\documentclass[openany,notoc]{tufte-book}
\hypersetup{colorlinks}
\usepackage{microtype}
\usepackage{booktabs}
\usepackage{graphicx}
\usepackage[normalem]{ulem}
\setkeys{Gin}{width=\linewidth,totalheight=\textheight,keepaspectratio}
\usepackage{fancyvrb}
\fvset{fontsize=\normalsize}

\usepackage{xspace}
\newcommand{\monthyear}{\ifcase\month\or January\or February\or March\or April\or May\or June\or July\or August\or September\or October\or November\or December\fi\space\number\year}

\usepackage{units}
\newcommand{\hlred}[1]{\textcolor{Maroon}{#1}}
\newcommand{\hangleft}[1]{\makebox[0pt][r]{#1}}

\newcommand{\tuftebs}{\symbol{'134}}
\providecommand{\XeLaTeX}{X\lower.5ex\hbox{\kern-0.15em\reflectbox{E}}\kern-0.1em\LaTeX}

\newcommand{\doccmddef}[2][]{\hlred{\texttt{\tuftebs#2}}\label{cmd:#2}\ifthenelse{\isempty{#1}}
{
\index{#2 command@\protect\hangleft{\texttt{\tuftebs}}\texttt{#2}}
}
{
\index{#2 command@\protect\hangleft{\texttt{\tuftebs}}\texttt{#2} (\texttt{#1} package)}
\index{#1 package@\texttt{#1} package}\index{packages!#1@\texttt{#1}}
}}
\newcommand{\doccmd}[2][]{
\texttt{\tuftebs#2}%
\ifthenelse{\isempty{#1}}
{
\index{#2 command@\protect\hangleft{\texttt{\tuftebs}}\texttt{#2}}
}
{
\index{#2 command@\protect\hangleft{\texttt{\tuftebs}}\texttt{#2} (\texttt{#1} package)}
\index{#1 package@\texttt{#1} package}\index{packages!#1@\texttt{#1}}
}}

\usepackage{makeidx}
\makeindex 

\usepackage{subcaption}
\usepackage{marginfix}
\usepackage{amsmath}
\usepackage{eso-pic}
\DeclareMathOperator*{\argmin}{arg\,min}
\DeclareMathOperator*{\argmax}{arg\,max}
\definecolor{cb1}{HTML}{e41a1c}
\definecolor{cb2}{HTML}{377eb8}
\definecolor{cb3}{HTML}{4daf4a}
\definecolor{cb4}{HTML}{984ea3}
\definecolor{cb5}{HTML}{ff7f00}
\definecolor{cb6}{HTML}{fec44f}
\definecolor{cb7}{HTML}{a65628}
\definecolor{cb8}{HTML}{f781bf}
\DeclareUnicodeCharacter{200E}{}

\title{The Atlas for the\\ \noindent Aspiring Network\\ \noindent Scientist}
\author[Michele Coscia]{Michele Coscia}
\publisher{IT University of Copenhagen}

\begin{document}

\mainmatter

\AddToShipoutPictureBG*{\includegraphics[width=8.5in,height=11in]{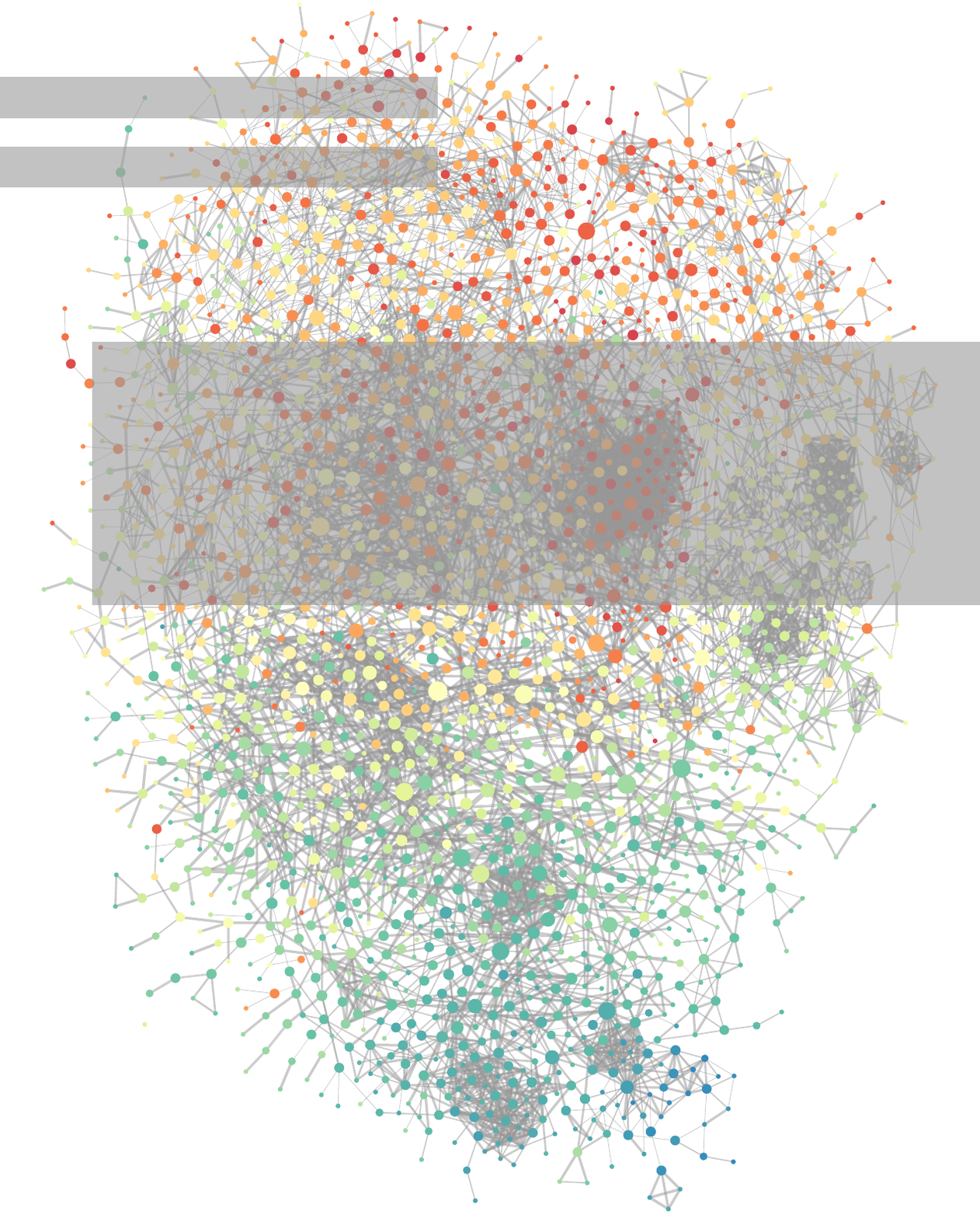}}

\maketitle


\newpage
\begin{fullwidth}
~\vfill
\thispagestyle{empty}
\setlength{\parindent}{0pt}
\setlength{\parskip}{\baselineskip}
Copyright \copyright\ 2025 \thanklessauthor

\par\smallcaps{\thanklessauthor\ is employed by the \thanklesspublisher, Rued Langgaards Vej 7, 2300 Copenhagen, Denmark}

\par\smallcaps{tufte-latex.googlecode.com}

\par Licensed under the Apache License, Version 2.0 (the ``License''); you may not use this file except in compliance with the License. You may obtain a copy of the License at \url{http://www.apache.org/licenses/LICENSE-2.0}. Unless required by applicable law or agreed to in writing, software distributed under the License is distributed on an \smallcaps{``AS IS'' BASIS, WITHOUT WARRANTIES OR CONDITIONS OF ANY KIND}, either express or implied. See the License for the specific language governing permissions and limitations under the License.\index{license}

\par\textit{First Edition, January, 2021}
\par\textit{Second Edition, January, 2025}
\end{fullwidth}

\tableofcontents 

\cleardoublepage

\chapter{Introduction}
Network science is a way to make sense of complex systems by modeling them as relations between their interacting components. It is immensely useful for two reasons: networks can model most -- if not all -- complex systems, and understanding complexity is the greatest challenge in front of humanity. This is, in my opinion, the reason why every scientist should be at least a little bit of a network scientist, and what motivated me to write this book.

\extrafloats{100}

\section{Why Complexity is the Key}
It is challenging to define what a complex system is\cite{ladyman2013complex}, but my way to understand complexity is by using this definition:

\begin{center}
A complex system is a system whose behavior cannot be reduced by analyzing its interacting parts.
\end{center}

This serves me well, because it can capture many key things about complex systems:

\begin{itemize}
\item Emergence: having properties that the parts do not have. There's no neuron that is conscious, no water molecule is wet, no person is a country.
\item Chaos\cite{packard1988adaptation}: changing a part can have hard-to-predict effects on the system. Removing a species from an ecosystem can lead to the extinction of another that did not have any relation -- direct or indirect -- with it, because it is difficult to foresee a chained series of reactions to that removal.
\item Self-Organization: the same part behaves differently depending on the context provided by (groups of) other parts. Think about the fact that all your cells have the same DNA, but a muscle cell is radically different from a bone cell.
\end{itemize}

And so on.

The diversity of these examples shows how complexity is all around us -- and so are networks. At some level, every aspect of reality seems to be made by interconnected parts. Societies are made by people entertaining multiple different types of relations with each other: artist citing each other's works, people making financial transactions, or developing friendships and enmities. The brains in their skulls are an intertwined web of neurons and synapses, but also machines to make inferences\cite{friston2010free}, which is another way to say connecting stimuli to each other. They're built with genes connected in a ballet of upregulating and downregulating dynamics. Genes are made with interacting proteins. Chemical compounds are atoms linked by bonds. Feynman's diagrams\cite{feynman1949theory} show elementary particles having all sorts of interesting relations. It's interactions all the way down.

If complexity is all around us, why did I say it is difficult to define? Part of the reason is because it is difficult to quantify. We have decided to describe reality with the language of math, to the point of believing that the math is actually the only thing that is real and objective\cite{tegmark2008mathematical} but, try as we might, we haven't been able to quantify complexity. If math is the language of science, then our understanding of complexity is pre-scientific, because we haven't found a way to fit complexity in our language. So we can't find the laws of complexity.

The solution, in my opinion, is a change in perspective. We need to develop a new language of complexity and go beyond our simple quantitative approach to science. We need to embrace complexity, not pigeonhole it into formulas. After all, the math is only useful when it hides the complexity away. One can be allured by the beauty of math and how well it describes reality, but math is only beautiful insofar it hides complexity, rather than explaining it. Take for instance the Standard Model of physics. This is a model that succinctly describes every elementary interaction (except gravity). However, if you were trying to use it to simulate a single iron atom in isolation you'd have a computationally intractable problem in your hands.

If you look at the formula behind the Standard Model of physics, you'll realize why:

\begin{center}
\begin{math}
-\frac{1}{2}\partial_{\nu}g^{a}_{\mu}\partial_{\nu}g^{a}_{\mu}
-g_{s}f^{abc}\partial_{\mu}g^{a}_{\nu}g^{b}_{\mu}g^{c}_{\nu}
-\frac{1}{4}g^{2}_{s}f^{abc}f^{ade}g^{b}_{\mu}g^{c}_{\nu}g^{d}_{\mu}g^{e}_{\nu}
+\frac{1}{2}ig^{2}_{s}(\bar{q}^{\sigma}_{i}\gamma^{\mu}q^{\sigma}_{j})g^{a}_{\mu}
+\bar{G}^{a}\partial^{2}G^{a}+g_{s}f^{abc}\partial_{\mu}\bar{G}^{a}G^{b}g^{c}_{\mu}
-\partial_{\nu}W^{+}_{\mu}\partial_{\nu}W^{-}_{\mu}-M^{2}W^{+}_{\mu}W^{-}_{\mu}
-\frac{1}{2}\partial_{\nu}Z^{0}_{\mu}\partial_{\nu}Z^{0}_{\mu}-\frac{1}{2c^{2}_{w}}
M^{2}Z^{0}_{\mu}Z^{0}_{\mu}
-\frac{1}{2}\partial_{\mu}A_{\nu}\partial_{\mu}A_{\nu}
-\frac{1}{2}\partial_{\mu}H\partial_{\mu}H-\frac{1}{2}m^{2}_{h}H^{2}
-\partial_{\mu}\phi^{+}\partial_{\mu}\phi^{-}-M^{2}\phi^{+}\phi^{-}
-\frac{1}{2}\partial_{\mu}\phi^{0}\partial_{\mu}\phi^{0}-\frac{1}{2c^{2}_{w}}M\phi^{0}\phi^{0}
-\beta_{h}[\frac{2M^{2}}{g^{2}}+\frac{2M}{g}H+\frac{1}{2}(H^{2}+\phi^{0}\phi^{0}+2\phi^{+}\phi^{-
})]+\frac{2M^{4}}{g^{2}}\alpha_{h}
-igc_{w}[\partial_{\nu}Z^{0}_{\mu}(W^{+}_{\mu}W^{-}_{\nu}-W^{+}_{\nu}W^{-}_{\mu})
-Z^{0}_{\nu}(W^{+}_{\mu}\partial_{\nu}W^{-}_{\mu}-W^{-}_{\mu}\partial_{\nu}W^{+}_{\mu})
+Z^{0}_{\mu}(W^{+}_{\nu}\partial_{\nu}W^{-}_{\mu}-W^{-}_{\nu}\partial_{\nu}W^{+}_{\mu})]
-igs_{w}[\partial_{\nu}A_{\mu}(W^{+}_{\mu}W^{-}_{\nu}-W^{+}_{\nu}W^{-}_{\mu})
-A_{\nu}(W^{+}_{\mu}\partial_{\nu}W^{-}_{\mu}-W^{-}_{\mu}\partial_{\nu}W^{+}_{\mu})
+A_{\mu}(W^{+}_{\nu}\partial_{\nu}W^{-}_{\mu}-W^{-}_{\nu}\partial_{\nu}W^{+}_{\mu})]
-\frac{1}{2}g^{2}W^{+}_{\mu}W^{-}_{\mu}W^{+}_{\nu}W^{-}_{\nu}+\frac{1}{2}g^{2}
W^{+}_{\mu}W^{-}_{\nu}W^{+}_{\mu}W^{-}_{\nu}
+g^2c^{2}_{w}(Z^{0}_{\mu}W^{+}_{\mu}Z^{0}_{\nu}W^{-}_{\nu}-Z^{0}_{\mu}Z^{0}_{\mu}W^{+}_{\nu}
W^{-}_{\nu})
+g^2s^{2}_{w}(A_{\mu}W^{+}_{\mu}A_{\nu}W^{-}_{\nu}-A_{\mu}A_{\mu}W^{+}_{\nu}
W^{-}_{\nu})
+g^{2}s_{w}c_{w}[A_{\mu}Z^{0}_{\nu}(W^{+}_{\mu}W^{-}_{\nu}-W^{+}_{\nu}W^{-}_{\mu})-
2A_{\mu}Z^{0}_{\mu}W^{+}_{\nu}W^{-}_{\nu}]
-g\alpha[H^3+H\phi^{0}\phi^{0}+2H\phi^{+}\phi^{-}]
-\frac{1}{8}g^{2}\alpha_{h}[H^4+(\phi^{0})^{4}+4(\phi^{+}\phi^{-})^{2}+4(\phi^{0})^{2}
\phi^{+}\phi^{-}+4H^{2}\phi^{+}\phi^{-}+2(\phi^{0})^{2}H^{2}]
-gMW^{+}_{\mu}W^{-}_{\mu}H-\frac{1}{2}g\frac{M}{c^{2}_{w}}Z^{0}_{\mu}Z^{0}_{\mu}H
-\frac{1}{2}ig[W^{+}_{\mu}(\phi^{0}\partial_{\mu}\phi^{-}-\phi^{-}\partial_{\mu}\phi^{0})
-W^{-}_{\mu}(\phi^{0}\partial_{\mu}\phi^{+}-\phi^{+}\partial_{\mu}\phi^{0})]
+\frac{1}{2}g[W^{+}_{\mu}(H\partial_{\mu}\phi^{-}-\phi^{-}\partial_{\mu}H)
-W^{-}_{\mu}(H\partial_{\mu}\phi^{+}-\phi^{+}\partial_{\mu}H)]
+\frac{1}{2}g\frac{1}{c_{w}}(Z^{0}_{\mu}(H\partial_{\mu}\phi^{0}-\phi^{0}\partial_{\mu}H)
-ig\frac{s^{2}_{w}}{c_{w}}MZ^{0}_{\mu}(W^{+}_{\mu}\phi^{-}-W^{-}_{\mu}\phi^{+})
+igs_{w}MA_{\mu}(W^{+}_{\mu}\phi^{-}-W^{-}_{\mu}\phi^{+})
-ig\frac{1-2c^{2}_{w}}{2c_{w}}Z^{0}_{\mu}(\phi^{+}\partial_{\mu}\phi^{-}-\phi^{-
}\partial_{\mu}\phi^{+})
+igs_{w}A_{\mu}(\phi^{+}\partial_{\mu}\phi^{-}-\phi^{-}\partial_{\mu}\phi^{+})
-\frac{1}{4}g^{2}W^{+}_{\mu}W^{-}_{\mu}[H^{2}+(\phi^{0})^{2}+2\phi^{+}\phi^{-}]
-\frac{1}{4}g^{2}\frac{1}{c^{2}_{w}}Z^{0}_{\mu}Z^{0}_{\mu}[H^{2}+(\phi^{0})^{2}+2(2s^{2}_{w}-
1)^{2}\phi^{+}\phi^{-}]
-\frac{1}{2}g^{2}\frac{s^{2}_{w}}{c_{w}}Z^{0}_{\mu}\phi^{0}(W^{+}_{\mu}\phi^{-}+W^{-
}_{\mu}\phi^{+})
-\frac{1}{2}ig^{2}\frac{s^{2}_{w}}{c_{w}}Z^{0}_{\mu}H(W^{+}_{\mu}\phi^{-}-W^{-}_{\mu}\phi^{+})
+\frac{1}{2}g^{2}s_{w}A_{\mu}\phi^{0}(W^{+}_{\mu}\phi^{-}+W^{-}_{\mu}\phi^{+})
+\frac{1}{2}ig^{2}s_{w}A_{\mu}H(W^{+}_{\mu}\phi^{-}-W^{-}_{\mu}\phi^{+})
-g^{2}\frac{s_{w}}{c_{w}}(2c^{2}_{w}-1)Z^{0}_{\mu}A_{\mu}\phi^{+}\phi^{-}-
g^{1}s^{2}_{w}A_{\mu}A_{\mu}\phi^{+}\phi^{-}
-\bar{e}^{\lambda}(\gamma\partial+m^{\lambda}_{e})e^{\lambda}
-\bar{\nu}^{\lambda}\gamma\partial\nu^{\lambda}
-\bar{u}^{\lambda}_{j}(\gamma\partial+m^{\lambda}_{u})u^{\lambda}_{j}
-\bar{d}^{\lambda}_{j}(\gamma\partial+m^{\lambda}_{d})d^{\lambda}_{j}
+igs_{w}A_{\mu}[-(\bar{e}^{\lambda}\gamma^{\mu}
e^{\lambda})+\frac{2}{3}(\bar{u}^{\lambda}_{j}\gamma^{\mu} 
u^{\lambda}_{j})-\frac{1}{3}(\bar{d}^{\lambda}_{j}\gamma^{\mu} 
d^{\lambda}_{j})]
+\frac{ig}{4c_{w}}Z^{0}_{\mu}
[(\bar{\nu}^{\lambda}\gamma^{\mu}(1+\gamma^{5})\nu^{\lambda})+
(\bar{e}^{\lambda}\gamma^{\mu}(4s^{2}_{w}-1-\gamma^{5})e^{\lambda})+
(\bar{u}^{\lambda}_{j}\gamma^{\mu}(\frac{4}{3}s^{2}_{w}-1-\gamma^{5})u^{\lambda}_{j})+
(\bar{d}^{\lambda}_{j}\gamma^{\mu}(1-\frac{8}{3}s^{2}_{w}-\gamma^{5})d^{\lambda}_{j})]
+\frac{ig}{2\sqrt{2}}W^{+}_{\mu}[(\bar{\nu}^{\lambda}\gamma^{\mu}(1+\gamma^{5})e^{\lambda})
+(\bar{u}^{\lambda}_{j}\gamma^{\mu}(1+\gamma^{5})C_{\lambda\kappa}d^{\kappa}_{j})]
+\frac{ig}{2\sqrt{2}}W^{-}_{\mu}[(\bar{e}^{\lambda}\gamma^{\mu}(1+\gamma^{5})\nu^{\lambda})
+(\bar{d}^{\kappa}_{j}C^{\dagger}_{\lambda\kappa}\gamma^{\mu}(1+\gamma^{5})u^{\lambda}_{j})]
+\frac{ig}{2\sqrt{2}}\frac{m^{\lambda}_{e}}{M}
[-\phi^{+}(\bar{\nu}^{\lambda}(1-\gamma^{5})e^{\lambda})
+\phi^{-}(\bar{e}^{\lambda}(1+\gamma^{5})\nu^{\lambda})]
-\frac{g}{2}\frac{m^{\lambda}_{e}}{M}[H(\bar{e}^{\lambda}e^{\lambda})
+i\phi^{0}(\bar{e}^{\lambda}\gamma^{5}e^{\lambda})]
+\frac{ig}{2M\sqrt{2}}\phi^{+}
[-m^{\kappa}_{d}(\bar{u}^{\lambda}_{j}C_{\lambda\kappa}(1-\gamma^{5})d^{\kappa}_{j})
+m^{\lambda}_{u}(\bar{u}^{\lambda}_{j}C_{\lambda\kappa}(1+\gamma^{5})d^{\kappa}_{j}]
+\frac{ig}{2M\sqrt{2}}\phi^{-}
[m^{\lambda}_{d}(\bar{d}^{\lambda}_{j}C^{\dagger}_{\lambda\kappa}(1+\gamma^{5})u^{\kappa}_{j})
-m^{\kappa}_{u}(\bar{d}^{\lambda}_{j}C^{\dagger}_{\lambda\kappa}(1-\gamma^{5})u^{\kappa}_{j}]
-\frac{g}{2}\frac{m^{\lambda}_{u}}{M}H(\bar{u}^{\lambda}_{j}u^{\lambda}_{j})
-\frac{g}{2}\frac{m^{\lambda}_{d}}{M}H(\bar{d}^{\lambda}_{j}d^{\lambda}_{j})
+\frac{ig}{2}\frac{m^{\lambda}_{u}}{M}\phi^{0}(\bar{u}^{\lambda}_{j}\gamma^{5}u^{\lambda}_{j})
-\frac{ig}{2}\frac{m^{\lambda}_{d}}{M}\phi^{0}(\bar{d}^{\lambda}_{j}\gamma^{5}d^{\lambda}_{j})
+\bar{X}^{+}(\partial^{2}-M^{2})X^{+}+\bar{X}^{-}(\partial^{2}-M^{2})X^{-}
+\bar{X}^{0}(\partial^{2}-\frac{M^{2}}{c^{2}_{w}})X^{0}+\bar{Y}\partial^{2}Y
+igc_{w}W^{+}_{\mu}(\partial_{\mu}\bar{X}^{0}X^{-}-\partial_{\mu}\bar{X}^{+}X^{0})
+igs_{w}W^{+}_{\mu}(\partial_{\mu}\bar{Y}X^{-}-\partial_{\mu}\bar{X}^{+}Y)
+igc_{w}W^{-}_{\mu}(\partial_{\mu}\bar{X}^{-}X^{0}-\partial_{\mu}\bar{X}^{0}X^{+})
+igs_{w}W^{-}_{\mu}(\partial_{\mu}\bar{X}^{-}Y-\partial_{\mu}\bar{Y}X^{+})
+igc_{w}Z^{0}_{\mu}(\partial_{\mu}\bar{X}^{+}X^{+}-\partial_{\mu}\bar{X}^{-}X^{-})
+igs_{w}A_{\mu}(\partial_{\mu}\bar{X}^{+}X^{+}-\partial_{\mu}\bar{X}^{-}X^{-})
-\frac{1}{2}gM[\bar{X}^{+}X^{+}H+\bar{X}^{-}X^{-}H+\frac{1}{c^{2}_{w}}\bar{X}^{0}X^{0}H]
+\frac{1-2c^{2}_{w}}{2c_{w}}igM[\bar{X}^{+}X^{0}\phi^{+}-\bar{X}^{-}X^{0}\phi^{-}]
+\frac{1}{2c_{w}}igM[\bar{X}^{0}X^{-}\phi^{+}-\bar{X}^{0}X^{+}\phi^{-}]
+igMs_{w}[\bar{X}^{0}X^{-}\phi^{+}-\bar{X}^{0}X^{+}\phi^{-}]
+\frac{1}{2}igM[\bar{X}^{+}X^{+}\phi^{0}-\bar{X}^{-}X^{-}\phi^{0}]
\end{math}
\end{center}

My thesis is that we need to understand complexity so we find a better language to describe reality.

We already know that, because no one uses that formula to do chemistry. We compartmentalized the fields of knowledge, because we know we can't describe societies via an explanation of quantum interactions of elementary particles, climbing the long and perilous ladder of fields in order of purity\footnote{\url{https://xkcd.com/435/}}. Some phenomena cannot be reduced to the underlying laws\cite{anderson1972more}. We \textit{do} need to toss away a good chunk of physics, add a bunch of new tools, to understand this new field called ``chemistry'', because the change in scale causes the emergence of new phenomena.

We have some starting pointers to understand how this compartmentalization of knowledge can help scientific investigation. This is what Hayek called ``division of knowledge''\cite{hayek1945use}, which is a much more powerful concept than Smith's classical division of labor\cite{smith1776wealth}. If I specialize as a chemist and hone my skills and tools to that specific task, I can be immensely more productive, because I am outsourcing all other knowledge discovery endeavors to other specialists. This is how societies grow their pool of knowledge efficiently. However, the result is that, now, no individual can really fully grasp a well-rounded picture of reality. The collective society can, but not its individual components. It is all deformed by the lens of their specialization.

The resulting irony is that we might not be able to get to the language of complexity because we need it in order to get it. To make an advancement in physics we need teams of tens or hundreds of people. The paper containing the discovery of the Higgs boson\cite{aad2012observation} has $5,154$ authors. The knowledge needed for it was so vast it could not fit in a single brain. Only a collective of interconnected brains could understand it -- and that is a complex system.

How do you meaningful coordinate a complex system to make an even vaster scientific discovery? You can only do it if you understand complexity -- which is what you need the collective of brains to do! It's a circular problem. We already know that simple interventions will cause unexpected local optima that are unsatisfactory. Science with its replication crisis\cite{ioannidis2005most}, its publish-or-perish misaligned incentives, its difficulty in dealing with misinformation\cite{bergstrom2021calling}, isn't going as smoothly as it could. Because we don't understand complexity. Because we need to understand complexity in order to understand complexity.

But I'll be damned if I don't give everything I have to make this understanding of complexity happen. And I think our best shot is via network science, because it is the field that gives us a way to talk about emergence. We need to know how the different fields -- physics, chemistry, biology, ... -- relate and transform into each other, which is necessary to reconstruct a picture of reality.

Connecting those fields means finding a shared language that can describe the \textit{relations} between the \textit{symbols} they use -- make a mental note of this, it'll come back later. A shared language can then be universal and move across layers and fields. If you understand intelligence as a way of how information is aggregated and manipulated in each part given its interactions, then this description is independent of what the parts actually are, as long as they can perform the same function. You can use this network theory of intelligence to describe not only how individual brains learn, but how collectives made of brains learn.

In summary, I hope I'm wrong when I say that we need to understand complexity in order to understand complexity. I hope network science can bootstrap our understanding. By teaching you network science with this book, I'm trying my darnedest to prove myself wrong. I want you -- the collective of all the $25$ people who'll read this thing -- to understand complexity and save science and society in the process. No pressure.

\section{The Creation Myth of Network Science}
I just put a big weight on the shoulders of network science and network scientists. That is because I think this is a promising field, because networks provide a versatile way of representing virtually anything that is made of interacting parts. To understand why, it is necessary to explore the surface of network science's history -- and deconstruct its creation myth, which is always a fun thing to do.

Normally, the creation myth of network science starts with Euler's solution to the famous K\"{o}nigsberg Bridge Problem\cite{euler1741solutio}. The problem asks whether it is possible to cross all bridges of K\"{o}nigsberg (now Kaliningrad) without crossing the same bridge twice. Euler realized that the problem didn't require reasoning about any quantities: it was exclusively a problem about the qualitative relationships between islands and bridges, whether you could use the latter to reach the former. So he abstracted quantities away and created the mathematical object of the graphs, which only has relations between the various parts.

This, as every single creation myth ever, is obviously false. Not only because Euler himself credited this idea of qualitative topology to Leibniz in the introduction of his paper\cite{schich2019cultural}. Not only because Euler's actual solution has no graph at all but it's instead combinatorial in nature. Not even because it's kind of ridiculous to say anyone invented graphs -- it'd be similar to say someone invented sticks, sure someone must have written about them first, but did they really invent them?

These are three great reasons why the creation myth of network science on Euler's desk is false, but there's one even more fundamental. That is, at best, a creation myth of \textit{graph theory}, but \textit{network science} is an entirely different beast. If graphs were born on Euler's desk in $1736$, why does $99.9\%$ of network science happened from the 1930s on?

Modern network science is a gift from sociology. Before sociology, graphs were seen as exact and deterministic mathematical objects, worthy of exploration through the manipulation of \textit{abstract symbols}. Sociologists saw the value in using these mathematical objects -- symbols -- to investigate a statistical and stochastic reality. This was the first -- fundamental and necessary -- explosion in possibilities for a true network science. If we want to have a better creation myth for network science, we should replace Euler's God-figure with Helen Hall Jennings -- who invented the sociograms\cite{moreno1938statistics} that were the real beginning of network science.

Jennings had two problems in the 1930s, though: she could only collect and manipulate data manually, and lacked a unified language to represent \textit{all} of reality as an analyzable symbol. She could do ad-hoc representations, tailored for a specific problem, but lacked the representation power necessary to use networks as the tool to understand complexity in general.

What Jennings needed was the computer. The value of the computer is not that it can perform lots of operations quickly, although that certainly helps. One can prove theorems and lemmas without computers. Rather, the revolution of the computer is in its symbolic language. The computer's zeroes and ones can be a universal language to represent reality -- that shared language I alluded to before. Computers seem to be able to allow you to manipulate anything: with spreadsheets you can tame problems in logistics, with XML you can map semantic concepts, with media players you can appreciate art and videos\footnote{It has been legendarily said that VLC, one of the most popular multimedia software, can open anything -- even a can of tuna.}. And yet, inside computers you just have a mass of zeroes and ones.

The power of the computer is its ability of seeing everything -- anything -- as a symbol. Once everything is a symbol, you can understand its relations with other symbols\footnote{One of the facts that never fails to blow my mind is the realization that a piece of software is, after all, just a very cleverly composed number. Thus you \textit{can} sum Adobe Photoshop to Google Chrome, although the result won't probably make much sense. That is also why there exist such a thing as an ``illegal number'' (\url{https://en.wikipedia.org/wiki/Illegal_number})}. In this sense, the true revolution of the computer was not pioneered by Babbage and von Neumann\cite{von1945first}. What they did was an immensely useful mechanical invention, but the revolution we needed was a logical invention for the manipulation of symbols and their interactions. This was gifted us by the first programmer Ada Lovelace\cite{lovelace1842sketch}, and by Wittgenstein's sharp observations of the relations between symbols and reality\cite{wittgenstein1921tractatus}.

The second half of the XX century was the moment when we started connecting symbols together in the same place: in the memory of a computer. Ironically, computers facilitated the emergence of even more networks that were latently waiting to express their potential. For instance, the invention of the Internet via ARPANET -- and of e-mail -- codified explicitly the relationships between centers of command and of knowledge creation that existed across the world. But it was arguably Berners-Lee\cite{berners1989information} -- following in the footsteps of Bush\cite{bush1945we} -- that truly understood how to piece symbols together in computers.

Once hypertexts birthed the Web, it was just a formality to get the papers published before modern network science could kick into gear. We finally had both the tools and the data to really understand how universal networks were, and developing the language we could use to talk about them. Thus, in a sense, the XX century planted the seed for the XXI: the great awakening of the world to the pervasive presence of networks -- and complexity -- in everything we do.

\section{How This Book Approaches Complexity}
There are a ton and a half other books about network science out there. I'd bet that a good chunk of them are better than mine. I do not advocate that this is the only book you should read to understand network science and to be a network analyst. In fact, I advocate the opposite: read diverse takes and use diverse structures to think about networks. I see my book as complementary to those out there.

Given that I am belatedly giving the credit to Jennings and sociologists for the creation of network science, one should consider social network analysis as the foundational approach. In that, Wasserman's and Faust's classic is a must read\cite{wasserman1994social}.

Post-$1998$ network science arguably took off because the innovations from sociology were picked up by physicists\cite{watts1998collective}\cite{barabasi1999emergence}, who inspired other fields with their quest for universal laws. It is no wonder that there exists a plethora of books\cite{caldarelli2012networks}\cite{caldarelli2007scale}\cite{latora2017complex} and review articles\cite{castellano2009statistical} taking the physics angle on network science. Among these, a few certainly stand out. Barab\'{a}si's book\cite{barabasi2016network} towers in accessibility and clarity, while Newman's work is probably the most complete and in-depth\cite{newman2018networks}. Physicists also have a good track record in publishing books for the wider audience\cite{barabasi2003linked}\cite{watts2004six}, not necessarily with a scientific background in mind.

Given the reliance on computational tools and the need of processing large amounts of data, the computer science angle should not be ignored\cite{menczer2020first}. A great favorite of mine combines the computer science methodology with applications in economics\cite{easley2010networks}. If you need yet another proof of the breadth of approaches in network science, consider that another major book on the topic was authored by a chemist\cite{estrada2012structure}.

The natural question now is: if there are already so many network science books and they are all great, what is the need of this one you're reading? Is it just for updating with the newest developments in the field? Not really. I hope you noticed that, when presenting the other network science books, I never introduced them as books written by a network scientist. This was not by accident. My impression is that these books are aimed at introducing people from a variety of disciplines into the skills and tools of network science, rather than examining network science from within.

There are now PhD programs for network scientists\footnote{\url{https://www.networkscienceinstitute.org/phd}}, but they are only a handful years old, meaning that there are only a few graduates coming out of them and they do not have yet the time or the experience to write a network science book. Worse still, I believe there are even fewer master and bachelor programs in network science, if any. This means that every book you can find on network science is a sort of ``something/network science'', with that something being sociology, physics, computer science, archaeology, or other.

This book has the -- probably overambitious -- aim of being no ``slash something''. It wants to be a pure breed: just a network science book. In other words, the difference between this and the other books is that this book considers ``network science'' not as something one attaches to another discipline, but rather it is a discipline in itself. People can -- and should! -- be trained from scratch in it.

I believe my background is as close as it could be to the right mix that network analysis requires. I am a digital humanist, a field pioneered by Busa\cite{busa1974index} which focuses on the digital processing of content produced by humans. This is to say: the computer-mediated manipulation and analysis of symbols representing different facets of reality -- which are not necessarily mathematical -- and their connections. If this sounds familiar, it is because this is the exact characterization of network science as the key or representing and understanding complexity that I adopted in this introduction.

By the time I started a PhD, there was no one offering it in network science -- or digital humanities -- so, formally, I am a computer scientist as well. However, from day one, I immersed myself in network science literature in all its facets -- physics, computer science, sociology -- armed with my digital humanities toolbox. I designed new network science algorithms and at the same time studied Dante's \textit{Inferno} as a complex network\cite{cappelli2011social}; I scouted for laws in complex systems whilst fighting Mexican drug traffic\cite{coscia2012knowing}. Everything I did was in an attempt to be an all-round network scientist. And this is the same hat I'm wearing as the author of this book.

If I do so, it is because I am intimately convinced that network science is truly a special field. This is not only because, as I opened this introduction, relations are what I consider being a fundamental way to understand reality. That consideration is only the beginning: it caused network science to have its complex and multifaceted origin story -- combining all the fields that I've been mentioning so far. By birthing out of many different scientific -- and non-scientific! -- disciplines, network science is truly a method to grasp emergence.

Connecting to the first section of this introduction, it should now be clear why I consider network science important, and a truly \textit{network} science book necessary: it is our best shot at building a collective understanding of all human knowledge, and such attempt needs to be approached with the proper humility of those who are not expert in anything else but gluing together the pieces created by the real experts.

That said, I don't want to oversell the importance of network science. If what I said is really true, it means that we can represent any -- or at least most -- aspects of reality as mathematical symbols and we can manipulate them with the mathematical tools of computer and network science. Which means that a complete understanding of them is necessarily out of reach. Not just because, as Poincar\'{e} would put it, ``the head of the scientist, which is only a corner of the universe, could never contain the universe entire''\cite{poincare1913foundations}. Rather, because G\"{o}del taught us that there is a strong bound of what is tractable in a formal system\cite{godel1931formal}. At some point, even when you represent the entirety of reality as interconnected mathematical symbols, you will need to jump out and look at the loops from the outside\cite{hofstadter1979godel}. No book can really give you a scientific road map on how to do so.

\section{What is in This Book?}
This is all fine and dandy but, at the end of the day, what does this book \textit{contain}?

At a general level, it contains the widest possible span of all that is related to network science that I know. It is the result of \sout{twelve} sixteen\footnote{Boy, writing a version $2$ of this book really makes me feel even older.} years of experience that I poured on the field. Virtually any concept that I used or that I simply came to know in these fifteen years is represented in at least a sentence in this book.

As you might expect, this is a lot to include and would not fit a book, not even a $\sim800$ pages like this one. By necessity, many -- if not all -- of the topics included in this book are treated relatively superficially. I would not say that this book would provide you what you need to know to be a network scientist. But it would \textit{point} you to what you need to know\footnote{\textit{Connecting the symbols} of network science, maybe?}. To borrow from Rumsfeld\footnote{\url{https://archive.defense.gov/Transcripts/Transcript.aspx?TranscriptID=2636}}: the book provides little to no \textit{known knowns}, but it will provide you with all the \textit{known unknowns} in network science -- so that your \textit{unknown unknowns} are aligned with those of everyone else. After internalizing this book, you will know what you don't know; you will be handed all the tools you need to ask meaningful questions about network science in $2025$. You can go to the other books or to any other article, and find the answers. Or you can figure out the answer yourself.

That is why I decided to call this book an ``Atlas''. It is the map you need to set foot among networks and start exploring. An atlas doesn't do the exploration for you, but you can't explore without an atlas. This is the book I wished I had fifteen years ago.

At a more specific level, the book is divided in fourteen parts.

\begin{itemize}
\item[Part I] provides the basics for the background knowledge to be a good network scientist. You need to know probability theory, statistics, machine learning, and linear algebra. That's the basic mental toolbox to handle networks. There are no networks in this part, so if you already have a good foundation in these fields you can skip it.
\item[Part II] teaches you what a graph is and how many features to the simple mathematical model were added over the years, to empower our symbols to tame more and more complex real world entities. Finally, it pivots perspectives to show an alternative way of manipulating networks, via matrices.
\item[Part III] is a carousel of all the simplest analyses you can operate on a graph. These are either local node properties -- how many connections a node has --, or global network ones -- how many connections on average I have to cross to go from one node to another. We see that some of these are easy to calculate on the graph structure, while others are naturally solved by linear algebra operations. Shifting perspectives is something you need to get used to, if you want to make it as a network scientist.
\item[Part IV] uses some of the tools presented in the previous part to build slightly more advanced analyses. Specifically, it focuses on the question: which nodes are playing which role in the network? And: can we say that a node is more important than another? If you want to answer these questions, you need to relate the entire network structure to a node, i.e. to use fully what Part \ref{par:properties} trained you to do.
\item[Part V] teaches you the main approaches for the creation of synthetic network data. It explores the main reasons why we want to do it. Sometimes, it is because we need to test an algorithm and we need a benchmark. Alternatively, we can use these models to reproduce the properties of real world networks we investigated in the previous parts, to see whether we understand the mechanisms that make them emerge.
\item[Part VI] starts considering not just network structures, but events on networks. That is, your nodes and your edges represent real world entities that actually do something, rather than simply connecting to each other. Specifically, Part \ref{par:sis} deals with things spreading through a network: when a node is affected by something, it has a chance to pass it to its neighbors via its connections. This something might be a disease in epidemiological models, or a behavior in sociological ones.
\item[Part VII] evolves the idea of dynamic events on networks. In Part \ref{par:sis}, we assumed that the network structure was unchanging: it was a timeless snapshot of reality and all nodes and connections are eternally the same. In Part \ref{par:lp}, we acknowledge that things change: if two nodes are not connected today, they might connect tomorrow. We investigate which techniques allow us to make a good guess on which connections we will observe in the future, on the basis of the ones we are observing now.
\item[Part VIII] performs another leap: up until now, we mostly inhabited an ideal world. We assumed we could model phenomena and cleanly gather insights. Here, we have our first impact with the real world. How do networks look like when you gather them via experiments and/or observations? Often, they don't look at all like the ones from your models. The only expectation that reality meets is its inability to meet expectations. This part trains you in the art of cleaning real world data to obtain something passable that can be fed to your neat theories and analyses.
\item[Part IX] opens the Pandora's Box of the level of analysis that is the most interesting and probably the one with which you will struggle most of the time: the mesoscale. The mesoscale is what lies between local node properties and global network statistics. This includes -- but is not limited to -- questions such as: does my network have a hierarchical structure? Is there a densely connected core surrounded by a sparsely connected periphery? Do nodes consider other nodes' properties to decide whether to connect to them?
\item[Part X] continues the exploration of the mesoscale. It needs to be split off Part \ref{par:meso} because there is one mesoscale analysis that has dominated all other subfields in network science: community discovery. Community discovery is the network equivalent of what clustering is for machine learning: the task of dividing nodes in a network into groups. Nodes in the same groups are densely connected to each other, more so than with nodes in different groups. Or so people would lead you to believe. The fact that there are literally thousands of papers proposing different algorithms to tackle this problem should be a hint at the fact that things might not be as simple.
\item[Part XI] takes a steep turn into the realm of computer science. It deals with graph mining: a collection of techniques that allow you to discover patterns in your graph structure, even if you are not sure about what these patterns might look like or hint at. It is what we would call ``bottom-up'' discovery. This is where you'll find a deep dive on graph neural networks, which is one of the fastest moving subfields of network science at the moment.
\item[Part XII] comprises a branch of network science that deals with analyzing the network structure as a whole. This is not simply providing summary statistics about the entire network like you'd find in Part \ref{par:properties}, but trying to embed the entire structure into a more advanced analysis. For instance, you'd deal with networks defining a complex non-Euclidean space.
\item[Part XIII] includes a few tips and tricks for an aspect of network science that is rarely covered in other books: how to browse/explore your network data and how to communicate your results. Specifically, I will show you some best practices in visualizing networks. I am a visual thinker and, sometimes, patterns and ideas about those patterns emerge much more clearly when you see them, rather than scouting through summary statistics. Moreover, network science papers thrive on visual communication and a good looking network has an amazing chance of ending up in the cover of the journal you're publishing on. It is a mystery to me why you would not spend some time in making sure that your network figures are at least of passable quality. Moreover, even if we are all primed to think dots and lines when it comes to visualizing a graph, you should be aware of the situations in which there are different ways to show your network.
\item[Part XIV] is a final collection of miscellanea that can help you to venture out in the real world of network analysis. It contains a quick discussion of the applications of network science I find most interesting, and a repository of tools you might need to kickstart your career.
\end{itemize}

\section{What's New in Version Two}
This is the second edition of the book. I decided to update the book because there were many things I found unsatisfactory with the first edition. Here I'm letting you know what's new besides making the introduction slightly less tongue-in-cheek -- so that, if you already read version one, you know where to check for what you have missed.

The first thing I greatly expanded was the introductory part. In version one, I only had a chapter on probability theory. The book contained some linear algebra and machine learning, but those concepts were scattered around in an unorganized way. Part \ref{par:background} of this book is a brand new effort to concentrate all that basic knowledge in the same place, adding some statistics for good measure.

The second major change is the coverage of graph neural networks. In version one, I only had a single chapter that put together shallow and deep learning. The explanations were simplistic and obscure, a necessary evil to convey so much content in so little space. Now that single chapter has exploded into four chapters, from Chapter \ref{cha:mining-embeddings} to Chapter \ref{cha:mining-deep2}. Hopefully, now those concepts are explained in a clearer way and I'm doing them justice.

The final major change is about uncertainty. In the original version of the book, I had only a brief half page section lamenting the fact that network scientists don't really give uncertainty its credit -- trusting that their networks are a given truth without measurement error. I wanted to amend that by giving justice to all the work done handling uncertainty and probabilistic connections in network science. Chapter \ref{cha:uncertainty} is the result.

There are a number or minor additions. The most significant are: a new discussion of effective resistance (Section \ref{sec:rw-effectres}) which then enables to talk about how it is used for network statistics (Section \ref{sec:nvd-stats}) and applications (Section \ref{sec:app-polar}); an improved overview of simplicial complexes for the high-order chapter (Section \ref{sec:hod-simplicial}); and some extensions in the discussion of node similarities (Section \ref{sec:centr-similarity}).

There are also a bunch of minor fixes and corrections, some of them were already made for version one but never appeared in print.

\section{Acknowledgements}
You'd be a complete fool if you trusted me to get this amount of knowledge correct by myself. Any and all scientific endeavors are only as good as the attention they receive by their peers, both in terms of building on top of those results, but also in terms of catching and correcting mistakes. This is in line with what I've been saying in this introduction: network science is too vast and hard for me to grasp, so I need to rely on the help of the experts from all of its subfields. It takes a village -- or an extensive social network -- to write a textbook. Here I want to thank all those who helped me in this journey.

The person who stood up tall above everybody was Aaron Clauset, who gets my most sincere thanks. Aaron is the only one who reviewed almost the whole version one of the book, all the $650$ friggin' pages of it. All while rocking a few months old baby daughter. Aaron is a superhero and should have everyone's deep respect.

Another one who went beyond the call of duty was Andres Gomez-Lievano. Andres and I shared a desk for years and I cherish those as the most fun I had at work. Andres didn't stop at the chapters I asked him to review, but deeply commented on the philosophy and framing of this book. I can see in his comments the spark of the years we spent together.

My other kind reviewers were, in alphabetical order: Alexey Medvedev, Andrea Tagarelli, Charlie Brummitt, Ciro Cattuto, Clara Vandeweerdt, Fred Morstatter, Giulio Rossetti, Gourab Ghoshal, Isabel Meirelles, Laura Alessandretti, Luca Rossi, Mariano Beguerisse, Marta Sales-Pardo, Matt\'{e} Hartog, Petter Holme, Renaud Lambiotte, Roberta Sinatra, Yong-Yeol Ahn, and Yu-Ru Lin.

For version two, I have recruited the additional help of: Giovanni Puccetti, Matteo Magnani, Maria Astefanoaei, Daniele Cassese, and Paul Scherer.

All these people donated hours of their time with no real tangible reward, just to make sure my book graduated from ``incomprehensible mess'' to ``almost passable and not misleading''. Thank you.

With their work, some reviewers expressed their intent to support charitable organizations. Specifically, they mentioned TechWomen\footnote{\url{https://www.techwomen.org/}} -- to support the careers of women in STEM fields --, Evidence Action\footnote{\url{https://www.evidenceaction.org/}} -- to expand our de-worming efforts and reaping the surprisingly high societal payoff --, and Doctors without Borders\footnote{\url{https://www.doctorswithoutborders.org/}}. You should also consider donating to them.

If there's any value in this book, it comes from the hard work of these people. All the mistakes that remain here are exclusively due to my ineptitude in properly implementing my reviewers' valuable comments. I expect there must be many of such mistakes, ranging from trivial typos and clumsily written sentences, to more fundamental issues of misrepresentation. If you find some, feel free to drop me an email to \texttt{mcos@itu.dk}.

If, for some reason, you only have access to a printed version of this book -- or you found the PDF somewhere on the Internet, know that there is a companion website\footnote{\url{https://www.networkatlas.eu/}} with data for the exercises, their solutions, and -- hopefully in the future -- interactive visualizations. 

\part{Background Knowledge}\label{par:background}

\chapter{Probability Theory}\label{cha:prob}
Before even mentioning networks, I need to lay the groundwork for a basic understanding of a few concepts necessary to make you a good network analyst. These concepts are part of four broad subjects: probability theory, statistics, linear algebra, and machine learning. We start with probability theory here, because that is the foundation on top of which this pyramid is built. Then, we deal with more specialized statistical concepts in Chapter \ref{cha:stats}, with machine learning in Chapter \ref{cha:machine-learning}, and linear algebra in Chapter \ref{cha:la}.

These chapters deal with entire fields of human knowledge that are much more vast a complicated than the extremely simplified picture I present here. As with many other chapters, my coverage of the subject is the bare minimum I can get away with. The wisest course of action for you would be to skip this book part entirely and take entire courses on these subjects and then come back to this book and start directly with Part \ref{par:graph}. Alas, that's not an option for everybody, and that is why this book part covers the basics you need to know to enjoy the rest of the book.

If you want to dive deep into probability theory, there are good books on the subject you should check out\cite{feller1968introduction}\cite{durrett1996probability}. I will attempt, where possible, to give forward references to later topics in this book, to let you know why understanding probability theory is important for a network scientist.

\section{Frequentism and Bayesianism}\label{sec:prob-freq-vs-bayes}
Probability theory is the branch of mathematics that allows you to work with uncertain events. It gives you the tools to make inferences in cases of uncertainty.

Probability theory is grounded in mathematical axioms. However, there are different ways to interpret what we really mean with the term ``probability''. With a very broad brush, we can divide the main interpretations into two camps: the frequentist and the Bayesian. There are more subtleties to this, but since these are the two main approaches we will see in this book, there is no reason to make this picture more complex than it needs to be.

To understand the difference, let's suppose you have Mrs. Frequent and Mr. Bayes experimenting with coin tosses. They toss a coin ten times and six out of ten times it turns heads up. Now they ask themselves the question: what is the probability that, if we toss the coin, it will turn heads up again?

Mrs. Frequent reasons as follows: ``An event's probability is the relative frequency after many trials. We had six heads after ten tosses, thus my best guess about the probability it'll come out as heads is 60\%''. Note that Mrs. Frequent doesn't really believe that ten tosses gave him a perfect understanding of that coin's odds of landing on heads. Mrs. Frequent knows that he will get the answer wrong a certain number of times, that is what confidence intervals are for, but for the sake of this example we need not to go there.

``Hold on a second,'' Mr. Bayes says, ``Before we tossed it, I examined the coin with my Coin Examiner\textsuperscript{TM} and it said it was a fair coin. Of course my Coin Examiner\textsuperscript{TM} might have malfunctioned, but that rarely happens. We haven't performed enough experiments to say it did, but I admit that the data shows it might have. So I think the probability we'll get heads again is $51\%$''. Just like Mrs. Frequent, also Mr. Bayes is uncertain, and he has a different procedure to estimate such uncertainty -- in this case dubbed ``credible intervals'' -- which again we leave out for simplicity.

Herein lies the difference between a frequentist and a Bayesian. For a frequentist only the outcome of the physical experiment matters. If you toss the coin an infinite number of times, eventually you'll find out what the true probability of it landing on heads is. For a Bayesian it's all about degrees of beliefs. The Bayesian has a set of opinions about how the world works, which they call ``priors''. Performing enough new experiments can change these priors, using a standard set of procedures to integrate new data. However a Bayesian will never take a new surprising event at face value if it is wildly off its priors, because those priors were carefully obtained knowledge coherent with how the world worked thus far.

\begin{figure}
\centering
\includegraphics[width=.66\columnwidth]{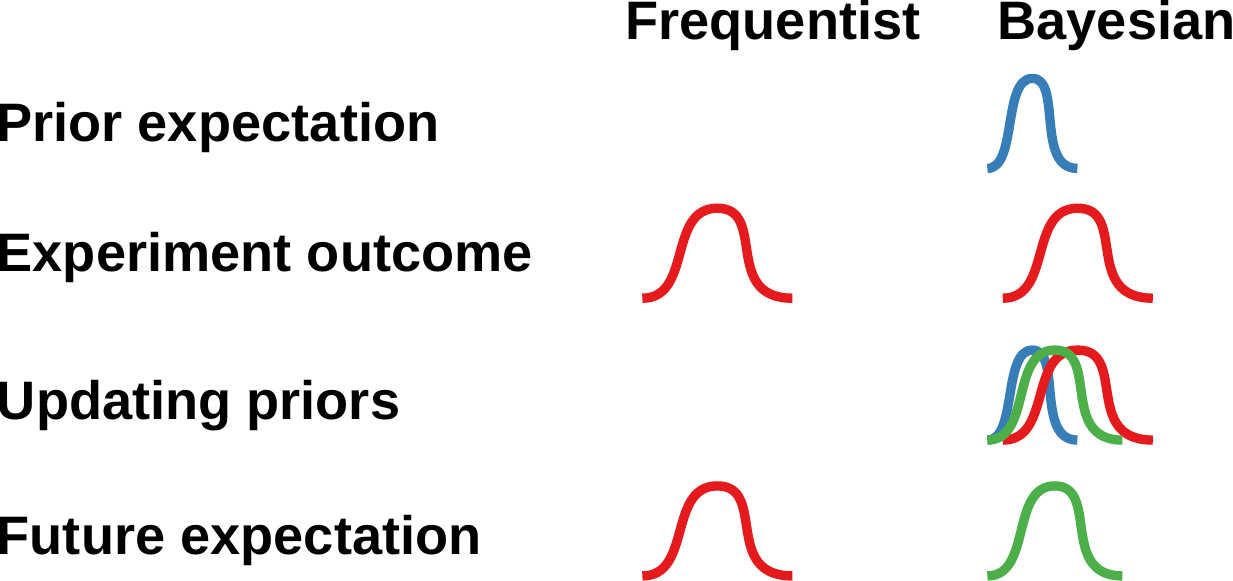}
\caption{Schematics of the mental processes used by a frequentist and a Bayesian when presented with the results of an experiment.}
\label{fig:freq-bayes}
\end{figure}

Figure \ref{fig:freq-bayes} shows the difference between the mental processes between a frequentist and a Bayesian. The default mode for this book is taking a frequentist approach. However, here and there, Bayesian interpretations are going to pop up, thus you have to know why we're doing things that way.

\section{Notation}
Probability theory is useful because it gives us the instruments to talk about uncertain processes. For instance, a process could be tossing a die. The first important thing is to understand the difference between \textit{outcome} and \textit{event}. An \textit{outcome} is a \textit{single} possible result of the experiment. A die landing on $2$ is an outcome. An \textit{event} is a \textit{set} of possible outcomes on which we're focusing. In our convention, we use $X$ to refer to outcomes. $X$ is a random variable and it can take many values and forms, and we don't know which of them it will be before actually running the process. As for events, they are the focus of all questions in probability theory: you can sum up probability theory as the set of instruments that allow you to ask and answer questions about events (sets of $X$) such as: ``What is the probability that $X$, the outcome of the process, is this and/or this but not that and/or that?''

Mathematically one writes such a question as $P(X \in S)$, where $S$ is a set of the values that $X$ takes in our question. $X$ is an outcome, $X \in S$ is an event. For instance, if we were asking about the event ``will the die land on an even number?'', $S = \{2, 4, 6\}$. So, $P(X \in S)$ asks what's the probability of the ``die lands on an even number'' event -- or for $X$ to take either of the $2, 4, 6$ values. Note that elements in $S$ are all possible alternatives: if we write $P(X \in \{2, 4, 6\})$, we're asking about the probability of landing on $2$ \textit{or} $4$ \textit{or} $6$. If you want to have the probability of two events happening simultaneously, you have to explicitly specify it with set notation: $P(X \in \{2,4,6\}) \cap P(X \in \{1\})$ asks the probability of landing on an even side and on $1$ at the same time.

\begin{figure}
\centering
\includegraphics[width=.45\columnwidth]{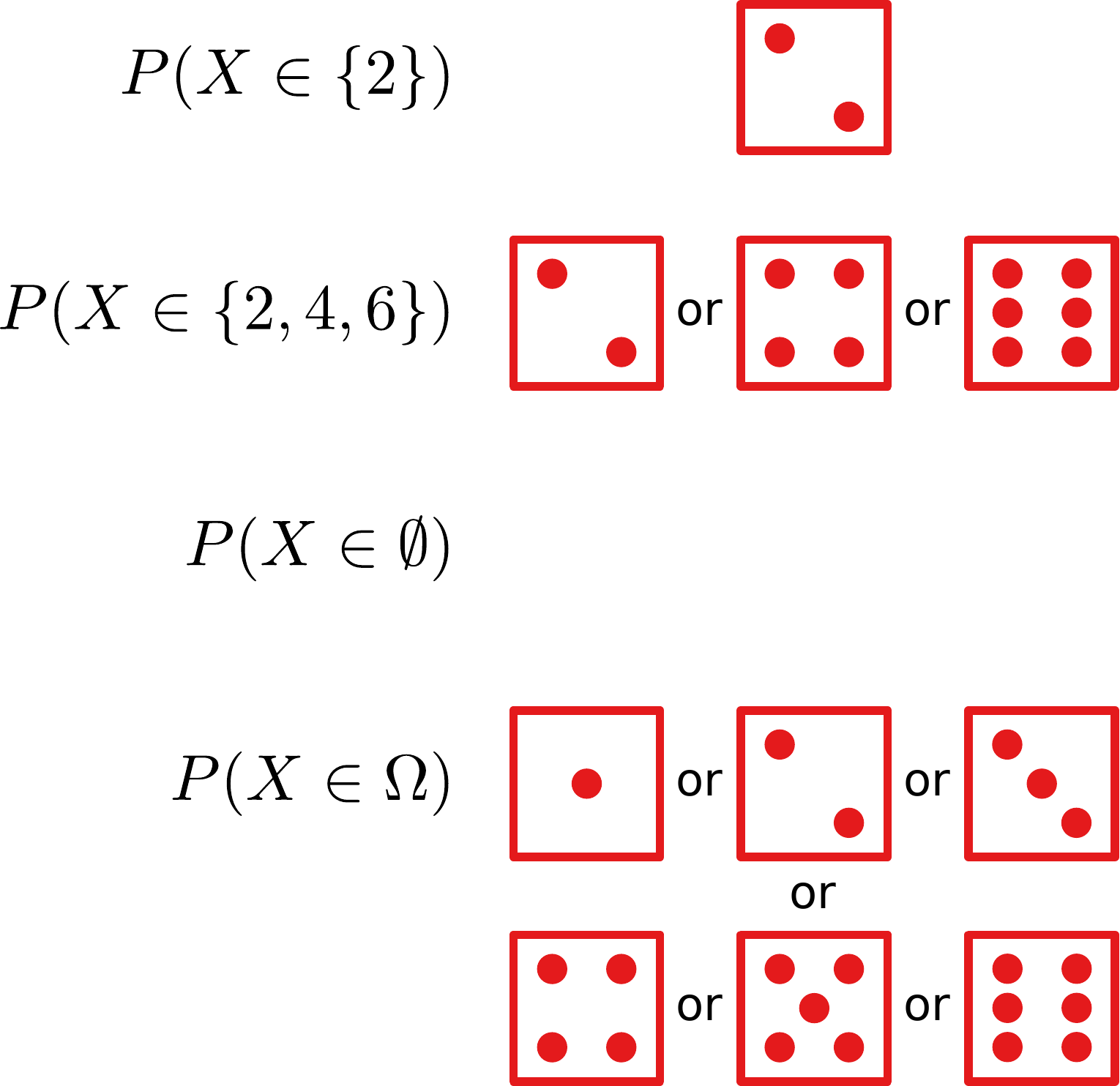}
\caption{A visual shorthand for understanding the mathematical notation of probabilities (left) and the possible outcomes of the ``tossing a die'' event.}
\label{fig:rnd-var-example}
\end{figure}

We also need to consider special questions. For instance, there is the case in which no event happens: $P (X \in \emptyset)$ (here $\emptyset$ refers to the empty set, a set containing no elements). The converse is also important: the probability of any event happening. In the case of the die, there are a total of six possible outcomes. Notation-wise, we define the set of all possible outcomes as $\Omega = \{1,2,3,4,5,6\}$. So this is represented as $P (X \in \{1, 2, 3, 4, 5, 6\})$, or $P(X \in \Omega)$. Figure \ref{fig:rnd-var-example} shows how the mathematical notation corresponds to our visual intuition.

To be more concise, we can skip the explicit reference to the variable $X$. For instance, we can codify the outcome ``the die lands on $3$'' with the symbol $3$. In this way, we can write $P(3)$ to refer to the probability of the die landing on $3$, $P(\{2,4,6\})$ for the probability of landing on an even number, $P(\{2,4,6\}) \cap P(1)$ for landing on an even number and on $1$, $P(\emptyset)$ for the probability of nothing happening, and $P(\Omega)$ for the probability of anything possible happening.

\section{Axioms}\label{sec:prob-axioms}
When building probability theory we need to establish a set of axioms: unprovable and -- hopefully -- self-evident statements that allow you to derive all other statements of the theory. Probability theory rests on three of such axioms.

First, the probability of an event is a non-negative number. Or: talking about a ``negative probability'' doesn't make any sense. Worst case scenario, an event $A$ is impossible, therefore $P(A) = 0$ -- for instance, this is the ``nothing happens'' case from the previous section when $A = \emptyset$. If $A$ is possible, $P(A) > 0$. In a borderless coin toss, there are only two possible outcomes: heads ($H$) or tails ($T$). The coin cannot land on the non-existing rim. Thus, the probability of landing on the rim is zero. It cannot be negative.

Second, certain events occur with probability equal to one. That is, if $A$ is an absolutely certain event, $P(A) = 1$. Using the notation from the previous section: $P(\Omega) = 1$, with $\Omega = \{H, T\}$ for a coin toss. Note that there isn't anything magical about the number $1$, we could have said that the maximum probability is equal to $42$, $\pi$, or ``meh''. It's just a convenient convention to define your units.

Third, the probability of happening for mutually exclusive events is the sum of their probabilities, or $P(\{H, T\}) = P(H) + P(T)$. A coin cannot land on heads and tails at the same time\footnote{Get those Schr\"{o}dinger coins out of my classroom!}, thus the probability that it lands on heads or tails is the sum of the probability of landing on heads and the probability of landing on tails.

As a corollary, you can also multiply probabilities. If $A$ and $B$ are \textit{independent} events, then $P(A)P(B)$ -- their multiplication -- tells you the probability of \textit{both} events happening. Independent events are events that have no relation to each other, such as you getting a promotion and the appearance of a new spot on the sun. For \textit{dependent} events, you need to take into account this dependence before applying the multiplication. In a fair die, $P(1) = P(2) = 1/6$, but we know that you can't get a $1$ if you are getting a $2$, so $P(1)P(2)$ is actually zero, not $1/36$. How to perform this check leads us to the world of conditional probabilities.

\section{Conditional Probability}
Events do not usually happen in isolation. Things that have happened in the past might influence what will happen in the future. There is a certain probability that the coin will land on heads: $P(H)$. But if I know something happened to the coin before the toss -- maybe I put some weights in it, event $W$ -- then the probability of heads will change. To handle this scenario, we introduce the concept of ``conditional probability''. In our scenario, the notation is $P(H|W)$. $P(H|W)$ is the probability of the coin landing on heads -- $H$ --  given that event $W$ happened.

This view of probability is particularly in line with the Bayesian interpretation, as what you call ``prior'' is really a synthesis of everything that happened in the past. That is not to say that a frequentist cannot understand conditional probabilities: they can, they just take the usual approach of simply observing what happen before/after something and be done with it.

\begin{figure}[b]
\centering
\includegraphics[width=.66\columnwidth]{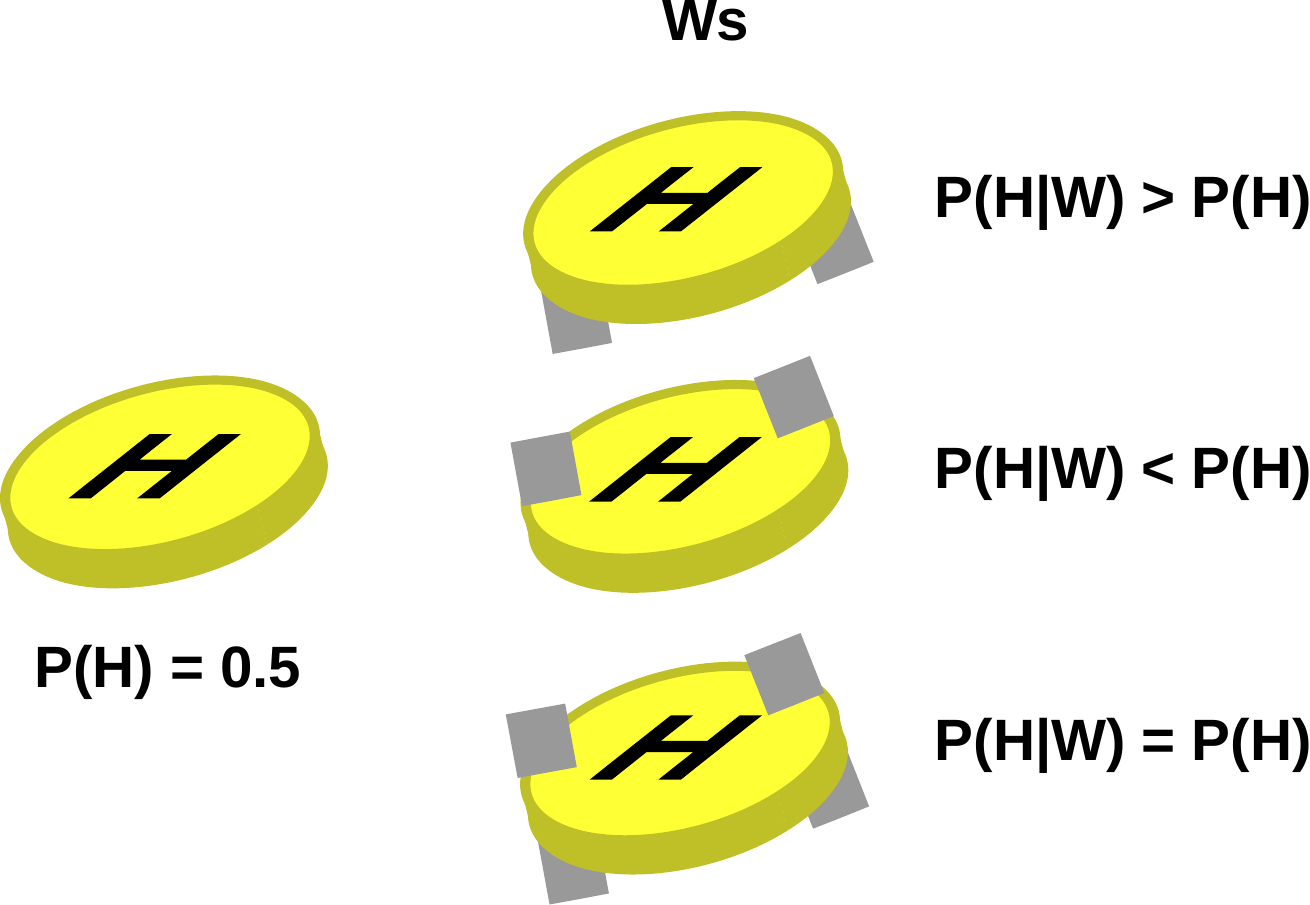}
\caption{The baseline probability of $H$ is $0.5$. When you add feet to the coin ($W$) the coin is more likely to land on the opposite side. Thus, $P(H|W) \neq P(H)$ and the two events are not independent -- unless you add feet on both sides as in the bottom example.}
\label{fig:conditional-p}
\end{figure}

Conditional probabilities enable you to make a nice set of inferences. Figure \ref{fig:conditional-p} shows the most basic ones. If you measure $P(H|W)$, you can figure out what event $W$ did to the coin. If $P(H|W) > P(H)$, it means that adding the weight to the coin made it more likely to land on heads. $P(H|W) < P(H)$ means the opposite: your coin is loaded towards tails. The $P(H|W) = P(H)$ case is equally interesting: it means that you added the weight uniformly and the odds of the coin to land on either side didn't change.

This is a big deal: if you have two events and this equation, then you can conclude that the events are independent -- the occurrence of one has no effect on the occurrence of the other\footnote{Note that here I'm talking about \textit{statistical} independence, which is not the same as \textit{causal} independence. Two events could be statistically dependent without being causally dependent. For instance, the number of US computer science doctorates is statistically dependent with the total revenue of arcades (\url{http://www.tylervigen.com/spurious-correlations}). This is what the mantra ``correlation does not imply causation'' means: correlation is mere statistical dependence, causation is causal dependence, and you shouldn't confuse one with the other. You should check \citep{pearl2018book} to delve deeper into this.}. This should be your starting point when testing a hypothesis: the null assumption is that there is no relation between an outcome (landing on heads) and an intervention (adding a weight). ``Unless,'' Mr. Bayes says, ``You have a strong prior for that to be the case.''

Reasoning with conditional probabilities is trickier than you might expect. The source of the problem is that, typically, $P(H|W) \neq P(W|H)$, and often dramatically so. Suppose we're tossing a coin to settle a dispute. However, I brought the coin and you think I might be cheating. You know that, if I loaded the coin, the probability of it landing on heads is $P(H|W) = 0.9$. However, you can't see nor feel the weights: the only thing you can do is tossing it and -- presto! -- it lands on heads. Did I cheat?

Naively you might rush and say yes, there's a $90\%$ chance I cheated. But that'd be wrong, because the coin already had a $50\%$ chance of landing on heads without any cheating. Thus $P(H|W) \neq P(W|H)$, and what you really want to estimate is the probability I cheated given that the coin landed on heads: $P(W|H)$. How to do so, using what you know about coins ($P(H)$) and what you know about my integrity ($P(W)$), is the specialty of Bayes' Theorem.

\section{Bayes' Theorem}\label{sec:prob-bayes}
Bayes' Theorem is an almost magical formula that allows you to estimate the probability of an event based on your priors. Keeping the example of cheating on a coin toss, we want to estimate the probability I cheated and rigged the coin so it lands on heads after we tossed it and it indeed landed on heads -- in mathematical notation: $P(W|H)$. To do so, you need to have priors. You need to know: what's the probability of heads for all coins in the world (whether they are rigged or not, $P(H)$), what's the probability I rigged the coin ($P(W)$), and what is the probability of obtaining heads on a rigged coin ($P(H|W)$). Without further ado, here's one of the most important formulas in human history:

$$ P(W|H) = \dfrac{P(H|W)P(W)}{P(H)}.$$

\begin{figure}
\centering
\includegraphics[width=\columnwidth]{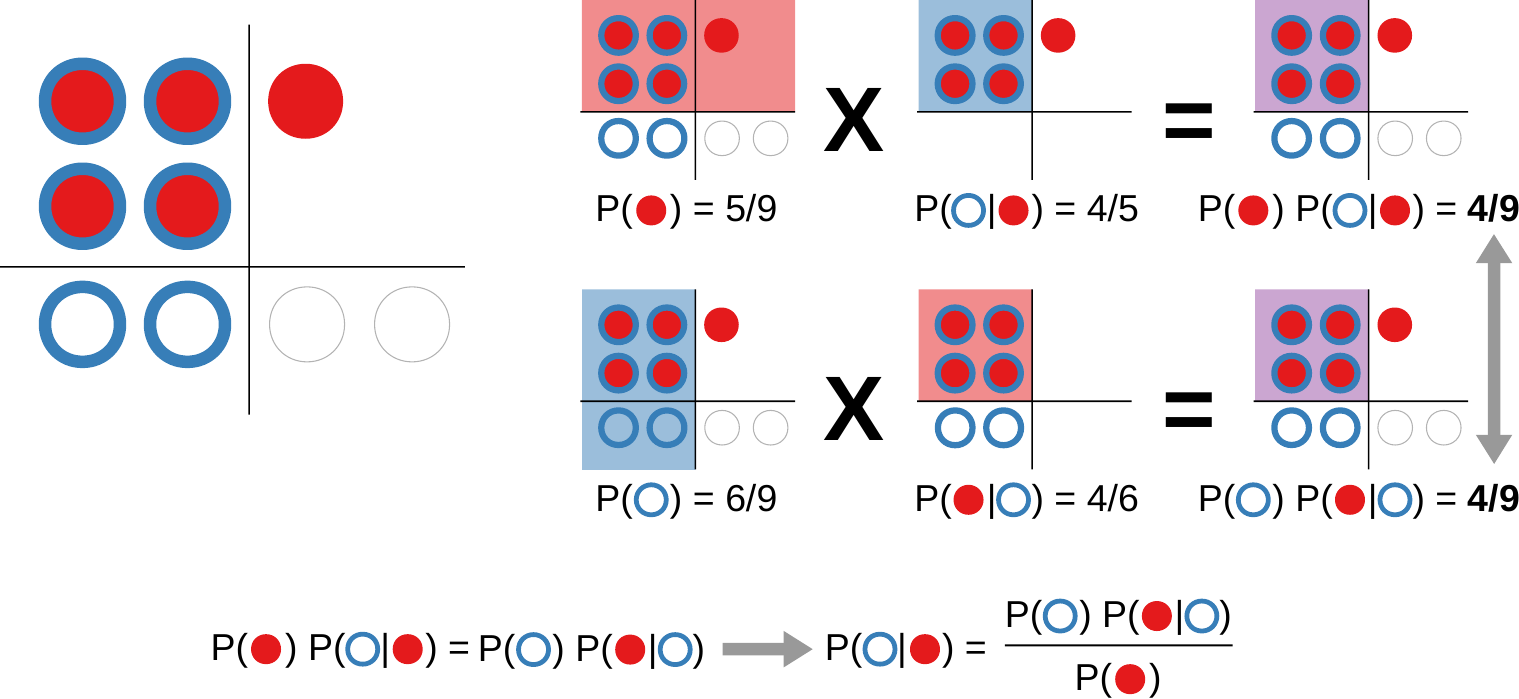}
\caption{The table on the left shows the occurrence of all possible events: red circles (5), blue borders (6), red circles with blue borders (4) and neither (2).}
\label{fig:bayes-theo}
\end{figure}

Figure \ref{fig:bayes-theo} shows a graphical proof of the theorem. When trying to derive $P(W|H)P(H)$, we realize that's identical to $P(H|W)P(W)$, from which Bayes' theorem follows.

I already told you that I'm a pretty good coin rigger ($P(H|W) = 0.9$). For the sake of the argument, let's assume I'm a very honest person: the probability I cheat is fairly low ($P(W) = 0.3$).

Now, what's the probability of landing on heads ($P(H)$)? $P(H)$ is trickier than it appears, because we're in a world where people might cheat. Thus we can't be naive and saying $P(H) = 0.5$. $P(H)$ is $0.5$ if rigging coins is impossible. It's more correct to say $P(H|-W) = 0.5$: a non rigged coin (if $W$ didn't happen, which we refer to as $-W$) is fair and lands on heads $50\%$ of the times. The real $P(H)$ is $P(H|-W)P(-W) + P(H|W)P(W)$. In other words: the probability of the coin landing on heads is the non rigged heads probability if I didn't rig it ($P(H|-W)P(-W)$) plus the rigged heads probability if I rigged it ($P(H|W)P(W)$).

The probability of not cheating $P(-W)$ is equal to $1 - P(W)$. This is because cheating and non cheating are mutually exclusive and either of the two \textit{must} happen. Thus we have $\Omega = \{W, -W\}$. Since $P(\Omega) = 1$ and $P(W) = 0.3$, the only way for $P(W,-W)$ to be equal to $1$ is if $P(-W) = 0.7$. 
 
This leads us to: $P(H) = P(H|-W)P(-W) + P(H|W)P(W) = 0.5 \times 0.7 + 0.9 \times 0.3 = 0.62$. Shocking.

The aim of Bayes' theorem is to update your prior about me cheating ($P(W)$) given that, suspiciously, the toss went in my favor ($P(W) \rightarrow P(W|H)$). Plugging in the numbers in the formula:

$$ P(W|H) = \dfrac{0.9 \times 0.3}{0.62} = 0.43.$$

A couple of interesting things happened here. First, since the event went in my favor, your prior about me possibly cheating got updated. Specifically, the event became more likely: from $0.3$ to $0.43$. Second, even if my success probability after cheating is very high, it is still more likely that I didn't cheat, because your prior about my lack of integrity was low to begin with.

This second aspect is absolutely crucial and it's easy to get it wrong in everyday reasoning. The textbook example is the cancer diagnosing machine. Let's say that $0.1\%$ of people develop a cancer, and we have this fantastic diagnostic machine with an accuracy of $99.9\%$: the vast majority of people will be diagnosed correctly (positive result for people with cancer and negative for people without). You test yourself and the test is positive. What's your chance of having cancer? $99.9\%$ accuracy is pretty damning, but before working on your last will, you apply Bayes' Theorem:

$$ P(C|+) = \dfrac{0.999 \times 0.001}{0.999 \times 0.001 + 0.001 \times 0.999} = 0.5.$$

The probability you have cancer is \textit{not} $99.9\%$: it's a coin toss! (Still bad, but not \textit{that} bad).\footnote{Of course, in the real world, if you took the test it means you thought you might have cancer. Thus you were not drawn randomly from the population, meaning that you have a higher prior that you had cancer. Therefore, the test is more likely right than not. Bayes' theorem doesn't endorse carelessness when receiving a bad news from a very accurate medical test.}

The real world is a large and scary environment. Many different things can alter your priors and have different effects on different events. The way a Bayesian models the world is by means of a Bayesian network: a special type of network connecting events that influence each other. Exploring a Bayesian network allows you to make your inferences by moving from event to event. I talk more about Bayesian networks in Section 
\ref{sec:extended-types}.

\section{Stochasticity}\label{sec:prob-stoch}
Colloquially, a stochastic process is one or more random variables that change their values over time. The quintessential stochastic process is Brownian motion. Brown observed very light pollen particles on water changing directions, following a \textit{stochastic} path that seemed governed purely by randomness. Interestingly, this problem was later solved by Einstein in one of his first contributions to science\cite{einstein1905molekularkinetischen}, working off important prior work\cite{bachelier1900theorie}. He explained the seemingly random changes of direction as the result of collision between the pollen and water molecules jiggling in the liquid.

\begin{figure}[t]
\centering
\begin{subfigure}[t]{.3\columnwidth}
\includegraphics[width=\textwidth]{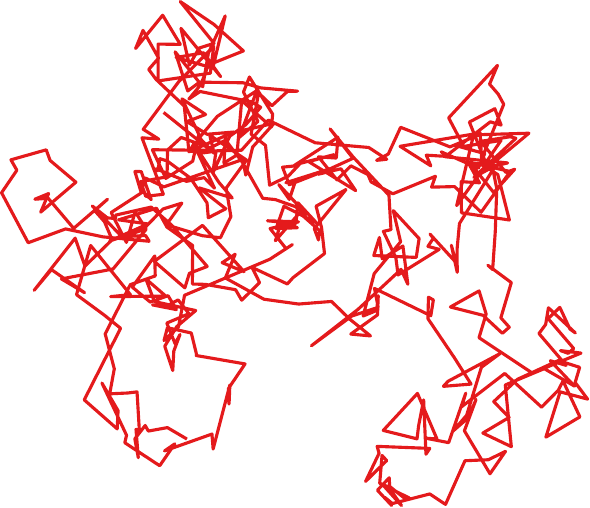}
\caption{}
\end{subfigure}
\qquad
\begin{subfigure}[t]{.2\columnwidth}
\includegraphics[width=\textwidth]{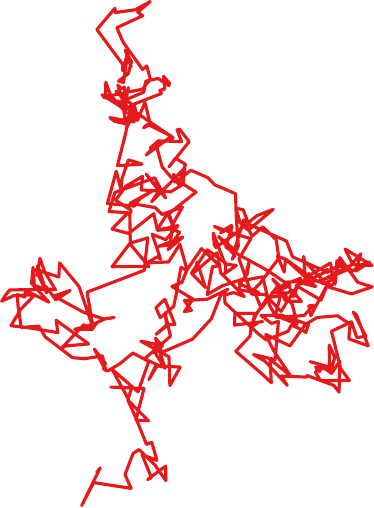}
\caption{}
\end{subfigure}
\qquad
\begin{subfigure}[t]{.29\columnwidth}
\includegraphics[width=\textwidth]{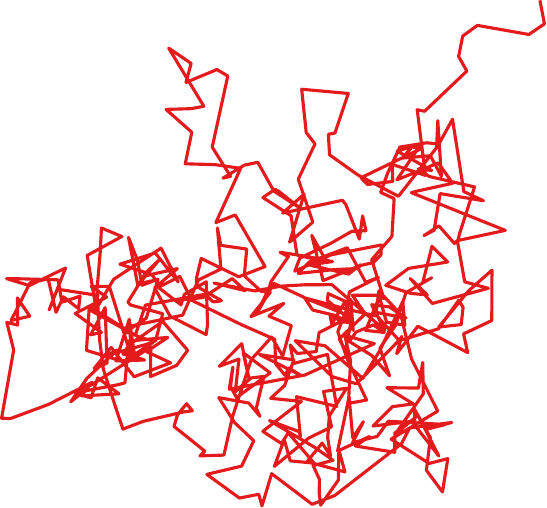}
\caption{}
\end{subfigure}
\caption{Three realizations of a Brownian stochastic motion on a two dimensional plane.}
\label{fig:brownian}
\end{figure}

When you have a stochastic process, there is an almost infinite set of results. The pollen can follow potentially infinite different paths. When you observe an actual grain, you obtain only one of those paths. The observed path is called a realization of the process. Figure \ref{fig:brownian} shows three of such realizations, which should help you visualize the intrinsic randomness of the change of direction.

Whenever you encounter the word ``stochastic'' in this book or in a paper, we're referring to a process governed by these dynamics. For instance, a stochastic matrix is a matrix whose rows and/or columns sum up to one. We call it stochastic, because such matrices are routinely used to describe stochastic processes. By having their rows to sum to one, you can interpret each entry of the row as the \textit{probability} of its corresponding event. The row in which you are tells you the current state of the process, the column tells you the next possible state, and the cell value tells you the probability of transitioning to each of the next possible states (column) given the current state (row). In other words, it is the probability of one possible realization of a single step in a stochastic process. In network science, you normally have stochastic adjacency matrices, which are the topic of Section \ref{sec:mat-mat-stochastic}.

\begin{figure}
\centering
\includegraphics[width=.45\columnwidth]{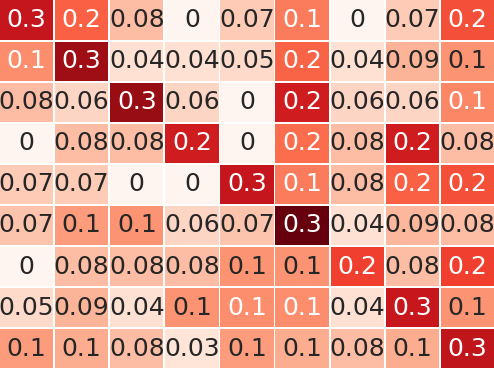}
\caption{A right stochastic matrix. }
\label{fig:stochastic-matrix}
\end{figure}

Figure \ref{fig:stochastic-matrix} is a stochastic matrix\footnote{Specifically, it is a right stochastic matrix: the rows sum to one, although there's a bit of rounding going on. In a left stochastic matrix, the columns sum to one.}. The rows tell you your current state and the columns tell you your next state. If you are in the first row, you have a $30\%$ probability of remaining in that state (the value of the cell in the first row and first column is $0.3$). You have a $20\%$ probability of transitioning to state two (first row, second column), $8\%$ probability of transitioning to state three, and so on.

\section{Markov Processes}\label{sec:prob-markov}
It should be clear now that, even if the next state is decided by a random draw, a stochastic process isn't necessarily uniformly random. In Brownian motion, the next position is determined by your previous position as well as a random kick. This observation is at the basis of a fundamental distinction between three flavors of stochastic processes, which are the most relevant for network science. The three flavors are: Markov processes, non-Markov processes, and higher-order Markov processes.

In a Markov process, the next state is exclusively dependent on the current state and nothing else. No information from the past is used: only the present state matters. That is why a Markov process is usually called ``memoryless''. The stochastic process I described when discussing Figure \ref{fig:stochastic-matrix} is a typical Markov process. The only thing we needed to know to determine the next state was the current state: in which row are we?

\begin{figure}
\centering
\includegraphics[width=.45\columnwidth]{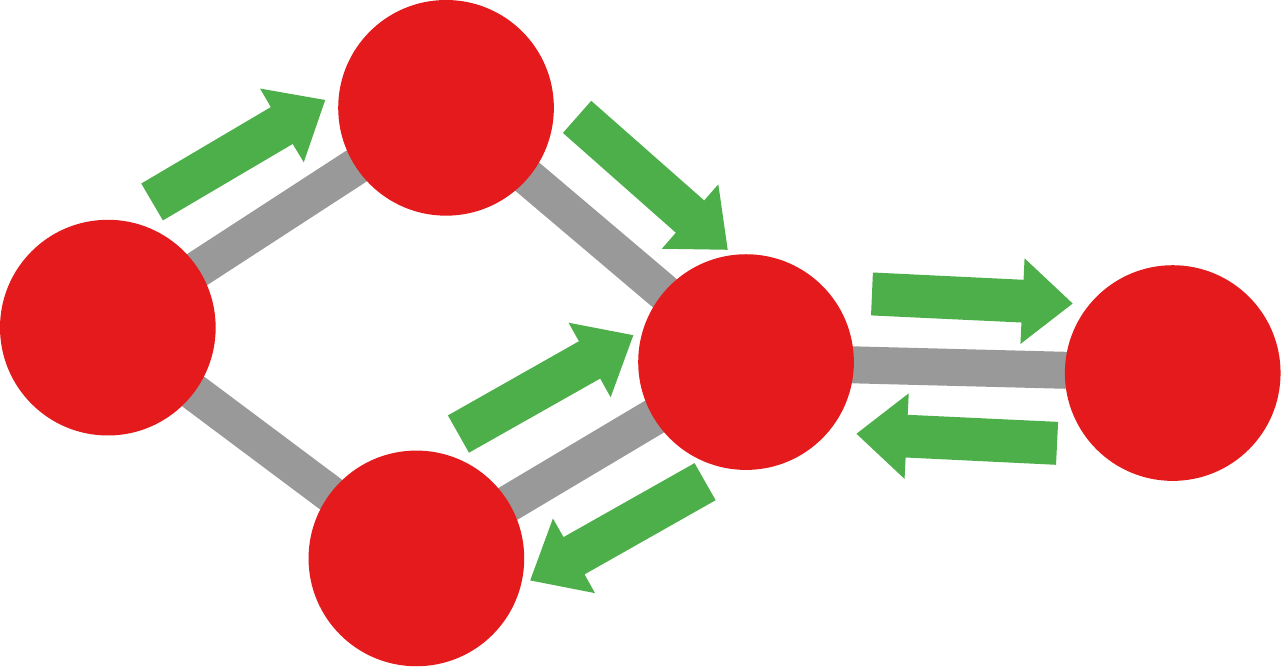}
\caption{A random walk. The green arrows show the state transitions.}
\label{fig:random-walk-example}
\end{figure}

The classical Markov process in network science is the random walk. A random walker simply chooses the next node it wants to occupy, and its options are determined solely by the node it is currently occupying. Rather surprisingly, random walks are one of the most powerful tools in network science and have been applied to practically everything. I'm going to introduce them properly in Chapter \ref{cha:rndwalks}, but they will pop up throughout the book -- for instance, in community discovery (Part \ref{par:cd}) and in network sampling (Chapter \ref{cha:sampling}). Figure \ref{fig:random-walk-example} shows an example of a random walk. As you can see, we start from the leftmost node. From that state, reaching the two rightmost ones is impossible because the nodes are not connected. Only when you transition to another state, new states become available.

A bit more formally, let's assume you indicate your state at time $t$ with $X_t$. You want to know the probability of this state to be a specific one, let's say $x$. $x$ could be the id of the node you visit at the $t$-th step of your random walk. If your process is a Markov process, the only thing you need to know is the value of $X_{t-1}$ -- i.e. the id of the node you visited at $t-1$. In other words, the probability of $X_t = x$ is $P(X_t = x | X_{t-1} = x_{t-1})$. Note how $X_{t-2}, X_{t-3}, ..., X_{1}$ aren't part of this estimation. You don't need to know them: all you care about is $X_{t-1}$.

On the other hand, a non-Markov process is a process for which knowing the current state doesn't tell you anything about the next possible transitions. For instance, a coin toss is a non-Markov process. The fact that you toss the coin and it lands on heads tells you nothing about the result of the next toss -- under the absolute certainty that the coin is fair. The probability of $X_t = x$ is simply $P(X_t = x)$: there's no information you can gather from your previous state.

Finally, we have higher-order Markov processes. Higher-order means that the Markov process now has a memory. A Markov process of order $2$ can remember one step further in the past. This means that, now, $P(X_t = x | X_{t-1} = x_{t-1}, X_{t-2} = x_{t-2})$: to know the probability of $X_t = x$, you need to know the state value of $X_{t-2}$ as well as of $X_{t-1}$. More generally, $P(X_t = x | X_{t-1} = x_{t-1}, X_{t-2} = x_{t-2}, ..., X_{t-m} = x_{t-m})$, with $m \leq t$.

\begin{figure}
\centering
\includegraphics[width=.45\columnwidth]{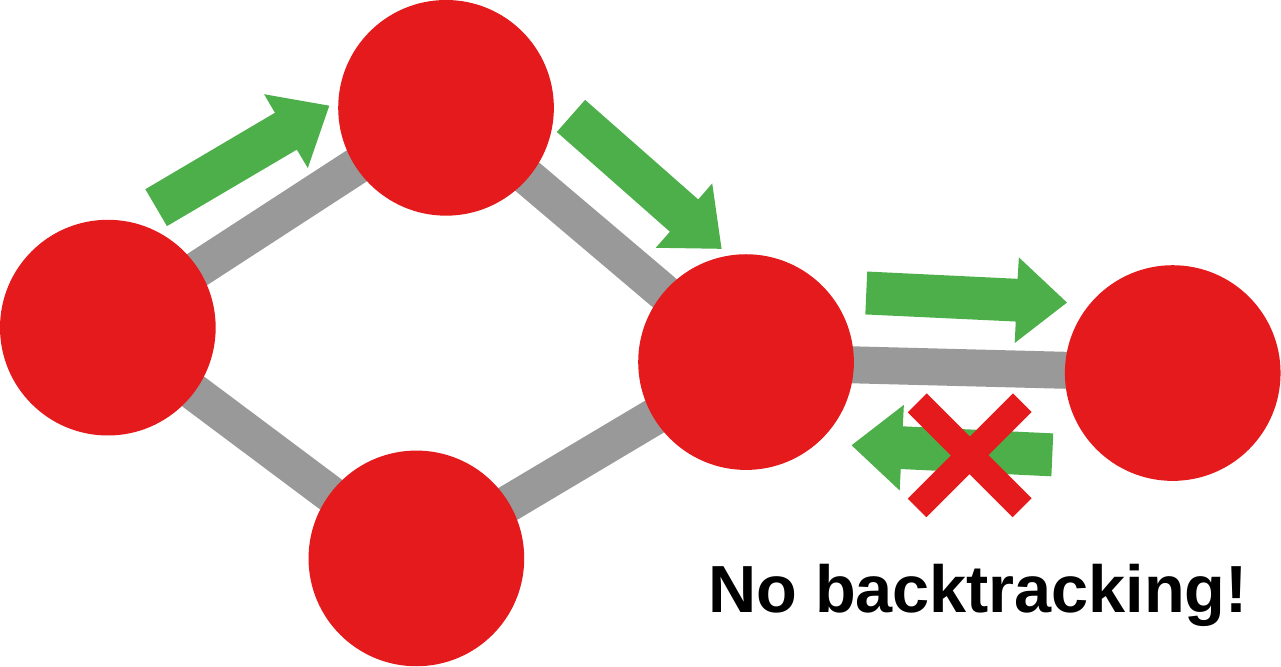}
\caption{A non-backtracking random walk. The green arrows show the state transitions.}
\label{fig:random-walk-nobacktrack}
\end{figure}

The classical network examples of a higher order Markov process is the non-backtracking random walk (Figure \ref{fig:random-walk-nobacktrack}). In a non-backtracking random walk, once you move from node $u$ to node $v$, you are forbidden to move back from $v$ to $u$. This means that, once you are in $v$, you also have to remember that you came from $u$. Higher order Markov processes are the bread and butter of higher order network problems, which is the topic of Chapter \ref{cha:hod}.

\section{Alternatives to Probability Theory}\label{sec:prob-alt}
When it comes to dealing with uncertainty, probability theory is not the only game in town. Here I briefly present two alternatives. These will be handy when we focus specifically on probabilistic networks in Chapter \ref{cha:uncertainty}.

\subsection{Dempster-Shafer's Theory of Evidence}\label{sec:prob-alt-dste}
In Dempster-Shafer's theory of evidence (DST) we take the key insight from Bayes and we turn it up to eleven. One thing that underlies Bayesian thought is that probabilities are subjective. If two people have different priors, say $A$ and $B$ such that $P(A) \neq P(B)$, then they will disagree on the probability of event $C$, which will depend on the priors. This doesn't happen with a frequentist framework, because there are no priors and $P(C)$ is based on objective data available to everyone. However, besides this subjectivity, Bayes still uses the axioms and the rules of probability theory.

DST is a generalization of probability theory, which moves from exact probabilities to probability intervals. Its central parts are \textit{beliefs} and \textit{plausibilities}, and these two things don't necessarily behave like probabilities\cite[-0.5in]{dempster1967upper}\cite{shafer1976mathematical}.

\begin{figure}
\centering
\includegraphics[width=.4\columnwidth]{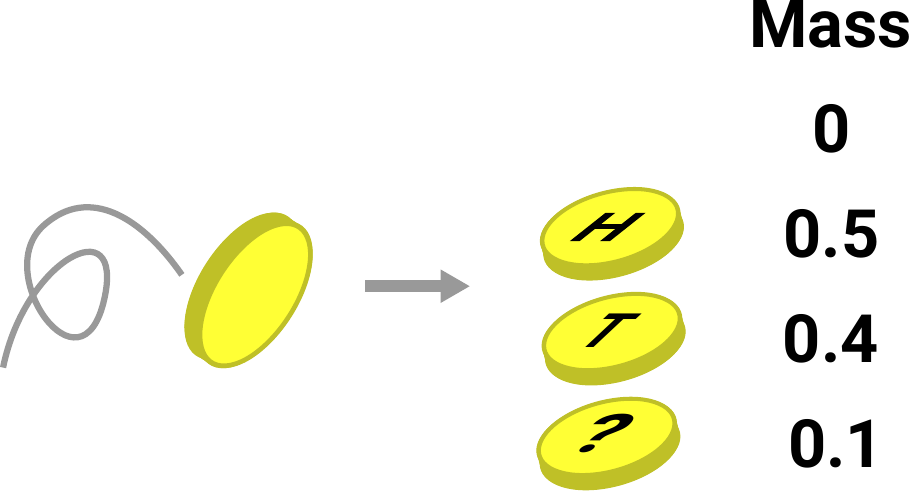}
\caption{An illustration of Mass in DST. After tossing a coin (left) each each subset of $\Omega$ obtains a value.}
\label{fig:prob-dst-mass}
\end{figure}

Starting with beliefs, the first thing you need is a \textit{degree of belief}, which estimates your ability to prove a set of beliefs\cite{pearl1990reasoning}. This degree of belief is quantified by a function which is conventionally called its Mass function. Figure \ref{fig:prob-dst-mass} shows a relatively simple example when tossing a coin -- slightly loaded on heads. The distinction between Mass in DST and classical probability is that it considers the case ``we don't know whether heads or tail'' as distinct from ``heads'' and ``tail''. In probability theory, you wouldn't make this distinction, because no other outcome than heads or tails can happen, even if for some reason you don't know the outcome. But in DST you want to model this, because we're talking about the ability of proving our statement, so we need to specifically take into account the situation in which we don't actually know the result -- e.g., if the coin rolled under the sofa and we can't see it. In that case, we don't have any evidence to say that the coin landed on heads or tail.

In summary, if $\Omega = \{H,T\}$, then $p(\Omega) = p(H) + p(T) = 1$, but $Mass(\Omega) \neq Mass(H) + Mass(T)$: $Mass(\Omega)$ is less trivial and actually informative -- it is the amount of uncertainty we have about the outcome of the event given the imperfection of our evidence. So the Mass function is basically giving all the available evidence a probability and obeys the following two rules:

\begin{enumerate}
\item $Mass(\emptyset) = 0$, and
\item $\sum \limits_{X \in 2^\Omega} Mass(X) = 1$.
\end{enumerate}

Here, $2^\Omega$ means all possible subsets of $\Omega$, which in my simple case are: $\emptyset$, $\{H\}$, $\{T\}$, and $\{H,T\}$. The first property means that it's impossible to prove that nothing happened -- we know we tossed the coin, it must have landed on something. That's why the ugly coin toss drawing on the left of Figure \ref{fig:prob-dst-mass} is necessary: to show we performed the tossing. The second property means that $\Omega$ contains all possible answers to our question -- so what's actually true must be a subset of $\Omega$.

\begin{figure}
\centering
\includegraphics[width=.6\columnwidth]{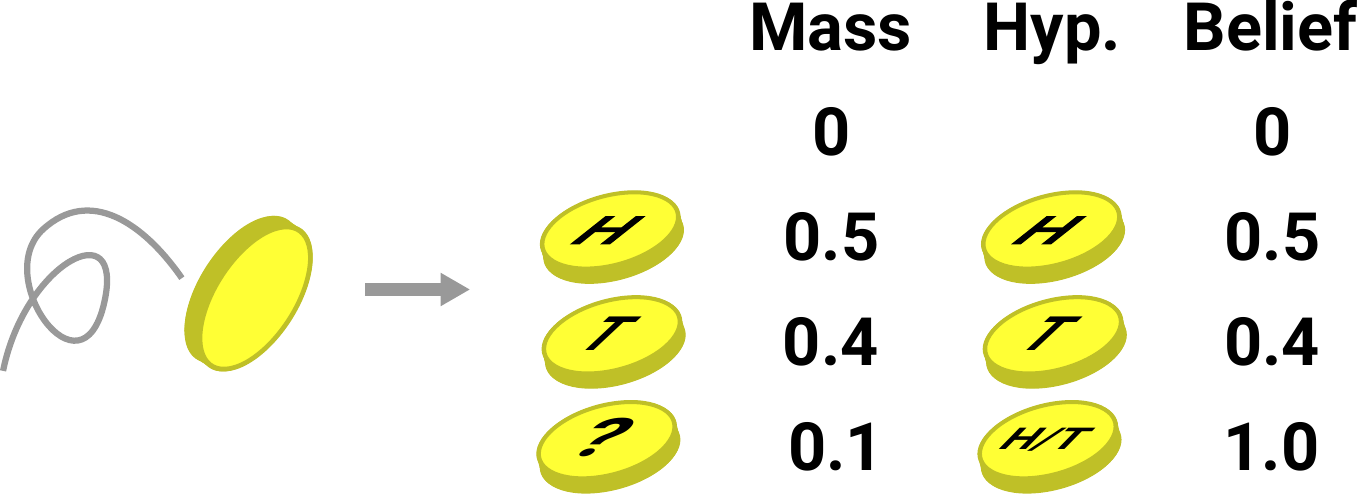}
\caption{An illustration of Belief in DST, based on the Mass from Figure \ref{fig:prob-dst-mass}. Note how Belief of a subset of size $1$ is equal to its Mass.}
\label{fig:prob-dst-belief}
\end{figure}

Now that you know the probabilities of all possible outcomes, you must make a hypothesis which is a set of potential outcomes. To estimate your ability to prove your hypothesis, you sum all the Mass values of all the subsets of your hypothesis. This is the Belief function, and you can see how it works in Figure \ref{fig:prob-dst-belief}. If your hypothesis has no support in the gathered evidence then its Belief value is zero, while if it is absolutely certain then Belief evaluates to one. So Belief tells you how likely your hypothesis -- or any of its subsets -- is to be proven given the available evidence.

\begin{figure}
\centering
\includegraphics[width=.7\columnwidth]{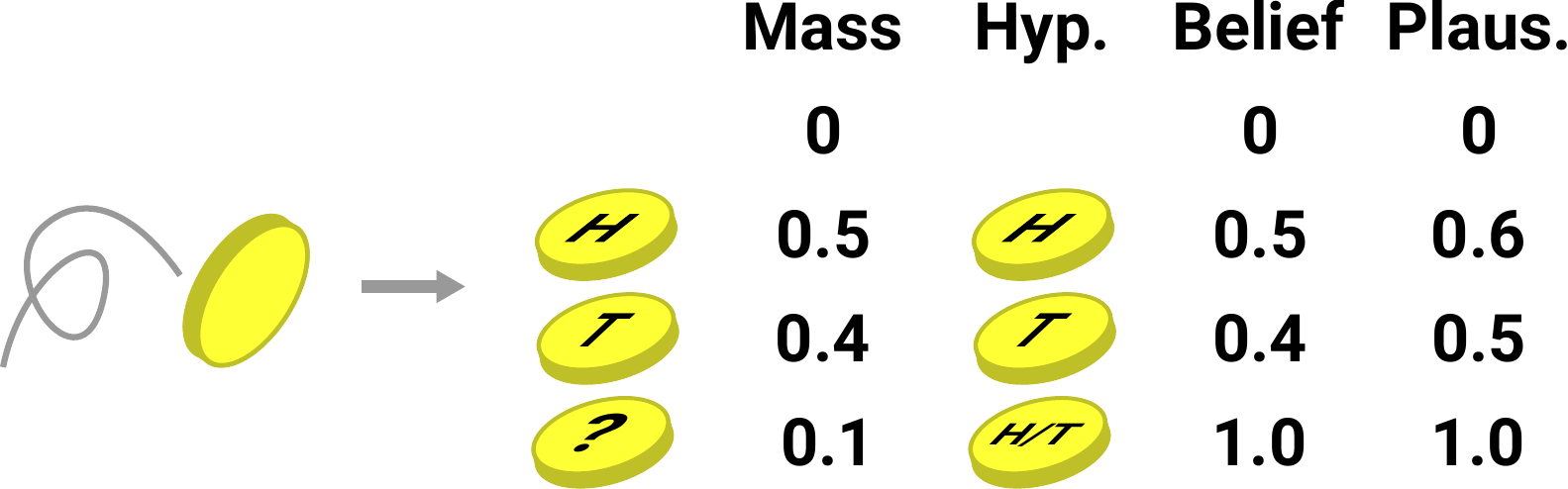}
\caption{An illustration of Plausibility in DST, based on the Mass and Belief Figure \ref{fig:prob-dst-belief}.}
\label{fig:prob-dst-plausibility}
\end{figure}

DST also allows to compute the Plausibility function, which is an upper bound of Belief. In practice, you can tally up all the evidence of your hypothesis not containing the truth and take the inverse of it. Figure \ref{fig:prob-dst-plausibility} shows how it works. Plausibility is basically estimating how much your hypothesis can survive an attempt to prove it false. A handy rule to remember is that $Plausibility(X) = 1 - Belief(\bar{X})$, where $\bar{X}$ is the complementary set of $X$ -- the plausibility of something is the opposite of your belief of that something being proven false.

We go through the trouble of defining these things because DST has some advantages. For instance, there are some operations you can do with the Belief and Plausibility functions -- which we are not going to see here -- but allow you to work with different hypotheses in conflict. Classical probability theory is ill suited to handle these cases. For instance, suppose that the ice cream shop has three flavors: chocolate, strawberry, and vanilla. We want to share an ice cream and my preference is $99\%$ chocolate and $1\%$ vanilla, while yours are $99\%$ strawberry and $1\%$ vanilla. We should obviously go for vanilla, and DST agrees, but if we took a probabilistic approach ignoring DST, you might instead say that vanilla is the least likely solution, and we would end up with either chocolate of strawberry, much to the distress of either of us. For instance by naively aggregating probabilities as $49.5\%$ for each strawberry and chocolate. To be fair, DST also ends up saying weird things occasionally, which has led researchers to formulate ways to turn it into computable functions that are different from the ones I explained\cite{sentz2002combination}.

Moreover, probability theory must assign a probability to an event, even if there is no evidence for it, while DST can simply give it zero Mass. For instance, if we have a potentially loaded die, in probability theory using a Bayesian approach we must start by assigning a prior probability of $1/6$ to all outcomes \textit{even if we have zero evidence for it}. In DST, you'd give them Mass zero instead, and Mass one to $\Omega$ and then start gathering evidence.

\subsection{Fuzzy Logic}\label{sec:prob-alt-fuzzy}
In probability theory, you only deal with boolean events, whose truth values can either be zero or one. Either something is false or it is true. The coin either landed on heads or on tails. In fuzzy logic, you work with something different. Things can have degrees of truthiness, which is to say we assign them a truth value between zero and one. If it is zero, we're certain that a statement is false, if it is one we're certain it is true, and if it is a value in between then there is some vagueness about whether it is true or false\cite{hajek2013metamathematics}.

For instance, at the moment of writing this paragraph I am $39$ years old. Is that young or old? Well, you could line up $100$ people and ask them this question. Maybe $60$ will say that I'm young, $39$ will say that I'm old (I'm so insecure I am disrespected even in my thought experiments), and $1$ will say something else. In probability theory you could model this as something like: there's a $39\%$ chance a random person will call me old (hey!). But in fuzzy logic you'd do something different. You could say that I belong to both the sets of young people and old people, with different strengths. I'm $60\%$ young and $39\%$ old -- I frankly don't know if that's an improvement over the alternative.

\begin{figure}
\centering
\includegraphics[width=.66\columnwidth]{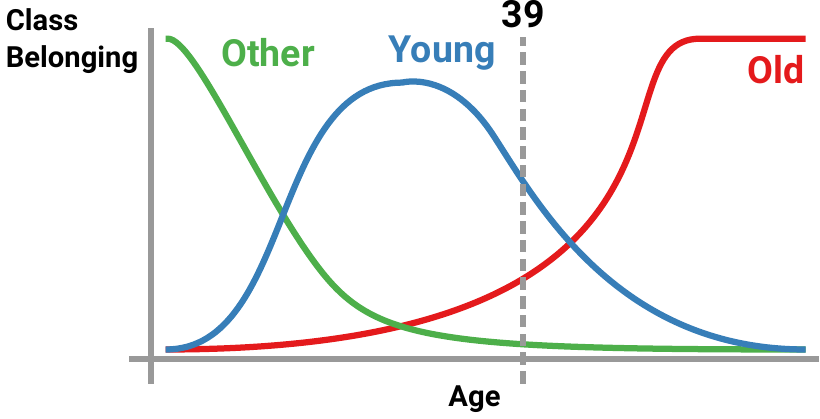}
\caption{An illustration of fuzzy logic. The degrees of belonging (y axis) to different age classes (line color) for a given age (x axis).}
\label{fig:prob-fuzzy}
\end{figure}

The consequences of this difference lead to different outcomes when working with fuzzy logic. We'll see a basic common example\cite{mamdani1974application}, but know that there are alternative ways to implement fuzzy logic\cite{takagi1985fuzzy}. Let's assume that the degree to which a person belongs to an age class depends on their age, following the function I draw in Figure \ref{fig:prob-fuzzy}.

For the age I highlight, probability theory can say the following things:

\begin{itemize}
\item The probability of picking somebody who could call me both young and old (assuming I ask multiple time and people can change their mind independently from what they said the first time) is $P(Y \cap O) = P(Y)P(O) = 0.234$;
\item The probability of picking somebody who will call me either young or old is $P(Y \cup O) = P(Y) + P(O) = 0.99$;
\item The probability of somebody not calling me young is $P(\bar{Y}) = 1 - P(Y) = 0.61$.
\end{itemize}

But in fuzzy logic we have:

\begin{itemize}
\item My belonging to the class of people who are both young and old is $P(Y \cap O) = \min(P(Y), P(O)) = 0.39$ -- it can't be any higher than my minimum belonging, because to fully belong to the young-old class I must be fully young and fully old;
\item My belonging to the class of people who are either young or old is $P(Y \cup O) = \max(P(Y), P(O)) = 0.6$ -- it can't be any lower than my maximum belonging, because if I am fully old then I am also fully-young-or-fully-old;
\item My belonging to the class of people who are not young is $P(\bar{Y}) = 1 - P(Y) = 0.61$.
\end{itemize}

\section{Summary}

\begin{enumerate}
\item Probability theory gives you the tools to make inferences about uncertain events. We often use a frequentist approach, the idea that an event's probability is approximated by the aggregate past tests of that event. Another important approach is the Bayesian one, which introduces the concept of priors: additional information that you should use to adjust your inferences.
\item Probabilities are non-negative estimates. The set of all possible outcomes has a probability sum of one. Summing two probabilities tells you the probability of either of two independent outcomes to happen.
\item The conditional probability $P(A|B)$ tells you the probability of an outcome $A$ given that you know another outcome $B$ happened. If $P(A|B) \neq P(A)$ then the two outcomes are not independent. Bayes' Theorem allows you to infer $P(A|B)$ from $P(B|A)$.
\item When we track the change over time of one or more random variables, we're observing a stochastic process. Markov processes are stochastic processes whose status exclusively depends on the status of the system in the previous time step.
\item There are alternative to probability theory when working with uncertainty. Dempster-Shafer's theory of evidence allows to work with the degrees of beliefs in specific hypotheses, while fuzzy logic allows for multiple things to be a little bit true at the same time.
\end{enumerate}

\section{Exercises}

\begin{enumerate}
\item Suppose you're tossing two coins at the same time. They're loaded in different ways, according to the table below. Calculate the probability of getting all possible outcomes:

\begin{tabular}{ll|rrrr}
$p_1(H)$ & $p_2(H)$ & H-H & H-T & T-H & T-T\\
\hline
0.5 & 0.5 &&&&\\
0.6 & 0.7 &&&&\\
0.4 & 0.8 &&&&\\
0.1 & 0.2 &&&&\\
0.3 & 0.4 &&&&\\
\end{tabular}

\item $60\%$ of the emails hitting my inbox is spam. You design a phenomenal spam filter which is able to tell me, with $98\%$ accuracy, whether an email is spam or not: if an email is not spam, the system has a $98\%$ probability of saying so. The filter knows $60\%$ of emails are spam and so it will flag $60\%$ of my emails. Suppose that, at the end of the week, I look in my spam box and see $963$ emails. Use Bayes' Theorem to calculate how many of those $963$ emails in my spam box I should suspect to be non-spam.
\item You're given the string: ``OCZ XJMMZXO VINRZM''. Each letter follows a stochastic Markov process with the rules expressed by the table at \url{http://www.networkatlas.eu/exercises/2/3/data.txt}. Follow the process for three steps and reconstruct the correct answer. (Note, this is a Caesar cipher\footnote{\url{https://en.wikipedia.org/wiki/Caesar_cipher}} with shift $7$ applied three times, because the Caesar cipher is a Markov process).
\item Suppose that we are examining a painting and we're trying to date it with the century when it was produced. Find out the Belief and Plausibility values for all hypotheses given the following Mass estimation (note that, by definition $\Omega = \{$XIV, XV, XVI$\}$ must have Belief and Plausibility equal to one):

\begin{tabular}{l|rrr}
Hypothesis & Mass & Belief & Plausibility\\
\hline
$\emptyset$ & $0.00$ & & \\
XIV & $0.16$ & & \\
XV & $0.04$ & & \\
XVI & $0.21$ & & \\
\{XIV, XV\} & $0.34$ & & \\
\{XV, XVI\} & $0.16$ & & \\
\{XIV, XVI\} & $0.08$ & & \\
$\Omega$ & $0.01$ & $1$ & $1$ \\
\end{tabular}


\end{enumerate}

\chapter{Statistics}\label{cha:stats}
In Chapter \ref{cha:prob} I explained the basic concepts of probability theory you need to understand to be a good network scientist. This chapter focuses on basic statistical concepts that are also necessary to analyze your networks. The disclaimer I put at the beginning of Chapter \ref{cha:prob} also holds here: statistics is much more vast and complicated than what I present here. This chapter is emphatically \textit{not} a substitute for a proper statistics textbook\cite{wheelan2013naked}\cite{mcelreath2018statistical}, which you could use to study this stuff further. You'll get a sense of how powerful statistics is\cite{huff1954lie}, in that it can allow you to support any point. As the saying goes: there are lies, damn lies, and statistics.

The main difference between probability and statistics is that you can do a lot of work in probability theory without actually looking at any data. When data takes the center stage, you enter in the world of statistics. Statistics covers more than simply describing your data: you should think in statistical terms also when collecting, cleaning, validating your data. But here we ignore all that and we focus on the tools that allow you to say something interesting about your data, assuming you did a good job collecting, cleaning, and validating it.

\section{Summary Statistics}\label{sec:stats-summary}

\subsection{Mean \& Median}
One common use of statistics is to give a quick description of what is in the data. The most classical task is to try and figure out what are the values you'd expect to find if you were to look directly at the data. For instance, if you want to know the height of the average human, you might want to calculate the mean height: $\mu(H) = \sum \limits_i H_i / |H|$, where $H_i$ is the height of one human. The mean in this case would tell you what you'd expect to see if you were to measure the height of a random person.

However, that works for height because height is normally distributed -- meaning that the average value is actually the mean and values farther from the mean are progressively more rare. We'll see what a normal distribution looks like in Section \ref{sec:prob-distr}. The same section will also tell you that not all variables distribute like that: in some cases the mean is actually not a great approximation of your average (or ``typical'') case. Wealth is like that: a tiny fraction of people own vastly more than the majority. In this case, the median gives you a better idea\cite{sheskin2003handbook}. The median tells you the value that splits the data in two equally populated halves: $50\%$ of the points are below the median and $50\%$ of the points are above.

\begin{figure}
\centering
\begin{subfigure}[t]{.3\columnwidth}
\includegraphics[width=\textwidth]{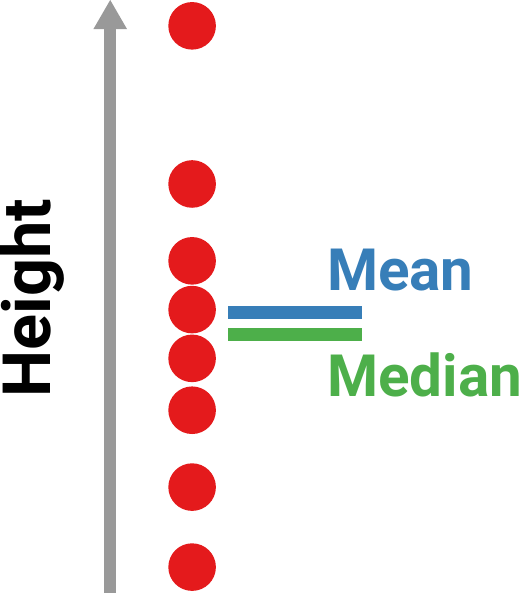}
\caption{}
\end{subfigure}
\qquad
\begin{subfigure}[t]{.3\columnwidth}
\includegraphics[width=\textwidth]{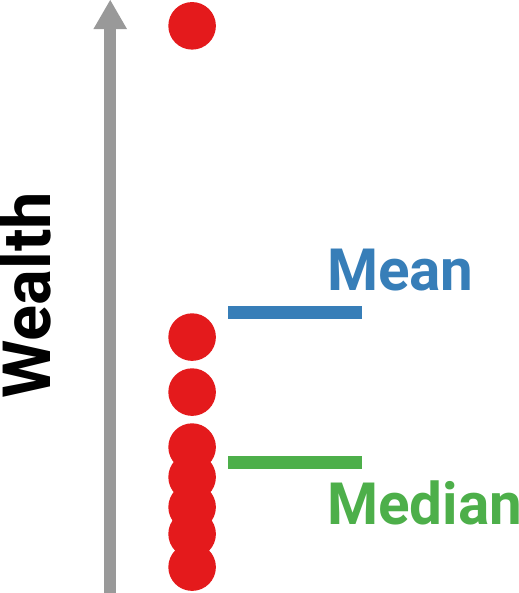}
\caption{}
\end{subfigure}
\caption{The mean (blue) and median (green) or two different variables (y axis): (a) normally distributed height; (b) skewed distribution of wealth.}
\label{fig:mean-median}
\end{figure}

Figure \ref{fig:mean-median} shows that the mean and the median can be quite different. In network science, we use the mean extensively -- even though maybe we shouldn't. When we'll talk about the number of connections a node has in a network (Chapter \ref{cha:degree}), we'll see we routinely take its ``average'' by calculating its mean. But connection counts in real networks typically do not follow a neat normal distribution. These distributions tend to look more like Figure \ref{fig:mean-median}(b) than Figure \ref{fig:mean-median}(a). So, perhaps the mean count of connections is not the most meaningful thing you can calculate.

This disconnect between the mean and the typical case is true for the arithmetic mean I show here, but there are other types of means -- such as geometric or harmonic -- which can take into account some special properties of the data\cite{fleming1986not}.

\subsection{Variance \& Standard Deviation}
When we deal with an average observation, we might want to know not only its expected value, but also how much we expect it to differ from the actual average value. Even if the average human height is, let's say, $1.75$ meters, we could think of two radically different populations. In the first, almost everyone is more or less $1.75$ and heights don't vary much. In the other, the opposite is true: the average is still $1.75$, but people could be anything between $1$ meter and $2.5$ meters. So the heights in this second population vary much more. We need to have a tool allowing us to distinguish these two populations. Since the difference is all about how much the heights \textit{vary}, we call this measure \textit{variance}. Variance  (and standard deviation) helps you quantify how dispersed your values are away from the mean.

\begin{figure}
\centering
\includegraphics[width=.33\textwidth]{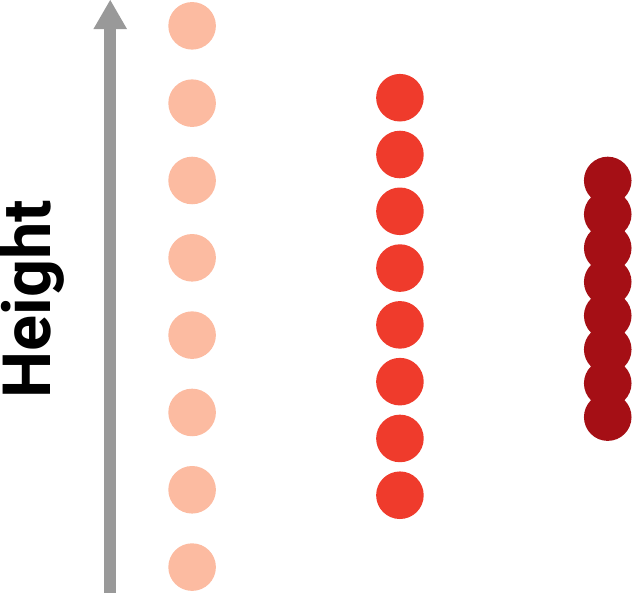}
\caption{The variances of three different populations, going from left (high variance, bright red) to right (low variance, dark red).}
\label{fig:stats-variance}
\end{figure}

Variance is almost literally the mean difference from the mean. The only tweak is that we take the square of this difference: $var(H) = \mu((H - \mu(H))^2)$. We take the square because we don't want values below the mean to cancel out values above the mean.  The standard deviation is simply the square root of the variance: $\sigma(H) = \sqrt{var(H)}$. The advantage of the standard deviation is that it ends up having the same units as the original variable. For example, for heights measured in cm, the variance will be in cm$^2$, but the standard deviation will be in cm again. Figure \ref{fig:stats-variance} shows what it looks like to have different variances for variables with the same mean.

The concept of variance will be important when we talk about data dimensionality reduction (Section \ref{sec:mat-factors}) and degree distributions (Section \ref{sec:degree-pl}). We will even see how to modify its definition to create a notion of network variance (Section \ref{sec:nvd-stats})

\subsection{Distribution \& Skewness}
Knowing how skewed your data is can be quite important -- in wealth distribution, how skewed the data is equals how screwed people are. Variance and standard deviation can help you quantify this. There are different formulas and different terms to talk about skewness\cite{von2005mean}, but for our purposes we limit ourselves to a bit of terminology.

\begin{figure}
\centering
\includegraphics[width=.8\columnwidth]{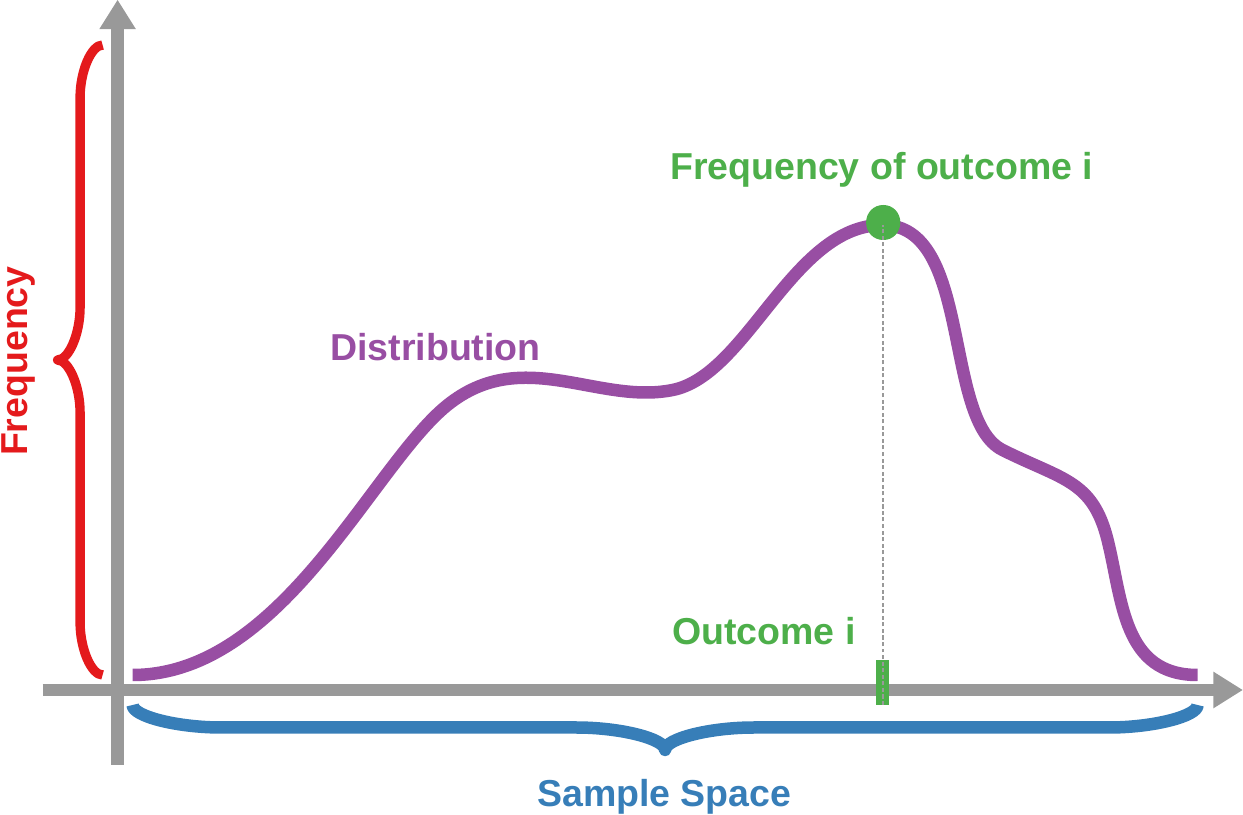}
\caption{A distribution, connecting every possible outcome in the sample space (x axis) to a frequency (y axis).}
\label{fig:p-distr}
\end{figure}

First: what actually is a distribution? I've used this term in an intuitive way without really defining it. Let's do it here. When you perform an experiment, or observe a stochastic process, you have many possible outcomes. For instance, you're measuring the heights of all people in a country and so your possible outcomes are all the possible heights a person can have. Sometimes, you're not interested on the frequency of measuring a specific outcome -- say $175$ cm. Sometimes, you want to study all possible outcomes together, to determine which is more likely, what you could expect when you measure more people, if there are maximums and minimums you don't expect to ever exceed, and so on. This is the task of a distribution. A distribution is a function that, for each outcome in the set of all possible ones (called the ``sample space''), tells you how many times you measured a given outcome. Figure \ref{fig:p-distr} shows a vignette on how to interpret a plot showing you a distribution.

\begin{figure}
\centering
\includegraphics[width=\columnwidth]{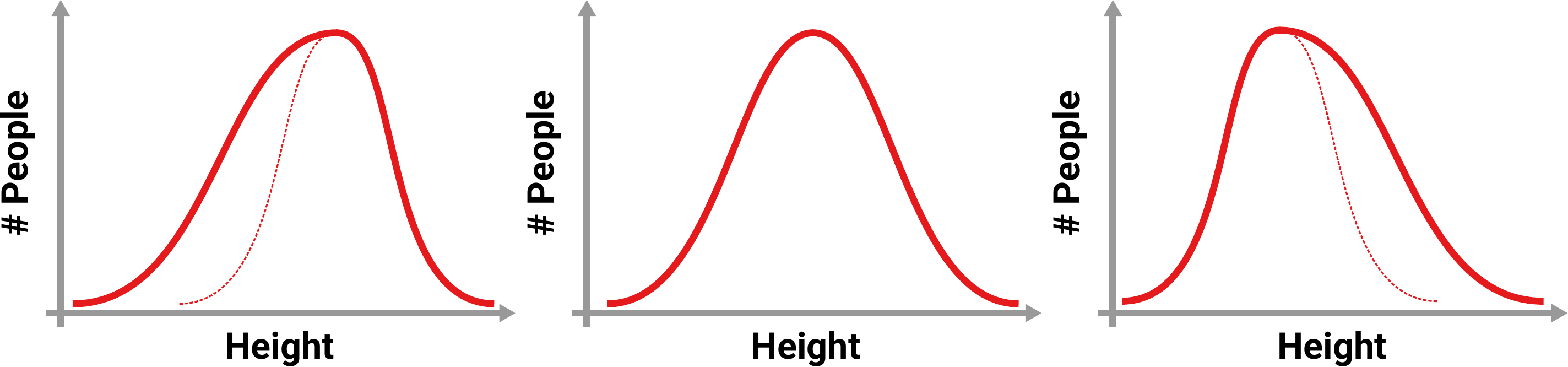}
\caption{Three distributions with different skewness: (a) the values lower than the average are more likely, (b) no skewness, (c) the values higher than the average are more likely.}
\label{fig:skewness}
\end{figure}

Skewness is a property of a distribution, it measures its symmetry. A symmetric distribution has no skewness, and any asymmetry will create a non-zero skewness value: positive if the skewedness is on the right and negative if it is on the left -- see Figure \ref{fig:skewness}.

\begin{figure}[t]
\centering
\includegraphics[width=.8\columnwidth]{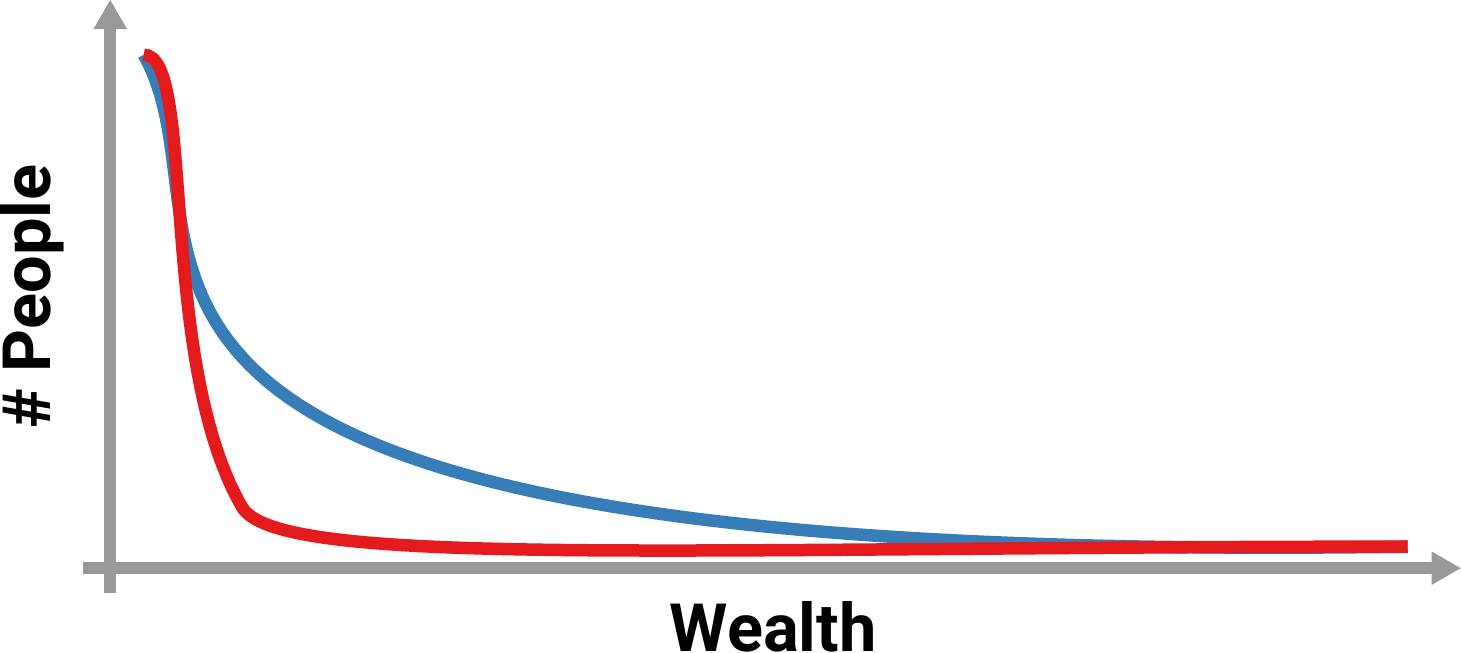}
\caption{A long tail (red) and a fat tail (blue).}
\label{fig:fat-vs-long}
\end{figure}

An important related concept to skewness is the heavy-tailed distribution. There are two types of heavy-tailed distributions that interest network scientists, because they're often observed in real world networks. They are the long tail and the fat tail\footnote{In case you were wondering: yes, statisticians body-shame distributions.}. In a long tail, you can find arbitrarily large outliers: that means the very highest value can be many times larger than the second-highest one. You might know these from the popular concept of the black swan\cite{taleb2007black}. In these kinds of distributions, observations can happen that are much more extreme than anything we have seen so far. You'll see a network example in Chapter \ref{cha:degree}, when we'll see it is common for nodes in real network to have a long tail in the number of connections attached to them. With a fat tail, you still have outliers that can be many times over the average value. However, these outliers are more common and less extreme. Figure \ref{fig:fat-vs-long} shows a graphical example.

If we're talking wealth, a long tail world is a world with a single Jeff Bezos and everybody else works in an Amazon warehouse. In a fat tail world, Jeff might not be quite as rich, but there are a few billionaire friends to keep him company.

\section{Important Distributions}\label{sec:prob-distr}
Chapter \ref{cha:degree} will drill in your head how important distributions are for network science, so it pays off to become familiar with a few of them. First, let's make an important distinction. There are two kinds of distributions, depending on the kinds of values that their underlying variables can take. There are discrete distributions -- for instance, the distribution of the number of ice cream cones different people ate on a given day. And there are continuous distributions -- for instance the distances you rode on your bike on different days. The difference is that the former has specific values that the underlying variable can take (you may have eaten two or three ice cream cones, but $2.5$ is not an option), the latter can take any real value as an outcome. In the first discrete case, we call the distribution a ``mass function''. In the second case, we call it a ``density function''.

Figure \ref{fig:distributions} shows some stylized representations of the most important distributions you should pay attention to, which are:

\begin{figure}
\centering
\includegraphics[width=.66\columnwidth]{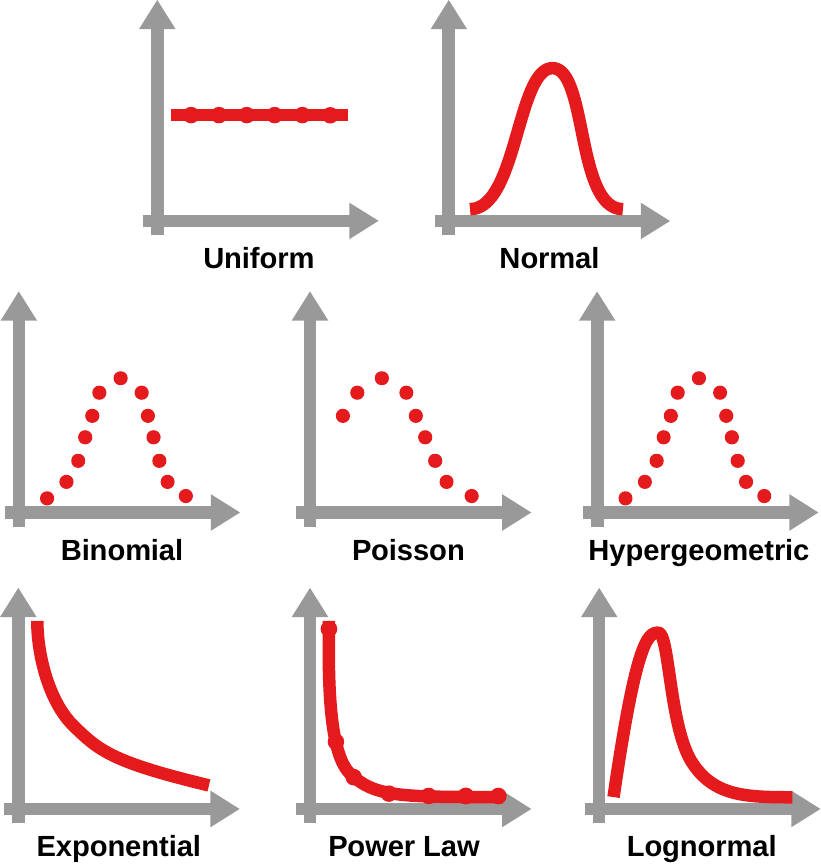}
\caption{A stylized representation of the most common distributions you'll encounter as a network scientist. Solid lines show continuous distributions, while dots show discrete ones -- note that some distributions can be both (dotted lines).}
\label{fig:distributions}
\end{figure}

\begin{itemize}
\item \textbf{Uniform}: in this distribution each event is equally likely. This distribution can be both discrete or continuous. In the discrete case, if you have $n$ possible events, each occurs with probability $p = 1/n$. You get a discrete uniform distribution if you look at the number on a ball extracted from an urn, where all balls in the urn are identified by distinct, progressive numbers without gaps. A continuous uniform distribution could be the amount of time you have to wait for the next drop to come from a leaky faucet that drips once a minute: if you haven't seen the last drop falling, the wait time could be any time between $0$ and $60$ seconds.
\item \textbf{Normal} (or \textbf{Gaussian}): this a very common distribution for continuous variables. The classical example is the distribution of people's heights: most people are of average height, and larger and larger deviations from the average get steadily less likely.
\item \textbf{Binomial}: this is a discrete distribution, in which you do $n$ experiments, each with success probability $p$, and you calculate the probability of having $n'$ successes. For instance, suppose you take five balls from an urn containing $50$ red and $50$ green balls -- each time putting the ball you extracted back into the urn. Let's say we count getting a green ball as a ``success'' here. The number of times you got $0$, $1$, $2$, $3$, $4$ or $5$ green balls over a bunch of trials would follow a binomial distribution. 
\item \textbf{Hypergeometric}: this is yet another discrete probability function. It is very similar to a binomial distribution. Where the binomial described the number of ``successes'' in an extraction-with-replacement urn game, the hypergeometric describes the more common case of extraction-without-replacement. When you extract a ball from the urn, you don't put it back. It is mathematically less tractable, but much more useful. This is used especially for the task of network backboning (Chapter \ref{cha:backboning}).
\item \textbf{Poisson}: this is another discrete distribution, which is the number of successes in a given time interval, assuming that each success arrives independently from the previous ones. For instance, the number of meteorites impacting on the moon each year will have a Poisson distribution. Interestingly, many examples commonly mentioned for explaining a Poisson distribution (number of admittances to a hospital in an hour, number of emails written in an hour, and so on) aren't actually Poisson distributions, because of the ``burstiness'' of human behavior\cite{barabasi2005origin}.
\item \textbf{Exponential}: the exponential distribution is a continuous distribution modeling cases in which the probability of something happening is not dependent of how much time has passed since you starting observing the phenomenon. For instance, if there's an epidemics out, the amount of time you have been infection-free bears no weight in determining your probability of being infected, if exposed. This is the reason why this distribution is sometimes called ``memoryless'' or that it ``doesn't age''.
\item \textbf{Power law}: a power law can be both a discrete or a continuous distribution. It describes the relationship between two quantities, the second quantity changes as a power of the first. One practical consequence is that, if you were given a power law plot without axis labels, you would not be able to tell where you are in the distribution, because the slope of the line always looks the same no matter how much you zoom in or out, or whether you're on the head or the tail. An example of discrete power law is Zipf's law\cite{newman2005power} recording the frequency of words in a document against their frequency rank. We'll see more than you want to know about power laws when talking about fitting degree distributions in Section \ref{sec:degree-pl}.
\item \textbf{Lognormal}: a lognormal distribution is the distribution of a continuous random variable whose logarithm follows a normal distribution -- meaning the logarithm of the random variable, not of the distribution. This is the typical distribution resulting from the multiplication of two independent random positive variables. If you throw a dozen 20-sided dice and multiply the values of their faces up, you'd get a lognormal distribution. It's very tricky to tell this distribution apart from a power law, as we'll see.
\end{itemize}

\begin{figure}
\centering
\includegraphics[width=\columnwidth]{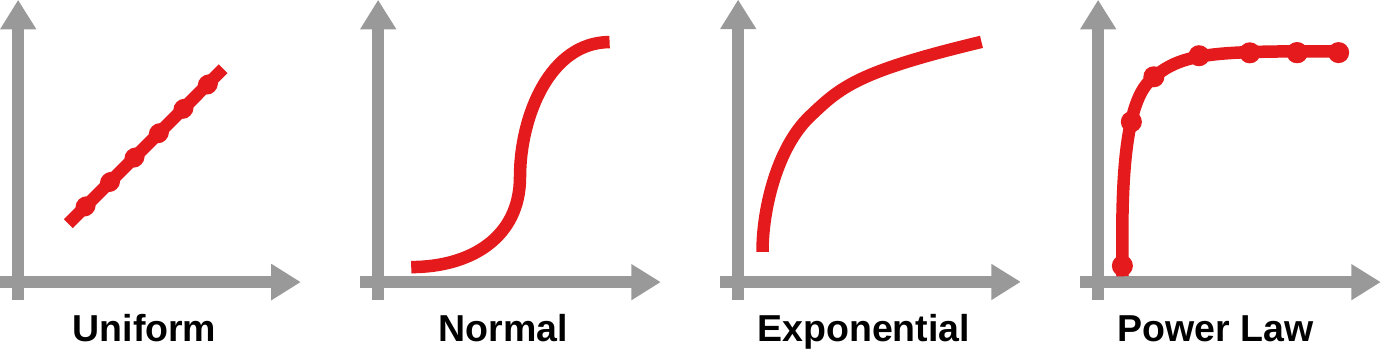}
\caption{A stylized representation of a few cumulative distributions. Same legend as Figure \ref{fig:distributions}.}
\label{fig:distributions-cdf}
\end{figure}

Sometimes, rather than looking at the mass/density functions, it's more useful to look at their cumulative versions. In practice, you want to ask yourself what is the number of -- say -- $x$ or \textit{fewer} successes. Each distribution changes in predictable ways, as Figure \ref{fig:distributions-cdf} shows.

For instance, a cumulative uniform distribution is a  line that goes straight up, because each event adds the same value to the cumulative sum. A cumulative normal distribution has an the shape of a flattened ``S''. In the power law case, as we'll see, we actually want to see the complement of the cumulative distribution ($1$ - CDF). This is, interestingly, also a power law.

\section{p-Values}\label{sec:stats-p}
One of the key tasks of statistics is figuring out whether what you're observing -- a natural phenomena or the result of an experiment -- can tell us something bigger about how the world works. The way this is normally done is making an hypothesis, for instance that a specific drug will cause weight loss. To figure out whether it is true, we need to prove that taking the drug actually does something rather than nothing. ``The drug does nothing'' is what we call the \textit{null hypothesis}, which is what we'd expect -- after all, most drugs don't cause weight loss. This is what we colloquially call ``burden of proof'': the person making the claim that something exists needs to prove that it does, because if we haven't proven that something exists yet there is no reason to believe it does.

We call what you want to prove -- ``the drug causes weight loss'' -- the \textit{alternative hypothesis}, because it's the alternative to the null hypothesis.

p-values are among of the most commonly used tools to deal with this problem\cite{wasserstein2016asa}. The interpretation of p-values is tricky and it is easy to get it wrong. The ``p'' stands for ``probability''. Suppose that you give your drug to a bunch of people and, after a few weeks, you see that their weight decreased by $5$kg. The p-value tells you the probability that you would be observing an effect this strong -- a loss of $5$kg -- if the null hypothesis was true -- i.e. if the drug actually did nothing. Lower p-values mean there is stronger evidence against the null hypothesis. What the p-value does not tell you (but might trick you into thinking it does) is:

\begin{itemize}
\item The p-value does \textbf{NOT} tell you how likely you are to be right;
\item The p-value does \textbf{NOT} tell you how strong the effect is;
\item The p-value does \textbf{NOT} tell you how much evidence you have against your hypothesis.
\end{itemize}

Etch these bullet points into your brain, because it is so easy to fool yourself. The last of them means that a high p-value does not mean that the null hypothesis is true. A high p-value just means that the observations we have are compatible with a world where the null hypothesis is true. But it could also mean that our sample is not big enough to draw firm conclusions. Given how tricky it is to get them right, some researchers have called for not using p-values altogether\cite{hubbard2008p}.

\begin{figure}
\centering
\includegraphics[width=.66\columnwidth]{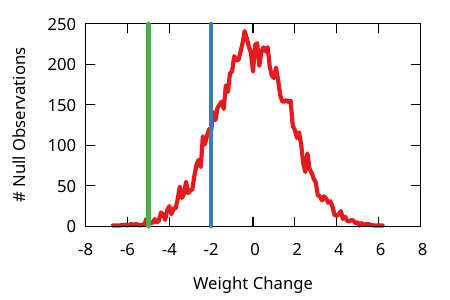}
\caption{In red we have the number of observations (y axis) with a given weight change (x axis) under the null hypothesis. In blue we have an observation of $2$kg weight loss after taking the drug ($p = 0.14$). In green we have an observation of $5$kg weight loss after taking the drug ($p = 0.003$).}
\label{fig:pvalues}
\end{figure}

Figure \ref{fig:pvalues} shows a graphical way to understand the p-value. You have a distribution of values that would be produced by the null hypothesis, you pit your measurements against those values, and the more unusual your observations look compared to those that would be produced by the null hypothesis, the lower the p-value. Exactly how to produce this null hypothesis distribution is not something we'll cover here, because it is not so close to the core of network science -- although we'll see something similar in Section \ref{seg:ergmodels-shuffling}.

One thing is worth mentioning, though. You will often see some magical p-value thresholds that people use as standards -- the most common being $p < 0.05$ and $p <0.01$. These p-value thresholds say something about the strength of the evidence we want to see before we are willing to reject the null hypothesis of no effect. Beware of these. Not only because -- as I said before -- they don't mean what you think they mean, but also because of Goodhart's Law\footnote{``When a measure becomes a target, it ceases to be a good measure.''}. If we say $p < 0.01$ is the gold standard we need to achieve to publish a paper, the natural tendency would be to try and repeat/modify experiments until we get $p < 0.01$. This is know in the literature as p-hacking\cite{head2015extent} and has led researchers to publish a flurry of false results.

If the p-value tells you the probability of observing a given result under the null hypothesis, then if you repeat your experiments $100$ times the probability that at least one of them will leave to a p-value $ \leq 0.01$ is actually $63\%$!\footnote{\url{https://xkcd.com/882/}} That is because the probability of getting a $p > 0.01$ is $99\%$ -- with this standard of evidence, one percent of the time, you will reject the null hypothesis by accident. You try $100$ times, so the formula to know how likely a $p \leq 0.01$ is becomes $1 - (0.99^{100})$. Sometimes, of course, you do need to run more than one test, and look at more than one p-value. A couple of common options to deal with this are applying  the Bonferroni\cite{bonferroni1936teoria}\cite{dunn1961multiple} or the Holm-Bonferroni\cite{holm1979simple} corrections to your p-values, systematically lowering the significance threshold to make up for ``cheating'' by running multiple tests.

p-values are important for network science because we will often have to create null models to figure out whether some property we observe in a network is actually interesting (Section \ref{seg:ergmodels-shuffling}). Another case is figuring out whether an edge has a weight significantly different from zero (i.e. it actually exists) or not (Chapter \ref{cha:backboning}).

\section{Correlation Coefficients}\label{sec:stats-corr}
So far we have worked with a single variable and we have done our best to describe how it distributes. However, more often than not, you have more than one variable and you want to know something interesting about how one relates to the other. For instance, you might want to describe how the weight of a person tends to be related to their height. 

In general, the taller a person is, the more we expect them to weigh. In this sense the two variables \textcolor{cb2}{vary} \textcolor{cb1}{together}. Therefore, we call this concept ``\textcolor{cb1}{co}\textcolor{cb2}{variance}\cite{rice2006mathematical}.'' The formula is pretty simple, it is the mean of the product of the deviations from the mean of both variables, so: $cov(H, W) = \mu((H - \mu(H))(W - \mu(W)))$.

\begin{figure}
\centering
\begin{subfigure}[t]{.3\columnwidth}
\includegraphics[width=\textwidth]{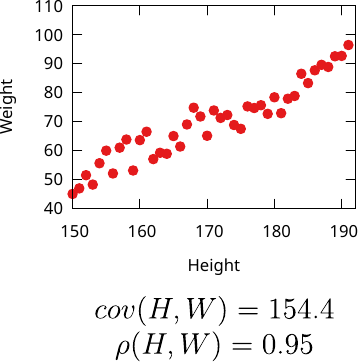}
\caption{}
\end{subfigure}
\quad
\begin{subfigure}[t]{.3\columnwidth}
\includegraphics[width=\textwidth]{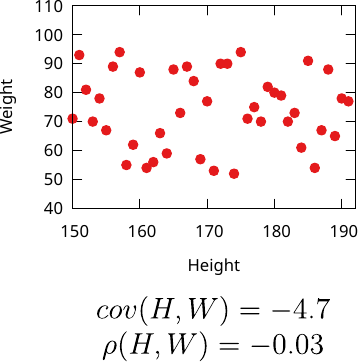}
\caption{}
\end{subfigure}
\quad
\begin{subfigure}[t]{.3\columnwidth}
\includegraphics[width=\textwidth]{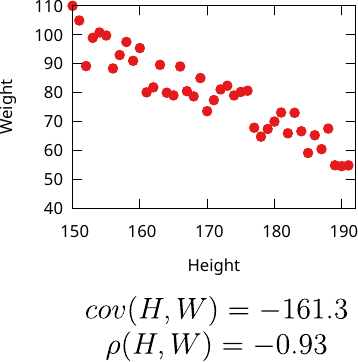}
\caption{}
\end{subfigure}
\caption{Three examples of datasets with (a) positive, (b) no, and (c) negative covariance between height and weight. Covariance ($cov(H, W)$) and correlation ($\rho(H, W)$) values below the scatter plots.}
\label{fig:cov-vs-pearson}
\end{figure}

If a person is both a lot taller than the mean and a lot heavier than the mean, the product of their height and weight deviations will be big, and they will contribute a large positive value to the covariance calculation. If a person is both a lot shorter than the mean and a lot lighter than the mean, the product of their height and weight deviations, both negative, will again be a large positive. So, they will also make the covariance turn out bigger. If someone is both tall and light, their positive height deviation and negative weight deviation will be multiplied to become a large negative number, pulling the covariance down. The covariance will be large in total if we observe many tall/heavy and short/light people, and not so many tall/light or short/heavy people (see Figure \ref{fig:cov-vs-pearson}, the covariance values are in the figure's caption, first row under $cov(H, W)$).

One problem is that the covariance depends on the units of your variables. That means the size of the covariance can be difficult to interpret -- is a covariance of $5.5$ high or low? To make sense of it, it helps to normalize it. This is what the Pearson correlation coefficient tries to do\cite{galton1877typical}. It divides the covariance by the variances of both variables. That means, for example, that if start measuring height in meters rather than centimeters, its covariance with weight will change, but its Pearson correlation with weight will stay the same. The Pearson correlation can only take values between $-1$ (perfect anticorrelation) and $+1$ (perfect correlation). 

\begin{figure}
\centering
\begin{subfigure}[t]{.3\columnwidth}
\includegraphics[width=\textwidth]{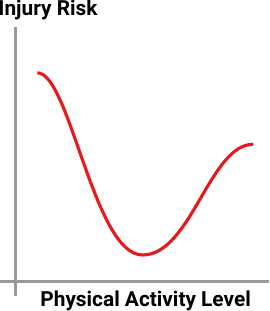}
\caption{}
\end{subfigure}
\quad
\begin{subfigure}[t]{.3\columnwidth}
\includegraphics[width=\textwidth]{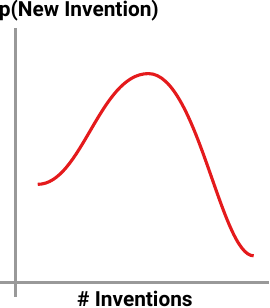}
\caption{}
\end{subfigure}
\caption{Two non monotone relationships (a) U-shaped: if you never exercise you're frail and prone to injury, and if you exercise too much you have more chances to injury yourself. (b) Inverted U-shaped (or A-shaped): if there are no inventions innovation is hard, if there are too many innovations there's nothing left to innovate.}
\label{fig:monotonicity}
\end{figure}

The Pearson correlation coefficient is nice and it is used for many things -- including in network science to predict links (Section \ref{sec:lp-other}), project bipartite networks (Section \ref{sec:projections-vectors}), estimate assortativity (Section \ref{sec:assortquant-plots}), ... you get the idea. There are a couple of problems with the Pearson correlation. The first is that it only estimates how monotone a relationship is: this means that it wants variables to always vary together in the same general direction. Figure \ref{fig:monotonicity} shows a couple of non-monotone relationships -- the classic ``U'' and ``inverted U'' shapes. In this case, even if there clearly is a relationship, Pearson will return zero and there isn't much you can do about it.

The second issue is that, even if the variables have a roughly monotone relationship, Pearson only measures it accurately if this relationship is \textit{linear}. Pearson will give us the wrong idea if increases in one variable are associated with changes in another variable, but less and less so. For example, we visit our friends less often if they live further away from us, but moving $20$ kilometers away if they lived next door before could make a big difference, whereas moving $20$ extra kilometers away if they already lived in another country is not so meaningful. 

\begin{figure}
\centering
\begin{subfigure}[t]{.4\columnwidth}
\includegraphics[width=\textwidth]{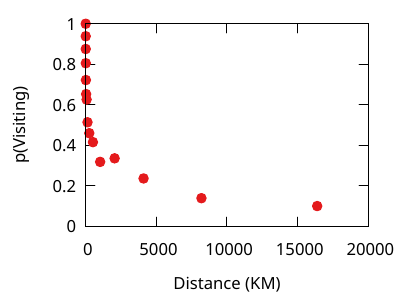}
\caption{}
\end{subfigure}
\qquad
\begin{subfigure}[t]{.4\columnwidth}
\includegraphics[width=\textwidth]{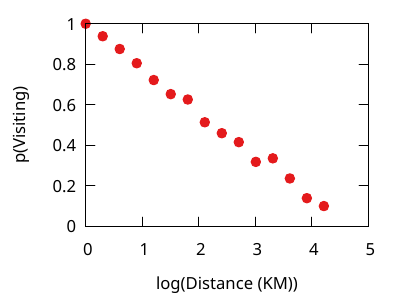}
\caption{}
\end{subfigure}
\caption{Data with skewed distributions across various orders of magnitude. (a) Linear plot. (b) Same data, with the x axis log-transformed.}
\label{fig:log-vs-linear}
\end{figure}

One neat trick you can often do to fix this is to do a log-transformation of your data before you calculate a correlation. In Figure \ref{fig:log-vs-linear}(a) you can see a pair of variables with an non-linear relationship -- the probability of deciding to visit a friend living at a given distance. Once you take the logarithm of the distance (Figure \ref{fig:log-vs-linear}(b)), a linear relationship with the visit probability comes to the surface. Often, you can log both variables.

Most of the times you can stop here, but when this fails you still have one tool that allows to calculate correlations of data related in just about any way. As long as a change in one variable monotonically correspond to a change in the other (so, no U- or A-shapes where the direction of the relationship changes midway), you can summarize this with the Spearman rank correlation\cite{spearman1904proof}. You do not assume anything at all about \textit{how much} each variable changes as the other changes. The Spearman rank correlation is still normalized to take values between $-1$ and $+1$  with the exact same meaning as Pearson.

How can you achieve this? It's actually rather simple. You still calculate a Pearson correlation. But rather than calculating it directly on the values of your variable, you do it on their \textit{ranks}. The observation with the highest value becomes a $1$, the second largest becomes a $2$, and so on. Then, you calculate the Pearson correlation of these ranks. That is why this is called the Spearman \textit{rank} correlation.

\begin{figure}
\centering
\begin{subfigure}[t]{.3\columnwidth}
\includegraphics[width=\textwidth]{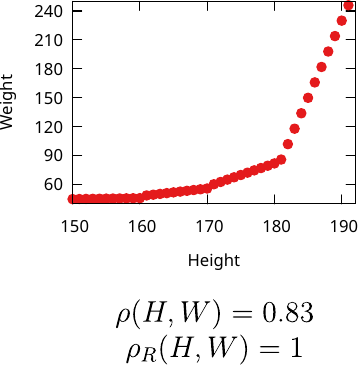}
\caption{}
\end{subfigure}
\quad
\begin{subfigure}[t]{.3\columnwidth}
\includegraphics[width=\textwidth]{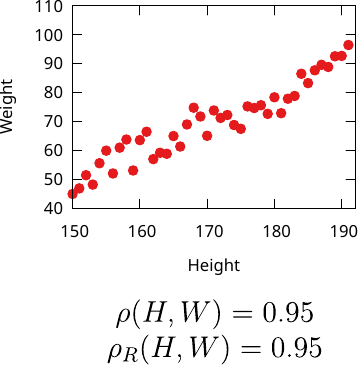}
\caption{}
\end{subfigure}
\quad
\begin{subfigure}[t]{.3\columnwidth}
\includegraphics[width=\textwidth]{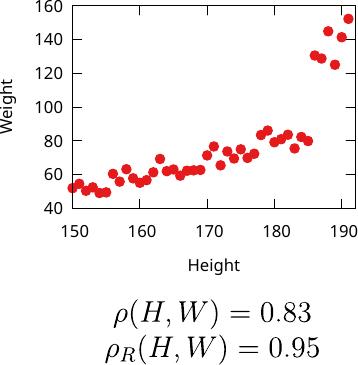}
\caption{}
\end{subfigure}
\caption{Three pairs of variables showing the difference between the Pearson ($\rho(H, W)$) and the Spearman ($\rho_R(H, W)$) correlation coefficients.}
\label{fig:spearman}
\end{figure}

Figure \ref{fig:spearman} provides some examples. You can see what a monotonic but not linear relationship looks like (Figure \ref{fig:spearman}(a)), that for data fitting the Pearson's assumptions Spearman returns comparable values (Figure \ref{fig:spearman}(b)), and the robustness of Spearman to outliers (Figure \ref{fig:spearman}(c)).

\section{Mutual Information}\label{sec:prob-mi}
Mutual Information (MI) is a key concept in information theory. It is another measure of how related two variables are. You can see it as a sort of a special correlation. Formally, MI quantifies how much information you obtain about one variable if you know the value of the other variable. For example, if we know how tall someone is, does this allow us to make a better guess at their weight? How much better? This ``amount of information'' is usually measured in bits.

To understand MI, we need to take a quick crash course on information theory\cite{cover1999elements}\cite{mackay2003information}, which starts with the definition of information entropy. It is a lot to take in, but we will extensively use these concepts when it comes to link prediction and community discovery in Parts \ref{par:lp} and \ref{par:cd}, thus it is a worthwhile effort.

\begin{figure}
\centering
\includegraphics[width=.8\columnwidth]{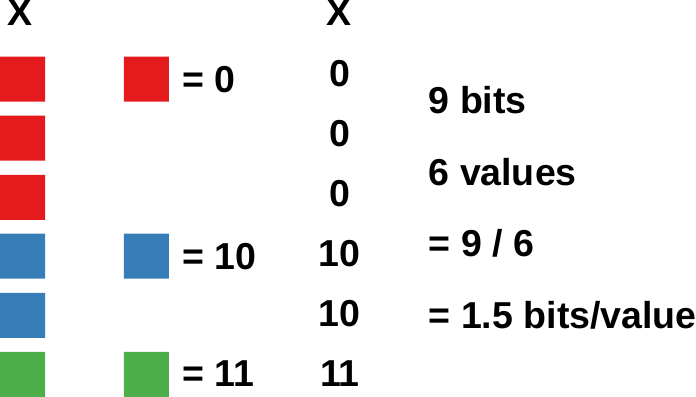}
\caption{A simple example to understand information entropy. From left to right: the vector $x$ has six elements taking three different values. We can encode each value with a sequence of zeros and ones. Doing so allows us to transmit $x$'s six elements using nine bits of information. This means that the number of bits per value is $1.5$.}
\label{fig:mi1}
\end{figure}

Consider Figure \ref{fig:mi1}. The figure contains a representation of a vector of six elements that can take three different values. The first thing we want to know is how many bits of information we need to encode its content.

We can be smart and use the shortest codes for the elements that appear most commonly, in this case the red square. Every time we see a red square, we encode it with a zero. If we don't see a red square, we write a one, which means that we need to look at a second bit to know whether we saw a blue or a green square. If it was a blue square, we write a zero, if it was green we write another one. With these rules, we can encode the original vector using nine bits, i.e. we use $1.5$ bits per element.

This is close to -- but not exactly -- the definition of information entropy. In information entropy, we weigh the probability of an event by its logarithm\cite{shannon1948mathematical}\cite{kolmogorov1965three}.

Consider flipping a coin. Once you know the result, you obtain one bit of information. That is because there are two possible events, equally likely with a probability $p$ of $50\%$. Generalizing to all possible cases, every time an event with probability $p$ occurs, it gives you $-\log_2(p)$ bits of information for... reasons\footnote{The amount of information of an event is a function that only depends on the probability $p$ of the event to happen, e.g. $i_a = f(p_a)$ for event $a$. If we have two events, $a$ and $b$, happening with probability $p_a$ and $p_b$, the event $c$ defined as $a$ and $b$ happening has probability $p_c = p_a p_b$. Now, each event also gives you an amount of information, namely $i_a$ and $i_b$. When $c$ happens, it means that both $a$ and $b$ happened, thus you got both pieces of information, or $i_c = i_a + i_b$. What we just said can be rewritten as $f(p_c) = f(p_a) + f(p_b)$, given the equation at the beginning. Since $p_c = p_a p_b$, then we can also rewrite the equation as $f(p_a p_b) = f(p_a) + f(p_b)$. The only function $f$ that we can possibly plug into this equation maintaining it true is the logarithm. Since probabilities are lower than $1$, the logarithm would be lower than zero, which would be nonsense -- you cannot get negative information. Thus we take the negated logarithm: $i_a = -\log(p_a)$.}. So, the total information of an event is the amount of information you get per occurrence times the probability of occurrence: $-p\log_2(p)$. Summed over all possible events $i$ in $x$: $H_x = -\sum \limits_i p_i\log_2(p_i)$, which is Shannon's information entropy -- how many bits you need to encode the occurrence of all events.

\begin{figure}[b]
\centering
\includegraphics[width=.8\columnwidth]{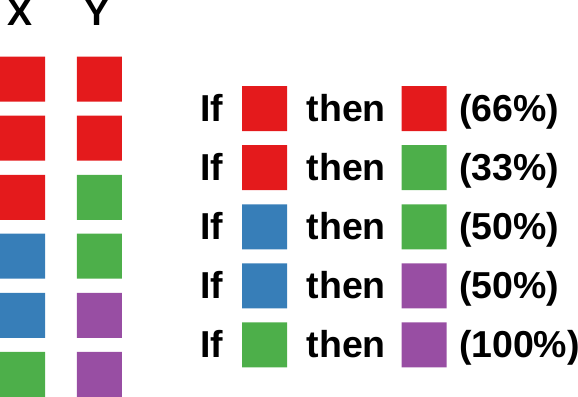}
\caption{An illustration of what mutual information means for two vectors. Vector $y$ has equal occurrences for its values (there is one third probability of any colored square). However, if we know the value of $x$ we can usually infer the corresponding $y$ value with a higher than chance confidence.}
\label{fig:mi2}
\end{figure}

Mutual information is defined for two variables. As I said, it is the amount of information you gain about one by knowing the other, or how much entropy knowing one saves you about the other. Consider Figure \ref{fig:mi2}. It shows the relationship between two vectors, $x$ and $y$. Note how $y$ has equally likely outcomes: each color appears three times. However, if we observe a green square in $x$, we know with $100\%$ confidence that the corresponding square in $y$ is going to be purple. This means that, knowing $x$'s values gives us information about $y$'s value. Mathematically speaking, mutual information is the amount of information entropy shared by the two vectors.

It would take $-\log_2(1/3) \sim 1.58$ bits to encode $y$ on its own (it is a random coin with three sides). However, knowing $x$'s values makes you able to use the inference rules we see in Figure \ref{fig:mi2}. Those rules are helpful: note how our confidence is almost always higher than $33\%$, which is the probability of getting $y$'s  color right without any further information. The rules will save you around $0.79$ bits, which is $x$ and $y$'s mutual information.

The exact formulation of mutual information is similar to the formula of entropy:

$$ MI_{xy} = \sum \limits_{j \in y} \sum \limits_{i \in x} p_{ij} \log \left( \dfrac{p_{ij}}{p_i p_j} \right), $$

where $p_{ij}$ is the joint probability of $i$ and $j$.  The meat of this equation is in comparing the joint probability of $i$ and $j$ happening with what you would expect if $i$ and $j$ were completely independent. If they are, then $p_{ij} = p_i p_j$, which means we take the logarithm of one, which is zero. But any time the happening of $i$ and $j$ is not independent, we add something to the mutual information. That something is the number of bits we save.

\section{Summary}

\begin{enumerate}
\item If you want to know what an average observation looks like in a variable, you might want to calculate the mean or the median.
\item The average observation will deviate from the mean: to estimate how far from the mean it will be on average, you can calculate the variance or the standard deviation.
\item Data can be skewed, meaning it distributes asymmetrically around the mean. There could be long or fat tails, depending how extreme and uncommon outliers are. The more skewed your data is, the more mean and median will disagree.
\item There are important distributions one should know: uniform, where all values are equally likely; normal, where values cluster around the mean; and various skewed distributions such as exponential, lognormal, and power law.
\item A cumulative distribution tells you the probability of observing a a values equal to or smaller than a given threshold.
\item The p-value tells you the probability of making a given observation under the null hypothesis (no change, no effect, nothing interesting is happening). A low p-value can prove the null hypothesis wrong but a high p-value does not prove the null hypothesis is true.
\item If you do multiple experiments, you need to correct the p-values you get otherwise you are going to misinterpret them.
\item When you have two variables, covariance tells you how much the two change together. Correlation coefficients are normalized covariances that do not change depending on the scale of the data. If your variables have a non-linear relationship, you need to use a correlation coefficient that can handle it.
\item Another approach to measure of how related two random variables are is mutual information. It tells you how many bits of information you gain about the status of one variable by knowing the other.
\end{enumerate}

\section{Exercises}

\begin{enumerate}
\item Calculate the mean, median, and standard deviation of the two variables at \url{http://www.networkatlas.eu/exercises/3/1/data.txt} (one variable per column).
\item Make a scatter plot of the variables used in the previous exercise -- with one variable on the x axis and the other on the y axis. Do you think that they are skewed or not? Calculate their skewness to motivate your answer.
\item Draw the mass function and the cumulative distribution of the following outcome probabilities:

\begin{tabular}{l|r}
Outcome & $p$\\
\hline
1 & 0.1\\
2 & 0.15\\
3 & 0.2\\
4 & 0.21\\
5 & 0.17\\
6 & 0.09\\
7 & 0.06\\
8 & 0.02\\
\end{tabular}

\item Which correlation coefficient should you use to calculate the correlation between the variables used in the exercise 2? Motivate your answer by calculating covariance, and the Pearson and Spearman correlation coefficients (and their p-values). Does the Spearman correlation coefficient agree with the Pearson correlation calculated on log-transformed values?
\item How many bits do we need to independently encode $v_1$ and $v_2$ from \url{http://www.networkatlas.eu/exercises/3/5/data.txt}? How much would we save in encoding $v_1$ if we knew $v_2$?
\end{enumerate}

\chapter{Machine Learning}\label{cha:machine-learning}
Machine learning is a collection of data analysis techniques that aim at discovering patterns in data. The main difference between statistics and machine learning is that statistics is ``top-down'' and machine learning is ``bottom-up''. By ``top-down'' I mean that, in statistics, you usually have a theory or hypothesis, you impose it on the data from above, and then you see if it fits. On the other hand, machine learning's ``bottom-up'' means that you instead scout for patterns and correlations in the data that you didn't necessarily know about, and you try to let them arise from the data.

Here's the usual disclaimer -- and I can tell you're already sick of it --: if you already know your machine learning, you can skip this chapter. If, instead, you want to truly learn machine learning, you should check out specialized texts\cite{alpaydin2020introduction}\cite{han2022data} that don't cram superficial explanations in a rushed chapter.

This chapter will not include one part of machine learning, which are evaluation measures: the functions telling you how well you did in your prediction task. There are a couple of reasons why. First, because they fit better in the link prediction part (Part \ref{par:lp}), since they are used mostly for that specific task in network science. The second reason is to highlight their difference with loss functions, which are instead treated here. Ostensibly, evaluation measures and loss functions do the same thing: given an objective, they tell you how close you are to it. However, loss functions are used for \textit{training} while evaluation measures are used for \textit{testing}. These two phases of machine learning -- as we'll see shortly -- are different, they should be as separate as possible, and that is why we should always have different measures and functions to drive them, which is why I create this big divide between evaluation measures and loss functions.

\section{General Structure}\label{sec:ml-general-infrastructure}
\begin{figure*}
\centering
\includegraphics[width=\columnwidth]{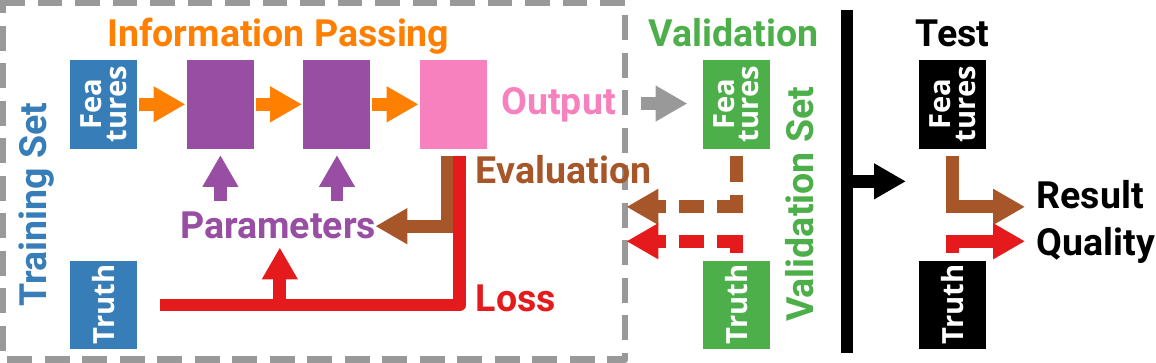}
\caption{A generic machine learning pipeline. Arrows indicate the flow of information. The red arrows and the truth boxes are only present for supervised learning.}
\label{fig:ml-general-infrastructure}
\end{figure*}

To explain machine learning, I'm going to rely on a generic architecture I depict in Figure \ref{fig:ml-general-infrastructure}. This is a bit of a simplification, but for a network science book it will do. It also looks daunting, but it's just a bunch of pipes, and I'm going to make a plumber out of you now. 

At a bird's eye view, you have three phases: \textbf{training}, \textbf{validation}, and \textbf{test}. Here, we deal only with the first two, leaving the test phase for Chapter \ref{cha:lp-experiment} about link prediction.

The objective of the \textbf{training} phase is to make your algorithm learn the patterns in your data. The way you do it is by passing the \textcolor{cb2}{data} into your algorithm governed by a bunch of \textcolor{cb4}{parameters}, possibly through a pipeline of several different operations -- \textcolor{cb5}{passing} the processed information --, and getting out an \textcolor{cb8}{output}. Then you have to figure out how good your output is and there are two ways to do it.

If you already have the right answers for a given problem, you can use them to drive the learning process: the algorithm should learn from that truth and you can estimate how much it is currently \textcolor{cb1}{getting it wrong}. This is \textit{supervised} learning, because we can supervise the algorithm and drive what it learns by using the correct answers.

Otherwise, you might just have a (bunch of) generic quality function(s) you want to \textcolor{cb7}{maximize/minimize}. This is \textit{unsupervised} learning, because the algorithm can roam freely through the parameter space in search of the patterns. Since there are no hard truth to reach in supervised learning, the evaluation functions could look rather different from the loss functions you use for training! One classical case is large language models: your evaluation could be how spooked humans are when reading the output, which is quite difficult to use as a loss function in the training phase -- given that you don't want to ask humans to evaluate millions of outputs.

\begin{figure}
\centering
\includegraphics[width=\textwidth]{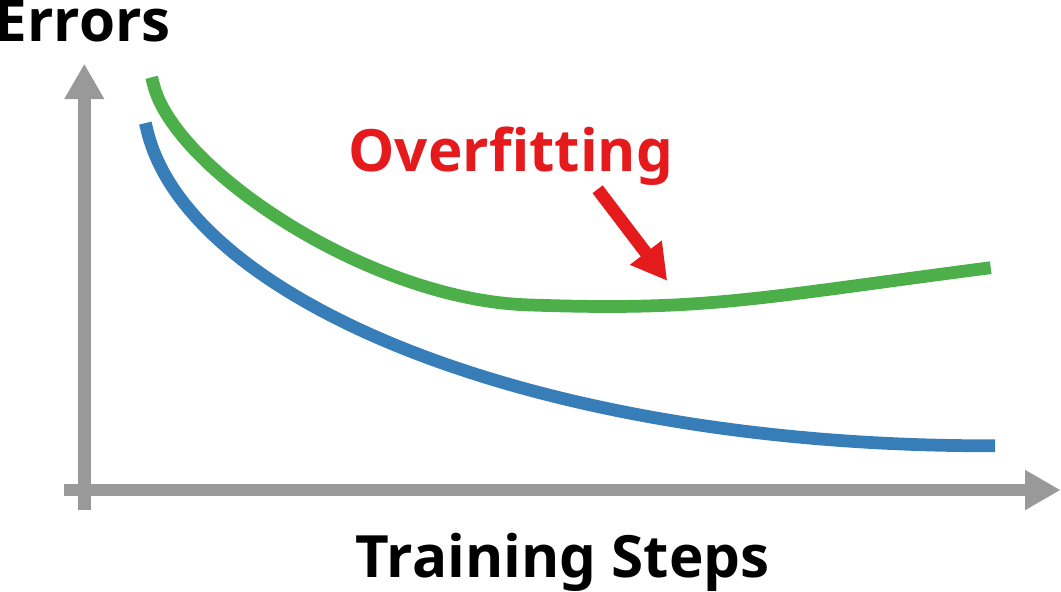}
\caption{Errors (y axis) as a function of training iterations (x axis) in the training (blue) and validation (green) sets. The red arrows highlights the moment in which we stop making progress on the validation set, a classical sign of overfitting.}
\label{fig:overfitting}
\end{figure}

Then there is the \textbf{validation} phase. This looks like the evaluation we did in the training phase. The key difference is that the \textcolor{cb3}{data} we use in the validation phase was never seen by the algorithm in the training phase. We do the validation step to reduce the problem of overfitting\cite{ying2019overview}. Figure \ref{fig:overfitting} shows a classical case of overfitting. It's possible for the algorithm to be getting better and better and better on the training data, but starting to give very poor predictions for new data. If we observe a much reduced quality / much higher errors in the validation set than in the training set, it means the algorithm cannot generalize and we should probably revise something about its architecture.

It is important to know that we never optimize parameters over the validation set data, which is only used once the training phase is over. That is why I box the training phase in the gray outline in Figure \ref{fig:ml-general-infrastructure}. If the validation set tells us we're overfitting, we revise the training phase in isolation and once we're done with training we try again to validate. If you were to accidentally leak information from the validation set in the training phase, you might think you're solving overfitting, but you're actually making the problem worse.

Once you're happy about having fixed your overfitting, you can move to the \textbf{test} phase. Note that no information ever makes it back from here: once you get to testing, it's do or die. Which means that the data in the test set must be separated from both training \textit{and} validation. The performance reports you get from the test phase are final.

There are alternatives to the supervised/unsupervised approach I described. One could do \textit{semi-supervised} learning\cite{chapelle2010semi}, where you have a small portion of annotated data which you can use to kickstart a first phase of supervised learning, before moving to the bulk of unlabeled data and use unsupervised learning that is guided by what you learned in the supervised phase. Another big paradigm is \textit{reinforcement} learning\cite{kaelbling1996reinforcement}\cite{henderson2018deep}, but that matters most for AI and agents and we skip it here. 

Ok, so now I gave you a lot of empty labels that need filling.

\begin{itemize}
\item By data we mean anything we can store in a bunch of numerical vectors. Think of describing a human being as its height, weight, running speed, ... There is no distinction in quality between the data in the \textcolor{cb2}{training set} or the one in the \textcolor{cb3}{validation set} -- except that they must be kept separate at all times. We'll see in Part \ref{par:mining} that it's tricky to do this with networks. In Chapter \ref{cha:mining-embeddings} we'll see methods to reduce complex network structures to numeric vectors, while in Chapter \ref{cha:mining-deep} we'll figure out how to do the machine learning directly on the network structure.
\item The \textcolor{cb4}{parameters} are the things regulating how the algorithm works. If you're trying to predict how weight influences height, your algorithm could be simply to multiply the height with a given number. That number is the parameter.
\item By \textcolor{cb5}{information passing} we mean the mechanism to inform a phase of your pipeline with the results of the previous phase -- or to sum up the results in an \textcolor{cb8}{output}. This is the job of activation functions, which we see in Section \ref{sec:ml-activation}.
\item The errors can be estimated using a loss function, the topic of Section \ref{sec:ml-loss}. Loss functions can help both if you have \textcolor{cb1}{errors on your true answers}, or you're trying to \textcolor{cb7}{minimize/maximize a given outcome}.
\end{itemize}

\section{Activation Functions}\label{sec:ml-activation}
Activation functions take the output of an arbitrary analysis and transform it so that you can do either of two things with it. You can either summarize all the information you have to make a choice, or you can pass on information for further processing. We need activation functions to ensure that the data passing through them assumes some convenient properties that will facilitate either of those two tasks\cite{nwankpa2018activation}. We'll see a couple of examples that fall in either class.

\subsection{Summarizing Information}
To summarize information, an activation function should take an arbitrary real number as an input and return a number that you can interpret, roughly as the probability or degree of belonging to a certain class. To understand what that means, let's take a look at a popular option, the softmax function\cite{bridle1990probabilistic}:

$$ softmax(X)_i = \dfrac{e^{x_i}}{\sum \limits_{x_j \in X} e^{x_j}}, $$

where $X$ is a vector of numbers and we want normalize its $i$th entry.

\begin{figure}
\centering
\includegraphics[width=\textwidth]{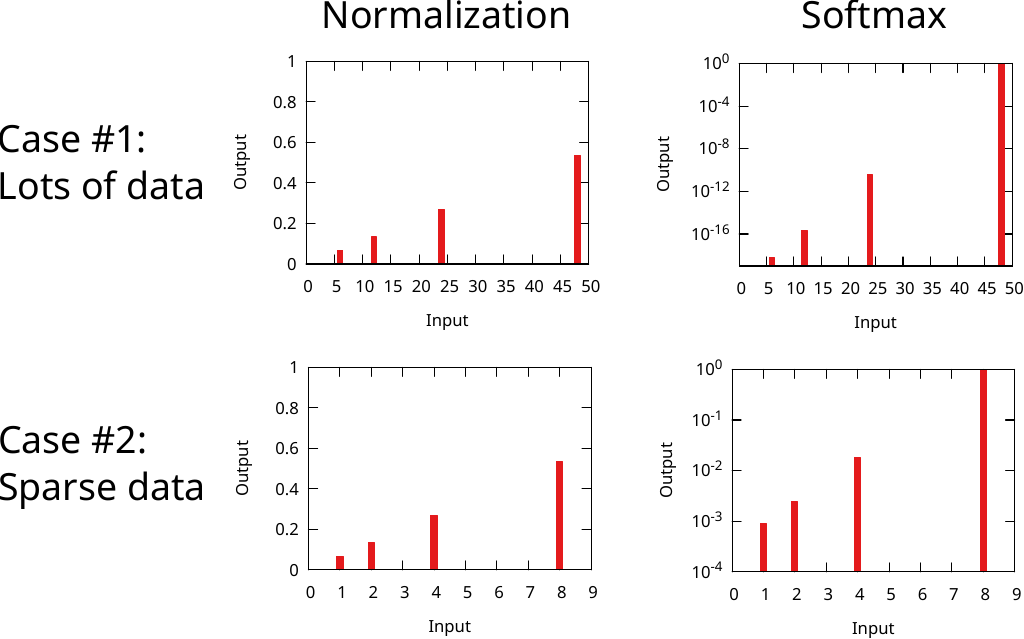}
\caption{Different activations for plain normalization (left column) and softmax (right columns) for different scenarios with data richness (top row) or sparsity (bottom row). Each plot shows the activation function value (y axis) for a given input (x axis).}
\label{fig:activation-softmax}
\end{figure}

The softmax function is useful because it can transform arbitrary vectors of real numbers into probabilities, which is convenient if you're doing a classification task with multiple possible outcomes. But why do we need such a complex function? Why all the exponentiations? Can't we just divide $X$ by its sum and call it a day? Well, let's compare what happens if we try to do that for a network task with both softmax and this normalization approach -- which I show in Figure \ref{fig:activation-softmax}.

Suppose you want to classify people into social circles or communities (Chapter \ref{cha:cd-partitions}). A totally legit way to do it is by counting the number of friends you have who we already sorted in a group. The more of your friends belong to community $c_1$ the more likely it is that you're part of $c_1$ yourself. If you have four communities, for a given person $v$ you might have these membership counts: $[48, 24, 12, 6]$ -- meaning you have $48$ friends in the first group, and so on. A simple normalization would give you a probability of $53\%$ of belonging to the first community, while softmax will be almost certain at more than $99.9\%$ -- this is the top row of Figure \ref{fig:activation-softmax}.

The reason is that simple normalization applies a frequentist approach, while softmax is more Bayesian (see Section \ref{sec:prob-freq-vs-bayes}): each friend in the first community is additional evidence you're part of that community. Consider what happens if we decrease all counts: $[8, 4, 2, 1]$. For the normalization approach nothing changes, because the proportions are the same. But this person has much fewer friends, so the Bayesian will admit more ignorance and give ``only'' a $97.8\%$ probability of belonging to the first community. This is the bottom row of Figure \ref{fig:activation-softmax} -- and pay attention to the differences in the y axis scales.

Another argument for softmax over normalization is that softmax can take a vector with negative numbers and still return probabilities. For instance passing $[4, 1, -1, -4]$ through softmax gives (approximately) $[0.946, 0.047, 0.006, 0.001]$. Passing the same vector to a plain sum normalization returns instead $[W, T, F, ?]$.

\begin{figure}
\centering
\includegraphics[width=\textwidth]{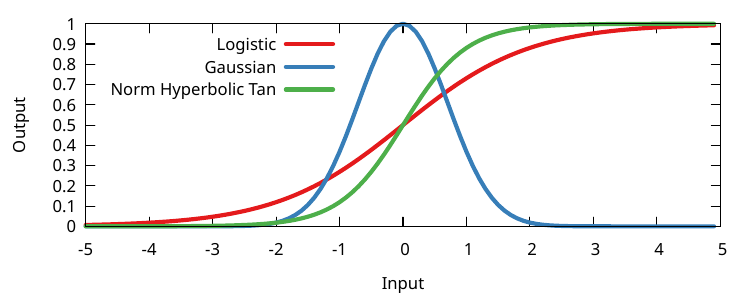}
\caption{Different activations functions (line color): the activation function value (y axis) for a given input (x axis).}
\label{fig:activation-bounded}
\end{figure}

Softmax is nice for classification tasks with multiple options, but more often than not you just want to know whether an output is or isn't a particular thing -- i.e. you're doing a binary classification. For instance, in link prediction (Chapter \ref{cha:lp-simple}), either a link exists or it doesn't. Doing softmax in this case is an overkill and you can use other popular activation functions such as the logistic\cite{gershenfeld1999nature} (or sigmoid), the Gaussian, and the hyperbolic tangent\footnote{Incidentally, ``hyperbolic tangent'' also describes well my writing style.}, which I show in Figure \ref{fig:activation-bounded}. Logistic is actually equivalent to softmax if softmax gets only two options. Note that, in the figure, I renormalize the hyperbolic tangent to be between $0$ and $1$, rather than its normal $-1$ to $+1$ interval.

For the logistic, the higher the value going in, the more likely it is to be an example of a positive class. For the Gaussian, we instead want to be close to a given value, usually zero -- which can be interpreted as ``the class is positive if its distance from a reference point is low''. The hyperbolic tangent is not interpretable as a probability, but you can still normalize it if you want -- as I did just to plot this figure.

\subsection{Passing Information}
\begin{figure*}[t]
\centering
\begin{subfigure}{.19\columnwidth}
\includegraphics[width=\textwidth]{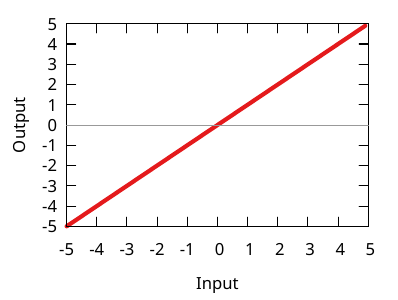}
\caption{Linear}
\end{subfigure}
\begin{subfigure}{.19\columnwidth}
\includegraphics[width=\textwidth]{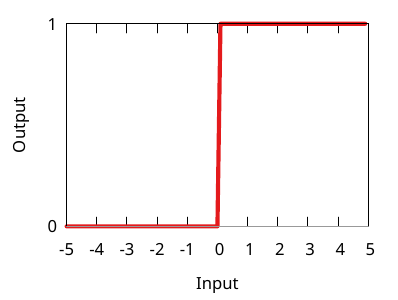}
\caption{Step}
\end{subfigure}
\begin{subfigure}{.19\columnwidth}
\includegraphics[width=\textwidth]{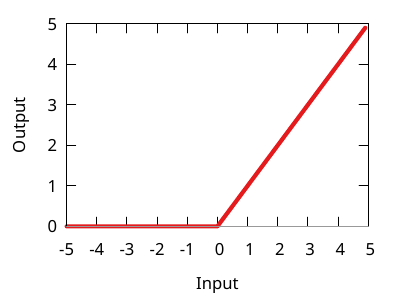}
\caption{ReLU}
\end{subfigure}
\begin{subfigure}{.19\columnwidth}
\includegraphics[width=\textwidth]{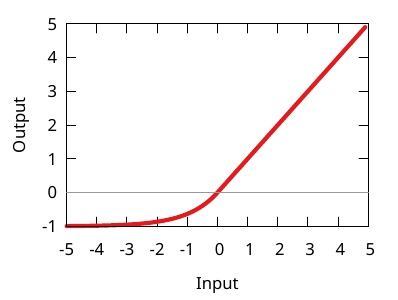}
\caption{ELU}
\end{subfigure}
\begin{subfigure}{.19\columnwidth}
\includegraphics[width=\textwidth]{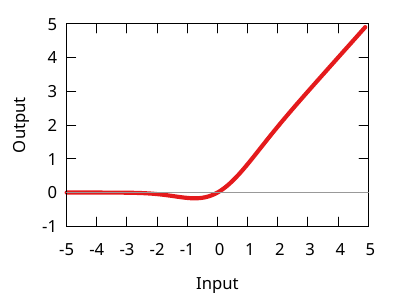}
\caption{GeLU}
\end{subfigure}
\caption{Five examples of activation functions passing information in a machine learning framework. Input of the function on the x axis and output on the y axis. I draw an horizontal gray line to show where the zero output is, to contextualize and compare the outputs.}
\label{fig:activation-passing}
\end{figure*}

Not to go onto another hyperbolical tangent, I'll keep this section barebone and mention that the activation functions we're dealing with here want to transform their input to make it more useful for a further analysis. In this case, activation functions should augment the informative character of their input and/or prevent problems down your analytic pipeline -- fancy stuff that has esoteric names used to evoke a sense of wonder and mystery such as ``vanishing gradients\cite[0.25in]{hochreiter1998vanishing}''. Let's focus on a handful of activation functions, which I show in Figure \ref{fig:activation-passing}.

The Linear activation function does nothing at all: what you pass in goes out. The Step activation function tests whether the input is positive or negative, and returns a binary output. If you want to go all fancy and impress people at cocktail parties, you can call it the Heaviside Step Function -- and enjoy the \textit{uuhh}s and \textit{aaah}s.

The Rectified Linear Unit\cite{fukushima1969visual} -- ReLU for its friends -- is the star of the bunch and probably the most used one\cite{ramachandran2017searching}, since it works so well in neural networks\cite{glorot2011deep}. If the input is negative it returns zero, and if it is positive ReLU is like the Linear function and it does nothing. In practice, $ReLU(x) = \max(0,x)$. ELU\cite{clevert2015fast} and GeLU\cite{hendrycks2016gaussian} are slight modifications of ReLU with more complex formulas, but they follow ReLU's approach. 

\section{Loss Functions}\label{sec:ml-loss}
\begin{figure}[t]
\centering
\includegraphics[width=\textwidth]{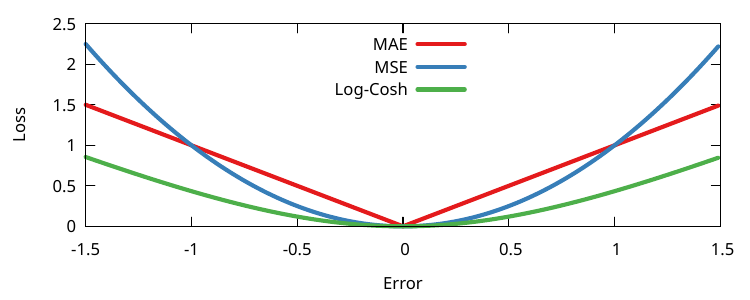}
\caption{Different loss functions (line color): the loss function value (y axis) for a given error (x axis).}
\label{fig:loss-functions}
\end{figure}

Whenever you're training your predictor, you're going to get things wrong. If you're never wrong during training, either your problem is so trivial you don't need machine learning, or something fishy is going on and there must be a mistake somewhere. The role of a loss function is to drive the training process to minimize its mistakes. Once you have a loss function, you can compare two potential updates in the training phase to pick the best one. Training means nothing else than trying something, calculate the loss function, try something else and, if the loss now is lower, keep the change and keep going in that direction. It follows that the loss function is as important as the activation function that propagates information. I show you a few examples of loss functions in Figure \ref{fig:loss-functions}.

The reason why you have many loss functions is because you might want to treat differently different types of errors. For instance, you can see how the MSE function in Figure \ref{fig:loss-functions} -- which I'm about to define -- shoots up for large errors. This means it is sensitive to outliers, which are expected to produce large errors. An alternative, such as Log-Cosh will give less weight to these outliers, producing a lower loss for the same error level\cite{jadon2020survey}.

Again, be reminded of the difference between loss and errors. Loss is something you care about minimizing during training, and loss functions should be generic enough not to lead to overfitting. Errors are an evaluation in test phase on which you cannot and should not minimize, only hope you get them as small as possible. When it comes to error functions, you can be as specific as you want, because you already passed the phase in which overfitting was a concern.

\subsection{Mean Errors}
Two classical examples are mean absolute error (MAE) and mean squared error (MSE). In MAE, if $y_i$ is the real outcome and $\bar{y}_i$ is what your method says, then you average the absolute value of the difference\cite{willmott2005advantages}:

$$MAE(\bar{Y}) = \sum \limits_i |y_i - \bar{y}_i| / |Y|,$$

where $\bar{Y}$ is your vector with all your answers and $Y$ is the vector with the corresponding correct answers.

It should be clear why we need the absolute value: if you were to make symmetric errors (for each overestimation you also make, for a different observation, an equal underestimation) your loss would be zero -- confusing symmetric errors with no errors at all. In MSE you solve the same issue of symmetric errors by taking not the absolute value, but the squared error\cite{bickel2015mathematical} (which is always positive, also for underestimates):

$$MSE(\bar{Y}) = \sum \limits_i (y_i - \bar{y}_i)^2 / |Y|.$$

\begin{figure}
\centering
\includegraphics[width=\textwidth]{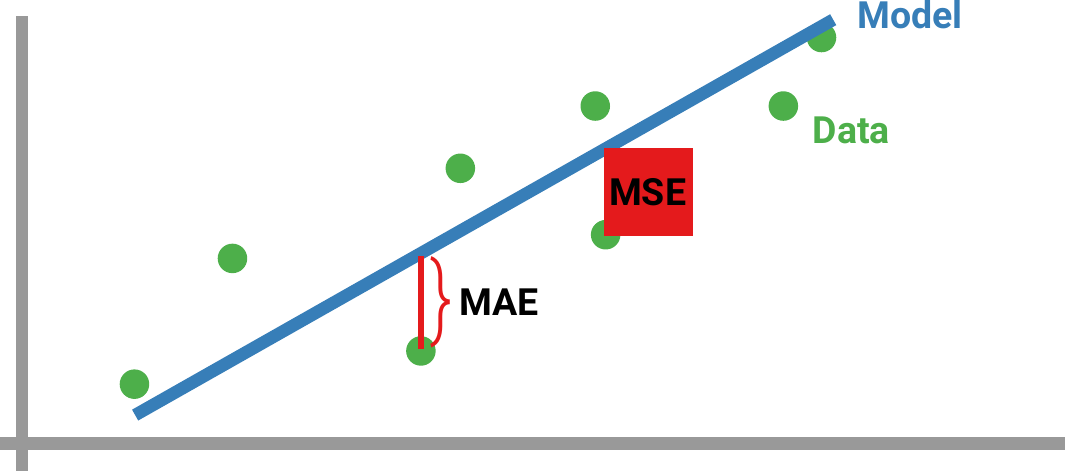}
\caption{Given a model (blue) for data points (green), we have two ways to interpret the errors (red).}
\label{fig:mae-vs-mse}
\end{figure}

If you find it distasteful to take the square because then you have an error in different units than your observation, you can always take the square root of the result and you have the root mean squared error -- just like you can take the root of the variance to get the standard deviation (Section \ref{sec:stats-summary}). Figure \ref{fig:mae-vs-mse} gives you a graphical representation of how the errors are calculated.

In Figure \ref{fig:mae-vs-mse}, for MAE, only the length of the segment matters. For MSE, you instead consider the entire area of the square.

\subsection{(Log) Likelihood}
The likelihood function is a function that tells you what is the probability of observing an event given a (set of) parameter(s) regulating such an event\cite{edwards1972likelihood}. Notation-wise, we use $\mathcal{L}$ to indicate the likelihood function, $x$ is the outcome of the event, and $\theta$ is the set of parameters.

\begin{figure}[t]
\centering
\includegraphics[width=\textwidth]{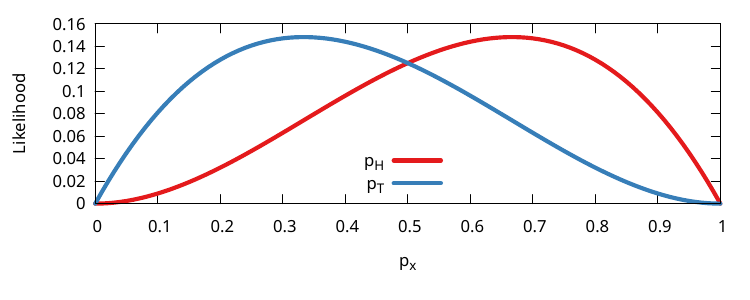}
\caption{The likelihood (y axis) of the parameter $p_H$ (x axis) for the $\{H, H, T\}$ event.}
\label{fig:log-likelihood-ex1}
\end{figure}

Let's make a simple example. Suppose that we are tossing a coin three times. The sides on which it lands, let's say heads twice and tails once, is our $x = \{H, H, T\}$. $\theta$ tells us the parameter regulating the coin. A coin is a fairly simple system, so it has only one parameter: the probability of the coin landing on heads -- which is $50\%$ when we know nothing about the coin and whether it is fair. So we can say $\theta = \{p_H = 0.5\}$. At this point we can estimate $\mathcal{L}(\theta, x)$ -- which, in our case, is $\mathcal{L}(\{p_H = 0.5\}, \{H, H, T\})$ -- which is the likelihood of the coin being fair given that we observe that given event. In this case, to know the $\mathcal{L}$ value given the event for any $p_H$ value we need to evaluate the formula $p_H^2(1-p_H)$ -- because we got two heads and one tails in entirely independent events and so we multiply the probabilities (Section \ref{sec:prob-axioms}). This is what Figure \ref{fig:log-likelihood-ex1} shows in red.

So the likelihood of $p_H = 0.5$ is $0.125$, because that's the value of $0.5^2(1-0.5)$. That's not the most likely value, since we got two heads. The most likely value is $p_H = 0.66$, or $0.2178$. Note that this is all symmetric to $p_T$ since $H$ and $T$ and independent, mutually exclusive, and one is one minus the other. The figure also shows the likelihood of $p_T$, in blue, for the same data, which is symmetric to $p_H$, being $p_T(1-p_T)^2$.

Note that $\mathcal{L}$ is NOT the probability that the coin is fair given the experiment result, because to reach that conclusion you should apply Bayes's theorem (Section \ref{sec:prob-bayes}). What you observe is $P(x|\theta)$ -- the probability of the event given the fairness --, but that is NOT equivalent to $P(\theta|x)$ -- the probability of the fairness given the event. So don't make this mistake when looking at the results of the likelihood function.

You can see from Figure \ref{fig:log-likelihood-ex1} why likelihood is useful: by looking at its value you can compare different parameter values and $\mathcal{L}$ will tell you which one is more likely to be accurate -- given the data you have. So you can use $\mathcal{L}$ to drive your the training phase. Of course, real world problems are much harder than figuring out how loaded a coin is, so it is not this trivial to fine tune your parameters with the likelihood function -- also because you might have many parameters and so you're trying to explore a multidimensional space, which here is monodimensional, since we only have $p_H$ to play with.

\begin{figure}[t]
\centering
\includegraphics[width=\textwidth]{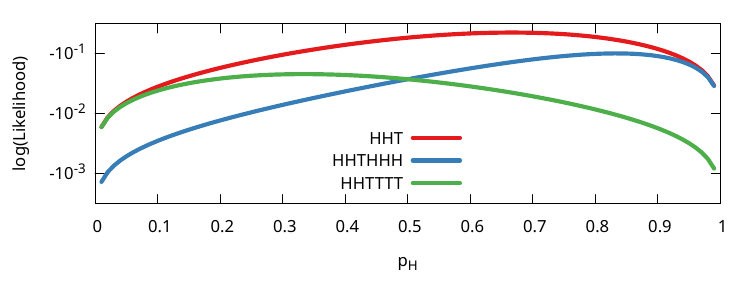}
\caption{The log-likelihood (y axis) of the parameter $p_H$ (x axis) for different events (line color).}
\label{fig:log-likelihood-ex2}
\end{figure}

For practical reasons, it is common to use $\log(\mathcal{L})$, i.e. to take the logarithm of the likelihood function rather than $\mathcal{L}$ itself. The reason is that, as you saw above, to calculate the likelihood we need to make a lot of multiplications. If we instead have log-likelihood we only have to do sums, because the logarithm of the product of two quantities is the sum of their logarithms. Sums are faster than products, and so this is convenient computationally. Figure \ref{fig:log-likelihood-ex2} shows a couple of log-likelihood functions for different events with our coin.

Since we're taking the logarithm of the likelihood function, which is always below $1$, the log-likelihood is always negative. In the figure, we can see that, when we get additional data, the log-likelihood function changes -- humping before for $p_H = 0.5$ if we get a lot of tails (meaning heads is less likely), or after if we keep getting heads.
 
\subsection{Cross Entropy}
\begin{figure}
\centering
\includegraphics[width=\textwidth]{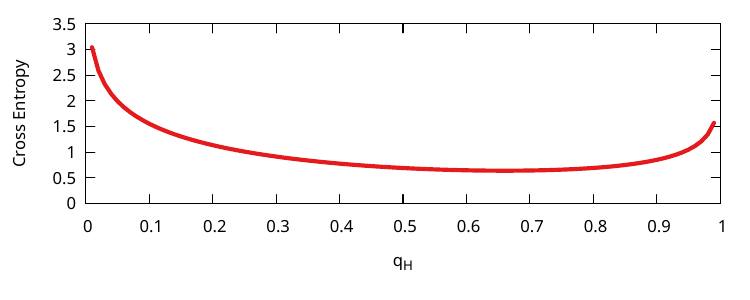}
\caption{The cross entropy (y axis) of our $q_H$ guess (x axis), given that the real $p_H = 0.66$.}
\label{fig:cross-entropy}
\end{figure}

In Section \ref{sec:prob-mi} I introduced the concept of information entropy, which is how many bits you need to encode the occurrence of all events. The formula is  $H_x = -\sum \limits_i p_i\log_2(p_i)$, with $p_i$ being the probability of event outcome $i$. If we have two different probability distributions, let's call them $p$ and $q$, we can calculate their cross entropy as $H_x = -\sum \limits_i p_i\log_2(q_i)$. This can be used as a loss function\cite{zhang2018generalized}\cite{mao2023cross}, because you can assume that $p$ and $q$ are describing the same events, but $p$ describes the real probabilities you're trying to predict and $q$ describes instead the probabilities you get out of your model. Figure \ref{fig:cross-entropy} shows how cross entropy works in our simplified example of guessing the $p_H$ of a loaded coin.

For the figure, I set $p_H = 0.66$. Then our classifier needs to spit out the $q_H$ it believes being the closest to the real $p_H$. The closest $q_H$ is to $0.66$, the lower cross entropy is. You can see how cross entropy shoots up the farther we get from our mark, while being tolerant of smaller mistakes -- the function is pretty flat around $0.66$.

\section{Dealing with Computational Complexity}\label{sec:ml-sampling}
One last important thing I should point out about machine learning is that it is normally done on a truckload of data. If you don't have a lot of observations, the appeal of machine learning is not that great. The more data points you have, the more likely it is you're going to discover whatever pattern lurks in them. Otherwise, the patterns might be overpowered by whatever other random fluctuation affects the observations. This introduces the problem of computational complexity. If you're going to do machine learning, you need to do it quickly, to process large amounts of information in a short time.

This is a problem because, even if the operations you perform are simple, you need to do them for each observation, potentially billions of times, which will take time. There are two main solutions to this issue: sampling and batching.

\subsection{Sampling}
In sampling, you decide not to perform your operation on all data points, but on a selection of them. That selection is a sample. The key problem to solve here is to get a representative sample: if all data points in your sample are ``weird'' in the same way, you might discover a pattern that is not present in your overall dataset. There are a few techniques to ensure that your sample is representative, but I'm not covering them here in details. That doesn't mean that you should avoid learning how to sample, since it's quite important\cite{wu2017sampling}. What I will say, though, is that sampling complex networks is its own variation of the problem, one we will look at in Chapter \ref{cha:sampling}.

\begin{figure}
\centering
\includegraphics[width=.4\textwidth]{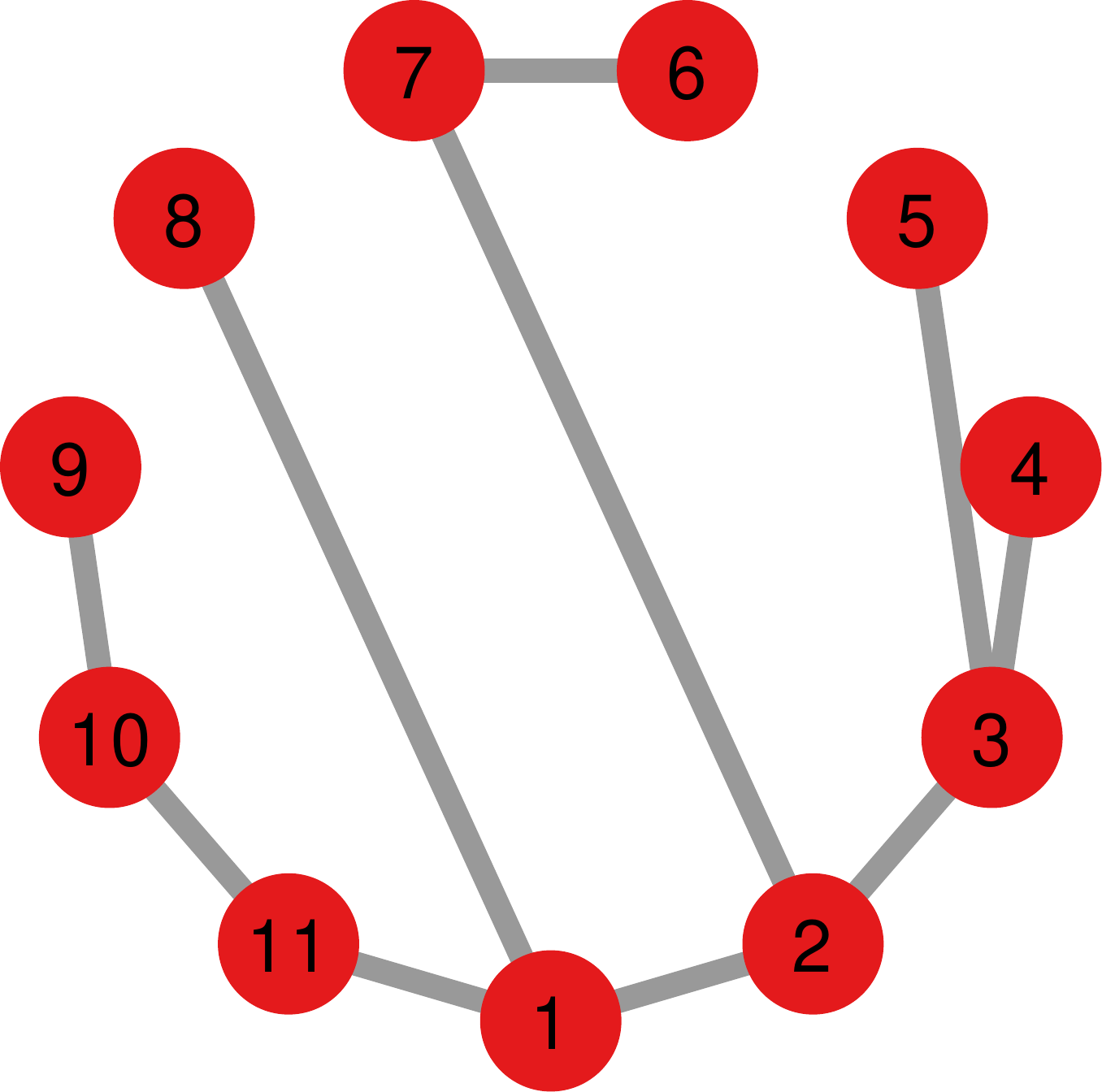}
\caption{A network with fewer connections than non-connections.}
\label{fig:ml-negative-sample}
\end{figure}

Of particular interest is a specific type of sampling called \textit{negative} sampling\cite{yang2020understanding}. Here a network example is helpful, let's look together at Figure \ref{fig:ml-negative-sample}. I know that I haven't formally introduced networks yet, but hopefully you can follow with some intuition.

Suppose we want to figure out what causes connections between nodes in Figure \ref{fig:ml-negative-sample}. We should collect various features and compare connections with non-connections. The problem is that the number of non-connections dwarfs the number of connections. We have $10$ connections, but -- trust me on this -- a whopping $45$ pairs of nodes that are not connected. And this is a super tiny network! Imagine what would happen if you had an actually large one. You won't be able to look at all the \textit{negative} instances -- the non-connections. That is why you need to \textit{sample} them, usually to have as many negative samples as observations -- so here you'd pick $10$ random non-connected node pairs. This is negative sampling.

\subsection{Batching}
Ok, now we've cut the potentially gigantic number of negative observations with negative sampling. Are we good? Not really. What you have now is twice the number of data points -- for each data point you have a companion generated via negative sampling. This can still be pretty huge, if you have a lot of data points -- as you should. What now?

Now, batching\cite{bjorck2018understanding}. In many common machine learning infrastructures, it is much faster -- and it requires less memory -- to run the training process on a small portion of the data at a time. Once you learn your parameters with this first chunk, you restart the learning process, updating the parameters with another portion of the data. Rinse and repeat until you have used all of your data. These chunks are the batches. Sometimes, it makes sense to use really small batches, and we have a cute pet name for them: minibatches\cite{goyal2017accurate}.

The advantages of batching are faster computation and less memory required. This might come at a cost of accuracy, so batching should be done with care -- specifically when choosing the batch size.

To remember the difference between batching and sampling, you can summarize them in your head as follows. If you only do sampling, you train once on part of the data. If you do only batching, you train multiple times on all your data, one chunk at a time.

Of course, you might end up doing both sampling and batching.

\section{Summary}

\begin{enumerate}
\item Machine learning is a set of techniques to discover patterns from data without necessarily knowing in advance what they might be. Generally, you have an algorithm with some parameters and you learn them through a training phase.
\item In the training phase, you might have the right answers you're interested in finding and your algorithm can learn from them. This is supervised learning. If you don't have them, you'll be doing unsupervised learning.
\item During training you might overfit, i.e. learn the odd peculiarities of the training set rather than the general patterns in the data. Having a separate validation set can help you to spot overfitting.
\item To update the parameters and producing an output you need to pass information about the data. This is the job of the activation functions: they take a generic signal and return one that satisfies some properties we care about -- e.g. to make it into an interpretable probability.
\item Loss functions drive your training phase by telling you how far from the objective your method is. They should not be confused with error functions, which are used in testing phase to tell you how wrong your final answers are -- and, since they are final, they cannot be changed, unlike training outputs.
\item Machine learning operates on large datasets. To increase efficiency you can sample your data points, learning only on a part of them. You can also do batching, updating your parameters in the training phase on one small chunk of your data at a time.
\end{enumerate}

\section{Exercises}

\begin{enumerate}
\item Generate a random vector with $100$ normally distributed random values (with zero average and standard deviation of one). Implement the softmax function. Plot the result with the original vector on the x axis and the softmax output on the y axis.
\item Generate a random vector with $100$ normally distributed random values (with zero average and standard deviation of one). Implement the ReLU function. Plot the result with the original vector on the x axis and the ReLU output on the y axis.
\item Generate a random vector with $100$ normally distributed random values (with zero average and standard deviation of one). Implement the MAE and MSE functions and compare their outputs when applied to the vector, by plotting each of them.
\item Plot the likelihood function for $p_H$ and $p_T$ for the events $\{H, H, T, H, T\}$.
\end{enumerate}

\chapter{Linear Algebra}\label{cha:la}
The final piece of groundwork we need to complete the foundations of network science is linear algebra. You will learn in Chapter \ref{cha:mat} that there are many different ways to represent a network as a matrix. Linear algebra is necessary to understand what sort of operations you can perform on those matrices, and what they mean.

Normally, in chapter preambles I provide you some generic academic references that cover the topic as a whole, pointing out that they are a much better and more complete resources than what I write. While that's certainly the case here as well\cite{strang1993introduction}\cite{meyer2000matrix}\footnote{\url{https://ocw.mit.edu/courses/mathematics/18-06-linear-algebra-spring-2010/}}, I'd say that my strongest recommendation by far does not come from academia, but from Youtube. The linear algebra series from 3blue1brown\footnote{\url{https://www.youtube.com/playlist?list=PLZHQObOWTQDPD3MizzM2xVFitgF8hE_ab}} is, quite possibly, the best and most intuitive introduction to linear algebra that I had the pleasure to consume.

\section{Vectors}\label{sec:la-vector}
The reason why linear algebra is called linear algebra is because it is a collection of algebraic techniques to solve systems of linear equations. However, we're not going to use this perspective, because using a \textbf{spatial} perspective is more intuitive and closer to what we're actually going to do with linear algebra -- which is to deal with networks.

\begin{figure}
\centering
\begin{subfigure}[t]{.4\columnwidth}
\includegraphics[width=\textwidth]{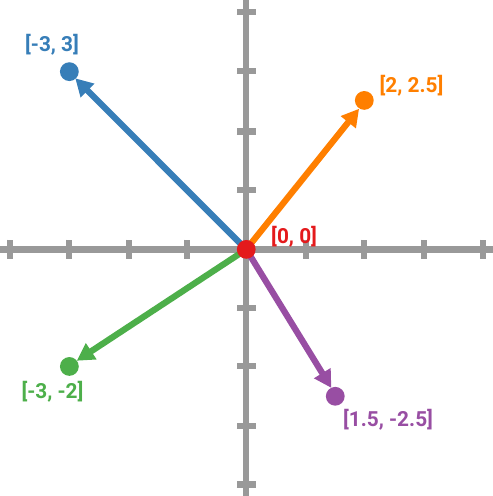}
\caption{}
\end{subfigure}
\qquad
\begin{subfigure}[t]{.4\columnwidth}
\includegraphics[width=\textwidth]{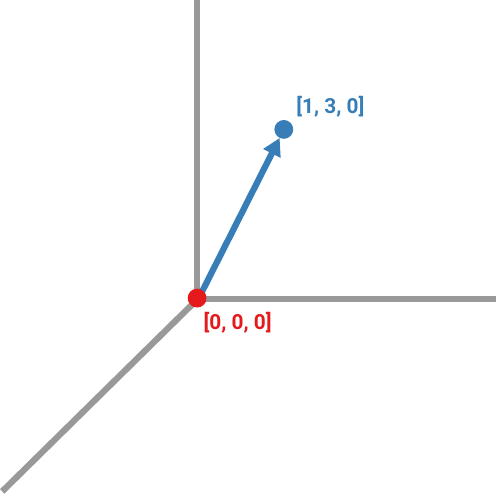}
\caption{}
\end{subfigure}
\caption{(a) Several vectors on a 2D space. (b) Two vectors on a 3D space.}
\label{fig:la-vector}
\end{figure}

The starting point of linear algebra is the vector. A vector is a list of numbers. So for instance $[3,1]$ is a vector. The number of entries in the vector is the dimension of the space it lives in. The example I just gave is a vector embedded in a two dimensional space, often we say it is a 2D vector, although this habit will get us in a bit of terminology trouble down the road. A three dimensional vector could be $[4,1,7]$. In the spatial interpretation of linear algebra, you can interpret these numbers as coordinates in space. By convention, the origin of the space, its central point, is a vector full of zeroes -- so $[0, 0]$ is the origin of a two dimensional space. While the vector is a point in space, sometimes it is convenient to think of it as an arrow -- so that's why I use both representations in Figure \ref{fig:la-vector}. The tail of the arrow is always at the origin and the point of the arrow is always at the coordinates specified by the vector.

From the figure you can see that it is a bit cumbersome to represent 3D vectors on a piece of 2D paper, so that's why I will stick to 2D examples when making figures. But there is no need to stop at 2D. In fact there is no need to stop at 3D either: all operations in linear algebra work the same in an arbitrary number of dimensions. They just become a little harder to picture in your head.

Notation-wise, normally a vector is written as $\overrightarrow{v}$, but I'm skipping the arrow and writing them as simple lowercase letters: $v$. This is a bit ambiguous, but in general it will be clear from the context when I'm talking about vectors or not.

\begin{figure}
\centering
\includegraphics[width=.4\textwidth]{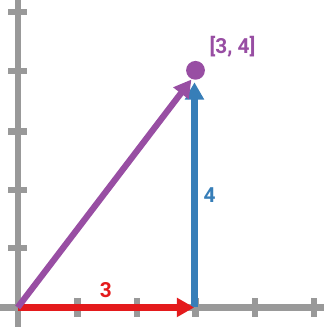}
\caption{Decomposing the vector into its dimension coordinates to apply Pythagoras's theorem to find its length.}
\label{fig:la-vector-length}
\end{figure}

One useful property of a vector is its length. The length of the vector is literally the length of the arrow we draw: it is the Euclidean (straight line) distance between the point identified by the vector and the origin. To calculate the Euclidean distance we can realize that a vector is nothing more than the hypotenuse of a special right triangle. Starting from the origin, we can first travel along the $x$ axis until we get to the first coordinate of the vector, then we walk up parallel to the $y$ axis to make up the second coordinate of the vector. So the $[3,4]$ vector in Figure \ref{fig:la-vector-length} is three steps along the $x$ axis and four steps parallel to the $y$ axis.

Pythagoras teaches us that the length of the hypotenuse is the square root of the sum of the squares of the catheti lengths\footnote{\url{https://en.wikipedia.org/wiki/Pythagorean_theorem}, Wikipedia will have to do, since Pythagoras never put a bibtex out...}. Putting it into mathematical form, our $[3,4]$ vector is $\sqrt{3^2+4^2} = 5$ long. More generally, any $d$ dimensional vector $v$ is $\sqrt{\sum \limits_{k=1}^{d} v_k^2}$ long.

What operations can you do with vectors? Fundamentally, we need two: \textbf{sum} and \textbf{multiplication}.

The \textbf{sum} of two vectors is pretty straightforward. Say you have $[1,3]$ and $[2,2]$. What do you think the result of their sum should be? It will be the element-wise sum of their entries. So $[1,3] + [2,1] = [3,4]$. Using a more abstract notation, ``element-wise'' means that, if you have two dimensional vectors $u$ and $v$, then their sum is $[u_1 + v_1, u_2 + v_2]$. You sum the first element of $u$ with the first element of $v$, then second with second, and so on (if you have more dimensions).

\begin{figure}
\centering
\includegraphics[width=.4\textwidth]{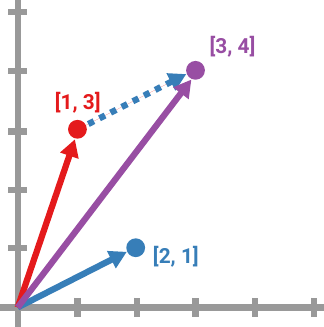}
\caption{An example of vector sum.}
\label{fig:la-vector-sum}
\end{figure}

The result of the sum of two vectors is another vector with the same dimensions. Visually, this means that we're taking the second vector and move its tail to the point of the first vector. The point where we end up is the new vector, the result of the sum. This is why it is sometimes easier to think of vectors as arrows rather than points, because this operation would be a bit less intuitive if we only have points. But arrows make it easy to see what we're doing, which is what Figure \ref{fig:la-vector-sum} shows.

This definition of the sum operation is consistent with how we defined vectors in the first place. Any vector could be considered the sum of itself to the origin point. If we have $[1,2]$, that is equivalent to $[0,0] + [1,2]$ -- i.e. you place the tail of $[1,2]$ on the point of $[0,0]$ and then you see where you end up. You can also think of vector sum as a sequence of movements in space. The first vector tells you the first movement. When you sum another vector to it, you take another movement, starting from where you ended up with the previous movement.

\begin{figure}
\centering
\includegraphics[width=.4\textwidth]{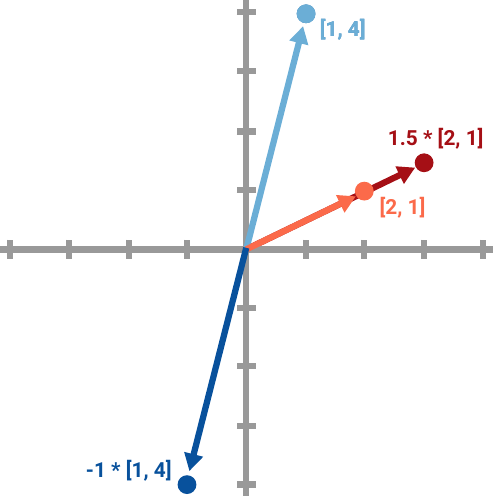}
\caption{Two examples of vector multiplications by a number (scalar). The light color version is the original vector, and the darker version is the result of the multiplication (the scaling).}
\label{fig:la-vector-mult}
\end{figure}

For the \textbf{multiplication}, for now we're only going to deal with the case of multiplying a number to a vector. As you can guess from the sum example, this means to multiply all the elements of the vector with that number: $2v = [2v_1, 2v_2]$. Visually, that means we extend the vector by that factor -- as you can see in Figure \ref{fig:la-vector-mult}. De facto, that means stretching the arrow's length while remaining on the same line -- we have to flip the direction if we're multiplying by a negative number.

Multiplying by $2$ means to end up with an arrow twice as long. Multiplying by a negative number means to reverse the direction before doing the stretching. And since stretching means to change the length of the arrow -- i.e. to ``scale'' it -- that's the reason sometimes one will refer to numbers as ``scalars''.

\section{Matrices}\label{sec:la-matrix}

\subsection{Basic Properties and Operations}
If vectors are lists of numbers, matrices are lists of lists of numbers. While you can write a vector as $[4,1,7]$, for a matrix you need to do something slightly more complicated:

$$
\begin{pmatrix}
    0 & 4 \\
    5 & 1 
\end{pmatrix}.
$$

Like vectors, matrices too have dimensions. The matrix above has two rows and two columns, so we say it is a $2 \times 2$ matrix. Since the number of rows and columns is the same, the matrix looks like a square and so we call it a square matrix. But nothing stops matrices from having a \textit{different} number of rows and columns. For instance, a $3 \times 2$ matrix is a non-square (rectangular) matrix, a totally valid object. Notation wise, $M_{ij}$ means to look at the value in the cell at the $i$th row and the $j$th column.

Before getting into what matrices are and what they are for, I want to list a few important things about matrices. First, square matrices have a \textit{diagonal}, which goes from the top left element down to the bottom right. Second, there's the concept of symmetry. A square matrix is \textit{symmetric} if you can mirror it along the diagonal and you get the same matrix. Mathematically, this means that is is always the case that $M_{ij} = M_{ji}$.

\begin{figure}
\centering
\begin{subfigure}[t]{.4\columnwidth}
\includegraphics[width=\textwidth]{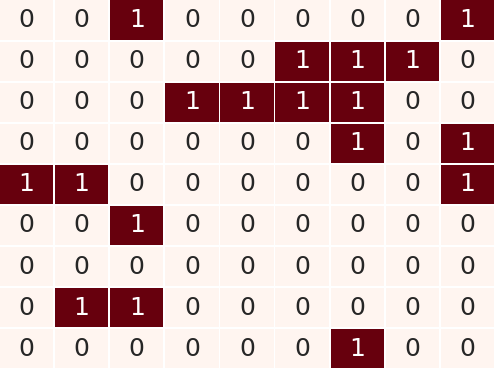}
\caption{$M$}
\end{subfigure}
\qquad
\begin{subfigure}[t]{.4\columnwidth}
\includegraphics[width=\textwidth]{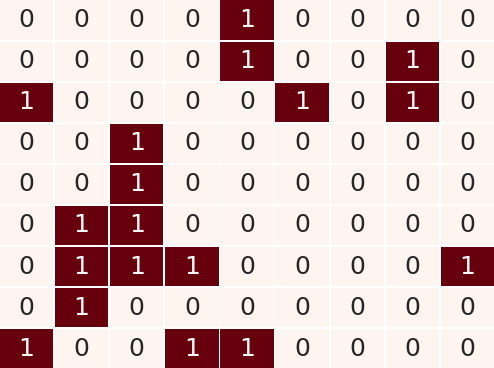}
\caption{$M^T$}
\end{subfigure}
\caption{A matrix and its transpose. Note how the $(i,j)$ entries of $M$ equal to one transposed to the $(j,i)$ entries of $M^T$, for instance $(1,3)$ to $(3,1)$.}
\label{fig:mat-transpose}
\end{figure}

Note how non-square matrices do not have a diagonal and cannot be symmetric. \textit{Transposing} a matrix $M$ means that all $M_{ij}$ values become $M_{ji}$ and viceversa. In this book, for convention, $M^T$ will be the transpose of $M$. Figure \ref{fig:mat-transpose} shows the case of a squared non symmetric matrix transpose. In practice, transposing is like placing a mirror on the diagonal.

If your matrix is squared and symmetric transposing has no effect: $M^T = M$. However, for non symmetric matrices, $M^T \neq M$. Moreover, for non square matrices, transposing an $n \times m$ matrix results into an $m \times n$ one, the dimensions flip.

\subsection{Vector-Matrix Multiplication}
To keep our spatial interpretation of linear algebra, in the case of matrices we need to jump right to the multiplication of a matrix with a vector. If you have matrix $M$ and vector $v$, then $Mv = w$: multiplying vector $v$ with matrix $M$ results a new vector $w$. The new vector $w$ will have different coordinates from $v$ -- with few interesting and useful exceptions we'll get to later. In practice, $M$ is moving $v$ so that it ends up in $w$. Alternatively, you can say that $M$ is changing the coordinate system of $v$: $w$ is still the same as $v$, but in a different coordinate system.

\begin{figure}[t]
\centering
\includegraphics[width=.4\textwidth]{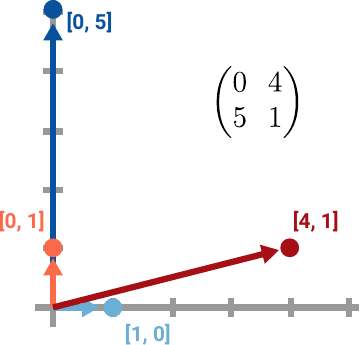}
\caption{A visualization of a matrix. The light color vectors are the original unit vectors, and their darker version is the result of the coordinate transformation applied by the matrix.}
\label{fig:la-matrix-vector-mult}
\end{figure}

So far this isn't helping much, it's still pretty abstract. What does the matrix I showed you before \textit{really means}? Each column of the matrix tells you how to change each coordinate in isolation. This is the same as stretching a vector that is equal to one in that given dimension, and zero everywhere else. So, since the first column of the matrix is $[0,5]$ then we know that the unit vector $[1,0]$ will end up in $[0,5]$. The second column tells you that the second unit vector $[0,1]$ will end up in $[4,1]$. Figure \ref{fig:la-matrix-vector-mult} shows you this transformation.

\begin{figure}
\centering
\includegraphics[width=.66\textwidth]{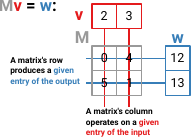}
\caption{A example of vector-matrix multiplication. Each entry of the result vector depends on the corresponding row of the matrix. Each entry of the input vector is handled by the corresponding column of the matrix.}
\label{fig:la-matrix-vector-mult-2}
\end{figure}

Once you know this, you know how to move any two vectors, because any vector is a combination of these two unit vectors. To know where any arbitrary vector $v = [v_1, v_2]$  ends -- to calculate $w$ --, you need to look at the matrix first by rows: the first row tells you the contribution of the matrix to the first entry of $w$. Then you look at columns: the first column tells you the effect of $v_1$, the second of $v_2$, and so on. To sum up in general terms, $Mv = w$ means that $w_1 = M_{1,1}v_1 + M_{1,2}v_2$ and $w_2 = M_{2,1}v_1 + M_{2,2}v_2$. You can see an example in Figure \ref{fig:la-matrix-vector-mult-2}.

It's important to remember that $M$ cannot change the coordinate system in any arbitrary way. We're still in \textit{linear} algebra, so $M$ can only change coordinates linearly. That is, if you have a line in a coordinate system, after applying $M$ that is still going to be a line -- maybe longer or shorter, maybe pointing in another direction, but it won't curve. We also want to maintain the origin -- in the case of 2D space $[0,0]$ -- in its place.

You might be wondering what happens when we use a non-square matrix to perform the transformation. What does it mean to multiply $[1,0]$ to a $3 \times 2$ matrix? Well, you just learned that the first column of the matrix gives you the coordinates of where $[1,0]$ lands. If a matrix is $3 \times 2$, it means that $[1,0]$ will land on a 3D space, because it will have three coordinates. So non-square matrices allow you to move between spaces with a different number of dimensions.

\begin{figure}
\centering
\includegraphics[width=.66\textwidth]{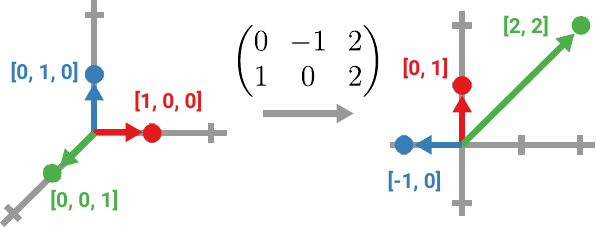}
\caption{A visualization of a non-square matrix. Each unit vector in the 3D space gets mapped to a new coordinate system in a 2D space.}
\label{fig:la-matrix-rect}
\end{figure}

In our $Mv = w$, besides $w_1$ and $w_2$, we also have $w_3 = M_{3,1}v_1 + M_{3,2}v_2$. More generally: $w_i = \sum \limits_{k=1}^{n} M_{ik}v_k$. This formula sneakily tells you one important thing: you cannot multiply any vector-matrix combination. The matrix must have the same number of columns as the number of entries in the vector. Otherwise you either have some entries of $v$ you can't transform, or portions of $M$ doing nothing. In general, multiplying a $d$ dimensional vector with a $n \times d$ matrix will result in an $n$ dimensional vector. Figure \ref{fig:la-matrix-rect} shows you this operation, and highlights how this coordinate change moves to a completely new 2D space from the original 3D one.

Possibly the most useful matrix for this book is the identity matrix, which we call $I$. This is defined as a matrix that has ones on its main diagonal -- the one going from top left to bottom right, and zero everywhere else:

$$
\begin{pmatrix}
    1 & 0 & \dots  & 0 \\
    0 & 1 & \dots  & 0 \\
    \vdots & \vdots & \ddots & \vdots \\
    0 & 0 & \dots  & 1
\end{pmatrix}.
$$

Can you guess what it does? Each column in this matrix is exactly the unit vector for that specific dimension. Since it tells you what the unit vector becomes after the transformation, what this means is that the unit vector will not change. If it does not change, nothing will! The identity matrix will preserve the coordinate system exactly as it is. Mathematically, for \textit{any} $v$: $Iv = v$.

\subsection{Matrix Multiplication}
It is often the case that you might want to multiply two matrices together, rather than a matrix and a vector. It's worth explaining what that operation means intuitively, because that will make it easy to understand why we achieve it the way we do.

\begin{figure}
\centering
\includegraphics[width=.5\textwidth]{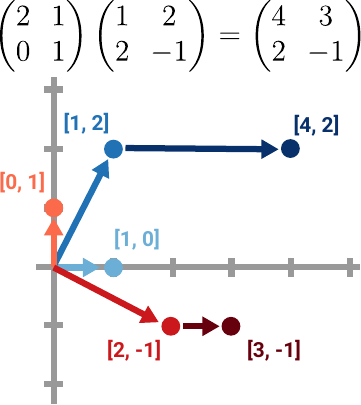}
\caption{A visualization of a matrix multiplication. The lightest color vectors are the original unit vectors. The midway dark version is the application of the first coordinate change, the darkest vectors apply the second coordinate change on top of that.}
\label{fig:mat-mult-viz}
\end{figure}

If a matrix is a change in the coordinate system, the multiplication of two matrices $A$ and $B$ is the effect of making first the $A$ transformation and then transform the result with $B$. The concatenation -- or composition -- of these two transformations is by itself another transformation, so it must be a matrix too, $C$. So, matrix multiplication is an operation that produces a matrix $C$ from two matrices $A$ and $B$. Figure \ref{fig:mat-mult-viz} shows an example. The first coordinate change moves $[1,0]$ to $1[1,2] + 0[2,-1] = [1,2]$, then the second change moves $[1,2]$ to $1[2,0] + 2[1,1] = [4,2]$, so $[4,2]$ must be the first column of the result. 

With formal notation, what we're saying here is that first we transform $v$ with $A$ as $Av$. Then we transform that result with $B$: $BAv$. That must equal $Cv$. So it is saying: $BAv = Cv$. And, at that point, you can simplify as $BA = C$, since this will work for any $v$. Note how we read this right to left: $BA$ means applying $A$ and then applying $B$ to the result. In most cases, $BA \neq AB$.

The advantage of thinking of this in terms of concatenation of transformations is that it immediately reduces it to the case of multiplying a vector to a matrix, since it's the same thing. The first column of $A$ tells you where $[1,0]$ lands, then the first column of $B$ tells you where that results lands in turn. If you expand this mathematically, you get that each $C_{ij}$ entry of $C$ is equal to the sum of the products of all entries in the $i$th row of $A$ and the $j$th column of $B$. Formally: $C_{ij} = \sum \limits_{k=1}^{m} A_{ik}B_{kj}$.

\begin{figure}[t]
\centering
\includegraphics[width=.5\textwidth]{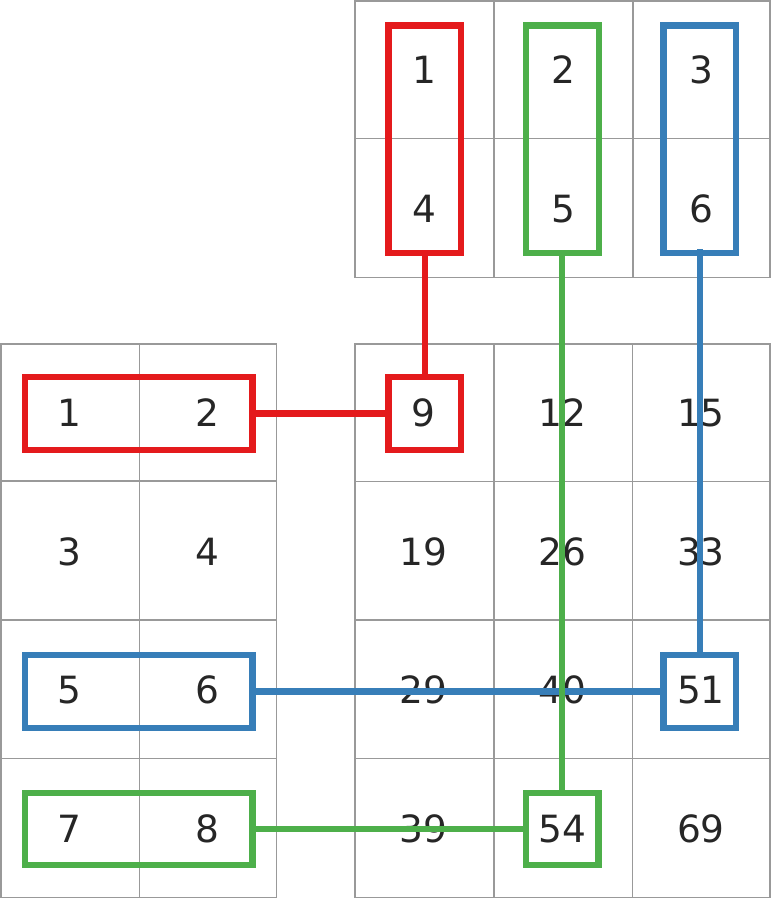}
\caption{An example of matrix multiplication. Each cell is the result of the combination of the rows/columns of the corresponding color, whose element-wise products are summed. So the red cell equals to $9$ because it is the sum of $1 \times 1$ (the product of the first elements) plus $2 \times 4$ (the product of the second elements).}
\label{fig:mat-mult}
\end{figure}

Just like in the case of vector-matrix multiplication, you also cannot multiply any two matrices together. $A$ and $B$ must have one dimension of equal size. So a $n \times m$ matrix can be multiplied by a $m \times x$ matrix to generate a $n \times x$ matrix. In this case, the common dimension ``disappears''. Figure \ref{fig:mat-mult} shows a graphical representation of the algorithm to multiply two non-square matrices -- which also generalizes to multiplying square matrices.

\section{Tensors}\label{sec:la-tensor}
Finally, we get to the concept of tensor. The tensor is the generalization of vectors and matrices. Here we clash a bit in terminology, because we already defined what a dimension is in linear algebra: the number of entries in a vector. Abusing terminology a bit, we can use the term ``dimension'' to mean another thing. You can think of vectors -- with any number of entries -- as one-dimensional lists of numbers, because they look like a line: $[1,3,6,3,8,2,\dots]$. On the other hand, matrices look like two-dimensional rectangles:

$$
\begin{pmatrix}
    1 & 3 & 4 & \dots  & 3 \\
    8 & 5 & 4 & \dots  & 9 \\
    \vdots & \vdots & \vdots & \ddots & \vdots \\
    1 & 2 & 7 & \dots  & 8
\end{pmatrix}.
$$

The tensor generalizes this intuition. A vector is a one dimensional tensor, it only needs one index to identify its entries $v_i$. A matrix is a two dimensional tensor, needing two indices for its entries $M_{ij}$. Then you can have three dimensional tensors, with three indices like $T_{ijk}$, four dimensional with four indices ($T_{ijkz}$) and so on. In fact, you can also generalize in a different direction. To indicate a number, you need no index. So a number is a zero dimensional tensor.

\begin{figure}
\centering
\includegraphics[width=.4\textwidth]{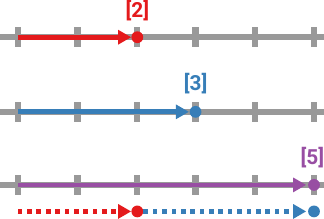}
\caption{An example of sum for two zero dimensional tensors.}
\label{fig:la-tensor-sum}
\end{figure}

This latter generalization also shows you how much it makes sense to make vector sums the way I explained in Section \ref{sec:la-vector}. The normal sum you're used to is defined the same way, for zero dimensional tensors. When we say $2+3=5$, what we mean is that we first move to the end of the first zero dimensional tensor to go from $0$ to $2$, then we apply the movement from the second zero dimensional tensor $3$ to end up in the final zero dimensional tensor $5$. Figure \ref{fig:la-tensor-sum} shows you exactly this.

Note that the dimension of the tensor has no relation with the dimension of the space in which the tensor lives! A vector is a one dimensional tensor, but can live in a 3D space, if it has three entries. Vice versa, a 2D space can contain a three dimensional tensor, for instance $[[[1,2],[2,4]],[[3,1],[2,5]]]$ is an example of such a tensor. Confusing, I know!

\section{Vector Products}\label{sec:la-dot}

\subsection{Dot Product}
The fact that both vectors and matrices are tensors suggests a profound thing: there is no qualitative distinction between a vector and a matrix. This is indeed true, and one cool repercussion is that vectors, just like matrices, are \textit{also} coordinate shifts. That is because any $d$ dimensional vector is also a $d \times 1$ rectangular matrix. So it can play a role in the vector-matrix and matrix-matrix multiplications I explained in Section \ref{sec:la-matrix}.

This is the dot product. Let's say you have two $d$ dimensional vectors: $v$ and $m$. We can decide that $m$ is actually a $d \times 1$ matrix and we use it to perform a vector-matrix multiplication with $v$. Since $m$ is a rectangular matrix, by now you know that it will transport you to a space with a different number of dimensions. In this case, you'll end up with only one dimension. A one-dimensional vector is a number.

\begin{figure}
\centering
\includegraphics[width=.4\textwidth]{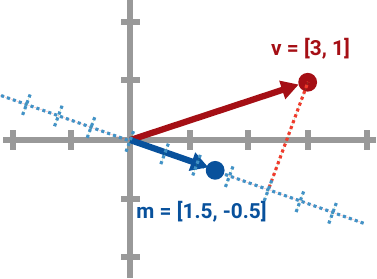}
\caption{An example of dot product. The blue dashed line is the number line defined by the dark blue vector $m$. The red dashed line shows the projection of the dark red $v$ vector onto the number line.}
\label{fig:la-dot-prod}
\end{figure}

What this means in our spatial perspective is that you're projecting $v$ onto the line defined by the direction pointed by $m$. This is because you're bringing $v$ into the 1D space of $m$, which is a number line. You will also have to stretch $v$'s projection proportionally to the length of $m$, just like entries in a matrix $M$ will stretch or squish the space if they are different than $1$. Figure \ref{fig:la-dot-prod} shows you how this looks like spatially.

Note how in Figure \ref{fig:la-dot-prod} the red vector $v$ lands on the fourth tick-mark of the number line determined by the direction of vector $m$. In fact, the dot product of $v = [3, 1]$ and $m = [1.5, -0.5]$ is exactly four, because it so happens that $m$ does not stretch anything on this number line -- it is its unit vector. However, if we were to do the dot product with another vector on that same number line -- say $m' = [3, -1]$ -- we will have to do some stretching. In this case, since $m'$ is twice as long as $m$, we will have to multiply the result of the projection by two. In fact, the dot product between $v$ and $m'$ is exactly $8$.

Mathematically, you're not doing anything differently than what you already learned in vector-matrix multiplication, only that now you have only one column to worry about. The dot product between $v$ and $m$ is still $w = \sum \limits_{k=1}^{n} m_{k}v_k$, which as you can see is a single number.

Notation-wise, what you're doing with the dot product is $m^Tv$. I want to point out one thing that will be useful down the road. You can dot product a vector with itself: $v^Tv$. This is normally known as the ``quadratic sum'' because, if you expand this mathematically, that's what it is: $v^Tv = \sum \limits_{k=1}^{n} v^2_{k}$. Looks familiar? It's the same sum to get the vector's length. So another way to calculate $v$'s length is $\sqrt{v^Tv}$. This actually generalizes to calculating the distance between two vectors: you simply need to take their difference and put it into the same formula. So the distance between $u$ and $v$ is $\sqrt{(u - v)^T(u - v)}$. The difference between $u$ and $v$ is simply the vector sum of $u$ to $-1v$, both things you know the definition of from Section \ref{sec:la-vector}.

\subsection{Outer Product}
There's another type of product that is useful in several part of this book. It is the cousin of the dot product: the outer product. When we decided to multiply two $d \times 1$ vectors, we decided we were collapsing the $d$ dimension to get a $1$ dimensional vector, a number. But that's not the only option. We could have instead collapsed the $1$ dimension and the result would be... can you guess it? A $d \times d$ matrix! In fact, if we decide to collapse the $1$ dimension, it actually doesn't matter whether $u$ and $v$ have the same dimension. Any two vectors can have an outer product. In this book I'll write the outer product as $u \otimes v$. 

Formally, this looks like:

$$\begin{aligned}\mathbf {u} \otimes \mathbf {v} =\mathbf {u} \mathbf {v} ^{T}={\begin{bmatrix}u_{1}\\u_{2}\\u_{3}\\u_{4}\end{bmatrix}}{\begin{bmatrix}v_{1}&v_{2}&v_{3}\end{bmatrix}}={\begin{pmatrix}u_{1}v_{1}&u_{1}v_{2}&u_{1}v_{3}\\u_{2}v_{1}&u_{2}v_{2}&u_{2}v_{3}\\u_{3}v_{1}&u_{3}v_{2}&u_{3}v_{3}\\u_{4}v_{1}&u_{4}v_{2}&u_{4}v_{3}\end{pmatrix}}.\end{aligned}$$

In practice, the outer product of two vectors $u$ and $v$ is a $|u| \times |v|$ matrix, whose $(i,j)$ entry is the multiplication of $u_i$ to $v_j$.

\subsection{Positive (Semi)Definite Matrices}
Positive definiteness is an interesting property for a matrix which will come in handy in Chapter \ref{cha:nvd}, so it's worthwhile to introduce it here. Suppose you have a vector $v$ of length $m$ and a $n \times m$ matrix $M$. It follows from the properties of matrix multiplication that you can always multiply them -- because a vector of length $m$ is a $m \times 1$ matrix. Hopefully, you know what the result would be: a vector of length $n$ -- remember: the common dimension ``disappears''. For the very same reason, you can always multiply a vector with the transpose of itself. $v^Tv$ is a legit operation, as we just saw when talking about the dot product, and produces a number. If we put together what we discovered in the previous two paragraphs: if $M$ is a square matrix, then $v^TMv$ is a scalar.

Now, some matrices $M$ are special. For these special matrices, it doesn't matter what you put in $v$, as long as it is a vector of real numbers: the result of $v^TMv$ is always going to be greater than zero. We call these special matrices ``positive definite''. Relaxing the concept a bit, if $v^TMv \geq 0$ for any real number $v$, then $M$ is positive semi-definite -- ``semi'' because we allow the result to be zero sometimes.

The identity matrix $I$ that I introduced a while ago is positive semi-definite. This means that any $v^TIv$ is going to be zero or positive. In fact, by properties of the identity matrix, $v^TIv = v^Tv$: the identity matrix does nothing, remember? So why did I mention this fact? Well, we can expand our formula of the Euclidean distance between two vectors as $\sqrt{(u - v)^TI(u - v)}$ without fear, because we know everything checks out. This gave some people some ideas. All it matters for a matrix to be a good fit in this formula and replace $I$ is that they have to be positive semi-definite. That's because the formula is under a square root and we don't want negatives there. The consequence is that any positive semi-definite matrix replacing $I$ will also define a special distance in a non-Euclidean space! (Admittedly, this excites me much more than it should)

This is the exact thing that you do, for instance, when calculating a Mahalanobis distance -- which is a smart Euclidean that takes into account the correlation between the vectors, see Section \ref{sec:nvd-nonnet}. The Mahalanobis distance is $((p - q)^Tcov(p,q)^{-1}(p - q))^{1/2}$, where $cov(p,q)^{-1}$ is the inverse of the covariance matrix between $p$ and $q$. The covariance matrix is a matrix whose element in the $i,j$ position is the covariance (Section \ref{sec:stats-corr}) between the $i$-th and $j$-th elements of $p$ and $q$. Surprise surprise, $cov(p,q)^{-1}$ is positive semidefinite.

This will be immensely useful in Chapter \ref{cha:nvd} when I will show how there are secret identity matrices hidden in many formulas that can be replaced with matrices representing your network, to expand many classical statistical measures into network statistical measures. It'll be uber cool, I pinky promise!

\section{Eigenvalues and Eigenvectors}\label{sec:la-eigen}
\begin{figure}
\centering
\includegraphics[width=.75\columnwidth]{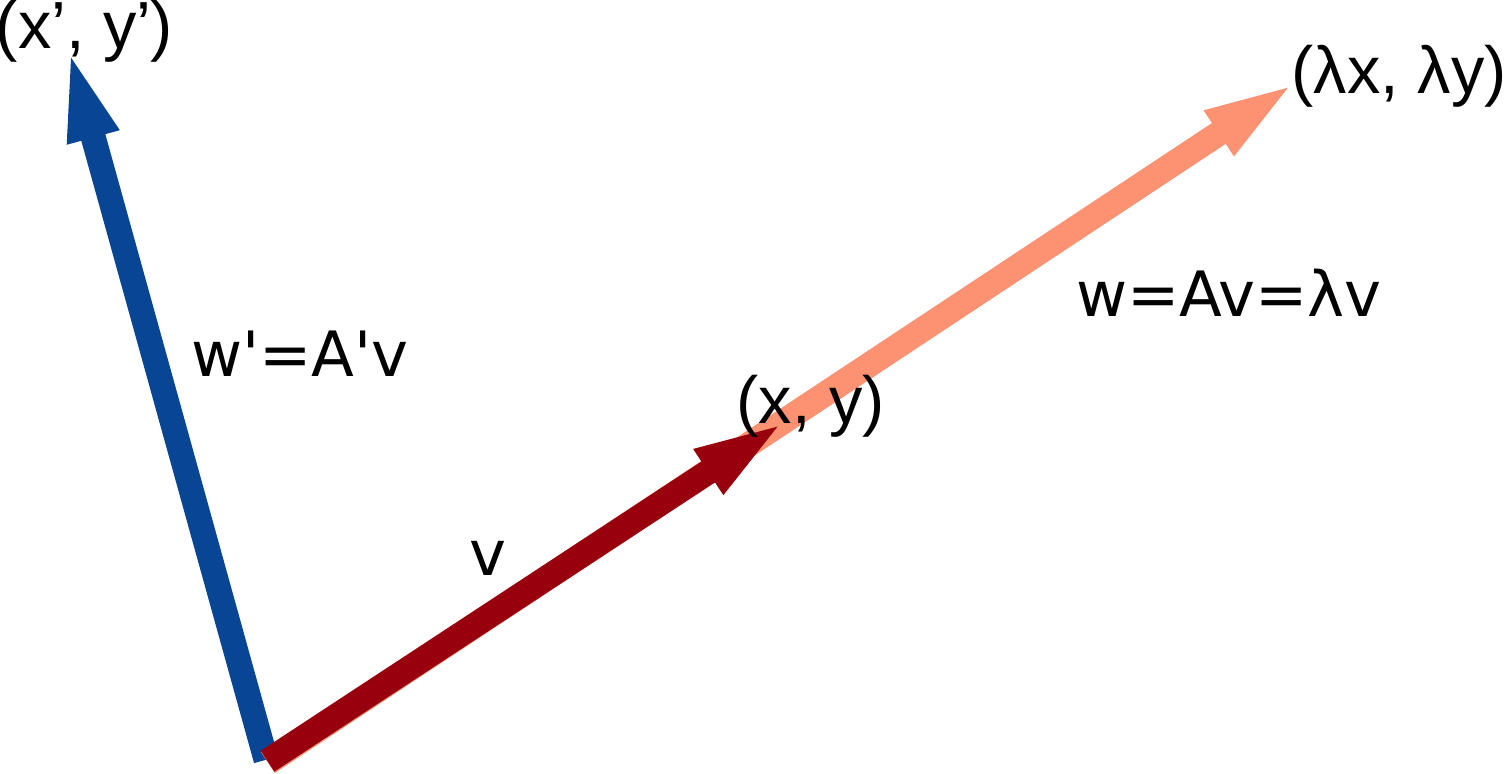}
\caption{A graphical depiction of an eigenvector.}
\label{fig:eigenvector}
\end{figure}

Two absolutely key concepts for network science are eigenvalues and eigenvectors. They are used almost everywhere in network science, so we need to define them here. Consider Figure \ref{fig:eigenvector}. Given a vector $v$, we learned we can apply an arbitrary matrix transformation $M$ to it -- as long as it has the correct dimensions. We then obtain a new vector $w = Mv$. Any transformation $M$ has special vectors: $M$ \textit{scales} these special vectors without altering their \textit{directions}. In practice, the transformation $M$ simply multiplies the elements of such vectors by the same scalar $\lambda$: $w = Mv = \lambda v$. We have a name for this: $v$ is $M$'s eigenvector and $\lambda$ is its associated eigenvalue. Mathematically, we represent this relation as $Mv = \lambda v$.

The formula we just introduced is the one for \textit{right} eigenvectors, because the vector multiplies the matrix \textit{from the right}. If $M$ is square, there are also \textit{left} eigenvectors, which multiply the matrix \textit{from the left}: $vM = v\lambda$. Right and left eigenvectors are different, have different values, but their corresponding eigenvalues are the same. From now on, when I mention eigenvectors, I refer to the \textit{right} eigenvectors. Right eigenvectors are the \textit{default}, and when I refer to \textit{left} eigenvectors I will explicitly acknowledge it.

A square matrix with $n$ rows and columns also has $n$ eigenvalues. By convention, we sort eigenvalues by their value, sometimes in increasing order, sometimes in decreasing order, depending on the application.

A key term you need to keep in mind is ``multiplicity''. The multiplicity of an eigenvalue is the number of eigenvectors to which it is associated. If you have an $n \times n$ matrix, but only $d < n$ distinct eigenvalues, some eigenvalues are associated to more than one eigenvector. Thus their multiplicity is higher than one.

\section{Matrix Factorization}\label{sec:mat-factors}
In some cases, you might want to express a matrix as the result of the multiplication of other matrices. This can be useful because the matrices you use to reconstruct your observed matrix might be made of pieces you can more easily interpret. We call this decomposition of a matrix ``factorization'', because we divide the matrix into its ``factors'', its building blocks. There are countless ways to factorize a matrix. Here we examine only the ones that you're most likely to encounter in network analysis. I divide them in two classes: the ones operating on regular bi-dimensional matrices, and the ones which work on scary and confusing multidimensional matrices (i.e. tensors).

\subsection{Matrix Decomposition}
One of the easiest ways to perform matrix factorization is what we call ``eigendecomposition''. A square matrix $M$ can always be decomposed as $M = \Phi \Lambda \Phi^{-1}$. Rather than being the left and right eyes of a really pissed frowny face, $\Phi$ is the matrix we obtain piling all eigenvectors next to each other, and $\Lambda$ is a diagonal matrix with the eigenvalues on its main diagonal and zeros everywhere else:

$$ \Lambda = 
\begin{pmatrix}
\lambda_0 & \dots & 0 \\
0 & \ddots & 0 \\
0 & \dots & \lambda_n. \\
\end{pmatrix}
$$

We are mostly interested in eigendecomposition as the special case of the more general Singular Value Decomposition (SVD) -- which can be applied to any matrix, even non-square ones. In SVD, we simply replace $\Phi$ and $\Lambda$ with generic matrices. In other words, we say that we can reconstruct $M$ with the following operation: $M = Q_1 \Sigma Q_2^T$. Like $\Lambda$, also $\Sigma$ is a diagonal matrix. The difference is that $\Sigma$ contains the singular values of $M$, rather than its eigenvalues. While there is only one valid $\Sigma$ to solve this equation -- that is why it is called ``singular'' -- there could be multiple $Q_1$ and $Q_2$ matrices that you could plug in, as long as they're both unitary matrices. A unitary matrix $Q$ is a matrix whose transpose is also its inverse: $Q Q^{-1} = I = Q Q^T$, with $I$ being the identity matrix. SVD is especially useful for estimating node distances on networks (Section \ref{sec:nvd-ge}).

Along with eigendecomposition, the two most common and useful matrix decomposition tools are the Principal Component Analysis (PCA) and the Non-Negative Matrix Factorization (NMF).

To understand PCA, suppose that your matrix is just a set of observations and variables. Each row of the matrix is an observation and each column is a variable. PCA, like NMF, is used to summarize this matrix of data. If two columns/variables are correlated it means they contain redundant information. Thus, you're after a way to describe your data in such a way that each variable has no redundant information. 

Figure \ref{fig:pca1} shows an example: each row is a day and each column is some measurement taken in that day -- the temperature, wind speed, the millimeters of rain/snow that fell that day, etc. You might expect that some of these variables might be correlated. For instance, it is very difficult to have a single millimeter of snow if the temperature is above a certain value. Rather than describing a day by all variables, you want to describe it by its similarity with an ``archetypal'' day: is this a snow day or a rain day?

\begin{figure*}
\centering
\begin{tabular}{l|rrrrr}
Day & Temp (\textdegree{}C) & Wind (km/h) & Sunlight (\%) & Rain (mm) & Snow (mm)\\
\hline
$1$ & $27$ & $10$ & $80$ & $2$ & $0$\\
$2$ & $26$ & $1.2$ & $95$ & $1$ & $0$\\
$3$ & $32$ & $7.6$ & $100$ & $0$ & $0$\\
$4$ & $12$ & $2.3$ & $12$ & $20$ & $0$\\
$5$ & $14$ & $3.8$ & $8$ & $25$ & $0$\\
$6$ & $6$ & $0.2$ & $24$ & $40$ & $1$\\
$7$ & $4$ & $0.1$ & $2$ & $8$ & $30$\\
$8$ & $2$ & $0.9$ & $4$ & $1$ & $40$\\
$9$ & $-1$ & $1.1$ & $4$ & $0$ & $80$\\
\end{tabular}
\caption{A table recording in a matrix the characteristics of some days.}
\label{fig:pca1}
\end{figure*}

This is the aim of PCA. Let's repeat the previous paragraph mathematically: you want to transform your correlated vectors in a set of uncorrelated, or orthogonal, vectors which we call ``principal components''. Each component is a vector that explains the largest possible amount of variance (Section \ref{sec:stats-corr}) in your data, under the condition of being orthogonal with all the other components. You can have as many components as you have variables, but usually you want much fewer -- for instance two, so you can plot the data. That is because the first component explains the most variance in the system, the second a bit less, and so on, until the last few components which are practically random. Thus you want to stop collecting components after you've taken the first $n$, setting $n$ to your delight. In Figure \ref{fig:pca2}, I collect the first two -- they're there for illustrative purposes so don't be shocked if you realize they're not really orthogonal.

\begin{figure}
\centering
\begin{tabular}{l|rr}
Day & PC1 & PC2\\
\hline
$1$ & $0.2$ & $-0.05$\\
$2$ & $-0.05$ & $0.1$\\
$3$ & $-0.1$ & $-0.1$\\
$4$ & $2.6$ & $-0.01$\\
$5$ & $2.9$ & $0$\\
$6$ & $3.2$ & $0.35$\\
$7$ & $0.3$ & $2.4$\\
$8$ & $0.1$ & $2.6$\\
$9$ & $-0.05$ & $2.8$\\
\end{tabular}
\caption{The first two principal components of the matrix in Figure \ref{fig:pca1}.}
\label{fig:pca2}
\end{figure}

\begin{figure}[b]
\centering
\includegraphics[width=.66\columnwidth]{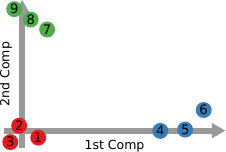}
\caption{A scatter plot with the first principal component of the matrix in Figure \ref{fig:pca1} on the x axis and the second component on the y axis.}
\label{fig:pca3}
\end{figure}

PCA is extremely helpful when performing data clustering. Suppose that we're looking only at the first two principal components of our matrix describing our days. We can make a two dimensional scatter plot of the system, with one point per day. It might look like Figure \ref{fig:pca3}. This seems successful, because we can clearly see three clusters: days dominated by the first component (in blue), days dominated by the second component (in green), and days which have low values in both (in red). When we look at the original data, we might recognize that the first class of days had high rain precipitation, the second high snow precipitation, and the third group was mostly sunny days. In this sense, PCA aided us in finding our archetypal days: the first component describes the archetypal rainy day, while the second component describes the archetypal snowy day.

PCA has no restrictions in the way it builds the principal components, besides the fact that all these components must be orthogonal with each other. This means that you might end up with components with negative values, as we do in Figure \ref{fig:pca2}. This might not be ideal. What does it mean for a day to be a ``negative rainy day''? PCA is interpretable, but sometimes the intepretation can be a bit... confusing.

Non-Negative Matrix Factorization solves this problem. Without going into technical details, NMF is PCA with the additional constraint that no component can have a negative entry -- hence the ``Non-Negative'' part in the name. At a practical level, if there were no negative entries in Figure \ref{fig:pca2}, then the two components in that figure could be results of NMF. This additional constraint comes at the expense of some precision: PCA can fit the data better because it does not restrict its output space. However, usually, NMF components are more easy to interpret.

Given their links to data clustering, both PCA and NMF are extensively used when looking for communities in your networks (Part \ref{par:cd}).

\subsection{Tensor Decomposition}
Let's assume you have a three dimensional tensor. You can think of a tensor as a cuboid and a slice of it is a matrix. If you find it difficult to picture this in your head, don't worry: you're not alone. That is why many researchers put effort into finding ways to decompose tensors in lower-dimensional representations that can sum up their main properties. This process is generally known as ``tensor decomposition''.

Tensor decomposition is a general term encompassing many techniques to express a tensor as a sequence of elementary operations (addition, multiplication, etc) on other, simpler tensors. For instance, you can represent a 3D tensor as a combination of three vectors, one per dimension. Or as a matrix and a vector. You want to do this to solve complex network analyses on multilayer networks -- whatever the hell this means, Section \ref{sec:mat-mat-mat} will enlighten you -- by taking the full dimensionality into account at the same time, rather than performing the analysis on each layer separately and then merge the results somehow. Examples of applications of tensor decomposition range from node ranking (Chapter \ref{cha:ranks}), to link prediction (Part \ref{par:lp}), to community discovery (Part \ref{par:cd}).

I am going to mention very briefly only two of these techniques: tensor rank decomposition and Tucker decomposition. You should look elsewhere for a more complete treatment of the subject\cite{kolda2009tensor}. There also exists a tensor SVD\cite{de2000multilinear}\cite{robeva2017singular}, but it is relatively similar to a special case of Tucker decomposition, so I will not cover it.

Tensor rank decomposition is the oldest of the two\cite{hitchcock1927expression} and has historically been referred to as PARAFAC\cite{harshman1970foundations} or CANDECOMP\cite{carroll1970analysis}. Let's say you have your nice 3D tensor $T$. This is a three dimensional matrix of dimensions $n \times n \times m$ -- for simplicity here we assume that the slices of the tensor are square matrices, thus two dimensions are the same, however the  method also works if all three dimensions are different. Tensor rank decomposition tells you that there is a way to decompose $T$ as the following combination:

$$ T \sim \sum \limits_k \lambda_k a_k \otimes b_k \otimes c_k.$$

Here, $T = a \otimes b \otimes c \rightarrow T_{ijk} = a_i b_j c_k$ which means that $\otimes$ represents the outer product. $a$ and $b$ are vectors of length $n$ and $c$ is a vector of length $m$. Finally, $\lambda_k$ is just a scaling factor that tells us how much to count the $k$th element of the sum. The convention is to call $\lambda_k a_k \otimes b_k \otimes c_k$ a component, while the vectors are the factors.

\begin{figure*}
\centering
\includegraphics[width=.95\columnwidth]{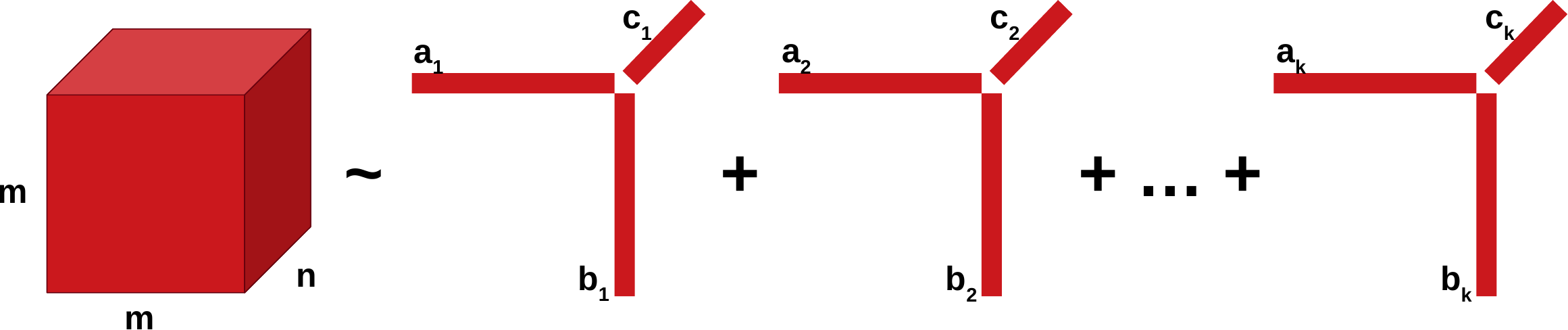}
\caption{A schema of tensor rank decomposition.}
\label{fig:parafac}
\end{figure*}

If you find difficult to understand what's going on by just looking at the formula, take inspiration from Figure \ref{fig:parafac}. What this operation does is to find the right set of one-dimensional vectors $a$, $b$ and $c$ such that, once they are scaled by factors $\lambda$, they can best represent the full tensor $T$. At that point, you are working in a lower dimensional space and all the rest of linear algebra starts making sense again.

How many components does this sum have? Or, in other words, how big should $k$ be to approximate $T$? That depends on the rank of the tensor. Unfortunately, calculating the rank of a tensor isn't as easy as calculating the rank of a matrix. The rank of a matrix is the number of columns (or rows) that are linearly independent from each other. The definition is the same for a tensor but, in this case, there is no straightforward algorithm to determine it\cite{kruskal1977three}. What happens is that, to find the rank of a tensor, you would literally apply the rank decomposition with different $k$ values and find the one that works the best.

Tucker decomposition\cite{tucker1966some} takes a different approach. It decomposes our tensor $T$ into a smaller core tensor and a set of matrices. If we keep our simplified case of a 3D tensor representing an adjacency matrix, mathematically speaking the Tucker factorization does:

$$ T \sim \mathcal{T} \times X \times Y \times Z.$$

Here, $\mathcal{T}$ is the core tensor, whose dimensions are smaller than $T$'s. $X$, $Y$, and $Z$ are matrices which have one dimension in common with $T$ and the other in common with $\mathcal{T}$ -- so that the matrix multiplication of them with $\mathcal{T}$ reconstructs a tensor with $T$'s dimensions.

\begin{figure}
\centering
\includegraphics[width=.66\columnwidth]{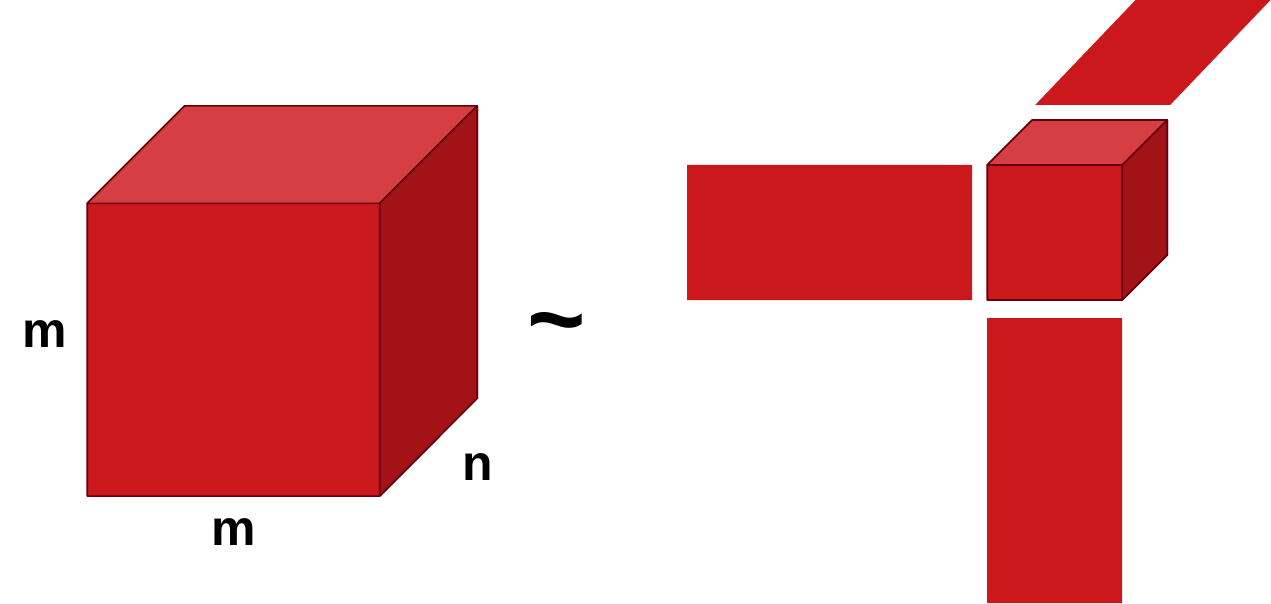}
\caption{A schema of Tucker decomposition.}
\label{fig:tucker}
\end{figure}

Again, for the visual thinkers, Figure \ref{fig:tucker} might come in handy. In Tucker decomposition you have the freedom to choose the dimensions of the core tensor $\mathcal{T}$. Smaller cores tend to be more interpretable, because they defer most of the heavy lifting to $X$, $Y$, and $Z$. However, they also tend to make the decomposition less precise in reconstructing $T$.

\section{Summary}

\begin{enumerate}
\item Vectors are lists of numbers, which specify the coordinates of a point in space. The number of entries in the list tells you the number of dimensions of its space. The length of a vector is its Euclidean distance to the origin.
\item Summing two vectors means to go to the point of the first vector and then move according to the coordinates of the second vector, but starting from where you landed rather than from the origin. So the result is another vector.
\item Matrices are coordinate transformations: multiplying a vector to a matrix means to find the coordinates of the vector in the new space transformed by the matrix. Multiplying two matrices gives another matrix, which performs both transformations one after the other.
\item Transposing means mirroring the matrix on its main diagonal. It can be used to flip the direction of edges in a directed network, or looking at two different modes of connections in a bipartite network.
\item A matrix's eigenvector is a vector that gets stretched by a factor (the eigenvalue) but does not change direction when multiplied by that matrix.
\item In Principal Component Analysis, we deconstruct a matrix in its ``principal components'': uncorrelated vectors that express most of the variation in the (correlated) values of the matrix. If no value in these vectors can be negative, we call it ``Non-Negative Matrix Factorization''.
\item Tensor decomposition is an operation expressing a multidimensional matrix as the result of the sum/product or other, simpler, tensors.
\end{enumerate}

\section{Exercises}

\begin{enumerate}
\item What is the length of the vector you obtain by summing $[0, 4]$ to $[5, 1]$?
\item Suppose:

$$
A = \begin{pmatrix}
    1 & 0 \\
    0 & 2 
\end{pmatrix}\ B = \begin{pmatrix}
    3 & 0 \\
    0 & -1 
\end{pmatrix}
$$

Are these two transformations commutative? Does applying $A$ first and $B$ second lead to the same transformation as applying $B$ first and $A$ second?

\item Calculate the eigenvalues and the right and left eigenvectors of the matrix from \url{http://www.networkatlas.eu/exercises/5/3/data.txt}. Make sure to sort the eigenvalues in descending order (and sort the eigenvectors accordingly). Only take the real part of eigenvalues and eigenvectors, ignoring the imaginary part.
\item Perform the eigendecompositions of the matrices from exercise $2$, showing that you can reconstruct the originals from their eigenvalues and eigenvectors.
\end{enumerate}

\part{Graph Representations}\label{par:graph}

\chapter{Basic Graphs}\label{cha:basic}

\section{Simple Graphs}\label{sec:basic-simple}
Every story should start from the beginning and, in this case, in the beginning was the graph\cite{bondy1976graph}\cite{west2001introduction}\cite{diestel2018graph}\cite{gross2005graph}. To explain and decompose the elements of a graph, I'm going to use the recurrent example of social networks. The same graph can represent different networks: power grids, protein interactions, financial transactions. Hopefully, you can effortlessly translate these examples into whatever domain you're going to work.

Let's start by defining the fundamental elements of a social network. In society, the fundamental starting point is you. The person. Following Euler's logic that I discussed in the introduction, we want to strip out the internal structure of the person to get to a node. It's like a point in geometry: it's the fundamental concept, one that you cannot divide up into any sub-parts. Each person in a social network is a node -- or vertex; in the book I'll treat these two terms as synonyms. We can also call nodes ``actors'' because they are the ones interacting and making events happen -- or ``entities'' because sometimes they are not actors: rather than making things happen, things happen to them. ``Actor'' is a more specific term which is not an exact synonym of ``node'', but we'll see the difference between the two once we complicate our network model just a bit\footnote{The understatement of the century.}, in Section \ref{sec:extended-multilayer}.

To add some notation, we usually refer to a graph as $G$. $V$ indicates the set of $G$'s vertices. Since $V$ is the set of nodes, to refer to the number of nodes of a graph we use $|V|$ -- some books will use $n$, but I'll try to avoid it. Throughout the book, I'll tend to use $u$ and $v$ to indicate single nodes.

So far, so good. However, you cannot have a society with only one individual. You need more than one. And, once you have at least two people, you need interactions between them. Again, following Euler, for now we forget about everything that happens in the internal structure of the communication: we only remember that an interaction is taking place. We will have plenty of time to make this model more complicated. The most common terms used to talk about interactions are ``edge'', ``link'', ``connection'' or ``arc''. While some texts use them with specific distinctions, for me they are going to be synonyms, and my preferred term will always be ``edge''. I think it's clearer if you always are explicit when you refer to special cases: sure, you can decide that ``arc'' means ``directed edge'', but the explicit formula ``directed edge'' is always better than remembering an additional term, because it contains all the information you need. (What the hell are ``directed edges''? Patience, everything will be clear)

Again, notation. $E$ indicates the set of $G$'s edges and $|E|$ is the number of edges -- some books will use $m$ as a synonym for $|E|$. Usually, when talking about a specific edge one will use the notation $(u,v)$, because edges are pairs of nodes -- unless we complicate the graph model. Now we have a way to refer to the simplest possible graph model: $G = (V, E)$, with $E \subseteq V \times V$. A graph is a set of nodes and a set of edges -- i.e. node pairs -- established among those nodes.

\begin{figure}
\centering
\begin{subfigure}{.075\columnwidth}
\includegraphics[width=\textwidth]{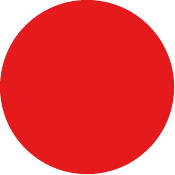}
\caption{}
\end{subfigure}
\qquad \qquad
\begin{subfigure}{.25\columnwidth}
\includegraphics[width=\textwidth]{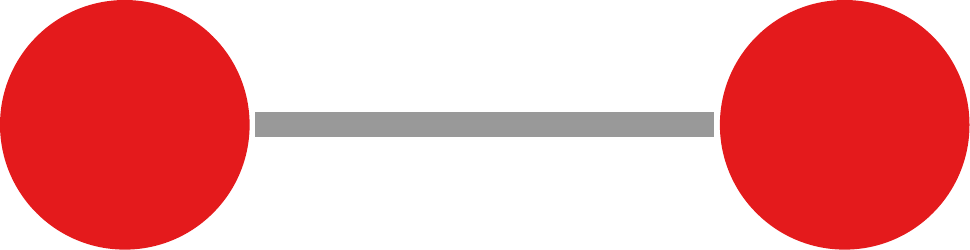}
\caption{}
\end{subfigure}
\qquad
\begin{subfigure}{.33\columnwidth}
\includegraphics[width=\textwidth]{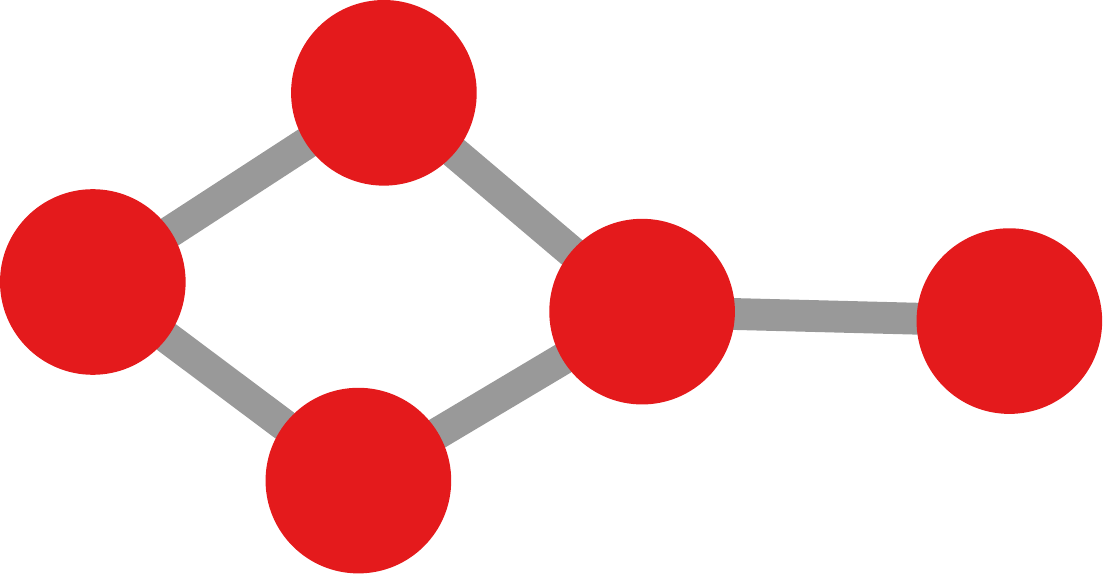}
\caption{}
\end{subfigure}
\caption{(a) A node. (b) An edge. (c) A simple graph.}
\label{fig:graph-elements}
\end{figure}

We're going to talk about how to visualize networks much later in Part \ref{par:netviz}, but it's better to introduce some visual elements now, otherwise how are we supposed to have figures before then? Nodes are usually represented as dots, or circles -- Figure \ref{fig:graph-elements}(a). Edges are lines connecting the dots -- Figure \ref{fig:graph-elements}(b). When all you have is nodes and edges, then you have a simple graph -- Figure \ref{fig:graph-elements}(c). Note that these visual elements are basic and widely used, but they are by no means the only way to visualize nodes and edges. In fact, when you want to convey a message about a network of non-trivial size, they're usually not a great idea.

\begin{figure}
\centering
\includegraphics[width=.66\columnwidth]{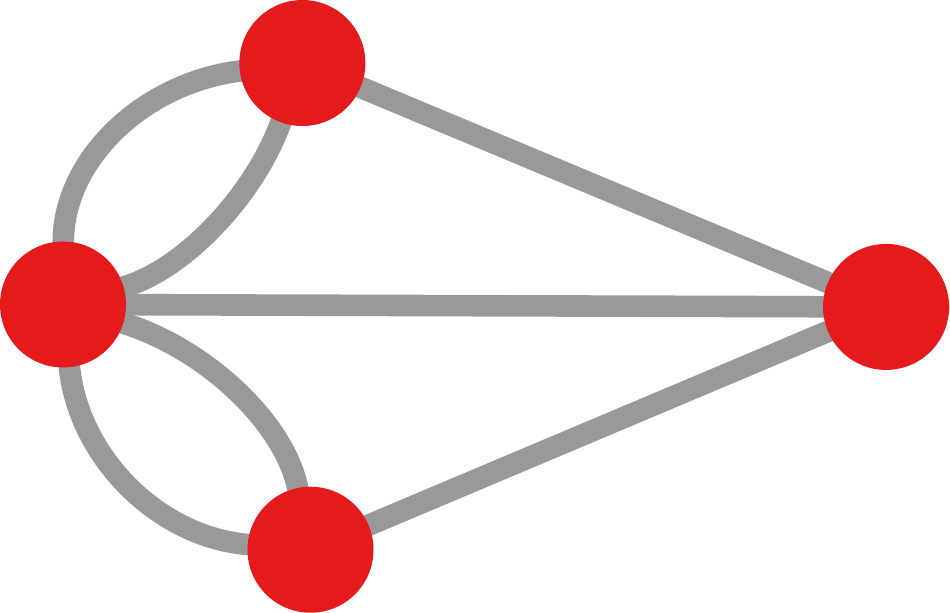}
\caption{The famous K\"{o}nigsberg graph Euler used.}
\label{fig:konig}
\end{figure}

The first famous graph in history is Euler's K\"{o}nigsberg graph, which I show in Figure \ref{fig:konig}. In the graph, each node represents a landmass and each edge represents a bridge connecting two landmasses. Since there were multiple bridges connecting the same landmasses, we have multiple edges between the same two nodes. This seemingly trivial fact is actually rather interesting.

``Simple graph'' means \textit{literally} simple: nothing more than nodes and edges -- no attributes, no possibility of having multiple connections between the same two nodes. If you add any special feature, it's not a simple graph any more. Under this light, we discover that Euler's first graph wasn't simple after all. It allowed for parallel edges: multiple edges between the same two nodes. Euler's first graph was a multigraph. That's so non-standard that we're not even going to talk about it in this chapter: you'll have to wait for the next one, specifically for Section \ref{sec:extended-multilayer}.

In our simple graph we also assume there are no self loops, which are edges connecting a node with itself. Our assumption is that we aren't psychopaths: everybody is friend with themselves, so we don't need to keep track of those connections.

\begin{figure}
\centering
\begin{subfigure}{.33\columnwidth}
\includegraphics[width=\textwidth]{figures/simple.pdf}
\caption{}
\end{subfigure}
\qquad \qquad
\begin{subfigure}{.33\columnwidth}
\includegraphics[width=\textwidth]{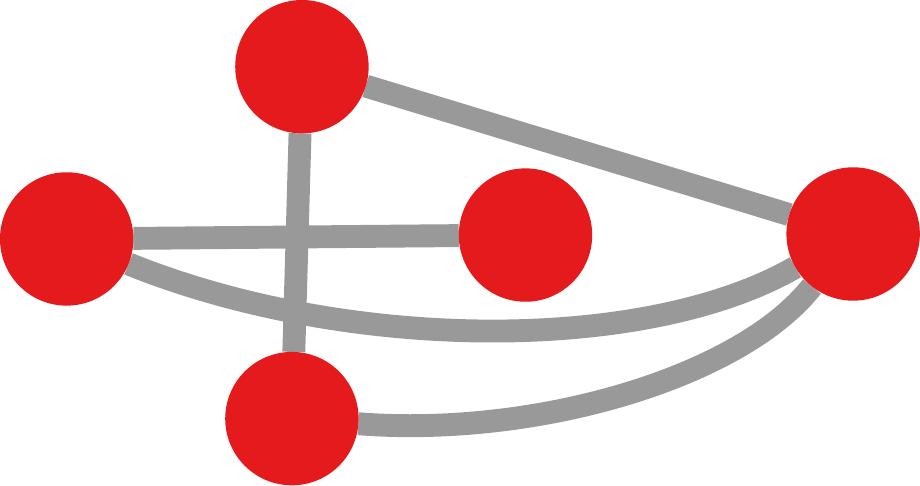}
\caption{}
\end{subfigure}
\caption{(a) A simple graph. (b) Its complement.}
\label{fig:complement}
\end{figure}

When you have a simple graph $G$, you can derive a series of special simple graphs related to $G$. For instance, you can derive the complement of $G$. This is equivalent to remove all of the original edges of $G$, and then connect all the unconnected pairs of nodes in $G$. Figure \ref{fig:complement} shows an example.

This operation basically views $G$ as a set of edges. If you take this perspective, you can define many operations on graphs as sets. Given two graphs $G'$ and $G''$, you can calculate their union, intersection, and difference, which are the union, intersection, and difference of their edge sets. The union of $G'$ and $G''$ is a graph $G$ that has the edges found in either $G'$ or $G''$; the intersection of $G'$ and $G''$ is a graph $G$ that has the edges found in both $G'$ and $G''$; and the difference of $G'$ and $G''$ is a graph $G$ that has the edges found in $G'$ but not in $G''$.

\begin{figure}
\centering
\begin{subfigure}{.33\columnwidth}
\includegraphics[width=\textwidth]{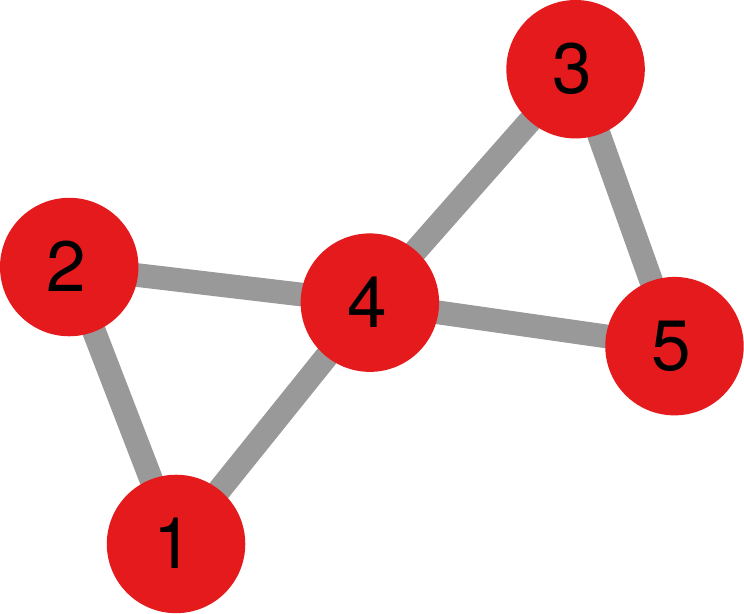}
\caption{}
\end{subfigure}\qquad\qquad
\begin{subfigure}{.25\columnwidth}
\includegraphics[width=\textwidth]{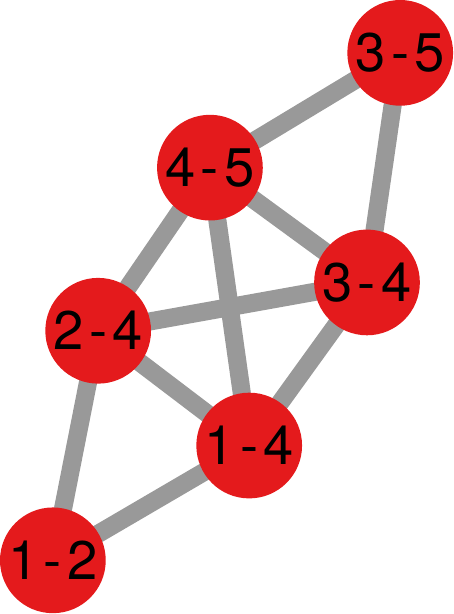}
\caption{}
\end{subfigure}
\caption{(a) A graph. (b) Its linegraph version.}
\label{fig:line-graph-ex}
\end{figure}

Another important special graph is the line graph\cite{whitney1932congruent}\cite{krausz1943demonstration}. The line graph of $G$ represents each of $G$'s edges as a node. Two nodes in the line graph are connected to each other if the edges they represent are attached to the same node in $G$. Figure \ref{fig:line-graph-ex} shows an example of line graph. We'll see how you can use line graphs to represent high order relationships in Chapter \ref{cha:hod}, to find overlapping communities in Chapter \ref{cha:ocd}, and to estimate similarities between networks in Chapter \ref{cha:netsimil}.

\section{Directed Graphs}\label{sec:basic-directed}
Simple graphs are awesome. They allow you to represent a surprising variety of different complex systems. But they are not the end all be all of network theory. There are many phenomena out there that cannot be simply reduced to a set of nodes interacting through a set of edges. Sometimes you really need to complicate stuff. In this and in the next section we're going to see two ways to enhance the simple graph models. They all work in the same way: by slightly modifying the definition of an edge. We're going to see even more fundamental reworkings of the simple graph model in Chapter \ref{cha:extended}.

The first thing we will do is realizing that not all relations are reciprocal. The fact that I consider you as my friend -- and I do, my dear reader -- doesn't necessarily mean that you also consider me as your friend -- wow, this book is getting very real very fast. We can introduce this asymmetry in the graph model. So far we said that $(u,v)$ is an edge and we implicitly assumed that $(u,v)$ is the same as $(v,u)$. Directed graphs\cite{harary1965structural} are graphs for which $(u,v) \neq (v,u)$.

\begin{figure}[t]
\centering
\begin{subfigure}{.32\columnwidth}
\includegraphics[width=\textwidth]{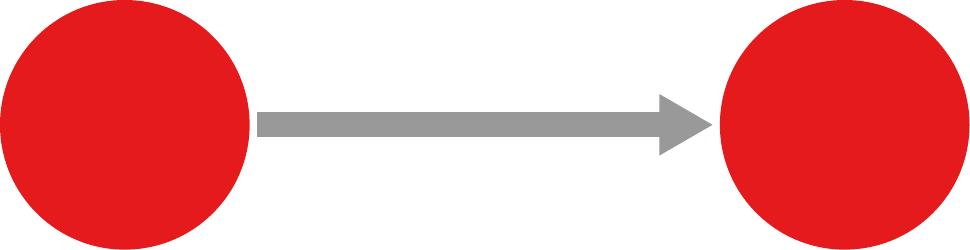}
\caption{}
\end{subfigure}
\qquad
\begin{subfigure}{.4\columnwidth}
\includegraphics[width=\textwidth]{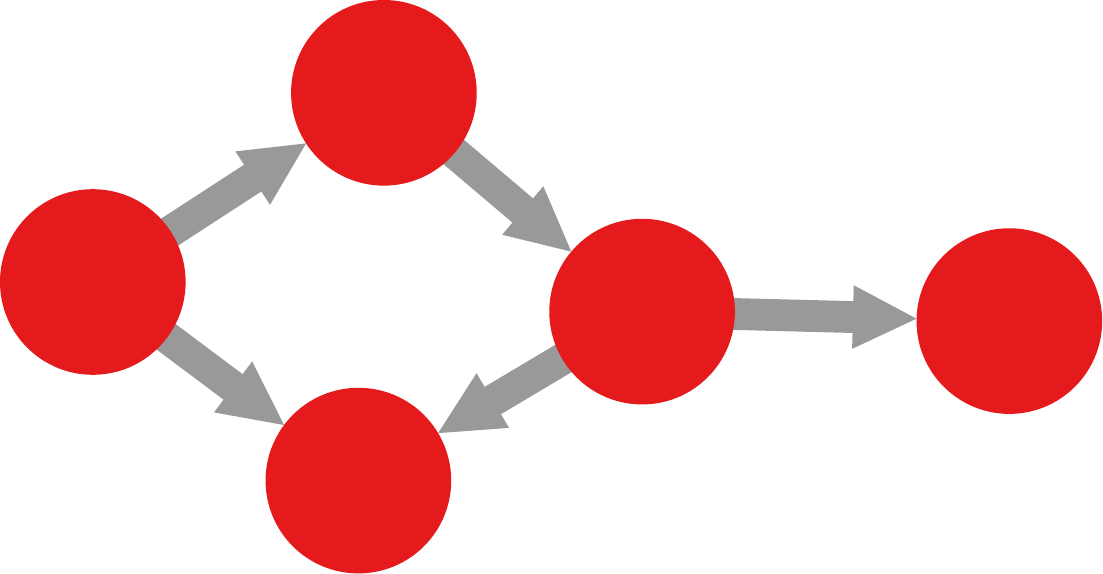}
\caption{}
\end{subfigure}
\caption{(a) A directed edge. (b) A directed graph.}
\label{fig:digraph-elements}
\end{figure}

In a message passing game, $(u,v)$ -- or $u \rightarrow v$ -- means that node $u$ can pass a message to node $v$, but $v$ cannot send it back to $u$. Directed graphs introduce all sorts of intricacies when it comes to finding paths in the network, a topic we're going to dissect in Chapter \ref{cha:paths}. The use of the arrow is a pretty straightforward metaphor to indicate the lack of reciprocity: relationships flow from the tail to the head of the arrow, not the other way around. It comes as no surprise, then, that we can use the arrow to indicate a directed edge, as we do in Figure \ref{fig:digraph-elements}(a). If $E$ contains directed edges, we have a directed graph -- Figure \ref{fig:digraph-elements}(b).

Note that, in a directed graph (or digraph) representation, an edge always has a direction. If two nodes have a reciprocal relationship, convention dictates that we draw two directed edges pointing in the two directions, to make such relationship explicit.

In general, when you have a directed graph $G$, you can calculate its reverse graph by flipping all edge directions.

\section{Weighted Graphs}\label{sec:basic-weighted}
Another way to make edges more interesting is realizing that two connections are not necessarily equally important in the network. One of the two might be much stronger than another. We are all familiar with the concepts of ``best friend'' and ``Facebook friend''. One is a much more tightly knit connection than the other.

For this reason, we can add weights to the edges\cite{barrat2004architecture}\cite{newman2004analysis}. A weight is simply an additional quantitative information we add to the connection. A possible notation could be $(u, v, w)$: nodes $u$ and $v$ connect to each other with strength $w$. So our graph definition now changes to $G = (V, E, W)$, where $W$ is our set of possible weights. $W$ is practically always included in the set of real numbers, and most of the times in the set of real positive numbers -- i.e. $W \subseteq R^+$. Now we have a weighted graph.

\begin{figure}
\centering
\includegraphics[width=.5\columnwidth]{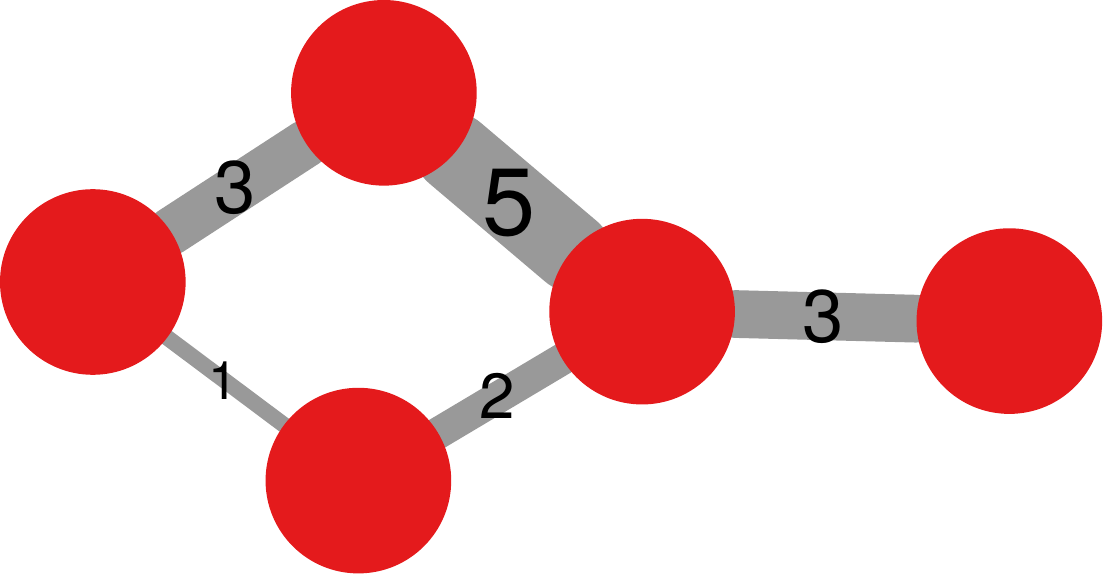}
\caption{A weighted graph. The weight of the edge dictates its label and thickness.}
\label{fig:weighted}
\end{figure}

Graphically, we usually represent the weight of a connection either by labeling the edge with its value, or simply by using visual elements such as the line thickness. I do both things in Figure \ref{fig:weighted}.

Edge weights can be interpreted in two opposite ways, depending on what the network is representing. They can be considered the \textit{proximity} between the two nodes or their \textit{distance}. This can and will influence the results of many algorithms you'll apply to your graph, so this semantic distinction matters. For instance, if you're looking for the shortest path (see Chapter \ref{cha:shortpath}) in a road network, your edge weight could mean different things. It could be a distance if it represents the length of the trait of road: longer traits will take more time to cross. Or it can be a proximity: it could be the throughput of the trait of road in number of cars per minute that can pass through it -- or the number of lanes. If the weight is a distance, the shortest path should avoid high edge weights. If the weight is a proximity, it should do its best to include them.

To sum up, ``proximity'' means that a high weight makes the nodes closer together; e.g. they interact a lot, the edge has a high capacity. ``Distance'' means that a high weight makes the nodes further apart; e.g. it's harder or costly to make the nodes interact.

Edge weights don't have to be positive. Nobody says nodes should be friends! Examples of negative edge weights can be resistances in electric circuits or genes downregulating other genes. This observation is the beginning of a slippery slope towards signed networks, which is a topic for another time (namely, for Section \ref{sec:extended-multilayer}, if you want to jump there).

The network in Figure \ref{fig:weighted} has nice integer weights. In this case, the edge weights are akin to counts. For instance, in a phone call network, it could be the number of times two people have called each other. Unfortunately, not all weighted networks look as neat as the example in Figure \ref{fig:weighted}. In fact, most of the weighted networks you might work with will have continuous edge weights. In that case, many assumptions you can make for count weights won't apply -- for instance when filtering connections, as we will see in Chapter \ref{cha:backboning}.

By far, the most common case is the one of correlation networks. In these networks, the nodes aren't really interacting directly with one another. Instead, we are connecting nodes because they are similar to each other, for some definition of similarity. For instance, we could connect brain areas via cortical thickness correlations\cite{bernhardt2011graph}, or currencies according to their exchange rate\cite{mizuno2006correlation}, or correlating the taxa presence in different biological communities\cite{friedman2012inferring}.

\begin{figure}
\centering
\includegraphics[width=\columnwidth]{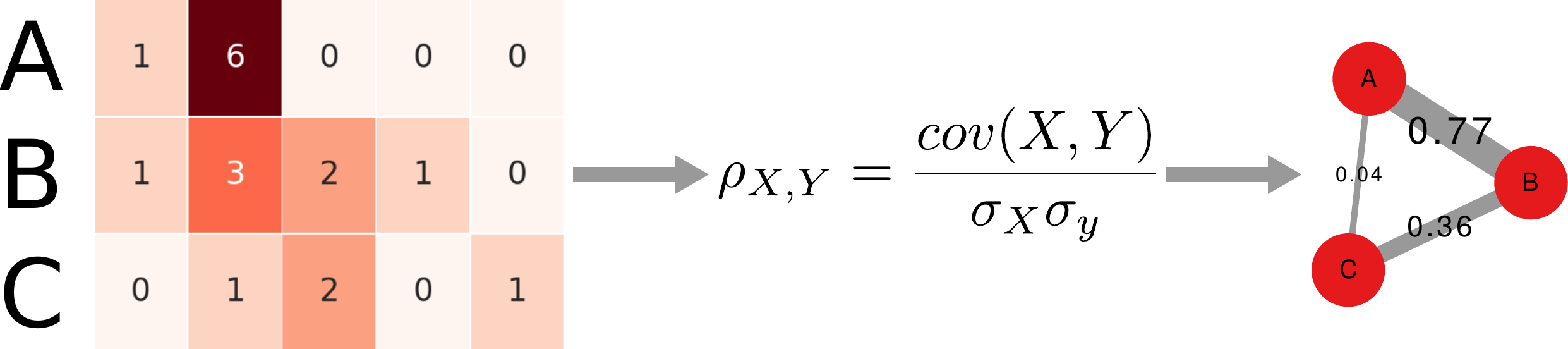}
\caption{A typical workflow for correlation networks: (left to right) from nodes represented as some sort of vectors, to a graph with a similarity measure as edge weigth.}
\label{fig:corr-net}
\end{figure}

These cases have more or less the same structure. I provide an example in Figure \ref{fig:corr-net}. In this case, nodes are numerical vectors, which could represent a set of attributes, for instance. We calculate a correlation between the vectors, or some sort of attribute similarity -- for instance mutual information (Section \ref{sec:prob-mi}). We then obtain continuous weights, which typically span from $-1$ to $1$. And, since every pair of nodes have a similarity (because any two vectors can be correlated, minus extremely rare degenerate cases), every node is connected to every other node. So, when working with similarity networks, you will have to filter your connections somehow, a process we call ``network backboning'' which is far less trivial that it might sound. We will explore it in Chapter \ref{cha:backboning}.

\section{Network Types}\label{sec:extended-types}
Now that you know more about the various features of different network models, we can start looking at different types of networks. I'm going to use a taxonomy for this section. I find this way of organizing networks useful to think about the objects I work with.

\subsection{Simple Networks}
The first important distinction between network types is between \textit{simple} and \textit{complex} networks. A simple network is a network we can fully describe analytically. Its topological features are exact and trivial. You can have a simple formula that tells you everything you need to know about it. In complex networks that is not possible, you can only use formulas to approximate their salient characteristics.

The difference between a simple network and a complex network is the same between a sphere and a human being. You can fully describe the shape of a sphere with a few formulas: its surface is $4 \pi r^2$, its volume is $\dfrac{4}{3} \pi r^3$. If you know $r$ you know everything you need to know about the sphere. Try to fully describe the shape of a human being, internal organs included, starting from a single number. Go on, I have time.

\begin{figure}
\centering
\begin{subfigure}[t]{.33\columnwidth}
\includegraphics[width=\textwidth]{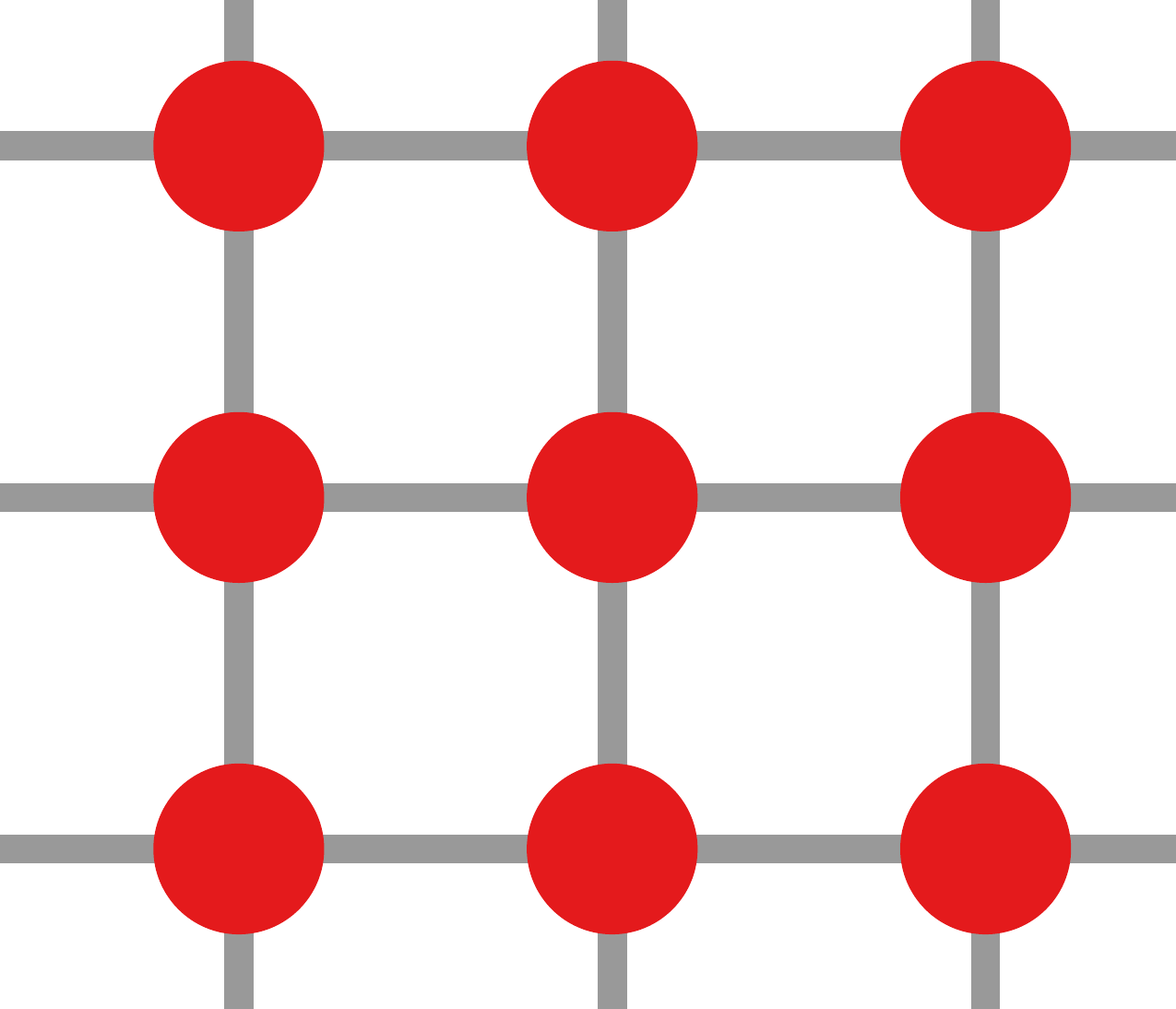}
\caption{}
\end{subfigure}
\qquad
\begin{subfigure}[t]{.33\columnwidth}
\includegraphics[width=\textwidth]{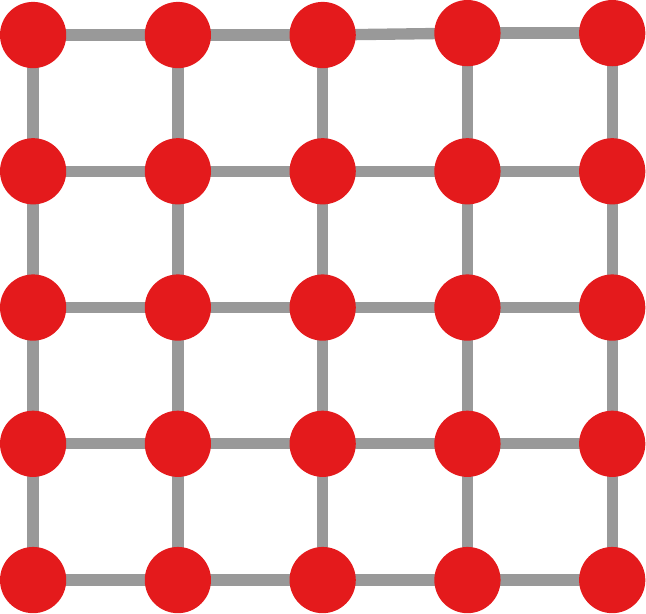}
\caption{}
\end{subfigure}
\caption{(a) An infinite lattice without boundaries. (b) A finite lattice with $25$ nodes and $40$ edges.}
\label{fig:lattice-example}
\end{figure}

What do simple networks look like? I think the easiest example conceivable is a square lattice. This is a regular grid, in which each node is connected to its four nearest neighbors. Such lattice can either span indefinitely (Figure \ref{fig:lattice-example}(a)), or it can have a boundary (Figure \ref{fig:lattice-example}(b)). Their fundamental properties are more or less the same. Knowing this connection rule that I just stated allows you to picture any lattice ever. That is why this is a simple topology.

\begin{figure}
\centering
\begin{subfigure}[t]{.17\columnwidth}
\includegraphics[width=\textwidth]{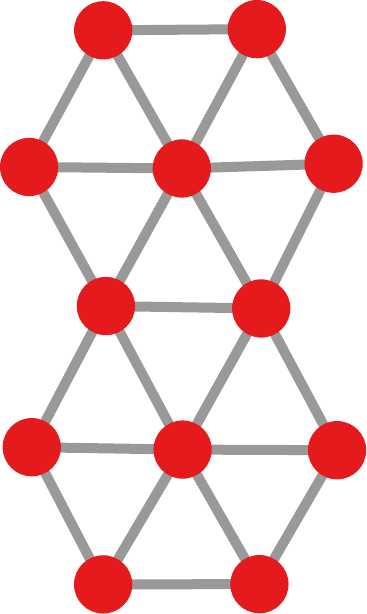}
\caption{}
\end{subfigure}
\quad
\begin{subfigure}[t]{.35\columnwidth}
\includegraphics[width=\textwidth]{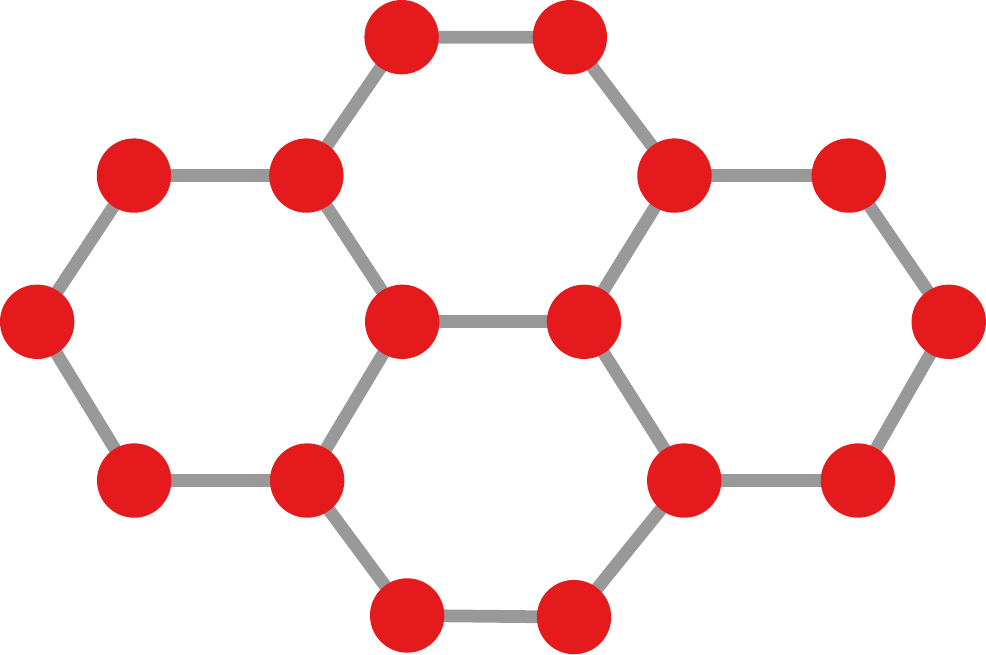}
\caption{}
\end{subfigure}
\quad
\begin{subfigure}[t]{.0275\columnwidth}
\includegraphics[width=\textwidth]{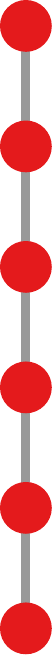}
\caption{}
\end{subfigure}
\quad
\begin{subfigure}[t]{.27\columnwidth}
\includegraphics[width=\textwidth]{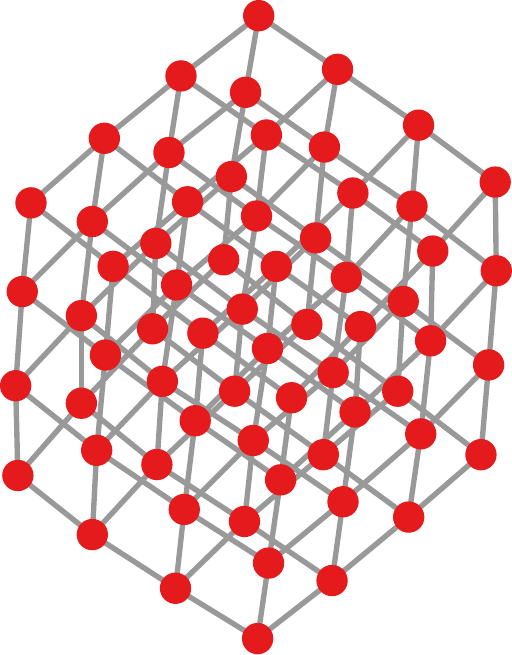}
\caption{}
\end{subfigure}
\caption{Different lattice types. (a) Triangular. (b) Hexagonal. (c) One dimensional. (d) Three dimensional cube.}
\label{fig:lattice-example2}
\end{figure}

Regular lattices can come in many different shapes besides square, for instance triangular (Figure \ref{fig:lattice-example2}(a)) or hexagonal (Figure \ref{fig:lattice-example2}(b)). They also don't necessarily have to be two dimensional as the examples I made so far: you can have 1D (Figure \ref{fig:lattice-example2}(c)) and 3D (Figure \ref{fig:lattice-example2}(d)) lattices -- the latter might be a bit hard to see, but it is a cube of with four nodes per side.

Even if deceptively simple, lattices can be extremely useful and are used as starting point for many advanced tasks. For instance, they are at the basis of the small-world graph generator (Section \ref{sec:physicsmodels-ws}) and of our understanding of epidemic spread in society (Chapter \ref{cha:epidemics}).

\begin{figure}
\centering
\begin{subfigure}[t]{.44\columnwidth}
\includegraphics[width=\textwidth]{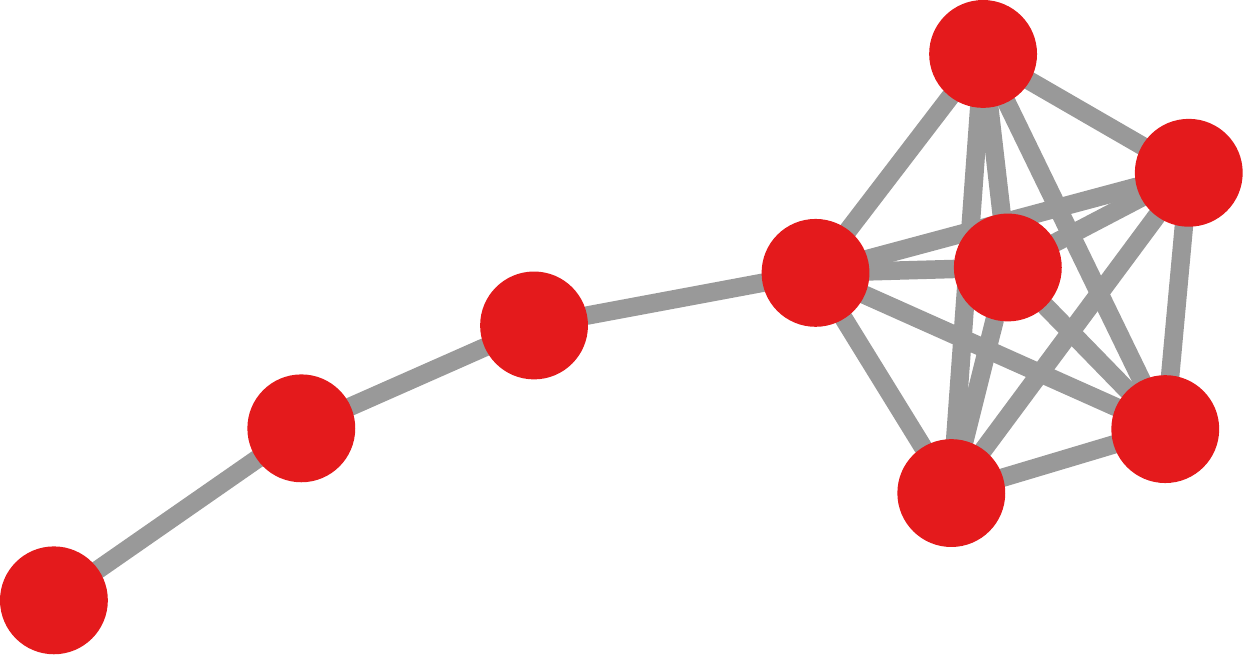}
\caption{}
\end{subfigure}
\quad
\begin{subfigure}[t]{.2\columnwidth}
\includegraphics[width=\textwidth]{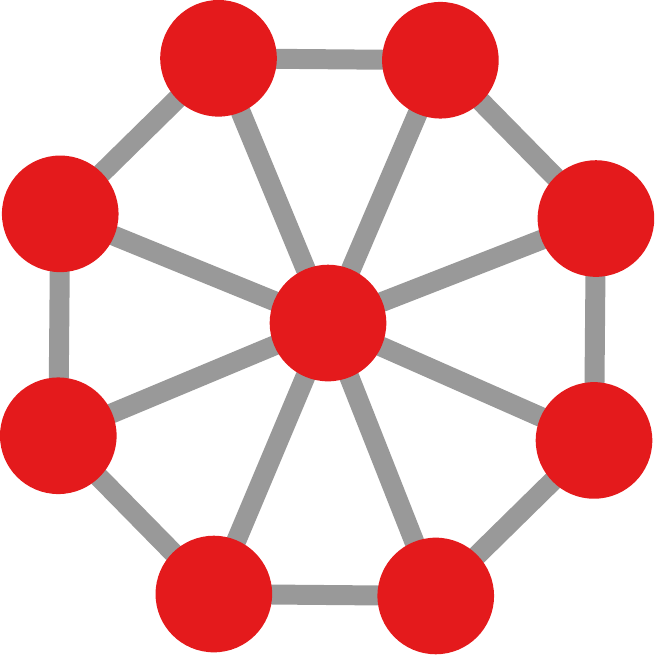}
\caption{}
\end{subfigure}
\quad
\begin{subfigure}[t]{.26\columnwidth}
\includegraphics[width=\textwidth]{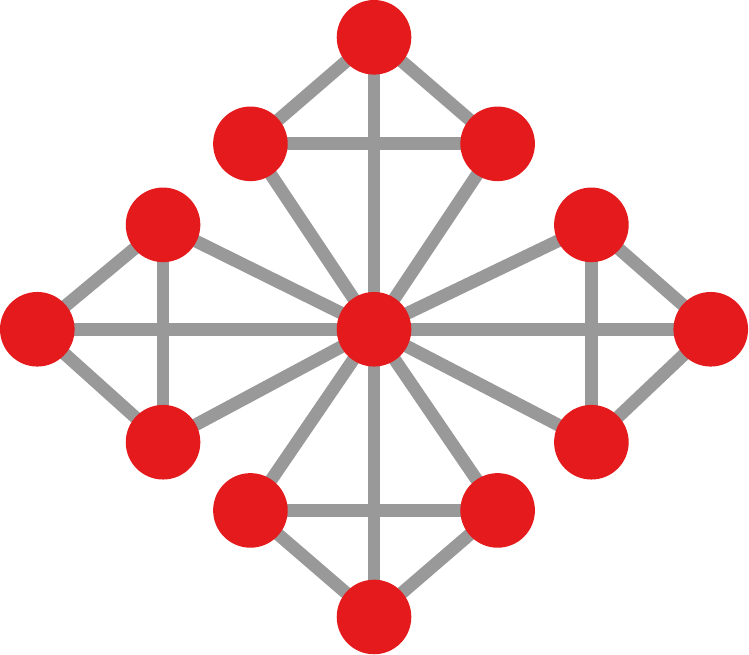}
\caption{}
\end{subfigure}
\caption{Different simple networks. (a) Lollipop graph. (b) Wheel graph. (c) Windmill graph.}
\label{fig:simple-examples}
\end{figure}

Lattices are not the only simple network out there. There is a wide collection of other network types. These are usually developed as the simplest illustrative examples for explaining new problems or algorithms. A few of my favorites (yes, I'm the kind of person who has favorite graphs) are the lollipop graph\cite{brightwell1990maximum} (a set of $n$ nodes all connected to each other plus a path of $m$ nodes shooting out of it, Figure \ref{fig:simple-examples}(a)), the wheel graph (which has a center connected to a circle of $m$ nodes, Figure \ref{fig:simple-examples}(b)), and the windmill graph (a set of $n$ graphs with $m$ nodes and all connections to each other, also all connected to a central node, Figure \ref{fig:simple-examples}(c)). Once you figure out what rule determines each topology, you can generate an arbitrary set of arbitrary size of graphs that all have the same properties.

\subsection{Complex Networks}
If simple networks were the only game in town, this book would not exist. That is because, as I said, you can easily understand all their properties from relatively simple math. That is not the case when the network you're analyzing is a complex network. Complex networks model complex systems: systems that cannot be fully understood if all you have is a perfect description of all their parts. The interactions between the parts let global properties emerge that are not the simple sum of local properties. There isn't a simple wiring rule and, even knowing all the wiring, some properties can still take you by surprise.

Personally, I like to divide complex networks into two further categories: complex network \textit{with fundamental metadata} and \textit{without fundamental metadata}. We saw that you can have edge metadata, the direction and weight. In Chapter \ref{cha:extended} we'll see you can have even more, attached to both nodes and edges. The difference I'm trying to make is that, if the metadata are fundamental, they change the way you interpret some or all the metadata themselves.

To understand \textit{non-fundamental} metadata, think about the fact that social networks, infrastructure networks, biological networks, and so on, model different systems and have different metadata attached to their nodes and edges. They can be age/gender, activation types, up- and down-regulation. However, the algorithms and the analyses you perform on them are the same, regardless of what the networks represent. They have nodes and edges and you treat them as such. You perform the Euler operation: you forget about all that is unnecessary so you can apply standardized analytic steps.

That is emphatically not true for networks with \textit{fundamental} metadata. In that case, you need to be aware of what the metadata represent, because they change the way you perform the analysis and you interpret the results. A few examples:

\begin{itemize}
\item \textit{Affiliation networks}. These are networks that, for instance, connect individuals to the groups they belong to. Here it is clear that one node type includes the other -- the group includes the individual. This is fundamentally different when you have node types at an equal level -- for instance if you connect people to the products they buy.
\item \textit{Interdependent networks}. These usually model some sort of physical system, for instance computers connected to the power plants they control. Edges express the dependencies of one node on the other they connect to. In this case, the removal of one node in one layer has immediate and non-trivial repercussions on all the layers depending into it, often with catastrophic consequences (see Section \ref{sec:epidemapps-interdependent}) -- which may not be true for other networks.
\item \textit{Correlation networks}. We saw a glimpse of these networks when we looked at weighted graphs. Here we have constraints on the edge weights, which can also be negative. The interpretation of such edge weights is different from what you would have in regular weighted networks. For instance, edges with very low weights are important here, because a strong negative correlation is interesting, even if its value ($-1$) is lower than no correlation at all ($0$).
\end{itemize}

A special mention for this class of networks should go to Bayesian networks\cite{jensen1996introduction}\cite{friedman1997bayesian}\cite{friedman1999learning}. In a Bayesian network, each node is a variable and directed edges represent dependencies between variables. If knowing something about the status of variable $u$ gives you information about the status of variable $v$, then you will connect $u$ to $v$ with a directed $(u,v)$ edge.

In the classical example, you might have three variables: the probability of raining, the probability of having the sprinklers on, and the probability that the grass is wet. Clearly, rain and sprinklers both might cause the grass to be wet, so the two variables point to them. Rain also might influence the sprinklers, because the automatic system to save water will not turn them on when it's raining, since it would be pointless. Obviously, the fact that the sprinklers are on will have no effect on whether it will rain or not. 

\begin{figure}
\centering
\begin{subfigure}{.38\columnwidth}
\includegraphics[width=\textwidth]{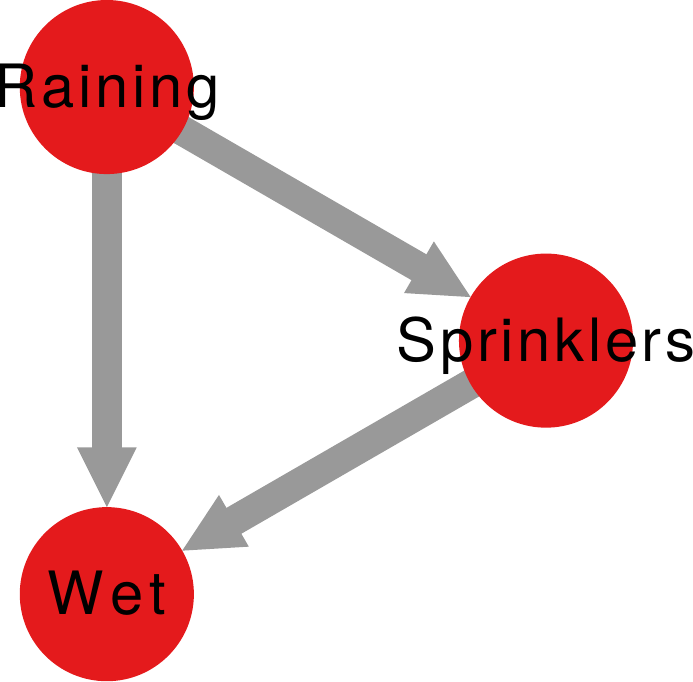}
\caption{}
\end{subfigure}\qquad
\begin{subfigure}{.36\columnwidth}
  \centering
  \begin{tabular}{l|cc}
    & \multicolumn{2}{c}{Rain}\\
    & T & F \\
    \hline
    & $0.20$ & $0.80$\\
    \multicolumn{3}{c}{}\\
  \end{tabular}
  \begin{tabular}{l|cc}
    & \multicolumn{2}{c}{Sprinkler}\\
    Rain & T & F \\
    \hline
    T & $0.01$ & $0.99$\\
    F & $0.20$ & $0.80$\\
    \multicolumn{3}{c}{}\\
  \end{tabular}
  \begin{tabular}{ll|cc}
    && \multicolumn{2}{c}{Wet}\\
    Rain & Sprinkler & T & F \\
    \hline
    T & T & $0.99$ & $0.01$\\
    T & F & $0.98$ & $0.02$\\
    F & T & $0.97$ & $0.03$\\
    F & F & $0.01$ & $0.99$\\
  \end{tabular}
\caption{}
\end{subfigure}
\caption{(a) A Bayesian network. (b) The conditional probability tables for the node states. The tables are referring to, from top to bottom: Rain, Sprinkler, Wet.}
\label{fig:bayesian}
\end{figure}

We can model this system with the simple Bayesian network in Figure \ref{fig:bayesian}(a) and the corresponding conditional probability tables in Table \ref{fig:bayesian}(b). Bayesian networks are usually the output of a machine learning algorithm. The algorithm will learn the best network that fits the observations. Then, you can use the network to predict the most likely probability of the state of a variable given a new observation of a subset of variables.

Simple examples like this might seem boring, but when you start having hundreds of variables you can find interesting patterns by applying some of the techniques you will learn later on. For instance, you might discover sets of variables that are independent of each other, even if, at first glance, it might be difficult to tell.

\begin{figure}
\centering
\begin{subfigure}[t]{.3\columnwidth}
\includegraphics[width=\textwidth]{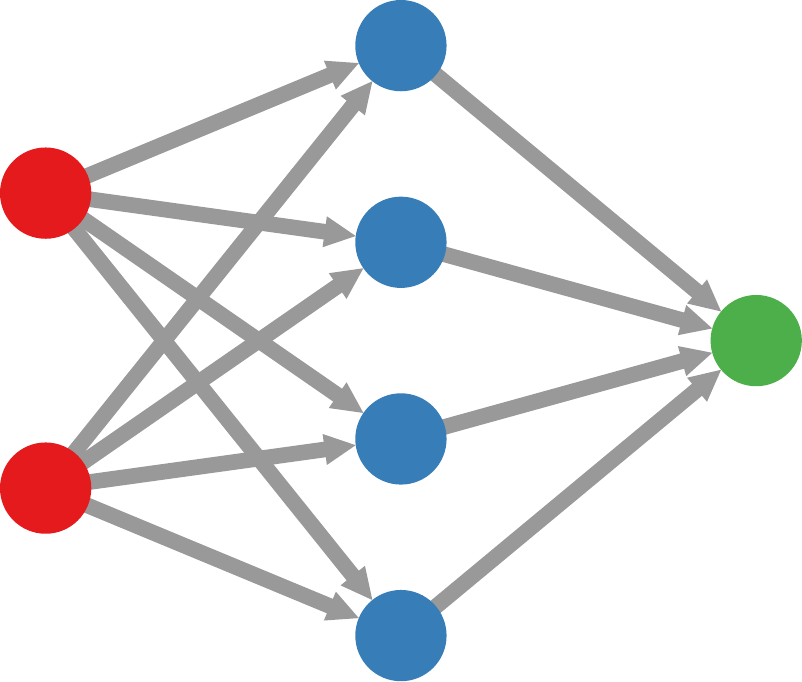}
\caption{Feedforward.}
\end{subfigure}
\quad
\begin{subfigure}[t]{.3\columnwidth}
\includegraphics[width=\textwidth]{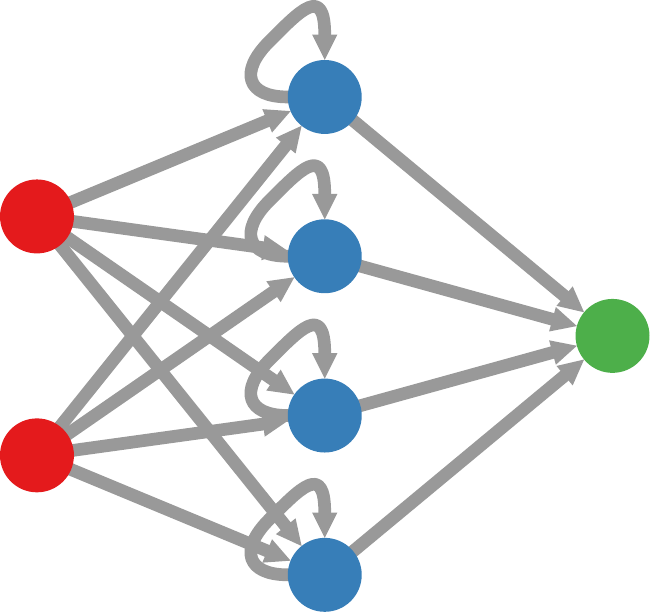}
\caption{Recurrent.}
\end{subfigure}
\quad
\begin{subfigure}[t]{.3\columnwidth}
\includegraphics[width=\textwidth]{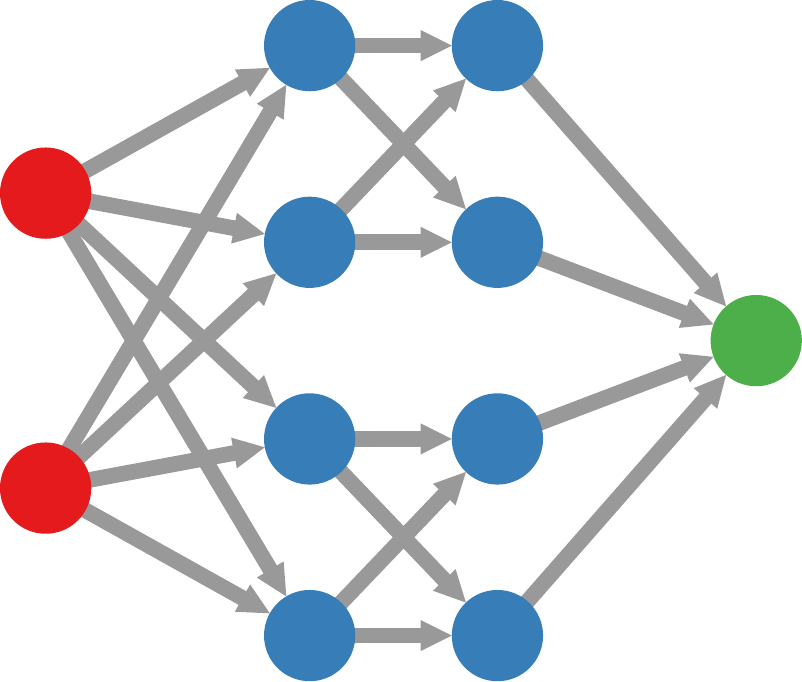}
\caption{Modular.}
\end{subfigure}
\caption{Different neural networks. The node color determines the layer type: input (red), hidden (blue), output (green).}
\label{fig:neural-nets}
\end{figure}

A not so distant relative of Bayesian networks are neural networks, the bread and butter of machine learning these days. Notwithstanding their amazing -- and, sometimes, mysterious -- power, neural networks are actually much more similar to simple networks than to complex ones. Differently from Bayesian networks, the wiring rules of neural networks -- of which I show some examples in Figure \ref{fig:neural-nets} -- are usually rather easy to understand.

The way they work is that the weight on each node of the output layer is the answer the model is giving. This weight is directly dependent on a combination of the weights of the nodes in the last hidden layer. The contribution of each hidden node is proportional to the weight of the edge connecting it to the output node. Recursively, the status of each node in the hidden layer is a combination of all its incoming connections -- combining the edge weight to the node weight at the origin. The first hidden layer will be directly dependent on the weights of the nodes in the input layer, which are, in turn, determined by the data.

What the model does is simply finding the combination of edge weights causing the output layer's node weights to maximize the desired quality function. 

\section{Summary}

\begin{enumerate}
\item The mathematical representation of a network is the graph: a collection of nodes -- the actors of the network --, and edges -- the connections among those actors. In a simple graph, no additional feature can be added, and there is only one edge between a pair of nodes.
\item If connections are not symmetric, meaning that if you consider me your friend I don't necessarily consider you mine, then we have directed graphs. In directed graphs, edges have a direction so relations flow one way, unless there is a reciprocal edge pointing back.
\item In weighted graphs, connections can be more or less strong, indicated by the weight of the edge, a numerical quantity. It doesn't have to be a discrete number, nor necessarily positive: for instance in correlation networks you can have negative continuous weights.
\item Weights can have two meanings: proximity -- the edge is the strength of a friendship --, or distance -- the edge is a cost to pay to cross from one node to another. Different semantics imply that some algorithms' results should be interpreted differently.
\item Simple networks are networks whose topology can be fully described with simple rules. For instance, in regular lattices you place nodes uniformly in a space and you connect them with their nearest neighbors.
\end{enumerate}

\section{Exercises}

\begin{enumerate}
\item Calculate $|V|$ and $|E|$ for the graph in Figure \ref{fig:graph-elements}(c).
\item Mr. $A$ considers Ms. $B$ a friend, but she doesn't like him back. She has a reciprocal friendship with both $C$ and $D$, but only $C$ considers $D$ a friend. $D$ has also sent friend requests to $E$, $F$, $G$, and $H$ but, so far, only $G$ replied. $G$ also has a reciprocal relationship with $A$. Draw the corresponding directed graph.
\item Draw the previous graph as undirected and weighted, with the weight being $2$ if the connection is reciprocal, $1$ otherwise.
\item Draw a correlation network for the vectors in \url{http://www.networkatlas.eu/exercises/6/4/data.txt}, by only drawing edges with positive weights, ignoring self loops.
\end{enumerate}

\chapter{Extended Graphs}\label{cha:extended}
The world of simple graphs is... well... simple. The only thing complicating it a bit so far was adding some information on the edges: whether they are asymmetric -- meaning $(u,v) \neq (v,u)$ -- and whether they are strong or weak. Unfortunately, that's not enough to deal with everything reality can throw your way. In this chapter, I present even more graph models, which go beyond the simple addition of edge information.

\section{Bipartite Graphs}\label{sec:extended-bip}
\begin{figure}[b]
\centering
\begin{subfigure}{.4\columnwidth}
\includegraphics[width=\textwidth]{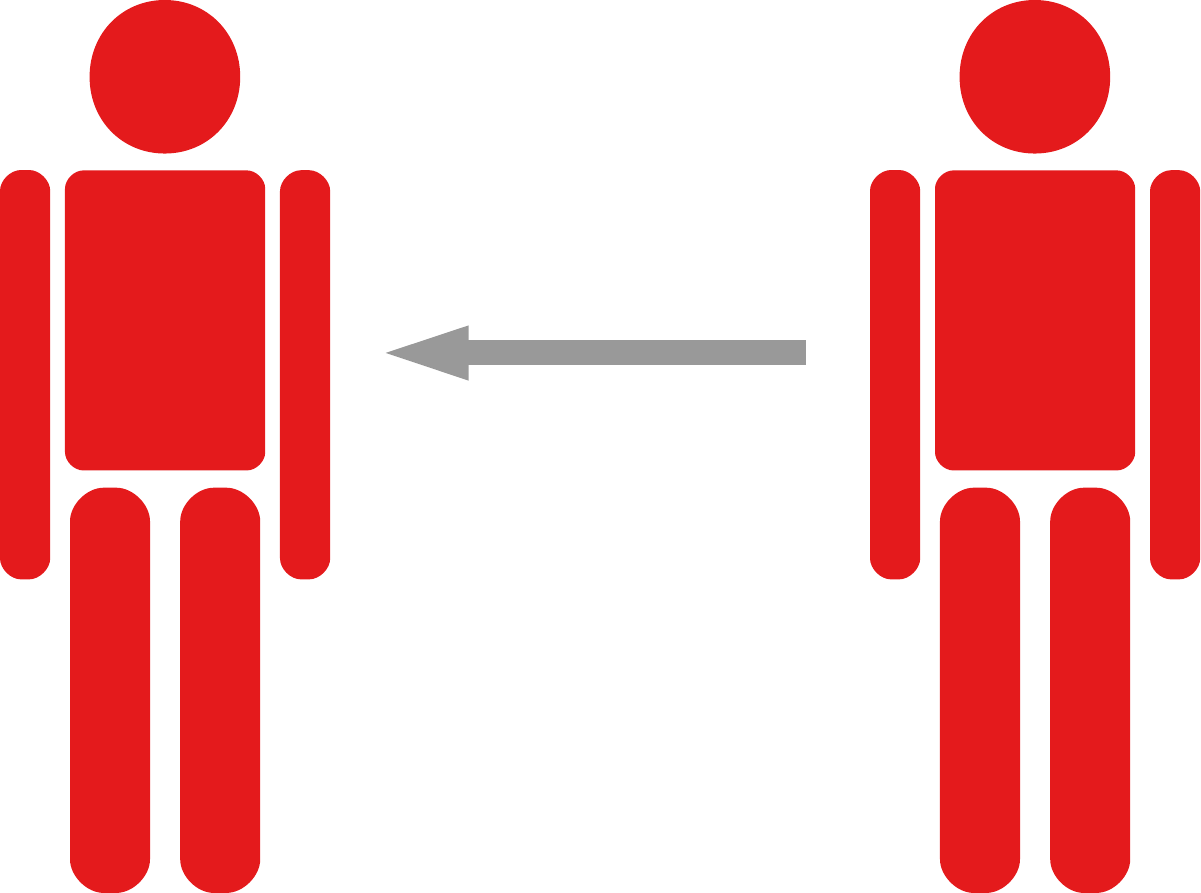}
\caption{}
\end{subfigure}
\qquad
\begin{subfigure}{.4\columnwidth}
\includegraphics[width=\textwidth]{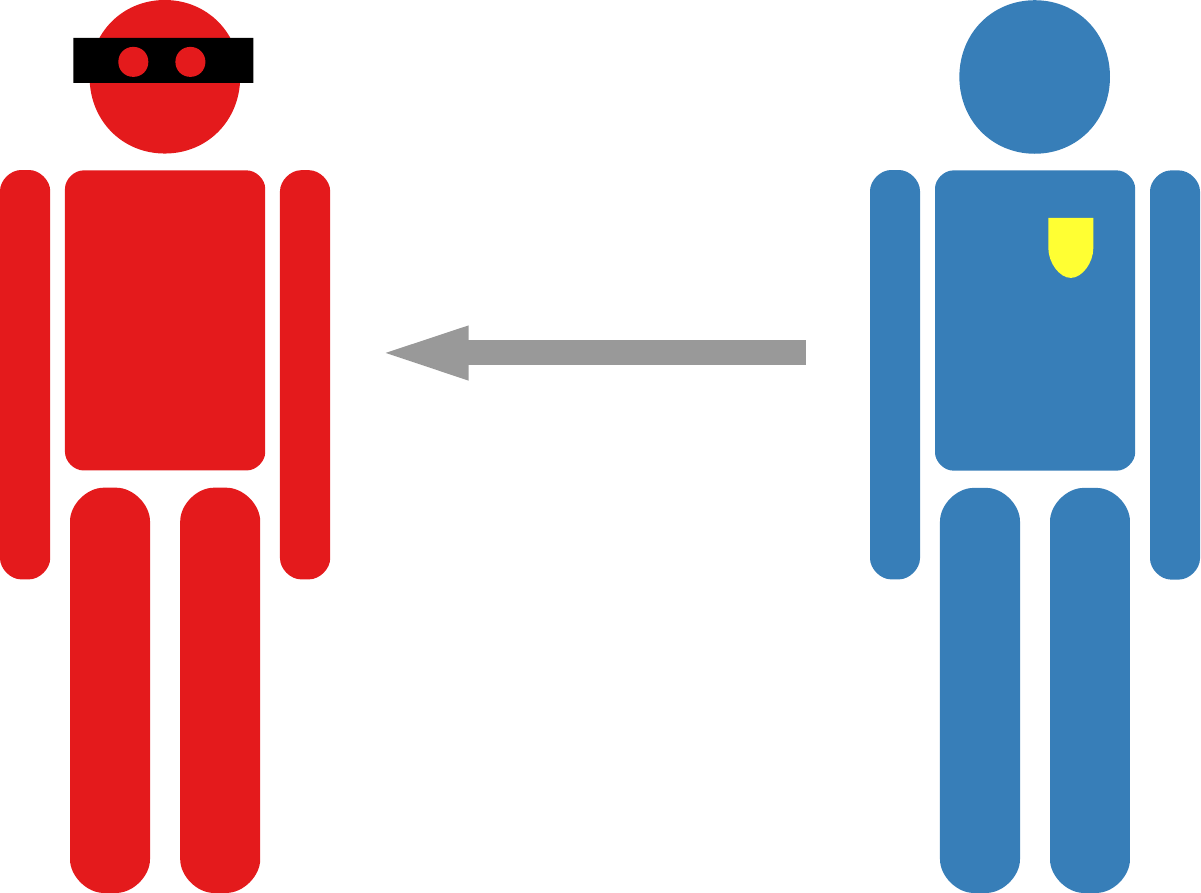}
\caption{}
\end{subfigure}
\caption{(a) A simple graph representing a social network with no additional constraints. (b) A cop-thief bipartite network: nodes can be either a cop or a thief, and cops can only catch (connect to) thieves.}
\label{fig:bipartite-example}
\end{figure}

So far we have talked about networks in which relations run between peers: nodes are all the same to us. But nodes might belong to two distinct classes. And connections can only be established between members of different classes. Figure \ref{fig:bipartite-example} provides an example. In a social network without node attributes nor types, anybody can be friend with anybody else and there isn't much to distinguish two nodes. But if we want to connect cops with the thieves they catch, then we are establishing additional connecting rules. Thieves don't catch each other. And, hopefully, cops aren't thieves. Another example could be connecting workers to the buildings hosting their offices.

\begin{figure}
\centering
\begin{subfigure}{.25\columnwidth}
\includegraphics[width=\textwidth]{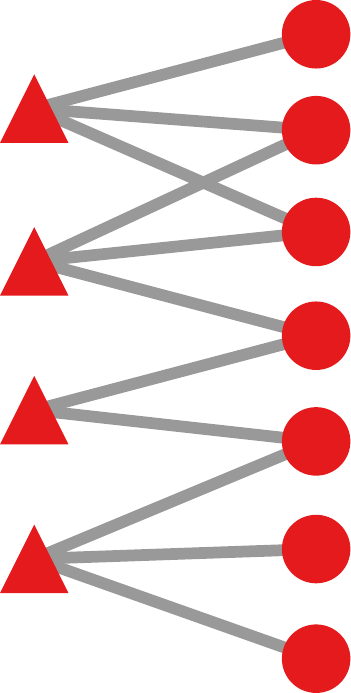}
\caption{}
\end{subfigure}
\qquad
\begin{subfigure}{.35\columnwidth}
\includegraphics[width=\textwidth]{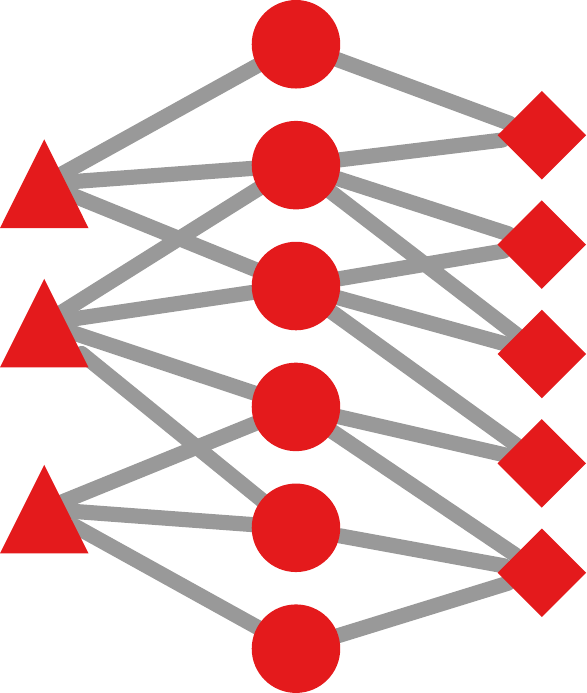}
\caption{}
\end{subfigure}
\caption{(a) An example of a bipartite network. (b) An example of a tripartite network.}
\label{fig:bipartite-tripartite}
\end{figure}

Stripping down the model to a minimum, bipartite networks are networks in which nodes must be part of either of two classes ($V_1$ and $V_2$) and edges can only be established between nodes of unlike type\cite{asratian1998bipartite}\cite{guillaume2004bipartite}. Formally, we would say that $G = (V_1, V_2, E)$, and that $E$ can only contain edges like $(v_1, v_2)$, with $v_1 \in V_1$ and $v_2 \in V_2$. Figure \ref{fig:bipartite-tripartite}(a) depicts an example.

Bipartite networks are used for countless things, connecting: countries to the products they export\cite{hidalgo2009building}, hosts to guest in symbiotic relationships\cite{muegge2011diet}, users to the social media items they tag\cite{lambiotte2006collaborative}, bank-firm relationships in financial networks\cite{marotta2015bank}, players-bands in jazz\cite{gleiser2003community}, listener-band in music consumption\cite{lambiotte2005uncovering}, plant-pollinators in ecosystems\cite{campbell2011network}, and more. You get the idea. Bipartite networks pop up everywhere.

However, by a curious twist of fate, the algorithms able to work directly on bipartite structures are less studied than their non-bipartite counterparts. For instance, for every community discovery algorithm that works on bipartite networks you have a hundred working on non-bipartite ones. The distinction is important, because the standard assumptions of non-bipartite community discovery do not hold in bipartite networks, as we will see in Part \ref{par:cd}.

Why would that be the case? Because practically everyone who works on bipartite networks projects them. Most of the times, you are interested only in one of the two node types. So you create a unipartite version of the network connecting all nodes in $V_1$ to each other, using some criteria to make the $V_2$ count. The trivial way is to connect all $V_1$ nodes with at least a common $V_2$ neighbor. This is so widely done and so wrong that I like to call it the Mercator bipartite projection, in honor of the most used and misunderstood map projection of all times. We'll see in Chapter \ref{cha:projections} why that's not very smart, and the different ways to do a better job.

Why stopping at bipartite? Why not go full $n$-partite? For instance, a paper I cited before actually builds a tri-partite network (Figure \ref{fig:bipartite-tripartite}(b) depicts an example): users connect to the social media they tag and with the tags they use. However, the gains you get from a more precise data structure quickly become much lower than the added complexity of the model. Even tripartite networks are a rarity in network science. A couple of examples are the recipe-ingredient-compound structure of the flavor network\cite{ahn2011flavor}, or the aid organization-country-issue structure\cite{coscia2013structure}.

\section{Multilayer Graphs}\label{sec:extended-multilayer}
In this section I describe models you can use to analyze multilayer networks. This should not be confused with the similarly-sounding, but actually completely different, multilevel network analysis\cite{lazega2015multilevel}. This is a whole different way to analyze social network data using multilevel analysis, of which I know little and I will attempt to cover in future versions of this book.

\subsection{One-to-One}
Traditionally, network scientists try to focus on one thing at a time. If they are interested in analyzing your friendship patterns, they will choose one network that closely approximates your actual social relations and they will study that. For instance, they will download a sample of the Facebook graph. Or they will analyze tweets and retweets.

However, in some cases, that is not enough to really grasp the phenomenon one wants to study. If you want to predict a new connection on Facebook, something happening in another social media might have influenced it. Two people might have started working in the same company and thus first connected on Linkedin, and then became friends and connected on Facebook. Such scenario could not be captured by simply looking at one of the two networks. Network scientists invented multilayer networks\cite{kivela2014multilayer}\cite{de2013mathematical}\cite{boccaletti2014structure}\cite{berlingerio2013multidimensional}\cite{dickison2016multilayer}\cite{magnani2011ml} to answer this kind of questions. 

\begin{figure}
\centering
\begin{subfigure}{.32\columnwidth}
\includegraphics[width=\textwidth]{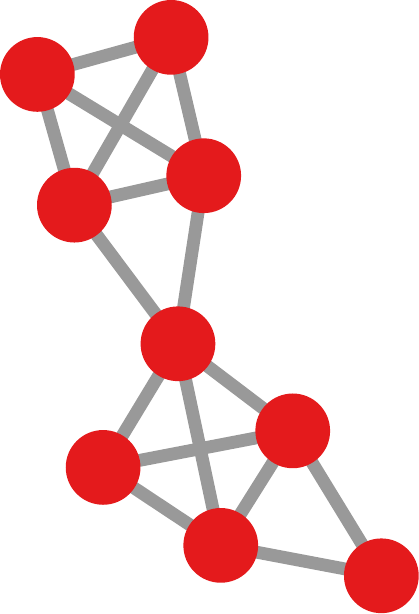}
\caption{}
\end{subfigure}
\begin{subfigure}{.32\columnwidth}
\includegraphics[width=\textwidth]{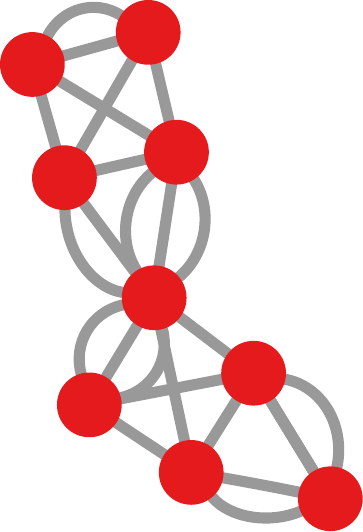}
\caption{}
\end{subfigure}
\begin{subfigure}{.31\columnwidth}
\includegraphics[width=\textwidth]{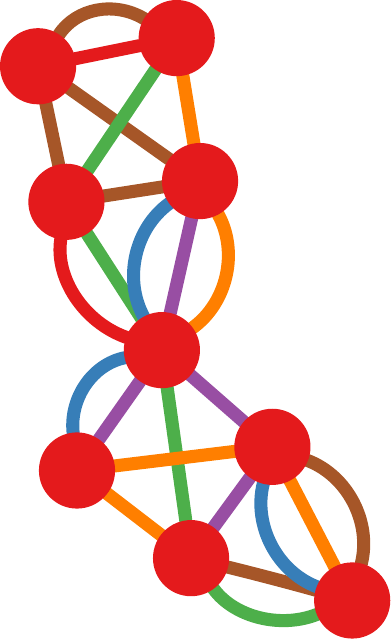}
\caption{}
\end{subfigure}
\caption{(a) A simple graph. (b) A multigraph, with multiple edges between the same node pairs. (c) A multilayer network, where each edge has a type (represented by its color).}
\label{fig:multigraph-multilayer}
\end{figure}

There are two ways to represent multilayer networks. The simpler is to use a multigraph. Remember Euler's parallel edges in the K\"{o}nigsberg graph from Figure \ref{fig:konig}? That's what makes a multigraph. Differently from a simple graph (Figure \ref{fig:multigraph-multilayer}(a)), in which every pair of nodes is forced to have at most one edge connecting them, in a multigraph (Figure \ref{fig:multigraph-multilayer}(b)) we allow an arbitrary number of possible connections.

If that is all, there wouldn't be much difference between multigraphs and weighted networks. If all parallel edges are the same, we could have a single edge with a weight proportional to the number of connections between the two nodes. However, in this case, we can add a ``type'' to each connection, making them \textit{qualitatively} different: one edge type for Facebook, one for Twitter, one for Linkedin (Figure \ref{fig:multigraph-multilayer}(c)), etc.

In practice, every edge type -- or label -- represents a different layer of the network. A pair of nodes can establish a connection in any layer, even at the same time. Each layer is a simple graph. In this book -- and generally in computer science -- the most used notation to indicate a multilayer network is $G = (V, E, L)$. $V$ and $E$ are the sets of nodes and edges, as usual. $L$ is the set of layers -- or labels. An edge is now a triple $(u, v, l) \in E$, with $u, v \in V$ as nodes, and $l \in L$ as the layer. This might seem similar to the notation used for weighted edges -- which was $(u, v, w)$. The key difference is that $w$ is a quantitative information, while $l$ is a qualitative one: a class, a type. We can make the two co-exist in weighted multigraphs, by specifying an edge as $(u,v,l,w)$.

The model that we introduce in Figure \ref{fig:multigraph-multilayer}(c) is but the simplest way to represent multilayer networks. This strategy rests on the assumption that there is a one-to-one node mapping between the layers of the network. In other words, the entities in each layer are always the same: you are always you, whether you manage your Facebook account or your Linkedin one. Such simplified multilayer networks are sometimes called multiplex networks.

Studies have shown how layers in a multiplex network could be complementary\cite{cardillo2013emergence}. This means that a single layer in the network might not show the typical statistical properties you would expect from a real world network -- the types of things we'll see in this book. However, once you stack enough layers one on top of the other, the resulting network does indeed conform to our structural expectations. In other words, multilayer networks have \textit{emerging} properties.

Multiplex networks, don't necessarily cover all application scenarios: sometimes a node in one layer can map to multiple nodes -- or none! -- in another. This is what we turn our attention to next.

\subsection{Many-to-Many}
\begin{figure}
\centering
\begin{subfigure}[t]{.4\columnwidth}
\includegraphics[width=\textwidth]{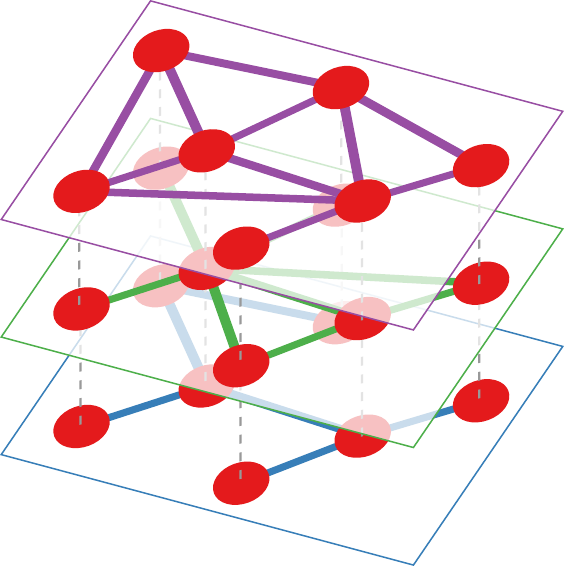}
\caption{}
\end{subfigure}\qquad
\begin{subfigure}[t]{.4\columnwidth}
\includegraphics[width=\textwidth]{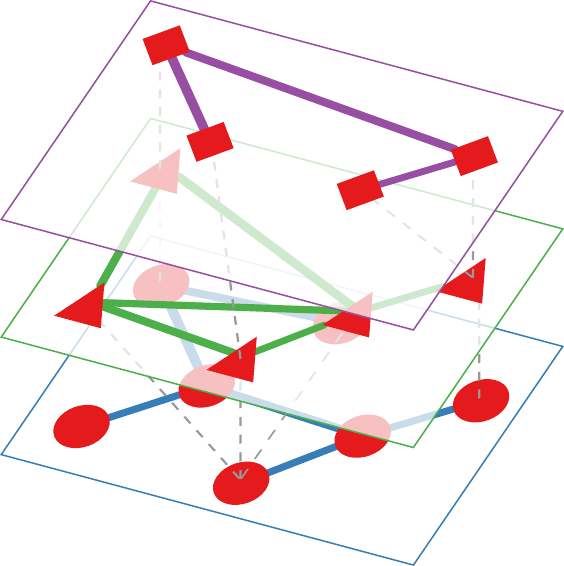}
\caption{}
\end{subfigure}
\caption{The extended multilayer model. Each slice represents a different layer of the network. Dashed grey lines represent the inter-layer coupling connections. (a) A multilayer network with trivial one-to-one coupling. (b) A multilayer network with complex interlayer coupling.}
\label{fig:multilayer}
\end{figure}

To fix the insufficient power of multiplex networks to represent true multilayer systems we need to extend the model. We introduce the concept of ``interlayer coupling''. In this scenario, the node is split into the different layers to which it belongs. In this case, your identity includes multiple personas: you are the union of the ``Facebook you'', the ``Linkedin you'', the ``Twitter you''. Figure \ref{fig:multilayer}(a) shows the visual representation of this model: each layer is a slice of the network. There are two types of edges: the intra-layer connections -- the traditional type: we're friends on Facebook, Linkedin, Twitter --, and the inter-layer connections. The inter-layer edges run between layers, and their function is to establish that the two nodes in the different layers are really the same node: they are \textit{coupled} to -- or \textit{dependent} on -- each other.

Formally, our network is $G = (V, E, L, C)$. $V$ is still the set of nodes, but now we split the set of edges in two: $E$ is the set of classical edges, the intra-layer one -- connections between different people on a platform --; and $C$ is the set of coupling connections, the inter-layer one, denoting dependencies between nodes in different layers.

Having a full set of coupling connections enables an additional degree of freedom. We can now have nodes in one layer expressing coupling with multiple nodes in other layers. In our social media case, we are now allowing you to have multiple profiles in one platform that still map on your single profile in another. For instance, you can run as many different Twitter accounts as you want, and they are still coupled with your Facebook account. To get a visual sense on what this means, you can look at Figure \ref{fig:multilayer}(b).

\begin{figure}
\centering
\includegraphics[width=.5\columnwidth]{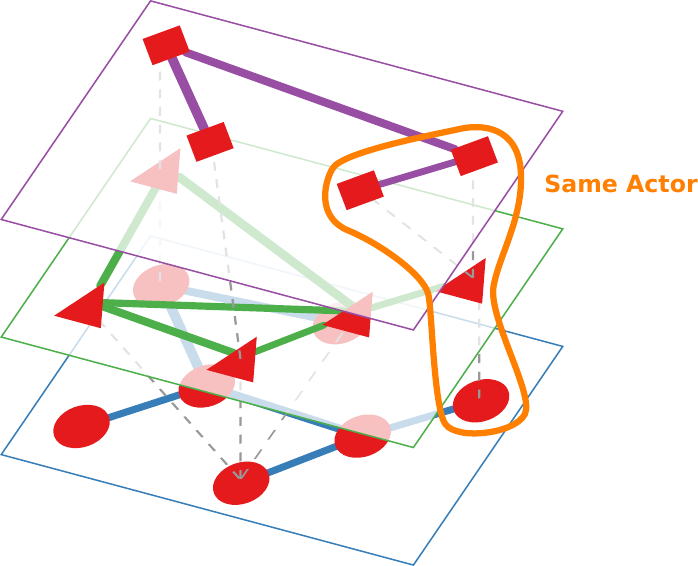}
\caption{An actor in a many-to-many coupled multilayer network. The orange outline surrounds nodes with coupling edges connecting them.}
\label{fig:actor}
\end{figure}

This new freedom comes to a cost. While in the one-to-one mapping it is easy to identify a node among layers, because all identities of a node are concentrated in a single point in a layer, in the many-to-many coupling that is not true any more. So we introduce the term ``actor'', which is the entity behind all the multiple identities across layers and within a layer. In practice, the actor is a connected component (see Section \ref{sec:paths-ccomps}), when only considering inter-layer couplings as the possible edges. If my three Twitter profiles all refer to the same person, with maybe two Flickr accounts and one Facebook profile, all these identities belong to the same actor: me. Figure \ref{fig:actor} should clarify this definition.

\begin{figure}
\centering
\begin{subfigure}[t]{.32\columnwidth}
\includegraphics[width=\textwidth]{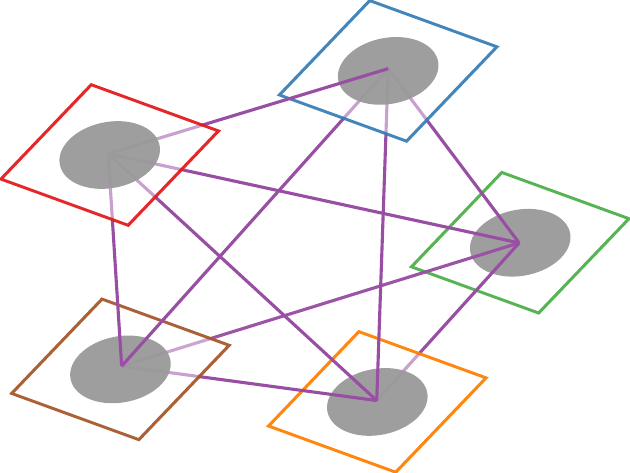}
\caption{Clique}
\end{subfigure}
\begin{subfigure}[t]{.32\columnwidth}
\includegraphics[width=\textwidth]{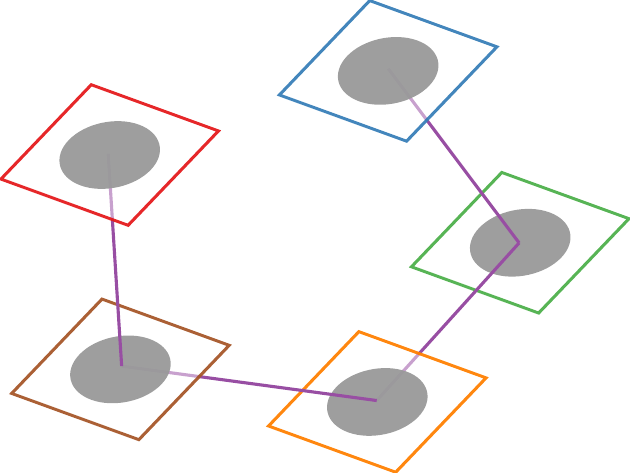}
\caption{Chain}
\end{subfigure}
\begin{subfigure}[t]{.32\columnwidth}
\includegraphics[width=\textwidth]{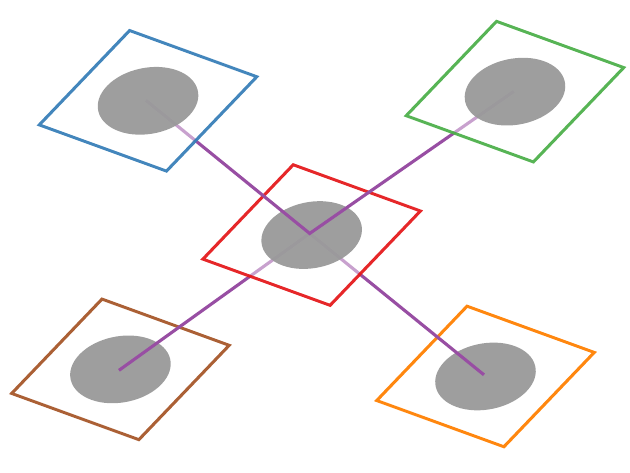}
\caption{Star}
\end{subfigure}
\caption{Different coupling flavors for your multilayer networks. Showing a network with a single actor and a single node per actor per layer (represented by the border-colored polygon). I color the coupling edges in purple.}
\label{fig:coupling-flavors}
\end{figure}

Note that there can be many ways to establish inter-layer couplings between the different nodes belonging to the same actor. As far as I know, when analyzing networks people usually use a ``cliquey'' approach: every node belonging to the same actor is connected to every other node as, for instance, in Figure \ref{fig:coupling-flavors}(a). This effectively creates a clique of inter-layer coupling connections -- for more information about what a clique is, see Section \ref{sec:density-cliques}.

However, this is usually too cumbersome to draw. So, for illustration purposes, the convention is to use a ``chainy'' approach (Figure \ref{fig:coupling-flavors}(b)): you sort your layers somehow, and you simply place a line representing your coupling connections piercing through the layers. We don't really have to stop there. One could imagine using a ``starry'' approach: defining one layer as the center of the system, and connecting all nodes belonging to that actor to the node in the central layer. To see what I mean, look at Figure \ref{fig:coupling-flavors}(c). Using different coupling flavors can be useful for computational efficiency: when you start having dozens or even hundreds of layers, creating cliques of layers can add a significant overhead.

Such many-to-many layer couplings are often referred to in the literature as ``networks of networks'', because each layer can be seen as a distinct network, and the interlayer couplings are relationships between different networks\cite{iacovacci2015mesoscopic}\cite{d2014networks}\cite{kenett2015networks}.

\subsection{Aspects}
Do you think we can't make this even more complicated? Think again. These aren't called ``complex networks'' by accident. To fully generalize multilayer networks, adding the many-to-many interlayer coupling edges is not enough. To see why that's the case, consider the fact that, up to this point, I considered the layers in a multilayer network as interchangeable. Sure, they represent different relationships -- Facebook friendship rather than Twitter following -- but they are fundamentally of the same type. That's not necessarily the case: the network can have multiple aspects.

For instance, consider time. We might not be Facebook friends now, but that might change in the future. So we can have our multilayer network at time $t$ and at time $t+1$. These are two aspects of the same network. All the layers are present in both aspects and the edges inside them change. Another classical example is a scientific community. People at a conference interact in different ways -- by attending each other talks, by chatting, or exchanging business cards -- and can do all of those things at different conferences. The type of interaction is one aspect of the network, the conference in which it happens is another.

I can't hope to give you here an overview of how many new things this introduces to graph theory. So I'm referring you to a specialized book on the subject\cite{bianconi2016multilayer}.

\subsection{Signed Networks}
Signed networks are a particular case of multilayer networks. Suppose you want to buy a computer, and you go online to read some reviews. Suppose that you do this often, so you can recognize the reviewers from past reviews you read from them. This means that you might realize you do not trust some of them and you trust others. This information is embedded in the edges of a signed network: there are positive and negative relationships.

Signed networks are not necessarily restricted to either a single positive or a single negative relationship -- e.g. ``I trust this person'' or ``I don't trust this person''. For instance, in an online game, you can have multiple positive relationships like being friend or trading together; and multiple reasons to have a negative relationship, like fighting each other, or putting a bounty on each other heads.

A key concept in signed networks is the one of structural balance. Since this is mostly related to the link prediction problem, I expand on this in Section \ref{sec:lp-multilayer-signed}.

Positive and negative relationships have different dynamics. For instance, in a seminal study looking at interactions between players in a massively multiplayer online game\cite{szell2010multirelational}, the authors studied the different degree distributions (Section \ref{sec:degree-distributions}) for each type of relationship. They uncovered that positive relationships have a marked exponential cutoff, while negative relationships don't. You'll become more accustomed to what a degree distribution is and all the lingo related to it in Chapter \ref{cha:degree}. For now, the meaning of what I just said is: there is a limit to the number of people you can be friends with, but there is no limit to the number of people that can be mad at you.

\section{Many-to-Many Relationships}\label{sec:extended-hyper}

\subsection{Hypergraphs}
\begin{figure}
\centering
\begin{subfigure}[t]{.4\columnwidth}
\includegraphics[width=\textwidth]{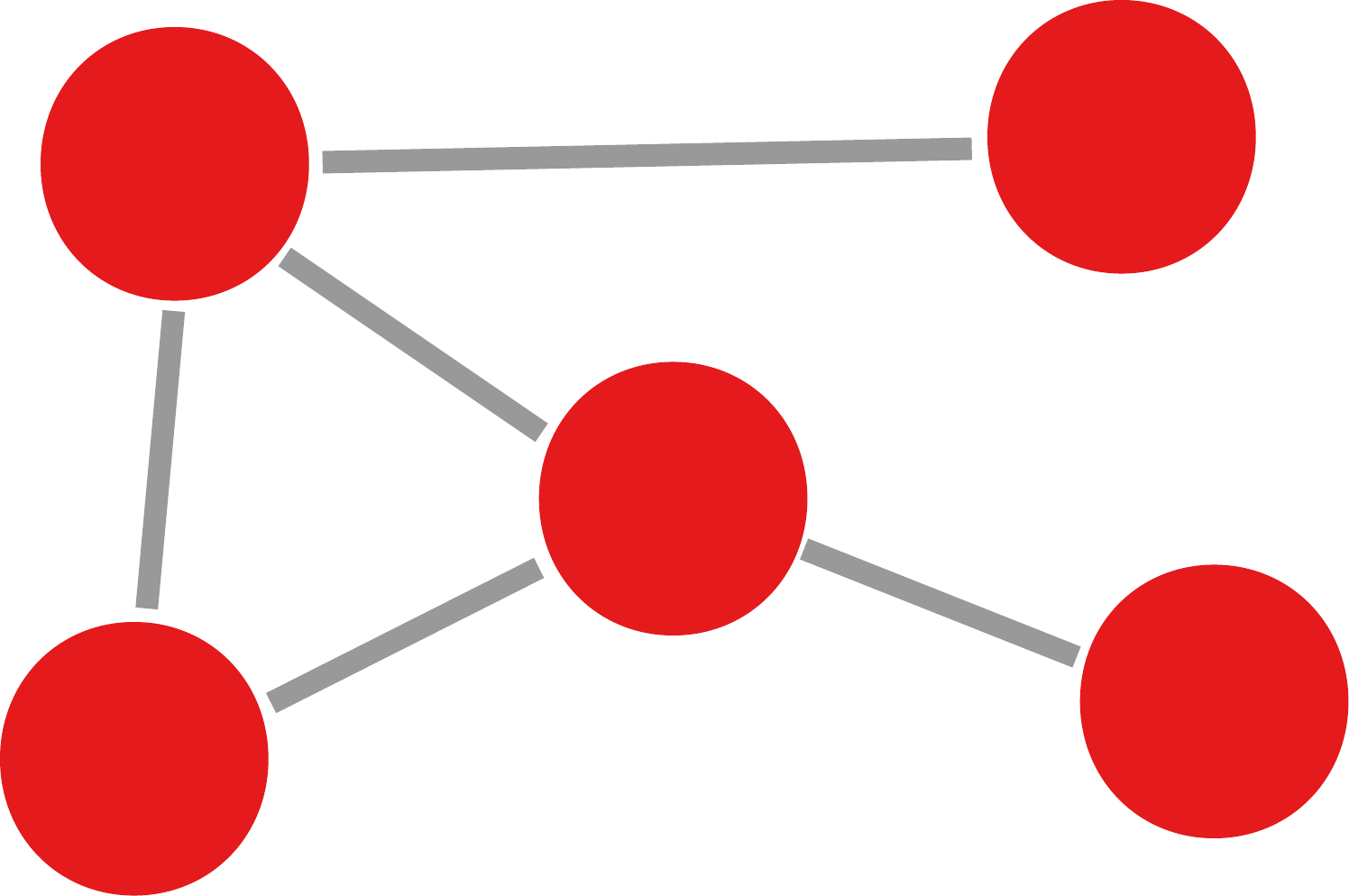}
\caption{}
\end{subfigure}
\qquad
\begin{subfigure}[t]{.4\columnwidth}
\includegraphics[width=\textwidth]{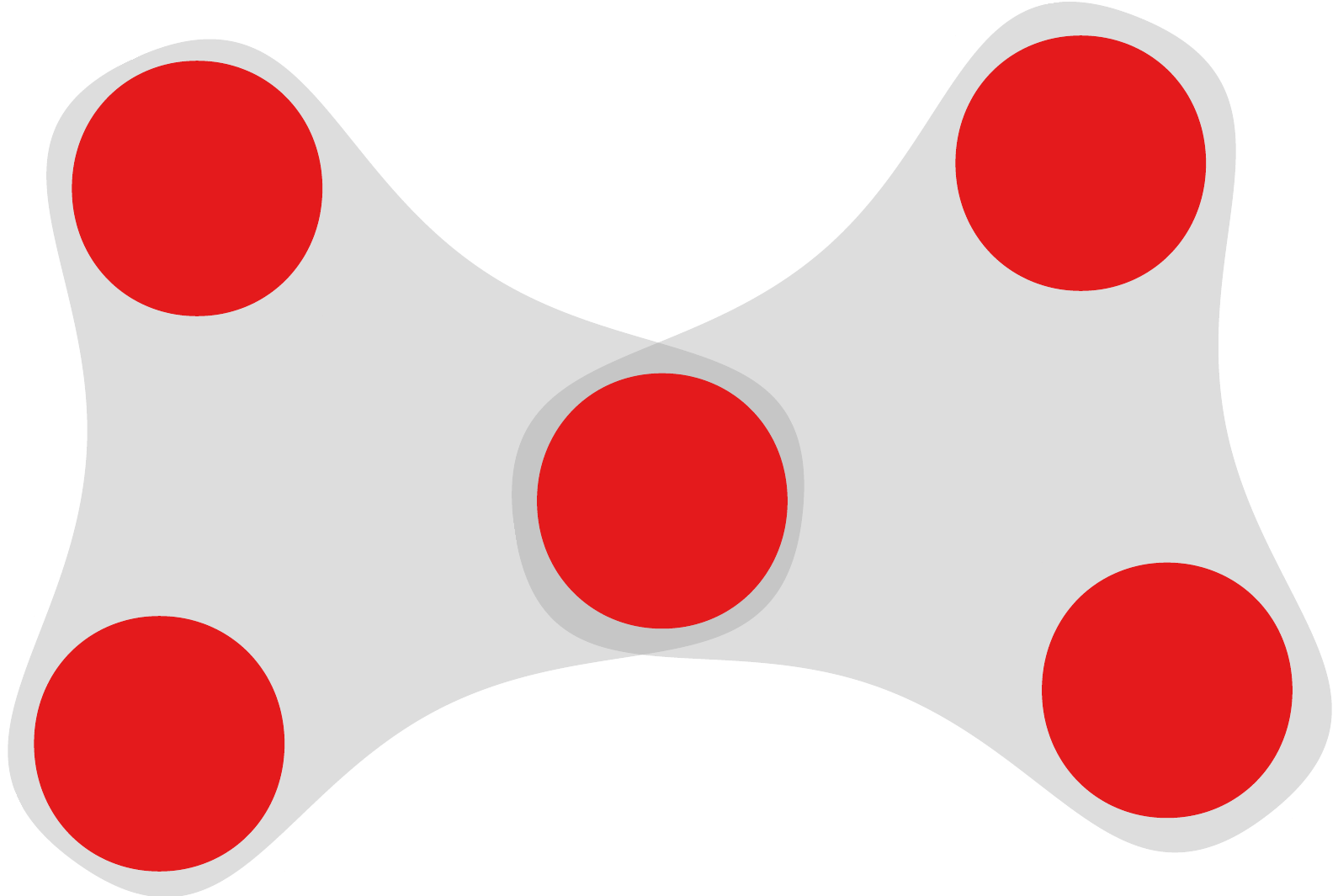}
\caption{}
\end{subfigure}
\caption{(a) Classical Graph. (b) Hypergraph.}
\label{fig:graph-hyper}
\end{figure}

In the classical definition, an edge connects two nodes -- the gray lines in Figure \ref{fig:graph-hyper}(a). Your friendship relation involves you and your friend. If you have a second friend, that is a different relationship. There are some cases in which connections bind together multiple people at the same time. For instance, consider team building: when you do your final project with some of your classmates, the same relationship connects you with all of them. When we allow the same edge to connect more than two nodes we call it a \textit{hyperedge} -- the gray area in Figure \ref{fig:graph-hyper}(b). A collection of hyperedges makes a \textit{hypergraph}\cite[-0.5in]{voloshin2009introduction}\cite{bretto2013hypergraph}.

To make them more manageable, we can put constraints to hyperedges. We could force them to always contain the same number of nodes. In a soccer tournament, the hyperedge representing a team can only have eleven members: not one more nor one less, because that's the number of players in the team. In this case, we call the resulting structure a ``uniform hypergraph'', and have all sorts of interesting properties\cite{hu2012algebraic}. In general, when simply talking about hypergraphs we have no such constraint.

It is difficult to work with hypergraphs\footnote{Source: I tried once.}. Specialized algorithms to analyze them exist, but they become complicated very soon. In the vast majority of cases, we will transform hyperedges into simpler network forms and then apply the corresponding simpler algorithms. 

There are two main strategies to simplify hypergraphs. The first is to transform the hyperedge into the simple edges it stands for. If the hyperedge connects three nodes, we can change it into a unipartite network in which all three nodes are connected to each other. In the project team example, the new edges simply represent the fact that the two people are part of the same team. The advantage is a gain in simplicity, the disadvantage is that we lose the ability to know the full team composition by looking at its corresponding hyperedge: we need to explore the newly created structures.

The second strategy is to turn the hypergraph into a bipartite network. Each hyperedge is converted into a node of type $1$, and the hypergraph nodes are converted into nodes of type $2$. If nodes are connected by the same hyperedge, they all connect to the corresponding node of type $1$. In the project team example, the nodes of type $1$ represent the teams, and the nodes of type $2$ the students. This is an advantageous representation: it is simpler than the hypergraph, but it preserves some of its abilities, for instance being able to reconstruct teams by looking at the neighbors of the nodes of type $1$. However, the disadvantage with respect to the previous strategy is that there are fewer algorithms working for bipartite networks than with unipartite networks.

\begin{figure}[b]
\centering
\includegraphics[width=.7\columnwidth]{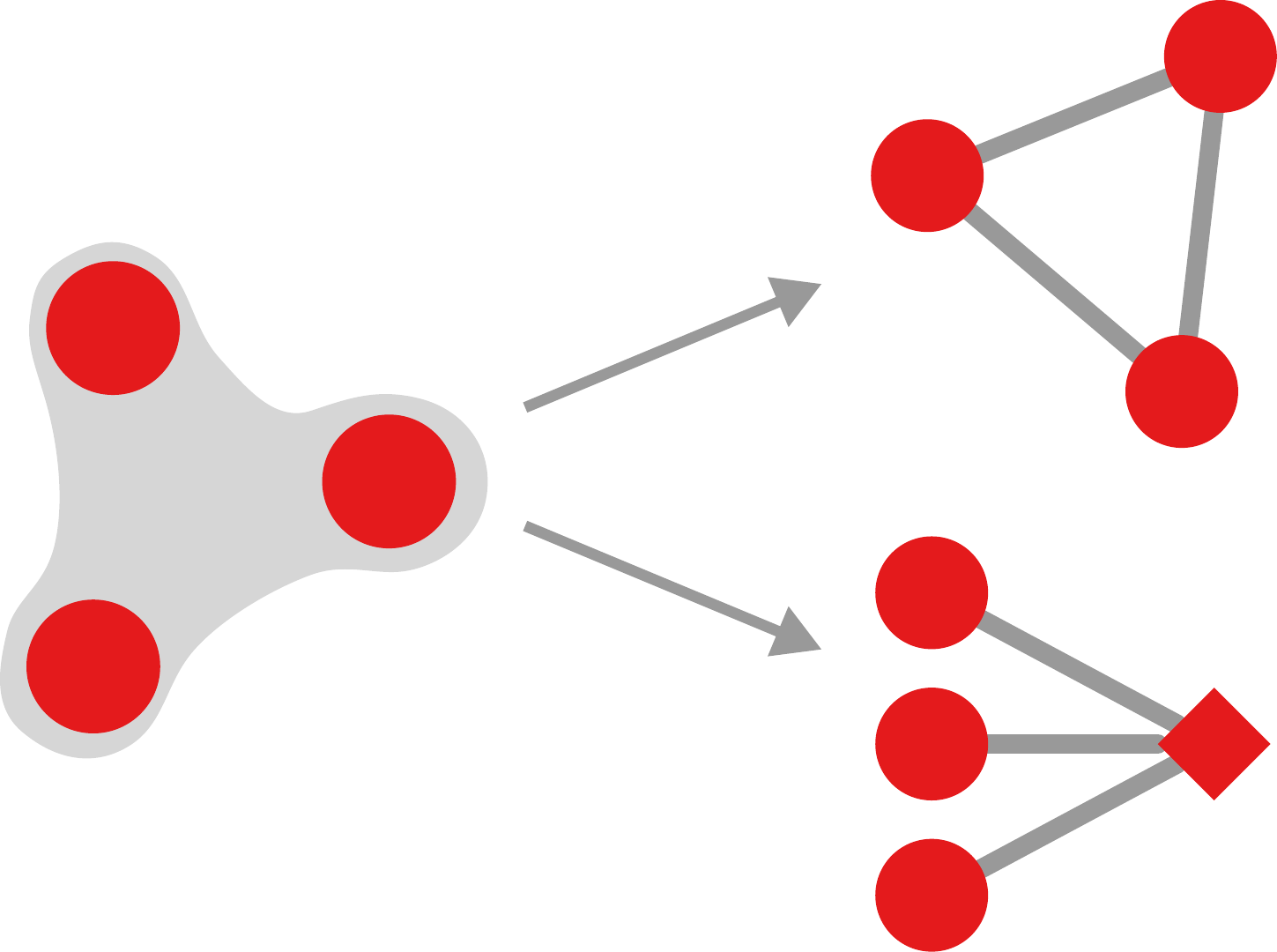}
\caption{The two ways to convert a hyperedge into simpler forms. A hyperedge connecting three nodes can become a triangle (top right), or a bipartite network (bottom right).}
\label{fig:hyperedge-conversion}
\end{figure}

Figure \ref{fig:hyperedge-conversion} provides a simple example on how to perform these two conversion strategies on a simple hyperedge connecting three nodes.

When it comes to notation, the network is still represented by the classical node and edge sets: $G = (V, E)$. However, the $E$ set now is special: its elements are not forced to be tuples any more. They can be triples, quartuplets, and so on. For instance, $(u, v, z)$ is a legal element that can be in $E$, with $u, v, z \in V$.

\subsection{Simplicial Complexes}
\begin{figure}
\centering
\includegraphics[width=.4\columnwidth]{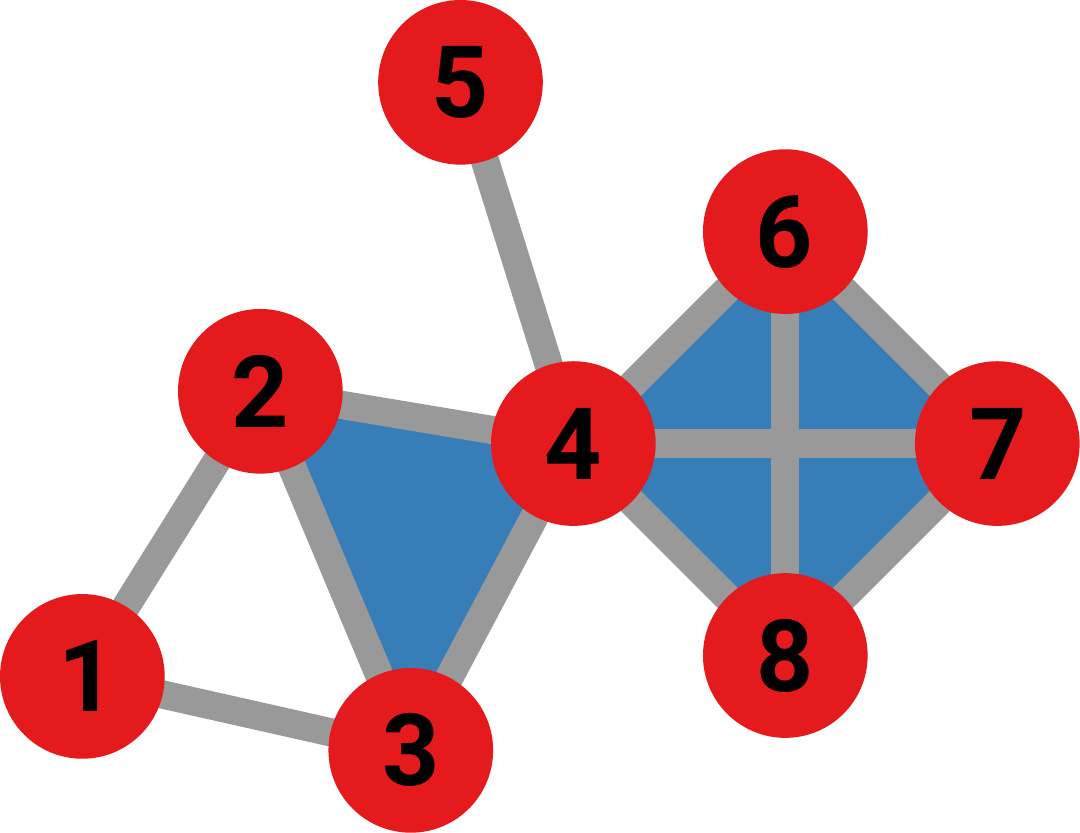}
\caption{An example of simplicial complex. The blue shades represent the two simplices in the complex.}
\label{fig:simplicial-example}
\end{figure}

Simplicial complexes\cite{salnikov2018simplicial}\cite{jonsson2008simplicial}\cite{bianconi2021higher} are related to hypergraphs. A simplicial complex is a graph containing simplices. Simplicial complexes are like hyperedges in that they connect multiple nodes, but they have a strong emphasis on geometry. Graphically, we normally represent simplicial complexes as fills in between the sides that compose the simplex -- as I show in Figure \ref{fig:simplicial-example}.

\begin{figure}
\centering
\begin{subfigure}[t]{.07\columnwidth}
\includegraphics[width=\textwidth]{figures/node.pdf}
\caption{}
\end{subfigure}
\quad
\begin{subfigure}[t]{.25\columnwidth}
\includegraphics[width=\textwidth]{figures/edge.pdf}
\caption{}
\end{subfigure}
\quad
\begin{subfigure}[t]{.2\columnwidth}
\includegraphics[width=\textwidth]{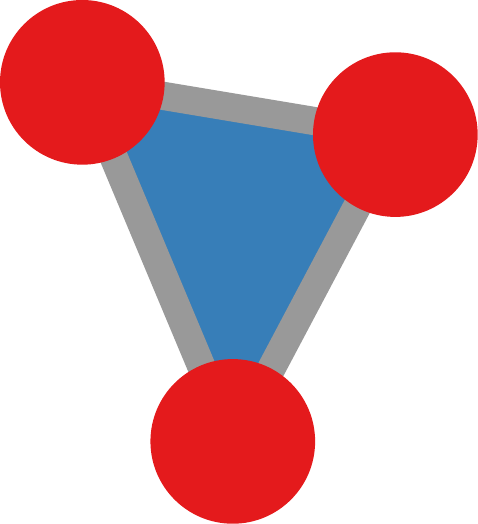}
\caption{}
\end{subfigure}
\quad
\begin{subfigure}[t]{.21\columnwidth}
\includegraphics[width=\textwidth]{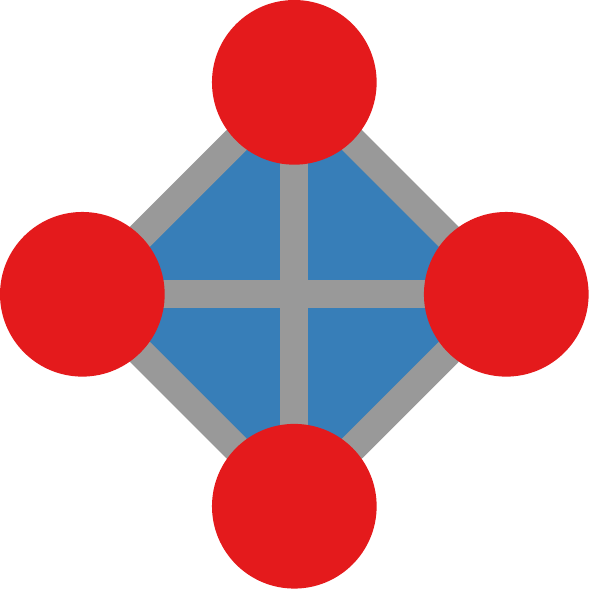}
\caption{}
\end{subfigure}
\caption{The smallest simplices a simplicial complex can have. (a) 0-simplex, (b) 1-simplex, (c) 2-simplex, (d) 3-simplex.}
\label{fig:simplicial-complexes}
\end{figure}

You can think of simplices as hyperedges on steroids. While a hyperedge including, say, four nodes is ``just'' a group of four nodes all interacting with each other, a simplex of four logically also contains all lower-level simplices, which are taken into account in the analysis. Notation-wise, a node is a 0-simplex, an edge is a 1-simplex, then a 2-simplex connects three nodes, and you can see where this is going. A simplex connecting $n+1$ nodes is a $n$-simplex. Figure \ref{fig:simplicial-complexes} shows you the first entries of the simplicial complex zoo.

What does it mean for a simplicial to contain all its lower level versions? Consider the 3-simplex in Figure \ref{fig:simplicial-complexes}(d). That simplex contains four 0-simplices -- the nodes --, six 1-simplices -- the edges --, and four 2-simplices. These are called the \textit{faces} of the simplex -- think of them as the faces of a solid in geometry, because that's what they are. On the other hand, a hyperedge with four nodes only contains those four nodes, it does not logically contain any three-node hyperedge -- unless that specific three-node hyperedge is explicitly coded as part of the data, but it is a wholly separate entity. In the paper writing example, four people writing a paper make a four-node hyperedge, but only a \textit{different} paper with three of those authors will generate a hyperedge contained in it -- while a 3-simplex will naturally contain all 2-simplices with no extra paper.

A simplicial complex -- a network with simplices -- has a \textit{dimension}: the largest dimension of the simplices it contains. The simplicial complex in Figure \ref{fig:simplicial-example} has dimension $3$. A \textit{facet} of a simplicial complex is a simplex that is not a face of any other larger simplex. The facets in the simplicial complex of Figure \ref{fig:simplicial-example} are: $[\{1,2\},\{1,3\},\{2,3,4\},\{4,5\},\{4,6,7,8\}]$. The sequence of facets of a simplicial complex fully distinguishes it from any other simplicial complex. Just like with uniform hypergraphs, we can also have pure simplicial complexes, which are complexes that contain only facets of the same dimension. Figure \ref{fig:simplicial-example} is not pure because it has facets of dimension three, two and one. Figure \ref{fig:simplicial-complexes}(d) is a 3-pure simplicial complex, because it only contains a simplex of dimension three. If you were to ignore all simplices and analyze the network without them, you'd be working with the \textit{skeleton} of the simplicial complex. In practice, any network is a skeleton of one or more simplicial complexes.

Simplicial complexes are one of the main ways to analyze high-order interactions in networks, and so we're going to look at them extensively in Chapter \ref{cha:hod}.

\section{Dynamic Graphs}\label{sec:extended-dynamic}
Most networks are not crystallized in time. Relationships evolve: they are created, destroyed, modified over time by all parties involved. Every time we use a network without temporal information on its edges, we are looking at a particular slice of it, that may or may not exist any longer.

For many tasks, this is ok. For others, the temporal information is a key element. Imagine that your network represents a road graph. Nodes are intersections, and edges are stretches of the street connecting them. Roadworks might cut off a segment for a few days. If your network model cannot take this into account, you would end up telling drivers to use a road that is blocked, creating traffic jams and a lot of discomfort. That is why you need dynamic -- or temporal -- networks\cite{yook2001weighted}\cite{barrat2004weighted}\cite{holme2012temporal}\cite{nicosia2013graph}\cite{masuda2016guidance}.

\begin{figure}
\centering
\begin{subfigure}[t]{.21\columnwidth}
\includegraphics[width=\textwidth]{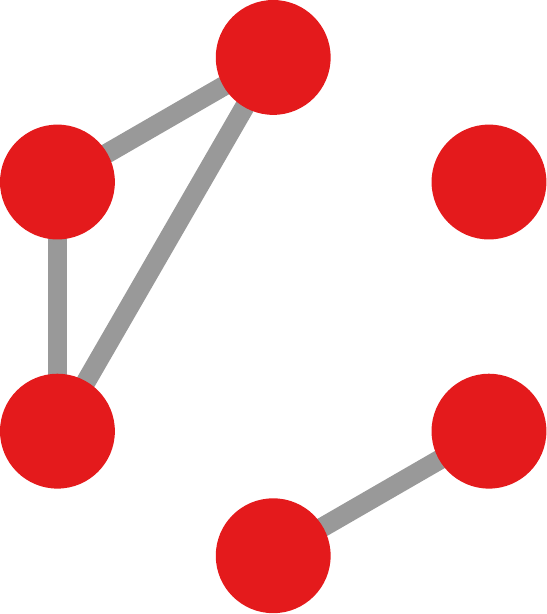}
\caption{$t = 1$}
\end{subfigure}
\quad
\begin{subfigure}[t]{.21\columnwidth}
\includegraphics[width=\textwidth]{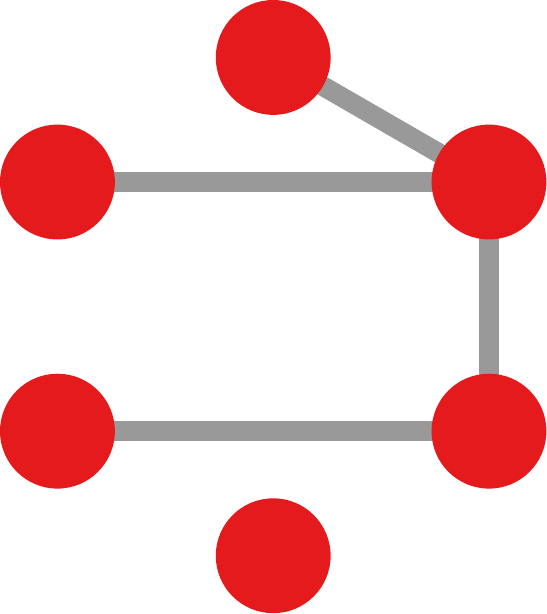}
\caption{$t = 2$}
\end{subfigure}
\quad
\begin{subfigure}[t]{.21\columnwidth}
\includegraphics[width=\textwidth]{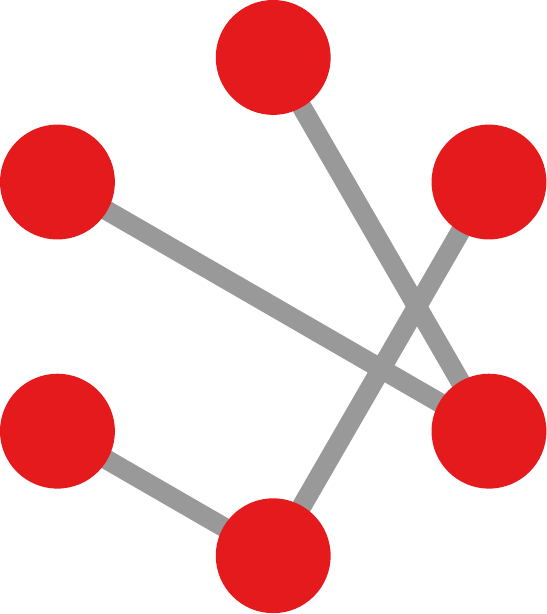}
\caption{$t = 3$}
\end{subfigure}
\quad
\begin{subfigure}[t]{.21\columnwidth}
\includegraphics[width=\textwidth]{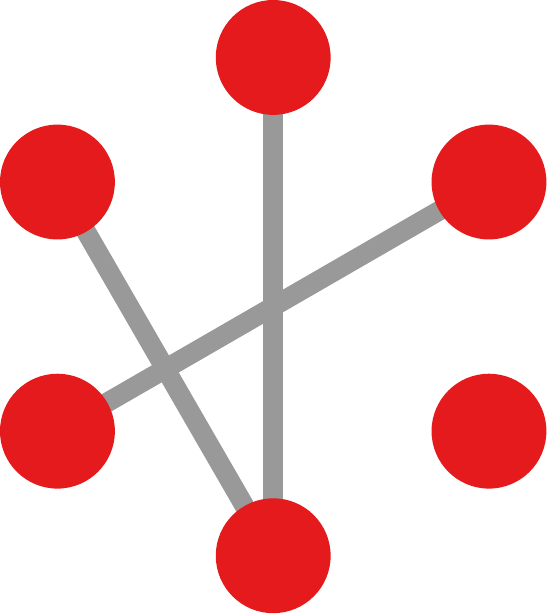}
\caption{$t = 4$}
\end{subfigure}
\caption{An example of dynamic network. Each figure represents the same network, observed at different points in time.}
\label{fig:dynamic}
\end{figure}

Consider Figures \ref{fig:dynamic}(a) to (d) as an example. Here, we have a social network. People are connected only when they are actually interacting with each other. We have four observations, taken at four different time intervals. Suppose that you want to infer if these people are part of the same social group -- or community. Do they? Looking at each single observation would lead us to say \textit{no}. In each time step there are individuals that have no relationships to the rest of the group. Adding the observations together, though, would create a structure in which all nodes are connected to each other. Taking into account the dynamic information allows us to make the correct inference. Yes, these nodes form a tightly connected group.

\begin{figure}
\centering
\includegraphics[width=.75\columnwidth]{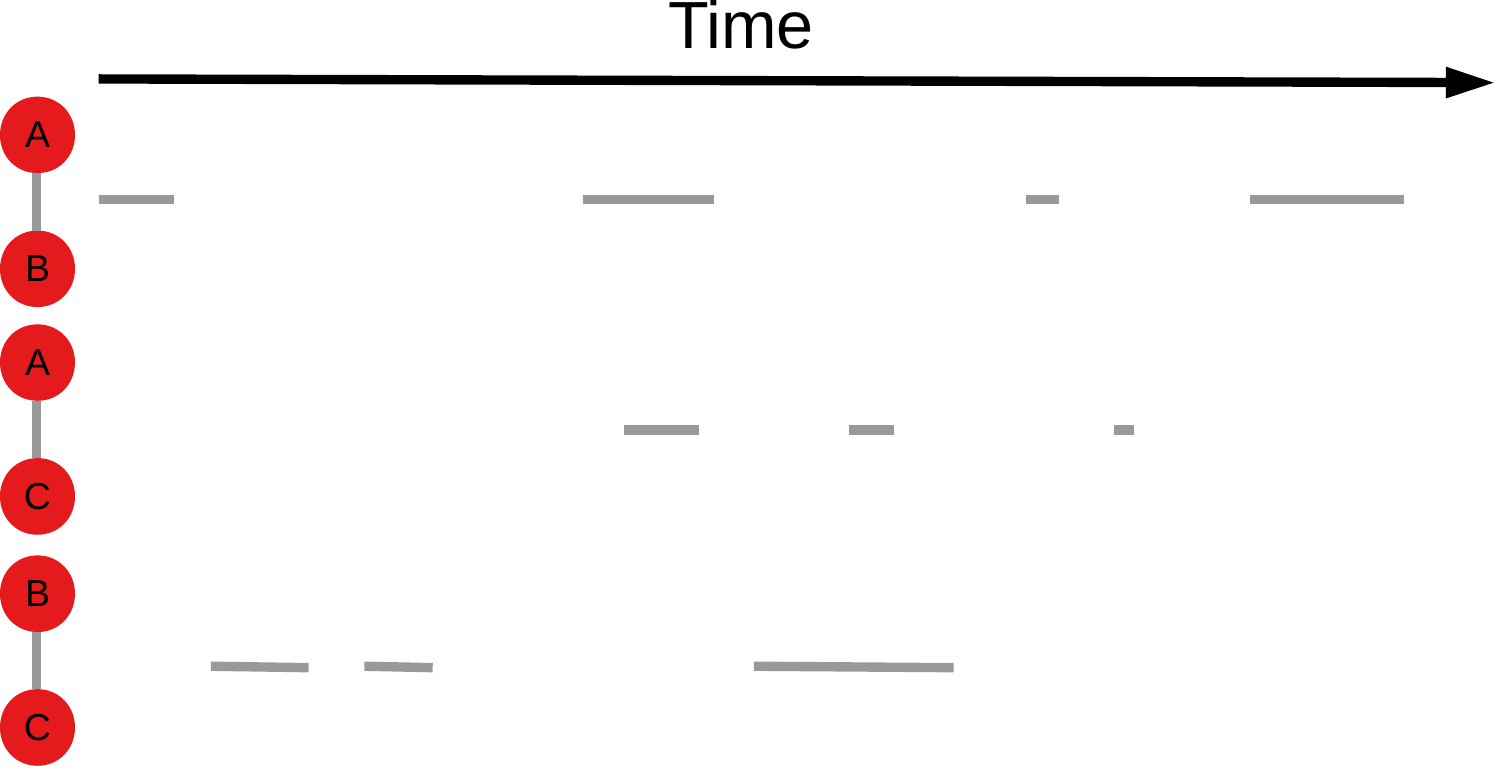}
\caption{An example of dynamic edge information. Time flows from left to right. Each row represents a possible potential edge between nodes $A$, $B$, and $C$. The moments in time in which each edge is active are represented by gray bars.}
\label{fig:dynamic-base}
\end{figure}

In practice, we can consider a dynamic network as a network with edge attributes. The attribute tells us when the edge is active -- or inactive, if the connection is considered to be ``on'' by default, like the road graph. Figure \ref{fig:dynamic-base} shows a basic example of this, with edges between three nodes.

More formally, our graph can be represented as $G = (G_1, G_2, ..., G_n)$, where each $G_i$ is the $i$-th snapshot of the graph. In other words, $G_i = (V_i, E_i)$, with $V_i$ and $E_i$ being the set of nodes and edges active at time $i$.

How do we deal with this dynamic information when we want to create a static view of the network? There are a four standard techniques.

\begin{itemize}
\item \textit{Single Snapshot} -- Figure \ref{fig:dynamic-windows}(a). This is the simplest technique. You choose a moment in time and your graph is simply the collection of nodes and edges active at that precise instant. This strategy works well when the edges in your network are ``on'' by default. It risks creating an empty network when edges are ephemeral and/or there are long lulls in the connection patterns, for instance in telecommunication networks at night.

\begin{figure*}[t]
\centering
\begin{subfigure}[t]{.47\columnwidth}
\includegraphics[width=\textwidth]{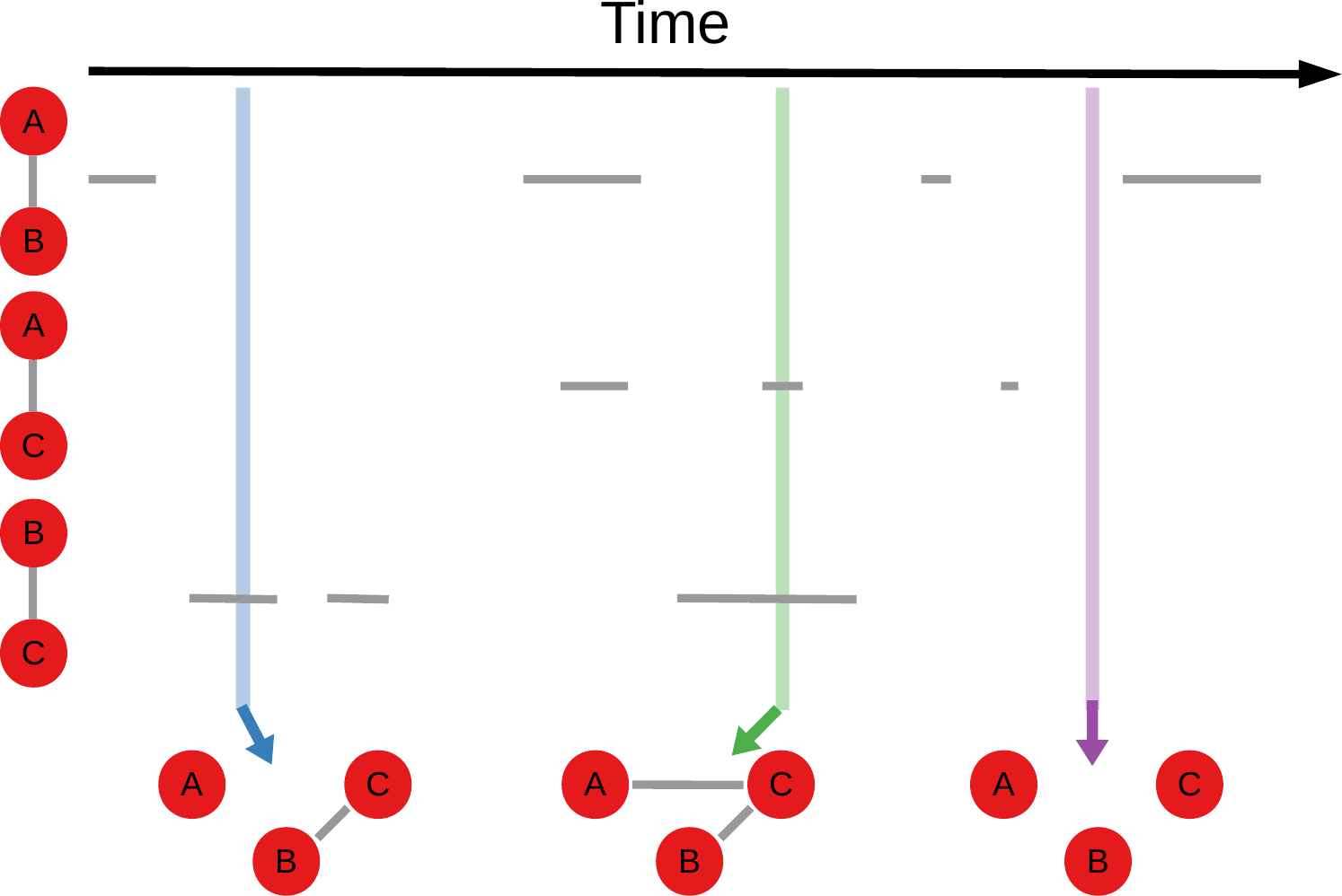}
\caption{Single Snapshot.}
\end{subfigure}
\quad
\begin{subfigure}[t]{.47\columnwidth}
\includegraphics[width=\textwidth]{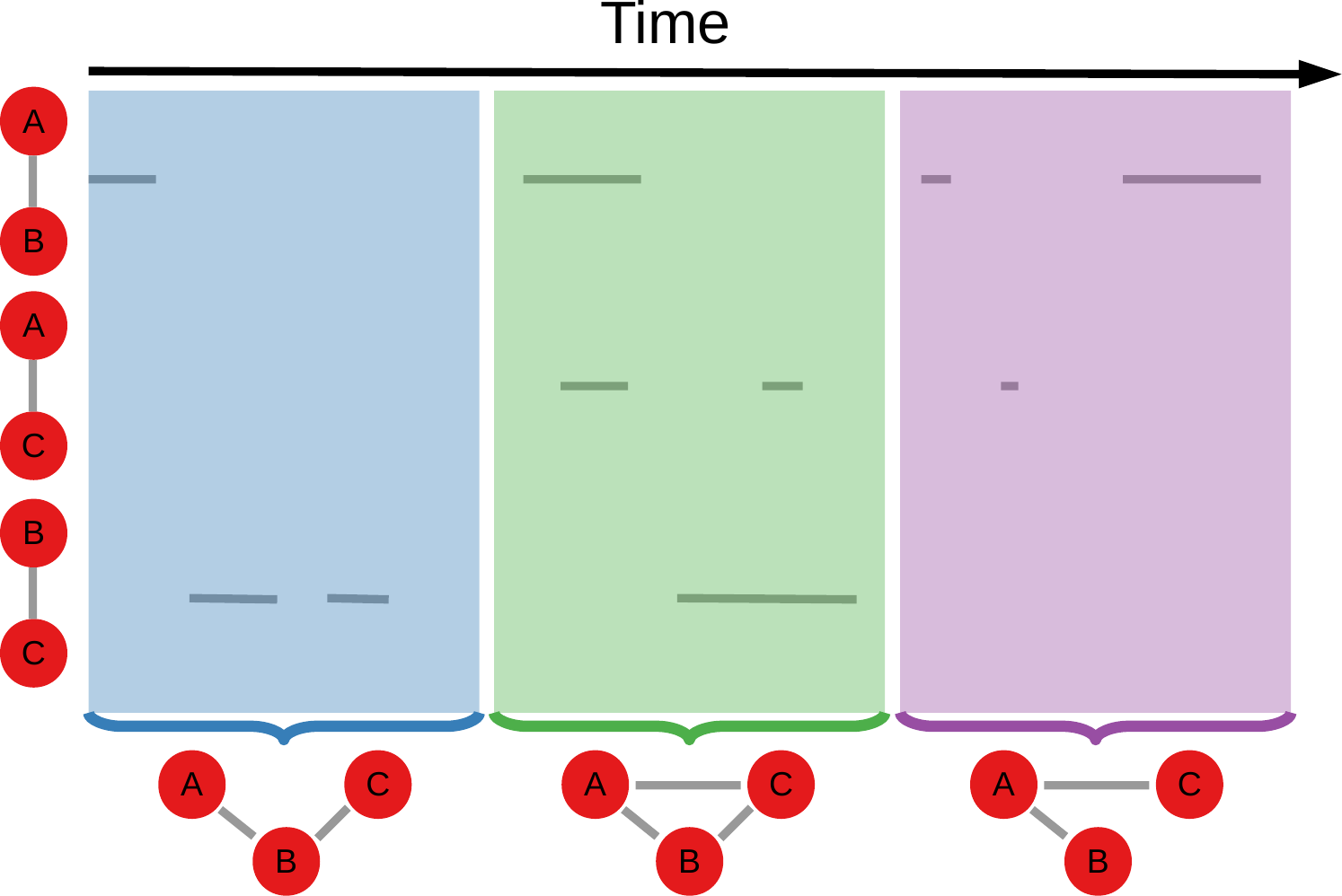}
\caption{Disjoint Windows.}
\end{subfigure}
\quad
\begin{subfigure}[t]{.47\columnwidth}
\includegraphics[width=\textwidth]{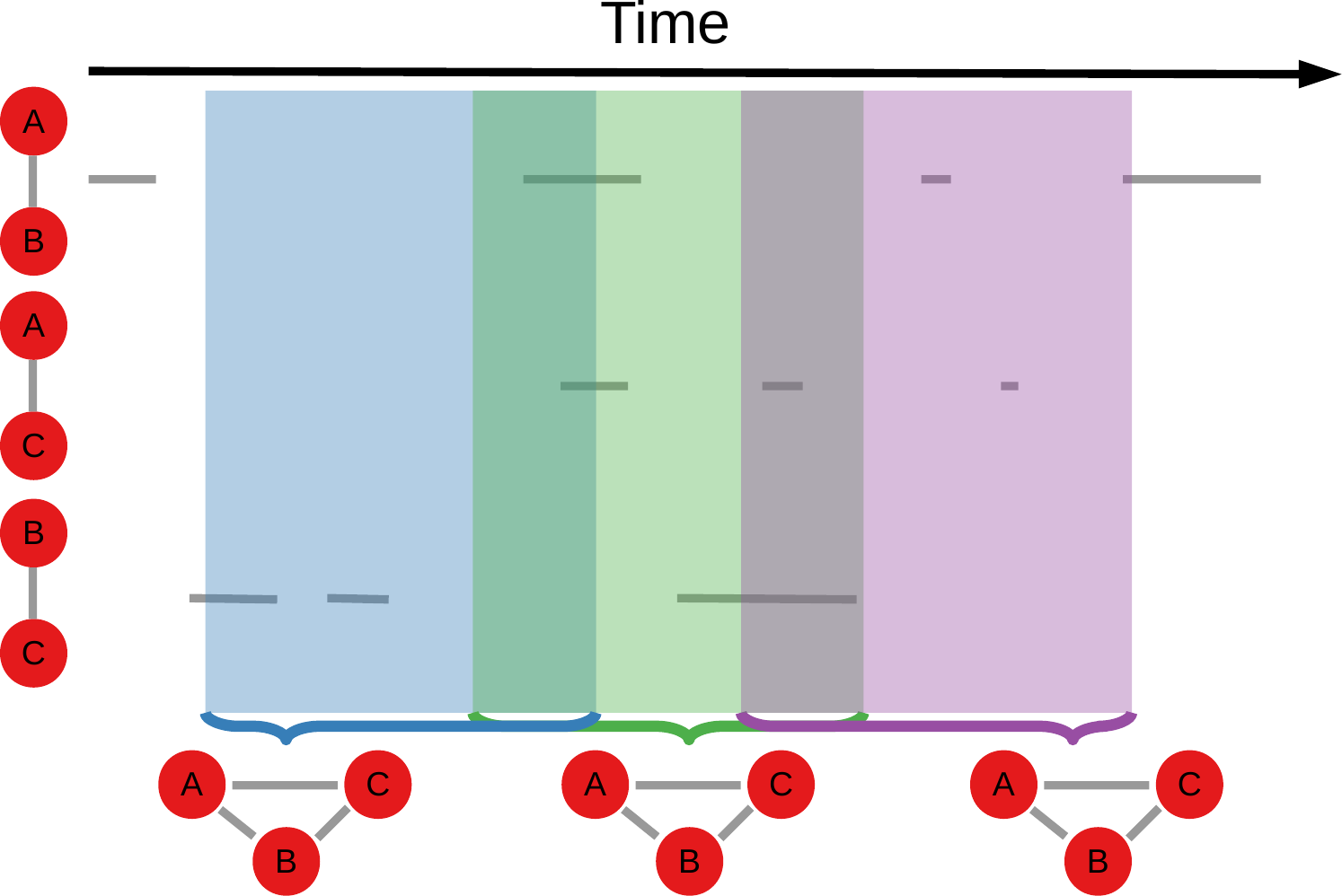}
\caption{Sliding Windows.}
\end{subfigure}
\quad
\begin{subfigure}[t]{.47\columnwidth}
\includegraphics[width=\textwidth]{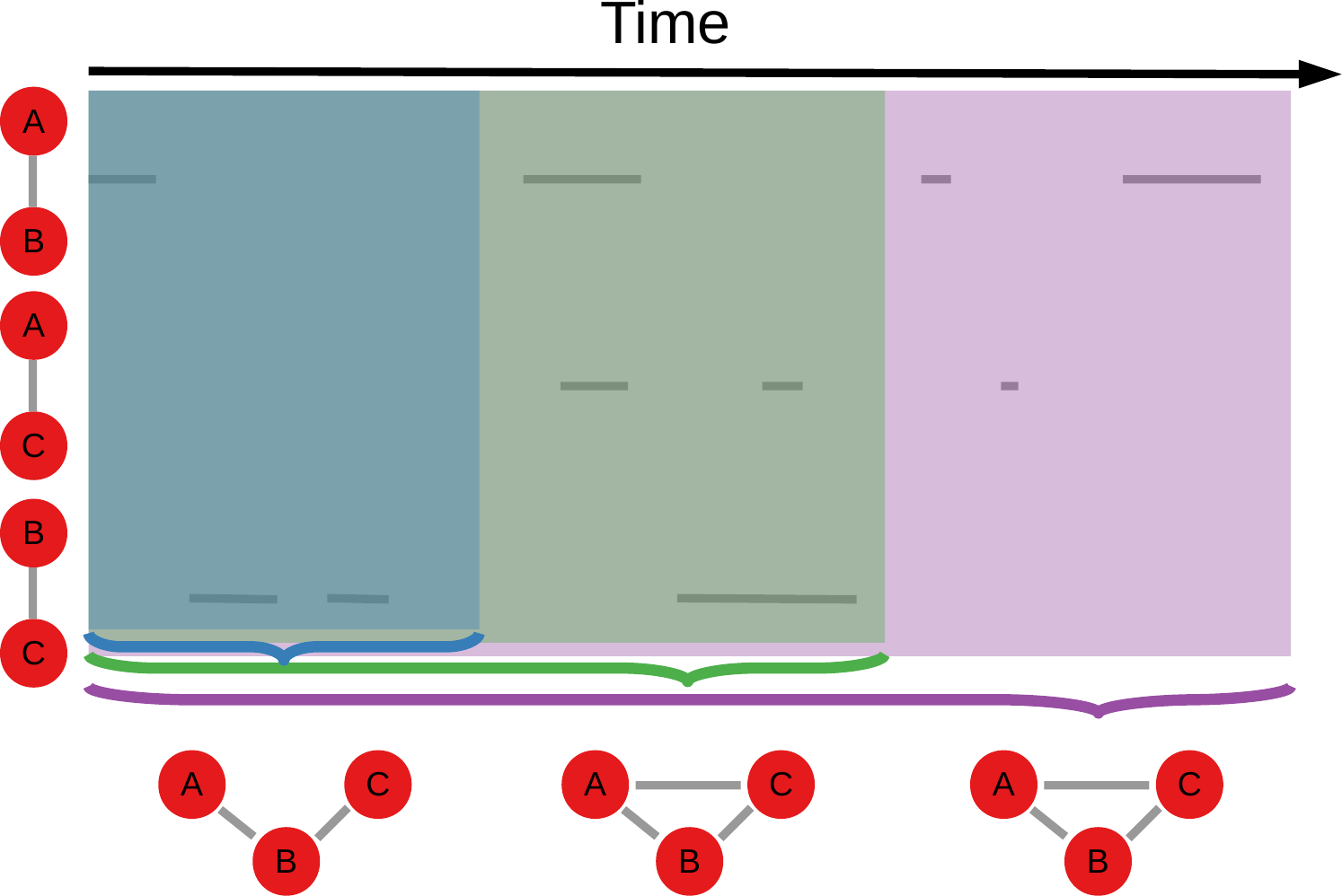}
\caption{Cumulative Windows.}
\end{subfigure}
\caption{Different strategies for converting dynamic edges into a graph view.}
\label{fig:dynamic-windows}
\end{figure*}

\item \textit{Disjoint Windows} -- Figure \ref{fig:dynamic-windows}(b). Similar to single snapshot. Here we allow longer periods of time to accumulate information. Differently from the previous technique, no information is discarded: when a window ends, the next one begins immediately. Works well when it's not important to maintain continuity.
\item \textit{Sliding Windows} -- Figure \ref{fig:dynamic-windows}(c). Similar to disjoint windows, with the difference that we allow the observation periods to overlap. That is, the next window starts before the previous one ended.  Works well when it is important to maintain continuity.
\item \textit{Cumulative Windows} -- Figure \ref{fig:dynamic-windows}(d). Similar to sliding windows, but here we fix the beginning of each window at the beginning of the observation period. Information can only accumulate: we never discard edge information, no matter how long ago it was firstly generated. Each window includes the information of all previous windows. Works well when the effect of an edge never expires, even after the edge has not been active for a long time.
\end{itemize}

Note how these different techniques generate radically different ``histories'' for the network in Figure \ref{fig:dynamic-windows}(a) to (d), even when the edge activation times are identical.

\section{Attributes on Nodes}\label{sec:extended-nodeattr}
Earlier I defined what a bipartite network is: a network with two node types and edges connecting exclusively nodes of unlike type. You could consider the node type as a sort of binary attribute on the node. Once you make the step of adding some metadata to the nodes, why stopping at just two values? And why constraining how edges can connect nodes depending on their attributes? Welcome to the world of node attributes!

\begin{figure}[!ht]
\centering
\begin{subfigure}[t]{.38\columnwidth}
\includegraphics[width=\textwidth]{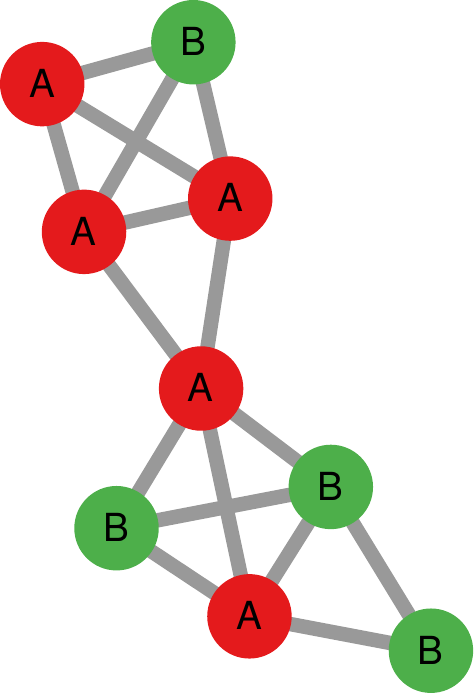}
\caption{}
\end{subfigure}
\qquad
\begin{subfigure}[t]{.38\columnwidth}
\includegraphics[width=\textwidth]{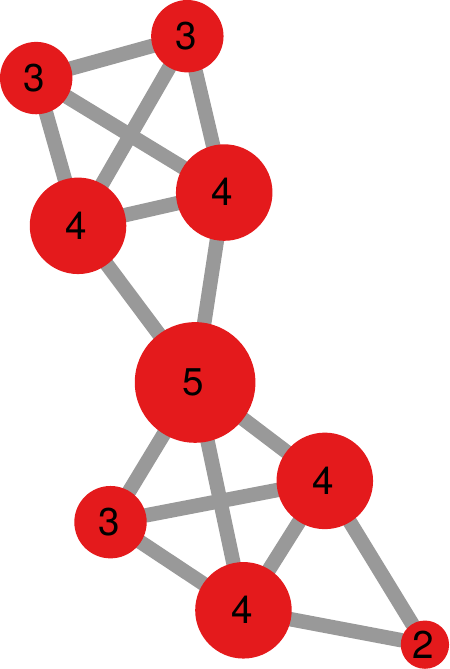}
\caption{}
\end{subfigure}
\caption{(a) A network with qualitative node attributes, represented by node labels and colors. (b) A network with quantitative node attributes, represented by node labels and sizes.}
\label{fig:nodeattrs}
\end{figure}

Here we do not have the requirement of only establishing edges between nodes with unlike attribute values. Moreover the attributes don't have to be binary. They also don't have to be qualitative at all (as in Figure \ref{fig:nodeattrs}(a)): they can be quantitative, as in Figure \ref{fig:nodeattrs}(b). For instance, the number of times a user logged into their social media profile. Finally, nodes can have an arbitrary number of attributes attached to them, not just one.

Consider for instance a trade network. The nodes in this network are the various countries. They connect together if one country exports goods to another. We can have multiple quantitative attributes on each country. For instance, it can be its GDP per capita, its population, its total trade volume. On the other hand, we can also put countries in different categories: in which world region are they located? Are they democracies or not? Of which trade agreement are they part of?

In this case, our graph changes form again: $G = (V, E, A)$. We can see each $v \in V$ not as a simple entity, but as a vector of attribute values: $v = (a_1, a_2, a_3, ...)$. In this representation, $a_1$ is the value for $v$ of the first attribute in $A$. $a_1$ can be a real, integer, or a category.

Node attributes are important because nodes might have tendencies of connecting -- or refusing to connect -- to nodes with similar attribute values. We'll explore this topic in the forms of ``homophily'' in Chapter \ref{cha:homophily} for qualitative attributes, and ``assortativity'' in Chapter \ref{cha:assortativity} for quantitative attributes. This is different from bipartite networks because in bipartite networks edges between nodes with the same attribute value are \textit{forbidden}, while in these cases edges are simply \textit{correlated} with attribute values. Moreover, bipartite networks are only defined for qualitative attributes, not quantitative.

To wrap up, no one forces you to use a single of these more complex graph models at a time. You can merge them together to fit your analytical needs. For instance, you can create this monster graph type: $G_n = (V_1, V_2, E, L, W, A)$: a bipartite graph with $V_1$ and $V_2$ nodes, each with attributes in $A$, which is weighted ($W$) multilayer with $|L|$ layers and -- for good measure -- is also a hypergraph, allowing edges in $E$ with more than two nodes. And, of course, you can observe it at multiple time intervals ($G_1$, $G_2$, ...). Yikes.

\section{Summary}

\begin{enumerate}
\item Bipartite networks are networks with two node types. Edges can only connect two nodes of different types. You can generalize them to be $n$-partite, and have $n$ node types.
\item In multigraphs we allow to have multiple (parallel) edges between nodes. We can have labeled multigraphs when we attach labels to nodes and edges. Labels on nodes can be qualitative or quantitative attributes.
\item If we only allow one edge with a given label between nodes we have a multiplex or multilayer network: the edge label informs us about the layer in which the edge appears.
\item Multilayer networks are networks in which different nodes can connect in different ways. To track which node is ``the same'' across layers we use inter-layer couplings. Couplings can connect a node in a layer to multiple nodes in another, making a many-to-many correspondence.
\item Signed networks are a special type of multilayer network with two layers: one positive (e.g. friendship) and one negative (e.g. enmity).
\item Hypergraphs are graphs whose (hyper)edges can connect more than two nodes at the same time. You can consider hyperedges as cliques or bipartite edges.
\item Simplicial complexes, like hypergraphs, allow nodes to connect in many-to-many relationships called simplices. Simplices are more powerful than hyperedges because a simplex of $4$ nodes logically contain all of its smaller simplices -- called ``faces''.
\item Dynamic graphs are graphs containing temporal information on nodes and edges. This information tells you when the node/edge was present in the network. There are many ways to aggregate this information to create snapshots of your evolving system.
\end{enumerate}

\section{Exercises}

\begin{enumerate}
\item The network in \url{http://www.networkatlas.eu/exercises/7/1/data.txt} is bipartite. Identify the nodes in either type and find the nodes, in either type, with the most neighbors.
\item The network in \url{http://www.networkatlas.eu/exercises/7/2/data.txt} is multilayer. The data has three columns: source and target node, and edge type. The edge type is either the numerical id of the layer, or ``C'' for an inter-layer coupling. Given that this is a one-to-one multilayer network, determine whether this network has a star, clique or chain coupling.
\item The network in \url{http://www.networkatlas.eu/exercises/7/3/data.txt} is a hypergraph, with a hyperedge per line. Transform it in a unipartite network in which each hyperedge is split in edges connecting all nodes in the hyperedge. Then transform it into a bipartite network in which each hyperedge is a node of one type and its nodes connect to it.
\item The network in \url{http://www.networkatlas.eu/exercises/7/4/data.txt} is dynamic, the third and fourth columns of the edge list tell you the first and last snapshot in which the edge was continuously present. An edge can reappear if the edge was present in two discontinuous time periods. Aggregate it using a disjoint window of size $3$.
\end{enumerate}

\chapter{Matrices}\label{cha:mat}
Graphs, with their fancy nodes and edges, are not the only way to represent a network. One can do so also by using matrices. In fact, ask some people and they will tell you that everything is a matrix. What's a number if not a zero-dimensional tensor (Section \ref{sec:la-tensor})? I mean, come on!

Unfortunately, I am not one of those people, so this chapter will contain only the bare minimum for you to smile and nod while talking to them. 

The reason of having this chapter is because sometimes operations are more natural to understand with the graph models, and sometimes they are just matrix operations. Which perspective is more useful -- graph vs matrix -- often depends on the perspective used by the researcher(s) discovering a given property of developing a given tool. So in the book I'll often switch back and forth between these two representations, and this chapter is your map not to get lost once I start rambling about the ``supra adjacency matrix'', whatever the hell that means.

In each of the sections of this chapter we will see a different way of making a matrix that has a meaningful correspondence to more and more advanced network structures. These will enable interesting operations via the linear algebra techniques I introduced in Chapter \ref{cha:la}.

\section{Adjacency Matrix}\label{sec:mat-mat-mat}

\subsection{Basics}
\begin{figure}[t]
\centering
\begin{subfigure}[t]{.41\columnwidth}
\includegraphics[width=\textwidth]{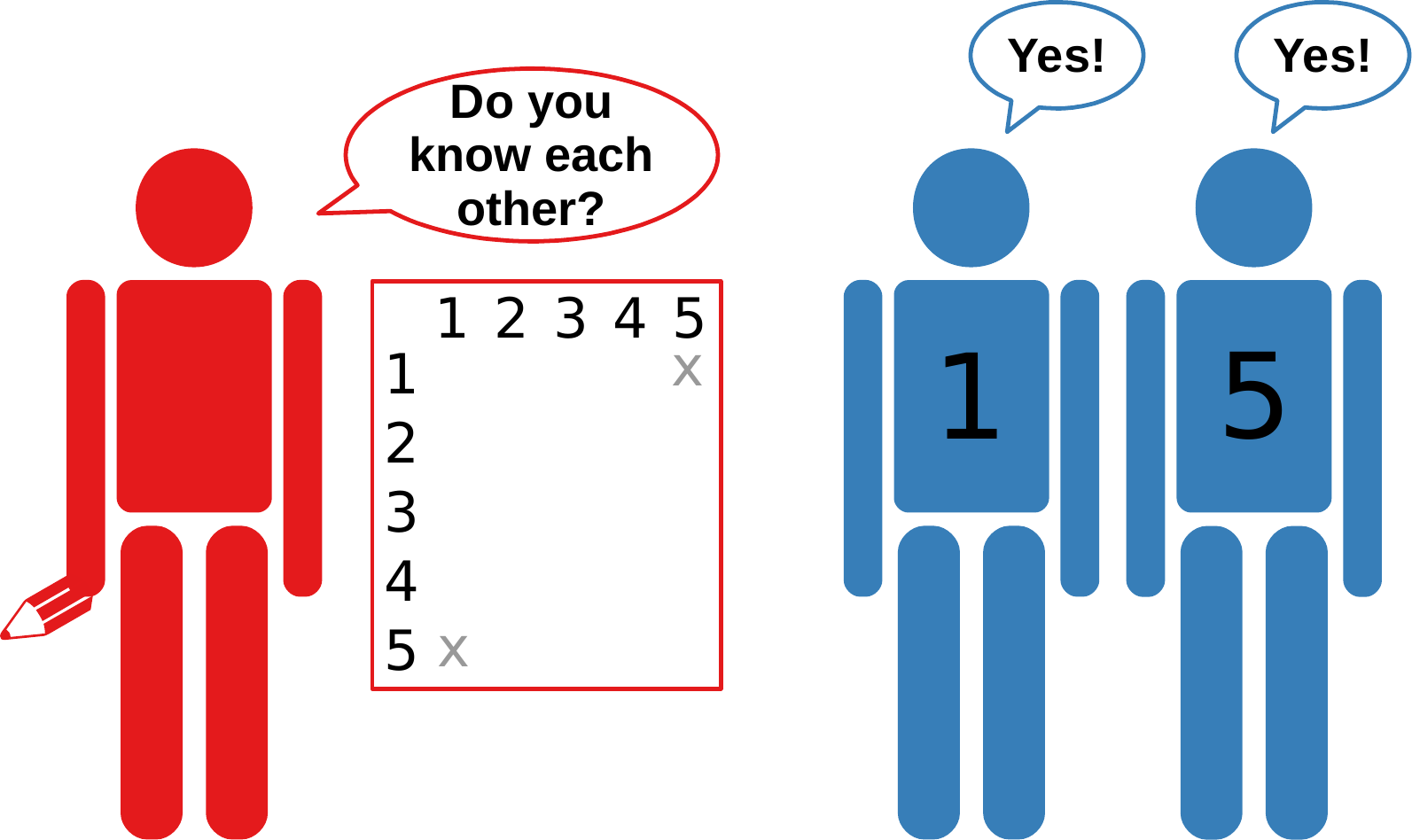}
\caption{}
\end{subfigure}
\begin{subfigure}[t]{.26\columnwidth}
\includegraphics[width=\textwidth]{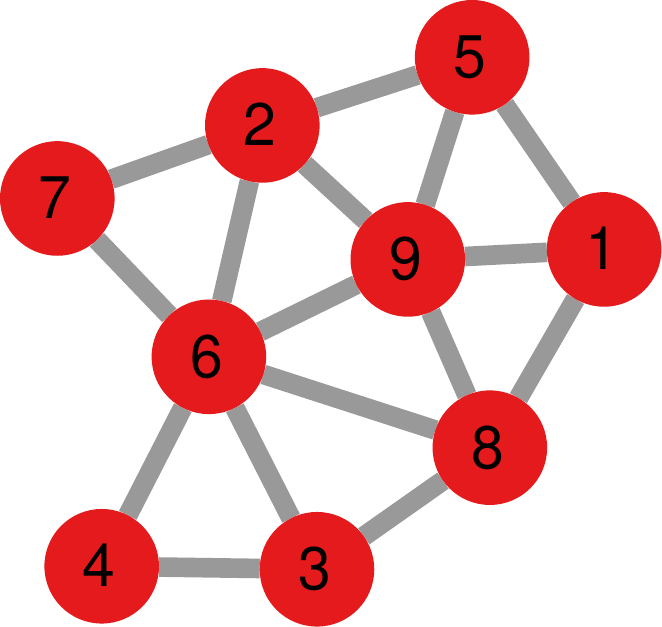}
\caption{}
\end{subfigure}
\begin{subfigure}[t]{.31\columnwidth}
\includegraphics[width=\textwidth]{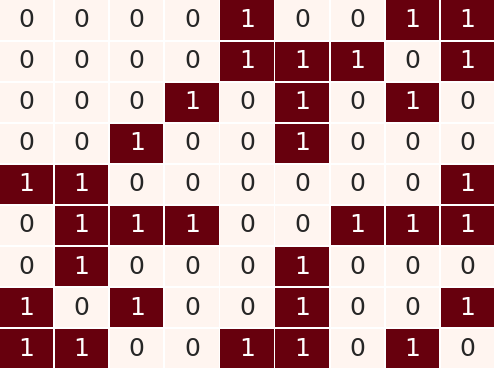}
\caption{}
\end{subfigure}
\caption{(a) A vignette of how one would construct an adjacency matrix. (b) An example graph. (c) The adjacency matrix of (b). Rows and columns are in the same order as the node ids (so the first row/column refers to node $1$, the second to node $2$, etc).}
\label{fig:adj-example}
\end{figure}

The adjacency matrix is a deceptively simple object. Suppose that you have a group of friends and you want to keep a tally of who's friend with whom. You can make a table with one friend per row and one friend per column. If two people say they know each other, you can just put a cross in the corresponding cell, as my example shows in Figure \ref{fig:adj-example}(a). Well, that's it. That's an adjacency matrix.

The adjacency matrix is the basic representation of a graph as a matrix. Each row/column corresponds to a node. Each cell represents an edge, set to one if the edge exists, and zero otherwise. If the graph is undirected, each edge sets two cells to one. If the edge connects nodes $u$ and $v$ both the $A_{uv}$ and the $A_{vu}$ entries are equal to one. Figure \ref{fig:adj-example}(b) shows a graph and Figure \ref{fig:adj-example}(c) shows its adjacency matrix -- in the graph view I labeled the nodes with the order as they appear in the adjacency matrix: the first row/column represents node $1$, the second row/column is for node $2$, and so on.

In Figures \ref{fig:adj-example}(b) and \ref{fig:adj-example}(c) we have the simplest graph possible: the unweighted undirected graph. In this case, the adjacency matrix carries a few properties. For instance, the graph has no self-loops -- edges connecting a node to itself. For this reason, the diagonal of the adjacency matrix contains zeros. We like to keep it that way, because we'll use the diagonal for all sorts of interesting stuff later on -- for instance later on when dealing with the graph Laplacian. The adjacency matrix is also square, meaning that it has the same number of rows and columns. Moreover, it is symmetric, meaning that $\forall u,v\ A_{uv} = A_{vu}$. The diagonal divides the matrix into two identical triangular halves.

You can calculate the complement of any graph by simply calculating $1 - A$, with $1$ being a matrix full of ones -- although you might want to fill its diagonal with zeros to avoid self loops, which are usually ignored in complement graphs.

A few interesting properties of binary adjacency matrices. The sum of the rows -- and of the columns in a symmetric matrix -- is equal to the node's number of connections, which we call the degree (and will be the topic of Chapter \ref{cha:degree}). If the graph is directed the rows/columns give you the the number of arrow tails and heads connected to the node. The sum of the entries of the matrix is $2$ times the number of edges (undirected) or the number of edges (directed).

\subsection{Directed, Weighted, Bipartite Adjacency}
\begin{figure}
\centering
\begin{subfigure}[t]{.4\columnwidth}
\includegraphics[width=\textwidth]{figures/matrix_extendendexample_02.png}
\caption{}
\end{subfigure}
\qquad
\begin{subfigure}[t]{.3\columnwidth}
\includegraphics[width=\textwidth]{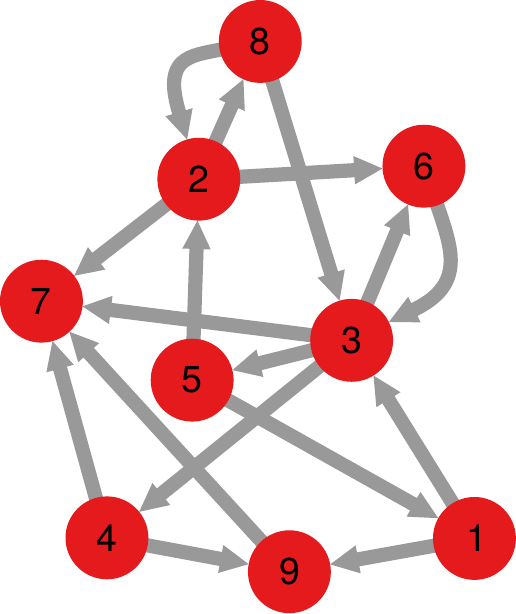}
\caption{}
\end{subfigure}
\caption{(a) A non-symmetric adjacency matrix. (b) The corresponding directed graph.}
\label{fig:adj-example-dir}
\end{figure}

We can adapt the adjacency matrix to deal with all the complications we introduced in the graph model in Chapters \ref{cha:basic} and \ref{cha:extended}. For instance, we can represent a directed graph by breaking the symmetry property we just enunciated. If $A_{uv}$ is allowed to be zero when $A_{vu}$ is one, then it means that we just introduced directionality in the matrix, as Figure \ref{fig:adj-example-dir} shows. Transposing $A$ in a directed graph means to reverse all edge directions. Note that different authors/papers might follow different conventions. Some will represent the $u \rightarrow v$ edge as $A_{uv}$ and some as $A_{vu}$. So make sure you identify the convention before you start working your way through the paper!

\begin{figure}
\centering
\begin{subfigure}[t]{.4\columnwidth}
\includegraphics[width=\textwidth]{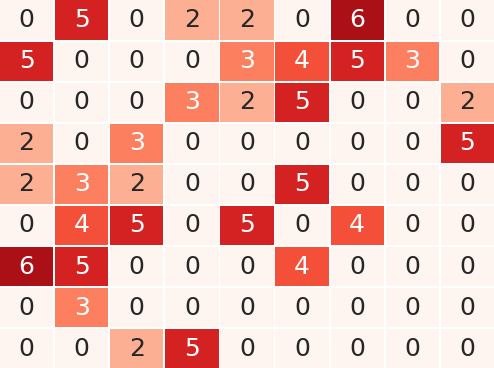}
\caption{}
\end{subfigure}
\qquad
\begin{subfigure}[t]{.45\columnwidth}
\includegraphics[width=\textwidth]{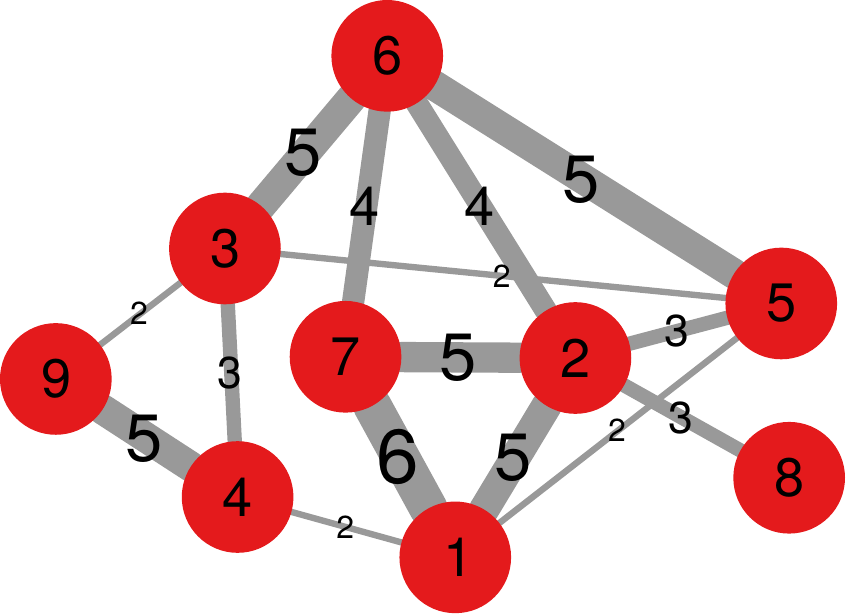}
\caption{}
\end{subfigure}
\caption{(a) A non-binary adjacency matrix. (b) The corresponding weighted graph.}
\label{fig:adj-example-wei}
\end{figure}

If we want edge weights to exploit the power of linear algebra also on weighted graphs (Section \ref{sec:basic-weighted}), we can allow values different than one for the cells representing edges. Now we can have an arbitrary real value in the cells, representing the connection's strength -- see Figure \ref{fig:adj-example-wei}.

\begin{figure}
\centering
\begin{subfigure}[t]{.38\columnwidth}
\includegraphics[width=\textwidth]{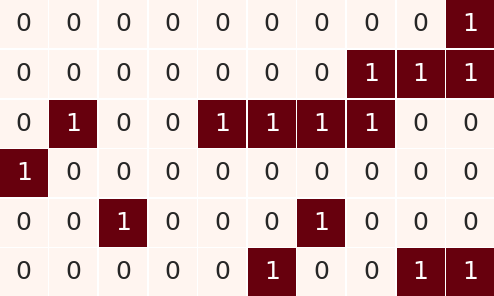}
\caption{}
\end{subfigure}
\quad
\begin{subfigure}[t]{.55\columnwidth}
\includegraphics[width=\textwidth]{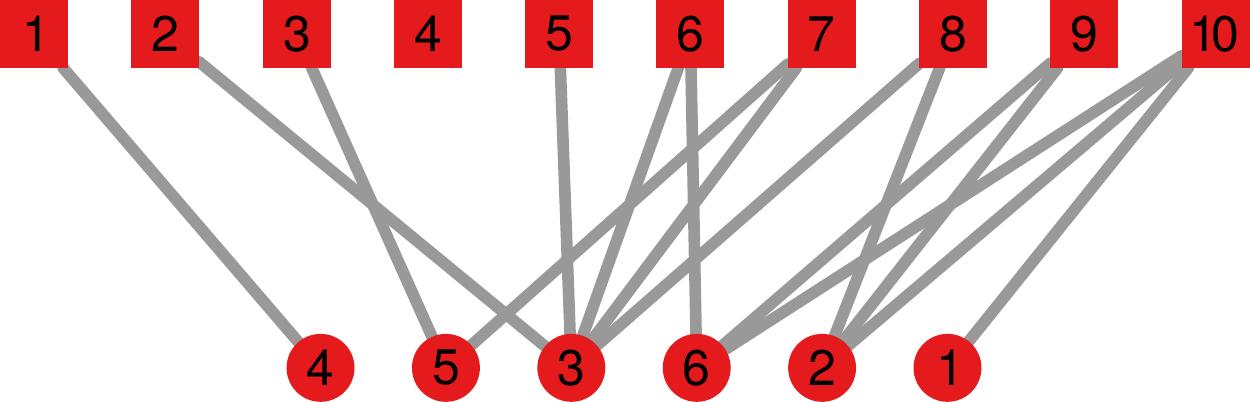}
\caption{}
\end{subfigure}
\caption{(a) A non-square adjacency matrix. (b) The corresponding bipartite graph.}
\label{fig:adj-example-bip}
\end{figure}

What else? We can make the adjacency matrix not square if we need to represent a bipartite network. The different numbers of rows and columns allow us to use one dimension to represent the nodes in $V_1$ and the other to represent the nodes in $V_2$. Figure \ref{fig:adj-example-bip} depicts an example. The downside is that we lose the power of the diagonal we had in the adjacency matrix -- which doesn't seem like a big deal now, because at the moment I'm being all hush hush about what this power really is.

\begin{figure}
\centering
\includegraphics[width=.45\columnwidth]{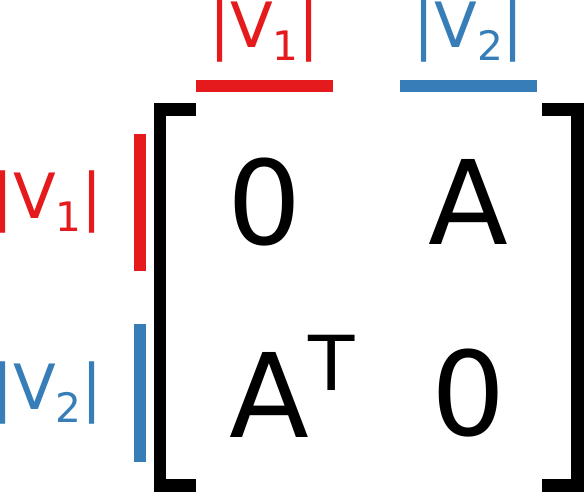}
\caption{A way to build a $(|V_1| + |V_2|) \times (|V_1| + |V_2|)$ square matrix starting from $A$, a non-square bipartite adjacency matrix.}
\label{fig:adj-example-bip-square}
\end{figure}

Of course, it's possible to have a square adjacency matrix for a bipartite network if $|V_1| = |V_2|$. You can also ``squarify'' a bipartite adjacency matrix by dividing it in four blocks. The blocks on the main diagonal contain zeros, while the blocks in the other diagonal contain the original adjacency matrix. Such a construct is a $(|V_1| + |V_2|) \times (|V_1| + |V_2|)$ matrix, and they can be useful. Figure \ref{fig:adj-example-bip-square} shows an example. Note that transposing the adjacency matrix of a bipartite network with $V_1$ and $V_2$ node types will change its shape, from being a $|V_1| \times |V_2|$ matrix to a $|V_2| \times |V_1|$ matrix.

\subsection{Multilayer Adjacency}
Finally we can -- and do -- represent even multilayer networks with matrices. We have two options to do so. We can either use \textbf{tensors} or the \textbf{supra adjacency matrix}.

I introduced \textbf{tensors} in Section \ref{sec:la-tensor} as generalized vectors. As a refresher, a vector can be seen as a monodimensional array: a list of values. A matrix could be said to be a two-dimensional array. A tensor is a multidimensional array: we can have as many dimensions as we want.

\begin{figure}
\centering
\begin{subfigure}[t]{.48\columnwidth}
\includegraphics[width=\textwidth]{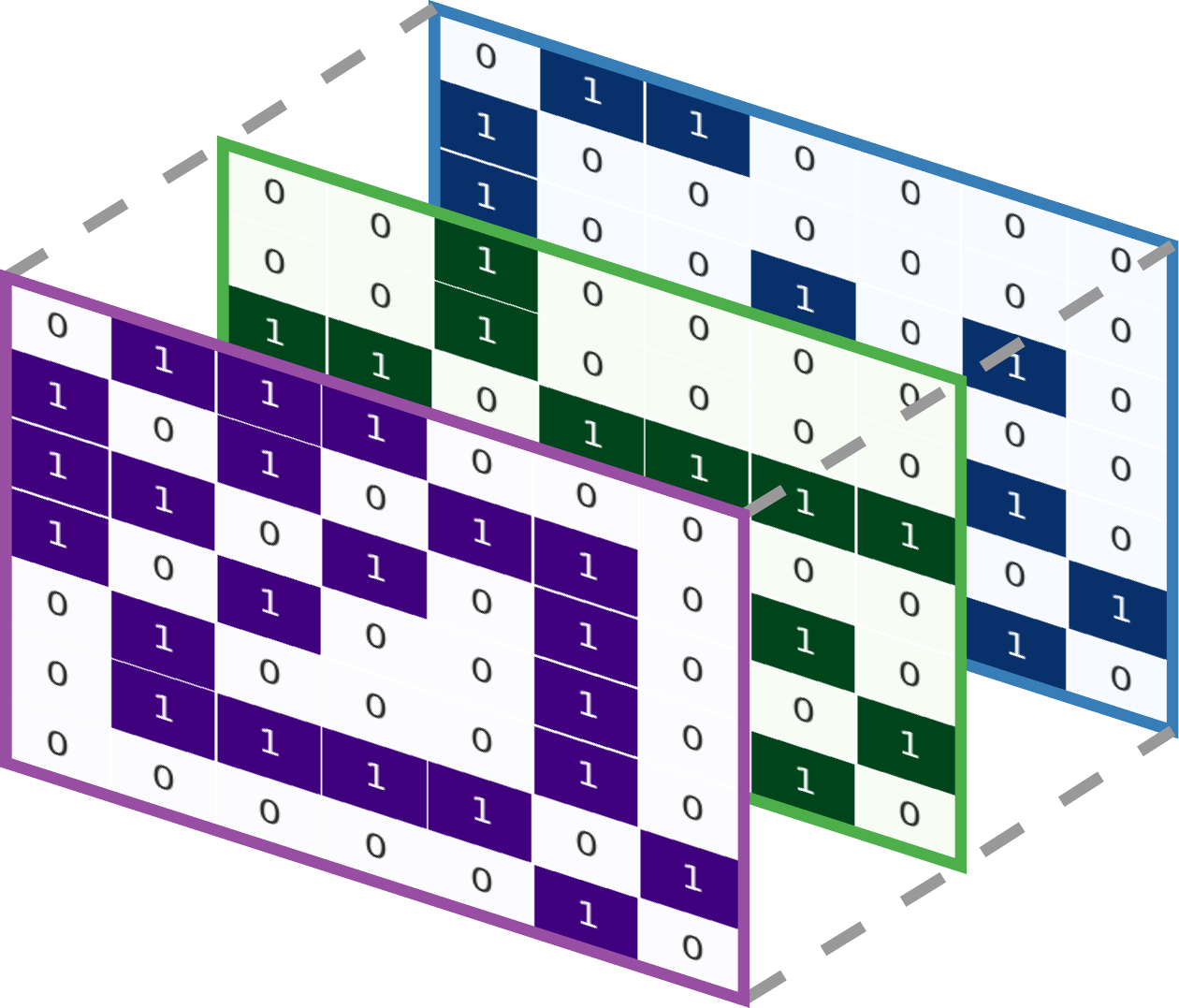}
\caption{}
\end{subfigure}
\quad
\begin{subfigure}[t]{.45\columnwidth}
\includegraphics[width=\textwidth]{figures/layers_combined_coupling_trivial.pdf}
\caption{}
\end{subfigure}
\caption{(a) A three dimensional tensor. (b) The corresponding multilayer graph.}
\label{fig:adj-example-multi}
\end{figure}

A one-to-one coupled multilayer network can be represented with a three-dimensional vector. The first two dimensions -- rows and columns -- are the nodes, and the third dimension is the layers. Mathematically, the $A_{uvl}$ entry in the tensor tells you the relationship between nodes $u$ and $v$ in layer $l$. Figure \ref{fig:adj-example-multi} provides an intuitive example. Note that we are assuming that the nodes are sorted in the same way across the third dimension, thus the inter-layer couplings (see Section \ref{sec:extended-multilayer}) are implicit. 

Tensors start getting into trouble when you want to deal with many-to-many couplings. In this case, my suggestion would be \sout{to find a different job} to use the \textbf{supra adjacency matrix}\cite{porter2018multilayer}. To build this matrix you apply a strategy that is relatively similar to the squarification of a bipartite adjacency matrix. In practice, you start by taking the simple adjacency matrices for each layer independently. You put them those as blocks in the diagonal of the supra adjacency matrix.

\begin{figure}
\centering
\begin{subfigure}[t]{.3\columnwidth}
\includegraphics[width=\textwidth]{figures/layers_combined_coupling_full.pdf}
\caption{}
\end{subfigure}\qquad
\begin{subfigure}[t]{.5\columnwidth}
\includegraphics[width=\textwidth]{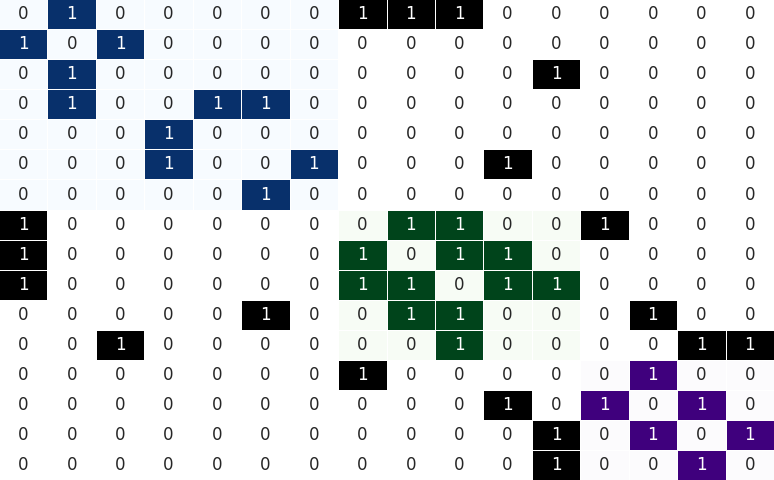}
\caption{}
\end{subfigure}
\caption{(a) A multilayer network with many-to-many interlayer couplings. Each slice represents a different layer of the network. Dashed grey lines represent the inter-layer coupling connections. (b) The supra adjacency matrix of (a).}
\label{fig:supra-adjacency}
\end{figure}

Then you can use the rest of the matrix to represent the inter-layer couplings. This can work because each entry $u$ in the supra adjacency matrix stands for a node in a given layer. A $A_{uv}$ entry outside the diagonal blocks tells you if the node-in-layer represented by the $u$th row in the matrix is coupled with the node-in-layer represented by the $v$th column. Figure \ref{fig:supra-adjacency} shows you a many-to-many multilayer network and its corresponding supra adjacency matrix.

As you can see, there is no problem if layers don't have the same number of nodes, as blocks are allowed to have different sizes and their off-diagonal parts are still able to represent the inter-layer couplings.

\section{Stochastic}\label{sec:mat-mat-stochastic}
\begin{figure}
\centering
\begin{subfigure}[t]{.27\columnwidth}
\includegraphics[width=\textwidth]{figures/graph_view.pdf}
\caption{}
\label{fig:adj-graph}
\end{subfigure}
\begin{subfigure}[t]{.345\columnwidth}
\includegraphics[width=\textwidth]{figures/matrix_firstexample_03.png}
\caption{}
\label{fig:adj-normal}
\end{subfigure}
\begin{subfigure}[t]{.345\columnwidth}
\includegraphics[width=\textwidth]{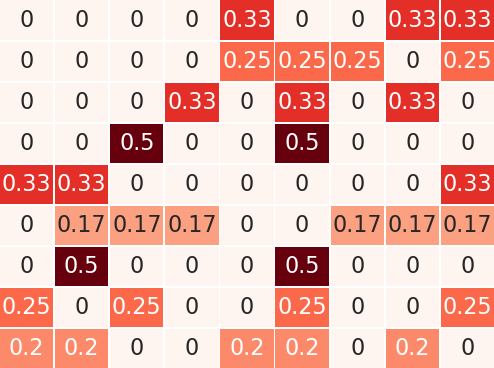}
\caption{}
\label{fig:adj-stochastic}
\end{subfigure}
\caption{(a) The original graph. (b) The adjacency matrix of (a). (c) The corresponding stochastic version.}
\label{fig:adj}
\end{figure}

Adjacency matrices are nice, but I think most of the times you'll see them transformed in various ways to squeeze out all the possible analytic juice. The simplest makeover we can give to the adjacency matrix is to convert it into a stochastic matrix. This means that we normalize it, dividing each entry by the sum of its corresponding row -- this means that each of its rows sums to one. If nodes $u$ and $v$ are connected, and $u$ has $5$ connections, the $A_{uv}$ entry will be $1 / 5 = 0.2$. Figure \ref{fig:adj} shows an example of this stochastic transformation.

What's the usefulness of the stochastic adjacency matrix? The first direct use we can make of it is to calculate transition probabilities. This is literally what it contains: each entry is the probability that a random walker (see Chapter \ref{cha:rndwalks}) on a given node (row) will cross that edge. Since non-edges have value zero, it is impossible to follow them. In Figure \ref{fig:adj}(a) we see that node $9$ has degree equal to five. This corresponds to having five entries set to one in Figure \ref{fig:adj}(b). If we close our eyes and pick one of these at random, each one has a probability of $0.2$ to be picked. That is the value in Figure \ref{fig:adj}(c).

Suppose we picked node $6$ and that we repeat the exercise. Picking one of node $6$'s neighbors at random has a probability of $0.17$, and we end up -- for instance -- on node $3$. We might want to know what was the likelihood of ending in node $3$ starting from node $9$ and doing exactly two random jumps. The probability is not simply the product of the two jumps -- as we would do naively for independent events (see Section \ref{sec:prob-axioms}) -- because there is an alternative route. We could have visited node $8$ first and \textit{then} moved to $3$. We have to keep track of all possible alternative paths, and this becomes really unwieldy when we start considering longer random walks.

Luckily, we don't have to do it. The stochastic matrix has the power of telling us what we want. It's literally its \textit{power}. Say $A$ is our stochastic matrix. We just saw how $A$ is just the probability of transitioning from one node to another. In other words, it gives us the probability of all transitions for random walks of length $1$ -- the length of a walk is the number of edges we crossed or the number of steps we took (we'll talk in depth about this in Chapter \ref{cha:paths}). Let's now write this matrix as $A^1$, which is the same thing as $A$. Let's say this again: $A^1$ is the probability of all transitions for random walks of length $1$. Could it be, then, that $A^2$ is the probability of all transitions for random walks of length $2$? And that $A^n$ is the probability of all transitions for random walks of length $n$? Yes, they are!

From the matrix multiplication crash course I gave you in Section \ref{sec:la-matrix} you know why: $A^2$'s $uv$ entry is, as the formula I wrote there shows, the sum of the multiplication of probabilities of all nodes $k$ that are connected to both $u$ and $v$, and thus can be used in a path of length $2$. Multiplying $A_{uk}$ to $A_{kv}$ means asking the probability of going from $u$ to $k$ \textit{and} from $k$ to $v$. Summing $A_{uk_1}A_{k_1v}$ to $A_{uk_2}A_{k_2v}$ means asking the probability of passing through $k_1$ \textit{or} $k_2$. See Chapter \ref{cha:prob} for a refresher on what multiplying and summing probabilities mean.

\begin{figure}
\centering
\begin{subfigure}[t]{.32\columnwidth}
\includegraphics[width=\textwidth]{figures/matrix_firstexample_04.png}
\caption{$A^1$}
\label{fig:stochastic-powers1}
\end{subfigure}
\begin{subfigure}[t]{.32\columnwidth}
\includegraphics[width=\textwidth]{figures/matrix_firstexample_06.png}
\caption{$A^2$}
\label{fig:stochastic-powers2}
\end{subfigure}
\begin{subfigure}[t]{.32\columnwidth}
\includegraphics[width=\textwidth]{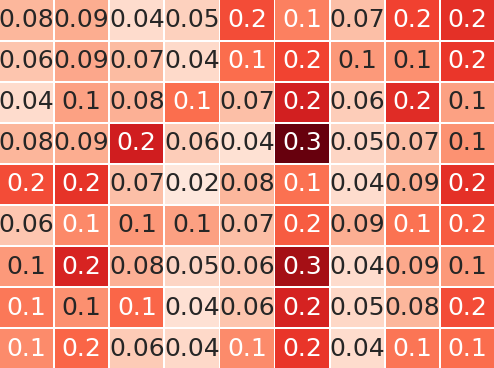}
\caption{$A^3$}
\label{fig:stochastic-powers3}
\end{subfigure}
\caption{Different powers of the stochastic adjacency matrix of the graph in Figure \ref{fig:adj}(b).}
\label{fig:stochastic-powers}
\end{figure}

Let's take a closer look at $A^2$ and $A^3$ in Figures \ref{fig:stochastic-powers}(b) and \ref{fig:stochastic-powers}(c). First, they are stochastic matrices, and Section \ref{sec:prob-stoch} taught you that the rows of a stochastic matrix always sum to $1$. So each entry in the matrix is still a transition probability. This time, though, it's not the transition probability of the direct connection, but of a path of length $2$ and $3$, respectively. 

Second, the diagonal is not zero any more. That is because, with a random walk of length $2$, there is a chance to select the same edge twice, and therefore returning to the point of origin. So the $A^2_{vv}$ entry tells you the likelihood of starting from $v$ and returning back to $v$ in two steps. That is because -- as the matrix multiplication section showed you mathematically -- $A^2_{vv}$ is the combination of the probabilities of going from any of $v$'s neighbors to $v$, weighted by the probability of having reached each of $v$'s neighbors from $v$ itself.

Third, while $A^2$ still has zero entries, $A^3$ does not. This is because some node pairs are farther than two edges away, so the probability of a random walker to reach them in two hops is zero -- check Figure \ref{fig:adj}(a) again if you don't believe me! On the other hand, no pair of nodes is farther than three hops away, and thus there is always a path of length three between any node pair, no matter how unlikely.

Finally, stochastic matrices -- either $A$ or any of its powers -- are not symmetric any more, even if the ``raw'' adjacency matrix of an undirected graph is. This is because the likelihood of ending in $u$ from $v$ isn't necessarily the same as the other way around: if $v$ has better connected neighbors, the random walkers are more likely to be led astray. For instance, the probability to go from node $4$ to $5$ in three random steps is $0.04$, while the other way around is $0.02$. That is because node $5$ has an extra connection, which can lead the walker to be unable to reach $4$ in two additional hops.

This last property means that, if you transpose a stochastic adjacency matrix, $A^T \neq A$ even for undirected graphs! Since you normalized by row sum, the $A_{uv}$ entry can be different from the $A_{vu}$: the only thing you know is that they're both non-zero.

There is one surprise hidden in the folds of $A^n$ for a suitably large $n$. This surprise is waiting for you in Section \ref{sec:rw-stationary}.

Note that there are two valid stochastic adjacency matrices for a bipartite network. If you normalize by row sum, the stochastic $A$ tells you the probability of going from a $V_1$ node to a $V_2$ node. Normalizing by column sum, which is equivalent of taking the stochastic $A^T$ (i.e., transposing $A$ beforehand), then the matrix tells you the probability of going from a $V_2$ node to a $V_1$ node. In fact, bipartite matrices allow us to make use of another linear algebra operation I talked about: the transpose.

Suppose you have a bipartite network and what you really want to know is not which node of type $V_1$ connects to nodes in type $V_2$, but how similar nodes in $V_1$ are, because they connect to the same $V_2$ nodes. This is practically the subject of Chapter \ref{cha:projections} but, to make it simple for this example, you want the probability of going from a $V_1$ node to another $V_1$ node, passing via $V_2$ nodes.

Dividing the bipartite adjacency matrix by its row sum gets you the probability to go from a $V_1$ node to a $V_2$ node -- as we just said. However, we don't know how to go back: in that case we should normalize by column sum! That could be achieved by normalizing $A^T$, $A$'s transpose, by its row sum. Since $A$ is a $|V_1| \times |V_2|$ matrix and $A^T$ is a $|V_2| \times |V_1|$ matrix, you can multiply one by the other: $AA^T$ is in fact a $|V_1| \times |V_1|$ which is exactly what you need.

But wait, what does $AA^T$ actually mean? What's the result of such an operation? Well, following Figure \ref{fig:mat-mult}, the $(v_1,v_2)$ cell is the sum of the probability of going from $v_1$ to any $V_2$ node times the probability of arriving to $v_2$ from any $V_2$ node. Which is what we wanted!

Without linear algebra, you'd have to represent the adjacency by sets of neighbors, and then calculate intersections and dividing various scalars in isolation. But transposes and matrix multiplications are such standard operations that many libraries will have implemented them very efficiently, with the result of being blazingly fast to calculate and extremely easy to incorporate in your code. Moreover, if you use sparse matrix representations you can also be memory efficient, meaning you can go big with your matrices!

Coming back to eigenvalues and eigenvectors (Section \ref{sec:la-eigen}), the largest eigenvalue of a stochastic adjacency matrix is always equal to one -- we also call it the ``leading'' eigenvalue, or $\lambda_1$. This takes a special value: any stochastic adjacency matrix you can come up with will always have $\lambda_1 = 1$. No eigenvalue will ever be greater.

As we saw in Figure \ref{fig:eigenvector}, each eigenvalue has a corresponding eigenvector. We call the eigenvector associated to the largest eigenvalue the ``largest'' or ``leading'' eigenvector, for convenience. The point of looking at eigenvectors is that there is a relationship between the $v$-th entry in the $i$-th eigenvector of an adjacency matrix and node $v$'s relationship with the entire graph.

For instance, the multiplicity of the largest eigenvalue of the stochastic adjacency matrix is important. It can happen that second largest eigenvalue $\lambda_2$ of $A$  could be equal to the first. And, actually, also the third, fourth, fifth, ... could be equal to the first. This is related to the first application of linear algebra to network analysis, which we will fully appreciate when it will be time to discuss about connected components (Section \ref{sec:paths-ccomps}).

\section{Incidence}
In general, an incidence matrix is a matrix telling you what are the relations between two classes of objects. For instance, you can have an incidence matrix telling you for which company a person works. Since the two classes might have a different number of members, incidence matrices are not necessarily square. In fact, you could say that the adjacency matrix of a bipartite network is an incidence matrix.

\begin{figure}
\centering
\begin{subfigure}{.45\columnwidth}
\includegraphics[width=\textwidth]{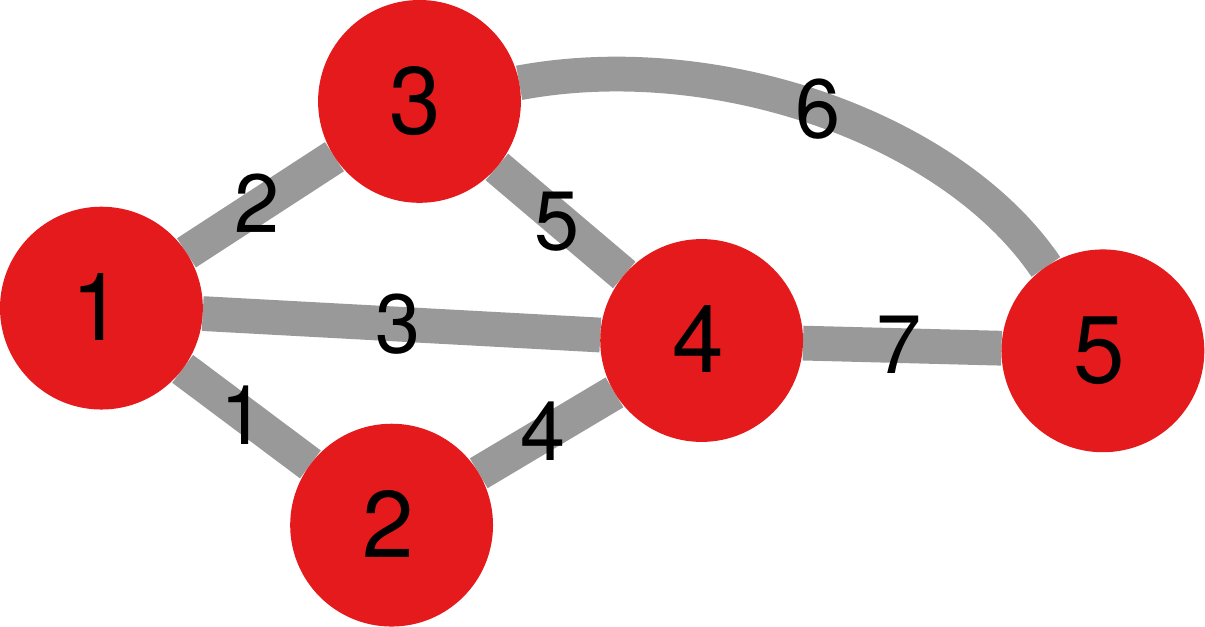}
\caption{}
\end{subfigure}
\qquad
\begin{subfigure}{.33\columnwidth}
\includegraphics[width=\textwidth]{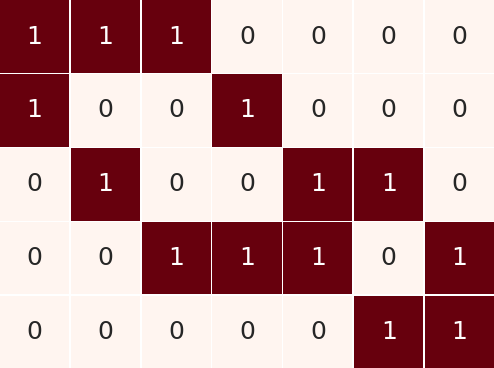}
\caption{}
\end{subfigure}
\caption{(a) A graph with nodes and edges labeled with their ids. (b) The incidence matrix of (a), with nodes on the rows and edges on the columns.}
\label{fig:incidence}
\end{figure}

However, when performing network analysis, there is one type of incidence matrix that is widely used, and thus ``owns'' this term. The vast majority of times, if you read a paper talking about the ``incidence matrix'', you'll see the same object: a matrix that has nodes on the rows, edges on the columns, and it has an entry equal to one if the node and the edge are connected to each other. Figure \ref{fig:incidence}(b) shows the incidence matrix of the graph at Figure \ref{fig:incidence}(a).

Incidence matrices have interesting properties, some more trivial than others. For instance, you know that, in the incidence matrix of a simple graph, each column sums to two because each edge only connects two nodes. Only in the incidence matrix of a hypergraph a column can sum to a number larger than $2$. You can use the incidence matrix to construct other special matrix representations. For instance, you can construct the adjacency matrix of the line graph of $G$  by calculating $B^TB - 2I$, assuming that $B$ is the incidence matrix, and $I$ is a $|E| \times |E|$ identity matrix.

An incidence matrix can also be oriented. In an oriented incidence matrix, the columns sum to zero. For every edge, one of the two non zero entries -- the nodes to which it is attached -- is equal to $1$ and the other is equal to $-1$. It doesn't really matter which of the two you pick, as long as you make sure all columns sum to zero. If $B$ is an oriented incidence matrix, you can use it to construct the Laplacian as $BB^T$. The Laplacian is a super cool matrix and I'll focus on it now.

\section{Laplacian}\label{sec:mat-mat-laplacian}

\subsection{Basic Laplacian}
\begin{figure}
\centering
\begin{subfigure}[t]{.32\columnwidth}
\includegraphics[width=\textwidth]{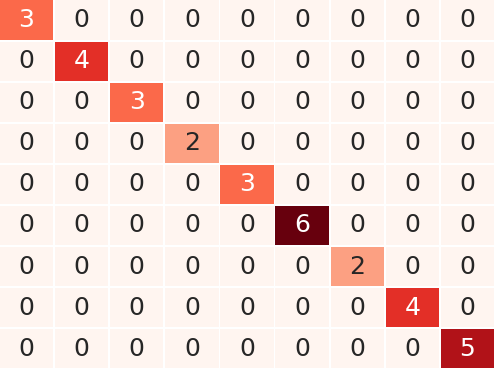}
\caption{}
\end{subfigure}
\qquad
\begin{subfigure}[t]{.27\columnwidth}
\includegraphics[width=\textwidth]{figures/graph_view.pdf}
\caption{}
\end{subfigure}
\caption{The degree matrix (a) of the sample graph (b).}
\label{fig:matrix-degree}
\end{figure}

The stochastic adjacency matrix is nice, but the real superstar when it comes to matrix representations of networks is the Laplacian. To know what that is, we need to introduce the concept of Degree matrix $D$ -- which is a very simple animal. It is what we call a ``diagonal'' matrix. A diagonal matrix is a matrix whose nonzero values are exclusively on the main diagonal. The other off-diagonal entries in the matrix are equal to zero. In $D$ the diagonal entries are the degrees of the corresponding nodes. Figure \ref{fig:matrix-degree} shows an example of a degree matrix.

\begin{figure}
\centering
\begin{subfigure}[t]{.32\columnwidth}
\includegraphics[width=\textwidth]{figures/matrix_firstexample_14.png}
\caption{$D$}
\label{fig:laplacian-d}
\end{subfigure}
\begin{subfigure}[t]{.32\columnwidth}
\includegraphics[width=\textwidth]{figures/matrix_firstexample_03.png}
\caption{$A$}
\label{fig:laplacian-a}
\end{subfigure}
\begin{subfigure}[t]{.32\columnwidth}
\includegraphics[width=\textwidth]{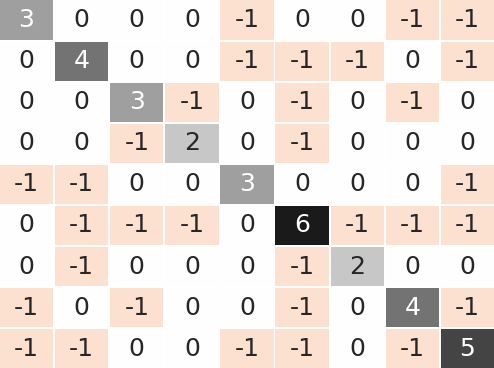}
\caption{$L$}
\label{fig:laplacian-l}
\end{subfigure}
\caption{The operation producing the Laplacian matrix $L$ (c), subtracting the adjacency matrix $A$ (b) from the degree matrix $D$ (a).}
\label{fig:laplacian}
\end{figure}

The Laplacian version of the adjacency matrix -- which we call $L$ -- is the result of a simple operation: $L = D - A$. In practice, we take the degree matrix $D$ and we subtract $A$ from it. $L$ is a matrix that has the node degree in the diagonal, $-1$ for each entry corresponding to an edge in the network, and zero everywhere else. Figure \ref{fig:laplacian} depicts the operation. Why is this matrix interesting? It was originally developed to represent something very physical: it captures the relation between voltages and currents between resistors\cite{kirchhoff1847ueber} -- represented by the edges of the graph.

However, you don't need to care about electric circuits to find a use for the Laplacian. $L$ has a number properties that are useful in general, regardless of what your graph represents. Some of the most obvious ones are that, since it has the degree of the node on the diagonal and $-1$ for each of the node's connection, the sums of all rows and columns are equal to zero.

The Laplacian of a connected undirected graph is part of the positive semi-definite club (Section \ref{sec:la-dot}), which will come in handy in Section \ref{sec:nvd-ge-lapl}. 

Just like for $A$, we are also interested in the eigenvectors of $L$. We need to make some adjustments, though. Instead of looking at the largest eigenvalues, we focus on the smallest ones -- meaning that now we use $\lambda_1$ to refer to the \textit{smallest} eigenvalue. This takes a special value like for $A$ but, for $L$, $\lambda_1 = 0$. Besides doing some of the same things you can do with the eigenvector of the adjacency matrix, the Laplacian has a few more tricks up its sleeve. For instance, one can use it to solve the normalized cut problem, which is useful for community discovery and we will discuss it in detail in Section \ref{sec:rw-mincut}.

Another connection between stochastic and Laplacian matrices is on multiplicity. The multiplicity of the smallest eigenvalue of the Laplacian plays the exact same role as the one of the largest eigenvalue of the stochastic matrix -- a role that you will appreciate in Section \ref{sec:paths-ccomps} when we will study connected components.

We will see what makes the Laplacian so important in Chapter \ref{cha:rndwalks}, when I'll show you how many things you can do with its quasi-mystical properties.

\subsection{Special Laplacians}
The way the Laplacian is built makes it non trivial how to generalize it for different types of networks. Here I focus on two cases: \textbf{directed} and \textbf{signed} graphs.

\begin{figure}
\centering
\begin{subfigure}[t]{.32\columnwidth}
\includegraphics[width=\textwidth]{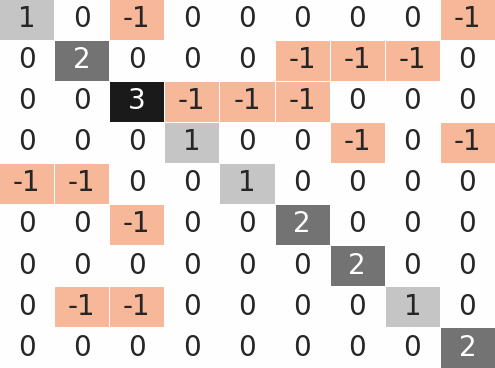}
\caption{Indegree Laplacian}
\end{subfigure}\qquad\qquad
\begin{subfigure}[t]{.32\columnwidth}
\includegraphics[width=\textwidth]{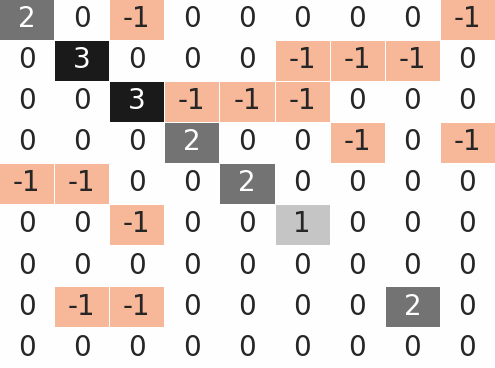}
\caption{Outdegree Laplacian}
\end{subfigure}
\caption{The two Lapacians for the directed graph in Figure \ref{fig:adj-example-dir}.}
\label{fig:laplacian-directed}
\end{figure}

For \textbf{directed} networks the problem comes from the fact that we have to put the degrees of the nodes in the diagonal. However, in a directed networks, Chapter \ref{cha:degree} will teach you that nodes have two degrees: the number of arrow heads pointing to the node (indegree) and the number of arrow tails coming out of the node (outdegree). Which one do we pick? In principle, we could pick either, creating two Laplacians: the outdegree Laplacian (Figure \ref{fig:laplacian-directed}(a)) or the indegree Laplacian (Figure \ref{fig:laplacian-directed}(b)).

However, this deceptively simple operation will make you lose some nice properties the Laplacian has. For instance, the Laplacian normally has real-valued eigenvalues. Unfortunately, as soon as you add a single directed link to the graph, the eigenvalues become complex, i.e. they contain a $i = \sqrt{-1}$ portion. Oops.

\begin{figure}[b]
\centering
\begin{subfigure}[t]{.35\columnwidth}
\includegraphics[width=\textwidth]{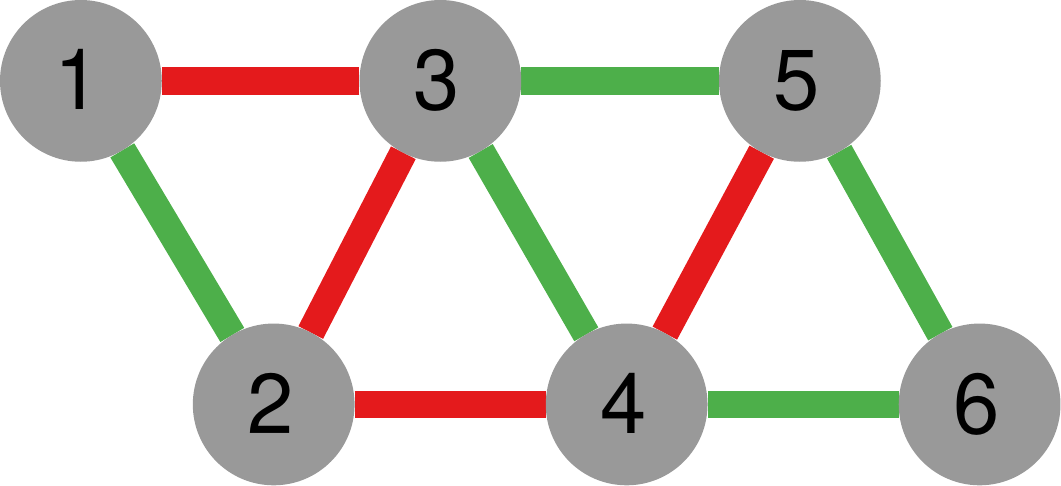}
\caption{}
\end{subfigure}\quad
\begin{subfigure}[t]{.275\columnwidth}
\includegraphics[width=\textwidth]{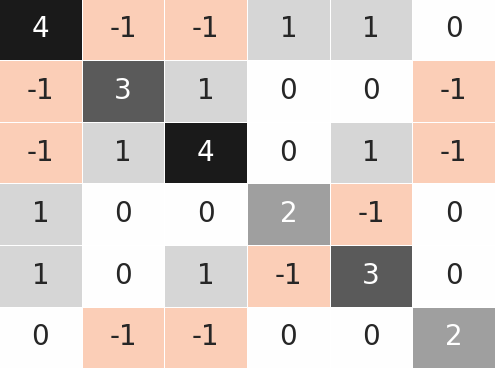}
\caption{}
\end{subfigure}\quad
\begin{subfigure}[t]{.275\columnwidth}
\includegraphics[width=\textwidth]{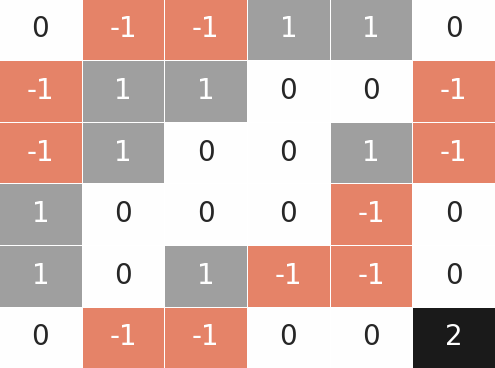}
\caption{}
\end{subfigure}
\caption{(a) A signed graph, with positive edges in green and negative edges in red, and its two Laplacians. (b) The signed Laplacian of (a). (c) The unsigned Laplacian of (a)}
\label{fig:laplacian-signed}
\end{figure}

\textbf{Signed} networks are a special case of multilayer networks I introduced in Section \ref{sec:extended-multilayer}. Also in this case you need to take some care, because you can't simply sum the adjacency matrix to make the degree matrix $D$. You need to sum the absolute value of $A$ to make $D$. If you do so, you get the signed Laplacian\cite{tian2024spreading} -- which I show in Figure \ref{fig:laplacian-signed}(b). You can also make an unsigned Laplacian -- Figure \ref{fig:laplacian-signed}(c) -- by ignoring this issue and summing positive and negative values alike\cite{shi2019dynamics}. However, that means you might end up with a zero entry on the diagonal.

The eigenvectors of the signed Laplacian are as cool as the ones of the regular one, and you'll briefly encounter them in Chapter \ref{cha:lp-multilayer}.

\section{Summary}

\begin{enumerate}
\item You can represent a graph with an adjacency matrix. The matrix has a row/column per node, and cells are equal to one if the two nodes are connected, zero otherwise.
\item Special graph types will have special adjacency matrices. A directed graph has an asymmetric adjacency, a bipartite graph has a non-square one, and multilayer graphs have tensors and supra adjacencies.
\item The stochastic adjacency matrix is a row-normalized (or column-normalized depending on the convention) adjacency matrix, whose rows (or columns) sum to one. It describes transition probabilities from one node to another.
\item The incidence matrix is a $|V| \times |E|$ matrix telling you if a given node is connected to a given edge.
\item The degree matrix $D$ is a matrix having the degree of the node in the main diagonal and zero everywhere else. If you subtract the adjacency matrix from the degree matrix you obtain the graph Laplacian, which is widely used in many network applications.
\item For directed networks, there are two Laplacians depending on whether you use the in- or out-degree for the diagonal. For signed networks, you can have two alternative Laplacians: signed and unsigned.
\end{enumerate}

\section{Exercises}

\begin{enumerate}
\item Calculate the adjacency matrix, the stochastic adjacency matrix, and the graph Laplacian for the network in \url{http://www.networkatlas.eu/exercises/8/1/data.txt}.
\item Given the bipartite network in \url{http://www.networkatlas.eu/exercises/8/2/data.txt}, calculate the stochastic adjacency matrix of its projection. Project along the axis of size $248$. (Note: don't ignore the weights)
\item Calculate the eigenvalues and the right and left eigenvectors of the stochastic adjacency of the network at \url{http://www.networkatlas.eu/exercises/8/2/data.txt}, using the same procedure applied in the previous exercise. Make sure to sort the eigenvalues in descending order (and sort the eigenvectors accordingly). Only take the real part of eigenvalues and eigenvectors, ignoring the imaginary part.
\item Generate the indegree and outdegree Laplacians of the directed graph at \url{http://www.networkatlas.eu/exercises/8/4/data.txt}. Calculate their eigenvalues as well as the eigenvalue of the undirected version of the graph.
\item Generate the signed and unsigned Laplacians of the signed graph at \url{http://www.networkatlas.eu/exercises/8/5/data.txt} -- the third column contains the sign. Calculate their eigenvalues as well as the eigenvalue of the version of the graph ignoring edge signs.
\end{enumerate}

\part{Simple Properties}\label{par:properties}

\chapter{Degree}\label{cha:degree}
So far we just described graph representations. We haven't actually done anything with them. Here, we start probing their properties, to say things about their nodes, edges, and structures. In this chapter we deal with the simplest possible statistics of a node: its degree. This is such a basic concept that I actually already mentioned it multiple times without defining it. It's hard not to, when talking about networks.

The intuition behind the degree is easy to understand in a social network. What is the simplest way for you to know how well you're doing in a society? Well, you could look around you, see the friends you have, and count them. This is the degree. I've got my decent couple of hundreds Facebook friends, like almost everyone else. But some people are superstars, and count the number of their acquaintances in the thousands. How many people does Brad Pitt know? Probably a couple orders of magnitude more than me. Those are the kinds of differences you can quantify by calculating the degree of all nodes in your network.

\begin{figure}
\centering
\includegraphics[width=.6\textwidth]{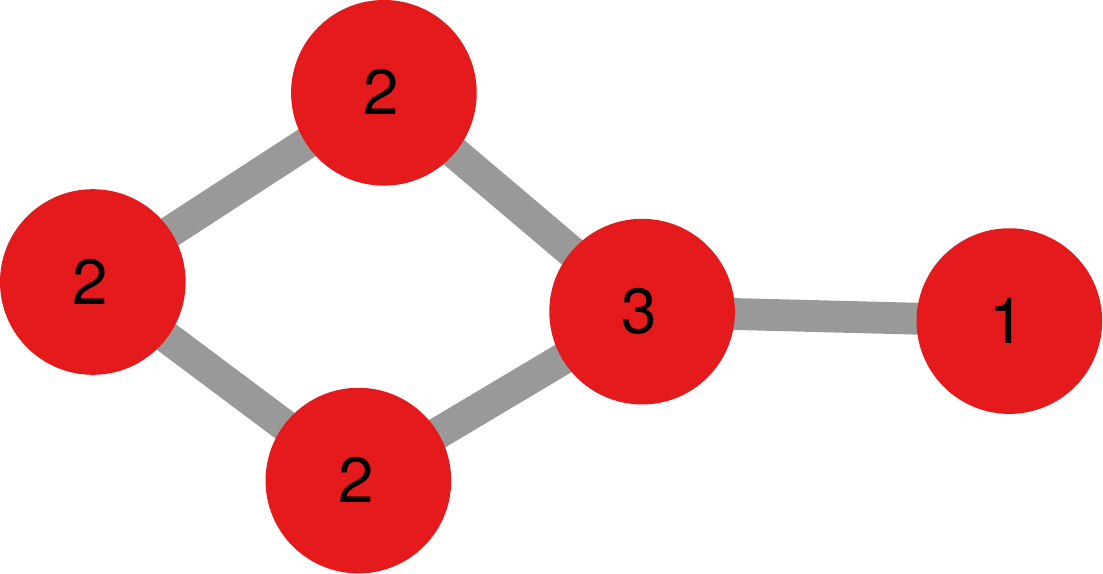}
\caption{A network where I labeled each node with its degree.}
\label{fig:degree}
\end{figure}

The degree of a node is simply the number of edges incident to it\cite{diestel2018graph}, as Figure \ref{fig:degree} shows. Or the number of its connections. Or -- given some assumption that I'll break later on -- the number of its neighbors. It is the first, most fundamental measure of structural importance of a node. A node of high degree ought to be an important node in the network. If it were to disappear, many nodes would lose a connection.

The degree is a property of a node. Let's call $k_v$ the degree of node $v$. We can aggregate the degrees of all nodes in a network to get a ``global'' information about its connectivity. The most common way to do it is by calculating the average degree of a network. This would be $\bar{k} = \sum \limits_{v \in V} k_v / |V|$, however it's much simpler to remember that $\bar{k} = 2|E|/|V|$. The average degree of a network is twice the number of edges divided by the number of nodes. Why twice? Because each edge increases by one the degree of the two nodes it connects.

In a social network, this is how many friends people have on average. What would that number be in your opinion? If we have a social network including two billion people, what's the average degree? It turns our that this number is usually ridiculously lower than one would expect, because -- as we'll see in Section \ref{sec:density-sparse} -- real networks are sparse\cite{broido2019scale}.

We call a node with zero degree, a person without friends, an isolated node, or a singleton. A node with degree one is a ``leaf'' node: this term comes from hierarchies, where nodes at the bottom -- the leaves of the tree -- can only have one incoming connection without outgoing ones. The sum of all degrees is $2|E|$, which implies that any graph can only have an even number of nodes with odd degree\cite{euler1741solutio} -- otherwise the sum of degrees would be odd and thus it cannot be two times something.

\begin{figure}
\centering
\begin{subfigure}[t]{.4\columnwidth}
\includegraphics[width=\textwidth]{figures/outline2.pdf}
\caption{}
\end{subfigure}
\qquad
\begin{subfigure}[t]{.4\columnwidth}
\includegraphics[width=\textwidth]{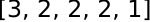}
\caption{}
\end{subfigure}
\caption{A graph and its degree sequence.}
\label{fig:degseq}
\end{figure}

A degree sequence is the list of degrees of all nodes in the network\cite{molloy1998size}\cite{bollobas2001degree}. Typically, we sort the nodes in descending degree, so you always start with the node with maximum degree and you go down until you reach the node with the lowest degree. Figure \ref{fig:degseq} shows an example.

Note that not all lists of integers are valid degree sequences. Some lists cannot generate a valid graph. The easiest case to grasp is if they contain an odd number of odd numbers. As we just saw, the degree sequence must sum to an even number ($2|E|$), thus a sequence summing to an odd number cannot describe a simple undirected graph\cite{sierksma1991seven}. We call all valid sequences ``graphic''.  We'll see that there are other, more subtle, requirements for a graphic sequence.

\section{Degree Variants}\label{sec:degree-variants}
Of course, the degree definition I just gave only makes sense in the world of undirected, unweighted, unipartite, monolayer networks. We had two whole chapters detailing when such a simple model doesn't work in complex real scenarios. We need to extend the definition of degree to take into account all different graph models we might have to deal with.

\subsection{Directed}
As we saw in Section \ref{sec:basic-directed}, edges can have a direction, meaning that the edge going from $u$ to $v$ doesn't necessarily point back from $v$ to $u$. Such is life. In directed graphs you can keep counting the degree as simply the number of connections of a node, but there is a more helpful way to think about it. You might want to distinguish the people who send a lot of connections -- but don't necessarily see them reciprocated --, and those who are the target of a lot of friends requests -- whether they accept them or not.

\begin{figure}
\centering
\begin{subfigure}[t]{.3\columnwidth}
\includegraphics[width=\textwidth]{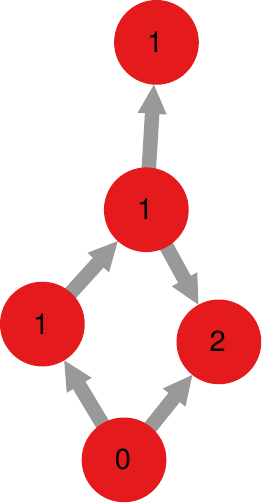}
\caption{}
\end{subfigure}
\qquad
\begin{subfigure}[t]{.3\columnwidth}
\includegraphics[width=\textwidth]{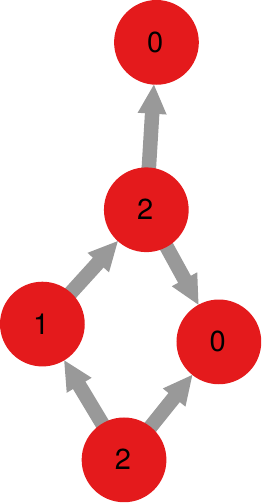}
\caption{}
\end{subfigure}
\caption{(a) A network where I labeled each node with its in-degree. (b) A network where I labeled each node with its out-degree.}
\label{fig:degree-directed}
\end{figure}

So we split the concept in two parts, helpfully named in-degree and out-degree\cite{harary1965structural}\cite{bang2008digraphs}. As one can expect, the in-degree is the number of \textit{in}coming connections. If we represent a directed edge as an arrow, the in-degree is the number of arrow heads attached to your node. See Figure \ref{fig:degree-directed}(a) for a helpful representation. The out-degree is the number of \textit{out}going connections, the number of arrow tails attached to your node. I show the out-degree of the nodes in my example in Figure \ref{fig:degree-directed}(b).

A directed graph's degree sequence is now a list of tuples. The first element of the tuple tells you the indegree, while the second element tells you the outdegree. Or you can have two sequences, but you need to make sure that the $n$th positions of the two sequences refer to the same node. If the two sequences are the same, meaning that every node has the same in- and out-degree, we have a ``balanced'' graph.

\subsection{Weighted}
Most of the time, people do not change the definition of degree when dealing with weighted networks. Many network scientists like how the standard definition works in weighted graphs, and keep it that way. The degree is simply the number of connections a node has.

\begin{figure}
\centering
\includegraphics[width=.6\textwidth]{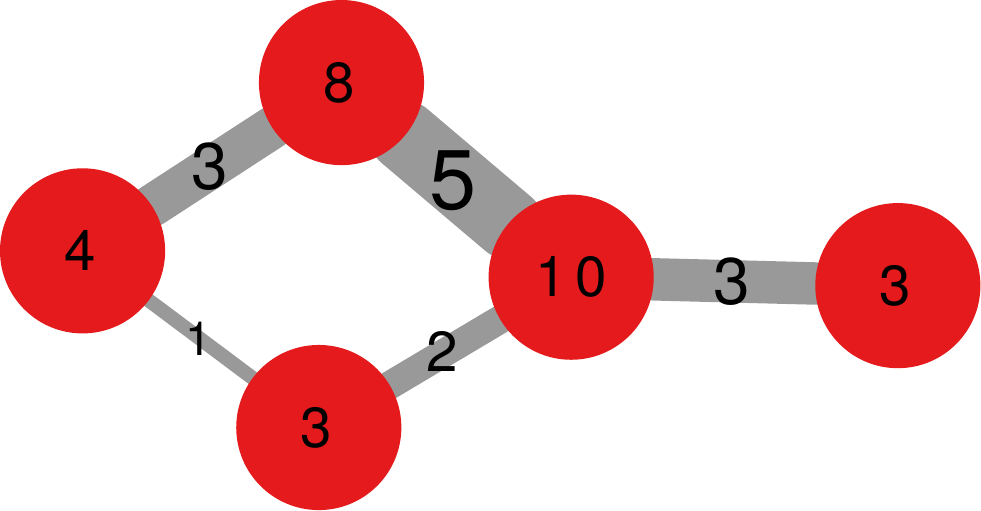}
\caption{A weighted network where I labeled each edge with its weight and each node with its weighted degree.}
\label{fig:degree-weighted}
\end{figure}

Other people don't\cite{barrat2004architecture}. In the case of weighted networks, one might be interested in the total weight incident into a node. We would call such quantity the ``weighted degree'' or ``node strength''\cite{opsahl2010node}. Node strengths are key concepts in investigating propagating failures on networks (Section \ref{sec:epidemapps-cascade}) and in some network backboning techniques (Section \ref{sec:bb-disfilter}).

Node strengths work exactly how you would expect them to do: to get $v$'s weighted degree you sum the weights of all the edges incident to $v$. Figure \ref{fig:degree-weighted} shows an example. The advantage of this definition is that it reduces to the classical degree definition if your network is unweighted -- that is to say that all of $G$'s edge weights are equal to one.

By separating the unweighted count of connections (degree) from the weighted sum of connections (weighted degree), we capture two distinct notions of connectivity. One can have a node with enormous strength but low degree -- a core router on the internet with few high-bandwidth connections -- and a ``peripheral'' router on your street -- which has a large number of low-bandwidth connections.

The reasons to do so are many. For instance, if you're looking a road graph, each edge represents a trait of road. It might be weighted with the number of cars passing through it per unit of time. Nodes, in this case, are road intersections. A weighted degree will tell you how many cars per unit of time want to clear that particular intersection. If the number is too high, you might be in trouble!

\subsection{Bipartite}
The bipartite case doesn't need too much treatment: the degree is still the number of connections of a node. It doesn't matter much that for $V_1$ nodes it is gained exclusively via connections to $V_2$ nodes and viceversa. However, there's a little change when one uses a matrix representation that it's worthwhile to point out. Assuming $A$ as a binary adjacency matrix (not stochastic), in the regular case the degree is the sum of the rows: the sum of first row tells you the degree of the first node, and so on.

\begin{figure}
\centering
\begin{subfigure}[t]{.4\columnwidth}
\includegraphics[width=\textwidth]{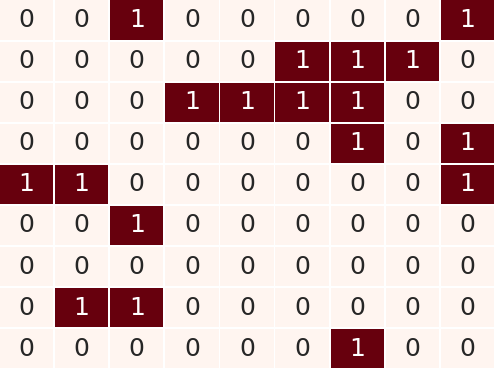}
\caption{$A$}
\end{subfigure}
\qquad
\begin{subfigure}[t]{.4\columnwidth}
\includegraphics[width=\textwidth]{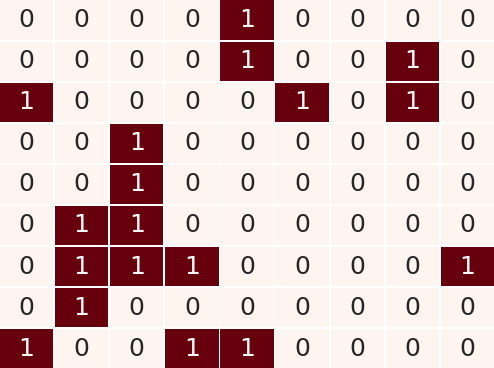}
\caption{$A^T$}
\end{subfigure}
\caption{Calculating the degree of a bipartite network via its adjacency matrix $A$ and its transpose $A^T$. The first $V_1$ node has degree equal to two (the sum of the first row is two). The first $V_2$ node has degree equal to one, which you can calculate either by summing the first column of $A$, or by summing the first row of $A^T$.}
\label{fig:bipartite-degree}
\end{figure}

In a bipartite network that will only tell you the degree of the $V_1$ nodes. You won't know anything about the $V_2$ nodes if you only look at row sums. You can fix the problem in two, equivalent, ways. You can either looking at the column sums, or you can look at the row sums of $A^T$, the transpose of $A$. $A^T$'s rows are $A$'s columns and vice versa, so the equivalence between these two approaches should be self-evident -- if it isn't, try to play with Figure \ref{fig:bipartite-degree}.

Just like directed graphs, also bipartite graphs have two degree sequences, one for $V_1$ nodes and the other for $V_2$ nodes. They both sum to the same value: $|E|$, implying that, in this case, you can have an odd number of odd degree nodes in each node type\cite{asratian1998bipartite}.

\subsection{Multigraph}
When I introduced the degree I said that it can be the number of a node's connections or the number of its neighbors. These two were assumed to be interchangeable, because each edge in a simple graph will bring you to a distinct neighbor. Say that $k_u$ is $u$'s degree, and $N_u$ the set of its neighbors. In a simple graph, $k_u = |N_u|$ -- assuming there are no self-loops or, if there are, that $N_u$ can contain $u$ itself.

That is not the case in a multigraph. Since we allow parallel edges, you can follow two distinct connections and end up in the same neighbor. So we need to solve this ambiguity. The way I saw most commonly accepted is to keep the degree ($k_u$) as the number of connections of a node. The number of neighbors of a node ($|N_u|$) will be just that: the number of neighbors. So, in a multigraph $k_u \neq |N_u|$ or, to be more precise, $k_u \geq |N_u|$.

\subsection{Multilayer}
The multilayer case is possibly the most complex of them all. At first, it doesn't look too bad. The degree is still the number of connections a node has. Then you realize that there are some connections you shouldn't count. For instance, no one -- that I know of -- counts the interlayer coupling connections as part of the degree. It's easy to see why: these are not connections that lead you to a neighbor in a proper sense. They lead you to... a different version of yourself.

\begin{figure}
\centering
\includegraphics[width=.33\textwidth]{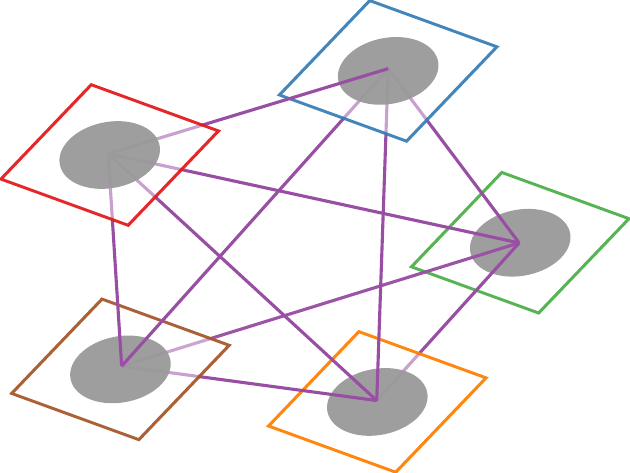}
\caption{Should we really say that the degree of this isolated node is ten just because there are five layers in the network and we couple them with each other? Eight out of ten cats say ``no''.}
\label{fig:coupling-degree}
\end{figure}

Even if we want to ignore this quirk, counting these connections won't really give you meaningful information. If you have a one-to-one multilayer network in which all nodes are part of all layers, they are all going to have the same number of inter-layer couplings. Sometimes, this number can be quite high. If you have five layers and you connect all identities of the same actor across layers, you effectively have a clique (see Section \ref{sec:density-cliques}) of inter-layer couplings. If you count those as part of the degree, this actor would have a degree starting from ten -- as I show in Figure \ref{fig:coupling-degree} --, which would be unreasonable. You could have fewer inter-layer coupling using different coupling strategies, but that wouldn't change the substance.

Since each layer is a network on its own, it is natural to want to have a measure telling us the degree of a node in a particular layer. So an actor can have many degrees: one per layer, and a general one, which we can define as the sum of each layer's degree. However, things can get complicated with a many-to-many mapping. In that case, the actor can ``own'' more than one node in a layer. Each node has its own degree, but how much do they contribute to the actor's degree? The answer might vary, depending on what you're interested in calculating.

\begin{figure}
\centering
\includegraphics[width=.75\textwidth]{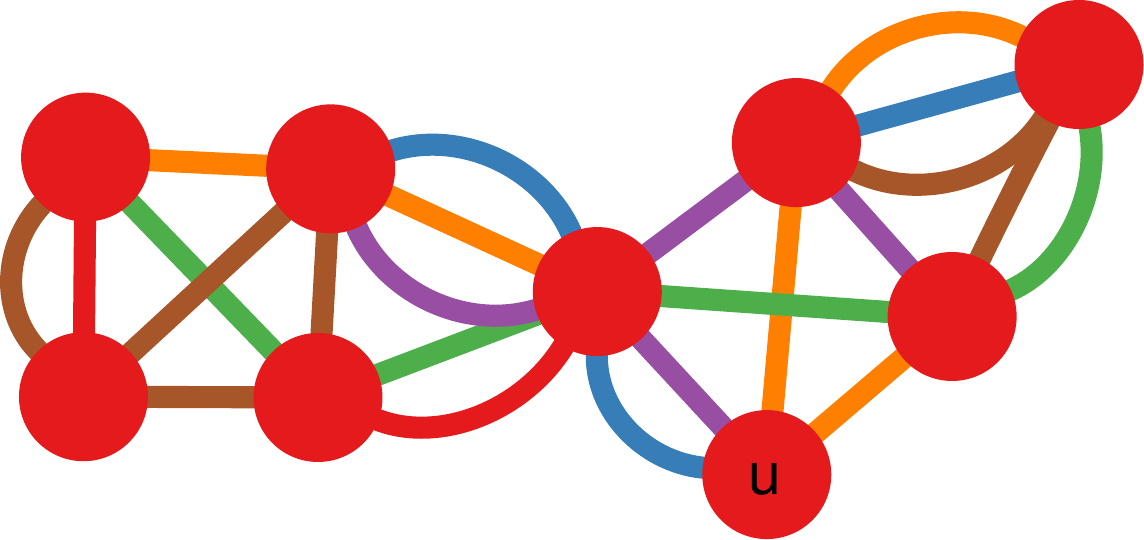}
\caption{A multigraph representation of a multilayer network with one-to-one mapping, where the edge color encodes layer in which it appears.}
\label{fig:dimension-relevance}
\end{figure}

One can also combine layers to do all sorts of interesting stuff. I'm going to give you some examples from a paper of mine\cite[-2.7in]{berlingerio2013multidimensional}, with the caveat that the space of actual possibilities is much vaster than this\cite{de2013mathematical}\cite{battiston2014structural}. What follows is also very related to the multilayer concept of ``versatility'': the ability of a node to be a relevant central node in different layers\cite{de2015ranking}\cite{de2014navigability}\cite{battiston2016efficient}.

Consider Figure \ref{fig:dimension-relevance}. Let's call $u$ the bottom node, the one with two orange edges, a blue and a purple one. We can see that its degree is four (four edges), and its neighbor set is of size three: $u$ has three neighbors, $|N_u| = 3$.

Now, we can count the size of the neighbor set per layer too, or $N_{u,l}$. In the orange layer $u$ has two neighbors ($|N_{u,l}| = 2$), in the blue and purple one it has only one ($|N_{u,l}| = 1$). There is a difference between the neighbors in the orange layer and the ones in the other layers. If $u$ wants to communicate with them, it has to use the orange layer: there is no alternative. On the other hand, if the blue layer were to disappear, $u$ could still use the purple one, and vice versa.

This observation is at the basis of the definition of the ``exclusive neighbor'' set, or $N^{XOR}$. Given a node $u$ and a layer $l$, the $N^{XOR}_{u,l}$ contains those neighbors of $u$ that can be reached exclusively via $l$. If there is an alternative path, those neighbors are not part of $N^{XOR}_{u,l}$. So $|N^{XOR}_{u,l}| = 2$, if $l$ is the orange layer, but $|N^{XOR}_{u,l}| = 0$ in the other two cases. So the exclusive neighbor gives us a rather intuitive measure: how many neighbors would $u$ lose if layer $l$ were to disappear?

We can use $N_{u,l}$ and $N^{XOR}_{u,l}$ to establish some generalized degree definitions, establishing the importance of $l$ for $u$. For instance, the Layer Relevance of $l$ for $u$ is the fraction of $u$'s neighbors that $u$ can reach through $l$, or $|N_{u,l}| / |N_u|$. In Figure \ref{fig:dimension-relevance} that's $2/3$ for the orange layer, and $1/3$ for both the blue and the purple layers. The exclusive variant of Layer Relevance is the fraction of $u$'s neighbors that $u$ can reach through $l$ and $l$ alone: $|N^{XOR}_{u,l}| / |N_u|$. In Figure \ref{fig:dimension-relevance} that's still $2/3$ for the orange layer, but it turns to zero for both the blue and the purple layers.

We can also have a normalized version of Layer Relevance such that it always sums to one for all nodes. In this version, for every pair of connected nodes $(u,v)$, each layer in which this connection appears does not contribute one to the sum, but $1/|L_{u,v}|$, where $|L_{u,v}|$ is the number of layers in which $u$ and $v$ are neighbors of each other. In my example, the normalized Layer Relevance of the orange dimension for $u$ is still $2/3$, but it turns to $1/6$ for the blue and the purple one, because they have to share the remaining $1/3$ of $u$'s neighbors.

\subsection{Hyper}
As one might expect, allowing edges to connect an arbitrary number of nodes -- rather than just two -- does unspeakable things to your intuition of the degree. We can still keep our usual definition: the degree in a hypergraph is the number of hyperedges to which a node belongs -- or: the number of its hyper-connections\cite{erdHos1983supersaturated}\cite{bretto2013hypergraph}. However, if you take any step further, all hell breaks loose. The number of neighbors has no relationship whatsoever with the number of connections: with a single hyperedge you can connect a node with the entirety of the network. Also the average degree is something tricky to calculate. Forget about $\bar{k} = 2|E|/|V|$: if a single hyperedge can connect the entire network, then $|E| = 1$, but $\bar{k} = |V|$.

Things are a bit less crazy for uniform hypergraphs -- where we force hyperedges to always have the same number of nodes. Which might explain why they're a much more popular thing to study, rather than arbitrary hypergraphs. I'll deal with the generalization of the degree for simplicial complexes in Section \ref{sec:hod-simplicial}, because it opens possibilities much more vast than the space I can allow them to have here.

\section{Degree Distributions}\label{sec:degree-distributions}
The degree of a node only gives you information about that node. The average degree of a network gives you information about the whole structure, but it's only a single bit of data. There are many ways for a network to have the same average degree. It turns out that looking at the whole degree distribution can shed light on surprising properties of the network itself. Since degree distributions can be so important, generating and looking at them is a second nature for a network scientist. As a consequence, there are a lot of standardized procedures you want to follow, to avoid confusing your reader by breaking them.

\begin{figure}
\centering
\includegraphics[width=.66\textwidth]{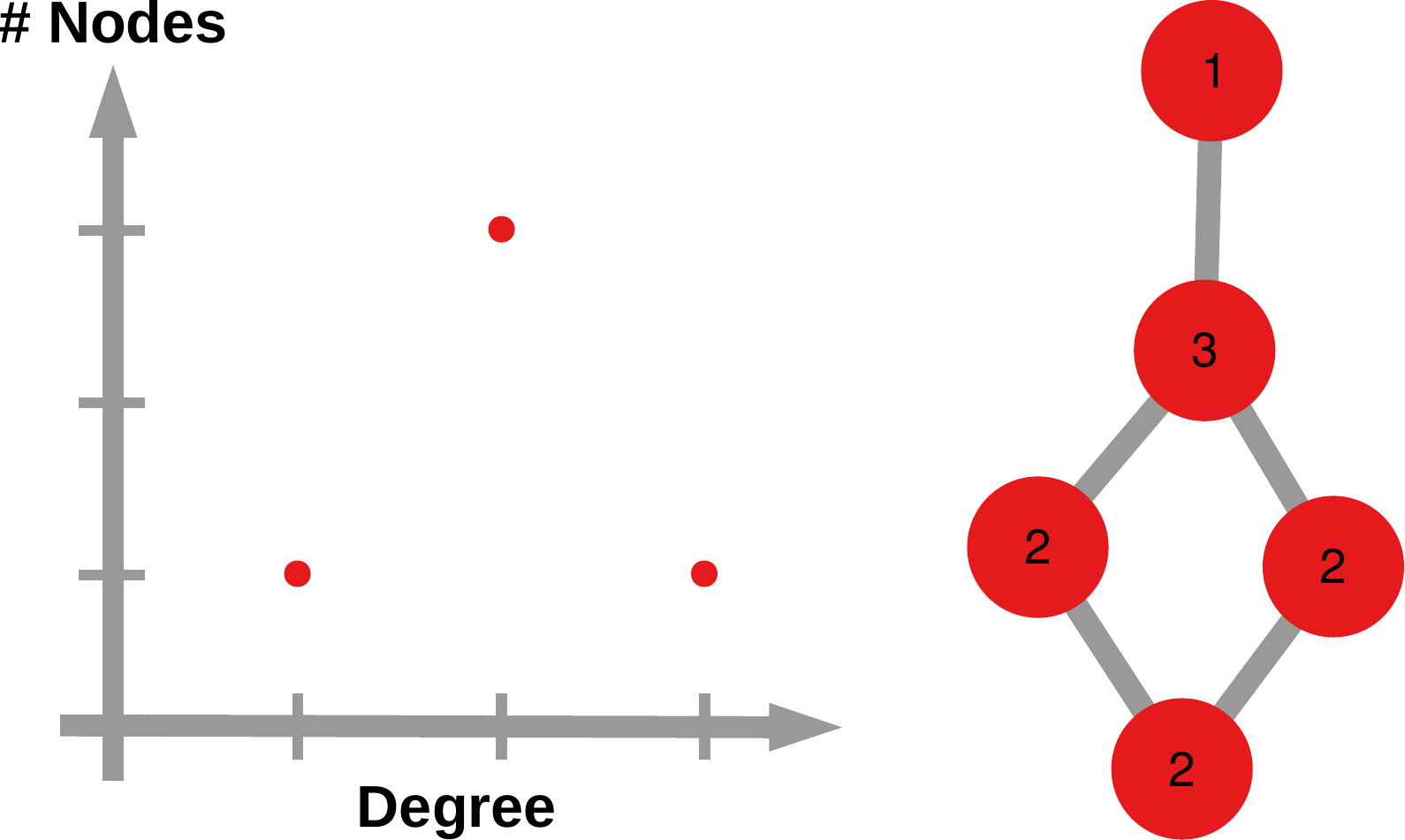}
\caption{The degree scatter plot (left) of the graph on the right.}
\label{fig:degree-histogram}
\end{figure}

Let's break down all the components of a good degree distribution plot. First, the basics. What's a degree distribution? At its most simple, it is just a degree scatter plot: the number of nodes with a particular degree. The degree should be on the x axis and the number of nodes on the y axis, just as I do in Figure \ref{fig:degree-histogram}. Commonly, one would normalize the y axis by dividing its values by the number of nodes in the network. At this point, the y axis is the probability of a node to have a degree equal to $k$, not simply the node count. That makes it easier to compare two networks with a different node count.

\begin{figure}
\centering
\begin{subfigure}{.45\columnwidth}
\includegraphics[width=\textwidth]{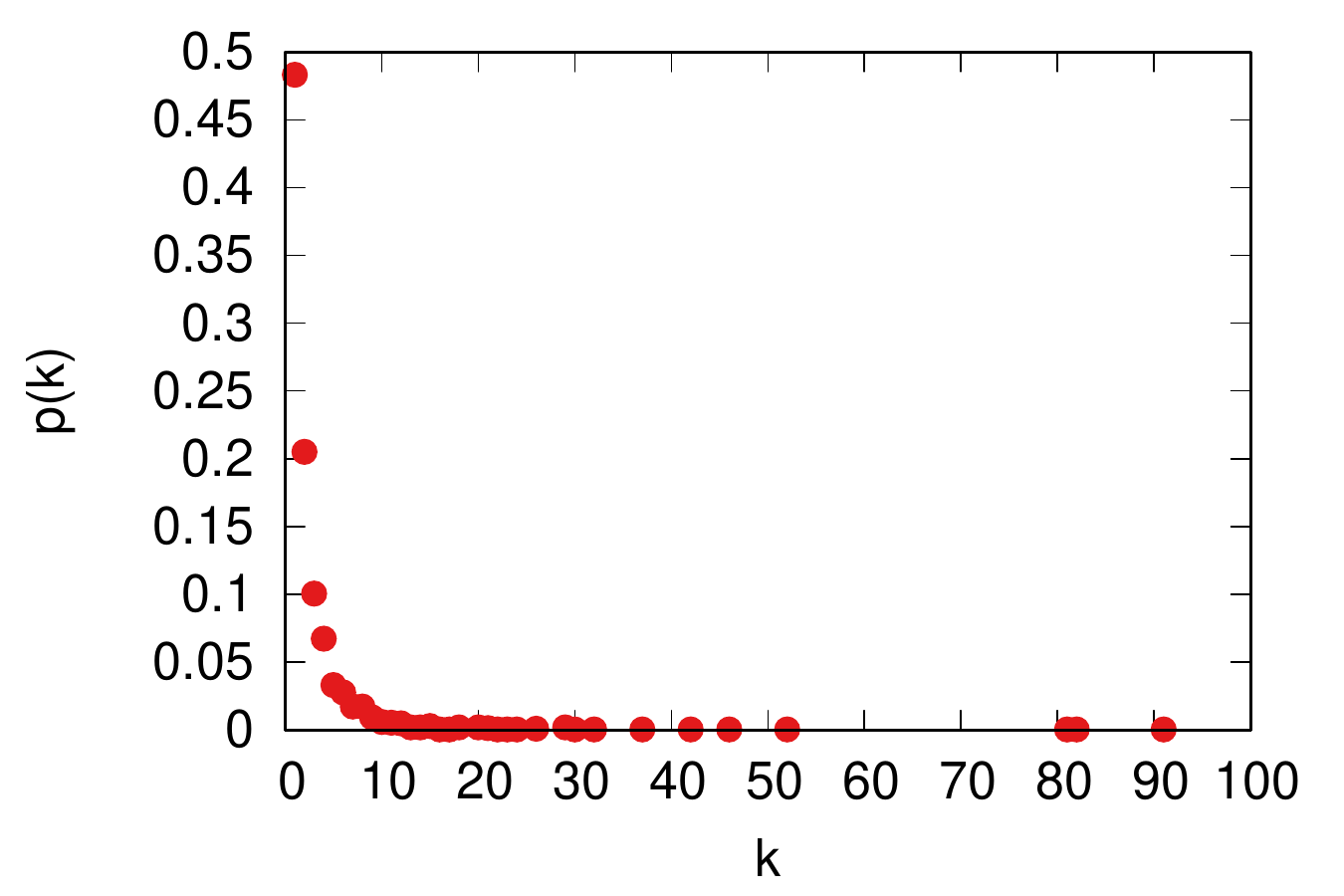}
\caption{}
\end{subfigure}
\qquad
\begin{subfigure}{.45\columnwidth}
\includegraphics[width=\textwidth]{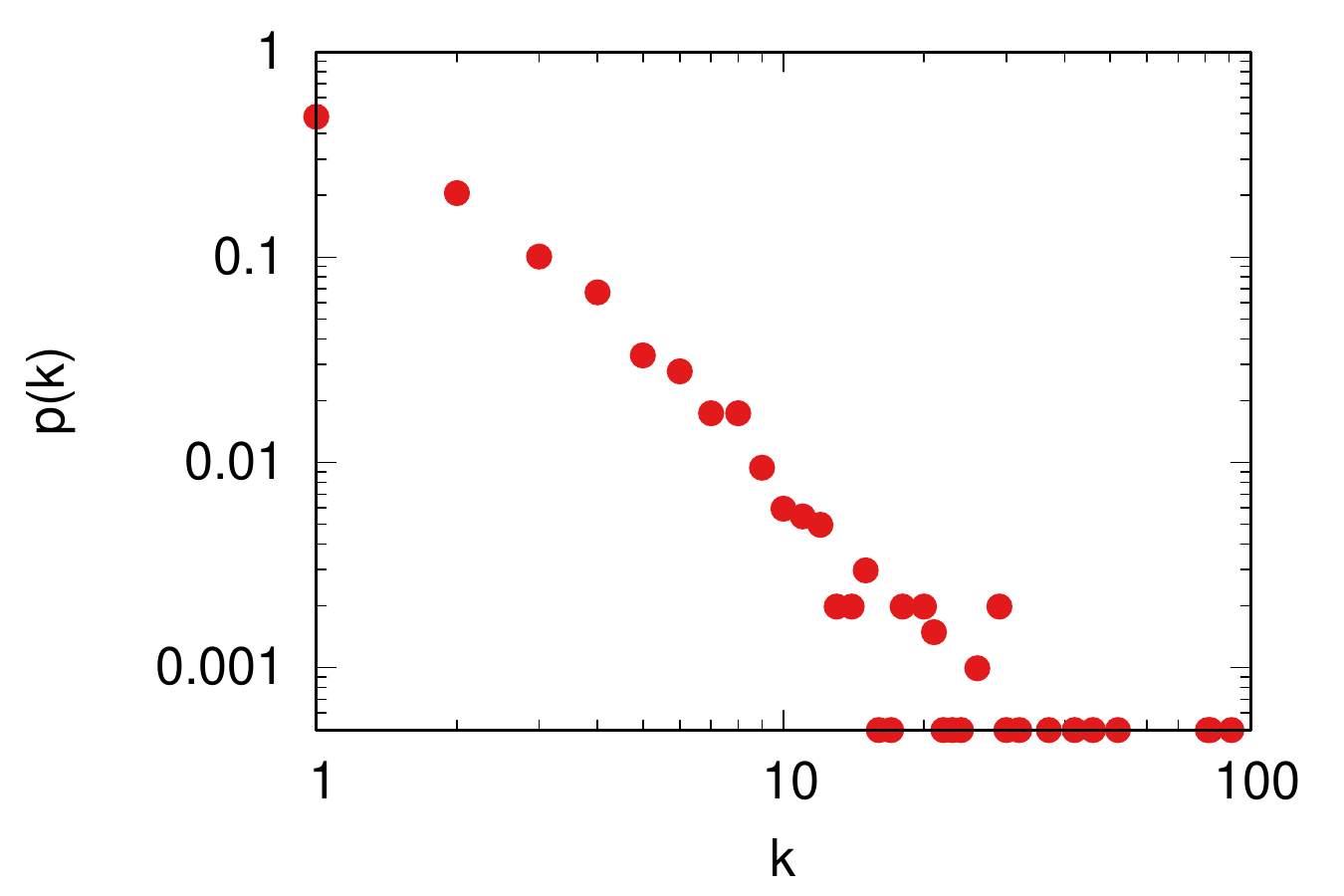}
\caption{}
\end{subfigure}
\caption{The degree distribution of the protein-protein interaction network. The distributions are the same, but in (a) we have a linear scale for the x and y axes, which is replaced in (b) by a log-log scale.}
\label{fig:degree-histogram2}
\end{figure}

Figure \ref{fig:degree-histogram2}(a) shows you the degree distribution of protein-protein interaction for the \textit{Saccharomyces Cerevisiae}, the beer bug. An interesting pattern is that there are lots of nodes with few interactions, and few nodes with many. As a consequence, we end up with all our datapoints concentrated in the same part of the plot, and it's difficult to appreciate both the low- and the high-degree structure. These degree patterns are more evident and easy to see when represented on a log-log scale, as Figure \ref{fig:degree-histogram2}(b) shows, which stretches out the low-degree area while compressing the high-degree one.

So, we just discovered that this protein-protein interaction network has something peculiar. The baseline assumption would be that nodes connect at random. If that were the case, we would expect the degree to distribute normally, in a nice bell-shape -- see Chapter \ref{cha:rndgraphs}. But Figure \ref{fig:degree-histogram2}(b) is not what a normal distribution looks like. The vast majority of nodes have a very low degree, and a few giant hubs have a degree much larger than average. Is this common?

\begin{figure}
\centering
\begin{subfigure}{.45\columnwidth}
\includegraphics[width=\textwidth]{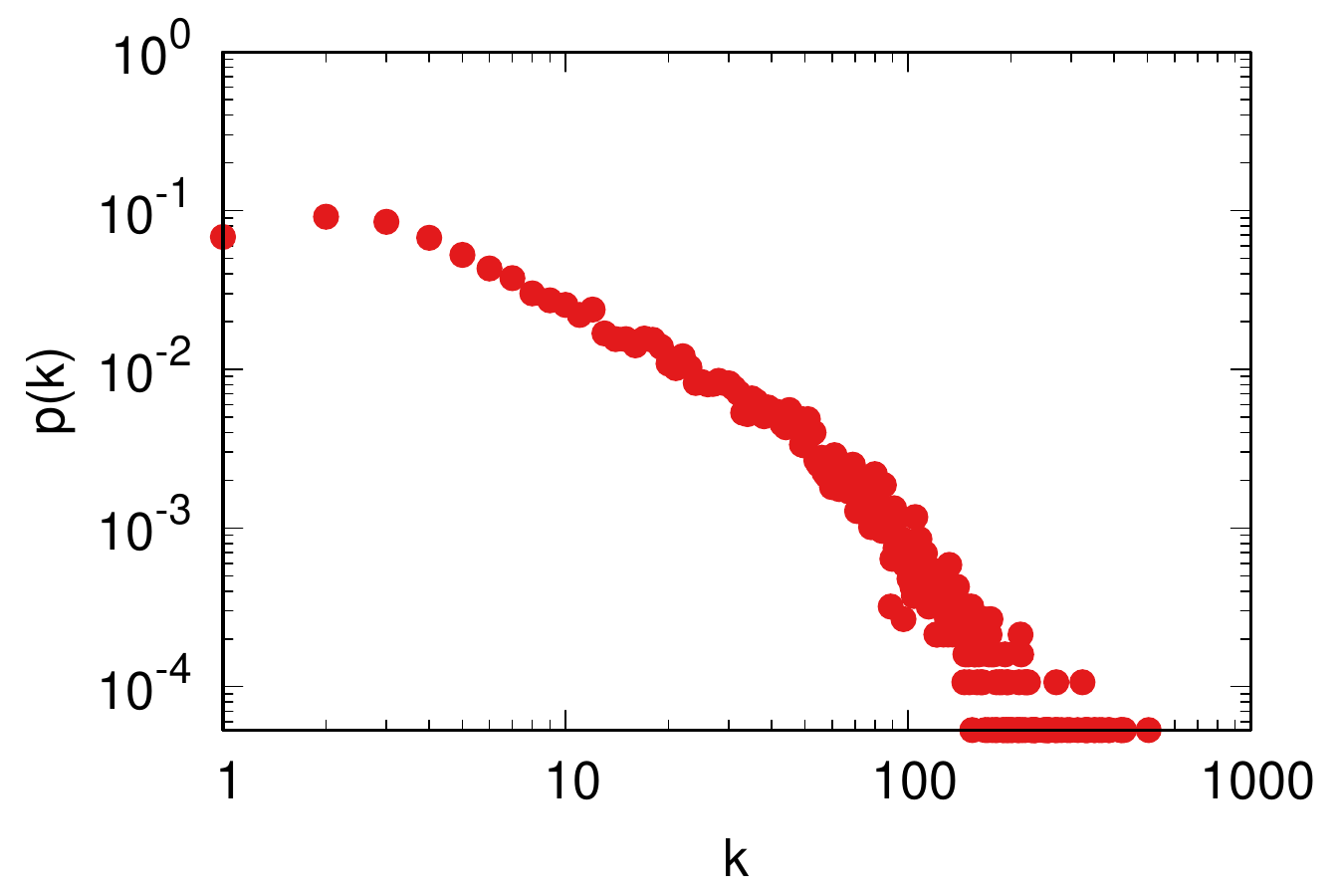}
\caption{}
\end{subfigure}
\qquad
\begin{subfigure}{.45\columnwidth}
\includegraphics[width=\textwidth]{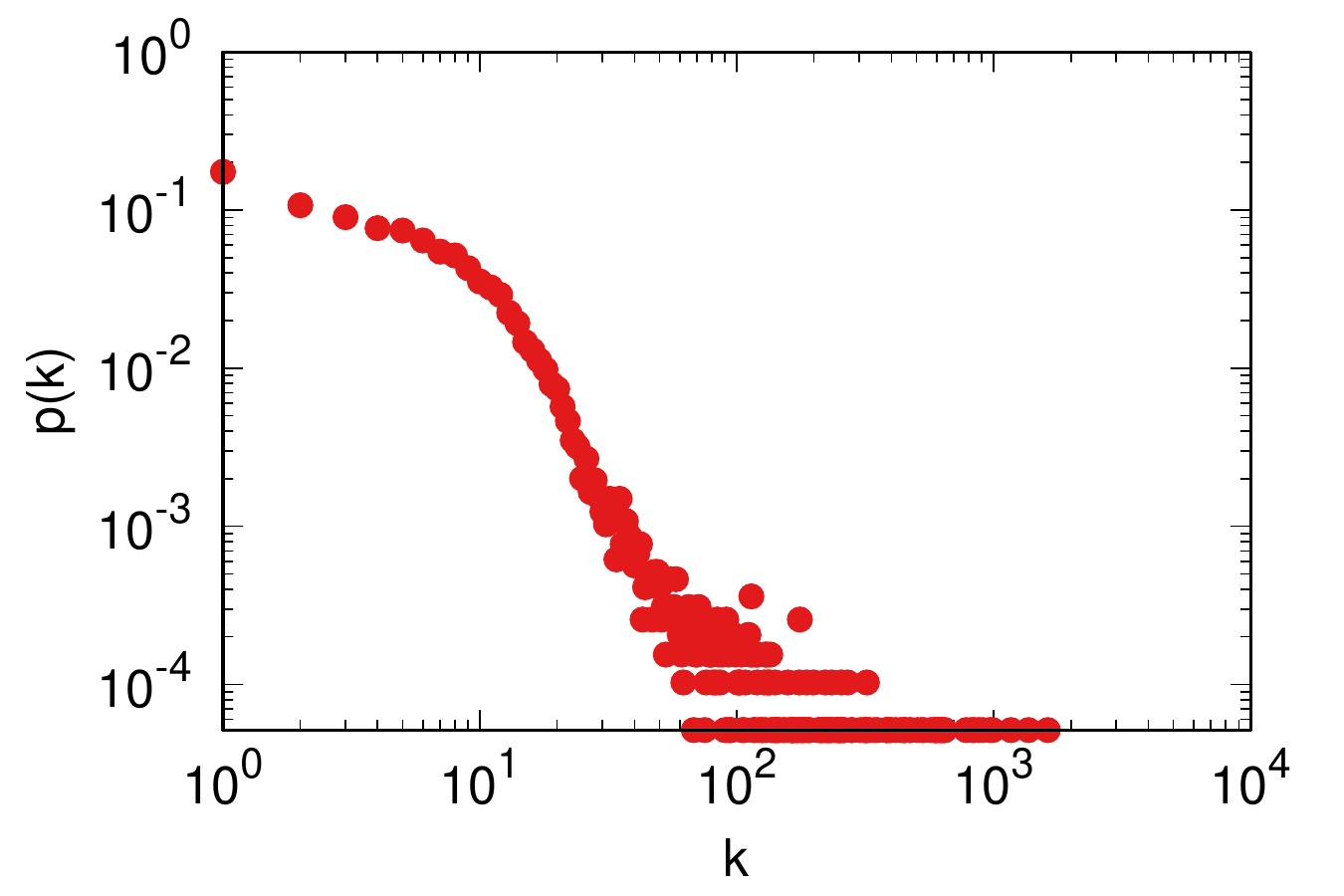}
\caption{}
\end{subfigure}
\qquad
\begin{subfigure}{.45\columnwidth}
\includegraphics[width=\textwidth]{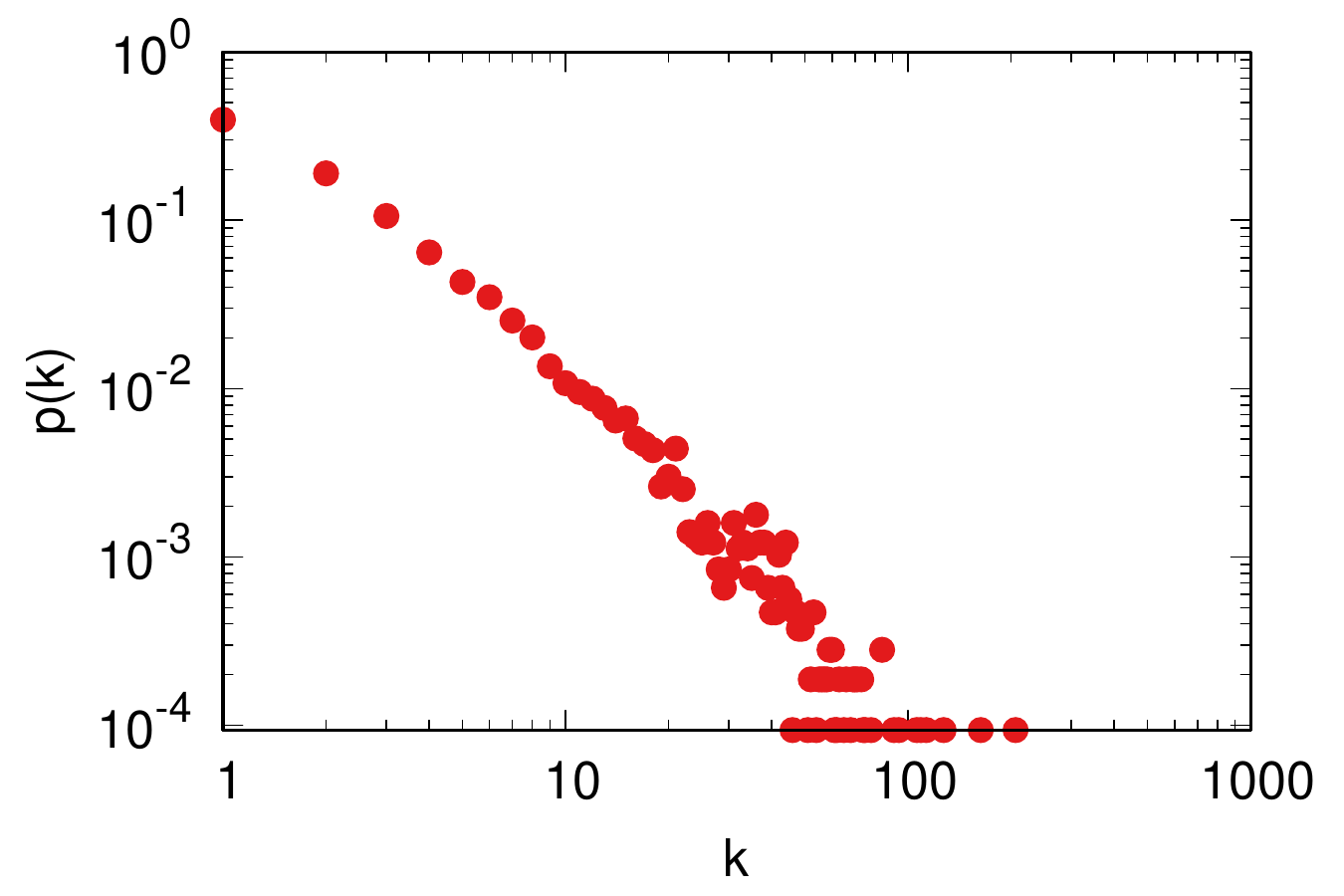}
\caption{}
\end{subfigure}
\qquad
\begin{subfigure}{.45\columnwidth}
\includegraphics[width=\textwidth]{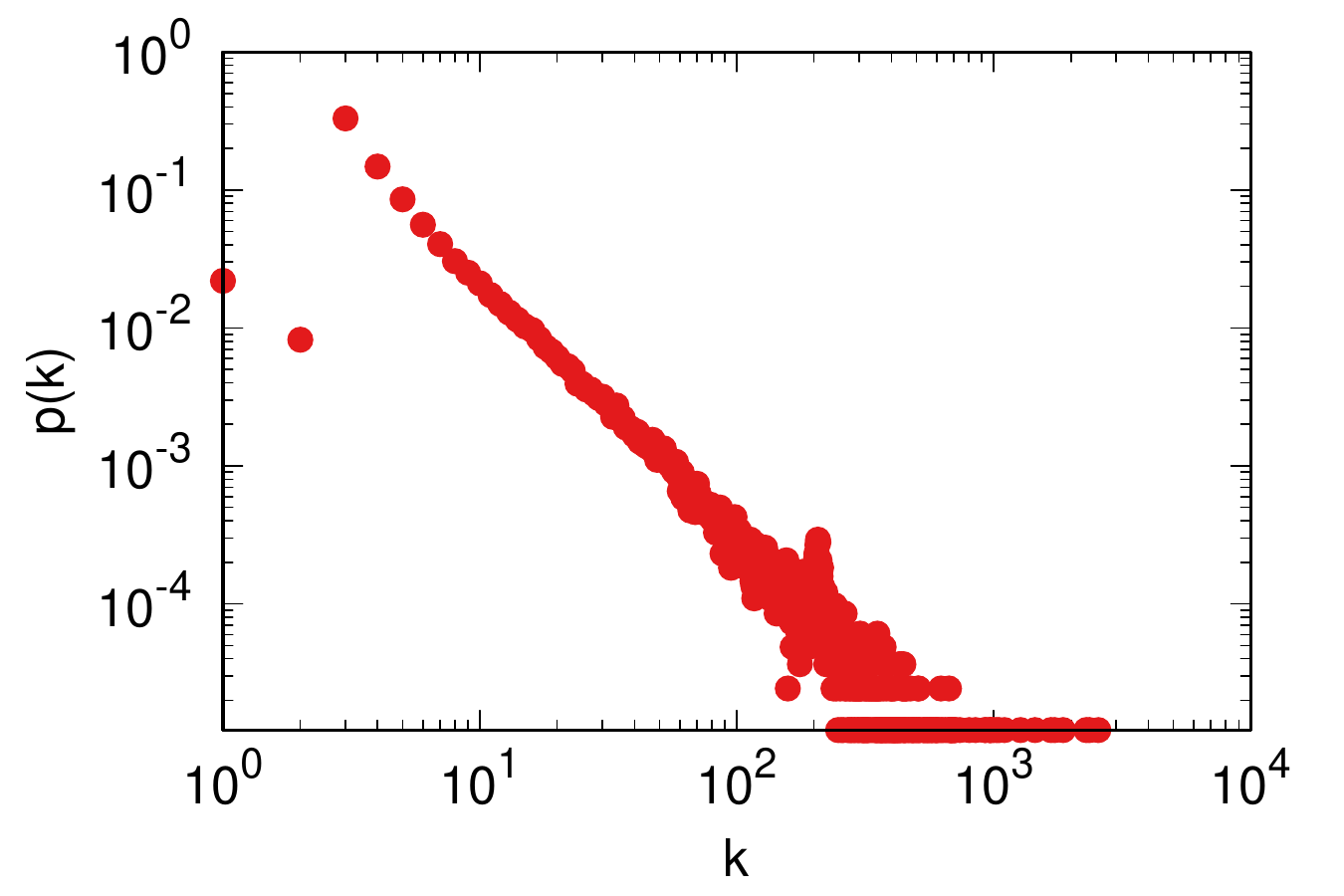}
\caption{}
\end{subfigure}
\caption{The degree distributions of many real-world networks: (a) coauthorship in scientific publication \citep{leskovec2007graph}; (b) coappearance of characters in the same comic book \citep{alberich2002marvel}; (c) interactions of trust between PGP users \citep{boguna2004models}; (d) connections through the Slashdot platform \citep{leskovec2009community}.}
\label{fig:degree-histogram3}
\end{figure}

Yes it is. Most real world networks would show such a broad distribution: email exchanges\cite{ebel2002scale}, synapses in the brain\cite{eguiluz2005scale}, internal cell interactions\cite{albert2005scale}. Take a look at the degree distribution zoo in Figure \ref{fig:degree-histogram3}. To put it simply: in most networks we have many orders of magnitude between the minimum and the maximum degree (x axis), and between the most and least popular degree value (y axis). This is not what scientists initially expected. And when things are not as we expected, we all get excited and start wonder why.

Before exploring these questions we need to finish our deep dive into how to generate and visualize a proper degree distribution. The disadvantages of the degree scatter plots is that they're a bit messy. The physicists in the audience would want a true functional form. But one cannot do that if we have such a broad scatter, especially for high degree values: the wide range of degree values carried only by a node in the network generate what we call a fat tail (Section \ref{sec:stats-summary}).

\begin{figure*}
\centering
\begin{subfigure}{.32\columnwidth}
\includegraphics[width=\textwidth]{figures/pgp_dd_log.pdf}
\caption{Regular scatter.}
\end{subfigure}
\begin{subfigure}{.32\columnwidth}
\includegraphics[width=\textwidth]{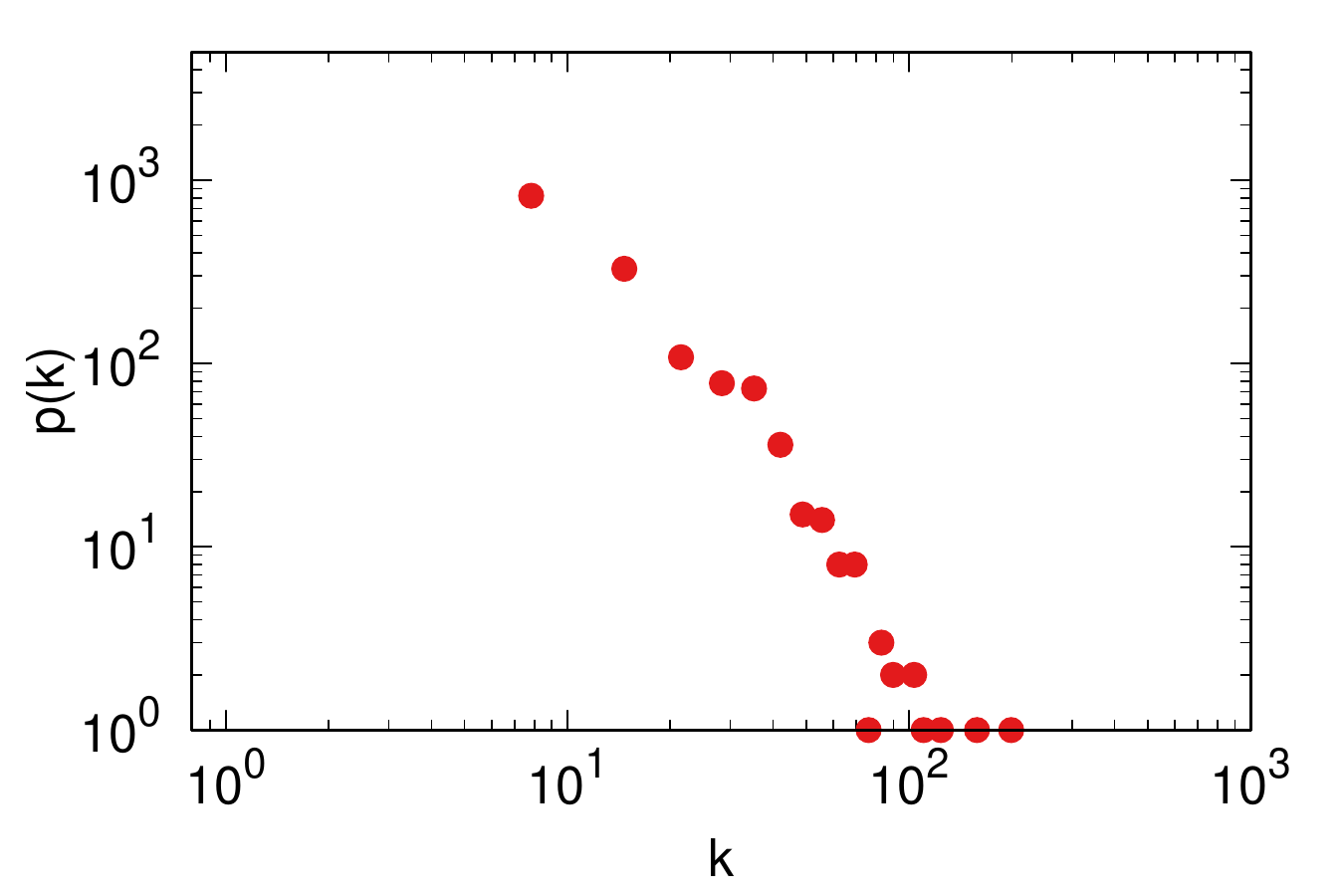}
\caption{Equal size binning.}
\end{subfigure}
\begin{subfigure}{.32\columnwidth}
\includegraphics[width=\textwidth]{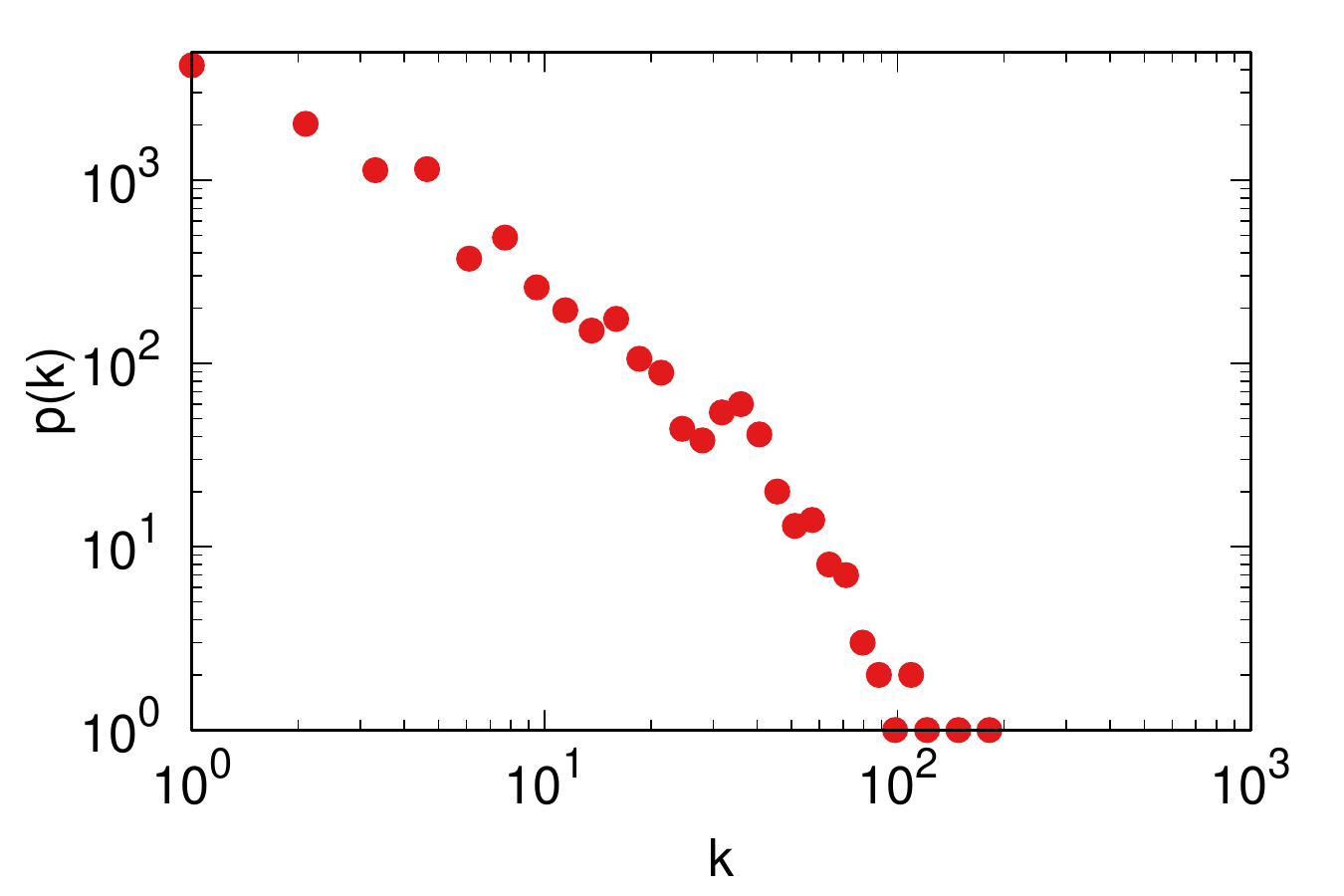}
\caption{Powerbinning.}
\end{subfigure}
\caption{The degree distribution of the PGP trust network.}
\label{fig:degree-histogram4}
\end{figure*}

There are two ways to do it. The first is to perform a power-binning of your x axis\cite[0.5in]{milojevic2010power}. Rather than drawing a point for each distinct degree value you have in your network, you can lump together values into larger bins. Using equally-sized bins -- with the same increment for the entire space -- doesn't work very well: for low degree values you're putting together very populated bins, while for high degree values usually the distribution is so dispersed that you aren't actually grouping together anything. See Figure \ref{fig:degree-histogram4}(b) for an example. In Figure \ref{fig:degree-histogram4}(b) we completely lost the head of the distribution -- the low degree values are all lumped together -- and, while less prominent, the fat tail is still there.

That's why you do power binning. You start with a small bin size, usually equal to $1$. Then each bin becomes progressively larger, by a constant multiplicative factor. At first, the bins are still small. But, as you progress, the bins start to be large enough to group a significant portion of your space\footnote[][0.5in]{An example of power binning, starting with size $1$ and increasing the bin size by $10\%$ at each step: $[1, 2, 3, 5, 6, 8, 9, 11, 14, ...,$ $1410, 1552, 1709, 1881, 2070, 2278, ...]$}. A good power bin choice can make the plot clearer, as the one in Figure \ref{fig:degree-histogram4}(c). In Figure \ref{fig:degree-histogram4}(c) we saved the head of the distribution and further reduced the fat tail.

One can do better than Figure \ref{fig:degree-histogram4}(c), that's why in network papers you rarely see power-binned distributions. An issue of power-binning is that it forces you to make a choice: to determine the bin size function. Having a choice is a double-edged sword: it opens you to the possibility of tricking yourself into seeing a pattern that is not there.

\begin{figure}
\centering
\includegraphics[width=.66\textwidth]{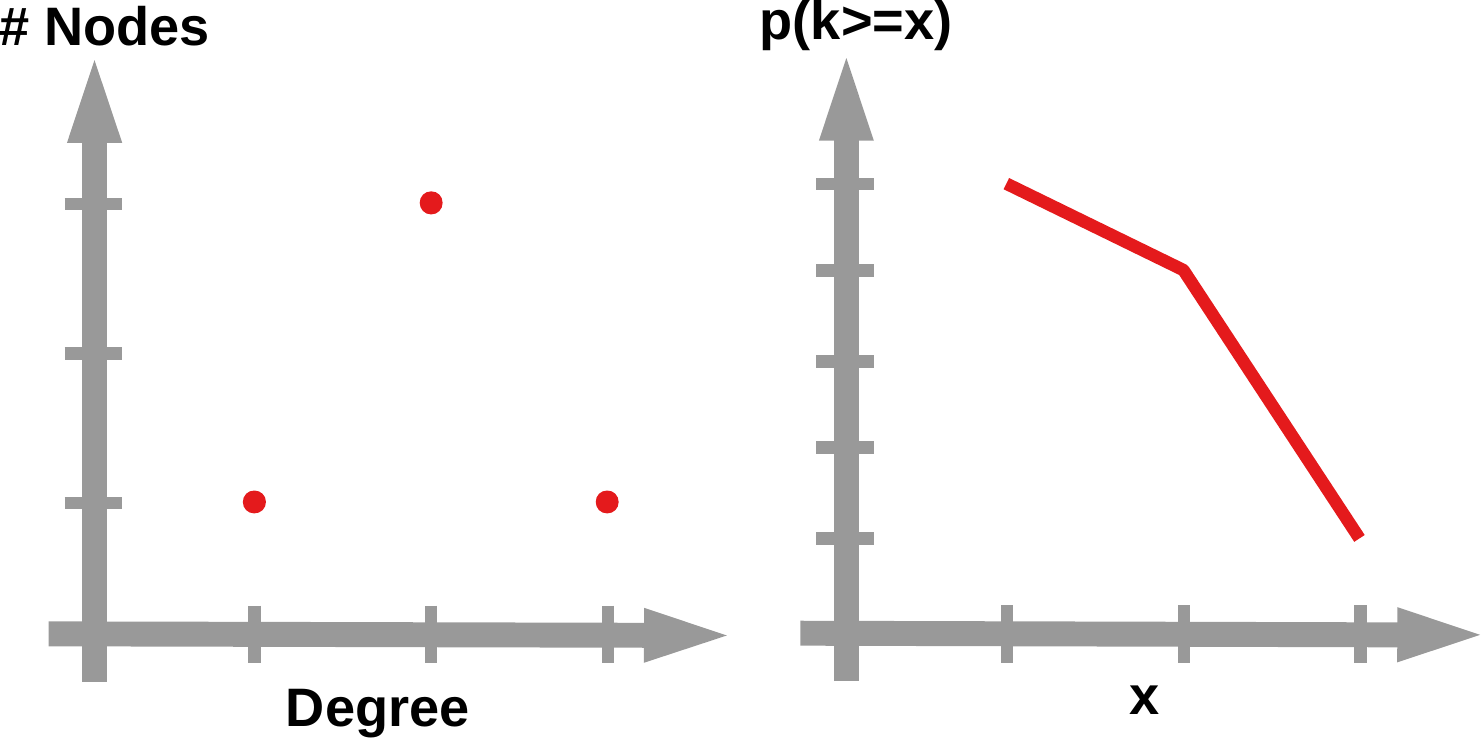}
\caption{The degree scatter plot (left) and its corresponding complement of the cumulative distribution (CCDF).}
\label{fig:degree-ccdf}
\end{figure}

The most common way to visualize degrees is by drawing cumulative distributions (CDF), or -- to be more aligned with the convention you'll see everywhere -- the complement of a cumulative distribution (CCDF). We can transform a degree histogram into a CCDF by changing the meaning of the y-axis. Rather than being the probability of finding a node of degree equal to $k$, in a CCDF this is the probability of finding a node of degree $k$ or higher. This is not a scattergram any more, but a function, which helps us when we need to fit it. Figure \ref{fig:degree-ccdf} shows an example, where we go from a degree histogram to its equivalent CCDF.

\begin{figure}
\centering
\begin{subfigure}{.45\columnwidth}
\includegraphics[width=\textwidth]{figures/protein_dd_log.pdf}
\caption{}
\end{subfigure}
\qquad
\begin{subfigure}{.45\columnwidth}
\includegraphics[width=\textwidth]{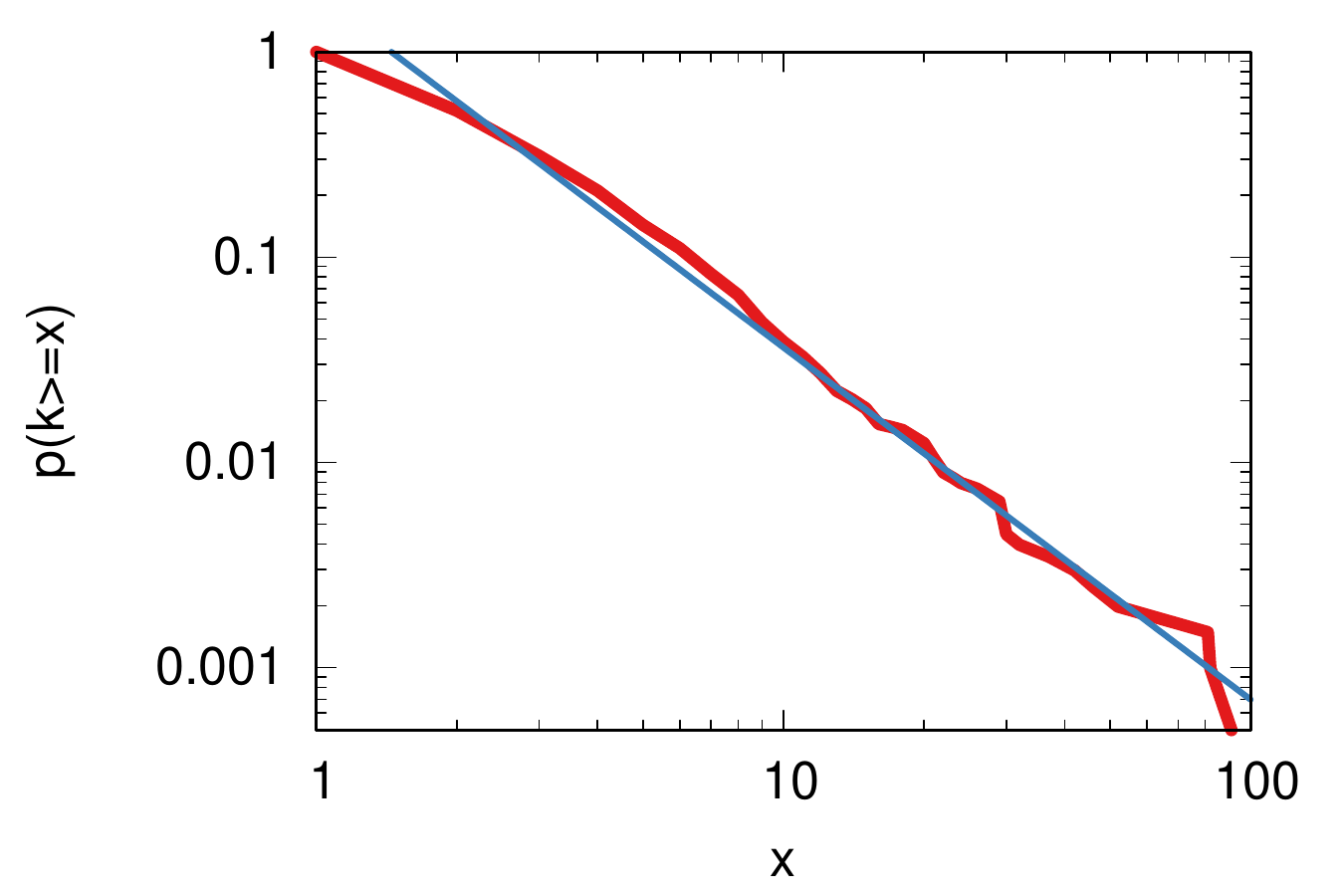}
\caption{}
\end{subfigure}
\caption{The degree distribution of the protein-protein interaction network. The distributions are the same and are both in log-log scale, but in (a) we have the degree histogram, and in (b) we show the CCDF version (with the best fit in blue).}
\label{fig:degree-ccdf2}
\end{figure}

We can see the relationship of our protein-protein network more clearly in Figure \ref{fig:degree-ccdf2}. It appears that, in log-log space, the relationship between degree and the number of nodes with a given degree is fixed. This relationship can be approximated with a straight line -- at least asymptotically: in Figure \ref{fig:degree-ccdf2}(b) you can see that the head doesn't really fit. Is this a coincidence, or does it have meaning? To answer this question we need to enter in the wonderful world of power-law degree distributions and scale free networks.

\section{Power Laws and Scale Free Networks}\label{sec:degree-pl}
\begin{figure}
\centering
\includegraphics[width=.66\textwidth]{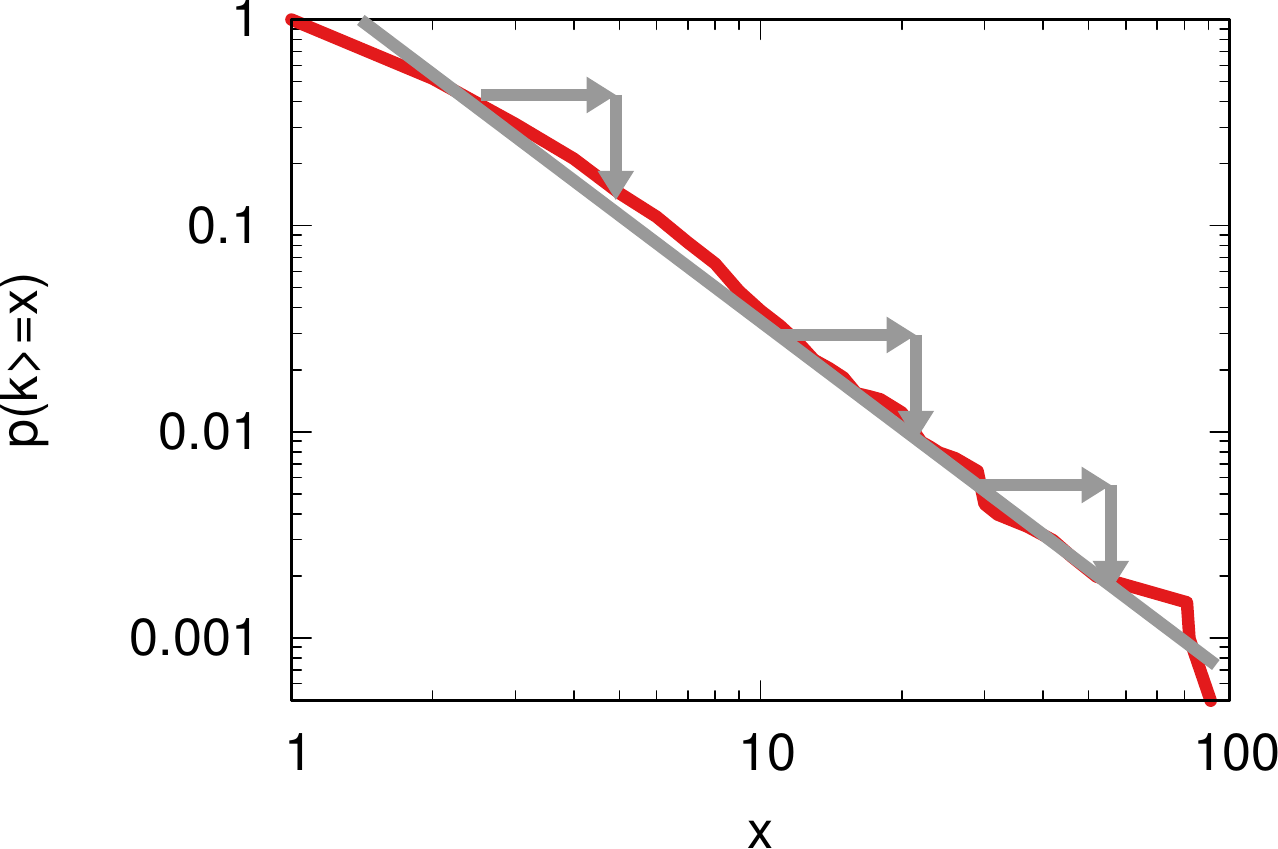}
\caption{An example of power law, showing how the red line always goes down by the same proportion as its right movement, no matter if we look a head, middle or tail.}
\label{fig:powerlaw}
\end{figure}

In statistics, a power law is a functional relationship between two quantities, where a relative change in one quantity results in a proportional relative change in the other quantity. The relation is independent of the initial size of those quantities: one quantity varies as a power of another. I show what I mean in Figure \ref{fig:powerlaw}: each time you move on the x-axis by a specific increment, you also always move on the y-axis by a fixed function of that x increment, no matter where you are in the distribution (head, tail, or middle).

\begin{figure}
\centering
\includegraphics[width=.6\columnwidth]{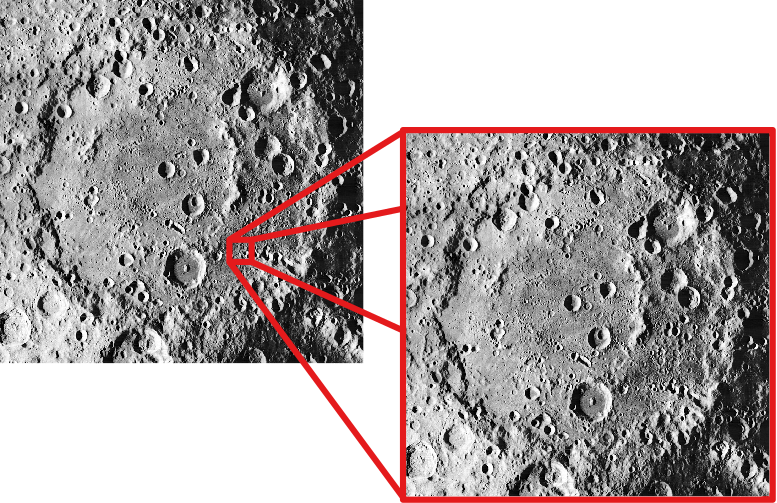}
\caption{The distribution of crater sizes in the moon is an example of quasi-power law, with many tiny ones and a huge one spanning the entire picture. If the moon were an ideal infinite plane and we could zoom in indefinitely, the picture we would get would be equivalent to the original one, i.e. the scale at which we're observing the moon would not influence the observation result.}
\label{fig:moon}
\end{figure}

You can find power laws in nature in many places: the frequencies of words in written texts, the distribution of earthquake intensities, etc... To grasp the concept you need a visual example, and my favorite is moon craters\cite{newman2005power}. You can see in Figure \ref{fig:moon} there are a lot of tiny craters caused by small debris and a huge one. This is fractal self-similarity: if the moon were an infinite plane, you could zoom in and out the picture and the size distributions would be the same. This is the scale invariance I'm talking about: no matter the zoom, the picture looks the same -- obviously in reality it doesn't, because the moon isn't an infinite plane, and you cannot zoom in infinitely many times (in fact, whether finite systems can actually generate power laws is a controversial topic\cite[0.1in]{stumpf2012critical}).

\begin{figure}[b]
\centering
\includegraphics[width=.66\textwidth]{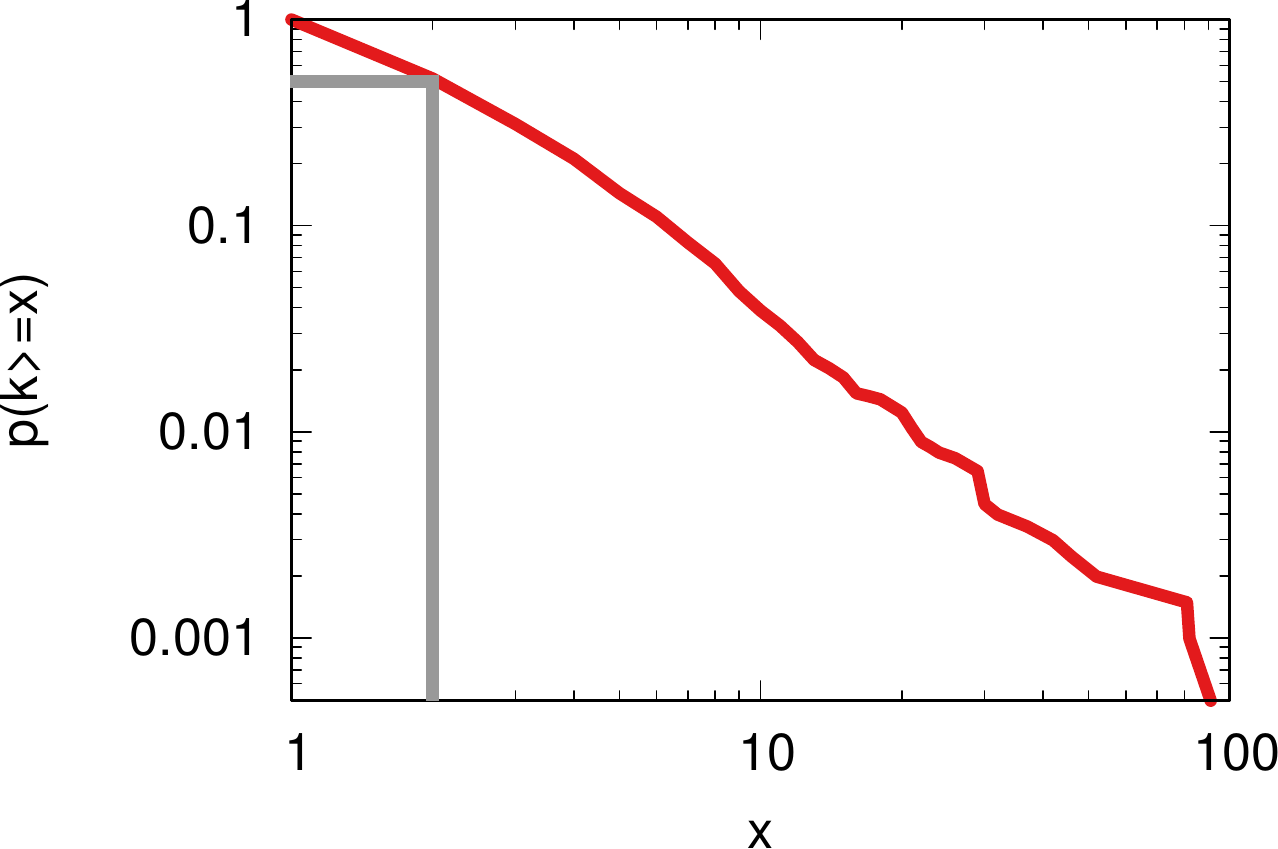}
\caption{An example of power law in a CCDF. The vertical gray bar shows that the point in the distribution is associated with degree equal to two. The horizontal gray bar shows that this degree correspond to a probability of around $0.5$. This means that half of the network has a degree equal to or greater than two. Or, in other words, that the other half of the network has degree equal to one.}
\label{fig:powerlaw2}
\end{figure}

This applies to networks too! There are many studies showing how some networks possess this sort of self-similar structure at different scales\cite{song2005self}\cite{serrano2008self} -- i.e. they are fractals. This is not necessarily the same thing as looking at the degree distribution\cite{kim2007fractality} -- although classifying a network as ``scale free'' by looking at its degree distribution is a common operation in the literature and it is also the stance I'll adopt from now on in the book.

In Figure \ref{fig:powerlaw2}, I show the usual CCDF of the protein-protein network: there we see that $50\%$ of the nodes have a degree of $2$ or more. This means that $50\%$ of the nodes have degree equal to one. A formula you'll see everywhere links the probability of a node having degree $k$ to $k$ to the power of a constant $\alpha$. Mathematically speaking, the scale free network master equation is:

$$ p(k) \sim k^{-\alpha}.$$

In this formula, we call $\alpha$ the scaling factor. Its value is important, because it determines many properties of the distribution. In general, if a real world network has a power law degree distribution, $\alpha$ tends to be low ($\alpha \sim 2$, and for a majority $\alpha < 3$, although you can find networks with higher $\alpha$s). This is rather unfortunate, because it means that the degree distribution has a well defined mean, but not a well defined variance (Section \ref{sec:stats-corr}). This implies that the average degree has meaning, but it's not very useful to do anything more than a superficial description of the network.

\begin{figure}
\centering
\includegraphics[width=.66\textwidth]{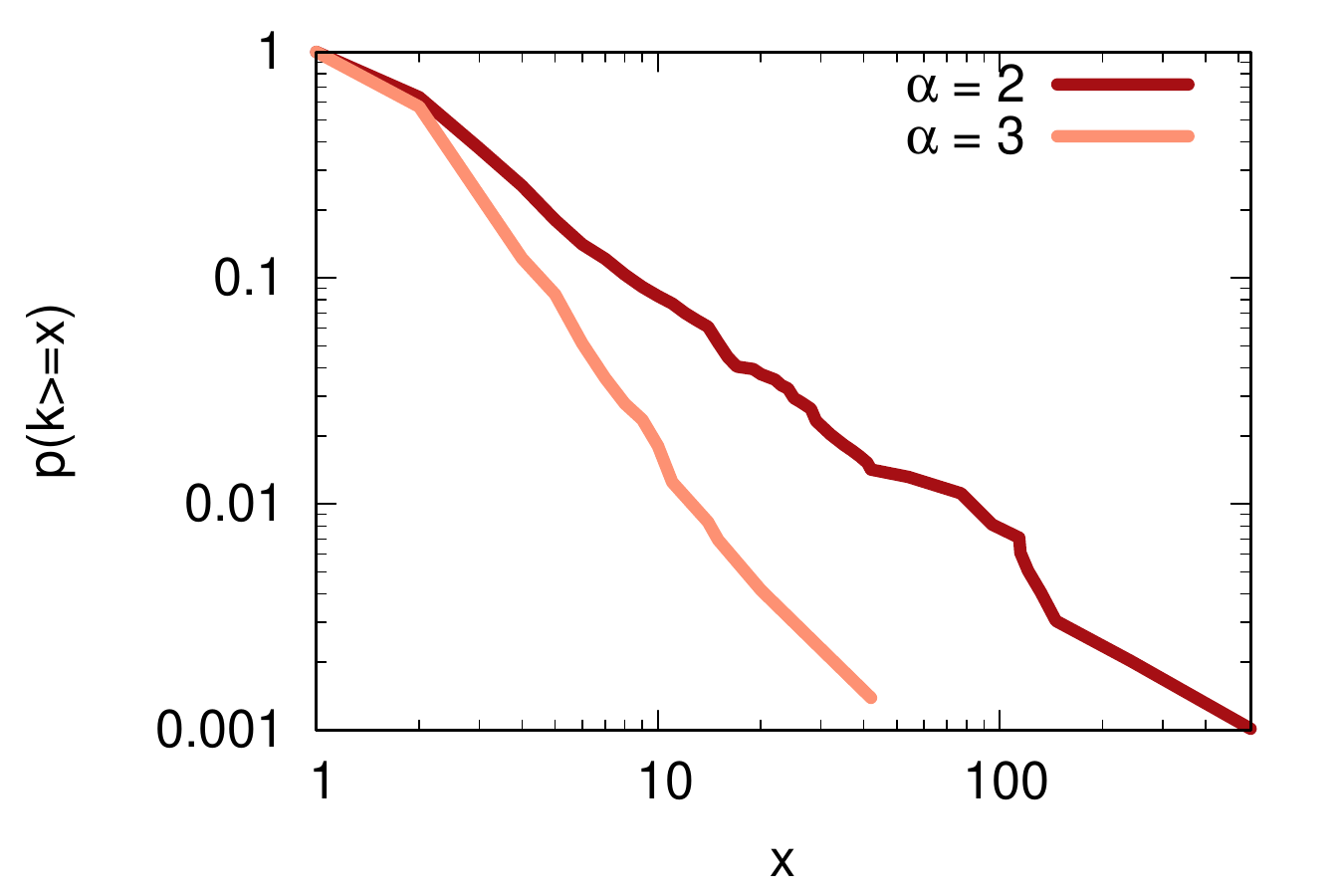}
\caption{The CCDF degree distributions of two random networks with different $\alpha$ exponents.}
\label{fig:scalefree-alpha1}
\end{figure}

\begin{figure}[b]
\centering
\begin{subfigure}{.41\columnwidth}
\includegraphics[width=\textwidth]{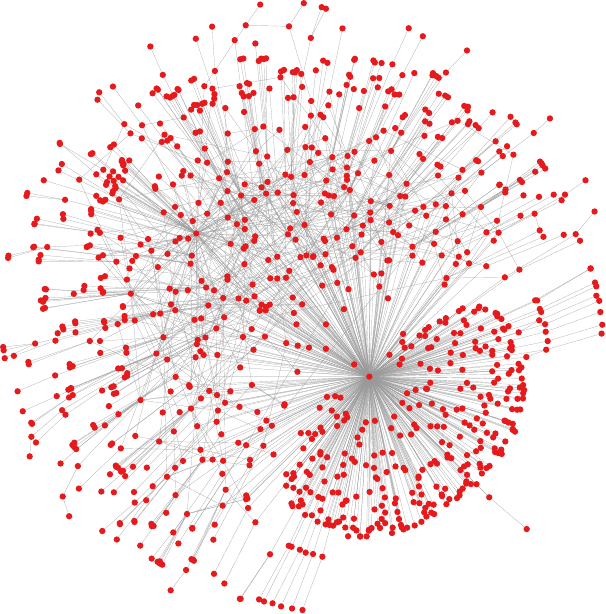}
\caption{$\alpha = 2$}
\end{subfigure}
\qquad
\begin{subfigure}{.49\columnwidth}
\includegraphics[width=\textwidth]{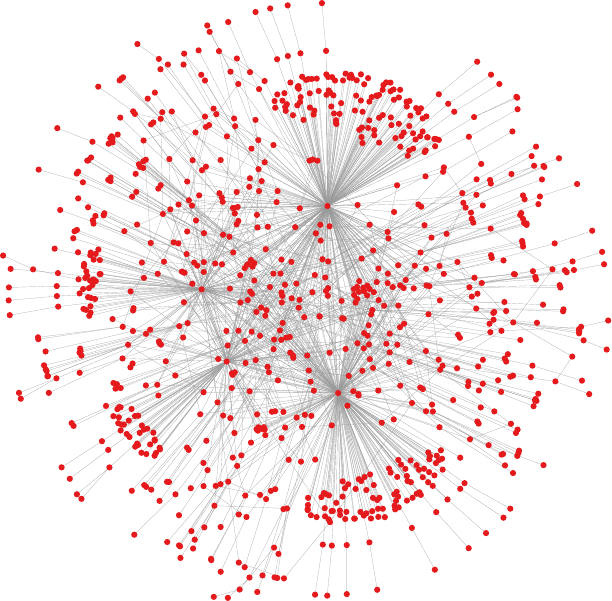}
\caption{$\alpha = 3$}
\end{subfigure}
\caption{An example of two networks with scale free degree distributions, with different $\alpha$ exponents.}
\label{fig:scalefree-alpha2}
\end{figure}

This is all well and good, but what does it mean exactly to have $\alpha = 2$ or $\alpha = 3$? How do two networks with these two different coefficients look like? I provide an example of their degree distributions in Figure \ref{fig:scalefree-alpha1}, and I show two very simple random networks with such degree distributions in Figure \ref{fig:scalefree-alpha2} -- obviously, systems this small are a very rough approximation. From Figure \ref{fig:scalefree-alpha1} you see that $\alpha$ determines the slope of the degree distribution, with a steeper slope for $\alpha = 3$. This means that the hubs in $\alpha = 3$ are ``smaller'', they do not have a ridiculously high degree.

Figure \ref{fig:scalefree-alpha2} confirms this: in Figure \ref{fig:scalefree-alpha2}(a) you see that, for $\alpha = 2$, you have only one obvious hub that is head and shoulders above the rest, practically connected to the entire network. In Figure \ref{fig:scalefree-alpha2}(b), instead, you still have a clear winner catching your eye (in the top), but it is much closer to the second best hub.

The average degree is heavily influenced by the outliers with thousands of connections. For instance, in Figure \ref{fig:powerlaw2} the average degree is equal to three, meaning that around $70\%$ of nodes are below average. This is well illustrated by the stadium example: you have a stadium with $79,999$ individuals sampled at random from the US population. If you calculate their average net worth you'll obtain a value -- it's difficult to be precise, but let's say it's around \$$100,000$. So their total net worth is $\sim 8$ billion dollars. However, the $80,000$th person entering the stadium is our outlier hub: Jeff Bezos. His net worth alone is $192$ billion dollars\footnote{Bear with me, I know it has probably doubled by the time you read this paragraph.}. The new average is $200$ billion divided by $80$ thousand people: $2.5$ million dollars. The average shifted dramatically: $2.5$ million is \textit{very} different from $100$ thousand. This is because net worth distributes broadly and thus has a crazy variance, which causes tremendous shifts in the average. This makes it incorrect to apply to this kind of distribution traditional statistics that are based on variance and standard deviation -- such as regression analysis, as we'll see in the next section.

\begin{figure*}
\centering
\includegraphics[width=\textwidth]{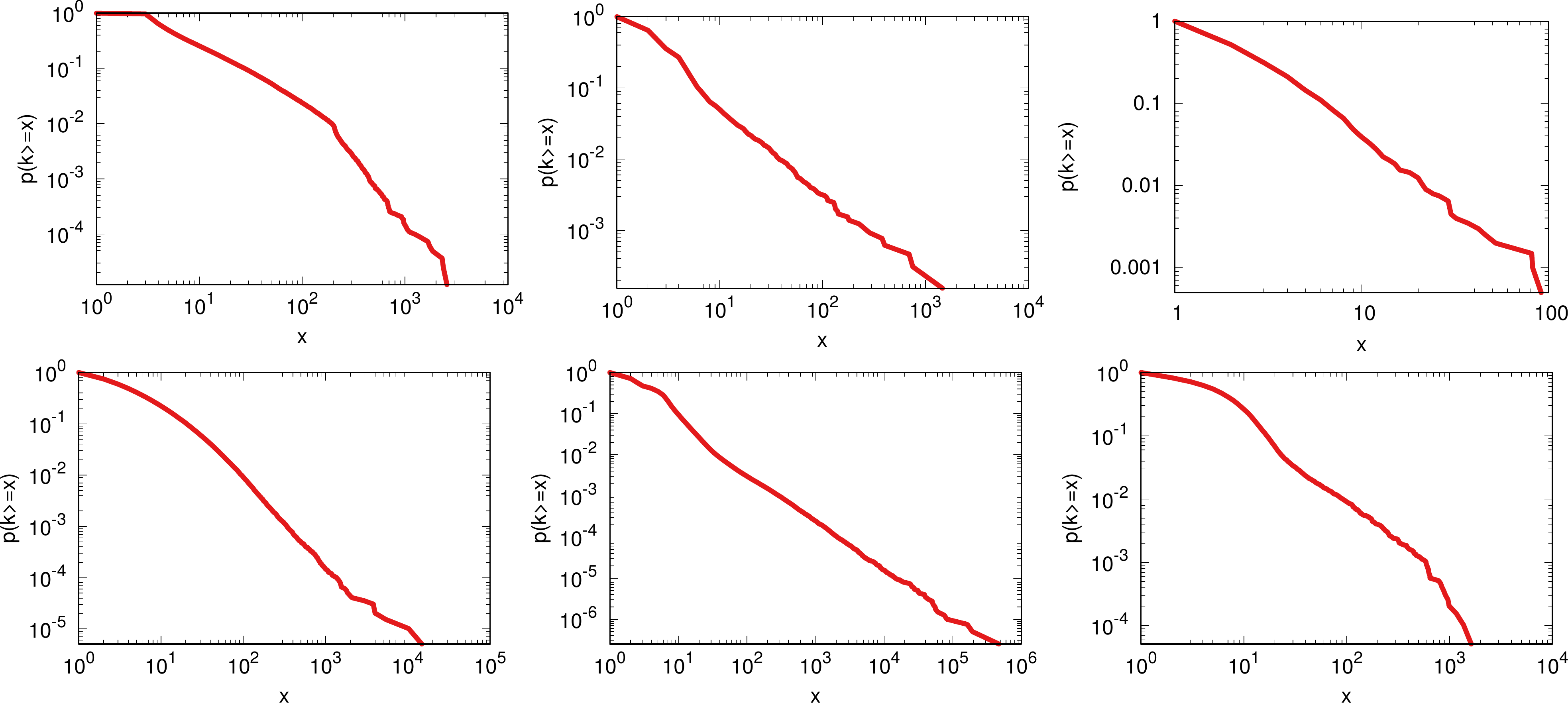}
\caption{A showcase of broad degree distributions from the same networks used in the examples in the previous section.}
\label{fig:powerlaw3}
\end{figure*}

Early works have found power law degree distributions in many networks, prompting the belief that scale free networks are ubiquitous. In fact, this seems true. Figure \ref{fig:powerlaw3} shows the CCDFs of many networks: protein interactions, PGP, Slashdot, DBpedia concept network, Gowalla, Internet autonomous system routers.

But we need to be aware of our tendency of seeing patterns when they aren't there -- after all, as Feynman says, the easiest person you can fool is yourself. So in the next section I'll give you an arsenal to defend yourself from your own eyes and brain.

\section{Testing Power Laws}\label{sec:degree-fit}

\begin{figure}
\centering
\begin{subfigure}{.45\columnwidth}
\includegraphics[width=\textwidth]{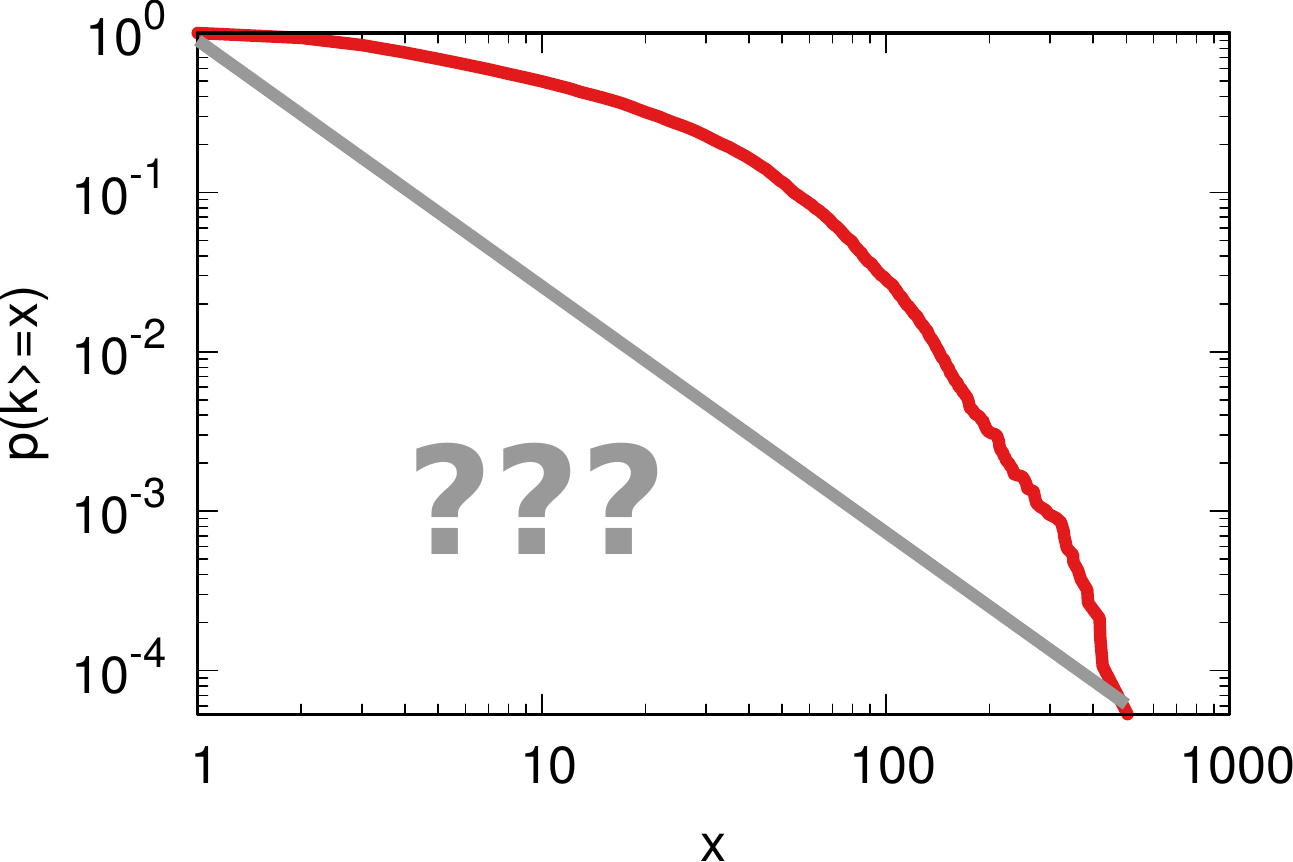}
\caption{}
\end{subfigure}
\qquad
\begin{subfigure}{.45\columnwidth}
\includegraphics[width=\textwidth]{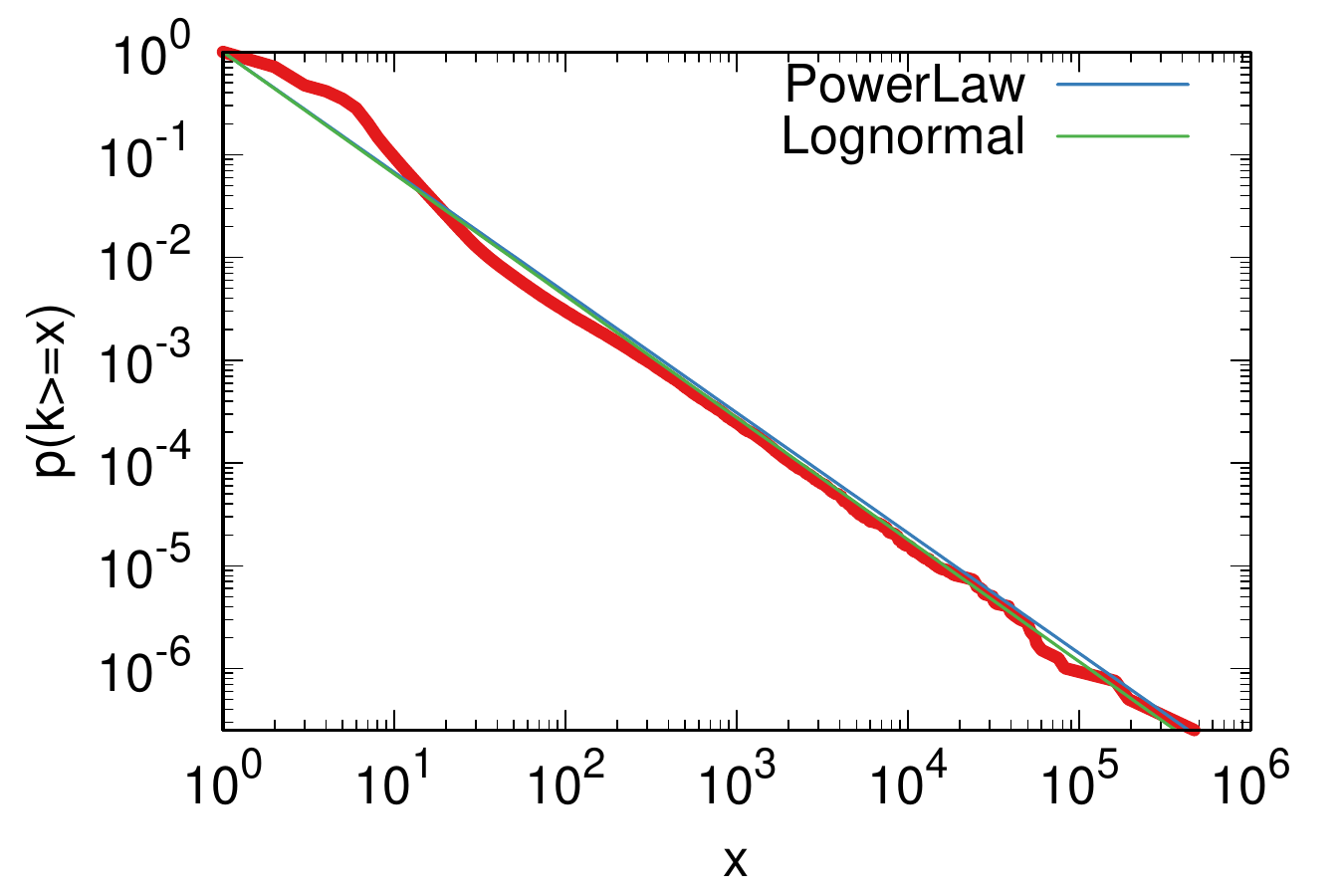}
\caption{}
\end{subfigure}
\caption{(a) An example of a CCDF that is most definitely NOT a power law, but that a researcher with a lack of proper training might be fooled into thinking it is. (b) Fitting a power law (blue) and a lognormal (green) on data (red) can yield extremely similar results.}
\label{fig:powerlaw4}
\end{figure}

Often, people will just assume that any degree distribution is a power law, calling ``power laws'' things that are not even deceptively looking like power laws. I've seen distributions as the one in Figure \ref{fig:powerlaw4}(a) passing as power laws and that's just... no. However, I don't want to pass as one perpetuating the myth that ``everything that looks like a straight line in a log-log space is a power law''. That is equally wrong, even if more subtle and harder to catch.

Seeing the plot in Figure \ref{fig:powerlaw4}(b), you might be tempted to perform a linear fit in the log-log space. This more or less looks like fitting the logged values with a $\log(p(x)) = \alpha \log(x) + \beta$. Transforming this back into the real values, the slope $\alpha$ becomes the scaling factor, and $\beta$ is the intercept, in other words: $\log(p(x)) = \alpha \log(x) + \beta$ is equivalent to $p(x) = 10^\beta x^{\alpha}$ -- assuming you logged to the power of ten.

A small aside: if you were to do this on the distributions from Figure \ref{fig:scalefree-alpha1}, you would expect to recover $\alpha \sim 2$ and $\alpha \sim 3$, because I told you I generated the degree distributions with those exponents. Instead, you will obtain $\alpha \sim 1$ and $\alpha \sim 2$, respectively. That is because, in Figure \ref{fig:scalefree-alpha1}, I showed you the \textit{CCDF} of the degree distribution, not the distribution itself. The CCDF of a power law is also a power law, but with a different exponent\cite{bauke2007parameter}. If you're doing the fit on the CCDF, you have to remember to add one to your $\alpha$ to recover the actual exponent of the degree distribution.

Back to parameter estimation. If you perform a simple linear regression, you'll get an unbelievably high $R^2$ associated to a super-significant $p$ value. Well, of course: you're fitting a straight line over a straight-ish line. Does that mean you're looking at a power law? Not really.

Just because something looks like a straight line in a log-log plot, it doesn't mean it's a power law. You need a proper statistical test to confirm your hypothesis. The reason is that other data generating processes, such as the ones behind a lognormal distribution, can generate plots that are almost indistinguishable from a power law. Figure \ref{fig:powerlaw4}(b) shows an example. You cannot really tell which of the two functions fits the data better.

What you need to do is to fit both functions and then estimate the likelihood (Section \ref{sec:ml-loss}) of each model to explain the observed data\cite{clauset2009power}. This can be done with, for instance, the \texttt{powerlaw} package\cite{alstott2014powerlaw}\footnote{\url{https://github.com/jeffalstott/powerlaw}} -- available for Python. However, be prepared for the fact that having a significant difference between the power law and the lognormal model is extremely hard.

In most practical scenarios, you'll have to argue that your network is a power law. How could you do it? Well, in complex networks power law degree distributions can arise by many processes, but one in particular has been observed time and time again: cumulative advantage. Cumulative advantage in networks says that the more connections a node has, the more likely it is that the new nodes will connect to it. For instance, if you write a terrific paper which gathers lots of citations this year, next year it will likely gain more citations than the less successful papers\cite{price1976general}.

This is the same mechanism behind -- for instance -- Pareto distributions and the $80$/$20$ rule. Pareto says that $80\%$ of the effects are generated by $20\%$ of the causes\cite{pareto1919manuale}. For instance, $20\%$ of people control $80\%$ of the wealth. And, given that it takes money to make money, they are likely to hold -- or even grow -- their share, given their ability to unlock better opportunities. In fact, the Pareto distribution is a power law. Similar to this is Zipf's Law, the observation that the second most common word in the English language occurs half of the time as the most common, the third most common a third of the time, etc\cite{estoup1916gammes}\cite{auerbach1913gesetz}\cite{zipf1935psycho}. In practice, the $n$th word occurs $1/n$ as frequently as the first, or $f(n) = n^{-1}$, which is a power law with $\alpha = 1$.

This is opposed to the data generating process of a lognormal distribution. To generate a lognormal distribution you simply have to multiply many random and independent variables, each of which is positive. A lognormal distribution arises if you multiply the results of many ten-dice rolls. You can see that there is no cumulative advantage here: scoring a six on one die doesn't make a six more likely on any other die -- nor influences subsequent rolls.

So, to sum up, to test for a power law you have to do a few things. First, make sure that your observations cannot be explained with an exponential. Confusion between a power law and some other distribution such as an exponential is hard. If you think a distribution might be an exponential, then it's definitely not a power law. Second, try to see if you can statistically prefer a power law model over a lognormal. In the likely event of you not being able to mathematically do so, you should look at your data generating process. If you have the suspicion that it could be due to random fluctuations, then you might have a lognormal. Otherwise, if you can make a convincing argument of non-random cumulative advantage, go for it.

There are a few more technicalities. Pure power laws in nature are -- as I mentioned earlier -- rare\footnote{Whether this holds true also for networks is the starting point of a surprisingly hot debate, see for instance \citep{broido2019scale} and \citep{voitalov2018scale}.}. Your data might be affected by two impurities. Your power law could be shifted\cite{johannesson2006afterglow}, or it could have an exponential cutoff\cite{clauset2009power}. In a shifted power law, the function holds only on the tail. In an exponential cutoff the power law holds only on the head.

\begin{figure}
\centering
\begin{subfigure}{.45\columnwidth}
\includegraphics[width=\textwidth]{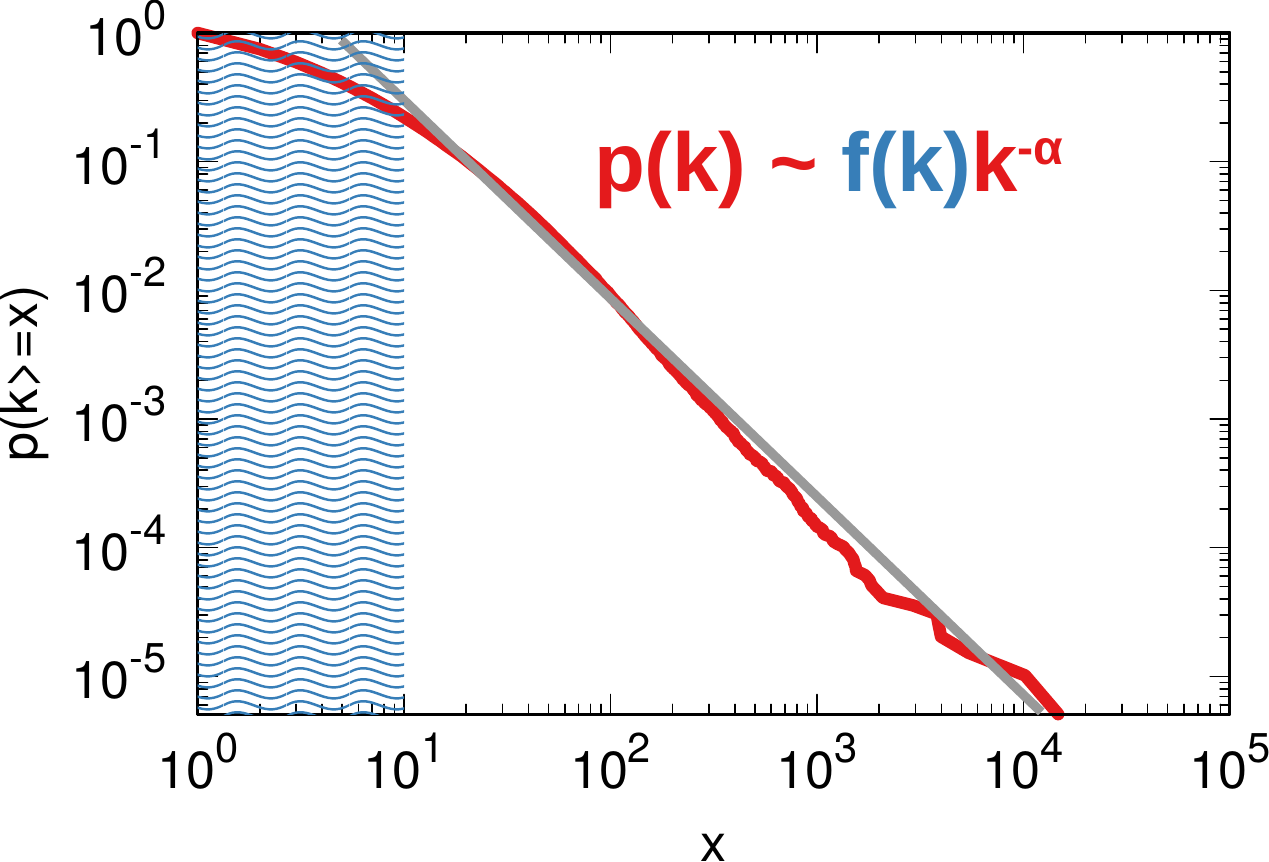}
\caption{}
\end{subfigure}
\qquad
\begin{subfigure}{.45\columnwidth}
\includegraphics[width=\textwidth]{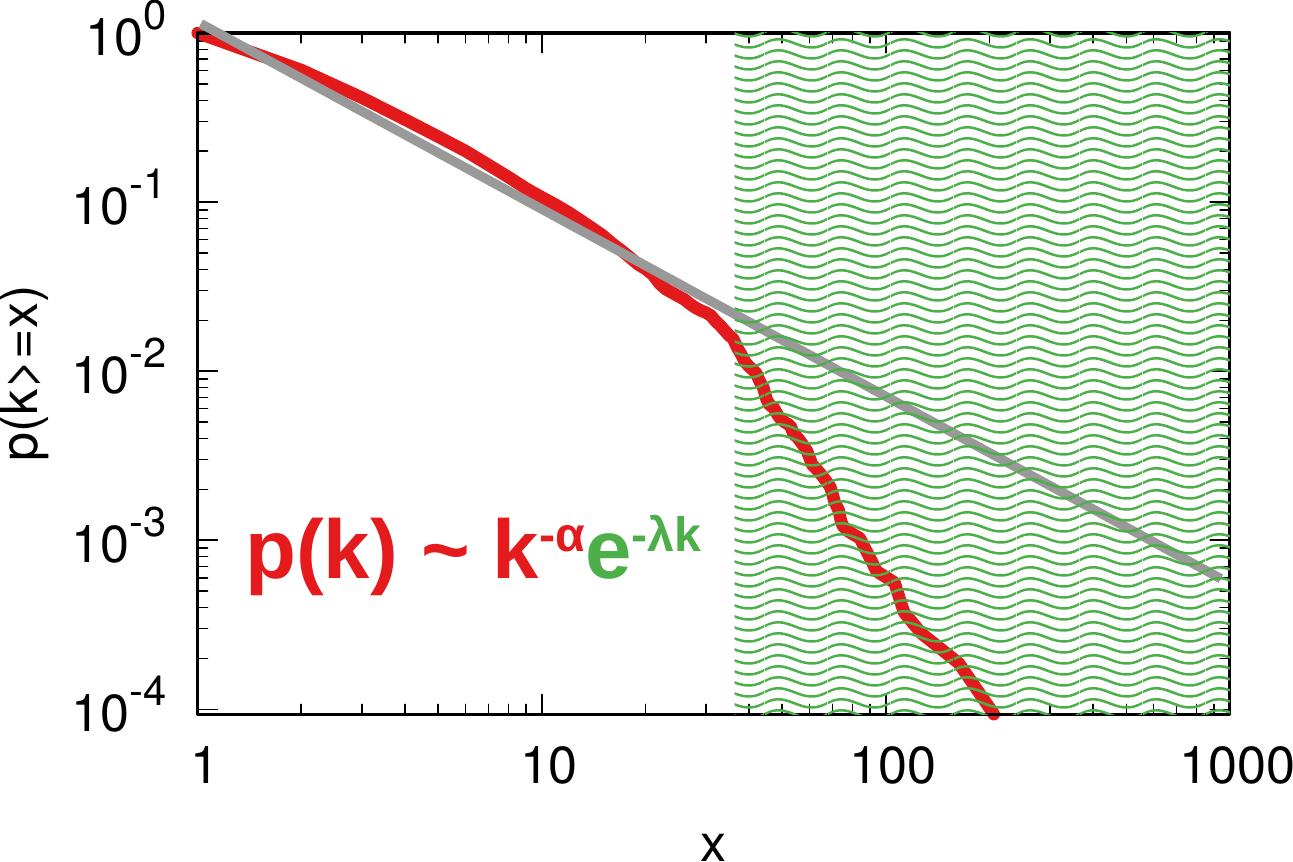}
\caption{}
\end{subfigure}
\caption{(a) An example of shifted power law. The area in which the power law doesn't hold is shaded in blue. (b) An example of truncated power law: a power law with an exponential cutoff. The area in which the power law doesn't hold is shaded in green.}
\label{fig:powerlaw5}
\end{figure}

Shifted power laws have an initial regime where the power law doesn't hold. Formally, the power law function needs a slowly growing function on top that will be overwhelmed by the power law for large values of $k$ -- as I show in Figure \ref{fig:powerlaw5}(a). So we modify our master equation as: $p(k) \sim f(k)k^{-\alpha}$, with $f(k)$ being an arbitrary but slowly growing. Slowly growing means that, for low values of $k$ it will overwhelm the $k^{-\alpha}$ term, but for high values of $k$, the latter would be almost unaffected. In power law fitting, this means to find the $k_{min}$ value of $k$ such that, if $k < k_{min}$ we don't observe a power law, but for $k > k_{min}$ we do.

Shifted power laws practically mean that ``Getting the first $k_{min}$ connections is easy''. If you go and sign up for Facebook, you generally already have a few people you know there. Thus we expect to find fewer nodes with degree $1$, $2$, or $3$ than a pure power law would predict. The main takeaway is that, in a shifted power law, we find fewer nodes with low degrees than we expect in a power law.

Truncated power laws are typical of systems that are not big enough to show a true scale free behavior. There simply aren't enough nodes for the hubs to connect to, or there's a cost to new connections that gets prohibitive beyond a certain point. This is practically a power law excluding its tail, that's why we call them ``truncated''. Mathematically speaking, this is equivalent to having an exponential cutoff added to our master equation: $p(k) \sim k^{\alpha}e^{-\lambda k}$. The exponential function is dominated by the power law function for low values of $k$, but it becomes dominant for high values of $k$. See Figure \ref{fig:powerlaw5}(b) for an example.

Truncated power laws practically mean that ``Getting the last connections is hard'': the biggest superstar on Twitter has a lot of followers, but relatively speaking they are not that many more as the second biggest superstar on Twitter. Thus its degree is not as big as we would expect. The main takeaway is that, in a truncated power law, the hubs have lower degrees than we expect in a power law.

At the end of the day, it doesn't matter too much if your network has an exponential, lognormal or power law degree distribution. On one thing the brotherhood of network scientists can agree: the vast majority of networks have broad degree distributions, spanning multiple orders of magnitude. Most nodes have below-average degree and hubs lie many standard deviations above the average. Even if they are not power laws at all, that's still pretty darn interesting.

\section{Summary}

\begin{enumerate}
\item The degree is the number of edges connected to a node and it's probably a node's most important and basic feature. It tells us how well connected and how structurally important the node is.
\item In more complex graph models (directed, weighted, bipartite, multilayer), the degree measure becomes itself more complex. For instance, in directed networks you have both in- and out-degree, depending on the direction of the edge.
\item The degree distribution is a plot telling you how many nodes have a specific degree value in the network, and it is one of the network's most important properties.
\item When plotting degree distributions, the standard choice is the complement of the cumulative distribution, shown in a log-log scale.
\item Many networks have a power law degree distribution, but rarely this is a pure power law: it is often shifted or truncated. Fitting a power law and finding the correct exponent is tricky and you should not do it using a linear regression: you should use specialized tools.
\item Moreover, determining whether the degree follows a power law is useful for modeling and theory, but it isn't crucial empirically. The interesting thing is that networks have broad and unequal degree distributions. You can describe them with statistics that are easier to get right than the tricky business of fitting power laws.
\end{enumerate}

\section{Exercises}

\begin{enumerate}
\item Write the in- and out-degree sequence for the graph in Figure \ref{fig:degree-directed}(a). Are there isolated nodes? Why? Why not?
\item Calculate the degree of the nodes for both node types in the bipartite adjacency matrix from Figure \ref{fig:bipartite-degree}(a). Find the isolated node(s).
\item Write the degree sequence of the graph in Figure \ref{fig:dimension-relevance}. First considering all layers at once, then separately for each layer.
\item Plot the degree distribution of the network at \url{http://www.networkatlas.eu/exercises/9/4/data.txt}. Start from a plain degree distribution, then in log-log scale, finally plot the complement of the cumulative distribution.
\item Estimate the power law exponent of the CCDF degree distribution from the previous exercise. First by a linear regression on the log-log plane, then by using the \texttt{powerlaw} package. Do they agree? Is this a shifted power law? If so, what's $k_{min}$? (Hint: \texttt{powerlaw} can calculate this for you)
\item Find a way to fit the truncated power law of the network at \url{http://www.networkatlas.eu/exercises/9/6/data.net}. Hint: use the \texttt{scipy.optimize.curve\_fit} to fit an arbitrary function and use the functional form I provide in the text.
\end{enumerate}

\chapter{Paths \& Walks}\label{cha:paths}
So far, we adopted a static vision of a network. We have a structure and we ask simple questions about the structure as it is. Does it have directed edges? What's the degree of the nodes? Those are interesting questions, but a network really shines when you use it for what it is for: exploring its connections.

To understand what I mean, let's come back to the social network example. Nodes are people and edges connect friends. Suppose you want to send a message to a person you are not connected to. Maybe you want to sell them a new shampoo. How do you do it? Well, you could tell the message to a friend, and instruct them to pass the message on. This means having a packet of information travel through the network, exploiting its edges.

There are many ways to cross a network using its edges, depending on which restrictions you want to put on your exploration. I'm going to define a few technical terms (following graph theory literature\cite{diestel2018graph}) but, if you find it simpler, you can call all of them ``paths'' plus a qualifier specifying what type of path it is. I'm going to present both conventions and you simply use the one you find most natural -- as everybody does in network analysis.

\begin{figure}
\centering
\begin{subfigure}{.45\columnwidth}
\includegraphics[width=\textwidth]{figures/walk.pdf}
\caption{}
\end{subfigure}
\qquad
\begin{subfigure}{.45\columnwidth}
\includegraphics[width=\textwidth]{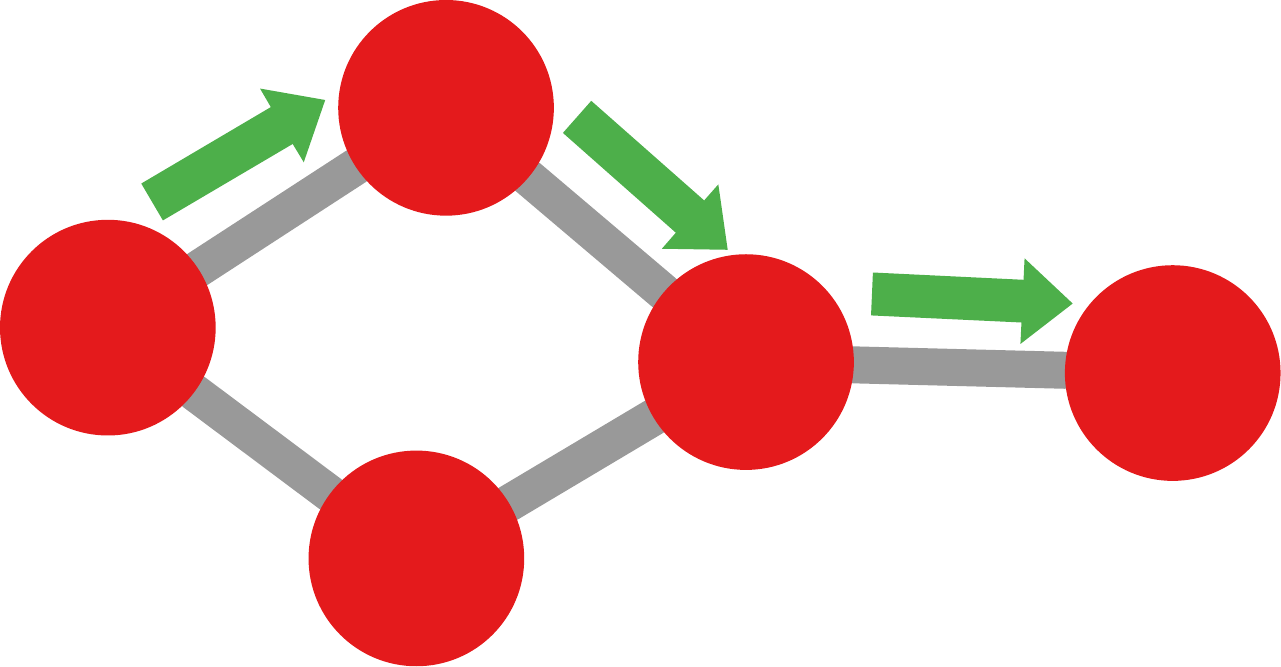}
\caption{}
\end{subfigure}
\caption{(a) An example of a walk of length six in the network, following the green arrows. (b) An example of a path of length three in the network.}
\label{fig:walk}
\end{figure}

The most basic way to explore a graph is by performing a \textbf{walk}, or ``path with repeating nodes''. A walk is a sequence of nodes with the property that each node in the sequence is adjacent to the node next to it. In a walk you're not imposing any rule in your exploration. You can go back and forth between the same two nodes by using the same edge as many times as you want. The \textbf{length} of a walk is the number of times you're using the edges in your walk. If you use the same edge $n$ times, this will increase the walk's length by $n$. Figure \ref{fig:walk}(a) shows an example of a walk of length $6$ in a network.

In a walk the choice of the next edge to explore is yours. You can have a slightly more constrained definition of a walk, where you put rules to choose the next edge to traverse. For instance, in the random walk you impose to make this choice completely at random. We already saw a way to calculate node exploration probabilities via a random walk using powers of the adjacency matrix in Section \ref{sec:mat-mat-mat}. We'll see that random walks are a phenomenally powerful way to explore your network's properties and are at the basis of countless methods: Chapter \ref{cha:rndwalks} will be but a superficial introduction.

When you impose even more constraints on your walks, then you can generate a \textbf{path}, or ``simple path'' (Figure \ref{fig:walk}(b)). This is a walk that does not repeat nodes nor edges. Again, you can put more qualifiers on your path to make it special. For instance, recalling the seven bridges problem, an Eulerian path is a path that travels through all edges of a connected graph -- since it is a path, not only it has to visit each edge, but it also has to do it exactly once. A cousin of the Eulerian path is the Hamiltonian path, which instead wants to visit each node -- not edge -- exactly once. More interestingly, you can try to find the \textit{shortest} path between two nodes. That will be the topic of Chapter \ref{cha:shortpath}.

Similarly to a walk, a path has a length as well. This is again defined as the number of edges the path crosses. Since no edge can be used twice in a path, this is also the number of distinct edges used.

\section{Walks and Matrices}\label{sec:paths-mat}
In Chapter \ref{cha:mat} I showed you that taking powers of the stochastic matrix is fun, because it tells us the probability of a random walker going from nodes $u$ to $v$. But looking at ye olde regular adjacency matrix can be insightful. If $A$ is binary and its diagonal is set to zero, then $A^n$ can tell us lots of interesting things.

\begin{figure}
\centering
\begin{subfigure}[t]{.23\columnwidth}
\includegraphics[width=\textwidth]{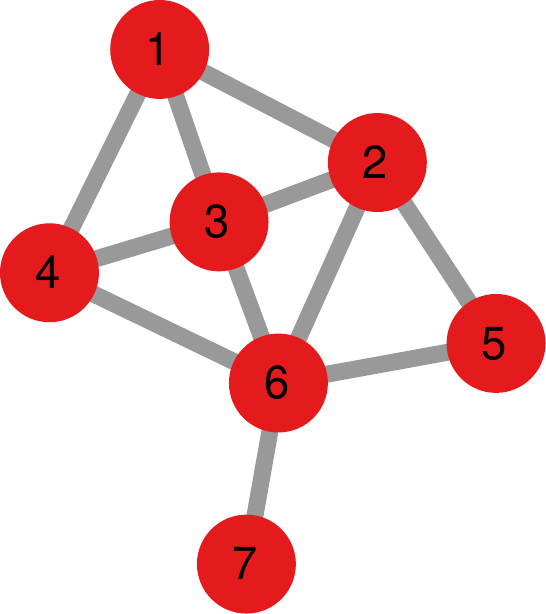}
\caption{$G$}
\end{subfigure}\qquad
\begin{subfigure}[t]{.32\columnwidth}
\includegraphics[width=\textwidth]{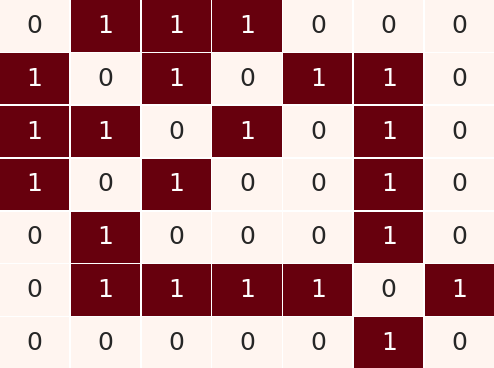}
\caption{$A^1$}
\end{subfigure}\qquad
\begin{subfigure}[t]{.32\columnwidth}
\includegraphics[width=\textwidth]{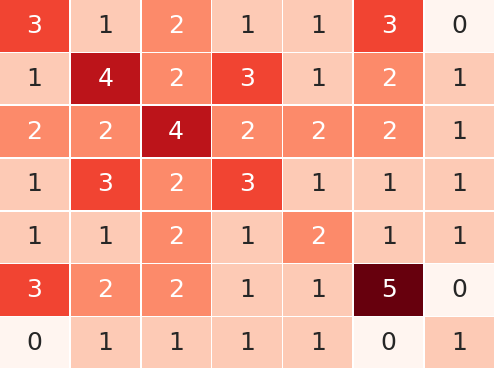}
\caption{$A^2$}
\end{subfigure}
\caption{(a) A graph. (b-c) Different powers of its binary adjacency matrix $A$.}
\label{fig:adj-powers-7}
\end{figure}

First, if $A^n_{uv} = 0$, then there are no walks of length $n$ going from $u$ to $v$. In Figure \ref{fig:adj-powers-7}(b) you see that $A^2_{1,7}$ is zero, because node $7$ is only connected to node $6$. Node $6$ isn't connected to node $1$ so there's no way to go from $1$ to $7$ in two hops. If, instead, $A^n_{uv} > 0$, then the $A^n_{uv}$ is exactly the number of such walks! There are two ways to go from node $1$ to node $3$ in two hops in Figure \ref{fig:adj-powers-7}(b): one via node $2$ and one via node $4$, since they're both connected to node $3$.

It doesn't end here: you can set the diagonal of $A$ to $1$ by summing to it the identity matrix $I$. Then, $(I + A)^n$ tells you the number of walks of length $n$ or less. The reason is that the diagonal represents self loops. If it is set to $1$, the walker in $u$ can choose to follow the self loop to $u$ an arbitrary number of times before reaching $v$.

Every time you calculate $A^n$, the resulting diagonal is interesting. Specifically, $A^n_{uu}$ is the number of loops or closed walks of length $n$ -- walks starting and ending in the same node -- to which $u$ participates. For $n = 2$ we have a special case: this value is equal to the degree -- or $A^2_{uu} = k_u$. Check the diagonal of Figure \ref{fig:adj-powers-7}(b) if you don't believe me! This means that you could consider $A^n_{uu}$ as a sort of generalized degree.

\section{Cycles}\label{sec:paths-cycles}
You can make a walk and a path in any graph, no matter its topology. There is a special path that you cannot always do, though. That is the cycle. Picking up the social network example as before, now you're not happy just by reaching somebody with your message. You want the message you originally sent to come back to you. Also, you don't want anybody to hear it twice. If you manage to do so, then you have found a cycle in your social network.

\begin{figure}
\centering
\begin{subfigure}{.25\columnwidth}
\includegraphics[width=\textwidth]{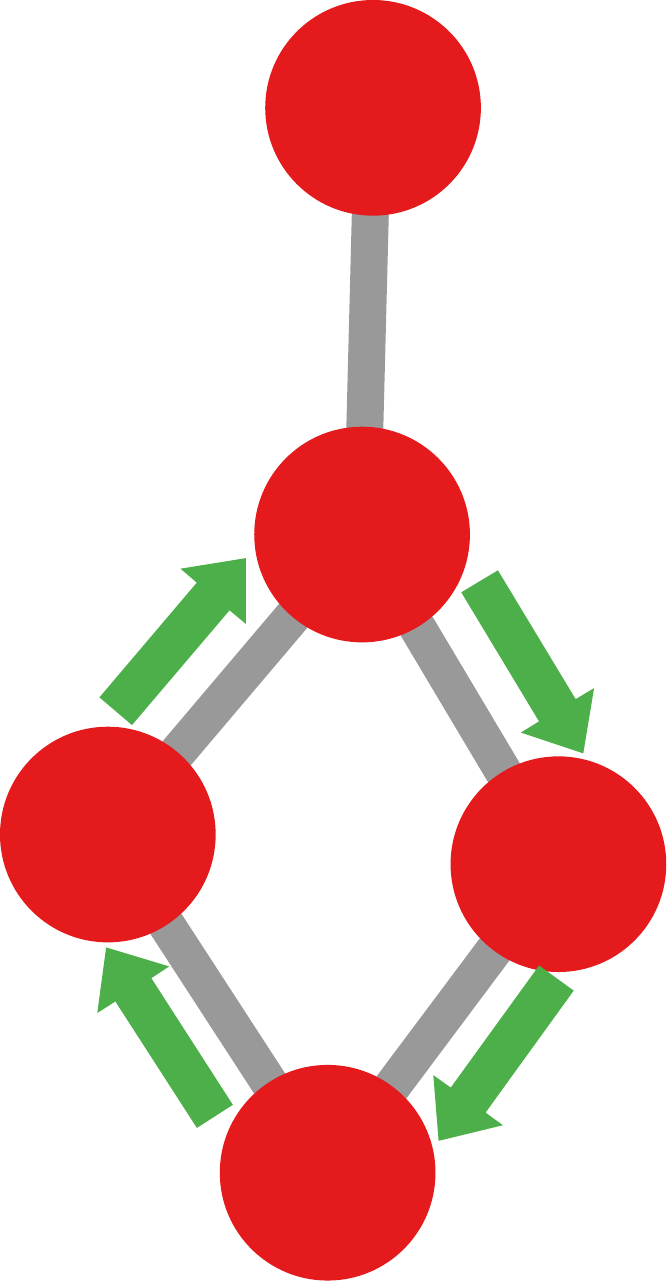}
\caption{}
\end{subfigure}
\qquad
\begin{subfigure}{.45\columnwidth}
\includegraphics[width=\textwidth]{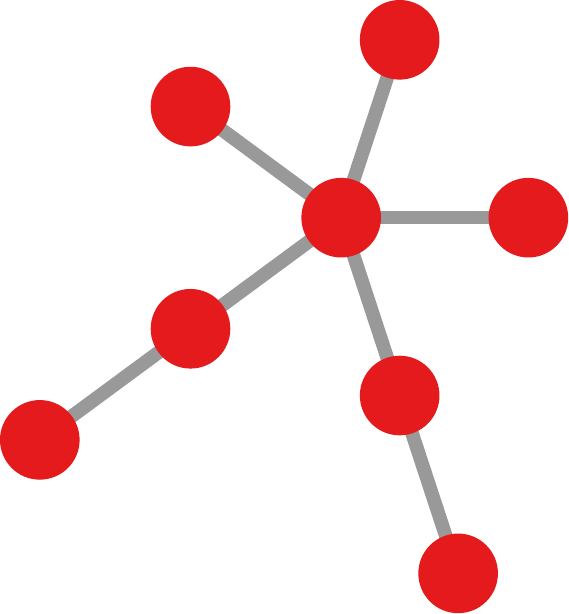}
\caption{}
\end{subfigure}
\caption{(a) An example of a cycle in the network, following the green arrows. (b) A tree.}
\label{fig:cycle}
\end{figure}

A \textbf{cycle} is a path that begins and ends with the same node. Note that I said ``path'', so we don't have any repeated nodes nor edges -- except the origin, of course. Figure \ref{fig:cycle}(a) shows an example of a cycle in the network. The cycle's length is the number of edges you use in your cycle. Given its topological constraints, that is also the number of its nodes.

Imposing cycles to be paths make them a non trivial object to have in your network. We can easily see why there might be nodes that participate in no cycles. If a node has degree equal to one, you can start a path from it, but you can never go back to complete a cycle. Doing so would force you to re-use the only connection they have. Thus a cycle is impossible for such nodes.

In fact, we can go further. We can imagine a network structure that has no cycles at all! I draw one such structure in Figure \ref{fig:cycle}(b). No matter how hard you squint, you're never going to be able to draw a cycle there. We have a special name for such structures: \textbf{trees}. Trees are simple graphs with no cycles. In a tree you cannot get your message back, unless somebody hears it twice. Given their lack of cycles, some even call them \textbf{acyclic graphs}.

\begin{figure}
\centering
\includegraphics[width=.45\textwidth]{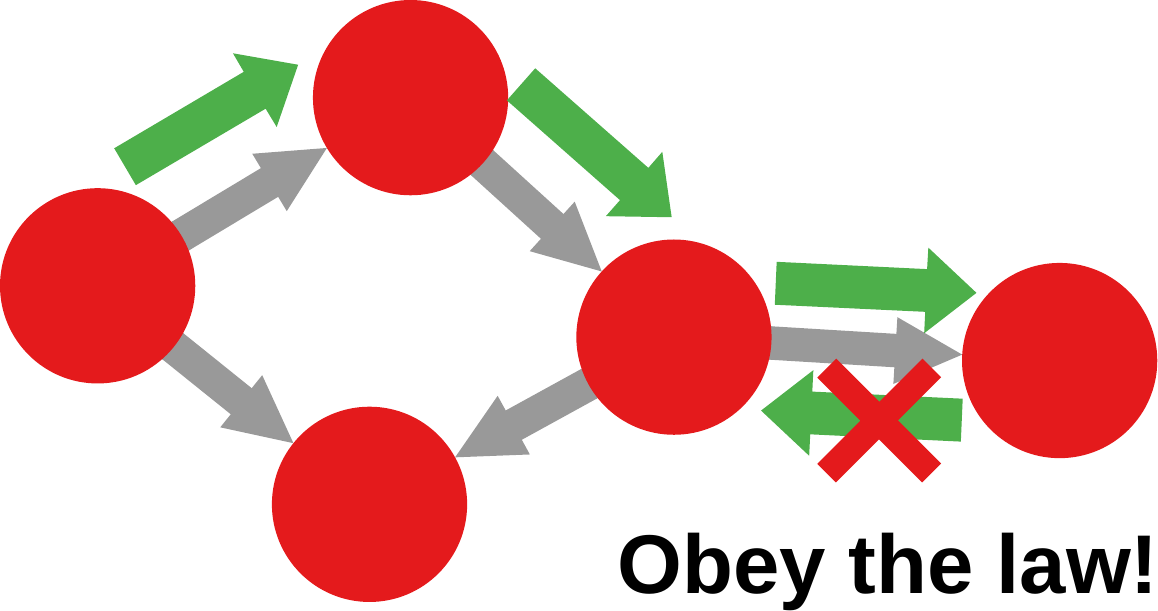}
\caption{An example of a directed walk, where we cannot explore an edge if we do not respect its direction.}
\label{fig:walk-directed}
\end{figure}

Directed networks add some spice to these concepts. In a directed social network, people are only willing to pass messages to their friends. If the friendship is not reciprocated, the receiver will not pass the message back to the sender. In practice, a walk -- and a path, and a cycle -- cannot use edges pointing to the direction opposite of the one they want to go. Figure \ref{fig:walk-directed} shows a naughty walk trying to do exactly that.

\begin{figure}
\centering
\begin{subfigure}{.19\columnwidth}
\includegraphics[width=\textwidth]{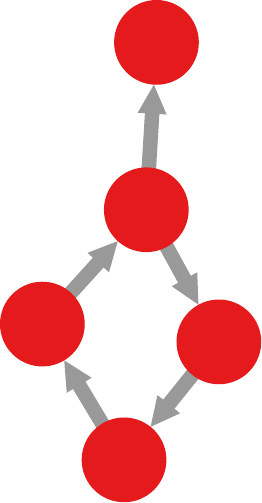}
\caption{}
\end{subfigure}
\qquad
\begin{subfigure}{.19\columnwidth}
\includegraphics[angle=90,origin=c,width=\textwidth]{figures/directed.pdf}
\caption{}
\end{subfigure}
\qquad
\begin{subfigure}{.19\columnwidth}
\includegraphics[width=\textwidth]{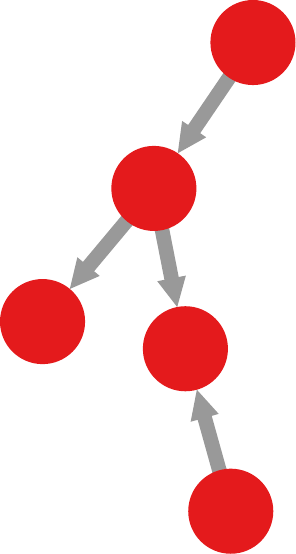}
\caption{}
\end{subfigure}
\qquad
\begin{subfigure}{.19\columnwidth}
\includegraphics[width=\textwidth]{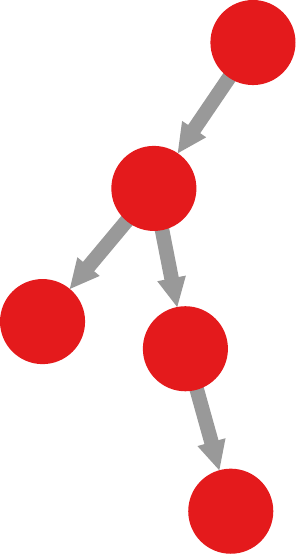}
\caption{}
\end{subfigure}
\caption{(a) A directed cyclic graph. (b) A directed acyclic graph. (c) A directed tree. (d) An arborescence.}
\label{fig:trees}
\end{figure}

In the undirected case, cycles create only two types of networks: those who have them (cyclic networks) and those who don't (acyclic networks). Instead, in the directed case, we can create a larger zoo of different directed structures. Figure \ref{fig:trees} showcases them. Figure \ref{fig:trees}(a) is the basic case: we have a cycle connecting the four nodes at the bottom, thus it is a \textbf{directed cyclic graph}. There's no such cycle present in Figure \ref{fig:trees}(b), thus we call it a \textbf{directed acyclic graph}.

There \textit{could} be a cycle in Figure \ref{fig:trees}(b), if we were to ignore edge directions. That is why we create a category for those directed graphs that would be acyclic if we were to ignore the edge directions. These are \textbf{directed trees}, and Figure \ref{fig:trees}(c) provides an example.

If you -- like me -- have even a mild case of self-diagnosed OCD, you'll probably be as irritated as I am about Figure \ref{fig:trees}(c). There's a natural flow to that directed tree, except for that little pesky edge at the bottom, going into the opposite direction. To restore sanity to the network world, we decided to create a final definition for directed graphs: \textbf{arborescences}. This is French for ``tree'', and in fact the two terms are often used interchangeably. But, technically speaking, an arborescence is a directed tree in which all nodes have in-degree of one, except the root. In an arborescence, the root is a special node: the only one with in-degree of zero. An arborescence must have one and only one root. Figure \ref{fig:trees}(d) fixes Figure \ref{fig:trees}(c) to be compliant to the definition of arborescence, and it is a work of art. So satisfying.

\section{Reciprocity}
Directed networks allow for a special type of path. In an undirected network without parallel edges, each path using two edges will necessarily bring you to a third node. You simply cannot use the same edge to go back to your origin. This is actually possible in a directed network. That is because the edge bringing you from $u$ to $v$ is not the same edge that brings you back to $u$ from $v$ -- by definition of what a directed edge is.

In the social network case, this is about replying messages, or considering as a friend somebody who also consider you as their friend. So these are cycles of length two, or containing two nodes and two edges.

In a social network, it is interesting to know the probability that, if I consider you my friend, you also consider me your friend -- which hopefully is $100\%$, but it rarely is so. This is an important quantity in network analysis, and we give it a name. We call it \textbf{reciprocity}, because it is all about reciprocating connections.

\begin{figure}
\centering
\includegraphics[width=.45\textwidth]{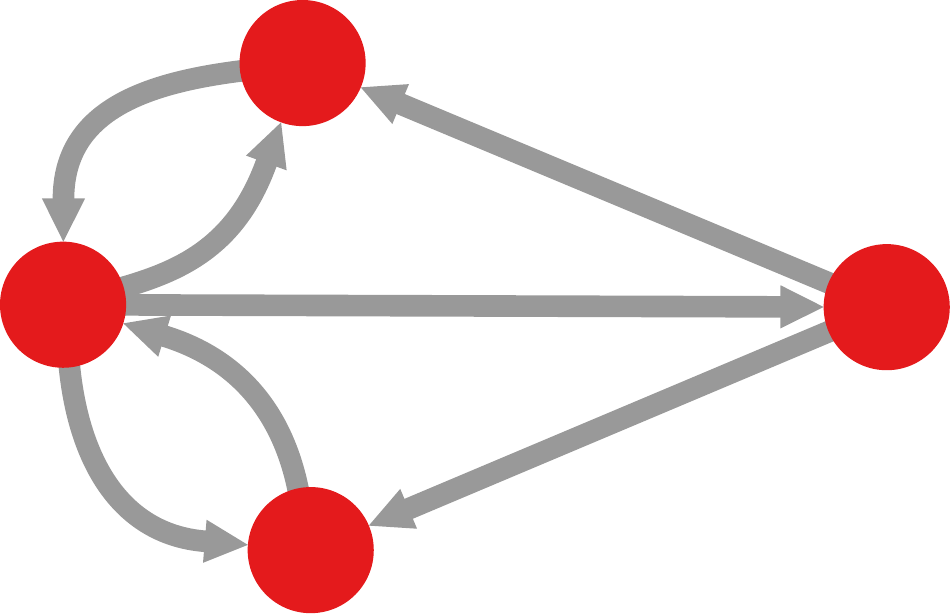}
\caption{An example of a directed network with some reciprocal edges.}
\label{fig:reciprocity}
\end{figure}

To calculate reciprocity we count the number of connected pairs of the network: pairs of nodes with at least one edge between them. In Figure \ref{fig:reciprocity}, we have five connected pairs. Then we count the number of connected pairs that have both possible edges between them: the ones reciprocating the connection. In Figure \ref{fig:reciprocity}, we have two of them. Reciprocity is simply the second count over the first one. So, for the example in Figure \ref{fig:reciprocity}, we conclude that reciprocity is $2 / 5$, or that the probability of a connection to be reciprocated is $40\%$. Sad.

\section{Connected Components}\label{sec:paths-ccomps}
Walks and paths can help you uncover some interesting properties in your network. Let's pick up our game of message-passing. In this scenario, we might end up in a situation where there is no way for a message to reach some of the people in the social network. The people you can reach with your message do not know anybody who can communicate to your intended targets. In this scenario, it is natural to divide people into groups that \textit{can} talk to each other. These are the network's ``components''.

\begin{figure}
\centering
\includegraphics[width=.5\columnwidth]{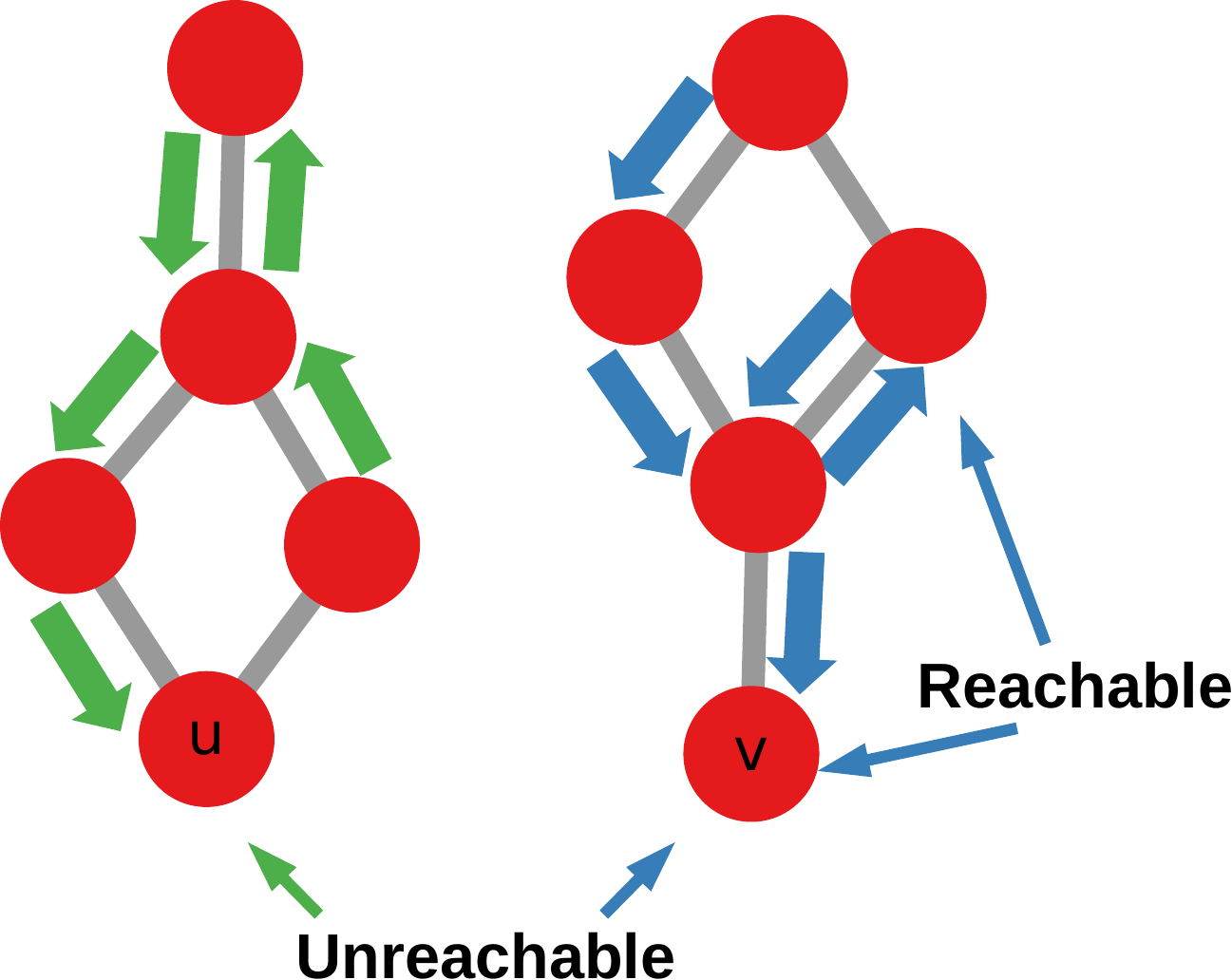}
\caption{A network with two connected components, each with five nodes. No matter how long you try, you can never find a path starting from $u$ and ending in $v$.}
\label{fig:ccomps}
\end{figure}

I can translate what I just said in terms of walks and paths. If two nodes cannot be connected by a walk, then they are on different connected components. Connected components are subgraphs whose nodes can be reached from one another by following the edges of the network. The network in Figure \ref{fig:ccomps} has two connected components: the nodes reachable with green-like walks, and the ones reachable by blue-like walks.\footnote{One nice thing about graph theorists is that they are less bad than average scientists in naming things. For instance, if you have a network made by different connected components and each of those components is a tree, then you can call that a ``forest''. 'cause it's made of trees. You get it? Anyone?}

A network with multiple connected components is usually bad news. The whole point of a network is to connect nodes together so that they are in the same shared structure. However, when you have multiple connected components, you effectively have two -- or more -- separate networks which cannot talk to each other. That's a bummer.

\begin{figure}
\centering
\includegraphics[width=.75\columnwidth]{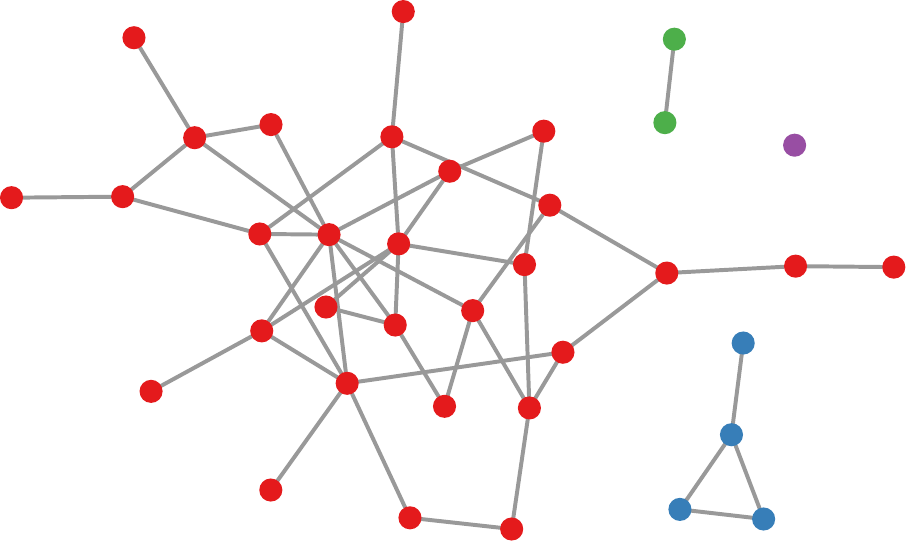}
\caption{A network with a giant connected component. The node color codes the component containing the node.}
\label{fig:gcc}
\end{figure}

As you might expect, real world networks tend to have multiple components. Reality always comes in the way of a good story. However, there is a silver lining. The vast majority of real networks host most of their nodes in a single connected component. In practice, networks have what we call a ``giant connected component'' (GCC). One of the components of the network is usually ridiculously larger than all the others\cite[-1in]{janson1993birth}\cite[-0.3in]{dorogovtsev2001giant}, as is the case in Figure \ref{fig:gcc}.

I mentioned earlier in this section that there is a relationship between some matrix operations and random walks. If you recall Section \ref{sec:mat-mat-mat}, raising the stochastic adjacency matrix to the power of $n$ tells you the probability of reaching a node with a random walk. So you might expect that there are also some matrix operations related to connected components.

Indeed, there are. We are interested in the eigenvalues of the stochastic adjacency matrix -- a more in-depth explanation of why this is the case will come in Section \ref{sec:rw-stationary}. In Section \ref{sec:mat-mat-stochastic} I said that the largest eigenvalue of the stochastic adjacency is equal to one. However, I also mentioned that the second eigenvalue $\lambda_2$ could also be equal to one -- and so could $\lambda_3$ and so on.  

\begin{figure}
\centering
\includegraphics[width=.75\columnwidth]{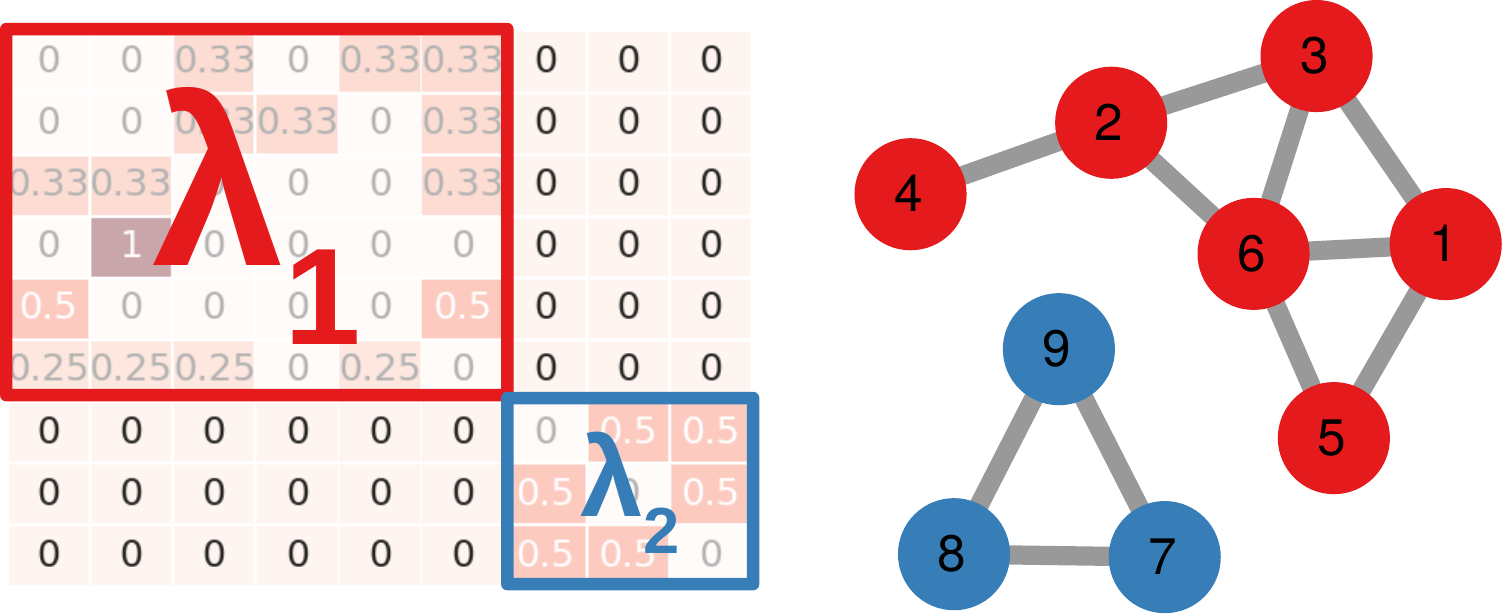}
\caption{The stochastic adjacency matrix of a disconnected graph looks like two different adjacency matrices pasted on the diagonal. Thus, they both have a (different) leading eigenvalue equal to one.}
\label{fig:eigenvector-leading}
\end{figure}

It turns out that the number of eigenvalues equal to one is the number of components in the graph. The reason is that you can consider the adjacency matrix as two adjacency matrices pasted into the same. They are disjoint matrices and each has as a maximum eigenvalue of one. I show an example in Figure \ref{fig:eigenvector-leading}.

\begin{figure}
\centering
\includegraphics[width=.66\columnwidth]{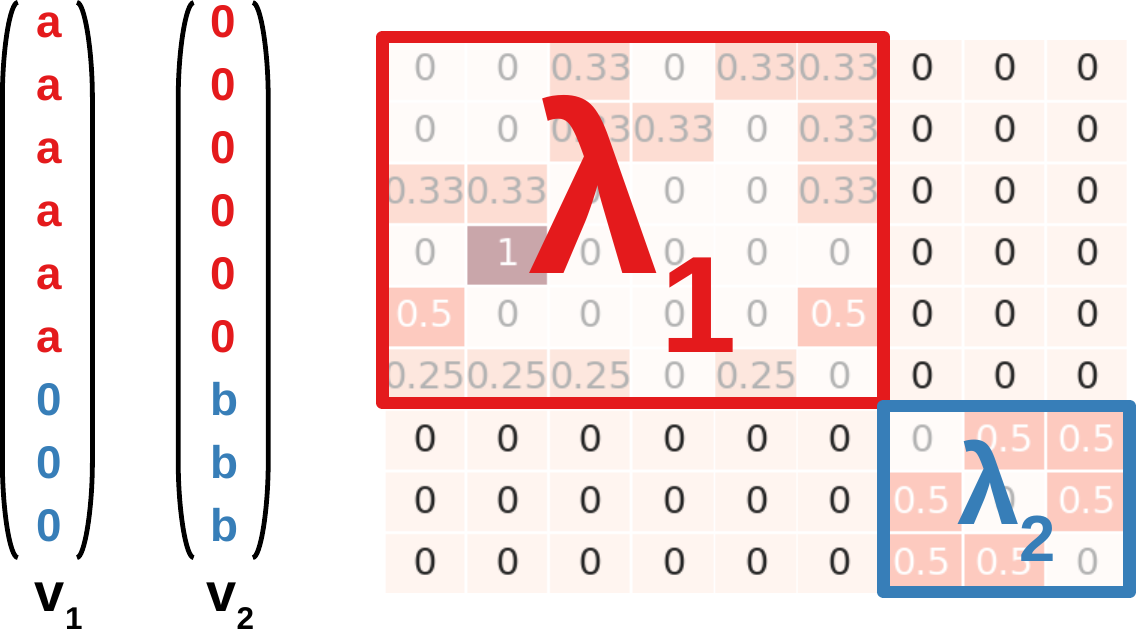}
\caption{If your graph has two components, the eigenvectors associated with the largest two eigenvalues of the stochastic adjacency matrix will tell you to which component the node belongs, by having a non-zero value.}
\label{fig:eigenvector-leading-2}
\end{figure}

If you can use the leading eigenvalues to count the number of connected components (Figure \ref{fig:eigenvector-leading}), the leading eigenvectors tell you to which component the nodes belong. If the network has two components, the nodes belonging to one will have a non-zero value in the eigenvector, while the nodes which do not belong to that component will have a zero -- see Figure \ref{fig:eigenvector-leading-2}. As you might have already deduced, if there is only one component then the leading eigenvector contains the same non-zero value. Similar properties hold for the Laplacian. The smallest eigenvectors of $L$ play the very same role as did the largest eigenvector of $A$: they are vectors telling us to which component the node belongs.

\subsection{Strong \& Weak Components}
So far, we saw that a measure of a network's usefulness is \textit{connectedness}. If there is no way to follow the edges of the network from node $u$ to node $v$, then $u$ has no way to influence -- or communicate to -- $v$. However, we dealt only with the case of undirected graphs. What if our edges are not symmetric, but have a direction?

\begin{figure}
\centering
\begin{subfigure}[t]{.33\columnwidth}
\includegraphics[width=\textwidth]{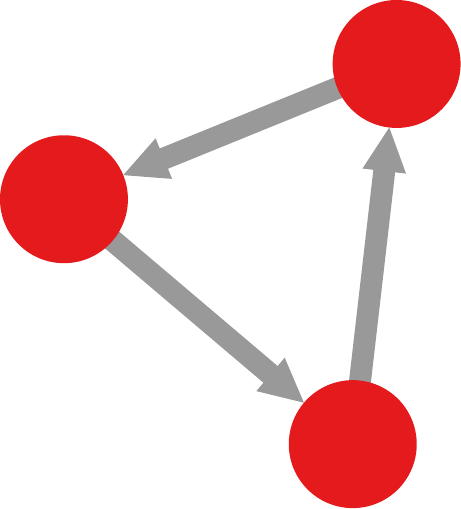}
\caption{}
\end{subfigure}
\qquad \qquad
\begin{subfigure}[t]{.4\columnwidth}
\includegraphics[width=\textwidth]{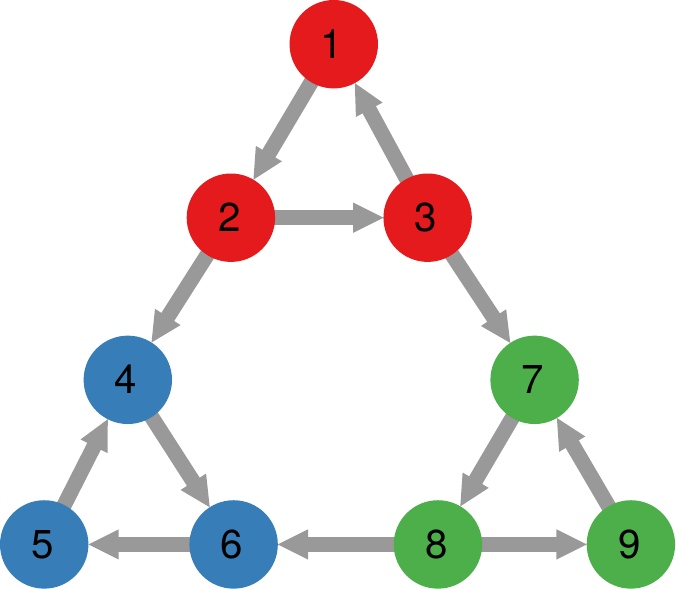}
\caption{}
\end{subfigure}
\caption{(a) A strongly connected component. (b) A network with multiple strongly connected components, coded by different node colors.}
\label{fig:scc}
\end{figure}

In that case we have two different scenarios. In the first scenario, which we call \textit{strong}, we want to ensure the same ability that the undirected network endowed us: $u$ must be able to contact $v$, and vice versa. Figure \ref{fig:scc}(a) shows an example of such a component. No matter where we choose to start our path, we can always go back. Therefore, any $u$ can reach any $v$, and vice versa. Since this strong requirement is satisfied, we call these ``strongly connected components'' (SCC).

It is not a surprise to reveal that strongly connected components contain cycles. By definition, if you can find a pair of nodes that you cannot join with a cycle -- meaning starting from $u$ and passing through $v$ makes it impossible to go back to $u$ -- then those nodes are not part of the same strongly connected component. SCCs are important: if you are playing a message-passing game where messages can only go in one direction, you can always hear back from the players in the same strongly connected component as you.

Popular algorithms to find strongly connected components in a graph are Tarjan's\cite{tarjan1972depth}, Nuutila's\cite{nuutila1994finding}, and others that exploit parallel computation\cite{hong2013fast}.

The definition of SCC leaves the door open for some confusion. Even by visually inspecting a network that appears to be connected in a single component, you will find multiple different SCCs -- as in Figure \ref{fig:scc}(b). In the figure, there is no path that respects the edge directions and leads from node $1$ to node $7$ and back. The best one could do is $1 \rightarrow 2 \rightarrow 3 \rightarrow 7 \rightarrow 8 \rightarrow 6 \rightarrow 5 \rightarrow 4$.

However, it \textit{feels} like this network should have one component, because we can see that there are no cuts, no isolated vertices. If we were to ignore edge directions, Figure \ref{fig:scc}(b) would really look like a connected component in an undirected network. This feeling of uneasiness led network scientists to create the concept of ``weakly connected components'' (WCC). WCCs are exactly what I just wrote: take a directed network, ignore edge directions, and look for connected components in this undirected version of it. Under this definition, Figure \ref{fig:scc}(b) has only one weakly connected component.

\subsection{In \& Out Components}
Not all weakly connected components are created equal. In large networks, one can find any sort of weird things. Suppose you are working in an office. The core of the office works on documents together, by passing them to each other multiple times and giving them the core's stamp of approval. This is by definition a strongly connected component.

But you're not part of the core of the office, you are in a weakly connected component. Your job is simply to receive a document, stamp it, and pass it to the next desk. Since you are in a WCC, you know you're never going to see the same document twice. That would imply that there is a cycle, and thus that you are in a strongly connected component with someone. However, what you see in the document can be radically different. The document might arrive to you with or without the core's stamp of approval. These two scenarios are quite different.

\begin{figure}
\centering
\begin{subfigure}[t]{.45\columnwidth}
\includegraphics[width=\textwidth]{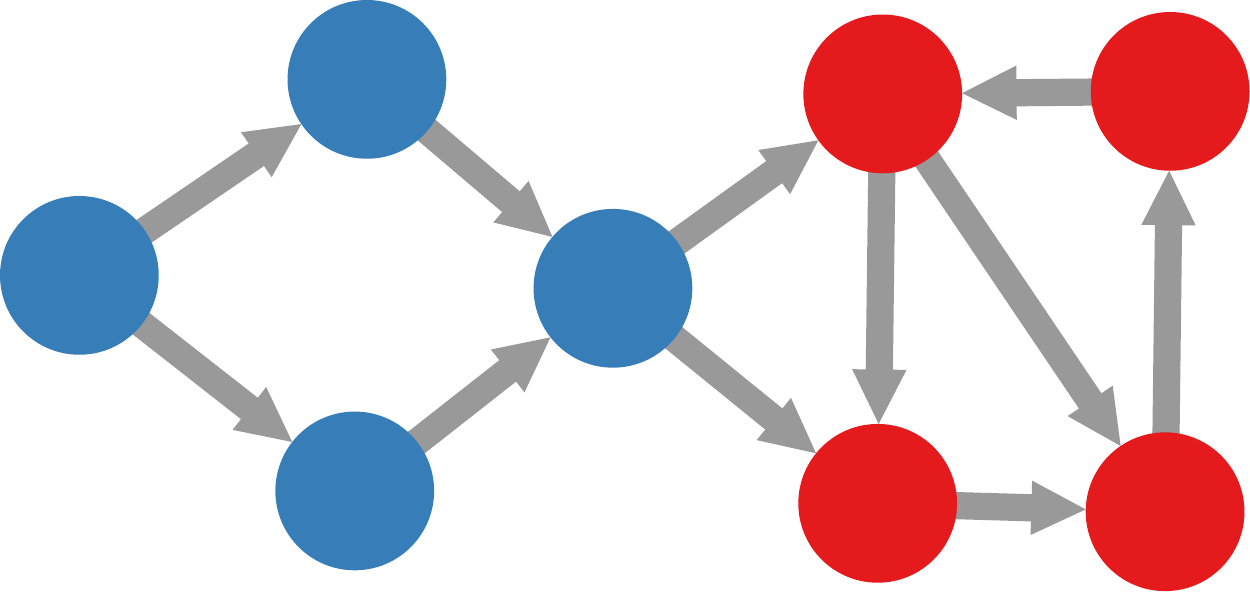}
\caption{}
\end{subfigure}
\qquad
\begin{subfigure}[t]{.45\columnwidth}
\includegraphics[width=\textwidth]{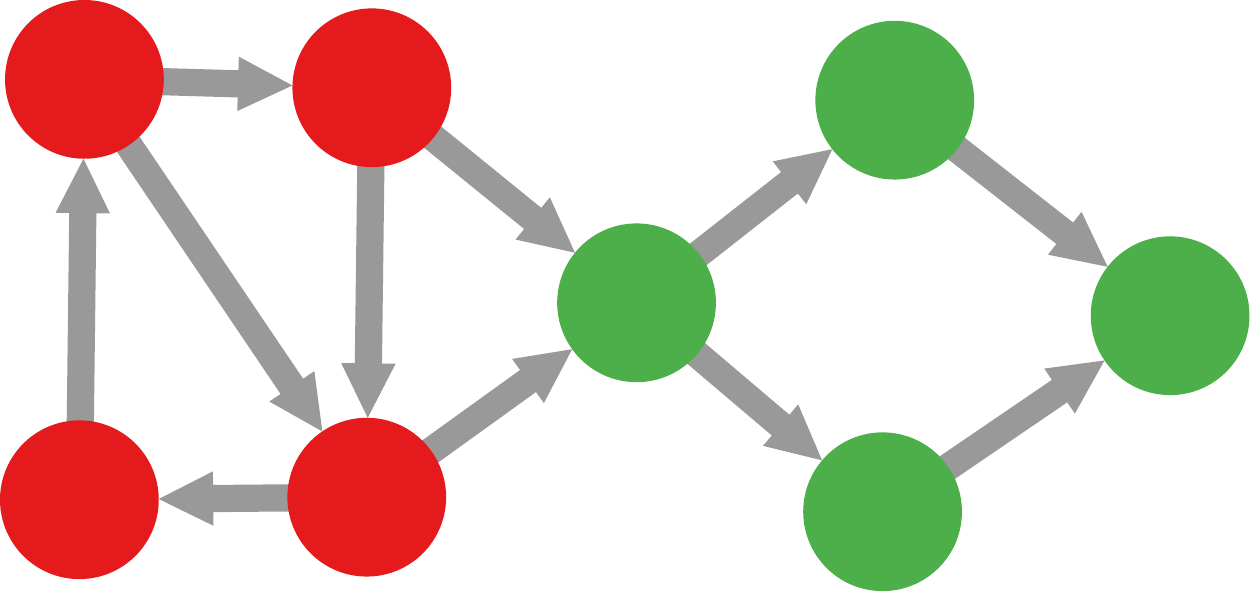}
\caption{}
\end{subfigure}
\caption{(a) An in-component (in blue), composed by the four leftmost nodes. (b) A out-component (in green), composed by the four rightmost nodes.}
\label{fig:comp-in-out}
\end{figure}

If you are in the first scenario, it means your WCC is positioned ``before'' the core. Documents pass through it and they are put \textit{in} the core. The flow of information originates from you or from some other member of the weakly connected component, and it is poured \textit{in}to the core. This is the scenario in Figure \ref{fig:comp-in-out}(a): you are one of the four leftmost nodes. In this paragraph I highlighted the word \textit{in} because we decided to call these special WCCs \textit{in}-components.

If you are in the second scenario, it means your WCC is positioned ``after'' the core. The core does its magic on the documents, and then \textit{out}puts them into your weakly connected component. The flow of information originates from the core and it is poured \textit{out} to your WCC. This is the scenario in Figure \ref{fig:comp-in-out}(b): you are one of the four rightmost nodes. In this paragraph I highlighted the word \textit{out} because we decided to call these special WCCs \textit{out}-components.

\section{Summary}

\begin{enumerate}
\item A walk is a sequence of nodes you can visit by following edges in the network. Its length is the number of edges you use. A path is a walk in which you never visit the same node or edge twice.
\item Cycles are paths which start and end in the same node. Acyclic graphs are graphs without cycles. An undirected acyclic graph is called a tree -- a graph with $|V|$ nodes and $|V| - 1$ edges. Otherwise, you can have directed acyclic graphs which are not trees.
\item A directed acyclic graph with $|V|$ nodes and $|V| - 1$ edges is a directed tree. If all nodes in a directed tree have in-degree of one, except one node with in-degree zero, then that directed tree is also an arborescence.
\item Reciprocal edges in directed networks are edges between two nodes pointing at each other. They allow cycles of length two. The number or reciprocated connections over the number of connected pairs is the reciprocity of the directed network.
\item A connected component is a set of nodes that can all reach each other by following walks on the edges. Real world networks usually have one giant connected component which contains the vast majority of nodes in the network.
\item You can count the number of connected components in a graph by counting the number of eigenvalues equal to one of its stochastic adjacency matrix. The non-zero entries in the corresponding eigenvectors tell you which nodes are in which connected component.
\item In directed networks you can have strong components: components of nodes that can reach each other respecting the direction of the edges. You can also have weak components, which ignore the edge direction.
\end{enumerate}

\section{Exercises}

\begin{enumerate}
\item Write the code to perform a random walk of arbitrary length on the network in \url{http://www.networkatlas.eu/exercises/10/1/data.txt}.
\item Find all cycles in the network in \url{http://www.networkatlas.eu/exercises/10/2/data.txt}. Note: the network is directed.
\item What is the average reciprocity in the network used in the previous question? How many nodes have a reciprocity of zero?
\item How many weakly and strongly connected component does the network used in the previous question have? Compare their sizes, in number of nodes, with the entire network. Which nodes are in these two components?
\end{enumerate}

\chapter{Random Walks}\label{cha:rndwalks}

\section{Stationary Distribution}\label{sec:rw-stationary}
\begin{figure}
\centering
\begin{subfigure}[t]{.32\columnwidth}
\includegraphics[width=\textwidth]{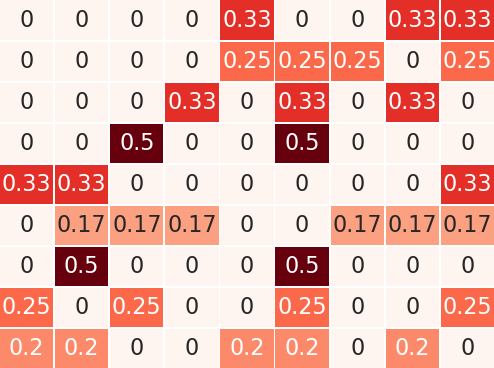}
\caption{$A^1$}
\label{fig:stochastic-powers1}
\end{subfigure}
\begin{subfigure}[t]{.32\columnwidth}
\includegraphics[width=\textwidth]{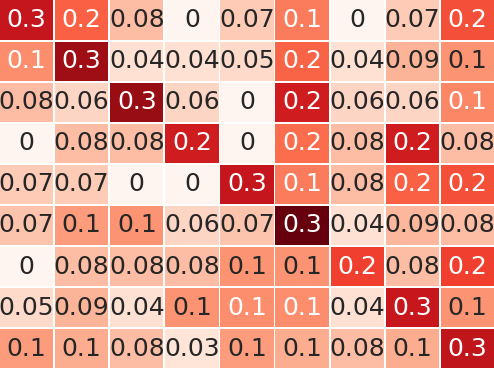}
\caption{$A^2$}
\label{fig:stochastic-powers2}
\end{subfigure}
\begin{subfigure}[t]{.32\columnwidth}
\includegraphics[width=\textwidth]{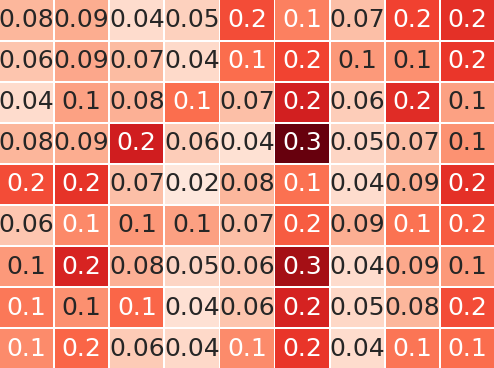}
\caption{$A^3$}
\label{fig:stochastic-powers3}
\end{subfigure}
\caption{Different powers of the stochastic adjacency matrix of the graph in Figure \ref{fig:adj}(b).}
\label{fig:stochastic-powers-again}
\end{figure}

Remember the stochastic adjacency matrix from Section \ref{sec:mat-mat-stochastic}? Figure \ref{fig:stochastic-powers-again} provides a refresher. Here we have the stochastic adjacency matrix of a graph, $A$, raised to different powers: $A^1$ (Figure \ref{fig:stochastic-powers-again}(a)), $A^2$ (Figure \ref{fig:stochastic-powers-again}(b)), and $A^3$ (Figure \ref{fig:stochastic-powers-again}(c)).

There's one curious thing if we look at the columns of the $A^1$ and $A^3$ matrices. In the first case, the minimum is zero and the maximum can be up to $0.5$. But in $A^3$ the minimum is higher than zero, and the maximum is just $0.3$. It seems that the values are somehow converging. So what happens if we take $A^{30}$ or -- gasp! -- $A^\infty$?

\begin{figure}
\centering
\includegraphics[width=.45\columnwidth]{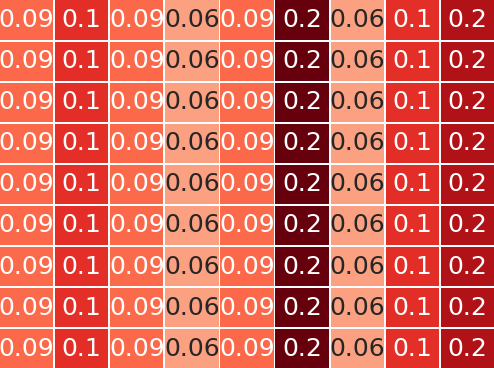}
\caption{The result when raising the stochastic $A$ to the power of $30$.}
\label{fig:stationary}
\end{figure}

Figure \ref{fig:stationary} shows the result. We see now that the columns are constant vectors. These numbers have a specific meaning. When we calculate $A^\infty$, what we're doing is basically asking the probability of being in a node after a random walk of infinite length. Since the length is infinite, it does not really matter from which node you originally started. That's why all rows of $A^\infty$ are the same -- remember that the row indicates the starting point while the column indicates the ending point.

This row vector -- you can pick any of them, since they're all the same -- is so important that we give it a name. We call it the ``stationary distribution'' -- or $\pi$, for short. $\pi$ tells us that, if you have a path of infinite length, the probability of ending up on a destination is only dependent on the destination's location and not on your point of origin. In practice, if you apply the transition probability ($A$) to the stationary distribution ($\pi$), you still obtain the stationary distribution: $\pi A = \pi$. Having a high value in the stationary distribution for a node means that you are likely to visit it often with a random walker -- by the way, this is almost exactly what PageRank estimates, plus/minus some bells and whistles, see Section \ref{sec:centr-eigen}.

Note that it is not necessary to calculate $A^\infty$ to know the stationary distribution. At least for undirected networks, $\pi$ is quite literally the normalized degree of the nodes: the degree divided by the sum of all degrees ($2|E|$). 

But... wait! This stationary distribution formula is oddly familiar: $\pi A = \pi$. Haven't we seen something similar to it? This kind of looks like our eigenvector specification ($Av = \lambda v$, see Section \ref{sec:la-eigen}), with a few odd parts. First, where's the eigenvalue? Well, we can always multiply a vector to $1$ and we won't change anything in the equation. So: $\pi A = \pi 1$. This is cool, because we already know that $1$ is the largest eigenvalue ($\lambda_1$) of a stochastic matrix. Second, the vector $\pi$ is \textit{on the left}, not \textit{on the right}. Putting these things together: the stationary distribution $\pi$ is the vector associated with the largest eigenvalue, if multiplied on the left of $A$. Therefore: $\pi$ is the leading left eigenvector.

If you're dealing with an undirected graph, there is a relationship between right and left eigenvectors. If you were to transpose the stochastic adjacency matrix, that is making it column-normalized instead of row-normalized, the left and right eigenvectors would swap. In different words: the left eigenvectors of $A$ are exactly the same as the right eigenvectors of $A^T$. Thus the vector of constant and $\pi$ are the right and left leading eigenvectors of $A$, and they swap roles in $A^T$.

\begin{figure}
\centering
\includegraphics[width=.75\columnwidth]{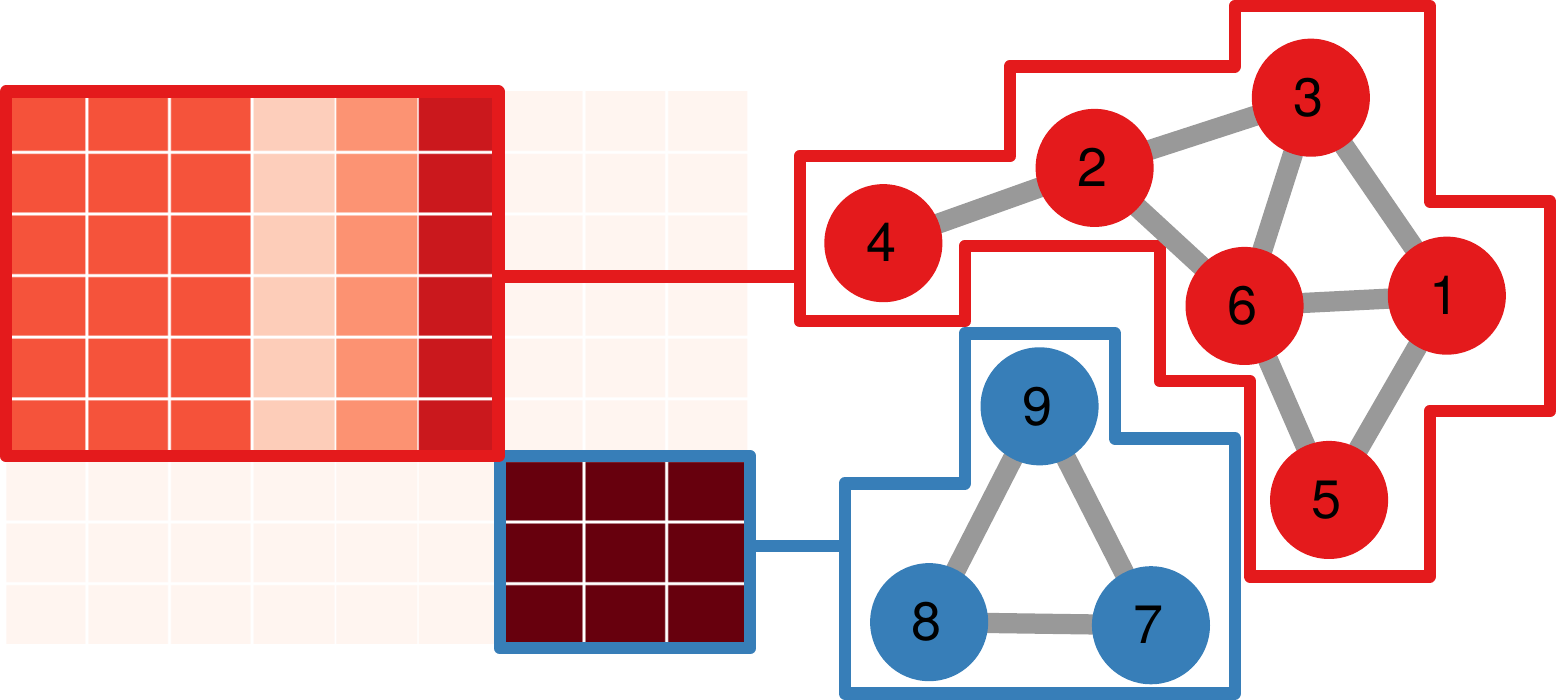}
\caption{The result when calculating the stationary distribution for an unconnected graph.}
\label{fig:stationary-disconnected}
\end{figure}

What do you do if your graph is not connected? No matter how many powers of $A$ you take, how infinitely long your walks are, some destinations are unreachable from some origins. We end up with two stationary distributions, one for one component, and one for the other. Figure \ref{fig:stationary-disconnected} shows an example. These two stationary distributions are not directly comparable one with the other. They are effectively telling you something about two different networks: one made by the first component, and the other composed by the second one.

This makes it clear why the eigenvector contains zeros for the entries corresponding to the nodes that are not part of the connected component we're looking at. The eigenvector contains the probability of ending up in a node after a random walk of infinite length. The probability of ending up in those nodes via a random walk -- no matter how long -- is zero, because there is no edge that you can use.

\section{Non-Backtracking Random Walks}\label{sec:rw-no-backtrack}
This is a good place to mention that there is an almost infinite number of different matrix representations you can  build for a graph. Each of those representations are useful to describe some sort of process on the graph. Since we're in the random walk section, I will mention another useful matrix representing random walks: the non-backtracking matrix. However, be aware that this is only one arbitrary choice about the many possible, you can have: cycle matrices\cite{mateti1976algorithms}\cite{harary1967graphs}, cut-set matrices\cite{vatn1992finding}, path and distance matrices, modularity matrices\cite{newman2006modularity}, to name a few.

\begin{figure}[b]
\centering
\includegraphics[width=.45\columnwidth]{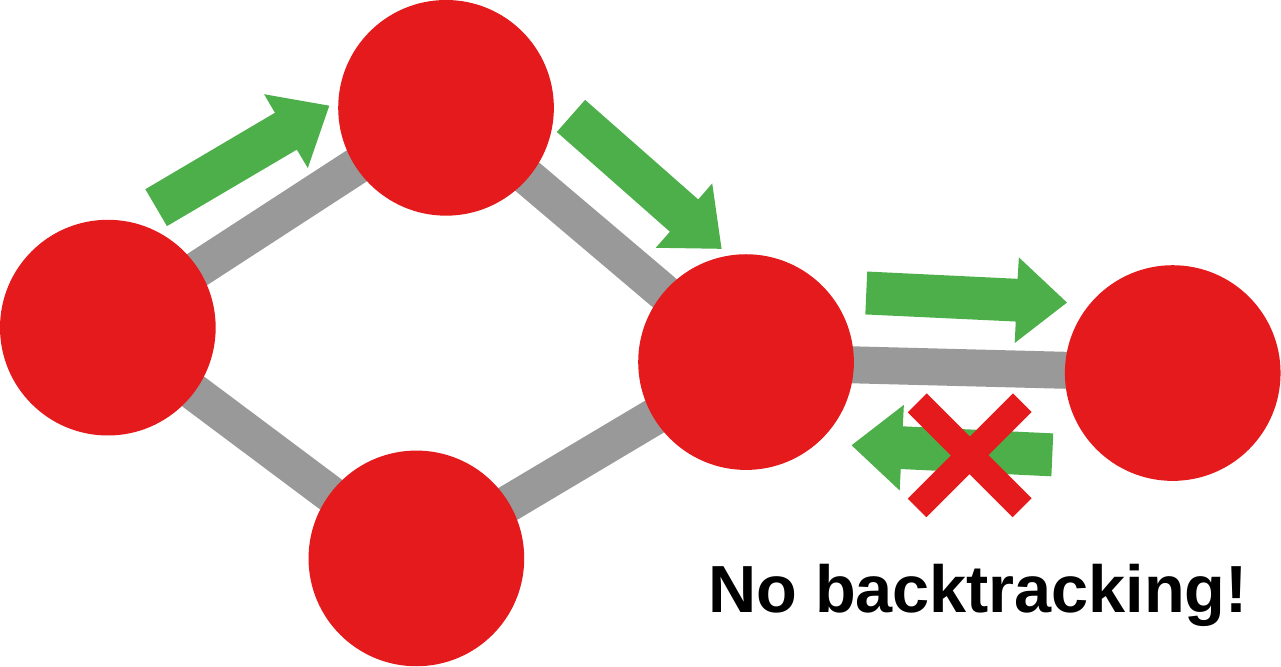}
\caption{A non-backtracking random walk. The green arrows show the state transitions.}
\label{fig:random-walk-nobacktrack2}
\end{figure}

So what's a non-backtracking matrix? As the name suggests, it's a matrix describing non-backtracking walks. A walk is non-backtracking when we forbid the walker to re-use the same edge twice in a row in its walk. Figure \ref{fig:random-walk-nobacktrack2} shows an example.

If we want to represent such a process with a matrix, we need to build it quite differently from an adjacency matrix. Rather than having a row and a column per node, we instead have two rows and columns per edge\cite{hashimoto1989zeta} -- or one row/column per edge direction if we have a directed graph, meaning that we treat an undirected graph as a directed one with perfect reciprocity. Each cell contains a one if we can use the edge direction for our non-backtracking walk, zero otherwise. Formally:

$$
NB_{uv,vz} = \begin{cases}
 1 & \text{if } u \neq z\\
 0 & \text{otherwise.}
\end{cases}
$$

\begin{figure}
\centering
\includegraphics[width=.8\columnwidth]{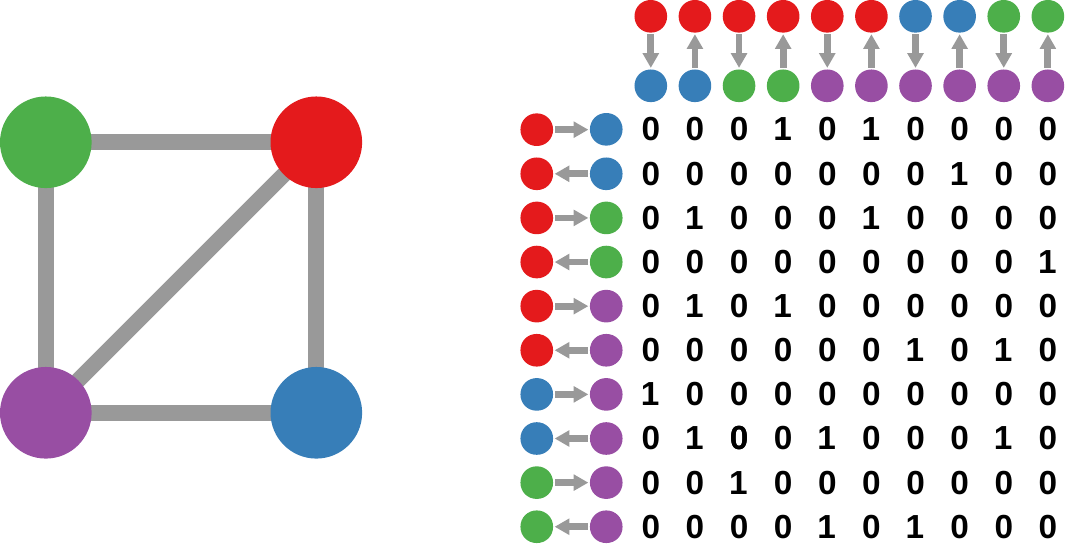}
\caption{A non-backtracking matrix.}
\label{fig:nobacktrack-matrix}
\end{figure}

So you see what's going on here: we can only transition to node $z$ from $v$ only if we got into $v$ via $u$ and $u$ is not the same node as $z$. If you're a graphical thinker, Figure \ref{fig:nobacktrack-matrix} might help you. The non backtracking matrix is not symmetric: if you go from red to blue (first column) you can go from blue to purple (first column, seventh row equals to $1$). But if you go from blue to purple (seventh column) you cannot go from red to blue (seventh column, first row equals to $0$). This breaks the symmetry.

As the figure shows, the non backtracking matrix has zero on the block diagonal, because the block diagonal contains all edges to themselves. Thus, a one in the diagonal is exactly a backtracking move: you used the $u,v$ edge to get to $v$ and then you use it again to get to $u$. Naughty backtracking walker!

On the other hand, if we go to the blue node from the red node, the only legal move is to go to the purple node. If we did the opposite move, from blue to red, we could reach either the green or the purple node. Using these rules, you can figure out each entry in the matrix. Of course, you can have a directed non-backtracking matrix, in which you need to respect the direction of the edge as well.

Non-backtracking matrices are useful for a bunch of applications: fixing eigenvector centrality degeneration\cite{martin2014localization} (Section \ref{sec:centr-eigen}); helping with community detection\cite{krzakala2013spectral} (Part \ref{par:cd}); describing percolation processes\cite{karrer2014percolation} (Chapter \ref{cha:epidemapps}); and even helping with counting motifs in graphs\cite{torres2019non} (Chapter \ref{cha:mining-base}).

\section{Hitting Time}\label{sec:rw-hitime}
The stationary distribution allows you to calculate the probability of visiting a node given an infinite length random walk. Another important thing you might be interested in discovering is how long you have to wait before a node gets visited by a random walker. Of course you have an intuition: if it is very likely to visit the node -- high value in the stationary distribution -- then probably it won't take long before we ``hit'' that node. But the two quantities, probability and average hitting time, are not the same. Especially since the hitting time of node $v$ depends from the starting point $u$.

We use $H_{u,v}$ to indicate the expected hitting time of $v$ for a random walk starting in $u$. How can we calculate $H_{u,v}$? Well, if we want to reach $v$ from $u$, first we have to go to a neighbor of $u$. Let's call it $z$. Once we are in $z$, it will take us $H_{z,v}$ steps to reach $v$, by definition. The probability of ending up in $z$ from $u$ is one over $u$'s degree ($k_u$) since we picked it at random. Thus the formula expands to:

$$ H_{u,v} = 1 + \dfrac{1}{k_u} \sum \limits_{z \in N_u} H_{z,v}.$$

Applying this formula naively wouldn't lead you very far, as you'll find yourself needing to calculate $H_{u,v}$ in order to find out $H_{u,v}$'s value. There is a way to find $H_{u,v}$ using -- what else? -- the eigenvectors and eigenvalues of the adjacency matrix\cite{lovasz1993random}. The exact mathematical derivation is not for the faint of heart and can be appreciated by the brave readers who will dare to look at this obscene footnote\footnote{So, here we go. The main formula can be rewritten in matrix form: $H = J + AH - F$, with $J$ being the matrix of ones, and $A$ the stochastic adjacency. What's that $F$, though? It's a diagonal matrix we have to remove from $H$ because, by definition, the diagonal of $H$ must be zero: the hitting time of the origin from the origin is zero, it is \textit{not} the time it takes for a random walker to go somewhere and coming back, which is what you'd get if you didn't take $F$ out. So did we just make it worse by adding something more we don't know? Actually, we can derive $F$. Let's rewrite the equation as $F = J + AH - H$. Let's multiply the stationary distribution to both sides: $F\pi = J\pi + H(A - I)\pi$ (I grouped the $H$ terms). Note that $A\pi = \pi$ by definition of a stationary distribution and that $I\pi = \pi$ by definition of an identity matrix. So the whole $H(A - I)\pi$ term disappears, leaving $F\pi = J\pi$. Again by definition of $\pi$, a matrix of ones times $\pi$ equals the scalar one, giving us $F\pi = 1$. So $F$ is a diagonal matrix with $1/\pi_u = 2|E|/k_u$ on its $u$th diagonal entry. We use $D$ to specify the diagonal matrix with the degree on the diagonal, giving us the equation $H(I - A) = J - 2|E|D^{-1}$. So we can derive $H$ easily now, right? Ahahah. Wrong. $(I - A)$ is singular, thus non-invertible: you can't calculate $(I - A)^{-1}$. So we need to multiply the left and the right side by something that will make $(I - A)$ disappear. That something is the special matrix $Z = (I - A + A^\infty)^{-1}$. This gives us $H = (I - A + A^\infty)^{-1}(J - 2|E|D^{-1})$. To cut a long story short, we can decompose $A$ using its eigenvectors, obtaining the formula in the main text. It took me one day to write this footnote. I'm not paid nearly enough for this.}, and ends with the following formula:

$$ H_{u,v} = 2|E| \sum \limits_{n = 2}^{|V|} \dfrac{1}{1 - \lambda_n} \left( \dfrac{w^2_{n,v}}{k_v} - \dfrac{w_{n,u}w_{n,v}}{\sqrt{k_uk_v}} \right),$$

where $|V|$ and $|E|$ are the number of nodes and edges. $\lambda$ and $w$  are eigenvalues and eigenvectors of a special decomposition of $A$, namely $N = \Delta^{1/2}A\Delta^{1/2}$. $\Delta = D^{-1}$ is the diagonal matrix with the inverse of the degree on the diagonal -- much like $D$ is the diagonal matrix with the degree on the diagonal, we met it when calculating the graph Laplacian. $\lambda_n$ is the $n$th eigenvalue -- sorted in descending order --; $w_{n,u}$ is the $u$th entry of the $n$th eigenvector.  Finally, $k_u$ is the degree of node $u$.

\begin{figure}[t]
\centering
\includegraphics[width=\columnwidth]{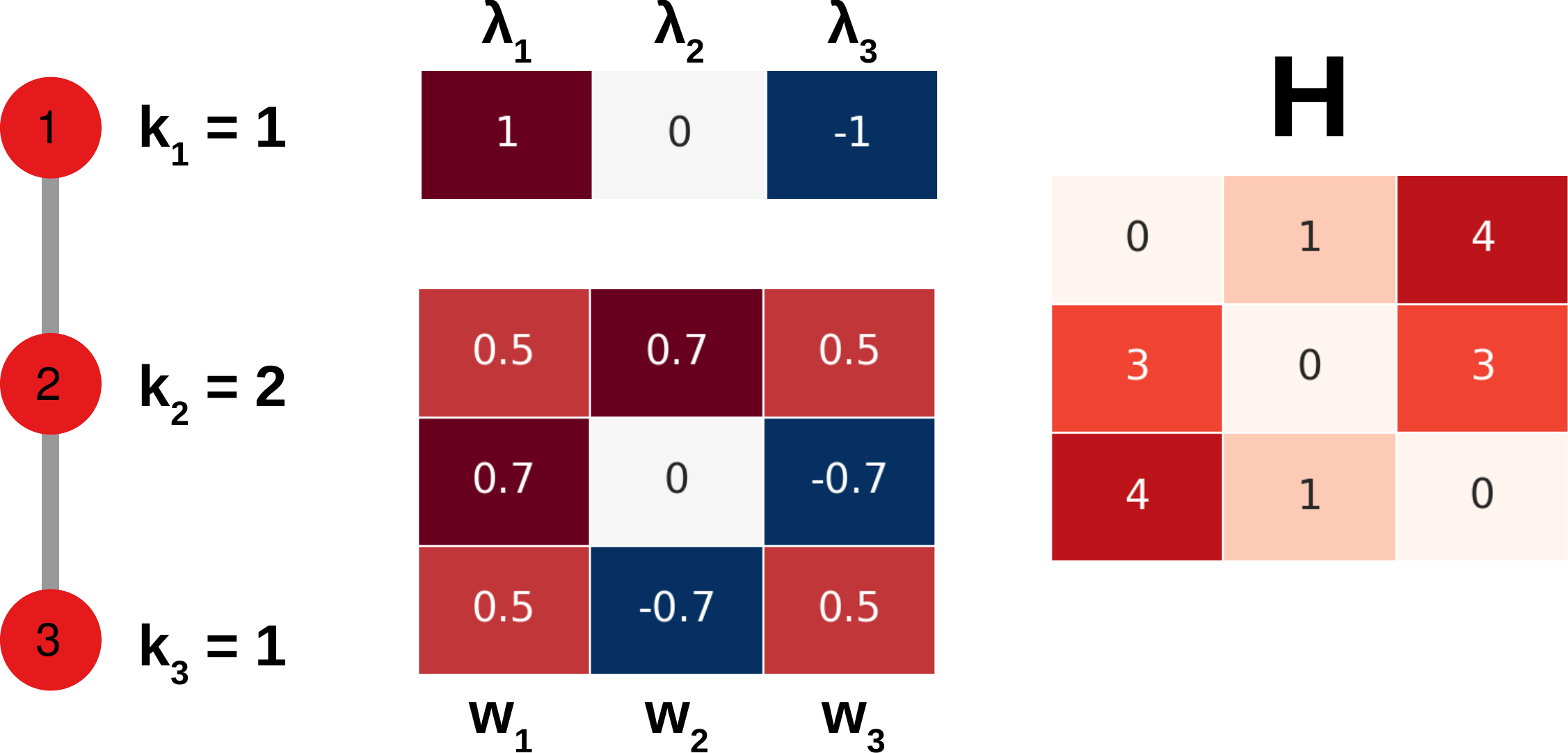}
\caption{The elements needed to calculate the hitting time of a graph. From left to right: the graph and its degree vector, the eigenvalues and eigenvectors of $N$, the resulting hitting time matrix $H$.}
\label{fig:hitime}
\end{figure}

The formula looks threatening, but it ought not to be. Let's look at the stupidest example possible, in Figure \ref{fig:hitime}. Here we have a simple chain graph. Its degrees $k$ are $(1, 2, 1)$, as the second node has two neighbors and the other nodes have only one. The eigenvalues of $N$ are $(1,0,-1)$ -- we know the largest must be one because $N$ is stochastic. $w$ shows the corresponding eigenvectors. Note how, in $w_1$, $w_{1,u} = (1 / \sqrt{2|E|})\sqrt{k_u}$, which you can derive following what you know about the stationary distribution of $A$ and the way we derived $N$ from $A$.

If you want to know $H_{1,3}$, you need to do two things. First, for $n = 2$, $\dfrac{1}{1 - \lambda_2} = 1$ and $\left( \dfrac{0.7^2}{1} - \dfrac{0.7 \times -0.7}{\sqrt{1}} \right) = 1$. Then, for $n = 3$, $\dfrac{1}{1 - \lambda_3} = 1/2$ and $\left( \dfrac{0.5^2}{1} - \dfrac{0.5 \times 0.5}{\sqrt{1}} \right) = 0$. So, for $n = 2$ the part on the right side of the sum evaluates to $1$ and for $n = 3$ it evaluates to zero. Thus the total sum is one. Multiplied to $2|E|$ you obtain four.

This is super intuitive. How long does it take to get from node $1$ to node $2$? Well, node $1$ has only one connection and it goes to node $2$, so it will always take one step. But to get from node $2$ to node $1$, you only have a $50\%$ chance of doing it in one step. The other $50\%$ of the times the random walker will go to node $3$. It will always come back after another step, and then we'll have another $50\%$ chance to go to node $1$. You sum that to infinity, and you get an expected hitting time of three.

From the formula and the example it is easy to see that $H$ is asymmetric, given that one of its parts is dependent on the degree of the destination $k_v$ and not on $k_u$, the degree of the origin. For this reason, another object of interest is the commute matrix $C$: the time it takes to go from $u$ to $v$ and back to $u$. This is $C$, and it is simply defined as: $C_{u,v} = H_{u,v} + H_{v,u}$, trivially symmetric.

There's one fun connection with the stationary distribution here. We call this the ``Random Target Lemma''. Suppose you start from node $u$ and you pick destinations $v$ at random. What's the expected hitting time? This is basically averaging $H_{u,v}$ over all possible destinations $v$. If you do that, you'll find out that the result is:

$$ \sum \limits_{v \in V} \pi_v H_{u,v} = \sum \limits_{n = 2}^{|V|} \dfrac{1}{1 - \lambda_n}.$$

Noticing something weird? The right hand side has no trace of $u$. This means that the average time to hit something doesn't depend on your starting point $u$, exactly like, in the stationary distribution, the probability of ending your random walk somewhere didn't depend on from where you started.

Note that this is a very short and incomplete treatment of the subject, just to let you know how to calculate hitting and commute times. You really should check out Lov{\'a}sz's paper (footnote 10) for a full treatment of the subject.

\section{Effective Resistance}\label{sec:rw-effectres}
The hitting time matrix $H$ is useful specifically when you want to exploit its asymmetry: you want to start from $u$ and know when you will hit $v$. However, if you're only interested in the symmetric commute time, then there is an easier way to calculate the commute matrix $C$. This calculation involves the effective resistance matrix $\Omega$ -- which is one of the awesomest matrices in network science and that's why you should know about it.

\begin{figure}
\centering
\begin{subfigure}[t]{.32\columnwidth}
\includegraphics[width=\textwidth]{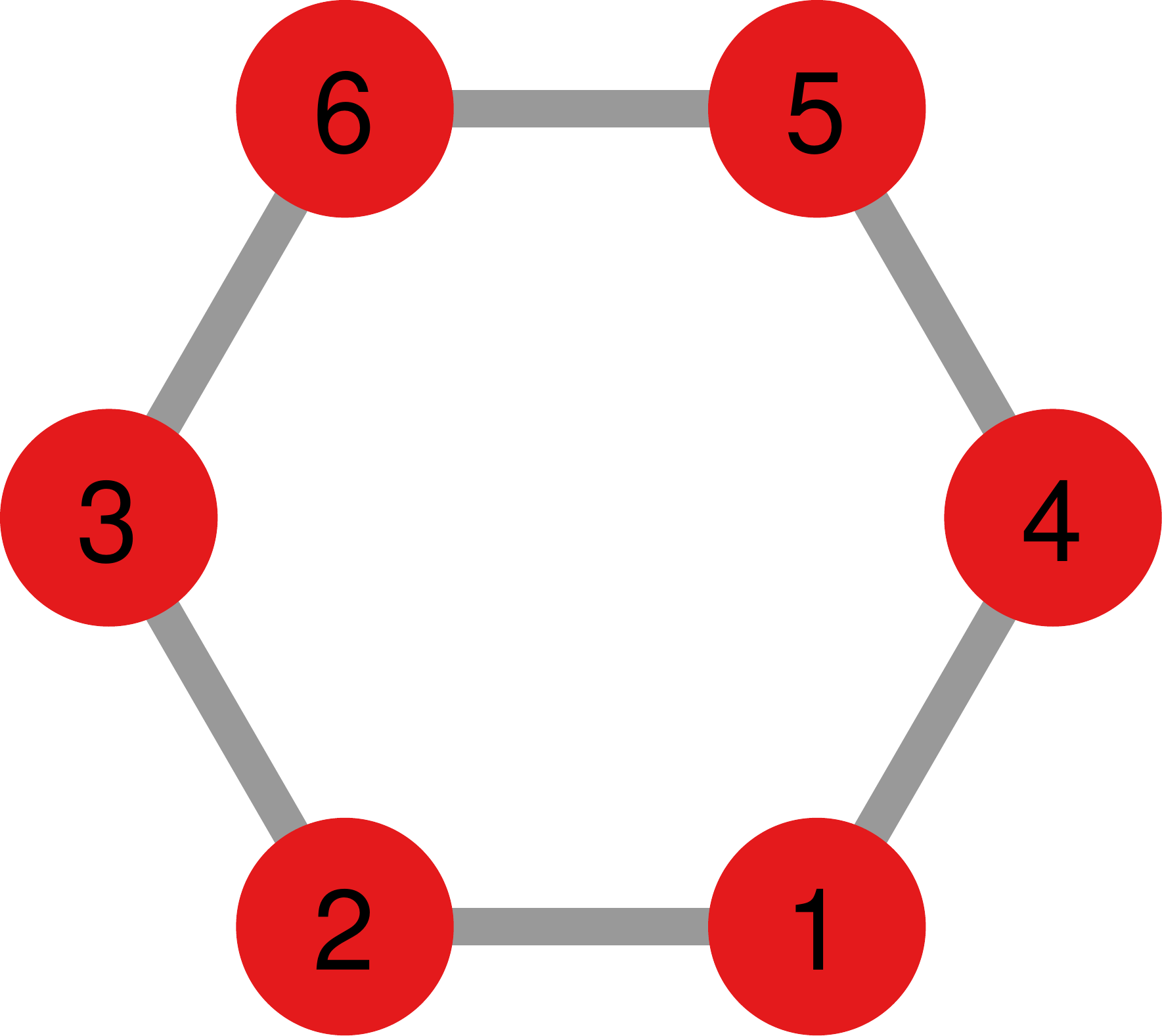}
\caption{}
\end{subfigure}
\qquad
\begin{subfigure}[t]{.4\columnwidth}
\includegraphics[width=\textwidth]{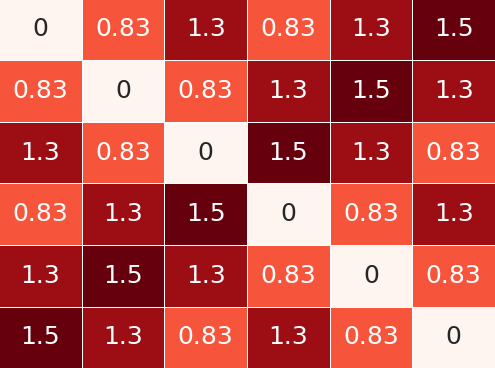}
\caption{}
\end{subfigure}
\caption{(a) A ring graph, (b) its effective resistance matrix.}
\label{fig:effres-1}
\end{figure}

Mathematically, $C = 2|E|\Omega$, which means that $\Omega$ is a sort of normalized commute time. It is the commute time, ignoring the overall size of the graph -- which is given by the $2|E|$ factor. You can see an example graph and its effective resistance matrix in Figure \ref{fig:effres-1}. The original definition of $\Omega$ is physical: you assume $G$ is an electrical network and each edge is a resistor of one Ohm. Then the effective resistance between nodes $u$ and $v$ $\Omega_{u,v}$ is the literal electric resistance you'd measure. If you remember, when I introduced the Laplacian (Section \ref{sec:mat-mat-laplacian}) I said that it was originally used to describe electric circuits, so you might expect it to pop up here. In fact, here's the formula to calculate effective resistance\cite{babic2002resistance}\cite{xiao2003resistance}: $ \Omega_{u,v} = \Gamma_{u,u} + \Gamma_{v,v} - 2\Gamma_{u,v}$, where:

$$ \Gamma = \left( L + \dfrac{1}{|V|} \mathbb{1} \right)^\dagger.$$

Here, $L$ is the Laplacian, $|V|$ is the number of nodes, and $\mathbb{1}$ is a $|V| \times |V|$ matrix filled with ones. In practice, inside the parentheses we have a matrix that is $L$ plus $1/|V|$ in all its entries. The $\dagger$ symbol means that we want to invert this matrix. If you remember, the Laplacian tells you how electricity flows in the network, so we need to invert it to estimate the resistances. The problem is that the Laplacian is a singular matrix, which means it cannot be inverted.

This is why we use $\dagger$ instead of $-1$: rather than inverting we take the Moore-Penrose pseudoinverse\cite{moore1920reciprocal}\cite{bjerhammar1951application}\cite{penrose1955generalized}, which is basically the inverse, but shhh don't tell the Laplacian or it will get mad. To get the Moore-Penrose pseudoinverse, the first step is to perform the singular value decomposition (SVD) of $L$. SVD is one of the many ways to perform matrix factorization (Section \ref{sec:mat-factors}). In SVD, we want to find the elements for which this equation holds: $Q_1 \Sigma Q_2^T = L$. The important part here is $\Sigma$, which is a diagonal matrix containing $L$'s singular values. We can easily build a $\Sigma^{-1}$ matrix, containing in its diagonal the reciprocals of $L$'s singular values. Then $Q_2 \Sigma^{-1} Q_1^T = L^\dagger$ is $L$'s Moore-Penrose pseudoinverse. It holds that $L L^\dagger L = L$ and that $L^\dagger L L^\dagger = L^\dagger$.

What makes effective resistance so awesome is that it defines a proper analogue to the Euclidean distance in a graph\cite{devriendt2022effective}. By far the most popular alternative is using shortest path distances -- we'll see how to calculate this distance in Chapter \ref{cha:shortpath}. However, differently from shortest paths, $\Omega$ is a proper metric, which means you can do a bunch of things (most of which we'll see in Chapter \ref{cha:nvd} when we'll talk about network distances) that would lead to mathematical nonsense if you were to use shortest paths instead.

\begin{figure}
\centering
\begin{subfigure}[t]{.32\columnwidth}
\includegraphics[width=\textwidth]{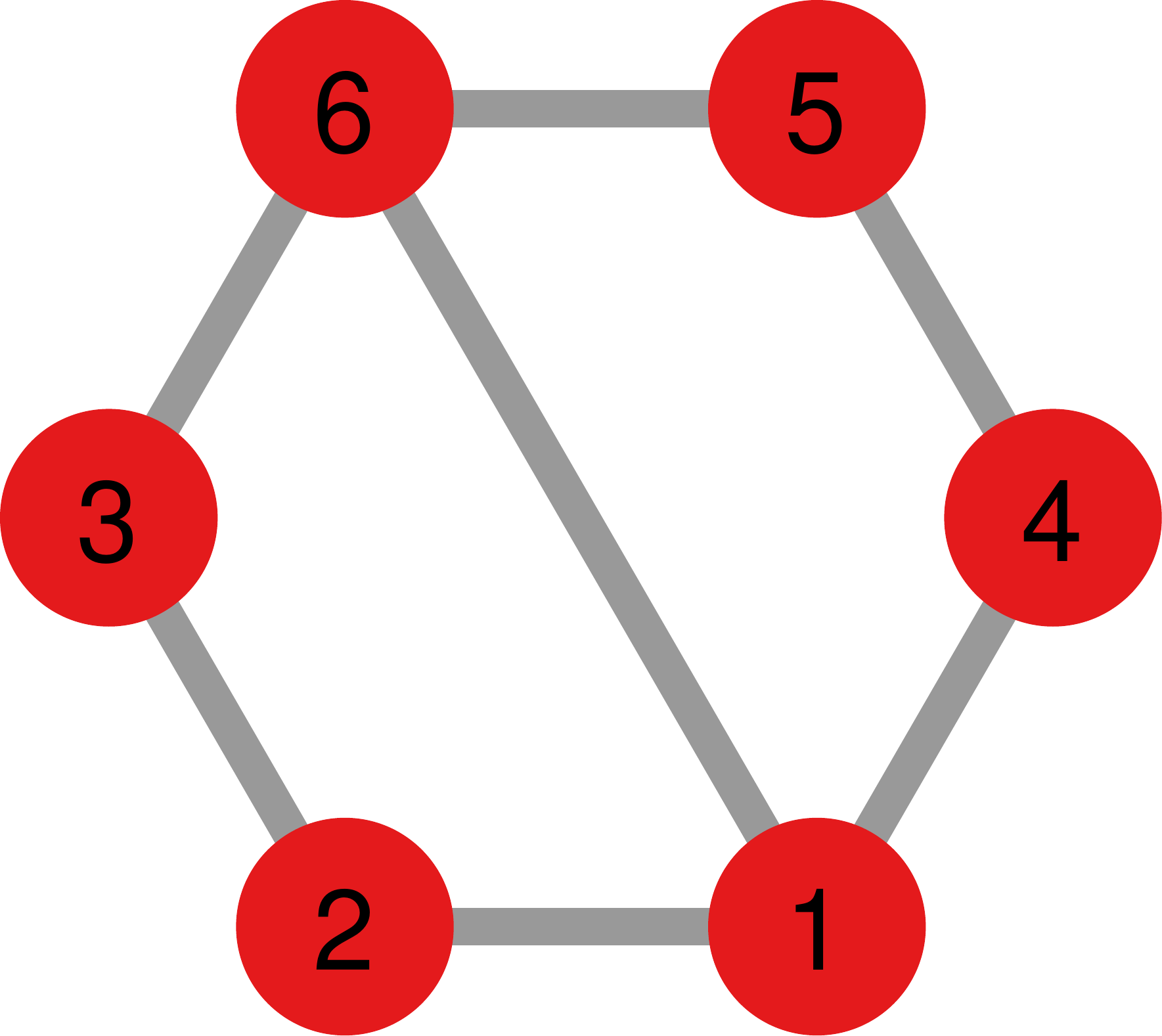}
\caption{}
\end{subfigure}
\qquad
\begin{subfigure}[t]{.4\columnwidth}
\includegraphics[width=\textwidth]{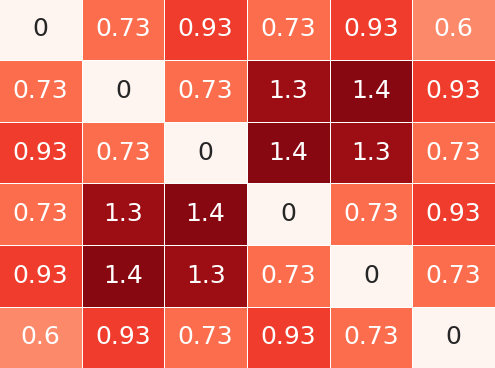}
\caption{}
\end{subfigure}
\caption{(a) A graph, (b) its effective resistance matrix.}
\label{fig:effres-2}
\end{figure}

Moreover, by being the result of a process of diffusion through the entirety of the graph -- current flows as it was presented -- $\Omega$ is also more resistant to random fluctuations. The removal or introduction of a single edge can radically change the shortest path distance between two nodes, but $\Omega$ will change less abruptly. Figure \ref{fig:effres-2} shows an example where I added an edge between nodes $1$ and $6$ to the graph in Figure \ref{fig:effres-1}. The shortest path distance reduces by a factor of three by adding a single edge, while effective resistance changes from $1.5$ to $0.6$ -- a factor of $2.5$.

The difference here is small because this is a tiny toy didactic example, but it can get quite significant. For instance, via a simulation, for a graph with $|V| = 1,000$ and $|E| = 3,000$ I could find nodes connected by a shortest path of length $7$. When connecting them, the shortest path decreases by a factor of $7$, but their effective resistance went down by a factor of less than $3$ (from $1.9$ to $0.65$). This is important, because network data is noisy and contains errors (Chapter \ref{cha:uncertainty}) and you don't want a small chance fluctuation to dramatically change your results.

\section{Mincut Problem}\label{sec:rw-mincut}
I already said in Section \ref{sec:paths-ccomps} that the smallest eigenvector of $L$ isn't so special after all. It plays the very same role as the largest eigenvector of $A$: it is a vector of constant, telling us to which component the node belongs. The reason why $L$ is interesting lies with the second smallest eigenvector (or the eigenvector associated to the smallest non-zero eigenvalue, if the graph has multiple components).

\begin{figure}
\centering
\begin{subfigure}[t]{.55\columnwidth}
\includegraphics[width=\textwidth]{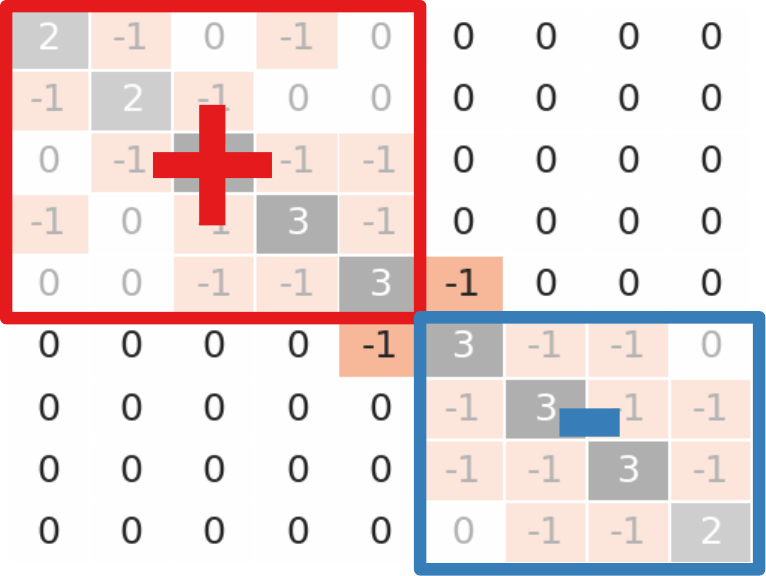}
\caption{}
\label{fig:laplacian-2nd-matrix}
\end{subfigure}
\quad
\begin{subfigure}[t]{.094\columnwidth}
\includegraphics[width=\textwidth]{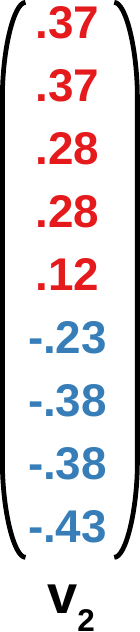}
\caption{}
\label{fig:laplacian-2nd-vector}
\end{subfigure}
\quad
\begin{subfigure}[t]{.204\columnwidth}
\includegraphics[width=\textwidth]{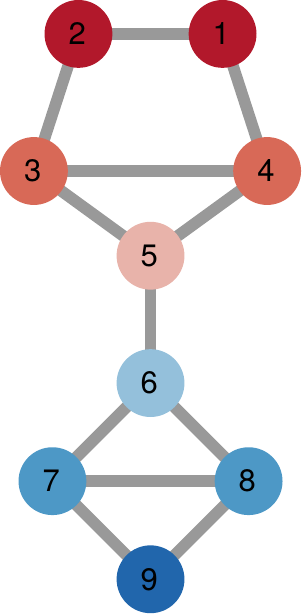}
\caption{}
\label{fig:laplacian-2nd-graph}
\end{subfigure}
\caption{(a) The Laplacian matrix of a graph. I show the 2-cut solution for this graph with the red and blue blocks. (b) The second smallest eigenvector of (a). (c) The graph view of (a) -- I color the nodes according to the value attached to them in the (b) vector.}
\label{fig:laplacian-2nd}
\end{figure}

One classical problem in graph theory is to find the minimum (or normalized) cut of a graph: how to divide nodes in two disjoint groups such that the number of edges running across groups is minimized. Turns out that the second smallest eigenvector of the Laplacian is a very good approximation to solve this problem\cite{fiedler1989laplacian}. How? Consider Figure \ref{fig:laplacian-2nd}. In Figure \ref{fig:laplacian-2nd}(a) I show the Laplacian matrix of a graph. I arranged the rows and columns of the matrix so that the 2-cut solution is evident: by dividing the matrix in two diagonal blocks there is only one edge outside our block structure that needs to be cut.

Now, why did I label the two blocks as ``+'' and ``-''? The reason lies in the magical second smallest eigenvector of the Laplacian -- also known as the Fiedler vector --, which is in Figure \ref{fig:laplacian-2nd}(b). We can see that the top entries are all positive (in red) and the bottom are all negative (in blue). This is where $L$ shines: by looking at the sign of the value of a node in its second smallest eigenvector we know in which group the node has to be to solve the 2-cut problem!

Not only that, but the values in Figure \ref{fig:laplacian-2nd}(b) are clearly in descending order. If we look at the graph itself -- in Figure  \ref{fig:laplacian-2nd}(c) -- and we use these values as node colors, we discover that there is much more information than that in the eigenvector. The absolute value tells us how embedded the node is in the group, or how far from the cut it is. Node $5$ is right next to it, while node $9$ is the farthest away.

\begin{figure*}
\centering
\begin{subfigure}[t]{.4\columnwidth}
\includegraphics[width=\textwidth]{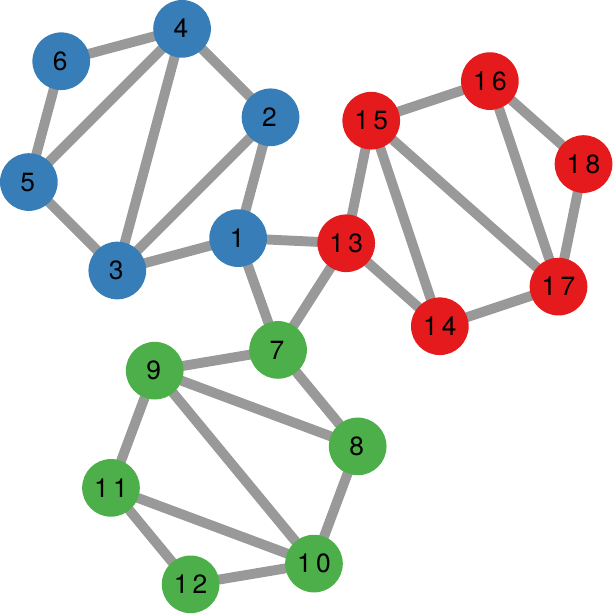}
\caption{}
\end{subfigure}
\begin{subfigure}[t]{.59\columnwidth}
\includegraphics[width=\textwidth]{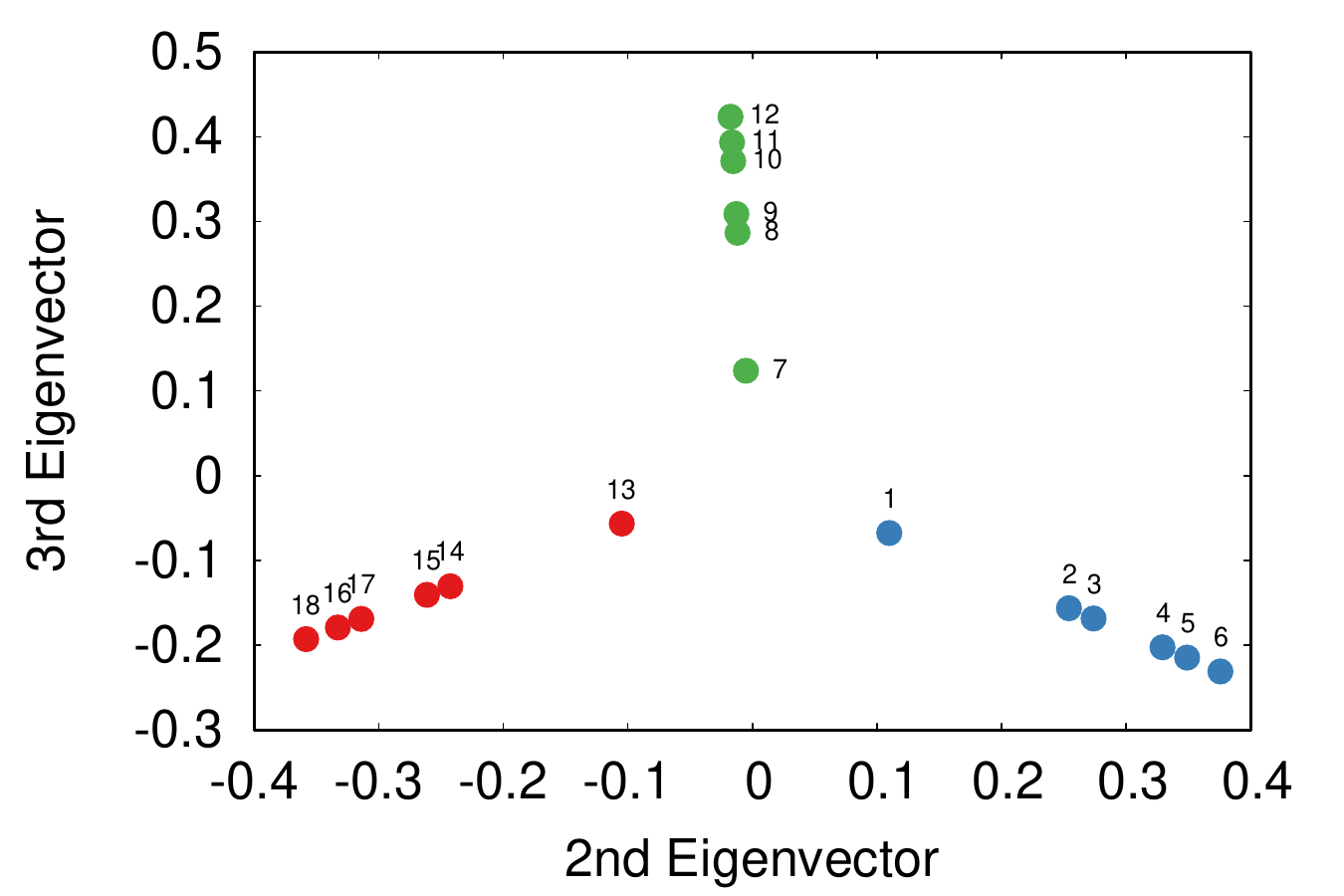}
\caption{}
\end{subfigure}
\caption{(a) A graph in which the node colors represent the best solution to the 3-cut problem. (b) The eigenspace of the Laplacian of the (a) graph. I plot the second smallest eigenvector in the x-axis, and the third smallest eigenvector on the y-axis. Data point color corresponds to the node color in (a). I label each data point with its corresponding node ID.}
\label{fig:laplacian-3cut}
\end{figure*}

\textit{Now} -- you're thinking -- \textit{you're itching to tell me you can use this eigenvector to solve all the $k$-cut problems, for any $k$ larger than two. But you can't, because the Fiedler vector is just a simple monodimensional vector}. Or can I? True: the second smallest eigenvector cannot solve, \textit{by itself}, the arbitrary $k$-cut problem, finding the minimum cuts to divide the graph in $k$ parts. That's why we have \textit{all the other eigenvectors} of the Laplacian.

Solving the 3-cut problem involves looking at the eigenvectors in a two dimensional space. I show an example of this in Figure \ref{fig:laplacian-3cut}. Figure \ref{fig:laplacian-3cut}(b) is a 2D representation of the second and third smallest eigenvectors of the Laplacian of the graph in Figure \ref{fig:laplacian-3cut}(a). We can see that there is a clear pattern: each node takes a position in this space on a different axis, depending on the block to which it belongs. Farther nodes on the axis are more embedded in the block, while nodes closer to the cuts are nearby the origin $(0,0)$. You can imagine that we could solve the 4-cut problem looking at a 3D space, and the $k$-cut problem looking at a $(k - 1)$D space. I can't show it right now because, although it's truly remarkable, the margin of my M\"{o}bius paper is too small to contain it.

Of course, at the practical level, real world networks are not amenable to these simple solutions. Most of the times, the best way to solve the 2-cut problem is to put in one group a node with degree equal to one and put all other nodes of the network in the other group. If you want to find non-trivial $k$-cuts of the network that are meaningful for humans... well... you have to do community discovery (and jump to Part \ref{par:cd}).

\section{Random Walks and Consensus}\label{sec:rw-consensus}
One thing that might be left in your head after reading the previous section is: why? Why do the eigenvectors of the Laplacian help with the mincut problem? What's the mechanism? To sketch an answer for this question we need to look at what we call ``consensus dynamics''. This is a subclass of studying the diffusion of something (a disease, word-of-mouth, etc) on a network -- which we'll see more in depth in Part \ref{par:sis}. This section is sketched from a paper\cite{schaub2019structured} that you should read to have a more in-depth explanation of the dynamics at hand. Consensus dynamics were originally modeled this way by DeGroot\cite{degroot1974reaching}.

In this section I'm going to use the stochastic adjacency matrix of the graph, but what I'm saying also holds for the Laplacian. The difference between the two -- as I also mention in Section \ref{sec:hod-algo} -- is that the stochastic adjacency matrix describes the discrete diffusion over a network. In other words, you have a clock ticking and nothing happens between one tick of the clock and the other. The Laplacian, instead, describes continuous diffusion: time flows without ticks in a clock, and you can always ask yourself what happens between two observations. Besides this difference, the two approaches could be considered equivalent for the level of the explanation in this section.

How does the stochastic adjacency help us in studying consensus over a network? Let's suppose that each node starts with an opinion, which is simply a number between $0$ and $1$. We can describe the status of a network with a vector $x$ of $|V|$ entries, each corresponding to the opinion of each node. One valid operation we could do is multiplying $x$ with $A$, the stochastic adjacency matrix, since $A$ is a $|V| \times |V|$ matrix.

What does this operation mean? Mathematically, from Section \ref{sec:la-matrix}, the result is a vector $x'$ of length $|V|$ defined as $x'_{v} = \sum \limits_{u=1}^{|V|} x_{u}A_{uv}$. In practice, the formula tells you that node $v$ is updating its opinion by averaging the opinion  of its neighbors. Non-neighbors do not contribute anything because $A_{uv} = 0$, and this is an average because we know that the rows of the adjacency matrix sum to $1$ -- thus each $x_{u}$ is weighted equally and $x'_{v}$ will still be between $0$ and $1$.

\begin{figure}
\centering
\includegraphics[width=.8\columnwidth]{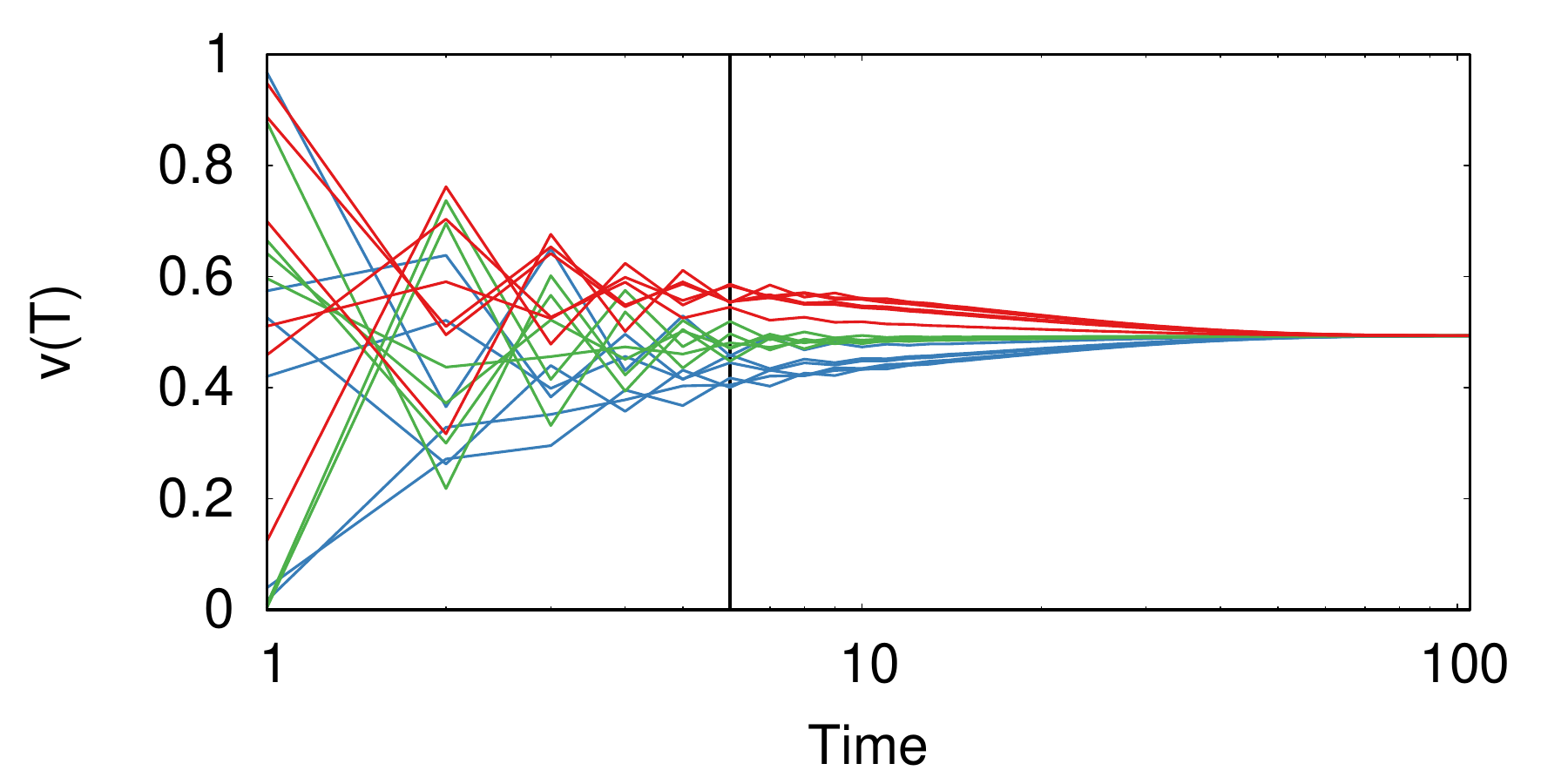}
\caption{The value of $x$ (y-axis) for each node in the graph from Figure \ref{fig:laplacian-3cut} over time (x-axis). At each time step, I update $x$'s values by multiplying them to $A$. The line color tells you to which community the node belongs. The black vertical line highlights step $6$.}
\label{fig:consensus}
\end{figure}

We can reach a consensus by multiplying $x$ with $A$ an infinite number of times. If you do that, $x$ will converge to a vector of constant -- which is the average of the initial $x$, in this case $0.5$ since values were extracted uniformly at random between $0$ and $1$. Figure \ref{fig:consensus} shows an example process on the network from Figure \ref{fig:laplacian-3cut}. Each node starts with a uniform random value and the line tracks this value over time.

Now the connection with eigenvectors: from Section \ref{sec:rw-stationary} you remember that the vector of constant is the leading eigenvector of $A$. This operation showed you why: if a network has separate connected components, it cannot reach a unique consensus; every connected component will reach its own consensus independently because there's no exchange of information across components.

The second eigenvector, instead, tells you \textit{how quickly} the nodes will reach the consensus. In the figure, the line color tells you the community of the node. You might notice that nodes bundle up with their community mates before reaching the network's final consensus. This is described by the inverse of the second eigenvalue of the Laplacian. For that network, it is $1 / \lambda_2 \sim 5.13$. This tells you that the nodes are expected to converge to their community's opinion between step $5$ and $6$, which is the step highlighted in Figure \ref{fig:consensus}. You might notice that one blue and one red node don't seem to converge to their community, but that's because they are nodes $1$ and $13$ and, as you can see from Figure \ref{fig:laplacian-3cut}, they are in between communities, i.e. they are close to the cut.

So the reason why the Fiedler vector allows you to solve the mincut is because it tells you how much it will take for the node to converge to the consensus. Nodes farther from the cut will take longer time. Moreover, you can use the sign to know on which side of the cut you are, because the nodes will first tend to converge to the value of their own community, which is the opposite of the value in the other community.

\section{Summary}

\begin{enumerate}
\item Raising the stochastic adjacency matrix to high powers will make it converge to the stationary distribution, which is the probability of ending in a node in the network after an infinite length random walk. This is also the leading left eigenvector, or the normalized degree.
\item A non-backtracking matrix is a matrix describing a non-backtracking random walk, with a row/column per edge telling you whether you can move from one edge to another. It has zero on its main diagonal.
\item The hitting time is the number of expected steps you need to take in a random walk to reach one node starting from another. It is related to a special eigenvector decomposition of the adjacency matrix.
\item The commute time of $u$ and $v$ is the hitting time from $u$ to $v$ plus the hitting time from $v$ to $u$. It is related to the effective resistance, which can be calculated by inverting the Laplacian. The effective resistance is a proper metric on a graph, less affected by random edge fluctuations than the shortest path distance.
\item The normalized cut problem aims to find the way to partition the network in $n$ balanced parts such that we ``cut'' the minimum number of edges (the ones flowing from one group to another). You can approximate the solution by looking at the $n - 1$ smallest eigenvectors of the Laplacian (skipping the first).
\item This is because the eigenvectors of the Laplacian tell you when a node will reach a consensus, and to which intermediate value it will converge before doing so. Nodes on the same side of a cut will converge to the same intermediate value.
\end{enumerate}

\section{Exercises}

\begin{enumerate}
\item Calculate the stationary distribution of the network at \url{http://www.networkatlas.eu/exercises/11/1/data.txt} in three ways: by raising the stochastic adjacency to a high power, by looking at the leading left eigenvector, and by normalizing the degree. Verify that they are all equivalent.
\item Calculate the non-backtracking matrix of the network used for the previous question. (The network is undirected)
\item Calculate the hitting time matrix of the network at \url{http://www.networkatlas.eu/exercises/11/3/data.txt}. Note: for various reasons, a naive implementation in python using numpy and scipy might lead to the wrong result. I would advise to try and do this in Octave (or Matlab).
\item Calculate the effective resistance matrix of the network at \url{http://www.networkatlas.eu/exercises/11/3/data.txt} and prove it is equal to the commute time divided by $2|E|$. Note: differently from above, the effective resistance matrix can be calculated in python without an issue. But the second part of the exercise might fail if not done in Octave (or Matlab).
\item Draw the spectral plot of the network at \url{http://www.networkatlas.eu/exercises/11/5/data.txt}, showing the relationship between the second and third eigenvectors of its Laplacian. Can you find clusters?
\end{enumerate}

\chapter{Density}

\section{Density \& Real Networks}\label{sec:density-sparse}
\begin{figure}
\centering
\begin{subfigure}[t]{.35\columnwidth}
\includegraphics[width=\textwidth]{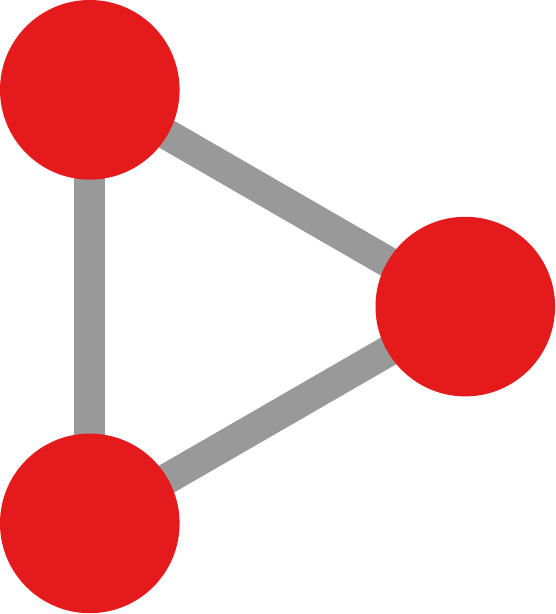}
\caption{}
\end{subfigure}
\qquad
\begin{subfigure}[t]{.55\columnwidth}
\includegraphics[width=\textwidth]{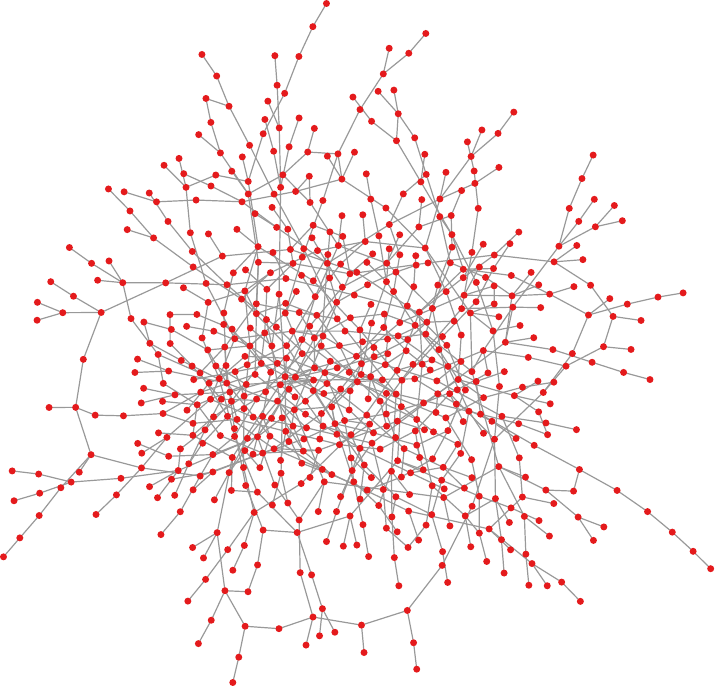}
\caption{}
\end{subfigure}
\caption{(a) A network with three nodes and average degree equal to two. (b) A network with around $650$ nodes and average degree equal to two.}
\label{fig:avgdegree}
\end{figure}

In Chapter \ref{cha:degree}, we saw that there is a quick way to get a sense of how well connected the nodes of your network are. You can calculate their average degree, that is to say the average number of edges each node uses to connect to its neighbors. However, in the same chapter, we also saw that the degree distribution is usually extremely broad, which makes the average degree an incomplete information. Moreover, depending on the size of your network, the same value of average degree could mean different things. A network with three nodes and average degree equal to two is fully connected. A network with the same average degree, but ten thousand nodes is quite sparse. See Figure \ref{fig:avgdegree} for a graphical example.

To quantify the difference between the two cases, network scientists defined the concept of network density. Informally, this is the probability that a random node pair is connected. Or, the number of edges in a network over the total possible number of edges that can exist given the number of nodes. We can estimate this latter quantity quite easily -- from now on, I'll just assume the network is undirected, unipartite and monolayer, for simplicity.

Here's the problem, rephrased as a question: how many edges do we need to connect $|V|$ nodes? Well, let's start by connecting one node, $v$, to every other node. We will need $|V| - 1$ edges -- we're banning self loops. Now we take a second node, $u$. We need to connect it to the other nodes in $V$ minus itself -- seriously, no self loops! -- and $v$, because we already added the $(u,v)$ edge at the previous step. So we add $|V| - 2$ edges. If you go on and perform the sum for all nodes in $v$, you'll obtain that the number of possible edges connecting $|V|$ nodes is $|V|(|V| - 1)/2$. In other words, you need $|V| - 1$ edges to connect a node in $V$ with all the other nodes, and you divide by two because each edge connects two nodes. In fact, the number of possible edges in a directed network is simply $|V|(|V| - 1)$, because you need an edge for each direction, as $(u,v) \neq (v,u)$.

Now we can tell the difference between the networks in Figures \ref{fig:avgdegree}(a) and \ref{fig:avgdegree}(b). The first one has three nodes. We would expect $3*2/2=3$ edges, and that's exactly what we have. Its density is the maximum possible: $100\%$. The network on the right, instead, contains $650$ nodes. Since the average degree of the network is also two, we know it also contains $650$ edges. This is a far cry from the $650*649/2=210,925$ we'd require. Its density is just $0.31\%$, more than three hundred times lower than the example on the left!

With all this talk about real world networks having low degree, we should expect them to be quite sparse. They are, in fact, even sparser than you think. A few examples -- the numbers are a bit dated, they refer to the moment in which these networks were studied in a paper\cite{barabasi2016network}.

The network connecting the routers forming the backbone of the Internet? It contains $|V| = 192,244$ nodes. So the possible number of edges is $|V|(|V| - 1)/2 = 18,478,781,646$. How many does it have, really? $609,066$, which is just $0.003\%$ of the maximum. How about the power grid? The classical studied structure has $4,941$ nodes, which means -- potentially -- $12,204,270$ edges. Yet, it only contains $6,594$ of them, or $0.05\%$. Well, compared to the Internet that's quite the density! Final example: scientific paper citations. A dataset from arXiV contains $449,673$ papers. The theoretical maximum number of citations is $101,102,678,628$. Physicists, however, are quite stingy: they only made $4,689,479$ citations, or $0.004\%$ of the theoretical maximum.

\begin{figure}
\centering
\includegraphics[width=.9\columnwidth]{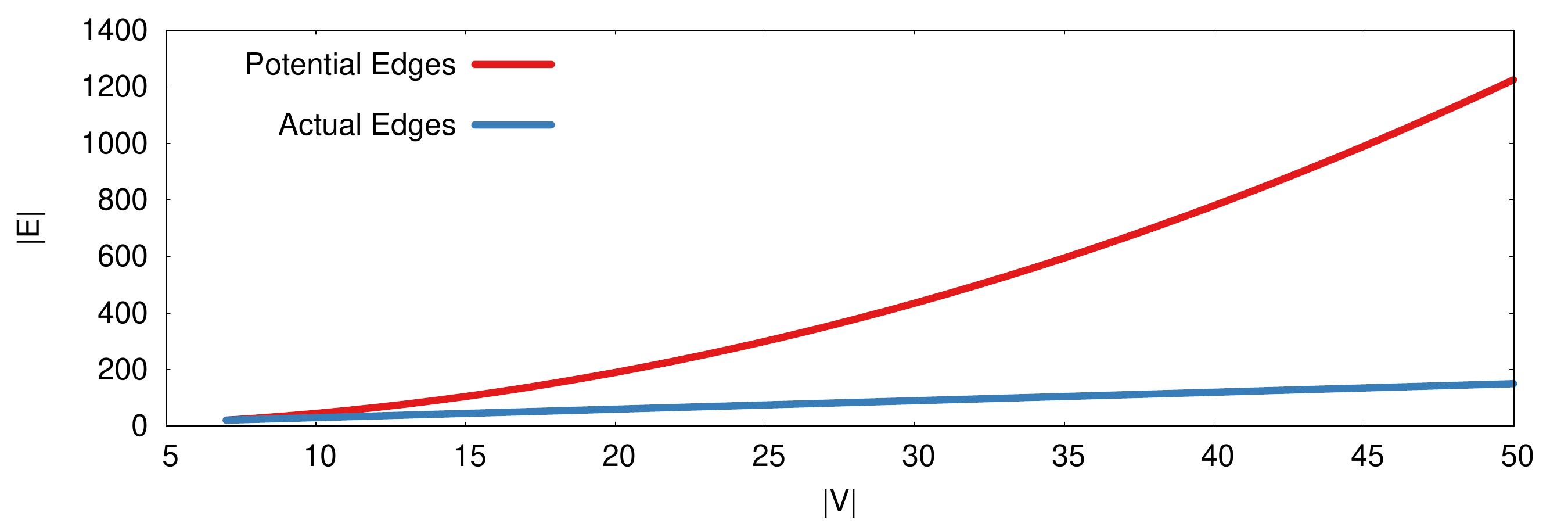}
\caption{The red line shows the number of possible edges ($|V|(|V| - 1)/2$) in a network with $|V|$ nodes (x axis). The blue line shows the number of actual edges, assuming the average degree being $\bar{k} = 3$.}
\label{fig:density-growth}
\end{figure}

You might have spotted a pattern there. The density of a network seems to go down as you increase the number of nodes. While not an ironclad rule, you might be onto something. The problem is that the denominator of the density formula grows quadratically. Each $v$ you add to $V$ allows $|V|(|V| -1)$ to grow pretty rapidly. Figure \ref{fig:density-growth} shows that as a red line. The numerator, instead, grows practically \textit{linearly} -- the blue line in Figure \ref{fig:density-growth}. Each added node will bring only few edges -- three in Figure \ref{fig:density-growth} --, as we know that the average degree of real world networks is low.

\section{Clustering Coefficient}\label{sec:density-clustering}

\subsection{Global Clustering}
Density doesn't solve all ambiguities you had in the case of the average degree. Two networks can have the same density and the same number of nodes, but end up looking quite different from each other. That is why the ever industrious network scientists created yet another measure to distinguish between different cases: the clustering coefficient.

\begin{figure}
\centering
\begin{subfigure}[t]{.55\columnwidth}
\includegraphics[width=\textwidth]{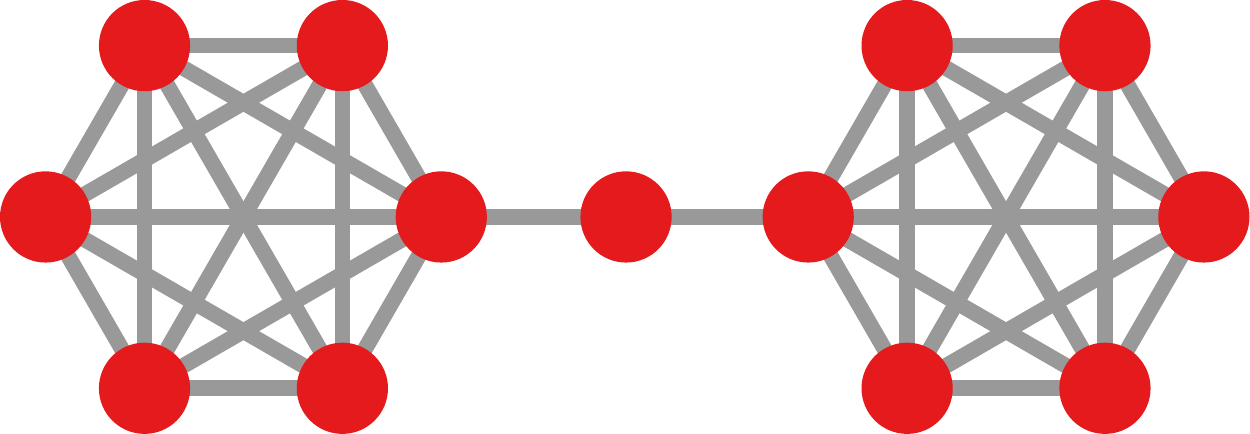}
\caption{}
\end{subfigure}
\qquad
\begin{subfigure}[t]{.3\columnwidth}
\includegraphics[width=\textwidth]{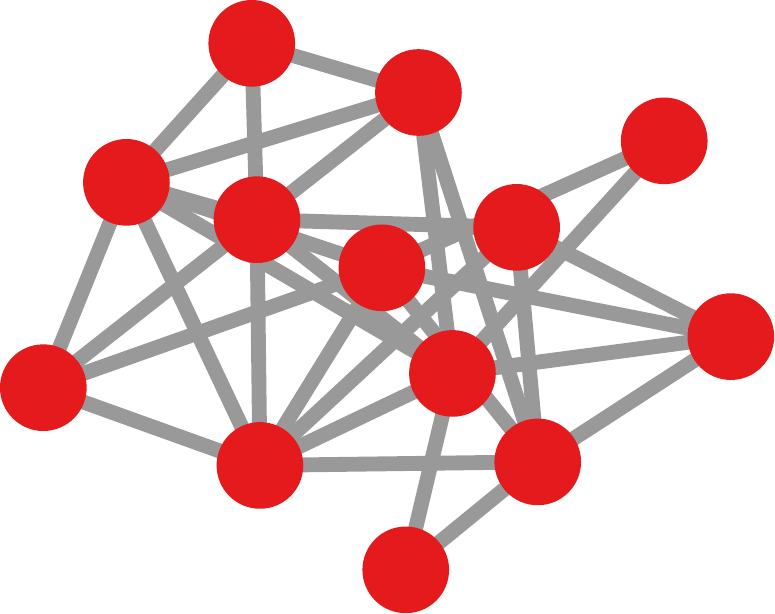}
\caption{}
\end{subfigure}
\caption{Both networks have $13$ nodes and $32$ edges. However, their topologies are different: in the example to the left the edges are more ``clustered'' together.}
\label{fig:clust-coef}
\end{figure}

Consider Figure \ref{fig:clust-coef}: it contains two networks with the same number of nodes and the same number of edges -- thus the same density. However, they look quite different. There's more a sense of ``order'' in Figure \ref{fig:clust-coef}(a). That is because, in Figure \ref{fig:clust-coef}(a), the edges are more ``clustered'' together. That is what the clustering coefficient aims at estimating quantitatively.

\begin{figure}
\centering
\begin{subfigure}[t]{.33\columnwidth}
\includegraphics[width=\textwidth]{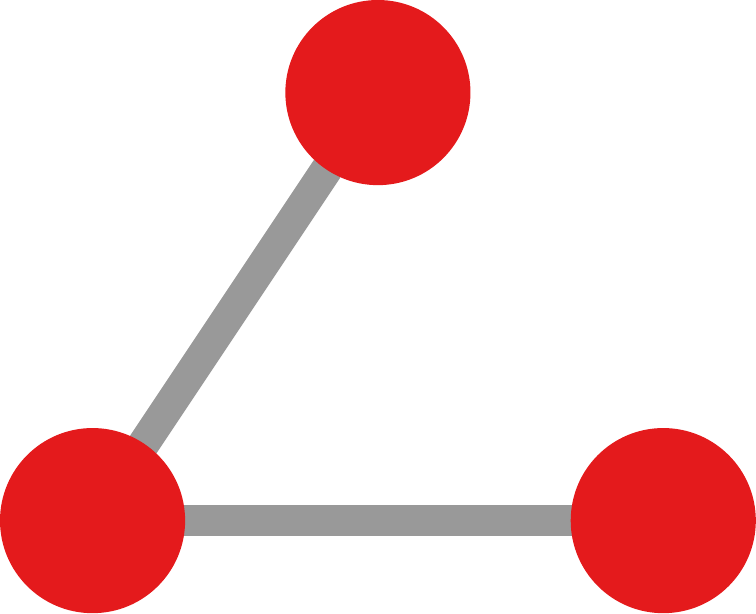}
\caption{Triad}
\end{subfigure}
\qquad
\begin{subfigure}[t]{.25\columnwidth}
\includegraphics[width=\textwidth]{figures/triangle.pdf}
\caption{Triangle}
\end{subfigure}
\caption{The two possible connected network patterns involving three nodes.}
\label{fig:clust-coef2}
\end{figure}

I can sum up the intuition behind the clustering coefficient as the old adage ``The friend of my friend is my friend''. If you have two friends it's overwhelmingly likely that they know each other, because they have something in common: you. You might invite them -- even by accident -- to the same event. This sort of dynamics in network is called ``triadic closure''. A set of three connected nodes is a triad -- Figure \ref{fig:clust-coef2}(a). If one member of the triad is connected to the other two, more often than not the triad will ``close'', meaning that the other two nodes will connect to each other to form a triangle. ``Triangle'' is the name we give to a specific network pattern: three nodes that are all connected to each other, and I show one in Figure \ref{fig:clust-coef2}(b).

A note for computer scientists: do not confuse the clustering coefficient with the operation you know as ``clustering'': the process of grouping similar rows of a matrix. The latter is a process that in network science is called community detection -- or discovery -- and will be the subject of Part \ref{par:cd}. The clustering coefficient is simply a number you return that describes quantitatively how ``clustered'' a network looks. A way to dispel the confusion is to use the term ``transitivity''. This takes inspiration from the transitive property: if $u$ is connected to $v$ and $v$ is connected to $z$, then $u$ is connected to $z$ too.

To sum up, the clustering coefficient answers the question: how often does a triad close down into a triangle? What's the likelihood that the ``friend of my friend is my friend'' rule holds in the network? To answer this question we have to count the number of triads and the number of triangles in the network. Then the global clustering coefficient (CC) is simply\cite{holland1971transitivity}: $CC = 3 \times \#Triangles / \#Triads$. Why do we have to multiply the number of triangles by three?

\begin{figure}
\centering
\begin{subfigure}[t]{.45\columnwidth}
\includegraphics[width=\textwidth]{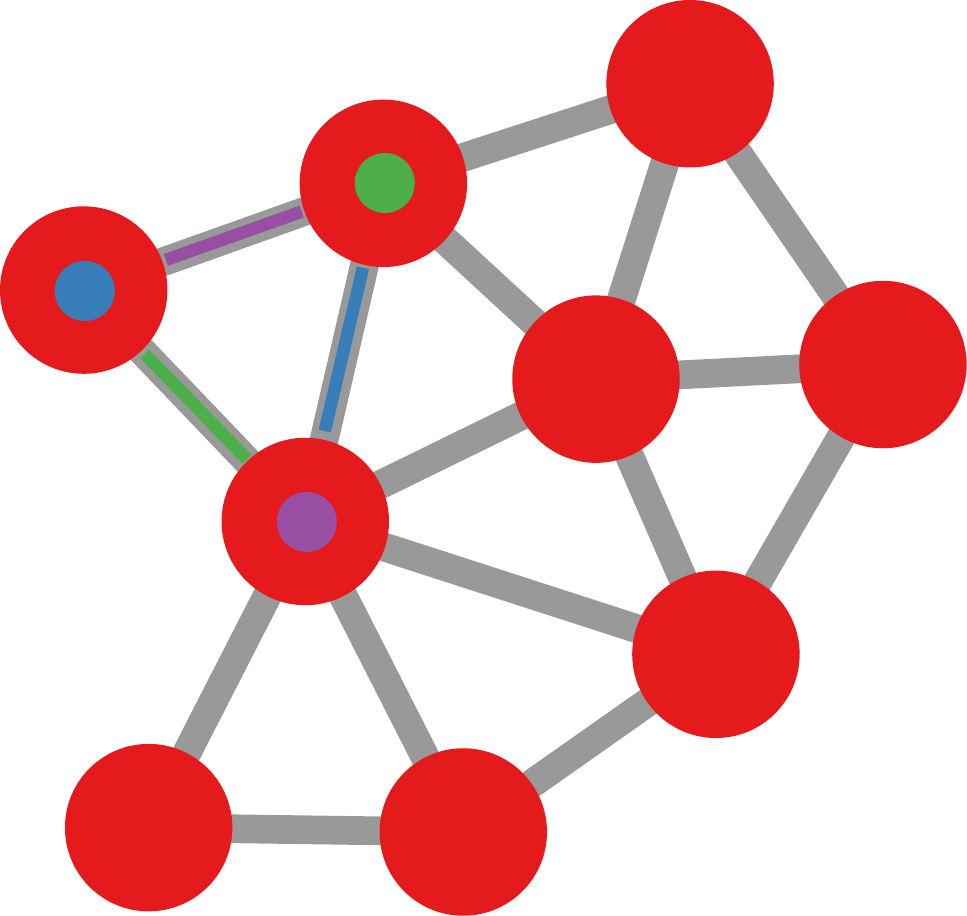}
\caption{}
\end{subfigure}
\qquad
\begin{subfigure}[t]{.45\columnwidth}
\includegraphics[width=\textwidth]{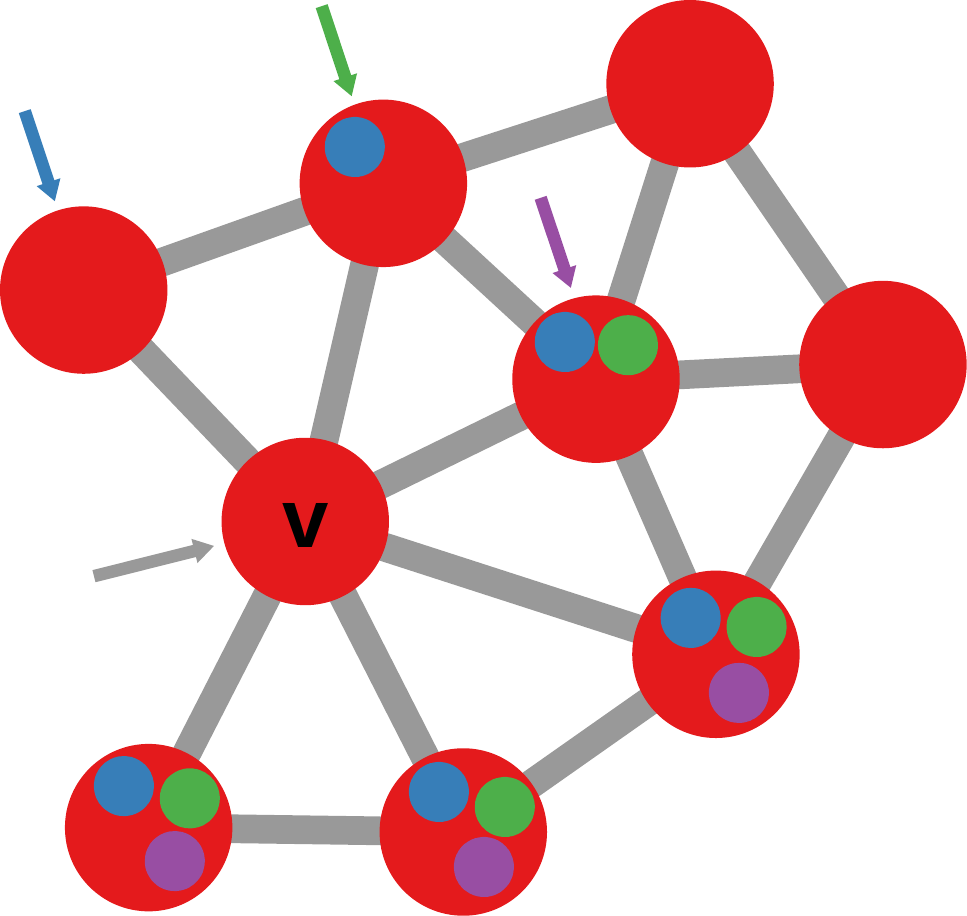}
\caption{}
\end{subfigure}
\caption{(a) Counting the number of triads closed by a triangle. From each node's perspective, a different triad is closed by the same triangle. (b) Counting the number of triads in the network.}
\label{fig:clust-coef3}
\end{figure}

Consider Figure \ref{fig:clust-coef3}(a). I highlighted a triangle in there. From the perspective of the node highlighted in blue. The triad is closed by the edge highlighted in blue. The same holds for the nodes highlighted in green and purple. That single triangle is closing three triads, that is the reason why we multiply the number of triangles in the network by three.

Counting triads is a bit more confusing, but in the end it's going to be easy to remember, because it connects with something you already know. I provide a representation of the process in Figure \ref{fig:clust-coef3}(b). In there, I count the number of triads centered on node $v$, meaning that we only count the triads that have $v$ connected to both of its members. This makes it easier, because we only have to look at $v$'s neighbors. I start by selecting its first neighbor, with the blue arrow. How many triads do $v$ and the blue neighbor generate? Well, one for each of the remaining neighbors of $v$, so I add a blue dot to each of the neighbors. When I move to the neighbor highlighted by the green arrow, I perform the same operation adding the green dot. I don't have to add one to the neighbor with the blue arrow, because I already counted that triad.

Sounds familiar? That is because this is the very same process you apply when you have to count the number of possible edges of a graph. The number of triads centered on node $v$ is nothing more that the number of possible edges among $k_v$ nodes, with $k_v$ being $v$'s degree. So, if we want to know the number of triads in a graph, we simply need to add $k_v(k_v - 1)/2$ for every $v$ in the graph.

Note that, as you might expect, the clustering coefficient takes different values for weighted\cite{onnela2005intensity}\cite{saramaki2007generalizations} and directed\cite{fagiolo2007clustering} graphs.

I primed you to expect that many statistical properties can be derived via matrix operations. This is true also for the clustering coefficient. It is done via the powers of the binary adjacency matrix -- see Section \ref{sec:paths-mat}. Triangles are closed paths of length $3$, while triads are paths of length $2$. The number of closed walks of length $3$ centered on $u$ is $A^3_{uu}$, while the number of walks of length $2$ passing through $u$ is $A^2_{uv}$, with $u \neq v$, which results in the formula:

$$ CC = \dfrac{\sum \limits_{u} A^3_{uu}}{\sum \limits_{u \neq v} A^2_{uv}}. $$

\begin{figure}[t]
\centering
\begin{subfigure}[t]{.45\columnwidth}
\includegraphics[width=\textwidth]{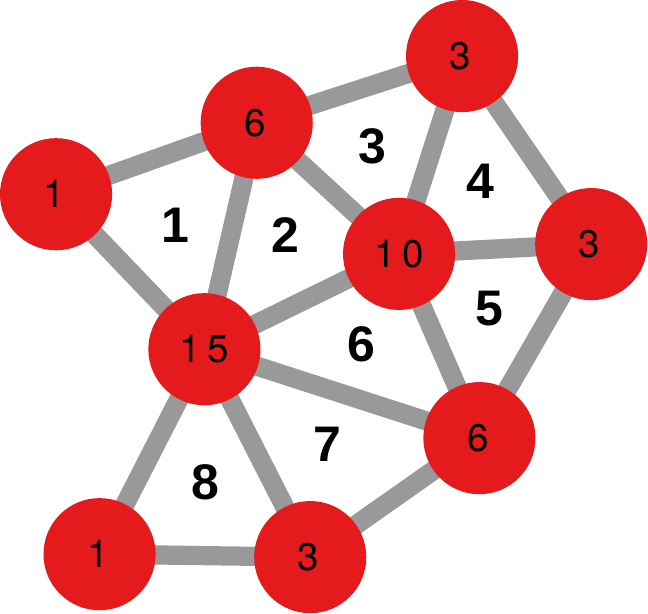}
\caption{}
\end{subfigure}
\qquad
\begin{subfigure}[t]{.45\columnwidth}
\includegraphics[width=\textwidth]{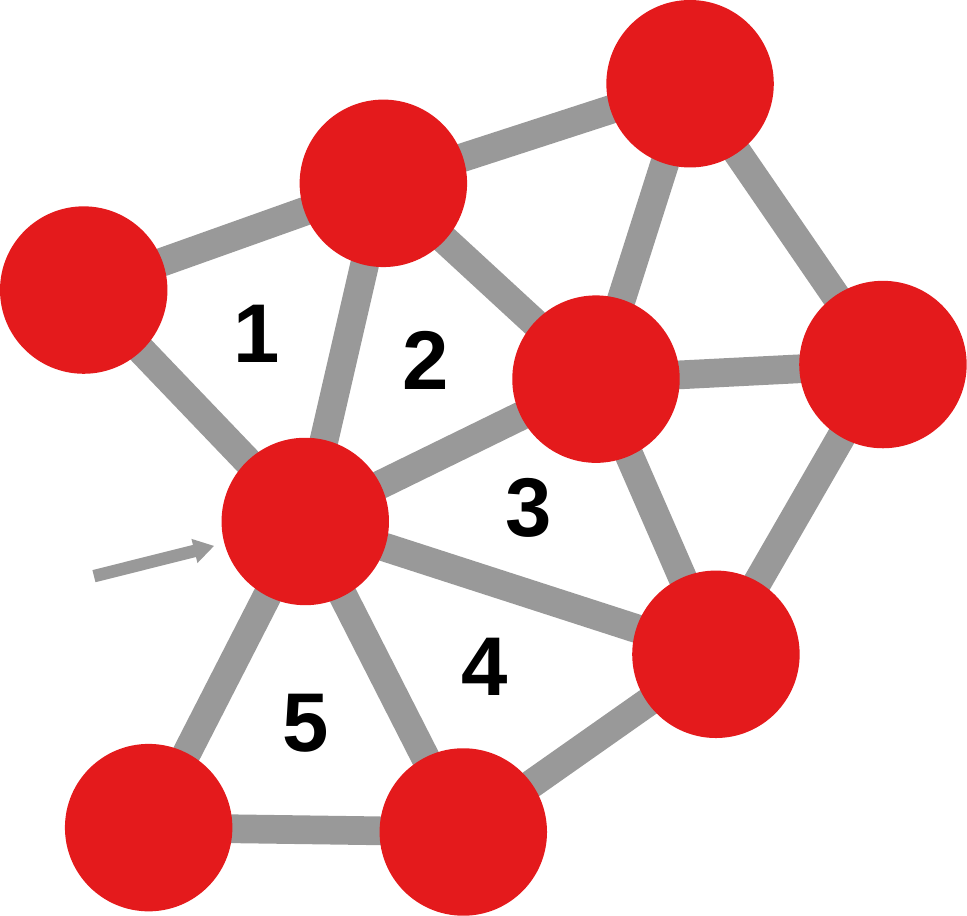}
\caption{}
\end{subfigure}
\caption{(a) Estimating the global clustering coefficient of a graph. Each node is labeled with the number of triads centered on it, and the numbers among the edges count the triangles. (b) Estimating the local clustering coefficient of a node, counting the triangles to which it belongs.}
\label{fig:clust-coef4}
\end{figure}

So, let's calculate the global clustering coefficient for the graph in Figure \ref{fig:clust-coef4}(a). We know how many triads there are in the graph. How many triangles are there? Here I made my life easier, because it is rather trivial to count the number of triangles in a planar graph -- a graph you can draw on a 2D plane without intersecting any edges. There are eight triangles and $48$ triads in the network. Thus the global clustering coefficient of the network is  $CC = 3 \times 8 / 48 = 0.5$. Half of the triads in the network close to form a triangle.

\subsection{Local \& Average Clustering}
The fact that the previous subsection was called ``global'' clustering should tip you off about the existence of a ``local'' clustering coefficient. Its definition is rather similar, but it is focused on a single node. It is the number of triangles to which the node belongs over the number of triads centered on it. We do not multiply by three the numerator, because we're focusing exclusively on the triads of $v$, that can then be closed only in one way.

Looking at Figure \ref{fig:clust-coef4}(b), let's try to calculate the local clustering coefficient of the node highlighted by the arrow. Again, we use our planar graph to have an easy time counting triangles: there are five that include node $v$. We already counted the number of triples before: it was $15$. Thus, the local clustering coefficient of $v$ is $CC_v = 5 / 15 = 0.\bar{3}$.

\begin{figure}
\centering
\includegraphics[width=.45\columnwidth]{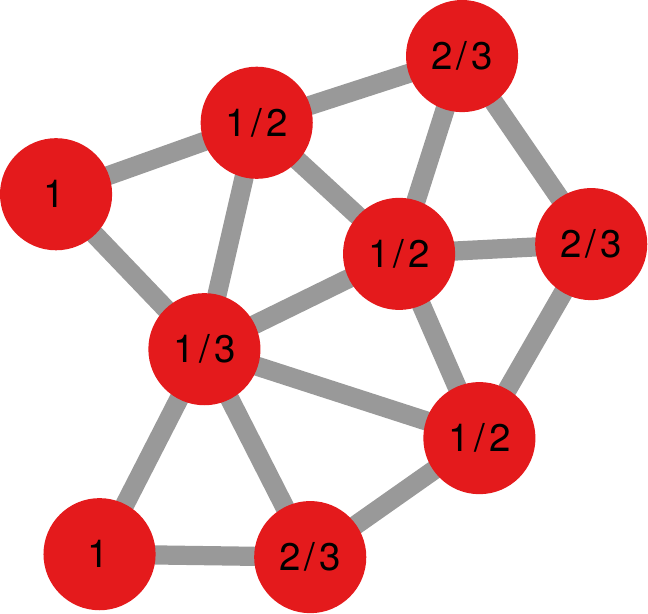}
\caption{I label each node in the network with its local clustering coefficient $CC_v$ value.}
\label{fig:clust-coef5}
\end{figure}

And, hopefully, that's it. Oh, who am I kidding. Of course there's more. Once you have a local clustering coefficient, one might get curious and desire to calculate the average local clustering coefficient of the network. This is simply $CC_{avg} = \dfrac{1}{|V|} \sum \limits_{v \in V} CC_v$. You would hope that $CC_{avg} = CC$. No such luck. For the network in Figure \ref{fig:clust-coef5}, we know that $CC = 0.5$. Unfortunately, if you average the $CC_v$ values I used to label each node, you'll find out that $CC_{avg} = 0.648$. So you have to remember that the global and the average clustering coefficient are two different things, and not to report one as the other.

I closed the previous section with a mantra -- ``real world networks are sparse'' --, so I want to do it again. However, there's a surprise here. If average degrees are low and networks are sparse, wouldn't you expect real world networks to have a low clustering too? Instead, the opposite holds: real world networks are clustered. The power grid example I used before? It has a $CC$ of $0.1032$, which is $150$ times higher than you would expect if its edges were distributed randomly. The scientific paper citations? It has $CC = 0.318$, more than $200$ times higher than expected.

This means that these systems might have few connections per node, but these connections tend to be clustered in the same neighborhood. Nodes tend to close triangles. This is especially true for social systems. In fact, the protein-protein network I used in Chapter \ref{cha:degree} has a clustering of $0.0236$, which is still higher than expected. But in this biological case, we only have a factor of $16$, a far cry from the factor of $200$ in the social paper citation system.

\subsection{Structural Holes \& Redundancy}
\begin{figure}[b]
\centering
\includegraphics[width=.45\columnwidth]{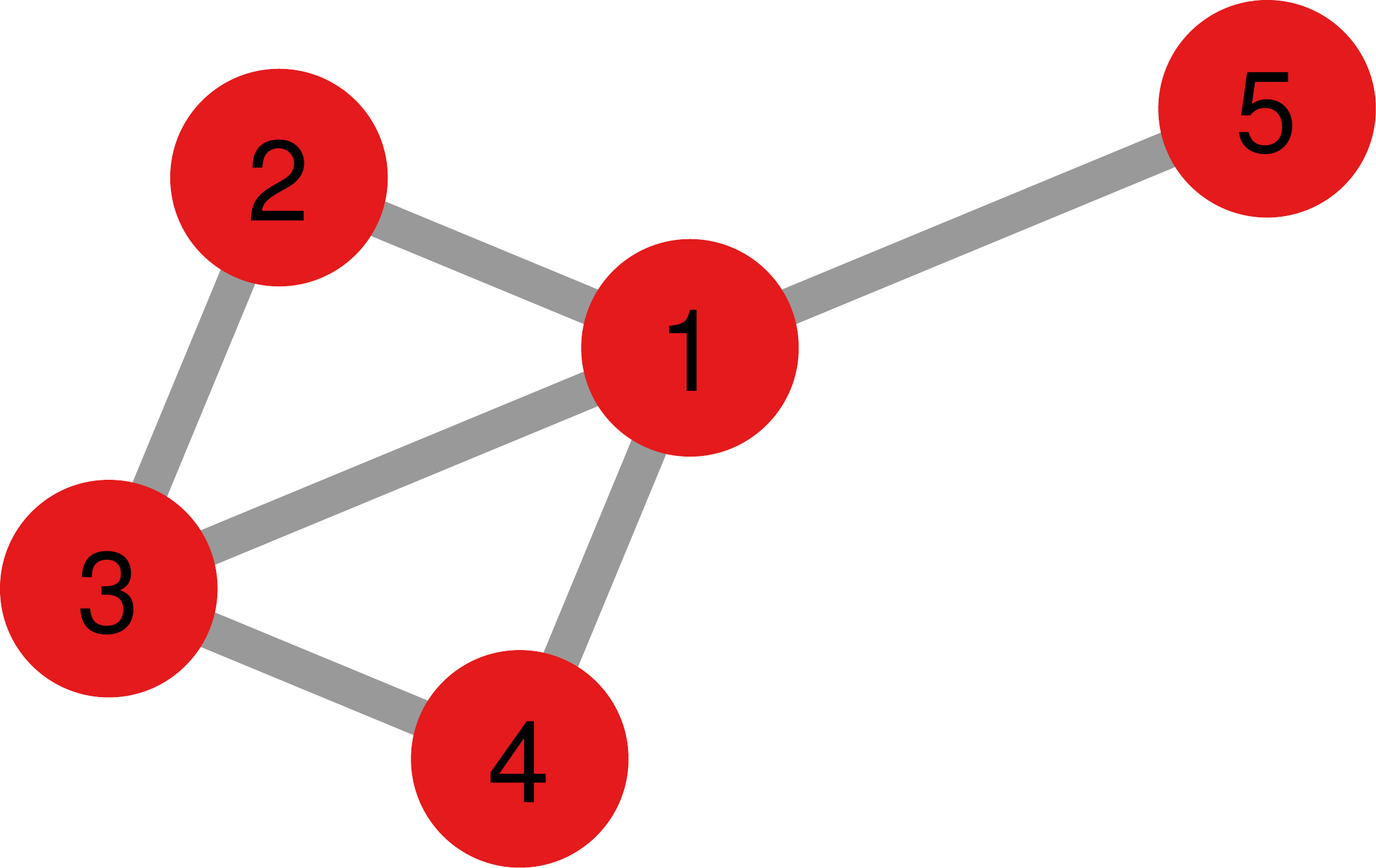}
\caption{A network with several structural holes, for instance between nodes: $2$ and $4$, $2$ and $5$, and $4$ and $5$.}
\label{fig:structural-holes}
\end{figure}

High global or local clustering coefficients in social networks imply that we expect triangles to close. It follows that, when they don't, we consider that weird. Every time there are common neighbors of a node $v$ that do not connect with each other we call that a structural hole\cite{burt1992structural}. Figure \ref{fig:structural-holes} shows a network with several structural holes. In practice, if $CC_v < 1$ then we know there are structural holes around $v$. A high $CC_v$ means there are a lot of ways to get around node $v$, so information can spread freely even if $v$ were to disappear. A low $CC_v$ means that $v$ is fundamental to spread information, since its common neighbors are hardly connected to each other.

Redundancy\cite{borgatti1997structural} is a measure that is relatively similar to the local clustering coefficient and can measure the presence of structural holes around a node. The redundancy of node $v$ is the average number of edges from a neighbor of $v$ to the other neighbors of $v$. Considering again Figure \ref{fig:structural-holes}, one can calculate the redundancy of node $1$. It has $4$ neighbors, so that's the denominator. Nodes $2$ and $4$ are connected with only one of node $1$'s neighbors. Node $3$ is connected with two neighbors of node $1$. Finally, node $5$'s degree with node $1$'s neighbors is zero. Putting everything together: $(1 + 1 + 2 + 0) / 4 = 1$.

De facto, local clustering is a normalized redundancy. If $R_v$ is the redundancy, then $CC_v = R_v / (k_v - 1)$ -- redundancy can be at most $k_v - 1$ because a neighbor of $v$ can only connect to that many other neighbors of $v$, it cannot connect with itself, hence the minus one.

\section{Cliques}\label{sec:density-cliques}
When it comes to clustering and density, you cannot do any better than having all possible edges among the nodes in your network. A network will contain many subsets of nodes for which this is true: you pick $k$ nodes in the network and all possible connections among them are present. This happens often in social networks: these complete graphs -- as we call them -- represent tightly knit groups of friends. They also get a special name: \textbf{cliques}.

\begin{figure}
\centering
\begin{subfigure}[t]{.19\columnwidth}
\includegraphics[width=\textwidth]{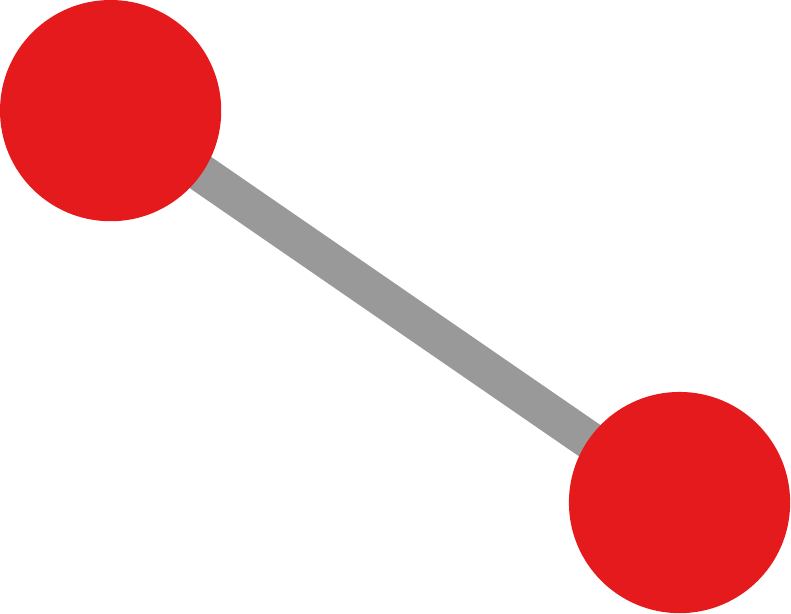}
\caption{2-clique.}
\end{subfigure}
\qquad
\begin{subfigure}[t]{.19\columnwidth}
\includegraphics[width=\textwidth]{figures/triangle.pdf}
\caption{3-clique.}
\end{subfigure}
\qquad
\begin{subfigure}[t]{.19\columnwidth}
\includegraphics[width=\textwidth]{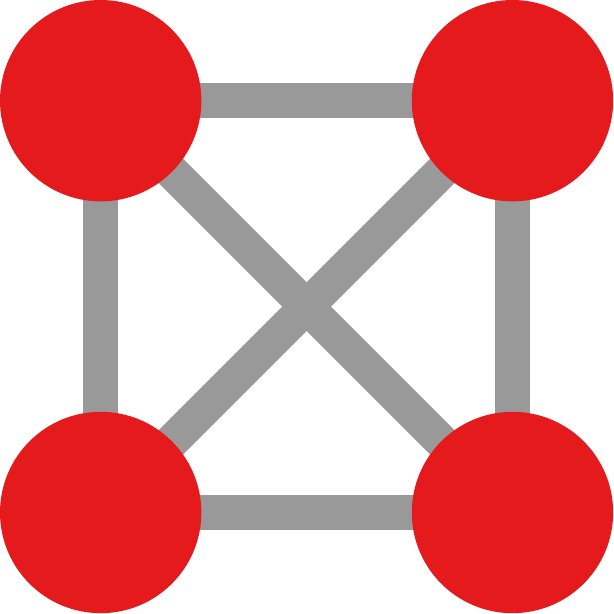}
\caption{4-clique.}
\end{subfigure}
\qquad
\begin{subfigure}[t]{.19\columnwidth}
\includegraphics[width=\textwidth]{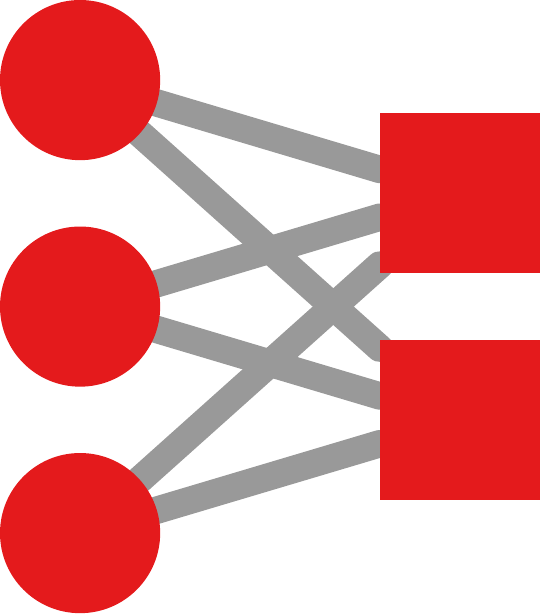}
\caption{3,2-clique.}
\end{subfigure}
\caption{The clique zoo.}
\label{fig:cliques}
\end{figure}

Formally, the definition of a clique is a subgraph of $k$ nodes and $k(k-1)/2$ edges -- assuming we're in the usual case of undirected unipartite networks. The most general way to refer to a clique of size $k$ is by calling it a $k$-clique. So an edge, which is a clique of size two, is a 2-clique (Figure \ref{fig:cliques}(a)). A clique with three nodes is a 3-clique (Figure \ref{fig:cliques}(b)), with four nodes we have a 4-clique (Figure \ref{fig:cliques}(c)) and, hopefully, you can generalize from that. A few cliques get special names too, given to their popularity. We already named the 3-clique in the previous section as a triangle.

The case of bipartite networks is a peculiar one worth mentioning. As we know from Section \ref{sec:extended-bip}, in bipartite graphs we're not allowed to connect nodes of the same type. So, when we define a clique as a subset of nodes where ``all possible connections among them are present'', we mean something radically different in a bipartite network. Here, we simply mean that all nodes of $V_1$ type are connected to all nodes of $V_2$ type. So we call this structure a \textbf{biclique}. Figure \ref{fig:cliques}(d) shows an example of biclique made of three nodes of type $1$ and two nodes of type $2$. We need to modify the way to refer to specific instances of a biclique: from $k$-clique to $n,m$-clique. So the one in Figure \ref{fig:cliques}(d) is a 3,2-clique.

\begin{figure}
\centering
\includegraphics[width=.36\columnwidth]{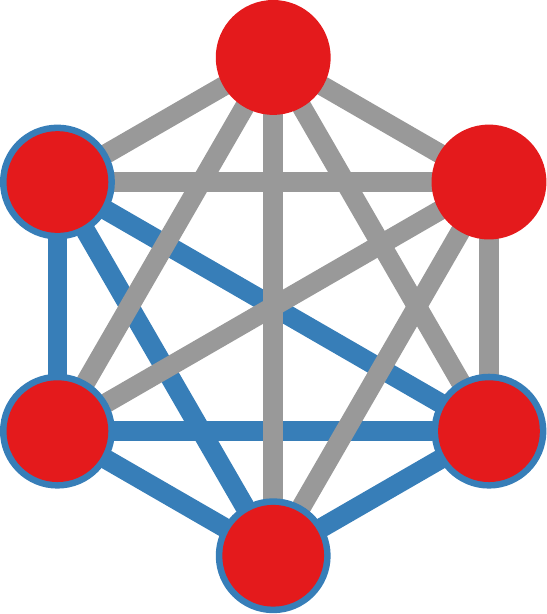}
\caption{An example of maximal 6-clique containing a non-maximal 4-clique, highlighted with the blue outline.}
\label{fig:cliques-maximal}
\end{figure}

It follows from the definition of a clique that any $k$-clique will contain many $(k-1)$-cliques, down to 2-cliques. If you have all possible edges between six nodes, you can pick any five (four, three, ...) of those six nodes and they're all connected to each other -- by definition. Figure \ref{fig:cliques-maximal} provides a depiction of this reasoning. We want to distinguish between cliques that are contained in other cliques, and cliques that aren't. We call the latter \textbf{maximal cliques}: the set of nodes to which you cannot add any other node and still make it a clique.

\section{Independent Sets}\label{sec:density-indsets}
\begin{figure}[b]
\centering
\includegraphics[width=.3\columnwidth]{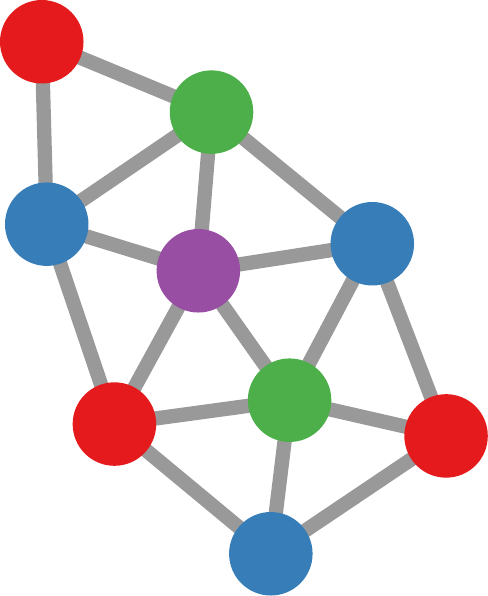}
\caption{A graph with several independent sets, represented by the color of the nodes.}
\label{fig:indset}
\end{figure}

Just like matter has anti-matter, also cliques have anti-cliques. By definition, a clique is a set of nodes all connected to each other. The anti-clique, which we call ``independent set'' is a set of nodes none of which are connected to each other\cite{godsil2013algebraic}. Figure \ref{fig:indset} shows a possible subdivision of a graph into independent sets.

Note that, in Figure \ref{fig:indset}, I force each node to have a color, i.e. to belong to at least an independent set. I could find a larger independent set, for instance the purple node could make an independent set with the two red nodes it isn't connected to -- since they are not connected to each other. The task of finding an independent set coverage of a graph is called graph coloring and it's a classical graph theory problem\cite{jensen2011graph}. Solving graph coloring tells you, for instance, how many colors you need to use for your map such that no two neighboring countries share the same hue -- in this case you represent countries as nodes and connect two countries if they share a land border. 

When it comes to independent sets, you should not confuse the \textit{maximal} independent set with the \textit{maximum} independent set. A maximal independent set, just like a maximal clique, is an independent set to which you cannot add any node and still make it an independent set. The green set in Figure \ref{fig:indset} is maximal because the two green nodes are connected to all other nodes in the graph.

On the other hand, the maximum independent set is simply the largest possible independent set you can have in your network. In Figure \ref{fig:indset}, the red set is the maximum independent set, or at least one of the many possible independent sets of size $3$. Finding the largest possible independent set is an interesting problem, because it tells you something about the connectivity of the graph. It also has applications in graph mining -- see Section \ref{sec:mining-single}.

\section{Summary}

\begin{enumerate}
\item We define a network's density as the number of its edges divided by the total possible amount of edges, which is a different number depending whether your network is directed or not.
\item Real world networks tend to be sparse, meaning that the density is usually lower than a few percentage points.
\item A way to estimate a local density by looking at the neighborhood of a node is the clustering coefficient: the number of common neighbors around node $u$ connected to each other. There are global, local, and average versions of the clustering coefficient, sometimes known as transitivity.
\item Differently from density, many real world networks have very high clustering.
\item A (sub)graph with density equal to one is a clique: a set of nodes all connected to each other. Bipartite networks have bicliques.
\item The opposite of a clique is an independent set: a group of nodes none of which connects to any other member of the group.
\end{enumerate}

\section{Exercises}

\begin{enumerate}
\item Calculate the density of hypothetical undirected networks with the following statistics: $|V| = 26, |E| = 180$; $|V| = 44, |E| = 221$; $|V| = 8, |E| = 201$. Which of these networks is an impossible topology (unless we allow it to be a multigraph)?
\item Calculate the density of hypothetical directed networks with the following statistics: $|V| = 15, |E| = 380$; $|V| = 77, |E| = 391$; $|V| = 101, |E| = 566$. Which of these networks is an impossible topology (unless we allow it to be a multigraph)?
\item Calculate the global, average and local clustering coefficient for the network in \url{http://www.networkatlas.eu/exercises/12/3/data.txt}.
\item What is the size in number of nodes of the largest maximal clique of the network used in the previous question? Which nodes are part of it?
\item What is the size in number of nodes of the largest independent set of the network used in the previous question? (Approximate answers are acceptable) Which nodes are part of it? 
\end{enumerate}

\part{Centrality}

\chapter{Shortest Paths}\label{cha:shortpath}
The degree (Chapter \ref{cha:degree}) is the most direct measure of importance of a node in a network. The more connections a node has, the more important it is. However, there are alternative ways to estimate the importance of a node. Sometimes, it doesn't matter how many connections you have, but how many people someone can reach by passing through you. Normally, the two are correlated -- more connections mean more possibilities -- but that's not always the case. We explore these differences in this part of the book.

Before we start ranking nodes in Chapter \ref{cha:ranks} we need to lay down some groundwork. A significant chunk of measures of node importance are based on the concept of shortest paths, which is what we're exploring in this chapter. We start by defining how to explore a graph and we build up from there.

\section{Graph Exploration}\label{sec:shortpath-exploration}
When we first encounter a graph, how do we know its topology and properties? Humans can ``see'' and parse simple graphs, but how does a computer do it? If we start from a node, how do we access information about its connections, its neighbors, and the connections among them? We need to perform a graph exploration -- or graph traversal. There are two main ways to do it: Depth First Search (DFS) and Breadth First Search (BFS).

\subsection{Depth \& Breadth First}

\begin{figure*}
\centering
\begin{subfigure}[t]{.45\columnwidth}
\includegraphics[width=\textwidth]{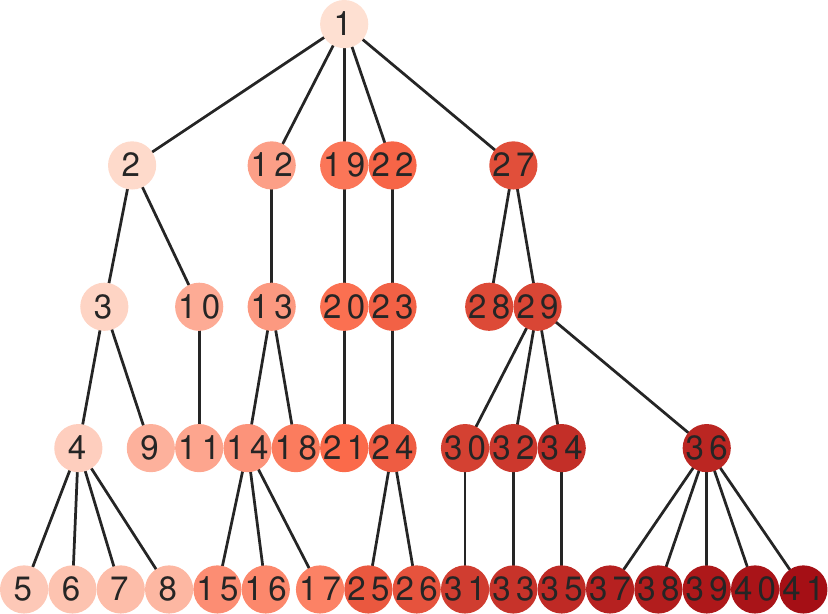}
\caption{DFS}
\end{subfigure}
\qquad
\begin{subfigure}[t]{.45\columnwidth}
\includegraphics[width=\textwidth]{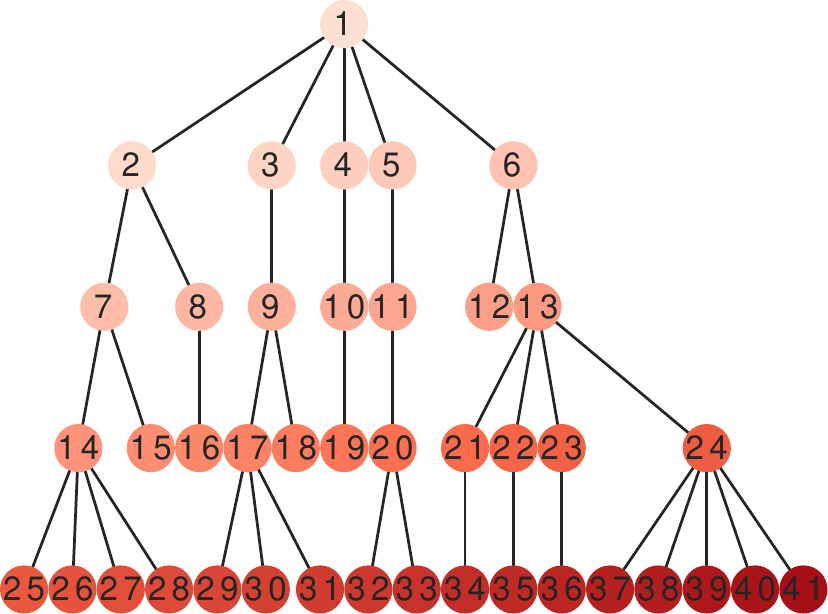}
\caption{BFS}
\end{subfigure}
\caption{Exploring a graph by DFS and BFS. In both cases, I label each node with the order in which it gets explored. I use the node color to encode the same information, from light (early explored) to dark (late explored).}
\label{fig:dfs-bfs}
\end{figure*}

In Depth First Search (DFS), you start by picking a root node. Then you put its neighbors in a Last-In-First-Out queue. You pick the last neighbor you added (you ``pop'' the queue) and you perform the same operation: you add its neighbors to the queue, making sure you don't add nodes you already explored. You continue until you have explored all nodes in the graph. Figure \ref{fig:dfs-bfs}(a) provides an example, showing the exploration order. Since you're using a Last-In-First-Out queue, the very first neighbor of the root node will be the last node to be explored -- unless you encounter it again as the neighbor of some other node down the exploration tree, of course.

Breadth First Search (BFS) is practically speaking the exact same algorithm as DFS, with a tiny change. Rather than putting the neighbors of the root node in a Last-In-First-Out queue, you put them into a First-In-First-Out queue. This changes fundamentally the way you explore the graph: you explore all neighbors of the root node before passing to the first neighbor of the first neighbor of the root node. Figure \ref{fig:dfs-bfs}(b) provides an example, showing the exploration order.

DFS tends to make a gradient over followed paths until it backtracks because it explored the entire neighborhood -- in the example from Figure \ref{fig:dfs-bfs}(a) it backtracks, for instance, from the eighth explored node back to the third explored node. BFS tends to make a gradient from the origin node to the farthest nodes in the network.

\subsection{Random Node/Edge Access}
In day to day computing, you might find yourself exploring the graph in two other ways. These are dependent on the way we store graphs on a computer's hard disk. We can call these two methods random edge access and random node access.

Random edge access is when you read the file containing your graph one line at a time, and each line contains an edge. We call this type of graph storing format an ``edge list'', because it's a list of one edge per line. In this case, you may or may not have sorted the edges in a particular way, but the baseline assumption is that they're in a random order. See Figure \ref{fig:edge-adj-list}(a) for an example.

\begin{figure}
\centering
\begin{subfigure}[t]{.1685\columnwidth}
  \begin{tabular}{l|l}
    Src & Trg \\
    \hline
    $1$ & $2$\\
    $2$ & $4$\\
    $5$ & $3$\\
    $1$ & $3$\\
    $4$ & $5$\\
  \end{tabular}
\caption{Edge list.}
\end{subfigure}
\qquad \qquad
\begin{subfigure}[t]{.33\columnwidth}
  \begin{tabular}{l|l}
    Node & Neighbors \\
    \hline
    $1$ & $2, 3$\\
    $2$ & $1, 4$\\
    $5$ & $3, 4$\\
    $3$ & $1, 5$\\
    $4$ & $2, 5$\\
  \end{tabular}
\caption{Adjacency list.}
\end{subfigure}
\caption{Two different ways to store graphs on disk.}
\label{fig:edge-adj-list}
\end{figure}

Random node access is the same, but the file records, in each line, the complete list of a node's neighbors. We call this type of graph storing format an ``adjacency list'', because it's a list of the adjacencies of one node per line. Also in this case, you may or may not have sorted the nodes -- and their neighbors -- in a particular way, but the baseline assumption is that they're in a random order. See Figure \ref{fig:edge-adj-list}(b) for an example.

\section{Finding Shortest Paths}
Recall that in Chapter \ref{cha:paths} we have defined the length of a walk and of a path as the number of edges one crosses to complete the walk/path. The concept of length is crucial when we want to talk about optimality -- which is to find the shortest (smallest length, fewest edges used) path between nodes $u$ and $v$.

The problem of exploring the graph via BFS or DFS is that they are not optimal, they cannot find the ``best'' (shortest) way to go from $v$ to $u$. Well, they can, but only in very specific cases under very specific assumptions. For instance, BFS finds shortest paths only for undirected unweighted graphs. This can be useful, for instance, to find the shortest path out of a maze\cite{even2011graph}\cite{moore1959shortest}\cite{zuse1972plankalkul}. But we still need a better, more general way.

\begin{figure}[t]
\centering
\includegraphics[width=.66\columnwidth]{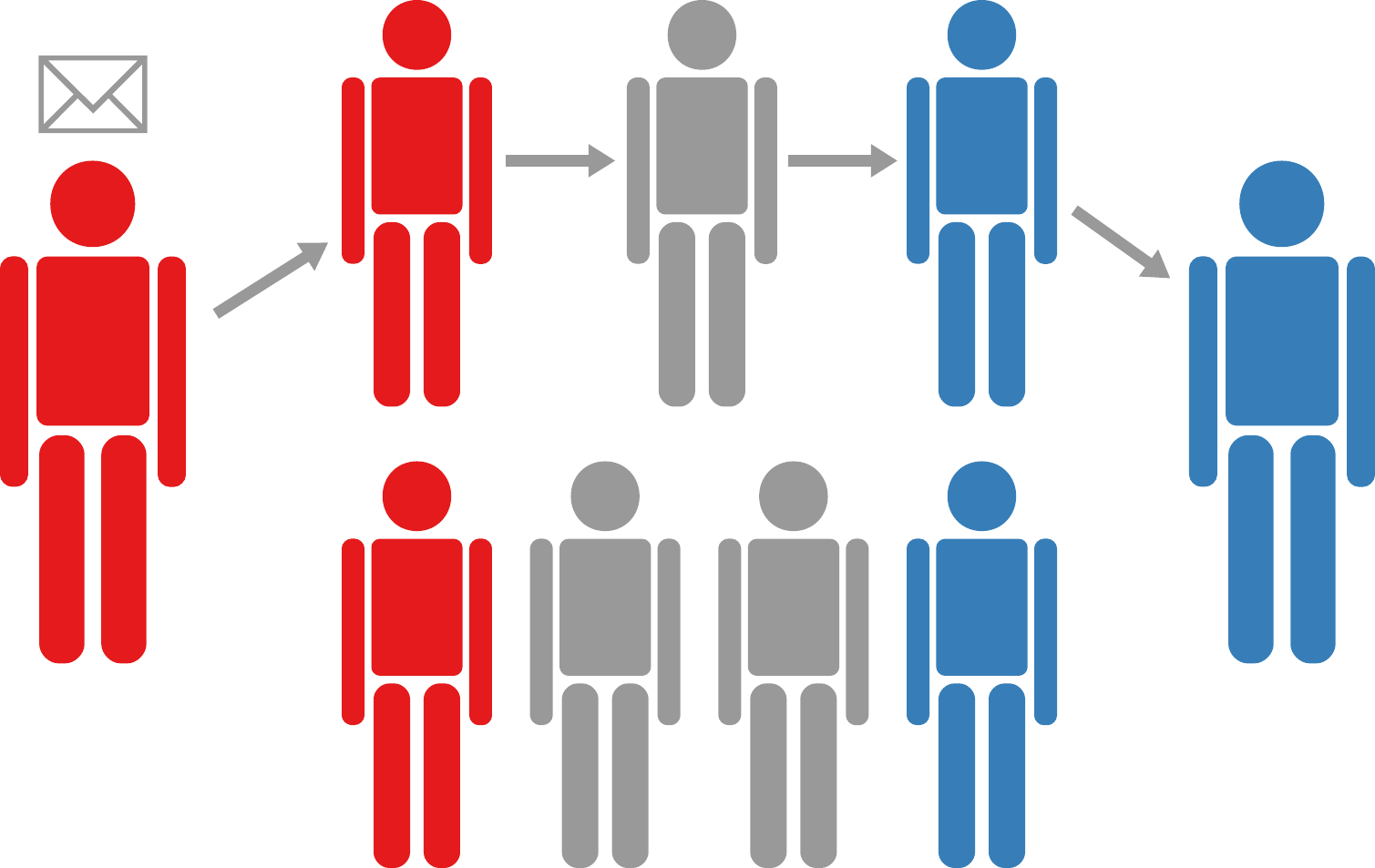}
\caption{A vignette representing the problem of delivering a letter through acquaintances: how do you know the best path is at the top since you're unaware of the existence of the people in gray?}
\label{fig:shortpath1}
\end{figure}

To see why finding ``best paths'' is important, suppose you have to deliver a letter to a person, as I show in Figure \ref{fig:shortpath1}. If you know them, no problem: you just give it to them. What if you don't? You might know one of their friends, and pass through them. Or you might know that one of your friends knows one of theirs. But if none of this is true, you have to know the shape of the entire social network -- the part in gray in the figure -- and discover what's the least amount of people you have to bother to get your letter to the recipient. This we call the ``shortest path problem'' in networks.

\begin{figure}[t]
\centering
\begin{subfigure}[t]{.45\columnwidth}
\includegraphics[width=\textwidth]{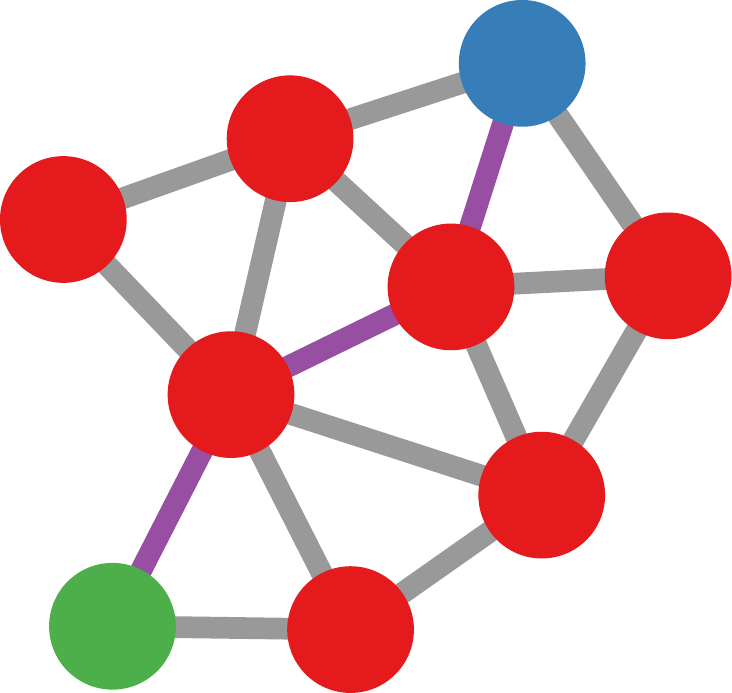}
\caption{}
\end{subfigure}
\qquad
\begin{subfigure}[t]{.45\columnwidth}
\includegraphics[width=\textwidth]{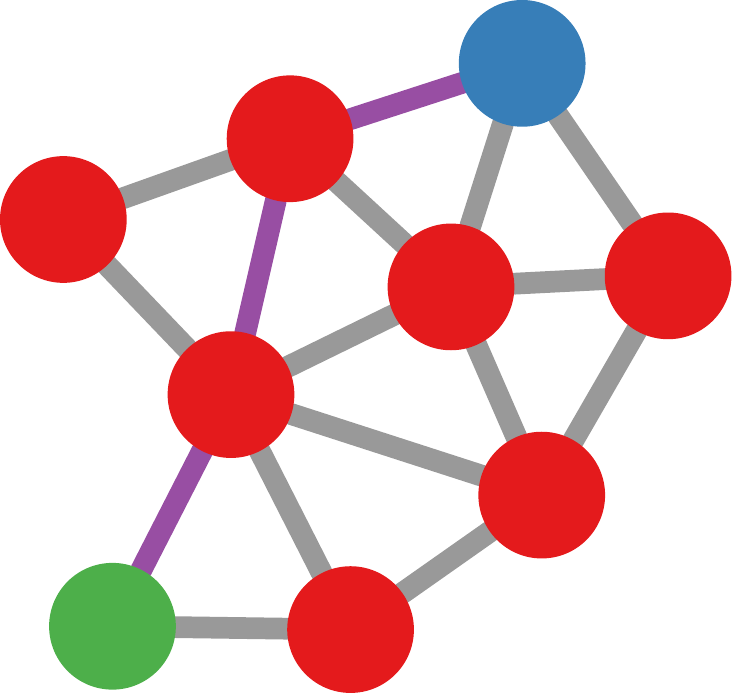}
\caption{}
\end{subfigure}
\caption{Finding the shortest path -- edges colored in purple -- between the start node (in blue) and the target node (in green). Note that (a) and (b) are both valid shortest paths which have the same length.}
\label{fig:shortpath2}
\end{figure}

The formal specification of the shortest path problem is the following. You're given a start node $v$ and a target node $u$. You have to find the path going from $v$ to $u$ crossing the fewest possible number of edges. Figure \ref{fig:shortpath2} provides a visualization of a shortest path between two nodes. In fact, the figure highlights a feature of this problem: it provides not one but two solutions. That is because, in unweighted undirected graphs, it is quite common to find multiple shortest paths between a given origin-destination pair. In other words, the solution to the shortest path problem is not necessarily unique.

\begin{figure}[b]
\centering
\begin{subfigure}[t]{.45\columnwidth}
\includegraphics[width=\textwidth]{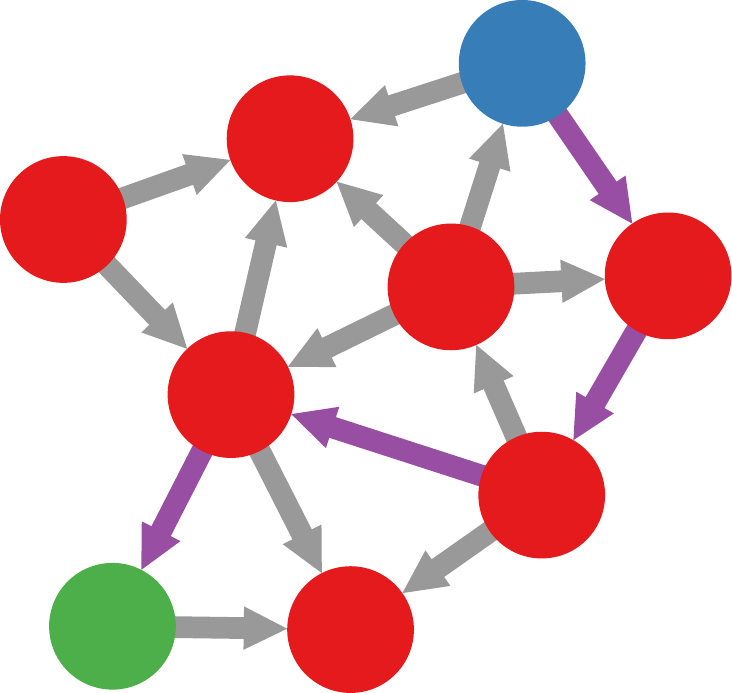}
\caption{}
\end{subfigure}
\qquad
\begin{subfigure}[t]{.45\columnwidth}
\includegraphics[width=\textwidth]{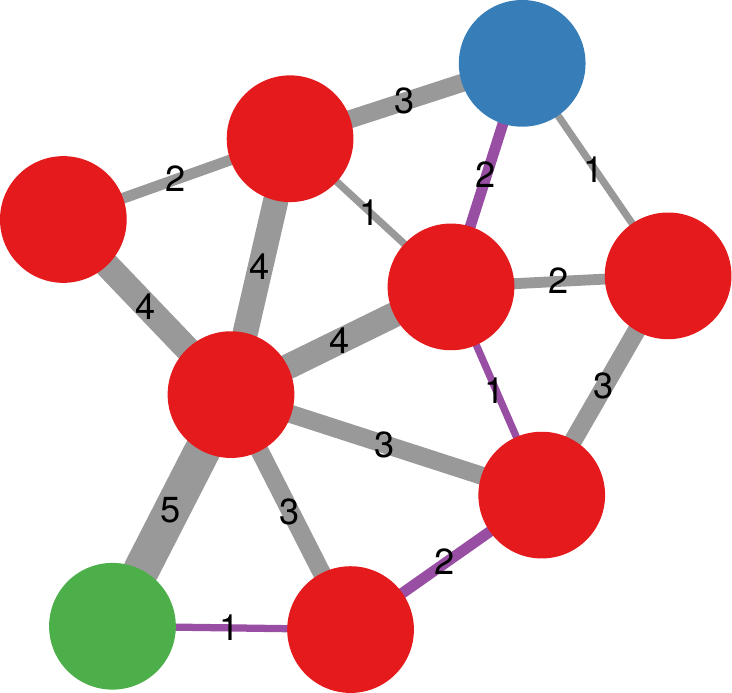}
\caption{}
\end{subfigure}
\caption{Finding the shortest path -- edges colored in purple -- between the start node (in blue) and the target node (in green). (a) Directed network. (b) Weighted network, where edge weights represent the cost of traversal.}
\label{fig:shortpath3}
\end{figure}

This is for the case of undirected, unweighted networks. If you have directed networks you obviously have to respect the edge directions -- see Figure \ref{fig:shortpath3}(a). If you have weighted networks, you might want to minimize the weight (as in Figure \ref{fig:shortpath3}(b)), assuming that the edge weight represents its traversal cost. If your edge weights represent proximities rather than costs, for instance they are the capacity of a trait of road as explained in Section \ref{sec:basic-weighted}, you'd do the opposite.

How do we find the shortest path? Depending on the properties of the graph (e.g. direct/undirected, weighted/unweighted), there are different algorithms for finding shortest paths. We also need to know if we just want to find a path from $v$ to $u$ (single-origin single-destination shortest path), or from $v$ to all other nodes (single-origin shortest path), or between every single pair of nodes.

\begin{figure}[t]
\centering
\includegraphics[width=.83\columnwidth]{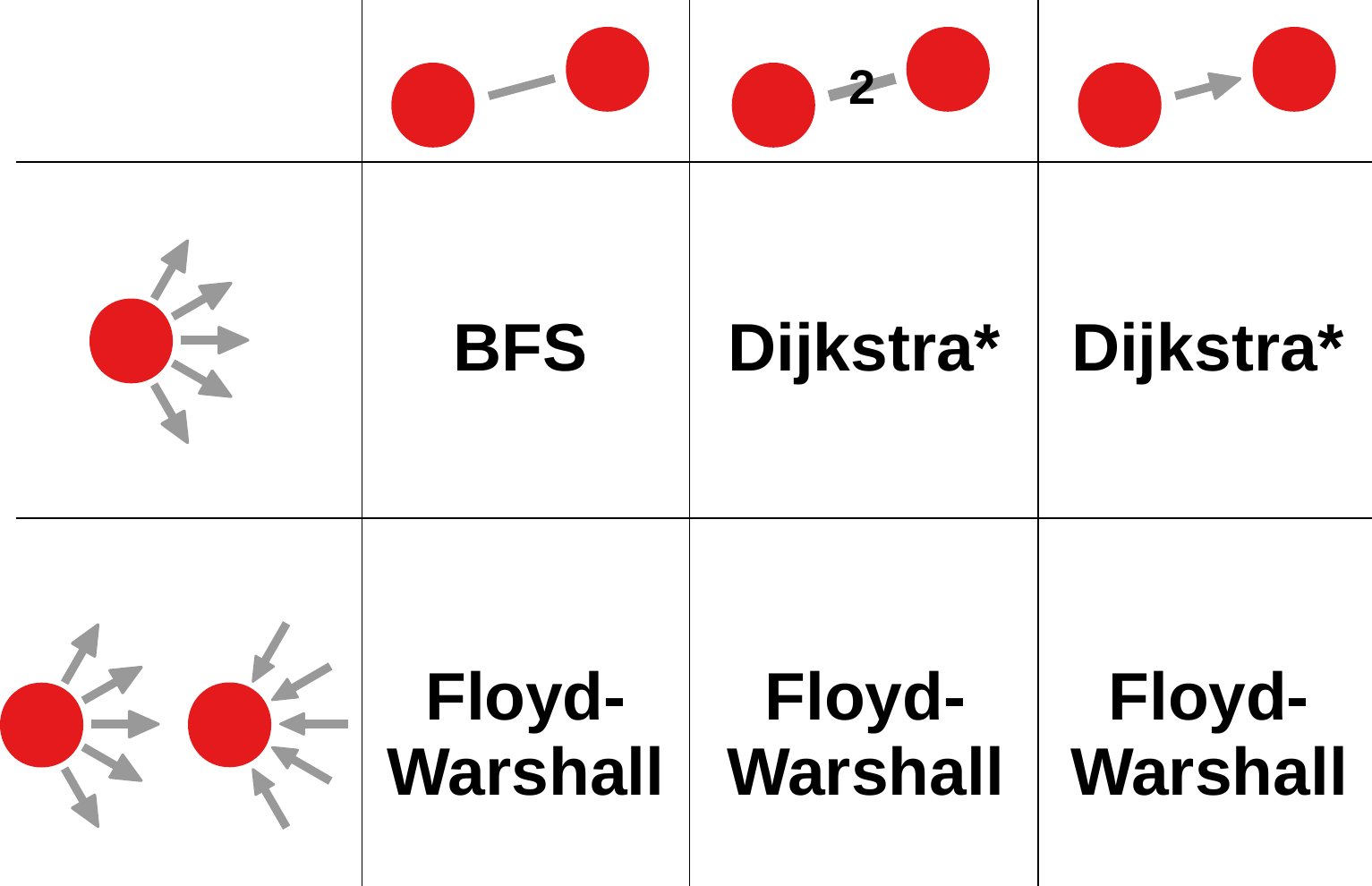}
\caption{A quick reference of the most well known algorithms used to solve specific shortest path problems. Columns (from left to right): simple, weighted, directed. Rows (top to bottom): single-origin, all pairs shortest path. Dijkstra is marked with a star because variants of the base algorithm can outperform it in special cases and they are used in most real world scenarios.}
\label{fig:shortpath4}
\end{figure}

Figure \ref{fig:shortpath4} provides you with a quick reference on which algorithm to use given each use case. For instance, as mentioned before, if you have an undirected unweighted network and you are interested in single-origin shortest paths, you can find them by performing a simple BFS exploration.

If you still have a single-origin in mind but your network contains directions and/or weights, you'll probably use one of the many flavors of the classical Dijkstra's algorithm\cite{dijkstra1959note}. Dijkstra's algorithm works as follows. You start by your origin, which you mark as you ``current node''. 

\begin{enumerate}
\item You look at all the unvisited neighbors of the current node and calculate their tentative distances through the current node.
\item Compare this tentative distance to the current assigned value and assign the smallest one.\footnote{For example, if the current node $u$ is at distance of $6$ from the source, and the edge connecting it with a neighbor $v$ has length $2$, then the distance to $v$ through $u$ is $6 + 2 = 8$. If you previously marked $v$ with a distance greater than $8$, you will change it to $8$. Otherwise you will keep the current value.}
\item When you are done considering all of the unvisited neighbors of the current node, mark the current node as visited. You will never check a visited node twice.
\item If the current node, the one you're marking as visited, is the destination node, you can stop. Otherwise, you can continue by selecting the unvisited node that is marked with the smallest tentative distance, as your new current node. Then go back to step 1.
\end{enumerate}

I cannot include in the book the best visual representation of the Dijkstra algorithm I know, because it is an animated GIF\footnote{\url{https://upload.wikimedia.org/wikipedia/commons/5/57/Dijkstra_Animation.gif}}.

Faster variations of the Dijkstra algorithm\cite{dial1969algorithm}\cite{ahuja1990faster}\cite{raman1997recent}\cite{thorup2000ram} use clever data structures and a few optimizations -- often under assumptions about the edge weights -- that are of no interest here.

The only other algorithm in the hall of fame of shortest path algorithms we consider here is Floyd-Warshall\cite{roy1959transitivite}\cite{warshall1962theorem}\cite{floyd1962algorithm}\footnote{Bernard Roy, who actually discovered the algorithm first, for mysterious reasons gets no naming rights.}. That is because it is the most used algorithm for the all-pairs shortest path problem, when you're not limiting yourself to a single origin and/or a single destination -- or to specific constraints on topology and/or edge weights. The algorithm uses recursive programming which, to this day, I still consider borderline magic.

Suppose you have a function $sp(u, v, K)$ that calculates the shortest path between $u$ and $v$ using only nodes in the set $K$. $K$ is a special set, it contains all nodes of the network with id equal to or lower than $K$. Obviously, if $K = 0$, then it is an empty set. Then $sp(u, v, 0)$ simply returns the weight of the edge between $u$ and $v$ -- if they are connected --, because we're not using any node in the path:

$$ sp(u, v, 0) = A_{uv}.$$

If $K > 0$ it means that we are adding a node as a possible member of the shortest path. When we do it, either of two things can happen: (i) adding the extra node allowed us to find a better (shorter) path, or (ii) it didn't. So:

$$ sp(u, v, K) = \min(sp(u, k, K) + sp(k, v, K),\ sp(u, v, K - 1)). $$

\begin{figure*}
\centering
\begin{subfigure}[t]{.23\columnwidth}
\includegraphics[width=\textwidth]{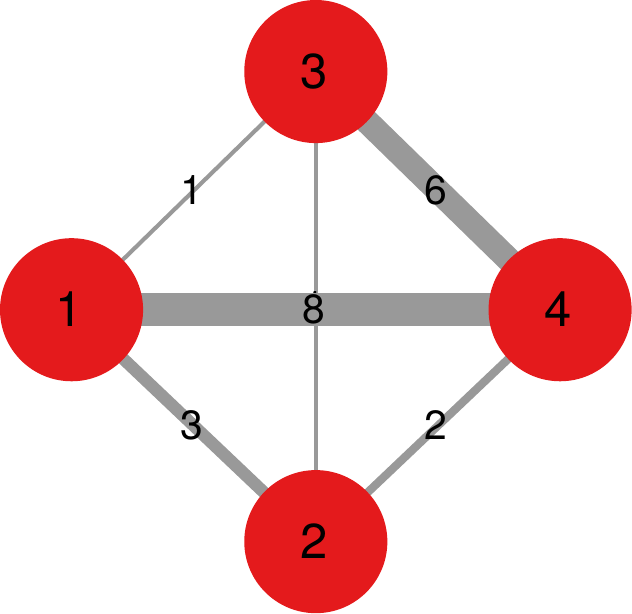}
\caption{Input}
\end{subfigure}
\quad
\begin{subfigure}[t]{.12\columnwidth}
\includegraphics[width=\textwidth]{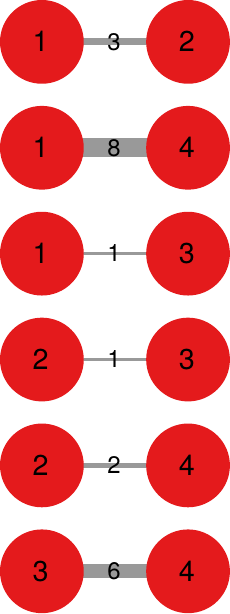}
\caption{$K = 0$}
\end{subfigure}
\quad
\begin{subfigure}[t]{.21\columnwidth}
\includegraphics[width=\textwidth]{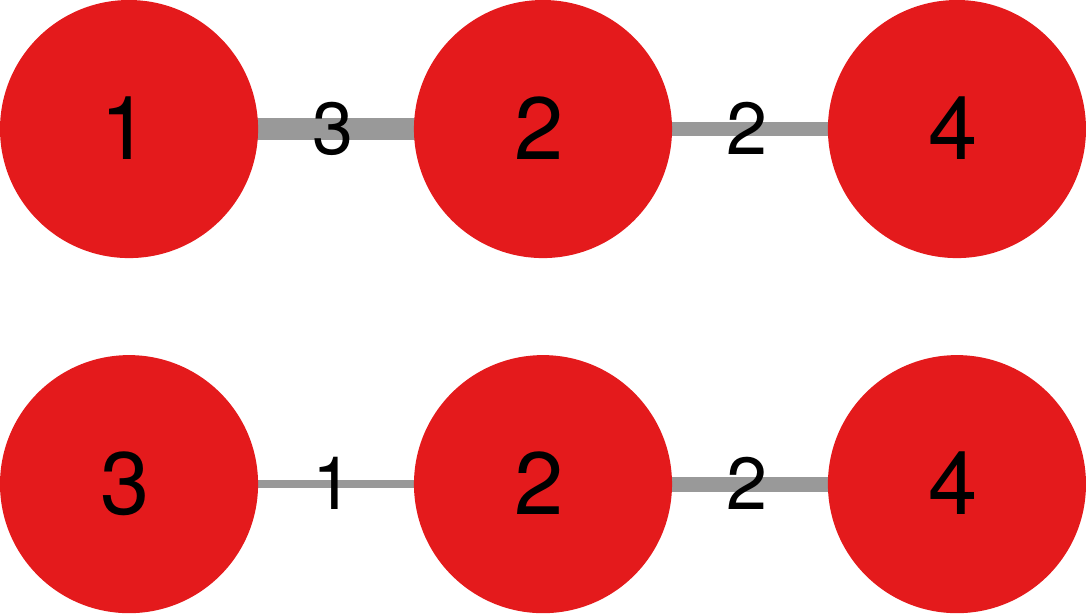}
\caption{$K = 2$}
\end{subfigure}
\quad
\begin{subfigure}[t]{.23\columnwidth}
\includegraphics[width=\textwidth]{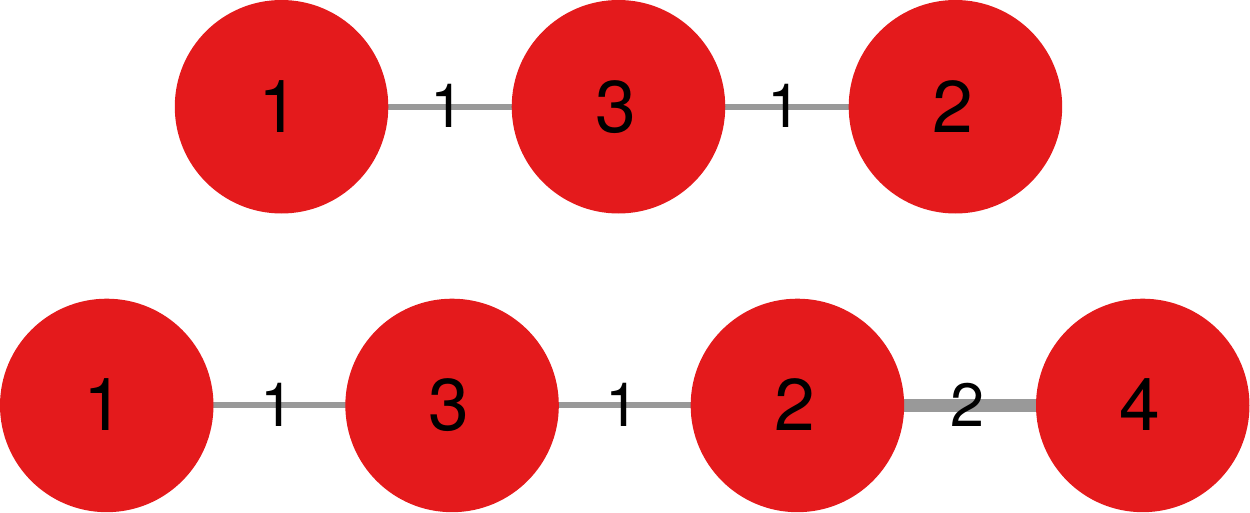}
\caption{$K = 3$}
\end{subfigure}
\caption{(a) The input for the Floyd-Warshall algorithm. (b-d) The temporary shortest paths at each step of the algorithm.}
\label{fig:floydwarsh}
\end{figure*}

Figure \ref{fig:floydwarsh} shows an example run. Figure \ref{fig:floydwarsh}(a) is an hypothetical input. At the first step, $K = 0$, we can only consider directly connected origins and destinations, setting the edge weights as the length -- Figure \ref{fig:floydwarsh}(b). For $K =1$ (not pictured) nothing happens: node $4$ cannot use node $1$ to go anywhere, because their edge is very costly, and nodes $2$ and $3$ have low cost connections to node $1$, but they are already directly connected by the minimum weight in the network. For $K = 2$ (Figure \ref{fig:floydwarsh}(c)) we're also allowed to use node $2$ for our paths. Both node $1$ and node $3$ use it to get to node $4$, given that their direct connection to node $4$ is costly. For $K = 3$ (Figure \ref{fig:floydwarsh}(d)) we can also use node $3$ in our paths. The path $1 \rightarrow 3 \rightarrow 2$ is the sum of two paths we already know from Figure \ref{fig:floydwarsh}(b): $1 \rightarrow 3$ an $3 \rightarrow 2$. It costs less than $1 \rightarrow 2$, so we select it. To go from node $1$ to node $4$ we sum two paths we already know: $1 \rightarrow 3$ (from Figure \ref{fig:floydwarsh}(b)) and $3 \rightarrow 2 \rightarrow 4$ (from Figure \ref{fig:floydwarsh}(c)). We discover then that the actual distance between the nodes $1$ and $4$ is four, rather than five -- as we though in Figure \ref{fig:floydwarsh}(c) -- or eight -- as we though in Figure \ref{fig:floydwarsh}(b).

\begin{figure}
\centering
\begin{subfigure}[t]{.3\columnwidth}
\includegraphics[width=\textwidth]{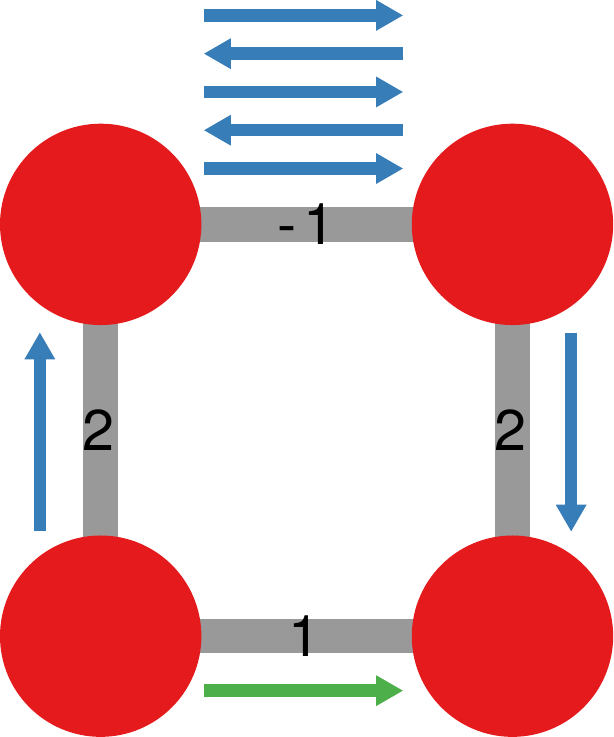}
\caption{}
\end{subfigure}
\quad
\begin{subfigure}[t]{.3\columnwidth}
\includegraphics[width=\textwidth]{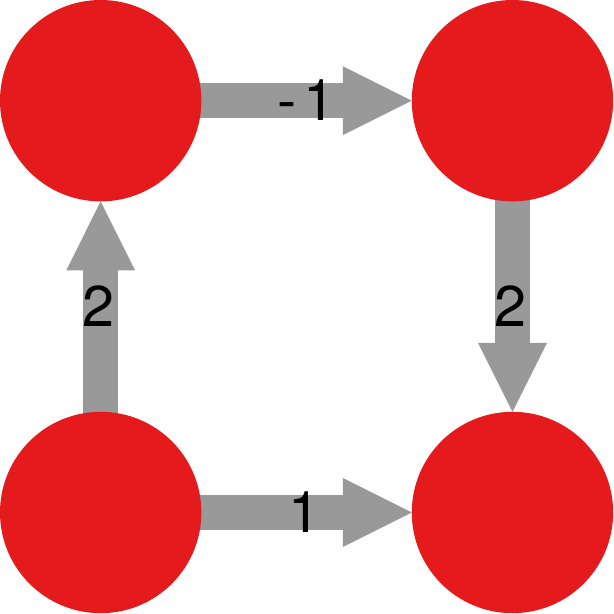}
\caption{}
\end{subfigure}
\quad
\begin{subfigure}[t]{.3\columnwidth}
\includegraphics[width=\textwidth]{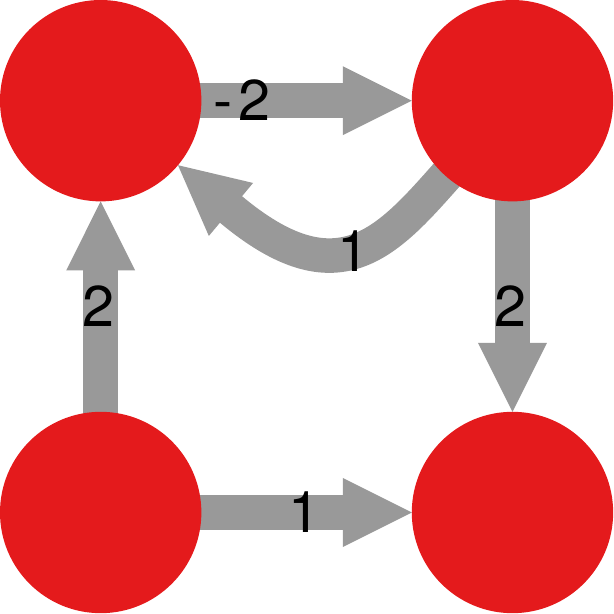}
\caption{}
\end{subfigure}
\caption{(a) A weighted network with negative weights which results in degenerate shortest paths -- in blue -- over preferred non-shortest paths -- in green. (b) A directed weighted network with negative weights but without the infinite negative weight problem. (c) A directed weighted network with negative cycles.}
\label{fig:negative-weights}
\end{figure}

A final word about negative weights. As presented earlier, there's no shame if your network contains them (see Section \ref{sec:basic-weighted}). However, you need to be careful when computing shortest paths. The reason is evident, as one can see from Figure \ref{fig:negative-weights}(a). The problem with negative weights is that we might think that it is trivial to find a shortest path (in green in the figure), but by going back and forth over a negative weight we can find an equivalent path. At that point, we can be stuck in an infinite loop of shorter and shorter paths without ever reaching the destination (in blue in the figure).

Directed networks can allow negative weights, because you're not allowed to follow the edge against its direction, as in Figure \ref{fig:negative-weights}(b). However, if there is a negative cycle -- see Figure \ref{fig:negative-weights}(c) -- you are in the same situation as before. A negative cycle is a cycle whose total edge weight sum is lower than zero.

If you're writing shortest path algorithms, you have to take care of these situations. Usually, you have to explicitly say that you're looking for \textit{paths}, not \textit{walks}. In paths, you cannot re-use the same edge twice (see Chapter \ref{cha:paths}), no matter how cool it would make your path length.

\section{Path Length Distribution}\label{sec:shortpath-avglength}
Just like with the degree, knowing the length distribution of all shortest paths in the network conveys a lot of information about its connectivity. A tight distribution with little deviation implies that all nodes are more or less at the same distance from the average node in the network. A spread out distribution implies that some nodes are in a far out periphery and others are deeply embedded in a core.

\begin{figure}
\centering
\includegraphics[width=.83\columnwidth]{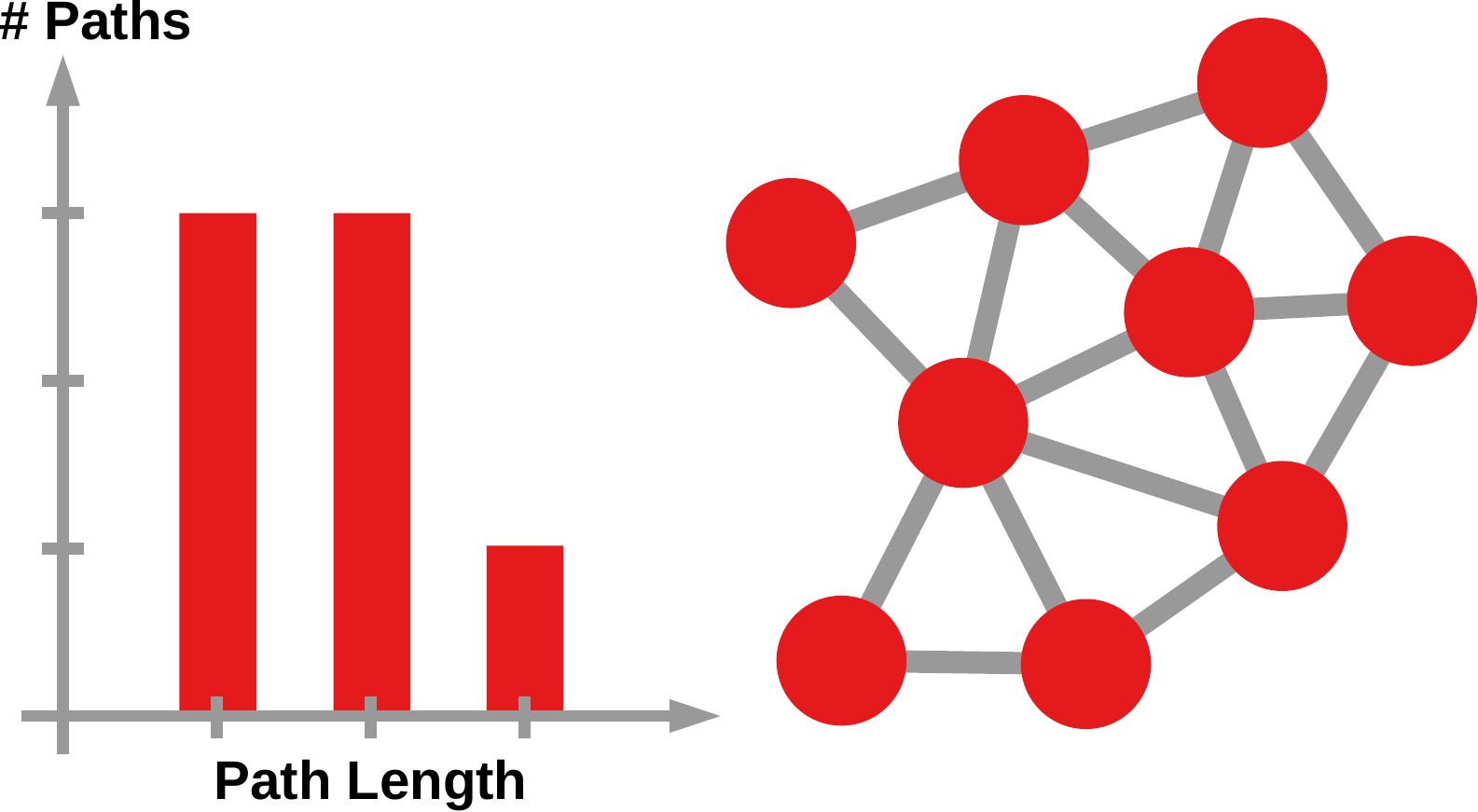}
\caption{The path length distribution (left) of a graph (right). Each bar counts the number of shortest paths of length one, two and three, which is the maximum length in the network.}
\label{fig:path-length-distr}
\end{figure}

To generate a path length distribution you perform the same operation you used to get the degree distribution: you have the path length on the x axis and the number of paths of a given length on the y axis. See Figure \ref{fig:path-length-distr} for an example. I'm not going to go on a tangent on log-log spaces and power laws like last time because usually path lengths distribute quasi-normally: you'll find a lot of classical bell shapes.

Some values in the distribution are fixed. For instance, the number of paths of length one is twice the number of edges, because each edge is used for two paths of length one ($u \rightarrow v$, and $v \rightarrow u$). It goes without saying that things are different in directed networks. The number of total shortest paths is $|V|(|V| - 1)$, because each origin has to reach each destination, minus one because we don't count the paths of length zero, from the origin to the origin.

\subsection{Diameter}
The rightmost column of the histogram in Figure \ref{fig:path-length-distr} is important. It records the number of shortest paths of maximum length. These are the ``longest shortest paths''. Since this is an important concept, such a long mouthful name won't do. We're busy people and we got places to be. So we use a different name for them or, to be more precise, to their length. We call it the \textbf{diameter} of the network.

Why do we care about the diameter? Because that's the worst case for reachability in the network. The diameter is the measure of the maximum possible separation between two nodes. A long diameter means that the problem of finding a shortest path for some pairs of nodes might be too hard because there are too many hypothetical paths and splits to consider. With a small diameter, everybody is reachable in one or two hops. With a large diameter, a full traversal of the graph might be impossible, especially if we only have local information about our neighborhood.

Let's go over a few values of diameter, just to get a grasp of the concept:

\begin{itemize}
\item Diameter = $1$ $\rightarrow$ You know everyone;
\item Diameter = $2$ $\rightarrow$ Your friends know everyone;
\item Diameter = $3$ $\rightarrow$ Your friends know someone who knows everyone;
\item ...
\end{itemize}

It's now easy to see that a network with diameter equal to three is easy to navigate. As the diameter grows, the number of people to rely on for a full traversal starts becoming unwieldy.

If your network has multiple connected components (Section \ref{sec:paths-ccomps}), we have a convention. Nodes in different components are unreachable, and thus we say that their shortest path length is infinite. Thus, a network with more than one connected component has an infinite diameter. Usually, in these cases, what you want to look at is the diameter of the giant connected component.

\subsection{Average}
The diameter is the worst case scenario: it finds the two nodes that are the most far apart in the network. In general, we want to know the typical case for nodes in the network. What we calculate, then, is not the longest shortest path, but the typical path length, which is the average of all shortest path lengths. This is the expected length of the shortest path between two nodes picked at random in the network.

\begin{figure}
\centering
\includegraphics[width=.75\columnwidth]{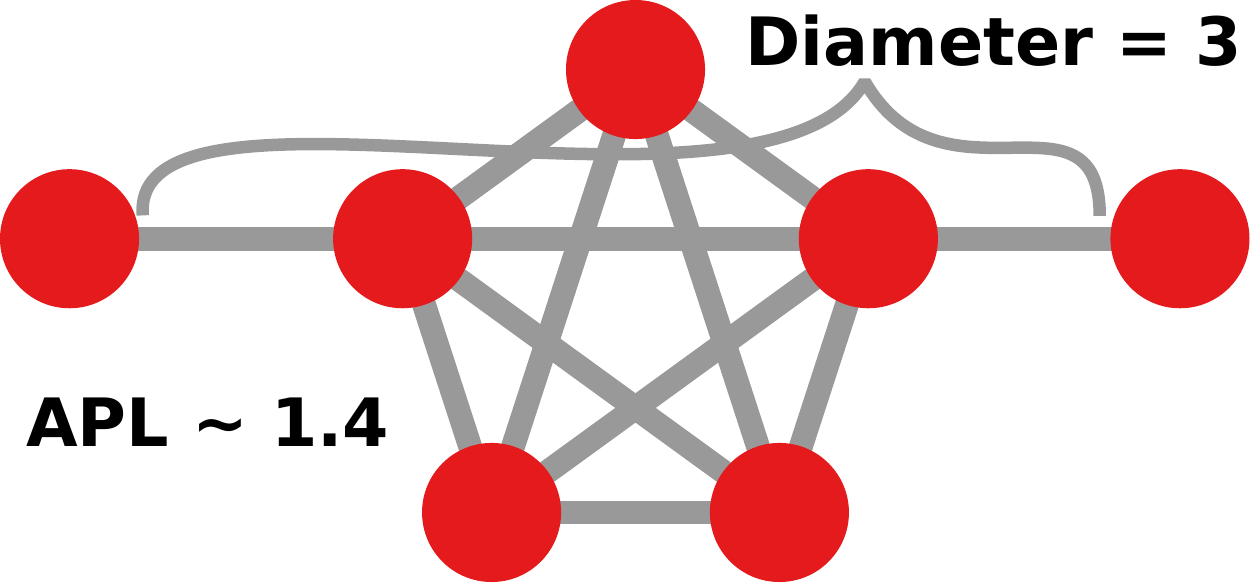}
\caption{The diameter and the APL in a graph can be quite different.}
\label{fig:diameter-vs-apl}
\end{figure}

If $P_{uv}$ is the path to go from $u$ to $v$ and $|P_{uv}|$ is its length, then the average path length of the network is $APL = \dfrac{\sum \limits_{u,v \in V} |P_{uv}|}{|V|(|V| - 1)}$. Figure \ref{fig:diameter-vs-apl} shows that, even in a tiny graph, the diameter and the APL can take different values, with the former being more than twice the length of the latter.

\begin{figure}
\centering
\includegraphics[width=.83\columnwidth]{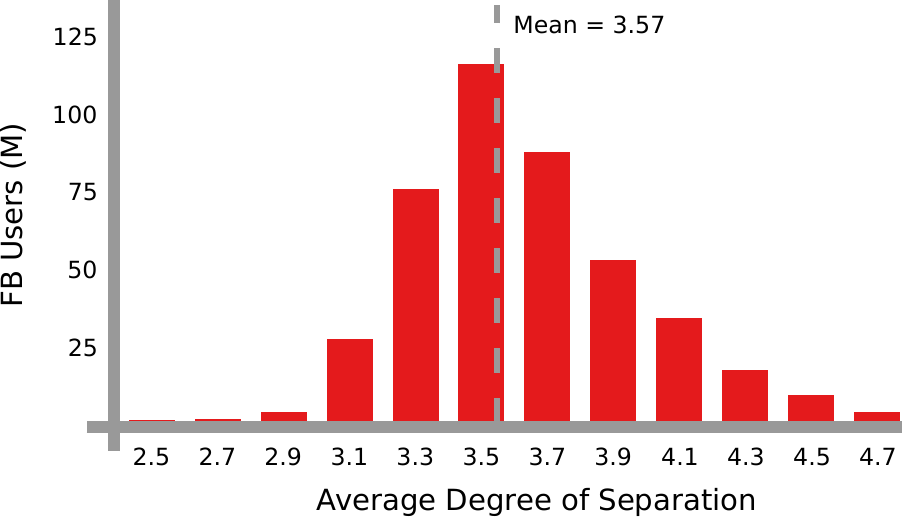}
\caption{The path length distribution for Facebook in $2012$.}
\label{fig:path-length-distr-facebook}
\end{figure}

With APL, we can fix the origin node. For instance, in a social network, you can calculate your average separation from the world. This would be an $APL_v$, the average path length for all paths starting at $v$. Then you can generate the distribution of all lengths for all origins. How does this $APL_v$ distribution look like for a real world network? One of the most famous examples I know comes from Facebook\cite{edunov2016three}. I show it in Figure \ref{fig:path-length-distr-facebook}. The remarkable thing is how ridiculously short the paths are even in such a gigantic network.

This is in line with classical results of network science, showing that the diameter and APL typically grow sublinearly in terms of number of nodes in the network\cite{newman2003structure}. In other words, there are diminishing returns to path lengths: each additional person contributes less and less to the growth of the system in terms of reachability. In fact, some researchers have found that adding people might even shrink the diameter\cite{leskovec2005graphs}\cite{leskovec2007graph}: as people join a social network, they create shortcuts and new paths that bring close together people that were previously far apart.

The most notorious enunciation of the surprising small average path length in large networks is the famous ``six degrees of separation''. This concept says that, on average, you're six handshakes away from meeting any person in the world, being a fisherman in Cambodia or an executive in Zimbabwe. People used this concept to describe the famous -- failed -- Milgram experiment.

In $1967$, Milgram published a paper\cite{milgram1967small} detailing the travels of a series of envelopes. He handed a destination address to people in the Midwest of the United States. The destination was in Boston, Massachusetts. The idea was that each recipient needed to attempt to have the letter reach its final destination. However, they could not mail it directly: they needed to hand it over to a person they knew on a first name basis. So they needed to figure out who in their acquaintances was most likely to know somebody (who knew somebody, who knew somebody, ...) in Massachusetts. Each handler of the envelope would have to write their name on it. When the envelope reached the destination, counting the names in it would give an approximation of the degrees of separation between the origin and destination individuals.

The number turned out to be $5.5$ on average, which gave fuel to the ``six degrees of separation'' urban legend. However, the experiment was arguably a failure given that, of the more than $400$ letters sent, less than a hundred actually arrived at the destination. The problem is that obviously there is no way to account for the fact that a letter might not successfully reach its target because some people in the chain were unreliable, rather than unconnected with the destination. Fascinating as it is, this theory might be wrong because the degrees of separation could be lower than six: people have proposed four\cite{backstrom2012four}, as we see in Facebook (Figure \ref{fig:path-length-distr-facebook}).

Diameter and average path length are only the two most famous and most used measures derived from the shortest path length distribution. There is a collection of other measures you might find in network science papers and books. Two other examples are the eccentricity of a node and the radius of a network. You can think of the eccentricity as a node-level diameter. It is the longest shortest path leading from node $u$ to the farthest possible node $v$ in the network. Thus, by definition the diameter is equal to the highest eccentricity among the nodes of the network. The radius of a network is, conversely, equal to the smallest eccentricity in the network.

\section{Spanning Trees \& Other Filtered Graphs}\label{sec:paths-spantree}
I conclude this chapter with a look at spanning trees and other ways to filter down a graph. These methods are usually deployed to reduce a network to its minimum terms and finding its fundamental structure in a way that is parsable by humans. They are also at the basis of some network backboning techniques (Chapter \ref{cha:backboning}).

A spanning tree of an undirected graph is a subgraph that: (i) is a tree (see Section \ref{sec:paths-cycles}), and (ii) it includes all of the vertices of the graph. In practice, it is that subgraph that can connect all nodes of the graph with the minimum number of edges, and no cycles.

\begin{figure}
\centering
\begin{subfigure}[t]{.3\columnwidth}
\includegraphics[width=\textwidth]{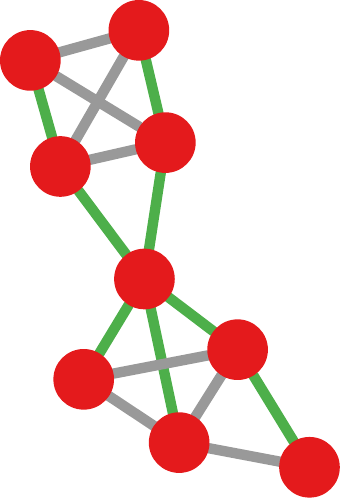}
\caption{}
\end{subfigure}
\quad
\begin{subfigure}[t]{.3\columnwidth}
\includegraphics[width=\textwidth]{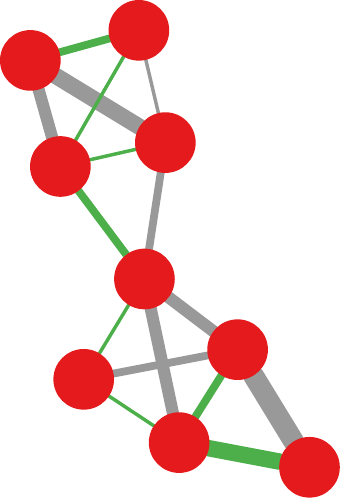}
\caption{}
\end{subfigure}
\quad
\begin{subfigure}[t]{.3\columnwidth}
\includegraphics[width=\textwidth]{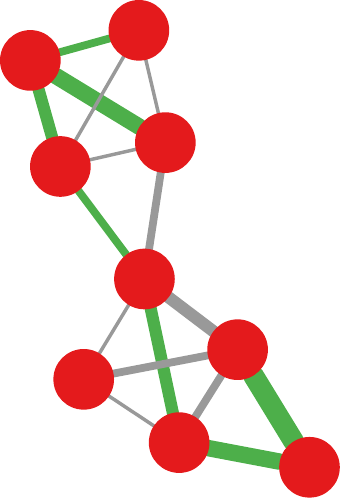}
\caption{}
\end{subfigure}
\caption{(a) A graph with one of its possible spanning trees highlighted in green. (b) The minimum spanning tree of a weighted graph, with the edge width proportional to its weight. (c) The maximum spanning tree of a weighted graph.}
\label{fig:spantree}
\end{figure}

Figure \ref{fig:spantree}(a) shows you an example of a spanning tree inside a graph. If your graph has multiple connected components you cannot find a single spanning tree, because you don't have a way to connect nodes in different components. However, you can make a spanning forest, by finding the spanning trees of each component separately.

Spanning trees are nice, but they get used mostly in weighted networks. In that case, you have to distinguish between weights as proximities and weights as distances (Section \ref{sec:basic-weighted}): is an edge with a high weight expressing the cost of going from $u$ to $v$, or is it saying how much $u$ and $v$ interact? In the first case we have a ``distance'' weight: we want to minimize costs. Imagine finding the tree connecting all your road intersections that minimizes driving distance -- the cost of an edge.

When your weights are distances you want a minimum spanning tree\cite{graham1985history}: the spanning tree among all spanning trees of a graph that has the minimum possible total edge weight. Figure \ref{fig:spantree}(b) shows an example. When your weights are proximities -- maybe because they tell you the capacity of the road -- then you want the maximum spanning tree: the spanning tree among all spanning trees of a graph that has the maximum possible total edge weight. Figure \ref{fig:spantree}(c) shows an example.

Of course, the algorithm to find the minimum and the maximum spanning tree is the same, you just flip the sign of the comparison. There's a good range of algorithms, from classical ones to more modern which use special data structures: Bor\r{u}vka\cite{boruuvka1926jistem}, Prim\cite{prim1957shortest}, Kruskal\cite{kruskal1956shortest}, Chazelle\cite{chazelle2000minimum}. They are usually all implemented in standard network analysis libraries.

Note that finding the minimum spanning tree doesn't really solve the traveling salesman problem\cite{lawler1985traveling}, although it sounds like it should. A quick recap: the traveling salesman problem is the quest to find the shortest possible route that visits each city and returns to the origin city, given a list of cities and the distances between each pair of cities. We can represent the problem as a weighted graph, with city distances as edge weights. The minimum spanning tree doesn't solve the problem: it creates a tree, which has no cycles. Thus, to get back to the origin city, you have to backtrack all the way through the tree -- not ideal.

\begin{figure}
\centering
\includegraphics[width=.3\columnwidth]{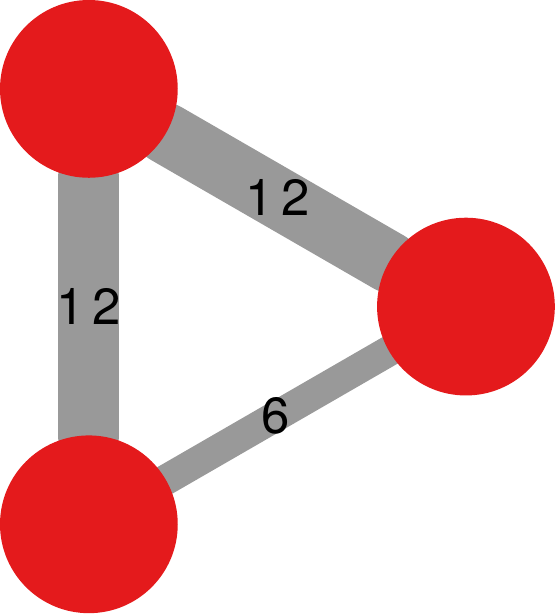}
\caption{An example of a weighted graph with a non unique minimum spanning tree.}
\label{fig:spantree2}
\end{figure}

Another thing to keep in mind is that rarely minimum/maximum spanning trees are unique: a weighted network can and will have multiple alternative minimum/maximum spanning trees. Consider the graph in Figure \ref{fig:spantree2}. Suppose we want to find its minimum spanning tree. The first choice is obvious: we use the edge of weight $6$. Then, we have to connect the final node. Each of the edges of weight $12$ is a valid addition to the tree: they will connect the node and the result will be a tree, an acyclic graph. So the graph has two valid minimum spanning trees.

There is an easy rule to remember to know whether a graph will have a unique minimum/maximum spanning tree or not. If each edge has a distinct weight then there will be only one, unique minimum spanning tree. As soon as you have two edges with the same weight, you open the door to the possibility of having more than one minimum spanning tree. In fact, in an unweighted graph where we assume that all edges have the same weight equal to one, then every spanning tree of that graph is minimum.

Spanning trees have some closely related cousins that are worthwhile mentioning. The first one is the planar maximally filtered graph\cite{tumminello2005tool}. As the name suggests, this is a technique to reduce any arbitrary graph into a planar version of itself, such that the edge weight sum is maximal (or minimal, depending on the meaning of your edge weights). Since a spanning tree is a tree, it means that it must have $|V| - 1$ edges. On the other hand, a planar maximally filtered graph must have $3(|V| - 2)$ or fewer edges.

\begin{figure}
\centering
\begin{subfigure}[t]{.3\columnwidth}
\includegraphics[width=\textwidth]{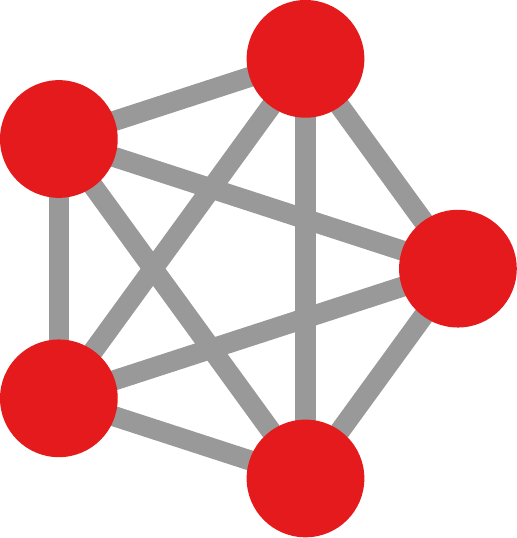}
\caption{}
\end{subfigure}
\qquad \qquad
\begin{subfigure}[t]{.39\columnwidth}
\includegraphics[width=\textwidth]{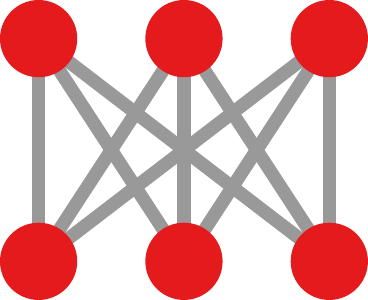}
\caption{}
\end{subfigure}
\caption{Two examples of non planar graphs that cannot be included in any planar maximally filtered graph. (a) A 5-clique; (b) a 3,3-biclique.}
\label{fig:nonplanar}
\end{figure}

Just like in the case of the tree, also in this case some motifs cannot appear. In a tree you cannot have cycles. In a planar graph you cannot have a motif that is impossible to draw as planar -- i.e. on a 2D surface without edge crossings --, for instance a 5-clique or a 3,3-biclique. Look at Figure \ref{fig:nonplanar} and try to draw those graphs in two dimensions without having any edge crossing another one. You'll find out that is not possible.

The second cousin of spanning trees is the triangulated maximally filtered graph\cite{massara2016network}. This was originally proposed as a more efficient algorithm to extract planar maximally filtered graphs from larger graphs. However, it also allows to specify different topological constraints, which are not necessarily making the graph planar.

\section{Classic Combinatorial Problems}
Graph exploration in general, and shortest paths in particular, are linked with some of the most famous problems discussed in computer science. We already saw one in Section \ref{sec:density-indsets} -- graph coloring: how many colors do I need to make sure that I don't give the same one to two connected nodes? Here I mention another, related to the classic Traveling Salesman Problem. In the Traveling Salesman Problem, we have a set of cities and we want to find the path that allows us to visit all cities by covering the minimum possible distance.

\begin{figure}
\centering
\includegraphics[width=.45\columnwidth]{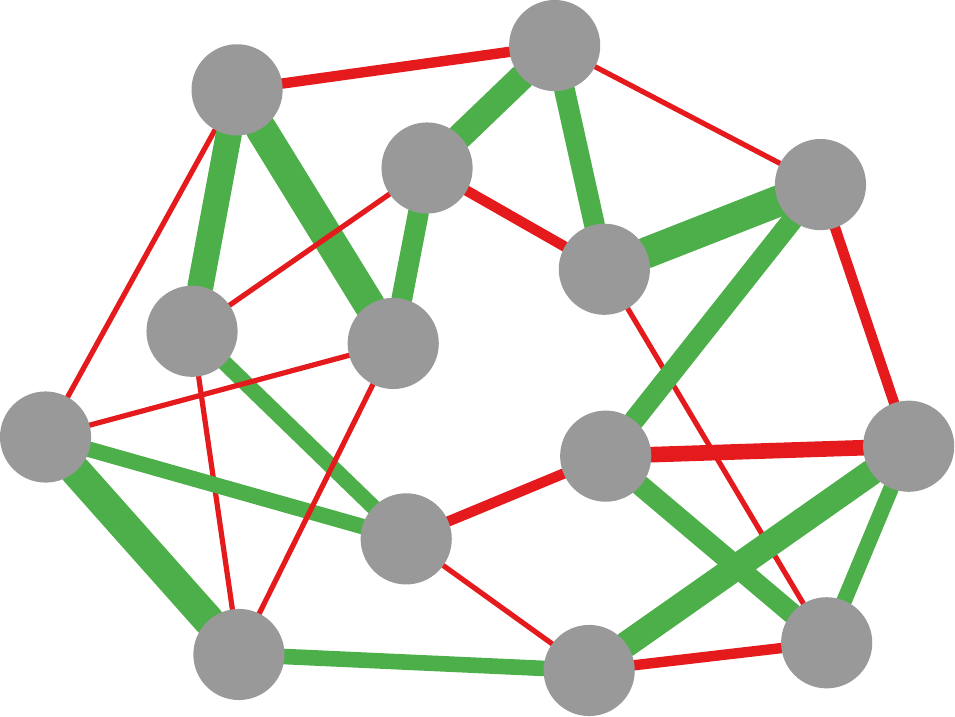}
\caption{A graph with two Hamiltonian cycles highlighted using the edge color. Red = minimum Hamiltonian; green = maximum Hamiltonian. The edge's thickness is proportional to its weight.}
\label{fig:min-hamiltonian}
\end{figure}

In this scenario, we are assuming that cities live in a two dimensional space and there is a path between any two cities. However, we could impose the existence of a road graph that makes some city-city connections impossible. In this case, we want to find the path of minimum cost in a graph that visits each node exactly once (i.e. the minimum Hamiltonian path -- see Chapter \ref{cha:paths}). Figure \ref{fig:min-hamiltonian} shows an example, with two Hamiltonian paths of different costs highlighted in red and green.

Such problems have a huge importance in computer science because they are classical examples of NP-hard problems. These problems have no known polynomial-time solution, meaning that we can usually only find approximate solutions in a reasonable time. Finding the best solution would require brute force algorithms whose time complexity make them unsuitable for problems of large size -- i.e. if your graph has more than a handful nodes.

Combinatorics and graphs have a much deeper relationship that this one, though. A vast number of problems in combinatorics can be represented as a graph problem, and often graphs are the best tool to solve them. Two other examples are the classic SAT problem, where we want to know if there is a true/false assignment so that a set of logical propositions is not contradictory; and vehicle routing, where we want to find the optimal set of routes for several vehicles to reach their destinations from their origins.

\section{Summary}

\begin{enumerate}
\item There are many ways to explore a graph structure. Breadth-First Search means to explore all neighbors of a node before exploring their neighbors; Depth-First Search means to explore a neighbor's neighbors before moving on to the next direct neighbor; random node and edge access means to explore one node or edge at a time ignoring the graph's topology.
\item Shortest paths are the paths connecting two arbitrary nodes in the network using the minimum possible number of edges. In directed networks you have to respect the edge's direction, in weighted networks you have to minimize (or maximize, depending on the problem definition) the sum of the edge weights.
\item The most common algorithms to solve shortest path finding are Dijkstra (if you have a fixed origin and destination) or Floyd-Warshall (if you are calculating the shortest paths between all pairs of nodes in the network).
\item Two important network connectivity measures are the diameter and the average path length. The diameter is the length of the longest shortest path. The average path length is the average length of all shortest paths in the network.
\item A minimum spanning tree is a tree connecting all nodes in the network which minimizes the sum of edge weights.
\end{enumerate}

\section{Exercises}

\begin{enumerate}
\item Label the nodes of the graph in Figure \ref{fig:spantree}(a) in the order of exploration of a BFS. Start from the node in the bottom right corner.
\item Label the nodes of the graph in Figure \ref{fig:spantree}(a) in the order of exploration of a DFS. Start from the node in the bottom right corner.
\item Calculate all shortest paths for the graph in Figure \ref{fig:spantree}(a).
\item What's the diameter of the graph in Figure \ref{fig:spantree}(a)? What's its average path length?
\end{enumerate}

\chapter{Node Ranking}\label{cha:ranks}
\begin{figure}
\centering
\includegraphics[width=.75\columnwidth]{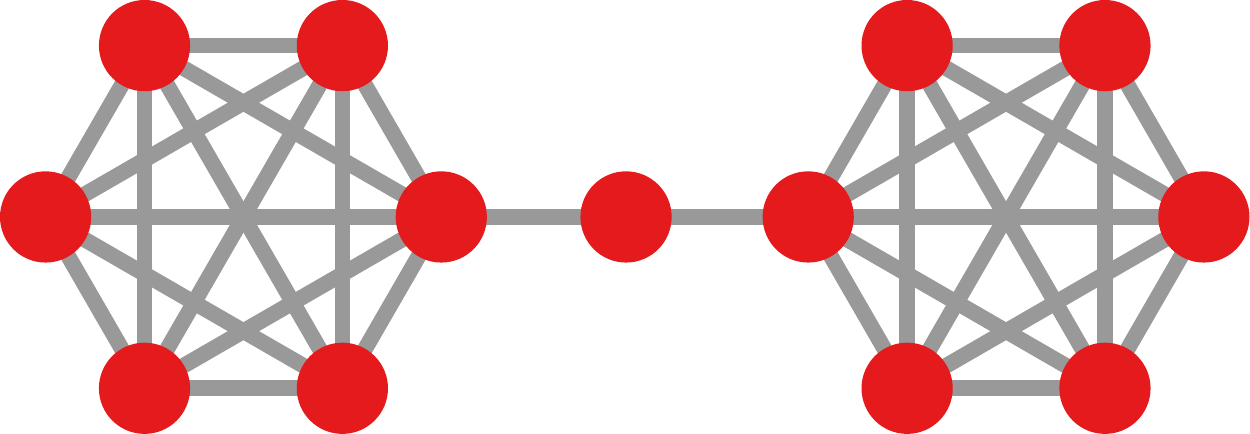}
\caption{An example of a social network in which the degree does not necessarily convey all the information about node importance.}
\label{fig:betweenness-example}
\end{figure}

The most direct way to find the most important nodes in the network is to look at the degree. The more friends a person has, the more important she is. This way of measuring importance works well in many cases, but can miss important information. What if there is a person with only few friends, but placed in different communities -- just like in Figure \ref{fig:betweenness-example}? The removal of such person will create isolated groups, which are now unable to talk to each other. Shouldn't this person be considered a key element in the social network, even with her puny degree?

Many networks scientists agree that she should, and developed different centrality measures accordingly. Here we focus on a few examples.

\section{Closeness}
If we want to know the closeness centrality\cite{bavelas1948mathematical} of a node $v$, first we calculate all shortest paths starting from that node to every possible destination in the network: $P$. Each of these paths $P_{vu}$ has a length, which is the number of edges you need to cross to get to your destination. Let's call it $|P_{vu}|$ -- the length to go from $v$ to $u$. We sum these distances in a total distance measure: $\sum_u |P_{vu}|$. We take the average of this value by dividing it by the number of all possible destinations $u$, which is the number of nodes in the network minus one (the origin): $\sum_u |P_{vu}| / (|V| - 1)$. Then, since the measure is called \textit{closeness}, we don't want to look at it directly. Closeness is the opposite of distance. So we actually want the opposite of what we just calculated, or $(|V| - 1) / \sum_u |P_{vu}|$. If this looks familiar, that's because it is. The closeness centrality of $v$ is nothing more than its inverse average path length (see Section \ref{sec:shortpath-avglength}), or $1/APL_v$.

\begin{figure}
\centering
\includegraphics[width=.5\columnwidth]{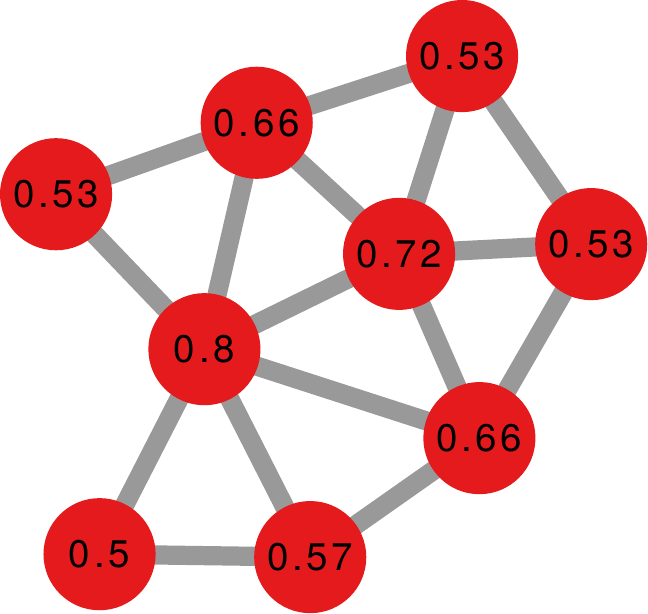}
\caption{A sample network. Node labels represent the closeness centrality value for the node.}
\label{fig:centrality-closeness}
\end{figure}

Let's look at an example -- in Figure \ref{fig:centrality-closeness}. Let's consider the node in the bottom left, labeled with $0.5$. That is its closeness centrality. How do we get to that value? First, we start with the nodes directly connected to it. The shortest paths to get to them is to follow the direct connections, thus only one edge is crossed. Both neighbors contribute $|P_{vu}| = 1$. Moving on, the two neighbors allow our $v$ node to access to four mode nodes. These four nodes require to cross an additional edge, thus they contribute $|P_{vu}| = 2$. We are left with two more nodes that require a third edge to be reached: $|P_{vu}| = 3$. So to recap, the total distance of this node is $1 + 1$ (the two direct neighbors) $+ 2 + 2 + 2 + 2$ (the four nodes at distance two) $+ 3 + 3$ (the final two nodes at distance three) $ = 16$. We then take the average ($16 / (9 - 1)$) and convert this into a closeness: $(9 - 1) / 16 = 0.5$.

The advantage of closeness centrality is that it has a spatial intuition: the closer you are on average to anybody, the more central you are. Exactly like standing in the middle of a room makes you closer on average to each member of the crowd in a party than standing in a corner. Empirically, in the vast majority of networks I analyzed, closeness centrality is distributed on a classical bell shape, i.e. normally. If you use closeness centrality, most of your nodes will have an average importance. This is not realistic for many networks: we know that degree distributions are very skewed -- the vast majority of nodes are unimportant, while only a few selected superstars take all the glory.

\begin{figure}
\centering
\includegraphics[width=.5\columnwidth]{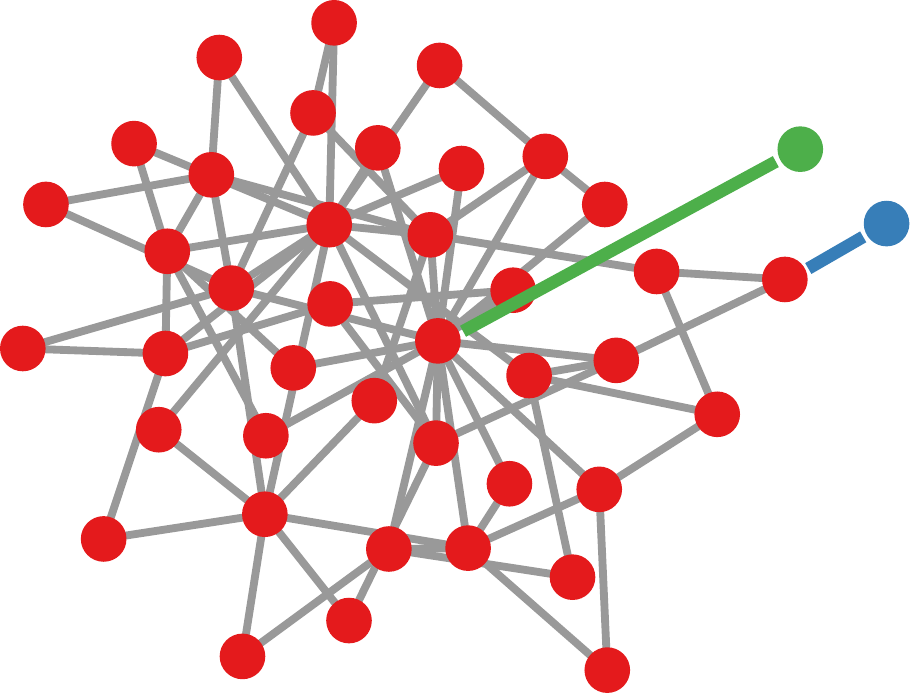}
\caption{The closeness centrality lottery. The blue and green nodes are new to the network and only have one edge to attach. The green node is lucky, and connects to a central hub. The blue node is unlucky and connects to a peripheral node. Thus nodes with the very same low degree end up with radically different closeness centrality values.}
\label{fig:closeness-lottery}
\end{figure}

Why does closeness centrality behave so differently from the degree? How can two nodes with very low degree -- for instance equal to one -- have different closeness centrality values so that they end up distributing normally instead of on a skewed arrangement? One possible explanation is that edge creation is a lottery. The many nodes with degree equal to one that you have in broad degree distributions can get lucky with their choice of neighbor. Sometimes, like in the case of the green node in Figure \ref{fig:closeness-lottery}, the neighbor is a hub. The green node's closeness centrality will then be high, because it is just one extra hop away from the hub itself -- which is very central. Sometimes the new node will attach itself to the periphery -- like the blue node in Figure \ref{fig:closeness-lottery} --, and thus have a very low closeness centrality.

\section{Betweenness}\label{sec:centr-betw}
\begin{figure}
\centering
\includegraphics[width=.5\columnwidth]{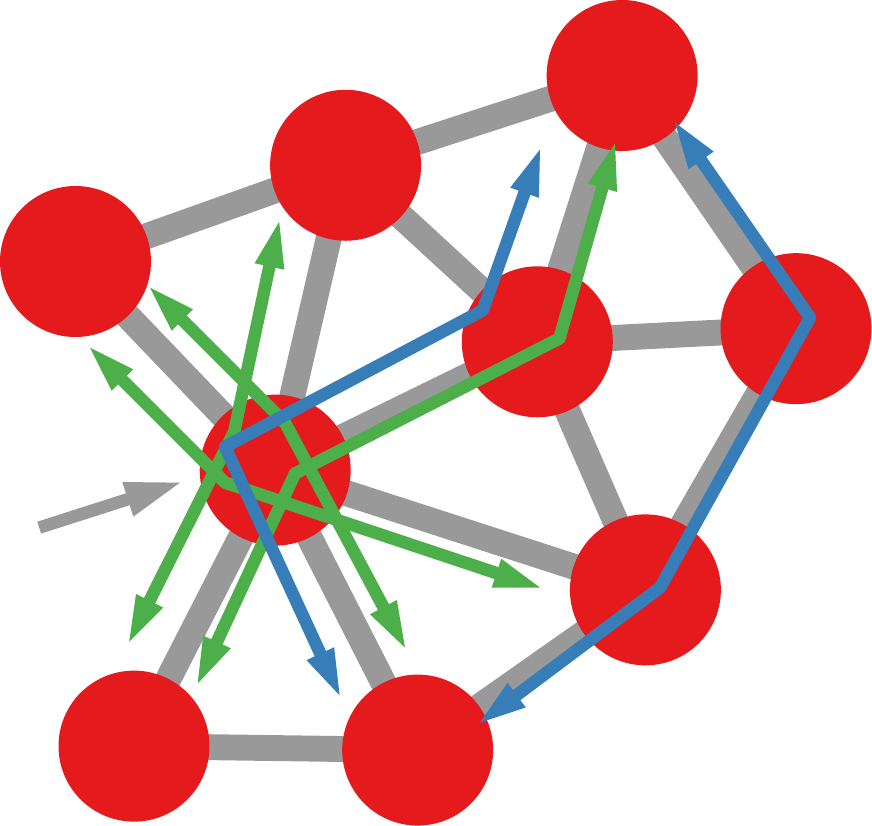}
\caption{An example on how to calculate the betweenness centrality of the node marked with the gray arrow. The shortest paths passing through it -- in green -- contribute to its betweenness centrality. If there are $n$ alternative paths not passing through the node, then the path contributes only $1 / n$ to the node's centrality -- I show an example in blue.}
\label{fig:centrality-betweenness}
\end{figure}

Network scientists developed betweenness centrality\cite{anthonisse1971rush}\cite{freeman1979centrality} to fix some of the issues of closeness centrality. Differently from closeness, with betweenness we are not counting distances, but paths. We still calculate all shortest paths between all possible pairs of origins and destinations. Then, if we want to know the betweenness of node $v$, we count the number of paths passing through $v$ -- but of which $v$ is neither an origin nor a destination. In other words, the number of times $v$ is \textit{in between} an origin and a destination. If there is an alternative way of equal distance to get from the origin to the destination that does not use $v$, we discount the contribution of the path passing through $v$ to $v$'s betweenness centrality. I provide an example in Figure \ref{fig:centrality-betweenness}.

The total number of paths that can pass through a node -- excluding the ones for which it is the origin or the destination -- are $(|V| - 1)(|V| - 2)$ in a directed network, and $(|V| - 1)(|V| - 2)/2$ in an undirected one.

One intuitive way to think about betweenness centrality is asking yourself: how many paths would become longer if node $v$ would disappear from the network? How much is the network structure dependent on $v$'s presence? Since real world networks have hubs which are closer to most nodes, the shortest paths will use them often. As a result, betweenness centrality distributes over many orders of magnitude, just like the degree. Unlike the degree, it takes into account more complex information than simply the number of connections.

The concept underlying betweenness centrality can be extended to go beyond nodes. You can use it to gauge the structural importance of edges. The definition is the same: the betweenness of an edge is the (normalized) count of shortest paths using the edge. If applied to connections, we call this measure ``edge betweenness''. This is a key concept especially for the field of community discovery, as we will find out in Part \ref{par:cd}.

\begin{figure}
\centering
\includegraphics[width=.6\columnwidth]{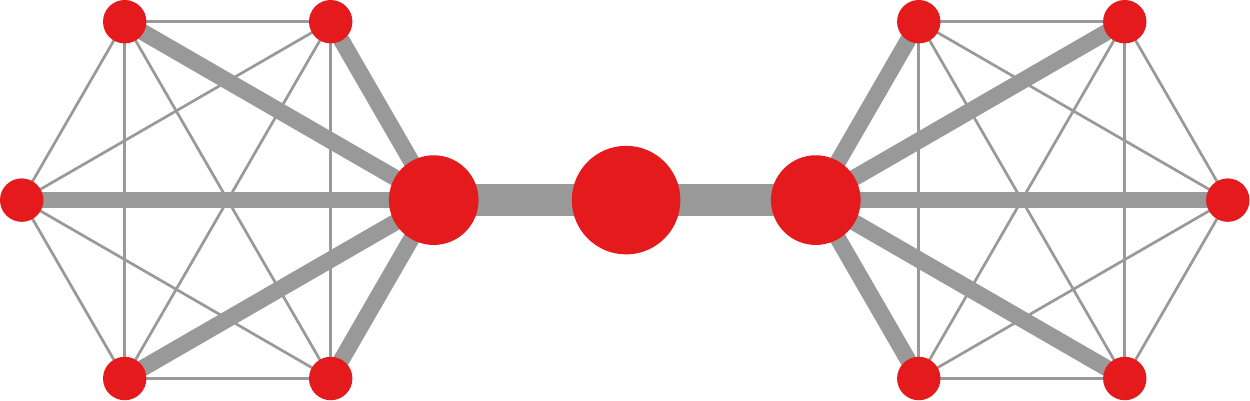}
\caption{In this network the edge thickness is proportional to its edge betweenness value. The node size is proportional to the node betweenness.}
\label{fig:centrality-betweenness2}
\end{figure}

Figure \ref{fig:centrality-betweenness2} shows an example of edge betweenness centrality. The intuition here is the same: the edge betweenness is the number of paths that would get longer if the edge were to disappear. Note, though, that if we remove the edge from the network, almost all of the edge betwenness centralities will have to be recalculated. It is very hard to figure out the second order effects of the edge disappearance on the shortest paths of the network. The edge betweenness of some edges might increase by one, but it is not easy to understand which ones.

In some cases -- and actually this might be very likely -- the network will be broken up into multiple components, meaning that no edge will increase its betweenness and, instead, many will lose part of their centrality. That is because now there will be many node pairs that cannot reach each other any more. Consider again Figure \ref{fig:centrality-betweenness2}: once we remove one of the two most central edges, no node in one clique can reach the nodes in the other one. All those paths passing through the removed edge are lost forever. The surviving edge between the two most central ones will have a much reduced edge betweenness centrality: it cannot be used to move between cliques any more. This consideration holds true not only for the edge betweenness, but also for the node betweenness.

A relaxed version of betweenness centrality does not use shortest paths, but random walks (Chapter \ref{cha:rndwalks}). This simulates the spreading of information into the network. The definition is similar: this ``flow'' centrality is the number of random walker passing through the node during the information spread event\cite{newman2005measure}\cite{brandes2005centrality}. Just like with the regular betweenness centrality, also in this case you can take an edge-centric approach, and count the number of random walks going through a specific edge. This has been used, for instance, to solve the problem of community discovery\cite{fortunato2004method}\cite{latora2005vulnerability}.

\section{Reach}\label{sec:centr-reach}
Reach centrality is only defined for directed networks. The local reach centrality of a node $v$ is the fraction of nodes in a network that you can reach starting from $v$\cite{mones2012hierarchy}. From this definition, one can see why it doesn't make much sense for undirected networks. If your network has a single connected component, then all nodes have the same reach centrality, which is equal to one. That is also the case if your directed network has only one strongly connected component. In a strongly connected component there are no ``sinks'' where paths get trapped, thus every node can reach any other node.

\begin{figure}
\centering
\includegraphics[width=.35\columnwidth]{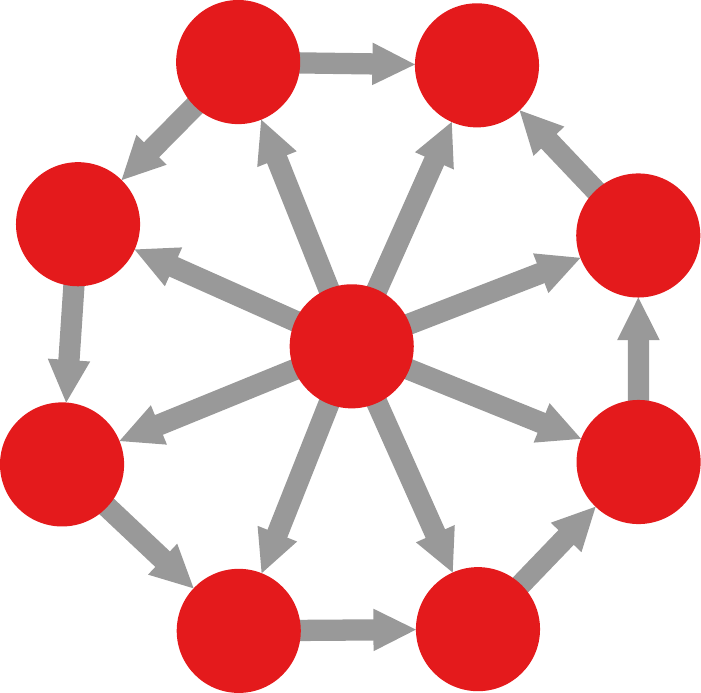}
\caption{A wheel graph with a flipped edge. The central node has the maximum reach centrality.}
\label{fig:centrality-reach}
\end{figure}

However, when you have multiple (or no) strongly connected components, reach centrality tells you how much you can command in your network if you're node $v$. Consider Figure \ref{fig:centrality-reach}. There is a clear boss in this figure: the central node. Following its edges, you can reach the entirety of the network. Thus its reach centrality is equal to one. On the other hand, the node on the top right has an out-degree of zero. You cannot reach anything from it, thus its reach centrality is zero. Reach centralities progressively increase if you follow the wheel clockwise, as more and more of the network gets reachable from the nodes' perspective.

Calculating the reach centrality is trivial. You start from node $v$ and you explore the graph with a BFS strategy. Once you cannot explore any more, you stop. The number of nodes you touched divided by the number of nodes in the network is your reach centrality. This is linear in the number of edges in the graph. 

Reach centrality is a key concept we use to detect the hierarchical structure of networks, as we will see in Chapter \ref{cha:hier}.

\section{Eigenvector}\label{sec:centr-eigen}
Betweenness centrality shares with closeness a drawback: computational complexity. Both measures require to calculate all shortest paths in the network. For large structures, this becomes unfeasible. The reason fully lies in the shortest paths calculation, which is very computationally expensive. One could approximate the node's importance for connectivity by looking not at shortest paths, but at random walks. 

This is different from the flow centrality I explained at the end of Section \ref{sec:centr-betw} because, in this case, we're not simulating a spreading event. Instead, we're looking at infinite length random walks and we're not bounded by origin-destination pairs of spreading events. Here, we are interested in knowing the expected probability of ending up in a node when we perform a random walk in the network. That is, we start from a random node and we keep choosing to traverse random edges. What's the likelihood of ending up in node $v$?

Calculating all shortest paths takes $|V|^3$ operations. By using clever linear algebra, running infinite length random walks could take only $|V|^2$. In fact, we already saw how to calculate the probability of ending in a node after an infinite length random walk: it is the stationary distribution, as I discussed in Section \ref{sec:rw-stationary}. By replacing the expensive step at the basis of betweenness centrality with simple random walks, you can obtain phenomenal speedups.

We call methods based on this technique ``Eigenvector Centralities'', as the stationary distribution is the leading left eigenvector of the stochastic adjacency matrix. If you take the straight up stationary distribution, you obtain what we call the eigenvector centrality\cite{seeley1949net}\cite{newman2016mathematics}.

\subsection{PageRank}
By far, the most famous approach in this category is PageRank\cite{page1999pagerank}: the algorithm that Google invented in $1998$ to rank webpages in their nascent search engine. PageRank is nothing more than calculating a stationary distribution over a directed adjacency matrix. PageRank differs from eigenvector centrality in one tiny -- but rather salient -- aspect.

If you remember Section \ref{sec:rw-stationary} you'll recall a small issue with the stationary distribution. If your network is not connected, meaning that it has more than one connected component (Section \ref{sec:paths-ccomps}), you will obtain multiple incomparable stationary distributions. This is bad news for the use case of PageRank: you want to use it to sort out webpages, and the users want to see a single ranking, not one per connected component!

\begin{figure}
\centering
\begin{subfigure}[t]{.4\columnwidth}
\includegraphics[width=\textwidth]{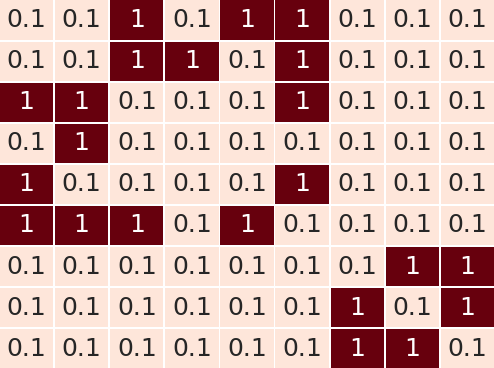}
\caption{}
\end{subfigure}
\quad
\begin{subfigure}[t]{.4\columnwidth}
\includegraphics[width=\textwidth]{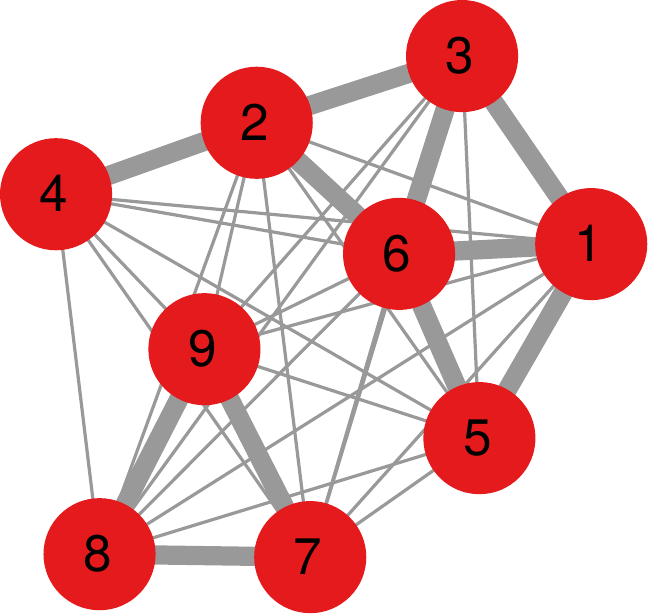}
\caption{}
\end{subfigure}
\caption{The practical implementation of the PageRank's teleportation trick: adjacency matrix (a) and resulting graph (b).}
\label{fig:stationary-teleport}
\end{figure}

Google's solution for the PageRank algorithm was to give the walker a teleportation device. At each step, the walker has a minuscule chance to request a teleportation, which then might land it on a different component. This mathematical trick is embarrassingly easy to implement. It is equivalent to the creation of ghost edges with very little weight connecting the entire graph. On matrix notation, this is the same as adding a tiny constant to the (not yet normalized) adjacency matrix: $A^* = A + \epsilon$. Figure \ref{fig:stationary-teleport} shows this teleportation trick in practice.

However, PageRank is not immune from downsides. PageRank is very close to the degree. How closely the degree approximates the PageRank depends on the value of our teleportation parameter $\epsilon$: in the literature, we call $1 - \epsilon$ the ``damping factor''. The magic value of $\epsilon$ is $0.15$, that is what Brin and Page used originally. If we set $\epsilon = 0$, PageRank is equivalent to $\pi$, and therefore to the degree\cite{fortunato2006approximating}\cite{ghoshal2011ranking}.

Of course, nowadays Google uses a much more complex algorithm to sort the results. Most of the tricks are either secret or too specialized to include here. However, there are a few tweaks of note. For instance, a very popular variant of PageRank is the personalized PageRank\cite{haveliwala2002topic}. In practice, one can split the network to a multilayer one depending on the topic of the hyperlink (e.g. its keywords) and calculate a set of PageRanks, one per topic. There are other possible ways to define a multilayer PageRank\cite{halu2013multiplex}.

\subsection{Katz}
Another popular variant of eigenvector centrality is Katz centrality\cite{katz1953new}. At a philosophical level, the difference between the two is that Katz says that nodes that are farther away from $v$ should count less when estimating $v$'s importance. So it matters whether $v$ is reached at the first step of the random walk, rather than at the second, or at the hundredth. For eigenvector centrality when you meet $v$ in a random walk makes no difference, for Katz it does.

If we were to write the eigenvector centrality not as an eigenvector, but as a sum, we would end up with something that looks a bit like this:

$$ EC_v = \sum \limits_{k=1}^{\infty} \sum \limits_{u \in V} (A^k)_{uv},$$

which means that $v$'s importance is the sum of the probabilities of getting from any $u$ to $v$ in $k$ steps, with $k$ going to infinity. Katz simply adds a term, $\alpha$, which is lower than one. He plugs it in the formula as follows:

$$ KC_v = \sum \limits_{k=1}^{\infty} \sum \limits_{u \in V} \alpha^k (A^k)_{uv}.$$

Since $0 < \alpha < 1$, as $k$ grows the contribution of $(A^k)_{uv}$ becomes more and more insignificant. Which is what Katz wants: longer walks contribute less to $v$'s centrality.

\subsection{UBIK}
A paper of mine presents UBIK, which is the lovechild between Katz centrality and the personalized PageRank I presented before\cite{coscia2013you}. The weird acronym is short for ``you (U) know Because I Know'' and we developed it with networks of professional in mind (like Linkedin). Each professional has skills, which allow her to perform her job. However, sometimes she is required to do something using a skill she doesn't have. In many cases, she might be able to perform the task anyway because she can ask for help in her social network. Think about any time you asked a friend to fix something in your script, or scrape some data, or patch a leaking water pipe.

\begin{figure}
\centering
\includegraphics[width=\columnwidth]{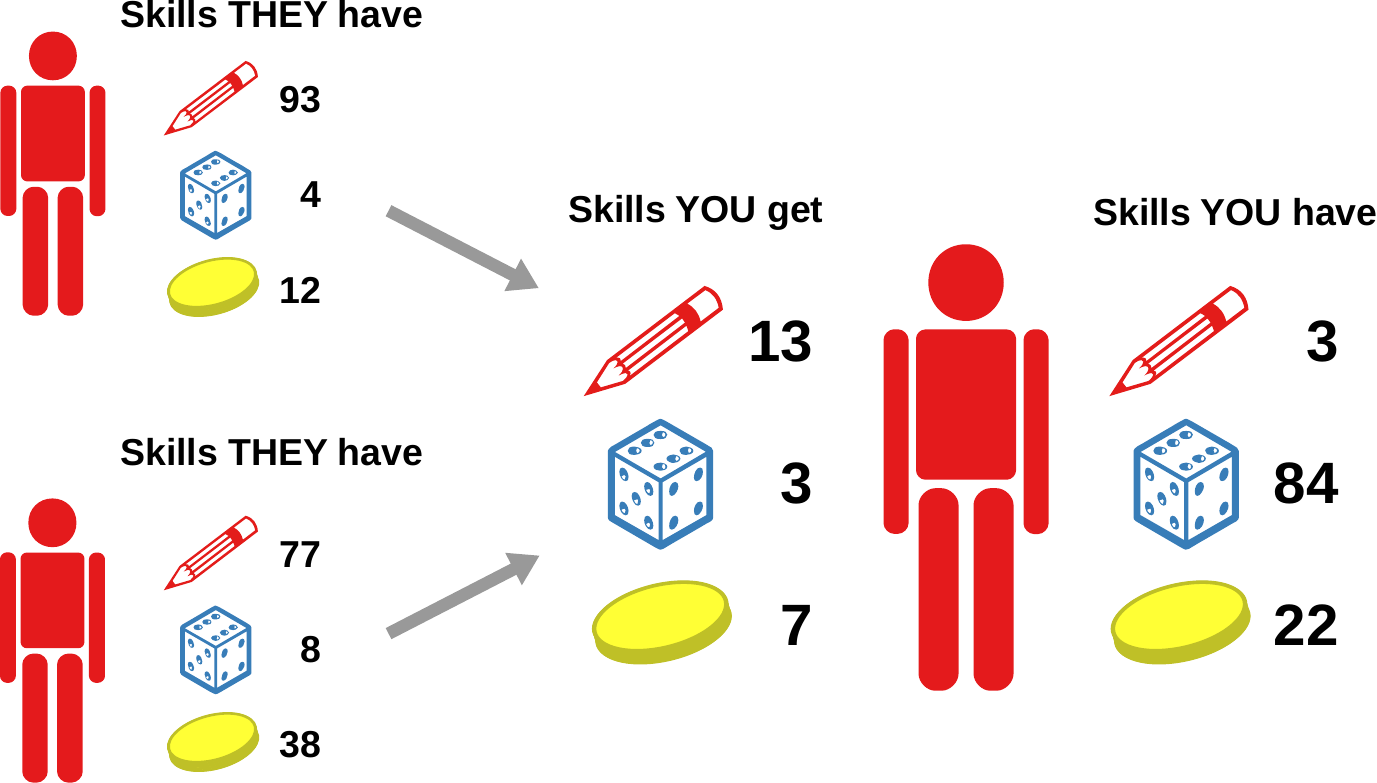}
\caption{How UBIK works. Each node (you) has different proficiency for different skills, here we have three: writing (red), statistical analysis (blue), and finance (yellow). But what each node really can do is the sum of what it knows plus a combination of what its neighbors know. You get from your social network some skills, in this case the sum of their skills raised to the power of $-1/l$, where $l$ is the degrees of separation plus one. So, in this case, the neighbors provide $13$ writing skill points, because $(93+77)^{1/2} \sim 13$.}
\label{fig:ubik}
\end{figure}

Of course, if the task you need to perform requires only knowledge you have, you can do it quickly. Every level of social interaction you add will slow you down. If you're a computer scientist you can think about this as memory layers. What you know is in your brain, your cache: it is ready to run on your CPU. What your friends know is the main memory, the RAM. The main memory is slower than the cache, because the data need to travel from the memory to the cache before it can be used. Your friends' friends are like a hard disk, and the friends of the friends of your friends are the Internet: a limitless amount of distributed information that is hard to search and collect. Figure \ref{fig:ubik} shows a vignette of this process.

So in UBIK we take the stance that a person's knowledge is the sum of her own knowledge plus some combination of her social networks and, ultimately, of mankind. This lofty philosophical picture boils down to just adding a few bells and whistles to Katz centrality. First, we don't use a simple graph, but a multilayer network. Different types of friends might have different levels of willingness or reactivity when asked to help. A colleague is just down the corridor, a close friend might want to do anything for you, that person you dated once during college maybe will pick up the phone if you call. So we have different adjacency matrices $A$ with a different topology and a different coefficient favoring or hampering the centrality contribution.

Second, rather than giving a single centrality score to each node, we have multiple. Each node gets a different score for each skill. You might be a dragon when it comes to do multivariate regression analysis, but unable to make yourself understood in an email. As a consequence, the initial condition is also different. The skills aren't distributed equally in the network. The nodes don't all start from the same level in all skills. Each node has its own personal story, and might start with higher scores in some skills and lower in others.

\begin{figure}
\centering
\begin{subfigure}[t]{.33\columnwidth}
\includegraphics[width=\textwidth]{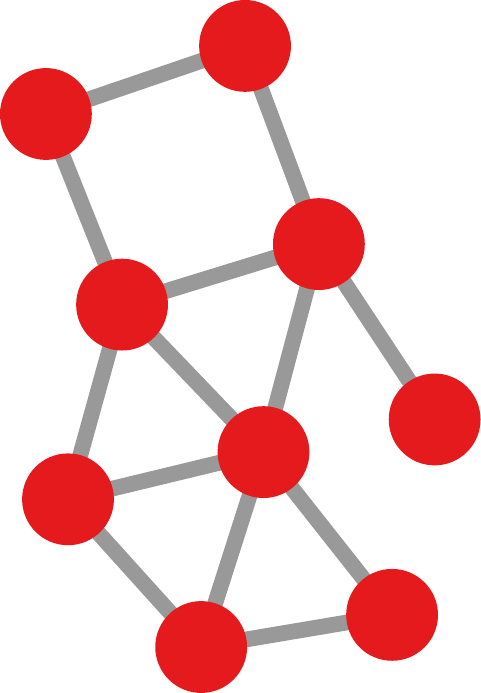}
\caption{Katz}
\end{subfigure}
\quad
\begin{subfigure}[t]{.33\columnwidth}
\includegraphics[width=\textwidth]{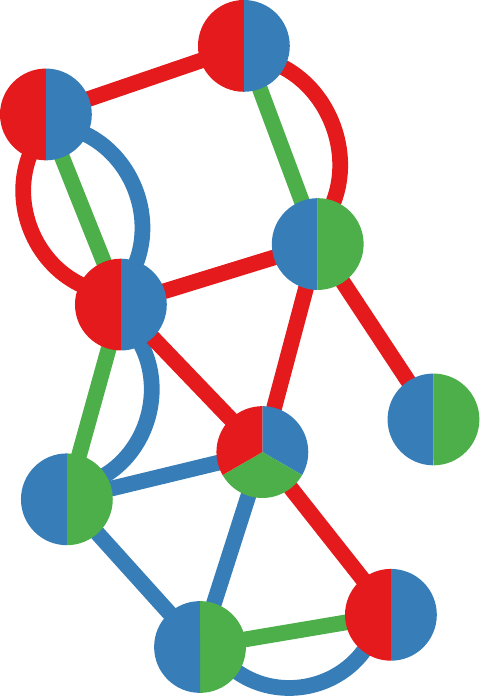}
\caption{UBIK}
\end{subfigure}
\caption{The difference in input between Katz and UBIK centralities. The node's color determines its initial condition: how much of a skill/centrality it possesses (different colors represent different skills/centralities). The edge's color determines its layer. Note how all nodes and all edges in Katz have the same color.}
\label{fig:ubik-vs-katz}
\end{figure}

So the difference between UBIK and Katz is basically in the input data. Where Katz works with a single layer network, a single centrality measure, and a uniform initial condition, UBIK uses a multilayer network, with multiple centrality scores, initialized differently for different nodes. Figure \ref{fig:ubik-vs-katz} depicts this difference. Then the process is practically the same: direct neighbors have a big effect on your centrality scores and, as you go to more and more degrees of separation, the contributions fade away to zero. Sure, UBIK has to do this multiple times for each skill and needs the extra parameters to distinguish between different layers but, at the end of the day, UBIK is a glorified Katz centrality.

Where UBIK shines is in the analysis of so-called ``expertise networks'': web-based communities of experts helping each other with problems related to their professions\cite{zhang2007expertise}\cite{adamic2008knowledge}. One could also use it to investigate the question whether team formation is a process that happens better if it is organized from the top -- like in organizations -- or spontaneously from the bottom, like it happens for instance in large open source software projects\cite{bird2008latent}.

\subsection{Alpha}
Another variant of eigenvector centrality is Bonacich's Alpha centrality. If Katz wanted to penalize long walks, Bonacich wants to add an external source of importance to the node's centrality\cite{bonacich2001eigenvector}. Practically, we are saying that, to know how important a node is in a network, we don't have to look exclusively at the topology of the network. The node might get its importance from somewhere else. A Web without Google would be poorer even if \texttt{google.com} would not be the most central node in the hyperlink network.

If, as we saw before, we can express the vanilla eigenvector centrality as an infinite sum:

$$ EC_v = \sum \limits_{k=1}^{\infty} \sum \limits_{u \in V} (A^k)_{uv},$$

then we can express Alpha centrality as the same sum, plus am external source of non-network importance:

$$ EC_v = (1 - \alpha) e_v  + \sum \limits_{k=1}^{\infty} \sum \limits_{u \in V} \alpha (A^k)_{uv}.$$

Differently from Katz, $\alpha$ doesn't change as the length $k$ increases: it just regulates how much weight we give to the traditional part of the eigenvector centrality. If $\alpha = 0$, then $100\%$ of the node's importance comes from the vector $e$. Each entry $e_v$ of $e$ is the external importance of node $v$. In the Web network, Google's $e_v$ would be through the roof. On the other hand, if $\alpha = 1$, this reduces to the classical eigenvector centrality.

\section{HITS}
HITS\cite{kleinberg1999web}\cite{kleinberg1999authoritative} is an algorithm designed by Jon Kleinberg and collaborators to estimate a node's centrality in a directed network. It is part of the class of eigenvector centrality algorithms from Section \ref{sec:centr-eigen}, but it deserves its own section due to its interesting characteristics. Differently from other centrality measures, HITS assigns \textit{two} values to each node. In fact, one can say that HITS assigns nodes to one of two roles -- we will see more node roles in Chapter \ref{cha:centr-roles}. The two roles are ``hubs'' and ``authorities''.

\begin{figure}
\centering
\begin{subfigure}[t]{.45\columnwidth}
\includegraphics[width=\textwidth]{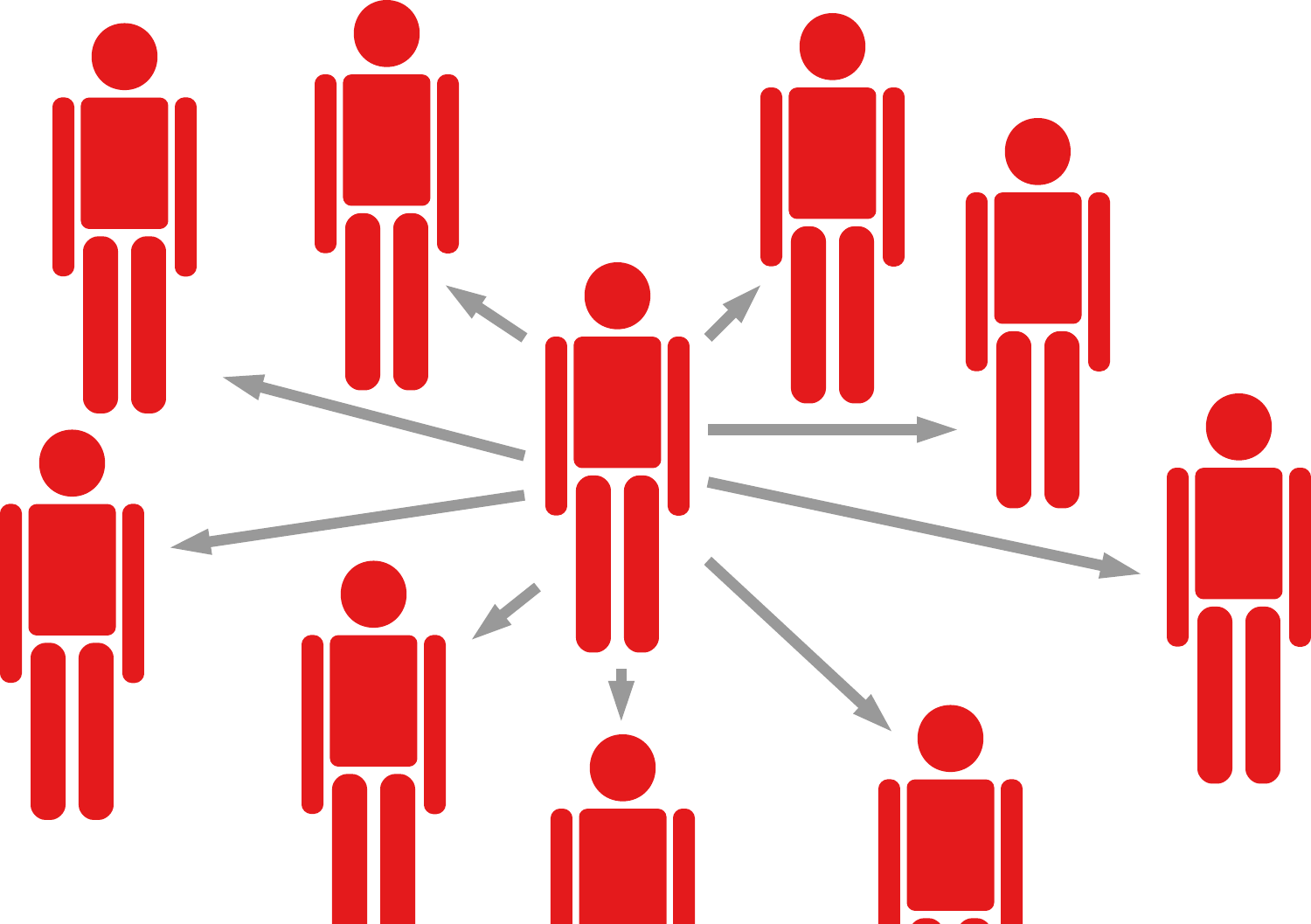}
\caption{Hub.}
\end{subfigure}
\qquad
\begin{subfigure}[t]{.45\columnwidth}
\includegraphics[width=\textwidth]{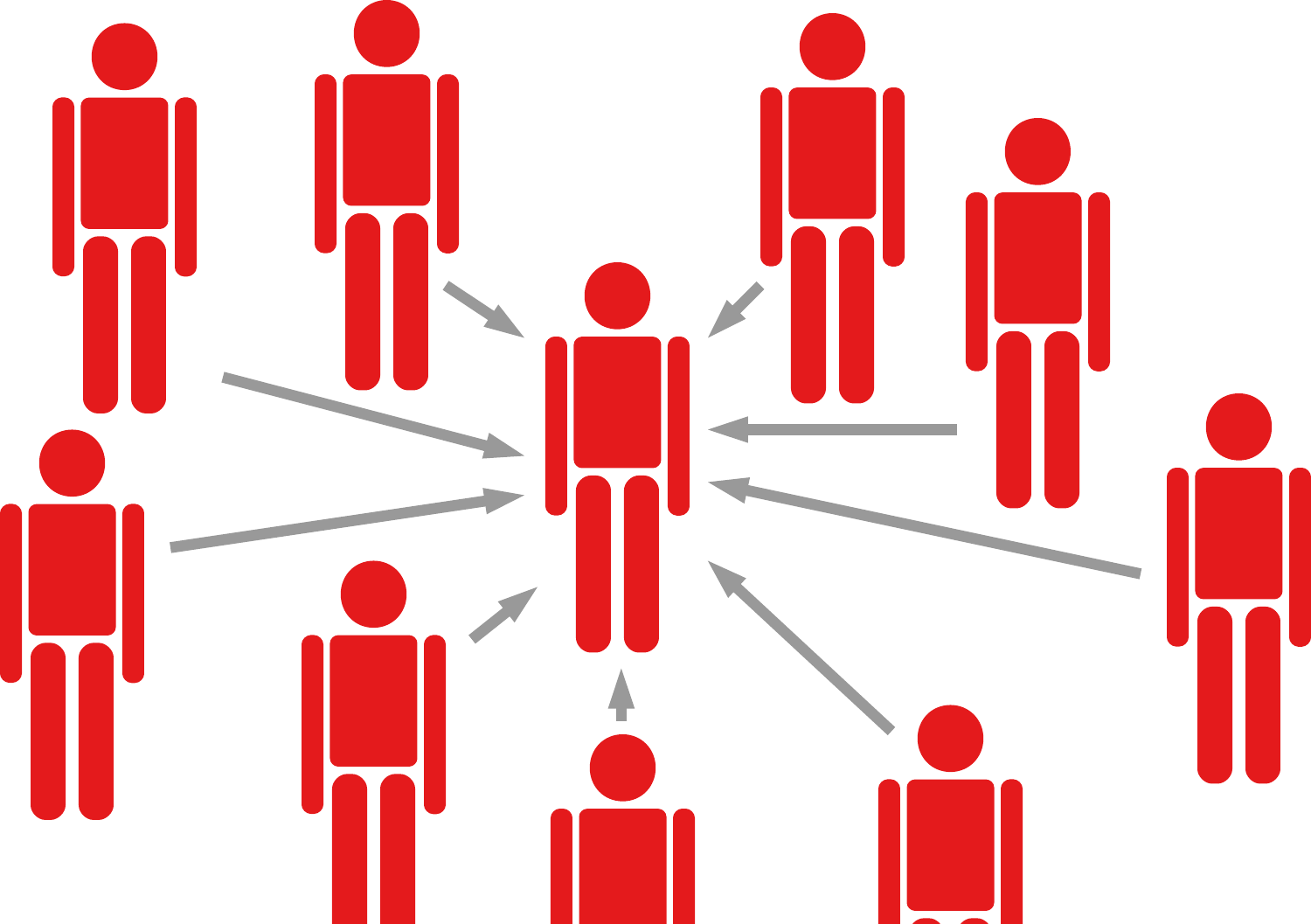}
\caption{Authority.}
\end{subfigure}
\caption{Hubs and authorities in directed networks.}
\label{fig:roles-hits}
\end{figure}

In a sense, both hubs and authorities are central nodes in the network. However, when you're dealing with directed networks, there are two ways in which a core member of a community can play its role. A core member might be a person who maybe does not know many things, but knows the people who know them. You will go to this member with a question and she will \textit{point to} someone who knows the answer. We call such linking resource a ``hub''. Figure \ref{fig:roles-hits}(a) provides an illustration.

The converse role of a hub is an authority. This is in principle the exact opposite of a hub -- although it's possible for a node to be partly a hub and partly an authority at the same time --: this person might not know many people in the social circle, but she has mastered her own topic of specialization. Everybody knows that, and so she is \textit{pointed by} everyone when someone asks about that particular topic. This happens because she is an ``authority'' on the subject. Figure \ref{fig:roles-hits}(b) provides an illustration.

Hubs and authorities are an instance in which the quantitative approach of the centrality measures and the qualitative approach of the node roles meet. There is a way to estimate the degree of ``hubbiness'' and ``authoritativeness'' in a network. This is what the HITS algorithm does. The underlying principle is very simple. A good hub is a hub that points to good authorities. A good authority is an authority which is pointed by good hubs. These recursive definitions can be solved iteratively -- or, more efficiently, with clever linear algebra -- and they eventually converge.

\begin{figure*}
\centering
\begin{subfigure}[t]{.3\columnwidth}
\includegraphics[width=\textwidth]{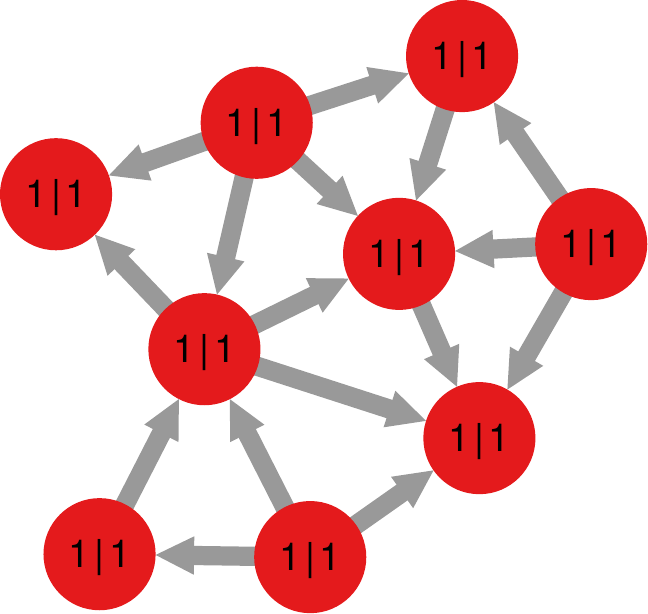}
\caption{0th iteration.}
\label{fig:hits-0}
\end{subfigure}
\quad
\begin{subfigure}[t]{.3\columnwidth}
\includegraphics[width=\textwidth]{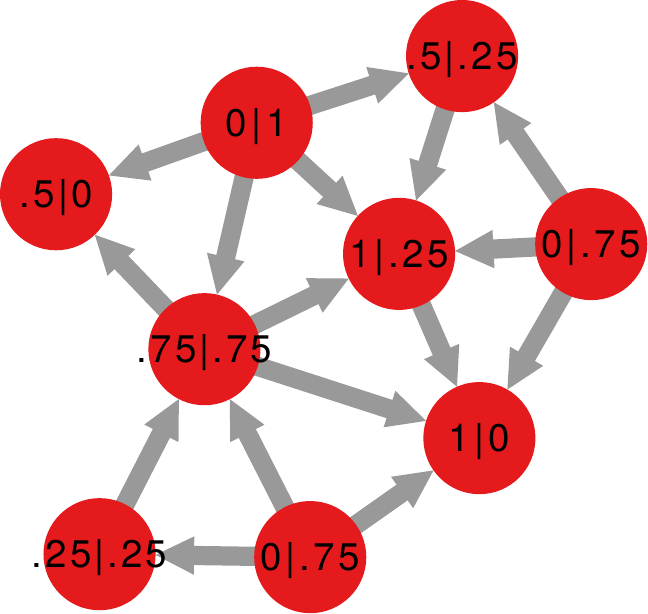}
\caption{1st iteration}
\label{fig:hits-1}
\end{subfigure}
\quad
\begin{subfigure}[t]{.3\columnwidth}
\includegraphics[width=\textwidth]{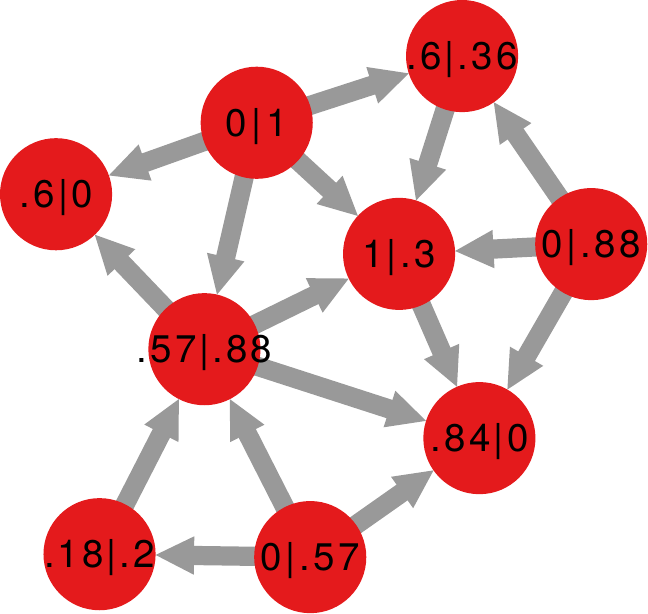}
\caption{n-th iteration.}
\label{fig:hits-n}
\end{subfigure}
\caption{A sample progression of the HITS algorithm to estimate hub and authority scores. Node labels are their authority (left) and hub (right) scores, separated by a pipe.}
\label{fig:hits}
\end{figure*}

Figure \ref{fig:hits} shows the progress when calculating the measure. Before the first iteration we assume that each node is equal. They thus have all the same hub score and authority score, equal to one -- Figure \ref{fig:hits}(a). At each iteration, we sum all the incoming hub scores of a node to determine its authority score. At the same time, we sum all the outgoing authority scores of a node to obtain its new hub score. We then normalize so that the maximum hub and authority score is one. At the first iteration -- Figure \ref{fig:hits}(b) -- hub and authority scores are equivalent to a normalized out- and in-degree, respectively.

After a sufficient number of iterations -- Figure \ref{fig:hits}(c) -- the scores stabilize. We can see that nodes with the same in-degree can have different authority scores -- the same holds for hub scores. Consider the two nodes with in-degree four: one has the maximum score of one, while the other has a score of $0.84$. This is because the more authoritative node has, on average, incoming connections from more reputable hubs.

HITS is an important algorithm in the computer science portion of the network analysis community. It was modified and extended in a number of ways, notably to work on multilayer networks\cite{kolda2006tophits}, enabling topic-dependent hub-authority scores. SALSA\cite{lempel2001salsa} is also a related method.

\section{Harmonic}
PageRank solves the problem of networks with multiple connected components. This is a common problem to have: all centrality measures based on shortest paths or random walks are ill defined when your network has pairs of unreachable nodes. This includes closeness, betweenness, reach, ... practically everything. But PageRank and its teleportation trick is not the only way to deal with multiple components.

The crux of the issue is that you cannot compare the closeness centrality of two nodes from different connected components, because one might have a higher closeness simply because its connected component is smaller. One approach to fix this issue is to consider the component size in the measure. We want the desirable property of saying that a central node in a large component is more important than a central node in a smaller component. Lin's centrality\cite{lin1976foundations} achieves this by multiplying the closeness centrality of a node by the size of its connected component, which -- incidentally -- just means to square the numerator:

$$LC_v = (|V_v| - 1)^2 / \sum_{u \in V_v} |P_{vu}|,$$

with $V_v \subseteq V$ here being the set of nodes part of the component in which $v$ resides.

Harmonic centrality represents another alternative which has been discussed in many slight different variations and scenarios\cite{marchiori2000harmony}\cite{rochat2009closeness}\cite{boldi2014axioms}\cite{cohen2007spatially}\cite{pan2011path}. In practice, you calculate the harmonic mean of all distances -- even those between unreachable nodes. Thus:

$$HC_v = \sum_u \dfrac{1}{|P_{vu}|}.$$

The harmonic centrality handles unreachable nodes properly, based on the assumption that $1 / \infty = 0$.

\section{k-Core}\label{sec:centr-kcore}
When it comes to node centrality, one common term you'll hear thrown around is one of ``core'' node. This is usually a qualitative distinction -- see Chapter \ref{cha:coreperiph}, but sometimes we need a quantitative one. With k-core centrality we look for a way to say that a node is ``more core'' than another. A k-core in a network is a subset of its nodes in which all nodes have at least $k$ connections to each other\cite{seidman1983network}. A connected component of a network is always a 1-core: each node in the component has at least one connection to the rest of the component. In a 2-core, each node must have at least two connections to the other nodes in the 2-core.

\begin{figure*}
\centering
\begin{subfigure}[t]{.19\columnwidth}
\includegraphics[width=\textwidth]{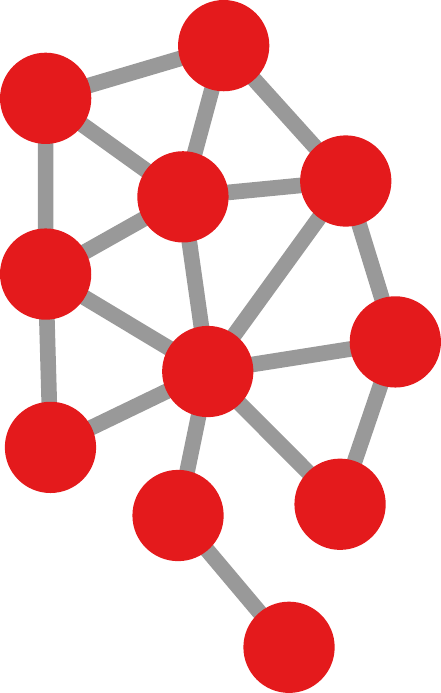}
\caption{}
\label{fig:kcore-0}
\end{subfigure}
\begin{subfigure}[t]{.19\columnwidth}
\includegraphics[width=\textwidth]{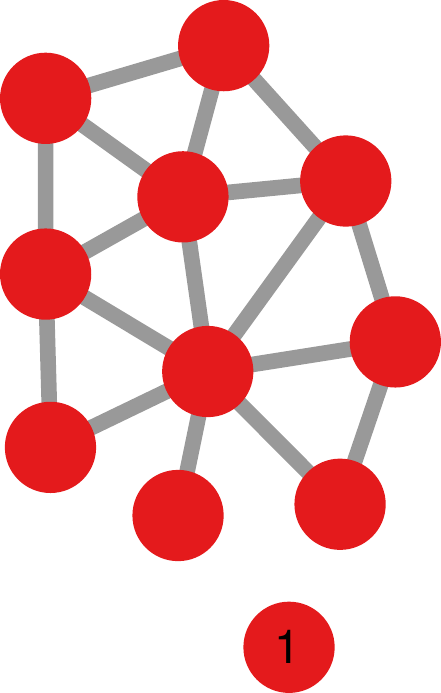}
\caption{}
\label{fig:kcore-1}
\end{subfigure}
\begin{subfigure}[t]{.19\columnwidth}
\includegraphics[width=\textwidth]{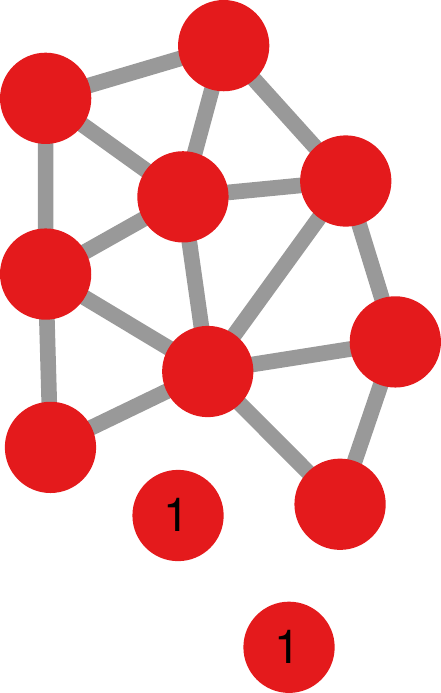}
\caption{}
\label{fig:kcore-2}
\end{subfigure}
\begin{subfigure}[t]{.19\columnwidth}
\includegraphics[width=\textwidth]{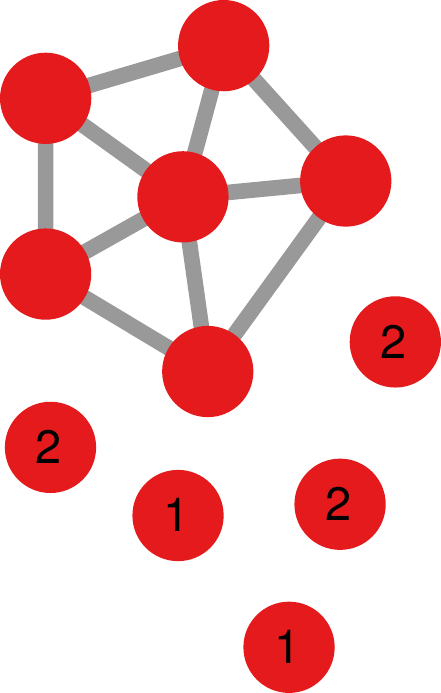}
\caption{}
\label{fig:kcore-3}
\end{subfigure}
\begin{subfigure}[t]{.19\columnwidth}
\includegraphics[width=\textwidth]{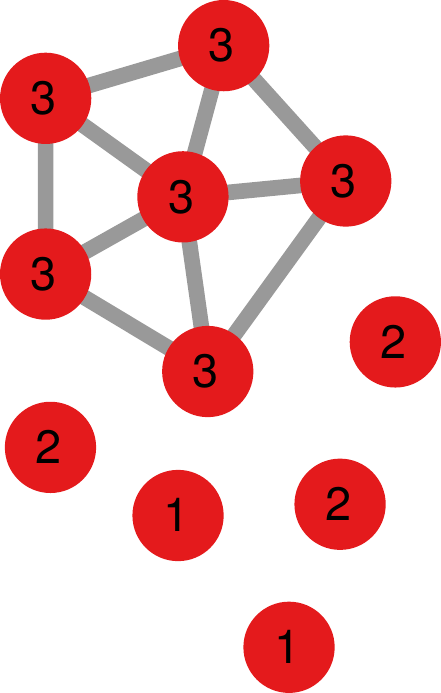}
\caption{}
\label{fig:kcore-4}
\end{subfigure}
\caption{The steps to determine the k-core value for each node in the network. (a) The starting network. (b-e) The steps of the algorithm.}
\label{fig:kcore}
\end{figure*}

One can easily identify the k-core of a network via the k-core decomposition algorithm\cite{batagelj2003m}. In Figure \ref{fig:kcore} we represent a stylized version of it. Figure \ref{fig:kcore}(a) shows the original network.

The first step is identifying the nodes with degree one. They are labeled as part of the 1-core of the network, and removed from the structure. -- Figure \ref{fig:kcore}(b). We need to apply this step recursively: there could be nodes that originally had degree two, but now have lower degree because we removed one of their neighbors (or both!). Also these nodes are part of the 1-core of the network -- Figure \ref{fig:kcore}(c).

Once the minimum degree in the network is higher than one we can proceed to the next phase. In this phase we identify the nodes that are part of the 2-core of the network. These are, unsurprisingly, the nodes with degree two -- and all nodes whose degree lowers to two or less once we remove their neighbors during this step (Figure \ref{fig:kcore}(d)). We continue the procedure to detect 3-, 4-, ..., k-cores until there are no remaining nodes in the network -- Figure \ref{fig:kcore}(e).

Note that the k-core decomposition approach is only the most famous among many similar which define a structure of interest and use it to define a centrality measure. Among popular examples we find D-cores\cite{giatsidis2011d} (for directed networks), k-shells\cite{carmi2007model}, k-coronas\cite{goltsev2006k}, and more.

\section{Centralization}\label{sec:centr-centr}
This chapter is all about knowing which nodes are central in a network. So it is a node-centric chapter. To wrap it up, let's change the perspective a little. Let's see how node centrality can say something about your network as a whole. This would be the \textit{centralization} of the network. A network is centralized when there is one node in it that is so much more central than everything else. Consider a star, where one node is in the middle, it is connected to every other node in the network and there are no other connections between its neighbors. A network cannot get more centralized that that\cite{freeman1978centrality}.

\begin{figure}
\centering
\begin{subfigure}[t]{.4\columnwidth}
\includegraphics[width=\textwidth]{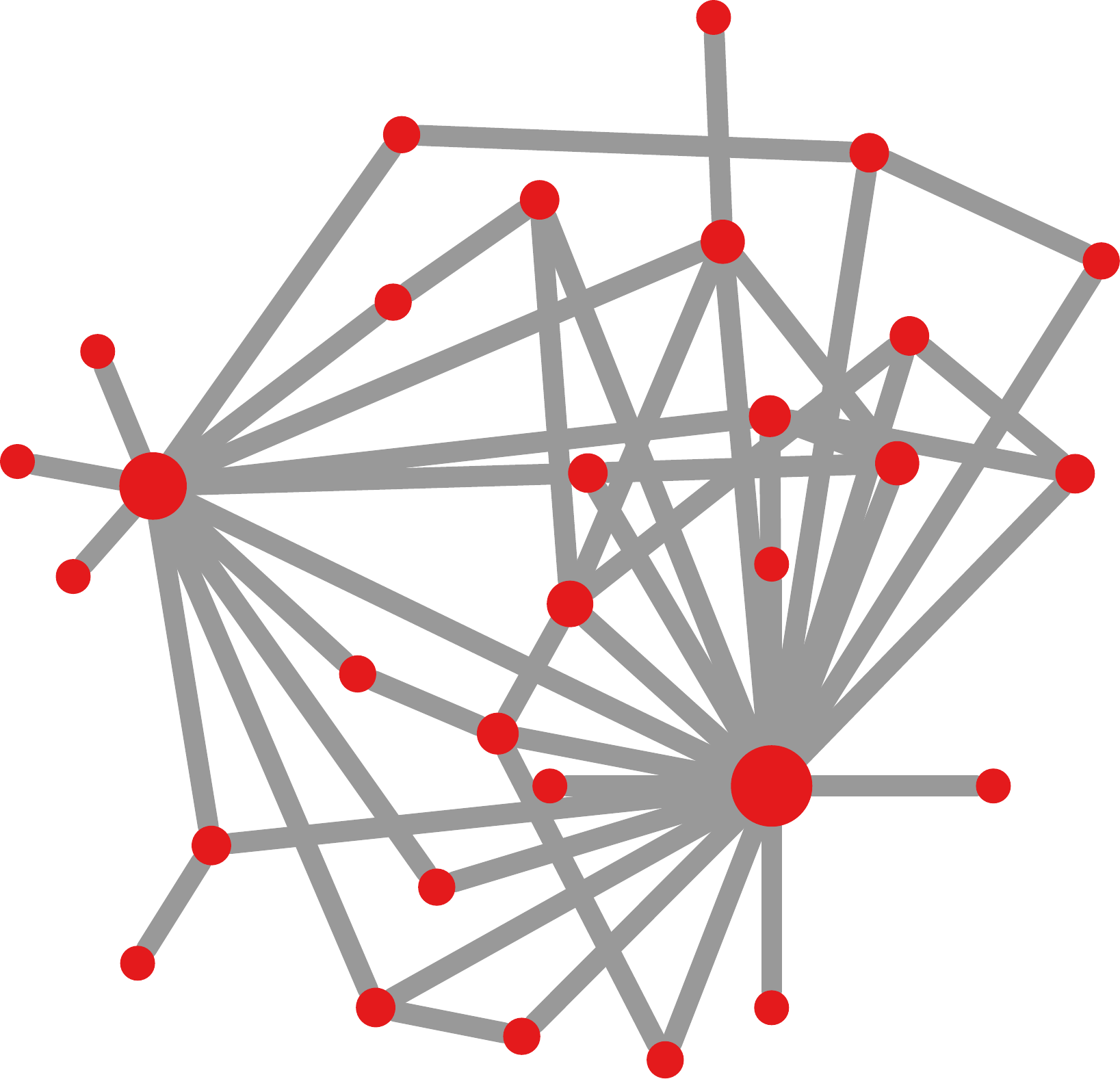}
\caption{$2.98$}
\end{subfigure}
\qquad
\begin{subfigure}[t]{.4\columnwidth}
\includegraphics[width=\textwidth]{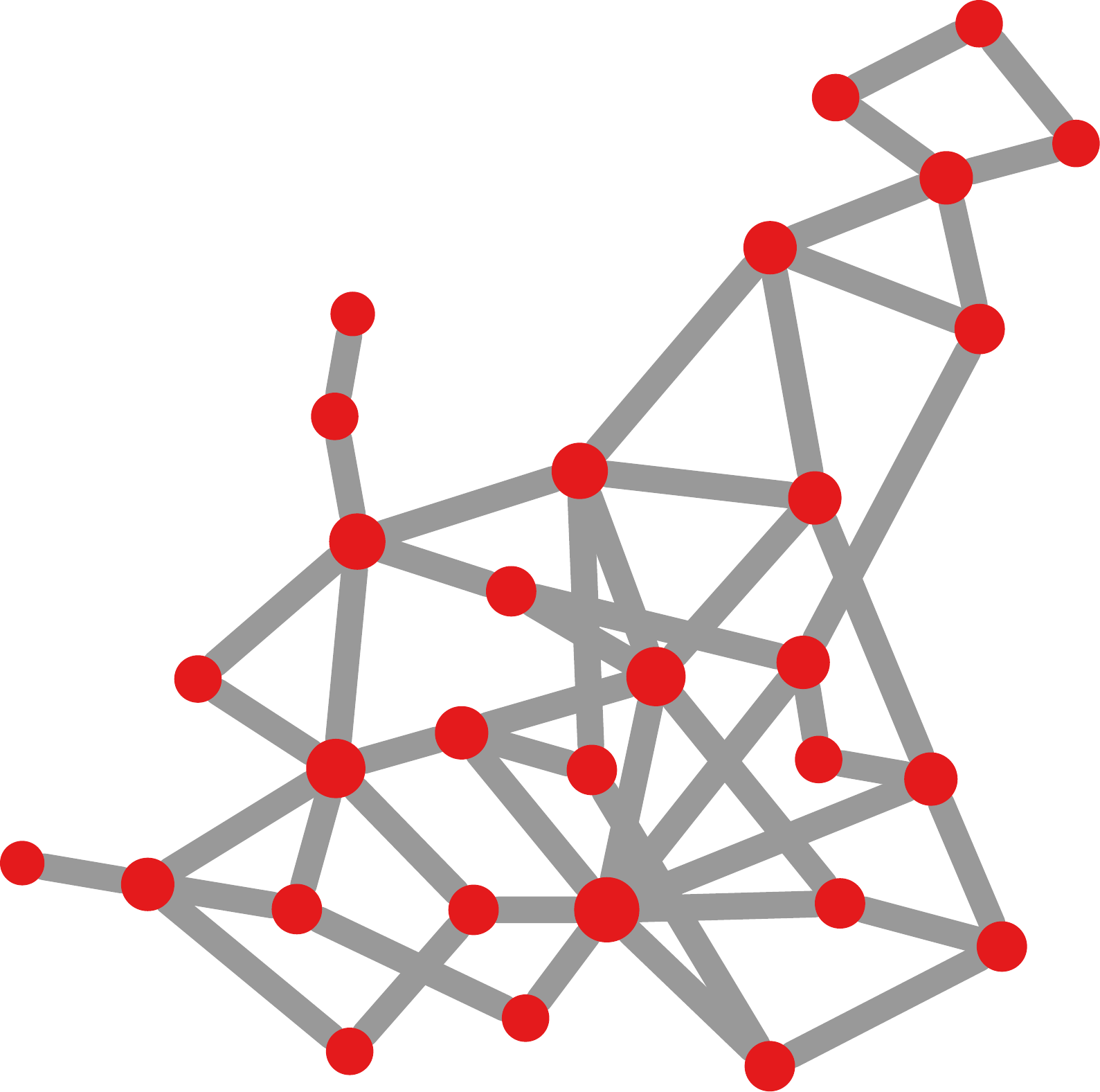}
\caption{$3.30$}
\end{subfigure}
\caption{Two networks with different levels of centralization. The node size is proportional to its degree. Entropy values in the captions.}
\label{fig:centralization-entropy}
\end{figure}

There are two main ways to detect centralization. The first approach is information-theoretic. You can look at the degree distribution and transform it into a probability distribution. This distribution answers the question ``What is the probability that an edge will be attached to this node?''. For instance, if you have the degree sequence $[4,2,2,1,1]$, you divide every entry by the sum: $[0.4,0.2,0.2,0.1,0.1]$. Now you can calculate Shannon's information entropy. As Figure \ref{fig:centralization-entropy} shows, a more centralized network will generate lower entropies\cite{dehmer2011history} given that there are few nodes connected to everything.

One issue with this approach is that it's not immediately obvious how centralized a network is by simply looking at the entropy value. Larger networks will generate higher entropies without necessarily being more centralized, because you need more bits to encode more entries -- check Section \ref{sec:prob-mi} for a refresher. You could normalize the entropy with the maximum entropy you'd get with $|V|$ entries -- the number of nodes in your network -- which is $\log |V|$. However, it's not that simple. The maximum entropy for a graph of $30$ nodes like the ones in Figure \ref{fig:centralization-entropy} is $3.40$. This means that the apparently centralized graph in Figure \ref{fig:centralization-entropy}(a) still has $87\%$ of the maximum entropy, which would lead you to conclude that it's decentralized. So you also need to figure out what is the \textit{minimum} possible entropy given a graph, which can be tricky\cite{cambie2022resolution}\cite{cambie2024extremal}. 

Note that Shannon's entropy is not the only option if you want to go this way: there's also Von Neumann's entropy\cite{passerini2008neumann}\cite{simmons2017quantum}. However, it's not the easiest thing in the world to calculate or interpret\cite{han2012graph}\cite{minello2019neumann}. And it tends to return similar values to Shannon's entropy anyway, so I personally stick with the version of entropy I explained.

One can also characterize the entropy of network ensembles\cite{bianconi2013statistical}\cite{anand2009entropy} -- i.e. what is the general entropy of all graphs generated with a given set of rules (which we'll see in Part \ref{par:synthnet}). This leads to interesting discoveries, such as that most naturally occurring graphs are part of families with lower entropy, which is not what you'd expect given the first expectation about entropy in the physical world -- entropy tends to be maximized.

If you don't want to go the information-theory route, you can calculate any centrality measure and see how skewed the maximum is with everything else. The procedure follows two steps. First, you sum the centrality differences between the most central node in the network and all other nodes. Then you calculate what would be the largest theoretical sum of differences in networks of comparable size. Usually the maximum is obtained by a star graph with the same number of nodes of your original network. The ratio between the two is the degree of centralization.

\begin{figure}[b]
\centering
\includegraphics[width=.8\columnwidth]{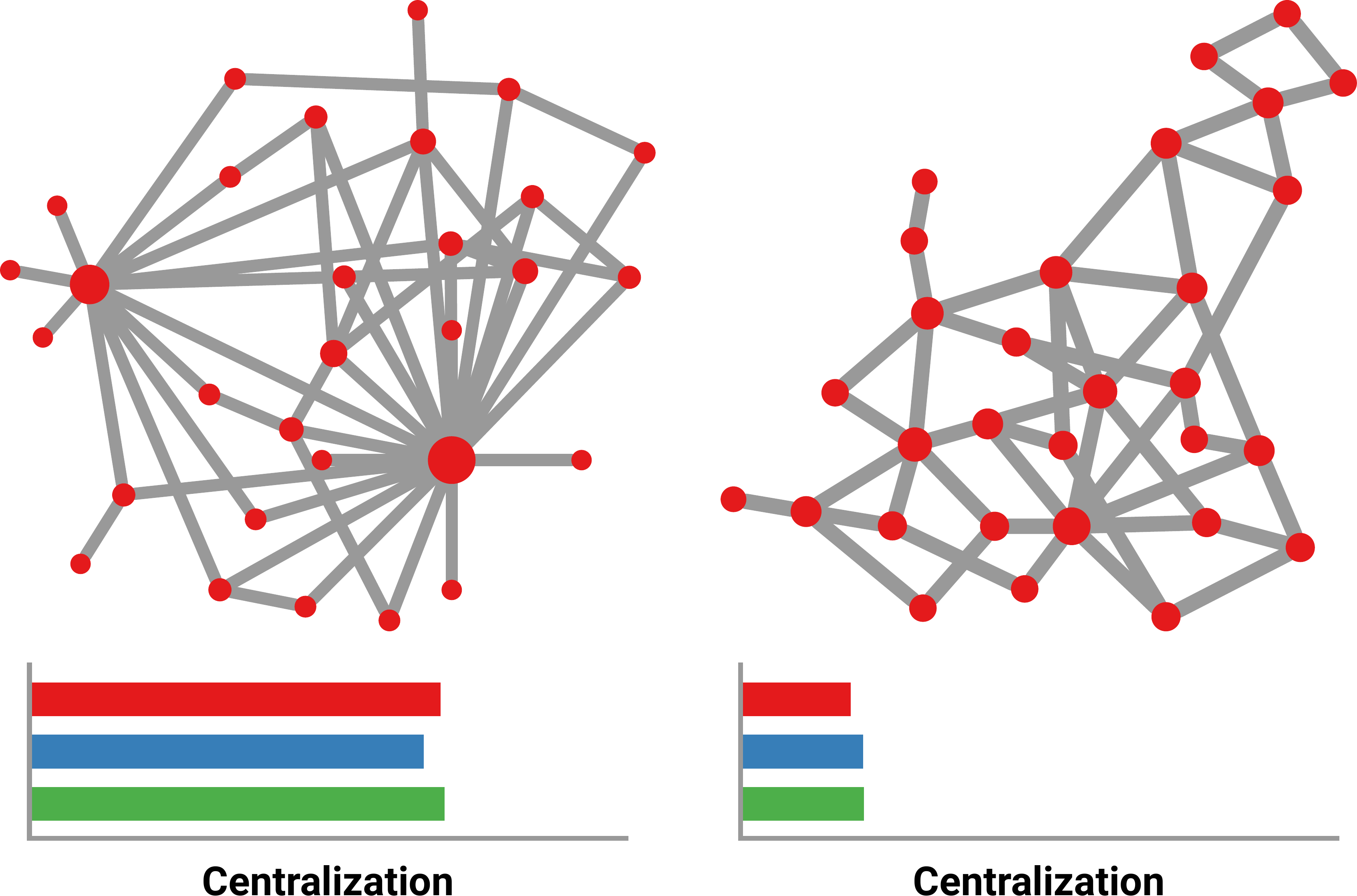}
\caption{The bar chart below the network indicates its three centralization values according to degree (red), betwenness (blue) and closeness (green) centrality.}
\label{fig:centralization-classical}
\end{figure}

Note that, depending on the centrality measure you picked, you're going to obtain different results. Figure \ref{fig:centralization-classical} shows a case using degree, betweenness, and closeness centrality. 

While the centralization measures agree that the network on the left is more centralized than the network on the right, the levels of centralization differ. For the network on the left, degree centrality overestimates centralization when compared to betweenness centrality. For the network on the right, the opposite happens.

There are variants of centralization measures. For instance, you can allow the graph to have multiple central points, not just the one with the maximum centrality. Once you add this degree of freedom, you can also ask yourself: given that the graph has multiple centers, are these centers close together, or are they scattered far apart? In the former case the graph is more centralized than in the latter.

\section{Summary}

\begin{enumerate}
\item We've seen many alternatives to the degree to estimate a node's importance. Many are based on shortest paths. The first measure is closeness centrality, answering the questions: how far is on average a node from every other node in the network? Betweenness centrality, instead, asks: how many shortest paths would become longer if this node were to disappear?
\item Alternatively, you can look at random walks, since they're less computationally expensive to calculate. You can calculate a family of eigenvector centralities, of which PageRank is one of the most famous examples.
\item HITS is another famous eigenvector centrality measure for directed networks, which divides nodes in two classes: hubs, who dominate out-degree centrality; and authorities, who dominate in-degree centrality.
\item Harmonic centrality is a version of closeness centrality which solves the issue of networks with multiple connected components. In such networks, there are pairs of nodes that cannot reach each other, thus other approaches based on shortest paths and random walks wouldn't work.
\item k-Core decomposition also works with networks with multiple components. It recursively removes nodes from the network with increasing degree thresholds. At iteration $k$, we say that surviving nodes are part of the $k$th core.
\item Regardless of your centrality measure, you can estimate how centralized your network is by comparing the highest observed centrality with the theoretical maximum centrality of a network with the same number of nodes: a star graph.
\end{enumerate}

\section{Exercises}

\begin{enumerate}
\item Based on the paths you calculated for your answer in the previous chapter, calculate the closeness centrality of the nodes in Figure \ref{fig:spantree}(a).
\item Calculate the betweenness centrality of the nodes in Figure \ref{fig:spantree}(a). Use to your advantage the fact that there is a bottleneck node which makes the calculation of the shortest paths easier. Don't forget to discount paths with alternative routes.
\item Calculate the reach centrality for the network in \url{http://www.networkatlas.eu/exercises/14/3/data.txt}. Keep in mind that the network is directed and should be loaded as such. What's the most central node? How does its reach centrality compare with the average reach centrality of all nodes in the network?
\item What's the most central node in the network used for the previous exercise according to PageRank? How does PageRank compares with the in-degree? (for instance, you could calculate the Spearman and/or Pearson correlation between the two)
\item Which is the most authoritative node in the network used for the previous question? Which one is the best hub? Use the HITS algorithm to motivate your answer (if using \texttt{networkx}, use the \texttt{scipy} version of the algorithm).
\item Based on the paths you calculated for your answer in the previous chapter, calculate the harmonic centrality of the nodes in Figure \ref{fig:spantree}(a).
\item Calculate the k-core decomposition of the network in \url{http://www.networkatlas.eu/exercises/14/7/data.txt}. What's the highest core number in the network? How many nodes are part of the maximum core?
\item What's the degree of centralization of the network used in the previous question? Compare the answer you'd get by using, as your centrality measure, the degree, closeness, and betweenness centrality.
\end{enumerate}

\chapter{Node Roles}\label{cha:centr-roles}
Sometimes, the differences between nodes can be estimated quantitatively. A person is measurably more or less connected in a social network. That is what we described in the previous chapter: if you can estimate the importance of the node (number of connections, centrality, etc), you do so by calculating the corresponding quantitative measure (degree, betweenness, etc).

Sometimes you cannot put a number to what you're trying to describe. What the person is doing in the social network does not have a quantity, but a quality: she is playing a specific role, which does not have a countable result. This could be explicitly represented in your data as a node attribute (Section \ref{sec:extended-nodeattr}): for instance, in a corporate network, nodes might be explicitly labeled as managers, executive, technicians, etc. But one thing is clear: in most cases, nodes perform different roles in the network.

If you don't have explicit qualitative data, you might want to put a label on the nodes based on the structural network data. Rather than being a characteristic of the person by itself, the node role is determined by her position in the network.

There are many ways to define node roles, dependent on the aspect of the network you want to describe. The main split in the literature is on the type of procedure you're following: unsupervised or supervised. In unsupervised role learning, no node in your data has a role and you're making up your own definition of roles depending on what's meaningful to you. This is the classical approach, which I dissect in Sections \ref{sec:centr-roles-classic} and \ref{sec:centr-similarity}. In supervised node learning, you already have partially labeled data and you want to figure out what are the underlying rules determining the node roles. This is the theme of Section \ref{sec:centr-roles-conv}.

\section{Classic Node Classification}\label{sec:centr-roles-classic}
In this section we introduce the concept of node roles by picking network communities as our focus, just to give an example. If your focus is different, you will probably define different roles -- for instance, you could look at paths in a directed network\cite{cooper2010role}. In fact there are countless centrality measures developed to identify specific node roles in complex networks\cite{borgatti2006graph}.

\begin{figure}
\centering
\includegraphics[width=.75\columnwidth]{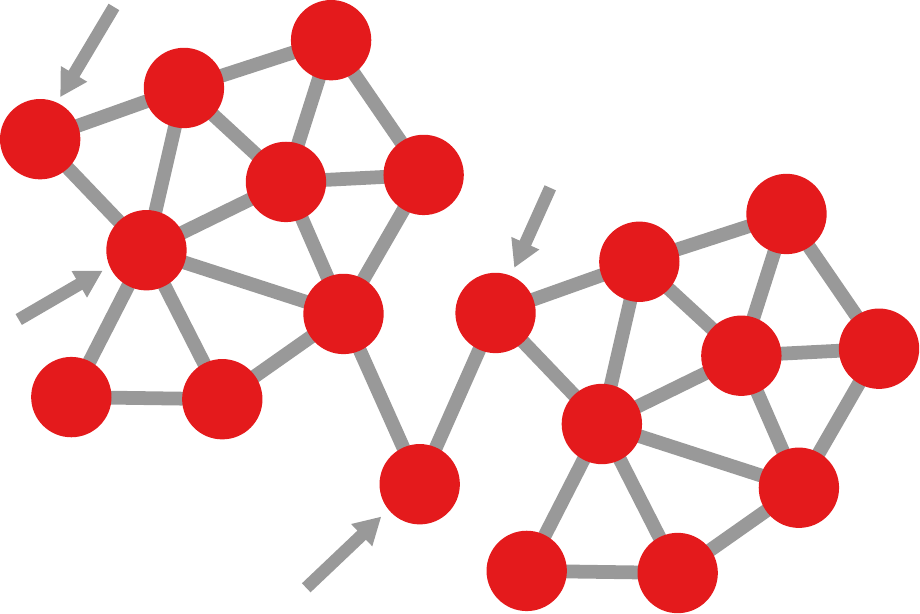}
\caption{Two hypothetical social circles: groups of nodes connected to each other on the left and on the right, with an intermediary in the middle. Are the nodes highlighted by the gray arrows all performing the same ``role'' in the network?}
\label{fig:roles-example}
\end{figure}

Let's consider the case of social circles. A social circle is a group of people interacting with each other because of shared interests and/or characteristics. We can say that Figure \ref{fig:roles-example} shows two connected social circles. The communities are not completely homogeneous: they have structure. They have a boundary, members that are more or less central, and outsiders connecting to them. We could define four roles in this network: brokers, gatekeepers, core, and periphery (the latter two not to be confused with the core-periphery mesoscale structure that we will see in Chapter \ref{cha:coreperiph}).

\begin{figure}
\centering
\begin{subfigure}{.45\columnwidth}
\includegraphics[width=\textwidth]{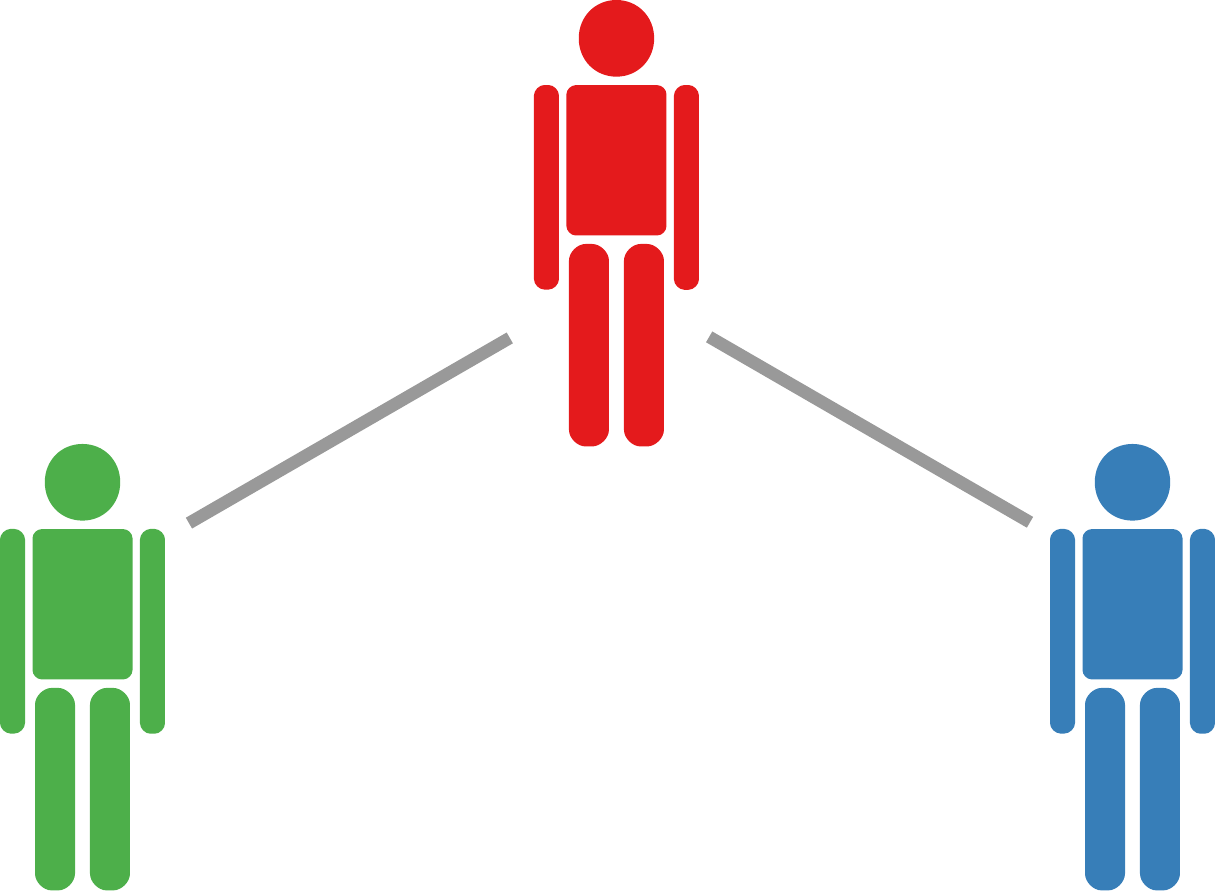}
\caption{}
\end{subfigure}
\qquad
\begin{subfigure}{.45\columnwidth}
\includegraphics[width=\textwidth]{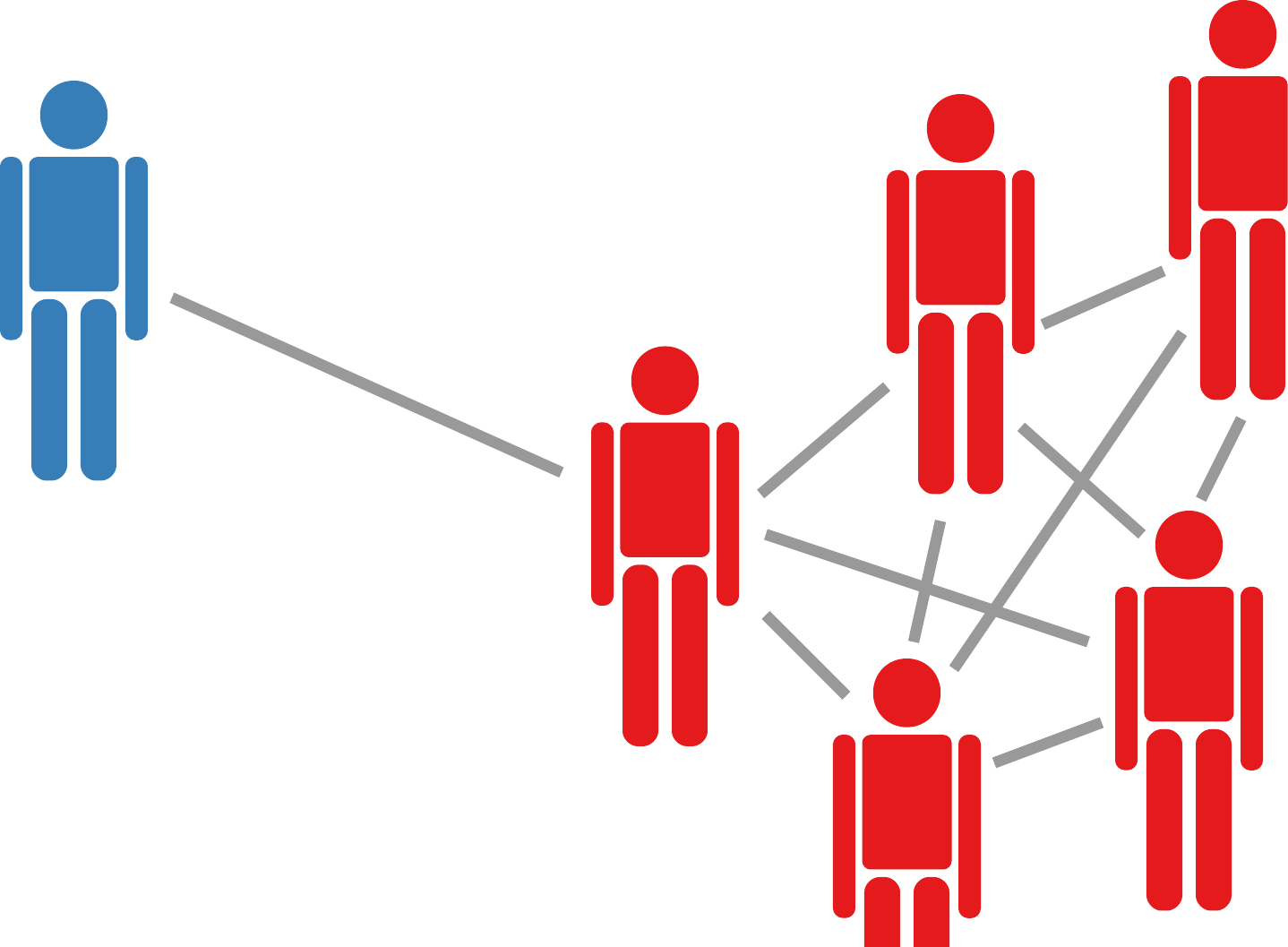}
\caption{}
\end{subfigure}
\caption{(a) An example of a broker. Color indicates the membership to a social circle. The red node isn't part of the two social circles it connects, so it brokers information between them. (b) An example of a gatekeeper. The blue person is not part of the red community. The member of the community to which it connects is managing the information access to the community. Thus, it is gatekeeping it.}
\label{fig:roles-1}
\end{figure}

\textbf{Broker}. Suppose we have two social circles. If the two communities do not share any member it means that they cannot communicate. However, sometimes you have people who are not part of either community -- because they only have few connections to each of them -- but they still have friends in both. These nodes can enable communication to happen between the communities, and so they are performing the role of information brokers. Figure \ref{fig:roles-1}(a) provides an illustration.

\textbf{Gatekeeper}. It is rare for a social circle to be completely isolated from the rest of society. Some of its members still have connections with people outside the community. If they do, they are managing how the community relates to society: both in the flow of information getting inside the community from the outside, and in what the community sends outside. These nodes are the gatekeepers. Figure \ref{fig:roles-1}(b) provides an illustration.

\begin{figure}
\centering
\includegraphics[width=.75\columnwidth]{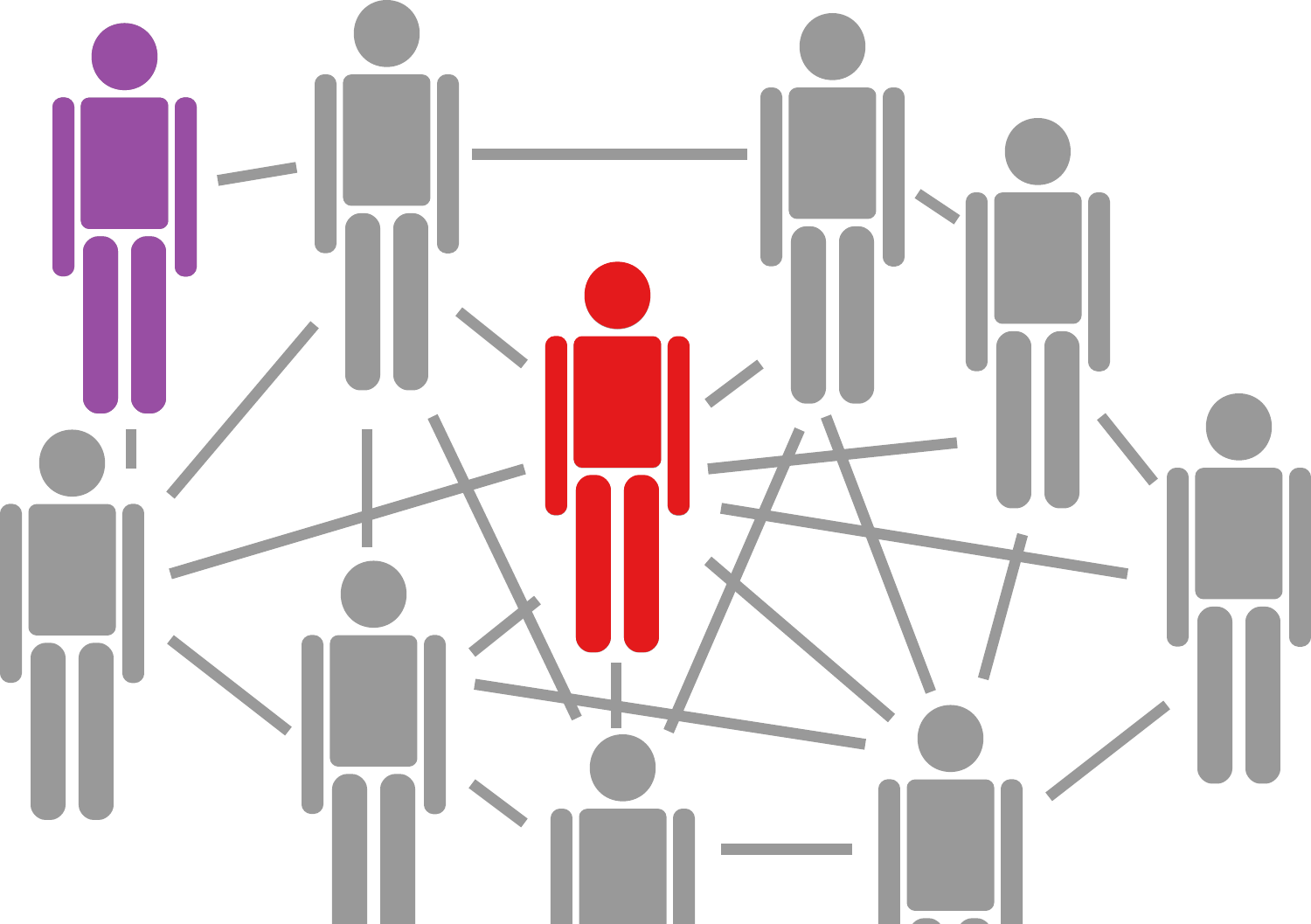}
\caption{Core and periphery nodes in a community. The red element of the social circle is very embedded in it: she is a core member. On the other hand, the purple person only has two friends in the social circle and no other relation. She is in the periphery.}
\label{fig:roles-core}
\end{figure}

\textbf{Core}. Some members of the community have a more central role than others. They connect exclusively with other members of the community, without establishing relations with outsiders. They also have many connections. They are the heart of the social circle, and thus composing its core. Figure \ref{fig:roles-core} provides an illustration.

\textbf{Periphery}. The other side of the coin of core members. A peripheral member does not have many connections in the community. Differently from brokers and gatekeepers, this is not because they also have connections to the external world. They just do not have many relations, and all they have are in their own community. They are thus peripheral to it. Figure \ref{fig:roles-core} provides an illustration.

Rolx\cite{henderson2012rolx} is one of the best known machine learning approaches for the extraction of node roles in complex networks -- a topic we'll greatly expand on in Part \ref{par:mining} of the book. The way it works is by representing nodes as vectors of attributes. Attributes can be, for instance, the degree, the local clustering, betweenness centrality, and so on. In practice, you decide which node features are relevant to determine the roles your nodes should be playing in the network. This means that, selecting the right set of features, you can recover all the roles I discussed so far -- core, periphery, broker, gatekeeper.

Rolx works in the way you would expect from a standard machine learning framework. The features are represented in a space where redundancies are eliminated, for instance by running principal component analysis (Section \ref{sec:mat-factors}). This space is then fed to a classifier, which tries to find the salient differences between different vector prototypes. These classes are the different roles a node can play.

\begin{figure}
\centering
\includegraphics[width=.75\columnwidth]{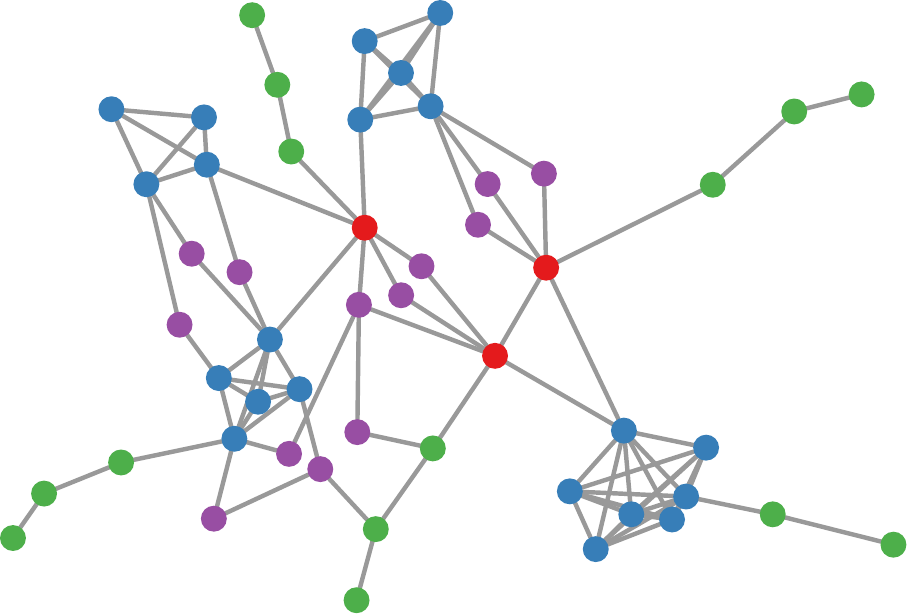}
\caption{An example of Rolx output. Node color encodes the node's role.}
\label{fig:rolx}
\end{figure}

Figure \ref{fig:rolx} shows an example of Rolx's output. The network represents co-authorship in a scientific network. Nodes are scientists connected if they collaborated on a paper. Rolx is able to find the different roles played by different authors. Specifically, authors found four roles of interest:

\begin{itemize}
\item Bridges (red): these are the hubs keeping the network together;
\item Tightly knit (blue): these are the authors who have a reliable group of co-authors, and are usually embedded in cliques;
\item Pathy (green): authors who are part of long stretches;
\item Mainstreram (purple): everything else.
\end{itemize}

Note that, with Rolx, you can also estimate how much each role tends to connect with nodes in a similar role. For instance, by their very nature, bridges tend to connect to nodes with different roles, while tightly knit nodes band together. This is related to the concepts of homophily and disassortativity, which we'll explore in Chapter \ref{cha:homophily}.

\section{Node Similarity}\label{sec:centr-similarity}

\subsection{Structural Equivalence}
When two nodes have the same role in a network they are, in a sense, similar to each other. Researchers have explored this observation and derived measures of ``node similarity''. We can also call this ``Structural Equivalence'', as two nodes with similar roles in similar areas of the network are keeping the network together in the same way. In fact, \textit{structural equivalence} is the stricter test of node similarity, which we can relax to obtain alternative measures.

In this part of the book we have assumed a structural view of nodes, unless otherwise specified. What this means is that, for betweenness centrality or the k-core algorithm, nodes don't have metadata, or internal statuses, or attributes. The only way to tell the difference between one node and another is by looking at their degrees and the nodes they connect to.

\begin{figure}
\centering
\begin{subfigure}{.4\columnwidth}
\includegraphics[width=\textwidth]{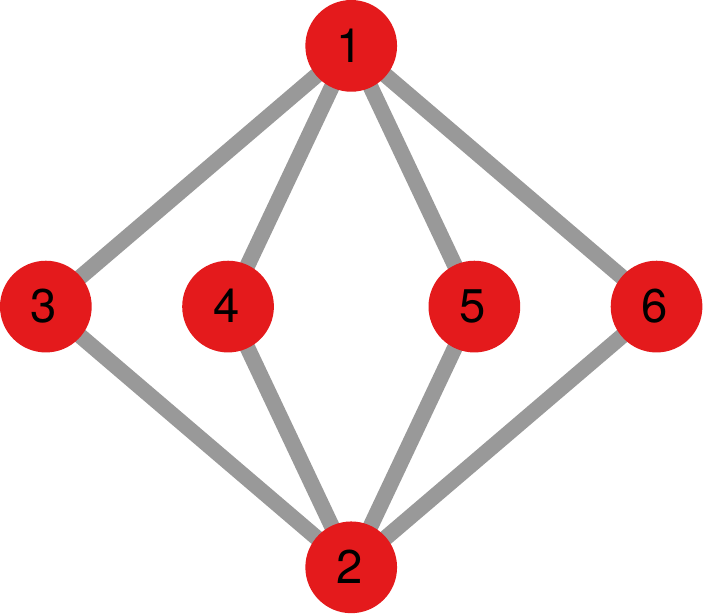}
\caption{}
\end{subfigure}
\qquad
\begin{subfigure}{.4\columnwidth}
\includegraphics[width=\textwidth]{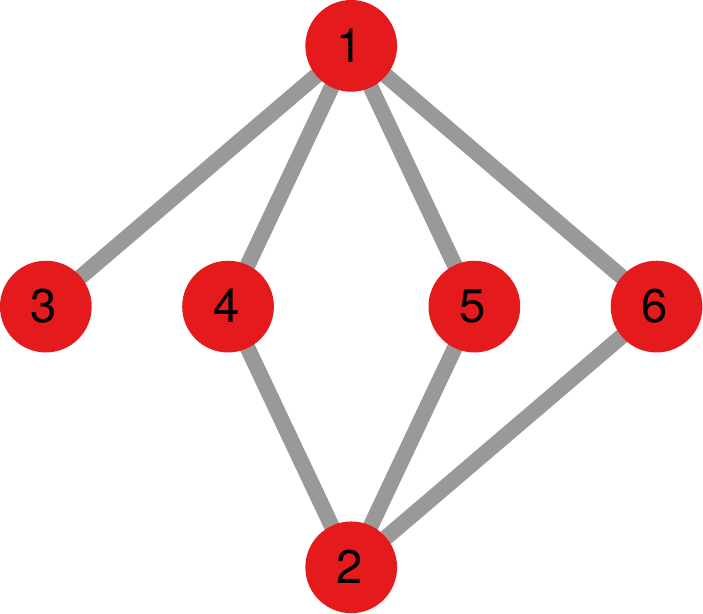}
\caption{}
\end{subfigure}
\caption{(a) An example of two structurally equivalent nodes (nodes $1$ and $2$). (b) Here, nodes $1$ and $2$ are not structurally equivalent, because node $1$ has a neighbor that node $2$ does not.}
\label{fig:str-equiv}
\end{figure}

This is important to point out in this section, because it helps understanding the definition of structural equivalence. For two nodes to be structurally equivalent they have to be connected to the same neighbors\cite{hanneman2005introduction}. If they do, they are indistinguishable from one another, therefore they cannot be any more similar. Consider Figure \ref{fig:str-equiv}(a): nodes $1$ and $2$ have the same neighbors and no other additional one. If I were to flip their IDs, you would not be able to tell. There is no extra information for you to do so, because all you have is their neighbor set.

On the other hand, we can tell the difference between nodes $1$ and $2$ in Figure \ref{fig:str-equiv}(b). That is because we know that node $1$ also connects to node $3$, which node $2$ does not. So the two nodes are not structurally equivalent. You can use any vector similarity measure to estimate structural equivalence. For instance, you can calculate the number of common neighbors or the Jaccard similarity of the neighbor sets between $u$ and $v$. In Figure \ref{fig:str-equiv}(b), nodes $1$ and $2$ have three common neighbors out of four possible, thus their structural equivalence is $0.75$.

\begin{figure}
\centering
\begin{subfigure}{.4\columnwidth}
\includegraphics[width=\textwidth]{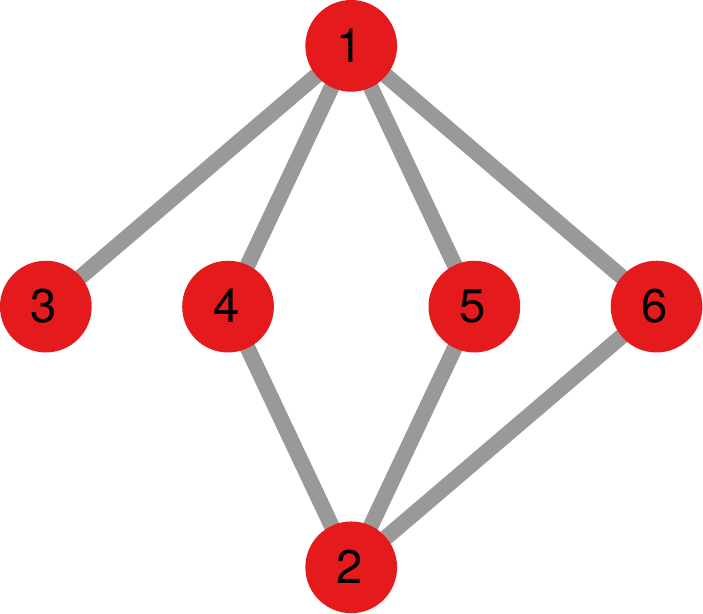}
\caption{}
\end{subfigure}
\qquad
\begin{subfigure}{.4\columnwidth}
  \centering
  \begin{tabular}{c|c}
    Node $1$ & Node $2$ \\
    \hline
    $0$ & $0$\\
    $0$ & $0$\\
    $1$ & $0$\\
    $1$ & $1$\\
    $1$ & $1$\\
    $1$ & $1$\\
  \end{tabular}
\caption{}
\end{subfigure}
\caption{(a) A simple graph. (b) The adjacency vector representations of node $1$ and node $2$.}
\label{fig:str-equiv2}
\end{figure}

Alternatively, one could use cosine similarity\cite{salton1989automatic}, Pearson correlation coefficients, or inverse Euclidean distance. In all these cases, you have to transform the neighbor set into a numerical vector. If you sort the nodes consistently, each node can be represented as a vector of zeros and ones. Zeros correspond to nodes not connected to $u$, while ones are $u$'s neighbors. These are the rows in the adjacency matrix corresponding to the nodes, as Figure \ref{fig:str-equiv2} shows. You can input the vectors corresponding to $u$ and $v$ to any of the mentioned measures and obtain their structural equivalence. For instance, the Pearson correlation coefficient of nodes $1$ and $2$ in Figure \ref{fig:str-equiv2} is around $0.7$.

\subsection{Automorphic Equivalence}
\textit{Automorphic equivalence} is a more relaxed version of structural equivalence. To understand it, we need to introduce the concepts of \textit{isomorphism} and \textit{automorphism}. We call two graphs ``isomorphic'' if they have the same topology: the graph in Figures \ref{fig:str-equiv}(a) and \ref{fig:str-equiv}(b) would be isomorphic if you were to add an edge in Figure \ref{fig:str-equiv}(b) between nodes $2$ and $3$. As a mnemonic trick: ``iso'' = same, and ``morph'' = ``shape'' -- two isomorphic graphs have the same shape. We'll see how to determine whether two graphs are isomorphic in Section \ref{sec:mining-isomorph}.

\begin{figure}
\centering
\begin{subfigure}{.4\columnwidth}
\includegraphics[width=\textwidth]{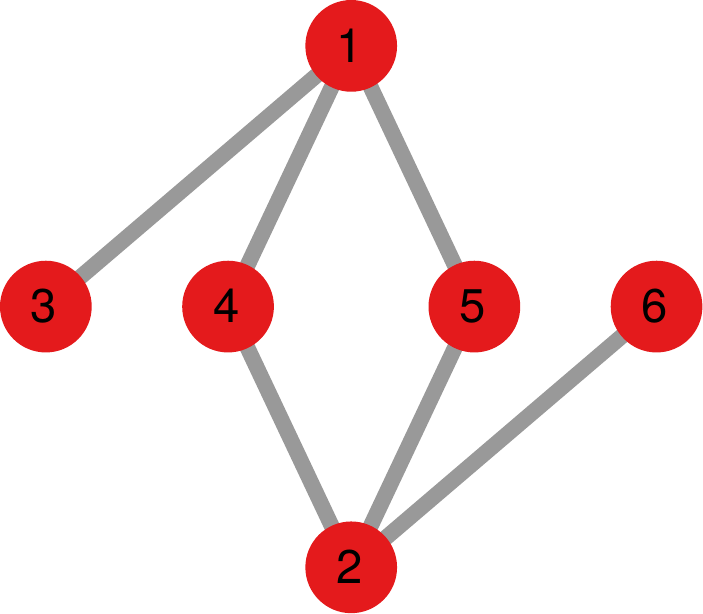}
\caption{}
\end{subfigure}
\qquad
\begin{subfigure}{.4\columnwidth}
\includegraphics[width=\textwidth]{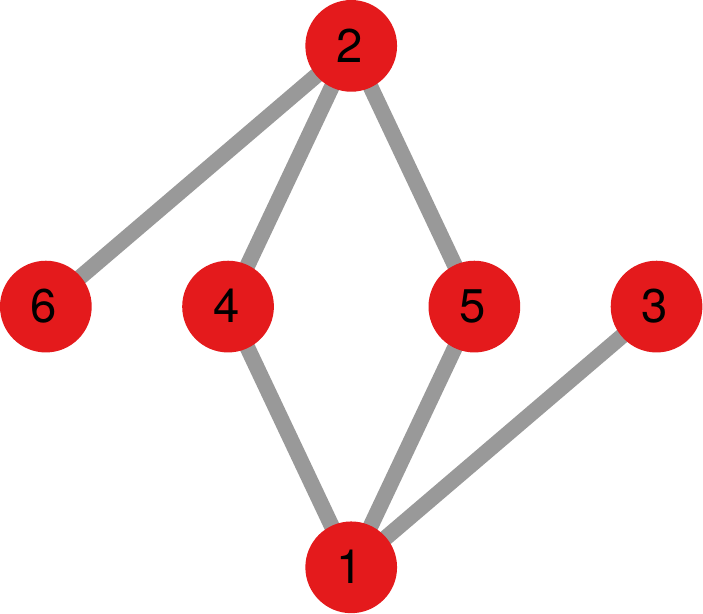}
\caption{}
\end{subfigure}
\caption{An example of relabeling of the graph to highlight the automorphic equivalence between nodes $1$ and $2$.}
\label{fig:str-equiv3}
\end{figure}

A graph is ``automorphic'' if it is isomorphic with itself. This means that you can shuffle all node IDs of your graph such that you preserve the neighborhoods of all nodes -- of course we care about non-trivial automorphisms, you can always find an automorphism by not swapping anything. If node $1$ was connected only to nodes $2$ and $3$ in $G$, you can only swap its ID with another node that only has two connections and those connections lead to nodes that have swapped their IDs with nodes $2$ and $3$. The graph in Figure \ref{fig:str-equiv3}(a) is automorphic because we can swap around labels respecting this rule -- as I do in Figure \ref{fig:str-equiv3}(b).

So, for automorphic equivalence, two nodes are equivalent if you can perform this re-labeling. To understand what this means consider nodes $1$ and $2$ in Figure \ref{fig:str-equiv2}(a). They are NOT automorphically equivalent because, if we swap their labels, there is no further relabeling we can do to render the graph isomorphic. We're not allowed to swap nodes $3$ and $6$, because they have a different number of neighbors. In Figure \ref{fig:str-equiv3}(a), instead, nodes $1$ and $2$ are automorphically equivalent. We can swap their labels, which forces us to swap node $3$ and $6$'s labels too. The resulting graph, in Figure \ref{fig:str-equiv3}(b), is identical to the original.

In this case, nodes $1$ and $2$ are not structurally equivalent, because they both have neighbors that the other node doesn't have, but they are automorphically equivalent, because you can perform the re-labeling. Every structurally equivalent pair of nodes is also automorphically equivalent, but two automorphically equivalent nodes might not be structurally equivalent.

In an automorphic graph, there must be pairs of automorphic equivalent nodes by definition. In non-automorphic graphs, some nodes can still be automorphic equivalent if you can perform a local relabeling. You could imagine Figure \ref{fig:str-equiv3}(a) to be embedded in a larger non-automorphic graph, but that would not affect the automorphism between nodes $1$ and $2$.

While structural equivalence focused on nodes that had literally the same neighbors in the network, automorphic equivalence focuses on nodes that belong to the same local structure type, even if they don't have the exact same set of neighbors.

\subsection{Regular Equivalence}
\begin{figure}[t]
\centering
\includegraphics[width=.3\columnwidth]{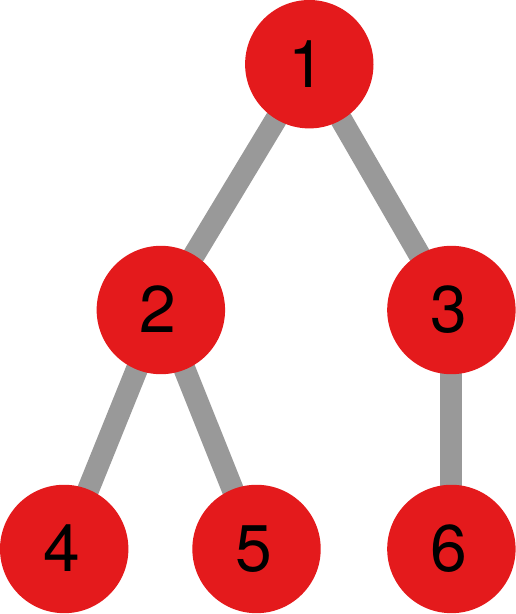}
\caption{A graph with three regularly equivalent classes of nodes.}
\label{fig:str-equiv4}
\end{figure}

The most relaxed variant of node similarity is \textit{regular equivalence}. In regular equivalence, nodes can be equivalent to each other if they have connections to equivalent nodes. This is most easily understood in hierarchies. Consider Figure \ref{fig:str-equiv4}. In the figure, we can find three equivalence classes containing, respectively: $\{1\}$, $\{2,3\}$, and $\{4,5,6\}$. The third class is defined by those nodes connected to nodes of class two but without connections to class one. Class two connects to both class one and class three, even if the number of connections to the members of class three can vary. Class one is the mirror of class three: it connects to nodes in class two, but to no node in class three.

How can you quantify regular equivalence? One way is to realize this is a recursive problem: similar nodes connect to similar nodes\cite{jeh2002simrank}\cite{blondel2004measure}. So, if we say our similarity score is $\sigma$, a matrix with a value per node pair, then we can say that two nodes $u$ and $v$ are similar if their neighbors are similar. In mathematical form:

$$ \sigma_{uv} = \alpha \sum \limits_{ij} A_{ui}A_{vj}\sigma_{ij},$$

so you are iterating over all pairs of $u$'s neighbors ($i$) and $v$'s neighbors ($j$) and using their similarity ($\sigma_{ij}$) to estimate $u$ and $v$'s similarity. If you write this in matrix form, you get that $\sigma = \alpha A \sigma A$, which looks like an eigenvector problem -- and, in fact, it is. To get the real similarity, you need to apply this formula many times, say $l$ times. That is why you need an $\alpha$ in front: by applying this formula $l$ times you are using paths of length $l$ to estimate the similarity, but you care more about nodes that are closer to $u$ and $v$ and not those that are $l$ hops away. If $\alpha < 1$, the $\alpha^l < \alpha^{l-1}$ and you get smaller contributions from nodes that are father away. This should be familiar to you, because that's the same logic as Katz centrality we just saw in Section \ref{sec:centr-eigen}.

\begin{figure}
\centering
\includegraphics[width=.3\columnwidth]{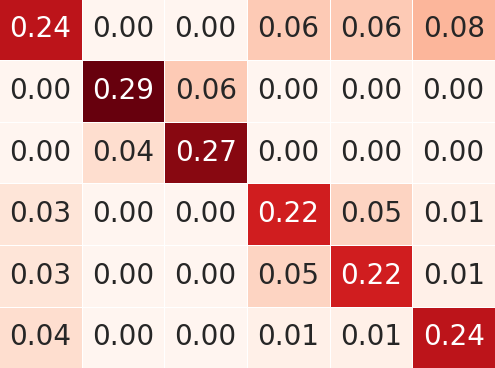}
\caption{The regular equivalence of the nodes from the network in Figure \ref{fig:str-equiv4}. The rows and columns of the matrix are sorted according to the numerical id of the node.}
\label{fig:str-equiv5}
\end{figure}

If you want to boost the similarity of a node with itself, you can always use the identity matrix: $\sigma = \alpha A \sigma A + I$. However, you should be careful because this measure as written here reduces to something similar to structural equivalence. If you run this formula on the network in Figure \ref{fig:str-equiv4} you get that nodes $1$ , $4$, $5$ and $6$ are all similar -- in intuitive terms, you'd get that CEOs are similar to interns because they both connect to middle managers. Figure \ref{fig:str-equiv5} shows you the similarity matrix you'd get, and you can see node $1$ is as similar to node $4$ as nodes $2$ and $3$ are to each other. So you need to ensure that your initial classes are respected somehow.

To sum up the difference between structural, automorphic, and regular equivalence, consider familial bonds. Two women with the same husband and the same children are structurally equivalent: they have the same relationships with the same people (in fact, they would be the same person -- although this is not a necessary requirement for structural equivalence). Two women can be automorphically equivalent if they have the same number of husbands and the same number of children. Finally, to be regularly equivalent to a married woman with children you have to be a married woman with children, even if you have a different number of relations -- the woman with fewer husbands will definitely lead a less frustrating life.

There are many other measures of node similarity. Researchers usually define new ones to better solve a problem called ``link prediction'', under the assumption that two similar nodes are more likely to connect to each other. We will see more node similarity measures than you want to know in Chapter \ref{cha:lp-simple}. Node similarity can be used to estimate network similarity, which is the topic of Chapter \ref{cha:netsimil}.

\section{Node Embeddings}\label{sec:centr-roles-conv}
So far we've been pretty rigid in the way we wanted to classify nodes into roles. Either we explicitly defined the roles with strict rules, or we adopted the similarity approach, finding which node plays a similar structural role to which other node. This is a sort of ``zero-dimensional'' approach, where everything collapses in a single label. One could use instead node embeddings, which determine the role of a node with a vector of numbers and then classifies the node with it. Recently, the most common way to discover such roles has become the use of graph neural networks. I will explain more in detail how they work much later in the book -- if you think what follows is a bunch of gobbledygook, you should check out Part \ref{par:mining}. Here I will just show the general shape of the problem they are solving and how it can be used to infer node roles.

\begin{figure}[t]
\centering
\includegraphics[width=.8\columnwidth]{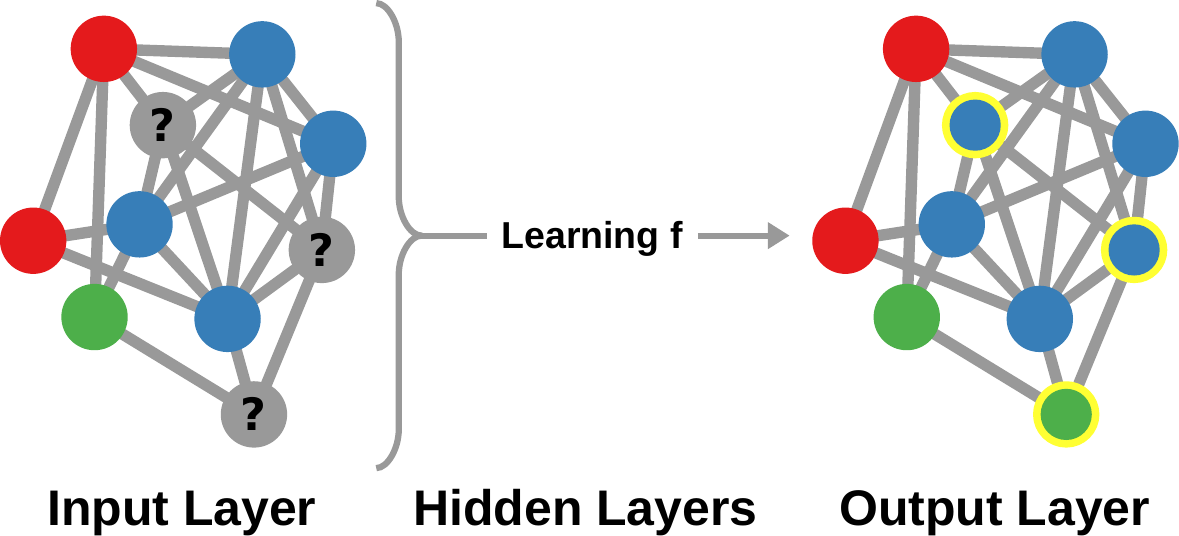}
\caption{A general schema for graph convolutional learning. The gray nodes in the input network are unclassified. By learning the function $f$ behind the classification of nodes, we can classify the rest of the network (yellow outline in the output layer).}
\label{fig:convolutional}
\end{figure}

When it comes to detect node roles, graph neural networks use machine learning techniques (Chapter \ref{cha:machine-learning}) to learn a function classifying nodes\cite{zhou2018graph}\cite{wu2019comprehensive}. Figure \ref{fig:convolutional} shows a general schema for graph neural networks. We start from some training data. This could be the graph itself with some nodes labeled and some not, or a graph or a collection of graphs with labeled nodes. We pass this graph as the input to the hidden layers. The role of the hidden layers is to learn a function $f$ that can explain why each node has a specific label/value. Once $f$ is learned, you will obtain the labels of the non-classified nodes, or you'll be able to classify a new, previously unseen, graph.

Typically, $f$ assumes that each node can be represented by a vector, called state. We're going to see in Chapter \ref{cha:mining-deep} more than you want to know on how to make smart $f$s.

\begin{figure}[b]
\centering
\includegraphics[width=.8\columnwidth]{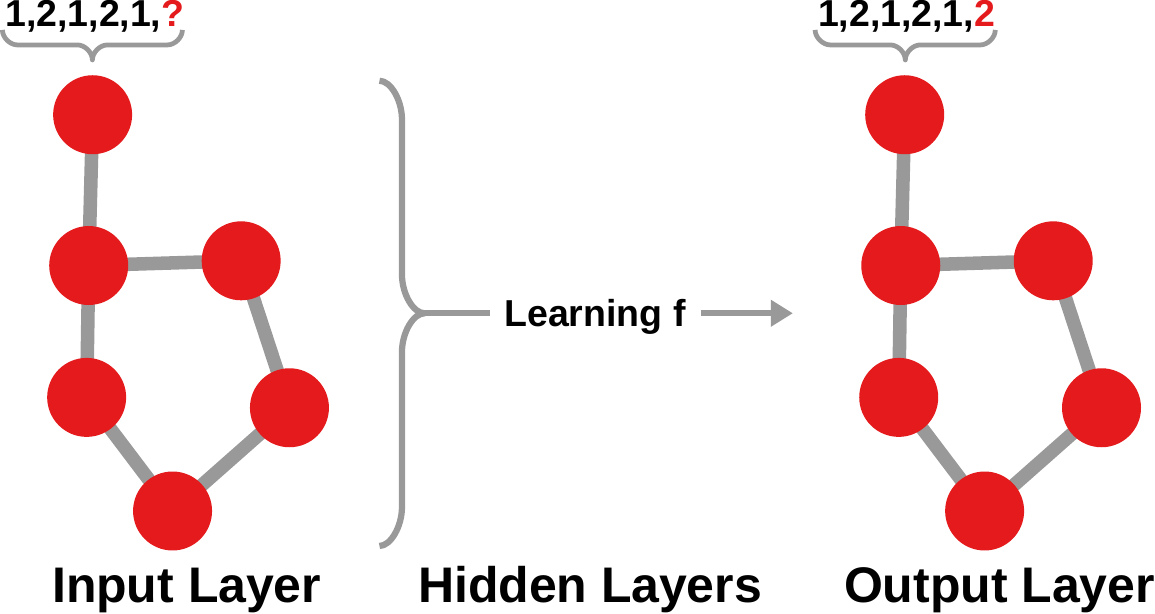}
\caption{A schema for spatial-temporal neural networks. We have an activation timeline for each node (here showing only one). The task is predicting the activation state in the next timestep.}
\label{fig:spatialtemporal}
\end{figure}

But modifying the way you learn $f$ is not the only possible variant. You could also modify the input and output layers, to change the task itself. For instance, you could try to predict spatiotemporal networks\cite{li2017diffusion}\cite{yu2018spatio}\cite{yan2018spatial}\cite{jain2016structural}: by having as input a dynamic network with changing node states, you could predict a future node state. Figure \ref{fig:spatialtemporal} shows a simplified schema for the task. Think, for instance, about traffic load: given the way traffic evolves, you want to be able to predict how many cars will hit a specific road straight (edge) or intersection (node).

Other possible applications are the generation of a realistic network topology (Chapter \ref{cha:csmodels}), the prediction of a link (Chapter \ref{cha:lp-simple}), or summarizing the graph (Chapter \ref{cha:mining-summarization}). Given that these are not related to node roles, I'll deal with such applications in the proper chapters. 

\section{Summary}

\begin{enumerate}
\item Going beyond node centrality, we can attach to nodes qualitative roles, rather than quantitative estimations of their importance. These qualitative roles are dependent on the node's position in the network topology. Traditionally, this is an ``unsupervised'' learning task, in which you don't know any node role and you're substantially inventing your own definition.
\item There are many ways to define roles, some well known are brokers -- in between communities --, gatekeepers -- on the border of a community --, and more. A popular algorithm to detect node roles is Rolx.
\item We can detect nodes playing the same role in a topology by estimating their structural similarity: their tendency of connecting to the same set of neighbors.
\item Node role could also be a supervised learning problem, where you have some node roles in your data and you want to discover the latent rules that determine them. Graph neutral networks are a popular way to do so. In this technique, we don't have (necessarily) a definition of what the role is, but some already labeled data on which we can train the algorithm.
\end{enumerate}

\section{Exercises}

\begin{enumerate}
\item For the network at \url{http://www.networkatlas.eu/exercises/15/1/data.txt}, I precomputed communities (\url{http://www.networkatlas.eu/exercises/15/1/comms.txt}). Use betweenness centrality to distinguish between brokers (high centrality nodes equally connecting to different communities) and gatekeepers (high centrality nodes connecting with different communities but preferring their own).
\item Use the network from the previous question to distinguish between core community nodes (high degree nodes with all their connections going to members of their own community) and peripheral community nodes (low degree nodes with all their connections going to members of their own community).
\item Calculate the structural equivalence of all pairs of nodes from the network used in the previous question. Which two nodes are the most similar? (Note: there could be ties)
\end{enumerate}

\part{Synthetic Graph Models}\label{par:synthnet}

\chapter{Random Graphs}\label{cha:rndgraphs}
To explore the properties of a network -- the degree distribution, the clustering, the centrality of its nodes -- there is one fundamental requirement. You have to have a network. If you don't have a network, you're going to look pretty silly when you try to analyze it\footnote{Just like searching in a dark room for a black cat that isn't there (\url{https://en.wikipedia.org/wiki/Black_cat_analogy}).}. At the very beginning of network analysis, there was a widespread lack of data. Thus, some of the most brilliant mathematical minds in the field determined different ways to create synthetic network data by defining network models. These models are the subject of this part of the book.

There are fundamentally three reasons to generate synthetic data today. The first is explanatory in nature. After you analyze a bunch of real world networks, you may realize they all seem to have a common property. Maybe they have a broad degree distribution (Section \ref{sec:degree-pl}), incredibly high clustering (Section \ref{sec:density-clustering}), or they contain communities (Part \ref{par:cd}). And you ask yourself: how did these properties arise? You might want to apply very simple rules, to see if they can reproduce the property of interest. If you succeed, you might be closer to explain their origins. This is what I explore in Chapter \ref{cha:physicsmodels}.

The second reason is to have a way to test your algorithms and analyses. If that's your aim, you want to generate fake networks that are as similar as possible to real world networks. They must have the same properties, especially the ones that are important for testing your method. This is what I explore in Chapter \ref{cha:csmodels}.

The third and final class focuses on description. As I just said, you might find a network with a peculiar property. You might ask yourself not how this property arose, but if it is something you'd expect any network to have, given other characteristics on its topology. In other words, you want to estimate the statistical significance of that observation. It's pretty hard to talk about statistical significance when you have a single observation -- a single network. So you might want to generate random networks with the same properties of your observed one and see if they all have that characteristic of interest. Chapter \ref{cha:ergmodels} is all about this task.

\section{Building Random Graphs}
Before diving deep into these more complex models, I need to spend some time with the grandfather of all graph models. It is the family of network generating processes created by Paul Erd\H{o}s and  Alfr\'{e}d R\'{e}nyi in their seminal set of papers\cite{erdHos1959random}\cite{erdos1960evolution}\cite{erdos1964random} (some credit goes also to Gilbert\cite{gilbert1959random} for a few variants of the model).

These are simply known colloquially as ``Random graphs''. I can divide them fundamentally in two categories: $G_{n,p}$ and $G_{n,m}$ models. The way they work is slightly different, but their results are mathematically equivalent. The difference is there simply for convenience in what you want to fix: $G_{n,p}$ allows you to define the probability $p$ that two random nodes will connect, while $G_{n,m}$ allows to fix the number of edges in your final network, $m$.

Ok, but... what \textit{is} a random graph? We're all familiar to the concept of random number. You toss a die, the result is random. But what does ``random'' mean in the context of a graph? What's randomized here? For this chapter, I will answer these questions assuming uncorrelated random graphs -- meaning that you can mentally replace ``random'' with ``statistical independence''. This is not strictly speaking necessary: in the same way that you can study the statistics of correlated coins, you can study correlated random graphs. However, that would make for a nasty math, and it isn't super useful for the aim of this book.

\begin{figure}
\centering
\includegraphics[width=.75\columnwidth]{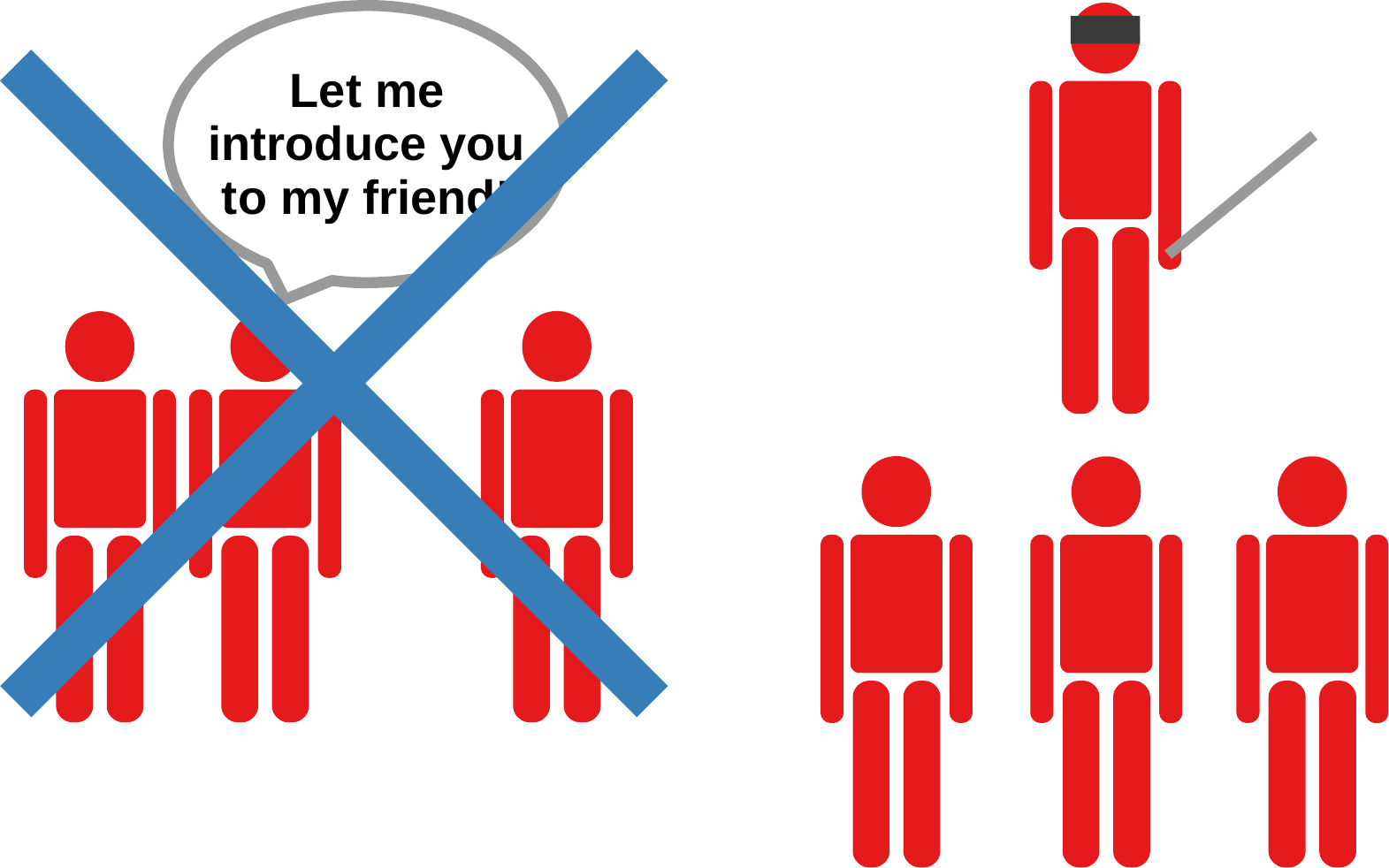}
\caption{In a social setting, friends introduce each other, thus edges are correlated: having a common friend increases the chance of being connected (see example on the left). In random graphs, edges are independent: blindfolded people establish the connections, as in the example on the right.}
\label{fig:random-vignette}
\end{figure}

For network scientists, ``random'' applies to the edges. In a social setting, connections are not independent: it is more likely for you to know people your friends know, because they can introduce you. In random graphs, this is strictly forbidden, as I show in the vignette in Figure \ref{fig:random-vignette}. If you are getting into a random graph and deciding where to put your new connection, you're going in completely blind. This means that you don't know anything about the connections that are already there.

If you're tired to read a book, this is the perfect occasion for a physical exercise. Here's a process you can follow to make your own random network. First, take a box of buttons and pour it on the floor. Yup, you heard me: just make a mess. Then take a yarn, cut some strings, and drop them on the buttons. The buttons are now your nodes, and the strings the edges. Congratulations! You have a random network! Time to calculate!

\begin{figure}
\centering
\includegraphics[width=.75\columnwidth]{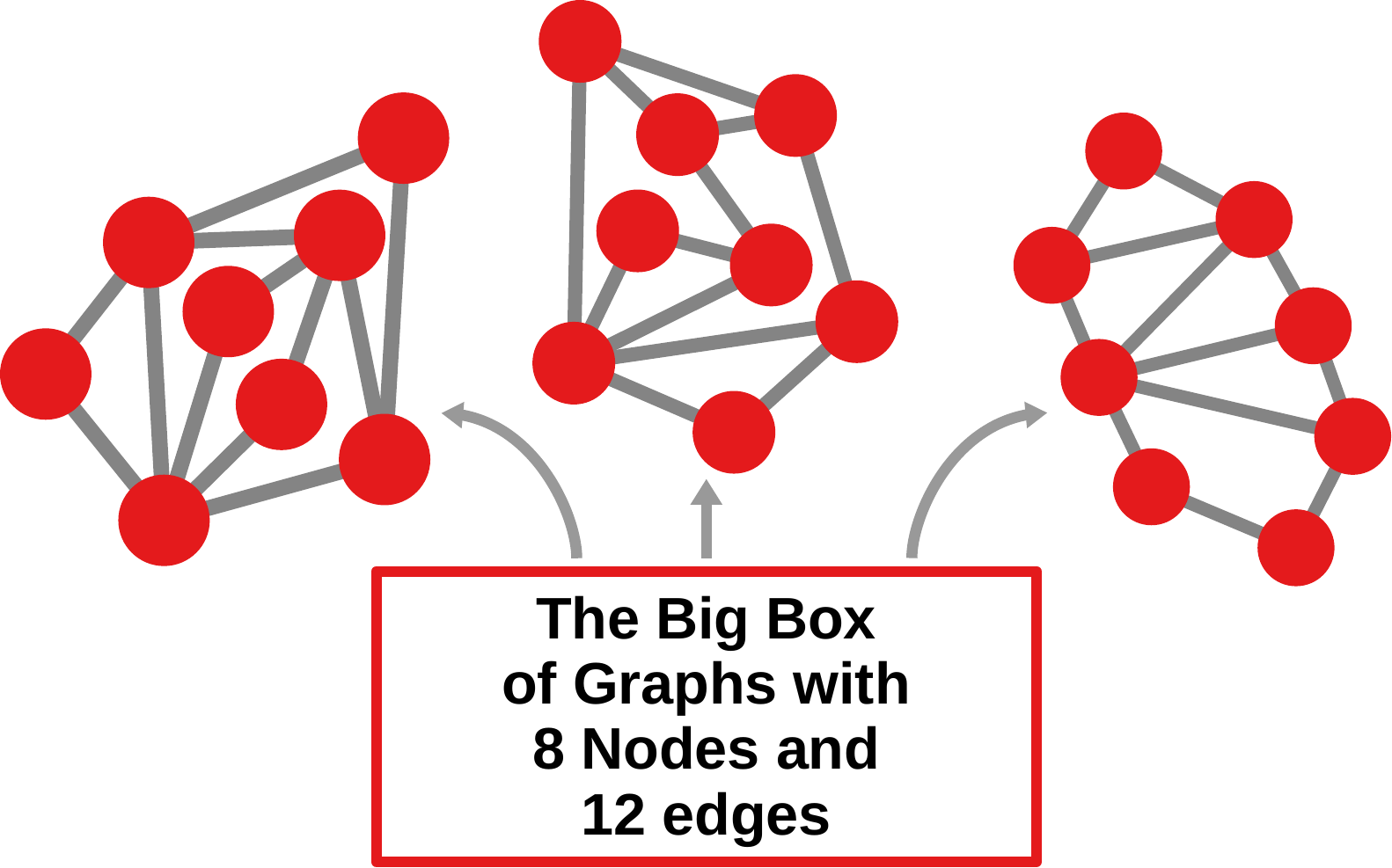}
\caption{How a $G_{n,m}$ model works: the $n$ and $m$ parameters determine the box from which you will extract your graph. The box contains all possible graphs with $n$ nodes and $m$ edges.}
\label{fig:random-gnm}
\end{figure}

Less facetiously, this process can illustrate clearly the distinction between $G_{n,p}$ and $G_{n,m}$ models. In $G_{n,m}$ you first fix the characteristics of the graph you want. You decide first how many nodes the graph should have -- which is the $n$ parameter --, say $8$. This is the number of buttons. Then you fix the number of edges it should have -- which is the $m$ parameter --, say $12$. This is the number of yarn strings you cut out. With those in mind, you go and take all possible graphs with $n$ nodes and $m$ edges, and you pick one at random -- as Figure \ref{fig:random-gnm} shows. All the graphs, by the power of having the right number of nodes and edges, are equally likely to be the result of our operation. Alternatively, you could simply extract $m$ random node pairs and connect them. The result will be the same.

In the $G_{n,p}$ variation we still say how many nodes we want: $n$. However, rather than saying how many edges we want, we just decide what's the probability that two nodes are connected to each other: $p$. It can be fifty-fifty, a coin toss. Then we consider all possible pairs of nodes, and for each one we toss the coin. If it lands heads we connect the nodes, if it lands tails we do not. $G_{n,p}$ is the perfect example of what we mean by ``random graph''. Given a pair of nodes we toss the coin and if it lands on heads we connect them. Another pair, same procedure. The two tosses are independent: the fact that one landed on heads does not influence the other. The coin is also the same, so each edge has an equal probability to appear.

\begin{figure}
\centering
\includegraphics[width=.75\columnwidth]{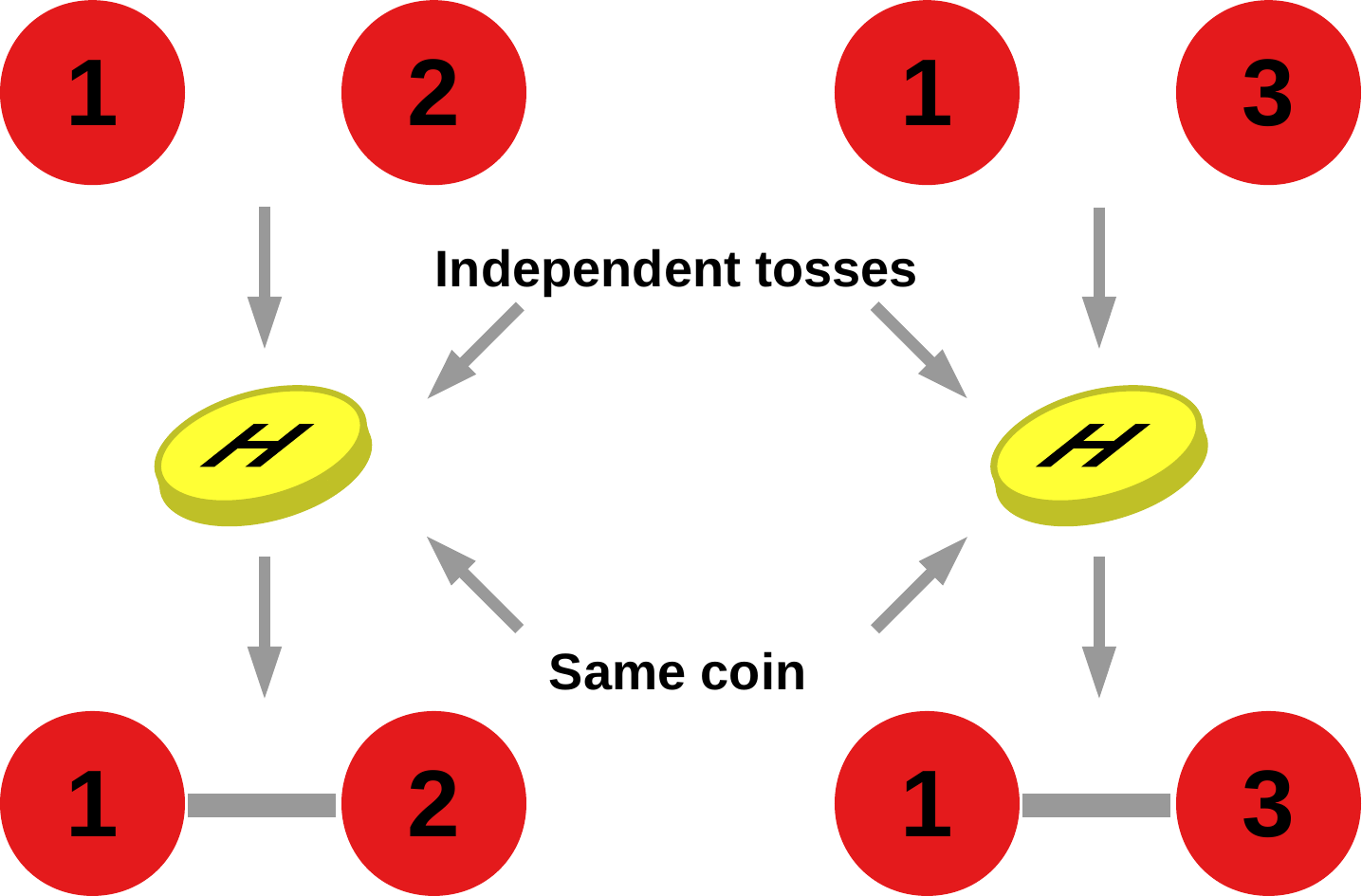}
\caption{How a $G_{n,p}$ model works: the $p$ parameter determines whether a node pair connects. The coin is not loaded, so it always has the same probability of landing on heads. The tosses are independent of each other, so the result of one doesn't affect the result of the other.}
\label{fig:random-gnp}
\end{figure}

Figure \ref{fig:random-gnp} depicts the process. Note that throwing coins for each node pair isn't exactly the most efficient way to go about generating a $G_{n,p}$ -- although I invite you to try. A few smart folks determined an algorithm to generate $G_{n,p}$ efficiently\cite{batagelj2005efficient}.

Since $G_{n,m}$ and $G_{n,p}$ generate graphs with the same properties, it means that $p$ and $m$ must be related. Since the graphs have $n$ nodes, we can derive the number of edges ($m$) from $p$. $p$ is applied to each pair of nodes independently. We know how many pairs of nodes the graph has, which is $n(n-1)/2$. Thus we have an easy equation to derive $m$ from $p$: the number of edges is the probability of connecting any random node pair times the number of possible node pairs, or

$$ p\dfrac{n(n-1)}{2} = m.$$

This is useful if you use $G_{n,p}$ but you want to have, more or less, control on how many edges you're going to end up with.

By the way this gives you an idea of what's the typical density of a random $G_{n,p}$ graph. The density (see Section \ref{sec:density-sparse}) is the number of links over the total possible number of links. We just saw that the number of links in a random graph is $p\dfrac{n(n-1)}{2}$ and the total possible number of links is $\dfrac{n(n-1)}{2}$. One divided by the other gives you $p$. So, if you want to reproduce the sparseness of real world networks, you can do that at will. Just tune the $p$ parameter to be exactly the density you want.

In the following sections we explore each property of interest of random graphs, to see when they model the properties of real world networks well, and when they don't. The latter is the starting point of practically any subsequent graph model developed after Erd\H{o}s and  R\'{e}nyi.

\section{Degree Distribution}\label{sec:rndgraphs-degdistr}
What's the expected degree of a node in a $G_{n,p}$ network with $9$ nodes? Well, we start by looking at the first potential connection and toss a coin. We do that for every possible node in the network. Since this is independent, if the probability of connection is $50\%$ and we make $8$ tosses -- one per potential neighbor --, on average we expect $8 \times 0.5$ head flips: $4$, plus minus a small random fluctuation.

\begin{figure}
\centering
\begin{subfigure}{.45\columnwidth}
\includegraphics[width=\textwidth]{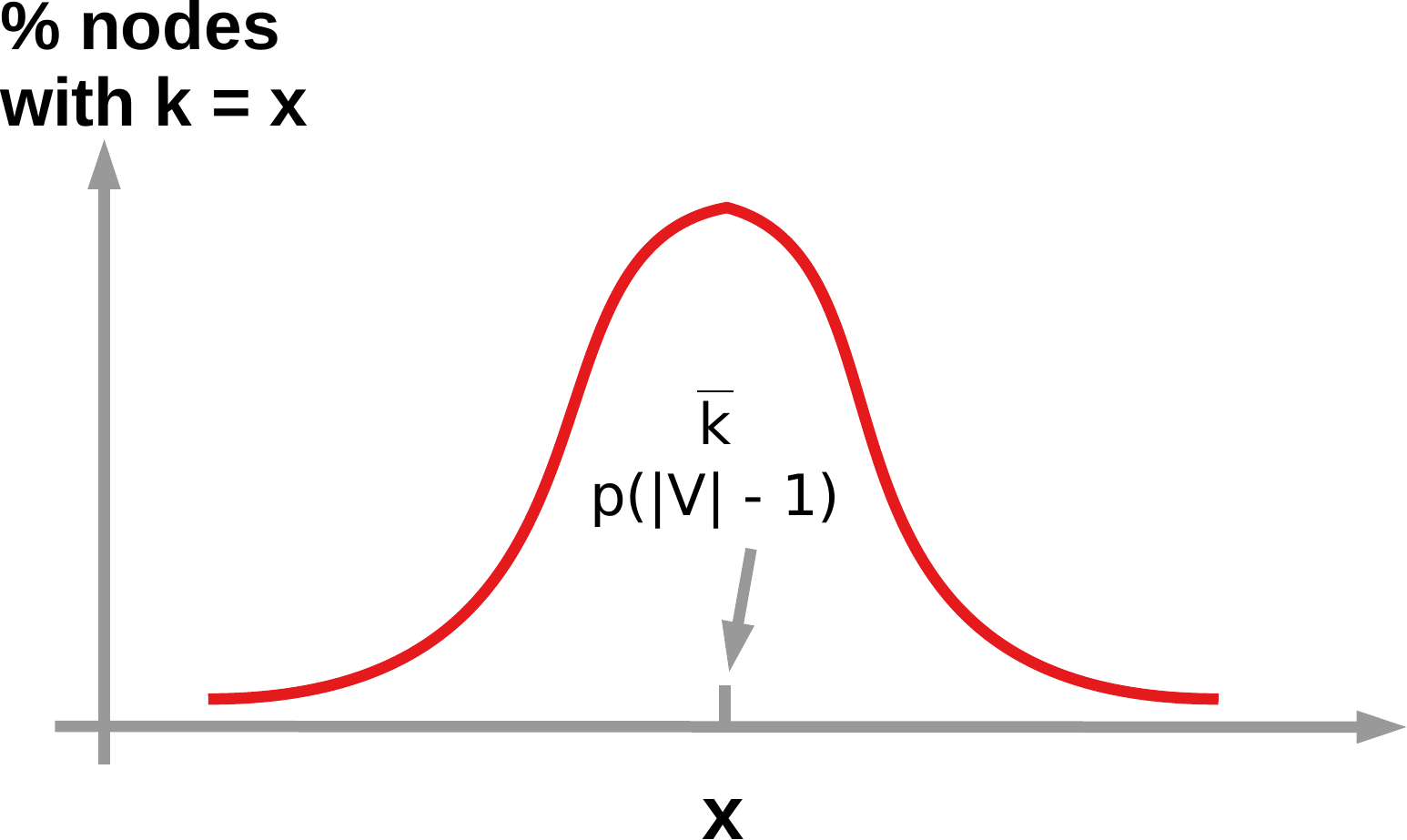}
\caption{}
\end{subfigure}
\qquad
\begin{subfigure}{.45\columnwidth}
\includegraphics[width=\textwidth]{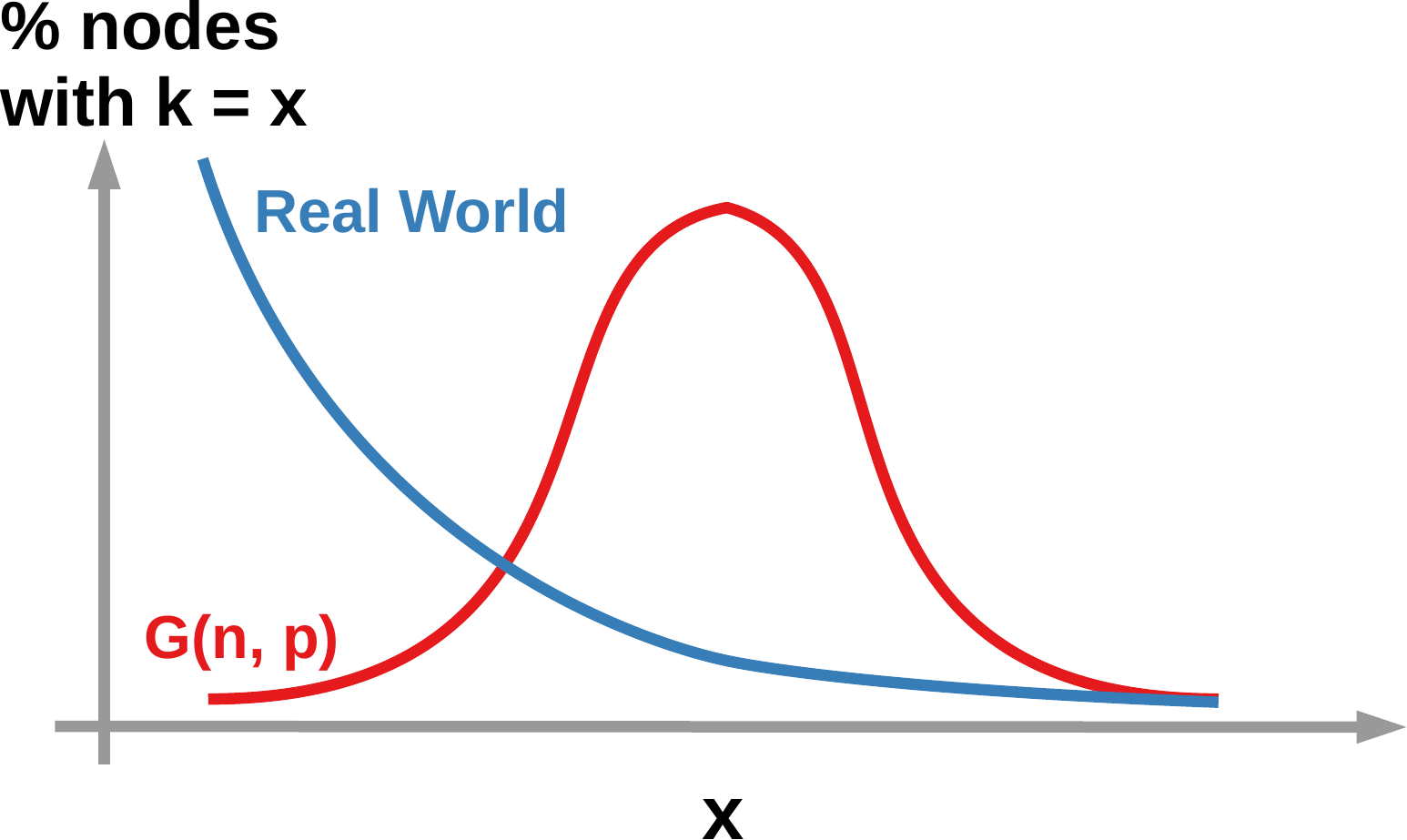}
\caption{}
\end{subfigure}
\caption{(a) The typical degree distribution of $G_{n,p}$ networks. (b) Comparing a $G_{n,p}$ degree distribution with one that you would typically get from a real world network.}
\label{fig:random-degdistr}
\end{figure}

A process like that generates a binomial distribution, where most nodes have a specific degree: $p$ times the number of nodes minus one -- because we avoid creating self loops. This is the average degree of the network. Figure \ref{fig:random-degdistr}(a) shows an example of a random degree distribution. Very few nodes have a much lower or much higher degree, because of how rarely you will get many heads or tails in a row from a fair coin.

Although the result is a binomial, many papers studying the degree distributions of random graphs use a Poisson distribution instead. They are practically identical, so this choice doesn't really matter -- specifically a binomial becomes equivalent to a Poisson when the number of trials is very large and the expected number of successes remains fixed (see Section \ref{sec:prob-distr}). We use a Poisson because the parameters regulating it make it easier to calculate the things that interest us -- they all depend on a single parameter: $\bar{k}$, the average degree.

The random graph's degree distribution is at odds with most real world networks which, as we saw in Section \ref{sec:degree-pl}, have many nodes with low degree. Moreover, the outliers with many connections -- the hubs -- in real networks have a much higher degree than the highest degrees you'll find in a random network. See Figure \ref{fig:random-degdistr}(b) for an example. So this is the first pain point of random networks: their degree distributions don't match the skewed ones we observe in the real world. Where is the real world broad degree distribution coming from? That's a question for Section \ref{sec:physicsmodels-ba}.

\section{Connected Components}
Besides broad degree distributions, we're interested in giant components, because all real world networks seem to have them. Remember that giant components are components that include the vast majority of -- or even all -- the nodes of the network, see Section \ref{sec:paths-ccomps} for a refresher.

\begin{figure}
\centering
\begin{subfigure}{.33\columnwidth}
\includegraphics[width=\textwidth]{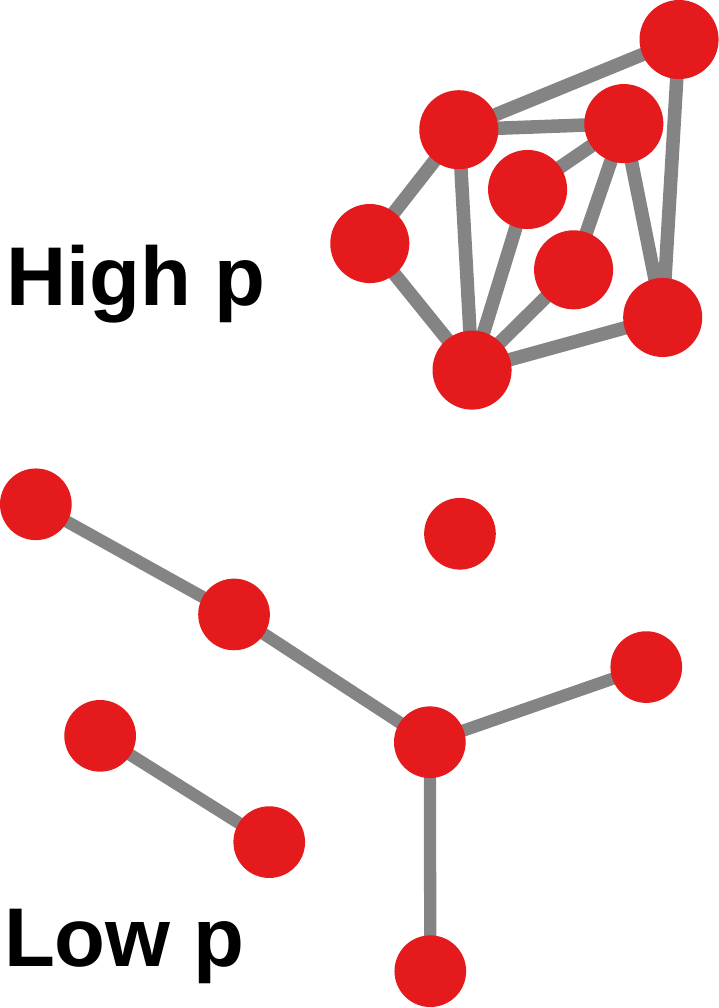}
\caption{}
\end{subfigure}
\qquad
\begin{subfigure}{.4\columnwidth}
\includegraphics[width=\textwidth]{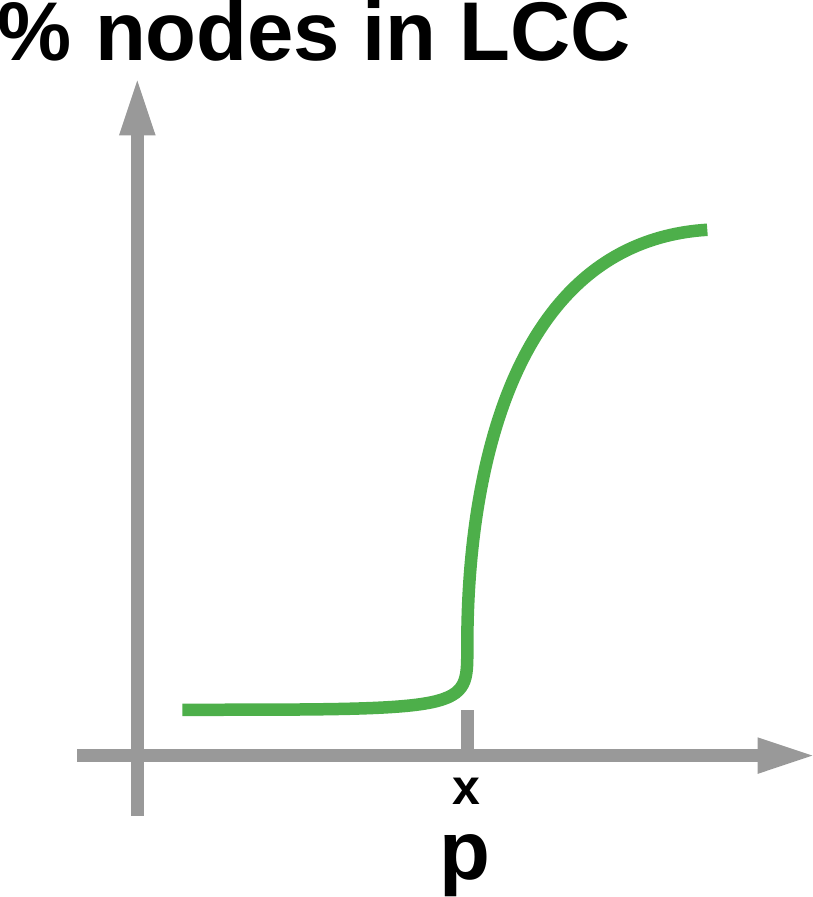}
\caption{}
\end{subfigure}
\caption{(a) The effect of $p$ on $G_{n,p}$'s connectivity. (b) The evolution of the number of nodes in the largest connected component of $G_{n,p}$ as $p$ changes.}
\label{fig:random-gcc}
\end{figure}

In a $G_{n,p}$ graph, the presence of a giant component is dependent on $p$. As it is easy to see from Figure \ref{fig:random-gcc}(a), if $p$ is high there are a lot of edges, if it is low there are few and the graph could be disconnected. One could make a plot, showing for which values of $p$ we have how many nodes in the largest connected component. One could expect a linear relationship, but that is not what we see: there is a special value of $p$ for which we observe a phase transition -- which I show in Figure \ref{fig:random-gcc}(b) --: if $p$ is lower than that value there is no giant component, if $p$ is higher then most nodes are in the GCC\cite{erdHos1961strength}\cite{erdos1966existence}\cite{achlioptas2009explosive}\cite{d2010local}.

Can we determine this magical value of $p$? Logically, a node cannot be part of any component if it doesn't have an edge. If the average degree is less than one, many nodes won't have edges. Thus they cannot be part of the largest connected component. When the average degree is higher than one, the giant component appears. Thus the magical value of $p$ is $1/|V|$.

So this is the value beyond which we start to see the largest connected component gobbling up the majority of the network's nodes. But there is another question. Is there a value of $p$ such that \textit{all} nodes are part of the giant component? This is equivalent of asking: when do we have fewer than one node without connections to the giant component? We start with the probability of connecting a node $u$ with a node $v$. This is $p$. It follows that the probability of $u$ and $v$ \textbf{not} to connect is $1-p$. Generalizing this, having no connection to a component with $|V|$ nodes is the same of landing tails for $|V|$ times in a row: $(1-p)^{|V|}$. The number of nodes without a connection to a component with $|V|$ nodes is equal to make $|V|$ attempts -- since we have $|V|$ nodes in our network --, all with that probability of success. The result is $|V|(1-p)^{|V|}$.

Our original question was knowing when there are no nodes outside the GCC. This is equivalent to say that we want to have fewer than one node outside GCC. We just said that the number of nodes outside the GCC is $|V|(1-p)^{|V|}$. Thus we want to know for which $p$ we have $|V|(1-p)^{|V|} < 1$. Which is $p = \ln|V| / |V|$.\footnote{The full mathematical derivation rests on the assumption that $(1-x/n)^n \sim e^{-x}$ if $n$ is large. At that point the derivation follows: $|V|(1-p)^{|V|} = |V|\left(1 - \dfrac{|V|p}{|V|}\right)^{|V|} = |V|e^{-|V|p}$. If $|V|e^{-|V|p} = 1$, as we said we want in the text, then:\\$|V|e^{-|V|p} = 1$\\$e^{-|V|p} = |V|^{-1}$\\$e^{|V|p} = |V|$\\$|V|p = \ln|V|$\\$p = \ln|V| / |V|$.}

Note that $\ln|V| / |V|$ tends to be a rather small number as $|V|$ grows, given that it pits a logarithmic growth in the numerator against a linear one in the denominator. Thus we discover that random graphs tend to have giant components as they grow larger, which is exactly what we see happening in real world networks. One point to team Erd\H{o}s-R\'{e}nyi!

\section{Average Path Length}

\begin{figure}
\centering
\includegraphics[width=.75\columnwidth]{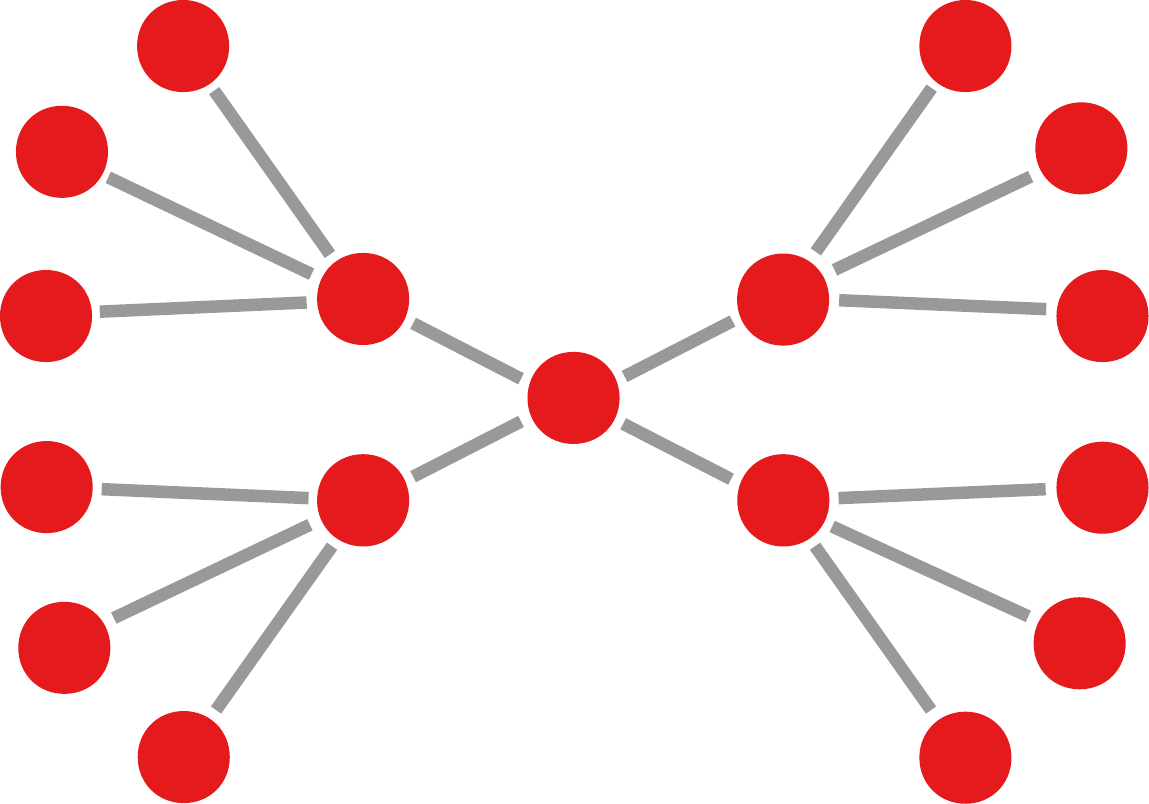}
\caption{A simplification of what happens to the average path length in $G_{n,p}$ random graphs. If the average degree ends up being four, a random node -- at the center -- will have four neighbors. Each of them, will contribute on average three more neighbors when considering paths of length two.}
\label{fig:random-apl}
\end{figure}

In a $G_{n,p}$ network, the number of nodes directly connected to a node is the average degree. This is the usual expectation as the connection probability is fixed to $p$, as we saw in Figure \ref{fig:random-degdistr}(a). In $G_{n,p}$ we can easily calculate the number of nodes at two hops from $v$. This is expected to be the average degree squared because, on average, each neighbor gives you access to its average degree number of neighbors -- see Figure \ref{fig:random-apl} for an example. This goes on for any number of hops $l$. The number of nodes at $l$ hops away is $\bar{k}^l$ -- remember that $\bar{k}$ indicates the average degree of the network.

If we know when $\bar{k}^l = |V|$, then we know the average path length of $G_{n,p}$: $\ln|V| / \ln(|V|p)$\cite{de1978contacts}\footnote{Time for another mathematical derivation!\\$\bar{k}^l = |V|$\\$l\ln\bar{k} = \ln|V|$\\$l = \ln|V| / \ln\bar{k}$.\\Since in a $G_{n,p}$ network $\bar{k} = |V|p$, you get the final derivation in the text.}. Note that this is relatively short, unless $p$ is really low. Which is another thing that $G_{n,p}$ graphs have in common with real world networks. It seems that these random graphs are not too shabby when it comes to reproducing real world properties!

\section{Clustering}
What's the clustering of a $G_{n,p}$ network? Remember Section \ref{sec:density-clustering}: the local clustering $CC_v$ of a node $v$ is number of triangles over the number of triads centered in $v$. The number of triads, as we saw then, is the number of possible edges among neighbors: $\bar{k}(\bar{k}-1)/2$. We can use $\bar{k}$ instead of $k_v$, because any $v$ in a $G_{n,p}$ graph is expected to have an average degree. That's the denominator of $CC_v$.

The numerator of $CC_v$ is the number of triangles centered of $v$. This is the probability an edge exists ($p$), times the number of possible edges among neighbors. Again, on the basis of Section \ref{sec:density-clustering}, we know that the number of possible edges among neighbors is $\bar{k}(\bar{k}-1)/2$.

\begin{figure}
\centering
\includegraphics[width=.75\columnwidth]{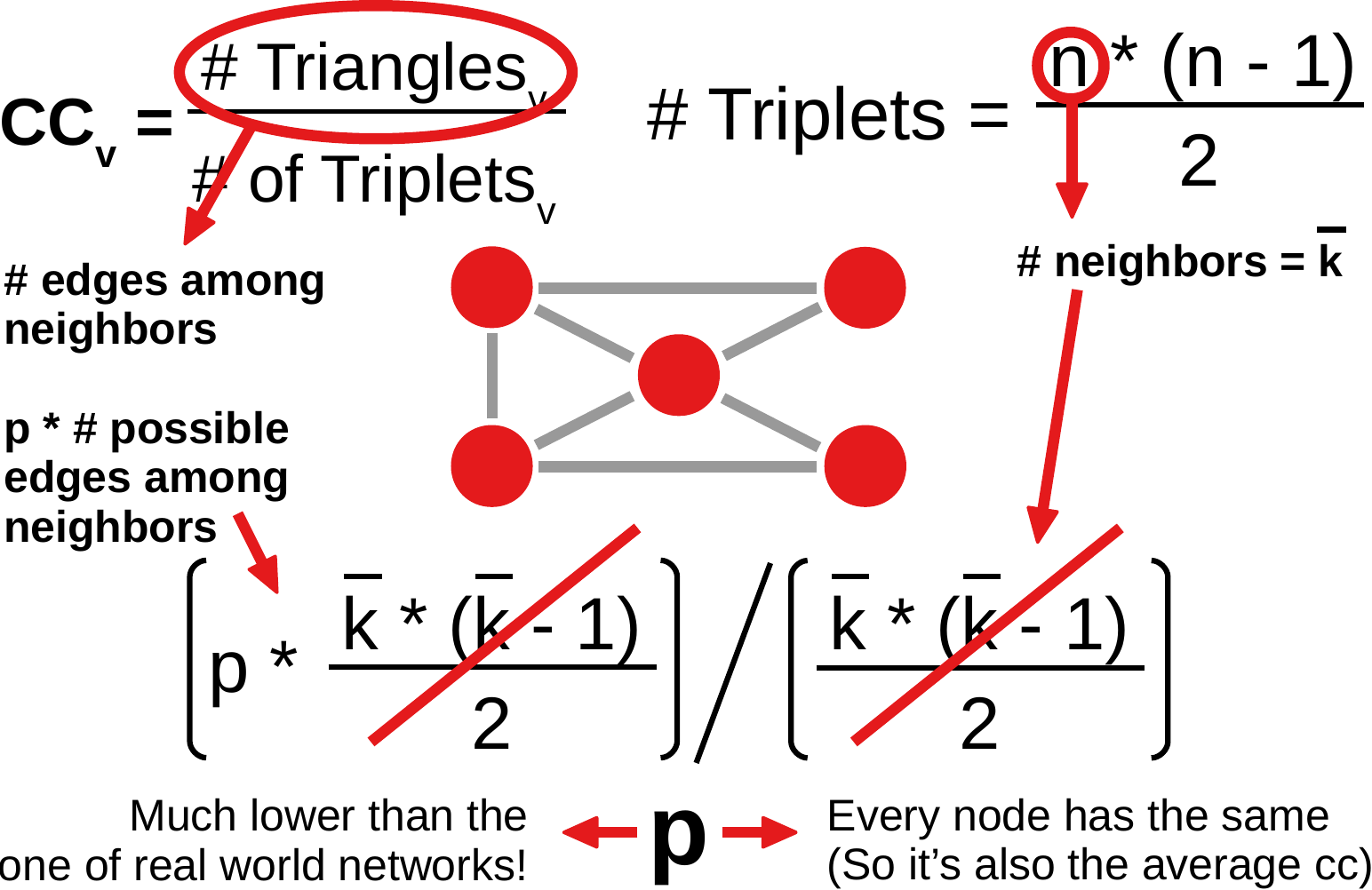}
\caption{The derivation of the clustering coefficient of a random $G_{n,p}$ network.}
\label{fig:random-cc}
\end{figure}

If we say, for simplicity, that $\bar{k}(\bar{k}-1)/2 = x$, then our $CC_v$ formula looks like: $CC_v = px/x = p$. This means that the clustering coefficient doesn't depend on any node characteristic, and it's expected to be the same -- equal to $p$ -- for all nodes. Figure \ref{fig:random-cc} provides a graphical version of this derivation.

Compared to real world networks, $p$ is usually a very low value for the clustering coefficient. In real world networks, it's more likely to close a triangle than to establish a link with a node without common neighbors. Thus the clustering coefficient tends to be higher than simply the probability of connection. This is a second pain point of Erd\H{o}s-R\'{e}nyi graphs when it comes to explain real world properties, after the lack of a realistic degree distribution, as we saw in Section \ref{sec:rndgraphs-degdistr}. Such a low clustering usually implies also the absence of a rockstar feature of many real world networks: communities.

To recap, $G_{n,m}$ and $G_{n,p}$ models correctly estimate the emergence of giant components and the relatively small diameters of real world systems. However they fail three of the most crucial tests: degree distribution, clustering, and communities. This is cause enough for researchers to push forward in the quest for better graph models.

You can find a deeper treatment of random graph models in Bollob\'{a}s's seminal work\cite{bollobas1998random}.

\section{Summary}

\begin{enumerate}
\item Random graph models are useful for testing your algorithms, explain how specific properties might arise in real world networks, and test whether an observed network is really as special as you think it could be, or if its properties are due to random chance.
\item The oldest and most venerable random graph model is the $G_{n,p}$ (or $G_{n,m}$) model, where we fix the number of nodes $n$ and the probability $p$ of connecting a random node pair, and we extract edges uniformly at random.
\item These random graphs have a binomial degree distribution, which is very different from the broad degree distributions of real world networks.
\item There is a phase transition when it comes to the largest connected component: if your random graph has an average degree higher than one, you'll have a connected component including most of the nodes; if the average degree is lower, you won't have such component.
\item Random graphs have a short average path length just like real world networks typically have. However, they have a much lower clustering coefficient than what you find in the wild. 
\end{enumerate}

\section{Exercises}

\begin{enumerate}
\item Consider the network in \url{http://www.networkatlas.eu/exercises/16/1/data.txt}.  Generate an Erd\H{o}s-R\'{e}nyi graph with the same number of nodes and edges. Plot both networks' degree CCDFs, in log-log scale. Discuss the salient differences between these distributions.
\item Generate a series of Erd\H{o}s-R\'{e}nyi graphs with $1,000$ nodes and an increasing $p$ value, from $.00025$ to $.0025$, with increments of $.000025$. Make a plot with the $p$ value on the x axis and the size of the largest connected component on the y axis. Can you find the phase transition?
\item Generate a series of Erd\H{o}s-R\'{e}nyi graphs with $p = .02$ and increasing number of nodes, from $200$ to $1,400$ with increments of $200$. Make a plot with the $|V|$ value on the x axis and the average path length on the y axis. Since the graph might not be connected, only consider the largest connected component. How does the APL scale with the number of nodes?
\item Generate an Erd\H{o}s-R\'{e}nyi graph with the same number of nodes and edges as the network used for question $1$. Calculate and compare the networks' clustering coefficients. Compare this with the connection probability $p$ of the random graph (which you should derive from the number of edges and number of nodes using the formula I show in this chapter).
\end{enumerate}

\chapter{Understanding Network Properties}\label{cha:physicsmodels}
We now move on to the class of network models developed primarily to explain network properties. The two most famous examples in this class are the small world model proposed by Watts and Strogatz and the preferential attachment model, independently discovered in many variants multiple times across many decades but usually attributed to Albert and Barab\'{a}si. The first aims at explaining the high level of clustering and small diameter of real world networks. The second focuses on power law degree distributions.

The interest in these models is twofold. First, it is a historic interest: these models were the first developed in the new wave of network science in the late 90s. Their impact in the development of the field was huge -- the original papers both accumulated more than $46$ thousand citations. Second, they give an idea of what are some of the original guiding principles of a certain flavor of network science: the hunt for universal patterns that apply to any network representation of a complex system. In practice, nowadays you'd seldom use these vanilla models, as they have been superseded by more sophisticated -- albeit often less mathematically tractable -- ones.

\section{Clustering}\label{sec:physicsmodels-cm}
Before the Stone Age, a caveman society was very simple. You had tribes living in their own caves. The tribes were very small, they were families. Everybody knew everyone else in their cave, but between caves there was almost no communication. Maybe there could have been one weak link if the two caves were close enough.

\begin{figure}
\centering
\begin{subfigure}{.22\columnwidth}
\includegraphics[width=\textwidth]{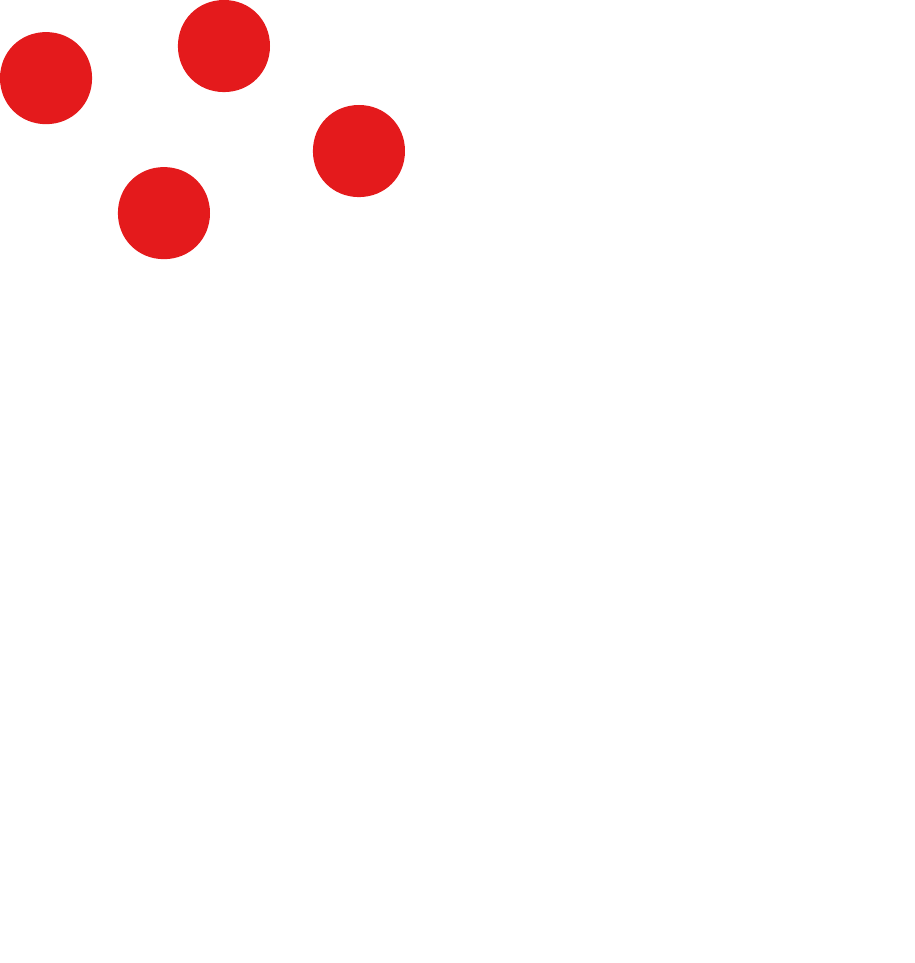}
\caption{}
\end{subfigure}\quad
\begin{subfigure}{.22\columnwidth}
\includegraphics[width=\textwidth]{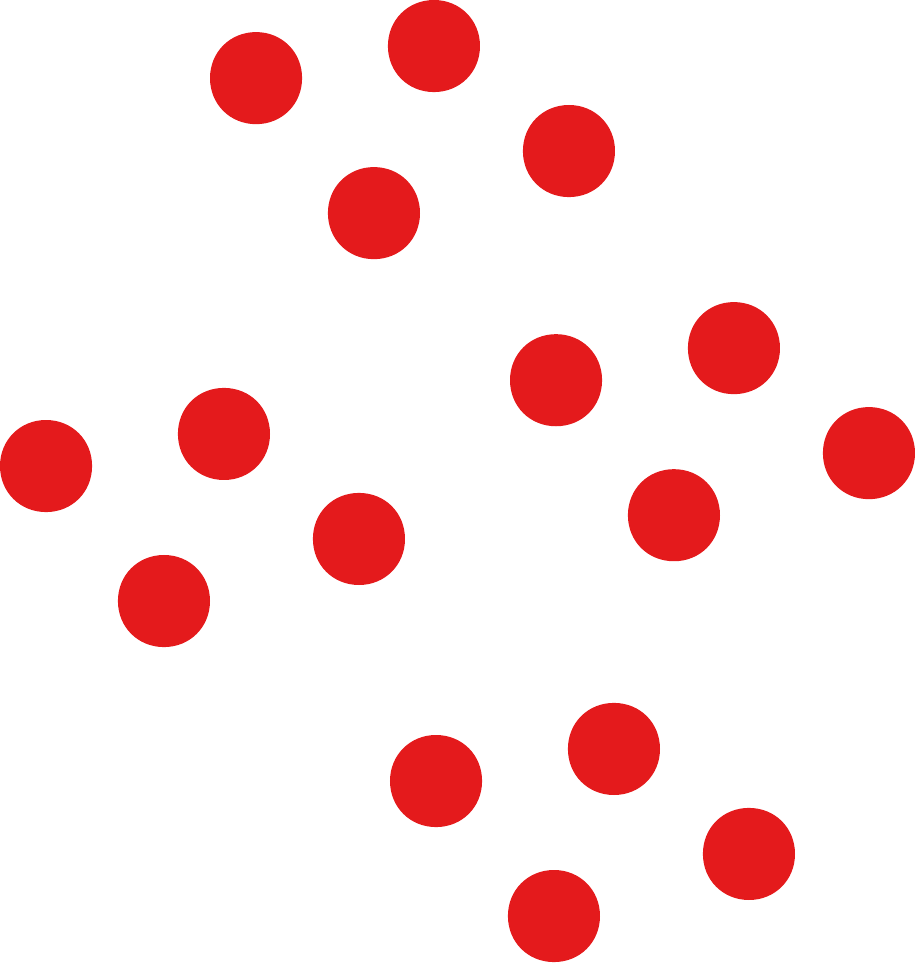}
\caption{}
\end{subfigure}\quad
\begin{subfigure}{.22\columnwidth}
\includegraphics[width=\textwidth]{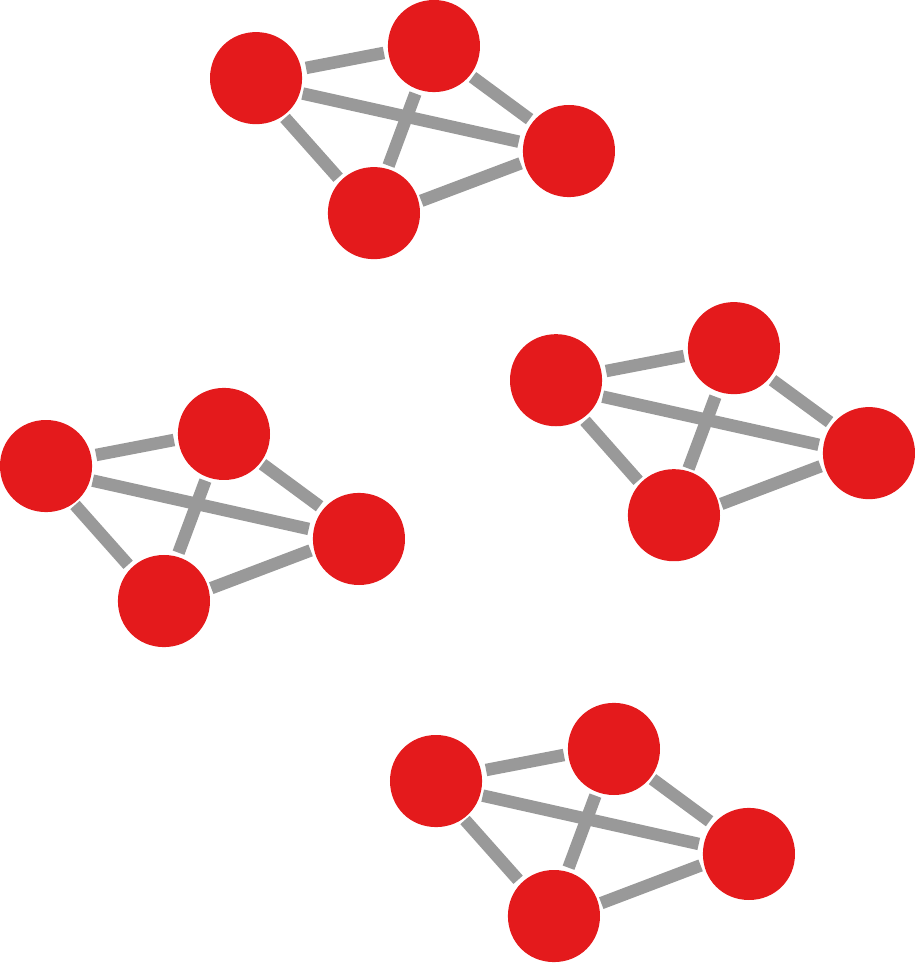}
\caption{}
\end{subfigure}\quad
\begin{subfigure}{.22\columnwidth}
\includegraphics[width=\textwidth]{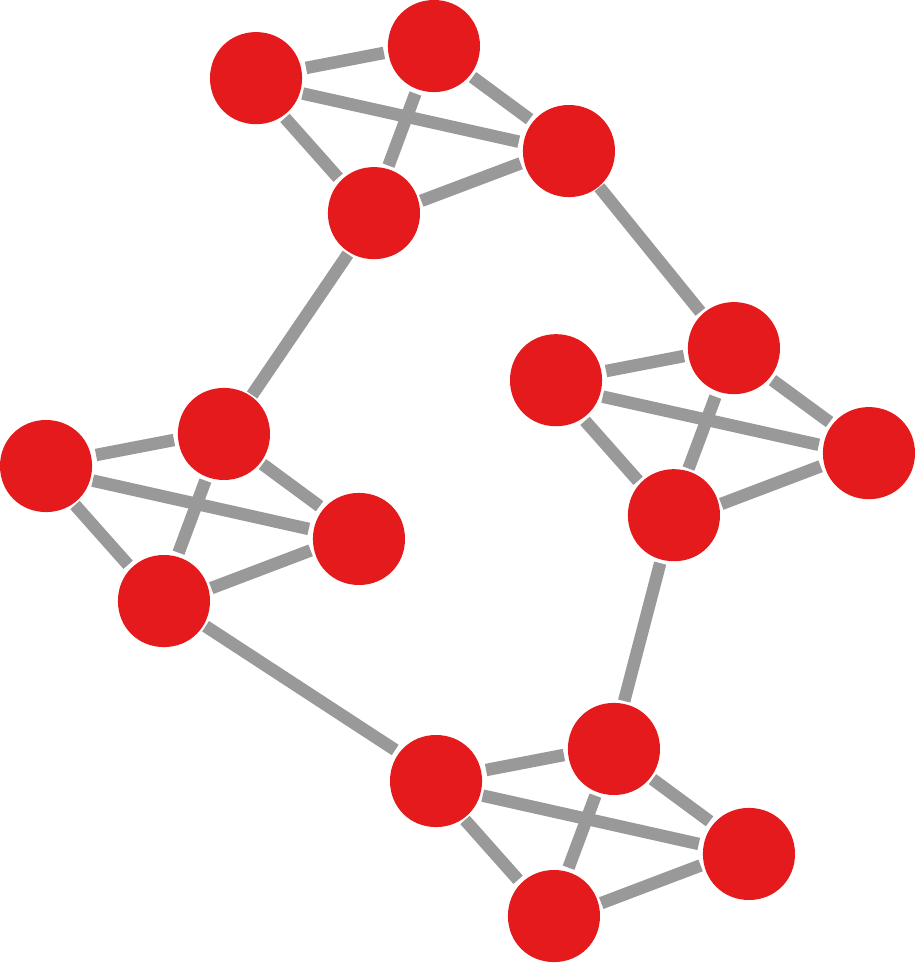}
\caption{}
\end{subfigure}
\caption{(a) First step of cavemen: decide the size of the cave. (b) Second step: decide the number of caves. (c) Third: make each cave in a clique. (d) Finally connect the nearest caves via random cave members.}
\label{fig:cavemen}
\end{figure}

This metaphor was the starting point for Watts in developing his ``cavemen'' model\cite{watts1999networks}. The cavemen model is part of the family of simple networks (see Section \ref{sec:extended-types}). It takes two parameters: the cave size (Figure \ref{fig:cavemen}(a)) and the number of caves (Figure \ref{fig:cavemen}(b)). The cave size is the number of people living in each cave. A cave is a clique: as said, everyone in the cave knows every cavemate (Figure \ref{fig:cavemen}(c)). Each cave ``elects'' a random member which will connect to a random member of the nearest cave on the left, and another member to connect to the nearest cave to the right (Figure \ref{fig:cavemen}(d)). That's it: the cavemen model.

By construction, the cavemen model has only one component, because the caves are always connected with their nearest neighbors. However, the cavemen model is worse than $G_{n,p}$ in approximating realistic diameters. To go from one cave to the farthest one in the network it takes a really long path. The degree distribution is also weird: in the example of Figure \ref{fig:cavemen}, all nodes inside a cave have the same degree (equal to three) and the nodes in between caves all have degree equal to four. Needless to say, this system with only two distinct degree values isn't found anywhere in natural networks.

So why do we want this type of graph? Well, differently from $G_{n,p}$, cavemen gives us clustering and communities. Having such well separated groups -- the caves -- makes it an ideal dataset to test whether your community discovery algorithm is working or if it is returning random results. You can't get anything clearer than a clique with just two edges pointing outwards.

\section{Path Lengths}\label{sec:physicsmodels-ws}
The small-world model is a more famous model developed by the same author\cite{watts1998collective}. It also models high clustering, but its primary target was to explain small diameters, which were discovered in real world social networks by Milgram, as I showed in Section \ref{sec:shortpath-avglength}. In a small world we start from you. You are standing in a certain point in space. Then there are other people, also in their spots. You can only communicate with people that are nearby you, because they are the ones who can listen to you. Their sets of listeners overlap with yours, because you're all nearby each other. But, since they are not exactly occupying your position, there are some folks whom you can reach and they cannot, and vice versa. They can talk to an extra neighbor. And so can their neighbors, to infinity.

\begin{figure}
\centering
\includegraphics[width=.75\columnwidth]{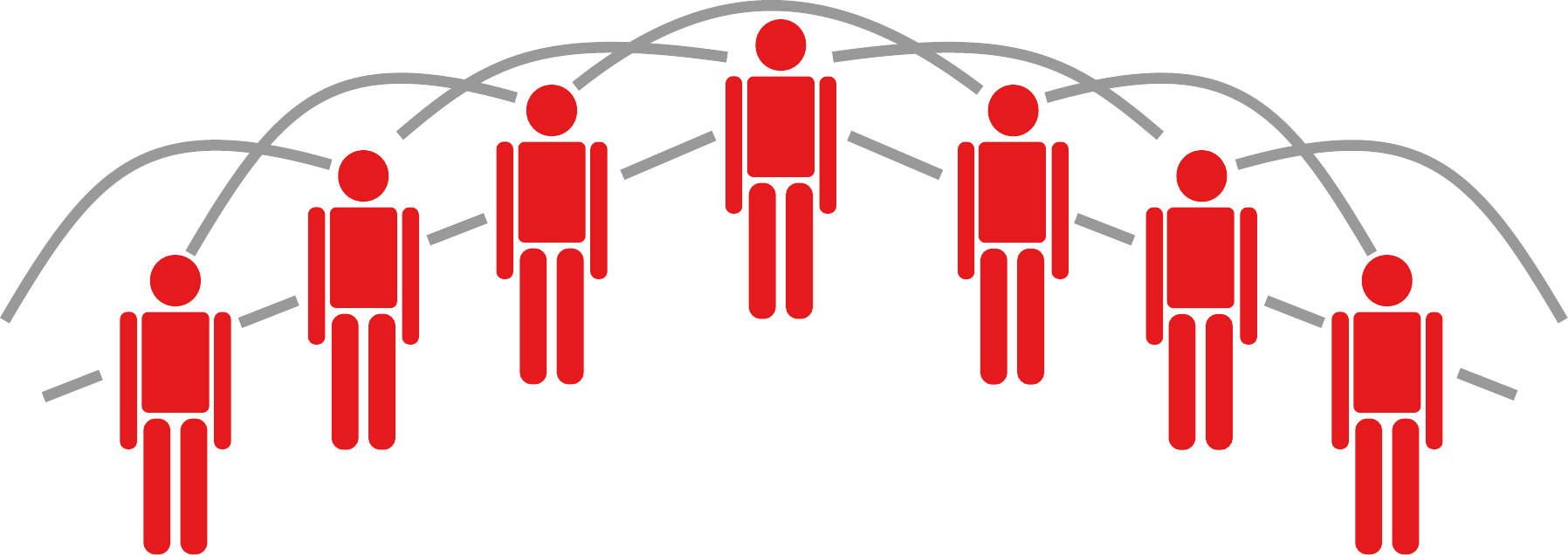}
\caption{The first step of a small world model: each individual can talk to their four most immediate neighbors. }
\label{fig:smallworld1}
\end{figure}

This creates a regular network, a lattice, where each node is the same as each other node. This is a simple network. Figure \ref{fig:smallworld1} provides a simple visualization. To use a more formal language, in the first step of a small world model you put nodes into a single dimensional space and connect them with their $k$ nearest neighbors -- with $k$ being the first parameter of the model that you can specify.

\begin{figure}
\centering
\includegraphics[width=.75\columnwidth]{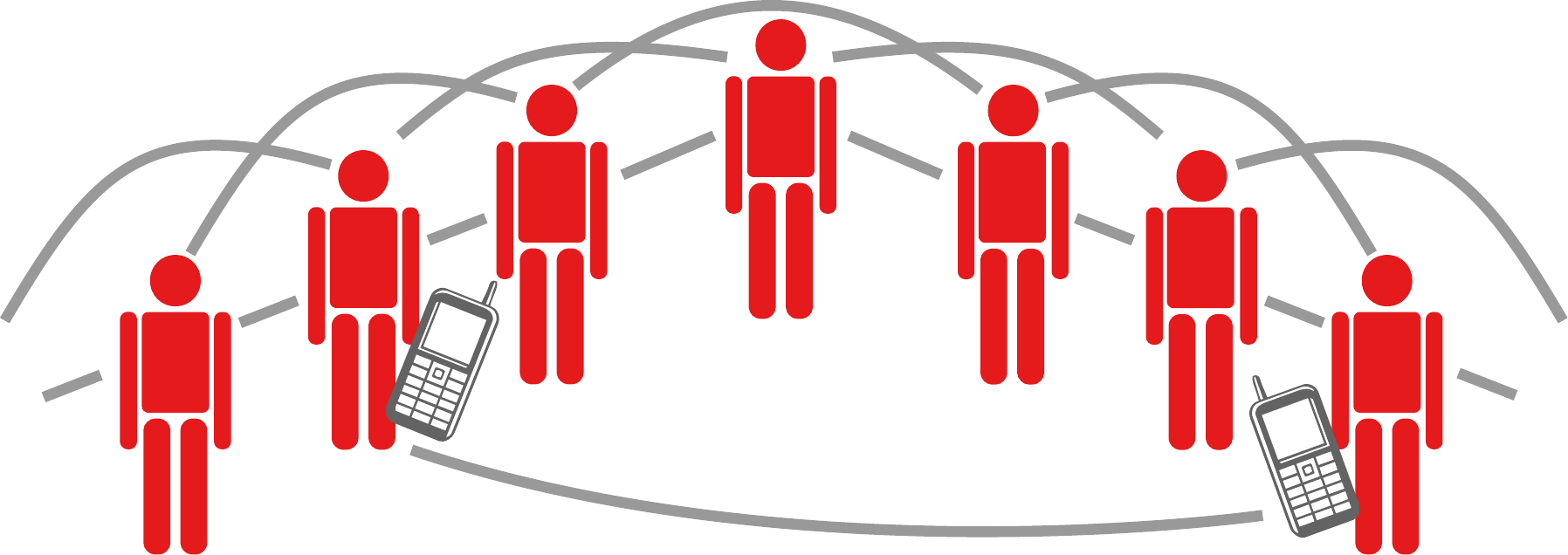}
\caption{The second step of a small world model: random individuals can talk to someone in the network, no matter their physical locations.}
\label{fig:smallworld2}
\end{figure}

However, sometimes, two folks can talk at distance, maybe because they have each other phone number. And so a shortcut is made, as I show in Figure \ref{fig:smallworld2}. Formally, you establish a rewiring probability $p$ -- the second parameter of the model. For each edge you toss a coin: if it lands on heads you delete the edge and you rewire it by picking a random destination. In variants of the model\cite{newman1999renormalization} you do not delete the original edge: for each existing edge you have a certain probability to pick an additional random destination for one of the two connected nodes.

\begin{figure}
\centering
\begin{subfigure}{.3\columnwidth}
\includegraphics[width=\textwidth]{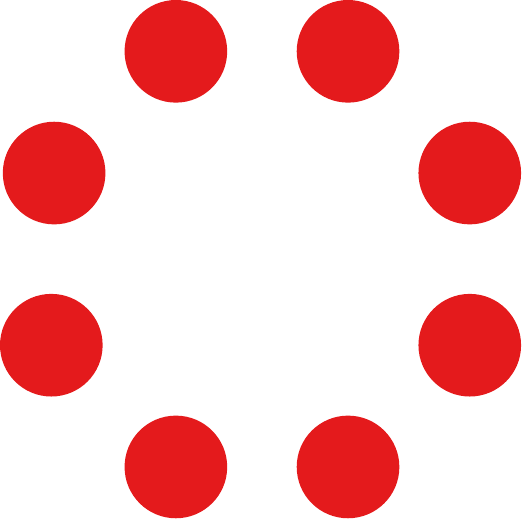}
\caption{}
\end{subfigure}\quad
\begin{subfigure}{.3\columnwidth}
\includegraphics[width=\textwidth]{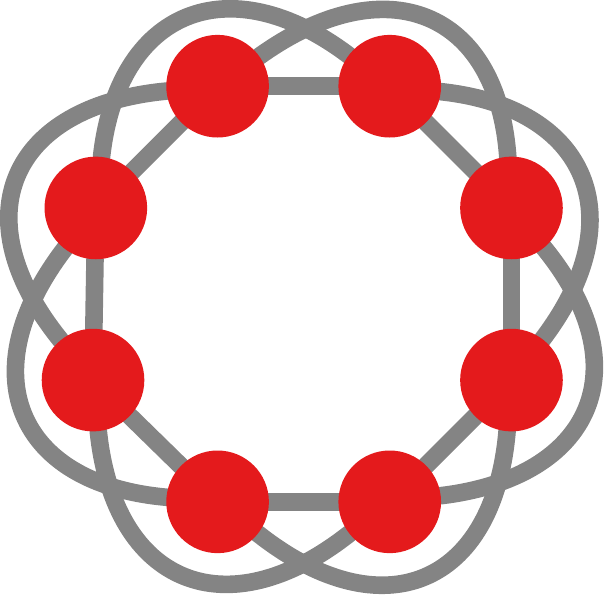}
\caption{}
\end{subfigure}\quad
\begin{subfigure}{.3\columnwidth}
\includegraphics[width=\textwidth]{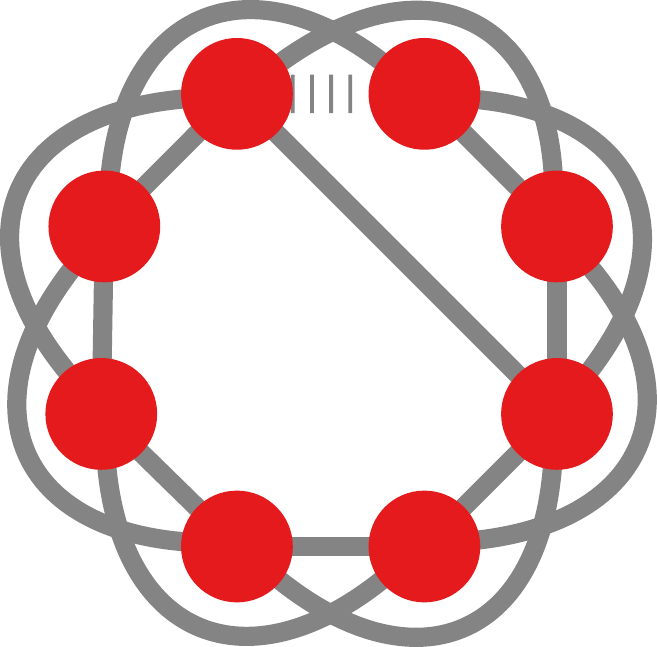}
\caption{}
\end{subfigure}
\caption{The steps of the small world mode: (a) Place nodes in a one-dimensional ring space. (b) Connect each node with its $k$ nearest neighbors. (c) Randomly rewire edges according to the parameter $p$.}
\label{fig:smallworld3}
\end{figure}

Figure \ref{fig:smallworld3} shows the full process. Again, by construction we don't need fancy math to prove that this model generates a single component. Since each node connects to a few nearest neighbors we have one component by default. In the rewiring model you could divide the network in separated components by unlucky rewiring draws, but this is pretty unlikely, especially for high $k$ values. Anyhow, given how unlikely this is, we can say that the small world model properly recovers the giant connected component property of real world networks.

Differently from cavemen, this time we have short paths. They are regulated by the rewiring probability $p$. Rewiring creates bridges that span across the network. Even a tiny bridge probability can connect parts of the network that are very far away. Thousands of shortest paths will use it and will be significantly shorter.

Still, the degree distribution of a small world model is very weird. If $p$ is low, it looks like a cavemen graph, because almost all nodes will have the same number of neighbors. If $p$ is high it means that we are practically rewiring every edge in the graph. At that point, randomness overcomes every other feature of the model, and the result would be almost indistinguishable from a $G_{n,p}$ model. We have high clustering because each connection that we don't rewire will create triangles with some of the neighbors of the connected nodes\footnote{Unless you set $k = 2$, then the clustering would be zero. But why would you set $k = 2$ in a small world model? People are weird.}. 

\begin{figure}
\centering
\includegraphics[width=.75\columnwidth]{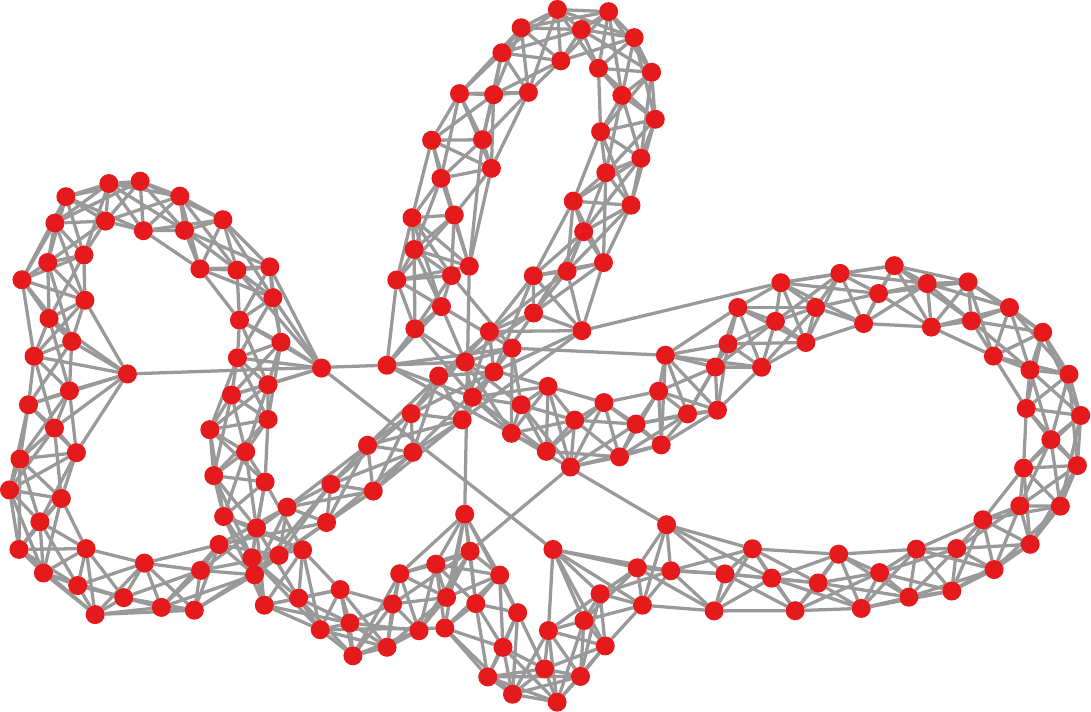}
\caption{A small world graph with $200$ nodes, average degree $\bar{k}$ equal to $8$, and rewiring probability $p = 0.01$.}
\label{fig:smallworld4}
\end{figure}

However, high clustering does not necessarily mean that you are going to have communities. In fact, a small world model typically doesn't have them. The triangles are distributed everywhere uniformly in the network. There are no discontinuities in the density, no differences between denser and sparser areas. This is especially evident for high $k$ and low $p$: as I show in Figure \ref{fig:smallworld4}, small world networks with such parameter combinations just look like odd snakes without clear groups of densely connected nodes. This is a precondition to have communities, and so you cannot find them in a small world model.

\section{Degree Distribution}\label{sec:physicsmodels-ba}
If we want to reproduce the degree distribution of a scale free network we need a process that can generate a power law degree distribution. One of the most popular approaches is cumulative advantage\cite{cole1974social}. This is a fundamentally dynamic model. You have an initial condition and then you keep adding one element at a time. Each element you add does not contribute to the preexisting ones uniformly at random, but \textit{prefers} to contribute to specific older elements, according to a rule you determine.

The preferential attachment model starts from the assumption that the rich get richer. For instance, suppose you have one coin and you invest in the stock market. If you're lucky, after a while, you will have another coin. Consider instead somebody who has a lot of coins. Not only she can match your returns, she can probably do better, because she can have a diversified portfolio which is resilient to market shocks and black swans -- highly improbable but also massively impactful events, like the one at the basis of the mortgage crisis\cite{taleb2007black}. Moreover, she can probably pay better advisers, and capital -- according to Piketty\cite{piketty2014capital} -- just has better returns at scale. In the time it takes for you to make a coin, she makes hundreds. Being already rich makes her proportionally richer than you.

\begin{figure}
\centering
\includegraphics[width=.75\columnwidth]{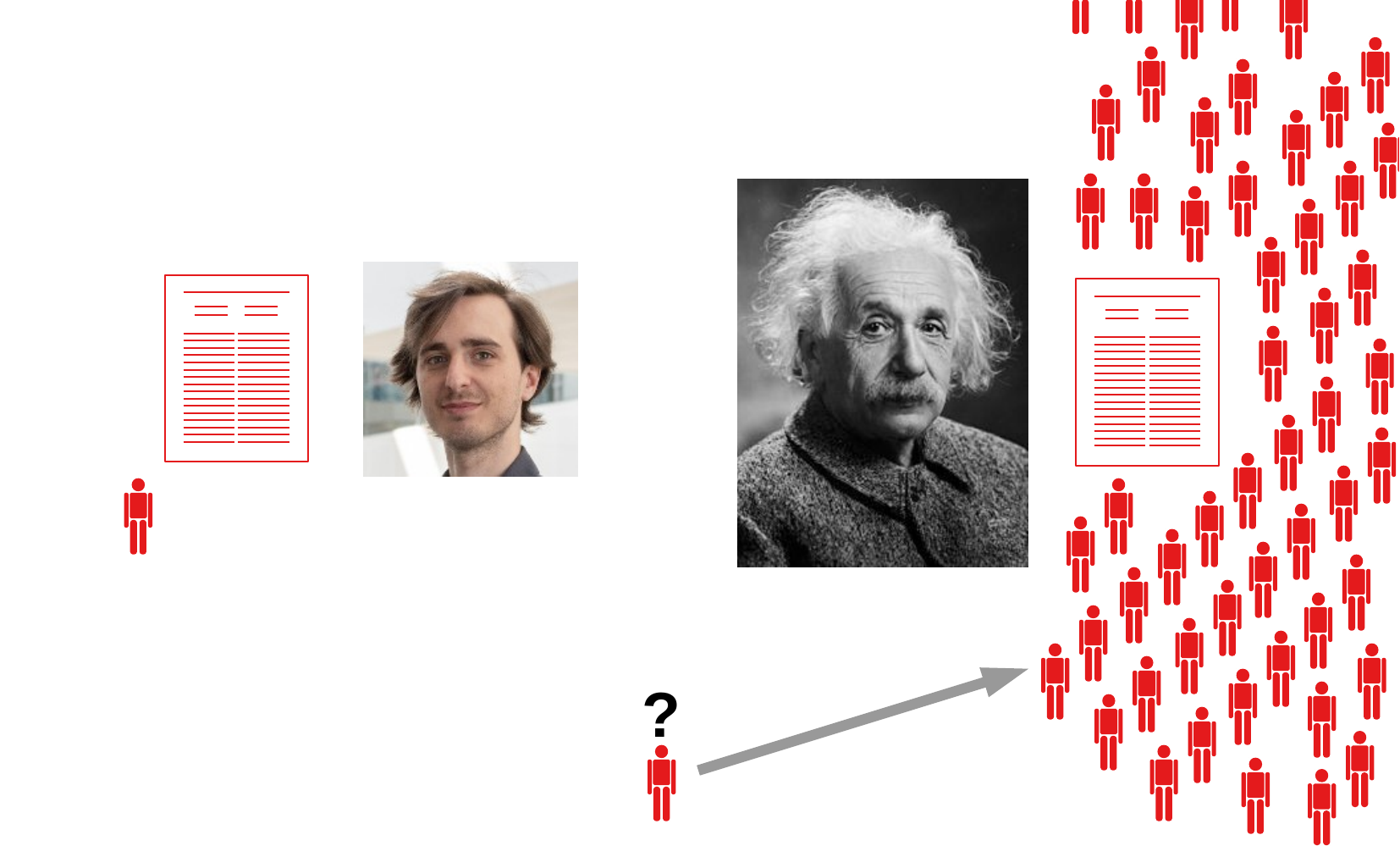}
\caption{We both have crazy hair, so the only difference between Einstein and me is that he got an unfair starting advantage which accumulates over time. Obviously.}
\label{fig:prefattach1}
\end{figure}

The textbook case of cumulative advantage is in scientific publishing\cite{merton1968matthew}. In terms of networks, consider a citation network. I can write a paper, and maybe at some point somebody will read it and cite it. On the other hand, we might have an actual researcher with a paper that has been cited hundreds of thousands of times. If we have a newcomer to the citation network, what is she more likely to see, and therefore to cite? The paper everybody knows. And so she will add to the pool, further increasing the odds that the paper will be seen by another newcomer. Figure \ref{fig:prefattach1} shows a vignette depicting this process.

As I said, this reasoning is at the basis of a dynamical \textbf{preferential attachment} model\cite[1\baselineskip]{simon1957models}\cite{price1976general}\cite{barabasi1999emergence}. In preferential attachment you start from an initial (set of) node(s) and you keep adding more. Each time you add a new node, it will connect to $m$ of the old ones, where $m$ is a parameter of the model. It will connect at random, but \textit{preferentially} to nodes with higher degree: the higher a node's degree, the more likely it is it will gather more connections. Let's follow the process step by step.

\begin{figure}[b]
\centering
\begin{subfigure}{.46\columnwidth}
\includegraphics[width=\textwidth]{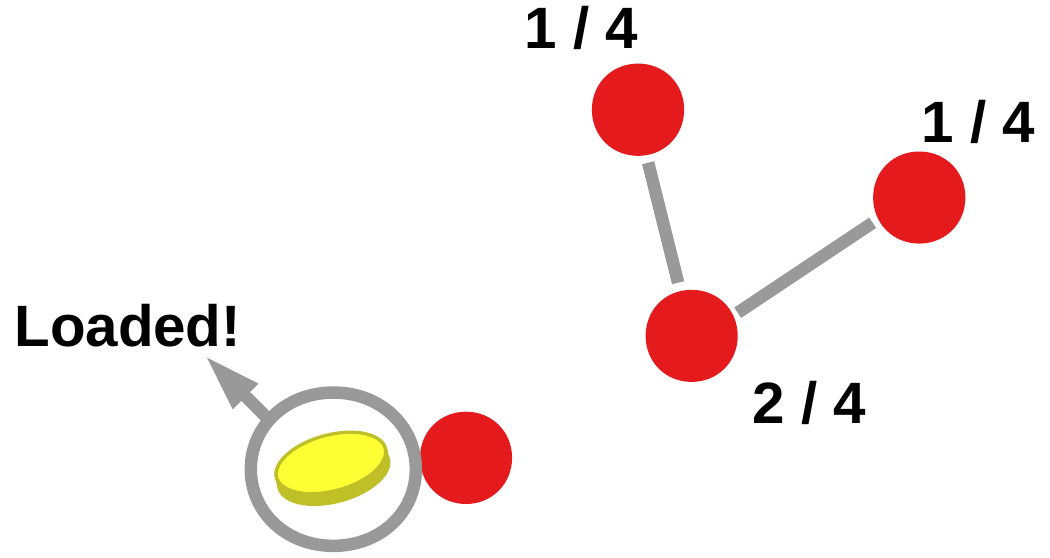}
\caption{}
\end{subfigure}\qquad
\begin{subfigure}{.39\columnwidth}
\includegraphics[width=\textwidth]{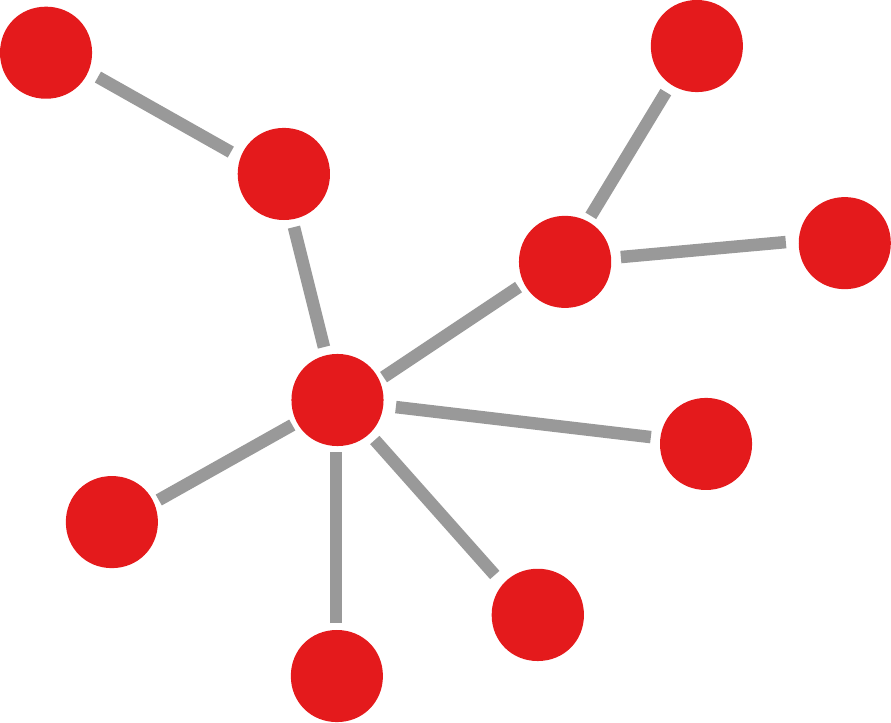}
\caption{}
\end{subfigure}
\caption{(a) Adding a fourth node in a preferential attachment model creates uneven probabilities of attachment (floating next to the nodes that are already part of the network). (b) A possible end result of the preferential attachment after adding ten nodes.}
\label{fig:prefattach2}
\end{figure}

The initial condition, meaning the topology of the network from which you start, is not specified by the model. You can have practically what you want: a clique, a chain, a star. The idea is that, after enough steps, what you started from doesn't matter. Of course, this seed has to have some characteristics that make it compatible with your model. For instance, since you are going to add a new node with $m$ connections, it means that the initial condition has to have at least $m$ nodes. An accepted convention is to have them be connected in a clique, so that they all have the same degree. One downside of this flexible initial condition is that it makes the model a bit less tractable mathematically, as you don't really have a formula describing an arbitrary graph. There are some variants of the model that fix this issue, for instance the linearized chord diagram\cite{bollobas2001degree}.

When you add your first new node, since you only have $m$ initial nodes in the seed and you have to place $m$ connections, there isn't much choice in deciding to whom you connect it. The new node will connect to all nodes in the seed. When you add more nodes, you flip a coin $m$ times to decide who gets the edges from the new node. By the time you're adding the third node, you have more than $m$ nodes to choose from, and they do not have the same degree: some have degree $m$, others have degree $m + 1$. You still flip a coin to decide where the edges go, but now it's a loaded one -- see Figure \ref{fig:prefattach2}(a), where I fix $m = 1$. The new edges are more likely to go to the nodes with more connections. If you keep repeating the process, you end up with something looking like Figure \ref{fig:prefattach2}(b). 

As I said earlier, the law determining the connection probabilities floating next to the nodes in Figure \ref{fig:prefattach2}(a) is the number of connections the node already has over twice the number of edges. In practice, newcomers prefer to attach to high degree nodes. This advantage accumulates over time: if Einstein is the highest degree node, he is the most likely to get a new edge, which makes it even more likely for him to get the next edge, and so on. You see that newcomers have an ever decreasing chance to get the new connections.

This is not the only way to create a cumulative advantage. In fact, the model has some defects. Preferential attachment requires the newcoming nodes to have global information about all the existing nodes' degree. This might be unrealistic in some cases -- you may not know the number of citations of every paper when you are making a citation. It is also not a necessary feature to generate a cumulative advantage.

\begin{figure}
\centering
\begin{subfigure}{.28\columnwidth}
\includegraphics[width=\textwidth]{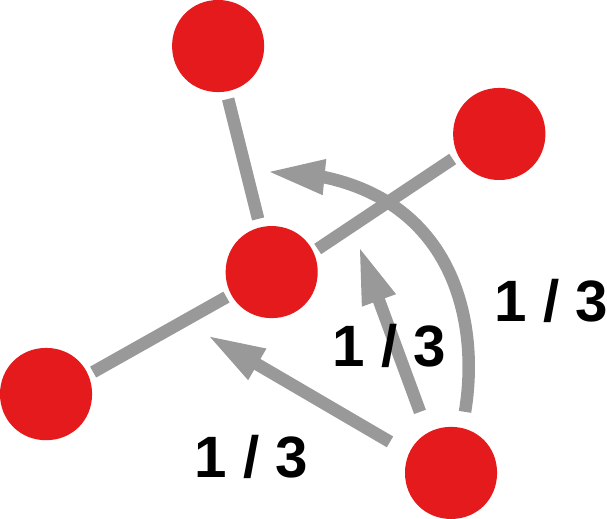}
\caption{}
\end{subfigure}\qquad
\begin{subfigure}{.25\columnwidth}
\includegraphics[width=\textwidth]{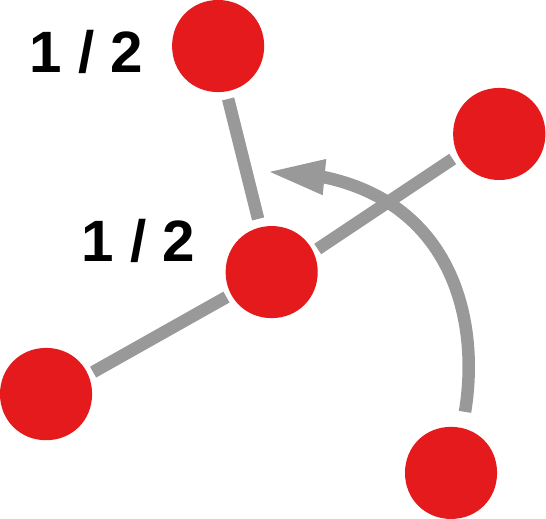}
\caption{}
\end{subfigure}\qquad
\begin{subfigure}{.3\columnwidth}
\includegraphics[width=\textwidth]{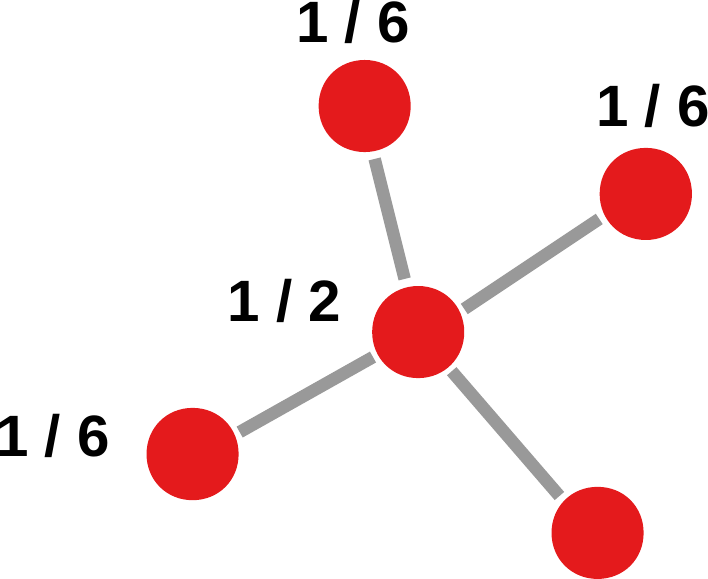}
\caption{}
\end{subfigure}
\caption{Adding a new node with the link selection model. (a) Pick an existing link at random. (b) Pick one of the two nodes connected by that link. (c) Connect to it. The probabilities of connecting to each node in this process are the ones floating next to it.}
\label{fig:linkselection}
\end{figure}

An alternative to preferential attachment is \textbf{link selection}\cite{dorogovtsev2002evolution}. In link selection, the newcoming node selects a link at random from the ones that exist in the network (Figure \ref{fig:linkselection}(a)). Then, it connects with one of the two nodes connected by that edge -- choosing uniformly at random between the two (Figure \ref{fig:linkselection}(b)). Cumulative advantage arises because nodes with more links are more likely to be selected, thus getting more links on average (Figure \ref{fig:linkselection}(c)). No matter which link you select, the central hub is connected to it.

\begin{figure}
\centering
\begin{subfigure}{.29\columnwidth}
\includegraphics[width=\textwidth]{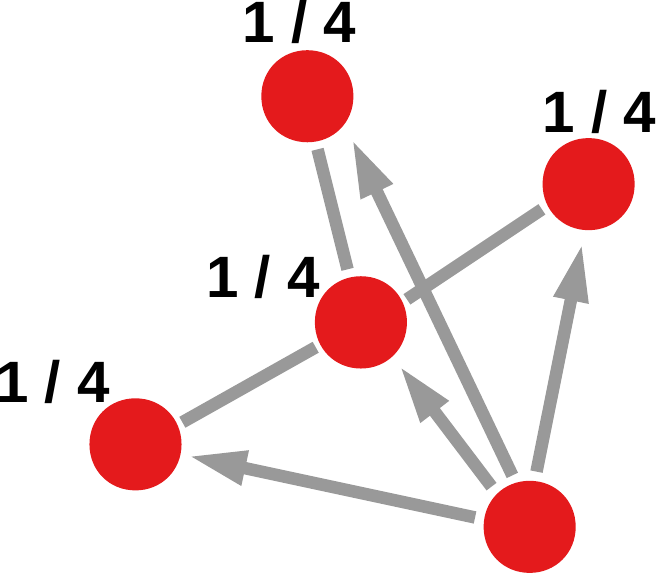}
\caption{}
\end{subfigure}\quad
\begin{subfigure}{.25\columnwidth}
\includegraphics[width=\textwidth]{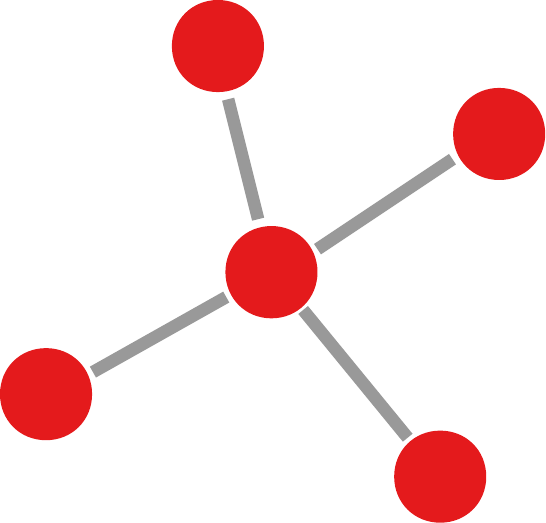}
\caption{}
\end{subfigure}\quad
\begin{subfigure}{.3\columnwidth}
\includegraphics[width=\textwidth]{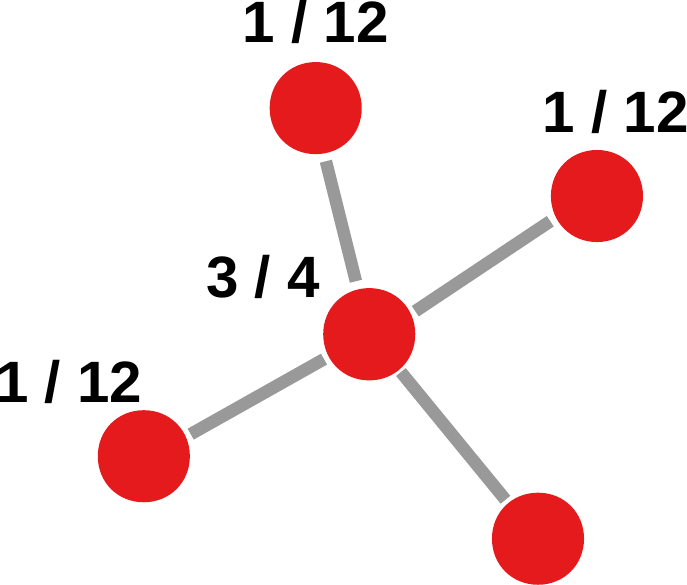}
\caption{}
\end{subfigure}
\caption{Adding a new node with the copying model. (a) Pick an existing node at random. (b) Copy one of its connections. (c) The probabilities of connecting to each node in this process are the ones floating next to it.}
\label{fig:copyingmodel}
\end{figure}

A third alternative from preferential attachment and link selection is the \textbf{copying model}. Just like in link selection, the newcoming node has no information about the network, it just picks something uniformly at random. Differently from the link selection model, here it picks another existing node, rather than a link (Figure \ref{fig:copyingmodel}(a)). It then copies one of its connections (Figure \ref{fig:copyingmodel}(b)). You can see again how it's more likely to connect to the hub: the hub has more neighbors, thus it is more likely to select one of its neighbors. Moreover, the neighbor of a hub is likely to be low degree, increasing the chances of selecting the hub in the copying step (Figure \ref{fig:copyingmodel}(c)). The copying model is based on an analogy on how webmasters create new hyperlinks to pre-existing content on the web\cite{kleinberg1999web}.

It is easy to see why a network generated with either of these three models has a single connected component. Since you always connect a new node with one that was already there, there is no step in which you have two distinct connected components. Thus, any cumulative advantage network following any of these models will have all its nodes in the same giant connected component, as it should.

\begin{figure}
\centering
\includegraphics[width=.8\columnwidth]{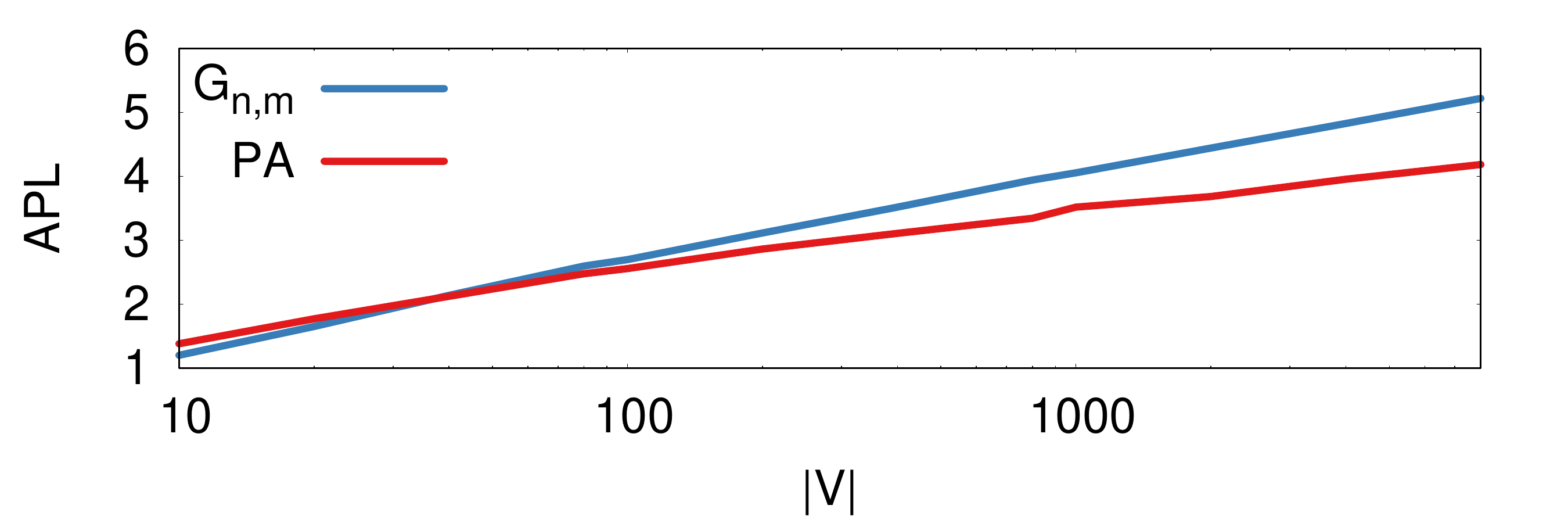}
\caption{The average shortest path length (y axis) for increasing number of nodes (x axis) for $G_{n,m}$ (blue) and preferential attachment (red) models, with the same average degree.}
\label{fig:ba-apl}
\end{figure}

These networks also have short diameters and average path lengths. Mechanically it is easy to see why. Hubs with thousands of connections can be used as shortcuts to traverse the network. Mathematically, the diameter of a preferential attachment network grows as $\dfrac{\log |V|}{\log \log |V|}$, thus very slowly, slower than a random graph. Figure \ref{fig:ba-apl} shows some simulations, comparing the average path length of a $G_{n,m}$ and a preferential attachment network with the same number of nodes and the same number of edges, as their size grows.

These models also reproduce power law degree distributions -- that's what they were developed for. In fact, you can calculate the exact degree distribution exponent for the standard preferential attachment model, which is $\alpha = 3$. This is independent from the $m$ parameter, meaning that you cannot tune it to obtain different exponents (obviously, you might get different exponents because of the randomness of the process, but as $|V| \rightarrow \infty$, then $\alpha \rightarrow 3$). If you want to reproduce a real world network with $\alpha = 2$, you cannot use the basic preferential attachment model.

However, there is a peculiar aspect about their degree distributions that is worth considering. As you saw from all examples, the cumulative advantage applies especially to ``old'' nodes. The earlier the node entered in your structure, the more likely it is to become a hub. This is especially easy to see in preferential attachment: the first node getting the second edge in Figure \ref{fig:prefattach2} already has an advantage that newcomers are unlikely to match.

\begin{figure}
\centering
\includegraphics[width=.66\columnwidth]{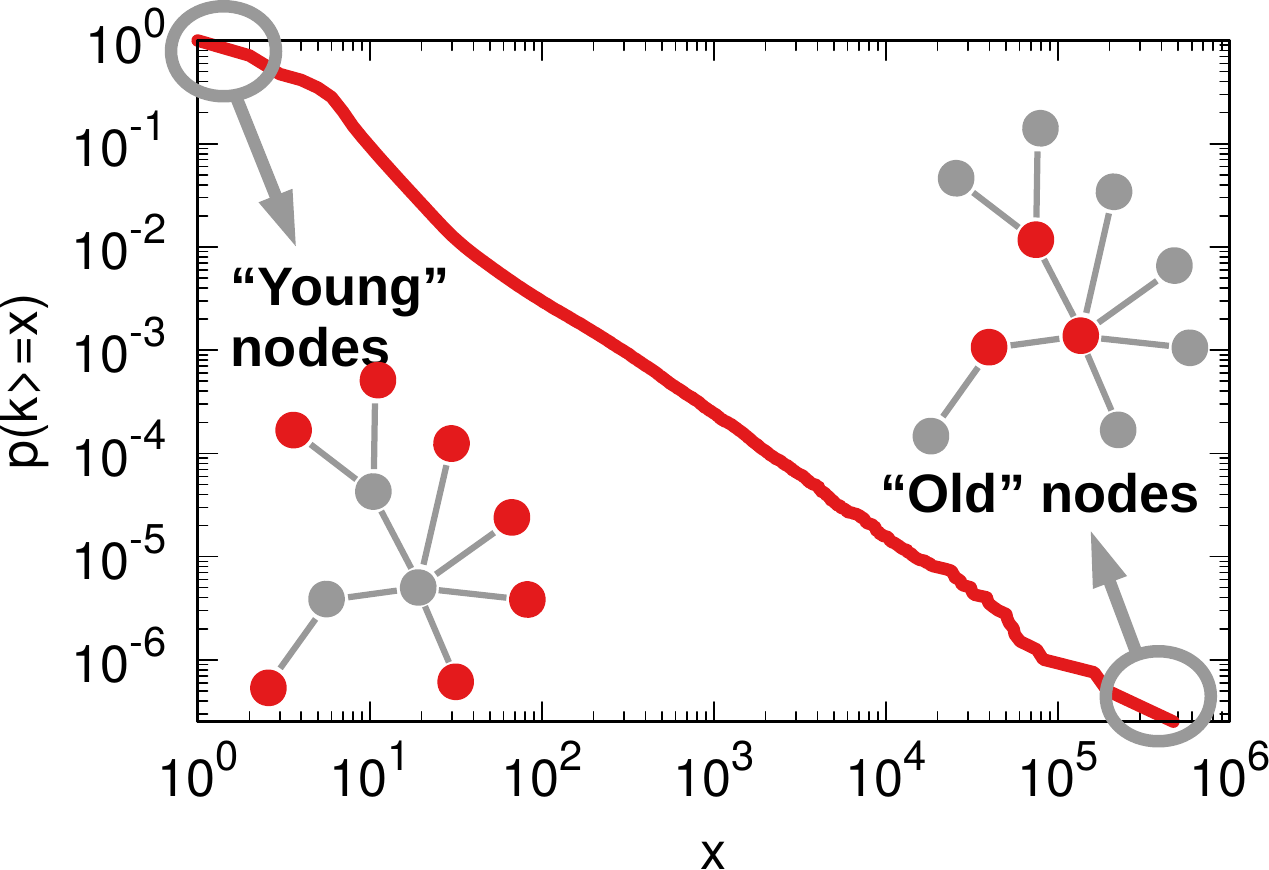}
\caption{The age effect in the degree distribution of the preferential attachment model.}
\label{fig:prefattach3}
\end{figure}

If you look at a degree distribution of a preferential attachment model, you'll find the old nodes in the tail -- the hubs -- and the head is going to be mostly composed by ``young'' nodes. I show an example in Figure \ref{fig:prefattach3}. Put it another way, there is a positive correlation between a node's age and its degree. This is particularly interesting because that is not something we observe in real world systems\cite{adamic2000power}. For instance, you could argue that, on the web, the number one website at the moment is \texttt{google.com}. However, \texttt{google.com} isn't the oldest website in the world. The web was already four years old when \texttt{google.com} was created. Moreover, readers a hundred years from now\footnote{I'm very optimistic about the success of this book.} might not even know what \texttt{google.com} is, because something replaced it.

There are ways to tweak the classical preferential attachment model to fix some of its issues. For instance, one way is to balance popularity and similarity\cite{papadopoulos2012popularity}. To determine where the connections of a new node attach to, the classical preferential attachment uses exclusively the popularity of the already present nodes: the more connections a node has, the more it'll gather. However, nodes will want to connect to other nodes that are similar to them -- we'll see this real world tendency when we'll discuss about homophily in Chapter \ref{cha:homophily}. Thus, if you have metadata about how similar two nodes are, you can create a model where this similarity score is as important as a node's popularity to determine which nodes will connect to it when they first arrive in the network.

From the examples made so far, you probably figured out that there's another thing missing: clustering. In particular, so far I made simple examples where a newcoming node will add only one edge to the network (the parameter $m$ is equal to one). If we add one node and one edge at a time it is impossible to create triangles.

You could set the parameter $m > 1$ to add two or more edges per new node, but that helps only to a certain point: it's not so likely to strike two already connected nodes thus creating a triangle. The preferential attachment model has a higher clustering than a random $G_{n,p}$ one, but not by much. It is still a far cry from the clustering levels you see in real world networks. Moreover, this is only true for $m > 1$. In that case, you add more than one edge per newcomer node, which means you end up losing the head of your distribution. The network will not contain a single node with degree equal to one.

Thus the clustering in these cumulative advantage models is much lower than real world networks, and there are no communities -- because everything connects to hubs which make up a single core. There are some extensions of the model which try to include clustering\cite{holme2002growing}. At every step of this model you have a choice. You either add a node with its links, or you just add links between existing nodes without adding a new one. The probability of taking that step regulates the clustering coefficient of the network.

However, triangles close randomly, thus we have no communities just like in the small world model. If we want to look at models which generate more realistic network data, we have to look at the ones I discuss in the next chapter.

\section{Summary}

\begin{enumerate}
\item To explain the high clustering and small diameter in real world networks we could use the small world model by Watts and Strogatz. In it, we place nodes on a regular distance in a low dimensional space and connect them to $k$ of their neighbors, ensuring high clustering. We then create few shortcuts connecting pairs of nodes at random with probability $p$, ensuring a small diameter.
\item The small world model has no communities, which you could generate with a caveman graph: a ring of cliques. However, the caveman graph has a long diameter.
\item To explain power law degree distributions you could use a preferential attachment model. In it, you grow the network one node at a time. Each node brings $m$ random connections. The probability of connecting to a node $u$ already present in the graph is proportional to $u$'s degree. This model has low diameter, but also low clustering.
\item Alternative models recreating power law degree distributions are the link selection and the copying models.
\item Another unrealistic effect of the preferential attachment model is the correlation between a node's age (how long ago we added it to the network) and its degree. Such correlation might not exist in real world networks.
\end{enumerate}

\section{Exercises}

\begin{enumerate}
\item Generate a connected caveman graph with $10$ cliques, each with $10$ nodes. Generate a small world graph with $100$ nodes, each connected to $8$ of their neighbors. Add shortcuts for each edge with probability of $.05$. The two graphs have approximately the same number of edges. Compare their clustering coefficients and their average path lengths.
\item Generate a preferential attachment network with $2,000$ nodes and average degree of $2$. Estimate its degree distribution exponent (you can use either the \texttt{powerlaw} package, or do a simple log-log regression of the CCDF).
\item Implement the link selection model to grow the graph in \url{http://www.networkatlas.eu/exercises/17/3/data.txt} to $2,000$ nodes (for each incoming node, copy $2$ edges already present in the network). Compare the number of edges and the degree distribution exponent with a preferential attachment network with $2,000$ nodes and average degree of $2$.
\item Implement the copying model to grow the graph in \url{http://www.networkatlas.eu/exercises/17/4/data.txt} to $2,000$ nodes (for each incoming node, copy one edge from $2$ nodes already present in the network). Compare the number of edges and the degree distribution exponent with networks generated with the strategies from the previous two questions.
\end{enumerate}

\chapter{Generating Realistic Data}\label{cha:csmodels}
Both the small world and the preferential attachment models are useful because they give us ideas on how some real world network properties arise. The small world model tells us that small diameters happen because a clustered network might have some random shortcuts. The preferential attachment model tells us that broad degree distributions arise because of cumulative advantage: having many links is the best way to attract more links.

Yet, neither of them is able to reproduce all the features of a real world network. If we want to do so, we have to sacrifice the explanatory power of a model. We have to fine tune the model so that we force it to have the properties we want, regardless of what realistic process made them emerge in the first place. This is the topic of this chapter.

\section{Configuration Model}\label{sec:csmodels-conf}
The easiest way to ensure that your network will have a broad degree distribution is to force it to have it. No fancy mechanics, no emerging properties. You first establish a degree sequence and then you force each node to pick a value from the sequence as its degree. This simple idea is at the basis of the configuration model\cite{newman2003structure}. In fact, the configuration model is more general than this. You can use it to match the degree sequence of \textit{any} real world graph, regardless of the simplicity or complexity of its actual degree distribution.

\begin{figure}
\centering
\includegraphics[width=.83\columnwidth]{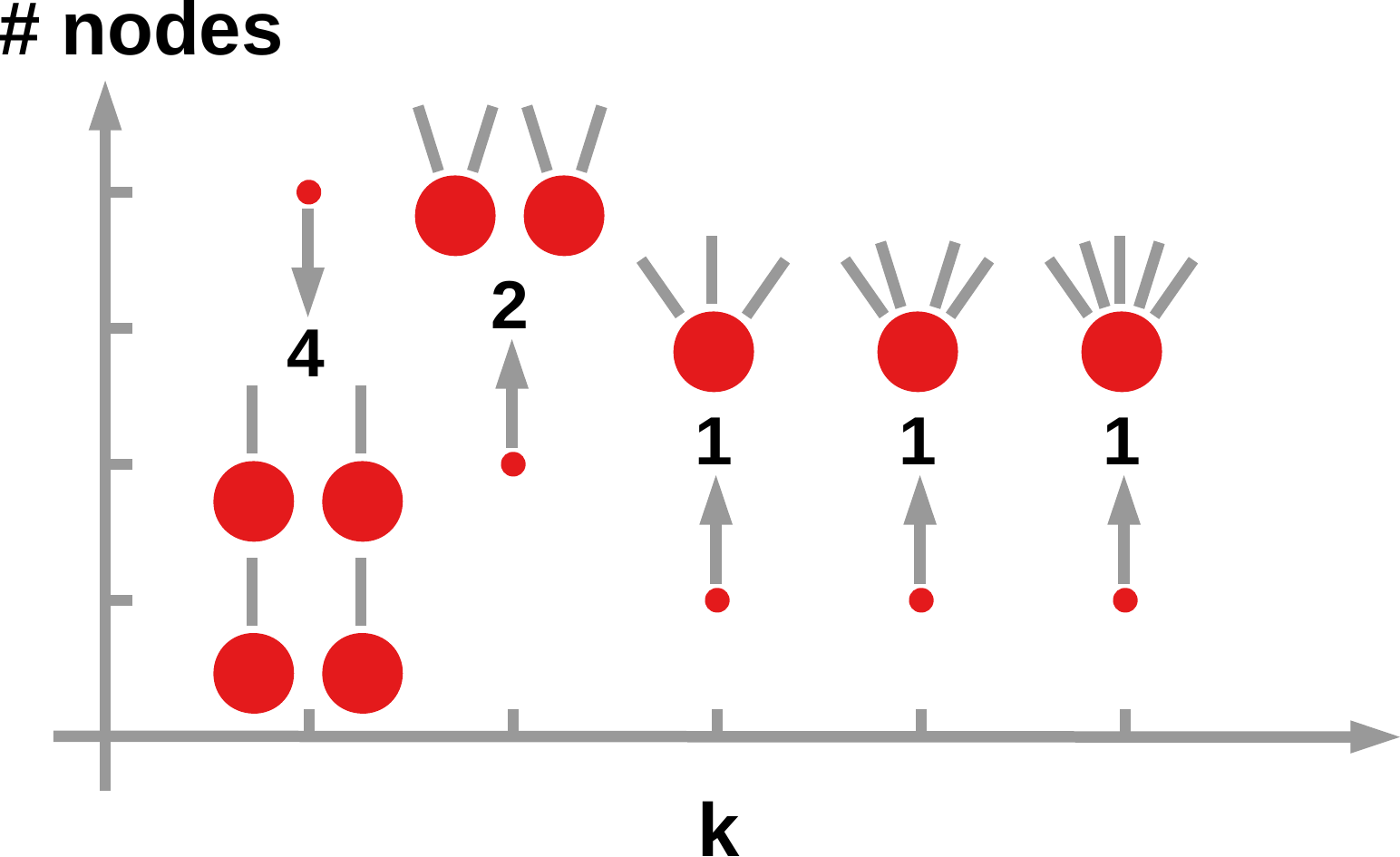}
\caption{In a configuration model, you start from the degree histogram to determine how many nodes have how many open ``edge stubs''.}
\label{fig:confmodel}
\end{figure}

The configuration model starts from the assumption that, if we want to preserve the degree distribution, we can take it as an input of our network generating process. We know exactly how many nodes have how many edges. So we forget about the actual connections, and we have a set of nodes with ``stubs'' that we have to fill in. Figure \ref{fig:confmodel} shows an example.

There's a relatively simple algorithm to generate a configuration model network, the Molloy-Reed approach\cite{molloy1995critical}\cite{newman2001random}. First, as we saw, you create a degree sequence, in which each value represents a node's degree. This sequence has a few constraints: the most important is that it has to sum to an even number. If it were to sum to an odd number, you'd have a node which cannot assign its last stub to any other neighbor -- i.e. the sequence is not ``graphic''.

Second, each node gets a unique identifier. As third step, you generate a list of these identifiers. You repeat each identifier in this list as many times as its assigned degree. For instance, if you know that node $1$ has degree four, the list will contain four $1$s. If node $2$ has degree twenty, you add twenty $2$s.

Finally, you create the actual connections. You pick two elements at random from this list, which are two node identifiers. If the identifiers are different and the two nodes are not already connected to each other, you connect them. The two conditions are necessary to avoid the creation of self loops and parallel edges -- if you're ok with either, you can skip these checks. Note that you might end up with a few unassigned edges, but usually these are an insignificant number which will not affect the degree distribution too much.

\begin{figure}[t]
\centering
\includegraphics[width=.83\columnwidth]{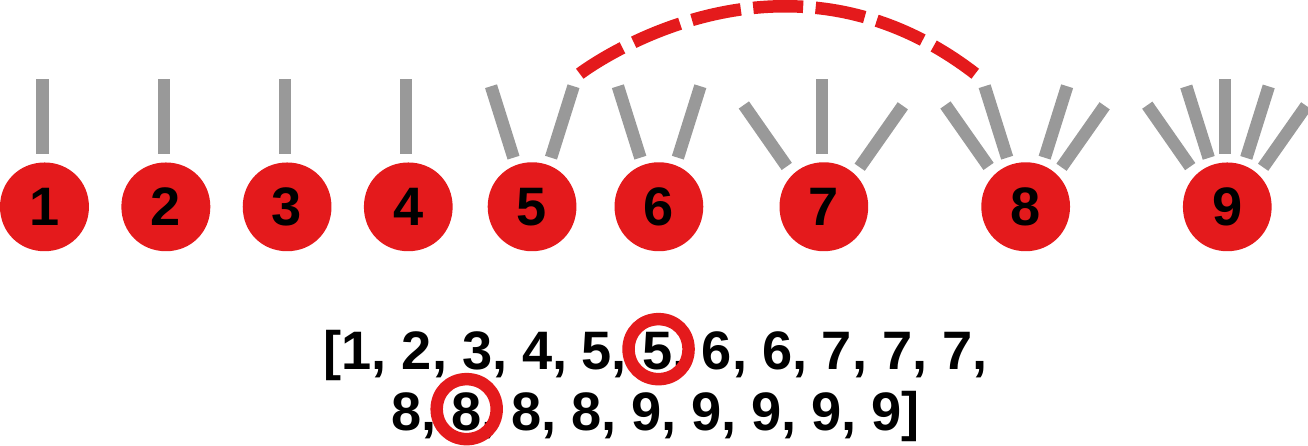}
\caption{A depiction of the algorithm to generate a network following the configuration model: the edge stubs on the nodes with their identifiers (top) and the list with node ids from which we pick nodes (bottom). The red circled ids are the ones picked, so we connect node $5$ to node $8$.}
\label{fig:confmodel2}
\end{figure}

Note that, each time we pick two ids from the list, we remove them from it. This ensures that each node will be picked only as many times as its degree -- or fewer times if we cannot find any legal connections at the end of the process. Figure \ref{fig:confmodel2} shows a depiction of a connection step in the configuration model.

The Molloy-Reed approach is not the only way to generate a random graph with a given degree distribution. For starters, there are closely related alternatives like the Chung-Lu model\cite{chung2002connected}. An alternative is the double swap edge algorithm which I describe in Section \ref{seg:ergmodels-shuffling}. As you'll see, in that case one doesn't need to worry about self-loops or parallel edges. The two algorithms are in different chapters because of their different aims. If you simply need a realistic graph with a given degree distribution for testing an algorithm or do asymptotic mathematics, the Molloy-Reed configuration model is good enough. But, if you want to compare a random graph to data, the differences are crucial\cite[0.25in]{fosdick2018configuring} and that's why the edge swap algorithm is in the chapter dedicated to evaluating statistical significance.

You can add a few features to the configuration models that were not trivial to add to either small world or preferential attachment. For instance, you can make a directed version of it. The only thing you need is to generate two degree distributions: one for the in-degree and one for the out-degree. This makes the connection step a bit more complex, as you have to pick the source from one list and the destination from another, but it is not a big deal.

You can see that this model will be spot on in replicating the vast majority degree distributions you pass to it. The way to estimate the difference between two distributions is by performing a Kolmogorov-Smirnov test\cite{kolmogorov1933sulla}\cite{smirnov1948table}. The test identifies the point of maximum separation between two distributions, along with a $p$-value telling you how likely it is that the two distributions are indistinguishable from each other.

The configuration model tends to generate giant connected components just like a random $G_{n,p}$ graph would, although this heavily depends on the $\alpha$ parameter of your power sequence\cite{aiello2000random} (see Section \ref{sec:degree-fit} for a refresher about the meaning of $\alpha$). Deriving the expected average shortest path length is a bit trickier, but it can be done\cite{chung2002average} and it is realistically short in most cases.

The clustering coefficient of a configuration model also depends on $\alpha$. For the majority of realistic values of $\alpha$ -- between $2$ and $3$ --, the clustering coefficient of a configuration model tends to zero, which is very unrealistic. There are a few valid values of $\alpha$ generating a properly high clustering, but these are rare enough that researchers needed to modify the configuration model to explicitly include the generation of triangles\cite{newman2009random}\cite{miller2009percolation}.

This is usually achieved by generating a joint degree sequence. Rather than simply specifying the degree of each node, we now have to fix two values. The first is the number of triangles to which the node belongs. The second is the number of remaining edges the node has that are not part of a triangle. One can see that we're still specifying the degree, because the number of triangles is simply half the number of edges we're adding to that node: each node in a triangle connects to other two nodes. If you know the number of triangles and the total degree of each node you know the clustering coefficient of the network (Section \ref{sec:density-clustering}), thus you can generate a sequence that will have the desired clustering coefficient.

\begin{figure}
\centering
\includegraphics[width=\columnwidth]{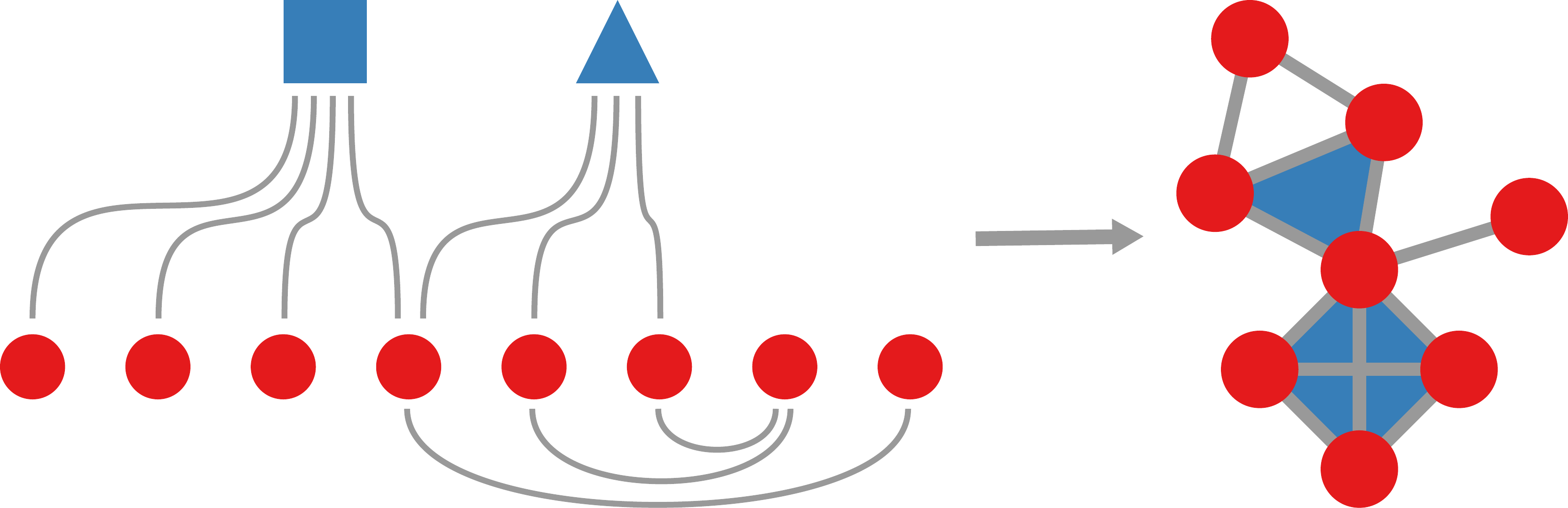}
\caption{A simplicial configuration model with $8$ nodes (in red) and two simplices (in blue). The nodes have the following open stubs (from left to right): $\{1,1,1,3,2,2,2,1\}$.}
\label{fig:confmodel3}
\end{figure}

An alternative -- and more flexible -- way to build higher order structures in your configuration model is to allow it to generate simplicial complexes. I introduced simplicial complexes in Section \ref{sec:extended-hyper}: these are graphs including relations between multiple nodes, rather than just normal edges, which are binary relationships. In a simplicial configuration model, besides nodes with open stubs, you also have simplices, each with the number of stubs determined by their dimension (from $2$ up)\cite{courtney2016generalized}\cite{young2017construction}. Figure \ref{fig:confmodel3} shows an example.

You can connect each node stub with either another node stub, or with a simplex, until there are no more open stubs. Note that, in this model, the number of open stubs is not the degree any more. Every time you connect a node to a simplex, the node's degree increases by the dimensionality of the stub minus one (thus by $2$ if you connect it to a triangular complex, by $3$ if you connect it to a square, and so on). Thus controlling the exact degree of a node is trickier.

\begin{figure}
\centering
\begin{subfigure}{.3\columnwidth}
\includegraphics[width=\textwidth]{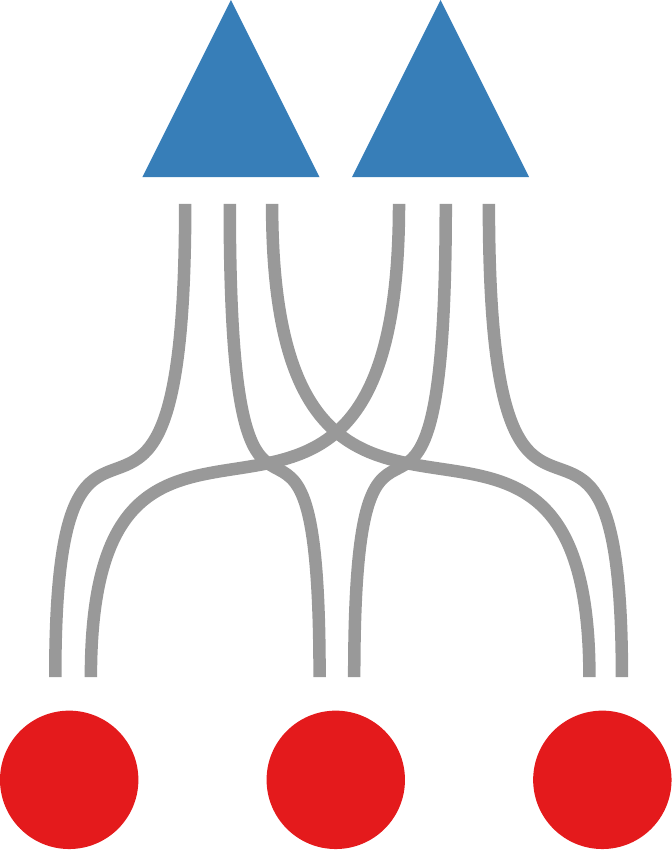}
\caption{}
\end{subfigure}\qquad
\begin{subfigure}{.3\columnwidth}
\includegraphics[width=\textwidth]{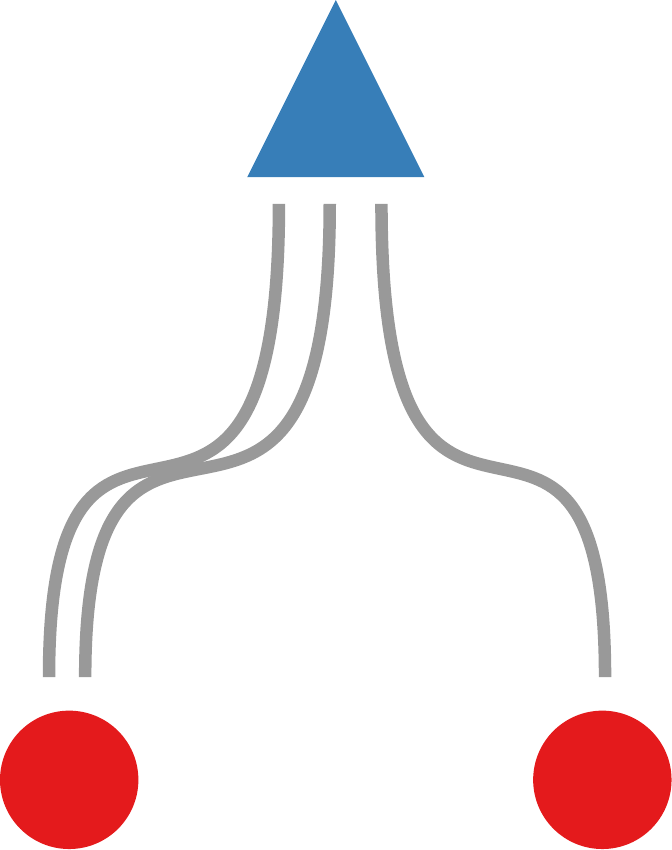}
\caption{}
\end{subfigure}
\caption{Two forbidden moves in the simplicial configuration model.}
\label{fig:confmodel4}
\end{figure}

Another complication is that you have more topolgical constraints. Forget about simply generating a graphical degree sequence: you now have a set of forbidden moves. A node can be part of multiple simplices, but you cannot make two distinct simplices using the very same nodes (Figure \ref{fig:confmodel4}(a)). Nor you can use two stubs from the same node in the same simplex (Figure \ref{fig:confmodel4}(b)). The first case is forbidden because you'd be creating two indistinguishable simplices -- and therefore only one instead of two. In the second case, you wouldn't be constructing a simplex at all!

The result is a network that has exactly the desired number of simplices. It might have, however, more triangles than you expect, because you can connect three nodes independently. There's an example of this mishap in Figure \ref{fig:confmodel3}: one of the two triangles is not filled, because it's made by three independent edges, rather than being a three-way relationship, like the blue-filled simplex.

You don't have to necessarily end up with a simplicial network or a hypergraph: once you have placed the shapes you can ``forget'' that they are higher order structures, and consider you network as a simple graph.

\section{Communities}\label{sec:csmodels-comms}
The configuration model can be tuned to include a high clustering, but it will still close triangles randomly in the network. This means that the number of triangles is correct, but their distribution in the network is random. As I mentioned multiple times, this is not the case for real world networks. The triangles tend to correlate and form denser areas we call communities -- specifically, assortative communities, the concept of community is a bit more complex than that and requires to read Part \ref{par:cd}, for now let's just roll with this simplification. Thus, the configuration model is still inadequate to fully reproduce a quasi-real network. Doing so is the task of a few models: stochastic block models, GN and LFR benchmark, and Kronecker graphs.

\subsection{Stochastic Block Models}
The easiest way to create communities in your graph model is to make a simple observation. Since communities are dense areas -- with nodes connecting to each other -- separated by sparse areas, it just means that nodes have two different probabilities to connect to each other. One probability, say $p_{in}$, determines the probability of a node $u$ to connect to another node inside the same community. Another probability, $p_{out}$, determines the likelihood of connecting to a node from a different community. Obviously, if we want dense communities, $p_{out} < p_{in}$.

$p_{in}$ and $p_{out}$ are the first two parameters of a Stochastic Block Model\cite{holland1983stochastic} (SBM). To fully specify the model you need a few additional ingredients. First, you have to specify $|V|$, the number of nodes in the graph. Then you have to plant a community partition. In other words, for each node -- from one to $|V|$ -- you have to specify to which community it belongs. Otherwise we don't know how to use the $p_{in}$ and $p_{out}$ parameters.

It's easy to see that, if $p_{in} = p_{out}$, then the SBM is fully equivalent to a $G_{n,p}$ model: each node has the same probability to connect to any other node in the network. However, if we respect the constraint that $p_{out} < p_{in}$, the resulting adjacency matrix will be block diagonal. Most of the non zero entries will be close to the diagonal, whose blocks are exactly the communities we planted in the first place!

\begin{figure}
\centering
\begin{subfigure}{.45\columnwidth}
\includegraphics[width=\textwidth]{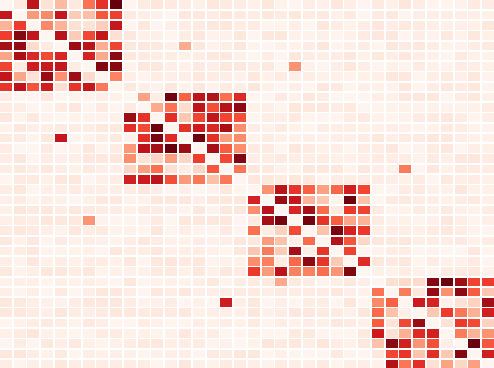}
\caption{}
\end{subfigure}\quad
\begin{subfigure}{.45\columnwidth}
\includegraphics[width=\textwidth]{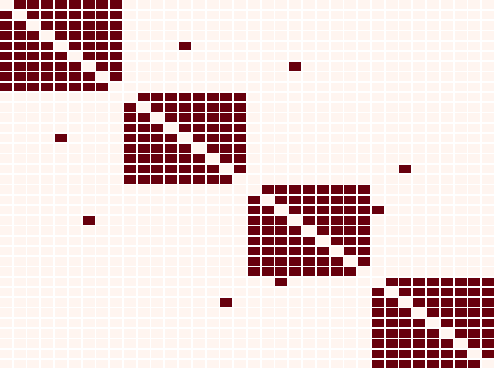}
\caption{}
\end{subfigure}
\caption{(a) A matrix containing, for each entry, the probability of the two nodes to connect to each other. (b) A realization of the SBM using the matrix from (a) as input. Each dark entry is an edge in the network.}
\label{fig:sbm}
\end{figure}

One could go even deeper and determine that each pair of nodes $u$ and $v$ can have its own connection probability. This would generate an input matrix for the SBM that looks like the one in Figure \ref{fig:sbm}(a). Figure \ref{fig:sbm}(b) shows a likely result of the SBM using Figure \ref{fig:sbm}(a) as an input. It's easy to see why we call this matrix ``block diagonal''. The blocks are.... on the diagonal, man.

In one swoop we obtained what we were looking for: both communities and high clustering. The very dense blocks contribute a lot to the clustering calculation, more than the sparse areas around the communities can bring the clustering down. One observation we will come back to is that, in real world networks, the community sizes distribute broadly, just like the degree: there are few giant communities and many small ones. This can be embedded in SBM, since we're free to determine the input partition as we please.

If you set $p_{out}$ to a relatively high value, you might make your communities harder to find, but you gain something else: smaller diameters. You're also free to set $p_{in} < p_{out}$ in which case you'd find a \textit{disassortative} community structure, where nodes tend to dislike nodes in the same community. See Section \ref{sec:homophily-assortativity} to know more about what disassortativity means.

Many real world networks will have overlapping communities sharing nodes. The standard SBM cannot handle this: in the input phase we can only put a node in a single community. However, smart folks have created Mixed-Membership Stochastic Block Models\cite{airoldi2008mixed}, in which nodes are allowed to span across communities. Another important variation of SBM is the degree-correlated SBM\cite{karrer2011stochastic}, which allows you to fix the degree distribution just like the configuration model does.

\subsection{GN Benchmark}
The GN benchmark is a modification of the cavemen graph and one of the first network models designed to test community discovery algorithms\cite{girvan2002community}. The first defining characteristic of this model is setting some of the parameters of the cavemen graph as fixed. In the benchmark, we have only four caves and each cave contains $32$ nodes. Differently from the caveman graph, the caves are not cliques: each node has an expected degree of $16$, thus it can connect at most to half of its own cave.

\begin{figure}
\centering
\begin{subfigure}{.22\columnwidth}
\includegraphics[width=\textwidth]{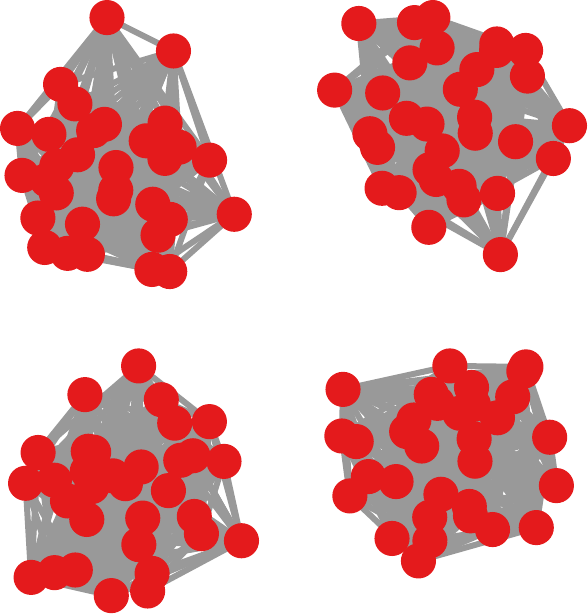}
\caption{$\mu = 0$}
\end{subfigure}\quad
\begin{subfigure}{.22\columnwidth}
\includegraphics[width=\textwidth]{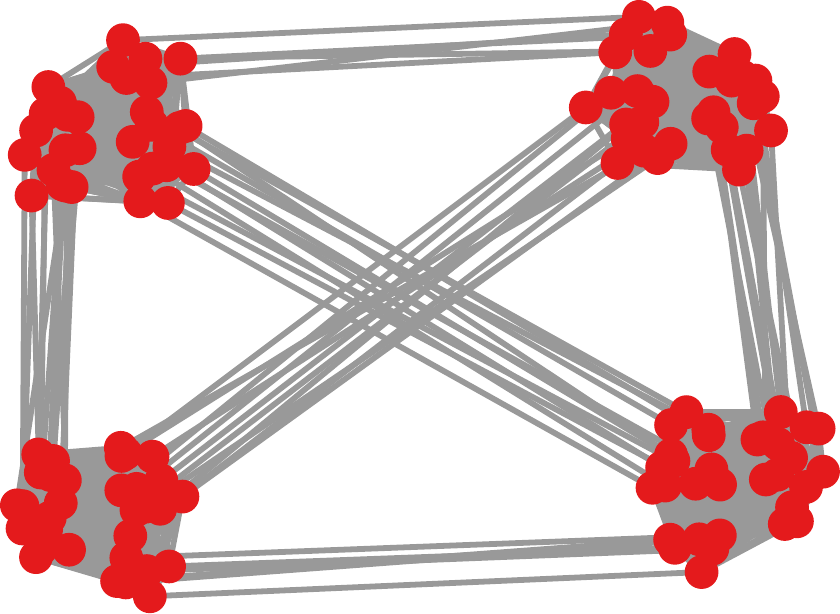}
\caption{$\mu = 0.0625$}
\end{subfigure}\quad
\begin{subfigure}{.22\columnwidth}
\includegraphics[width=\textwidth]{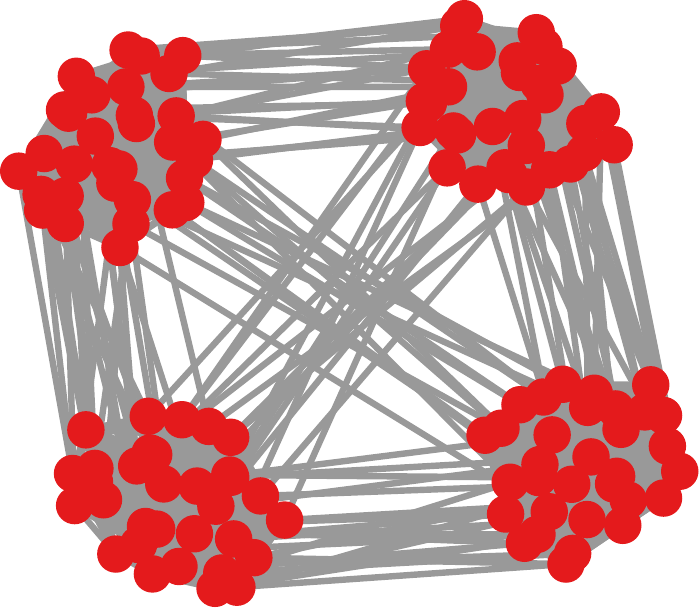}
\caption{$\mu = 0.125$}
\end{subfigure}\quad
\begin{subfigure}{.22\columnwidth}
\includegraphics[width=\textwidth]{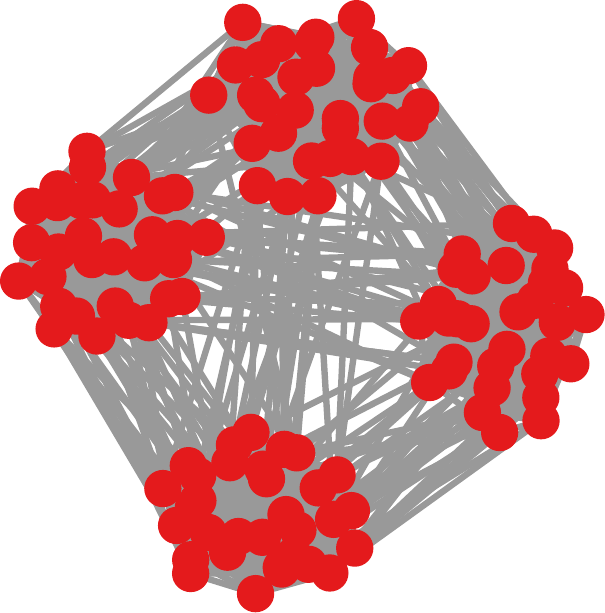}
\caption{$\mu = 0.1875$}
\end{subfigure}
\caption{Results from the GN benchmark for increasing values of $\mu$.}
\label{fig:gn-model}
\end{figure}

The GN benchmark then introduces a parameter, usually called ``mixing'' parameter, or $\mu$. This is the share of edges that a node has to nodes that are not part of its own cave. You can use $\mu$ to introduce the amount of noise you want in the community structure. If $\mu = 0$, then all nodes will exclusively connect with fellow members of the same cave. This results in four isolated connected components with no paths among them. If $\mu = 0.5$, half of the edges of a node point outside its community, to a random other cave. You can see the effect of $\mu$ in the sequence from Figure \ref{fig:gn-model}.

The GN benchmark isn't particularly realistic. It has a fixed number of nodes, rather low when compared with real world networks. Its degree distribution is binomial, which rarely happens in the real world. It is also rare to have equally-sized communities. The LFR benchmark fixes all these issues.

\subsection{LFR Benchmark}
The LFR Benchmark was developed to serve as a test case for community discovery algorithms\cite{lancichinetti2008benchmark}. The objective is to generate a large number of benchmarks to test a new algorithm such that we know the ``true'' allegiance of each node. Once an algorithm returns us a possible node partition, we can compare its solution with the true communities.

Since you want to have networks with lots of realistic properties, some of which are difficult to reproduce organically, the LFR benchmark takes lots of input parameters. If you want an LFR network, you have to specify:

\begin{itemize}
\item The $\alpha$ exponent of the power law degree distribution of the graph;
\item The $\beta$ exponent of the power law size distribution of the communities in the graph;
\item The $|V|$ number of nodes in your graph;
\item The $\bar{k}$ average degree of the nodes -- you can also set $k_{min}$ and $k_{max}$ as the minimum and maximum degree, respectively;
\item $s_{min}$ and $s_{max}$, as the minimum and maximum community size;
\item The $\mu$ mixing parameter, regulating the fraction of edges going outside their planted communities;
\item [Optional] $o_n$: the fraction of nodes overlapping between multiple communities, if using the overlapping variant of LFR;
\item [Optional] $o_m$: the number of communities to which an overlapping node belongs.
\end{itemize} 

\begin{figure}
\centering
\includegraphics[width=.66\columnwidth]{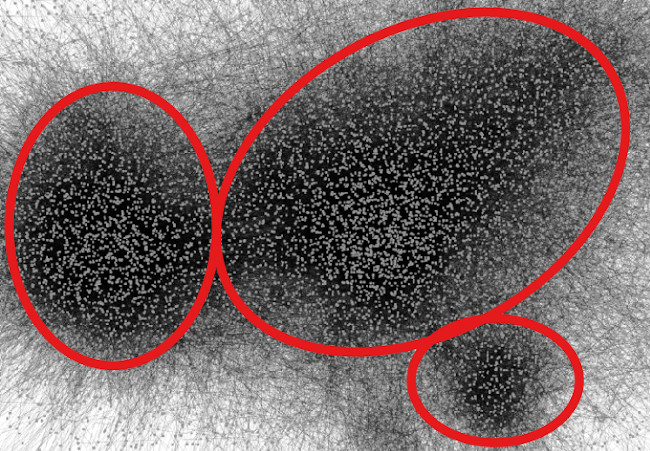}
\caption{A messy real world network in which I highlighted with red outlines the three largest communities.}
\label{fig:lfr-commsizedistr}
\end{figure}

As you can see, the LFR assumes that both the degree distribution and the size of your communities distribute like a power law. The latter is regulated by the $\beta$ parameter, with one gigantic community including the majority of nodes and many trivial communities of size $s_{min}$. In Figure \ref{fig:lfr-commsizedistr} I show a real world network and its three largest communities, showing how their sizes rapidly decline. This is a rather realistic assumption, although there are obvious exceptions. You can take care of such exceptions by forcing the maximum community size to a known -- and lower -- $s_{max}$ size. You can also set the minimum community size if you don't want to have too many trivial communities in your network, using the $s_{min}$ parameter.

\begin{figure}
\centering
\begin{subfigure}{.18\columnwidth}
\includegraphics[width=\textwidth]{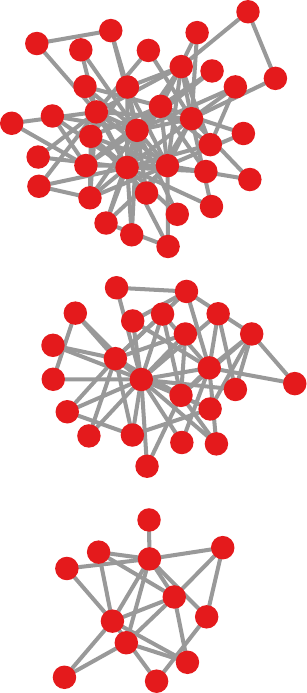}
\caption{$\mu = 0$}
\end{subfigure}\quad
\begin{subfigure}{.26\columnwidth}
\includegraphics[width=\textwidth]{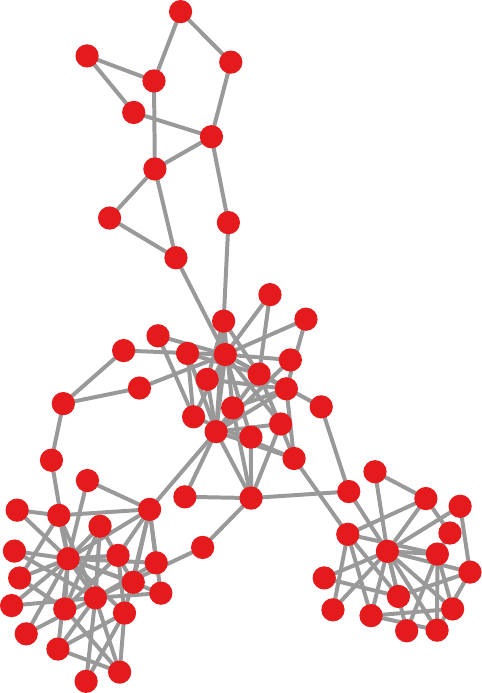}
\caption{$\mu = 0.05$}
\end{subfigure}\quad
\begin{subfigure}{.23\columnwidth}
\includegraphics[width=\textwidth]{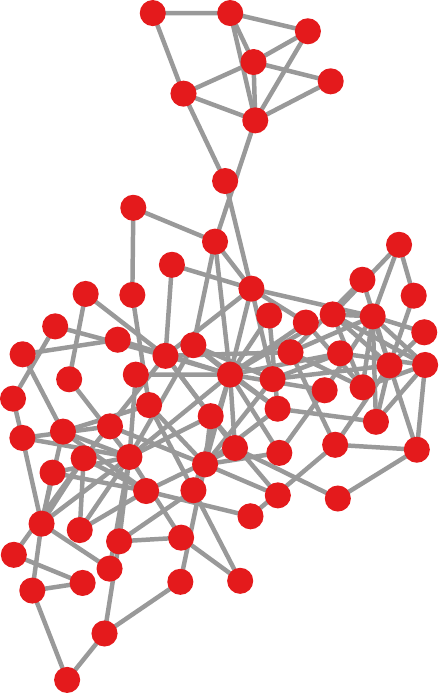}
\caption{$\mu = 0.2$}
\end{subfigure}\quad
\begin{subfigure}{.23\columnwidth}
\includegraphics[width=\textwidth]{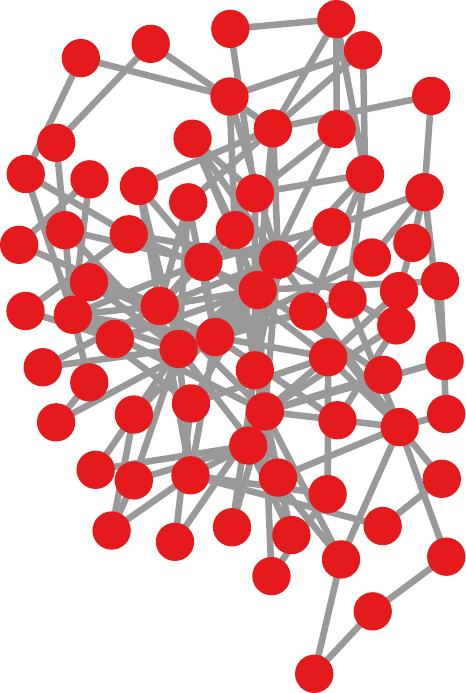}
\caption{$\mu = 1$}
\end{subfigure}
\caption{Examples of LFR benchmarks at different $\mu$ levels.}
\label{fig:lfr-mu}
\end{figure}

The mixing parameter $\mu$ regulates how hard it is to find communities in the network. If $\mu = 0$ all edges run between nodes part of the same community, i.e. each community becomes a distinct connected component of the network (Figure \ref{fig:lfr-mu}(a)). In this scenario, recovering community information is trivial. If $\mu = 1$ then nodes in the same community do not connect at all (Figure \ref{fig:lfr-mu}(d)).  In this scenario, recovering the community wiring is impossible. Usually, you want to set $\mu$ to a reasonably low non-zero value. From Figures \ref{fig:lfr-mu}(b-c) you can see why this is the case: even the seemingly low value of $\mu = 0.2$ generates hard to distinguish communities (Figures \ref{fig:lfr-mu}(c)).

\begin{figure*}
\centering
\begin{subfigure}{.25\columnwidth}
\includegraphics[width=\textwidth]{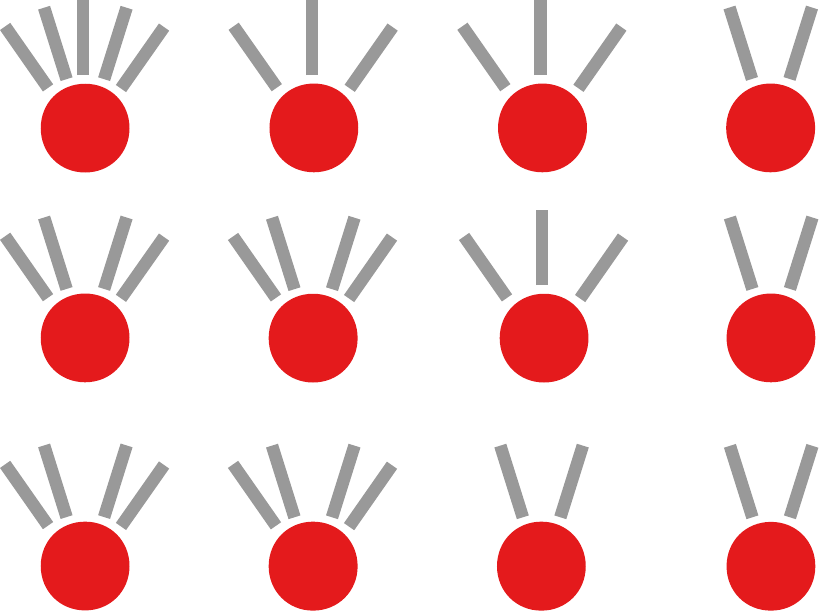}
\caption{Step \#1}
\end{subfigure}\quad
\begin{subfigure}{.25\columnwidth}
\includegraphics[width=\textwidth]{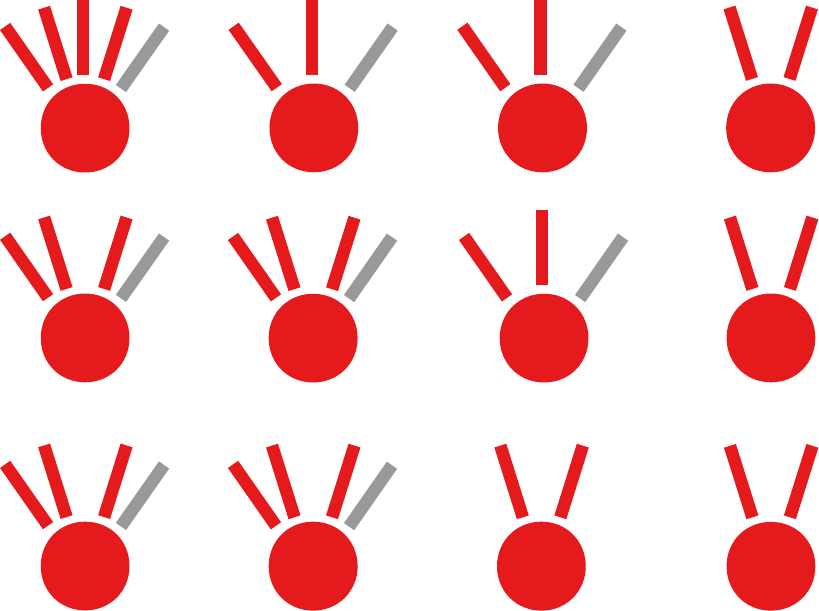}
\caption{Step \#2}
\end{subfigure}\quad
\begin{subfigure}{.25\columnwidth}
\includegraphics[width=\textwidth]{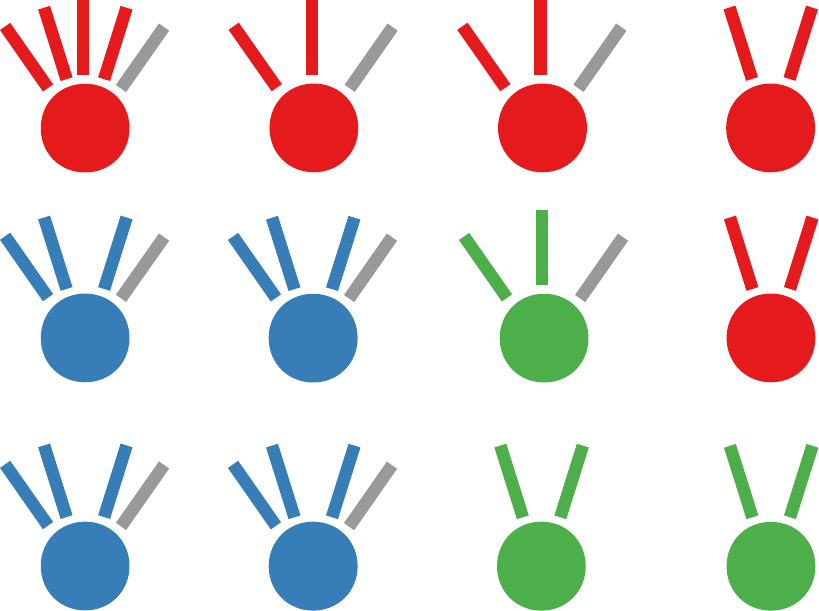}
\caption{Steps \#3 \& \#4}
\end{subfigure}\quad
\begin{subfigure}{.17\columnwidth}
\includegraphics[width=\textwidth]{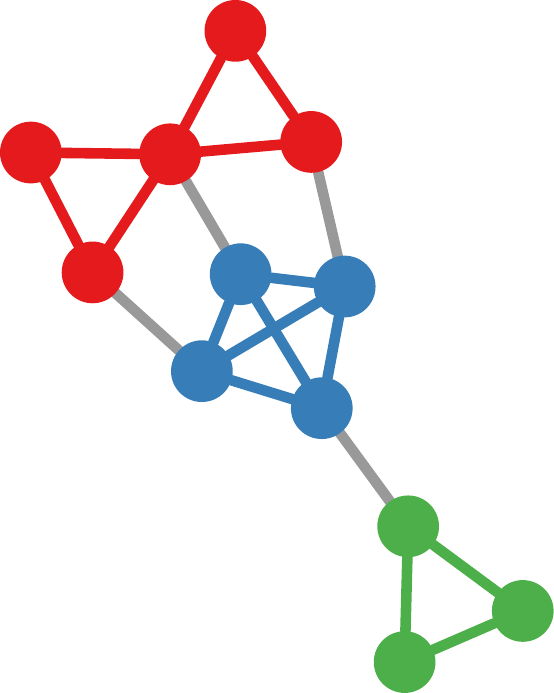}
\caption{Step \#5}
\end{subfigure}
\caption{A run through a simple LFR model.}
\label{fig:lfr-run}
\end{figure*}

The basic LFR algorithm, which generates disjoint communities, works using the following strategy:

\begin{enumerate}
\item Generate the degree distribution of the network, with exponent $\alpha$ and whose minimum and maximum degree are $k_{min}$ and $k_{max}$ (Figure \ref{fig:lfr-run}(a));
\item Mark a fraction $\mu$ of its edges as connecting outside the community and the rest as connecting inside the community (Figure \ref{fig:lfr-run}(b));
\item Generate the community size distribution, with exponent $\beta$ and whose minimum and maximum size are $s_{min}$ and $s_{max}$;
\item Assign each node $v$ to a community $c$ at random, ensuring that $k_v(1 - \mu) < s_c$, i.e. that the community will have enough nodes for $v$ to connect to them\footnote{If $k_v(1 - \mu) > s_c$, once you gave to $v$ $s_c$ internal edges, you still have internal edges to assign to $v$ that cannot be attached to nodes inside $c$, because $v$ is already connected to all its community mates, thus breaking the model. Remember that $s_c$ is the size of community $c$, in number of nodes.} (Figure \ref{fig:lfr-run}(c));
\item Apply a modified configuration model to establish the edges. Each node $v$ connects to $k_v(1 - \mu)$ random nodes in its community $c$, and to $k_v\mu$ random nodes outside $c$ (Figure \ref{fig:lfr-run}(d)).
\end{enumerate}

Note that, in light of step \#4, you have some constraints in your choice of parameter. Specifically $k_{min} < s_{min}$ and $k_{max} < s_{max}$, otherwise the nodes with minimum and maximum degree will never belong to any community. Even with such constraints, sometimes the combination of parameters will require the generation of an impossible graph, so the LFR benchmark will always be some sort of approximation of your desires. In practice, differences are going to be relatively tiny and insignificant.

Since we're plugging in a power law degree distribution and communities, it is obvious that LFR benchmarks will reproduce these characteristics of real world networks well -- although now you're actually \textit{forced} to have a power degree distribution, which in some cases you might not want. They also respect clustering and small diameters, making them the most realistic model we have.

\subsection{Kronecker Graphs}
\begin{figure}
\centering
\begin{subfigure}{.25\columnwidth}
\includegraphics[width=\textwidth]{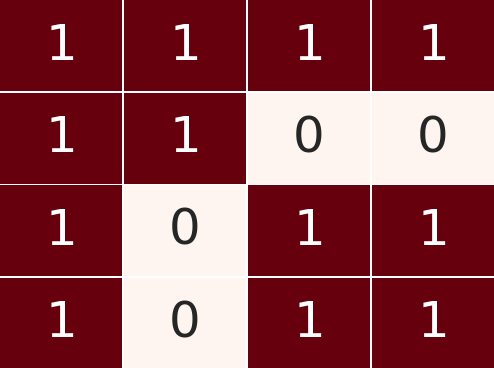}
\caption{}
\end{subfigure}\quad
\begin{subfigure}{.25\columnwidth}
\includegraphics[width=\textwidth]{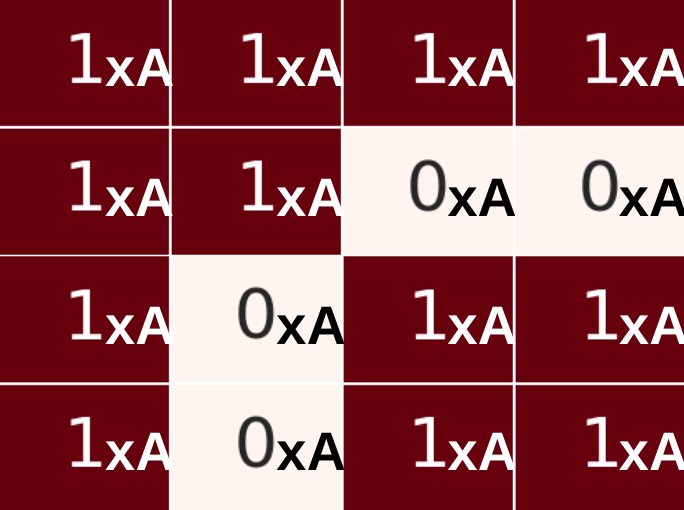}
\caption{}
\end{subfigure}
\caption{An example of Kronecker product. (a) A matrix $A$. (b) The operation we perform to obtain $A \otimes A$.}
\label{fig:kronecker-ex}
\end{figure}

The idea of a Kronecker graph originates from the Kronecker product operation. The Kronecker product is the matrix equivalent of the outer product of two vectors. The outer product of two vectors $u$ and $v$ is a $|u| \times |v|$ matrix, whose $i,j$ entry is the multiplication of $u_i$ to $v_j$. The Kronecker product is the same thing, applied to matrices. Figure \ref{fig:kronecker-ex} shows an example. To calculate $A \otimes B$, we're basically multiplying each entry of $A$ with $B$\cite{zehfuss1858gewisse}. 

When it comes to generating graphs, the matrix we're multiplying is the adjacency matrix. We usually multiply it with itself. So we're calculating $A \otimes A$, as I show in Figure \ref{fig:kronecker-ex}(b). This generates a new squared matrix, whose size is the square of the previous size. We can multiply this new adjacency matrix with our original one once more, for as many times as we want. We stop when we reach the desired number of nodes\cite{leskovec2005realistic}\cite{leskovec2007scalable}.

\begin{figure}[h!]
\centering
\begin{subfigure}{.25\columnwidth}
\includegraphics[width=\textwidth]{figures/kronecker_01.png}
\caption{}
\end{subfigure}\quad
\begin{subfigure}{.25\columnwidth}
\includegraphics[width=\textwidth]{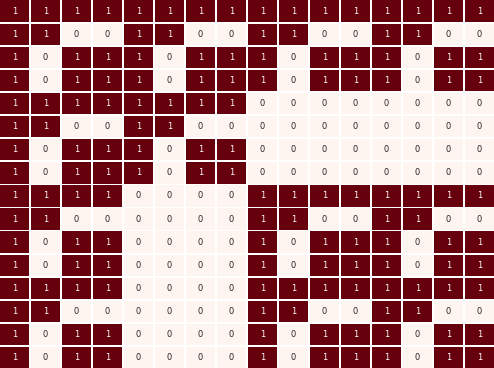}
\caption{}
\end{subfigure}\quad
\begin{subfigure}{.25\columnwidth}
\includegraphics[width=\textwidth]{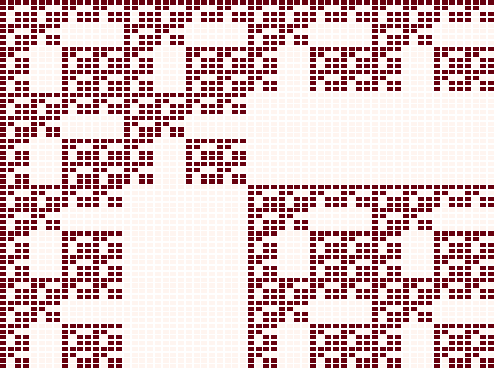}
\caption{}
\end{subfigure}
\caption{An example of Kronecker graph. (a) The seed adjacency matrix. (b) Kronecker product of (a) with itself. (c) Kronecker product of (b) with (a).}
\label{fig:kronecker}
\end{figure}

Figure \ref{fig:kronecker} shows the progression of the Kronecker graph. Figure \ref{fig:kronecker}(a) is our seed graph which we multiply to itself (Figure \ref{fig:kronecker}(b)) twice (Figure \ref{fig:kronecker}(c)).

One small adjustment that is customary to do when generating a Kronecker graph is to fill the diagonal with ones instead of zeros. If you remember my linear algebra primer, this means we consider every node to have a self-loop to itself. This is because we want the Kronecker graph to be a block-diagonal matrix, with lots of connections around the diagonal. This is required if we want them to show a sort of community partition.

\begin{figure}[b]
\centering
\begin{subfigure}{.25\columnwidth}
\includegraphics[width=\textwidth]{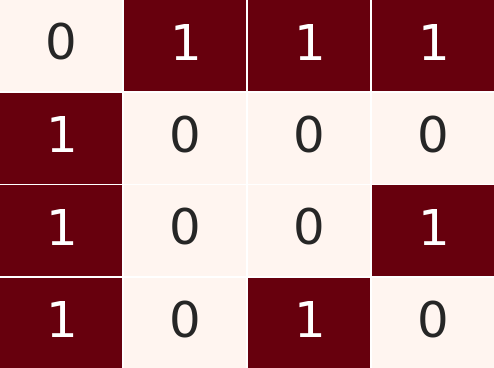}
\caption{}
\end{subfigure}\quad
\begin{subfigure}{.25\columnwidth}
\includegraphics[width=\textwidth]{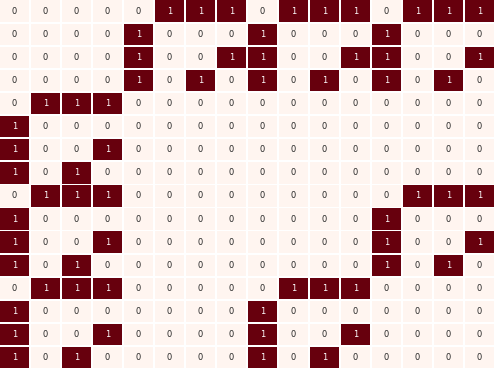}
\caption{}
\end{subfigure}\quad
\begin{subfigure}{.25\columnwidth}
\includegraphics[width=\textwidth]{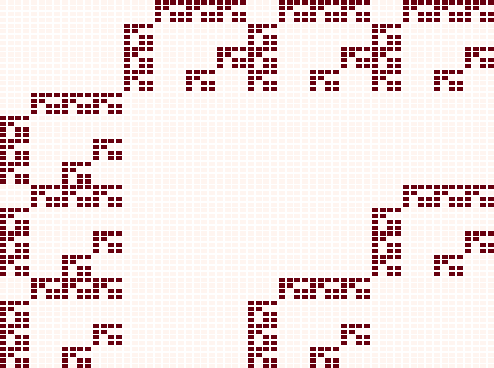}
\caption{}
\end{subfigure}
\caption{An example of Kronecker graph, similar to the one in Figure \ref{fig:kronecker}, but without setting the main diagonal to one.}
\label{fig:kronecker2}
\end{figure}

By how the Kronecker product is defined you can see that, if the seed matrix had an empty diagonal, we would not get a block diagonal matrix after applying the Kronecker product. Figure \ref{fig:kronecker2} shows an example, in which you can see the devastating effect on the graph's density of leaving the main diagonal empty. We can always reset the main diagonal to zero once we're done with the Kronecker products.

The question underlying generating graphs with an iterative Kronecker product is: why? Well, for starters, Kronecker graphs are fractals. Personally, I don't need any other reason that that. Look at Figure \ref{fig:kronecker}(c): if you tell me it doesn't speak to your heart then I question whether you're really human. If you're not an incurable fractal romantic like me, the deceptively simple process that generates Kronecker graphs solves all the issues we want from a graph generating process. In some cases, it is even better than LFR.

\begin{figure}
\centering
\begin{subfigure}{.25\columnwidth}
\includegraphics[width=\textwidth]{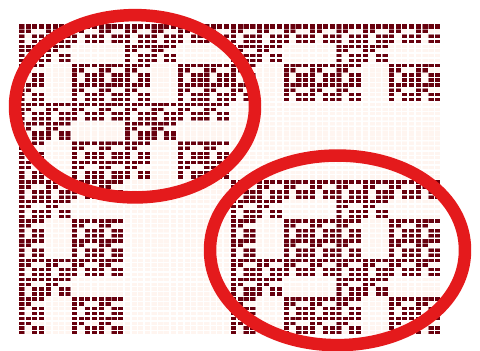}
\caption{}
\end{subfigure}\quad
\begin{subfigure}{.25\columnwidth}
\includegraphics[width=\textwidth]{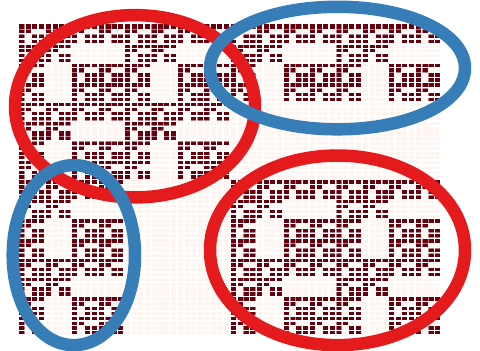}
\caption{}
\end{subfigure}\quad
\begin{subfigure}{.25\columnwidth}
\includegraphics[width=\textwidth]{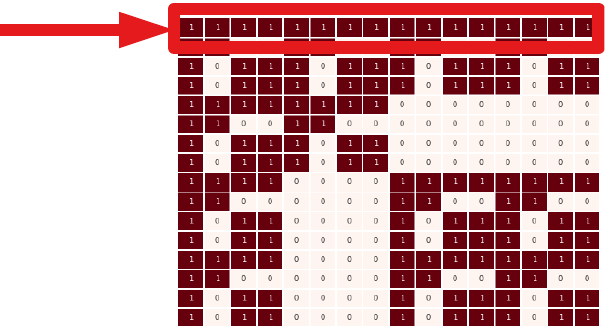}
\caption{}
\end{subfigure}
\caption{Some properties of Kronecker graphs. (a) Communities -- circled in red --; (b) Communities (red) with their overlap (blue); (c) Small diameter -- as the highlighted node in red is connected to every other node in the network making the diameter equal to two.}
\label{fig:kronecker3}
\end{figure}

Kronecker graphs have high clustering and communities (Figure \ref{fig:kronecker3}(a)), even hierarchical and overlapping (Figure \ref{fig:kronecker3}(b)). They even have a hint of core-periphery structures, which I'll present fully in Chapter \ref{cha:coreperiph}. They are small world (Figure \ref{fig:kronecker3}(c)), and with a power-lawish degree distribution (usually shifted because they have few low degree nodes). LFR is preferred because it leaves space for the randomness of real world noise, but Kronecker graphs have the advantage of being more simple to understand and implement.

\section{Random Geometric Graph}
Another property you might want to preserve in your graph is the spatial structure. Many networks live on a physical space, and this physical space constraints the edge generating process. If two nodes are too far way from each other, they cannot connect. For instance, if two cities are at the antipodes of the globe, there might not be a plane able to fly directly from once city to the other. We can model these constraints using random geometric graphs\cite{penrose2003random}\cite{dall2002random}.

The concept is simple. You first decide the dimensionality of your space: is it a 2D plane, three dimensional, or n-dimensional? Then you generate $|V|$ points in this space, by extracting them uniformly at random. Finally, you connect two points if they are at $r$ distance -- or less -- from each other. Every point will be at distance zero from itself but, for the sake of simplicity, we ignore self-loops. Note that you are free to decide how to calculate the distance between points: you're not forced to use the Euclidean.

\begin{figure}
\centering
\begin{subfigure}{.25\columnwidth}
\includegraphics[width=\textwidth]{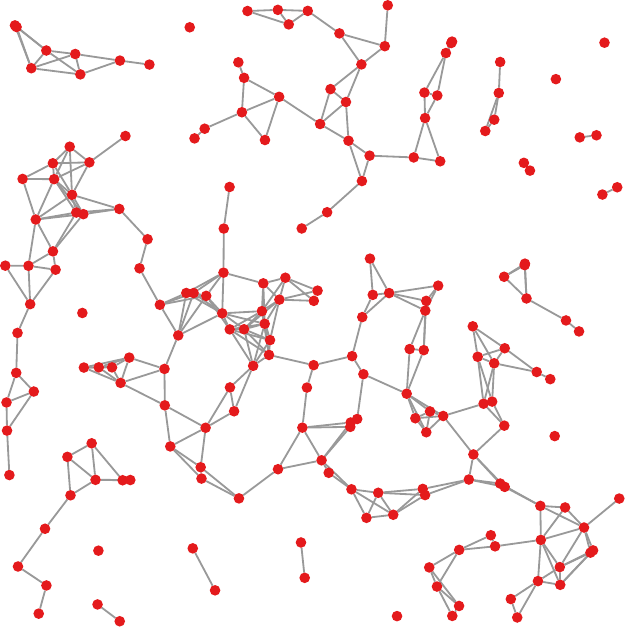}
\caption{$r = 0.08$}
\end{subfigure}\quad
\begin{subfigure}{.25\columnwidth}
\includegraphics[width=\textwidth]{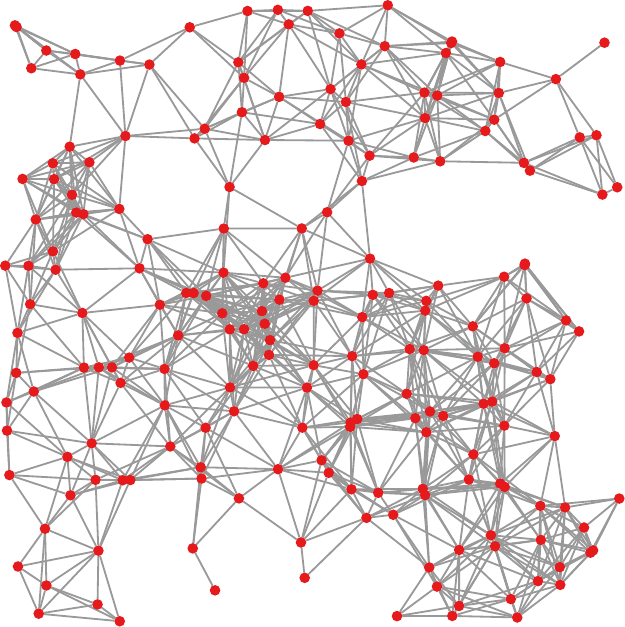}
\caption{$r = 0.14$}
\end{subfigure}\quad
\begin{subfigure}{.25\columnwidth}
\includegraphics[width=\textwidth]{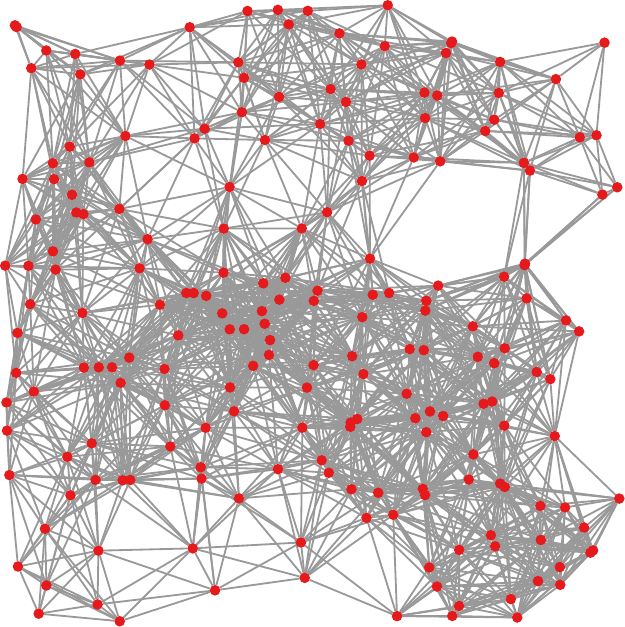}
\caption{$r = 0.2$}
\end{subfigure}
\caption{Random geometric graphs with $200$ nodes. X and Y positioning of nodes are the same for all figures, but the $r$ parameter increases from left to right, generating a higher number of longer edges.}
\label{fig:rnd-geo-graph}
\end{figure}

This is a rather simple way of generating random graphs. A random geometric graph is fully described by a handful of parameters: the number of nodes $|V|$ -- which is the number of points you extracted --; the maximum distance $r$; the number of dimensions; and the measure you used to calculate point-point distances. Figure \ref{fig:rnd-geo-graph} shows a few results. In the figure, I used a 2D plane and the Euclidean distance measure. The networks have the same number of nodes -- in fact their coordinates are the same --, and I simply play with the $r$ parameter.

There is a simple naive algorithm to generate random geometric graphs. First, you extract $|V|$ uniform random tuples -- depending on your chosen dimensionality, for a 2D plane they'd be pairs. Then you calculate the pairwise distance between all of them and connect the nodes if the distance is lower than $r$. There are smarter algorithms\cite{funke2019communication}, especially designed to avoid computing all pairwise distances -- and exploiting parallel processing.

Assuming that you use the Euclidean distance, you can derive the probability of having a given number of isolated vertices or a value of clustering coefficient by looking at the parameters you used to generate the network. In general, giant connected components appear easily in these types of graphs, provided that $|V|e^{-\pi r^2 |V|} < 1$. This magic value derives from the fact that the expected degree of a node is $\pi r^2 |V|$, given that it will connect to all nodes in a circle around it -- which has an area of $\pi r^2$. For instance, in Figure \ref{fig:rnd-geo-graph}(a), $r = 0.08$ and thus $|V|e^{-\pi r^2 |V|} \sim 3.6$, which allows for a few isolated nodes; while in Figure \ref{fig:rnd-geo-graph}(c) this value is $\sim 2 \times 10^{-9}$, which makes the presence of a single connected component almost certain.
 
You can also have probabilistic random geometric graphs\cite{waxman1988routing}. The difference is that you do not always connect nodes at distance lower than $r$ with probability $1$, but rather with some probability $p < 1$.

\section{Graph Generative Networks}\label{sec:csmodels-ggn}
One thing that all models discussed so far have in common is that they are engineered to have specific properties. These are the properties we think are salient in real world networks: broad degree distributions, community structures, etc. But what if we are wrong? Maybe some of these properties are not the most relevant things about a network we want to model. Moreover: what if there are other properties that we aren't seeing? Edges are dependent on each other, but these dependencies can be complex and it's difficult to put them in simple measures we can then optimize. Here I will describe a couple of approaches in isolation. However, to make more sense of the neural network lingo, you should look up Section \ref{sec:mining-deep2-generative}.

The field of graph generative networks\cite{li2018learning}\cite{de2018molgan}\cite{bojchevski2018netgan} aims at tackling this problem. Here we want to generate networks that look like specific real world networks, without us knowing what ``looking like'' actually means. In other words, we want the generative process to ``learn'' how a real world network looks like, so that it can generate synthetic versions at will.

The most trivial way you can do this is by feeding the adjacency matrix of your graph -- or a suitably modified version of it -- to an algorithm that can learn the dependencies between edges. You can think of this approach as an SBM process without inferring the communities beforehand. SBM wants to preserve the community structure and, on this basis, learns edge probabilities that only depend on the community affiliation. Here, we want to preserve the general network properties, thus each edge probability is dependent on the entirety of the adjacency matrix.

However, this approach has two problems. First, it will only generate a graph with the same number of nodes as the input, while you might want to vary the size of your synthetic networks. Second, it can only learn from a single graph at a time. Sometimes, you might want to model a class of graphs.

\begin{figure}[b]
\centering
\includegraphics[width=.7\columnwidth]{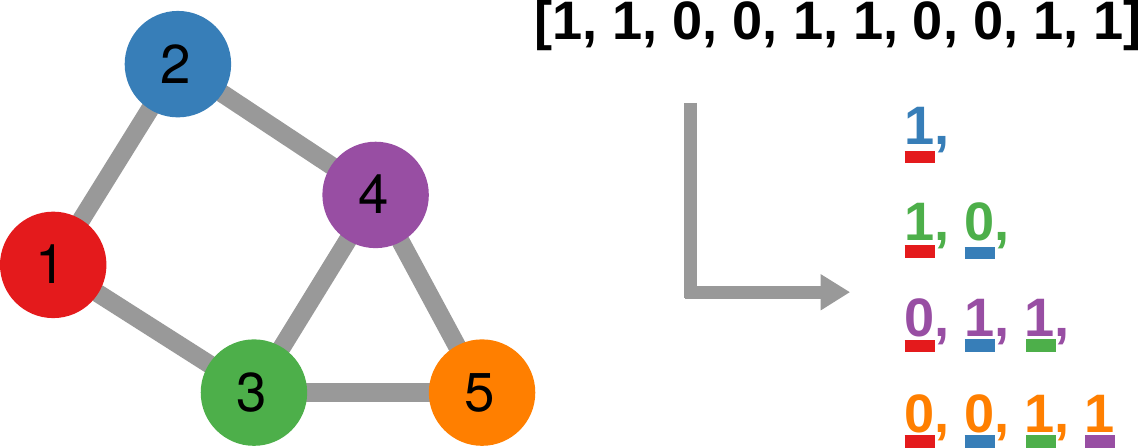}
\caption{A graph and its sequence representation in GraphRNN. Each element in the sequence belongs to a node (character color) and records whether that node connects to a specific node preceding it in the sequence (underline color).}
\label{fig:graph-gen-net1}
\end{figure}

These limitations are solved in a variety of ways. Just to give an example, GraphRNN\cite{you2018graphrnn} allows for two moves: a graph-level update and an edge-level update. In the first step, GraphRNN adds a new node into the network. Every time a new nodes is added, the edge-level update is triggered, determining to which nodes the new node connects. This is achieved by representing the graph as a sequence. For each node, in order, we list to which of the previous nodes it connects. For instance, the sequence $[1,1,0,0,1,1,0,0,1,1]$ corresponds to the graph in Figure \ref{fig:graph-gen-net1}.

To see why, it is useful to break down the sequence in sections, each one referring to a node, as the figure does. The first node has no element in the sequence, because it has no preceding nodes. The second node contributes only one element to the sequence: $0$ if it doesn't connect to the first node, $1$ otherwise. The third node contributes two values, one for its edge with the first node and one for the second node, and so on.

\begin{figure*}
\centering
\includegraphics[width=.9\columnwidth]{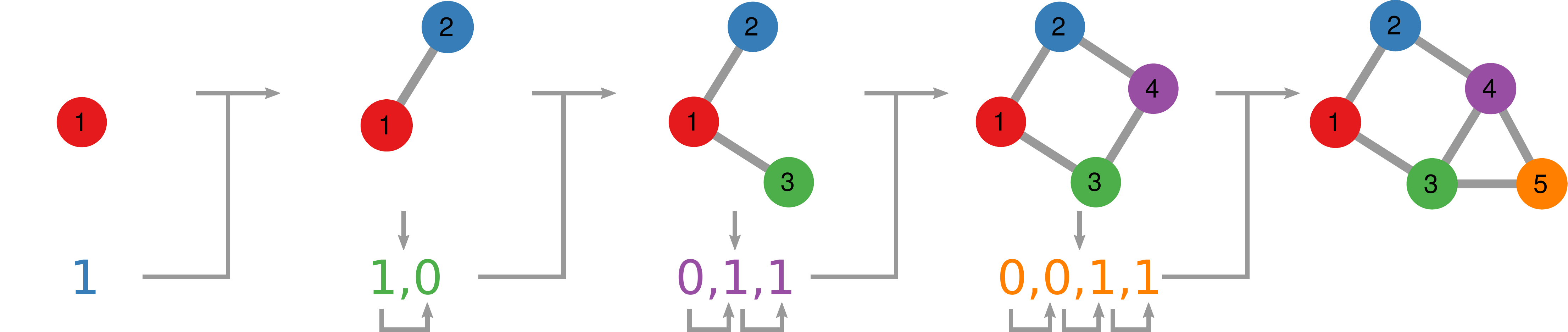}
\caption{The GraphRNN workflow. From left to right, we progressively add new nodes, in the form of an extension of the sequence representing the connections.}
\label{fig:graph-gen-net2}
\end{figure*}

In practice, GraphRNN expands the sequence, by adding the $n$th node (graph-level update) as a new subsequence of length $n-1$ (edge-level update). Figure \ref{fig:graph-gen-net2} shows a simple iteration. These updates are implemented via autoregressive models (see Section \ref{sec:mining-deep2-generative}).

\section{Summary}

\begin{enumerate}
\item The configuration model is a way to have a synthetic network with an arbitrary degree distribution. However, if you don't allow for parallel edges or self loops, the degree distribution is likely only going to be approximated.
\item Stochastic block models can recreate a community structure by taking as input a node partition and the probabilities of connecting to nodes inside the same community and between communities.
\item The GN and LFR benchmark were created to test community discovery algorithms. The GN benchmark creates equal size communities and normal degree distributions, while LFR is able to return power-law degree distributions and communities of varying size.
\item Kronecker graphs are generated from a simple seed matrix to which you recursively apply the Kronecker product, creating high clustering networks with shifted power law degree distributions and communities.
\item Alternatively, you can make a random geometric graph, by placing nodes uniformly on an n-dimensional space and connecting nodes to all their closest neighbors, at a maximum distance that you can set as parameter.
\item Finally, you can learn a neural network representation from your original (set of) network(s), which will be able to generate more synthetic networks with comparable properties to your original one(s).
\end{enumerate}

\section{Exercises}

\begin{enumerate}
\item Generate a configuration model with the same degree distribution as the network in \url{http://www.networkatlas.eu/exercises/18/1/data.txt}. Perform the Kolmogorov-Smirnov test between the two degree distributions. 
\item Remove the self-loops and parallel edges from the synthetic network you generated in the previous question. Note the \% of edges you lost. Re-perform the Kolmogorov-Smirnov test with the original network's degree distribution.
\item Generate an LFR benchmark with $100,000$ nodes, a degree exponent $\alpha = 3.13$, a community exponent of $1.1$, a mixing parameter $\mu = 0.1$, average degree of $10$, and minimum community size of $10,000$. (Note: there's a \texttt{networkx} function to do this). Can you recover the $\alpha$ value by fitting the degree distribution?
\item Use \texttt{kron} function from \texttt{numpy} to implement a Kronecker graph generator. Plot the CCDF degree distribution of a Kronecker graph with the following seed matrix multiplied $4$ times (setting the main diagonal to zero once you're done):

$$
A = 
\begin{pmatrix}
1 & 1 & 1 & 0 \\
1 & 1 & 1 & 0 \\
1 & 1 & 1 & 1 \\
0 & 0 & 1 & 1 \\
\end{pmatrix}
$$

\end{enumerate}

\chapter{Evaluating Statistical Significance}\label{cha:ergmodels}
One of the big issues when it comes to analyzing complex networks is that, usually, you only have one network to base your observations on. Therefore, whenever you observe a given property -- power law degree distribution, reciprocity, clustering -- you don't have the statistical power to claim that what you're observing is interesting. You need to have multiple versions of your network, a null model, to test your observation. If keeping everything fixed about a network minus the property of interest gives you something indistinguishable from your observation, then you know that the particular feature arose at random. There is no fundamental non-random force behind it.

To do so, you need to generate a (set of) synthetic graph(s), which is why I put this chapter in this part of the book. In other words, you consider your observed network as part of a family of networks, which all have the same fixed properties. Then you ask if the one you did not fix is also a typical characteristic of this family of networks. If it is, then it's not an interesting discovery. If it isn't, the deviation of the network from its family is interesting.

As far as I know, there are two ways to generate this network family: the easy way -- network shuffling, Section \ref{seg:ergmodels-shuffling} --, and the right way -- the Exponential Random Graph approach, Section \ref{seg:ergmodels-ergm}.

\section{Network Shuffling}\label{seg:ergmodels-shuffling}
Network shuffling is a way to generate synthetic networks that is based on directly manipulating your observed network. At a fundamental level, it is a process of rewiring edges for a given number of times, until we think that we are sufficiently far from the starting point. I call this the ``easy'' way because, as we'll see in a moment, the process is usually straightforward. On the other hand, this method is significantly less rigorous than the Exponential Random Graph model (ERGM), and thus should be used only when you don't need the statistical power ERGM can give you.

\begin{figure}[t]
\centering
\begin{subfigure}{.3\columnwidth}
\includegraphics[width=\textwidth]{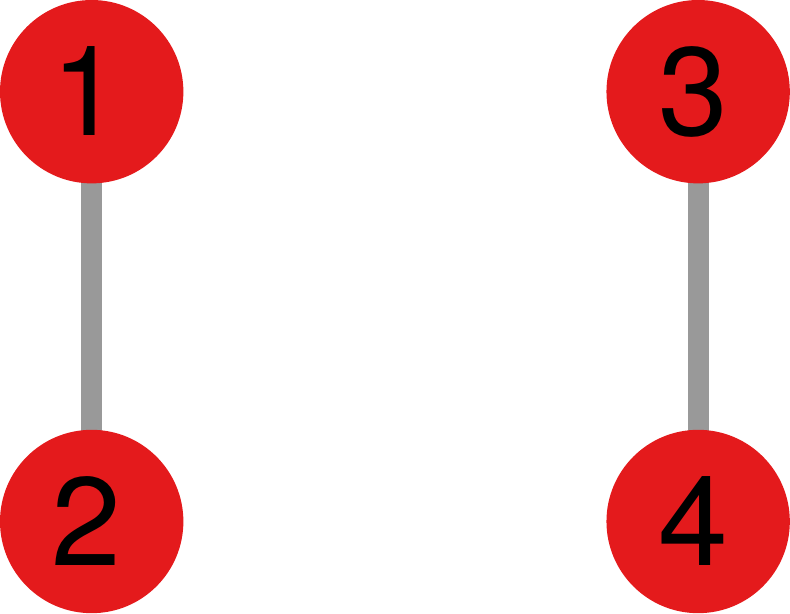}
\caption{}
\end{subfigure}\qquad\qquad
\begin{subfigure}{.3\columnwidth}
\includegraphics[width=\textwidth]{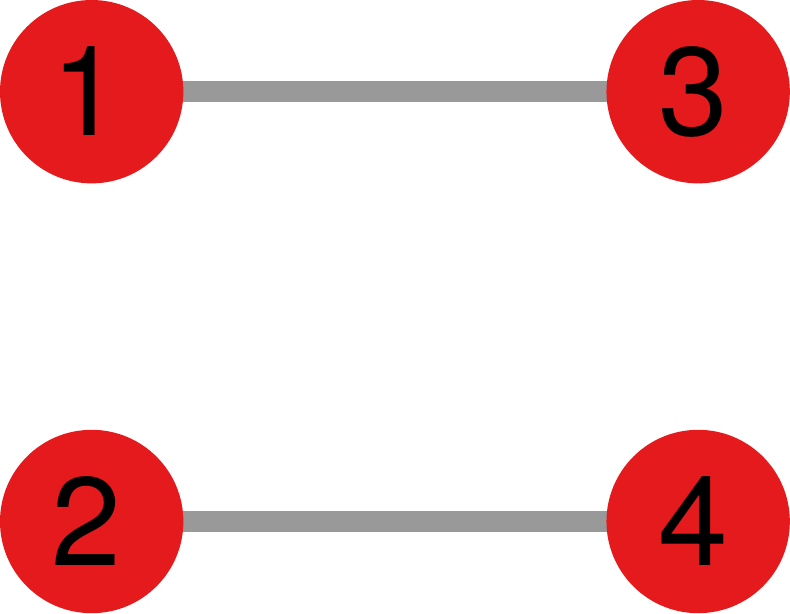}
\caption{}
\end{subfigure}
\caption{The edge swap procedure.}
\label{fig:edgeswap}
\end{figure}

The fundamental basis of the network shuffling model is the edge swap operation. Figure \ref{fig:edgeswap} depicts it in all its simplicity. You pick two pairs of connected nodes, in this examples node $1$ is connected to node $2$, and node $3$ is connected to node $4$. Then, you flip the edge around, deleting $(1, 2)$ and replacing it with $(1, 3)$, and deleting $(3, 4)$ replacing it with $(2, 4)$. If you do this enough times, the resulting network will be quite different from your original one. However, it will still have the same number of nodes, the same number of edges, and the same degree distribution -- also in case of a directed network, provided that you always swap edges in the correct direction.

This procedure is usually performed in network games, where edges are rewired with some objective in mind\cite{alon2013basic}. Note that the number of swaps to perform before stopping is a non trivial quantity to evaluate\cite{bottazzi2010measuring}.

An attentive reader will surely notice that this result is practically the same as the one you would obtain from a configuration model. However the two approaches have completely different objectives. The configuration model wants to simply generate a network with more or less the same degree distribution. The networks generated by shuffling are significantly more similar to the original network than the ones obtained from a configuration model, because they need to be compared to it.

This becomes more obvious once you explore the differences with the configuration model more in depth. First, edge swapping is always possible, while at some point in the configuration model process you might have to create self-loops or parallel edges. You cannot create self-loops in network shuffling unless there were already self loops in the original network. You can easily avoid parallel edges by checking that your node pairs are not connected -- for instance, you can reject the operation in the example in Figure \ref{fig:edgeswap} if either the $(1, 3)$ or the $(2, 4)$ edges are already in the network.

Second, what I explained is only the simplest way to perform network shuffling. You can add a few more features that are hard to embed in a configuration model. For instance, you can keep fixed the number of connected components, by rejecting an edge swap if it would disconnect more nodes, or join two different components. You can keep the clustering fixed, by making sure to keep the number of triads and triangles constant. You can preserve the communities, by only allowing edge swaps inside the clusters. And so on.

If all you want to do is to have a randomized version of your original network, then you're done. But, since I mentioned the problem of determining the statistical significance of your observations, let's push on. How would you use the networks generated via edge swap for such a task? I usually apply the following procedure:

\begin{enumerate}
\item Fix all reasonable properties of the network (at least number of nodes, edges, and degree distribution) except the one of interest;
\item Generate a large set of shuffled networks, with independent shuffles -- the number of shuffles depends on the number of edges;
\item Calculate the property of interest in the observed and in the generated networks;
\item Estimate the distribution of the property in the shuffled networks and calculate how far from the expectation the observed value is.
\end{enumerate}

\begin{figure}
\centering
\begin{subfigure}{.45\columnwidth}
\includegraphics[width=\textwidth]{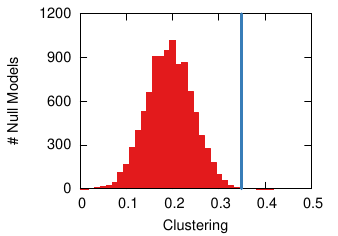}
\caption{}
\end{subfigure}\quad
\begin{subfigure}{.45\columnwidth}
\includegraphics[width=\textwidth]{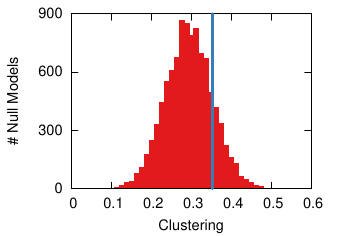}
\caption{}
\end{subfigure}
\caption{The edge swap statistical test. The plot shows how many null models (y-axis) scored a given value of the property of interest (in this case clustering, x-axis). The blue vertical bar shows the observation.}
\label{fig:edgeswap2}
\end{figure}

Usually, this ends up with a plot looking like Figure \ref{fig:edgeswap2}(a) or \ref{fig:edgeswap2}(b). The histogram shows how many null models scored a value of the property of interest in a given interval. The blue vertical bar shows the value for the observed network.

In the easiest scenario, the null model will show a nice normal or pseudo-normal distribution, making the estimation of statistical significance easier. That's the case for the figures I show, which means I can simply calculate how many standard deviations from the average my observation is. In the case of Figure \ref{fig:edgeswap2}(a) the observation is significantly higher than expectation, given it's three standard deviations away from the average. That is not true for Figure \ref{fig:edgeswap2}(b): in that case, we cannot reject the null explanation. The observation is less than a standard deviation away from null expectation.

By the way, the thing that I call ``number of standard deviations above (or below) average'' is know as z-score. You can automatically convert from the z-score into a p-value (Section \ref{sec:stats-p}), provided that you know whether you're interested in a one-sided or a two-sided test. The one-sided test means that your success is exclusively on one side of the distribution -- e.g. you want to score more than average, you're not interested whether your score is significantly below average\footnote{When discussing discoveries in physics, you'll hear often the term ``five sigma'' ($5\sigma$) thrown around. This means a z-score equal to $5$. In turn, this can be converted to a (one-sided) p-value lower than $10^{-6}$, way lower than the $p < 0.01$ you'll see in other fields. For $p < 0.01$, you're looking at a z-score a bit higher than $2.3$. I'm simplifying a lot here, since this is not -- and never will be -- a statistics book.} (or vice versa).

Of course, if the null model distribution is not pseudo-normal, estimating the statistical significance is a bit trickier. We don't need to go into that, because we're about to learn how to perform this task in the ``right'' way, using ERGMs.

\section{Exponential Random Graphs}\label{seg:ergmodels-ergm}
As introduced in this chapter, ERGM is a technique to generate a set of graphs that have the same properties of an observed network. ERGM is also know in the literature as p* model\cite{anderson1999p}\cite{robins2007introduction}. The observed network is seen as the result of a stochastic (random) process with a set of parameters. ERGM creates other networks using the same process and the same parameters. The problem we need to solve to generate an exponential random graph ensemble of networks is figuring out which parameter values to use. This is usually achieved through maximizing the likelihood function I introduced in Section \ref{sec:ml-loss}.

\begin{figure}
\centering
\begin{subfigure}{.23\columnwidth}
\includegraphics[width=\textwidth]{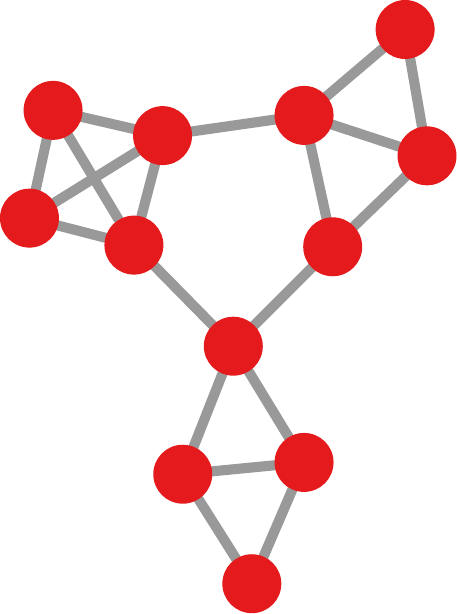}
\caption{}
\end{subfigure}\quad
\begin{subfigure}{.19\columnwidth}
\includegraphics[width=\textwidth]{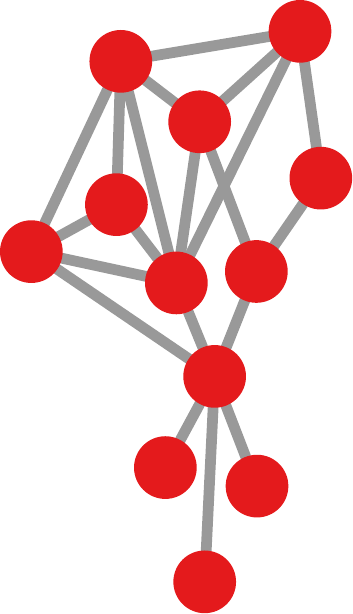}
\caption{}
\end{subfigure}\quad
\begin{subfigure}{.225\columnwidth}
\includegraphics[width=\textwidth]{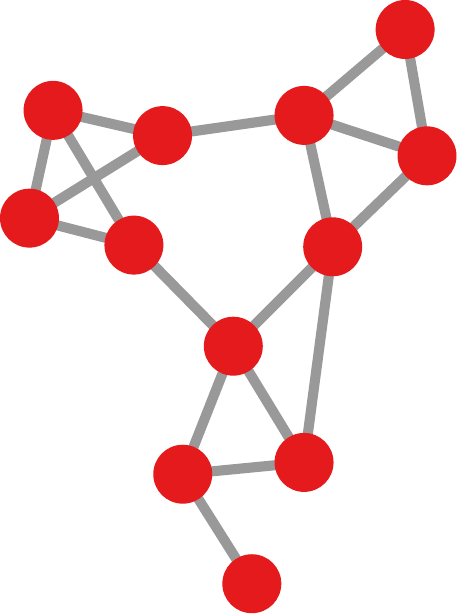}
\caption{}
\end{subfigure}\quad
\begin{subfigure}{.225\columnwidth}
\includegraphics[width=\textwidth]{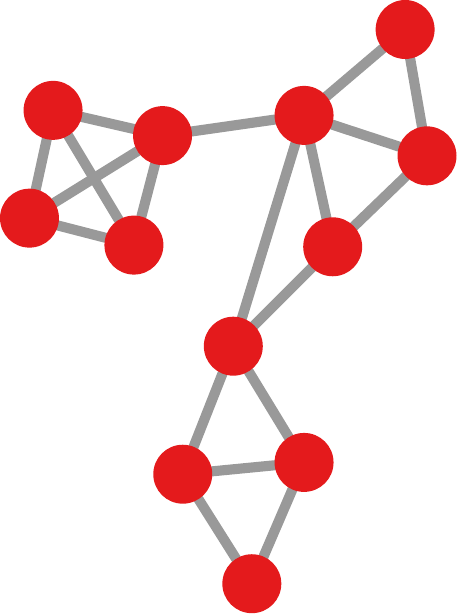}
\caption{}
\end{subfigure}
\caption{An illustrative example of the ERGM process. (a) Observed network. (b) Unlikely random network. (c) Random network more likely to be in the same family of (a) than (b). (d) Most likely random network. The parameters used to generate it are more likely to be the ones of the family of random graphs to which (a) belongs.}
\label{fig:ergm}
\end{figure}

The sketch of the solution is the following. Suppose you're observing a graph $G$. There is an immensely large set of other random $G$ graphs we could have observed: they are those we could generate with a random process with a given set of parameters (same number of nodes, edges, ...). However, some of these random graphs are more likely than others to be observed -- namely, the ones most similar to $G$. Knowing this, we can identify the parameters values these likely $G$s have in common. Once we find these values, we can generate an arbitrary number of graphs with them. We can use them as new synthetic graphs for our purposes, or we can test them against the observed $G$ to verify whether they also share with it some other property we did not fix. Figure \ref{fig:ergm} provides a vignette of this process.

If you talk statistics, the following process might give you an inkling about how ERGMs work. In this scenario, we consider an edge as a random variable. In the simplest case of unweighted network, this will be a binary variable, equal to one if the edge is present, and to zero if it isn't. You hypothesize what sort of process might be the one determining the edge presence in your network. This is sort of similar to estimating a logistic regression. You have a binary outcome (edge present/absent) and a set of variables that might be able to predict its value.

\begin{figure*}[b]
\centering
\begin{subfigure}{.23\columnwidth}
\includegraphics[width=\textwidth]{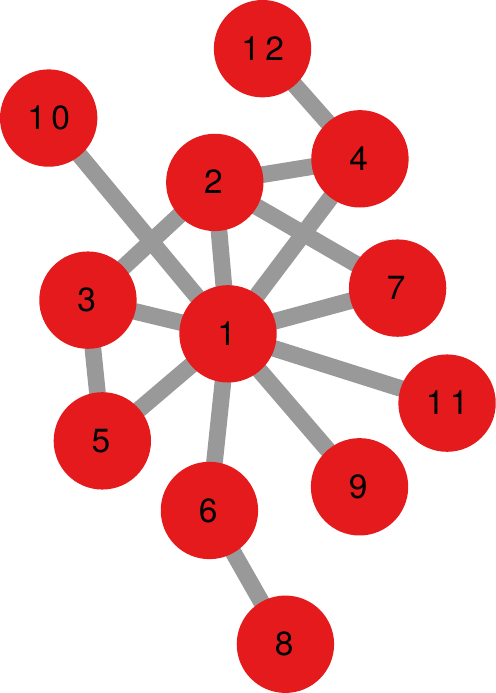}
\caption{}
\end{subfigure}
\begin{subfigure}{.23\columnwidth}
  \centering
  \begin{tabular}{c|cc}
    $A_{u,v}$ & $k_u$ & $k_v$ \\
    \hline
    $1$ & $9$ & $4$\\
    $1$ & $9$ & $3$\\
    $1$ & $9$ & $3$\\
    $1$ & $9$ & $2$\\
    $1$ & $9$ & $2$\\
    $1$ & $9$ & $2$\\
    $0$ & $9$ & $1$\\
    ... & ... & ...\\
  \end{tabular}
\caption{}
\end{subfigure}
\begin{subfigure}{.23\columnwidth}
  \centering
  \begin{tabular}{c|cc}
    $\overline{A_{u,v}}$ & $k_u$ & $k_v$ \\
    \hline
    $.92$ & $9$ & $4$\\
    $.87$ & $9$ & $3$\\
    $.87$ & $9$ & $3$\\
    $.8$ & $9$ & $2$\\
    $.8$ & $9$ & $2$\\
    $.8$ & $9$ & $2$\\
    $.7$ & $9$ & $1$\\
    ... & ... & ...\\
  \end{tabular}
\caption{}
\end{subfigure}
\begin{subfigure}{.23\columnwidth}
\includegraphics[width=\textwidth]{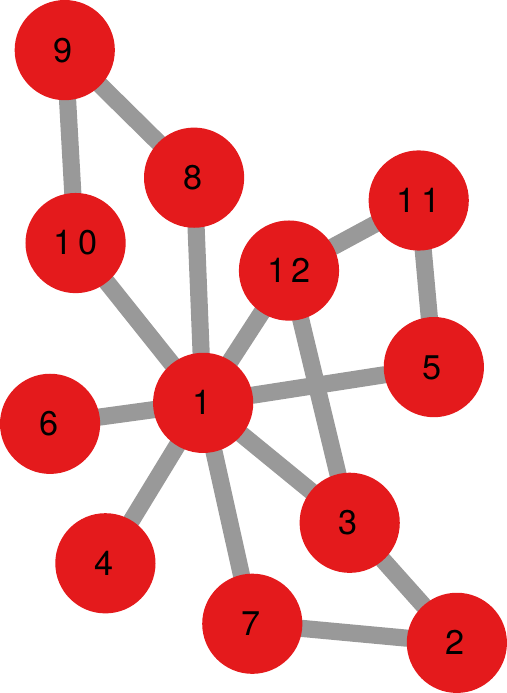}
\caption{}
\end{subfigure}
\caption{Step-by-step example of a simple ERGM proces. (a) Observed network. (b) Observed edge table (only first seven rows). $A_{u,v}$ is one if the edge is present, zero otherwise; $k_u$ and $k_v$ are the degrees of the two nodes. (c) Result of the logit regression (only first seven rows). $\overline{A_{u,v}}$ is the estimated probability of the edge existing. (d) An extracted ERGM from the edge probabilities in (c).}
\label{fig:ergm2}
\end{figure*}

Let's make an example. Figure \ref{fig:ergm2}(a) represents our observed graph. I generated it with a configuration model with the degree sequence $(9, 4, 3, 3, 2, 2, 2, 1, 1, 1, 1, 1)$, but the ERGM doesn't know that. The edge presence, the outcome, is our adjacency matrix $A$. So the edge between $u$ and $v$ is $A_{u,v}$. I now make an hypothesis: the degree of a node influences its likelihood of getting a connection. Or, in mathematical terms, $A_{u,v} = \beta_1 k_u + \beta_2 k_v$. The degrees of $u$ and $v$ ($k_u$ and $k_v$) can be used to predict the probability of existence of an edge. This is equivalent of running a logit regression on an edge table like the one in Figure \ref{fig:ergm2}(b): we're trying to predict the binary $A_{u,v}$ variable using the degrees of $u$ and $v$.

Once the logit model is done, for each $(u,v)$ pair we have a probability of its existence: $\overline{A_{u,v}}$ (Figure \ref{fig:ergm2}(c)). We can now flip a loaded coin for each node pair and add the edge in case of success. $\overline{A_{u,v}}$ is determined by the $\beta_1$ and $\beta_2$ parameters. Since they are the result of a logit model estimation, they are the ones most likely to describe the family of random graphs from which we extracted the observed $G$. By using Figure \ref{fig:ergm2}(c) to generate a new graph (e.g. the one in Figure \ref{fig:ergm2}(d)), we're sure to extract a graph from the same family that generated the original one -- at least when it comes to its degree distribution.

So far, I simplified the process for the sake of intuition. For instance, I assumed that the likelihood of an edge only depends on a node's characteristic -- in this case the degree. This is not necessarily the case in the general ERGM. You can plug in all complex structures you can express mathematically. For instance, you can ensure triadic closure to preserve the clustering coefficient or other, more complex, motifs. I also assumed the functional form of the model, namely a linear one. The degrees of the two nodes interact linearly to give us the result. That might not be the case.

We can describe the full model making no such assumptions as:

$$ Pr(A = A') = \dfrac{1}{B} \exp \left( \sum \limits_{g} \beta_g g_{A'} \right).$$

There's a bit to unpack here:

\begin{itemize}
\item $Pr(A = A')$ is the probability that the adjacency matrix $A$ we extract is equal to a given adjacency matrix $A'$, dependent on $A'$'s characteristics.
\item $\exp$ is the exponential function. We use it to define the probability because exponentials come from maximum entropy distributions. We want to use a function that can have the highest possible entropy while still having a positive and definite mean -- which is necessary to define a probability (see Section \ref{sec:prob-axioms}). ``Highest possible entropy'' simply means that whatever statistic we haven't incorporated in the model will be ``as random as possible''.
\item $g$ is a graph configuration. It can be any pattern, for instance a triangle, or a clique of four nodes, or even just an edge.
\item $\beta_g$ is the parameter corresponding to this particular configuration. In our previous example, it is the thing telling you how much the degree of a node influences the connection probability. This is the knob you have to use to maximize your quality function. The ``right'' $\beta_g$ value is the one best describing your data.
\item $g_{A'}$ is a function applying pattern $g$ to $A'$. It tells you whether the pattern is in the network. Mathematically: $g_{A'} = \prod \limits_{A'_{uv} \in A'} A'_{uv}$, which means that $g_{A'} = 1$ if and only if all parts of $g$ are in $A'$.
\item $B$ is simply a normalization parameter needed to ensure that the rest of the equation is a proper probability distribution -- i.e. that it sums to one.
\end{itemize}

To use human language: the probability of observing an adjacency matrix $A$ is the probability of extracting a random $A'$ from all possible adjacency matrices, weighted by how well $A'$'s topological properties fit the $\beta$ parameters we observed in our original graph, over the patterns $g$ that interest us -- any other pattern not in $g$ is assumed to appear entirely at random.

Such a model can have lots of $\beta$ parameters. That is why usually there is an additional step of parameter reduction, through what we call ``homogeneity constraints''. For instance you could have a parameter for each node, telling us how likely that node is to reciprocate connections. However, the homogeneity assumption says that -- most likely -- all nodes in the same network have more or less the same tendency of reciprocating connections. Thus, rather than having $|V|$ reciprocity parameters, you have a single, network-wide, one.

This functional form is a general version of more specific ones that were studied in the literature in the eighties: $p_1$ models\cite{holland1981exponential} and Markov graphs\cite{frank1986markov}.

To understand a bit more the magic behind the formula I just presented, let's consider a few special cases. Given their general form, ERGMs include many of the network models we saw so far. For instance, we can represent a $G_{n,p}$ model, by noting that, in this case, edges are all independent to each other. Without the homogeneity assumption, we would have a parameter for each pair of nodes, giving us the equation:

$$ Pr(A = A') = \dfrac{1}{B} \exp \left( \sum \limits_{u,v} \beta_{u,v} A'_{u,v} \right). $$

In this case, the graph pattern $g$ is a single $u,v$ edge. Since $g_{A'}$ is equal to one if $A'$ contains pattern $g$, it reduces to $A'_{u,v}$, which is one if $A'$ contains the $u,v$ edge. Further, we have a different $\beta_{u,v}$ ($\beta_g$) per edge. Edges present in our observed network will have corresponding high $\beta_{u,v}$ parameters, while absent edges will have $\beta_{u,v}$ values close to zero. This in turn implies that an $A'$ is likely to be extracted if it has edges attached to high $\beta_{u,v}$ values.

However, as we said, this is too many parameters. If we apply the homogeneity assumption, we will just say that any pair of nodes has the same probability $p$ of connecting -- which is the same assumption of the $G_{n,p}$ model. This means that we get rid of a lot of parameters, which are substituted by the single parameter $p$:

$$ Pr(A = A') = \dfrac{1}{B} \exp \left( \sum \limits_{u,v} p A'_{u,v} \right).$$

Since $\sum \limits_{u,v} A'_{u,v}$ is simply the number of edges $|E'|$, this simplifies to:

$$ Pr(A = A') = \dfrac{1}{B} \exp \left(p|E'|\right),$$

which is exactly a $G_{n,p}$ model: a graph whose edges are all equally likely to be observed. We can also simulate a stochastic blockmodel by not reducing all connection probabilities to $p$ but by having multiple $p$s for each block (and for inter-block connections). If you have a directed graph you can represent reciprocity with the probability $p_1$ of a node to reciprocate the connection, adding a term to the $G_{n,p}$ model:

$$ Pr(A = A') = \dfrac{1}{B} \exp \left(p|E'| + p_1 R(A')\right).$$

Here $R(A')$ is the number of reciprocated ties. You can make edges dependent on node attributes as researchers do in $p_2$ models\cite{lazega1997position}\cite{van2004p2}. Finally, you can also plug higher-order structures in the model, for instance:

$$ Pr(A = A') = \dfrac{1}{B} \exp \left(p|E'| + \tau T(A')\right),$$

where -- under the homogeneity assumption -- $T(A')$ is the number of triangles in $A'$. This way, you can also control the transitivity of the graph.

These models can be very difficult to solve analytically for all but the simplest networks. Modern techniques rely on Monte Carlo maximum likelihood estimation\cite{snijders2002markov}\cite{snijders2006new}. We don't need to go too much into details, but these work similarly to any Markov chain Monte Carlo method\cite{gilks1995markov}. However, if your network is dense, your estimation might need to take an exponentially large number of samples to estimate your $\beta$s\cite{bhamidi2008mixing}. There are ways to get around this problem by expanding the ERG model\cite{chandrasekhar2014tractable}, but by now we're already way over my head and I don't think I can characterize this fairly.

\begin{figure}
\centering
\includegraphics[width=.8\columnwidth]{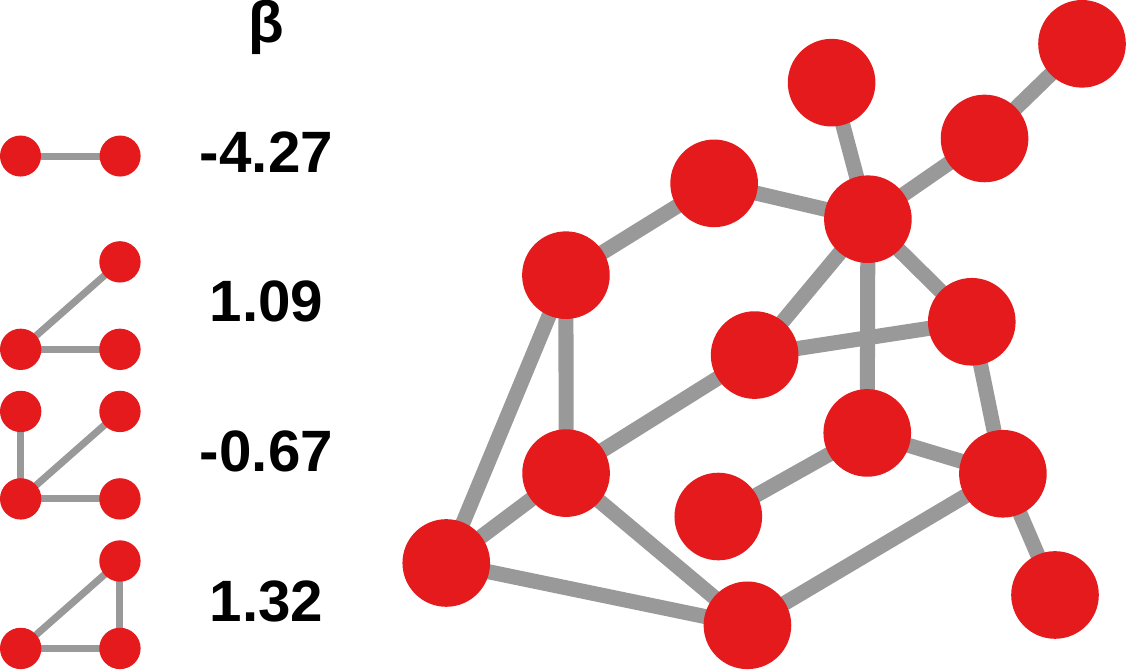}
\caption{On the left we have the estimated parameters from the observation for four patterns, with positive values indicating a ``more than chance'' occurrence of the pattern, and negative values a ``less than chance''. On the right we have a likely network extracted from the set of ERGM with the given parameters.}
\label{fig:ergm3}
\end{figure}

How does all of this look like in practice? The result of your model might look like something from Figure \ref{fig:ergm3}. Here we decide to have four parameters: a single edge (this is always going to be present in any ERGM), a chain of three nodes, a star of four nodes, and a triangle. Each motif has a likelihood parameter: the higher the parameter the more likely the pattern. Negative values mean that the pattern is less likely than chance to appear.

The negative value for simple edges means that the network is sparse: two nodes are unlikely to be connected. The positive value for the triangle means that triangles tend to close: when you have a triad, it is more likely than chance to have the third edge. The other two configurations are not significantly different from zero (you can't tell because I omitted the standard errors, but trust me on that). Thus we should not emphasize their interpretation too much.

On the right side of the figure you can see a potential network that is very likely to be extracted by this ERGM. In fact, I cheated a bit, because that is the network on which I fitted the model. It is the famous graph mapping the business relationship between Florentine families in the Renaissance\cite{padgett1993robust}.

In this chapter I presented only the simplest of the ERGM forms. Recent research has shifted to more sophisticated models. A few of those are:

\begin{itemize}
\item Longitudinal ERGMs\cite{cranmer2011inferential}, which are specialized to deal with networks that are evolving over time, for instance co-sponsorship of bills in the US Congress -- two representatives might co-sponsor a bill in one year, but not in another;
\item Similarly, TERGMs\cite{hanneke2010discrete} introduce the temporal aspect in ERGMS. This contains the ``separable'' TERGMs\cite{krivitsky2014separable} which works on discrete models, rather than modeling the evolution as happening on a continuous time flow;
\item ERGMs that can take into account edge weights, initially only continuous weights\cite{desmarais2012statistical}, but subsequently also discrete ones\cite{krivitsky2012exponential};
\item ERGMs for multilayer networks\cite{caimo2020multilayer}.
\end{itemize}

ERGMs have been successfully applied in many fields. For instance, they help in cases in which longitudinal network data collection is
unfeasible -- e.g. informal face to face contacts in certain business clusters\cite{balland2016dynamics}, or inside firms\cite{broekel2013explaining}. In economics they are particularly useful because of their ability to estimate structural network parameters, extending conventional analyses that use, for instance, gravity models\cite{broekel2014modeling}. To make an example, in migration a gravity model would say that the number of migrants from country $u$ to $v$ is directly proportional to the size -- in number of inhabitants -- of the two countries, and inversely proportional to their distance. ERGMs allow you to model more complex interdependencies\cite{windzio2018network}.

\section{Summary}

\begin{enumerate}
\item Network shuffling is a way to create a null version of your network, by performing edge swapping. In edge swapping, you pick two pairs of connected nodes and you rewire the edges to connect nodes from the other pair.
\item Once you generate thousands of null versions of your network, you can test a property of interest and obtain an indication of how statistically significant your observation is, by counting the number of standard deviations between the observation and the null average.
\item In Exponential Random Graphs you use a series of characteristics of the network of interest as a predictor of the presence of an edge between two nodes.
\item Once you know the relationship between these parameters and the presence of an edge, you can randomly extract graphs that are likely results of such predictors.
\item There is a trade off between the number of parameters you can use and the complexity of the extraction process. Many different heuristics have been proposed to sample the space of all possible ERGMs.
\end{enumerate}

\section{Exercises}

\begin{enumerate}
\item Perform $1,000$ edge swaps, creating a null version of the network in \url{http://www.networkatlas.eu/exercises/19/1/data.txt}. Make sure you don't create parallel edges. Calculate the Kolmogorov-Smirnov distance between the two degree distributions. Can you tell the difference?
\item Do you get larger KS distances if you perform $2,000$ swaps? Do you get smaller KS distances if you perform $500$? 
\item Generate $50$ $G_{n,m}$ null versions of the network in \url{http://www.networkatlas.eu/exercises/19/3/data.txt}, respecting the number of nodes and edges. Derive the number of standard deviations between the observed values and the null average of clustering and average path length. (Consider only the largest connected component) Which of these two is statistically significant?
\item Repeat the experiment in the previous question, but now generate $50$ Watts-Strogatz small world models, with the same number of nodes as the original network and setting $k=16$ and $p=0.1$.
\end{enumerate}

\part{Spreading Processes}\label{par:sis}

\chapter{Epidemics}\label{cha:epidemics}
So far, we've seen some dynamics you can embed in your network. In Section \ref{sec:extended-dynamic} I showed you how to model graphs whose edges might appear and disappear, while in the previous book part we've seen models of network growth: nodes arrive steadily into the network and we determine rules to connect them such as in the preferential attachment model. This part deals with another type of dynamics on networks. Here, edges don't change, but nodes can transition into different states.

The easiest metaphor to understand these processes is disease. Normally, people are healthy: their bodies are in a homeostatic state and they go about their merry day. However, they also constantly enter into contact with pathogens. Most of the times, their immune systems are competent enough to fend off the invasion. Sometimes this does not happen. The person transitions into a different state: they now are sick. Sickness might be permanent, but also temporary. People can recover from most diseases. In some cases, recovery is permanent, in others it isn't.

These are all different states in which any individual might find themselves at any given time. Like individuals, nodes too can change state as time goes on. This book part will teach you the most popular models we have to study these state transitions. In this chapter we look at three models we defined to study the progression of diseases through social networks\cite{pastor2015epidemic}. Note that such models can easily represent other forms of contagion, for instance the spread of viruses in computer and mobile networks\cite{wang2009understanding}.

We're going to complicate these models in Chapter \ref{cha:triggers}, to see how different criteria for passing the diseases between friends affect the final results. Then, in Chapter \ref{cha:epidemapps}, we'll see how the same model can be adapted to describe other network events, such as infrastructure failure and word-of-mouth systems to aid a viral marketing campaign.

Another complication is the one introduced by simplicial spreading. If you remember (Section \ref{sec:extended-hyper} is there if you don't), simplicial complexes contain these simplices, linking together multiple nodes in higher-order structures (triangles, 4-cliques, etc). In some cases, to be infected you might need to be part of such a complex structure. For instance, peer-pressure might not work well if you're only connected by an edge. However, if you're part of a simplicial complex of three nodes, that might be enough to trigger you. The combined pressure from your two friends overcomes your resistance. I'll expand on this in Section \ref{sec:hod-simplicial}, when we'll do a deep dive into this kind of high order interactions.

\section{SI}
Sickness can be fatal for the individual and extremely debilitating for entire societies. If tomorrow an epidemic sends to bed $90\%$ of the population at the same time -- even if it doesn't kill anyone -- it can grind the planet to a halt\footnote{This chapter was drafted before COVID-19 happened, and it shows.}. For this reason, humans have a strong incentive to study contagion dynamics at large, to predict whether such a situation might occur in the future. Researchers have developed simple models to describe the dynamics of diseases\cite{kermack1927contribution}\cite{clayton1993statistical}\cite{omran2005epidemiologic}. These are usually known as ``Compartmental models'' -- although I've been calling ``state'' what is traditionally known as ``compartment''.

The model divides individuals into two states. The first state is called Susceptible. A person in the Susceptible state is... well... susceptible to contract a disease. This marks healthy people that show no symptoms and are functioning properly. The second state is called Infected. People in the Infected state -- you'll never believe it -- have contracted the disease. We use $S$ to indicate the set of individuals in the Susceptible state, and $I$ to indicate the set of individuals in the Infected state.

The model allows for only one possible transition between states. The only thing that can happen in this model is the transition from the Susceptible to the Infected state: $S \rightarrow I$. In this world, the only possible action is for a healthy person to contract the disease. Nothing else is allowed.

\begin{figure}
\centering
\includegraphics[width=.66\columnwidth]{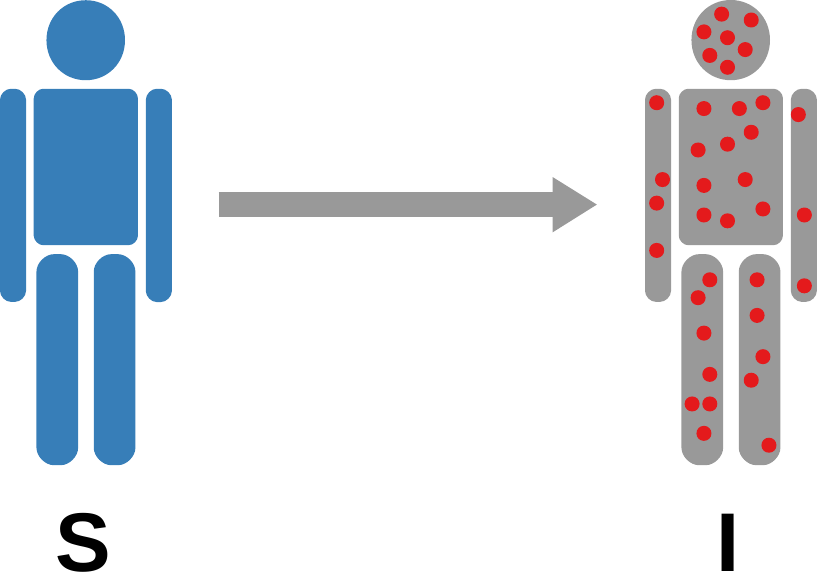}
\caption{The schema underlying the SI Model: two possible states and one possible transition.}
\label{fig:si}
\end{figure}

Given that there are only two states ($S$ and $I$) and only one transition ($S \rightarrow I$), we call this the SI Model. Figure \ref{fig:si} shows the schema fully defining the model. In practice, SI models diseases with no recovery. An example would be some variants of the herpes virus. Love goes by, herpes is forever.

There is one assumption underlying the traditional SI Model: \textbf{homogenous mixing} -- keep this in mind because it's important. In homogenous mixing, we assume that each susceptible individual has the same probability to come into contact with an infected person. This is simply determined by the current fraction of the population in the infected state. Once the susceptible individual meets an infected, there is a probability that they will transition into the $I$ state too. This probability is a parameter of the model, traditionally indicated by $\beta$. If $\beta = 1$, any contact with an infected will transmit the disease, while if $\beta = 0.2$, you have an $20\%$ chance to contract the disease.

\begin{figure}
\centering
\includegraphics[width=\columnwidth]{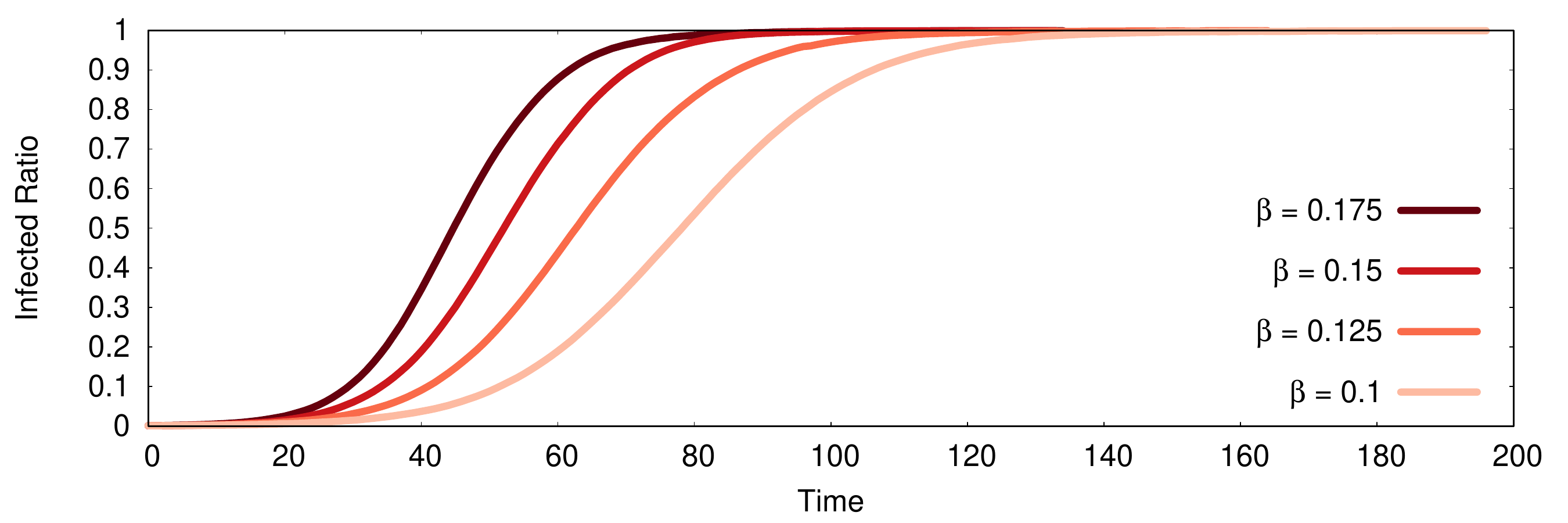}
\caption{The solution of the SI Model for different $\beta$ values. The plot reports on the y axis the share of infected individuals ($i = |I|/(|I|+|S|)$) at a given time step (x axis).}
\label{fig:si2}
\end{figure}

Once you have $\beta$ you can solve the SI Model. Usually, the way it's done is assuming that at the first time step you have a set of one or more patient zeros scattered randomly into society. Then, you track the ratio of people in the $I$ status as time goes on, which is $|I|/(|I|+|S|)$. This usually generates a plot like the one in Figure \ref{fig:si2}.

SI models have the same signature. At first, the ratio of infected individuals grows slowly, because there are few people in the $I$ state. Then, as soon as $I$ expands a little, we see an exponential growth, as more and more people have a chance to meet an infected individual. After a critical point, the growth of $I$ slows down, because there aren't many people left in $S$ to infect.

Eventually, all SI models stop when every single individual is in the set $I$ and so no one else can transition. All SI Models, no matter the value of $\beta$ will end up with a complete infection, where $S$ is empty and $I$ contains the entirety of society. The only thing $\beta$ affects -- as you can see from Figure \ref{fig:si2} -- is the speed of the system: when the exponential growth of $I$ starts to kick in and when $S$ gets emptied out.

We can re-tell the story I've just exposed in mathematical form. In our SI model, the probability that an infected individual meets a susceptible one is simply the number of susceptible individuals over the total population, because of the homogenous mixing hypothesis: $|S|/|V|$ (remember $|V|$ is our number of nodes). There are $|I|$ infected individuals, each with $\bar{k}$ meetings (the average degree). Thus the total number of meetings is $\bar{k} \dfrac{|I||S|}{|V|}$. Since each meeting has a probability $\beta$ of passing the disease, at each time step there are $\beta \bar{k} \dfrac{|I||S|}{|V|}$ new infected people in $I$.

We can simplify the equation a bit, because $|I|/|V|$ and $|S|/|V|$ are related. They sum to one, since $S$ and $I$ are the only possible states in which you can have a node. So, if we say $i = |I|/|V|$, that is, the fraction of nodes in $I$, then $|S|/|V| = 1 - i$. So our formula becomes: $i_{t+1} = \beta \bar{k} i_t (1 - i_t)$\footnote{Note that this is a differential equation, so you need to integrate it to actually find the share of infected nodes at $t+1$. Also, I'm only including the addition to $i_{t+1}$, not its full composition. So, pedantically, the correct formula should be $i_{t+1} = i_t + \beta \bar{k} i_t (1 - i_t)$, but that would make the discussion harder to follow -- and it would not change the results we are interested in here. This warning applies to all formulas with the time subscript.}, where $t$ is the current time step. If we integrate over time, we can derive the fraction of infected nodes depending solely on the time step\cite{barabasi2016network}:

$$ i = \dfrac{i_0 e^{\beta \bar{k} t}}{1 - i_0 + i_0 e^{\beta \bar{k} t}}. $$

This is the mathematical solution to the SI model with homogenous mixing, generating the plot in Figure \ref{fig:si2}. You can see why you have an initial exponential growth at the beginning and a flat growth at the end. If $i_0 \sim 0$, then the denominator is $1$ and the numerator is dominated by the $e^{\beta \bar{k} t}$ factor: exponential growth (very slow at the beginning because multiplied with the small $i_0$). When $i_0 \sim 1$, both the denominator and the numerator reduce to $e^{\beta \bar{k} t}$, which means that, in the end, $i \sim i_0 \sim 1$, so there's no growth.

Why did we go to the trouble of all this math? Because, at this point, we have to tear down the homogenous mixing hypothesis. The formulas will allow to see the difference better.

\begin{figure}
\centering
\begin{subfigure}[t]{.37\columnwidth}
\includegraphics[width=\textwidth]{figures/lattice_infinite.pdf}
\caption{}
\end{subfigure}
\qquad
\begin{subfigure}[t]{.27\columnwidth}
\includegraphics[width=\textwidth]{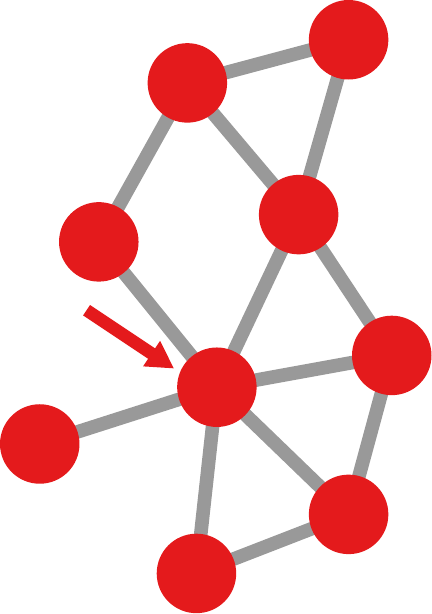}
\caption{}
\end{subfigure}
\caption{The underlying assumption of an SI model: that social networks look like a uniform lattice (a). Instead, the degree is distributed differently (b), with hubs -- pointed by the red arrow -- having many more connections and, thus, infection chances.}
\label{fig:si3}
\end{figure}

Homogenous mixing is based on the assumption that the more people are infected, the more likely you're going to be infected. In practice, it assumes everybody is the same. In homogenous mixing, the global social network is a lattice: a regular grid where each node is connected only to its immediate neighbors. Figure \ref{fig:si3} shows an example of square lattice (Section \ref{sec:extended-types}): each node connects regularly to four spatial neighbors. On a lattice, the infection spreads like water filling a surface\footnote{This is a useful mental image: \url{https://upload.wikimedia.org/wikipedia/commons/a/a6/SIR_model_simulated_using_python.gif}.}.

We know that real networks are not neat regular lattices. The degree is distributed unevenly, with hubs having thousands of connections -- see Section \ref{sec:degree-pl}. When the infection hits such a hub, it will accelerate faster through the network. In fact, it is extremely easy to infect an hub early on. Hubs have more connections, thus they are more likely to be connected to one of your patient zeros. Those same connections make them super-spreaders: once infected, the hub will allow the disease to reach the rest of the network quickly. In fact, when searching information in a peer-to-peer network, your best guess is always to ask your neighbor with highest degree\cite{adamic2001search}.

To treat the SI model mathematically you have to first group nodes by their degree. Rather than solving for $i$ -- the fraction of infected nodes --, you solve for $i_k$: the fraction of infected nodes of degree $k$. The formula for a network-aware SI model is similar as the one we saw for the vanilla SI model:

$$i_{k,t+1} =  \beta k f_k (1 - i_{k,t}).$$

The two differences are that: (i) we replace the average degree $\bar{k}$ with the actual node's degree $k$, and (ii) rather than using  $i_{k,t}$ we use $f_k$ -- a function of the degree $k$. This is because real world networks typically have degree correlations: if you have a degree $k$ the degree of your neighbors is usually not random (see Section \ref{sec:assortquant-plots} for more). If it were random, then we could simply use $i_{k,t}$, because the number of infected individuals around you should be proportional to the current infection rate. But it isn't: in presence of degree correlations, if you have $k$ neighbors then there exists a function $f_k$ able to predict how many neighbors they have. Thus the likelihood of a node of degree $k$ of having infected neighbors is specific to its degree, and not (only) dependent on $i_{k,t}$.

\begin{figure}
\centering
\includegraphics[width=\columnwidth]{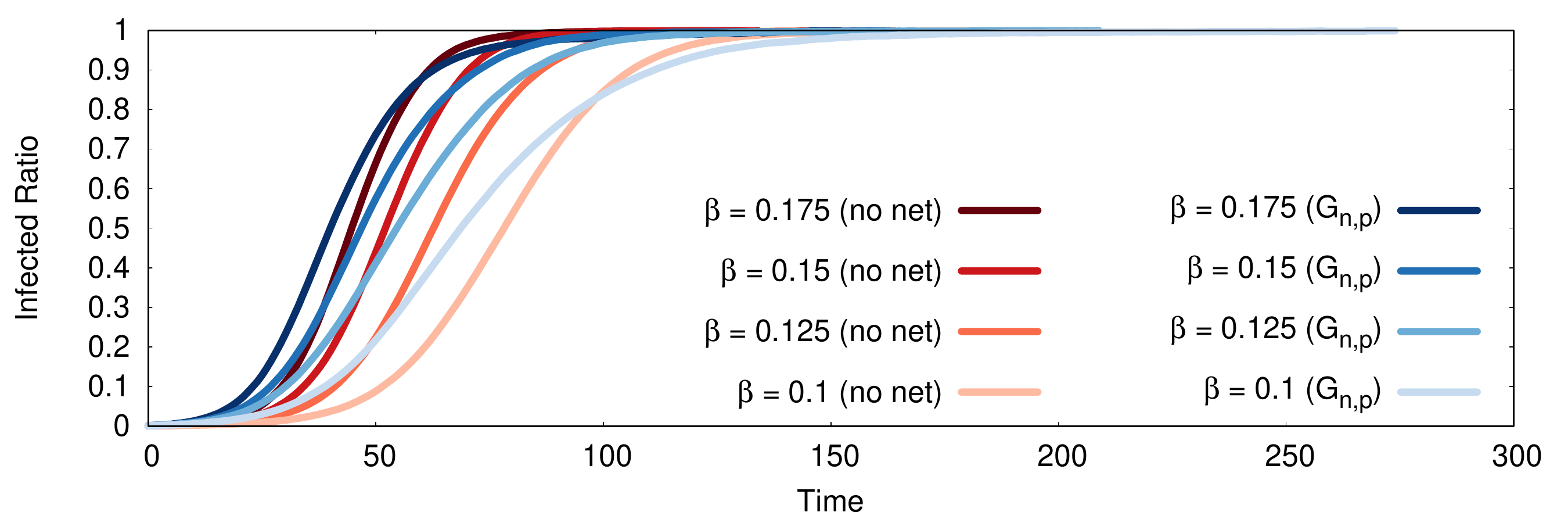}
\caption{The solution of the SI Model for different $\beta$ values in homogeneous mixing (reds) and $G_{n,p}$ graphs (blues). The plot reports on the y axis the share of infected individuals ($i = |I|/(|I|+|S|)$) at a given time step (x axis).}
\label{fig:si4}
\end{figure}

If you do the proper derivations\cite{pastor2001epidemic}, you'll discover that in a $G_{n,p}$ network the dynamics have the same functional form to the ones of the homogeneous mixing, as Figure \ref{fig:si4} shows. In $G_{n,p}$ the exponential rises faster at the beginning -- due to the few outliers with high degree -- and tails off slower at the end -- due to the outliers with low degree -- but the rising and falling of the infection rates is still an exponential. It also depends, obviously, on the average degree you give to the lattice and to the $G_{n,p}$ graph.

Is that it? Did I really throw Greek letters at you for such an underwhelming discovery? Of course not. Remember that $G_{n,p}$ is a poor approximation of social networks, \textit{especially} when it comes to degree distributions. Let's look at what happens when you have a network with a power law degree distribution.

\begin{figure}
\centering
\includegraphics[width=\columnwidth]{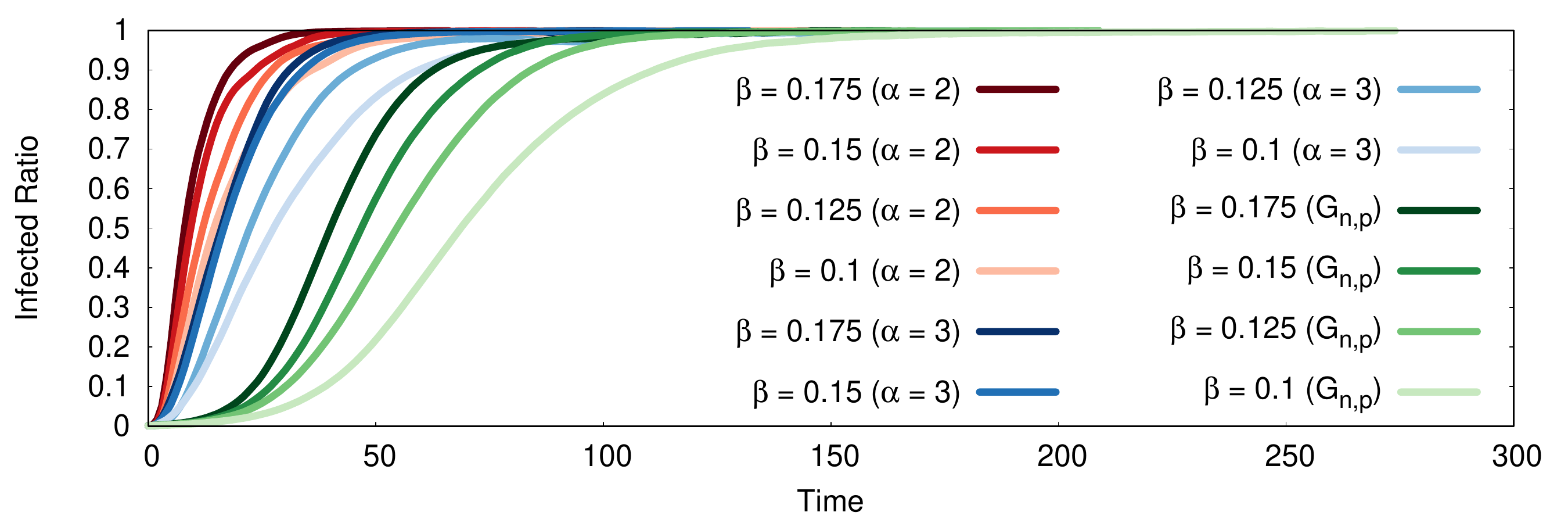}
\caption{The solution of the SI Model for different $\beta$ values for networks with power law degree distribution with exponent $\alpha = 2$ (reds), $\alpha = 3$ (blues) and $G_{n,p}$ graphs (greens). The plot reports on the y axis the share of infected individuals ($i = |I|/(|I|+|S|)$) at a given time step (x axis).}
\label{fig:si5}
\end{figure}

Figure \ref{fig:si5} shows you the results of a bunch of simulations on networks with different degree distributions. The slowest infections (in green) happen for $G_{n,p}$ graphs. When looking at a power law random network, the thing that matters the most is the exponent of the degree distribution, $\alpha$ (for a refresher on its meaning, see Section \ref{sec:degree-pl}).

If $\alpha = 3$ we have not-so-large hubs. These hubs contribute enormously to the speed of infection: it is easy to catch them and, once you do, the disease spreads faster. You can see how the exponential growth regimes, for the blue data series, are much steeper than in the green $G_{n,p}$ cases. Even for $\beta = 0.1$, a mildly contagious disease, a power law degree distribution with $\alpha = 3$ gets infected faster than a $G_{n,p}$ network with a much more aggressive disease (with $\beta = 0.175$, almost twice as infectious!).

The same comparison applies when pitting the $\alpha = 3$ case with $\alpha = 2$ networks. In the latter case, there's not even a recognizable exponential warm up any more (in red in Figure \ref{fig:si5}). You know that, no matter where you started, you're going to hit the largest hub of the network at the second time step $t = 2$, because it is connected to practically every node. And, since it is connected to practically every node, at $t = 3$ you'll have almost the entire network infected.

In fact, I ran the simulations from Figure \ref{fig:si5} on imperfect and finite power law models. Theoretically, if you had a perfect infinite power law network, infection would be \textit{instantaneous for any non-zero value} of $\beta$. Meaning that, no matter how infectious a disease is, with $\alpha = 2$ it will infect the entire network almost immediately. And things get even more complicated when you add to the mix the fact that networks evolve over time\cite{valdano2015analytical}. Scary thought, isn't it?

\section{SIS}
Just like in the SI model, also in the SIS model nodes can only either be Susceptible or Infected\cite{hethcote1989three}. However, the SIS model adds a transition. Where in SI you could only get infected without possibility of recovery ($S \rightarrow I$), in SIS you can heal ($I \rightarrow S$).

Thus the SIS model requires a new parameter. The first one, shared with SI, is $\beta$: the probability that you will contract the disease after meeting an infected individual. Once you're infected, you also have a recovery rate: $\mu$. $\mu$ is the probability that you will transition from $I$ to $S$ at each time step. High values of $\mu$ mean that recovering from the disease is fast and easy. Note that recovery puts you back to the Susceptible state, thus you can catch the disease again in the future.

\begin{figure}
\centering
\includegraphics[width=.66\columnwidth]{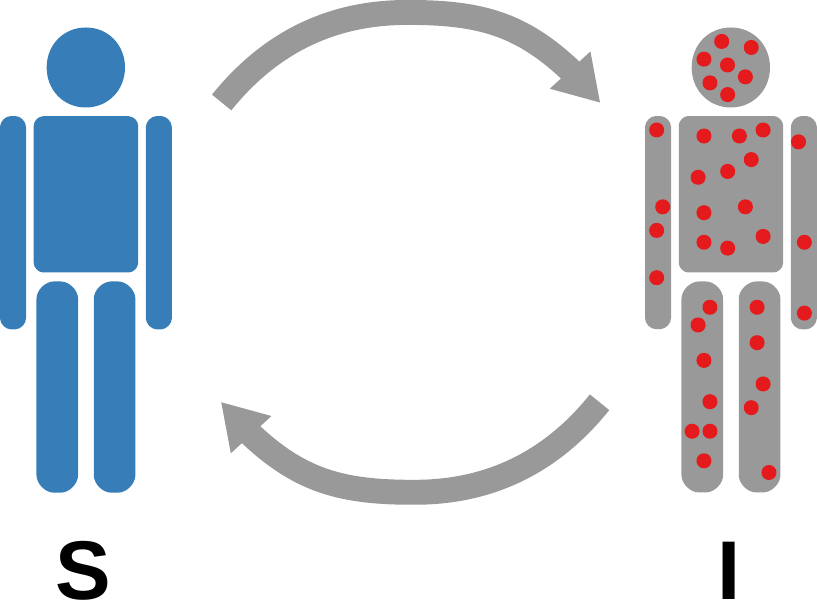}
\caption{The schema underlying the SIS Model: two possible states and two possible transitions.}
\label{fig:sis}
\end{figure}

Figure \ref{fig:sis} shows the schema fully defining the model. In practice, SIS models disease with recovery and relapse. An example would be the general umbrella of the flu family. Once you heal from a particular strain of the flu you're unlikely to fall ill again under the same strain. However, you can easily catch a similar strain, thus cycling each year between the $S$ and $I$ states.

The presence of $\mu$ changes the outcome of the model. SI models always reach full saturation: eventually, every node will end up in status $I$. For SIS models that is not true, because a certain fraction of nodes -- $\mu$ -- heal at each time step. The interplay between the recovery rate $\mu$, the infection rate $\beta$, and the average degree $\bar{k}$ determines the asymptotic size of $I$: the share of infected nodes as time approaches infinity ($t \rightarrow \infty$). To see how, let's look at the math again.

\begin{figure}
\centering
\includegraphics[width=.8\columnwidth]{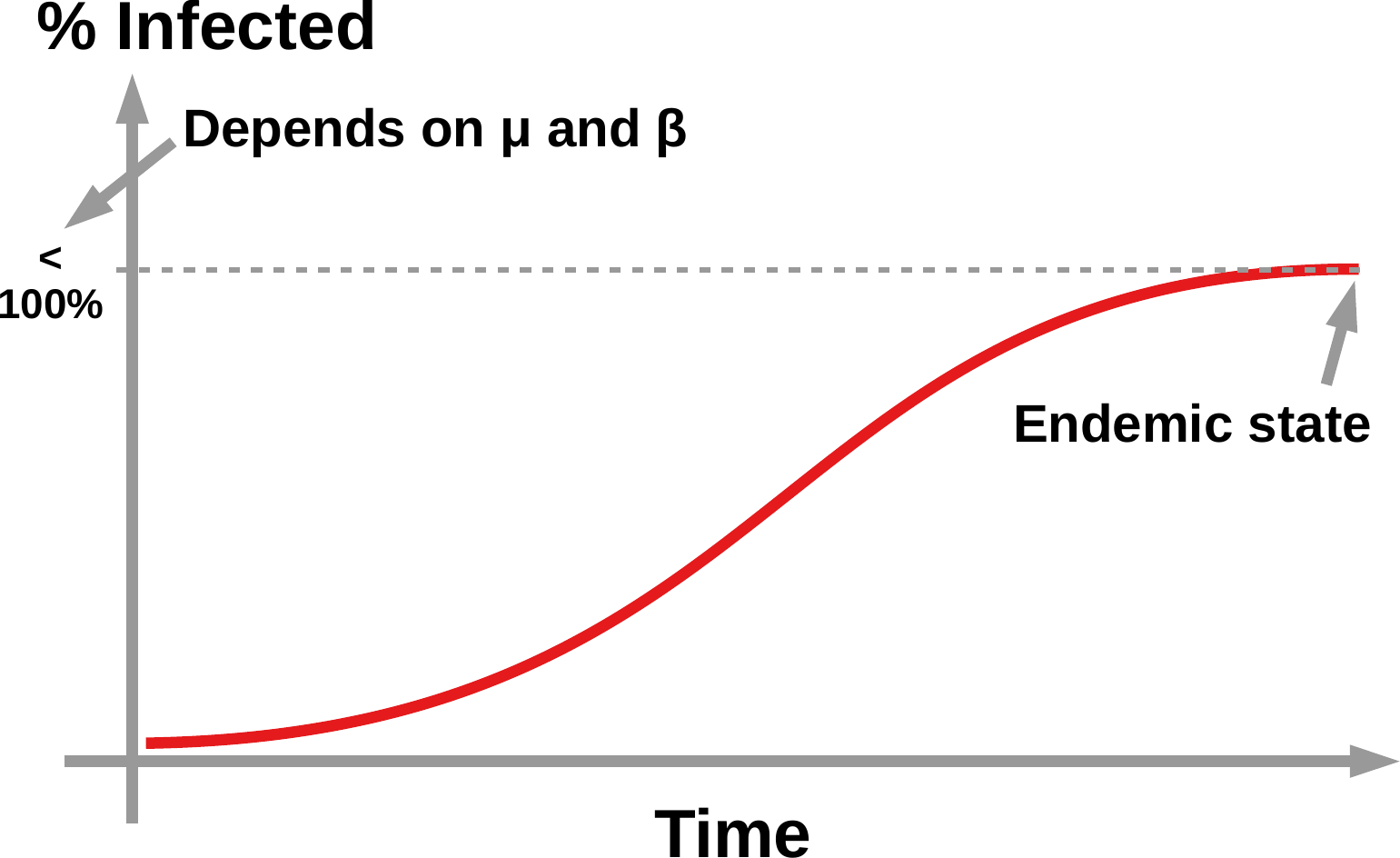}
\caption{The typical evolution of an endemic SIS model: the equilibrium state is the one in which a constant fraction $i < 1$ contracted the disease. The rate at which infected people recover and the infection rate are perfectly balanced, as all things should be.}
\label{fig:sis-endemic}
\end{figure}

The SI model could be described by the formula $i_{t+1} = \beta \bar{k} i_t (1 - i_t)$. If, at each time step, a fraction $\mu$ of the infected nodes $i$ recovers, we just have to remove it from $i$. Thus, the SIS model is simply $i_{t+1} = \beta \bar{k} i_t (1 - i_t) - \mu i_t$. This should raise your eyebrow. As $i_t$ grows, so does $\mu i_t$, obviously. Eventually, $\beta \bar{k} i_t (1 - i_t) = \mu i_t$: that is when the share of infected nodes $i$ doesn't grow any more. We reached the endemic state where the number of people recovering is perfectly balanced by the new infected. Figure \ref{fig:sis-endemic} depicts this situation.

\begin{figure}
\centering
\includegraphics[width=\columnwidth]{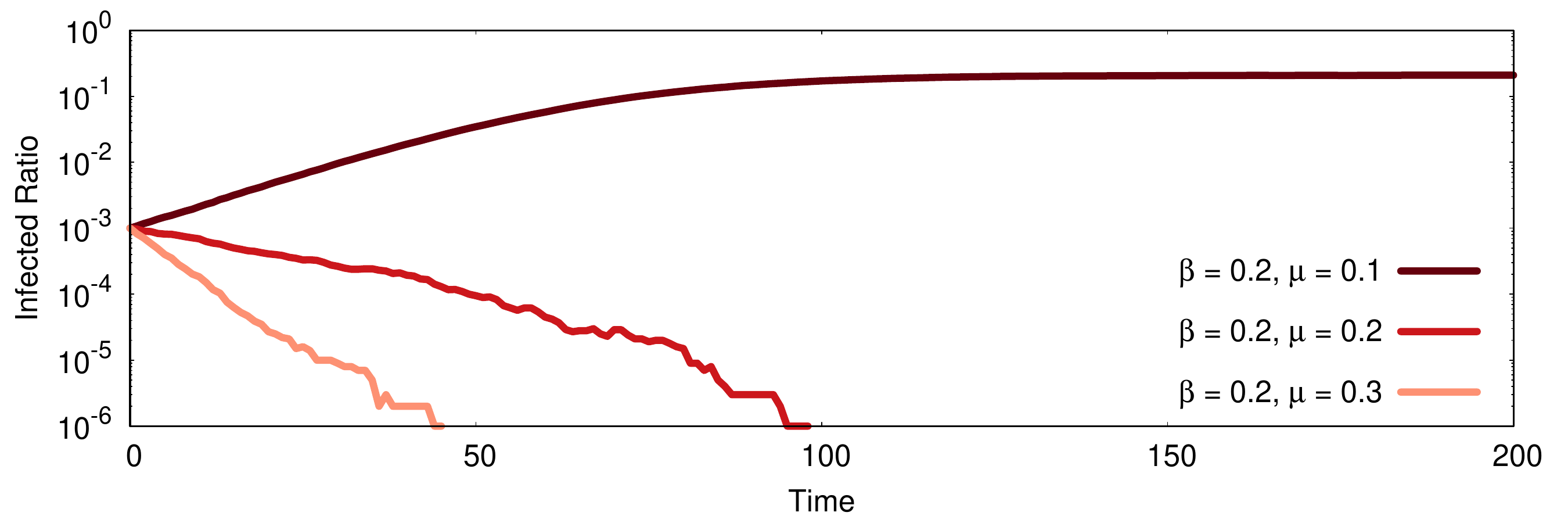}
\caption{The solution of the SIS Model, keeping $\beta$ fixed but varying $\mu$. The plot reports on the y axis the share of infected individuals ($i = |I|/(|I|+|S|)$) at a given time step (x axis).}
\label{fig:sis-endemic2}
\end{figure}

Is it possible that $\beta \bar{k} i_t (1 - i_t) < \mu i_t$? Meaning: is there a situation when people are recovering faster than new infected pop up? \textit{Yes}! We \textit{can} get rid of a disease in the SIS model. If you do the proper derivations\cite{pastor2001epidemic2}, you discover that the magic value of $\mu$ for that to happen is $\beta \bar{k}$. If $\mu < \beta \bar{k}$, recovery isn't fast enough to escape the endemic state, and you're in the situation we saw in Figure \ref{fig:sis-endemic}. But, if $\mu \geq \beta \bar{k}$, eventually the endemic state is the one for which $i = 0$! Congratulations! No more infected people. You defeated the disease. The evolution of the outbreak looks like what I sketch in Figure \ref{fig:sis-endemic2}.

In my simulations for Figure \ref{fig:sis-endemic2} I set $\bar{k} = 1$, to make the comparisons easier. You can see that, when $\mu \geq \beta$, eventually the disease dies out. The case for which $\mu < \beta$ reaches the endemic state, showing that the disease will persist in the population. For $\mu = \beta$, the disease dies out, although not as quickly as for $\mu > \beta$.

The relationship between $\mu$ and $\beta$ is so important that we can study the evolution of infections in a network according to their ratio $\lambda = \beta / \mu$. Since we ignore the degree when calculating $\lambda$, we know that $\lambda$ depends exclusively by the pathogen's characteristics. We just saw in Figure \ref{fig:sis-endemic2} that, in some cases, the SIS model predicts a non-zero endemic state -- there are always at least $i > 0$ infected individuals -- and, in other cases, the disease dies out -- thus $i = 0$. So there must be a critical value of $\lambda$ that make us transition between the endemic and non-endemic state.

In $G_{n,p}$ networks with homogeneous mixing this critical $\lambda$ value depends on the average degree of the network $\bar{k}$. Specifically, if $\lambda > 1 / (\bar{k} + 1)$ then the disease will be endemic. If $\lambda$ is below that threshold, then the pathogen will eventually disappear. Note that, in a $G_{n,p}$ graph, $\bar{k}$ is always positive and equal to $p|V|$ (see Section \ref{sec:rndgraphs-degdistr} for a refresher). This mean that you can find a value of $\lambda$ below the critical endemic threshold: the only way for $1 / (\bar{k} + 1)$ to be equal to zero would be if $\bar{k} = \infty$, which is clearly nonsense. The average degree in a $G_{n,p}$ graph cannot be infinite. Thus any $G_{n,p}$ graph will be resistant to a disease with a $\lambda$ lower than $1 / (\bar{k} + 1)$

Surprising absolutely no one, when we drop the homogeneous mixing assumption and we look at a preferential attachment network the situation changes radically. Here, the critical value is $\bar{k} / \bar{k^2}$: the average degree over the average squared degree -- note that we square the degrees and \textit{then} we take the average, we don't simply raise the average degree to the power of $2$. Here's the problem: the average degree in a preferential attachment network is low. But the network contains large hubs with a ridiculously high degree: squaring it eclipses the small contributions from the peripheral nodes that kept $\bar{k}$. In other words, as you add more and more nodes to the network, $\bar{k}$ remains constant and low -- because you're adding peripheral nodes with low degree -- but $\bar{k^2}$ grows fast. Each of those new nodes tend to add to the degree of the largest hubs, because of preferential attachment -- shooting $\bar{k^2}$ in the stratosphere.

The consequence? Well, if $\bar{k}$ stays constant -- or even decreases -- and $\bar{k^2}$ grows relatively to it, the critical threshold $\bar{k} / \bar{k^2}$ tends to zero. If we say that you have an endemic value if $\lambda > \bar{k} / \bar{k^2}$ and $\bar{k} / \bar{k^2} = 0$, then any disease, no matter $\beta$ and $\mu$, will be endemic in a network with a power law degree distribution. Oops.

\begin{figure}
\centering
\includegraphics[width=\columnwidth]{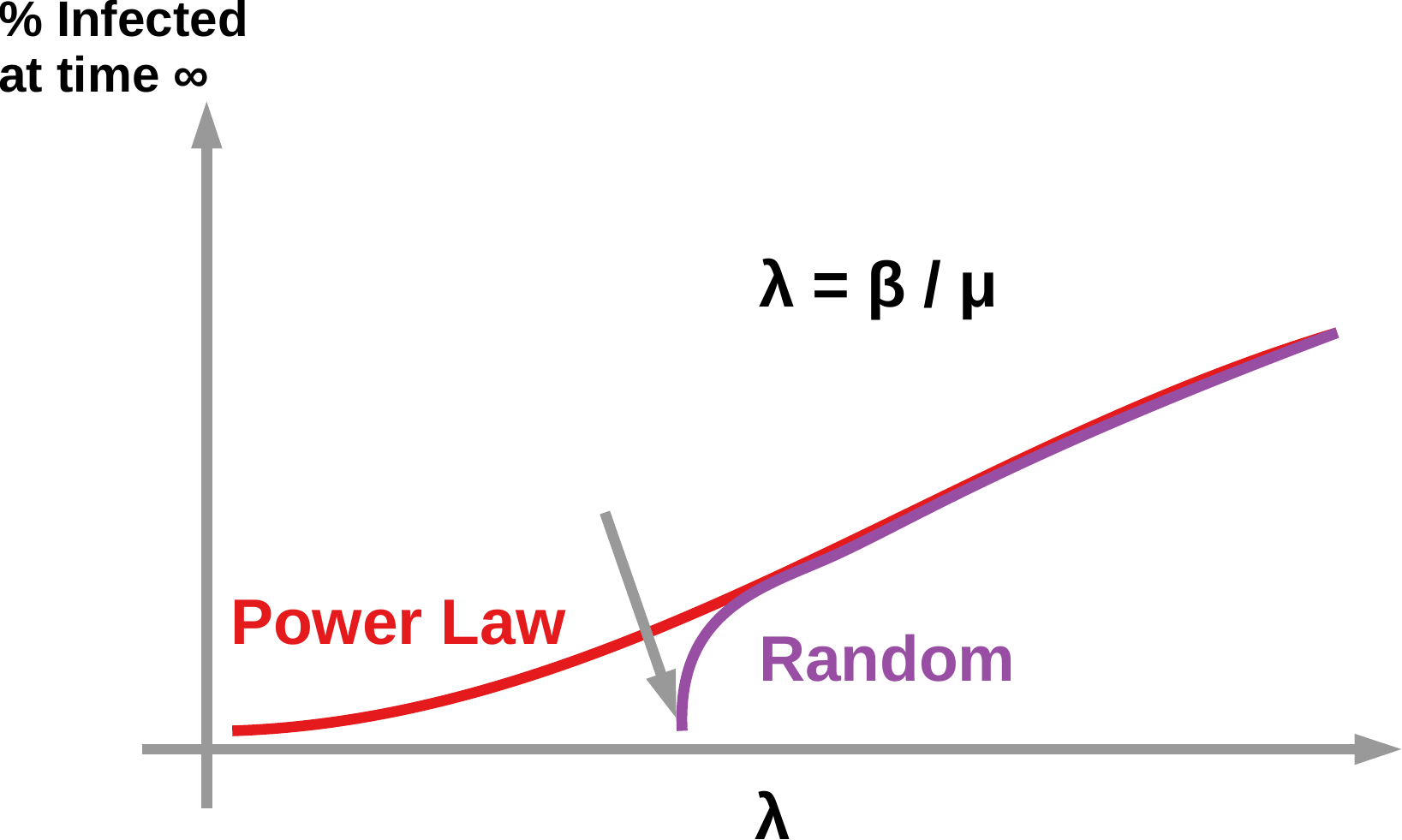}
\caption{The solution of the SIS Model for $\lambda$. As $\lambda$ grows (x axis), I show the share of infected individuals $i$ at the endemic state ($t \rightarrow \infty$).}
\label{fig:sis-lambda}
\end{figure}

I sum up the situation in Figure \ref{fig:sis-lambda}: in a power law random network, no matter $\lambda$, the pathogen can always be endemic, even if it's not very infectious. In a $G_{n,p}$ network you see that for some values of $\lambda$ greater than zero you do not have endemic infections, thus the network's topology has a big effect on the dynamics of the epidemic. Heavy tailed degree distributions, which are ubiquitous in reality, are closer to the power law line than to the $G_{n,p}$ line, meaning that we should expect to see a similar behavior in real networks.

\section{SIR}\label{sec:epidemics-sir}
The next step in modeling epidemics on networks is by considering those diseases you can catch only once in your lifetime. Think about the bubonic plague. If you have the bad luck of encountering the \textit{Yersinia pestis}, there are only two possible outcomes. Either you die, or you survive. If you survive, your immune system is now trained to recognize the bacterium and will not allow you to be infected again. In either case, you are Removed from the outbreak.

\begin{figure}
\centering
\includegraphics[width=.8\columnwidth]{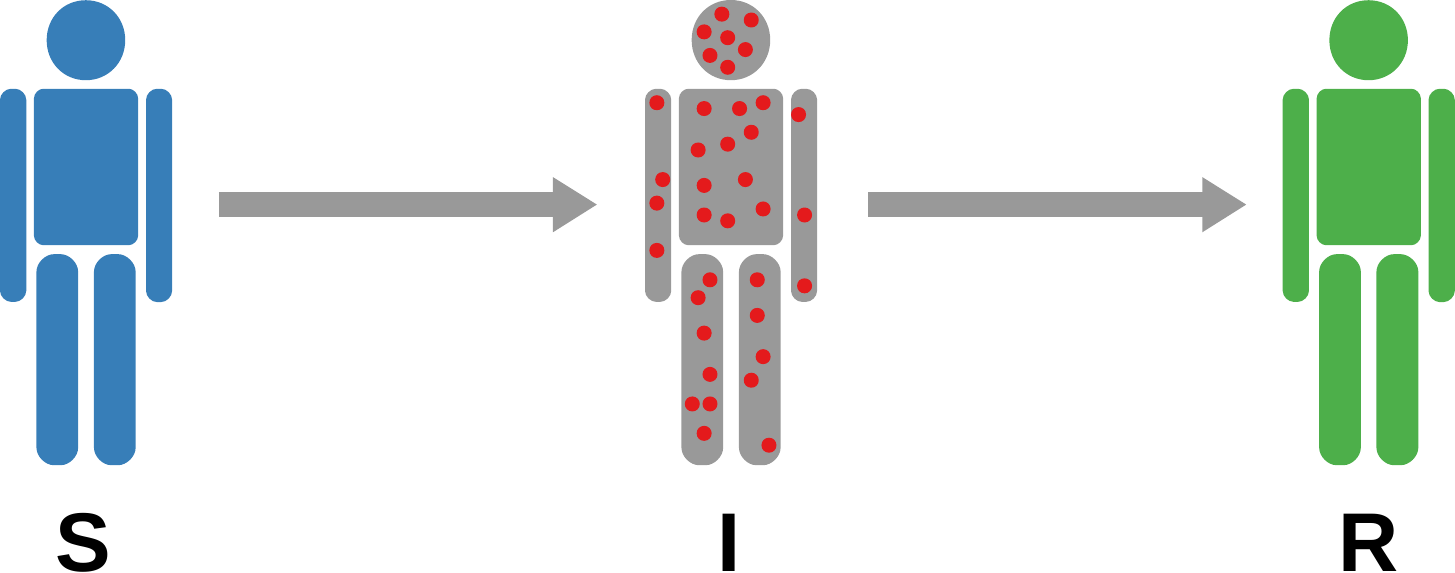}
\caption{The schema underlying the SIR Model: three possible states and two possible transitions.}
\label{fig:sir-schema}
\end{figure}

Removed is exactly the state we add to the SI model in the SIR model. Now the only two possible state transitions are $S \rightarrow I$ -- when you contract the disease -- and $I \rightarrow R$ -- when you heal or die. Figure \ref{fig:sir-schema} shows the schema fully defining the model.

\begin{figure}
\centering
\includegraphics[width=.8\columnwidth]{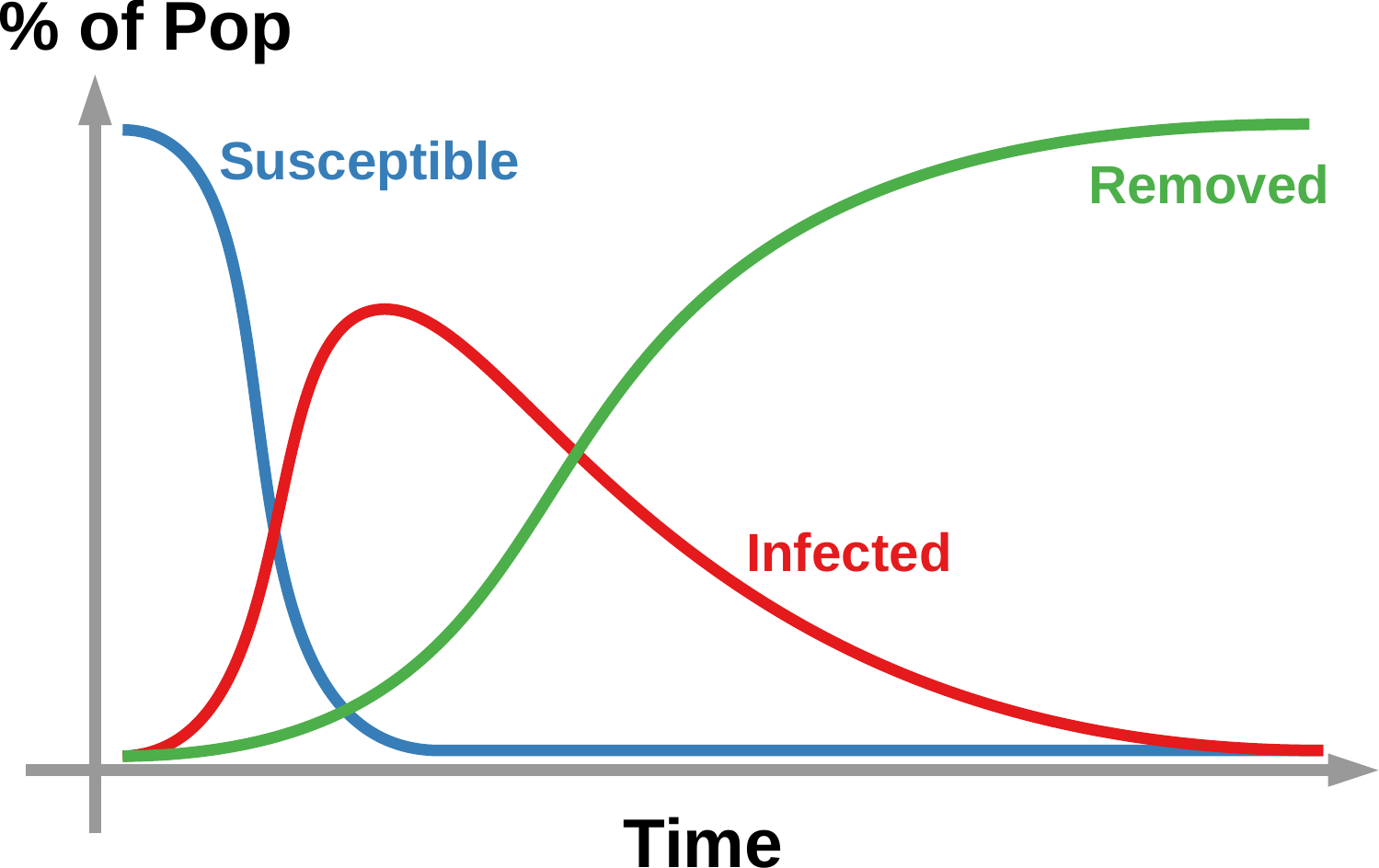}
\caption{The typical evolution of an SIR model: after an initial exponential growth of the infected, the removed ratio takes over until it occupies the entire network.}
\label{fig:sir}
\end{figure}

The defining characteristic of a SIR model is its lack of endemicity. Either the disease kills everybody, or every individual still alive has had the disease and healed. Figure \ref{fig:sir} shows such a typical evolution. At the beginning, everybody is susceptible. Then, people start getting infected, so $I$ grows. $R$ cannot start growing immediately, as $I$ is still too small for the recovery parameter $\mu$ to significantly contribute to $R$ size. As $I$ grows, though, there are enough infected individuals that start being removed. Eventually every $I$ individual transitions to $R$.

Note that the evolution of Figure \ref{fig:sir}, where eventually there are only people in the $R$ state, isn't necessarily the only possible. If you're lucky, $I$ empties out before $S$, meaning that the disease dies out in $R$ before every susceptible individual has had the privilege of sneezing.

Mathematically speaking, the evolution of the recovery ratio $r = |R|/|V|$ is the simplest possible. At each time step, a fraction $\mu$ of $I$ transitions into $R$. In SIR, just like in SIS, $\mu$ is the recovery parameter. So $r_{t+1} = \mu i_t$. The evolution of $i$ is a bit trickier, but it boils down to making sure of removing the nodes in $|R|$ from the potential pool of infected.

Of course, in the quest of making models more and more accurate to fit the actualy dyamics of infections, you don't have to stop with the SIR model. In the literature you can find: SEIR, adding an ``Exposed'' status before the infection triggers in an individual\cite{li1995global}\cite{li1999global}; you can have an immune status M; relapsing to susceptibility in a SIRS model after being removed\cite{li2014analysis}; and, of course, combining everything together in a warm and fuzzy pile of states, in the MSEIRS model (I wish I was kidding). As you might expect, the math becomes fiendishly complicated and it's just not worth delving into that for an introductory chapter to network epidemics such as this one.

This chapter is also by necessity just a superficial sketch of network epidemics. There's plenty more research on endemic and epidemic states and their relationship with network topology\cite{wang2003epidemic}\cite{durrett2010some}\cite{castellano2010thresholds} that you can check if you find the topic fascinating.

\section{Summary}

\begin{enumerate}
\item In simple contagion epidemics models, nodes are in specific states given their exposure to the disease and can transition in different states according to simple contact rules.
\item In SI models there are two states: Susceptible and Infected. Nodes transition from $S$ to $I$ with a certain probability $\beta$ if they have at least an $I$ neighbor.
\item All SI models end up with the entire network in the $I$ state. $\beta$ determines how quickly this happens. Networks with a degree distribution characterized by a low $\alpha$ exponent are infected more quickly.
\item SIS models are like SI models, but nodes can transition back to $S$ state with a stochastic probability $\mu$ at each time step.
\item The $\lambda = \beta / \mu$ ratio determines whether the disease will be endemic or it will die out. In power law random networks, no matter $\lambda$, the disease will always be endemic. 
\item In SIR models, nodes in state $I$ recover at a $\mu$ rate rather than moving back to $S$. Eventually, all nodes will move to the $R$ state, and no disease can be endemic.
\end{enumerate}

\section{Exercises}

\begin{enumerate}
\item Implement an SI model on the network at \url{http://www.networkatlas.eu/exercises/20/1/data.txt}. Run it 10 times with different $\beta$ values: $0.05$, $0.1$, and $0.2$. For each run (in this and all following questions) pick a random node and place it in the Infected state. What's the average time step in which each of those $\beta$ infects $80\%$ of the network?
\item Run the same SI model on the network at \url{http://www.networkatlas.eu/exercises/20/2/data.txt} as well. One of the two networks is a $G_{n,p}$ graph while the other has a power law degree distribution. Can you tell which is which by how much the disease takes to infect $80\%$ of the network for the same starting conditions used in the previous question?
\item Extend your SI model to an SIS. With $\beta = 0.2$, run the model with $\mu$ values of $0.05$, $0.1$, and $0.2$ on both networks used in the previous questions. Run the SIS model, with a random node as a starting Infected set, for 100 steps and plot the share of nodes in the Infected state. For which of these values and networks do you have an endemic state? How big is the set of nodes in state I compared to the number of nodes in the network? (Note, randomness might affect your results. Run the experiment multiple times)
\item Extend your SI model to an SIR. With $\beta = 0.2$, run the model for $400$ steps with $\mu$ values of $0.01$, $0.02$, and $0.04$ and plot the share of nodes in the Removed state for both the networks used in Q1 and Q2. How quickly does it converge to a full $R$ state network?
\end{enumerate}

\chapter{Complex Contagion}\label{cha:triggers}
You may or may not have noticed that, in the previous chapter, all our models of epidemic contagion shared an assumption. Every time a susceptible individual comes in contact with an infected individual, they have a chance to become infected as well. If that doesn't happen, the healthy person is still in the susceptible pool. The next time step represents a new occasion for them to contract the disease. And so on, \textit{ad infinitum}.

Without that assumption, the models wouldn't be mathematically tractable. For instance, if each node gets only one chance to be infected, you can easily see how it is not given that a SI model would eventually infect the entire network. In fact, it takes any $\beta < 1$ to make that impossible. The first time you fail to infect somebody you won't get the chance to try again.

SI, SIS, and SIR models are useful and generated tons of great insights. But this limitation allows them to model only rather specific types of outbreak. We usually consider them models of simple contagion. There are fundamentally two ways to make such models more complex and realistic. They involve changing two things: (i) the triggering mechanism, which is the condition regulating the $S \rightarrow I$ transition, and (ii) the assumption that each individual gets infinite chances to infect their neighbors.

We deal with the triggering mechanisms in Section \ref{sec:triggers-triggers} and infection chances in Section \ref{sec:triggers-resistance}. We also explore the possibilities of interfering with the outbreak in Section \ref{sec:triggers-intervention}, dedicated to epidemic interventions.

\section{Triggers}\label{sec:triggers-triggers}

\begin{figure}
\centering
\begin{subfigure}{.45\columnwidth}
\includegraphics[width=\textwidth]{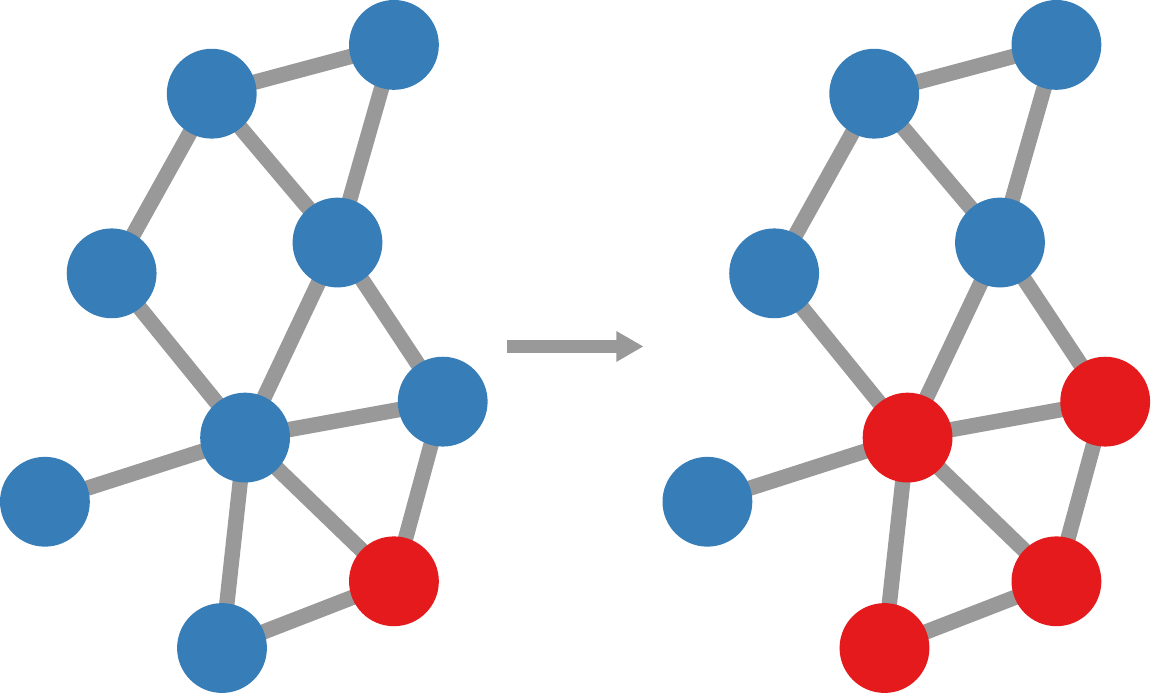}
\caption{}
\end{subfigure}\quad
\begin{subfigure}{.45\columnwidth}
\includegraphics[width=\textwidth]{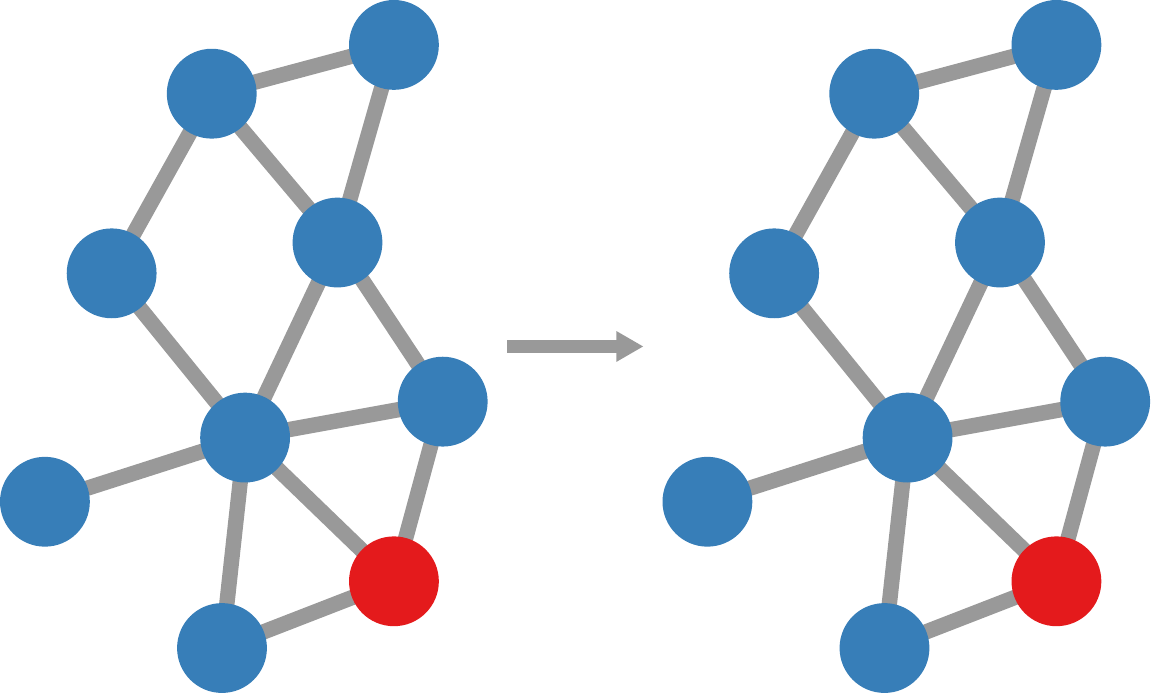}
\caption{}
\end{subfigure}
\caption{A simple introduction to complex contagion. (a) A simple contagion where any contact can and will transmit the disease. (b) A complex contagion where you need two contacts to contract the disease.}
\label{fig:complex-contagion}
\end{figure}

As I mentioned before, in the simple contagion we explored in Chapter \ref{cha:epidemics}, one contact with an infected node is enough for you to have a chance to be infected. The main difference between simple and complex contagion is that, in the latter, you require reinforcement. You can consider Figure \ref{fig:complex-contagion} as the simplest possible introduction to this concept. If we require a node to enter into contact with two infected individuals rather than one, the figure shows that the outbreak from a single seed is impossible.

In reality, complex contagion is more nuanced than this. A single contact may or may not infect you, but if you have multiple contacts your likelihood to transition into the $I$ state grows. There are fundamentally two types of reinforcement we can consider, which are subtypes of complex contagion: Cascade and Threshold. Before looking at them, though, let's consider an easy extension of simple contagion since, in a sense, it can be turned into the simplest possible reinforcement mechanism.

\subsection{Classical}
In classical reinforcement you have an independent probability of being infected for each of your neighbors that are infected. Note that this is different from the simple contagion of Chapter \ref{cha:epidemics}: in there, you get $\beta$ chance to transition regardless whether you have one or more infected neighbors. Here, more infected neighbors mean more chances of infection.

If you have $n$ sick friends, and you visit them one by one, at each visit you toss a coin. To calculate the probability you are going to be infected, it is easier to calculate the probability of not being infected by any contact, and then invert it. If our parameter $\beta$ tells us the probability of being infected by a single contact, then $(1 - \beta)$ is the probability of not being infected. Since the coin tosses are all independent, the probability of never being infected by any of the $n$ contacts is $(1 - \beta)^n$. So the probability that at least one contact will infect us is $1 - ((1 - \beta)^n)$.

\begin{figure}
\centering
\includegraphics[width=.8\columnwidth]{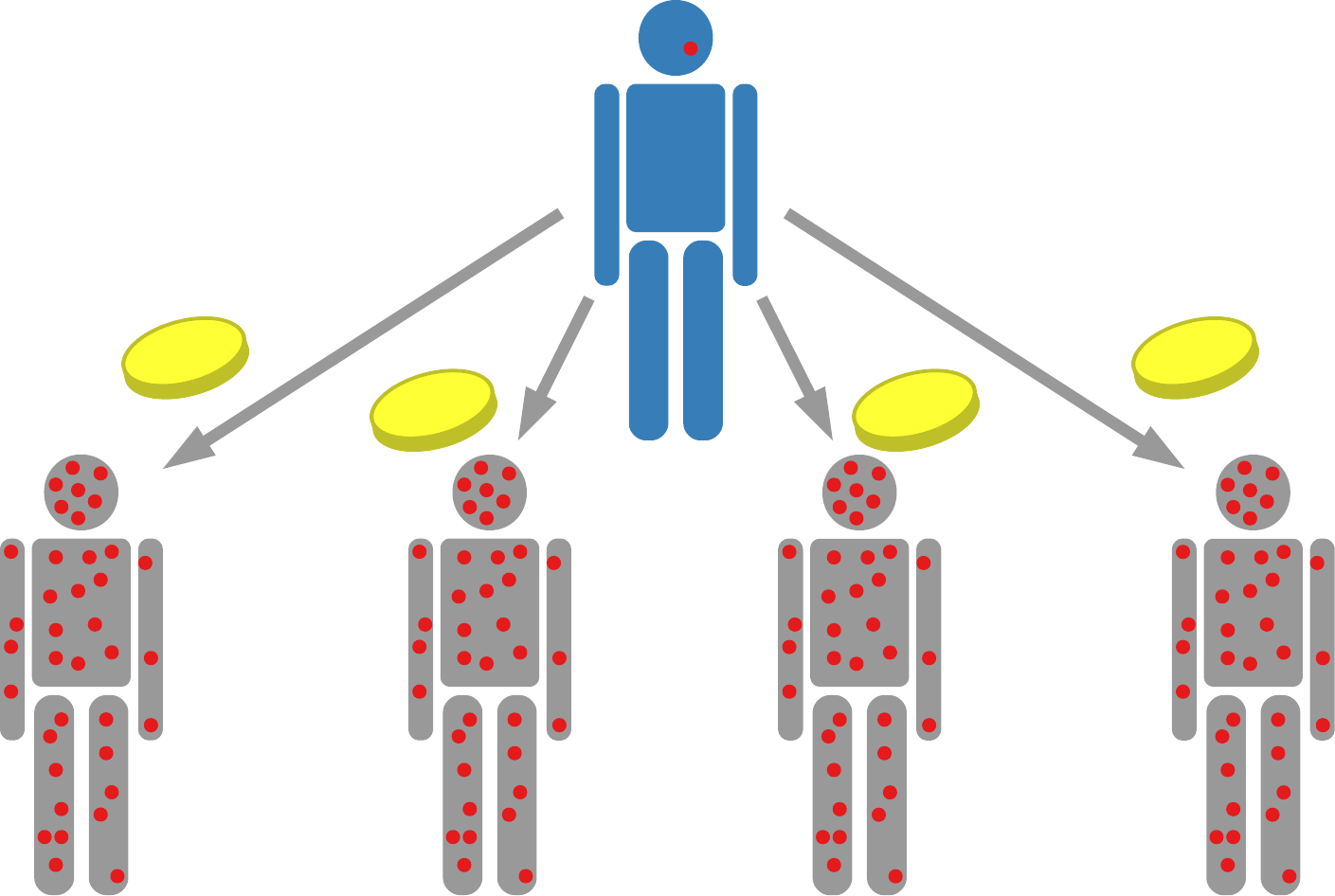}
\caption{An example of classical contagion. The healthy individual performs an independent check for contagion -- a coin toss -- with each of their neighbors.}
\label{fig:complex-classical}
\end{figure}

Figure \ref{fig:complex-classical} shows a vignette of this process. The healthy individual has four neighbors, all of which are infected. Thus she has to make four independent coin tosses, each of which has $\beta$ chance to succeed. Thus, the more infected neighbors the more likely she will contract the disease. The difference with a simple SI model without reinforcement is that in the simple SI model you always toss a single coin at each time step, no matter how many infected neighbors you have -- as long as you have at least one. So, at each time step, in simple SI the infection probability is $\beta$ if you have $1$ or $n$ infected neighbors. In classical complex SI, you have $1 - ((1 - \beta)^n)$ probability of being infected. The whole difference between the two models is that the latter depends on $n$, the number of your friends that are infected.

The vignette makes clear why, in the classical model, it's easy to infect hubs: they have more neighbors. More neighbors mean that they toss their coins much more often. This is what generates the super-exponential -- theoretically instantaneous -- outbreak growth in power law models with large hubs.

\subsection{Threshold}
The threshold model is a sophisticated version of the introductory example I made with Figure \ref{fig:complex-contagion}. To be infected, you need multiple infected neighbors. The threshold model adds a parameter, let's call it $\kappa$. If more than $\kappa$ of your neighbors are infected, then they pass the infection to you\cite{granovetter1978threshold}.

For instance, if $\kappa = 4$, you need four infected friends to have a chance to be infected. Note that you can still inject in this model the $\beta$ chance of infection, by saying that, once you clear the $\kappa$ threshold, you have a chance $\beta < 1$ to contract the disease. In this latter case, if $\kappa = 1$, this model is the same as the simple SI without reinforcement: as long as you have at least one infected neighbor, you toss your $\beta$ coin.

\begin{figure}
\centering
\includegraphics[width=.8\columnwidth]{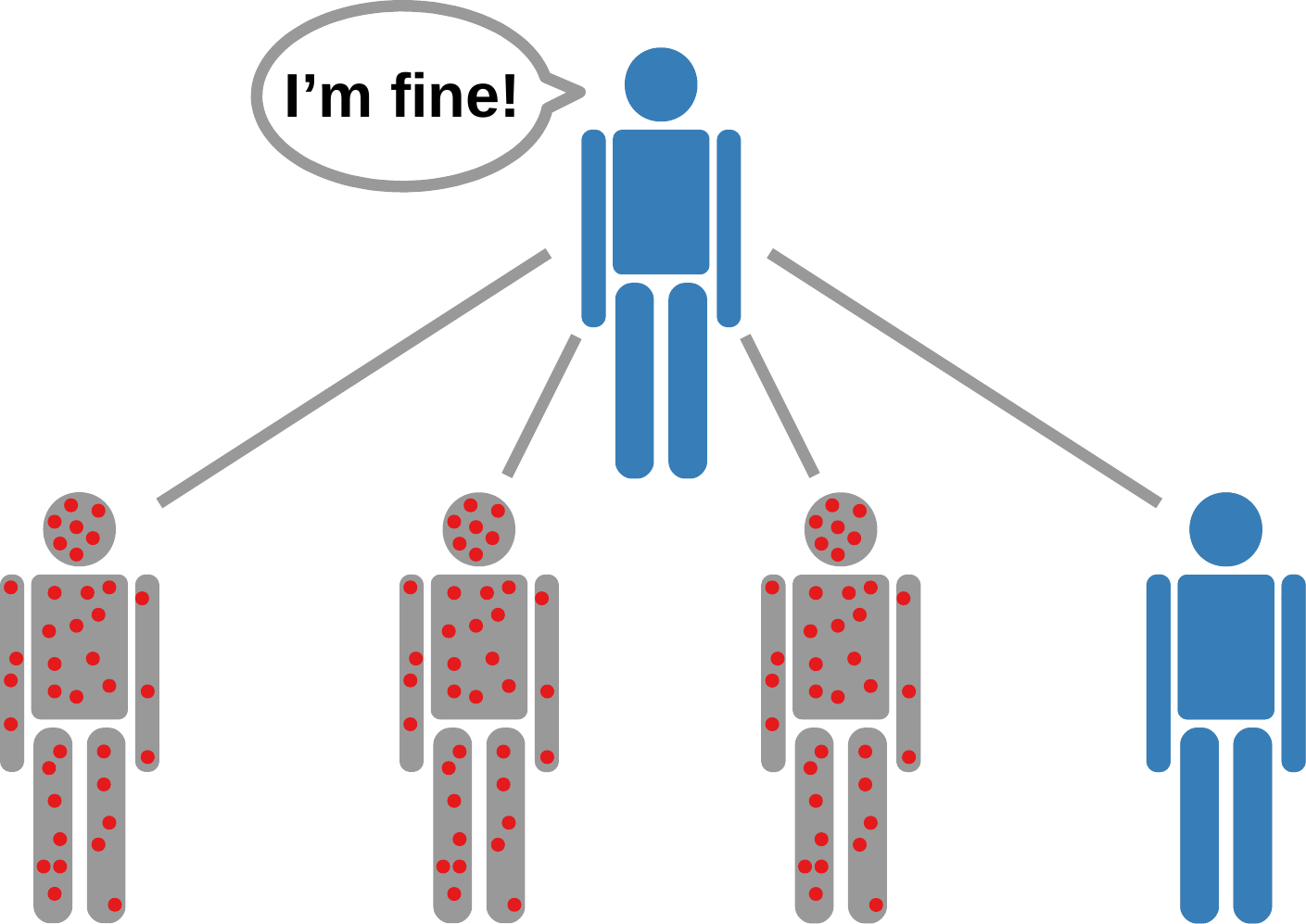}
\caption{An example of threshold contagion. The healthy individual checks how many infected neighbors she has. If they're less than $\kappa$, she's fine. In this example, $\kappa = 4$.}
\label{fig:complex-threshold}
\end{figure}

Figure \ref{fig:complex-threshold} shows a vignette of this process. If we were in classical contagion, the hub at the top would toss three coins with $\beta$ chance of getting infected at each check. In threshold contagion, since $\kappa = 4$, the probability of her being infected is zero. Threshold models usually find an easy time to infect hubs, because we usually set $\kappa$ to be low. Any hub will have more infected friends than that. Any $\kappa > 1$ renders peripheral nodes safe, since most of them have only one connection.

The threshold model is where epidemiology starts to blend in with sociology. Rather than modeling the spread of a virus, the threshold assumption works best when explaining the spread of a behavior. The assumption is that individuals' behavior depends on the number of other individuals already engaging in that behavior\cite{granovetter1983threshold}. This can be used to explain racial segregation\cite{granovetter1988threshold} and customer demand\cite{granovetter1986threshold}. We're going to dive in deep on the racial segregation angle when we'll deal with homophily in social networks in Chapter \ref{cha:homophily}. The customer demand angle explains why variations of the threshold models are one of the favorite instruments of researchers involved in studying viral marketing. We'll see more of that later on in this chapter.

You can spice up the threshold model by allowing $\kappa$ to be a node-dependent parameter, rather than a global one. This means that each node $v$ has a different $\kappa_v$ activation threshold. Some might be convinced to change their behavior by a single individual contact. Or, to keep our epidemic metaphor, they might have a weak immune system, prone to concede defeat to the disease after the first exposure. Others require a high $\kappa_v$: their defining characteristic is being stubborn, whether in their head or in their antibodies. When it comes to social behavior, individuals' thresholds may be influenced by many factors: social economic status, education, age, personality, etc.

\subsection{Cascade}
The cascade model is a straightforward variant of the threshold model, in fact they were developed together by the same researchers. However, their differences are important enough to mention them separately. In the cascade model you also need reinforcement from more than one neighbor to transition to the $I$ state. However, while in the threshold model this was governed by an \textit{absolute} parameter $\kappa$, here we use a \textit{relative} one.

In other words, in the cascade model you need a fraction of your neighbors to be infected in order for you to be infected\cite{watts2002simple}. In the cascade model, the size of your neighborhood influences your likelihood to transition. In the threshold model it didn't: whether you have one or one hundred neighbors, you always need the same $\kappa$ number of them in the $I$ state to consider transitioning.

So if you have four friends, but you need a fraction $\beta > .75$ to transition, if only one, two, or three of them are infected you're fine. In such a condition, you need all of your neighbors to be infected in order to get sick. Only when the fourth neighbor is infected you'll be triggered. Here, I use the same notation $\beta$ I had in the classical contagion, but note that its meaning is slightly different. $\beta$ is not the probability of infection given a contact, but the share of neighbors in set $I$ required for you to transition to $I$. If in classical contagion a low $\beta$ means low infection chance, in cascade a low $\beta$ means that it's more likely to get infected, as you need fewer neighbors to make you transition.

Why do I separate the cascade and the threshold model? Because of hubs. We saw that, in the threshold model, infecting hubs was easy. Since they have lots of connections, the likelihood of them having at least $\kappa$ infected friends is high. In the cascade model the opposite holds. It's harder to infect hubs in a cascade than in a threshold model, because -- for a hub -- the $\beta$ fraction of neighbors required to be infected usually includes hundreds or thousands of nodes. So, in a threshold model, hubs are the primary spreaders of the disease. In a cascade model, they're the last bastion of defense. Once the hubs fall, there's no more chance for salvation.

Again, you're allowed to vary $\beta$ to account for the heterogeneity of gullibility. With proper, rather complicated, fine tuning of $\kappa_v$ in the threshold model and $\beta_v$ in the cascade model, you can render them equivalent. However, it's useful to know that there is a simple way to model contagion in the two cases where you want to simulate a high or low resistance of hubs to the new spreading behavior.

\begin{figure}
\centering
\begin{subfigure}{.31\columnwidth}
\includegraphics[width=\textwidth]{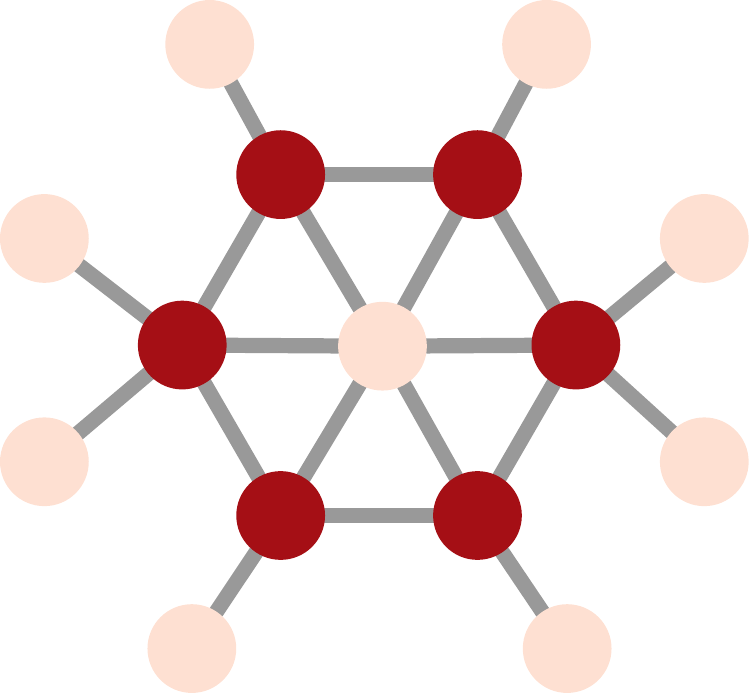}
\caption{$\phi = 0$}
\end{subfigure}\quad
\begin{subfigure}{.31\columnwidth}
\includegraphics[width=\textwidth]{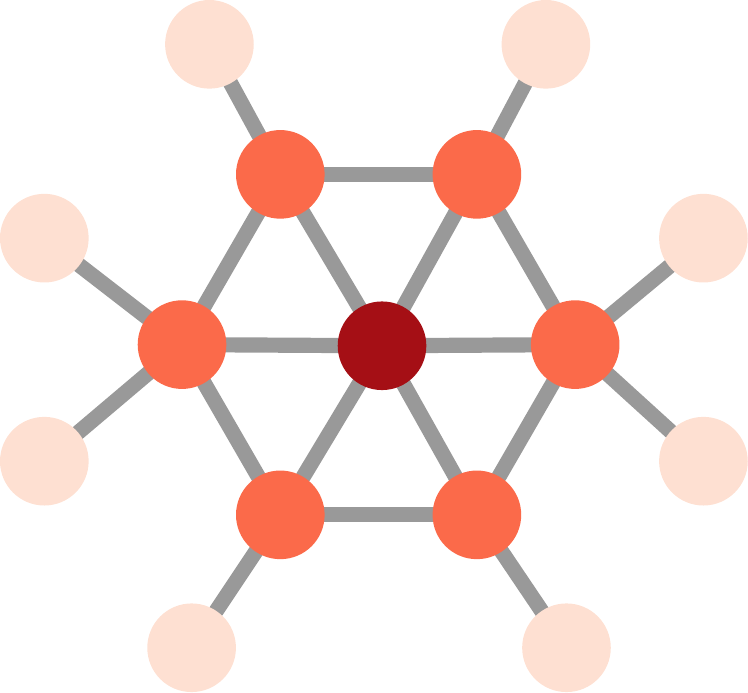}
\caption{$\phi > 0$}
\end{subfigure}\quad
\begin{subfigure}{.31\columnwidth}
\includegraphics[width=\textwidth]{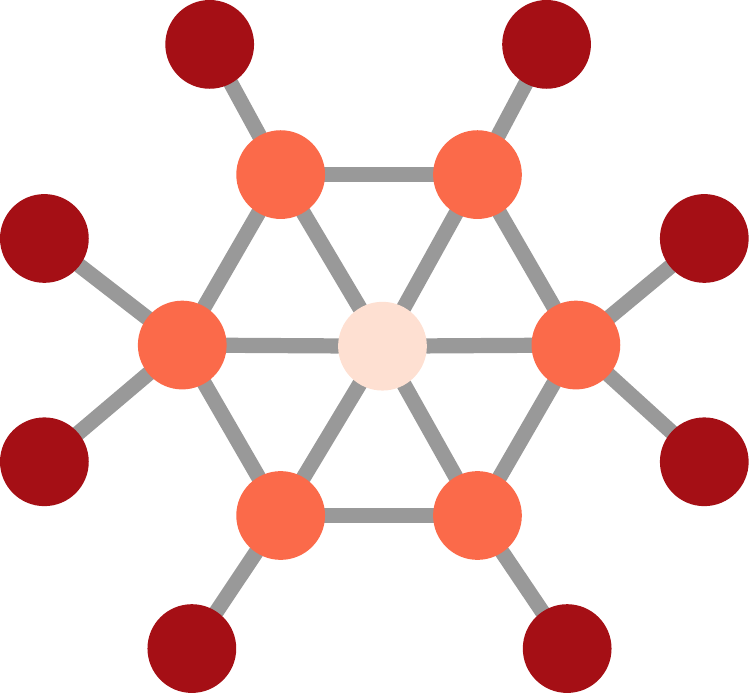}
\caption{$\phi < 0$}
\end{subfigure}
\caption{The three classes of complex contagion as regulated by the $\phi$ parameter. The node color represent the activation time of the node from dark (early) to bright (late).}
\label{fig:netgps}
\end{figure}

Separating the cascade and threshold models in different compartments would make you think they obey completely different rules and they are just different phenomena. This needs not to be the case. The separation is mostly done out of a pedagogical need. In fact, there is a universal model of spreading dynamics concentrating on hubs\cite{barzel2013universality}. We don't need to delve deep into the details on how this model works, but it mostly hinges on a parameter: $\phi$. $\phi$ determines the interplay between the degree of a node and its propensity of being part of the epidemics. If $\phi = 0$, the likelihood of contagion of the node is independent with its degree. If $\phi
 > 0$ we are in the threshold scenario: hubs have a stronger impact on the network. With $\phi < 0$, as you might expect, the opposite holds. Figure \ref{fig:netgps} shows a vignette of the model.

Sprinkling a bit of economics into the mix, you can relate the threshold or the cascade parameter with the utility an actor $v$ gets from playing along or not. Each individual calculates their cost and benefit from undertaking or not undertaking an action. There is a cost in adopting a behavior before it gets popular, and in not doing so after it did\cite{schelling1973hockey}. Being aware of these effects makes for very effective strategies to make your own decisions while you're in doubt. You can establish a Schelling point which determines whether or not you're going to undertake an action\footnote{\url{https://www.lesswrong.com/posts/Kbm6QnJv9dgWsPHQP/schelling-fences-on-slippery-slopes}}, which effectively means you consciously set your own $\kappa_v$. However, this is getting dangerously close to a weird blend of economics, philosophy, and game theory. If you're interested in learn more, you'd be best served by closing this book and looking elsewhere\cite{gintis2014bounds}.

\section{Limited Infection Chances}\label{sec:triggers-resistance}
So far we have mainly looked at diseases spreading through a network of contacts as a bad thing that we want to minimize. If you want to look at the opposite problem -- how to spread things faster and faster through a social network -- without looking like a sociopath, you need to slightly change the perspective. You can concoct a scenario in which, for instance, you want to sell a product, thus you want people to talk about it and convince each other. In practice, you want them to \textit{infect} themselves with the \textit{idea} that the product is good\cite{domingos2001mining}\cite{wang2011information}.

The obvious strategy would be to target hubs, since they have more connections. However, this heavily depends on your triggering model, and hubs come with a disadvantage. First, by being prominent, hubs are targeted by many things, thus they have a very high barrier to attention. Second, they have many connections: if the triggering mechanism requires reinforcement, most of their connections might not get it, thinning out the intervention. A third and final problem might be that you have only one shot at convincing a person. If you fail, it's game over forever. If a hub fails, you might not have a second shot to get to all their peripheral nodes.

\begin{figure}
\centering
\begin{subfigure}{.35\columnwidth}
\includegraphics[width=\textwidth]{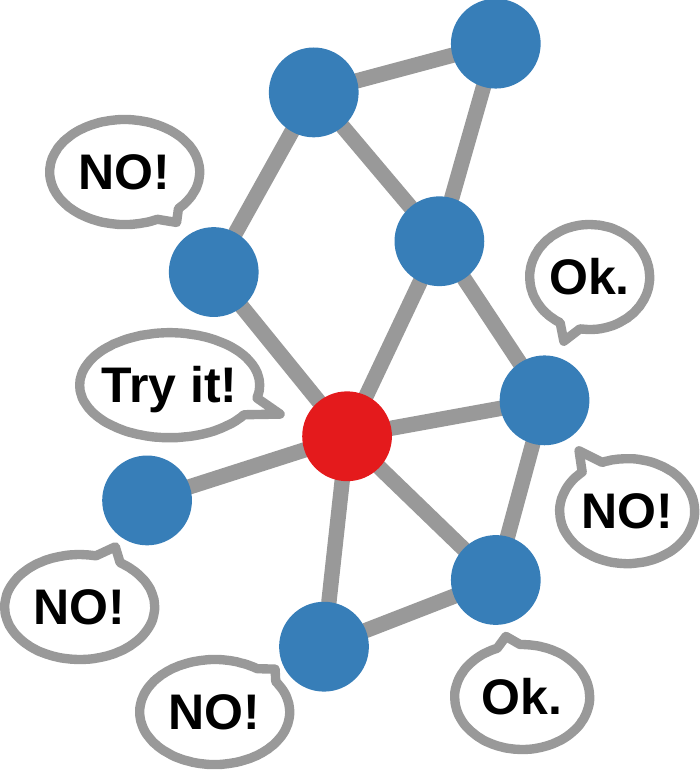}
\caption{$t = 1$}
\end{subfigure}
\begin{subfigure}{.3\columnwidth}
\includegraphics[width=\textwidth]{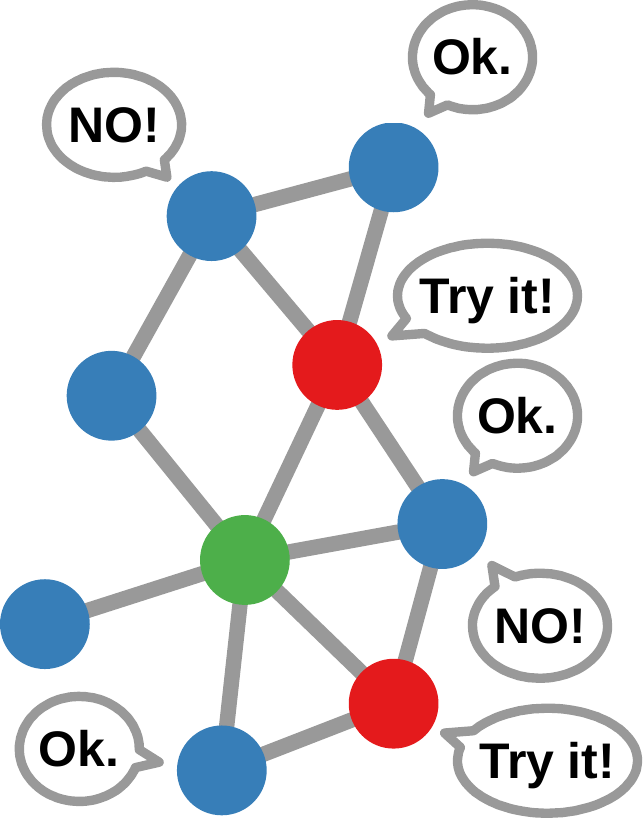}
\caption{$t = 2$}
\end{subfigure}
\begin{subfigure}{.29\columnwidth}
\includegraphics[width=\textwidth]{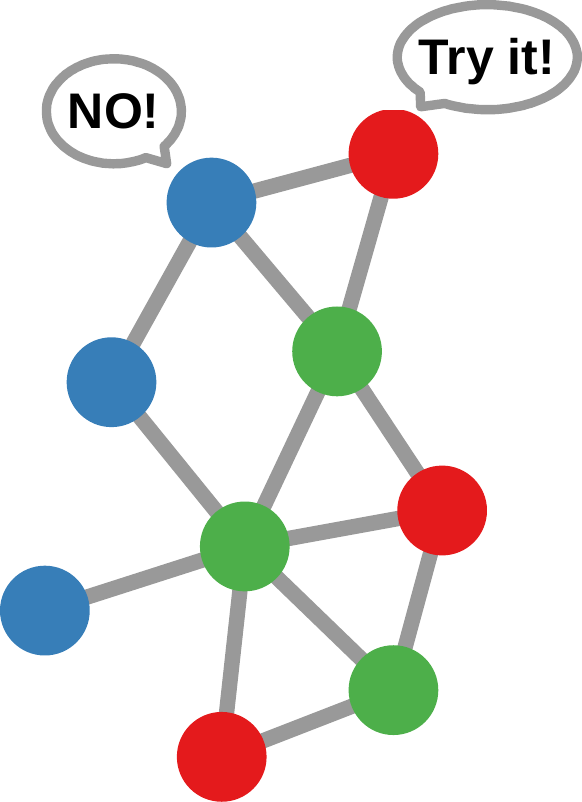}
\caption{$t = 3$}
\end{subfigure}
\caption{An example of independent cascade. The node's state is encoded by the color: blue = susceptible, red = infected and contagious, green = infected but not contagious.}
\label{fig:independent-cascade}
\end{figure}

You can think of this model as a modified SI or SIR, as I show in Figure \ref{fig:independent-cascade}. Suppose that an infected neighbor makes you transition from $S$ to $I$ at time $t$. At the next time step $t + 1$ you will attempt to do the same to your $S$ neighbors. However, at $t + 2$ you transition to a non-contagious stage, where you won't attempt to convert your neighbors any more. You will be in $I$ forever, this is the reason why this is an SI model, but you won't propagate the disease. Or, you could see yourself as transitioning to $R$, with $\mu = 1$: every $i$ individual will always transition to $R$ immediately. The difference is that $R$ individuals are still infected, they just cannot pass the disease.

Note how the node at the bottom in Figure \ref{fig:independent-cascade}(a) resisted the hub at time $t = 1$, but in Figure \ref{fig:independent-cascade}(b) gave up on the second attempt by another node at time $t = 2$. At the same time, the rightmost node has to give two answers because of two independent attempts from its two neighbors. At time $t = 3$ (Figure \ref{fig:independent-cascade}(c)) we see only one persuasion attempt, given that the other infectious nodes are connected to already infected ones. The cascade ends with unconvinced nodes, due to the lack of any possible further move in $t = 4$.

It should be clear by now that, once you tried the first time to convince me to buy a product, any further attempts won't work if you didn't convince me immediately. Maybe another person will persuade me, just not you. This is a crucial difference with regard to SI models. We know that any SI model will eventually fill the entire network. The independent cascade model\cite{goldenberg2001using} won't: the nodes we choose to start the infection with are very important to maximize the reach of our message.

In the simplest model, node $u$ has a probability $p_{u,v}$ of convincing node $v$. However, the past history of attempts to convince $v$ might influence this probability, that's why you should get to hubs when you're the most sure you're going to convince them. So we can modify that probability as $p_{u,v}(S)$, with $S$ being the set of nodes who already tried to influence $v$\cite{kempe2003maximizing}\cite{kempe2005influential}\footnote{Note that, if we only use $p_{u,v}$ we call this \textit{independent} cascade model, because the previous attempts do not influence future attempts. When we introduce $p_{u,v}(S)$ the cascades are not independent any more. Specifically, for the paper I'm citing, we have \textit{decreasing} cascades because, the more people try, the hardest it is to convince $v$, i.e. $p_{u,v}(S) < p_{u,v}(S \cup z)$. If we did the opposite, $p_{u,v}(S) > p_{u,v}(S \cup z)$, then this model would be practically equivalent to the threshold model: the more infected neighbors you have, the more likely you're going to turn.}. The process ends when all infected nodes exhausted all their chances of convincing people so no more moves can happen.

So you get the problem: find the set of cascade initiators $I_0$ such that, when the infection process ends at time $t$, the share of infected nodes in the network $i_t$ is maximized. Kempe et al. solve the problem with a greedy algorithm. We start from an empty $I_0$. Then we calculate for each node its marginal utility to the cascade. We add the node with the largest utility, meaning the number of potential infected nodes, to $I_0$ and we repeat until we reach the size we can afford to infect. Of course, each node we add to $I_0$ changes the expected utility of each other node, because they might have common friends, thus we cannot simply choose the $|I_0|$ nodes with the largest initial utility.

\begin{figure}
\centering
\begin{subfigure}{.3\columnwidth}
\includegraphics[width=\textwidth]{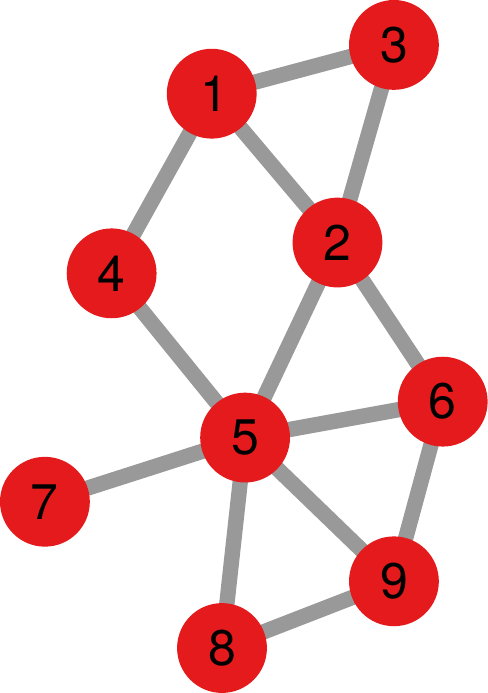}
\caption{}
\end{subfigure}\quad
\begin{subfigure}{.36\columnwidth}
  \begin{tabular}{c|c|c}
    Time & Node & Action \\
    \hline
    $1$ & $1$ & $a_1$\\
    $2$ & $1$ & $a_2$\\
    $2$ & $2$ & $a_1$\\
    $3$ & $1$ & $a_3$\\
    $3$ & $2$ & $a_2$\\
    $3$ & $3$ & $a_1$\\
    $3$ & $6$ & $a_1$\\
  \end{tabular}
\caption{}
\end{subfigure}\quad
\begin{subfigure}{.225\columnwidth}
\includegraphics[width=\textwidth]{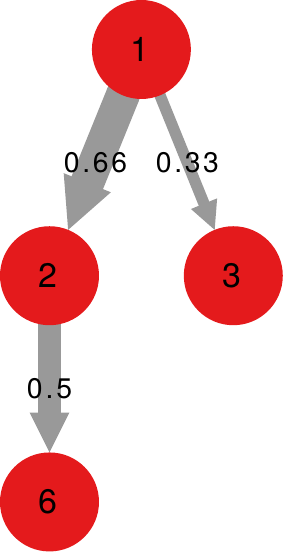}
\caption{}
\end{subfigure}
\caption{(a) The underlying social network. (b) The actions nodes made. (c) A possible inferred influence graph.}
\label{fig:gurumine}
\end{figure}

There are many improvements for this algorithm, focused on improving time efficiency, lowering the expected error, and integrating different utility functions. However, things get more interesting when you start adding metadata to your network. For instance, Gurumine\cite{goyal2010learning} is a system that lets you create influence graphs, as I show in Figure \ref{fig:gurumine}. You start from a social network (Figure \ref{fig:gurumine}(a)) and a table of actions (Figure \ref{fig:gurumine}(b)). You know when a node did what.

You can use the data to infer that node $v$ does action $a_1$ regularly after node $u$ performed the same action. In the example, for two actions $a_1$ and $a_2$ you see node $2$ repeating immediately after node $1$. Since these two nodes are connected, maybe node $1$ is influencing node $2$. You can use that to infer $p_{u,v} = 0.66$ (Figure \ref{fig:gurumine}(c)) -- or, if you're really gallant, to infer $p_{u,v}(S)$ by looking at all neighbors of $v$ performing $a_1$ before it.

Note that node $6$ performed the same action at the same time as node $3$. Node $6$ could only be influenced by node $2$. For node $3$ we prefer inferring that node $1$ did it, because we know that it influenced node $2$ too, so that's the most parsimonious hypothesis. The size in number of nodes of these cascades can be approximated by -- you guessed it -- a power law\cite{bakshy2011everyone}. 

When running such models on real data you can find funny things. For instance, I ran it with some co-authors on LastFm data, a social network recording which user listened to which musical artist at which time\cite{pennacchioli2013three} -- the artist is considered the ``action''. In doing so, we discovered that we could build these influence graphs and describe their trade offs. For instance, the more intensely a user was influenced by a prominent friend -- meaning that they listened the new artist a lot -- the fewer friends the influencer hit. In other words: the stronger you want to influence people, the fewer people you can influence.

\begin{figure}
\centering
\includegraphics[width=.8\columnwidth]{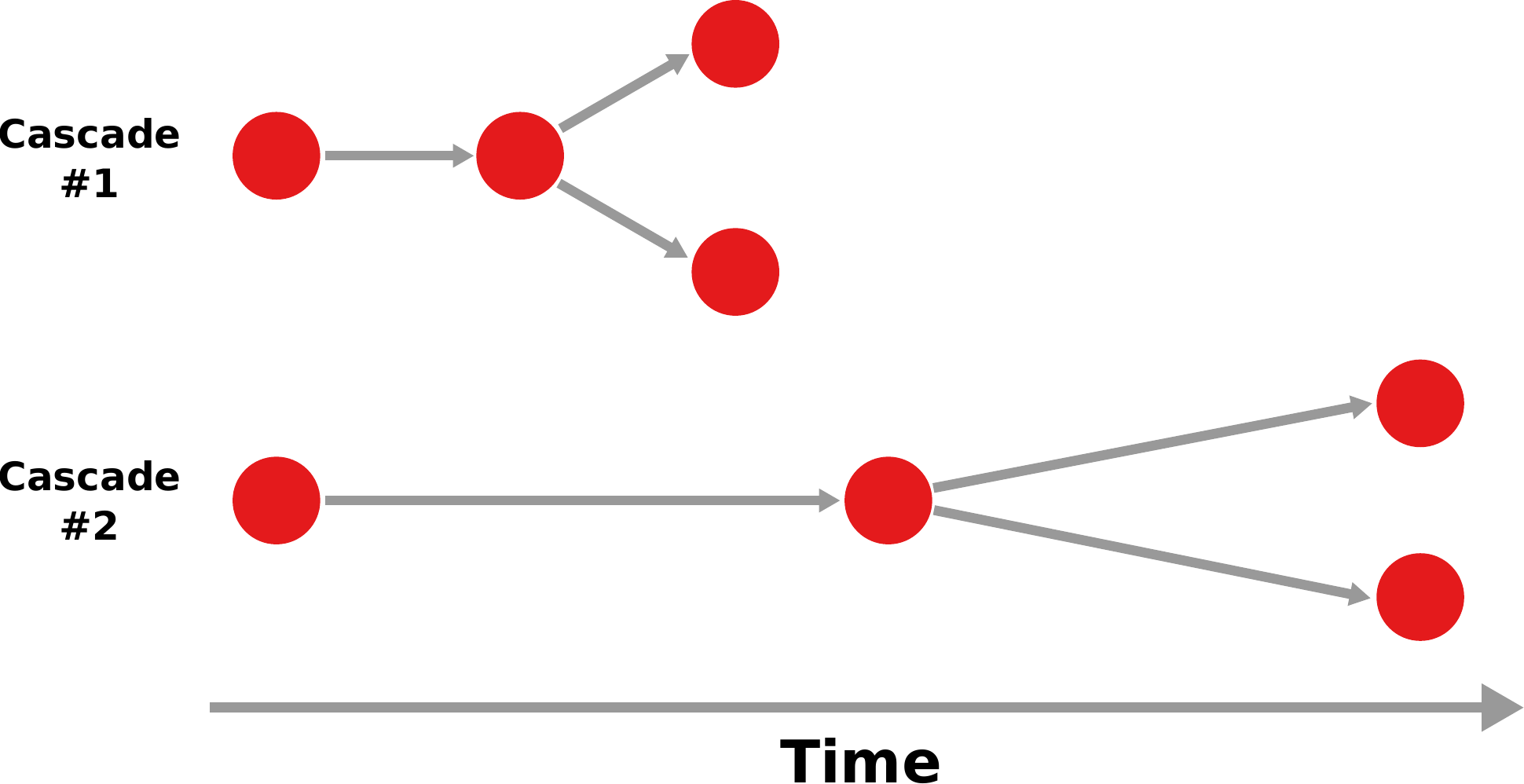}
\caption{Anatomies of two different cascades. Time flows from left to right. A node at a given position on the x axis denotes when they share the content on their profile. An arrow indicates from where they re-shared an item, i.e. who influenced them.}
\label{fig:adamic-cascades}
\end{figure}

Similar studies on Facebook tried to find which early signs we can use to predict the size of a cascade. A cascade is when I share something on my profile, then other people share it too and so on until it hits the news. Counterintuitively, the answer seem to have very little to do with the actual content of the idea per se, but with the speed with which it triggers other people\cite{cheng2014can}\cite{cheng2016cascades}. For instance, in Figure \ref{fig:adamic-cascades} we have two hypothetical cascades with the same number of shares and the same topological pattern: one re-sharer then two. However, the fact that the first cascade happened faster is enough for us to infer that it's much more likely to end up being much larger than the second, slower, one. I'm going to explore more in depth this idea about memes spreading when talking about classical results in network analysis in Chapter \ref{cha:history}.

You can further complicate models by having competing ideas spreding into the network. There are some people who are complete enthusiasts about iPhones, while others really hate them. The love/hate opinions are both competing to spread through the network. You can see them, for instance, as a physical heating/cooling process which will eventually make nodes converge to a given temperature\cite{ma2008mining}. The classic survey of viral marketing applications of network analysis\cite{leskovec2007dynamics} is a good starting point for diving deeper into the topics only skimmed in this section.

\section{Interventions}\label{sec:triggers-intervention}
Once we have a disease spreading through a social network, we might be interested in using our knowledge to prevent people to become sick. In practice, if this were a SIR model, we want to flip some people directly from the $S$ to the $R$ state, without passing by $I$. This is equivalent to vaccinate them and, if done properly, would stop the epidemics in its tracks. You can try an online game with this premise and see how much of a network you can save from an evil disease\footnote{\url{https://github.com/digitalepidemiologylab/VaxGame}}.

The first question is: who should we vaccinate? The answer is rather obvious once you run your simulation numbers: the hubs. If the disease attacks the hubs, it will spread to the entirety of the network almost instantly. This assumes that its degree exponent is $\alpha < 3$ and we know that, unfortunately, this is true for the majority of social systems we know.

However, now we have a second question: how do we find hubs? We might not have a complete -- or even a partial! -- picture of our social network. Luckily, this book has prepared you to figure out a way to find hubs even if you know nothing about the network's topology. You can exploit the fact that hubs have lots of connections. The simplest and unreasonably effective vaccination strategy is to pick a node at random in the network and vaccinate one of its friends\cite{cohen2003efficient}. Statistically speaking, the friend of our random sampled individual is more likely to have a higher degree than our first choice.

\begin{figure}
\centering
\includegraphics[width=.66\columnwidth]{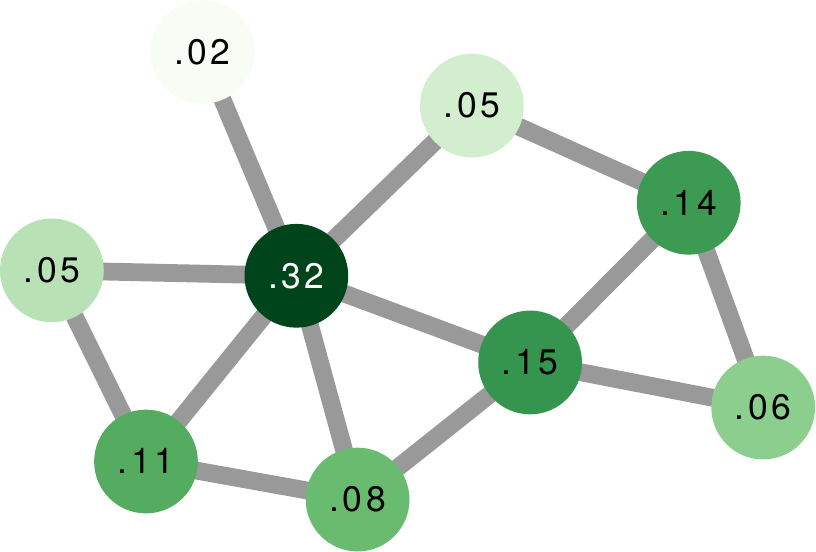}
\caption{The probability of vaccinating a node with the ``vaccinate-a-friend'' strategy.}
\label{fig:immunization}
\end{figure}

To see why, consider Figure \ref{fig:immunization}. Here we apply the ``vaccinate-a-friend'' strategy and report the probability of choosing each node. Note that this is done completely blindly, we don't know anything about the topology of this network. If we were to vaccinate the randomly sampled node, we would have only one chance out of nine to find the hub, given that the example has nine nodes. However, the probability of vaccinating the hub with our strategy is almost three times as high. This is related to a curious network effect on hubs, known as the ``Friendship Paradox'', which we'll investigate further in Section \ref{sec:assortquant-paradox}.

Of course, this strategy makes a number of assumptions that might not hold in practice. For instance, we only consider a simple SIR model,  without looking at the possibility of complex contagion. Luckily, there is a wealth of research relaxing this assumption and proposing ad hoc immunization strategies that can work in realistic scenarios\cite{chen2015node}. One of the most historically important approaches in this category is Netshield\cite{tong2010vulnerability}.

\begin{figure}[t]
\centering
\includegraphics[width=.8\columnwidth]{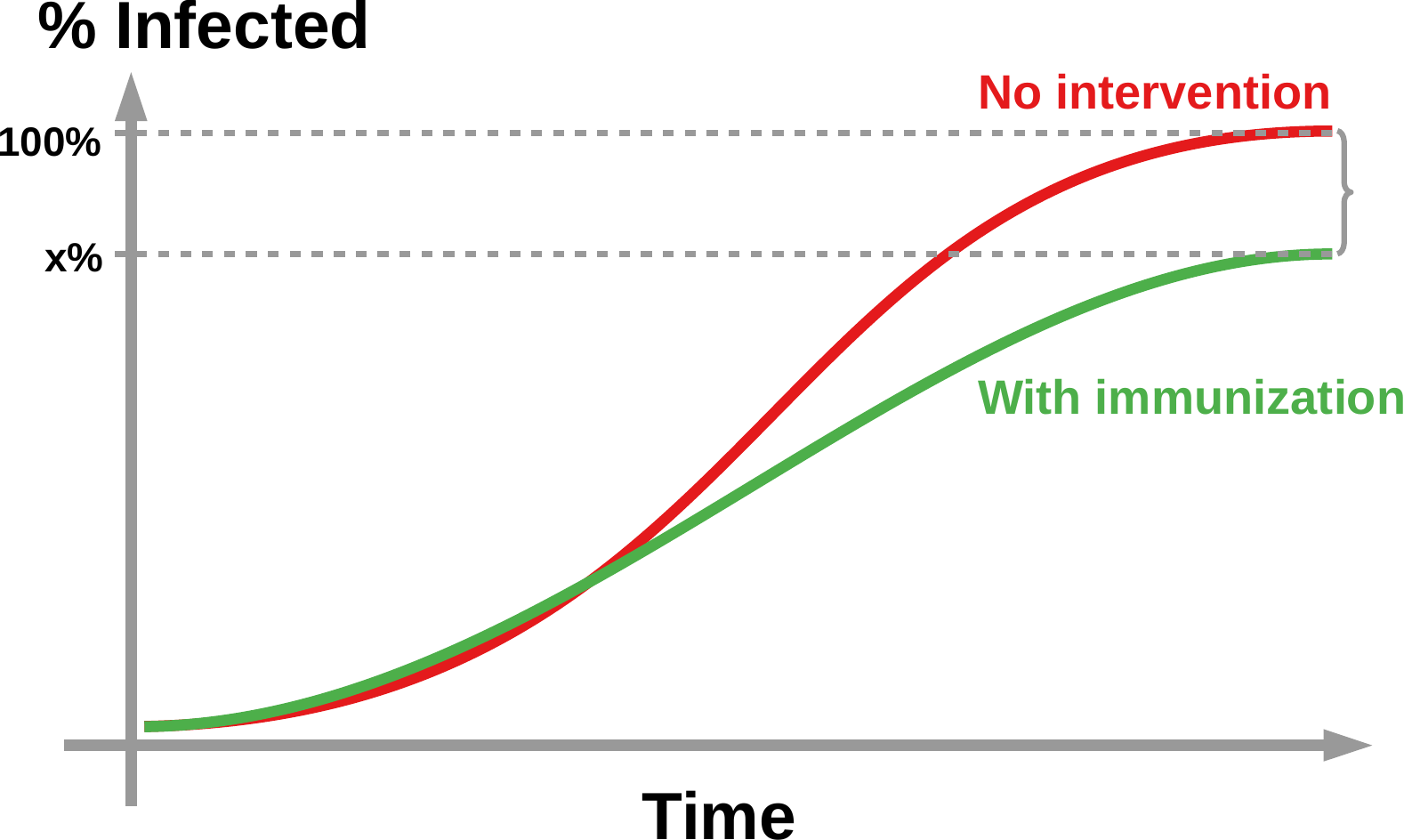}
\caption{The first criterion of immunization success: the share of infected nodes at the end of the outbreak is lower than $100\%$ in a SI model.}
\label{fig:immunization2}
\end{figure}

How do we know if we did a good job? How can we evaluate the impact of an intervention? There are two things we want to look at. First, we look at the size of the final infected set and simply subtract the predicted infected share without immunization with the one with immunization. The higher the difference the better. Figure \ref{fig:immunization2} gives you a sense of this. An SI model without immunization reaches saturation when all nodes are infected. A smart immunization strategy can make sure that the outbreak stops at a share lower than $100\%$.

\begin{figure}[t]
\centering
\includegraphics[width=.8\columnwidth]{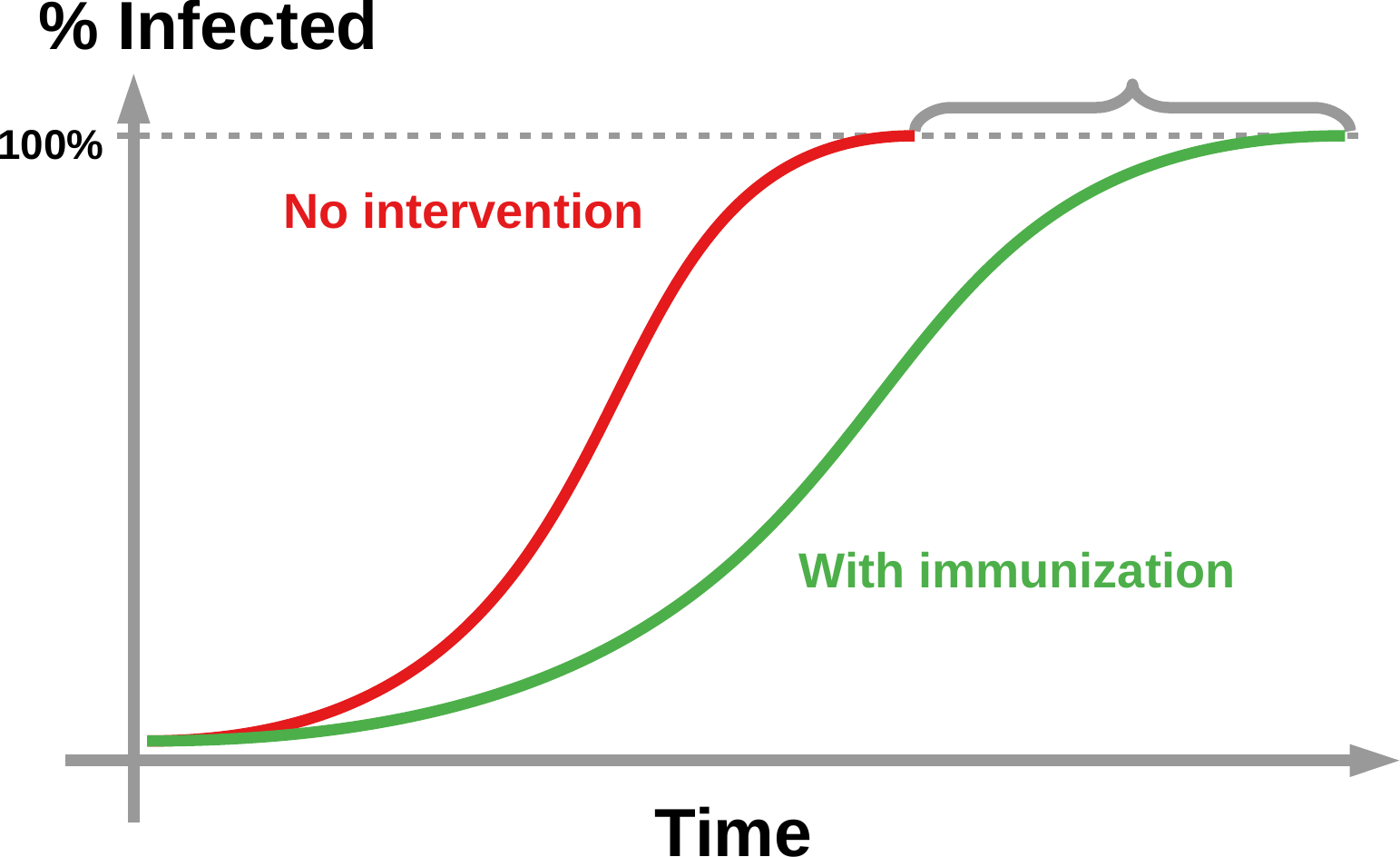}
\caption{The second criterion of immunization success: temporary immunity can delay propagation.}
\label{fig:immunization3}
\end{figure}

A second criterion might be just delaying the inevitable. Once immunized, the nodes can revert to the $S$ state, and therefore to $I$, after a certain amount of time. This time can be used to develop a real vaccine or might be a feature in itself, preventing having too many people transitioning to $I$ at the same time. Figure \ref{fig:immunization3} provides an example. In this case, we might want to either calculate the time $t$ at which the system reaches saturation, or compute the area between the two curves as a more precise sense of the delay we imposed.

We can combine the two criteria at will. By immunizing nodes, we make the disease unable to reach saturation at $100\%$ infection AND we delay its spread in the network. Thus the two scenarios are not mutually exclusive.

Obviously, here I only assumed the perspective of limiting the outbreak of a disease. If you're in the viral marketing case you can invert the perspective: your interventions wants to favor the spread of the idea in the social network. In this case, the second scenario makes more sense: even if the idea was bound to reach everyone eventually, if it does so faster it can have great repercussion. Think about the scenario of condom use to prevent HIV infections. You want to convince as many people as fast as you can, even if eventually your message was going to reach everybody anyway.

\section{Controllability}\label{sec:triggers-control}
A related problem is the classical scenario of the controllability of complex networks\cite{liu2011controllability}\cite{gao2014target}\cite{yan2015spectrum}. Here the task is slightly different: nodes can change their state freely and there can be an arbitrary number of states in the network. What we want to ensure is that all -- or most -- nodes in the network end up in the state we desire. To do so, we need to identify driver nodes: the smallest possible subset of nodes we have to manipulate so that they will influence the other nodes to switch to the state we want them to assume.

There already is a branch of mathematics dedicated to figure out how to control simple engineered or natural systems, unoriginally named control theory\cite{lee1967foundations}\cite{francis1987course}. However, we define complex systems exactly on their nature of being difficult to predict, as their parts interact with each other and thus let non-obvious properties and behaviors emerge.

\begin{figure}
\centering
\begin{subfigure}{.25\columnwidth}
\includegraphics[width=\textwidth]{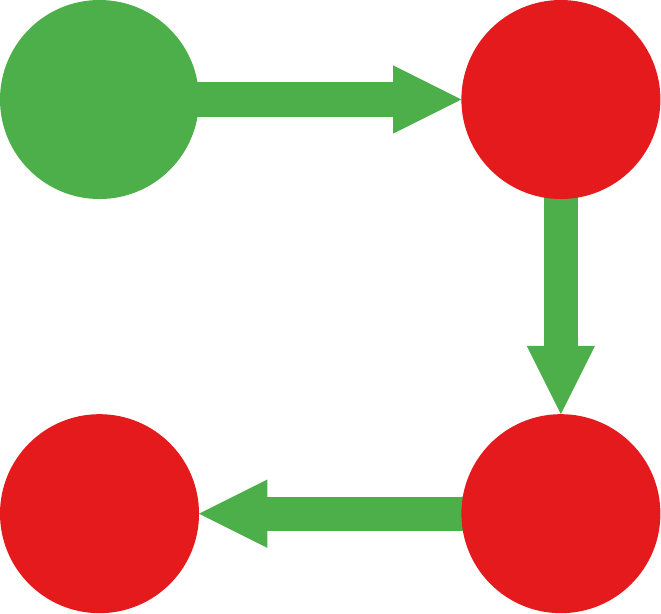}
\caption{}
\end{subfigure}\qquad
\begin{subfigure}{.25\columnwidth}
\includegraphics[width=\textwidth]{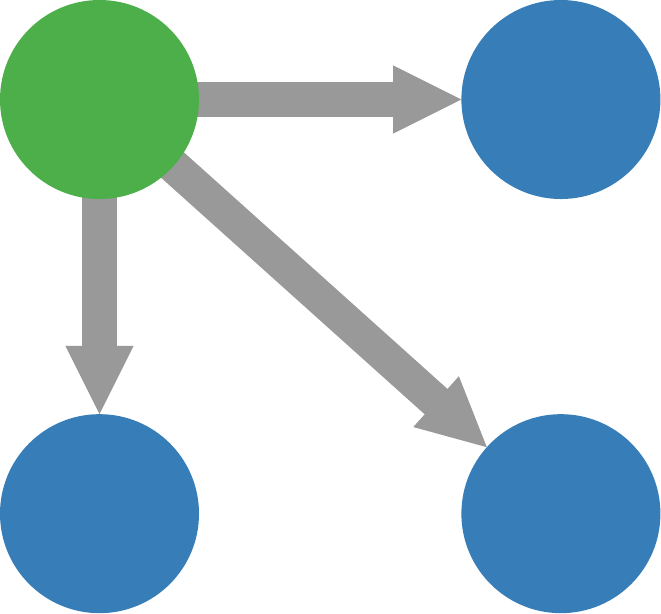}
\caption{}
\end{subfigure}\qquad
\begin{subfigure}{.25\columnwidth}
\includegraphics[width=\textwidth]{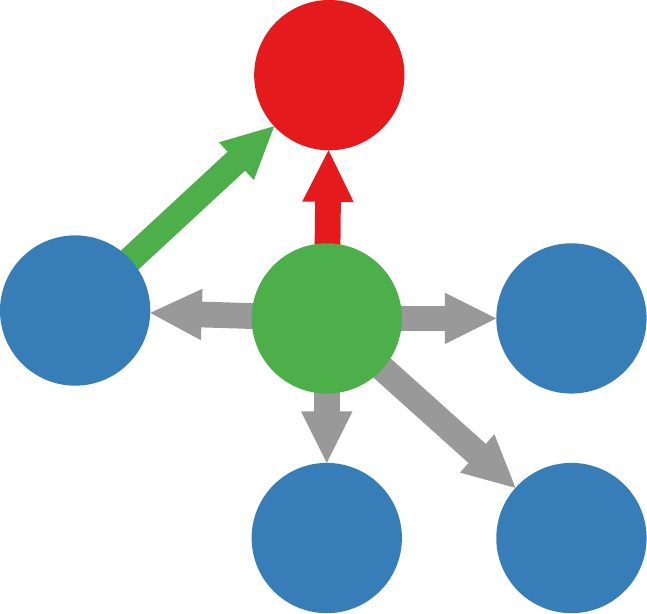}
\caption{}
\end{subfigure}
\caption{Different controllability scenarios. The nodes we're forced to select as drivers are in green, the ones we could or could not choose are in blue, the ones we don't need as drivers are in red. Similarly for links, the links we need to have to ensure controllability are in green, the ones we could or could not have are in gray, and the ones we could remove from the network without hampering controllability are in red.}
\label{fig:controllability}
\end{figure}

In complex systems, controllability is a bit more complicated. Figure \ref{fig:controllability} shows a few simple examples. In Figure \ref{fig:controllability}(a), a chain, we only need to control the origin of the chain and the rest of the system will fall into place. In Figure \ref{fig:controllability}(b), somewhat surprisingly, one needs to control at least two among the blue nodes besides the hub in green to ensure control the system. One also needs to control three of the four blue nodes in Figure \ref{fig:controllability}(c). Unfortunately, the mathematical details to reach this conclusion are beyond the scope of this book, and I invite you to read the papers cited at the beginning of this section if you're interested in them.

As you might expect from a complex network paper, the final conclusion is that sparse networks with a power law degree distribution characterized by a low $\alpha$ exponents are extremely difficult to control: they require a large number of driver nodes. In another twist going against our intuition, driver nodes tend not to be hubs. You would expect nodes connecting to the majority of the network to be natural choices to control the system, yet it seems that peripheral nodes have their role.

There are numerous applications of controllability in complex systems, for instance in networks modeling the brain\cite{power2011functional}.

\section{Summary}

\begin{enumerate}
\item In simple contagion at each timestep you have the same chance of getting infected if you have one or more infected neighbors. In complex contagion more infected neighbors reinforce the infection chances. Different models work with different triggering mechanisms.
\item Classically, you can have an independent $\beta$ probability to transition for each infected neighbor. In the threshold model you need at least $\kappa$ neighbors, independently of your degree. In the cascade model you need at least a fraction of neighbors.
\item This changes the behavior of hubs: in the threshold model it is easy to have at least $\kappa$ infected contacts because hubs have so many neighbors, but for the very same reason it is difficult to reach the relative limit in the cascade model.
\item You can estimate which type of infection model a real world outbreak follows by estimating the universality class of the spreading, via its parameter $\phi$: $\phi > 0$ is similar to a threshold model (positive correlation between degree and chance of infection), $\phi < 0$ is similar to a cascade model (negative correlation between degree and chance of infection).
\item In viral marketing models of word-of-mouth you don't have infinite chances to infect a node. The problem becomes identifying the set of initial infected seeds so that you maximize the number of infected nodes in the network.
\item One effective strategy to prevent a global outbreak is to immunize the friends of randomly chosen nodes. This strategy works because the randomly picked neighbors of randomly picked nodes are more likely to be hubs.
\end{enumerate}

\section{Exercises}

\begin{enumerate}
\item Modify the SI model developed in the exercises of the previous chapter so that it works with a threshold trigger. Set $\kappa = 2$ and run the threshold trigger on the network at \url{http://www.networkatlas.eu/exercises/21/1/data.txt}. Show the curves of the size of the $I$ state for it (average over $10$ runs, each run of $50$ steps) and compare it with a simple (no reinforcement) SI model with $\beta = 0.2$.
\item Modify the SI model developed in the previous exercise so that it works with a cascade trigger. Set $\beta = 0.1$ and compare the $I$ infection curves for the three triggers on the network used in the previous exercise (average over $10$ runs, each run of $50$ steps).
\item Modify the simple SI model so that nodes become resistant after the second failed infection attempt. Compare the $I$ infection curves of the SI model before and after this operation on the network used in the previous exercise, with $\beta = 0.3$ (average over $10$ runs, each run of $50$ steps).
\item Run a classical SIR model on the network used in the previous exercise, but set the recovery probability $\mu = 0$. At each timestep, before the infection phase pick a random node. Pick one random neighbor in status $S$, if it has one, and transition it to the $R$ state. Compare the $I$ infection curves with and without immunization, with $\beta = 0.1$ (average over $10$ runs, each run of $50$ steps).
\end{enumerate}

\chapter{Catastrophic Failures}\label{cha:epidemapps}
In Section \ref{sec:triggers-resistance} we saw an interesting thing: when limiting the infection chances nodes get, the disease might be unable to reach some parts of the network. This is great when fighting a disease, but networks model much more than social systems hosting pathogens. The roads you use every day for your commute are part of a network. The power grid is a network. The beloved cat pictures you look at every day flow through edges of the interwebz. The fact that something might prevent them to reach you is alarming and deserves to be studied.

\begin{figure}
\centering
\begin{subfigure}{.3\columnwidth}
\includegraphics[width=\textwidth]{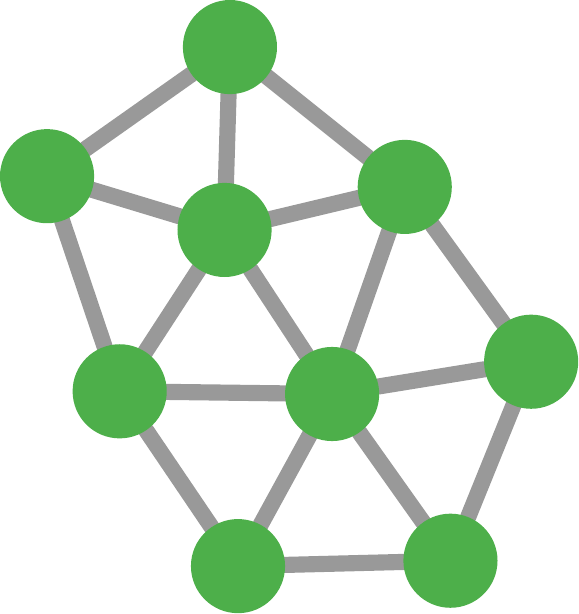}
\caption{}
\end{subfigure}\quad
\begin{subfigure}{.3\columnwidth}
\includegraphics[width=\textwidth]{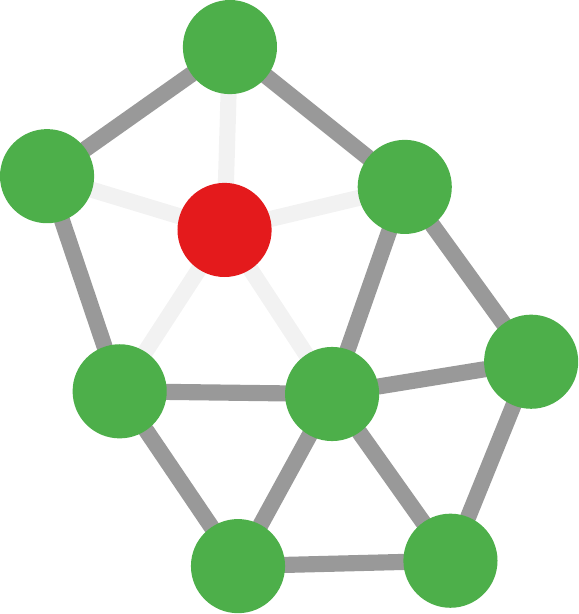}
\caption{}
\end{subfigure}\quad
\begin{subfigure}{.3\columnwidth}
\includegraphics[width=\textwidth]{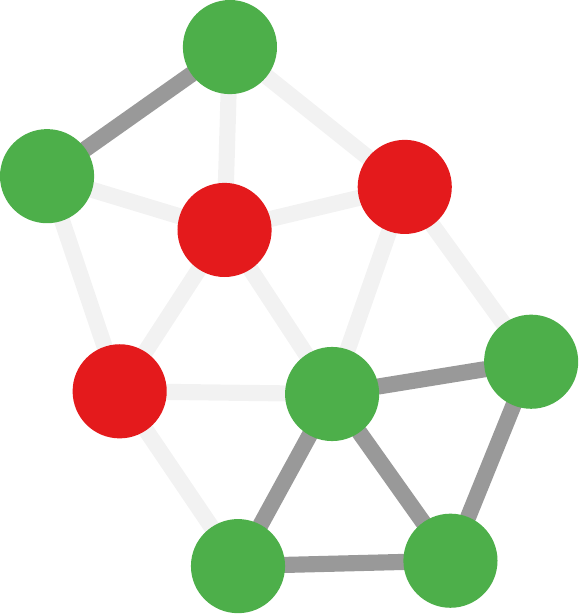}
\caption{}
\end{subfigure}
\caption{The effect of node failures on the connectivity of a network. Node color: green = active; red = failing. (a) Starting condition, all nodes active. (b) First failure. (c) Propagating failure disconnects the two nodes on top from the main component.}
\label{fig:robustness}
\end{figure}

In this chapter we do exactly that. We again slightly change the perspective of our epidemic model to study the conditions under which networks break down. Rather than propagating a disease, we propagate failures. In their standard status, nodes are active and fulfill their duties. See Figure \ref{fig:robustness}(a) for an example.  However, for random or deliberate reasons they might transition into an $R$ status: they might fail. The fundamental question is: how does the network react to such failures? Can information still flow through the Internet if some routers go down? How many blocked roads does it take for cars not to be able to drive around town?

Networks are usually resilient to small failures: a single node going down does not affect communication in the structure (see Figure \ref{fig:robustness}(b)). Our criterion to say whether a network is still fulfilling its purpose is the share of nodes part of its largest component. If there still is a path between all or most nodes in the network, even if it becomes longer, the network still works. However, when nodes start breaking down in multiple components and getting isolated -- as in Figure \ref{fig:robustness}(c) -- then the network is failing.

We start by looking at random failures in Section \ref{sec:epidemapps-random} to move then to deliberate attacks in Section \ref{sec:epidemapps-targeted}. We then put some dynamics on failures by considering correlated cascade failures in Section \ref{sec:epidemapps-cascade}, giving a special attention to a specific case of multilayer structures: interdependent networks (Section \ref{sec:epidemapps-interdependent}).

This chapter is related to the mathematical problem of percolation theory\cite{stauffer2014introduction}, which has then been adapted to the network scenario\cite{albert2000error}\cite{crucitti2004error}\cite{gao2016universal}. Note that I'm using the power grid example mostly as a way to give color to the math -- and for traditional reasons. However, power grids failures don't necessarily follow such percolation approach. The model here propagates failures linearly and locally, but real failures in power lines are neither, due to the underlying laws governing electrical flows.

You can use these methods in other scenarios, but only if you're sure that the assumptions made here are respected in the phenomenon you're studying.

\section{Random Failures}\label{sec:epidemapps-random}
In this section we look at random failures. In this case, nodes can spontaneously break for uncorrelated and not deliberate reasons. Think about normal wear and tear. Any power generator can only take so much. Moreover, slight differences in the manufactory process, or in the model, can give different failure rates. Thus it is difficult to predict when one component will fail. The error rate will appear to be more or less random.

How does a network respond to these random failures? We're assuming that all its nodes are in the same Giant Connected Component (GCC), so that power can flow freely through the grid. When will the network lose its giant component? In other words: when will we need to rely on local generators rather than on the entire grid?

\begin{figure}
\centering
\includegraphics[width=.8\columnwidth]{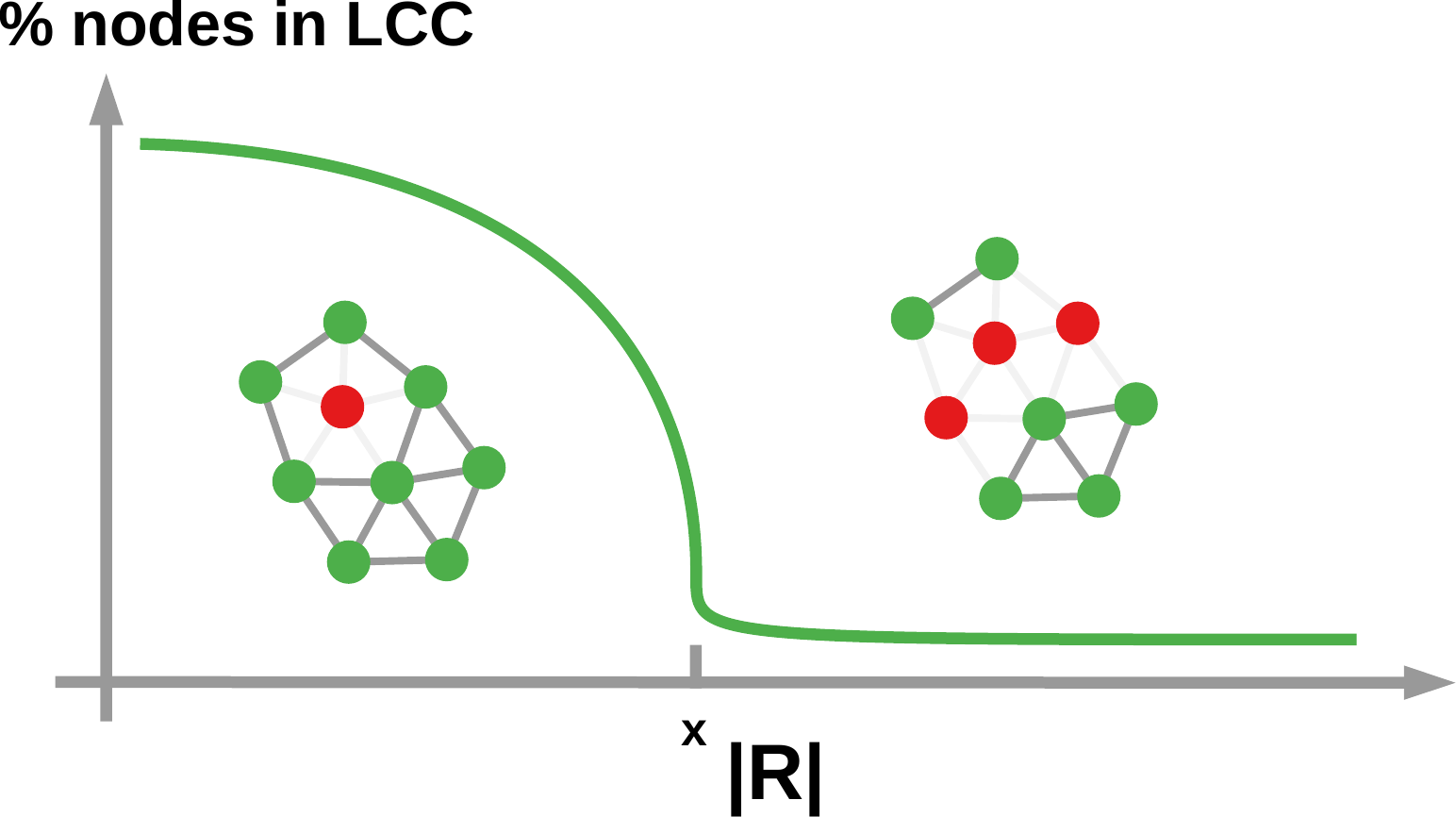}
\caption{The probability of being part of the largest connected component as a function of the number of failing nodes in a random $G_{n,p}$ graph.}
\label{fig:rndfail-rndgraph}
\end{figure}

The answer depends on the original topology of the network. Let's start by considering the case of a random $G_{n,p}$ network. What we want to see is how much the probability of a node being part of the GCC changes as we put more and more nodes in the failure state $R$. I ran a few simulations and they generate a plot similar to what Figure \ref{fig:rndfail-rndgraph} shows.

Now, if this seems the very same plot as one you've already seen, calm down: you're not taking crazy pills. You have indeed seen something like this. It was back to Figure \ref{fig:random-gcc}(b), when we were talking about the probability of a node being part of the GCC in a $G_{n,p}$ model. In that case, the function on the x-axis was the probability $p$ of establishing an edge between two nodes. In fact, the two are practically equivalent: if you have a $G_{n,p}$ graph with failures it is as if you're manipulating $n$ and $p$.

What Figure \ref{fig:rndfail-rndgraph} says is that a $G_{n,p}$ network will withstand small failures: a few nodes in $R$ will not break the network apart. However, the failure will start to become serious very quickly, until we reach a critical value of $|R|$ beyond which the GCC disappears and the network effectively breaks down. Just like the appearance of a GCC for increasing $p$ in a $G_{n,p}$ model is not a gradual process, so are random failures. At some point there is a \textit{phase transition}, from having to not having a GCC.

\textit{Ah} -- you say -- \textit{but we're not amateurs at this. Who would engineer a random power grid network? For sure it won't be a $G_{n,p}$ graph}. Good point. In fact that's true: the power grid's degree distribution is skewed. For the sake of the argument -- and the simplicity of the math -- let's check the resilience to random failures of a network with a power law degree distribution.

\begin{figure}
\centering
\includegraphics[width=.8\columnwidth]{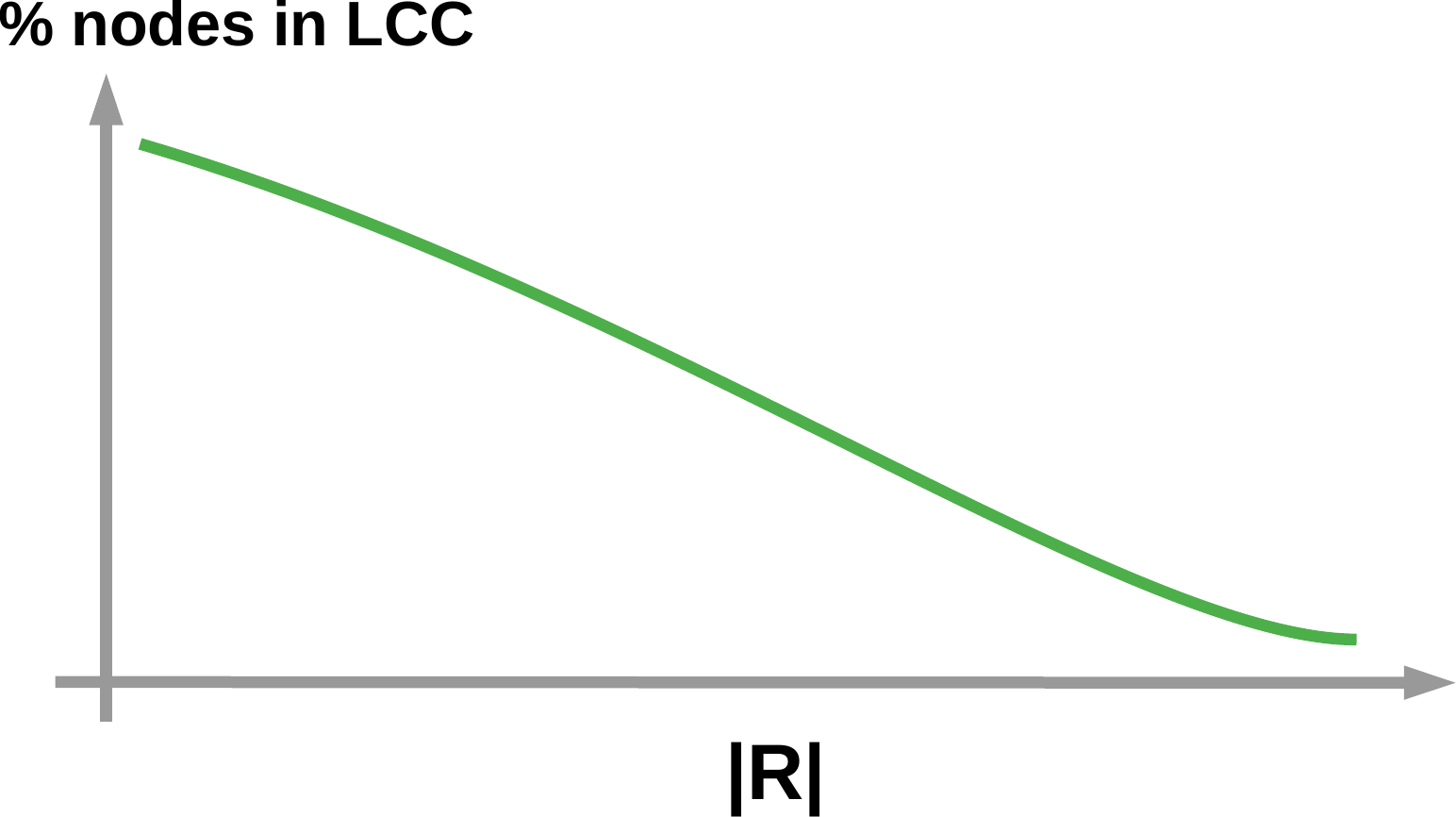}
\caption{The probability of being part of the largest connected component as a function of the number of failing nodes in a network with a skewed degree distribution.}
\label{fig:rndfail-scalefree}
\end{figure}

Good news everybody! Power law random networks are more resilient than $G_{n,p}$ networks to random failures. The typical signature of a power law network under random node disappearances looks something like Figure \ref{fig:rndfail-scalefree}. In the figure you see no trace of the phase transition. The critical value under which the GCC disappears is much higher than in the $G_{n,p}$ case. Of course the size of the largest connected component goes down, because you're removing nodes from the network. However, the nodes that remain in the network still tend to be able to communicate to each other, even for very high $|R|$.

Why would that be the case? The reason is always the power law degree distribution. If you remember Section \ref{sec:degree-pl}, having a heavy tailed degree distribution means to have very few gigantic hubs and a vast majority of nodes of low degree. When you pick a node at random and you make it fail, you're overwhelmingly more likely to pick one of the peripheral low degree ones. Thus its impact on the network connectivity is low. It is extremely unlikely to pick the hub, which would be catastrophic for the network's connectivity.

Since, by now, you must be a ninja when it comes to predict the effect of different degree exponents on the properties of a network with a power law degree distribution, you might have figured out what's next. The exponent $\alpha$ is related to the robustness of the network to random failures. An $\alpha = 2$, remember, means that there are fewer hubs and their degree is higher. If $\alpha > 3$, the hubs are more common and less extreme.

\begin{figure}
\centering
\includegraphics[width=.8\columnwidth]{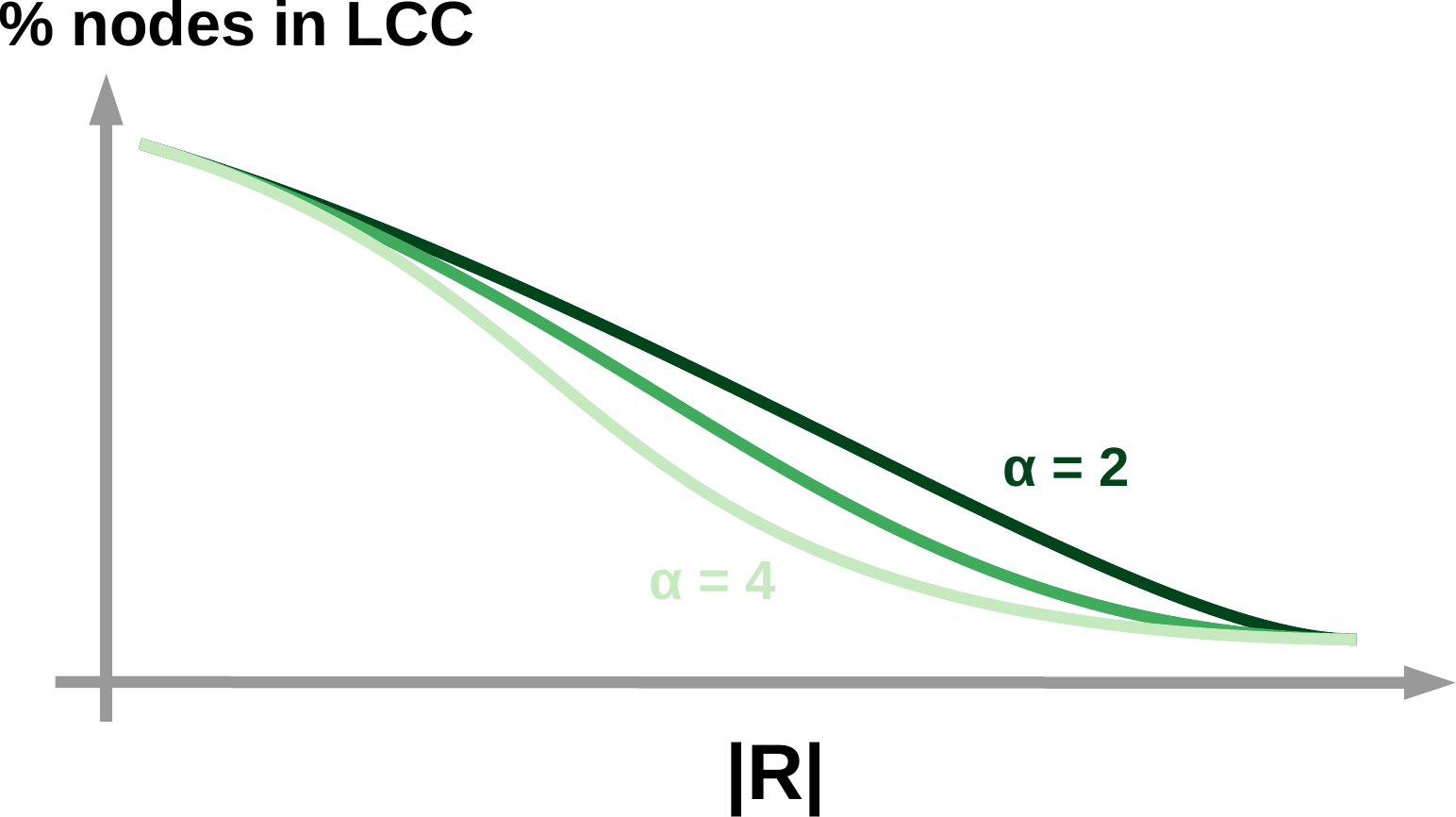}
\caption{The probability of being part of the largest connected component as a function of the number of failing nodes in a network for different $\alpha$ exponents of its power law degree distribution.}
\label{fig:rndfail-scalefree2}
\end{figure}

More common hubs equals higher likelihood of picking them up in a random extraction. Thus the failure functions for different $\alpha$ values follow the pattern I show in Figure \ref{fig:rndfail-scalefree2}. The lower your $\alpha$  the fewer hubs, the more resilient the network. By now you probably start to get an inkling on why network scientists are so obsessed about finding that their networks are scale free. If they are, then there are tons of properties you can infer by just knowing its degree exponent and relatively simple math. In this part we already saw two: robustness to random failures (here) and  outbreak size and speed in SI and SIS models (in Chapter \ref{cha:epidemics}).

By the way, so far we've been looking at random \textit{node} failures, i.e. a generator blowing up in the power grid. \textit{Edge} failures can be equally common: think about road blocks. However, the underlying math is rather similar and the functions describing the failures are not so different than the ones I've been showing you so far\cite{callaway2000network}\cite{cohen2000resilience}. For this reason we keep looking at node failures.

\section{Targeted Attacks}\label{sec:epidemapps-targeted}
So far we've assumed the world is a nice place and, when things break down, they do so randomly. We suspect no foul play. But what if there was foul play? What if we're not observing random failures, but a deliberate attack from a hostile force? In such a scenario, an attacker would not target nodes at random. They would go after the nodes allowing them to maximize the amount of damage while minimizing the effort required.

This translates into prioritizing attacks to the nodes with the highest degree. Taking down the node with most connections is guaranteed to cause the maximum possible amount of damage. What would happen to our network structure?

\begin{figure}
\centering
\includegraphics[width=\columnwidth]{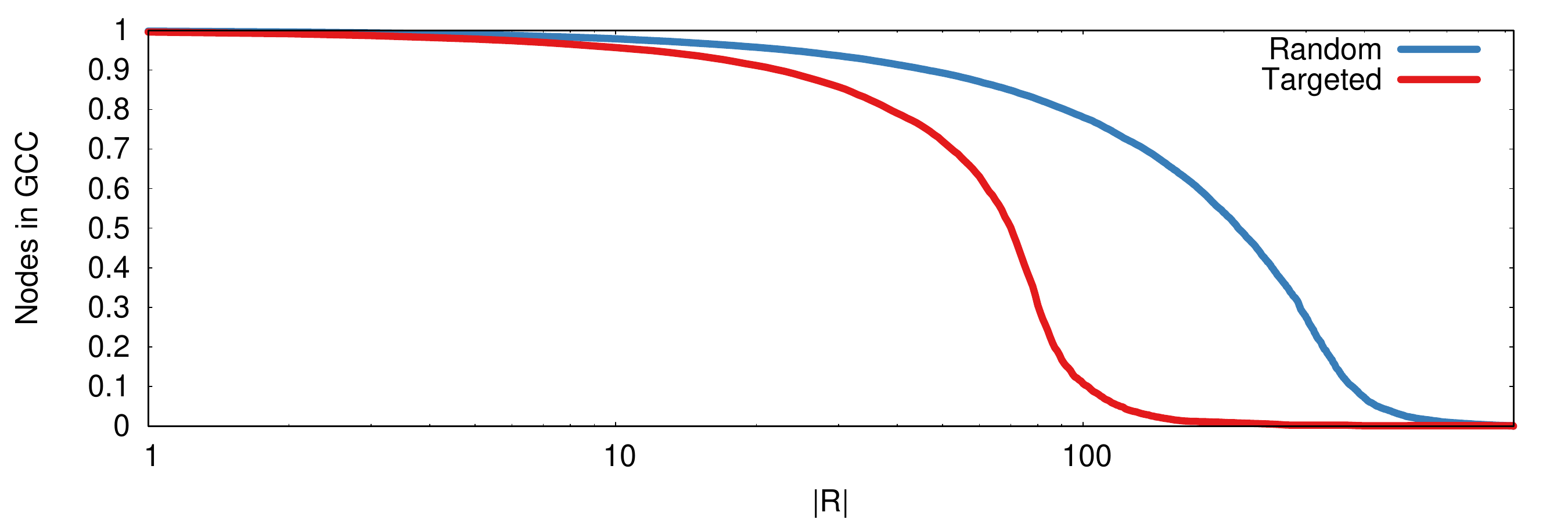}
\caption{The probability of being part of the largest connected component as a function of the number of failing nodes in a $G_{n,p}$ network, for random (blue) and targeted (red) failures.}
\label{fig:attack-rndgraph}
\end{figure}

Let's start again by considering a $G_{n,p}$ network. I ran a few simulations and Figure \ref{fig:attack-rndgraph} shows the result. We knew that $G_{n,p}$ networks aren't particularly good under random failures. It turns out that targeted attacks don't change the scenario much. Sure, the critical threshold is a bit lower, but the failure function is fundamentally the same.

Why? Remember that a $G_{n,p}$ model generates a normal degree distribution. This means that hubs are less common and their degree isn't much different from the average degree of all other nodes. If you pick up nodes randomly, you are likely to pick a node with higher-than-average degree and, even if you don't, whatever you pick isn't much different.

\begin{figure}
\centering
\includegraphics[width=\columnwidth]{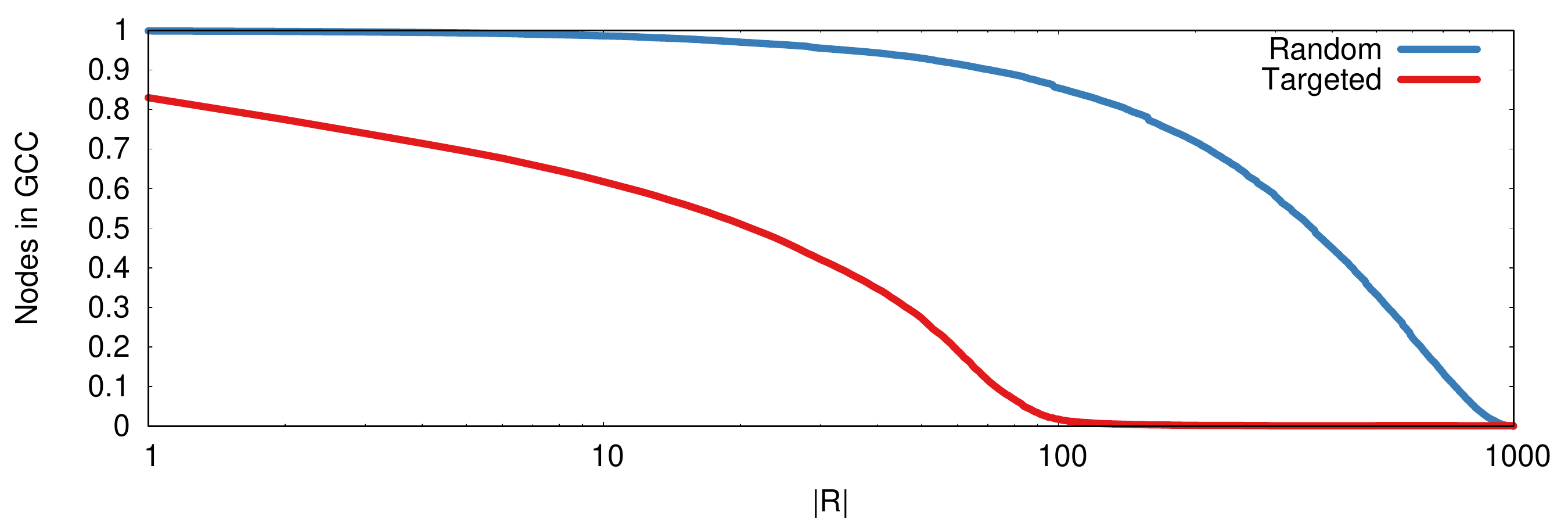}
\caption{The probability of being part of the largest connected component as a function of the number of failing nodes in a degree skewed network, for random (blue) and targeted (red) failures.}
\label{fig:attack-scalefree}
\end{figure}

The case is oh-so-much different when we turn our attention to networks with power law degree distributions. Since they have large hubs, prioritizing them for your attack will have devastating effects, as Figure \ref{fig:attack-scalefree} shows. Removing even a single node brings down the GCC size by almost $20\%$ in this case. To make a similar damage to a $G_{n,p}$ network, you have to remove around $40$ nodes.

In fact, networks with power law degree distributions break down \textit{more} easily than $G_{n,p}$ equivalents, when under a targeted attacks. As a consequence, different topologies should be used for different failure scenarios. If we're talking about random failures, your should plan your network to be scale free. If you want to defend from hostile takeovers, you probably want something similar to a random $G_{n,p}$ graph or, even better, a mesh-like network\cite{baran1964introduction}.

\begin{figure}
\centering
\includegraphics[width=\columnwidth]{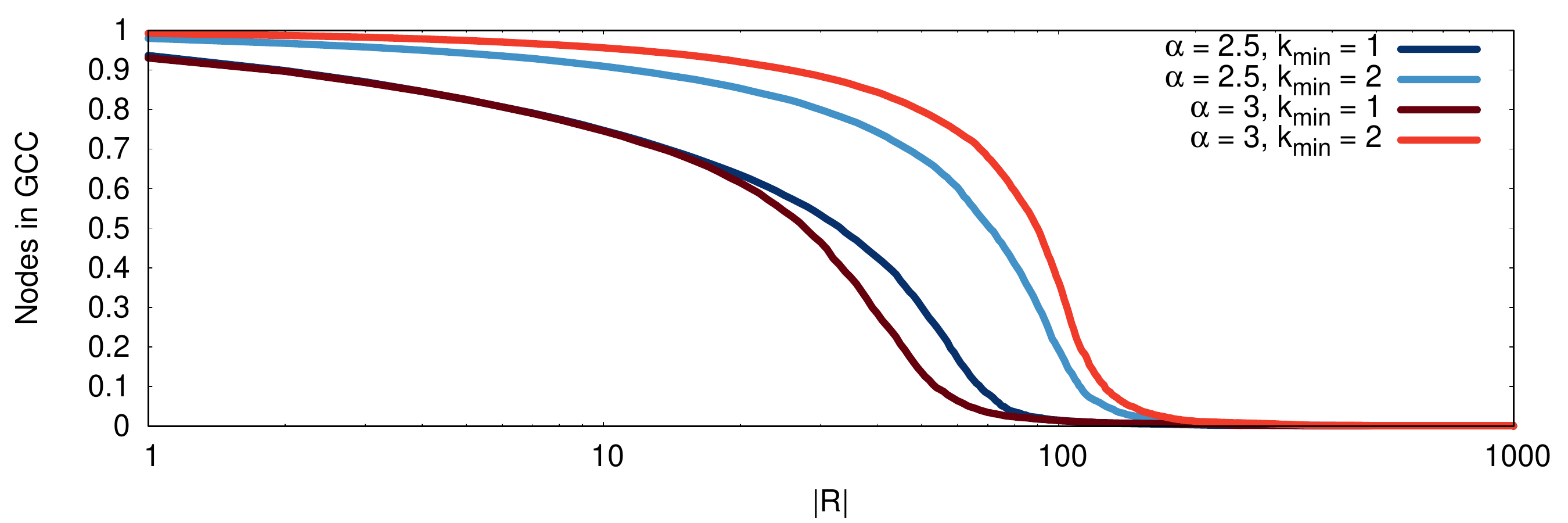}
\caption{The probability of being part of the largest connected component as a function of the number of failing nodes in a degree skewed network, for different $\alpha$ and $k_{min}$ combinations.}
\label{fig:attack-scalefree2}
\end{figure}

As you might expect, the $\alpha$ exponent of the power law degree distribution has something to do with the fragility of a network to deliberate attacks. However, it is a non-linear relationship, which also depends on the minimum degree of the network $k_{min}$\cite{cohen2001breakdown}\cite{bollobas2004robustness}. Figure \ref{fig:attack-scalefree2} shows the results of a few simulations. If $k_{min}$ is low, higher $\alpha$ exponents tend to make your network rather fragile, so it's better to have $\alpha = 2.5$ rather than $\alpha = 3$. However, if we increase $k_{min}$, then the opposite holds true: higher $\alpha$ actually make your network stronger.

\section{Chain Effects}\label{sec:epidemapps-cascade}
So far in this chapter we've relying on an unreasonable assumption: failures don't propagate. We said that a power generator goes boom randomly and studied what this means for the structure of the power grid. However, it is important to note that energy demand does not go down just because there was a failure. People will still turn their light bulbs on. Therefore, whatever power that generator was providing has to come from somewhere else. However, if that generator was there, there was a reason. Maybe the other nodes in the network cannot satisfy the additional demand. Thus there is a high chance that they will themselves go out of business.

In this scenario, the failure of one node propagates in a cascade and causes more correlated failures. This sort of snowball effect can turn into an avalanche and shut the entire network down. And it has happened, many times\cite{dobson2007complex}, also in structures that have nothing to do with power grids such as airline schedules\cite{fleurquin2013systemic}.

The models we use to simulate such propagating failures are yet another family of variations of the threshold model from Granovetter (Section \ref{sec:triggers-triggers}), for instance the Failure Propagation model\cite{dobson2005loading}. You can define failure propagation as being literally equivalent to the threshold model, by having a node $V$ failing if a fraction $f_v$ of its neighbors are failing. However, things become more interesting when you take into account more information.

All nodes start in the state $S$. They are characterized by a current load and by a total capacity. Think of this as road intersections: the load is how many cars pass on average through the intersection and the capacity is the maximum amount of cars that can pass before congestion happens.

\begin{figure*}
\centering
\begin{subfigure}{.24\columnwidth}
\includegraphics[width=\textwidth]{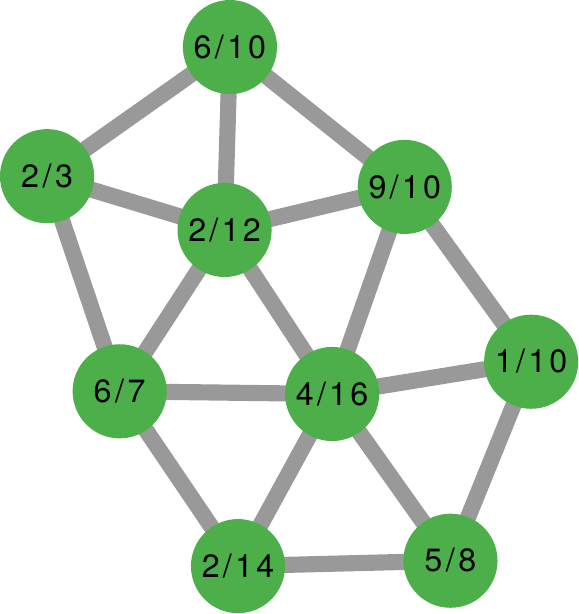}
\caption{$t = 0$}
\end{subfigure}
\begin{subfigure}{.24\columnwidth}
\includegraphics[width=\textwidth]{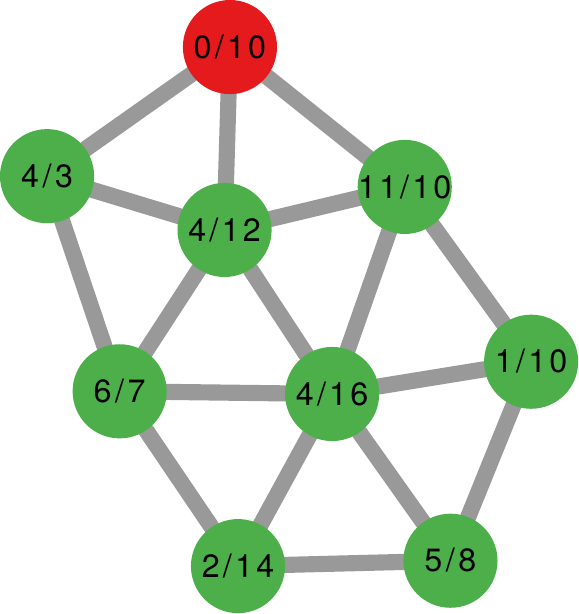}
\caption{$t = 1$}
\end{subfigure}
\begin{subfigure}{.24\columnwidth}
\includegraphics[width=\textwidth]{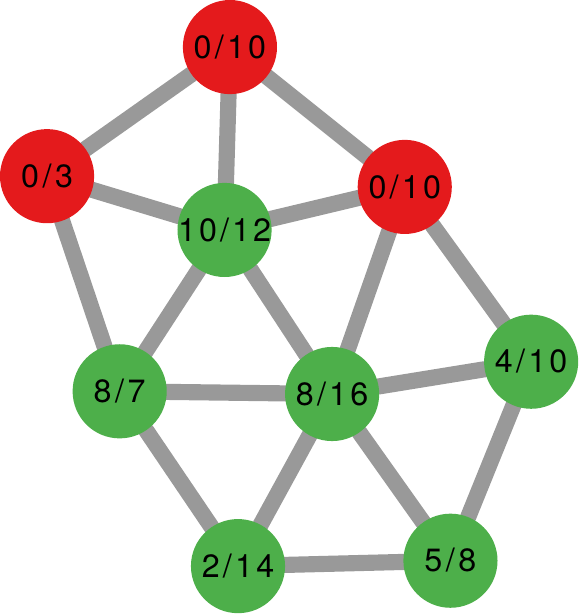}
\caption{$t = 2$}
\end{subfigure}
\begin{subfigure}{.24\columnwidth}
\includegraphics[width=\textwidth]{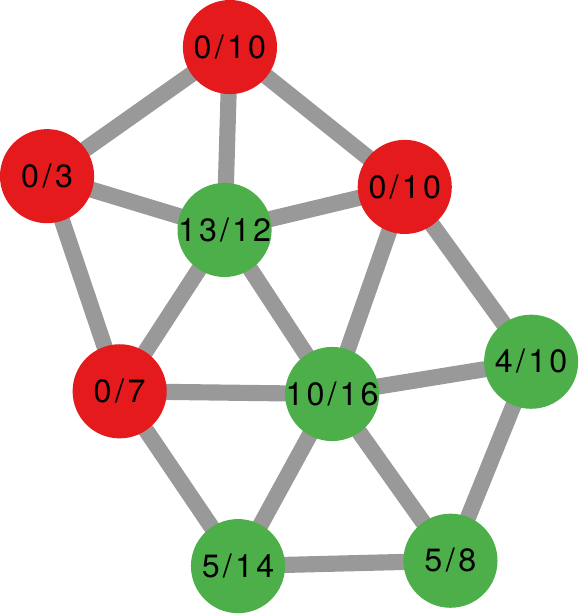}
\caption{$t = 3$}
\end{subfigure}
\begin{subfigure}{.24\columnwidth}
\includegraphics[width=\textwidth]{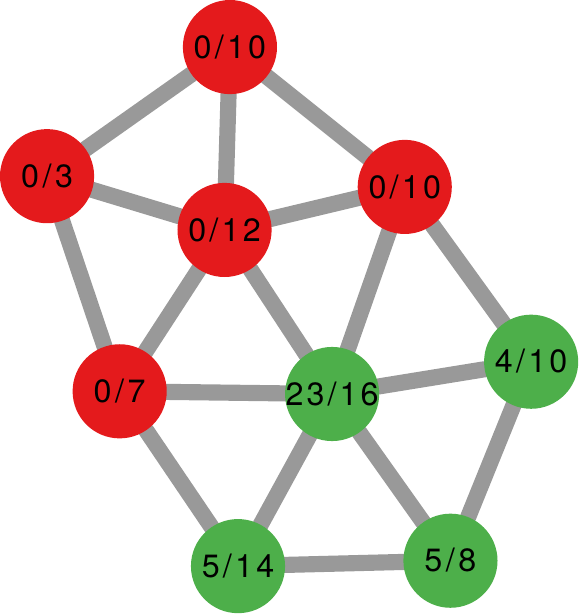}
\caption{$t = 4$}
\end{subfigure}
\begin{subfigure}{.24\columnwidth}
\includegraphics[width=\textwidth]{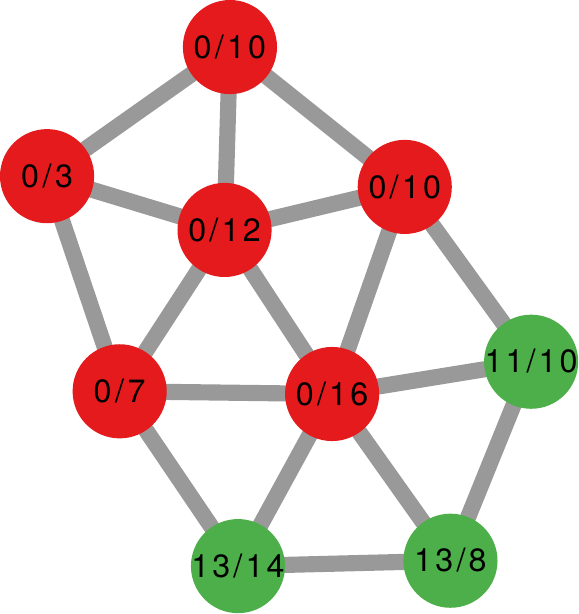}
\caption{$t = 5$}
\end{subfigure}
\begin{subfigure}{.24\columnwidth}
\includegraphics[width=\textwidth]{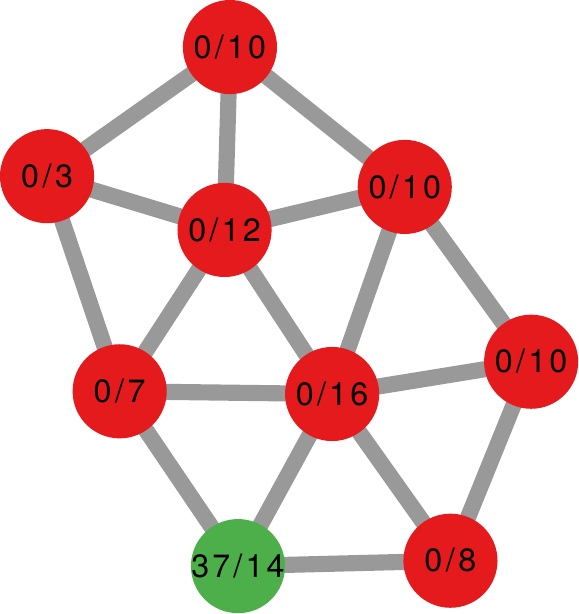}
\caption{$t = 6$}
\end{subfigure}
\begin{subfigure}{.24\columnwidth}
\includegraphics[width=\textwidth]{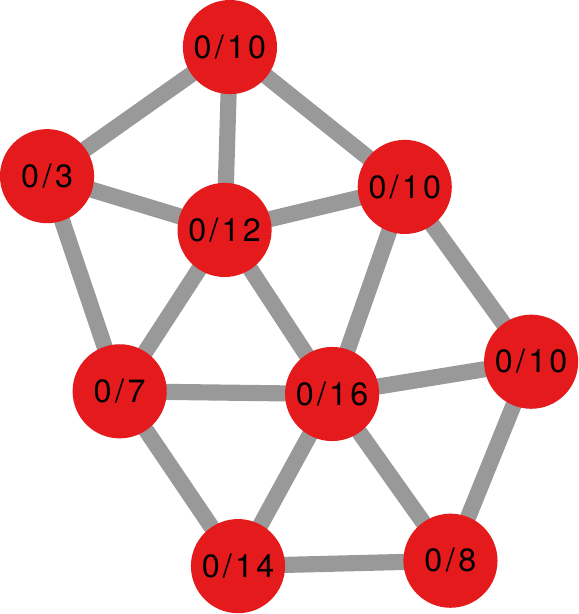}
\caption{$t = 7$}
\end{subfigure}
\caption{A simulated failure propagation. Each node is labeled with ``load / capacity''. Green nodes are active, red nodes have failed. If load $>$ capacity, the node will fail at the next time step.}
\label{fig:failure-propagation}
\end{figure*}

At time $t = 1$ we shut down a node in the network. Maybe the traffic light failed and so no one can pass through until we repair it. This means that the node transitions to state $I$. People still need to do their errands, so we have to redistribute the load of cars that wanted to pass through that intersection through alternative routes: the neighbors of that node. However, that means that their load will increase. If the new load exceeds the capacity of the node, also this node shuts down due to congestion. So its load has also to be redistributed to its neighbors and so on and so forth.

Figure \ref{fig:failure-propagation} shows an example of failure propagation with this load-capacity feature. You can see that the network was built with some slack in mind: its normal total load is $37$ -- the sum of all loads of all nodes -- for a maximum capacity of $90$ -- the sum of all nodes' capacities. Yet, shutting down the top node whose load was only $6$ and redistributing the loads causes a cascade that, eventually, brings the whole network down.

\begin{figure}
\centering
\includegraphics[width=.8\columnwidth]{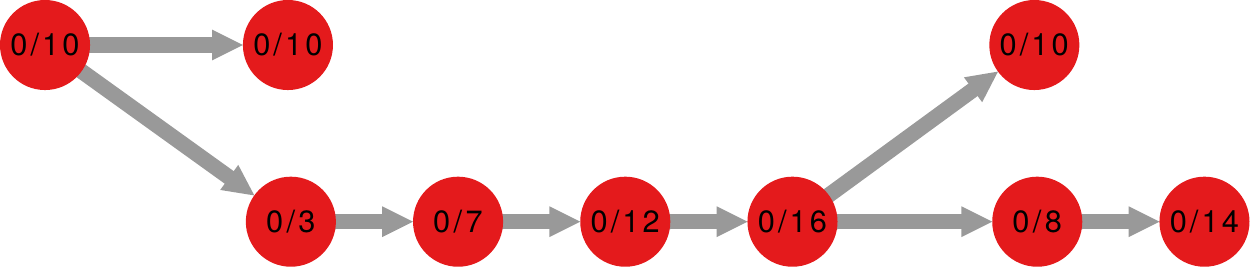}
\caption{The branch model representing the same cascade failure of Figure \ref{fig:failure-propagation}.}
\label{fig:failure-propagation-tree}
\end{figure}

One could represent the failure cascade from Figure \ref{fig:failure-propagation} as the branches of a tree. The first node failing is the root. We then connect each node to the nodes it causes to fail. The final structure would be something like Figure \ref{fig:failure-propagation-tree}.

Using this perspective has its own advantages. It makes the failure propagation model more amenable to analysis. The final size of the failure cascade depends on the average degree of nodes in this tree $\bar{k}$. The critical value here is $\bar{k} = 1$. If, on average, the failure of a node generates another node failure -- or more -- the cascade will propagate indefinitely, until all nodes in the network will fail. If, instead, $\bar{k} < 1$, the failure will die out, often rather quickly.

It's easy to see why if you have the mental picture of a domino snake: each domino falling will cause the fall of another domino, until there's nothing standing. If, however, there is as much as a single gap in this chain, the rest of the system will be unaffected.

Quick show of hands: how many of you expect the size of a failure cascade to be a power law? Good, good: by now you learned that every goddamn thing in this book distributes broadly. The exponent of the cascade size is also related to the $\alpha$ exponent of your degree distribution. With a power degree exponent $\alpha > 3$, networks behave like $G_{n,p}$ graphs, but for $\alpha < 3$ then the cascade size will grow with exponent $\alpha / (\alpha - 1)$.

\section{Interdependent Networks}\label{sec:epidemapps-interdependent}
There is an additional thing you have to consider when describing cascading failures in real world structures. So far, we have considered our networks as living in isolation. A failure in the power grid propagates only through the power grid. In our interconnected world that is not the case. In fact, once you realize how fragile networked systems can be, why would you rely on such systems without additional fail-safes? For instance, you might want to control what's happening on a power grid with an automatic computerized controller, so that it can try to isolate the failure and prevent it from propagation.

How are you going to do that with a single computer? You need a network of terminals close to the action. How are you going to provide the power they need? Through the power grid itself. So now you have two interdependent networks: the power grid needs computers to work and the computers need power to work. Interdependent networks don't behave like isolated networks\cite{gao2012networks}. Researchers have studied propagating failures in such interdependent systems and found out that \textit{even if the two networks are resilient to random failures in isolation, the inter-dependencies cause them to be fragile to failures propagating back and forth between them}\cite{buldyrev2010catastrophic}. Ouch.

\begin{figure}[t]
\centering
\includegraphics[width=.66\columnwidth]{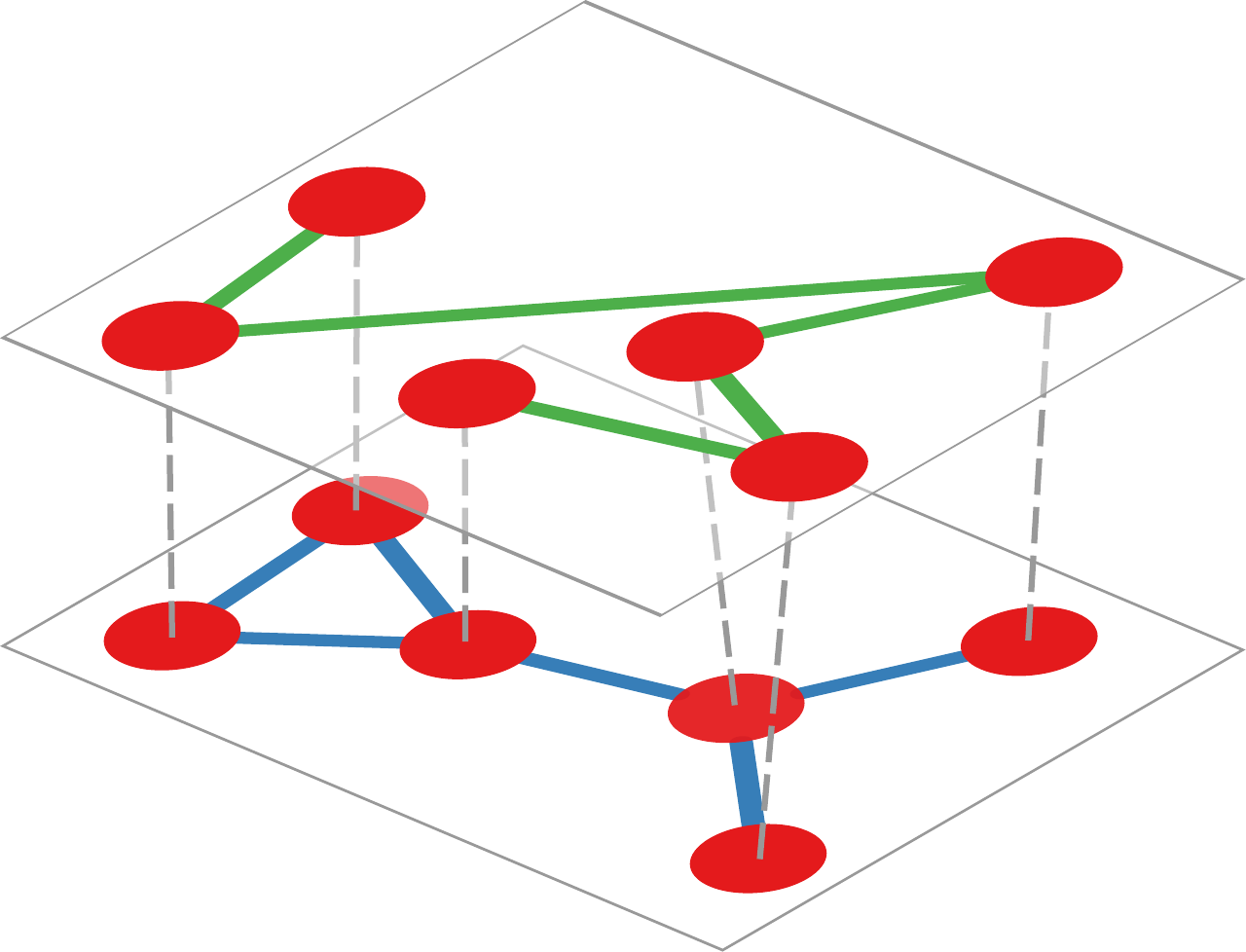}
\caption{A multilayer network representing inter-dependencies between two layers: the nodes in one layer depend on the nodes on the other layer connected to them via an inter-layer coupling.}
\label{fig:interdependent-failure}
\end{figure}

You model this problem using multilayer networks (Section \ref{sec:extended-multilayer}). Figure \ref{fig:interdependent-failure} shows an example. The blue layer represents the power stations and the green layer represents the computers. A power station needs the coupled computer to work and vice versa. The network can work via connected components in both layers: if nodes get isolated in one layer, the nodes on the other layer coupled to different components get disconnected. A power station needs to know the statuses of its neighboring stations: if they are on a different computer component it cannot know it, so their links deactivate.

\begin{figure*}[t]
\centering
\begin{subfigure}{.3\columnwidth}
\includegraphics[width=\textwidth]{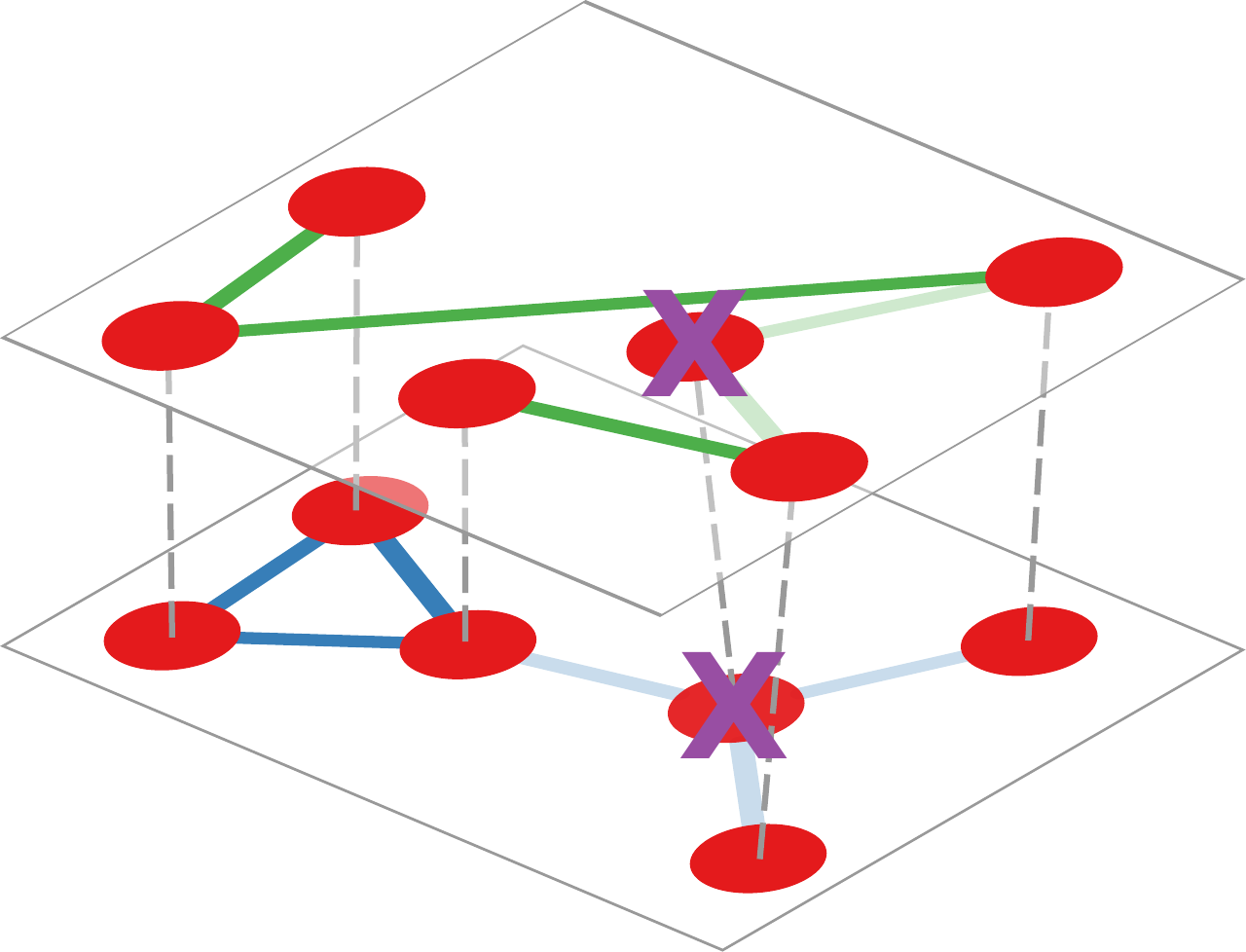}
\caption{$t = 1$}
\end{subfigure}\quad
\begin{subfigure}{.3\columnwidth}
\includegraphics[width=\textwidth]{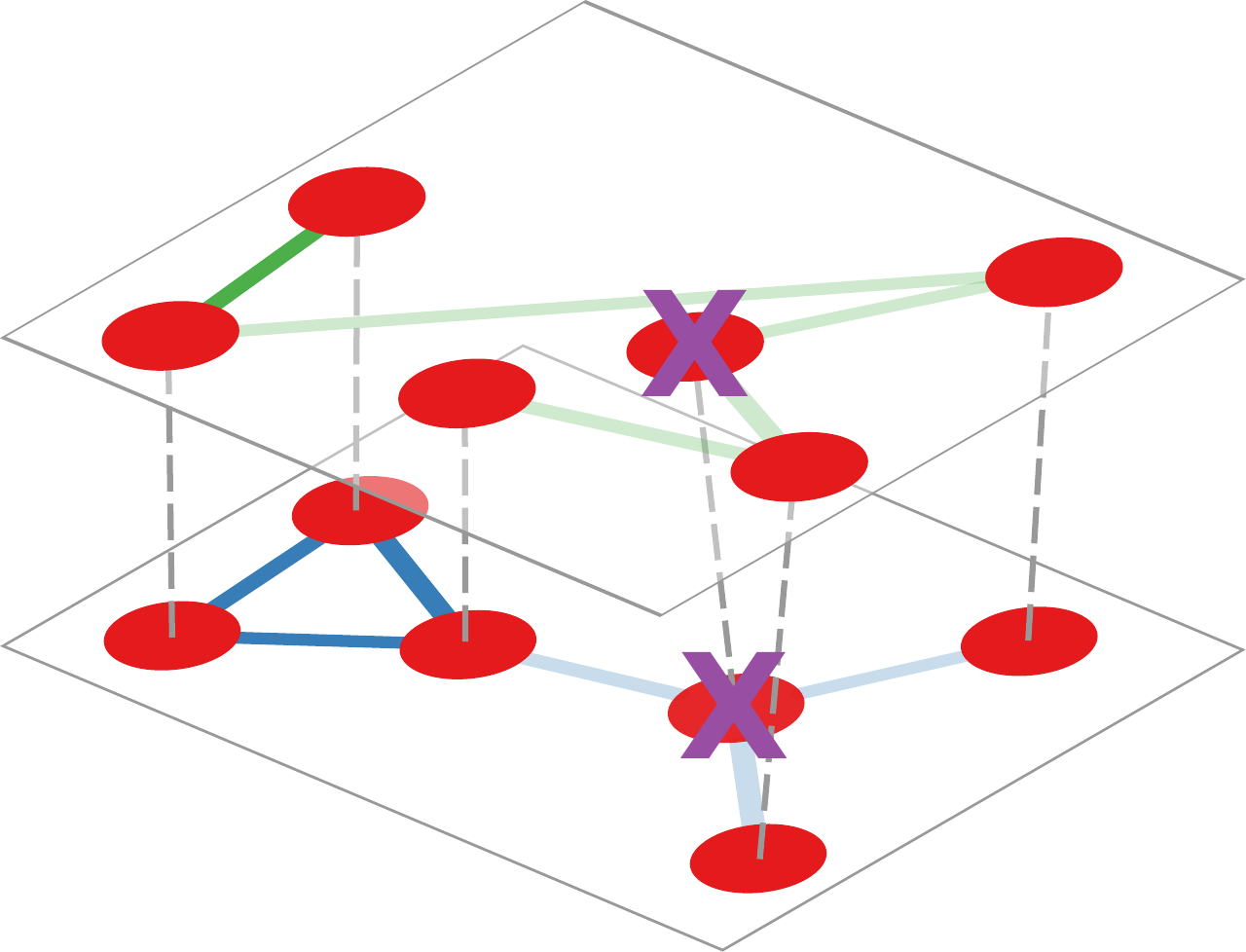}
\caption{$t = 2$}
\end{subfigure}\quad
\begin{subfigure}{.3\columnwidth}
\includegraphics[width=\textwidth]{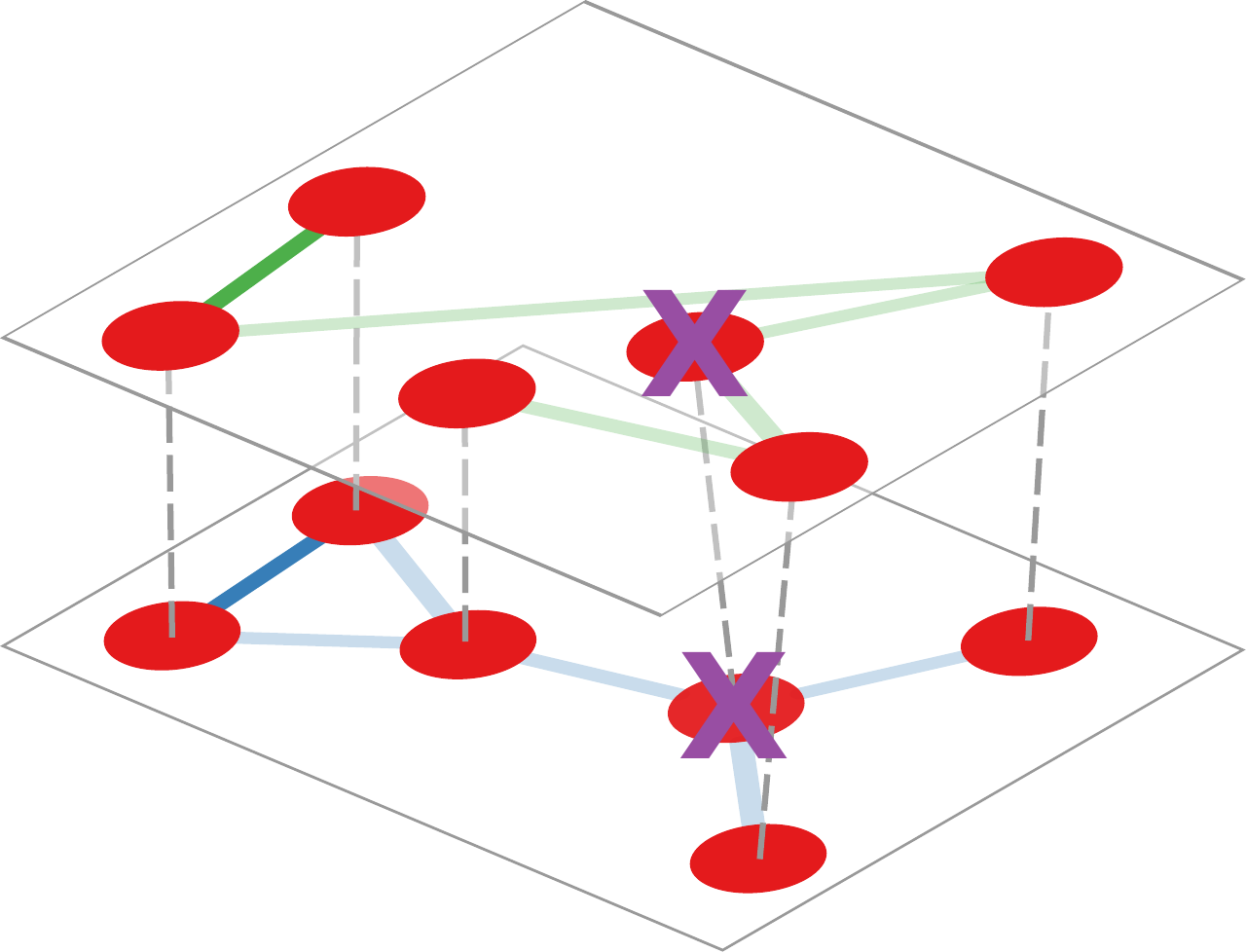}
\caption{$t = 3$}
\end{subfigure}
\caption{Propagating failures in interdependent networks. The purple cross shows the original failing nodes. At each time step, nodes depending on nodes in different components lose their connections (indicated by a faded edge color).}
\label{fig:interdependent-failure2}
\end{figure*}

If a node in one layer fails (Figure \ref{fig:interdependent-failure2}(a)) it breaks its connections and causes its coupled node to lose its connections too. Now we have multiple connected components in one layer, so the links in the other layer going across components start to fail as well (Figure \ref{fig:interdependent-failure2}(b)). These failures propagate in a chain reaction until we end up in a situation where practically every node in both layers is isolated, and the network almost completely failed (Figure \ref{fig:interdependent-failure2}(c)).

When describing failures in single layer networks, we asked ourselves what's the value of $|R|$ such that the network breaks down. In other words: what's the fraction of initially failing nodes that will make the GCC disappear. We can ask the same question here, realizing that, in interdependent networks, this $|R|$ value is much lower than the corresponding one for single layer networks. In fact, \textit{if you were to calculate the critical $|R|$ value for each layer separately you would obtain a result much higher than the one for the interdependent network as a whole}. Meaning that, if you were to analyze the layers independently, you'd grossly \textit{under}estimate the risk of a catastrophic failure propagating through the entire network.

Remember when, in Section \ref{sec:epidemapps-random}, I said that networks with a heavy tail in their degree distribution are particularly robust to random failures? The reason was that large hubs keep the network together and there are very few of them, so it's unlikely to pick them up at random. Well... In two interdependent networks it is likely that hubs in one layer will couple to nodes with a lower degree in the other. Guess what: that makes coupled power law networks fragile to random failures. In fact, they're more fragile than random $G_{n,p}$ graphs, contrarily to what was the case before.

\begin{figure}[t]
\centering
\includegraphics[width=.8\columnwidth]{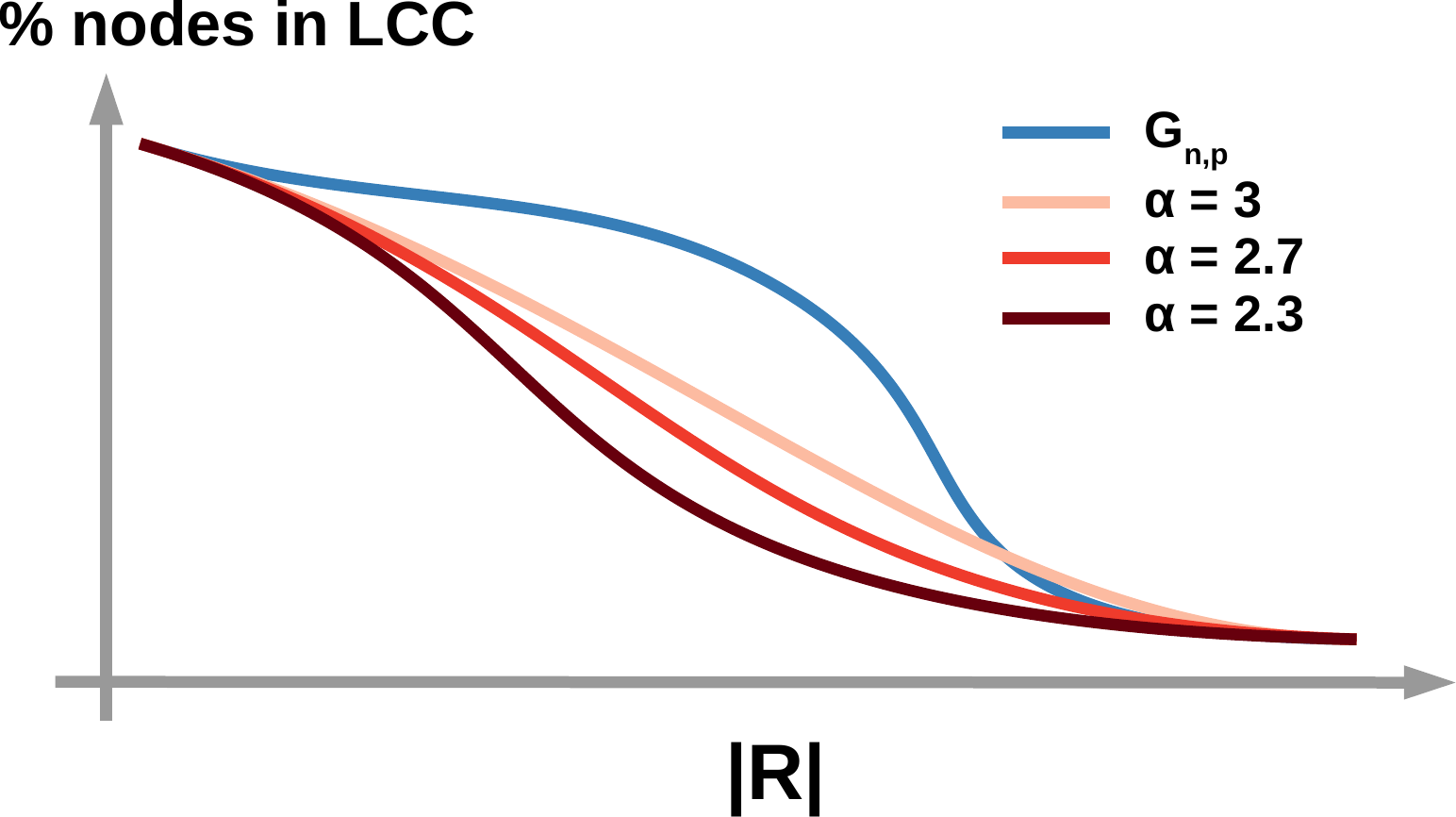}
\caption{The relationship between the degree exponent $\alpha$ of coupled power law networks and the fraction of nodes in $R$ state (x axis) needed to destroy the GCC (shades of red). In blue the equivalent plot for coupled random $G_{n,p}$ graphs.}
\label{fig:interdependent-failure3}
\end{figure}

Figure \ref{fig:interdependent-failure3} shows a schema of fragility for different coupled network topologies. Moving from random failures to targeted attacks still makes interdependency bad: failures will spread more easily than in isolated networks\cite[-1in]{huang2011robustness}.

Fixing this issue is not easy. First, one needs to estimate the propensity of the network to run the risk of failure. This has been done in two-layer multiplex networks\cite[-0.9in]{burkholz2016systemic}. One would think that the best thing to do is to create more connections between nodes, to prevent breaking down in multiple components. However that's not a trivial operation, as these networks are embedded in a real geographical space, where creating new power lines might not be possible. However, there are also theoretical concerns that show how more connections could render the network more fragile, as it would give the cascade more possible pathways to generate a critical failure\cite[-1.6in]{brummitt2012suppressing}. A better strategy involves so-called ``damage diversification'': mitigating the impact of the failure of a high degree node\cite[-0.95in]{burkholz2016damage}.

Note that we assumed that power law networks are randomly coupled: hubs in one layer will pick a random node to couple to in the other layer. As a consequence, they'll likely to pick a low degree node. Other papers study the effect of degree correlations in inter-layer coupling\cite[-1.35in]{min2014network}: what if hubs in one layer tend to connect to hubs in the other layer? If such correlations were perfect, we'd obtain again the robustness of power law networks to random failures. These correlations are luckily observed in real world systems\cite[-1.2in]{parshani2011inter}\cite[-0.5in]{reis2014avoiding}, showing how they're not as fragile as one might fear. Phew.

\section{Summary}

\begin{enumerate}
\item $G_{n,p}$ networks are fragile to random failures: beyond a critical number of removed nodes the giant connected component will disappear. Networks with skewed degree distributions are instead robust because they rely on few hubs which are unlikely to be picked by random failures.
\item In targeted attacks we take down nodes from the most to least connected. Power law random networks are very fragile and break down quickly under this scenario.
\item When one node fails, all its load needs to be redistributed to non-failing nodes. This can and will make the failure propagate on the network in a cascade event which might end up bringing the entire network down.
\item In interdependent networks we have a multilayer network whose nodes in one layer are required for the functioning of nodes in the others.  Depending on the degree correlations among layers, failures can propagate across layer and bring down power law networks even under random accidents. 
\end{enumerate}

\section{Exercises}

\begin{enumerate}
\item Plot the number of nodes in the largest connected component as you remove $2,000$ random nodes, one at a time, from the network at \url{http://www.networkatlas.eu/exercises/22/1/data.txt}. (Repeat 10 times and plot the average result)
\item Perform the same operation as the one from the previous exercise, but for the network at \url{http://www.networkatlas.eu/exercises/22/2/data.txt}. Can you tell which is the network with a power law degree distribution and which is the $G_{n,p}$ network?
\item Plot the number of nodes in the largest connected component as you remove $2,000$ nodes, one at a time, in descending degree order, from the networks used for the previous exercises. Does the result confirm your answer to the previous question about which network is of which type?
\item The network at \url{http://www.networkatlas.eu/exercises/22/4/data.txt} has nodes metadata at \url{http://www.networkatlas.eu/exercises/22/4/node_metadata.txt}, telling you the current load and the maximum load. If the current load exceeds the maximum load, the node will shut down and equally distribute all of its current load to its neighbors. Some nodes have a current load higher than their maximum load. Run the cascade failure and report how many nodes are left standing once the cascade finishes.
\end{enumerate}

\part{Link prediction}\label{par:lp}

\chapter{For Simple Graphs}\label{cha:lp-simple}
\begin{figure}
\centering
\includegraphics[width=.8\columnwidth]{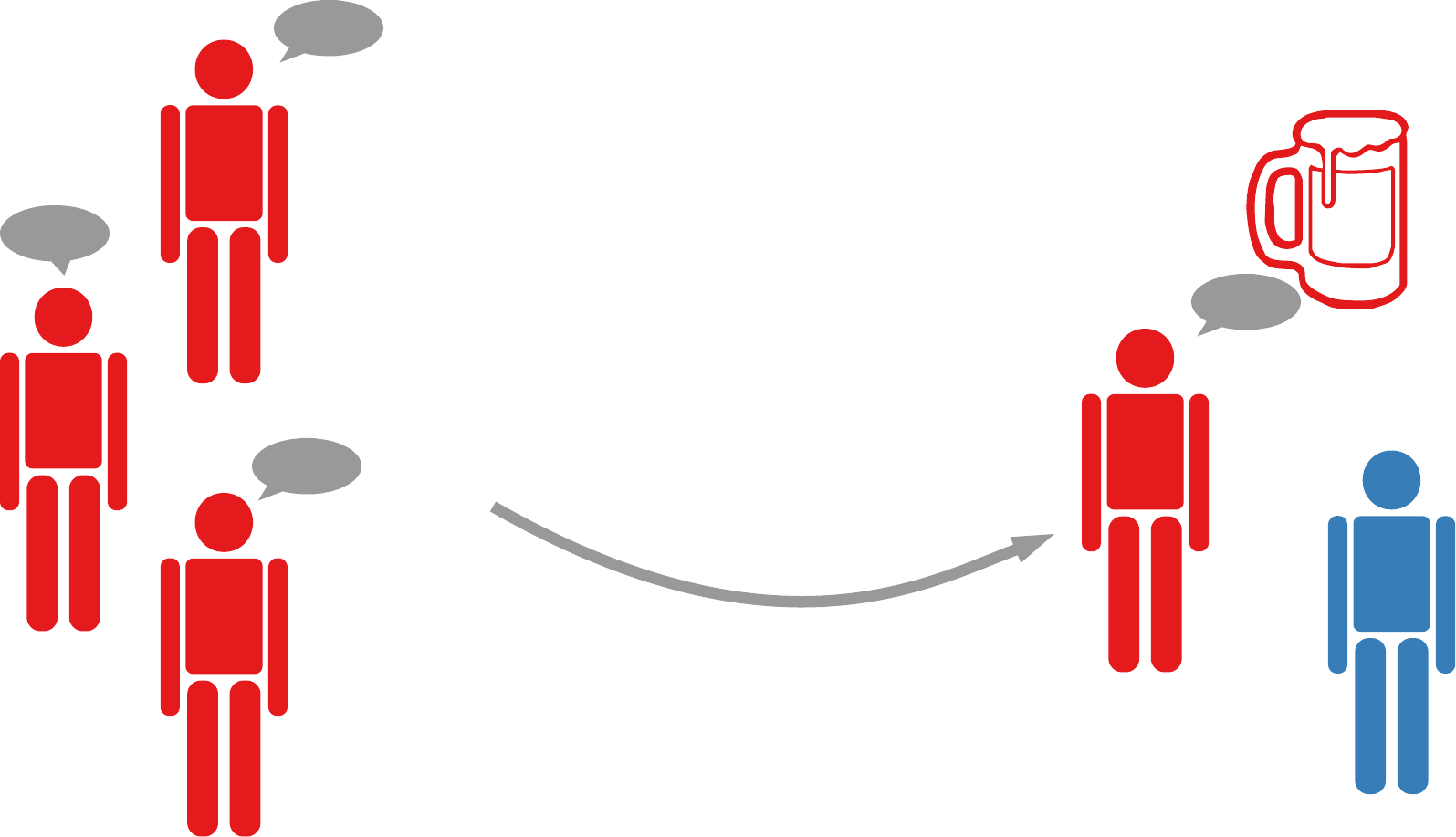}
\caption{A vignette explaining the link prediction aim. at a party, red individuals know each other and express their relationships (or links) by talking. One member might detach from the group for any reason and, in doing so, she exposes herself to the possibility of establishing a new link with the blue individual. Link prediction is all about finding the true reason that might make this happen.}
\label{fig:linkprediction}
\end{figure}

Link prediction is the branch of network analysis that deals with the prediction of new links in a network. In link prediction, you see the network as fundamentally dynamic, it can change its connections. Suppose you're at a party. You came there with your friends, and you're talking to each other, using the old connections. At some point, you want to go and get a drink so you detach from the group. On the way, you could meet a new person, and start talking to them. This creates a new link in the social network. Link prediction wants to find a theory to predict these events -- in this case, that alcohol is the main cause of new friendships at parties, or so I'm told --, as I show the vignette in Figure \ref{fig:linkprediction}.

In other words, given a network with nodes and edges, we want to know which link is the most likely to appear in the future. Or, if we think we're seeing an incomplete version of the network, we ask ourselves which edges are currently missing from the structure.

Link prediction happens in three steps. The starting point is your desire to place a new link in the network: which edge will appear next? The first thing you do is to observe the current links. On the basis of this observation you formulate a hypothesis on how nodes decide to link in the network. Finally, you operationalize this hypothesis: if nodes are created via process $x$, you apply $x$ to the current status of the network and that will tell you which link is most likely to appear next.

In this chapter, we are going to focus on the simplest possible case for link prediction: predicting new links in a simple graph. We delve deep into the classical approaches to link prediction, which are the simplest and most used in the literature\cite[-0.5in]{getoor2005link}\cite{liben2007link}\cite{lu2011link}\cite{wang2015link}\cite{martinez2017survey}: Preferential Attachment, Common Neighbor, Adamic-Adar, Hierarchical Random Graph models, Resource Allocation, and Graph Evolution Rules. We also briefly mention other approaches, to give justice to a gigantic subfield of network analysis in computer science.

We will deal with multilayer link predictions later, in Chapter \ref{cha:lp-multilayer}. Once we understand how to assign a score to every possible future link, it's time to estimate whether we did a good job. This will be covered in Chapter \ref{cha:lp-experiment}.

\section{Preferential Attachment}

\begin{figure}
\centering
\includegraphics[width=.75\columnwidth]{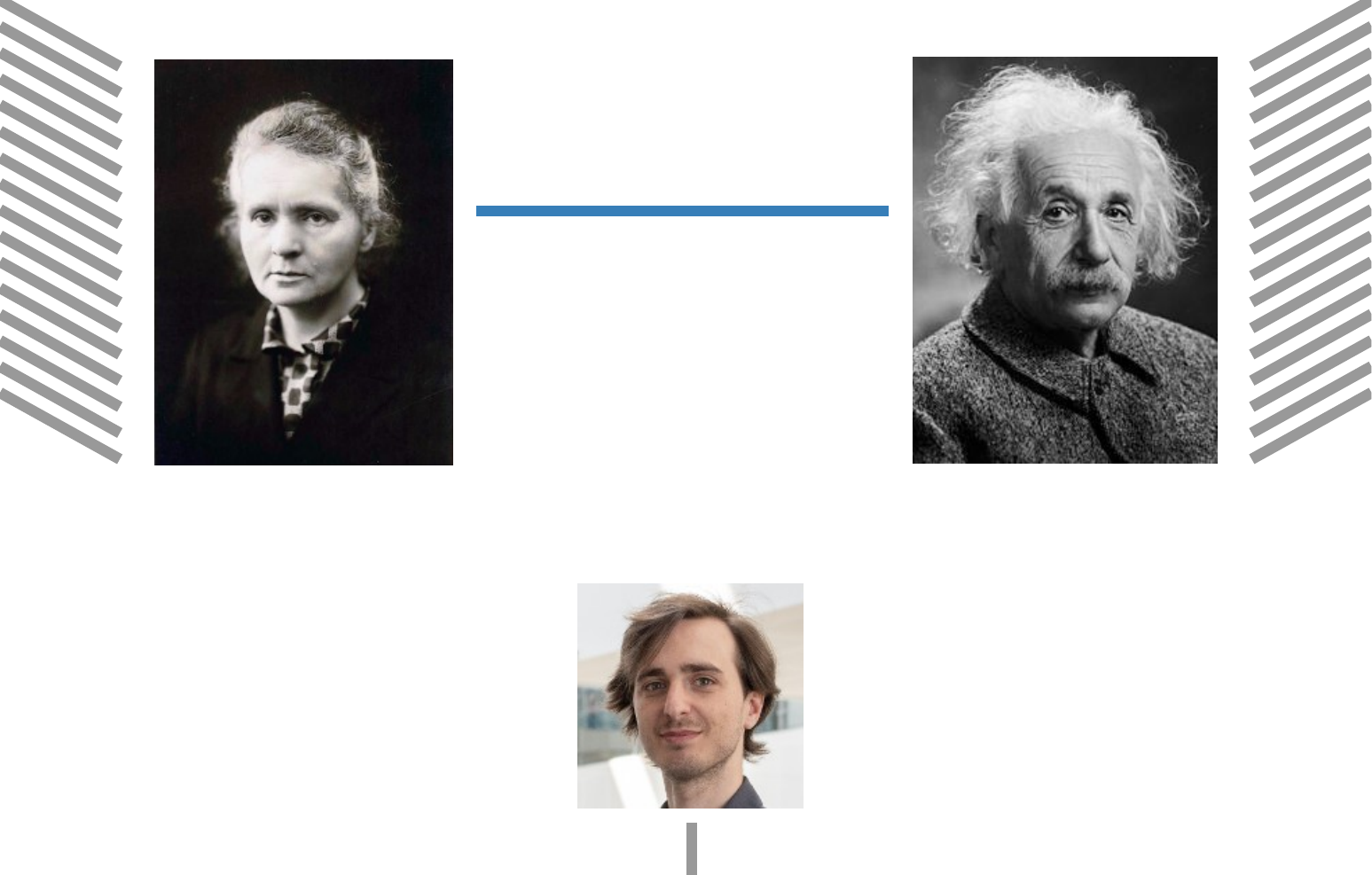}
\caption{An example of a Preferential Attachment link prediction. Two hubs (Einstein and Curie) have a lot of connections (in gray), while a third author only has few. For PA, the most logical link to predict is in blue between the hubs, because rich get richer, and thus they will attract the new connections.}
\label{fig:lp-pa}
\end{figure}

Let's start with Preferential Attachment (PA). Consider scientific publishing. We have three authors: two of them -- Einstein and Curie -- have a lot of collaborators, while the third -- me -- has only few -- see Figure \ref{fig:lp-pa}. If we have to make a guess of what collaboration is more likely to happen next, which one would we expect? It's more likely to see the two high degree hubs to connect, because they're more prominent, and thus visible to each other.

If we want to predict links, we have to formulate a hypothesis and then translate it into a score of $u$ connecting to $v$ for any pair of $u, v$ nodes: $score(u, v)$. In PA, the hypothesis is that ``rich get richer'', nodes with lots of edges will attract more edges\cite{newman2001clustering}\cite{barabasi2002evolution}. So we look for pairs of nodes that have attracted so far the most edges. Our PA model would consider it strange if they are not connected to each other. Our best guess is that they will connect soon. In practice, the probability of connecting two nodes is directly proportional to their current degree: $score(u, v) = k_u k_v$, where $k_u$ and $k_v$ are $u$'s and $v$'s degrees, respectively.

\section{Common Neighbor}\label{sec:lpsimple-cn}

\begin{figure}
\centering
\includegraphics[width=.75\columnwidth]{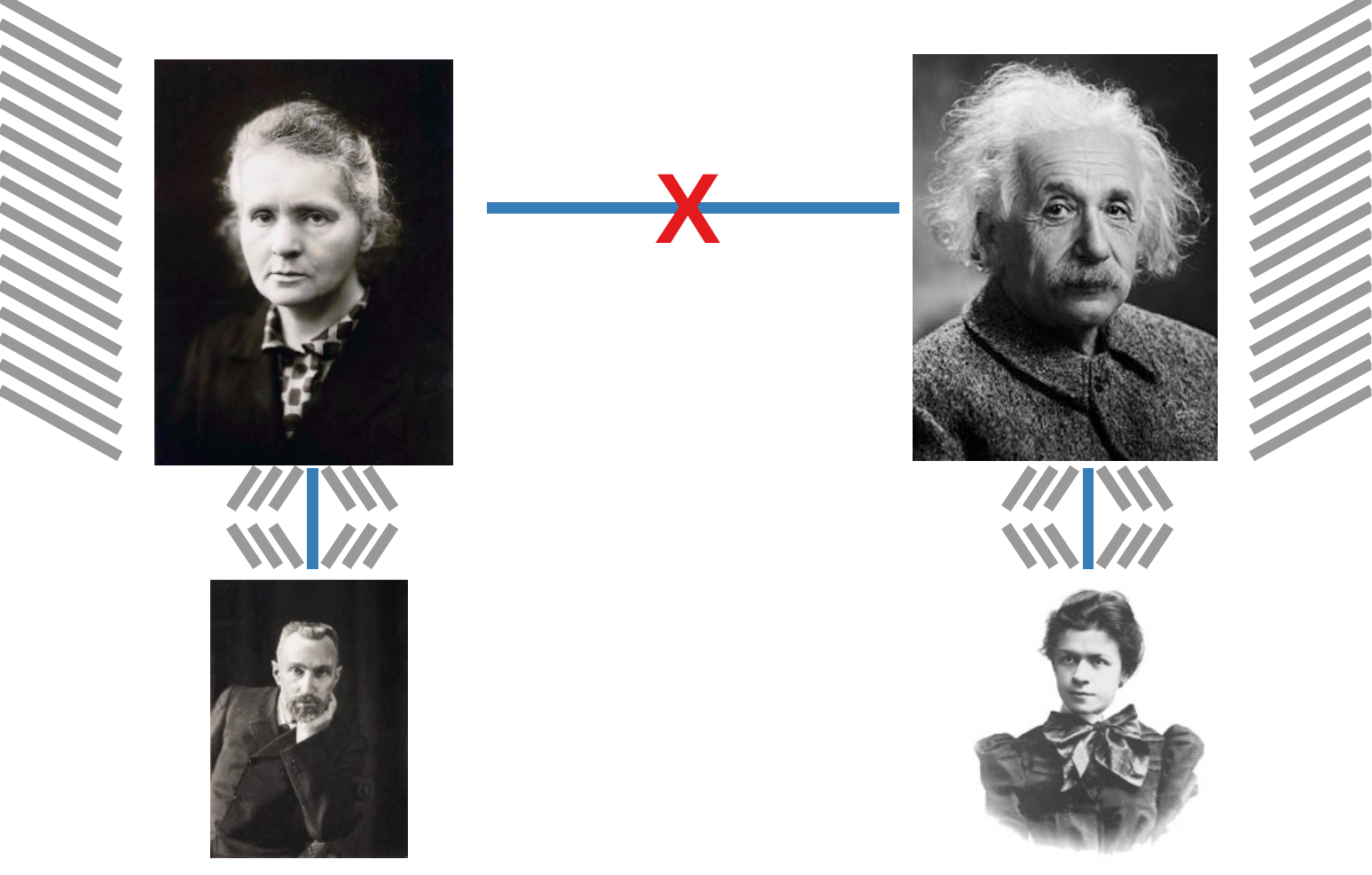}
\caption{An example of a Common Neighbor link prediction. The hubs (Einstein and Curie) do not share any connection, thus it's less likely they will connect to each other. But they share a  lot of connections with other lower degree nodes. CN will give the possibility of linking to them a boost.}
\label{fig:lp-cn}
\end{figure}

The preferential attachment example in Figure \ref{fig:lp-pa} has one defect: its prediction is wrong, Curie and Einstein never collaborated. This is because PA fails to consider the social element: it is more likely to collaborate not only if one is good at collaborating, but also if the two people are likely to meet. Given that Curie and Einstein are from slightly different fields, it is difficult for a meeting between the two to stick into a collaboration. On the other hand, they might have a lot of common collaborators with other people in the same field: Curie shared a Nobel prize with her husband Pierre Curie, and Einstein owes a great debt to Mari\'{c} -- see Figure \ref{fig:lp-cn}. Neither Pierre nor Mari\'{c} had as many collaborations as Curie or Einstein, but for the Common Neighbor (CN) model the thing that matters most is the number of neighbors they share with them.

Common Neighbor's basic theory is that triangles close: the more common neighbors $u$ and $v$ have, the more triangles we can close with a single edge connecting $u$ to $v$\cite{newman2001clustering}. So the likelihood of connecting two nodes is proportional to the number of shared elements in their neighbor sets: $score(u, v) = |N_u \cap N_v|$, where $N_u$ and $N_v$ are the set of neighbors of $u$ and $v$, respectively. A variant controls for how many neighbors the two nodes have: the same number of common neighbors weighs more if it's the total set of connections the two neighbors have. This is the Jaccard variant: $score(u, v) = |N_u \cap N_v| / |N_u \cup N_v|$.

\section{Adamic-Adar}\label{sec:lpsimple-aa}

\begin{figure}
\centering
\includegraphics[width=.75\columnwidth]{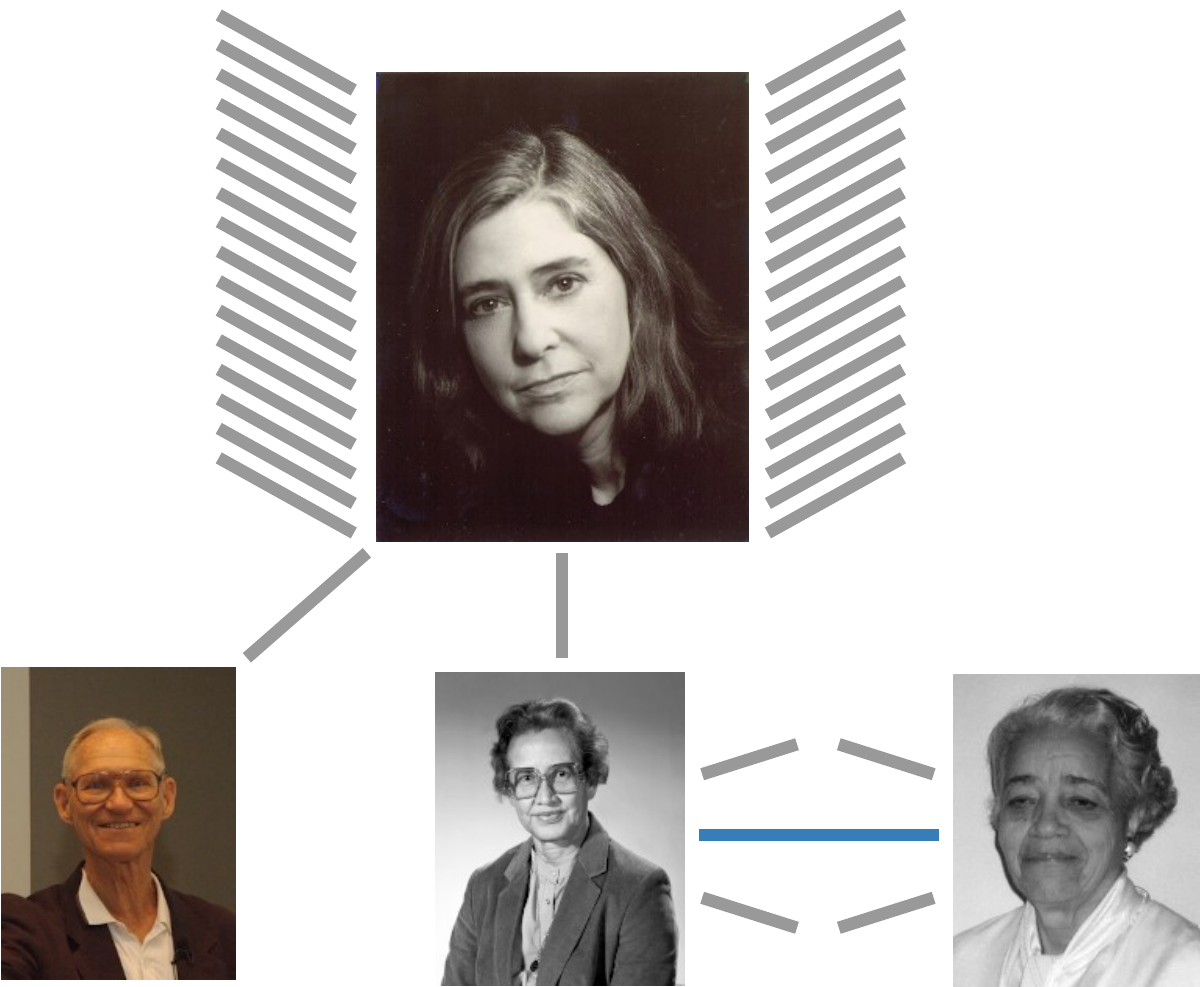}
\caption{An example of a Adamic-Adar link prediction. As a hub, Hamilton (top) does not have enough time to introduce all possible pairs of her many collaborators. Thus it's less likely Johnson (middle bottom) can connect to Boehm (left bottom) than she can connect to Vaughan (right bottom), with whom she shares common connections with fewer collaborators than Hamilton.}
\label{fig:lp-aa}
\end{figure}

Common Neighbor has a problem which is bandwidth. Even if you have Hamilton as a collaborator, she has collaborated with so many other people that the likelihood of Johnson to connect to Boehm -- both Hamilton's collaborators -- is low, because Hamilton does not have enough bandwidth to make the introduction. On the other hand, few common collaborators, if they have few connections, can represent a stronger attraction between two people. In this case, they are more likely to make the introduction, because they have the time to do so. Thus they make the new collaboration happen, as is the case for Johnson and Vaughan -- see Figure \ref{fig:lp-aa}.

In Adamic-Adar (AA)\cite{adamic2003friends} we say that common neighbors are important, but the hubs contribute less to the link prediction than two common neighbors with no other links, because the hubs do not have enough bandwidth to make the introduction. In AA, our score function discounts the contribution of each node with the logarithm of its degree: $score(u, v) = \sum \limits_{z \in N_u \cap N_v}\dfrac{1}{\log k_z}$. The formula says that, for each common neighbor, instead of counting one -- as we do in Common Neighbor when we look at the intersection --, we count one over the common neighbor's degree (log-transformed).

\section{Resource Allocation}\label{sec:lpsimple-ra}
The Resource Allocation index\cite{zhou2009predicting} is almost identical to Adamic-Adar. It stems from the very same principle: nodes have bandwidth. The likelihood of $u$ connecting to $v$ is proportional to the amount of resources $u$ can send to $v$ and vice versa. Thus we have $score(u, v) = \sum \limits_{z \in N_u \cap N_v}\dfrac{1}{k_z}$. The only difference with Adamic-Adar is that the scaling is assumed to be linear rather than logarithmic. Thus, Resource Allocation punishes the high-degree common neighbors more heavily than Adamic-Adar. You can see that the difference between $k_z$ and $\log k_z$ is practically nil for low values of $k_z$, but balloons when $k_z$ is high.

One could make a more complex version of the Resource Allocation index by assuming that the bandwidth of each node and of each link is not fixed. Thus the amount of resources $u$ sends can change, and the amount of resources that can pass through the $(u,v)$ link can also be different from the one passing through other edges.

\section{Hierarchical Random Graphs}\label{sec:lpsimple-hrg}
\begin{figure}
\centering
\includegraphics[width=.9\columnwidth]{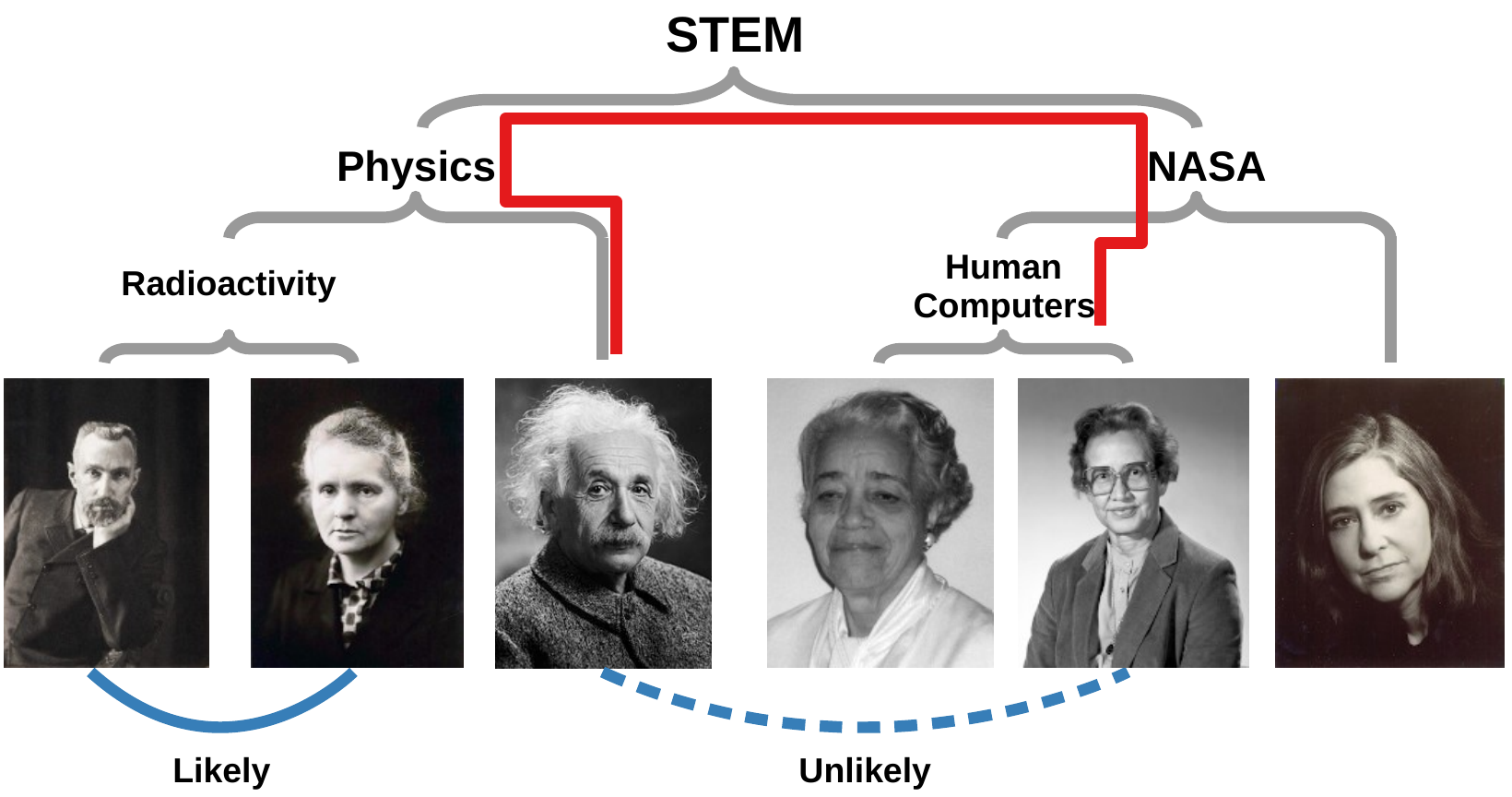}
\caption{An example of a Hierarchical Random Graph link prediction. The hierarchy fits the observed connections, showing that researchers in the same field are more likely to connect. Then HRG looks at pairs of nodes in the same part of the hierarchy that are not yet connected, and gives them a higher score.}
\label{fig:lp-hrg}
\end{figure}

With the Hierarchical Random Graph (HRG) model\cite{clauset2008hierarchical} we start to look at different approaches to link prediction. Its main difference with what we saw so far is that in HRG we're not just looking at pairs of nodes and their neighbors, but at the entire network. First we look at all connections and we create a hierarchical representation of it that fits the data. In practice, we want to group nodes in the same part of the hierarchy if they have a high chance of connecting. In our recurring example, the field of study of the scientist is a good way to group nodes. Then we say that it is more likely for nodes in the same part of the hierarchy to connect in the future if they haven't done so yet. Making a long path through this hierarchy to establish a new connection is less likely -- see Figure \ref{fig:lp-hrg}.

In HRG we're basically saying that communities matter: it is more likely for nodes in the same community to connect. Thus we fit the hierarchy and then we say that the likelihood of nodes to connect is proportional to the edge density of the group in which they both are. If the nodes are very related, the group containing both nodes might be a semi-clique with almost maximum density; if the nodes are far apart, the group containing both nodes might be just the entire network. In a schematic way: $score(u, v) = |E_c| / e(|E_c|)$, where $c$ is the community to which $u$ and $v$ belong, $|E_c|$ is the number of edges it contains, and $e(|E_c|)$ is the number of edges we expect it to contain, under a null random configuration model assumption (see Section \ref{sec:csmodels-conf}).

\section{Association Rules}\label{sec:lp-germ}
Just like HRG, GERM\cite{berlingerio2009mining}\cite{bringmann2010learning} is a peculiar approach to link prediction that has almost nothing in common with the standard approach of simply evaluating node-node similarity. GERM is short for Graph Evolution Rule Mining and it is rarely considered in link prediction surveys because it's a bit harder to implement and its only known implementation is proprietary software. But the approach deserves to be mentioned, given its cleverness.

GERM looks at any possible network motif (see Section \ref{sec:mining-motifs}) and counts how many times each appears in the network. This is a spectacularly hard problem, and I'll tell you how to perform it in the part of this book dedicated to graph mining (Chapter \ref{cha:mining-base}). For now, let's just assume that an oracle told us how many times each pattern occurs in our network.

\begin{figure}
\centering
\includegraphics[width=.55\columnwidth]{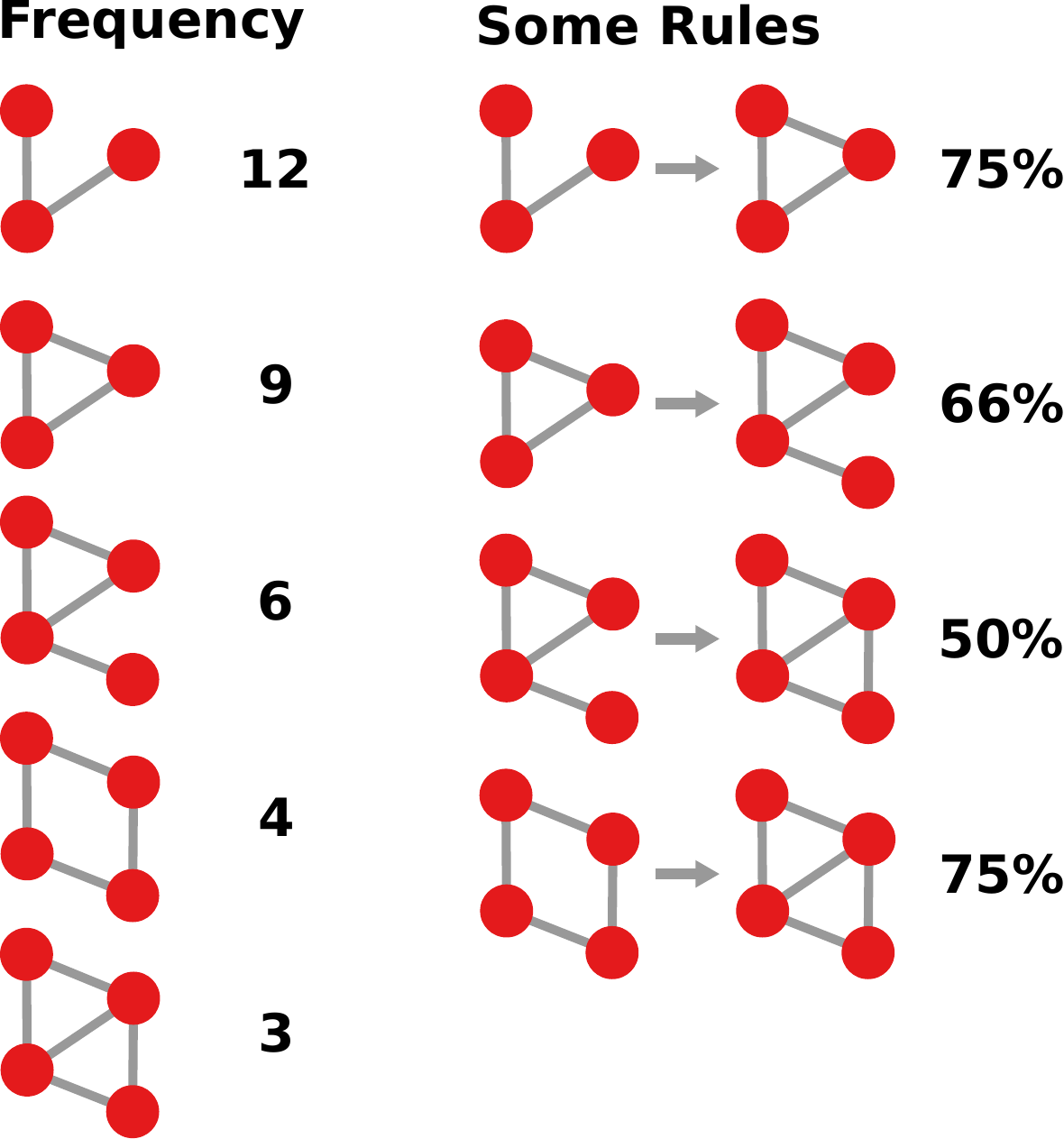}
\caption{An example of GERM. The frequency of each pattern is on the left. The graph evolution rules with their relative frequency is on the right. Note from the last two rules how the same consequent can be predicted with different levels of confidence by two different antecedents.}
\label{fig:germ}
\end{figure}

The idea is to identify all graph patterns that are a simple extensions of other, simpler patterns. By ``simple extension'' I mean that the consequent should have at most one additional edge -- and, possibly, one additional node -- added to its antecedent. This is the hard part. Once you detect all possible antecedent-consequent pairs, you can generate the rules, as Figure \ref{fig:germ} shows.

Suppose you have two patterns $G'$ and $G''$, which differ only by one edge -- with $G''$ including $G'$, for instance the top two patterns in Figure \ref{fig:germ}'s left column. Given their frequencies, you know that, $75\%$ of the times you see $G'$, you'll also see $G''$. So you can infer, with $75\%$ confidence, that $G'$ evolves into $G''$.

A crucial difference between GERM and whatever we saw so far is that it doesn't directly assign a similarity score to any two particular nodes. Each of the methods listed so far has a $score(u,v)$ for each $u,v$ pair. In GERM, rather than iterating over each pair of unconnected nodes to estimate their score, we iterate over rules and identify which pairs should be connected.

\begin{figure}
\centering
\includegraphics[width=\columnwidth]{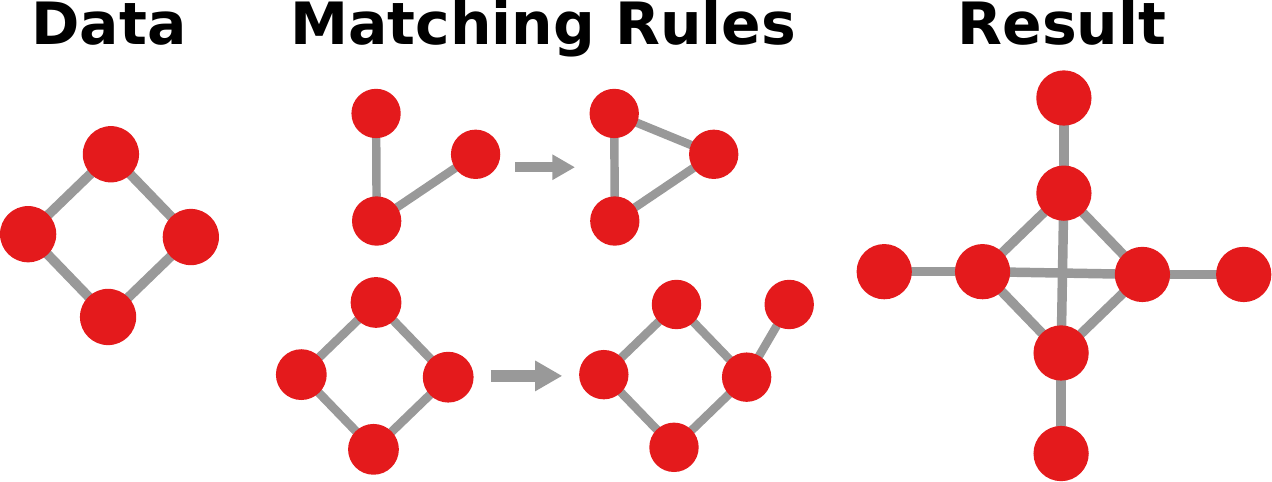}
\caption{An example of link prediction in GERM. The pattern matches two rules -- in the middle --, which we apply to all combinations of it to obtain, at the next time step, the extended pattern on the right.}
\label{fig:germ2}
\end{figure}

To understand the process, consider Figure \ref{fig:germ2}. Imagine our starting network is on the left. Suppose that we found two rules whose antecedents match the data. The first (on top) is a classic triad closure pattern that we see in many real world networks (see Section \ref{sec:density-clustering} for a refresher on clustering). The second (on the bottom) says that these ``square'' patterns attract a fifth node.

With the power of these two rules we know we can close the two open triads we have and add a new neighbor to each node in the original data. The end result is on the right. We now have more open triads and could apply the rules again, which would in turn create more square patterns and so on and so forth. In fact, one could use GERM not only as a link predictor but also as a graph generator (and put it in Chapter \ref{cha:physicsmodels}).

Note that I made two simplifications to GERM that the original papers don't make. First, in Figure \ref{fig:germ2}, I assumed that each rule applies with the same priority. I ignored its frequency and its confidence. Of course, that would be sub-optimal, so the papers describe a way to rank each candidate new edge according to the frequency and confidence of each rule that would predict it.

Second, in all my examples I always assumed that the rules add a new edge in the next time step and that all edges in the antecedent are present at the same time. In reality, GERM allows to have more complex rules spanning multiple time steps. You could have a rule saying something like: you have a single edge at time $t = 0$, you add a second edge at time $t = 1$ creating an open triad, and \textit{then} you close the triangle at time $t = 2$. This triad closure rule spans three time steps, rather than only two.

\begin{figure}
\centering
\includegraphics[width=.7\textwidth]{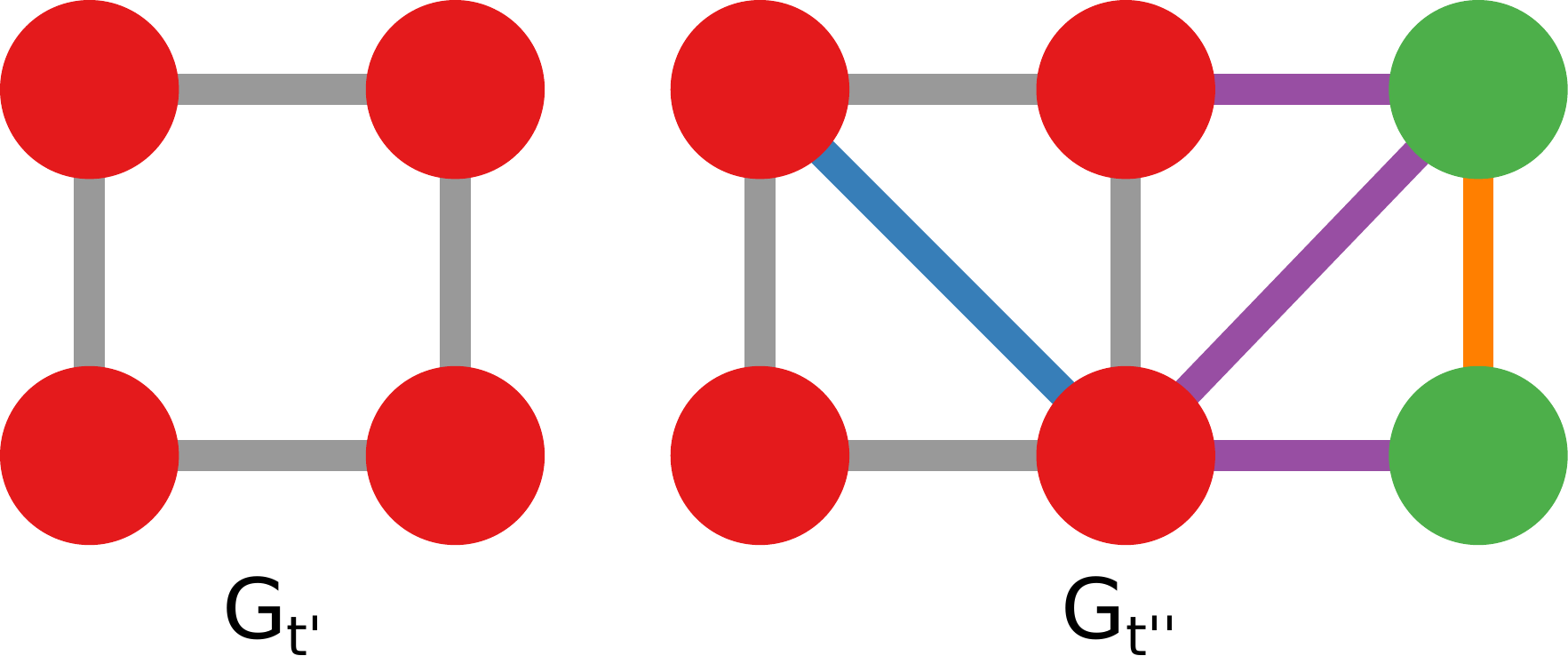}
\caption{Two observations of the graph $G$ at time $t'$ (left) and $t''$ (right). The node color encodes its type (red = ``old'', green = ``new''). The edge color encodes its type: gray = original link; blue = a new ``old-old'' link between two old nodes; purple = a new ``old-new'' link between an old node and a new node; orange = a new ``new-new'' link, between two nodes which were not in the graph at time $t'$.}
\label{fig:germ3}
\end{figure}

GERM has a final ace up its sleeve. We can classify new links coming into a network into three groups: old-old, old-new, and new-new. We base these groups according to the type of node they attach to. I show an example in Figure \ref{fig:germ3}. An ``old-old'' link appearing at time $t + 1$ connected two nodes that were already present in the network at time $t$. These are two ``old'' nodes. You can expect what  an ``old-new'' link is: a link connecting an old node with a node that was not present at time $t$ -- a ``new'' node. New nodes can also connect to each other in a ``new-new'' link. If the network represents paper co-authorships, this would be a new paper published by two or more individuals who have never published before.

Every method we saw so far -- and the ones we'll see -- predict exclusively ``old-old'' links. They work by creating a score between nodes $u$ and $v$ and, if either of those nodes were not present at time $t$, then their score is undefined. GERM is the only method I know that is able to predict also old-new and new-new links. Look again at Figure \ref{fig:germ2}: the result of GERM's prediction has more nodes than the original graph. That is because GERM predicted a few old-new links.

There's nothing stopping GERM to, in principle, predict also new-new links. However, estimating its precision in doing so is tricky. Thus the papers presenting the algorithm did not explore that dimension.

\begin{figure}
\centering
\includegraphics[width=.8\columnwidth]{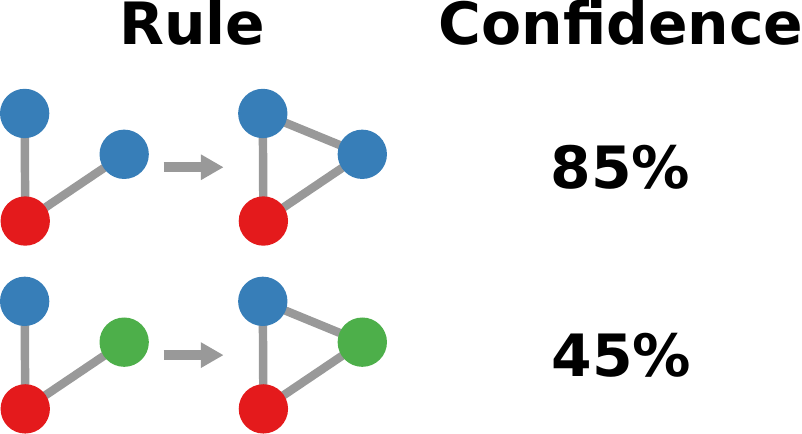}
\caption{An example of link prediction in GERM considering also node types. The node color here represent the node's label.}
\label{fig:germ4}
\end{figure}

Finally, note how all the rules I mentioned so far have no information on the nodes. This is a simplification I made which can be thrown away with ease. We can have labels on the nodes, so that we will obtain different link prediction scores for the same edges depending on the characteristics of the nodes. Figure \ref{fig:germ4} provides an example. Only one qualitative information can be represented at a time in this scenario. Also, implementing quantitative node attributes -- such as the degree -- is non-trivial, as one would ideally want to encode the fact that the attribute values $1$ and $2$ should be considered ``more similar'' to each other than $1$ and $1,000$. Extensions taking care of these limitations might be possible, but I'm not aware of one.

This is not a function unique to GERM, though. Many link prediction methods can be extended to take into consideration node attributes as well. In fact, this is also a key ingredient in some network generating processes. Node attributes are used, for instance, when modeling exponential random graphs, as we saw in Section \ref{seg:ergmodels-ergm}. In this case, differently than GERM, quantitative attributes represent no issue.

\section{Other Approaches}\label{sec:lp-other}
Link prediction is a vast subfield of network analysis. It is not quite as vast as community discovery (Part \ref{par:cd}) is, but we're not that far off. I cannot give justice to all methods out there. I chose the ones for the previous sections because they are the simplest and most didactic examples that everyone knows. In this section, I group together the most prominent examples of ``all the rest''.

I don't have the mental firepower to give you refined intuitions of the following methods as I did so far. Thus this is going to be just a big lump of concepts and formulas, necessarily superficial. If I did it any other way, this would not be a network analysis book, but a link prediction book. If you want to specialize in the field and really understand all of these methods -- and more! -- please go and read the review papers I cited at the beginning of the chapter, and a few newer ones\cite{ghasemian2020stacking}. Understood? Good.

And now: God helps us all, we're going in.\bigskip

\textit{Graph Embeddings}. This is a very big family of methods which has blasted into scene recently -- not only in link prediction, but in network analysis as a whole. We're going to see in details what a ``graph embedding'' is in Chapter \ref{cha:mining-embeddings}. For now, suffice to say that it is a way to represent a node as a vector of values. Depending on how you construct this vector of values, you could argue that similar nodes tend to connect to each other. On this basis, you can use any vector similarity measure to predict links in your network.

This is a dumbed-down version of the general approach shared by most of graph embedding link predictors. As you might expect, this general template has been applied in multiple ways to tackle different challenges. For instance, embeddings can feed their node representation to a deep neural network\cite{zhang2018link}; they can easily go beyond purely structural methods, because node attributes are just another entry in the vector that can be fed to a machine learning framework\cite{liao2018attributed}; finally, they have been used to predict relations in semantic graphs\cite{ji2015knowledge}\cite{trouillon2016complex}\cite{dettmers2018convolutional}, which encode relationships between different entities such as ``king marries queen''.

The list goes on and on and it would deserve a specialized book on the subject, but you get the picture so let's move on. \bigskip

\textit{Katz}\cite{katz1953new}. We already saw Katz's name when we talked about centrality (in Section \ref{sec:centr-eigen}). His idea of centrality was one where you get a centrality contribution from your neighbors, their neighbors, the neighbors of their neighbors, and so on. With each additional degree of separation, Katz established a penalty: the farther away a node is, the less it contributes to your centrality.

We can apply a similar strategy to derive our $score(u,v)$. Two nodes are strongly related if there are many short paths between them. So one would calculate all paths between $u$ and $v$ and sum their count. Of course, short paths contribute more because they represent a closer relationship. Thus the formula would be something like: $score(u,v) = \sum \limits_{l=1}^{\infty} \alpha^l|P_{u,v}^l|$.

The assumption is that, the more short paths are between $u$ and $v$, the more the two nodes are related. This is regulated by the $0 < \alpha < 1$ parameter: a lower $\alpha$ penalizes long paths more -- because they have a high $l$. Here, $\alpha$ plays the exact same role it did in Katz centrality. If we choose a very small $\alpha$, the Katz score is practically equivalent to counting common neighbors, as $\alpha^2 \sim 0$.

\begin{figure}[t]
\centering
\includegraphics[width=.66\columnwidth]{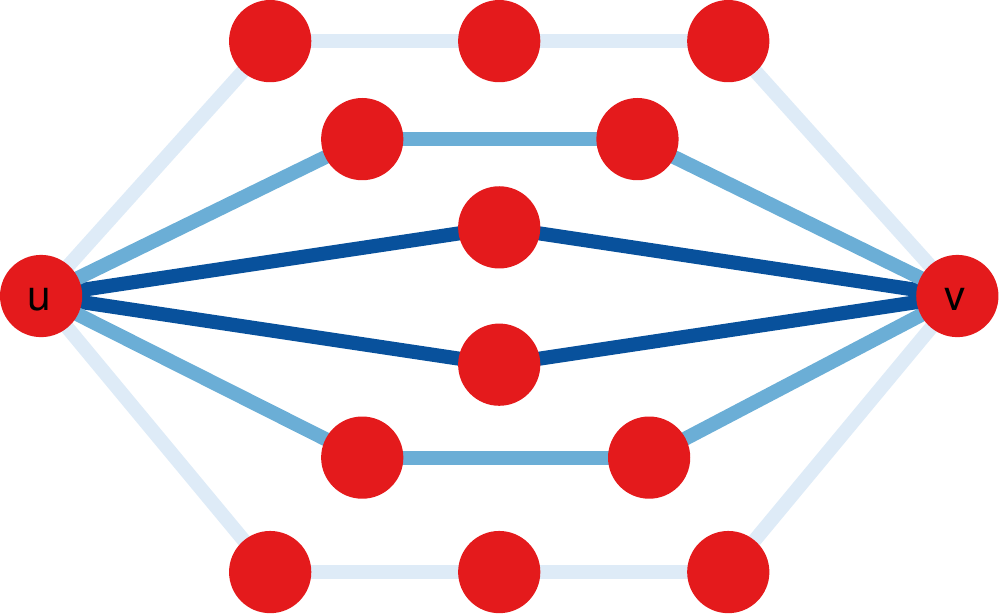}
\caption{An example of the Katz correction for using paths as a score for link prediction. The shade of blue of a path is proportional to its contribution to $score(u,v)$, with darker paths contributing more.}
\label{fig:katz-lp}
\end{figure}

Figure \ref{fig:katz-lp} shows an example of this correction, with longer paths fading to almost no contribution. \bigskip

\textit{Hitting Time}. We all tried to forget the shenanigans of Section \ref{sec:rw-hitime}, when I defined the hitting time: the expected number of steps it takes for a random walker to reach $v$ from $u$. Alas, we're reminded of it now, as there is a way to use $H_{u,v}$ as a basis for $score(u,v)$. After all, if $u$ and $v$ have very low hitting times, doesn't it mean that they are very related, and thus likely to connect?

\begin{figure}
\centering
\begin{subfigure}{.3\columnwidth}
\includegraphics[width=\textwidth]{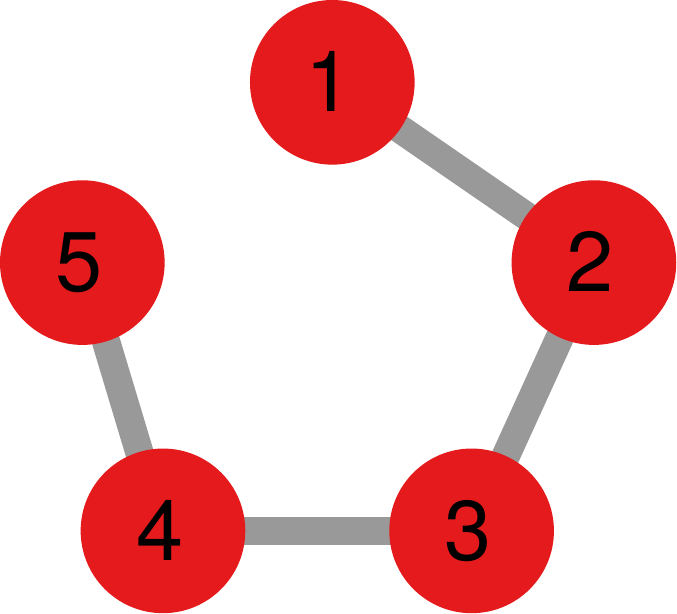}
\caption{}
\end{subfigure}\qquad
\begin{subfigure}{.4\columnwidth}
\includegraphics[width=\textwidth]{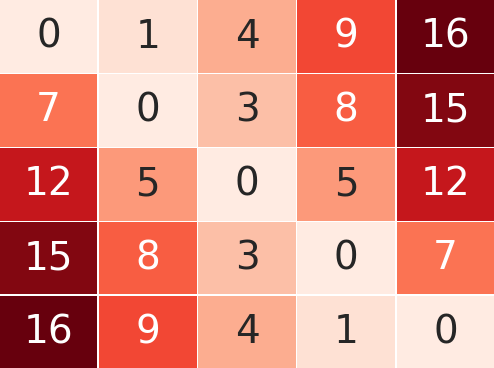}
\caption{}
\end{subfigure}
\caption{(a) A chain of five nodes. (b) The corresponding hitting time matrix $H$.}
\label{fig:hitime-lp}
\end{figure}

The easiest way to use hitting time as a score is to simply negate it ($score(u,v) = -H_{u,v}$) or negate the commute time ($score(u,v) = -(H_{u,v} + H_{v,u})$). For instance, in Figure \ref{fig:hitime-lp}, nodes $2$ and $4$ are more likely to connect ($score(2,4) = -(H_{2,4} + H_{4,2}) = -16$) than nodes $1$ and $5$ ($score(1,5) = -(H_{1,5} + H_{5,1}) = -32$) because they are closer. However, this has the problem of greatly favoring connections to hubs, since their hitting times as destinations are quite low. Thus some authors normalize them by multiplying $H_{u,v}$ with the stationary distribution of the target ($score(u,v) = -H_{u,v}\pi_v$). Hubs have higher stationary distribution values, thus they are penalized more.

Another problem of using $H_{u,v}$ is that the hitting time might increase even if $u$ and $v$ are nearby in the graph, simply because the graph has many vertices and edges that can lead the random walkers astray. To counteract this problem, a solution could be allowing the random walker to restart from $u$. This is practically equivalent to calculate the PageRank, with the difference that you fix the origin of the random walker. For this reason, since you random walker has a root ($u$), it is usually called ``rooted PageRank''. \bigskip

\textit{SimRank}\cite{jeh2003scaling}. As the name suggests, SimRank is based on an idea of node similarity. The more similar two nodes are, the more likely they are to connect. Similarity here is defined recursively: two nodes are similar if they are connected to similar neighbors. This is a definition in line with the philosophy of the regular equivalence we saw in Section \ref{sec:centr-similarity}. Thus, if we say that $score(u,u) = 1$, we can define all other scores as:

$$ score(u,v) = \gamma \dfrac{\sum \limits_{a \in N_u} \sum \limits_{b \in N_v} score(a,b)}{k_u k_v}.$$

$\gamma$ is a parameter you can tune. This is surprisingly similar to the hitting time approach. The expected value of a SimRank score is $\gamma^l$, where $l$ is the length of an average random walk from $u$ to $v$. \bigskip

\textit{Vertex similarity}\cite{leicht2006vertex}. The name of this approach should tip you off regarding its relationship with SimRank. However, it's actually much closer to the Jaccard variant of common neighbor. In fact, the only difference with Jaccard is the denominator. While Jaccard normalizes the number of common neighbors by the total possible number of common neighbors -- which is the union of the two neighbor sets -- this approach builds an expectation using a random configuration graph as a null model. This is a definition in line with the philosophy of the structural equivalence we saw in Section \ref{sec:centr-similarity}.

In practice, $score(u,v) = |N_u \cap N_v| / (k_u k_v)$. This is because two nodes $u$ and $v$ with $k_u$ and $k_v$ neighbors are expected to have $k_u k_v$ common neighbors (multiplied by a constant derived from the average degree which would not make any difference as it is the same for all node pairs in the network).

The same authors in the same paper also make a global variant of this measure. Their inspiration is the Katz link prediction, where they again provide a correction for a random expectation in a random graph with the same degree distribution as $G$. I won't provide the full derivation, which you can find in the paper, but their score is:

$$ score(u,v) = 2|E| \lambda_1 D^{-1} \left(I - \dfrac{\phi A}{\lambda_1} \right)^{-1} D^{-1}.$$

The elements in this formula are the usual suspects: $|E|$ is the number of edges, $\lambda_1$ is the leading eigenvalue of the adjacency matrix $A$ (not the stochastic, as that would be equal to one), $D$ is the degree matrix, and $I$ is the identity matrix. The only odd thing is $0 < \phi < 1$, which is a parameter you can set at will. This is similar to the parameter of Katz: smaller $\phi$ give more weight to shorter paths. \bigskip

\textit{Local and superposed random walks}\cite{liu2010link}. These two methods are a close sibling to the hitting time approach. To determine the similarity between $u$ and $v$, we place a random walker on $u$ and we calculate the probability it will hit node $v$. Note that, if we were to do infinite length random walks, this would be the stationary distribution $\pi$. This would be bad, as you know that this only depends on the degree of $v$, not on your starting point $u$. For this reason, the authors limit the length of the random walk, and also add a vector $q$ determining different starting configurations -- namely, giving different sources different weights.

To sum up, the local random walk method determines $score(u,v) = q_u\pi_{u,v} + q_v\pi_{v,u}$. The superposed variant works in the same way, with the difference that the random walker is constantly brought back to its starting point $u$. This tends to give higher scores to nodes closer in the network. \bigskip

\textit{Stochastic block models}\cite{guimera2009missing}. We saw the stochastic block models (SBM) as a way to generate graphs with community partitions (Section \ref{sec:csmodels-comms}) -- and we will see them again as a method to detect communities (Section \ref{sec:cd-partition-sbm}). In fact, any link prediction approach, in a sense, is a graph generating model. Given the close relationship of SBMs with community discovery, this class of solutions is particularly related to the Hierarchical Random Graph approach.

\begin{figure}
\centering
\begin{subfigure}{.33\columnwidth}
\includegraphics[width=\textwidth]{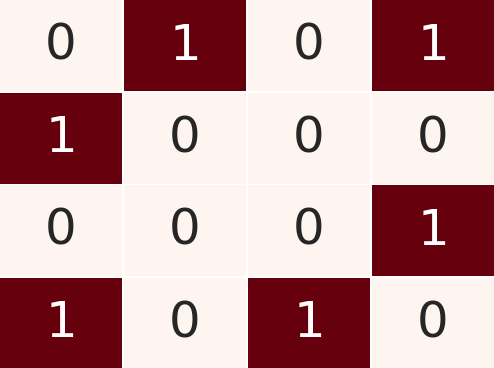}
\caption{}
\end{subfigure}\qquad
\begin{subfigure}{.33\columnwidth}
\includegraphics[width=\textwidth]{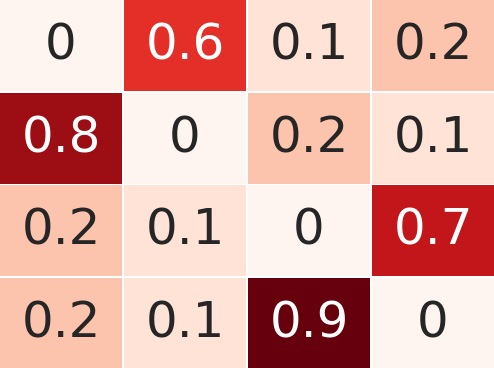}
\caption{}
\end{subfigure}
\caption{(a) The adjacency matrix of a simple graph. (b) One of the possible connection probability matrices that could generate the graph in (a). Each cell reports the probability of observing the edge.}
\label{fig:lp-sbm}
\end{figure}

Suppose you're observing a graph, in the form of its adjacency matrix (Figure \ref{fig:lp-sbm}(a)). Given an adjacency matrix, we can infer a matrix of connection probabilities (Figure \ref{fig:lp-sbm}(b)), telling us the likelihood of observing each edge. We don't need to know how this works -- we'll study this process in details when it comes to use SBMs for community detection -- but for now suffice to say we use the Expectation Maximization algorithm\cite{moon1996expectation}.

This matrix will give you a probability of observing a connection that isn't there (yet). You can use that probability as your $score(u,v)$ for your link prediction. The cool thing about this approach is that, differently from the ones we saw so far, it can also tell you when an observed link is likely to be spurious, because it is associated with a low probability.

This is a family of solutions, because you can bake in different assumptions on how your stochastic blockmodel works. For instance, you can have mixed-membership ones where nodes are allowed to be part of multiple blocks. Or you can have versions working with multilayer networks.\bigskip

\textit{Probabilistic models}\cite{friedman1999learning}\cite{heckerman2007probabilistic}\cite{yu2007stochastic}. In this subsection I group not a single method, but an entire family of approaches to link prediction. They have their differences, but they share a common core. In probabilistic models, you see the graph as a collection of edges and attributes attached to both nodes and edges. The hypothesis is that the presence of an edge is related to the values of the attributes.

In practice, the hypothesis is that there exist a function taking as input the attribute values of each node and edge. This function models the observed graph. Then, depending on the values of the attributes for nodes $u$ and $v$, the function will output the probability that the $u,v$ edge should appear -- and with which attributes. \bigskip

\textit{Mutual information}\cite{tan2014link}. In Section \ref{sec:prob-mi} I introduced the concept of mutual information: the amount of information one random variable gives you over another. This can be exploited for link prediction. If you remember how it works, you'll remember that MI allows you to calculate the relationship between two non-numerical vectors, which is not really possible using other correlation measures -- not without doing some non-trivial bending of the input. In link prediction, this advantage is crucial: you can define your function as $score(u,v) = MI_{uv}$, where the ``events'' that allow you to calculate $MI_{uv}$ are the common neighbors between nodes $u$ and $v$. \bigskip

\textit{CAR Index}\cite{cannistraci2013link}. In this index you favor pairs of nodes that are part of a local community, i.e. they are embedded in many mutual connections. This is a variant of the idea of common neighbor: each shared connection counts not equally, but proportionally more if it also shares neighbors with $u$ and $v$. This basic idea can be implemented in multiple ways, depending on which of the traditional link prediction methods we want to extend. For instance, if we extend vanilla common neighbors, you'd say that:

$$ score(u,v) = \sum \limits_{z \in N_u \cap N_v} 1 + \dfrac{|N_u \cap N_v \cap N_z|}{2}.$$

\begin{figure}
\centering
\includegraphics[width=.6\columnwidth]{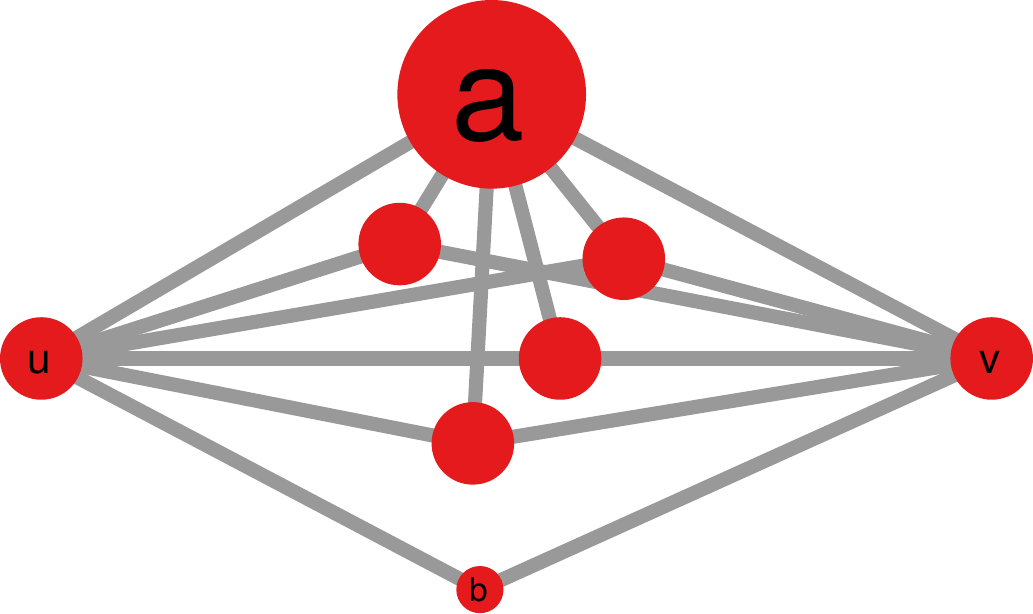}
\caption{Comparing the CAR index contribution to $score(u,v)$ for nodes $a$ and $b$. Node size is proportional to the contribution.}
\label{fig:car-index-lp}
\end{figure}

Note how here we simply added a second intersection to the basic common neighbors, the one with the common neighbor $z$. Figure \ref{fig:car-index-lp} shows an example. Node $a$ in this case contributes much more than node $b$, because it shares four common neighbors with $u$ and $v$. Node $b$, on the other hand, lies outside this local community.

One could use this CAR approach to create a new family of measures. For instance, we can have a CAR version of the resource allocation approach:

$$ score(u,v) = \sum \limits_{z \in N_u \cap N_v} \dfrac{|N_u \cap N_v \cap N_z|}{|N_z|}.$$

Here, we normalize by $z$'s bandwidth. \bigskip

\textit{Katz Tensor Factorization}\cite{dunlavy2011temporal}. This approach doesn't really add a new idea to link prediction, but it is worthwhile mentioning for a few reasons. It is an application of the Katz criterion we saw earlier. However, it implements it via tensor factorization. If you recall Section \ref{sec:mat-mat-mat}, we could view particular tensors as ``3D'' matrices. In practice, they are a collection of adjacency matrices. In that section I introduced the idea to represent multilayer networks, where each adjacency is a layer of the network. Here, instead, each layer is a temporal snapshot of the network.

The advantage of this approach is that, like GERM (Section \ref{sec:lp-germ}), one could predict links not necessarily at time $t + 1$, but also at $t+ 2$, $t +3$, etc... This is possible because the adjacency matrix at any time $t$ is simply a slice of the tensor. Thus, there's nothing fundamentally different between predicting a link in the slice $t + 1$ or in the slice $t + 3$ -- your precision might be a bit lower because of the strongly sequential nature of link appearance, but it's still possible to provide a good guess. \bigskip

\textit{Rank aggregation techniques}\cite{dwork2001rank}. When it comes to link prediction, you can go full meta. And I never drop an opportunity to go full meta. You can take all -- yes, I mean all -- the methods listed so far. Each method will rank unobserved edges differently. Meaning that they have an ordered list of preferences as to which new link should be the next one observed. You can use a rank aggregation method to create a final list that uses all the information from all the methods. Rank aggregation is the general process of having two ordered list and producing a single ordered list that is the one agreeing the most with both your inputs.

There is a classical way to solve the issue, which is the Borda's method\footnote{\url{https://en.wikipedia.org/wiki/Borda_count}}. This is what happens in those electoral systems where citizens will vote not for one candidate, but will rank their top $n$ candidates. Each candidate receives a number of points proportional to its rank, and the candidate with most points win.

A more sophisticated aggregation methods uses the Kendall $\tau$. The Kendall $\tau$ counts the number of pairwise disagreements: pairs of edges that have the opposite rankings in the two lists. If in one list $u_1,v_1$ is ranked higher than $u_2,v_2$ while the opposite holds in the other list, then you have a disagreement -- this is sort of similar to the Spearman rank correlation\cite{bonett2000sample}.

\section{Summary}

\begin{enumerate}
\item In link prediction we want to take an observed network and infer the most likely connections to appear in the future, or the ones that might already be there but for some reason we aren't seeing yet.
\item The most common approach is to compute a score for each pair of unconnected nodes by using some theory about the topology of the network. For instance, we can say that usually triangles close and thus count the number of common neighbors between two nodes.
\item Other approaches model mesoscale structures of the network such as communities, or find overexpressed graph patterns in the network and rank node pairs on whether they are likely to make more of these patterns appear in the network.
\item Many approaches from other branches of network science can be used to predict links, for instance ranking algorithms (Katz), random walk hitting time, stochastic blockmodels, mutual information, etc. 
\item Nothing stops you from using all the link prediction methods at once and then aggregate their results. Really, it's a free country.
\end{enumerate}

\section{Exercises}

\begin{enumerate}
\item What are the ten most likely edges to appear in the network at \url{http://www.networkatlas.eu/exercises/23/1/data.txt} according to the preferential attachment index?
\item Compare the top ten edges predicted for the previous question with the ones predicted by the jaccard, Adamic-Adar, and resource allocation indexes.
\item Use the mutual information function from \texttt{scikit-learn} to implement a mutual information link predictor. Compare it with the results from the previous questions.
\item Use your code to calculate the hitting time (from exercise 3 of Chapter \ref{cha:rndwalks}) to implement a hit time link predictor -- use the commute time since the network is undirected. Compare it with the results from the previous questions.
\end{enumerate}

\chapter{For Multilayer Graphs}\label{cha:lp-multilayer}
Link prediction takes a distinctive new flavor when your input is a multilayer network. In single layer networks, all you have to do is asking the question: ``who will be the next two people to become friends with each other?'' The question becomes harder and more interesting in multilayer networks. Here you don't want to only know \textit{which} two people will connect next, but also \textit{how}. Are they going to be best buds? Work colleagues? Lovers? Enemies? You see that this new dimension adds a lot of spice to the problem. You don't want to make a friend suggestion on Facebook for two people with a strong connection prediction if you also knew that the connection type between the two would be an enmity link -- and researchers in social media studies have looked at this problem\cite{tang2015negative}.

Multilayer link prediction is the topic of this chapter. We start from the simplest case where we have only two layers in the network with a very precise semantic: link prediction in signed networks (Section \ref{sec:lp-multilayer-signed}). We then move on to the generalized case of an arbitrary number of layers with no clear semantic relationship with each other (Section \ref{sec:lp-multilayer-general}).

\section{Signed Networks}\label{sec:lp-multilayer-signed}

\subsection{Social Balance Theory}
There are two reasons why signed networks represent the simplest case of link prediction when your network has multiple different types of connections. First, signed networks are a subtype of multilayer networks with strong constraints on the edges. You can only have two edge types: positive and negative. Moreover, these edge types are exclusive: if you have a positive edge between $u$ and $v$, you cannot have also a negative one -- unless the network is directed and the edge direction flows in the opposite way. This reduces the search space for a link predictor.

The second reason why signed networks are easier to predict is because the positive/negative sign of an edge gives it a precise meaning. We have strong priors as to which structures we can see in a signed network. These are discussed in what we call ``Social Balance Theory''\cite{heider2013psychology}\cite{antal2005dynamics}. According to the theory, positive and negative relationships are balanced. For instance, if I have two friends, they are more likely to like each other. On the other hand, if I have an enemy, I expect my friends to be enemies of them as well.

\begin{figure}
\centering
\begin{subfigure}[t]{.21\columnwidth}
\includegraphics[width=\textwidth]{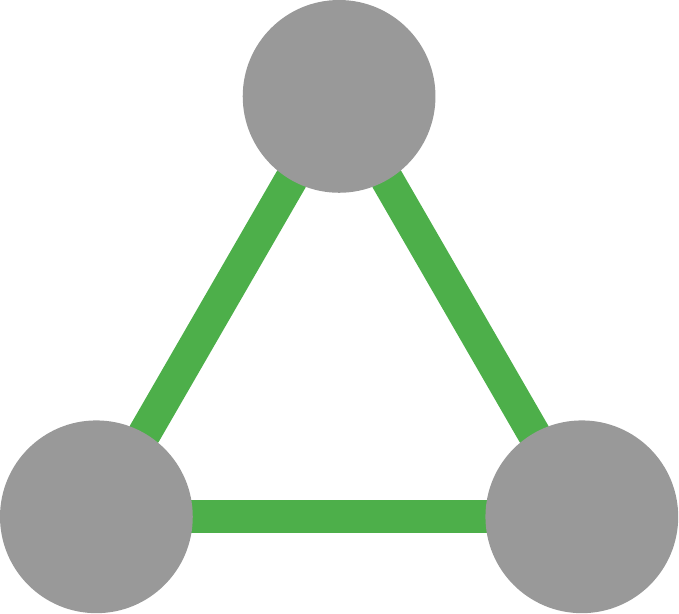}
\caption{}
\end{subfigure}
\quad
\begin{subfigure}[t]{.21\columnwidth}
\includegraphics[width=\textwidth]{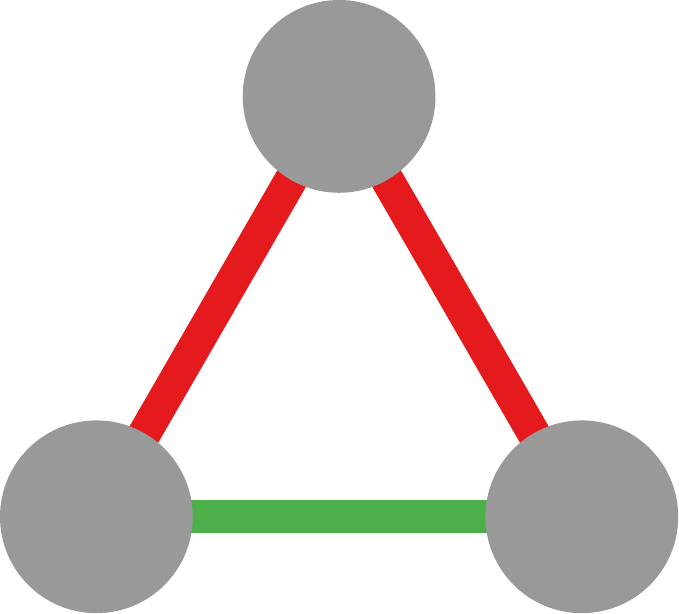}
\caption{}
\end{subfigure}
\quad
\begin{subfigure}[t]{.21\columnwidth}
\includegraphics[width=\textwidth]{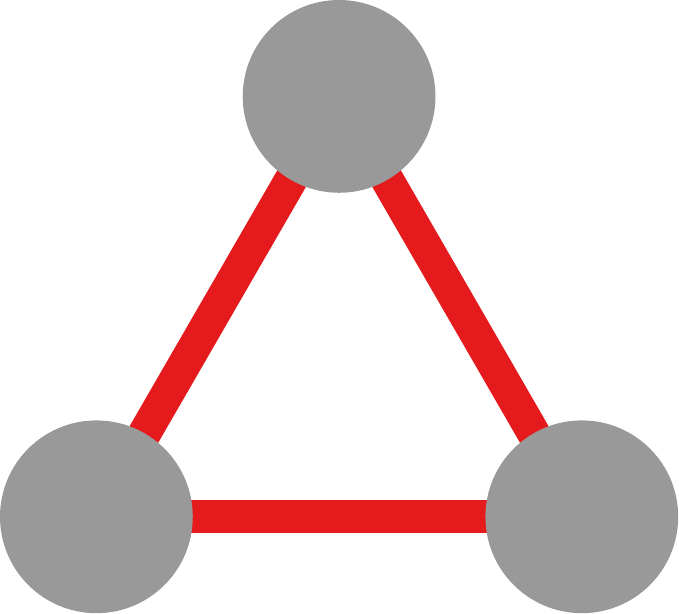}
\caption{}
\end{subfigure}
\quad
\begin{subfigure}[t]{.21\columnwidth}
\includegraphics[width=\textwidth]{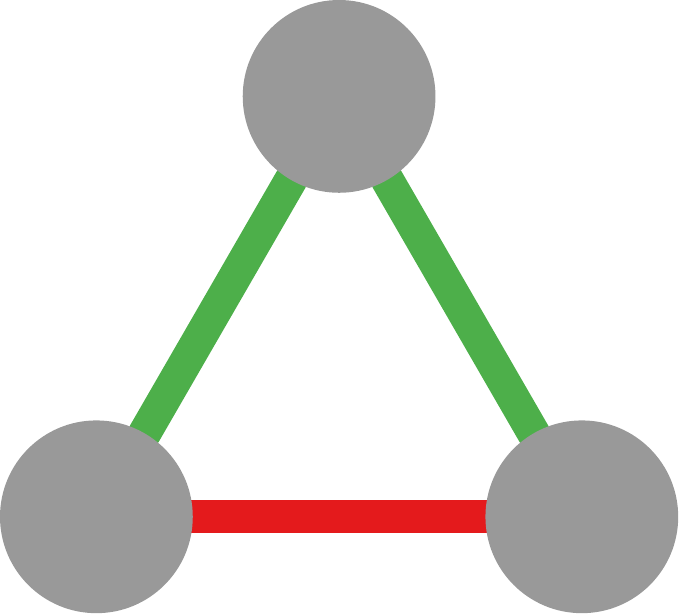}
\caption{}
\end{subfigure}
\caption{The four possible types of triangles when considering a mutually exclusive pair of positive (in green) and negative (in red) relationships. (a) and (b) are balanced triangles because they have an odd number of positive relationships. (c) and (d) are classically considered unbalanced, although, under certain circumstances, (c) can be considered a balanced or neutral configuration.}
\label{fig:socialbalance}
\end{figure}

Social balance theory looks predominantly at triangles -- although there are ways to look at longer cycles\cite[9\baselineskip]{chiang2011exploiting}. It divides them in two classes: balanced and unbalanced, see Figure \ref{fig:socialbalance}. Balanced triangles can be understood with common sense: the friend of my friend is my friend, the enemy of my friend is my enemy. Unbalanced triangles are relationships that we expect to change in the future. In fact, the prediction is that balanced triangles are overexpressed in real networks over our expectation -- and unbalanced triangles are underexpressed --, and that is generally observed.

One note about the all-negative triangle (Figure \ref{fig:socialbalance}(c)): in some views it is not considered unbalanced, and it is in fact more commonly found in real networks than the other unbalanced triangle (Figure \ref{fig:socialbalance}(d)). A typical case of balanced all-negative triangle is campanilism in Tuscany: the worst enemy of a person from Pisa is a person from Livorno. The second worst enemy for both of them is a person from Lucca. And people from Lucca hate indiscriminately both Pisa and Livorno. And this has gone on for centuries. Pretty balanced.

\begin{figure}
\centering
\includegraphics[width=.66\columnwidth]{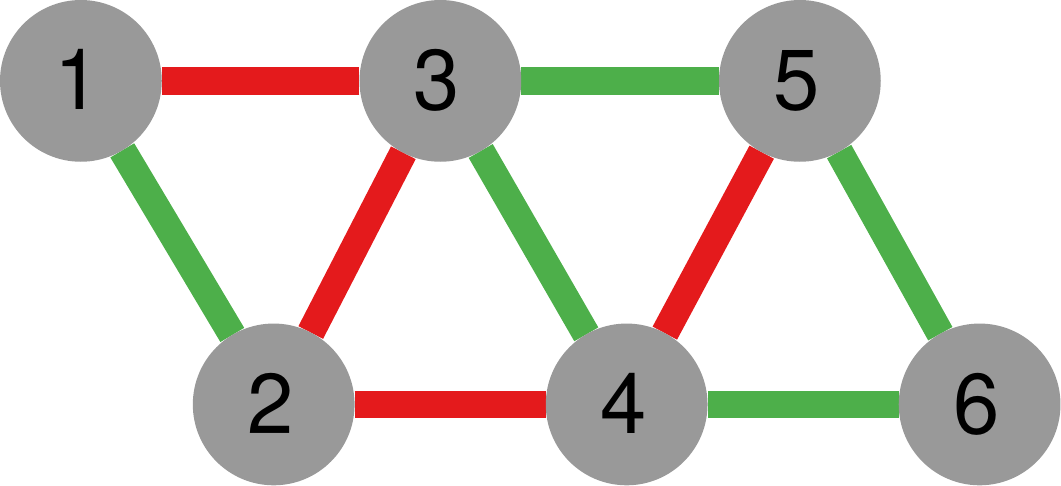}
\caption{A graph with two unbalanced triangles, the ones including edge $(4,5)$.}
\label{fig:frustration}
\end{figure}

You can calculate a summary statistics telling how much, on average, your whole network is balanced. There are many ways to do this, but I think the most popular one is called \textit{frustration}\cite{harary1959measurement}. In frustration, you count the number of edges whose removal -- or negation -- would result in a perfectly balanced network. You can normalize this over the total number of edges in the network. Figure \ref{fig:frustration} shows an example network with two unbalanced triangles: $(3,4,5)$ and $(4,5,6)$. Both triangles would turn balanced if we were to flip the sign of edge $(4,5)$ -- or, alternatively, frustration would dissipate if we were to remove the edge altogether. Thus, the frustration of this graph is $1 / 9$, since it contains $9$ edges.

\begin{figure}
\centering
\begin{subfigure}[t]{.4\columnwidth}
\includegraphics[width=\textwidth]{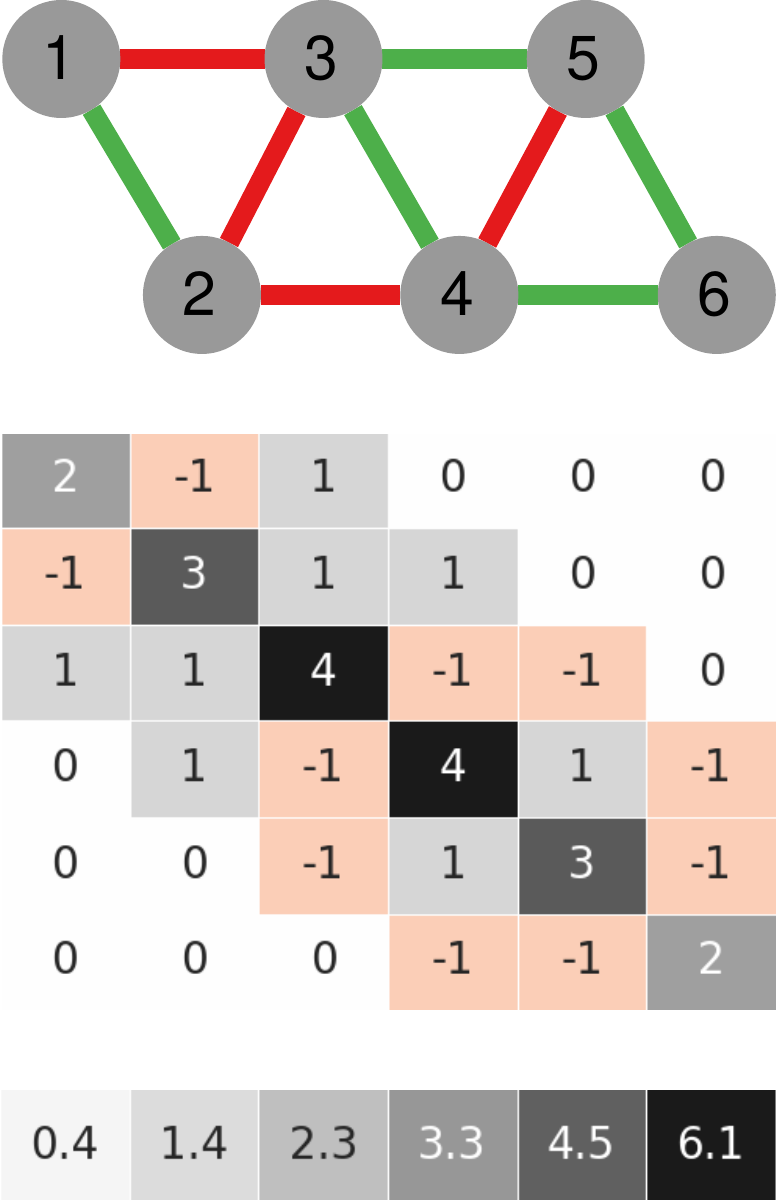}
\caption{}
\end{subfigure}
\qquad\qquad
\begin{subfigure}[t]{.4\columnwidth}
\includegraphics[width=\textwidth]{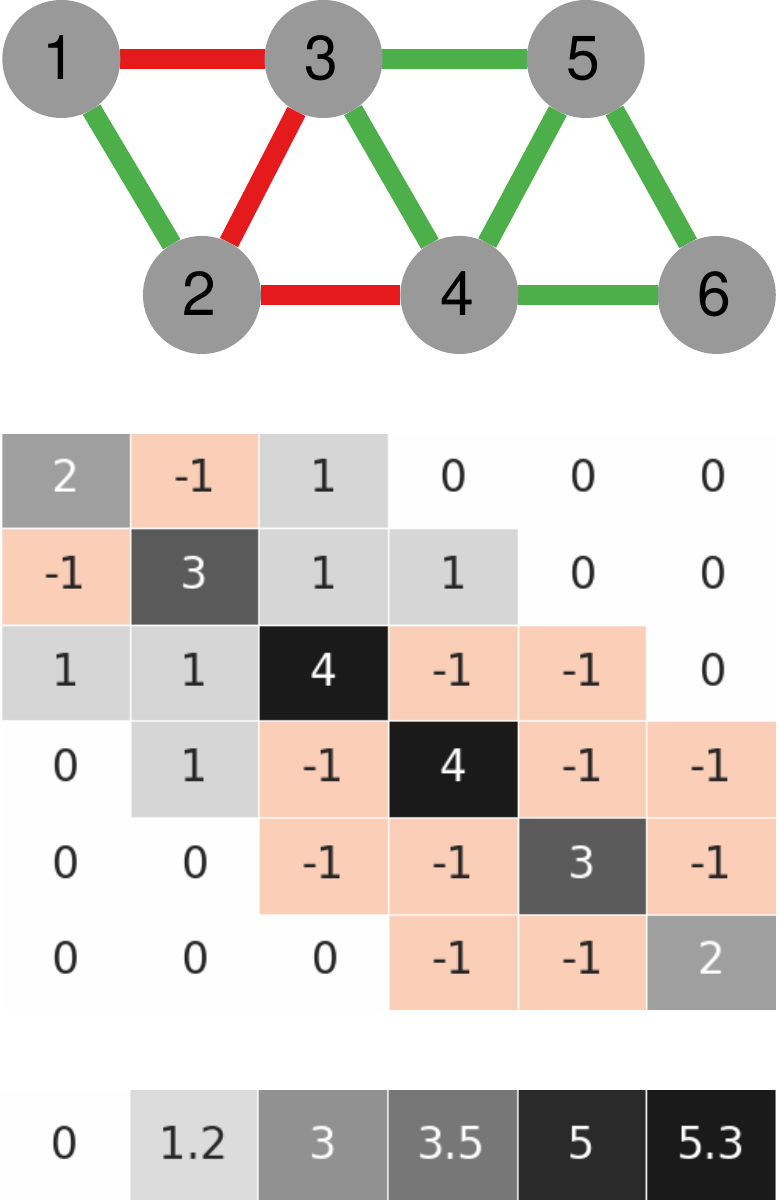}
\caption{}
\end{subfigure}
\caption{From top to bottom: signed graph, its signed Laplacian, the sorted eigenvalues of the signed Laplacian. (a) Unbalanced graph. (b) Balanced graph.}
\label{fig:frustration-eigenvalue}
\end{figure}

Frustration is a bit computational complex to calculate, but there are heuristics you can use to speed up your calculations\cite{aref2019balance}. One relatively simple check you can do to figure out whether your signed network is balanced is by looking at the smallest eigenvector of its signed Laplacian -- I told you how to calculate the signed Laplacian in Section \ref{sec:mat-mat-laplacian}. The smallest eigenvector of a balanced graph will be zero. If your signed graph is not balanced, the smallest eigenvector of its signed Laplacian will be higher than zero. In fact, how much higher than zero it is puts an upper bound to the frustration value\cite{belardo2014balancedness} -- that is to say, frustration can only be lower than or equal to the smallest eigenvector of the signed Laplacian. Figure \ref{fig:frustration-eigenvalue} shows you an example of balanced and an example of unbalanced graph with their eigenvectors.

This relates to link prediction when we consider evolving signed networks. If we find a configuration with three nodes connected by two positive edges, it is overwhelmingly more likely that, in the future, the triangle will close with a positive relationship (Figure \ref{fig:socialbalance}(a)) rather than with a negative one (Figure \ref{fig:socialbalance}(d)). On the other hand, if we find a positive and a negative relationship, we expect the triangle to close with a negative edge (Figure \ref{fig:socialbalance}(b)). The case with an initial condition of two negative edges is more difficult to close, but we prefer to close it with a positive edge (Figure \ref{fig:socialbalance}(b)) than with a negative one (Figure \ref{fig:socialbalance}(c)).

So you see that you can perform signed link prediction by first predicting the pair of nodes that will connect, calculating a $score(u,v)$ with any of the methods presented in the previous chapter. Then you will decide the sign of the link, by using social balance theory.

\subsection{Social Status Theory}
There is a competing theory to social balance, which is the status theory\cite{leskovec2010signed}. This arises from a different interpretation of the sign. A positive sign in a social setting might mean that the user originating the link feels to be lower status than -- and thus giving social credit to -- whomever receives the link. Conversely, a negative link is a way for a higher status node to shoot down a lower status one. Note that here we started talking about the direction of an edge, meaning that we have more than four types of triangles. In fact, we have $32$.

\begin{figure}
\centering
\begin{subfigure}[t]{.2\columnwidth}
\includegraphics[width=\textwidth]{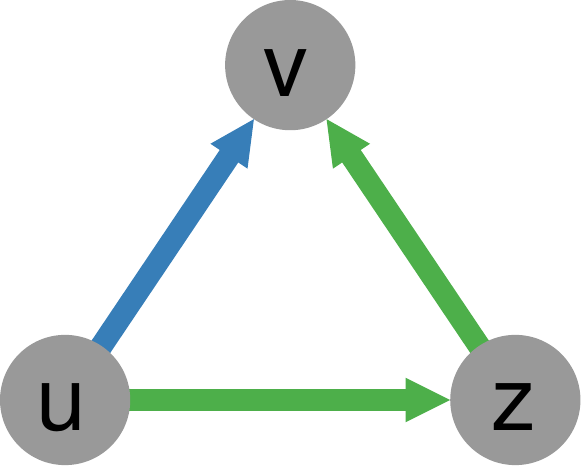}
\caption{}
\end{subfigure}
\quad
\begin{subfigure}[t]{.2\columnwidth}
\includegraphics[width=\textwidth]{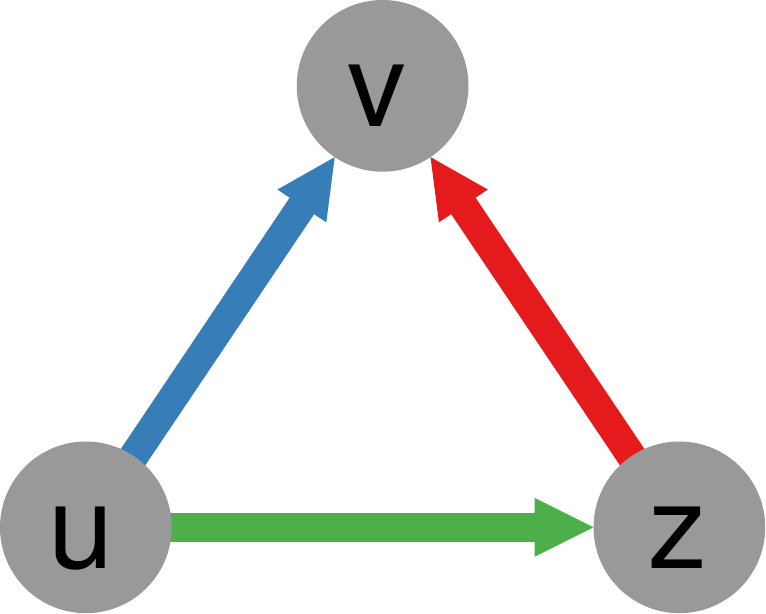}
\caption{}
\end{subfigure}
\quad
\begin{subfigure}[t]{.2\columnwidth}
\includegraphics[width=\textwidth]{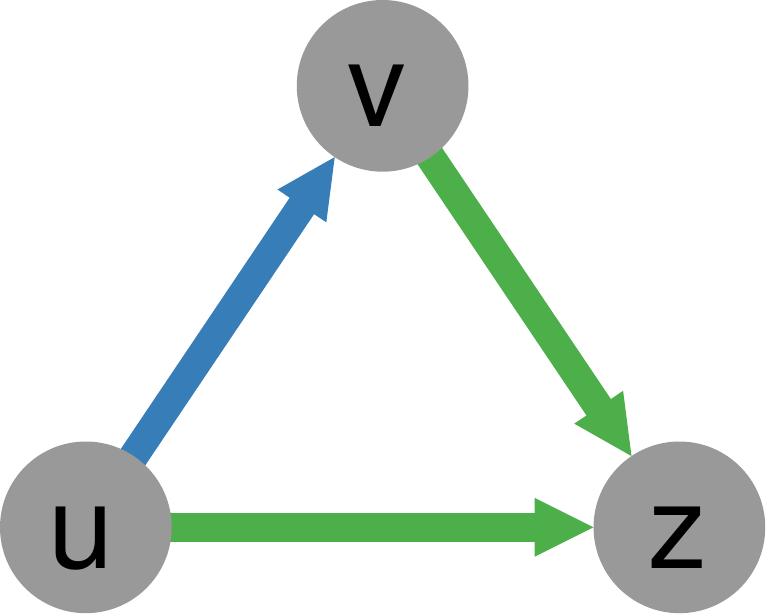}
\caption{}
\end{subfigure}
\quad
\begin{subfigure}[t]{.2\columnwidth}
\includegraphics[width=\textwidth]{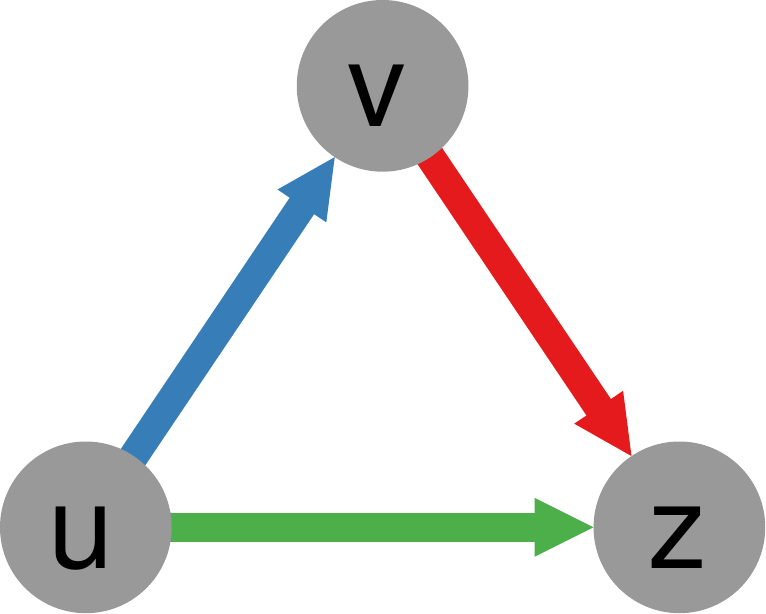}
\caption{}
\end{subfigure}
\quad
\begin{subfigure}[t]{.2\columnwidth}
\includegraphics[width=\textwidth]{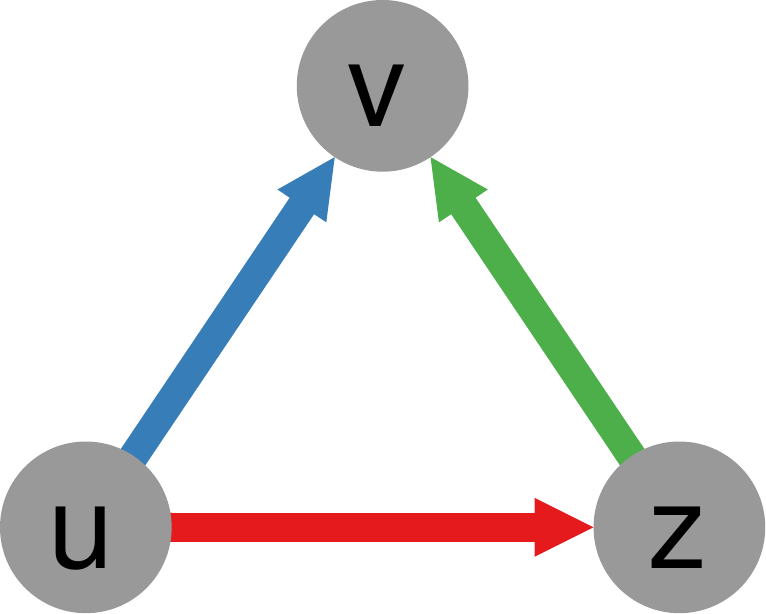}
\caption{}
\end{subfigure}
\quad
\begin{subfigure}[t]{.2\columnwidth}
\includegraphics[width=\textwidth]{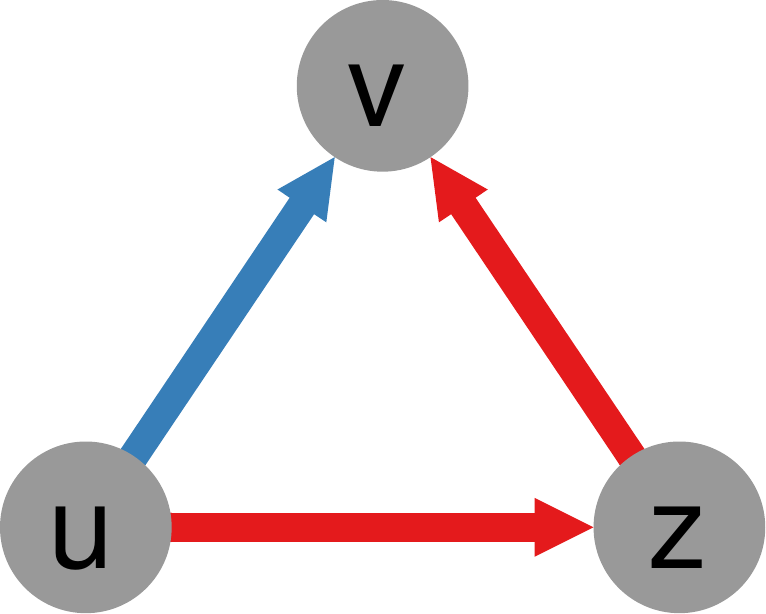}
\caption{}
\end{subfigure}
\quad
\begin{subfigure}[t]{.2\columnwidth}
\includegraphics[width=\textwidth]{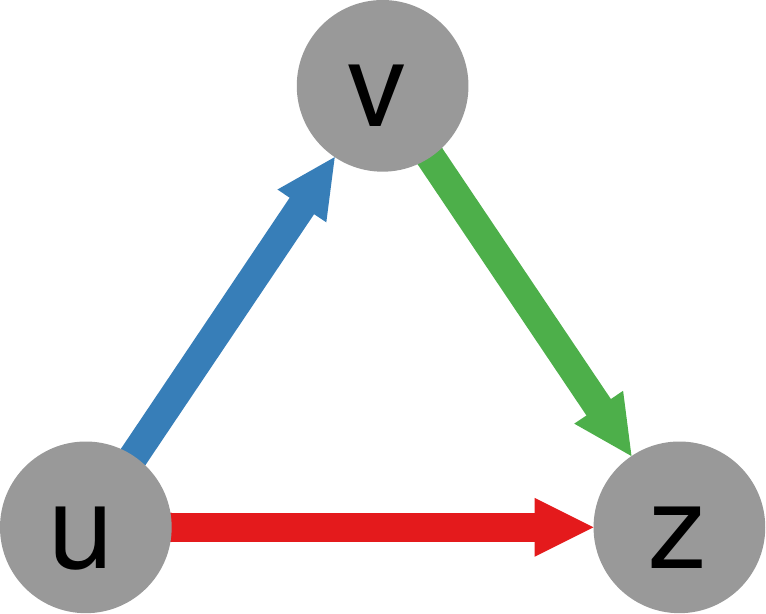}
\caption{}
\end{subfigure}
\quad
\begin{subfigure}[t]{.2\columnwidth}
\includegraphics[width=\textwidth]{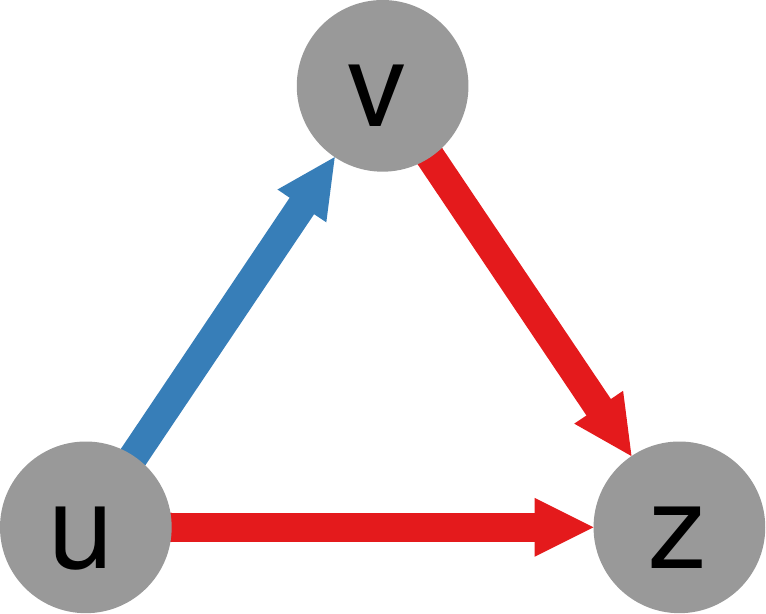}
\caption{}
\end{subfigure}
\quad
\begin{subfigure}[t]{.2\columnwidth}
\includegraphics[width=\textwidth]{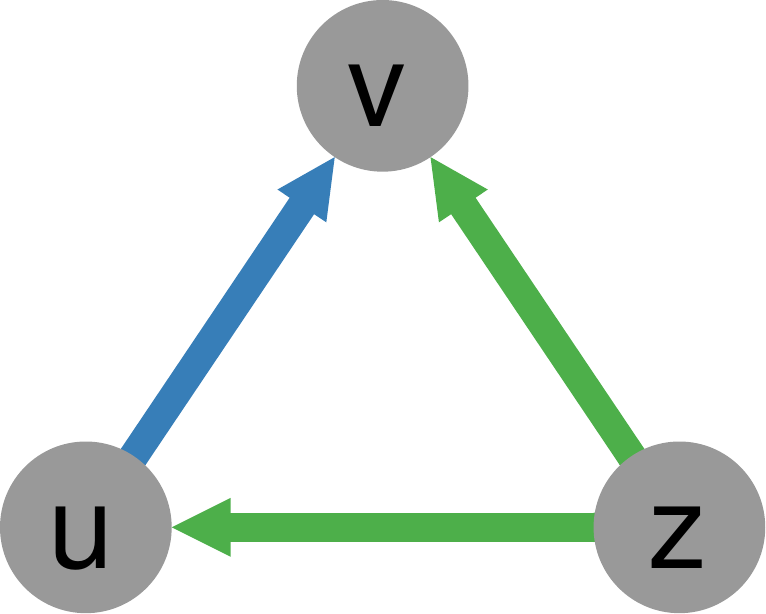}
\caption{}
\end{subfigure}
\quad
\begin{subfigure}[t]{.2\columnwidth}
\includegraphics[width=\textwidth]{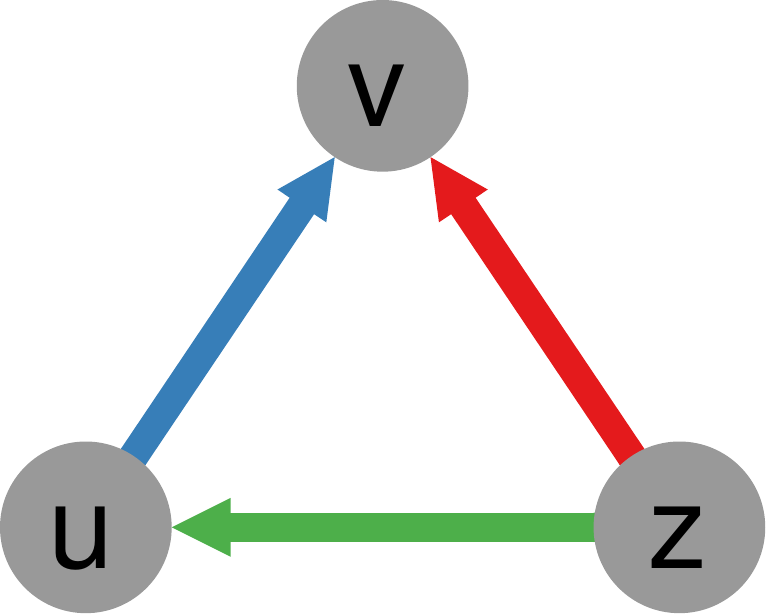}
\caption{}
\end{subfigure}
\quad
\begin{subfigure}[t]{.2\columnwidth}
\includegraphics[width=\textwidth]{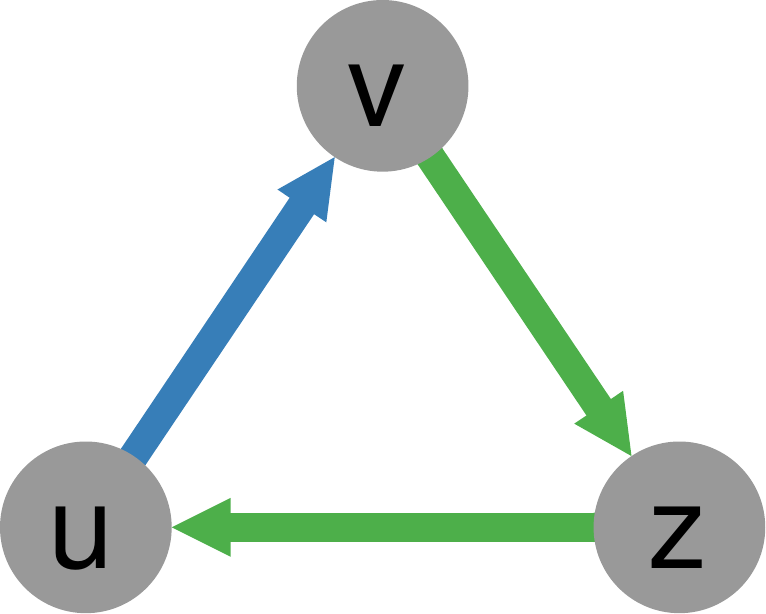}
\caption{}
\end{subfigure}
\quad
\begin{subfigure}[t]{.2\columnwidth}
\includegraphics[width=\textwidth]{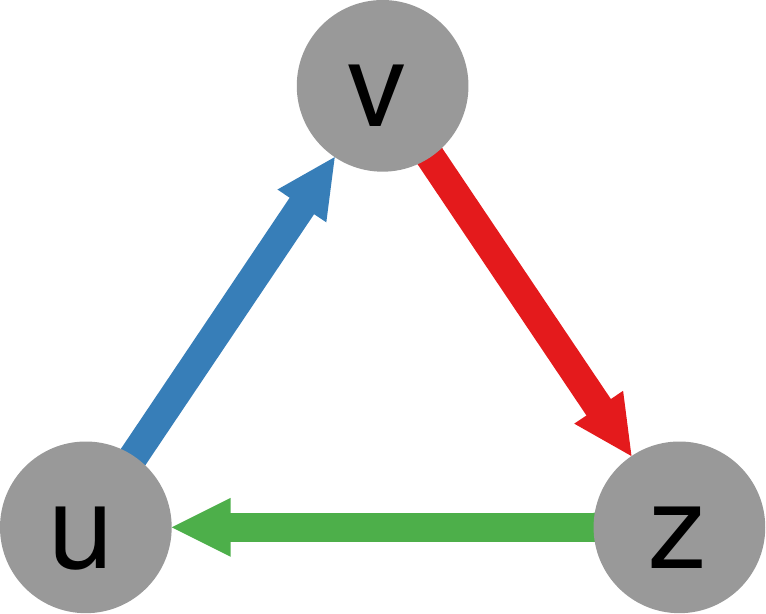}
\caption{}
\end{subfigure}
\quad
\begin{subfigure}[t]{.2\columnwidth}
\includegraphics[width=\textwidth]{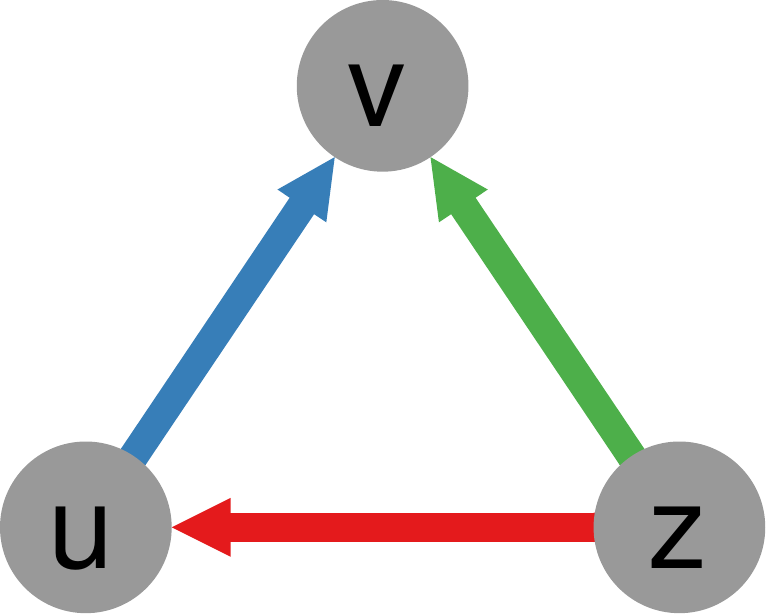}
\caption{}
\end{subfigure}
\quad
\begin{subfigure}[t]{.2\columnwidth}
\includegraphics[width=\textwidth]{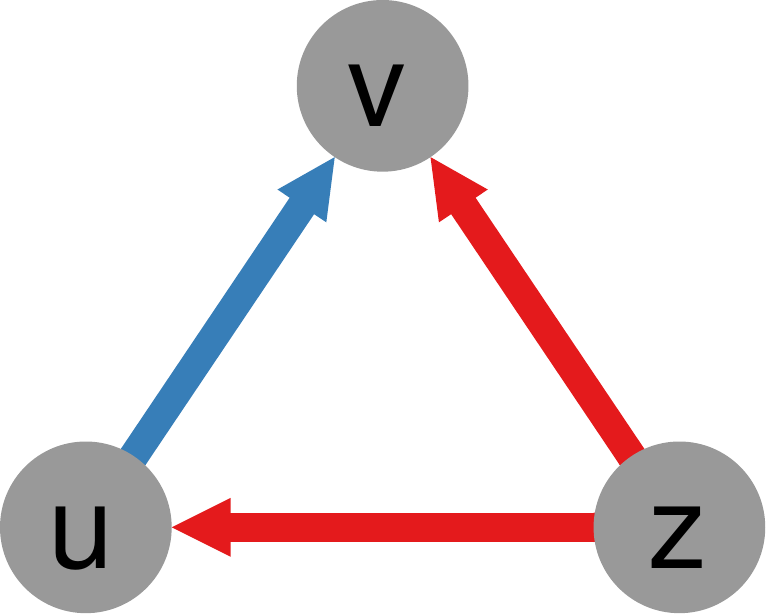}
\caption{}
\end{subfigure}
\quad
\begin{subfigure}[t]{.2\columnwidth}
\includegraphics[width=\textwidth]{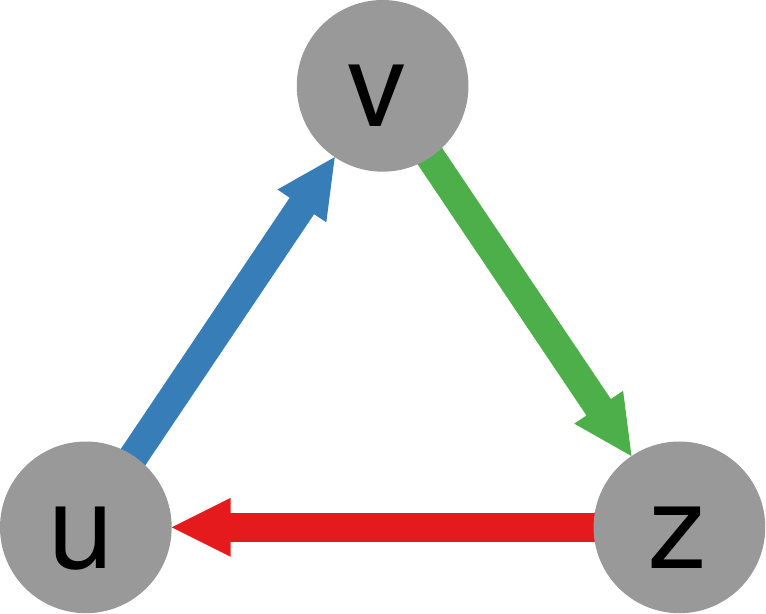}
\caption{}
\end{subfigure}
\quad
\begin{subfigure}[t]{.2\columnwidth}
\includegraphics[width=\textwidth]{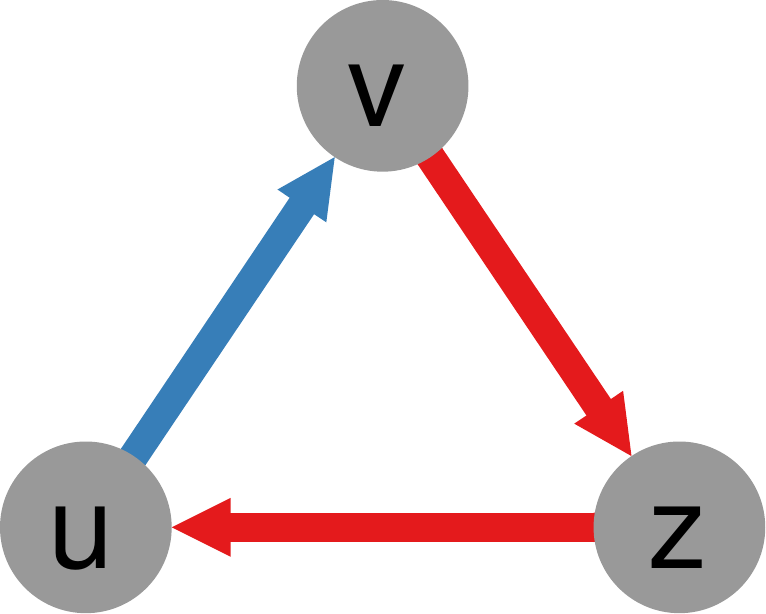}
\caption{}
\end{subfigure}
\caption{The $16$ templates of directed signed triangles in social status networks. The color of the edge determines its status: green = positive, red = negative, blue = the edge we are trying to predict -- can be either positive or negative.}
\label{fig:socialstatus}
\end{figure}

Figure \ref{fig:socialstatus} shows the $16$ main configurations of these triangles. The closing edge connecting $u$ to $v$ can be either positive or negative, generating $32$ final possible configurations. Status theory generates predictions that are more sophisticated and -- sometimes -- less immediately obvious. Some cases are easy to parse. For instance, consider Figure \ref{fig:socialstatus}(a). The objective is to predict the sign of the $(u,v)$ edge. In the example, $u$ endorses $z$ as higher status. $z$ endorses $v$. If $v$ is on a higher level than $z$, and $z$ is on a higher level than $u$, then it's easy to see how $v$ is also on a higher level than $u$. Thus the edge will be of a positive sign. In fact, this specific configuration is grossly over represented in real world data.

The situation is not as obvious for other triangles. For instance, the one in Figure \ref{fig:socialstatus}(i). Here we have the $z$ node endorsing both $u$ and $v$. We don't really know anything about their relative level, only that they are both on a higher standing with respect to $z$. The paper presenting the theory makes a subtle case. The edge connecting $u$ to $v$ is more likely to be positive than the generative baseline on $u$, but less likely to be positive than the receiving baseline of $v$. So, suppose that $50\%$ of edges originating from $u$ are positive, while $80\%$ of the links $v$ receives are positive. The presence of a triangle like the one in Figure \ref{fig:socialstatus}(i) would tell us that the probability of connecting $u$ to $v$ with a positive link is higher than $50\%$, but lower than $80\%$.

Given this sophistication, and the fact that social status works with more information than social balance -- namely the edge's direction --, it is no wonder that there are cases in which social status vastly outperforms social balance. For instance, the original authors apply social status to a network of votes in Wikipedia. Here the nodes are users, who are connected during voting sessions to elect a new admin. The admin receives the links, positive if the originating user voted in favor, negative if they voted against. Triangles in this network connect with the patterns predicted by social status theory.

\subsection{Atheoretical Sign Prediction}
\begin{figure*}
\centering
\includegraphics[width=\columnwidth]{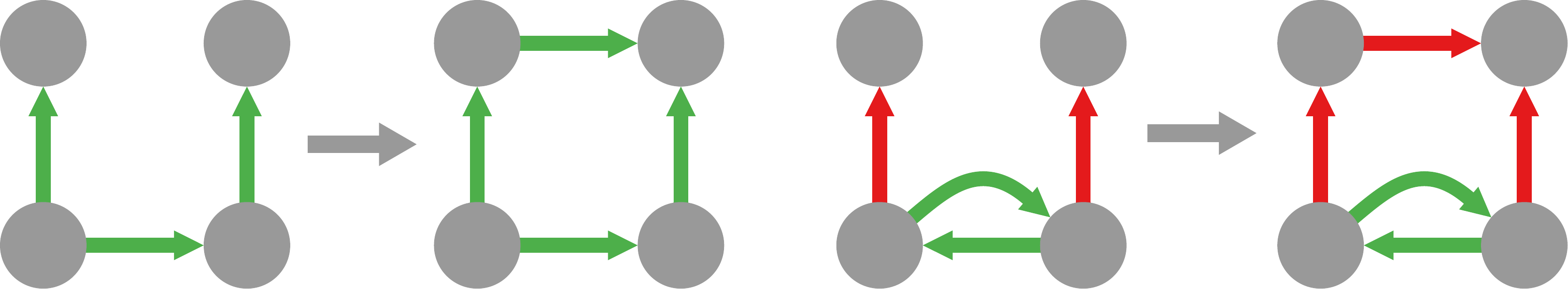}
\caption{Two possible graph association rules extracted from a directed signed network.}
\label{fig:germ-signed}
\end{figure*}

Of course, calling onto us the powers we unlocked in Section \ref{sec:lp-germ}, we can apply a strategy similar to GERM to extract graph association rules also in this scenario\cite{bachi2012classifying}. Figure \ref{fig:germ-signed} shows two possible association rules extracted from two different datasets. Looking at the figure, two advantages for this strategy emerge.

First, using a variation of GERM we free ourselves from the tyranny of the triangles. We can look at an arbitrary set of rules, not necessarily involving three nodes and triadic closure, which may not apply for all networks.

Second, what social balance and status have in common is that they will make the same prediction no matter the network you're going to have as input. They establish universal laws that might apply in general, but overlook specific laws that might apply for the phenomenon we're observing right now. For instance, voting in Wikipedia might be very different from trusting someone's opinion in a product review database. Consider the rule on the right in Figure \ref{fig:germ-signed}. That is a voting pattern in Wikipedia. It makes sense from a social status point of view: the node receiving the last negative link should expect to receive it, because it already received one, so it received a signal of being of low status.

But that rule makes little sense when moving to the scenario of trusting reviews. What the negative link means here is that the originator of the edge doesn't trust the recipient of the edge. However, we already know that the node originating the last link disagrees with the bottom two nodes, who trust each other. So we would expect it to trust -- to send a positive rather than a negative link -- to the node in the top right. In fact, while the pattern is widely popular in the Wikipedia network, it doesn't appear in the social review dataset.

\section{Generalized Multilayer Link Prediction}\label{sec:lp-multilayer-general}
So far we have only considered the case of two possible edge types. Moreover, these two types have a clear semantic: one type is positive, the other is negative. Both assumptions make the link prediction problem easier: there are few degrees of freedom and we move in a space constrained by strong priors. It is now time to drop these assumptions and face the full problem of multilayer link prediction as the big boys we are.

Generalized multilayer link prediction is the task of estimating the likelihood of observing a new link in the network, given the two nodes we want to connect and the layer we want to connect them through. Nodes $u$ and $v$ might be very likely to connect in the immediate future, but they might do so in any, some, or even just a single layer. Thus, we extend our score function to take the layer as an input: from $score(u,v)$ to $score(u,v,l)$.

\subsection{Layer Independence}
As you might expect, there are tons of ways to face this problem. The most trivial way to go about it is to apply any of the single layer link prediction methods from Chapter \ref{cha:lp-simple} to each layer separately. Then, you can create a single ranking table by merging all these predictions\cite{pujari2015link}.

\begin{figure}
\centering
\begin{subfigure}{.5\columnwidth}
\includegraphics[width=\textwidth]{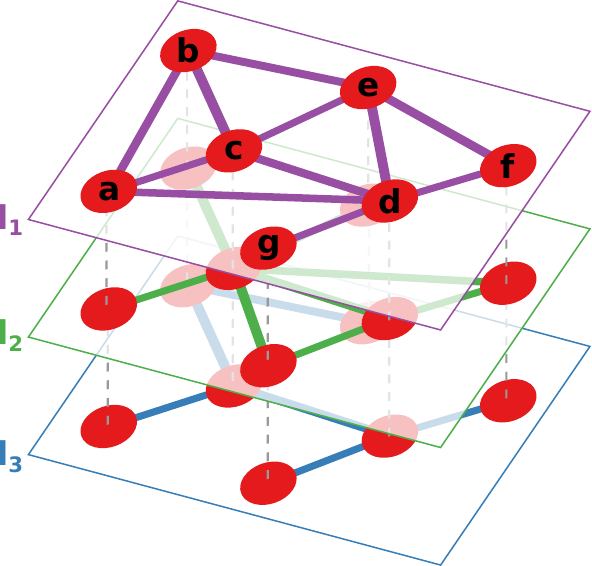}
\caption{}
\end{subfigure}\quad
\begin{subfigure}{.25\columnwidth}
  \begin{tabular}{l|l|l}
    Nodes & Layer & CNs \\
    \hline
    a, b & $l_2$ & $1$\\
    a, b & $l_3$ & $1$\\
    b, d & $l_1$ & $2$\\
    b, d & $l_2$ & $1$\\
    b, d & $l_3$ & $1$\\
    a, g & $l_1$ & $1$\\
    a, g & $l_2$ & $1$\\
    a, g & $l_3$ & $0$\\
  \end{tabular}
\caption{}
\end{subfigure}
\caption{The easiest way to perform multilayer link prediction. Given the input network, perform single layer link prediction on each of the layer separately. In this case, we count the number of common neighbors between pairs of nodes. We then predict the one with the overall highest score.}
\label{fig:multilayer-lp-1}
\end{figure}

Figure \ref{fig:multilayer-lp-1} depicts an example for this process. Note that here I use a rather trivial approach to aggregate, by comparing directly the various scores. One could also apply to this problem the rank aggregation measures presented in the previous chapter. In this way, you could also aggregate different scores using different criteria: common neighbors, preferential attachment, and so on.

This is practically a baseline: it will work as long as we have an assumption of independence between the layers. As soon as having a link in a layer changes the likelihood of connecting into another layer, we expect to grossly underperform.

\subsection{Blending Layers}
\begin{figure}
\centering
\includegraphics[width=.45\columnwidth]{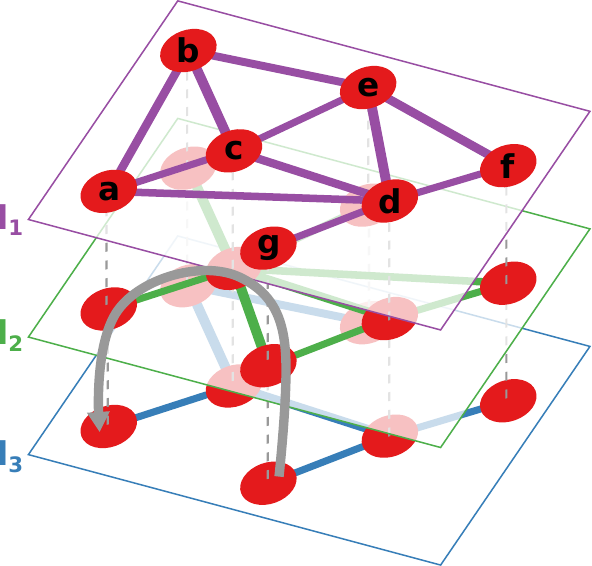}
\caption{A slightly more sophisticated way to perform multilayer link prediction. Given the input network, perform the link prediction procedure on the full structure. In this case, the gray arrow simulates a random walker going from node $g$ in layer $l_3$ to node $a$ in the same layer, passing through node $c$ in layer $l_2$. The mutlilayer random walker contributes to the $score(g, a, l_3)$.}
\label{fig:multilayer-lp-2}
\end{figure}

A slightly more sophisticated alternative is to consider the multilayer network as a single structure and perform the estimations on it. For instance, consider the hitting time method. This is based on the estimation of the number of steps required for a random walker starting on $u$ to visit $v$. We can allow the random walker to, at any time, use the inter layer coupling links exactly as if they were normal edges in the network. At that point, a random walker starting from $u$ in layer $l_1$ can and will visit node $v$ in layer $l_2$. The creation of our connection likelihood score is thus well defined for multilayer networks. Figure \ref{fig:multilayer-lp-2} depicts an example for this process.

These paths crossing layers are often called meta-paths. The information from these meta-paths can be used directly as we just saw, informing a multilayer hitting time. Or we can feed them to a classifier, which is trying to put potential edges in one of two categories: future existing and future non-existing links. Any classifier can perform this job once you collect the multilayer information from the meta-path: naive Bayes, support vector machines (SVM), and others\cite{jalili2017link}.

Other extensions to handle multilayer networks have been proposed\cite{davis2011multi}. These studies show that multilayer link prediction is indeed an interesting task, as there is a correlation between the neighborhood of the same nodes in different layer. The classical case involves the prediction of links in a social media platform using information about the two users coming from a different platforms\cite[0.25in]{hristova2016multilayer}. Such layer-layer correlations are not limited to social media, but can also be found in infrastructure networks\cite{kleineberg2016hidden}.

\subsection{Multilayer Scores}
The last mentioned strategy is better, but it still doesn't consider all the wealth of information a multilayer network can give you. To see why, let's dust off the concept of layer relevance we introduced in Section \ref{sec:degree-variants}. That is a way to tell you that a node $u$ has a strong tendency of connecting through a specific layer. If a layer exclusively hosts many neighbors of $u$, that might mean that it is its preferred channel of connection.

\begin{figure}
\centering
\includegraphics[width=.45\columnwidth]{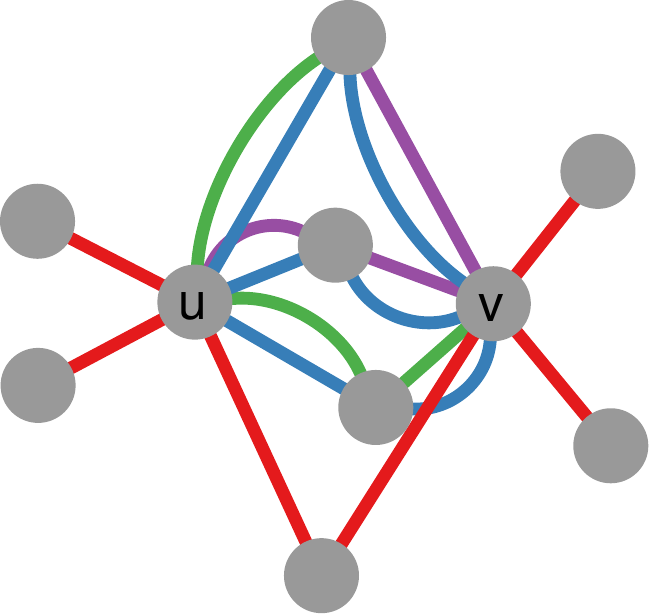}
\caption{The multilayer neighborhood of nodes $u$ and $v$. Edge color indicates the layer to which the connection belongs.}
\label{fig:multilayer-lp-3}
\end{figure}

This suggests that other naive ways to estimate node-node similarity for our score should be re-weighted using layer relevance\cite{rossetti2011scalable}. Consider Figure \ref{fig:multilayer-lp-3}. We see that the two nodes have many common neighbors in the blue layer. They only have one common neighbor in the red layer. However the blue layer, for both nodes, has a very low exclusive layer relevance. There is no neighbor that we can reach using exclusively blue edges. In fact, in this case, the exclusive layer relevance is zero.

The opposite holds for the layer represented by red edges. There are many neighbors for which red links are the only possible choice. In this particular case, we might rank the red layer as more likely to host a connection for nodes $u$ and $v$. In practice, this boils down to multiplying the layer relevance to the common neighbor score. Such weighting schema can be applied to most of the link prediction strategies we saw so far.

A related approach tries to estimate not whether a link will exist in the future, but its strength. If we have an unweighted multilayer network, we might still be able to estimate how strong a connection is. By looking at the various layers connecting two nodes, one could estimate such tie strength\cite{pappalardo2012well}.

Another approach to multilayer link prediction is the usage of tensor factorization. We briefly mentioned tensor factorization at the end of the previous chapter, for single layer link prediction. In that case, the third dimension of our tensor was representing time. In this case, we can apply the same technique by changing the meaning of this third dimension. Rather than using it to represent time, we can use it to represent the layer in which the edge appear. The same technique can now be applied, to discover in which layer new edges are likely to pop up.

And, since we're mentioning flexible methods that can be applied in multiple scenarios, why don't we dust off GERM again? We already saw how graph association rules can be extracted in signed networks without batting an eye. There is, in principle, no issue in extending the algorithm to deal with multilayer networks\cite{coscia2020multiplex} -- as long as you can properly and efficiently solve the graph isomorphism problem (Section \ref{sec:mining-isomorph}) for labeled multigraphs.

\begin{figure}[t]
\centering
\includegraphics[width=.8\columnwidth]{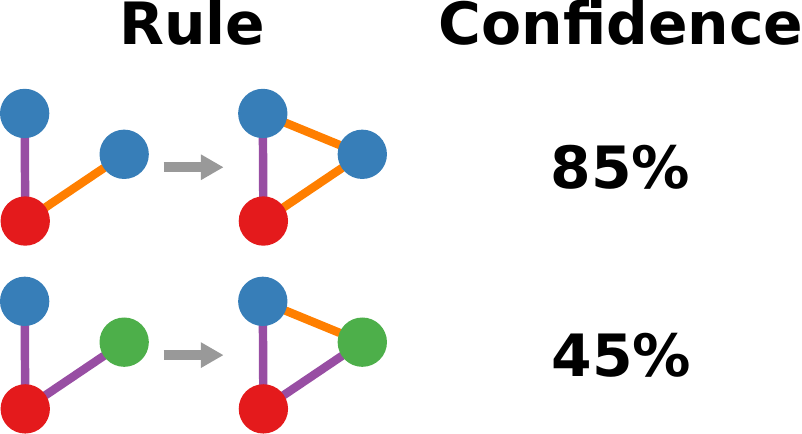}
\caption{Two possible graph association rules extracted with the multilayer version of GERM. Node color represents the node's label, while edge color represents the edge's layer.}
\label{fig:multilayer-lp-germ}
\end{figure}

Figure \ref{fig:multilayer-lp-germ} shows multilayer association rules in all their glory. In this case I encode also node attributes -- because why not? -- rendering the rules extracted by multilayer GERM extremely multifaceted. By collecting all the rules I showed so far in this book part, you realize that there are really a lot of ways to close a triangle in complex networks!

By using the edge labels to represent the layer in which an edge appears, we lose one of the powers of GERM. Namely, we are not able to make predictions at time steps farther than one. Remember that, with GERM, we could predict that a link will appear at time $t + 2$. This is not the case any more here, because the way we were able to do that was by encoding the edge arrival time in its label. But here we're using the edge label to indicate the layer in which it appears. This is an acceptable price to pay if we're able to perform multilayer link prediction.

Finally, as in the single layer case, there are some promising approaches using multilayer network embedding to predict links in multilayer networks, both in the signed case\cite{wang2017signed} and in the multilayer proper case\cite{bordes2013translating}\cite{zhang2018scalable}\cite{matsuno2018mell}.

\subsection{Heterogeneous Networks}
\begin{figure}
\centering
\includegraphics[width=.8\columnwidth]{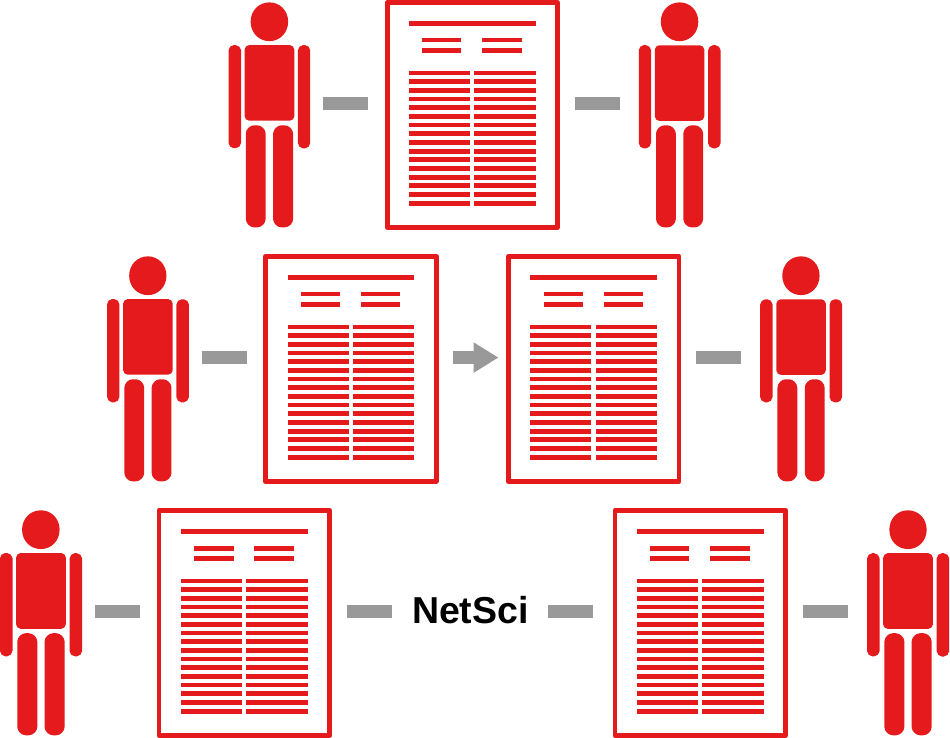}
\caption{Some examples of metapaths in a heterogeneous network with multiple node types, in a scientific publication scenario. From top to bottom, connecting authors because: they co-author a paper (both nodes of type authors are connected to the same node of type paper); they cite each other (a node of type author connects to a node of type paper citing another paper-type node); or they publish in the same venue.}
\label{fig:multilayer-lp-metapaths}
\end{figure}

In computer science, specifically in data mining and machine learning, multilayer networks are often called ``heterogeneous'' networks, because they have edges of different, heterogeneous, types. Heterogeneous link prediction is one of the main tasks tackled in this subfield. This is actually where metapaths were firstly developed\cite{sun2011co}. Figure \ref{fig:multilayer-lp-metapaths} shows examples of possible metapaths in a co-authorship network. These metapaths form the input of a classifier, which will then spit out the most likely new metapaths involving specific nodes.

Other common approaches use a ranking factor graph model\cite{dong2012link}, which searches for common general patterns shared by the various layers of the network; or consider link prediction as a matching problem\cite{kong2013inferring}.

By the way, the converse of what I said about GERM and tensor factorization applies also to heterogeneous link predictions. There is research showing how you can use this class of approaches to predict \textit{when} an edge will appear, rather than its type\cite{sun2012will}.

I should also mention that link prediction, community discovery, and generating synthetic networks are sides of the same weirdly triangular coin. This holds also for multilayer networks. There are efforts to create models generating multilayer networks than can then be applied to predict new links on already existing real-world multilayer networks\cite{de2017community}\cite{roxana2019edge}.

In this chapter, as in all chapters of this book, I presented only the most prominent methods to tackle the issue at hand, and the ones I'm most familiar with. The study of a deeper review work\cite{sun2013mining} is necessary if you want to make a living off solving multilayer link prediction.

\section{Summary}

\begin{enumerate}
\item In multilayer link prediction, besides predicting the appearance of a new edge, you also need to guess in which layer the new edge will appear. It's not only about \textit{whether} two nodes will connect, it's also about \textit{how} they will connect.
\item A simplified version of multilayer link prediction involve signed networks. In this case, real world networks have a preference for balanced structures, which you can use to predict the sign of the relationship. 
\item An alternative approach is by using status theory. Whether you should use social balance or social status depends on what your network represents, for instance trust would follow balance, while voting would follow status.
\item For generalized multilayer link prediction the common approach is to create multilayer generalizations of single layer predictors, with some strategy to aggregate multilayer information.
\item Other approaches rely on multilayer extensions of graph mining and on the use of metapaths: paths connecting nodes across layers.
\end{enumerate}

\section{Exercises}

\begin{enumerate}
\item You're given the undirected signed network at \url{http://www.networkatlas.eu/exercises/24/1/data.txt}. Count the number of triangles of the four possible types.
\item You're given the directed signed network at \url{http://www.networkatlas.eu/exercises/24/2/data.txt}. Does this network follow social balance or social status? (Consider only reciprocal edges. For social balance, the reciprocal edges should have the same sign. For social status they should have opposite signs)
\item Consider the multilayer network at at \url{http://www.networkatlas.eu/exercises/24/3/data.txt}. Calculate the Pearson correlation between layers (each layer is a vector with an entry per edge. The entry is $1$ if the edge is present in the layer, $0$ otherwise). What does this tell you about multilayer link prediction? Should you assume layers are independent and therefore apply a single layer link prediction to each layer?
\end{enumerate}

\chapter{Designing an Experiment}\label{cha:lp-experiment}
As the name of the problem suggests, link prediction is fundamentally a task that involves making claims about the future. Evaluating the performance of an oracle in getting things right is harder than it might seem. There are surprising ways to get it wrong. Luckily, making predictions is the bread and butter of machine learning. Thus we have a large set of best practices we can follow. In Chapter \ref{cha:machine-learning} we saw the general architecture of a machine learning framework. This chapter is a crash course on how to adapt that general pipeline to the specific problems of link prediction. This is mostly taken from the literature\cite{yang2015evaluating}, which you should check out to get a deeper view on the problem.

The chapter is divided in two parts. First, we have to figure out how to perform the test (Section \ref{sec:lp-experiment-test}). Since link prediction is about the future, one option would be to just wait until new data comes in. This is often not ideal, because you  don't really know when you'll get new information, and you want to publish your paper \textit{right now}. So you have to work with the data you have. Once you make your prediction, you have to then evaluate how well you perform (Section \ref{sec:lp-experiment-eval}).

\section{Train/Test Sets}\label{sec:lp-experiment-test}

\subsection{The Basics}
When it comes to evaluate your prediction algorithm, you have to distinguish between the training and the test datasets. The training dataset is what your model uses to learn the patterns it is supposed to predict. For instance, if you're doing a common neighbor link predictor, the training dataset is what you use to count the number of shared connections between two nodes. Once you're done examining the input data, you have generated the results of the $score(u,v)$ function for all possible pairs of $u,v$ inputs.

The test dataset is a set of examples used to assess performance. When your model is done learning on the training dataset, it is unleashed on the test dataset and will start making predictions. Every time it gets it right you increase its performance, every time it gets it wrong you decrease it.

\begin{figure}
\centering
\begin{subfigure}{.33\columnwidth}
\includegraphics[width=\textwidth]{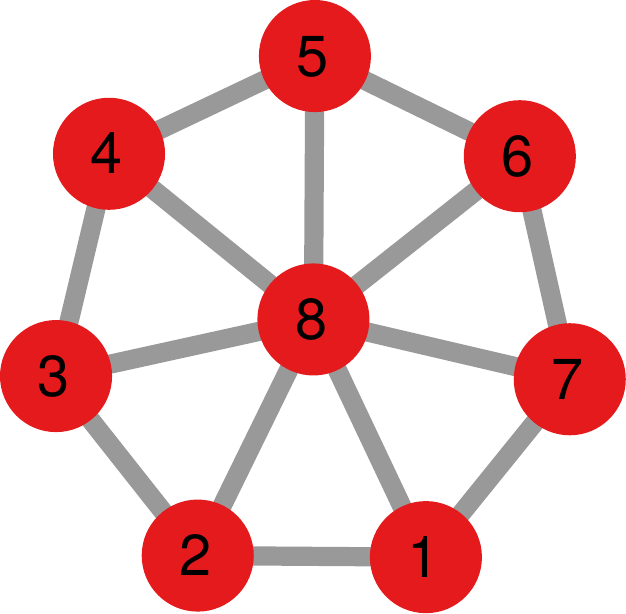}
\caption{Train}
\end{subfigure}\quad
\begin{subfigure}{.25\columnwidth}
  \begin{tabular}{l|r}
    Edge & Score \\
    \hline
    $1, 2$ & $1$\\
    $1, 3$ & $2$\\
    $1, 4$ & $1$\\
    $1, 5$ & $1$\\
    $1, 6$ & $2$\\
    $2, 4$ & $2$\\
    $\ldots$ & $\ldots$\\
  \end{tabular}
\caption{Scores}
\end{subfigure}\quad
\begin{subfigure}{.33\columnwidth}
\includegraphics[width=\textwidth]{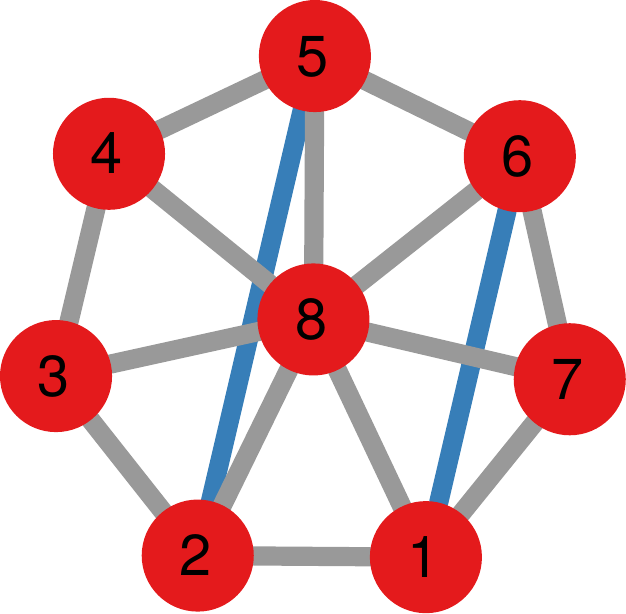}
\caption{Test}
\end{subfigure}
\caption{An example of train and test sets for a network. The information (a) we use to build the score table (b), using the common neighbor approach. I highlight the test edges (c) in blue.}
\label{fig:lp-train-test-0}
\end{figure}

Figure \ref{fig:lp-train-test-0} shows an example of the difference between the two sets. Figure \ref{fig:lp-train-test-0}(a) is the training dataset: it contains all the information we can use to infer our scores. Figure \ref{fig:lp-train-test-0}(b) are the scores we calculate based on the data from Figure \ref{fig:lp-train-test-0}(a). Specifically, for each pair of nodes I calculate the number of common neighbors shared by the two nodes. Figure \ref{fig:lp-train-test-0}(c) shows the test set in blue. These are the actual new edges that appeared in the network. I include the training set in gray because it makes it easier to put the new edges into context -- besides, when performing a prediction you need to make sure you remember what was in the training set and discard any prediction you might have done about those edges. For instance, in this case, we need to throw away the $(1,2)$ edge prediction.

In machine learning there is also what you'd call a ``validation'' set (Section \ref{sec:ml-general-infrastructure}), but there is nothing special about the validation set for link prediction. Just like in any other machine learning scenario, if you worry about overfitting in your link prediction, you should have a validation set. Since there is nothing special about link prediction and validation sets, I simply refer you back to Chapter \ref{cha:machine-learning}.

The fundamental tenet of machine learning also holds for link prediction: you can never ever ever use the data that trained your model to test it. In other words, training and test sets have to be \textit{disjoint}. If you test your method on the same data that trained it, you're going to grossly overestimate your actual performance in the real world. You have only learned about your training set and nothing about the general forces that shaped it the way it is.

What this means in link prediction is that you cannot claim to have predicted a link that was already in your data. You have to focus only on those pairs of nodes that were not connected in the training set. That is why in Figure \ref{fig:lp-train-test-0}(c) the edges that were already in the training set are gray rather than blue: we won't make predictions on those, because we already know they exist.

So now the problem is: how do you do that? If you have a network, how do you divide it into training and test sets?

\begin{figure}[t]
\centering
\includegraphics[width=.66\columnwidth]{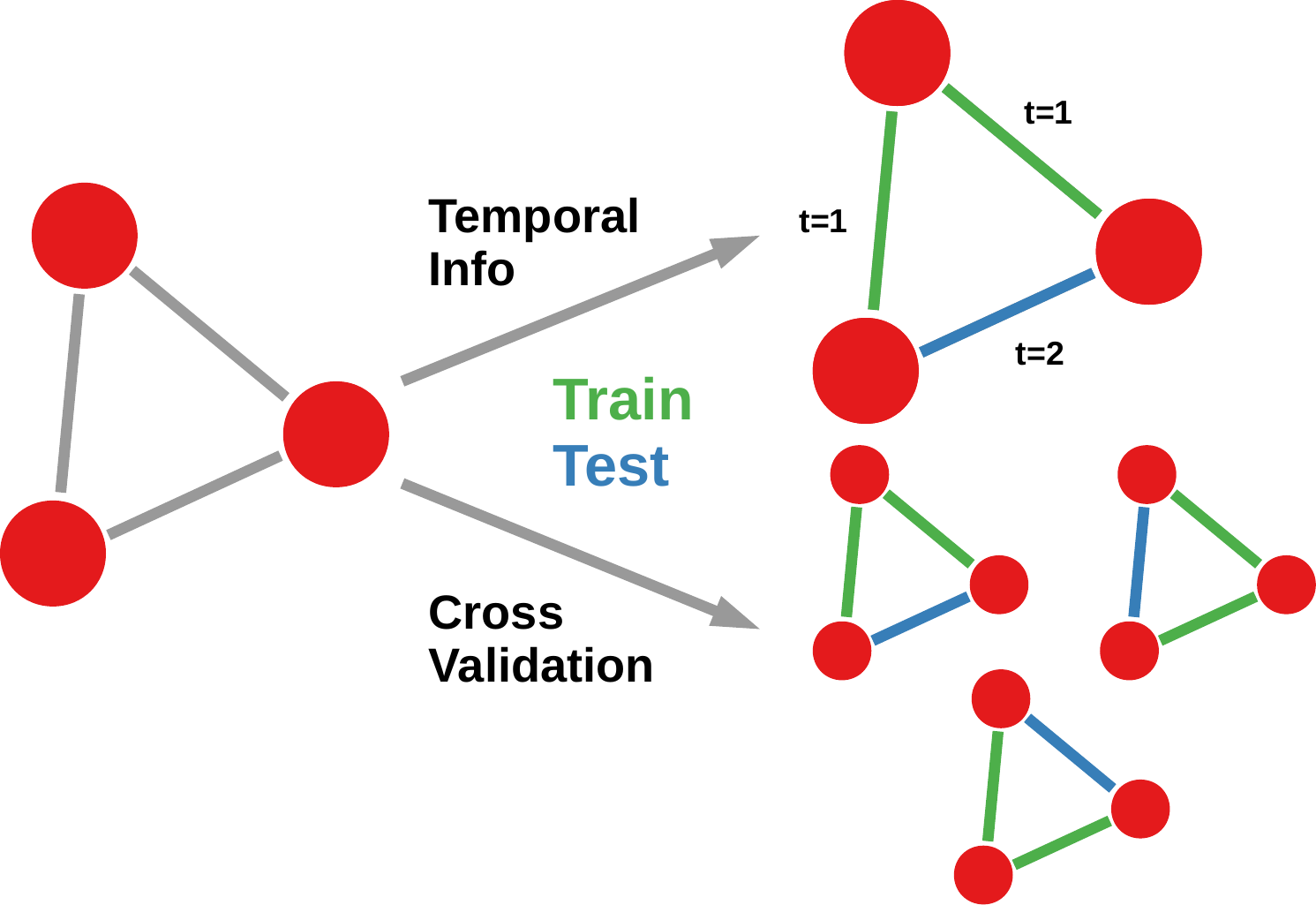}
\caption{Two approaches to build your train and test sets for link prediction. On the left you have the input data. On the right, the partition of links into the two sets: train (green) and test (blue).}
\label{fig:lp-train-test}
\end{figure}

You have two options, as Figure \ref{fig:lp-train-test} shows. If you have temporal information on your edges you can use earlier edges to predict the later ones. Meaning that your train set only contains links up to time $t$, and the test set only contains links from time $t + 1$ on. If you don't have the luxury of time data, you have to do k-fold cross validation: divide your dataset in a train and test set (say $90\%$ of edges in train and $10\%$ in test) and then perform multiple runs of train-test by rotating the test set so that each edge appears in it at least once\cite{kohavi1995study}.

\subsection{Specific Issues}
Link prediction comes with a few peculiarities that might not be a problem in other machine learning tasks. I focus on two: size of the search space and sparsity of positives.

The first refers to the fact that, as we saw in Section \ref{sec:density-sparse}, real networks are sparse. I made the example of the Internet backbone: with its $192,244$ nodes, the number of possible edges is $|V|(|V| - 1)/2 = 18,478,781,646$. However, it only contains $609,066$ actual edges. This means that, if you were to use its current state as a train set, the score function would have to compute a result for $18,478,781,646 - 609,066 = 18,478,172,580$ potential edges. That is an unreasonable burden, both for computation time and memory storage.

For this reason, when you perform link prediction you will often sample your outputs. You will not calculate $score(u,v)$ for every possible $u,v$ pair, but you will sample the pairs according to some expectation criterion. Such criterion can be as hard to pin down as the link prediction problem itself.

\begin{figure}
\centering
\begin{subfigure}{.34\columnwidth}
\includegraphics[width=\textwidth]{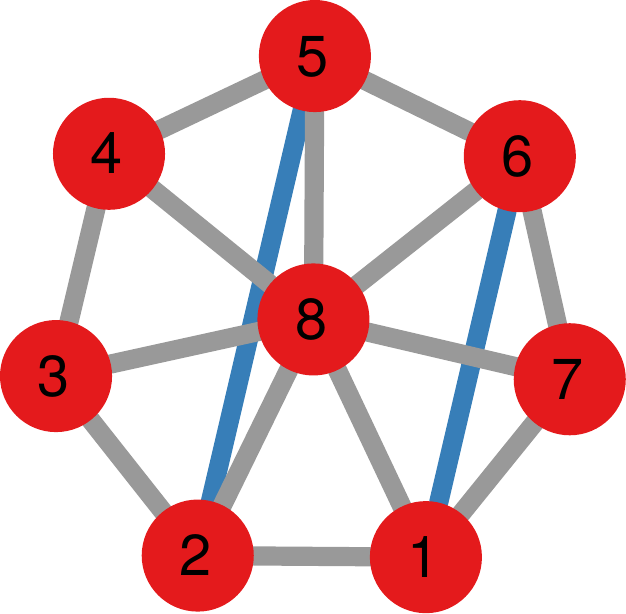}
\caption{Test}
\end{subfigure}\qquad
\begin{subfigure}{.3\columnwidth}
  \begin{tabular}{l|rr}
    Edge & Prediction & Correct?\\
    \hline
    $1, 3$ & $0$ & $1$\\
    $1, 4$ & $0$ & $1$\\
    $1, 5$ & $0$ & $1$\\
    $1, 6$ & $0$ & $0$\\
    $2, 4$ & $0$ & $1$\\
    $2, 5$ & $0$ & $0$\\
    $2, 6$ & $0$ & $1$\\
    $\ldots$ & $\ldots$\\
  \end{tabular}
\caption{Scores}
\end{subfigure}
\caption{Estimating the performance of the ``always negative'' predictor on our test set.}
\label{fig:lp-always-negative}
\end{figure}

The second problem is intertwined with the first. Suppose that the Internet backbone adds edges at a $5\%$ rate per time step. That means that, if at time $t$ you had $609,066$ edges, at time $t + 1$ you will observe $609,066 \times .05 \sim 30,453$ new edges. As we just saw, the number of potential edges is just above 18B. Putting these two facts together lets us reach an absurd conclusion: we can build a link prediction method that will tell us that no new link will ever appear. If we do so, we would be right $99.999\%$ of the times. We would make 18B correct predictions -- no edge -- and we would get it wrong only 30k times. The accuracy of the ``always negative'' predictor in Figure \ref{fig:lp-always-negative} is $\sim 85\%$: not bad!

However that's... kind of not the point? We're in this business because we want to predict new links. Returning a negative prediction for all possible cases is not helpful. The usual fix for this problem is building your test set in a balanced way\cite{lichtenwalter2010new}\cite{lichtnwalter2012link}. Rather than asking about all possible new edges, you create a smaller test set. Half of the edges in the test set is an actual new edge, and then you sample an equal number of non-edges. This would make our Internet test set containing 60k edges, not 18B. We called this sampling technique ``negative sampling'' in Section \ref{sec:ml-sampling}.

\section{Evaluating}\label{sec:lp-experiment-eval}
Let's assume that we have competently built our training and test set. We made our model learn on the former. We now have two things: prediction -- the result of the model -- and reality -- the test set. We want to know how much these two sets overlap. Since you know what edges are in the test set, this is a supervised learning task (Section \ref{sec:ml-general-infrastructure}) and we can focus on the loss/quality functions (Section \ref{sec:ml-loss}) specific for this scenario. Even more narrowly, link prediction is a binary task: you give a yes/no answer that is either completely correct or wrong. So there are four possible cases:

\begin{itemize}
\item True Positives (TP): you predict a link that really appeared;
\item False Positives (FP): you predict a link that didn't appear;
\item True Negatives (TN): you correctly didn't predict a link that, in fact, didn't appear;
\item False Negatives (FN): you didn't predict a link that appeared.
\end{itemize}

These are simple counts on which we can build several quality measures. Two basic combinations of these counts are the True Positive Rate (TPR) and False Positive Rate (FPR). TPR -- also known as sensitivity or recall -- is the ratio between true positives and all positives: $TPR = TP / (TP + FN)$. It tells you how many times you got it right over the maximum possible number of times you could. Or, what's the share of correct results you found.

FPR is defined similarly: $FPR = FP / (FP + TN)$. This is the share of your wrong answers over all the possible instances of a negative prediction.

\subsection{Confusion Matrix}
Humans like single numbers, because seeing a number going up tingles our pleasure centers (wait, what? You don't feel inexplicable arousal while maximizing scores? I question whether you're in the right line of work...). However, we should beware of what we call ``fixed threshold metrics'', i.e. everything that boils down a complex phenomenon to a single number. Usually, to reduce everything to a single measure you have to make a number of assumptions and simplifications that may warp your perception of performance.

\begin{figure}
\centering
\includegraphics[width=.8\columnwidth]{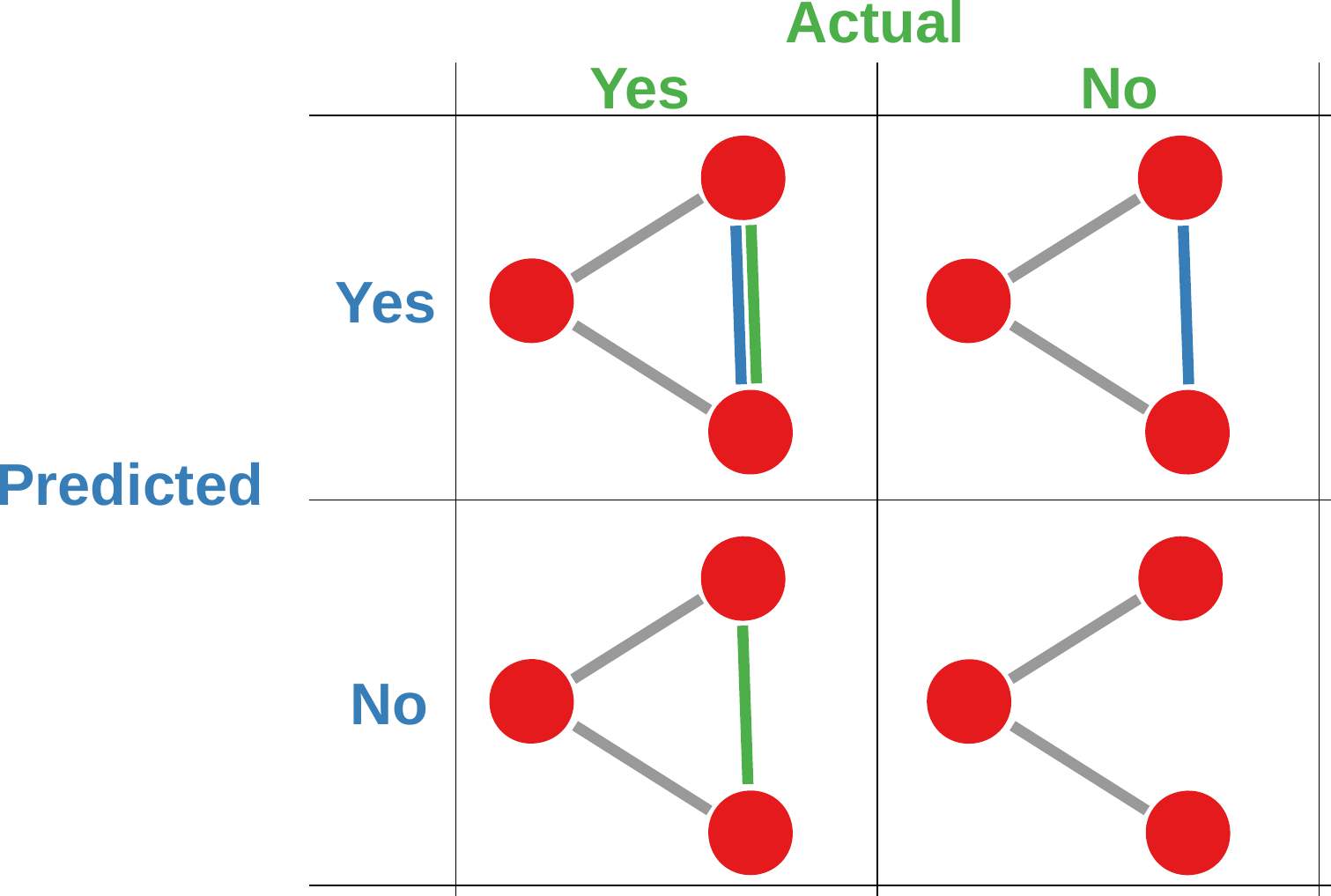}
\caption{The schema of a confusion matrix for link prediction. From the top-left corner, clockwise: true positives, false positives, true negatives, false negatives.}
\label{fig:confusion}
\end{figure}

That is why one of the first thing you should look at is a confusion matrix. A confusion matrix is simply a grid of four cells, putting the four counts I just introduced in a nice pattern\cite{stehman1997selecting}. You can see an example in Figure \ref{fig:confusion}. Confusion matrices are nice because they don't attempt to reduce complexity, but at the same time you see information in an easy-to-parse pattern.

\begin{figure}
\centering
\begin{subfigure}{.33\columnwidth}
\includegraphics[width=\textwidth]{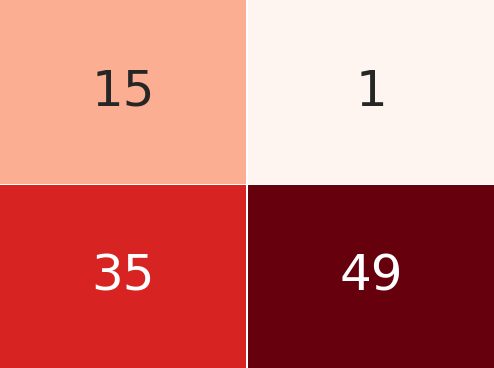}
\caption{}
\end{subfigure}
\qquad
\begin{subfigure}{.33\columnwidth}
\includegraphics[width=\textwidth]{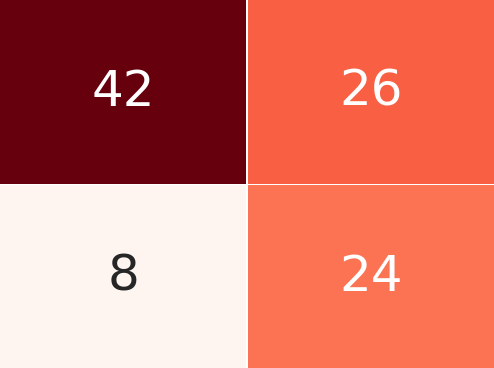}
\caption{}
\end{subfigure}
\caption{Two distinct confusion matrices for different predictors.}
\label{fig:confusion2}
\end{figure}

By looking at two confusion matrices you can say surprisingly sophisticated things about two different methods. The one in Figure \ref{fig:confusion2}(a) does a better job in making sure a positive prediction really corresponds to a new link: there are very few false positives (one) compared to the true positives ($15$). The one in Figure \ref{fig:confusion2}(b) minimizes the number of false negatives, with the downside of having a lot of false positives.

By combining the cells of a confusion matrix, you can easily derive measures like TPR or FPR, or many others. They are simple operations on the rows and columns.

If you didn't balance your test set, the confusion matrix can end up being irrelevant, as the vast majority of your observations will end up in the true negative cell, obliterating all the rest.

Another disadvantage of the confusion matrix is that you have to pick a threshold in your score. In other words, you predict the appearance of a link if it obtains a score higher than the specific threshold, otherwise you don't. This is in itself a problematic choice, thus it is common to show the evolution of your accuracy as you change that threshold. For high values of the threshold you only report high confidence predictions, which become less and less confident as you decrease the threshold. This is the topic explored in the rest of the chapter.

\subsection{ROC Curves \& AUC}
\begin{figure*}[t]
\centering
\begin{subfigure}{.4\columnwidth}
\includegraphics[width=\textwidth]{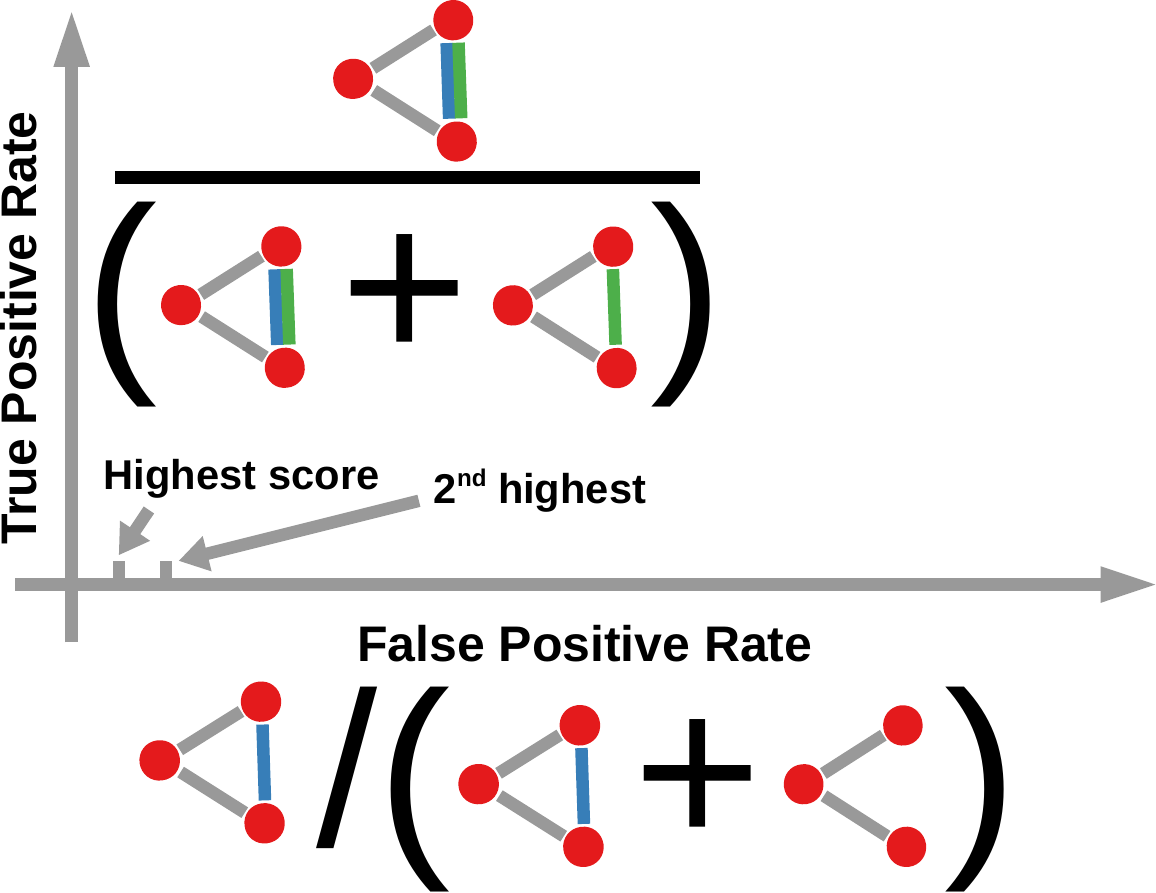}
\caption{}
\end{subfigure}
\quad
\begin{subfigure}{.45\columnwidth}
\includegraphics[width=\textwidth]{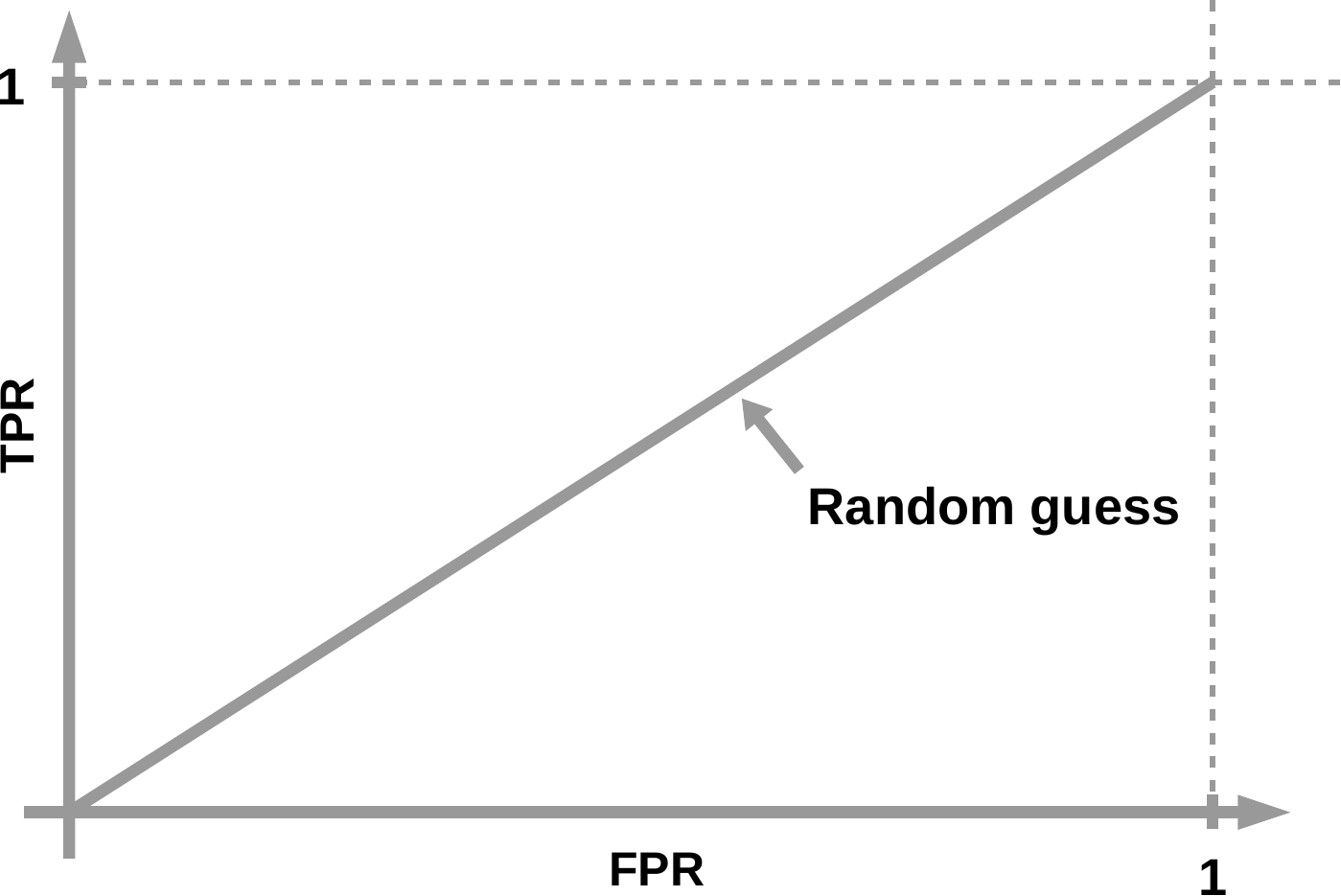}
\caption{}
\end{subfigure}
\caption{Schema of ROC curves.}
\label{fig:roc0}
\end{figure*}

The classic evaluation instrument for classification tasks is the Receiver Operating Characteristic (ROC) curve. This is a plot, with the false positive rate on the x-axis and the true positive rate on the y-axis (see Figure \ref{fig:roc0}(a))\cite{hanley1982meaning}\cite{fawcett2006introduction}. We sort all our predictions by their score such that we look at the highest scores first. Then we keep track of the evolution of TPR and FPR. 

In a ROC curve, the $45$ degree line corresponds to the random guess (Figure \ref{fig:roc0}(b)). Suppose $80\%$ of possible links did not appear and $20\%$ did, and there are a total of $20$ new links. If we make ten random guesses, we'll get eight false positives and two true positives. The two true positives represent $10\%$ of all the positives, so TPR = $2 / 20 = 0.1$. On the other hand, we know that there are $80$ negatives. Since we got eight false positives, FPR = $8 / 80 = 0.1$. This shows that FPR and TPR grow at the same rate for a random predictor.

\begin{figure}
\centering
\includegraphics[width=.9\columnwidth]{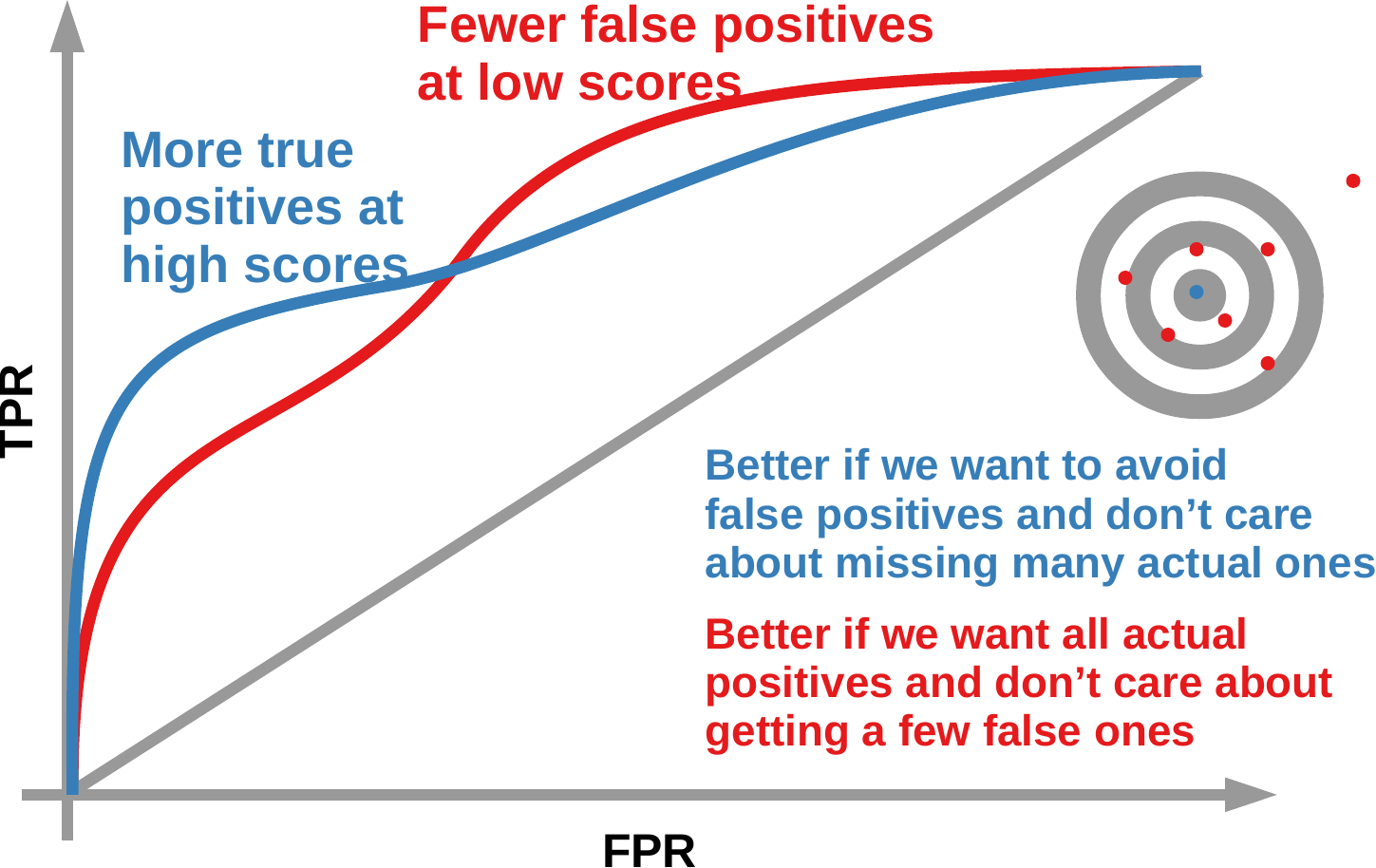}
\caption{An example of ROC curves. The gray line corresponds to random guesses. The blue and red lines correspond to two different predictors, with different behaviors at different score levels.}
\label{fig:roc}
\end{figure}

What we want to see is that our best guesses are more likely to be true positives, and thus contribute to the y-axis more than they do to the x-axis. Just like in the confusion matrix, there are multiple ways for this to happen. We can be very precise at high scores, or at all scores on average. The two classifiers in Figure \ref{fig:roc} will be used in different scenarios with different requirements.

ROC curves are great -- you might even say that they ROC -- but, at the end of the day, you might want to know which of the two classifiers is better on average. ROC curves can be reduced to a single number, a fixed threshold metric. Since we just said that the higher the line on the ROC plot the better, one could calculate the Area Under the Curve (AUC). The more area under that curve, the better your classifier is, because for each corresponding FPR value, your TPR is higher -- thus encompassing more area.

You don't need to know calculus to estimate the area under the curve, because it's such a standard metric that any machine learning package will output it for you. The AUC is $0.5$ for the random guess: that's the area under the $45$ degree line. An AUC of $1$  -- which you'll never see and, if you do, it means you did something wrong -- means a perfect classifier.

Note that ROC curves and AUCs are unaffected if you sample your test set randomly, namely if you only test potential edges at random from the set of all potential edges -- I discussed before how this is a common thing to do because of the unmanageable size of the real test set. However, that is not true if you perform a non-random sampling. This means choosing potential edges according to a specific criterion. If your criterion is ``good'', meaning that your sampling method is correlated with the actual edge appearance likelihood, you're going to see a different -- lower -- AUC value. That is because, if you don't sample, the vast number of easy-to-predict false negatives increases your classifier's accuracy.

\subsection{Precision \& Recall}
\begin{figure}
\centering
\includegraphics[width=\columnwidth]{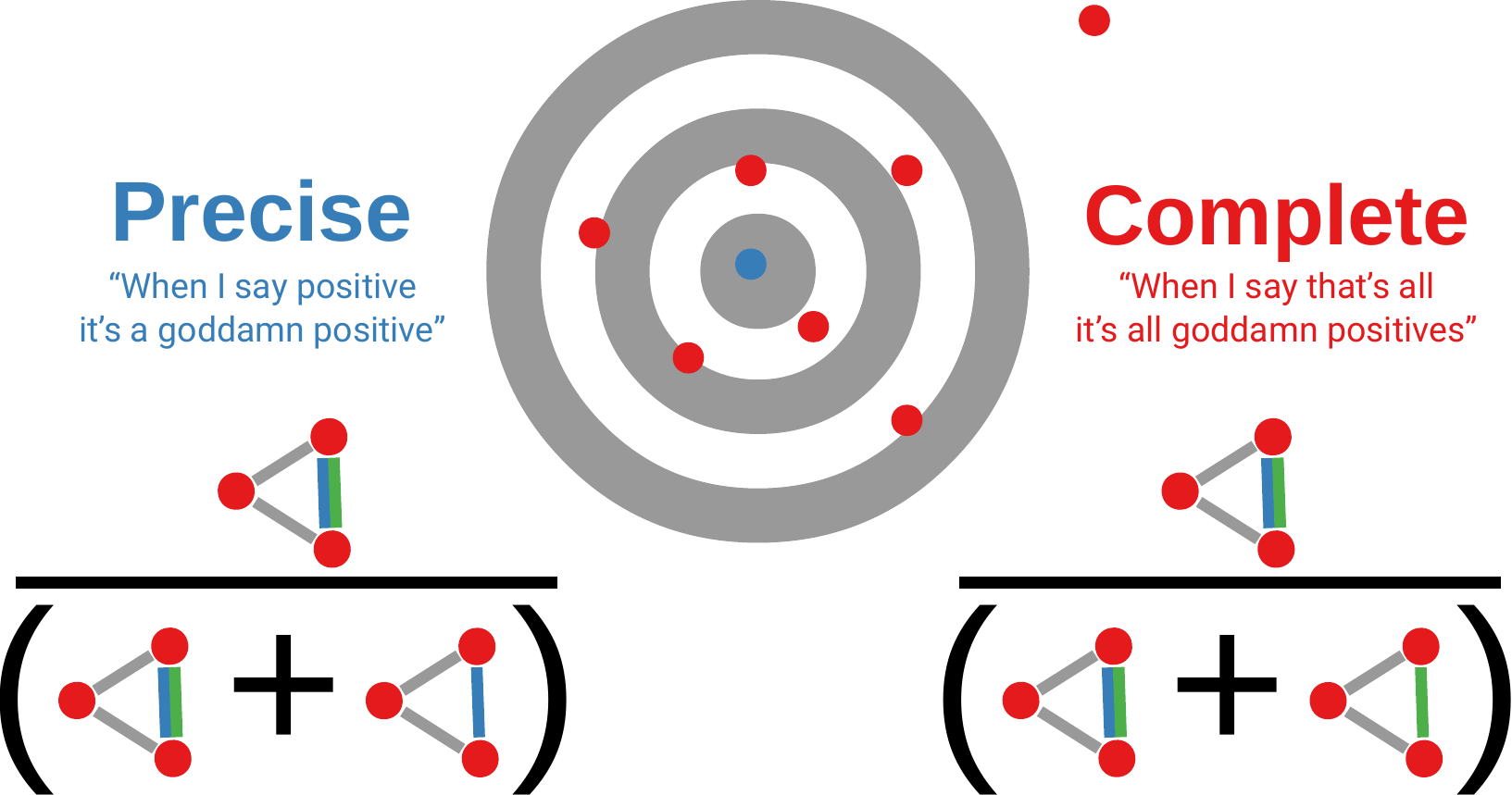}
\caption{A representation of precision and recall.}
\label{fig:precision-recall}
\end{figure}

Another way of putting a number to evaluate the quality of the prediction is to look at Precision and Recall\cite{powers2011evaluation}. Precision means that, when we predict that a link exists, it exists (even if we fail to predict actual links). Recall means that there are very few existing links we do not predict, even if we might have predicted many that didn't actually exist. So Precision is true positives over all predicted positives (including false positives): $TP / (TP + FP)$. Recall is another name for the True Positive Rate: $TP / (TP + FN)$. Figure \ref{fig:precision-recall} shows a visual example.

You can do a few things with precision and recall. First, you can transform them into fixed threshold metrics. This is done by calculating what we call ``$Precision @ n$'', defining $n$ as the number of predictions we want to make. For instance, in $Precision@100$ we only consider as an actual prediction the $100$ pairs of nodes that have the highest scores. Everything else is classified as ``no link''.

You can also combine precision and recall to generate a derived score, balancing them out. This is known as the $F1$-score, which is their harmonic mean: $F1 = 2(Precision \times Recall) / (Precision + Recall)$. This is a single number, like AUC, capturing both types of errors: failed predictions and failed non-predictions.

\begin{figure}
\centering
\includegraphics[width=\columnwidth]{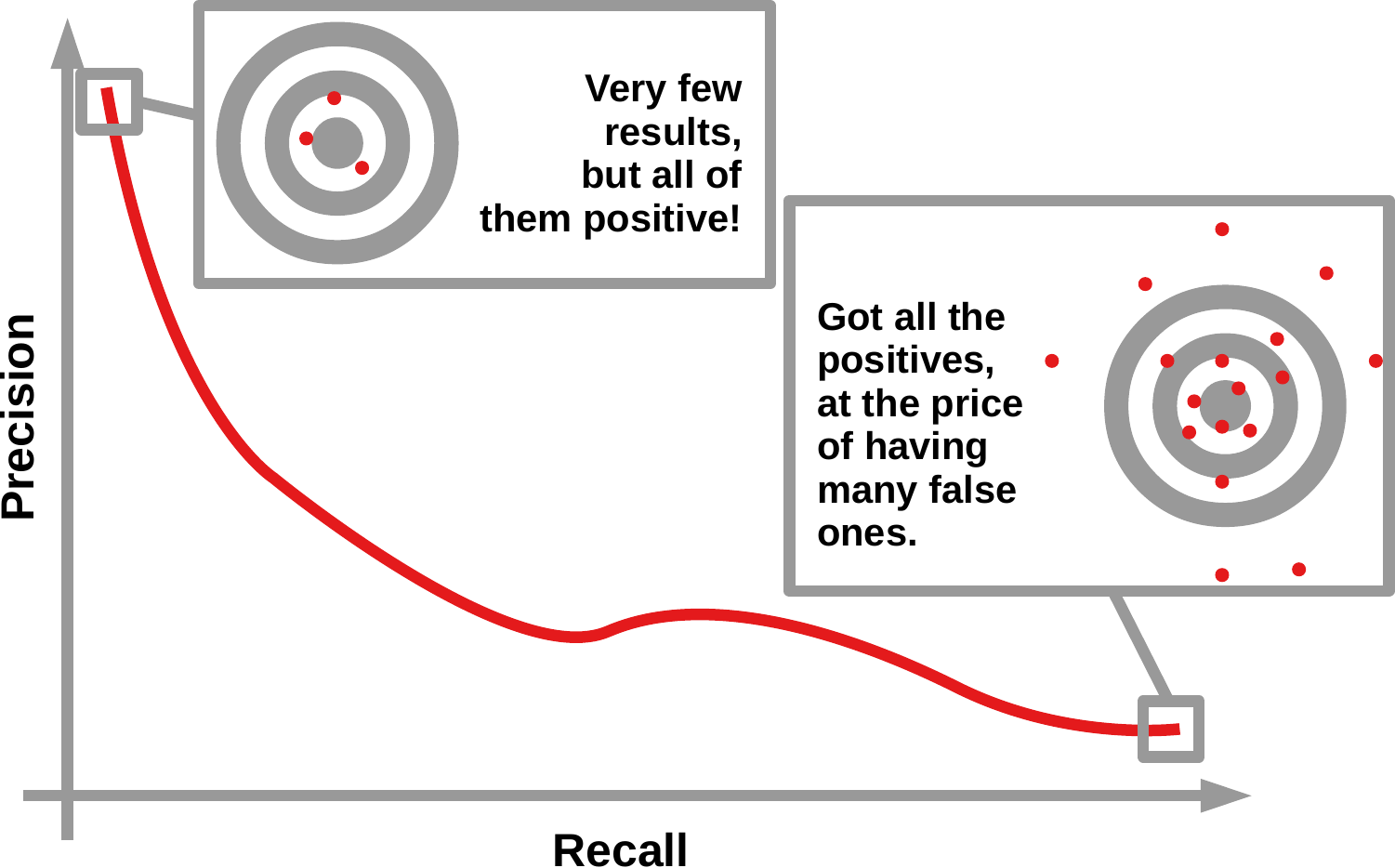}
\caption{A representation of precision-recall curves.}
\label{fig:pr-curve}
\end{figure}

A powerful way to use precision and recall is by using them as an alternative to ROC curves. The so-called Precision-Recall curves have the recall on the x-axis and the precision on the y-axis (see Figure \ref{fig:pr-curve}). They tell you how much your precision suffers as you want to recover more and more of the actual new edges in the network. Recall basically measures how much of the positive set your recover. But, as you include more and more links in that set, you're likely to start finding lots of false positives. That will make your recall go up, but precision go down.

In a Precision-Recall plot, the random classifier is a horizontal line. If you have $10$ positive samples in a dataset of $100$ entries, then the horizontal line is at $10 / 100 = 0.1$. That is because, if you're making random predictions, you always have a $10\%$ chance of getting it right. It is a horizontal line because you can achieve any recall value, as long as you keep trying. You can make $100$ predictions, which by definition will get you all positives -- and a recall of $1$ -- while the precision will still be $0.1$.

A final way to use precision for evaluating link prediction methods is to use the \textit{prediction power}\cite{cannistraci2013link}. This is a measure that compares the precision of your classifier with the one you would obtain from a random classifier returning random links without looking at the network topology. If we say that your precision is $P$ and the random precision is $P_r$, then the prediction power $PP$ is

$$ PP = 10 \log_{10} \dfrac{P}{P_r}.$$

This is a decibel-like logscale: a $PP = 1$ implies your predictor is ten times better than random, while $PP = 2$ means you are one hundred times better than random. You can also create $PP$-curves by having on the x-axis the share of links you remove from your training set. By definition, the random predictor is an horizontal line at $0$. The more area your $PP$ curve can make over the horizontal zero, the more precise your predictor is.

In closing, I should also mention another popular measure: accuracy. This is simply $(TP + TN) / (TP + TN + FP + FN)$: the number of times you got it right over all the attempts. The lure of accuracy is its straightforward intuition. However, it hides the difference between type I and type II errors -- false positives and false negatives -- and thus it should be handled with care.

\section{Summary}

\begin{enumerate}
\item To evaluate the quality of a link prediction you need to train your algorithm and then test it. To do so, you need to divide the data in mutually exclusive train and test sets.
\item If your data has temporal information you can decide a cutoff date to divide the two sets. Otherwise you have to perform cross validation: divide the data in ten blocks and rotate one block as test set using the other nine as training, until you tested on all data.
\item Since real networks are sparse, there are more non-edges than edges. Thus a link prediction always predicting non-edge would have high performance. That is why you should balance your test sets, having an equal number of edges and non-edges.
\item A classical evaluation strategy is the ROC curve, recording your true positive rate against your false positive rate. The more area under this curve (AUC) you have the better your prediction performance.
\item Precision is the ability of returning only true positive results at the price of missing some. Recall is the ability of returning all positive results, at the price of returning also lots of false positives. You can draw precision-recall curves, again with the objective of maximizing their AUC.
\end{enumerate}

\section{Exercises}

\begin{enumerate}
\item Divide the network at \url{http://www.networkatlas.eu/exercises/25/1/data.txt} into train and test sets using a ten-fold cross validation scheme. Draw its confusion matrix after applying a jaccard link prediction to it. Use 0.5 as you cutoff score: scores equal to or higher than 0.5 are predicted to be an edge, anything lower is predicted to be a non-edge. (Hint: make heavy use of \texttt{scikit-learn} capabilities of performing KFold divisions and building confusion matrices)
\item Draw the ROC curves on the cross validation of the network used at the previous question, comparing the following link predictors: preferential attachment, jaccard, Adamic-Adar, and resource allocation. Which of those has the highest AUC? (Again, \texttt{scikit-learn} has helper functions for you)
\item Calculate precision, recall, and F1-score for the four link predictors as used in the previous question. Set up as cutoff point the ninetieth percentile, meaning that you predict a link only for the highest ten percent of the scores in each classifier. Which method performs best according to these measures? (Note: when scoring with the \texttt{scikit-learn} function, remember that this is a binary prediction task)
\item Draw the precision-recall curves of the four link predictors as used in the previous questions. Which of those has the highest AUC?
\end{enumerate}

\part{The Hairball}\label{par:hairball}

\chapter{Bipartite Projections}\label{cha:projections}
Reality does not usually match expectations. Let's consider three examples:

\begin{enumerate}
\item Degree distributions;
\item Epidemics spread;
\item Communities.
\end{enumerate}

Many papers have been written on how power law degree distributions are ubiquitous\cite{barabasi1999emergence}\cite{barabasi2003scale}\cite{albert2005scale}\cite{barabasi2009scale}. Chances are that any and all the networks you'll find on your way as a network analyst do not have even a hint of a power law degree distribution. In the best case scenario you are going to have shifted power laws, or exponential cutoffs -- if you're lucky -- (for a refresher on these terms, see Section \ref{sec:degree-fit}).

\begin{figure}[b]
\centering
\begin{subfigure}[t]{.4\columnwidth}
\includegraphics[width=\textwidth]{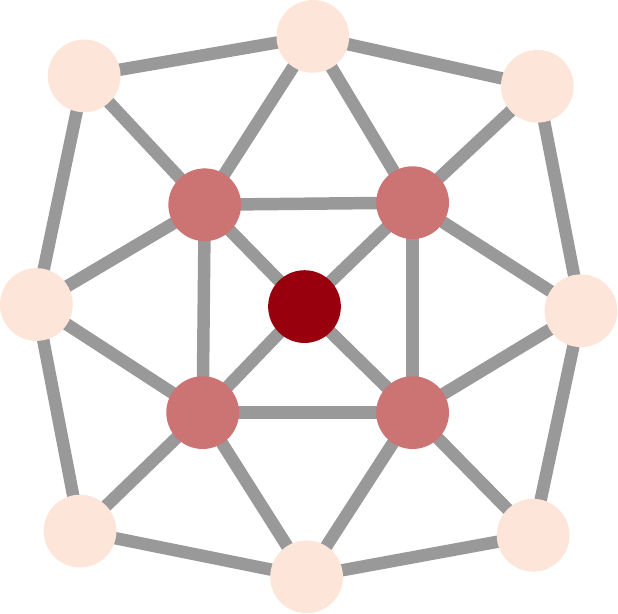}
\caption{}
\end{subfigure}
\quad
\begin{subfigure}[t]{.4\columnwidth}
\includegraphics[width=\textwidth]{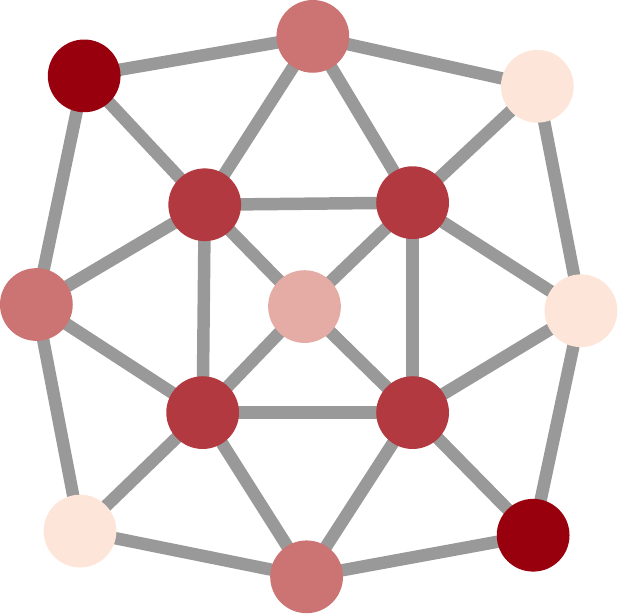}
\caption{}
\end{subfigure}
\caption{(a) Theory-driven mechanically explained activation times, represented by the node color (from dark to bright). (b) Real data swamped with noise, which only mildly conforms to the network topology.}
\label{fig:hairball-sis}
\end{figure}

My second example is epidemics spread -- Figure \ref{fig:hairball-sis}. As we saw in Part \ref{par:sis}, SIS/SIR models tell us exactly when the next node is going to be activated. In practice, data about real activation times has (a) high levels of noise, (b) many exogenous factors that have as much power in influencing how the infection spreads as the network connections have.

\begin{figure}
\centering
\begin{subfigure}[t]{.4\columnwidth}
\includegraphics[width=\textwidth]{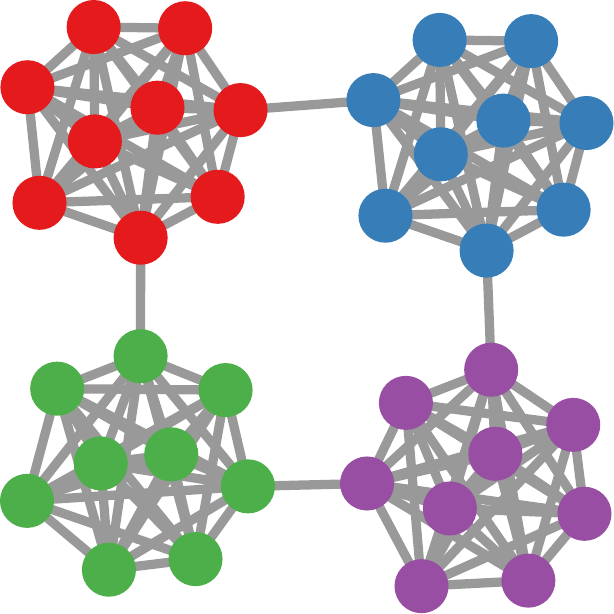}
\caption{}
\end{subfigure}
\quad
\begin{subfigure}[t]{.4\columnwidth}
\includegraphics[width=\textwidth]{figures/hairball_comms.jpg}
\caption{}
\end{subfigure}
\caption{(a) Well-separated groups internally densely connected. (b) The ubiquitous and mighty hairball.}
\label{fig:hairball-cd}
\end{figure}

Third, and more famously, communities. We are not going to dive deeply into the topic only until Part \ref{par:cd}. But, very superficially, when it comes to community discovery, the vast majority of papers propose a very naive standard definition of what constitute communities in a network: ``Groups of nodes that have a very large number of connections among them and very few to nodes outside the group''. Many papers claim that most networks have this kind of organization -- references provided in Part \ref{par:cd}. $99\%$ of networks will instead look like a blobbed mess. We have not one but three names for this useless visualization of an (apparently) useless network structure: ridiculogram -- a term which you can find sneaking around in some papers\cite{holme2011atmospheric} and attributed to Marc Vidal --; spaghettigraph -- a term I'm fond of due to my Italian origins; and hairball -- the term I'll use from now on in the book. See Figure \ref{fig:hairball-cd} for an example.

\begin{figure*}[b]
\centering
\begin{subfigure}[t]{.3\columnwidth}
\includegraphics[width=\textwidth]{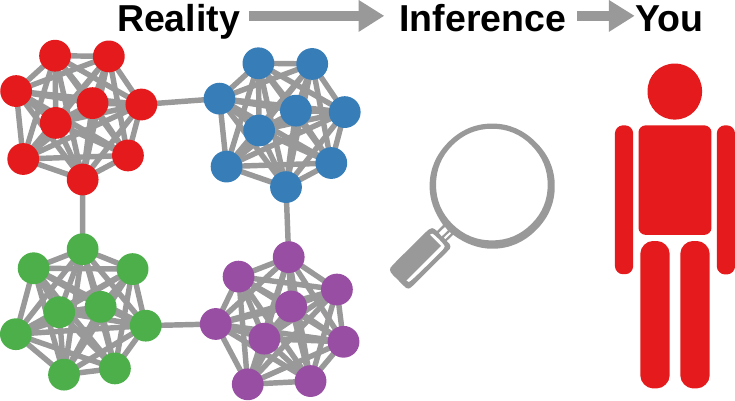}
\caption{}
\end{subfigure}
\quad
\begin{subfigure}[t]{.3\columnwidth}
\includegraphics[width=\textwidth]{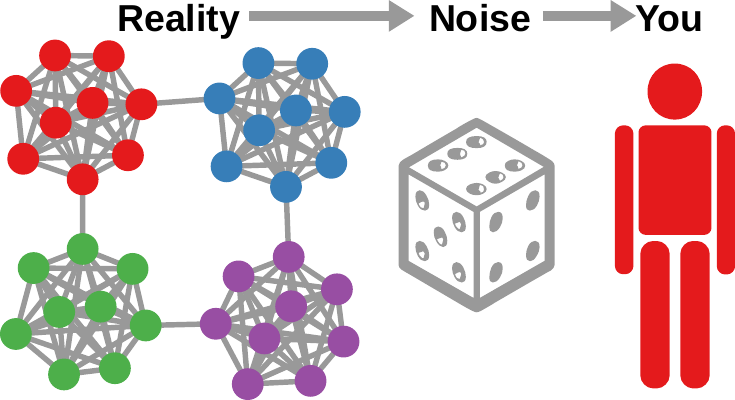}
\caption{}
\end{subfigure}
\quad
\begin{subfigure}[t]{.3\columnwidth}
\includegraphics[width=\textwidth]{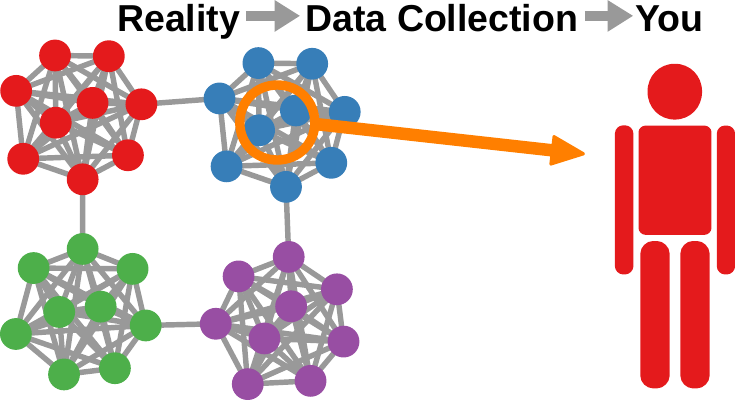}
\caption{}
\end{subfigure}
\caption{The typical breeding grounds for hairballs: (a) Indirect observation, (b) Noise in the data, (c) Incomplete samples.}
\label{fig:hairball-origins}
\end{figure*}

There are a few ways in which hairballs arise, which are the focus of this book part. First, many networks are not observed directly: they are inferred (Figure \ref{fig:hairball-origins}(a)). If the edge inference process you're applying does not fit your data, it will generate edges it shouldn't. Second, even if you observe the network directly, your observation is subject to noise (Figure \ref{fig:hairball-origins}(b)), connections that do not reflect real interactions but appear due to some random fluctuations. Finally, you might have the opposite problem: you're looking at an incomplete sample (Figure \ref{fig:hairball-origins}(c)), and thus missing crucial information.

In the chapters of this book part, we tackle each one of these problems to see some examples in which you can avoid giving birth to yet another hairball. Chapter \ref{cha:backboning} deals with network backboning: how to clear out noise from your edge observations. Chapter \ref{cha:sampling} focuses on the problem of network sampling: if you have a huge network in front of you, how do you extract a part of it so that your sample is representative?

Here, we start by tackling the first problem: how to deal with indirectly observed networks. Most of the times, you want to connect things because they are somehow similar, or they do similar things, or they relate to similar things. For instance, you want to connect users because they watch the same movies on Netflix. The most natural way to represent these cases is with bipartite networks (see Section \ref{sec:extended-bip}): in my example, a network connecting each user to the movie they watched. However, you don't want a bipartite network, you want a normal, down-to-earth, honest-to-god unipartite network. What can you do in this case?

Project! Bipartite projection means that you have a bipartite network with nodes of type $V_1$ and $V_2$, and you want to create a unipartite network with only nodes of type $V_1$ (or $V_2$). In my Netflix example, all you observe is people watching movies. As I said before, this is a bipartite network: nodes of type $V_1$ are people, nodes of type $V_2$ are movies, and edges go from a person to a movie if the person watched the movie. However, the holy grail is to know which movies are similar, to make recommendations to similar users.

In the following sections we explore the different ways in which one can project a bipartite network. They all boil down to the same strategy: we use a different criterion to give the projected edges a weight, we establish a threshold, and drop the edges below this minimum acceptable weight.

\section{Simple Weights}

\begin{figure}[b]
\centering
\includegraphics[width=.66\columnwidth]{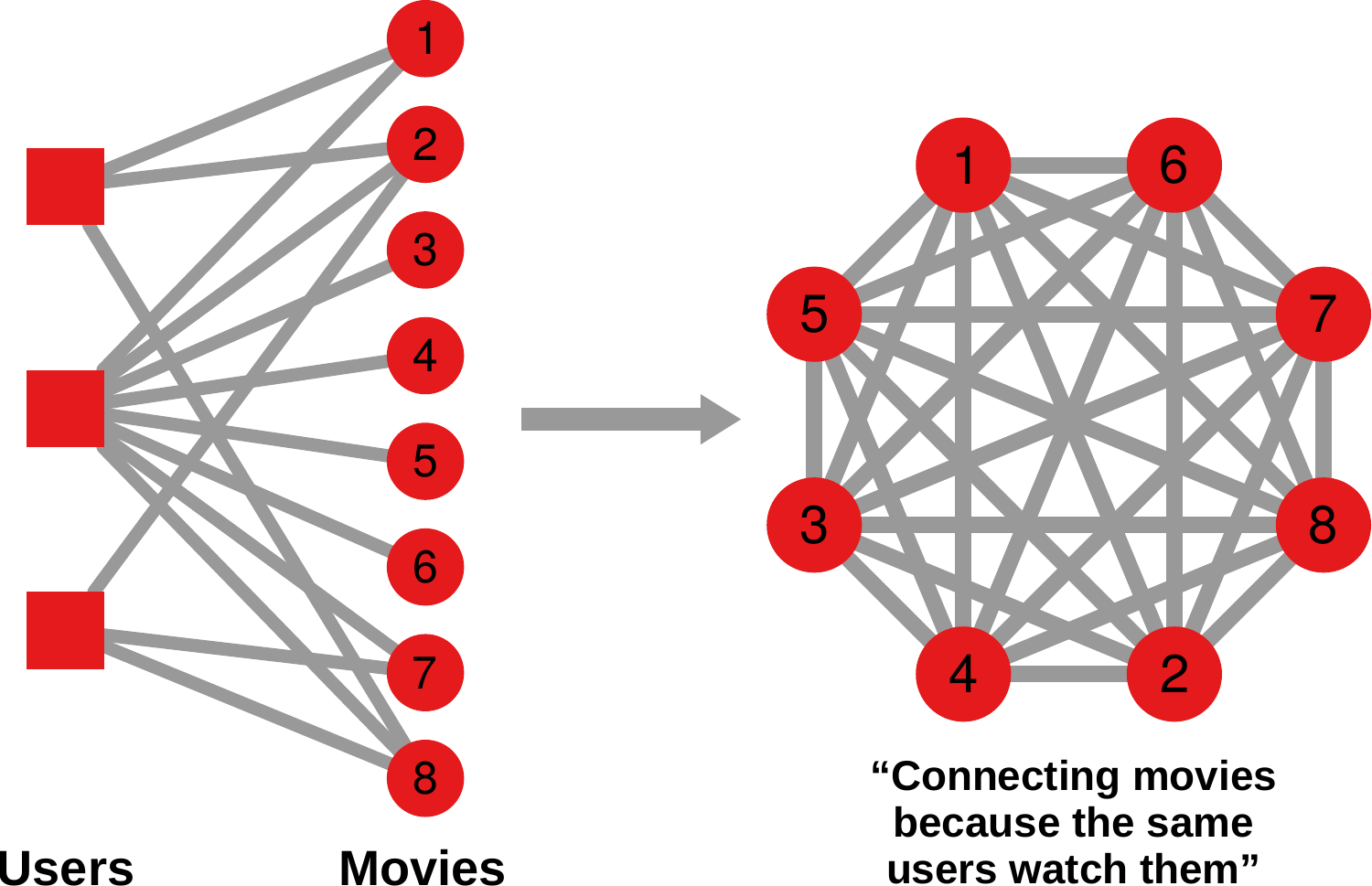}
\caption{An example of naive bipartite projection, where we connect nodes of one type if they have a common neighbor.}
\label{fig:projection-naive}
\end{figure}

Let's stick with our Netflix example\footnote{Note that, hereafter, I ignore the fact that in Netflix you could also rate the movie, i.e. that the bipartite network is weighted. In my example, I treat the bipartite network as unweighted.}. Naively, you might think that you can connect movies because the same people watched them\cite{newman2001scientific} -- as in Figure \ref{fig:projection-naive}. The problem is that -- as we saw -- degree distributions are broad. This means that there are going to be some users in your bipartite user-movie network with a very high degree. These are power users, people who watched everything. They are a problem: under the rule we just gave to project the bipartite networks, you'll end up with all movies connected to each other. A hairball. The key lies in recognizing that not all edges have the same importance. Two movies that are watched by three common users are more related to each other than two movies that only have one common spectator.

\begin{figure}
\centering
\includegraphics[width=.66\columnwidth]{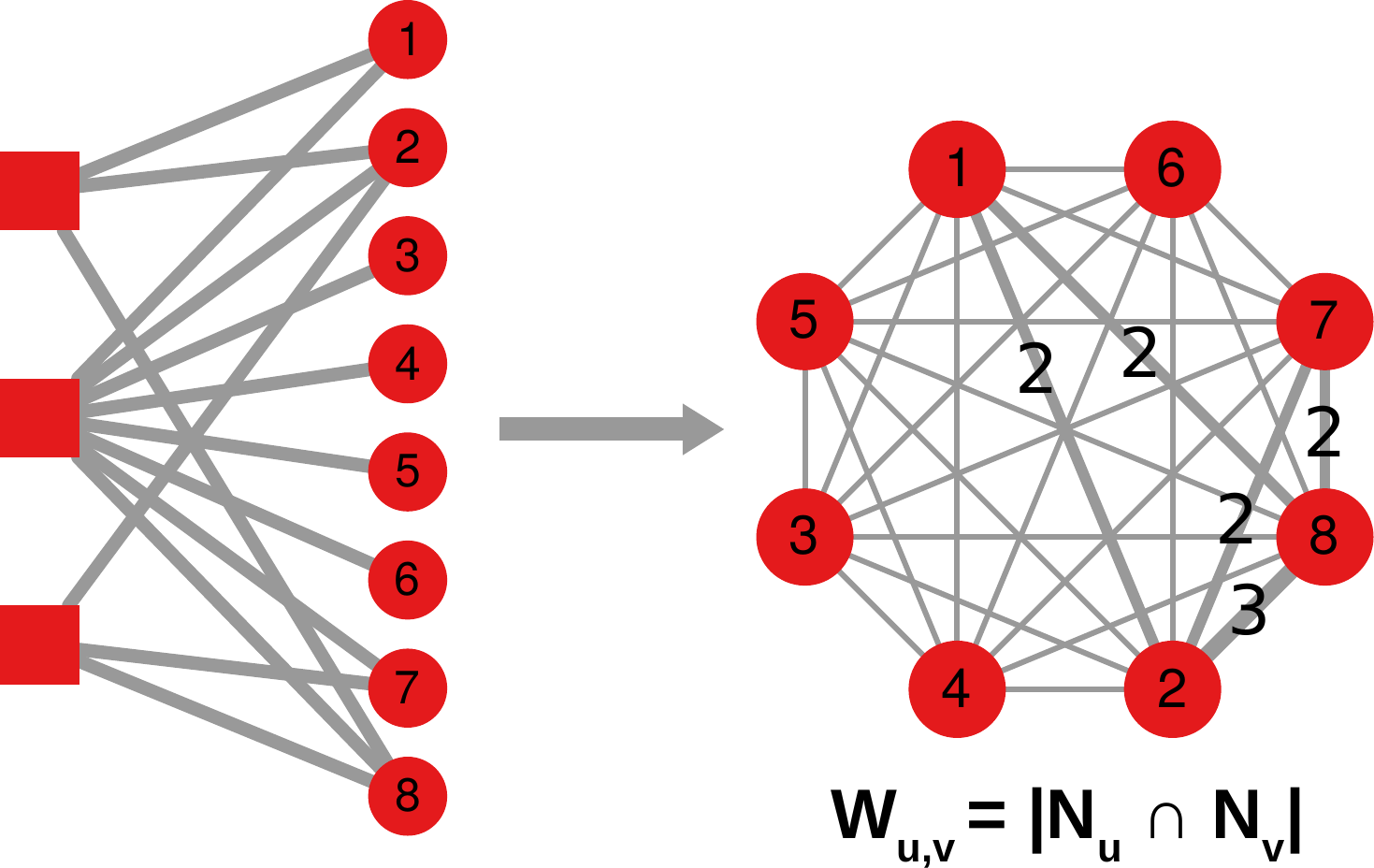}
\caption{An example of Simple Weight bipartite projection, where we connect nodes of one type with the number of their common neighbors.}
\label{fig:projection-simple}
\end{figure}

The easiest way to take this information into account is to perform \textbf{simple weighting}. For each pair of nodes you identify the number of common neighbors they have, and that's the weight of the edge -- see Figure \ref{fig:projection-simple}. In practice, you don't simply require that movies are connected if there is at least one person who has watched both of them. You connect movies with a weighted link, and the weight is the number of people who watched them both: $w_{u,v} =  |N_u \cap N_v|$. This weighting scheme is similar to Common Neighbors in link prediction (Section \ref{sec:lpsimple-cn}), and of course you can do a Jaccard correction by normalizing it with the size of the union of the neighbor sets: $w_{u,v} =  |N_u \cap N_v| / |N_u \cup N_v|$.

If you like to think in terms of matrices (Chapter \ref{cha:mat}), this is equivalent to multiplying the bipartite adjacency matrix with its transpose. Of course, you need to pay attention to the dimension onto which you're projecting. If $A$ is a $|V_1| \times |V_2|$ matrix, then $AA^T$ is a $|V_1| \times |V_1|$ matrix, while $A^TA$ is a $|V_2| \times |V_2|$ one. When multiplying binary matrices, the result in cell $A_{uv}$ is the number of common entries set to one between the $u$th and the $v$th rows, which is exactly the number of common neighbors between nodes $u$ and $v$. The diagonal will tell you the degree of the node, which you can simply set to zero.

This approach can be integrated with a second step\cite{saracco2017inferring}. In this second step, one wants to evaluate the statistical significance of the edge weights you obtained by counting the number of common neighbors. This is sort of the same thing as first projecting and then performing network backboning -- a task we'll see in Chapter \ref{cha:backboning}. The main difference is that this backboning is specially defined to clean up the result of bipartite projections. One can define a series of null bipartite network models, either via exponential random graphs (Section \ref{seg:ergmodels-ergm}) or configuration model (Section \ref{sec:csmodels-conf}). These null models will give birth to a bunch of null projections, which will give an expected weight for all possible edges in the unipartite network. Then, you can keep in your projection only those links significantly exceeding random expectation.

\section{Vectorized Projection}\label{sec:projections-vectors}
There are many criticisms of the simple counts as a weighting approach. Here we see the one called saturation problem\cite{li2005weighted}. Another issue is the bandwidth problem, which I explain in Section \ref{sec:projections-hyper}, along with the projection methods designed to fix it.

Some authors noticed that the simple count scheme has what they call a ``saturation'' problem. As an illustration, consider the following example: suppose you are an author and you collaborated with another scientist on a new paper. The contribution of that new paper to your similarity is not linear. If in your previous history you only had a single other paper with this person, then the new paper is your second collaboration. This is a strong contributor: it represents $50\%$ of your entire scientific output. If, instead, this was your hundredth collaboration, this new paper only adds little to your connection strength. Giving the same weight in these two different scenarios is not a good proxy to estimate the similarity in the original network.

We can exploit edge weights to solve the saturation problem. Edge weights are something that simple counting cannot handle easily, and if you try to handle them by doing a weighted simple counting, you probably end up doing something similar to what I present in this section anyway. In this scenario, you don't want to count each common $V_2$ neighbor equally. You need your adjacency matrix to contain non-zero values different than one. For simplicity, I'm going to make the following examples with a binary matrix anyway, also to show that these techniques can handle this simpler scenario as well.

Our sophisticated needs imply that we need to change the way we look at the problem. As the title of this section suggests, we are considering \textbf{nodes as vectors}. Specifically, consider the binary adjacency matrix of our bipartite network. Each row is a node of type $V_1$. Each entry tells us whether it is connected to a node of type $V_2$. So we can see a node as a vector of zeroes and ones.

\begin{figure}
\centering
\includegraphics[width=\columnwidth]{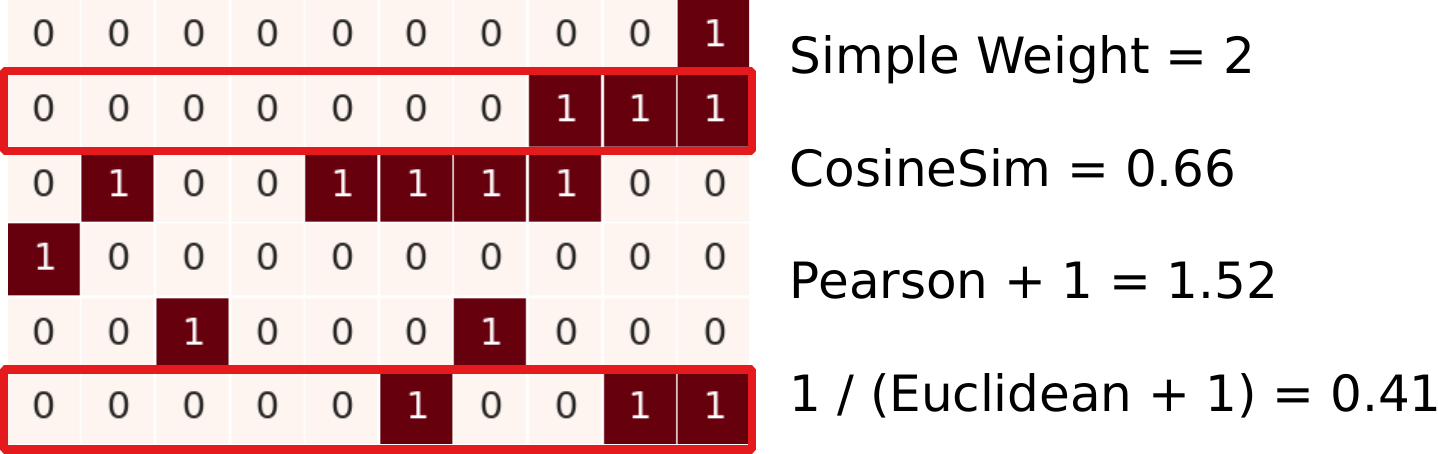}
\caption{An example of vectorized bipartite projection, where we connect nodes of one type with the inverse of some vector distance measure of their rows in the bipartite adjacency matrix.}
\label{fig:projection-vector}
\end{figure}

Once we do -- as Figure \ref{fig:projection-vector} shows --, we discover that we can apply a large number of distance metrics between two numerical vectors. If these numerical vectors represent two $V_1$ nodes, then the distance between them must be -- inversely -- related to how similar they are. Popular choices to establish the strength of the connection between these two nodes are the Euclidean distance, cosine similarity and Pearson correlation -- but the list could be much longer and you can get inspiration from the set of vector distance measures implemented in any statistical library\footnote{\url{https://docs.scipy.org/doc/scipy/reference/spatial.distance.html}}.

One nice thing about many of these measures, besides properly handling edge weights, is that they handle also common zeroes. In simple weighting, you only count common neighbors. However, two nodes might be similar also based on the neighbors they \textit{don't} connect to. This is elegantly handled in the Pearson correlation, for instance. Such indirect effects are not always good: for instance they are a problem when performing link prediction\cite{barzel2013network}.

You need to be aware of a few problems with this approach. First, it's not always immediately obvious how to translate a distance into a similarity while preserving its properties. You cannot always take the inverse, or multiply by minus one, or doing one minus the distance. Each of these solutions might work with some measures, but catastrophically fail with others.

The second issue is more subtle. None of these measures were really developed with network data in mind. So they might not work because they don't take into account what the edge creation process of the bipartite network looks like. They are not going to necessarily solve the issues simple weighting has, because they're still prone to fall into the trap of large hubs and very skewed degree distributions.

\section{Hyperbolic Weights}\label{sec:projections-hyper}
The second problem is similar to the saturation one, but cannot be solved by looking at edge weights. If you're in a CERN paper, you coauthor with hundreds of people, but you don't really know all of them. In practice, we're acknowledging that ``bandwidth'' is finite: having too many coauthors implies having only a superficial relationship with all of them. This bandwidth argument is not new, we saw a similar one when we introduced link prediction methods like Adamic-Adar in Section \ref{sec:lpsimple-aa} and Resource Allocation in Section \ref{sec:lpsimple-ra}.

\begin{figure}
\centering
\includegraphics[width=.66\columnwidth]{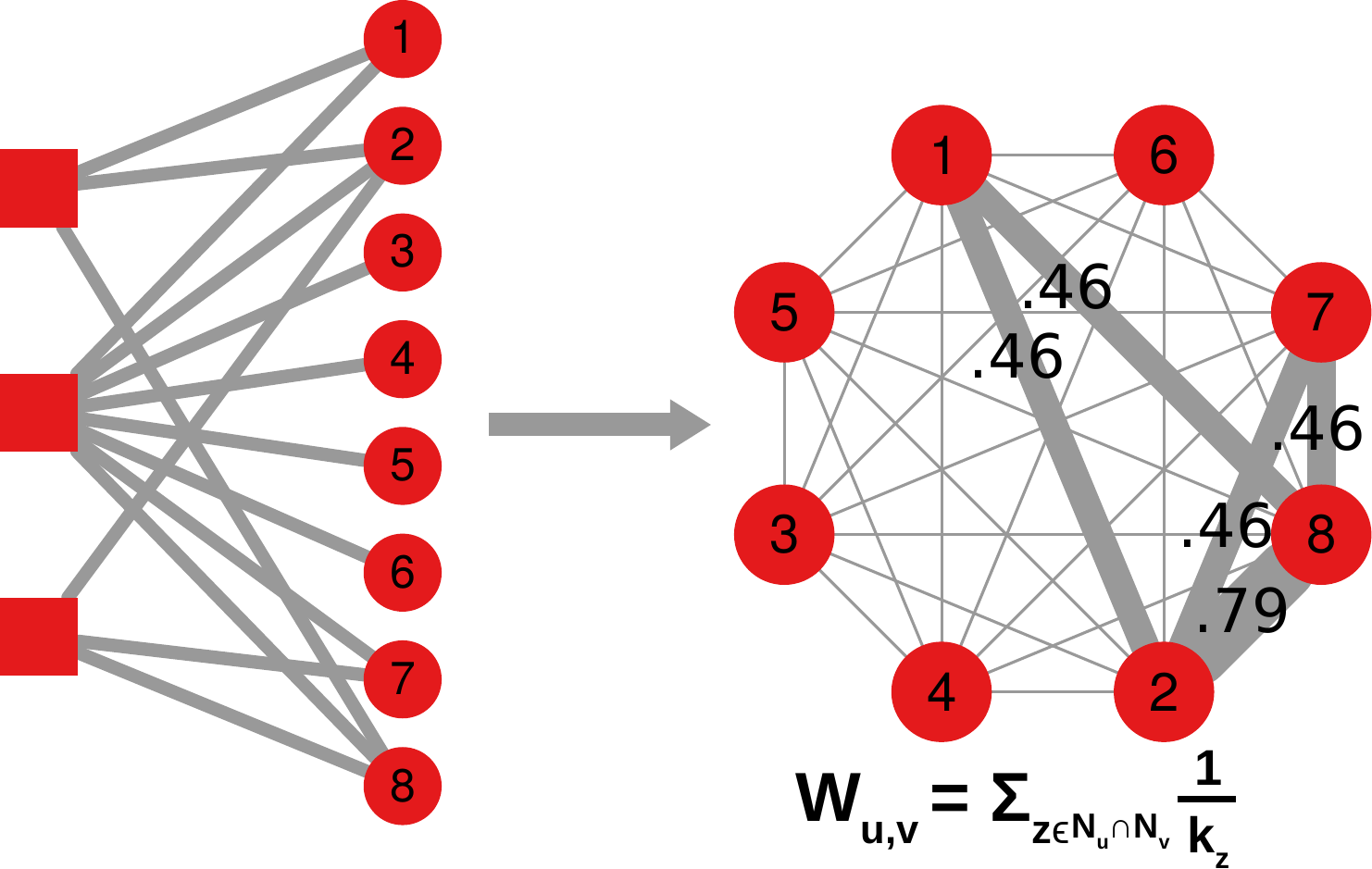}
\caption{An example of Hyperbolic Weight bipartite projection, where each common neighbor $z$ contributes $k_z^{-1}$ to the sum of the edge weight.}
\label{fig:projection-hyper}
\end{figure}

In \textbf{hyperbolic weight} we recognize that hubs contribute less to the connection weight than non-hubs\cite{newman2001scientific2}. Such a weight scheme is similar to the link prediction strategies I just mentioned: each common neighbor $z$ contributes $k_z^{-1}$ rather than $1$ to the weight of the edge connecting the two nodes: $w_{u,v} =  \sum \limits_{z \in N_u \cap N_v} \dfrac{1}{k_z - 1}$. The final result in this example is similar to simple weight -- see Figure \ref{fig:projection-hyper} --, but it exaggerates the differences, so that thresholding becomes easier.

Note that the minus one in the denominator -- which we do because $u$ never checks its similarity with itself -- means that we're effectively ignoring all papers with only one author. And, if an author only wrote with herself, she won't appear in the network. This makes sense at some level -- how can you connect with anybody else if you never collaborate? -- but it also implies that there is going to be no information in the diagonal of the resulting adjacency matrix.

Again, this projection is simple to implement as a matrix operation. Rather than multiplying the bipartite adjacency matrix with its transpose, you multiply it with the transpose of its degree normalized stochastic version. If you do so, rather than counting the common ones, you sum up all the $1 / k_z$ entries. Again, pay attention to the dimension over which you project, because normalizing by row sum or by column sum will change the result.

\section{Resource Allocation}\label{sec:projections-probs}
\begin{figure}
\centering
\includegraphics[width=.66\columnwidth]{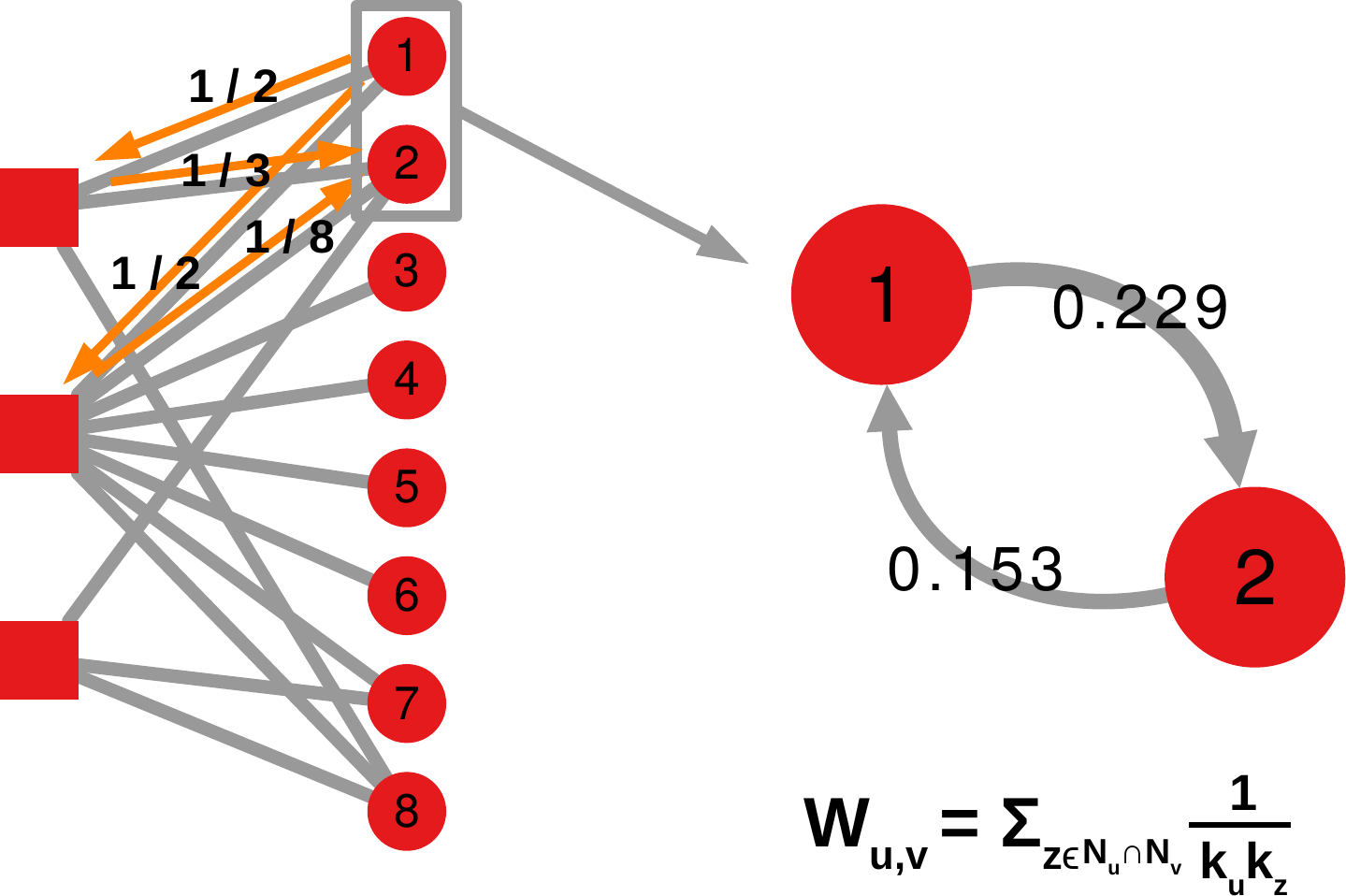}
\caption{An example of Resource Allocation bipartite projection, where each common neighbor $z$ contributes $(k_u k_z)^{-1}$ to the sum of the edge weight. When connecting node $1$ to node $2$, from node $1$'s perspective the edge weight is $(1/2 * 1/3) + (1/2 * 1/8)$, because the two common neighbors have degree of $3$ and $8$, respectively, and node $1$ has degree of two. However, from node $2$'s perspective, the edge weight is $(1/3 * 1/3) + (1/3 * 1/8)$, because node $2$ has three neighbors.}
\label{fig:projection-resources}
\end{figure}

In \textbf{resource allocation} we do the same thing as hyperbolic weight, but considering two steps instead of one. Rather than only looking at the degree of the common neighbor, we also look at the degree of the originating node\cite{zhou2007bipartite}. In the paper-writing example, not only it is unlikely to be strongly associated with a co-author in a paper with hundreds of authors, it is also difficult to give attention to a particular co-author if you have many papers with many other people. So each common neighbor $z$ that node $u$ has with node $v$ contributes not $k_z^{-1}$ as in hyperbolic weights, but $(k_u k_z)^{-1}$ -- see Figure \ref{fig:projection-resources}. The weight is then:

$$w_{u,v} =  \sum \limits_{z \in N_u \cap N_v} \dfrac{1}{k_u k_z}.$$

This generates the unipartite weight matrix $W$.

This strategy also works for weighted bipartite networks. If $B$ is your weighted bipartite adjacency matrix, the entries of $W$ are:

$$w_{u,v} =  \sum \limits_{z \in N_u \cap N_v} \dfrac{B_{uv}}{k_u k_z}.$$

In practice, you replace the $1$ in the numerator with the edge weights connecting $z$ to $v$ and $u$. Moreover, we can also have node weights, noticing that some nodes might have more resources than others. Suppose that you have a function $f$ giving each node in the network a resource weight. After you perform the resource allocation projection, each node will have a new amount of resources $f' = Wf$.

Note that, in this case, $W$ is not symmetric: in the scenario with a single common neighbor $z$, $u$'s score for $v$ would be $(k_u k_z)^{-1}$, while $v$'s score would be $(k_v k_z)^{-1}$. If $k_u \neq k_v$, then the scores are different. In many cases, this provides a better representation of the network than one ignoring asymmetries. You might be the most similar author to me because I always collaborated with you, but if you also contributed to many other papers with other people, then I might not be the author most similar to you.

$W$ has a well-defined diagonal: $w_{u,u} =  \sum \limits_{z \in N_u} \dfrac{1}{k_u k_z} = \dfrac{1}{|N_u|} \sum \limits_{z \in N_u} \dfrac{1}{k_z}$. In fact, this diagonal is the maximum possible similarity value of the row: only a node $v$ with the very same neighbors and nothing else can have a weight $w_{u,v} = w_{u,u}$. 

In some other cases you might consider having a directed projection an inconvenience, because you really want an undirected network as a result. You can make the result of resource allocation symmetric by always choosing the minimum or maximum between $w_{u,v}$ and $w_{v,u}$, or simply their average: $(w_{u,v} + w_{v,u}) / 2$. Also self-loops can be annoying sometimes. If you have no use for them, you can manually set $W$'s diagonal to zero.

The resource allocation as presented so far is only one of the many possible variants following the same idea. The one I explained so far is known as ProbS and uses $k_u$, the degree of the origin of the two-step random walk, as the normalizing factor. A variant known as HeatS\cite{zhou2010solving} uses instead $k_v$, the destination of the random walk. Thus, the weight of the $u,v$ connection is now $w_{u,v} =  \sum \limits_{z \in N_u \cap N_v} \dfrac{1}{k_v k_z}$.

Surprising absolutely no one, some authors decided to combine ProbS and HeatS in a single Hybrid framework\cite{lu2011information}. The combination is exactly what you would expect: $w_{u,v} =  \sum \limits_{z \in N_u \cap N_v} \dfrac{1}{k_u^\lambda k_v^{(1 - \lambda)} k_z}$. This introduces a parameter in the equation: $\lambda$. This should be a number between zero and one, determining how much importance the degree of the origin has compared to the degree of the destination. For $\lambda = 0$ you have HeatS, for $\lambda = 1$ you have ProbS, and for $\lambda = 1/2$ you have the middle point between HeatS and ProbS.

In matrix terms, ProbS is the same as the hyperbolic projection (Section \ref{sec:projections-hyper}), but now you normalize differently. In hyperbolic, you multiply the adjacency matrix $A$ with its degree-normalized transpose. In ProbS you multiply the stochastic with a stochastic version of the transpose. Meaning, in hyperbolic first you normalize then you transpose, in ProbS first you transpose and then you normalize. Finally, HeatS is the transpose of ProbS.

\section{Random Walks}\label{sec:projections-ycn}
In \textbf{Random Walks}, we take the resource allocation to the extreme. Rather than looking at 2-step walks, we look at infinite length random walks. Which means that the strength between $u$ and $v$ is the probability of visiting $v$ starting from $u$. If we have infinite random walks, this means that we can use the stationary distribution to estimate the edge weight: $w_{u,v} = \pi_v A_{u,v}$, where $A$ is a transition probability matrix (recording the probability of the path $u \rightarrow z \rightarrow v$, for any $z$)\cite{yildirim2014using}. Note that $A$ here is different than a simple binary adjacency matrix, as it encodes the probabilities of all random walks of length two. This means that its interpretation is slightly different than what I originally presented for $\pi$ in Section \ref{sec:centr-eigen}.

\begin{figure}
\centering
\includegraphics[width=.66\columnwidth]{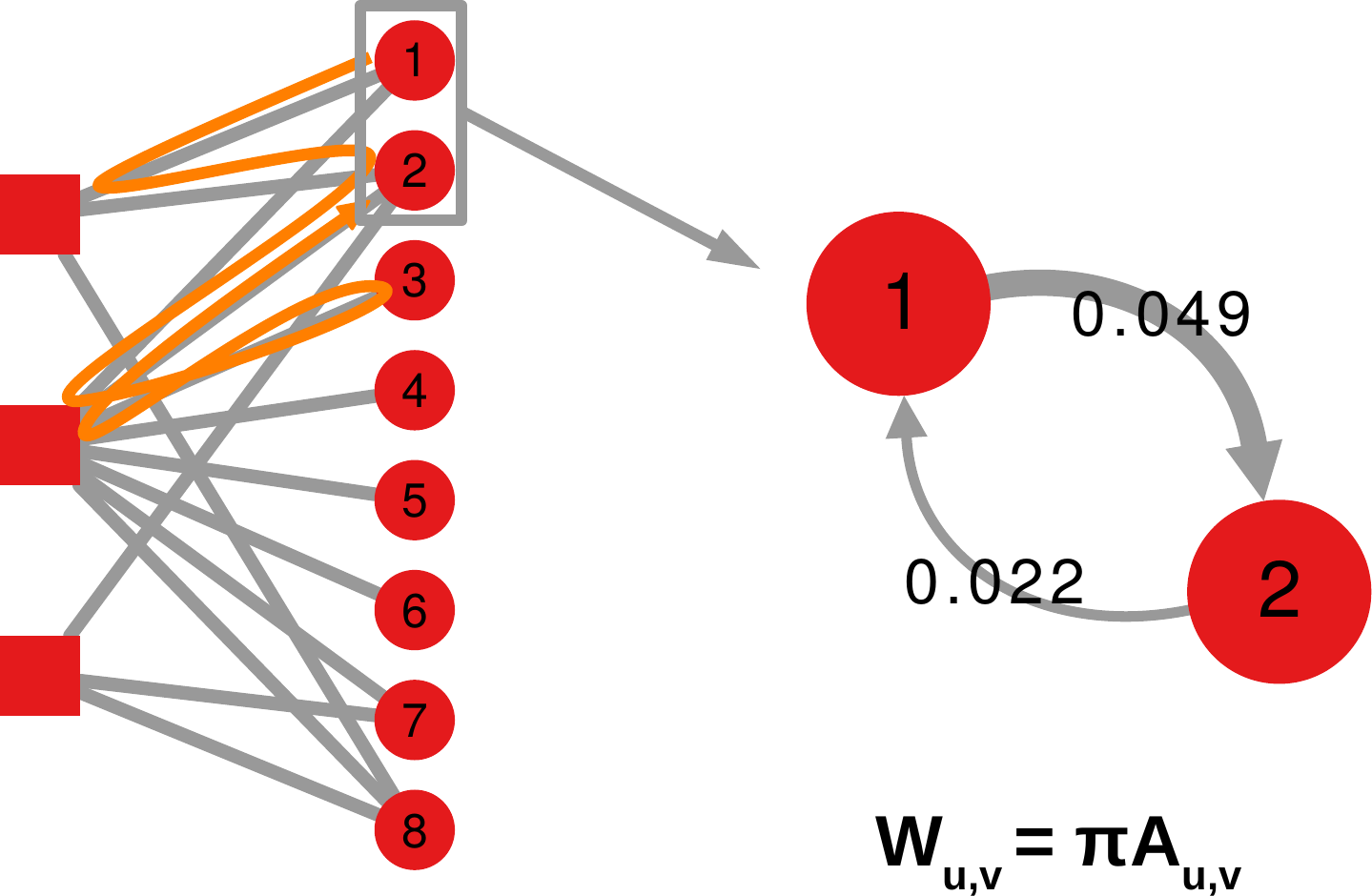}
\caption{An example of Random Walks bipartite projection, where the connection strength between $u$ and $v$ is dependent on the stationary distribution $\pi$, telling us the probability of ending in $v$ after a random walk.}
\label{fig:projection-rw}
\end{figure}

In matrix terms, you take the result ProbS' multiplication, which is a square $|V_1| \times |V_1|$ matrix, and you multiply it with its stationary distribution.

As in the resource allocation case, this means that the measure is not symmetric, and the differences between nodes now are more extreme than before: the $1 \rightarrow 2$ edge weight is now more than twice as $2 \rightarrow 1$, while in resource allocation it was just about 50\% higher. See Figure \ref{fig:projection-rw} for an example. Another parallelism between these two approaches is the presence of a well-defined diagonal, which you can use in case you're not afraid of self-loops (I am).

\section{Comparison in a Practical Scenario}
I showed you how these different methods approach the projection process and the different results they obtain in a toy example. Do these differences in simple scenarios translate to big practical differences in real-world cases? To answer this question, let's just take a superficial look at the projections I get using a bipartite network extracted from Twitter. In the network, the nodes of type $V_1$ are websites, and nodes of type $V_2$ are Twitter users. I connect a Twitter user to a website if the user included the URL of the website in one of her tweets.

\begin{figure*}
\centering
\begin{subfigure}[t]{.24\columnwidth}
\includegraphics[width=\textwidth]{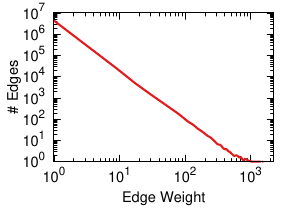}
\caption{Simple}
\end{subfigure}
\begin{subfigure}[t]{.24\columnwidth}
\includegraphics[width=\textwidth]{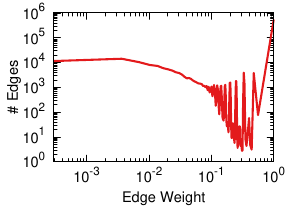}
\caption{Jaccard}
\end{subfigure}
\begin{subfigure}[t]{.24\columnwidth}
\includegraphics[width=\textwidth]{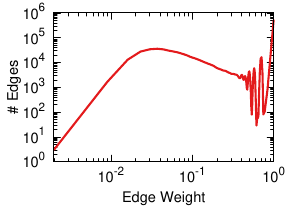}
\caption{Cosine}
\end{subfigure}
\begin{subfigure}[t]{.24\columnwidth}
\includegraphics[width=\textwidth]{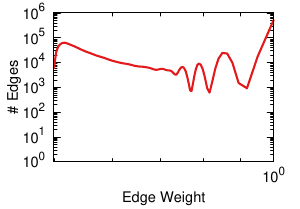}
\caption{Pearson}
\end{subfigure}
\begin{subfigure}[t]{.24\columnwidth}
\includegraphics[width=\textwidth]{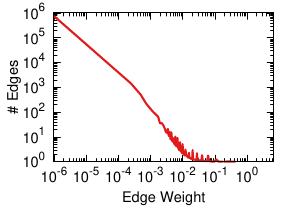}
\caption{Hyperbolic}
\end{subfigure}
\begin{subfigure}[t]{.24\columnwidth}
\includegraphics[width=\textwidth]{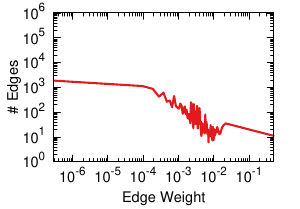}
\caption{ProbS}
\end{subfigure}
\begin{subfigure}[t]{.24\columnwidth}
\includegraphics[width=\textwidth]{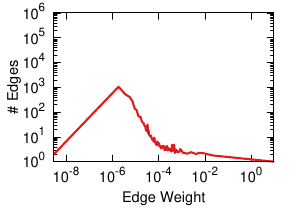}
\caption{Hybrid ($\lambda = 0.5$)}
\end{subfigure}
\begin{subfigure}[t]{.24\columnwidth}
\includegraphics[width=\textwidth]{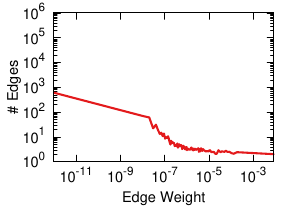}
\caption{Random Walks}
\end{subfigure}
\caption{The distributions of edges weights in the projected Twitter network for eight different projection methods. The plot report the number of edges (y axis) with a given weight (x axis).}
\label{fig:projection-realcases}
\end{figure*}

I project the network so that I have a unipartite version with only $V_1$ nodes: websites are connected if the same users tweet about them. This is a sort of website similarity index. Now let's see how different the space of edge weights looks like if we use different approaches. This is what Figure \ref{fig:projection-realcases} is all about.

The figure shows that the space of the edge weights looks pretty different according to different projection methods. For instance, the simple projection (Figure \ref{fig:projection-realcases}(a)) shows a power law distribution of edge weights, with more than four million edges with weight equal to one and one edge with weight equal to $2,252$, while the average weight is $2.14$. This is very much not the case for other projection strategies such as Jaccard (Figure \ref{fig:projection-realcases}(b)), where there is no trace of a power law. And, in many cases such as cosine and Pearson (Figures \ref{fig:projection-realcases}(c-d)), the highest edge weight is actually the most common value, rather than being an outlier such as in the hyperbolic projection (Figure \ref{fig:projection-realcases}(e)).

\begin{figure*}
\centering
\begin{subfigure}[t]{.3\columnwidth}
\includegraphics[width=\textwidth]{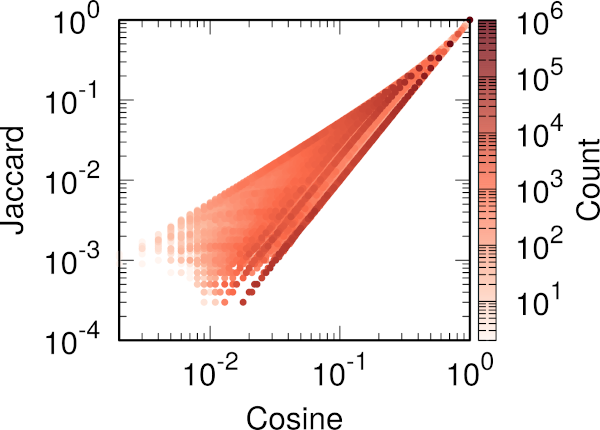}
\caption{}
\end{subfigure}
\qquad
\begin{subfigure}[t]{.3\columnwidth}
\includegraphics[width=\textwidth]{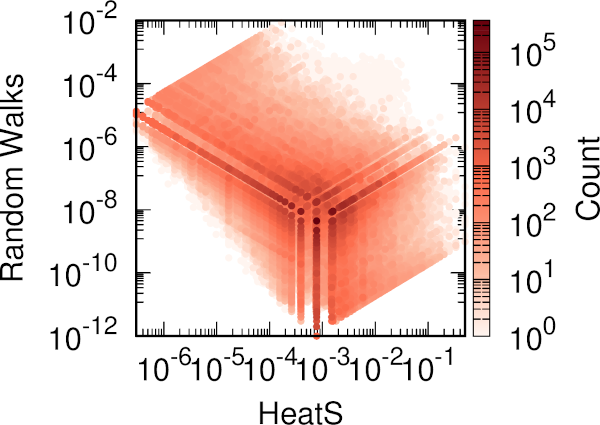}
\caption{}
\end{subfigure}
\qquad
\begin{subfigure}[t]{.3\columnwidth}
\includegraphics[width=\textwidth]{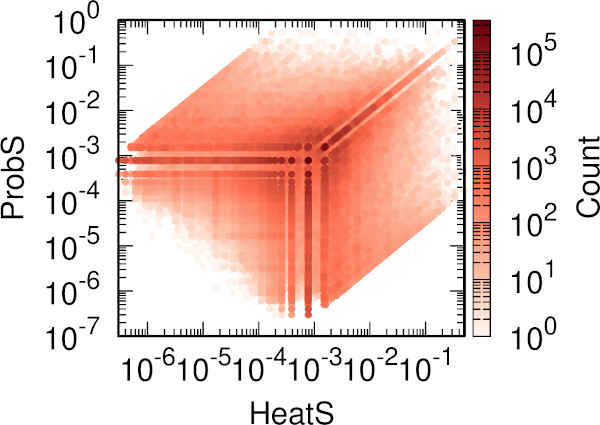}
\caption{}
\end{subfigure}
\caption{The comparison between the edge weights according to different network projections. Each point is an edge. The x-y coordinates encode its weight in the two different projections. The color encodes how many edges share she same x-y score. (a) Cosine vs Jaccard; (b) HeatS vs Random Walks; (c) HeatS vs ProbS.}
\label{fig:projection-realcases2}
\end{figure*}

Is the difference exclusively in the shape of the distribution, or do these approaches disagree on the weights of specific edges? To answer this question we have to look at a scattergram comparing the edge weights for two different projection strategies. This is what I do in Figure \ref{fig:projection-realcases2}.

I picked three cases to show the full width of possibilities. In Figure \ref{fig:projection-realcases2}(a), I compare the cosine projection against the Jaccard one. This is the pair of projections that, in this dataset, agree the most. Their correlation is $>0.94$. Looking at the figure, it is easy to see that there isn't much difference. You can pick either method and you're going to have comparable weights. The opposite case compares two method that are anti-correlated the most. This would be HeatS and the random walks approach, in Figure \ref{fig:projection-realcases2}(b). They correlation in a log-log space is a staggering $-0.7$. From the figure you can probably spot a few patterns, but the lesson learned is that the two methods build fundamentally different projections.

Ok, but these are extreme cases. How does the average case looks like? To get an idea, I chose a particular pair of measures: HeatS and ProbS (Figure \ref{fig:projection-realcases2}(c)). You might expect the two to be more similar than the average method: after all, one is the transpose of the other. You'd be very wrong. In this dataset, HeatS and ProbS are actually anti correlated, at $-0.34$ in the log-log space. HeatS and ProbS would be positively correlated if the nodes of type $V_1$ with similar degrees connect to the same nodes of type $V_2$. But that is not the case in this specific Twitter dataset. Here, it is not true that the people sharing lots of URLs share the same URLs.

At this point, you might be asking yourself how do you choose the projection method that is most suitable for your application. The general guideline is to study what each method does and see if it aligns with your expected edge generation process. However, I feel it's a bit too early to ask this question. That is because network projection is rarely the only thing you're going to do. Almost all these methods return the same set of non-zero weighted edges. They also return extremely dense projections, as a single common node is enough to create an edge in the projection.

In fact, the Twitter data I just used has $\sim 15k$ users and $\sim 14k$ domains, with $\sim 175k$ edges connecting them. The undirected projections return $\sim 5.3M$ edges, meaning a density of $2 \times 5.3M / 14k^2 \sim 5\%$, or an average degree of $\sim 713$. This is usually way too much for an intelligible network. That is why, if we want to avoid hairballs and related problems, these techniques -- while necessary -- are not usually sufficient. The process to get rid of hairballs has two steps: first one performs the bipartite projection, and then she applies a threshold to throw away low-weighted edges. The next chapter expands on how to perform this second step properly.

\section{Summary}

\begin{enumerate}
\item Most network analysis algorithms work with unipartite networks, but many phenomena have a natural bipartite representation. To transform a bipartite network into a unipartite network you need to perform the task of network projection.
\item In network projection you pick one of the two node types and you connect the nodes of that type if they have common neighbors of the other type. Normally you'd count the number of common neighbors they have (simple weighted) and then evaluate their statistical significance.
\item Real world bipartite networks have broadly distributed degrees which might make your projection close to a fully connected clique. Then you need a smart weighting scheme to aid you in removing weak connections.
\item You could use standard vector distances (cosine, euclidean, correlation) but we have specialized network-aware techniques. For instance, considering nodes as allocating resources to their neighbors, inversely proportional to the number of neighbors they have (hyperbolic).
\item In resource allocation, you also have nodes sending resources, but you take two steps instead of one: you're not discounting only for the degree of nodes of type one, but also for the degree of nodes of type two.
\item Finally, you can also do resource allocation with infinite length random walks by looking at the stationary distribution. The resulting edge weights from all these techniques can create very different network topologies.
\end{enumerate}

\section{Exercises}

\begin{enumerate}
\item Perform a network projection of the bipartite network at \url{http://www.networkatlas.eu/exercises/26/1/data.txt} using simple weights. The unipartite projection should only contain nodes of type $1$ ($|V_1| = 248$). How dense is the projection?
\item Perform a network projection of the previously used bipartite network using cosine and Pearson weights. What is the Pearson correlation of these weights compared with the ones from the previous question?
\item Perform a network projection of the previously used bipartite network using hyperbolic weights. Draw a scatter plot comparing hyperbolic and simple weights.
\end{enumerate}

\chapter{Network Backboning}\label{cha:backboning}
Network backboning is the problem of taking a network that is too dense and removing the connections that are likely to be not significant -- or ``strong enough''. If you ever found yourself in a situation thinking ``there are too many edges in this network, I'm going to filter some out'', then you performed network backboning. Even if it is rarely explicitly labeled like that, network backboning is one of the most common tasks performed in network analysis.

There are many reasons why you would want to backbone your network. First, this is a book part about hairballs. If your network is a hairball, meaning that the tangle of connections is too dense to reach any meaningful conclusion, you might want to sparsify your network. Graph sparsification\cite{spielman2004nearly}\cite{satuluri2011local} -- sometimes called ``pruning'' in combinatorics\cite{harabor2011online} and neural networks\cite{liu2018rethinking} -- could be an alternative name for backboning, but it is often used in a more narrow context, namely the second application field of backboning: your network simply has too many connections to be computationally tractable and so you need to filter out the ones that are unlikely to affect your computation. Finally, a third scenario might be the presence of noise: you don't know whether the edges you're observing are real connections and you need a statistical test to determine that. This overlaps with the notion of measurement error -- which is the topic of Chapter \ref{cha:uncertainty} --, but it is slightly different. The techniques presented in this chapter aim to remove connections that are likely to be noisy, while in Chapter \ref{cha:uncertainty} we will see how to deal with such connections without removing them, mainly via the use of probabilistic networks.

When wearing its ``graph sparsification'' hat, network backboning could be confused with graph summarization: the task of taking a large complex network and reducing its size so that we can describe it better. However, there is a crucial difference between the two tasks: one of the central objectives of graph summarization is to reduce the number of nodes of the network as much as possible, often even merging them into ``meta nodes''. This is exactly the opposite of what backboning wants to do: in network backboning you do not merge nodes and you want to keep as many as possible in your network. The reason is that you want to let the strong connections emerge, but you want to preserve all the entities in your data. If you remove nodes from your network, you cannot describe them directly any more in your analysis. In a nutshell, network backboning wants to allow you to perform node- and global-level analyses, while graph summarization only focuses on empowering meso-level analysis (Part \ref{par:meso}) where you lose sight of the single individual nodes. For this reason, graph summarization has its own chapter (Chapter \ref{cha:mining-summarization}) in a totally different part of this book.

There are several network backboning methods which aim to tackle this problem. I'll look at a few techniques divided in two macro categories: structural approaches (naive thresholding, Doubly Stochastic, High Salience Skeleton, convex network reduction), and statistical ones (Disparity Filter, Noise Corrected). These are, to the best of my knowledge, the most used and are the ones that I'm the most familiar with. There are other backboning methods, many of which are based on the same ``urn extraction'' procedure we'll see in depth when we talk about the noise-corrected backbone: for bipartite networks\cite{tumminello2011statistically}, using Polya urns\cite{marcaccioli2019polya}, and more. Another common approach is to create a null version of the observed network and testing the edge weights against such expectation\cite{radicchi2011information}, in line with the disparity filter we'll see. 

Note that finding the maximum spanning tree, planar maximally filtered graphs, and the triangulated maximally filtered graphs could also be considered a way to perform structural network backboning, and they were covered in Section \ref{sec:paths-spantree}.

The vast majority of methods in this area of research work on weighted networks. You could, in principle, apply some of them to unweighted networks as well, but you might fall off the use cases that were taken in consideration when developing such algorithms.

\section{Naive}
The reason not many network researchers mention the backboning problem is because they usually apply a limited set of naive strategies and do not recognize it as a problem in itself. In fact, there is an easy naive solution that most researchers apply without a second thought. If we have a weighted network and we want to keep the ``strongest connections'', we sort them in decreasing order of intensity. We decide a threshold, a minimum strength we accept in the network. Everything not meeting the threshold is discarded.

\begin{figure}
\centering
\includegraphics[width=.8\columnwidth]{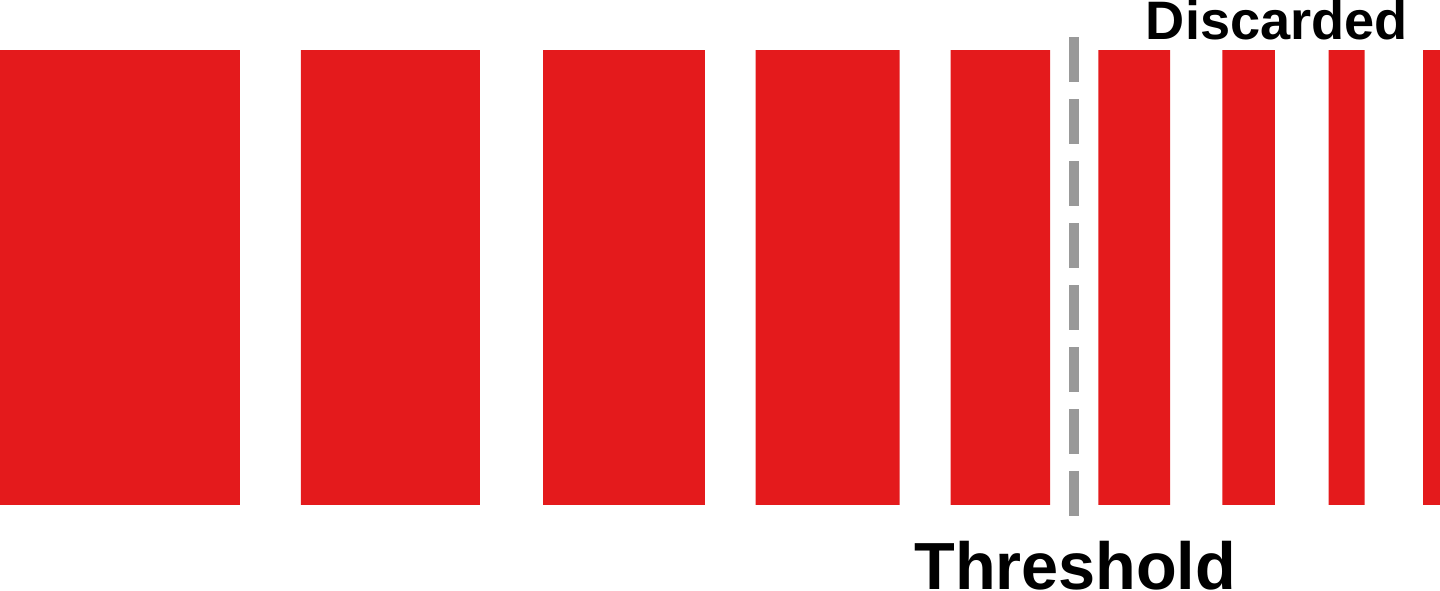}
\caption{A vignette of the naive thresholding procedure. Each red bar is an edge in the network. The bar's width is proportional to the edge's weight. Here, I sort all edges in decreasing weight order. I then establish a threshold and discard everything to its right.}
\label{fig:backboning-naive}
\end{figure}

Figure \ref{fig:backboning-naive} provides a vignette of this procedure. There are two problems with the naive strategy. 

\subsection{Broad Weight Distributions}
\begin{figure}
\centering
\includegraphics[width=.66\columnwidth]{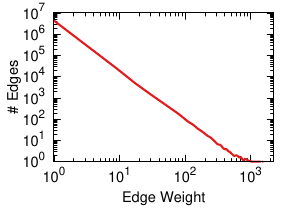}
\caption{The distribution of edge weights in the projected Twitter network using the simple projection strategy. The plot reports the number of edges (y axis) with a given weight (x axis).}
\label{fig:backboning-naive-problems}
\end{figure}

The first problem is that, in real world networks, edge weights distribute broadly in a fat-tail highly skewed fashion, much like the degree (Section \ref{sec:degree-pl}). Let's take a quick look again at the edge weight distribution we got using the simple projection in the previous chapter for our Twitter network. I show the distribution again in Figure \ref{fig:backboning-naive-problems}.

In this network, $82\%$ of the edges have weight equal to one. The smallest possible hard threshold would remove $82\%$ of the network, without allowing for any nuance. Moreover, since we have a fat tailed edge weight distribution, it is hard to motivate the choice of a threshold. Such a highly skewed distribution lacks of a well-defined average value and has undefined variance (Section \ref{sec:stats-corr}). You cannot motivate your threshold choice by saying that it is ``$x$ standard deviations from the average'' or anything resembling this formulation.

\subsection{Local Edge Weight Correlations}
The second problem is that edge weights are usually correlated. Nodes that connect strongly tend to connect strongly with everybody. In our sample Twitter network, let's consider a user $u$ who only had shared a single URL. If $u$ connects to any $v$, there can be only one possible edge weight in the simple projection: one. All edges around $u$ will have weight equal to one. On the other hand, if $u$ had shared thousands of URLs, it will likely connect to another user with similar sharing patterns, because statistically speaking they have high odds of sharing at least few of the same URLs, even if it happens by chance. Thus many edges around $u$ will have high weights.

\begin{figure}
\centering
\includegraphics[width=.66\columnwidth]{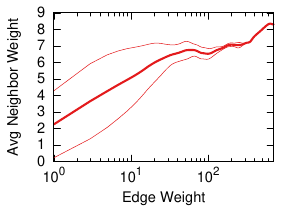}
\caption{The average weight of edges sharing a node with a focus edge (y axis) against the weight of the focus edge (x axis). Thin lines show the standard deviation. One percent sample of the Twitter network.}
\label{fig:backboning-naive-problems2}
\end{figure}

This is what I mean when I say that the weight of an edge is correlated with the weights of the edges of the nodes it connects. Figure \ref{fig:backboning-naive-problems2} shows how this correlation looks like in the Twitter network. The higher an edge weight, the higher on average the weights of edges sharing a node with it. Here, the correlation is $\sim 0.69$. The figure has the edge weights in log scale, since they are broadly distributed. The correlation of the average neighbor weight against the logarithm of the edge weight is $\sim 0.84$.

\begin{figure}
\centering
\begin{subfigure}[t]{.4\columnwidth}
\includegraphics[width=\textwidth]{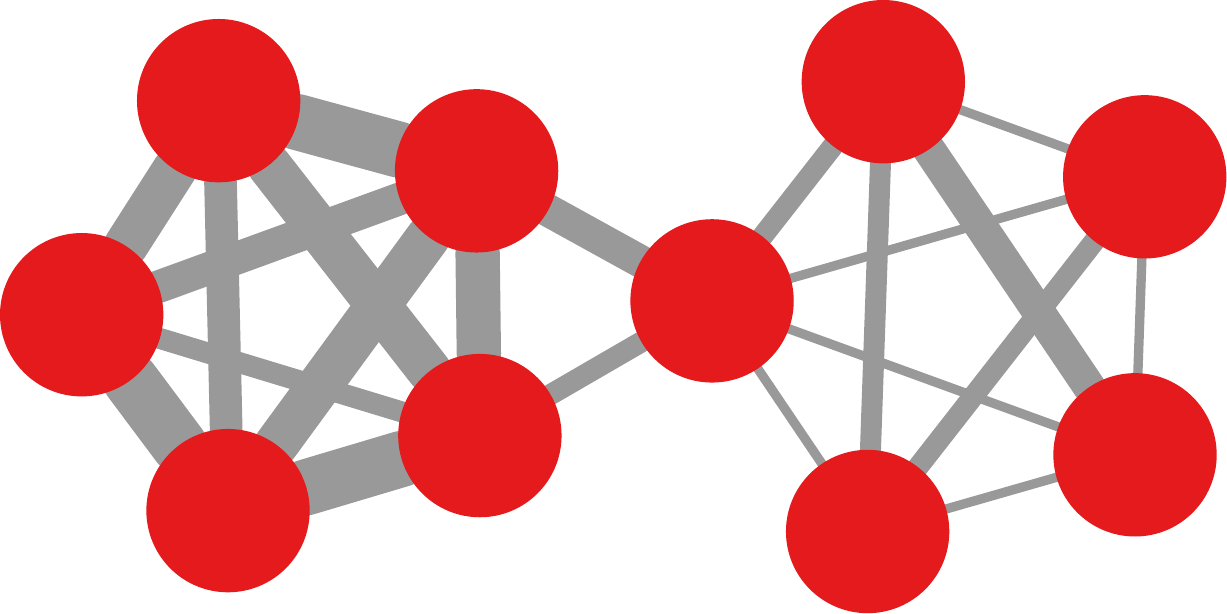}
\caption{}
\end{subfigure}
\quad
\begin{subfigure}[t]{.4\columnwidth}
\includegraphics[width=\textwidth]{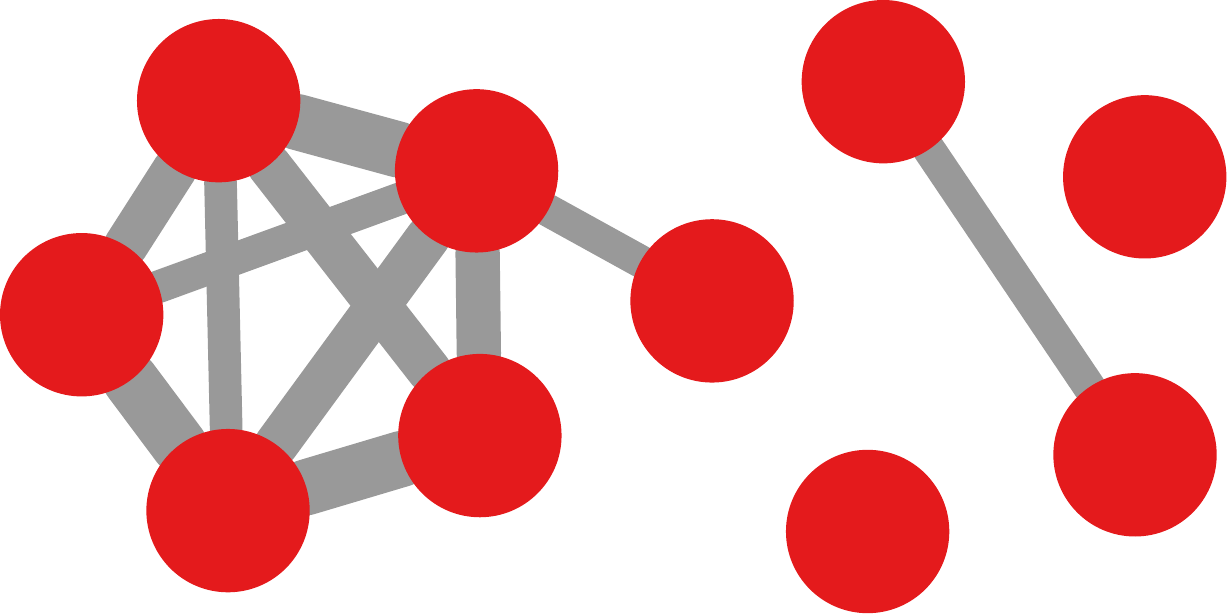}
\caption{}
\end{subfigure}
\caption{Establishing a hard threshold in a network with correlated edge weights. (a) I represent the edge weight with the width of the line. (b) I eliminate all the edges with a weight lower than a given threshold, equal for all edges.}
\label{fig:backboning-naive-problems3}
\end{figure}

This means that there are areas of the network with high edge weights and areas with low weights. If we impose the same threshold everywhere, some nodes will retain all their connections and others will lose all of theirs, without making the structure any clearer. Figure \ref{fig:backboning-naive-problems3} provides a vignette of this issue. In the figure, we completely destroy the topological information in the rightmost clique, while at the same time being unable to sparsify the leftmost clique.

An alternative ``naive'' strategy you could apply is to simply pick the top $n$ strongest connections for each node. This would not be affected by the issues I mentioned. However, by applying it you're effectively determining the minimum degree of the network to be $n$. This is a heinous crime against the God of power law degree distributions and, if you commit it, you will be tormented by scale free demons in network hell for all eternity.

\section{Doubly Stochastic}
The next approach we look at is the \textbf{doubly stochastic} strategy. Remember what a stochastic matrix is: it is the adjacency matrix normalized such that the sum of the columns is $1$ (Section \ref{sec:mat-mat-stochastic}). A \textit{doubly} stochastic matrix is a matrix in which the sums of both rows \textit{and} columns are equal to one. You can transform an adjacency matrix into its corresponding doubly stochastic by alternatively normalizing rows and columns until they both sum to $1$.

\begin{figure}
\centering
\begin{subfigure}[t]{.4\columnwidth}
\includegraphics[width=\textwidth]{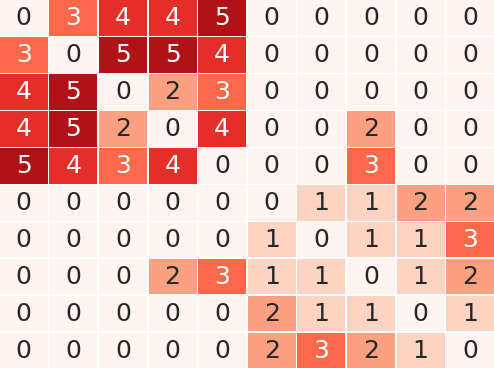}
\caption{}
\end{subfigure}
\quad
\begin{subfigure}[t]{.4\columnwidth}
\includegraphics[width=\textwidth]{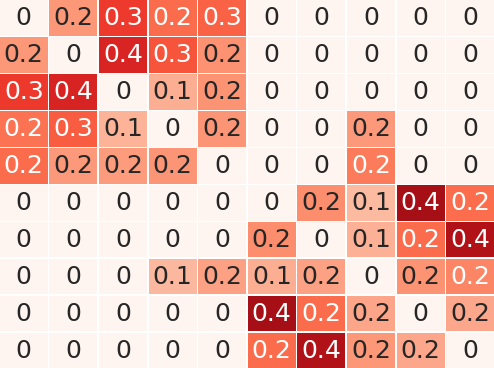}
\caption{}
\end{subfigure}
\caption{An example of Doubly Stochastic network backboning. (a) The adjacency matrix has areas of the network with different edge weight scales. (b) Its doubly stochastic counterpart has no such correlations.}
\label{fig:backboning-ds}
\end{figure}

After you perform such normalization, the scale of all edges is the same, and you break local correlations -- as I show in Figure \ref{fig:backboning-ds}. You can now threshold the edges without fearing for the issues we mentioned before\cite{slater2009two}. The original paper proposing this technique has specific guidelines on how to perform this thresholding. You should pick the threshold that allows your graph to be a single connected component. However, in many cases you might have different analytic needs. Thus you can specify your own threshold.

The downside of this approach is that not all matrices can be transformed into a doubly stochastic. Only strictly positive matrices can\cite{sinkhorn1964relationship}\cite{sinkhorn1967concerning}, meaning that the matrix cannot contain zero elements. Since real world networks are sparse, they actually contain lots of zeros.

So this solution cannot be always applied, although, in practice, my experience is that failure to convergence is the exception rather than the rule. The easy solution of adding a small $\epsilon$ to the matrix to get rid of zero entries does not always make sense. As $\epsilon \rightarrow 0$, meaning that $A + \epsilon \rightarrow A$, the normalization of $A + \epsilon$ does not converge.

Note also that a doubly stochastic matrix must be square. This is easy to see: if all rows sum to one, then the sum of all entries in the matrix must be the number of rows. On the other hand, if all columns sum to one, the sum of all the entries of the matrix must be the number of columns. Thus, the number of rows and the number of columns are the same number. This cheeky proof means that you cannot apply the doubly stochastic backboning to bipartite networks, unless $|V_1| = |V_2|$.

Doubly stochastic matrices have other fun properties. If you remember Section \ref{sec:rw-stationary}, the leading left eigenvector of a stochastic adjacency matrix is the stationary distribution, while the leading right eigenvector is a constant -- assuming the graph is connected. In a doubly stochastic matrix, both the left and the right eigenvectors are equal to a constant or, in other words, the stationary distribution of a doubly stochastic matrix is constant. This isn't really a necessary thing to know while doing network backboning, but I though it was cool, so do with this information what you will.

\section{High-Salience Skeleton}
The intuition behind the \textbf{high salience skeleton} (HSS) is that a network is a structure facilitating the exchange of information or goods. Thus, some connections are more important than others because they keep the network together in a single component. The main imperative is to allow all nodes to reach all other nodes in the most efficient and high-throughput way possible. Thus you need to interrogate each node and ask them what are the most efficient paths from their perspective. This cannot be done repurposing measures such as edge betweenness -- whose objective is also telling us how structurally important an edge is (Section \ref{sec:centr-betw}) -- because these measures adopt a ``global'' point of view: they are the salient connections for the network \textit{as a whole}, but they might leave some nodes poorly served.

To build an HSS we loop over the nodes and we build their shortest path tree: a tree originating from a node, touching all other nodes in the minimum number of hops possible and maximum amount of edge weight possible. In practice we start exploring the graph with a BFS and note down the total edge weight of each path. When we reach a node that we already visited we consider the edge weights of the two paths and the one with the highest one wins.

\begin{figure}
\centering
\includegraphics[width=.66\columnwidth]{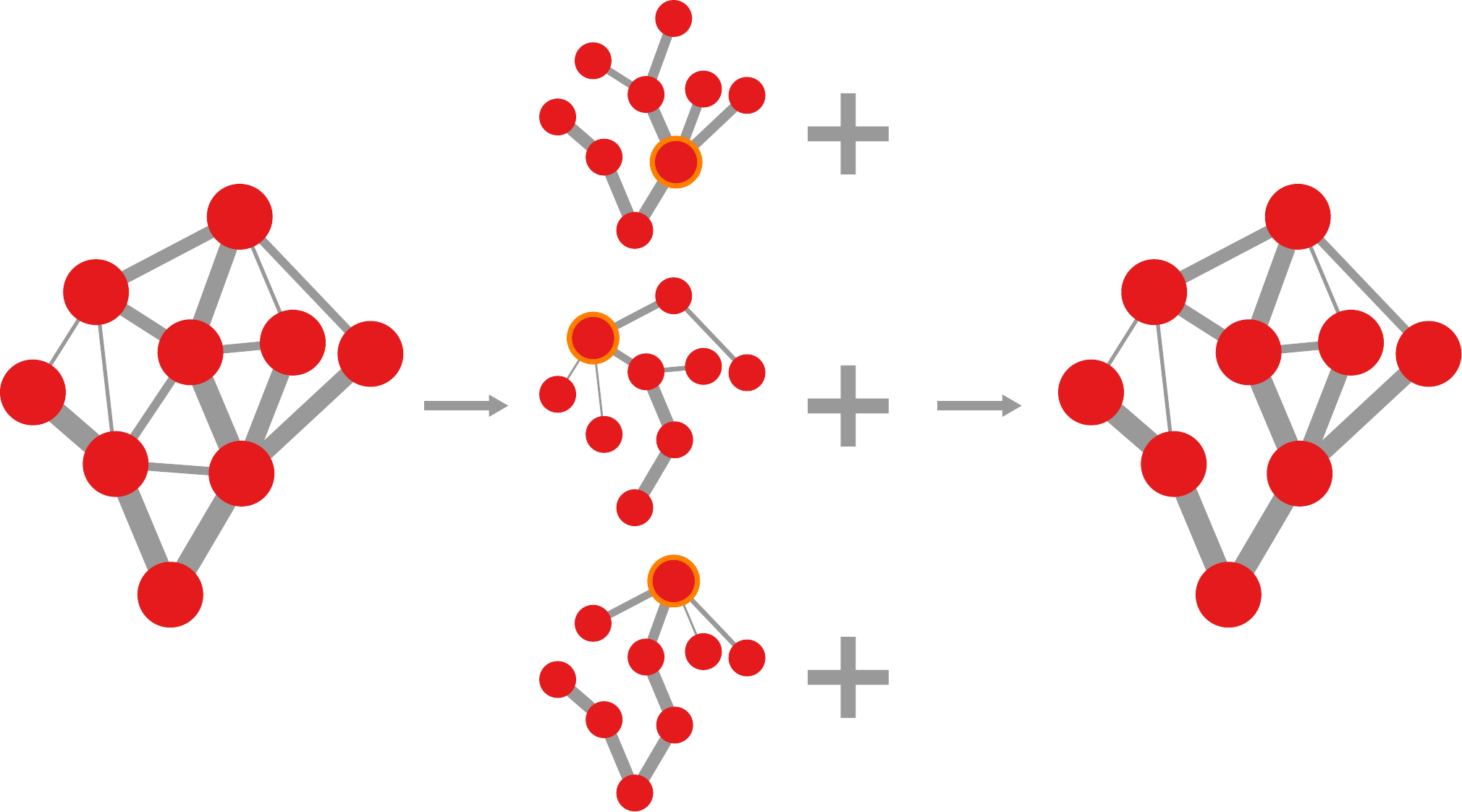}
\caption{An example of High Salience Skeleton network backboning. The original graph is used to create a shortest path tree for each node in the network. In each tree, I highlight the focus node with the orange outline. The trees are then summed, and the result is new edge weights for the original graph that can be thresholded.}
\label{fig:backboning-hss}
\end{figure}

Note that we have the constraint of the structure originating from a node to be a tree. Thus it cannot contain a triangle. Consider Figure \ref{fig:backboning-hss} as an example. In the bottom example, we might want to save two edges at the same time. Our origin node, the one at the top of the network connects strongly with one node which also connects strongly to the node on the left. However, we cannot have both edges in the shortest path tree, as that would create a cycle. The final salience skeleton is allowed to have triangles and cycles, because it is the sum of all the shortest path trees.

We perform this operation for all nodes in the network and we obtain a set of shortest path trees. We sum them so that each edge now has a new weight: the number of shortest path trees in which it appears\cite[0.1in]{grady2012robust}. The network can now be thresholded with these new weights.

Forbidding the creation of cycles in shortest path trees causes the main difference with the edge betweenness measure (Section \ref{sec:centr-betw}). One could think that the edges are simply sorted according to their contributions to all shortest paths in the network, but that is not the case. By forcing the substructures to be trees, we are counting the edges that are salient from each node's local perspective, rather than the network's global perspective. The authors in the paper show the subtle difference between shortest path tree counts and edge betweenness, also showing how a hypothetical skeleton extracted using edge betweenness performs more poorly.

\begin{figure}
\centering
\includegraphics[width=.66\columnwidth]{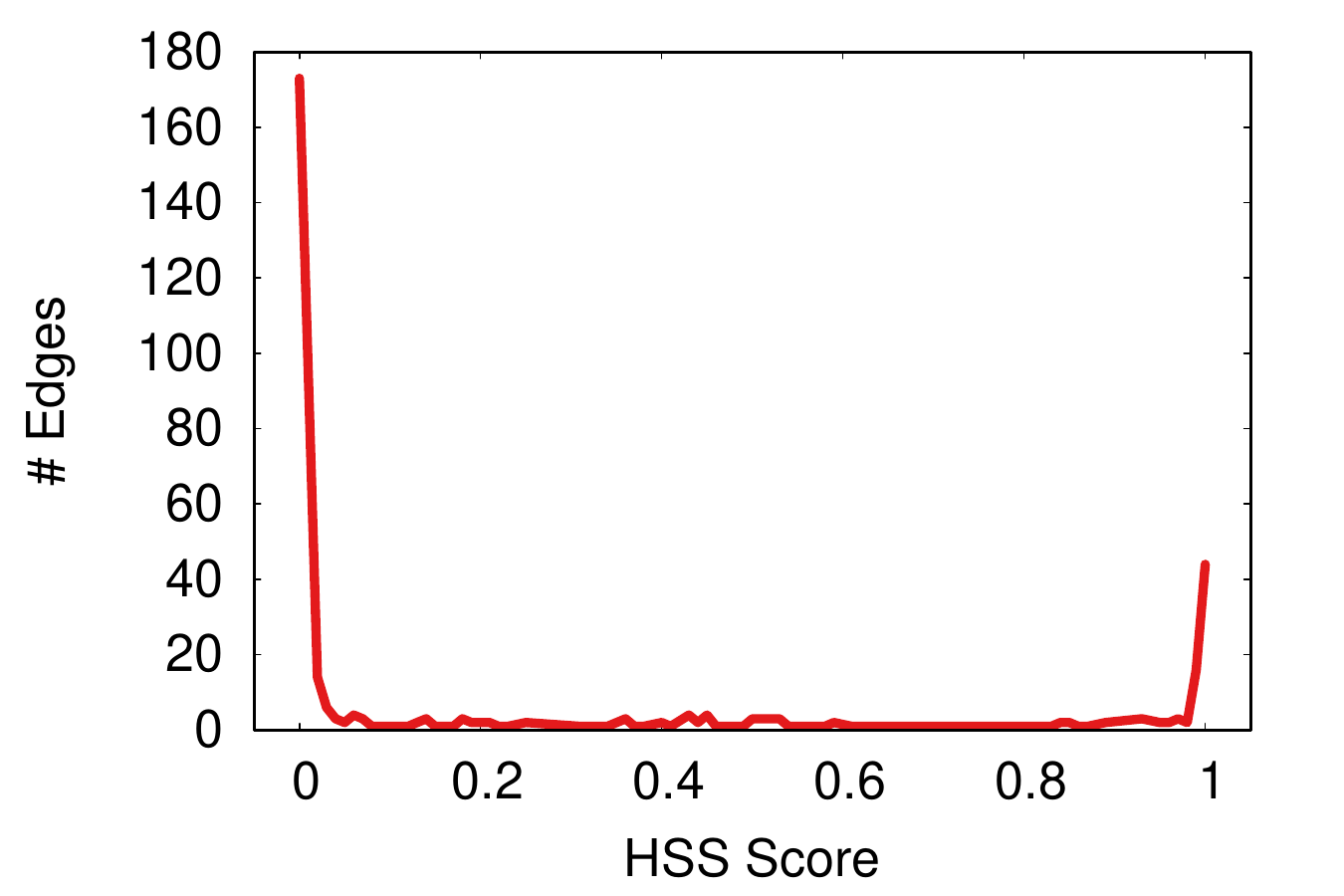}
\caption{A typical ``horns'' plot for the edge weight distribution in HSS. The plot reports how many edges (y axis) are part of a given share of shortest path trees (x axis).}
\label{fig:backboning-hss2}
\end{figure}

The HSS makes a lot of sense for networks in which paths are meaningful, like infrastructure networks. However, it requires a lot of shortest path calculations -- which makes it computationally expensive. Moreover, the edges are either part of (almost) all trees or of (almost) none of them. Figure \ref{fig:backboning-hss2} shows an example of this edge weight distribution, showcasing the typical ``horns'' shape of the HSS score attached to the original edges. You can see clearly that there are two peaks: one at zero -- the edge is in no shortest path tree --; the other at one -- the edge is part of all shortest path trees.

This can be nice, because it means HSS can be almost parameter free: the thresholding operation does not have many degrees of freedom. On the other hand, when there are few edges with weights close to one your skeleton might end up being too sparse and it is difficult to add more edges without lowering the threshold close to zero.

\section{Convex Network Reduction}\label{sec:bb-convex}
\begin{figure}
\centering
\begin{subfigure}[t]{.45\columnwidth}
\includegraphics[width=\textwidth]{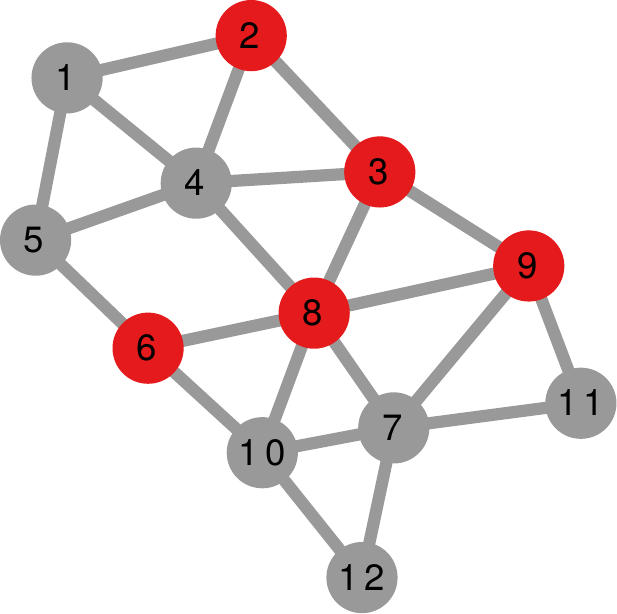}
\caption{}
\end{subfigure}
\qquad
\begin{subfigure}[t]{.45\columnwidth}
\includegraphics[width=\textwidth]{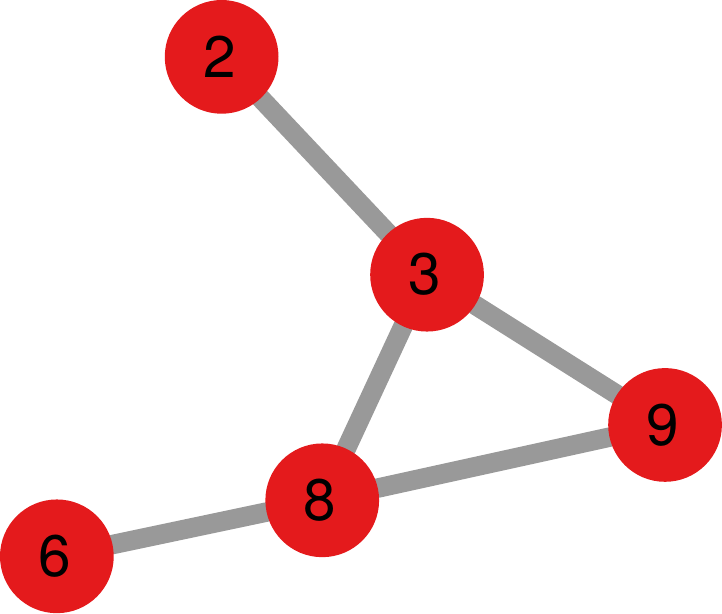}
\caption{}
\end{subfigure}
\caption{An example of induced graph. (a) The original graph. I highlight in red the nodes I pick for my induced graph. (b) The induced graph of (a), including only nodes in red and all connections between them.}
\label{fig:sampling-induced}
\end{figure}

A subgraph of a network $G$ is convex if it contains all shortest paths existing in the main network $G$ between its $V' \subseteq V$ nodes\cite{harary1981convexity}. We can expand this concept of convexity to apply to a full network $G$. To do so, we need to introduce the concept of ``induced'' subgraph: an induced subgraph is a graph formed from a subset of the vertices of the graph and all of the edges connecting pairs of vertices in that subset. Figure \ref{fig:sampling-induced} shows an example of the inducing procedure.

A network is convex if all its connected induced subgraphs are convex. No matter which set of nodes you pick: as long as they are part of a single connected component, they are all going to be convex. This might look like a weird and difficult to understand concept, but you can grasp it with the help of elementary building blocks you already saw in this book.

\begin{figure}
\centering
\begin{subfigure}[t]{.4\columnwidth}
\includegraphics[width=\textwidth]{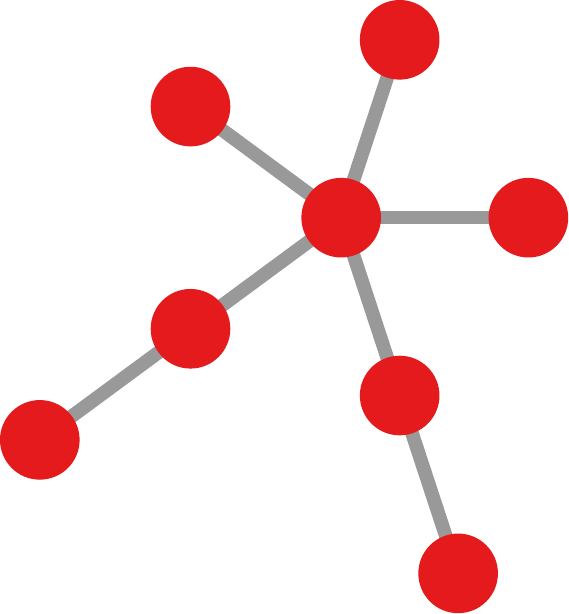}
\caption{}
\end{subfigure}
\qquad
\begin{subfigure}[t]{.3\columnwidth}
\includegraphics[width=\textwidth]{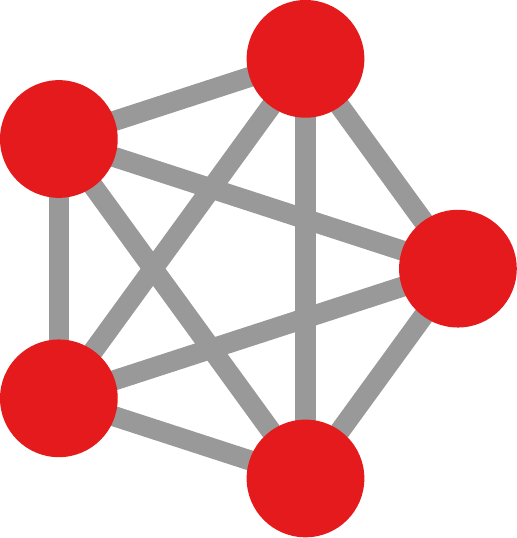}
\caption{}
\end{subfigure}
\caption{Two examples of convex networks. (a) A tree. (b) A clique.}
\label{fig:bb-convex-1}
\end{figure}

Figure \ref{fig:bb-convex-1} shows the two basic alternatives for a convex network. In a tree -- Figure \ref{fig:bb-convex-1}(a) -- any set of connected nodes is a convex subgraph. There are no other edges in $G$ you can use to make shortcuts, because they'd create a cycle and trees cannot contain a cycle. A clique -- Figure \ref{fig:bb-convex-1}(b) -- is a convex network as well: all possible connections are part of $G$, so picking any subset $V'$ of nodes will also result in a clique. Since all nodes are connected to each other in a clique, you have all the shortest paths between them, making it convex.

You can build an arbitrary convex network by stitching together trees and cliques. In practice, it's just stitching together cliques, because the ``tree-like'' parts are nothing more than 2-cliques.

One could use the concept of convex networks to create a skeleton of a real world network\cite{vsubelj2018convex}. Convex networks are almost impossible in nature, because adding a single edge to a tree or removing a single edge from a clique completely destroys convexity. However, one could make a real world network into a convex network by finding the minimal set of edges to remove to reduce the network into a tree of cliques. This is a possible way of backboning your network.

\section{Disparity Filter}\label{sec:bb-disfilter}
In this and the following section, we're slightly turning the perspective on network backboning. You could consider these as a different subclass of the problem. They all apply a general template to solve the problem of filtering out connections, which relate to the ``noise reduction'' application scenario of network backboning. Up until now, we adopted a purely structural approach which re-weights edges according to some topological properties of the graph. Here, instead, given a weighted graph, we adopt a template composed by three main steps: (1) define a null model based on node distribution properties; (2) compute a p-value (Section \ref{sec:stats-p}) for every edge to determine the statistical significance of properties assigned to edges from a given distribution; (3) filter out all edges having p-value above a chosen significance level, i.e. keep all edges that are least likely to have occurred due to random chance.

\begin{figure}
\centering
\includegraphics[width=.33\columnwidth]{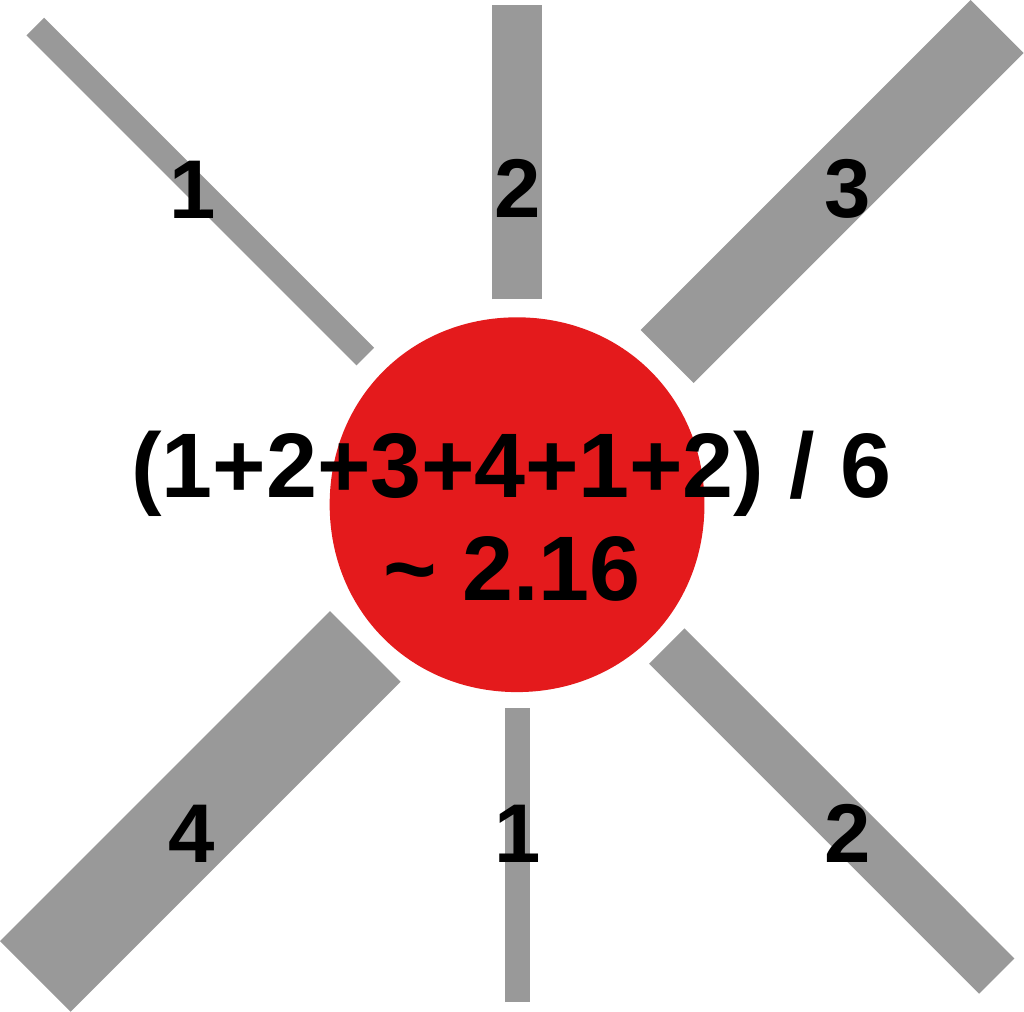}
\caption{A schematic simplification of Disparity Filter network backboning. The node determines its customized threshold by building an expectation of its average connection strength. Every edge weight higher than this expectation in a statistically significant way is kept.}
\label{fig:backboning-df}
\end{figure}

The \textbf{disparity filter} (DF) is the first example in this class of solutions. It takes a node-centric approach. Each node has a different threshold to accept or reject its own edges. This is done by modeling an expected typical ``node strength'', for instance the average of its edge weights. Then we keep only those edges which are higher than the expected edge weight for this node, making sure that this difference is statistically significant\cite[0.5in]{serrano2009extracting}. Figure \ref{fig:backboning-df} depicts a simplification of the method.

More precisely, the disparity filter defines $u$'s p-value for an edge $u,v$ of weight $w_{u,v}$ as:

$$ p((u,v), u) = \left( 1 - \dfrac{w_{u,v}}{\sum \limits_{v' \in N_u} w_{u,v'}} \right)^{(|N_u| - 1)},$$

where $N_u$ is the set of $u$'s neighbors (thus, $|N_u|$ is a fancy way to represent $u$'s degree). The original paper also shows how to calculate the expected edge weight and its variance, from which you can derive this p-value, but the procedure is a bit too convoluted to be included here. All you need, really, is the p-value. You can easily see that, if $|N_v| \neq |N_u|$ and/or $\sum \limits_{v' \in N_u} w_{u,v'} \neq \sum \limits_{v' \in N_v} w_{v,v'}$, the p-values for the same edge $u,v$ will be different depending whether we focus on $u$ or on $v$.

\begin{figure}
\centering
\includegraphics[width=.66\columnwidth]{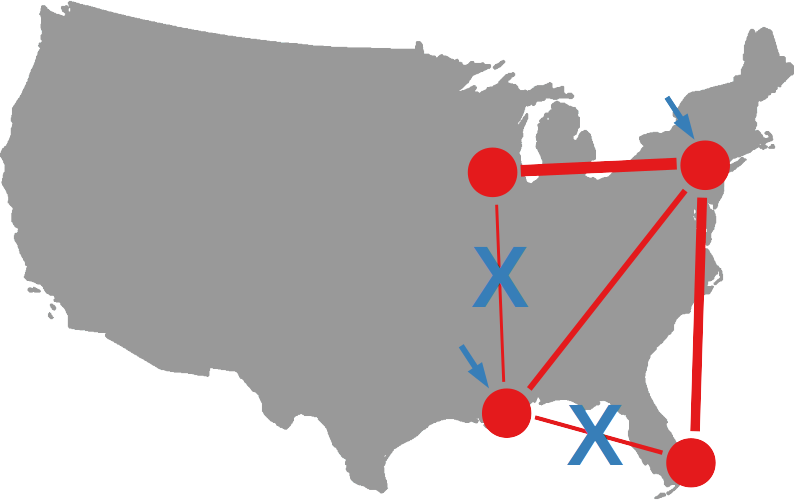}
\caption{An example of hub dominance in the DF filtering schema. Edge thickness is proportional to the weight. We check the same edge from both perspectives (blue arrows).}
\label{fig:backboning-df2}
\end{figure}

The disparity filter doesn't take into account that some nodes have inherently stronger connections. For instance, consider a mobility network, tracking commuters between cities in the United States. Figure \ref{fig:backboning-df2} provides an example. New York has a lot of people and thus will have strong mobility links with any place in the US. In the disparity filter, edges are checked twice from both nodes' perspectives: few of New York's links are stronger than its average, but almost all of them are the strongest in the perspective of the smaller towns to which New York connects.

We check New York against a small town in the south, for instance Franklington in Louisiana. Let's say that New York's connections, on average, involve $10$k travelers. The traveler traffic with Franklington involves only $1$k. This is way less than New York's average so, when we check this edge from New York's perspective, we mark it for deletion. However, on average, Franklington's connections involve only $500$ travelers. Thus, when we check the edge from Franklington's perspective, we will find it significant and so we will keep it. 

Since you need one success out of the two attempts to keep the edge, you end up with strong hubs connected to the entire network, and few peripheral connections (hub-spoke structure, or core-periphery, with no communities). In other words, the disparity filter tends to create networks with high centralization (Section \ref{sec:centr-centr}), broad degree distributions, and weak communities. In many cases, that is fine. For some other scenarios, we might want to consider an alternative.

In summary, this means that the disparity filter ignores the weights of the neighbors of a node when deciding whether to keep an edge or not. There is a collection of alternatives\cite{dianati2016unwinding}\cite{gemmetto2017irreducible} that take this additional piece of information into account and are thus less biased.

In this section I explained the disparity filter only in the case of undirected networks. You can apply the same technique also for directed networks. In this case, you need to make sure that you're properly accounting for direction in your p-value calculation: the edge must be significant either when compared to the out-connections of the node sending the edge, or when compared to the in-connection weights of the node receiving it.

\section{Noise-Corrected}
\begin{figure}
\centering
\includegraphics[width=.5\columnwidth]{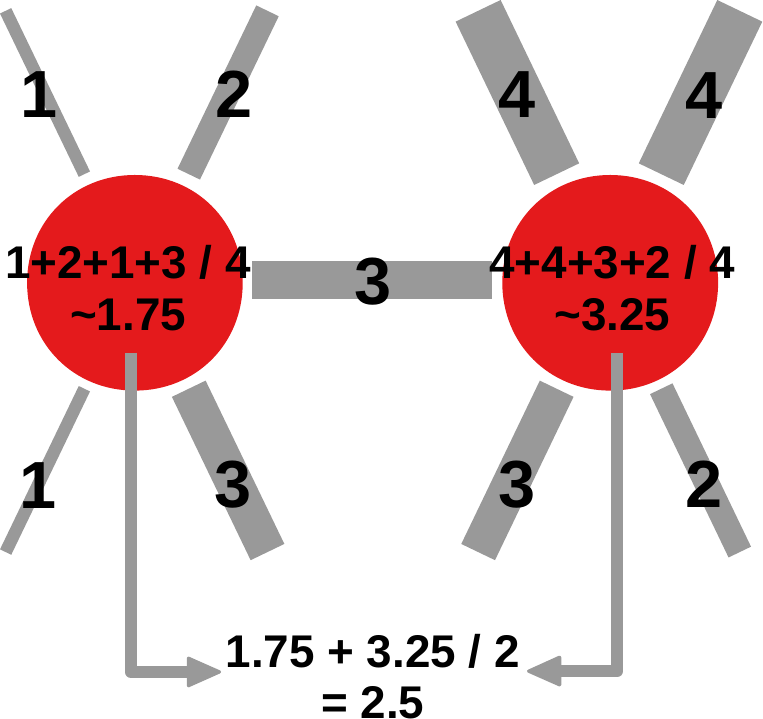}
\caption{A schematic simplification of Noise Corrected network backboning. The edge determines its customized threshold by building an expectation of the average connection strength of its two nodes. If its weight is higher than the expectation in a statistically significant way then the edge is kept.}
\label{fig:backboning-nc}
\end{figure}

The \textbf{noise-corrected} (NC) approach attempts to fix the issues of the disparity filter\cite{coscia2017network}. In spirit, it is very similar to it. However, the focus is shifted towards an edge-centric approach: each edge has a different threshold it has to clear if it wants to be included in the network. The assumption is that an edge is a collaboration between the nodes. It has to surpass the weight we expect given both nodes' typical connection strength. Again, we have to make sure that this difference is statistically significant. Figure \ref{fig:backboning-nc} depicts a simplification of the method.

Formally speaking, the p-value of NC is calculated by looking at the CDF of a binomial distribution. The observed value (number of successes) is the weight of the edge $w_{u,v}$, the number of trials is the total sum of edge weights in the network $\sum \limits_{u,v} w_{u,v}$, and the probability of success is given by:

$$ p_{u,v} = \dfrac{\sum \limits_{v' \in N_u} w_{u,v'} \times \sum \limits_{u' \in N_v} w_{u',v}}{\left( \sum \limits_{u',v'} w_{u',v'} \right)^2}.$$

So, in practice, we're looking at the probability of having a weight higher than $w_{u,v}$ in a binomial distribution with $\sum \limits_{u,v} w_{u,v}$ trials and a probability of success $p_{u,v}$. Given that we use a binomial as a null model, you can see that NC works only for discrete counts as edge weights, because the binomial is a discrete distribution (Section \ref{sec:prob-distr}). Moreover, all the elements here ($w_{u,v}$, $\sum \limits_{u,v} w_{u,v}$, and $p_{u,v}$) are the same in the perspective of $u$ and $v$, thus this measure is $u,v$ specific, differently from the disparity filter. Of course, if your network is directed, $w_{u,v} \neq w_{v,u}$ and you'll get a different null expectation for either direction of the edge, because the $u,v$ edge is different from the $v,u$ edge.

In the mobility network example I used before, a way to understand the difference between DF and NC is that, in NC, we require both nodes to agree to keep the edge. So, in this case, the edge between Franklington and New York will not be kept, because New York voted for deletion. If you attempt to create backbones with the same number of edges, Franklington will end up connecting to its local neighborhood, because those edges are more likely to be agreed upon by all the smaller towns nearby our focus.

\begin{figure}
\centering
\includegraphics[width=.66\columnwidth]{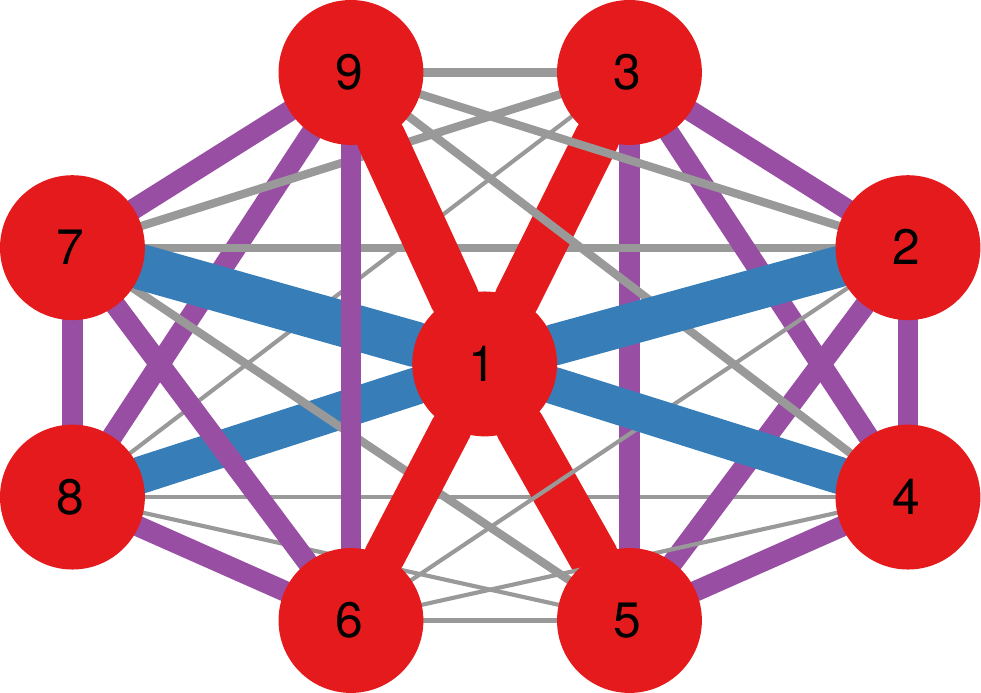}
\caption{Different choices between DF and NC backboning. Edge width is proportional to its weight. Edge color: red = selected by both DF and NC; blue = DF only; purple = NC only; gray = neither.}
\label{fig:backboning-nc2}
\end{figure}

Figure \ref{fig:backboning-nc2} abstracts from our geographical example to show the crux of the difference between DF and NC. DF favors the centralization around a hub. NC favors the horizontal peripheral connections which are the basis of the community structure of the network.

As you might expect, these methods exist because they give very different results. It is up to you to decide which of their assumptions best fits the network you are analyzing and the type of things you want to say about the network. A naive threshold fixes the same obstacle for all nodes no matter how strong, favoring the connections of the hub; HSS can include weaker links if they're the only path to a node; DF is similar to naive, but can recognize important weak edges; and NC overweights peripheries and communities: it is the most punishing method for the central hubs.

\section{Summary}

\begin{enumerate}
\item Backboning is the process of removing edges in a network. Reasons to do so span from a simple need of getting a sparser network, to facilitate computation on large networks, to the removal of connections that are not statistically significant. Most methods are developed assuming weighted networks, although you could apply some of them to unweighted networks as well.
\item One cannot simply establish a fixed threshold and remove all edges with a weight lower than the threshold. Edge weights usually distribute broadly and are correlated in different parts of the network, both factors that make the naive threshold approach not reasonable.
\item In doubly stochastic backboning, you transform the adjacency matrix in a doubly stochastic matrix (whose rows and columns sum to one) to break the local edge correlations. Such transformation is not always possible.
\item In high-salience skeleton, you calculate the short path tree for each node and you re-weight the edges counting the  number of trees using them. Then you keep the most used edges. This is usually computationally expensive.
\item In disparity filter and noise corrected, you create a null expectation of the edge weight and keep only the ones whose weight is significantly higher than the expectation. This expectation is node-centric in the disparity filter and edge-centric in noise-corrected.
\end{enumerate}

\section{Exercises}

\begin{enumerate}
\item Plot the CCDF edge weight distribution of the network at \url{http://www.networkatlas.eu/exercises/27/1/data.txt}. Calculate its average and standard deviation. NOTE: this is a directed graph!
\item What is the minimum statistically significant edge weight -- the one two standard deviations away from the average -- of the previous network? How many edges would you keep if you were to set that as the threshold?
\item Can you calculate the doubly stochastic adjacency matrix of the network used in the previous exercise? Does the calculation eventually converge? (Limit the normalization attempts to 1,000. If by 1,000 normalizations you don't have a doubly stochastic matrix, the calculation didn't converge)
\item How many edges would you keep if you were to return the doubly stochastic backbone including all nodes in the network in a single (weakly) connected component with the minimum number of edges?
\end{enumerate}

\chapter{Uncertainty \& Measurement Error}\label{cha:uncertainty}
To any person who has ever worked with real world data, it should come as no surprise that datasets are often disappointing. They contain glaring errors, incomprehensible omissions, and a number of other issues that make them borderline useless if you don't pour hours of effort into fixing them. The infamous 80-20 rule\cite{pareto1919manuale} that holds for literally everything (including degree distributions, as we saw in Section \ref{sec:degree-fit}) will tell you that $80\%$ of data science is just about cleaning data, and only $20\%$ about shiny and fun analysis techniques. This obviously applies to network data as well. You'll find edges in your networks with incorrect weights and perhaps they shouldn't even be there, and you'll have plenty of missing or unobserved connections. And don't take my word for it: there are some works outside network science proper that also cite measurement error as one of the many things you should think about when working with networked data\cite{advani2014empirical}.

Admittedly, techniques to clean network data would deserve an entire book part but, frankly, there are surprisingly few techniques to deal with this problem. In fact, a survey paper about measurement error in network data\cite{wang2012measurement} points out that measurement error is routinely considered only a problem about missing data, rather than the more general framing as uncertainty. Network data cleaning is thus the lovechild of network backboning and link prediction, but that's a rather barren marriage -- as far as I know. In fact, one of the few papers I know\cite{peixoto2018reconstructing} delivers the truth in a brutal and deadpan way: ``[in network analysis] the practice of ignoring measurement error is still mainstream''. I hope this chapter will contribute to make things change.

To give you an idea of the significance of the measurement error blind spot in network science, consider the Zachary Karate Club network. As I'll explain in details in Section \ref{sec:utilities-graphs}, everyone in our field is madly in love with this toy example. The paper presenting the network\cite{zachary1977information} has been cited more than six thousand times -- and not everybody using this network cites it. The fun thing about this graph is that we actually don't know whether it has $77$ or $78$ edges. The bewildering thing about this graph is that \textit{almost no one even mentions this problem}!

I start the chapter with a few methods that model measurement error to return a classical adjacency matrix, on which we can operate with classical algorithms. The rest of the chapter, instead, deals with probabilistic networks, structures whose edges have a probability of existing. Then, we want to define techniques to extract various things we care about -- such as the degree distribution or the shortest paths -- from a network containing edges that we're not certain they are there.

To wrap up this introduction, let's mention one obvious thing someone might want to do when they know their networks are affected by measurement error: to correct these errors! We'll see some techniques in this chapter, but in general one can use generic algorithm to find missing edges (link prediction, in Chapter \ref{cha:lp-simple}), to throw away spurious edges (network backboning, in Chapter \ref{cha:backboning}), or to find node duplicates. In the latter case, we can use network techniques to find such nodes, e.g. with node similarity (Section \ref{sec:centr-similarity}). However, you could also approach this task with non-network techniques such as entity resolution\cite{getoor2012entity} -- e.g. figuring out that two nodes with different names ``Sofia Coppola'' and ``Sophia Coppola'' are actually referring to the same entity. This is a natural language processing task and therefore not of interest here.

\section{Error Estimation \& Correction}
To be more concrete about errors in network, we can talk about some specific cases in which we have certified experience of such imperfections. For instance, there is a great deal of uncertainty when it comes to detect interactions between proteins in living organisms\cite{wodak2009challenges}. Some researchers have worked trying to reconstruct the physical backbone of the Internet\cite{clauset2005accuracy}, but as technology evolves this task gets harder and harder. Finally, social networks are not immune to this problem\cite{marsden1990network}: people might answer to surveys incorrectly -- because their memory is faulty --, and researchers can do it as well -- as in the case of Zachary.

\begin{figure}
\centering
\includegraphics[width=.75\textwidth]{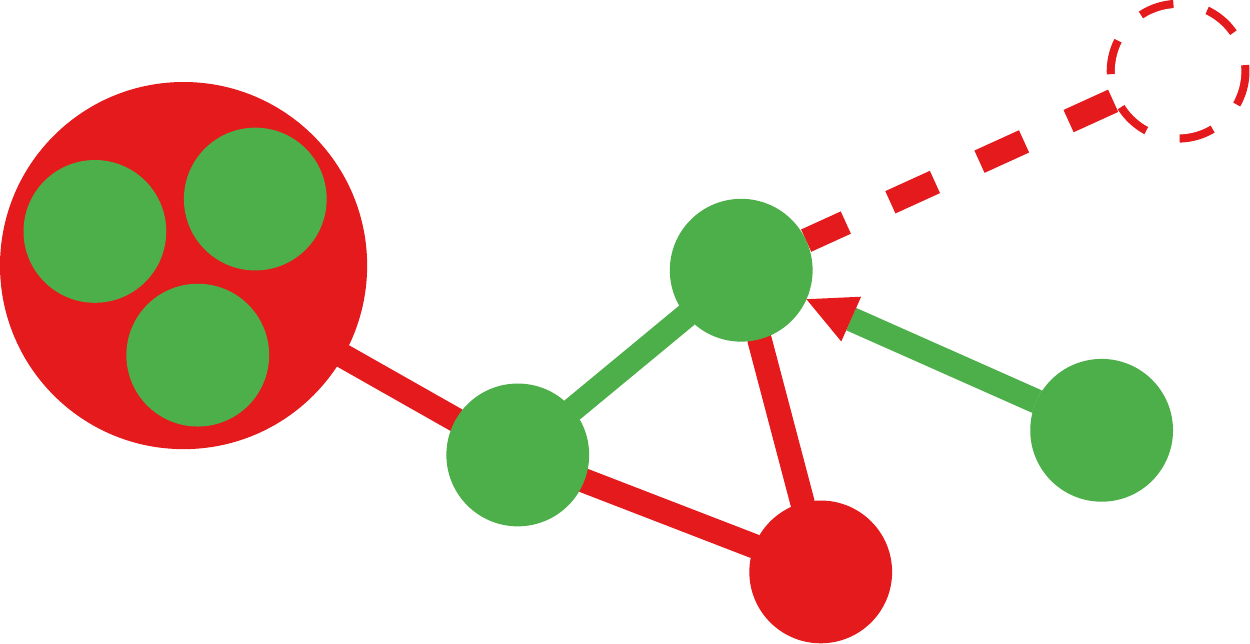}
\caption{A network with all sorts of errors. Correct features are in green, errors are in red.}
\label{fig:network-errors}
\end{figure}

Let's organize all the ways in which measuring a network can go wrong. I have a graphical representation in Figure \ref{fig:network-errors} of what could be happening.

\begin{itemize}
\item Missing nodes/edges: failing to report entities or connections between them that should have been there. This can be considered akin to having false negatives in your data. In Figure \ref{fig:network-errors} you can see a node with a dashed red outline that we missed, and its corresponding red dashed edge -- we missed that too;
\item Additional nodes/edges: reporting entities or connections that don't actually exist. This is instead more similar to have false positives. In Figure \ref{fig:network-errors}, those would be the solid red nodes and edges;
\item Duplicated nodes: an entity might exist, but it has been mistakenly split in two nodes. For instance, if you are making a network of actors, you might not know that the same actor might act under different pseudonyms. In Figure \ref{fig:network-errors}, the solid red node could be a duplicate of one of the green nodes;
\item Aggregated nodes: it is possible to mistakenly think of a group of distinct entities as a single one. For instance, a music group might be reported as a single entity, even if we are interested in individual artists. In Figure \ref{fig:network-errors}, we have a big node with three green nodes in them: we didn't see the green nodes and mistakenly thought they were the red one (who appears to be very shocked by this);
\item Misestimated node/edge features: the edge might exist, but in a weighted network we might be unsure about the actual strength of the relationship; in a directed network we may have gotten the direction wrong; or nodes and/or edge attributes might be incorrect. In Figure \ref{fig:network-errors}, there is a directed edge that does exist, but its arrowhead is red, because we got the direction wrong.
\end{itemize}

So what do you do when these things happen in your network? We're mostly focusing here on presence/absence of edges, because that's the case in which network techniques are most useful. As mentioned before, some of these problems can be tackled with non-network techniques, such as when you need to disambiguate nodes.

The basic way to go about estimating (and correcting) measurement error is by measuring network data multiple (random and independent) times\cite{newman2018network}. This is a way to reconstruct a primary error estimate, i.e. to diagnose how good or bad our data collection is. Before going to networks, let's consider what you do with normal data, for instance estimating a quantity.

\begin{figure}
\centering
\includegraphics[width=.66\textwidth]{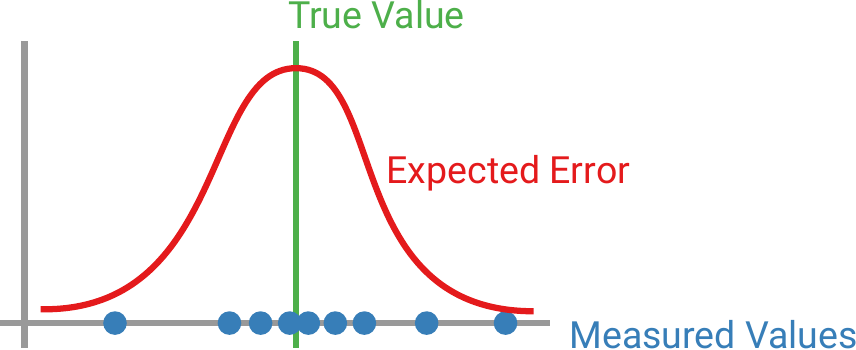}
\caption{Estimating a true quantity (in green) with different measurements (in blue). In red we have the distribution of the expected errors. The x-axis represents the measurement value, the y-axis the likelihood of observing a given measurement value.}
\label{fig:gaussian-error}
\end{figure}

Figure \ref{fig:gaussian-error} shows the usual scenario: you make a lot of measurements, none of them is the exact value you look for, but you can assume a Gaussian error. Once you know what sort of errors you might expect, you can do what one calls ``maximum likelihood\footnote{The ``likelihood'' I mention here is the log-likelihood loss function I introduced in Section \ref{sec:ml-loss}.} estimation'': given the measurements I have and the expected distribution of the errors, what is the most likely true value? One powerful algorithm to deal with this problem is Expectation Maximization\cite{moon1996expectation} (EM) -- we already mentioned it for link prediction (Section \ref{sec:lp-other}) and we will encounter it again when talking about community discovery.

The same framework -- maximum likelihood estimation via Expectation Maximization -- can be done also for networks. There is nothing magical differentiating estimating an adjacency matrix $A$ from a single number. It's just a little bit more complicated. One needs to introduce additional parameters -- dealing with different types of errors I before dubbed false positives (extra edges) and false negatives (missing edges). The additional complexity is not a problem for the EM algorithm, and has the nice advantage that this framework is basically performing link prediction and network backboning at once.

Inferring directly the most likely adjacency matrix $A$ is not necessarily the best thing you can do. If you focus on a specific measure of the network -- say, the clustering coefficient -- you can get its expected value with this method. The reason why this is better is because, this way, we can use alternative $A$s that might be slightly less likely than the most likely $A$, but still pretty likely. As a consequence, rather than having a single number that is the clustering coefficient of the most likely $A$, you get a distribution of potential values, each weighted by how likely their corresponding $A$ is. The distribution would represent a more accurate and useful information than the simple estimate -- since the most likely $A$ could still be pretty unlikely in absolute terms, and thus its clustering coefficient might be very wrong.

This all sounds awesome, but there are a couple of problems with this approach. First, we need to make an assumption about the errors on our measurement. Different networks will call for different error models for networks. Those models are available\cite{butts2003network}\cite{guimera2009missing}, but you're still going to be uncertain whether you have picked the right error model for your network. 

Second, it is very very unlikely you're going to measure your network multiple times. The most likely scenario is that you will either measure it once, or that you're going to work with a network you find in the literature, also measured once and without any possibility of making a second measurement. Finally, it is extremely likely that, even if you can measure the same network twice, your measurements are not going to be independent. In a social setting, if people forget one of their social relationships, they're likely to forget it again if they are asked a second time. And that's a problem\cite{yuan2015graph}.

Luckily there are ways out. For instance, in the paper I cited in the introduction of this chapter\cite{peixoto2018reconstructing}, the author can build a clever Bayesian framework to find the most likely false positive and false negative edges. One still has to have a model, but instead of modeling the \textit{errors} as we did above, we actually model the \textit{signal}. We assume that the observed network fits a specific model and then we can use the model to figure out which edges are more or less likely to be unobserved or spurious.

\begin{figure}
\centering
\begin{subfigure}{.45\columnwidth}
\includegraphics[width=\textwidth]{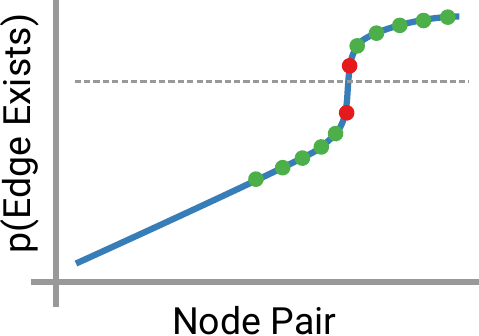}
\caption{}
\end{subfigure}
\qquad
\begin{subfigure}{.45\columnwidth}
\includegraphics[width=\textwidth]{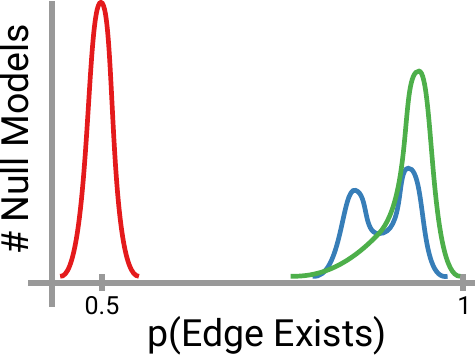}
\caption{}
\end{subfigure}
\caption{(a) The probability (y axis) of a node pair (x axis) being connected given a network model. Node pairs are sorted from least to most likely. The gray horizontal dashed line represents the probability threshold for accepting/rejecting an edge hypothesis. Red points represent errors in the data according to the model. (b) Number of null models (y axis) producing a probability for an edge existing (x axis) for different network models: $G_{n,p}$ (red), configuration model (blue), stochastic blockmodel (green).}
\label{fig:bayesian_errors}
\end{figure}

You can use any of the network generative methods I showed in Part \ref{par:synthnet}. Any of those will allow to get edge probabilities. If you do things properly, you should see something like Figure \ref{fig:bayesian_errors}(a). For each node pair you get a probability of them being connected. At some point, you'll find a discontinuity, transitioning from very low to very high existence probabilities. That's where you can put your threshold (above that threshold you expect the edge to exist, below you expect it not to exist. Any unconnected node pair above this threshold (in red in the figure) is likely to be a missing edge. Any connected node pair below this threshold is likely to be a spurious edge (also in red in the figure). In green in the figure you see cases in which theory and data agree: expected edges actually there (above the dashed line) and unexpected edges that are indeed absent from the data (below the line).

However, if the model you choose is not a good representation of the data you have, your estimates are going to be way off. Figure \ref{fig:bayesian_errors}(b) shows an example of this. Each model run will give you a distribution of probabilities for an edge existing. If the model is pretty bad -- say a $G_{n,p}$ model (Chapter \ref{cha:rndgraphs}) -- you'll probably end up with a 50-50 chance of having or not having the edge, which is pretty useless. Better models fitting the data better, like the configuration model (Section \ref{sec:csmodels-conf}) or stochastic blockmodels (SBMs, Section \ref{sec:csmodels-comms}), will give more useful estimates. In the figure we're testing for an edge we know it's there, and thus these more accurate models have most of their guesses close to $p = 1$.

Of course, it's not a given that these models will work better for your data. Different data will call for different models. So if you don't find a good model fitting your data, you're back at square one. The original paper uses a sophisticated Hierarchical Degree Corrected SBM\cite{peixoto2017nonparametric} (HDCSBM). HDCSBM is great, because it can model both (hierarchical) communities and the degree distribution. Still, it won't be able to fit every single network and it may not be the best choice in any case, so you're still left with the task of finding the best model for your network.

\section{Probabilistic Networks}
In the previous section we saw how to estimate errors and assign probabilities to edges. The papers I cited so far give you an idea on how to work with edge probabilities, assuming a model of the network or of the errors. In some cases, you might have neither. One example could be that the edges in your network come already with their own probabilities, and you don't know the model that generated them. In this case, you have to use a probabilistic network.

\begin{figure}
\centering
\includegraphics[width=.4\columnwidth]{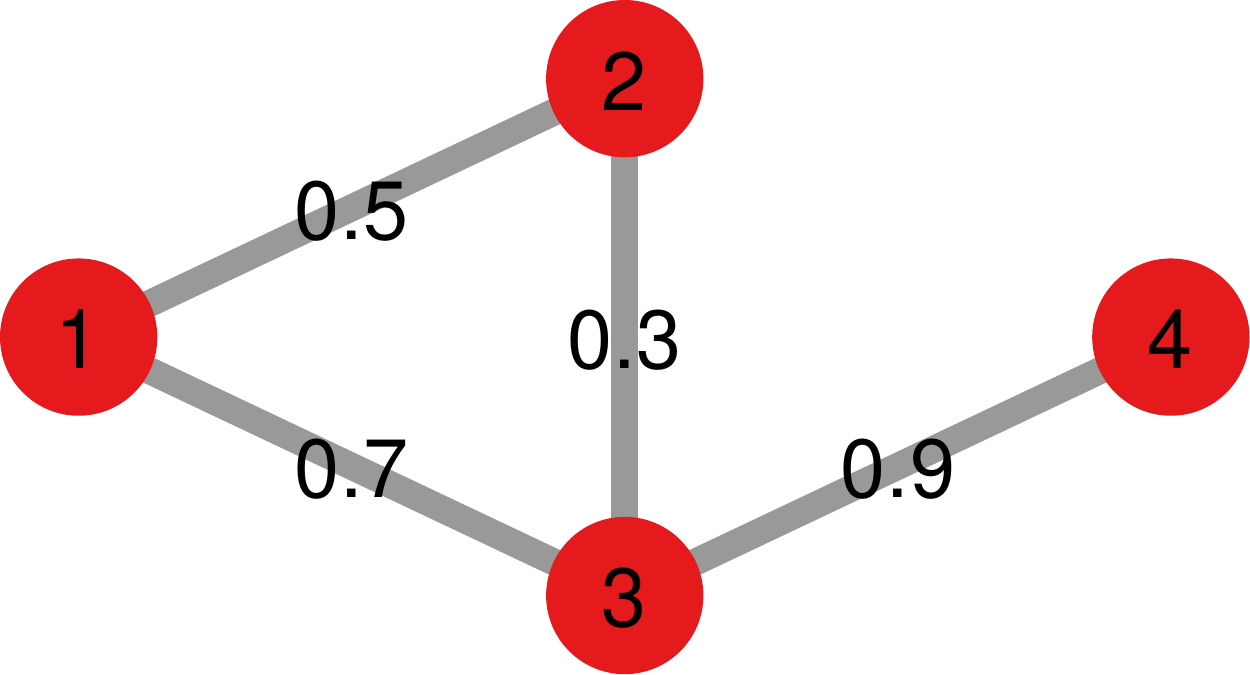}
\caption{A probabilistic network. The edge label indicates the edge's probability of existing.}
\label{fig:probabilistic-network}
\end{figure}

A probabilistic network is a graph defined as $G = (V, E, \Pi)$. In this case, $\Pi$ is a vector that assigns a probability of existence $p_{u,v}$ to each observed $u,v$ edge in $E$. Normally, the probabilities in $\Pi$ are all independent from each other -- we're going to work with this assumption here, but it's not the only one: in some cases the edge probabilities should be dependent, as for instance in some cases edge failures might affect the local structure of the network\cite{yuan2012efficient}. Figure \ref{fig:probabilistic-network} provides an example. 

Of course the models can be more complex than simply attaching a probability to edges (what about nodes\cite{tong2014efficient}? What about edge weights\cite{bonchi2014core}?), but to preserve our mental sanity we'll go with this simpler example and then it's up to you if you want to go down this specific rabbit hole. It's worth reiterating a difference between probabilistic network analysis and network backboning: in the latter we want to remove spurious edges, here we want to use all the information we have and integrate it in the analysis, no matter how unlikely an edge is to exist.

Probabilistic networks have been used in many different fields from mobility sensors\cite{gao2017energy}, to protein-protein interactions in biology\cite{srihari2013survey}. They have been used to extend the k nearest neighbor algorithm to work with probabilities\cite{potamias2010k} and, as we'll see, to estimate centrality measures for road networks\cite{fushimi2019new}, and communities\cite{kollios2011clustering}\cite{liu2012reliable} (of course).

The naive approach to deal with probabilistic networks would be to realize that each edge $(u,v)$ can either exist or not. So you could create two ``possible worlds'', which is to say two alternative networks, one with the edge and one without. The first has likelihood $p_{u,v}$ (because the edge exists) and the other has likelihood $1 - p_{u,v}$ (because the edge doesn't exist). These likelihoods act as a sort of weight. For instance, if you're absolute certain about the existence of an edge ($p_{u,v} = 1$) the network in which that edge doesn't exist isn't considered in the degree distribution calculation, because that possible world happens with probability zero. 

\begin{figure*}[ht!]
\centering

\begin{subfigure}{.2\columnwidth}
\includegraphics[width=\textwidth]{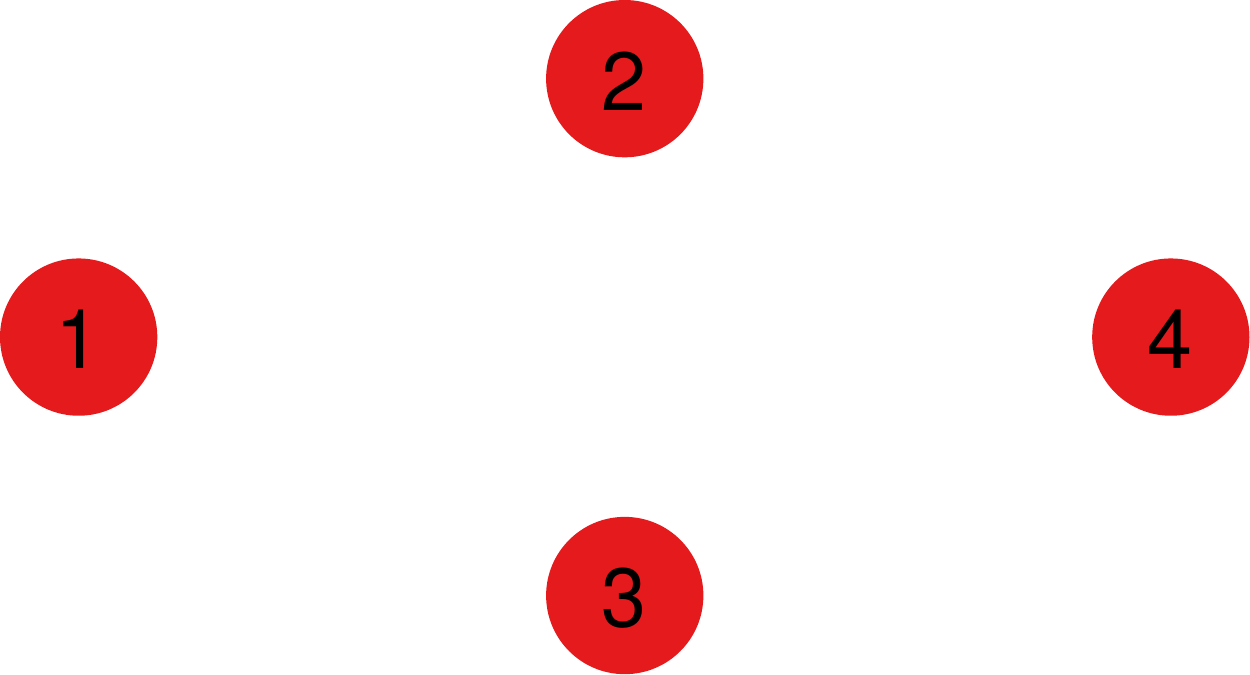}
\caption{$p = 0.0105$}
\end{subfigure}\quad
\begin{subfigure}{.2\columnwidth}
\includegraphics[width=\textwidth]{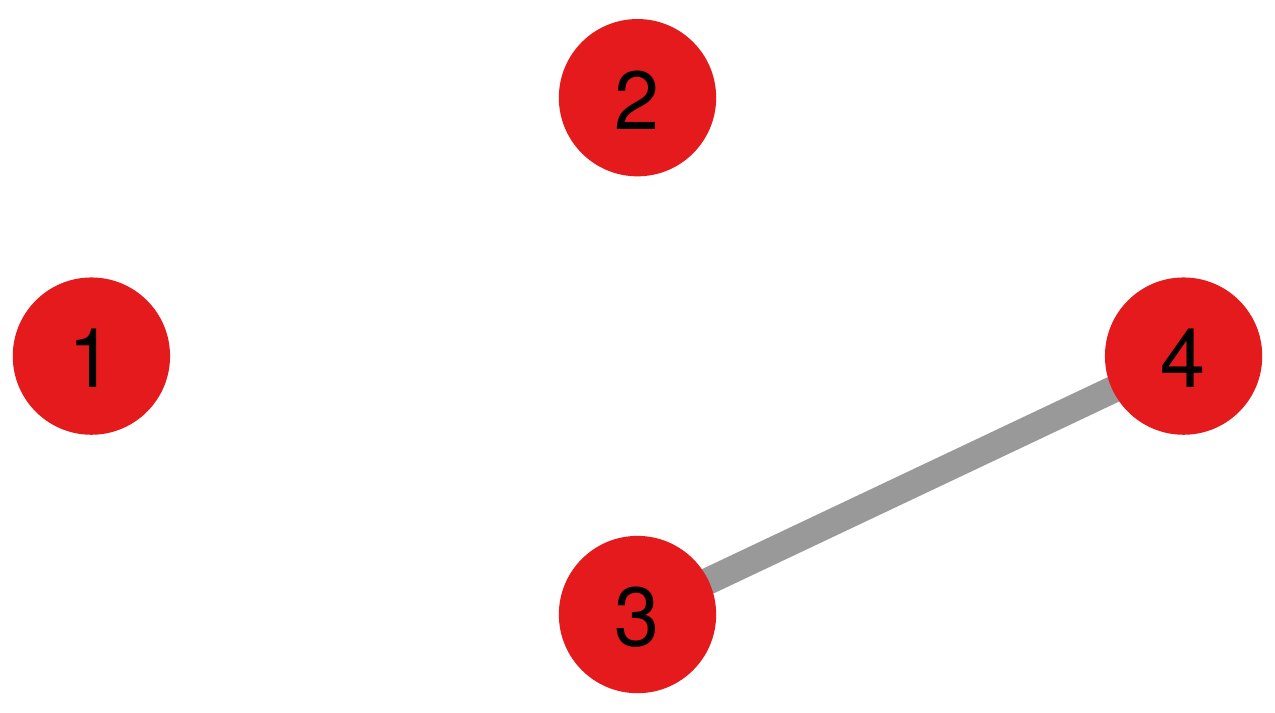}
\caption{$p = 0.0945$}
\end{subfigure}\quad
\begin{subfigure}{.2\columnwidth}
\includegraphics[width=\textwidth]{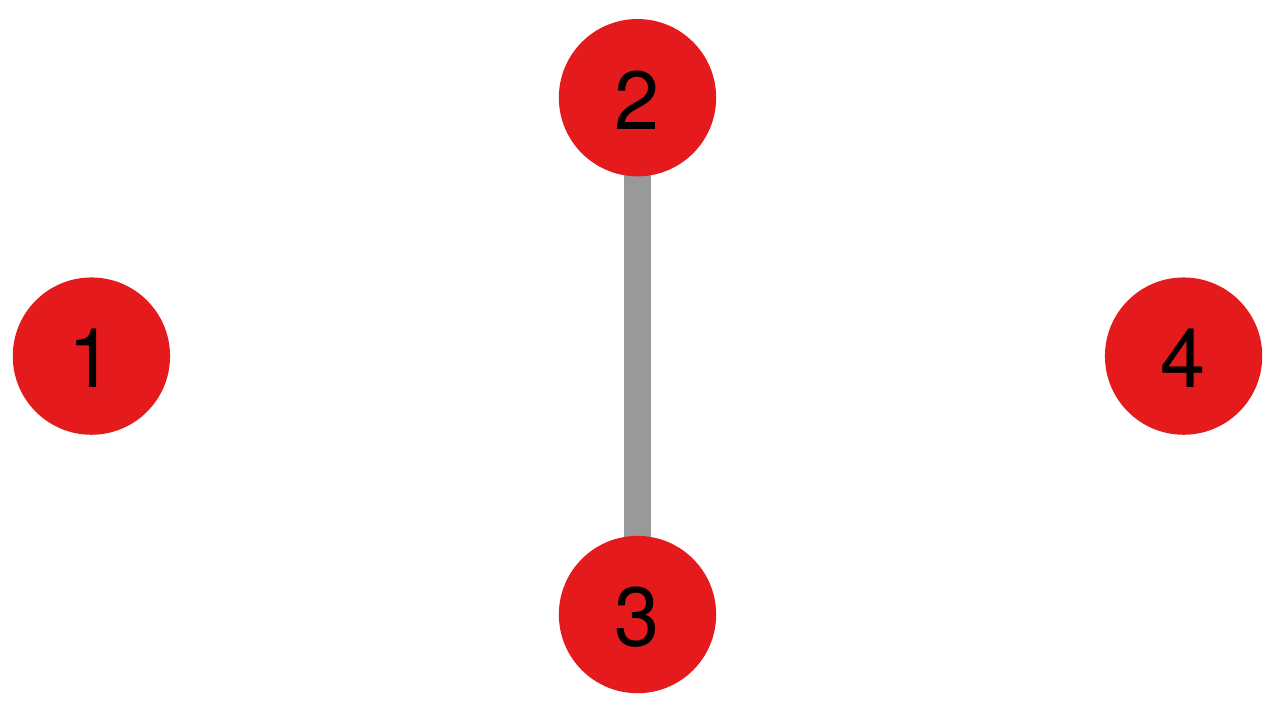}
\caption{$p = 0.0045$}
\end{subfigure}\quad
\begin{subfigure}{.2\columnwidth}
\includegraphics[width=\textwidth]{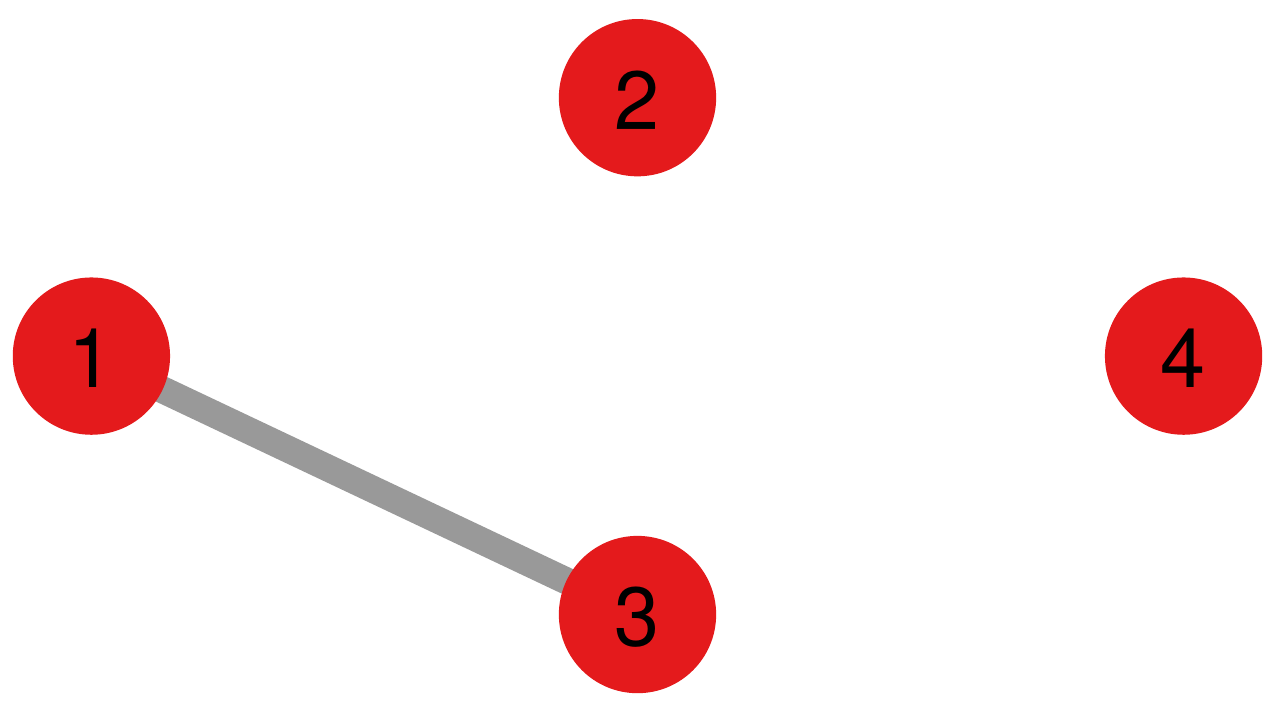}
\caption{$p = 0.0245$}
\end{subfigure}

\begin{subfigure}{.2\columnwidth}
\includegraphics[width=\textwidth]{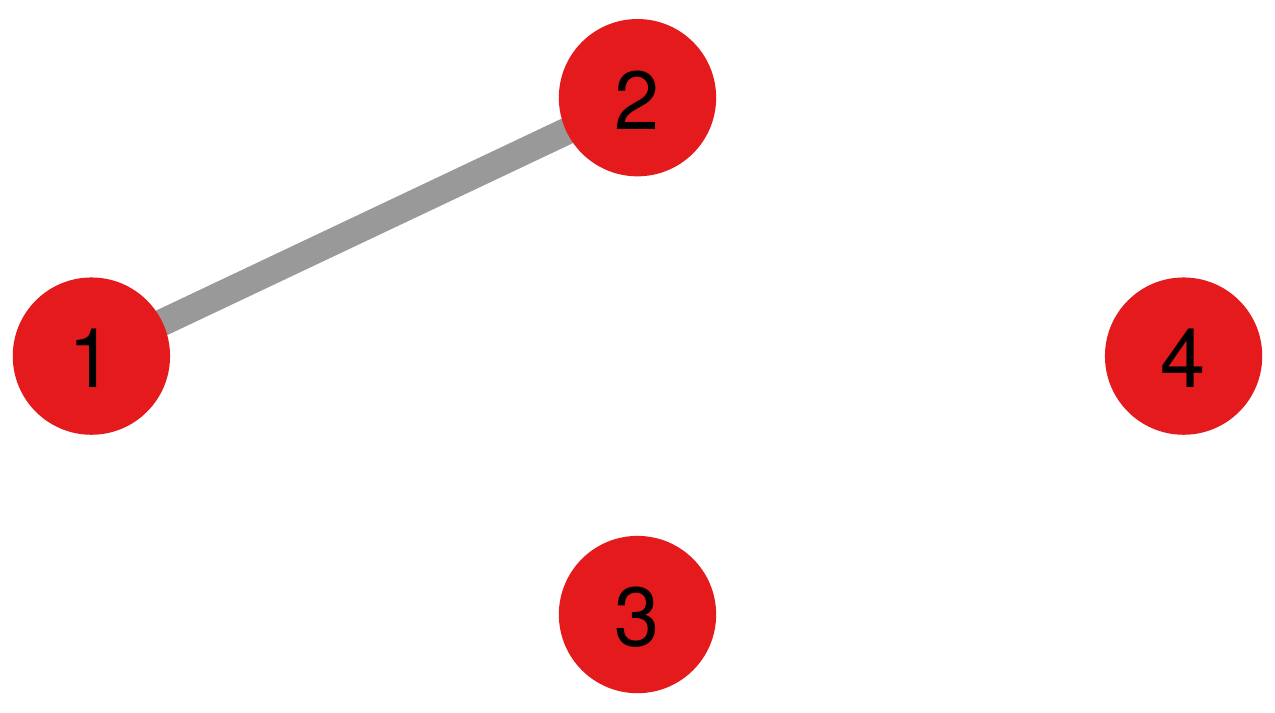}
\caption{$p = 0.0105$}
\end{subfigure}\quad
\begin{subfigure}{.2\columnwidth}
\includegraphics[width=\textwidth]{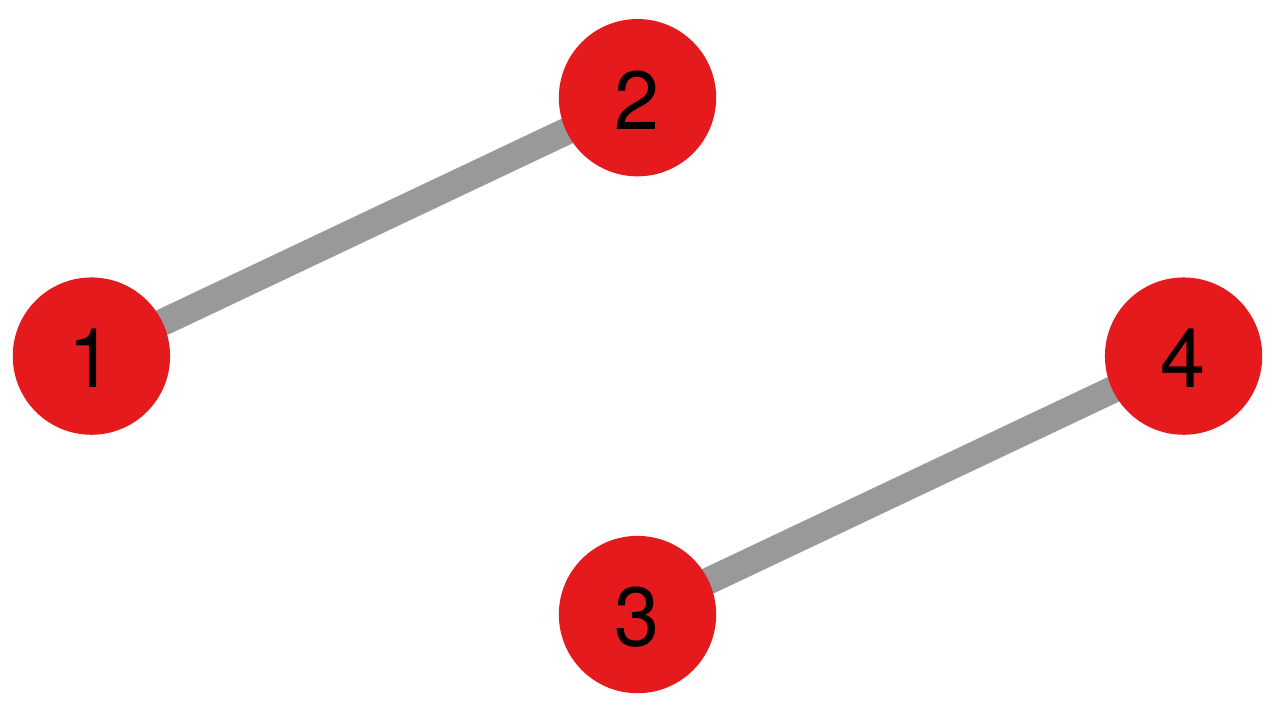}
\caption{$p = 0.0945$}
\end{subfigure}\quad
\begin{subfigure}{.2\columnwidth}
\includegraphics[width=\textwidth]{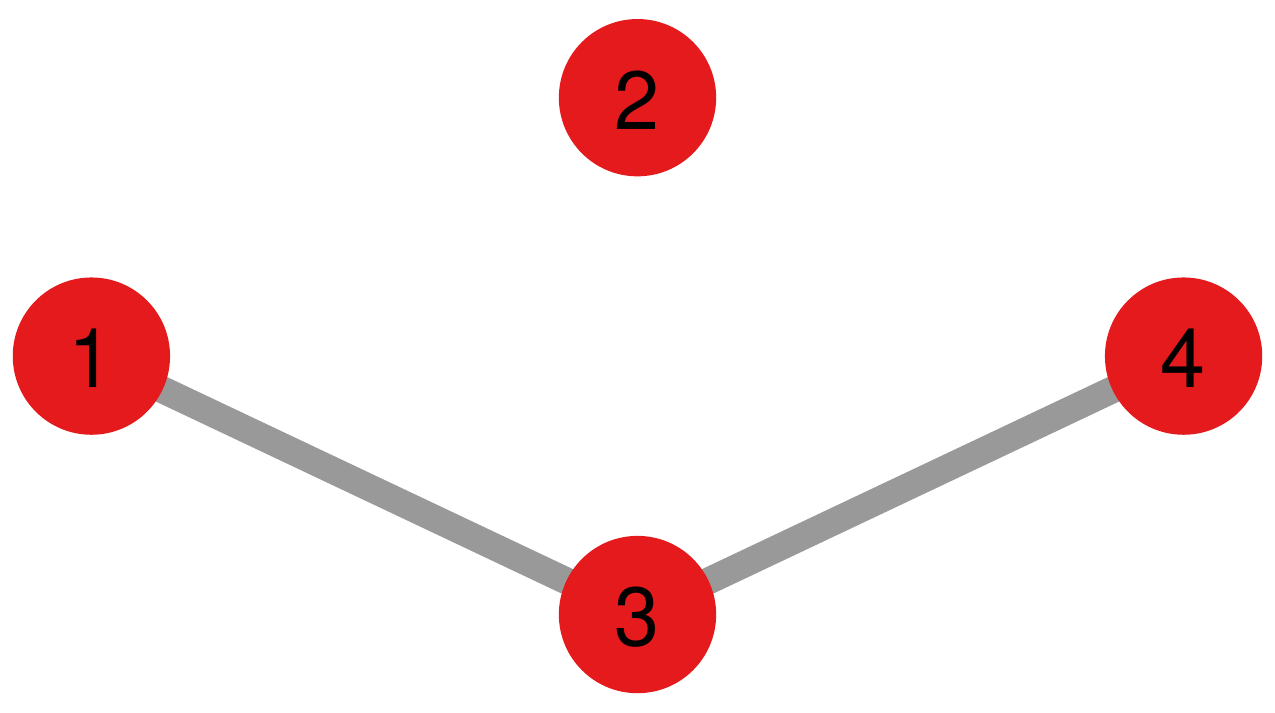}
\caption{$p = 0.2205$}
\end{subfigure}\quad
\begin{subfigure}{.2\columnwidth}
\includegraphics[width=\textwidth]{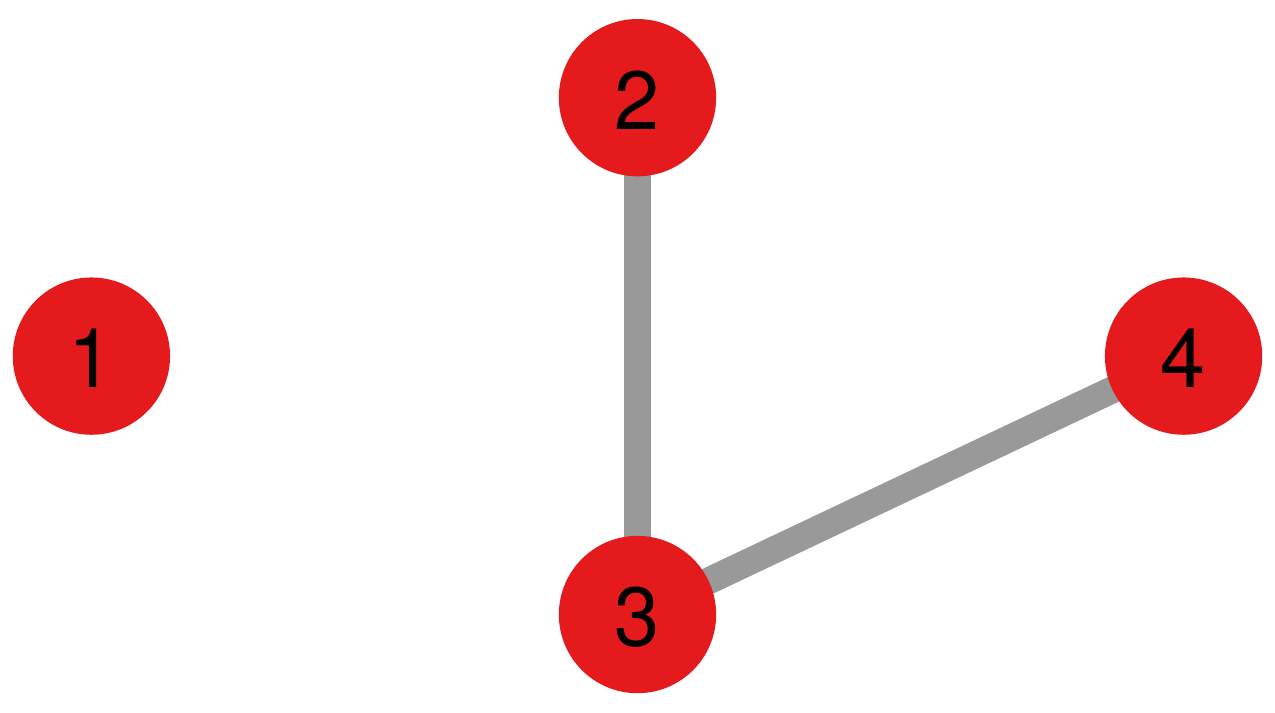}
\caption{$p = 0.0405$}
\end{subfigure}

\begin{subfigure}{.2\columnwidth}
\includegraphics[width=\textwidth]{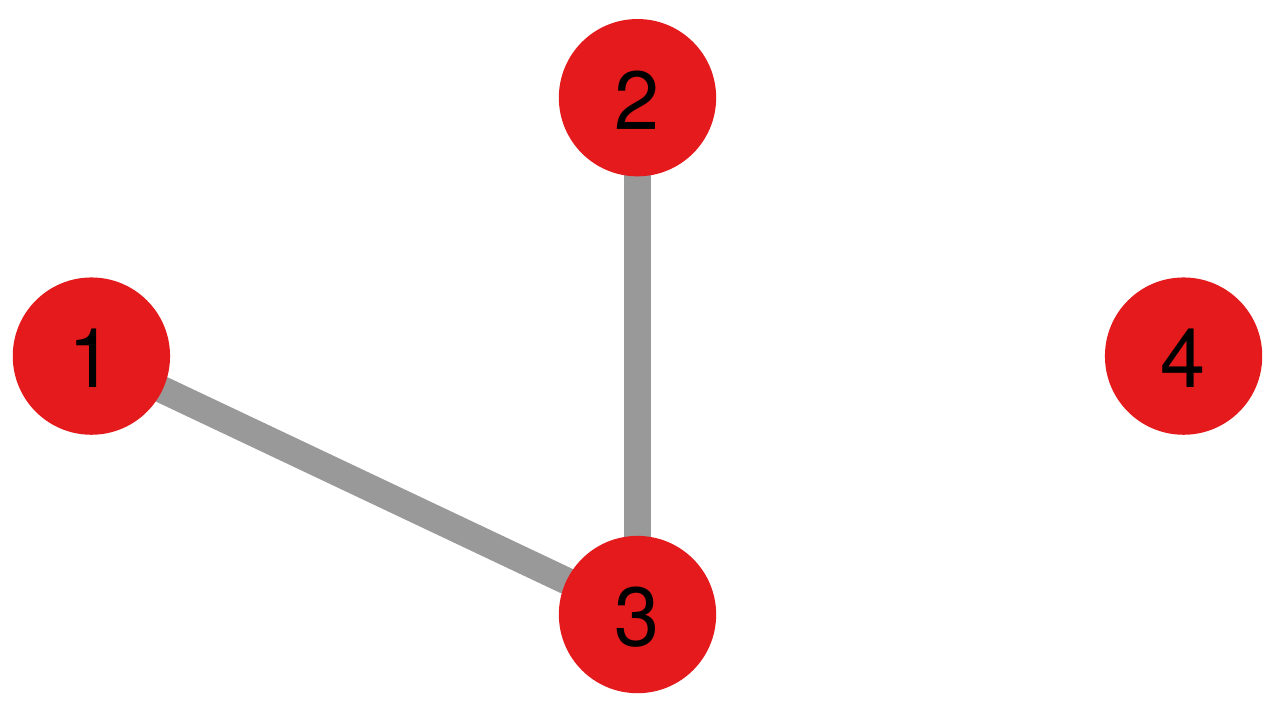}
\caption{$p = 0.0105$}
\end{subfigure}\quad
\begin{subfigure}{.2\columnwidth}
\includegraphics[width=\textwidth]{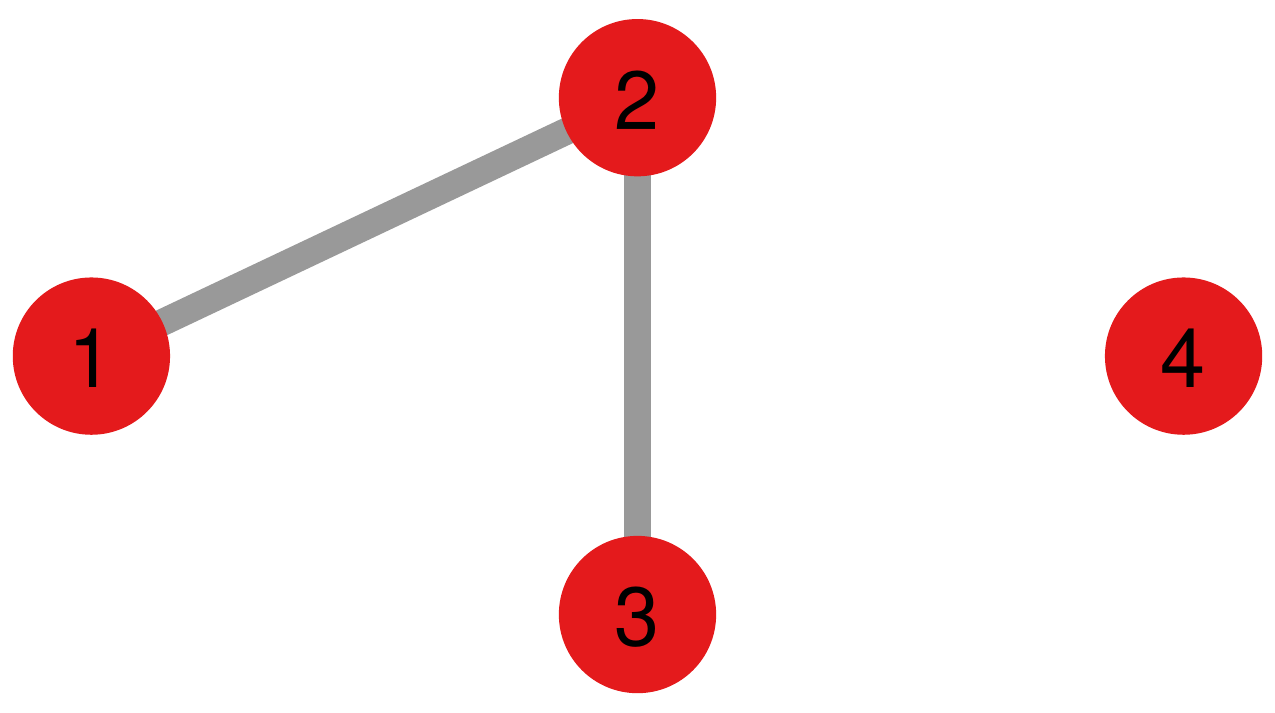}
\caption{$p = 0.0045$}
\end{subfigure}\quad
\begin{subfigure}{.2\columnwidth}
\includegraphics[width=\textwidth]{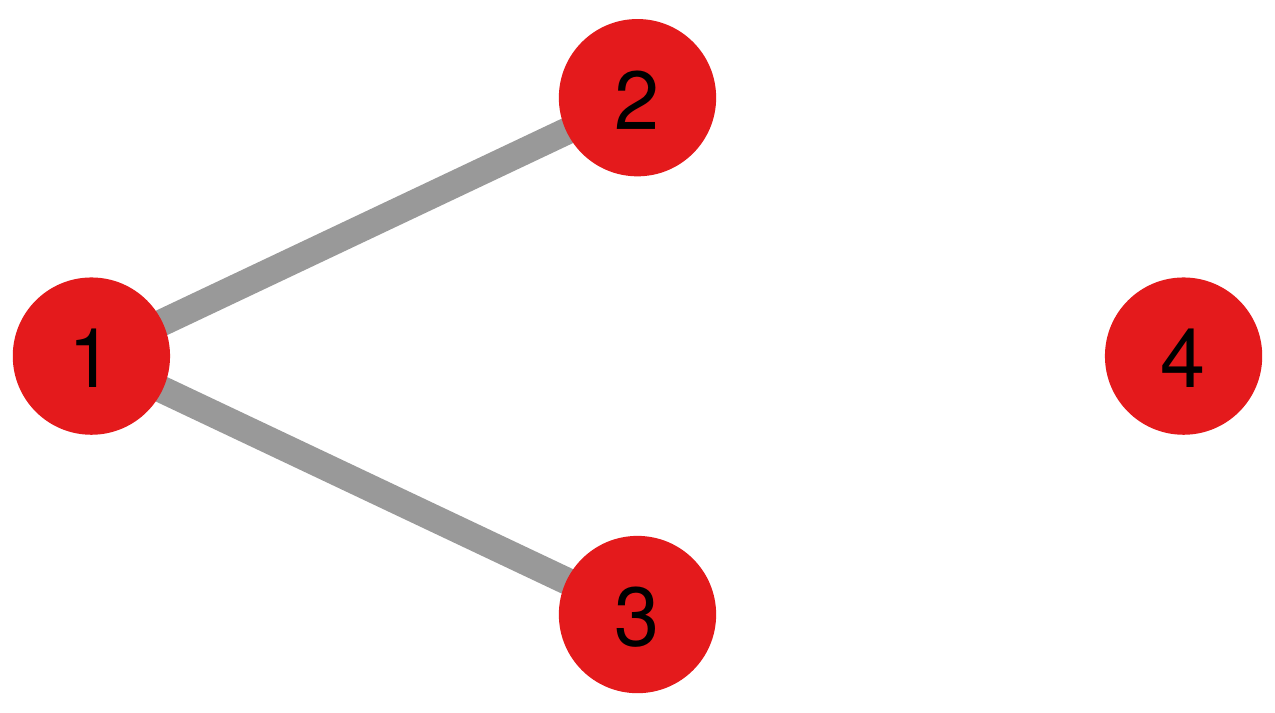}
\caption{$p = 0.0245$}
\end{subfigure}\quad
\begin{subfigure}{.2\columnwidth}
\includegraphics[width=\textwidth]{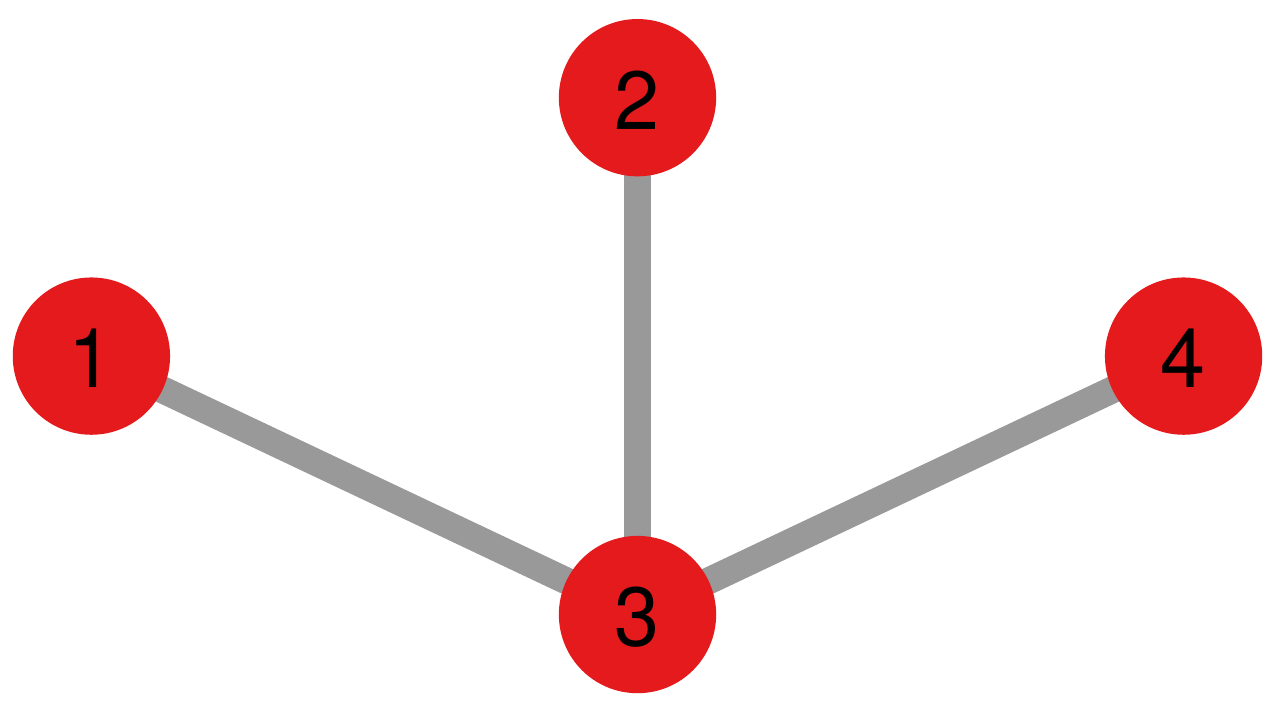}
\caption{$p = 0.0945$}
\end{subfigure}

\begin{subfigure}{.2\columnwidth}
\includegraphics[width=\textwidth]{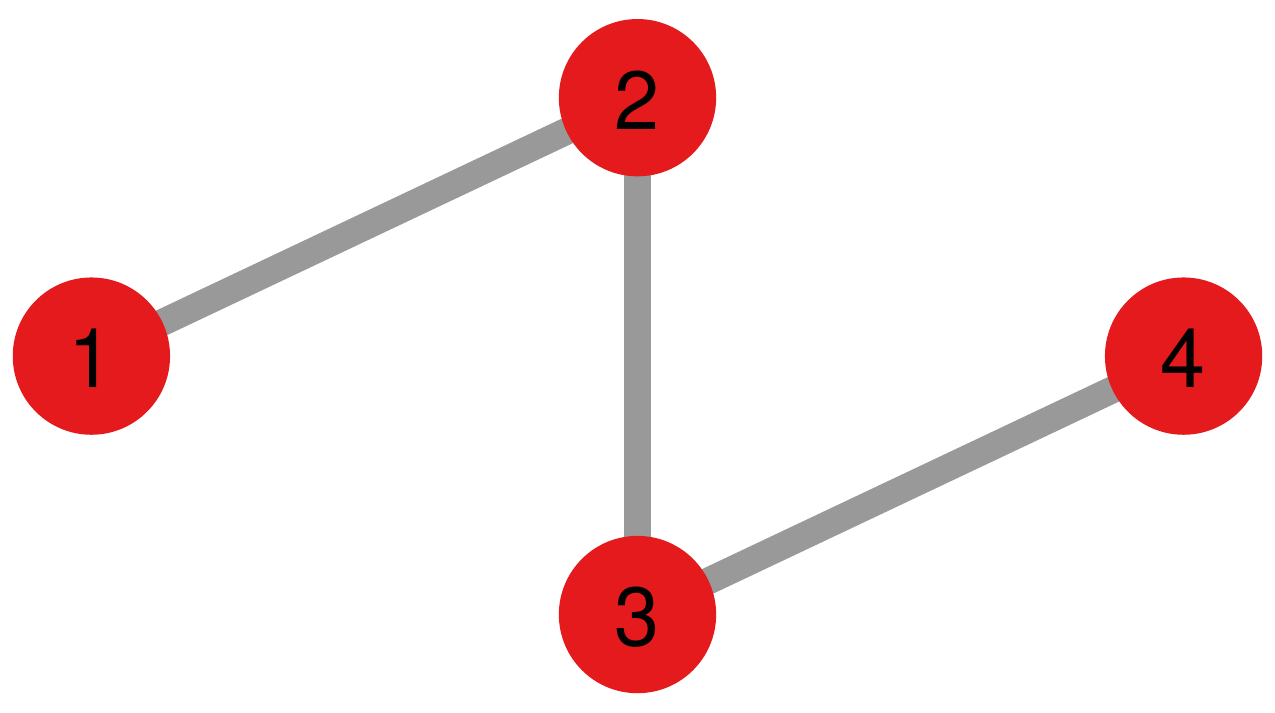}
\caption{$p = 0.0405$}
\end{subfigure}\quad
\begin{subfigure}{.2\columnwidth}
\includegraphics[width=\textwidth]{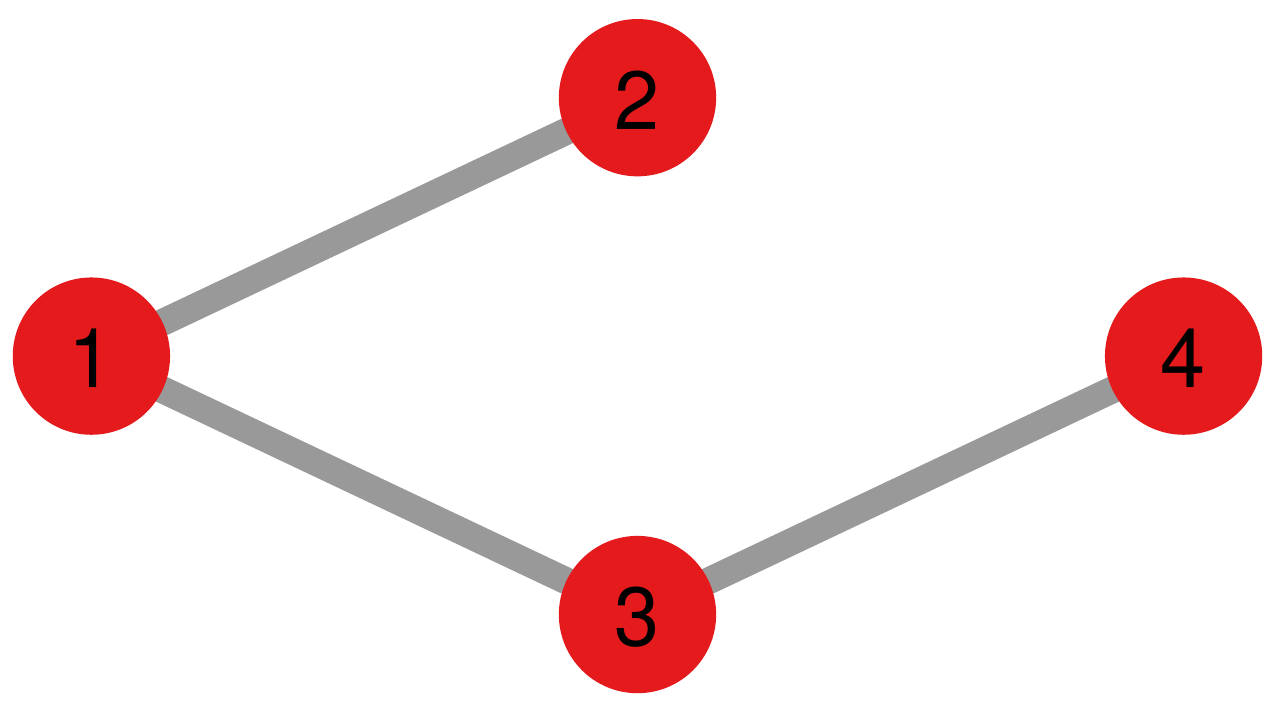}
\caption{$p = 0.2205$}
\end{subfigure}\quad
\begin{subfigure}{.2\columnwidth}
\includegraphics[width=\textwidth]{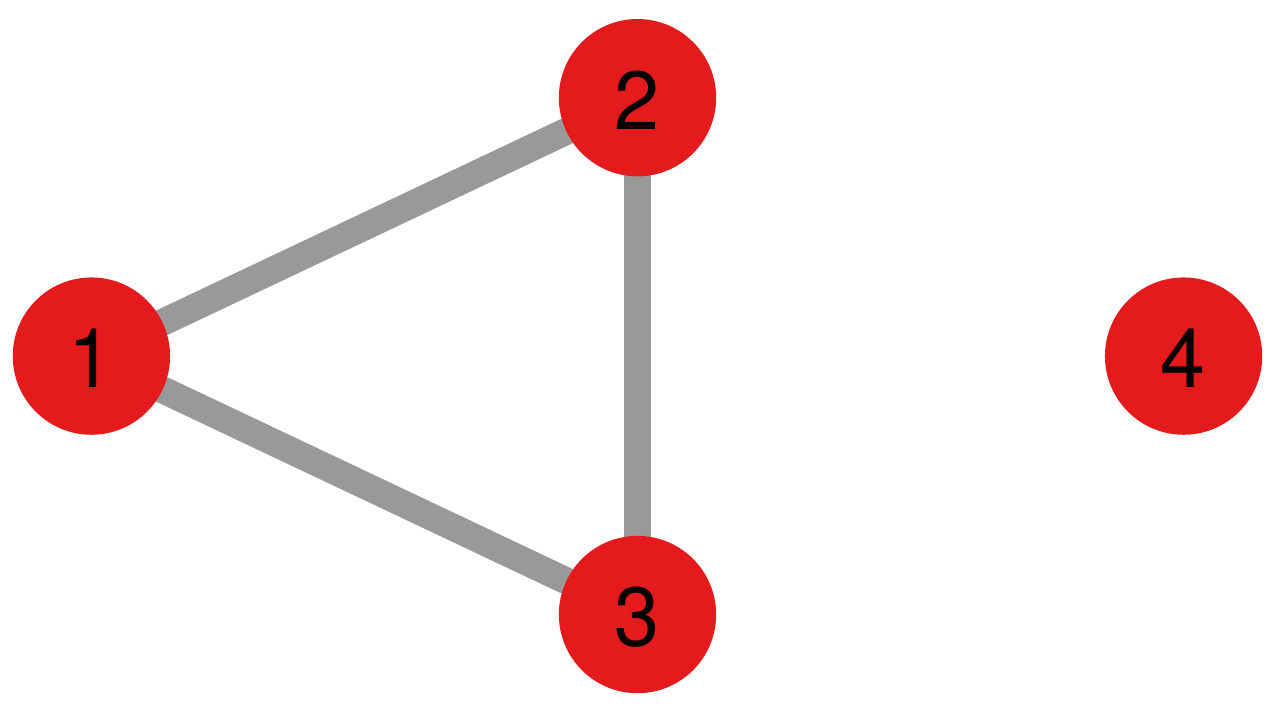}
\caption{$p = 0.0105$}
\end{subfigure}\quad
\begin{subfigure}{.2\columnwidth}
\includegraphics[width=\textwidth]{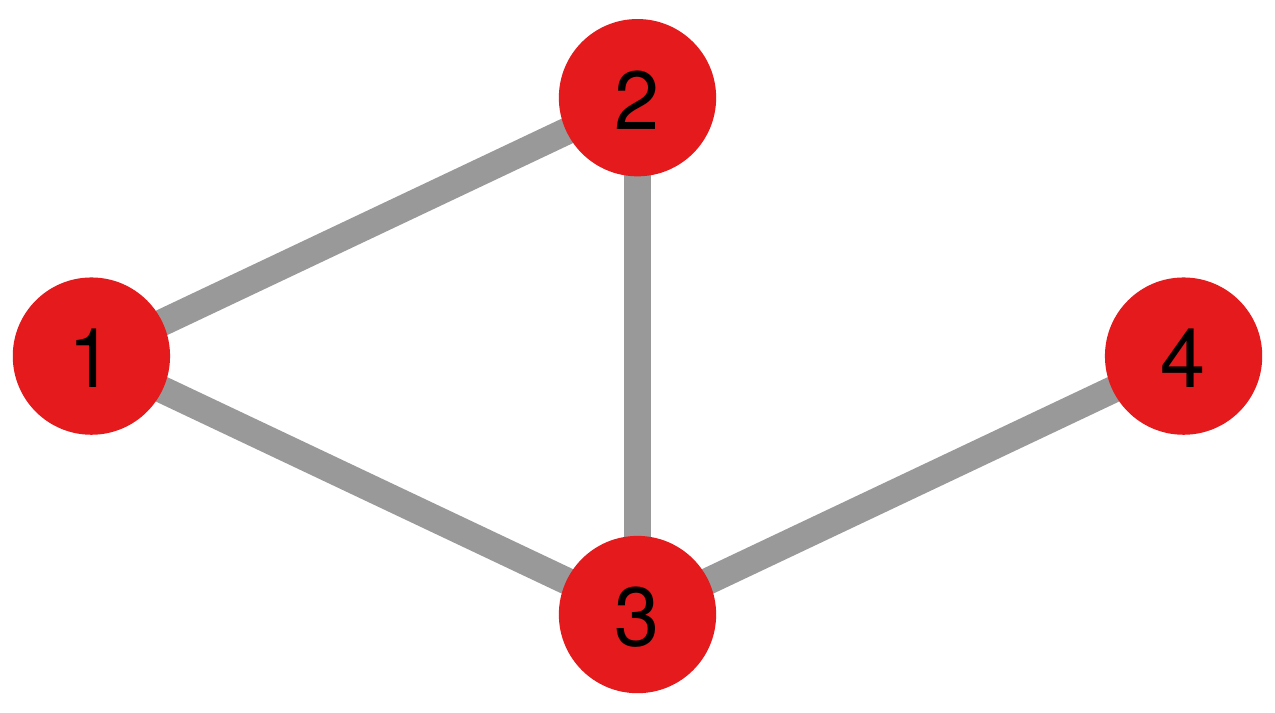}
\caption{$p = 0.0945$}
\end{subfigure}

\caption{Each possible world from the probabilistic network in Figure \ref{fig:probabilistic-network}. The subcaptions report the probability of the possible world.}
\label{fig:probabilistic-network-pws}
\end{figure*}

The problem with this naive approach is that you need to do this for each edge. In a network with $|E|$ edges, you have $2^{|E|}$ possible worlds, with each possible combination of edges existing or not existing. This can (and will) get out of hand pretty quickly. Even the simple network in Figure \ref{fig:probabilistic-network}(a) will generate the monstrous amount of possible worlds I show in Figure \ref{fig:probabilistic-network-pws} -- and here I omit all possible words with $p = 0$, for instance the one connecting only node $1$ to node $4$.

Looking at Figure \ref{fig:probabilistic-network-pws}, it is immediately obvious that one cannot compute all possible worlds for networks beyond a trivial size. Even if one could do it, it would still be a bad idea. What do you do when your analysis is by itself already expensive to calculate on a single network -- as is the case, for instance, for the calculation of the average shortest path length? One solution would be to sparsify\cite[1in]{parchas2018uncertain}: if you have fewer probabilistic edges to begin with, you have much fewer possible worlds.

There is one way to sparsify that picks the edges to keep by minimizing entropy, which will give you the most representative possible worlds for the measure you want to calculate. The reason why you want to minimize entropy is that because you want to sample possible worlds using a Monte Carlo approach (we briefly mentioned it in Section \ref{seg:ergmodels-ergm}): in this case, the lower the entropy the more accurate results you will get by sampling fewer possible worlds. However, this sampling approach is not general purpose: the edges you want to keep are going to be different depending on your objective -- for instance clustering coefficient rather than average path length.

This leads to focus on specific problems we want to solve and see how specialized approaches can lead us to deal with uncertain graphs.

\section{Specific Probabilistic Network Solutions}\label{sec:uncertainty-pnets-classical}
I will make a small deep dive into the following classical problems and one of their possible solutions in probabilistic networks:

\begin{itemize}
\item Node degree (for the classical version, see Chapter \ref{cha:degree});
\item Ego networks (which I formally introduce only in Section \ref{sec:homophily-ego});
\item Betweenness centrality (see Section \ref{sec:centr-betw} for a refresher);
\item Connected components (which we deal with in Section \ref{sec:paths-ccomps});
\item Finding the densest subgraph (similar to k-core decomposition -- Section \ref{sec:centr-kcore}).
\end{itemize}

Some of these things we'll only see later in the book, so I'll give a brief context for the time being.

\subsection{Node Degree}
If you wanted to know the degree distribution of the network in Figure \ref{fig:probabilistic-network}, what would you do? The naive approach would be to determine the expected degree of each node: go over all the possible worlds and take the average of the degrees of the node, weighted by how likely the world is to exist. 

\begin{figure}
\centering
\begin{subfigure}{.2\columnwidth}
\includegraphics[width=\textwidth]{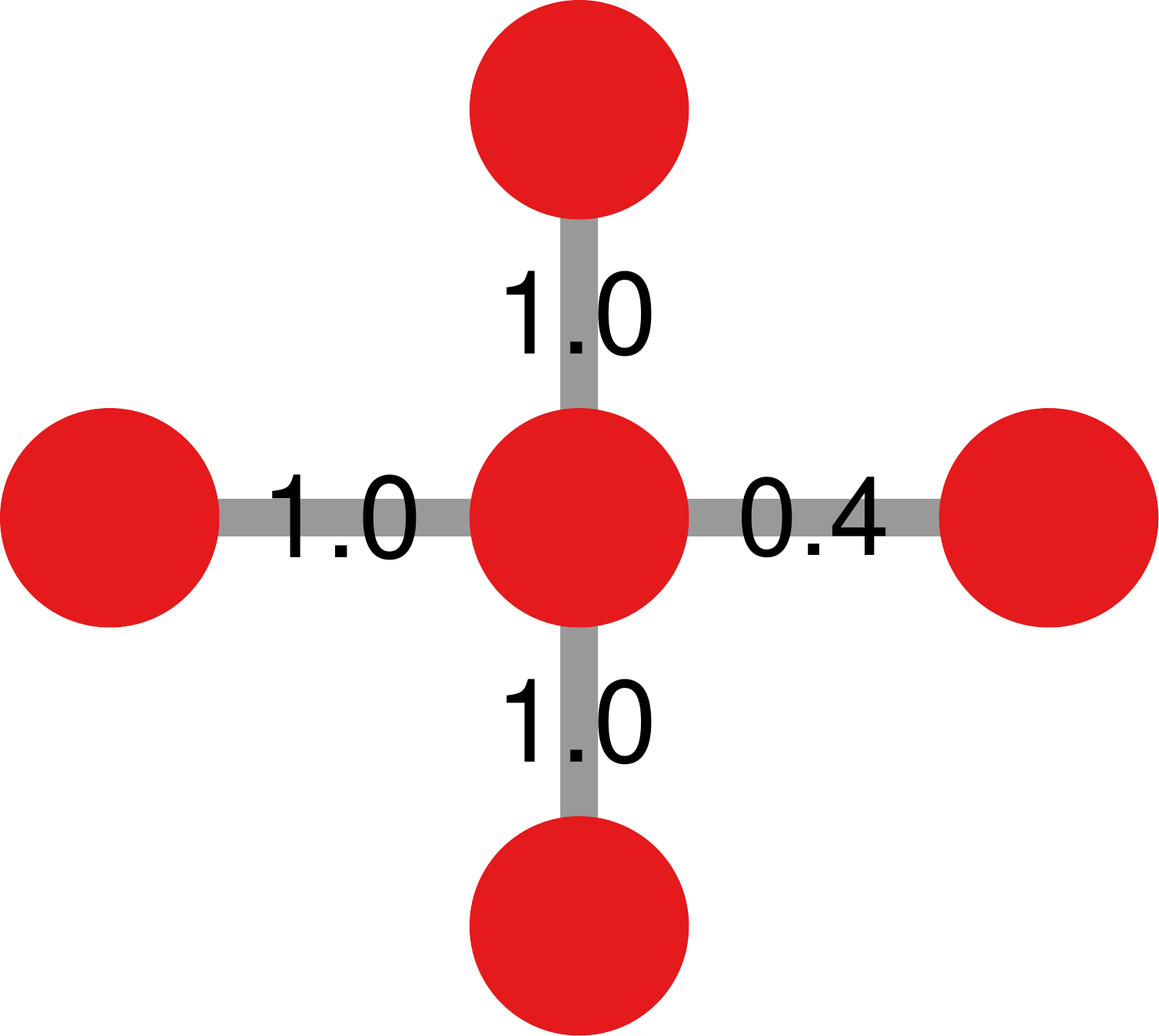}
\caption{}
\end{subfigure}
\quad
\begin{subfigure}{.2\columnwidth}
\includegraphics[width=\textwidth]{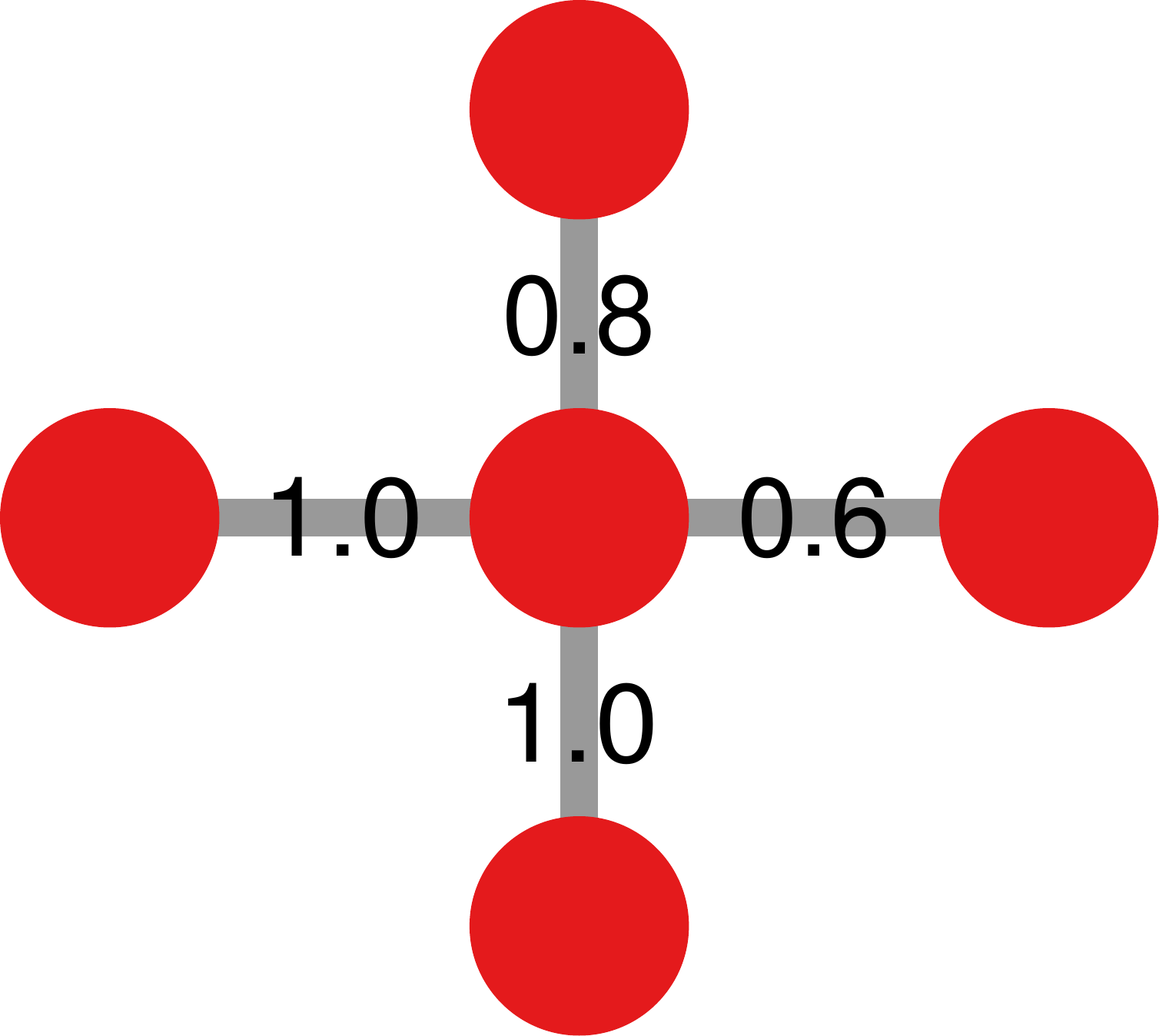}
\caption{}
\end{subfigure}
\quad
\begin{subfigure}{.2\columnwidth}
\includegraphics[width=\textwidth]{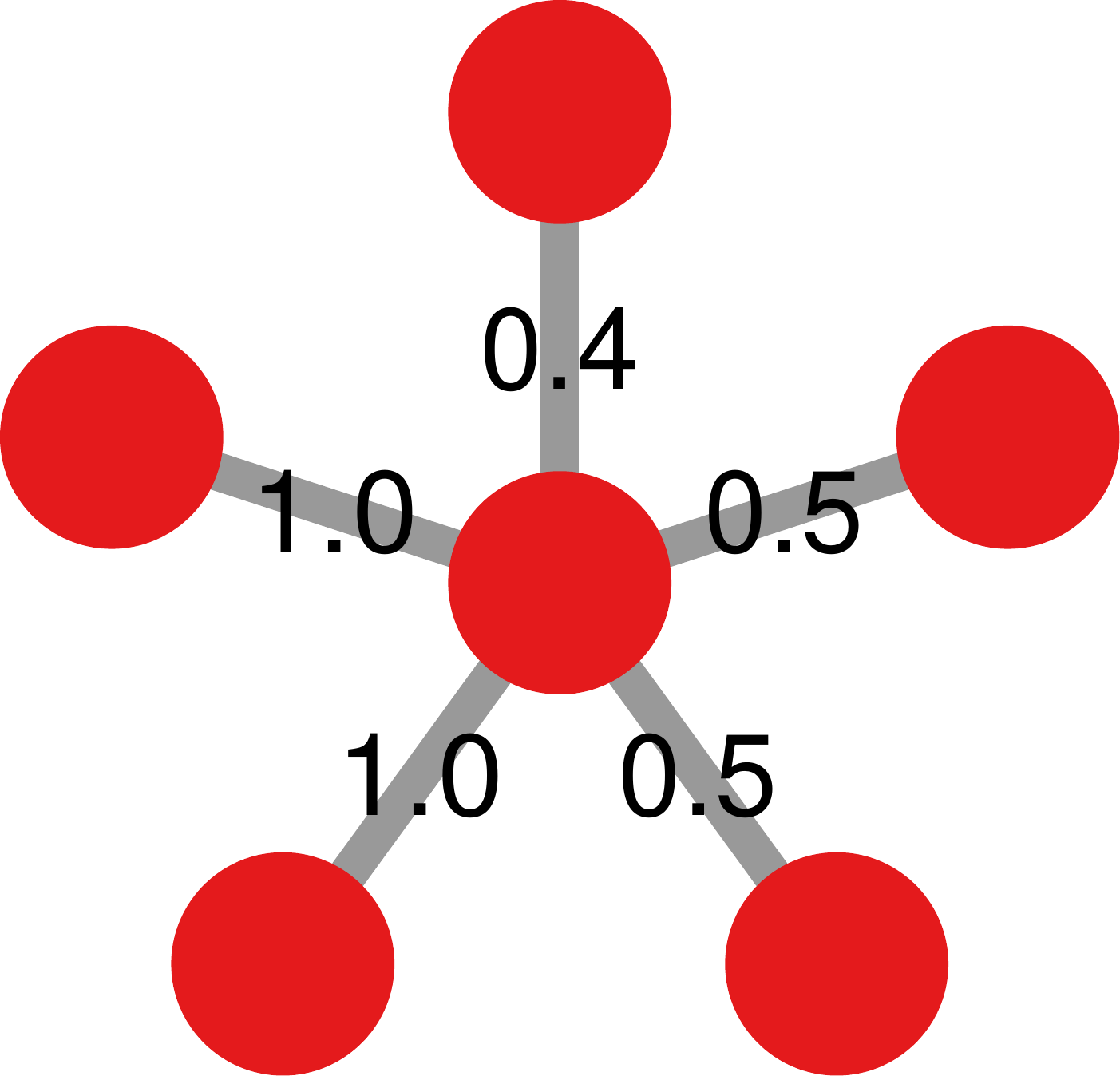}
\caption{}
\end{subfigure}
\quad
\begin{subfigure}{.2\columnwidth}
\includegraphics[width=\textwidth]{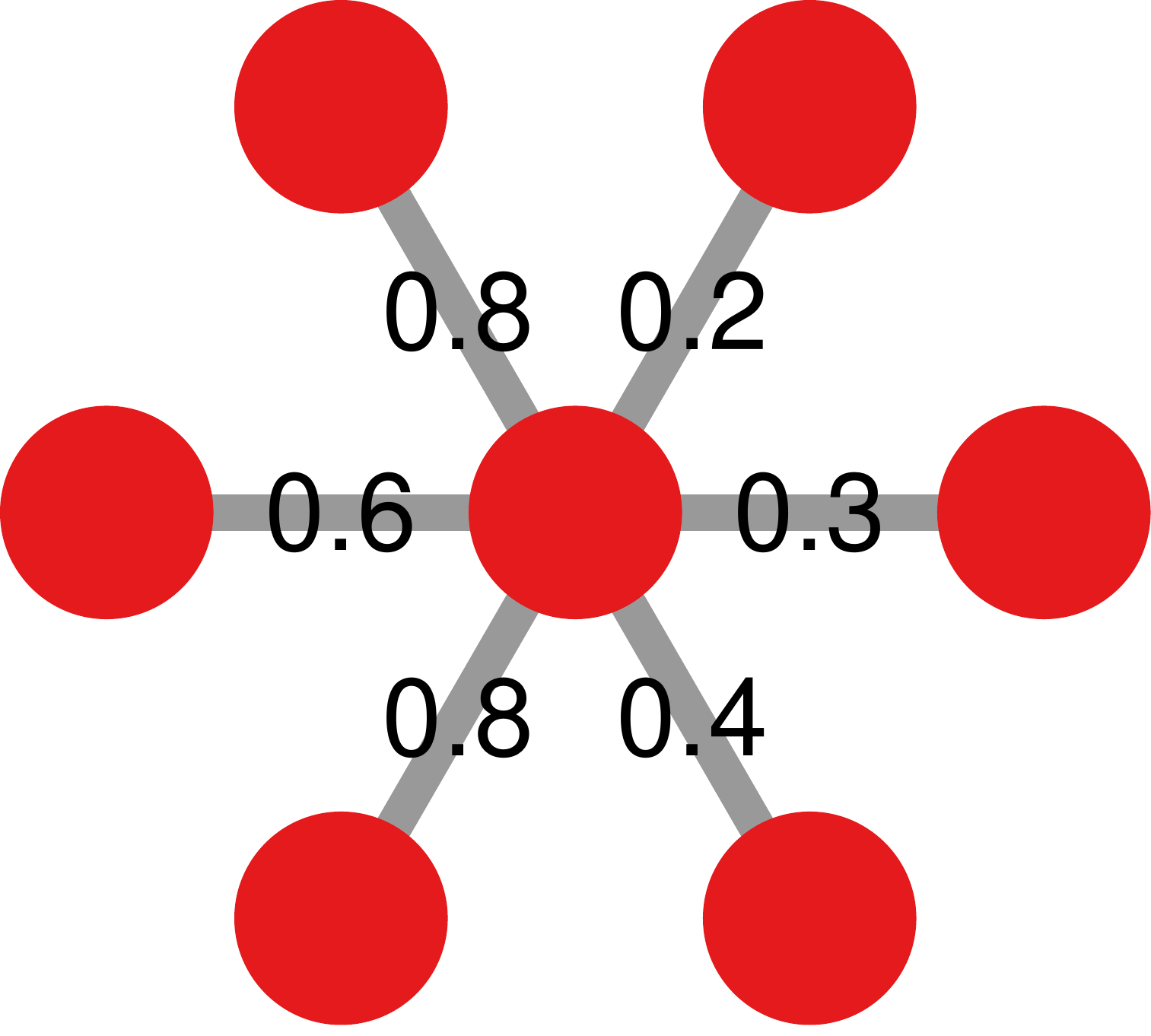}
\caption{}
\end{subfigure}
\caption{Probabilistic networks. The edge label reports an edge's probability of existing.}
\label{fig:probnet-expected-degree}
\end{figure}

However, that would be a bit silly: the degree is a count, it shouldn't be a continuous number. The interpretability of what you'd do by averaging would go out of the window. To see why, consider Figure \ref{fig:probnet-expected-degree}. The nodes in the figure have the same expected degree ($3.4$ -- what the hell does it mean for a node to have degree equal to $3.4$?), but their connection topologies are radically different. 

\begin{figure}
\centering
\begin{subfigure}{.4\columnwidth}
\includegraphics[width=\textwidth]{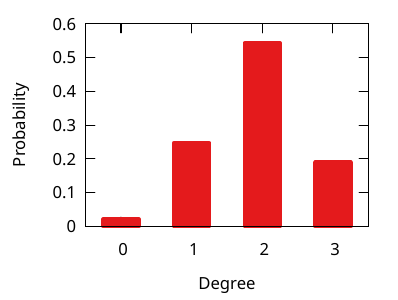}
\caption{}
\end{subfigure}
\qquad
\begin{subfigure}{.4\columnwidth}
\includegraphics[width=\textwidth]{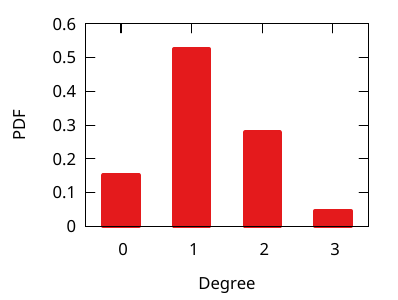}
\caption{}
\end{subfigure}
\caption{(a) The probability (y axis) for node $3$ in Figure \ref{fig:probabilistic-network} to have a given degree (x axis). (b) The probabilistic degree PDF for the entire network in Figure \ref{fig:probabilistic-network}.}
\label{fig:probabilistic-network-dd}
\end{figure}

It is much better to give each node a probability distribution for each possible degree value which allows you to differentiate between the nodes in Figure \ref{fig:probnet-expected-degree}. In Figure \ref{fig:probabilistic-network-dd}(a), I focus on node $3$ of Figure \ref{fig:probabilistic-network}. Once you do all your weighted averages for all the possible worlds with their likelihoods, you'll end up with the correct degree distribution in Figure \ref{fig:probabilistic-network-dd}(b). To get to Figure \ref{fig:probabilistic-network-dd}(b) the method needs to take into account the whole probability distribution for the node, rather than a simple point estimate\cite{kaveh2019comparing}.

One way to do it is by counting how many nodes had a specific degree in each of the worlds in Figure \ref{fig:probabilistic-network-pws} and then weight the probability of those discrete values with the likelihood of the world. Similar approaches can work, e.g., to perform core decomposition in probabilistic networks\cite{bonchi2014core}, since to perform core decomposition you need to know the degrees of the nodes (see Section \ref{sec:centr-kcore}).

\subsection{Ego Networks}
\begin{figure}[t!]
\centering
\includegraphics[width=\columnwidth]{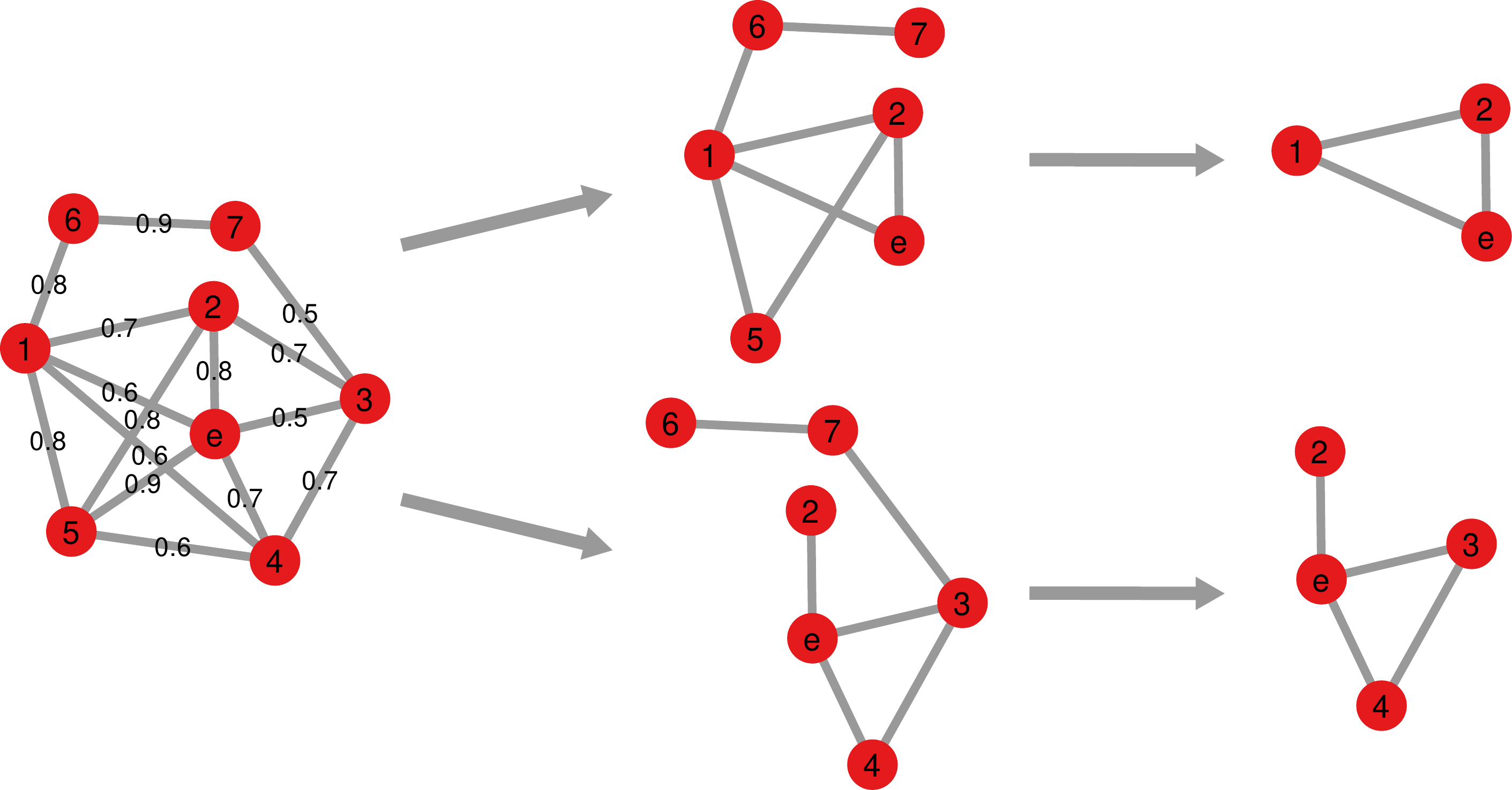}
\caption{The process of extracting an ego network from a probabilistic network by first realizing the possible worlds and then extracting their ego networks. Ego marked with the ``e'' label.}
\label{fig:probnet-ego-1}
\end{figure}

\begin{figure}[t!]
\centering
\includegraphics[width=\columnwidth]{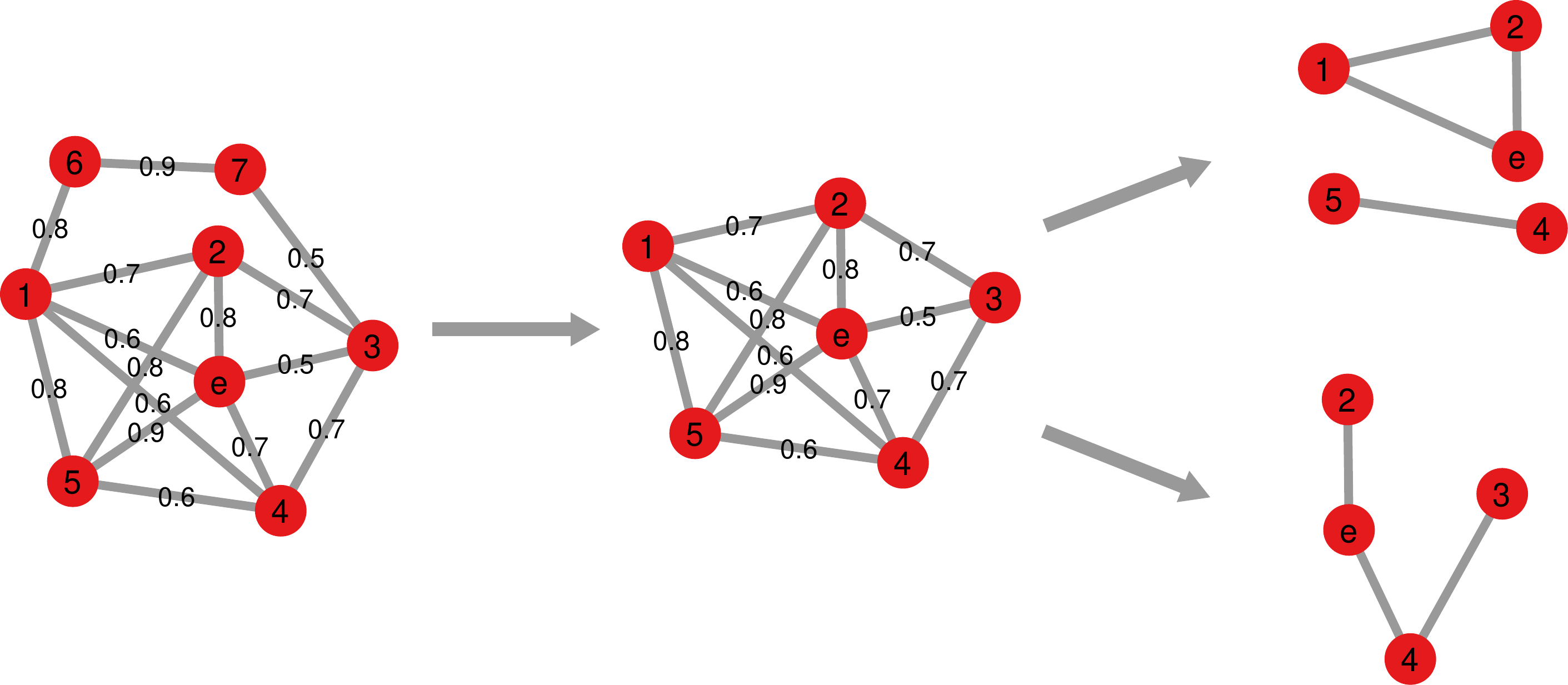}
\caption{The process of extracting an ego network from a probabilistic network by first extracting a probabilistic ego network and then realizing its possible worlds. Ego marked with the ``e'' label.}
\label{fig:probnet-ego-2}
\end{figure}

As we will see in Section \ref{sec:homophily-ego}, an ego network is a subset of a graph. The subset contains one node -- the ``ego'' --, all of its direct neighbors and all of the connections between these nodes (ego + neighbors). When dealing with probabilistic networks, there are fundamentally two ways of extracting an ego network\cite{kaveh2021defining}. You can generate first the possible worlds and then get the ego networks from each world separately (Figure \ref{fig:probnet-ego-1}) or you can extract the ego network directly from the probabilistic network and then apply specialized algorithms to calculate your measure of interest on the probabilistic ego network (Figure \ref{fig:probnet-ego-2}).

You might think this is a ``Well, duh'' moment, but I invite you to consider a few fun things that could happen. In an ego network, by definition, the ego is connected to all of its neighbors. Logically, ego networks must have a single connected component. If you go for the first option ``possible worlds first, ego networks later,'' both of these properties are guaranteed in your result. However, if you extract an ego network as a probabilistic network, when you then realize the various possible worlds you are likely to get a few neighbors that are not connected to the ego and you might even get disconnected components!

More in general, the results you'd get from these two approaches are highly correlated, but they're not the same. So this choice matters. And having disconnected ego networks or an ego not connected to all of its neighbors can be a valid way to analyze your ego network, depending on the requirements you need to satisfy to answer your research question. So there isn't necessarily a ``best option'' between the two.

\subsection{Betweenness Centrality}\label{sec:uncertainty-pnets-betweenness}
Not everything is more difficult when working with probabilistic networks. Sometimes, you're lucky. Betweenness centrality is one of those cases. You remember from Section \ref{sec:centr-betw} that betweenness centrality is the number of shortest paths passing through a node. In probabilistic networks, you have to weight each path by its probability of existing.

Since a path is a sequence of edges, and since each edge has a probability of existing, it follows that a long path is -- under reasonable assumptions -- less likely to exist than a short path. A path can only exist if all of the edges composing it exist. Therefore, its probability of existing is the product of all its edge probabilities. Here we assume that each edge's existence probability is independent from all other edges, which is what people normally assume when working with probabilistic networks. And the probability of two independent events happening is the product of their probabilities -- the probability of getting heads twice is the probability of getting heads once, squared -- see Section \ref{sec:prob-axioms} for a refresher.

Imagine a probabilistic network where every single edge has 50/50 odds of existing. Any path of length $3$ in this network exists with probability $p = (1 / 2)^3 = 0.125$. But a path of length $4$ has $p = (1 / 2)^4 = 0.0625$ and one of length $5$ has $p = (1 / 2)^5 = 0.03125$. As you can see, these probabilities keep halving. At some point, we can decide that the contributions from paths of length $l$ is negligible, and simply ignore them. The resulting betweenness will be practically the same\cite{kaveh2021probabilistic}. This speeds up computation quite a bit. On the other hand, the number of possible paths will go up, which might counteract the vanishing probabilities. Finding the right balance is something that probably depends on the probability distributions on your edges and there is no silver bullet.

\subsection{Connected Components}
In a deterministic network, when two nodes are part of the same connected component they are reachable: there exists a path following the edges of the network leading you from one node to another (see Section \ref{sec:paths-ccomps}). This is not necessarily true in a probabilistic network. Once you think about this in terms of probabilistic networks, it is not hard to see why.

\begin{figure}
\centering
\begin{subfigure}{.45\columnwidth}
\includegraphics[width=\textwidth]{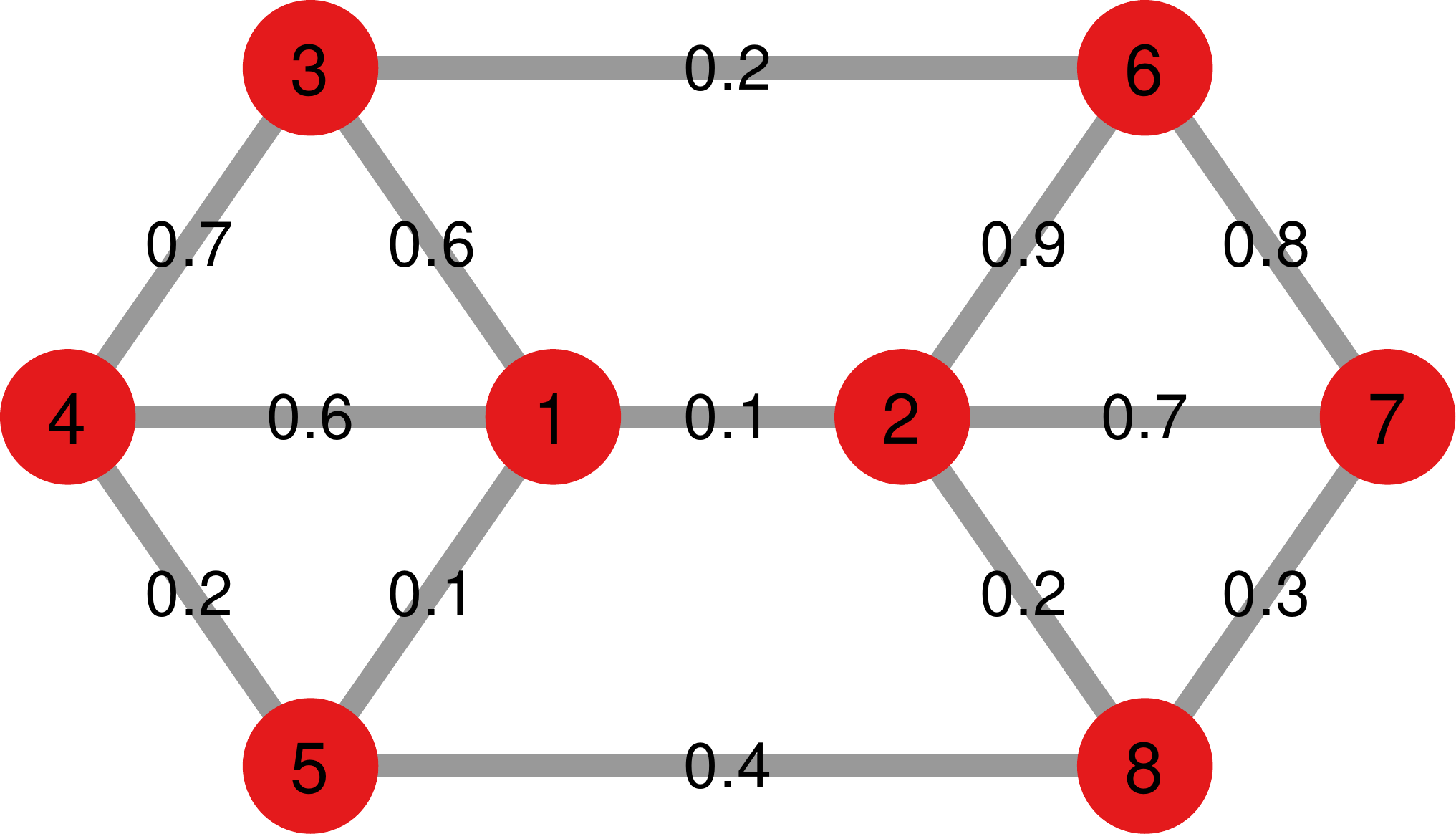}
\caption{}
\end{subfigure}
\quad
\begin{subfigure}{.45\columnwidth}
\includegraphics[width=\textwidth]{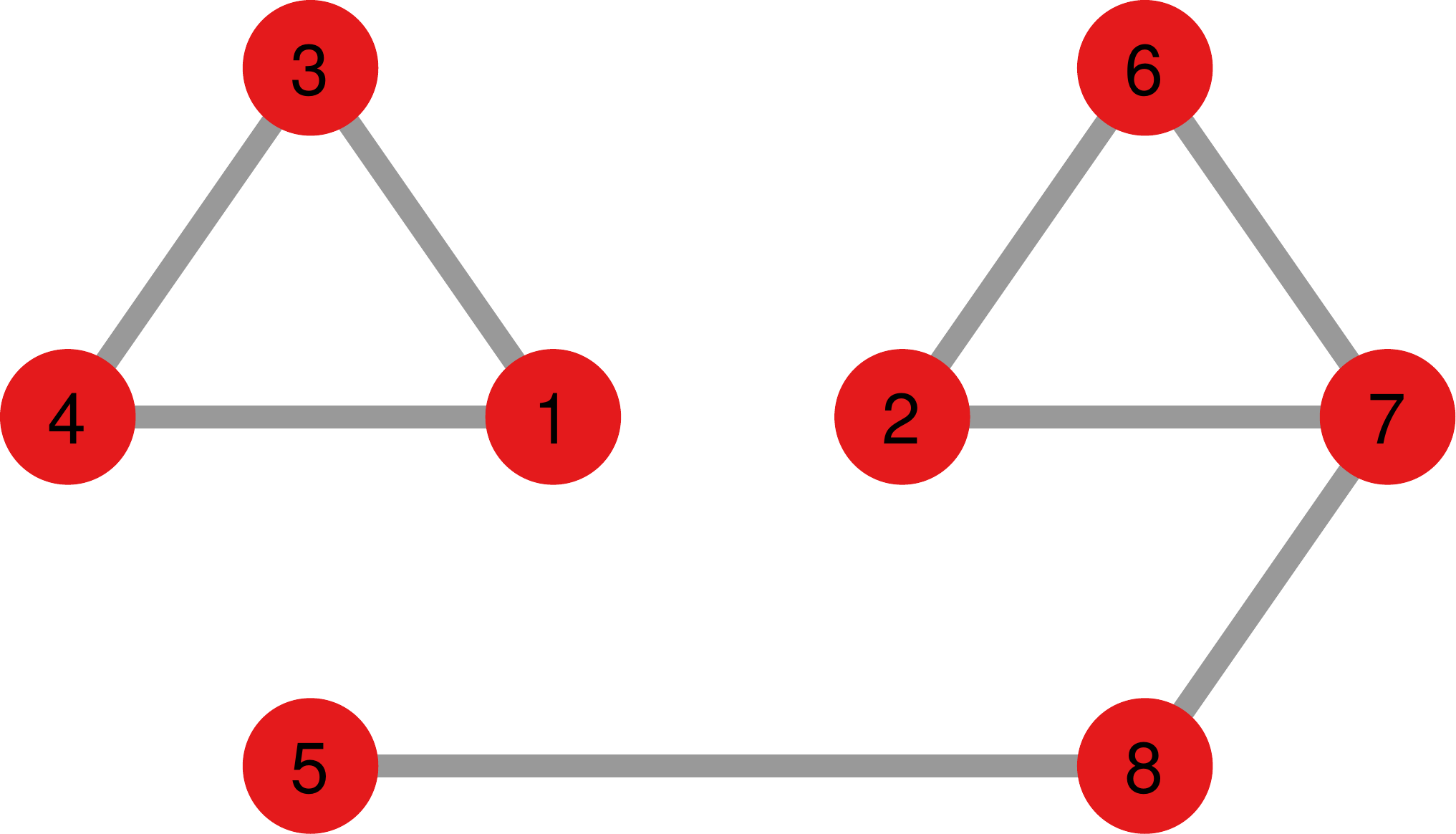}
\caption{}
\end{subfigure}
\caption{(a) A probabilistic network with edge probabilities as the edge's labels. (b) One of the possible worlds of (a).}
\label{fig:probabilistic-reliability}
\end{figure}

Consider Figure \ref{fig:probabilistic-reliability}(a). This is a probabilistic network in which nodes $1$ and $2$ are not only in the same component, they are directly connected to each other. However, many of the possible worlds of Figure \ref{fig:probabilistic-reliability}(a) have nodes $1$ and $2$ in different components -- Figure \ref{fig:probabilistic-reliability}(b) is but one example.

In probabilistic networks, we can estimate a probability that two nodes are reachable -- i.e. they are part of the same connected component. We call this probability `reliability''. The reliability of node pairs $(1,2)$ is simply the sum of the probabilities of all possible worlds in which they are in the same connected component\cite{jin2011discovering}\cite{khan2014fast}.

\subsection{Densest Subgraph}
In many applications, we might want to extract the densest subgraph of a network\cite{goldberg1984finding}. As the name implies, the densest subgraph of a network is a subset of its nodes and the edges between them such that its density is the highest than the one of any other subgraphs in the network. You cannot find a subgraph denser than the one you extracted. Of course there are trivial solutions -- for instance any connected node pair is a densest subgraph because it has density one -- but one can specify a minimum number of nodes they are interested in, or a different measure to optimize such as the average degree.

Of course, there are ways to solve this problem in many different scenarios, for instance in streaming graphs as the data comes in little by little\cite{bahmani2012densest}, but here we're interested in probabilistic networks specifically. The idea is, as usual, to try and define a measure of expected density for a subgraph, which is its density across all possible worlds weighted by their probability of existing. There are many methods to do so, and a few notable ones can also avoid making the assumption that the edge existence probabilities are independent\cite{zou2013polynomial}.

One related problem is to find not the densest, but any specific subgraph instead of a large graph. This is something we will see in depth when we talk about subgraph mining (Chapter \ref{cha:mining-base}). For now, let's just say that you might have a specific pattern in mind and you want to know how many times it appears in the data. When you have a probabilistic network, you could instead ask for all patterns that are more likely to exist than a certain probability\cite{zou2010discovering}\cite{papapetrou2011efficient}.

\begin{figure}[b!]
\centering
\begin{subfigure}{.04\columnwidth}
\includegraphics[width=\textwidth]{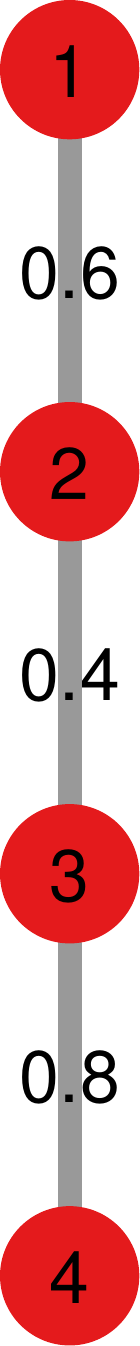}
\caption{}
\end{subfigure}
\qquad\qquad
\begin{subfigure}{.04\columnwidth}
\includegraphics[width=\textwidth]{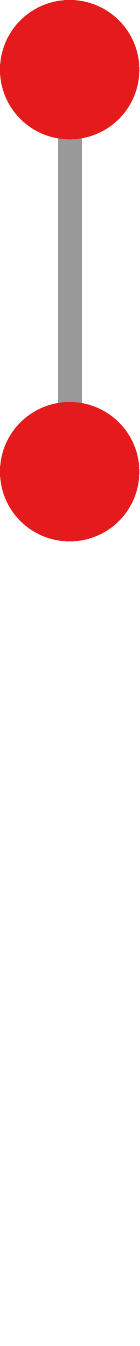}
\caption{}
\end{subfigure}
\qquad\qquad
\begin{subfigure}{.04\columnwidth}
\includegraphics[width=\textwidth]{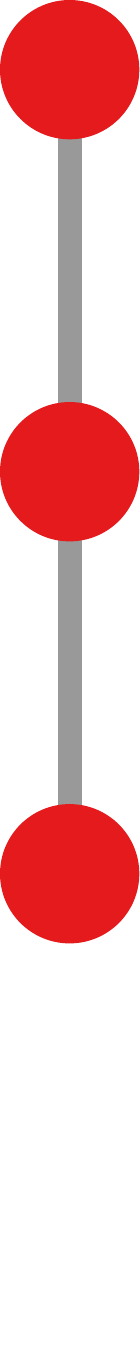}
\caption{}
\end{subfigure}
\qquad\qquad
\begin{subfigure}{.04\columnwidth}
\includegraphics[width=\textwidth]{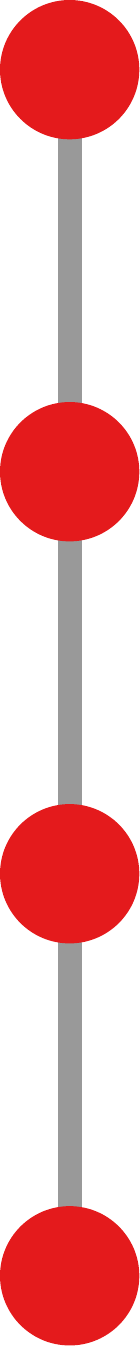}
\caption{}
\end{subfigure}
\caption{(a) A probabilistic network with edge probabilities as the edge's labels. (b-d) Some subgraphs that can be extracted from (a). They exist with probabilities: (b) $p = 0.8$, (c) $p = 0.32$, (d) $p = 0.192$ -- the probability of a pattern is the maximum combined probability of the set of edges that can form the pattern.}
\label{fig:probabilistic-patterns}
\end{figure}

The potentially huge search space -- a large graph has a lot of potential subgraphs -- can be efficiently reduced. Let's take a look at Figure \ref{fig:probabilistic-patterns}(a). 

This probabilistic network is extremely simple -- it's only a chain with four nodes and three edges. Even this ridiculously simple network already has a surprisingly high number of possible subgraphs. However, Figures \ref{fig:probabilistic-patterns}(b-d) shows a subset of them, and it also shows that the probabilities of these patterns existing goes down quite dramatically with their size (also discussed in Section \ref{sec:uncertainty-pnets-betweenness}). A larger pattern has a strictly lower probability of existing than a smaller pattern -- basically the product of all its edges' probabilities, if these probabilities are all independent. If you say that you only want patterns with $25\%$ probability of existing or more, the number of potential subgraphs you need to check in Figure \ref{fig:probabilistic-patterns}(a) becomes more sane.

\section{Alternatives to Probability Theory}
As I discuss in Chapter \ref{cha:prob}, probability theory is not the only game in town when confronted with uncertainty. One could use Dempster-Shafer's theory of evidence (DST) or fuzzy logic (Section \ref{sec:prob-alt-fuzzy} -- go back and refresh your understanding of these two approaches, or the rest of this section won't make much sense). DST has been used to classify\cite[-0.2in]{dhaou2019belief} and predict\cite{mallek2019evidential} edges. Besides shortest path distance\cite{simas2015distance} -- as we will see in a moment -- fuzzy logic has also been applied to the problem of link prediction\cite{moradabadi2017link} and to estimate the centrality of nodes\cite{hu2015centrality}. There is also uncertainty theory\cite{liu2010uncertainty}, which has been used for transportation planning\cite{gao2016uncertain} and shortest paths\cite{sheng2020uncertain}, but I won't treat it here.

\begin{figure}
\centering
\begin{subfigure}{.275\columnwidth}
\includegraphics[width=\textwidth]{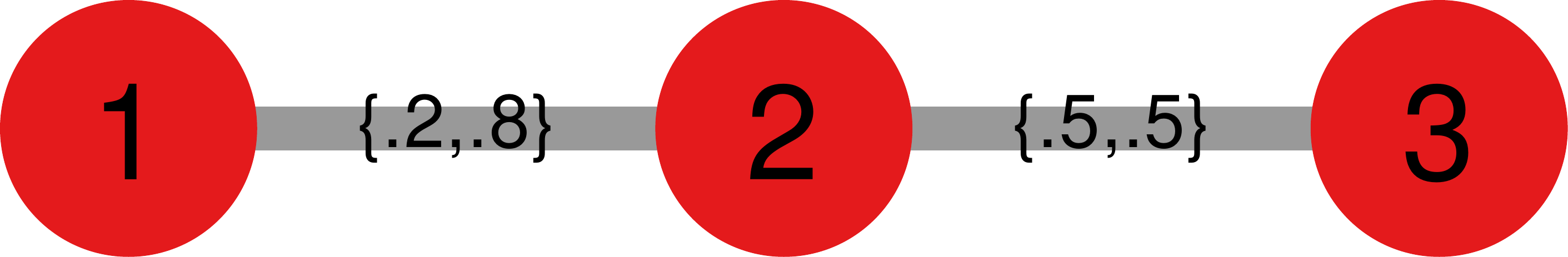}
\caption{}
\end{subfigure}
\quad
\begin{subfigure}{.275\columnwidth}
\includegraphics[width=\textwidth]{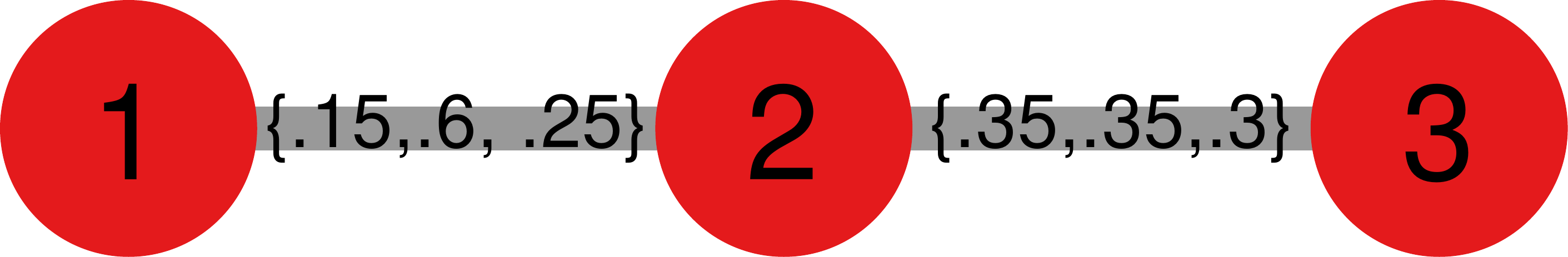}
\caption{}
\end{subfigure}
\quad
\begin{subfigure}{.275\columnwidth}
\includegraphics[width=\textwidth]{figures/prob_dst_fuzzy_1.pdf}
\caption{}
\end{subfigure}
\caption{The edge labels represent the uncertain edge weights, using different models. (a) Probability theory: the probability of the edge being of weight $1$ and $2$. (b) DST: Mass values of edge weights $\{1\}$, $\{2\}$, and $\{1,2\}$. (c) Fuzzy logic: degree of belonging to the classes of weight $1$ and $2$.}
\label{fig:prob-dst-fuzzy}
\end{figure}

For the sake of simplicity, I'll focus on how DST and fuzzy logic will represent a problem differently than probability theory and lead to different results. I'll limit the discussion to the estimation of the path distance between two nodes. Figure \ref{fig:prob-dst-fuzzy} shows three representations of the same network with uncertain edge weights. So what is the distance between nodes $1$ and $3$ -- counting the weights of the edges?

Let's start with probability theory -- Figure \ref{fig:prob-dst-fuzzy}(a). There are four possible worlds out there: both weights could be $1$ (with $p = 0.2 \times 0.5 = 0.1$), then they could be $1$-$2$ (with $p = 0.2 \times 0.5 = 0.1$), $2$-$1$ (with $p = 0.8 \times 0.5 = 0.4$), or both $2$ (with $p = 0.2 \times 0.5 = 0.4$). So probability theory will tell you that the distance between node $1$ and node $3$ has a probability distribution: it could be $2$ with $p = 0.1$, it could be $3$ with $p = 0.5$, or it could be $4$ with $p = 0.4$. Splendid.

Onto DST, then. Looking at Figure \ref{fig:prob-dst-fuzzy}(b) we see that we now have also to deal with the fact that we might not be able to prove that the edge weight is either $1$ or $2$. We know those are the only two possibilities, though, so we need to estimate the belief and plausibility of the weight being $2$ -- for brevity's sake I'll do the math only for the first edge. The Belief of the $1,2$ edge having weight $2$ is $0.6$ -- equal to its Mass since this set has no subsets. The Plausibility is $1 - Belief(1) = 0.85$. Following the literature\cite[-0.1in]{szucs2009route}, we can define the edge bounds as follows: $1 + [1 \times 0.6, 1 \times 0.85] = [1.6, 1.85]$ -- i.e. the minimum weight we think the edge has is $1.6$ and the maximum is $1.85$.

Why did we aggregate this way? As I mentioned, we take for granted that the edge weight must be at least $1$, so that's why we start with ``$1 +$''. Then, Belief($2$) -- which is $0.6$ -- tells us how much we can prove there is an additional cost of $1$. So the total cost cannot be anything less than $1.6$. Belief($1$), instead, tells us how much we can prove the edge weight is not $2$ ($0.15$) which means we can prove that the weight cannot be any higher than $1.85$.

Applying the same reasoning to the $2,3$ edge leads us with an interval of $[1.35,1.65]$. The final cost of the full path must then be within the interval $[1.6, 1.85] + [1.35, 1.65] = [2.95, 3.5]$. Our available evidence tells us that the path costs at least $2.95$ to traverse, but not more than $3.5$.

Just like with DST and probability theory, there are multiple ways to use fuzzy logic\cite{klein1991fuzzy}, so here I'll just pick one as an example of the different results different framework can give. Since the edge weights are fuzzy, the length of the shortest path $1 \rightarrow 2 \rightarrow 3$ is fuzzy as well. It can belong to three classes: length $2$ (if both edges have weight $1$), length $3$ (if one edge has weight $2$), and length $4$ (when both edges have length $2$). To find out to which degrees the path belongs to these length classes we need to work backwards. First we ask the distance of $3$ with itself, which can only be $0$, by definition.

Then we need the distance between node $2$ and node $3$, which is $1$ with degree $0.5$ and $2$ with degree $0.5$ -- because those are the classes of the $2,3$ edge and node $3$ contributes zero. We write this as $\{1/0.5, 2/0.5\}$.

We apply the same reasoning to get the length of $1 \rightarrow 2 \rightarrow 3$. The $1,2$ edge contributes $\{1/0.2, 2/0.8\}$ to what we already had from the edge $2,3$. We need to add up all the possible combinations and taking the minimum belongings. The path can be of length $2$ only if both edges weights are $1$, so its belonging in the minimum of the two (see Section \ref{sec:prob-alt-fuzzy}) which is $0.2$. The path can be of length $3$ if the weights are either $2-1$ (which is $min(0.8,0.5) = 0.5$) or $1-2$ (which is $min(0.2,0.5) = 0.2$). In fuzzy logic, we need to take the maximum between the two, which is $0.5$. Finally, to be of length $4$ both edges need to belong to the weight $2$ class, and that's $min(0.8,0.5) = 0.5$. So the final result is $\{2/0.2, 3/0.5, 4/0.5\}$.

As you can see the three theories give three very different results. The differences are not only quantitative, but also qualitative: you get results that are truly incompatible with each other -- a distribution, an interval, a set of classes.

\section{Summary}

\begin{enumerate}
\item Network data is not immune to measurement error. There could be missing or extra edges/nodes, and incorrectly reported features such as edge direction, weight, or any sort of edge/node attribute.
\item To correct measurement errors you need to have a model of your errors. If you know the type of errors you are getting, then you can use statistical methods to infer the most likely true value, based on the measurements you observe. This, however, requires to measure your network multiple times -- which is unusual in network data.
\item One can sidestep the requirement of measuring the network multiple times by having an accurate model of the generating process of the network. Then one can generate many synthetic versions of the observed data, as if they were multiple measurements. The downside being that one now has to have an accurate model of the network, which can be tricky.
\item If one has data about the probability of existence of an edge, then they can work directly with probabilistic networks -- network whose edges have probability values. To analyze them, one has to create all the possible networks -- which are $2^{|E|}$ because each edge can exist or not exist. Then they can calculate a measure of interest in all possible worlds and average it with the probabilities of each possible world.
\item Since there are so many possible worlds, there are specialized techniques to reduce the computational complexity and estimate the probability distribution of given network measures such as the degree, betweenness centrality, connected components, and more.
\item One can work with alternative approaches to deal with uncertainty and create different types of networks with uncertain data, which will give different interpretations (and results) with the same data.
\end{enumerate}

\section{Exercises}

\begin{enumerate}
\item Consider the network at \url{http://www.networkatlas.eu/exercises/28/1/data.txt}. This is an undirected probabilistic network with four columns: the two connected nodes, the probability of the edge existing and the probability of the edge non existing. Generate all of its possible worlds, together with their probability of existing. (Note, you can ignore the fourth column for this exercise)
\item Calculate the probabilistic degree distribution of the network used in exercise 1. (Note, you can ignore the fourth column for this exercise)
\item Estimate the reliabilities of node $4$ from the previous network with each of the other nodes in the network. (Note, you can ignore the fourth column for this exercise)
\item Calculate the length of the shortest path between node $2$ and node $4$ in the previous network using fuzzy logic, assuming that the third column reports the probability of the edge to have weight $1$ and the fourth column reports the probability of the edge to have weight $2$.
\end{enumerate}

\chapter{Network Sampling}\label{cha:sampling}
\begin{figure}
\centering
\begin{subfigure}[t]{.45\columnwidth}
\includegraphics[width=\textwidth]{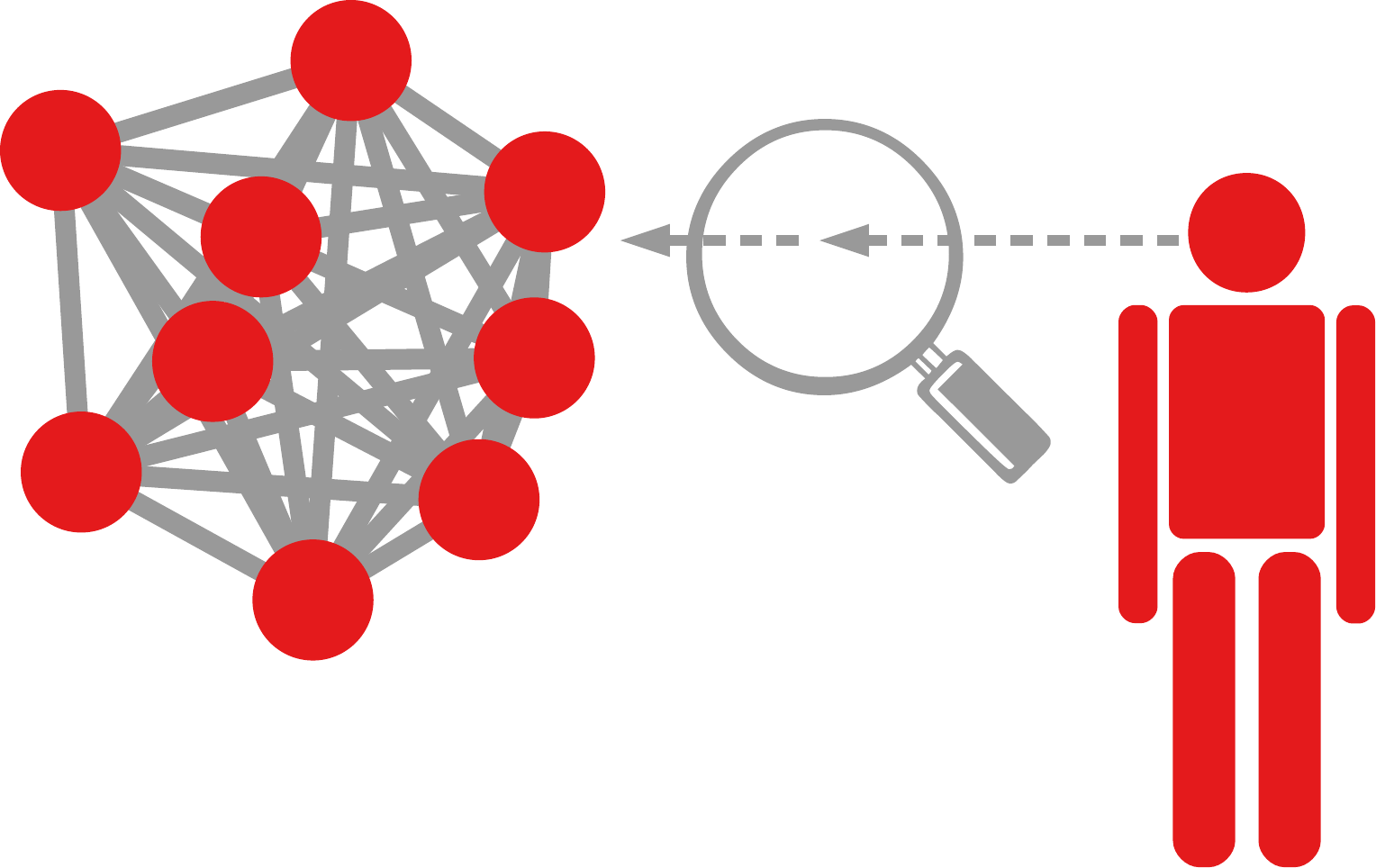}
\caption{}
\end{subfigure}
\qquad
\begin{subfigure}[t]{.45\columnwidth}
\includegraphics[width=\textwidth]{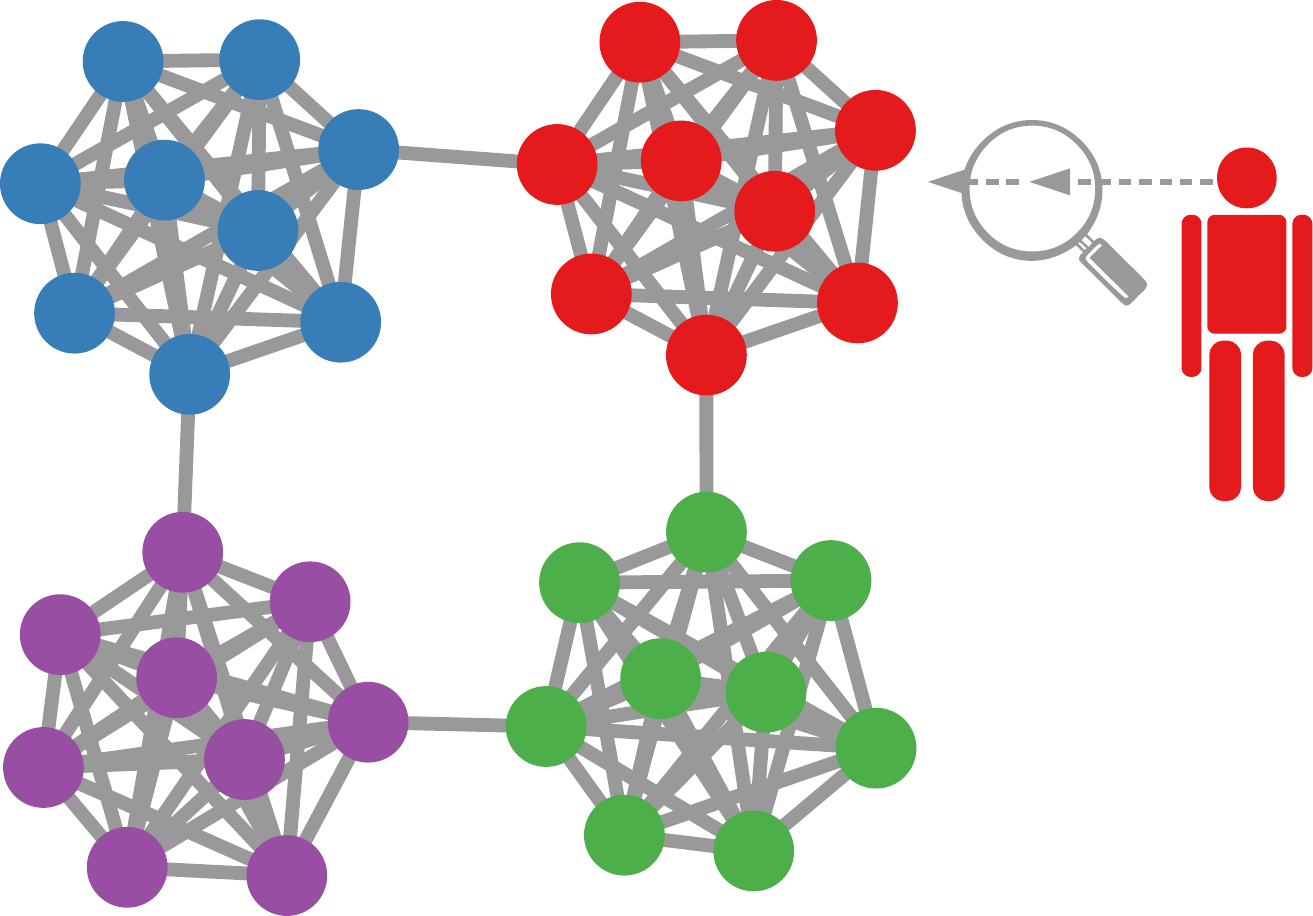}
\caption{}
\end{subfigure}
\caption{A representation of the sampling conundrum. (a) The sample you're able to observe looks like a hairball, with everything connected with everything else. (b) Zooming out to the whole structure shows a different story, with a clear community structure we could not observe due to the improper sample.}
\label{fig:sampling-issue}
\end{figure}

Sometimes, having a good edge induction or network backboning technique still doesn't help you. Sometimes you're observing a network directly and it's just a hairball. In these cases, it's useful to make a step back and consider that the act of observation in itself is not neutral. We decide what to focus on, whether we do it because of our interests, or simply because of data availability. If we could zoom out, we would see the structure, as Figure \ref{fig:sampling-issue} shows. When you're unable or unwilling to look at the entire network you have to perform network sampling.

``Network sampling'' means to extract from your network a smaller version of it. This smaller version, the sample, should be a representative subset of the data. By ``representative'' we mean that the property you're interested in studying should be more or less the same in the sample as in the network at large. For instance, if you're interested in estimating the clustering coefficient, extracting the only triangle from a large network which otherwise has none wouldn't be a good sampling. The sample's clustering is one, while the network at large has a clustering approaching to zero. Put it in other words, a proper network sampling will ensure that the tiny sliver you observe is carrying the properties of the whole structure you're interested in. Sampling can be extremely important also in new emerging scenarios, as it is fundamental for some types of graph neural networks to be able to deal with gigantic graphs (Section \ref{sec:mining-deep2-practical}).

\begin{figure}
\centering
\includegraphics[width=.75\columnwidth]{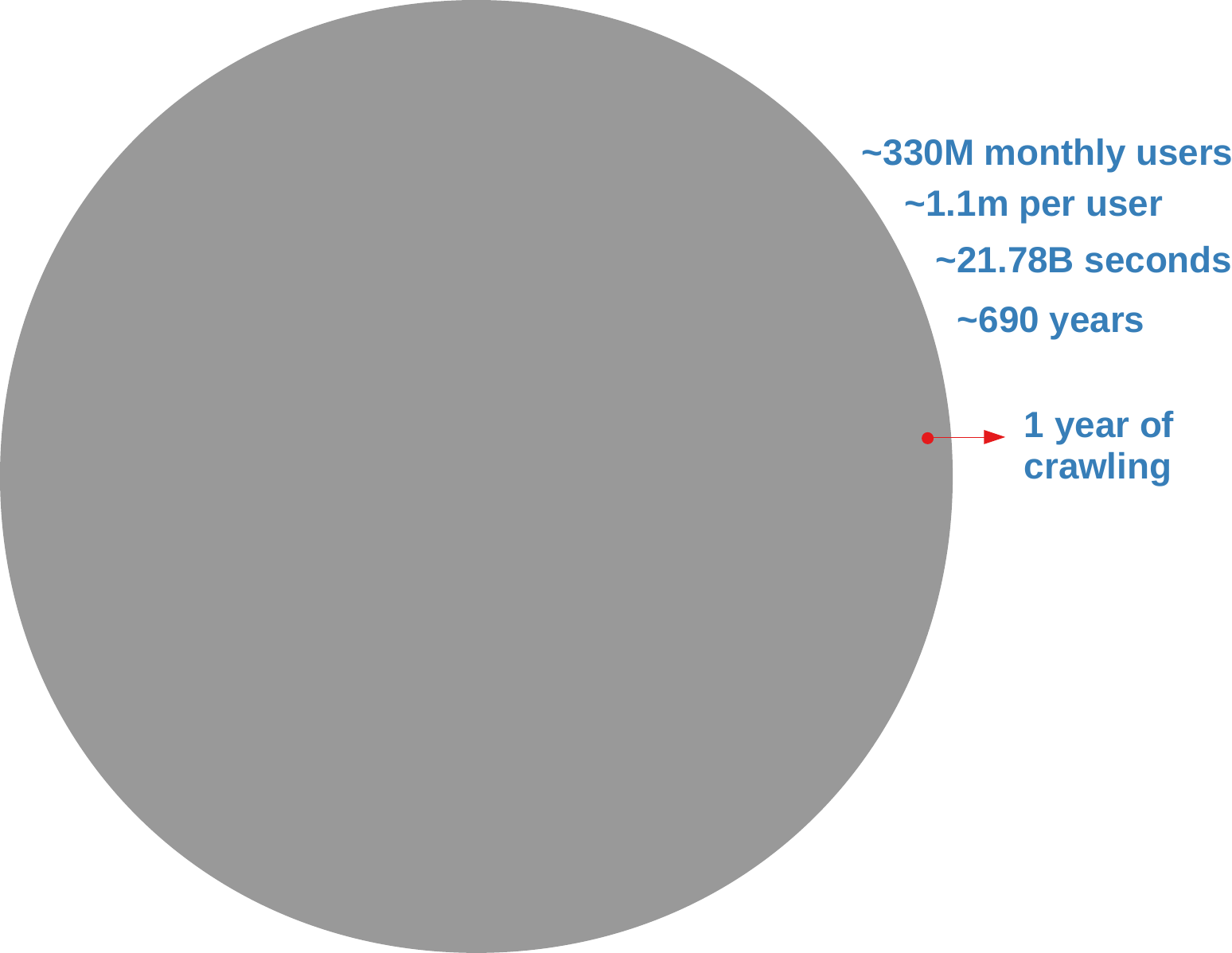}
\caption{The gray circle represents the set of users in Twitter. Given the platform's API constraints, a non-stop one-year crawl of the Twitter network would yield the set of nodes encompassed by the red circle.}
\label{fig:sampling-issue2}
\end{figure}

To put in perspective how bad the problem is, consider Twitter. As of writing this paragraph, Twitter has more than $300$ million active users. According to its API, it takes a bit more than a minute on average to fully know the connections of a user. This means that it takes more than $20$ billion seconds to crawl the entirety of Twitter, or just a bit less than $700$ years. If you were to crawl constantly for one year, you'd get a bit more than $0.1\%$ of Twitter. If you're a visual thinker, Figure \ref{fig:sampling-issue2} shows a depiction of the fact I just narrated. You can understand that what ends up in your $0.1\%$ has to be the best possible representation of the whole, and thus it has to be chosen carefully.

We already saw some ways to explore a graph: BFS and DFS (Section \ref{sec:shortpath-exploration}). They are reasonable ways to explore a graph, but their underlying assumption is that, eventually, they will cover the entire network. Here we focus on a slightly different perspective. We don't want the entire network: we want to prevent biases to creep into our sample. 

We can classify network sampling strategies -- in the broadest terms possible -- as induced and topological techniques. These are the focus of the next sections. What I'm writing is based on review works on network sampling\cite{gjoka2010walking}\cite{gjoka2011practical}\cite{dasgupta2012social}\cite{blagus2015empirical}. I'm going to mostly focus on the case in which the sampled network is stable, or it is evolving too slowly to make any significant difference during the sampling procedure. There are specialized methods to sample graphs when this assumption is not true. For instance streaming or evolving graphs, whose properties might significantly change as you explore them\cite{stutzbach2006sampling}\cite{rasti2008evaluating}\cite{ahmed2014network}.

Nowadays, you rarely want to sample a large graph that you fully own. We have enough computing and storing capabilities to process humongous structures. The case is different when you rely on an external data source. Most of the times, such data source will be a large social media platform. In this scenario, one has to apply double carefulness. API-based sampling is affected by fundamental issues. Works in the past have shown that one has to be careful when working with data sources that potentially yield non-representative samples of the phenomenon at large\cite{morstatter2013sample}\cite{morstatter2014biased}.

Note that you are not the only person in the world performing network sampling. In most cases, you're going to work with data that has already been sampled by somebody else and you have no control over how they extracted that sample from reality. This is true also if you're convinced that you are at the data source itself, for instance the API of the social media platform. But then you should ask yourself a few questions. Who has decided to use the platform? Who is active and who is present but inactive? What data does the provider make available? In such cases, you might need to carefully consider what you do with the data and/or decide to perform a network completion process (Section \ref{sec:sampling-completion}) -- if it is possible at all.

\section{Induced}
Induced sampling works with a guiding principle. You specify a set of elements that must be in your sample. Then, you collect all information that is connected to the elements you selected\cite{leskovec2006sampling}.

This is related, but not the same thing as, the concept of induced subgraph, a graph formed from a subset of the vertices of the graph and all of the edges connecting pairs of vertices in that subset -- see Section \ref{sec:bb-convex}. When performing an induced subgraph, you only focus on nodes, and you won't obtain new nodes from your induction procedure. When performing induced samples, instead, you usually want to add nodes to your sample as well, besides edges.

Differently from simply making an induced graph, you can do induced sampling in two ways: by focusing on nodes or by focusing on edges.

\subsection{Node Induced}
If you focus on nodes, it means that you are specifying the IDs of a set of nodes that must be in your sample. Then, usually, what you do is collecting all their immediate neighbors. The issue here is clearly deciding the best set of node IDs from which to start your sampling. There are a few alternatives you could consider.

The first, obvious, one is to choose your node IDs completely at random. Random sampling is a standard procedure in many other scenarios, and has its advantages. If the properties you're interested in studying are normally distributed in your population, a large enough random sample will be representative. However, when it comes to real world networks, such expectation might not be accurate. For two reasons.

\begin{figure}[t]
\centering
\begin{subfigure}[t]{.53\columnwidth}
\includegraphics[width=\textwidth]{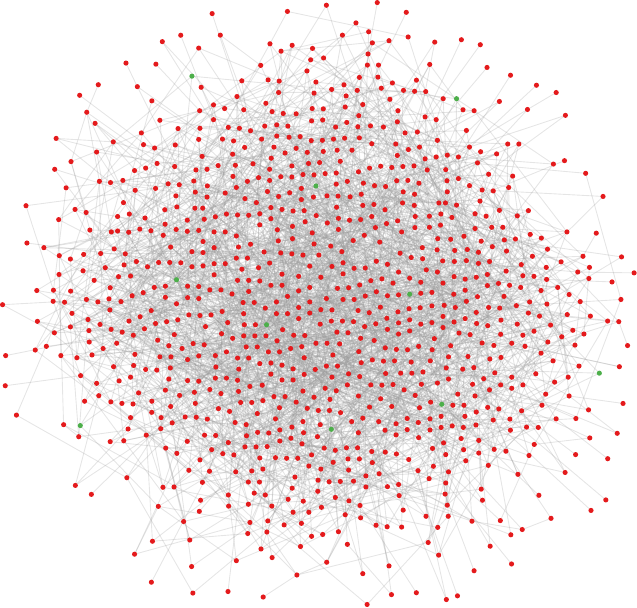}
\caption{}
\end{subfigure}
\qquad
\begin{subfigure}[t]{.37\columnwidth}
\includegraphics[width=\textwidth]{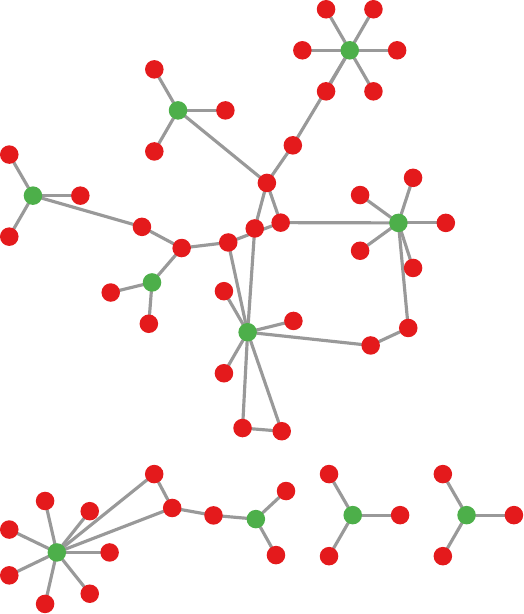}
\caption{}
\end{subfigure}
\caption{(a) A graph with a thousand nodes. I select uniformly at random $1\%$ of the nodes, in green. I then induce a graph with the selected nodes, all their neighbors, and all connections between them. (b) The resulting node-induced graph.}
\label{fig:sampling-induced-nodes}
\end{figure}

First, if your network is large -- and if your network isn't large why the heck are you sampling it? -- choosing node IDs at random might end up reconstructing a disconnected sample. The likelihood of two random nodes -- or their neighbors -- being connected is stupidly low. Figure \ref{fig:sampling-induced-nodes} shows an example of this issue. Even with a very generous $1\%$ random node sampling -- which, in the Twitter example I made earlier, would mean three million nodes! -- the resulting node-induced graph breaks down in multiple components. This might not be a problem but, usually, large social networks are connected. Thus ending up with a disconnected network, by definition, will mean that you don't have a representative sample.

Second, one of the properties most network scientists are interested in is the degree distribution. The degree distribution is emphatically not distributed normally in your population (Section \ref{sec:degree-pl}). Thus, a random node-induced sample is unlikely to fairly represent the hubs in your network.

Standard solutions for these two issues are simple. One can weight their samples. Nodes are more likely to be extracted and be part of the sample if they have a higher degree or PageRank. However, this requires knowing this information in advance, which is not feasible if you're crawling your network from an API system.

\subsection{Edge Induced}
Another way to generate induced samples is to focus on edges rather than nodes. This means selecting edges in a network and then crawl their immediate neighbors. There are a few techniques to do so. One is the obvious extension of random node induced sampling: random edge induced sampling. You select edges at random and you collect all their direct neighbors. Two more sophisticated approaches are Totally Induced Edges Samples (TIES)\cite{ahmed2011network} and Partially Induced ones (PIES)\cite{ahmed2012space}.

The idea behind edge sampling is that it counteracts the downward bias when it comes to the degree. In a network with a heavy-tailed degree distribution, most nodes have a low degree. Thus, if you pick one at random, it's overwhelmingly likely that it will be a low degree node. On the other hand, most edges are attached to large hubs. Thus, if you pick an edge at random, it is likely that a hub will be attached to it. This is a similar consideration of the vaccination strategy we saw in Section \ref{sec:triggers-intervention}.

There is an obvious downside to the edge sampling technique. You cannot easily use it when interfacing yourself with a social media API system. Very rarely such systems will allow you to start your exploration from a randomly selected edge. Thus, in one way or another, you're always going to perform some form of node-induced sampling.

\section{Topological Breadth First Search Variants}
The alternative to induced sampling is topological sampling. In topological sampling you also start from a random seed, but then you start exploring the graph. You're not limited to the immediate neighborhood of your seed as in the induced sampling, but you can get arbitrarily far from your starting point. That is why one of the key differences between induced sampling and topological sampling is the seed set size. In induced sampling you have to have the largest possible seed set, while in topological sampling you can start from a single seed and explore from there.

One of the key advantages of topological sampling is that it works well with API systems. There is also research showing that topological sampling is, in general, less biased than induced sampling\cite{lee2006statistical}. If used for sampling purposes, DFS and BFS graph exploration fall into this category.

There are fundamentally two families of topological sampling. The first is a modification of the BFS approach. The idea is to perform a BFS, but then adding a few rules to prevent some of the issues affecting that strategy. This is what we focus on in this section. The second big family is based on random walks and it will be the topic of the next section. Note that this division is largely arbitrary, as there is cross-pollination between these two categories, but it is a useful way to organize this chapter.

\subsection{Snowball}
\begin{figure}
\centering
\includegraphics[width=.75\columnwidth]{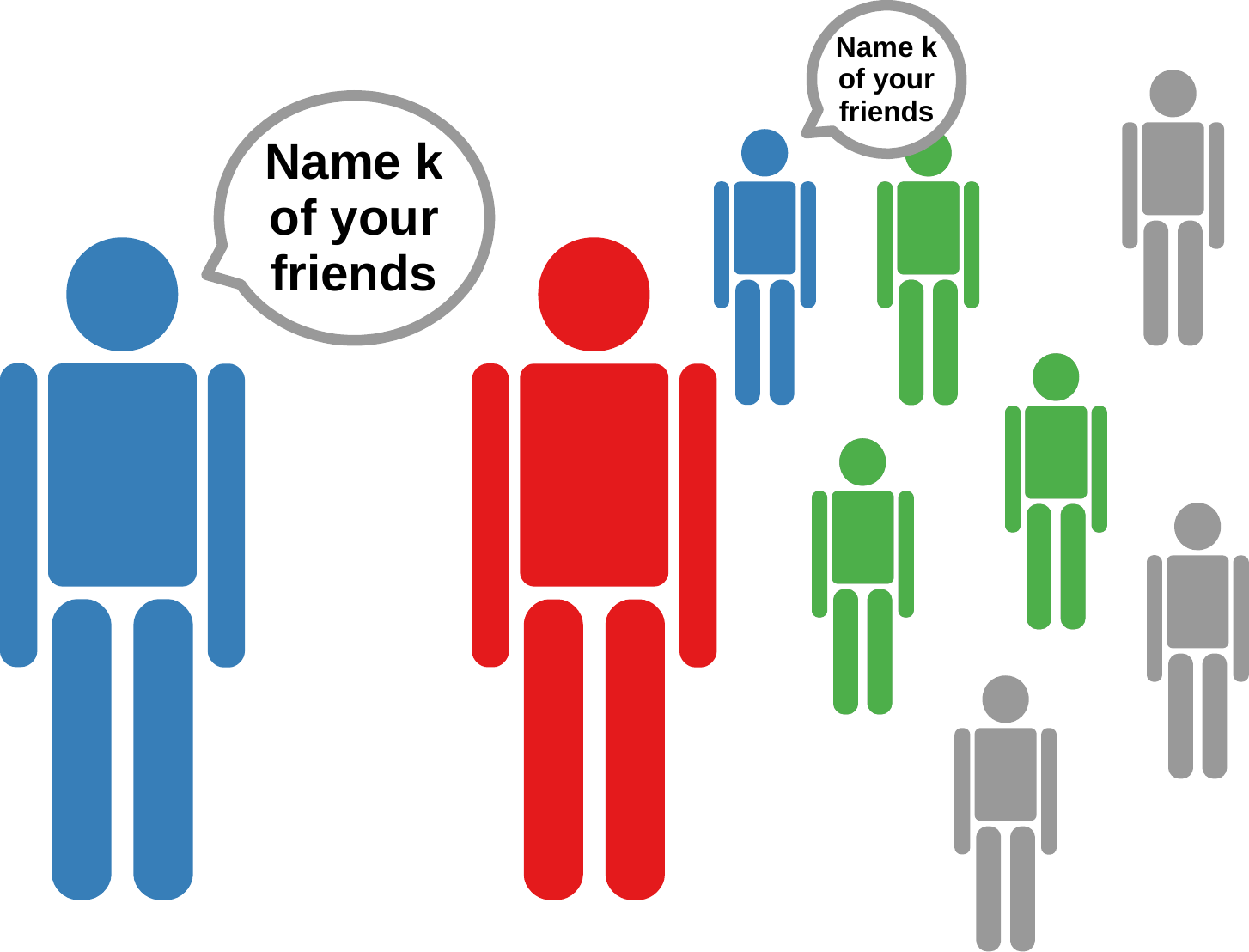}
\caption{Snowball sampling. Your sampler (blue) starts from a seed (red) and asks for $k = 3$ connections. Red names their green friends, but not the gray ones. The interviewer then recursively asks the same question to each of the newly sampled green individuals. If no one ever mentions the gray ones, those are not sampled and won't be part of the network.}
\label{fig:sampling-snowball}
\end{figure}

In \textbf{Snowball} sampling we start by taking an individual and asking her to reveal $k$ of her connections\cite[0.1in]{goodman1961snowball}\cite{biernacki1981snowball}. She might have more than $k$ friends, but we only take $k$. Then, we use these new individuals and we ask them the same question: to name $k$ friends. We do so with a BFS strategy. In practice, Snowball is BFS, but imposing a cap in the number of connections we collect at a time: $k$. Figure \ref{fig:sampling-snowball} depicts the process. Note that $k$ is not the maximum degree of the network, because a node might be mentioned by more than $k$ neighbors, if they have them.

Snowball has some advantages. It is cheap to perform in the real world, where the cost of identifying nodes is high, because the nodes identify themselves as a part of the survey process. This is less relevant for social media, where node discovery is relatively easy. Snowball has a smaller degree bias: with the  ``nominate-a-friend'' strategy we're likely to encounter hubs. However, their degree is somewhat capped, since they can only name $k$ of their friends, rather than the full list. This generates weird degree distributions with a sharp cutoff, which aren't very realistic. 

When it comes to sampling from social media, Snowball has a surprising advantage. It works well with pagination: in API systems, when you ask the connections of a node, you rarely get all of them. Social media \textit{paginate} results, so you only get $k$ connections at a time. With Snowball you can easily decide the maximum number of pages you want.

\subsection{Forest Fire}
In \textbf{Forest Fire}, like in Snowball, the base exploration is a BFS. However, once we get all neighbors of a node, we do not explore them all. Instead, for each of them, we flip a coin and we explore the node only with probability $p$. The advantage of Forest Fire is usually linked with a proper estimation of the clustering coefficient of the network, since with a BFS we would overestimate it -- because we fully explore the neighborhood of nodes\cite[-0.3in]{leskovec2006sampling}.

\begin{figure}
\centering
\includegraphics[width=.75\columnwidth]{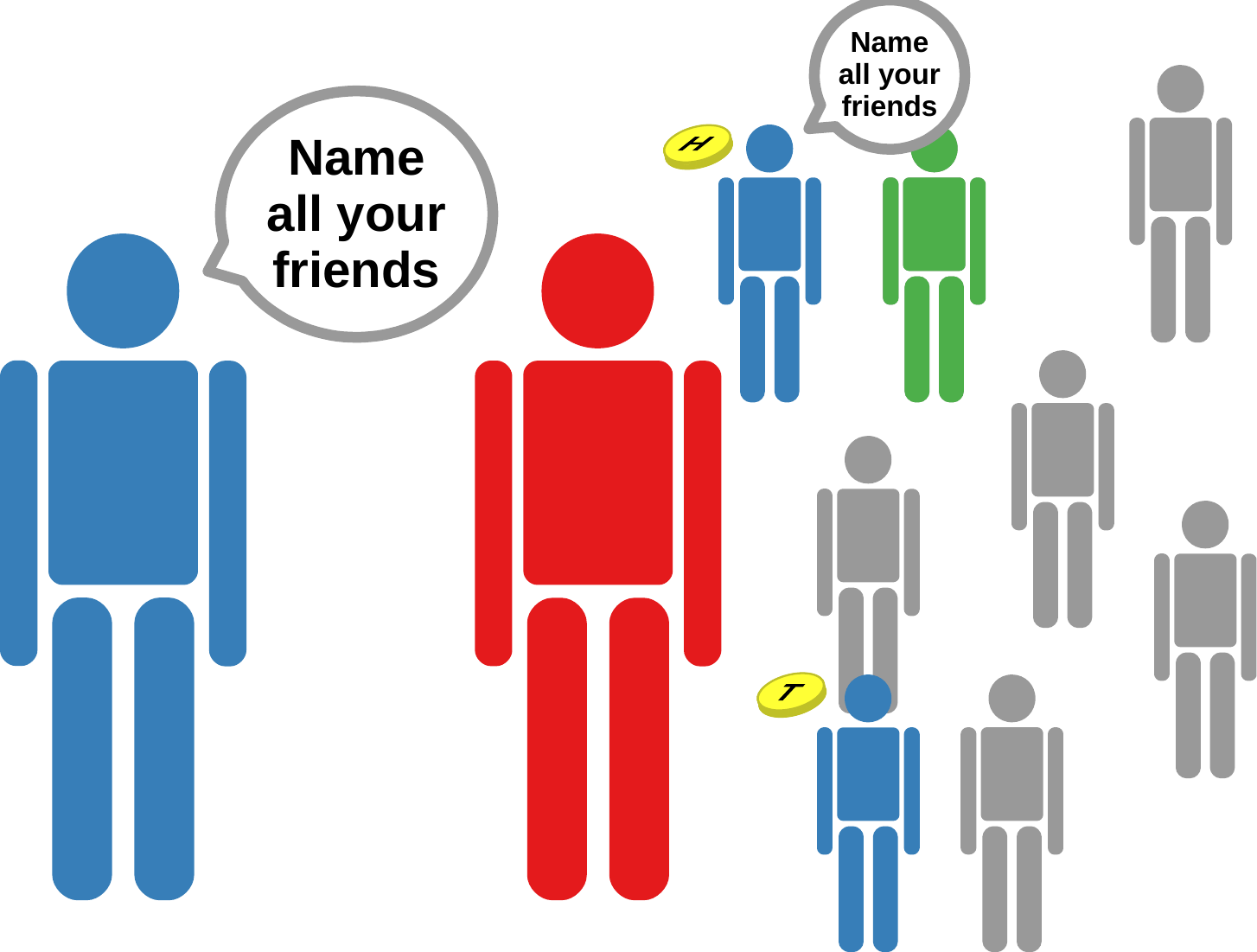}
\caption{Forest fire sampling. Your sampler (blue) starts from a seed (red) and asks for all the connections a node. If the probability test succeeds, the neighbor turns green and is also explored. If it fails, the neighbor remains gray and is not explored further.}
\label{fig:sampling-ff}
\end{figure}

Figure \ref{fig:sampling-ff} provides an example. After sampling a node and getting all its neighbors, we continue the BFS exploration. But, before sampling the neighbors of a neighbor, we flip a coin. If the test fails, we skip the neighbor and we go to the next one. Usually, one won't try to visit again the neighbors that have been skipped.

Forest Fire has an interesting relationship with your sampling budget. Usually, you're in a scenario in which you have a limited amount of resources to gather your network -- normally, the time it takes to perform the crawl. Assuming your network is sufficiently large, all sampling methods seen so far will eventually use up all your budget. However, if you set $p$ sufficiently low, you might end up in a situation where your Forest Fire crawl ends before you used up your budget. In this case you have to decide whether you want to stop your crawl and forgo the rest of your budget, or you're allowed to re-visit skipped nodes. The decision should be made depending on what's most important to keep: if the sample's topological properties are paramount, you cannot re-visit skipped nodes and you will have to make peace with having wasted part of your budget.

\section{Random Walk}\label{sec:sampling-rw}
The random walk sampling family does exactly what you would expect it to do given its name: it performs a random walk on the graph, sampling the nodes it encounters. After all, if random walks are so powerful and we can use them for ranking nodes (Section \ref{sec:centr-eigen}) or projecting bipartite networks (Section \ref{sec:projections-ycn}), why can't we use them for sampling too? I'll start by explaining the simplest approach and its problems, moving into sophisticated variants that address its downsides.

\subsection{Vanilla}
\begin{figure}
\centering
\includegraphics[width=.75\columnwidth]{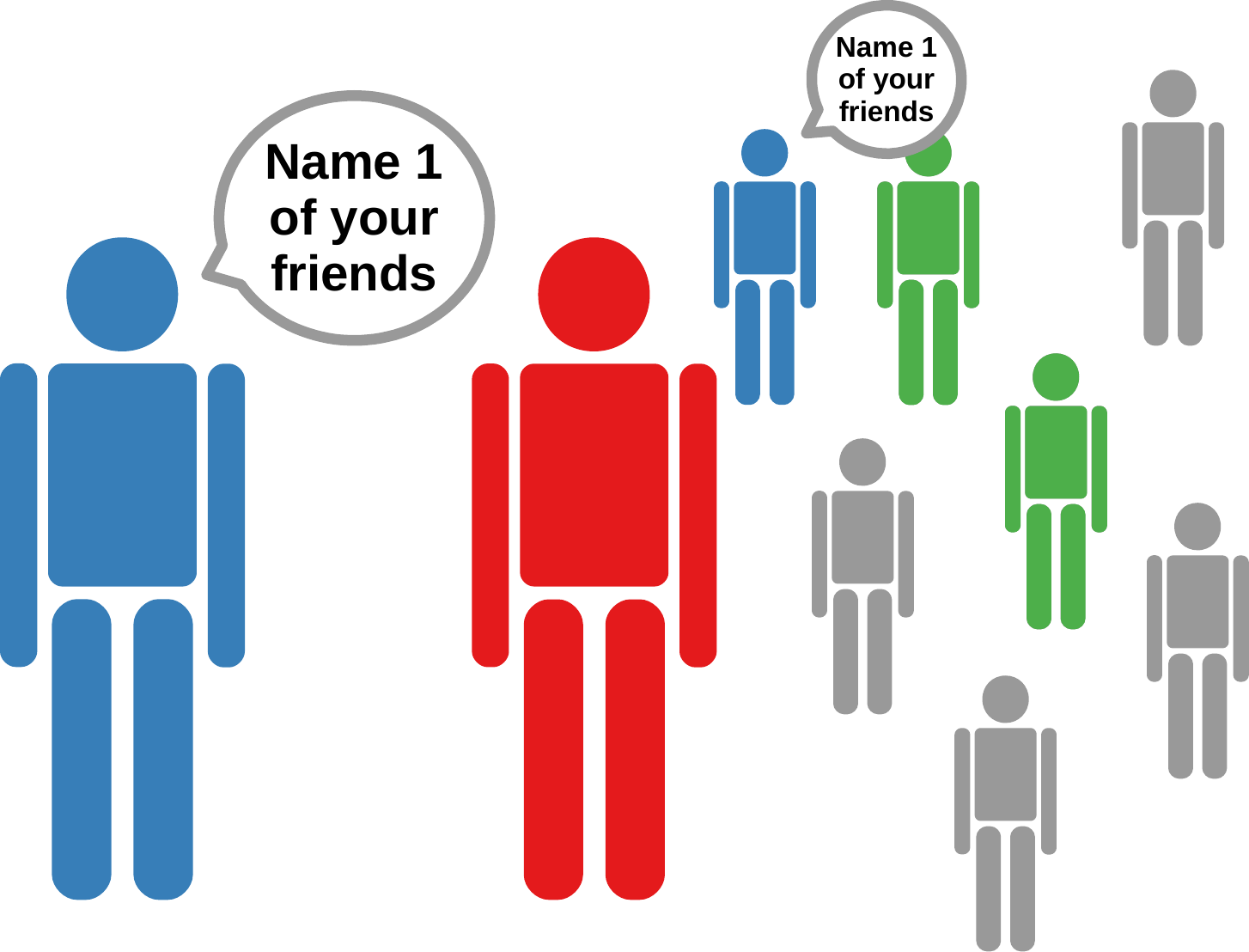}
\caption{Random walk sampling. Your sampler (blue) starts from a seed (red) and asks for all the connections a node (green + gray). One of the neighbors is picked at random and becomes the new seed (green) and, when asked, will name another green node to become the new seed.}
\label{fig:sampling-rw}
\end{figure}

In \textbf{Random Walk} (RW) sampling, we take an individual and we ask them to name one of their friends at random. Then we do the same with her and so on. Figure \ref{fig:sampling-rw} shows the usual vignette applied to this strategy.

This is an easy approach which can be very effective, but it has problems. First, you might end up trapped in an area of the network where you already explored all nodes, thus unable to find new ones. This can be easily solved by allowing a random teleportation probability, just like PageRank does to avoid being stuck in a connected component of the network.

More importantly, RW sampling has a degree bias. Remember the stationary distribution (Section \ref{sec:rw-stationary}): the probability of ending in a node with a random walk is known and constant no matter where we started. And the stationary distribution has a 1-to-1 correspondence to the degree. This means that high degree nodes are very likely to be sampled, while low degree nodes not so much. Thus, with RW, your sample is not representative -- at least when it comes to representing nodes with all degrees fairly.

Note that not all biases are entirely bad, some are useful\cite{tsugawa2020benefits}. Specifically, we could compare this upward degree bias with the downward degree bias of node induced sampling. Arguably, if we have to be biased, at least let's oversample the important nodes in the network, rather than the unimportant ones. This philosophy is implemented by the Sample Edge Counts (SEC) method\cite{maiya2011benefits}. SEC ranks the neighbors of all the sampled nodes according to their degree and then explores the neighbor with the highest edge count towards already explored nodes.

In this vein, one could avoid the limit of performing a single random walk at a time. A simple extension of RW sampling is m-dependent Random Walk (MRW)\cite{ribeiro2010estimating}. This involves performing $m$ random walks at once. The random walkers are not independent: we choose which of the $m$ random walker will take the next step by looking at the degree of the nodes they are currently visiting. Thus, if there are three random walkers and they are currently on nodes with degrees $3$, $2$, and $1$, we will continue from the first random walker with $3/(3 + 2 + 1) = 0.5$ probability.

\subsection{Metropolis-Hastings}
One way in which we could fix the issues of random walk sampling is by performing a ``random'' walk. Meaning that we still pick a neighbor at random to grow our sample, but we become picky about whether we really want to sample this new node or not. 

In the \textbf{Metropolis-Hastings Random Walk} (MHRW), when we select a neighbor of the currently visited node, we do not accept it with probability $1$. Instead, we look at its degree. If its degree is higher than the one of the node we are visiting, we have a chance of rejecting this neighbor and trying a different one. This probability is the old node's degree over the new node's degree. The exact formula for this decision is $p = k_v / k_u$, assuming that we visited $v$ and we're considering $u$ as a potential next step\cite{stutzbach2009unbiased}\cite{krishnamurthy2008few}.

\begin{figure}
\centering
\includegraphics[width=.75\columnwidth]{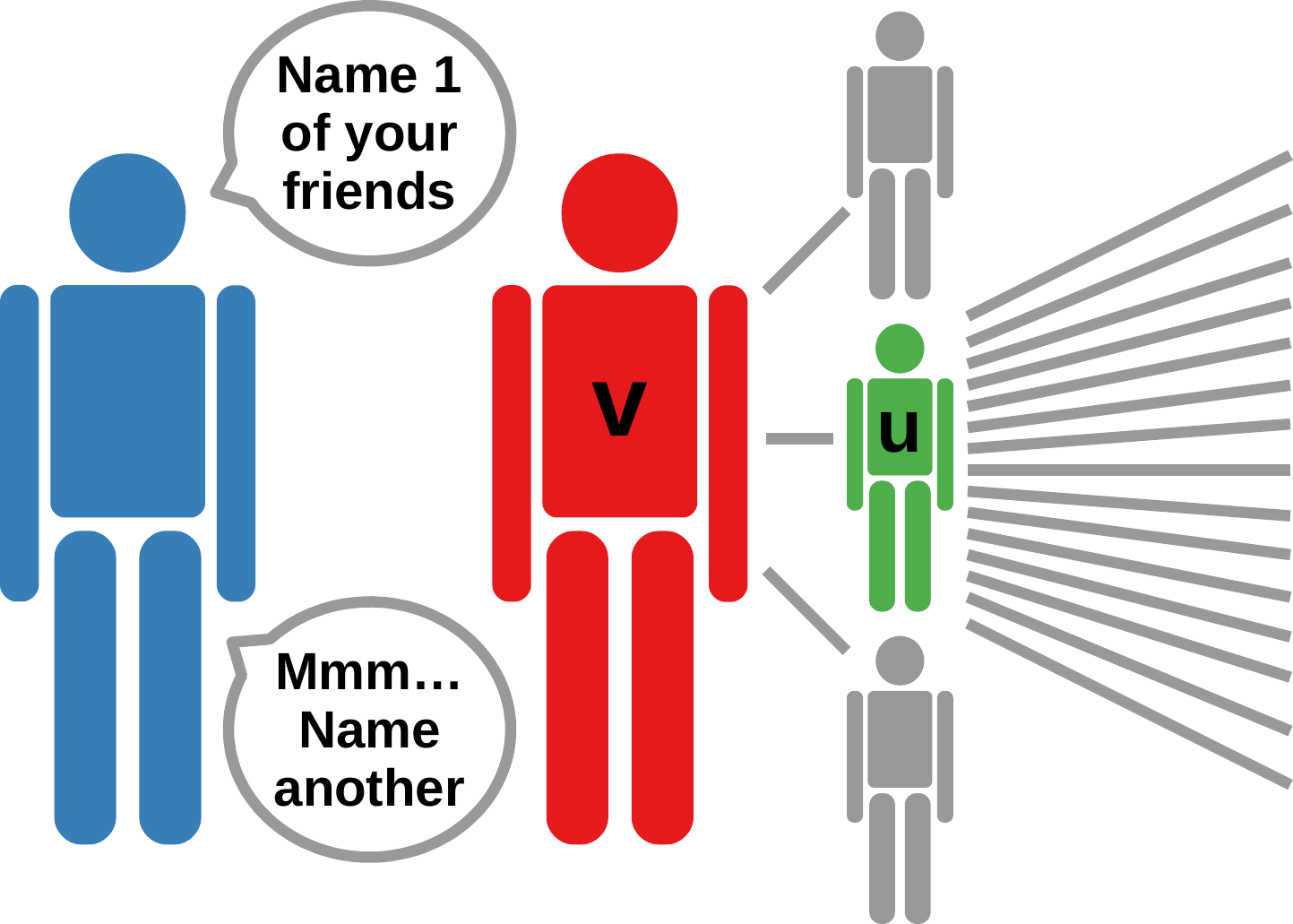}
\caption{Metropolis-Hastings Random Walk sampling. Your sampler (blue) starts from a seed (red) and asks for all the connections a node (green + gray). One of the neighbors is picked at random and we attempt to make it the new seed (green). However, since $u$ has so many connections, it is likely that the sampler will ask for a different neighbor.}
\label{fig:sampling-mhrw}
\end{figure}

If the current node $v$ has degree of $3$, and its $u$ neighbor has degree of $100$, the probability of transitioning to $u$ is only $3\%$ -- note that this is \textit{after} we selected $u$ as the next step of the random walk, thus the visit probability is actually lower than $3\%$: first you have a $1/k_v$ probability of being selected and \textit{then} a $k_v / k_u$ probability of being accepted. If we were, instead, to transition from $u$ to $v$, we would always accept the move, because $100/3 > 1$, thus the test always succeeds. In practice, we might refuse to visit a neighbor if its degree is higher than the currently visited node. The higher this difference, the less likely we're going to visit it. A random walk with this rule will generate a uniform stationary distribution. Figure \ref{fig:sampling-mhrw} shows the mental process of our Metropolis-Hastings sampler.

\subsection{Re-Weighted}
In Re-Weighted Random Walk (RWRW) we take a different approach. We don't modify the way the random walk is performed. We extract the sample using a vanilla random walk. What we modify is the way we look at it. Once we're done exploring the network, we correct the result for the property of interest\cite{salganik2004sampling}\cite{rasti2009respondent}. Say we are interested in the degree. We want to know the probability of a node to have degree equal to $i$. We correct the observation with the following formula: 

$$ p_{i} = \dfrac{ \sum \limits_{v \in V_{i}} i^{-1}}{\sum \limits_{v' \in V} x_{v'}^{-1}}.$$

\begin{figure}
\centering
\includegraphics[width=.75\columnwidth]{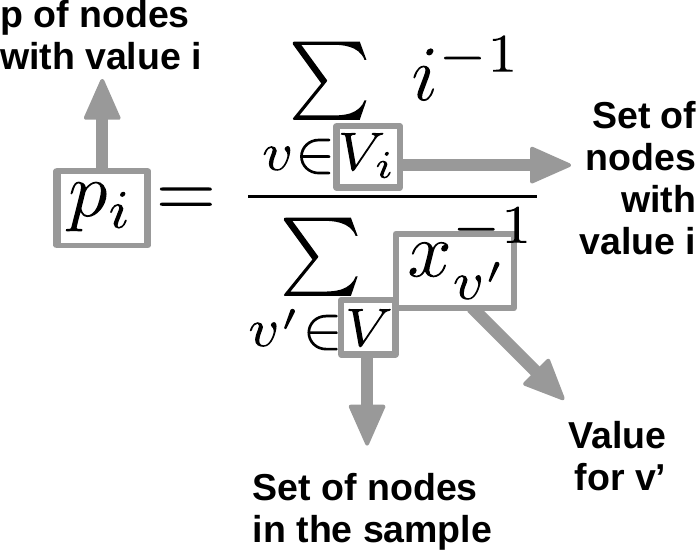}
\caption{The Re-Weighted Random Walk formula, estimating the probability $p_i$ of observing the $i$ value in a property of interest, using the set of sampled nodes $V_i$ with that particular value in the total set of $v$ sampled nodes.}
\label{fig:sampling-rwrw}
\end{figure}

The formula tells us the probability of a node to have degree equal to $i$ ($p_i$). This is the sum of $i^{-1}$ -- the inverse of the value -- for all nodes in the sample with degree $i$ ($V_i$), over $1 /$ degree ($x_{v'}^{-1}$) of all nodes in the sample ($V$). This is also known as Respondent-Driven Survey\cite{bernard2013social}, because it is used in sociology to correct for biases in the sample when the properties of interest are rare and non-randomly distributed throughout the population. Figure \ref{fig:sampling-rwrw} attempts to break down all parts of the formula.

Let's make an example. Suppose you want to estimate the probability of a node to have degree $i = 2$. First, you perform your vanilla random walk sample. Say you extracted $100$ nodes. Twenty of those nodes have degree equal to two. So your numerator in the formula will be the sum of $i^{-1} = 1/2$ for $|V_{i}| = 20$ times: $20 * 1/2 = 10$. If we assume that there were $50$ nodes of degree $1$, $10$ of degree $3$, $8$ of degree $4$, $7$ of degree $5$, and $5$ of degree $6$, our denominator would be:

$$(50/1) + (20/2) + (10/3) + (8/4) + (7/5) + (5/6),$$

which is $67.5\bar{6}$. Hopefully, you can spot what I did there. To bring the formula together, we discover that $p_2 = 10 / 67.5\bar{6} \sim 0.148$. So RWRW is telling us that the overall probability of a node having degree equal to $2$ is not $20\%$ as we would have inferred from the -- biased -- random walk sample. It is actually lower, it is $14.8\%$.

Note that the formula reported here only works when the variable of interest is discrete, i.e. $i$ is an integer, like in the case of the degree. You can still apply RWRW sampling even if the variable you want to study is continuous, for instance the local clustering coefficient. However, you'll have to perform a kernel density estimate\cite{rosenblatt1956remarks}\cite{parzen1962estimation}\footnote{\url{https://en.wikipedia.org/wiki/Kernel_density_estimation}}, and I'm not particularly keen of going into that nest of vipers. You're on your own, have a blast.

RWRW works particularly well when there are hidden populations who might actively try not to be sampled\cite{heckathorn2017network}. For instance, it has been successfully applied to the sampling of drug users\cite{bell2017comparison}.

RWRW has a crucial downside. While it is excellent to estimate the distribution of a property in a network, it will still return a biased vanilla random walk sample. So, if what you need was the sample rather than the estimation of a simple measure, you're out of luck. You cannot use this method to have a representative sample.

\subsection{Neighbor Reservoir Sampling}
Neighbor Reservoir Sampling\cite{lu2012sampling} (NRS) is one of those methods blending between the two families of sampling I talked about. It happens in two phases. In the first phase, NRS builds its set of core nodes and connections. Starting from the seed we provide as an input, NRS performs a normal random walk, including in the sample all nodes and edges it finds during this exploration.

However, the majority of NRS's budget is spent in the second phase. Once we have a core, we start modifying it. Suppose that, after the first phase, you sampled nodes in a set $V'$. In this second phase, you make a loop. At each iteration $i$ of the loop, you pick two nodes at random: $u$ and $v$. Node $v$ is a member of $V'$, the set of explored nodes. Node $u$ is not a member of $V'$ -- meaning that you haven't explored it yet, but it is a neighbor of a node in $V'$.

Our objective is to add $u$ to $V'$ and to remove $v$ from $V'$ at the same time. We can do it only if two conditions are met. First condition: we want our sample to be a single connected component. We cannot remove $v$ if that would break the graph into multiple components -- adding $u$ isn't going to add new components, because we only consider $u$s that are connected to a node in $V'$ ensuring connectivity. Note that $u$ and $v$ usually are not connected to each other.

The second condition is a random test. We extract a uniform random number $0 < \alpha < 1$. We perform the switch if and only if $\alpha < |V'| / i$, where $i$ is set to be equal to $|V'|$ at the beginning and it is increased by one at each attempt. In practice, this has a few consequences. By swapping $u$ and $v$, we ensure that the size of our sample stays constant, i.e. $|V'|$ doesn't change. Moreover, at the beginning, since $i \sim |V'|$ all initial attempts to modify $V'$ succeed. As we progress, the chances of accepting a new node in the set vanish.

This isn't really a random walk nor a BFS, because the random neighbor selected can be from any node in $V'$. So you can see how hard it is to fit it into a neat category.

NRS ensures a realistic clustering coefficient distribution. How can that happen? The trick lies in the connectivity test. We only perform the $u$-$v$ swap if it doesn't break the network into distinct connected components. Which means that not all $v$s have the same probability of being removed from the sample. The $v$ with higher clustering have higher probability to be replaced, because by removing them it is more likely that the graph will stay connected. High clustering coefficient means that their neighbors are connected to each other (see Section \ref{sec:density-clustering}), thus making it more likely there are alternative paths to keep the network together.

\begin{figure}[t]
\centering
\includegraphics[width=.75\columnwidth]{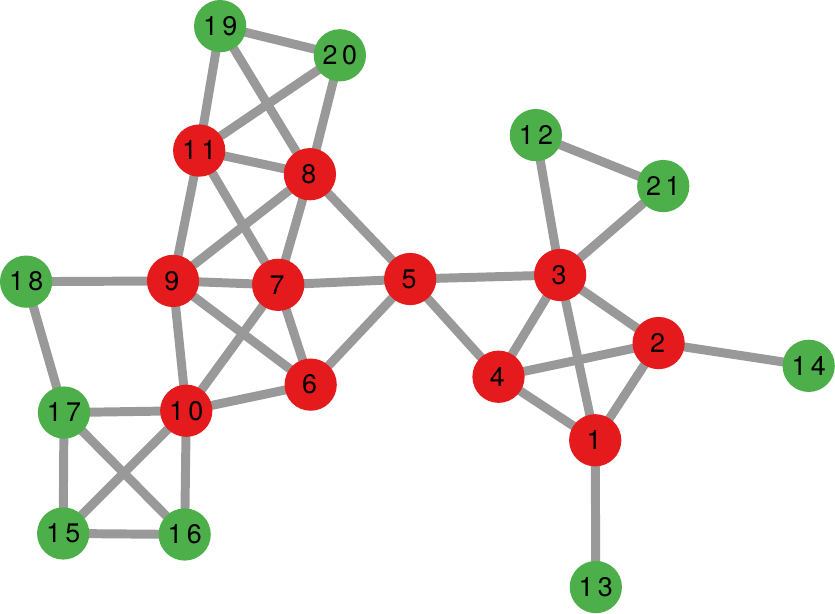}
\caption{Neighbor Reservoir sampling. Nodes in the explored set $V'$ are in red. Neighbors of $V'$ -- the reservoir -- are in green.}
\label{fig:sampling-nrs}
\end{figure}

To see why it's the case, consider Figure \ref{fig:sampling-nrs}. NRS will pick a node in green and a node in red at random. It will then remove the red node and add the green node. However, it will always refuse to perform the operation if the red node you pick is node $5$. Removing that node will create two connected components, which is unacceptable. Other unlucky $u$-$v$ draws are forbidden too. For instance, you cannot perform the swap if you pick nodes $3$ and $12$.

\begin{figure*}[t]
\centering
\begin{subfigure}{.32\textwidth}
\includegraphics[width=\textwidth]{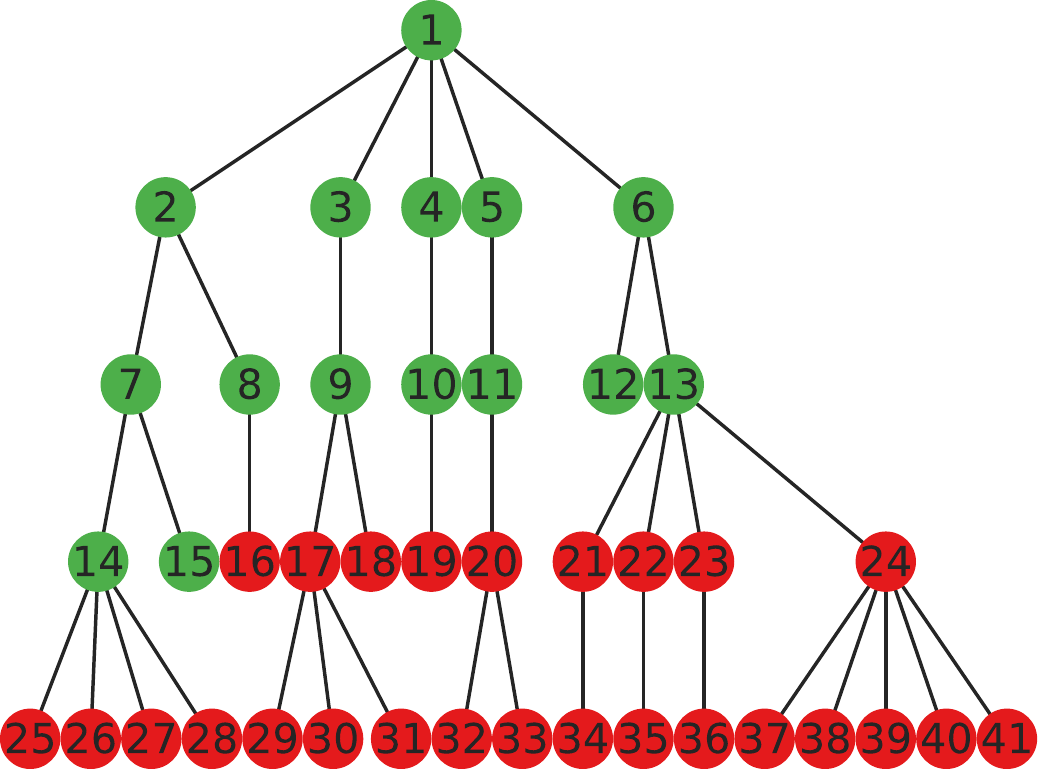}
\caption{BFS}
\end{subfigure}\hfill
\begin{subfigure}{.32\textwidth}
\includegraphics[width=\textwidth]{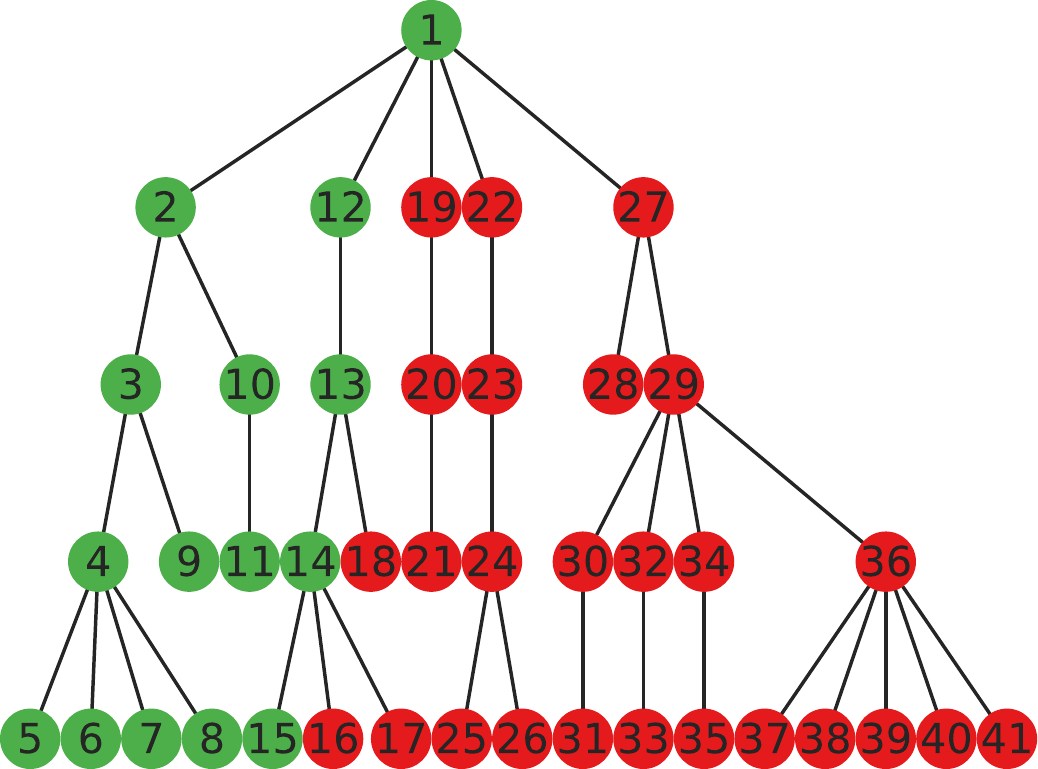}
\caption{DFS}
\end{subfigure}\hfill
\begin{subfigure}{.32\textwidth}
\includegraphics[width=\textwidth]{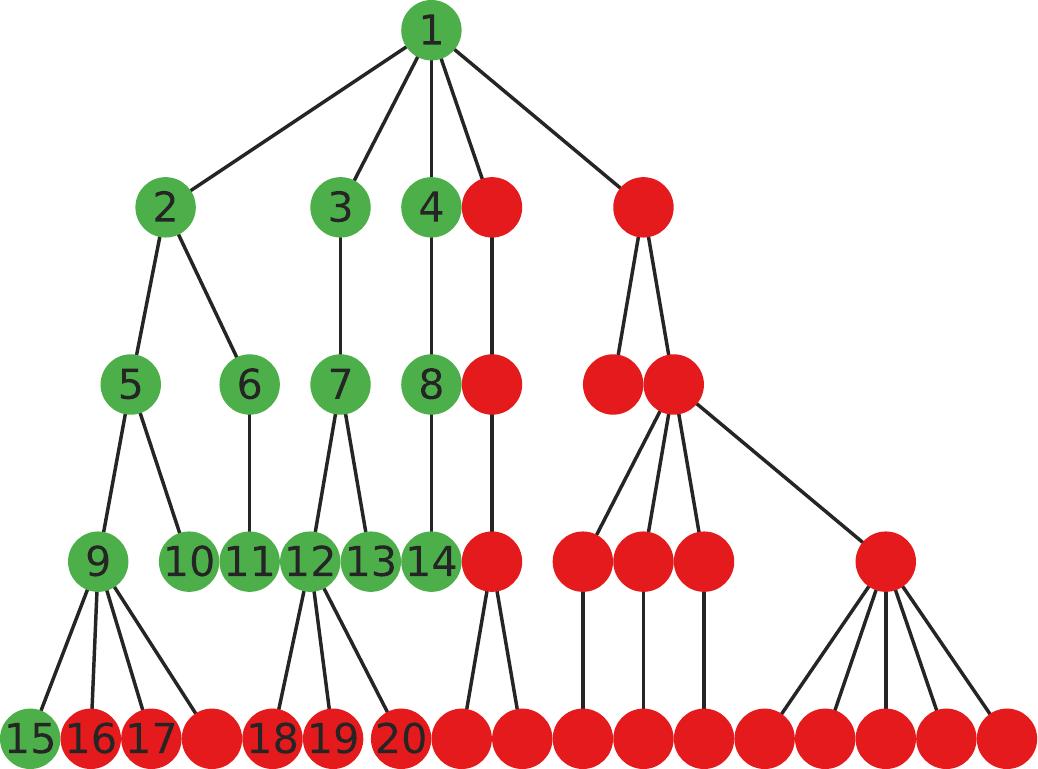}
\caption{Snowball}
\end{subfigure}\hfill
\begin{subfigure}{.32\textwidth}
\includegraphics[width=\textwidth]{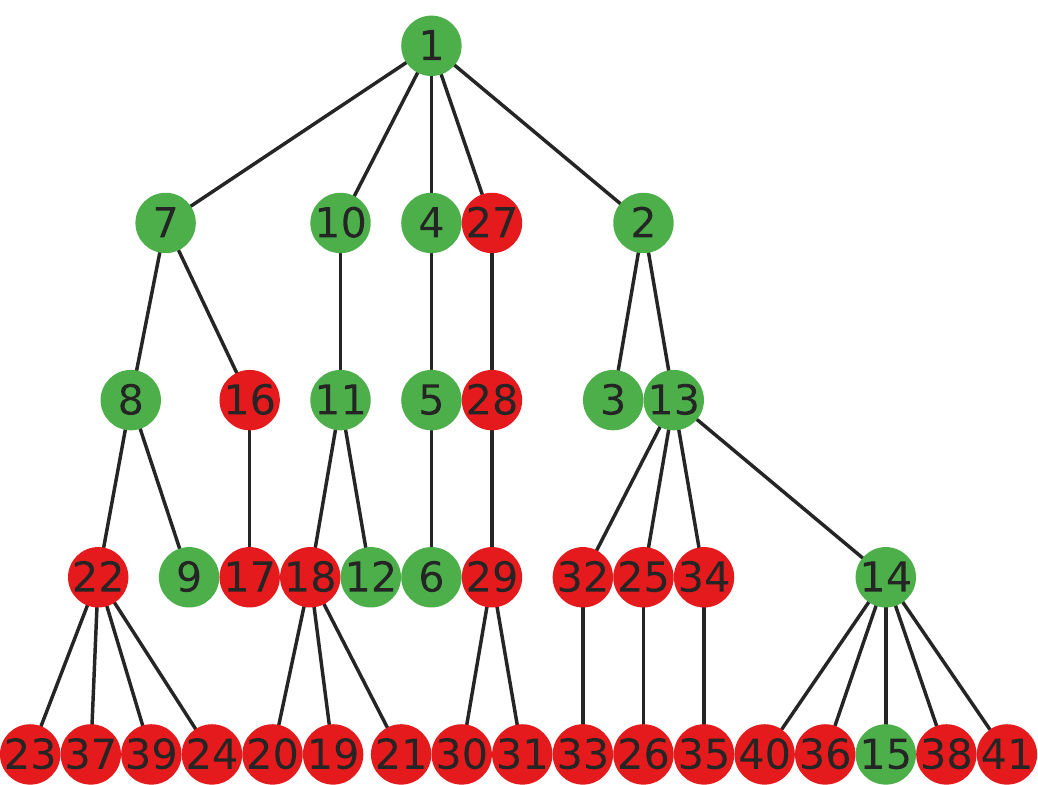}
\caption{Random Walk}
\end{subfigure}\hfill
\begin{subfigure}{.32\textwidth}
\includegraphics[width=\textwidth]{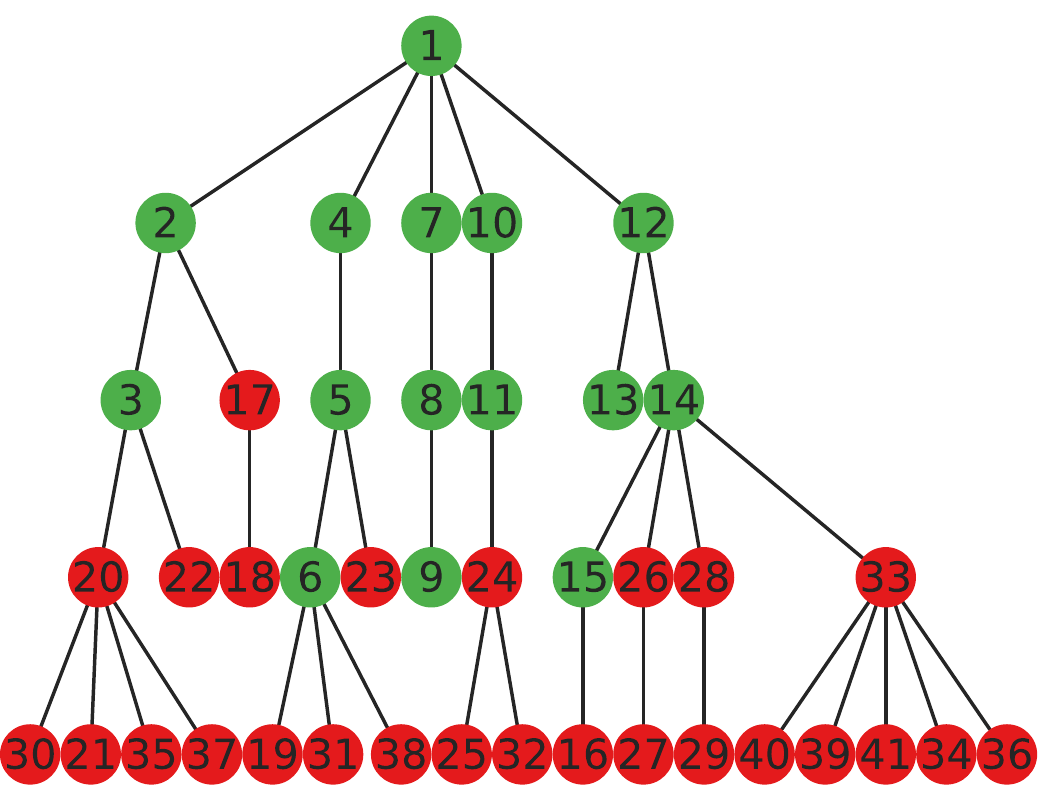}
\caption{MHRW}
\end{subfigure}\hfill
\begin{subfigure}{.32\textwidth}
\includegraphics[width=\textwidth]{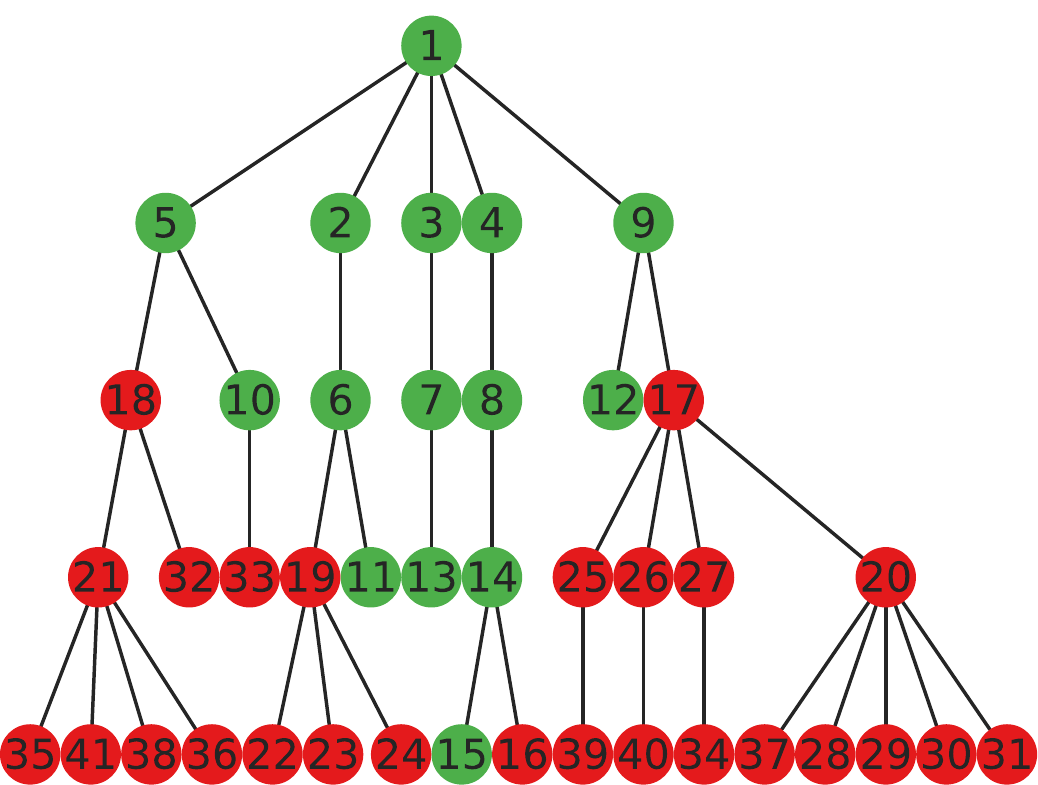}
\caption{Forest Fire}
\end{subfigure}\hfill
\caption{Examples of how different network sampling strategies explore a given network. Each node is labeled with the order in which it is explored. The node color shows whether the node was sampled (green) or not (red), assuming a budget of $15$ units and a constant cost of $1$ unit per node. Snowball assumes $n = 3$ (unlabeled nodes are not explored due to this parametric restriction), while Forest Fire has a burn probability of $.5$.}
\label{fig:alg-det-ex}
\end{figure*}

Figure \ref{fig:alg-det-ex} shows an example of how some of these different strategies would explore a simple tree. I don't show RWRW, because the samples it extracts are indistinguishable from the vanilla random walk ones. I also don't include NRS, because it's too subtle to really be appreciated in a figure like this one.

\section{Sampling Issues}
When talking about Snowball sampling I mentioned the issue of pagination. To recap: social media APIs will rarely give you all connections of a user when you ask for them. Rather, they will send you a list of $k$, chosen with some criterion that is opaque to you (likely in the order they are stored in their internal database). If you want the remaining ones, you have to ask again. You're always getting $k$ connections at a time. Each request is a ``page'', with $k$ being the page size.

It can be tricky to know how pagination will affect crawl time. Imagine two different API policies. The first returns big pages -- say $100$ edges per page -- but requires a long waiting time between queries -- say two seconds. The second policy returns small pages -- ten edges per page -- but more often -- you only need to wait one second between queries. We can calculate the edge throughput of these two policies. In this case, the one with big pages returns more edges per unit of time, on average: $50$ edges per second versus $10$ edges per second. However, how will these policies behave on a real world network?

\begin{figure}
\centering
\includegraphics[width=.6\columnwidth]{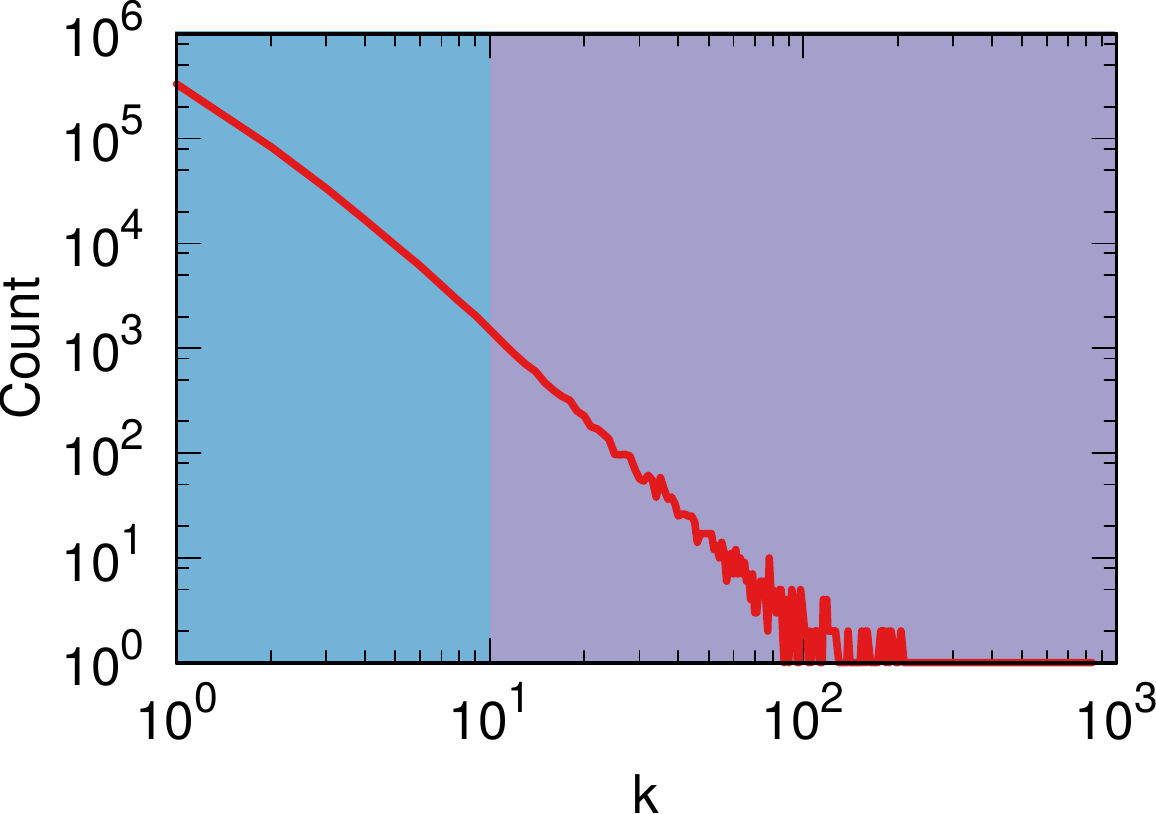}
\caption{A power law degree distribution, showing the count of nodes (y-axis) with a given degree (x-axis). The colors in the plot represent in which cases the first API policy described in the text is faster than the second (purple) and when the second is faster than the first (blue).}
\label{fig:pagination-paradox}
\end{figure}

As we saw in Section \ref{sec:degree-pl}, real world networks have broad degree distributions, like the one we show in Figure \ref{fig:pagination-paradox}. For some of these nodes, the second policy is better: if they have $10$ or less edges, we can fully explore them with a single query, thus we're going faster because we have lower waiting times between nodes. In Figure \ref{fig:pagination-paradox}, we color in blue the part of the degree distribution for which this holds true. If the node has a degree higher than $10$, then the first policy is better, because it requires to perform fewer queries, even if they are spaced out more in time. In Figure \ref{fig:pagination-paradox}, we color in purple the part of the degree distribution for which this holds true.

The second policy is in theory $5$ times slower than the first ($10$ edges/sec versus $50$ edges/sec, on paper), however it will allow you to crawl this network in half of the time\cite[1in]{coscia2018benchmarking}. This is because, in a broad degree distribution, we have way more nodes with low degree -- for which the second policy is faster. In Figure \ref{fig:pagination-paradox}, out of $500$k nodes, $492$k have degree of $10$ or less. 

You could conclude that the best API policy possible is the one that gives you only one node at a time, imposing no waiting time between requests. This is true only in theory. In practice there are a few things you need to consider, which you can lump into the issue of network latency. First, it still takes time for the information to travel from the server to your computer. This is not exactly the speed of light, so the requests will never be truly instantaneous. Second, a server which gets hit too frequently with too many requests will also naturally slow down, often in unpredictable ways. Thus some level of pagination and waiting time will always be part of an API system. Which means that there are going to be trade-offs when reconstructing the underlying network.

Pagination is often not the only thing you need to worry about. Other challenges might be sampling a network in presence of hostile behavior\cite{bortnikov2009brahms}. For instance, some hostile nodes will try to lie about their connections and it's your duty to reconstruct the true underlying structure. Or not: there are reasonable and legit reasons to lie about one's connection, for instance to protect one own privacy.

In another scenario, you might not be interested in the topological properties of the full network. What's interesting for you is just estimating the local properties of one -- or more -- nodes. In that case, specialized node-centric strategies can be used\cite{papagelis2013sampling}.

\section{Network Completion}\label{sec:sampling-completion}
Network completion is a related -- but not identical -- problem. Like sampling, it wants to establish a topological strategy for the exploration of a network. Different from sampling, its aim is not to extract a smaller version of the full dataset. Rather, we want to complete the sample. The idea is that you downloaded from somewhere a sample of a network, but you are able to process a larger dataset. Rather than starting collecting data from scratch, you can use what you have as a seed and try to complete it.

The question now is: what's the most efficient way to do so? What strategy would give you the most information about the full structure with the least amount of effort? Specifically, you want to obtain the highest possible number of new nodes by asking the lowest amount possible of new queries. You could simply apply any of the network sampling strategies I explained so far. However, there are dedicated techniques developed to solve this problem.

Note that here you don't know the strategy originally used to collect the sample you're given. If you knew that it was a Metropolis-Hastings random walk you'd probably use a different strategy than if it was a standard BFS. But, since you don't know this piece of information, you need a general strategy working regardless of the shape of the initial sample.

Naively, you might think to just go and probe the nodes with the highest degree. However, there are a few considerations to make. First, since -- by definition -- your sample is incomplete, you don't really know the true degree of a node. You only know how many neighbors it has in your sample. Second, since the node has a high degree in your sample, there's some chance you already explored all its neighbors, thus probing it won't help you. 

The first technique, MaxReach\cite{soundarajan2016maxreach}, estimates the true degree of a node and its clustering coefficient using the information gathered so far. It does so with a technique similar to Re-Weighted Random Walk. The difference is that, in RWRW, we only want to know how many nodes have a given degree $i$. In MaxReach, we want to also know which nodes have that given degree value. At this point, the score of a node is the difference between its estimated degree and its degree in the sample. Nodes with higher scores are probed earlier. After each probe, since we gathered more information in the sample, MaxReach will recalculate the degree estimates.

$\epsilon$-wgx\cite{soundarajan2017varepsilon} is a more recent alternative.

\section{Summary}

\begin{enumerate}
\item Network sampling is a necessary operation when the network you need to analyze is too large and/or you need to gather data one node/edge at a time from a high latency source (e.g. the API of a social media platform). Sometimes the decision is not up to you and all you can access is a sample made by somebody else.
\item The main objective is to extract a sample that is representative of the network at large for the property you're interested in studying. For instance, it has to have a comparable degree distribution if you want to infer its shape (e.g. whether it is a power law).
\item Sampling methods can be induced or topological. In induced sampling you extract a random sample of nodes/edges and you collect all that it is attached to it. In topological sampling you explore the structure one node at a time.
\item Variants of BFS explorations are: Snowball, in which we impose a maximum number $k$ of explored neighbors of a node; and Forest Fire, in which we have a probability of rejecting some edges.
\item Variants of random walk exploration are: Metropolis-Hastings, where we have a probability of refusing visiting high degree nodes; and Re-Weighted, where we correct the statistical properties of the sample after we collected it.
\item When sampling from real API systems one has to be careful that the throughput in edges per second is not necessarily a good indicator of how quickly you can gather a representative sample. Due to pagination, high-throughput sources might return smaller samples.
\item A related problem is network completion: given an incomplete sample of a network, find the best strategy to complete the sample in the least number of queries possible.
\end{enumerate}

\section{Exercises}

\begin{enumerate}
\item Perform a random walk sampling of the network at \url{http://www.networkatlas.eu/exercises/29/1/data.txt}. Sample 2,000 nodes (1\% of the network) and all their connections (note: the sample will end up having more than 2,000 nodes).
\item Compare the CCDF of your sample with the one of the original network by fitting a log-log regression and comparing the exponents. You can take multiple samples from different seeds to ensure the robustness of your result.
\item Modify the degree distribution of your sample using the Re-Weighted Random Walk technique. Is the estimation of the exponent of the CCDF more precise?
\item Modify your random walk sampler so that it applied the Metropolis-Hastings correction. Is the estimation of the exponent of the CCDF more precise? Is MHRW more or less precise than RWRW?
\end{enumerate}

\part{Mesoscale}\label{par:meso}

\chapter{Homophily}\label{cha:homophily}
``Mesoscale'' is the term we use to indicate network analyses that operate at the level that lies in between the global and the local one. At the global level, we have analyses that sum up the topological characteristics of a network with a single number. For instance, the exponent of the degree distribution, the global clustering coefficient, or the diameter. At the local level, we sum up individual node characteristics with a single number. They can be its degree, its local clustering coefficient, or closeness centrality.

At the mesoscale we want to describe groups of nodes. How do they relate to each other? How does their local neighborhood look like? There are many different meso-level analyses you can perform. This part of the book groups almost all of them together, leaving one out: community discovery. Community discovery is, by far, the most popular meso-level analysis of complex networks. Given its size, it deserves a part on its own, which will be the next one. Here, we're talking about all the meso-rest.

We start with homophily: the love (\textit{philia}) of the similar (\textit{homo}). ``Birds of a feather flock together'' is a popular way of saying. It originates from the fact that many species of birds flock with individuals of their own kind and coordinate when moving around. This phrase has been adopted in sociology to exemplify the concept of homophily: people will tend to associate with people with similar characteristics as their own. In a social network, homophily implies that nodes establish edges among them if they have similar attributes\cite{mcpherson2001birds}. If you have a particular taste in movies, and there are two potential friends, you are more likely to choose the one with similar tastes as yours, because you have more things to talk about and have less potential for conflict.

Many factors influence and favor homophily, and they are not necessarily exclusively explained by individual preference: sometimes homophily is a property emerging from access. It's not only the fact that you don't \textit{like} the different, but rather that you cannot \textit{access} the different. In other words, homophily is not only the result of our preferences, but also of social constructs. That is why the term ``homophily'' is problematic, and we use it only because of historic reasons.

\begin{figure}
\centering
\includegraphics[width=\columnwidth]{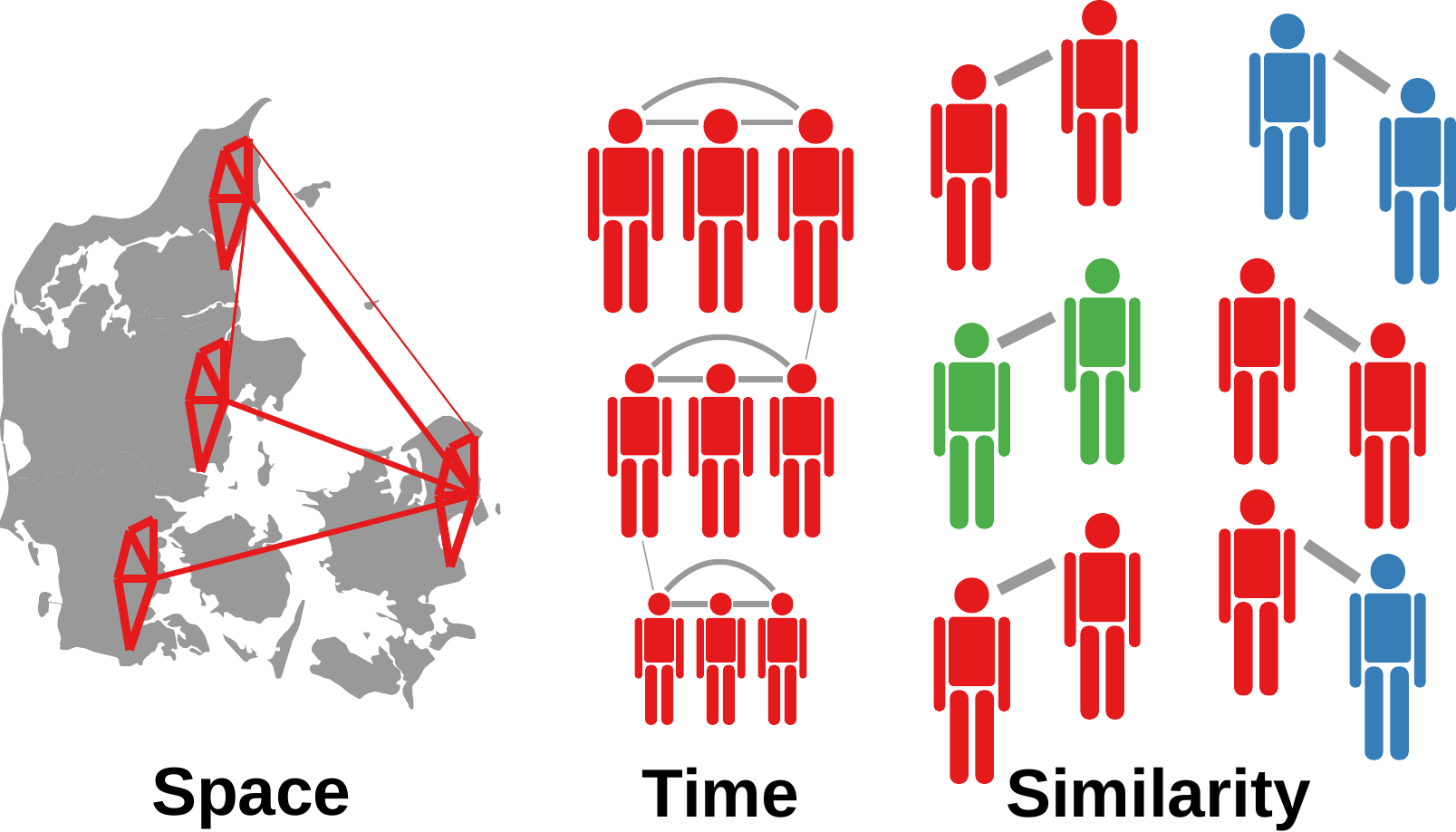}
\caption{Some examples of homophily driven by spatial, temporal, and attribute similarity.}
\label{fig:homophily}
\end{figure}

With that said, let me go through a carousel of examples of homophily, some of which I represent in Figure \ref{fig:homophily}. There are so many observed examples in real world social networks that I have to push their references down to the next page otherwise my Latex won't compile. So have a picture of my cat Ferris. He is, incidentally, a great example of homophily, in that he hates everything that is not himself.

\begin{figure}
\centering
\includegraphics[width=.75\columnwidth]{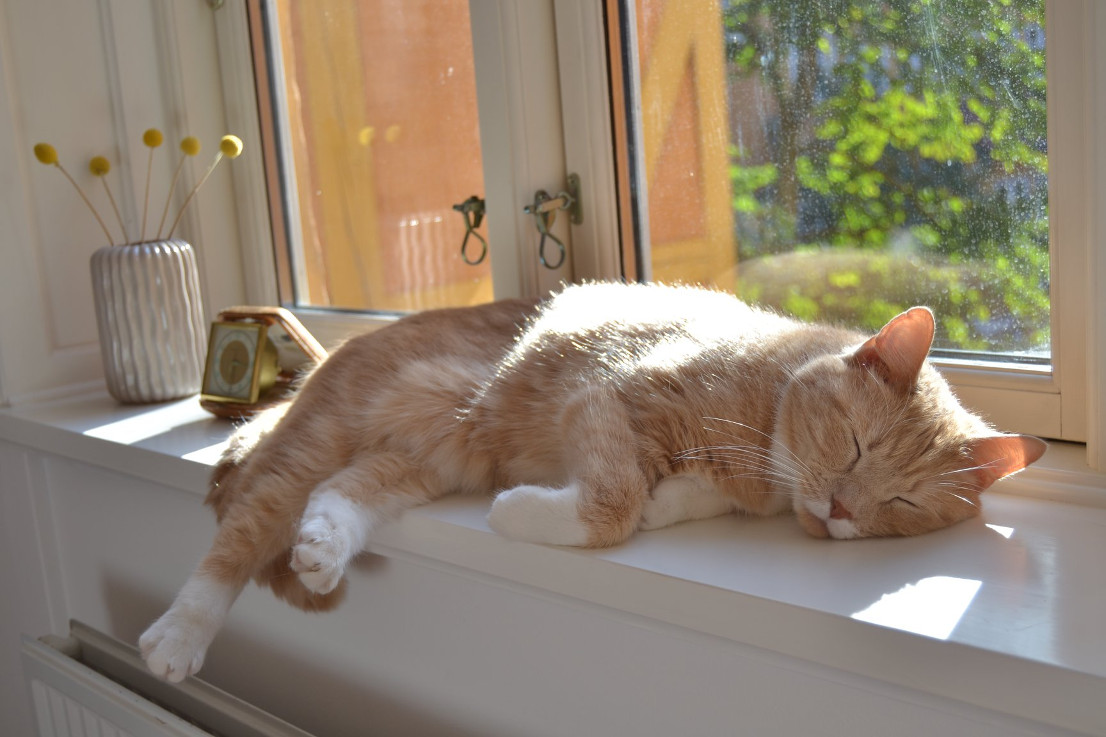}
\caption{My cat Ferris. In the picture, I color in orange the parts of the cat that are orange.}
\label{fig:ferris}
\end{figure}

First, gender\cite[-3.1in]{stehle2013gender} and race\cite{gonzalez2007community} are glaring examples. School children are more likely to make friends with people sharing their gender or race\cite{joyner2000school}. We observe this in adults too: in marriage ties it is so overwhelmingly likely to date\cite{mcclintock2010does} or marry\cite{raley2015growing} someone of the same race that sociologists don't study this fact any more because it's so boringly obvious. In this, we're truly similar to other animals we often look down to\cite{jiang2013assortative}. Rather than asking whether romantic ties show homophily, it's more interesting to use the degree of homophily of romantic ties to compare societies.

\begin{figure}
\centering
\includegraphics[width=.66\columnwidth]{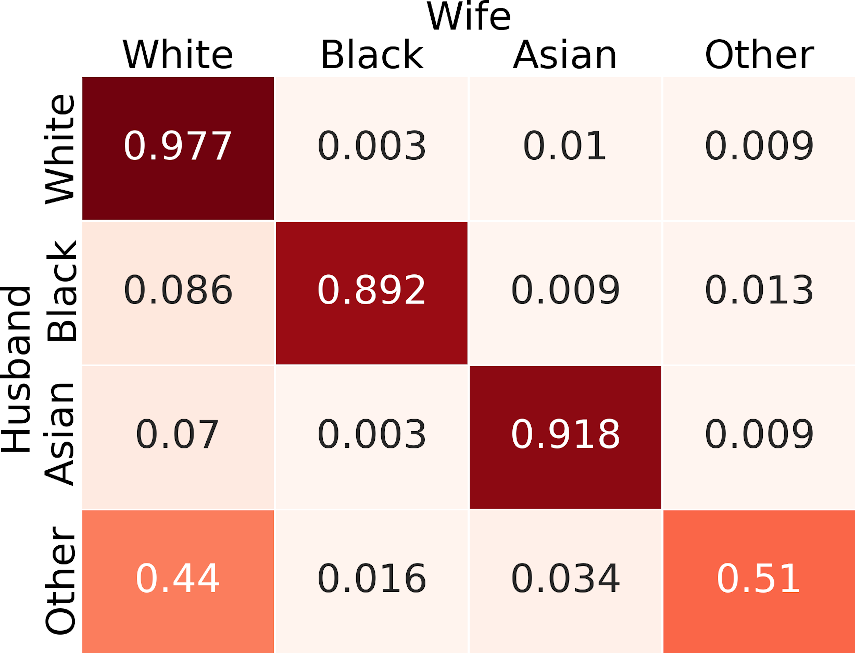}
\caption{The mixing matrix of interracial marriage in the US: share of husbands per race with a wife of a given race (Census Bureau).}
\label{fig:homophily-marriage}
\end{figure}

In Figure \ref{fig:homophily-marriage} you see an example of mixed marriage in the United States. To that diagonally dominated matrix, you have to add the consideration that the United States is probably one of the most diverse countries in the world. Imagine how this would look like elsewhere! 

\extrafloats{36}

Another example is spatial homophily: living in the same place makes it easier to have stronger connections, a factor that overcomes other correlates such as race\cite[-3.9in]{scellato2011socio}\cite[-3.2in]{sailer2012social}\cite{wong2006spatial}. A sub-type of spatial homophily is mobility homophily: going to the same places influences the likelihood of connecting socially\cite{wang2011human}\cite{toole2015coupling}. The reverse is also true -- as it might seem obvious --: being friends increases the likelihood to go to the same places\cite{cho2011friendship}. The connection between geographical space and social space is very strong, showing how, even in presence of (almost) limitless communication ranges, social ties still decay with distance\cite{onnela2011geographic}\cite{coscia2015evidence}\cite{deville2016scaling}.

Another factor of homophily is time, meaning that being in the same age range favors connections. Think about school friends: $38\%$ of a person's friends are within a 2-year age gap -- this figure comes from McPherson's paper.

In the rest of the chapter we are going to explore some techniques to study the mesoscale, such as the usage of ego networks. We're going to quantify homophily and see some of the consequences in network dynamics.

\section{Ego Networks}\label{sec:homophily-ego}
\begin{figure}
\centering
\begin{subfigure}{.425\textwidth}
\includegraphics[width=\textwidth]{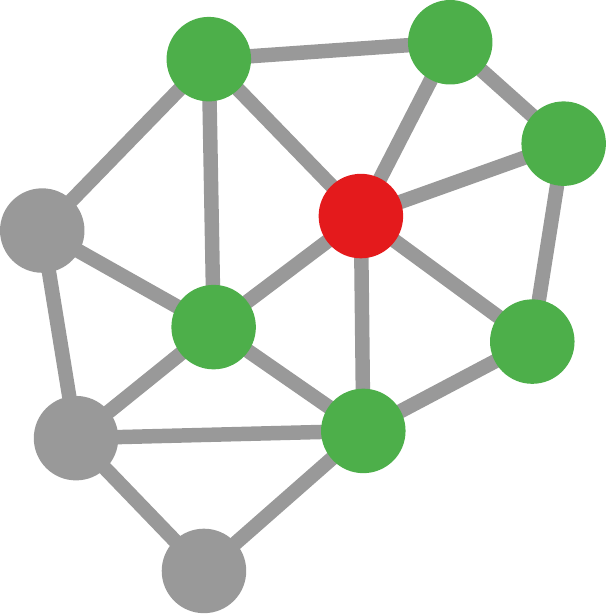}
\caption{}
\end{subfigure}\qquad
\begin{subfigure}{.425\textwidth}
\includegraphics[width=\textwidth]{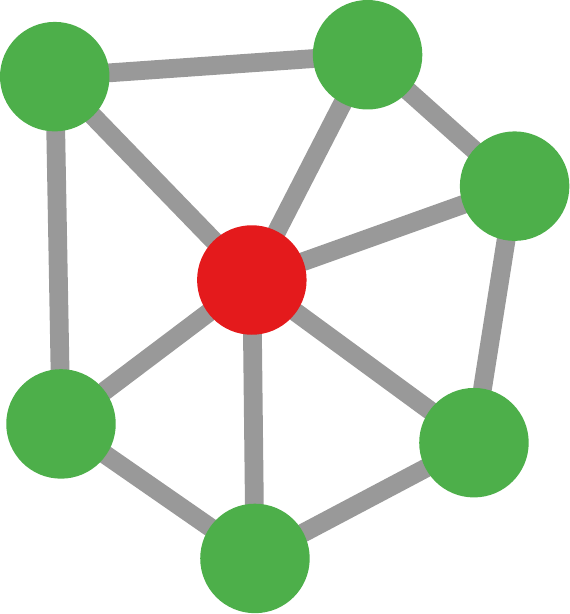}
\caption{}
\end{subfigure}
\caption{The procedure to extract an ego network from a larger network. (a) We select the ego (in red) and all its neighbors (in green). (b) We create a view only using red and green nodes, and all connections among them.}
\label{fig:ego}
\end{figure}

Ego networks are a common technique to explore the meso-level around a node\cite{crossley2015social}. ``Ego'' in Latin means ``I'', the self. An ego network is a subset of a larger network. You first have to identify your ``ego'': the node on which the ego network is centered. Then, you select all of its neighbors and the connections among them. The resulting network formed by all the nodes and edges you selected is an ego network. Figure \ref{fig:ego} provides an example of this procedure.

Once you have an ego network, you can start investigating its ``global'' properties such as the degree distribution or its homophily, and these are not properties of the global network as a whole, but of the local neighborhood of the ego, the ego network, which lives in the mesoscale. Ego networks are frequently used in social network analysis\cite{borgatti2009network}\cite{leskovec2012learning}, for instance to estimate a person's social capital\cite{borgatti1998network}.

A consequence of this procedure is that we know that an ego node is connected to all nodes in its ego network. This is unfortunate in some cases, depending on our analytic needs. For instance, all ego networks have a single connected component and will have a diameter of two. If those forced properties are undesirable, one can extract an ego network and then remove the ego and all its connections. 

\section{Quantifying Homophily}\label{sec:homophily-assortativity}
When it comes to homophily, we want to have an objective way to quantify how much it is driving a network's connections. This means that the nodes connect to other nodes depending on the value of one of their attributes. There are two possible scenarios. The attribute driving the connections could be quantitative (e.g. age) or qualitative (e.g. gender). When the attribute is quantitative, you can use the same technique to estimate the degree assortativity, which we cover in the next chapter.

Here we focus on the case of a qualitative attribute. Let's start by making a simple scenario: biological sex. In humans, this is -- barring rare and exceptional cases -- a categorical binary attribute.

\begin{figure}
\centering
\includegraphics[width=.75\columnwidth]{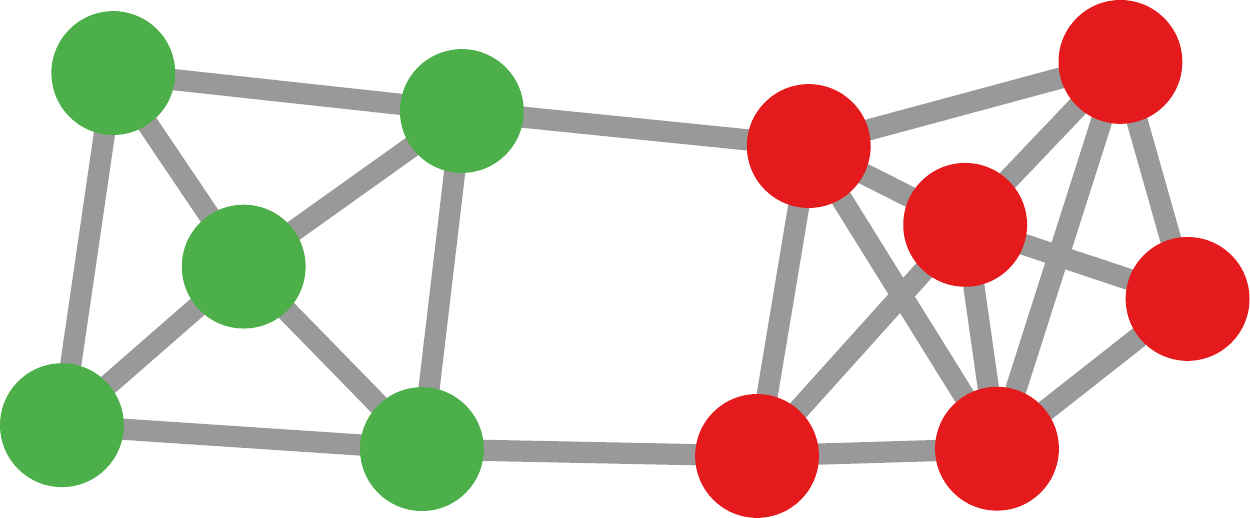}
\caption{A toy example to test our measures of homophily. We represent the categorical binary attribute with node color, with two possible values: red and green.}
\label{fig:homophily2}
\end{figure}

In this scenario, you can estimate the probability of an edge to connect alike nodes, and compare it to the probability of connection in the network. Consider Figure \ref{fig:homophily2}. We have $20$ edges connecting nodes with the same color over $22$ total edges in the graph. Therefore, the observed probability of edges between alike nodes is $20 / 22$. In the graph we have $11$ nodes, thus the number of possible edges is $|V|(|V| - 1) / 2 = 55$ (with $|V| = 11$). So the probability of having a connection between any node pair is $22 / 55$.  Thus we see that the probability of an edge being between alike nodes is more than twice what we would have expected: $(20 / 22) / (22 / 55) \sim 2.27$. Values higher than one imply homophily, while values lower than one mean that nodes tend to connect with similar nodes less than we expect -- i.e. the network is disassortative, nodes don't like to connect to similar nodes, another totally valid thing that can happen often in social networks (in Section \ref{sec:homophily-contagion} I call this concept ``heterophily'').

This approach breaks down if you have more than two possible values for your attribute, and also if some values are more popular than others. In these cases, you might conclude that there is assortativity in a non-assortative network, simply because you're assuming the incorrect null model of equal attribute value popularity.

In this case, you should use a different approach\cite{newman2002assortative}\cite{newman2003mixing}. You want to look at the probability of edges connecting alike nodes per attribute value $i$, and then compare it to the probability of an edge to have at least one node with attribute value $i$. The formula is:

$$ r = \dfrac{\sum \limits_{i} e_{ii} - \sum \limits_{i} a_{i}b_{i}}{1 - \sum \limits_{i} a_{i}b_{i}},$$

where $e_{ii}$ is the probability of an edge to connect two nodes which both have value $i$, $a_i$ is the probability that an edge has as origin a node with value $i$, and $b_i$ is the probability that an edge has as destination a node with value $i$. In an undirected network, the latter two are equal: $a_i = b_i$. This formula takes values between $-1$ (perfect disassortativity) and $1$ (perfect assortativity: each attribute is a separate component of the network).

\begin{figure}
\centering
\includegraphics[width=.8\columnwidth]{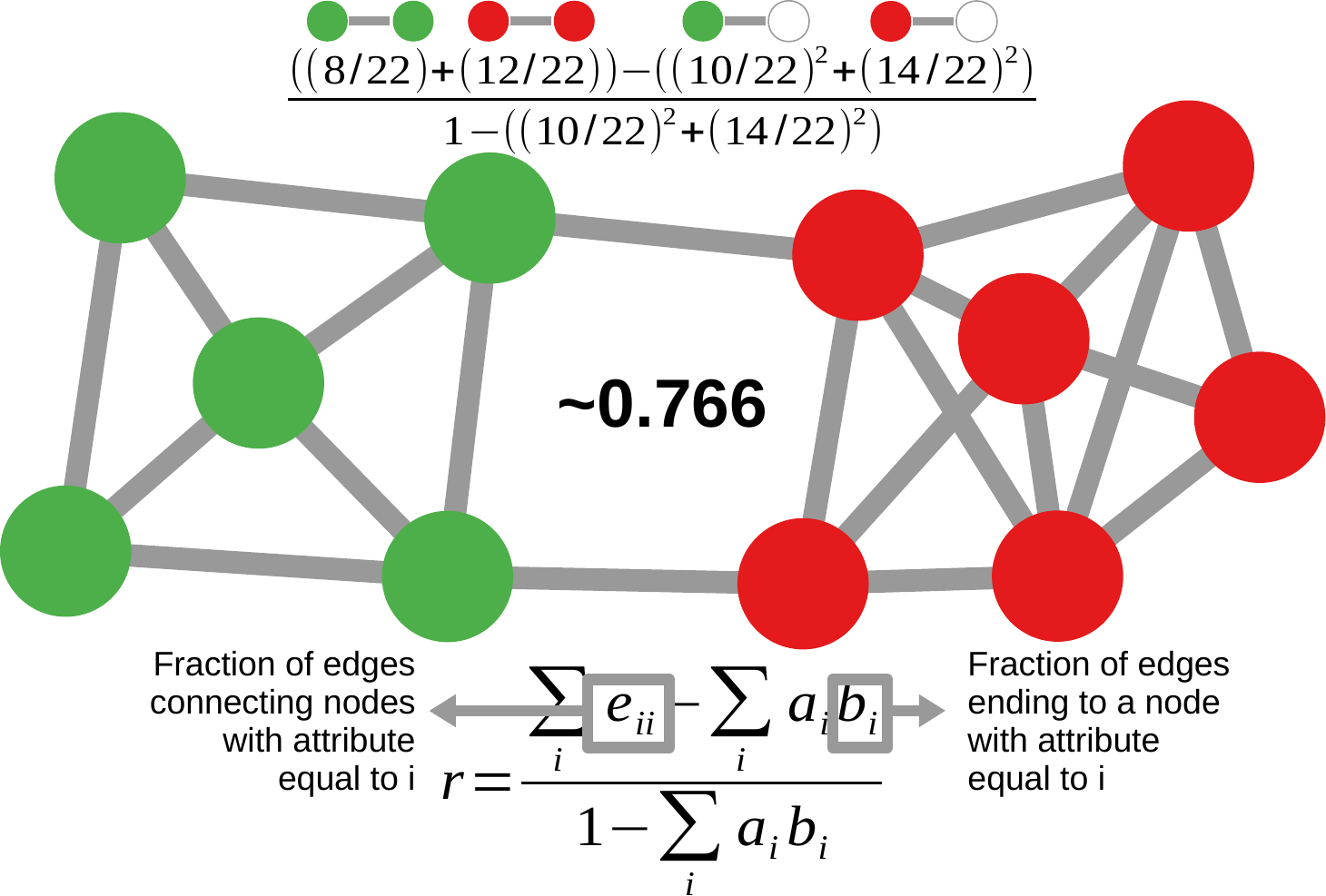}
\caption{How to calculate homophily using the formula in the text.}
\label{fig:homophily3}
\end{figure}

In Figure \ref{fig:homophily3} we have two values $i$: red and green. There are $22$ edges in the graph: eight green-green edges -- thus the probability is $8/22$ -- and $12$ red-red edges -- thus the corresponding $e_{ii}$ value is $12/22$. Ten edges originate (or end) in a green node: $a_i = b_i = 10 / 22$; and $14$ originate (or end) in a red node: $a_i = b_i = 14 / 22$. The final value of homophily is $\sim 0.766$. This value is interpretable as a sort of Pearson correlation coefficient, which means that $0.766$ is pretty high.

\section{Strength of Weak Ties}
Is homophily a good thing? In some aspects yes. A person who is surrounded by people with similar tastes and behaviors is happy. But suppose this person is looking for a job. It is very difficult, in presence of high homophily, for a message to arrive to the job seeker, because she only has close ties who cannot broker to her new information from the outside -- assuming that the network has a strong assortative community structure.

\begin{figure}
\centering
\includegraphics[width=.75\columnwidth]{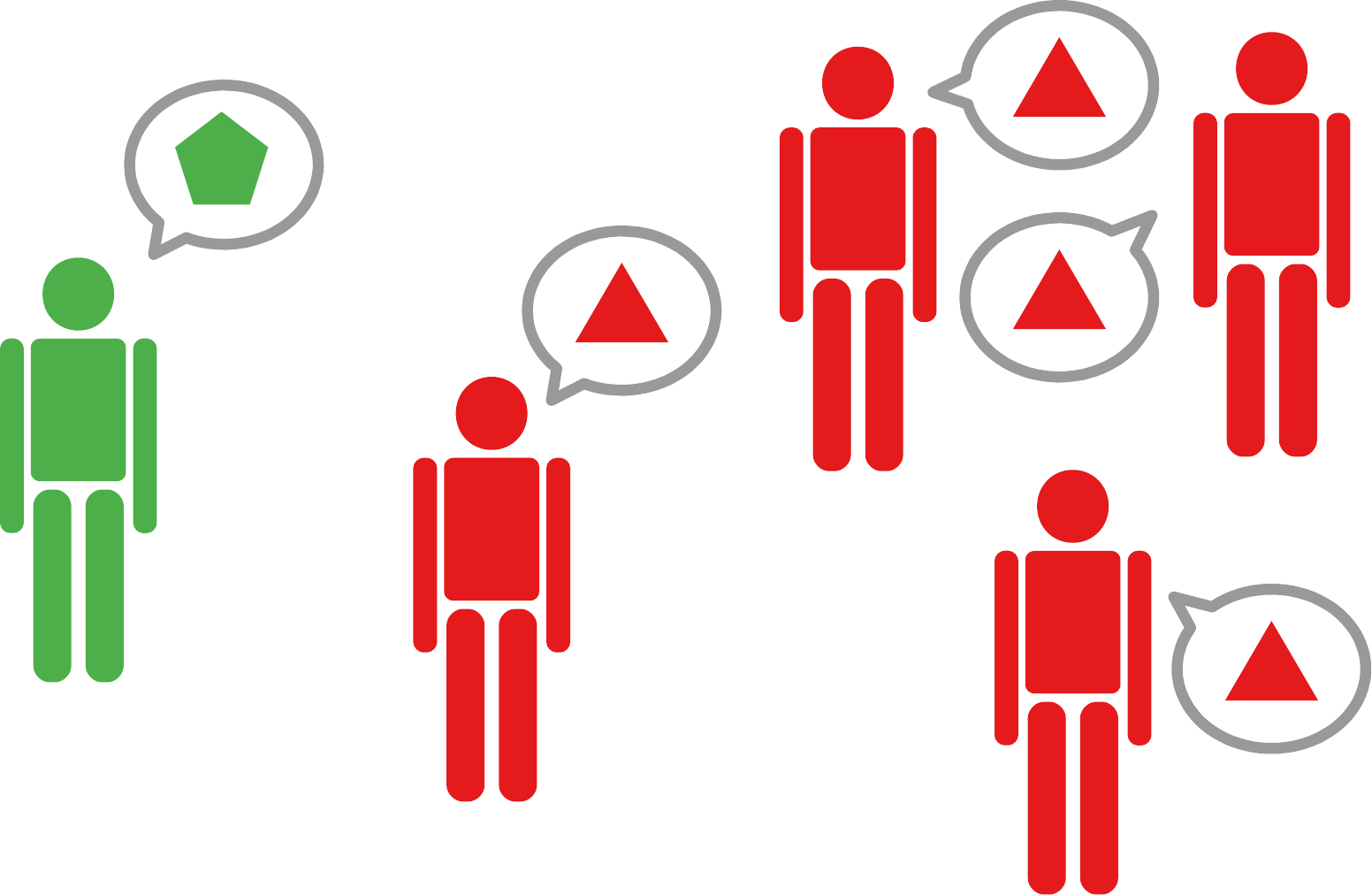}
\caption{An example of strength of weak ties. The green individual is part of a different community and thus only weakly linked with the red community. However, by being exposed to different information, she can bring it to the community she is not part of, but connected to it via a weak tie.}
\label{fig:strength-weak-ties}
\end{figure}

The ties that bind different communities with different people are the so-called ``weak ties'' and they have been shown to be fundamental in the job market by Granovetter\cite{granovetter1977strength}\cite{granovetter1983strength}. To put it simply: it's rarely your closest friends who make you find a job, but that far acquaintance with whom you rarely speak, because your close friends usually access the same information as you do and so cannot tell you anything new. Figure \ref{fig:strength-weak-ties} shows an example of the weak ties effect.

Note that Granovetter divides ties in three types: weak, strong, and absent. The terminology should not fool you. In this case, we are not referring to the edge's weight (Section \ref{sec:basic-weighted}). This is rather a categorical difference, more akin to multilayer networks (Section \ref{sec:extended-multilayer}). A weak tie is established between individuals whose social circles do not overlap much. A strong tie is the opposite: an edge between nodes well embedded in the same community. The absent tie is more of a construct in sociology, which lacks a well-defined counterpart in network science. It can be considered as a potential connection lurking in the background. For instance, there is an absent tie between you and that neighbor you always say ``hello'' to but never interact beyond that. You could consider an absent tie as one of the most likely edges to appear next, if you were to perform a classical link prediction (Part \ref{par:lp}).

You can see now that you can have strong, weak, and absent ties in an unweighted network. We can, of course, expect a correlation between being a weak tie and having a low weight. However, we can construct equally valid scenarios in which there is an anti-correlation instead. For instance, we could weight the edges by their edge betweenness centrality (Section \ref{sec:centr-betw}). A weak tie must have a high edge betweenness, because by definition it spans across communities and thus all the shortest paths going from one community to the other must pass through it.

Note that, notwithstanding their usefulness in favoring information spread, weak ties are not the only game in town in a society. The competing concept of the ``strength of strong ties'' shows that strong ties are important as well. They are specifically useful in times of uncertainty: ``Strong ties constitute a base of trust that can reduce resistance and provide comfort\cite[-0.25in]{krackhardt2003strength}''.

\section{Social Contagion}\label{sec:homophily-contagion}
Homophily can lead to a surprising number of counter-intuitive social dynamics. This section is intimately linked with Part \ref{par:sis}, where we looked at spreading events in networks. Here, we explore some more social explanations behind behavioral change in networks, mostly fueled by the right combination of homophily and heterophily (the love of the different).

\begin{figure}
\centering
\includegraphics[width=.8\columnwidth]{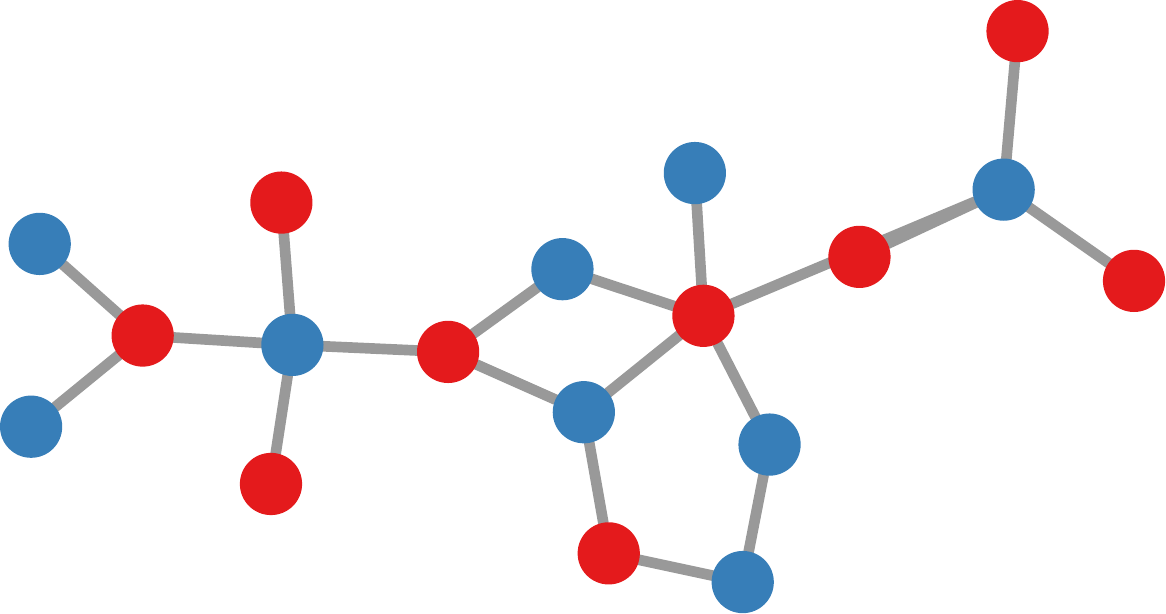}
\caption{A dating network. The node color encodes the gender (red = female, blue = male).}
\label{fig:dating-network}
\end{figure}

What's heterophily? Some things in social networks are very disassortative. For instance, consider sexual partners. When looking at some attributes, it is a network driven by homophily: people try to find mates with similar characteristics. They like the same music, movies, food. On the other hand, other attributes are very disassortative, for instance gender. Notwithstanding notable exceptions, the majority of edges in this network are between unlike genders -- as Figure \ref{fig:dating-network} shows.

If we live in a network governed by homophily, we know that connections are driven by the characteristics of the nodes. In some cases that is the only possible explanation. For instance, race is given: one cannot change their race and race homophily means that one's race influences which social connections are more or less likely to be established.

But consider the other side of the coin: if we observe a strong homophily, it could be because our social connections are influencing us into adopting behaviors we would not otherwise. For instance, drug use. One can decide whether to use drugs, and will be more likely to do so if the majority of their friends are drug users. It turns out that the right network topology can create an illusion of majority. Even if the majority of people do not use drugs, we can draw a network in which everybody thinks that the opposite is true\cite{lerman2016majority}!

\begin{figure}
\centering
\includegraphics[width=.33\columnwidth]{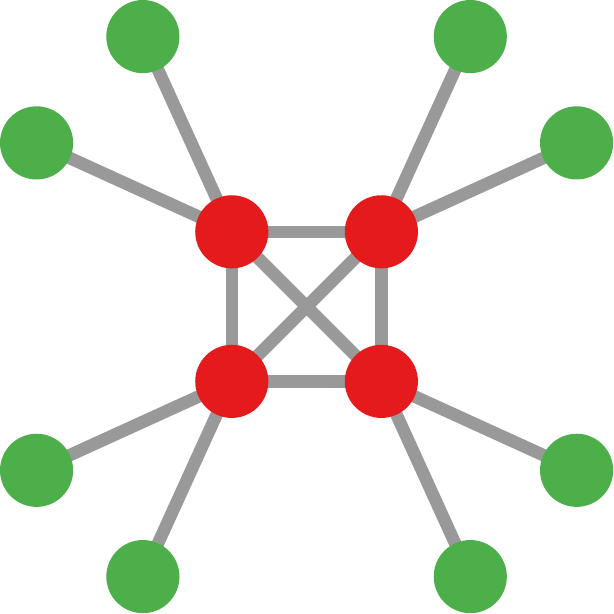}
\caption{The majority illusion in a toy network. Nodes in red are drug users, nodes in green are not. For every node, its neighbors include a majority of drug users.}
\label{fig:majority-illusion}
\end{figure}

You can look at Figure \ref{fig:majority-illusion} to see an example of this counter intuitive result. Or, you can play a simple game showing how to build networks fooling people into thinking everybody is binge-drinking\footnote{\url{http://ncase.me/crowds-prototype/}}.

Since humans are social animals and tend to succumb to peer pressure, homophily can be a channel for behavioral changes. In a health study, researchers looked at health indicators from thousands of people in a community over $32$ years. They saw that behavior and health risks that should not be contagious actually are. For instance obesity: if you have an obese friend, the likelihood of you becoming obese increases by $57\%$ in the short term\cite{christakis2007spread}. This is like the Susceptible-Infected epidemic models we saw, even if obesity is not a biological virus. It is rather a social type of virus.

Same with smoking, although in this case it worked the opposite: people were quitting in droves\cite{christakis2008collective}. This is due to social pressure and homophily: a behavior you might not adopt by yourself is brokered by your social circle, which you trust because it is made by people like you -- it speaks to your identity.

\begin{figure}
\centering
\includegraphics[width=\columnwidth]{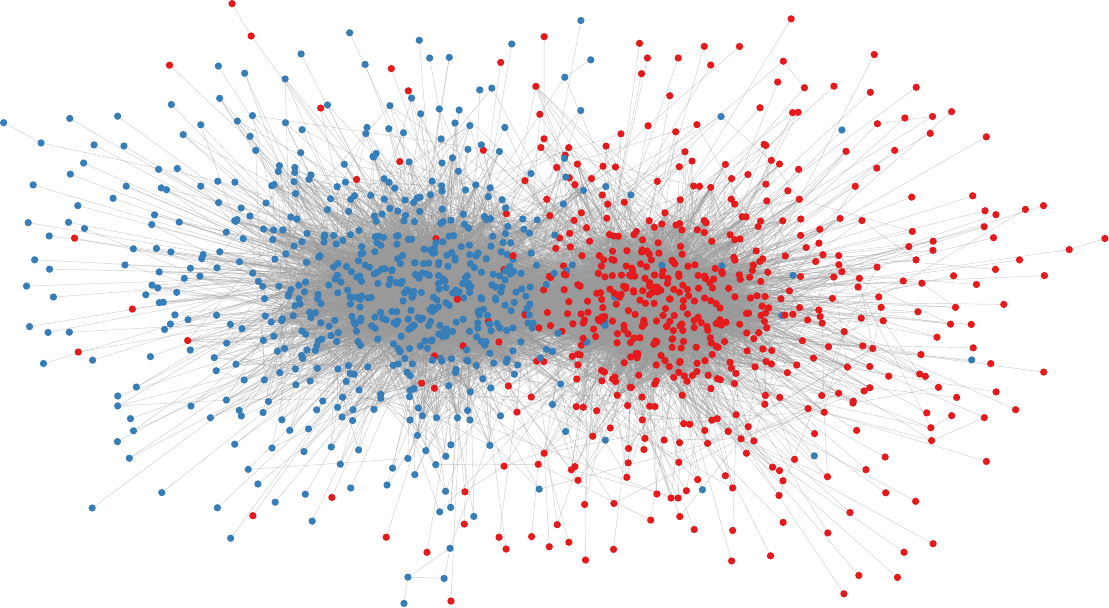}
\caption{The network of political blogs. Each node is a blog. Node's color encodes its political leaning (blue = democrat, red = republican). Two nodes are connected if either blog links to the other.}
\label{fig:polblogs}
\end{figure}

Another paper shows strong homophily in political blogs\cite{adamic2005political}. In Figure \ref{fig:polblogs} we see a visualization of how people writing online about politics connect to each other. A common political vision is the clear driving force behind the creation of an hyperlink from one blog to another.

Homophily arises very strongly even with mild preferences. One classical example is segregation. The famous ``Parable of Polygons\footnote{\url{https://ncase.me/polygons/}}'' starts from a simple assumption: people want to live with at least some similar people next to them. Even if they do not seek a majority of alike neighbors the end result is very clustered. Try to make an experiment and set the threshold to $40\%$, which means that people are happy being in the \textit{minority}. You'll still end up with segregation. There's no network in this interactive example, but one could easily introduce one by allowing nodes to rewire their friendship preferences according to the same rules. Experiments building relation graphs via RFID tags show that these dynamics may shape the topology of networks of face-to-face interactions\cite[-1in]{cattuto2010dynamics}.

\begin{figure}
\centering
\begin{subfigure}{.225\textwidth}
\includegraphics[width=\textwidth]{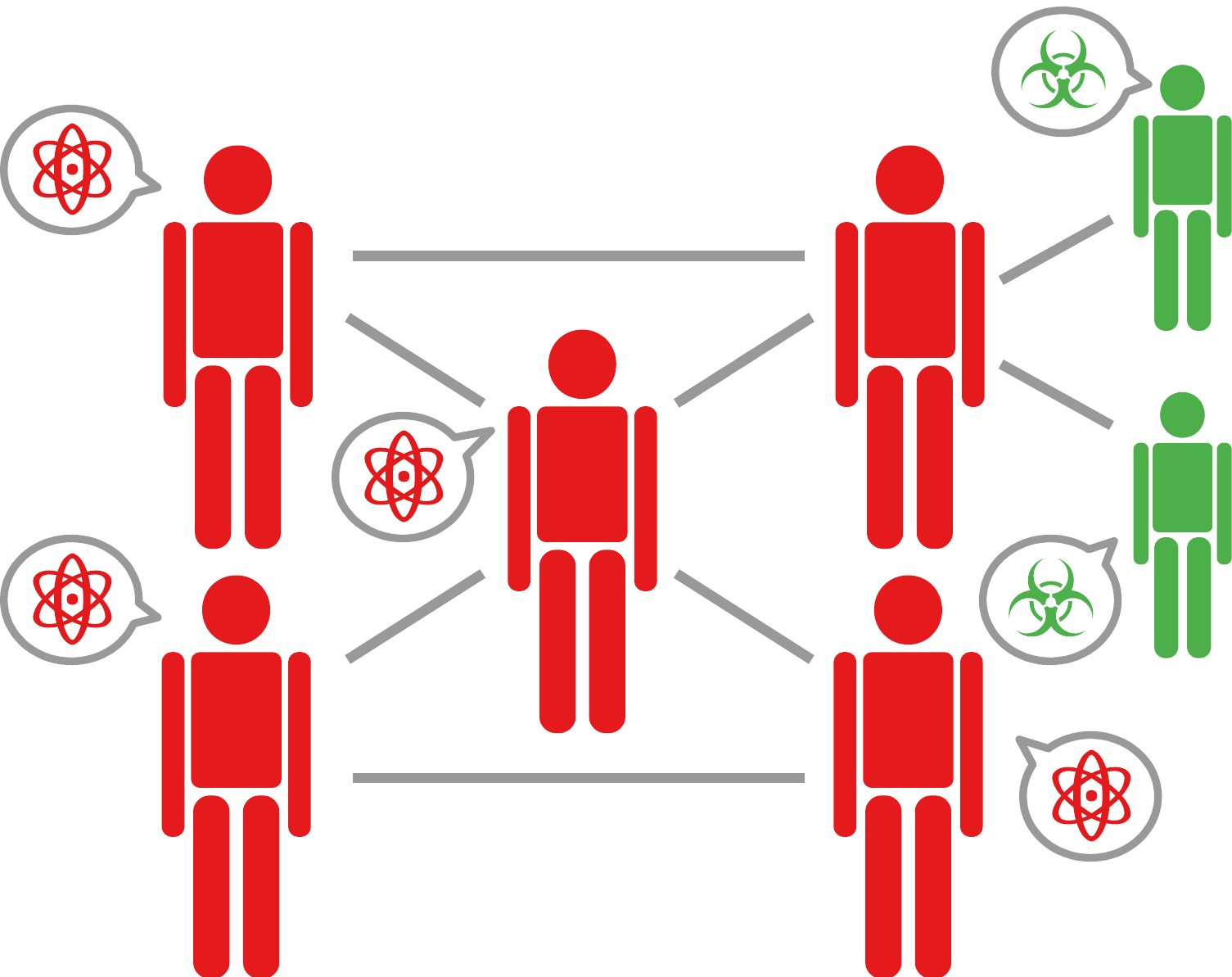}
\caption{}
\end{subfigure}\quad
\begin{subfigure}{.225\textwidth}
\includegraphics[width=\textwidth]{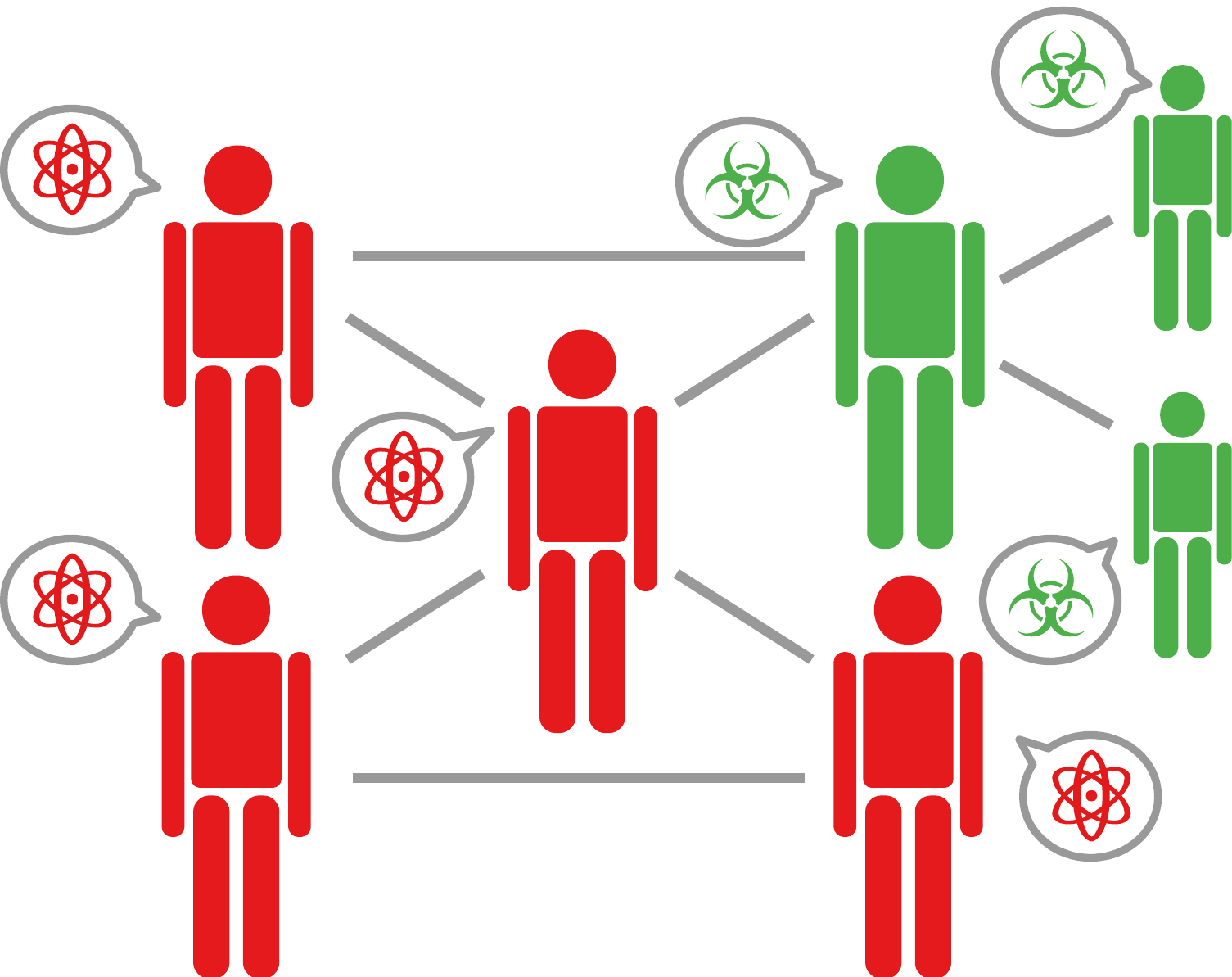}
\caption{}
\end{subfigure}\quad
\begin{subfigure}{.225\textwidth}
\includegraphics[width=\textwidth]{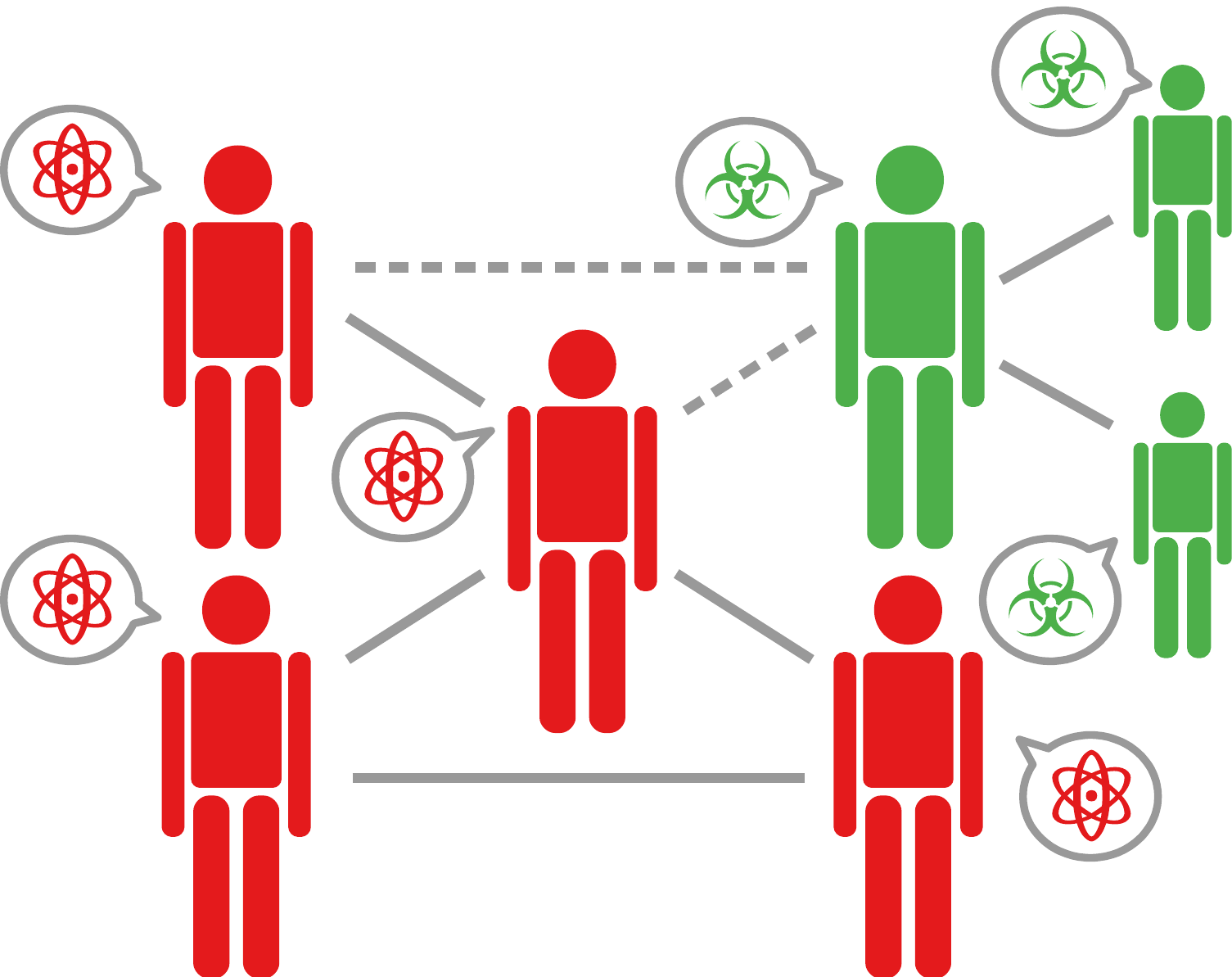}
\caption{}
\end{subfigure}\quad
\begin{subfigure}{.225\textwidth}
\includegraphics[width=\textwidth]{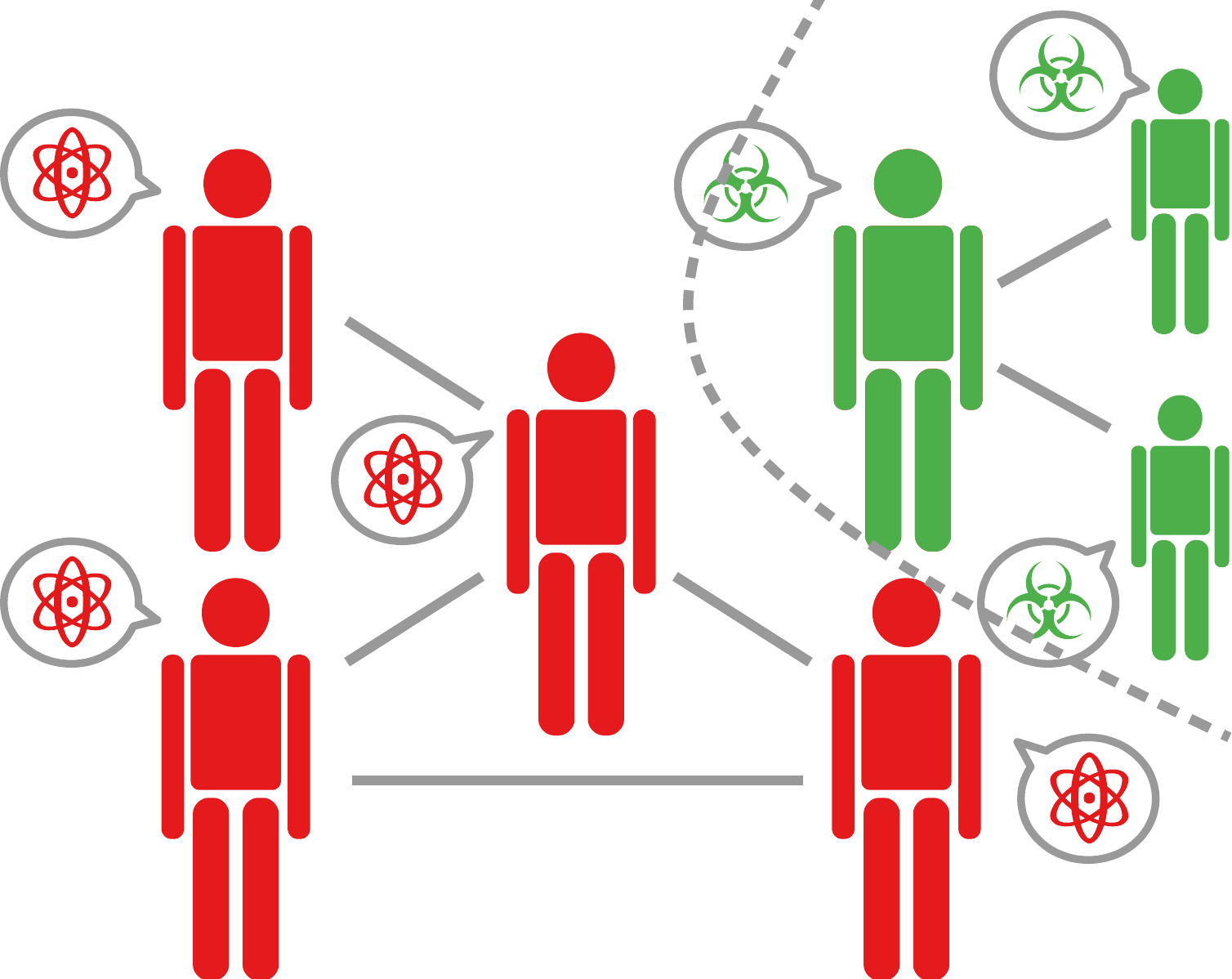}
\caption{}
\end{subfigure}
\caption{Homophily driving echo chambers. Science-oriented people (in red) rescind connections from conspiracy theorists (green) creating communities which have no possibility of communicating.}
\label{fig:echo-chamber}
\end{figure}

We are venturing now in new territory. So far we have seen homophily as a constructive force, meaning that people with similar characteristics link to each other. But with segregation we're doing something different. We're seeing homophily as a \textit{destructive} force: polygons are moving away if their expectation of uniformity isn't met. In network terms, people who are connected and discover differences in their characteristics might decide to rescind their connection.

Recently, researchers have started investigating this effect: rather than preferring to connect to similar strangers, we preferably rescind connections from dissimilar friends. Suppose you're on Facebook and you share a lot of content on scientific topics. One of the members of your community has outside connections, which could be convincing them of something like anti-vaccination. This person starts sharing anti-vax content, and the rest of the community is likely to rescind its connections. Which ends up creating groups that cannot connect any more people with different ideas, and thus reinforce each other convictions without any debate\cite[-5.05in]{del2017modeling}. Figure \ref{fig:echo-chamber} shows a vignette of this process.

This is particularly problematic, as there is a large body of research showing how easy it is for misinformation to spread through social media\cite{del2016spreading} and how strong online echo chambers can be\cite{del2016echo}\cite{bakshy2015exposure}. In fact, a sufficiently determined single actor can magnify their impact online, as the challenge in creating and operating difficult-to-detect bot nets is easy to overcome\cite{shao2017spread}\cite{ferrara2016rise}. There is suggestive research showing how this might already have happened\cite{bessi2016social}.

\section{Summary}

\begin{enumerate}
\item Homophily or assortativity is the tendency of nodes in a network to connect with nodes that are similar to them in some attribute. For instance, people tend to be friends in a social network with other people of similar age or same race.
\item A way to study this meso scale property is by creating ego networks: you pick one node as ego and then you create a network view including only its neighbors and the connections among them.
\item There are measures to estimate attribute assortativity, usually interpreted as a correlation coefficient taking values from $+1$ (perfect assortativity) to $-1$ (perfect disassortatvitiy).
\item Disassortativity is the opposite of assortativity: nodes tend to connect to other nodes with different attributes from their own. For instance, the dating network tends to be disassortative by gender.
\item Homophily interacts with network process. Links lowering homophily connect nodes with different attributes, which can favor information spread (the ``strength of weak ties'').
\item In other cases, nodes can be fooled into seeing a minority attribute as always the majority option in their friends: the majority illusion.
\end{enumerate}

\section{Exercises}

\begin{enumerate}
\item Load the network at \url{http://www.networkatlas.eu/exercises/30/1/data.txt} and its corresponding node attributes at \url{http://www.networkatlas.eu/exercises/30/1/nodes.txt}. Iterate over all ego networks for all nodes in the network, removing the ego node. For each ego network, calculate the share of right-leaning nodes. Then, calculate the average of such shares per node.
\item What is the assortativity of the leaning attribute?
\item What is the relative popularity of attribute values ``right-leaning'' and ``left-leaning''? Based on what you discovered in the first exercise, would you say that there is a majority illusion in the network?
\end{enumerate}

\chapter{Quantitative Assortativity}\label{cha:assortativity}
In the previous chapter we talked about homophily, the love of the similar. It is our tendency of liking the people who are similar to us: similar race, similar places we hang around, similar movies we watch. These are all \textit{qualitative} attributes. In this chapter, we make the jump towards \textit{quantitative} homophily.

Many node attributes are quantitative: age, number of friends, etc. We can still estimate the level of homophily in a network based on these attributes. In this case, we perform a small change in terminology. We use the term ``assortativity'' instead. This change is largely arbitrary, but can help you in differentiating between the concepts. Just like ``(qualitative) homophily = (quantitative) assortativity'', we have ``(qualitative) heterophily = (quantitative) disassortativity''.

Shifting our attention to quantitative attributes means we can use slightly different tools to estimate homophily, since there is a clear sorting in the attribute values and an intuition of similarity. If two nodes of values $1$ and $2$ connect, it is true that they have a different attribute value, but it still counts more towards assortativity than, say, connecting a node of value $1$ to a node of value $100$.

The set of possible quantitative attributes can be vast. For this chapter, I'm going to focus mainly on one example, which is the most studied case: the degree. However, don't be fooled: any technique for the estimation of degree assortativity can be employed to estimate any other quantitative attribute's assortativity. However, by focusing on the degree, I can introduce other fun assortativity-related network effects, such as the friendship paradox.

\section{Degree Correlations}\label{sec:assortquant-plots}
The degree is the most studied example of assortativity because it is directly related to the edge creating process. In a degree-assortative network we see that hubs connect preferentially to hubs, while peripheral nodes connect preferentially to other peripheral nodes. This is like a dating network, where celebrities hook up with each other much more than you would expect if the dating network would be fair\footnote{Which is \textit{clearly} the only reason why I've been unsuccessful in getting a date with Jennifer Lawrence. No other possible explanation.}.

\begin{figure}
\centering
\includegraphics[width=.8\columnwidth]{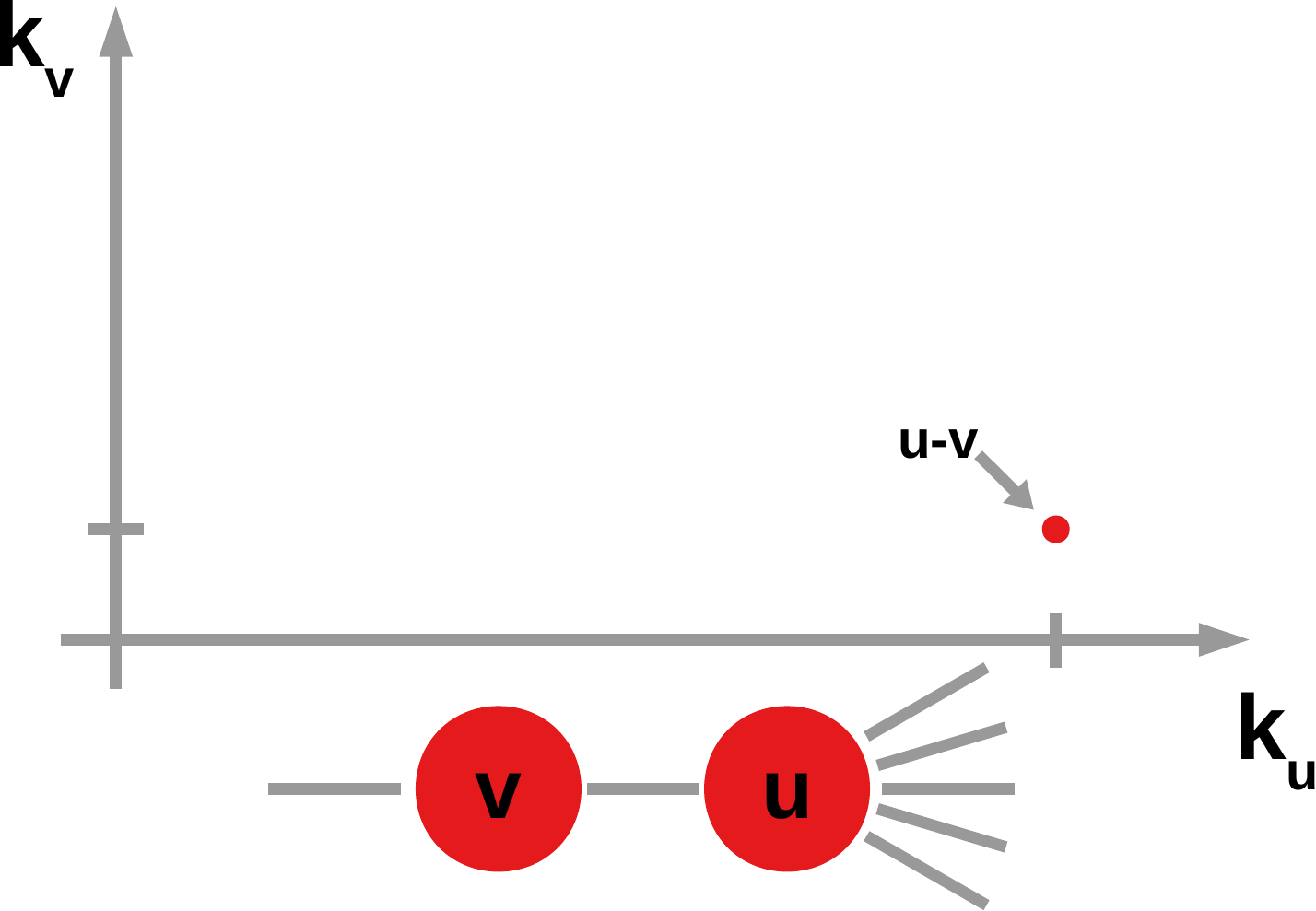}
\caption{A scatter plot we can use to visualize degree assortativity. For each edge, we have the degree of one node on the x axis and of the other node on the y axis.}
\label{fig:assortativity}
\end{figure}

In a disassortative network, hubs connects to periphery, as it happens for instance in protein networks\cite{uetz2000comprehensive}. This is more similar to the classical preferential attachment, where newcoming low-degree nodes will connect more often to older and high-degree hubs.

\begin{figure}[b]
\centering
\begin{subfigure}{.32\textwidth}
\includegraphics[width=\textwidth]{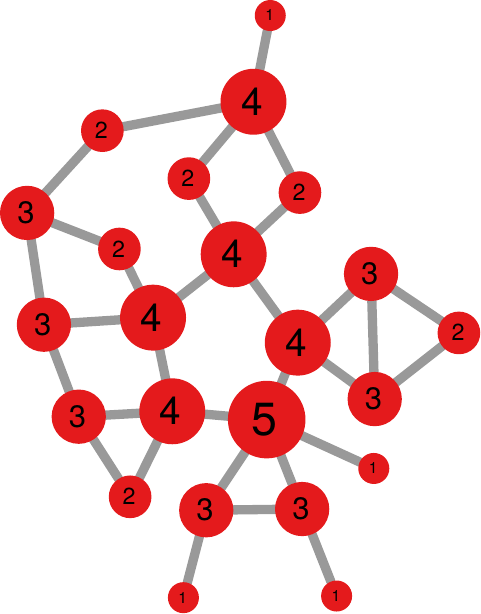}
\caption{}
\end{subfigure}\qquad
\begin{subfigure}{.5\textwidth}
\includegraphics[width=\textwidth]{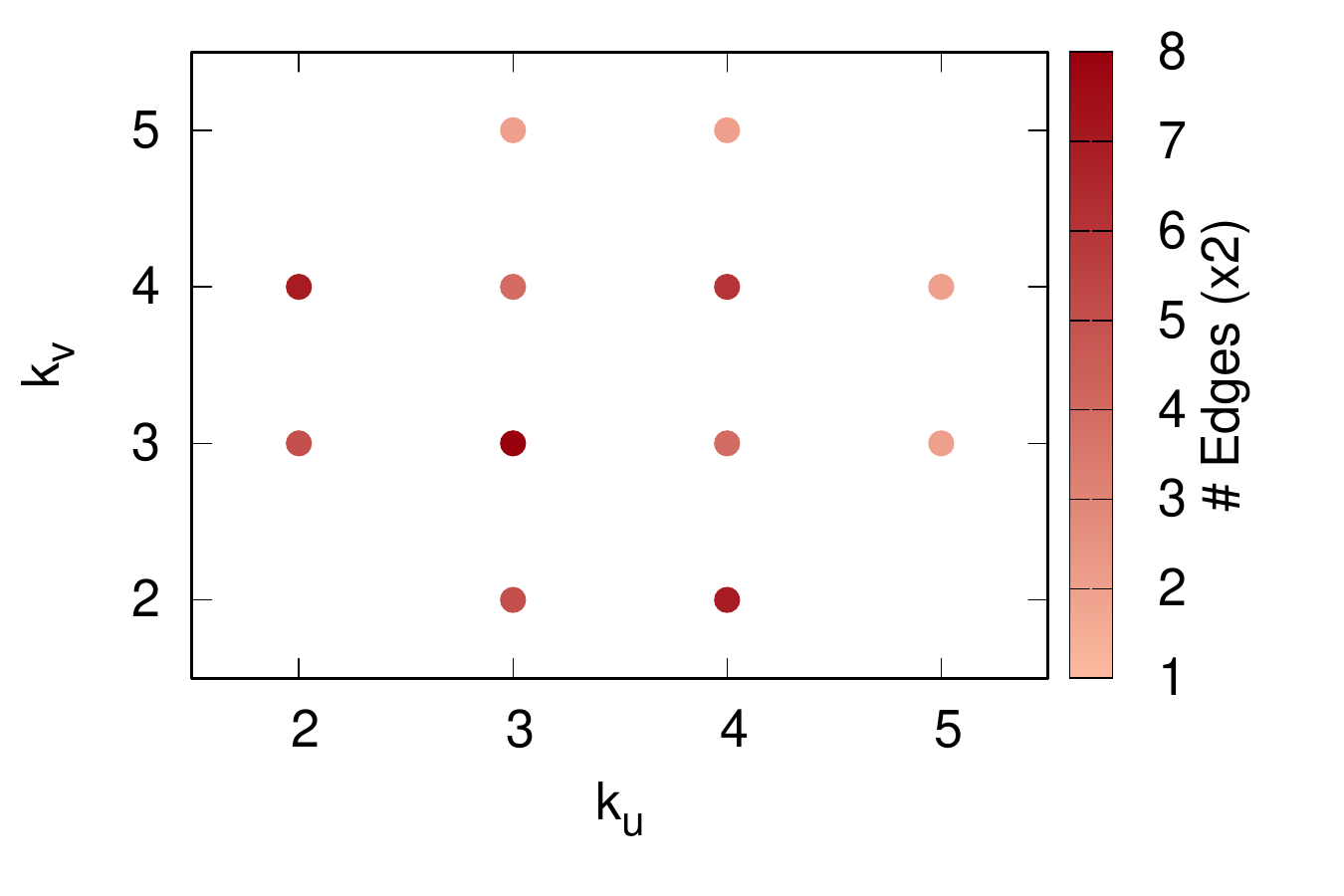}
\caption{}
\end{subfigure}
\caption{(a) A network with nodes labeled with their degree. (b) A scatter plot we can use to visualize (a)'s degree assortativity. Each point is a possible degree combination of an edge. The data point color tells you how many edges have that particolar degree combination. Usually, this count should also be log-transformed.}
\label{fig:assortativity2}
\end{figure}

A way to visualize degree assortativity is to consider each edge as an observation. We create a scattegram, recording the degree of one node on the x-axis and of the other node on the y-axis. So each point in this scatter plot is an edge of the network. Remember that the degree in real world networks follows a skewed distribution spanning many orders of magnitude. So, usually, these plots will have a log-log scale. Figure \ref{fig:assortativity} shows the skeleton of such a scatter plot. For an example network, such scatter would look like the one in Figure \ref{fig:assortativity2}.

Such a visualization suggests us a way to compute a possible index of degree assortativity. This is the first of two options you have if you want to quantify the network's assortativity. You iterate over all the edges in the network and put into two vectors the degrees of the nodes at the two endpoints. Note that each edge contributes two entries to this vector -- unless your network is directed. So, if your network only contains a single edge connecting nodes $1$ and $2$, your two vectors are $x = [k_1, k_2]$ and $y = [k_2, k_1]$, with $k_v$ being the degree of node $v$. Then, assortativity is simply the Pearson correlation coefficient of these two vectors.

There is only one way to achieve perfect degree assortativiy. In such a scenario, each node is connected only to nodes with the exact same degree. This is true only in a clique. Thus, a perfectly degree assortative network is one in which each connected component is a clique.

\begin{figure}
\centering
\includegraphics[width=\columnwidth]{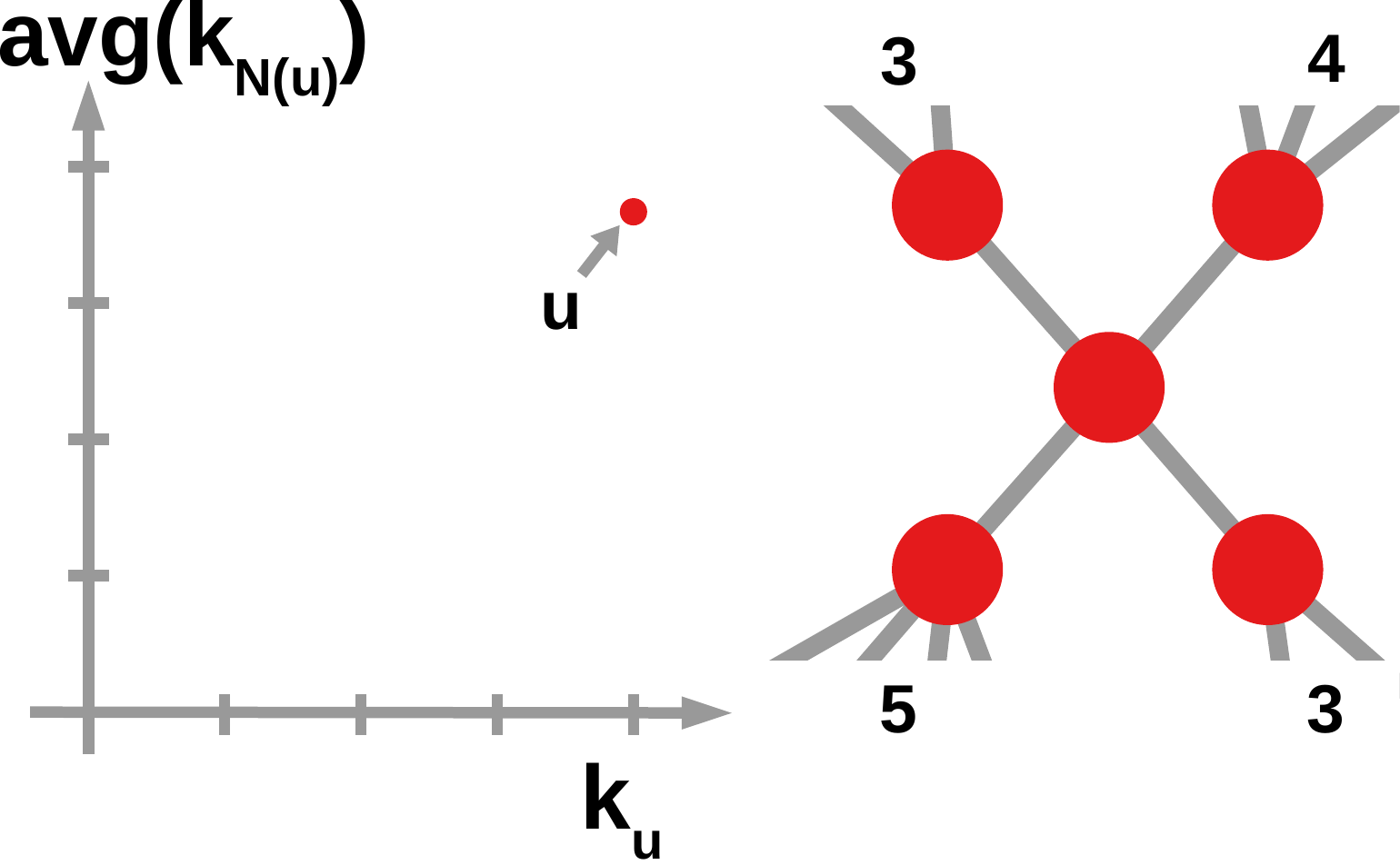}
\caption{A second strategy to visualize degree assortativity. The scatter plot has a point for each node in the network, reporting its degree (x axis) against the average degree of its neighbors (y axis).}
\label{fig:assortativity3}
\end{figure}

Figure \ref{fig:assortativity3} shows the second strategy to estimate degree assortativity in a network. Rather that plotting each edge, we plot each node. We compare a node's degree with the average degree of its neighbors. In a degree assortative network, we expect to see a positive correlation: the more connections the node has, the more connections, on average, its neighbors have\cite{pastor2001dynamical}.

Again, you shouldn't forget to log-transform the degree values, given the broad degree distributions of most real world networks. This also means that you should perform a power fit. The exponent of such a fit tells you whether the network is degree assortative (if it's positive), disassortative (if it's negative), or non assortative (if it's statistically indistinguishable from zero).

\begin{figure}
\centering
\begin{subfigure}{.45\textwidth}
\includegraphics[width=\textwidth]{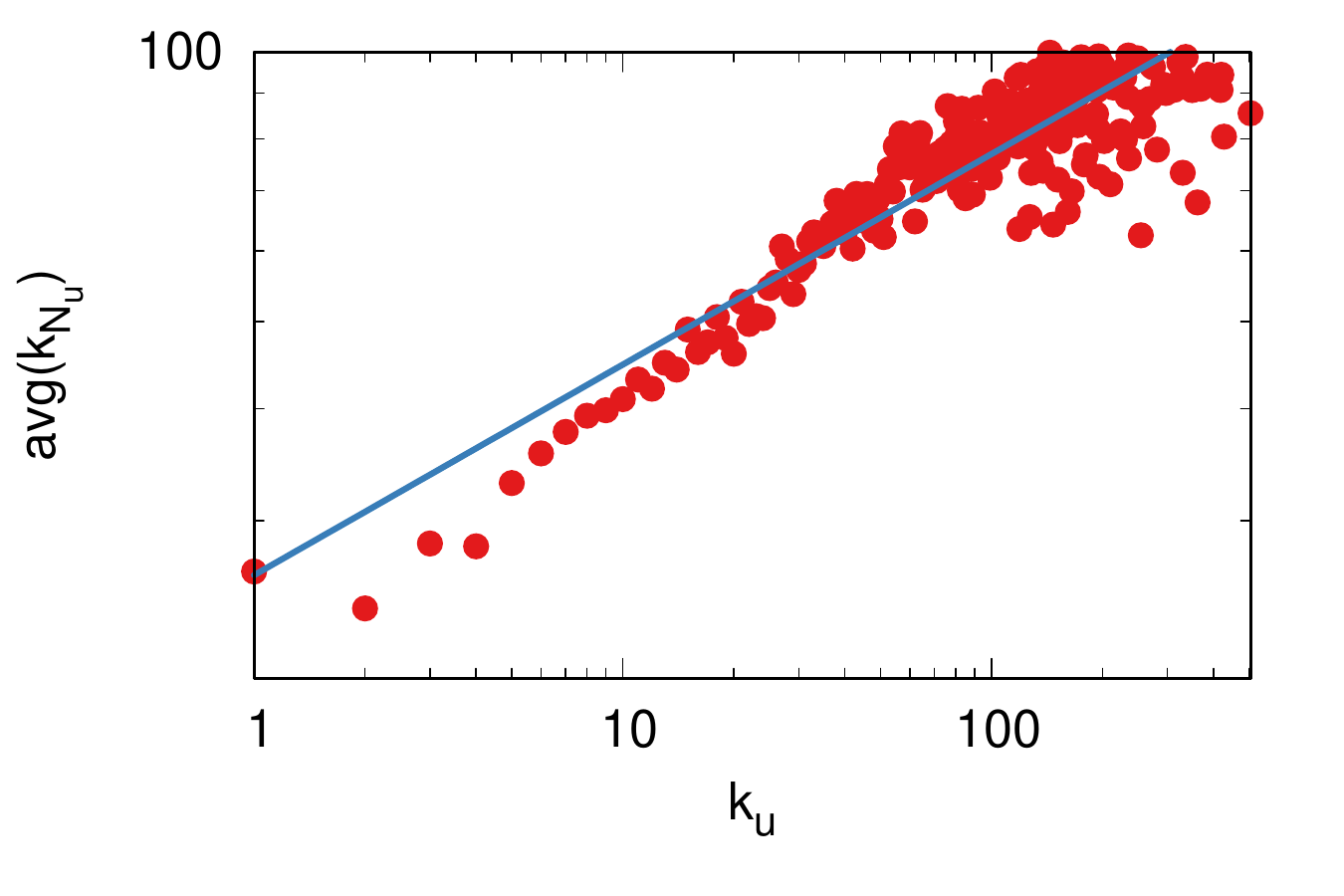}
\caption{}
\end{subfigure}
\begin{subfigure}{.45\textwidth}
\includegraphics[width=\textwidth]{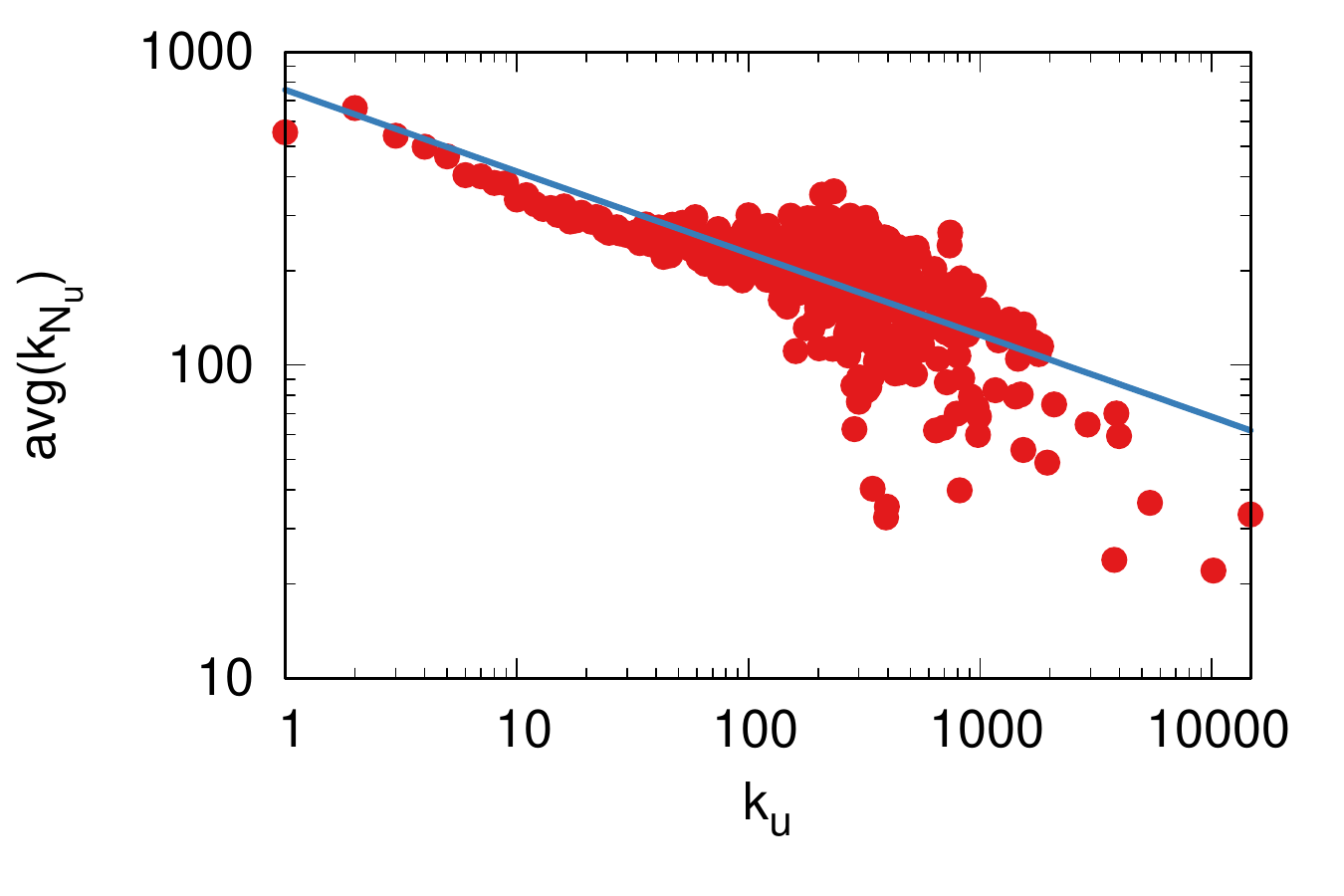}
\caption{}
\end{subfigure}\\
\begin{subfigure}{.45\textwidth}
\includegraphics[width=\textwidth]{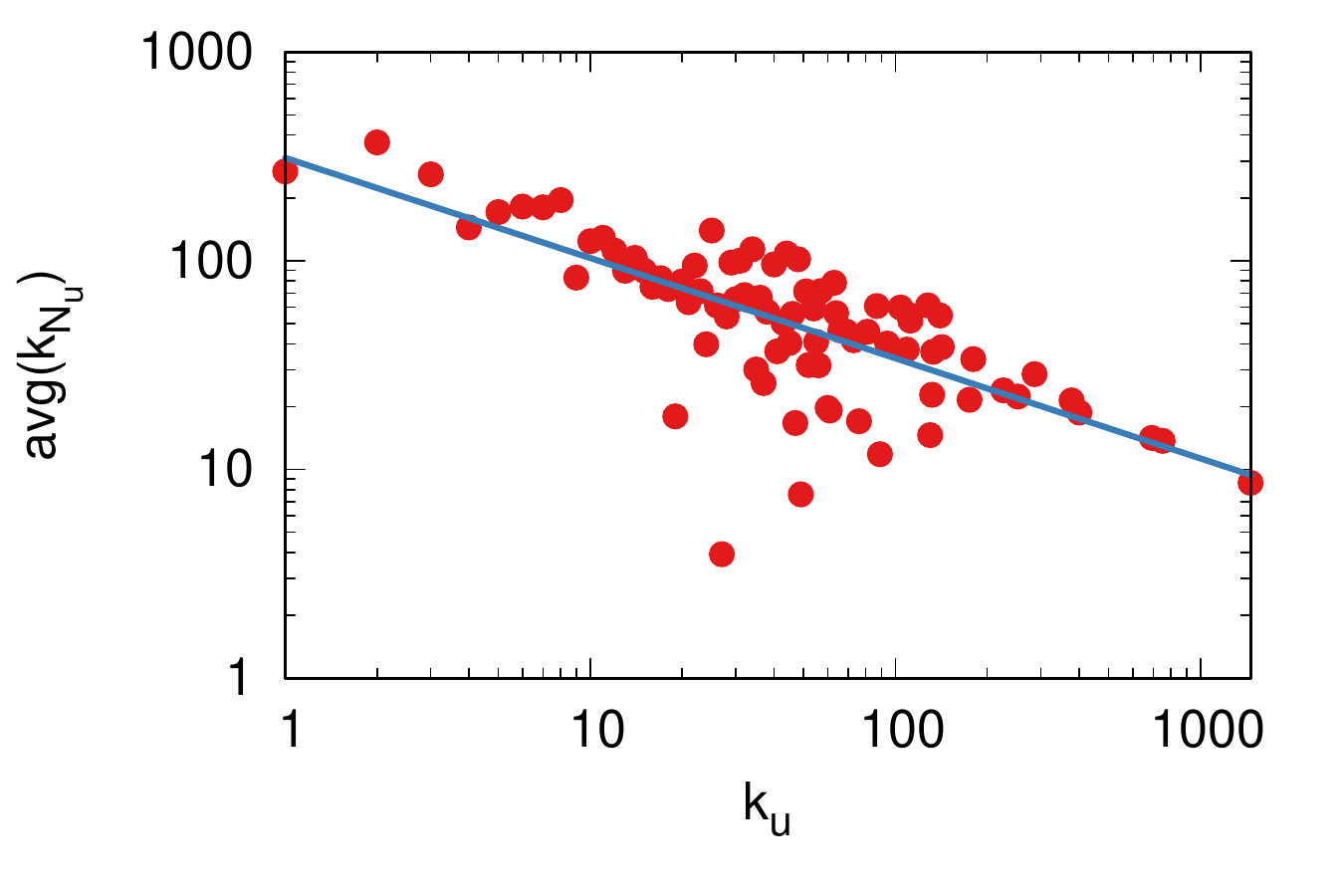}
\caption{}
\end{subfigure}
\begin{subfigure}{.45\textwidth}
\includegraphics[width=\textwidth]{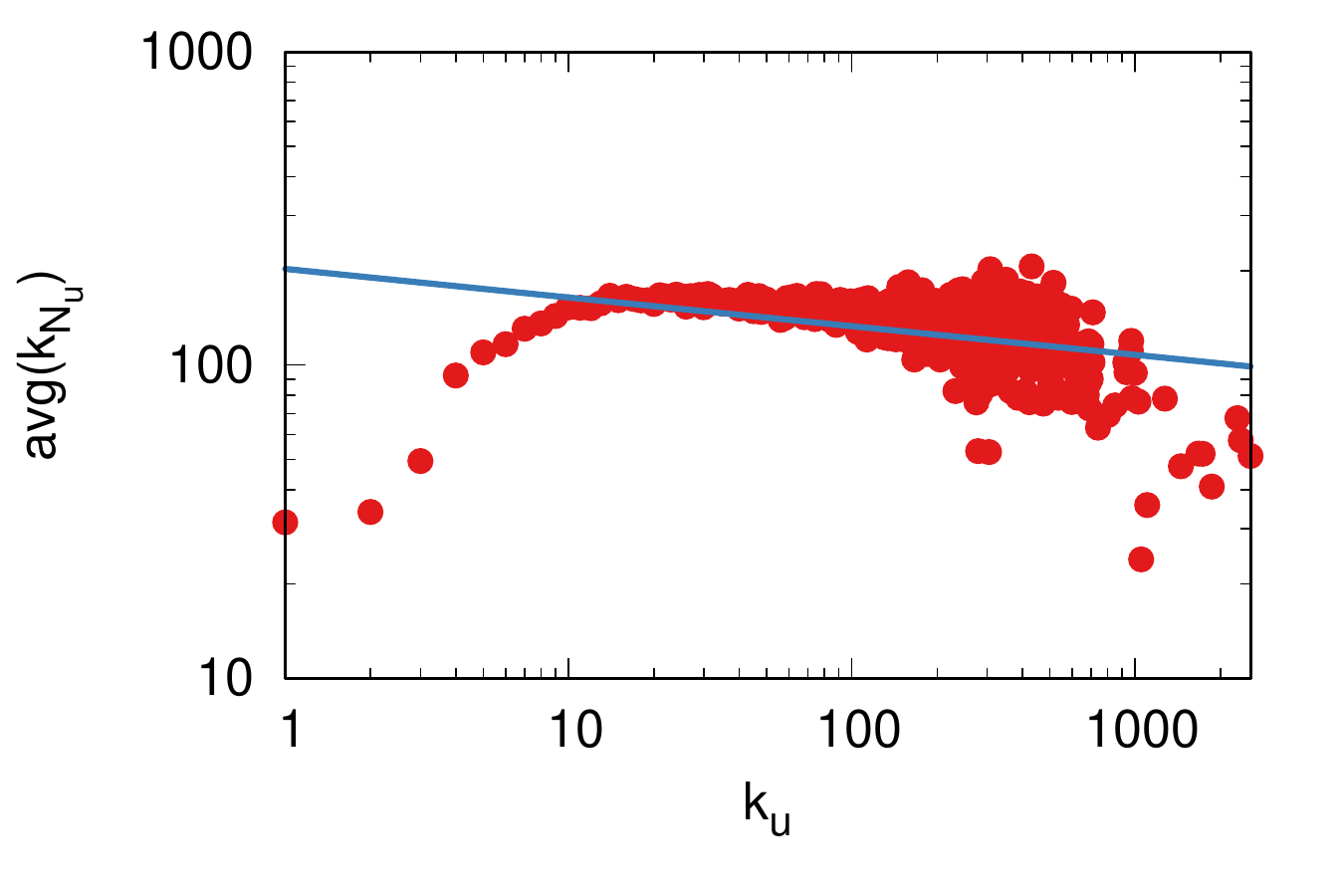}
\caption{}
\end{subfigure}
\caption{A collection of assortativity plots for four real world networks: (a) co-authorship in scientific publishing, (b) P2P network, (c) Internet routers, (d) Slashdot social network.}
\label{fig:assortativity4}
\end{figure}

Figure \ref{fig:assortativity4} shows few examples of assortativity in real world networks. Coauthorship is assortative: if I have a lot of coauthors, on average, they have a lot of coauthors too. Gowalla is disassortative: users with few friends likely attach to hubs. Slashdot is still disassortative, but it is the closest we could find to show what a neutral network looks like: one where my degree doesn't tell me anything about my neighbors' degree.

Note that, for real world networks, we usually aggregate in the same data point all nodes with the same degree. Thus what we're plotting is actually the average degree of all neighbors of all nodes of degree $k$. Otherwise, we would have an unreadable cloud of points at low degree values, since most nodes in real world networks have a degree equal to one or two.

Degree assortativity is a super important property for your network. Degree correlations radically change many network dynamics\cite{boguna2003absence}\cite{vazquez2003resilience}\cite{sorrentino2006synchronization}\cite{posfai2013effect}. This is especially true when you have multilayer networks and we're looking at assortativity inter-layer besides intra-layer\cite{wang2015evolutionary}. Meaning: if I am a hub in one layer, am I also a hub in the other layers? We touched the topic in Section \ref{sec:epidemapps-interdependent}. If the answer to this question is yes, failure-resistant layers become failure-prone\cite{gao2011robustness}\cite{shao2011cascade}.

There is a curious tension between degree assortativity and other common statistical properties of real world networks. For instance, we just saw that scientific collaboration is an assortative network. However, we also know it has a broad degree distribution.

The two properties clash against each other: assortativity means that hubs connect to hubs, but in a network with a heavy-tailed degree distribution there are few huge hubs and many one-degree nodes. The likelihood of connecting a hub to many small nodes seems too high. In fact, if we were to generate a random version of the co-authorship network respecting its degree distribution -- for instance via a configuration model (Section \ref{sec:csmodels-conf}) --, we would obtain a degree disassortative network\cite{boguna2004cut}. This makes quite interesting networks that both have a skewed degree distribution and are degree assortative! They have some non-trivial machinery driving their nodes' connections that cannot be captured by simple models.

The network generators I discussed in Part \ref{par:synthnet} weren't really designed with assortativity in mind. As we just saw, the configuration model is naturally disassortative, as is preferential attachment and anything which imposes a power law degree distribution. By definition, random graphs such as $G_{n,p}$ are non-assortative.

However, if you start with any synthetic network, there are algorithms to rewire the edges such that the degree distribution will be preserved, but you will obtain an assortative (or disassortative) network\cite{boguna2003class}\cite{xulvi2004reshuffling}\cite{xulvi2005changing}. 

\begin{figure}
\centering
\begin{subfigure}{.3\textwidth}
\includegraphics[width=\textwidth]{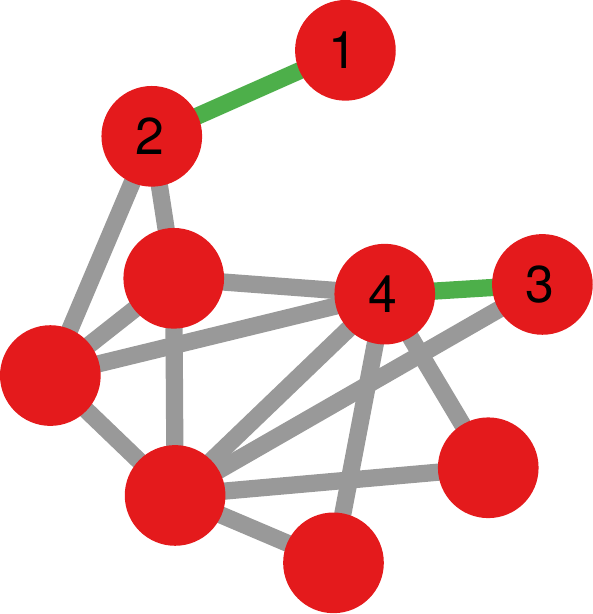}
\caption{}
\end{subfigure}\quad
\begin{subfigure}{.3\textwidth}
\includegraphics[width=\textwidth]{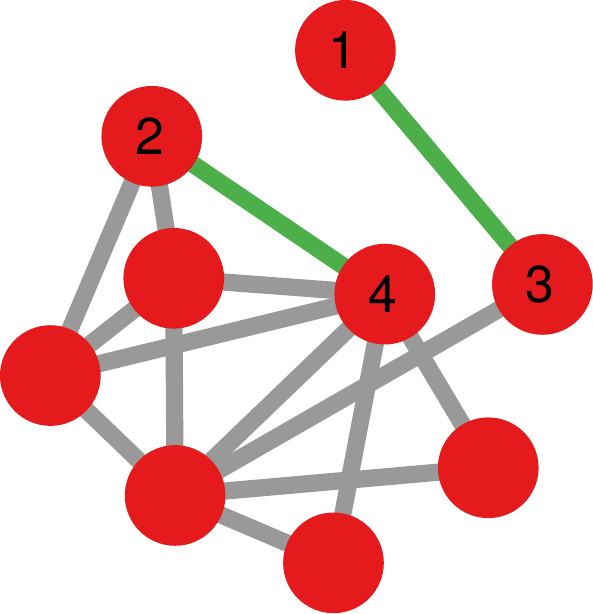}
\caption{}
\end{subfigure}\quad
\begin{subfigure}{.3\textwidth}
\includegraphics[width=\textwidth]{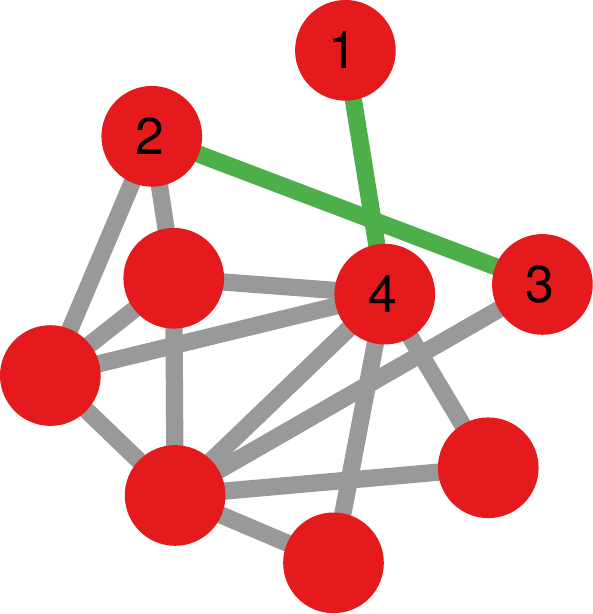}
\caption{}
\end{subfigure}
\caption{The (dis)assortativity inducing model. (a) Select two pairs of connected nodes (in green the edges we select). (b) Assortativity inducing move. (c) Disassortativity inducing move.}
\label{fig:assortativity-model}
\end{figure}

This happens through edge swap, as Figure \ref{fig:assortativity-model} shows. First, you select two connected node pairs. Then you sort them according to their degree, in the figure the order is nodes $4, 2, 3, 1$ ($k_4 = 6$, $k_2 = 3$, $k_3 = 2$, $k_1 = 1$). The next move depends on whether you want to induce assortativity or disassortativity. In the first case, you connect the two nodes with the highest degree to each other, and the two with lowest degree to each other (Figure \ref{fig:assortativity-model}(b)). In the second case, you do the opposite: connect the highest degree node with the lowest, and the two middle ones (Figure \ref{fig:assortativity-model}(c)).

Note that this swap doesn't always change the topology nor alter the characteristics of the network. For instance, if all nodes have the same degree, the move would not affect assortativity. But, after enough trials in a large enough network, you'll see that these operations will have the desired effect.

Degree assortativity, as I discussed it so far, is defined for undirected networks. There are straightforward extensions for directed networks\cite{foster2010edge}. The standard strategy is to look at four correlation coefficients: in-degree with in-degree, in-degree with out-degree (and vice versa), and out-degree with out-degree.

Obviously, everything I wrote so far on degree assortativity also applies to any other quantitative attribute you might have on your nodes. For instance, if you have a social network, it applies to a person's age, height, weight, and so on. You can build the scattergrams and calculate the best fit to figure out if your social circle sort themselves according to their height or income.

\section{Friendship Paradox}\label{sec:assortquant-paradox}
\begin{figure}
\centering
\includegraphics[width=.75\columnwidth]{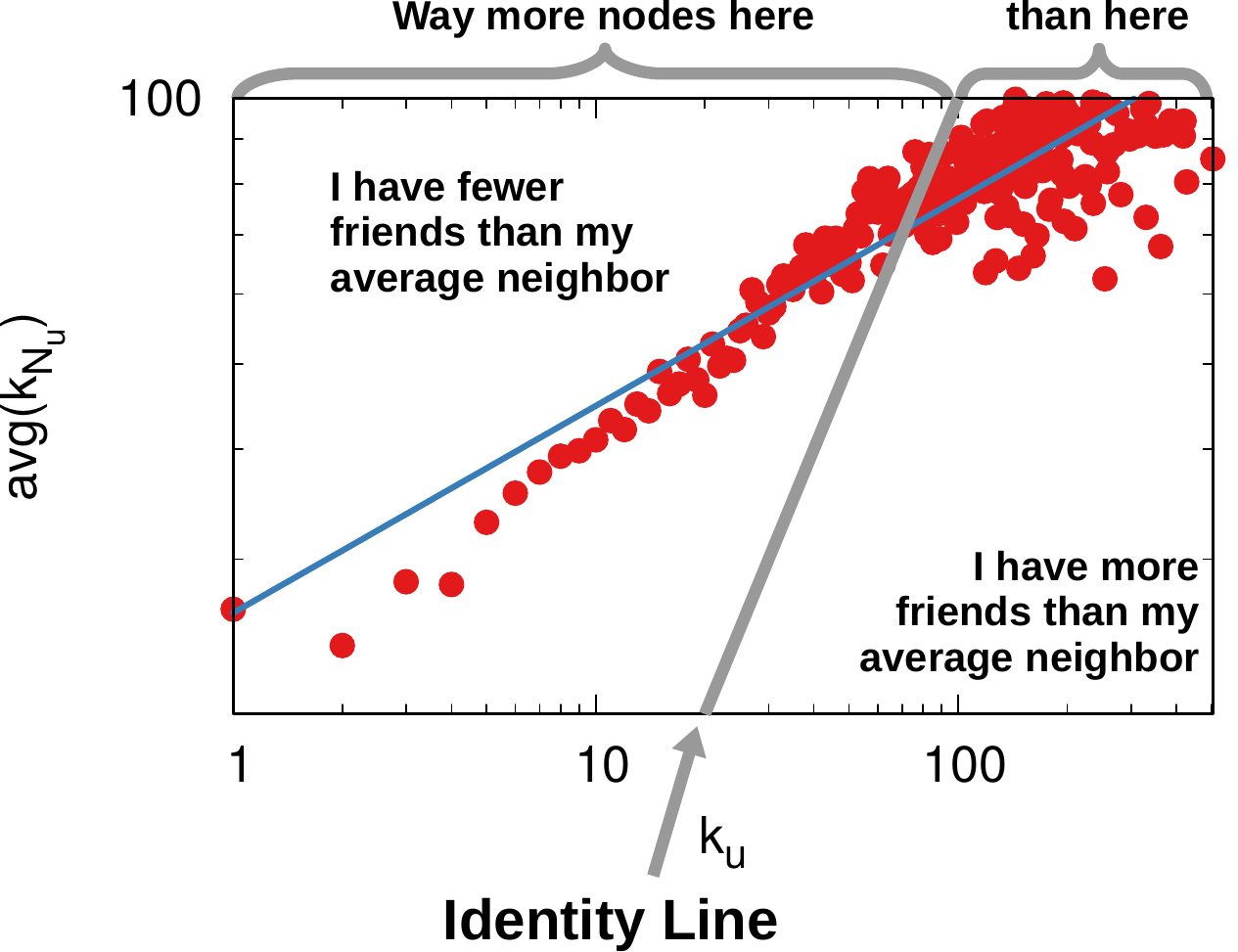}
\caption{The friendship paradox. The plot shows a node's degree (x axis) against the average degree of its neighbors (y axis). The blue line is the best fit, while the gray line is the identity line. Nodes above the identity line have fewer friends than their friends' average.}
\label{fig:friend-paradox}
\end{figure}

You might want to make a double take on Figure \ref{fig:assortativity4}, because it contains a not-so-obvious but intriguing -- and enraging -- message. In Figure \ref{fig:friend-paradox} I focus on one of those assortativity plots, the one about scientific collaborations. I add an additional line to the plot: the identity line. This line runs through the part of the plane where the x axis and the y axis have the same value.

In other words, the identity line divides the space in two. Above the identity line we have all the nodes for which, on average, the neighbor degree is higher than the node's degree. Below the identity line it's the opposite: the node's degree is higher than the neighbors degree.

At first glance, the situation seems balanced. There are as many points above the identity line as there are below. However, remember that we're aggregating all nodes with the same degree value in a single point. We know that the degree has a broad distribution, because we visualized it for the co-authorship network before. Therefore, there are way more nodes above the identity line than below.

That's the friendship paradox: your friends are, on average, more popular than you\cite{feld1991your}\cite{zuckerman2001makes}! This means that, for the average node, its degree is lower than the average degree of their neighbors. This is actually pretty obvious once you think about it: a node with degree $k$ appears in $k$ other nodes' averages, and hence is ``over-counted'' by an amount equal to how much larger it is than the network's mean degree. A high degree node appears in many more node neighborhoods than does a low degree node, and hence it skews many local averages. The only way to escape such paradox is by having a network whose degree distributes mostly regularly: for instance small-world networks (Section \ref{sec:physicsmodels-ws}) are usually immune to the friendship paradox because most nodes have the same degree (the probability of rewiring is low).

The friendship paradox sounds pretty depressing, but we actually already made use of it in a rather uplifting scenario. The effective ``vaccinate-a-friend'' scheme I discussed in Section \ref{sec:triggers-intervention} is nothing else than a practical application of this network property.

\section{Distribution of Quantitative Attributes}
Quantitative assortativity does not stop at the degree. There are examples of papers analyzing quantitative attributes discovering all sort of interesting phenomena. I'm going to give you one example I know well, as it came from a paper I wrote\cite{barone2018birds}.

\begin{figure}
\centering
\includegraphics[width=.66\columnwidth]{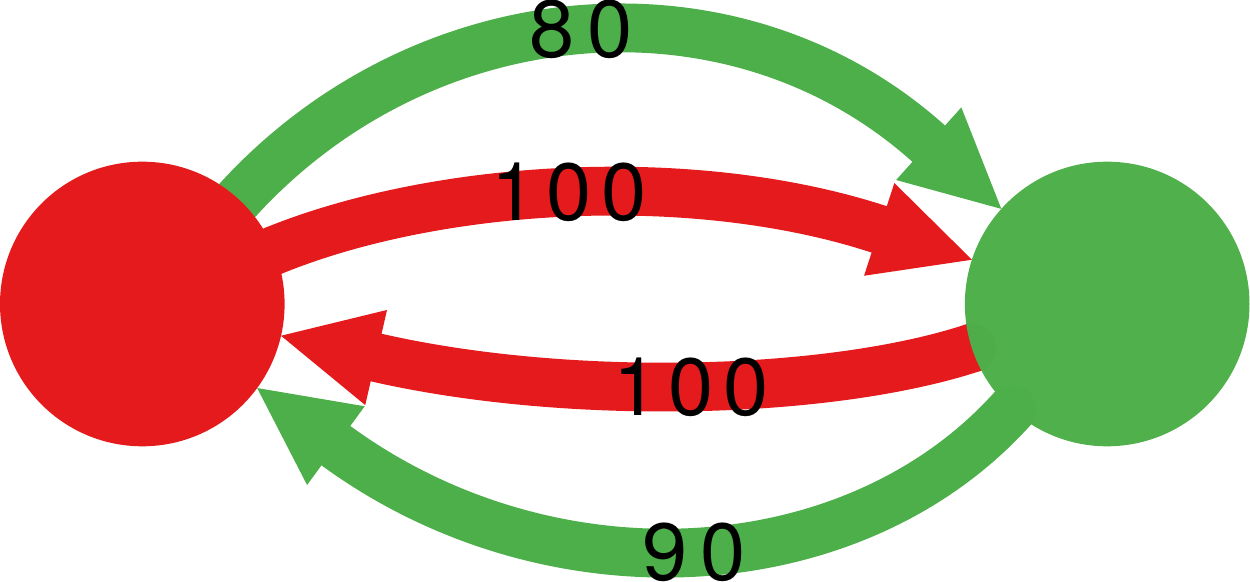}
\caption{The data model of the business to business network. The edge color tells corresponds to the node making the claim about the transaction. For instance, the green node reports selling $90$ to the red node and buying $80$ back.}
\label{fig:assort-b2b}
\end{figure}

In the paper, I analyze a business to business network, connecting businesses if they are customers or suppliers of each other. Each edge is a B2B transaction and both businesses have to report it, as Figure \ref{fig:assort-b2b} shows. Thus, if they report a different amount, we know someone is lying. I create a measure of trustworthiness, which is the average value of mismatch a business has, weighted by how trustworthy its neighbors are.

This trustworthiness score is a quantitative attribute. It is strongly correlated with the likelihood that the business was in fact cheating on their taxes, as I have information whether the audited businesses were fined and, if they were, how much they had to pay.

\begin{figure}
\centering
\includegraphics[width=.66\columnwidth]{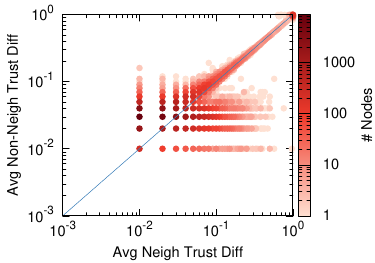}
\caption{The average trustworthiness score difference between neighbors (x axis) and non-neighbors (y axis). The blue line shows the identity, and the point color the number of nodes with the given value combination.}
\label{fig:assort-b2b-2}
\end{figure}

Simulations show that, with this correction, in a randomly wired network the score should be disassortative. Instead, in the real observed network, the score is assortative. Figure \ref{fig:assort-b2b-2} shows the relation: there are more nodes above the identity line than below -- it might appear that the opposite is true, but you need to take into account the color of the dots. Being above the identity line means that the trust score difference between neighbors is lower than between non-neighbors: a sign of assortativity.

Given the connection with the actual tax fines, the assortativity analysis can make us conclude something tangible about business connections. In this case, that fraudulent and untrustworthy businesses band together. If I know your customer/suppliers are scammers, I should update my priors on whether you're a scammer as well.

A more famous example of quantitative attribute assortativity is related to the friendship paradox that I just described in the previous section, and it is ten times more enraging. We humans are social animals. Notwithstanding introverted cavemen like myself, in general our level of happiness is correlated with the number of friends we have. In fact, just like the degree, happiness is assortative in social networks: happy people tend to befriend each other\cite{bollen2011happiness}.

What I just stated is that more friends imply more happiness. And the friendship paradox tells us that our friends have more friends than us. Do I mean to tell that, like with friendship, there is also a happiness paradox? Why, yes there is\cite{bollen2017happiness}. If you ask people about their level of happiness in a social network, you will find out that the average happiness level of one's friends tends to be higher than their own happiness level. That is probably why you think everyone is having such a great time on social media. Everyone but you. It's not you, it's the system. Luckily, the researchers behind this discovery have a few guidelines on how to unplug from social media toxicity and live a more fulfilling life\cite{bollen2018network}.

In general, everything that correlates with degree -- be it happiness, income, or tax fraud -- will get its own paradox for free.

\section{Summary}

\begin{enumerate}
\item Assortativity works not only on qualitative, but also on quantitative attributes. The most studied case is degree assortativity: the tendency of high degree nodes to connect to other high degree nodes (degree assortative) or to low degree nodes (degree disassortative).
\item You can calculate degree assortativity -- or any quantitative assortativity -- by correlating the attribute values at the endpoints of each edge. Alternatively, you can correlate a node's attribute value with the average of their neighbors.
\item Graph generators usually are unable to provide degree assortative networks, but there are postprocessing techniques that can induce either degree assortativity or degree disassortativity.
\item By a cruel mathematical property of degree distributions, social systems are affected by the friendship paradox: the average person has fewer friends than their friends on average.
\item Even crueler, since in social system happiness is usually correlated with the number of friends a person has, the friendship paradox also implies than the average person is less happy than their friends on average. So it's not just you.
\end{enumerate}

\section{Exercises}

\begin{enumerate}
\item Draw the degree assortativity plots of the network at \url{http://www.networkatlas.eu/exercises/31/1/data.txt} using the first (edge-centric) and the second (node-centric) strategies explained in Section \ref{sec:assortquant-plots}. For best results, use logarithmic axes and color the points proportionally to the logarithmic count of the observations with the same values.
\item Calculate the degree assortativity of the network from the previous question using the first (edge-centric Pearson correlation) and the second (node-centric power fit) strategies explained in Section \ref{sec:assortquant-plots}.
\item Prove whether the network from the previous questions is affected or not by the friendship paradox.
\end{enumerate}

\chapter{Core-Periphery}\label{cha:coreperiph}
When you obtain a new network dataset and you plot it for the first time, in the vast majority of cases you will see a blobbed mess. This is usually due to the fact that raw network data is usually a hairball, and you need to backbone it, or perform other data cleaning tasks, as I detailed in Part \ref{par:hairball}. However, in some cases, there is an unobjectionable truth. It might be that, deep down, your network really is a hairball.

Many large scale networks have a common topology: a very densely connected set of core nodes, and a bunch of casual nodes attaching only to few neighbors. This should not be surprising. If you create a network with a configuration model and you have a broad degree distribution, the high degree nodes have a high probability of connecting to each other -- see Section \ref{sec:csmodels-conf}. The surprising part is that the cores of some empirical networks are even denser than what you'd anticipate by looking at the degree distribution of the network\cite{zhou2004rich}!

Since everything that departs from null expectation is interesting, this phenomenon in real world networks has attracted the attention of network scientists. They gave a couple of names to this special meso-scale organization of networks: core-periphery\cite{holme2005core}\cite{csermely2013structure}, with the core sometimes dubbed as ``rich club''\cite{colizza2006detecting}.

We already saw the concept of core and periphery when we discussed node roles and k-core centrality in Section \ref{sec:centr-kcore}. Such method is not included here because it finds a fundamentally different type of core-periphery structure, which is more similar to a hierarchical decomposition of the network estimating the centrality of the nodes\cite{gallagher2020clarified}. In this chapter I discuss ways in which you can detect a core-periphery structure that is less hierarchical and more ``hub-and-spoke'', in which we want to determine which nodes are in the core and which others are in the periphery. I also discuss the tension between the ubiquity of core-periphery structures and the equally common and (only apparently) contradictory presence of communities in complex networks. Finally, I'll connect core-periphery structures with a few real world dynamics that might be able to generate them.

\section{Models}
There are many ways to extract core-periphery structures from your networks. However, two methods dominate in the literature, especially in sociology. These are the discrete and the continuous model\cite{borgatti2000models}.

\subsection{Discrete}
Core-periphery networks emerge when all nodes belong to a single group. Some nodes are well connected while others, while still being part of that group, are not. In a pure idealized core-periphery network the nodes can be classified strictly into two classes. The core nodes are the one with a high degree of interconnectedness. The periphery nodes are the rest, the ones that are only sparsely connected in the network.

\begin{figure}
\centering
\includegraphics[width=.5\columnwidth]{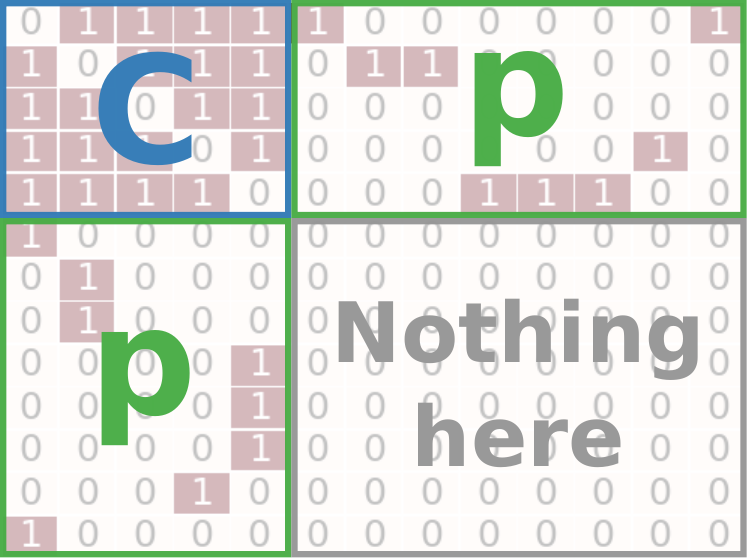}
\caption{A toy example of the Discrete Model for core-periphery structures. This adjacency matrix shows, highlighted in blue, a dense area of the network with many connections. In green, a sparser area: the periphery. Connections only go to (or from) a core member, meaning that in the main diagonal in the peripheral area there are no entries larger than zero.}
\label{fig:cp-discrete}
\end{figure}

There can be only two types of connections: between core nodes -- which is the most common edge type, since the core is densely connected -- and between a core-periphery pair. Peripheral nodes do not connect to each other. In the adjacency matrix, which I show in Figure \ref{fig:cp-discrete}, there's a big area with no connections. This is known as the ``Discrete Model''. It is a very strict one, and rarely real world networks comply with this standard. A perfect discrete model in which the core is composed by a single node is a star.

If you want to detect the core-periphery structure using the discrete model, you have a simple quality measure you want to maximize. This is $\sum \limits_{uv} A_{ij}\Delta_{uv}$, with $A$ being the adjacency matrix, and $\Delta$ a matrix with a value per node pair. An entry in $\Delta$ is equal to one if either of the two nodes is part of the core.

Since $A$ is immutable, your quest is to find the best $\Delta$ such that the sum is maximized, i.e. you are capturing all edges established between a core node and all other nodes in the network. In a network following the perfect discrete model, this sum is equal to $|E|$ the number of edges in a network.

Of course, you cannot compare the ``coreness'' of two different networks unless they have the same number of nodes and edges, because this measure will take different expected values. That is why, sometimes, you can simply calculate the Pearson correlation coefficient between $A$ and $\Delta$.

Without going into details of specialized algorithms, one can find the best $\Delta$ using classical randomization algorithms. Options are genetic algorithms, simulated annealing, or basin hopping.

\subsection{Continuous}
Reality rarely conforms with strict expectations. Having only two classes in which to put nodes is exceedingly restrictive. What if nodes can be sorted in three classes? What if a semi-periphery exists? This is an enticing opportunity, until you realize that you could also ask: why three classes? Why not four? Why not five? Why not... you get the idea.

\begin{figure}
\centering
\includegraphics[width=.5\columnwidth]{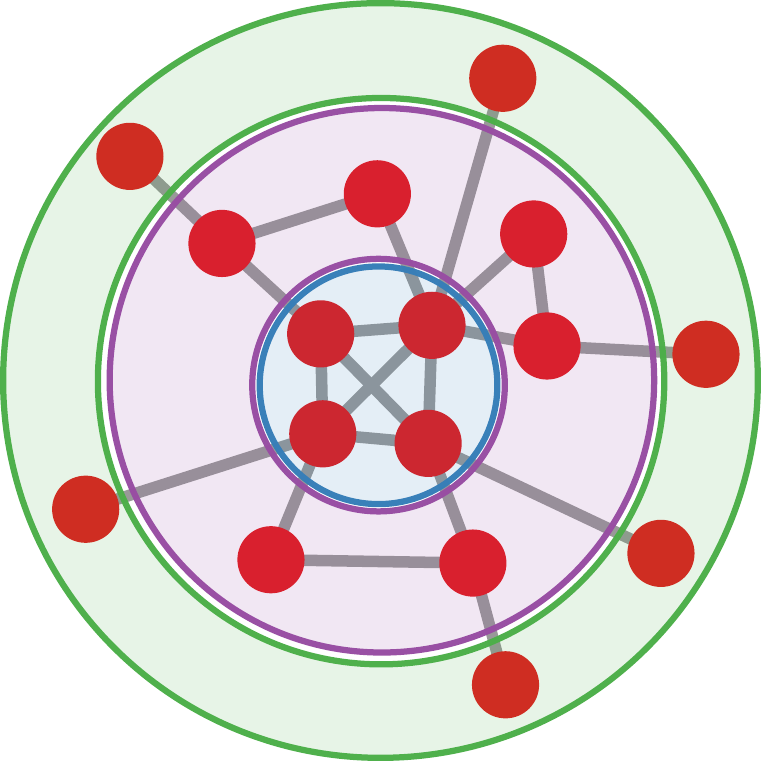}
\caption{A toy example of the Continuous Model for core-periphery structures. This network shows a densely connected core (blue), a pure periphery whose nodes do not connect to each other (green), and an intermediate stage which is not dense enough to be part of the core, but whose nodes still connect to each other (purple).}
\label{fig:cp-continuous}
\end{figure}

This is where the Continuous Model comes into play. Looser than the Discrete Model, in the Continuous Model we have an arbitrary number of classes for the nodes beyond the core. The intermediate classes can interconnect with each other and with the periphery proper, just in a looser way than the core proper. I show an example in Figure \ref{fig:cp-continuous}.

To be more precise, rather than creating an arbitrary number of classes, we assign to each node a ``coreness'' value. Mathematically speaking, the difference with the discrete model is tiny. The quality measure we want to optimize is still $\sum \limits_{uv} A_{uv}\Delta_{uv}$. However, now the entries of $\Delta$ are not binary any more. Instead we have $\Delta_{uv} = c_u c_v$, with $c_u$ being the coreness value of node $u$.

For instance, we could say that, in Figure \ref{fig:cp-continuous}, the nodes in the blue circle have coreness equal to $1$, the ones in purple have coreness equal to $0.5$, and the rest has coreness equal to $0$. If you force nodes to have a coreness of either $1$ or $0$, you're back in the discrete model.

How could one establish the $c_u$ values for the Continuous Model? I already mentioned the k-core centrality algorithm from Section \ref{sec:centr-kcore} in this chapter. This could be a good choice, because we know that nodes with low k-core centrality have a low degree, while nodes with a high k-core value tend to connect to each other in a core.

Of course, that is an a priori approach: it fixes your $\Delta$, which may or may not fit your data well. The alternative is to use a technique similar to the ones I mentioned before, to build your $\Delta$ via the $c$ vector in such a way that your quality measure is maximized.

\subsection{Other Approaches}
The continuous model is powerful, but it doesn't really tell you much on how you should build your $c_i$ values. Rombach et al.\cite{rombach2014core} propose a way to build such vector, introducing two parameters, $\alpha$ and $\beta$. $\beta$ determines the size of the core, from the entirety of the network to an empty core. $\alpha$ regulates the $c$ score difference between the core classes. If a node $u$ at a specific core level has a score $c_u$, the node $v$ at the closest highest core level will have $c_v = \alpha + c_u$ -- or, really, any function taking $\alpha$ as a parameter.

\begin{figure}
\centering
\begin{subfigure}{.3\textwidth}
\includegraphics[width=\textwidth]{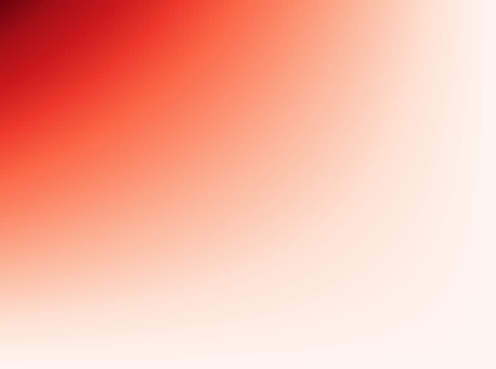}
\caption{$\Delta_{uv} = c_u c_v$}
\end{subfigure}\quad
\begin{subfigure}{.3\textwidth}
\includegraphics[width=\textwidth]{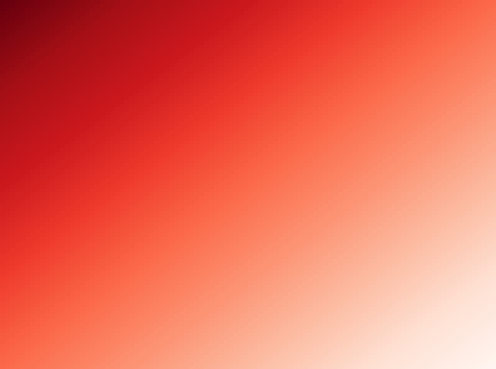}
\caption{$\Delta_{uv} = c_u + c_v$}
\end{subfigure}\quad
\begin{subfigure}{.3\textwidth}
\includegraphics[width=\textwidth]{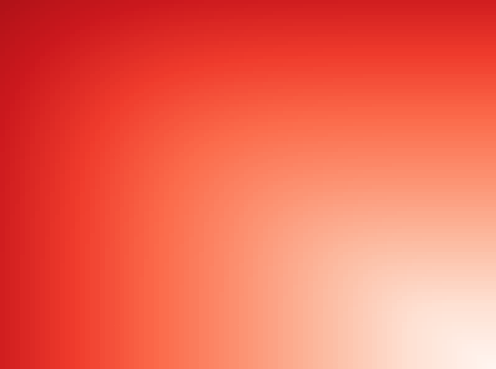}
\caption{$\Delta_{uv} = \sqrt[5]{c_u^5 + c_v^5}$}
\end{subfigure}
\caption{Some matrix masks you can use to build $\Delta$. The darkness of the color is directly proportional to how much the core value combination contributes to $\Delta$. Matrices are sorted so that the top-left corner shows the value for $c_u = c_v = 1$ and the bottom-right corner shows the value for $c_u = c_v = 0$.}
\label{fig:cp-delta-matrices}
\end{figure}

Another freedom you can take is to build $\Delta$ differently. In the standard continuous model, we build it via multiplication: $\Delta_{uv} = c_u c_v$. An alternative is to build it via a p-normalization: $\Delta_{uv} = \sqrt[p]{c_u^p + c_v^p}$. The higher $p$, the more weight you're putting into a classic discrete model core. Figure \ref{fig:cp-delta-matrices} shows the different effects of different criteria to build $\Delta$.

Other approaches in the literature make use of the Expectation Mazimization or the Belief Propagation algorithms\cite{zhang2015identification}.

All methods discussed so far detect the core via a statistical inference or the development of a null model. Thus there are issues of scalability when you need to infer the parameters of the model to make the proper inference. Alternative methods exploit the fact that the core is densely connected and nodes have a high degree. Thus, the expectation is that a random walker would be trapped for a long time in the core\cite{ma2015rich}. By analyzing the behavior of a random walker, one could detect the boundary of the core.

What about multilayer networks (Section \ref{sec:extended-multilayer})? Does it make sense to talk about a core in a network spanning multiple interconnected layers? The analysis of some naturally occurring multilayer networks, for instance the brain connectome\cite{battiston2018multiplex}\cite{guillon2019disrupted}, suggest that it is. However, I would say that at the moment of writing this paragraph, a systematic investigation of core-periphery in multilayer networks, along with a general method to detect them, is an open problem in network science.

\section{Tension with Communities}
There is a tension between core-periphery structures (CP) and the classical community discovery (CD) assumption: in CP there isn't space for communities, given that there's only one dense area and everything connects to it. In CD, there's little space for peripheries, and there are multiple cores.

You can see this mathematically. For simplicity, let's consider the discrete model: $\sum \limits_{uv} A_{ij}\Delta_{uv}$. There is a strong correlation between being an high degree node and being in the core: after all, the nodes in the core are highly connected. We'll see that a community is a set of nodes densely connected to each other. Thus, nodes in a community have a relatively high degree and should be considered part of the core. So, for all nodes deeply embedded in a community, $\Delta_{uv} = 1$.

\begin{figure}
\centering
\begin{subfigure}{.35\textwidth}
\includegraphics[width=\textwidth]{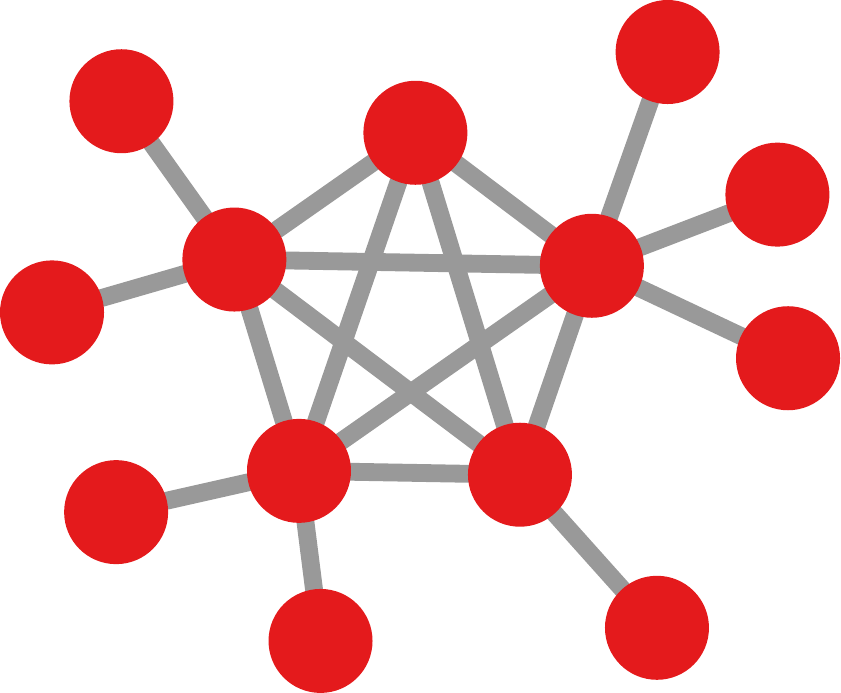}
\caption{}
\end{subfigure}\qquad
\begin{subfigure}{.34\textwidth}
\includegraphics[width=\textwidth]{figures/outline8.pdf}
\caption{}
\end{subfigure}
\caption{(a) A classical core-periphery structure. (b) A classical community structure.}
\label{fig:cp-vs-cd}
\end{figure}

However, a traditional community is also sparsely connected to nodes outside the community. This means that, if nodes $u$ and $v$ are in different communities, likely $A_{uv} = 0$. But we just saw that their $\Delta_{uv}$ should be $1$ because they have high degree! All of that score is wasted! Maximizing the quality function would imply to put all nodes in the same core, sacrificing the defining characteristic of a core: the fact that nodes in it should tend to connect to each other. Figure \ref{fig:cp-vs-cd} shows the difference between the two archetypal meso-scale organizations.

This is problematic since we have evidence that core-periphery structures are ubiquitous, and so are communities. There are a couple of explanations we can use to restore our sanity.

\begin{figure}
\centering
\includegraphics[width=\columnwidth]{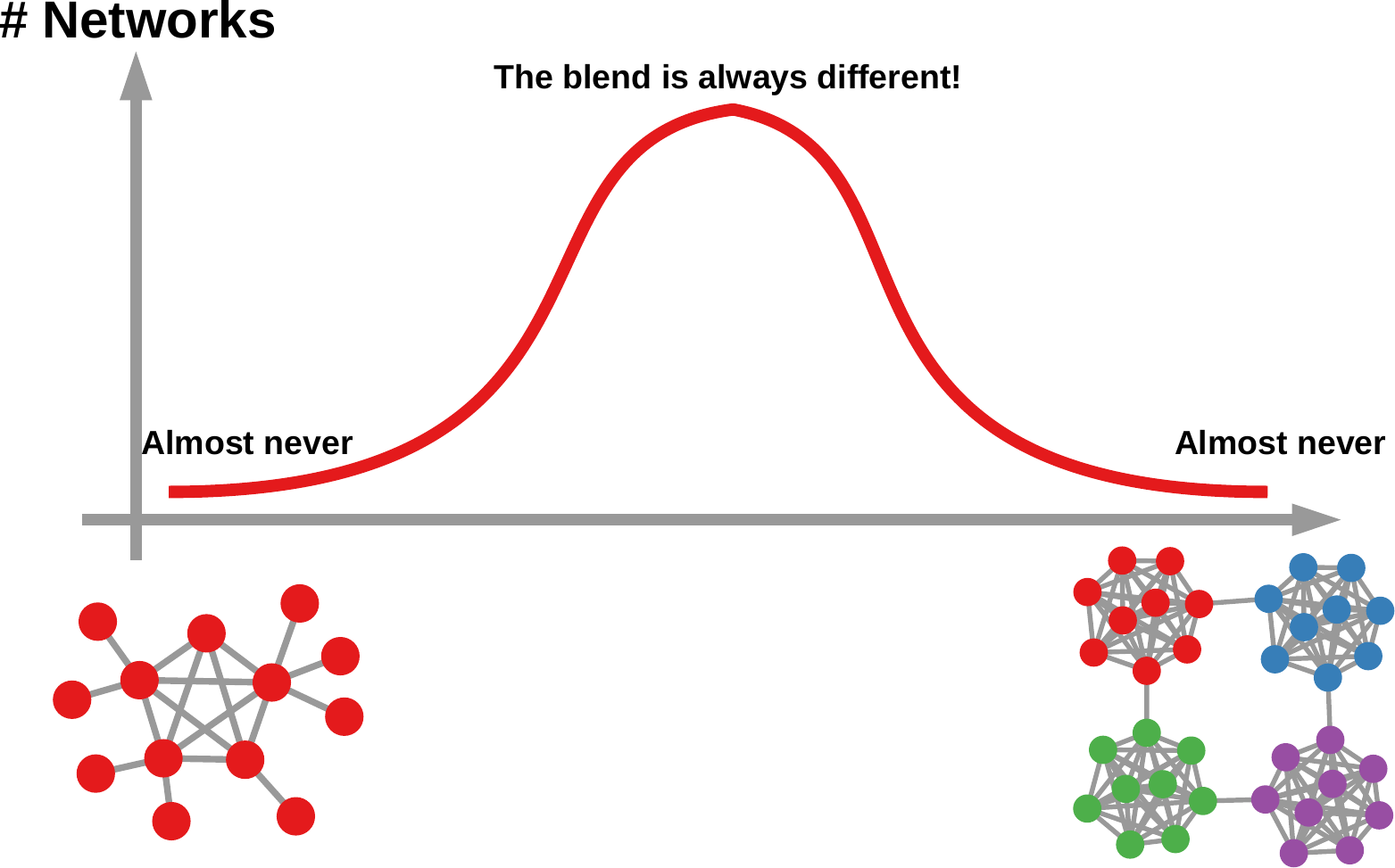}
\caption{The number of networks with a pure core-periphery network and with a pure community structure is actually tiny.}
\label{fig:cp-vs-cd2}
\end{figure}

The first explanation is realizing that every network lives on a core-periphery to community structure continuum. The real world networks we observe distribute through this continuum in such a way that perfect instances are extremely rare -- as Figure \ref{fig:cp-vs-cd2} shows. You'll very rarely find a natural discrete model, exactly as rarely as finding a real world network organizing like a caveman graph (Section \ref{sec:physicsmodels-cm}) -- the quintessential community structure. As a consequence, one way to solve this conundrum is to admit that a network might have multiple cores. Thus, one first performs community discovery to find the multiple cores and then applies the core-periphery detection algorithm independently on each community\cite{kojaku2018core}.

In most cases, networks are in a middle way. Another important thing to keep in mind is that this continuum is not monodimensional -- although on this page I had to squeeze it on a line. The way in which each network blends core-peripheries with communities is always different and difficult to fully characterize.

There are many mechanisms that makes this the case. One is related to overlapping communities (Chapter \ref{cha:ocd}). These communities allow nodes to belong to multiple groups at the same time. These nodes in between communities can be considered a special core of the network\cite{yang2014overlapping}, bringing together the community structure with the core-periphery organization.

Alternative explanations use the power of random walkers to explain core-peripheries\cite{della2013profiling}, an approach that is also commonly used in community discovery. Other models attempt to embed nodes into a spatial dimentions, showing how this can create core-periphery structures\cite{hojman2008core}. It is following this example that I'll try to draw a connection between core-periphery and some well studied aspects of economics in the next section.

\section{Emergence from Social Behavior}
The emergence of core-periphery structures can be observed in many systems. I'll make one example from economic geography.

Consider Hotelling's law\cite{hotelling1929stability}. Suppose that you are on a beach with two ice cream vendors of equal quality. People want ice cream and, since the two vendors offer the same quality, the customers will go to the closest vendor. Therefore, a rational vendor would move their stand so that it can capture the people in the middle. The other vendor would do the same. The solution is an equilibrium in which vendors concentrate in the middle, even though that means increasing the walk length for every customer.

\begin{figure*}
\centering
\begin{subfigure}{.21\columnwidth}
\includegraphics[width=\textwidth]{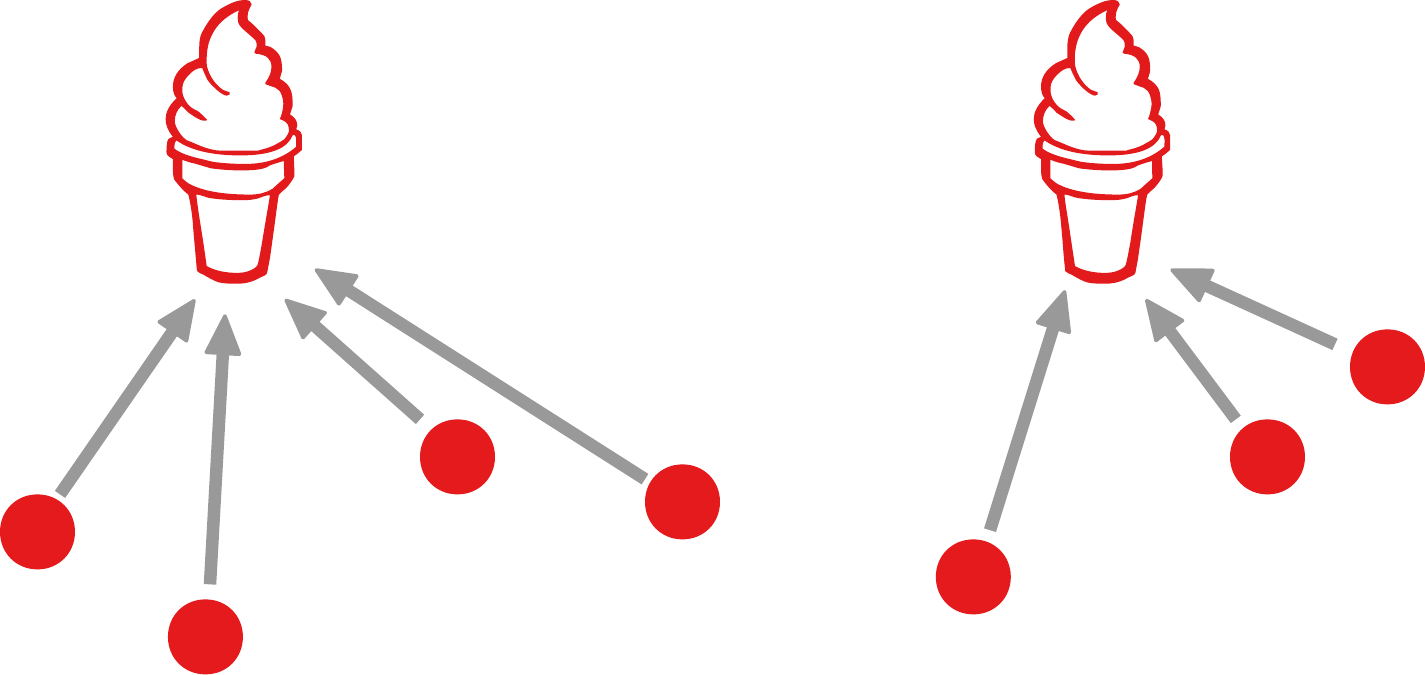}
\caption{$t = 1$}
\end{subfigure}\qquad
\begin{subfigure}{.21\columnwidth}
\includegraphics[width=\textwidth]{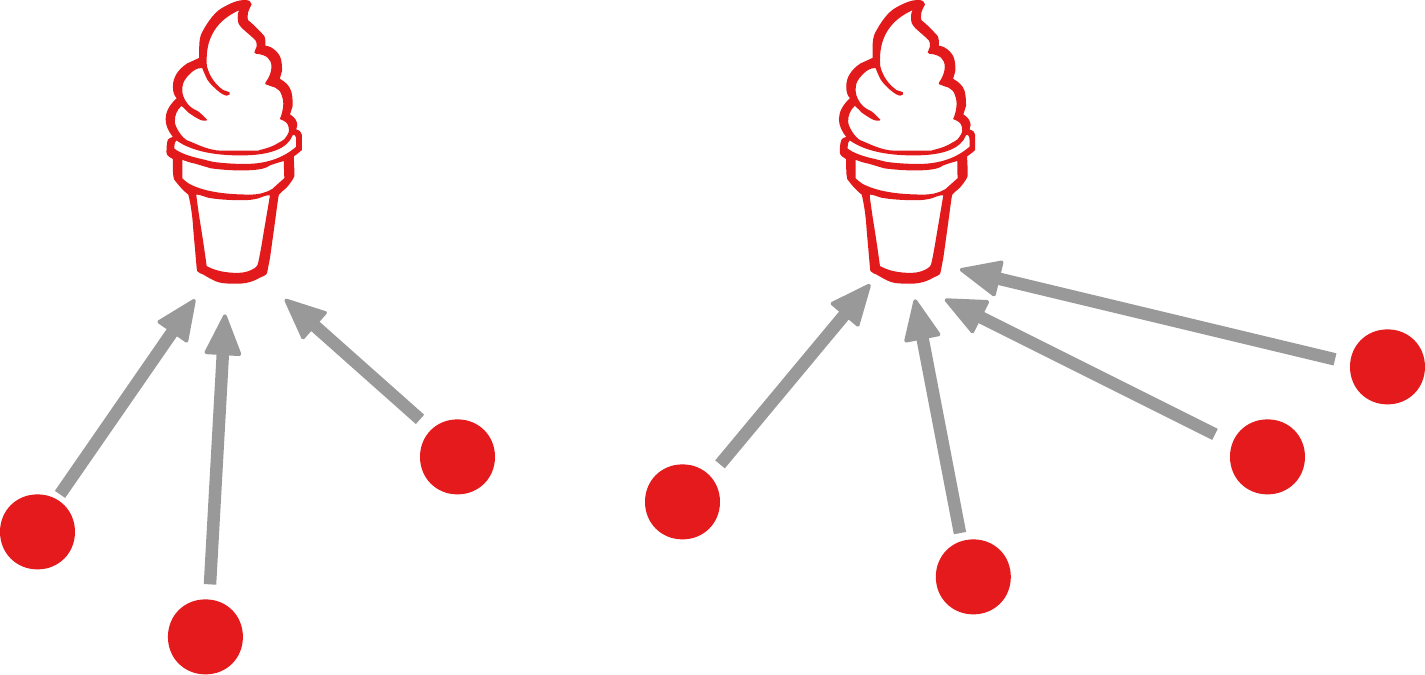}
\caption{$t = 2$}
\end{subfigure}\qquad
\begin{subfigure}{.21\columnwidth}
\includegraphics[width=\textwidth]{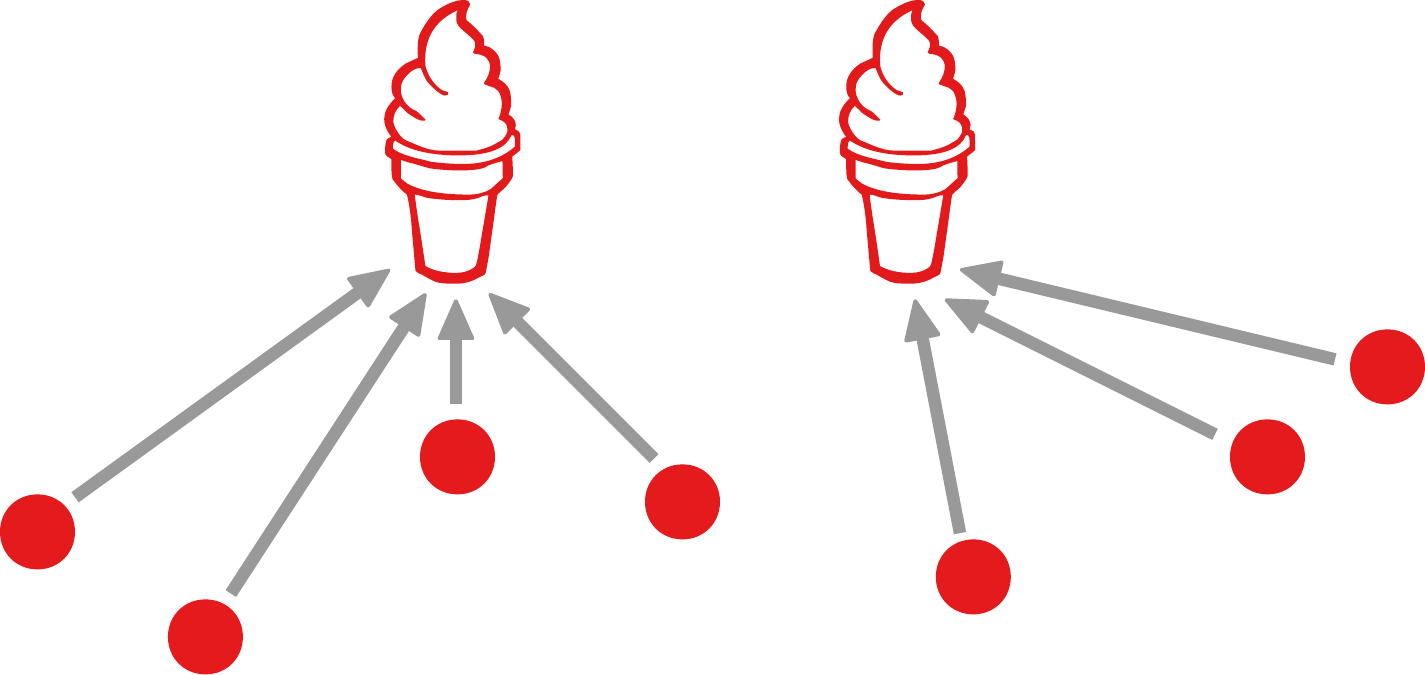}
\caption{$t = 3$}
\end{subfigure}\qquad
\begin{subfigure}{.21\columnwidth}
\includegraphics[width=\textwidth]{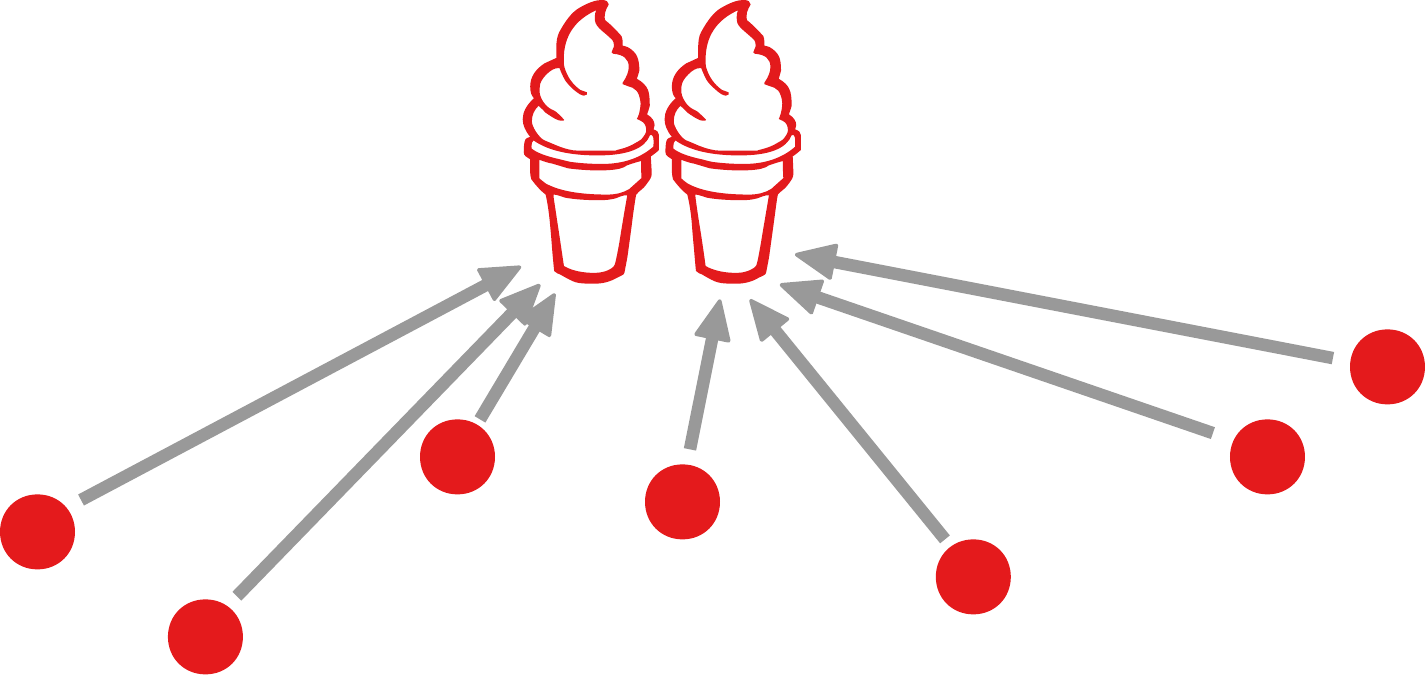}
\caption{$t = 4$}
\end{subfigure}
\caption{A depiction of Hotelling's Law. Each customer (circle) walks to the closest ice-cream stand. As times goes by, the stands get closer and closer, to capture the people in the middle.}
\label{fig:cp-hotelling}
\end{figure*}

Figure \ref{fig:cp-hotelling} captures the law in all of its enraging, negative-sum, glory. The situation hardly changed for the businesses, while the customers are served less efficiently.

This is observed in reality, at multiple levels. Consider a federal government structure, where state governments coalesce into state capitals that need to be visited by their peripheries. In turn, federal capitals provide even higher level services and thus attract people from each state.

This is related to ekistics, the science of human settlements\cite{doxiadis1968ekistics}. In ekistics, Doxiadis shows how improvements in human transportation have introduced a cumulative advantage for the best connected city centers. These are swallowing up their surroundings, sparsifying periphery-periphery connections and attracting all wealth and connections for themselves, creating a centralized rich club. 

\begin{figure}
\centering
\begin{subfigure}{.35\columnwidth}
\includegraphics[width=\textwidth]{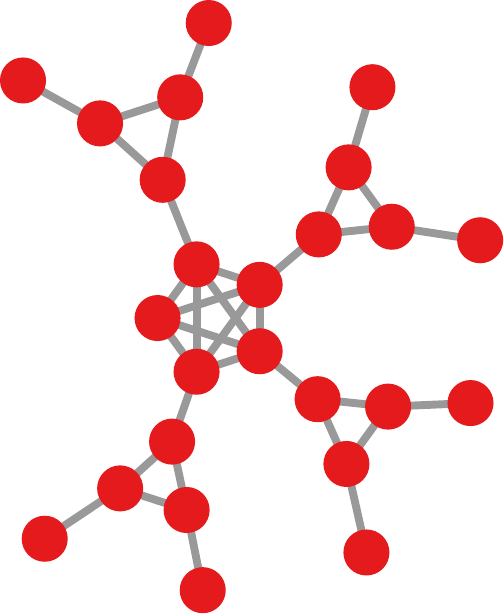}
\caption{}
\end{subfigure}\qquad
\begin{subfigure}{.55\columnwidth}
\includegraphics[width=\textwidth]{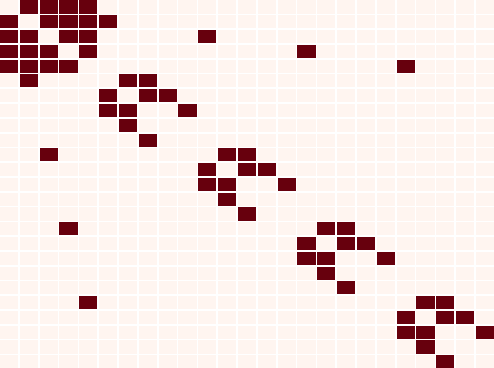}
\caption{}
\end{subfigure}
\caption{A toy example of a network built accordingly to Doxiadis' ekistics. (Left) Graph view: the most central 5-clique is the union of $5$ settlements that make arise the best connected metropolis. The smaller 3-cliques represent other competing cities connecting to their peripheries, which are bound to be eventually absorbed by the core. (Right) Adjacency matrix view.}
\label{fig:centralplace}
\end{figure}

If we look at the types of networks generated with such theory in mind -- for instance Figure \ref{fig:centralplace} --, we realize they do not conform very much to the strict core-periphery structure imposed by the Discrete Model, and they are not quite the same as the Continuous Model, due to their multi-core nature.

\begin{figure}
\centering
\includegraphics[width=.8\columnwidth]{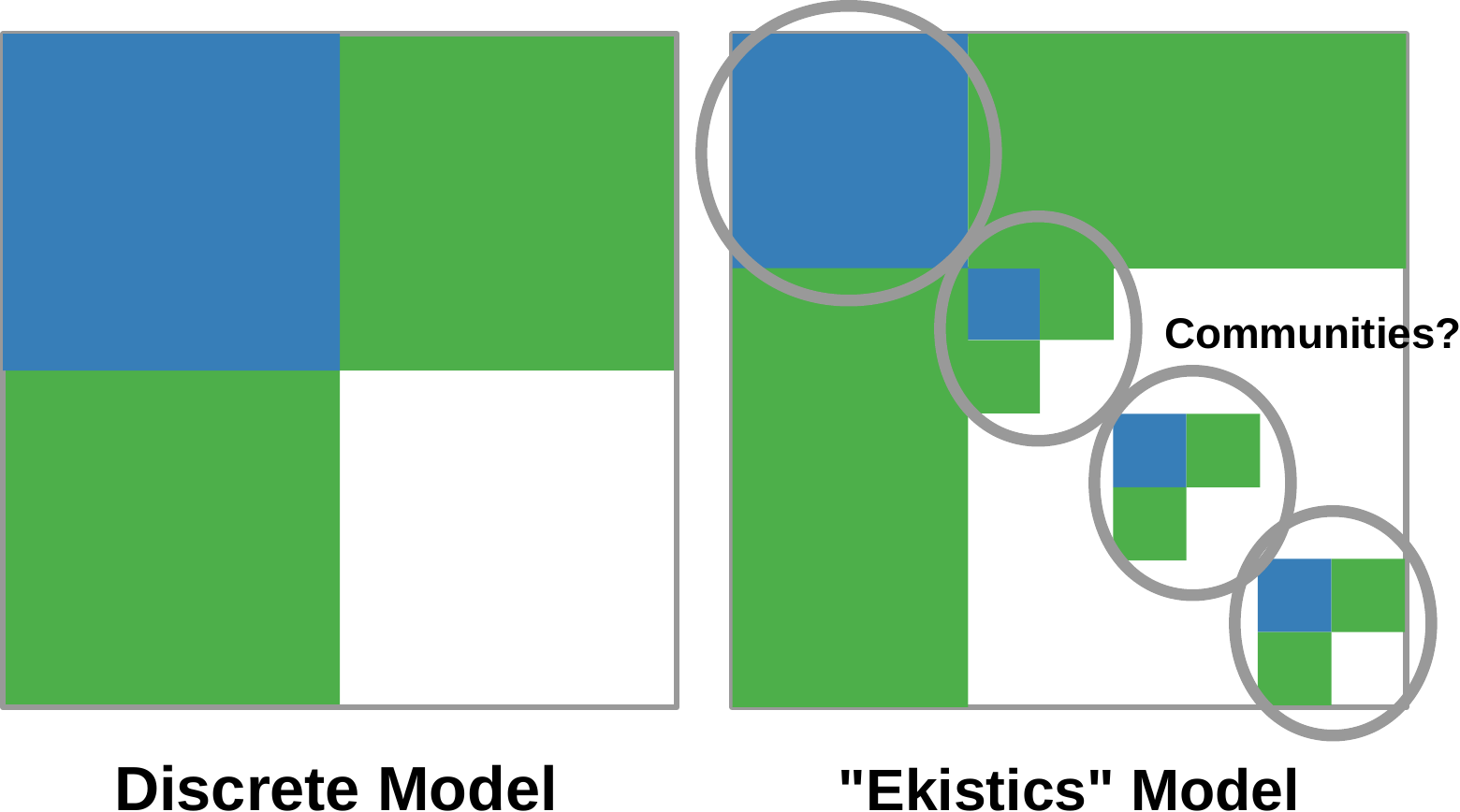}
\caption{Two schematic representations of ideal core-periphery structures: the discrete model on the left and the ekistics model on the right. The blue area represents a densely connected set of nodes, while the green area is a sparser periphery.}
\label{fig:centralplace2}
\end{figure}

However, their schematic structure brings us some sort of closure when it comes to the tension between core-periphery and community discovery. The networks do have a core, and also some sort of communities, coalescing around the secondary cores. Figure \ref{fig:centralplace2} shows a possible model representing a temporary core-periphery organization -- which ekistics predicts to eventually collapse into a single core. These networks are very similar to the Kronecker graphs we saw in Section \ref{sec:csmodels-comms}, in all their fractal glory.

The Central Place Theory in economic geography\cite{christaller1966central} is an alternative and less pessimistic interpretation of geographical core-periphery structures. It says that settlements function as ``central places'' providing services to surrounding areas. The more sophisticated the service the harder it is to provide, and thus it requires more skill integration and thus more centralization. If we keep centralizing sophisticated services, we end up with a high-level central core, several secondary lower level cores and various peripheries, in a fractal way. However, the conceptual jump to core-periphery structures is larger here, because to explain central place we need to introduce this service-providing system -- with sophistication and skills --, while ekistics uses the simpler cumulative advantage mechanics that we already know are a part of many networked systems.

\section{Nestedness}\label{sec:coreper-nestedness}
Core-periphery structures are a generalization of a specific meso-scale organization of complex systems that is relevant in multiple fields: nestedness\cite{lee2016network}. A nested system is one where the elements containing few items only contain a subset of the items of elements with more items.

In terms of networks, an ideal nested system has a hub which is connected to all nodes in the network. The node with the second largest degree is connected to a subset of the neighbors of a hub. The third largest degree node is connected, in turn, to a subset of the neighbors of the second node. It rarely connects to nodes not connected to its antecedent in this hierarchy\cite{bascompte2003nested}.

Nestedness was originally developed in ecology studies\cite{mac1967theory}\cite{patterson1987principle}\cite{bastolla2009architecture} -- specifically biogeography. It originated from an observation of related ecosystems. Suppose you have a mainland with a set of species. Then you have an archipelago, with many islands at an increasing distance from the coast. The closest island contains all the species able to cross water. The second island contains species that can cross water and have a slightly higher range. The third island contains only species with a further increased range, and so on. Clearly, the species which did not make it to the second island are unable to get to the third. Thus the third island is a perfect subset of the second.

I should point out that this explanation of nestedness is not the only one in literature: other authors suggest that nestedness could arise simply from the degree distribution\cite{payrato2019breaking}, and that there are fewer nested system than we originally thought.

\begin{figure}
\centering
\includegraphics[width=\columnwidth]{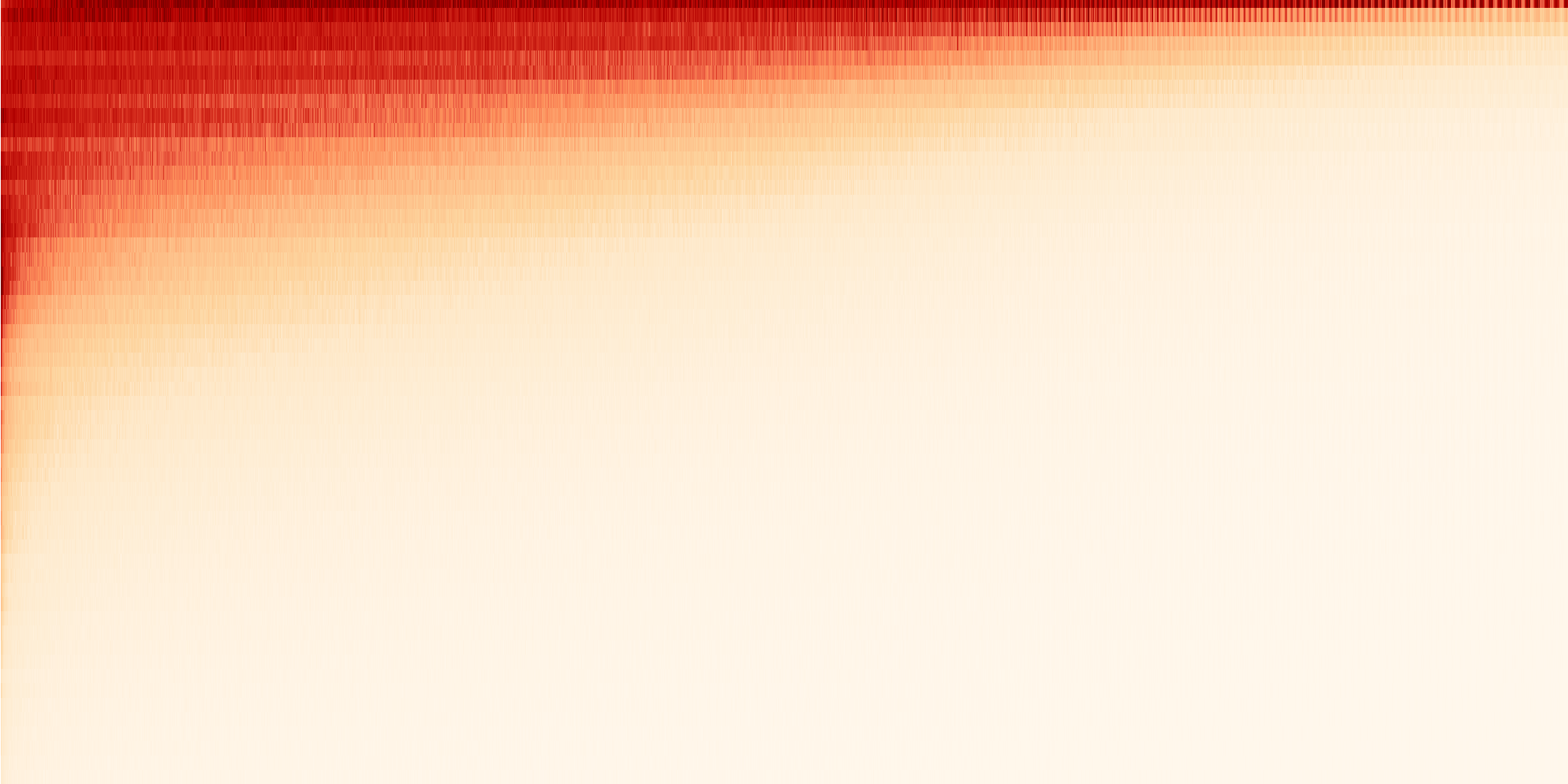}
\caption{A matrix view of a nested network. The matrix has species on the columns and ecosystems on the rows.}
\label{fig:nestedness}
\end{figure}

A typical nested network is bipartite. One node type, in our ecology case, is the species, and the other is the ecosystem. If you were to plot the adjacency matrix of such network, you'd end up with a picture that looks like Figure \ref{fig:nestedness}. To highlight the nested pattern, you want to plot the matrix sorting rows and columns by their sum in descending order. If you do so, most of the connections end up in the top-left of the matrix. That is why we often call nested matrices ``upper-triangular''.

Just like in the discrete model, it's also rare for a real world network to be perfectly nested. Thus, researchers developed methods to calculate the degree of nestedness of a matrix\cite{atmar1993measure}\cite{guimaraes2006improving}\cite{jonhson2013factors}. I'm going to give you a highly simplified view of the field.

These measures are usually based on the concept of temperature, using an analogy from physics. A perfectly nested matrix has a temperature of zero, because all the ``particles'' (the ones in the matrix) are frozen where they should be: in the upper-triangular portion. Every particle shooting outside its designated spot increases the temperature of the matrix, until you reach a fully random matrix which has a temperature of $100$ degrees.

A key concept you need to calculate a matrix's temperature is the isocline. The isocline is the ideal line separating the ones from the zeroes in the matrix. You should try to draw a line such that most of the ones are on one side and most of the zeros are on the other. Usually, there are two ways to go about it.

The parametric way is to figure out which function best describes the curve in your upper triangular matrix: a straight line, a parabolic, a hyperbolic curve, the ubiquitous and mighty power-law. Then you fit the parameters so that the isocline snugs as close as possible to said border.

The non-parametric way is to simply create a jagged line following the row sum or the column sum. If your row (or column) sums to $50$, then you expect a perfectly nested matrix to have $50$ ones followed only by zeros. So your isocline should pass through that point.

\begin{figure}
\centering
\begin{subfigure}{.45\columnwidth}
\includegraphics[width=\textwidth]{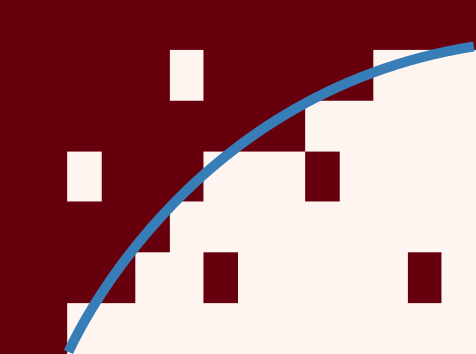}
\caption{}
\end{subfigure}\qquad
\begin{subfigure}{.45\columnwidth}
\includegraphics[width=\textwidth]{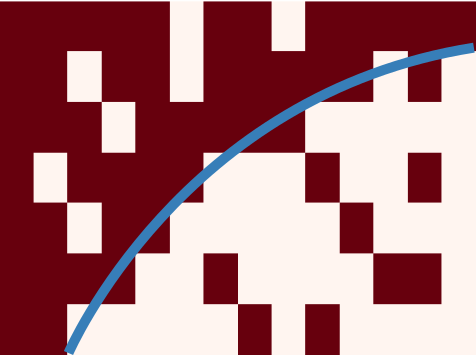}
\caption{}
\end{subfigure}
\caption{Two matrices at a different level of nestedness: (a) high nestedness (low temperature), (b) low nestedness (high temperature). The line in blue is the best isocline.}
\label{fig:nestedness2}
\end{figure}

Once you have your isocline, you can simply calculate how many mistakes you made: how many ones are on the side of the zeroes, and vice versa? Figure \ref{fig:nestedness2} shows an example of the different levels of nestedness two toy matrices can have.

I'm mentioning nestedness in this book because I have encountered it a suprisingly high amount of times in my research, in complex systems that have nothing to do with ecology -- economics for instance.

Some colleagues built a bipartite network connecting countries to the products they can export successfully in the global market\cite{bustos2012dynamics}\cite{cristelli2015heterogeneous}. The most competitive export-oriented economies (Japan, Germany, ...) are able to have an export edge in almost any product, no matter how complex. As you go down the ladder of competitiveness, the countries are able to export a subset of what their immediate higher ranked country can. When you get to the least diversified economies in the world, you end up with countries that are able to export only the products that every country in the world can make.

I personally found nested patterns in a supermarket matrix, with customers and products as the two node types\cite{pennacchioli2014retail}. A connection goes from a customer to the product they bought in significant quantities. Nestedness emerges as there are different customer types, organized in a continuum: from those who buy everything they need in the shop we studied, to those who only buy immediate necessities. 

Note that I constantly used adjacency matrix pictures to convey the ideas behind core-peripehry and communities -- from Figure \ref{fig:cp-discrete} to Figure \ref{fig:nestedness2}. This is because one could unify core-peripehry and communities in a single model using a mixing matrix, which is at the basis of using stochastic blockmodels (Section \ref{sec:csmodels-comms}) to find communities (Section \ref{sec:cd-partition-sbm}) and/or cores in networks.

\section{Summary}

\begin{enumerate}
\item A core-periphery structure is a meso-level organization of a complex network in two parts: one set of nodes densely interconnected with each other (the core) and a sparsely connected set of nodes with just one or few connections per node to the core (the periphery).
\item The discrete model allows to detect such structures by penalizing periphery-periphery connections. Other approaches, such as the continuous model, are more flexible and allow for varying degrees of ``coreness''.
\item Core-periphery structures are ubiquitous just as community structures are, yet a pure core-periphery structure is incompatible with the general notion of community. In reality, these two meso-scale organizations of complex networks co-exist on a spectrum.
\item Many real world dynamics may be at the basis of a core-periphery structure. For instance, geographical agglomeration -- i.e. the creation of a localized core -- makes sense when combining skills to provide complex services in an economy.
\item Nestedness in ecology and economics is another classical core-periphery structure for bipartite networks. In an archipelago, islands with the most species have all species, while islands with few species only have the species that are present in all islands.
\end{enumerate}

\section{Exercises}

\begin{enumerate}
\item Install the \texttt{cpalgorithm} library (\texttt{sudo pip install cpalgorithm}) and use it to find the core of the network using the discrete model (\texttt{cpa.BE}) and Rombach's model (\texttt{cpa.Rombach}) on the network at \url{http://www.networkatlas.eu/exercises/32/1/data.txt}. Use the default parameter values. (Warning, Rombach's method will take a while) Assume that Rombach's method puts in the core all nodes with a score higher than 0.75. What is the Jaccard coefficient between the cores extracted with the two methods?
\item The network at \url{http://www.networkatlas.eu/exercises/32/2/data.txt} has multiple cores/communities. Use the Divisive algorithm from \texttt{cpalgorithm} to find the multiple cores in the network.
\item \url{http://www.networkatlas.eu/exercises/32/3/data.txt} contains a nested bipartite network. Draw its adjacency matrix, sorting rows and columns by their degree.
\end{enumerate}

\chapter{Hierarchies}\label{cha:hier}
We usually represent organizations with networks. Each person in the company is a node. Directed edges connect nodes. They usually flow from the superior to the subordinate, from the coordinator to its team. These networks have a particular structure. In an ideal organization, there is a single head. If this were a corporation, that would be the CEO. The head commands a small group, its top level executives. They, in turn, have their own team of managers. The managers command their own groups, and so on and so forth, until we get to the foot soldiers.

Hierarchical networks arise not only in social systems\cite{clauset2015systematic}, but also -- and especially -- in biological ones\cite{ravasz2002hierarchical}\cite{ravasz2003hierarchical}\cite{yu2006genomic}\cite{vazquez2004topological}\cite{csermely2013structure2}\cite{johnson2017looplessness}.

This is some sort of core-periphery structure (Chapter \ref{cha:coreperiph}), with a few distinctions.

First, in core-periphery there's still some degree of horizontal connections. People at the same level are able to connect to each other. In a perfect hierarchy that is not the case: horizontal connections are banned. You can assign each worker to a level, and workers can only connect to lower level -- and being connected by higher levels.

Second, usually hierarchical networks are directed, while core-periphery normally doesn't care all that much about the direction of the edges.

We can then consider hierarchies as some sort of special case of core-periphery structures. In this chapter we'll explore the concept of hierarchical networks, as it can be interpreted in multiple ways. We're then going to introduce key concepts -- and the main methods using such concepts -- to estimate the ``hierarchicalness'' of a directed network.

\section{Types of Hierarchies}
When talking about ``hierarchies'' in complex networks, researchers mainly refer to three related but distinct concepts. We can classify them in three categories: order, nested, and flow hierarchy\cite{mones2012hierarchy}. I'll present the three of them in this section, noting how this chapter will then only focus on flow hierarchy. Order and nested hierarchies are covered elsewhere in this book with different names.

\subsection{Order}
In an order hierarchy, the objective is to determine the order in which to sort nodes. We want to place each node to its corresponding level, according to the topology of its connections. Usually, this is achieved by calculating some sort of centrality score. The most central nodes are placed on top and the least central on the bottom.

\begin{figure}
\centering
\begin{subfigure}{.45\columnwidth}
\includegraphics[width=\textwidth]{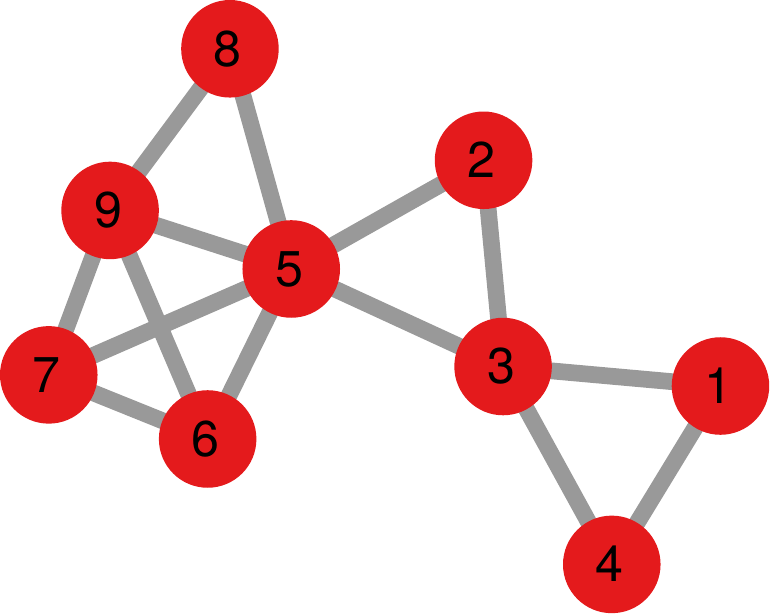}
\caption{}
\end{subfigure}\qquad
\begin{subfigure}{.45\columnwidth}
\includegraphics[width=\textwidth]{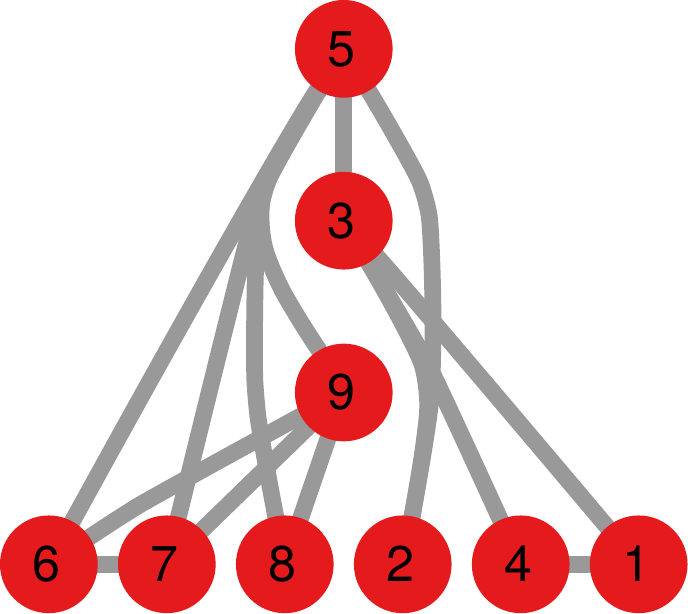}
\caption{}
\end{subfigure}
\caption{(a) A toy network. (b) Its order hierarchy. I place nodes in descending order of betweenness centrality from top to bottom.}
\label{fig:hier-order}
\end{figure}

Figure \ref{fig:hier-order} provides an example of detecting an order hierarchy in a toy network, using betweenness centrality as the guiding principle. Node $5$ has the highest betweenness centrality, followed by node $3$ and then node $9$. All other nodes have the same betweenness centrality -- equal to zero.

Perhaps, the most famous example in this class is SpringRank\cite{de2018physical}. SpringRank only works for directed networks. It sees an $u \leftarrow v$ edge as a sign that node $u$ is above $v$ in the hierarchy. SpringRank tries to put all nodes in a specific height -- $h_u$ and $h_v$ for nodes $u$ and $v$ respectively. As the name suggests, it sees each edge as a spring. Therefore, for any positioning $h_u$ and $h_v$ of the nodes, it can calculate the energy in the spring, which is $H_{uv} = \frac{1}{2}(h_u-h_v-1)^2$ -- this is the literal amount of energy in a literal spring of that literal size, it's a purely physical approach. Once you have the energy for a pair of nodes, you can calculate the energy for the entire network, which is the sum of the energies of all edges that exist or, mathematically:

$$ H_G = \sum \limits_{u,v \in V} A_{uv}H_{uv} = \dfrac{1}{2}\sum \limits_{u,v \in V} A_{uv}(h_u-h_v-1)^2,$$

the $A_{uv}$ term is the adjacency matrix and will cancel to zero the energies of edges that don't exists, because in that case $A_{uv} = 0$. Then SpringRank finds a few clever ways to minimize this quantity, which involve the pseudoinverse of the Laplacian we discussed in Section \ref{sec:rw-effectres}.

In general, solutions to the order hierarchy result in a continuous value of each node in the network that puts it on the y axis on a line. This is quite literally equivalent to finding a \textit{ranking} of nodes in the network. One can easily see that we already covered this sense of hierarchical organization of complex networks. The order hierarchy is nothing more than a different point of view of node centrality. Thus, I refer to Chapter \ref{cha:ranks} for a deeper discussion on the topic.

\subsection{Nested}
Nested hierarchy is about finding higher-order structures that fully contain lower order structures, at different levels ultimately ending in nodes. In the corporation example, the largest group is the corporation itself, encompassing all workers. We can first subdivide the corporation into branches, if it is a multinational, they could be regional offices. Each office can be broken down into different departments, which have teams and, finally, the workers in each team.

\begin{figure}
\centering
\begin{subfigure}{.45\columnwidth}
\includegraphics[width=\textwidth]{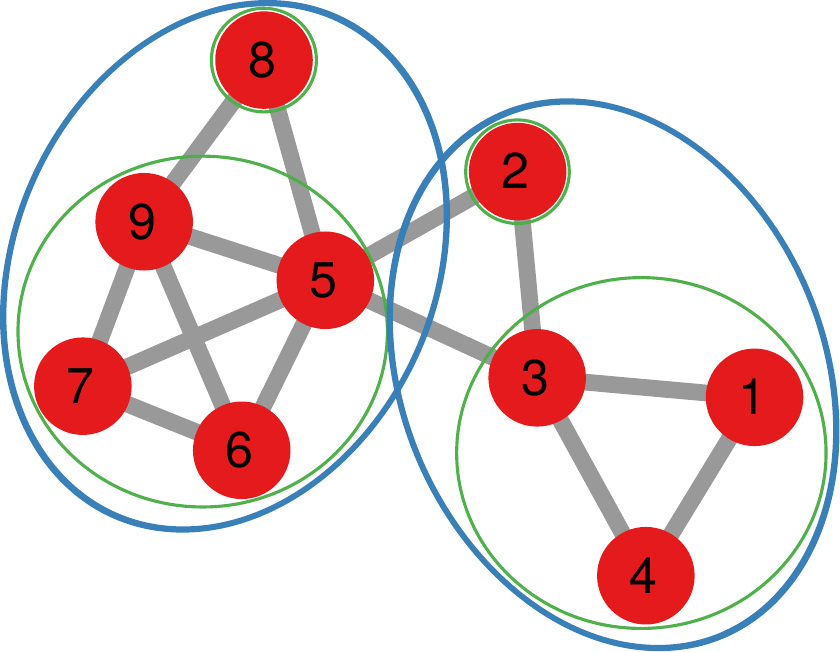}
\caption{}
\end{subfigure}\qquad
\begin{subfigure}{.45\columnwidth}
\includegraphics[width=\textwidth]{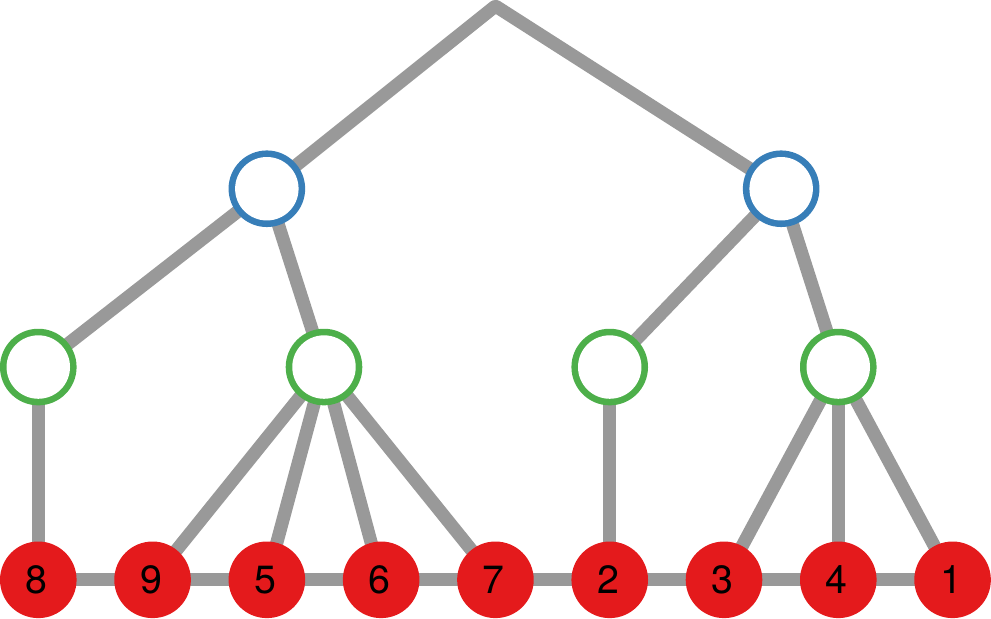}
\caption{}
\end{subfigure}
\caption{(a) A toy network. Colored circles delineate nested substructures. (b) Its nested hierarchy, according to the highlighted substructures. Each node and substructure is connected to the substructure it belongs to.}
\label{fig:hier-nested}
\end{figure}

Figure \ref{fig:hier-nested} provides an example of detecting a nested hierarchy in a toy network. This is usually done by detecting smaller and smaller densely connected units in the network. Note how, in this case, the hierarchy does not place nodes on levels, but organizes the detected substructures.

This is equivalent to performing hierarchical community discovery on complex networks. Thus, I refer to Chapter  \ref{cha:hcd} for further reading.

\subsection{Flow}
In a flow hierarchy, nodes in a higher level connect to nodes at the level directly beneath it, and can be seen as managers spreading information or messages to the lower levels. We call it a ``flow'' hierarchy because you can see the highest level node as the origin of a flow, which hits first the nodes at the level directly beneath it, and so on until it reaches the leaves of the network: the nodes at the bottom layer.

\begin{figure}
\centering
\begin{subfigure}{.45\columnwidth}
\includegraphics[width=\textwidth]{figures/hierarchy_order.pdf}
\caption{}
\end{subfigure}\qquad
\begin{subfigure}{.45\columnwidth}
\includegraphics[width=\textwidth]{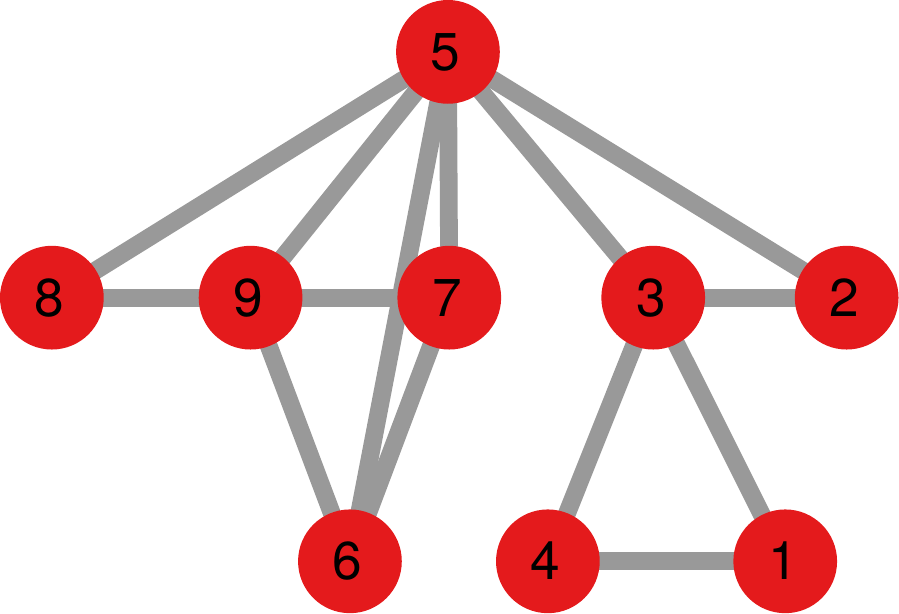}
\caption{}
\end{subfigure}
\caption{(a) A toy network. (b) Its flow hierarchy.}
\label{fig:hier-flow}
\end{figure}

Figure \ref{fig:hier-flow} provides an example of detecting a flow hierarchy in a toy network. Note how it tends to have high centrality nodes on top, like the order hierarchy, but it creates a substantially different organization. Nodes directly connected to a given level tend to belong to the level immediately beneath it, no matter their different centrality values. One could think that the order hierarchy is a special case of flow hierarchy, but that is incorrect: in a flow hierarchy all nodes belonging to a level need to connect to the level directly below and above, while that's not the case for the order hierarchy. Moreover, the order hierarchy is just something we add on top of a node-level measure (centrality), while the flow hierarchy is a meso-level analysis: it describes how groups of nodes -- the ones at different hierarchical levels -- relate to each other.

Since this concept is not covered elsewhere in this book, I focus on flow hierarchies for the rest of this chapter. There are four main approaches to detect hierarchies and quantify the hierarchicalness of real world directed networks. They are: cycle-based Flow Centrality, Global Reach Centrality (GRC), Agony, and Arborescence.

From now on, we always assume that the network we're analyzing is directed.

\section{Cycles}\label{sec:hier-cycles}
I introduced the concept of cycles in Section \ref{sec:paths-cycles}: special paths that start and end in the same node. Cycles are natural enemies of hierarchies. There is no way to have a perfect hierarchy if you have a cycle in your network. In a perfect hierarchy, you can always tell who's your boss. Your boss will never take orders from you, nor from your peer, and even less so from any of your underlings.

However, if you have a cycle, that is not true. A cycle means that your boss gives you an order, you pass it down to one of your underlings and, somehow, they give it back to your boss. This is clear nonsense and should be avoided at all costs.

\begin{figure}
\centering
\begin{subfigure}{.45\columnwidth}
\includegraphics[width=\textwidth]{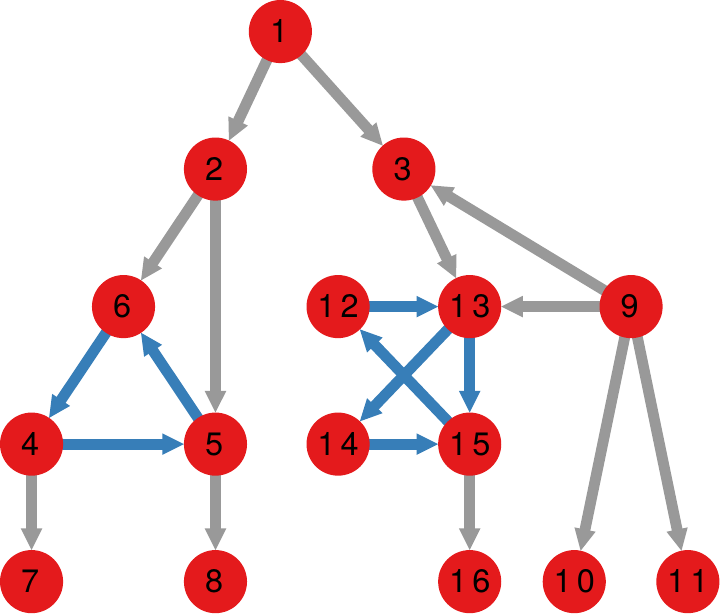}
\caption{}
\end{subfigure}\qquad
\begin{subfigure}{.45\columnwidth}
\includegraphics[width=\textwidth]{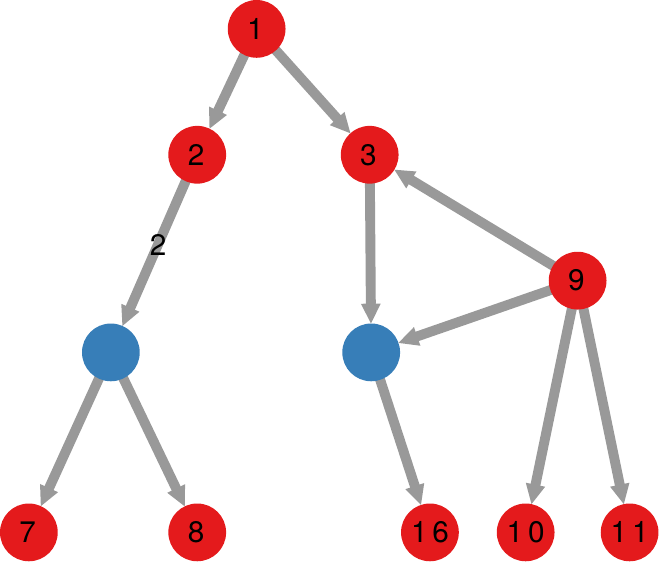}
\caption{}
\end{subfigure}
\caption{(a) A directed network. In the figure, I highlight in blue the edges partaking in cycles. (b) The condensed version of (a), where all nodes part of a strongly connected component are condensed in a node (colored in blue).}
\label{fig:hier-fh}
\end{figure}

Thus, the simplest way to estimate the hierarchicalness of a directed network is to count the number of edges involved in a cycle. The fewer edges are part of a cycle in a network, the more hierarchical it is\cite{luo2011detecting}. You can see in Figure \ref{fig:hier-fh}(a) that the somewhat hierarchical network has only a handful of edges involved in cycles.

You can calculate the flow hierarchicalness of a network by simply condensing the graph. Graph condensation follows two steps. First, you reduce all the graph's strongly connected components each to a single node. Then you connect that node to all nodes that were connected to a node part of that component. When condensing the graph in Figure \ref{fig:hier-fh}(a), you'll obtain the graph in Figure \ref{fig:hier-fh}(b). Note that, if you're merging two edges, you need to keep track of this information, for instance in the edge weight. The $(2,5)$ and $(2,6)$ edges collapse in a single edge outgoing form node $2$, which now must have a weight of two.

The ratio between the sum of the edge weights in the condensed graph and the number of edges in the original graph is the flow hierarchy of your network. The original graph in Figure \ref{fig:hier-fh}(a) had $20$ edges. Since the condensed graph in Figure \ref{fig:hier-fh}(b) has $11$ edges with a total weight sum of $12$ (because of the edge of weight two coming out of node $2$), we can conclude that the network's flow hierarchy is equal to $12 / 20 = 0.6$.

A consequence of this definition is that any directed network composed by a single strongly connected component has a hierarchicalness of zero by definition.

\begin{figure}
\centering
\begin{subfigure}{.35\columnwidth}
\includegraphics[width=\textwidth]{figures/hierarchy_wheel.pdf}
\caption{}
\end{subfigure}\qquad
\begin{subfigure}{.55\columnwidth}
\includegraphics[width=\textwidth]{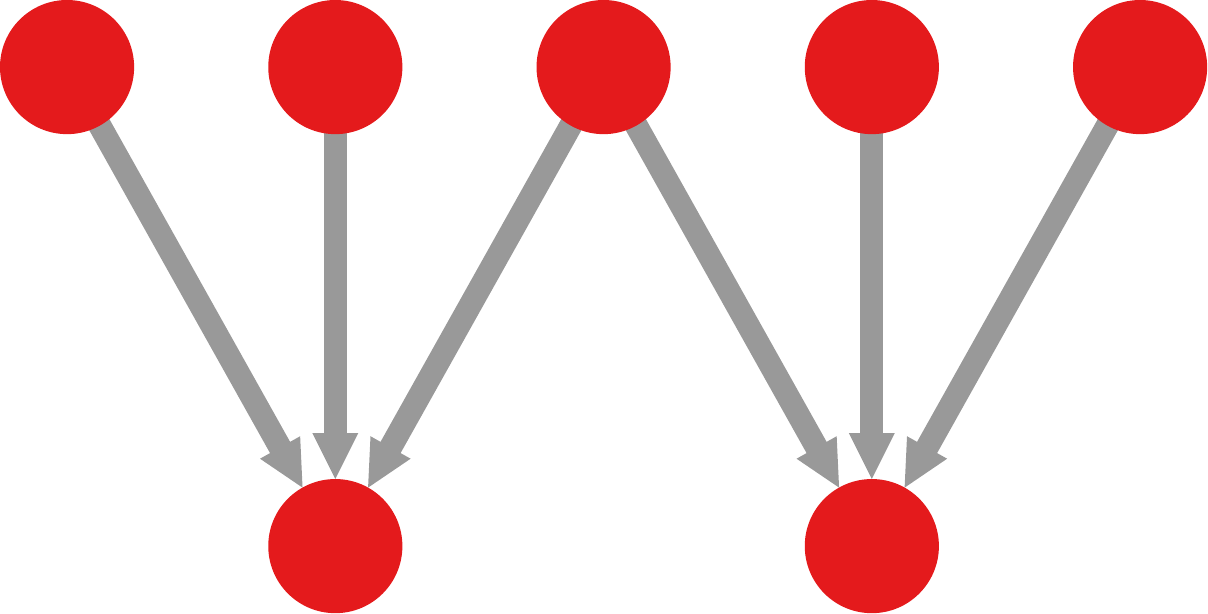}
\caption{}
\end{subfigure}
\caption{Two directed acyclic graphs not conforming to our intuition of a perfect hierarchy. (a) A wheel graph with one flipped edge. (b) A ``hierarchy'' with more bosses than workers.}
\label{fig:hier-fh2}
\end{figure}

Flow hierarchy has a major flaw: it's too lenient. In fact, it says that any and all directed acyclic graphs are perfect hierarchies. It is very easy to construct toy examples of less-than-ideal structures that this cycle-based flow hierarchy will consider perfect. Figure \ref{fig:hier-fh2} shows two of such examples.

In Figure \ref{fig:hier-fh2}(a) we have no cycles because I flipped one edge. However, arguably, one should not be able to go from a perfect hierarchy to less than $50\%$ hierarchicalness by simply flipping one direction. Figure \ref{fig:hier-fh2}(b) shows another case where there are more bosses than workers: you don't need to have a master in management to see that this is no way to run an organization! Yet, since they are both directed acyclic graphs, for this cycle-based definition of hierarchy, both toy examples are ideal hierarchies.

\section{Global Reach Centrality}
I introduced the concept of reach centrality back in Section \ref{sec:centr-reach}: the reach centrality of node $v$ in a directed network is the fraction of nodes it can reach using directed paths originating from itself. This measure is often dubbed ``local'' reach centrality, because the same authors defined a ``global'' reach centrality\cite{mones2013hierarchy} (GRC). GRC is not a measure for nodes any more: it is a way to estimate the hierarchicalness of a network.

The intuition behind GRC is simple. A network has a strong hierarchy if there is a node which has an overwhelming reaching power compared to the average of all other nodes. Or, to put it in other words, if there is an overseer that sees all and knows all. To calculate GRC you first estimate the local reach centrality of all nodes in the network. You then find the maximum value among them, say $LRC_{MAX}$. Then:

$$ GRC = \dfrac{1}{|V| - 1} \sum \limits_{v \in V} LRC_{MAX} - LRC_v.$$

In practice, you average out its difference with all reach centrality values in the network. This is an effective way of counteracting the degeneracy of cycle-based hierarchy measures. In both toy examples from Figure \ref{fig:hier-fh2}, GRC is well behaved, returning values of $0.555$ and $0.22$, respectively.

\begin{figure}
\centering
\includegraphics[width=.75\columnwidth]{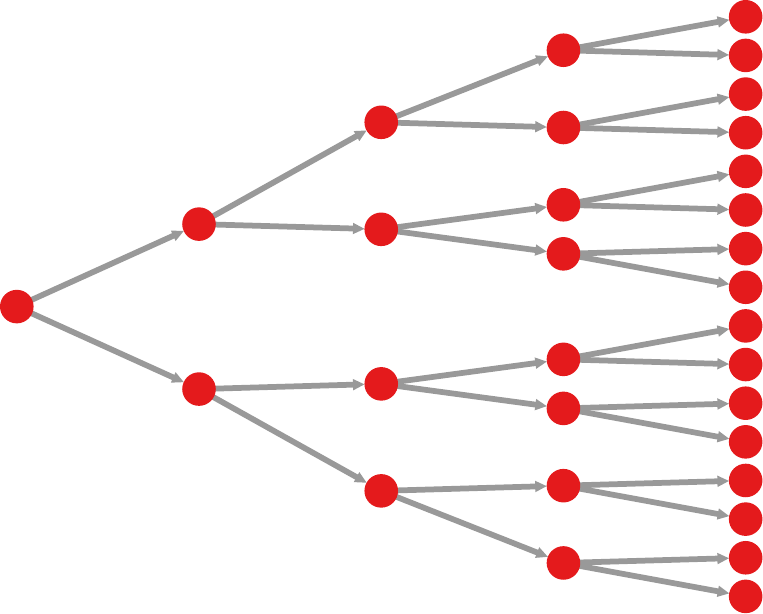}
\caption{A network we could consider a perfect hierarchy, for which GRC fails to give a perfect score.}
\label{fig:hier-grc}
\end{figure}

However, GRC has a blind spot of its own. Since we're averaging the differences between the most central node against all others, we know we will never get a perfect GRC score if there is more than one node with non-zero local reach centrality. Consider Figure \ref{fig:hier-grc}. I don't know about you, but to me it looks like a pretty darn perfect hierarchy. Yet, we know that the two nodes connected by the root don't have a zero local reach centrality. In fact, the GRC for that network is $0.89$.

So, if cycle-based flow hierarchy is too lenient -- every directed acyclic graph is a perfect hierarchy --, GRC is too strict: even flawless hierarchies might fail to get a $100\%$ score. If for cycle-based flow hierarchy the perfect hierarchy is a DAG, for GRC the only perfect hierarchy is a star: a central node connected to everything else, and no other connections in the network.

\section{Arborescences}
We introduced the concept of arborescence in Section \ref{sec:paths-cycles}, as a stricter definition of a directed acyclic graph. To sum up: an arborescence is a directed tree in which all nodes have in-degree of one, except the root, which has in-degree of zero. For instance, Figure \ref{fig:hier-grc} is an arborescence. Given their properties, arborescences seem particularly well suited to inform us about hierarchies. Every arborescence is a perfect hierarchy: all nodes have a single boss, there are no cycles, and there is one node with no bosses -- the CEO.

That is why I used them to create my own hierarchicalness score\cite{coscia2018using}. I take an approach similar to the one developed from the cycle-based flow hierarchy. In fact, the first step is almost identical: take your directed graph and condense all its strongly connected components into a single node. The only difference is that here we ignore edge weights. This gives us a directed acyclic graph version of the original network. Note that there are alternative methods to reduce a generic directed network to a DAG\cite{lu2016exploring}\cite{sun2017breaking}, which can preserve more edges.

To transform it in an arborescence, we need to remove all edges going ``against'' the flow. We cannot allow any node to have an in-degree larger than one. So all edges contributing to a larger-than-one in-degree have to be removed. We cannot remove them at random: we need to remove the ones pointing towards the root and keep the ones pointing away from it. I use closeness centrality to determine which edges to keep: the ones coming from the node with the lowest closeness centrality. This is a bit counter-intuitive: the closer a node is to the root, the lower its closeness centrality is. This is due to the fact that these nodes have more possible paths originating from them, and they tend to be longer because they can reach a larger portion of the network.

Once all edges breaking the arborescence requirements are eliminated, we can count how many connections survived. This is also equivalent to the final step of the cycle-based flow hierarchy. The more edges we needed to remove to obtain an arborescence, the less the original network was resembling a perfect hierarchy. In the example from Figure \ref{fig:hier-fh}, we would preserve nine edges out of $20$, giving us an arborescence score of $9 / 20 = 0.45$. Differently from the cycle-based measure, the arborescence score is not fooled by the examples in Figure \ref{fig:hier-fh2}, returning a score of $0.5$ and $0.33$, respectively.

\begin{figure}
\centering
\begin{subfigure}{.45\columnwidth}
\includegraphics[width=\textwidth]{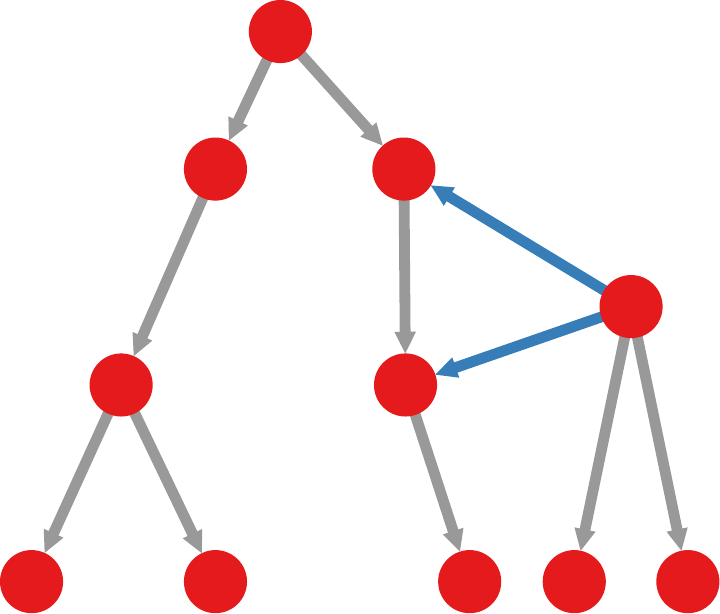}
\caption{}
\end{subfigure}\qquad
\begin{subfigure}{.45\columnwidth}
\includegraphics[width=\textwidth]{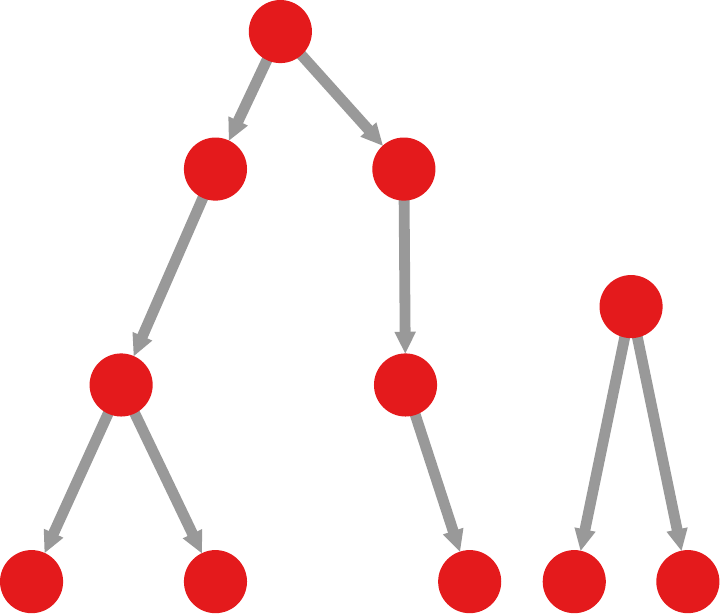}
\caption{}
\end{subfigure}
\caption{The additional step to go from cycle-based hierarchy to arborescence. (a) I highlight in blue the two edges going ``against the flow''. (b) The final result, reduced from Figure \ref{fig:hier-fh}(a): an arborescence forest.}
\label{fig:hier-arb}
\end{figure}

Note from Figure \ref{fig:hier-arb} that, technically speaking, this technique reduces an arbitrary directed network into an arborescence forest, not an arborescence. This is another difference with the cycle-based method, as condensing a graph will never break it into multiple weakly connected components. Arborescence is a very punitive measure, much more than cycle-based flow hierarchy, but less so than GRC.

\section{Agony}
In the agony measure we start from the assumption that we can partially order nodes into levels. The CEO lives at the top of the hierarchy (level $1$), its immediate executive are at level $2$, the managers beneath them are at level $3$, and so on. Let's say that $l_v$ tells us the level of node $v$. In this scenario, a perfect hierarchy only has edges going from a node in a lower (more important) level to a node in a higher level. If $l_u < l_v$ then a $u \rightarrow v$ edge is ordinary and expected. On the other hand, a $u \leftarrow v$ edge will cause ``agony'': something isn't as it is supposed to be.

How much agony does it cause? Well, this is proportional to the level difference between the nodes. You won't be shocked if the CEO accepts orders from another top executive, but you'll go to the madhouse if she does what the most recently hired intern says. In the original paper\cite{gupte2011finding}, the authors define the agony of the $u \leftarrow v$ edge as: $l_v - l_u + 1$. The $+1$ is necessary because, if we were to exclude it, we could put all nodes in the same level and obtain zero agony, which would defeat the purpose of the measure.

Ultimately, this reduces to calculating the result of:

$$ A(G,l) = \sum \limits_{(u,v) \in E} \max(l_v - l_u + 1, 0).$$

Every time $l_u < l_v$, we contribute zero to the sum. Note that agony requires you to specify the $l_u$ value for all nodes in the network. This is not usually something you know beforehand. So the problem is to find the $l_u$ values that will minimize the agony measure. There are efficient algorithms to estimate the agony of a directed graph\cite{tatti2015hierarchies}.

\begin{figure}
\centering
\begin{subfigure}{.45\columnwidth}
\includegraphics[width=\textwidth]{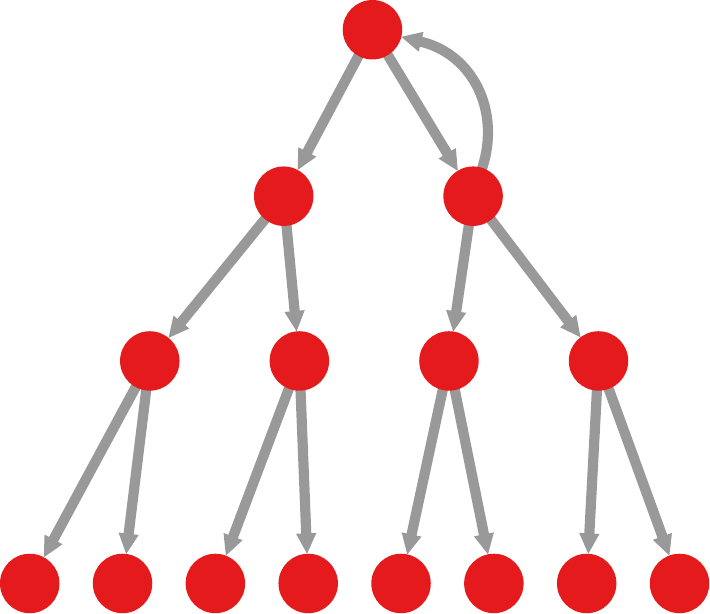}
\caption{}
\end{subfigure}\qquad
\begin{subfigure}{.45\columnwidth}
\includegraphics[width=\textwidth]{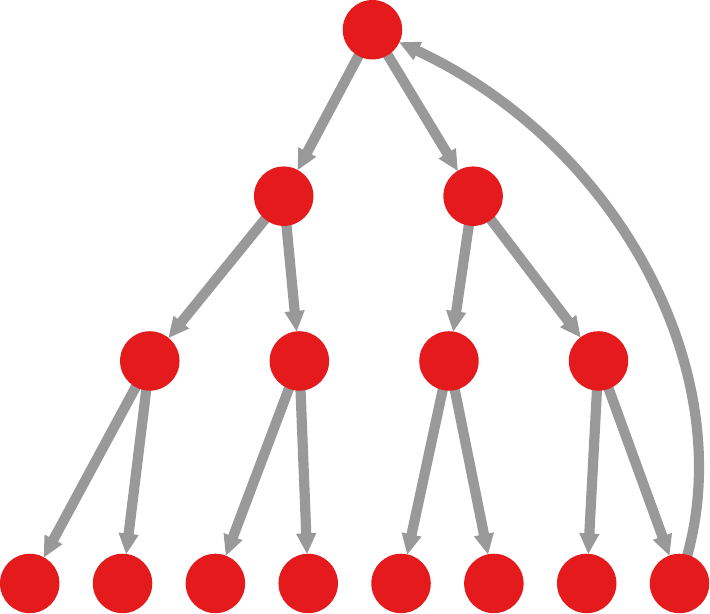}
\caption{}
\end{subfigure}
\caption{Two hierarchies with different values of agony. The vertical positioning of each node determines its level, from top ($l_u = 1$) to bottom ($l_u = 4$).}
\label{fig:hier-agony}
\end{figure}

Consider Figure \ref{fig:hier-agony}. In both cases, we have only one edge going against the flow. Agony, however, ranks these two structures differently. In Figure \ref{fig:hier-agony}(a), the difference in rank is only of one, thus the total agony is $2$. In Figure \ref{fig:hier-agony}(b), the difference in rank is $3$, resulting in a higher agony. Also a cycle-based flow hierarchy measure takes different values, as Figure \ref{fig:hier-agony}(b) involves more edges in a cycle (four edges, versus just two in Figure \ref{fig:hier-agony}(a)).

Ultimately, the resting assumption of agony is the same of the cycle-based flow hierarchy. Agony considers any directed acyclic graph as a perfect hierarchy. Thus it will give perfect scores to the imperfect hierarchies from Figure \ref{fig:hier-fh2}.

\section{Drawing Hierarchies}
To wrap up this chapter, note that all these methods have a various degree of graphical flavor to them. Meaning that you can use them to create a picture of your hierarchy, which might help you to navigate the structure. The most rudimentary method is the cycle-based flow hierarchy, because it just reduces the graph to a DAG, which doesn't help you much.

\begin{figure}
\centering
\begin{subfigure}{.4\columnwidth}
\includegraphics[width=\textwidth]{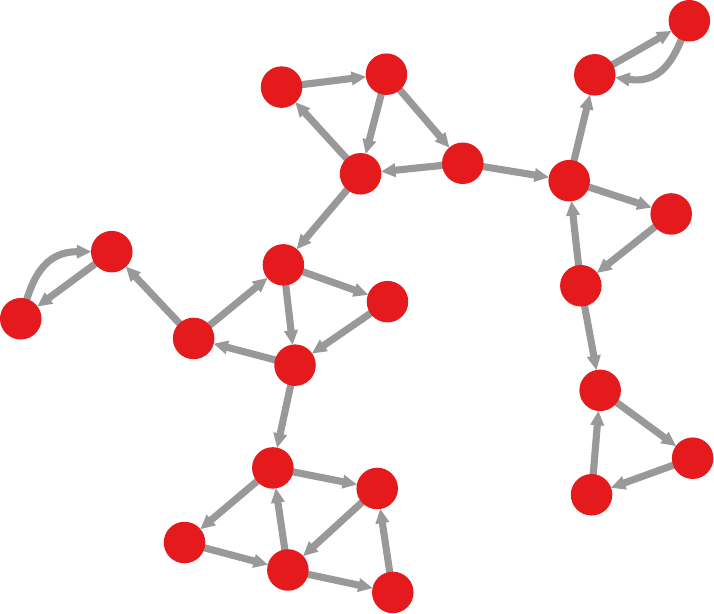}
\caption{}
\end{subfigure}\quad
\begin{subfigure}{.14\columnwidth}
\includegraphics[width=\textwidth]{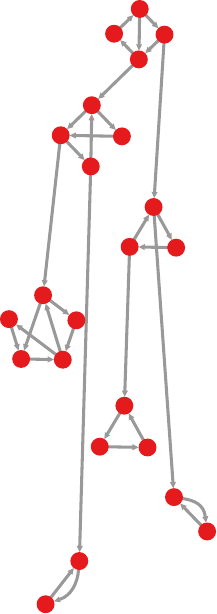}
\caption{}
\end{subfigure}\quad
\begin{subfigure}{.39\columnwidth}
\includegraphics[width=\textwidth]{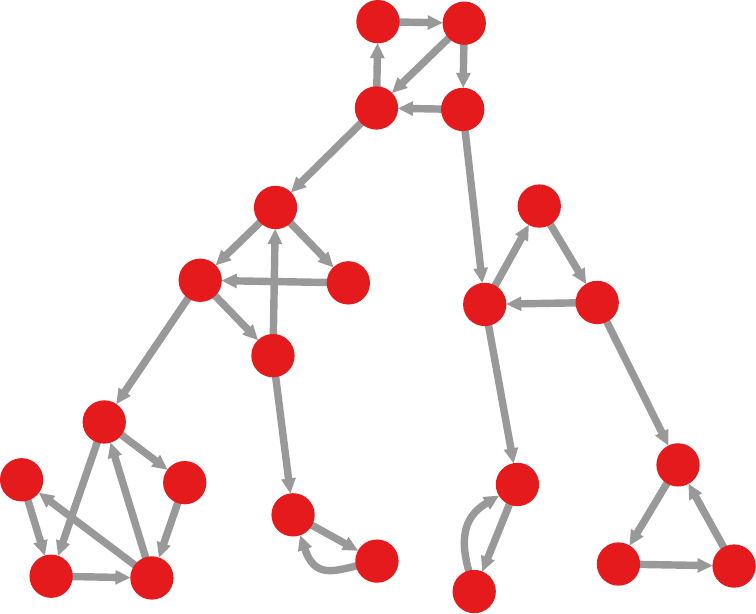}
\caption{}
\end{subfigure}
\caption{(a) A directed graph. (b) The graph from (a), layout according to local reach centrality -- with the most central nodes on top and the least central on the bottom. (c) The graph from (a) layered according to its arborescence scheme or its agony levels -- the two are equivalent.}
\label{fig:hier-layout}
\end{figure}

Global reach centrality is better, as you can place nodes on a vertical level according to their local reach centrality value. In Figure \ref{fig:hier-layout}(b) I apply a reach centrality informed layout to the directed graph from Figure \ref{fig:hier-layout}(a). The reach centrality layout is quite rudimentary, as the resulting picture looks more akin to an order hierarchy than a flow hierarchy. In the figure, I had to do a bit of manual work to make it look more like a flow hierarchy, which you might not be able to do for larger graphs. Since the method allows you to find flow hierarchies, this mismatch could be confusing. However, at least it allows you to find out the root of the hierarchy, the node(s) with the highest reach centrality, which the previous method could not do.

The arborescence approach is a further step up. Since it reduces the network to an arborescence, one can draw the resulting condensed graph, identifying not only the root of the hierarchy, but at which level each node lies. The same can be said for agony: it assigns each node to a level, thus you can plot the network by layering nodes vertically according to their assigned rank. Figure \ref{fig:hier-layout}(c) shows a possible layout informed by arborescence (or agony).

\section{Summary}

\begin{enumerate}
\item There are many different ways to intend the meaning of ``hierarchy'' in complex networks. Order hierarchy is like centrality: sorting nodes according to their importance. Nested hierarchy is like communities: grouping nodes in teams and teams of teams.
\item Here we look at flow hierarchy: a structural organization where we have nodes working at different levels and information always flows in one direction, from nodes in a higher level to nodes in the directly lower level. We assume networks are directed.
\item There are many ways to estimate the hierarchicalness of a network. A perfect hierarchy cannot have cycles, which are nodes at a lower level linking against the flow to higher levels. One can simply remove cycles, or calculate how much ``agony'' a connection brings to the structure.
\item We can identify the head of the hierarchy as the node with the highest reach. Alternatively, arborescences are prefect hierarchies -- directed acyclic graphs with all nodes having in-degree of one, except the head of the hierarchy having in-degree of zero.
\end{enumerate}

\section{Exercises}

\begin{enumerate}
\item Calculate the flow hierarchy of the network at \url{http://www.networkatlas.eu/exercises/33/1/data.txt}. Generate $25$ versions of the network with the same degree distributions of the observed one (use the directed configuration model) and calculate how many standard deviations the observed value is above or below the average value you obtain from the null model.
\item Calculate the global reach centrality of the network at \url{http://www.networkatlas.eu/exercises/33/1/data.txt} (note: it's much better to calculate all shortest paths beforehand and cache the result to calculate all local reaching centralities). Is there a single head of the hierarchy or multiple? How many?
\item The arborescence algorithm is simple: condense the graph to remove the strongly connected components and then remove random incoming edges from all nodes remaining with in-degree larger than one, until all nodes have in-degree of one or zero. Implement the algorithm and calculate the arborescence score.
\item Perform the null model test you did for exercise $1$ also for global reach centrality and arborescence. Which method is farther from the average expected hierarchy value?
\end{enumerate}

\chapter{High-Order Dynamics}\label{cha:hod}
Networks are a great tool to represent a complex system in a simple and elegant way. They embed most of their subtleties into a coherent structure. However, if you only look at the structure you might miss part of the story. This is especially true if you're interested in knowing how different agents act in it.

For instance, consider air travel. The airlines give you the structure, by deciding from where their planes take off and to where they land. And, if you're interested only in studying how different airports connect together, that is all you need to know. But you might instead want to study the behavior of the passengers taking those flights. In this case, the structure itself might be misleading. If you ever made some air travel, you know that, in most cases and especially for long trips, you would take connecting flights. Meaning that you will hop through one or more airports before you reach your intended destination from your origin.

But, if all you look at is the structure of flights, you're not going to be able to recover such information. When a traveler boards in $u$ and jumps off in $v$, if all you have is the structure, you only know that they are going to either stay in $v$ or go to one of $v$'s neighbors, maybe even back to $u$. If you want to really model the traveler's behavior, you need some sort of \textit{memory}, you need to know they arrived from $u$ into $v$. The next step is not dependent exclusively on the fact we're in $v$.

\begin{figure}
\centering
\includegraphics[width=.75\columnwidth]{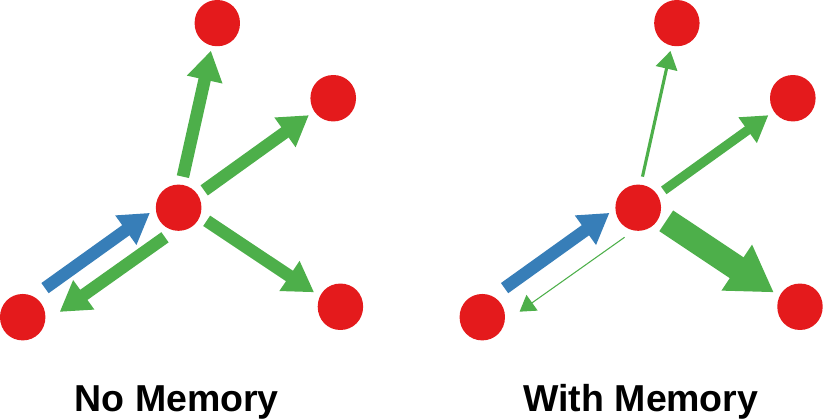}
\caption{An example of possible high order dynamics. The blue arrow shows the first step, while the green arrows show the potential second steps.}
\label{fig:highord-example}
\end{figure}

Figure \ref{fig:highord-example} shows an example of this. The network might have equal weights on the edges, because from the central airport there is an equal number of flights or passengers in each link. Thus, all we can say is that all steps from the central node are equally likely (left). However, from observed data, we might see that some second steps -- from our given origin -- are much more likely than others (right). In this case, we cannot trust the unweighted edges at face value: we need to take this additional information into account. 

When you start talking about memory, you're talking about higher order dynamics -- for a refresher on the terminology, see Chapter \ref{cha:prob}. In this chapter, we'll explore three ways of embedding high-order interactions in your network. The first two are structural, the latter is algorithmic. The first is by using simplicial complexes. These don't have memory, but have a way to encode many-to-many relationships, which can handle a large variety of high order interactions. Then we're moving to explicitly encoding memory into your network analysis workflow. You can either modify your network data, embedding the higher order dynamics into the structure; or you can modify your algorithm.

\section{High Order with Simplicial Complexes}\label{sec:hod-simplicial}
I defined the terminology and some basic facts about simplicial complexes in Section \ref{sec:extended-hyper}, and showed how to generate one in Section \ref{sec:csmodels-conf}. We're focusing on complexes here even though, technically, some of the things you can do with simplicial complexes you can also do with hypergraphs. But simplicial complexes are more flexible and completely contain hypergraphs.

\subsection{Simple High Order Statistics}
\begin{figure}
\centering
\includegraphics[width=.4\columnwidth]{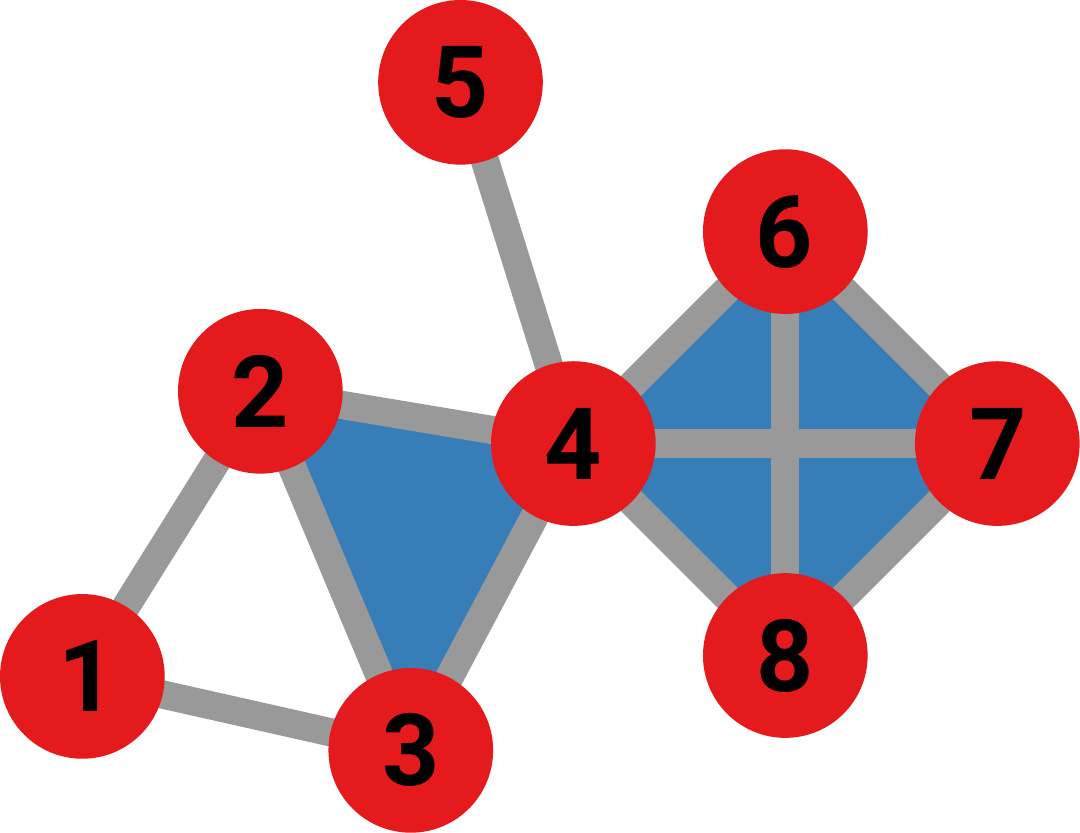}
\caption{An example of simplicial complex. The blue shades represent the two simplices in the complex.}
\label{fig:simplicial-example-hod}
\end{figure}

There are a bunch of simple analyses you can do with simplicial complexes. The first is generalizing the degree. In a network without simplices, the degree is only a property of a node. But in a simplicial complex, each face has a generalized degree\cite{bianconi2016network}. With $k_{d,m}$ we can indicate the number of $d$ dimensional simplices incident on an $m$-face -- with $m < d$. Consider Figure \ref{fig:simplicial-example-hod}. The $k_{2,0}$ of node $4$ is four. That's because we're looking at $d = 2$, i.e. 2-simplices -- also known as triangles --, on an $m = 0$ face -- which is a node. Node $4$ gets three 2-simplices from its 3-simplex connecting it to nodes $6$, $7$, and $8$, and another 2-simplex with nodes $2$ and $3$. On the other hand, the $k_{2,1}$ of edge $(4,8)$ is two. The 1-simplex -- i.e. edge -- $(4,8)$ participates in two 2-simplices -- i.e. triangles -- with nodes $6$ and $7$.

There is a special generalized degree that we call ``incidence''. This is $k_{d,d-1} - 1$, i.e. the number of $d$ dimensional simplices incident on a face that is one dimensional step below them, so a $d-1$ face. Incidence is important to define manifolds, but before we do that we also need to expand the concept of connected component in a simplicial complex.

\begin{figure}
\centering
\includegraphics[width=.5\columnwidth]{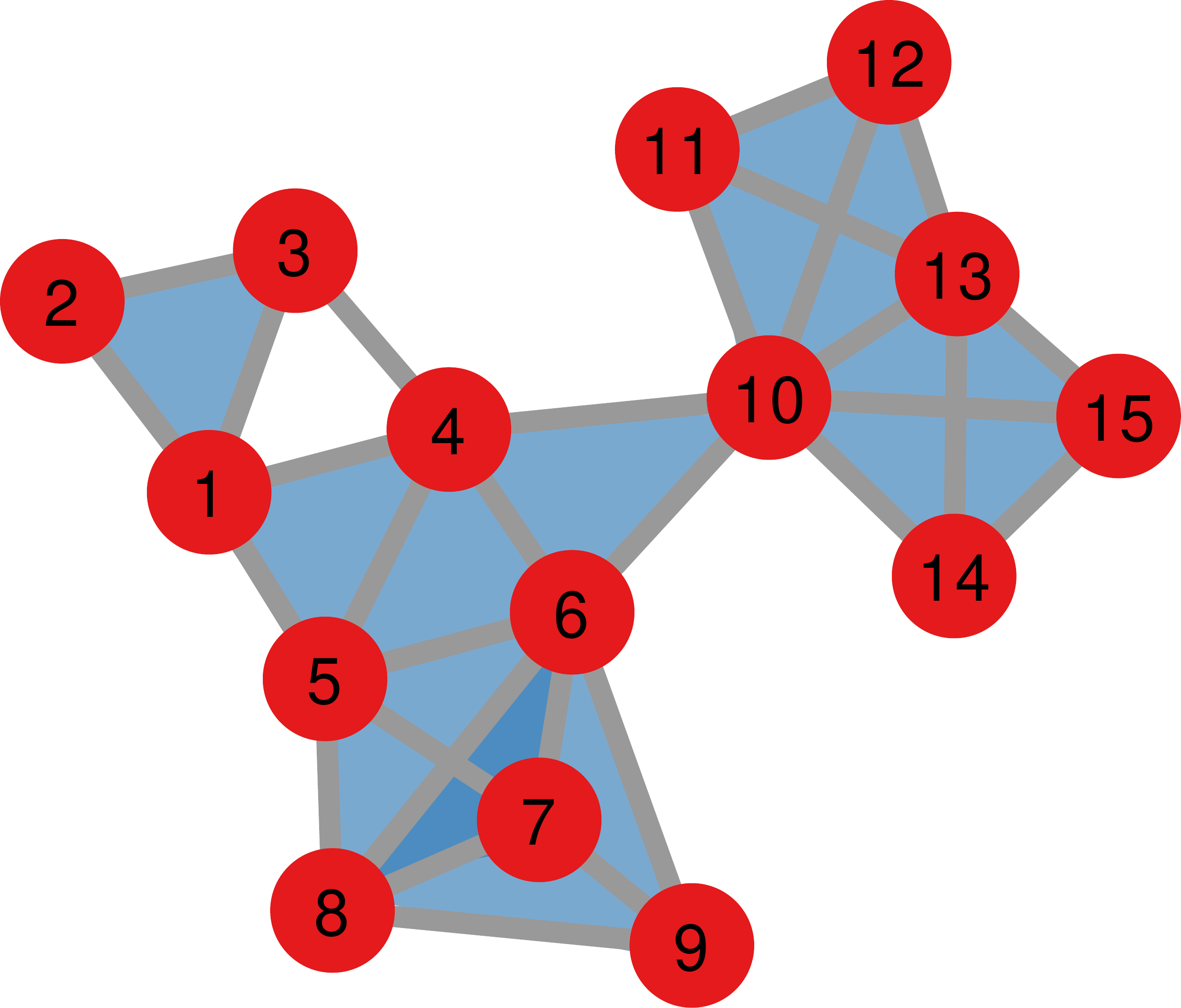}
\caption{A simplicial complex. The blue shades represent the two simplices in the complex. A darker shade indicates simplices that overlap.}
\label{fig:simplicial-ccs}
\end{figure}

A $d$-connected component is the set of adjacent facets of at least dimension $d$. Two facets of dimension $d$ are connected if there is a sequence of simplices of at least dimension $d$ that allows you to go from one to the other. A 0-connected component is the same as the notion of connected component in a normal network, since 0-simplices are vertices and thus any edge connecting two 0-simplices will make them part of the same 0-connected component. The 1-connected component is a little more tricky. Consider Figure \ref{fig:simplicial-ccs}.

The simplicial complex has three 1-components: $\{1,4,5,6,7,8,9,10\}$, $\{10,11,12,13,14,15\}$, and $\{1,2,3\}$. That is because nodes $1$, $3$, and $4$ make up a triangle but not a simplex, so there is no simplex of dimension at least $1$ connecting those 1-simplices together -- there are only 0-simplices, the nodes themselves. The complex only has a 2-component: $\{5,6,7,8,9\}$. There are other facets of dimension $2$, but they are only connected by a 1-simplex, so they do not make up a 2-component.

\begin{figure}
\centering
\begin{subfigure}{.4\columnwidth}
\includegraphics[width=\textwidth]{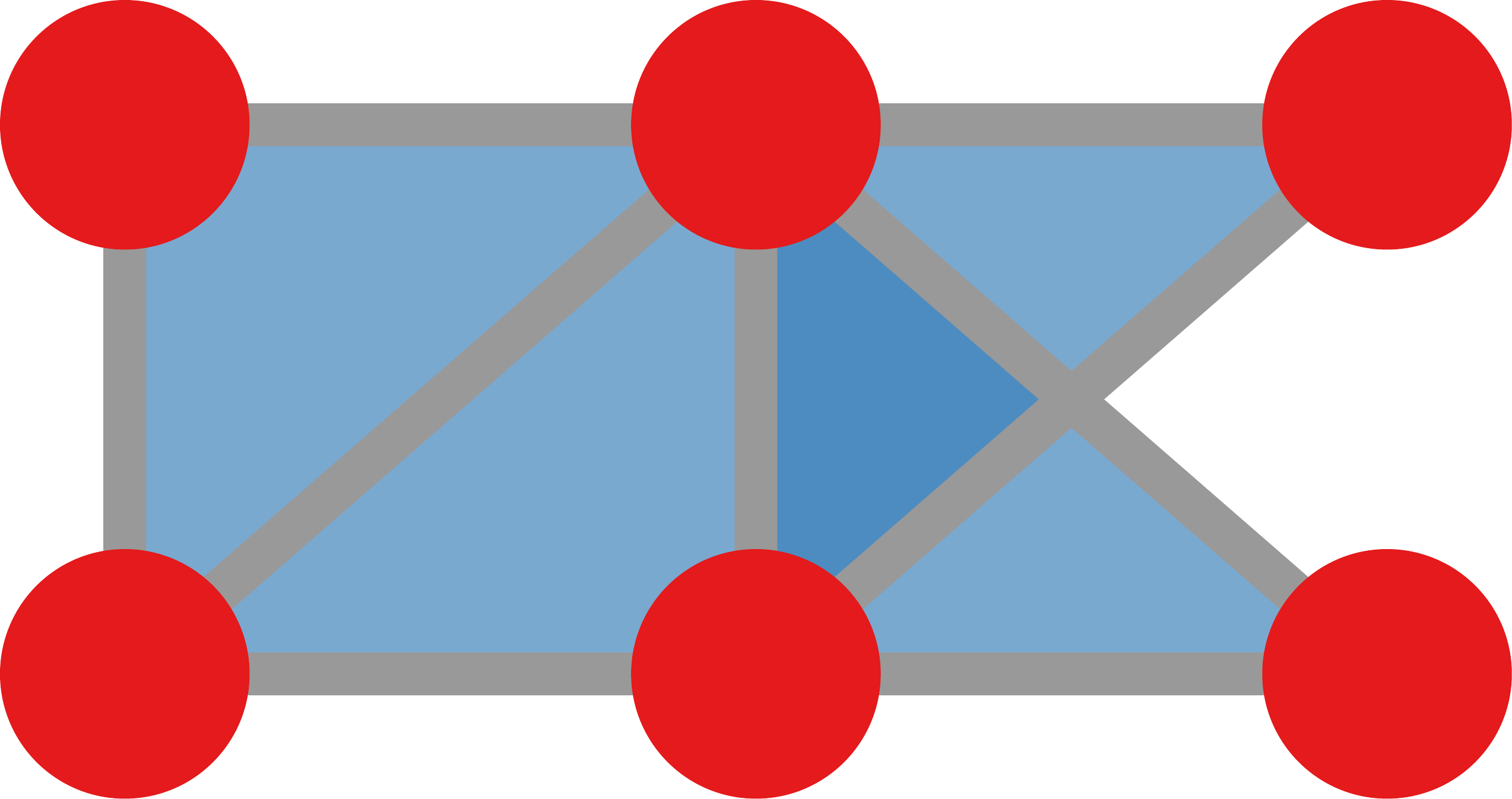}
\caption{}
\end{subfigure}\qquad\qquad
\begin{subfigure}{.4\columnwidth}
\includegraphics[width=\textwidth]{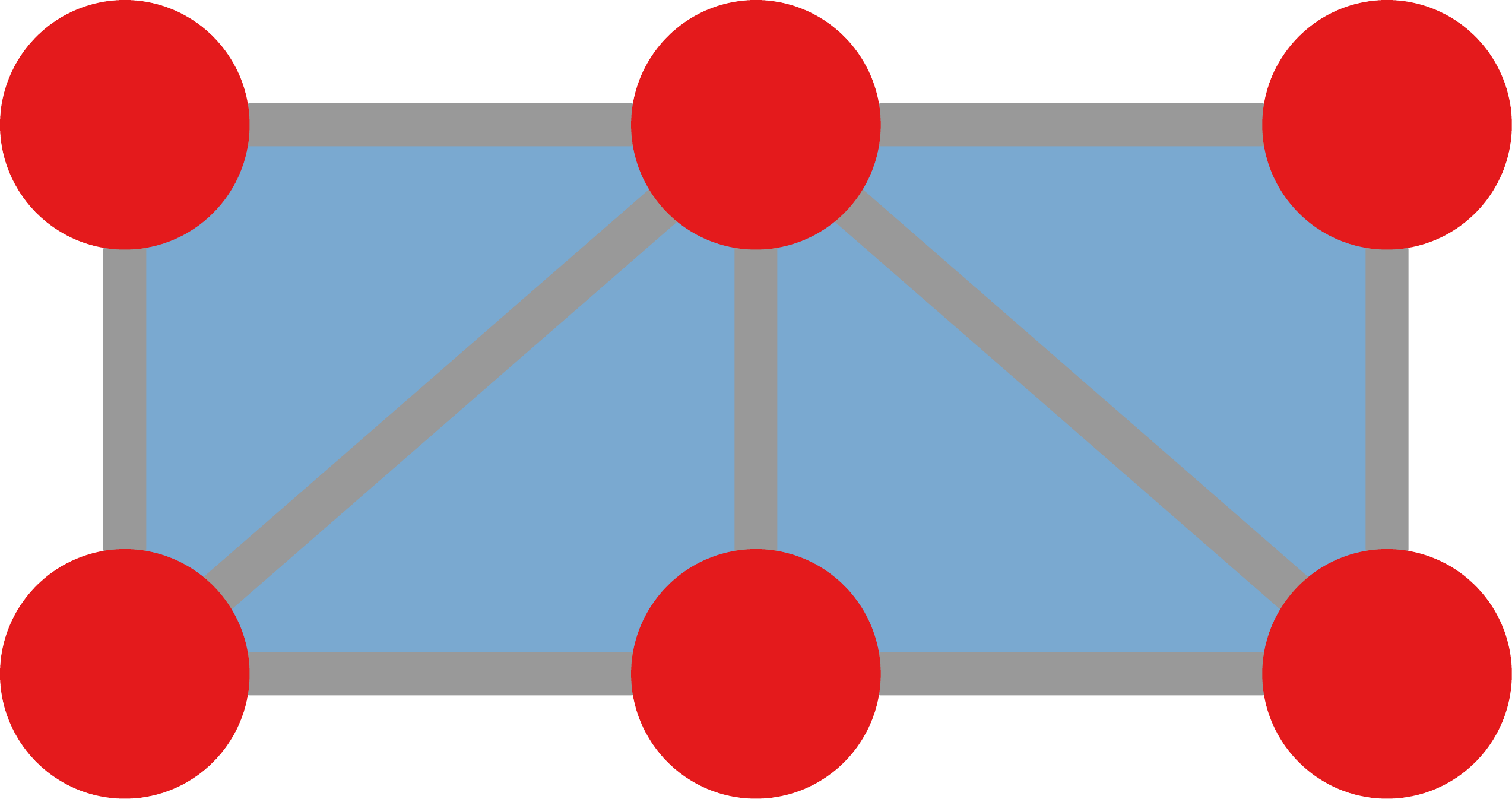}
\caption{}
\end{subfigure}
\caption{Simplicial complexes with simplices shaded in blue. (a) Not a manifold. (b) A manifold.}
\label{fig:simplicial-manifold}
\end{figure}

Now we can define a manifold. A manifold is a simplicial complex that is $d$ connected and all its $d$ faces have incidence of either $0$ -- i.e. they are not connected -- or $1$ -- i.e. they only have a single incident face. Figure \ref{fig:simplicial-manifold}(a) shows a case that is not a manifold, because there is a face with incidence larger than one, while Figure \ref{fig:simplicial-manifold}(b) shows a manifold.

\subsection{High Order Processes on Simplicial Complexes}
The reason why you want to work with simplicial complexes is not because of slightly fancier network measures: it is because you want to study processes where these high order interactions can make a difference. Of course, there are too many to report on them all, so I picked three that can hopefully give you a taste of the possibilities in front of you. We're going to see briefly the problems of \textbf{synchronization}, \textbf{percolation}, and \textbf{epidemics}.

\begin{figure}
\centering
\includegraphics[width=.8\columnwidth]{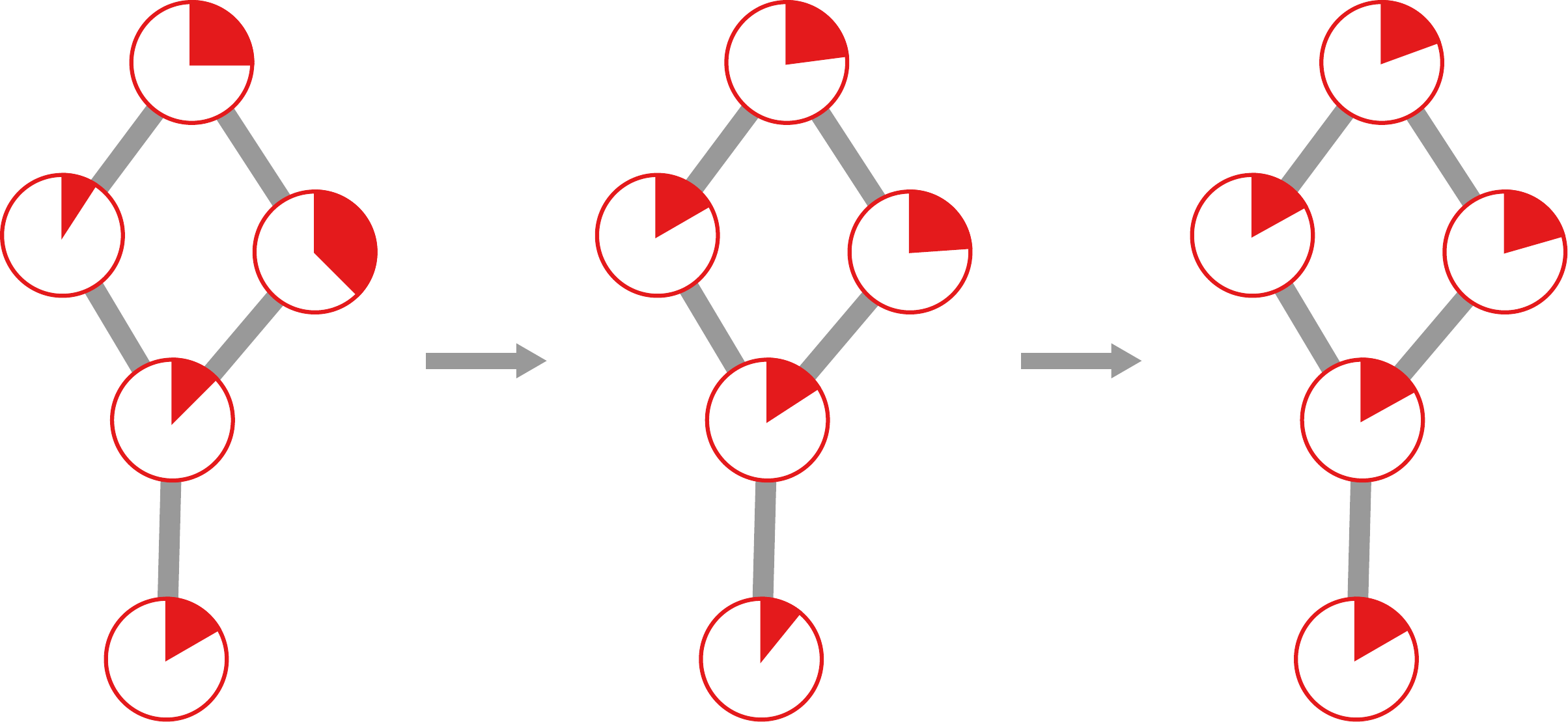}
\caption{Synchronization on a network. The shaded portion of each node is proportional to its firing frequency. From left to right we progress over time, with nodes changing their frequencies according to the ones of their neighbors.}
\label{fig:kuramoto}
\end{figure}

Networks have been used to study \textbf{synchronization} for a long time. One classical example is thinking of nodes as neurons who are firing at random intervals. Each node fires with a frequency that is drawn uniformly at random. However, nodes that are connected will influence each other. At each pulse, their frequencies will get closer and closer together -- a process regulated by a synchronization speed parameter $\sigma$. Figure \ref{fig:kuramoto} shows a simple example, where nodes get synchronized over time.

\begin{figure}[t]
\centering
\begin{subfigure}{.4\columnwidth}
\includegraphics[width=\textwidth]{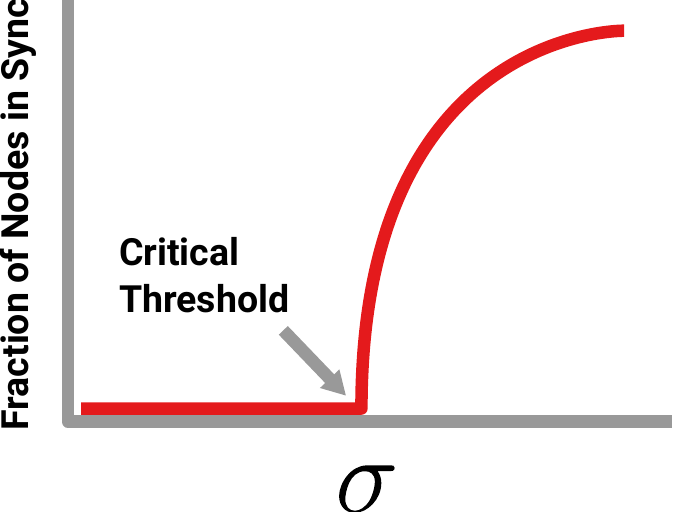}
\caption{}
\end{subfigure}\qquad\qquad
\begin{subfigure}{.4\columnwidth}
\includegraphics[width=\textwidth]{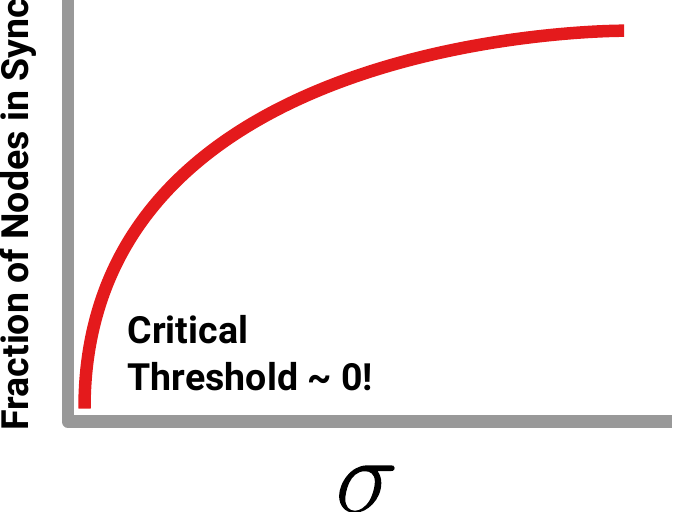}
\caption{}
\end{subfigure}
\caption{Different synchronization dynamics: fraction of nodes at the same frequency (y axis) at a given $\sigma$ value (x axis). (a) No many-to-many interactions. (b) With many-to-many interactions.}
\label{fig:kuramoto-synch}
\end{figure}

This process has a phase transition. If $\sigma$ is too low, no or very few nodes will synchronize. Beyond a critical value of $\sigma$, the fraction of nodes that is firing at the same rate will start to increase abruptly, until it reaches almost the entirety of the network. Figure \ref{fig:kuramoto-synch}(a) shows how the fraction of synchronized nodes evolves for increasing values of $\sigma$.

When you run this model on a simplicial complex, you enable many-to-many interactions: now nodes in a simplex will all influence each other at the same time. There are many ways to define such a model\cite{skardal2019abrupt}\cite{millan2020explosive}\cite{ghorbanchian2021higher}, but one key result is that the critical $\sigma$ threshold tends to go to zero -- as Figure \ref{fig:kuramoto-synch}(b) shows. In practice, when you enable many-to-many interactions, you will always see at least some synchronization in the system, even for small synchronization strength $\sigma$, which you would not see it if many-to-many interactions were to be disabled.

With \textbf{percolation}, I intend the catastrophic network failures I discussed in Chapter \ref{cha:epidemapps}. In that case, we saw that nodes or edges could fail, and their failures could propagate (= percolate) through the network, causing a chain reaction. In simplicial complexes, you can have simplices failing as well. The classical case of node or edge failure is a $d = 1$ type of failure. But you can have also simplices failing at $d = 2$, which means you have a certain probability triangles will fail\cite{bianconi2018topological}. And it goes without saying that you can experience failures at higher dimensions $d$.

Finally, in \textbf{epidemics} you can expand the SI and related models I explained to you in Chapters \ref{cha:epidemics} and \ref{cha:triggers}. In the basic model, you have a certain probability of transitioning $\beta$ if you have a contact with an infected individual. When you have simplices, you can have a different transition probability if you are involved in a simplex with an infected individual. This models well the case when you enter a room and you don't particularly interact with anyone, but you're in the same space with someone infected. That's a many-to-many interaction, given that there could be multiple (potentially infected) people in the room.

So now you have, say, a $\beta_1$ for the normal face-to-face interactions -- which are a $d = 1$ simplex --, a $\beta_2$ for a many-to-many interaction with two people -- because that's a $d=2$ simplex --, and so on\cite{landry2020effect}\cite{iacopini2019simplicial}... What we normally do when we ignore many-to-many interactions is that we only study the fraction of infected individuals for different values of $\beta_1$. This basically means we assume $\beta_2 = 0$, and we get a given function telling us how much we're screwed given how infective at the face-to-face level a disease is.

\begin{figure}
\centering
\includegraphics[width=.8\columnwidth]{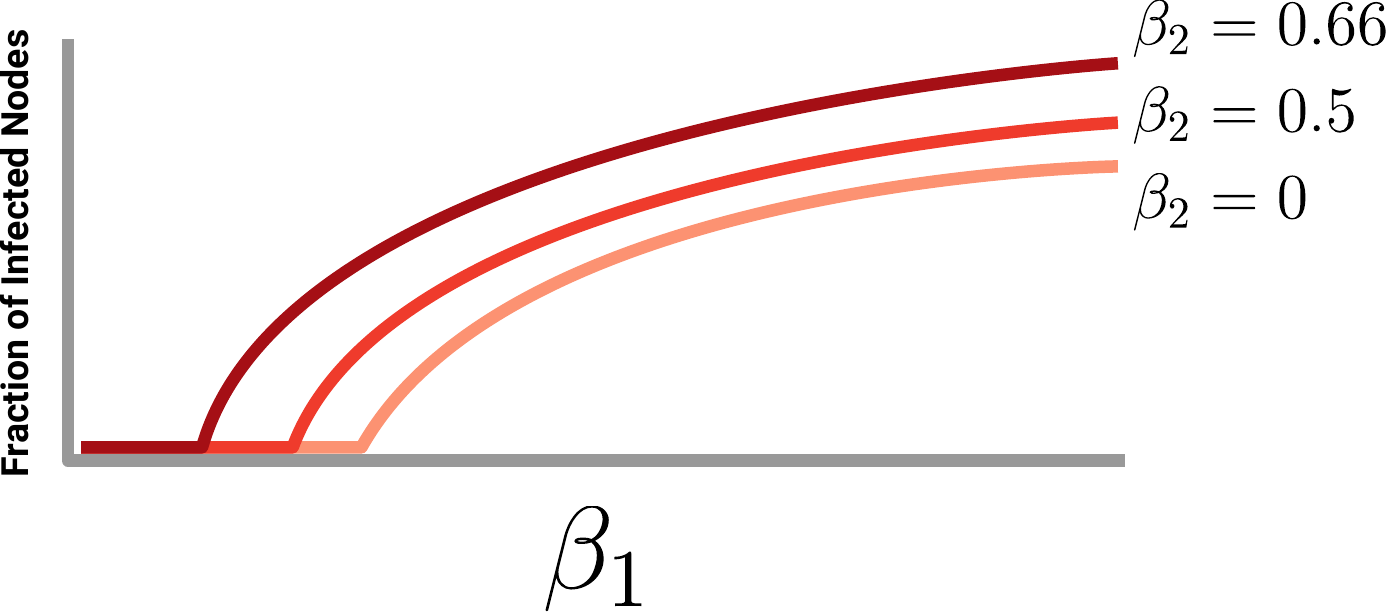}
\caption{The fraction of infected nodes (y axis) for a given value of $\beta_1$ (x axis) and $\beta_2$ (line color).}
\label{fig:simplicial-contagion}
\end{figure}

However, Figure \ref{fig:simplicial-contagion} shows you that, if $\beta_2 > 0$, then the contagion will happen \textit{faster} and will infect \textit{more} nodes than we would expect for the \textit{same value} of $\beta_1$.

\subsection{Generating High Order Networks}
\begin{figure}[t]
\centering
\begin{subfigure}{.46\columnwidth}
\includegraphics[width=\textwidth]{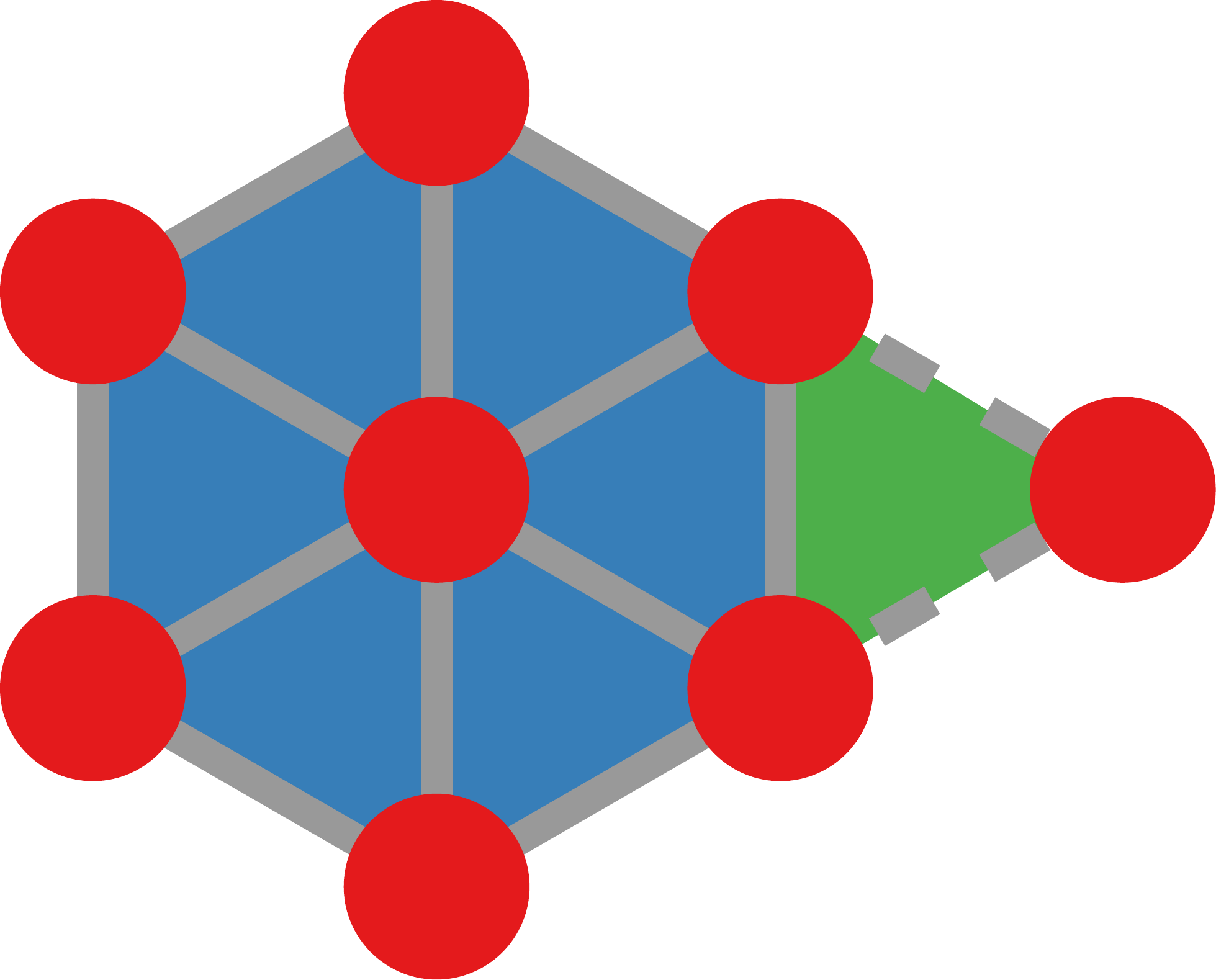}
\caption{}
\end{subfigure}\qquad\qquad
\begin{subfigure}{.33\columnwidth}
\includegraphics[width=\textwidth]{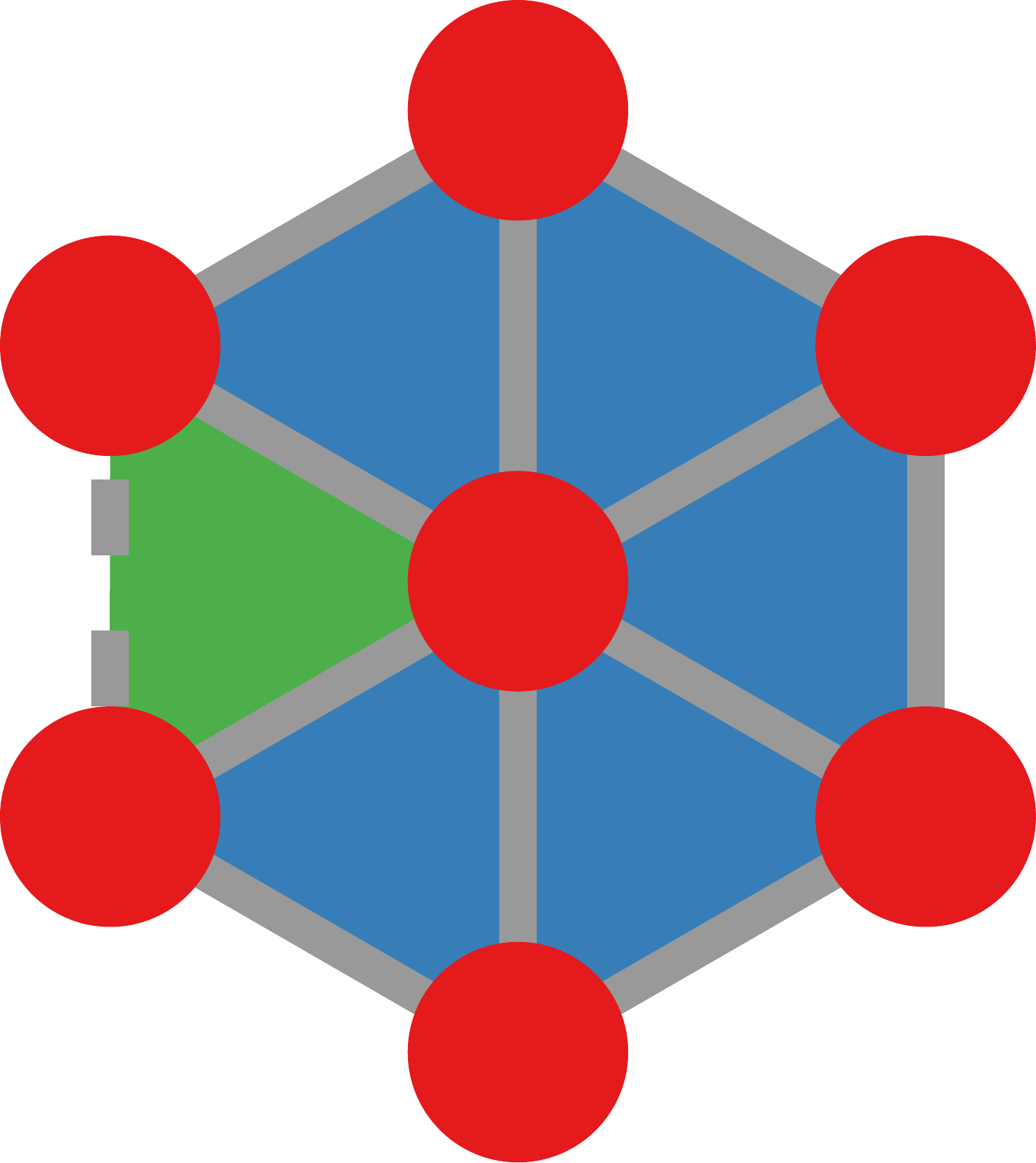}
\caption{}
\end{subfigure}
\caption{Two rules to grow a manifold. The blue shade is for the simplices already in the manifold, the added simplex is in green. (a) Adding a simplex to an unsaturated link. (b) Saturating two links.}
\label{fig:manifold-growth}
\end{figure}

Another reason to love manifolds -- besides enabling the interesting high order dynamics we just saw -- lies in their generative power. For the rest of the section we assume a manifold with $d = 2$. In any $d$-manifold, you can classify links as either saturated or unsaturated. A link is saturated if it is part of exactly $d$ simplices. If not, it is unsaturated. Once you classify the links you have, you can grow the manifold in two ways. First, you can pick a single unsaturated link with probability $1 / |E_u|$, with $E_u$ being the set of saturated links, and saturate it by gluing a triangle to it -- which implies that you need to also add a node with which to close the triangle. Second, with probability $p$ -- which is a parameter you can choose -- you can find two unsaturated links and add the simplex to saturate them -- this does not add a node. Figure \ref{fig:manifold-growth} shows how these two moves look like.

\begin{figure}[t]
\centering
\includegraphics[width=.8\columnwidth]{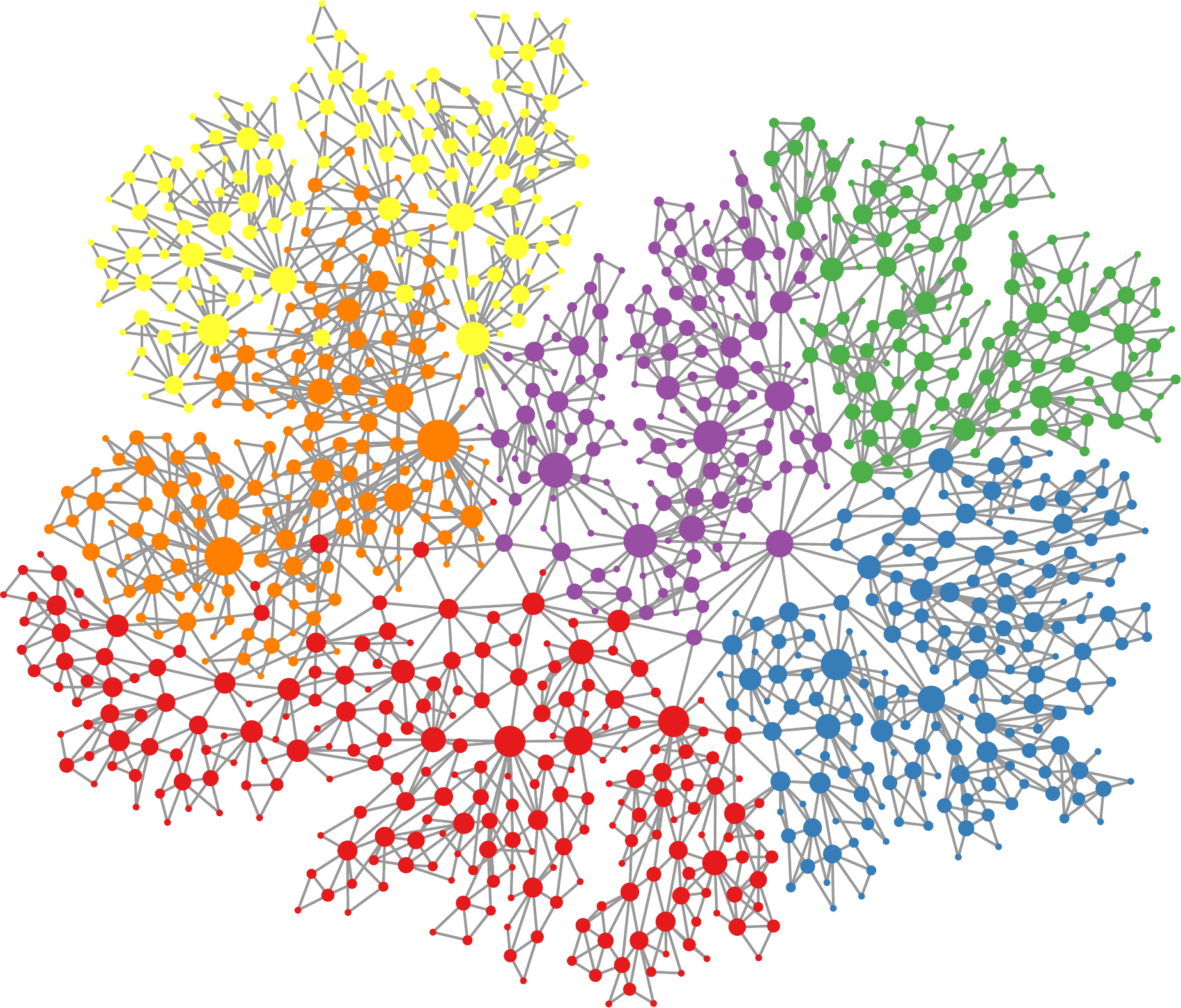}
\caption{A manifold grown with the rules from Figure \ref{fig:manifold-growth}. The node color is its community. The node size is proportional to its degree.}
\label{fig:manifold-generated}
\end{figure}

If you do this, you can grow a manifold\cite{wu2015emergent}. The reason why this is cool is because, with properly defined parameters $d$ and $p$, you can reproduce a lot of real world properties in your manifold -- which was the holy grail of Chapter \ref{cha:physicsmodels}. For some specific $d$ and $p$ values -- for Figure \ref{fig:manifold-generated} I set $p = 0.5$ and $d = 2$ -- you obtain a small world network, with a broad degree distribution, high clustering, and high modularity as well. This hints at the fact that these sort of high order interactions might play a role in the formation of many real world networks.

The cool thing about generating manifolds this way is that, with a slightly more advanced model, you can have a single network generating process that generates a wide variety of other models. With suitable choices of $d$ and $p$, you can have $G_{n,p}$ model, or a preferential attachment one, and you get clustering and communities for free.

Just like edges, simplices can also have weights, and these weights can be used meaningfully to enhance the network generating process\cite{courtney2017weighted}. Even more interestingly, we can assign an energy value to each node -- an arbitrary non-negative number. The nodes pass their energy to the face they belong to. This energy acts like a weight but, since it comes from the node, will lead to different dynamics when growing the network -- again regulated by a parameter of your choice. Depending on the values of this parameter, your manifold's properties will approach the ones of different quantum states: Fermi-Dirac, Boltzmann, or Bose-Einstein -- which, admittedly, sounds unbelievably cool, but I will stop here because my quantum physics isn't exactly up to the standards required to discuss this. The excellent book by Bianconi\cite{bianconi2021higher} is what you should use to quench your knowledge thirst.

Additionally, there are models that can generate multilayer simplicial complexes\cite{sun2021higher}.

Suppose that you find simplices cool, but you already have your network -- so generating a manifold is not an option, However, your network is not a simplicial complex. What do you do? Technically, you can make any arbitrary network into a simplicial complex. You can consider every clique in the network as a simplex and perform the analyses I described on that structure.

\begin{figure}
\centering
\includegraphics[width=.66\columnwidth]{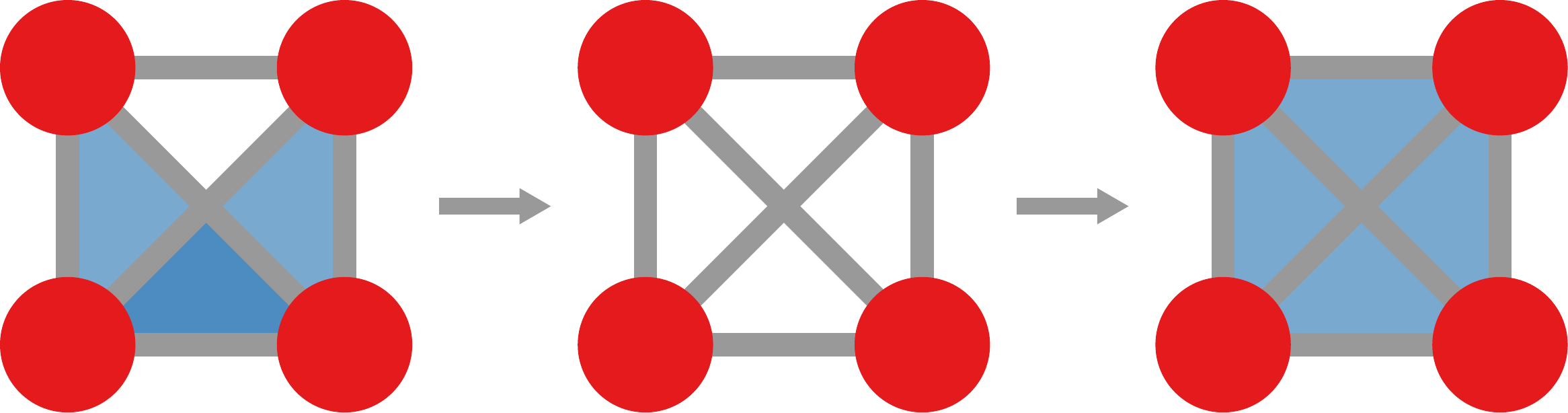}
\caption{From left to right: a simplicial complex with two 2-simplices, its skeleton, the simplicial complex reconstructed by assuming each clique in a network is a simplex.}
\label{fig:simplices-commutativity}
\end{figure}

You should be careful about one thing, though. As I told you in Section \ref{sec:extended-hyper}, you can do the opposite operation: every simplicial complex can be reduced to a normal complex network by ignoring the simplices and treating them like cliques. However, these two operations -- simplices to cliques and cliques to simplices -- are \textit{not commutative}. If you apply them one after the other, you're not going to go back to your original simplicial complex -- bar weird coincidences. Figure \ref{fig:simplices-commutativity} shows you a case in which a clique of four nodes was actually made up by two 2-simplices rather than being a 3-simplex, and therefore the reconstruction by cliques leads to a different simplicial complex.

\section{Embedding Memory into the Structure}\label{sec:hod-structure}
A natural way to have higher order memory in your network analysis is by embedding it into the structure itself. This is a powerful approach, because it allows you to use any non-high-order algorithm you want. You have the entirety of the network analysis toolbox at your disposal. The price you have to pay is that you need to keep track of your operation. You need to reconstruct the original structure if you want to properly interpret your results.

This practically boils down to performing a pre-processing on your data structure and a post-processing on your results. Here we focus mainly on the pre-processing as, hopefully, how to post-process the results should be straightforward. Unfortunately, the pre-process is not going to be as simple as I make it to be in Figure \ref{fig:highord-example}. You cannot simply re-weight your edges, because the re-weighting would be dependent on the current position of the agent in the network. Thus you'd have to have a re-weighting for every node in the network. Worse still, if you do that you're just implementing second-order dynamics: if you need to go to a higher order than that (say, you need to remember the last two nodes through which you passed) you're in no better position than before.

I'm going to specifically focus on a few papers in this line of research, but hopefully you could see how the general approach in this category of high-order analysis works.

\subsection{High Order Network}
\begin{figure}
\centering
\begin{subfigure}{.4\columnwidth}
\includegraphics[width=\textwidth]{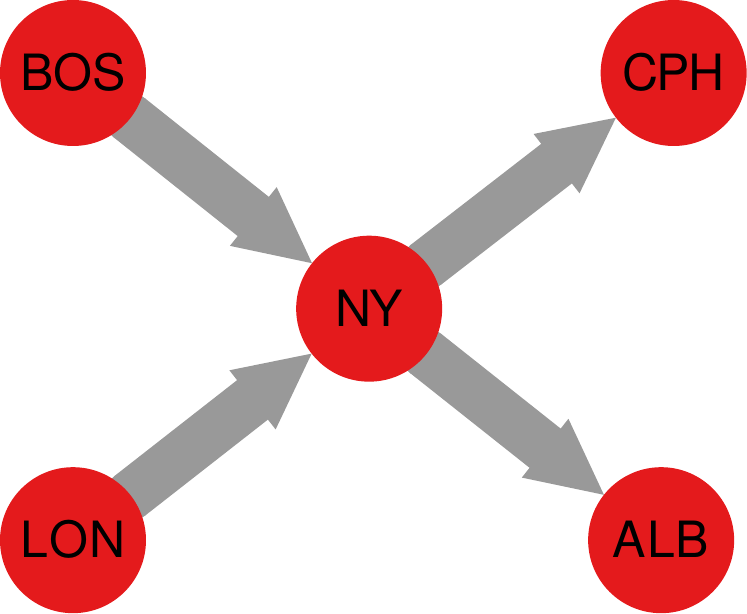}
\caption{}
\end{subfigure}\qquad\qquad
\begin{subfigure}{.4\columnwidth}
\includegraphics[width=\textwidth]{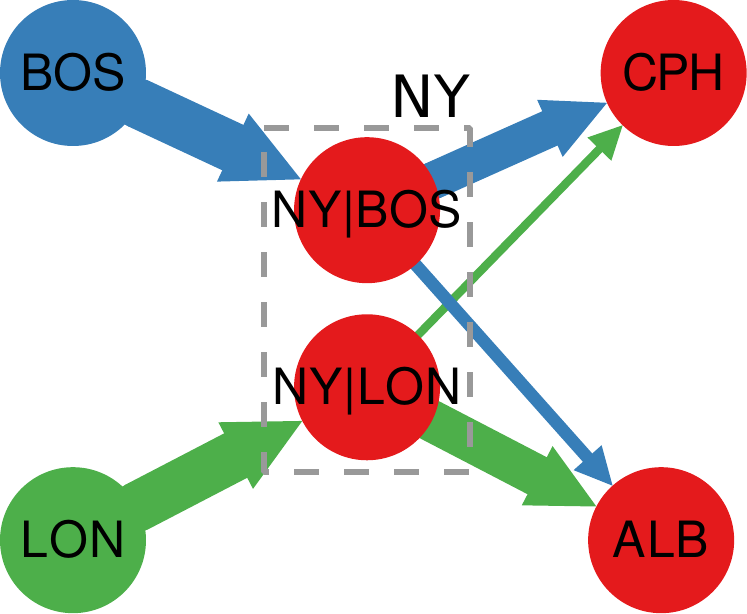}
\caption{}
\end{subfigure}
\caption{(a) A simple directed network of airplane travel. (b) Its corresponding High Order version, introducing conditional nodes depending on the origin of the flow.}
\label{fig:hon1}
\end{figure}

What we want to build here is a High Order Network\cite{xu2016representing} (HON). Let's go back to our airline travel example. In the original network, nodes are airports and edges represent flows of passengers between them. Now, the crucial assumption we're making here is that, if I'm landing in New York in a plane coming from Boston, this is fundamentally a different travel than the one which makes me land in the same airport, but from London. Thus, in the HON representation, we split the New York node in two meta nodes. One of the two meta nodes captures the passenger flow from Boston, the other from London. Figure \ref{fig:hon1} shows an example of this procedure.

What happens here is that now we have a way to represent different flows. A passenger from London might be much more likely to want to go to Alberta, while the Bostonian would instead go to Copenhagen. Thus we can re-part the outgoing edge weight of New York into those two meta nodes, to make them a more accurate representation of the data.

\begin{figure}
\centering
\includegraphics[width=\columnwidth]{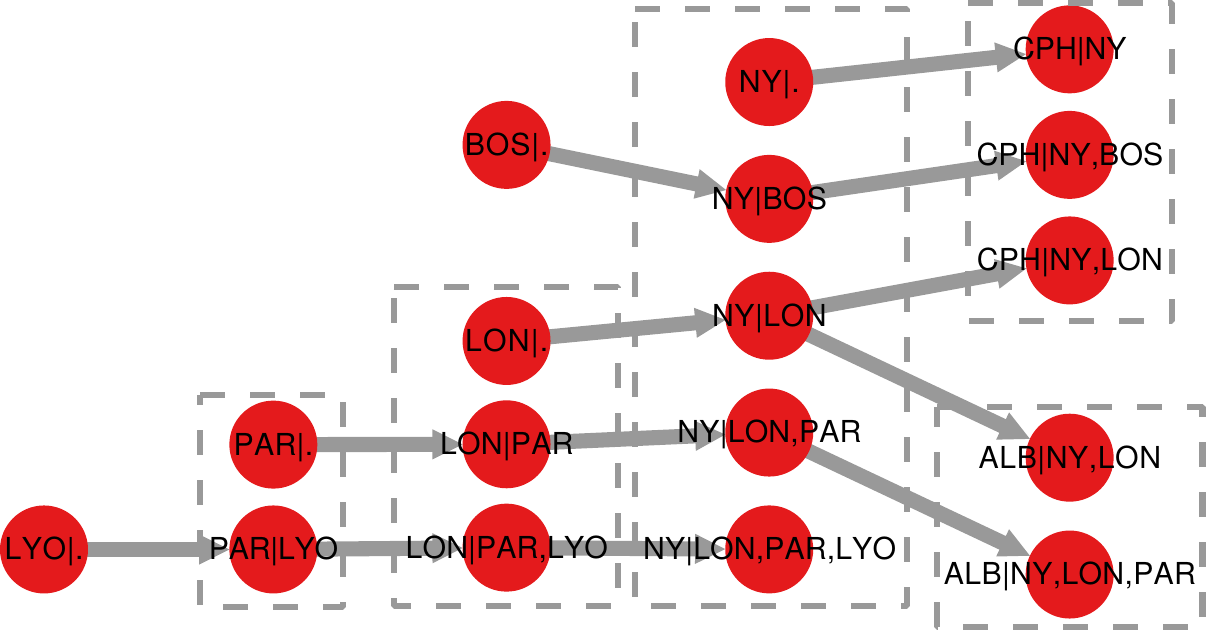}
\caption{A small example of a full High Order Network with complex dependencies up to an order of four. the dashed outlines group all the meta nodes originating from the same original node.}
\label{fig:hon2}
\end{figure}

Of course, this represents only a single step in the creation of the HON structure. The origin nodes that brought us to New York were themselves the product of another high order transition. Thus they are also split into several meta nodes. The general HON structure looks like the one in Figure \ref{fig:hon2}.

You might have noticed that we use the conditional probability notation introduced in Chapter \ref{cha:prob}. Each node represents the transition probability of getting into node $v$ given that we come from node $u$ -- and, possibly, given that we reached $u$ from another node, and so on. While airline travel might have short dependency chains -- rarely a trip involves more than two transfers -- other networks show dependency chains up to length five: the global shipping network, for instance. When I label a node $v$ as, for instance, $v|.$, it means that this specific node has no dependencies: it represents passengers who are starting their first step in $v$.

If you think HON structures looks like a big and complex Bayesian network -- see Section \ref{sec:extended-types} -- don't worry: you're not alone.

There are potential downsides of using a HON representations. First, as you can see from Figure \ref{fig:hon2}, the structure can become really unwieldy. The original network only had $7$ nodes and $6$ edges, and ballooned into having $17$ nodes and $16$ edges. Therefore you need to find the right trade off between the complexity of your HON representation and the analytic gains it gives you. In the original paper, for instance, the authors limit themselves to an order of five, even if the shipping network they analyze might have even longer dependencies. The increase in complexity simply wasn't worth it.

Finally, HON networks tend to transform into weakly connected graphs, or even not connected. The original network from Figure \ref{fig:hon2} had a single connected component, while its HON representation splits into $5$ connected components. If your algorithm cannot handle multiple connected components, you might be in trouble.

\subsection{Memory Network}
\begin{figure}
\centering
\begin{subfigure}{.33\columnwidth}
\includegraphics[width=\textwidth]{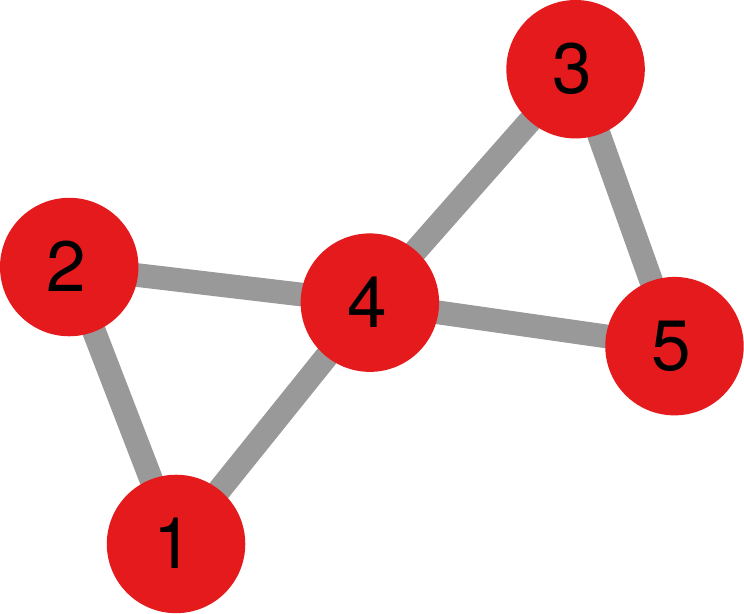}
\caption{}
\end{subfigure}\qquad\qquad
\begin{subfigure}{.25\columnwidth}
\includegraphics[width=\textwidth]{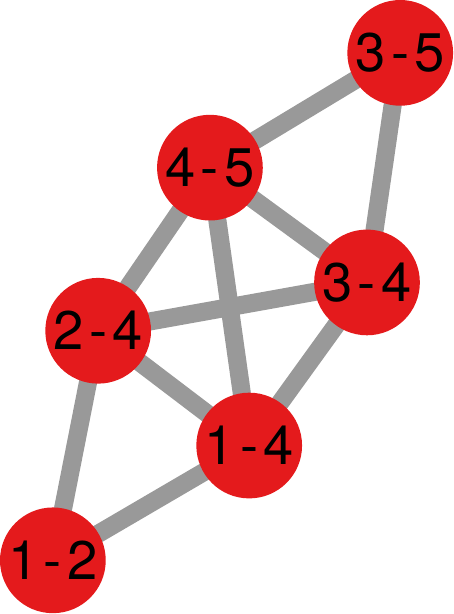}
\caption{}
\end{subfigure}
\caption{(a) A graph. (b) Its linegraph version.}
\label{fig:linegraph}
\end{figure}

Alternatives to HON exist. For instance, researchers built what they call a ``memory network\cite{rosvall2014memory}''. To model second-order dynamics we can create a line graph. If you recall Section \ref{sec:basic-simple}, a line graph is a transformation of the original graph. The edges of the original graph become the nodes of the network and they are connected to each other if the original edges shared a node in the network. Figure \ref{fig:linegraph} is a reproduction of Figure \ref{fig:line-graph-ex} to remind you of the building procedure. Line graphs will pop up also in overlapping community discovery (Section \ref{sec:ocd-link-linegraph}).

You can model third order dynamics by making a more complicated version of a line graph, in which nodes stand in for paths of length three. You can see how memory graphs are a generalization of line graphs, and they start looking extremely similar to the HON model. In fact, once you have built your memory network with the desired order, then the simple memoryless Markov processes on the memory network describe the high-order processes, of the order you used to build the network in the first place. The adjacency matrix of a memory network of second order is a non-backtracking matrix (Section \ref{sec:rw-no-backtrack}), provided that the memory network has no self-loops. Non-backtracking random walks are another example of high order network analysis.

The difference between memory networks and HONs is that HONs are a bit more flexible, because they allow you to have nodes of any order mixed together in the same structure. In the memory network, all nodes represent transitions of the same order.

\section{Embedding Memory into the Algorithm}\label{sec:hod-algo}
The alternative approach to deal with high order dependencies is to leave your structure alone and to embed the high order logic directly into your algorithm. In practice, you ``hide'' both the pre- and post-process from the previous section into your analysis. This way, you don't have to deal with the complexity yourself.

There is a bit more diversity in this category of solutions, due to the many different valid ways one could incorporate high orders into network analysis.

\subsection{Motif Dictionary}
The first approach I consider is the one building motif dictionaries. In this approach, one realizes that there are different motifs of interest that have an impact on the analysis. For instance, one could focus specifically on triangles. Once you specify all the motifs you're interested in, you take a traditional network measure and you extend it to consider these motifs.

This move isn't particularly difficult once you realize that low-order network measures still work with the same logic. It's just that they exclusively focus on a single motif of the network: the edge. An edge is a network motif containing two nodes and a connection between them. Once you realize that a triangle is nothing else but a motif with three nodes and three edges connecting them, then you're in business.

To make this a bit less abstract, let's consider a specific instance of this approach\cite{benson2016higher}\cite{yin2017local}\cite{tsourakakis2017scalable}. In the paper, the idea is to use motifs to inform community discovery, the topic of Part \ref{par:cd}. I'm going to delve deeper into the topic in that book part, for now let's just say that we're interested in solving the 2-cut problem (Section \ref{sec:rw-mincut}): we want to divide nodes in two groups such that we minimize the number of edges connecting nodes in different groups.

In the classical problem we want to make a normalized cut such that the number of edges flowing from one group to the other is minimum, considering some normalization factor. So we have a fraction that looks something like this:

$$ \phi(S) = \dfrac{|E_{S,\bar{S}}|}{\min\left(|E_S|, |E_{\bar{S}}|\right)}, $$

where $S$ is one of the two groups -- i.e. a set of nodes on one side of the cut --, $\bar{S}$ is its complement -- i.e. $\bar{S} = V - S$, the set of nodes on the other side of the cut. $E_x$ is the set of edges in the set $x$, and $E_{x,y}$ is the set of edges established between a node in $x$ and a node in $y$. In practice, let's find $S$ such that $\phi(S)$ is minimum. We do so by minimizing the number of edges between $S$ and its complement, normalized by their sizes (so that we don't find trivial solutions by cutting off simply a dangling leaf node).

But here we say: \textit{No! Not the number of edges! We are interested in higher order structures! We want to minimize the number of triangles between groups!} How would that look like? Exactly the same. We just count not the number of the edges, but the number of arbitrary motifs $M$ spanning between $S$ and non-$S$:

$$ \phi(S, M) = \dfrac{|M_{S,\bar{S}}|}{\min\left(|M_S|, |M_{\bar{S}}|\right)}.$$

\begin{figure}
\centering
\begin{subfigure}{.2\columnwidth}
\includegraphics[width=\textwidth]{figures/triangle.pdf}
\caption{}
\end{subfigure}\qquad
\begin{subfigure}{.5\columnwidth}
\includegraphics[width=\textwidth]{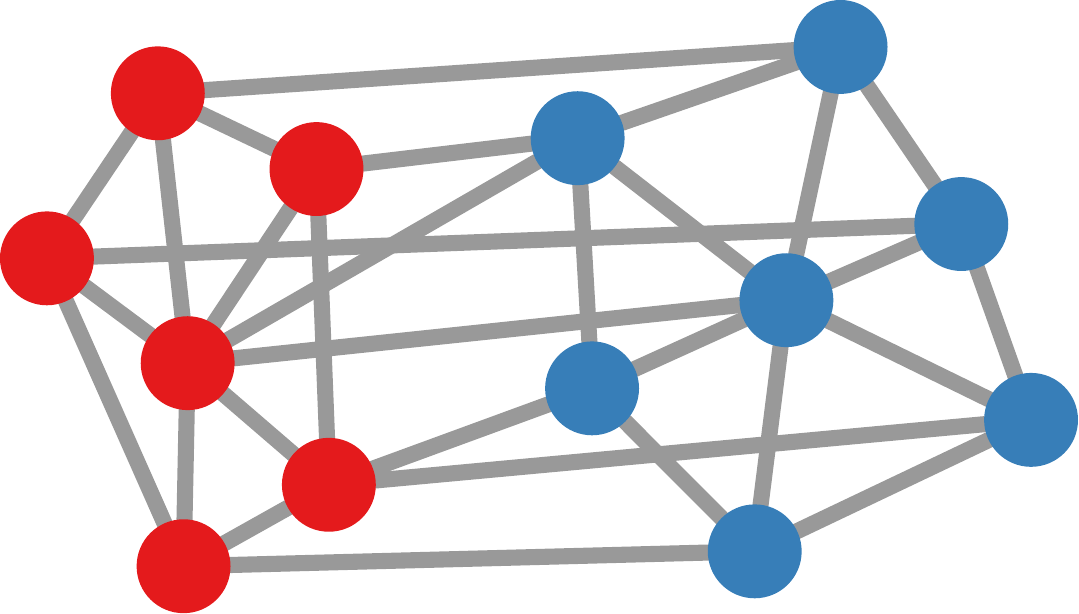}
\caption{}
\end{subfigure}
\caption{(a) The motif we want to minimize the cut for. (b) A high order normalized cut solution, which I represent via the node's color.}
\label{fig:highord-conductance}
\end{figure}

Boom. By finding an $S$ minimizing this specific $\phi(S, M)$ we just made a high-order normalized cut. This can be done exactly like finding the ``normal'' normalized cut, by examining the eigenvectors of a specially constructed Laplacian\cite{benson2015tensor}. Figure \ref{fig:highord-conductance} shows an example. In the figure, the two groups still have tons of edges going from one group to another, this is hardly a solution for a regular normalized cut. But there are few triangles between $S$ and $\bar{S}$, thus this is a proper solution for the higher order normalized cut. The shape of this solution can be applied to many other problems.

\subsection{Processes with Memory}
In this category of solutions you find all the approaches who take the equation of the Markov process and add a temporal factor. For instance, suppose that we're observing random walks (Chapter \ref{cha:rndwalks}). We have a vector of probability $p_k$ that tells us the status of the random walkers at time $k$, namely the probability of the walkers to be in a node ($p_k$ is a vector of length $|V|$). Now, if we were doing a normal random walk and perform one step, we'd know that the next step is simply given by the stochastic adjacency matrix, i.e. $p_{k+1} = p_k D^{-1}A$ (assuming $A$ is the normal adjacency matrix, and $D$ the degree diagonal matrix). This is a \textit{discrete} random walk on a graph.

But what if we want higher order dynamics? What if we want to look at the result of the process after $t$ steps? Well, the idea is to add the temporal information to the equation: $p_{k+t} = p_k T_t$. So what's $T_t$? $T_t$ is a matrix telling us the transition probability from $u$ to $v$ after $t$ steps. If you followed the linear algebra math in Chapter \ref{cha:rndwalks} you know that, to perform a random walk of length $t$, you can simply raise $D^{-1}A$ to the power of $t$: $(D^{-1}A)^t$.

The limitation here is that time ticks discretely, i.e. the random walker makes one move per timestep. In many cases, you might want to simulate the passage of time as a continuous flow\cite{schaub2012encoding}. If you want to do it, you need to derive a different $T_t$. First, rewrite $D^{-1}A$ as $-D^{-1}L$. This changes the random walk from discrete to continuous\cite{lambiotte2008laplacian}. This allows us to let time flow and take the result of the random walk at time $t$: $T_t = e^{-tD^{-1}L}$. If we set $t = 1$, we recover exactly the transition probabilities of a one-step random walk. But now we're free to change $t$ at will, to get second, third, and any higher order Markov processes.

\subsection{Other Approaches}
There is a whole bunch of other approaches to introduce high order dynamics into your network structure. More than I can competently cover. I provide here some examples with brief, but overly simplistic, explanations.

One way is to use tensors (Section \ref{sec:la-tensor}). In practice, you create a multidimensional representation of the topological features of the network. Each added dimension of the tensor represents an extra order of relationships between the nodes\cite{chertok2010efficient}\cite{duchenne2011tensor}. Then, by operating on this tensor, you can solve any high order problem: in the papers I cite the problem the authors focus on is graph matching.

Another general category of solutions is a collection of techniques to take high order relation data and transform it into an equivalent first order representation\cite{ishikawa2009higher}\cite{fix2011graph}. The first order solution in this structure then translates into the high order one, much like in the HON and memory network approach. The cited techniques are also generally applied to the problem of finding a high order cut in the network.

Being able to study high order interactions can help you making sense of many complex systems. For instance, they have been used to explain the remarkable stability of biodiversity in complex ecological systems\cite{grilli2017higher}. Other application examples include the study of infrastructure networks\cite{wegner2013higher} -- how to track the high order flow of cars in a road graph --, and aiding in solving the problem of controlling complex systems\cite{muhammad2006control} -- which we introduced in Section \ref{sec:triggers-control}.

\section{Summary}

\begin{enumerate}
\item Classical network analysis is single-order: only the direct connections matters. But many phenomena they represent are high-order: team collaborations are many-to-many relationships or, in a flight passenger network, your next step in the trip is influenced by all the steps you took in the recent past.
\item We can use simplicial complexes to model how many-to-many relationships make synchronization easier, create new potential failure types in infrastructures, and make epidemics more infective than we would normally expect.
\item Generating networks using simplicial complex dynamics can reproduce many real world properties, hinting at many-to-many interactions being a potential mechanism behind natural interacting processes.
\item There are two approaches to handle high-order processes with memory: you can modify the network structure so that it explicitly codes its memory; or you give to the algorithm the task of remembering previous moves.
\item In High Order Networks you split each node to represent all the paths that lead to it. In memory networks you instead model second order dynamics with a line graph, third order dynamics with the line graph of the line graph, and so on.
\item Embedding memory in the analysis can be done by either building dictionaries of temporal motifs, adding a temporal factor in the classical Markov equation (e.g. for random walks), and other approaches.
\end{enumerate}

\section{Exercises}

\begin{enumerate}
\item Calculate the distribution of the $k_{2,0}$ and $k_{2,1}$ degrees of the network at \url{http://www.networkatlas.eu/exercises/34/1/data.txt}. Assume every clique of the network to be a simplex.
\item Run an SI model with simple contagion on the simplicial complex from the previous exercise. The seed node is node $0$. Run it for all possible combinations of $\beta_1$ from $0.1$ to $0.9$ (in $0.1$ increments) and for $\beta_2 = \{0.0, 0.25, 0.5\}$. Make $100$ independent runs and average the results. Visualize the ratio of infected nodes after $3$ steps for each process -- like in Figure \ref{fig:simplicial-contagion}.
\item Generate the two-order line graph of the network at \url{http://www.networkatlas.eu/exercises/34/3/data.txt}, using the average edge betweenness of the edges as the edge weight.
\item Assume that the edge weight is proportional to the probability of following that edge. Which 2-step node transitions became more likely to happen in the line graph compared to the original network? (For simplicity, assume that the probability of going back to the same node in 2-steps is zero for the line graph)
\end{enumerate}

\part{Communities}\label{par:cd}

\chapter{Graph Partitions}\label{cha:cd-partitions}
We have reached the part of network analysis that has probably received the most attention since the explosion of network science in the early $90$s: community discovery (or detection). To put it bluntly, community discovery is the subfield of network science that postulates that the main mesoscale organization of a network is its partition of nodes into communities. Communities are groups of nodes that are very related to each other. If two nodes are in the same community they are more likely to connect than if they are in two distinct communities. This is an overly simplistic view of the problem, and we will decompose this assumption when the time comes, but we need to start from somewhere.

So the community discovery subfield is ginormous. You might ask: ``Why?'' Why do we want to find communities? There are many reasons why community discovery is useful. I can give you a couple of them. First, this is the equivalent of performing data clustering in data mining, machine learning, etc. Any reason why you want to do data clustering also applies to community discovery. Maybe you want to find similar nodes which would react similarly to your interventions. Another reason is to condense a complex network into a simpler view, that could be more amenable to manual analysis or human understanding.

More generally, decomposing a big messy network into groups is a useful way to simplify it, making it easier to understand. The reason why there are so many methods to find communities -- which, as we'll see, rarely agree with each other -- is because there are innumerable ways to simplify a network.

It is difficult to give you a perspective of how vast this subfield of network science is. Probably, one way to do it is by telling you that there are so many papers proposing a new community discovery algorithm or discussing some specific aspect of the problem, that making a review paper is not sufficient any more. We are in need of making review papers of review papers of community discovery. This is what I'm going to attempt now.

I think that we can classify review works fundamentally into four categories, depending on what they focus on the most. Review papers on community discovery usually organize community discovery algorithms by:

\begin{itemize}
\item \textbf{Process}. In this subtype of review paper, the guiding principle is how an algorithm works\cite{fortunato2010community}\cite{fortunato2016community}\cite{parthasarathy2011community}\cite{porter2009communities}\cite{newman2004detecting}\cite{danon2005comparing}\cite{gulbahce2008art}. Does it use random walks rather then eigenvector decomposition? Does it use a propagation dynamic or a Bayesian framework? Does it aim at describing the network as is, or at modeling what underlying latent process could have generated what we observe\cite{peixoto2023descriptive}?
\item \textbf{Performance}. Here, all that matters is how well the algorithm works in some test cases\cite{leskovec2010empirical}\cite{yang2016comparative}\cite{harenberg2014community}\cite{orman2009comparison}\cite{lancichinetti2009community}. The typical approach is to find many real world networks, or creating a bunch of synthetic benchmark graphs (usually using the LFR benchmark -- see Section \ref{sec:csmodels-comms}) and rank methods on how well they can maximize a quality function.
\item \textbf{Definition}. More often than not, the standard community definition I gave you earlier (nodes in the same community connect to each other, nodes in different communities rarely do so) isn't exactly capturing what a researcher wants to find. We'll explore later how this definition fails. Some review works acknowledge this, and classify methods according to their community definition\cite{schaeffer2007graph}\cite{coscia2011classification}\cite{malliaros2013clustering}. Different processes might be based in different definitions, so there's overlap between this category and the first one I presented, but that is not always the case.
\item \textbf{Similarity}. Finally, there are review works using a data-driven approach to figure out which algorithms, on a practical level, return very similar communities for the same networks\cite{vinh2010information}\cite{dao2018estimating}\cite{dao2018community}\cite{ghasemian2019evaluating}. This is similar to the performance category, with the difference that we're not interested in what performs \textit{better}, only in what performs \textit{similarly}.
\end{itemize}

The \textbf{process} approach is the most pedagogical one, because it focuses on us trying to understand how each method works. Thus, it will be the one guiding this book part the most. However, I'll pepper around the other approaches as well, when necessary. This book part will necessarily be more superficial than what one of the excellent surveys out there can do in each specific subtopic of community discovery, so you should check them out.

In this chapter I'll limit myself discussing the most classical view of community discovery, the one sheepishly following the classical definition. We're going to take a historical approach, exploring how the concept of network communities came to be as we intend it today. Later chapters will complicate the picture. Buckle up, this is going to be a wild ride.

\section{Stochastic Blockmodels}\label{sec:cd-partition-sbm}

\subsection{Classical Community Definition}
When people started mapping complex systems as networks, they realized that the edges didn't distribute randomly across nodes. We already saw that deviating from random expectation is a source of interest when we talked about degree distributions (Section \ref{sec:degree-distributions}). On top of the rich-get-richer effect, researchers realized that edges distributed unevenly among different groups of nodes. Especially, but not only, in social networks, there were lumps of connections, separated by very sparse areas.

\begin{figure}
\centering
\begin{subfigure}{.45\columnwidth}
\includegraphics[width=\textwidth]{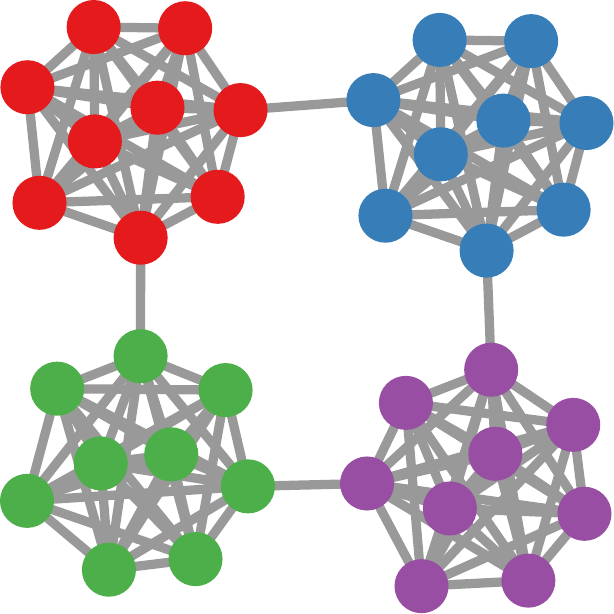}
\caption{}
\end{subfigure}\quad
\begin{subfigure}{.45\columnwidth}
\includegraphics[width=\textwidth]{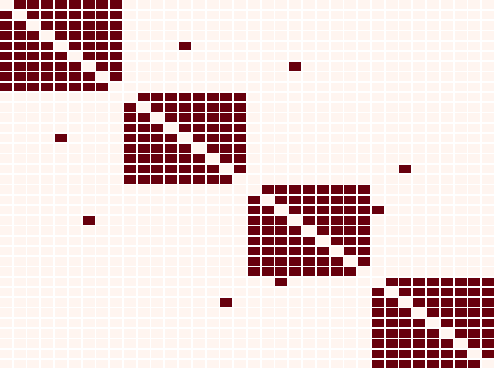}
\caption{}
\end{subfigure}
\caption{(a) A representation of a classical ideal community structure. I encode with the node color the community to which the node belongs. (b) The adjacency matrix of (a).}
\label{fig:community-classic}
\end{figure}

The ideal scenario is something resembling Figure \ref{fig:community-classic}(a). The natural next step was trying to see if we could separate these lumps of connections into coherent groups: communities. This created the first and most commonly accepted definition of a network community:

\begin{center}
\textit{Communities are groups of nodes densely connected to each other and sparsely connected to nodes outside the community.}
\end{center}

I would call this the classical definition of a network community. This definition can be attacked and deconstructed from multiple parts but, for now, let's accept it. Note that, for now, we assume that a node can only be part of a single community. We use the term ``partition'' to refer to the assignment of nodes to their community.

In the early community discovery days\cite{holland1983stochastic} -- before we even had coined the term ``community'' --, the main approach was using stochastic blockmodels. We already introduced them in Section \ref{sec:csmodels-comms} as a method to generate a synthetic graph. How do we apply them to the problem of finding communities? The first step in our quest is changing the perspective over the graph from Figure \ref{fig:community-classic}. Figure \ref{fig:community-classic}(b) shows you its corresponding adjacency matrix. The diagonal blocks are the ones referred by the name ``blockmodel''.

\subsection{Maximum Likelihood}

\begin{figure}
\centering
\begin{subfigure}{.425\columnwidth}
\includegraphics[width=\textwidth]{figures/block_communities.png}
\caption{}
\end{subfigure}\qquad
\begin{subfigure}{.425\columnwidth}
\includegraphics[width=\textwidth]{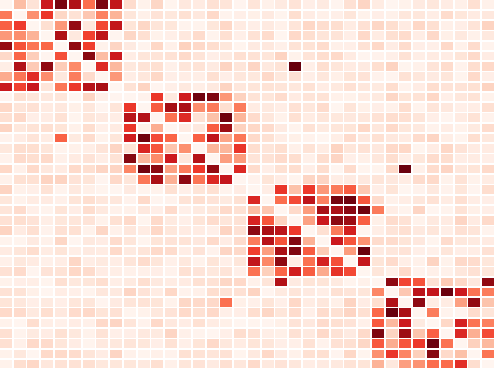}
\caption{}
\end{subfigure}
\caption{(a) An observed adjacency matrix. (b) A possible SBM representation, darker cells indicate edges with higher probability. Finding the most likely (b) explaining (a) is the task of SBM-based community detection.}
\label{fig:community-sbm}
\end{figure}

What we want to do now is to find the SBM model that most accurately reconstructs the adjacency matrix in Figure \ref{fig:community-classic}(b). To do so, we will need to plant a community structure in the SBM model. If the resulting network is similar to the original one, it is very likely that the partition we planted corresponds to the ``real'' partition in the original graph. As the vignette in Figure \ref{fig:community-sbm} shows, we do so by employing the principle of maximum likelihood\cite{king1998unifying}\cite{jaynes2003probability} -- for details about the likelihood function, see Section \ref{sec:ml-loss}. I'm going to give you a very incomplete and simplified view of what that means, prompting you to read more if you're interested in the topic.

Maximum likelihood estimation means to estimate the parameters of a model that are more likely to generate your observed data. Suppose that you have your parameters in a vector $\theta$. The likelihood function $\mathcal{L}$ tells you how good of a job you made using $\theta$ to approximate your observed adjacency matrix $A$. In practice, you're after those special parameters $\hat{\theta}$ such that $\mathcal{L}_{\hat{\theta}, A}$ is maximum. Mathematically:

$$ \hat{\theta} = \argmax \limits_{\theta \in \Theta} \mathcal{L}_{\theta, A},$$

where $\Theta$ is the space of all possible parameters, and $A$ is your given adjacency matrix.

So let's make an extremely simplified example of what that means when using SBM to find communities. This sketched example has a planted partition, only two parameters, and it exclusively focuses on the simplest definition of community -- the one I gave you earlier: simple non-overlapping assortative communities. I'm imposing these limitations for pedagogical purposes: in reality, SBMs are much more powerful than this.

\begin{figure}
\centering
\begin{subfigure}{.4\columnwidth}
\includegraphics[width=\textwidth]{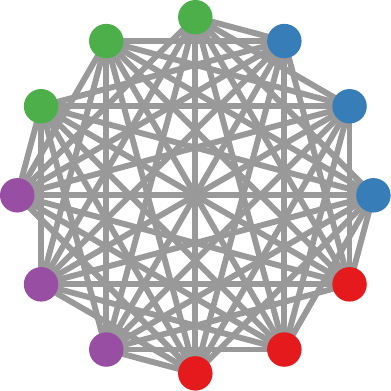}
\caption{}
\end{subfigure}\qquad
\begin{subfigure}{.425\columnwidth}
\includegraphics[width=\textwidth]{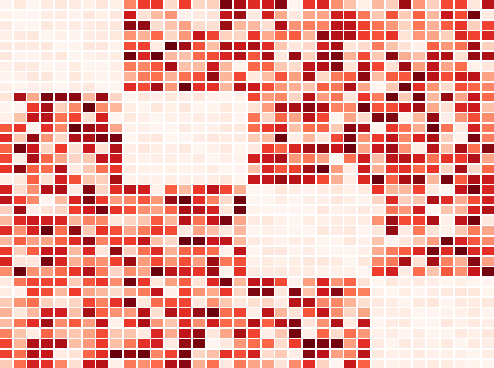}
\caption{}
\end{subfigure}
\caption{(a) A network with a disassortative community structure. (b) A possible disassortative SBM representation, darker cells indicate edges with higher probability.}
\label{fig:community-sbm-disass}
\end{figure}

I already mentioned in Section \ref{sec:csmodels-comms} why: SBM can find \textit{disassortative} communities, communities in which the nodes tend to connect to their community mates less than chance. The classical definition I gave you earlier is only for assortative communities. You can compare the disassortative community structure from Figure \ref{fig:community-sbm-disass}(a) with the classical assortative one in Figure \ref{fig:community-classic}(a). SBMs don't break a sweat in this scenario: you can simply flip the parameters and -- presto! -- you switch from the classical model in Figure \ref{fig:community-sbm}(b) to the disassortative model in Figure \ref{fig:community-sbm-disass}(b). I haven't even started explaining you the first community discovery method and we already found the first hole in the classical definition!

However, let's take it slow. To understand the SBM basics it is better to start with a cartoonishly simplified example and then introduce the features you can add to SBMs every time it'll be relevant in the book.

As mentioned, in the simplest SBM, we only have three parameters. The first two are the probabilities of two nodes connecting if they are -- or are not -- in the same community: $\theta_1$ and $\theta_2$, respectively. The third (hyper)parameter $\theta_3$ is the planted partition: it contains all node pairs that are in the same community.

Let's have the simplest possible $\mathcal{L}$ function. Let's define a helper variable defined for a pair of nodes: 

$$
l_{\theta, A, u, v} =
\begin{cases}
 \theta_1 - 1, & \text{if } A_{uv} = 1 \text{ and } (u,v) \in \theta_3\\
 \theta_2 - 1, & \text{if } A_{uv} = 1 \text{ and } (u,v) \not\in \theta_3\\
 -\theta_1, & \text{if } A_{uv} = 0 \text{ and } (u,v) \in \theta_3\\
 -\theta_2, & \text{if } A_{uv} = 0 \text{ and } (u,v) \not\in \theta_3.
\end{cases}
$$

In practice, if two nodes are connected, we subtract one from their probability of connections in the SBM -- since $\theta_1 > \theta_2$, this means we reward the case in which we put the two nodes in the same community. If they are not, we penalize ourselves proportionally to how sure we were they were connected: in this case we get the lowest penalty if we assumed them not being in the same community.

Then, we go over every node pair to check whether we assigned a high probability of edge existence to two nodes that were indeed connected, and we aggregate the scores:

$$ \mathcal{L}_{\theta, A} = \sum \limits_{u,v \in A} l_{\theta, A, u, v}.$$

Let's fix $\theta_1 = 1$ and $\theta_2 = 0.01$ for simplicity, since we're only interested in the partition ($\theta_3$). If we apply $\mathcal{L}$ to the perfect partition, the one following the colors in Figure \ref{fig:community-classic}(a), we get $\mathcal{L}_{\theta, A} = -8.78$ (I'm ignoring the diagonal because I don't allow self loops). Why?

We get zero contribution from the blocks in the diagonal, since they're fully connected and also in the same community (case $1$ in the formula). We get $4 \times (\theta_2 - 1)$ contribution from the edges connecting nodes in different communities (case $2$ in the formula) -- note that I don't double count because I assume the network is undirected. We get zero contribution from case $3$, as there are no disconnected nodes that we put in the same community. Finally, we get $482 \times -\theta_2$ contribution from the $482$ disconnected node pairs in different communities (case $4$).

Any other partition would return a lower likelihood. For instance, having only two communities, each fully including two blocks, has (following the four cases): $\mathcal{L}_{\theta, A} = 0 + (2 \times (\theta_2 - 1)) + (160 \times -\theta_1) + (322 \times -\theta_2) = -165.2$.

One of the problems of this approach is that the space of all possible partitions ($\theta_3$) is huge. How do we know which one is best? Besides, how do we even know what's the correct number of communities? A reliable way to estimate maximum likelihoods in presence of unknowns such as these, is the Expectation Maximization algorithm\cite{dempster1977maximum}.

Note that here I showed you an intuitive, but cartoonish, likelihood function. This will fail in most, if not all, real world networks. There are smarter alternatives out there, taking into account the degree distributions\cite{karrer2011stochastic}, and more that we will see in later chapters. There are also better heuristics to find the best partition\cite{peixoto2014efficient}.

Don't be deceived by the fact that the SBM approach I just described is clunky. I introduced it as a historic approach because it was the one used by sociologists before the network science renaissance of the late nineties. However, as I mentioned, SBMs are more sophisticated than this.

Many community detection methods use different approaches, and some can be shown as equivalent to SBMs. One such key algorithm is modularity maximization. However, since this is closely related with evaluating the quality of a partition, I'm going to defer presenting modularity maximization methods to Section \ref{sec:cd-eval-mod}.

Another approach uses the spectrum of the graph\cite{von2007tutorial}\cite{newman2006finding}. Well-separated communities will show large gaps in the eigenvalues. The eigenvectors are simply a coordinate system that you can use to place nodes in a multidimensional space and then use a standard clustering algorithm to find communities, such as k-means. I already covered the main concepts for this approach in Section \ref{sec:rw-mincut}, so I won't go into details here. Hereafter, we explore alternatives to the problem of discovering communities.

\section{Random Walks}\label{sec:cd-partition-rw}
A common approach for detecting communities looks at network processes. Random walks (Chapter \ref{cha:rndwalks}) are a popular choice. These walks tell us a lot about the community structure of the network. Consider the example in Figure \ref{fig:community-classic}(a). In the network, there are eight ``border nodes'' with a connection to a different community, out of $36$ nodes. And only one edge out of nine they have points outside the community. So we know that the probability for a random walker to get out of the community it is visiting is very low.

\begin{figure}
\centering
\includegraphics[width=.45\columnwidth]{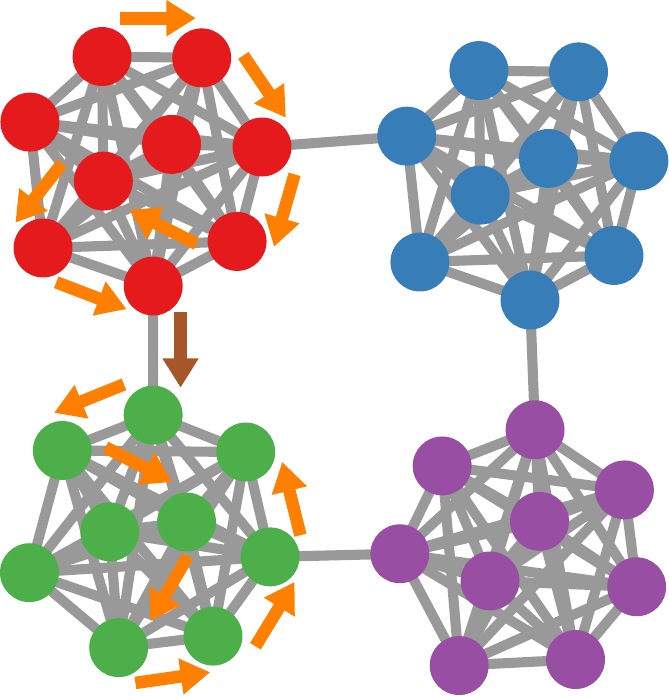}
\caption{A random walk in a network with communities. The node color indicates to which community a node belongs. The arrows show each step of the random walker. The orange arrows are transitions between nodes in the same community. The brown arrow is a transition between nodes in different communities.}
\label{fig:community-rndwalk}
\end{figure}

Appearing together in a random walk is a strong indication that two nodes are in the same community. Figure \ref{fig:community-rndwalk} shows why. When we perform our random walk, only one step out of the $13$ we took crossed communities. Let's do some napkin math. The probability of being a ``border node'' is $8 / 36 = 0.\bar{2}$. The probability of picking the one edge going outside of the community if you are in a border node is $1 / 9 = 0.\bar{1}$, because the node has degree nine and only one edge going out. So the probability of transitioning between communities in a random walk is $(8 / 36) \times (1 / 9) \sim 0.025$. Pretty low!

The guiding principle of detecting communities with random walks is that a random walk is likely to be trapped inside a community and to keep visiting nodes belonging to the same community. This has been exploited in a number of different ways\cite[-0.85in]{dongen2000cluster}\cite[-0.55in]{pons2006computing}\cite{zlatic2010topologically}\cite{satuluri2009scalable}\cite{weinan2008optimal}. These are only a few examples of the many papers using this approach -- you're going to hear this excuse from me a lot, to save myself from citing literally everything and making this a book about community discovery.

\begin{figure}
\centering
\includegraphics[width=.66\columnwidth]{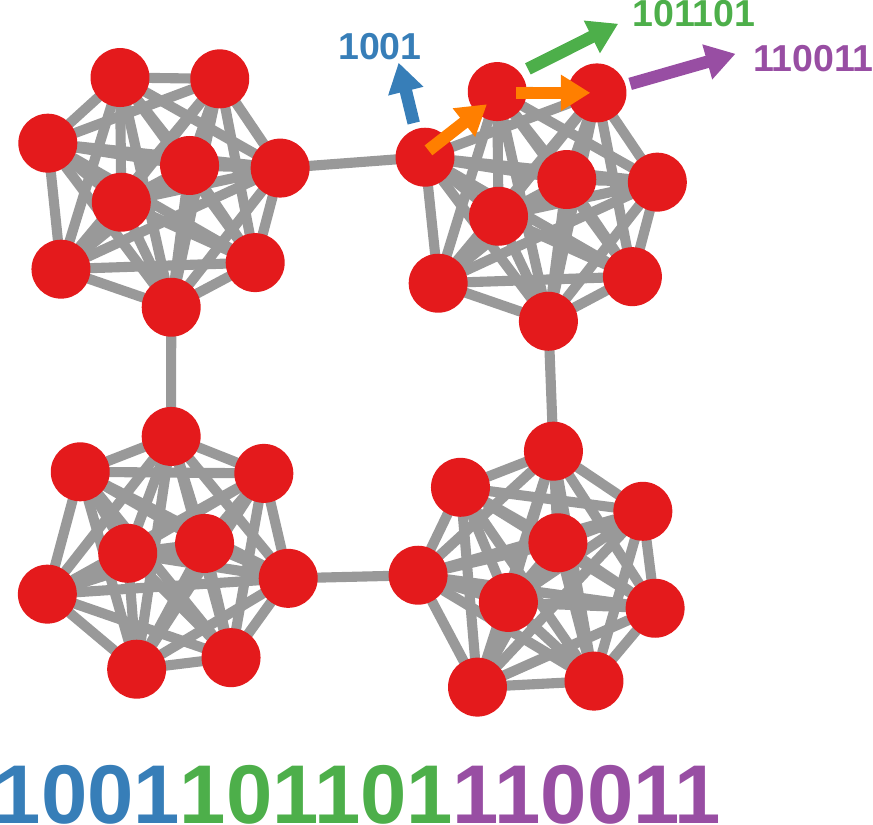}
\caption{A binary node ID schema you'd use to encode a random walk. Each colored arrow points to the ID of the three nodes involved in the orange walk.}
\label{fig:community-infomap}
\end{figure}

The most known and best performing of these approaches is usually considered the map equation approach, or Infomap\cite{rosvall2008maps}\cite{bohlin2014community}. The map equation is what you use to encode the random walk information with the minimum possible number of bits -- i.e. minimizing the ``code length''. Suppose that you give each node a binary ID. Since we have $36$ nodes, we need around $5$ bits, but we can save a little if we give shorter codes to central nodes (they are going to be visited more often). Then the cost of describing the random walk is simply the length of the code of the node multiplied by the number of times we're going to see it in a random walk, which is given by the stationary distribution (Section \ref{sec:rw-stationary}). In the example from Figure \ref{fig:community-infomap}, the orange walk is fully described by the bits in the figure: the node id sequence.

\begin{figure}
\centering
\includegraphics[width=.8\columnwidth]{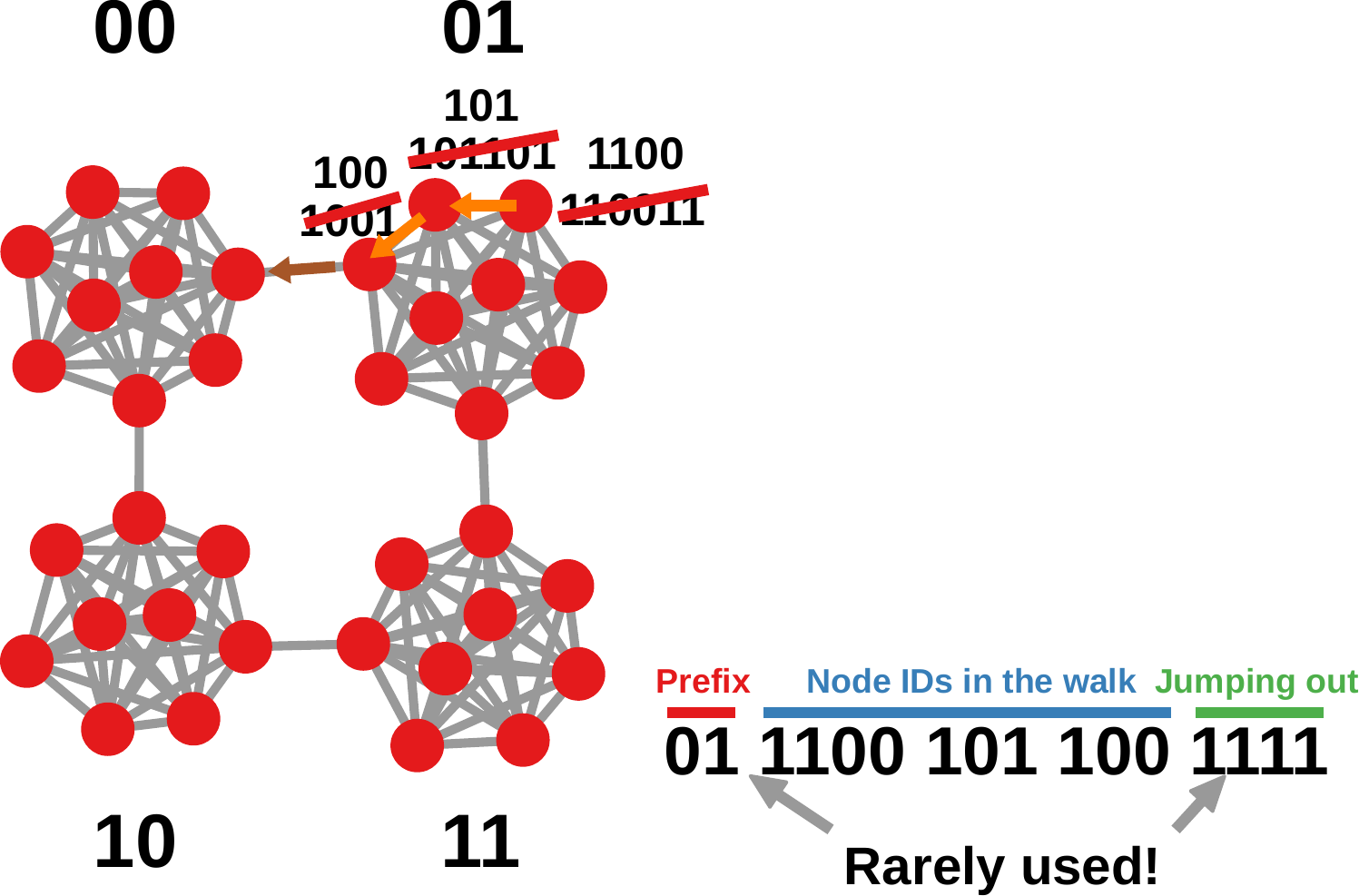}
\caption{The large two-digit codes are the IDs of the communities. Each node gets a new shorter ID, given that IDs need to be community-unique, rather than network-unique. Now the random walk uses the prefix (in red) to indicate in which community it is, then the new shorter node IDs (in blue) and finally adds an extra ID to indicate it's jumping out of a community (in green).}
\label{fig:community-infomap2}
\end{figure}

Infomap saves bits by using community prefixes. Nodes in the same community get the same prefix. So now we need fewer bits to uniquely refer to each node, because we can prepend the community code the first time and then omit it as long as we are in the same community. Since a community contains, in this case, $9$ nodes instead of $36$, we can use shorter codes. We need to add an extra code that allows us to know we're jumping out of a community. This is an overhead, but the assumption here is that a random walker will spend most of its time in the community, so this community prefix and jump overhead is rarely used. Figure \ref{fig:community-infomap2} shows this re-labeling process.

\begin{figure}
\centering
\includegraphics[width=\columnwidth]{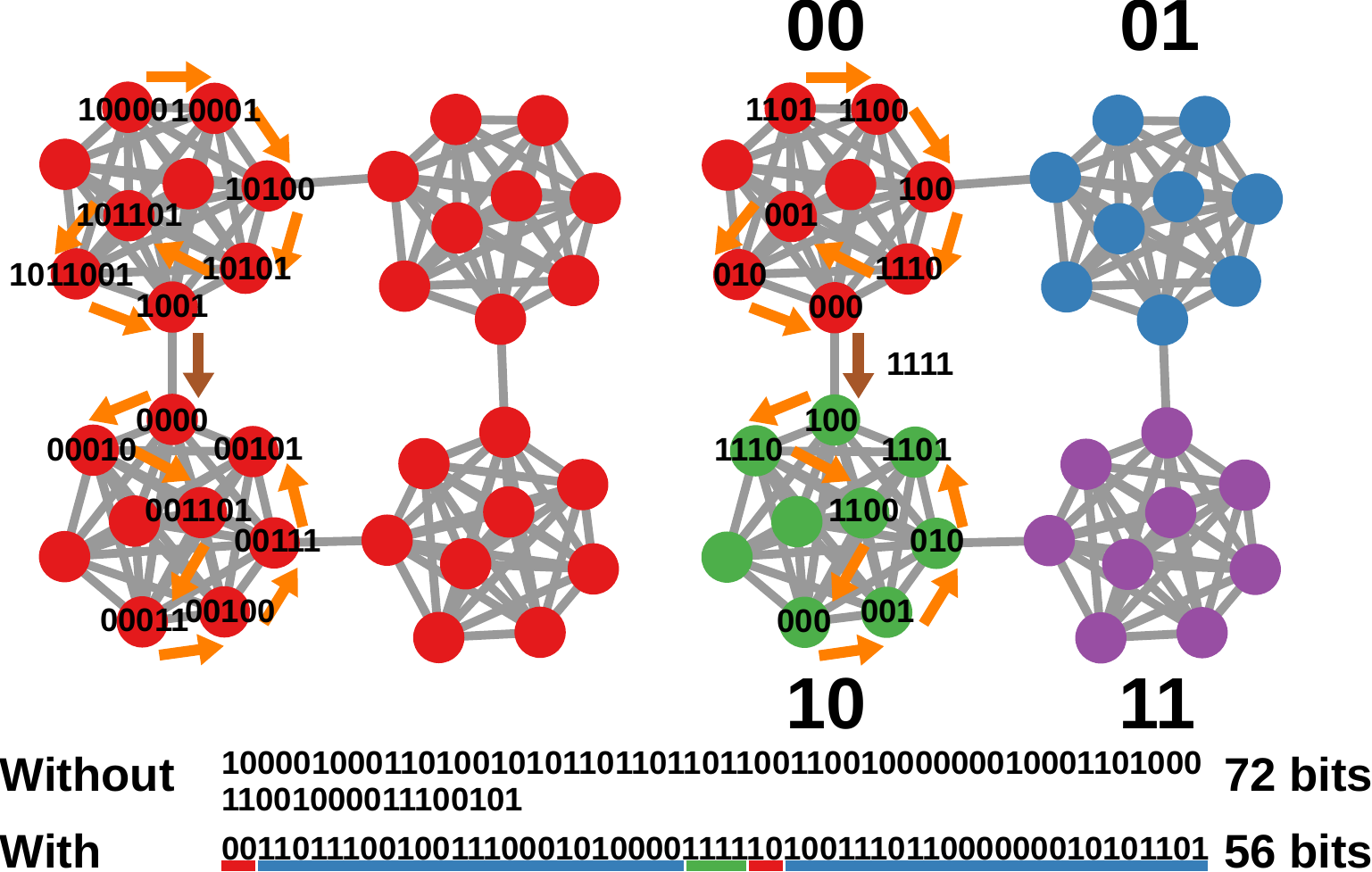}
\caption{The cost of encoding a random walk with the naive scheme (left) compared with the Infomap scheme (right). In the Infomap scheme, I underline in red the community prefixes, in green the inter-community jump overhead, and in blue the node ID encoding.}
\label{fig:community-infomap3}
\end{figure}

If the partition is good, we can compress the random walk information by a lot. Consider the example in Figure \ref{fig:community-infomap3}. Without communities we have no overhead, but we need to fully encode our $36$ nodes. The path in orange is simply the sequence of node IDs and can be stored in $72$ bits. If we have community partitions, we add the community prefixes and the jump overhead (for the community jump in brown), but the node IDs are shorter. The encoding of the same walk is $56$ bits, and we can see that the overhead parts are tiny compared with the rest.

If my explanation still makes little sense, you can try out an interactive system showing all the mechanics of the map equation approach\footnote{\url{http://www.mapequation.org/apps/MapDemo.html}}. Infomap has been adapted to numerous scenarios. Many involve hierarchical, mutlilayer and overlapping community detection, which we will explore in later chapters. Other modifications include adding some ``memory'' to the random walkers\cite{schaub2012encoding}\cite{rosvall2014memory} -- effectively using higher order networks (Chapter \ref{cha:hod}).

This means that the walker is not randomly selecting destinations any more, but it follows a certain logic. Consider a network of flight travelers. If you fly from New York to Los Angeles, your next leg trip isn't random. You're much more likely, for instance, to come back to New York, since you were in LA just for a vacation or visiting family.

Some approaches do not use vanilla random walks, but also consider the information encoded in node attributes in the map equation\cite{smith2016partitioning}.

Since these methods are based on a fundamentally random process -- random walks -- they tend to be non-deterministic. This means that running the same algorithm on the same input twice might return different results.

\section{Label Percolation}\label{sec:cd-partition-lp}
Random walks are running a dynamic process to detect communities, meaning that we're performing some sort of event on the network to uncover the community structure. Another very popular dynamic approach is having nodes deciding for themselves to which community they belong by looking at their neighbors' community assignments.

We use node labels to indicate to which community each node belongs. We start with a network whose node labels are scattered randomly. Then each node looks at its neighbors and adopts the most common labels it sees (if there is a tie, it will choose a random one among the most popular). As a result, the labels will percolate through the network until we reach a state in which no more significant changes can happen. The assumption is that, in a community, nodes will end up surrounded by nodes with the same label. That is why this class of solutions is usually know as ``label percolation'' (or propagation). 

\begin{figure}
\centering
\begin{subfigure}{.425\columnwidth}
\includegraphics[width=\textwidth]{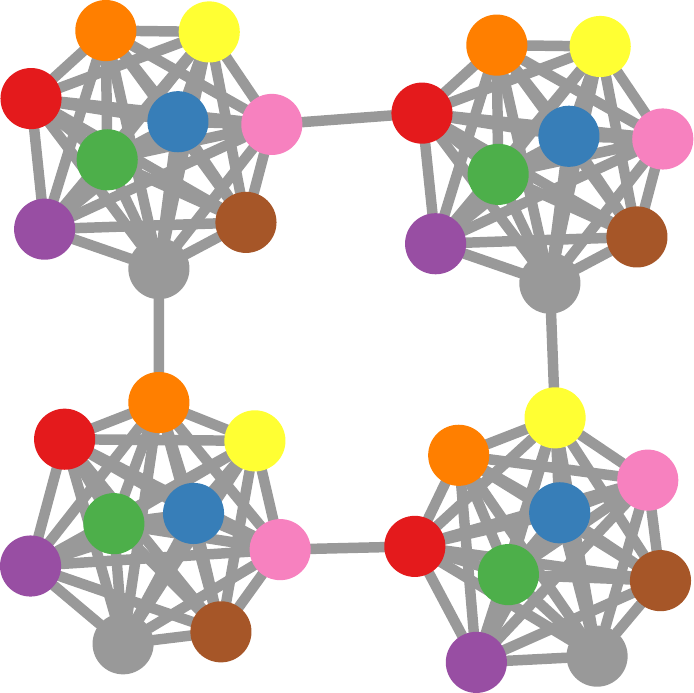}
\caption{}
\end{subfigure}\qquad
\begin{subfigure}{.425\columnwidth}
\includegraphics[width=\textwidth]{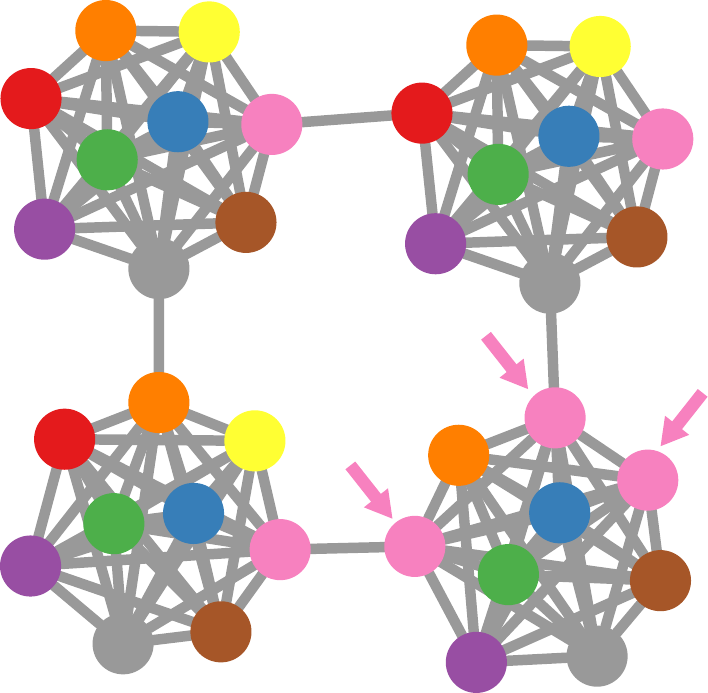}
\caption{}
\end{subfigure}
\caption{(a) Starting condition of label propagation algorithm, with labels distributed randomly in the network. (b) After some iterations, randomly, some neighbors will adopt the same label. The highlighted pink nodes will take over the community at the next time step.}
\label{fig:community-lp}
\end{figure}

At the beginning, nodes will switch their labels randomly. However, by chance, some nodes will eventually adopt the same label. If they are in the same community, all of a sudden this label is the only one with two nodes in the cluster. It starts becoming the majority label for many nodes in the cluster, and thus it will be eventually adopted by everybody, as Figure \ref{fig:community-lp} shows.

This approach is fairly straightforward and computationally simple. In fact, one of the claims to fame of such an algorithm is its time complexity: it runs linearly in terms on the number of edges in the network. You just have to iterate over your edge list a few times before convergence.

The most important dimension along which the many papers implementing label propagation community detection differ is in the strategy they employ for the nodes to look around their neighborhood. We can classify them in three classes: asynchronous\cite{raghavan2007near} -- which is also the original formulation of the label propagation principle --, semi-synchronous\cite{cordasco2010community}, and synchrnous\cite{xie2011slpa}.

The asynchronous case uses the labels that the nodes had at the previous iteration. For instance, at the $i$th iteration a node will decide which label to adopt by looking at the majority label between its neighbors at the $(i-1)$th iteration. If some neighbors have, in the mean time, updated their label, this information is ignored, and will be used only at the $(i+1)$th iteration.

In the synchronous approach this is not the case: you always use the most up-to-date information you have. If some of your neighbors already changed their label, you look the their $i$th iteration label. Otherwise, you look at their $(i-1)$th iteration label. The semi-synchronous case is, as you might expect, a combination of the two.

Note that, just like in the random walk case, the label propagation algorithms are typically non-deterministic. The random choices nodes make when breaking ties among the most popular labels around them can lead to differences in the detected communities.

\section{Temporal Communities}\label{sec:cd-partition-evo}

\subsection{Evolutionary Clustering}
So far I've framed the community discovery problem as essentially static. You have a network and you want to divide it into densely connected groups. However, we saw in Section \ref{sec:extended-dynamic} that many of the graphs we see are views of a specific moment in time. Networks evolve and you might want to take that information into account. A couple of good review works\cite{cai2016survey}\cite{rossetti2018community} focus on dynamic community discovery and can help you obtaining a deeper understanding of this problem. Let's explore what can happen to your communities over time.

\begin{figure}
\centering
\includegraphics[width=.8\columnwidth]{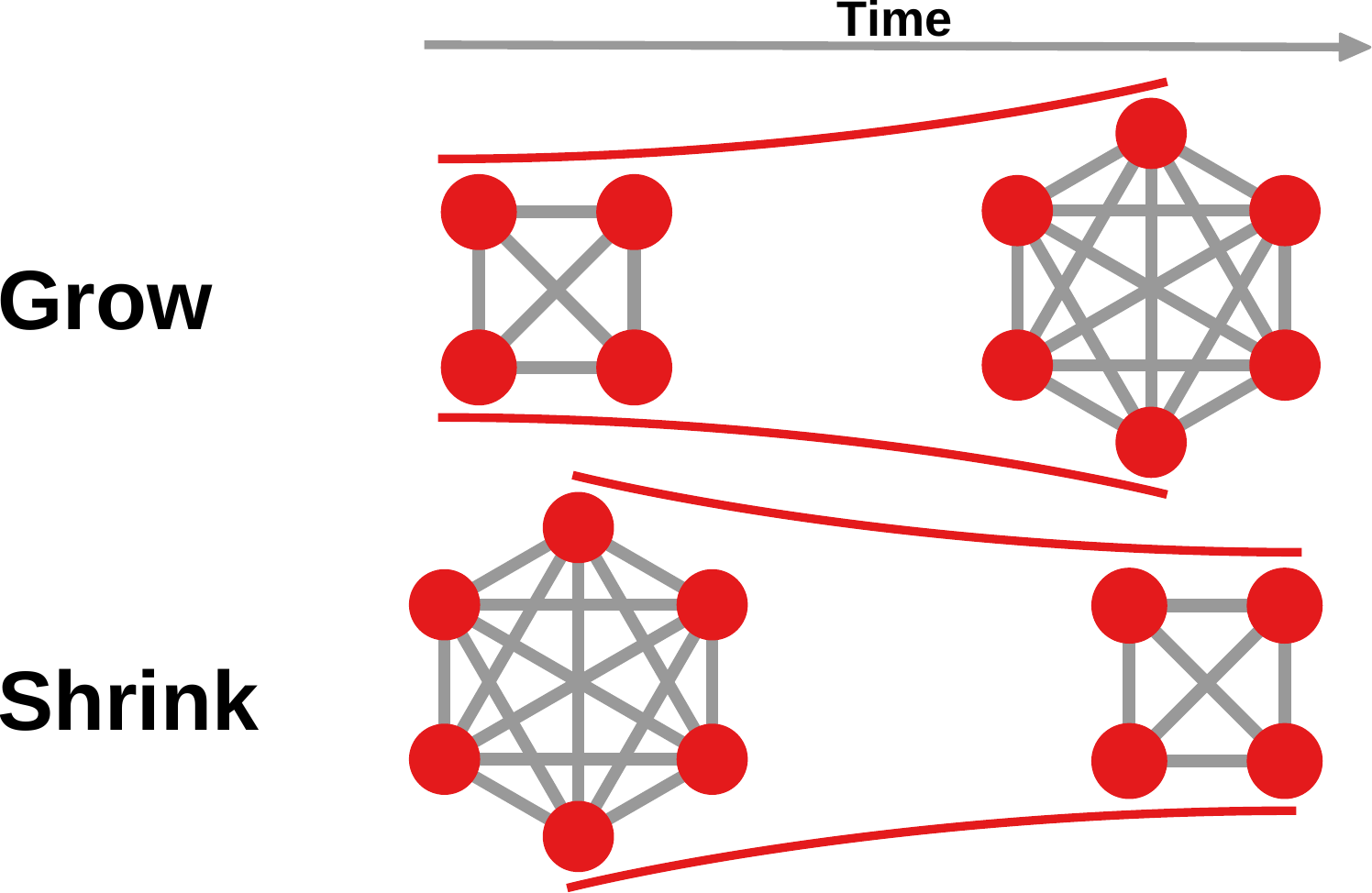}
\caption{Two things that can happen to your communities in an evolving network: growing and shrinking.}
\label{fig:community-evo-1}
\end{figure}

One possibility is that the community will grow: it will attract new nodes that were previously unobserved. The other side of the coin is shrinking: nodes that were part of the community disappear from the network. Figure \ref{fig:community-evo-1} shows visual examples of these events.

\begin{figure}[t]
\centering
\includegraphics[width=.8\columnwidth]{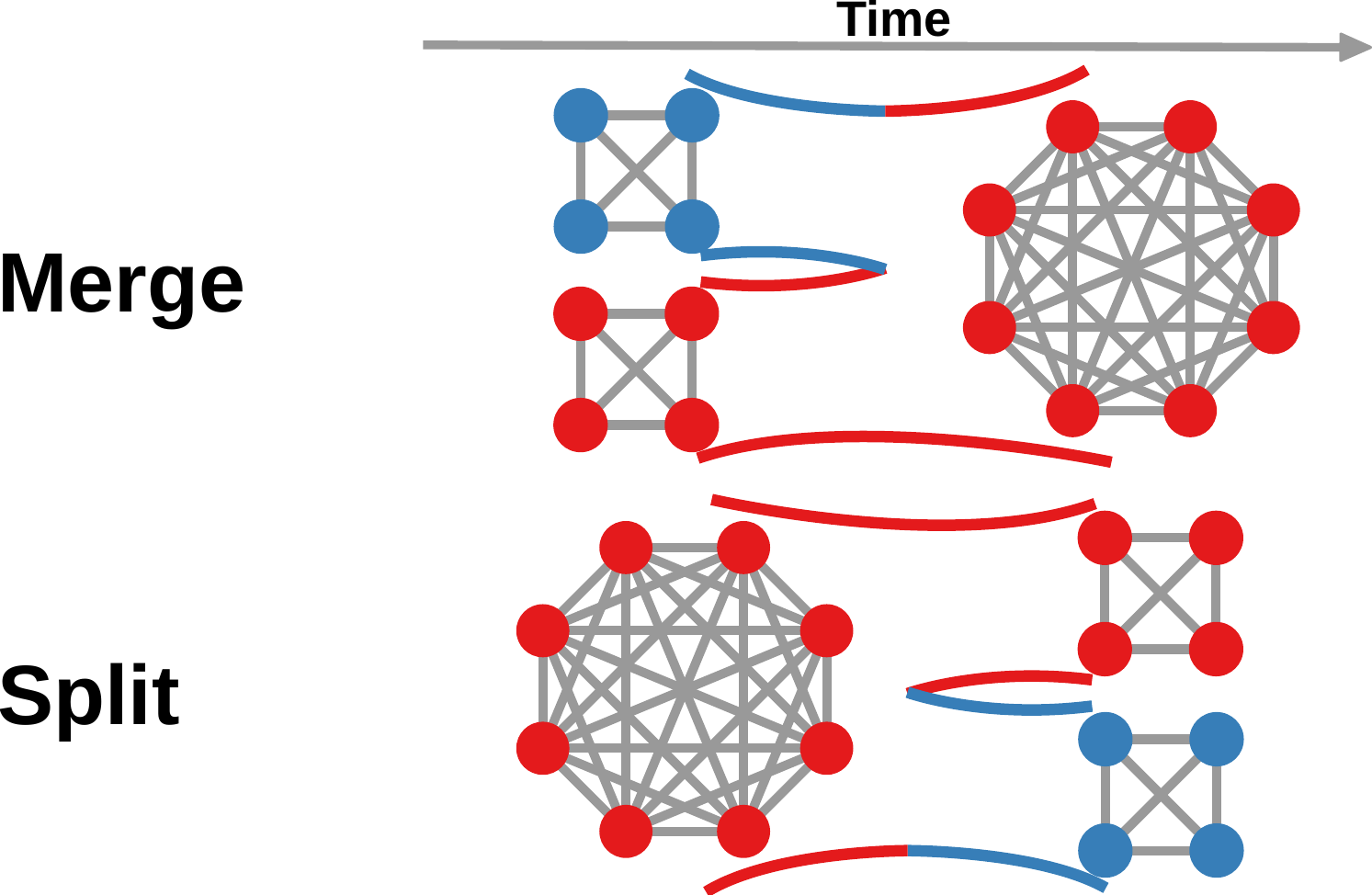}
\caption{Two things that can happen to your communities in an evolving network: merging and splitting.}
\label{fig:community-evo-2}
\end{figure}

Another way for a community to grow is by merging with other communities. The difference with the previous case is that in the growth case the added nodes to the community were not previously observed. In this case they were, and they were classified in a different community. ``Grow'' happens with the addition of nodes, while ``merge'' usually happen with the addition of edges. Again, we can have the opposite case: a community which splits into two or more new communities, due to the loss of edges (rather than nodes, as it was the case in the ``shrink'' scenario). Figure \ref{fig:community-evo-2} shows visual examples of these events.

\begin{figure}[t]
\centering
\includegraphics[width=.8\columnwidth]{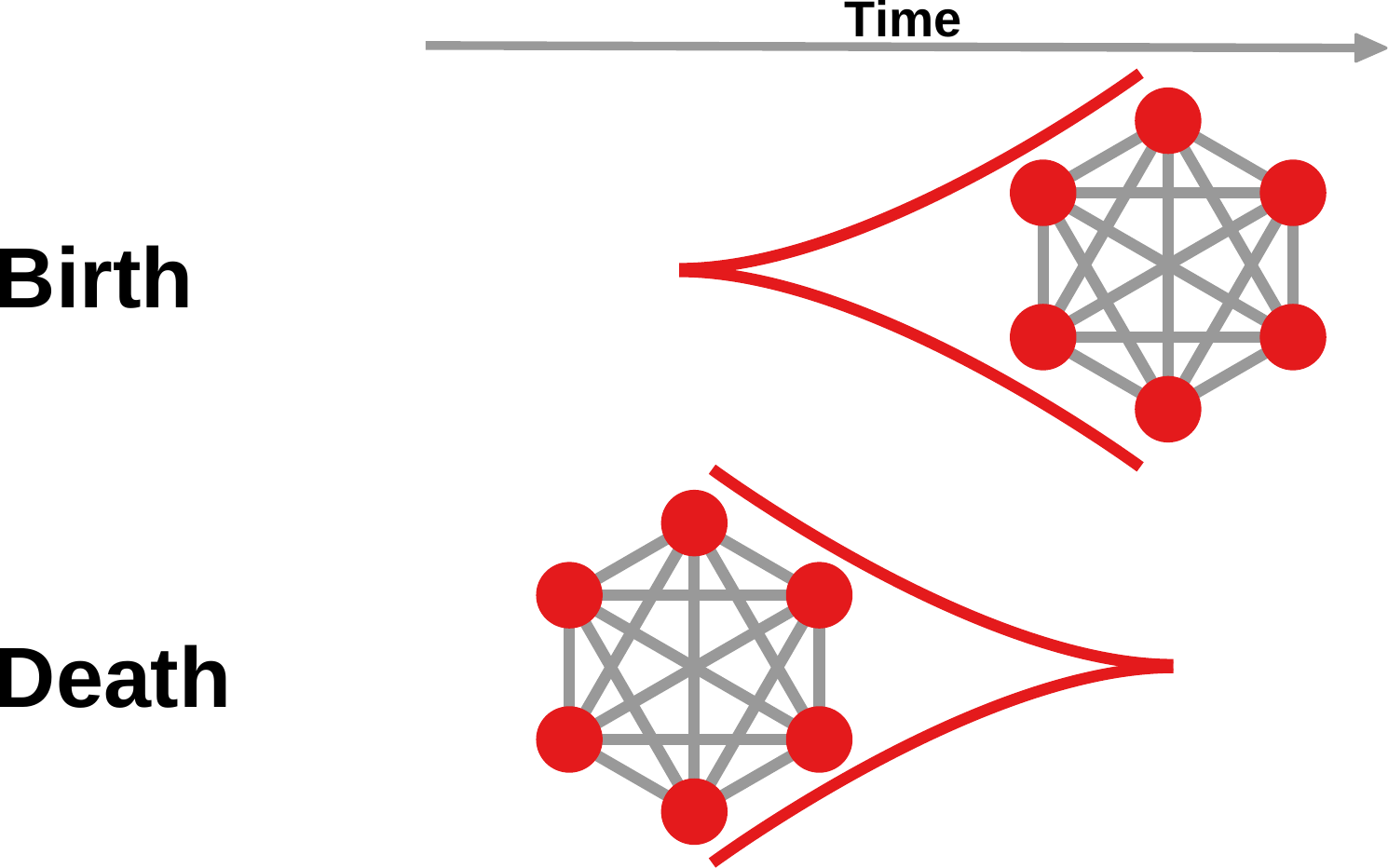}
\caption{Two things that can happen to your communities in an evolving network: birth and death.}
\label{fig:community-evo-3}
\end{figure}

Communities can also arise from nothing. This is like ``grow'', except that none of the nodes forming the community were previously observed in the network. The converse is community death: every node which was part of it disappears from the network. Figure \ref{fig:community-evo-3} shows visual examples of these events.

Note that, as everything else in this chapter, also this description is cartoonish. What happens in real world networks is way messier. Communities grow, shrink, split, and merge at the same time. Merges could also be partial, with communities ``stealing'' nodes from each other, resulting in an evolution that is not nearly as neat as the one I depicted here. Don't assume that you're going to be able to say, unequivocally, something like ``community $C$ split in $C_1$ and $C_2$ at time $t$''.

How do you detect communities in an evolving graph? One possible approach could be to define a series of network snapshots, apply a community discovery algorithm to each of these snapshots, and then combine the results into a single clustering\cite{hopcroft2004tracking}\cite{asur2009event}\cite{palla2007quantifying}. This is fine in some scenarios, but is generally not advisable, as it makes the quiet assumption that snapshots are sort of independent.

It becomes challenging to link the community discovery of each snapshot to the next. For each community at time $t$ you're trying to find the community at time $t + 1$ that is the most similar to it, and say that the most recent community is an evolution of the older one. A possible similarity criterion would be calculating the Jaccard coefficient.

A better solution is performing evolutionary clustering\cite{chakrabarti2006evolutionary}. This means that we add a second term to whatever criterion we use to find communities in a snapshot -- a procedure sometimes called ``smoothing''. Suppose you're using Infomap. The aim of the algorithm, as I presented earlier, is to encode random walkers with the lowest number of bits. Let's say this is its quality function -- which is known as code length ($CL$).

In evolutionary clustering you don't just optimize $CL$. You have $CL$ as a term in your more general quality function $Q$. The other term in $Q$ is consistency. For simplicity sake, let's just assume it is some sort of Jaccard coefficient between the partitions at time $t$ and the partition at time $t - 1$. To sum up, a very simple evolutionary clustering evaluates the partition $p_t$ at time $t$ as:

$$ Q_{p_t} = \alpha CL_{p_t} + (1 - \alpha) J_{p_t, p_{t - 1}}.$$ 

\begin{figure}
\centering
\begin{subfigure}{.32\columnwidth}
\includegraphics[width=\textwidth]{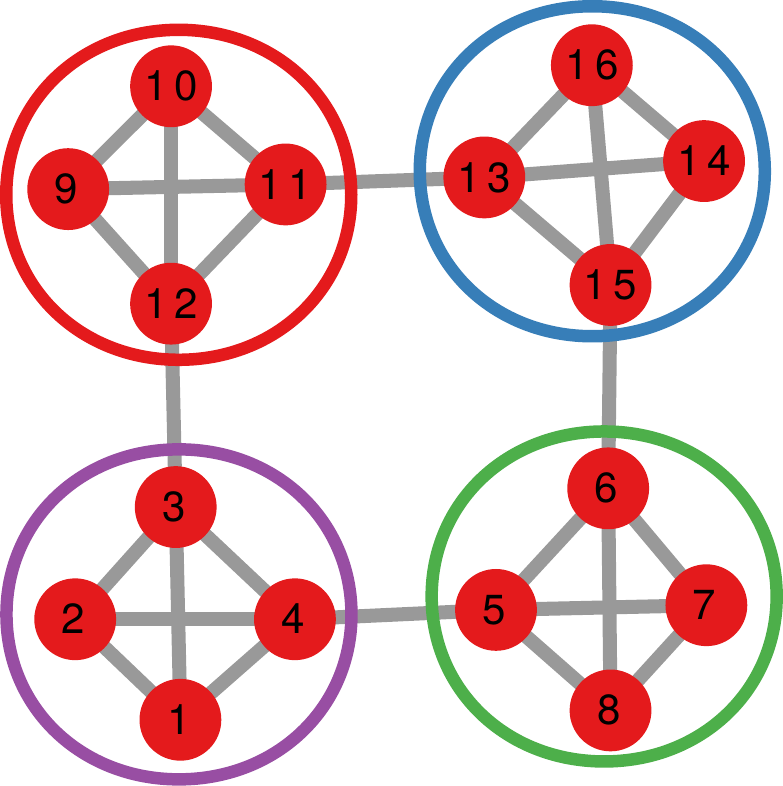}
\caption{}
\end{subfigure}
\begin{subfigure}{.32\columnwidth}
\includegraphics[width=\textwidth]{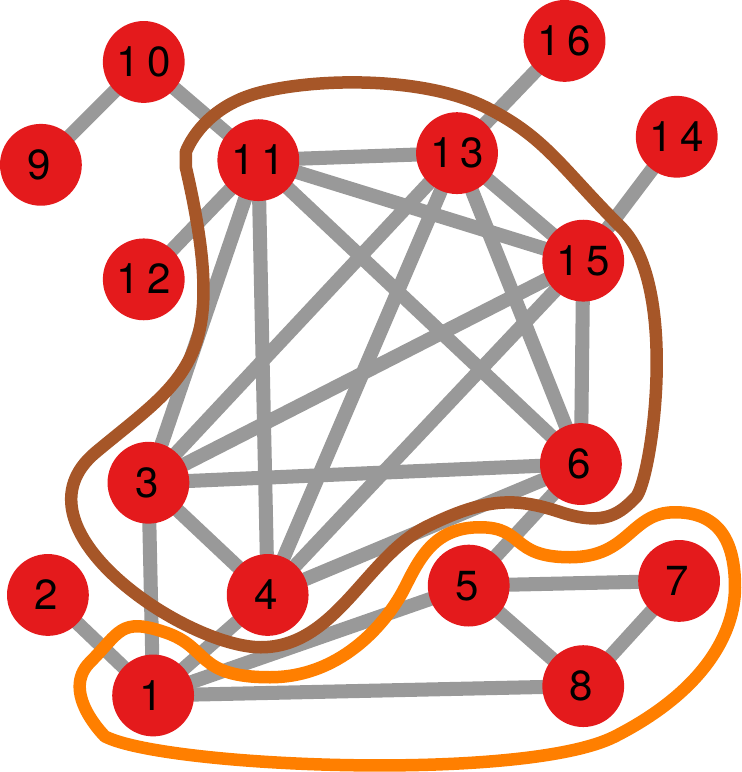}
\caption{}
\end{subfigure}
\begin{subfigure}{.32\columnwidth}
\includegraphics[width=\textwidth]{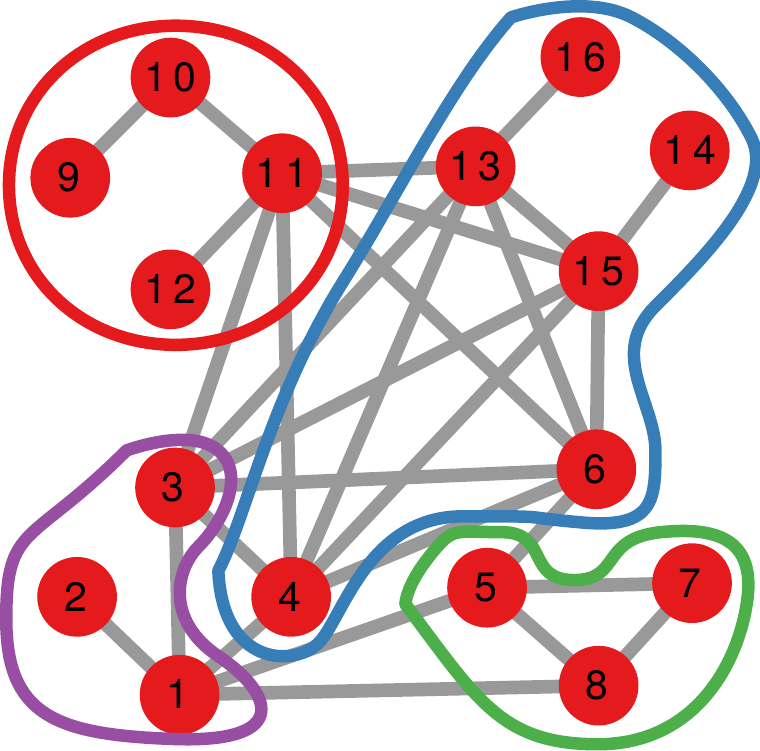}
\caption{}
\end{subfigure}
\caption{(a) The community partition of a graph at time $t$. (b) A partition of the graph at time $t + 1$ exclusively optimizing the code length, using Infomap. (c) A partition of the graph at time $t + 1$ balancing a good code length and consistency with the partition at time $t + 1$.}
\label{fig:community-evo4}
\end{figure}

Here, $\alpha$ is a parameter you can specify which regulates how much weight you want to give to your previous partitions. For $\alpha = 1$ you have standard clustering, while for $\alpha = 0$ the new information is discarded and you only use the partition you found at the previous time step. Figure \ref{fig:community-evo4} shows you that maximizing $CL_{p_t}$ might yield significantly different results than maximizing a temporally-aware $Q_{p_t}$ function.

This is only one -- the simplest -- of the many ways to perform smoothing, which the other review works I cited describe more in details. However, all these methods (and the ones that follow) have something in common: they are all at odds with the classical definition of community that I gave you earlier. That is because, at time $t + 1$, we're not simply trying to group nodes in the same community according to the density of their connections. Eventually, we're going to end up with a partition with many edges running between communities, which is against the traditional definition of community. Together with the ability of SBMs to find disassortative communities, these are yet more cracks appearing in the classical community detection assumption of assortative communities.

Smoothing is not necessarily applied to adjacent snapshots: you can have a longer memory looking at $t - 2$, $t- 3$, and so on\cite{goldberg2011tracking}\cite{morini2017revealing}. In alternative approaches, you can skip the smoothing altogether. You can identify a ``core-node'' which is the center of the community and will identify it for all snapshots. You then find communities around that node\cite{chen2010detecting}\cite{wang2008commtracker}.

\subsection{Other Approaches}
Alternatives to evolutionary clustering exist. You could find an optimal partition only for the very first snapshot of your network. As you receive a new snapshot, rather that starting from scratch and then smoothing, you can adapt the old communities to the new network, whether you do it via global optimization\cite{miller2010continuous}, or using a specific set of rules to update the old communities\cite{rossetti2017tiles}\cite{sun2010community}\cite{tang2008community}.

Another approach consists in defining a dynamic null model: a null version of your evolving network which has no communities\cite{bassett2013robust}, much like a random graph. Then you look at deviations from this expected null model in the network as the potential sources of dynamic communities.

You can use SBMs in this case as well -- you can use SBMs for \textit{any case}, really. SBMs tend to be more principled than evolutionary clustering approaches, because they model temporal communities directly\cite{peixoto2015inferring}, rather than chasing communities around as your network evolves. As an example, you could use a tensor representation in which each slice of the tensor is a snapshot of the network. Since a tensor is nothing more than a high dimensional matrix, and SBMs understand matrices, you can make a tensor-SBM\cite{tarres2019tensorial}. One nice thing about the SBM approach is that it allows to estimate from data the timescale at which the community structure changes -- which is inferred by the coupling between snapshots. This is nice because then you don't need to decide yourself the granularity of the temporal observation. Basing your inferences on data rather than taking a guess is always a plus!

A final approach is not to consider the different snapshots as separate, but taking the entire structure of the network as input all at once, as if it were a single structure. For instance, you can split each node $v$ into many meta-nodes $v_{t_1}, v_{t_2}, ...$ connected to each other by special edges\cite{sun2007graphscope}\cite{gauvin2014detecting}\cite{peel2015detecting}\cite{ghasemian2016detectability}\cite{viard2016computing}. This is similar to performing multi-layer community discovery, which we'll see later.

\section{Local Communities}\label{sec:cd-partition-local}
In some cases, you are not interested in grouping every node into a community. I'm not just referring to allowing nodes to be part of no communities -- a feature included in many algorithms, regardless of their guiding principle. Sometimes, you want to find local communities: you're interested in knowing the communities around a specific (set of) node(s), regardless of the rest of the network. This makes sense if the network is very large and some nodes are just too far to ever influence the results on your specific objectives. Or you cannot analyze it fully because it would take too much memory. Or it might take too much time to access the entire network, imposing you to sample it (see Chapter \ref{cha:sampling}).

This is usually done by exploring the graph one node at a time, putting nodes into different bins according to their exploration status -- and their community affiliation. For instance, you start from a seed node $v_0$, which by definition is part of your local community $\mathcal{C}$. All of its neighbors are part of the unexplored node set $\mathcal{U}$.

\begin{figure}
\centering
\begin{subfigure}{.22\columnwidth}
\includegraphics[width=\textwidth]{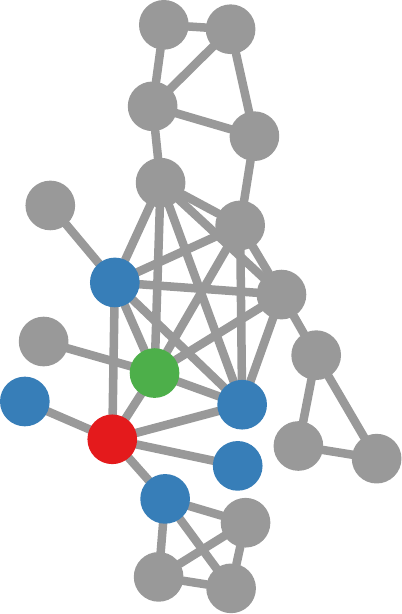}
\caption{}
\end{subfigure}\quad
\begin{subfigure}{.22\columnwidth}
\includegraphics[width=\textwidth]{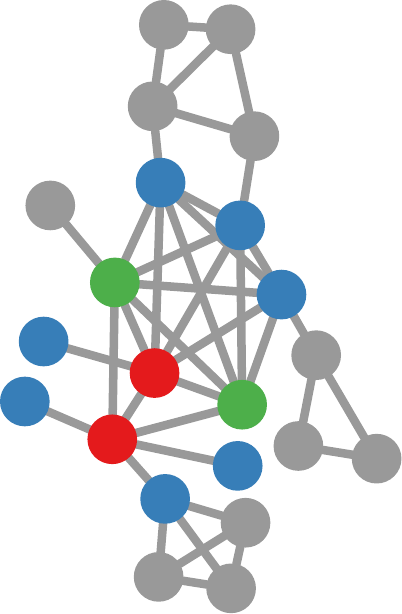}
\caption{}
\end{subfigure}\quad
\begin{subfigure}{.22\columnwidth}
\includegraphics[width=\textwidth]{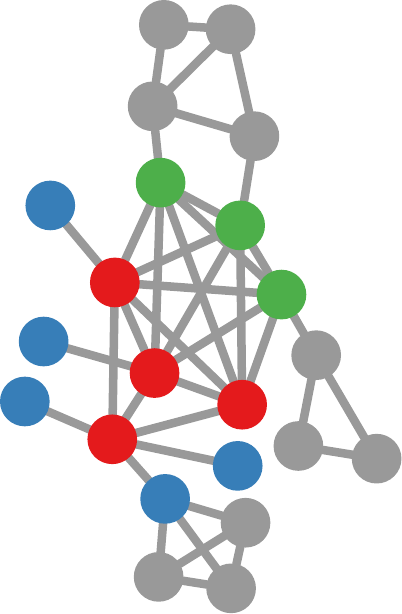}
\caption{}
\end{subfigure}\quad
\begin{subfigure}{.22\columnwidth}
\includegraphics[width=\textwidth]{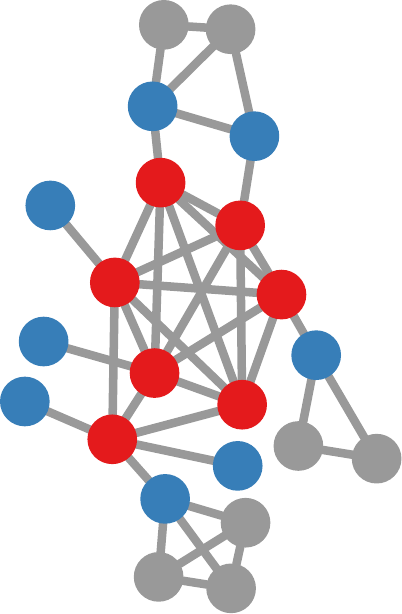}
\caption{}
\end{subfigure}
\caption{The process of discovering local communities. Red: nodes in $\mathcal{C}$. Blue: nodes in $\mathcal{U}$. Green: nodes maximizing the number of edges in $\mathcal{C}$ and therefore being added to the community.}
\label{fig:community-local}
\end{figure}

You iterate over all members of $\mathcal{U}$, trying to find the one that would maximize some community quality function. For instance, it could be simply the number of edges connecting inside $\mathcal{C}$ -- with ties broken randomly. We add to $\mathcal{C}$ the $v_1$ node with the most edges to the local community. Then, $\mathcal{U}$ is updated with the new neighbors, the ones $v_1$ has but $v_0$ did not.

We continue until we have reached our limit: we explored the number of nodes we wanted to test, or we ran out of time, or we actually explored all nodes in the component to which $v_0$ belongs. Figure \ref{fig:community-local} shows some steps of the process. As you can see, we can terminate after we explore a certain set of nodes. At that point, we detected the local community of node $v_0$, without exploring the entire network. We did not explore the blue nodes -- although we know they exists -- and we're absolutely clueless about the existence of the grey nodes.

The algorithm I just described is one\cite{clauset2005finding} of the many possible\cite{bagrow2008evaluating}\cite{luo2008exploring}\cite{papadopoulos2009bridge}. All these algorithms are variations of this exploration approach. More alternatives have been proposed, for instance using random walks like Infomap to explore the network\cite{jeub2015think}\cite{de2014mixing}. You can explore the literature more in depth using one of the survey papers I cited at the beginning of the chapter.

\section{Using Clustering Algorithms}\label{sec:cd-partition-practical}
I barely started scratching the complex landscape of different approaches to community discovery. We're going to have time in the next chapters to explore even more variations. However, a question might have already dawned on you. \textit{If there are such different approaches to detecting communities, how do I find the one that works for me?} And, \textit{how do I maximize my chances of finding high quality communities?} As you might expect, the answers to these questions are difficult and often tend to be subjective.

Let's start from the second one: designing a strategy to ensure you find close-to-optimal communities. In machine learning, we discovered a surprising lesson. If you want to improve accuracy, designing the most sophisticated method in the world usually helps only up to a certain point. Having many simple methods and averaging the results could potentially yield better results.

\begin{figure*}
\centering
\begin{subfigure}{.22\columnwidth}
\includegraphics[width=\textwidth]{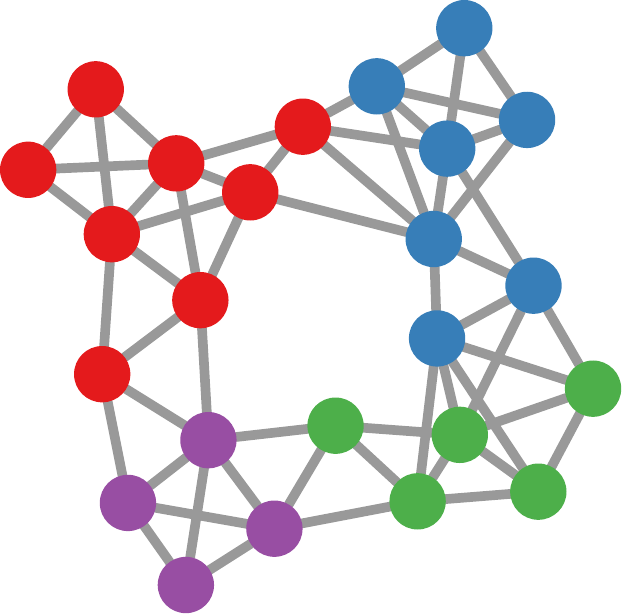}
\caption{Method \#1}
\end{subfigure}\quad
\begin{subfigure}{.22\columnwidth}
\includegraphics[width=\textwidth]{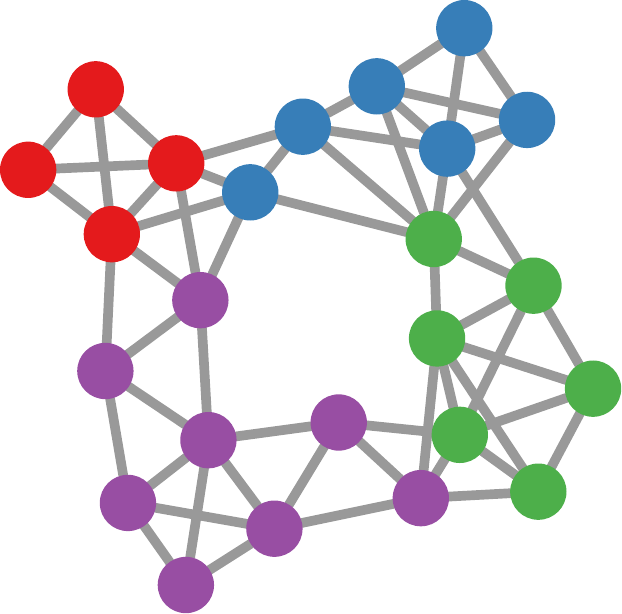}
\caption{Method \#2}
\end{subfigure}\quad
\begin{subfigure}{.22\columnwidth}
\includegraphics[width=\textwidth]{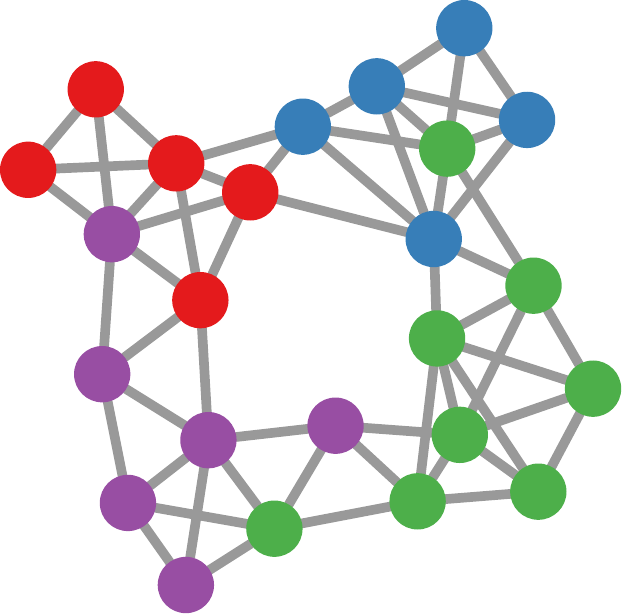}
\caption{Method \#3}
\end{subfigure}\quad
\begin{subfigure}{.22\columnwidth}
\includegraphics[width=\textwidth]{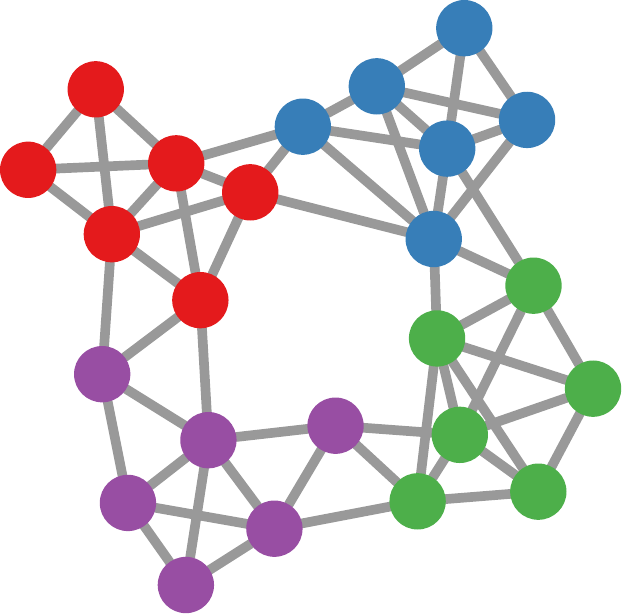}
\caption{Consensus}
\end{subfigure}
\caption{An example of consensus clustering. Node color represents the community. (a-c) The results of three independent methods. (d) Their majority vote.}
\label{fig:community-ensemble}
\end{figure*}

This observation is at the basis of what we know as ``consensus clustering'' (or ensemble clustering)\cite{strehl2002cluster}. This strategy has been applied to detecting communities\cite{lancichinetti2012consensus} in the way you'd expect. Take a network, run many community discovery algorithms on it, average the results. Figure \ref{fig:community-ensemble} shows an example of the procedure. Note how none of the methods (Figure \ref{fig:community-ensemble}(a-c)) found the best communities, which is their consensus: Figure \ref{fig:community-ensemble}(d). Note also how the third method finds rather absurd and long stretched sub-optimal communities. However, its evident blunders are easily overruled by the consensus between the other two methods, and its tiebreaker improves the overall partition.

This is a fine strategy, but you should not apply it blindly. You have to make sure the ensemble of community detection algorithms you're considering is internally consistent. In particular, the methods should have a coherent and compatible definition of what a community is. Limiting ourselves to the small perspective on the problem from this chapter -- ignoring all that is coming next -- combining a dynamic community discovery with a static local community discovery will probably not help.

Even mashing together superficially similar algorithms might result in disaster. For instance, the flow-based communities Infomap returns aren't extremely compatible with the density optimization algorithms we'll see in the next chapter, even if the community definitions on which they are based don't look too dissimilar.

So, how do you go about choosing which algorithms to include in your ensemble? In a paper of mine\cite[0.2in]{coscia2019discovering}, I explore the relationship between around $70$ algorithms, comparing how similar their resulting partitions are. This results in an algorithm similarity network, which has distinct communities: groups of algorithms returning potentially interchangeable results that are significantly different from algorithms in a different group.

\begin{figure*}
\centering
\includegraphics[width=\columnwidth]{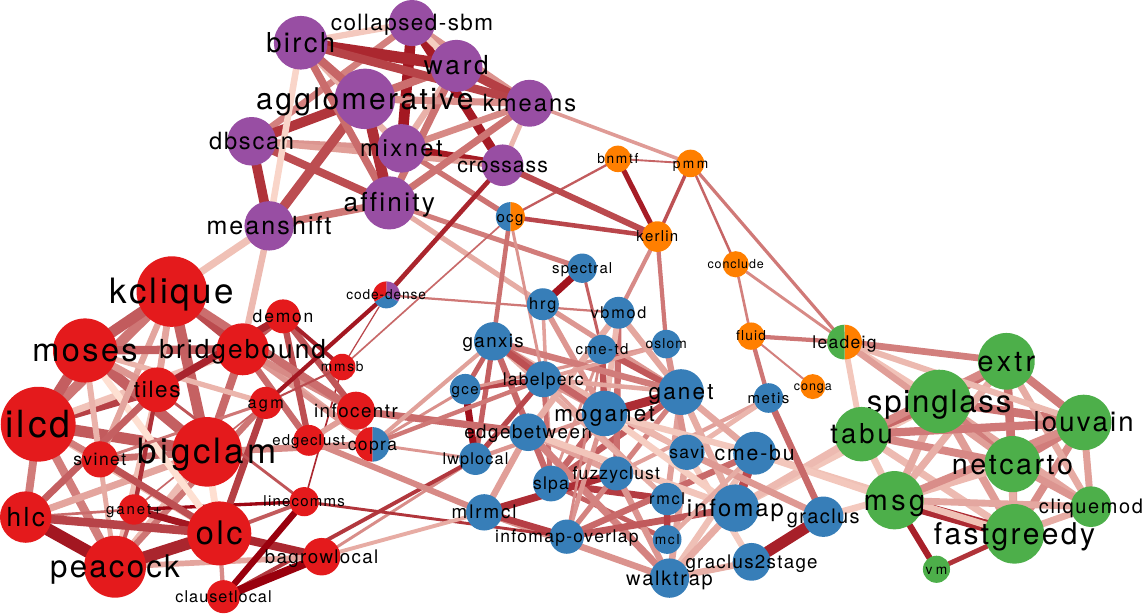}
\caption{The community detection algorithm similarity network. Each node is a community discovery algorithm. They are connected if the two algorithms return similar partitions. I use the node color to represent the algorithm's community. Multicolored nodes belong to multiple communities.}
\label{fig:community-asn}
\end{figure*}

Figure \ref{fig:community-asn} shows the result. Note that, in there, I label each algorithm with a tag. Chapter \ref{cha:utilities} contains a map from the tag to a resource where to recover the specific algorithm.

Now what? Well, we could... aehm... detect... communities... on this -- I love being meta. These communities of community discovery can drive you in choosing your ensemble set. To find them, I use a version of Infomap allowing nodes to be part of multiple communities. This is the so-called overlapping community discovery, the topic of Chapter \ref{cha:ocd}. Most algorithms allowing overlapping communities are in the red community in the figure -- along with the local clustering approach I presented in Section \ref{sec:cd-partition-local}.

The blue community contains Infomap and the label propagation approaches I explained earlier (Sections \ref{sec:cd-partition-rw} and \ref{sec:cd-partition-lp}). This shows the close relationship between the two. The purple community includes methods departing from the classical ``internal density'' definition, to use a ``neighbor similarity'' approach. These will be covered in detail in Section \ref{sec:bcd-clustering}.

A very popular approach is using a quality measure to evaluate how good a partition is, and then finding a smart strategy to optimize it. Most methods applying this strategy are in the green community, which is the one we're exploring in the next chapter.

Before moving to it, let me highlight an important lesson that we learn from this algorithm similarity network. Its communities \textit{are} well-defined. This means that there are different and mutually exclusive notions of what a community is. This is yet another proof that the naive definition of community commonly accepted without criticism must be only one of the many possible. In fact, we can go deeper than this. The notion that there is a golden partition of the network is a utopia. As I mentioned at the beginning of the chapter, community discovery is more useful than that: it decomposes a network and simplifies it. Since there are innumerable ways -- and reasons why -- to simplify a network, then there are innumerable approaches to community discovery, and they are all valid even if they don't chase the mythical golden pot of ``true'' communities at the end of the rainbow.

\section{Summary}

\begin{enumerate}
\item According to the classical definition, communities in complex networks are groups of nodes densely connected to each other and sparsely connected to the rest of the network. They are one of the most common mesoscale organizations of real world networks.
\item One of the oldest approaches in community detection is to assume a planted partition of nodes in the network and then finding the distributions of nodes in communities that has the highest likelihood of explaining the observed connections.
\item A random walker would tend to be trapped in a community, because most of the neighbors of a node are part of its same community. By the same principle, we can detect communities by letting nodes assume the community label that is most common among their neighbors.
\item Networks evolve and so do communities. One can track the evolution of communities by having a two-part community quality function. One part tells us how well we're partitioning the network, the other tells us how compatible the new communities are with the old ones.
\item Sometimes your input network is too big or you have no interest in partitioning all of it. In that case, you can perform local community detection, detecting communities only in the neighborhoods of one or more query nodes.
\item There are hundreds of community detection algorithms. To choose one, you need to know what type of communities it returns. Alternatively, you can perform ensemble clustering, averaging out the results of multiple algorithms.
\item The classical community definition is assortative. Disassortative communities can exist, where nodes don't like to connect to members of the same group. Temporal communities are also not always assortative. This shows that there are more types of communities than the one assumed by the classical definition and they are all valid objectives you can follow to simplify your network.
\end{enumerate}

\section{Exercises}

\begin{enumerate}
\item Find the communities in the network at \url{http://www.networkatlas.eu/exercises/35/1/data.txt} using the label propagation strategy. Which nodes are in the same community as node 1?
\item Find the local communities in the same network using the same algorithm, by only looking at the 2-step neighborhood of nodes 1, 21, and 181.
\item Suppose that the network at \url{http://www.networkatlas.eu/exercises/35/3/data.txt} is a second observation of the previous network. Perform the label propagation community detection on it and use the Jaccard coefficient to determine how different the communities containing nodes 1, 21, and 181 are.
\end{enumerate}

\chapter{Community Evaluation}\label{cha:cd-eval}
How do you know if you found a good partition of nodes into communities? Or, if you have two competing partitions, how do you decide which is best? In this chapter, I present to you a battery of functions you can use to solve this problem. Why a ``battery'' of functions? Doesn't ``best'' imply that there is some sort of ideal partition? Not really. What's ``best'' depends on what you want to use your communities for. Different functions privilege different applications. So we need a quality function per application and you need to carefully choose your evaluation strategy to match the problem definition you're trying to solve with your communities.

Think about ``evaluating your communities'' more as a data exploration task than a quest to find the ultimate truth. Since there is no one True partition -- and not even one True definition of community as I suggested in the previous chapter --, there also cannot be one True quality function. You have, instead, multiple ways to see different kinds of communities, some of which might be more or less useful given the network you have and the task you want to perform.

In the first two sections, I start by focusing on functions that only take into account the topological information of your network. In this case, the only thing that matters are the nodes and edges -- at most we can consider the direction and/or the weight of an edge.

In the latter two sections I move to a different perspective. First, we consider the network as essentially dynamic and we use communities as clues as to which links will appear next, under the assumption the communities tend to densify: it is much more likely that a new link will appear between nodes in the same community. Finally, we look at metadata that could be attached to nodes, which might be providing some sort of ``ground truth'' for the actual communities in which nodes are grouped into in the real world.

\section{Modularity}\label{sec:cd-eval-mod}

\subsection{As a Quality Measure}
When it comes to functions evaluating the goodness of a community partition using exclusively topological information, there is one undisputed queen: modularity\cite{newman2006modularity}. You shouldn't be fooled by its popularity: modularity has severe known issues that limits its usefulness. We'll get to those in the second half of this section.

Modularity is a measure following closely the classical definition of community discovery. It is all about the internal density of your communities. However, you cannot simply maximize internal density, as the partition with the highest possible density is a degenerate one, where you simply have one community per edge -- two connected nodes have, by definition, a density of one.

\begin{figure}
\centering
\begin{subfigure}{.45\columnwidth}
\includegraphics[width=\textwidth]{figures/outline8.pdf}
\caption{}
\end{subfigure}\quad
\begin{subfigure}{.45\columnwidth}
\includegraphics[width=\textwidth]{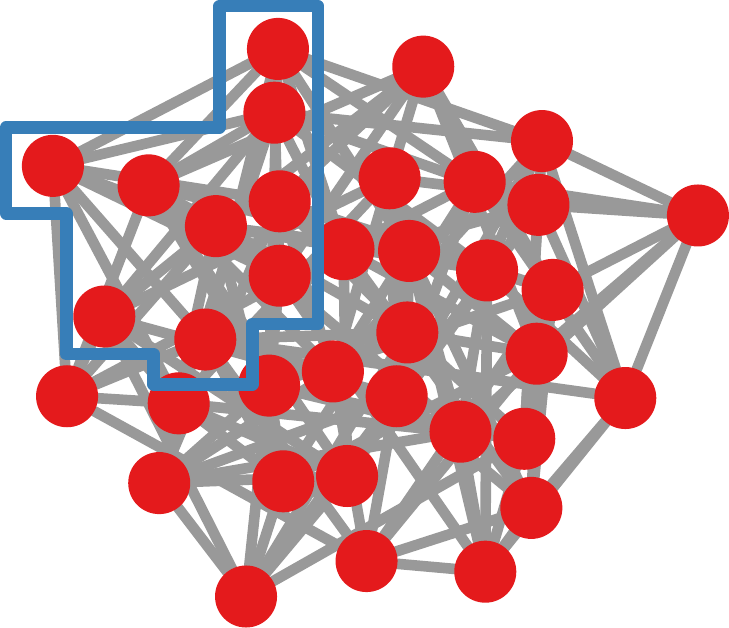}
\caption{}
\end{subfigure}
\caption{(a) A network with a community structure. The node colors represent the community partition. (b) A configuration model of (a), preserving the number of nodes, edges, and degree distribution. The blue outline identifies the nodes that were grouped in the blue community in (a).}
\label{fig:community-mod-null}
\end{figure}

Modularity solves this issue by comparing the observed network with a random expectation. For instance, consider the network in Figure \ref{fig:community-mod-null}(a). If we were to create a randomized version of it, it'd look like the graph in Figure \ref{fig:community-mod-null}(b). In Figure \ref{fig:community-mod-null}(b), each node has the very same degree that it has on the left. However, the edges are shuffled around. It is clear that this random network has no community structure. The difference between the two networks is that the communities of nine nodes have many more links inside them that any grouping of nine nodes in Figure \ref{fig:community-mod-null}(b).

At an abstract level, modularity is the comparison between the observed number of edges inside a community and the expected number of edges. The expectation is based on a null model of a random graph with the same degree distribution as the observed graph (i.e. a configuration model, Section \ref{sec:csmodels-conf}). In Figure \ref{fig:community-mod-null}, we see that this number is positive: there are more edges in the community structure network than in its randomized version. The blue community in Figure \ref{fig:community-mod-null}(a) contains $36$ edges -- all communities in the figure do. Picking those nodes from Figure \ref{fig:community-mod-null}(b) results in finding only $17$ edges among them.

\begin{figure*}
\centering
\includegraphics[width=\columnwidth]{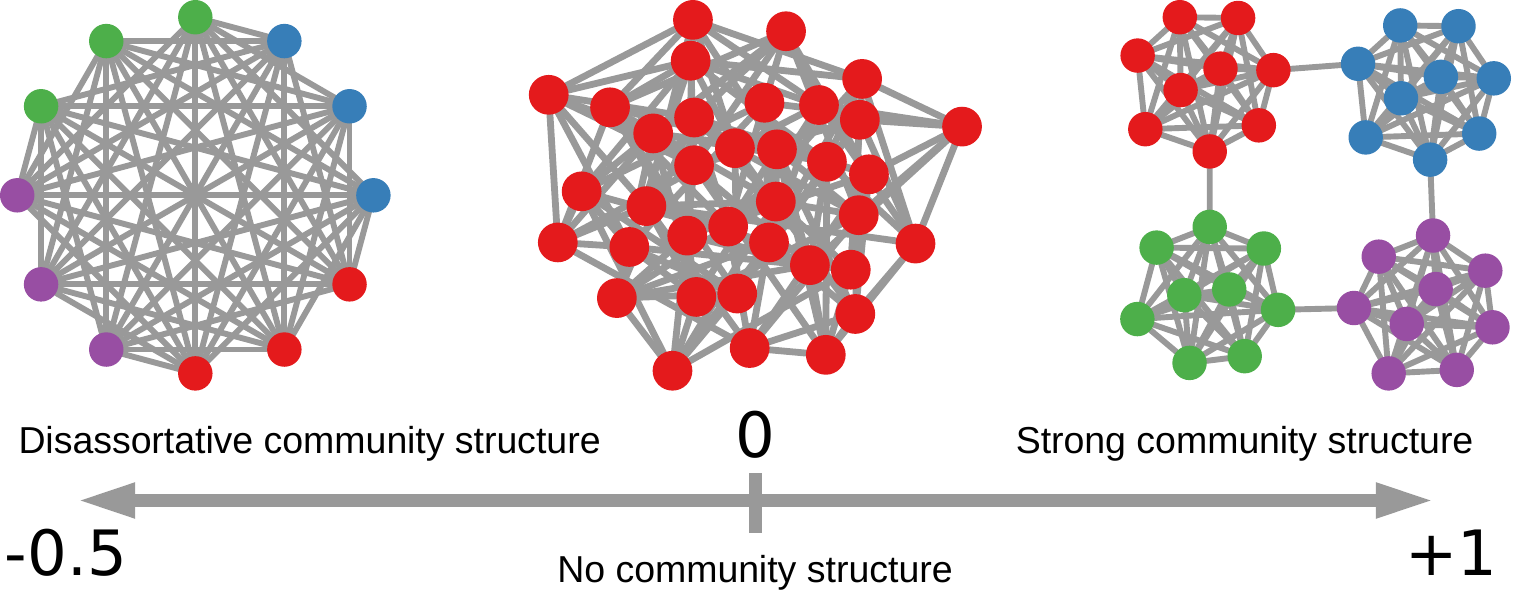}
\caption{The domain of modularity, with example partitions returning a given value, from $-0.5$ (disassortative communities) to $+1$ (assortative communities, the most common case), passing via $0$ (random graph with no communities).}
\label{fig:community-mod-domain}
\end{figure*}

The domain of the modularity function is thus defined between $+1$ and $-0.5$, as Figure \ref{fig:community-mod-domain} shows. A positive modularity happens when our partition finds nodes whose number of edges exceeds null expectation. When expectation exactly matches the number of edges in our community partition, modularity is zero. You can achieve negative modularity by trying to group nodes together that connect to each other less than chance. This can be a reasonable scenario: for instance, if you have disassortative communities (see Section \ref{sec:homophily-assortativity}). Note that, in the leftmost graph in Figure \ref{fig:community-mod-domain}, nodes of the same color do not connect with each other.

The question is: how do we build this expectation, mathematically? This boils down to estimating the connection probability between any two nodes $u$ and $v$. If this were a true random graph ($G_{n,p}$) it'd be easy: the connection probability is $p = 2|E|/|V|$. But we have the constraint of keeping the degree distribution. Each node $v$ has a number of connection opportunities equal to its degree. The number of possible wirings we can make in the network is twice the number of edges. In a configuration model, the probability of connecting $u$ and $v$ is $(k_u k_v) / 2|E|$.

Now we have all we need to build the modularity formulation:

$$ M = \dfrac{1}{2|E|} \sum \limits_{u,v \in V} \left[ A_{uv} - \dfrac{k_v k_u}{2|E|} \right] \delta(c_v, c_u),$$

where $A$ is our adjacency matrix, and $\delta$ is the Kronecker delta: a function return one if $u$ and $v$ are in the same community ($c_u = c_v$), zero otherwise.

Modularity's formula is scary looking, but it ought not to be. In fact, it's crystal clear. Let me rewrite it to give you further guidance:

$$ M =
   \mathbin{\textcolor{cb1}{\dfrac{1}{2|E|}}}
   \sum \limits_{\mathbin{\textcolor{cb2}{u,v \in V}}}
   \left[ \mathbin{\textcolor{cb3}{A_{uv}}} -
   \dfrac{\mathbin{\textcolor{cb4}{k_v k_u}}}{\mathbin{\textcolor{cb5}{2|E|}}} \right]
   \mathbin{\textcolor{cb7}{\delta(c_v, c_u)}}
,$$

which translates into: \textcolor{cb2}{for every pair of nodes} \textcolor{cb7}{in the same community} subtract from \textcolor{cb3}{their observed relation} the expected number of relations given \textcolor{cb4}{the degree of the two nodes} and \textcolor{cb5}{the total number of edges in the network}, then \textcolor{cb1}{normalize} so that the maximum is $1$.

Modularity and Stochastic Blockmodels are related. Optimizing the community partition following modularity is proven to be equivalent to a special restricted version of SBM\cite{newman2016equivalence}. Specifically, you need to use the degree-correlated SBM -- since it fixes the degree distribution just like the configuration model does (which is the null model on which modularity is defined). Then, you must fix $p_{in}$ and $p_{out}$ -- the probabilities of connecting to nodes inside and outside their community -- to be the same for all nodes.

In general, you can use both to evaluate the quality of your partition, but there are subtle differences. SBM is by nature generative: it gives you connection probabilities between your nodes. Modularity doesn't. On the other hand, modularity has this inherent test against a null graph which you don't really have in SBMs. In fact, you can easily extend modularity in such way that you can talk about a statistically significant community partition, one that is sufficiently different from chance\cite{karrer2008robustness}.

\begin{figure}
\centering
\begin{subfigure}{.3\columnwidth}
\includegraphics[width=\textwidth]{figures/outline8.pdf}
\caption{$M = 0.723$}
\end{subfigure}\quad
\begin{subfigure}{.3\columnwidth}
\includegraphics[width=\textwidth]{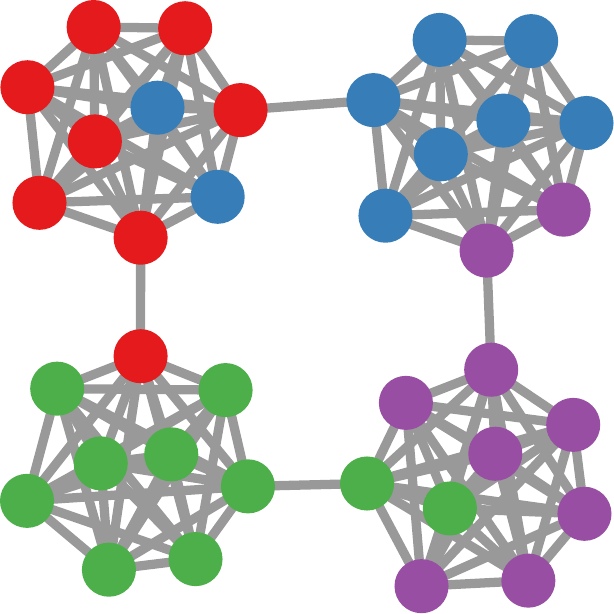}
\caption{$M = 0.411$}
\end{subfigure}
\quad
\begin{subfigure}{.3\columnwidth}
\includegraphics[width=\textwidth]{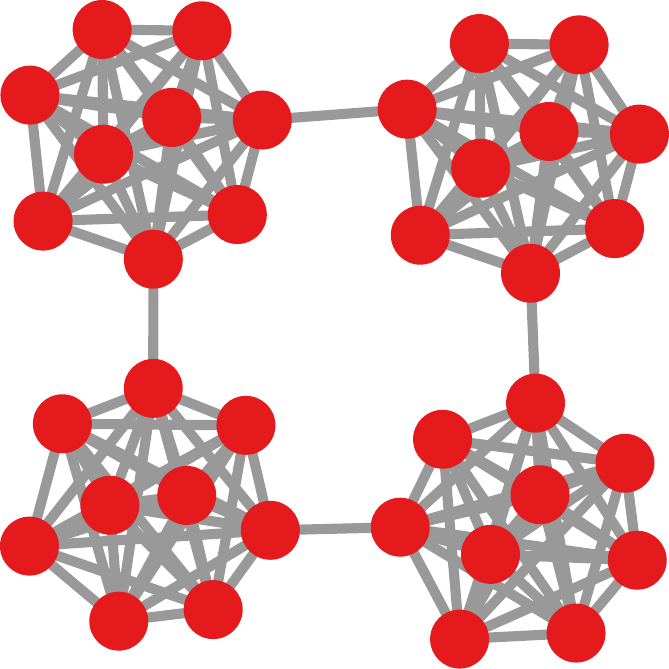}
\caption{$M = 0$}
\end{subfigure}
\caption{A network with a community structure. The node colors represent the community partition. (a) Optimal partition. (b) Sub optimal partition. (c) Partition grouping all nodes in the same community.}
\label{fig:community-mod-ex}
\end{figure}

Modularity also gives us an intuition about whether a partition is better than another, without the need of calculating the likelihood, which is a more generic tool that was not developed with networks in mind -- in fact, I introduced it in Section \ref{sec:ml-loss}. We can compare partitions and see that a higher modularity implies a better partition, as Figure \ref{fig:community-mod-ex} shows. Moving nodes outside the optimal partition lowers modularity (compare the scores in the captions of Figures \ref{fig:community-mod-ex}(a) and \ref{fig:community-mod-ex}(b)). If we do not do community discovery and return a single partition (Figure \ref{fig:community-mod-ex}(c)), modularity will be equal to zero.

\subsection{As a Maximization Target}
\begin{figure}
\centering
\begin{subfigure}{.4\columnwidth}
\includegraphics[width=\textwidth]{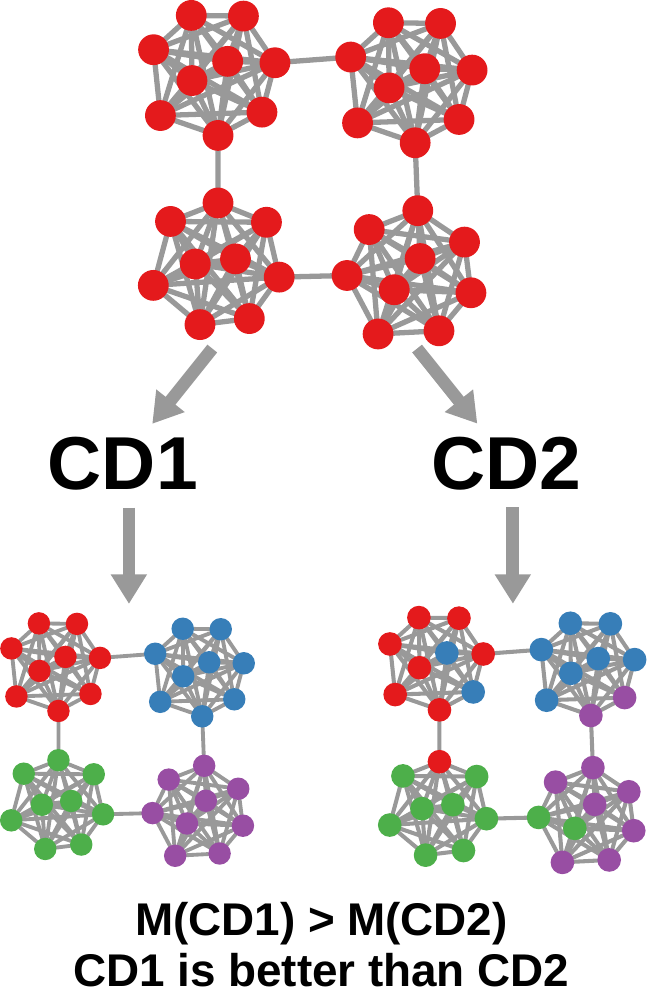}
\caption{}
\end{subfigure}\quad\quad
\begin{subfigure}{.4\columnwidth}
\includegraphics[width=\textwidth]{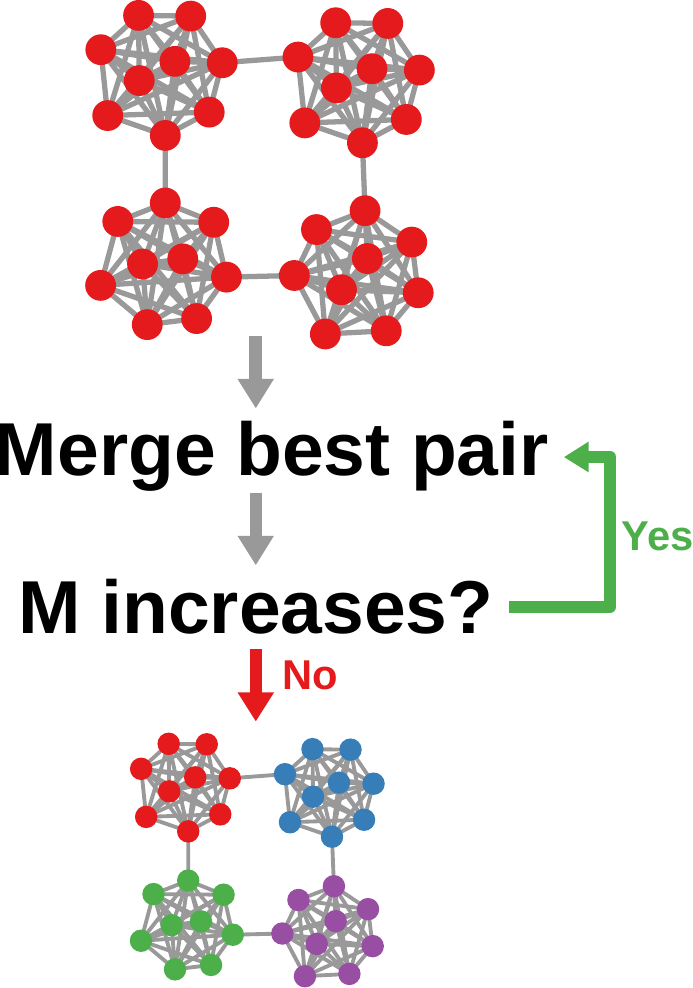}
\caption{}
\end{subfigure}
\caption{(a) The workflow of using modularity as a quality criterion for your partitions. (b) The workflow of using modularity as an optimization target to find the best community partition.}
\label{fig:community-mod-max}
\end{figure}

As I mentioned earlier, modularity can be used in two ways. So far, we've seen the use case of evaluating your partitions. You start from a graph, you try two algorithms (or the same algorithm twice) and you get two partitions. The one with the highest modularity is the preferred one -- see Figure \ref{fig:community-mod-max}(a).

The alternative is to directly optimize it: to modify your partition in a smart way so that you'll get the highest possible modularity score  -- see Figure \ref{fig:community-mod-max}(b). For instance, your algorithm could start with all nodes in a different community. You identify the node pair which would contribute the most to modularity and you merge it in the same community. You repeat the process until you cannot find any community pair whose merging would improve modularity\cite{newman2004fast}\cite{clauset2004finding}. Most approaches following this strategy return hierarchical communities, recursively including low level ones in high level ones, and I cover them in detail in Chapter \ref{cha:hcd}.

But there are other ways to optimize modularity. One strategy is to progressively condense your network such that you preserve its modularity\cite{arenas2007size}. Or using modularity to optimize the encoding of information flow in the network, bringing it close to the Infomap philosophy\cite{nematzadeh2014optimal}. Another approach is using genetic algorithms\cite{pizzuti2008ga}\cite{pizzuti2012multiobjective} or extremal optimization\cite{duch2005community}: an optimization technique similar to genetic algorithms, which optimizes a single solution rather than having a pool of potential ones.

Other approaches include, but are not limited to:

\begin{itemize}
\item Progressively merging cliques, under the assumption that a clique is a structure that has the highest possible modularity\cite{yan2009detecting};
\item Performing the merging of communities I describe earlier allowing multiple communities to merge at the same time and then refining the results by allowing single nodes to move at the end\cite{schuetz2008efficient};
\item Using simualted annealing\cite{guimera2005functional}, integrated by using spinglass dynamics\cite{reichardt2006statistical}. This technique can also take into account whether your network is signed\cite{traag2009community} -- i.e. it has positive and negative connections (see Section \ref{sec:extended-multilayer}), a special and simpler case of multilayer community discovery, which I'll cover in details in Chapter \ref{cha:mcd};
\item Using Tabu search, another optimization technique related to simulated annealing and working mostly using local information\cite{arenas2008analysis};
\item Even including geospatial terms in the definition, when your nodes live into an actual geometric space\cite{expert2011uncovering}.
\end{itemize}

As you can gather from the number of references, we network folks really like to optimize our modularities.

\subsection{Expanding Modularity}
Unfortunately, modularity is not the end-all be-all of community detection as it initially appeared to be. There are several issues with it. We start by looking at the less problematic -- but still annoying -- ones. If you go back to the formula, you'll recognize that it is \textit{a little bit too simple}.

The standard definition of modularity works exclusively with undirected, unweighted, disjoint partitions. We'll take care about extending modularity to cover the overlapping case in Chapter \ref{cha:ocd}. For now, let's see what we can do when our graphs have directed edges and/or weighted ones.

The most straightforward way to extend modularity when your graphs have directed edges is simply modifying your expected number of edges between two nodes\cite{leicht2008community}. If in the undirected case we simply used the degree for both nodes $u$ and $v$, now we have to use their in- and out-degree alternatively. So the expectation turns from $(k_u k_v) / 2|E|$ into $(k_u^{in} k_{v}^{out}) / |E|$ for the $v \rightarrow u$ edges. Modularity thus becomes:

$$ M = \dfrac{1}{2|E|} \sum \limits_{u,v \in V} \left[ A_{uv} - \dfrac{k^{out}_v k^{in}_u}{|E|} \right] \delta(c_v, c_u).$$

Since we're here, why stopping at directed unweighted graphs? Let's add weights! Say that the $(u,v)$ edge has weight $w_{uv}$, and that $w_{u}^{out}$ is the sum of all edge weights originating from $u$ (with $w_{u}^{in}$ defined similarly for the opposite direction). Then:

$$ M = \dfrac{1}{2|E|} \sum \limits_{u,v \in V} \left[ w_{uv} - \dfrac{w^{out}_v w^{in}_u}{\sum \limits_{u,v \in V} w_{uv}} \right] \delta(c_v, c_u).$$

Note that this simple move makes optimizing modularity a tad more complicated -- you should check out the original paper to see why.

This is all well and good, since there aren't competing definitions of directed/weighted modularity. What's that? I'm being told there are. Oh boy. For instance, an alternative is to look at a directed network as if it were a bipartite network\cite{guimera2007module}, where each node $v$ can be seen as two nodes $v^{in}$ and $v^{out}$.

It has also being pointed out that, while this generalized modularity gives out different results than the standard modularity, it actually doesn't really distinguish the $u \rightarrow v$ and $v \rightarrow u$ cases very well\cite{kim2010finding}. In this case, the proposed solution is to use the PageRank of nodes $u$ and $v$ as an expectation of their connection strength. And, once you do that, you open the floodgates of hell, as any directed measure can be now used to determine your expectation, generating hundreds of different modularity versions, each with its own community definition.

\subsection{Known Issues}
But the issues raised so far are only child's play. Let's take a look at the \textit{real} problematic stuff when it comes to modularity. There are three main grievances with modularity. The first is that random fluctuations in the graph structure and/or in your partition can make your modularity increase\cite{guimera2004modularity}. However, I already mentioned that modularity can be extended to take care of statistical significance.

A harder beast to tame is the infamous resolution limit of modularity. To put it bluntly, modularity has a preferred community size, relative to the size of the graph. This means that a partition that a human would consider the natural partition of the network could be rejected by modularity maximization as it is not at the preferred resolution\cite{fortunato2007resolution}. Empirically, it has been shown that modularity maximization approaches tend to find $\sqrt{|E|}$ communities in the network -- a number of communities that seems to be common for many other partitioning algorithms\cite{ghasemian2019evaluating}.

\begin{figure}
\centering
\includegraphics[width=.8\columnwidth]{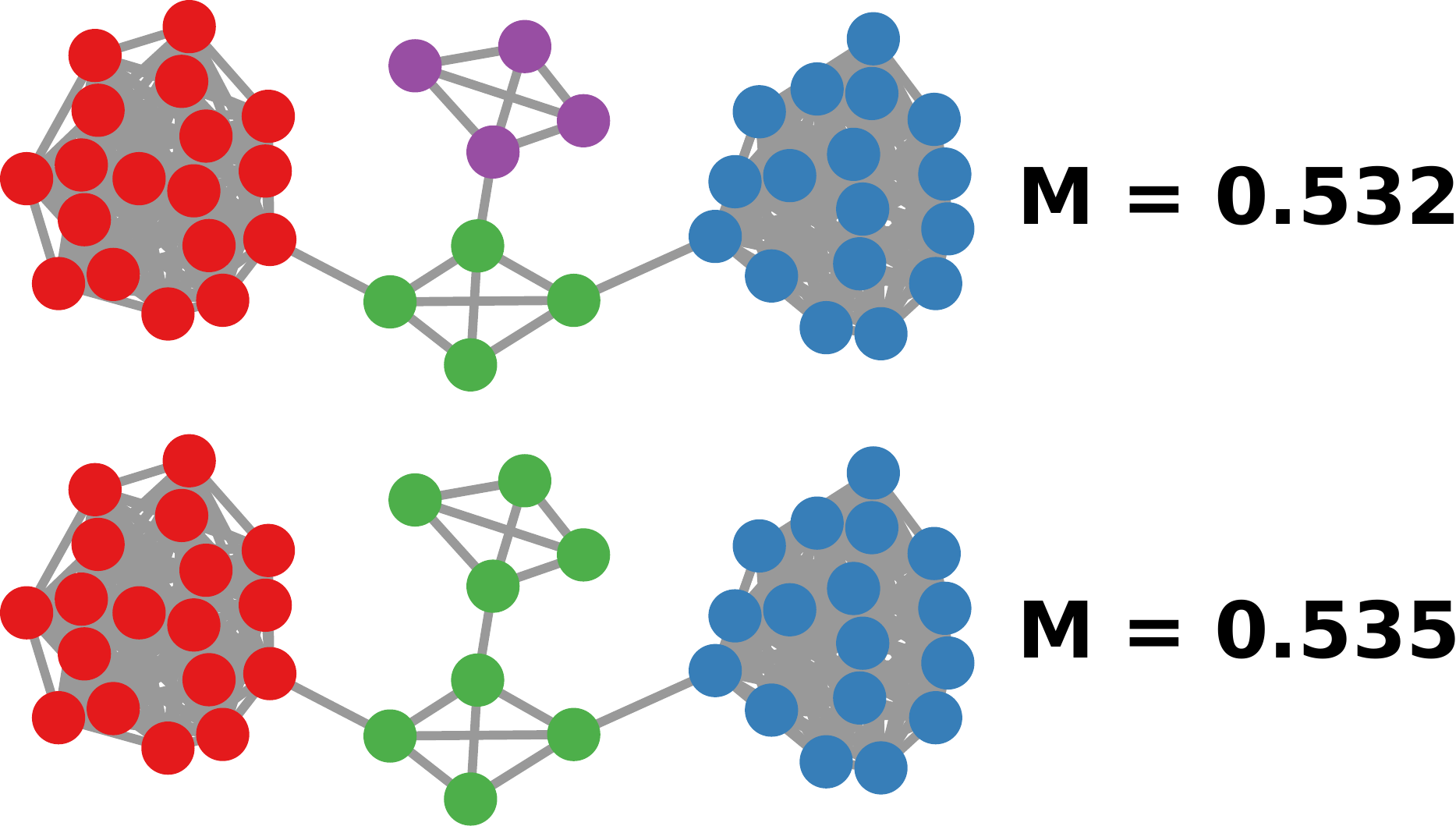}
\caption{The resolution limit of modularity. For the same network, I propose two different partitions in communities, using the node color.}
\label{fig:community-mod-resolution}
\end{figure}

For instance, consider the network and partition in Figure \ref{fig:community-mod-resolution} (top). Modularity is positive, thus this is a good partition. However, the partition to the bottom of Figure \ref{fig:community-mod-resolution} is better, even if a human would probably disagree. This is because, when we have small communities relative to the number of the edges of the network, for modularity it is better to merge them, even if they are clearly and intuitively distinct. This is the resolution limit of modularity: it accepts partitions only of a comparable size with the size of the network.

Mathematically speaking, it's not too hard to extend modularity so that it can work at multiple resolutions. The common strategy is to add a resolution parameter\cite{arenas2008analysis}\cite{huang2011towards}. This can be interpreted as adding a bunch of self-loops to each node, such that the number of edges in a small community can still be considerable, due to the presence of such self loops. However, now you're not only optimizing modularity, you also have to search for the optimal value of this parameter. Uff.

\begin{figure}
\centering
\includegraphics[width=.8\columnwidth]{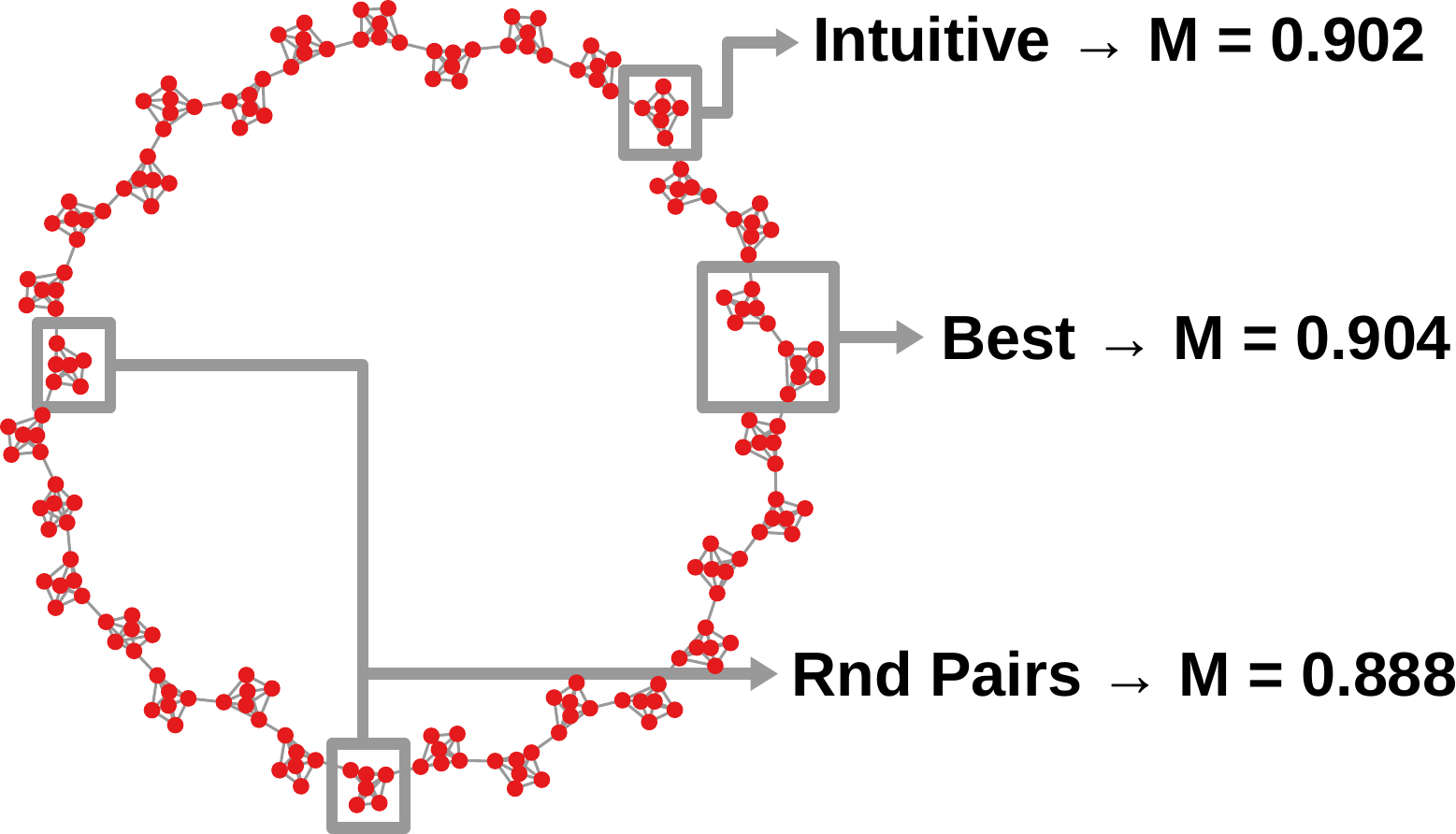}
\caption{A ring of cliques, showing another side of the resolution limit problem of modularity.}
\label{fig:community-mod-resolution2}
\end{figure}

This can get really tricky. Consider Figure \ref{fig:community-mod-resolution2} as another example. Here we have a ring of cliques, a classical caveman model. How would you partition this network? It seems natural to just have one community per clique. \textit{Silly human}, modularity says, \textit{the best partition is instead merging two neighboring cliques}. You look at modularity, puzzled by this sentence. But she is not done: \textit{however, we could also put random clique pairs in the same community, even if they're not adjacent. That's a good partition as well}.

\textit{Go home, modularity}, you now say, \textit{you're drunk}.

Joking aside, this is related to another well known problem of modularity: the field of view limit\cite{schaub2012markov}. Modularity cannot ``see'' long range communities. If your communities are very large and span across multiple degrees of separation, modularity will overpartition them. This means that, if a proper community has nodes whose shortest connecting path is more than a few edges long, you might end up splitting them in different groups. This field of view limit is shared with other community discovery approaches using Markov processes (random walks). Even vanilla Infomap tends to overpartition such communities. In fact, in the paper I cited in this paragraph, the authors show how you could interpret the modularity formula as a one-step Markov process.

All of this is to say that optimizing modularity is NP-hard and the heuristics have a hard time finding the best partitions because the space of possible solutions is crowded by high values that look very different. This is the third grievance with modularity: the degeneracy of good solutions\cite{good2010performance}\cite{lancichinetti2011limits}. It's easy to get stuck in local maxima even when the partition you're returning makes no sense. A classical solution is to summon consensus clustering\cite{zhang2014scalable} -- we saw it in Section \ref{sec:cd-partition-practical}. Hopefully, strategies based on perturbation will converge to different local maxima, and the mistakes will cancel each other out, leading you to a global maximum.

\section{Other Topological Measures}
Given these issues, it's no wonder that researchers have looked elsewhere for alternative quality measures. The ones I'm mentioning are by no means perfect, and they have been scrutinized less than modularity, so the absence of known issues should be taken with a grain of salt.

The likelihood measure introduced for SBMs is an obvious candidate as modularity alternative, and I wrote about it in details in Section \ref{sec:cd-partition-sbm}. The obvious downside here is that it needs your community partition to be a ``generative'' one: it has to give you a model of connection probabilities among nodes. Without that, you cannot estimate how likely your model is to generate the observed network.

There is also a quality measure lurking behind Infomap (Section \ref{sec:cd-partition-rw}). The ``code length'' Infomap is trying to minimize is the number of bits you need to encode the random walks using your partition. There is, in principle, no issue in generating a community partition using something else than Infomap, and then testing it with code length. Thus that is also a valid quality measure. You have to be careful, because the standard code length is not normalized. Two networks with different sizes in number of nodes will have a different expected code lengths. Thus a better partition in a larger network could have a worse (higher) code length than a bad partition in a small network.

A direct evolution of modularity which aims at being a more general version of it is stability\cite{lambiotte2008laplacian}\cite{delvenne2010stability}\cite{delvenne2013stability}. In modularity, you see the graph as static. Nodes $u$ and $v$ contribute to modularity only insofar as they are directly connected or not. In stability, you see your graph as a flow. You take into account the amount of time it would take to reach $u$ from $v$ and vice versa. In practice, modularity is equivalent to stability when you only look at immediate diffusion.

There's a battery of other quality measures, conveniently grouped in a review paper\cite{leskovec2010empirical}. I'm going to cover a few here. In all cases, I use a generic $f(C)$ to refer to the function taking the community $C$ as an input and returning its evaluation of the quality of that community.

\subsection{Conductance}
The idea behind conductance is that communities should not be conductive: whatever falls into -- or originates from -- them should have a hard time getting out. In practice, this translates in comparing the volume of edges pointing outside the cluster\cite{shi2000normalized}\cite{kannan2004clusterings}\cite{leskovec2008statistical}\cite{leskovec2009community}. Here we assume that $C \subseteq V$ is a set of nodes grouped in a community. Mathematically speaking, let's define two sets of edges. The first set of edges, $E_C$ is the number of edges fully inside the community $C$. That means $E_C = \{ (u,v) : u \in C, v \in C \}$. The second set of edges is the boundary of $C$: the edges attached to one node in $C$ and one node outside $C$: $E_{B,C} = \{ (u,v) : u \in C, v \not\in C \}$. We can now define conductance as:

$$ f(C) = \dfrac{|E_{B,C}|}{2|E_C| + |E_{B,C}|}. $$

\begin{figure}
\centering
\begin{subfigure}{.4\columnwidth}
\includegraphics[width=\textwidth]{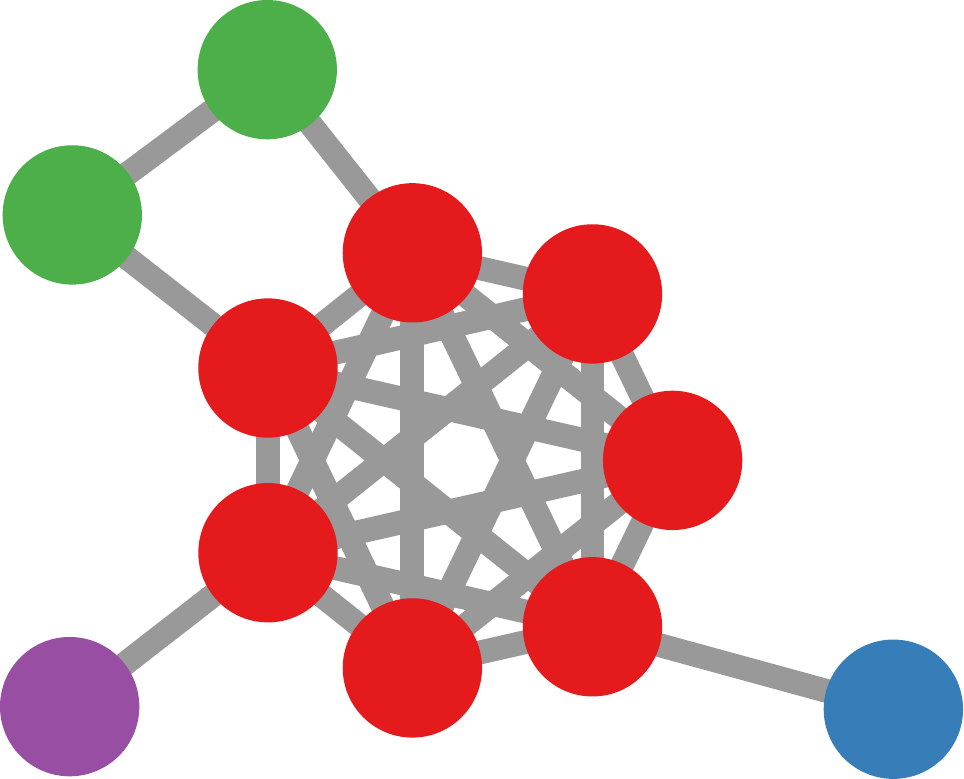}
\caption{}
\end{subfigure}\quad\quad
\begin{subfigure}{.4\columnwidth}
\includegraphics[width=\textwidth]{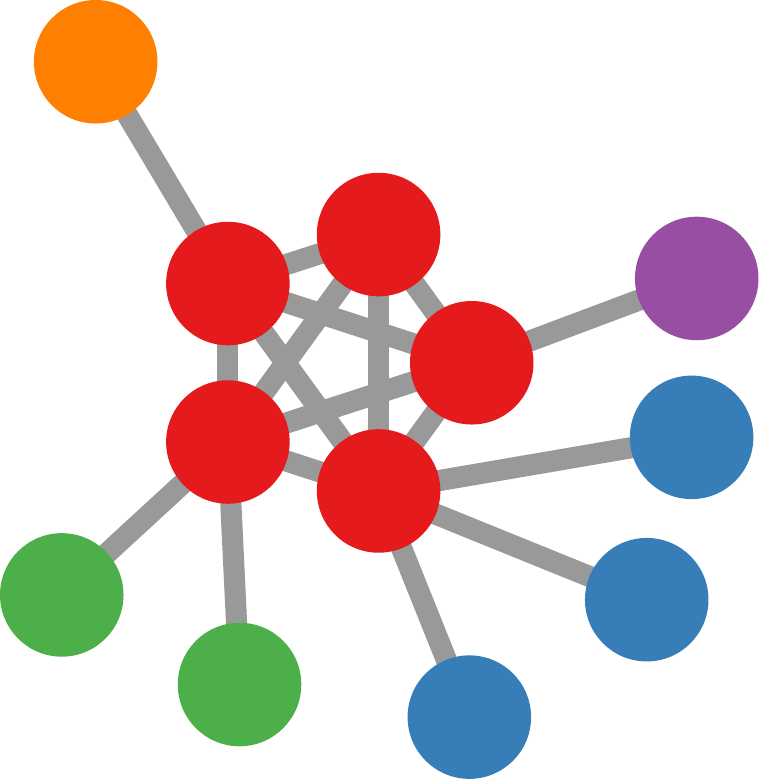}
\caption{}
\end{subfigure}
\caption{Two examples of communities at different conductance levels. I represent the community as the node color. In the text, I focus on the red community.}
\label{fig:community-conductance}
\end{figure}

Note that we want to minimize this function, namely we want to find the partition of $G$ such that the average conductance across all communities is minimal. Figure \ref{fig:community-conductance} shows two examples of communities with different levels of conductance.

Figure \ref{fig:community-conductance}(a) (in red) is a low conductance community. It is a clique of seven nodes, thus we know that $|E_C| = 7 \times 6 / 2 = 21$. It has four edges in its boundary ($|E_{B,C}| = 4$), which gives us $4 / ((2 \times 21) + 4) \sim 0.087$. Figure \ref{fig:community-conductance}(b) (also in red), on the other hand, has a higher conductance. Being a clique of five nodes, $|E_C| = 5 \times 4 / 2 = 10$. From the figure we see that $|E_{B,C}| = 7$, giving us a conductance of $7 / ((2 \times 10) + 7) \sim 0.26$.

Note how conductance doesn't care too much about internal density, as one would expect from the classical definition. It cares, instead, about external sparsity: making sure that the community is as isolated as possible from the rest of the network. Here, both communities are cliques, the denset possible structure. But, since the one in Figure \ref{fig:community-conductance}(b) also has a lot of connections to the rest of the network, the resulting conductance is almost three times higher than the value we get from the community in Figure \ref{fig:community-conductance}(a).

Finally, be aware that you cannot build a community discovery algorithm that simply minimizes conductance -- the same way you'd try to maximize modularity. That is because there's a trivial community with zero conductance: the one including all nodes in your network.

\subsection{Internal density}
The other side of the conductance coin is the internal density measure. This is exactly what you'd think it is: how many edges are inside the community over the total possible number of edges the community could host\cite{radicchi2004defining}. Borrowing $E_C$ from the previous section:

$$ f(C) = \dfrac{|E_C|}{|C|(|C| - 1) / 2}. $$

So you can see that, in this case, both communities in Figure \ref{fig:community-conductance} have an internal density of $1$, since they're cliques. Thus, internal density is unable to distinguish between them, which we would like since community Figure \ref{fig:community-conductance}(b) is clearly ``weaker'', given its high number of external connections.

\begin{figure}
\centering
\includegraphics[width=.8\columnwidth]{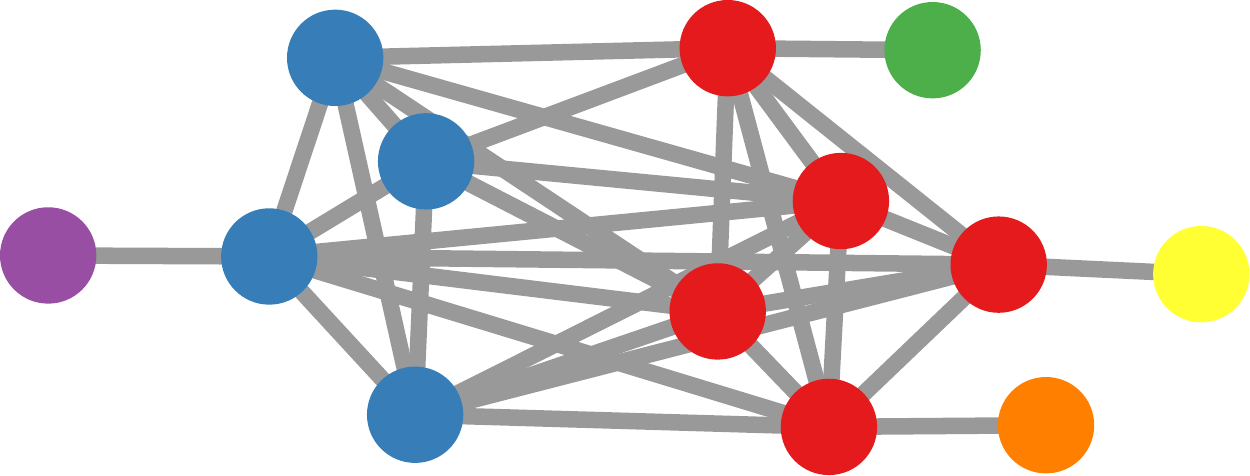}
\caption{The best internal density partition of this core community. I encode the node's community with its color.}
\label{fig:community-intdens}
\end{figure}

You can appreciate the paradoxical result of internal density by looking at Figure \ref{fig:community-intdens}. Here, one might be tempted to merge the red and blue communities, since their nodes are so densely connected to each other and to not much else. Yet, the red nodes are a 5-clique and the blue nodes are a 4-clique, while the red and blue nodes are not a 9-clique. Thus, the best way to maximize internal density is to split these clearly very related nodes.

Neither conductance nor internal density fully capture the classical definition of community discovery I provided in the previous chapter. The definition wants communities to be both internally dense and externally sparse. Each of the two measures only satisfies one of the two requirements. Thus, if you're using either of them to evaluate your communities, you're practically having a different definition of what a community \textit{is}.

Just like conductance, don't try to blindly maximize internal density, as you're only going to find cliques in your networks.

\subsection{Cut}
Originally, we define the cut ratio as the fraction of all possible edges leaving the community. The worst case scenario is when every node in $C$ has a link to a node not in $C$. There are $|C|$ nodes in $C$ and $(|V| - |C|)$ nodes outside $C$, so there can be $|C|(|V| - |C|)$ such links. Thus:

$$ f(C) = \dfrac{|E_{C,B}|}{|C|(|V| - |C|)}. $$

This is usually what gets minimized when solving the mincut problem (Section \ref{sec:rw-mincut}). Again, this is a measure easy to game. That is why we often modify it to be a ``normalized'' mincut:

$$ f(C) = \dfrac{|E_{B,C}|}{2|E_C| + |E_{B,C}|} + \dfrac{|E_{B,C}|}{2(|E| - |E_C|) + |E_{B,C}|}. $$

The most attentive readers already noticed that the first term in this equation is conductance. The second term is also a conductance of sorts. If the first term is the conductance from the community to the rest of the network, the second term is the conductance from the rest of the network to the community. The two are not the same, because the number of edges in $C$ is $|E_C|$, while the number of edges outside $C$ is $|E| - |E_C|$.

\subsection{Out Degree Fraction}
The out degree fraction (ODF), as the name suggests, looks at the share of edges pointing outside the cluster. It follows a strategy similar to conductance. The difference lies in a normalizing factor. While conductance normalizes with the total number of edges in the community, in the out degree fraction you normalize node by node. In the original paper\cite{flake2000efficient}, the authors present a few variants of the same philosophy.

In the Maximum-ODF, you simply pick the node which has the highest number of edges pointing outside the community (relative to its degree) as your yardstick:

$$ f(C) = \max \limits_{u \in C} \dfrac{|{(u,v) : v \not \in C}|}{k_u}. $$

The idea here is that, in a good community partition, there shouldn't be \textit{any} node with a significant number of edges pointing outside the community. We can tolerate if a node has a large \textit{number} of edges pointing out, only if the node is a gigantic hub with a humongous degree $k_u$.

Requiring that there is absolutely no node with a large out degree fraction might be a bit too much. So we also have a relaxed Average-ODF:

$$ f(C) = \dfrac{1}{|C|} \sum \limits_{u \in C} \dfrac{|{(u,v) : v \not \in C}|}{k_u}. $$

In this case, we're ok if, on average, nodes tend not to connect relatively much to neighbors outside the cluster. If there is one node doing so, the presence of many other nodes without external connections will overwhelm it.

Finally, Flake et al. in their paper propose a further variant of the same idea:

$$ f(C) = \dfrac{1}{|C|} |\{u : u \in C, |{(u,v) : v \in C}| < k_u / 2\}|. $$

For each node $u$ in $C$, we count the number of edges pointing outside the cluster. If it's more than half of its edges, we mark the node as ``bad'', because it connects more outside the community than inside. A node shouldn't do that! The measure tells you the share of bad nodes in $C$, which is something you want to minimize.

\begin{figure}
\centering
\includegraphics[width=.8\columnwidth]{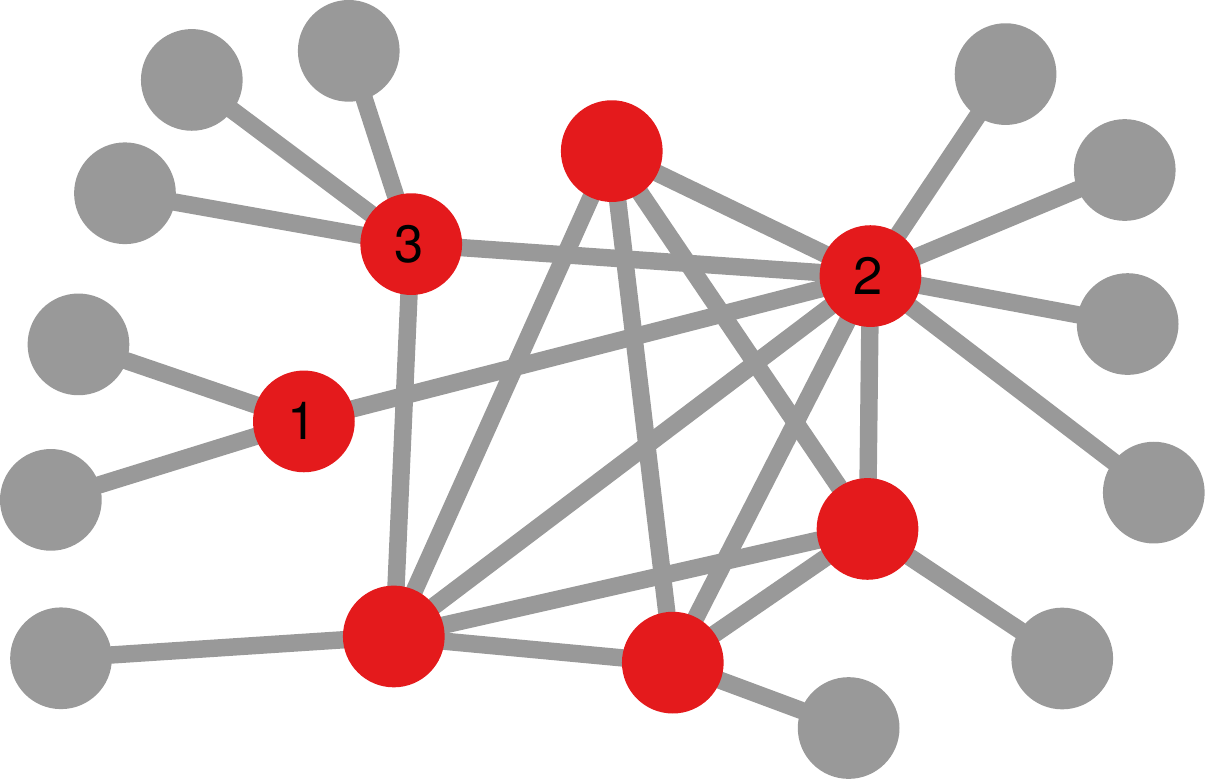}
\caption{An example of community (in red).}
\label{fig:community-odf}
\end{figure}

Figure \ref{fig:community-odf} shows an example community, which we can use to understand the difference between the various ODF variants. In the Maximum-ODF, we're looking for the node with the relative highest out degree. That is node $1$ as its degree is just three, and two of those edges point outside the community. Thus, the Maximum-ODF is $f(C) = 2/3$. Both nodes $2$ and $3$ have a higher out-community degree, but they also have a higher degree and thus they don't count at all for the community quality. You can see how Maximum-ODF is a blunt tool which disregards lots of information.

For Average-ODF we have (clockwise starting from node $1$): $f(C) = (2/3 + 3/5 + 0 + 4/10 + 1/5 + 1/5 + 1/6) / 7 \sim 0.319$. This is awfully close, but not quite, conductance -- which is $12/(2\times(13)+12) \sim 0.316$. Finally, we only have two nodes with more links going outside the community than inside: these are nodes $1$ and $3$. Thus, Flake-ODF is $f(C) = 2/7$.

\section{Link Prediction}
If you have a temporal network, you gain a new way to test the quality of your communities. After all, communities are dense areas in the network, thus they tell you something about where you expect to find new links. In a strong assortative community partition, there are more links between nodes in the same community than between nodes in different communities. Otherwise, your communities would be weak -- or there won't be communities at all\footnote{Or so the classical definition of community says. I already started tearing it apart, and I'll continue doing so, but in this specific test you base your assumption on this classical definition. If you have a different definition of community, don't use this test.}.

Thus you can use your communities to have a prior about where the new links will appear in your dynamic network. This sounds familiar because it is: it is literally the definition of the link prediction problem (Part \ref{par:lp}). In this approach of community evaluation, you use the community partition as your input. You use it to estimate the likelihood of connection between any pair of nodes in the network, and then you can design the experiment (Chapter \ref{cha:lp-experiment}) and use any link prediction quality measure as your criterion to decide which community partition is better. The higher your AUC, the better looking your ROC curve, the better your partition is.

The classical way to create a $score(u,v)$ is having a simple binary classifier: $1$ if $u$ and $v$ are in the same community, $0$ otherwise. This is a bit clunky, so you usually want to add a bit of information: how well embedded are the nodes in the network? This also works in the case of overlapping community discovery (Chapter \ref{cha:ocd}), when nodes can be part of multiple communities. In that case, $score(u,v)$ can be the number of communities they have in common. This seemingly innocuous operation has some rather interesting repercussions, which we will see dubbed as the ``overlap paradox'' in Section \ref{sec:ocd-paradox}.

A method that works naturally well to be evaluated via link prediction is finding communities via SBMs. This is because, in the general SBM, every pair of nodes receives a $p_{in}$ or a $p_{out}$ connection probability given the planted partition. Only in the simplest SBM techniques these values are the same for every pair of nodes. In more sophisticated approaches they are personalized for each node pair, and thus serve as a natural $score(u,v)$ function.

Remember that, when looking at communities, you could have a disassortative community partition, where nodes tend to connect to other nodes \textit{outside} their own community. You can still use this approach, now penalizing in your $score(u,v)$ function nodes part of the same community. You could even do more fun stuff, by creating a community-community similarity score, in which nodes that are in more interconnected communities receive a higher $score(u,v)$ value.

You know from Chapter \ref{cha:lp-experiment} that you can still evaluate your link prediction also in presence of static network data, via k-fold cross validation. This would allow you to evaluate your communities as input for link prediction even lacking a temporal network, making this a more general evaluation tool.

\section{Normalized Mutual Information}\label{sec:cd-eval-nmi}
Your network might not be temporal, but you could have additional information about the nodes, besides to which other nodes they connect (Section \ref{sec:extended-nodeattr}). In this context, node attributes are usually referred to as ``node metadata''. There is a widespread assumption in community discovery: if you have good node metadata, some of them have information about the true communities of the network. Nodes with similar values, following the homophily assumption (Chapter \ref{cha:homophily}), will tend to connect to each other. Therefore there should be some sort of agreement between the community partition of the network and the node metadata\cite{yang2015defining}.

\begin{figure}
\centering
\includegraphics[width=.8\columnwidth]{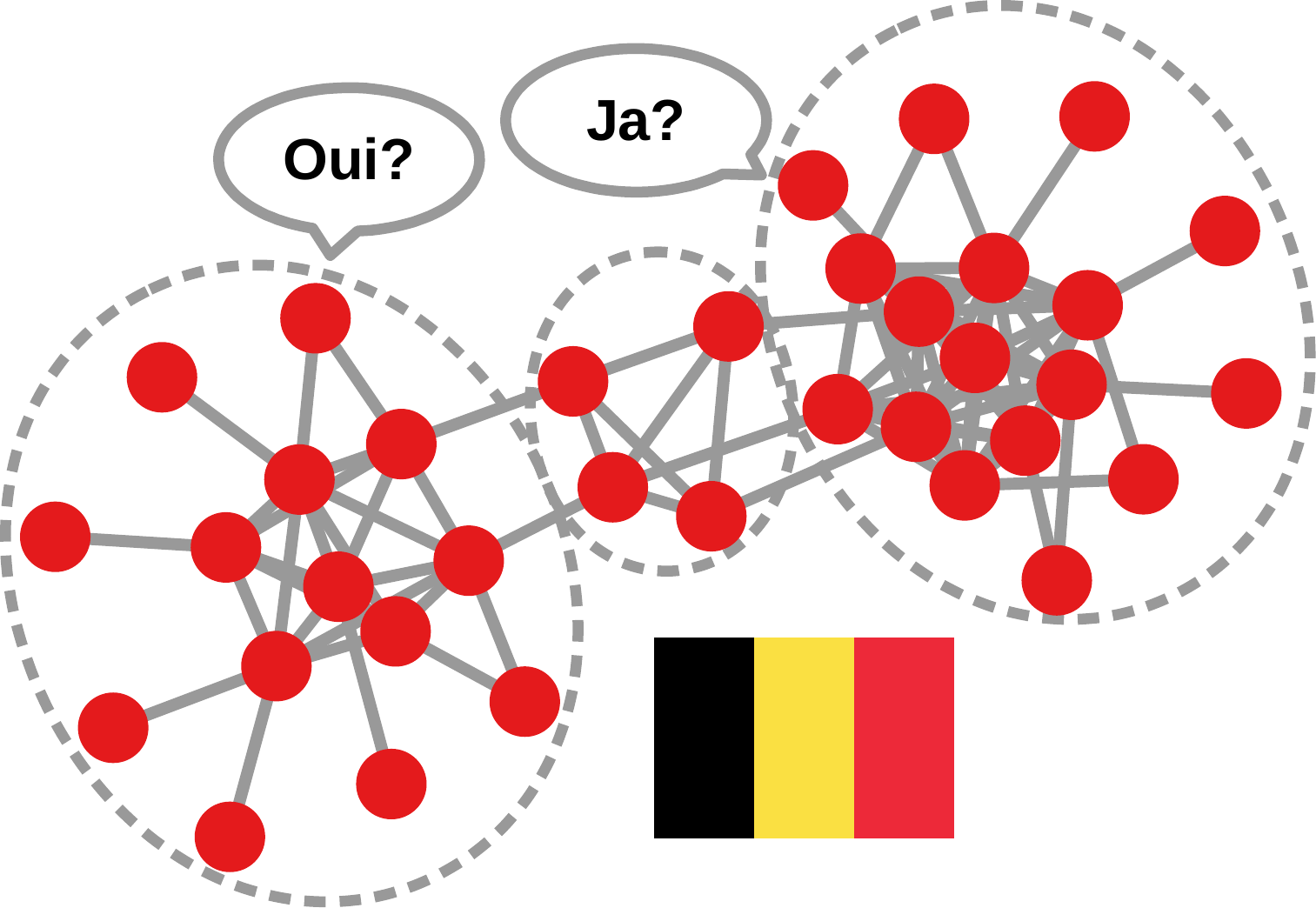}
\caption{A simplification of the Belgian cellphone call graph, highlighting three communities (with the dashed gray outline).}
\label{fig:community-nmi}
\end{figure}

For instance, a classical paper\cite{blondel2008fast} analyzed a network whose nodes were cellphones, connected together if they made a significant number of calls to each other. The network showed three well-separated communities. Figure \ref{fig:community-nmi} shows an extreme simplification of that (very large) graph.

Why was that the case? Why were there gigantic communities? It all becomes clear when I tell you that the country they studied was Belgium, where roughly half of the population is French-speaking and the other half Dutch-speaking, and so they do not call each other. The intersection in the middle is the capital Brussels, where the two populations have to interact. Knowing which language you speak should have almost a one-to-one correspondence with the network community in this case.

How would you calculate such agreement? We can re-use a concept we encountered early on: mutual information (Section \ref{sec:prob-mi}). To recap briefly: you can consider each node in the network as being an entry in two vectors. In the first vector, the node is associated with its metadata: the language the person speaks or whether she lives in Brussels. In the second vector, the node is associated to its community.

\begin{figure}
\centering
\includegraphics[width=.8\columnwidth]{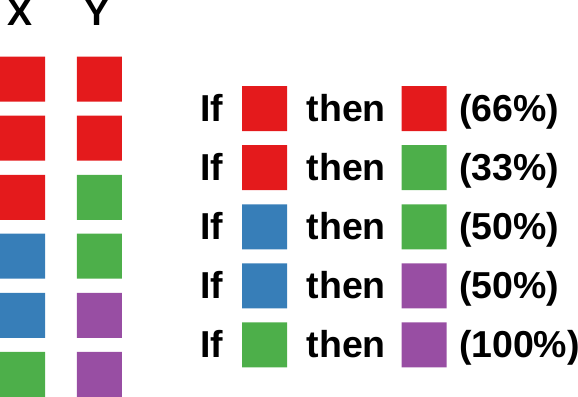}
\caption{An illustration of what mutual information means for two vectors. Vector $y$ has equal occurrences for its values (there is one third probability of any colored square). However, if we know the value of $x$ we can usually infer the corresponding $y$ value with a higher-than-chance confidence.}
\label{fig:mi2-2}
\end{figure}

Mutual information tells you the number of bits of information you gather about one vector by knowing the other vector. Figure \ref{fig:mi2-2} (a reprisal of Figure \ref{fig:mi2}) should help you understanding what mutual information means: having a set of rules that allow you to infer the values in one vector by knowing the other with better-than-chance odds. The Rand Index\cite{rand1971objective}\cite{hubert1985comparing} provides a similar measure, by counting the number of node pairs agreeing between the community partition and the ground truth, without the information theoretic properties fo mutual information.

In Section \ref{sec:lp-other} we used mutual information for link prediction, meaning that we didn't care much about comparing different networks, as everything happened in the same network. However, when evaluating community partitions, you need a standardized yardstick to know whether a network has communities more tightly knit than another -- or if it has communities at all! But mutual information is dependent on the amount of bits you need to encode the vectors in the first place. A longer vector needs more bits to be encoded. Thus, the same value of mutual information can mean different things when you have $100$ nodes or $100,000$.

That is why we often use a \textit{normalized} mutual information (NMI). This is a simple normalization that forces mutual information to take a value between zero and one. This is generally achieved by dividing mutual information by some combination of the entropy of the two vectors (the community partition and the node metadata)\cite{strehl2002cluster}\cite{press2007numerical}.

While normalizing mutual information so that it's comparable across networks is nice, that is not the full story. Remember that our end here is knowing whether there is a relationship between the communities we found and the node attributes. The problem of mutual information is that it is always non-zero. This means that there will be always a little mutual information between two vectors, even if they are both completely random! 

\begin{figure}
\centering
\includegraphics[width=.8\columnwidth]{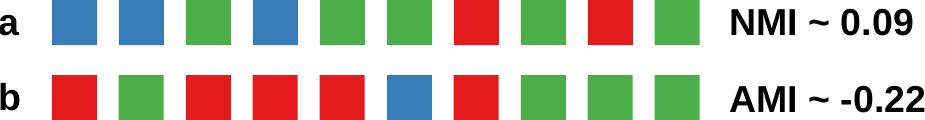}
\caption[][-0.3in]{Two random vectors with a positive NMI even if generated completely independently from one another.}
\label{fig:community-ami}
\end{figure}

Consider the vectors in Figure \ref{fig:community-ami}. I generated them by extracting ten random elements, with three possible values. This is done uniformly at random and with independent draws -- pinky promise! Yet, if you calculate their NMI values, you're going to obtain around $0.09$: a non-zero mutual information from vectors that literally have nothing to do with each other. This is not good.

That is why researchers developed a new normalization for mutual information: Adjusted mutual information (AMI)\cite{meilua2007comparing}\cite{vinh2010information}. In this case, we subtract from mutual information the amount of bits we would expect to obtain about a vector by pure chance. We can do the same thing for the Rand index I mentioned before, generating the Adjusted Rand Index\cite{steinley2004properties} (ARI). In this, AMI and ARI are similar to modularity: you're comparing the observed value with the one you'd get from some sort of null model. AMI and ARI are defined to be equal to zero when you get nothing more than you'd expect by just tossing coins. At this point, any positive value starts getting interesting. These indexes can be negative, AMI is for the two vectors in Figure \ref{fig:community-ami}. A negative AMI means that your clustering isn't good.

It can also mean another thing. You see, so far I grounded this section on a key assumption: that node metadata go hand in hand with the network structure. That... is not always the case. Researchers have seen how much the two notions can diverge\cite{hric2014community}. Node metadata and structural network communities are rarely the same thing. Nodes can share attribute values and not being connected to each other due to a variety of reasons.

In fact, the assumption that communities and node metadata go hand in hand rests on shaky ground. It seems to give some sort of importance and status to the node metadata because it calls it ``meta'' data. But, at the end of the day, in real observed networks metadata is just data. Real metadata is like the planted community in an LFR model (Section \ref{sec:csmodels-comms}), but when you do data gathering there's no such a thing as metadata: what you find is often -- if not always -- incomplete, irrelevant, or wrong to a certain degree.

We can call this a ``data'' problem. Which is yet another issue with the classical definition of communities. Wanting to find a structural way to group nodes with the same attributes -- especially when we don't know their values and we want to infer them -- is a totally valid aim on which to base your community definition. It's just that it doesn't correlate well with the notion of communities made by densely connected nodes. In this scenario, we might even have disconnected communities, made by multiple components without paths leading from one node in the community to another node in the same community. This is absolutely verboten in the classical view of community discovery.

\begin{figure}
\centering
\includegraphics[width=.66\columnwidth]{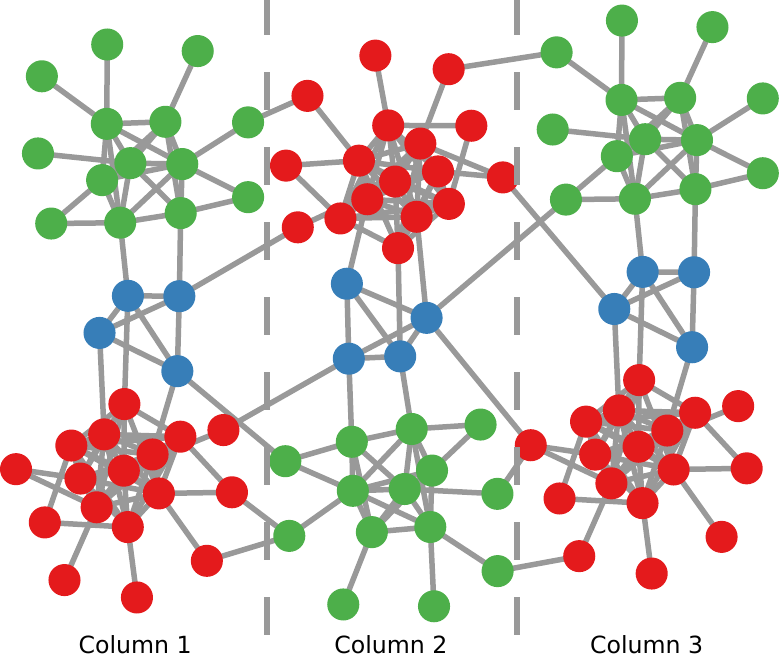}
\caption{A network with a community structure made by conflicting attributes. One node attribute is the color, while the other is its horizontal positioning. Nodes connect most likely with nodes of the same color \textit{and} in the same ``column''.}
\label{fig:community-metadata-issues}
\end{figure}

The assumption that we can infer the ``real'' communities from node attributes rests on data availability: we have some metadata and we assume that the network follows it. But the network could be wired following multiple different attributes and the interactions between them. Figure \ref{fig:community-metadata-issues} shows a simplified example of that. The communities found by node colors are ``good'' to approximate one attribute, but they are terrible for the classical notion of community.

There's another problem with ground truth and community discovery, a more theoretical one. As many other facets of life, in community discovery there is no free lunch\cite{peel2017ground}. This means that community discovery is a large problem with many different network types and valid community definitions. There is no single algorithm who is going to work reliably better than average in all these scenarios.

We are going to see yet more ways in which the classical definition of communities break. But this is a good moment to have a brief pause and collect our thoughts. The real definition of community depends on what the network represents: if you're looking at a social network some definitions of communities make sense, but if you're looking at an infrastructure network they do not. Communities depend on what you're looking for: whether you're trying to approximate a real world property or compress your network. And they also depend on what's your criterion of success: nobody says you cannot use modularity, or NMI, or anything else, as long as it is a motivated choice.

I want you to learn a lesson, my dear reader. You didn't have to go and look for a definition of community: the real definition was inside you all along.

\section{Summary}

\begin{enumerate}
\item The most common function used to evaluate community partitions is modularity. Modularity compares the number of edges inside the communities you detected with the expected number of edges in a configuration model which has, by definition, no communities.
\item You can also use modularity for something more than evaluating the communities you found: it can be an optimization target. Your algorithm will operate on your communities until it cannot find any additional move that would increase modularity.
\item Standard modularity is defined for undirected and unweighted graphs. There are extensions of the measure to deal with directed and/or weighted graphs, however such extensions are not unique: there are multiple competing versions.
\item Modularity has been extensively studied and we know it has several issues. The main one is resolution limit, where communities have to be of similar size. There is also degeneracy: many partitions are close to optimal, even if they are very different from each other.
\item Many other quality functions have been defined. Conductance aims at minimizing edges flowing out of a community. Internal density aims at maximizing the edges inside the community.
\item One could use communities as the basis of link prediction, since nodes in the same community are expected to connect to each other. Thus a better partition is one that would be more accurate in predicting new links.
\item Normalized mutual information is another way to evaluate your partition when you have metadata about your nodes -- if you assume that communities should be used to recover latent node attributes. Be aware, though, that not always nodes with similar attributes connect to each other. 
\end{enumerate}

\section{Exercises}

\begin{enumerate}
\item Detect communities of the network at \url{http://www.networkatlas.eu/exercises/36/1/data.txt} using the asynchronous and the semi-synchronous label propagation algorithms. Which one does return the highest modularity?
\item Find the communities of the network at \url{http://www.networkatlas.eu/exercises/36/2/data.txt} using label propagation and calculate the modularity. Then manually create a new partition by moving nodes $25, 26, 27, 31$ into their own partition. Recalculate modularity for this new partition. Did this move increase modularity?
\item Repeat exercise $1$, but now evaluate the difference in performance of the two community discovery algorithms by means of conductance, cut size, and normalized cut size.
\item Assume that \url{http://www.networkatlas.eu/exercises/36/4/nodes.txt} contains the ``true'' community partition of the nodes from the network at \url{http://www.networkatlas.eu/exercises/36/1/data.txt}. Determine which algorithm between the asynchronous and the semi-synchronous label propagation achieves higher Normalized Mutual Information with such gold standard.
\end{enumerate}

\chapter{Hierarchical Community Discovery}\label{cha:hcd}
\begin{figure}
\centering
\includegraphics[width=.8\columnwidth]{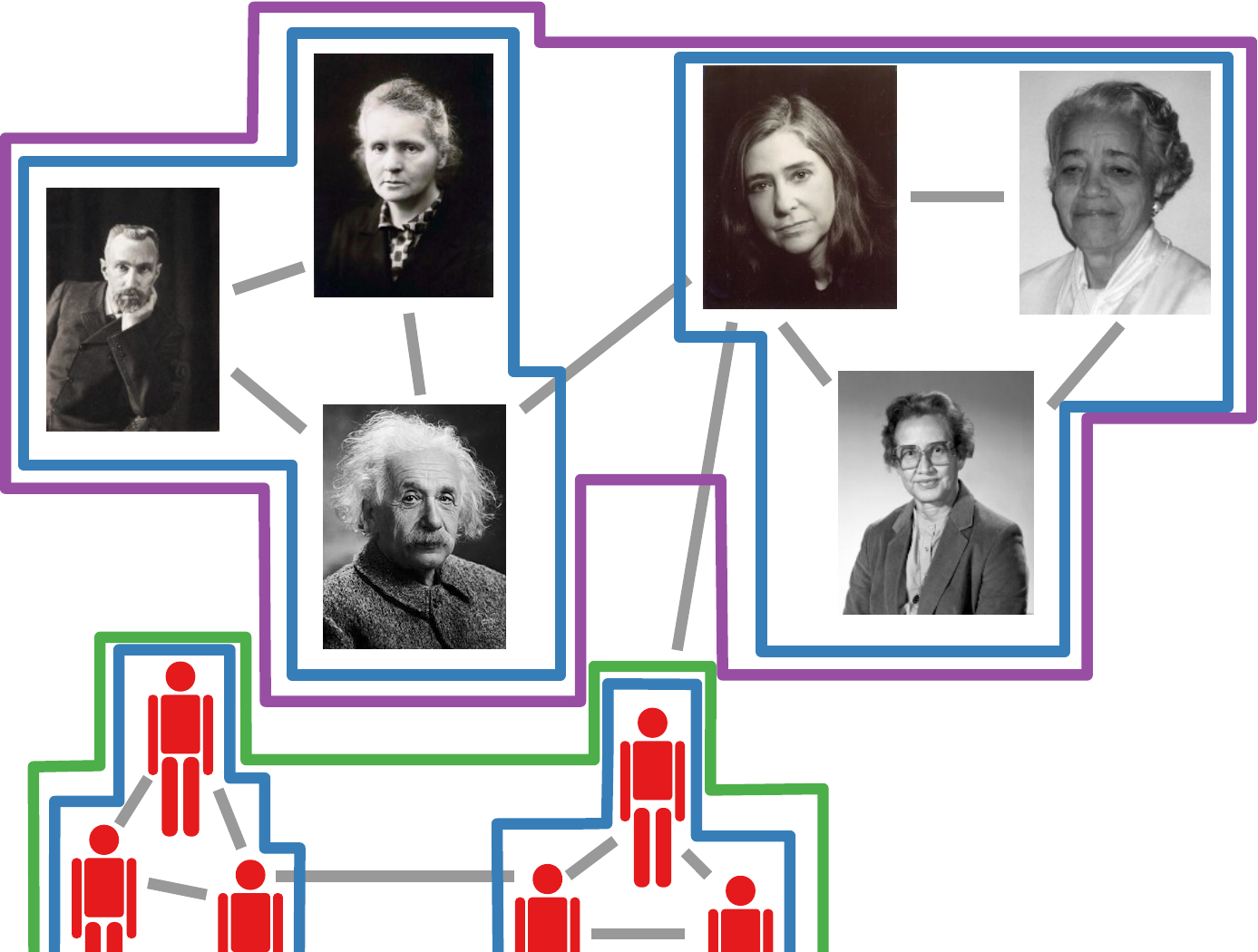}
\caption{Two possible valid partitions of this network. Gray lines are the edges. The blue outlines identify strong, narrow communities. The purple line outlines the scientific community, opposed to the green outline: the laymen community.}
\label{fig:hcd-example}
\end{figure}

When talking about the issues of modularity as a quality measure for network partitions, we touched on an important subject which deserves to be explored more deeply. When doing community discovery, you might have a situation where the network can be divided in different ways and they're all valid partitions. For instance, in Figure \ref{fig:hcd-example}, the obvious assortative partition (blue outlines) will divide scientists into their fields, and laymen into their own communities. However, it is also reasonable to enlarge the definition of a field and say that there is a scientific community, which incorporates all of its subfields (purple outline), and a non-scientific community, which incorporates all non-scientific subcommunities (green outline).

To generalize the issue, once you find tightly knit communities, you might realize that some communities are more related to each other than others. And so there is a second level on the partition you can impose on the network. This means that, after finding communities of nodes, you want to find communities of communities. And communities of communities of communities. And so on. This is the ``hierarchical''  community discovery problem: how to create a hierarchy of communities that best describes the structure of your network.

I'm going to present some general approaches to hierarchical community discovery. As usual, be aware that there are more than I can cover here\cite{papadimitriou2008hierarchical}\cite{huang2010shrink}. However, once you know them, it's easy for you to see that you can redefine many standard community discovery algorithms to find hierarchical communities. For instance, the Infomap algorithm I described in the previous chapter has a natural hierarchical version\cite{rosvall2011multilevel}.

\section{Recursive Approaches}\label{sec:hcd-recursive}
You can transform any community discovery algorithm into a hierarchical community discovery algorithm by applying it recursively to coarsened views of your network. The easiest way to do so is by following this simple meta algorithm:

\begin{enumerate}
\item Apply your algorithm to $G$ and find the optimal communities;
\item Condense your graph by collapsing all nodes belonging to community $C$ into a meta-node $C$.;
\item Connect all $C$s with each other, according to how many edges there were between the nodes they include;
\item You now have a new graph $G'$, so you can go back to step $1$.
\end{enumerate}

\begin{figure}
\centering
\begin{subfigure}{.225\columnwidth}
\includegraphics[width=\textwidth]{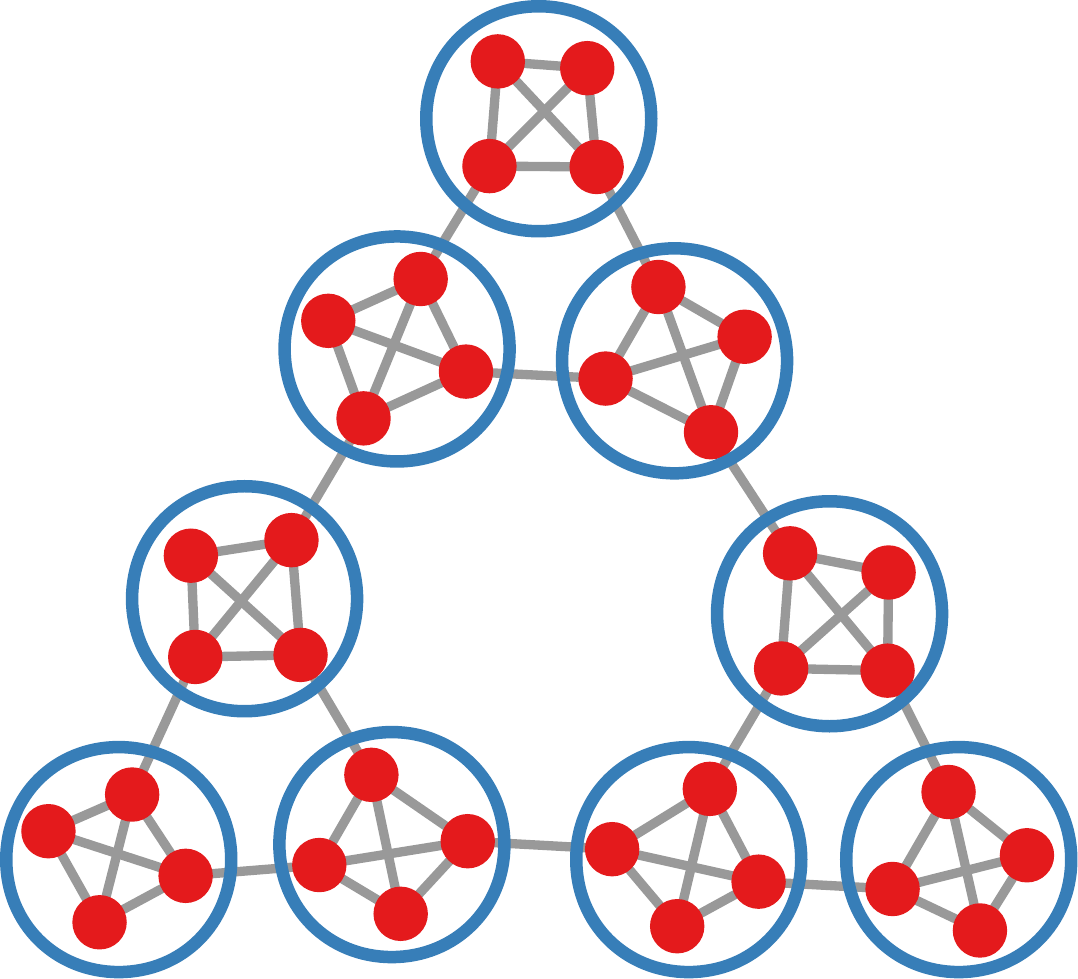}
\caption{}
\end{subfigure}\quad
\begin{subfigure}{.225\columnwidth}
\includegraphics[width=\textwidth]{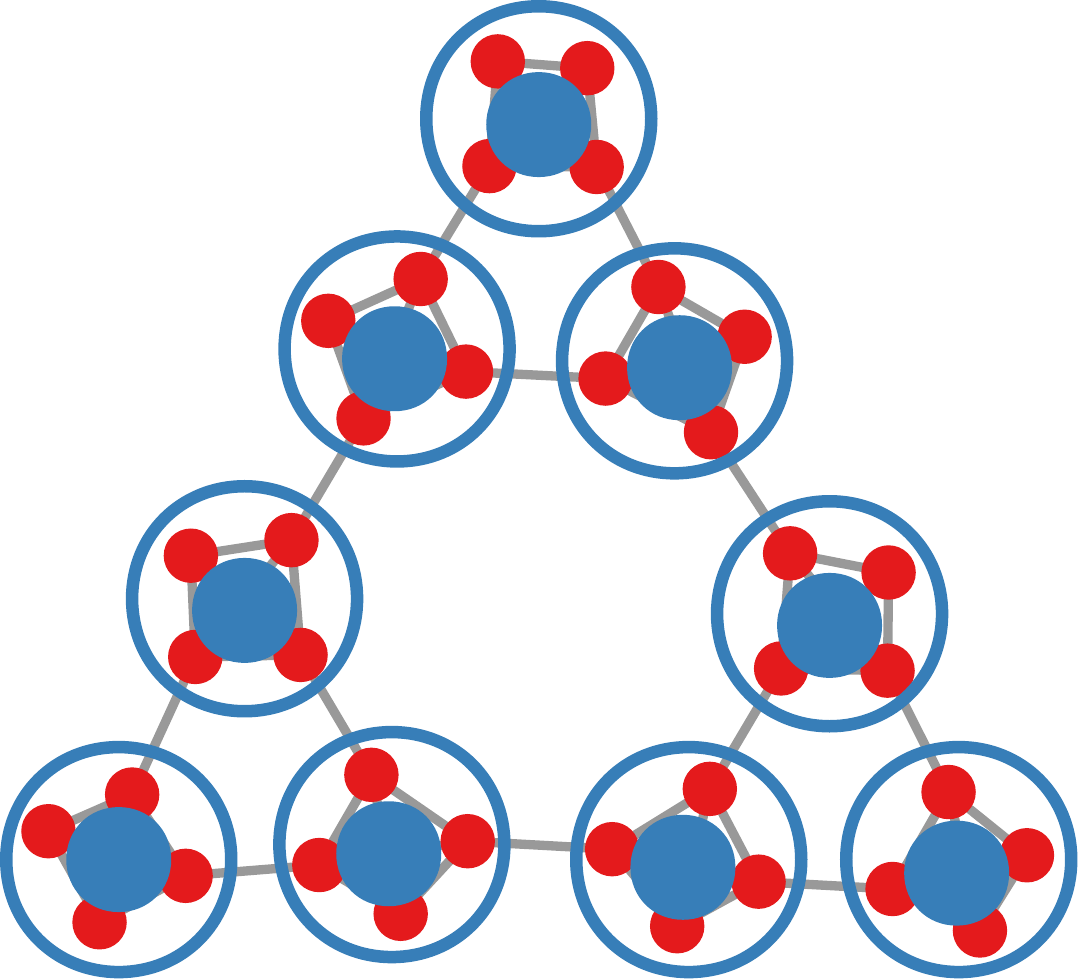}
\caption{}
\end{subfigure}\quad
\begin{subfigure}{.225\columnwidth}
\includegraphics[width=\textwidth]{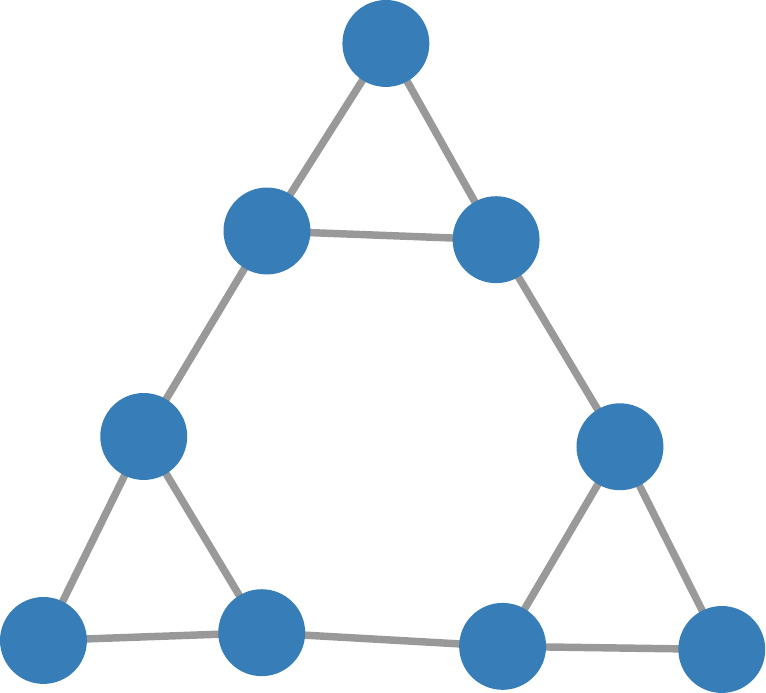}
\caption{}
\end{subfigure}\quad
\begin{subfigure}{.225\columnwidth}
\includegraphics[width=\textwidth]{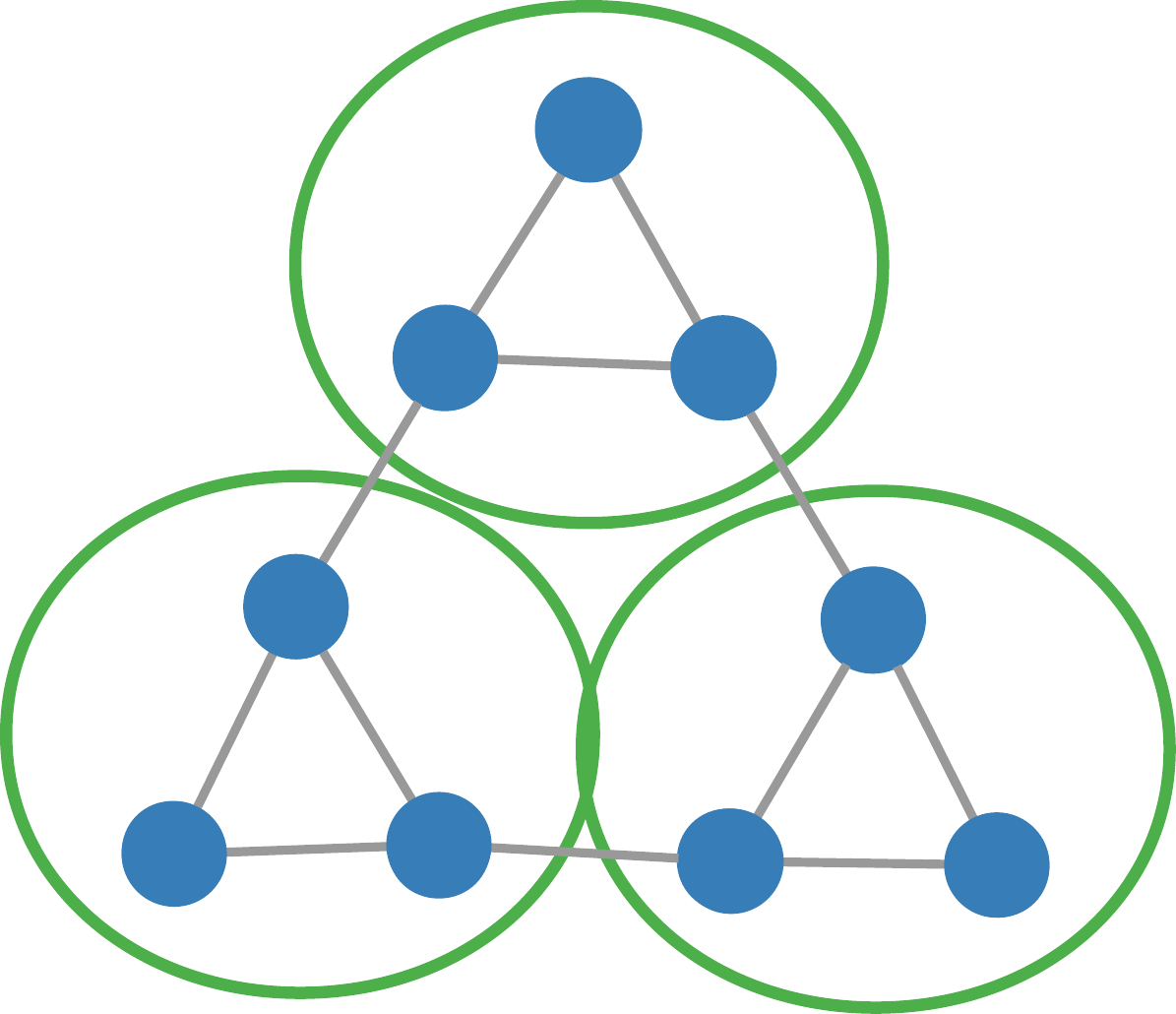}
\caption{}
\end{subfigure}
\caption{(a) The optimal node partition. (b) Collapsing each community into a meta-node. (c) Connecting meta-nodes made by nodes which were originally connected to each other. (d) Finding the second level partition.}
\label{fig:hcd-merge1}
\end{figure}

Figure \ref{fig:hcd-merge1} shows a graphical example of this meta algorithm. You can modify step $3$ in case your algorithm was an overlapping algorithm, which allows communities to share nodes. In that case, you can count the number of shared nodes between the communities\cite{coscia2014uncovering}, rather than the number of edges connecting them.

However, such approach is just a hack on top of non-hierarchical community discovery. In reality, we want to have a hierarchy-aware approach that was built with this feature in mind from the beginning, rather than adding it as an afterthought. There are two meta-approaches for baking in hierarchies in your community discovery: merging and splitting. 

\subsection{Merging}
In the merging approach, you start from a condition where all your nodes are isolated in their own community and you create a criterion to merge communities. This is a bottom-up approach. It is similar to the meta-algorithm from earlier, but it's not really the same. Let's take a look at how it works, highlighting where the differences with the meta-algorithm are.

The template I'm using to describe this approach is the Louvain algorithm\cite{blondel2008fast}. This is one of the many heuristics used to recursively merge communities with the aim of maximizing modularity\cite{sales2007extracting}\cite{peixoto2014hierarchical}, which happens to be among the fastest and most popular.

\begin{figure}
\centering
\includegraphics[width=.45\columnwidth]{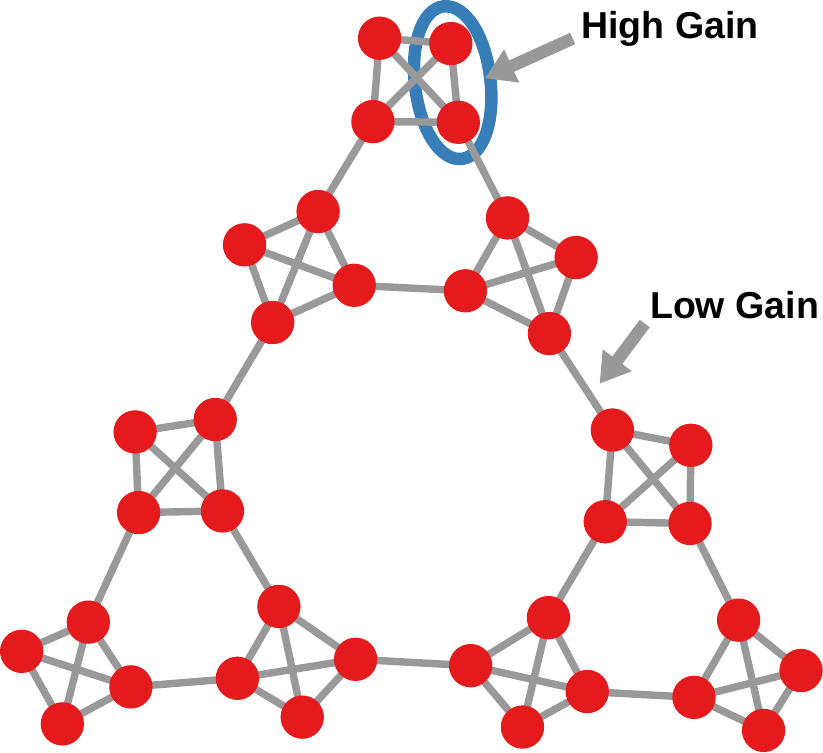}
\caption{An example of the first step of the Louvain algorithm. All in-clique edges (like the representative I highlight in blue) are merged, while all out-clique edges (like the representative I point to with a gray arrow) are ignored.}
\label{fig:hcd-louvain}
\end{figure}

The Louvain algorithm starts with each node in its own community. It calculates, for each edge, the modularity gain one would get if they were to merge the two nodes in the same community. Then it merges all edges with a positive modularity gain. Now we have a different network for which the expensive modularity gains need to be recomputed. However, this network is smaller, because of all the edge merges. You repeat the process until you have all nodes in the same community. Figure \ref{fig:hcd-louvain} shows an example of this process.

The Louvain algorithm is particularly smart and optimized to find the best merges by minimizing the amount of computation needed. Its first step is expensive, because for every edge you have to know what's the modularity gain of merging the nodes. However, once you start merging, it's fast because you only need to update the gains of the nodes directly connected to the new partition. Finally, there is no need to go all the way: we know that putting all nodes in the same community has modularity zero, so at some point there are no moves that can improve modularity, and we can stop. The algorithm inspiring it\cite{clauset2004finding}, for instance, only made one merge per modularity gain and thus had to perform the expensive modularity gain calculation more often for larger networks.

\begin{figure}
\centering
\includegraphics[width=\columnwidth]{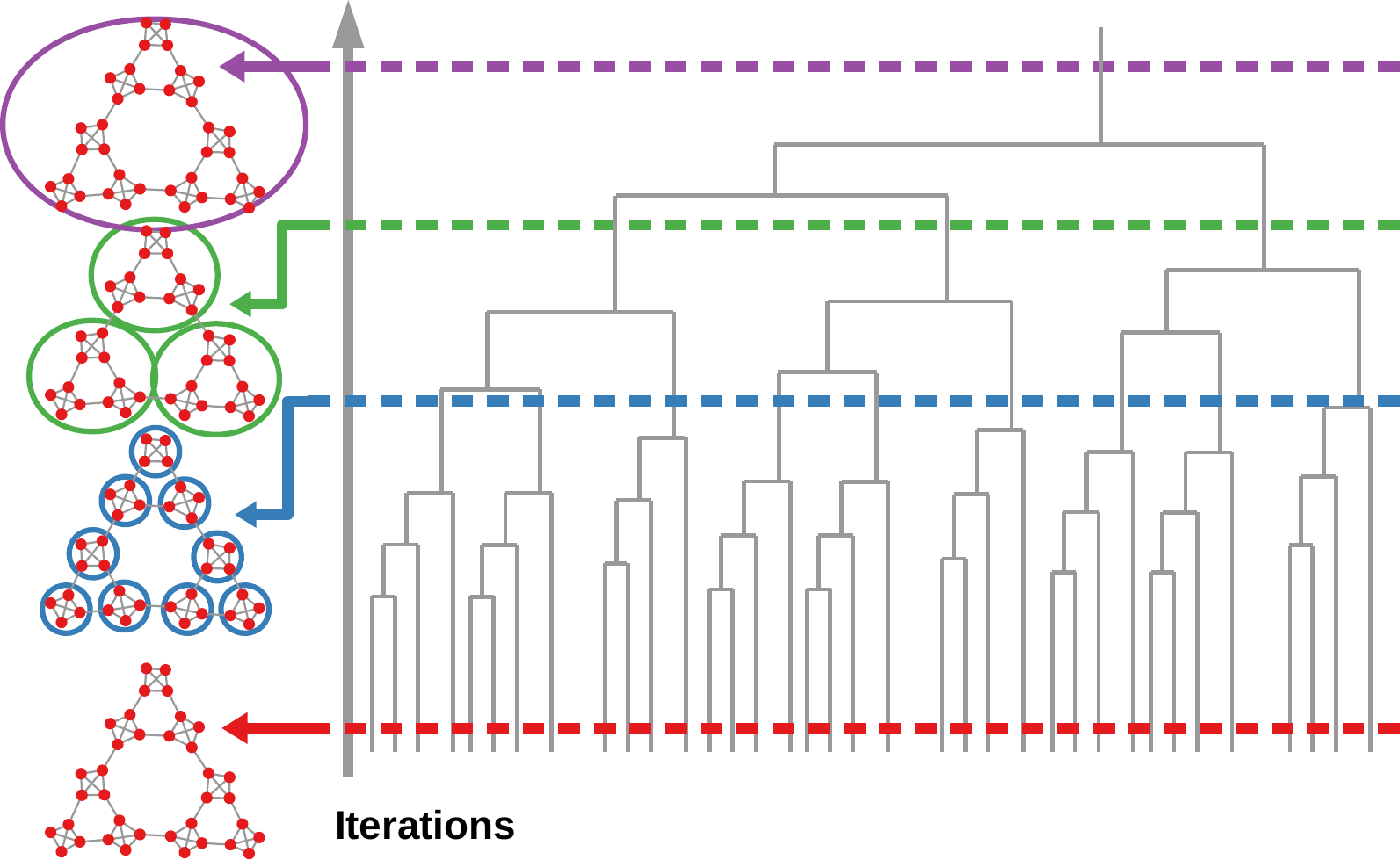}
\caption{The dendogram building from the bottom up typical of a ``merging'' approach in hierarchical community discovery.}
\label{fig:hcd-bottomup}
\end{figure}

What the algorithm does, in practice, is building a dendogram of communities from the bottom up. Each iteration brings you further up in the hierarchy. We start with no partition: each node is in its own community. And then we progressively make larger and larger communities, until we have only one. Figure \ref{fig:hcd-bottomup} shows an example of this approach. This is the crucial difference between the merging approach and what I discuss previously. In the meta algorithm, you don't perform all the merges, you make lots of them at once when you run your step $1$ to find the initial communities.

\subsection{Splitting}
In the splitting approach, you do the opposite of what I described so far. You start with all nodes in the same community and you use a criterion to split it up in different communities. For instance by identifying edges to cut. This is a top-down approach.

Historically speaking, the first algorithm using this approach used edge betweenness as its criterion to split communities\cite{girvan2002community}\cite{newman2004finding}. That is not to say there aren't valid alternatives as your splitting criterion, including -- but not limiting to -- edge clustering\cite{radicchi2004defining} and information centrality\cite{fortunato2004method}. However, given its historical prominence, I'm going to allow the edge betweenness Girvan-Newman algorithm to have its place under the limelight.

\begin{figure}
\centering
\includegraphics[width=\columnwidth]{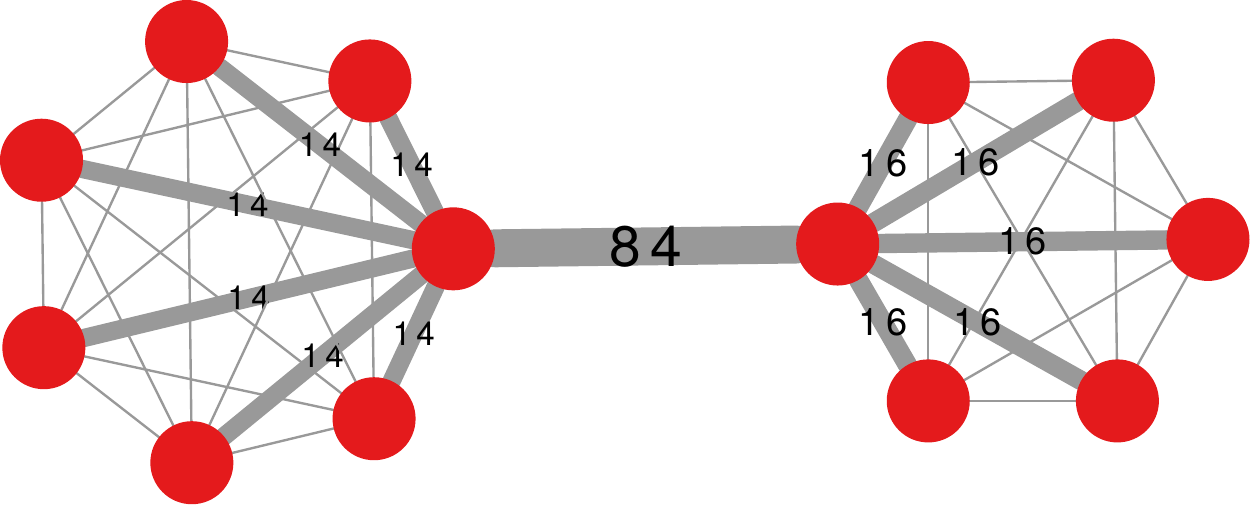}
\caption{Two cliques connected by an edge. I label links with edge betweenness higher than one with the number of shortest paths passing through them.}
\label{fig:hcd-eb}
\end{figure}

The first step of the algorithm is to calculate the edge betweenness of each edge in the network, that is the normalized number of shortest paths passing through it (Section \ref{sec:centr-betw}). The assumption is that edges between assortative communities will have a systematically higher edge betweenness value than edges inside the communities. Figure \ref{fig:hcd-eb} shows an example. All edges inside the communities have a low value because there are many alternative paths you can take, since all nodes are connected to everybody else in the community. On the other hand, if you want to go from one node in one community to a node in another, there's only one edge you can use. As a result, its edge betweenness value skyrockets.

The second step of the algorithm is to cut the edge with the highest edge betweenness. The final aim is to break the network down into multiple components. Each component of the network is a community.

Unfortunately, after each edge deletion you have to recalculate the betwennesses. Every time you alter the topology of the network you change the distribution of its shortest paths. This makes edge betweenness extremely computationally heavy. Calculating the edge betweenness for all edges takes an operation per node and per edge ($O(|V||E|)$) and you have to repeat this for every edge you delete, resulting in a crazy complexity of $O(|V||E|^2)$. You cannot apply this naive algorithm to anything but trivially small networks.

You can now see the parallels with the Louvain method I described earlier. The difference is that you are exploring the dendogram of communities from the top down, rather than bottom up. Each iteration brings you further down in the hierarchy. At the very top you start with a network with a single connected component. As you delete edges, you find different connected components. As you continue, you end up with more and more. At the last iteration, each node is now isolated.

\begin{figure}
\centering
\includegraphics[width=\columnwidth]{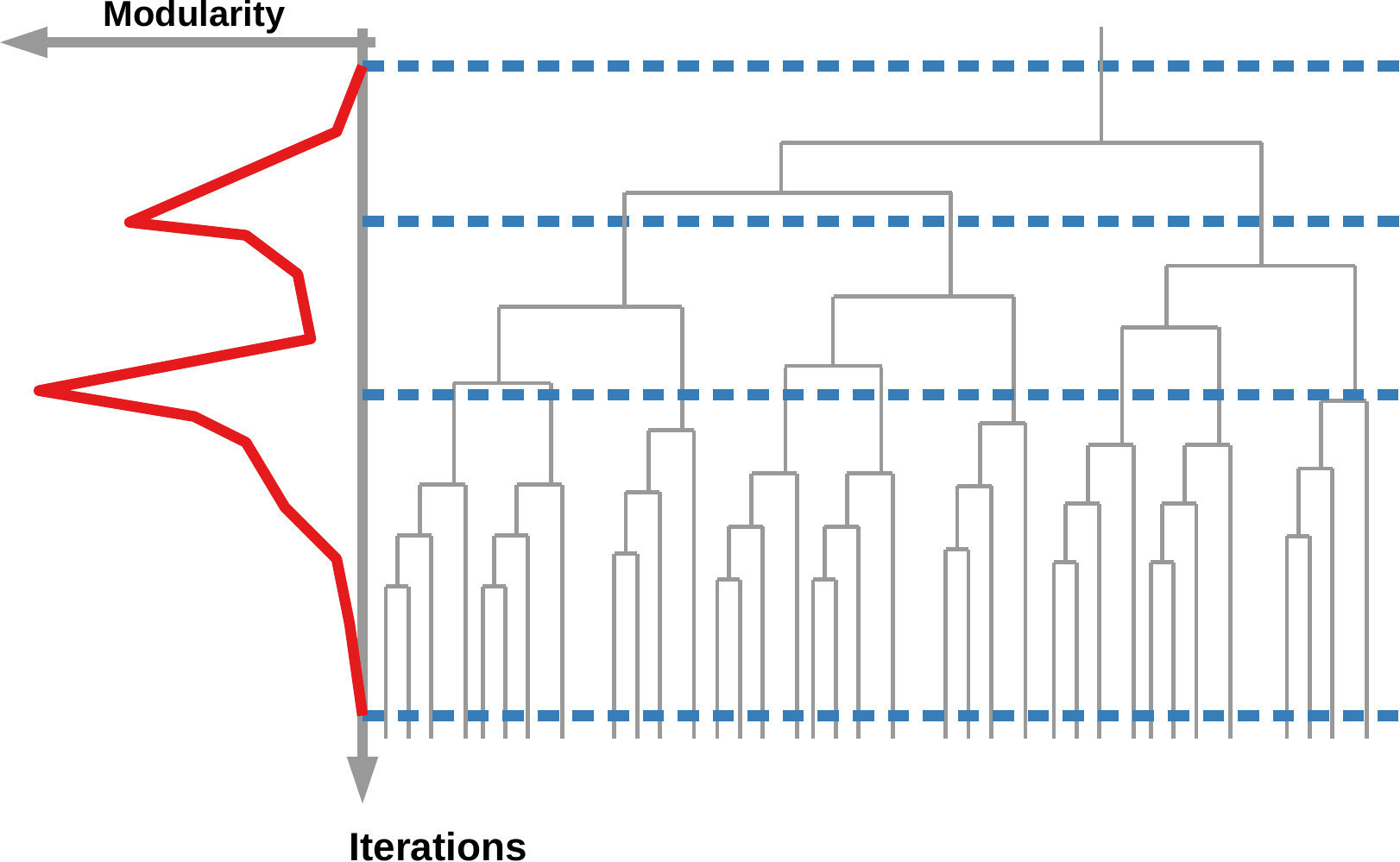}
\caption{The dendogram building from the top down typical of a ``splitting'' approach in hierarchical community discovery. The left panel shows the modularity values of each possible cut.}
\label{fig:hcd-topdown}
\end{figure}

Differently from the Louvain algorithm, in the Girvan-Newman method you do not calculate modularity gains as you explore the dendogram. Thus, the algorithm will normally perform all the possible splits and returns you the full structure, rather than the cut that maximizes modularity. Thus you will have to calculate the modularity of each split yourself, something similar to what you see in Figure \ref{fig:hcd-topdown}.

Higher modularities are better partitions, thus better cuts. The good cuts will appear as peaks in the modularity profile of the dendogram. Thus they are the natural points for us to cut it and get the best partition. Multiple peaks are a clue of a hierarchical organization, because they identify good partitions with a very different number of communities.

The aforementioned edge clustering and information centrality variants use the same algorithm, changing the criterion to determine which edge to cut. In edge clustering the assumption is that edges with high clustering are embedded in communities, because all their neighboring nodes are connected to each other. Note that in this case we also modify the definition of local clustering coefficient so that it applies to edges rather than nodes. The guiding principle is to cut the edges with the lowest clustering first. This is computationally more efficient than using edge betweenness, because when you cut an edge you only change the clustering of the neighboring edges: in edge betweenness all values need to be recomputed.

The information centrality variant has no such benefit, because it simply uses a different definition of edge betweenness. Specifically it uses the edge current flow centrality\cite{brandes2005centrality}. It is still more computationally efficient, because it uses random walks rather than shortest paths, which will treat your CPU with more respect.

\section{Hierarchical Random Graphs: Part 2}
This section is a throwback to Section \ref{sec:lpsimple-hrg}, where I introduced the usage of Hierarchical Random Graphs to solve the problem of link prediction. If you remember, the idea was to divide the graph into a hierarchical organization, under the assumption that nodes part of the same hierarchical branch are more likely to connect to each other. If we start from this assumption, then the natural consequence is also that these nodes \textit{already are} densely connected. Thus they are proper communities!

Note that, however, HRG is more flexible than that: it can also uncover a disassortative community structure where nodes are \textit{less} likely to connect to their community mates.

In Section \ref{sec:lpsimple-hrg} I simply said that ``we create a hierarchical representation of the observed connections that fits the data,'' which is a rather mysterious non-explanation of how we actually group nodes into communities. The way the algorithm works is by creating a dendrogram to fit the data. In the dendrogram, the $|V|$ leaf nodes are the nodes of the network. The other nodes of the dendrogram are so-called ``internal nodes'', and we need $|V| - 1$ of them to properly build the full structure.

\begin{figure}
\centering
\includegraphics[width=.66\columnwidth]{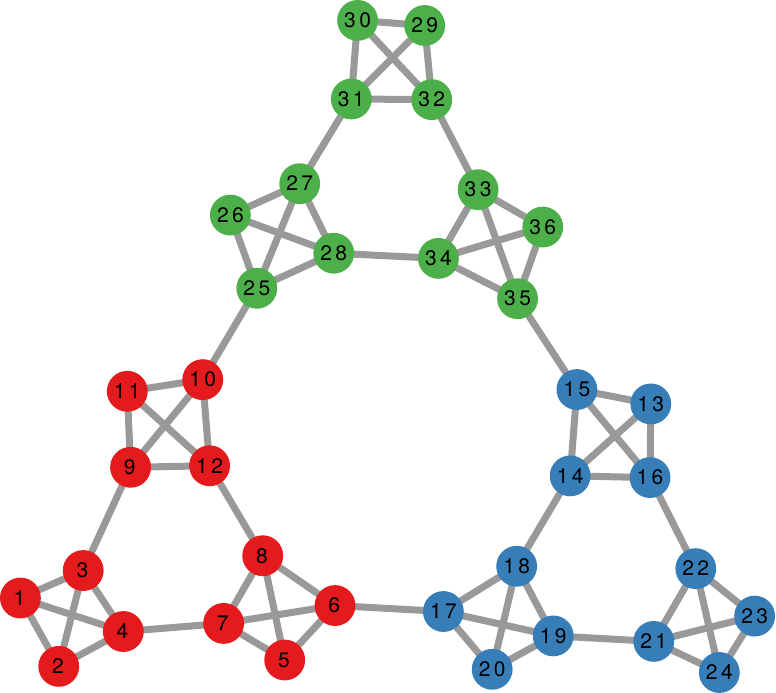}
\caption{A graph with hierarchical communities (node color according to the community partition at one level of the hierarchy).}
\label{fig:hrg-comms1}
\end{figure}

\begin{figure*}
\centering
\includegraphics[width=\columnwidth]{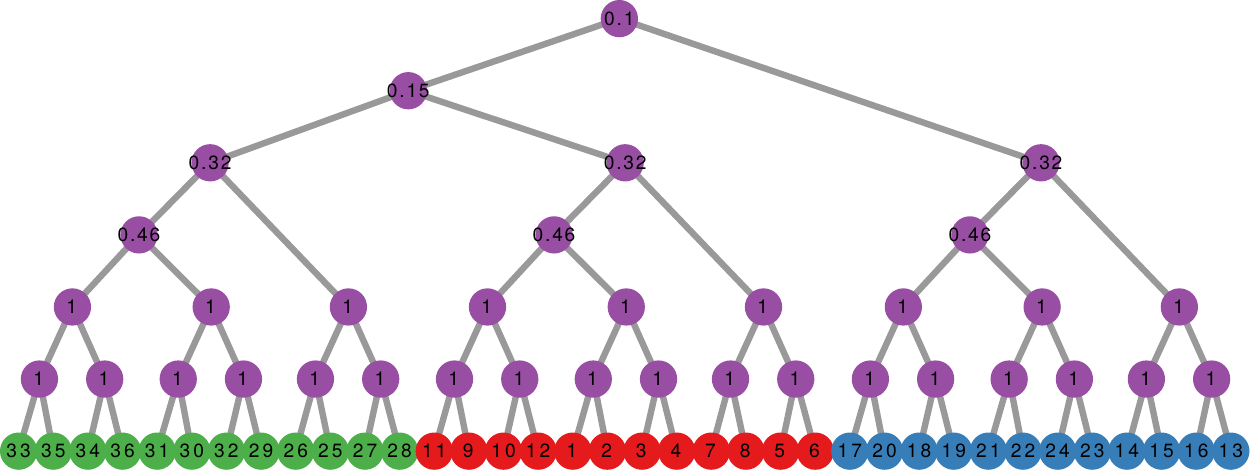}
\caption{A likely HRG representation of Figure \ref{fig:hrg-comms1}. Purple nodes are internal nodes. I label them according to their $p_i$ probability, which is the number of edges between nodes in the branches over all possible number of edges between them.}
\label{fig:hrg-comms2}
\end{figure*}

Figure \ref{fig:hrg-comms1} shows a network with a hierarchical community structure. Figure \ref{fig:hrg-comms2} is a possible HRG representation of Figure \ref{fig:hrg-comms1}. Each internal node $i$ has an associated probability $p_i$. This is the probability of connecting two nodes $u$ and $v$ that are in the branch attached to $i$. So our aim is to assign to all internal nodes the proper $p_i$ probabilities and to shuffle the leaf nodes properly so that the dendrogram we end up with is the one most likely to describe the real data. 

If this sounds familiar, you're not wrong. This looks like a special hierarchical version of stochastic blockmodels. You are basically assigning to each node pair a different $p_i$ probability of connecting, depending on which is their most proximate common internal node ancestor $i$.

At this point, finding the dendrogram that most likely fits the data is just a choice of how you want to explore the space of all possible dendrograms. In the original paper, authors use a Markov chain Monte Carlo method, where each dendrogram is sampled proportionally to its likelihood value.

Note that the dendrogram in Figure \ref{fig:hrg-comms2} is not the only possible good description of Figure \ref{fig:hrg-comms1}. For instance, I could have grouped nodes $1$ and $3$ together, at the lowest level, rather than $1$ and $2$. The resulting dendrogram would have been equally likely, and would generate an equally good hierarchical representation.

\section{Directed Communities}
HRGs allow us to transition to a slightly different way of interpreting hierarchical community discovery. HRGs still group communities into super communities, but takes a more statistical approach rather than applying a recursive merge/split operation. Pushing further this idea, we can get to the scenario of directed community discovery: to find communities in directed networks. Here we totally reject the idea that we should group communities into super communities and we simply establish that the communities we found organize in a hierarchy. Some communities are at the bottom because they prevalently point to other communities -- here pointing to is interpreted as being below in the hierarchy.

\begin{figure}
\centering
\begin{subfigure}{.4\columnwidth}
\includegraphics[width=\textwidth]{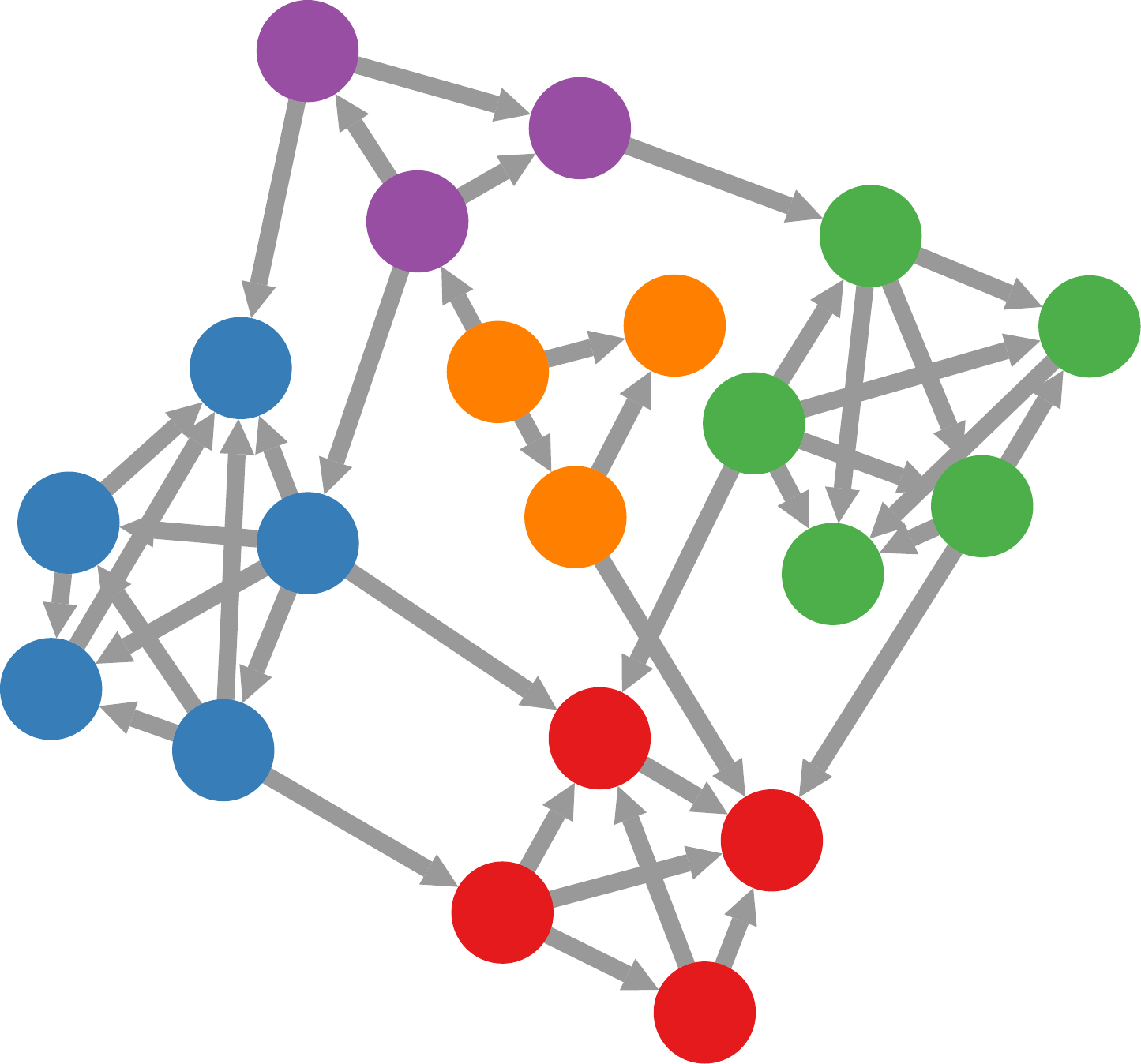}
\caption{}
\end{subfigure}\quad
\begin{subfigure}{.4\columnwidth}
\includegraphics[width=\textwidth]{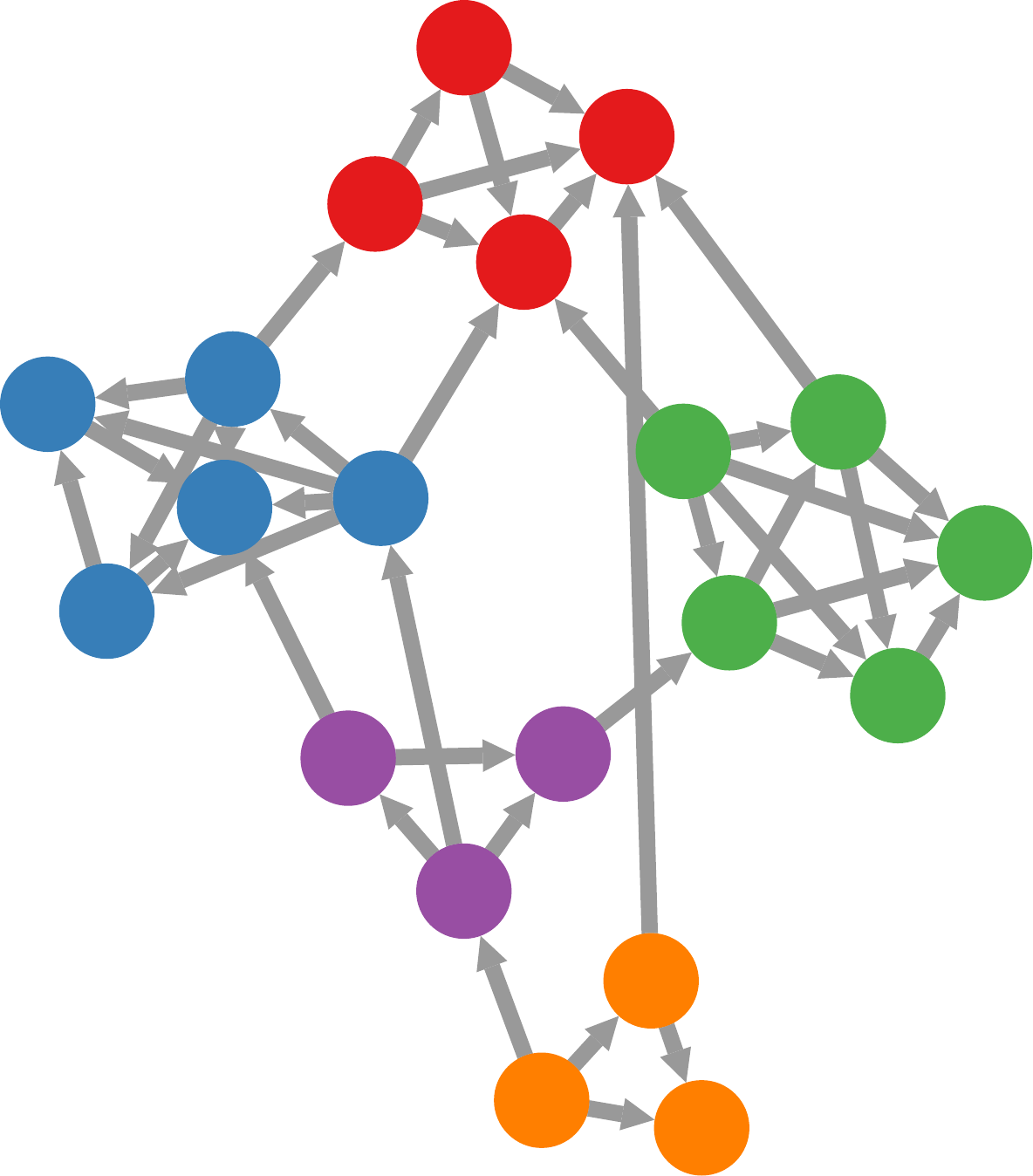}
\caption{}
\end{subfigure}
\caption{(a) A classical community partition for a network. The node color identifies the node's community. (b) A hierarchy-aware directed community partition, using the vertical dimension to show the community's ranking.}
\label{fig:directed-cd}
\end{figure}

One approach\cite{peixoto2022ordered} is relatively similar to HRGs: we're still doing a degree corrected SBM. However, as Figure \ref{fig:directed-cd}, our objective is not to merge communities, but to figure out how to arrange them vertically so that most links point upwards. Note how in Figure \ref{fig:directed-cd}(b) all edges going from one community to another always point upwards. The algorithm will find the vertical positioning of the communities most appropriate to guarantee this feature the best way it can. As a result, you can consider the top community as the one dominating the network, in a hierarchical fashion.

\begin{figure}
\centering
\begin{subfigure}{.4\columnwidth}
\includegraphics[width=\textwidth]{figures/neuralnet_feedforward.pdf}
\caption{}
\end{subfigure}\quad
\begin{subfigure}{.4\columnwidth}
\includegraphics[width=\textwidth]{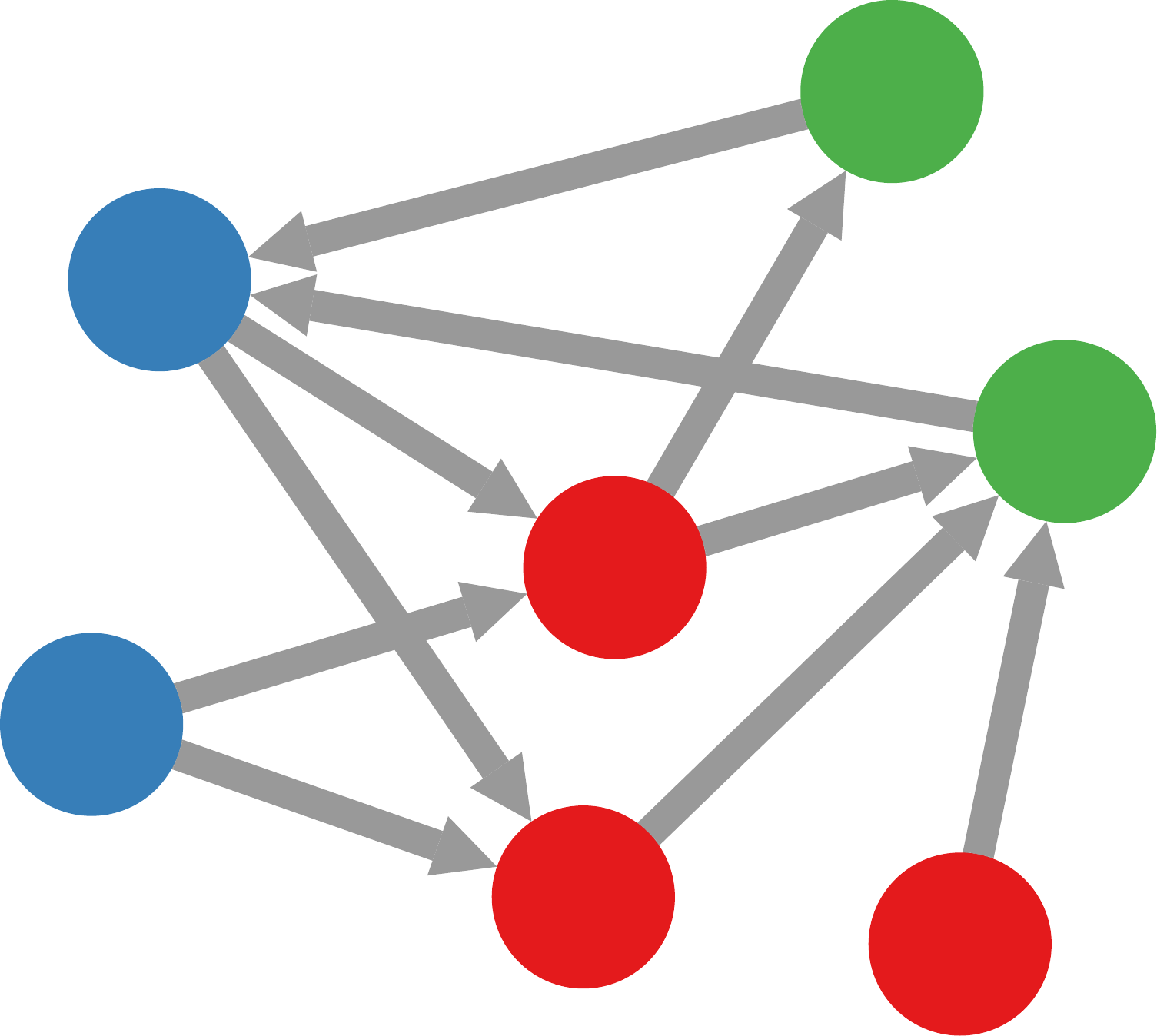}
\caption{}
\end{subfigure}
\caption{The node color identifies the node's community. (a) A pattern-based directed community partition. (b) A flow-based directed community partition.}
\label{fig:directed-cd-patterns}
\end{figure}

Of course, this is not the only way to intend directed community discovery. I already showed you in Section \ref{sec:cd-eval-mod} that you could use (one of the many competing versions of) directed modularity to do the job. In general, directed networks open the possibility of finding communities with a logic that is different from the usual one based on density. We might want to identify groups of nodes that follow a certain pattern\cite{lai2011partitioning}. For instance, Figure \ref{fig:directed-cd-patterns}(a) shows that nodes can be grouped not because they connect to each other -- in the classical density case -- but because they are pointed by (or point to) the same nodes. In this case, red points to blue and blue points to green.

This does not necessarily imply the hierarchical directionality we just saw. You can have a headless flow-based pattern like in Figure \ref{fig:directed-cd-patterns}(b) that still satisfies the definition I just gave, without allowing you to find any hierarchical organization. In this case, blue points to red, red to green and green to blue, so there's no head to this cycling pattern.

\section{Density vs Hierarchy}
Hierarchical community discovery poses some issues with our traditional community definition based on density. If there is a partition with maximum internal density in the network, then a partition at a different hierarchical level must have a lower density -- by definition. Is that still a valid partition of the network? When you perform hierarchical community discovery you'd say yes, and that would be totally valid. There are valid analytic scenarios where you can divide up a social network at multiple levels. For instance the example I made at the beginning, dividing it up into a layman and scientific community, and then breaking down the scientific community in fields and subfields. But that is in direct contradiction with our classical community definition.

\begin{figure*}[t]
\centering
\begin{subfigure}{.45\columnwidth}
\includegraphics[width=\textwidth]{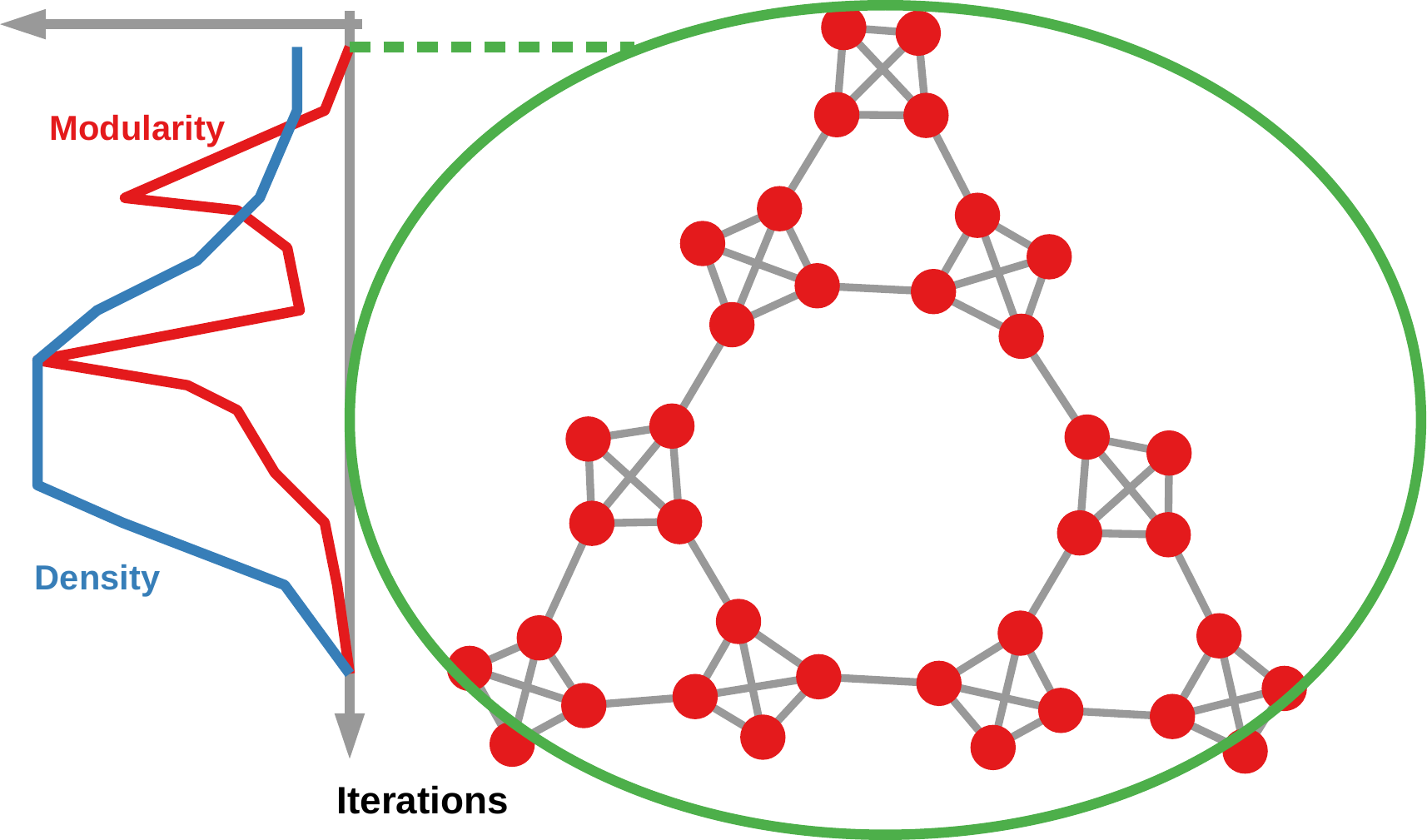}
\caption{}
\end{subfigure}\quad
\begin{subfigure}{.45\columnwidth}
\includegraphics[width=\textwidth]{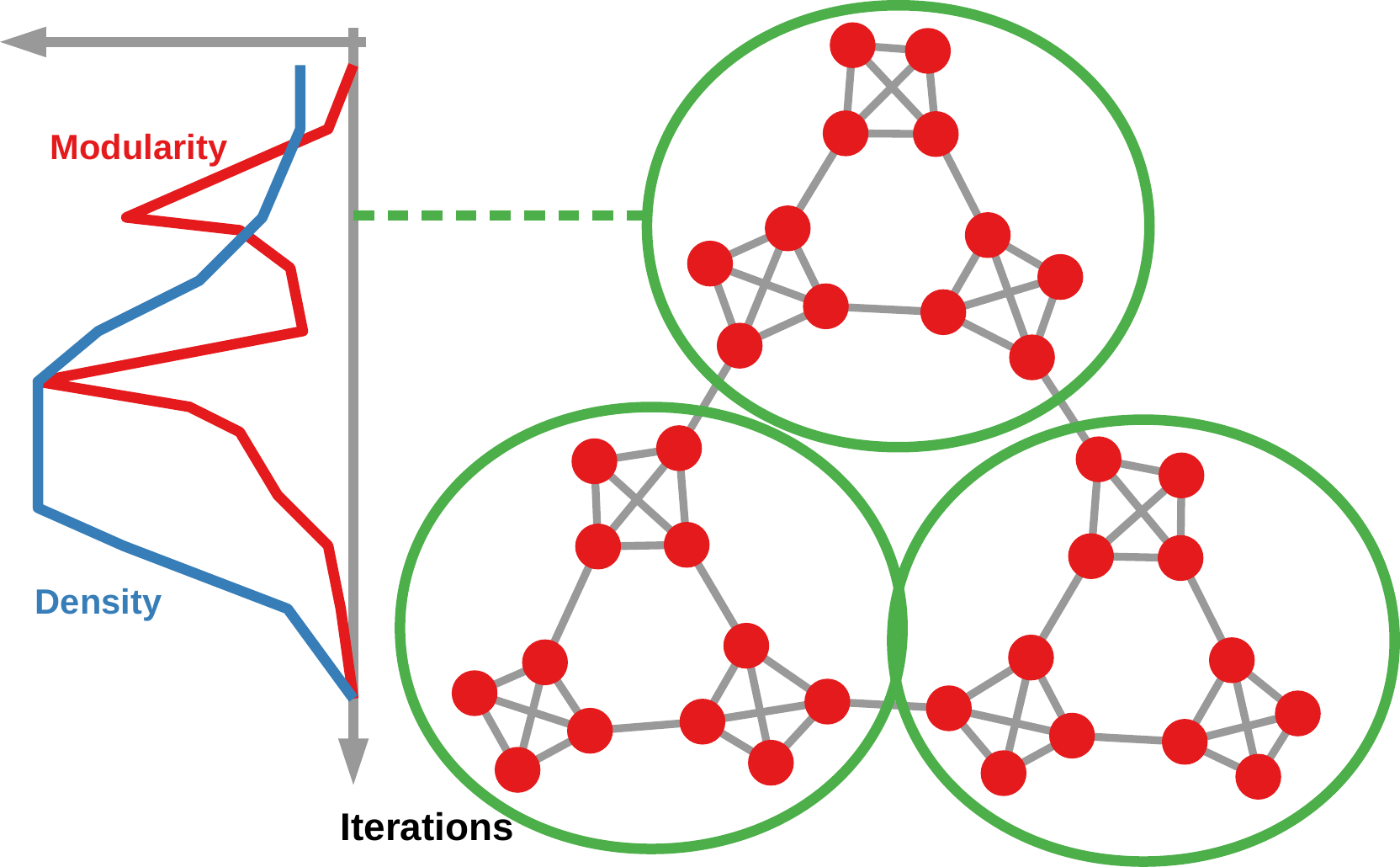}
\caption{}
\end{subfigure}\quad
\begin{subfigure}{.45\columnwidth}
\includegraphics[width=\textwidth]{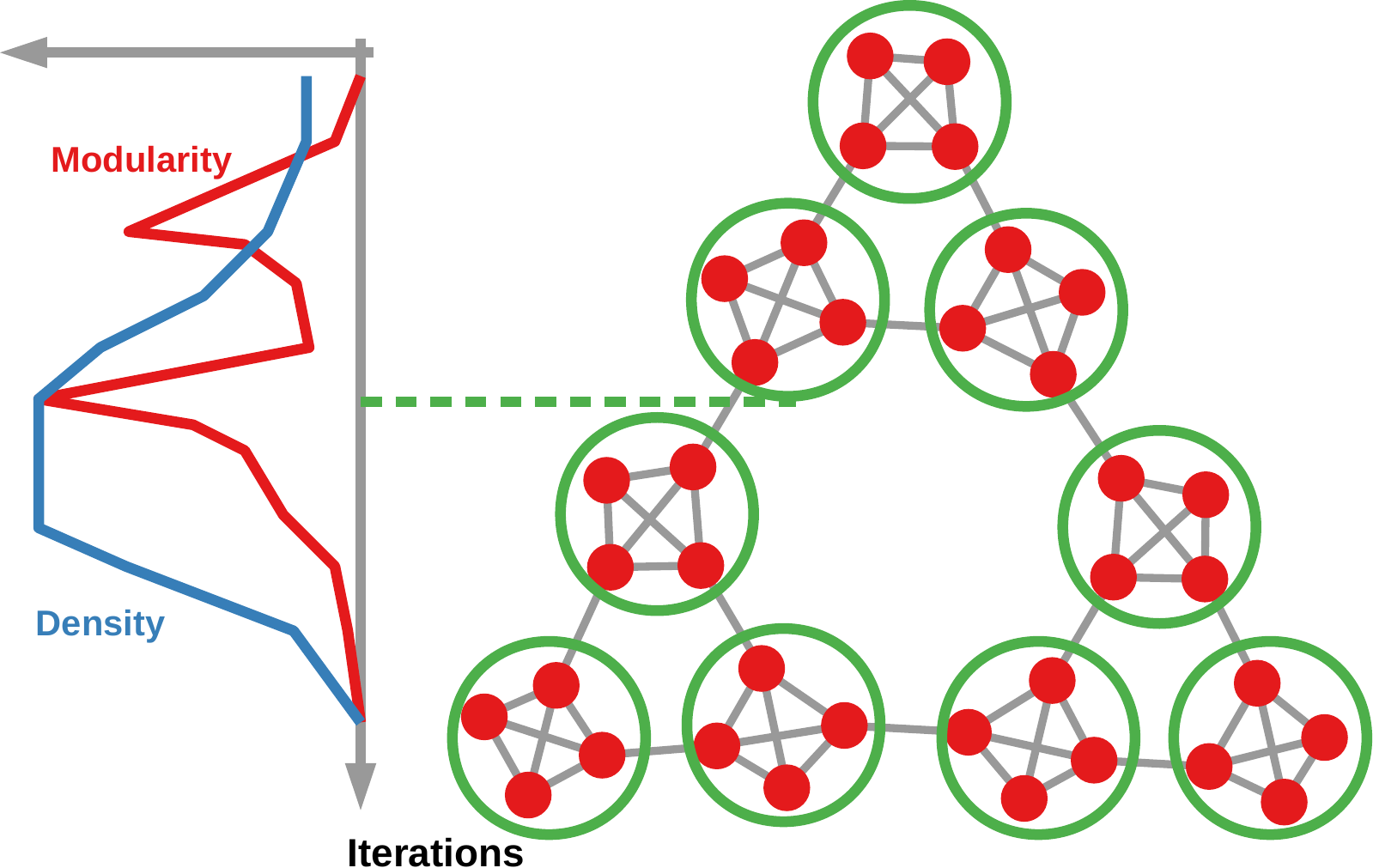}
\caption{}
\end{subfigure}\quad
\begin{subfigure}{.45\columnwidth}
\includegraphics[width=\textwidth]{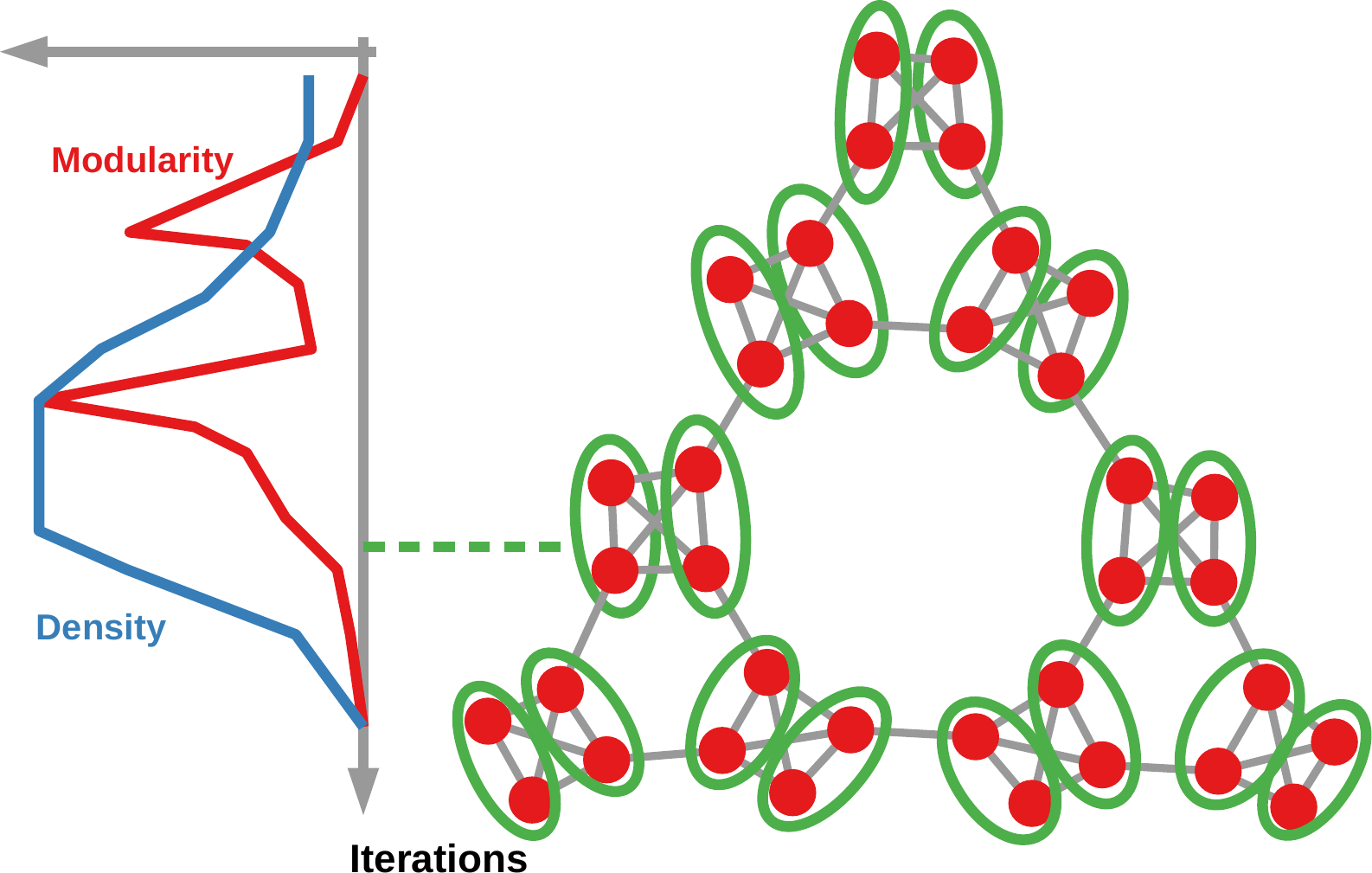}
\caption{}
\end{subfigure}
\caption{The contrast between modularity and density at different cuts in the hierarchical community organization: (a) every node in the same community; (b) sub-optimal high-level partition; (c) optimal low-level partition; (d) maximal density but low modularity partition.}
\label{fig:hcd-paradox}
\end{figure*}

This is particularly tricky since even modularity, which should be defined as in direct correspondence with this density-based definition, actually disagrees with it. In other words, the density profile changes differently from the modularity profile. When we group everything in a community, there's some density even if modularity is zero (Figure \ref{fig:hcd-paradox}(a)). At the top hierarchical level we have high modularity but low internal density (Figure \ref{fig:hcd-paradox}(b)). At the best partition we have agreement (Figure \ref{fig:hcd-paradox}(c)). But density is still high even with low modularity for a partition that puts together connected node pairs (Figure \ref{fig:hcd-paradox}(d)).

\section{Summary}

\begin{enumerate}
\item You can find communities at different scales in a network. Meaning that there are communities of nodes, communities of communities, communities of communities of communities, and so on. The process to find such structures is hierarchical community discovery.
\item You can find hierarchical communities with either a top-down or a bottom-up approach. In the bottom-up or merging approach, you start with each node in its own community and then you recursively merge communities optimizing a quality function.
\item In the top-down or splitting approach, you start with the network encapsulated in a single community and you recursively split it following some guiding principle, e.g. removing the edges with the highest betweenness.
\item The third alternative is to model your network as a hierarchical system, and then find the hierarchical organization that is the most likely explanation of your observed data.
\item Not all communities at all levels of the hierarchy maximize the internal density and/or the external sparsity of your communities. Thus, even if totally valid, hierarchical communities defy our classical definition of communities based on edge density.
\end{enumerate}

\section{Exercises}

\begin{enumerate}
\item Use the edge betweenness algorithm to find hierarchical communities in the network at \url{http://www.networkatlas.eu/exercises/37/1/data.txt}. Since the algorithm has high time complexity, perform only the first 10 splits. What is the split with the highest modularity?
\item Change the splitting criterion of the algorithm, using the inverse edge weight rather than edge betweenness. Since this is much faster, you can perform the first 20 splits. Do you get higher or lower modularity relative to the result from exercise 1?
\item Use the maximum edge weight pointing to a community as a guiding principle to merge nodes into communities using the bottom-up approach. (An easy way is to just condense the graph by merging the two nodes with the maximum edge weight, for every edge in the network)
\item Using the algorithm you made for exercise 3, answer these questions: What is the latest step for which you have the average internal community edge density equal to 1? What is the modularity at that step? What is the highest modularity you can obtain? What is the average internal community edge density at that step? 
\end{enumerate}

\chapter{Overlapping Coverage}\label{cha:ocd}
Let's go back for a moment to the classical definition of communities in networks: 

\begin{center}
\textit{Communities are groups of nodes densely connected to each other and sparsely connected to nodes outside the community.}
\end{center}

\begin{figure}
\centering
\includegraphics[width=.8\columnwidth]{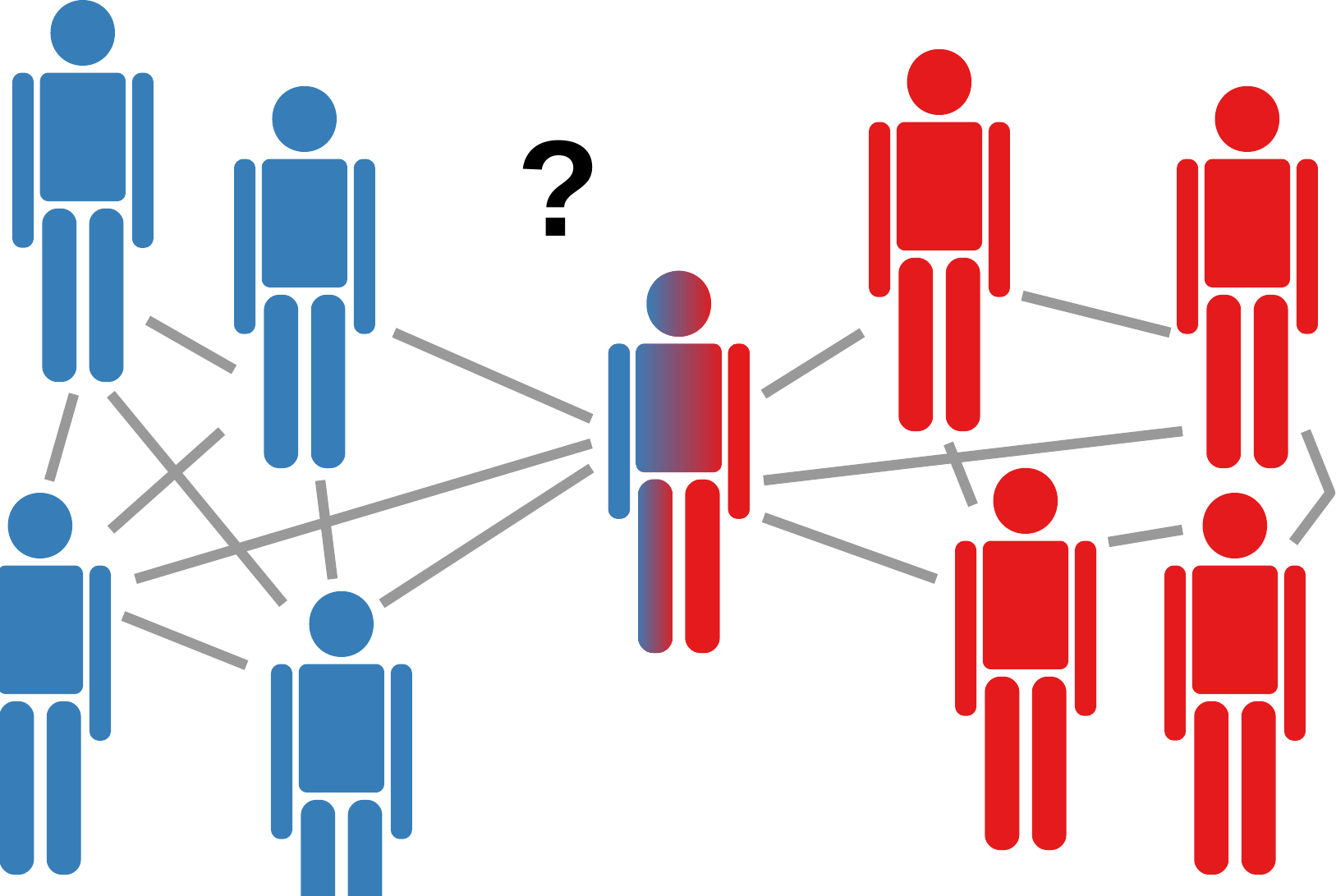}
\caption{A social network with an individual having an equal number of relationships distributed among two communities.}
\label{fig:ocd-example}
\end{figure}

This seems to imply that communities are a clear cut case. Nodes have a majority of connections to other nodes in their community. However real world networks do not have to conform to this expectation, and in fact often they don't. There are numerous cases in which nodes belong to multiple communities: to which community does the center person in Figure \ref{fig:ocd-example} belong? The red or the blue one?

The classical community definition forces us to make a choice. Regardless of the choice we make -- red or blue -- it wouldn't be a satisfying solution. The more reasonable answer is ``she belongs to both''. For instance, a person can very well be part of one community because it is composed by the people they went to school with. And she can be part of a work community too, of people she works with. Some of these people could be the same, but usually they are not.

The problem is that none of the methods seen so far allow for such a consideration. For instance, the basic stochastic blockmodels only allows you to plant a node in a community, not multiple. Modularity also has issues, because of the Kronecker delta: since this is going to be $1$ for multiple communities for a node, there will be double-counting and the formula breaks down.

This is where the concept of \textit{overlapping} community discovery was born. We need to explicitly allow for overlapping communities: communities that can share nodes. There are many ways to do this, which have been reviewed in several articles\cite{xie2013overlapping}\cite{amelio2014overlapping} dedicated especially to this sub problem of community detection (itself a sub problem of network analysis: it's communities all the way down).

Here we explore a few of the most popular approaches.

\section{Evaluating Overlapping Communities}
Before we delve deep into overlapping community discovery, let's amend Chapter \ref{cha:cd-eval} to this new scenario. We can have a few options when we try to evaluate how well we divided the network into overlapping communities.

Normalized mutual information expects you to put nodes into a single category. However, there are ways to make it accept an overlapping coverage\cite{lancichinetti2009detecting}\cite{mcdaid2011normalized}. The obstacle is that NMI wants to compare the vector of metadata with the vector containing the community partition. The vector can only have one value per node but, in an overlapping coverage, it can have multiple values. Thus we don't compare the vectors directly. We compare two bipartite matrices.

\begin{figure}
\centering
\begin{subfigure}{.5\columnwidth}
\includegraphics[width=\textwidth]{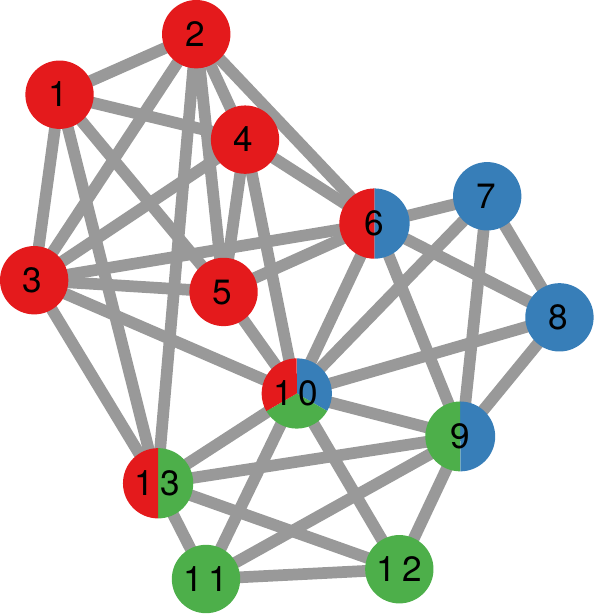}
\caption{}
\end{subfigure}\quad
\begin{subfigure}{.42\columnwidth}
\includegraphics[width=\textwidth]{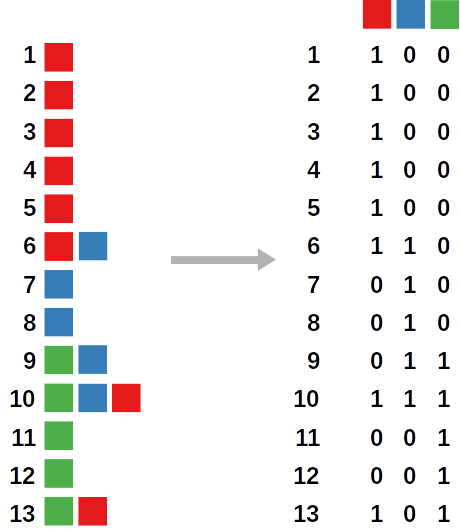}
\caption{}
\end{subfigure}
\caption{(a) A network with three overlapping communities, encoded by the node's color. (b) Transforming the overlapping coverage into a binary affiliation matrix, which we can use as input for the overlapping version of NMI}
\label{fig:ocd-onmi}
\end{figure}

Suppose you found $\mathcal{C}$ communities, and you have $\mathcal{A}$ node attributes. You can describe the overlapping coverage in communities with a $|V| \times \mathcal{C}$ binary matrix, whose $u,C$ entry is equal to $1$ if node $u$ is part of community $C$. The node attribute matrix is similarly defined. Figure \ref{fig:ocd-onmi} shows an example of this procedure. Now you can calculate the mutual information between the two matrices by pairing the columns such that we assign to each column on one side the ones on the other side that is the most similar to it.

We can normalize this mutual information in different ways. In fact, the papers I cited earlier propose six alternatives, providing different motivations for each of those. These overlapping NMIs share with their original counterpart the issue of non-zero values for independent vectors -- although they try to mitigate the issue with different strategies.

Some researchers have pointed out a few biases in the overlapping extensions of NMI and similar measures\cite{gates2019element}. They propose a unified framework that can evaluate disjoint, overlapping, and even hierarchical communities. This is achieved by creating a node-community bipartite affiliation graph and then project it so that you obtain a new node-node unipartite graph -- just like the original network you started with. However, now, the relations between the nodes are due to their common cluster affiliations. The similarity between two community coverages is estimated by looking at the similarity of the stationary distributions of the projected affiliation graphs.

An alternative measure, called Omega Index, attempts to be the overlapping equivalent of the adjusted mutual information: the one correcting for chance\cite{collins1988omega}. This is an extension of the adjusted Rand index (Section \ref{sec:cd-eval-nmi}) for non-disjoint clusters. The Rand index is simply the number of times two partitions agree over all possible pairs of nodes. The adjusted-for-chance version establishes the probability of agreeing by chance and uses that to normalize the index. The solution again passes through a procedure similar to what we explained for the overlapping version of NMI -- but note that there is no universal null model to correct for, and the one you assume has a severe impact on the results you'll be seeing\cite{gates2017impact}.

What about modularity? Of course there should be a way to extend it to work with multiple clusters! How hard can that be? It isn't at all. In fact, it is so easy that there are multiple conflicting ways to extend modularity for the overlapping case. Because woe to the scientific community that can agree on something.

Solutions span from replacing the binary Kronecker delta with a continuous node similarity measure based on the product\cite{nepusz2008fuzzy}\cite{shen2009quantifying} or the average\cite{zhang2007identification} of ``belonging'' coefficients (i.e. how much a node really belongs to a community); to simply calculating the average modularity of all communities\cite{lazar2010modularity}; to a version incorporating both overlap \textit{and} directed edges\cite{nicosia2009extending}.

Mentioning ``belonging'' coefficients allows me to make a distinction here. You can perform overlapping community discovery in two different ways. The first is by saying that nodes fully belong to multiple communities -- i.e. that all the communities they belong to are equal for them. There is no way to say that a node is ``more'' part of one community or another. This is in contrast with fuzzy clustering\cite{bezdek2013pattern}, in which nodes have such belonging coefficients and thus can tell you whether they really feel like they're strongly part of a community or not.

\begin{figure}
\centering
\includegraphics[width=.66\columnwidth]{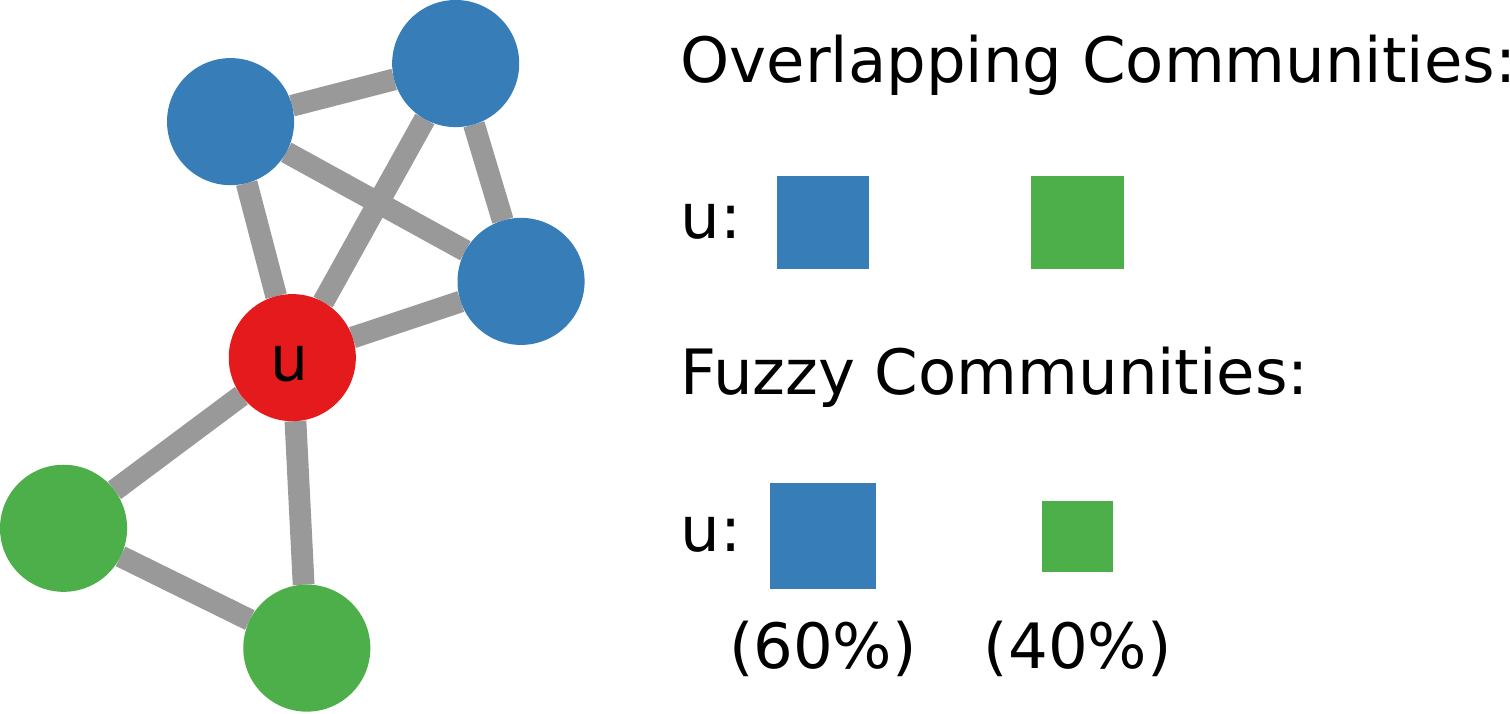}
\caption{Comparing overlapping and fuzzy clustering for node $u$: the size of the square is proportional to $u$'s ``belonging'' coefficient, in share of number of $u$'s edges connected to the community.}
\label{fig:ocd-vs-fuzzy}
\end{figure}

Figure \ref{fig:ocd-vs-fuzzy} depicts this distinction. Fuzzy communities are more difficult to calculate, but they are also more precise. However, you should be careful in not relying on the coefficients you get from fuzzy clustering too much. Is there really a difference in saying that $49\%$ of a node ``belongs'' to a community, versus saying that $51\%$ of the node belongs to it? In some cases it might be, but you need to have trust in the fact that your data allows you such level of confidence.

\section{Adapted Approaches}
On the basis of having an overlapping version of modularity, overlapping community discovery needs not to be a separated problem with specialized solutions. We already have delineated a procedure to solve the problem: trying to maximize a target function. So we can take all the algorithms which maximize modularity, and make them maximize overlapping modularity instead.

This move can be applied to multiple other algorithms. For instance, we have an adaptation of Infomap\cite{esquivel2011compression}. If you remember Section \ref{sec:cd-partition-rw}, in Infomap we're looking for a smart way to encode random walks. We do so by using short binary codes to identify nodes inside modules and special codes to indicate when the random walk crosses between communities.

\begin{figure}[t]
\centering
\includegraphics[width=.66\columnwidth]{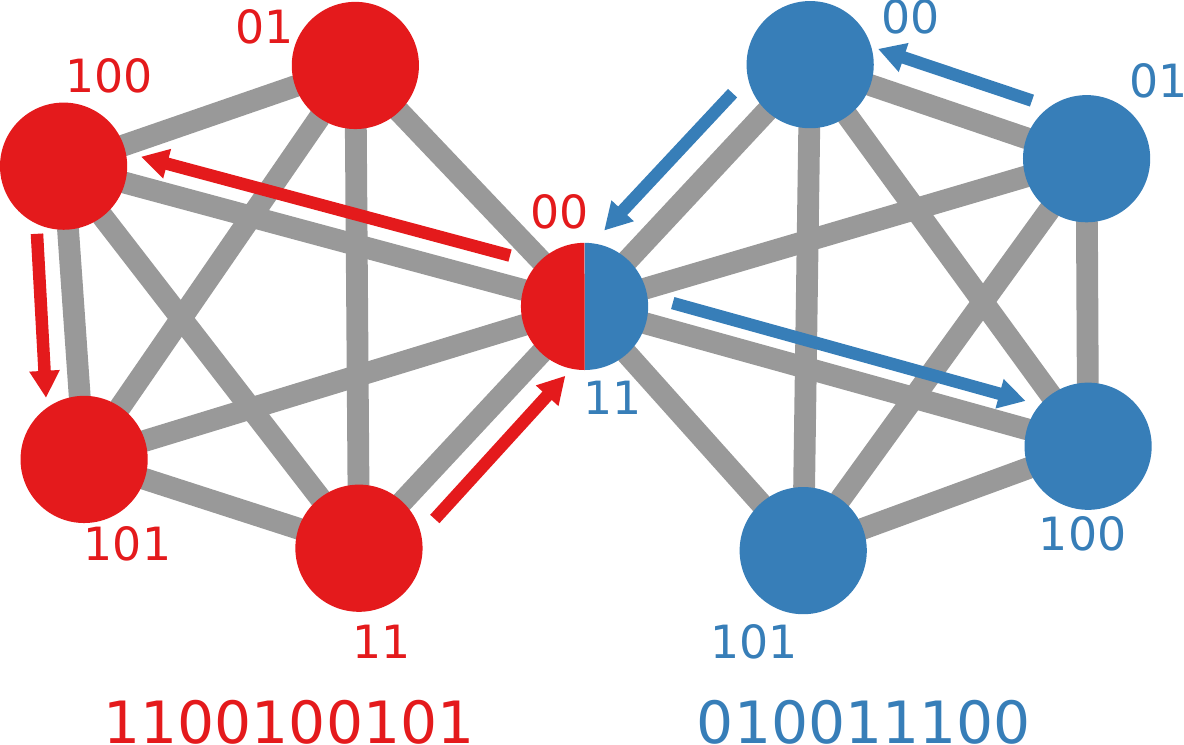}
\caption{The encoding of two random walks in the overlapping version of Infomap. Note that in neither the red nor the blue path we're crossing community boundaries, so we don't use the community crossing code.}
\label{fig:ocd-infomap}
\end{figure}

In such strategy, the communities are disjoint, because if a node is part of two communities you would have to use the code for crossing between communities when you visit it. However, there is a way to make this encoding compatible with overlapping communities. That is giving to a node part of multiple communities a different code per community. Figure \ref{fig:ocd-infomap} shows an example.

For instance, if the Reykjavik airport is part of both the North American and the European community, it will have two codes. We would use one code if the random walk approaches Reykjavik from a North American airport, and the other code if it approaches Reykjavik from a European one. You would then use the community crossing code only if you actually transition between clusters in the next step.

Finally, let's consider label propagation. There is a way to extend the classical label propagation algorithm to allow for overlapping communities\cite{gregory2010finding}. The idea is that nodes will not just adopt the single most common label among their neighbors. They will adopt all labels, each weighted by a ``belonging coefficient'', which is the weighted average of the belonging coefficient of that label across all neighbors.

Labels that are below a specific belonging coefficient threshold are removed from the node at each step, preventing the nodes to converge to a single global status where all nodes belong to all communities at the same level of belonging.

\begin{figure}
\centering
\begin{subfigure}{.2\columnwidth}
\includegraphics[width=\textwidth]{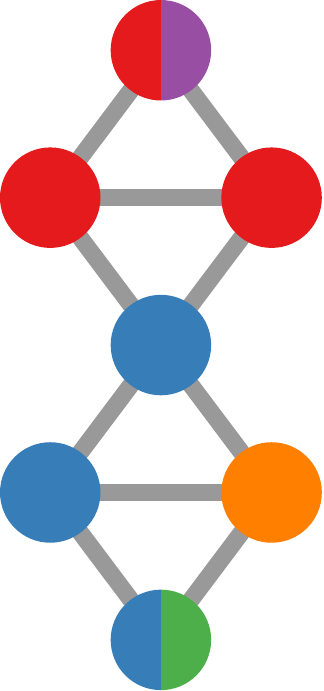}
\caption{}
\end{subfigure}\qquad
\begin{subfigure}{.2\columnwidth}
\includegraphics[width=\textwidth]{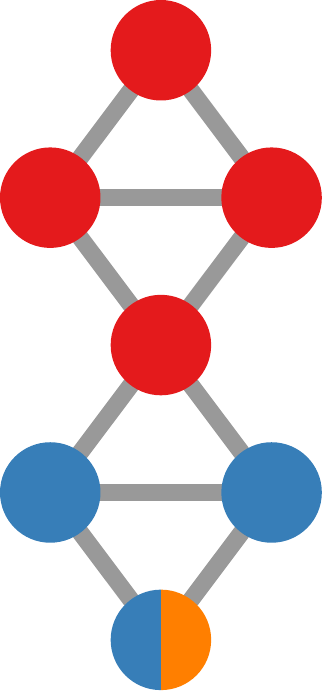}
\caption{}
\end{subfigure}\qquad
\begin{subfigure}{.2\columnwidth}
\includegraphics[width=\textwidth]{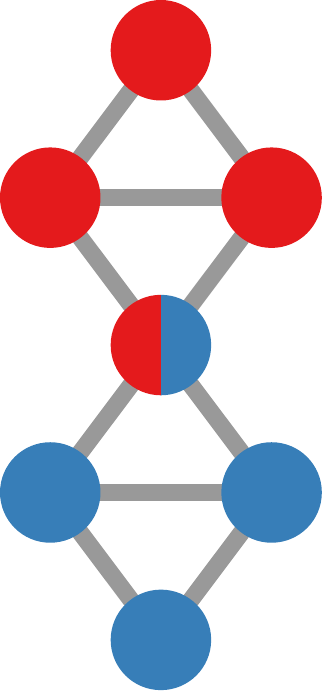}
\caption{}
\end{subfigure}
\caption{A simple run of the overlapping label propagation. The node's color is a pie chart representing the belonging coefficient of the node to a label.}
\label{fig:ocd-lp}
\end{figure}

Figure \ref{fig:ocd-lp} shows a small run of this principle. Let's suppose that we set our minimum belonging coefficient to $0.5$. In the first step (from Figure \ref{fig:ocd-lp}(a) to \ref{fig:ocd-lp}(b)), the two fully red nodes stay fully red, because they both receive $0.5$ of the red label, $0.33$ of the blue and $0.16$ of the purple label. The only label clearing the $0.5$ threshold is the red one, thus they become fully red. The half red half purple node becomes red because that's the only label around it. The central node is also red, receiving $0.5$ red, $0.25$ blue and $0.25$ orange. From Figure \ref{fig:ocd-lp}(b) to \ref{fig:ocd-lp}(c) though, the central node will correctly split between red and blue, because they both contribute half of its neighborhood.

\section{Explicit Structural Approaches}\label{sec:ocd-struct}
In the class of structural approaches we find methods that have a definition of what an overlapping community should look like, and try to find such a structure in the network. The idea is different from modularity maximization, because it is not primarily driven by the optimization of a function. I add the word ``explicit'' because these approaches exclusively look at the structure as it is, and try to find the communities there. In the next section we'll see ``latent'' structural approaches which assume that the observed structure is the result of a hidden one driving the connections.

The explicit structural approach is historically the oldest solution to overlapping community discovery. I can think of fundamentally two subclasses: the famous clique percolation approach, and node splitting.

\subsection{Clique Percolation}
Clique percolation starts from the observation that communities should be dense. What is the densest possible subgraph? The clique. In a clique, all nodes are connected to all other nodes. So the problem of community discovery more or less reduces to the problem of finding all cliques in the network. However, this is a bit too strict: there are subgraphs in the network that, while being very dense and close to being a clique, are not fully connected. It would be a pity to split them into many small substructures.

Thus researchers developed the more sophisticated $k$-clique percolation algorithm\cite{derenyi2005clique}. Clique percolation says that communities must be cliques of at least $k$ nodes, with $k$ being a parameter you can freely set. In the first step, the algorithm finds all cliques of size $k$, whether they are maximal or not. Then, it attempts to merge two communities in the same community if the two communities share at least a $k-1$ clique.

\begin{figure}
\centering
\begin{subfigure}{.33\columnwidth}
\includegraphics[width=\textwidth]{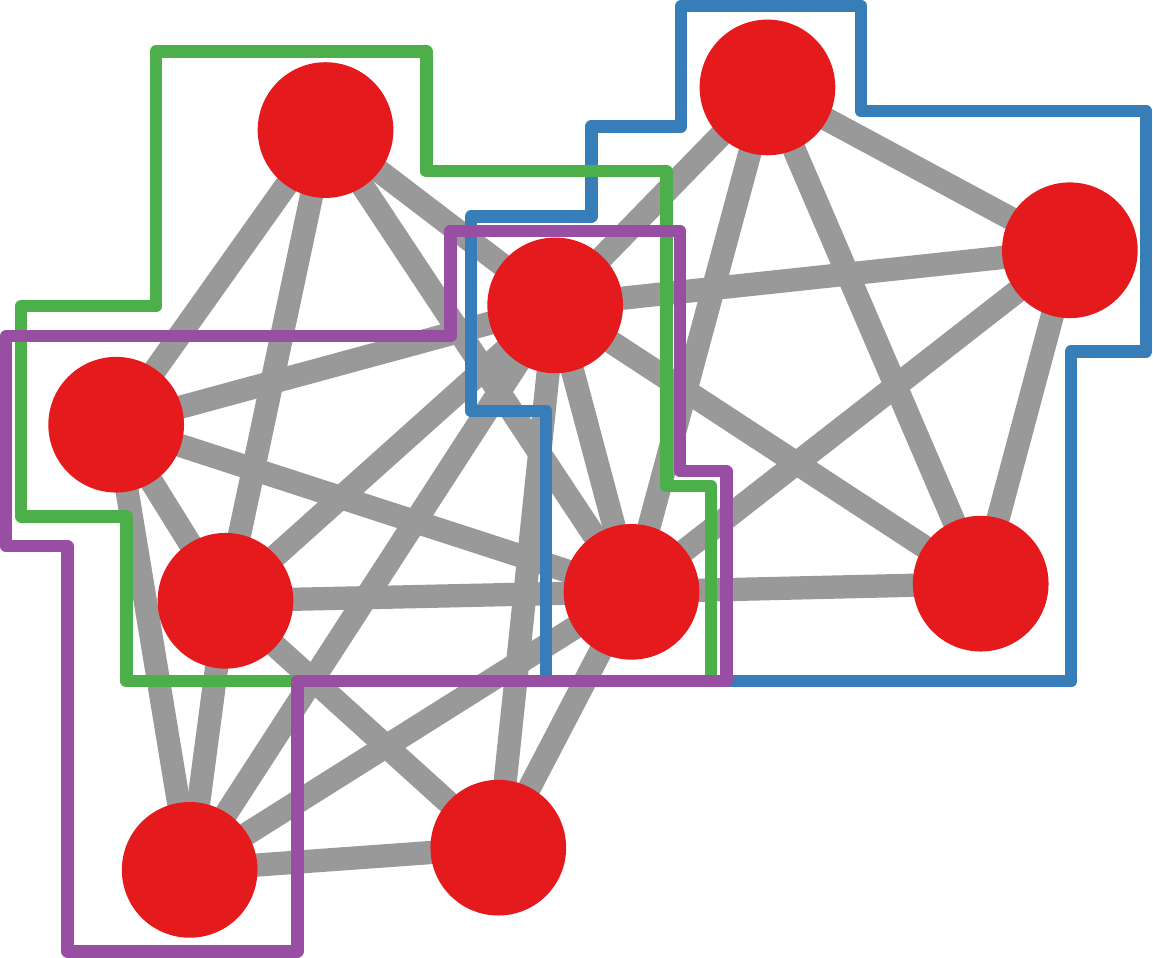}
\caption{}
\end{subfigure}\qquad
\begin{subfigure}{.33\columnwidth}
\includegraphics[width=\textwidth]{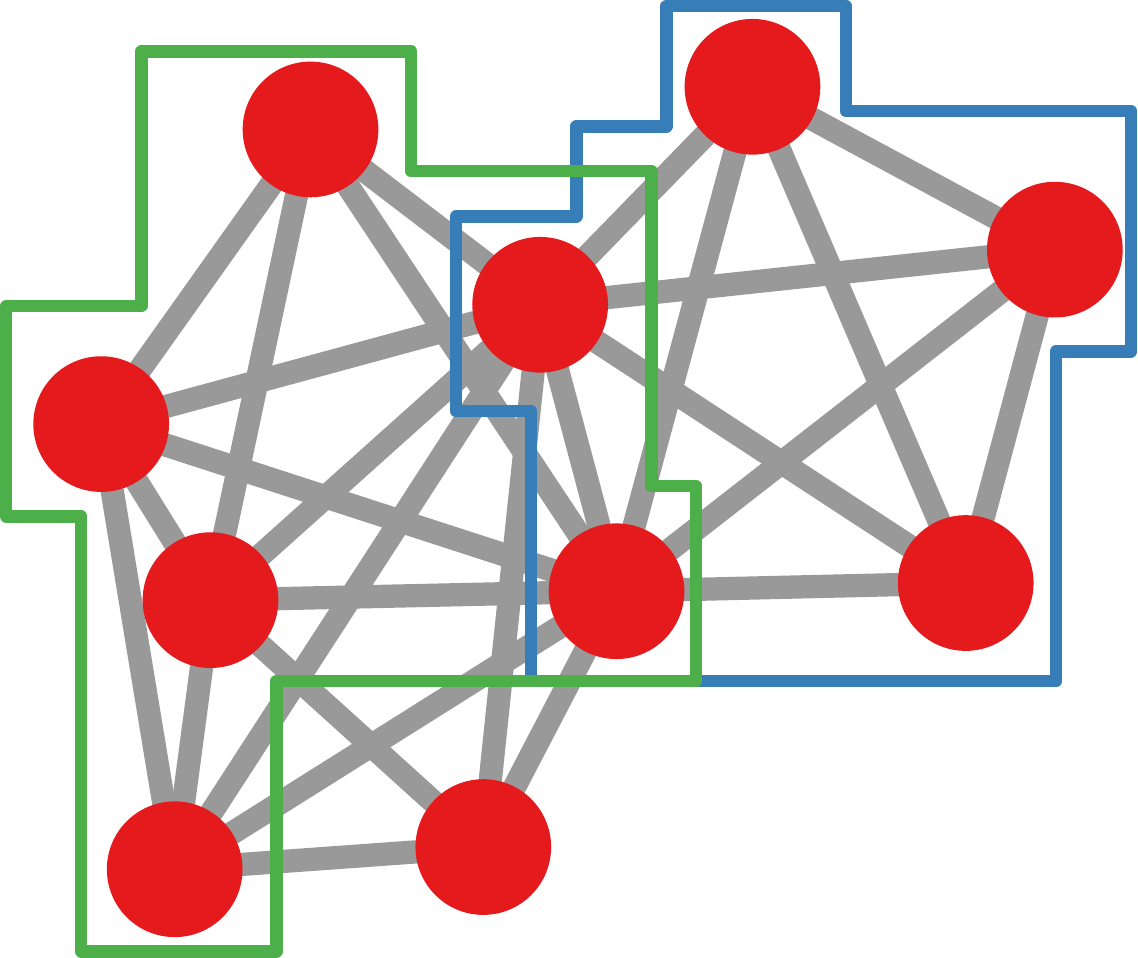}
\caption{}
\end{subfigure}\qquad
\begin{subfigure}{.33\columnwidth}
\includegraphics[width=\textwidth]{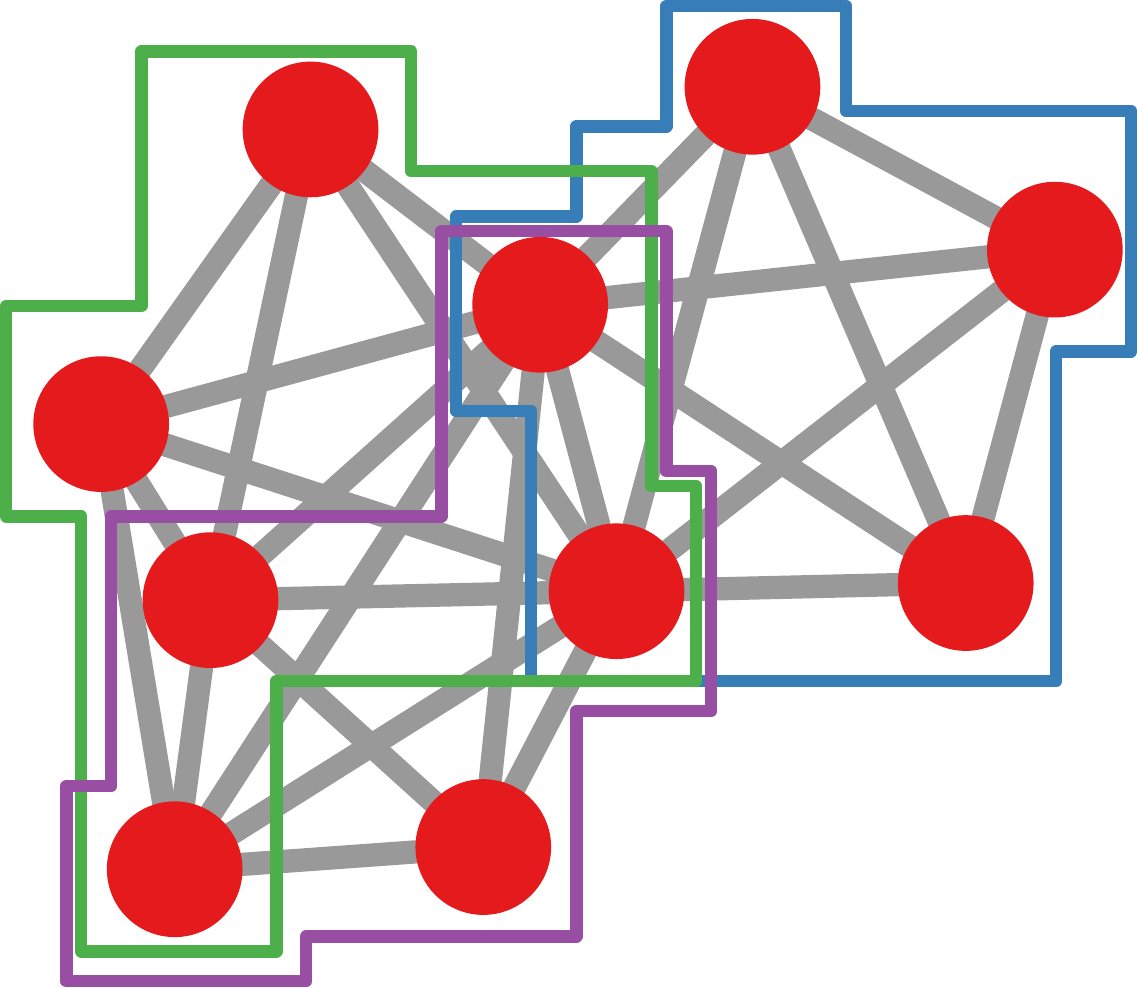}
\caption{}
\end{subfigure}\qquad
\begin{subfigure}{.33\columnwidth}
\includegraphics[width=\textwidth]{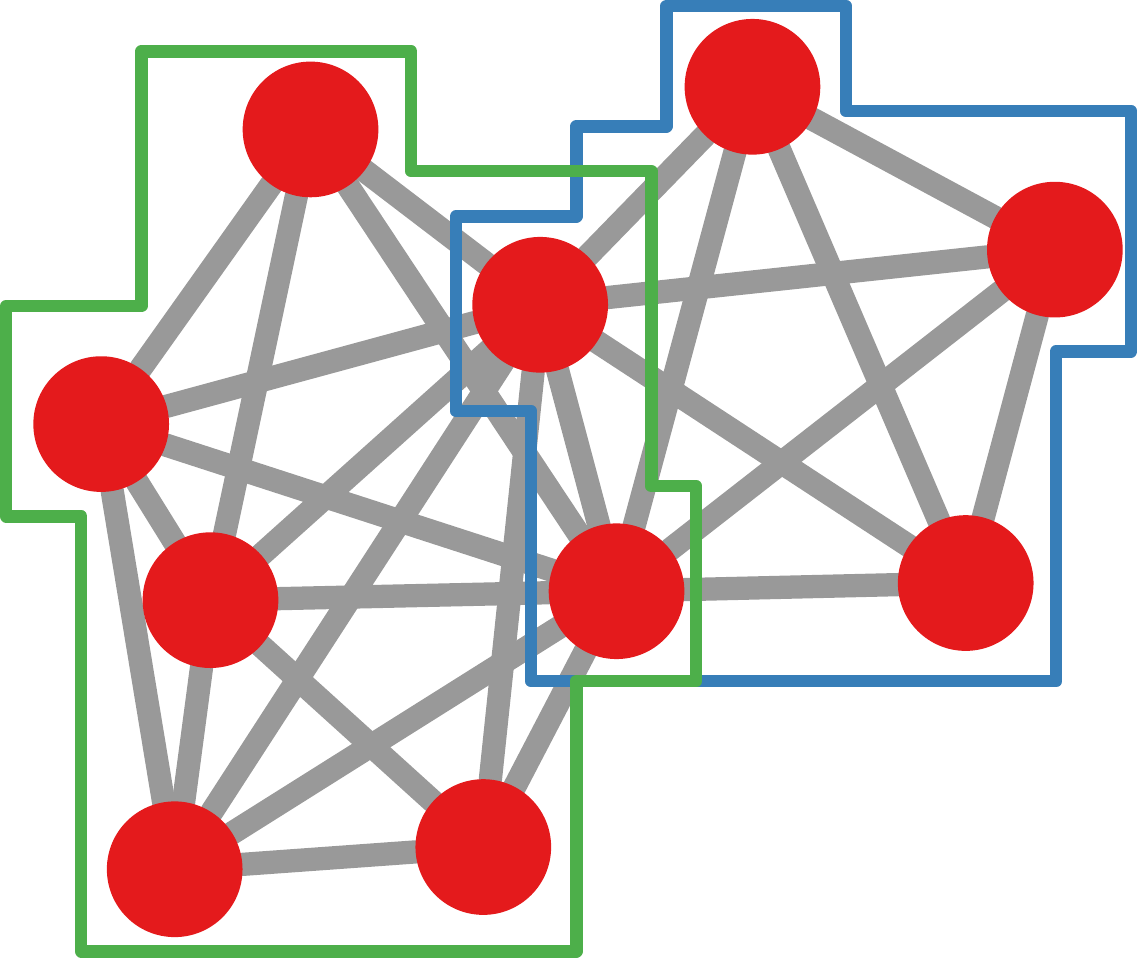}
\caption{}
\end{subfigure}
\caption{An example of clique percolation. 5-Cliques are highlighted by outlines. Green cliques percolate with purple cliques because they share $k-1 = 4$ nodes.}
\label{fig:cliqueperc}
\end{figure}

For instance, consider the example in Figure \ref{fig:cliqueperc}, setting the parameter $k = 5$. The blue and green 5-cliques only share two nodes, so it cannot be a 4-clique. But the green and purple do share a 4-clique, so they are merged (top row). And there is another purple 5-clique that can now be merged with the green community (bottom row).

This is generally implemented via the creation of a clique graph\cite{evans2010clique}. The nodes of a clique graph are the cliques in the original graph. We connect two cliques if they share nodes. For instance, if we only connect cliques sharing $k -1$ nodes, then we can efficiently find all communities by finding all connected components in the clique graph.

This algorithm works well in practice. It has been used to study overlapping friendship patterns in school systems\cite{gonzalez2007community} -- due to classroom being quasi-cliques: pupils have rare but significant friendships across classes --, and in metabolic networks\cite{zhang2006identification}. However, it has a couple of downsides.

First, finding all cliques in a network is computationally expensive. One could fix this problem by setting $k$ to be relatively high. If we set $k = 5$ we know that nodes with degree three or less cannot be in any community, because they need at least four edges to be part of a 5-clique. Since most networks have broad degree distributions (Section \ref{sec:degree-pl}), this means that we can safely ignore the vast majority of the network, thus reducing the number of operations we need to find communities. This is a suboptimal solution because it implies that one will not classify most nodes into communities. For this reason, there are developments of this algorithm\cite{kumpula2008sequential}\cite{reid2012percolation} that are a bit more computationally efficient.

Second, it has limited coverage for sparse networks. That means that it might end up being unable to classify nodes in networks because they are not part of any clique. If you set your $k$ relatively low, e.g. $k = 4$, all nodes with degree equal to one cannot be part of any community. This is because a node with degree equal to one cannot be part of a 3-clique. Thus it will never be merged into any 4-clique, which are the basis of our communities.

\subsection{Node Splitting}
Another approach is to simply recognize that a node is part of multiple communities if it has different identities. This is extremely similar to the approach of overlapping Infomap. In that case we represented the two identities of the node by giving it two different codes: one per community to which it belongs. Here we literally split it in two. We modify the structure of the network in such a way that, when we are done, by performing a normal non-overlapping community discovery we recover the overlapping clusters. In the resulting structure we have multiple nodes all referring to the same original one.

If we want to split nodes, we need to answer two questions: which nodes do we split and how. First we identify the nodes most likely to be in between communities. If you remember the definition of betweenness, you'll recollect that nodes between communities are the gatekeepers of all shortest paths from one community to the other. So they are the best candidates to split. There are many ways to perform the split, but I'll focus on the one that involves calculating a special betweenness: pair betweenness\cite[-0.7in]{gregory2007algorithm}\cite{gregory2009finding}. Pair betweenness is a measure for a pair of edges: the number of shortest paths that use both of them.

\begin{figure}
\centering
\begin{subfigure}{.4\columnwidth}
\includegraphics[width=\textwidth]{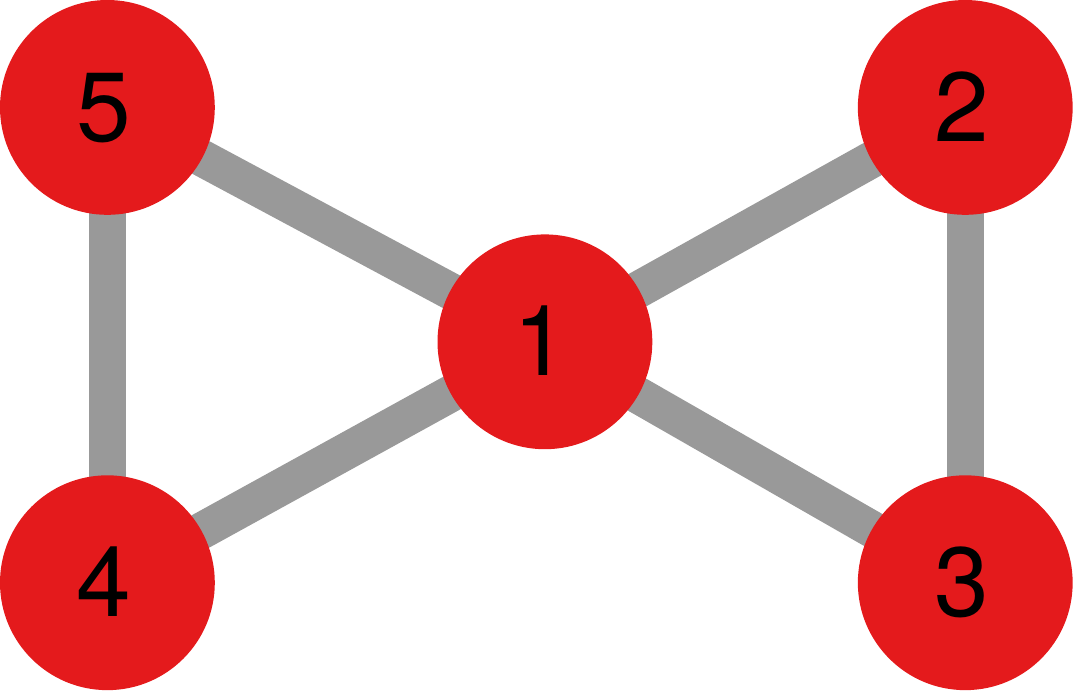}
\caption{}
\end{subfigure}\qquad
\begin{subfigure}{.4\columnwidth}
\includegraphics[width=\textwidth]{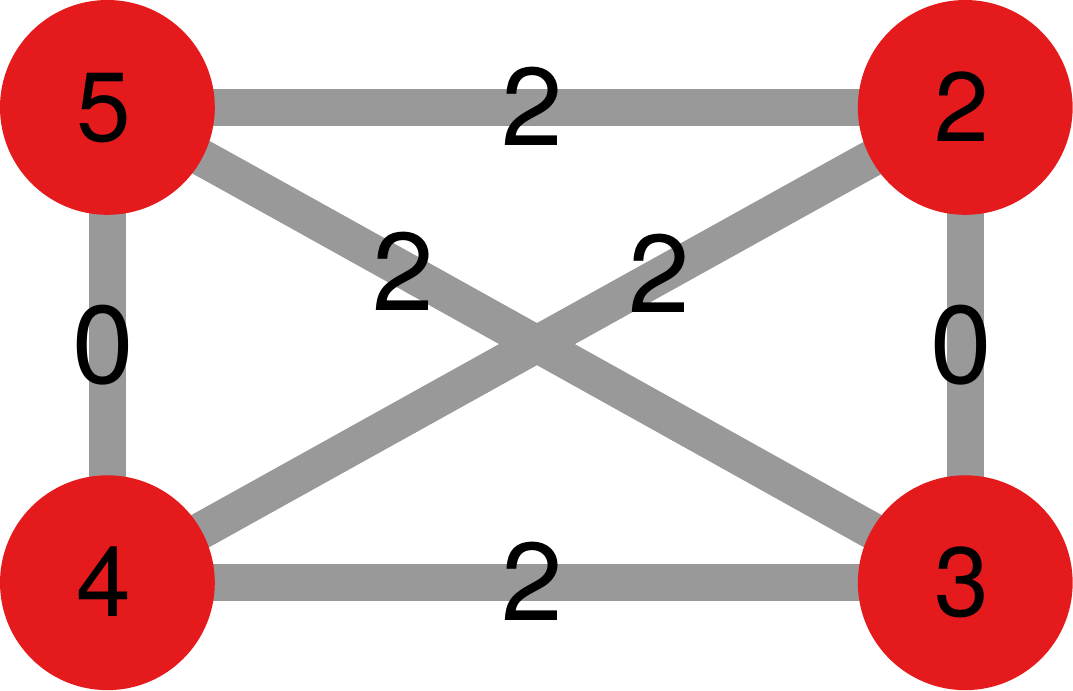}
\caption{}
\end{subfigure}
\caption{Attempting to find the best node to split in (a). Selecting node $1$ as the candidate, we build a pair betweenness graph (b). I label each edge in (b) with its pair betweenness.}
\label{fig:conga1}
\end{figure}

For instance, consider the graph in Figure \ref{fig:conga1}(a). The most central node is node $1$. To try and split it, we build its split graph. Meaning that we remove node $1$ and we connect all nodes that were connected by $1$. Each edge has a weight: the number of shortest paths in the original graph that passed through node $1$. In this case, there are two shortest paths using the $(4,1)$ and $(1,3)$ edges: the one going from node $4$ to node $3$ and the one going from node $3$ to node $4$. We can represent the pair betweenness of all neighbors of node $1$ with a weighted clique (Figure \ref{fig:conga1}(b)). 

\begin{figure*}[t]
\centering
\begin{subfigure}{.3\columnwidth}
\includegraphics[width=\textwidth]{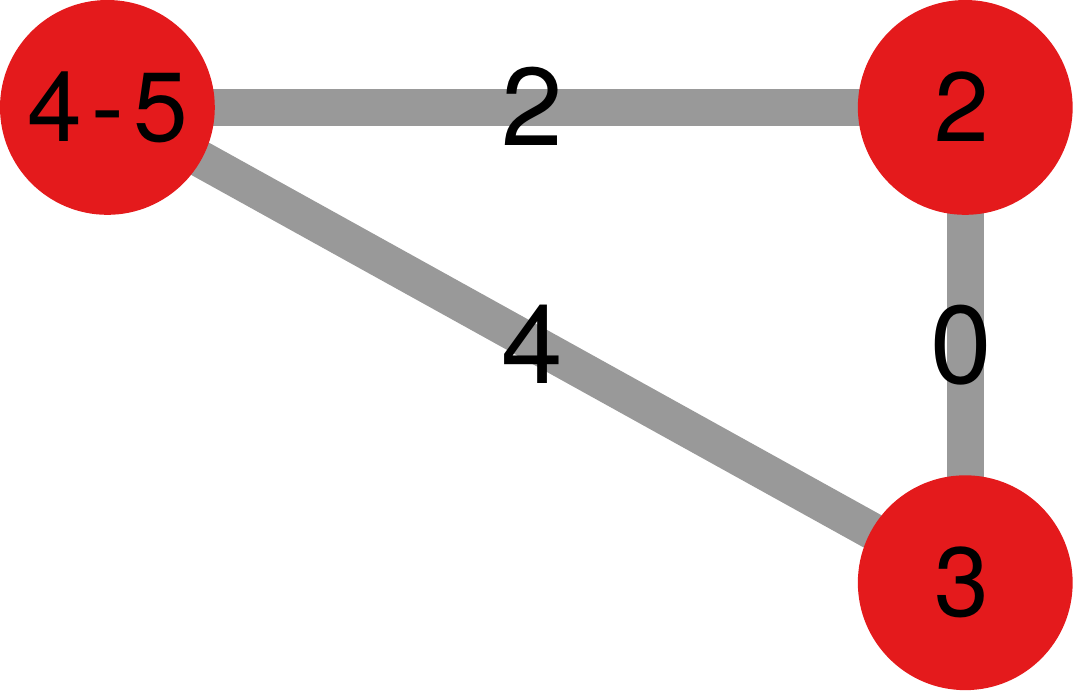}
\caption{}
\end{subfigure}\qquad
\begin{subfigure}{.3\columnwidth}
\includegraphics[width=\textwidth]{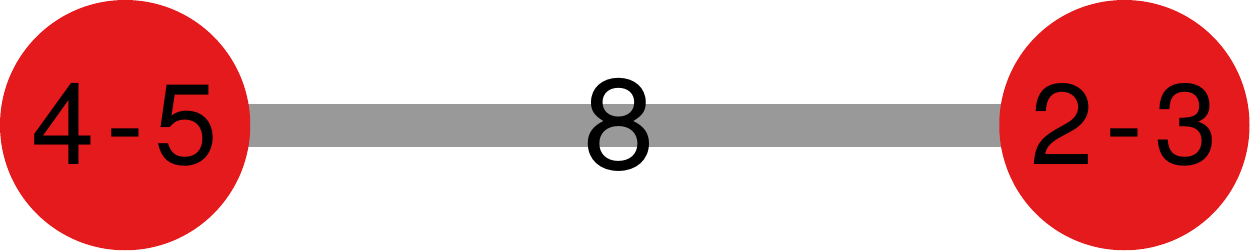}
\caption{}
\end{subfigure}\qquad
\begin{subfigure}{.3\columnwidth}
\includegraphics[width=\textwidth]{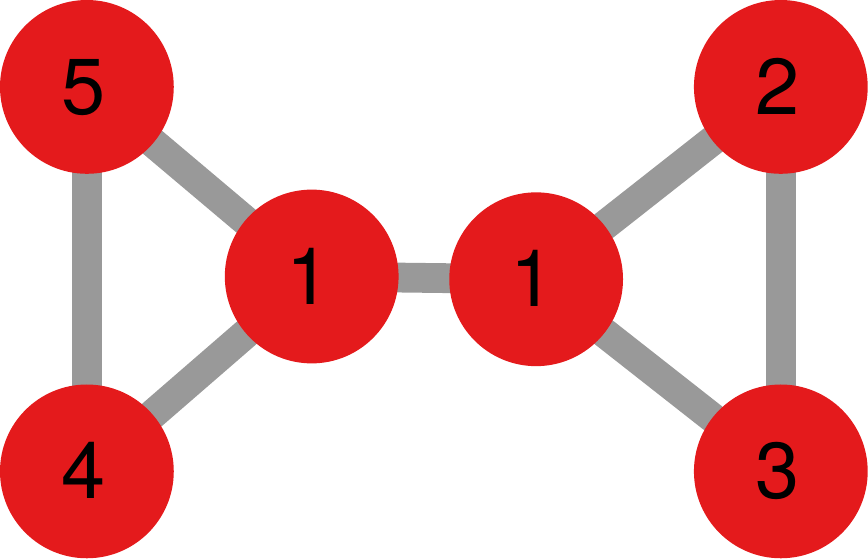}
\caption{}
\end{subfigure}
\caption{(a-b) Merging the nodes connected by the weakest pair betweenness edges. (c) The resulting split in the original graph.}
\label{fig:conga2}
\end{figure*}

To find the split we use a simple algorithm: we identify the edges with the lowest pair betweenness and we merge the nodes connected by those edges (Figures \ref{fig:conga2}(a-b)). At each merge, we sum up the pair betweennesses of all edges that got merged together by the merging of the node. Once we have one remaining edge, the resulting split is the best one. The reason is that edges with low pair betweenness are likely to be in the same community. Once you identify the split (Figure \ref{fig:conga2}(c)), it is easy to find disjoint communities and then merge them into overlapping.

\section{Latent Structural Approaches}\label{sec:ocd-mmsbm}
In this section we have a collection of methods that make an assumption: the observed community division of the network is the result of a latent structure. In this latent structure, we have nodes assigned to communities. Then, the probability of observing an edge between two nodes is proportional to the number of communities the two nodes have in common in the latent structure. The two classes of approaches in this category I consider are Mixed Membership Stochastic Blockmodels (MMSB) and the community affiliation graph.

\subsection{Mixed Membership Stochastic Blockmodels}
The description of the latent structural approaches I just wrote should turn on a light bulb in your head. The idea that the probability of connecting two nodes is dependent on a latent community partition is not new: it is exactly the starting point of the stochastic blockmodel approach. In SBM, we assume there is a partition of nodes and we assign a higher probability of connection between nodes in the same partition than the one between nodes in different partitions. The little problem we need to solve now is how to make this mathematical machinery work when we want to allow nodes to be part of multiple communities. That solution constitutes the Mixed Membership Stochastic Blockmodels\cite{airoldi2008mixed}, an object that I already mentioned in Section \ref{sec:csmodels-comms}.

The trick here is that we represent each node's membership as a vector. The vector tells us how much the node belongs to a given community. Then, we also have a community-community matrix, that tells us the probability of a node belonging to community $c_1$ to connect to a node belonging to a community $c_2$. These are the two ingredients that replace the simple community partition in the regular SBM. From this moment on, you attempt to find the set of community affiliation vectors and the community-community probability matrix that are most likely to reproduce your observed data, exactly as you do in SBM.

Just like we saw in Section \ref{sec:cd-partition-evo}, we can have dynamic MMSB, adding time to the mix\cite[-1.8in]{fu2009dynamic}\cite[-0.9in]{xing2010state}\cite[-0.2in]{ho2011evolving}\cite{xu2013dynamic}: the community affiliation vectors and the community-community matrix can change over time. There is also a hierarchical (Chapter \ref{cha:hcd}) variant of MMSB, allowing a nested community structure\cite{sweet2014hierarchical}.

\subsection{Community Affiliation Graph}
Affiliations graphs have been often used to describe the overlapping community structure of real world networks\cite{noh2005growing}. In a community affiliation graph you assume that you can describe your observed network with a latent bipartite network. In this bipartite network, the nodes of one type are the nodes of your observed network. The other type, the latent nodes, represent your communities. Nodes are connected to the communities they belong to. This is the community affiliation graph, because it describes the affiliations to communities of your nodes.

\begin{figure}
\centering
\begin{subfigure}{.35\columnwidth}
\includegraphics[width=\textwidth]{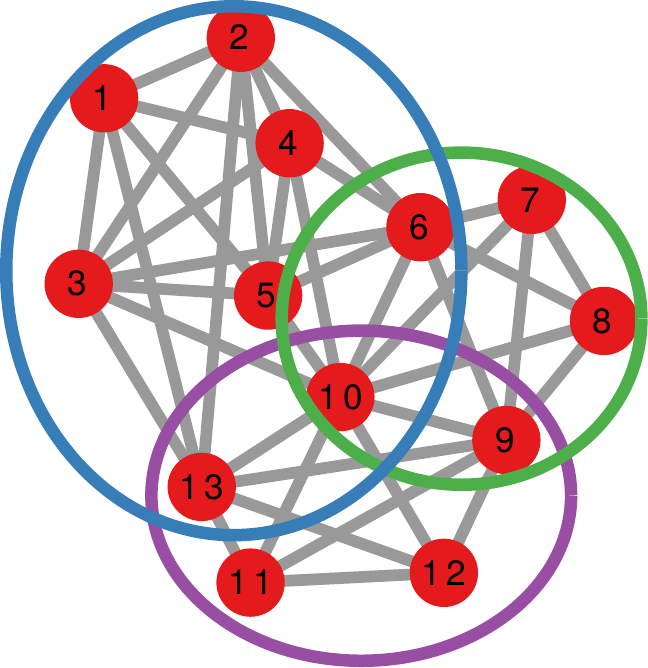}
\caption{}
\end{subfigure}\qquad
\begin{subfigure}{.55\columnwidth}
\includegraphics[width=\textwidth]{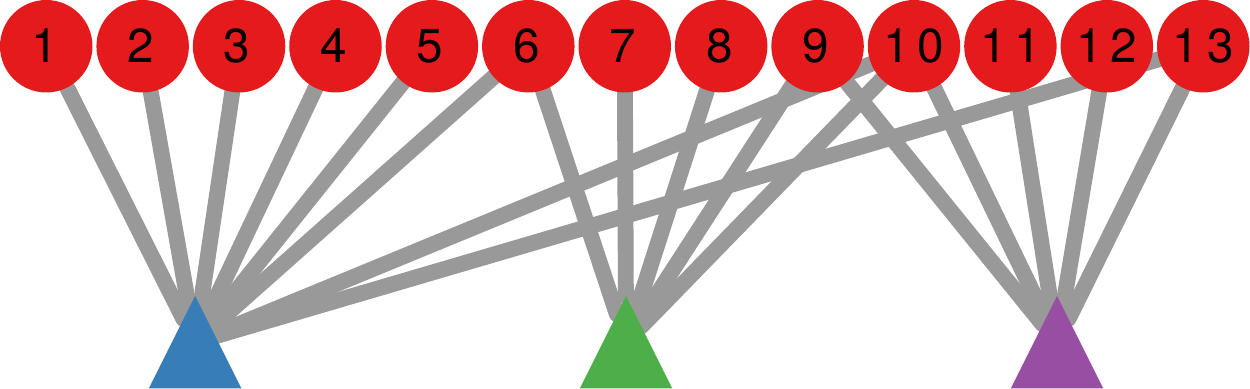}
\caption{}
\end{subfigure}
\caption{(a) A graph with overlapping communities indicated by the colored outlines. (b) Its corresponding community affiliation graph. The community latent nodes are triangular and their color corresponds to the color used in (a).}
\label{fig:bigclam}
\end{figure}

Figure \ref{fig:bigclam} shows a representation of a community affiliation graph. Of course, you can build such a graph easily once you already know to which communities the nodes belong. The hard part is finding out the best representation. There are a few ways to do so, usually relying on the expectation maximization algorithm that is also at the basis of the MMSB. One such approach is BigClam\cite{yang2013overlapping}, which uses non-negative matrix factorization as the guiding principle (see Section \ref{sec:mat-factors} for a refresher).

\section{Clustering Links}\label{sec:ocd-link}
An alternative approach is to look at edges. Why looking at edges? Because people can be in multiple communities, as we saw: work mates, school mates, etc. However, a link usually is created for one single reason: you met that person in one situation and that's the defining characteristic of your relationship. You originally met $u$ as a work colleague, and $v$ as a school mate. So one can cluster links with a non-overlapping method and say that a person is part of all the communities to which their edges belong.

There can be many similarity measures and link communities can be found by adapting and optimizing such functions to the link community case.

\subsection{Line Graphs}\label{sec:ocd-link-linegraph}
To cluster the edges rather than the nodes we can transform the network into its corresponding line graph\cite{evans2009line}. In a line graph, as we saw in Section \ref{sec:basic-simple}, the edges become nodes and they are connected if they're incident on the same node. A way to do so is to generate a weighted line graph.

\begin{figure}
\centering
\begin{subfigure}{.33\columnwidth}
\includegraphics[width=\textwidth]{figures/conga1.pdf}
\caption{}
\end{subfigure}\qquad
\begin{subfigure}{.45\columnwidth}
\includegraphics[width=\textwidth]{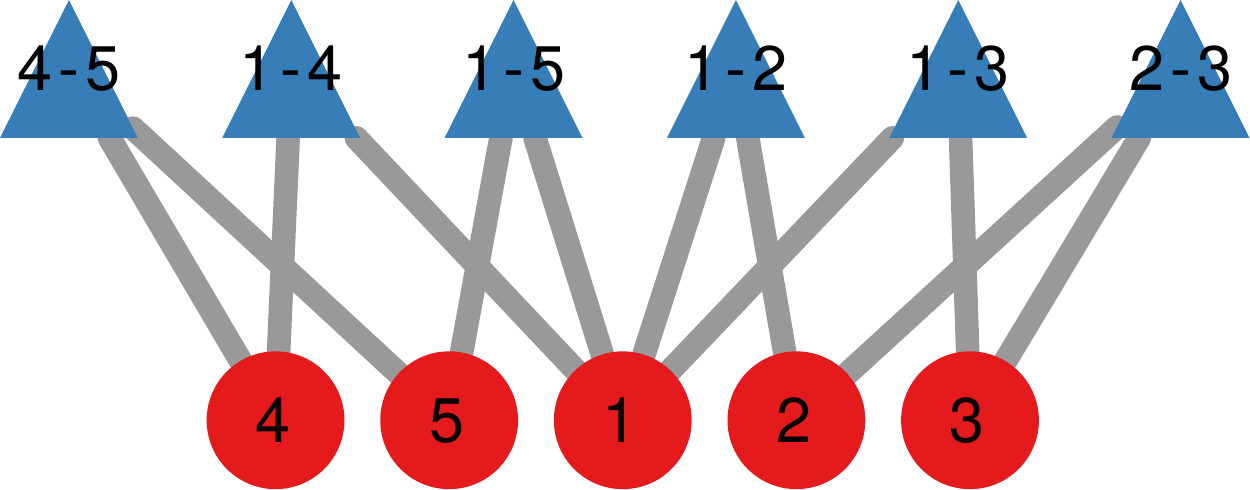}
\caption{}
\end{subfigure}
\caption{(a) A simple graph. (b) A bipartite version of (a) connecting each node to its edges.}
\label{fig:ocd-linegraph1}
\end{figure}

To create a line graph you first transform the network into bipartite connecting the nodes to the edges they are connected to, as Figure \ref{fig:ocd-linegraph1} shows. Then you project this network over the edges. The most important thing to define is how to weight the edges in the line graph. Different weight profiles will steer the community discovery on the line graph in different directions.

You could use any of the weighting schemes I discussed in Chapter \ref{cha:projections}, but the researchers proposing this method also have their suggestions. The reason you might need a special projection is because you want nodes that are part of an overlap to give their edges lower weights, because their connections are spread out in different communities.

\begin{figure}
\centering
\begin{subfigure}{.4\columnwidth}
\includegraphics[width=\textwidth]{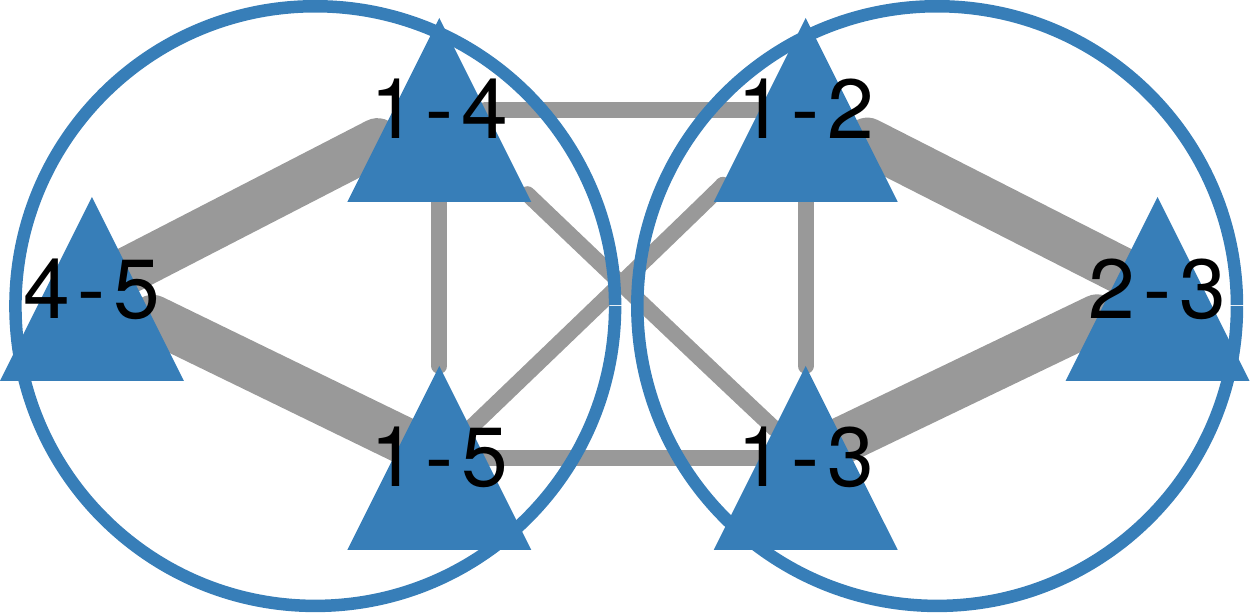}
\caption{}
\end{subfigure}\qquad
\begin{subfigure}{.33\columnwidth}
\includegraphics[width=\textwidth]{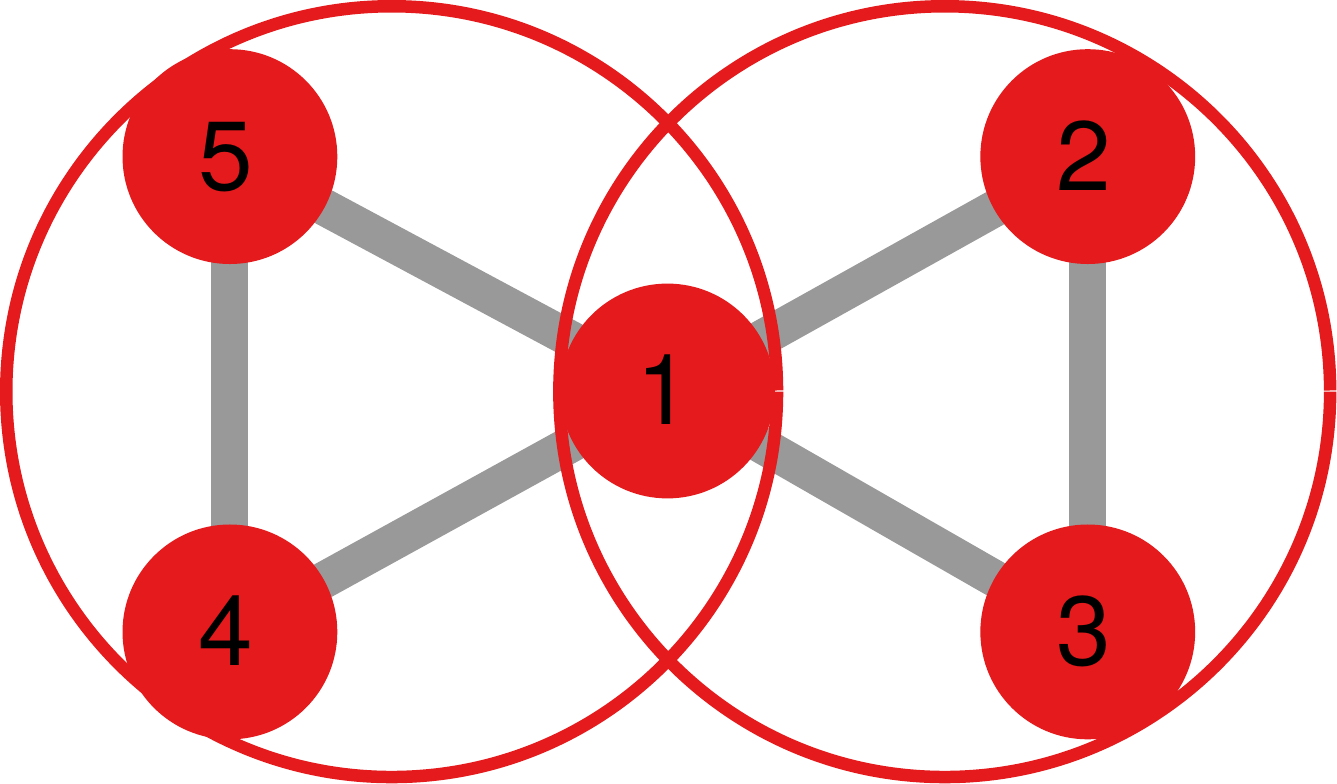}
\caption{}
\end{subfigure}
\caption{(a) The blue circles are disjoint communities in the line graph. (b) The red circles are the corresponding overlapping communities in the original graph.}
\label{fig:ocd-linegraph2}
\end{figure}

At that point, a disjoint community discovery will downplay the weak edges and find communities with strong edge weights, as Figure \ref{fig:ocd-linegraph2}(a) shows. Once we bring back the communities to the other side of the projection, as I do in Figure \ref{fig:ocd-linegraph2}(b), we have overlapping ones.

Other approaches in this class exist, including random walks on line graphs\cite{deng2015finding}.

\subsection{Hierarchical Link Clustering}
In Hierarchical Link Clustering\cite{ahn2010link} (HLC), the first step is to calculate a measure of similarity between two edges. We only calculate it for edges sharing one node: edges $(u,k)$ and $(v,k)$ share node $k$. If two edges share no node, their similarity is zero. If the edges share a node, their similarity is the Jaccard coefficient of the neighborhoods of the two non-shared nodes:

$$ S_{(u, k), (v, k)} = \dfrac{|N_u \cap N_v|}{|N_u \cup N_v|}.$$

\begin{figure}
\centering
\includegraphics[width=.45\columnwidth]{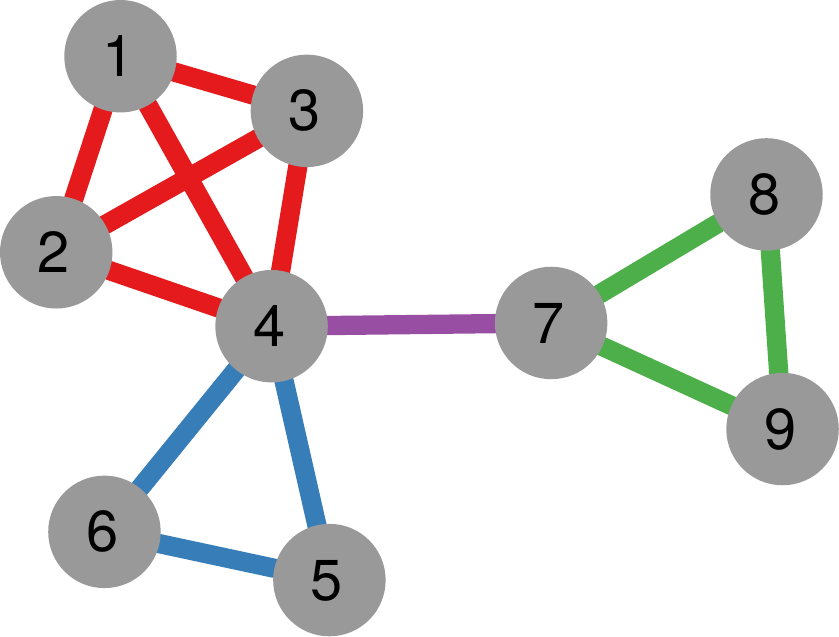}
\caption{A graph and its best link communities. The color of the edge represents its community. Nodes are part of all communities of their links. For instance, node $4$ belongs to three communities: red, blue, and purple.}
\label{fig:hlc}
\end{figure}

The edges with the highest $S$ value are merged in the same community. For instance, in Figure \ref{fig:hlc}, edges $(1, 2)$ and $(1, 3)$ have a high $S$ value: the neighborhoods of nodes $2$ and $3$ are identical, thus $S_{(1, 2), (1, 3)} = 1$. On the other hand, edges $(4,7)$ and $(7,8)$ only have one node in the numerator, thus: $S_{(4, 7), (7, 8)} = 1 / 6$.

Then, the merging happens recursively for lower and lower $S$ values, building a full dendrogram, as we saw in Chapter \ref{cha:hcd} for hierarchical community discovery. We then need a criterion to cut the dendrogram. We cannot use modularity, because these are link communities, not node communities.

The original authors develop a new quality measure called ``partition density''. For each link community $c$, we have $|E_c|$ as the number of edges it contains, and $|V_c|$ as the number of nodes connected to those edges. Its density $D_c$ is

$$ D_c = \dfrac{|E_c| - (|V_c| - 1)}{|V_c|(|V_c| - 1) / 2  - (|V_c| - 1)},$$

which is the number of links in $c$, normalized by the maximum number of links possible between those nodes ($|V_c|(|V_c| - 1) / 2$), \textit{and} its minimum $|V_c| - 1$, since we assume that the subgraph induced by $c$ is connected. Note that, if $|V_c| = 2$, we simply take $D_c = 0$. All $D_c$ scores for all $c$s in your link partition are aggregated to find the final partition density, which is the average of $D_c$ weighted by how many links are in $c$: $D = \dfrac{1}{|E|} \sum \limits_c |E_c| D_c$.

\subsection{Ego Networks}
Assuming that links exists for one primary reason works usually well, but it is a problematic assumption. Let's look back at the case of work and school communities. What would happen if you were to end up working in the same company and play in the same team of a former schoolmate? Is it still fair to say that the link between the two of you exists for only one predominant reason?

Modeling truly overlapping communities can get rid of this problem. There are many ways to do it, but we'll focus on one that is easy to understand. The starting observation is that networks have large and messy overlaps. However, just like in the assumption of clustering links, here we realize that the neighbors of a node usually are easier to analyze. It is easy for a node to look at a neighbor and say: ``I know this other node for this reason (or set of reasons)''.

\begin{figure}[t]
\centering
\includegraphics[width=.8\columnwidth]{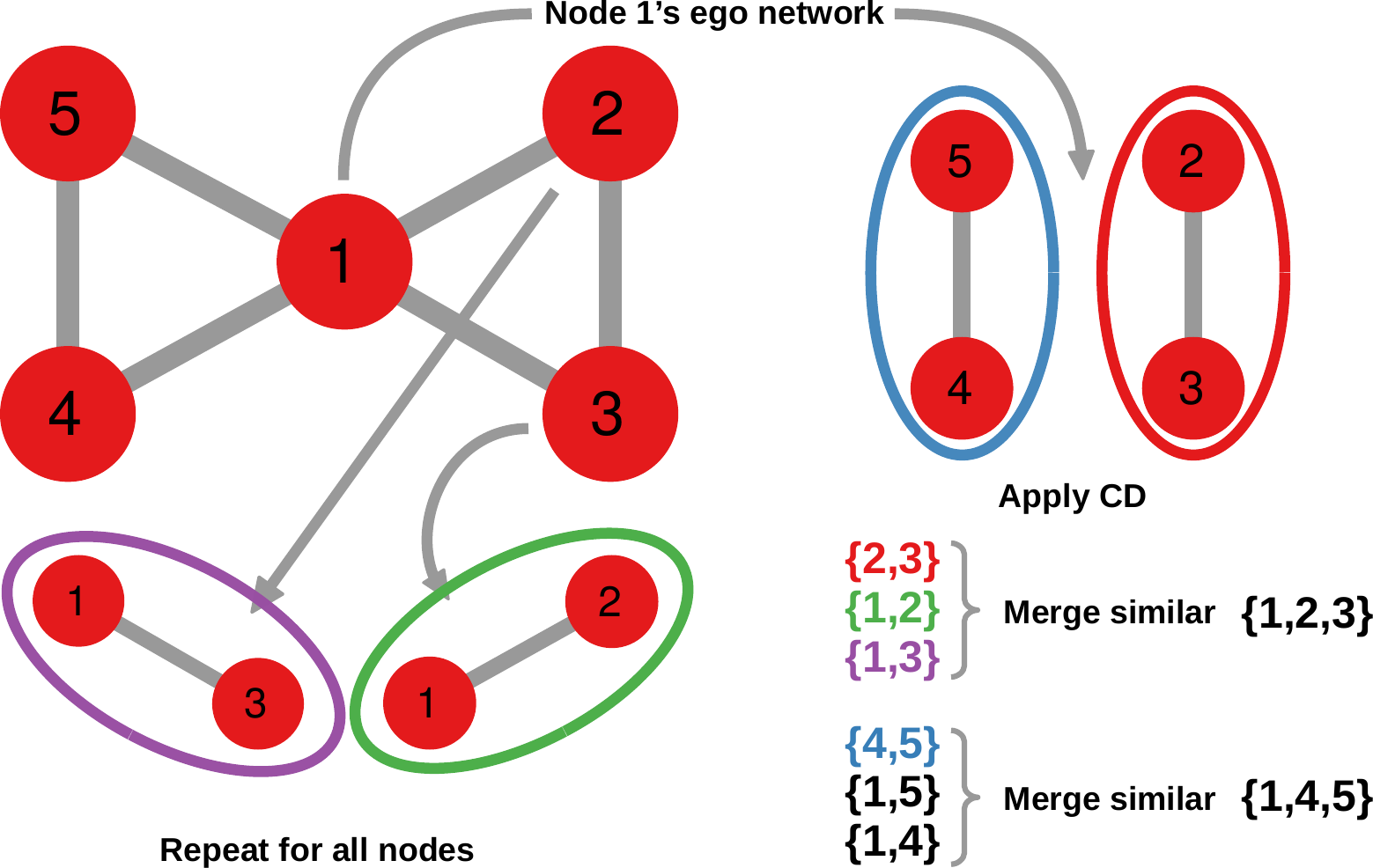}
\caption{The process of community discovery via the breaking down of the network into ego networks.}
\label{fig:ocd-demon}
\end{figure}

The procedure\cite{coscia2012demon} works as follows, and I use Figure \ref{fig:ocd-demon} to guide you. First, we extract the ego network of a node, removing the ego itself. This creates a simpler network to analyze. In the figure, I start by looking at node $1$ on the top right. This is a graph with two connected components: one connecting nodes $2$ and $3$, the other connecting nodes $4$ and $5$.

Then, we apply a disjoint community discovery algorithm to the ego network. The ego network is easier to analyze and often has easily distinguishable communities. In the example, these are the blue and red outlines, which find two trivial communities. We repeat the process for all nodes in the network, extracting all ego networks: in the figure I show the ego networks of nodes $2$ and $3$, with the communities I find in those cases, the green and purple outlines, respectively. Note that I omit the ego networks for nodes $4$ and $5$, but you can hopefully see where this is going.

Once we're done, we have a set of communities, and we can merge them according to some criterion. In the case of the figure, we merge communities if they share at least one node that is not part of too many communities. So we merge $\{2,3\}$ to $\{1,2\}$ and $\{1,3\}$ on the basis of them sharing nodes $2$ and $3$. Same reasoning for merging $\{4,5\}$ to $\{1,4\}$ and $\{1,5\}$. We don't merge $\{1,2,3\}$ and $\{1,4,5\}$, because the only node they have in common is $1$, which is in common with all communities in the network.

The original paper uses label propagation as the community discovery algorithm to apply to each network, but this needs not to be the case. We can instead apply a naive overlap algorithm. This move allows us to solve the problem of the assumption we mentioned before: we're allowing the links to have multiple origins, thus we don't necessarily force each link to be present for exclusively one reason.

\section{Other Approaches}
An interesting and more general approach to overlapping community discovery is the Order Statistics Local Optimization Method \cite{lancichinetti2011finding} (OSLOM). In reality, OSLOM is a bit of a Swiss army knife, in the sense that it isn't limited by finding overlapping communities: it can deal with edge weights and directions, hierarchical and evolving communities. Its basic philosophy is extremely similar to modularity: it builds an expected number of edges in a community by means of a configuration model.

Differently from modularity, OSLOM attempts to establish the statistical significance of the partition. That is, it asks how likely it is to observe the given community subgraphs in a configuration model. The less likely a vertex is to be included in a community in a null model, the more likely it is that we should add it to the community. Thus, these $p$ values can be used as a mean of ranking the next move, a move being adding a node to a community.

One can see how OSLOM is also a hierarchical community discovery method of the merge type: it assumes all nodes being on their own community at the beginning, and then it progressively merges them. Differently from classical hierarchical CD, a node is still a merge candidate even after it has been added to a community, allowing overlap. Moreover, OSLOM can be used as a post-processing strategy, to refine the communities you already found using another method. In fact, one could use OSLOM to transform a hard disjoint partition into an overlapping coverage.

\section{The Overlap Paradox}\label{sec:ocd-paradox}
Let's go back to the beginning of this chapter. Let's reiterate the definition of a community in a network:

\begin{center}
\textit{Communities are groups of nodes densely connected to each other and sparsely connected to nodes outside the community.}
\end{center}

As we've been seeing, overlapping communities make a paradox arise in our definition of community. If we have no overlap the ``denser inside'' and ``sparser outside'' assumption works well. However, if we add overlap, we cannot have it both ways.

\begin{figure}
\centering
\begin{subfigure}{.4\columnwidth}
\includegraphics[width=\textwidth]{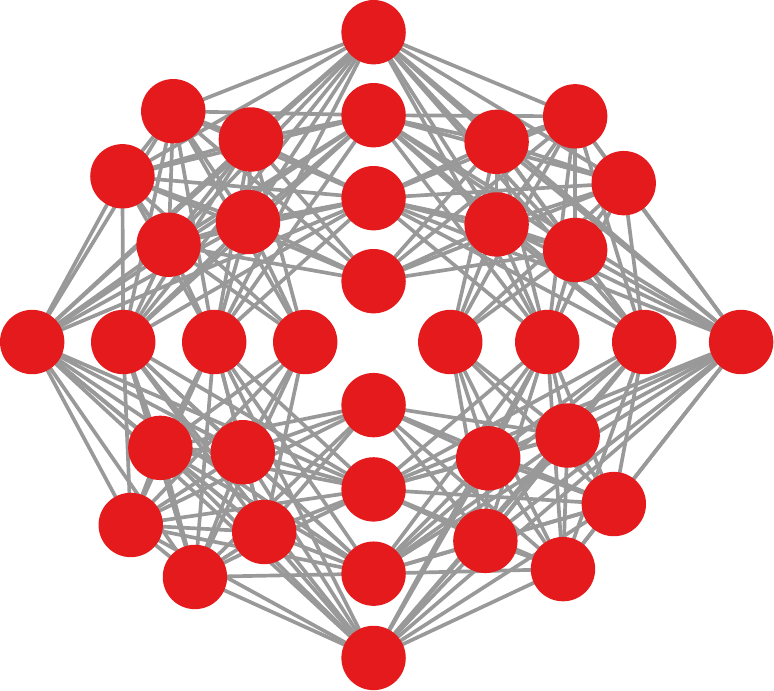}
\caption{}
\end{subfigure}\qquad
\begin{subfigure}{.4\columnwidth}
\includegraphics[width=\textwidth]{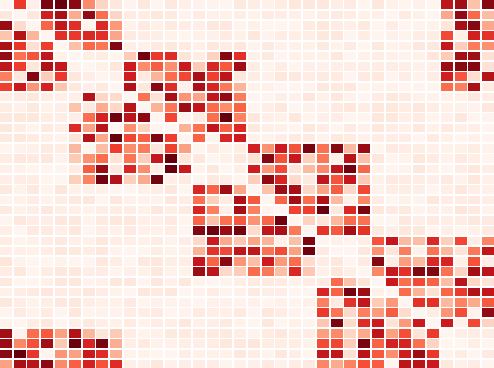}
\caption{}
\end{subfigure}
\caption{A graph (a) and its stochastic blockmodel (b).}
\label{fig:overlap-exclusive}
\end{figure}

If there are nodes in between communities either of two things will happen. The nodes in the overlap could connect with all nodes in both communities but not to each other, to maintain the ``external sparsity'' condition. But doing so contradicts the ``internal density'' part, because the overlap nodes do no connect to each other even though they belong to the same community. Figure \ref{fig:overlap-exclusive} provides an example for this scenario.

In Figure \ref{fig:overlap-exclusive}(a) we have a graph made by four 5-cliques and four sets of four nodes overlapping between two neighboring cliques. The overlap nodes don't connect to each other. Figure \ref{fig:overlap-exclusive}(b) shows how a stochastic blockmodel would interpret such a structure. You can clearly see that there are ``holes'' in the communities where the overlap nodes should be. If the overlap nodes don't connect to each other, they have low connection probability, which contradicts the fact that they are part of the same community.

\begin{figure}
\centering
\begin{subfigure}{.4\columnwidth}
\includegraphics[width=\textwidth]{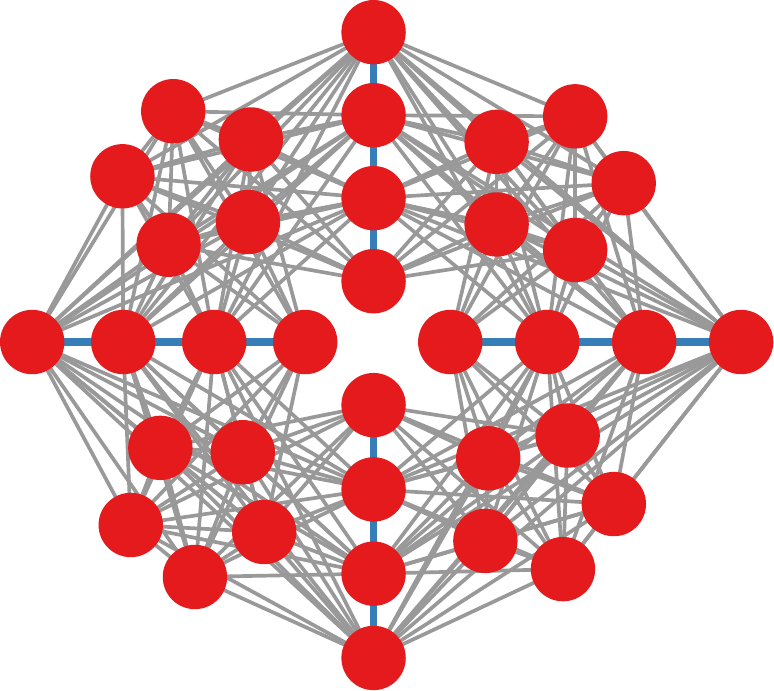}
\caption{}
\end{subfigure}\qquad
\begin{subfigure}{.4\columnwidth}
\includegraphics[width=\textwidth]{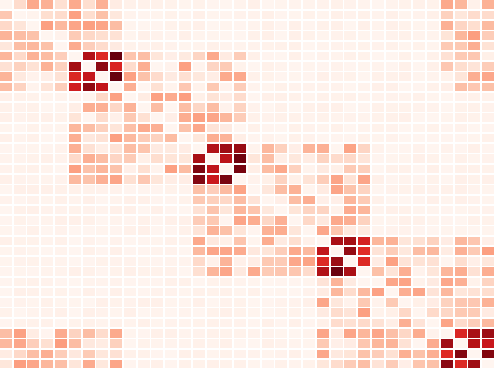}
\caption{}
\end{subfigure}
\caption{(a) The same graph as the one in Figure \ref{fig:overlap-exclusive}, but the added edges are highlighted in blue. (b) The corresponding stochastic blockmodel.}
\label{fig:overlap-inclusive}
\end{figure}

If, on the other hand, we maintain the ``internal density'' condition, since these nodes share not one but two communities, then they are more likely to connect to each other than nodes sharing only one community. In doing so, we end up with the opposite problem: breaking external sparsity. The overlap, which by definition is between the two communities, is denser than the community itself\cite{yang2012community}!

In Figure \ref{fig:overlap-inclusive}(a) we have such a scenario, with a graph similar to the one from Figure \ref{fig:overlap-exclusive}(a). But here all the overlap nodes are connected to each other, and they are connected more strongly than non-overlap nodes, given that they share more communities with each other. The corresponding stochastic blockmodel (Figure \ref{fig:overlap-inclusive}(b)) now shows that the communities themselves look weaker than the overlap.

This is another reason why our golden rule, the standard definition of communities in complex networks, isn't as shiny as we originally thought.

\section{Summary}

\begin{enumerate}
\item In real world networks, communities can overlap, meaning that nodes can be part of multiple communities at the same time. For instance, you're part of both the community of your high school friends, and of your university colleagues.
\item Many quality measures like modularity or mutual information cannot deal with overlapping communities. Thus we have several extensions that allow them to take into account node-sharing communities.
\item Disjoint community discovery algorithms can be adapted to find overlapping communities, for instance by means of fuzzy clustering: each node is given a ``belonging coefficient'' for each community, and can have multiple coefficients larger than zero.
\item Explicit structural approaches define the structure of an overlapping community and attempt to find it in the network. For instance, by percolating cliques, or splitting nodes so that each of their copies can belong to different communities.
\item Alternatively, one can divide edges into communities rather than nodes. In such approaches, the nodes belong to all communities to which their connections belong.
\item Overlapping communities put our classical community definition in crisis: if it is true that the more communities two nodes share the more likely they are connected, then the overlap of multiple communities is denser than the communities themselves, i.e. there are more links going outside communities than inside.
\end{enumerate}

\section{Exercises}

\begin{enumerate}
\item Use the k-clique algorithm to find overlapping communities in the network at \url{http://www.networkatlas.eu/exercises/38/1/data.txt}. Test how many nodes are part of no community for $k$ equal to 3, 4, and 5.
\item Compare the k-clique results with the coverage in \url{http://www.networkatlas.eu/exercises/38/2/comms.txt}, by using any variation of overlapping NMI from \url{https://github.com/aaronmcdaid/Overlapping-NMI}. For which value of $k$ do you get the best performance?
\item Implement the ego network algorithm: for each node, extract its ego minus ego network and apply the label propagation algorithm, then merge communities with a node Jaccard coefficient higher than 0.1 (ignoring singletons: communities of a single node). Does this method return a better NMI than k-clique percolation for $k = 3$?
\end{enumerate}

\chapter{Bipartite Community Discovery}\label{cha:bcd}
So far we have extended the community discovery problem by adding or discussing features of the output: do we want communities to share nodes or not? Do we want to have some sort of hierarchical optimization? In this and in the next chapter we instead discuss advancements on the other direction: what if our input is special? Here we deal with the case of bipartite networks, leaving multilayer networks for the next chapter.

We start by briefly amending modularity to the bipartite case. The bulk of the chapter is dedicated to alternative and more specialized ways to find bipartite communities. As usual, you can find a specialized survey of bipartite community detection methods\cite{alzahrani2016community}.

\section{Evaluating Bipartite Communities}
Since it has been a looming presence across this entire book part, let's start again with modularity, the elephant in the room of community discovery. Network scientists in the community detection business love modularity. If there is a scenario in which modularity doesn't work, they panic and start amending it to hell, until it works again. We've seen this with directed and overlapping community discovery, and we're seeing it again.

There are a couple of alternatives when it comes to define a modularity that works for bipartite networks. If you remember the original version of the modularity, it hinges on the fact that we want the partition to divide the network in communities that are denser than what we would expect given a null model -- the configuration model. Thus, extending modularity means to find the right formulation of a null model for bipartite networks\cite{barber2007modularity}\cite{guimera2007module}.

This is not that difficult, the only thing to keep in mind is that the expected number of edges in a bipartite network is different than in a regular network. So, while in the traditional modularity the configuration model connection probability was $\dfrac{k_u k_v}{2|E|}$, here it is instead $\dfrac{k_u k_v}{|E|}$, with the added constraints that $u$ and $v$ needs to be nodes of unlike type. The sum of modularity is made only across pairs of nodes of unlike types, otherwise we would have negative modularity contributions from nodes that cannot be connected, which would make the modularity estimation incorrect.

To see why this is the case, suppose that we're checking $u$ and $v$ and they are of the same type. Since they are of the same type and we're in a bipartite network, they cannot connect to each other, so $A_{uv} = 0$. But they are both part of the network, thus $k_u \neq 0$ and $k_v \neq 0$. Thus $\dfrac{k_u k_v}{|E|} > 0$, meaning that $A_{uv} - \dfrac{k_u k_v}{|E|} < 0$. Negative modularity contribution.

Once you have a proper bipartite modularity you can use any of the modularity maximization algorithms to find modules in your network, or even specialized ones\cite{beckett2016improved}.

\section{Via Projection}
As we saw in previous sections, bipartite networks have nodes of two different types, and edges are established exclusively between nodes of different types. In Netflix, we have users watching movies. It's natural to want to find communities in these networks. You want to know which movies are similar to each other so you can suggest them to users that are similar to each other.

In this case there is an easy obvious strategy. You take the bipartite network, you project it to unipartite using one of the techniques we saw in Chapter \ref{cha:projections} -- simple or hyperbolic weighting, random walks, etc -- and then you apply a normal unipartite community discovery to the result. Then you can project on the other set of nodes and find the other communities.

There are a couple of issues with this strategy. The first is that by projecting you're losing information. You connect movies with a weighted edge, which carries a quantitative information. But the bipartite network had qualitative information: a structure of users watching different things. That information is lost.

This is related to the second issue: once you project your network on your two types of edges and find their communities, you have movies grouped together because watched by the same users. But you don't know who those users are. Same with communities of users: which are their common movies? You have to go back to the bipartite network to know. A way to solve this issue is to use the dual projection approach\cite{everett2013dual}. In dual projection, you project the bipartite networks into its two unipartite versions and then you analyze them at the same time with specialized techniques. This dual projection approach has been applied to community discovery\cite{melamed2014community}, with encouraging results.

\begin{figure}
\centering
\includegraphics[width=.66\columnwidth]{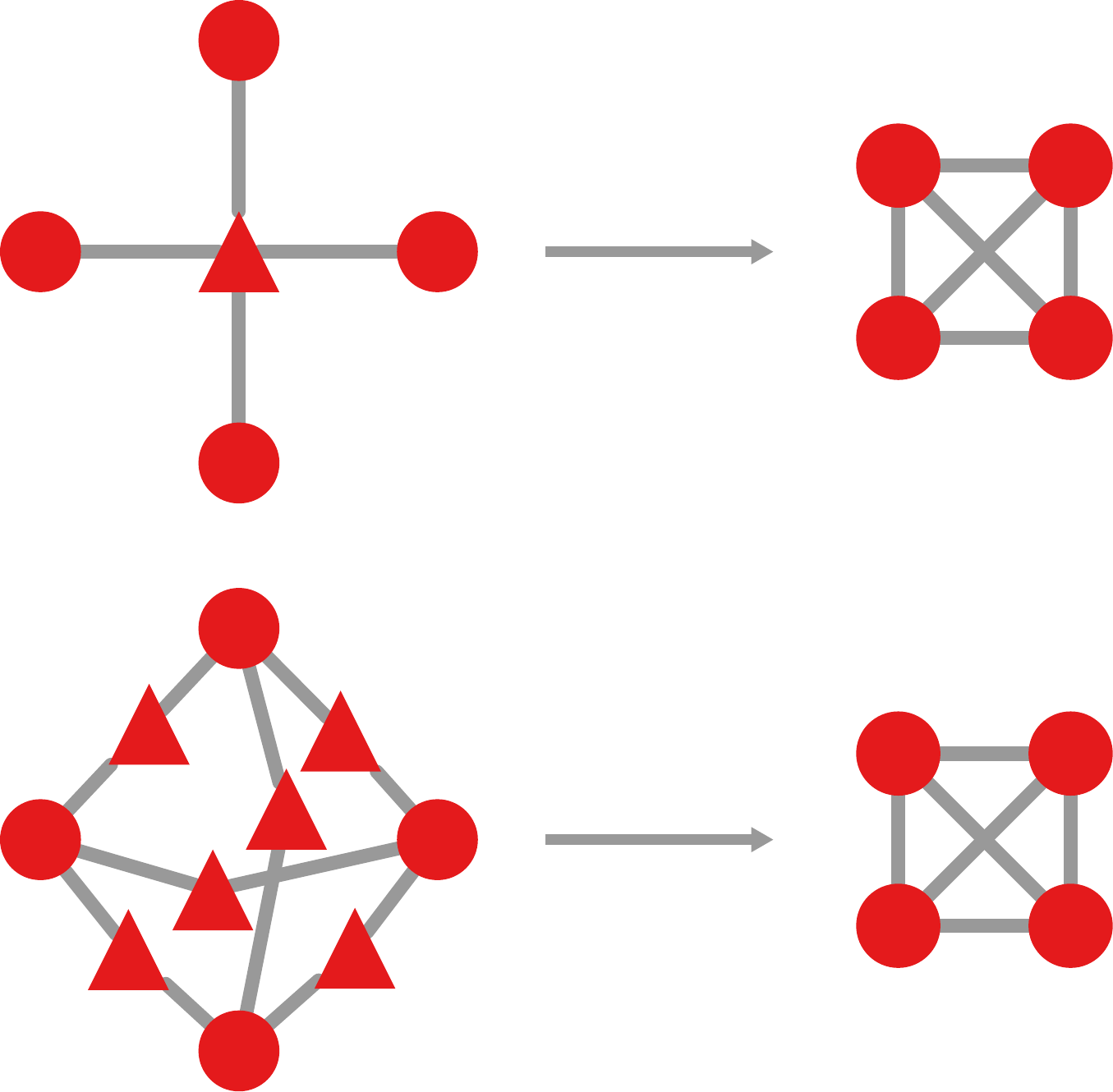}
\caption{Examples showing different bi-structures projecting to the same uni-structures. Both a 1,4-clique (top) and six 1,2-cliques (bottom) project into a 4-clique.}
\label{fig:bi-k-difference}
\end{figure}

There is an additional, more subtle, issue with this strategy. There are a number of different structures in bipartite networks that projects into the same unipartite graphs. This means that a proper bipartite community discovery algorithm and its hypothetical unipartite version will return different results, even if someone runs the mirror algorithm on the projection of the bipartite graph. Figure \ref{fig:bi-k-difference} shows two different bipartite networks that project to the same result. As the figure shows, projecting always means losing information. If we were to perform such projection we wouldn't be able to distinguish between the two cases Figure \ref{fig:bi-k-difference} shows, while a proper bipartite community detection algorithm could.

\section{Direct Bipartite Module Detection}\label{sec:bcd-direct}

\subsection{Bi-Clique Percolation}
The solution is to perform the community discovery directly on the bipartite structure. Here, we use the concept of bi-clique we saw earlier. Remember that a clique is a set of nodes in which all possible edges are present. A bi-clique is the same thing, considering that some edges in a bipartite network are not possible. For instance, a 5-clique in a unipartite network is a graph with five nodes and ten edges. In a bipartite network, a 2,3-clique has two nodes of type $1$, three nodes of type $2$, and all nodes of type $1$ are connected to nodes of type $2$ -- six edges in total. This is the starting point of bipartite community discovery.

\begin{figure}[t]
\centering
\includegraphics[width=.83\columnwidth]{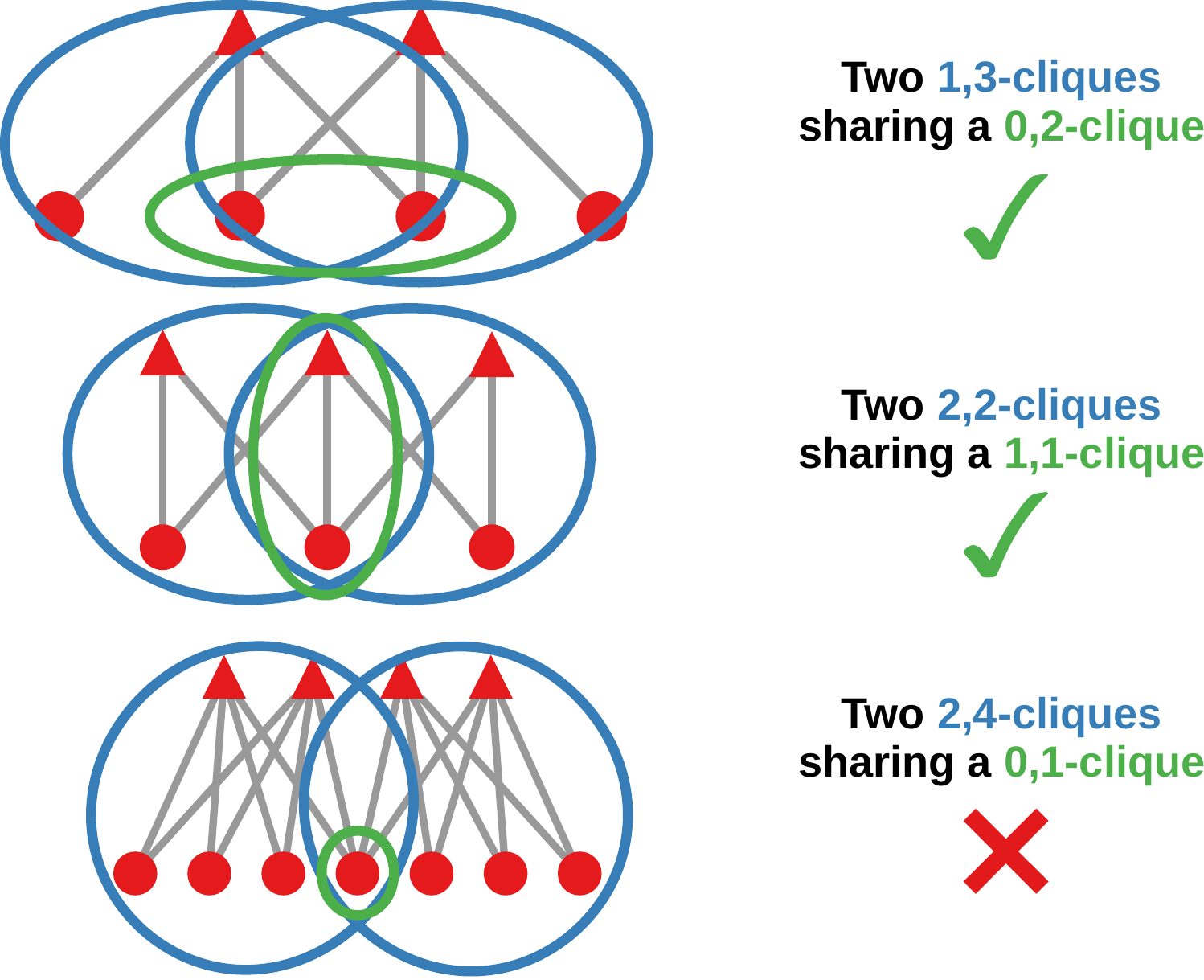}
\caption{Examples of bi-clique percolation. The top two examples will lead to percolation because they satisfy the $n-1$,$m-1$ constraint. The last example on the bottom does not.}
\label{fig:bicliqueperc}
\end{figure}

Bi-clique percolation works in the same way as the unipartite k-clique percolation algorithm -- described in Section \ref{sec:ocd-struct} --, with the added headache of having two numbers of nodes to keep track, because now it's a biclique. Communities are n,m-cliques (again, $n$ and $m$ are parameters) and they get merged if they share an $(n-1)(m-1)$-clique. Figure \ref{fig:bicliqueperc} shows some examples of bi-clique percolation. Note that any combination of $n$ nodes of the same type can be a 0,n-clique, thus allowing percolation.

The bi-clique percolation algorithm inherits from its predecessor the ability of returning overlapping communities. Thus in this case we get not only bipartite communities, but these communities can also share nodes.

\subsection{Adapting Classical Approaches}
As we already got used to see, many of the classical approaches to community discovery can be adapted to take into account complications in the network structure. Bipartite community discovery is no exception. So let's see what we can do to transform random walks, label propagation,  and stochastic blockmodels to the bipartite case.

There are a couple of ways to exploit random walks and find bipartite communities\cite{kheirkhahzadeh2016efficient}\cite{alzahrani2014community}. We already considered one: simply project the network as unipartite and perform the random walks normally. The issue is that the abundance of links will lower the power of random walkers, because the network will be too dense and the boundaries between communities will be difficult to find. That is why you might also want to perform network backboning (Chapter \ref{cha:backboning}).

Alternatively, you could perform high-order random walks. We saw in Chapter \ref{cha:hod} that we can add a parameter to a random walker, telling it how much time it passes between one step and another. You can effectively model 2-steps random walks this way, which will allow you to find the communities for nodes of one type -- and then of the other type, by repeating the process with a different starting point.

\begin{figure}
\centering
\begin{subfigure}{.33\columnwidth}
\includegraphics[width=\textwidth]{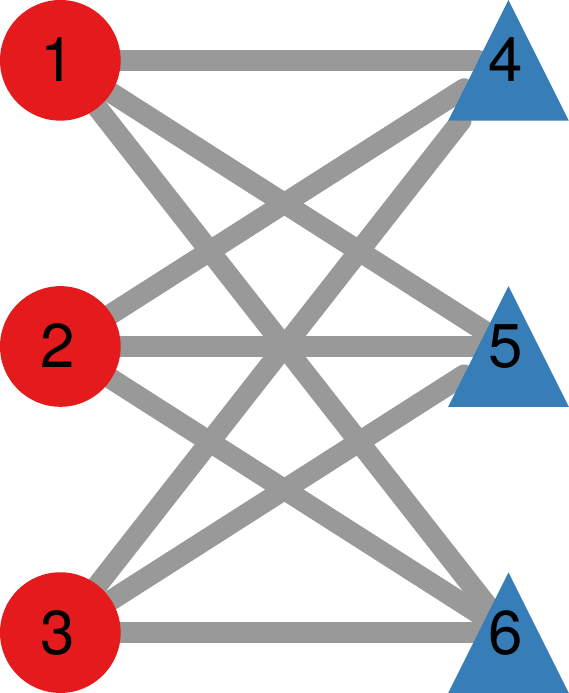}
\caption{}
\end{subfigure}\qquad
\begin{subfigure}{.33\columnwidth}
\includegraphics[width=\textwidth]{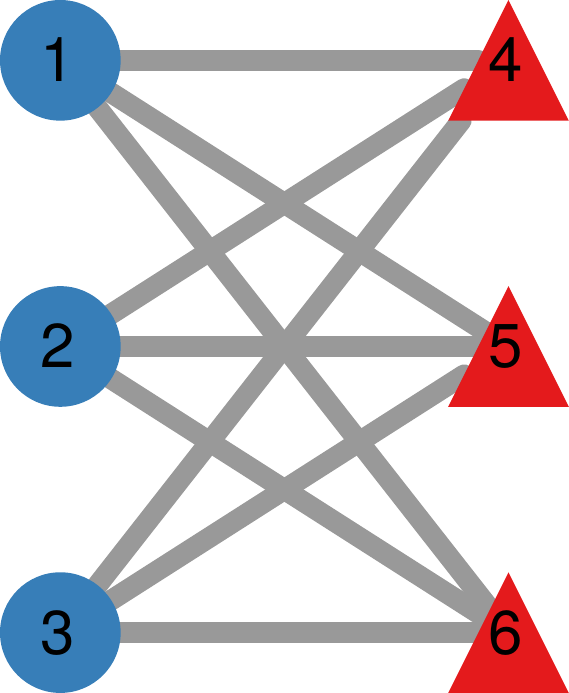}
\caption{}
\end{subfigure}
\caption{(a) Communities from a synchronous label propagation at an hypothetical time $t$ (the node color is the community label). Each node of circle type has a blue label majority in its neighborhood, and each triangle has a red majority. (b) At time $t + 1$ the labels oscillate according to the label majority at time $t$. The system will oscillate forever between (a) and (b) without converging.}
\label{fig:label-oscillation}
\end{figure}

Classical label percolation community discovery can be tricky. The first reason is that synchronous label percolation has one problem known as a label oscillation. Figure \ref{fig:label-oscillation} provides an example. After a few iterations, it might be that a community of nodes has a majority label on one side and a different majority label on the other side. The algorithm will be then stuck oscillating the labels at each time step. In such a scenario, it will never converge\cite{vsubelj2011robust}.

One could solve the issue in various ways. First, one could simply use asynchronous updating. However, traditional asynchronous updating will update nodes in a random order. This might impact the stability of the resulting partition, because the order in which nodes are updated matters. Two subsequent runs of the algorithm could yield very different results. Moreover, it might impact convergence time, because there are many node orders which will still result in label oscillation. The most sure way to prevent label oscillation is to update first all nodes of one type and then all nodes of the other.

There are a few other ways to prevent oscillation. First, one could integrate label propagation with the modularity approach\cite{liu2010community}. In this scenario, one doesn't run label propagation until convergence, but only for a few steps. Then they would refine the communities by maximizing bipartite modularity. Alternatively, one could put constraints on how we allow labels to propagate. For instance, we could force communities to be of comparable sizes in number of nodes or edges\cite{barber2009detecting}.

Finally, we can adapt stochastic blockmodels to the bipartite case, creating a biSBM\cite{larremore2014efficiently}. This has some similarities with the MMSB we saw for overlapping community discovery in Section \ref{sec:ocd-mmsbm}. First, we don't look directly at the $|V_1| \times |V_2|$ biadjacency matrix $B$. It is more convenient to look at its adjacency matrix equivalent:

$$ A = 
\begin{pmatrix}
0 & B \\
B^T & 0 \\
\end{pmatrix}
$$

The zeros on the main diagonal mean that nodes of the same type cannot connect to each other, enforcing the bipartite structure (see Section \ref{sec:mat-mat-mat}).

Then, just like in the overlapping case, we can have a special community-community matrix that tells us the probability of nodes in two distinct communities to connect to each other. The special condition here is that communities grouping nodes of the same type will have zero probability of connecting to each other, respecting the bipartite constraint.

Once we have these two special structures in place, one can proceed finding the most likely blockmodel that explains the observed data, which is the one with the best community partition, with the same strategies as in vanilla SBM. Note that this method can be trivially extended to multi-partite networks, modifying the fundamental structures accordingly.

The biSBM clusters the two modes separately, so you get a mixing matrix that tells you how the groups in the $V_1$ nodes interact with the groups in the $V_2$ nodes. In contrast, bipartite modularity and some other approaches will produce mixed groups, which contain nodes from both $V_1$ and $V_2$. This makes biSBM a co-clustering method, like the ones we'll see in Section \ref{sec:bcd-clustering}. The difference with those methods is that they find communities discovery via neighbor similarity, which is not the philosophy of biSBM.

A related method uses matrix factorization\cite{zhang2015community}. It starts by noticing that zeroes in $A$ have different meanings. The zeroes in the main diagonal block represent impossible connections, connections that shouldn't be penalized. The zeroes in the off-diagonal block instead represent edges that \textit{could} exist. Thus, the authors define a mask matrix $M$, with the same dimensions as $A$, with zeros on the main diagonal blocks and ones in the off diagonal blocks. By factorizing the product of $M$ and $A$ together with our best guess at the community organization of $A$, we obtain a function we can maximize to find the best community partition, knowing that we're only penalizing zeroes corresponding to connections that could exist.

\subsection{Redefining the Clustering Coefficient}
One reasonable way to find communities in unipartite networks is by exploiting the clustering coefficient. Nodes with high clustering coefficients are supposedly well embedded in a community, because all their neighbors are connected to each other. Thus, an algorithm cutting low local clustering coefficient areas could perform well. If you recall Section \ref{sec:hcd-recursive}, one such strategy is to derive an edge clustering measure from the local node clustering, and then use it to determine which edges to cut to find a hierarchical community structure.

Our problem here is that the local clustering coefficient in bipartite networks is zero for all nodes. This is because in bipartite networks there cannot be triangles, which are at the basis of the computation of the local clustering coefficient (Section \ref{sec:density-clustering}). A triangle, by definition, connects three nodes together. However, in bipartite networks, the edge closing the triangle cannot exist, because it would connect two nodes of the same type.

\begin{figure}
\centering
\begin{subfigure}{.3\columnwidth}
\includegraphics[width=\textwidth]{figures/triangle.pdf}
\caption{}
\end{subfigure}\qquad
\begin{subfigure}{.33\columnwidth}
\includegraphics[width=\textwidth]{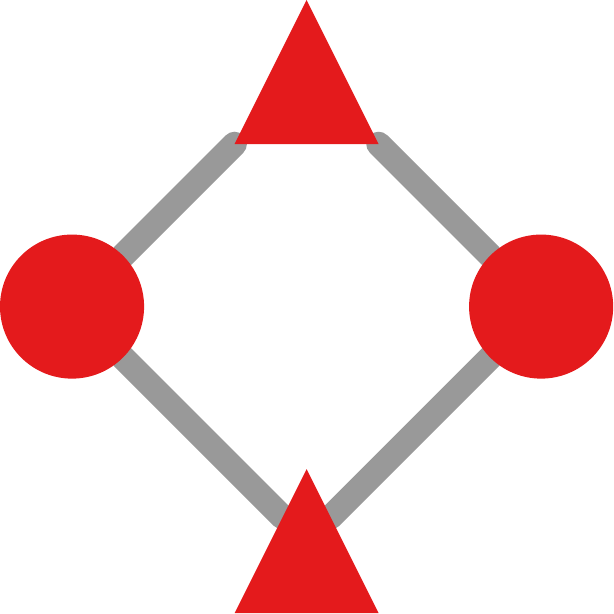}
\caption{}
\end{subfigure}
\caption{(a) The structure at the basis of the clustering coefficient of unipartite networks: the triangle. (b) Its equivalent in bipartite networks: the square.}
\label{fig:bipartite-clustercoef}
\end{figure}

This needs not to worry us. We can redefine the clustering coefficient to make sense in a bipartite network. In a unipartite network, the triangle is the smallest non-trivial cycle, the one that does not backtrack using the same edge, as you can see in Figure \ref{fig:bipartite-clustercoef}(a). We can also have a smallest non-trivial cycle in bipartite networks. It involves four nodes, as Figure \ref{fig:bipartite-clustercoef}(b) shows. So we can say that the local clustering coefficient of a node in a bipartite network is the number of times such cycles appear in its neighborhood, divided by the number of times they could appear given its degree\cite{zhang2008clustering}.

Let us assume that we want to know the local square clustering coefficient of node $z$. If we say that nodes $u$, $v$, and $z$ are involved in $s_{uvz}$ squares, then contribution of nodes $u$ and $v$ to the square clustering coefficient of $z$ is:

$$C4_{u,v}(z) = \dfrac{s_{uvz}}{s_{uvz} + (k_u - \eta_{uvz}) + (k_v - \eta_{uvz})}, $$

with $\eta_{uvz} = 1 + s_{uvz}$. In practice, the number of possible squares (in the denominator) is the number of actual squares plus how many additional squares you could have given $u$'s and $v$'s free edges, edges not involved in any square. Here, $u$ and $v$ are the nodes of the same type.

\begin{figure}
\centering
\includegraphics[width=.66\columnwidth]{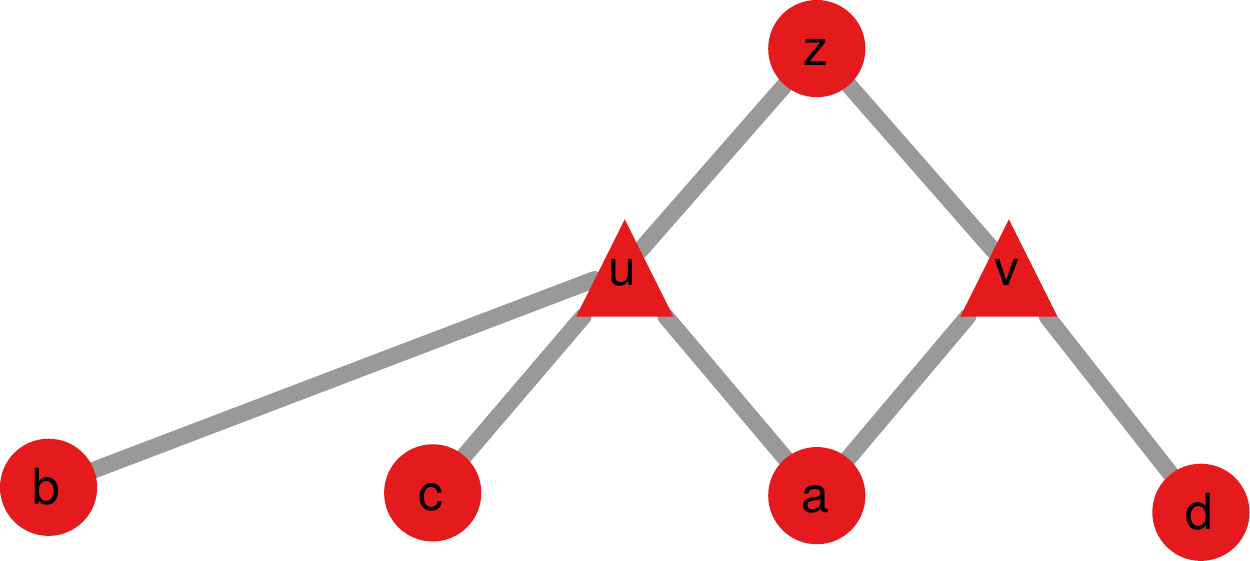}
\caption{An example of bipartite network on which we can calculate the local clustering coefficient of node $z$.}
\label{fig:square-cc}
\end{figure}

Consider Figure \ref{fig:square-cc}. In the figure, $s_{uvz} = 1$, because nodes $u$, $v$, and $z$ are involved in one square. As a consequence, $\eta_{uvz} = 2$. Since $k_u = 4$ and $k_v = 3$, we know that there could be $s_{uvz} + (k_u - \eta_{uvz}) + (k_v - \eta_{uvz}) = 1 + (4 - 2) + (3 - 2) = 4$ squares between the nodes: $uvza$ (which exists), $uvzb$, $uvzc$, and $uvzd$ (which do not exist). Since $z$ has no additional neighbors, its local clustering coefficient is thus $1 / 4$.

Once you have a properly defined local node clustering measure, you can use it to derive the corresponding edge clustering measure. Then, you can apply the same edge splitting algorithm we explored in the hierarchical community discovery case to find hierarchical bipartite communities.

\section{Neighbor Similarity}\label{sec:bcd-clustering}
A wholly different category of approaches tries to look at the adjacency matrix of a bipartite graph under a different perspective. The point of a community in a bipartite network is not that the nodes connect densely to each other, but that they connect to the same neighbors -- nodes of the other type. Thus it's not much about internal density as it is about structural similarity (Section \ref{sec:centr-similarity}). Some approaches use some sort of common neighbor approach\cite[-0.5in]{daminelli2015common}.

\begin{figure}
\centering
\begin{subfigure}{.3\columnwidth}
\includegraphics[width=\textwidth]{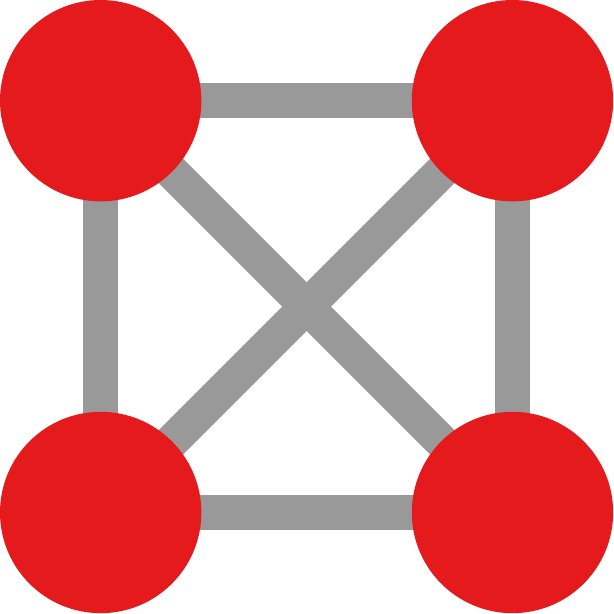}
\caption{}
\end{subfigure}\qquad
\begin{subfigure}{.33\columnwidth}
\includegraphics[width=\textwidth]{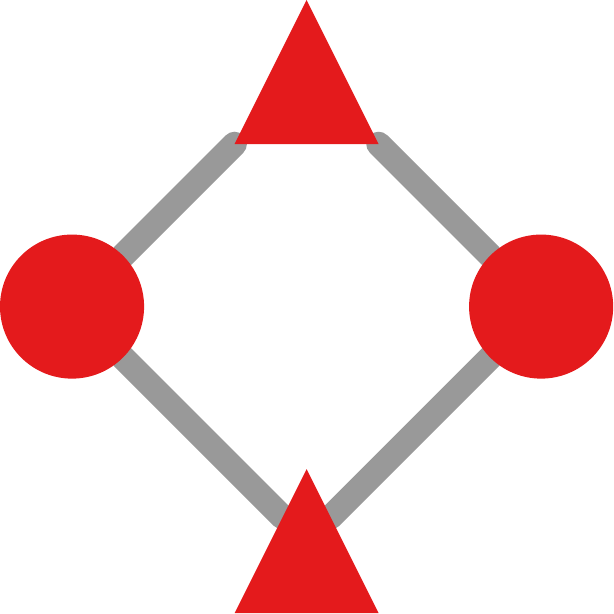}
\caption{}
\end{subfigure}
\caption{(a) A classical unipartite community. (b) A bipartite community.}
\label{fig:bipartite-neighsim}
\end{figure}

Figure \ref{fig:bipartite-neighsim} shows a way to perform such a mental pivot. In a regular unipartite network -- Figure \ref{fig:bipartite-neighsim}(a) -- we're looking at a community as a \textit{set of nodes that connect to each other}. In a bipartite network -- Figure \ref{fig:bipartite-neighsim}(b) -- we're looking at a community as a \textit{set of nodes that connect to the same nodes}. Thus, in a bipartite community, we don't require two nodes that are part of the same community to connect to each other.

This is a fully valid definition of community that can be translated to unipartite networks, which generates another way to look at the community discovery problem. I have showed this as a valid ``community'' of community discovery algorithms in Section \ref{sec:cd-partition-practical}.

The change in perspective might seem small at first, but it opens up a sea of possibilities. When you look at an adjacency matrix as a simple set of feature vectors, you can perform data clustering on it. Meaning that you can see a node as a point in a multidimensional space, a space with as many dimensions as there are nodes of the other type in the network. Then you can pick any of your favorite machine learning algorithms, spanning from k-means\cite{steinhaus1956division}\cite{macqueen1967some}\cite{lloyd1982least} to dbscan\cite{ester1996density}, and interpret their clusters as communities.

You should also be encouraged to look at bi-clustering (or co-clustering) techniques\cite{dhillon2001co}\cite{dhillon2003information}\cite{kluger2003spectral}. The advantage of such techniques is that they will cluster the rows and the columns of your matrix at the same time. In this way, you don't have to manually map the clusters you found by looking at the $|V_1| \times |V_2|$ matrix with the ones you found in the $|V_2| \times |V_1|$ matrix.

\begin{figure*}[t]
\centering
\begin{subfigure}{.4\columnwidth}
\includegraphics[width=\textwidth]{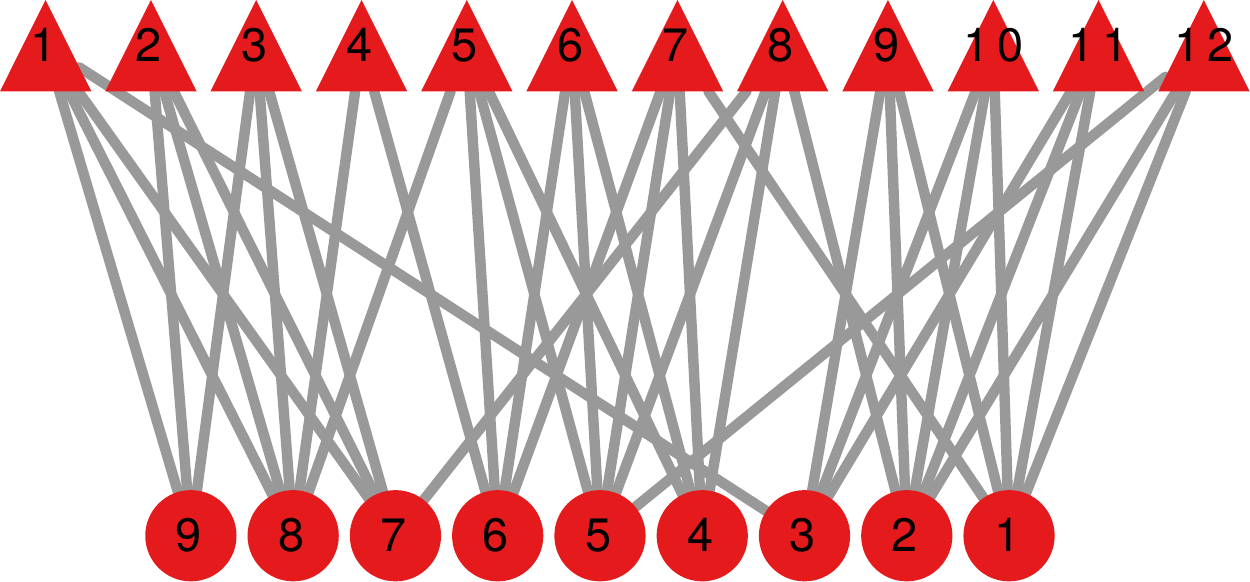}
\caption{}
\end{subfigure}\quad
\begin{subfigure}{.25\columnwidth}
\includegraphics[width=\textwidth]{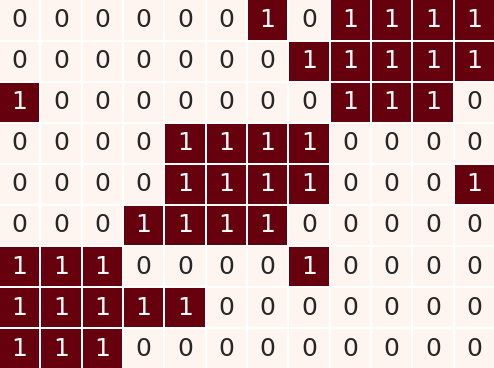}
\caption{}
\end{subfigure}\quad
\begin{subfigure}{.25\columnwidth}
\includegraphics[width=\textwidth]{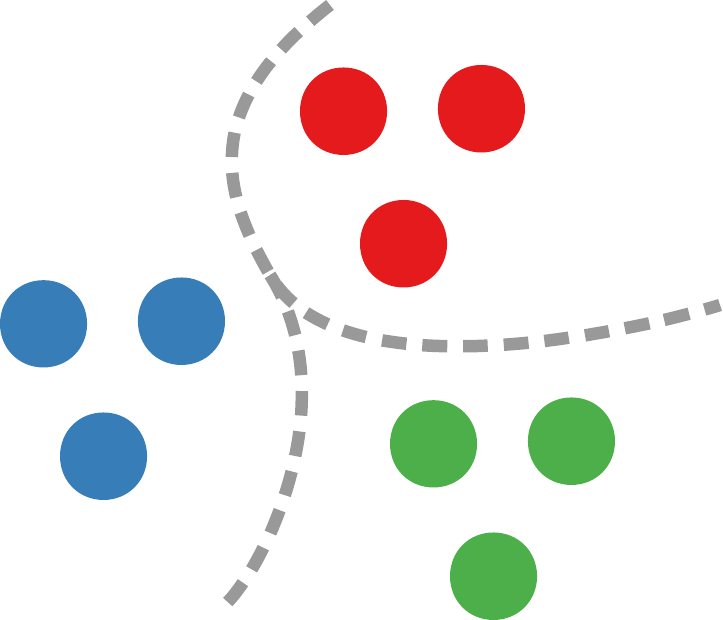}
\caption{}
\end{subfigure}
\caption{(a) A bipartite network. (b) Its adjacency matrix. (c) A 2D spatial representation of the circular nodes, using their adjacencies to determine the position. Node color is its cluster, as identified by spatial clustering (dashed line).}
\label{fig:bipartite-clustering}
\end{figure*}

The way these methods find clusters is usually by estimating the pairwise distance (Euclidean or otherwise) between data points. Then clusters are sets of points that lump together in this complex space. Hopefully, boundaries between clusters are clear, as few points are equidistant from multiple cluster centers. In community discovery, your data points are the nodes. Figure \ref{fig:bipartite-clustering} shows an example.

Besides old data clustering techniques, there is a lot of excitement for the application of neural network approaches to community discovery\cite{bruna2017community}. However, usually, adjacency matrices are too sparse and constrained to provide a proper input to neural networks. Thus, most of the deep learning attacks to community detection use more sophisticated representations of the relations between nodes in the graph, in the form of graph embeddings, which we'll explore more in depth later on (in Chapter \ref{cha:mining-embeddings}).

To wrap up this chapter, let's recall again our definition of communities in complex networks:

\begin{center}
\textit{Communities are groups of nodes densely connected to each other and sparsely connected to nodes outside the community.}
\end{center}

\begin{figure}
\centering
\includegraphics[width=.5\columnwidth]{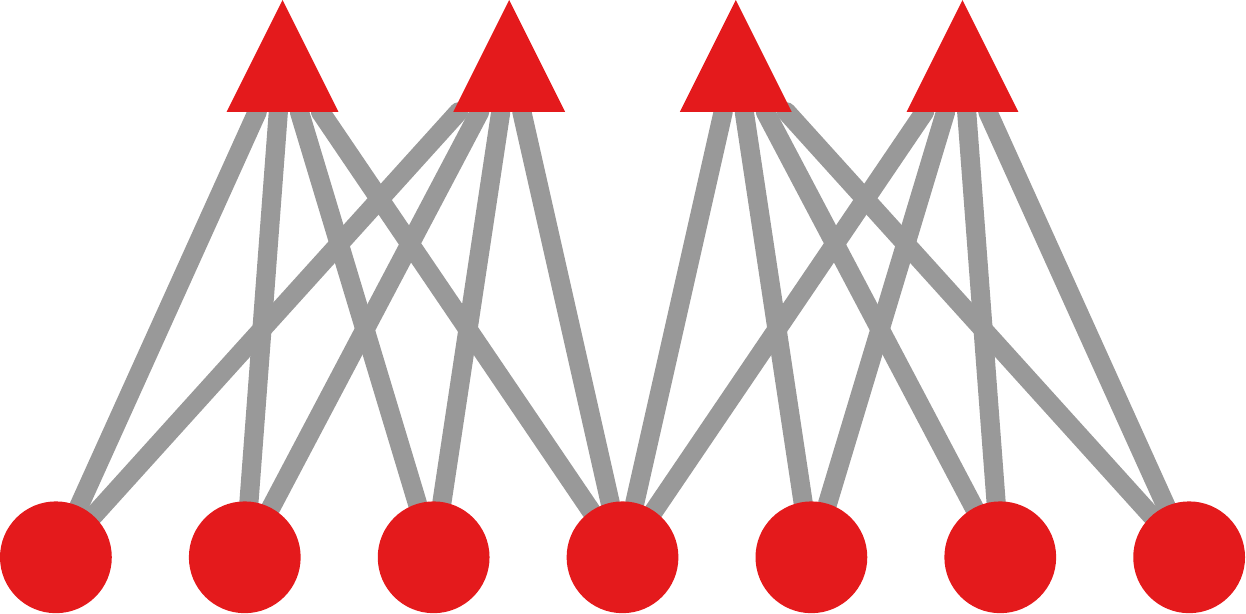}
\caption{A possible bipartite community.}
\label{fig:biclique-density}
\end{figure}

The bipartite community discovery introduces another issue with the standard definition of community based on density. On the one hand, nodes of the same type cannot connect in a bipartite graph -- so they have density of zero --, but they can and will be part of the same community. Figure \ref{fig:biclique-density} has many missing links, but it is still a valid bipartite community. On the other hand, we are looking at a community definition that is more based on neighborhood similarity than on internal density. So again this criterion of internal density is a bit flaky.

\section{Summary}

\begin{enumerate}
\item In bipartite networks, nodes of the same type cannot connect to each other. However, they could still be in the same community, because they have lots of common neighbors. Thus we need to adapt modularity and other community quality measures to take this into account.
\item One way to perform bipartite community discovery is by projecting the bipartite network into unipartite form and then perform community discovery there. However, the resulting network will be too dense and we will lose information in the projection.
\item We can adapt clique percolation by percolating bi-cliques. We can perform random walks by making them perform two steps at a time. We can also adapt label propagation by performing it asynchronously and refining its results.
\item We can also redefine the local clustering coefficient, so that it is based not on the number of triangles around a node (which cannot exist in a bipartite network), but on the number of its surrounding squares.
\item Alternatively, one could simply infer node similarity measures by looking at their neighbors, or perform simple data clustering. However, all these strategies show how our classical definition of communities in networks cannot really capture the bipartite case.
\end{enumerate}

\section{Exercises}

\begin{enumerate}
\item The network at \url{http://www.networkatlas.eu/exercises/39/1/data.txt} is bipartite. Project it into unipartite and find five communities with the Girvan-Newman edge betweenness algorithm (repeat for both node types, so you find a total of ten communities). What is the NMI with the partition proposed at \url{http://www.networkatlas.eu/exercises/39/1/nodes.txt}?
\item Now perform asynchronous label propagation directly on the bipartite structure. Calculate the NMI with the ground truth. Since asynchronous label propagation is randomized, take the average of ten runs. Do you get a higher NMI?
\item Consider the bi-adjacency matrix as a data table and perform bi-clustering on it, using any bi-clustering algorithm provided in the \texttt{scikit-learn} library. Do you get a higher NMI than in the previous two cases?
\end{enumerate}

\chapter{Multilayer Community Discovery}\label{cha:mcd}
The last chapter of community discovery, at least for this book, focuses on multilayer networks. In multilayer networks, nodes can belong to different layers and thus they can connect for different reasons. In multilayer networks we want to find communities that span across layers. For example, we want to figure out communities of friends even if your friends are spread across multiple social media platforms.

There are a few review works you can check out to have a more in-depth exploration of the topic\cite{kim2015community}\cite{hanteer2019community}. Here, I go over briefly the main approaches and peculiar problems of community discovery in multilayer networks.

\section{Flattening}
\begin{figure}
\centering
\begin{subfigure}[t]{.4\columnwidth}
\includegraphics[width=\textwidth]{figures/layers_combined_coupling_trivial.pdf}
\caption{}
\end{subfigure}\qquad
\begin{subfigure}[t]{.4\columnwidth}
\includegraphics[width=\textwidth]{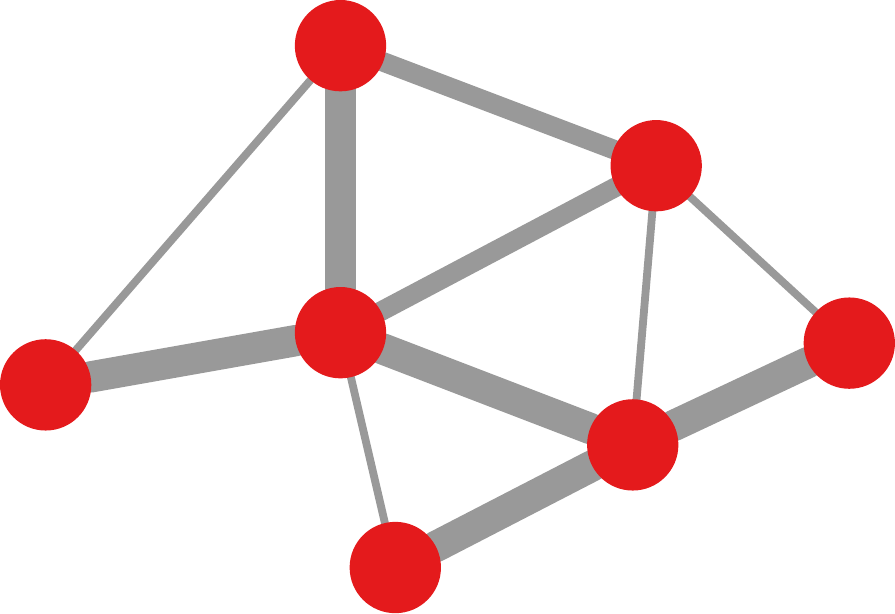}
\caption{}
\end{subfigure}
\caption{(a) A multilayer network. (b) A weighted flattening of (a). An edge's width is proportional to its weight.}
\label{fig:multilayer-cd}
\end{figure}

Similar to the bipartite case, there is a relatively simple solution. You can flatten the network by collapsing nodes across layers\cite{berlingerio2011finding}, meaning that you reduce the multilayer network to a network with a single layer and weighted edges.  The weights on the edges depend on the multilayer connections. In practice, you're collapsing a qualitative information -- in which layer a connection appears -- into a quantitative one -- an edge weight. This assumes that every edge type is equally important. Then you can perform a normal mono-layer community discovery. Figure \ref{fig:multilayer-cd} shows an example.

There are a few choices for your edge weights. The simplest one could be to simply count the number of layers in which the connection between the nodes appear. However, you might want to take into account some interplay between the layers. For instance, you can count the number of common neighbors that two nodes have and use that as the weight of the layer, under the assumption that a layer where two nodes have many common neighbors should count for more when discoverying communities. Or you could use ``differential flattening\cite{kim2017differential}'': flatten the multilayer graph into the single layer version of it such that its clustering coefficient is maximized. 

As with the bipartite case, simplistic solutions create problems. The issue is the same: we lose information. When we translate a layer into a weight, we don't know any more which type of links we're looking at. Some types of links might contribute differently to the communities. Moreover, flattening is not always possible, or at least not straightforward. If a node in one layer is coupled with multiple nodes in another layer, how do we represent it in the flattened network?

\section{Layer by Layer}
There is another solution that is slightly more sophisticated than flattening a multilayer network but that at the same time doesn't require you to go fully multidimensional in your analysis. It is performing community discovery separately on each layer of your network and then somehow combine your results. This is sort of like the approach of dynamic community discovery where you perform your detection on each snapshot of the network and then you aggregate the results (Section \ref{sec:cd-partition-evo}).

In fact, one could see each snapshot of the network as a layer -- or vice versa: each layer as a snapshot. Thus anything we do on an evolving network we could also do on a multilayer one. And -- why not? -- we could even have \textit{evolving} multilayer networks, putting the two together\cite{bazzi2016community}. The solution is not ideal, though, because there are many assumptions you have on a dynamic networks that you might break on a multilayer network -- and vice versa. For instance, in a dynamic graph you have some sort of continuity assumption: one snapshot should be, in principle, similar to the next. There is no requirement of similarity between layers: in fact, they can even be strongly anti-correlated.

For these reasons, you need some specialized community discovery approaches. In one approach, one could build a matrix where each row is a node and each column is the partition assignment for that node in a specific layer\cite{tang2012community}. This is then a $|V| \times |C|$ matrix. Then one could perform kMeans on it, finding clusters of nodes that tend to be clustered in the same communities across layers.

A similar approach\cite{berlingerio2013abacus} uses frequent pattern mining, a topic we'll see more in depth in Section \ref{sec:mining-transact}. For now, suffice to say that we again perform community discovery on each layer separately. Each node can then be represented as a simple list of community affiliations. We then look for sets of communities that are frequently together: these are communities sharing nodes across layers.

\begin{figure*}
\centering
\begin{subfigure}{.3\columnwidth}
  \begin{tabular}{c|ccc}
    Node & L$1$ & L$2$ & L$3$\\
    \hline
    $1$ & \textcolor{cb1}{C$1$L$1$} & \textcolor{cb3}{C$1$L$2$} & \textcolor{cb5}{C$1$L$3$}\\
    $2$ & \textcolor{cb1}{C$1$L$1$} & \textcolor{cb3}{C$1$L$2$} & \textcolor{cb5}{C$1$L$3$}\\
    $3$ & \textcolor{cb1}{C$1$L$1$} & \textcolor{cb3}{C$1$L$2$} & \textcolor{cb5}{C$1$L$3$}\\
    $4$ & \textcolor{cb2}{C$2$L$1$} & \textcolor{cb3}{C$1$L$2$} & \textcolor{cb5}{C$1$L$3$}\\
    $5$ & \textcolor{cb2}{C$2$L$1$} & \textcolor{cb3}{C$1$L$2$} & \textcolor{cb7}{C$2$L$3$}\\
    $6$ & \textcolor{cb2}{C$2$L$1$} & \textcolor{cb3}{C$1$L$2$} & \textcolor{cb8}{C$3$L$3$}\\
    $7$ & \textcolor{cb1}{C$1$L$1$} & \textcolor{cb4}{C$2$L$2$} & \textcolor{cb8}{C$3$L$3$}\\
    $8$ & \textcolor{cb2}{C$2$L$1$} & \textcolor{cb4}{C$2$L$2$} & \textcolor{cb8}{C$3$L$3$}\\
    $9$ & \textcolor{cb2}{C$2$L$1$} & \textcolor{cb4}{C$2$L$2$} & \textcolor{cb8}{C$3$L$3$}\\
    $10$ & \textcolor{cb1}{C$1$L$1$} & \textcolor{cb4}{C$2$L$2$} & \textcolor{cb7}{C$2$L$3$}\\
  \end{tabular}
\caption{}
\end{subfigure}\qquad
\begin{subfigure}{.3\columnwidth}
  \begin{tabular}{l|r}
    MLComm & SLComms\\
    \hline
    MLC$1$ & \textcolor{cb1}{C$1$L$1$}, \textcolor{cb3}{C$1$L$2$}, \textcolor{cb5}{C$1$L$3$}\\
    MLC$2$ & \textcolor{cb2}{C$2$L$1$}, \textcolor{cb3}{C$1$L$2$}\\
    MLC$3$ & \textcolor{cb4}{C$2$L$2$}, \textcolor{cb8}{C$3$L$3$}\\
  \end{tabular}
\caption{}
\end{subfigure}\qquad
\begin{subfigure}{.3\columnwidth}
  \begin{tabular}{l|r}
    MLComm & Nodes\\
    \hline
    MLC$1$ & $1$, $2$, $3$\\
    MLC$2$ & $4$, $5$, $6$\\
    MLC$3$ & $7$, $8$, $9$\\
  \end{tabular}
\caption{}
\end{subfigure}
\caption{(a) The communities found in each layer of each node. (b) The merged multilayer communities. (c) The final node-community affiliation.}
\label{fig:abacus}
\end{figure*}

Figure \ref{fig:abacus} shows an example. In Figure \ref{fig:abacus}(a) we have the communities found for each layer for each node. Then we decide that we want to merge communities if they have at least three nodes in common, i.e. they appear in at least three rows of the table.

Figure \ref{fig:abacus}(b) shows the multilayer communities mapping and there are many interesting things happening. First, we only want maximal sets, meaning that we aren't interested in returning C$1$L$1$ by itself if we also find it in a larger set of communities. Second, we are ok if a community gets merged in different sets -- i.e. the multilayer communities can overlap --: C$1$L$2$ is part of two maximal sets, MLC$1$ and MLC$2$. Figure \ref{fig:abacus}(c) shows the final output: the multilayer community affiliation. A node is part of a multidimensional community if it is part of all communities composing it.

Node $10$ is an example of a final interesting thing: it is part of no multidimensional community because its affiliation is a weird combination of communities. We can decide to let it be without community affiliation, or to allow it to be part only of its non-multilayer communities.

There are other algorithms solving the same problem and inspired by frequent pattern mining\cite{silva2012mining}\cite{zeng2006coherent}.

The last solution for this section is inspired by ensemble clustering. Again, we have a community per layer. Then we use the same strategy I outlined in Section \ref{sec:cd-partition-practical}: we consider each community partition in each layer as a valid clustering of the same underlying relationship via different datasets\cite{tagarelli2017ensemble}. We then find the ``true'' clustering, which is the partition that is the closest one to the combination of all partitions.

Aggregating communities across layers has some benefits. For instance, it might solve the resolution problem of modularity\cite{taylor2016enhanced} that I discussed in Section \ref{sec:cd-eval-mod}. However, all these methods have the downside of relying more or less on the same assumption: that the layers are correlated to each other. While this might not be a bad assumption to start with\cite{hristova2014keep}, disassortative layers exist and might represent a problem.

\section{Multilayer Adaptations}

\subsection{Multilayer Modularity}
I already mentioned how obsessed networks scientists are with modularity, so you know what's coming next: multilayer modularity\cite{mucha2010community}. Suppose we're using the Louvain method, which grows communities node by node. If we found a triangle in a layer, can we extend it by taking a node in a different layer? Intuitively yes, the edge should count because the node is the same. However, if we were to represent this as a flat network, the new node is not densely connected to the rest of the triangle: a node couples only with itself, not with its community fellows. So the coupling edges have to count in some special way.

\begin{figure}
\centering
\includegraphics[width=.83\columnwidth]{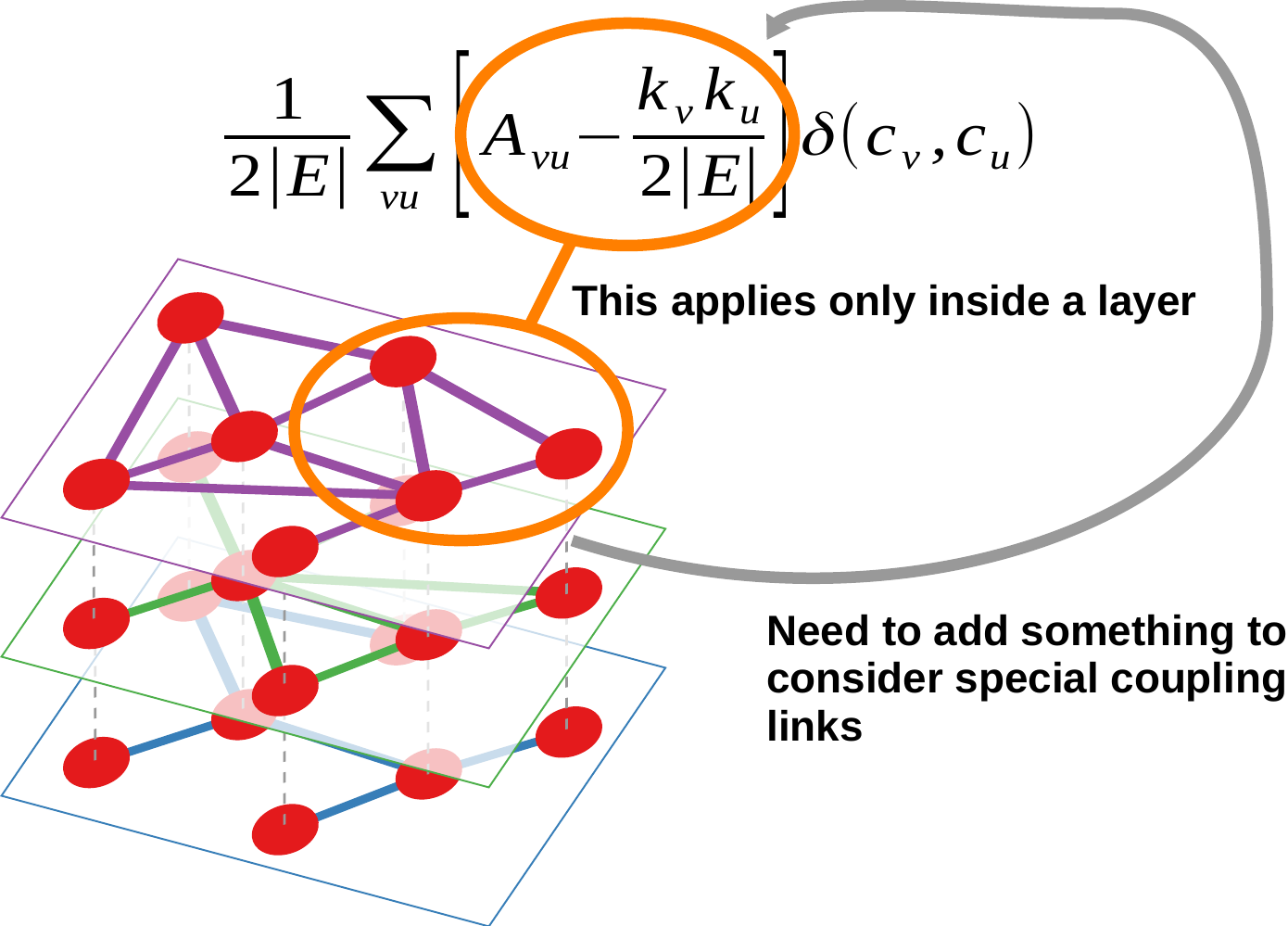}
\caption{The issues in applying modularity to the multilayer case. The part looking at the community density only applies inside a single layer, thus we need to add something to it to consider the special coupling edges.}
\label{fig:modularity-mono}
\end{figure}

In practice, standard modularity works well in each layer separately. Consider Figure \ref{fig:modularity-mono}: in modularity, the part testing for the density of the community is $A_{uv} - \dfrac{k_u k_v}{2|E|}$. If we use this same part for the inter-layer coupling, we would end up with a case in which the community cannot be expanded across layers, because there are only sparse connections between layers. A node couples only with itself in a different layer, not connecting to its community members, making a multi-layer community sparser than it actually is. So we need to add something that will allow us to count the coupling links, so that we don't end up with the trivial result of all mono-layer communities.

The full formulation of multilayer modularity is the following:

$$ \dfrac{1}{2(|E| + |C|)} \sum \limits_{vusr} \left[ \left(A_{vus} - \gamma_s \dfrac{k_{vs}k_{us}}{2|E_s|}\right) \delta_{sr} + C_{vsr} \delta_{uv} \right] \delta(c_{us},c_{vr}).$$

\begin{figure}
\centering
\includegraphics[width=.83\columnwidth]{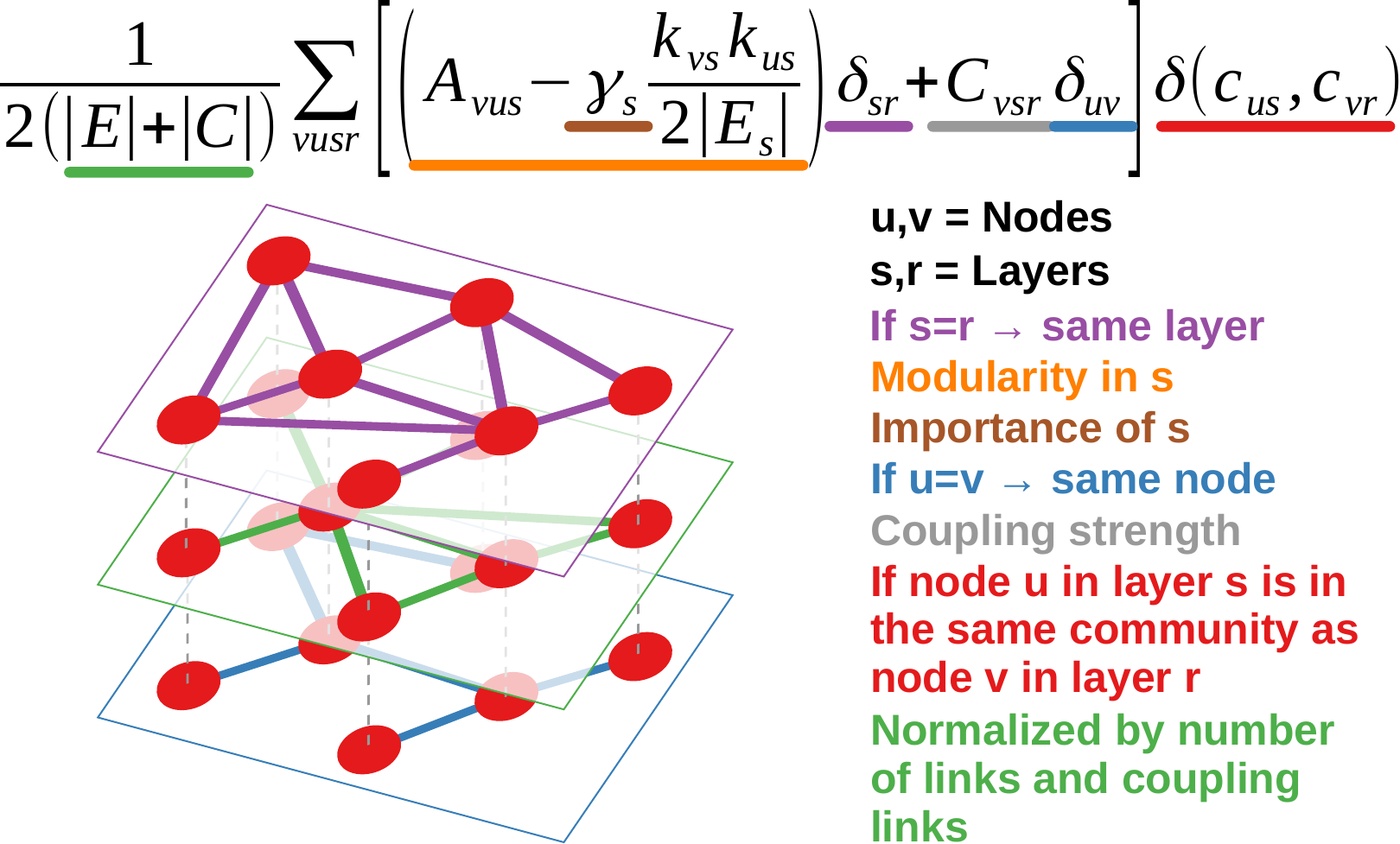}
\caption{The adaptation of modularity to the multilyer setting. Each part of the formula is underlined with a color corresponding to its interpretation.}
\label{fig:modularity-multi}
\end{figure}

Let's break it down -- and you can check Figure \ref{fig:modularity-multi} for a graphical representation and you can always go back to Section \ref{sec:cd-eval-mod} to read more about the notation of classical modularity and compare it with this formulation. $E$ and $C$ are the sets of (intra-layer) edges and (inter-layer) couplings. $A_{vus}$ is our multilayer adjacency tensor, it is equal to $1$ if nodes $u$ and $v$ are connected in layer $s$, and it is $0$ otherwise. $k_{us}$ and $k_{vs}$ are the degrees of $u$ and $v$ in layer $s$, respectively. $|E_s|$ is the number of edges in $s$.

The major complication in multilayer modularity is that we have many Kronecker deltas ($\delta$). The first is $\delta_{sr}$: its role is to make sure that standard modularity is applied inside a layer (when $s = r$, $\delta_{sr} = 1$, because $s$ and $r$ are the same layer). The second is $\delta_{uv}$ and it checks whether we are looking at two nodes that are coupled across layers: $\delta_{uv}$ is $1$ if $u = v$. Note that $\delta_{sr}$ and $\delta_{uv}$ are mutually exclusive. If $s = r$ we are in the same layer and so it must be that $u \neq v$, because we don't have self loops. On the other hand, if $u = v$ then $s \neq r$, because by definition coupling connections go across layers, thus $s$ and $r$ must refer to different layers. Both deltas can be zero if we're looking at uncoupled nodes in different layers. The final delta, $\delta(c_{us}, c_{vr})$ is the same as in standard modularity, it is equal to $1$ only if we are looking at nodes inside the same community, i.e. $c_{us} = c_{vr}$.

With $\gamma_s$ we can regulate how important each layer $s$ is for the community. In practice, $\gamma$ is a vector of weights, one per layer $s$ of the network.

$C_{vsr}$ is the strength of the coupling link, which is a parameter just like $\gamma$ is: you can decide how strong the layer couplings should be. It matters only when we're looking at the same node connected by a coupling link across layer ($u = v$), and so $\delta_{uv}$ is $1$. In this case, nothing else matters, because $\delta_{sr}$ is $0$ (because $s \neq r$), so the standard modularity part cancels out.

Just like in standard modularity, only nodes in the same community contribute to the sum, so when node $u$ in layer $s$ is in the same community as node $v$ in layer $r$ (meaning that $\delta(c_{us}, c_{vr}) = 1$). This is normalized by the number of edges ($|E|$) across all layers plus all coupling links ($|C|$).

\begin{figure}
\centering
\begin{subfigure}{.4\columnwidth}
\includegraphics[width=\textwidth]{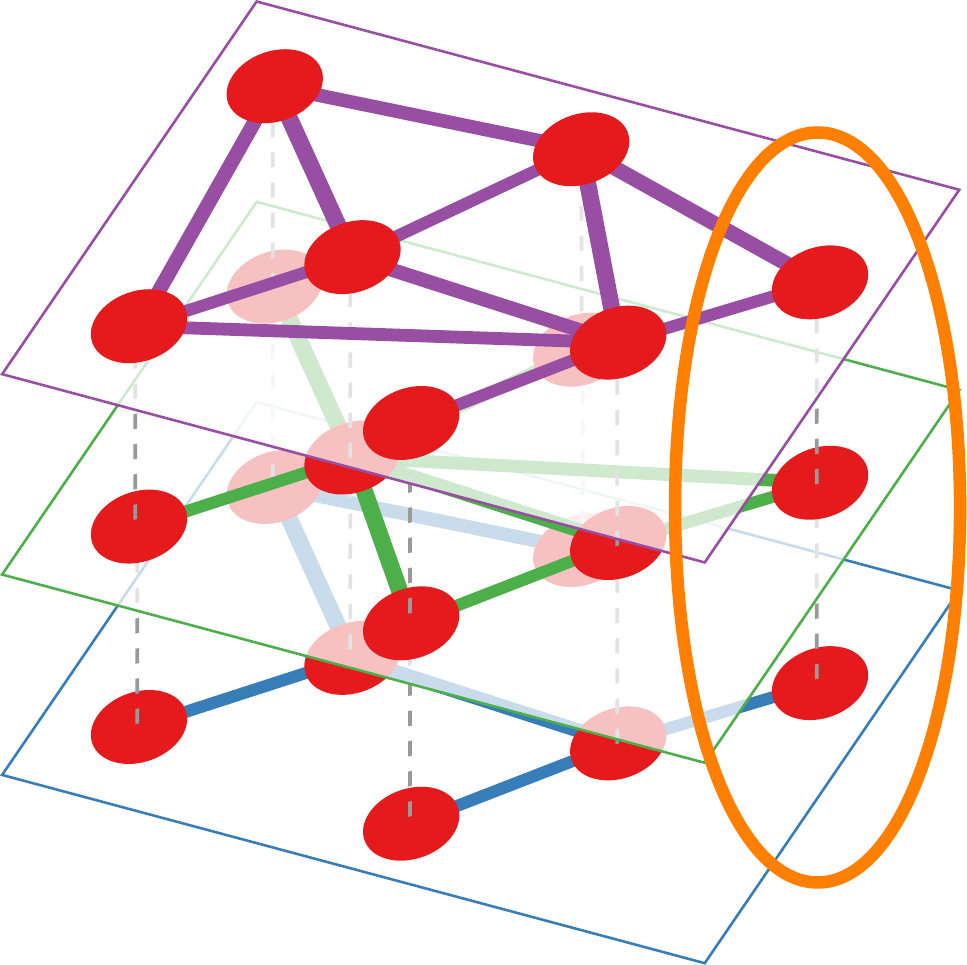}
\caption{}
\end{subfigure}\qquad
\begin{subfigure}{.4\columnwidth}
\includegraphics[width=\textwidth]{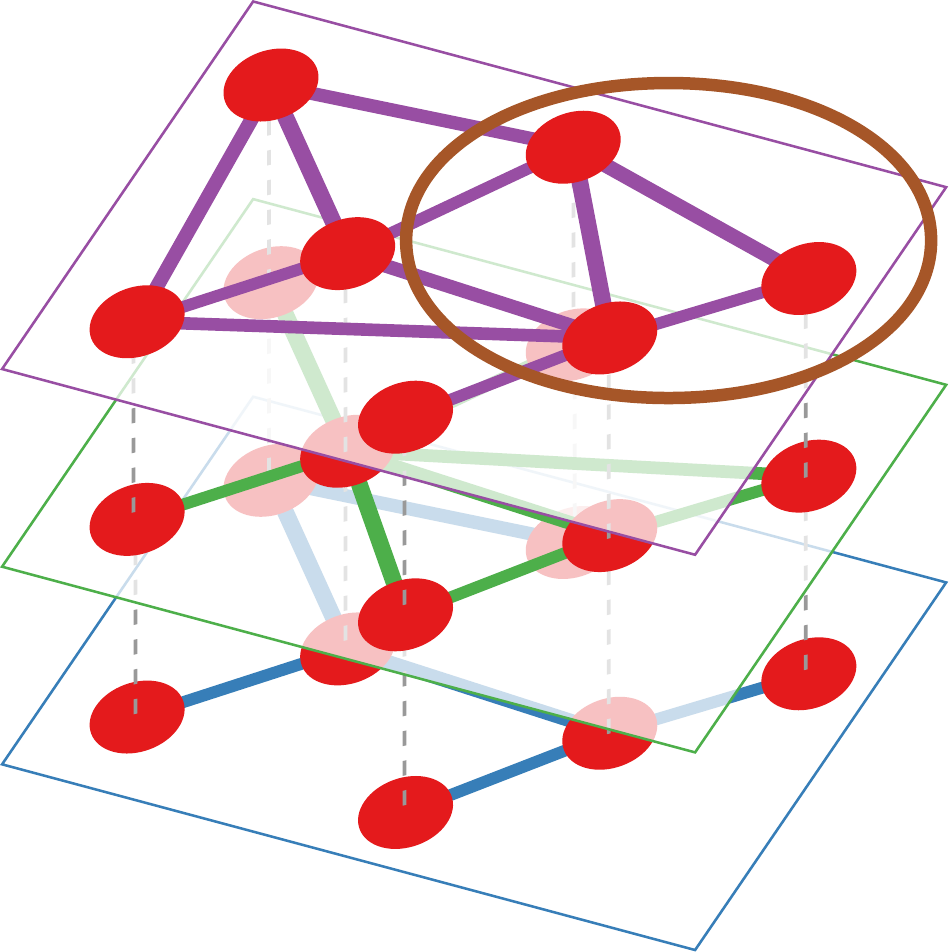}
\caption{}
\end{subfigure}
\caption{The effect of the $C_{vsr}$ parameter on multilayer modularity. (a) High values imply pillar communities. (b) Low values imply flat communities (right).}
\label{fig:modularity-multi-c}
\end{figure}

If you decide that your inter layer couplings are very strong, you'll end up with ``pillar communities'' where nodes tend to favor grouping with themselves across layers: the inter layer couplings trump any intra-layer regular edge. If your inter layer couplings are weak (low $C_{vsr}$) then you'll end up with ``flat communities'' as nodes prefer to group with other nodes in the same layer. I show an example in Figure \ref{fig:modularity-multi-c}.

\begin{figure}
\centering
\begin{subfigure}{.4\columnwidth}
\includegraphics[width=\textwidth]{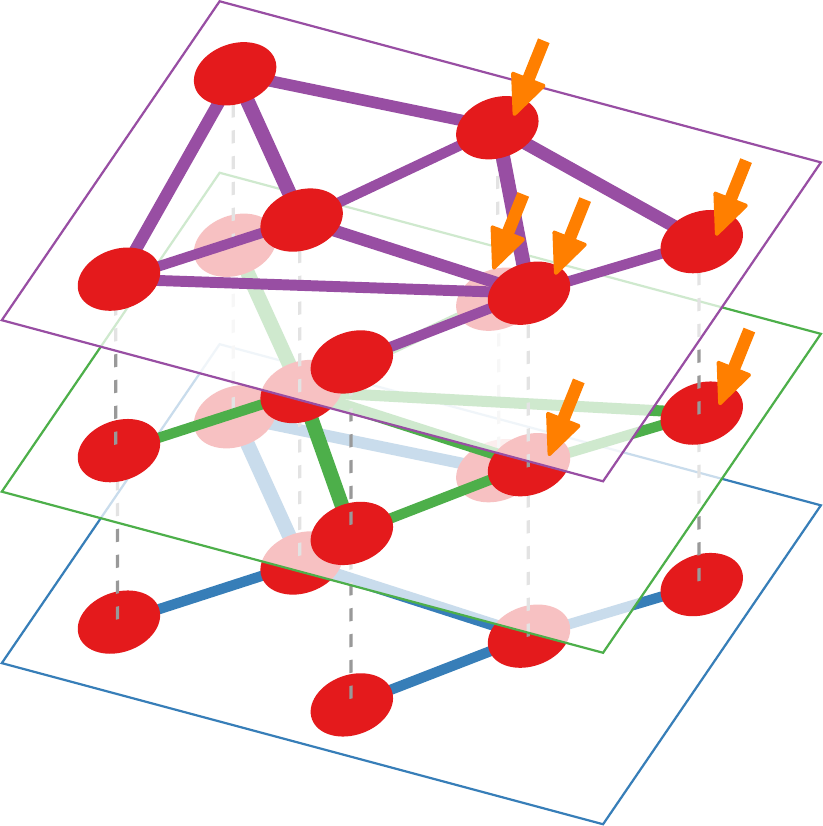}
\caption{}
\end{subfigure}\qquad
\begin{subfigure}{.4\columnwidth}
\includegraphics[width=\textwidth]{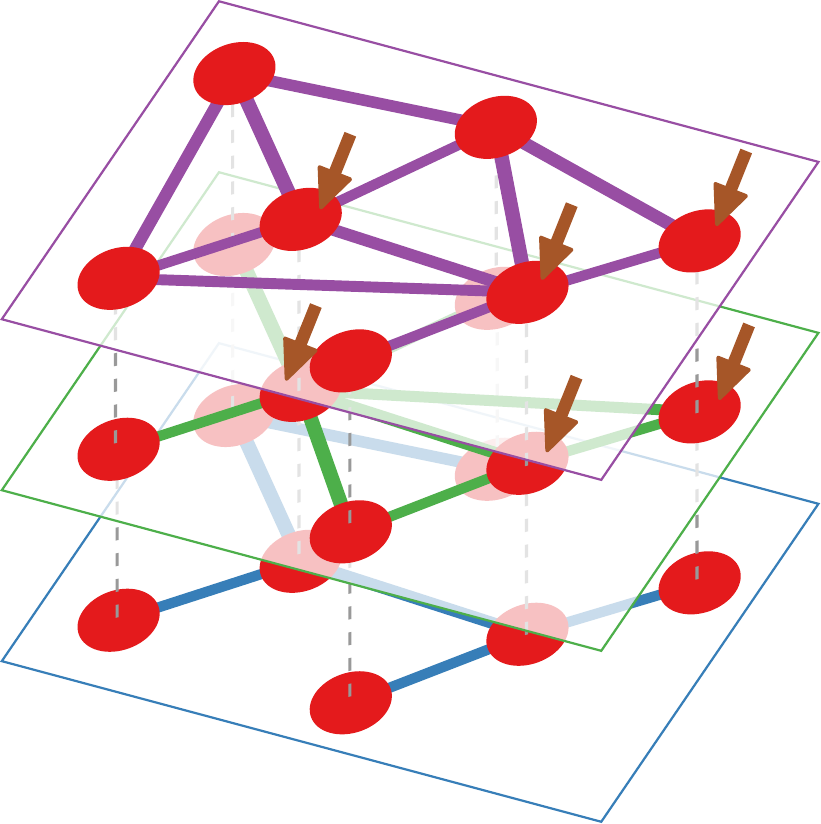}
\caption{}
\end{subfigure}
\caption{The effect of the $\gamma_s$ parameter on multilayer modularity. (a) High $\gamma$ for the top (purple) community and low for the mid (green) community. (b) Low $\gamma$ for the top (purple) community and high for the mid (green) community.}
\label{fig:modularity-multi-gamma}
\end{figure}

Instead, $\gamma$ allows you to indicate some layers as more important than others, as I show in Figure \ref{fig:modularity-multi-gamma}. If the purple layer is more important than the green one, multilayer modularity will group in the community a node that is not connected with the two nodes in the green layer. If we flip the $\gamma$ values to make green more important than purple, the situation is reversed, and modularity will return different communities.

An alternative way to adapt modularity maximization to multilayer networks is to adapt the Louvain algorithm (see Section \ref{sec:hcd-recursive}) to handle networks with multiple relation types\cite{jutla2011generalized}.

\subsection{Other Approaches}
Modularity is not the only algorithm that can be adapted to multilayer networks. Following the same strategy we applied for bipartite networks, we can adapt the kitchen sink of community discovery to multilayer structures.

\begin{figure}[b]
\centering
\includegraphics[width=.4\columnwidth]{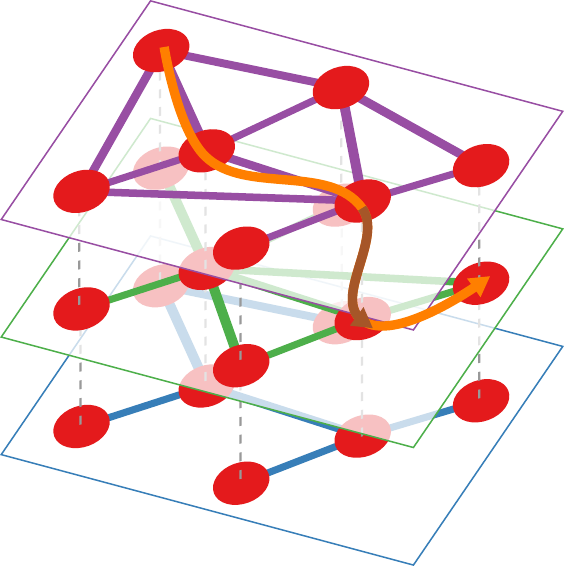}
\caption{In multilayer random walk community discovery, the random walker (orange) has a certain probability to perform a layer jump (brown).}
\label{fig:mcd-infomap}
\end{figure}

The random walks approach can be multilayer\cite[-0.5in]{kuncheva2015community}\cite{de2015identifying}. In practice, we have a random walker that normally selects edges in the same layer, but it has a special rule that sometimes allows it to go through a coupling edge, and then resume its normal random walk. Figure \ref{fig:mcd-infomap} shows an example of such a move. Many ways have been proposed to expand random walks to multilayer networks, and a special set of algorithms focuses on local-first community discovery. We center our focus on a specific (set of) query nodes and we find the communities surrounding them\cite[-0.25in]{jeub2017local}\cite[1.3in]{interdonato2017local}.

Similarly, one could propagate labels across layers with special rules\cite{boutemine2017mining}, and thus adapt the fast label propagation algorithm to find multilayer communities. First, we cannot use synchronous label propagation: like in the bipartite case, also for multilayer networks we could be stuck with label oscillation (Section \ref{sec:bcd-direct}), this time across layers. Second, the authors define a quality function that regulates the propagation of labels. This is done because there might be layers that are relevant for a community and layers that are not. We do not want a community, which is very strong in some layers, to ``evaporate away'' just because in most layers the nodes are not related.

\begin{figure}
\centering
\begin{subfigure}{.4\columnwidth}
\includegraphics[width=\textwidth]{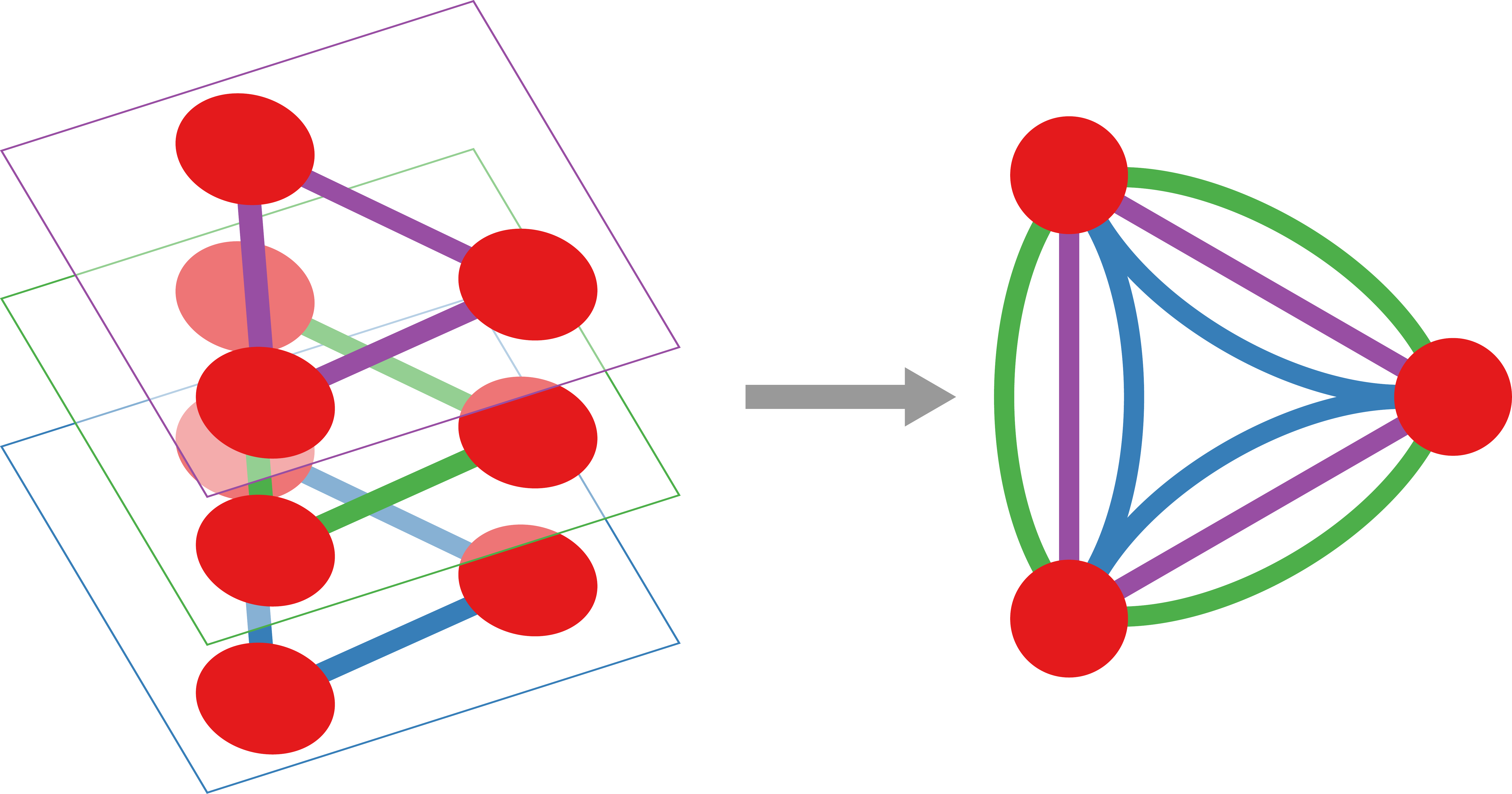}
\caption{}
\end{subfigure}\qquad
\begin{subfigure}{.4\columnwidth}
\includegraphics[width=\textwidth]{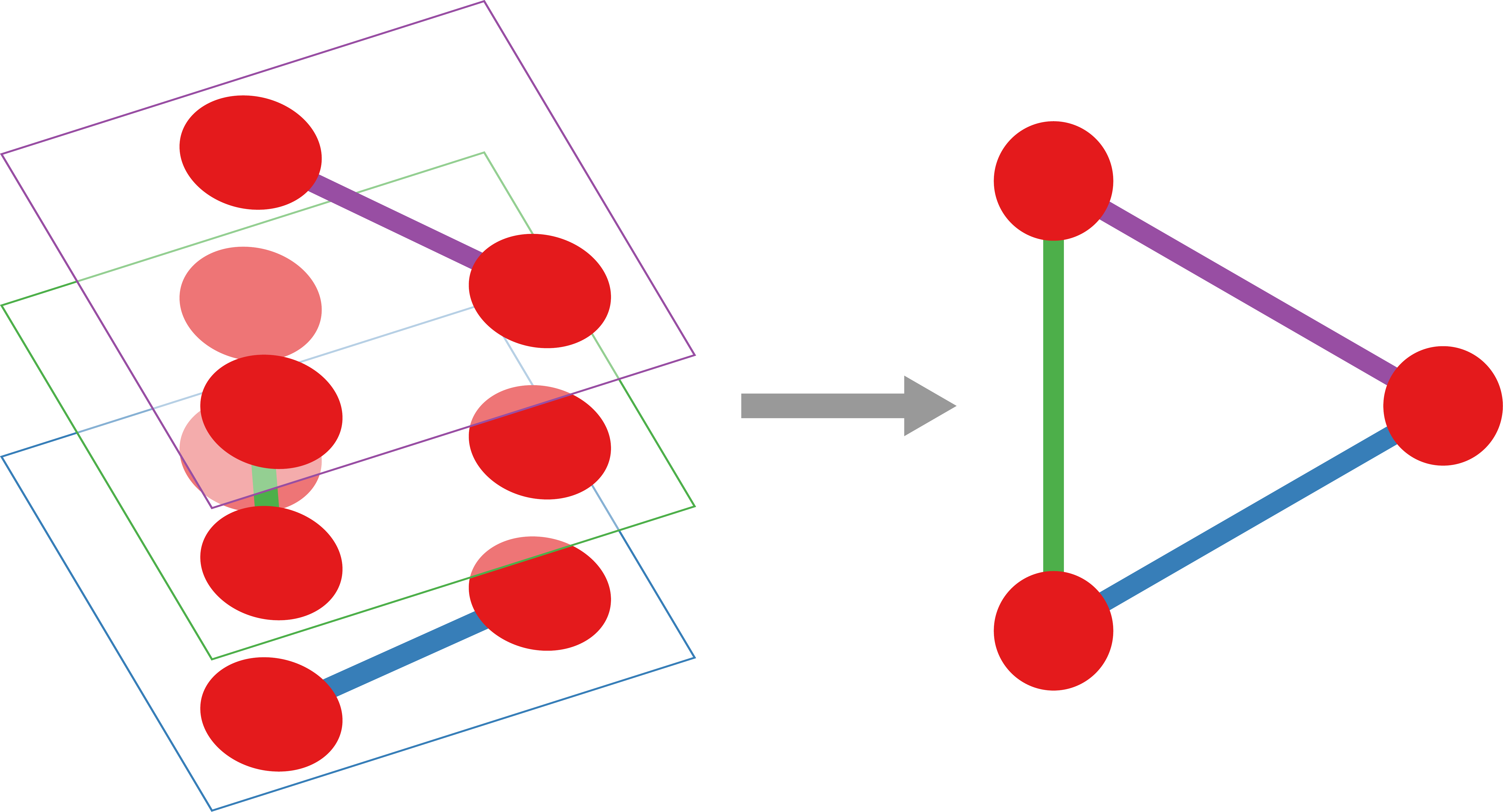}
\caption{}
\end{subfigure}
\caption{(a) A 3-3-AND-clique. (b) A 3-3-OR-clique.}
\label{fig:multilayer-density}
\end{figure}

Next on the menu is k-clique percolation\cite{afsarmanesh2016finding}. In this scenario, we need to redefine a couple of concepts, particularly what a clique is in a multilayer network, and how we determine when two multiplex cliques are adjacent. For the first case, we need to talk about k-l-cliques: a set of $k$ nodes all connected through a specific set of $l$ layers. Moreover, there are two ways for nodes to be all connected via the layers: all pairs of nodes could be connected in all layers at the same time, or they could be connected in only one layer at a time. The first type of clique is an k-l-AND-clique, the second type is a k-l-OR-clique. Figure \ref{fig:multilayer-density} shows an example.

\begin{figure}
\centering
\begin{subfigure}{.4\columnwidth}
\includegraphics[width=\textwidth]{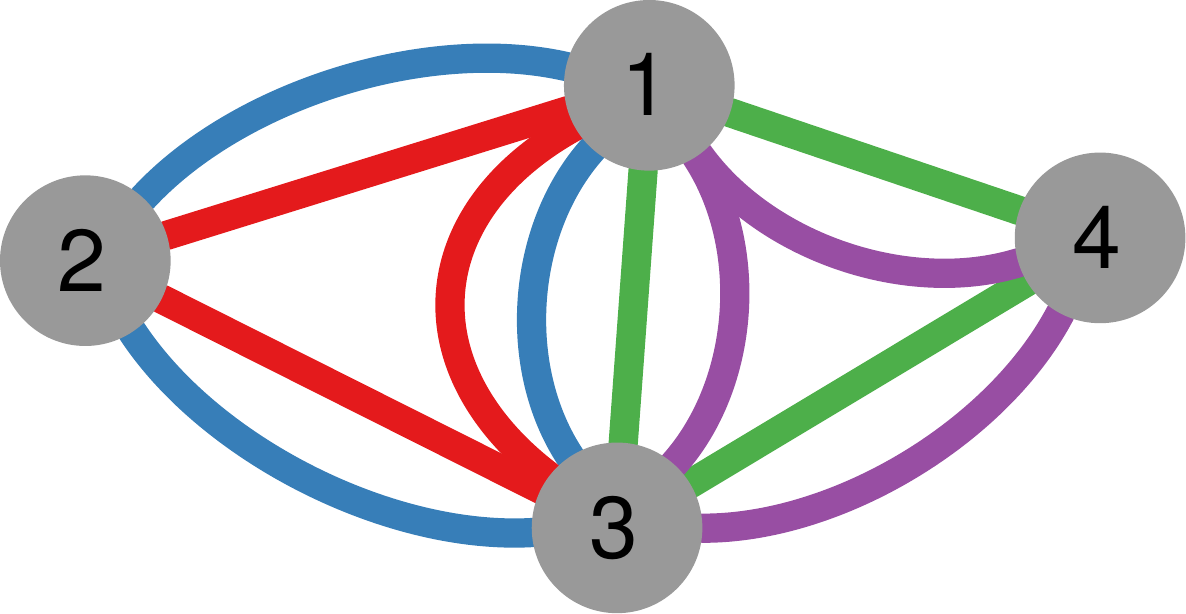}
\caption{}
\end{subfigure}\qquad
\begin{subfigure}{.4\columnwidth}
\includegraphics[width=\textwidth]{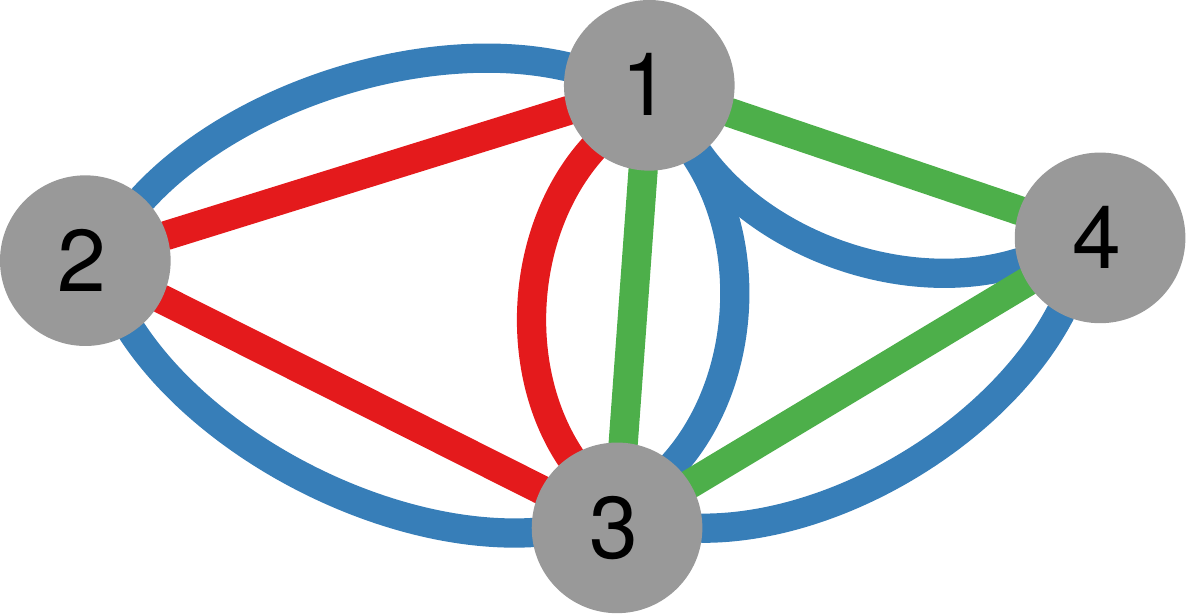}
\caption{}
\end{subfigure}
\caption{Two 2,2-cliques sharing a 1,2-clique. The edge color represents the edge layer. If $m=1$ (a) does NOT percolate because the rightmost clique does not share a layer with the leftmost clique; (b) DOES percolate, since the two cliques share the blue layer.}
\label{fig:multilayer-clique-perc}
\end{figure}

It becomes clear that two k-l-cliques might share $(k-1)$ nodes across different layers. In such a case, we need some care in defining a parameter to regulate percolation. We need a minimum number $m$ of shared layers to allow the percolation. If the two cliques do not share at least $m$ layers, even if they share $k-1$ nodes they are not considered adjacent. Figure \ref{fig:multilayer-clique-perc} shows an example.

The final adaptation we consider is the stochastic blockmodels\cite[-3in]{de2017community}\cite[-2in]{stanley2016clustering}. Just like we saw for overlapping and bipartite SBMs, we need to add an additional matrix into our expectation maximization framework. For overlapping and hierarchical SBMs it was a community-community matrix telling us how strongly communities connect to each other. In this case, instead, we have a layer-layer matrix telling how likely it is for two layers to have the same edges.

This is neat, because it allows us to model assortative, disassortative, and non-assortative layer relationships. In the first case, being connected in a layer increases the chances of being connected in the other layer: it is more likely to be friends on Facebook if we are also friends in real life. The second case is the opposite: a relation in one layer makes it harder to be connected on another: if you attack me in an online game it's less likely that we'll be friend. The non-assortative case covers the scenario in which being connected in a layer gives us no information on whether we're connected in the other.

All these cases -- the special layer-layer jump probability in random walks, the quality function in label propagation, and the layer-layer probability matrix in SBM -- allow us to select the relevant layers for a multilayer community. Thus, we can keep a node group together even if in many layers the nodes don't connect to each other, probably because those layers were disassortative with the layers in which the community appears.

Still an adaptation of already existing methods, but more properly multilayer, is the approach of multilink similarity\cite{mondragon2018multilink}. This is heavily inspired by the hierarchical link clustering we saw in the overlapping case (Section \ref{sec:ocd-link}). The objective is the same: to define a link-link similarity measure that we can use to progressively merge link communities in a hierarchical fashion.

The similarity measure for multilayer networks is:

$$ S_{(u, k), (v, k)} = \epsilon z^{\beta_{uk,vk}} + (1 - \epsilon) |P_{uv\bar{k},2}|. $$

Here, $\epsilon$ and $z$ are parameters between $0$ and $1$ you can set. The real work is made by $\beta_{uk,vk}$ and $P$. $\beta_{uk,vk}$ takes values between $0$ and $1$, and it is one minus the share of layers in common connecting $uk$ and $vk$. So, for instance, if the node pairs $uk$ and $vk$ connect in mutually exclusive layers -- i.e. no layers in common -- then $\beta_{uk,vk} = 1$. On the other hand, if they connect in exactly the same layers and no other layer, then $\beta_{uk,vk} = 0$. Since our parameter $z$ is between $0$ and $1$, the whole term tells us how much we weight in the link-link similarity the absence of layers, because mutually exclusive layer set will just be $z^1 = z$.

$|P_{uv\bar{k},2}|$ is instead the count of paths of length $2$ between $u$ and $v$ that do not pass through $k$, plus the number of layers in which $u$ and $v$ connect directly to each other. This is normalized by the lowest degree between $u$ and $v$, excluding all connections going to $k$. Thus, in practice, the parameter $\epsilon$ regulates the weight we want to give to the number of the shared layers of edges $uk$ and $vk$, over the local multilayer clustering of nodes $u$ and $v$. If $\epsilon = 1$ we only care about clustering together node pairs that connect through the same sets of layers.

\begin{figure}
\centering
\includegraphics[width=.4\columnwidth]{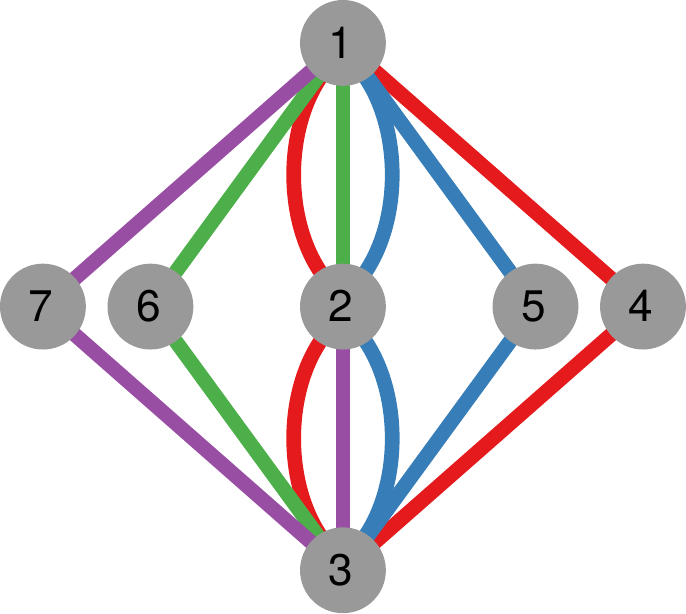}
\caption{A multilayer network, with the edge color encoding the layer in which it appears.}
\label{fig:multilink-similarity}
\end{figure}

Let us consider Figure \ref{fig:multilink-similarity} and attempt to estimate the similarity of node pairs $(1, 2)$ and $(1, 3)$. The pairs share two out of four possible layers, thus $\beta_{uk,vk} = 0.5$. There are also four other paths going from $1$ to $3$ that do not use node $2$. Both node $1$ and $3$ have degree equal to four -- remember we don't count the connections going through node $2$ --, thus $|P_{uv\bar{k},2}| = 4 / 4 = 1$. If we were to set $z = 0.6$ and $\epsilon = 0.4$, then $ S_{(u, k), (v, k)} = (0.4 \times 0.6^{0.5}) + ((1 - 0.4) \times 1) \sim 0.91$.

Other adaptations I'm not going to discuss in details are an extension of local community discovery to multilayer network\cite{hmimida2015community}. This is based on redefining the simple concepts of degree and neighborhood for the multilayer case, and then apply a classical local exploration approach, as the one we saw in Section \ref{sec:cd-partition-local}.

Another class of solutions include representing multilayer graphs in a lower dimensional space with a technique known as ``Grassmann manifold''\cite{dong2013clustering}.

\section{Multilayer Density}
We have talked about how to find communities in multilayer networks. Implicitly, we're resting on our definition, which is based on density. But what actually \textit{is} multilayer density? This turns out to be an ambiguous concept.

Let's go back to Figure \ref{fig:multilayer-density} for a moment. Is a group of nodes ``multilayer densely'' connected when they are connected in all layers (Figure \ref{fig:multilayer-density}(a))? Or is it that you need to look at all connections across layers (Figure \ref{fig:multilayer-density}(b))? To use a different perspective, let's represent multilayer networks with a labeled multigraph. In our first example, we have a multigraph with connections in all labels. In the second example we sill have a triangle, so the multigraph is dense. But the two concepts are not the same. Which of the two are we looking at?

This is another case when one has to make their own judgment. The answer depends on the type of analysis, and the type of data, you are looking at. In some cases, you want connections in all layers. In some others, you are ok with looking at all layers to find communities. You cannot rely on a fixed definition of communities based on density, because it cannot apply to all scenarios.

Thus, you need to have measures to determine when you are in one case and when you are in another. I proposed a couple in a paper of mine\cite{berlingerio2011finding2}. We decided to call them ``redundancy'' and ``complementarity''. Note that this redundancy here has nothing to do with the cousin of the local clustering coefficient we talked about in Section \ref{sec:density-clustering}.

Redundancy is the easiest of the two. To consider a set of nodes to be densely connected in a multilayer network, we require that the edges appear in \textit{all} the layers we are interested in. If we have a community $c$ containing a set of nodes, and we test it over the set of layers $L$, redundancy is simply the share of actual edges over all the edges we would need to connect every pair of nodes through every layer:

$$ \rho_c = \sum \limits_{u,v \in P_c} \dfrac{|\{l : \exists (u,v,l) \in E\}|}{|L| \times P_c}, $$

where $P_c$ is the set of all node pairs in $c$.

Complementarity is a little trickier, because it is the intersection of three concepts: variety, exclusivity, and homogeneity. Variety is through how many different layers nodes in community $c$ connect to each other. This is simply the number of layers in $c$ divided by the number of layers in the network: $\dfrac{|L_c| - 1}{|L| - 1}$. Exclusivity counts how many pairs of nodes connect
in just one layer within $c$: if $\overline{P_{c,l}}$ is the number of node pairs in $c$ which connect only in layer $l$, the exclusivity is $\dfrac{\sum_{l \in L} \overline{P_{c,l}}}{|P_c|}$. Finally, homogeneity estimates how uniformly the edges in $c$ distribute across layers, which is $1 - \dfrac{\sigma_c}{\sigma_c^{\max}}$. Here, $\sigma_c$ is the standard deviation of the distribution of the edges in $c$ across layers, and it is normalized by its theoretical maximum.

\begin{figure}
\centering
\begin{subfigure}{.45\columnwidth}
\includegraphics[width=\textwidth]{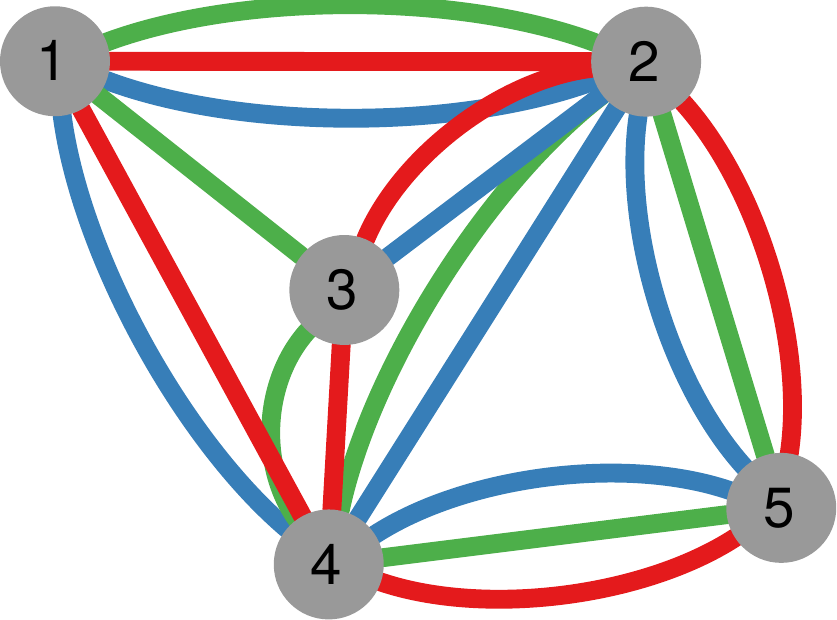}
\caption{}
\end{subfigure}\qquad
\begin{subfigure}{.35\columnwidth}
\includegraphics[width=\textwidth]{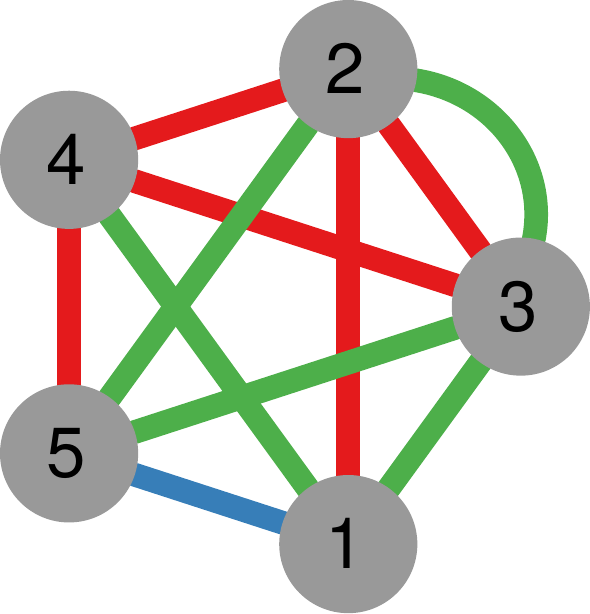}
\caption{}
\end{subfigure}
\caption{Two examples of different types of multilayer density. The edge color encodes the layer in which the edge appears. (a) A high redundancy case. (b) A high complementarity case.}
\label{fig:red-vs-compl}
\end{figure}

Let's see some examples to put all these Greek letters to good use. Let's consider Figure \ref{fig:red-vs-compl}, assuming that the network has a total of three layers. In Figure \ref{fig:red-vs-compl}(a) we have a high redundancy case. Since the community includes all layers of the network, the numerator of redundancy is simply the count of edges: $18$. The denominator is $3$ (the number of layers) times the number of node pairs in the community, which is $10$, since we have $5$ nodes. Thus the redundancy is $18 / 30 = 0.6$.

Figure \ref{fig:red-vs-compl}(b) is instead a high complementary case. Variety is $1$ by definition, since the community contains all layers of the network. Exclusivity is $9 / 10$, because there is one pair of nodes (nodes $2$ and $3$) which is connected in two layers, and thus it is not counted. Finally, the standard deviation of the distribution of the edges in $c$ across layers is the standard deviation of the vector $[5, 1, 5]$, since there are five edges in the red and green layer, and only one in the blue layer. This is $\sim 1.88$, which is exactly two thirds of the maximum possible, leaving us with a total homogeneity of $0.33$. Thus, complementarity is $1 \times 0.9 \times 0.33 = 0.297$, penalized by the low representation of the blue layer in the community.

\section{Mopping Up Community Discovery}
We have finally reached the end of this extremely simplified trip through community discovery. It all started with a simple, nothing-up-my-sleeve definition of what communities are in complex networks:

\begin{center}
\textit{Communities are groups of nodes densely connected to each other and sparsely connected to nodes outside the community.}
\end{center}

Yet, as we progressed in this journey, we realized that there is a gazillion ways in which such definition needs to be stretched, it is not the full story, or simply does not work. To sum up our list of grievances:

\begin{itemize}
\item The standard definition implies assortative communities: nodes in the same communities tend to connect to each other more than random. However, disassortative communities are a thing as well: nodes in a disassortative community tend to connect to nodes in different communities (Section \ref{sec:cd-partition-sbm}).
\item If our network is evolving, communities are evolving too. The information from past communities should be taken somehow into account, making the communities at time $t + 1$ a compromise between their density and their similarity with the communities at time $t$ (Section \ref{sec:cd-partition-evo}).
\item Communities can be local (Section \ref{sec:cd-partition-local}), meaning that we might prevent ourselves from discovering all members of a community and thus leaving out some parts of the network that would make the community denser.
\item Measures defined with the idea of maximizing density and external sparsity have counter-intuitive behavior, for instance modularity has resolution limit, degeneracy of good solutions, and limits in its field of vision (Section \ref{sec:cd-eval-mod}).
\item We want communities to be interpretable and/or to tell us something about the real world properties of the entities we are grouping. This can be achieved by finding the communities maximizing the normalized mutual information of some other data we have about nodes. While this is a worthwhile task for many real world applications, it can -- and most often does -- clash with the internal density requirement (Section \ref{sec:cd-eval-nmi}). This is because node ``meta''data is just data, and it doesn't necessarily have any relevance to the edge creation process of your network.
\item Many networks have a hierarchical community structure, where we can find communities of communities. However, by definition, these communities must be more sparsely connected that the communities forming them. This should not make them any less valid, as they are a useful tool to explain many natural structures (Chapter \ref{cha:hcd}).
\item It is equally a fact that many real world networks have overlapping communities: nodes can be part of multiple communities at the same time. But if it is true that the more communities two nodes have in common the more likely they are to connect, we end up with networks in which the overlap of communities is denser than the communities themselves, which contradicts our original definition (Chapter \ref{cha:ocd}).
\item We can find communities in bipartite networks as well. However, by definition, these communities will be somewhat sparse, given that we forbid connections between nodes of the same type. We need to modify our definition of density accordingly (Chapter \ref{cha:bcd}).
\item However, there are proper ways to find bipartite communities -- and even regular unipartite communities -- by adapting classical data clustering algorithms from data mining. These algorithms will simply take as input the adjacency matrix of the graph as if it was an attribute table. The meaning of the communities found this way would change, though: these are not any more nodes densely connected to each other, but rather nodes connected to the same neighbors (Section \ref{sec:bcd-clustering}).
\item Finally, as we saw in this chapter, we need to adapt density to the multilayer case as well. However, this is necessarily an ambiguous operation with multiple valid alternatives, as multilayer density can be intended both in a ``redundancy'' and in a ``complementary'' sense.
\end{itemize}

The moral of this story is that you can intend ``communities'' in complex networks in a thousand different ways. Performing community discovery is not a neutral operation you would do like adding two numbers. It is a complex problem that starts from the question: what is a community \textit{for me}? Or for \textit{my data}? What am I \textit{actually} looking for? Just picking an algorithm because someone tells you so, or because it is the most used, is guaranteed to misfire.

This is also why, if you encounter somebody who uncritically tells you the classical definition or uses it in their paper without at least a mention of these caveats, you should tell them they're wrong, otherwise Michele would have written these 100 pages for nothing.

\section{Summary}

\begin{enumerate}
\item Multilayer community discovery is the adaptation of community discovery for networks with multiple edge types. The simplest approach is to flatten the multilayer structure by collapsing all layers into a weighted simple graph.
\item Alternatively, you can find communities using a non-multilayer algorithm on each of your layers separately. Then, you would merge the results.
\item One can adapt modularity to multilayer networks by adding a term that takes into account the inter-layer connection strength, binding the nodes in the communities across layers. There are similar adaptations for random walks and label percolation approaches.
\item The concept of ``multilayer density'' is intrinsically ambiguous. One could intend it as the requirement of all nodes connected through all layers at the same time, or in a single different layer for each node pair.
\end{enumerate}

\section{Exercises}

\begin{enumerate}
\item Take the multilayer network at \url{http://www.networkatlas.eu/exercises/40/1/data.txt}. The third column tells you in which layer the edge appears. Flatten it twice: first with unweighted edges and then with the count of the number of layers in which the edge appears. Which flattening scheme returns a higher NMI with the partition in \url{http://www.networkatlas.eu/exercises/40/1/nodes.txt}? Use the asynchronous label percolation algorithm (remember to pass the edge weight argument in the second case).
\item Using the same network, perform label percolation on each layer separately. Build the $|V| \times |C|$ table, perform kMeans (setting $k = 4$) on it to merge the communities. Does this return a higher NMI when compared with the ground truth?
\item Calculate the redundancy of each community you found for the previous exercises.
\end{enumerate} 

\part{Graph Mining}\label{par:mining}

\chapter{Frequent Subgraph Mining}\label{cha:mining-base}
Graph mining is a category of network analysis including the application of machine learning and data mining techniques to the analysis of complex networks. You got your machine learning primer back in Chapter \ref{cha:machine-learning}, which you should use as a reference in case you find something here you don't understand. This chapter focuses on one of the traditional ways to do graph mining: to find frequent patterns in the graph. In further chapters, I will then cover more recent techniques which connect graph mining with the developments in neural networks: first by transforming the graph in a less complex format hat can be understood by traditional neural network architectures (Chapters \ref{cha:mining-embeddings} and \ref{cha:mining-rwemb}); and then by creating new neural network architectures that can exploit the complex structure of a graph directly (Chapter \ref{cha:mining-deep}).

The idea in this chapter is to describe your graph by counting all the different motifs that compose it -- sometimes called graphlets\cite{shervashidze2009fast}. I will start with a brief introductory section connecting machine learning and network science, and then delve deeper into why and how we perform frequent subgraph mining.

\section{Machine Learning with Graphs}\label{sec:machine-learning-example}
The idea of applying machine learning to network science is to train a model to find something interesting about your network. To train your model you need to give it some features in some sort of numerical form. This is important because one key thing you need to do during training (and then during the evaluation) is to \textit{quantify} how well the model is doing. The problem with networks is that they are structures, and not numbers. It is difficult to quantify something with them. You need to transform the structure into a (set of) number(s) -- and do it in a smart way.

\begin{figure}
\centering
\includegraphics[width=\textwidth]{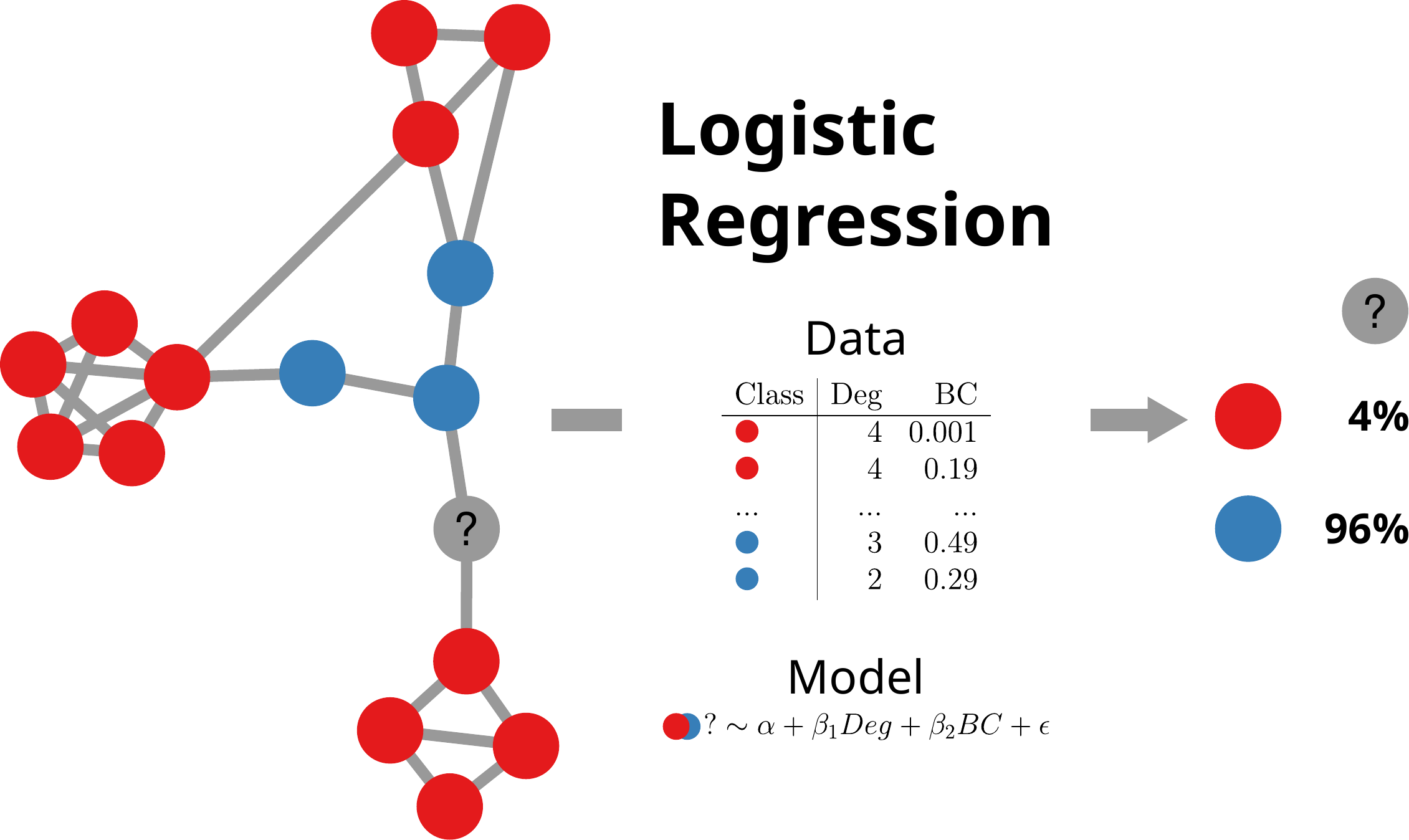}
\caption{An example of classical machine learning on graphs. The node colors represent the node's classes.}
\label{fig:ml-graphs-holistic}
\end{figure}

Traditionally, you would just collect a bunch of characteristics of your network (or nodes, or edges) and feed them to a classifier. In Figure \ref{fig:ml-graphs-holistic}, we give the nodes' degree and betweenness centrality to a simple logistic regression model. The model can figure out that blue nodes have high centrality but low degree, so that unclassified node we have in grey is probably blue. This is how Rolx worked, all the way back in Section \ref{sec:centr-roles-classic} -- but here we take a more general approach than simply inferring node structural roles.

We can shift the perspective from learning something about the nodes to learning something about the network in its entirety. The same approach of computing statistics for a classifier can be used here as well, for instance to estimate the similarity between two networks (Section \ref{sec:netsimil-simil}). You can use the degree distribution, the global clustering coefficient, etc. You can also use a ``bag of nodes'' approach, in which a network is described by the list of individual properties of its nodes. Finally, and more relevantly for this chapter, you can enumerate all subgraph motifs and their relative occurrences to compare two networks, boiling down a complex network to the counts of its substructures.

\section{Network Motifs}\label{sec:mining-motifs}
Our experience with modeling real world networks tells us that they are not random: they are an expression of complex dynamics shaping their topology. Networks will tend to have overexpressed connection patterns. Nodes and edges will form different shapes much more -- or less -- often than what you'd expect if the connections were random. For instance, the clustering coefficient analysis tells us that you're going to find more triangles than expected given the number of nodes or edges.

Frequent subgraph mining is the branch of network science that attempts to find these overexpressed patterns, when they are more complex than a simple triangle. Want to know whether a square with a dangling edge appears more often than chance? You have to perform subgraph mining! In frequent subgraph mining we have a wealth of techniques to systematically and efficiently enumerate all possible subgraphs and finding the ones that occur more often in your networks.

\begin{figure}[b]
\centering
\begin{subfigure}{.22\columnwidth}
\includegraphics[width=\textwidth]{figures/triangle.pdf}
\caption{}
\end{subfigure}\qquad
\begin{subfigure}{.23\columnwidth}
\includegraphics[width=\textwidth]{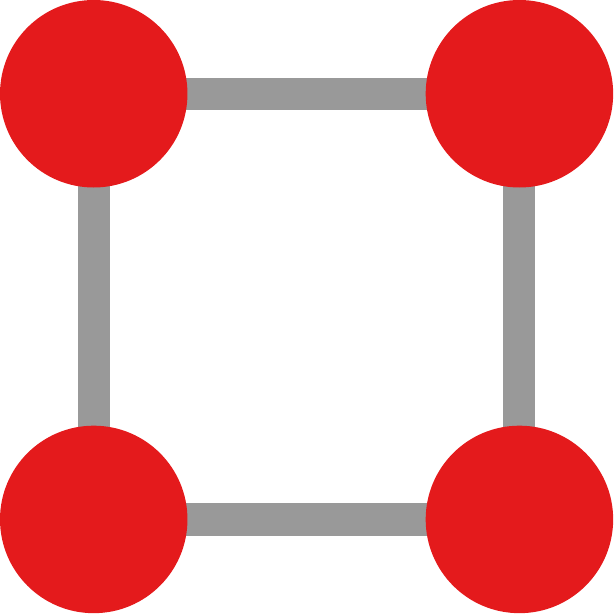}
\caption{}
\end{subfigure}\qquad
\begin{subfigure}{.39\columnwidth}
\includegraphics[width=\textwidth]{figures/simple.pdf}
\caption{}
\end{subfigure}
\caption{Three examples of motifs in complex networks.}
\label{fig:motifs}
\end{figure}

We start by defining the building blocks of complex networks. These are network motifs\cite{milo2002network}\cite{shen2002network}\cite{alon2007network}\cite{onnela2005intensity}. A network motif is a subgraph of your original network with a given topology. A triangle is a motif, a square is a motif, the five nodes with the connection pattern in Figure \ref{fig:motifs}(c) is a motif.

Generally, one wants to know which motifs are relevant for a network and which ones aren't. So the standard technique is to follow a relatively simple procedure. First, you count how many times each motif appears in your network. Second, you define a null model of your network, keeping its relevant properties fixed -- maybe just the degree distribution. Third, you count the expected number of occurrences of the motifs in the null model. Finally, you compare with your observation, so that you can build an idea of the statistical significance of the motif.

We use network motifs for many different applications. For instance, and I won't get tired of bringing this up, we use them for community discovery\cite{arenas2008motif}. Of course nobody forces you to have exclusively the simple motifs I depict in Figure \ref{fig:motifs}. One can extend the concept of network motifs to encompass temporal networks\cite{kovanen2011temporal}\cite{paranjape2017motifs} -- so time-evolving motifs --, and multilayer networks\cite{battiston2017multilayer}\cite{de2015structural}.

You might have noticed that the examples I show in Figure \ref{fig:motifs} are all very small. They include only a handful of nodes and edges. There's a reason for that. Finding motifs in a large network is a hard problem. There are clever techniques to enumerate specific small motifs which are reasonably fast\cite{pinar2017escape}\cite{torres2019non}. However, in general, one has to solve the scary graph isomorphism problem, which is the topic of the next section.

\section{Graph Isomorphism}\label{sec:mining-isomorph}

\subsection{General Idea}
Colloquially speaking, we can state the graph isomorphism problem as follows: given two graphs, decide whether they are the same graph. Two graphs are the ``same'' if they have the same topology. More formally, graph isomorphism is the search of a function which maps each node of a graph to each node of the other graph, such that they have the same neighbors -- identically mapped nodes\cite{mckay1981practical}.

\begin{figure}
\centering
\begin{subfigure}{.29\columnwidth}
\includegraphics[width=\textwidth]{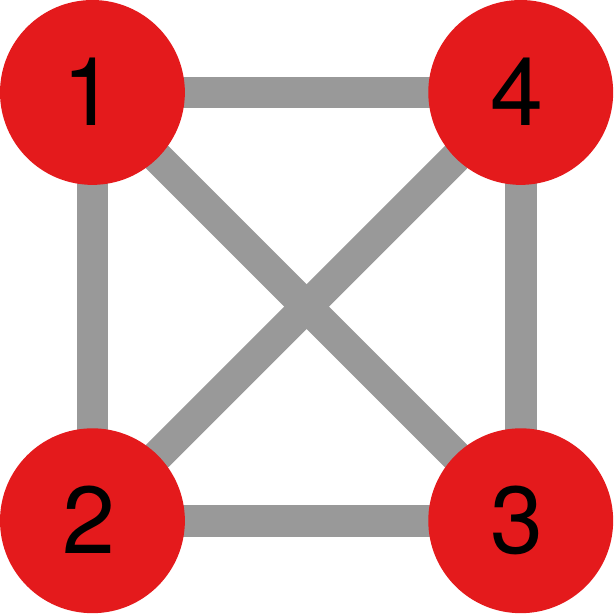}
\caption{}
\end{subfigure}\qquad
\begin{subfigure}{.29\columnwidth}
\includegraphics[width=\textwidth]{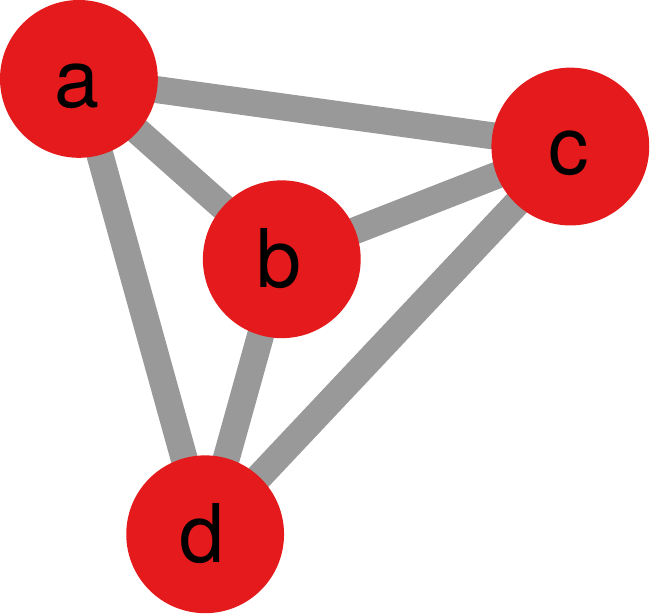}
\caption{}
\end{subfigure}\qquad
\begin{subfigure}{.2\columnwidth}
  \centering
  \begin{tabular}{c|c}
    $G1$ & $G2$ \\
    \hline
    $1$ & $c$\\
    $2$ & $a$\\
    $3$ & $d$\\
    $4$ & $b$
  \end{tabular}
\caption{}
\end{subfigure}
\caption{(a, b) Two graphs, with their nodes labeled with their ids. (c) A function connecting the node ids from graph (a) to the node ids of graph (b).}
\label{fig:isomorphism}
\end{figure}

Are the graphs in Figure \ref{fig:isomorphism}(a-b) isomorphic? Table \ref{fig:isomorphism}(c) attempts to answer positively: it relabels nodes from Figure \ref{fig:isomorphism}(a) into nodes from Figure \ref{fig:isomorphism}(b). Since all nodes are connected to their identically labeled neighbors, the answer is yes, the graphs are isomorphic -- in fact they're both 4-cliques.

This example is simple enough, but the problem gets very ugly very soon when we start considering non-trivial graphs. Subgraph isomorphism is an NP-complete problem\cite{aaronson2017p}: a type of problem where a correct solution requires you to try all possible combinations of labeling. This grows exponentially and requires a time longer than the age of the universe even for simple graphs of a few hundreds nodes.

That is why we're looking for efficient algorithms to solve graph isomorphism. Recently, a claim of a quasi-polinomial algorithm shook the world\cite{babai2016graph} -- well, parts of it. However, this is more of a theoretical find, which cannot be used in practice. Given how hard the problem is, we can either try to solve it quickly by paying the price of getting it wrong sometimes, or more slowly but having an exact solution. I'll give an example for both strategies, since they are both quite important for the rest of this book part.

\subsection{Approximate Solutions}
\begin{figure}
\centering
\includegraphics[width=.8\textwidth]{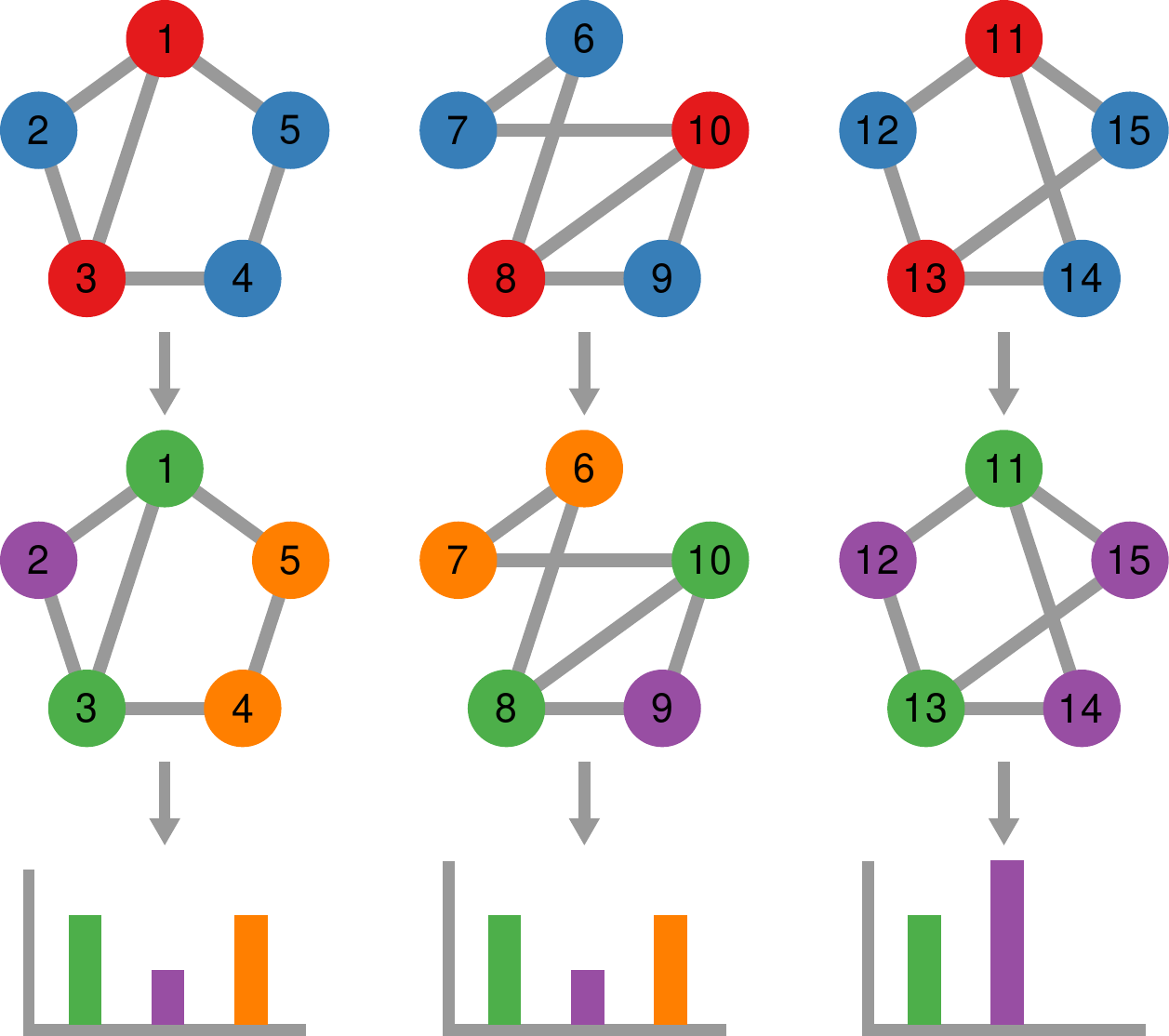}
\caption{The Weisfeiler-Lehman algorithm. The node colors represent the node's classes, based on the degree. In the first step (top) red means degree $3$ and blue means degree $2$ for all networks. In the second step each color changes according to a consistent rule within a given network. Finally, we have the histogram of node colors.}
\label{fig:weisfeiler-lehman-success}
\end{figure}

One of the most well-known ways to approximate a solution to graph isomorphism is to use the Weisfeiler-Lehman algorithm\cite{weisfeiler1968reduction}. Part of the reason why Weisfeiler-Lehman is popular comes from the fact that it approximates the behavior of message-passing neural networks -- which we'll see in Chapter \ref{cha:mining-deep} --, providing a nice connection between isomorphism and deep learning on graphs. The Weisfeiler-Lehman approach is intuitively simple, and I show an example in Figure \ref{fig:weisfeiler-lehman-success}.

We start by giving a color to a node -- classically, each node with the same degree will get the same color. Then, each node will get a new color by combining its color with the colors of all its neighbors. For instance, in the second step of Figure \ref{fig:weisfeiler-lehman-success}, the left and middle networks encode three new colors. Green means ``I'm red and I'm connected with two blue nodes and one red node'', purple means ``I'm blue and I'm connected with two red nodes'', and orange means ``I'm blue and I'm connected with one red and one blue node''. The network on the right is different, because it is not isomorphic with the other two, so its colors are determined with different rules -- which I'm sure you can figure out.

In technical computer science terms, these sentences I used to determine the new colors are called a ``hash function''. A hash function gives a unique output to each different input, that is to say we end up with a new color that is uniquely determined by all the colors that went in. At some point, there are no more changes, the graph reaches stability and there is no point in continuing. Note that, for the network on the right in Figure \ref{fig:weisfeiler-lehman-success} that actually happened at step one, so step two wasn't necessary -- I included it for aesthetic purposes. At this point a network can be described by a the color histogram of its nodes, which is the number of nodes with a given color -- the bottom row of Figure \ref{fig:weisfeiler-lehman-success}. Two networks with the same color histograms are isomorphic: they have the same number of nodes with the same colors. Indeed, the networks on the left and in the middle are isomorphic to each other, while the one on the right isn't.

\begin{figure}
\centering
\includegraphics[width=.66\textwidth]{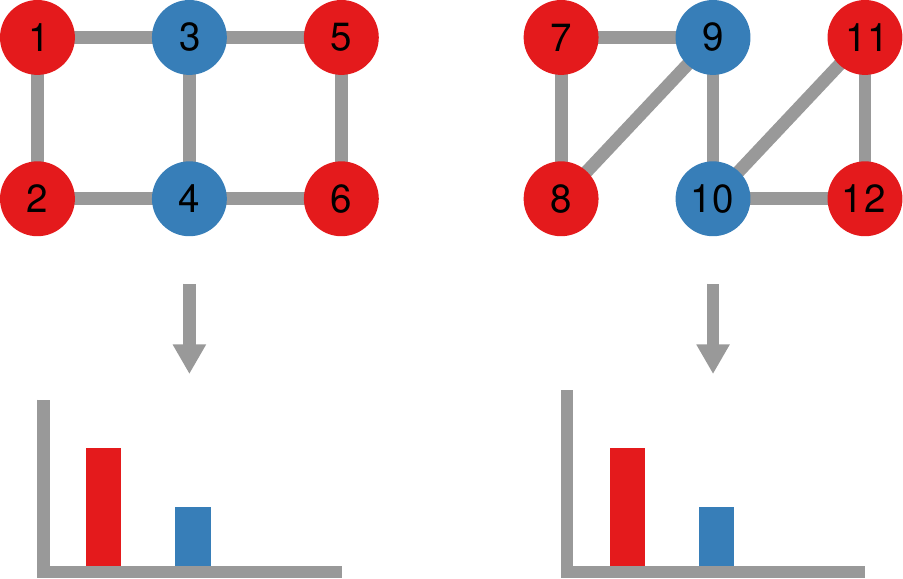}
\caption{The Weisfeiler-Lehman algorithm. The node colors represent the node's classes, based on the degree. In both cases, red means ``one red and one blue neighbor'' and blue means ``two red and one blue neighbor''.}
\label{fig:weisfeiler-lehman-fail}
\end{figure}

This works well for most networks, and it is guaranteed to finish after at most $|V|$ steps\cite{babai1979canonical}. However, you're a smart reader and you know you're reading a section titled ``approximate solutions'', so you know there's a catch. The test sometimes fails. You can see in Figure \ref{fig:weisfeiler-lehman-fail} a case in which two non-isomorphic graphs result in the same color histogram. You can tell that the graphs are not isomorphic, because one has two triangles and the other has none.

There are ways to fix this issue, for instance by aggregating not only the information from direct neighbors, but from nodes at k-hops away\cite{cai1992optimal}. However, as you might expect, this greatly increases the computational complexity of the approach.

Note that here I use simple colors based on the degree for simplicity. Nothing would change in my explanation if you were to have more complex labels -- for instance some sort of node attributes. As long as you hash those attributes into a label uniquely identify by a specific set of attribute values, the algorithm works the same and will produce the same results shown in this section.

Another popular solution in this class is by using DFS codes, which I'm going to treat extensively in Section \ref{sec:mining-transact} due to their convenient connections to itemset mining. This property of DFS codes make them more useful in many practical contexts than Weisfeiler-Lehman. Specifically, we'll see how you can save computations with DFS codes if you are growing motifs by adding nodes and edges, while Weisfeiler-Lehman doesn't allow you to do so.

\subsection{Exact Solutions}
If one needs to solve graph isomorphism exactly, to the best of my knowledge, the current practical state of the art to solve graph isomorphism is the VF2 algorithm\cite{cordella2001improved}\cite{cordella2004sub} -- recently evolved to VF3\cite{carletti2017introducing}\cite{carletti2017challenging}. Since I'm just going to give the general idea of the backtracking process it implements, the sophisticated differences between the various approaches aren't all that important. Most of the power of these algorithms come in a series of heuristics they can use to make good guesses, but here I'm just interested in talking about the general idea.

``Heuristics'' means to do your darnedest not to actually run the algorithm itself, or to do it in a way that you expect it to finish more quickly. For instance, the first step of VF2 is to make all easy checks that don't really require much thought. As an example, two graphs cannot be isomorphic if they have a different number of nodes or a different number of edges. If this check fails, you can safely deny the isomorphism. Once all these heuristic all have confirmed that the graphs could be isomorphic, then you have to begrudgingly actually look at the network, which is to say that you apply the following procedure:

\begin{itemize}
\item Step \#1: match one node in $G1$ with a node in $G2$;
\item Main loop: try to match the n-th node in $G1$ with the n-th node in $G2$. If the match is unsuccessful, recursively step back your previous matches and try a new match;
\item End loop, case $1$: you explored all nodes in $G1$ and $G2$, then the graphs are isomorphic;
\item End loop, case $2$: you have no more candidate match, then the graphs are not isomorphic.
\end{itemize}

\begin{figure}
\centering
\begin{subfigure}{.22\columnwidth}
\includegraphics[width=\textwidth]{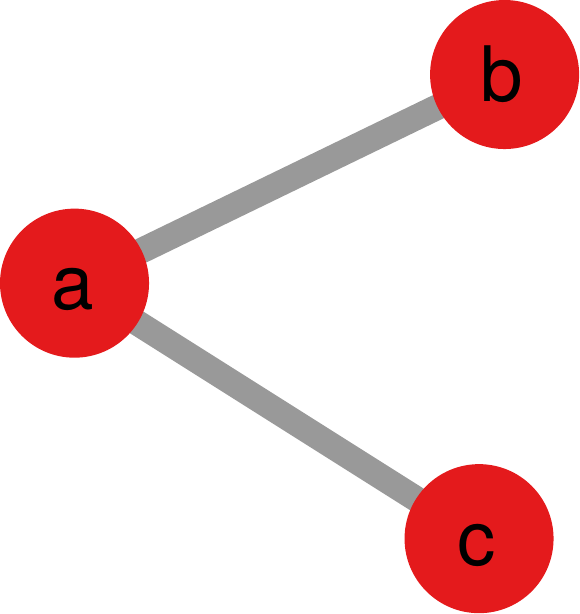}
\caption{}
\end{subfigure}\qquad
\begin{subfigure}{.2\columnwidth}
\includegraphics[width=\textwidth]{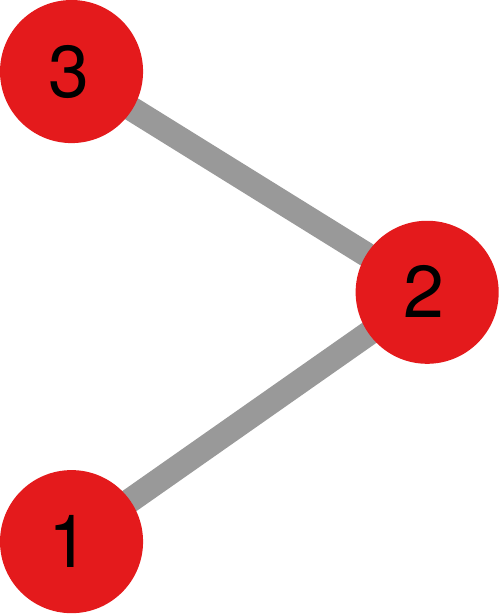}
\caption{}
\end{subfigure}\qquad
\begin{subfigure}{.43\columnwidth}
\includegraphics[width=\textwidth]{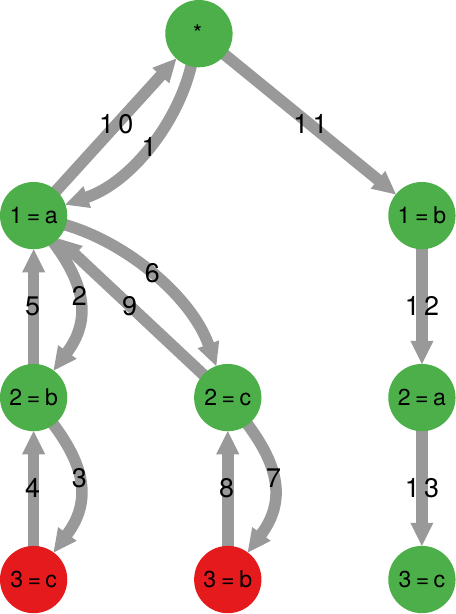}
\caption{}
\end{subfigure}
\caption{(a, b) Two graphs, with their nodes labeled with their ids. (c) The inner data structure used by the VF2 algorithm to test for isomorphism. I label each node with the attempted match. The node color tells the result of the match (green = successful, red = unsuccessful). I label the edges to follow the step progression of the algorithm.}
\label{fig:isomorphism2}
\end{figure}

Most of the heavy lifting is made in the main loop, when checking whether a match is successful or not, but most of the cleverness is in step \#1: prioritizing good matches that are more likely to succeed -- for instance never trying to match two nodes with different degrees. An illustrated example with a simple graph would probably be helpful. Consider Figure \ref{fig:isomorphism2}. VF2 attempts to explore the tree of all possible node matching (Figure \ref{fig:isomorphism2}(c)). It starts from the empty match -- the root node.

The first attempted match always succeeds, as any node can be matched to any other node -- in this case matching node $1$ with node $a$. For the second match to succeed, we need that the two matched nodes are connected to each other. Since node $a$ connects to node $b$ and node $1$ connects to node $2$, then the $1=a$ and $2=b$ match is a success.

However, attempting to match $3=c$ fails, because while node $2$ is connected to node $3$, node $b$ (matched with $2$) isn't connected to $c$ (matched to $3$). Thus VF2 backtracks: it undoes the last matches and starts from the last successful match -- provided that there are possible matches to try. In this case there aren't , so it backtracks again.

Trying to set $2=c$ and $3=b$ fails again, for the same reason as before. So VF2 has to give up also on the $1=a$ match and start from scratch. Luckily, there's another possible move: $1=b$. When we go down the tree all matches are successful, until we touched all nodes in the graph. At that point, we can safely conclude the two graphs are isomorphic. Note that Figure \ref{fig:isomorphism2}(c) doesn't include the branches that VF2 never tries in this case, for instance the $1=c$ branch.

You can see here what I meant by smart heuristics that can help making the algorithm faster. The $1=a$ starting move was particularly stupid, because node $1$ and node $a$ don't have the same degree, so they could never have matched. With the heuristic of ``only try to match nodes with the same degree'' you could have saved a lot of computation by starting directly with the $1=b$ move -- and that's what VF2 does.

As expected, multilayer networks provide another level of difficulty. One can perform graph isomorphism directly on the full multilayer structure\cite{kivela2017isomorphisms}, or give up a bit of the complexity and represent them as labeled multigraphs\cite{ingalalli2016sumgra}\cite{micale2019fast}.

\section{Transactional Graph Mining}\label{sec:mining-transact}
So far we've been dealing with network motifs on a ``top-down'' approach. We have some motifs of interest and we ask ourselves whether they are overexpressed or underexpressed. This implies that you have to start with your motifs already in mind. This might not be possible. Sometimes, you need a ``bottom-up'' approach: you want an algorithm telling you the frequencies of all possible simple network motifs. This is usually the task of frequent subgraph mining.

We split frequent subgraph mining in two: transactional and simple graph mining. Transactional graph mining was developed first, because single graph mining introduces some non-trivial problems. It's best to start by explaining transactional graph mining, and we'll deal with the additional obstacles of single graph mining later (in Section \ref{sec:mining-single}).

\subsection{Frequent Itemset Mining}
\begin{figure}
\centering
\includegraphics[width=.75\columnwidth]{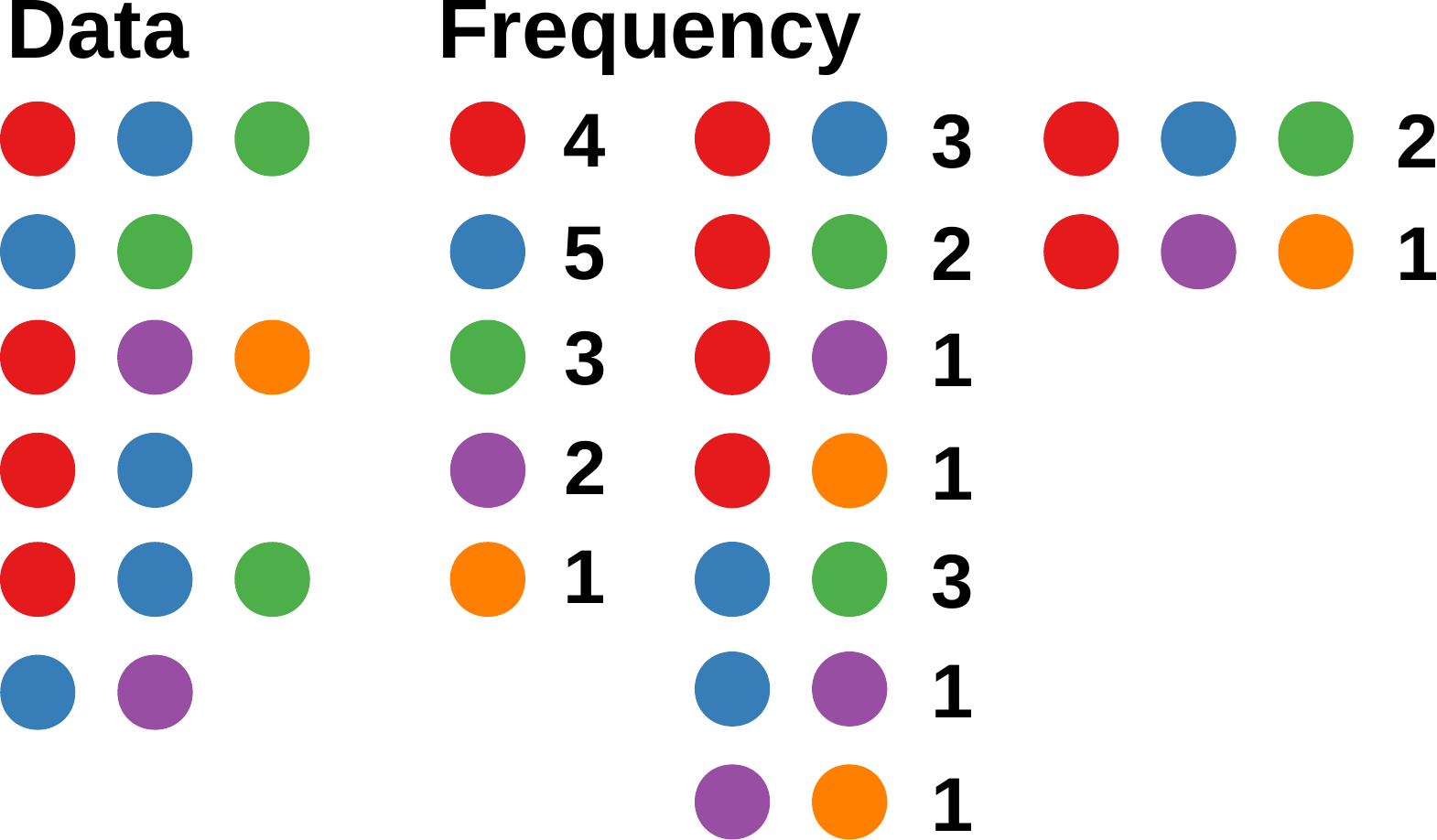}
\caption{An example of frequent itemset mining. The original data is on the left, one line per set of items (itemset). We calculate the frequency of each itemset, including all subsets (right).}
\label{fig:freqitems}
\end{figure}

Transactional graph mining is inspired by frequent itemset mining, a classical problem in data mining\cite[-0.2in]{han2000mining}\cite{pei2000closet}\cite{han2007frequent}. In frequent itemset mining, your input is a collection of sets. Each set includes different objects. The objective is to find the subsets that appear more often in your collection of sets. The number of times each subset appear is called support. We increase support by one each time we find a subset inside a set in the collection. Figure \ref{fig:freqitems} provides an example.

From Figure \ref{fig:freqitems} you can see that this problem explodes in complexity very soon. Even with just five itemsets and five items, the number of possible subsets can get very high. Thus the crucial problem in frequent itemset mining is how to efficiently explore the search space. There are many algorithms to do it, but I'll focus on the old and legendary Apriori\cite{agrawal1994fast}, given its simplicity and its didactic potential.

The first thing you do is giving up on the idea of finding \textit{all} subsets. You only want to find the \textit{frequent} ones. Thus you establish a support threshold: if a subset fails to occur in that many sets, then you don't want to see it. This allows you to prune the search space. If subset $A$ is not frequent, then none of its extensions can be: they have to contain it so they can be at most as frequent as $A$ is\footnote{That is, the support function is anti-monotonic: it can only stay constant or shrink as your set grows in size.}. Thus, once you rule out subset $A$, none of its possible extensions should even be considered, since none can be frequent. This usually allows to perform much fewer tests than the possible ones, and still return all frequent subsets.

For instance, in Figure \ref{fig:freqitems}, the orange circle only occurs once. If the support threshold is $2$, we know we don't need to check the red-orange, purple-orange, and red-purple-orange subsets. With one check, we prevented three.

\begin{figure}
\centering
\includegraphics[width=.75\columnwidth]{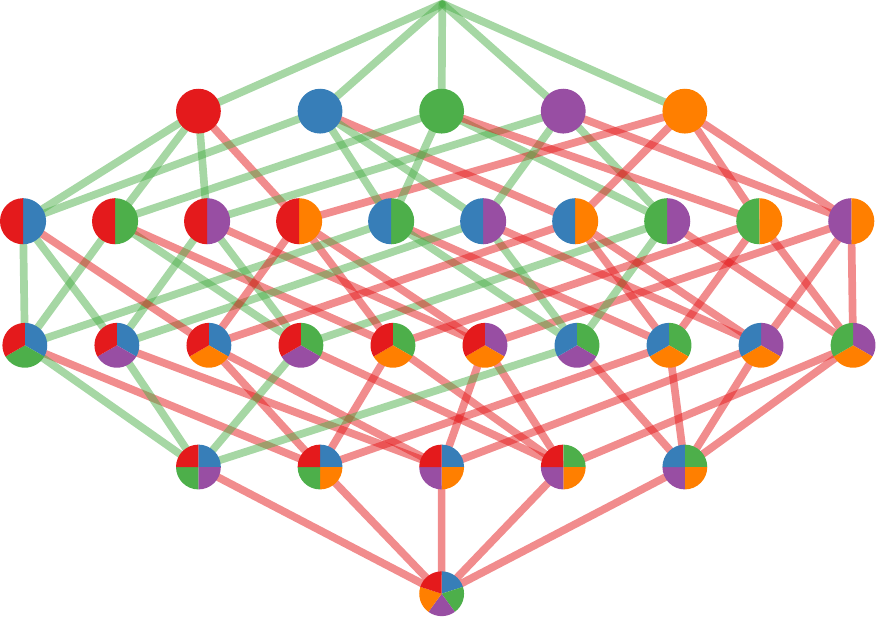}
\caption{Apriori's search space. Each node represents a subset and bears the colors of the items it contains. Given that the orange item is not frequent, Apriori marks as red the links it needs not to follow, because they lead to a subset containing the infrequent orange item, which cannot make the support threshold.}
\label{fig:apriori}
\end{figure}

Actually, we prevented many more. Figure \ref{fig:apriori} shows the entire search space in a dataset with five different items. As you can see, if we have an item which does not pass the support threshold, the search space crumbles. Apriori explores this graph and marks all nodes with an orange item as infrequent, checking only the itemsets without red connections. This doesn't even take into account the other infrequent subsets from Figure \ref{fig:freqitems}.

\begin{figure}
\centering
\includegraphics[width=.25\columnwidth]{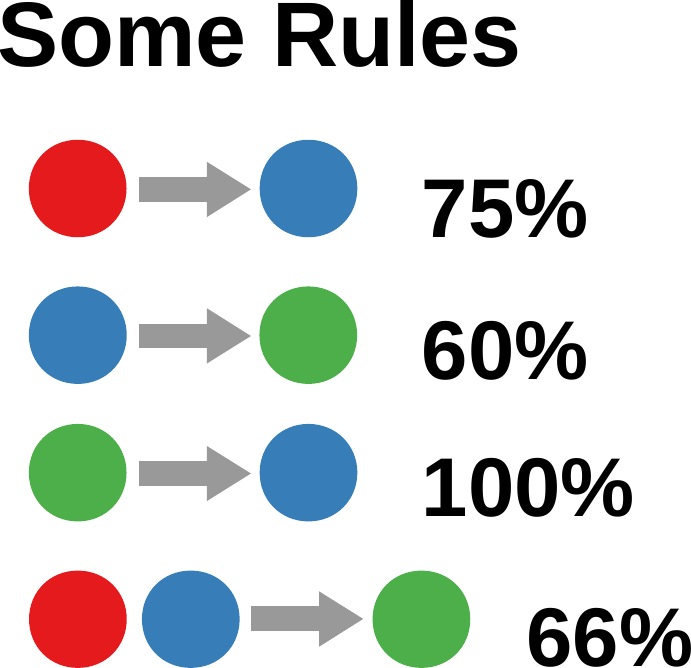}
\caption{An example of association rule mining. Assume the frequencies of each itemset are the ones from Figure \ref{fig:freqitems}. We generate rules recording the relative frequency of observing two itemsets. Note that these frequencies are not symmetric! While the green item only occurs $60\%$ of the times a blue item occurs, every time green occurs we also have the blue item.}
\label{fig:assrules}
\end{figure}

As a small aside, note that, once you know the frequencies of all sets, you can build what we call ``association rules''. What you want to do is to find all rules in the form of: ``If a set contains the objects $A$, then it is likely to also contain object $b$''. Figure \ref{fig:assrules} shows a simple example of the problem we're trying to solve.

Suppose that, in your data, you see $100$ instances of sets containing objects $a_1$, $a_2$, and $a_3$. And let's say that, among them, $80$ also contain object $b$. Then you can say, with $80\%$ confidence, that the following rule applies: $\{a_1, a_2, a_3\} \rightarrow b$. The $\{a_1, a_2, a_3\}$ part is the antecedent of the rule, while $b$ is the consequent.

You can also correct your confidence for chance, if you know $b$'s overall frequency in the data, and the size of the dataset. This is the ``lift'' measure. Let's say that, in our example, $b$ appears in $120$ sets. Also, our dataset contains a total of $400$ sets. The lift of the rule is the relative frequency (support) of $\{a_1, a_2, a_3, b\}$ ($80/400$) over the product of the support of the antecedent and the consequent ($100/400 \times 120/400$). This latter quantity gives us the probability of the antecedent and the consequent to co-appear randomly. Doing the math: $.2 / (.25 \times .3) = 2.\bar{6}$. This means that the rule appears more than twice as much as we would expect if there was no relationship between the antecedent and the consequent. A lift lower than one indicates items appearing less often than they would do at random: a sign that the rule is unlikely to be interesting.

Lift is one of those Swiss army knives that has been independently invented in multiple fields. For instance it is also known as Relative Risk in in statistics\cite{zhang1998s}, and Revealed Comparative Advantage in trade economics\cite{balassa1965trade}.

When you replace itemsets with motifs, you obtain the GERM algorithm: that is why I introduced it as graph association rule mining in the link prediction chapters (Sections \ref{sec:lp-germ} and \ref{sec:lp-multilayer-general}). How to go from itemsets to network motifs is the topic of the rest of this section.

\subsection{From Itemsets to Network Motifs}
\begin{figure}
\centering
\includegraphics[width=\columnwidth]{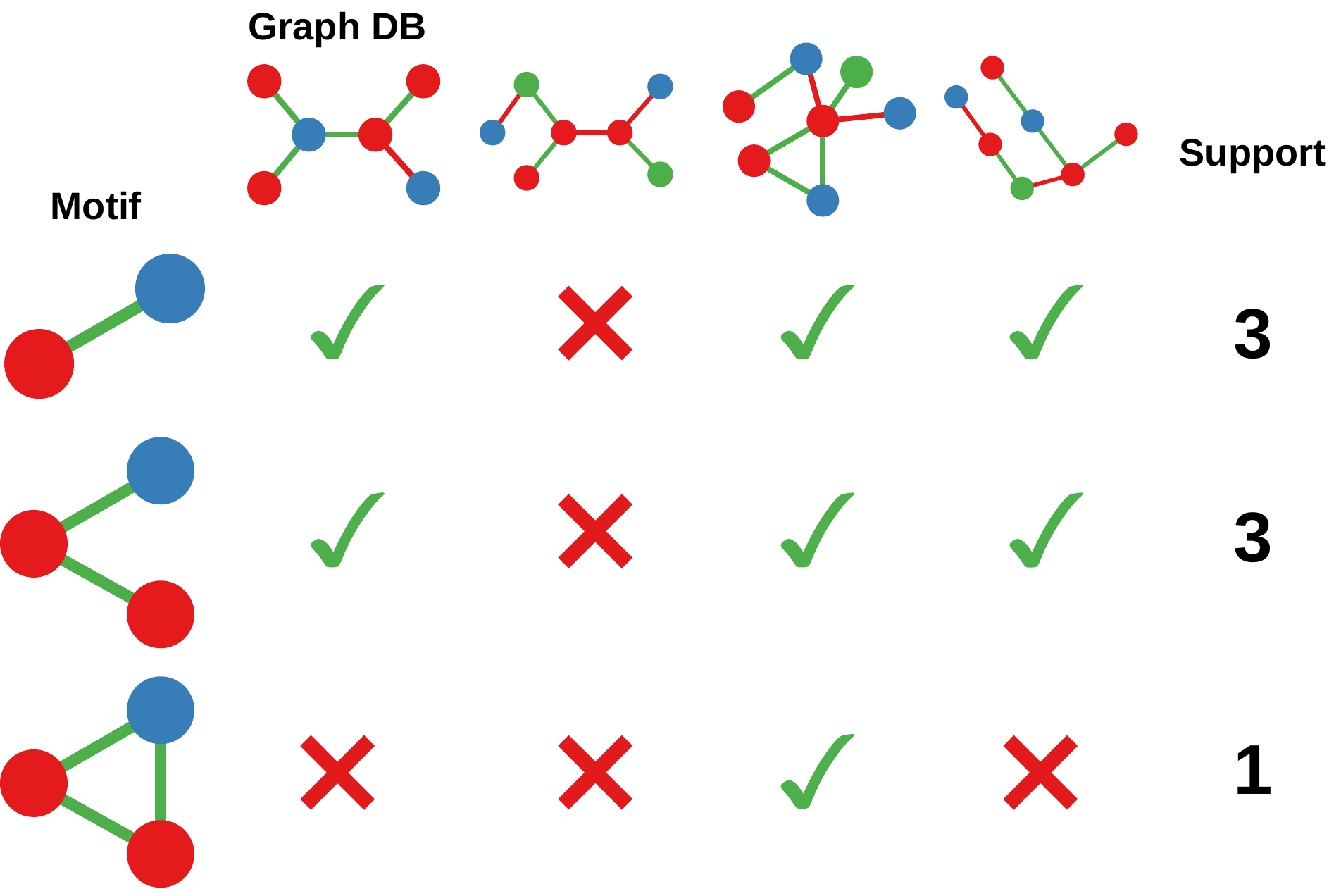}
\caption{For each of the patterns on the left we check whether a graph in the database (on top) contains it (green checkmark) or not (red cross). The number of graphs in the database containing the motifs is its support.}
\label{fig:graphmining}
\end{figure}

Transactional graph mining applies all this machinery to graphs. Rather than looking at simple sets of items, we look at graph patterns: triangles, 4-cliques, bi-cliques... any possible combination of nodes and edges. So we have a database of many different graphs and we ask in how many graphs the motif appears. This is our definition of support, as Figure \ref{fig:graphmining} shows.

The big problem is how to enumerate all possible graphs efficiently. We want to avoid trying to count the occurrences of a graph pattern $G''$ if it contains a pattern $G'$, which we already counted and found not frequent enough. Since $G''$ is an extension of $G'$, we already know it cannot be frequent enough: a larger graph can at most be as frequent as the least frequent of its subgraphs.

There are many ways to do this. An incomplete list of approaches includes FFSM\cite{huan2003efficient}, Gaston\cite{nijssen2005gaston}, Moss\cite{borgelt2002mining}, etc. You can find relevant literature for a historic quantitative comparison of these methods\cite[2.5in]{worlein2005quantitative}. Just like for frequent itemset mining, I'll focus on a specific method, gSpan\cite{yan2002gspan}\cite{yan2003closegraph}, given its historical and didactic relevance.

Graphs are more complex structures than itemsets, so building a search space like the one Apriori creates (Figure \ref{fig:apriori}) is tricky. If you cannot explore the search space like Apriori does, it's even harder to prune it by avoiding exploring patterns you already know they are not frequent. gSpan solves the problem by introducing a graph lexicographic order (which I'm going to dumb down here, the full details are in the paper).

\begin{figure}
\centering
\includegraphics[width=\columnwidth]{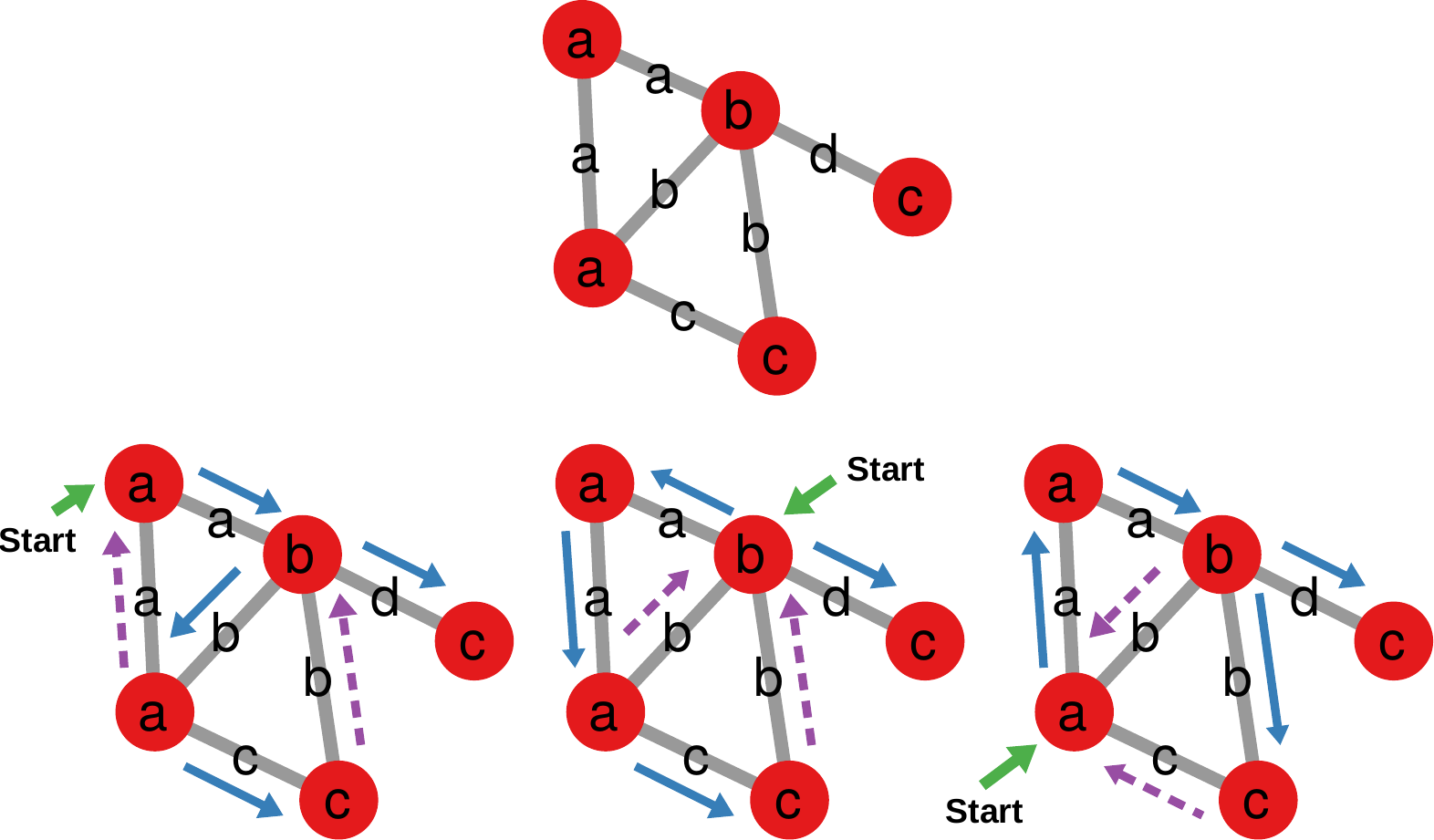}
\caption{Three possible DFS explorations of the graph on top. Blue arrows show the DFS exploration and purple dashed arrows indicate the backwards edges (pointing to a node we already explored). Remember that DFS backtracks to the last explored node, not to where the backward edges points.}
\label{fig:dfscode}
\end{figure}

Suppose you have a graph, as in Figure \ref{fig:dfscode} (top). You can explore it using a DFS strategy. Actually, you can have many different DFS paths: you can start from node $a$, from node $b$, ..., then you can move through a different edge any time. We can encode each DFS exploration with a DFS code: a sequence of quintuples (id of source node, id of target node, label of source node, label of edge, label of target node). Figure \ref{tab:dfscode} shows the DFS codes for the three explorations in Figure \ref{fig:dfscode}. Note that, every time we explore a node with backward edges, we insert them in the code immediately, before continuing with the DFS exploration.

\begin{figure}
\centering
\begin{tabular}{c|ccc}
Order & DFS1 & DFS2 & DFS3\\
\hline
$0$ & $\mathbin{\textcolor{cb2}{(0,1,a,a,b)}}$ & $\mathbin{\textcolor{cb2}{(0,1,b,a,a)}}$ & $\mathbin{\textcolor{cb2}{(0,1,a,a,a)}}$\\
$1$ & $\mathbin{\textcolor{cb2}{(1,2,b,b,a)}}$ & $\mathbin{\textcolor{cb2}{(1,2,a,a,a)}}$ & $\mathbin{\textcolor{cb2}{(1,2,a,a,b)}}$\\
$2$ & $\mathbin{\textcolor{cb4}{(2,0,a,a,a)}}$ & $\mathbin{\textcolor{cb4}{(2,0,a,b,b)}}$ & $\mathbin{\textcolor{cb4}{(2,0,b,b,a)}}$\\
$3$ & $\mathbin{\textcolor{cb2}{(2,3,a,c,c)}}$ & $\mathbin{\textcolor{cb2}{(2,3,a,c,c)}}$ & $\mathbin{\textcolor{cb2}{(2,3,b,b,c)}}$\\
$4$ & $\mathbin{\textcolor{cb4}{(3,1,c,b,b)}}$ & $\mathbin{\textcolor{cb4}{(3,0,c,b,b)}}$ & $\mathbin{\textcolor{cb4}{(3,0,c,c,a)}}$\\
$5$ & $\mathbin{\textcolor{cb2}{(1,4,b,d,c)}}$ & $\mathbin{\textcolor{cb2}{(0,4,b,d,c)}}$ & $\mathbin{\textcolor{cb2}{(2,4,b,d,c)}}$\\
\end{tabular}
\caption{The DFS codes for the DFS explorations in Figure \ref{fig:dfscode}. DFS exploration edges in blue. For backward edges (purple) the node id of the source is higher than the node id of the target.}
\label{tab:dfscode}
\end{figure}

Once you have all DFS codes for a graph, you can find the \textit{minimum} DFS code, by simply sorting them alphanumerically. The minimum DFS code -- in our example the third one (DFS3 in Table \ref{tab:dfscode}) -- is a canonical representation for a graph. Two graphs with the same minimum DFS code are isomorphic, and if you encounter a non-minimum DFS code in your search space you can safely ignore it. It is necessary to remind you what I said in Section \ref{sec:mining-isomorph}: DFS codes provide an approximate solution to the graph isomorphism problem, so you might get it wrong sometimes. However, it is an occurrence rare enough not to impact you in practice most of the times.

We're ok paying this price of approximation because DFS codes allow you to reduce the graph mining problem to a frequent itemset problem. What before I called items $a_1$, $a_2$, and $a_3$ are now items like $(0,1,a,a,a)$, $(1,2,a,a,b)$, and $(2,0,b,b,a)$. If you always find them together with the additional $(2,3,b,b,c)$, you have built a graph association rule! So, with this canonical graph representation, you can solve the frequent subgraph mining problem with any frequent itemset algorithm (like Apriori).

\section{Single Graph Mining}\label{sec:mining-single}
Transactional graph mining is great because it's a very similar problem to frequent itemset mining and has some interesting applications. For instance, your graph database could contain thousands of different chemical compounds, and you want to find the most common substructures among all those molecules. However, there's  a great deal of network data that doesn't fit this mold.

For instance, if you want to mine all frequent patterns in a social network, you typically have a single large graph. In those cases, you cannot have as definition of support the ``number of graphs in which the pattern appears'' as in the transactional setting. That number is always going to be either zero -- the pattern doesn't appear -- or one. What you want to do, instead, is to count the number of times the patterns appear in the single network.

\begin{figure}
\centering
\includegraphics[width=.5\columnwidth]{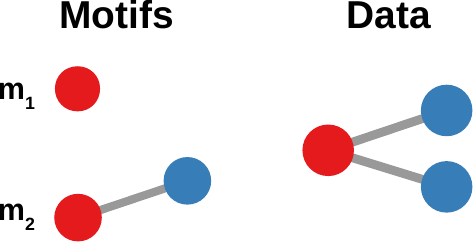}
\caption{Two motifs (left) and our graph data (right). Motif $m_1$ appears only once. How many times does motif $m_2$ appears?}
\label{fig:single-gm}
\end{figure}

However, such naive support definition won't work. To see why, consider the example in Figure \ref{fig:single-gm}. It's obvious that the red node motif appears only once: there's only one red node. However, naively, we could say that motif $m_2$ appears twice. This is unacceptable, because it breaks the anti-monotonicity rule: a motif can only occur at most as much as the least frequent of its sub-motifs. Since $m_2$ contains $m_1$, it cannot appear more often than $m_1$.

If we were to accept it to have higher support than its submotifs, we could not prune the search space using Apriori's strategy, meaning that we would have to explore an exponentially growing set of possibilities. This would make single graph mining impractical in all but trivial scenarios.

So the main quest for single graph mining is the one for an anti-monotonic support definition that is sufficiently easy to compute -- otherwise we don't gain much time -- and that hopefully makes intuitive sense. I'm going to show you three alternatives: using ego networks (Section \ref{sec:homophily-ego}), harmful overlap, and the minimum image support.

\subsection{Ego Networks}
One option is to bring back the problem into familiar territory. One can split the single graph in many different subgraphs and then apply any transactional graph mining technique. For example, one could take the ego networks of all nodes in the network. The support definition would then be the number of nodes seeing the pattern around them in the network.

\subsection{Harmful Overlap}
Another option starts from recognizing that the entire problem of non-monotonicity is due to the fact that motif $m_2$ appears twice only because we allow the re-use of parts of the data graph when counting the motif's occurrences. In Figure \ref{fig:single-gm}, we use the red node in the data graph twice to count the support of $m_2$. In practice, the two patterns supporting $m_2$ overlap: they have the single red node in common. We could forbid such overlap: we don't allow the re-use of nodes when counting a motif's occurrences. With such a rule, $m_2$ would appear only once in the data graph. If we applied the rule, we would have an anti-monotone support definition\cite{kuramochi2005finding}: larger motifs would only appear fewer times or as many times as the smaller motifs they contain.

\begin{figure}
\centering
\includegraphics[width=\columnwidth]{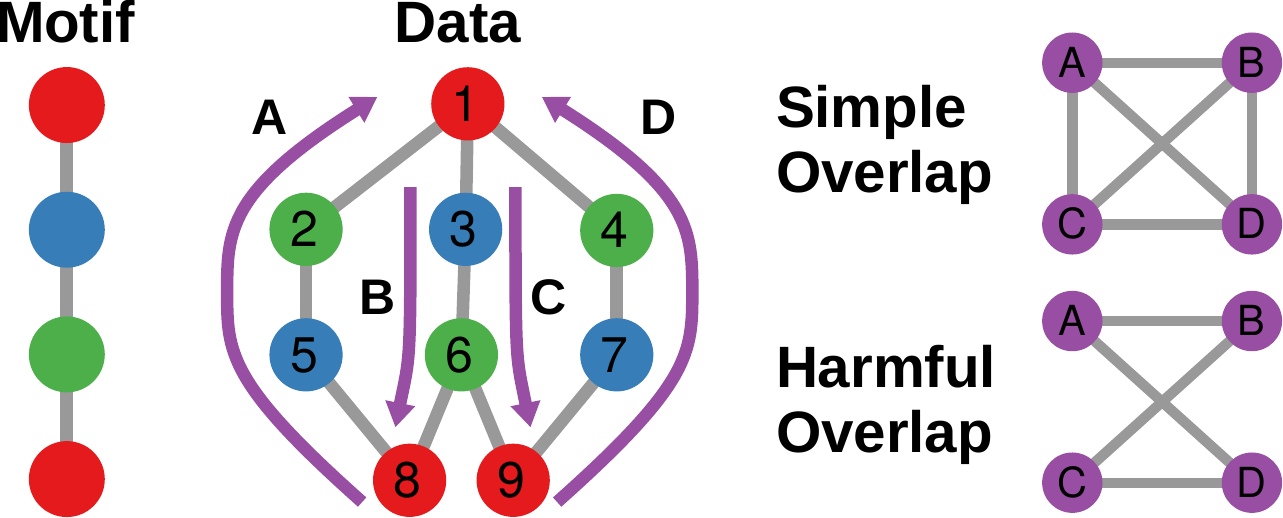}
\caption{From left to right: a pattern, the graph dataset, and its corresponding simple and harmful overlap graphs. I label each occurrence of the motif with a letter, which also labels the corresponding node in the overlap graph.}
\label{fig:single-gm-overlap}
\end{figure}

To see how, consider Figure \ref{fig:single-gm-overlap}. The motif appears four times, but each of these four occurrences share at least one node. We can create an ``overlap graph'' in which each node is an occurrence in the data graph, and we connect occurrences if they share at least one node. If we forbid overlaps, we only want to count ``complete'' and non-overlapping occurrences. This is equivalent of solving the maximum independent set problem (see Section \ref{sec:density-cliques}) on the overlap graph: finding the largest set of nodes which are not connected to any other member of the set. In this case, we have four independent sets all including a single node -- because the overlap graph is a clique -- and thus the pattern occurs only once.

This is the simple overlap rule and it is usually too strict. There are some overlaps between the occurrences that do not ``harm'' the anti-monotonicity requirement for the support definition\cite{fiedler2007support}. To find harmful overlaps you need to do two things. First, you look at which nodes are in common between two occurrences. For instance, in Figure \ref{fig:single-gm-overlap}, $A \cap B = \{1, 8\}$, $A$ and $B$ share nodes $1$ and $8$. Second, you need to make sure that these nodes that are in common between the two occurrences are not required to map the same nodes at the same time. In the case of $A$ and $B$ they are, because the only way to map the top red node in the motif is to use node $8$ for $A$ and $1$ for $B$. This is a harmful overlap.

The non-harmful overlaps are the ones in which this doesn't happen. For instance $A$ and $C$ are not overlapping harmfully: the only node in common between $A$ and $C$ is node $1$. However, we do not need node $1$ to map the same node in the motif in $A$ and $C$: when we use node $1$ in $A$ we use node $9$ in $C$, when we use node $1$ in $C$ we use node $8$ in $A$. Node $1$ is also the only node in common between $A$ and $D$, but in this case the overlap is harmful, because we use it to map the same node in the motif: the bottom red node.

The simple and harmful overlap build different overlap graphs, but then they count occurrences in the same way, using the maximum independent set problem. In the example from Figure \ref{fig:single-gm-overlap} this leads to different support values: for the harmful support, the motif occurs twice in the network -- you have two independent sets of size two ($A,C$ and $B,D$).

\subsection{Minimum Image Support}
The problem of simple and harmful overlap is that they have to solve the maximum independent set problem for every motif we search in a possibly very large overlap graph, which is a hard problem. Thus, researchers proposed a new definition which skips this computation: the minimum image support\cite{bringmann2008frequent}. In this definition, what matters is that we do not re-use the same node in the network to play the same role in the motif.

\begin{figure}
\centering
\begin{subfigure}{.06\columnwidth}
\includegraphics[width=\textwidth]{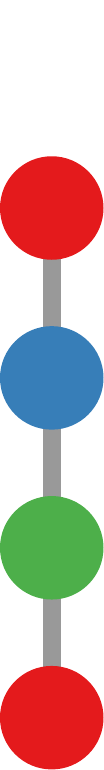}
\caption{}
\end{subfigure}\qquad
\begin{subfigure}{.6\columnwidth}
  \huge
  \begin{tabular}{cccc|c}
    $A$ & $B$ & $C$ & $D$ & Count \\
    \hline
    $8$ & $1$ & $1$ & $9$ & $3$\\
    $5$ & $3$ & $3$ & $7$ & $3$\\
    $2$ & $6$ & $6$ & $4$ & $3$\\
    $1$ & $8$ & $9$ & $1$ & $3$\\
  \end{tabular}
\caption{}
\end{subfigure}
\caption{(a) The motif (b) The image table for the minimum image support definition, with the motif's nodes as rows and all the occurrences of the motif as columns. Each cell records the node id we use for the mapping.}
\label{fig:single-gm-mis}
\end{figure}

In practice, we look at which node in the network we use to map each node in the motif. To do so, we build an ``image'' table. Figure \ref{fig:single-gm-mis} shows the image table for the example in Figure \ref{fig:single-gm-overlap}. In the table, we record the node in the network playing the role of a specific node in the motif. Thus, the support of the motif is the minimum number of distinct row values -- two identical values in a row stand for an incompatible pair of occurrences.

The aforementioned Moss method is able to deal with multigraphs, thus it can be used for some multilayer graph mining as well -- assuming your multilayer network can be represented as a multigraph. Otherwise, Muxviz\cite{de2015muxviz} allows for multilayer motif counting, but employs a naive support definition and thus cannot be used for graph mining, due to the break of the anti-monotonicity requirement. 

\section{Summary}

\begin{enumerate}
\item Machine learning can give you powerful insights about your networks, but first you need to find a way to transform the complex structure of a network into numerical values that can be handled by machine learning algorithms. One way to do so is to count network motifs.
\item Network motifs are small simple graphs that you can use to describe the topology of a larger network. For instance, you can count the number of times a triangle or a square appears in your network.
\item To do so, you need to solve the problem of ``graph isomorphism'', which is the task of determining whether two graphs have the exact same topology. Graph isomorphism is a computationally heavy problem to solve, which can be tackled quickly in an approximate way, or slowly in an exact way.
\item Frequent subgraph mining is the graph equivalent of frequent itemset mining: to efficiently find all the graph motifs that appear in your network, avoiding to perform the expensive graph isomorphism problem for patterns that you already discovered not being frequent.
\item Frequent subgraph mining comes in two flavors. Transactional mining analyzes many small networks and counts the number of networks containing the motif we're counting. Single graph mining analyzes a single large graph and counts the number of times the motif appears.
\item Unfortunately, simply counting motif appearances in a single graph cannot support an efficient exploration of the search space, because larger motifs might appear more often than smaller motifs, i.e. the counting function is not-monotonic.
\item We have different ways of counting the frequency of a motif in a single graph that are anti-monotonic. They are all based on the concept that we should not count twice patterns that are overlapping, for different definitions of what ``overlapping'' means.
\end{enumerate}

\section{Exercises}

\begin{enumerate}
\item Test whether the motifs in \url{http://www.networkatlas.eu/exercises/41/1/motif1.txt}, \url{http://www.networkatlas.eu/exercises/41/1/motif2.txt}, \url{http://www.networkatlas.eu/exercises/41/1/motif3.txt}, and \url{http://www.networkatlas.eu/exercises/41/1/motif4.txt} appear in the network at \url{http://www.networkatlas.eu/exercises/41/1/data.txt}.
\item How many times do the motifs from the previous question appear in the network? \url{http://www.networkatlas.eu/exercises/41/1/motif2.txt} is included in \url{http://www.networkatlas.eu/exercises/41/1/motif3.txt}: is the latter less frequent the former as we would require in an anti-monotonic counting function?
\item Suppose you define a new type of clustering coefficient that is closing \url{http://www.networkatlas.eu/exercises/41/1/motif3.txt} with \url{http://www.networkatlas.eu/exercises/41/1/motif4.txt}. What would be the value of this special clustering coefficient in the network?
\end{enumerate}

\chapter{Shallow Graph Learning}\label{cha:mining-embeddings}
In the next chapters we're going to fully explore the recent and quickly evolving field of Graph Neural Networks (GNNs). I already mentioned some basics of neural networks in Section \ref{sec:extended-types} and ideas on what they can do on graphs in Section \ref{sec:centr-roles-conv}. Now it's time to explain exactly how you can use neural networks on graphs to perform those -- and more -- tasks.

There are a few good books if you want to get into this topic more in depth\cite{hamilton2020graph}\cite{bronstein2021geometric}, which you should check out. Some of the surveys in the literature\cite{goyal2018graph} also come with code\footnote{\url{https://github.com/palash1992/GEM}} that you can use to follow along the explanations.

For the time being, we're focusing on building ``embeddings''. A small terminology note: building embeddings and using them in GNNs can go under different terms depending on where the focus is. For instance, GNNs can be considered a special case of the more general geometric learning that goes beyond the use of Euclidean space in data analysis\cite{bronstein2017geometric}. We'll see more on the use of complex non-Euclidean geometries in Chapter \ref{cha:nvd}. Another term you could see is ``collective classification'' field: the attempt of classifying nodes by looking at how they relate to the rest of the network\cite{sen2008collective}.

Finally, one thing you need to know to read these chapters critically. In the next couple of chapters, I am adopting a purely structural approach to build node embeddings. This means that they are built exclusively by looking at their connections. It is only in Chapter \ref{cha:mining-deep} that I'm going to give you a fuller picture that takes the node attributes into account as well. So don't assume that looking at the connections is the only way to build a node embedding! 

\section{What is a Node Embedding?}\label{sec:mining-embeddings-what}
\begin{figure}
\centering
\begin{subfigure}{.28\columnwidth}
\includegraphics[width=\textwidth]{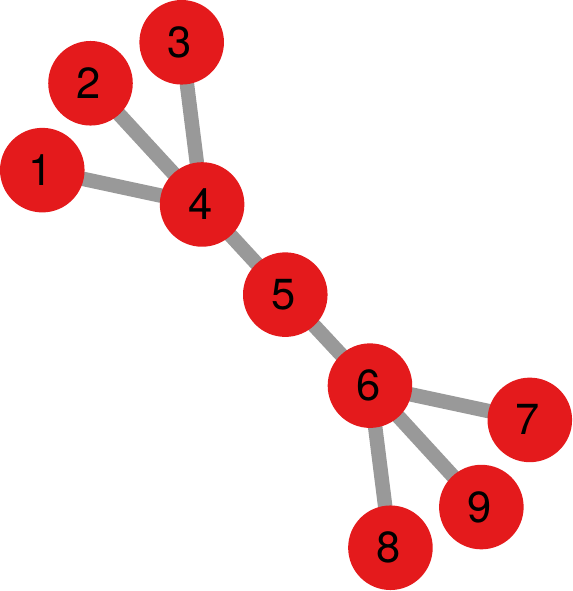}
\caption{}
\end{subfigure}
\begin{subfigure}{.3\columnwidth}
  \footnotesize
  \begin{tabular}{l|r}
    Node & Embedding \\
    \hline
    $1$ & $\{0.88, 0.69\}$\\
    $2$ & $\{0.88, 0.69\}$\\
    $3$ & $\{0.88, 0.69\}$\\
    $4$ & $\{1, 1\}$\\
    $5$ & $\{0.48, 0.66\}$\\
    $6$ & $\{0, 0\}$\\
    $7$ & $\{0.42, 0.1\}$\\
    $8$ & $\{0.42, 0.1\}$\\
    $9$ & $\{0.42, 0.1\}$\\
  \end{tabular}
\caption{}
\end{subfigure}
\begin{subfigure}{.34\columnwidth}
\includegraphics[width=\textwidth]{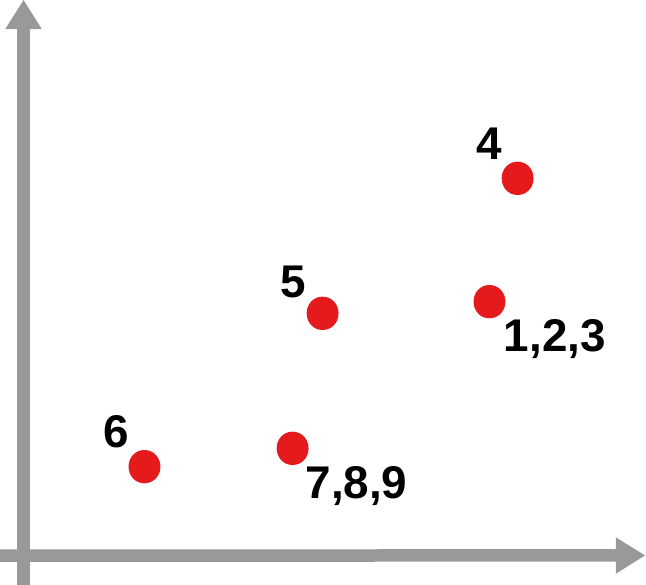}
\caption{}
\end{subfigure}
\caption{(a) An example graph. (b) One of the possible embeddings of (a), assigning a two dimensional vector to each node. (c) The scatter plot representation of (a)'s embeddings.}
\label{fig:embeddings-example}
\end{figure}

A node embedding is a vector of numbers that describes the node's role in the complex network structure. To get a node embedding you need a function that maps each node of your network to a vector of numbers. Figure \ref{fig:embeddings-example} shows a stylized example of what a node embedding is. We transform the original graph (Figure \ref{fig:embeddings-example}(a)) into a set of two-dimensional numerical vectors for each of its nodes (Figure \ref{fig:embeddings-example}(b)). These vectors have a spatial relationship reflecting some of the graph's properties (Figure \ref{fig:embeddings-example}(c)).

\subsection{Properties of Good Embeddings}
\begin{figure}
\centering
\includegraphics[width=.8\textwidth]{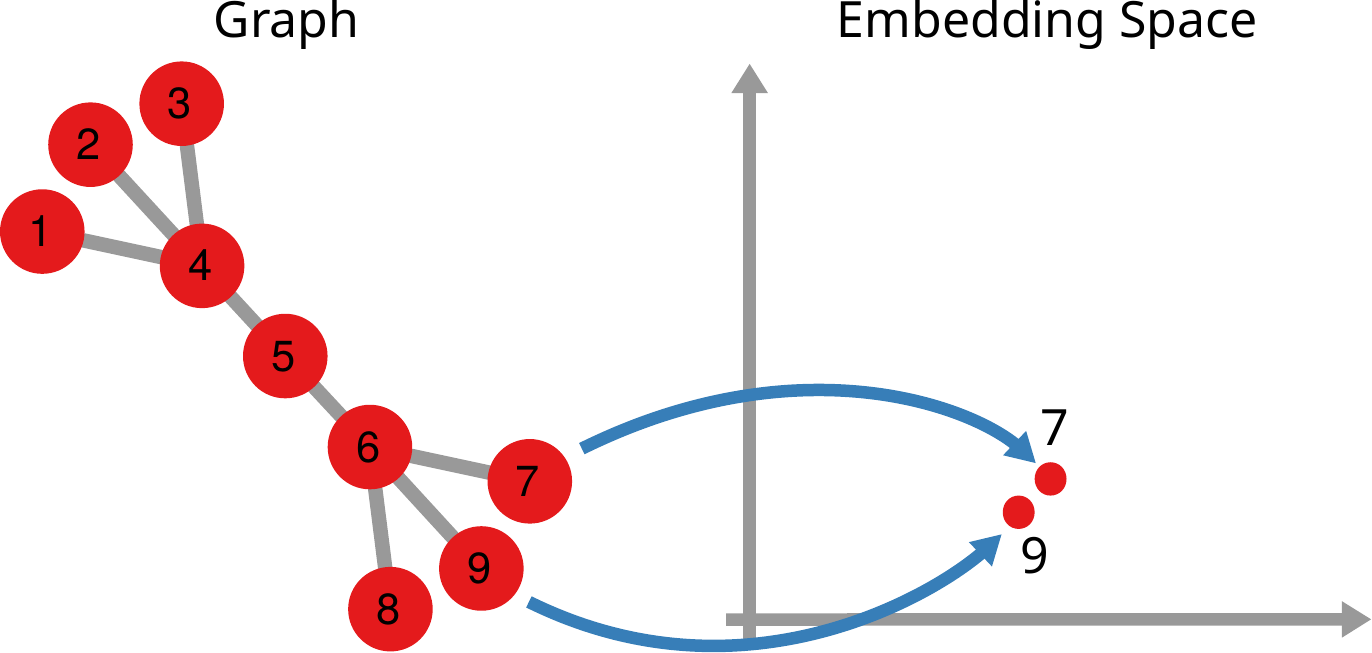}
\caption{A graph and its embedding space, where the embedding function represented by the blue arrows places structurally equivalent nodes nearby in a 2D embedding space.}
\label{fig:embeddings-similarity}
\end{figure}

Not all functions turning a node into a vector of numbers are good. To be useful, a node embedding should satisfy at least a few basic properties:

\begin{enumerate}
\item An embedding should be small, meaning that it should have a low dimensionality, few entries. If your network has $|V|$ nodes, then the vector representing each node should have fewer than $|V|$ entries.
\item An embedding should be dense, you don't want most of your entries in most of your node embeddings to be zeroes -- otherwise, why having those dimensions in the first place, if they don't give you much extra information?
\item An embedding should be permutation invariant: we shouldn't be getting different embedding depending on the order in which we're exploring the nodes.
\item An embedding should be a faithful representation of the topology of your network. Nodes that are ``similar'' should be represented by vectors at a low distance -- as Figure \ref{fig:embeddings-similarity} shows.
\end{enumerate}

The aim of the embedding function is to embed your nodes into a low dimensional space in a way that respects at least those properties. You can think of your nodes as the points of a scatter plot: points that are spatially close to each other are similar. If this low dimensional representation is any good, you can then analyze this scatter plot and discover interesting properties of your nodes. This is helpful, because usually the scatter plot is (i) easier to analyze than a graph, and (ii) a more common data structure than a graph on which you can apply a more diverse set of algorithms that were not developed with graphs in mind.

The real meat that makes embeddings interesting is the last constraint -- which connects similarity of embeddings with the similarity of the nodes. But before we dive into that, there might be a question lingering in your mind: why are we bothering with graph embeddings? We can already represent easily a node with a vector. In fact, this is something you taught me since Chapter \ref{cha:mat}! A node is nothing more than a row in the adjacency matrix of the graph. Thus it is a vector. Why can't we use that as our ``embedding''?

The reason is that $A$ fails each of the first three constraints:

\begin{enumerate}
\item It is not low-dimensional. If you slice the adjacency matrix, your nodes will be represented by a vector of length $|V|$, which doesn't save you any dimensions.
\item It is binary and sparse, therefore not information-dense. Most algorithms that you want to apply on your embeddings don't work well with this sort of input data.
\item It is not permutation invariant. You will get a different embedding depending on how you sort the rows and columns of your matrix. So you could get two isomorphic graphs with different embeddings!
\end{enumerate}

\begin{figure}
\centering
\begin{subfigure}{.34\columnwidth}
\includegraphics[width=\textwidth]{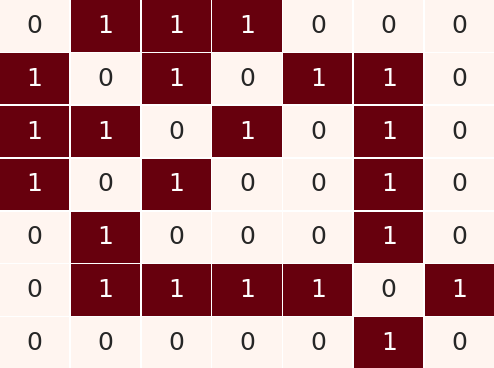}
\caption{}
\end{subfigure}
\qquad
\begin{subfigure}{.34\columnwidth}
\includegraphics[width=\textwidth,angle=180]{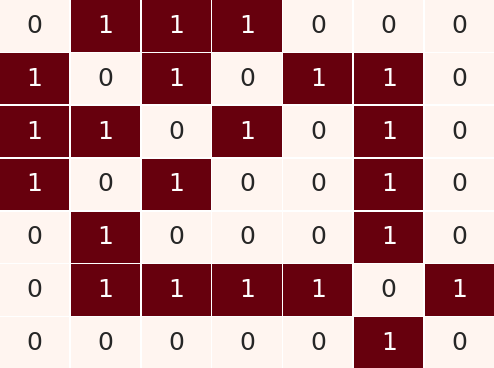}
\caption{}
\end{subfigure}
\caption{(a) An adjacency matrix. (b) The adjacency matrix of a graph isomorphic to the one represented by the adjacency matrix in (a).}
\label{fig:embeddings-adj-permutation}
\end{figure}

I show you an example of the last problem in Figure \ref{fig:embeddings-adj-permutation}. You can tell that the two graphs defined by the two matrices I show there are isomorphic, because one matrix is the rotation of the other -- quite literally, I was too lazy to make another figure from scratch, so I just rotated it in latex.

In fact, you want something even stronger than permutation invariance. You want permutation equivariance: not only the embeddings of a node should not be changed by the order in which you visit the graph, but nodes in the exact same position in the graph should always get the same embeddings.

\subsection{Embeddings and Node Similarity}
Let's now focus on the most interesting of the constraints I mentioned: similar nodes should get similar embeddings. This means that we need to decide how to calculate the similarity of nodes and of embeddings.

The similarity of embeddings is the easiest, because they are just numerical vectors, so let's start from there. We can store all our embeddings in a matrix, let's call it $Z$. $Z$ is a $|V| \times d$ matrix, where each node $v \in V$ gets a $d$ dimensional vector. One can estimate the similarity between two node embeddings $Z_u$ and $Z_v$ by their dot product $Z_u^TZ_v$. Great.

\begin{figure}
\centering
\begin{subfigure}{.3\columnwidth}
  \footnotesize
  \begin{tabular}{l|r}
    Node & Embedding \\
    \hline
    $1$ & $\{0.01, 0.01\}$\\
    $2$ & $\{0.01, 0.01\}$\\
    $3$ & $\{0.01, 0.01\}$\\
    $4$ & $\{0.01, 1\}$\\
    $5$ & $\{0.33, 0.5\}$\\
    $6$ & $\{0.05, 0.9\}$\\
    $7$ & $\{0.05, 0.05\}$\\
    $8$ & $\{0.05, 0.05\}$\\
    $9$ & $\{0.05, 0.05\}$\\
  \end{tabular}
\caption{}
\end{subfigure}
\begin{subfigure}{.35\columnwidth}
\includegraphics[width=\textwidth]{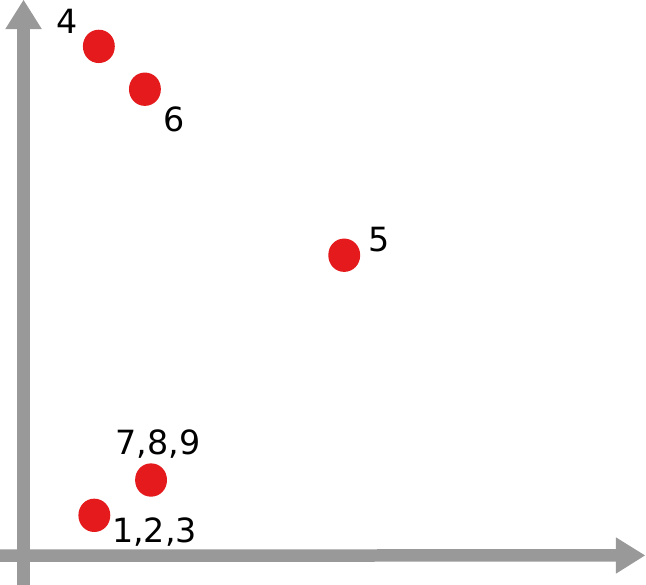}
\caption{}
\end{subfigure}
\caption{(a) A different valid embedding of the graph in Figure \ref{fig:embeddings-example}(a), assigning a two dimensional vector to each node. (b) The scatter plot representation of (a)'s embeddings.}
\label{fig:embeddings-example2}
\end{figure}

What about node similarity? The whole point of making embeddings was that we cannot easily quantify a network structure, so aren't we lost in recursion? Well, actually there are many ways to estimate node similarity: we saw a few when we talked about different roles and structural equivalences in Chapter \ref{cha:centr-roles}. The fact that there are no unique ways to define node similarity shouldn't scare you, but empower you: you can define different embeddings depending on your notion of node similarity, which depends on what you want to do with your embeddings. In other words, you can build and optimize embeddings for many specific tasks -- Figure \ref{fig:embeddings-example2} shows you a perfectly valid alternative embedding for the network from Figure \ref{fig:embeddings-example}.

The example I show in Figure \ref{fig:embeddings-example} is one you'd use if you wanted your embeddings to help you with community discovery or some spreading event on the network. However, it would be poor when it comes to estimate, for instance, structural equivalence (Section \ref{sec:centr-similarity}), which Figure \ref{fig:embeddings-example2} instead captures.

Note that you have an additional degree of freedom: not only you can decide which function you use to create the embedding, you can also decide the shape of the space in which you're creating the embedding. For simplicity, in Figures \ref{fig:embeddings-example2}(a) and \ref{fig:embeddings-example2}(b) I use an Euclidean space: each dimension has equal importance and the distance between two points is determined by the length of the straight line between the points. This is not the only choice. For instance, some researchers use a hyperbolic space\cite{krioukov2010hyperbolic}\cite{bianconi2017emergent}\cite{kitsak2020link}, where the distance between two points is not determined by a straight line, but by the branch of a hyperbole.

\subsection{Classical Embeddings}
\begin{figure}
\centering
\includegraphics[width=.3\textwidth]{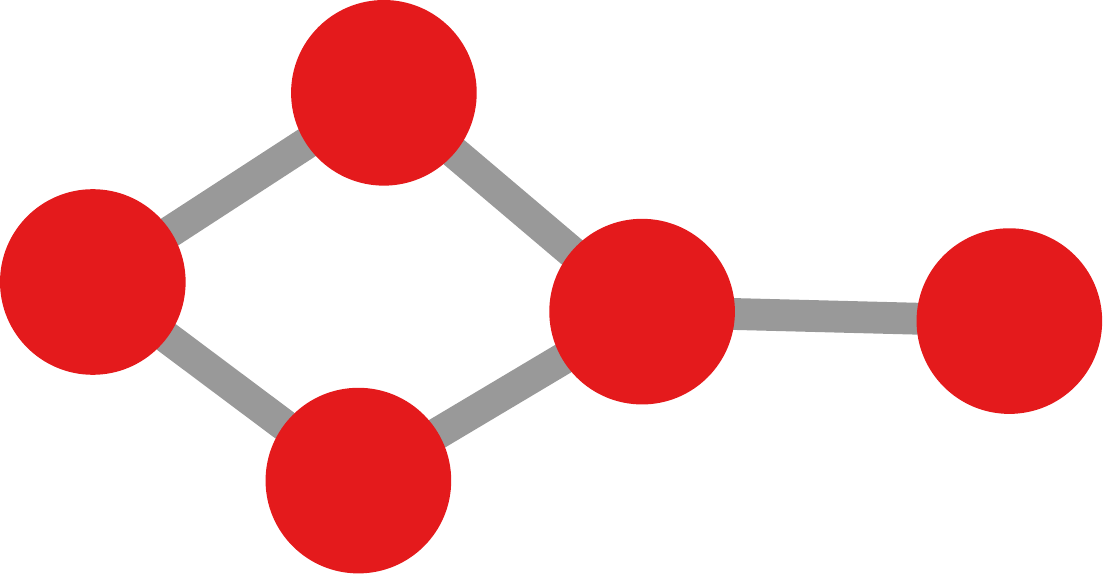}
\caption{A simple graph whose nodes are located on a 2D space interpreting their 2D embeddings as spatial coordinates.}
\label{fig:embeddings-layout-simple}
\end{figure}

At this point in the chapter, you might feel a bit daunted by the concept of embeddings. It ostensibly looks simple: to embed is to transform a node in a vector of numbers. But you might think that, in order to encode the complexity of what a node does in a potentially large structure you might have to apply ridiculously sophisticated and advanced algorithms. In this subsection, I want to convince you that this isn't true. In fact, you have seen me using node embeddings -- without me telling you that I was -- ever since Section \ref{sec:prob-markov}! Let me show you the first case of node embeddings you saw in the book, stripped away of some unnecessary complexity: Figure \ref{fig:embeddings-layout-simple} is a simplified reproduction of Figure \ref{fig:random-walk-example} -- literally the first network you saw in the book.

I can hear you thinking: \textit{Michele, what the hell are you talking about?} But bear with me for a second. In Figure \ref{fig:embeddings-layout-simple} I need to decide where to put the nodes on a 2D plane. That means I need to figure out a pair of $x$ and $y$ coordinates so that the network looks understandable. The way I decide the values of $x$ and $y$ for each node depends on their connections in the network. This is equivalent to say that I use the structure of the network to assign to each node a 2D numerical vector. In turn, this also means exactly that I embed the network structure in a 2D space.

It would be ugly if there were edge crossings or the nodes overlapped, so the network structure constrains how I build these vectors. Every time you try to display your network in the clearest form possible, you are creating 2D (sometimes 3D) node embeddings. In fact, there are a bunch of specialized algorithms that layout your networks -- we'll see them in Chapter \ref{cha:layouts}. They are a relatively simple way to create node embeddings -- but, crucially, it is a totally \textit{valid} way to do it.

\section{From Neural Networks to Graph Neural Networks}\label{sec:mining-embeddings-nn}
\begin{figure}
\centering
\begin{subfigure}[t]{.3\columnwidth}
\includegraphics[width=\textwidth]{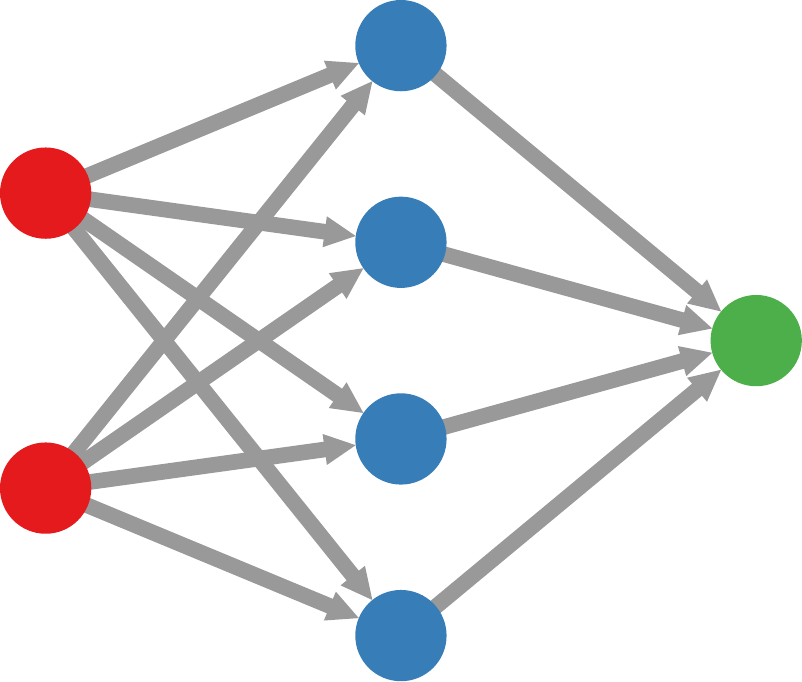}
\caption{Feedforward.}
\end{subfigure}
\quad
\begin{subfigure}[t]{.3\columnwidth}
\includegraphics[width=\textwidth]{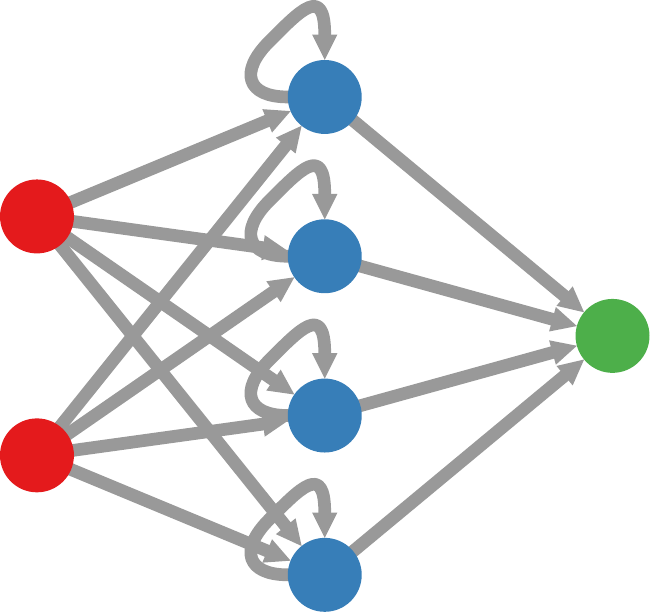}
\caption{Recurrent.}
\end{subfigure}
\quad
\begin{subfigure}[t]{.3\columnwidth}
\includegraphics[width=\textwidth]{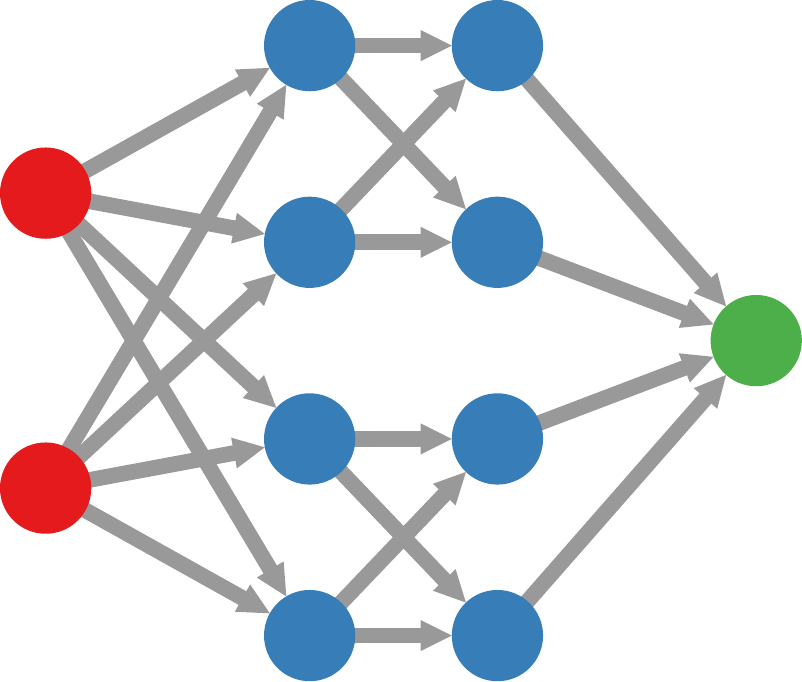}
\caption{Modular.}
\end{subfigure}
\caption{Different neural networks. The node color determines the layer type: input (red), hidden (blue), output (green).}
\label{fig:neural-nets-2}
\end{figure}

The main reason people nowadays can't get enough of making new ways of building node embeddings is because node embeddings are a perfect way to boil down the complexity of a graph in a way that classical neural network architectures can understand it. To see what I mean, we shall take another look at Figure \ref{fig:neural-nets}, which I reproduce here as Figure \ref{fig:neural-nets-2}.

From the figure, you can see the general architecture of a neural network. The input goes into the input layer, it is processed by one or more hidden layers, until the neural network produces an output in the output layer. Now, as I mentioned in Chapter \ref{cha:mining-base}, the problem here is that the input layer is a vector of numbers, but networks aren't vectors of numbers. If they were, analyzing them would be easy and this book wouldn't exist. So we need to find a way to transform them into something a neural network understands. Which is embeddings, as I mentioned.

We generally call this approach ``shallow learning'' because we use the network only to build the embeddings. However, once we have the embeddings, we forget about the network and we don't use it any more -- there are no more edges in the vectors of numbers we feed into the algorithm. As I already outlined, when we get to Chapter \ref{cha:mining-deep}, we'll deal with ``deep learning''. In this case, we use the edges themselves as the neural network, which then become a central part of the learning process. This will require us to define new graph neural network algorithms, rather than re-using generic neural network approaches that were not defined specifically for graphs.

\begin{figure}
\centering
\includegraphics[width=0.45\columnwidth]{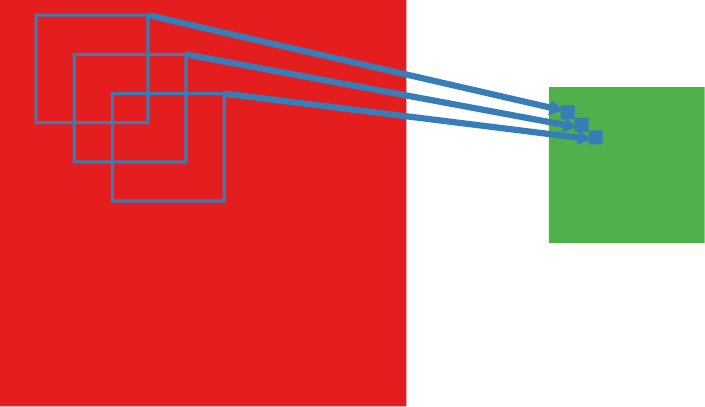}
\caption{A convolution operation. Color determines the layer of the neural network: red = input, blue = hidden, green = output.}
\label{fig:convolution}
\end{figure}

However, I like to think that the two approaches aren't so different. This is just my personal view but, if you squint hard at it, many of the classical neural network approaches are actually graph neural networks. In Section \ref{sec:mining-rwemb-basic} I'll give an example from natural language processing, but here I'll focus on another classical neural network approach: image convolution. Figure \ref{fig:convolution} shows a depiction of what a simple convolution operation looks like at an abstract level.

The figure shows what one normally does in image convolution: we chunk the image in a collection of (overlapping) pixels. Then the neural network learns -- across multiple layers -- relationships between adjacent chunks of pixels. Eventually it will figure out that these chunks here look like an eye, those over there look like a mouth, and so on until it classifies the image as a face.

\begin{figure}
\centering
\includegraphics[width=0.45\columnwidth]{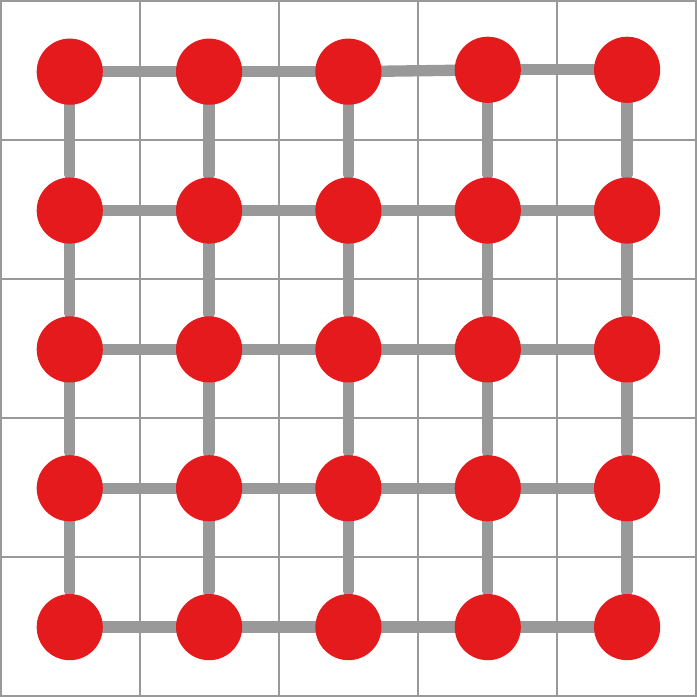}
\caption{A way of thinking about image convolution as a graph neural network problem. Each square in the grid represents a pixel. Each pixel can be though of as a node connected to its adjacent pixels.}
\label{fig:convolution-gnn}
\end{figure}

But consider a perspective shift, which I represent in Figure \ref{fig:convolution-gnn}. Here, each square is a (collection of) pixels. What the neural network does is to look at each pixel's neighbors and use them to update the pixel's representation. This is literally the same as having a simple network -- a regular grid of nodes -- as the input of your neural network architecture. So, in this simplified example, there are two differences between a neural network and a graph neural network. A non-graph neural network:

\begin{enumerate}
\item Only accepts simple graphs as inputs; and
\item It requires the simple graphs to be \textit{ordered} -- we know which nodes come ``before'' and ``after'' the current one, for instance the ones representing the pixel on the left and on the right.
\end{enumerate}

A complex network doesn't have such a canonical node ordering. So, to move from neural networks to graph neural networks, we ``simply'' have to figure out how to handle arbitrarily complex unordered graphs.

We now take a look at specialized ways to build graph embeddings. I decided to focus on two main families of approaches. Here we focus on spectral embeddings. Then, in the next chapter, we'll see a random walk embeddings, which has proven to be a popular and powerful way of building embeddings.

\section{Spectral Embeddings}
This is the oldest category of graph embeddings techniques. The objective of the researchers working in this early definition of the problem was simple dimensionality reduction. In other words, they were mainly interested in having small vectors representing the topology around a node without having to look at the entirety of its neighborhood -- that is to say, its adjacency vector. In practice, they were trying to have some sort of network-aware Principal Component Analysis (Section \ref{sec:mat-factors}).

Let's repeat the general idea of embedding nodes: if nodes $u$ and $v$ are connected to each other by a strong link $A_{uv}$, then their embeddings $Z_u$ and $Z_v$ should be similar. In spectral embeddings you want to find a good loss function that approximates well your objective which can be optimized by solving an eigenvector problem. In practice, you want to find the right matrix representation of your graph corresponding to that specific loss function.

\begin{figure}
\centering
\begin{subfigure}[t]{.135\columnwidth}
\includegraphics[width=\textwidth]{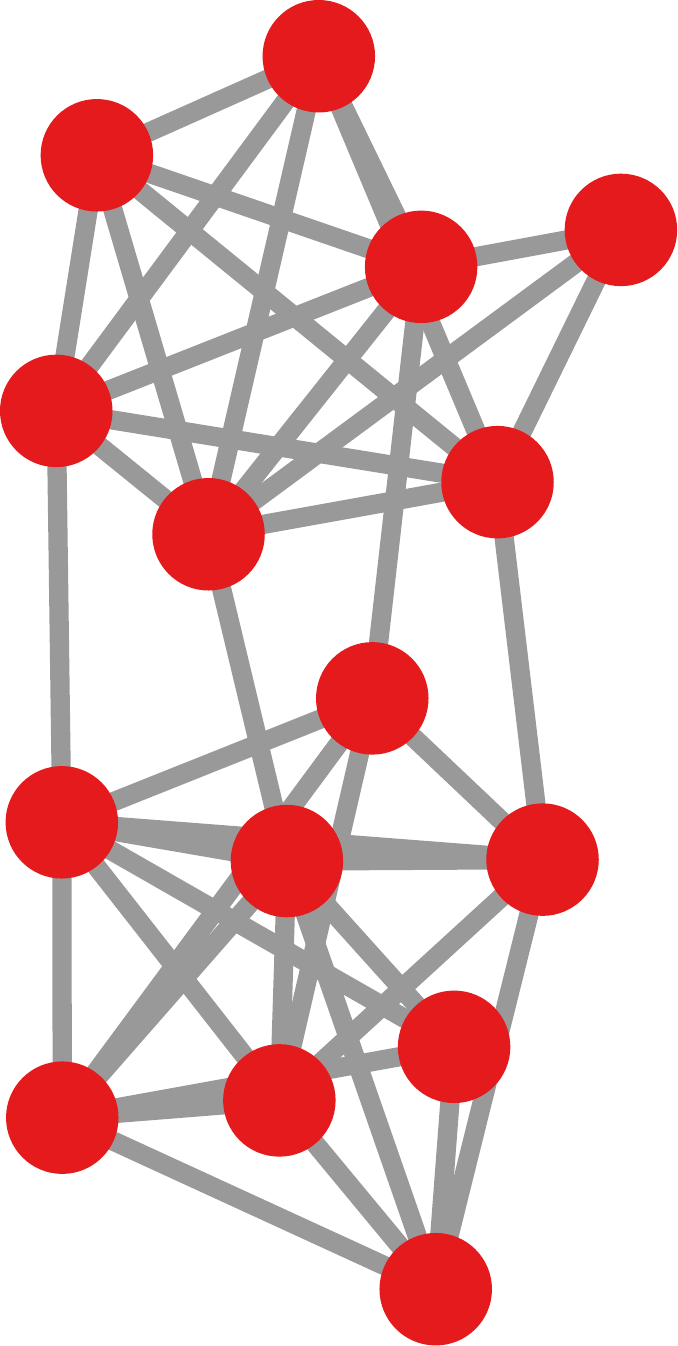}
\caption{}
\end{subfigure}
\quad
\begin{subfigure}[t]{.385\columnwidth}
\includegraphics[width=\textwidth]{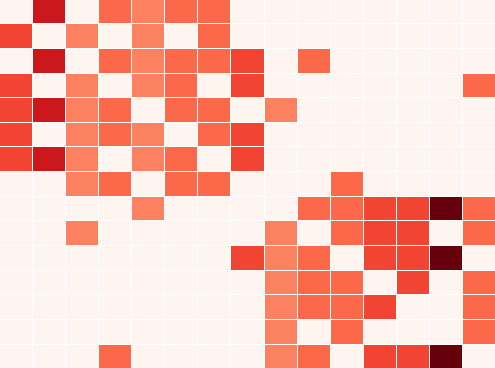}
\caption{}
\end{subfigure}
\quad
\begin{subfigure}[t]{.385\columnwidth}
\includegraphics[width=\textwidth]{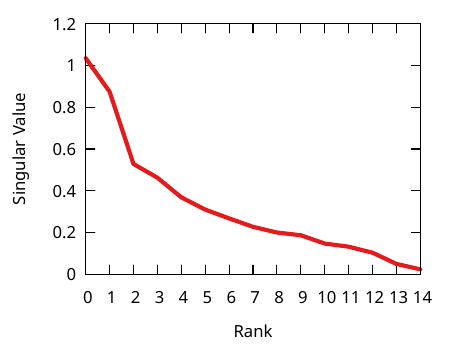}
\caption{}
\end{subfigure}
\caption{(a) A graph. (b) Its $D^{-1/2}AD^{-1/2}$ matrix. (c) The singular values of (b), sorted on the x axis in descending order.}
\label{fig:spectral-embeddings}
\end{figure}

These matrix representations are sometimes transformations of $A$, but most often of the Laplacian $L$. Let's look at a simple example to warm up. Figure \ref{fig:spectral-embeddings}(a) shows a graph. Here, I choose to transform $A$ into $D^{-1/2}AD^{-1/2}$ -- see Figure \ref{fig:spectral-embeddings}(b)). The first step to get the embedding is to calculate the Single Value Decomposition (SVD) of that transformed matrix -- see Section \ref{sec:mat-factors}.

Remember that SVD produces three outputs -- the singular values and two matrices with the singular vectors. For now, I only show you the singular values in Figure \ref{fig:spectral-embeddings}(c), which give you an idea of the explanatory power of a given singular vector. You can see that, to describe $D^{-1/2}AD^{-1/2}$ the first two single vectors are more powerful than the rest. This is equivalent in saying that you can represent each node with a 2D embedding, composed by the values corresponding to that node in the first two single vectors.

\begin{figure}
\centering
\includegraphics[width=0.5\columnwidth]{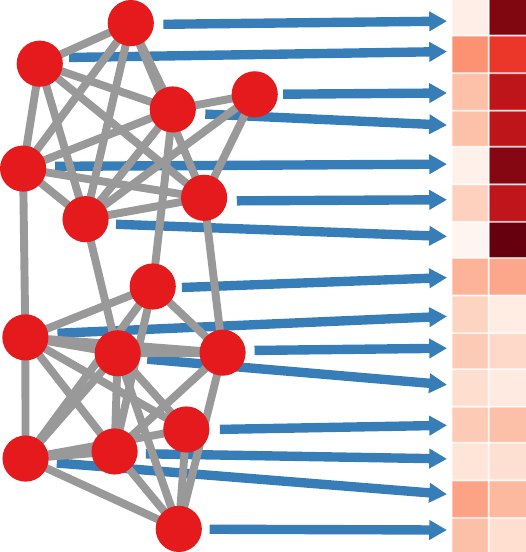}
\caption{On the right you have the two singular vectors corresponding to the top two singular values. Each node corresponds to a row in this matrix.}
\label{fig:spectral-embeddings-2}
\end{figure}

Figure \ref{fig:spectral-embeddings-2} shows you how those 2D embeddings look like, highlighting the difference between nodes in one community and the ones in the other. In practice, if you decided you want 2D embeddings, you can take the first two singular values and those will make your $Z$ embedding matrix.

This could be familiar to you, it is practically equivalent to solving min-cut by using the eigenvectors of the Laplacian -- something we saw all the way back in Section \ref{sec:rw-mincut}. So that's why $L$ is a popular solution here. But building spectral embeddings is more general than just finding the mincut solution. A community structure is not necessary to create spectral embeddings. You can do that with arbitrary networks, provided you have a good loss function. Different loss functions will lead to slightly different transformations of $L$ before its decomposition. As far as I can tell, this is the main thing differentiating spectral embedding approaches.

Let's take a look at a couple of examples. There are other approaches, most notably ways to include node attributes in the construction of the embedding\cite{zhang2021spectral}. But, once you got a few basic examples laid down, you're equipped to think about spectral embeddings techniques in general.

\subsection{Locally Linear Embedding}
In Locally Linear Embedding\cite{roweis2000nonlinear}, assume that you have your $Z$ embeddings. Then, the loss function you want to minimize is:

$$ \sum \limits_u \left | Z_u - \sum \limits_v Z_v A_{uv} \right |^2. $$

Here, the straight bars mean that we are taking the length of the vector $Z_u - \sum \limits_v Z_v A_{uv}$ and then we square it. In practice, your loss is the difference between $Z_u$ and $Z_v$, weighted by how strongly they are connected in the graph ($A_{uv}$). If $u$ and $v$ are connected by a high $A_{uv}$ link strength, we better have a low $Z_u - Z_v$ difference to cancel it out! It turns out that finding $Z$, the matrix containing the embeddings $Z$ for all our nodes can be solved as an eigenvector problem. Specifically, you can take the smallest eigenvectors of the matrix $(I - A)^T(I - A)$ -- but discarding the actual smallest one --, with $I$ being the identity matrix.

\subsection{Laplacian Eigenmaps}
Laplacian Eigenmaps\cite{belkin2002laplacian} change the loss function, which means the objective is still the eigenvector decomposition of a matrix, but the matrix itself is built differently. In this case, the loss function is:

$$\dfrac{1}{2} \sum \limits_u \left | Z_u - Z_v \right |^2 A_{uv},$$

which gives more weight to the $Z_u - Z_v$ difference than before -- because it squares it before multiplying it with the $A_{uv}$ link strength. Also in this case you can solve the problem by taking the smallest eigenvector of the normalized Laplacian $D^{-1/2}LD^{-1/2}$, with $D$ being the degree diagonal matrix of the graph $G$.

\section{Pooling}\label{sec:mining-embeddings-pool}
You might have noticed that so far I have exclusively talked about generating node embeddings. However, in many cases, you don't want to have node embeddings. Maybe you want to have edge embeddings to do link prediction -- trying to figure out if there is a key function distinguishing the embedding of edges that exists from edges that don't, which will tell you which non-existing links are likely to exist. Or you might want to have an embedding for the entire network. A classical case is trying to classify molecules to predict some of their characteristics. A molecule is a network of atoms, so you want to embed the entire structure.

To do this, you need to perform what we call ``pooling''. The role of a pooling function is to take a bunch of embeddings and return one that is a good representation of all of them. Pooling for edges is relatively straightforward, since we only have two node embeddings to deal with and not much complexity. For edge $(u,v)$, you have embeddings $Z_u$ and $Z_v$. To get to a $Z_{uv}$ embedding you can average them ($Z_{uv} = (Z_u + Z_v) / 2$) or multiply them element-wise ($Z_{uv} = Z_u \times Z_v$), or take the L1 or L2 norm, you get the idea. To give you an example, I take the node embeddings from Figure \ref{fig:embeddings-example}(b) and I create a bunch of edge embeddings out of them in Figure \ref{fig:edge-pooling}.

\begin{figure}
\centering
\begin{tabular}{l|rrrr}
Edge & AVG Pool & Prod Pool & L1 Norm & L2 Norm \\
\hline
$(1, 4)$ & $\{0.94,0.84\}$ & $\{0.88,0.69\}$ & $\{1.88,1.69\}$ & $\{1.33,1.21\}$\\
... & ... & ... & ... & ... \\
$(4, 5)$ & $\{0.74,0.83\}$ & $\{0.48,0.66\}$ & $\{1.48,1.66\}$ & $\{1.11,1.20\}$\\
$(5, 6)$ & $\{0.24,0.33\}$ & $\{0.00,0.00\}$ & $\{0.48,0.66\}$ & $\{0.48,0.66\}$\\
$(6, 7)$ & $\{0.21,0.05\}$ & $\{0.00,0.00\}$ & $\{0.42,0.10\}$ & $\{0.42,0.10\}$\\
... & ... & ... & ... & ... \\
\end{tabular}
\caption{A few different ways of pooling the node embeddings from Figure \ref{fig:embeddings-example}(b) into edge embeddings.}
\label{fig:edge-pooling}
\end{figure}

Pooling information for graphs is... a bit more complicated. The structure of a single edge is trivial. The structure of a graph is not. It could have an arbitrary number of nodes and edges, distributed in an arbitrary way. A good pooling function needs to take this into account to build the final embedding. We're going to see a few methods briefly, and you can read some surveys out there\cite{grattarola2022understanding}\cite{liu2022graph} to gain a deeper understanding.

\subsection{Flat Pooling}
The algorithms that do flat pooling for graphs all think: \textit{nah, Michele is wrong, pooling graphs ain't that different from pooling edges}. In this strategy, ``flat'' means that all vectors are aggregated in a manner that doesn't really take the network into account. So you could apply any of the functions that I mentioned before to all the node embeddings of a graph.

Slightly more sophisticated methods try to weight some of these embeddings differently depending on some characteristic that can be discovered with a neural network, using the attention mechanism\cite{itoh2022multi} -- which I haven't explained to you, yet. Another approach\cite{zhang2018end} sorts the node embeddings in increasing order and then performs a convolution until you get to the desired number of dimensions -- solving the issue that networks with a different number of nodes would generate a graph embedding with different dimensions.

\begin{figure}
\centering
\includegraphics[width=0.75\columnwidth]{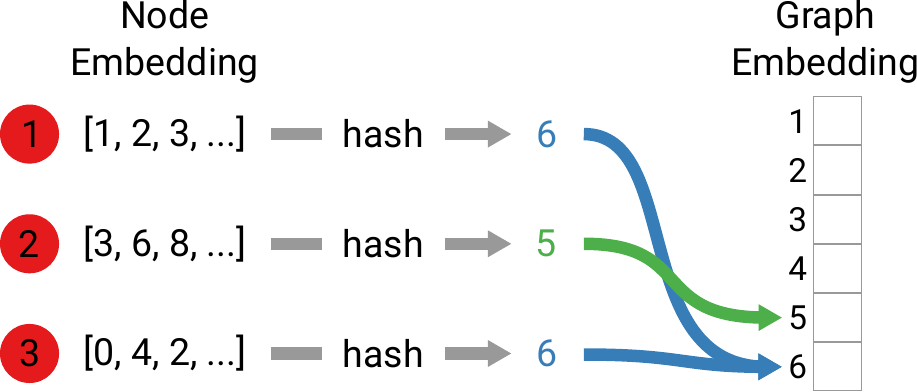}
\caption{The pooling approach determining the index of the graph embedding to be updated by hashing the node embedding.}
\label{fig:pooling-hash}
\end{figure}

Other clever methods involve hashing the node embedding so that it becomes an index and then update the corresponding index in the graph embedding\cite{duvenaud2015convolutional}. Figure \ref{fig:pooling-hash} shows how this happens in practice. Nodes $1$ and $3$ happen to get the same hash and so update the same part of the graph embedding. This is fine and expected -- if you have a graph embedding with $256$ entries, but a graph with two million nodes, you're expecting a ton of hash collisions.

The downside of this approach is that, if you have a small graph, the resulting graph embedding will be sparse. You can see it in the figure, where more than half of the resulting embedding is not updated.

There are more approaches in this class, but you get the idea: here the structure of the network doesn't matter that much. Let's see a couple of examples where instead it matters a lot.

\subsection{Node Clustering Pooling}
\begin{figure}
\centering
\begin{subfigure}[t]{.45\columnwidth}
\includegraphics[width=\textwidth]{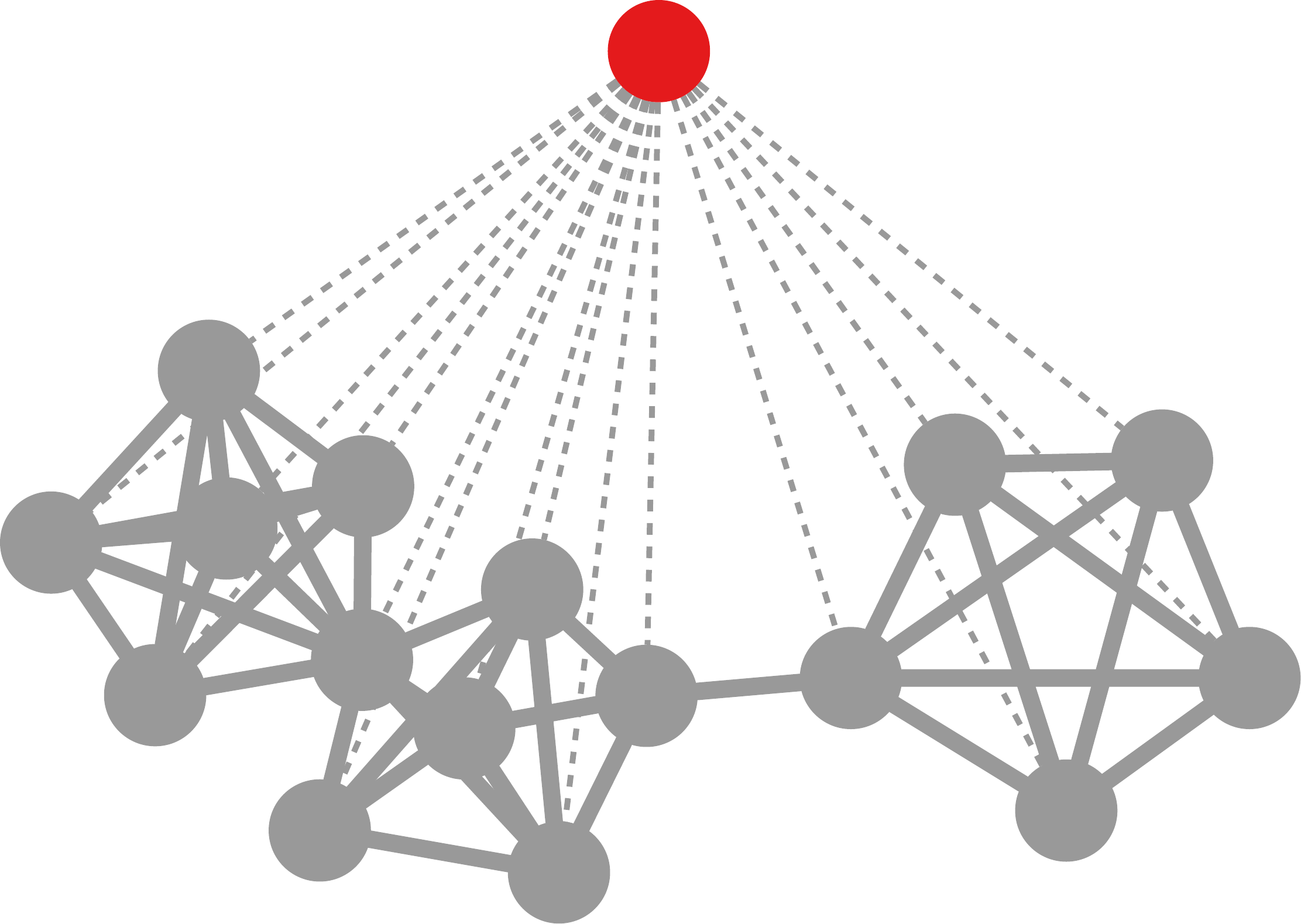}
\caption{}
\end{subfigure}
\quad
\begin{subfigure}[t]{.45\columnwidth}
\includegraphics[width=\textwidth]{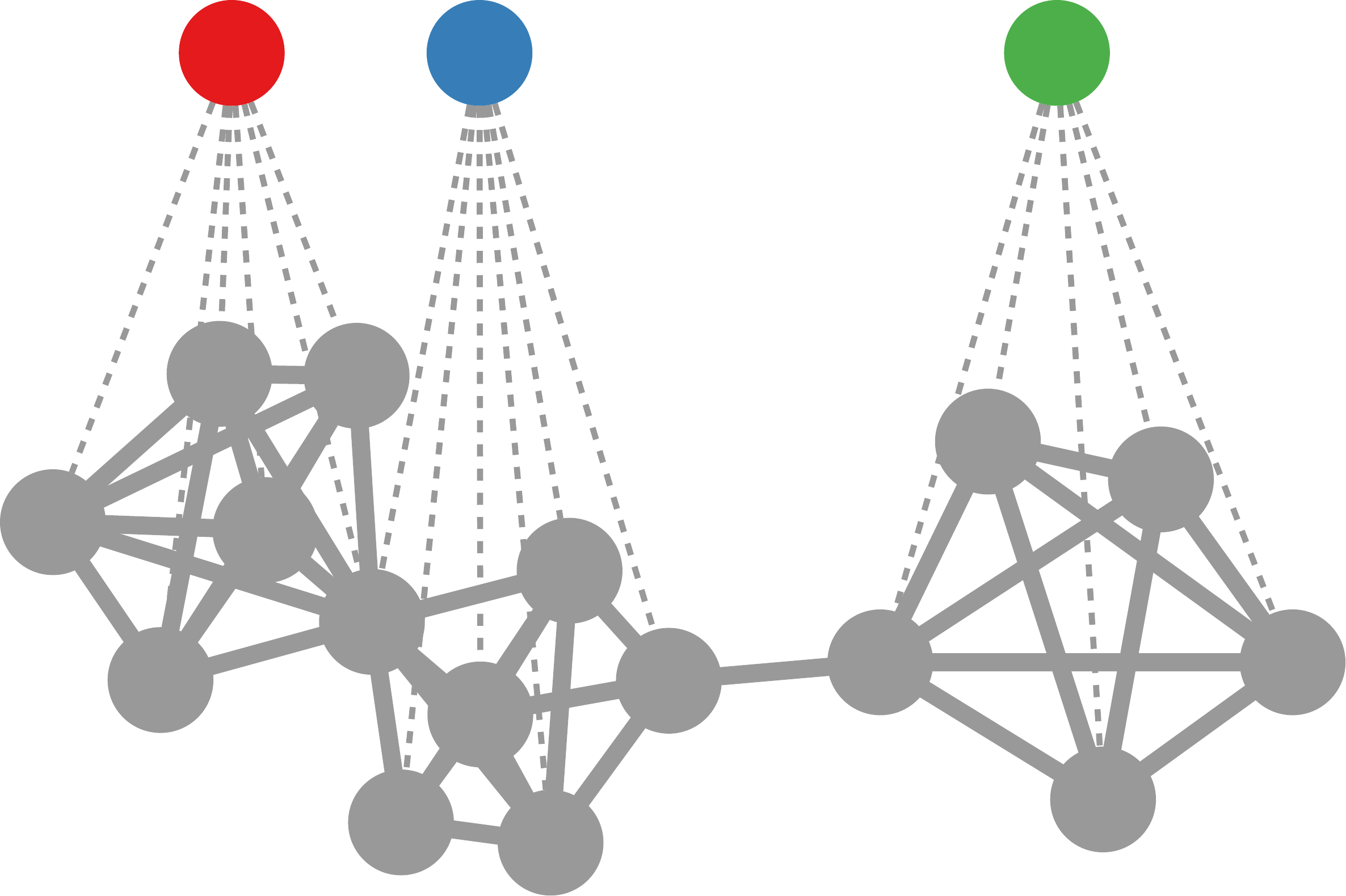}
\caption{}
\end{subfigure}
\caption{(a) Flat pooling via a virtual node (in red) connected with all nodes in the network by virtual edges (thin dashed lines). (b) Node clustering pooling with a fixed number of virtual nodes (in red, blue, and green) connected to dense subgraphs.}
\label{fig:node-clustering-pooling}
\end{figure}

You can see the flat pooling approach I presented in the previous section using a different perspective. Averaging all node embeddings is equivalent to say that there exists a ``virtual node'' connected to all nodes in the network, whose embedding is the average of all its neighbors\cite{li2015gated} -- i.e. all the nodes of the network. That will give you the same result as averaging all node embeddings. You can see this perspective in Figure \ref{fig:node-clustering-pooling}(a). 

Once you adopt this perspective, you see immediately how to make a smarter choice. What if we decide to have not one but a fixed number of virtual nodes? In that way, we can have a fixed size graph embedding. If we choose those nodes wisely, we get a good embedding that is representative of the structure of the network. This is what we're doing in this section, and Figure \ref{fig:node-clustering-pooling}(b) shows you how to visually think in these terms.

Specifically, we choose virtual nodes via node clustering\cite{defferrard2016convolutional}\cite{simonovsky2017dynamic}\cite{bianchi2020spectral}\cite{yuan2020structpool}, connecting each virtual node with a dense subgraph of the network -- as we expect those node embeddings to be coherent with each other. The criterion to choose to which nodes a virtual node connects seems awfully familiar: it's quite literally community discovery (Part \ref{par:cd}). In fact, many special things you can do to find smarter communities might help you in finding better virtual nodes.

\begin{figure}[b!]
\centering
\includegraphics[width=.66\columnwidth]{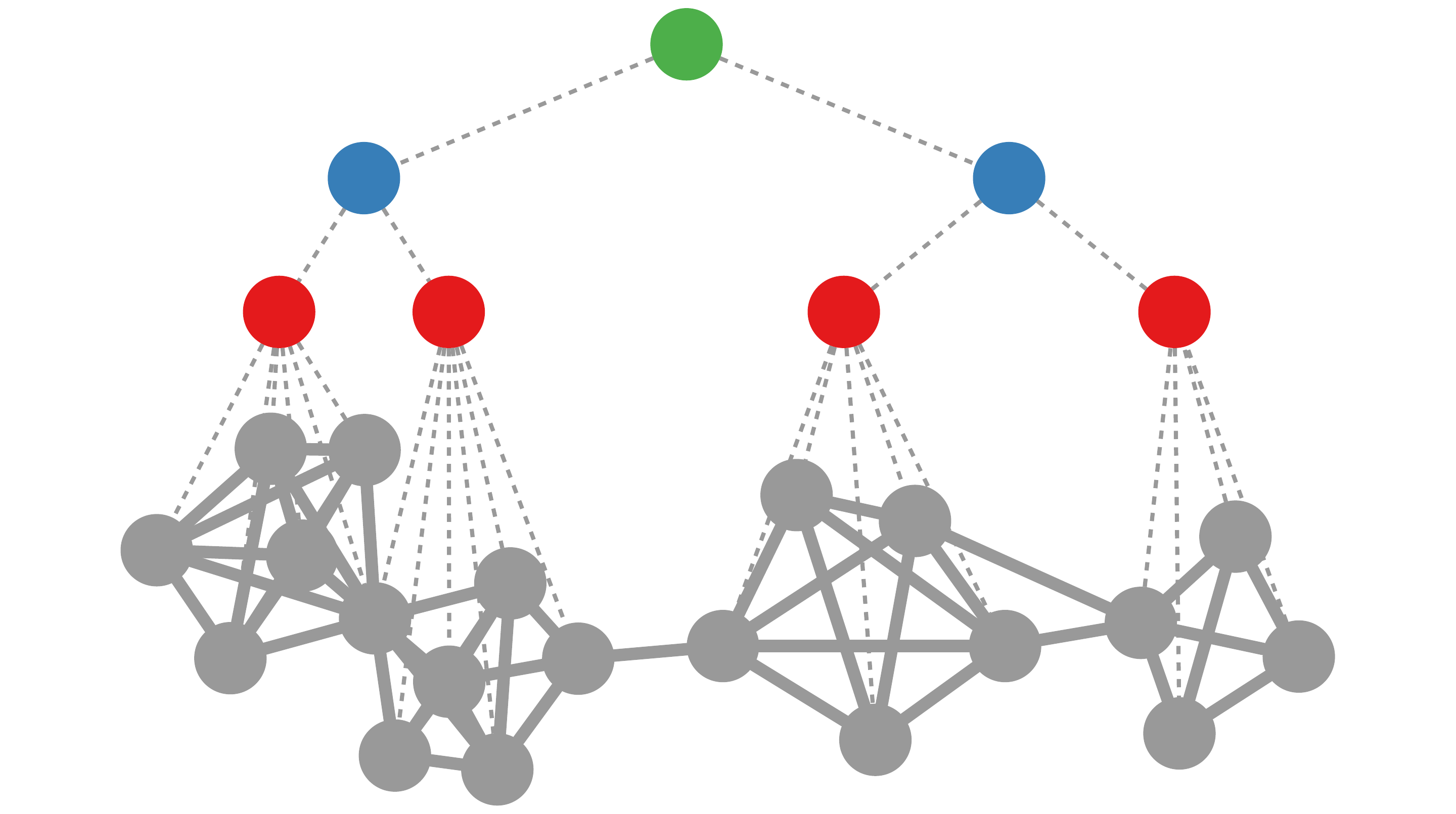}
\caption{A hierarchical node clustering pooling.}
\label{fig:node-clustering-pooling-hier}
\end{figure}

For instance, virtual nodes can share neighbors. You can see that the red and blue virtual nodes in Figure \ref{fig:node-clustering-pooling}(b) share one, because it is a node well connected to two clusters. If you want to allow that, you can pick an overlapping community discovery algorithm (Chapter \ref{cha:ocd}). And you don't have to give up the idea of having a single embedding for the entire graph. If you adopt a hierarchical community discovery mindset (Chapter \ref{cha:hcd}), you can make virtual nodes aggregating information from virtual nodes -- see Figure \ref{fig:node-clustering-pooling-hier}.

This is essentially what DiffPool\cite{ying2018hierarchical} does, with the difference that DiffPool doesn't use an off-the-shelf hierarchical community discoverer, but learns the hierarchical structure together with the node embeddings. This is advantageous, because the properties you want to capture with the embeddings might not be all that correlated with the structure. For instance, if you want spectral embeddings, you should coarsen your graph with a technique that preserves it spectrum\cite{deng2019graphzoom}.

\subsection{Node Drop Pooling}
Another way to eventually end up with a fixed-size embedding is to apply a three step strategy:

\begin{enumerate}
\item Score nodes by how relevant they are for the structure;
\item Select the least relevant nodes;
\item Coarsen the graph by dropping the nodes you just selected.
\end{enumerate}

\begin{figure}
\centering
\includegraphics[width=.66\columnwidth]{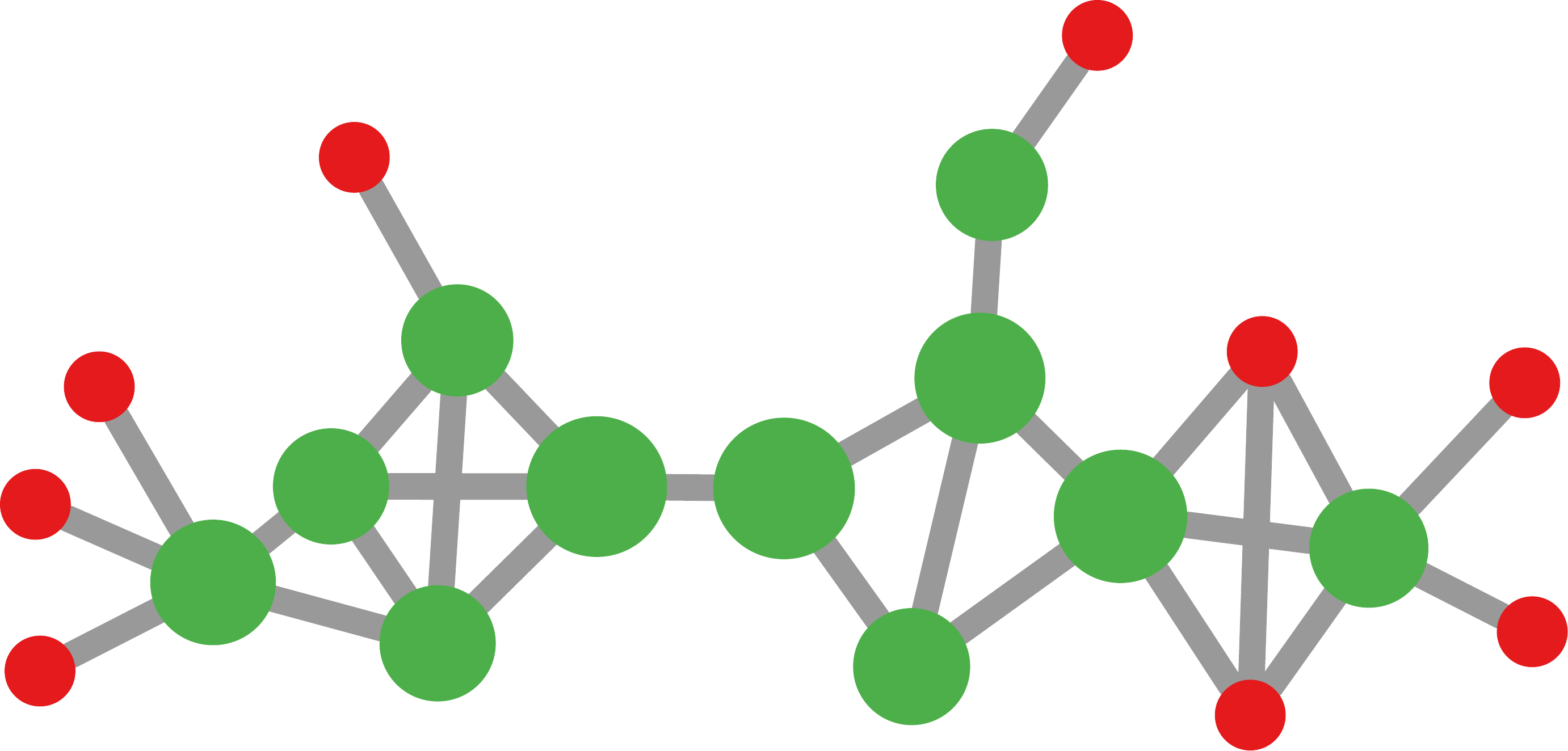}
\caption{A step of node drop pooling. The size of each node is proportional to its betweenness centrality. Nodes in red have betweenness centrality of zero, so we will drop them.}
\label{fig:node-drop-pooling}
\end{figure}

You stop when you have the desired number of nodes. Step \#2 is usually trivial, you just select the $k$ nodes with the lowest scores. Most of the work is done in step \#1: we need to find good scoring rules. In Figure \ref{fig:node-drop-pooling} I show an example of a simple scoring rule, by dropping nodes with the lowest betweenness centrality.

A general difference in how to calculate the scores is whether you're looking at the graph holistically in a single pass\cite{gao2019graph}, or whether you do multiple passes. In the latter case, a typical approach is to do two passes, to score the relevance of nodes for their local structure (e.g. a community) and then globally for the whole network\cite{gao2021topology}. For instance, a bridge nodes might be completely irrelevant locally because it's sparsely connected to the community, but crucial globally because it is the sole node keeping the community connected with the rest of the network.

While most attention is devoted to step \#1, occasionally step \#3 can be important. For instance, there are approaches that do not only drop the selected nodes. They will also drop some of the non-selected nodes, under the assumption that the selected nodes are ``weird'' by definition, and we want to build the graph embedding with a sample that is ``representative'', whatever representative means for our current analysis\cite{li2020graph}.

\section{Summary}

\begin{enumerate}
\item A node embedding is a vector of numbers representing a node. It should be small (with length $d << |V|$), dense, not dependent on the order in which you explore the nodes, and representative of the node attributes and position in the network -- similar nodes should have spatially close embeddings.
\item Depending on how you build them, embeddings can have multiple meanings and facilitate different analyses. For instance, you can use embeddings to determine node communities, or identify structurally equivalent nodes.
\item We want good node embeddings because they allow us to use any machine learning techniques to learn patterns in a complex network, which cannot be fed directly into those types of algorithms.
\item One way to build embeddings is to formulate a loss function that can be minimized by finding the eigenvectors of a corresponding transformation of the adjacency (or Laplacian) matrix of the graph. This is spectral embeddings.
\item To get edge embeddings you can perform simple combinations of the embeddings of the nodes they connect. To get graph embeddings you need to pool information from all the nodes in the network.
\item Pooling options for graph are: flat, where you don't look too hard at the structure of the graph; node clustering, where you collapse nodes into virtual nodes in a smart way; and node drop, where you recursively remove nodes that aren't salient for the network.
\end{enumerate}

\section{Exercises}

\begin{enumerate}
\item Build a 2D node embedding for the network at \url{http://www.networkatlas.eu/exercises/42/1/data.txt}. You should build it by taking the first two singular values of the  $D^{-1/2}AD^{-1/2}$ matrix.
\item Compare the embeddings you obtained from the previous exercise with the ones you get by taking the second and third eigenvectors of $(I - A)^T(I - A)$. Which one has the lowest loss according to $\sum \limits_u \left ( Z_u - \sum \limits_v Z_v A_{uv} \right )^2$?
\item Compare the embeddings you obtained from the previous two exercises with the ones you get by taking the first two eigenvectors of $D^{-1/2}LD^{-1/2}$. Which one has the lowest loss according to $\dfrac{1}{2} \sum \limits_u \left ( Z_u - Z_v \right )^2 A_{uv}$?
\item Make 2D edge embeddings by pooling each of the three node embeddings from the previous exercises. Use the product as the pooling function. Plot them on a scatter plot, coloring existing vs non existing edges with two different colors.
\item Make a graph embedding of the network by hierarchically aggregating the node embeddings in two layers: first from the nodes into two virtual nodes corresponding to the two clusters of the network (which you have to find somehow) and then aggregating the two virtual nodes into a single final virtual node. Show the difference between these embeddings and flat embedding.
\end{enumerate}

\chapter{Random Walk Embeddings}\label{cha:mining-rwemb}
In this chapter we continue looking at how to build node and graph embeddings. This time, we focus on doing so via random walks. I split this chapter out from the previous one, because there are a ton of random walk embedding methods. The reason is that making embedding via random walks hits three seducing spots. First, it is easy to do, there isn't much easier than doing a random walk. Second, it is flexible: there are many ways to bias your random walks, which can lead to meaningfully different embeddings. Finally, random walk embeddings work really well, they are very expressive. Let's take a look at them.

\section{Basic Idea}\label{sec:mining-rwemb-basic}
The general approach of the embedding methods in the random walk category is to generate embeddings of a node by performing several -- you guessed it -- random walks. The embedding of node $v$ is created by starting a bunch of random walks from $v$ and noting down which nodes appear in these random walks.

\begin{figure}
\centering
\includegraphics[width=0.66\columnwidth]{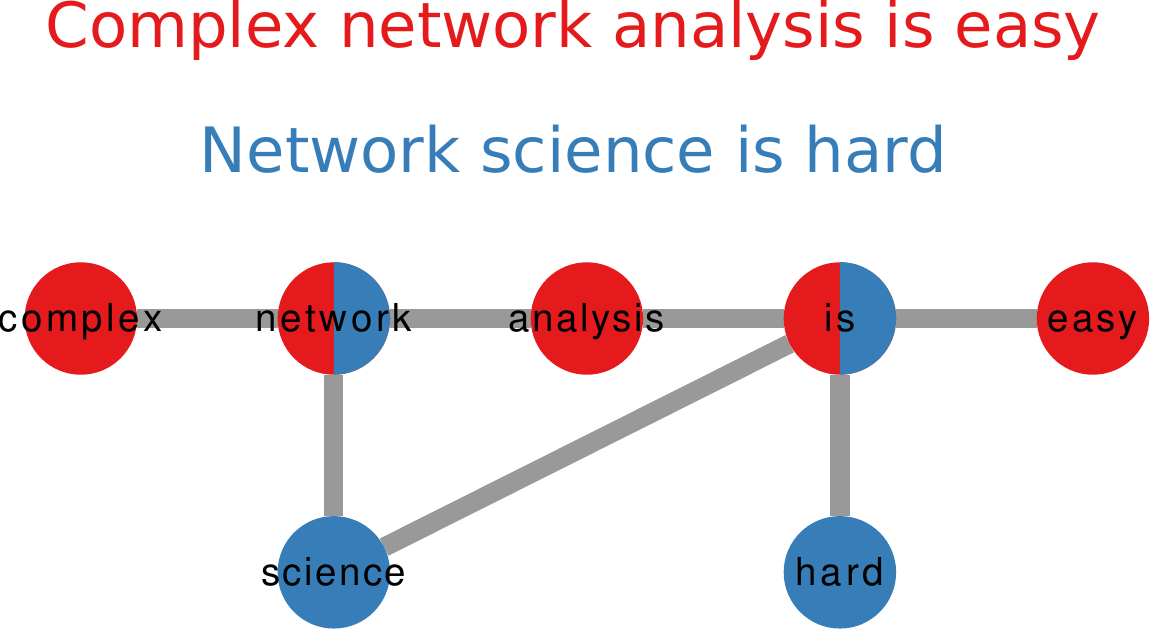}
\caption{Representing two sentences as a graph, showing the connection between Word2Vec and network science.}
\label{fig:word2vec}
\end{figure}

We know this is useful, because it works in natural language processing (NLP). In the famous Word2Vec approach\cite{mikolov2013efficient}\cite{mikolov2013distributed} we want to make embeddings of words. To embed the word $v$ we note down the other words used in sentences where we use $v$. Two words, $u$ and $v$ have similar embeddings if they are used around the same words -- that means they must share some semantic or syntactic similarity. The reason why I'm describing Word2Vec is that this NLP technique can be translated directly into network terms. In fact, if you were a network supremacist (I know I am), you might even say that Word2Vec is a network algorithm, and the NLP people are quiet about it -- just like in Section \ref{sec:mining-embeddings-nn} we saw that image convolution is a network algorithm specifically designed for simple grids. Let's look at Figure \ref{fig:word2vec}.

In Word2Vec you are basically saying that each word is a node and each sentence is a random walk. The random walk happens on a network we don't have, the one connecting each word to each other because of their semantic and syntactic relationships. But not having the network doesn't matter, because all you need to create the embeddings are the random walks. Similarly to the random walk approach in community discovery (Section \ref{sec:cd-partition-rw}), we base this on the assumption that random walks of two similar nodes will hit the same neighbors.

Once you realize this, you can apply Word2Vec to your networks and get node embeddings. This is quite literally what DeepWalk\cite{perozzi2014deepwalk} does. You simply perform a bunch of random walks -- normally a certain number of random walks starting at a given node, with a given length, and so on. The more two words co-appear in a random walk, the more similar they are. Therefore, whatever your $Z$ embeddings are, they need to be a good approximation of the co-appearances frequencies.

\begin{figure}
\centering
\includegraphics[width=0.8\columnwidth]{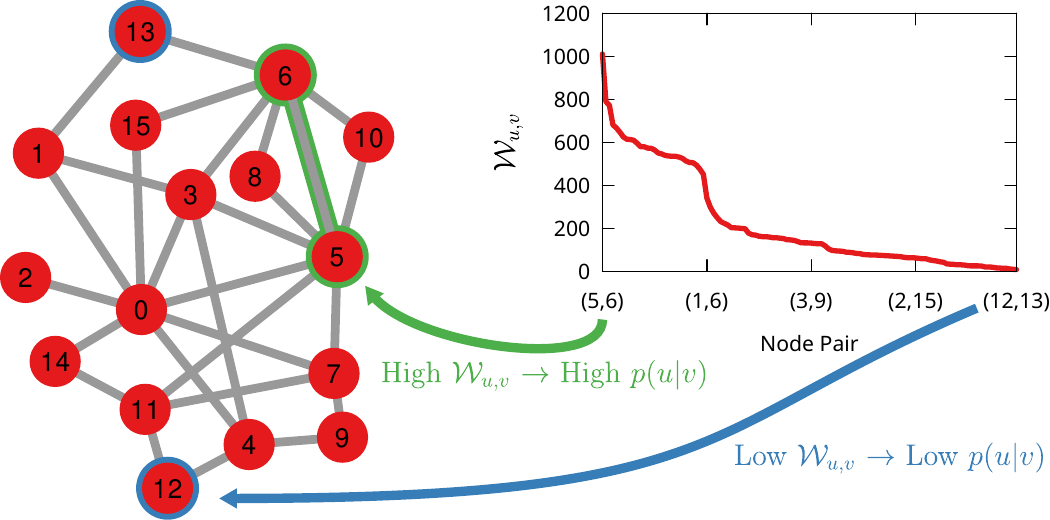}
\caption{(Left) A graph. (Right) The graph's frequency of node co-appearance in a random walk (y axis) per node pair (x axis) sorted in descending order. Node pairs on the left of the frequency plot should get as similar embedding as possible.}
\label{fig:deepwalk}
\end{figure}

You can write this in mathematical form, all you need is to find the right loss function (Section \ref{sec:ml-loss}) -- which in this case is the log-likelihood. I'm going to simplify DeepWalk a bit, to help intuition. Let's assume $\mathcal{W}$ is the set of our random walks. We can say that $|\mathcal{W}_{u,v}|$ is the number of random walks in which $u$ and $v$ co-appear. Then, we can consider this log-likelihood function:

$$ \mathcal{L} = - \sum \limits_{u,v \in \mathcal{W}}  |\mathcal{W}_{u,v}| \log\left(p(u|v)\right), $$

where $p(u|v)$ is our estimation of the probability of observing node $u$ in a walk that touched node $v$. This is a likelihood function because we're summing logs of probabilities, which is equivalent to multiply the probabilities themselves -- see Section \ref{sec:ml-loss} for a refresher with a simple case with coin tosses. We multiply each probability with $|\mathcal{W}_{u,v}|$ because we want a probability to weigh more if the two nodes co-appear in a random walk a lot of times. If that happens -- meaning $|\mathcal{W}_{u,v}|$ is high -- we want $p(u|v)$ to be high as well. This is what Figure \ref{fig:deepwalk} shows for a random network.

Ok, so the question left unanswered is: how do we estimate $p(u|v)$? That's the job of our embeddings $Z$. If we find a really good $Z$, we can get high $p(u|v)$ for each high $|\mathcal{W}_{u,v}|$. We already know that the embedding similarity of $u$ and $v$ is $Z_u^TZ_v$, but the unfortunate thing is that this isn't a probability. No problem, we have the softmax function (Section \ref{sec:ml-activation}) whose superpower is to transform anything into a probability. So:

$$ p(u|v) = softmax(Z_u, Z_v) = \dfrac{e^{Z_u^TZ_v}}{\sum \limits_{w \in V} e^{Z_u^TZ_w}}. $$

Note that this is not computationally efficient. In the softmax function, for the denominator, we need to loop over all the nodes in the network. To avoid that, we can approximate the softmax normalization via negative sampling, i.e. estimating the similarity of two nodes that co-appear in a random walk with a sample of the similarities of the nodes that don't co-appear with them. Since we're sampling, we can choose a suitably low number of samples, reducing computational complexity. Since this is all about random walks, you can perform the negative sampling with a random walk as well (Section \ref{sec:sampling-rw}), which basically means to pick nodes in the negative sample with a probability proportional to their degree.

While using negative sampling technically gives you a function that is not softmax, it approximates it well enough to be considered the same\cite{goldberg2014word2vec}.

Note that approaches in this family require you to set a lot of parameters. You need to decide how many walks per node you want to do, how long the random walks should be, etc. I don't know about you, but I don't like to think and to make lots of decisions. Some methods cover your back and can learn the optimal values for those parameters for you\cite{abu2018watch}.

\section{Variants on Simple Graphs}
If we wanted to improve over DeepWalk, what would we do? For now, let's keep things simple and assume we only deal with simple graphs -- graphs with a single edge type. We deal with heterogeneous / multilayer network later. We have a few options, the most common being to deploy high order random walks. This is popular because, depending on how the high order is implemented, you can get node embeddings with radically different meanings and properties.

\subsection{Node2Vec}
\begin{figure}
\centering
\includegraphics[width=0.4\columnwidth]{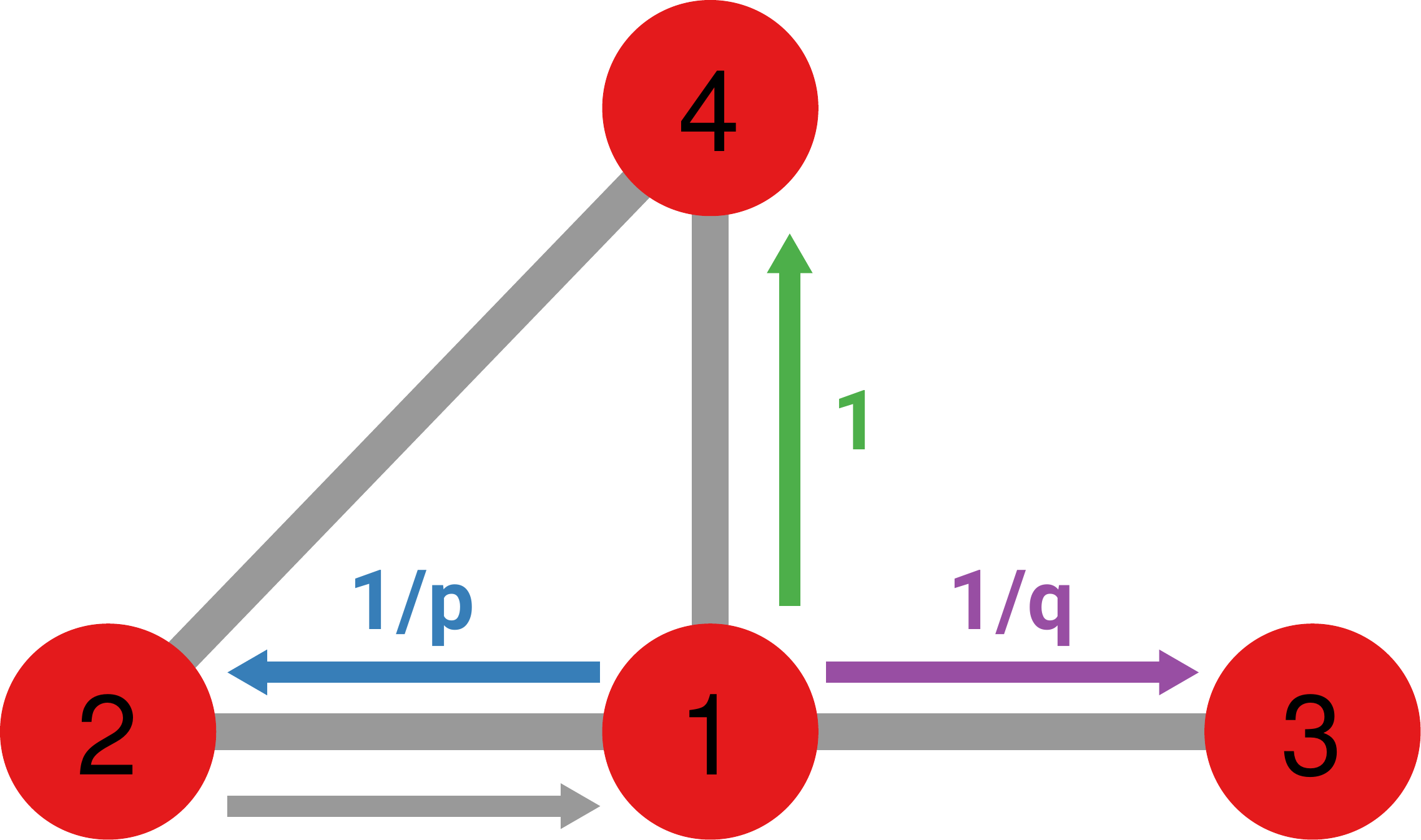}
\caption{The biased random walk implemented in Node2Vec. The gray arrow is the step we just took. The colored arrows show the reweighting of the various possibilities, with the $p$ and $q$ parameters. Blue for backtrack, green for common neighbor, purple for not shared neighbors.}
\label{fig:node2vec}
\end{figure}

Node2vec\cite{grover2016node2vec} notices that DeepWalk uses uniform random walks, picking the next step of the walk completely at random. However, there is something to gain by performing higher order random walks (Chapter \ref{cha:hod}). Figure \ref{fig:node2vec} shows you an example.

In a high order random walk what matters is not only the node you're currently in, but also the node you come from. Node2Vec uses two parameters to exploit this information, $p$ and $q$. In practice, it tries to modulate between a BFS-like strategy and a DFS-like one (Section \ref{sec:shortpath-exploration}). In Figure \ref{fig:node2vec} you just followed the gray arrow to go from node $2$ to node $1$. How do you pick the next step?

Parameter $p$ tells us how much we hate to backtrack. The step bringing you back to node $2$ is weighted $1/p$. So if $p$ is high, $1/p$ is low, and so it is unlikely we backtrack. But if $p$ is below $1$, it is more likely to backtrack, which is what a BFS exploration would do. Parameter $q$, instead, tells us how much we hate to explore parts of the network we haven't seen yet. Among all of node $1$'s neighbors, those that are not also neighbor with node $2$ will get a $1/q$ weight to be explored. With low $q$ we're trying to get as far from node $2$ as possible -- which is how DFS would behave. With high $q$ we're more likely to explore a common neighbor between nodes $1$ and $2$.

DeepWalk is equal to Node2Vec when we set $p = q = 1$, so each neighbor of node $1$ is treated equally. But setting $p$ and $q$ with various combinations leads to different random walks and, as a consequence, different embeddings $Z$ which privilege some information about the network over some other.

\subsection{HARP}
HARP\cite{chen2018harp} is a meta-strategy that one can apply to improve any other random walk embedding method, including DeepWalk and Node2Vec. The idea is to employ a smarter way to initialize the weights of the function summarizing your nodes -- before you start performing your random walks. The way HARP does it is basically as a sort of ``reverse pooling''. If you remember Section \ref{sec:mining-embeddings-pool}, node clustering pooling takes node embeddings and combines them hierarchically to obtain a graph embedding. Here, instead, we coarsen the graph hierarchically \textit{before} we calculate the embeddings. We then calculate the embeddings on the coarsened graph. Then we refine the embeddings for each node on the original graph.

\begin{figure}
\centering
\includegraphics[width=0.8\columnwidth]{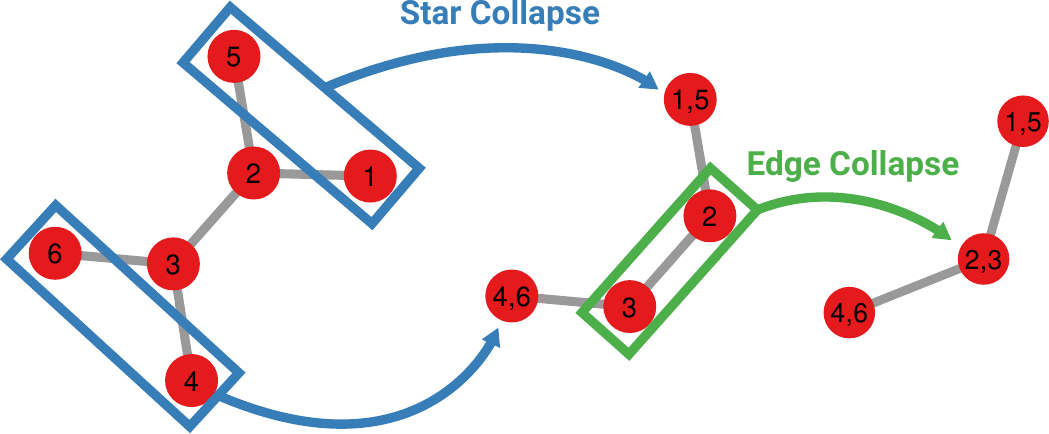}
\caption{The star and edge collapse processes in HARP. Nodes in the blue outlines merge during the star collapse phase, nodes in the green outlines merge in the edge collapse phase.}
\label{fig:rwemb-harp}
\end{figure}

The coarsening is done with two approaches: edge and star collapsing, which I show in Figure \ref{fig:rwemb-harp}. The idea is to preserve both first order relationship (edges) and second order relationships (paths of length two).

First one does star collapsing, where pairs of structurally equivalent nodes of degree one are merged. This means that, if you have $k$ of them, you perform $\log_2 k$ merges. Then, you can do edge collapsing. First, you create set $E' \subset E$ such that no two edges in $E'$ share a node. Then you collapse each edge in $E'$ into a single node. You have to do star collapsing first because, if you did not, you'd have to perform $k$ merges here instead of $\log_2 k$. So this ordering of operations is mostly a convenience to reduce computational complexity. On the other hand, the order in which one explores the graph matters for the final result, as the same graph can generate different $E'$ sets depending on how you explore it. However, this turns out not to be too consequential, as the resulting embeddings normally aren't that sensitive to this choice.

An alternative approach tries to include in the random walk the global structure of the graph by raising the stochastic matrix to several powers and using these $k$ step transition matrix to build the embedding\cite{cao2015grarep}. You can also integrate node attributes in your representation\cite{yang2016revisiting}\cite{pan2016tri}.

\subsection{LINE}
\begin{figure}
\centering
\includegraphics[width=\columnwidth]{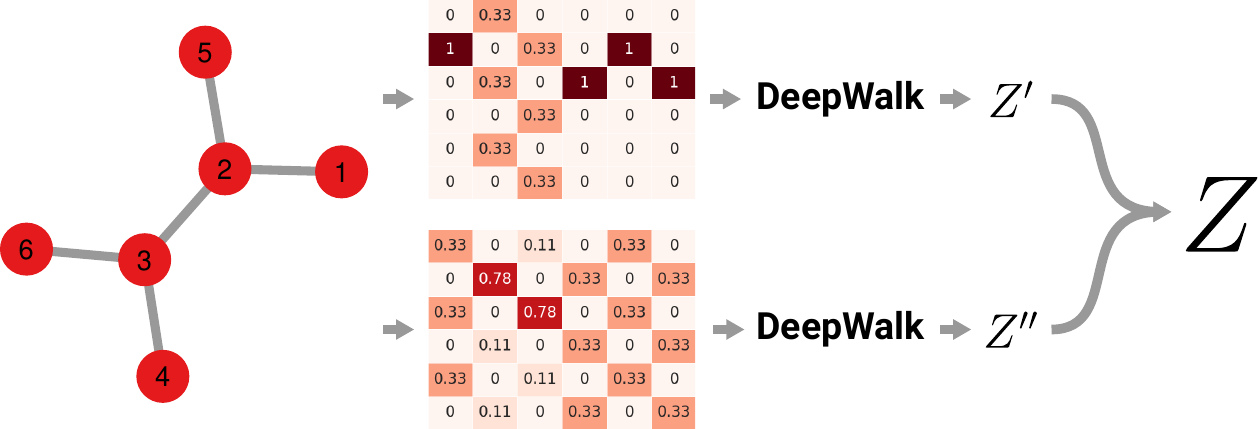}
\caption{The general shape of LINE's framework. From left to right the graph has two matrix representations for first and second order, on which we perform the random walks.}
\label{fig:rwemb-line}
\end{figure}

LINE\cite{tang2015line} is another approach worth mentioning. Just like in HARP and Node2Vec, LINE realizes that one needs to take into account both first and second order relationships between nodes, a feature that is absent from DeepWalk. I depict the general idea of LINE in Figure \ref{fig:rwemb-line}.

LINE represents a graph with two matrices: one with the first order and one with the second order proximities. LINE has its own definition of what these matrices should be, but simplifying a bit one could think that the first order proximities can be represented with the stochastic adjacency matrix $A$, since it gives you the transition probability via a random walk. Then, the second order proximities could be the squared stochastic adjacency matrix, $A^2$, which gives the probability of moving between the nodes with a random walk of length two.

Once you have these two matrices, you can use the same loss function of DeepWalk, twice. In the first pass, you create embeddings minimizing the loss over the first order proximities, and in the second pass you minimize the loss over the second order proximities. Then, you combine the embeddings.

\subsection{Struct2Vec}
Struct2Vec\cite{ribeiro2017struc2vec} follows the guiding principle of structural similarity. The idea is to guide the random walker to explore nodes that are structurally similar. The key here is that there could be structurally similar nodes that are very far apart in the network -- and therefore unlikely to appear in a naive random walker. How do we fix the problem?

\begin{figure}
\centering
\begin{subfigure}{.3\columnwidth}
\includegraphics[width=\textwidth]{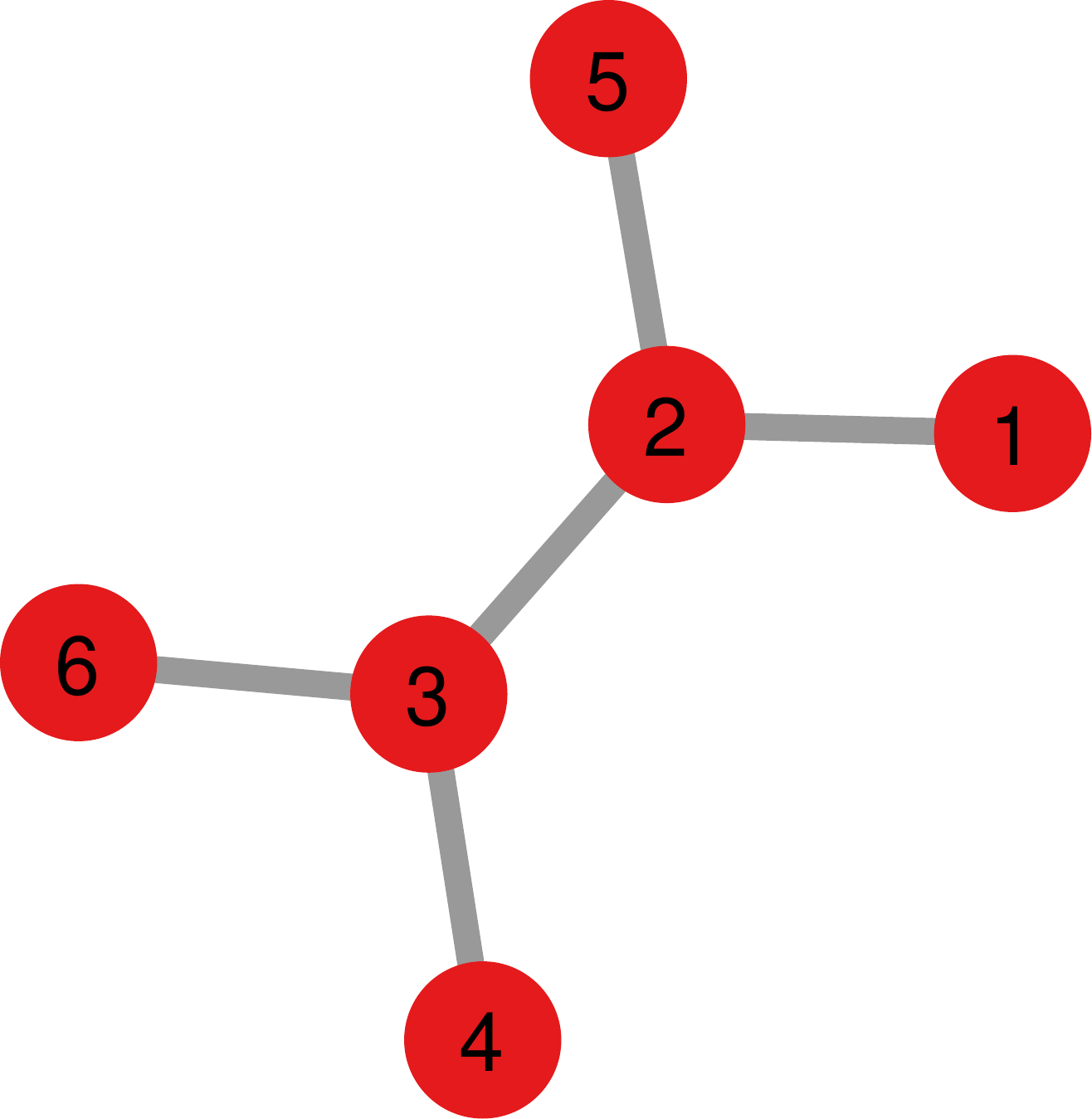}
\caption{}
\end{subfigure}\quad
\begin{subfigure}{.65\columnwidth}
\includegraphics[width=\textwidth]{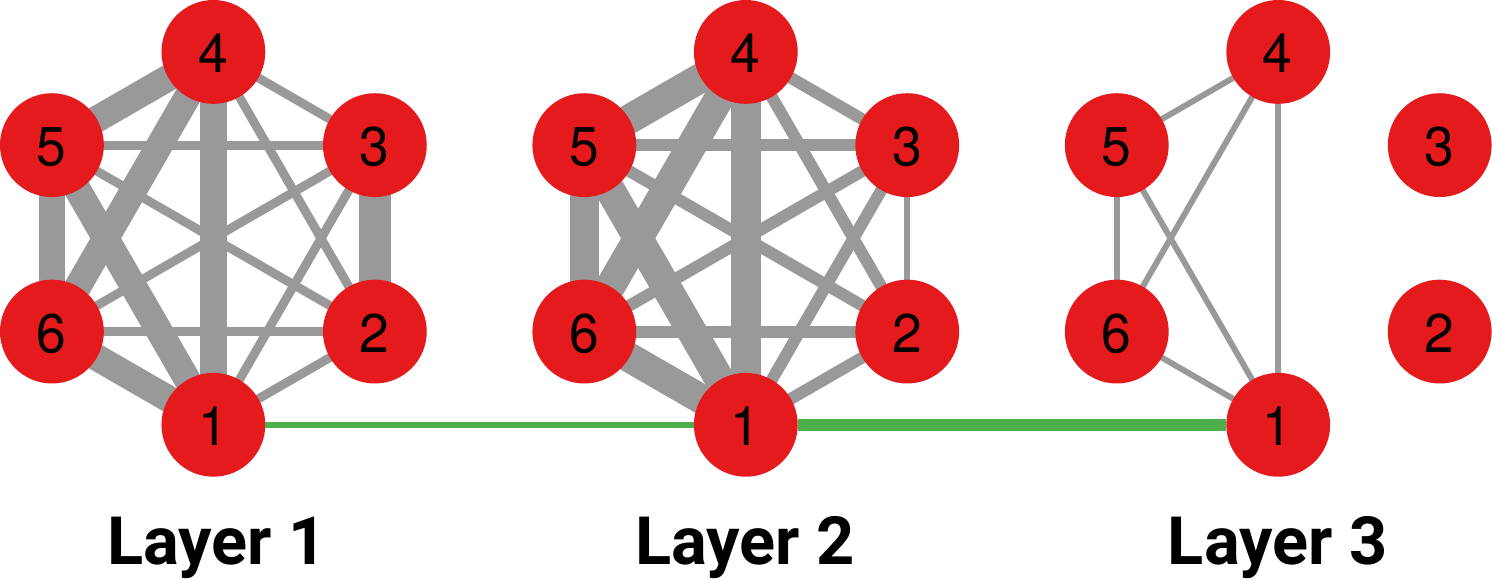}
\caption{}
\end{subfigure}
\caption{(a) The original graph. (b) The multilayer network Struct2Vec builds. I only show one set of interlayer couplings, in green, for node $1$. The line thickness is proportional to the edge's (and coupling's) weight.}
\label{fig:struct2vec}
\end{figure}

The idea here is to build a multilayer graph $G'$ out of your original $G$. Figure \ref{fig:struct2vec} shows how it looks like. This multilayer graph has $k$ layers, where $k$ is the diameter of $G$. Each layer contains the same nodes as $G$. In each layer, all pairs of nodes in $G$ are connected -- although some edges are allowed to have a weight of zero and so appear non-existent, I explain later why that is the case. The embeddings are built using DeepWalk on $G'$ rather than $G$.

You might think: \textit{what's the point of doing random walks on a giant clique?} Well, it all comes down to the weights of those edges, which guide the random walker to explore structurally similar nodes. A $u,v$ edge in layer $i$ has a weight proportional to the structural similarity of $u$ and $v$ at distance $i$. In practice, you take the set of all nodes at $i$ distance to $u$ and you compare it with the set of all nodes at $i$ distance to $v$. Struct2vec compares them using the degrees of these nodes, enforcing a structural similarity function that says that two nodes are $i$-structurally similar if the nodes at distance $i$ from them have similar degree sequences. Note that, in Figure \ref{fig:struct2vec}, some nodes appear disconnected: they are actually connected with an edge weight of zero, because of their low structural similarity.

$G'$ is more useful than $G$ because, when the random walk transitions to any layer that is not the first, it can explore nodes that are far apart more easily. Once the random walker reaches the last layer $k$, in one step it can transition between the two farthest apart nodes in $G$ -- because that layer represents the diameter.

The random walker can transition to a different layer via inter layer couplings -- in Figure \ref{fig:struct2vec} I show only two of them, in green. These couplings connect node $u$ to itself in the neighboring layers -- so $u$ in layer $i$ connects to itself in layers $i + 1$ and $i - 1$. The strength of the coupling is proportional to $u$'s average similarity to all other nodes in layer $i$. If $u$ is similar to all layer $i$ nodes, then it's more informative to go up one layer to $i+1$.

\section{Heterogeneous Graph Embeddings}
Graph embedding techniques, regardless of their chosen approach, need to be adapted to be able to handle heterogeneous\cite{chang2015heterogeneous} and multilayer networks, networks where nodes and/or edges can be of multiple different types. For instance, one could adopt the metapath approach\cite{dong2017metapath2vec} we saw in Section \ref{sec:lp-multilayer-general} when talking about link prediction in multilayer networks. The problem with heterogeneous networks is that there are some node types that are more dominant -- i.e. connected -- than others. Their representations would then be extremely noisy. In metapath2vec, the problem is solved by switching one's attention from nodes to metapaths.

\begin{figure}
\centering
\includegraphics[width=.66\columnwidth]{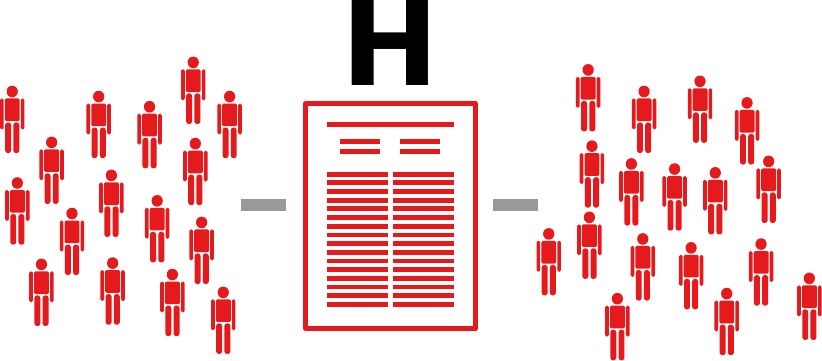}
\caption{A troubling set of connections in an heterogeneous network: a paper with hundreds of co-authors.}
\label{fig:metapaths-embed}
\end{figure}

Figure \ref{fig:metapaths-embed} shows a stylized depiction of the issue. The paper reporting the discovery of the Higgs boson has $5,154$ co-authors. Every pair of co-authors is a valid path in the co-authorship network. As a result, the embedding of the node representing the paper is extremely noisy, as it could be visited by $10^7$ walks of length $2$, not even counting the ones that could lead you back to your origin node. But by instead focusing on each of the metapaths (Section \ref{sec:lp-multilayer-general}), we will have a much cleaner signal.

Another way to deal with heterogeneous networks is to infer different embeddings for different edge types\cite{shi2018easing}. Let's say you have two nodes of different types: a conference venue $c$ and a topic $t$. They are connected to the same node $a$, an author, who published in that conference and in that topic, but not at the same time. Thus, $c$ and $t$ are not connected and not similar to each other. Then $a$ will be equidistant from them. However, when we focus on the ``author-topic'' edge type, $a$ will be closer to $t$, and when we focus on the ``author-conference'' edge type, $a$ will be closer to $c$.

Knowledge Graph Embedding\cite{nickel2011three} is particularly popular and well-suited for these techniques and that's why I dedicate some space to it. This is the application of graph embedding techniques on knowledge graphs. Knowledge graphs are graphs of entities related to each other by a certain semantic. \textit{De facto}, these are special heterogeneous networks. Examples of knowledge graphs are, for instance, DBPedia\cite{lehmann2015dbpedia}, Wikidata, and more. What makes them special is their massive size and the rich semantics behind the connections.

\begin{figure}[t]
\centering
\begin{subfigure}{.3\columnwidth}
\includegraphics[width=\textwidth]{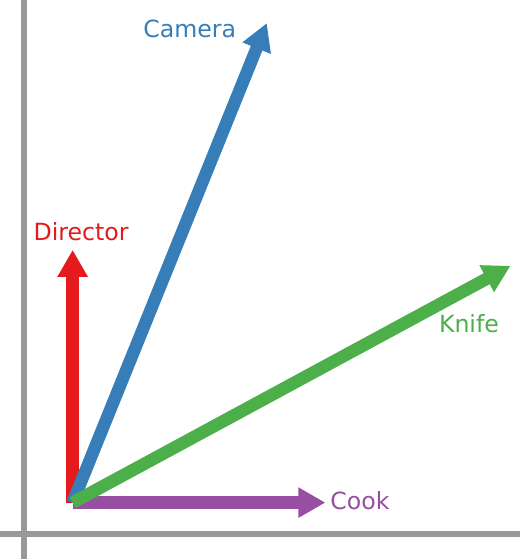}
\caption{}
\end{subfigure}\quad
\begin{subfigure}{.3\columnwidth}
\includegraphics[width=\textwidth]{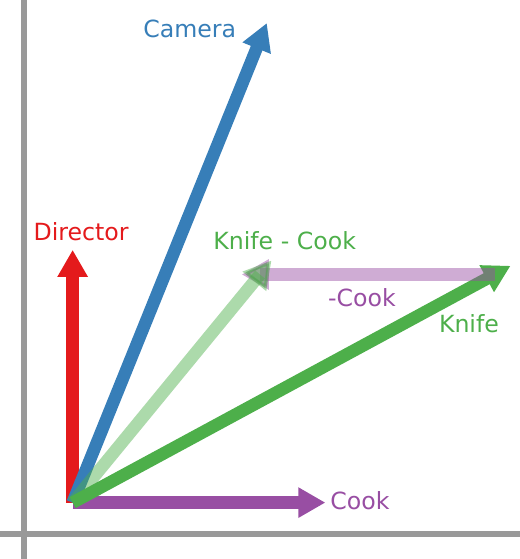}
\caption{}
\end{subfigure}\quad
\begin{subfigure}{.3\columnwidth}
\includegraphics[width=\textwidth]{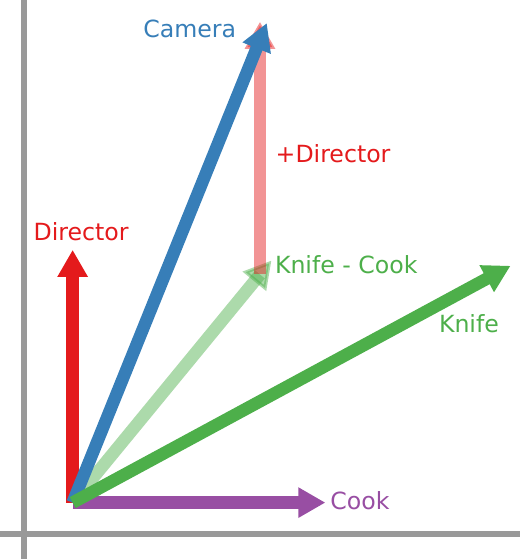}
\caption{}
\end{subfigure}
\caption{(a) Some node embeddings we learned from a knowledge graph. (b-c) Manipulating node embeddings by adding-subtracting their vector representations can uncover new knowledge.}
\label{fig:knowgraph-embed}
\end{figure}

In such cases, graph embedding techniques acquire a special meaning. We are now establishing relations between concepts, building a way to determine that, for instance, knives are to cooks what cameras are to movie directors. This allows us to create automatically new connections, previously unknown relationships, in the knowledge graph. Figure \ref{fig:knowgraph-embed} shows an example.

In the figure, I create the new abstract concept of ``knife minus cook''. By adding ``movie director'' to such an abstract concept, I discover the camera-director relationship as equivalent to knife-cook. Thus the ``knife minus cook'' is a useful concept, which we might use for all professions.

Such operations can be done directly from text using Natural Language Process techniques. What makes this special, is that the embeddings were created not by looking at natural text, but at the structure of a knowledge graph. To do this, we need specialized tools\cite[1in]{wang2014knowledge}\cite{lin2015learning}\cite{schlichtkrull2018modeling}. As usual, you can find more details on this problem in dedicated survey papers\cite{nickel2015review}\cite{wang2017knowledge}.

\section{Applications}\label{sec:mining-embeddings-app}
So, what can you do with node embeddings? To cut a story short: practically everything. We already saw a couple of applications, in this chapter and in others. Without repeating myself too much, I'll spend only a couple of words about the applications we've already seen in the book:

\begin{itemize}
\item Community discovery (Part \ref{par:cd}): if you build embeddings encoding the similarity of neighbordhoods of nodes (like in Figure \ref{fig:embeddings-example}), then you can feed them into any data clustering algorithm such as kMeans or DBSCAN and get communities;
\item Structural equivalence (Section \ref{sec:centr-similarity}): if you build embeddings encoding the structural similarity of nodes (like in Figure \ref{fig:embeddings-example2}) the same data clustering algorithms will identify structurally equivalent nodes;
\item Link prediction\cite{ou2016asymmetric} (Part \ref{par:lp}): by comparing the embeddings of edges that exist against those of edges that do not, you can figure out which edges should exists, and be predicted;
\item Node role detection (Chapter \ref{cha:centr-roles}): you can build embeddings to conform to your expectation of specific roles in the networks, and classify all nodes based on that.
\end{itemize}

Here I want to unpack the topic of network visualization, since I teased it in Section \ref{sec:mining-embeddings-what}. We're going to do a deep dive on network visualization in Part \ref{par:netviz}, but hopefully Section \ref{sec:mining-embeddings-what} did a good job to make you understand why embeddings are useful in this case. One could reduce each node into two or three dimensions, and then simply use them as the $(x, y, z)$ coordinates to display them spatially. After all, this is what is already being done with traditional dimensionality reduction techniques: it is the whole selling point of the quasi-magical t-distributed stochastic neighbor embedding\cite{maaten2008visualizing} (t-SNE) or UMAP\cite{mcinnes2018umap}.

\begin{figure}[b]
\centering
\begin{subfigure}{.5\columnwidth}
\includegraphics[width=\textwidth]{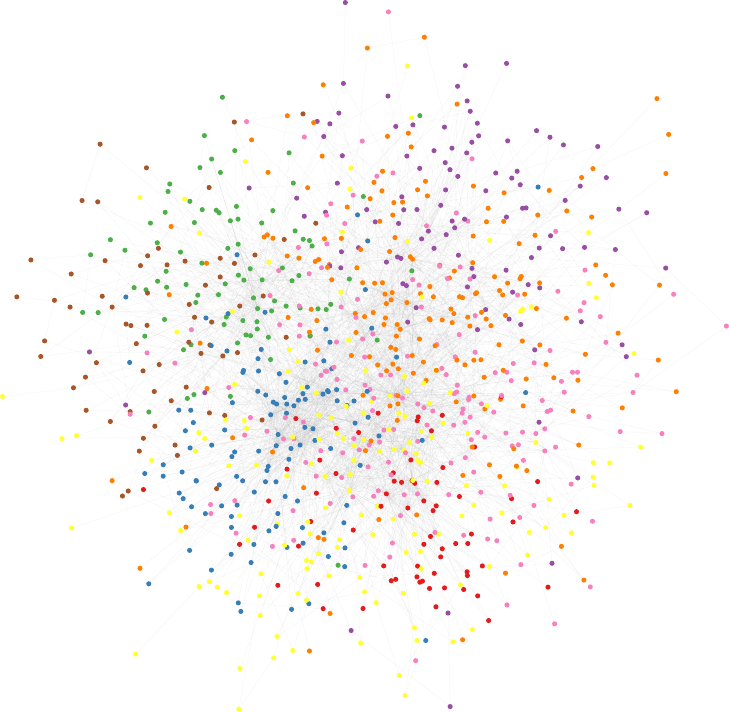}
\caption{}
\end{subfigure}\quad
\begin{subfigure}{.4\columnwidth}
\includegraphics[width=\textwidth]{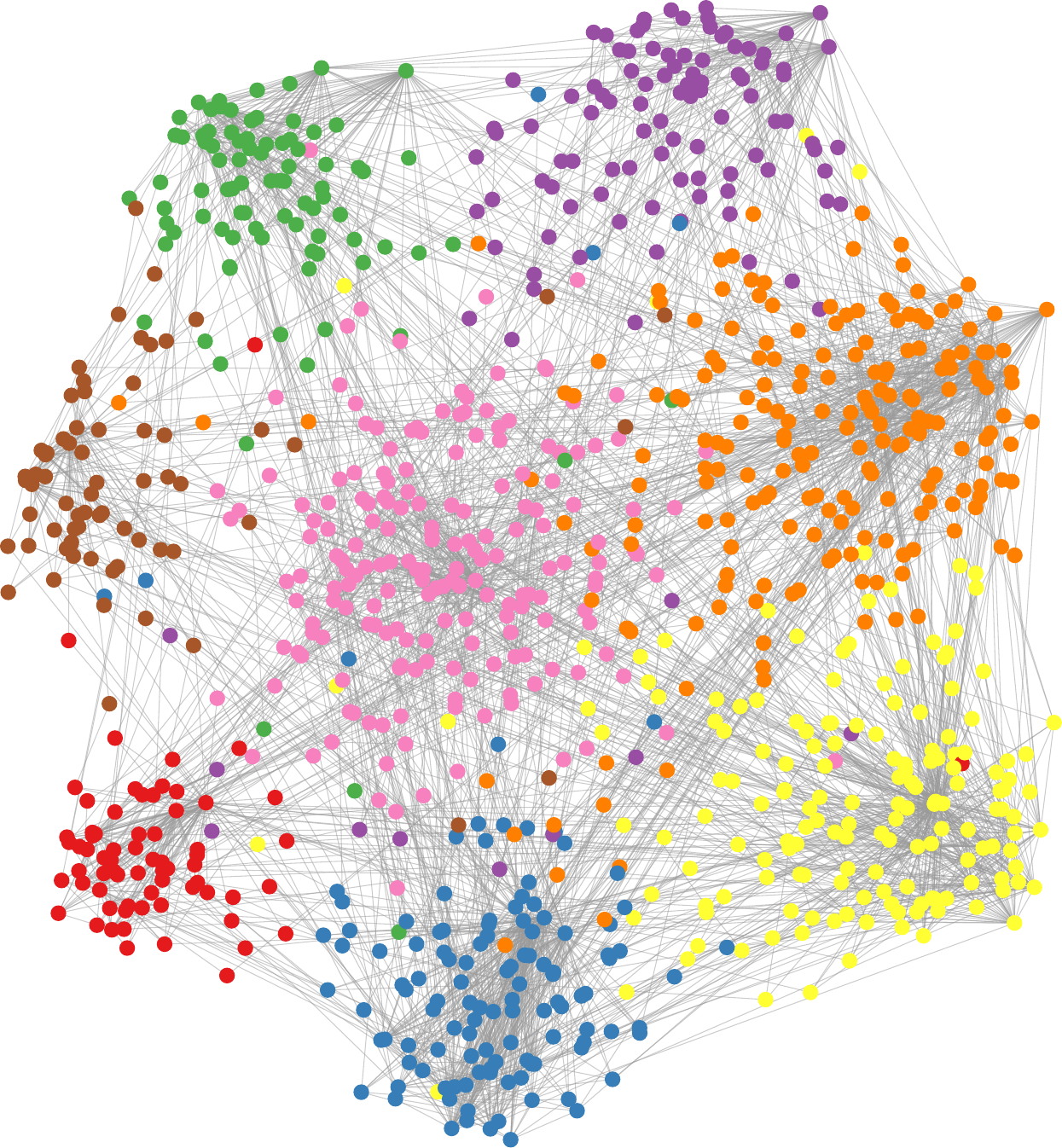}
\caption{}
\end{subfigure}
\caption{An LFR benchmark. The node color represents its community affiliation. (a) Force directed layout (Section \ref{sec:layouts-force}). (b) layout based on node embeddings via random walks.}
\label{fig:embeddings-viz}
\end{figure}

In fact, the standard approach includes two steps\cite{tang2016visualizing}. First, you use your favorite graph embedding technique to reduce all nodes to vectors of length $d$. Then, you apply a dimensionality reduction technique to reduce those $d$ dimensions to two and you plot the result. Figure \ref{fig:embeddings-viz} shows what happens to a relatively simple LFR benchmark with a high mixing parameter -- i.e. communities which tend to share lots of inter-community edges. In Figure \ref{fig:embeddings-viz}(a) you see what a classical network layout would do, while Figure \ref{fig:embeddings-viz}(b) shows you a combination of Node2Vec embeddings, reduced with UMAP. In the latter case, the community structure is more evident. Imagine what you could do if you could make a more advanced node embedding, taking into account node attributes, edge weights, and more!

Another natural application of node embeddings we haven't seen so far is graph summarization. We're going to examine the problem more in details in Chapter \ref{cha:mining-summarization}. For now, suffice to say that, if two nodes have very similar embeddings, we could just collapse them into the same node. The benefit would be to have a smaller structure to analyze -- less memory and time consumption for your algorithm -- while still preserving the general properties of the network as a whole.

\begin{figure}[b]
\centering
\begin{subfigure}{.3\columnwidth}
\includegraphics[width=\textwidth]{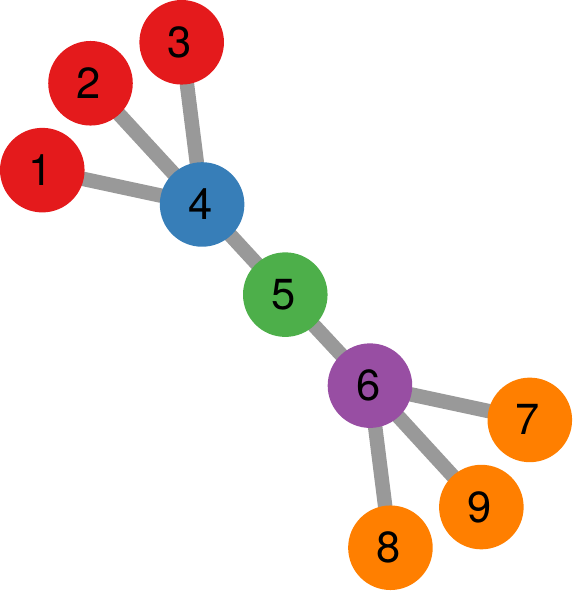}
\caption{}
\end{subfigure}\quad
\begin{subfigure}{.3\columnwidth}
\includegraphics[width=\textwidth]{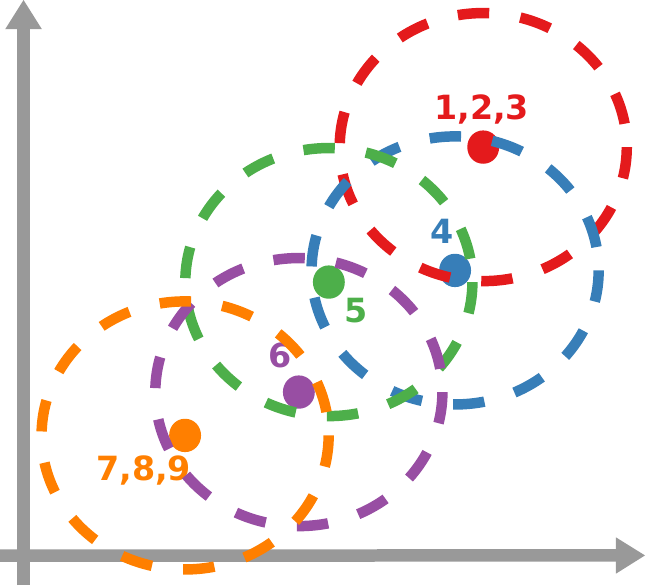}
\caption{}
\end{subfigure}\quad
\begin{subfigure}{.3\columnwidth}
\includegraphics[width=\textwidth]{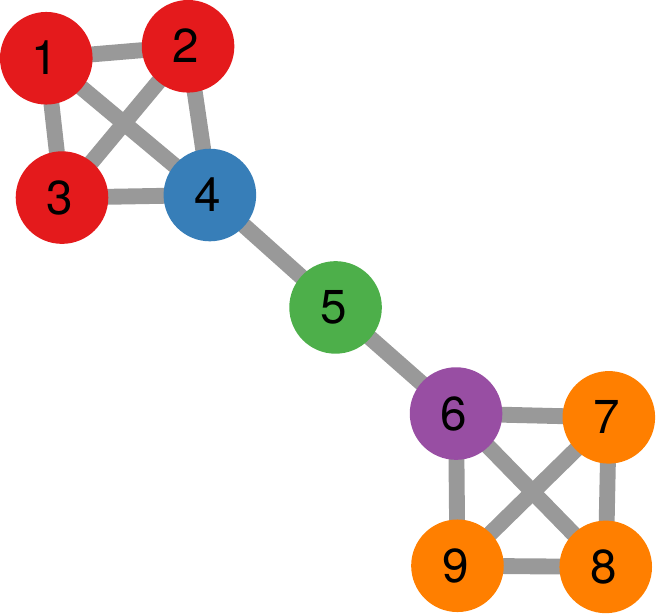}
\caption{}
\end{subfigure}
\caption{(a) An example graph. (b) One of the possible embeddings of (a). The dashed circles encapsulate the radius inside which we consider the nodes as connected together. (c) The version of graph (a) reconstructed through its two dimensional embedding.}
\label{fig:embeddings-summary}
\end{figure}

One common way is to do the following. First, take your graph and calculate its node embeddings. These, as we saw, are spatial representations in $d$ dimensions. Then, calculate the pairwise distance between all these points. You can use the Euclidean distance or whatever floats your boat. At this point, you can establish a certain distance $k$. Nodes that are closer than $k$ get connected together, otherwise they remain disconnected. This is a graph reconstructed from the embeddings. You can compare the reconstructed graph with the original one. The more similar they are, the better the embedding worked. Figure \ref{fig:embeddings-summary} shows an example of this procedure.

You can consider closer points as the same node and summarize your graph this way. For instance, nodes $7,8,9$ have the same embedding and thus they could be considered to be the same node, just like nodes $1,2,3$.

Other classical applications of graph embeddings are node ranking\cite{park2019estimating}, solving classical combinatorial problems like the traveling salesman problem\cite{khalil2017learning}, and network alignment\cite{heimann2018regal}, the problem of figuring out which nodes from two distinct networks might refer to the same real world entity. This is still limited to the realm of network analysis for network analysis' sake, but we know we can use networks -- and, therefore, graph embeddings -- to solve many more problems. Some include natural language processing\cite{lin2017structured}, computer vision\cite{yan2018spatial}, and bioinformatics\cite{zitnik2018modeling}, to cite a few pointers.

\section{Limitations}\label{sec:mining-embeddings-limits}
The classical node embeddings approaches had a lot of issues: they didn't capture structural similarity, they ignored node and edge attributes, etc. These issues were fixed with various new methods, which we have explored in the past two chapters. There remains one fundamental issue with the majority of the most popular node embedding approaches. Throwing some jargon at you, the issue is that these methods are not inductive, but transductive. What does that mean?

\begin{figure}
\centering
\includegraphics[width=\columnwidth]{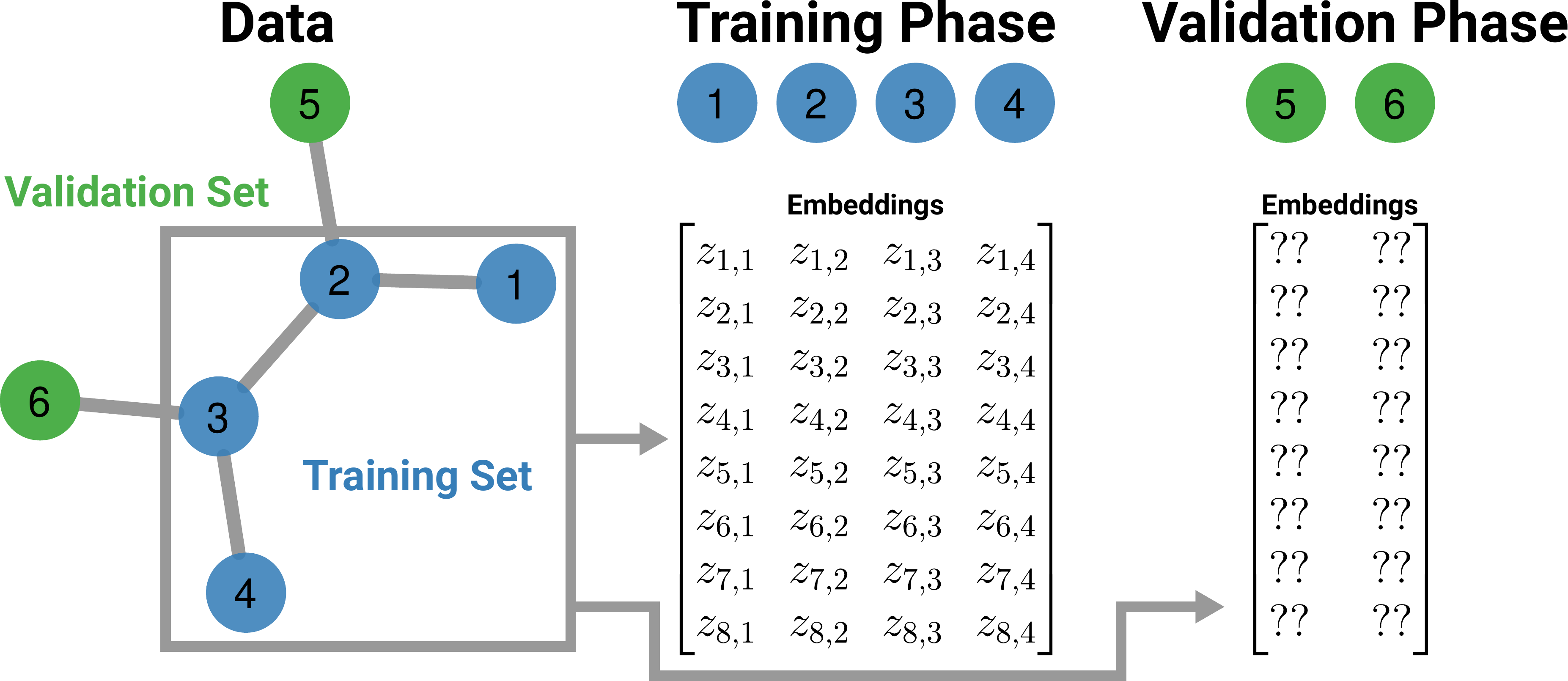}
\caption{The original network is divided in training nodes (in blue) and validation nodes (in green). We create the embedding (grey outline) using only the training nodes and their connections.}
\label{fig:rwemb-transductivity}
\end{figure}

Figure \ref{fig:rwemb-transductivity} shows an example. Here we're in a fairly classical scenario, where we try to train our algorithm by dividing nodes into a training set $V_t$ and a validation set $V_v$, to figure out whether we're overfitting. If you remember Section \ref{sec:ml-general-infrastructure}, these two sets are disjoint: all nodes in the validation set are not part of the training set, or $V_t \cap V_v = \emptyset$. How does that affect your node embeddings?

If you're using a spectral approach, you can only calculate the eigenvectors on the matrix including only the nodes in the training set. This means $Z$ is a $V_t \times d$ matrix. In the Figure \ref{fig:rwemb-transductivity}, we have eight dimensional embeddings, so $Z$ is $4 \times 8$. Here you have no embeddings for the nodes in the validation set!

You might be tempted to perform validation on the full network. But adding the nodes from the validation set means recalculating \textit{all} the embeddings, even those of the nodes in the training set -- meaning you just trained on embeddings that are different than the ones you use in the validation. Oops. The same holds for random walk embeddings: the presence/absence of new nodes will change the random walk content, by making some node pairs all of a sudden much more (or less) likely to appear together in a random walk.

This issue is not limited to the mere practicality of making training and validation sets, though. This limitation also makes it technically hard to deal with a network that evolves over time. Every time some nodes/edge can come in -- or out -- of your structure, all your embeddings will change and need to be recalculated from scratch. There are ways to go around this issue\cite{li2017attributed}, but in most cases we'll need to dive into different techniques designed specifically for dynamic graphs\cite{kazemi2020representation}\cite{liu2023deep}, or to solve a specific subproblem.

In the next chapter we'll see how deep learning can give us more natural inductive methods that do not suffer from this issue.

\section{Summary}

\begin{enumerate}
\item The basic idea of random walk embeddings is to perform many random walks on a network and to minimize the loss function describing those random walks. Two nodes have similar embeddings if they co-appear often in the same random walk.
\item You can get different and meaningful embeddings by modifying the random walk strategy. You can have high order random walks exploring the graph either in depth or in width, you can collapse the network in various ways to explore structural similarities, and more.
\item With heterogeneous graphs your strategy needs to take into account that nodes and edges of different types might require special rules to be traversed.
\item Knowledge graphs are a special case of heterogeneous networks: they express relations between real world concepts. In this scenario, embeddings can help us to uncover previously unknown meanings -- i.e. groups of nodes in the knowledge graph.
\item Node embeddings can be applied to many problems we saw in the book: community discovery, link prediction, and more. Visualization is another natural one: by building 2D embeddings we can interpret them as spatial coordinates to display our network.
\item Many common node embedding methods can only build and use embeddings for nodes that are already present in the graph. If you add new nodes -- because of a temporal graph or a train-validation split -- you need to recompute all embeddings from scratch.
\end{enumerate}

\section{Exercises}

\begin{enumerate}
\item Perform $10,000$ random walks of length $6$ in the network at \url{http://www.networkatlas.eu/exercises/43/1/data.txt}. Build embeddings with $d = 32$ using Word2Vec (I suggest using the \texttt{gensim} implementation). Reduce them to two dimensions using \texttt{sklearn.manifold.TSNE}. Visualize the network by using the 2D embeddings as spatial coordinates. Use \url{http://www.networkatlas.eu/exercises/43/1/nodes.txt} to determine the node colors.
\item Use the Word2Vec embeddings reduced to 2D with tSNE you found in the previous exercise to find $4$ clusters in the data using any clustering algorithm (if kMeans, set $k=4$, otherwise you can use DBSCAN and find the eps parameter giving you $4$ clusters). What is the NMI (use the \texttt{sklearn} function to calculate it) of the clusters with the ground truth you can find at \url{http://www.networkatlas.eu/exercises/43/1/nodes.txt}?
\item Compare the NMI score from the previous exercise to the one you would get from a classical community discovery like label propagation. Note: both methods are randomized, so you could perform them multiple times and see the distributions of their NMIs.
\item Use the Word2Vec embeddings reduced to 2D with tSNE you found in the previous exercise as a link prediction score (if $Z_u$ and $Z_v$ are node embeddings, then the score for edge $u,v$ is $Z^T_uZ_v$). Draw the ROC curve of your predictions, assuming that the true new edges are the ones you can find in \url{http://www.networkatlas.eu/exercises/43/4/newedges.txt}.
\item Is the AUC you get from the previous question better or worse than the one you'd get from a classical link prediction like Jaccard, Resource Allocation, Preferential Attachment, or Adamic-Adar?
\end{enumerate}

\chapter{Message-Passing \& Graph Convolution}\label{cha:mining-deep}
In this chapter we start dealing with the development of deep learning methods that work on a graph. The difference with what we've seen so far is that the embedding methods from the previous chapters encode the graph's structure so that we can use non-graph deep learning methods to learn something about the graph. Here we scrap that approach and try to adapt the deep learning methods to be able to work and learn directly on the graph. Which is why the organization of this chapter and the next has one section per deep learning approach, explaining how that specific deep learning approach can be performed on a graph.

For these two chapters to make most sense and to understand most of the lingo, you'd be required to know more about what deep learning is. This is outside the scope of this book, as it deserves a (couple of) book(s) on its own. For this reason, I suggest you some readings\cite{deng2014deep}\cite{lecun2015deep}\cite{goodfellow2016deep}, which might help you figuring out better what is going on.

A note about terminology. Formally speaking, graph convolutional networks and other more sophisticated approaches such as graph attention networks are all part of the family of message-passing graph neural networks, which is the most general model and it includes the two. \textit{However}, to take things step by step and make them more digestible, I'll slightly abuse the terminology. For now on, ``message-passing graph neural network'' in this book refers not to the \textit{general} class, but to the \textit{basic} class. This makes it easier to introduce graph convolutional networks for the special thing they do. But, technically speaking, there is no difference between graph convolutional networks and message-passing graph neural networks, because graph convolutional networks are a type of message-passing graph neural networks.

\section{Message-Passing Graph Neural Networks}\label{sec:deep-mpgnn}
Before I start throwing math at you, I'll start with a simple graphical example of what a message-passing graph neural network (MPGNN) is. The reason is that this type of architecture is the basis of all the variants that will follow. Understanding this will provide the foundation to figure out what happens in graph convolution networks and in other models. The example I give here is pretty cartoonish, but is should help understanding. It is based on the early attempts to define graph neural networks\cite[-1in]{scarselli2005graph}\cite{gori2005new}\cite{merkwirth2005automatic}, which have been constantly refined over time\cite{gilmer2017neural}.

One key component in this framework is that the nodes start with some feature matrix, which we will call $H^0$. $H^0$ is a $|V| \times d$ matrix, associating each node with a vector of length $d$ of features. If your graph doesn't have node attributes, you can always use some structural properties as attributes, for instance their degree.

The key process of a message-passing graph neural network is... message passing. At each iteration -- which we call layer -- we do three things:

\begin{enumerate}
\item Each node formulates a \textit{message} based on its current node attributes and passes it to all its edges;
\item Each node receives the messages from its neighbors and \textit{aggregates} them;
\item Each node \textit{updates} its own attributes based on the aggregated messages.
\end{enumerate}

\begin{figure}[b]
\centering
\includegraphics[width=\columnwidth]{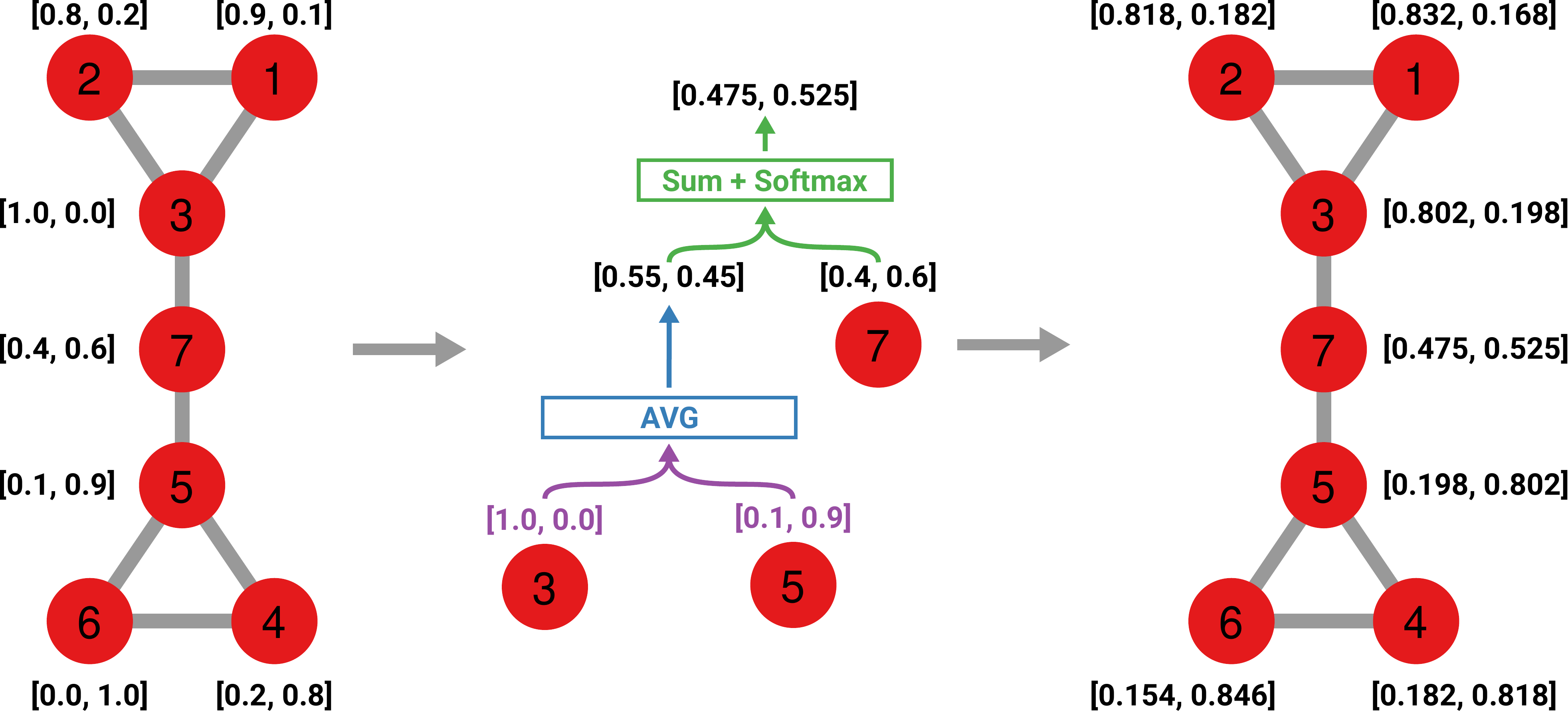}
\caption{A potential message-passing graph neural network architecture. From left to right: a network with 2D node attributes; the message-passing procedure for node $7$; the result of the message-passing for all nodes. The different colors in the message-passing procedure highlight the three steps: message composition and passing in purple, aggregation in blue, and update in green.}
\label{fig:mpgnn}
\end{figure}

Figure \ref{fig:mpgnn} shows this process graphically for a node, and the final result for the whole network.

Note how I emphasized the words ``message'' ``aggregate'' and ``update''. That is because those are the functions that power the special juice of MPGNNs. Depending on how you compose the messages, aggregate them, and use them to update a node, you can lead your graph neural network to achieve different objectives. In the figure I decided that the message composition was just the identity, i.e. the node attributes are passed ``as they are''. The aggregation function was the element-wise average, so nodes $3$ and $5$ aggregate to $[(1+0.1)/2, (0+0.9)/2] = [0.55, 0.45]$. Then the updating function was sum and then softmax -- to keep the property that the node attributes for each node sum to $1$. So node $7$ updates as $softmax([0.4 + 0.55, 0.6 + 0.45]) = [0.475, 0.525]$. Of course you can use any arbitrary combination of functions here.

\begin{figure}
\centering
\begin{subfigure}{.23686\columnwidth}
\includegraphics[width=\textwidth]{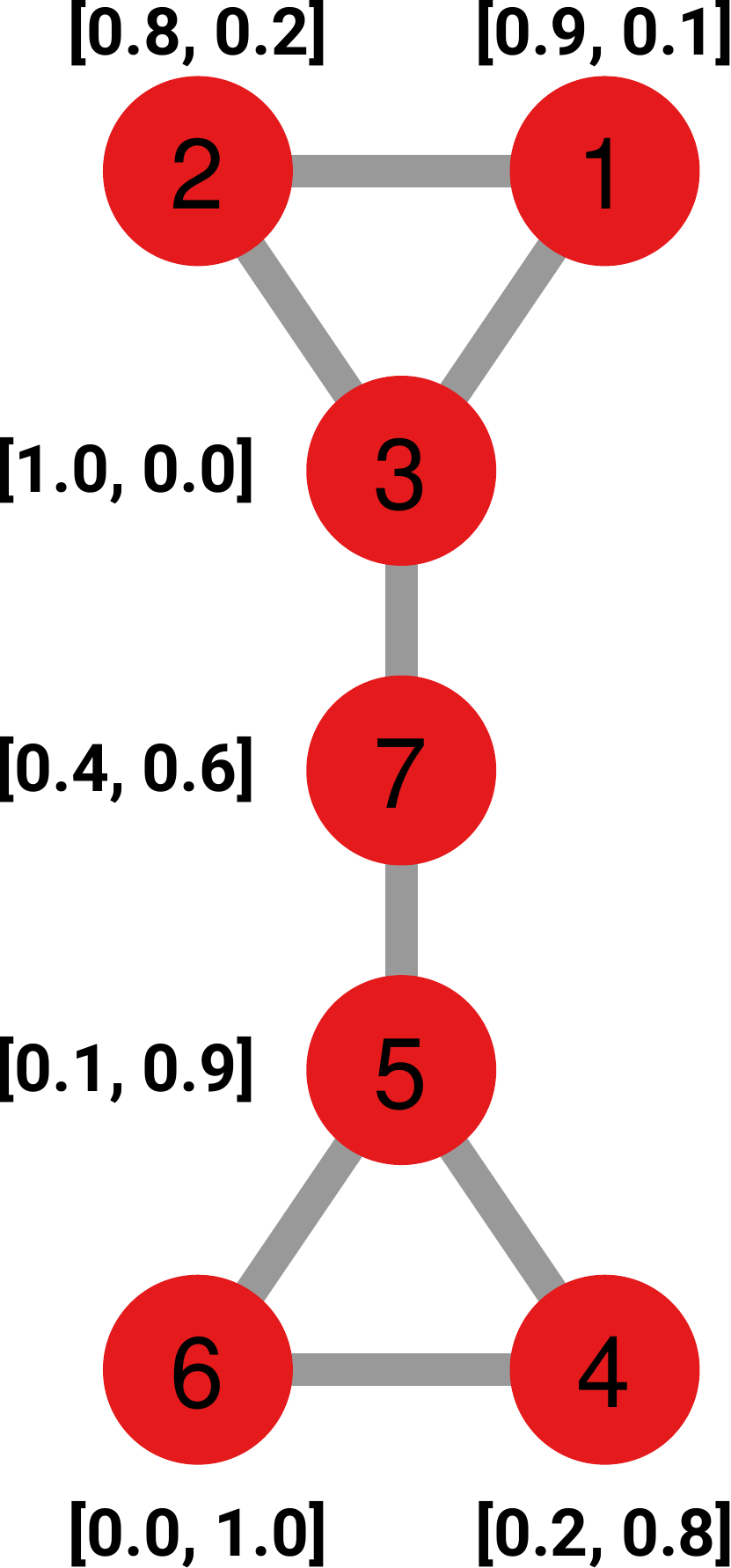}
\caption{$H^0$}
\end{subfigure}\quad
\begin{subfigure}{.34\columnwidth}
\includegraphics[width=\textwidth]{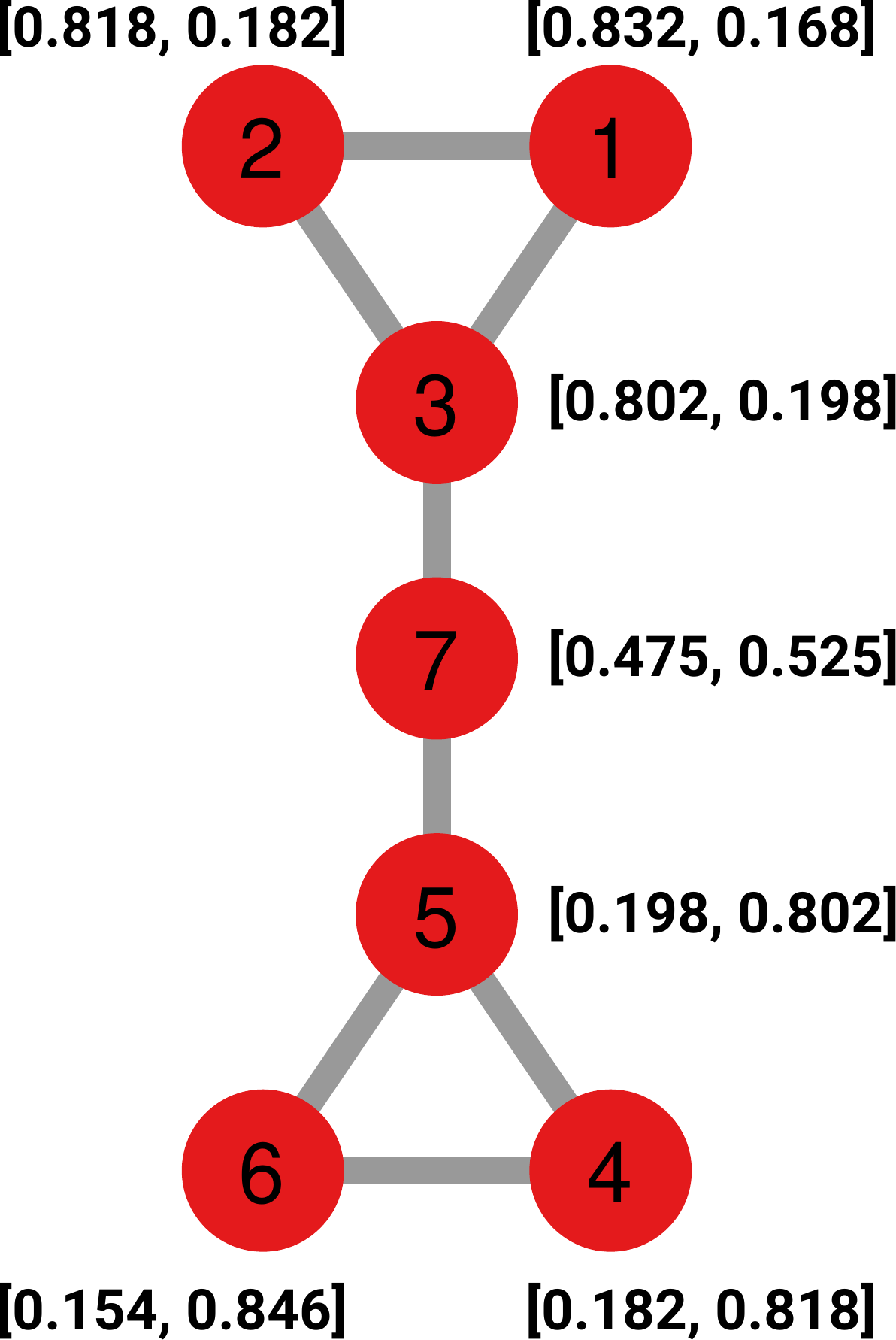}
\caption{$H^1$}
\end{subfigure}\quad
\begin{subfigure}{.34\columnwidth}
\includegraphics[width=\textwidth]{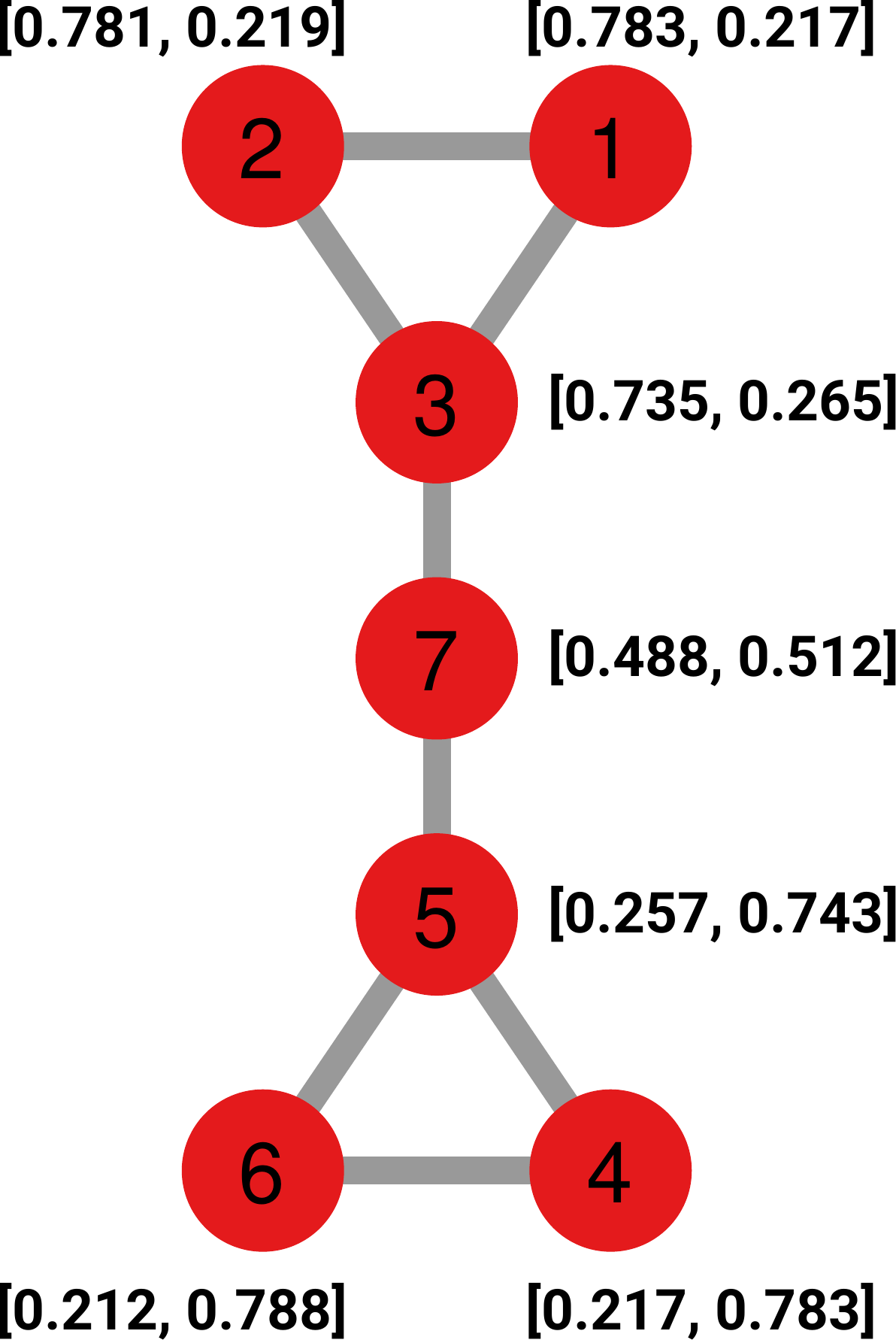}
\caption{$H^2$}
\end{subfigure}
\caption{(a) The original network. (b) The result of applying the first layer of message-passing. (c) The result of applying the second layer.}
\label{fig:mpgnn-layers}
\end{figure}

One round of messaging, aggregating, and updating constitutes one layer of the MPGNN. In the figure, that results in building a new attribute matrix. We call it $H^1$, because it is the result of the first layer applied to $H^0$. The power of MPGNNs is that you can pile up multiple layers. Figure \ref{fig:mpgnn-layers} shows you how we get $H^2$ from the second layer, which boils down to re-applying the same aggregate and update functions to $H^1$.

You can see how this can be powerful: in $H^2$ we're starting to delineate very clearly not only the clustered structure of the network, but also which nodes are in which position in this clustered structure, whether on the border or deeply embedded in the community.

Why do we want to have multiple layers? Because in this way, for each node $v$, we can pool information from $v$'s extended neighborhood. At the first layer, we only get information from its direct neighbors. At the second layer, however, those neighbors have embedded some information from \textit{their} neighbors, which is now passed to $v$ as well. At layer $l$, $H^l_v$ encodes information at $l$ hops from $v$. More interestingly, this ``information'' that is passed is not merely about the features, but it depends on the structure of the network as well. $H^1$ is not dependent only on $H^0$ but also on the structure of $G$! The same $H^0$ on a different $G$ will produce a different $H^1$. Isn't that neat?

Before I move on to the sections showing where this neatness ends, a few concluding remarks. So far we have worked with simple graphs. However, you know that networks come in various flavors, and your MPGNN should take that into account. For instance, if you're working with multilayer networks, you need your aggregate and update functions to run over the various different edge types\cite{marcheggiani2017encoding}\cite{teru2020inductive}. However, a more flexible approach is to realize that the edge type is nothing more than a feature of the edge. So we can treat this as any other feature, and including it in the aggregate function that we use to pass the message to the nodes\cite{sinha2019clutrr}.

Note also that there's no one stopping from having edges passing messages to each other, or to summarize a graph embedding by updating at each layer the graph's embedding with a combination of its node and edge embeddings\cite{battaglia2018relational}\cite{barcelo2020logical}.

\section{Limits of the Message-Passing Approach}\label{sec:deep-mpgnn-limits}

\subsection{Relations to Isomorphism Test}
It's worth to make a quick pause and point back to Section \ref{sec:mining-isomorph}, where I explained the Weisfeiler-Lehman isomorphism test. In that section I told you that this isomorphism test is important because it approximates the behavior of MPGNNs. If you go and read this section again, you'll realize that Weisfeiler-Lehman is exactly a message-passing network, with specific aggregating and updating functions. In fact, there are proofs showing that this style of message-passing network can distinguish -- at best -- the same structures of Weisfeiler-Lehman\cite{xu2018powerful}. That is to say that, in the cases where Weisfeiler-Lehman fails the isomorphism test, a MPGNN would fail to distinguish structures -- returning the same $H^l$ for different $G$s. Bummer.

\begin{figure}
\centering
\begin{subfigure}{.5\columnwidth}
\includegraphics[width=\textwidth]{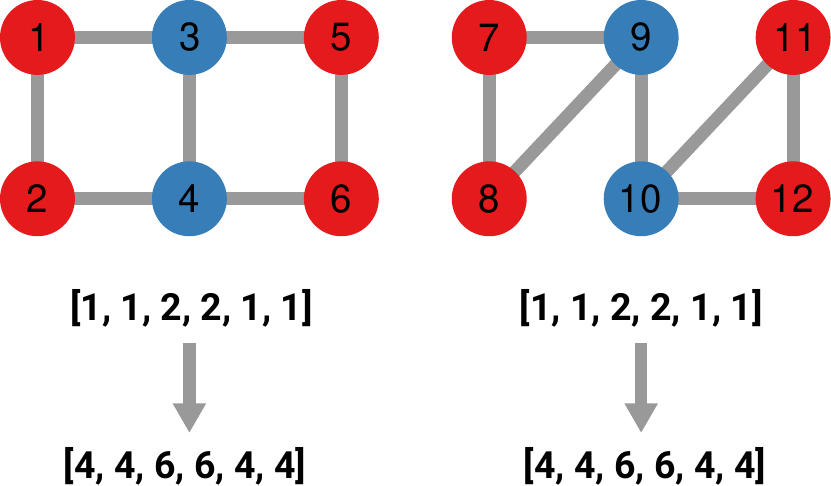}
\caption{}
\end{subfigure}\quad
\begin{subfigure}{.44\columnwidth}
\includegraphics[width=\textwidth]{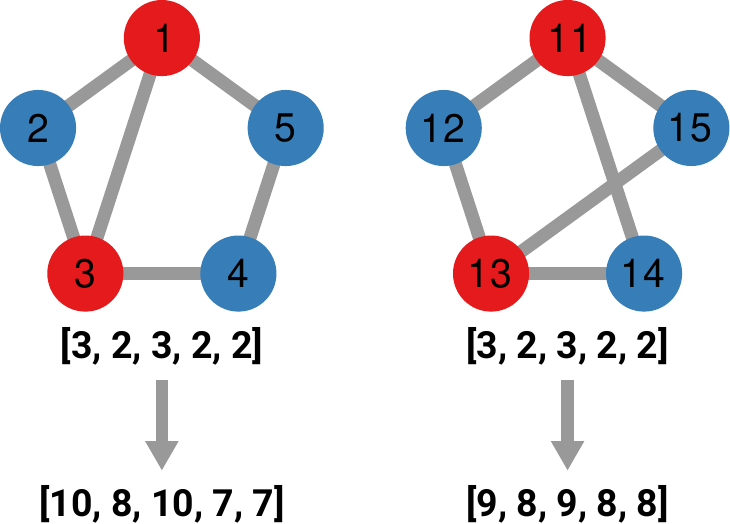}
\caption{}
\end{subfigure}
\caption{Two non-isomorphic networks with 1D embeddings passing through one layer of a simple MPGNN. (a) Leading to the same result. (b) Leading to different results.}
\label{fig:mpgnn-weisfeiler}
\end{figure}

The full proof is sophisticated and a bit much to include in this book, but I'll give you a superficial intuition about it. Consider Figure \ref{fig:mpgnn-weisfeiler}(a). Here I show you the networks I used in Figure \ref{fig:weisfeiler-lehman-fail} to show you a case in which Weisfeiler-Lehman gets it wrong: it says the graphs are isomorphic when they aren't. I use as simple 1D embedding the node degrees. The aggregate and update functions of the MPGNN are the sum. As you can see, once we pass through a layer, both networks generate the same embeddings, even though the structures are different and one would expect them to lead to different embeddings.

On the other hand, in Figure \ref{fig:mpgnn-weisfeiler}(b) we have the graphs I used in Figure \ref{fig:weisfeiler-lehman-success} to show when Weisfeiler-Lehman gets it right. As you can see, using the same 1D embeddings and the sum as the aggregate and update functions, the different structures generate different embeddings, as one would expect.

There are ways to fix this problem. A simple one involves extending $H^0$ with additional features, specifically features that associate each node with a unique ID\cite{murphy2019relational}. In this way, the MPGNN can identify when two nodes share a neighbor, which will allow it to distinguish structures Weisfeiler-Lehman cannot. It's relatively simple to do this: just concatenate to $H^0$ the $|V| \times |V|$ identity matrix $I$. This is all fine and dandy, but the problem is that now we lose the permutation invariance requirement, since now we're forcing the graph to have a specific node ordering. This can be solved by aggregating information from all possible orderings, basically running the MPGNN for all possible permutation of nodes. While this will lead to permutation invariance, one should note that trying all permutations of nodes is something you can do only for small graphs with few nodes.

\begin{figure}
\centering
\begin{subfigure}{.33\columnwidth}
\includegraphics[width=\textwidth]{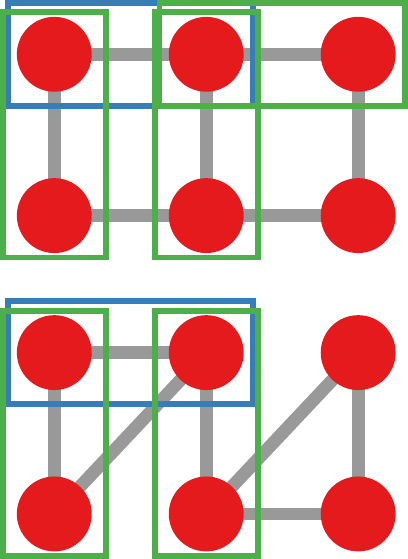}
\caption{}
\end{subfigure}\qquad
\begin{subfigure}{.33\columnwidth}
\includegraphics[width=\textwidth]{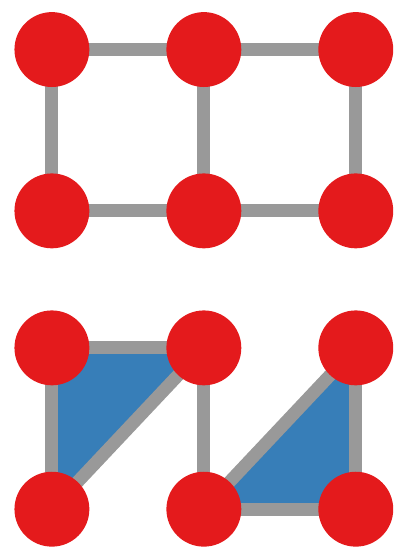}
\caption{}
\end{subfigure}
\caption{(a) A 2-MPGNN where I identify with a green outline all node pairs adjacent to the node pair in the blue outline. (b) A MPGNN with simplicial complexes of three nodes, highlighted with a blue fill.}
\label{fig:mpgnn-highorder}
\end{figure}

More tractable solutions exploit high order dynamics (Chapter \ref{cha:hod}). For instance, in the k-MPGNN approach\cite{morris2019weisfeiler} one realizes that what I just described is merely a 1-MPGNN, where nodes get messages individually from their neighbors. But we can imagine a 2-MPGNN where each pair of nodes receives messages from its neighboring pairs of nodes -- pairs with which they share a node. Then you can have as many k-MPGNN as you want, with each set of $k$ nodes communicating with all other sets of $k$ nodes with which they share $k - 1$ nodes. Figure \ref{fig:mpgnn-highorder}(a) shows an example, where you can see the two graphs should lead to different embeddings, as the node pair in blue gets updates from a different number of adjacent node pairs -- in green -- in the two cases.

Another high-order tactic is to use simplicial complexes\cite{bodnar2021weisfeiler}. To cut a long story short, the idea here is that messages not only pass through edges, but also through simplicial complexes, with their own customized aggregate and update functions. This solves many problems because, as you can see in Figure \ref{fig:mpgnn-highorder}(b), even if Weisfeiler-Lehman thinks that the two graphs are isomorphic, a simplicial MPGNN does not, because it sees that one has two simplicial complexes that the other does not. This approach of using high order structures connects to a more general ``topological'' learning, of which graph neural networks are a special case.

\subsection{Smoothing Problems}
There's another issue with MPGNNs. As I told you, more layers allow you to pool information from farther away in the network. One thing you might be tempted to do is to pool information from the entire network. That is to say, you could decide to have $l$ layers so that $l$ is the diameter of the network. This way you know that, in the end, even the farthest away pair of nodes will exchange some information. However, you start to run into troubles, a specific kind of trouble we call ``smoothing''.

\begin{figure}
\centering
\begin{subfigure}{.3\columnwidth}
\includegraphics[width=\textwidth]{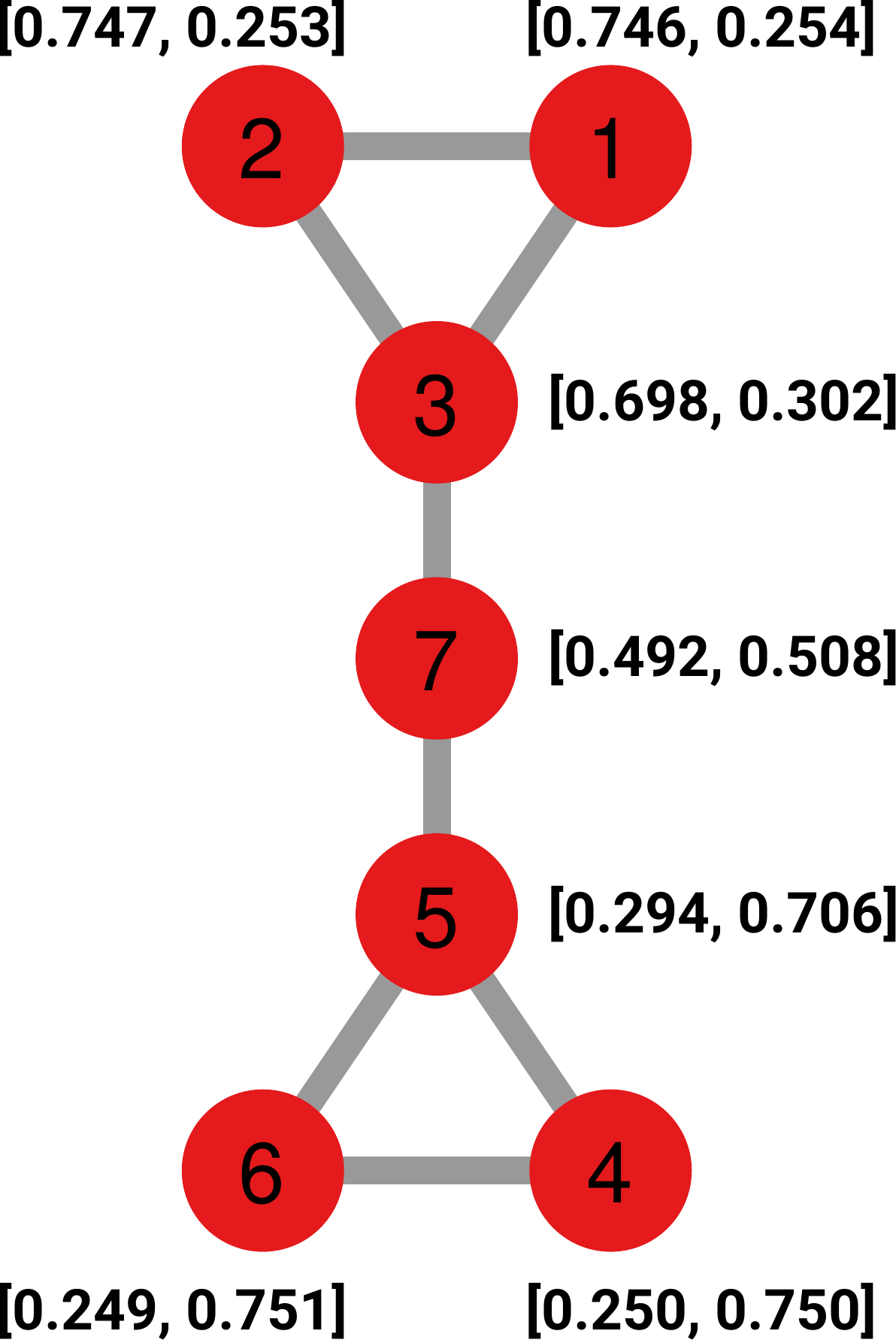}
\caption{$H^3$}
\end{subfigure}\quad
\begin{subfigure}{.3\columnwidth}
\includegraphics[width=\textwidth]{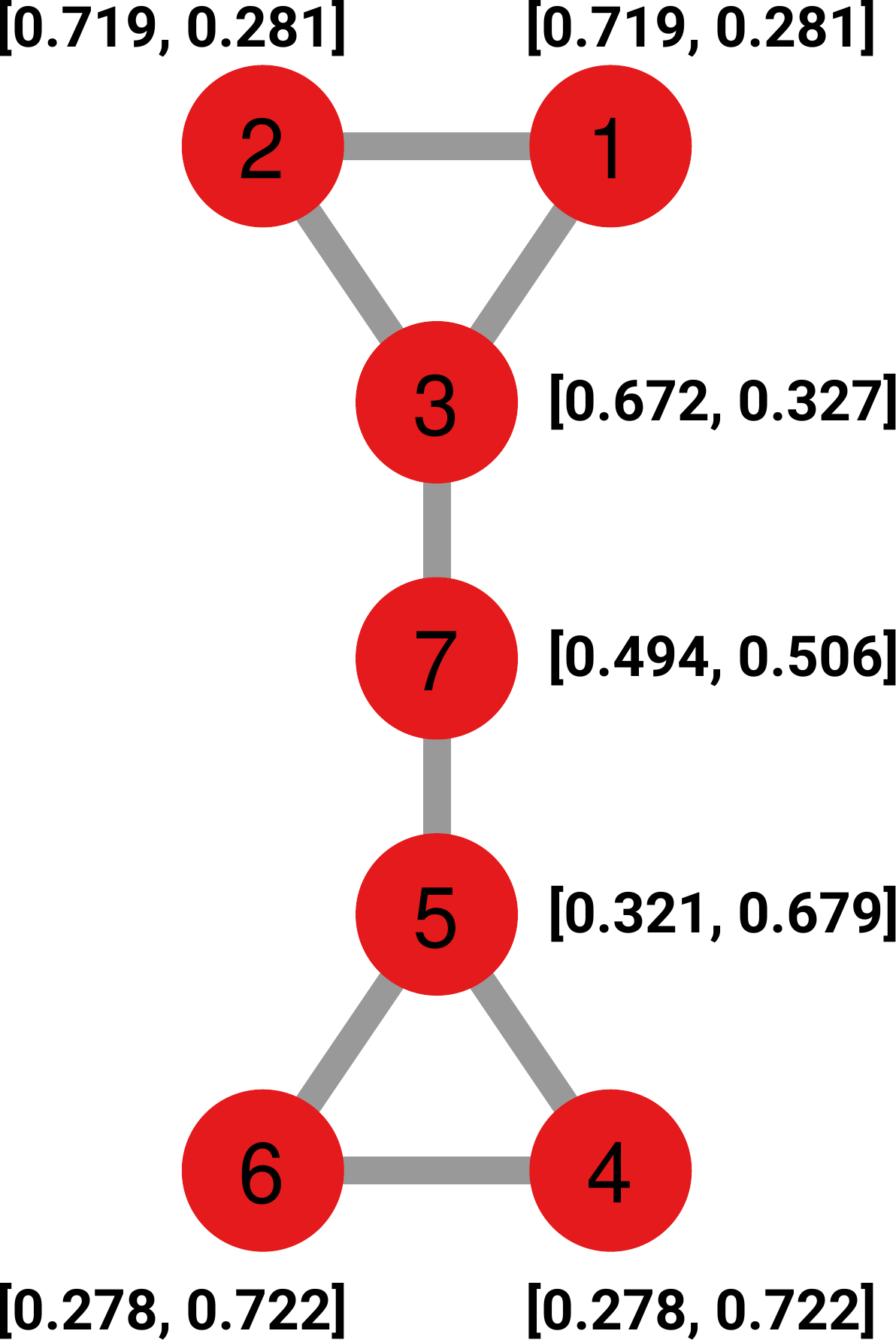}
\caption{$H^4$}
\end{subfigure}\quad
\begin{subfigure}{.3\columnwidth}
\includegraphics[width=\textwidth]{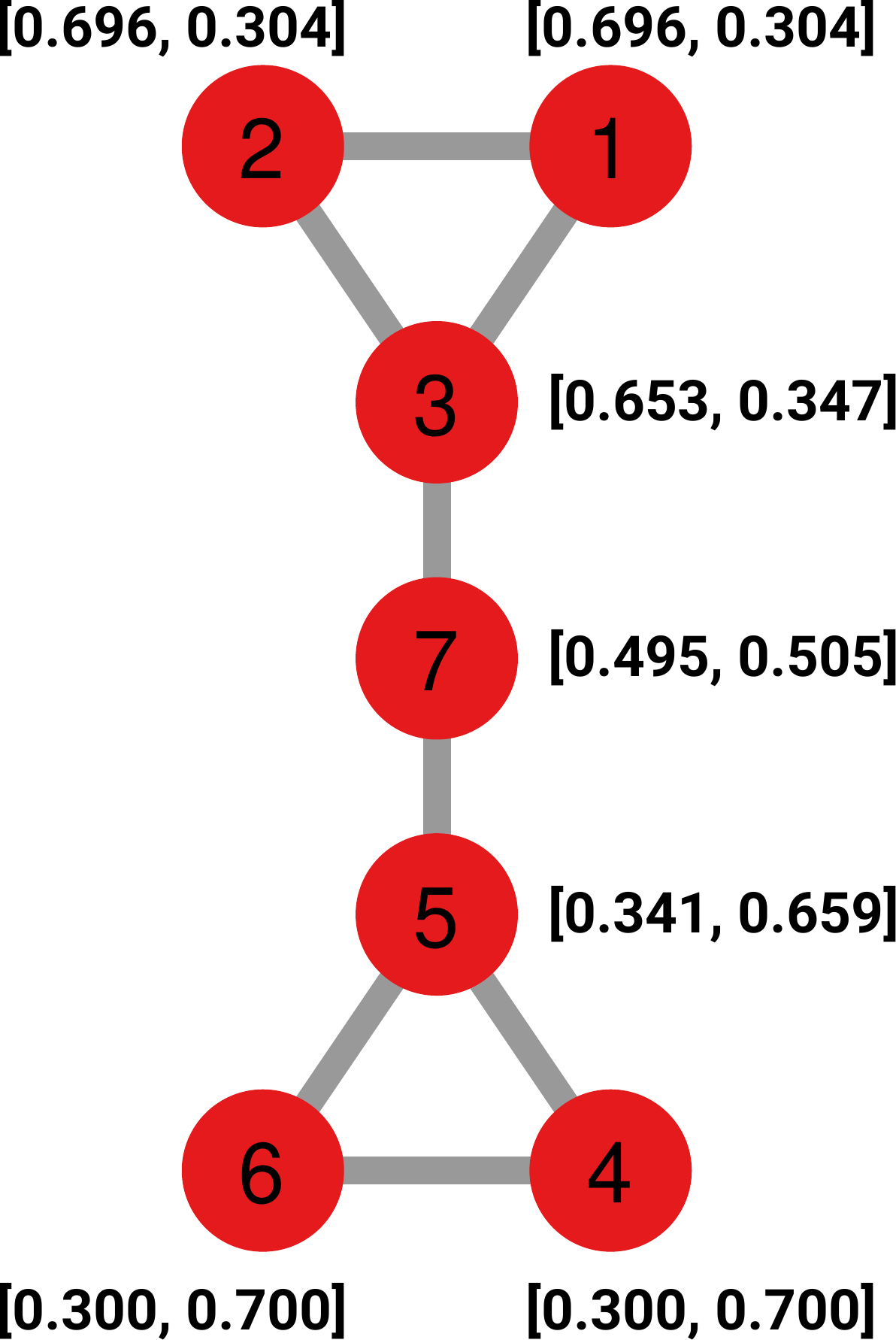}
\caption{$H^5$}
\end{subfigure}
\caption{The result of applying several more layers to the network in Figure \ref{fig:mpgnn-layers}: (a) third, (b) fourth, and (c) fifth layer.}
\label{fig:mpgnn-smoothing}
\end{figure}

To understand smoothing, let's take the process we started in Figure \ref{fig:mpgnn-layers} and let's keep unfolding it. In Figure \ref{fig:mpgnn-layers} we stopped at the second layer, so in Figure \ref{fig:mpgnn-smoothing} I start from the third and I continue for a few layers. What do you observe?

All embeddings are getting more and more similar to each other. That is, the MPGNN is losing its capability to distinguish the nodes in the network. A few more layers, and each node will get the same embeddings. In MPGNN, you either stop early or you run long enough to become a vector of constants. You knew this was going to happen, because this is exactly the same dynamics I showed you back in Section \ref{sec:rw-consensus}, when we talked about reaching a consensus in the network. From that section you know that the eigenvalues of the Laplacian will tell you when you will reach the consensus, so you should stop far before that point.  

Well, actually there are other things you can do to prevent smoothing. One of them is to use the skip-connection approach. The simplest way to do it is by not blindly accepting the message arriving from the neighbors. For each node $v$, you can run your update function on the message $v$ receives plus $v$'s representation in the previous layer\cite{veit2016residual}. In practice, $v$ remembers what it was and only updates a little bit with the new information. This makes information propagate more slowly in the MPGNN which, given that our problem is over-smoothing, it's a feature, not a bug. The way you combine the new message with the old representation is up to you, you can figure out which function works best for your case\cite{pham2017column}. Since you're using a node's embedding to update itself, this is a type of recurrent neural network\cite{cho2014learning}. 

Another neat thing you can do is the so-called jumping knowledge network\cite{xu2018representation}. In this approach, the final embedding for node $v$ in your MPGNN is not the final layer, but a concatenation of $v$'s intermediate embeddings from all layers. If you had $d$ dimensional embeddings and $l$ layers, the final embedding for $v$ has not length $d$, but length $ld$.

You could also add layers that do not pass messages among nodes\cite{you2020design}, such as ones coming from more classical neural network designs.

\subsection{Over-Squashing}
\begin{figure}[t]
\centering
\includegraphics[width=.8\columnwidth]{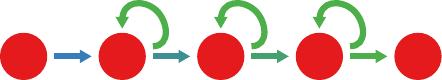}
\caption{An example of oversquashing. Horizontal arrows represent the passed message. Loopy arrows represent the aggregate step of the receiving node. The color of the arrow represent what's being passed.}
\label{fig:oversquashing}
\end{figure}

So, adding too many layers will cause smoothing, which is a bad news. The even worse news is that it won't even help with the original problem, which is that there are some nodes in the periphery whose messages are going to be largely ignored. Even if you have $l$ layers in a network with diameter $l$, when the message from the peripheral node arrives to its most distant nodes, it has changed so much it became irrecognizable. Figure \ref{fig:oversquashing} shows you why: the original message -- the pure blue color representing its content -- has passed through many ``aggregate'' steps -- in green --, each of them changing it with the current node's attributes. 

It's like a game of telephone that has ran for too long. By the end, the message passed to the final recipient node is completely green: the intermediate nodes have erased all the blue information contained in the original message. Central hubs with high betweenness are the culprit and can ``squash'' virtually all messages with their own attributes, resulting in less expressive embeddings at the end\cite{topping2021understanding}.

\section{Convolution}
It is now time to go into the dreaded math I foreshadowed at the beginning of the chapter. There are many reasons why this is a good idea. One of them is to realize that most of what I said so far is nothing more than a bunch of matrix multiplications. This gives you an idea why there has been an explosion of popularity with neural network approaches: it's not because these are fancy algorithms. Matrix multiplication ain't fancy but it works, and it has the convenient property that you can do it blazingly fast on a GPU. It is aligning the neural network approaches with how the hardware works that led to the gains.

\subsection{Message-Passing as Matrix Multiplications}
\begin{figure}
\centering
\includegraphics[width=.5\columnwidth]{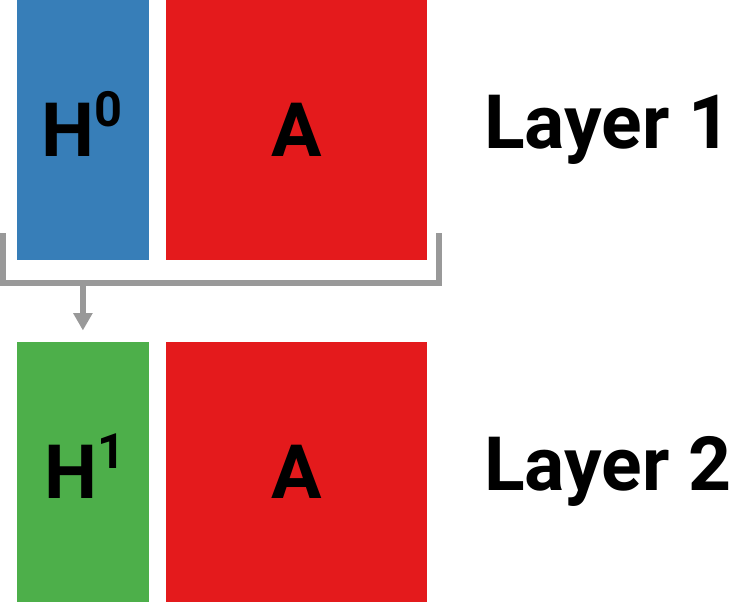}
\caption{The basic architecture of a MPGNN in matrix form. The blue rectangle represents the matrix for node features, red rectangle for the adjacencies, and green rectangle for the hidden features.}
\label{fig:gcn-basics}
\end{figure}

Ok, so how is a MPGNN really a matrix multiplication? Remember the key ingredient: we have a $|V| \times d$ matrix $H^0$ with $d$ dimensional node features. Then we want each node $v$ to look at the features of all its neighbors. What matrix can tell you the neighbors of a node? The good ol' adjacency matrix $A$. So the most barebone graph neural network possible can be expressed simply as:

$$ H^l = AH^{l-1}, $$

that is to say: to get the values for layer $l$, you take those from layer $l - 1$ and you aggregate them by multiplying $H^{l-1}$ by the adjacency matrix $A$. You know from Section \ref{sec:la-matrix} that the result of multiplying a $|V| \times d$ matrix ($H^0$) to a $|V| \times |V|$ matrix ($A$) is going to be another $|V| \times d$ matrix -- so $H^1$ will be just another node attribute matrix. Figure \ref{fig:gcn-basics} shows an abstract representation of how this happens.

Note how in the figure I use different colors for $H^0$ and $H^1$, stressing that the former is data while the latter is a representation of what we've learned in the first layer. This process -- and equation -- describes the aggregation step of the MPGNN I described at the beginning of Section \ref{sec:deep-mpgnn} -- except that here the aggregation function is the sum and not the average.

There are some issues with this formulation which you can solve by implementing the update step of the MPGNN. If $H^k = H^{l-1}A,$ then each node $v$ completely forgets its $l - 1$ features -- which is a problem especially for the first layer, since in that case the $H^0$ features we're forgetting are literally node $v$'s actual real features. To fix this issue we add self loops in the update step, assuming that each node $v$ is a neighbor of itself. This is also a matrix operation! It is equivalent to adding the identity matrix $I$ to $A$. So now we have:

$$ H^l = (I + A)H^{l-1}. $$

Using a different perspective, this is equivalent to sum to the result $H^{l-1}$ to $H^{l-1}A$. This would lead to $H^l = H^{l-1} + AH^{l-1}$, and we can group the $H^{l-1}$ terms, since $H^{l-1}I = H^{l-1}$ -- with $I$ being the identity matrix.

Great, but we're not done reproducing the basic MPGNN I showed you at the beginning of the chapter. The first thing we need to solve is that, in my original MPGNN, the aggregation function was not the sum, but the average. Why might we prefer the average over the sum? Well, if you have a lot of layers, constantly summing features will lead them to eventually blow up in scale. All values will diverge to infinity. You might want to keep each $H^l$ within the same scale. You can actually see this problem happening in the example from Figure \ref{fig:mpgnn-weisfeiler}, with the resulting embedding being much larger than the original one.

You might think to solve this problem by using the adjacency matrix, normalized by the degree -- yet another thing you can do with matrix multiplications: $D^{-1}A$. However, sometimes a more sophisticated approach can help you. For instance, the symmetric normalization $D^{-1/2}AD^{-1/2}$ still produces a stochastic matrix, but here you're normalizing the values coming from your neighbors, taking into account how many neighbors they themselves have\cite{kipf2017semi}. So:

$$ H^l = \hat{D}^{-1/2}(I + A)\hat{D}^{-1/2} H^{l-1}, $$

with $\hat{D} = D + I$, i.e the degree matrix which takes into account the fact that you added self loops to $A$. Note that you don't have to always normalize: in fact, normalizing by the degree reduces the number of structures you can distinguish (Section \ref{sec:deep-mpgnn-limits}) because now all of a sudden you can't distinguish nodes with different degrees any more.

We're so close to reproduce the initial MPGNN, we are only missing one step: the final activation function in the update step. You might want to apply any activation function $\sigma$ to the result -- see Section \ref{sec:ml-activation} for some options. The reason is that matrix multiplications can only allow you to learn linear functions -- this is linear algebra, after all! However, more often than not, you might want to learn nonlinear functions. An activation function such as ReLU or the sigmoid can allow you to inject nonlinearity in the system. The result is:

$$ H^l = \sigma\left(\hat{D}^{-1/2}(I + A)\hat{D}^{-1/2} H^{l-1} \right). $$

\begin{figure}
\centering
\begin{subfigure}{.2\columnwidth}
\includegraphics[width=\textwidth]{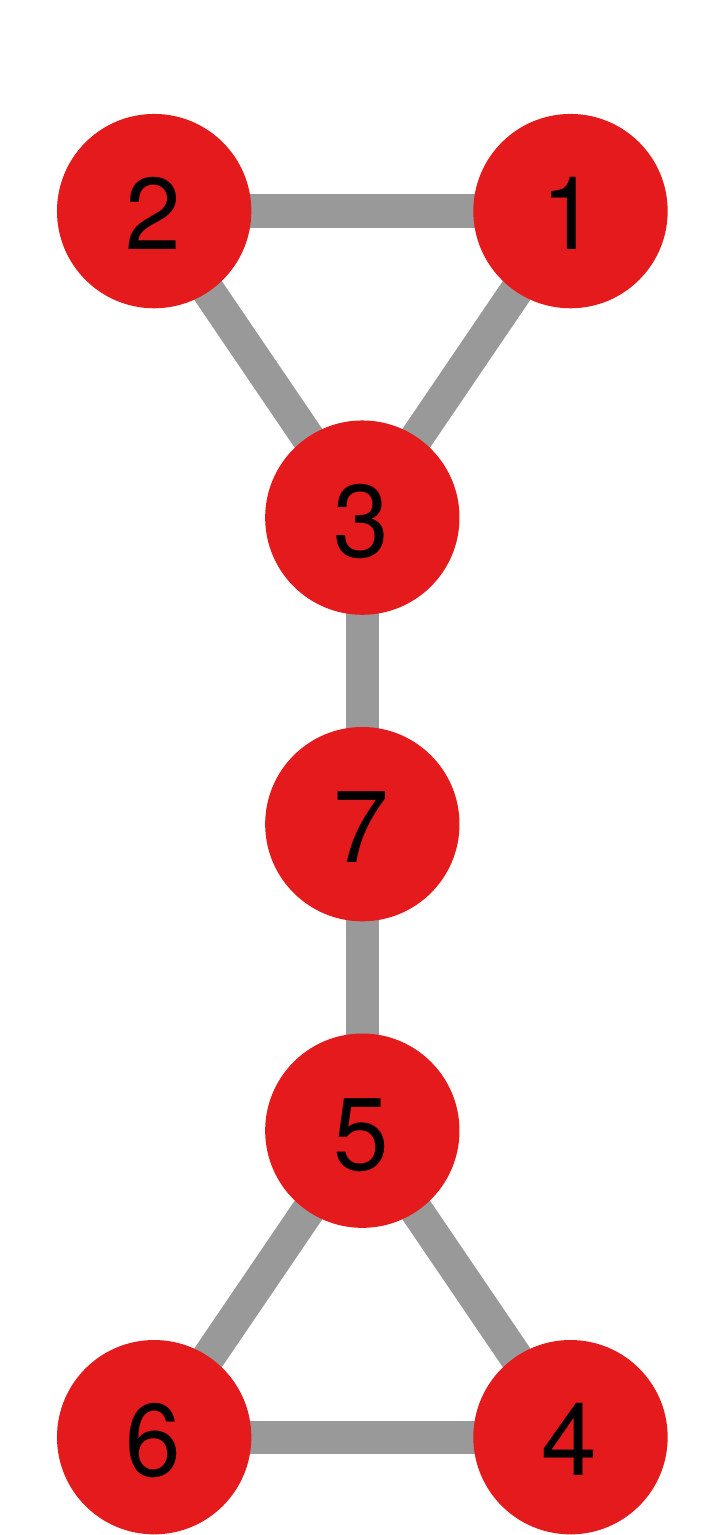}
\caption{}
\end{subfigure}\qquad\qquad
\begin{subfigure}{.2\columnwidth}
\includegraphics[width=\textwidth]{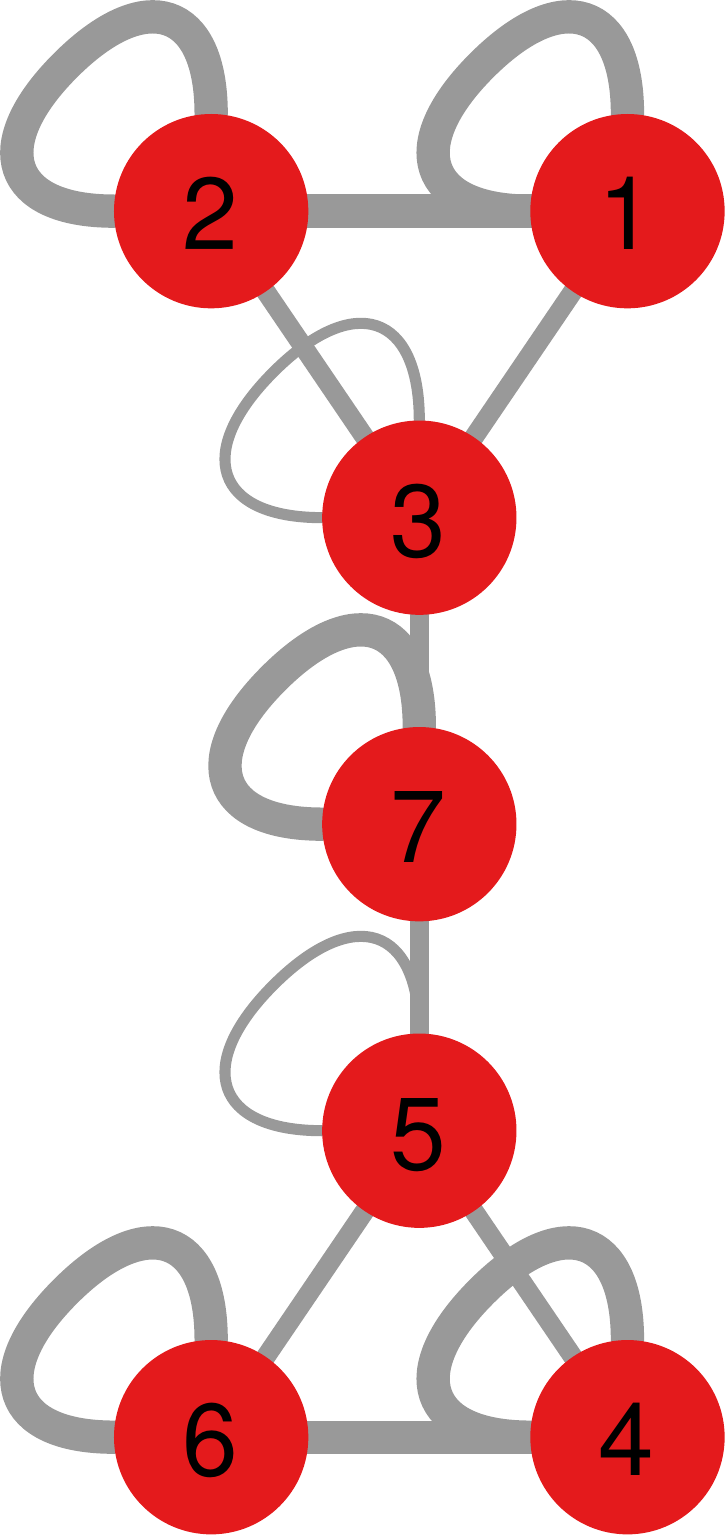}
\caption{}
\end{subfigure}
\caption{(a) The original network. (b) The network after the $\hat{D}^{-1/2}(I + A)\hat{D}^{-1/2}$ operation. The thickness of an edge is proportional to the weight it gets as a result of the transformation.}
\label{fig:mpgnn-vs-gcn}
\end{figure}

This is a big pile of linear algebra already, but we can still demystify it a bit with a graphical representation. That formula is equivalent to the operation I depict in Figure \ref{fig:mpgnn-vs-gcn}.

Now we're getting somewhere, but this somewhere still isn't very far. Up until this moment we have reproduced the MPGNN framework. It is now time to really go deep.

\subsection{From Message-Passing to Convolution}
The key ingredient to really start with deep learning on graphs looks innocuous. We just add another matrix multiplication to the framework. We call this mysterious family of matrices $W$, and we have a different $W^l$ matrix for each layer $l$. So:

$$ H^l = \sigma\left(\hat{D}^{-1/2}(I + A)\hat{D}^{-1/2} H^{l-1} W^{l}\right). $$

This is starting to become a handful, isn't it? Yet, it's still matrix multiplications all the way down. And we know what each piece means, except $W$ and what its role is, so let's take it step by step. Fundamentally, $W$ does three things:

\begin{enumerate}
\item By multiplying it to the other matrices it weighs the results -- that's why it's called $W$, $W$ for Weights.
\item It changes the dimensions of the embeddings, as we're free to decide one of its dimensions.
\item It allows to the neural network to learn in a supervised task.
\end{enumerate}

In other words, $W$ is the primary ingredient of the secret sauce that allows to move from MPGNN to actual Graph Convolutional Networks (GCNs). The first point is the least interesting: multiplying $W$ to any matrix will change the matrix, over- or under-weighting parts of it -- unless $W$ is the identity matrix.

\begin{figure}
\centering
\includegraphics[width=.5\columnwidth]{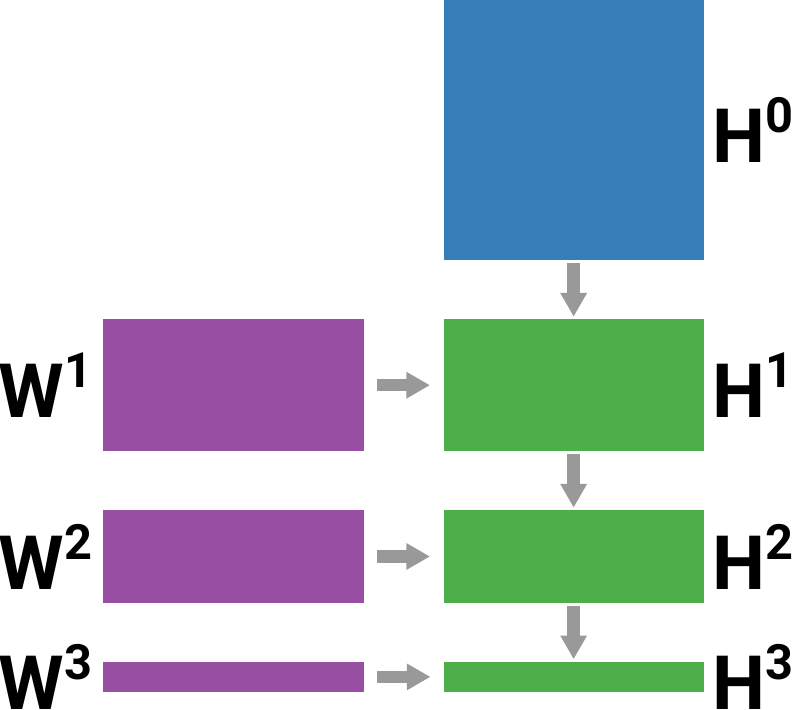}
\caption{The dimensionality reduction that one can apply to $H^{l}$ by multiplying $H^{l-1}$ to $W^l$. The length of the sides of the rectangles are proportional to the dimensions of their matrices.}
\label{fig:gcn-w}
\end{figure}

How do we achieve the second point? Well, $W$ is a $d \times d'$ matrix. We're multiplying it to a $|V| \times d$ matrix, so you know that you're going to end up with a $|V| \times d'$ matrix in the end. Unless $d = d'$, you will end up with a set of node features that is different than the original set. Stripping away complexity that is not necessary to get to the point, Figure \ref{fig:gcn-w} shows you how this dimensionality change happens. 

Each $W^l$ must share one dimension with $H^l$ and will determine the shape of $H^{l+1}$. In most cases, you want $d' < d$, for instance because the original set was sparse and you want it to be denser. Often, for the last layer, you want $d' = 1$, so that you end up with a single number per node, and use it to classify the nodes.

If these two things -- weighting and dimensionality reduction -- were all that $W$ does, you could still technically remain within the bounds of MPGNNs. If we only have fixed weights, it is as if our graph was weighted from the beginning. And as long as you always keep $d = d'$, we still have our familiar MPGNN. But the burning question you might have, and the one which will mark our permanent departure to GCNs is: what's actually inside $W$? We know what $H$, $A$, $I$, and $D$ contain. But $W$ seems mysterious.

To understand $W$ we need to unpack its third role in the GCN: to allow the neural network to learn. In this scenario, we have some sort of ground truth we want to predict, so we're performing a supervised learning task. Let's assume that we have this ground truth in matrix $T$ -- for simplicity assuming it is merely a $|V| \times 1$ vector. Ideally, we want the final embeddings $H^l$ to also be $|V| \times 1$, and to be identical to $T$ -- plus/minus a constant multiplicative factor. If that's the case, we can perfectly predict the $T$ values with $H^l$.

A MPGNN cannot do this because it doesn't have $W$ and so it can't learn. That is to say, with given starting node features $H^0$, you're going to get the same $ H^l$ as output. It might be good to predict $T$, but switch $T$ with $T'$ -- a different ground truth, a different set of new node features we want to predict -- and the MPGNN won't be able to adapt to this new task. But a GCN can, because it has $W$. At each layer $l$, a GCN can calculate how different $H^l$ is to any arbitrary $T$. This difference will be estimated using any loss function (Section \ref{sec:ml-loss}) we think is appropriate.

Then the GCN can perform one trick, which is so powerful we gave it a cool name: backpropagation\cite{linnainmaa1970representation}\cite{rumelhart1986learning}. The GCN can backpropagate the errors. Basically this means that, after each layer $l$ has computed the value of its loss function, it can call up layer $l - 1$ and tell it: ``you dummy! Look what a bad job you did! Do it better so that my loss function is lower!'' And layer $l - 1$, very embarrassed, will get to work to minimize the loss function. How can it do it? By changing the only thing it can change: $W^{l-1}$.

Backpropagation is a bit more complicated than this: it is a clever way to figure out how the loss function is changing given the changes in the parameters -- in technical terms you call the landscape of such changes ``gradients'' and you want to ``descend'' them, i.e. to find the minimum value of the loss function. One way to do it is by stochastic gradient descent\cite{robbins1951stochastic}. However, you only care about these details if you're doing machine learning, so we skip some nitty-gritty here. This is more or less the general picture you need as a network scientist.

\begin{figure}
\centering
\includegraphics[width=.4\columnwidth]{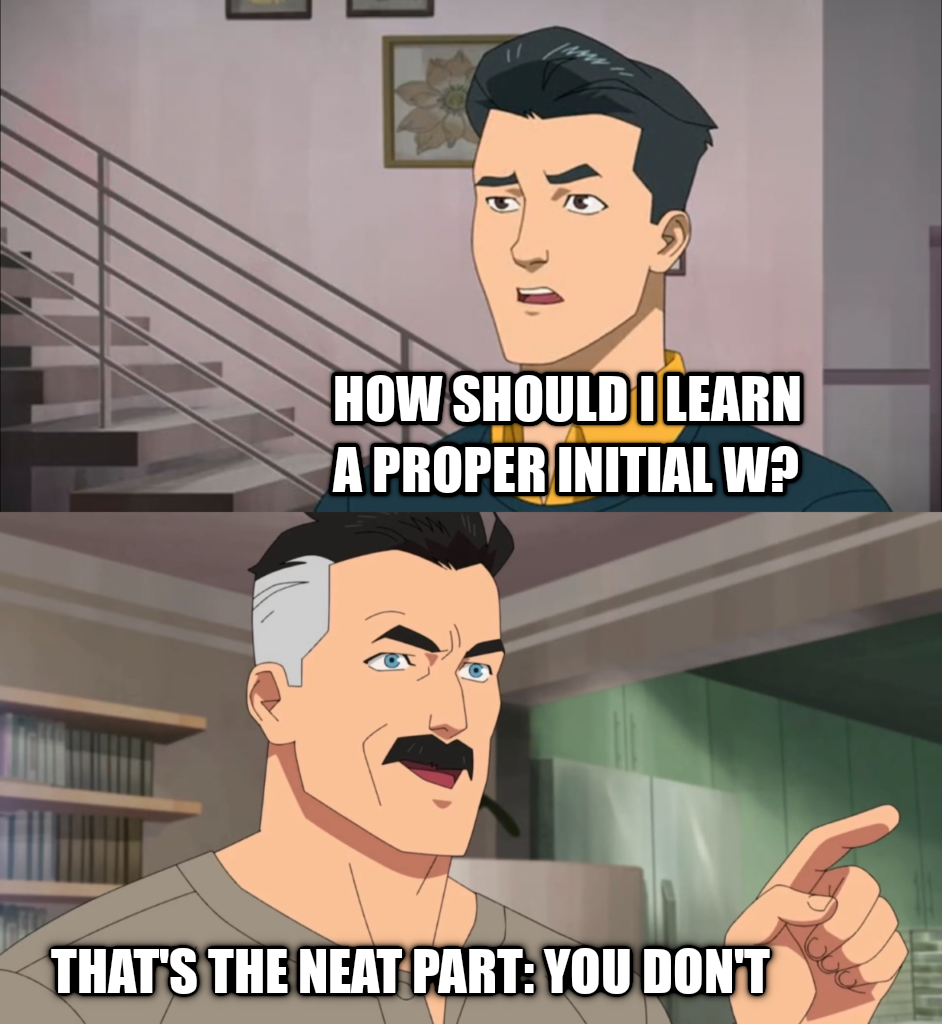}
\caption{The initialization algorithm for $W$ in a graph convolutional network.}
\label{fig:gcn-w-meme}
\end{figure}

There might still be one last thing worrying you. Backpropagation is cool, but it only tells you how to update $W$ to go from a high to a low loss. It doesn't tell you how to initialize $W$. To initialize $W$ you need to apply the strategy I depict in Figure \ref{fig:gcn-w-meme}.

That is to say: you can start with a random $W$ and backpropagation will lead any random $W$ to predict $T$ with minimal loss\cite{velickovic2019deep}. Sure, some people will tout the idea of pre-training, which is to do a training phase to get a great $W$ before you actually run your GCN. But that's not necessarily going to help.

\begin{figure}
\centering
\includegraphics[width=\columnwidth]{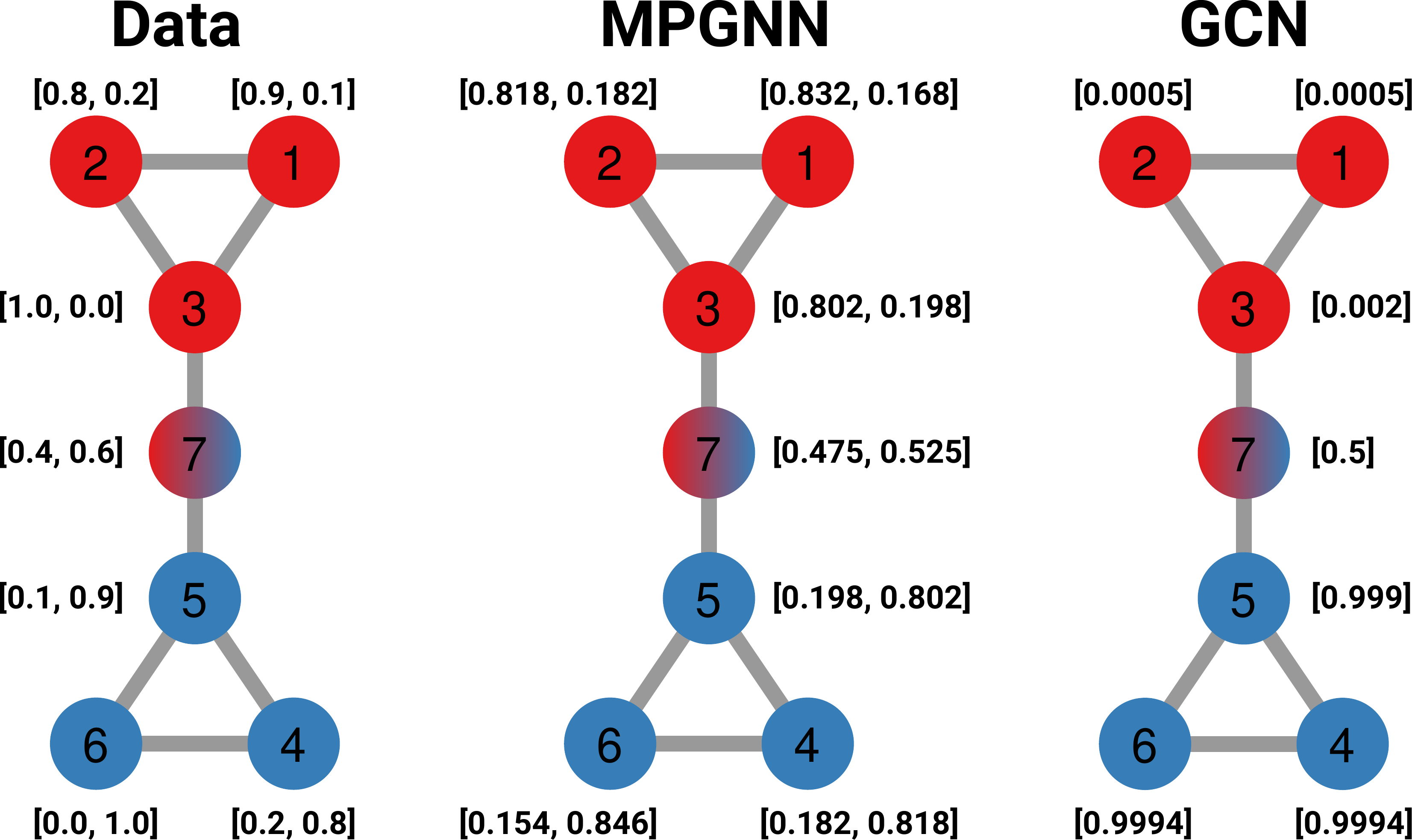}
\caption{(Left) The initial data: nodes are papers connected by undirected edges if the papers cite one another. Each node starts with two attributes. The node's color tells us the field of the paper it represents, with node $7$ belonging to both the red and the blue class. (Right) The numbers next to the node represent the node's embeddings obtained from the last layer of the MPGNN and GCN.}
\label{fig:gcn-vs-mpgnn1}
\end{figure}

To give you an example of the flexibility of GCNs against MPGNNs consider Figure \ref{fig:gcn-vs-mpgnn1}. Here we assume the nodes are papers, connections are citations, and nodes have features connecting them to the words in their abstracts. Our $T$ here might be a binary matrix, telling us whether a paper is about a specific subfield. This is a fairly classical test case of graph neural networks. And MPGNNs can do well here, as the figure shows. Since papers citing each other also have fairly similar abstracts, the messages will be able to detect this synergy between the network's structure and the node embeddings.

Note how the GCN can produce 1D embeddings even if we started with 2D ones, which here works fine because we only have two classes to predict, so we can place the node in a spectrum. One difference is how much more confident the GCN is with respect to the MPGNN: the scores are much closer to zero and to one, which means the GCN is sure about the label assignment. Also, the GCN doesn't distinguish between nodes that are structurally equivalent, such as nodes $4$ and $6$, even if they started with different embeddings.

But these are mere details: the MPGNN here worked equally as well as the GCN. But suppose that the same network -- same nodes, same citations, same abstracts -- now wants to predict a different $T'$: a binary vector telling us whether the paper was authored by Jure Leskovec. A MPGNN is condemned to run the same process, and would probably perform poorly -- as Jure is an eclectic researcher. A GCN, instead, can make its $W$ into an efficient Leskovec-detector -- a ``lesk2vec'' if you may\footnote{I'm begrudgingly crediting Sune Lehmann for the best joke of the chapter.} --, and spot him in no time.

\begin{figure}
\centering
\includegraphics[width=\columnwidth]{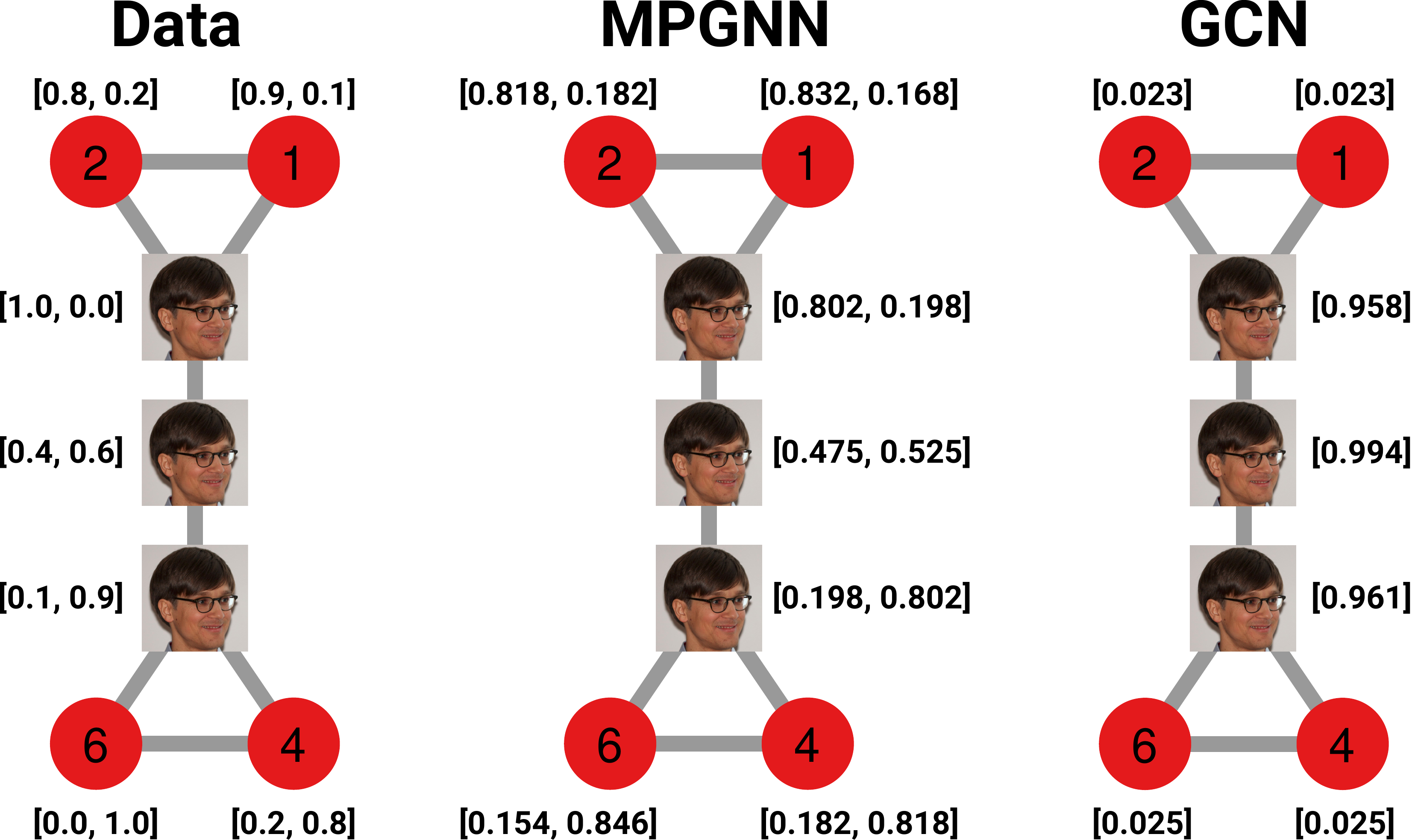}
\caption{(Left) The same network as Figure \ref{fig:gcn-vs-mpgnn1}, however this time we want to predict the node's picture rather than its color. (Right) The embeddings produced by the MPGNN and the GCN.}
\label{fig:gcn-vs-mpgnn2}
\end{figure}

Figure \ref{fig:gcn-vs-mpgnn2} shows you that there is no connection between the MPGNN embeddings and whether the node is a Leskovec node, while the GCN only gives high values to the Leskovec node and low values to all other nodes -- a radically different embedding even if the starting point was exactly the same. Backpropagation lead it to correctly modify $W$ to adapt to this new $T'$.

As a final note, so far I talked about deep graph neural networks for supervised tasks, where we have a ground truth we want to predict. This is not the only option. There are unsupervised approached that can be driven by a maximization of mutual information between local and global graph properties\cite{velivckovic2018deep}.

\section{Spectral Convolution}\label{sec:deep-mpgnn-spectral}
Before moving on to even more sophisticated methods, I want you to consider a slight shift in perspective. If you look again at the most recent pile of matrix multiplications that define a GCN, you'll notice that it is only dependent on $A$, the adjacency matrix:

$$ H^l = \sigma\left(\hat{D}^{-1/2}(I + A)\hat{D}^{-1/2} H^{l-1} W^{l}\right). $$

Even if $H^{l-1}$ aggregates information from potentially many steps away, to determine the new embedding of node $v$ we're still operating exclusively on the information that reached the direct neighbors of $v$. This is what we call a ``spatial'' approach\cite{scarselli2008graph}. This is powerful and has many advantages. One of them is that the information to determine the status of a node is essentially local: you don't need to know the entire topology of the graph to learn a specific node's embedding. There are many popular GCN methods that are spatial\cite{atwood2016diffusion}\cite{niepert2016learning}.

This, by the way, solves the transductivity problems of shallow learning we saw in Section \ref{sec:mining-embeddings-limits}. If you get a new node in your network, you can determine its new embeddings by only running the local part of the graph neural network.

However, spatial approaches are not the only game in town, just like they weren't when we dealt with shallow learning. In shallow learning, random walk-based methods are also spatial, and I showed you there are alternatives, namely by using a spectral approach. In GCNs we can have a spectral approach as well\cite{bruna2014spectral}\cite{liao2019lanczosnet}. If in the spatial approach we learn the embedding of $v$ only by looking at $v$'s neighbors, in the spectral approach $v$'s updated embedding depends on the entire graph as a whole.

This is because we see the embedding as a signal that is processed by the entire graph. To make sense of this sentence you need to understand a few things about signal processing, the Fourier transform, and its discrete version on graphs. This is a huge topic, very complex, so I'm going to give you an extremely simplified and incomplete version of what's going on, and you'd do better reading more advanced texts on the subject\cite{shuman2013emerging}.

\begin{figure}[t]
\centering
\includegraphics[width=.4\columnwidth]{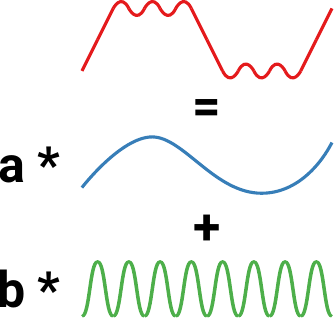}
\caption{The Fourier transform. The original signal in red can be thought as the weighted sum of frequencies (in blue and green).}
\label{fig:fourier-transform}
\end{figure}

Basically, the Fourier transform is a way to decompose any periodic signal into a weighted sum of frequencies that make it up. If, for any moment in time, you sum the value of each frequency with the weight you learned by the Fourier transform, you'll get the exact value of the signal. Figure \ref{fig:fourier-transform} shows you how to think about this visually.

In the graph Fourier transform, you want to do the same, but your signal is $H^{l-1}$. How to do it? Well, one key part of the regular Fourier transform is to calculate the difference between the value of the signal at a point and the values of the signal in the neighboring points. This is done with the Laplace operator and you should have alarm bells ringing in your head right now. The Laplacian does exactly the same thing on a graph: the $-1$ entry for each edge gives you exactly the difference of a node with its neighbor.

\begin{figure}
\centering
\includegraphics[width=.66\columnwidth]{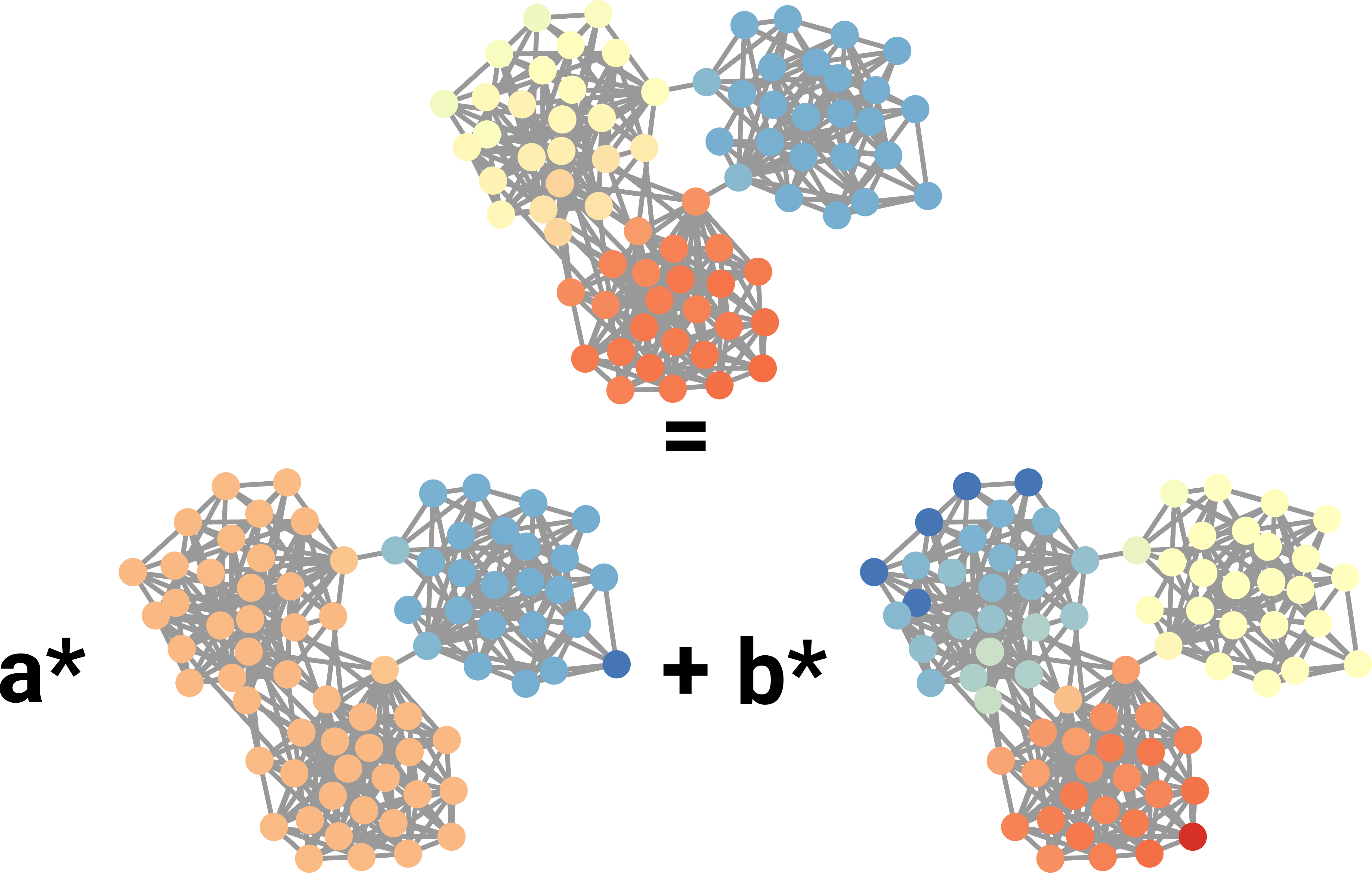}
\caption{The Fourier transform. The original signal (top) in red can be thought as the weighted sum of eigenvectors of the Laplacian (bottom).}
\label{fig:graph-fourier-transform}
\end{figure}

The way to use $L$ in the graph Fourier transform is to exploit the following equivalence: $L = \Phi \Lambda \Phi^{-1}$ -- this is the eigendecomposition we saw in Section \ref{sec:mat-factors}. What are $\Phi$ and $\Lambda$? $\Lambda$ is $L$'s eigenvalue diagonal matrix:

$$ \Lambda = 
\begin{pmatrix}
\lambda_0 & \dots & 0 \\
0 & \ddots & 0 \\
0 & \dots & \lambda_n \\
\end{pmatrix}
$$

where $0 = \lambda_0 < \lambda_1 \leq \lambda_2 \leq \dots \leq \lambda_{|V| - 1}$ are the eigenvalues of $L$ sorted in increasing order (as usual, we assume $G$ is connected). $\Phi$ is the matrix of $L$'s eigenvectors, in the same order as $\Lambda$. We do this because in this way we have established that the ``frequencies'' we were talking about in the normal Fourier case are the eigenvalues of $L$. As a consequence, we can reconstruct any signal on $G$ as the weighted sum of the eigenvalues of $L$, as Figure \ref{fig:graph-fourier-transform} depicts.

This means that you can use $\Phi$ as a convolution operator instead of $A$, and $\Phi$ will give you information about how the entire graph at once processes a single  node $v$'s embeddings. The obvious downside here is that calculating $\Phi$ is computationally expensive. So it is no surprise that people have worked on ways to perform this update locally\cite{defferrard2016convolutional}. Then there are ways to render this approach more sophisticated and expressive\cite{levie2018cayleynets}\cite{bianchi2021graph}.

\section{Summary}

\begin{enumerate}
\item Deep learning on graphs uses the message-passing graph neural network (MPGNN) framework: each node aggregates messages from its neighbors' features and uses them to update its own features. One can repeat this aggregation-update operation many times, each of which being a layer in the MPGNN.
\item Limitations of this approach come from the fact that there are structures MPGNNs aren't able to distinguish, that piling up too many layers will usually result in all nodes having the same embeddings, and that messages from peripheral nodes are usually lost.
\item A MPGNN can be described as a series of matrix operations in linear algebra, for instance: $H^{l}=\sigma\left(\hat{D}^{-1/2}(I + A)\hat{D}^{-1/2} H^{l-1}\right)$, which means to add self loops, normalize edge weights, then pass messages through edges, and update with a non-linear $\sigma$ function such as ReLU.
\item In graph convolutional networks (GCNs), you add another matrix family $W^l$ to the multiplication. $W^l$ can weight how much a node's embedding depends on each of its neighbors, and it can reduce the dimensions of your embeddings.
\item $W^l$ can also allow you to predict specific node features you're interested in. At the beginning, $W^l$ is random, but training it via backpropagation can lead to find the $W^l$ that minimizes your loss with any prediction target you might have.
\item One can go from a spatial approach with $A$ to a spectral approach with $\Phi$ -- the matrix of eigenvectors of the Laplacian -- performing a graph Fourier transform, processing each embedding via the entire graph rather than only the immediate neighborhood.
\end{enumerate}

\section{Exercises}

\begin{enumerate}
\item Implement the MPGNN described in Section \ref{sec:deep-mpgnn}. You need to define an ``aggregate'' function which takes a list of nodes and their embeddings and returns the element-wise mean of those embeddings. You need to define an ``update'' function which takes two vectors, sums them, and return their softmax. Finally, you need a message-passing function which loops over all nodes of the network, applies aggregate to its neighbors, and applies update with the result of the aggregation and the node's embedding. Run a single layer of it on the network at \url{http://www.networkatlas.eu/exercises/44/1/network.txt}, with node features at \url{http://www.networkatlas.eu/exercises/44/1/features.txt}. Do you get the same results as Figure \ref{fig:mpgnn}?
\item Run the MPGNN you designed in the previous exercise. Make a scatter plot of the node embeddings using the two dimensions as $x$ and $y$ coordinates at the first, fifth, tenth, and twentieth layer. What do you observe?
\item Implement a MPGNN as a series of matrix operations, implementing $H^{l}=\sigma\left(\hat{D}^{-1/2}(I + A)\hat{D}^{-1/2} H^{l-1}\right)$, with $\sigma$ being softmax. Apply it to the network at \url{http://www.networkatlas.eu/exercises/44/3/network.txt}, with node features at \url{http://www.networkatlas.eu/exercises/44/3/features.txt}. Compare its running time with the MPGNN you implemented in the first exercise, running each for $20$ layers and making several runs noting down the average running time.
\end{enumerate}

\chapter{Deep Graph Learning Models}\label{cha:mining-deep2}

\section{Attention}
In standard machine learning, ``attention'' is a strategy that allows you to have embeddings that adapt to the situation\cite{bahdanau2014neural}. The classical case is natural language processing. Each word must have an embedding, but problems arise when you have words with multiple meanings. You have an example in this book part, with the awkwardness of the word ``network'', which we use in this book to refer to an object -- for instance a \textit{social} network -- that is rather different from a \textit{neural} network. A world where the same word ``network'' can have more than one embedding would be wonderful. Attention allows you to do that.

\begin{figure}
\centering
\includegraphics[width=\columnwidth]{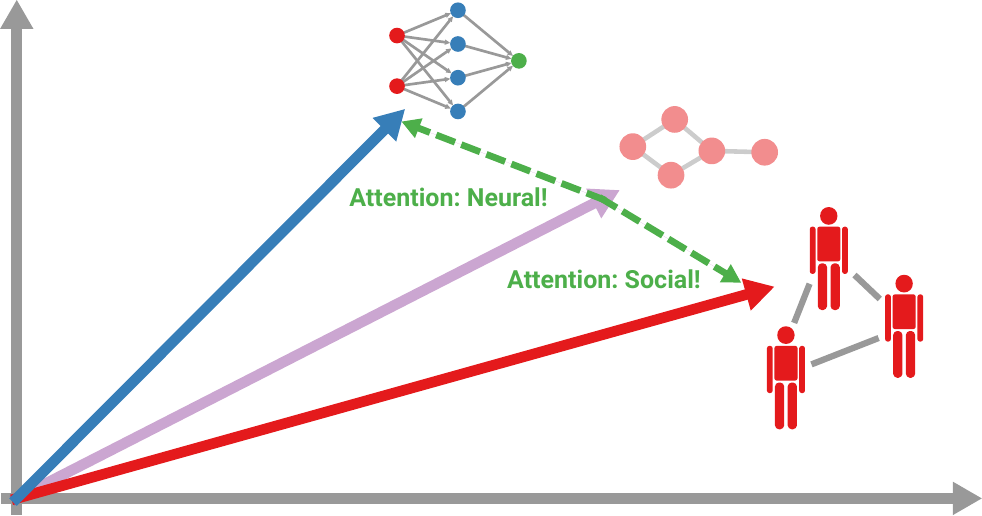}
\caption{The mechanism of attention. Vectors represent the embeddings of words, pointing to their location in a 2D embedding space. The dashed green arrow translates the original generic embedding in purple to more specialized concepts, depending on which word preceded ``network''.}
\label{fig:nn-attention}
\end{figure}

In practice, with attention, the presence of a word in the same sentence can change the embedding of the words that follow. That makes sense: if I write ``social'' before ``network'' you immediately know what I'm talking about, and you know it is different than the network I'd be talking about had I written ``neural'' instead. You can see this happening graphically in Figure \ref{fig:nn-attention}: attention is nothing more than taking a generic concept pointing to the general idea of ``network'' and multiplying a different vector depending on which words preceded ``network''. The result is to point in a slightly different direction, which is the more concrete embedding of which meaning of the word we are actually referring to.

Mathematically this means that the $H^{l}$ embeddings we use to calculate the $H^{l+1}$ ones are not really really the $H^{l}$ embeddings: they are $\alpha^l(H^l)$, i.e. the result of applying the attention function $\alpha$ to them at layer $l$.

The same principle that works in NLP works in networks too. When you receive messages from your neighbors, you might want to pay attention more to some over others. For instance, in a paper citation network, you might have hubs that are cited by everyone, or interdisciplinary papers whose citations are ambiguous -- because they can come from multiple fields, and so might confuse you when trying to predict fields with citations. In those cases, you want to pay less attention to them.

In Graph Attention Networks\cite{velivckovic2017graph}\cite{brody2021attentive} (GATs) the way to implement this is by looking at the $D^{-1}A$ transformation that we have in MPGNNs. This is sort of an attention mechanism: it is the trivial one where each neighbor get the same amount of attention. In this simplified example, each neighbor of node $v$ gets exactly $1/k_v$ attention, with $k_v$ being $v$'s degree. In this framework, $v$'s attention only depends on $v$'s characteristics, specifically its degree.

GATs unlock a new degree of freedom allowing $v$'s attention to depend also on the characteristics of the neighboring node $u$ as well\cite{tailor2021adaptive}. GCNs could sort of do the same, because the symmetric $\hat{D}^{-1/2}(I + A)\hat{D}^{-1/2}$ normalization also depends on the neighbor $u$'s degree, but this attention is always fixed to be $1 / \left(\sqrt{k_v}\sqrt{k_u}\right)$. Instead, GATs try to learn -- again with backpropagation -- a different attention value for each of $v$'s neighbors. Since we introduced this learnable function as $\alpha$, the GAT formula is:

$$ H^l = \sigma \left( H^{l-1} \alpha^l W^l \right). $$

\begin{figure}[t]
\centering
\begin{subfigure}{.2\columnwidth}
\includegraphics[width=\textwidth]{figures/mpgnn_naked.pdf}
\caption{}
\end{subfigure}\qquad\qquad
\begin{subfigure}{.2\columnwidth}
\includegraphics[width=\textwidth]{figures/mpgnn_gcn.pdf}
\caption{}
\end{subfigure}\qquad\qquad
\begin{subfigure}{.2\columnwidth}
\includegraphics[width=\textwidth]{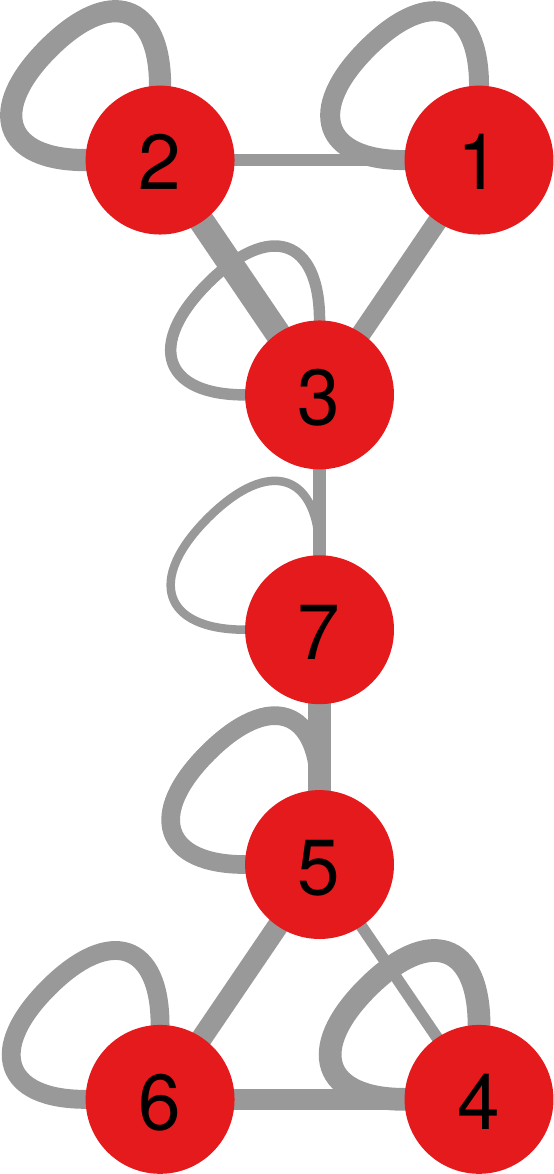}
\caption{}
\end{subfigure}
\caption{(a) The original network. (b) The network after the $\hat{D}^{-1/2}(I + A)\hat{D}^{-1/2}$ operation, leading to a GCN. The thickness of an edge is proportional to the weight it gets as a result of the transformation. (c) The network after an arbitrary transformation determined by a learned $\alpha$ function, leading to a GAT.}
\label{fig:gcn-vs-gat}
\end{figure}

There is quite some freedom in how to define $\alpha$ for GATs. The key thing is that $\alpha$ takes as input both $H^{l-1}$ and $W^l$, which is what allows it to be so flexible. So, in summary, GCNs are a special type of GATs: they are GATs that have a fixed non-learnable $\alpha$. Therefore, GATs are a generalization of GCNs. In Figure \ref{fig:gcn-vs-gat} I call back to Figure \ref{fig:mpgnn-vs-gcn}. GCNs allow us to go from Figure \ref{fig:gcn-vs-gat}(a) to Figure \ref{fig:gcn-vs-gat}(b), preventing some issues that would be caused by using only $A$. GATs allow us to go from Figure \ref{fig:gcn-vs-gat}(b) to Figure \ref{fig:gcn-vs-gat}(c), realizing we're not forced to always use the same transformation, but we can actually earn the best possible one to minimize our loss.

Using attention, you realize why you might want to have multiple layers. Once you change the embedding of the word ``network'' because you saw the word ``social'' before, then every word following ``network'' might want to pay more (or less) attention to it.

Further refinements of this model are able to tackle heterogeneous networks with different node and edge types\cite{wang2019heterogeneous}, deal with data embedded in space and time\cite{zhang2019spatial}, with signed networks\cite{huang2019signed} and more.

\section{Transformers}
Ever since a seminal paper came out\cite{vaswani2017attention}, transformers became all the rage in neural networks, even though I've been playing with them since the 80s\footnote{I'm now told that puppet robots don't count in machine learning, although I'm pretty sure Megatron is too evil not to be the product of some sort of rogue AI.}. In fact, the T in the popular Chat-GPT algorithm actually stands for Transformer. So, of course the graph neural network literature took a good look at them.

Conceptually, the important part of a transformer is not all that difficult. It is merely a parallel attention mechanism. Instead of having a single $\alpha^l$ function per layer, you have many: $\alpha^l_1$, $\alpha^l_2$, ... , $\alpha^l_k$. These are all trained independently, so they are free to learn different things. Then, the representation you get can be any combination of these newly learned representations -- more often than not it is a linear combination and/or you stitch them together by concatenating the various matrices. If you want to learn the lingo, each of the $\alpha^l_k$ independent attention is called an attention ``head''.

\begin{figure}[t]
\centering
\includegraphics[width=\columnwidth]{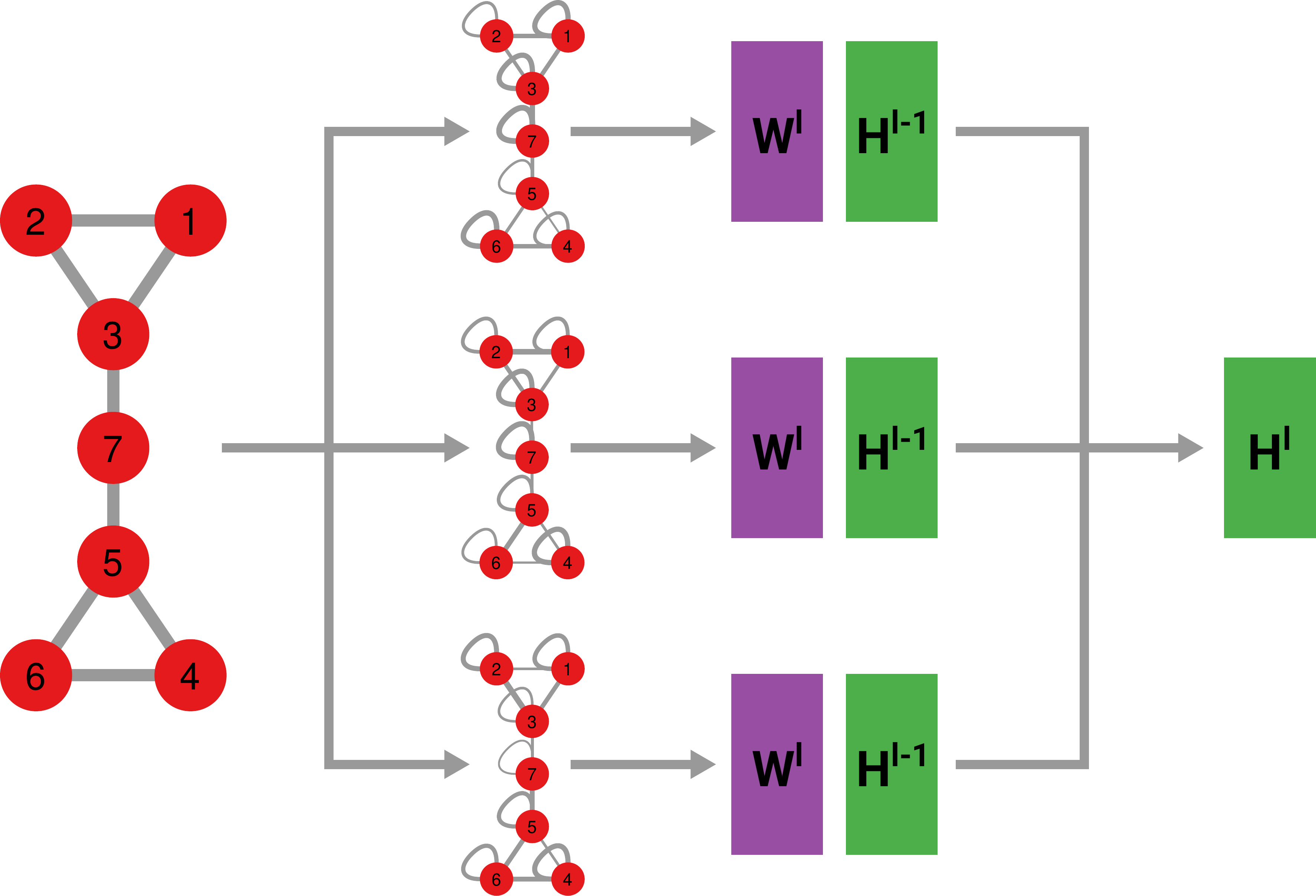}
\caption{What happens in the layer $l$ of a graph transformer.}
\label{fig:gnn-transformer}
\end{figure}

One can apply this strategy exactly as written to GATs, leading to graph transformers. In fact, the paper proposing GATs for the first time -- which I described in the previous section -- already defines graph transformers. Figure \ref{fig:gnn-transformer} shows you what happens in each layer $l$ of a graph transformer. The various attention heads are trained independently, they each operate on the $H^{l-1}$ and $W^l$ matrices, then the result is combined to produce $H^l$.

This can be seen as a further generalization away from GCNs. In GATs we decided we wanted to learn the attention weights with a function, rather than having fixed ones like in GCNs. In graph transformers we ask: why limiting ourselves to learning a single attention function? And so we learn an arbitrary number of them.

Of course there are different ways to make more sophisticated graph transformers\cite{yun2019graph}\cite{min2022transformer}. And, following Section \ref{sec:deep-mpgnn-spectral}, one can ditch the spatial approach to GATs and graph transformers to embrace the spectral one\cite{kreuzer2021rethinking}.

\section{Deep Generative Graph Models}\label{sec:mining-deep2-generative}
One cool thing you can do with graph neural networks is to learn how to generate a graph that looks like a real one. This is arguably something we already discussed in Part \ref{par:synthnet}, so this section could fit in there -- and I gave you a teaser of some deep generative approaches in Section \ref{sec:csmodels-ggn}. However there are a couple of reasons why this section is here. First, it relies on advanced graph neural network concepts that we could not introduce before Part \ref{par:synthnet}. More importantly, the mechanism here is rather different. Rather than fixing a handful of key parameter and let a fixed algorithm -- such as preferential attachment -- to do the rest, here we actually learn from many graphs how the algorithm should wire the network. Exponential random graphs and the configuration model get closer to this approach, but they too only learn from a single graph.

\subsection{Variational Autoencoders}
To understand variational autoencoders\cite{kingma2013auto} we need to start by breaking down their name into its component parts: we start by understanding what an ``encoder'' is, what it means for it to be ``auto'', and what ``variational'' means.

\begin{figure}
\centering
\includegraphics[width=.8\columnwidth]{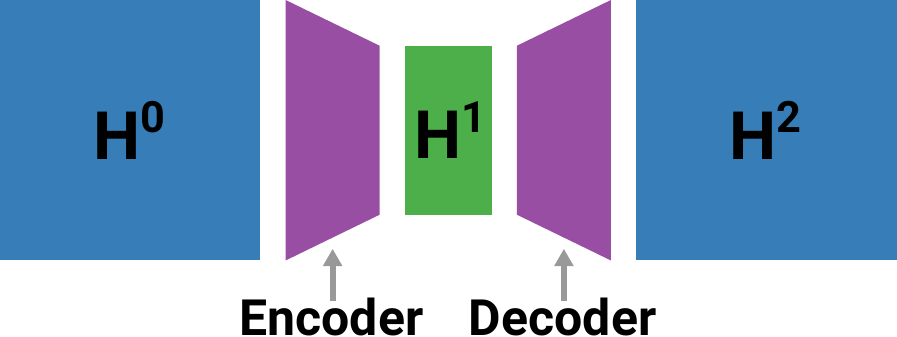}
\caption{The structure of an autoencoder: (left to right) the input data (blue) is encoded (purple) in an embedding (green) then decoded (purple) to a second layer embedding (blue) which is as similar as possible to the original data.}
\label{fig:autoencoder}
\end{figure}

The \textit{encoder} part is the easiest. An encoder is a function that takes and input and produces a code -- a more succinct representation of the same data that went in. You've already seen a bunch of encoders up until now. Every time you produce a node embedding you're encoding it into a different representation. An \textit{auto}encoder aims to reconstruct the original data by passing it through an encoder and then recovering it. This means we need a decoder that reconstructs the original data. Figure \ref{fig:autoencoder} shows you the general structure of an autoencoder.

Both the encoder and the decoder can be learned, so that the loss in reconstructing the original data is minimal. You might think it's pointless to try and reconstruct the original data -- after all you have it, so why bother regenerating it? -- but the value is that now your encoder and decoder have learned the key features of the data. At this point, you don't need the original data any more and you can generate as much data as you want.

\begin{figure}[t]
\centering
\includegraphics[width=\columnwidth]{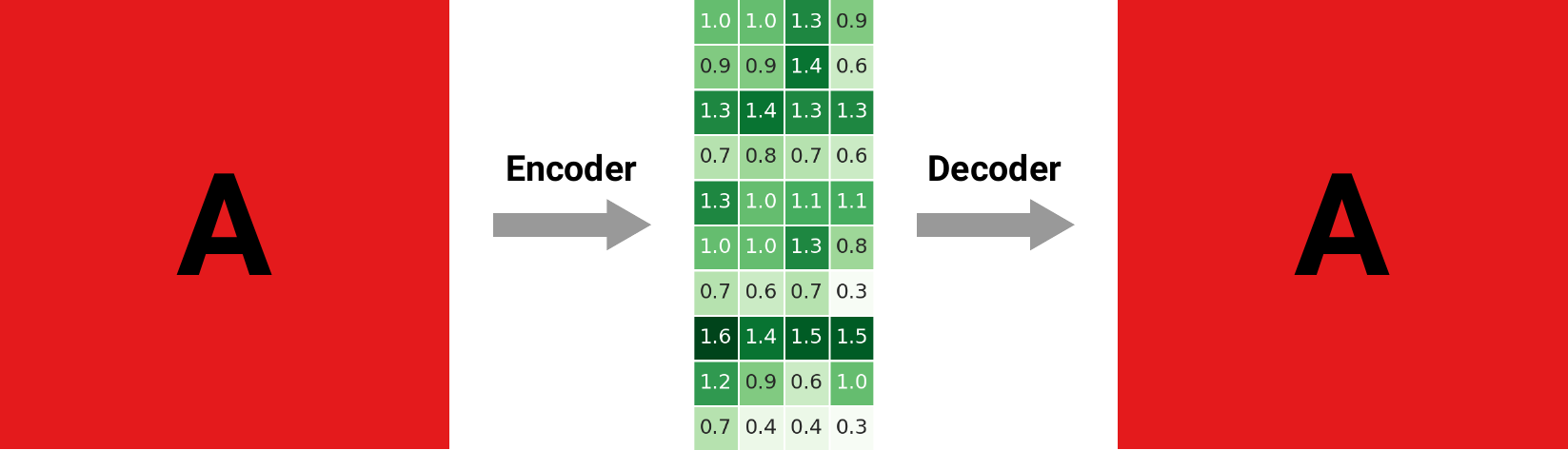}
\caption{An autoencoder working on an adjacency matrix (in red). The encoder learns the $H^1$ embeddings (in green) and the decoder reconstructs the adjacency matrix.}
\label{fig:vae-novar}
\end{figure}

Why would we need a \textit{variational} part? That's what introduces randomness. If you always reconstruct the input features $H^0$, then you're going to always get the same embeddings from the autoencoder, as you can see in Figure \ref{fig:vae-novar} -- where I unpack the values inside $H^1$. In practice, the objective of an autoencoder is to learn directly the embeddings and use them to reconstruct the original adjacency matrix $A$.

However, what you can do instead is to learn not the embeddings themselves, but the \textit{distribution} of their values. For instance, you can assume the embeddings distribute normally, and then you use two neural networks to learn the mean and standard deviations of those distributions. This means that you treat the embeddings as random variables, which can be more succinctly described by the parameters of their distribution. Once you have learned such parameters -- in our case the mean and standard deviation -- you can generate random embeddings with those parameters. You're going to get a new random embedding for each node which looks like the original one -- because it has the same average and standard deviation -- but it's different enough that can lead you to reconstruct a different $A$.

\begin{figure*}[t]
\centering
\includegraphics[width=\columnwidth]{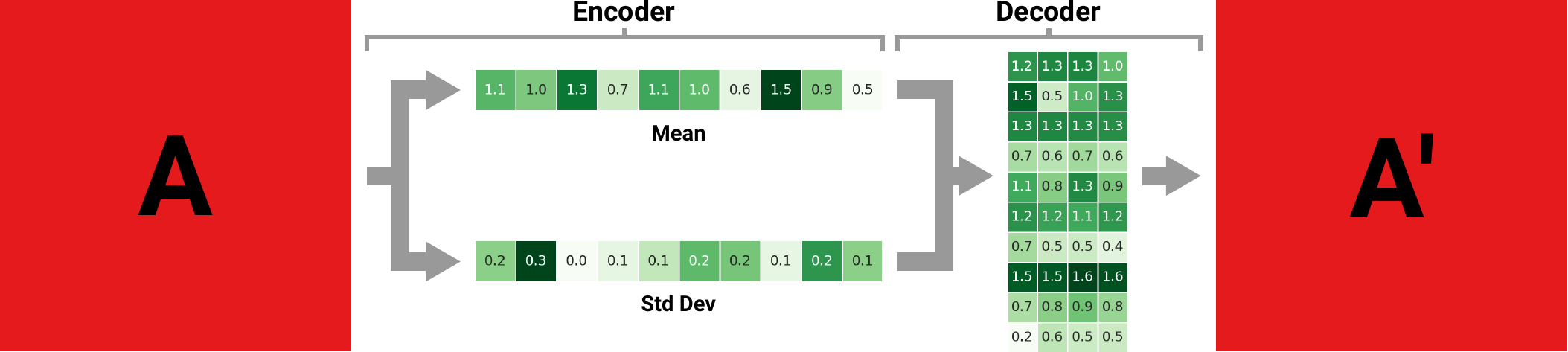}
\caption{A variational autoencoder working on an adjacency matrix (in red). The encoder learns the mean and standard deviations behind the true $H^1$ embeddings (in green) and the decoder first samples a random embedding with those means and standard deviations and the uses it to reconstruct the adjacency matrix.}
\label{fig:vae-var}
\end{figure*}

You can see in Figure \ref{fig:vae-var} how the variational part makes the encoder and decoder more complex. The encoder cannot simply reconstruct the embeddings, but needs to learn their distributions. The decoder has the task of sampling from these distributions and then use the sampled embeddings to reconstruct the data.

When it comes specifically to graphs, the main approaches are whether you want to learn the parameters of the distributions for each node separately\cite{kipf2016variational}\cite{grover2019graphite} -- and so reconstruct node embeddings from their own means and standard deviations -- or if you want to pool everything together and learn directly at the level of the entire graph\cite{simonovsky2018graphvae} -- reconstructing the entire adjacency matrix all at once.

Depending on how you build the embeddings, the autoencoder can learn different things. For example, you can try to use the autoencoder to reconstruct the information present in random walks\cite{cao2016deep}, and if you go high-order you can learn the first and second order relationships between nodes at the same time\cite{wang2016structural}.

\subsection{Generative Adversarial Networks}
The idea behind Generative Adversarial Networks (GANs) is relatively simple\cite{goodfellow2014generative}\cite{creswell2018generative}: we employ a two step strategy. First, we create a generator -- of any type, with any algorithm -- that can generate some realistic-looking data. Then we pass it to a classifier, whose task is to learn how to distinguish real data from randomly generated one. The two are the adversaries: the generator is trying to fool the classifier, and the classifier is trying to spot the generator's outputs.

Graph versions of GANs are what you expect them to be. Both the generator and the classifier need to be some sort of neural network architecture that can generate adjacency matrices and use them as input for the classification\cite{de2018molgan}\cite{bojchevski2018netgan}. 

\subsection{Autoregressive Models}
Autoregressive models can be used in combination with VAEs and GANs to augment them -- because they can be used as decoder for a VAE or as the generator for a GAN. But they can also be their own thing, just generating the networks without a subsequent step in the VAE or GAN architecture.

Their main objective is to relax a key assumption VAEs and GANs make: that edges are generated independently. We know that's not true: high clustering coefficients mean that the friend of my friend is more likely to be my friend. This means that edge generation is not independent. Autoregressive models solve the problem by reducing the generation of the adjacency matrix as the generation of a linear sequence of zeroes and ones. The next entry on the sequence is dependent on the sequence we have generated so far -- so here's the dependency.

I already explained how one of these methods work -- without using too much deep learning lingo -- in Section \ref{sec:csmodels-ggn}, where I showed you GraphRNN\cite{you2018graphrnn}. In this case, we're still operating on an edge-by-edge basis -- even if we use all the previously generated edges to make a new one. An alternative approach looks at general descriptive statistics of a graph -- such as degree distribution, the spectrum of the Laplacian, and more -- and use those as points of comparison to generate the new graph\cite{liao2019efficient}.

The advantage of this approach is that it can be generic -- working on properties that any graph has -- but it can also be specific. If you know what these graphs represent and what peculiar statistics they might have -- for instance, some physical and chemical properties that must be respected when generating graphs representing molecules\cite{jin2018junction} --, you can integrate that information in the generation process.

\section{Alternative Message-Passing Techniques}
\begin{figure}
\centering
\begin{subfigure}{.33\columnwidth}
\includegraphics[width=\textwidth]{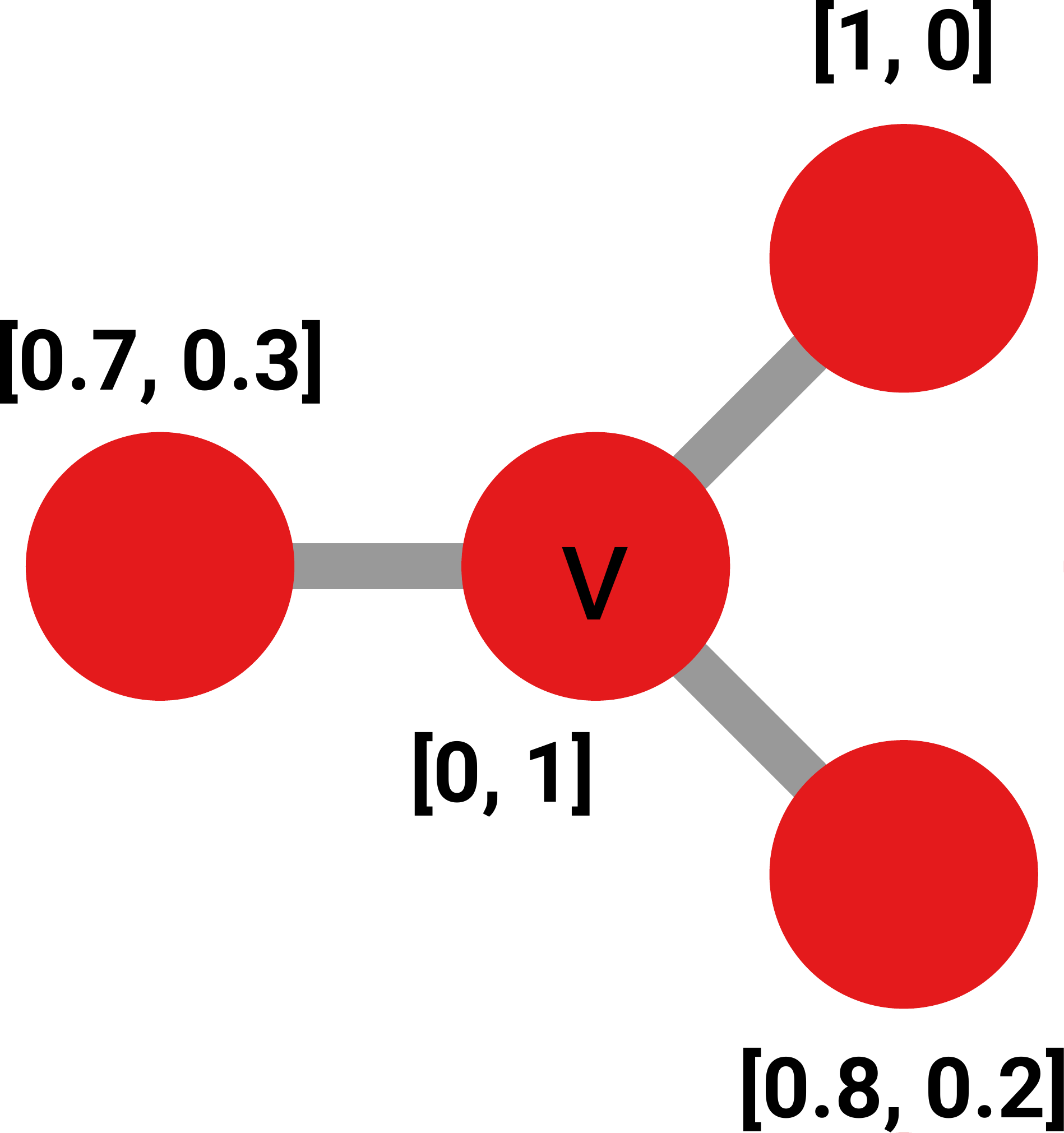}
\caption{}
\end{subfigure}\qquad\qquad
\begin{subfigure}{.33\columnwidth}
\includegraphics[width=\textwidth]{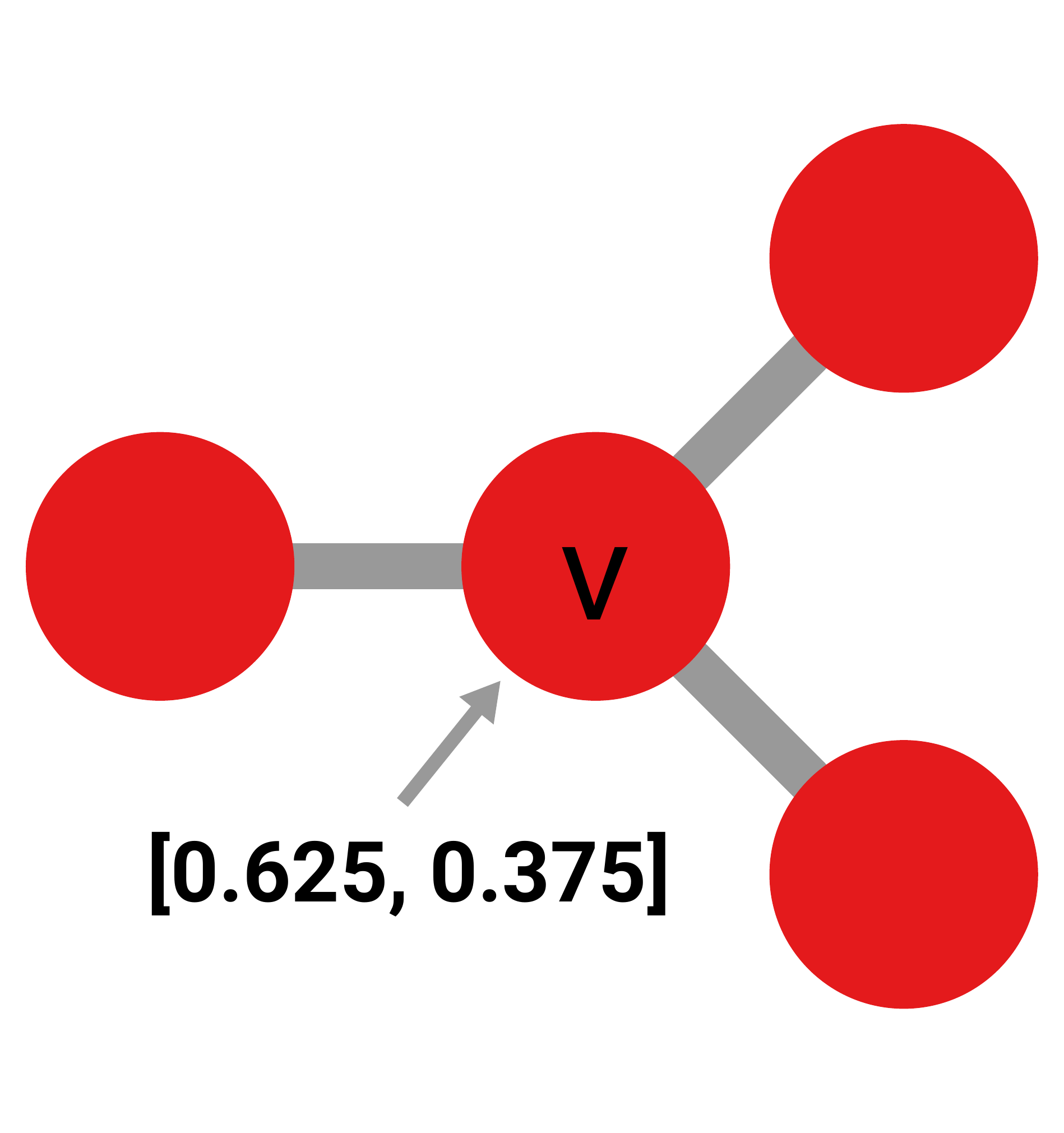}
\caption{}
\end{subfigure}
\caption{(a) A network with 2D node embeddings next to their corresponding nodes. (b) The resulting embedding for node $v$, using average as the aggregation function.}
\label{fig:mpgnn-aggregation-alternatives-1}
\end{figure}

So far I've only discussed the same type of message-passing technique. This is the case where the messages from all nodes are aggregated in discrete steps and summed up in a single-valued summary. For instance, consider Figure \ref{fig:mpgnn-aggregation-alternatives-1}. Here node $v$ aggregates the messages from the neighbors by averaging them with its own embedding. There are a couple of ways to challenge this approach.

\begin{figure}[b]
\centering
\begin{subfigure}{.33\columnwidth}
\includegraphics[width=\textwidth]{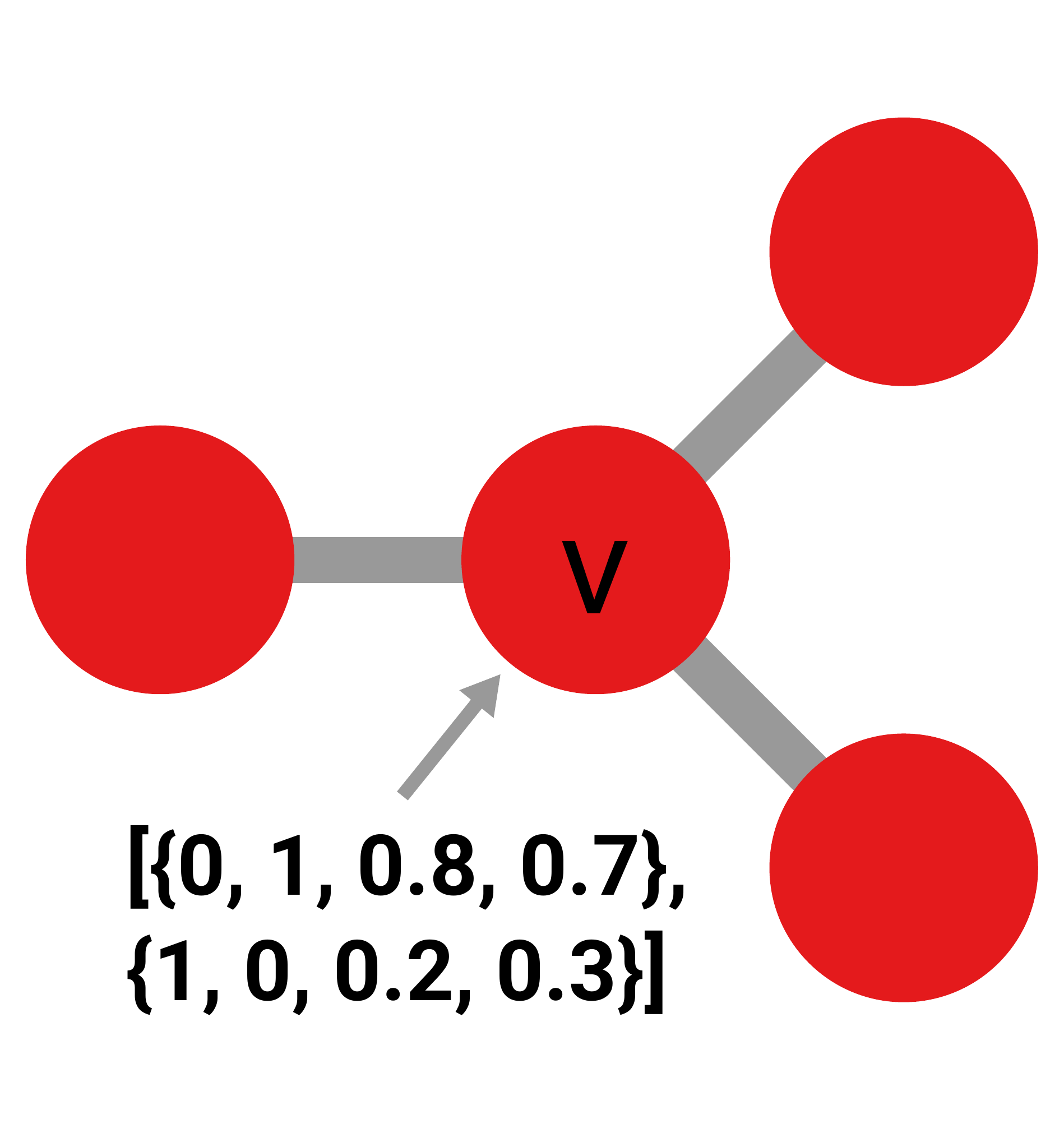}
\caption{}
\end{subfigure}\qquad\qquad
\begin{subfigure}{.33\columnwidth}
\includegraphics[width=\textwidth]{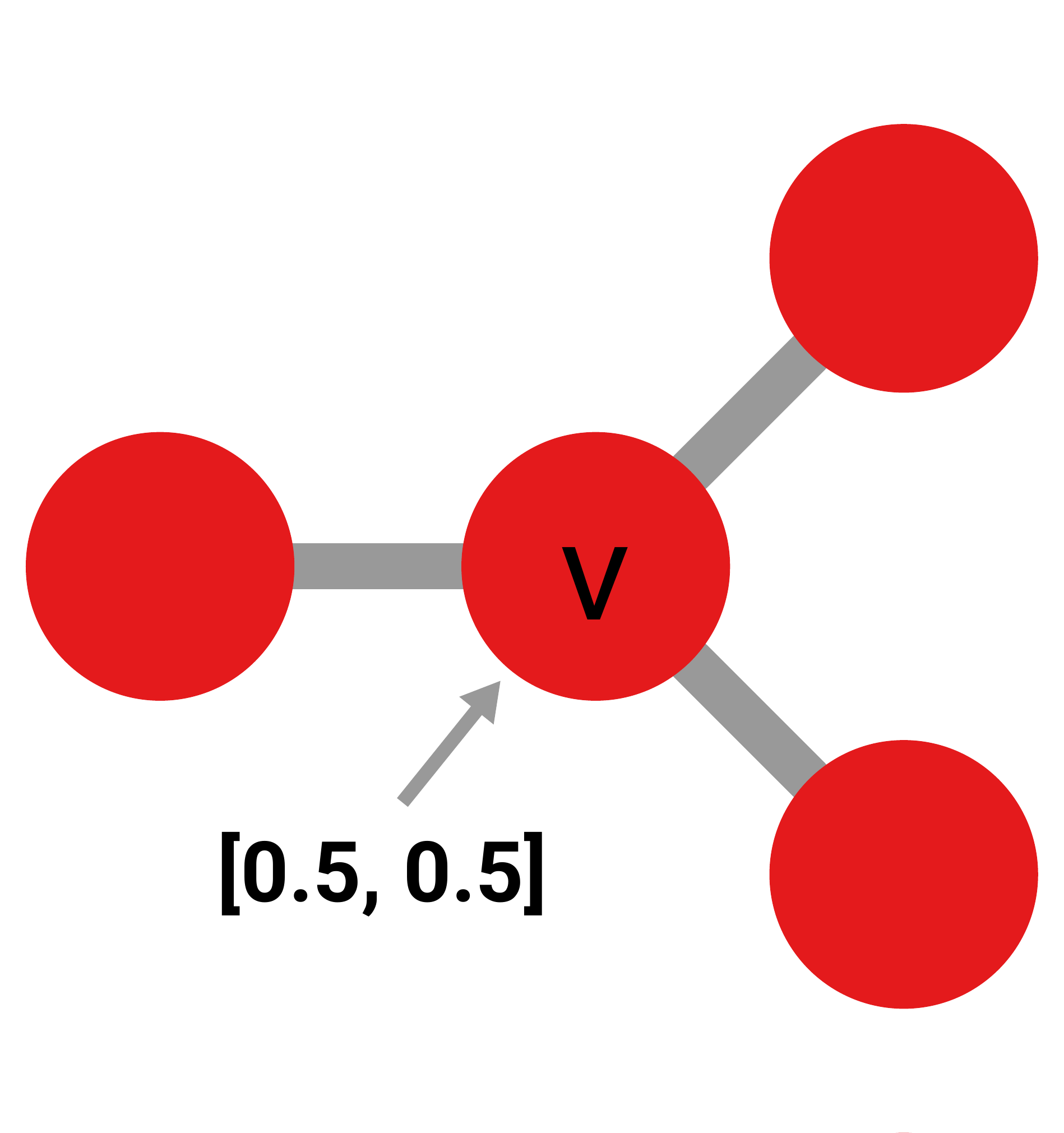}
\caption{}
\end{subfigure}
\caption{(a) The resulting embedding for node $v$ from Figure \ref{fig:mpgnn-aggregation-alternatives-1}(a), using a set aggregation function. (b) Same as (a), but by using a continuous aggregation function.}
\label{fig:mpgnn-aggregation-alternatives-2}
\end{figure}

The first is by using set aggregators. In practice, this replaces the average function we just used. You take the three embeddings as a set and you use a function that, given a set, produces an output. One example of such function is the hash we used for the Weisfeiler-Lehman isomorphism test. To give you an idea of how it might work, Figure \ref{fig:mpgnn-aggregation-alternatives-2}(a) shows you a stupid set aggregator. Here we simply say that the result of the message-passing for node $v$ is a list of two unordered sets. Unordered means that, in this case, $\{0, 1, 0.8, 0.7\}$ would be identical to $\{0.7, 1, 0, 0.8\}$ -- and any possible permutation. We need this property because the message-passing still needs to be a permutation-invariant operation -- it doesn't matter the order in which we look at the neighbors.

There are, of course, smarter way to define set aggregators\cite[-0.7in]{zaheer2017deep}\cite{qi2017pointnet}. And, with Janossy pooling, you can actually use a function that is not permutation invariant, and then average it over all permutations\cite{murphy2018janossy}. Normally, there will be too many permutations to be able to do this exhaustively. So you can randomly sample some permutations, or deploy a canonical ordering of nodes -- so there is only one order and this operation becomes permutation invariant.

The second alternative is to use normal aggregation functions, but refusing to run them in a discrete way. You can simulate a continuous flow of information and then discretize it at an arbitrary moment. This is a technique that we call continuous message passing\cite{chamberlain2021grand}\cite{chamberlain2021beltrami}\cite{bodnar2022neural}\cite{frasca2020sign} and has relations with the more generic form of geometric deep learning -- of which graph neural networks are a special, discrete, case.

To give an oversimplified example, in Figure \ref{fig:mpgnn-aggregation-alternatives-2}(b) I only pass half of the messages from the neighbors, which results in node $v$ preserving more of its own original embedding.

\section{Practical Considerations}\label{sec:mining-deep2-practical}

\subsection{Computational Efficiency}
In practically all cases, you don't want to implement deep graph neural network architectures by naively applying the operations I explained to you so far. Consider the message-passing model: a node can only send the same message to the MPGNN architecture. If you have a hub with thousands of connections, you're going to repeat the same operation thousands of times, to achieve the same result. In a GCN, since you're doing a single matrix multiplication, you don't have this problem, but you have another one: if your graph is huge, so are your matrices, and they might not fit in memory.

\begin{figure}
\centering
\begin{subfigure}{.45\columnwidth}
\includegraphics[width=\textwidth]{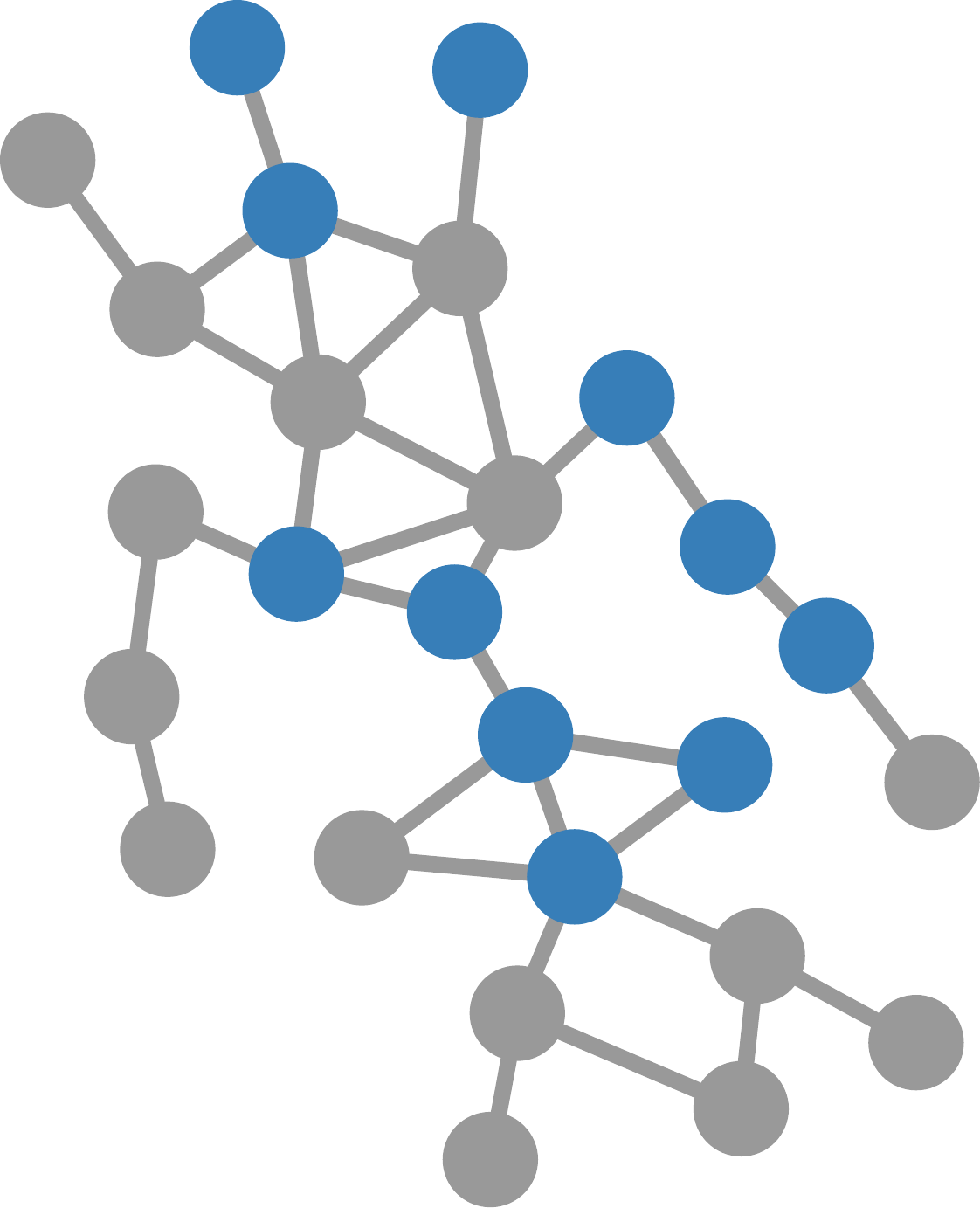}
\caption{}
\end{subfigure}
\begin{subfigure}{.45\columnwidth}
\includegraphics[width=\textwidth]{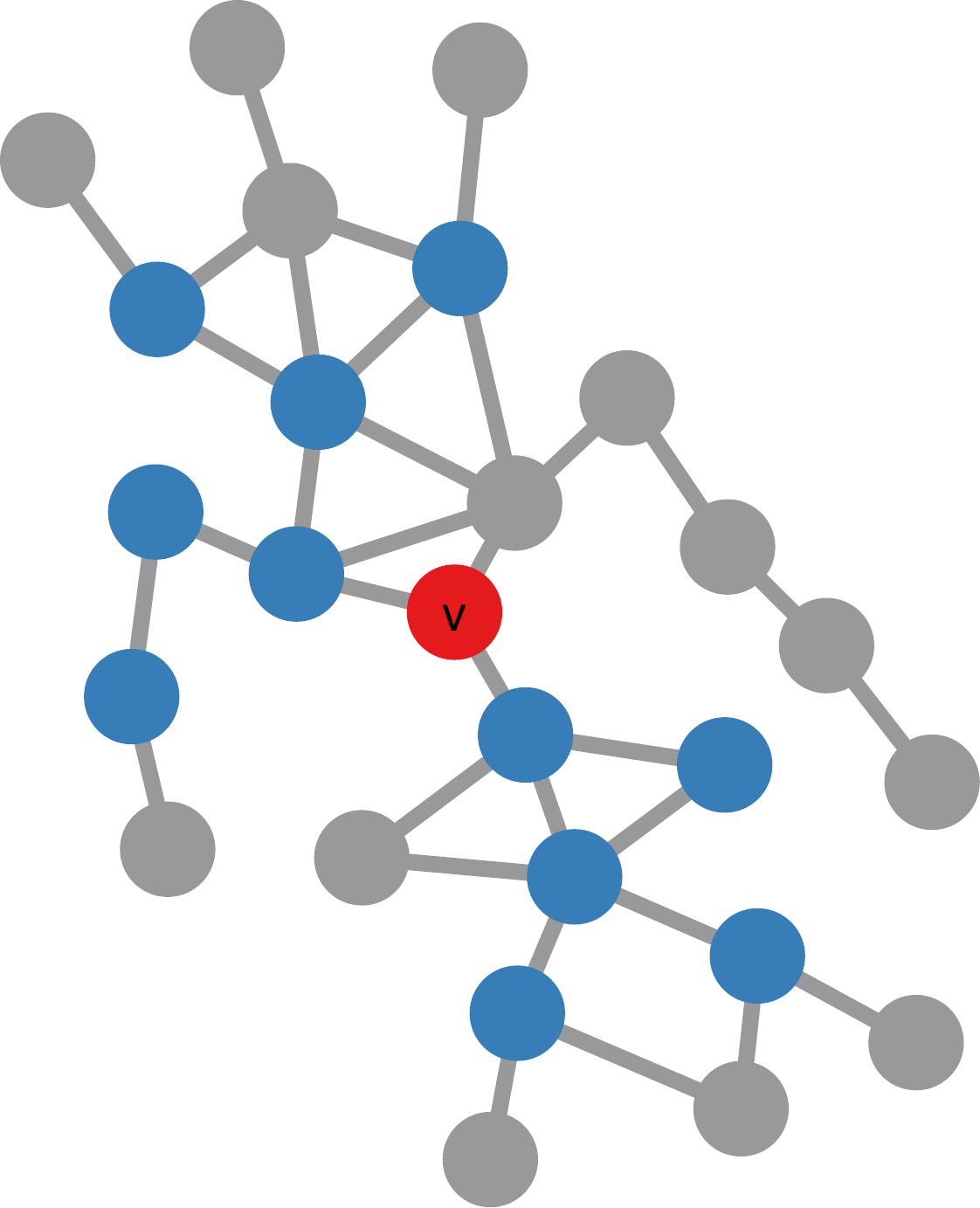}
\caption{}
\end{subfigure}
\caption{Two network minibatching approaches. The gray nodes are not sampled and the blue nodes are sampled. (a) Random sampling, (b) Minibatching around the red focus node.}
\label{fig:minibatching-sage}
\end{figure}

One thing you can do is to apply the sampling and mini-batching strategies\cite{ying2018graph}\cite{hamilton2017inductive}\cite{zeng2019graphsaint}\cite{chiang2019cluster} (Section \ref{sec:ml-sampling}). In a graph you need specific ways of sampling. You know the intricacies of network sampling from Chapter \ref{cha:sampling}. As a refresher, Figure \ref{fig:minibatching-sage}(a) shows that sampling nodes randomly will most likely lead to disconnected samples where any message-passing strategy won't have anything to work with.

In the figure we were even lucky that the randomness lead to somewhat connected patches, but each patch cannot communicate with the others and we have wasted the time to sample and to compute the messages to essentially receive no information whatsoever. One minibatching strategy -- which I show in Figure \ref{fig:minibatching-sage}(b) -- will get the messages for node $v$ from a given number of nodes from $v$'s neighborhood, and then recursively sample the neighbors of the neighbors and so on, until we reached the desired depth. In the figure, we want to receive messages from up to two neighbors, up to a distance of three steps.

\subsection{Regularization}
As I mentioned in Section \ref{sec:ml-general-infrastructure}, one of the greatest woes of machine learning is overfitting. Since you're learning from a lot of data bottom-up without imposing a theory top-down, you might end up just learning the special quirks of the data you use for training. In neural networks there are a few techniques to avoid overfitting, which we normally call ``regularization''. A classical strategy is called ``dropout'': you take your matrix $W$ and you randomly set to zero some entries or entire rows\cite{wan2013regularization}\cite{srivastava2014dropout}.

When working with a graph neural network, you can do a special variation of this: edge dropout\cite{schlichtkrull2018modeling}. Since you also have $A$ besides $W$, you can do dropout in $A$ as well. This means to randomly remove a few edges between layers, which will make it harder for your framework to overfit to the original $A$.

\subsection{Augmentation}
\begin{figure}[t]
\centering
\includegraphics[width=.2\columnwidth]{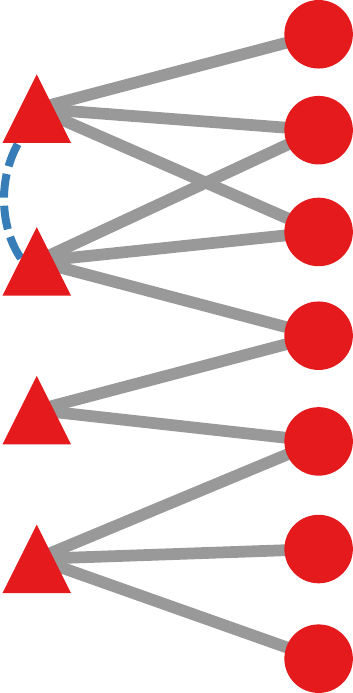}
\caption{A bipartite network, augmented with the virtual edge in the blue dashed line.}
\label{fig:gnn-augmentation}
\end{figure}

Augmentation is something you'd do if your input graph is too sparse, which means information might have a hard time propagating through the layers. Curiously, in this case you'd do the exact opposite of the regularization strategy we just discussed: you'd try to add virtual edges -- or virtual nodes\cite{gasteiger2018predict}. It might be difficult sometimes to add proper virtual edges to actually improve the situation rather than adding noise, but in some cases things can be easier. For instance, if you deal with a bipartite network, it's kind of natural to add virtual edges between nodes of the same type, if they have a lot of common neighbors -- as Figure \ref{fig:gnn-augmentation}, where the nodes sharing two neighbors are connected with a virtual edge.

\subsection{Applications}
When deeply embedded in this book part, it is easy to lose oneself in the mountain of ever-evolving details that make up one of the most prolific publication grounds of the last half decade or so. One should not lose sight that GNNs are always means to an end. I already mentioned a few applications of shallow embedding learning in Section \ref{sec:mining-embeddings-app}, so I think it is high time we refresh our minds with a few successful applications of GNNs out there.

Popular application of deep learning include in general those tasks that are difficult to approach from a purely theoretical standpoint. They are mostly scenarios when you'd normally make judgment calls like ``I'll know it when I see it''. The GNN can try to do it for you. For instance we have:

\begin{itemize}
\item Recommender systems: mostly this is link prediction in a bipartite network of users consuming some sort of product (movie, book, song, etc). This is normally done by learning which user embeddings are more likely to connect with which product embedding.
\item Misinformation detection on social media: hoping to spot differences in the embeddings of real and fake news, or accounts that spread misinformation from accounts that do not\cite{monti2019fake}\cite{han2020graph}.
\item Network medicine: identifying organisms that could serve as new antibiotics by figuring out how they would interact with given pathogens\cite{stokes2020deep}.
\end{itemize}

And more.

\section{Summary}

\begin{enumerate}
\item Graph Attention Networks (GATs) allow you to weight the contribution of each of node $v$'s neighbors with a learnable function, which can help you further minimizing your loss function. This makes them a more general version of GCNs, which have fixed attention weights.
\item Graph transformers are GATs with multiple independently initialized and learned attention functions. The results of these functions are then combined to produce the layer's output.
\item Graph variational autoencoders can help you generate random graphs. They first learn the parameters behind the distributions of the embeddings and then use these parameter to generate realistic random embeddings, from which they can reconstruct a new random adjacency matrix.
\item In Generative Adversarial Networks we have a generator and a detector. The generator tries to make synthetic data as similar to the real data as possible. The detector tries to distinguish synthetic data from real one. This can lead to better and better generators.
\item Autoregressive models allow you to represent and learn dependencies between edges when generating a graph, by treating it as a sequence of zeros and ones whose next entries depend on the previously generated ones.
\item You can have more sophisticated message-passing techniques that operate on sets rather than numbers, and run the message-passing process in a continuous way rather than in discrete steps.
\item Minibatching by sampling the nodes' neighborhood can help with computational efficiency, removing random edges can help avoiding overfitting, and adding virtual edges can help when your network is too sparse for effective message-passing.
\end{enumerate}

\section{Exercises}

\begin{enumerate}
\item Implement a simple attention mechanism by replacing the $\hat{D}^{-1/2}(I + A)\hat{D}^{-1/2}$ term from the function you developed in the exercises of the previous chapter. The new $\alpha$ term comes from the edge weights in the third column of the network at \url{http://www.networkatlas.eu/exercises/45/1/network.txt}. The features are at \url{http://www.networkatlas.eu/exercises/45/1/features.txt}. Then run it on that network. (For the purposes of this and the following exercises, you can use a completely random $W$)
\item Implement a simple transformer by repeating the attention operation from the previous exercise with the alternative weights in the third, fourth, and fifth column of \url{http://www.networkatlas.eu/exercises/45/1/network.txt}. Combine them with a final layer averaging all the attention heads.
\item Implement a simple graph variational autoencoder. Learn the embeddings with the transformer you just implemented on the network from the previous exercise. Then calculate each node embedding's mean and standard deviation. Then generate ten random embeddings with the same average and standard deviations.
\item Implement a basic continuous message-passing. Each node averages its own embedding with the average of its neighbors times a factor $k$. Try $k = 0.5$, $k = 1$, and $k = 1.33$ with the network from the previous exercises.
\end{enumerate}

\part{Holistic Network Analysis}

\chapter{Graph Summarization}\label{cha:mining-summarization}
Graph summarization is a data mining class of algorithms that take an input graph and reduce its size -- summarizing it -- returning a smaller graph as an output. The output graph should respect the salient characteristics of the input graph, so that analyses performed on the output return results that can be used to reconstruct what the whole input graph would return\cite{liu2018graph}.

There are many reasons why one would want to perform graph summarization. The main ones are four: 

\begin{itemize}
\item Algorithmic speedup: this is the classic motivation\cite{feder1995clique}. If your original graph has, like Facebook, $\sim 10^9$ nodes, even the most elementary algorithms will take a long time to run. This might be a problem if, for instance, you want to perform online analysis and return results in real time. If you manage to reduce the size of your network to $\sim 10^6$ nodes, this would buy you a lot of time and reactivity.
\item Storage facilitation: hard disks might be cheap nowadays, but they ain't free. Again using the Facebook example, it might not be a problem storing $\sim 10^9$ nodes, but if all those nodes perform several activities per day, you might start to get into trouble. Moreover, you will have to hit some physical limits: even if you can create a system with several petabytes of storage capability, if you want to \textit{use} those petabytes of data you have to move them around, and all of a sudden the speed of light seems so slow.
\item Noise reduction: noise creeps up inside your data at every twist and turn. Storing and using your full network as it is measured might not be a good idea. Graph summarization can help you to smooth out the noise using information theoretic techniques, reconstructing the underlying signal.
\item Visualization: as we will see in Chapter \ref{cha:layouts}, plotting a network is hard and takes a lot of time. No one will ever visualize directly a $\sim 10^9$ node network. If you summarize it so that all its visual features are respected, you might then be able to have something meaningful to show.
\end{itemize}

Before going on the specific methods, I need to clarify what this chapter is about and what it isn't about. This chapter focuses on four main approaches to summarize graphs:
 
\begin{itemize}
\item Aggregation (Section \ref{sec:mining-summarization-aggr}): collapsing nodes -- and the edges connecting them -- into super nodes;
\item Compression (Section \ref{sec:mining-summarization-compr}): finding a set of rules enabling you to encode your network by using fewer bits than its adjacency matrix;
\item Simplification (Section \ref{sec:mining-summarization-simpl}): tossing ``unimportant'' nodes and edges;
\item Influence-based (Section \ref{sec:mining-summarization-infl}): finding the smallest possible graph which is still able to describe propagation events in the same way as the original one.
\end{itemize}

They all have one thing in common: they attempt to reduce a graph by lowering the number of nodes and edges such that the output is another -- smaller -- graph which represents the entire structure, only simplified.

In this sense, graph summarization is \textit{not} network sampling, and I will also not use space in this chapter for other similar methods such as low-rank approximations.

Graph summarization is fundamentally distinct from network sampling (Chapter \ref{cha:sampling}) even if both branches start from the same point: networks are too large and we cannot take them all in at once. However, in network sampling, you explore a \textit{part} of the network and you operate on the observed structure \textit{directly}. Neither is true in graph summarization. In summarization you want to analyze and understand the \textit{whole} structure, and you do so \textit{indirectly} by manipulating it. There are methods that attempt summarization-by-sampling\cite{najork2009less}\cite{ko2020incremental}, but I'm not going to cover them.

Conversely, low-rank approximations seek to reduce the data size with low reconstruction error\cite{sun2007less}. This is practically a Principal Component Analysis technique that is specialized to work on the adjacency matrices of a large sparse graph, rather than on generic attribute tables. Thus one could consider this more akin to matrix factorization techniques, for which I invite you to refer to Section \ref{sec:mat-factors}.

Finally, a word about how to evaluate your graph summary. Many methods come with their own quality measure that they are trying to optimize. Thus you could use one of these measures to decide whether you have obtained a good summary or not. For instance, in the compression class we're trying to minimize the number of bits needed to describe the graph.

However, Part \ref{par:cd} of this book -- on community discovery -- should have drilled something in your head: the real quality criterion you want to run is dependent on what you want to do with your graph summary. Thus the ideal test should be something as closely related to your final analysis task as possible. For instance, you might want to preserve some properties of interest -- e.g. the clustering coefficient. The more you depart from the original value, the more poorly your method is performing.

\section{Aggregation}\label{sec:mining-summarization-aggr}
In the aggregation approach, the idea is to take the original structure and start aggregating nodes -- or edges -- into superstructures that stand in for the observed ones. This process is normally guided by some function that determines the quality loss of each aggregation operation. The function must be designed with some application in mind (e.g. community discovery, link prediction, or others).

\begin{figure}
\centering
\begin{subfigure}{.3\columnwidth}
\includegraphics[width=\textwidth]{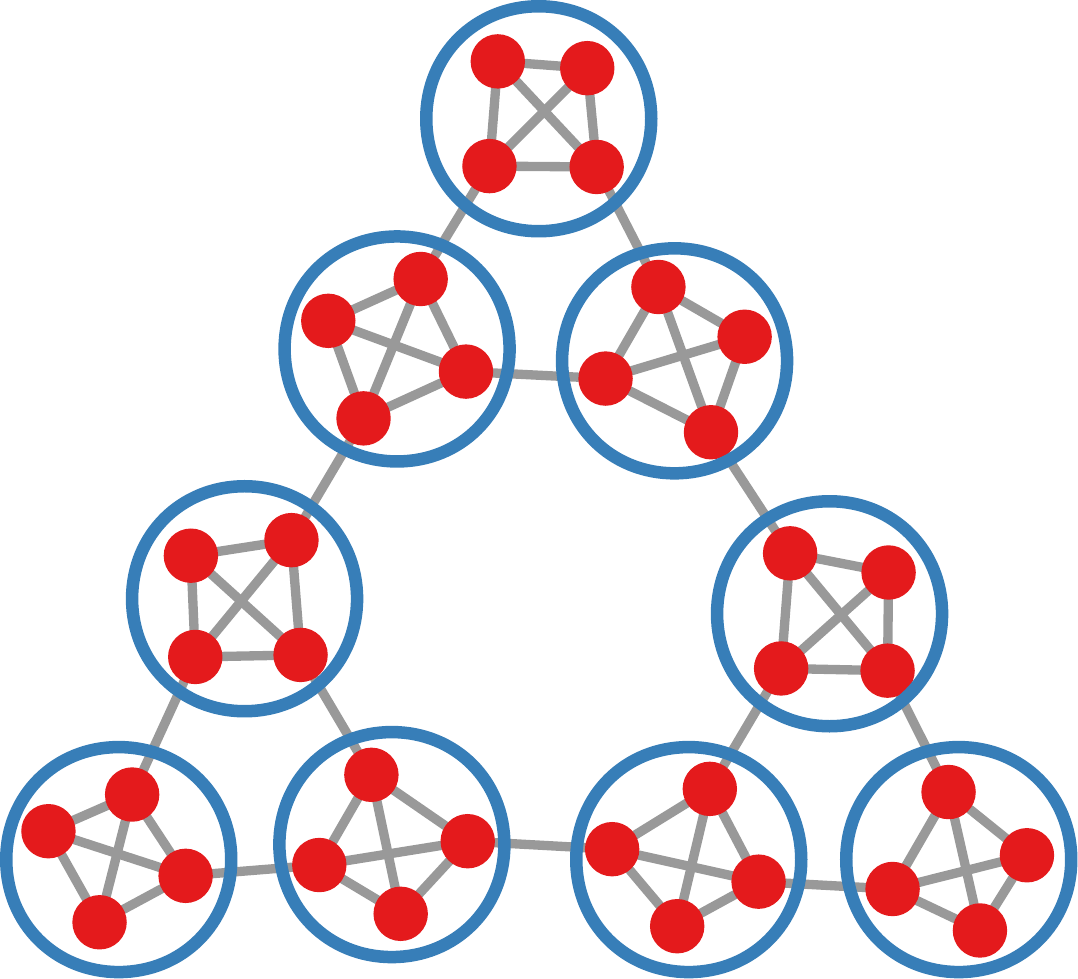}
\caption{}
\end{subfigure}\quad
\begin{subfigure}{.3\columnwidth}
\includegraphics[width=\textwidth]{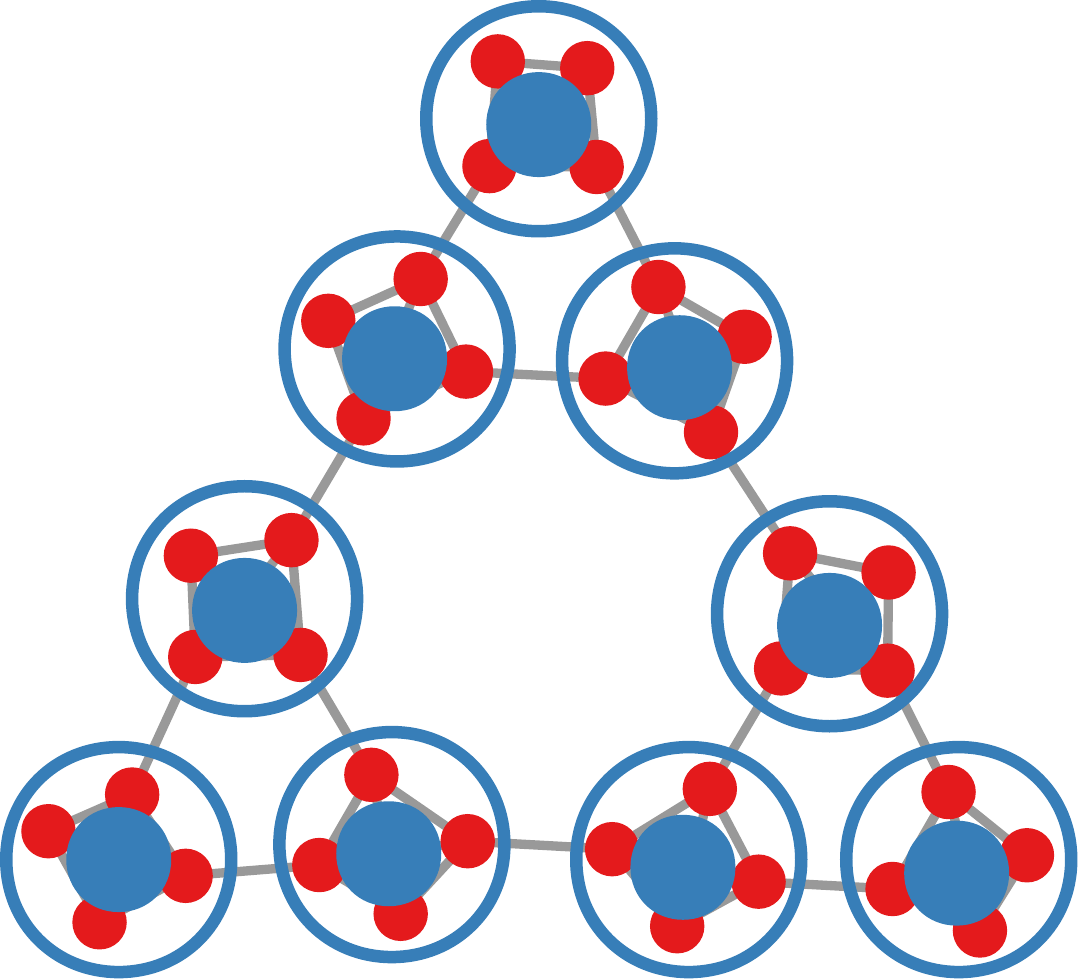}
\caption{}
\end{subfigure}\quad
\begin{subfigure}{.3\columnwidth}
\includegraphics[width=\textwidth]{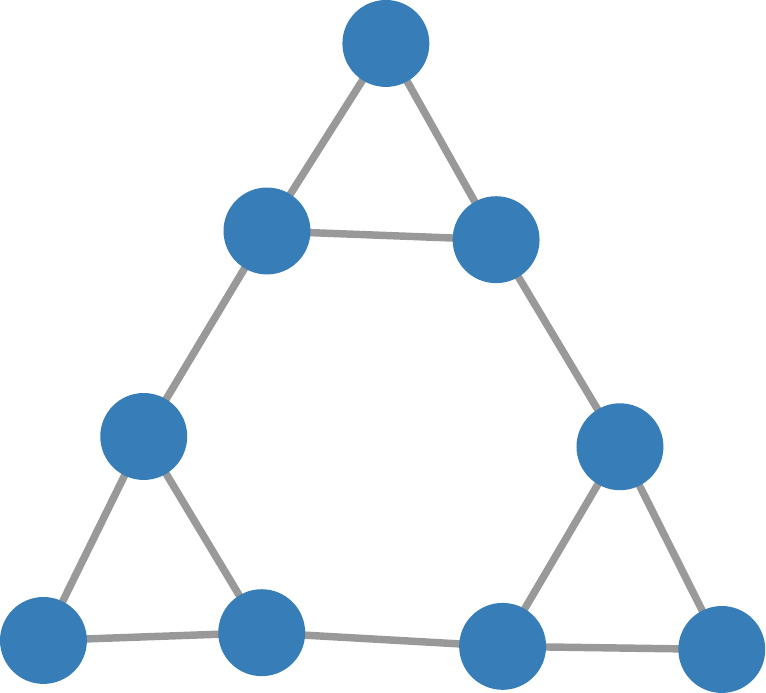}
\caption{}
\end{subfigure}
\caption{(a) An input graph with highlighted cliques. (b) Aggregation step: we consider each clique as a ``supernode'' (in blue). (c) The summarized graph.}
\label{fig:aggregation}
\end{figure}

We already saw a flavor of this approach when we discussed hierarchical community discovery in Section \ref{sec:hcd-recursive}. Figure \ref{fig:aggregation} is a reprise of Figure \ref{fig:hcd-merge1} and is a great example of the aggregation approach. In it, we use a community discovery technique to highlight densely connected modules, and we then collapse each into a ``super node''.

Another example from the past comes from Section \ref{sec:hier-cycles}, where we discussed graph condensation, another summarization technique. In the case of condensation, it involves neither communities nor cliques -- clique-reduction\cite{ishikawa2009higher} is another aggregation method --, but strongly connected components in directed networks.

There's also an approach from a future section of this book. When visualizing networks, one thing you need to decide is the positioning of the nodes onto a 2D plane. This is the problem of finding a good layout for your graph, and I'll talk in depth about this in Chapter \ref{cha:layouts}. In that chapter, you'll see that one of the biggest problems is that nodes sometimes snuggle together a bit too closely, overlapping with each other. One could identify such nodes that tend to occupy the same position in space and simply aggregate them into a super node\cite{gansner2008efficient}, and use this information to bundle up edges as well\cite{gansner2011multilevel} -- edge bundling is a classic visualization improvement I'll discuss in Section \ref{sec:layouts-bends}.

Of course, community discovery, graph condensation, or visualization were not originally developed with summarization in mind. Thus it is possible to design node aggregation methods that are specialized for summarization, even if inspired by other related approaches. One is Grass\cite{lefevre2010grass}\cite{riondato2017graph}. In Grass one performs the node aggregation in such a way that the errors in reconstructing the original adjacency matrix are minimized.

\begin{figure*}
\centering
\begin{subfigure}{.25\columnwidth}
\includegraphics[width=\textwidth]{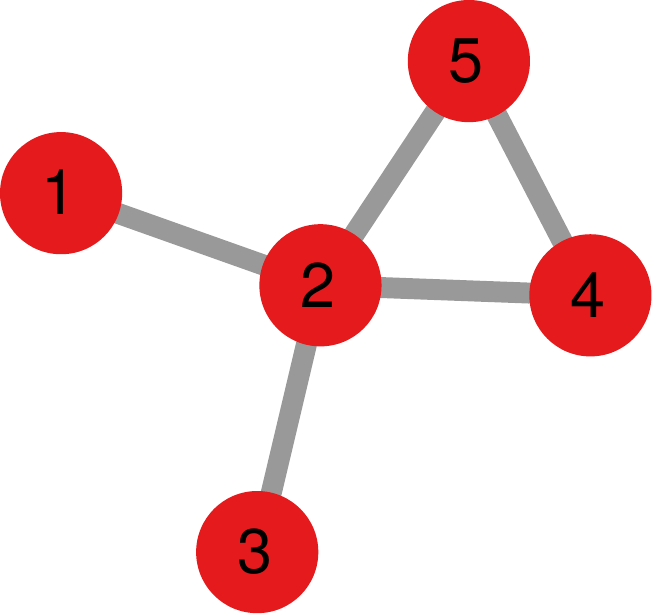}
\caption{}
\end{subfigure}\qquad
\begin{subfigure}{.2\columnwidth}
\includegraphics[width=\textwidth]{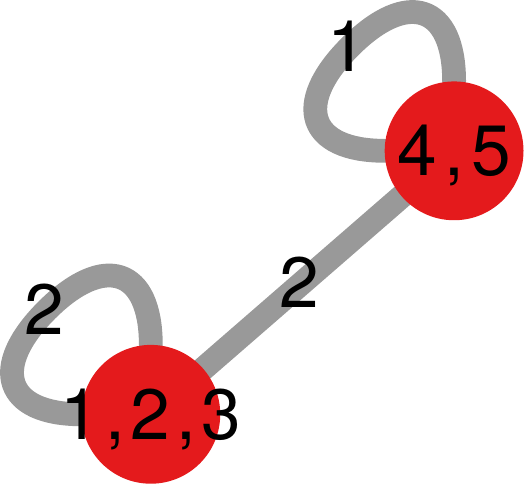}
\caption{}
\end{subfigure}\qquad
\begin{subfigure}{.3\columnwidth}
  \begin{tabular}{c|ccccc}
    & $1$ & $2$ & $3$ & $4$ & $5$ \\
    \hline
    $1$ & $0$ & $2/3$ & $2/3$ & $1/3$ & $1/3$\\
    $2$ & $2/3$ & $0$ & $2/3$ & $1/3$ & $1/3$\\
    $3$ & $2/3$ & $2/3$ & $0$ & $1/3$ & $1/3$\\
    $4$ & $1/3$ & $1/3$ & $1/3$ & $0$ & $1$\\
    $5$ & $1/3$ & $1/3$ & $1/3$ & $1$ & $0$\\
  \end{tabular}
\caption{}
\end{subfigure}
\caption{(a) An input graph. (b) Aggregation of (a). Node labels report the nodes collapsed into the super node. Edge labels record the number of edges inside or between super nodes. (c) The adjacency matrix of (a) as reconstructed via (b).}
\label{fig:grass}
\end{figure*}

Suppose you condensed the graph in Figure \ref{fig:grass}(a) into the graph in Figure \ref{fig:grass}(b). Now all you have is Figure \ref{fig:grass}(b), but you might want to know what is the probability that nodes $1$ and $4$ are connected. You can reconstruct Figure \ref{fig:grass}(a)'s adjacency matrix via Figure \ref{fig:grass}(b)'s -- and keeping track of the original number of edges inside and between each super node.

For instance, if two nodes $u$ and $v$ are in the same super node $a$, then their expected connection probability is $|E_a| / (|V_a| (|V_a| - 1))$, namely the number of edges collapsed inside $a$ over all edges $a$ could contain. Vice versa, if $u$ and $v$ are in different super nodes $a$ and $b$, then their connection probability is $|E_{ab}| / (|V_a||V_b|)$, again: number of edges between $a$ and $b$ over all the possible edges that there could be. Thus, the reconstructed adjacency matrix of the original graph is the one in Figure \ref{fig:grass}(c). The quality function guiding this process is mutual information: the higher the mutual information between the original and the reconstructed matrix, the better the aggregation is.

Other approaches compress structurally equivalent nodes\cite{toivonen2011compression} -- see Section \ref{sec:centr-similarity} for a refresher on structural equivalence.

Respecting the adjacency of nodes is not necessarily the only reasonable guiding principle for your aggregation. As I mentioned previously, one might want to perform summarization to aid visualization. In this case, one might want to just simplify complex motifs that would tangle up your visualization\cite{dunne2013motif}.

Since we're shifting perspectives, we might as well keep shifting them. So far, we have assumed that aggregation involves the collapse of nodes into super nodes. However, we could very well collapse \textit{edges} instead. In this case, the edge is aggregated into what we call a ``compressor'', or virtual node. The idea is that high degree nodes, especially those embedded in very dense parts of the network, are at the center of a lot of redundant information.

\begin{figure}
\centering
\begin{subfigure}{.3\columnwidth}
\includegraphics[width=\textwidth]{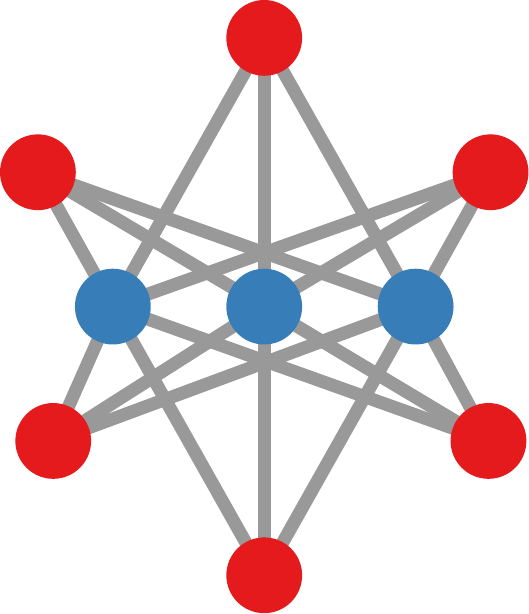}
\caption{}
\end{subfigure}\qquad
\begin{subfigure}{.4\columnwidth}
\includegraphics[width=\textwidth]{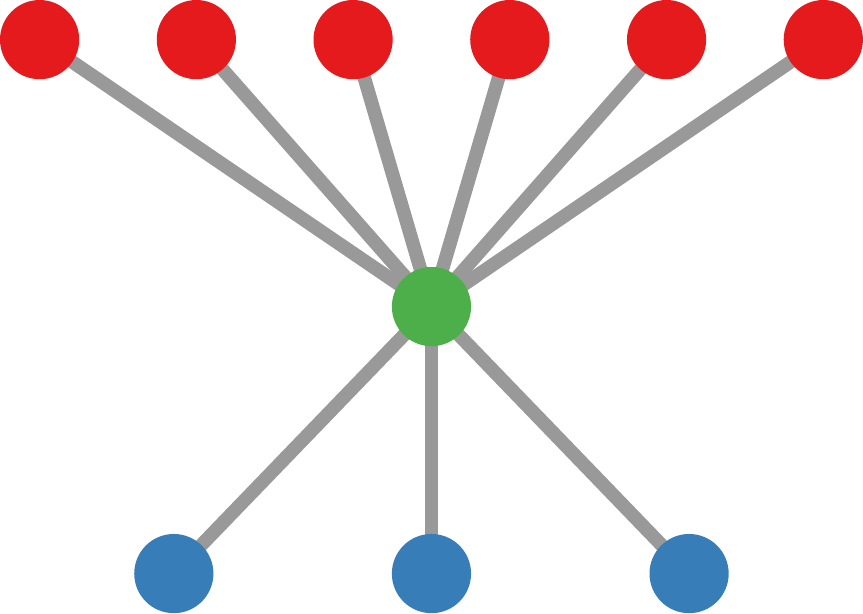}
\caption{}
\end{subfigure}
\caption{(a) An input graph. Nodes in red have low degree, nodes in blue have high degree. (b) Its summarization via edge aggregation into a compressor (in green).}
\label{fig:dedens}
\end{figure}

Take Figure \ref{fig:dedens}(a) as an example: its structure can be summarized with a very simple formula -- all red nodes connect to all blue nodes. We can compress this information in a node that represent all red-blue connections. The resulting graph -- in Figure \ref{fig:dedens}(b) -- describes exactly the same structure, but does so with nine edges instead of $18$, at the price of adding a single node. This looks a bit like the community affiliation graph we saw in Section \ref{sec:ocd-mmsbm}, and in fact I'll have you notice that -- in this case -- we're practically just compressing a 6,3-clique.

\section{Compression}\label{sec:mining-summarization-compr}
If you're familiar with zipped archives and programs like WinRAR, you already know what the guiding principle of this branch of graph summarization is. The idea here is to compress the original structure so that we minimize the number of bits we need to encode it. Suppose you're compressing a text in ASCII format. Each letter will cost you seven bits. However, by looking at the text, you realize that many ASCII characters never appear. Thus you can re-encode all characters using shorter codes: if you only use $40$ distinct characters, you only need a bit more than five bits per character, reducing a 1MB text into $\sim 760kB$. You can do better than that, realizing that most of the times the character ``h'' follows specific other characters and so on. In practice, you're modeling your text with a model $M$. Encoding $M$ takes some bits, but it saves many more. This is following the same philosophy as the Infomap community discovery approach we saw in Section \ref{sec:cd-partition-rw}.

Translating this into graph-speak, you want to construct a model $M$ of your graph $G$ so that the length $L$ describing both is minimal, or: $\min L(G,M) = L(M) + L(G|M)$. An example\cite{navlakha2008graph} creates $M$ using a two-step code: (i) each super node in the summary is connected, in the original graph, to all nodes in all super nodes adjacent to it in the summary; and (ii) we correct every edge mistake with an additional instruction.

\begin{figure}
\centering
\begin{subfigure}{.4\columnwidth}
\includegraphics[width=\textwidth]{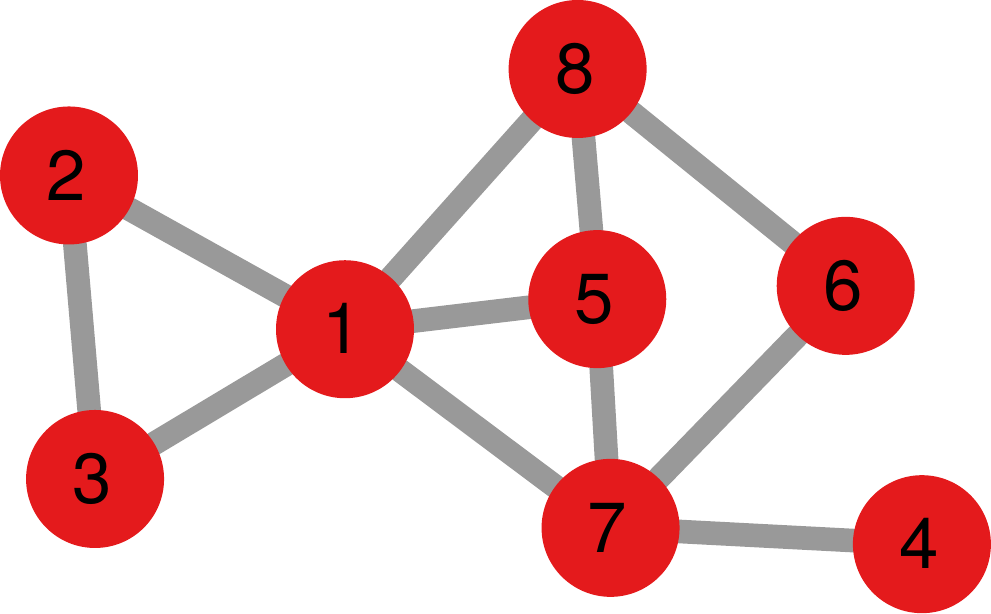}
\caption{}
\end{subfigure}\quad
\begin{subfigure}{.425\columnwidth}
\includegraphics[width=\textwidth]{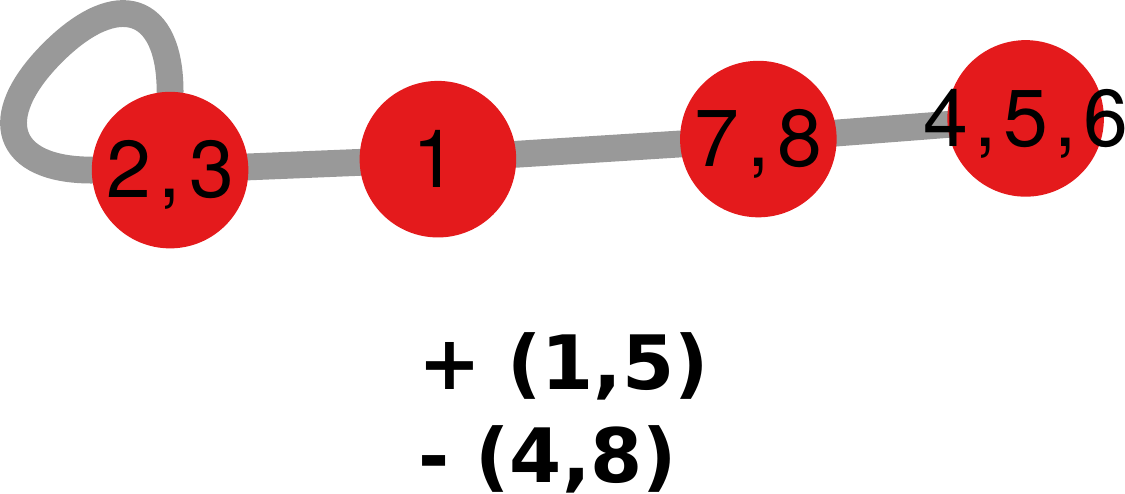}
\caption{}
\end{subfigure}
\caption{(a) An input graph. (b) Its summarization via minimization of description length. I label each super node with the list of nodes it contains. On the bottom, the additional rules we need to reconstruct (a).}
\label{fig:summary-mdl}
\end{figure}

Figure \ref{fig:summary-mdl} shows the approach. The original graph in Figure \ref{fig:summary-mdl}(a) can be compressed in the graph in Figure \ref{fig:summary-mdl}(b). However, the summary is not perfect. It assumes the existence of an edge that does not really exist, and it misses another edge that exists. We add these two rules to the model $M$ and now the summary is a perfect reconstruction of Figure \ref{fig:summary-mdl}(a). In Figure \ref{fig:summary-mdl}(b) we say that nodes $7$ and $8$ are connected to nodes $4$, $5$, and $6$. This is mostly accurate, but not completely correct: we need the additional rule that nodes $4$ and $8$ do not connect to each other.

The objective now is to find the best combination of summary and additional rules that uses as little information as possible, given or take a margin of error you can set as parameter. There are many information-theoretic approaches in this category\cite{boldi2004webgraph}\cite{ahnert2013power}\cite{koutra2014vog}.

A natural domain of application for compression-based summarization is in the description of evolving networks. In practice, one has many snapshots of the same network and they are trying to reconstruct what the whole network looks like\cite{shah2015timecrunch}. The idea is to find the model that is able to best represent all the snapshots you collected.

You might feel like aggregation and compression are basically the same category. After all, if you look at Figure \ref{fig:summary-mdl}, what you're seeing is basically an aggregation strategy. The fundamental difference between the two categories is the existence of the model $M$. In aggregation, there is no $M$: we simply brute force our way through the graph to save every node or edge we can, regardless whether we're uncovering common patterns or not. The existence of $M$ in the compression category, instead, forces us only to perform an aggregation if it results in a leaner and more elegant $M$. As a consequence, $M$ itself is an important result of the procedure, because it contains information about the common patterns you can find in your original structure.

\section{Simplification}\label{sec:mining-summarization-simpl}
In this class we group solutions that are the lovechildren of network sampling (Chapter \ref{cha:sampling}) and backboning (Chapter \ref{cha:backboning}). The idea here is not to create ``super nodes'' like in the previous two classes. Here, we look at the original structure. However, we simplify it by removing nodes and edges that we consider ``unimportant''. Sampling and backboning techniques can be considered simplification strategies that are purely structural: they only use information coming from the graph's topology.

\begin{figure}
\centering
\begin{subfigure}{.5\columnwidth}
\includegraphics[width=\textwidth]{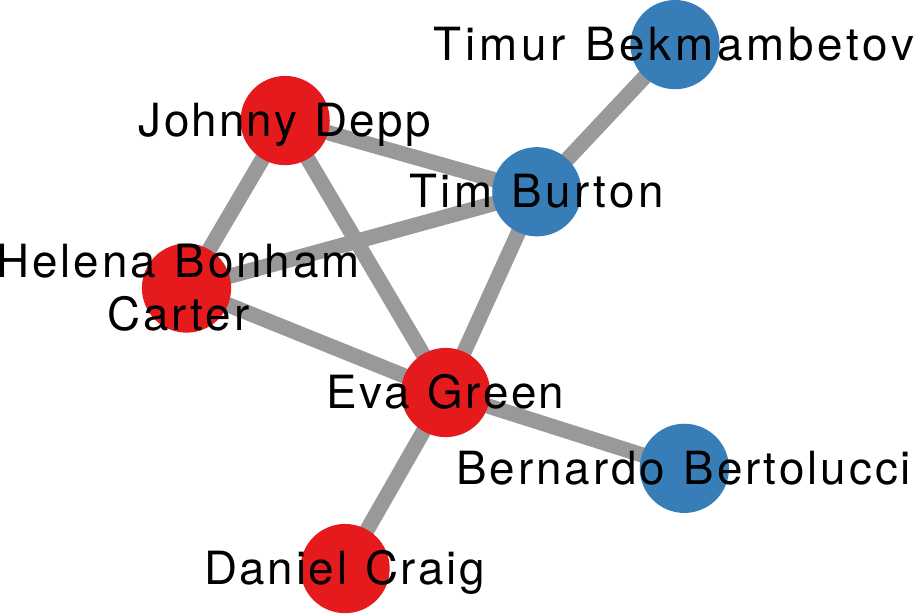}
\caption{}
\end{subfigure}\quad
\begin{subfigure}{.4\columnwidth}
\includegraphics[width=\textwidth]{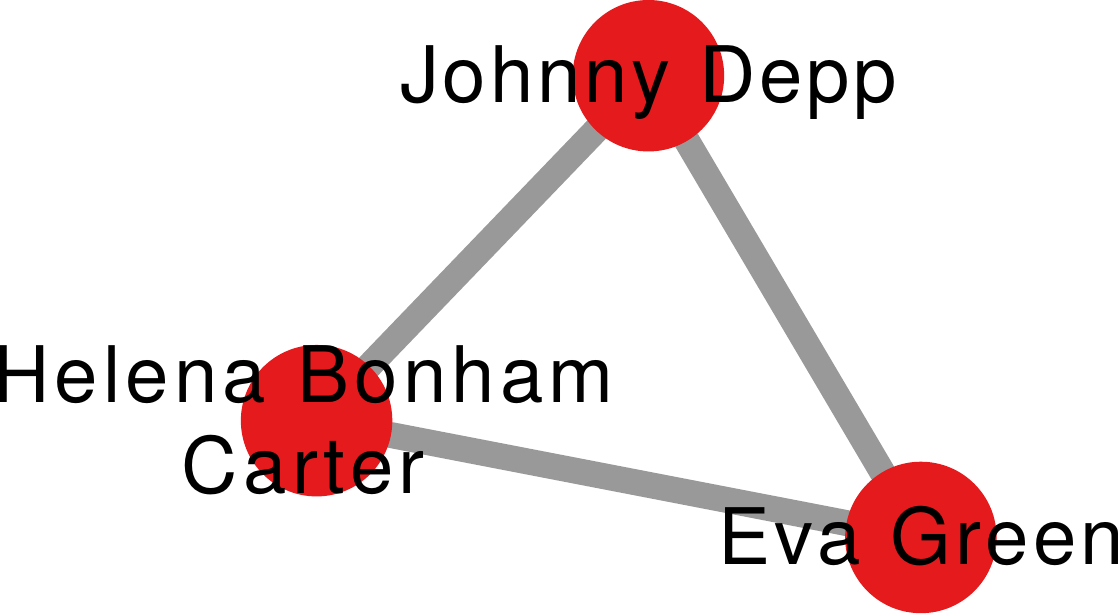}
\caption{}
\end{subfigure}
\caption{(a) An input graph: directors (in blue) and actors (in red) connected if they collaborated with each other. (b) Its simplification via the selection of only actor-type nodes directly connected to Tim Burton.}
\label{fig:ontovis}
\end{figure}

Those are not the only valid approaches. If the graph also has metadata attached to nodes -- or edges -- we can exploit them. For instance, Ontovis\cite{shen2006visual} allows the simplification of the graph via the specification of a set of attribute values we're interested in studying. For instance, the graph in Figure \ref{fig:ontovis}(a) can be simplified into the one in Figure \ref{fig:ontovis}(b), if we're only interested in knowing the relationships between actors (node type value) working with Tim Burton (topology attribute).

Ontovis finds the best way to simplify the graph, primarily focusing on its visual characteristics when plotted in 2D: it is first and foremost a visualization-aiding tool. Ontovis focuses on node attributes, but one could also switch their focus to edges\cite{li2009egocentric}.

Note, also, that another difference with sampling is that, in graph simplification, we are not really interested in preserving any specific property of the original graph. This is, instead, a core focus of network sampling.

\begin{figure*}
\centering
\includegraphics[width=\columnwidth]{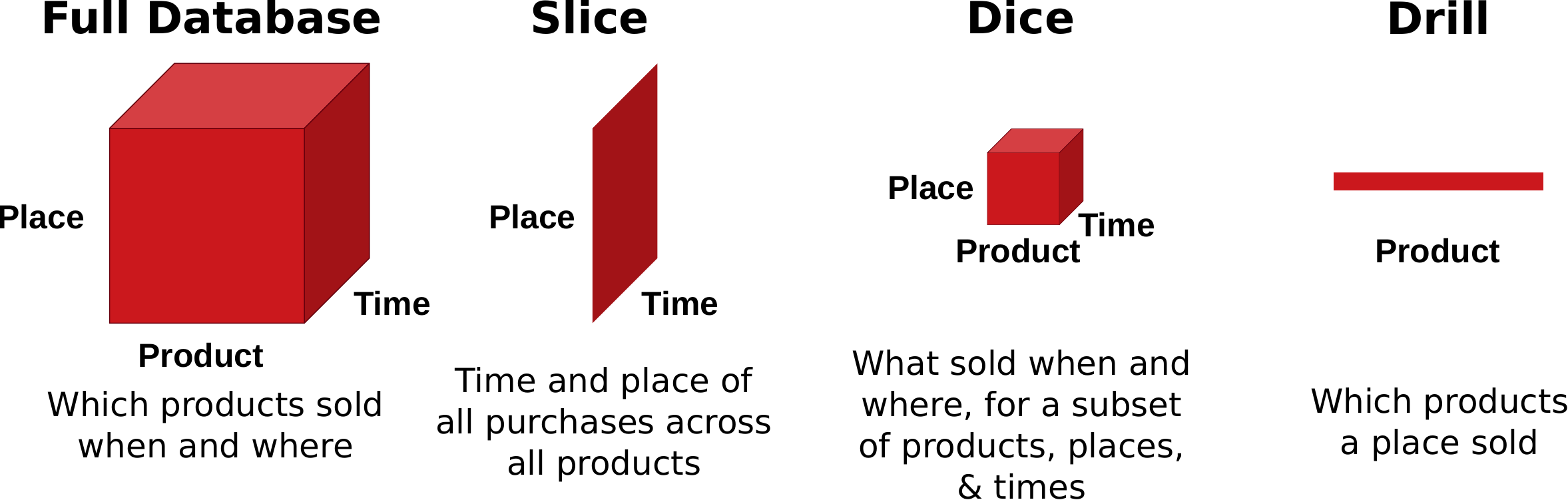}
\caption{An example of operations on an OLAP database.}
\label{fig:olap}
\end{figure*}

When not necessarily focusing on visualization, the simplification approach uses a database metaphor to help the user navigate between different ``views'' of the network data. A classical database infrastructure is OLAP\cite{chaudhuri1997overview}, which stands for OnLine 
Analytical Processing. In an OLAP database you have data that is inherently multidimensional, for instance sales can happen in different shops, at different times, via different product categories, and customer classes, etc. OLAP allows you to represent this with a ``data cube'' that you can slice and dice to aggregate the dimensions. Figure \ref{fig:olap} shows you a visual representation of what different operations look like -- and mean.

Graph OLAP\cite{tian2008efficient}\cite{chen2008graph}\cite{zhao2011graph} is fundamentally the same thing applied to networks, using node/edge attributes and characteristics to drive the simplification procedure. While traditional graph OLAP works best with categorical node attributes, there are also ways to dice and slice your graph using numeric attributes\cite{zhang2010discovery}, allowing you to also consider node properties such as the degree.

Another related category of approaches is ``graph sketches''\cite{ahn2012graph}\cite{liberty2013simple}\cite{ghashami2016efficient}.

\section{Influence Based}\label{sec:mining-summarization-infl}
In influence-based summarization one is interested in having a summarized graph in which influence events follow the same dynamics as in the original graph. That is: if something is flowing through the graph, percolating node to node, the percolation in the summarized graph follows the same dynamics as in the original graph.

There are a few ways to wrap your head around this concept. I feel that I can provide two very different approaches for you, that should serve different types of understanding graph dynamics. The first rests on a data-driven approach. Suppose you have a social network, where nodes are people. Some users are early adopters of a new product. As they perform an action, some of their friends will see it and will imitate them. This means that you can detect ``tribes'' of people reacting with the same timing to the same stimuli. This is basically like doing community discovery, with the difference being that you're not maximizing internal density, but simply the synchronization of an action. What I described is, for instance, what GuruMine detects, and you can use Section \ref{sec:triggers-resistance} as a refresher.

\begin{figure}
\centering
\begin{subfigure}{.4\columnwidth}
\includegraphics[width=\textwidth]{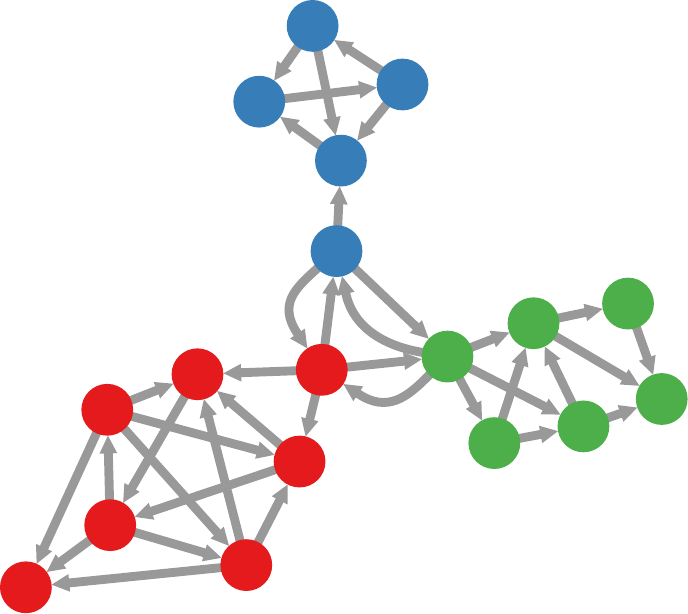}
\caption{}
\end{subfigure}\qquad
\begin{subfigure}{.33\columnwidth}
\includegraphics[width=\textwidth]{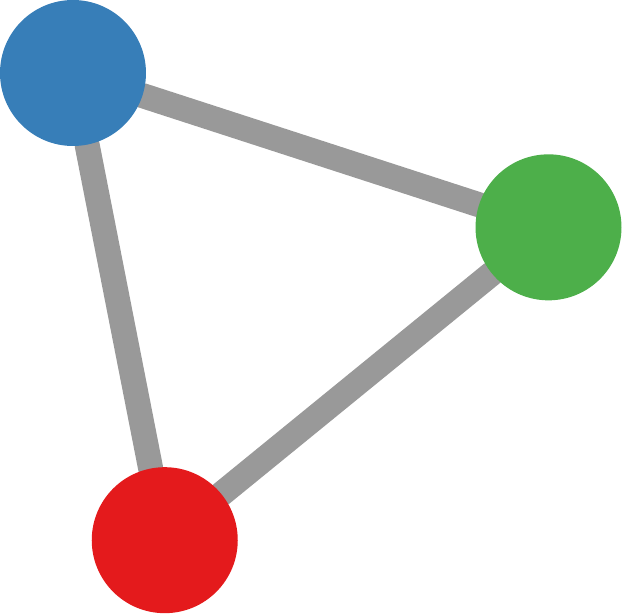}
\caption{}
\end{subfigure}
\caption{(a) An influence graph. The edge direction tells you who influences whom. The node color tells you the detected ``tribes''. (b) The summary graph, compressing all tribes into a node, to preserve influence dynamics.}
\label{fig:summary-gurumine}
\end{figure}

Visually, this would look like Figure \ref{fig:summary-gurumine}. The central hubs influence each other, and each is responsible for influencing their branch. Thus one could summarize the graph as a clique of interacting tribes. Of course, you don't have to use GuruMine for this. In some cases, researchers have used special adaptations to estimate community-level influence\cite{mehmood2013csi}.

There's a completely different way to interpret summarization by influence preservation. We have seen that the spectrum of the Laplacian can be used to partition a graph -- solving the cut problem (Section \ref{sec:rw-mincut}). This is related to diffusion processes: the reason why the eigenvectors of the Laplacian help you with cutting is because they are a sort of simulation of a diffusion process, and the edges to cut are the bottlenecks though which things cannot flow efficiently. For now, let's take this for granted, but we'll see more about this relationship in Section \ref{sec:nvd-ge}, where we'll talk about using the Laplacian to estimate distances between sets of nodes by releasing a flow from the nodes in the origin to the nodes in the destination.

If we want to summarize the graph to preserve these diffusion properties, we can use the Laplacian to guide our process. What we're after, in the simplest possible terms, is a smaller Laplacian matrix, with fewer rows and columns, that has the same eigenvectors\cite[-1in]{purohit2014fast}.

\begin{figure}
\centering
\begin{subfigure}{.4\columnwidth}
\includegraphics[width=\textwidth]{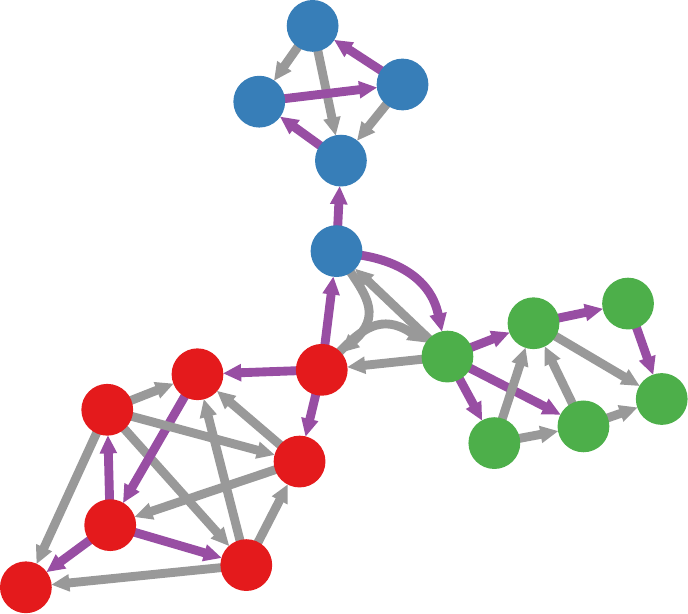}
\caption{}
\end{subfigure}\qquad
\begin{subfigure}{.4\columnwidth}
\includegraphics[width=\textwidth]{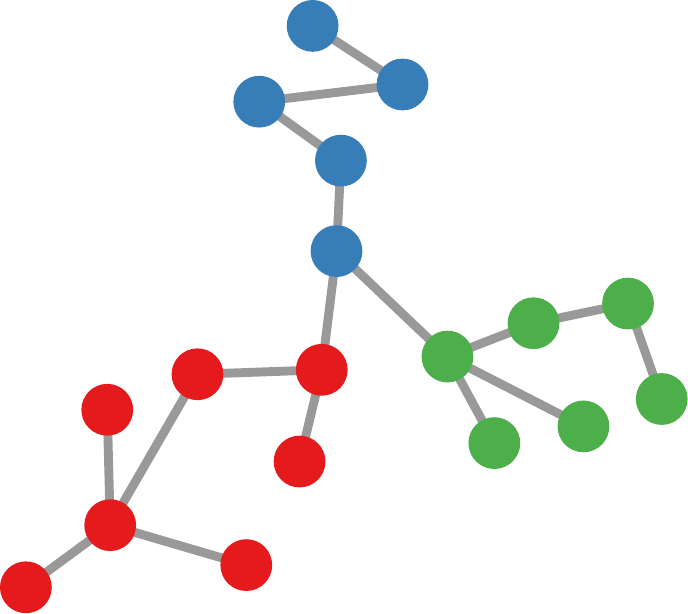}
\caption{}
\end{subfigure}
\caption{(a) An influence graph. The edge color represents the path taken by an hypothetical spreading event. (b) The summary graph, including only the edges used in the spreading process.}
\label{fig:summary-spine}
\end{figure}

Among other interesting approaches there is SPINE\cite{mathioudakis2011sparsification}. In SPINE, one analyzes many influence events in the network. Then, SPINE only keeps in the network the edges that are able to better explain the paths of influence you observe. You might realize that an edge is never used to transport information, and thus you could remove it from the structure without hampering your explanatory power. Figure \ref{fig:summary-spine} shows an example of this.

\section{Summary}

\begin{enumerate}
\item Graph summarization is the task of reducing the size of your graph so that you can facilitate different operations, such as analysis, storage, data cleaning, and/or network visualization.
\item There are many ways to perform summarization. One is to do so via aggregation: you find coherent modules in your network and you collapse them into the same node, aggregating all incoming and outgoing edges.
\item A second category is compression: you similarly try to aggregate the graph, but this time you record your operation in a model. The model also has to be encoded, and thus you have to find the simplest possible model that best represents your original graph.
\item The third approach is simplification. This is especially relevant for visualization: you want to simplify the graph so that motifs that would clutter its representation are reduced and do not cross each other.
\item Finally, you might have influence in mind: some process is spreading on your network and you want to keep only the connections that are most likely responsible for that process to percolate through the nodes.
\item Making a summary of a chapter about summarization is delightfully meta and I'm having a hell of a good time.
\end{enumerate}

\section{Exercises}

\begin{enumerate}
\item Perform label percolation community discovery on the network at \url{http://www.networkatlas.eu/exercises/46/1/data.txt}. Use the detected communities to summarize the graph via aggregation.
\item The table at \url{http://www.networkatlas.eu/exercises/46/2/diffusion.txt} contains the information of which node (first column) influenced which other node (second column). Use it to summarize the graph by keeping only the edges used by the spreading process.
\item Summarize the summary you generated answering question \#1 with the data from question \#2. Do you still obtain a connected graph?
\end{enumerate}

\chapter{Node Vector Distance}\label{cha:nvd}
In Chapter \ref{cha:shortpath} we saw a way to determine the distance between two nodes: the number of edges you need to cross in the graph to go from one to the other. Alternatively, one could use the hitting time (Section \ref{sec:rw-hitime}): how long it will take for a random walk to hit both nodes. However, sometimes you don't want the distance between a pair of nodes. Sometimes you want to ask: what is the distance between a \textit{group} of nodes and another, given that some nodes might weigh more than others?

\begin{figure}
\centering
\includegraphics[width=.45\columnwidth]{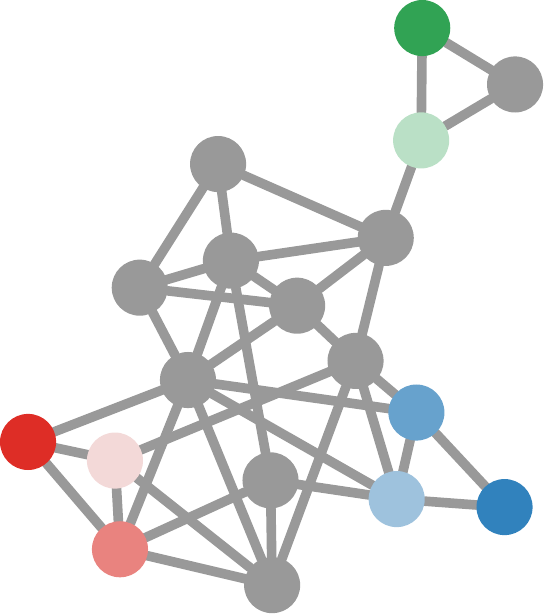}
\caption{A graph with different highlighted groups of nodes. The intensity of the color is proportional to how much weight is on the node.}
\label{fig:nvd-example}
\end{figure}

Figure \ref{fig:nvd-example} provides an intuitive example to understand this question. Is the total red hue distributed closer to the blue color, or to the green? How much does it matter that the darker nodes carry more weight in estimating such distance?

I call this problem the Node Vector Distance, and it has many applications:

\begin{itemize}
\item In computer vision\cite{peleg1989unified}\cite{rubner2000earth}, we represent an image as a graph of points of interest, with different values, proportional to how much light or color is in them. Two images can then be compared by estimating how much ``light'' we have to transport from the interest points of one to the interest points of the other. Small amounts will indicate that the images are similar.
\item In economics\cite{hausmann2014atlas}\cite{hidalgo2007product}, you can represent products as nodes, connected if there is a significant number of countries that are able to co-export significant quantities of them. A country occupies the products in this network it can export. From one year to another, the country will change its export basket, by shifting its industries to different products. How dynamic is the country's export basket?
\item In epidemics\cite{colizza2006role}\cite{ganesh2005effect}\cite{pastor2001epidemic}, a disease occupies the nodes in a social network it has infected. Across time, the disease will move from a set of infected individuals to another. Similarly, in viral marketing, product adoption can be modeled as a disease.
\end{itemize}

All these cases can be represented by the same problem formulation. You have a network $G$. Then you have two vectors: an origin vector $p$ and a destination vector $q$. Both $p$ and $q$ tell you how much value there is in each node. $p_u$ tells you how much value there is in node $u$ at the origin, and $q_v$ tells you how much value there is in node $v$ at the destination.

All you want to do is to define a $\delta(p, q, G)$ function. Given the graph and the vectors of origin and destination, the function will tell you how far these vectors are. There are many ways to do so, which are organized in a survey paper\cite{coscia2020node}, on which this chapter is based. Before we jump into the network distances, it is probably wise to have a refresher on non-network distances, since it will allow us to introduce concepts that will be helpful later.

\section{Non-Network Distances}\label{sec:nvd-nonnet}
How to estimate node vector distances on networks is a new and difficult problem. Let's take it easy and first have a quick refresher on the many ways we can estimate distances of vectors \textit{without} a network. The easiest way to do it is by assuming a vector of numbers just represents a set of coordinates in space. If you're on Earth, with three numbers you can establish your latitude, longitude and altitude. That is enough to place you on a position in a three dimensional space. Another person might be at a different latitude, longitude and altitude than  you. What is the distance between you and your friend? Easy! You throw a straight rope between you and your friend and its length is the distance between you. This is the Euclidean distance.

\begin{figure}
\centering
\includegraphics[width=.5\textwidth]{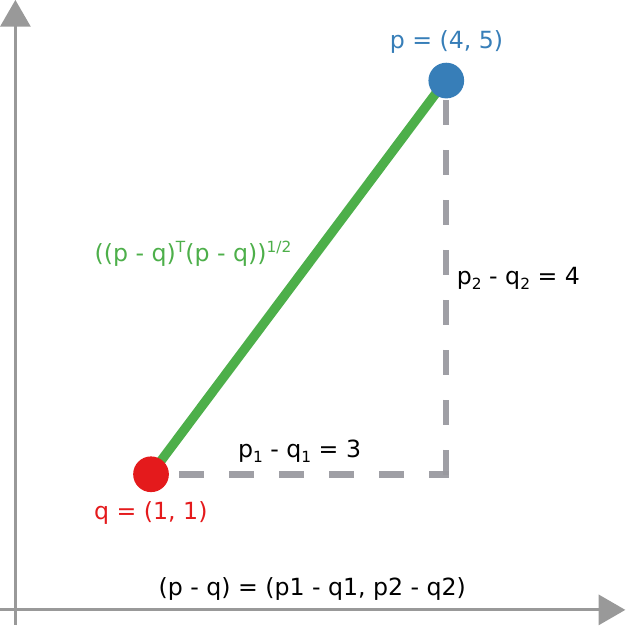}
\caption{Euclidean distance in $m=2$ dimensions. We build the special $p - q$ vector to have, at its $i$th entry, the difference between the $i$th entries of $p$ and $q$. }
\label{fig:euclidean2}
\end{figure}

In Section \ref{sec:la-dot} I did my very best to connect this intuitive idea in real life with linear algebra operations. The pain you felt back then should pay off now. To sum up, if $p$ and $q$ are the vectors defining your two positions in space, the Euclidean distance is $((p - q)^T I (p - q))^{1/2}$, where $I$ is the identity matrix. Figure \ref{fig:euclidean2} should help refreshing your intuition behind the Euclidean distance.

You can put any matrix in this formulation instead of $I$, as long as they are positive semi-definite matrices. Why would you want to put any other matrix in there? Remember that matrices are nothing more than spatial transformations: they tell you how to bend and warp your space. So putting something else than $I$ will make the $(p - q)$ vector bend in special ways. What are these ways? Well, the role of $I$ in the Euclidean distance is to tell us that each dimension of the vector makes the exact same contribution to the distance. So the vectors $(0,1,0)$ and $(0,0,1)$ are exactly equidistant from $(1,0,0)$.

However, in some cases, we might notice that some dimensions are correlated: they change together. So, if the vectors differ in a direction opposite from what we would expect, this change should count more than the one we would expect. We know we can calculate the covariance of two variables (Section \ref{sec:stats-corr}), so we can store the relations between the dimensions of $p$ and $q$ in their covariance matrix $cov(p,q)$ and, since we want to count more when we have a change going in the opposite direction from the expected, we invert the matrix: $((p - q)^T cov^{-1}(p, q) (p - q))^{1/2}$. This is the Mahalanobis distance.

\begin{figure}[b]
\centering
\includegraphics[width=.5\textwidth]{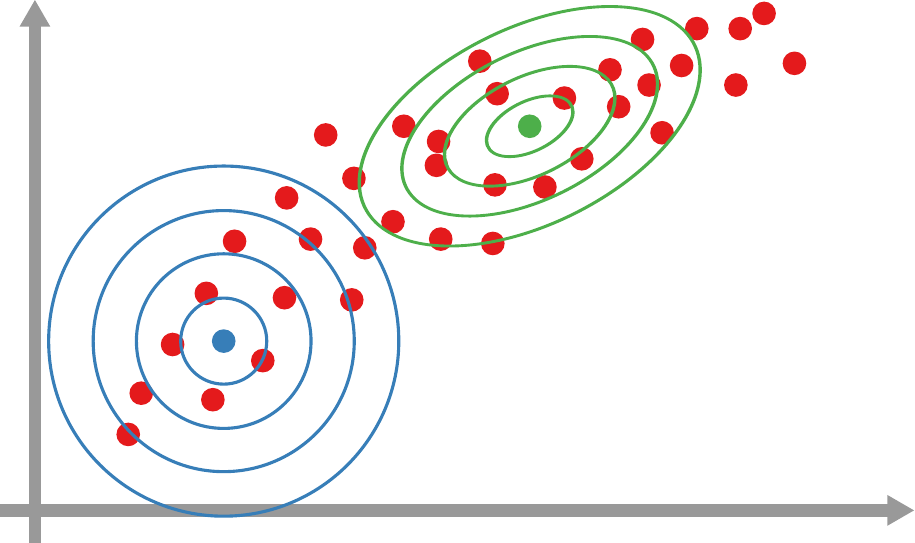}
\caption{An example of two correlated variables. The blue concentric circles represent Euclidean distances centered on the blue point. The green concentric ellipses represent Mahalanobis distances centered on the green point.}
\label{fig:mahalanobis}
\end{figure}

Figure \ref{fig:mahalanobis} shows you an example. The two dimensions of the scatter plot are clearly correlated. This means that moving in a direction orthogonal to the correlation line counts for more distance covered: it is an unexpected move. That is what the Mahalanobis distance, in green, is capturing. The Euclidean distance is oblivious to this and, for it, all directions are equally important. Crossing the same number of lines means covering the same distance: while the direction you choose in the Euclidean case doesn't matter, it does in the Mahalanobis distance.

In the next section we'll see how we can use a graph's topology to weight the dimensions of our vector in such a way that the difference between $p$ and $q$ is constrained to happen in the space described by our network. This boils down to find a smart positive semi-definite matrix to put in the place occupied by $I$ or $cov^{-1}(p, q)$.

Are we done with non-network vector distances? Of course not: we haven't even started yet. I already mentioned a few other distances when I talked about network projections in Section \ref{sec:projections-vectors}. We have correlation distances, cosine distances, etc. I need not to go into details on how each of these measures work. I will only explain the cosine distance just to make a point: the Euclidean way of estimating distances is not the only proper way.

What do I mean by this? Consider the rope example I made before. For some distance measures, the length of the rope between you and your friend is not the most important thing. You can have two points requiring a longer rope to connect that could be closer to each other than points requiring a shorter rope. This is the case of the widely used cosine distance.

\begin{figure}
\centering
\includegraphics[width=.5\textwidth]{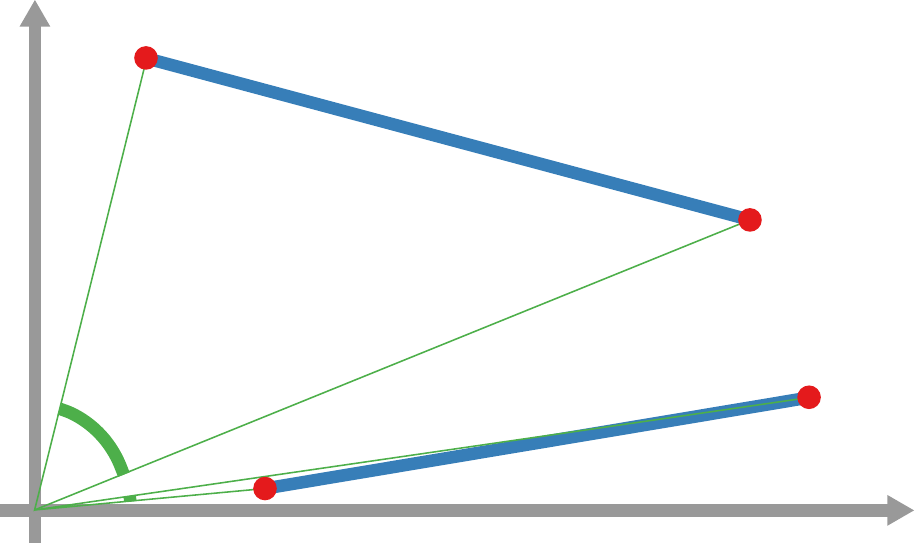}
\caption{The difference between the Euclidean distance between two points (in blue) and their cosine distance (thick green arc tracing the angle between the two points).}
\label{fig:cosine}
\end{figure}

In cosine distance you look at the \textit{angle} made by the vectors connecting the two points, as Figure \ref{fig:cosine} shows with a thick green line. The distance between them is one minus the cosine of that angle. This is useful, because the cosine is $1$ for angles of zero degrees and $0$ for angles at ninety degrees. Two points on the same straight line will have a distance of zero, even if they're infinitely farther apart on such a line. For instance, the two points at the bottom of Figure \ref{fig:cosine} are at a considerable Euclidean distance, but practically neighbors when it comes to cosine distance. Sometimes in life it doesn't matter where you are, as long as you're going in the same direction.

If you're curious about why we need the cosine distance, the reason is to deal with high-dimensional data. When you have lots of dimensions, data points tend to all roughly be at the same Euclidean distance and it's very difficult to tell differences between them\cite{beyer1999nearest}. This is what we normally call the ``curse of dimensionality.'' By refusing to calculate Euclidean distances between the points and considering only the angle they make with each other -- i.e. by using cosine distances -- we can tame the curse of dimensionality, at least partially.

The importance of the cosine distance is in reminding us that a distance measure does not have to necessarily respect the triangle inequality like the Euclidean distance does. The triangle inequality says that the distance between $a$ and $b$ is always lower than or equal to the distance between $a$ and $c$ and $b$ and $c$. This is how things work in the real physical world. But it is not the only way things \textit{can} work.

\section{Generalized Euclideans}\label{sec:nvd-ge}
\begin{figure}
\centering
\includegraphics[width=.5\textwidth]{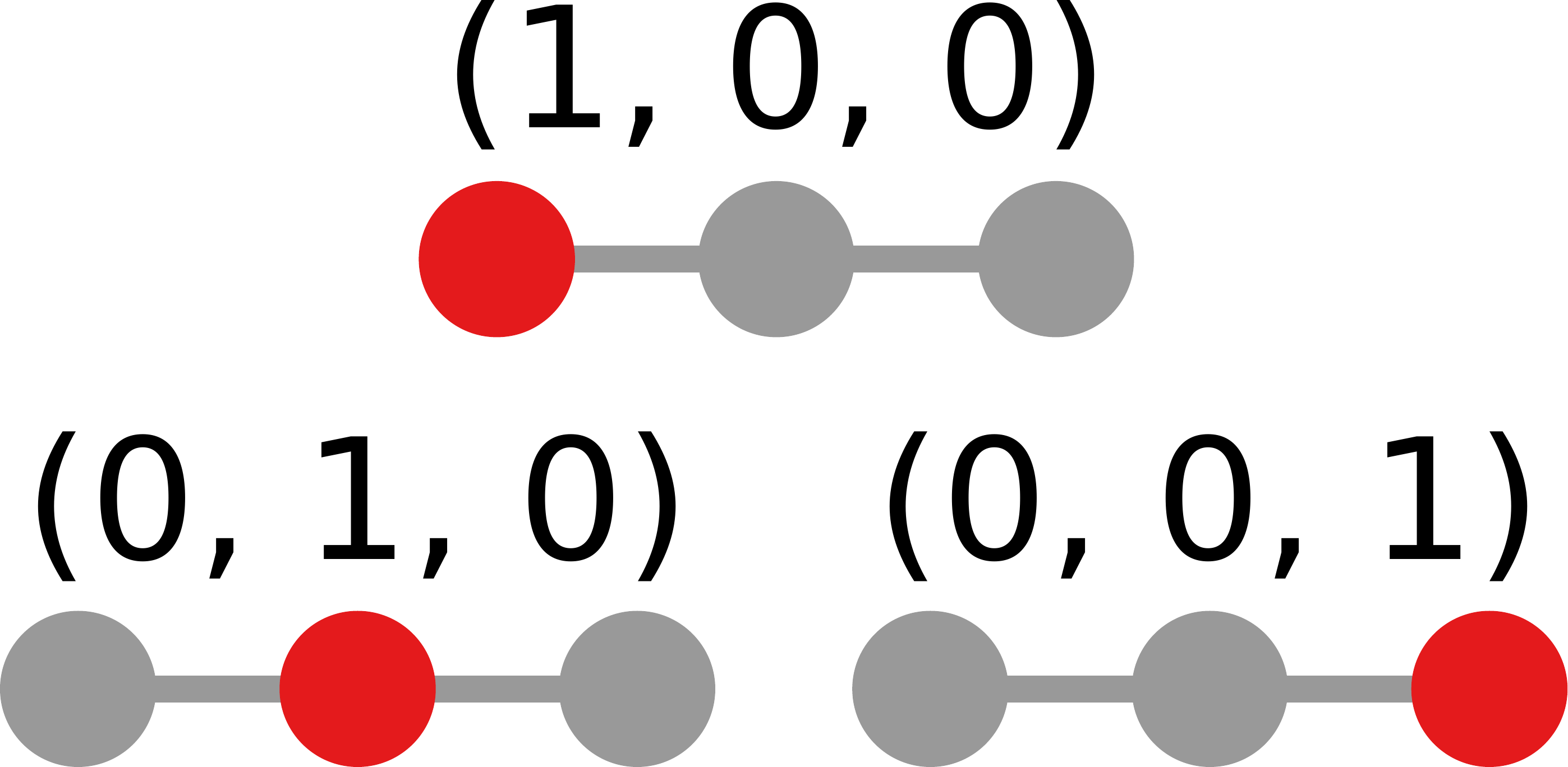}
\caption{The graph representation of three vectors. I highlight in red the node corresponding to the entry equal to $1$ in the vector.}
\label{fig:nvd-not-euclidean}
\end{figure}

In this section we're going to generalize the Euclidean distance by replacing the identity matrix $I$ with some matrix dependent on the topology of the network we're working with. In practice, this means that we are constraining the diffusion process to happen through the edges of the network. We don't want the diffusion to jump between any two pairs of nodes. If the nodes are far apart in the network, such jump should be considered differently than the one happening between two nodes that are directly connected. For instance, in Figure \ref{fig:nvd-not-euclidean}, I make the case in which the $(0,1,0)$ and $(0,0,1)$ vectors should not be considered equidistant from $(1,0,0)$, because the leftmost and rightmost nodes in this network are not connected, and they are thus farther apart.

With its reliance on $I$, the Euclidean distance considers these cases as equidistant. So we need to replace $I$ with something else. We'll look at three alternatives.

\subsection{Laplacian}\label{sec:nvd-ge-lapl}
In the Laplacian version\cite{coscia2020generalized}, we look at the Laplacian of the adjacency matrix. Remember from Section \ref{sec:mat-mat-laplacian}, that the Laplacian $L = D - A$, with $D$ being the degree matrix and $A$ the adjacency matrix. Since the smallest eigenvalue of $L$ is zero, $L$ is positive semi-definite.

We use the graph Laplacian because the Laplace operator describes mathematically statuses of equilibrium\cite{evans1997partial}. In practice, if you have a liquid on a container making waves -- i.e. being out of equilibrium -- the Laplace operator will tell you how the diffusion of the liquid will behave when transitioning to its equilibrium state, where there are no more waves.

Just like in the Mahalanobis case, $L$ like $cov$ tells us how close together nodes are. Thus, we need to invert it. If you recall Section \ref{sec:rw-effectres}, $L$ is singular and singular matrices -- by definition -- cannot be inverted. But we're not too picky about what ``inverting'' means, so we take the Moore-Penrose pseudoinverse.

So, to sum up, the Laplacian's $\delta$ function is:

$$ \delta(p, q, G) = ((p - q)^T L^\dagger (p - q))^{1/2},$$

with $L^\dagger$ being the pseudoinverse of the graph Laplacian of $G$.

\subsection{Markov Chain}
\begin{figure}[b]
\centering
\includegraphics[width=.5\textwidth]{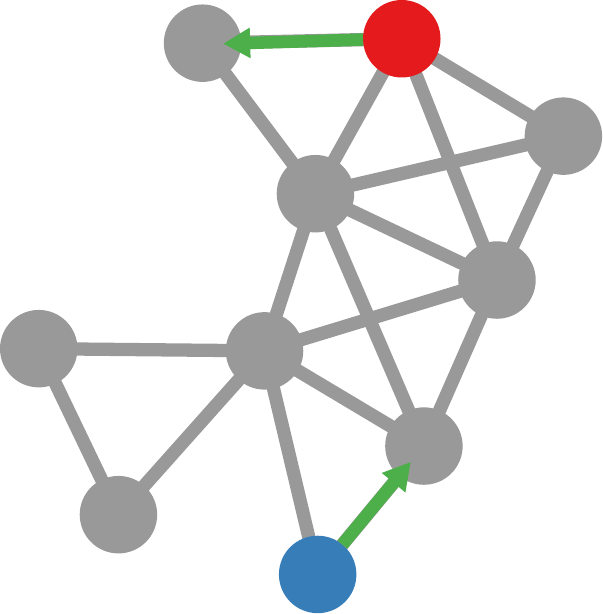}
\caption{To estimate the distance between the red and the blue node, we could release random walkers (in green) from them and calculate how much time it will take for them to arrive at the other node.}
\label{fig:nvd-mmc}
\end{figure}

Random walks are helpful to estimate node-node distances (Section \ref{sec:centr-eigen}). If we are in the situation of Figure \ref{fig:nvd-mmc}, we could estimate the distance between the red and blue node by simply asking how long it will take for a random walker to go from one node to the other. Here, we generalize this idea to groups of nodes.

In the Markov Chain distance we start from the assumption that, given a starting point $p$, by looking at $G$ we can construct an expected ``next step'', which is $E(q) = Ap$. This expected behavior follows a simple random walk (which is a Markov process, see Section \ref{sec:prob-markov}). In other words, we expect that $q$ should be the result of a one step diffusion of $p$ via random walks. For this to be the case, $A$ needs to be the stochastic adjacency matrix. Here, we also set the diagonal of the adjacency matrix to be equal to one before we transform it in its stochastic version. This is equivalent to add a self loop to all nodes in the network: we want to allow the diffusion process to stand still in the nodes it already occupies.

For each node $u$ in the network, we can calculate the expected occupancy intensity in the next time step by unrolling the previous formula: $E(q_u) = \sum \limits_{v \in V} A_{u,v} p_v$. This is helpful, because it allows us to calculate the standard deviation of this expectation, $\sigma_{u,v}$, which we can do by making a few assumptions on the distribution of this expectation (which are spelled out in the original paper\cite{coscia2020node}). For now, suffice to say that such deviation is $\sigma_{u,v} = (p_v A_{u,v} (1 - A_{u,v}))$.

Since we now have an expectation and a deviation, we can calculate the z-score of the observation, which is a measure of how many standard deviations your observation is distant from the expectation. We calculate a z-score for each node in the network and place it in its corresponding spot in a diagonal matrix:

$$ [Z]_{u,u} = \sum \limits_{v \in V} \sigma_{u,v}^2.$$

Now, the problem of this formulation is that it'd make this distance not symmetric, because to build $\sigma_{u,v}$ we only used $p$ as the origin of the diffusion. So, if $\sigma_{u,v}$ is the deviation of the diffusion from $v$ to $u$, we can also calculate a deviation of the diffusion from $u$ to $v$: $\sigma_{v,u}$. This is done as above, switching $p$'s and $q$'s places. Then the $u,u$ entry in $Z$'s diagonal is the sum of $\sigma_{u,v}^2$ and $\sigma_{v,u}^2$.

Finally we can write our distance as:

$$ \delta_{p, q, G} = ((p - q)^T Z^{-1} (p - q))^{1/2}.$$

In this distance measure you can tune the propagation duration. Maybe you don't want to make one-step random walks by using simply the stochastic matrix $A$. Maybe you want to have two steps random walks. In this case, you would use $A^2$. To ensure that the two farther apart nodes can still reach each other, you could consider to use $A^\ell$, with $\ell$ being the diameter of the network (Section \ref{sec:shortpath-avglength}). If you were to take $A^\infty$, then you'd be using the stationary distribution (Section \ref{sec:rw-stationary}). In that case, the topology of $G$ doesn't matter any more: the only thing that makes two nodes closer or farther is their degree.

\subsection{Annihilation}
Let's take that last thought a bit further. If you let your $p$ and $q$ vectors to diffuse via random walks for an infinite amount of time, they will distribute themselves to all nodes of $G$ proportionally to their degree, because they will both tend to approximate the stationary distribution. In the wavy water basin I mentioned before, $p$ and $q$ are simply two different waves conditions, while the stationary distribution is... well ... the stationary distribution: a waveless basin where the water is at the same level everywhere. Mathematically, this means that $\sum \limits_{k=0}^{\infty} A^kq$ and $\sum \limits_{k=0}^{\infty} A^kp$ are the same thing, or $ \sum \limits_{k=0}^{\infty} A^k(p - q) = 0$.

Now, the interesting bit is for which value of $k$ this is true or, put in another words, how fast will $p$ and $q$ cancel out. If they cancel each other out quickly, it means that they were already pretty similar to begin with. In fact, that equation would be true at $k = 0$ if $p = q$. So we're interested in the \textit{speed} of that equation. This is given us by the following formula:

$$ \delta_{p, q, G} = ((p - q)^T \sum \limits_{k=0}^{\infty} A^k (p - q))^{1/2}.$$

You shouldn't be scared of the infinite sum $\sum \limits_{k=0}^{\infty} A^k$: it converges and there is a proper solution to it, which is in the survey paper I've been citing in this chapter.

\section{Shortest Path Based}
The solutions based on shortest paths start from the assumption that the problem of establishing distances between sets of nodes can be generalized from solving the problem of finding the distance between pairs of nodes. This is a well understood and solved problem: using a shortest path algorithm -- for instance Dijkstra's\cite{dijkstra1959note} -- one can count the number of edges separating node $u$ to $v$. 

\begin{figure}[b]
\centering
\includegraphics[width=.5\textwidth]{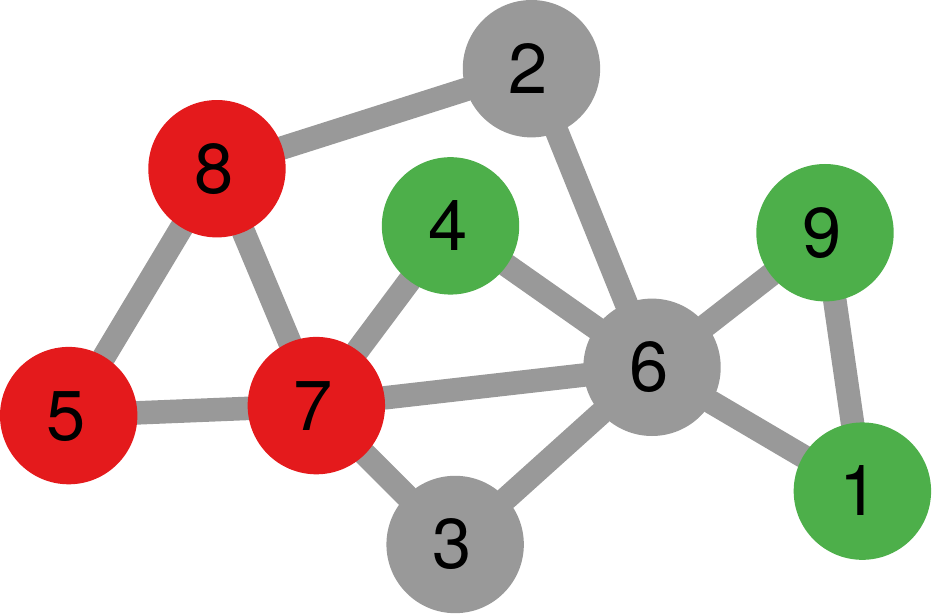}
\caption{A network where the red nodes represent the origins and the green nodes represent the destinations.}
\label{fig:nvd-spl-toy}
\end{figure}

If we take Figure \ref{fig:nvd-spl-toy} as an example, we'd start by collecting a bunch of distances: $[2,3,3]$ when starting from node $5$ or node $8$, and $[1,2,2]$ when starting from node $7$. We then want to aggregate these distances with a given strategy, to define several functions solving the problem.

Since this section deals with shortest paths, a useful convention is to refer to all possible paths between origins and destinations as $P_{p, q}$. This is to avoid to calculate all shortest paths between all pairs of nodes in the network, which is computationally expensive and not necessary, since we don't need the distances between nodes that have a zero value in both $p$ and $q$. A path length $|P_{u,v}| \in P$ is the minimum number of edges required to cross to move from node $u$ to node $v$.

There are two subcategories in this group: methods which try to optimize the paths from $p$ to $q$, and methods which do not. We start from the latter.

\subsection{Non-Optimized}
Here we show a set of possible aggregations of shortest path distances between the nodes in $p$ and $q$, by taking hierarchical clustering as an inspiration. There are of course more strategies than the ones listed here, but I can't really list them all -- and most haven't really been researched yet.

\begin{figure}
\centering
\includegraphics[width=.5\textwidth]{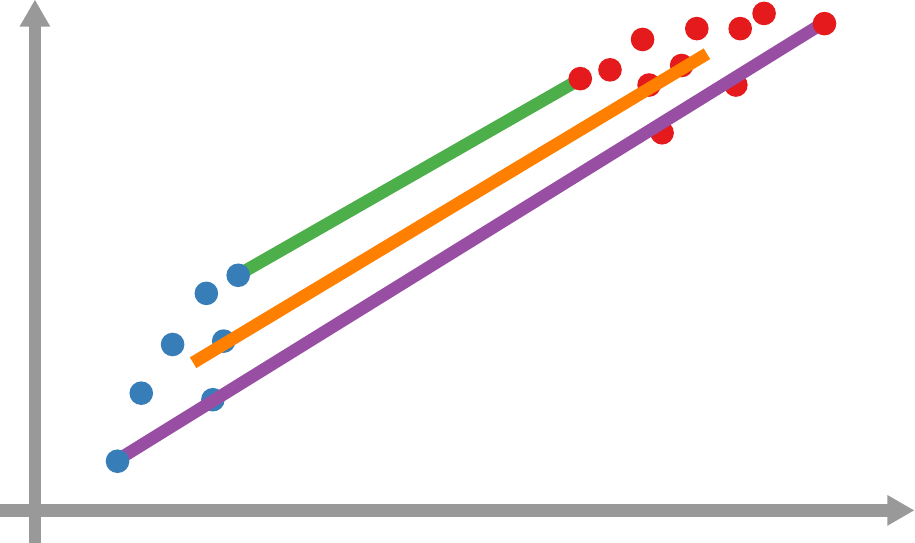}
\caption{Different linkage strategies to estimate the distances between clouds of points: single (green), complete (purple), and average (orange).}
\label{fig:nvd-linkage}
\end{figure}

When performing hierarchical clustering, there are three common ways to merge clusters according to their distance\cite{szekely2005hierarchical}: single, complete, and average linkage. Single linkage (green in Figure \ref{fig:nvd-linkage}) means that the distance between two clusters is the distance between their two closest points. On the other hand, complete linkage (purple in Figure \ref{fig:nvd-linkage}) considers the distance of the two farthest points as the cluster distance. In average linkage (orange in Figure \ref{fig:nvd-linkage}), one calculates the average distance between all pairs of points in the two clusters as the distance between the clusters.

Similarly, our aim is to reach the destination from the origin in the minimum distance possible. In the single linkage strategy, the ``cost'' of reaching a destination node is the distance of it from the closest possible origin node. First, we need to make sure that $\sum p = \sum q$. If that isn't the case, we rescale up the vector with the smallest sum so that this equation is satisfied. For instance, if $q$ had a lower sum, we transform it: $q' = (\sum p / \sum q) q$.

Then we start a loop. At each iteration, we want to find the pair of closest nodes ($\argmin \limits_{u,v} |P_{u,v}|$) that can exchange the largest possible value. How much value can two nodes exchange? A node can only give what they have, so we will exchange the minimum of the two values: $\min(p_u, q_v)$. The contribution of this move to the distance is $|P_{u,v}| \min(p_u, q_v)$: we move $\min(p_u, q_v)$ across $|P_{u,v}|$ edges. Once the value is exchanged, we update $p_u$ and $q_v$ to reflect the successful transaction: $p_u = p_u - \min(p_u, q_v)$ and $q_v = q_v - \min(p_u, q_v)$. Eventually, all values would have been transferred, because we ensured that the two vectors sum to the same value, and the iterations will stop.

The complete linkage uses the very same operation: the only difference is using at each step $\argmax \limits_{u,v} |P_{u,v}|$ instead of $\argmin \limits_{u,v} |P_{u,v}|$. This means that we preferentially exchange value between the node pairs that are \textit{farthest}, not closest.

The average linkage is conceptually simpler: it is the weighted average path distance between all $u \in p$ and all $v \in q$:

$$ \delta_{p, q, G} = \dfrac{\sum \limits_{\forall v \in q} \sum \limits_{\forall u \in p} p_u q_v |P_{u,v}|}{\sum p}.$$

Here it doesn't matter what we put in the denominator, since we already ensured that $p$ and $q$ sum to the same value.

\begin{figure}
\centering
\includegraphics[width=.5\textwidth]{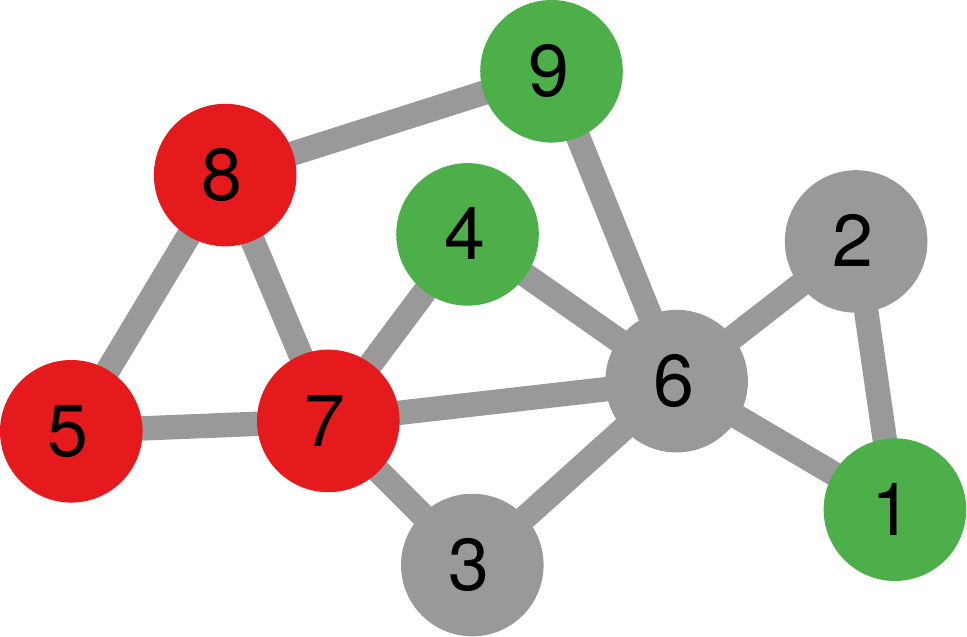}
\caption{A network where the red nodes represent the origins and the green nodes represent the destinations.}
\label{fig:nvd-spl-toy2}
\end{figure}

Looking at Figure \ref{fig:nvd-spl-toy2}, we can now estimate the different distances between red and green nodes. In single linkage, we try to find the shortest path to the closest destination from each origin. Origin $8$ goes to destination $9$ because they are directly connected, and so does origin $7$ with destination $4$. Origin $5$ has to take a path of length $3$ to reach destination $1$: $5 \rightarrow 7 \rightarrow 6 \rightarrow 1$. Thus, for single linkage, $\delta_{p, q, G} = 1 + 1 + 3 = 5$. If we normalize the vectors beforehand, each step counts for $1/3$, and thus the distance would be $5 / 3 = 1.\bar{6}$.

The average linkage looks at all nine shortest paths, and calculates an average. The total length of all shortest paths is $18$. It then normalizes with the total moved weight: $\sum p = 3$. Thus the average linkage estimates the distance as $18 / 3 = 6$. If we normalized the vectors, we would again count each path as contributing one third, i.e. $(18 / 3) / 3 = 2$.

In complete linkage, we perform a similar operation as in single linkage, but looking at the farthest destination for each origin. The farthest destination is $5 \rightarrow 1$, at three steps; then $8 \rightarrow 4$ and $7 \rightarrow 9$ at two steps each. Thus, complete linkage will return $3 + 2 + 2 = 7$ as distance. If we normalized the vectors, we would again count each path as contributing one third, i.e. $7 / 3 = 2.\bar{3}$.

\subsection{Optimized}
Here we try to be a bit smarter than the aggregation strategies we saw so far. In this branch of approaches, we try to optimize this aggregation such that the number of edge crossing is minimized.

If there are no further constraints in this optimization problem, we are in the realm of the Optimal Transportation Problem (OTP) on graphs\cite{mcgregor2013sketching}. In its original formulation\cite{monge1781memoire}, OTP focuses on the distance between two probability distributions without an underlying network. However, it has been observed how this problem can be applied to transportation through an infrastructure, known as the multi-commodity network flow\cite{hitchcock1941distribution}. Specifically, one has to simply specify how distant two dimensions in the vector are. The distance needs to be a metric, and the number of edges in the shortest path between two nodes satisfies the requirement.

In its most general form, the assumption is that we have a distribution of weights on the network's nodes, and we want to estimate the minimal number of edge crossings we have to perform to transform the origin distribution into the destination one. This is a high complexity problem, which has lead to an extensive search for efficient approximations\cite{assent2006approximation}\cite{erbar2017computation}\cite{essid2017quadratically}\cite{karakostas2008faster}\cite{maas2011gradient}\cite{pele2008linear}\cite{pele2009fast}\cite{solomon2016continuous}. For what concerns us, all these methods are equivalent: they all solve OTP and the difference between them is how they perform the expensive optimization step. Thus, they all return a very similar distance given $p$, $q$ and $G$ -- plus or minus some approximation due to their optimization strategy --, and fall in the same category.

More formally, in OTP we want to find a set of movements $M$ such that:

$$ M = \argmin \limits_{m_{p_u,q_v}} \sum \limits_{p_u} \sum \limits_{q_v} m_{p_u,q_v} d_{u,v},$$

where $p_u$ and $q_v$ are the weighted entries of $p$ and $q$, respectively; $m_{p_u,q_v}$ is the amount of weights from $p_u$ that we transport into $q_v$; and $d_{p_u,q_v}$ is the distance between them. Then:

$$ \delta_{p, q, G} = \dfrac{\sum \limits_{p_u} \sum \limits_{q_v} m_{p_u,q_v} d_{u,v}}{\sum \limits_{p_u} \sum \limits_{q_v} m_{p_u,q_v}}, $$

where the $m_{p_u,q_v}$ movements come from the $M$ we found at the previous step. The differences between the methods cited before almost exclusively lie in the strategy to find the optimal $M$. The thing left to determine in the $\delta$ formula is the distance function $d_{u,v}$ between pairs of nodes. As mentioned previously, we choose this to be the length of the shortest path in $G$ between $u$ and $v$, or: $d_{u,v} = |P_{u,v}|$. This is zero if $u = v$.

There \textit{could} be additional constraints to this optimized many-to-many distance. For instance, we could have the constraint that, while we are moving something from node $u$ to node $v$ via the edge that connects them, nothing else can pass through that edge. In practice, we are simulating an actual physical transportation system, in which edges and nodes have capacities. If we want to move two values at the same time through the same edge, we need to re-route one of them, because the edge -- or the node -- is occupied. How to find the optimal way to solve this problem is the realm of Multi-Agent Path Finding (MAPF)\cite[-1in]{goldreich2011finding}\cite[-0.3in]{yu2015pebble}.

\begin{figure}
\centering
\includegraphics[width=.5\textwidth]{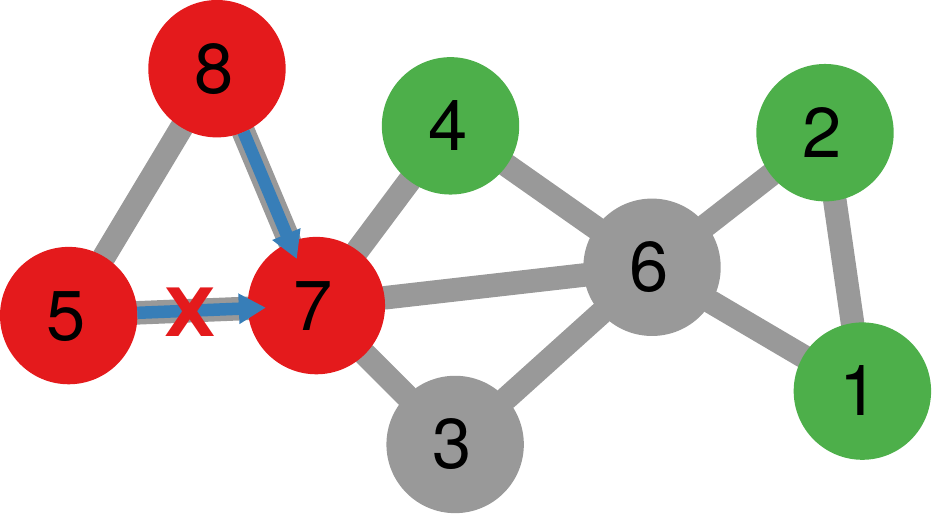}
\caption{Attempting to find a MAPF solution from red nodes to green nodes. Blue arrows show two attempted moves that cannot be executed at the same time.}
\label{fig:nvd-spl-toy-mapf}
\end{figure}

Figure \ref{fig:nvd-spl-toy-mapf} shows an example of this problem: we want to go from node $8$ to node $2$ and from node $5$ to node $1$. Unfortunately, the shortest paths for these two objectives both involve passing through node $7$ at the same time. This cannot happen, since node $7$ can host only a single walker at a time. Thus, either the walker in $8$ or the walker in $5$ needs to wait for a bit until node $7$ is clear again.

In general MAPF, you have multiple robots occupying one node at a time and they each have a specific node as their intended destination\cite{foerster2017multi}. This is slightly different from our problem, where each weight in $p$ can potentially reach any other destination in $q$. So one has to determine, before running MAPF, which $u \in p$ should go to which $v \in q$. One solution is to simply use the same strategy we used for single linkage in the non-optimized shortest path category: we look for the shortest path length $|P_{u,v}|$ carrying the largest possible weight $\min(p_u, q_v)$.

Moreover, since in MAPF robots cannot be in the same node at the same time, you still have a problem. Say that we assigned a robot to go from $u$ to $v$ in our preprocessing. If $p_u \neq q_v$, then either $u$ or $v$ has some unallocated weight. Thus we would need to add at least a second robot that can either start in $u$ or terminate in $v$. But this violates MAPF. The way we solve the issue is by running a sequence of MAPF sessions. In each session, we attempt to move all the weights that were left over during the previous session. We keep running smaller and smaller sessions until all weights have been allocated -- which we can guarantee by normalizing either $p$ or $q$ so that they sum to the same value, as we did in the non-optimized solutions.

There are many algorithms to solve MAPF\cite[-0.3in]{godoy2015adaptive}\cite{snape2011hybrid}\cite{wagner2015subdimensional}\cite{dobson2017scalable}\cite[1in]{ma2017multi}\cite{walker2017using}\cite{yakovlev2017any}\cite{walker2018extended}\cite{li2019multi}\cite{andreychuk2019multi}\cite{liu2019task}, each of them providing a different solution to NVD with our preprocessing strategy.

\begin{figure}
\centering
\includegraphics[width=.5\textwidth]{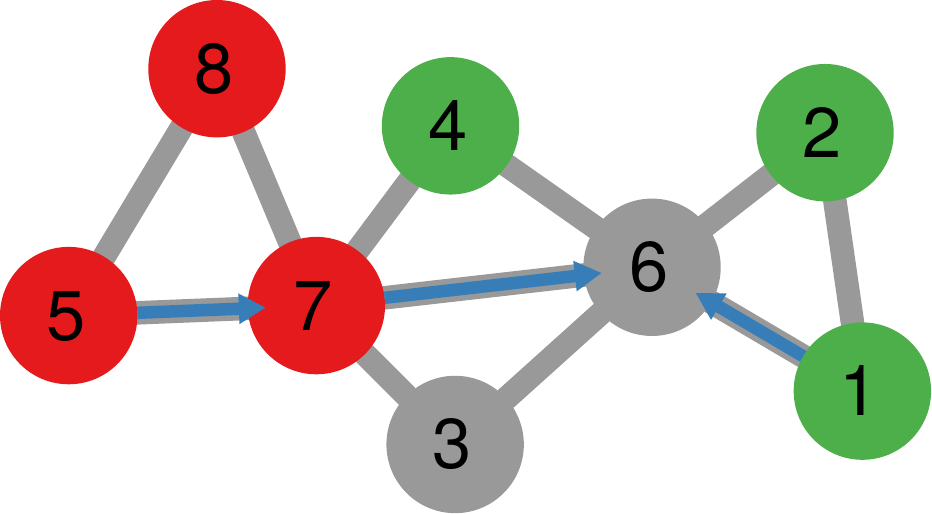}
\caption{Attempting to find a pursue solution from red nodes to green nodes. Blue arrows show attempted moves.}
\label{fig:nvd-spl-toy-pursue}
\end{figure}

Another variant to OTP is pursuit-evasion games. In these games, we populate a space with a set of robots. Some robots are pursuers and they aim at capturing the other robots, the evaders. In discrete pursuit-evasion (DPE) we force the robots to move through the nodes and edges of a graph, rather than in a Euclidean space\cite{parsons1978pursuit}. Many algorithms have been proposed to model different strategies and constraints both on the pursuer and on the evader side.

One can see how it is possible to adapt DPE to solve our distance problem. First, we set the pursuers as $p$ and the evaders as $q$. Then we run any DPE solving algorithm. Alternatively, both $p$ and $q$ are sets of pursuers and try to capture each other, with no evasion. Figure \ref{fig:nvd-spl-toy-pursue} is an example: here nodes $5$ and $1$ are trying to capture each other. Every time pursuers capture each other, the one carrying the $\min(p_u, q_v)$ weight disappears and the other one carries its own weight minus $\min(p_u, q_v)$. The amount of time/moves it takes for all weights to disappear is the distance between $p$ and $q$. Since $p$ and $q$ sum to the same value -- either by normalization or by the usual expansion strategy --, the system will terminate.

There are a number of solutions to DPE, satisfying a vast number of different constraints\cite{akhoondian2015graph}\cite{alspach2006searching}\cite{fomin2008annotated}\cite{luccio2007intruder}\cite{mejia2019solutions}\cite{stiffler2016pursuit}.

\section{Graph Fourier Transform}
There are other ways to solve the node vector distance problem in networks that do not start either from shortest paths nor from a generalization of Euclidean distances. They are generally linked to the spectrum of the graph, since the spectrum can be used to describe diffusion processes on the network, and the node vector distance is a type of diffusion process.

In the signal processing literature, a common scenario is one where the analyst has a battery of sensors, whose readings are correlated with each other. In order to extract the actual signal $\hat{s}$ from the noisy and correlated signal data $s$, these relationships between sensor outputs have to be taken into account. The relationships can be modeled with a network $G$ connecting related sensors. Then, the outputs are smoothed using the Graph Fourier Transform $\hat{s} = \Phi^T s$\cite{hammond2011wavelets}\cite{shuman2016vertex}.

I explained what the Graph Fourier Transform is in Section \ref{sec:deep-mpgnn-spectral}. For the purposes of this section, suppose that our signal -- the vector assigning each node to a value -- is $p$. Then the corrected signal is equal to: $\widehat{p} = \Phi^T p$ -- remember that $\Phi$ is the matrix of $L$'s eigenvectors. With this transformation, $\widehat{p}$ tells us how much each of the eigenvectors contributes to $p$. In other words, we are changing our representation from the ``spatial nodes'' ($p$) to the ``frequency modes'' ($\widehat{p}$). We can now weight the modes so that we take into account the topology of the graph. This is usually achieved by filtering the signal in the spectral domain, multiplying it with the diagonal matrix of the Laplacian's eigenvectors $\Lambda$.

Once we apply this transformation to both $p$ and $q$, we have encoded $G$'s topology in the vectors. The Euclidean distance between them is the node vector distance that we are looking for:

$$ \delta_{p, q, G} = Euclidean(p \Lambda \Phi^T , q \Lambda \Phi^T).$$

Note that this is not one, but a family of measures. One could replace the Euclidean distance with any other off-the-shelf measure (cosine, correlation, etc) to estimate the distance between the filtered $p$ and $q$, because they already contain $G$'s topology in their values.

These approaches can be used to establish the distance between two different signals on a graph. However, this is but one of the applications of graph signal processing. Other scenarios include signal cleaning\cite{hagmann2008mapping}, frequency analysis\cite{sandryhaila2014discrete}, sampling\cite{anis2014towards}, interpolation\cite{narang2013localized}, and trend filtering\cite{wang2016trend}, to cite a few. This also means that the transformation proposed here might not be the optimal one, and it is for sure not the only one.

\section{Statistics on a Network Space}\label{sec:nvd-stats}
While reading this chapter, you might have been wondering something. If we have defined an analogue to the Euclidean distance when the network represents the space in which our observations live, can we do the same thing for other statistical measures? The answer turns out to be yes. I know of two works in this direction. One redefines the concept of variance for node vectors, while the other works with a generalization of normalized co-variance -- also known as the Pearson linear correlation.

\subsection{Network Variance}
We introduced the concept of variance all the way back in Section \ref{sec:stats-summary}. We said it quantifies how much we expect any random observation to differ from the average (squared). One way to write this formula is\cite{devriendt2022variance}:

$$ var(x) = \dfrac{1}{2} \sum \limits_{u,v} x_u x_v d^2_{uv}, $$

assuming $u$ and $v$ are all possible pairs of observations in our distribution $x$ -- and so $x_u$ is the value of $x$ for observation $u$. Here, $d_{uv}$ is the Euclidean distance between observations $u$ and $v$. But what if $u$ and $v$ are nodes in a graph $G$? Then their distance should not be the Euclidean distance, but a graph distance. So, if for $d_{uv}$ we take a graph distance, we have just defined a notion of network variance.

Luckily, it is very easy to define a graph distance: $d_{uv}$ could simply be the shortest path between $u$ and $v$. A better option, though, is to use the effective resistance between the two nodes (Section \ref{sec:rw-effectres}). This is because the effective resistance is a proper metric and it uses random walks rather than shortest paths. The latter is desirable, because it makes effective resistance less sensitive to small changes in the structure -- such as the addition or the removal of an edge -- which can dramatically change the shortest path distance.

\begin{figure}
\centering
\begin{subfigure}[t]{.4\columnwidth}
\includegraphics[width=\textwidth]{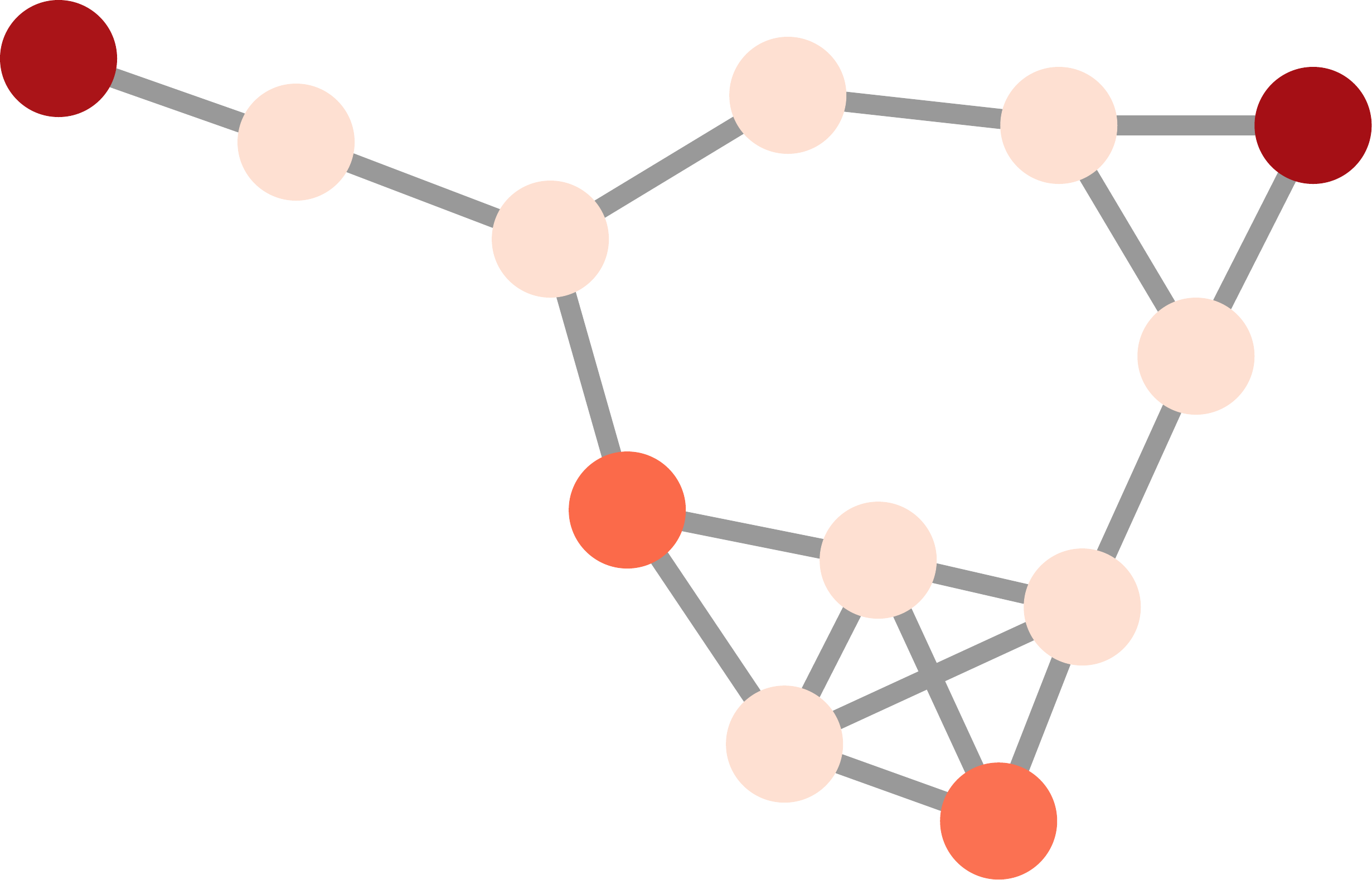}
\caption{$var(x) = 2.92$}
\end{subfigure}
\qquad
\begin{subfigure}[t]{.4\columnwidth}
\includegraphics[width=\textwidth]{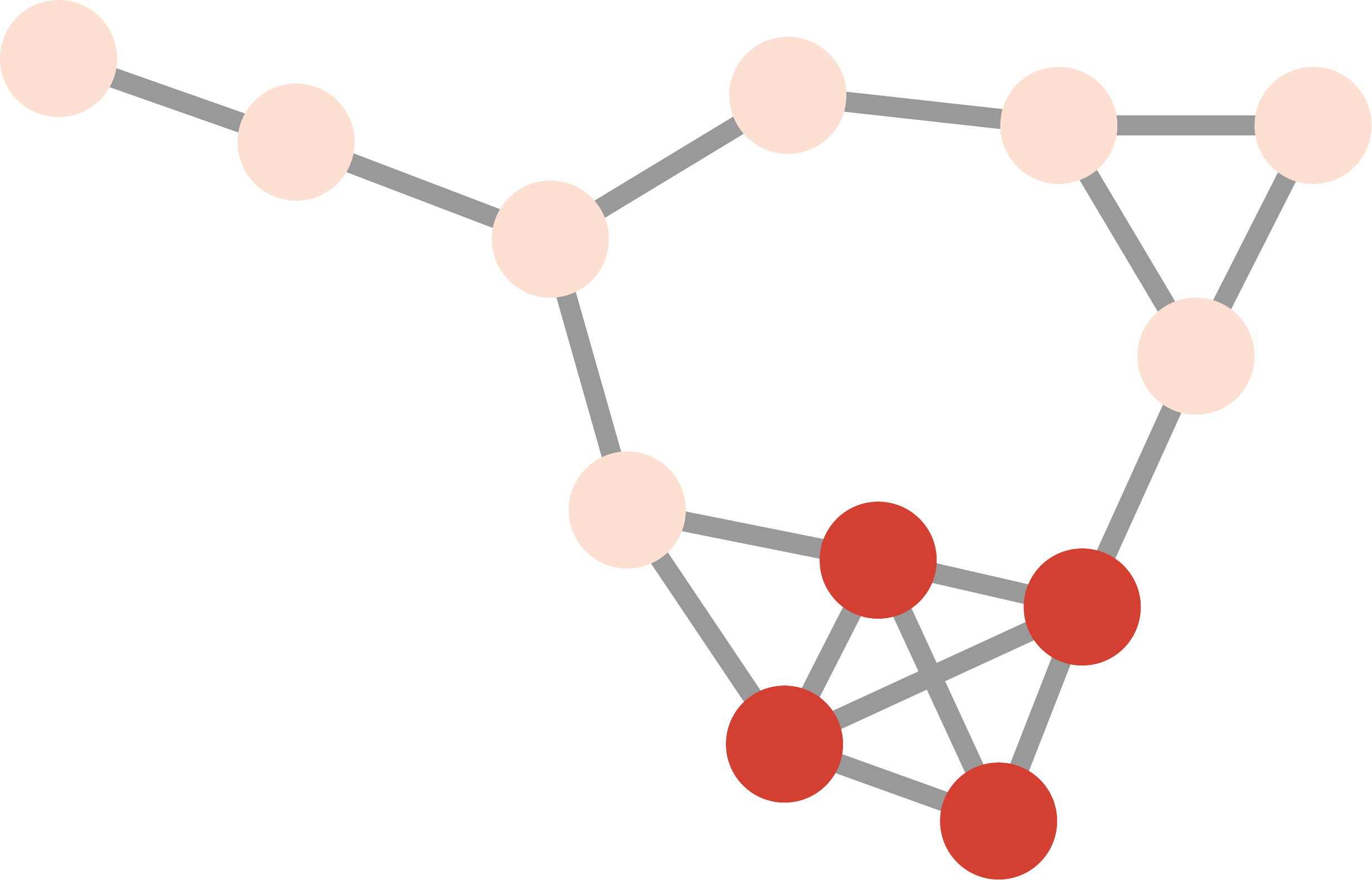}
\caption{$var(x) = 0.08$}
\end{subfigure}
\caption{Two different attributes on the same network. The darker the node, the higher the value on that node for a given attribute.}
\label{fig:network-variance}
\end{figure}

Figure \ref{fig:network-variance} shows an example, which should help you have a visual understanding of what it means to calculate a network variance. We have the same graph, but two different variables defined on its nodes. Intuitively, the variable I show in Figure \ref{fig:network-variance}(a) has a high variance, because it is scattered around the periphery of the graph. On the other hand, the variable I show in Figure \ref{fig:network-variance}(b) has a low variance, because it clusters in a small subpart of the graph. The network variance estimated via effective resistance confirms this intuition, giving a high value for Figure \ref{fig:network-variance}(a) and a low one for Figure \ref{fig:network-variance}(b) -- which I report in the subcaptions.

\subsection{Network Correlation}
We can use a similar trick to extend the familiar Pearson correlation coefficient (Section \ref{sec:stats-corr}) to variables defined on a network\cite{coscia2021pearson}. Let's suppose that we have two vectors, $x$ and $y$, each assigning a value to a node in the graph $G$. Each node $v$ in $G$ contributes something to the Pearson correlation, and that something is the relation between its corresponding values $x_v$ and $y_v$. If, across all $v$s, every time we have a high $x_v$ value we also have a high $y_v$ value, and vice versa, then the Pearson correlation is positive and high.

But, wait, if we are in a graph $G$, why should we only weigh $v$'s values to determine the correlation coefficient? The basic assumption that is at the basis of literally everything in network science is that the network structure influences the behavior of the nodes -- and vice versa. With this assumption we must conclude that $v$'s neighbors influence the $x_v$ value, and therefore they should play a role in the correlation.

\begin{figure}
\centering
\begin{subfigure}{.23\columnwidth}
\includegraphics[width=\columnwidth]{{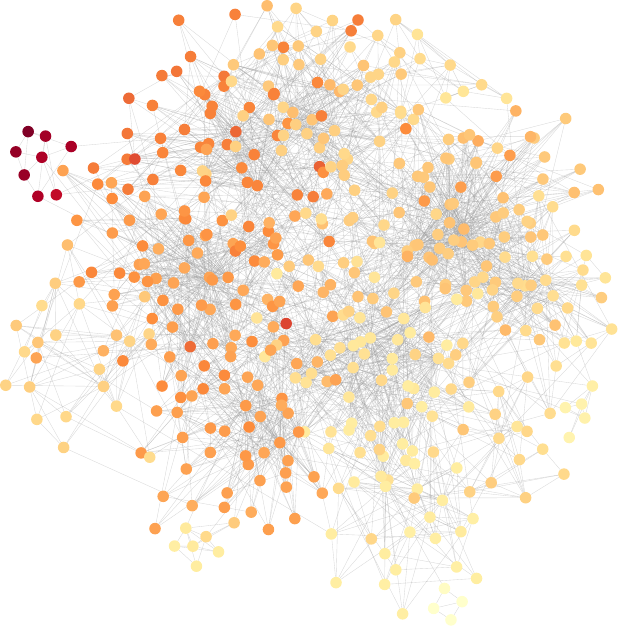}}
\caption{}
\end{subfigure}
\begin{subfigure}{.23\columnwidth}
\includegraphics[width=\columnwidth]{{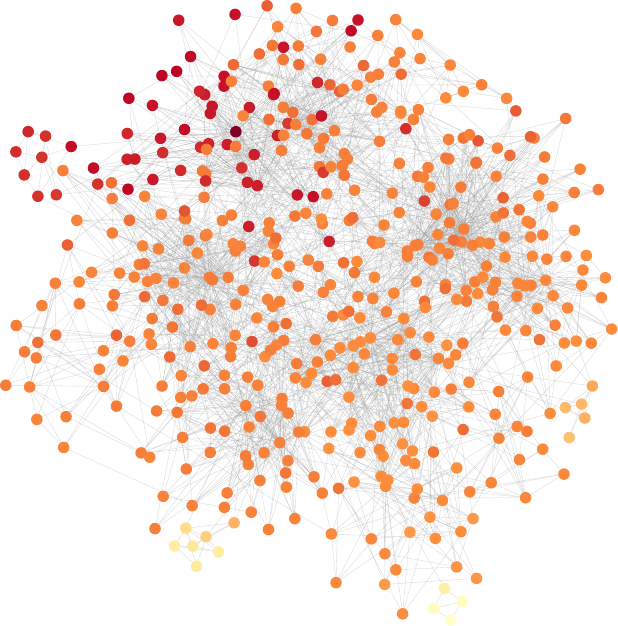}}
\caption{}
\end{subfigure}
\begin{subfigure}{.23\columnwidth}
\includegraphics[width=\columnwidth]{{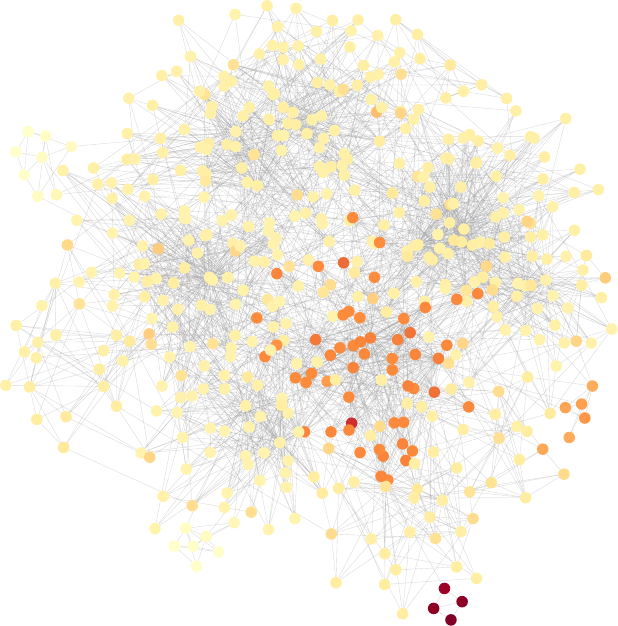}}
\caption{}
\end{subfigure}
\begin{subfigure}{.23\columnwidth}
\includegraphics[width=\columnwidth]{{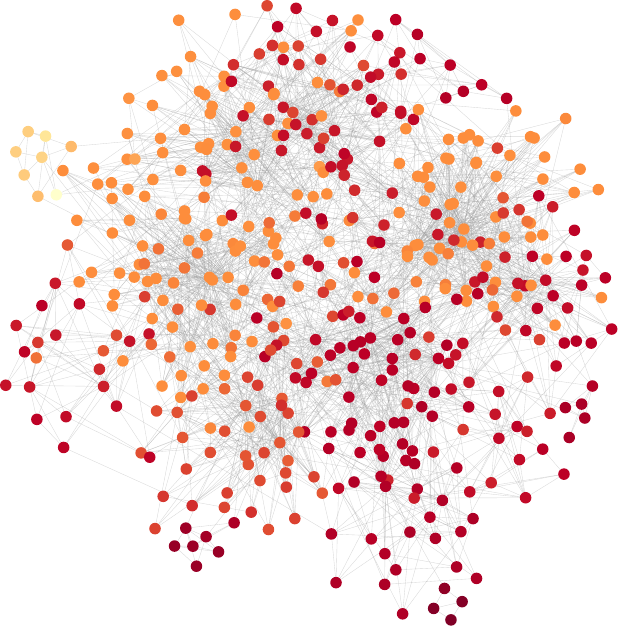}}
\caption{}
\end{subfigure}
\caption{Four different node vectors on the same network. The node's color represents the value in $x$, from dark red (high) to bright yellow (low). (a) A node vector $v$. (b) A node vector with a positive network correlation with $v$. (c) A node vector with no correlation with $v$. (d) A node vector with an anti-correlation with $v$.}
\label{fig:network-correlation-toy}
\end{figure}

Figure \ref{fig:network-correlation-toy} gives you a visual understanding of what a network correlation means for a node vector. For a given node vector $v$ in Figure \ref{fig:network-correlation-toy}(a), I show you node vectors with positive (Figure \ref{fig:network-correlation-toy}(b)), no (Figure \ref{fig:network-correlation-toy}(c)), and negative (Figure \ref{fig:network-correlation-toy}(d)) network correlations.

The neighbors' role should be inversely proportional to their distance from $v$: farther nodes exert a weaker influence -- indirectly through their neighbors, and neighbors of neighbors and so on. This distance is usually estimated via -- again -- the effective resistance matrix as it was the case for network variance. In this case one should not use the shortest path lengths, because they are not a proper metric on a network and they could lead to mathematical errors\cite{coscia2024pearson}.

\section{Summary}

\begin{enumerate}
\item The node vector distance problem is the quest for finding a way to estimate a network distance between two vectors describing the degree of occupancy of the nodes in the network. If at time $t$ I occupy nodes $1,2$, and at time $t + 1$ I occupy nodes $3,4,5$, how much did I move in the network?
\item The Euclidean distance can be used to estimate distances between vectors in a homogeneous space. Thus, a family of solution focuses on ``warping'' the space so that it is described by the topology of the network. At that point, you can use the Euclidean distance on such a warped space.
\item Another family of solutions uses shortest paths: you calculate all shortest paths between all nodes of origin and destination, and you aggregate the results somehow. Alternatively, you can try to find only those shortest paths minimizing the resulting distance.
\item You can add several constraints to the optimized shortest path strategy. For instance, you could model a real infrastructure network: nodes and edges have finite capacity.
\item Finally, you can use signal cleaning techniques. You can see your network as describing sets of sensors that return correlated results. Thus, two ``signals'' are far apart if they are reported by uncorrelated sensors, which are not connected to each other.
\item You are not limited to calculating distances between node vectors. One can also extend the basic concepts of variance and correlation to the case in which the vector lives in the complex space defined by a graph.
\end{enumerate}

\section{Exercises}

\begin{enumerate}
\item Calculate the distance between the node vectors in \url{http://www.networkatlas.eu/exercises/47/1/vector1.txt} and \url{http://www.networkatlas.eu/exercises/47/1/vector2.txt} over the network in \url{http://www.networkatlas.eu/exercises/47/1/data.txt}, using the Laplacian approach. The vector files have two columns: the first column is the id of the node, the second column is the corresponding value in the vector. Normalize the vectors so that they both sum to one.
\item Calculate the distance using the same data as the previous question, this time with the average linkage shortest path approach. Normalize the vectors so that they both sum to one.
\item Calculate the distance using the same vectors as the previous questions, this time on the \url{http://www.networkatlas.eu/exercises/47/3/data.txt} network, with both the average linkage shortest path and the Laplacian approaches. Are these vectors closer or farther in this network than in the previous one?
\item Calculate the variances of the vectors used in the previous exercises on both networks used in the previous exercises.
\end{enumerate}

\chapter{Topological Distances}\label{cha:netsimil}
In the previous chapter we learned how to estimate the distance between two vectors describing the occupancy of sets of nodes in the same network. In that problem, you get two vectors with $|V|$ entries, and you calculate their distance on the \textit{same} network topology. We called this ``node vector distance'', a certain type of ``network distance''. There are other ways one could interpret the term ``network distance''. I group them all in this chapter.

Specifically I talk about:

\begin{itemize}
\item Network similarity (Section \ref{sec:netsimil-simil}): how to tell if two graphs $G_1$ and $G_2$ have a similar topology;
\item Network alignment (Section \ref{sec:netsimil-alignment}): finding nodes in $G_1$ that are similar to nodes in $G_2$, so that we could couple them and consider $G_1$ and $G_2$ as two layers of a multilayer network;
\item Network fusion (Section \ref{sec:netsimil-fusion}): given multiple observations of a network, combine them to create a summary that is the most similar to all observations.
\end{itemize}

\section{Network Similarity}\label{sec:netsimil-simil}
By far, the most common and popular way to intend the term ``network distance'' is as the opposite of the similarity between two networks. The term ``network similarity'' is, unfortunately, rather ambiguous, and you might find papers dealing with very different problems but using the same terminology. For instance, one could intend ``network similarity'' as a measure of how similar two nodes are (see Section \ref{sec:centr-similarity}). Or one could be talking about ``similarity networks'', which are ways to express the similarities between different entities by connecting the ones that are the most similar to each other -- something you might do via bipartite projections (Chapter \ref{cha:projections}).

Here, we focus on a different problem. The idea here is simple: we have two networks $G_1$ and $G_2$ and we want to know how similar the two are. Namely, how easy it is to mistake $G_1$ as $G_2$ by looking at their edges. There are many ways to do this, and I'll try to give a general overview.

Most of the applications of these techniques are in biology and chemistry. The idea is to compare networks describing specific pathways. However, there are also more peculiar applications, for instance in malware detection\cite{runwal2012opcode} and image recognition\cite{sorlin2005reactive}. I am going to include the approaches used mostly for practical problems in computer science. However, there are many more distance measures that have a more distinctively ``mathy'' flavor. The bible for this kind of things is for sure the Encyclopedia of Distances\cite{deza2009encyclopedia}. Some examples of distance measures you can find there are the Chartrand-Kubicki-Schultz distance\cite{chartrand1998graph}, the rectangle distance\cite{borgs2006graph}, and many more others.

\begin{figure}
\centering
\includegraphics[width=.9\textwidth]{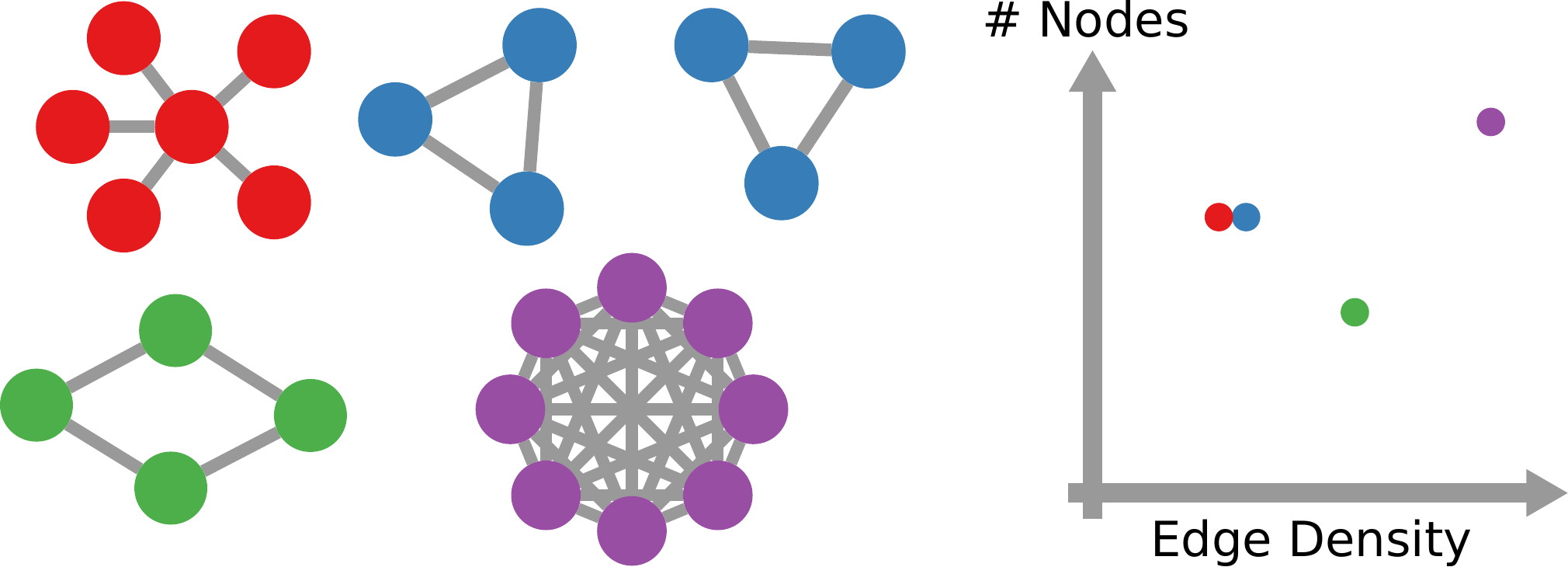}
\caption{On the left we have four graphs, each identified by the color of its nodes. On the right, I make a two dimensional projection by recording each graph's node count (y axis) and edge density (x axis). The similarity between two graphs is the inverse of their distance in this space.}
\label{fig:netdist-collapse}
\end{figure}

At a practical level, all the methods that follow have one thing in common. Comparing two networks using a handful of summary statistics means to define a low-dimensional space in which every network is a point. You can visualize that as a scatter plot: Figure \ref{fig:netdist-collapse} makes a super simple one where I decide to classify networks by their number of nodes and edge density. However, networks are a high-dimensional object, and the summary statistics commonly used in network science are usually non orthogonal -- differently from what you'd get from, for instance, Principal Component Analysis (Section \ref{sec:mat-factors}) --: in my case, from Section \ref{sec:density-sparse} you know that the number of nodes is usually negatively correlated with edge density.

The main issue is that we still don't know which set of network statistics is sufficient to cover the space of all possible networks. Whatever dimensions you use to organize your networks will collapse many -- possibly dissimilar -- networks into the same place in your scatter plot. This happens in Figure \ref{fig:netdist-collapse}, where a star (in red) is confused with a set of unconnected cliques (in blue). This is not necessarily a bad thing! If the summary statistics you chose are meaningful to you in some fundamental way, this is a feature. However, if you're hunting for ``universal'' patterns, this approach could mislead you.

\subsection{Global Property Comparison}
The most basic way to tell whether two networks are similar is by looking at their global properties\cite{li2011graph}. If two networks have the same degree distribution, the same average path length, the same clustering coefficient, the same average degree, and so on... Well, doesn't that mean that these two networks are.... the same?

This is a seducing option because, as we'll see, estimating the similarity between two networks by looking at their topology is computationally very hard. It is related to the graph isomorphism problem, and we saw that graph isomorphism is a though nut to crack in Section \ref{sec:mining-isomorph}. On the other hand, estimating many global properties is trivial and instantaneous in many cases, and well studied and optimized in others.

\begin{figure}
\centering
\begin{subfigure}{.36\columnwidth}
\includegraphics[width=\textwidth]{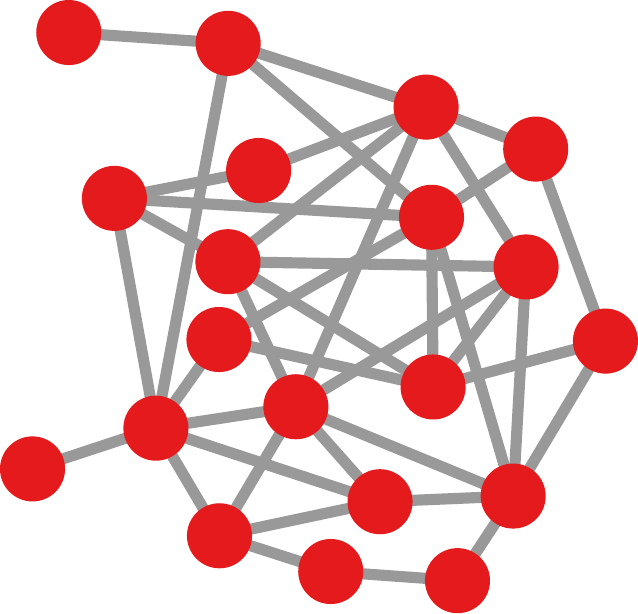}
\caption{}
\end{subfigure}\qquad
\begin{subfigure}{.33\columnwidth}
\includegraphics[width=\textwidth]{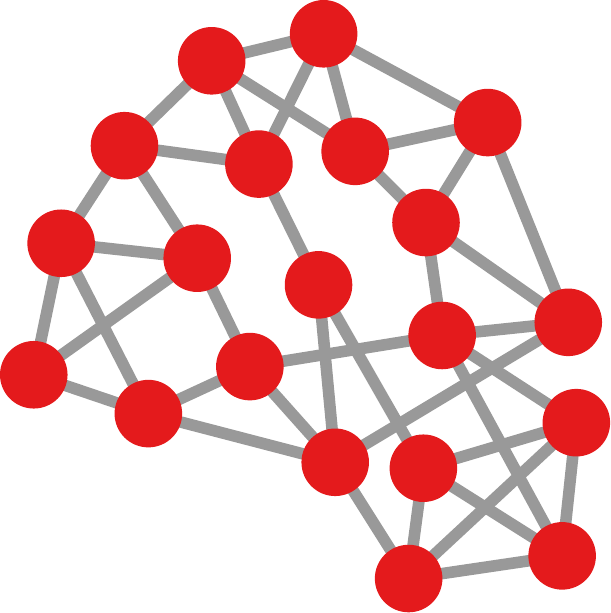}
\caption{}
\end{subfigure}
\caption{Two graphs of which we want to estimate the similarity.}
\label{fig:netsim-glob}
\end{figure}

Of course, you need to be extremely careful in considering two things. First, what are the global properties you're looking at? Second, how do you aggregate the differences between these properties to end up with a single measure of similarity? These are important questions, because you might end up considering as similar two networks that are very different. Consider Figures \ref{fig:netsim-glob}(a) and \ref{fig:netsim-glob}(b). The two networks have a lot in common: same number of nodes and edges (thus the average degree and density are the same as well). They have almost identical degree distributions, approximated by a Gaussian. They have the same diameter and a very similar average path length ($2.1$ vs $2.4$). Up until now, you'd consider them practically equivalent. And yet, they're still relatively different, as they were generated using two very different processes. Figure \ref{fig:netsim-glob}(a) is a $G_{n,m}$ random graph, while Figure \ref{fig:netsim-glob}(b) is a small-world graph. The crucial factor I forgot to check is the clustering coefficient, which is low for $G_{n,m}$ graphs ($0.17$ in this instance) and high for small-world networks ($0.41$ here).

\subsection{Pairwise Node Similarity}
A common approach is the estimation of all possible combinations of node similarities. This is a relatively popular way to attack the problem, which underlies many other techniques. The reason is that it is a natural way to think about network similarity: two networks are similar if they have the same nodes and these nodes connect to the same neighbors. Estimating all the pairwise node similarities is the first step to tell which nodes are the same. More often than not, that is the end goal of estimating network similarity: we might be less interested in how similar two networks are and more in which nodes from one network are the same nodes in the other. This is the problem of network alignment and we'll see it more in depth in Section \ref{sec:netsimil-alignment}.

We have seen dozens of ways to tell how similar two nodes are, both in Section \ref{sec:centr-similarity} and in Chapter \ref{cha:lp-simple}. The general idea is to try and estimate the structural equivalence of all nodes in the two graphs. Then you can either average all the node-node similarities you calculated, or find the best way to map nodes: for each $u_1$ in $G_1$ you find the best corresponding $u_2$ in $G_2$ such that, when you mapped all nodes, the average similarity is maximized.

\begin{figure}
\centering
\begin{subfigure}{.3\columnwidth}
\includegraphics[width=\textwidth]{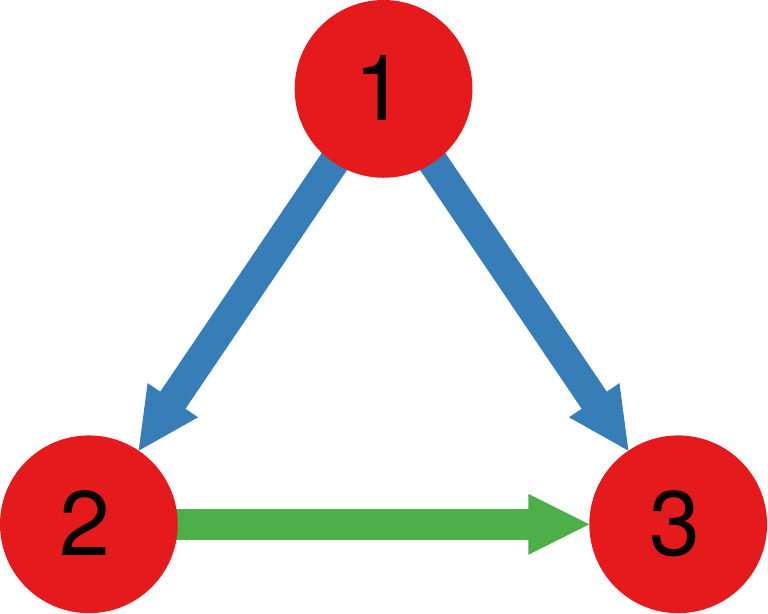}
\caption{}
\end{subfigure}\qquad
\begin{subfigure}{.3\columnwidth}
\includegraphics[width=\textwidth]{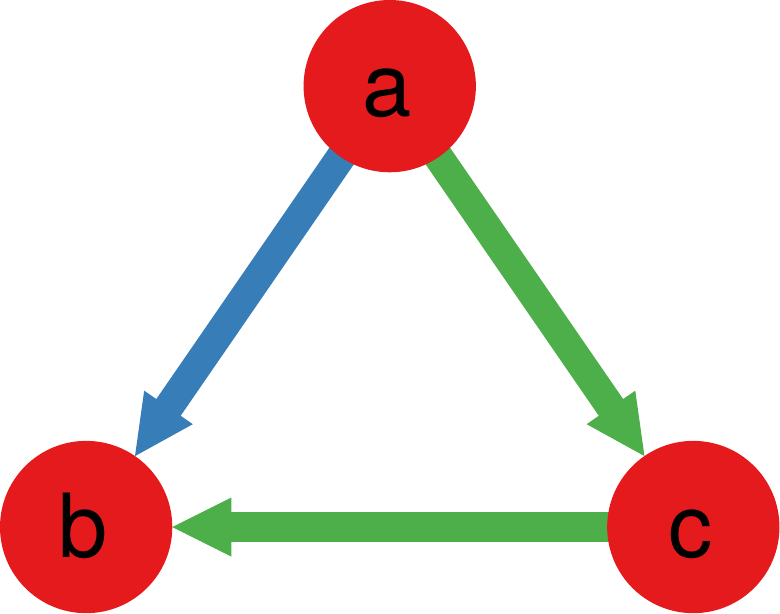}
\caption{}
\end{subfigure}
\caption{Two graphs of which we want to estimate the similarity. Edge color represents its type.}
\label{fig:simflooding}
\end{figure}

Just to get a better intuition on how this might work, consider Figure \ref{fig:simflooding}. We can estimate the networks' similarity by looking at the structural equivalence of their nodes. Nodes $1$ and $a$ are very similar: they both have outdegree of two and they point to the same neighborhood -- two nodes connected by a single green edge. The only difference is in the label of one of their edges. Nodes $2$ and $c$ are also of relatively high similarity, given their equal in- and out-degree with again the sole difference of the edge color. Nodes $3$ and $b$ are, on the other hand, almost structurally identical, with the sole difference being not between them, but between their neighbors. We can conclude, then, that the two graphs are extremely similar, since we just made a node mapping among very similar nodes.

This is a simplification of real approaches\cite{melnik2002similarity}\cite{zager2008graph}. The hard part is defining an efficient technique to find such mappings.

\subsection{Graph Edit Distance}
The strictest possible criterion to establish the similarity between two networks is by solving the graph isomorphism problem. If two graphs are literally the same, their similarity is equal to one. Of course, the graph isomorphism test is binary, thus it is too strict. A single edge difference would net you a zero similarity. We can transform this test into something more useful by counting the number of edge differences between the two graphs. This is akin to define a ``graph edit distance''.

The edit distance between objects $a$ and $b$ is an estimation of the number of edits you need to make on $a$ in order to transform it into $b$. Perhaps the most known and used edit distance is the string edit distance, of which the most famous is the Levenshtein distance\cite{levenshtein1966binary}: this tells you how far apart two strings are. Variants of it are widely used, for instance, by search engines and word processors: when you mistype a word, the software will look up what are the properly spelled words that are at the smallest edit distance from what you typed, and it will suggest them to you. This works well because, usually, you won't make more than one or two mistakes in typing something -- unless you're me and you're trying to retype ``Levenshtein'' from memory.

String and graph edit distances work with the same principles\cite{gao2010survey}\cite{riesen2009approximate}. In strings you are allowed to perform three operations: character insertion, deletion, and replacement. In graphs you have the same three operations, but you can apply them to either nodes or edges, for a total of six operations. Your nodes and edges might have labels, so you want to be able to flip the label values as well.

\begin{figure}
\centering
\begin{subfigure}{.4\columnwidth}
\includegraphics[width=\textwidth]{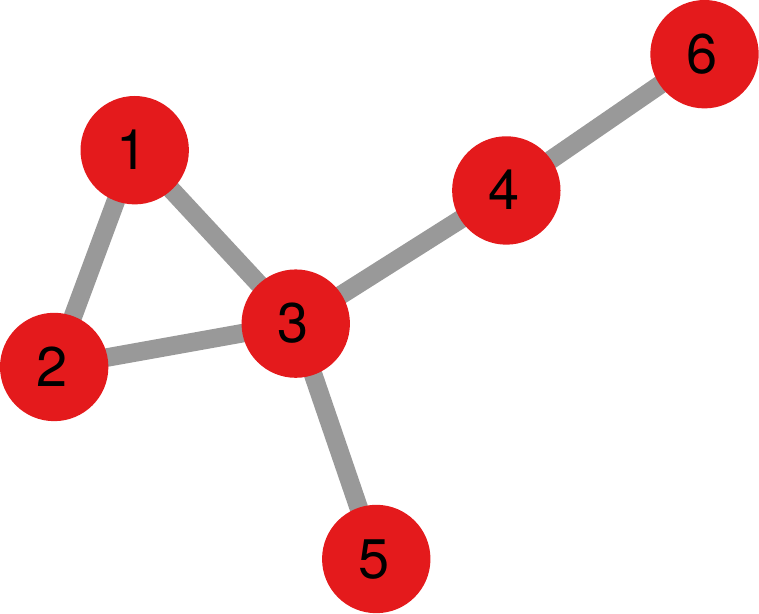}
\caption{}
\end{subfigure}\qquad
\begin{subfigure}{.33\columnwidth}
\includegraphics[width=\textwidth]{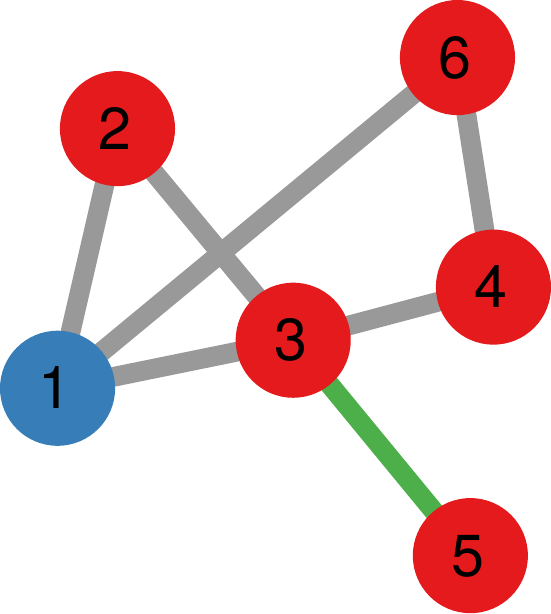}
\caption{}
\end{subfigure}
\caption{Two graphs of which we want to estimate the similarity. Node and edge color represents their type.}
\label{fig:graphedit}
\end{figure}

Figure \ref{fig:graphedit} can help you to visualize the process. Here, we want to know how many operations we need to go from the graph in Figure \ref{fig:graphedit}(a) to the graph in Figure \ref{fig:graphedit}(b). Starting from node $1$, we need to change its label (from red to blue) and to add the edge connecting it to node $6$. Node $2$ is fine, but node $3$ needs to replace its edge to node $5$ with one labeled in green. There are no more edits we need to do, so the distance between the two graphs is three.

Of course, the hard part of graph edit distance is finding the minimum set of edits, so there are a bunch of ways to go about it, ranging from Expectation Maximization to Self-Organizing Maps, to subgraph isomorphism\cite{bunke1997relation}. Special network types deserve special approaches, for instance in the case of Bayesian networks\cite{myers2000bayesian} (see Section \ref{sec:extended-types}) and trees\cite{reis2004automatic}\cite{pawlik2011rted}.

The prototypical graph edit distance metric\cite{balavz1986metric} is relatively simple to understand. It is based on the maximum common subgraph. Given two graphs $G_1$ and $G_2$, first you find the largest common subgraph $G_s$: the largest collection of nodes and edges that is isomorphic in both graphs. Then, the distance between $G_1$ and $G_2$ is simply the number of nodes and edges that remain outside $G_s$:

$$ \delta_{G_1, G_2} = |E_1 - E_s| + |E_2 - E_s| + ||V_1| - |V_2||,$$

with $V_x$ and $E_x$ being the set of nodes and edges of graph $G_x$. This is actually a metric, as it respects the triangle inequality. An evolution of this approach tries to find the maximum common edge subgraph\cite{raymond2002rascal}, which is found in the line graph representation of $G$.

The problem gets significantly easier when the networks are aligned. Two networks are aligned if we have a known node correspondence between the two. This means that we know that node $u$ in one network is the same as node $v$ in the other. How to align two networks is an interesting problem in and of itself, and we're going to look at it in Section \ref{sec:netsimil-alignment}. For now, we just take for granted that the two networks we're comparing are already aligned.

\begin{figure}
\centering
\begin{subfigure}{.29\columnwidth}
\includegraphics[width=\textwidth]{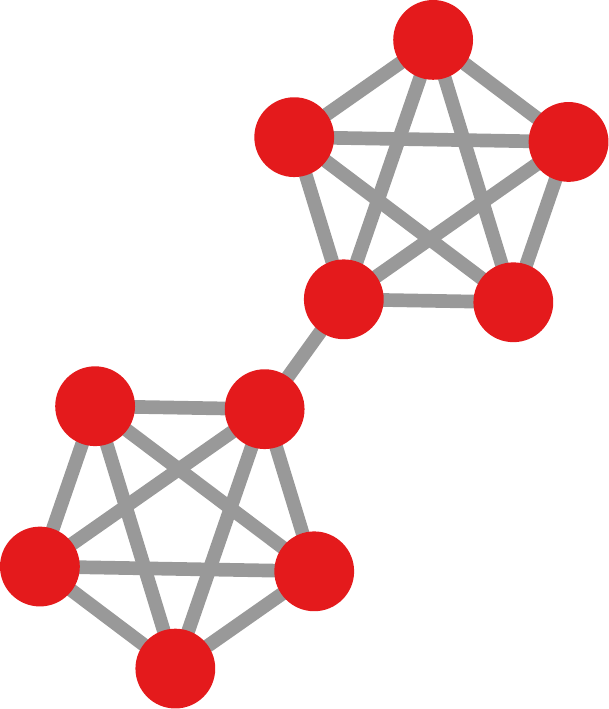}
\caption{}
\end{subfigure}\quad
\begin{subfigure}{.29\columnwidth}
\includegraphics[width=\textwidth]{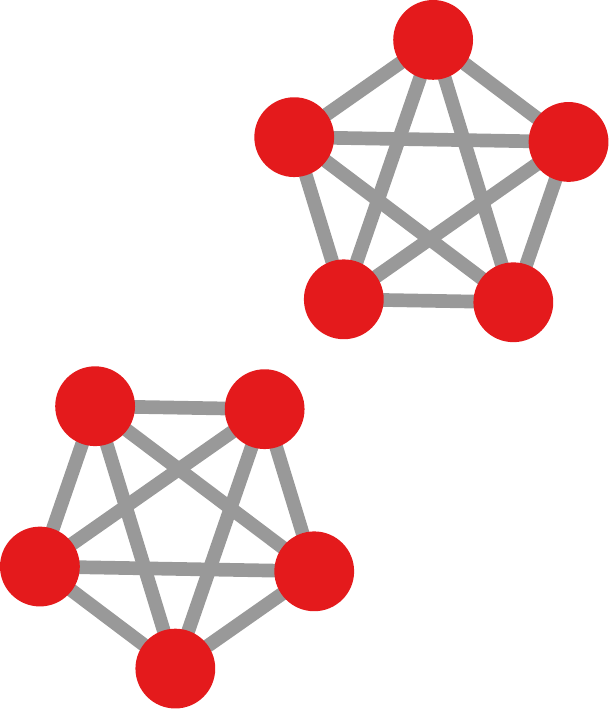}
\caption{}
\end{subfigure}\quad
\begin{subfigure}{.29\columnwidth}
\includegraphics[width=\textwidth]{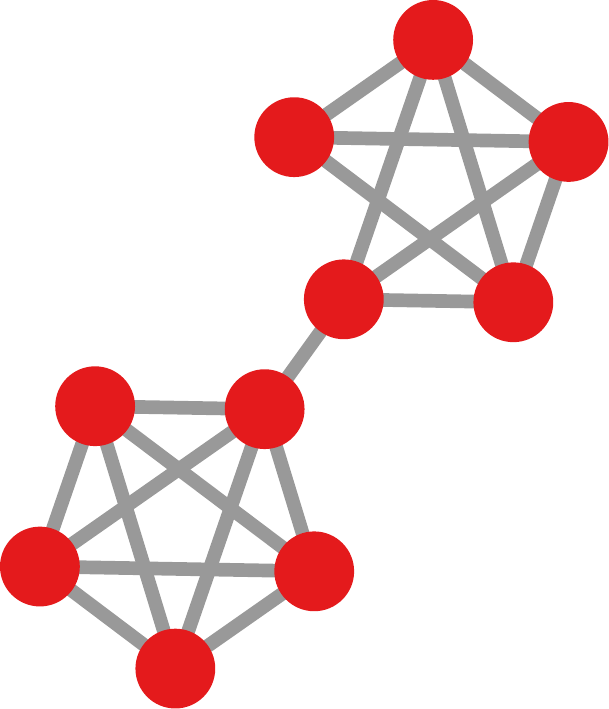}
\caption{}
\end{subfigure}
\caption{Three graphs of which we want to estimate the similarity. Note how (b) misses the edge connecting the cliques, and (c) misses an edge inside the top clique.}
\label{fig:deltacon}
\end{figure}

In this case, we don't have to go looking for maximum subgraphs. We can just iterate over all the nodes and edges in the two networks and note down every time we find an inconsistency: a node or an edge that is present in one network and absent in the other. Simply counting won't do much good, though, because some differences should count more than others, if they are significantly affecting the local or global properties of the network. Consider Figure \ref{fig:deltacon}: both Figure \ref{fig:deltacon}(b) and Figure \ref{fig:deltacon}(c) are just one edge away from Figure \ref{fig:deltacon}(a). However, since Figure \ref{fig:deltacon}(b) breaks down in multiple connected components, its difference should be counted as higher.

There are a few strategies to estimate these differences. Deltacon is based on some sort of node affinity estimation\cite{koutra2013deltacon}. One could also do vertex rank comparison\cite{papadimitriou2010web}: if the most important nodes in two networks are the same, then the networks must be similar, to some extent. The same authors propose other ways to estimate network similarity, for instance via shingling: reducing the networks to sequences and then applying a sequence comparing algorithm. These latter approaches are specialized to find changes in the same time-evolving network.

The big caveat for using graph edit distances is that they only work for specific data generating processes. For instance, remember the $G_{n,p}$ uniform random graphs from Chapter \ref{cha:rndgraphs}? Two $G_{n,p}$ graphs with the same $n$ and (low) $p$ are similar, in the sense that they are realizations of the same process. However, since edges are independent and the graphs are sparse, they will have almost no edge in common. As a consequence, their edit distance is large! So what you're looking for when using edit distances is for a generating process that has strong dependencies between edges: the fact that two nodes are connected implies the presence/absence of other edges in their neighborhood. You should discard graph edit distance measures as soon as you think that the edges inside your networks are independent from each other.

\subsection{Substructure Comparison}
Substructure comparison\cite{yan2005substructure}\cite{shang2010connected} is similar to graph edit distance. In this class of methods, you describe the network as a dictionary of motifs and how they connect to each other. Usually, you'd find the motifs by applying frequent subgraph mining (Chapter \ref{cha:mining-base}). In practice, graph edit distance is equivalent to a simple substructure comparison, where the only substructure you're focusing on is the edge. There is not much to say about this class, given its similarity with the previous one: the same considerations and warnings that applied there also apply here.

When it comes to applications of substructure similarity, the classical scenario is estimating compound similarities at the molecular level in a biological database\cite{hagadone1992molecular}. But there are more fun scenarios, such as an analysis of Chinese recipes\cite{wang2008substructure}.

\subsection{Holistic Approaches}
In the holistic category I group a series of approaches that are a mixture of the four previous strategies. Meaning that they use parts of all global properties, node similarities, and edit distance, to build a more general similarity measure for networks. The idea is to build a ``signature vector'': a numerical vector that describes the relevant aspects of the topology of $G$. Then, the similarity between two graphs is simply the similarity between their signature vectors. In practice, one could think this class to include a sort of ``graph embeddings'' (Chapter \ref{cha:mining-embeddings}).

\begin{figure}
\centering
\includegraphics[width=.9\textwidth]{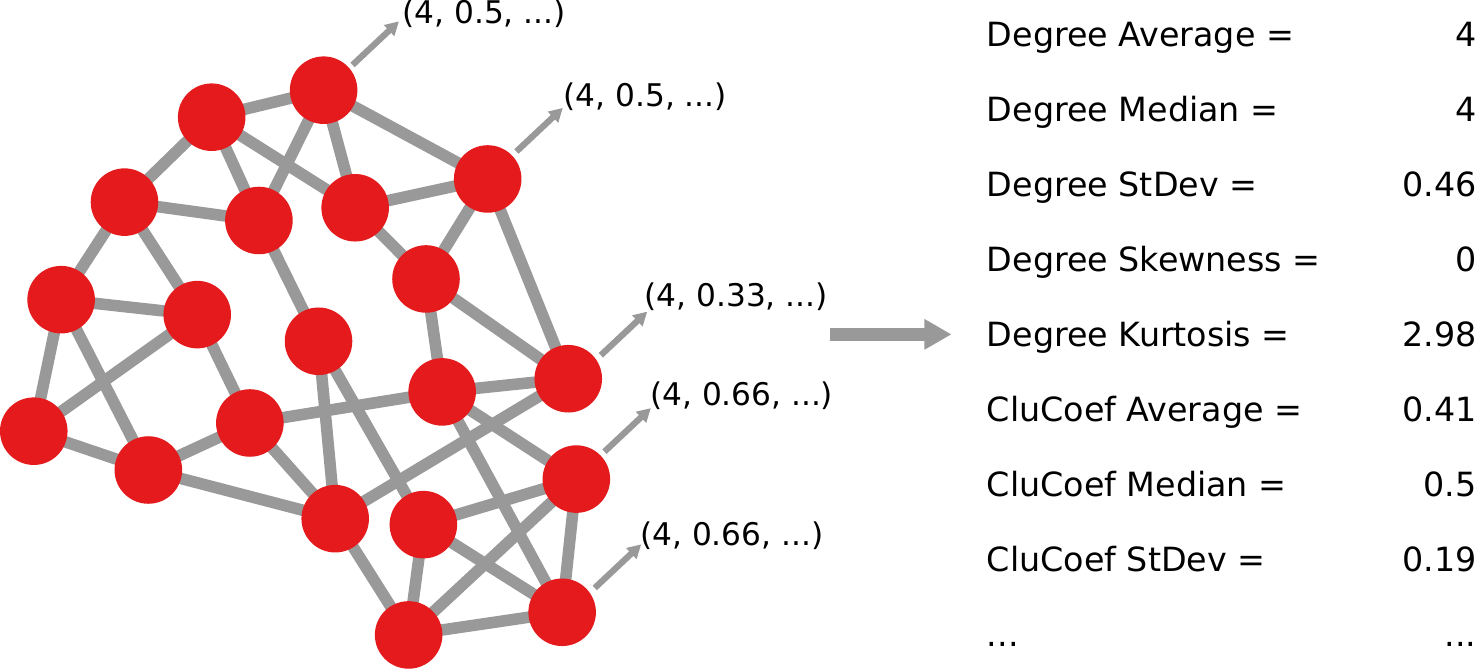}
\caption[][-2\baselineskip]{The workflow of NetSimile. The rightmost column of number is the beginning of the graph's signature vector.}
\label{fig:netsimile}
\end{figure}

NetSimile\cite{berlingerio2012netsimile} is one of the many algorithms in this class. I represent its workflow in Figure \ref{fig:netsimile}. First, NetSimile calculates seven features for each node of the graph: degree, local clustering coefficient, average neighbor degree, average neighbor local clustering coefficient, number of edges among neighbors, etc. Then, these features are aggregated across nodes, i.e. NetSimile calculates their summary statistics like average, standard deviation, etc. This is the signature vector of the graph, which can now be used to compare $G$ with any other graph. Any distance measure discussed so far in the book -- cosine, Euclidean, ... -- can be used to perform the comparison. The authors focus specifically on the Camberra distance\cite{lance1966computer}.

Similar approaches are graph hashes\cite{wang2012g}, designed for optimizing graph similarity searches in a graph database possibly containing thousands of graphs; and approaches that are more rooted in social theories\cite{berlingerio2013network}. The latter case is an evolution of NetSimile. Rather than including a laundry list of all the measures we think we can use to compare graphs, we pick the ones that are theoretically motivated. We define which are the criteria of similarity based on different theories, and we discard the rest. The objective is to be able to better interpret the similarity scores.

A close cousin of holistic approaches is the one of graph kernels\cite{gartner2003graph}\cite{vishwanathan2006fast}\cite{kang2012fast}. Just like in NetSimile and in graph embeddings, a graph kernel is the reduction of a complex high-dimensional graph into a vector of numbers. These vectors are then fed to a machine learning algorithm that is able to learn the shape of the space in which these vectors live and thus the similarity between them. Just like with many graph embeddings techniques, these kernel are usually created by means of some sort of random walk process.
 
\subsection{Information Theory}
A radically different approach works directly with the adjacency matrix of a graph. The idea here is to generalize the Kullback-Leibler divergence (KL-divergence) so that it can be applied to determining the distance between two graphs. The KL-divergence is a cornerstone of information theory and linked with the concept of information entropy -- see Section \ref{sec:prob-mi} for a refresher.

The KL-divergence is also known as ``relative entropy''. From Section \ref{sec:prob-mi}, you learned that the information entropy of a vector $X$ is the number of bits per element you need to encode it. Now, of course when you try to encode a vector, you try to be as smart as possible. You create a codebook that is specialized to encode that particular vector. If there is an element that appears much more often than the others, you will give it a short code: you will have to use it more often and, if it is shorter, every time you use it you will save bits. This is the strategy used by Infomap to solve community discovery -- see Section \ref{sec:cd-partition-rw}.

Now suppose you have another vector, $Y$. You want to know how similar $Y$ is to $X$. One thing you could do is to encode $Y$ using the code book you optimized to encode $X$. If $X = Y$, then the codebook is as good encoding $X$ as it is encoding $Y$: you need no extra bits. As soon as there are differences between $X$ and $Y$, you will start needing extra bits to encode $Y$, because $X$'s codebook is not perfect for $Y$ any more. The KL-divergence boils down to the number of extra bits you need to encode $Y$ using $X$'s codebook.

\begin{figure}
\centering
\begin{subfigure}{.4\columnwidth}
\includegraphics[width=\textwidth]{figures/entropy1.pdf}
\caption{}
\end{subfigure}\qquad
\begin{subfigure}{.4\columnwidth}
\includegraphics[width=\textwidth]{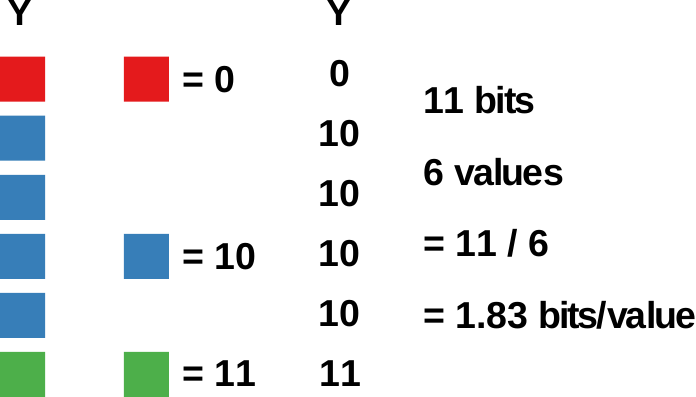}
\caption{}
\end{subfigure}
\caption{An example of the spirit of KL-divergence. The code we use for $X$ (a) requires additional bits to encode $Y$ (b).}
\label{fig:kldivergence}
\end{figure}

Figure \ref{fig:kldivergence} presents a rough outline of the idea behind the KL-divergence -- simplified to help intuition. The $X$ vector in Figure \ref{fig:kldivergence}(a) requires $1.5$ bits per element. Using its codebook to encode $Y$ in Figure \ref{fig:kldivergence}(b) increases the requirement to $11$ total bits instead of the original $9$.

In its original formulation, the KL-divergence is defined for pairs of vectors. However, one can expand it to allow it to consider different inputs\cite{galas2017expansion}. One can say that entries in the vectors are dependent on each other. Thus, if you consider a graph as a series of $|V|$ variables, one per node, you can express the pairwise dependencies as the edges of the graph. This approach has applications in chemistry\cite{mcclendon2012comparing}.

Another way of comparing networks by means of analyzing their adjacency matrices comes from comparing the eigenvectors of their Laplacians\cite{banerjee2012structural}. Similar networks will experience similar spreading patterns, which are reflected in their spectra. 

\section{Network Alignment}\label{sec:netsimil-alignment}
Earlier in the chapter, I mentioned what aligned networks are: two networks are aligned if we have a node to node mapping, i.e. for each node in network $G_1$ we have a corresponding node in $G_2$ representing the same entity. Many networks are naturally aligned. The most typical case of aligned networks are time-evolving networks. Two snapshots of a structure are simply two different networks: since we have the same ids on the nodes, we have the alignment for free. Another example could be networks describing brain scans: we divide the brain in different areas for all individuals, and these areas might interact differently between individuals. The areas are the nodes, thus their identities are known, while the interactions are the edges, which might change. In general, any multilayer network (Section \ref{sec:extended-multilayer}) could be seen as a collection of aligned networks: each layer is a network and the inter layer couplings are the mappings from one layer to another.

However, you might not be as lucky as in the case of evolving networks: sometimes you have two observations that you think you should be able to align, but you actually do not have neither consistent node ids, nor a reliable node mapping. For instance, you might have collected a bunch of data from different social media. You know that people have profiles in different platforms, but these platforms will use different and mutually incompatible identifiers. Thus, you will need to figure out who is who in all the networks you collected. This is the network alignment problem\cite{trung2020comparative}. A classical application of network alignment is the attempt to map the protein-protein interaction of different organisms\cite{kelley2003conserved}, discovering that many biological pathways are preserved across species.

\begin{figure}
\centering
\begin{subfigure}{.3\columnwidth}
\includegraphics[width=\textwidth]{figures/graphedit_dist1.pdf}
\caption{}
\end{subfigure}\qquad
\begin{subfigure}{.25\columnwidth}
\includegraphics[width=\textwidth]{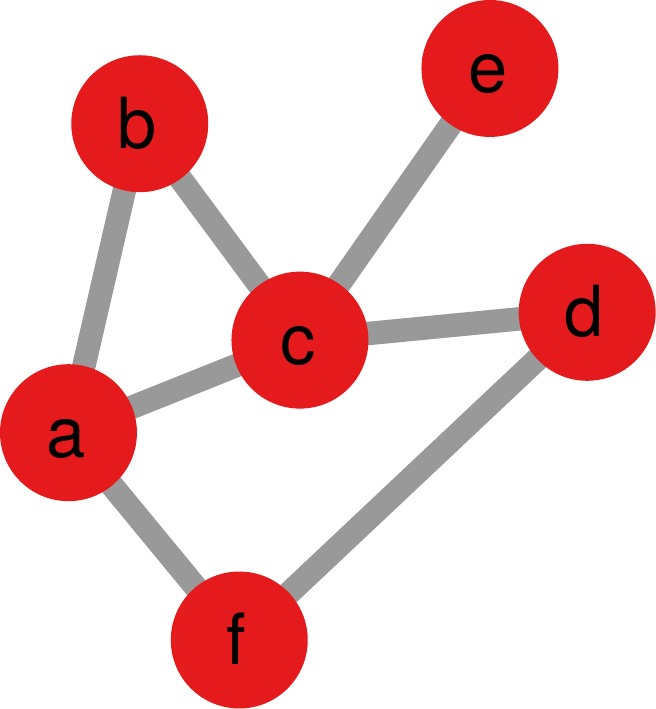}
\caption{}
\end{subfigure}\qquad
\begin{subfigure}{.3\columnwidth}
\includegraphics[width=\textwidth]{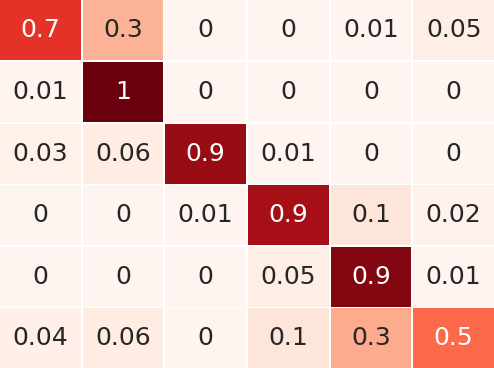}
\caption{}
\end{subfigure}
\caption{Two graphs of which we want to discover the alignment. (c) assigns to each node pair from (a) and (b) an alignment probability.}
\label{fig:alignment}
\end{figure}

Figure \ref{fig:alignment} shows an example. Given two graphs as input with $V_1$ and $V_2$ as their node sets, you want to produce a $|V_1| \times |V_2|$ matrix telling you the probability of each node from the first graph to be the node in the second graph. The way this matrix is built can rely on any structural similarity measure, we saw a few in different parts of this book. Then, the idea is to pick the cells in this matrix so that the sum of the scores is maximized and, at the same time, we match as many nodes as possible\cite{kuchaiev2011integrative}. If $|V_1| \neq |V_2|$ you will have to face a choice: either you do not map some nodes or you allow nodes from one network to map to multiple nodes in the other. This common approach can be extended, for instance, by calculating multiple versions of this matrix using different measures and then seeking a consensus matrix which is a combination of all the similarity measures.

One could also find just a few node mappings with extremely high confidence and then expand from that seed\cite{kollias2011network}, assuming that the neighborhoods around these high confidence nodes should look alike. Of course, a large portion of network alignment solutions rely on solving the maximum common subgraph problem: if you find isomorphic subgraphs in both networks, chances are that the nodes inside these subgraphs are the same, and thus should be aligned to each other\cite{klau2009new}. Other approaches rely on the fact that isomorphic graphs have the same spectrum, thus similar values in the eigenvectors of the Laplacian imply that the nodes are relatively similar\cite{patro2012global}.

Another approach uses a dictionary of networks motifs. Each node is described by counting the number of motifs it is part of. We can then describe the node as a numerical count vector. Two nodes with similar vectors are similar\cite{milenkovic2010optimal}. This approach has to solve the graph isomorphism problem as well, but it needs to do so only for small graph motifs rather than for -- supposedly -- large common subgraphs. This way, it can be more efficient.

\begin{figure}
\centering
\begin{subfigure}{.29\columnwidth}
\includegraphics[width=\textwidth]{figures/graphedit_dist1.pdf}
\caption{}
\end{subfigure}\qquad
\begin{subfigure}{.6\columnwidth}
\includegraphics[width=\textwidth]{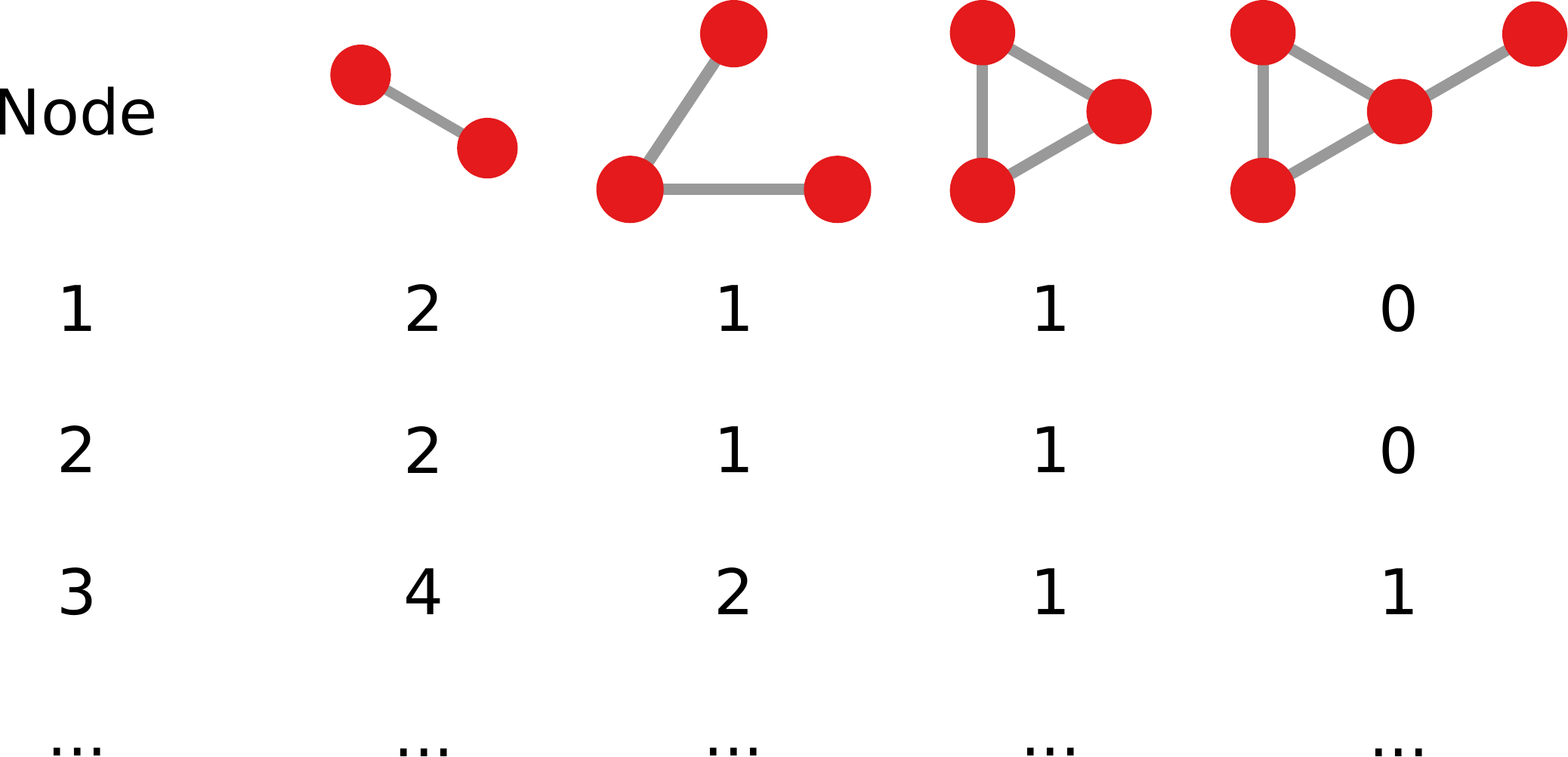}
\caption{}
\end{subfigure}
\caption{(a) A graph. (b) A node-motif table, counting how many times each node is part of a given motif.}
\label{fig:alignment2}
\end{figure}

Figure \ref{fig:alignment2} shows an example: here I choose a relatively small set of motifs, which generate a short vector. However, one could define as many motifs as they are relevant for a specific application, and obtain much more precise vectors describing the nodes. A final approach I mention is MAGNA, which uses a genetic algorithm approach: it tries aligning by exploring the search space of all possible node mappings, allowing the best matches to survive and evolve, and dropping the worst matches\cite{saraph2014magna}.

\section{Network Fusion}\label{sec:netsimil-fusion}
The final problem related to network distance/similarity is network fusion. Network fusion is a relatively old term and branch of computer science that up until recently had little to do with network science proper. It was introduced a few decades ago in the field of neural networks\cite{hansen1990neural}\cite{cho1995multiple}. The idea was that you trained a neural network on some data. The network has grown to be able to capture as much of the variation as possible. If you use the same algorithm to train on different data, you might end up with a similar, but not identical, configuration in your neural network. Some connections are stronger, others are weaker. Most of the variation between the same neural networks trained on different data is due to overfitting. Thus you want to build yet another neural topology, smoothing out all the noise. This is network fusion, because effectively you want to fuse together all the neural networks that you have trained. This might be an old idea, but is still an area of active research\cite{du2017fused}.

\begin{figure}
\centering
\begin{subfigure}{.3\columnwidth}
\includegraphics[width=\textwidth]{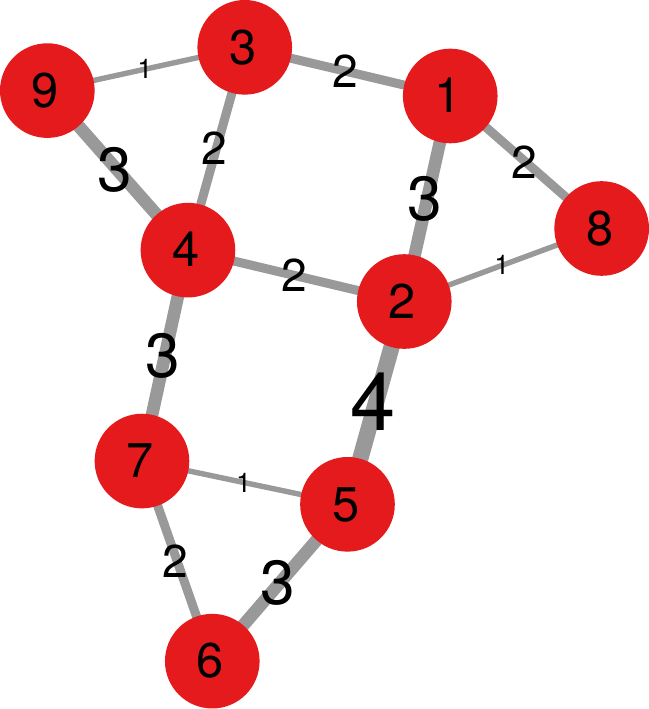}
\caption{}
\end{subfigure}\quad
\begin{subfigure}{.3\columnwidth}
\includegraphics[width=\textwidth]{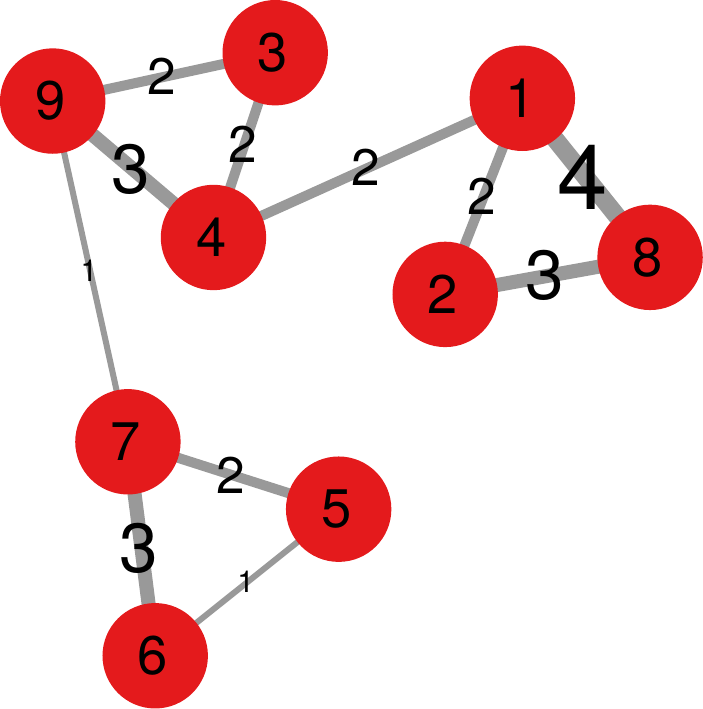}
\caption{}
\end{subfigure}\quad
\begin{subfigure}{.3\columnwidth}
\includegraphics[width=\textwidth]{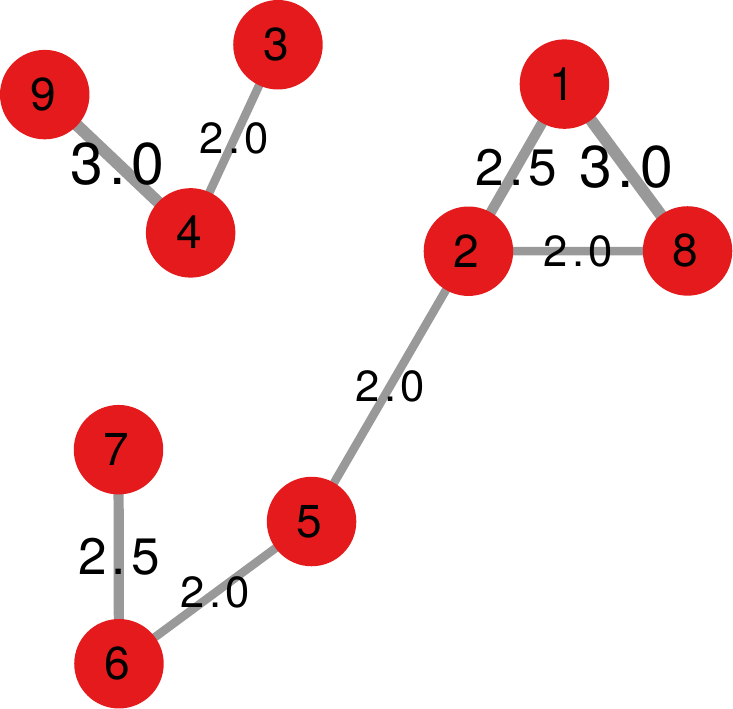}
\caption{}
\end{subfigure}
\caption{An example of network fusion: (c) is the result of the fusion of (a) and (b). Edge thickness proportional to its weight.}
\label{fig:fusion}
\end{figure}

Figure \ref{fig:fusion} is the easiest mental picture you need to understand the principle of network fusion. We have two aligned networks in Figure \ref{fig:fusion}(a) and Figure \ref{fig:fusion}(b). We decide that we want to fuse them together by calculating the average edge weight. We also decide that we keep a connection in the fused network only if its resulting average weight is higher than $2$. Figure \ref{fig:fusion}(c) is the result. Of course, real network fusion algorithms are much smarter and more sophisticated than this.

Slowly but surely, network fusion crept into network science and found applications that go beyond increasing the performance and applicability of neural networks. For instance, consider genomic data. You can collect samples of interactions from many individuals. These are similar, but not always the same. You might want to combine them to create a prototypical interaction network\cite{wang2014similarity}. Alternatively, it could be that some of these samples are incomplete, and you can use their fusion as the complete genomic data.

\section{Summary}

\begin{enumerate}
\item In this chapter we explore different ways to estimate the similarity/distance between the topologies of two networks: given two graphs, quantitatively estimate how much of their topologies are the same. Related applications are network alignment and fusion.
\item Network similarity can be done in many ways. One could compare the global properties of the network such as degree distribution and clustering; or aggregate all the node pairwise similarities; or estimate the number of edits needed to go from one network to the other.
\item One can combine all those approaches in a holistic one, determining which elements of the similarities between networks are more relevant, based on what the networks represent. Additionally, one could see adjacency matrices as signals and calculate the mutual information entropy between them.
\item Network alignment is the problem of finding a node-to-node mapping between two networks. We hypothesize that the two networks represent the connections between the same real world entities and we need to re-identify them by looking exclusively at the network topologies.
\item Network fusion is the process of taking multiple versions of the same network and reconstructing the underlying structure. The idea is that each observation might be noisy or incomplete, while their combination should represent the ideal structure.
\end{enumerate}

\section{Exercises}

\begin{enumerate}
\item Estimate the similarity between the networks at \url{http://www.networkatlas.eu/exercises/48/1/data1.txt}, \url{http://www.networkatlas.eu/exercises/48/1/data2.txt}, and \url{http://www.networkatlas.eu/exercises/48/1/data3.txt}, by comparing their average degree, average clustering coefficient, and density (average their absolute differences). Which pair of networks are more similar to each other?
\item Calculate the structural similarities of all pairs of nodes for all pairs of networks used in the previous question. Derive a network similarity value by averaging the node-node similarities. Since the networks are aligned, the node-node similarity is the Jaccard coefficient of their neighbor sets, and you should only calculate them for pairs of nodes with the same id. Which pair of networks are more similar to each other?
\item Calculate the graph edit distances between the networks used in the previous questions. Remember that the networks are aligned, thus you just need to iterate over nodes and compare their neighborhoods. Which pair of networks are more similar to each other?
\item Fuse the three networks together to produce a consensus network. You can keep an edge in the consensus network only if it appears in two out of three networks -- assume that their are aligned and that nodes with the same id are the same node.
\end{enumerate}

\part{Visualization}\label{par:netviz}

\chapter{Node Visual Attributes}\label{cha:netviz-node}
Data visualization is at the core of any data analysis undertaking -- call it statistics, data science, or whatever else. There are two reasons why: gathering insights in exploratory data analysis, and presenting your results.

\begin{figure*}
\centering
\begin{subfigure}{.5\columnwidth}
  \centering \footnotesize
\begin{tabular}{cc|cc|cc|cc}
\multicolumn{2}{c}{\textcolor{cb1}{I}}  & \multicolumn{2}{c}{\textcolor{cb2}{II}}  & \multicolumn{2}{c}{\textcolor{cb3}{III}}  & \multicolumn{2}{c}{\textcolor{cb4}{IV}}\\
x  & y  & x  & y  & x  & y  & x  & y\\
\hline
\textcolor{cb1}{$10$}  & \textcolor{cb1}{$8.04$}  & \textcolor{cb2}{$10$}  & \textcolor{cb2}{$9.14$}  & \textcolor{cb3}{$10$}  & \textcolor{cb3}{$7.46$}  & \textcolor{cb4}{$8$}  & \textcolor{cb4}{$6.58$}\\
\textcolor{cb1}{$8$}  & \textcolor{cb1}{$6.95$}  & \textcolor{cb2}{$8$}  & \textcolor{cb2}{$8.14$}  & \textcolor{cb3}{$8$} & \textcolor{cb3}{$6.77$}  & \textcolor{cb4}{$8$}  & \textcolor{cb4}{$5.76$}\\
\textcolor{cb1}{$13$}  & \textcolor{cb1}{$7.58$}  & \textcolor{cb2}{$13$}  & \textcolor{cb2}{$8.74$}  & \textcolor{cb3}{$13$}  & \textcolor{cb3}{$12.74$}  & \textcolor{cb4}{$8$}  & \textcolor{cb4}{$7.71$}\\
\textcolor{cb1}{$9$}  & \textcolor{cb1}{$8.81$}  & \textcolor{cb2}{$9$}  & \textcolor{cb2}{$8.77$}  & \textcolor{cb3}{$9$}  & \textcolor{cb3}{$7.11$}  & \textcolor{cb4}{$8$}  & \textcolor{cb4}{$8.84$}\\
\textcolor{cb1}{$11$}  & \textcolor{cb1}{$8.33$}  & \textcolor{cb2}{$11$}  & \textcolor{cb2}{$9.26$}  & \textcolor{cb3}{$11$}  & \textcolor{cb3}{$7.81$}  & \textcolor{cb4}{$8$} & \textcolor{cb4}{$8.47$}\\
\textcolor{cb1}{$14$}  & \textcolor{cb1}{$9.96$}  & \textcolor{cb2}{$14$}  & \textcolor{cb2}{$8.1$}  & \textcolor{cb3}{$14$}  & \textcolor{cb3}{$8.84$} & \textcolor{cb4}{$8$}  & \textcolor{cb4}{$7.04$}\\
\textcolor{cb1}{$6$}  & \textcolor{cb1}{$7.24$}  & \textcolor{cb2}{$6$}  & \textcolor{cb2}{$6.13$}  & \textcolor{cb3}{$6$}  & \textcolor{cb3}{$6.08$}  & \textcolor{cb4}{$8$}  & \textcolor{cb4}{$5.25$}\\
\textcolor{cb1}{$4$}  & \textcolor{cb1}{$4.26$}  & \textcolor{cb2}{$4$}  & \textcolor{cb2}{$3.1$}  & \textcolor{cb3}{$4$}  & \textcolor{cb3}{$5.39$}  & \textcolor{cb4}{$19$}  & \textcolor{cb4}{$12.5$}\\
\textcolor{cb1}{$12$}  & \textcolor{cb1}{$10.84$}  & \textcolor{cb2}{$12$}  & \textcolor{cb2}{$9.13$}  & \textcolor{cb3}{$12$}  & \textcolor{cb3}{$8.15$}  & \textcolor{cb4}{$8$}  & \textcolor{cb4}{$5.56$}\\
\textcolor{cb1}{$7$}  & \textcolor{cb1}{$4.82$}  & \textcolor{cb2}{$7$}  & \textcolor{cb2}{$7.26$}  & \textcolor{cb3}{$7$}  & \textcolor{cb3}{$6.42$}  & \textcolor{cb4}{$8$}  & \textcolor{cb4}{$7.91$}\\
\textcolor{cb1}{$5$}  & \textcolor{cb1}{$5.68$}  & \textcolor{cb2}{$5$}  & \textcolor{cb2}{$4.74$}  & \textcolor{cb3}{$5$}  & \textcolor{cb3}{$5.73$}  & \textcolor{cb4}{$8$}  & \textcolor{cb4}{$6.89$}
\end{tabular}
\caption{}
\end{subfigure}
\begin{subfigure}{.49\columnwidth}
\includegraphics[width=\textwidth]{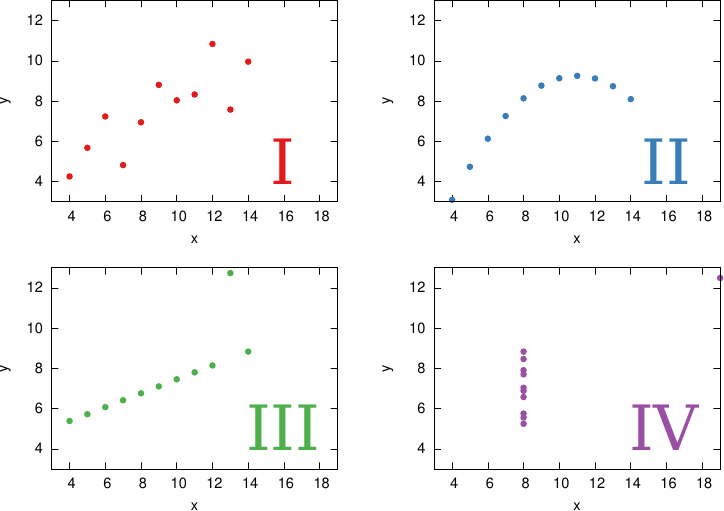}
\caption{}
\end{subfigure}
\caption{(a) Four datasets with $x$ and $y$ coordinates. (b) Data visualization of (a) in the form of a scatter plot.}
\label{fig:anscombe}
\end{figure*}

Let's start from gathering insights. It is sometimes -- not always! -- easier to spot patterns when you look at them with a proper visualization, rather than relying on summary statistics. The classical case for this position is the Anscombe quartet\cite{anscombe1973graphs}. In Figure \ref{fig:anscombe}(a) you have four datasets of two variables.

In all datasets, these variables have the same mean and standard deviation, the same correlation and even drawing a regression line between the two variables leads to the same result. You'd think that the four datasets are identical and no clear patterns distinguish them. However, simply \textit{seeing} how these datasets look like -- as I show in Figure \ref{fig:anscombe}(b) with a humble scatter plot -- immediately tells you that there's something interesting going on.

Then there is result communication. When you write a paper, you must state your results clearly and in an intuitive matter. In many cases, it is true that showing a picture of them is necessary. Thus, you need to be proficient in data visualization techniques, so that you won't accidentally trick your reader -- or you won't tricked yourself while reading a paper from a malicious miscommunicator.

The classical building blocks of network visualization are nodes and edges. Traditionally, we represent nodes as circles or dots, and edges as lines connecting dots. Then, nodes are scattered around so that the ones not connected to each other tend to be far apart. Instead, edges tend to be as short as possible, so connected nodes appear in close spatial proximity. This is so ingrained in network science visualization that you saw me using this approach throughout most of the examples I presented so far.

There are reasons why rules and best practices exist. They work in most scenarios. They also build familiarity: if you're exposed to the same strategies over and over again, you become literate and know immediately what's going on. In the majority of what follows I will align myself with these conventions. However, the first thing that needs to be highlighted is something that might appear obvious. Even if we represent them that way, nodes aren't dots, and edges aren't lines. The dot-line diagram is a \textit{map}, not the \textit{territory}. The reality that lurks behind a graph can take many forms and some will communicate better your intentions than others.

\begin{figure}
\centering
\includegraphics[width=.66\columnwidth]{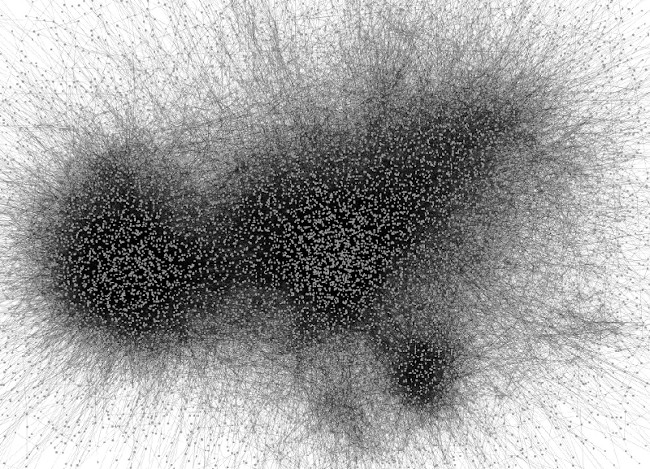}
\caption{Hello hairball, my old friend. I've come to talk with you again, because a vision softly creeping left its seeds while I was sleeping
and the vision that was planted in my brain still remains.}
\label{fig:hairball}
\end{figure}

In practice, my aim is to give you the best practices and then empower you to break them when you feel they get in the way of your network visualization. Our journey is a fight against the nemesis of every network scientist who dares visualizing her own networks: the hairball. A hairball, or spaghettigraph, is something that looks like the example in Figure \ref{fig:hairball}.

We already saw how the hairball can get in your way when you're analyzing network data in Part \ref{par:hairball}. Here, we're trying to defeat it in the realm of communicating to others what your network contains. If we trust the node-link diagram convention too much, we end up with visualizations that are cluttered like in Figure \ref{fig:hairball} and do not communicate much besides ``it's complicated''.

So, if there's something you will take away from this part, it is how to make hairballs less hairbally. We start in this chapter with node visual attributes, then we move on to edge visual attributes (Chapter \ref{cha:edgeviz}) and network layouts (Chapter \ref{cha:layouts}), with a small carousel of peculiar examples.

My software of choice is usually Cytoscape\cite{shannon2003cytoscape}\footnote{\url{https://cytoscape.org/}}, which is the one I'm most proficient with. Most of the examples in this part will be based on Cytoscape and can be achieved by using it without any real programming skill. A popular alternative would be Gephi\cite{bastian2009gephi}\footnote{\url{https://gephi.org/}}.

Finally, I should say that what follows is all practical knowledge of me messing up with Cytoscape for ten years and learning myself what looks good and what doesn't. I'm not an expert on data visualization and visual communication in general. If you want a more in-depth dive into proper visualization techniques -- which go beyond simple network visualization -- you're best served with one of the many awesome books and papers out there\cite{shneiderman1996eyes}\cite{cairo2012functional}\cite{meirelles2013design}\cite{tufte2014visual}\cite{munzner2014visualization}. You should also consider keeping an eye on some conferences on data visualizations such as IEEE Visualization Conference and the ACM Computer-Human Interaction conference.

\section{Size}
\begin{figure}
\centering
\includegraphics[width=.8\columnwidth]{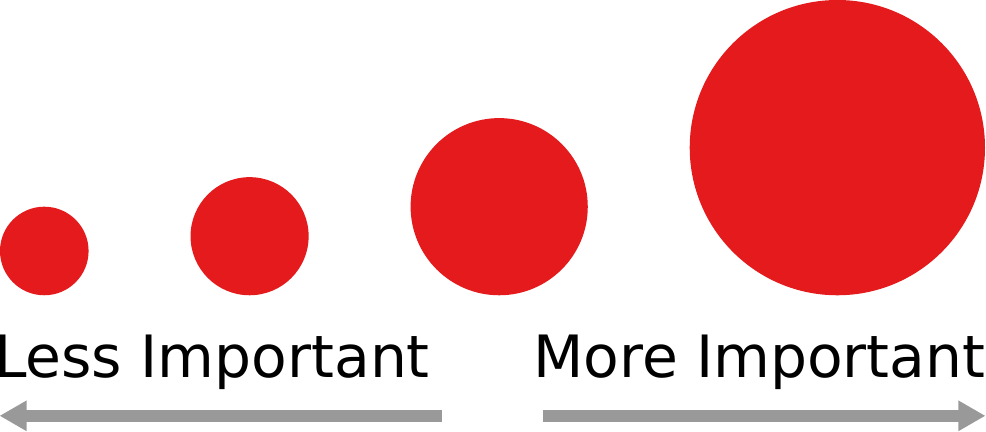}
\caption{The natural scale with which we can use node size -- meaning: its area -- to confer the idea of its importance.}
\label{fig:nodesize}
\end{figure}

The first thing you might want to modify about a node is its size. This should be used for \textbf{quantitative} attributes, measuring some sort of importance of the node. They can be directly calculated from the graph properties (such as number of connections, PageRank, etc) or they can be provided as quantitative metadata. For instance, in a network where nodes are traffic junctions, it could be the number of cars that can pass through a street crossing per unit of time. It seems natural to encode the node's importance directly on its size, as I show in Figure \ref{fig:nodesize}.

The reason for using sizes -- and other visual features as we will see -- is to facilitate perception of the quantities and hence facilitate inference and insight. You want to make quantitative distinctions so that the variables you're visualizing stand out. If you cannot tell the difference between two different node sizes because the variations are too subtle, you're not communicating anything to your viewer. Thus, you have to have enough diversity in your visual features: just the amount that the human eye can perceive. This is one of the reasons why it is so hard to create visualizations. Finding the right scale to encode a quantity into a size is hard, especially when you're dealing with continuous variables and you have to bin them yourself.

\begin{figure}
\centering
\begin{subfigure}{.32\textwidth}
\includegraphics[width=\textwidth]{figures/multigraph_simple.pdf}
\caption{}
\end{subfigure}
\begin{subfigure}{.32\textwidth}
\includegraphics[width=\textwidth]{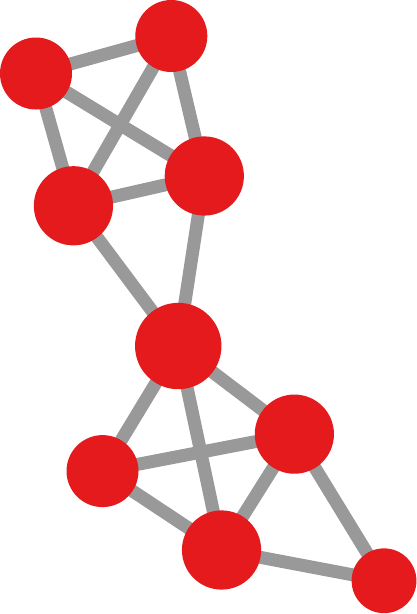}
\caption{}
\end{subfigure}
\begin{subfigure}{.32\textwidth}
\includegraphics[width=\textwidth]{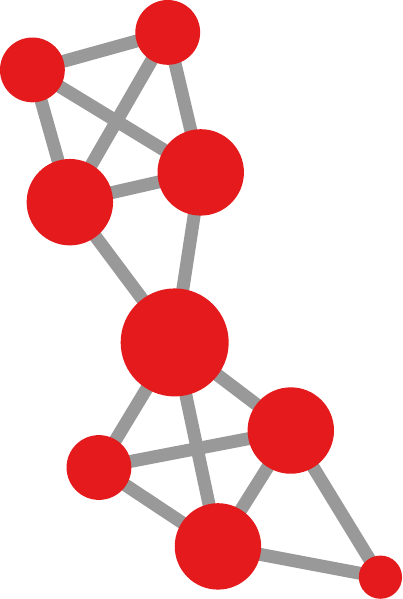}
\caption{}
\end{subfigure}
\caption{(a) A graph in which all nodes have the same size. (b) A graph in which the node's degree determines its size, with subtle variations. (c) Same as (b), but exaggerating the node size variation.}
\label{fig:nodesize2}
\end{figure}

Taking Figure \ref{fig:nodesize2} as an example, in Figure \ref{fig:nodesize2}(b) you cannot really tell who is boss by simply looking at the node sizes. As soon as we exaggerate the size difference -- Figure \ref{fig:nodesize2}(c) --, it becomes clearer and the visualization becomes more informative and, arguably, visually more pleasing. Differences have to jump to the eye: making subtle changes is not going to communicate much to the viewer. In other words, to facilitate the viewer's perception of the differences in the quantities mapped, your visualization has to have enough ``action'', differences, it has to say something.

\begin{figure*}
\centering
\begin{subfigure}{.32\textwidth}
\includegraphics[width=\textwidth]{figures/centrality_degree_degeneration.pdf}
\caption{}
\end{subfigure}\hfill
\begin{subfigure}{.32\textwidth}
\includegraphics[width=\textwidth]{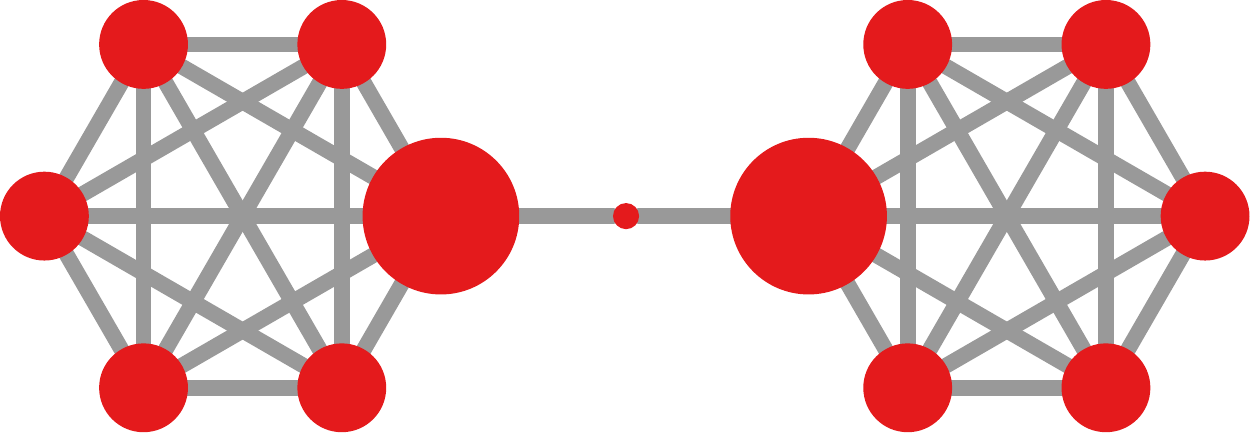}
\caption{}
\end{subfigure}\hfill
\begin{subfigure}{.32\textwidth}
\includegraphics[width=\textwidth]{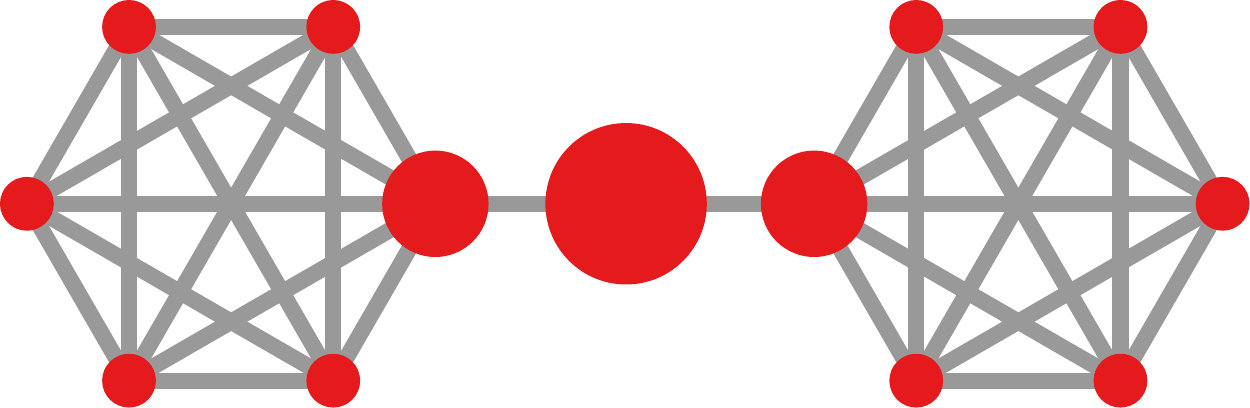}
\caption{}
\end{subfigure}
\caption{(a) A graph in which all nodes have the same size. (b) A graph in which the node's degree determines its size. (c) Same as (b), but using betweenness centrality instead of the degree for the nodes' size.}
\label{fig:nodesize3}
\end{figure*}

You cannot simply take away the message that any quantitative measure of node importance is an equally good choice for your node size. Some of those measures will not highlight what you want to highlight. For instance, the degree is not always the right choice. Consider Figure \ref{fig:nodesize3}(a): would you think to use the node's degree as a measure of its size? If you do, you end up with Figure \ref{fig:nodesize3}(b) where the node playing arguably the strongest role in keeping the network together almost disappears. A much better choice, in this case, is betweenness centrality (Figure \ref{fig:nodesize3}(c)).

It shouldn't surprise you -- after all the network analysis we've done -- to hear that many variables of interest in a network have very broad distributions. Degree and betweenness centrality, the two examples cited so far, follow (quasi) power laws, with few gigantic hubs and many nodes with minimum values. This means that linear size scales are not going to work very well: everything is going to be tiny and then BAM! One huge node, the hub.

\begin{figure*}
\centering
\begin{subfigure}{.425\textwidth}
\includegraphics[width=\textwidth]{figures/marvel_dd_log.pdf}
\caption{}
\end{subfigure}\qquad
\begin{subfigure}{.425\textwidth}
\includegraphics[width=\textwidth]{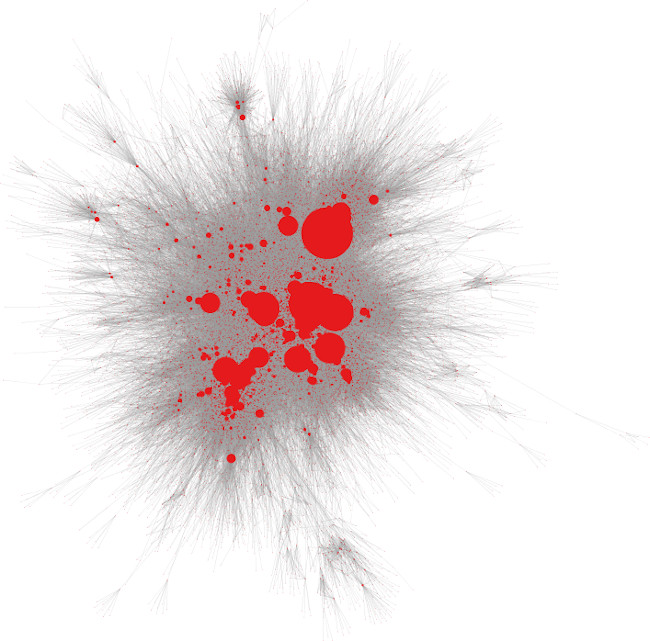}
\caption{}
\end{subfigure}
\caption{(a) Degree distribution of the Marvel social network example. (b) Visualizing the network with a linear node size map, where the degree directly determines the node size.}
\label{fig:marvel}
\end{figure*}

Consider Figure \ref{fig:marvel}. Here, we use the degree to determine the node size and we use a linear scale. Since this is an example of comic book characters, we expect the ones appearing with many other characters in the same comic book to be the most important. And they are: the largest nodes are the ones you would expect. But... they are too much the ones you expect. Their size swamps everything else. As a result, the visualization might be \textit{truthful}, but it's not \textit{informative}. It doesn't show you any new information. You already knew all you can gather from it.

To counteract this, you need to apply a quasi-logarithmic scaling. If you're creating your visualizations programmatically you can have an actual log scale, although you probably will still have to manually tweak it a bit to make the result more pleasing. The idea is to have diminishing returns to the contribution of the degree to the node size. The differences in size from the minimum degree, to the average degree -- which can be quite low -- are big but, from that point on, the contribution to the node's size plateaus.

\begin{figure}
\centering
\begin{subfigure}{.475\textwidth}
\includegraphics[width=\textwidth]{figures/marvel.jpg}
\caption{}
\end{subfigure}
\begin{subfigure}{.475\textwidth}
\includegraphics[width=\textwidth]{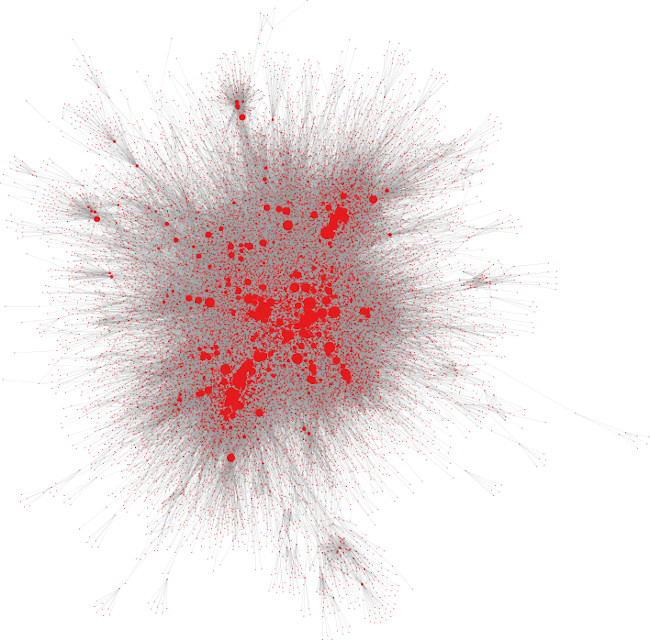}
\caption{}
\end{subfigure}
\caption{(a) The comic book social network using degree directly for node size. (b) Same network using a quasi-logarithmic scaling for the node size.}
\label{fig:marvel2}
\end{figure}

If you do so, you can find new clusters that were previously cluttered by the huge nodes, or that had a low degree and so they did not pop up. You can compare the two hairballs in Figures \ref{fig:marvel2}(a) and \ref{fig:marvel2}(b). Note that the visualization is still truthful: we're never going to make nodes with lower degree larger than nodes with higher degree. That would be bad and land you in the naughty corner. We're just making the visualization more useful.

This is probably a good place to stop and make a disclaimer. Even if eyes are the highest bandwidth sensors we have, it doesn't mean they are flawless. Nor that our monkey brain is able to use the information they gather in a perfect way. Human perception is flawed and you cannot expect that something a computer understands will appear obvious to your viewers as well. In the case of node size this takes the form of the confusion between radii and areas.

Unless otherwise specified by the software/program of choice, you are going to decide the \textit{radius} of the node when determining its size. This can be trouble if you don't handle this choice properly. The reason is that, when you increase the radius, you are substantially performing a linear increase: you think that, if the degree increases by one unit, you should increase the radius by one unit. Unfortunately, what a viewer will perceive is you changing the \textit{area} of the circle. The crux of the problem is that a radius is a one dimensional quantity, and it should never be used for controlling a two dimensional one such as an area -- which is what your readers perceive. You think you're increasing something linearly, but you're actually raising that increase by the power of two.

\begin{figure}[t]
\centering
\includegraphics[width=.66\columnwidth]{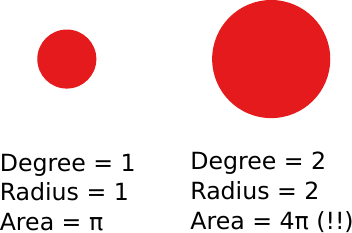}
\caption{A human perceives a node with radius one as being of size $\pi$, its area. So she will also perceive a node of radius two as being of size $4\pi$: double degree, but four times as large!}
\label{fig:nodesize4}
\end{figure}

Figure \ref{fig:nodesize4} shows you why you need to be well aware of the difference. What you think is a small increase can seem humongous to your reader.

\section{Color}
The second obvious feature to manage for your nodes is their color. If we routinely use node sizes for quantitative attributes, we primarily use node color for \textbf{qualitative} ones. The reason is that, while humans perceive size as quantitative, color hue is not perceived in the same way. Cleveland and McGill\cite{cleveland1984graphical} distinguish between different data types and how much different graphical features are effective for each data type. Color is good for nominal attributes -- categories that cannot be compared/sorted, like ``apple'' vs ``orange''. Color could be used for ordinal attributes, that are still categories but can be compared -- for instance days of the week, Monday comes before Tuesday. Color is terrible for quantitative attributes, for which areas are a more effective tool. As always, what follows is based on my experience and, if you want or need more in-depth explanations, you should check out the paper.

When it comes to network visualization, this implies that we put nodes into classes and we use colors to emphasize that different nodes are in different classes. Classical examples can be the node's community -- Part \ref{par:cd} --, or its role -- Chapter \ref{cha:centr-roles}. I already mentioned nodes can have metadata, and these metadata could be categorical. For instance, in a network connecting online shopping products because they are co-purchased together you could use the color to determine their category (outdoors, rather than kitchen, rather than electrical appliances).

\begin{figure}
\centering
\includegraphics[width=.8\columnwidth]{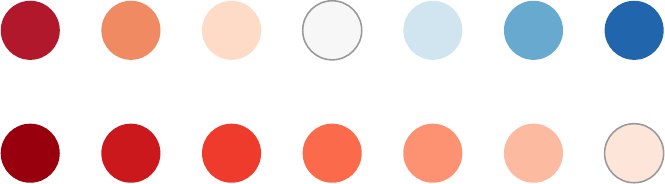}
\caption{(top) A gradient palette for diverging quantities and a meaningful middle point of the spectrum. (bottom) An intensity gradient, useful to go from zero to a maximum value without a meaningful middle point.}
\label{fig:nodegradient}
\end{figure}

You could still use node color for ordinal attributes, and maybe for quantities as well, provided that you have clear and intuitive bins. The way one would use colors for quantities is by implementing a gradient. A classical one is a blue-red spectrum for temperatures: this is a diverging scale that can be useful, e.g., if you have some sort of correlation data. You have a very precise and semantically meaningful middle point, and nodes can diverge in either of two directions, as I show in Figure \ref{fig:nodegradient} (top). Otherwise, if we're talking of a more classical intensity -- say how much money a customer spent in your online shop -- you want a simple sequential gradient, just like the one in Figure \ref{fig:nodegradient} (bottom).

There are many things you need to take into account when using colors. One of the trickiest ones is cultural associations\cite{silva2011using}. When you visualize something, your visualizations come after centuries -- if not millennia -- of other people using colors for different tasks. These usages ingrained in our mind a quick way to decode information. For instance, we associate red with danger, yellow with caution, green with ``good to go''. Black is death, and -- stereotypical -- blue is for boys and pink is for girls. But blue is also Democrat against red Republican if we're talking about elections in the US -- which, interestingly, is the opposite of the left-right wing spectrum for other countries in which red is on the left. It all depends on the context in which you're visualizing. Color can aid you in making your visualization quicker to decode, but if you're instead using it differently from a convention it can make things harder.

This doesn't even take into consideration the deficits in the physical perception by humans. Just to repeat myself -- we are very limited when it comes to distinguish colors. If you ask your laptop how many colors there are out there, a popular reaction would be counting the number of possible RGB combinations and to reply: $16$ millions! That would be very wrong for any human with a hint of common sense. In fact, what I would say is that you should never use more than nine colors in your visualization, and I'm sure that a few of my data designer friends are already gasping in horror to the extent of my liberalism. Nine, for them, is already way too much.

\begin{figure}
\centering
\begin{subfigure}{.6\textwidth}
\includegraphics[width=\textwidth]{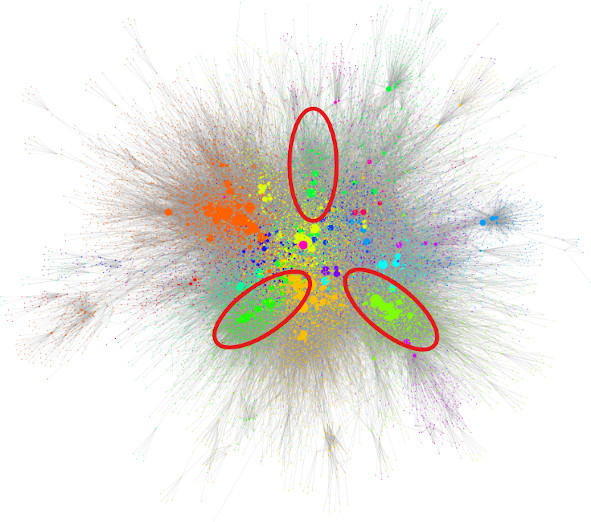}
\caption{}
\end{subfigure}
\begin{subfigure}{.3\textwidth}
\includegraphics[width=\textwidth]{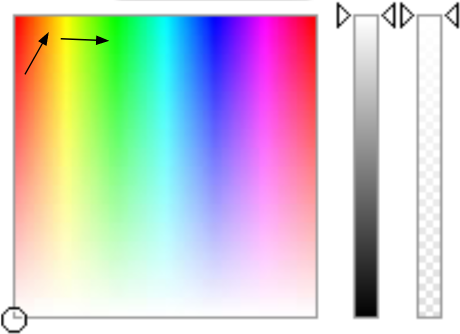}
\caption{}
\end{subfigure}
\caption{(a) The comic book social network using communities for the node's color. (b) An example of the RGB color space. The black arrows indicate equal distance movements in this space, connecting colors at different human perceptive distances.}
\label{fig:rgb}
\end{figure}

It's not just about the quantity of colors, though, it is also about how to choose and use them. How many colors would you say I used for the nodes in Figure \ref{fig:rgb}(a)? If you guessed $16$ -- which is the correct answer -- you're very lucky, or you have some Truman Capote levels of pattern recognition. In the network I highlighted three groups of nodes. These have different colors, believe me or not. Few -- if any -- people would be able to tell without scanning the figure for more than a handful of seconds. Requiring this level of effort from your viewer means to lose them.

Why does Figure \ref{fig:rgb}(a) fail? Because it assumes that RGB is a perceptive color space. In a perceptive color space, if you move by a given amount, you get to a color that will be perceived differently. This is wrong, as Figure \ref{fig:rgb}(b) shows: the two black arrows in it make two movements in the space and show that the same RGB space distance can connect either two virtually identical colors -- virtually identical for our monkey brains -- or two very distinct ones.

\begin{figure}
\centering
\includegraphics[width=.8\columnwidth]{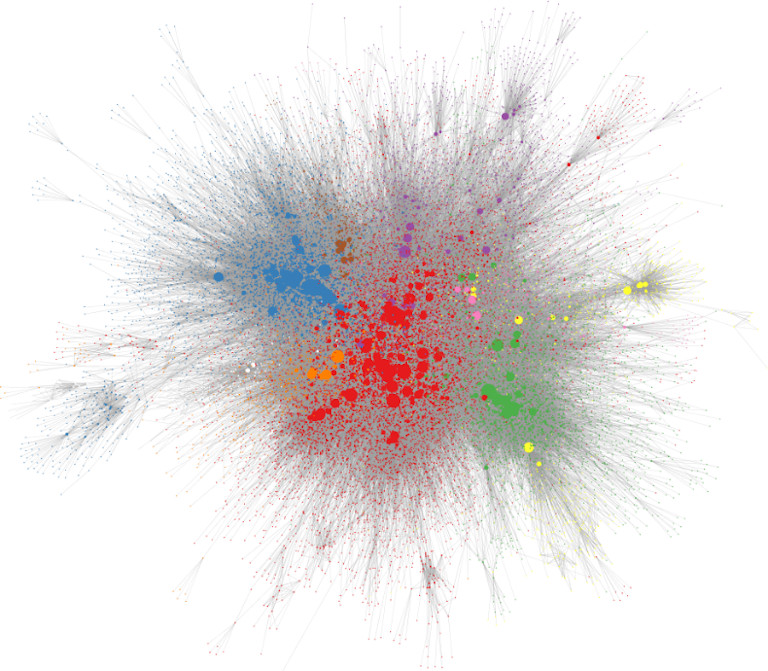}
\caption{A version of Figure \ref{fig:rgb}(a) with fewer colors and based on a more sane color space than RGB.}
\label{fig:colorbrewer}
\end{figure}

If there is one message I wish I could ingrain in you after reading this material is this one: RGB is a terrible terrible terrible color space for information visualization and nobody should use it for anything related to data design ever. There is some research backing color palettes that align better with human perception -- also including the case of people with color blindness\cite{brewer1994color}. A good resource you can use is the Color Brewer interactive tool\cite{harrower2003colorbrewer}, which will generate the palettes for you\footnote{\url{http://colorbrewer2.org/}}. Color Brewer is embedded in many software/programming packages that you might already use for your visualizations, including R\cite{neuwirth2014colorbrewer}, Cytoscape (since version $3.7.1$, for earlier version you need the Color Cast plugin), QGis, Python (Matplotlib and Seaborn, for instance), and Matlab.

Figure \ref{fig:colorbrewer} uses the Color Brewer space and fixes one of the many problems of Figure \ref{fig:rgb}(a). In Figure \ref{fig:colorbrewer} we use also fewer colors -- just nine -- which is always good.

Color Brewer and RGB are not the only possible color spaces you could use. If you are creating visualization for printing, you should use a CMYK color space. This is similar to RGB, but RGB is an \textit{additive} color space, while CMYK is \textit{subtractive}. Additive color spaces describe how different wavelengths of light add to each other, which is how computer screens work. Subtractive color spaces, instead, describe how ink combines on the page, which is why it'll show better how things will look in print. HSV and HSL are alternative color spaces which transform RGB to be more perceptually-relevant: we as humans don't really perceive colors as combinations of red-blue-green, but as variation in hue, saturation and lightness, which is what HSL stands for.

\begin{figure}
\centering
\includegraphics[width=.66\columnwidth]{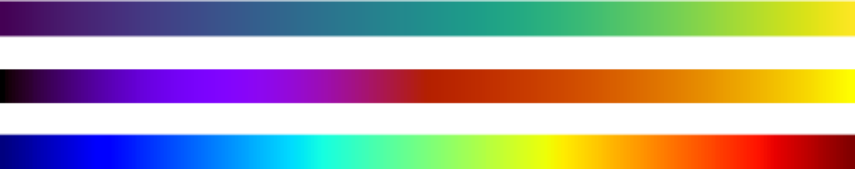}
\caption{Some linear color palettes you'll find available in different software. From top to bottom: viridis (Matplotlib), Gnuplot default, and jet (Matlab).}
\label{fig:palettes}
\end{figure}

\begin{figure}[t]
\centering
\includegraphics[width=.66\columnwidth]{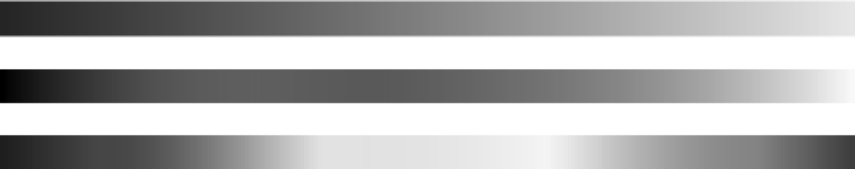}
\caption{The linear color palettes from Figure \ref{fig:palettes}, transformed in a grayscale. From top to bottom: viridis (Matplotlib), Gnuplot default, and jet (Matlab).}
\label{fig:palettes-bw}
\end{figure}

There are other options for linear color gradients -- which I show in Figure \ref{fig:palettes}. Some of these palettes were systematically compared across a series of tasks viewer might want to perform\cite[0.2in]{liu2018somewhere}, as well as different issues your readers might have with perceiving colors. For instance, to tackle the aforementioned issue of color blindness, you could transform these palettes into their correspondent black and white version and see how they look like to a person unable to distinguish colors but only relying on lightness. I do exactly this in Figure \ref{fig:palettes-bw} and show that, for instance, the jet palette in Matlab performs poorly because the two ends of the spectrum become indistinguishable. Of course, testing for color blindness and other human vision deficiencies is much more complex than this, and you should delve deeper in the literature I cited at the beginning of the chapter.

Note that here I used node colors for communities. It seems that I'm suggesting that you shouldn't find more than nine communities in your networks. That is not exactly what I'm saying. Of course, when it comes to the \textit{analysis}, you will find the number of communities that you will find. The sky is the limit there. It's when it comes to \textit{visualizing} them that you should never show more than nine. If you try to break that limit, you may as well not visualize anything. A famous motto in data visualization is: ``emphasizing everything means to emphasize nothing''. So you need to find a different solution, maybe showing smaller extracts of your data.

As with node sizes, also in node colors -- if you're applying a gradient -- you're best served using a (quasi)logarithmic scale to highlight differences better and make color variance more meaningful.

\section{Other Features}
To wrap up this chapter, let's see a few more things you can do to your nodes. They both stem from the same idea: your nodes represent something, and so you want to communicate this to your viewers.

The first strategy involves \textbf{node labels}. If you want the audience to know something, you simply tell them. You plaster some text on top of your nodes and you call it a day. In my opinion, this is a desperate move and it should be avoided if possible. Just as in movies, also in data visualizations it's better to ``show, not tell''. In other words, nobody wants to \textit{read} your network. They want it to speak to them.

\begin{figure}
\centering
\includegraphics[width=.75\columnwidth]{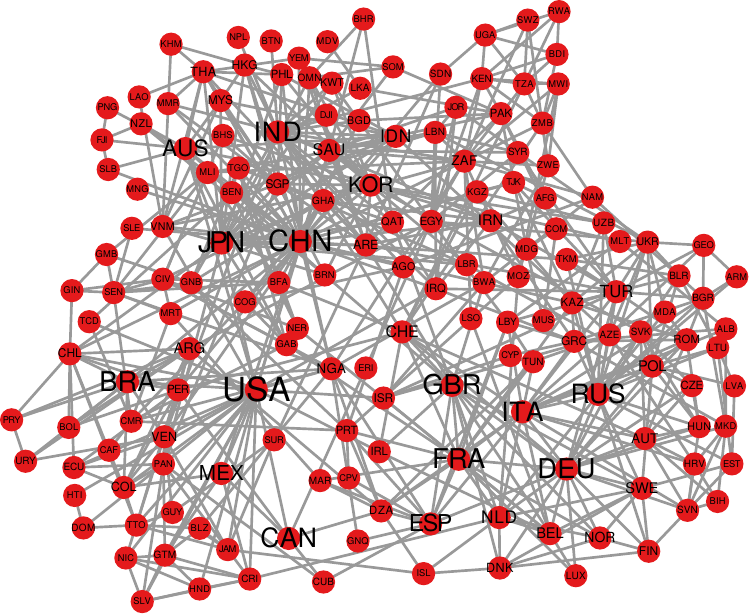}
\caption{A network with node labels conveying information about a node's importance. In this trade network, it is the country's Gross Domestic Product.}
\label{fig:nodelabels}
\end{figure}

That is not to say that sometimes a good choice of node labels can enhance your visualization. You can practically transform your network into a glorified word cloud. I don't love it, but I grudgingly admit that sometimes it works. An example could be the one in Figure \ref{fig:nodelabels} -- although in this case one should choose a less saturated color for the nodes, because the current red goes in the way of the readability of the label. My rule of the thumb is that the node label font size should have a one-to-one correspondence to the node size. It would look weird to have a gigantic label on top of a tiny node, and vice versa.

The second visual attribute you could play with is the node's border. This is an interesting one, because it could be used for quantitative and qualitative attributes at the same time. For the node border, you can both decide the color \textit{and} the thickness. Again, you should really ask yourself whether you really need to do it. Personally I almost never touch node borders -- I'd say that in $99\%$ of my visualization the border is invisible. If you already have node sizes and colors, adding a border of a different size and color would just cause information overload in your reader's brain. You should only do it if there are extremely clear patterns in your network, which involve no more than a handful distinct values, and that can be easily parsed.

\begin{figure}
\centering
\includegraphics[width=.75\columnwidth]{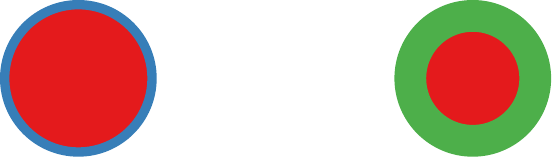}
\caption{Two nodes with same color and size, but with borders of different thickness and color.}
\label{fig:nodeborders}
\end{figure}

For instance, in Figure \ref{fig:nodeborders}, we could have two nodes of same size and color -- perhaps these are two plants in the same country (color) and employing the same number of people (size). However, they process different products (border color) and they have different throughput in number of products processed per day (border thickness).

Another strategy is more creative -- and for this reason you should apply tons of caution if you want to go this way. It involves xenographic\footnote{\url{https://xeno.graphics/}}. This translates to ``weird visualizations'', stuff that has very specific and almost unique use cases, and thus it's likely to choose a style that people haven't seen before. You can be creative with what you put on your nodes, as long as you don't abuse it and it has a meaningful relationship with your message.

One obvious way you can communicate differences in kind when it comes to a node would be to represent it not as a dot, but as a figure. The classical case is by transforming the node's shape. I already used this approach in this book for bipartite networks. A classic way to visualize them is to use one node shape for $V_1$ nodes, and another for $V_2$ nodes. For instance they can be circles vs squares. Another way is to use symbols. For instance, you could have a network of dogs and use a silhouette of the dog's appearance to encode its breed -- this is inspired by the beautiful ``Top Dog'' visualization\footnote{\url{https://www.informationisbeautiful.net/visualizations/best-in-show-whats-the-top-data-dog/}}.

My favorite use case, instead, keeps the node's shape constant, but transforms it into a chart in itself. This involves the often-maligned pie charts. Pie charts get a bad rep, often deservedly so, but can be rather useful in specific instances. You can use them both in the qualitative and in the quantitative use case, making them more versatile than either node color or size.

\begin{figure}
\centering
\begin{subfigure}{.55\textwidth}
\includegraphics[width=\textwidth]{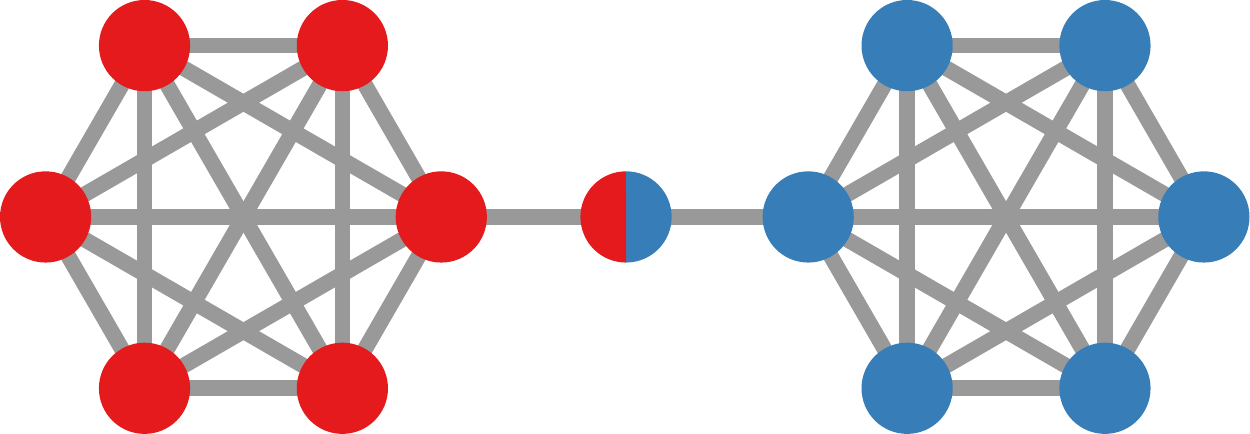}
\caption{}
\end{subfigure}\qquad
\begin{subfigure}{.25\textwidth}
\includegraphics[width=\textwidth]{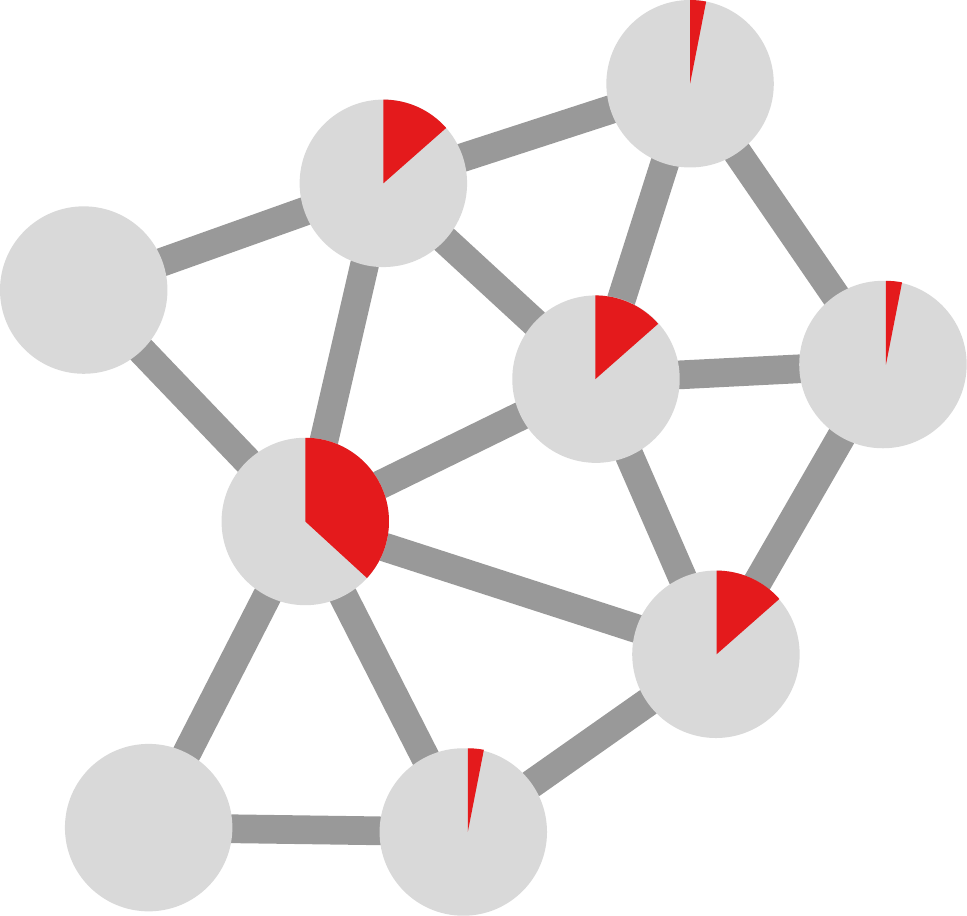}
\caption{}
\end{subfigure}
\caption{(a) Using pie charts on nodes to signify their allegiance to multiple communities. (b) Using pie charts to represent the relative centrality of each node.}
\label{fig:piecharts}
\end{figure}

We know that communities in networks can share nodes (Chapter \ref{cha:ocd}). If you're using the node color to encode the community, what do you do if a node belongs to more than one of them? You can use a pie chart for that! Figure \ref{fig:piecharts}(a) shows an example. This assumes that the node is not part of too many communities but, if you have enough communities in your network to break a pie chart, you shouldn't use colors to encode them to begin with.

In the quantitative case, pie charts are more limited, but still work in case you have ``quantitative classes''. With that, I mean that you have a quantitative attribute that can take very specific and very different values, such as a centrality. In that case, distinguishing between few very different pie charts is possible even for the human brain, as you can see in Figure \ref{fig:piecharts}(b).

\begin{figure}
\centering
\includegraphics[width=.75\columnwidth]{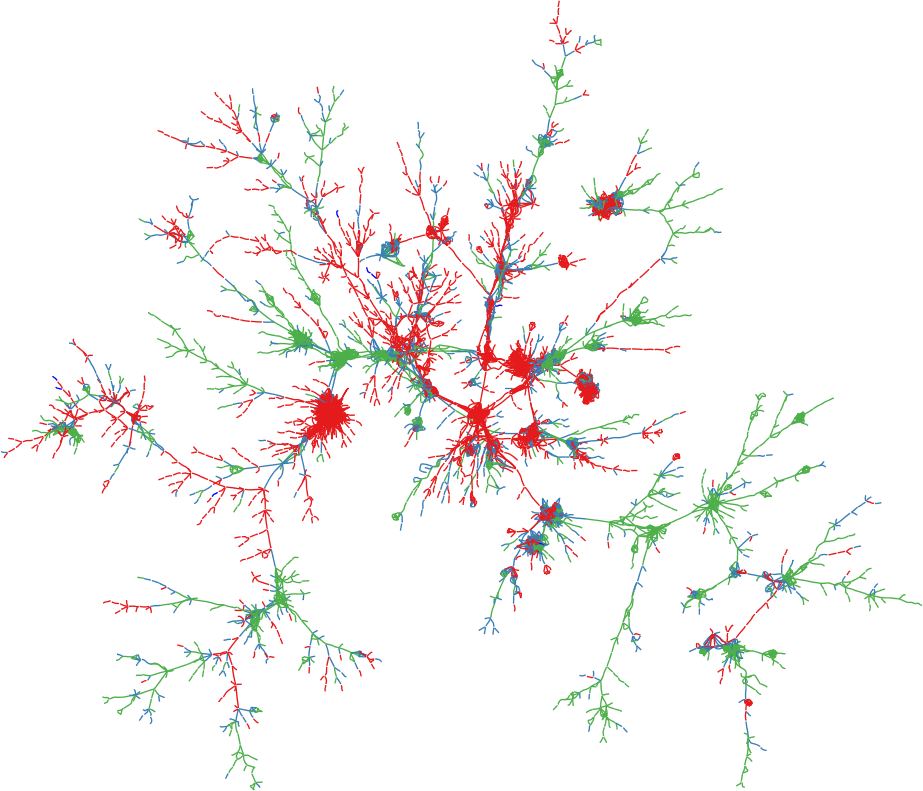}
\caption{A network with fully transparent nodes, where all the topological information is conveyed by the edge colors.}
\label{fig:nodetransparency}
\end{figure}

A final xenographic touch concerns playing with the alpha channel -- i.e, the opacity of the node. In this paragraph, I consider the extreme case of not showing the nodes at all. Normally, there's no point in having invisible nodes. After all, the nodes are what you want to see in a network, so why making it impossible to look at them? However, there are some use cases in which this rule can be broken. I present one in Figure \ref{fig:nodetransparency}. I'm not arguing that the figure is a \textit{good} visualization: what I'm saying is that being able to see the nodes would not make much of a difference, especially since they do not have attributes of interest. Rather, the visualization allows you to see where different types of edges create red and green clumps, and which edge type keeps the network together in which branches. And how to deal with edge visual attributes is exactly the topic of the next chapter.

\section{Summary}

\begin{enumerate}
\item The first visual attribute of nodes is their size. Usually, you want to show quantitative attributes via size -- the degree, the capacity, etc. Be aware that you should always manipulate the \textit{area} of the node, which is what your viewer perceives. If your software only allows you to control a node's \textit{radius}, keep in mind that your area will change quadratically for each linear change of the radius.
\item Second, you can control a node's color. Usually, this is for qualitative attribute, e.g. community affiliation. Use no more than nine distinct colors, from a perceptual-aware space (\textit{not} RGB rainbows!).
\item Gradients can be used for quantitative attribute: diverging ones for quantities with a clear midpoint -- e.g. correlations --, otherwise sequential ones for quantities going from zero to an arbitrary maximum.
\item You can augment your nodes with additional visual elements. Labels could be used -- sparingly -- and their size should be locked with the node's area size. You can use pie charts and icons to embed additional information on the nodes.
\end{enumerate}

\section{Exercises}

\begin{enumerate}
\item Import the network at \url{http://www.networkatlas.eu/exercises/49/1/data.txt}, calculate the nodes' degrees and use them to set the node size. Make sure you scale it logarithmically. This can be performed entirely via Cytoscape. (The solution will be provided as a Cytoscape session file)
\item Import the community information from \url{http://www.networkatlas.eu/exercises/49/2/nodes.txt} and use it to set the node color. (The solution will be provided as a Cytoscape session file)
\end{enumerate}

\chapter{Edge Visual Attributes}\label{cha:edgeviz}
When it comes to edge visual attributes, most of the things already mentioned for nodes in Chapter \ref{cha:netviz-node} still apply. So this is going to be mostly a recap, with a few additional warnings.

\section{Classical Visual Elements}

\subsection{Size}
\begin{figure}
\centering
\includegraphics[width=.6\columnwidth]{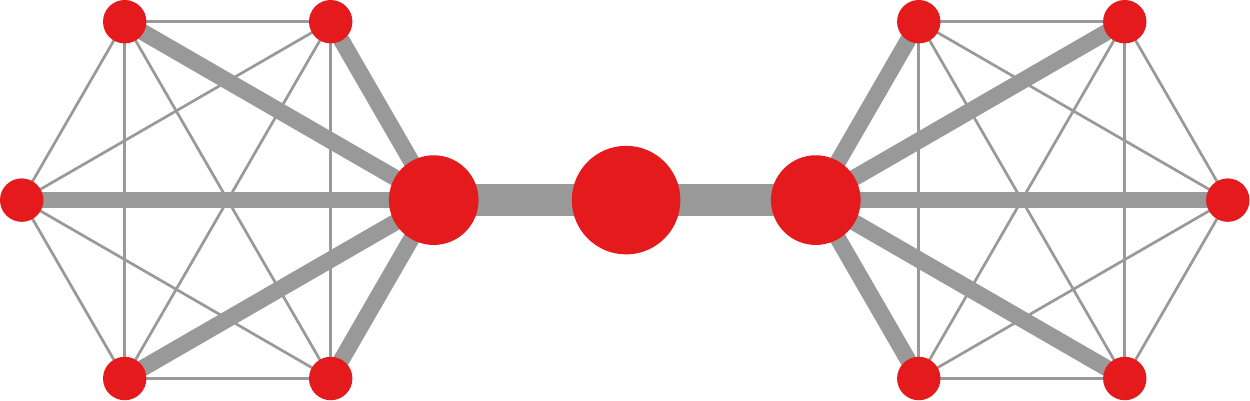}
\caption{In this network the edge thickness is proportional to its edge betweenness value. The node size is proportional to the node betweenness.}
\label{fig:edgesize}
\end{figure}

The equivalent for edges of node size is the thickness. As in the previous case, this is mostly for quantitative attributes on edges. The most trivial one is the edge's weight: heavy edges usually appear to be more thick. Another common use case is to put edge betweenness as the determinant of the edge thickness. This works well when used in conjunction with nodes sizes following the same semantics. It gives a sense of balance to the visualization, so you can see which edges are contributing to the node's centrality. Figure \ref{fig:edgesize} shows an example.

Just like node betwenness, also edge betweenness is unevenly distributed across edges. And, as you already saw, typically edge weights distribute equally broadly -- see Chapter \ref{cha:backboning} for a refresher. So you have to apply the same pseudo log scaling for edge thickness as you did for node size. Lines are considered one dimensional, so you shouldn't worry too much about the square area problem I mentioned for node sizes. It will start to be a problem only if your edges are so large that your eyes start interpreting lines as rectangles, at which point they're probably already too large!

\subsection{Color}
When it comes to colors, there are more differences between edges and nodes than we just saw for sizes\cite{szafir2017modeling}. The fundamental difference between edges and nodes is that there are so many more of the former than of the latter: typically twice or three times as many. Also, dots and circles are much easier to see than lines, especially thin lines. Since most edges will have low weights, most of them will be relatively thin, as we just discussed. Thus, seeing the edges is trickier.

There is another reason, which is even more practical. Very rarely, if at all, you will have good qualitative information about your edges. By their very nature of being the glue connecting things, edges are much more likely to have quantitative information attached to them. We usually classify things, not the glue between things.

\begin{figure}
\centering
\begin{subfigure}{.33\textwidth}
\includegraphics[width=\textwidth]{figures/multilayer.pdf}
\caption{}
\end{subfigure}\qquad
\begin{subfigure}{.5\textwidth}
\includegraphics[width=\textwidth]{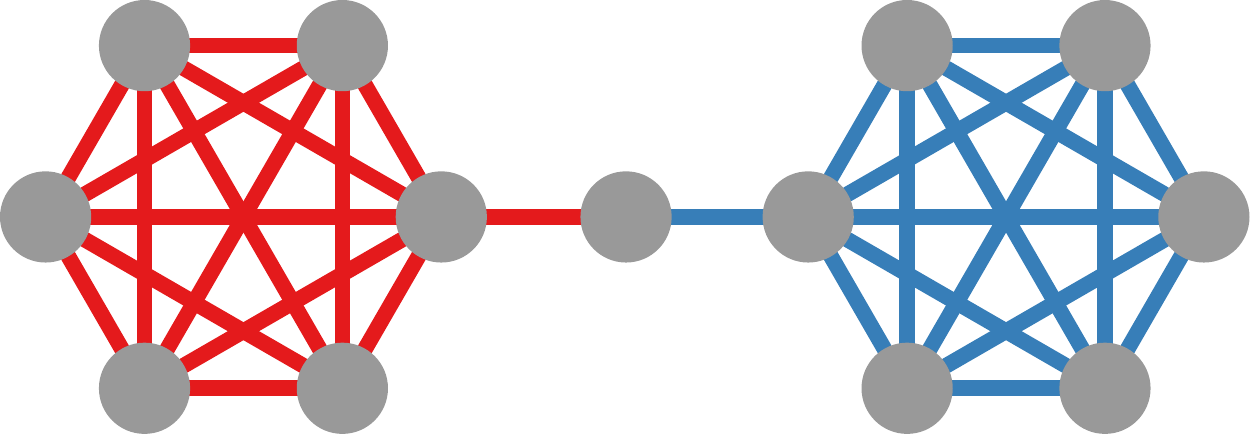}
\caption{}
\end{subfigure}
\caption{(a) Using edge colors to represent the edge's layer. (b) Using edge colors to represent the link community to which they belong.}
\label{fig:edgecolors}
\end{figure}

The most obvious exceptions are two. You can have qualitative information telling you to which layer and to which community an edge belongs, if you have multilayer networks (Section \ref{sec:extended-multilayer}) and/or link communities (Section \ref{sec:ocd-link}). For multilayer networks you can use edge colors to represent the layer if you adopt a multigraph visualization -- as I do in Figure \ref{fig:edgecolors}(a). However, this will get unwieldy pretty soon, as the number of nodes, edges and layers grows beyond an elementary size. In fact, it's usually best to use dedicated tools for the visualization of multilayer networks\cite{de2015muxviz} -- although the field of visualizing multilayer networks is still in its infancy.

For link communities, keep in mind the same warning I made regarding node communities. It is pointless to try and visualize more than a handful -- nine -- communities, even more so when it comes to links. Figure \ref{fig:edgecolors}(b) shows an example. Again, it's not that you should always find fewer than nine communities in your analysis. You can and should find however many there are in the network. It's \textit{visualizing} them that is a problem.

\begin{figure}
\centering
\begin{subfigure}{.55\textwidth}
\includegraphics[width=\textwidth]{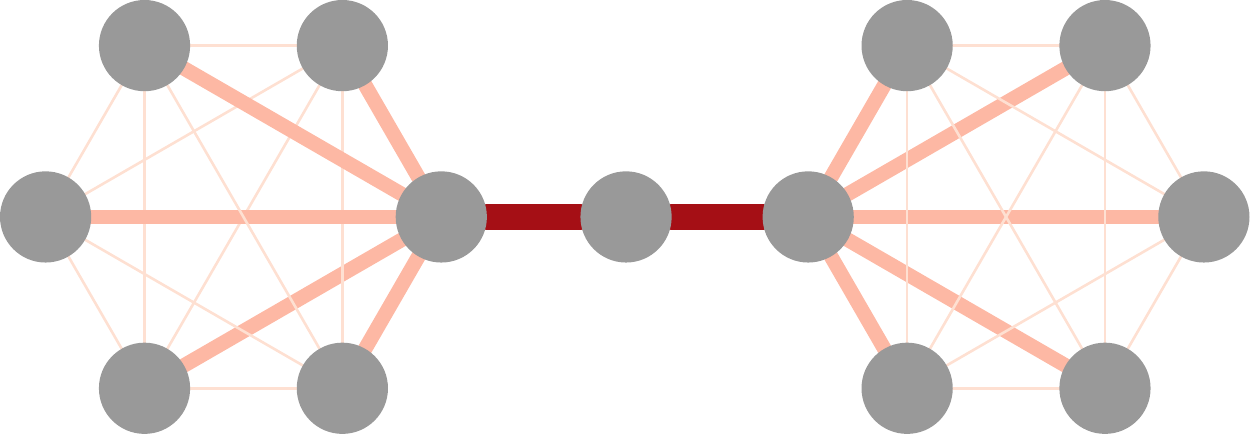}
\caption{}
\end{subfigure}\qquad
\begin{subfigure}{.25\textwidth}
\includegraphics[width=\textwidth]{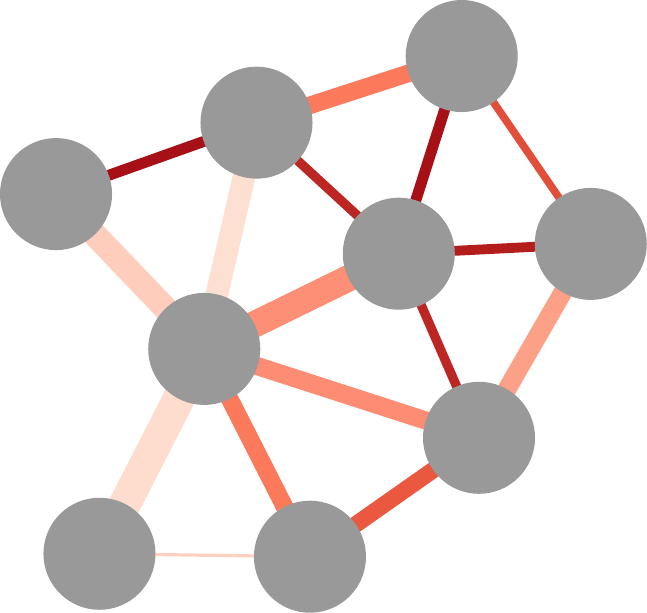}
\caption{}
\end{subfigure}
\caption{(a) Using edge colors to reinforce the message conveyed by the thickness: the edge's betweenness centrality. (b) Using edge colors for an orthogonal quantitative information: edge thickness is its weight, while the color represent the weight's significance.}
\label{fig:edgecolors2}
\end{figure}

As said, most often you will have quantitative information on your edges. So it is much more common to use gradients on the edges than it is on the nodes. You have gathered that I'm not a great fan of gradients from the previous chapter, but that's what we have to work with. The way I usually fix the problem is to use the same quantitative attribute for both thickness and color, so that the two can work in concert and reinforce each other. Two imperfect visual clues can sum and make each other clearer. This is the case for edge betweenness, determining both color and thickness of the edges in Figure \ref{fig:edgecolors2}(a).

That is not to say that there aren't good reasons to break this rule. One dimension I play with is usually the one of the edge weight's significance, for instance when doing network backboning -- see Chapter \ref{cha:backboning}. In that case, the color can work in contrast with the thickness. I find more natural to use thickness to represent how heavy an edge is, so to assign it to represent the weight. This is under the metaphor that large things generally weigh more. On the other hand, there is no inherent connection between the color of a thing and its weight. So I assign the color to represent the significance. Usually, larger weights tend to connect nodes which have higher average connection weights, so large links will tend to have paler colors than smaller links, which creates a nice contrast in Figure \ref{fig:edgecolors2}(b).

A final word of caution about the number of colors in your visualization. You might have noticed that Figures \ref{fig:edgecolors} and \ref{fig:edgecolors2} use edge colors but choose a single hue for the nodes. This is because in a network you have two visual elements -- circles and lines -- and allowing both of them to be colored differently effectively doubles the number of visual elements in your visualizations. Already Figure \ref {fig:edgecolors}(a) feels way too busy. If in the previous chapter I told you not to use more than nine colors in the visualization, here I'm giving you a more nuanced guideline: the sum of the distinct number of colors for edges and nodes must be nine or lower. Meaning that, if you have five colors for nodes, you shouldn't have more than four for edges.

\subsection{Transparency}
Transparency is another aspect in which edges diverge from nodes. In the previous chapter I mentioned that nodes should be fully visible, and provided only a single use case in which I believe transparency can add something to the visualization by removing the nodes from sight. When it comes to edges, I usually abuse transparency lavishly. Most commonly, I make transparencies work together with colors, to reinforce them. Significant links have darker colors \textit{and} are more opaque. The objective of playing with the alpha channel for edges is to create a visual hierarchy, where nodes come to the forefront and edges go to the background.

However, sometimes you can play with edge transparencies even if you don't have any attribute to attach to them at all! This is because of the sheer number of edges: in most real world networks, they are going to overlap to each other, no matter what. Thus edge transparency, even a fixed value, can highlight structure, because there are going to be more overlaps in dense areas of the network than in sparser ones. Thus, you can highlight such clusters even without any edge metadata.

\begin{figure}
\centering
\begin{subfigure}{.49\textwidth}
\includegraphics[width=\textwidth]{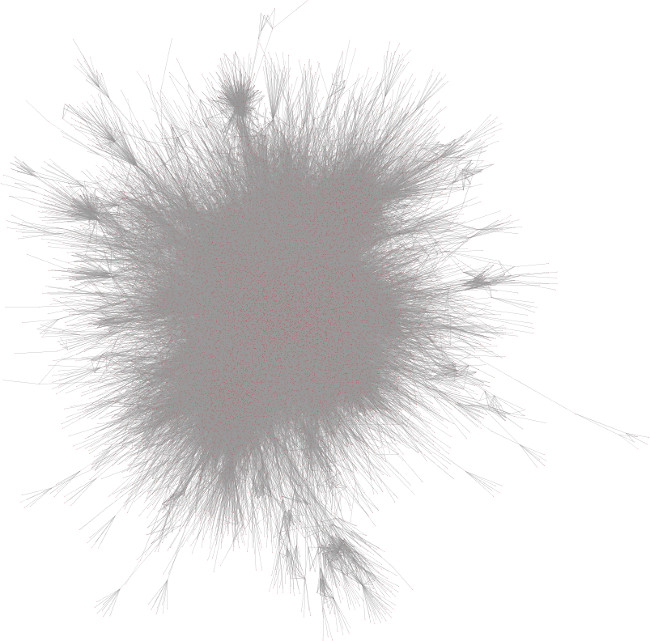}
\caption{}
\end{subfigure}
\begin{subfigure}{.49\textwidth}
\includegraphics[width=\textwidth]{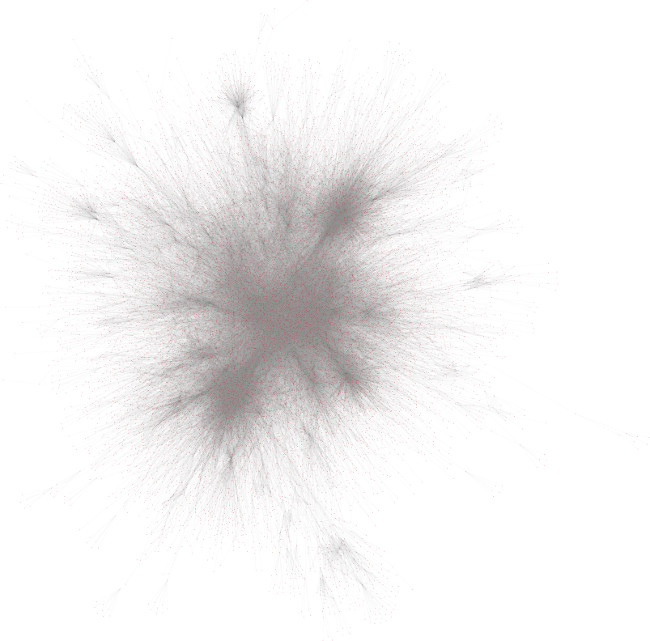}
\caption{}
\end{subfigure}
\caption{(a) A network with solid edges with $100\%$ opacity. (b) The same network, but this time all edges have a $37.5\%$ opacity, no matter their weight.}
\label{fig:edgetransparency}
\end{figure}

I do exactly that in Figure \ref{fig:edgetransparency}(b). Compared to Figure \ref{fig:edgetransparency}(a), the version with edge transparency looks less like a random smudge on the paper and shows a few structures of interest, even if I added literally zero bits of information by removing some edge opacity.

\subsection{Labels}
Moving on to edge labels: if I said that node labels have to be rarely used, then use edge labels even less than that. This is an extremely rare use case, you should avoid edge labels at (almost) all costs. Practically, they're only useful for scholastic examples, when explaining simple dynamics of super simple graphs. For instance, I used edge labels in this book for examples of weighted edges in weighted networks, with a grand total of five nodes and five edges. That's more or less stretching the use case of edge labels to the limit.

\section{Xenographic Elements}
Just like with nodes, also with edges you can be... edgy in how you visualize them. There are two fundamental aspects I'm going to mention here: shapes and bends.

The classical edge visualization is as a straight, solid line. This is what you should do in $99\%$ of the cases. However, in many cases, you might want to slightly change this shape. The most common shape change for edges is when you are working with directed connections. In this case, the convention is to add an arrow that indicates the direction of the edge. The arrow points from the originator of the edge to the target.

\begin{figure}
\centering
\begin{subfigure}{.4\textwidth}
\includegraphics[width=\textwidth]{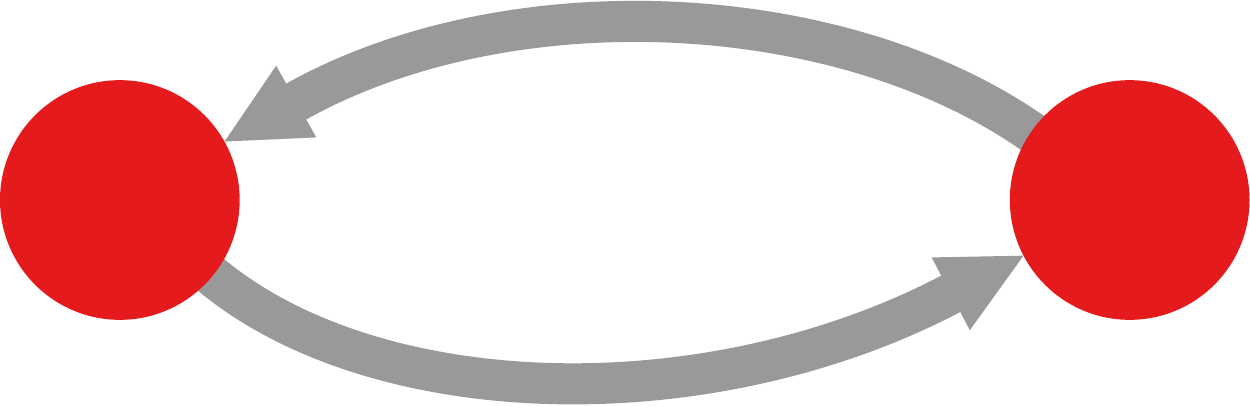}
\caption{}
\end{subfigure}\qquad
\begin{subfigure}{.4\textwidth}
\includegraphics[width=\textwidth]{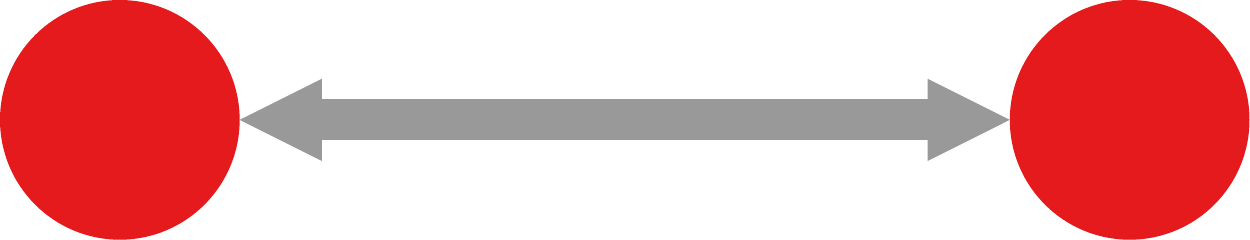}
\caption{}
\end{subfigure}
\caption{(a) The classical reciprocal edge visualization with two bent edges. (b) Merging the two edges in a single reciprocal edge.}
\label{fig:draw-recipr-edge}
\end{figure}

Directed networks are more challenging to visualize than you might think. The reason is not only that you're doubling the possible number of edges, which is true and it is an issue. But the real trouble is that now you might have a significant number of double edges between the same two nodes: $u \rightarrow v$ and $u \leftarrow v$. This might make your visualization a real mess. One convention you can implement is not to actually draw the two edges. What you can do is to draw a single edge and add to it a second arrow pointing in the opposite direction if that edge is reciprocal. Figure \ref{fig:draw-recipr-edge} shows how this strategy looks like.

Unfortunately, this is not an immediately obvious thing to do with standard network plotting software, so it might take some effort. Moreover, this visualization technique gets significantly more complicated in the case of weighted directed networks. If the two edges, $u \rightarrow v$ and $u \leftarrow v$, have two different weights, it is even less clear how you should handle visualizing them in a single line.

Line shapes can be manipulated in other ways, specifically by altering the line style. One can have dashed, dotted, wavy lines. These are clearly differences in qualities of the connection, and thus should only be used for qualitative attributes. My advice would be to rely on changing the edge line style exclusively for (i) very small networks, and (ii) just in those rare cases you need to print your visualization in black and white and thus cannot use color instead. It is clear that, once you have a lot of connections, they are going to inevitably be drawn one on top of the other. Distinguishing between a dashed and a dotted line if they overlap is nigh impossible.

The final odd thing you could do to your edges is to bend them, meaning that the $(u, v)$ edge is not a straight line from $u$ to $v$ any more, but it takes a ``detour''. Why would you want to do this? There are fundamentally two reasons. Figure \ref{fig:draw-recipr-edge}(a) provides an example: since there are two edges between the nodes, we want the visualization to be more symmetric and pleasant, and thus we bend the two edges.

More often, edge bends are used to make your network layout more clear. You bend edges to bundle together the ones going from nearby nodes to other nearby nodes. Since this is done mostly to clean up the visualization \textit{after} you already decided where the nodes should be placed, I will deal with this topic in the network layout chapter (Chapter \ref{cha:layouts}).

\section{Network Lifting}
Let's recap all the advice I gave you on node and edge visual attributes and see a case of applying each feature one by one to go from a meaningless hairball to something that conveys at least a little bit of information. Our starting point is the smudge of edges you already saw a couple of times: that's Figure \ref{fig:edgetransparency}(a). Note that this isn't really the starting point, because we already settled on a network layout, but that will be the topic of the next chapter.

The usual order I apply to my networks after I settled on a layout is the following:

\begin{enumerate}
\item Edge transparencies;
\item Edge sizes;
\item Edge colors;
\item Node sizes;
\item Node colors.
\end{enumerate}

So let's do this.

\textbf{Edge transparencies}. In this network, I do have quantitative information, that is the edge betweenness of each connection. However, I think that it's better if I limit that to the other edge visual features, so I fix the same edge transparency to all links. The result is Figure \ref{fig:edgetransparency}(b).

\begin{figure}
\centering
\begin{subfigure}{.49\textwidth}
\includegraphics[width=\textwidth]{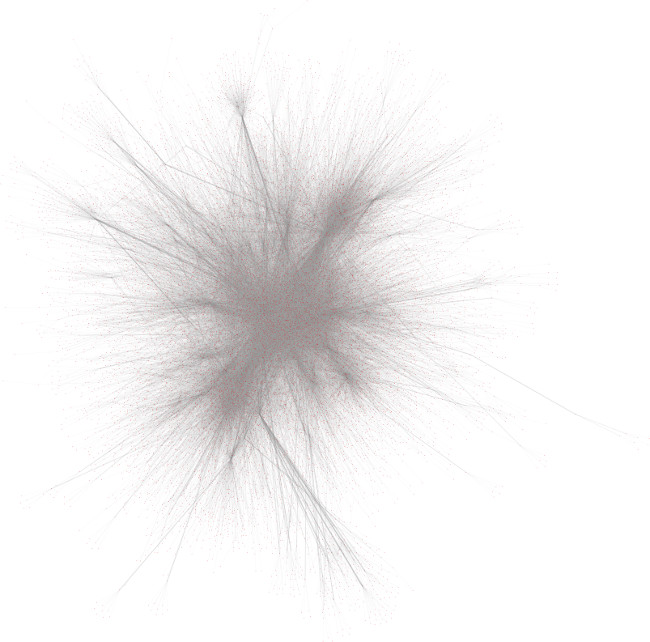}
\caption{}
\end{subfigure}
\begin{subfigure}{.49\textwidth}
\includegraphics[width=\textwidth]{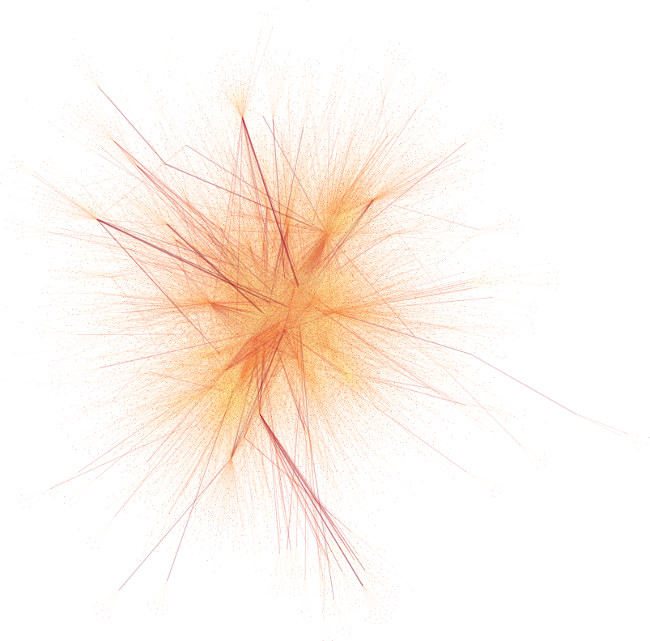}
\caption{}
\end{subfigure}
\caption{(a) Using edge betweenness for the link's thickness of Figure \ref{fig:edgetransparency}. (b) Using it for the link's color as well.}
\label{fig:pimp1}
\end{figure}

\textbf{Edge sizes \& colors}. We now move on to use edge betweenness. I merge the two steps of edge size and color into one, because using simply the thickness does not make a significant difference with the previous visualization. Compare Figure \ref{fig:edgetransparency}(b) with Figure \ref{fig:pimp1}(a) and see that not much has changed. So I apply a Color Brewer color gradient to the edges, resulting in Figure \ref{fig:pimp1}(b). Hopefully now you can see that there are a few very important long connections keeping the network together, connecting very central comic book characters to a sub universe they're almost exclusively part of.

\begin{figure}
\centering
\begin{subfigure}{.49\textwidth}
\includegraphics[width=\textwidth]{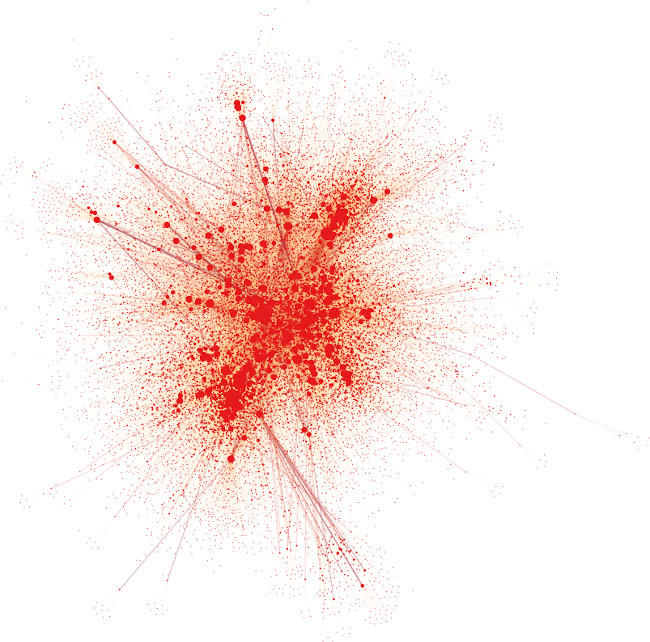}
\caption{}
\end{subfigure}
\begin{subfigure}{.49\textwidth}
\includegraphics[width=\textwidth]{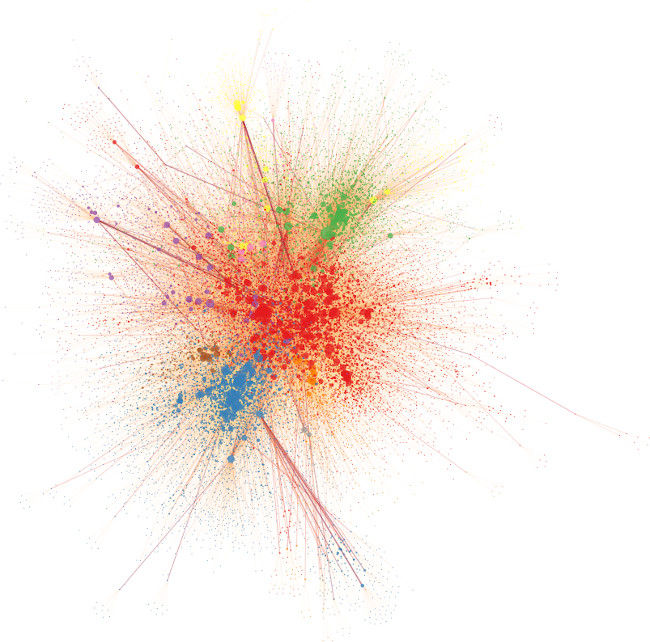}
\caption{}
\end{subfigure}
\caption{(a) Using node degree for the node's size of Figure \ref{fig:pimp1}. (b) Using communities for the node's color.}
\label{fig:pimp2}
\end{figure}

\textbf{Node sizes}. It's time to deal with nodes. There's something to be said for keeping them almost invisible, but that's not what we want to do here. This is a comic book network and so we want to know which characters are tightly connected to the universe of which other characters. So first we need to know who the important fellows are. We use node size for that. As outlined in the previous chapter, this is a job for the degree. The more connections a node has, the more important it is, the larger it should be. And that's what Figure \ref{fig:pimp2}(a) does. Note the pseudo-logarithmic node size scaling.

\textbf{Node colors}. Finally, we discover groups of characters using a community discovery algorithm -- see Part \ref{par:cd}. I tweak it so that it will return no more than nine communities. Again, I trust the qualitative color palette that Color Brewer provides, I'm partial to Set1 -- even if it is not exactly color blind friendly. See Figure \ref{fig:pimp2}(b) for the final result.

Figure \ref{fig:pimp2}(b) still has a long way to go before we can call it a good network visualization. But, compared to the starting point in Figure \ref{fig:edgetransparency}(a) we can definitely say more things about its structure. Which is exactly what network visualization is for.

A way to improve this picture would be to choose a better layout, and to apply some tweaks to it. That is the topic of next chapter.

\section{Summary}

\begin{enumerate}
\item Edge visual elements should be paired, whenever possible, with the same semantics as node visual elements. The thickness of an edge should be proportional to the same -- or a similar -- variable determining node size. If node size is its betweenness centrality, edge size should be its edge betweenness.
\item Differently from node colors, edge colors are used mostly for quantitative attributes. Usually, there are more edges than nodes in a network, thus it is harder to limit the number of edge colors. Anyhow, you should not have more than nine different colors in total, whether they are node or edge colors. Classical qualitative edge colors choices are layers or link communities.
\item You should try to collapse reciprocal edges in a directed network into a single edge with arrows pointing in both directions. You should use line style very sparingly, only for specific scenarios like black and white printing.
\item A classical workflow to improve your network visualization is to determine edge and node visual attributes in this order: edge transparency, edge width, edge color, node size, and node color.
\end{enumerate}

\section{Exercises}

\begin{enumerate}
\item Build on top of you visualization from exercise \#2 in Chapter \ref{cha:netviz-node}. Assign to edges a sequential color gradient and a transparency proportional to the logarithm of their edge betweenness (the higher the edge betweenness the more opaque the edge).
\item Import edge data from \url{http://www.networkatlas.eu/exercises/50/2/edges.txt} and use the attribute for the edge's width.
\end{enumerate}

\chapter{Network Layouts}\label{cha:layouts}
The network layout is the algorithm that decides where to place each node. This is usually a result of the other nodes to which it is connected. That is why I dub this process ``the art of scattering nodes around''. Such art has a queen: Mary Northway\cite{northway1940method}, the first person who realized -- in $1940$! -- that displaying nodes accurately is a must if one wants to parse social networks.

\begin{figure*}
\centering
\begin{subfigure}{.425\textwidth}
\includegraphics[width=\textwidth]{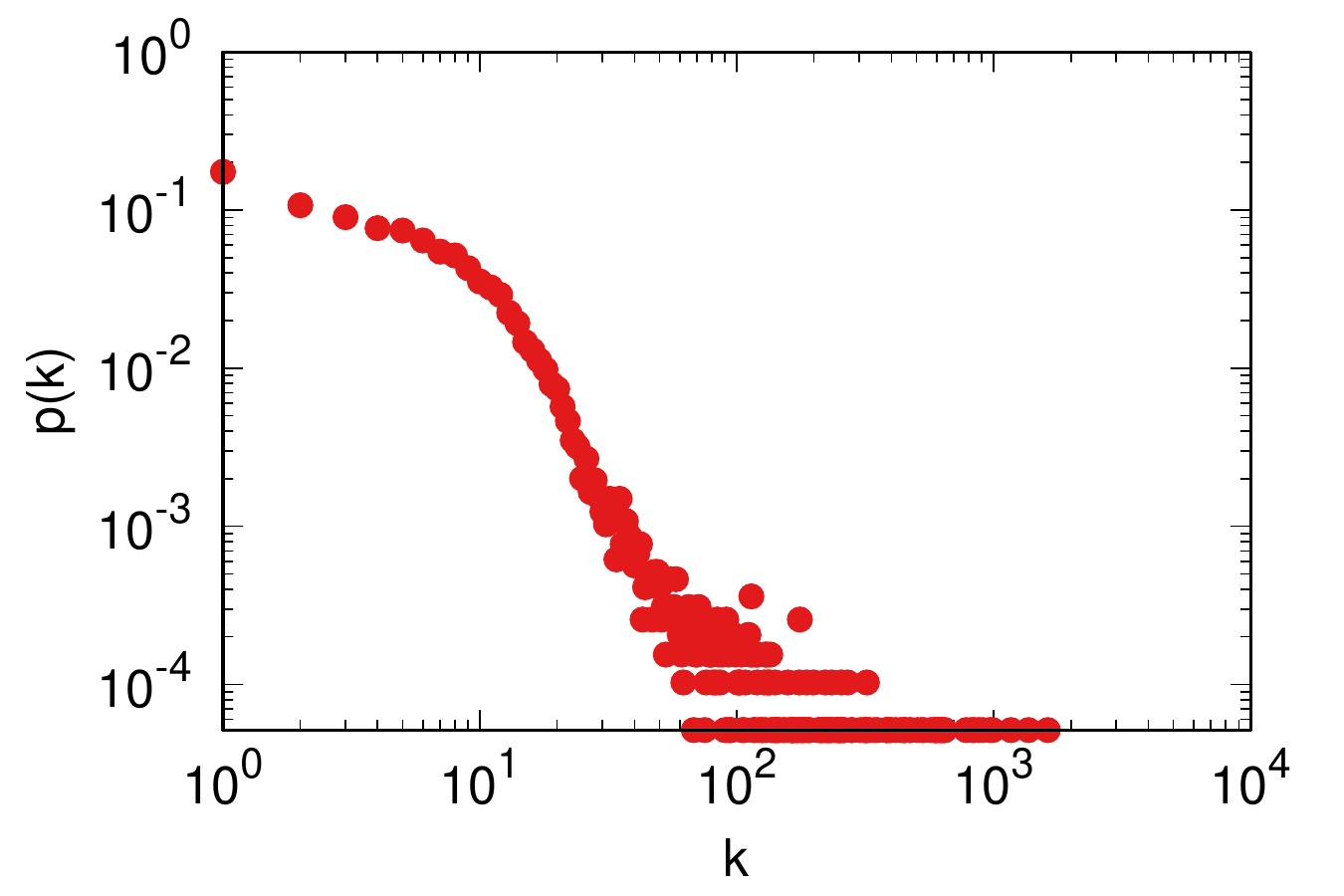}
\caption{}
\end{subfigure}\qquad
\begin{subfigure}{.425\textwidth}
\includegraphics[width=\textwidth]{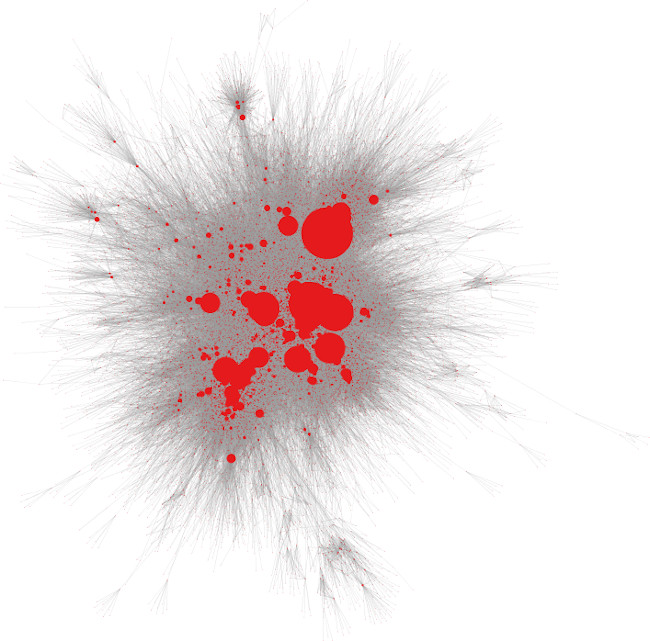}
\caption{}
\end{subfigure}
\caption{(a) In a scatter plot, changing the x-y coordinates of a point is forbidden, because that will change the data. (b) In a network, you can move nodes around, as long as you do it with a consistent set of rules to all nodes.}
\label{fig:netviz-freedom}
\end{figure*}

Changing graphical elements as we saw in the previous chapters is good, but network data has a peculiarity that other data types don't have. Networks are a particular data type that allow our representations an additional degree of freedom. This influences the network visualization. To see what I mean consider that, in a scatter plot, you cannot move the points around because that would change the data. But in a network you have connections. What counts is not the ``absolute'' position of a node, but its ``relative'' one. Figure \ref{fig:netviz-freedom} shows an example of what I mean.

The aim of this chapter is to present a few of the classical network layouts, trying to provide a rough guide on what should be used when -- although this is sometimes subtle and a matter of personal preferences.

There is a general issue you should be aware of: the vast majority of network layouts were developed with single layer networks in mind. However, they are commonly used for tasks that go beyond these types of networks. For instance, we will see that one common layout principle is the force directed one. People have been applying it to more complex structures such as hypergraphs\cite{makinen1990draw}\cite{arafat2017hypergraph}, or multilayer networks\cite{de2015muxviz}.

Unfortunately, there is not much visual research that I am aware of in these subfields. How to visualize a hypergraph or a multilayer network is a problem with its own challenges and they are both in need of finding their own conventions. The conventions I used when talking about these structures, specifically in Sections \ref{sec:extended-multilayer} and \ref{sec:extended-hyper}, are a good starting point. Otherwise, you should check out some recent surveys\cite{mcgee2019state}.

Another entry in the ``weird network types that we don't really know how to visualize'' is high order networks -- see Chapter \ref{cha:hod}. As far as I know, HoNVis\cite{tao2017honvis} is the only technique occupying the high order network visualization niche, so you might as well just read that paper.

\section{Force-Directed}\label{sec:layouts-force}
By far, the layout you will see in network science papers most often is the force directed layout. There are many variants of the same basic principle, but here I'll limit myself to explain the mechanics underlying most of them, and provide four examples you can find implemented in Cytoscape.

\begin{figure}
\centering
\begin{subfigure}{.3\textwidth}
\includegraphics[width=\textwidth]{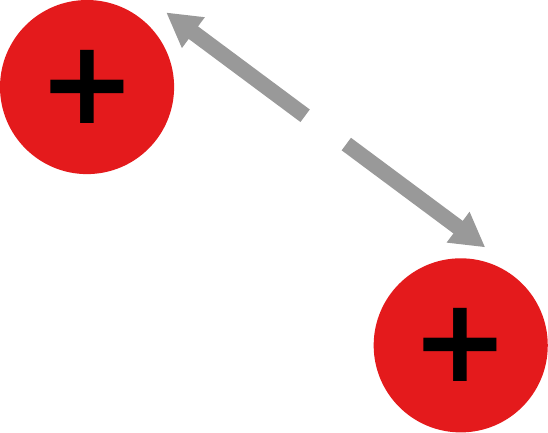}
\caption{}
\end{subfigure}\qquad
\begin{subfigure}{.3\textwidth}
\includegraphics[width=\textwidth]{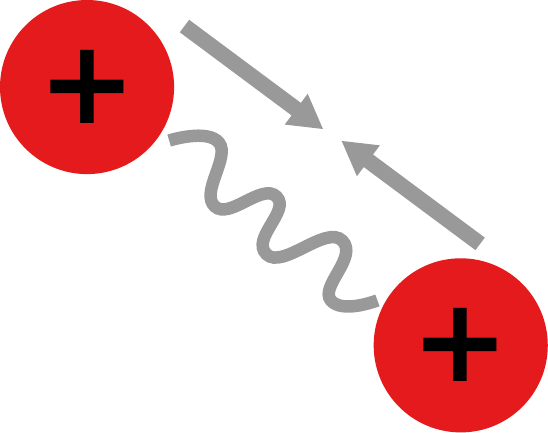}
\caption{}
\end{subfigure}
\caption{(a) Nodes behave like particles of the same charge repelling each other. (b) Edges behave like springs, bringing connected nodes together.}
\label{fig:forcedir}
\end{figure}

The basic principle underlying any force directed method is that nodes should try to repel each other so that they do not overlap, as you can see in Figure \ref{fig:forcedir}(a). However, connected nodes should be closer to each other, to represent relatedness using spatial proximity. The way this is usually implemented is to consider nodes to have the same magnetic sign -- creating the repulsive force. To bring connected nodes together, edges are -- rather than strings -- springs. The spring wants to be as short as possible and so it will pull connected nodes close together -- see Figure \ref{fig:forcedir}(b). Usually, springs are stronger than the repelling charge force at long distances, but have a minimum length, so that when the nodes are very close together they will not overlap. Of course, as soon as you add more nodes to the mix this gets pretty complicated, as nodes pull other nodes in different directions and so springs overstretch, even if they're stronger than charges.

Different force-directed network layouts include Prefuse\cite{heer2005prefuse}, yFiles organic\cite{wiese2004yfiles}, regular\cite{kamada1988simple} and compound spring embedded\cite{dogrusoz2009layout}, Fruchterman-Reingold\cite{fruchterman1991graph}, simulated annealing\cite{davidson1996drawing}, GME\cite{frick1994fast} and a bunch that I'm probably forgetting. They're all implemented in either Cytoscape or igraph. They usually differ in the strength and length they give to charges and springs, or in the strategy they apply to find the configuration in which charges and springs are the least stressed, i.e. the system has the minimal possible residual energy.

\begin{figure}
\centering
\begin{subfigure}{.475\textwidth}
\includegraphics[width=\textwidth]{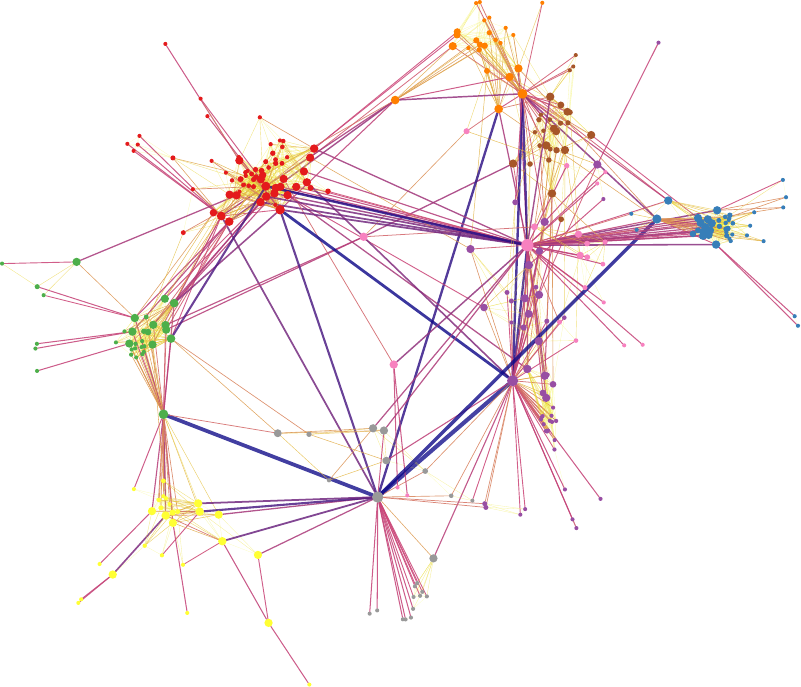}
\caption{Spring Embedded}
\end{subfigure}
\begin{subfigure}{.475\textwidth}
\includegraphics[width=\textwidth]{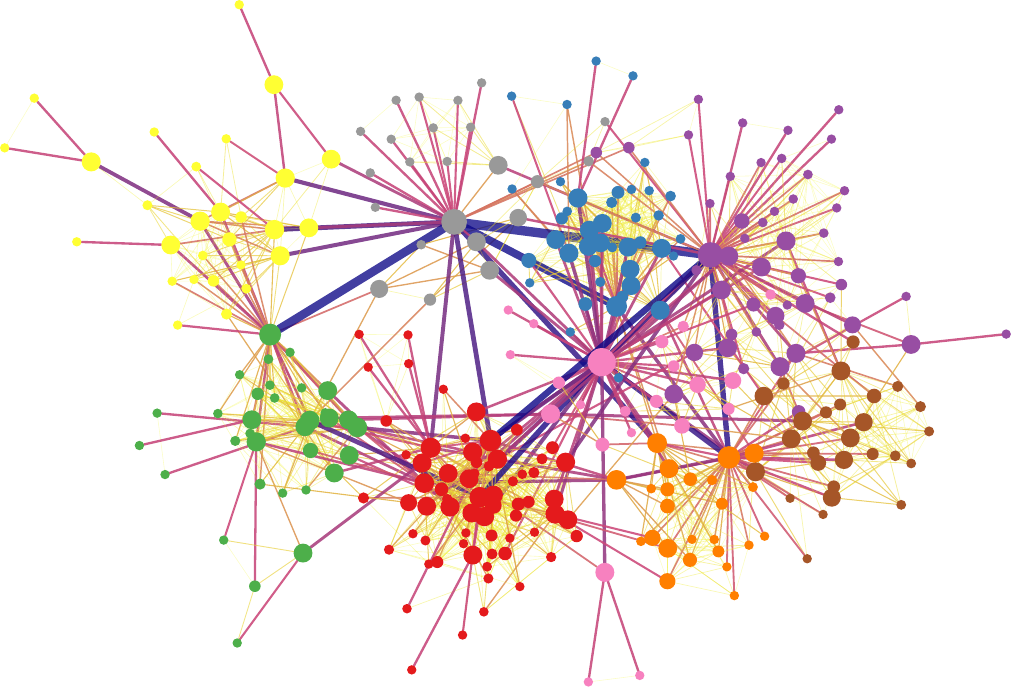}
\caption{Prefuse}
\end{subfigure}
\begin{subfigure}{.475\textwidth}
\includegraphics[width=\textwidth]{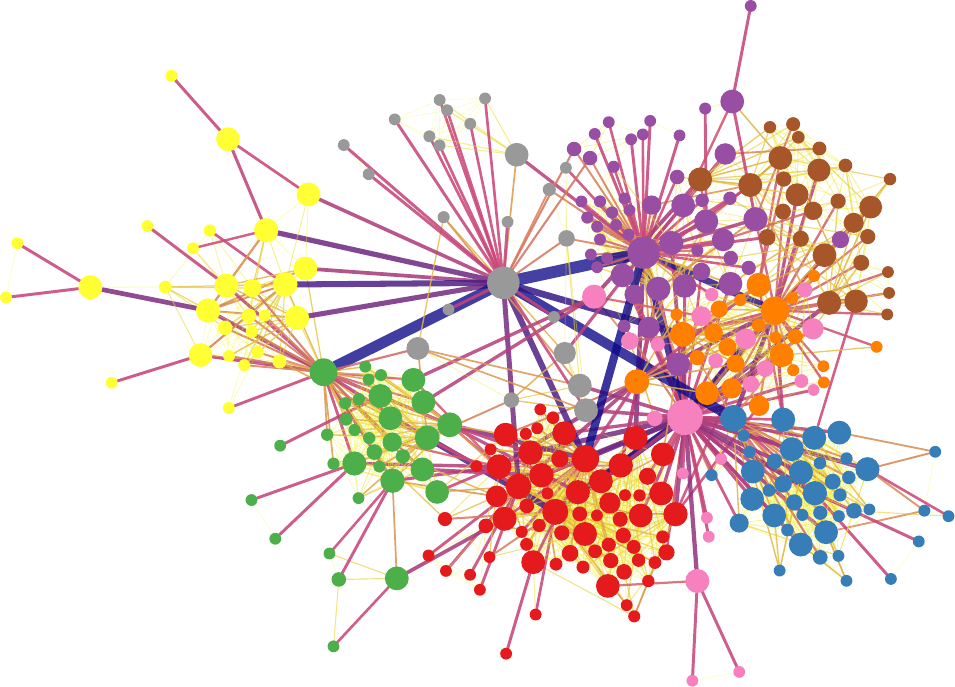}
\caption{yFiles organic}
\end{subfigure}
\begin{subfigure}{.475\textwidth}
\includegraphics[width=\textwidth]{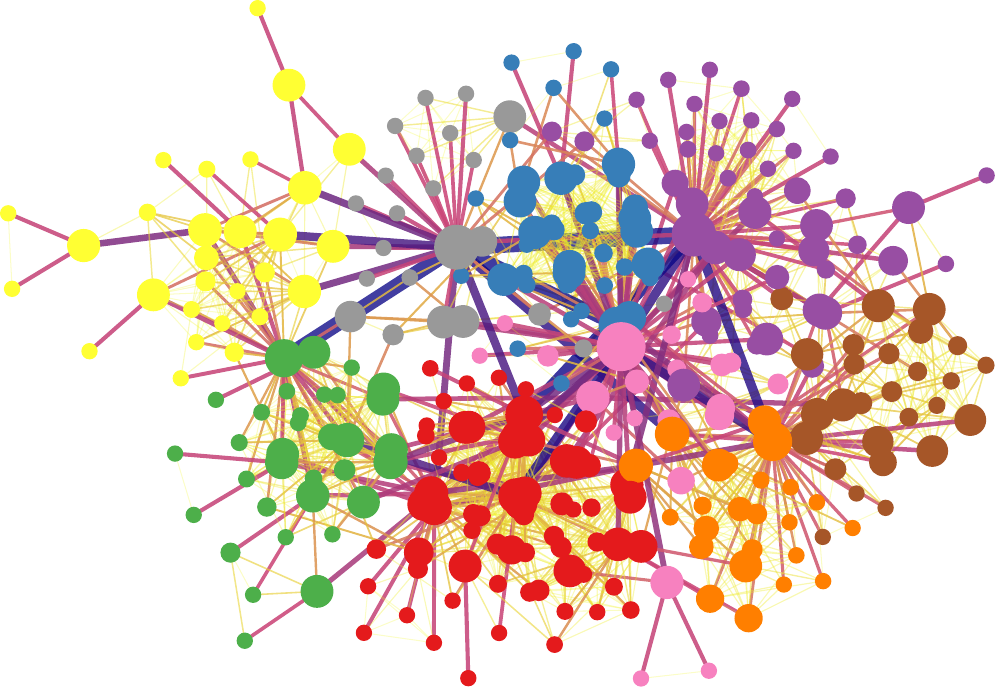}
\caption{Compound spring embedded}
\end{subfigure}
\caption{The same network displayed with different flavors of force directed layout. Node colors are their communities, node sizes are their degree. Edge colors and sizes are proportional to their betweenness centrality.}
\label{fig:forcedir2}
\end{figure}

Of these, I provide a few visualizations of the typical results you'd get in Cytoscape with its default parameters choices. They appear to be quite similar, with a few differences. From the strongest to the weakest charges: spring embedded (Figure \ref{fig:forcedir2}(a)), Prefuse (Figure \ref{fig:forcedir2}(b)), yFiles organic (Figure \ref{fig:forcedir2}(c)), and compound spring embedded (Figure \ref{fig:forcedir2}(d)).

Spring embedded (Figure \ref{fig:forcedir2}(a)) tends to shoot nodes in the stratosphere. It highlights clusters and central backbones better, but nodes tend to overlap. Note how the purple community here looks central. We'll see it changing its relative position in the other layouts, a sign that even different flavors of force directed can communicate different messages about the centrality of nodes and/or communities.

Prefuse (Figure \ref{fig:forcedir2}(b)) is the bread and butter of network visualization. It isn't particularly good, but works ok in most cases, and tends to be computationally less expensive than the other force directed alternatives. You will find Prefuse to be the default choice in many cases, in Cytoscape for instance. It deploys a classical balance between charges and springs. Note how the purple community isn't as central here as it was in Figure \ref{fig:forcedir2}(a).

yFiles organic (Figure \ref{fig:forcedir2}(c)) is very similar to Prefuse force directed. It tends to cluster nodes a bit more snuggingly, usually works a bit better for more complex networks than the one in Figure \ref{fig:forcedir2}. It is my default choice, unless the network is too big, since the yFiles organic algorithm has a higher computational complexity.

Compound spring embedded (Figure \ref{fig:forcedir2}(d)) is at the opposite end of the spectrum when compared to spring embedded. It forces nodes to be more or less equidistant from each other. It is a good option if the nodes are all more or less equivalent, but plays badly once you have diverse node sizes. You can see a lot of nodes overlapping in Figure \ref{fig:forcedir2}(d).

My rule of thumb here is that, the more complex interconnections you have the more you want your clusters -- if you have any -- to be separated. So you would choose the spring embedded layout. On the other hand, well balanced and separated clusters, with nodes more or less on equal footing, will mean going to the opposite end of the spectrum, to the compound spring embedded.

Note that Cytoscape's implementation of layout algorithms is not the most efficient. However, as most network plotting programs, it will accept you to pass x and y coordinates as attributes of your nodes, which you can use to display them. One trick is to calculate the layout using a more efficient script -- for instance igraph in R or C -- and then import the result in Cytoscape. Alternatively, remember that you can have a force directed style layout even without applying the force directed algorithm. Section \ref{sec:mining-embeddings-app} taught you how to use node embeddings to determine the placement of nodes on a 2D space.

Force directed is a good choice for sparse to medium-sparse networks. It works well when you have well defined clusters: any additional density coming from quasi-cliques is well handled because the nodes will bunch up in a ball together no matter what. Problems arise when the additional density comes from interconnections spanning across clusters. Also, it tends to fit your network onto a virtual sphere: its layouts tend to be circular. This is not always the best choice, as some networks are going to fit different shapes better.

\section{Other Node-Link Layouts}

\subsection{Hierarchical}
If force directed and its variants are a good default choice, they are not the only way to display your networks. As I concluded in the previous section, they have some pretty limited use cases. What happens when we go out of those use cases?

\begin{figure*}
\centering
\begin{subfigure}{.24\textwidth}
\includegraphics[width=\textwidth]{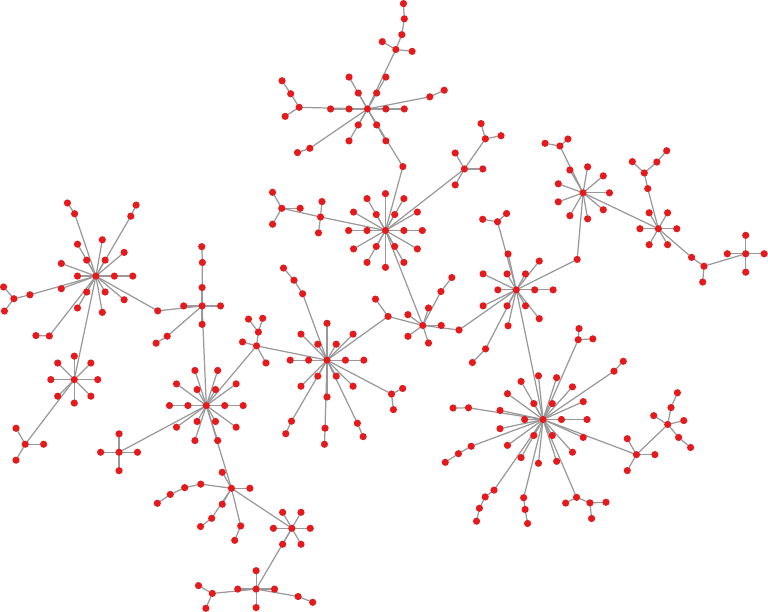}
\caption{yFiles organic}
\end{subfigure}\quad
\begin{subfigure}{.39\textwidth}
\includegraphics[width=\textwidth]{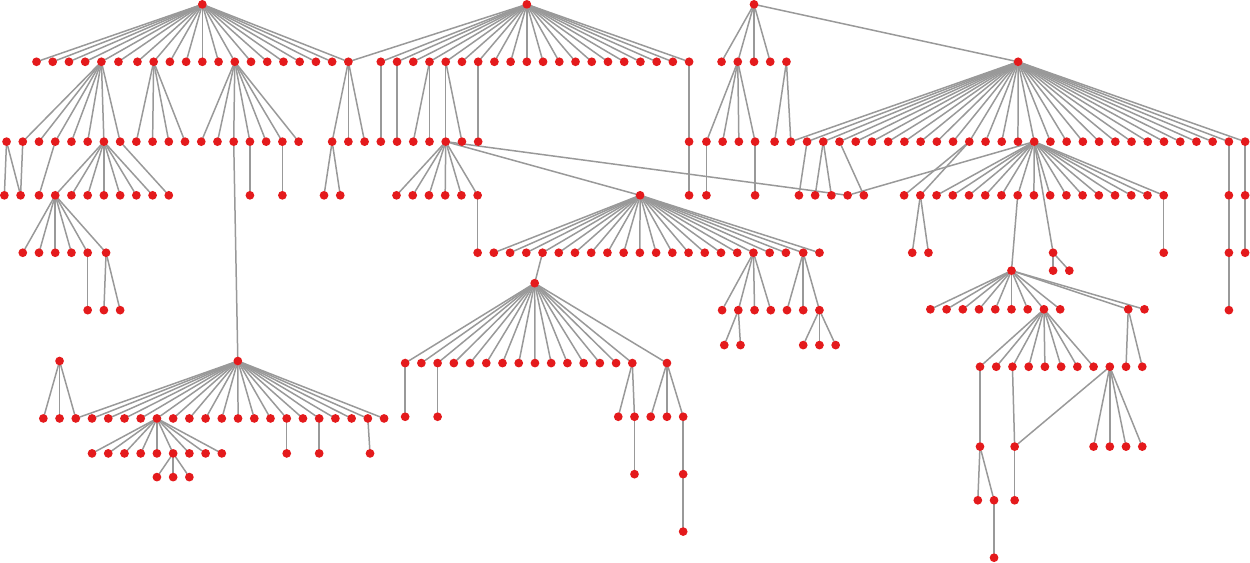}
\caption{yFiles tree}
\end{subfigure}\quad
\begin{subfigure}{.3\textwidth}
\includegraphics[width=\textwidth]{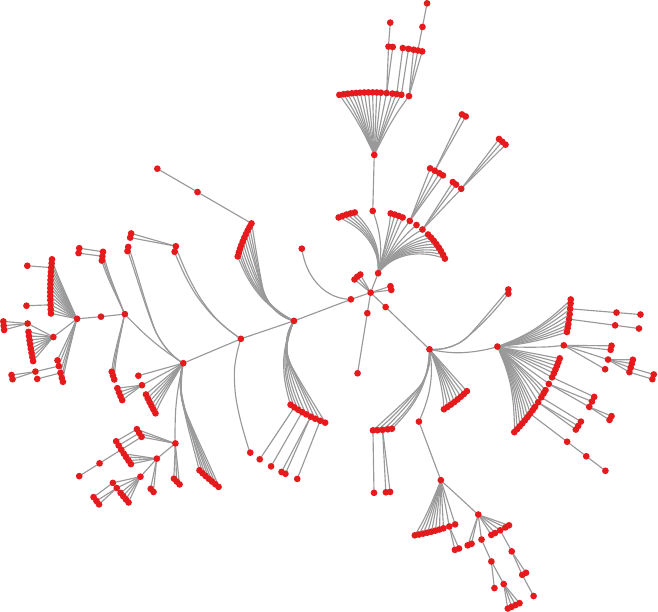}
\caption{yFiles radial}
\end{subfigure}
\caption{The same network displayed with different layouts, force directed and hierarchical.}
\label{fig:layout-hier}
\end{figure*}

The first scenario we consider is the one of extremely sparse networks. In that case you can obviously use force directed. However, it can arguably be not the better choice. Consider Figure \ref{fig:layout-hier}(a): even for this extremely sparse graph, the layout still manages to have many awkward node-edge overlaps. This is because very sparse networks are similar to trees, and a tree hardly fits the assumption of a circular layout. A tree has a root and leaves, thus a inherent top-down flow, which doesn't fit well the ``in-out'' flow of a circle (from the center to the circumference).

In this case you want to use a tree layout as I show in Figure \ref{fig:layout-hier}(b). A popular variant would be a radial layout -- Figure \ref{fig:layout-hier}(c). The radial layout is a compromise that still respects the tree-like structure of sparse graphs and, at the same time, has a force-directed ``in-to-out'' flavor to it.

\begin{figure}
\centering
\begin{subfigure}{.4\textwidth}
\includegraphics[width=\textwidth]{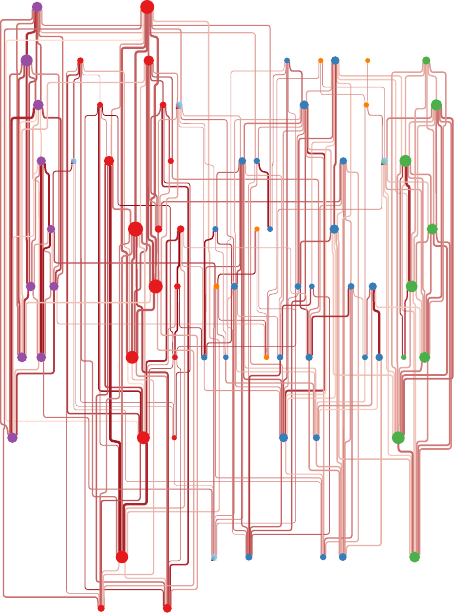}
\caption{}
\end{subfigure}\qquad
\begin{subfigure}{.3\textwidth}
\includegraphics[width=\textwidth]{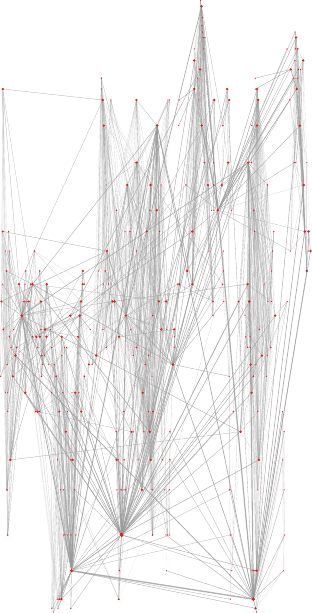}
\caption{}
\end{subfigure}
\caption{(a) A dense network still manageable with a hierarchical layout. (b) An example of a network too complex to be displayed with a hierarchical layout}
\label{fig:layout-hier2}
\end{figure}

As soon as your network becomes more dense than a quasi-tree, you should probably stop considering the hierarchical layouts. In some extreme cases they can work, if you want to highlight some specific messages. For instance, I would not use it for the network in Figure \ref{fig:layout-hier2}(a), but I can see how it can communicate something about the clusters of the network. The layout manages to put nodes in the same community in the same column, showing how some communities have stronger -- darker, thicker -- connections than others. However, a non-trivial number of nodes and edges would make you network visualization completely unintelligible, as it happens in Figure \ref{fig:layout-hier2}(b).

\subsection{Circular}
A second scenario to consider is the case of an extremely dense network. The network could be so dense, that the force directed is not able to pull nodes apart and show structure. In this case, the first step of the solution involves considering a layout that might not seem the best for the job, but has a few tricks up its sleeve: the circular layout. Which is exactly what it sounds: it places nodes on a circle, equidistant from one another.

The first part of the trick in using circular layouts is not to display the nodes in a random order, but choosing an appropriate one. Ideally, you want to place nodes in bunches such that most connections happen across neighbors. Usually this is achieved by identifying the nodes' attribute which groups them best. You can also run a custom algorithm deciding the order and then provide that as the attribute for the circular layout.

\begin{figure}
\centering
\includegraphics[width=.66\columnwidth]{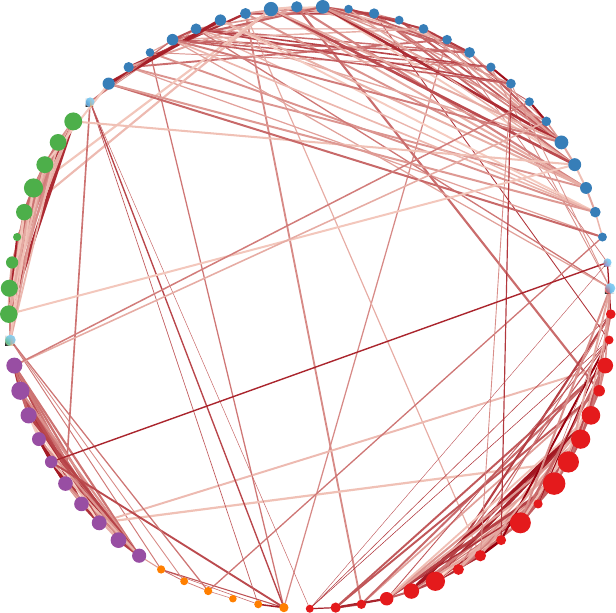}
\caption{A circular layout.}
\label{fig:layout-circle}
\end{figure}

Figure \ref{fig:layout-circle} shows an example. In the figure you can see that the layout still works: it shows how most connections remain within the communities, and clearly points at how many and where the inter-community connections are. In a force directed layout, these connections would stretch long and be forced in the background of denser areas, with the effect of being difficult to appreciate. However, the real kicker for circular layouts happens when you consider the use of an additional visual feature: edge bends.

\section{Edge Bends}\label{sec:layouts-bends}
The default choice in network layouts up until recently was to show edges as straight lines. However this is usually not pleasing visually. Go back to Figure \ref{fig:forcedir2}(a) and you will feel how clunky the long thick edges feel. People started playing with edge bends and discovered that they can solve a series of other problems, rather than being a simple cosmetic enhancement.

\begin{figure*}
\centering
\begin{subfigure}{.31\textwidth}
\includegraphics[width=\textwidth]{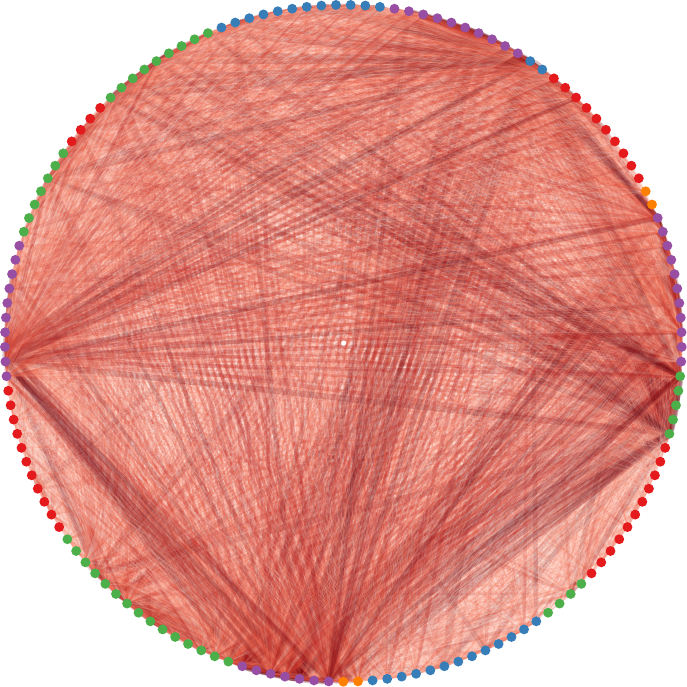}
\caption{}
\end{subfigure}\quad
\begin{subfigure}{.31\textwidth}
\includegraphics[width=\textwidth]{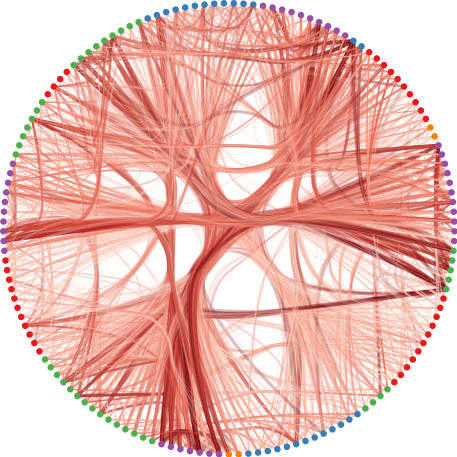}
\caption{}
\end{subfigure}\quad
\begin{subfigure}{.31\textwidth}
\includegraphics[width=\textwidth]{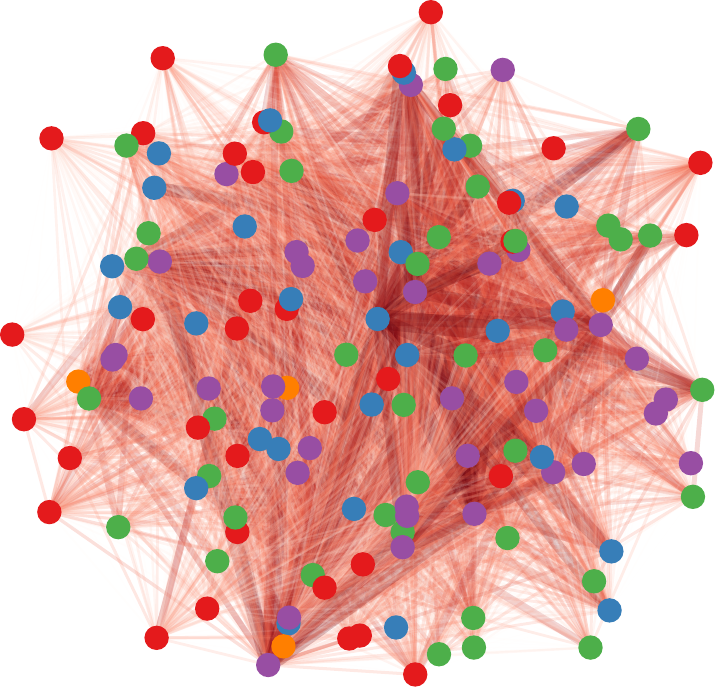}
\caption{}
\end{subfigure}
\caption{The same network displayed with different layouts: (a) circular with no bends, (b) circular with edge bundling, (c) Prefuse force directed.}
\label{fig:layout-circle2}
\end{figure*}

The first strategy to bend your edges is to bundle together the ones coming/going from/to more or less the same part of the network\cite{holten2006hierarchical}\cite{holten2009force}. This is extremely useful for the very dense networks in circular layouts I teased at the end of the previous section\cite{gansner2006improved}. Consider Figure \ref{fig:layout-circle2}(a): here the circular layout isn't helping us much in making sense of this extreme density. However, once we bundle edges in Figure \ref{fig:layout-circle2}(b), we obtain a much clearer idea of what is going where. Sure, the visualization is still complex and it is still hard to tell which nodes are connected to which other. However, that is a much better situation than the alternative visualization you would get from a force directed without edge bends. Figure \ref{fig:layout-circle2}(c) is my proof.

\begin{figure}
\centering
\begin{subfigure}{.5\textwidth}
\includegraphics[width=\textwidth]{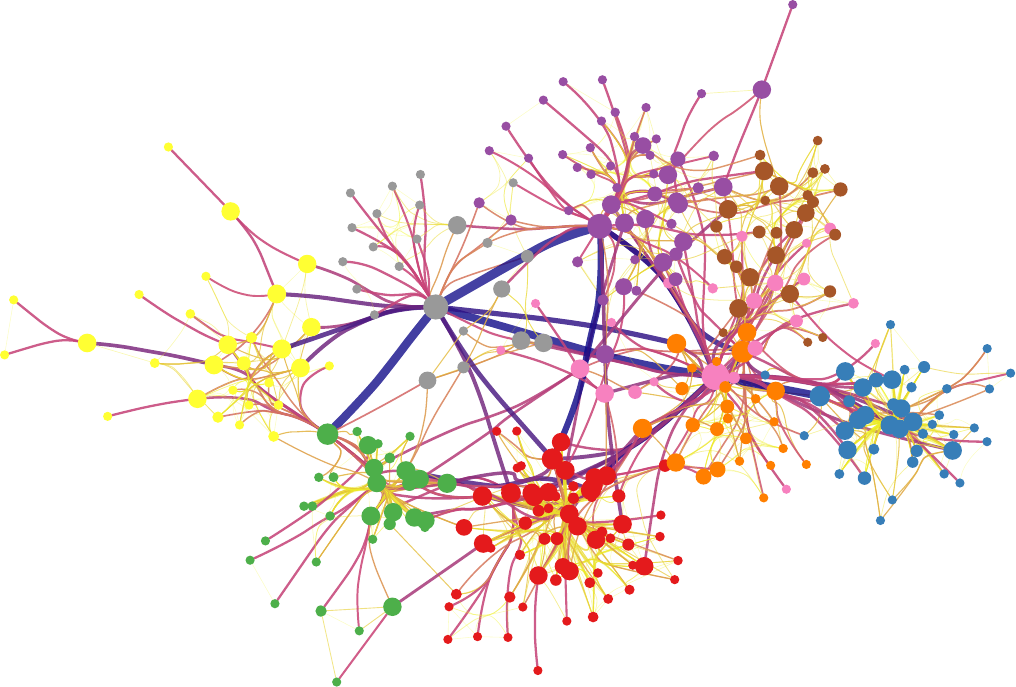}
\caption{}
\end{subfigure}\quad
\begin{subfigure}{.33\textwidth}
\includegraphics[width=\textwidth]{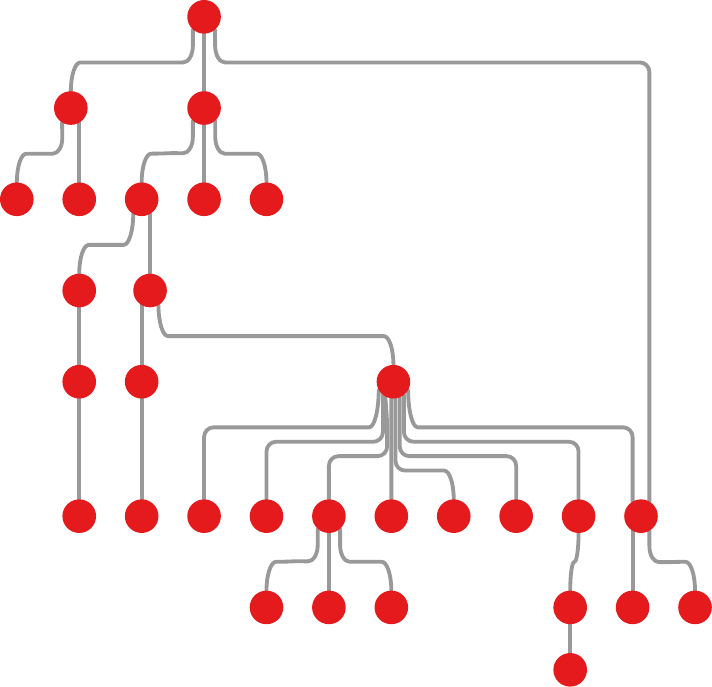}
\caption{}
\end{subfigure}
\caption{Edge bundles in (a) force directed and (b) hierarchical network layouts.}
\label{fig:edgebundle}
\end{figure}

You can use edge bundling in other layouts too. Usually this will reduce clutter and make your visualization clearer and easier to parse. For instance, in a force directed layout communities will reduce to flower bouquets where all connections collapse in the same point. See Figure \ref{fig:edgebundle}(a) for an example. This is as if the community is a hyperedge connecting all the nodes that it groups. This is not exactly the full topological information in the network, but oftentimes it is an acceptable approximation.

A hierarchical network layout benefits from edge bends as well -- see Figure \ref{fig:edgebundle}(b). This is because it is difficult to display hierarchies in a tight space, they tend to grow horizontally and vertically. With edge bends you can make your edges take a slightly longer route which increases the picture's readability. Of course, a bend can decrease the readability if you are then unable to clearly distinguish which edge goes where in a bundle. So there are techniques to make this distinction\cite{bach2016towards}.

Bending edges helps in a second scenario. When squeezing complex multidimensional structures, such as dense networks, onto a two dimensional plane, you'll often have no good placement choice for your nodes and edges. In many cases this is just awkward and displeasing to the eye, but in many others it can create problems and untruthful visualizations. The most dreaded case you should avoid at all cost is ghost edges.

\begin{figure}
\centering
\begin{subfigure}{.3\textwidth}
\includegraphics[width=\textwidth]{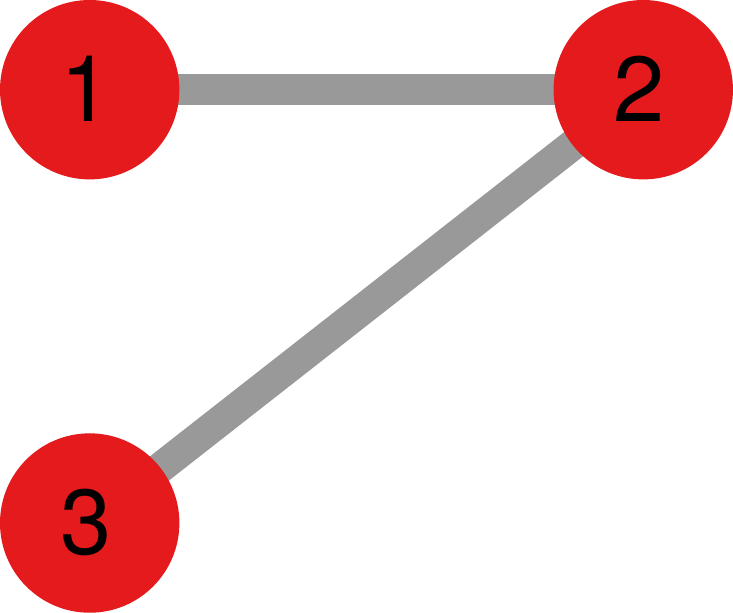}
\caption{}
\end{subfigure}\qquad
\begin{subfigure}{.3\textwidth}
\includegraphics[width=\textwidth]{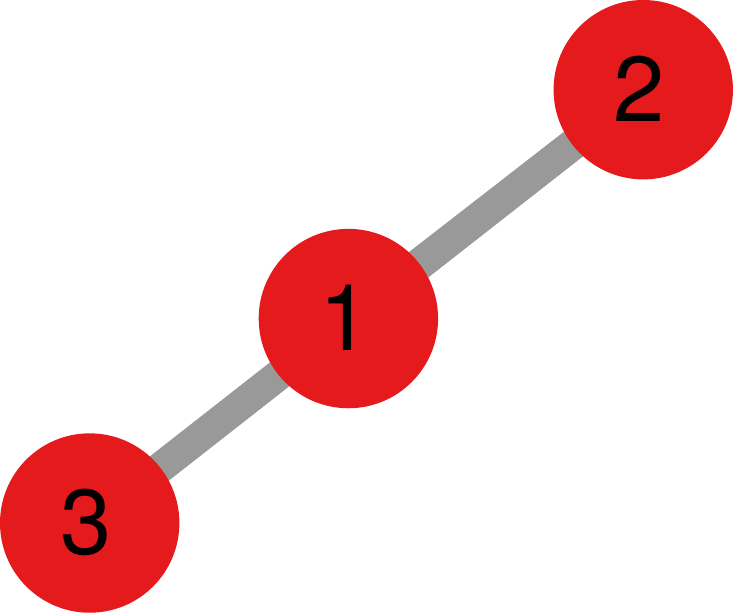}
\caption{}
\end{subfigure}
\caption{A scenario in which you might create ghost edges.}
\label{fig:ghostedge}
\end{figure}

Consider Figure \ref{fig:ghostedge}. Suppose we know that the network's topology involving those three nodes is on the left. Node $1$ is connected to node $2$ which is connected to node $3$ in a chain. However, these three nodes don't live in a vacuum. There are thousands of other nodes and edges around them, pulling and pushing them in different directions. After a few layouts, you might not notice that your three nodes ended up in the configuration to the right.

Seeing that configuration, a viewer would instantly assume that node $1$ is connected to both node $2$ and $3$, without a connection between the latter two. This, as we know, is wrong. Worse still, there's no way by looking at the configuration to the right to know what really is going on between these three nodes. It might be that node $1$ is not connected to either of those, and ended up there by accidents of your layout. Or that could be a squeezed triangle. Or it might be even true that the three nodes are not connected at all to each other, and that's just a long edge connecting two other nodes out of sight. There are so many ways to lie in a 2D network layout -- whether you do it accidentally or on purpose.

\begin{figure}
\centering
\includegraphics[width=.83\columnwidth]{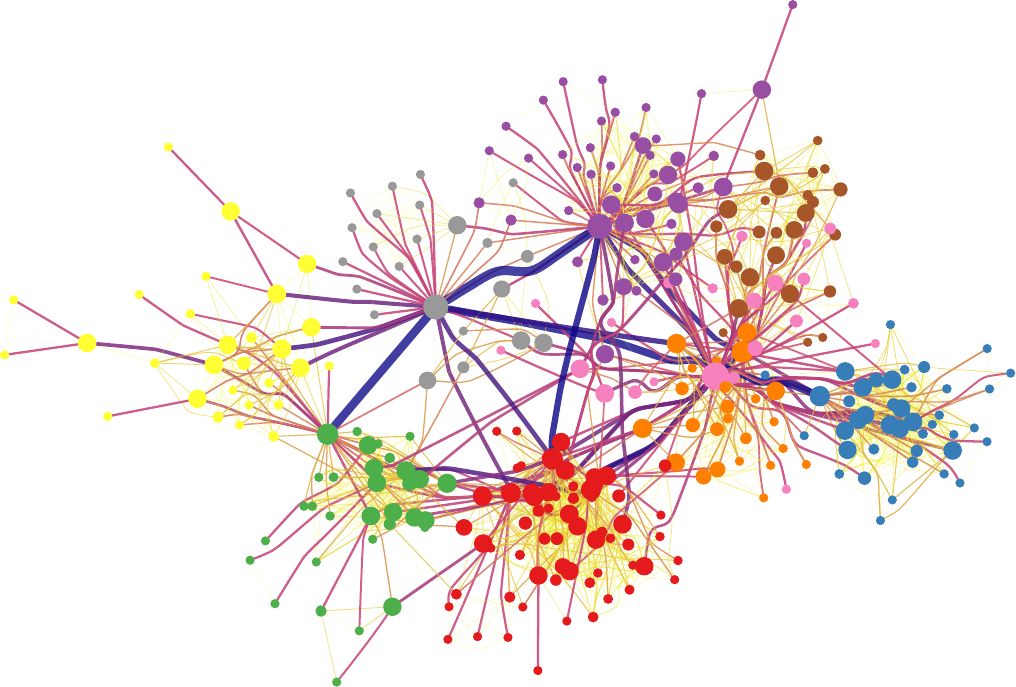}
\caption{A force directed layout with organic edge routing.}
\label{fig:edgeroute}
\end{figure}

Edge bends can partially save you. In particular, the trick is to use organic or orthogonal edge routing\cite{dwyer2006integrating}, implemented by yFiles. To know what it looks like, consider Figure \ref{fig:edgeroute}. In the figure, there are many cases in which the vanilla force directed layout would pass a straight edge across the nodes. As it happens, some of those cases were actually two edges with a node in between. The organic edge router would not change the edge shape in that case. But some edges get a dramatically evident bend, because the vanilla force directed would make you believe they connected nodes while, in reality, they did not.

\section{Alternative Layouts}
In this section, we break the assumption that nodes are circles and edges are lines. We try to find weird ways to summarize the network topology in a way that is more compact and compelling.

\subsection{Matrix Layouts}
What's the last resource for networks in which the density is too high even for a circular layout plus edge bundles? If your network is so dense, then you don't have a network: you have a matrix and you should visualize it as such. In these cases, what matters more is not really which area is denser than which other, but which blocks of nodes have connections with higher and lower weights.\footnote{If your network is this dense and \textit{also} unweighted, consider changing job.}

\begin{figure}
\centering
\begin{subfigure}{.45\textwidth}
\includegraphics[width=\textwidth]{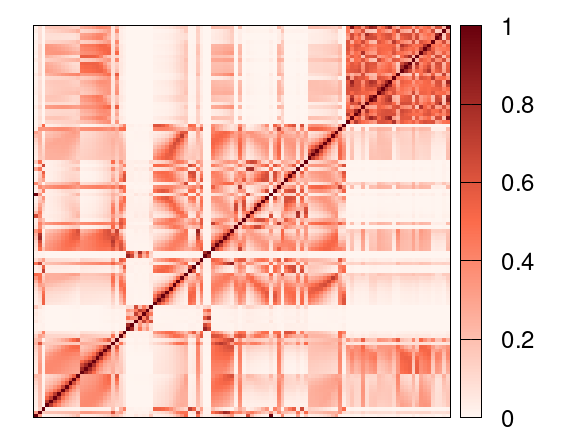}
\caption{}
\end{subfigure}\qquad
\begin{subfigure}{.45\textwidth}
\includegraphics[width=\textwidth]{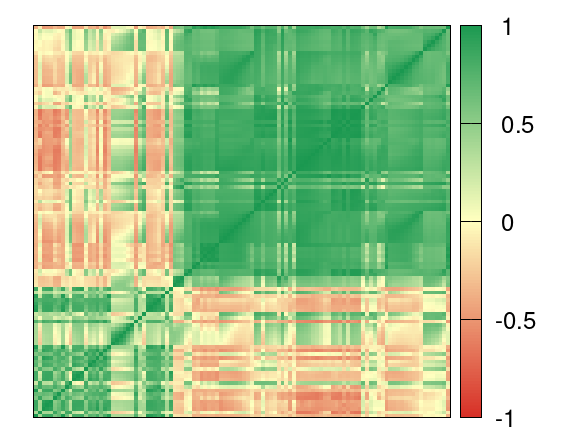}
\caption{}
\end{subfigure}
\caption{(a) A matrix view of a network with progressive edge weights. (b) A matrix view of a network with divergent edge weights, such as correlations.}
\label{fig:matrixview}
\end{figure}

Color scales should be chosen depending whether your edge weights are defined as progressive or divergent. The first case, in Figure \ref{fig:matrixview}(a), is the classical case of an edge indicating the intensity of the connection between two nodes, or the cost of edge traversal. The second case, in Figure \ref{fig:matrixview}(b), is a classical correlation network. This is a default visualization scenario, as correlations are always defined between any pair of nodes, and thus the network will be complete.

\begin{figure}
\centering
\includegraphics[width=\columnwidth]{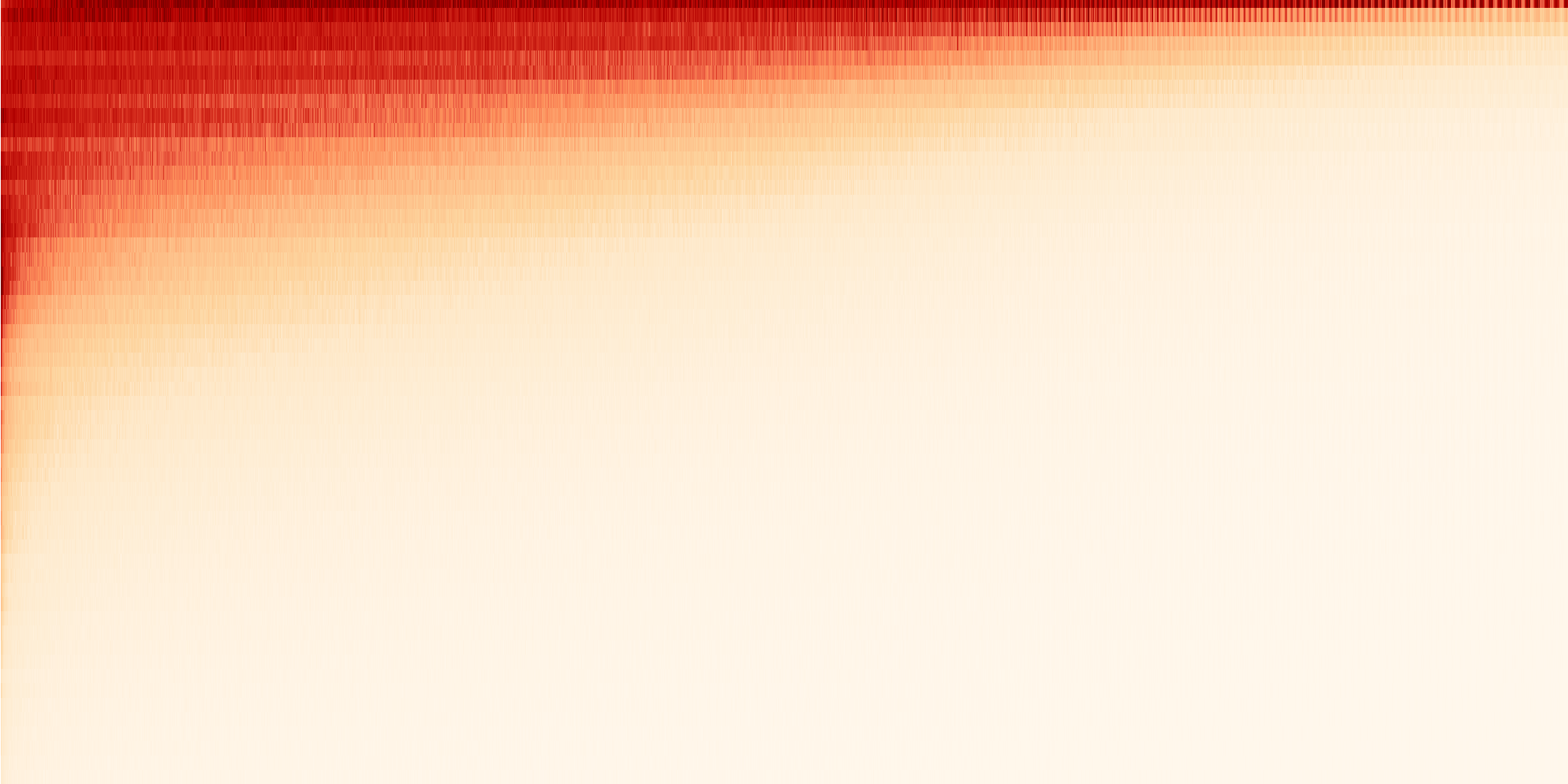}
\caption{A matrix view of a nested network.}
\label{fig:matrixview2}
\end{figure}

That is not to say that this is the only use case of a matrix visualization. Even for sparser networks, sometimes a matrix is worth a thousand nodes. Consider the case of nestedness (Section \ref{sec:coreper-nestedness}), a particular core-periphery structure for bipartite networks where you can sort nodes from most to least connected. The most connected node connects to every node in the network, while the least connected nodes only connects to the nodes that everyone connects to. This sort of linear ordering naturally lends itself to a matrix visualization. While a node-link diagram would make a mess of such a core, rendering the message difficult to perceive, a matrix view is deceptively simple, as I show in Figure \ref{fig:matrixview2}.

This case is a good example of the main problem in visualizing networks as matrices: the order of the rows/columns you choose is the most important thing. You should put your nodes in the sequence that highlights the crucial structural characteristics the best. In the nestedness case, nodes are sorted by degree (or total incoming weight sum). If your network has communities, you want to have nodes next to their community mates. This creates the classical block diagonal matrices. There are other criteria you might want to consider\cite{behrisch2016matrix}.

\subsection{Hive Plots}
The issue with all traditional node-link diagrams is that the position in space of a node is arbitrary: it does not reflect its properties, but it is just relative to its connections. As such, layouts are not reproducible, because a small change in the initial conditions in the placement of a single node will result in a completely different layout. Hive plots\cite{krzywinski2011hive} try to fix these issues by providing a way to have a deterministic node layout placement.

The idea is the following: first, the user determines a set of rules. The aim of these rules is to divide nodes into classes. Nodes in the same class will be grouped together on the same axis. Then, the user selects a specific measure, which determines the position of a node on that axis. The hive plot will then attempt to place the axes in such a way to minimize edge crossings. If there are connections between the nodes on the same axis, the axis will be duplicated in order to avoid confusing loops.

\begin{figure}
\centering
\begin{subfigure}{.5\textwidth}
\includegraphics[width=\textwidth]{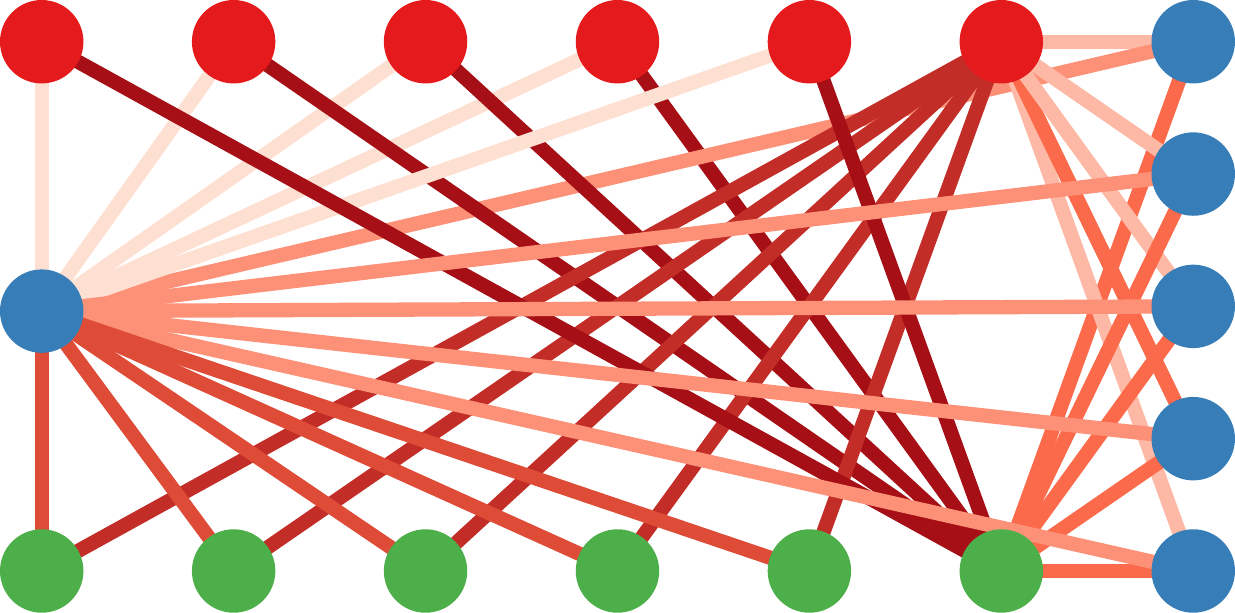}
\caption{}
\end{subfigure}\qquad
\begin{subfigure}{.4\textwidth}
\includegraphics[width=\textwidth]{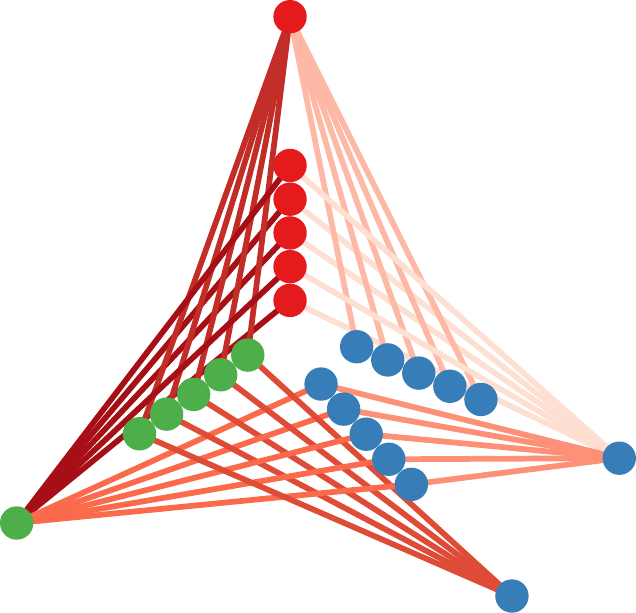}
\caption{}
\end{subfigure}
\caption{(a) A graph where I encode with colors the node and edge types. (b) The hive plot version of (a).}
\label{fig:hive}
\end{figure}

Figure \ref{fig:hive} shows a simple example. In Figure \ref{fig:hive}(a) we have three node types, which result in three axes in Figure \ref{fig:hive}(b). However, the blue nodes also connect to each other, and have more complex connecting patterns with the other groups. Thus, we duplicate the blue axis, meaning that we have a copy of it at slightly different angles. The blue-blue links flow between the copies, and we divided the connections between blue and other nodes as to minimize the number of crossings.

The position of nodes on the axes is determined by the degree. The nodes with high degree on each axis are the ones on top, separated by the rest. The other nodes have all the same degree, so they are grouped, although we separate them so that the nodes don't overlap with each other. You can choose which measure to use for the node positioning on the axis, you are not forced to use the degree.

\subsection{Graph Thumbnails}
Just like hive plots, also graph thumbnails\cite{yoghourdjian2018graph} aim to provide a deterministic layout, where two isomorphic graphs result in the same visualization. In graph thumbnails, we decide to give up the ability of analyzing local structures. We are not seeing each individual node: the visualization is a summary of the graph's global structure. The idea is to dissect a graph into its main core components in a hierarchical fashion. Each core is then visualized as a circle, whose color tells us its core level. You should use graph thumbnails when you need to compare a large number of graphs in a compact way and you care about the high-level organization of the graph as a whole, rather than the meso-level communities -- or the individual nodes.

The decomposition is done via the classical k-core detection -- see Section \ref{sec:centr-kcore}. Each connected component of a network is part of a 1-core. Then, there could be multiple k-cores around the network. Each k-core is represented as a circle, and it is nestled inside the ($k-1$)-core that contains it.

\begin{figure}
\centering
\begin{subfigure}{.3\textwidth}
\includegraphics[width=\textwidth]{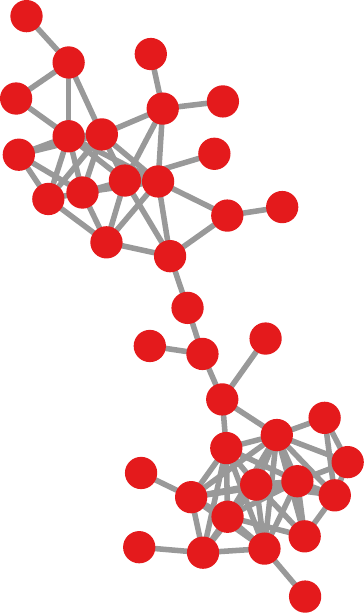}
\caption{}
\end{subfigure}\qquad
\begin{subfigure}{.4\textwidth}
\includegraphics[width=\textwidth]{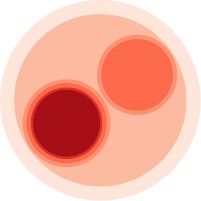}
\caption{}
\end{subfigure}
\caption{(a) A graph and its (b) graph thumbnail visualization. The color of the circle encodes the $k$ value of the k-core, while its size is (loosely) proportional to the number of nodes in that particular core.}
\label{fig:graphthumb}
\end{figure}

Figure \ref{fig:graphthumb} shows an example. The graph in Figure \ref{fig:graphthumb}(a) has two communities. All nodes in the graph are part of the 1-core because it's a single connected component. Some nodes are not part of the 2-core, but they are only the peripheral dangling ones: both communities are part of the same 2-core. The communities split when we consider the 3-core, that is why the second circle in the graph thumbnail in Figure \ref{fig:graphthumb}(b) contains two subcircles. The smallest community only contains a 4-core, while the largest goes up to a 6-core, explaining why the second circle goes to darker hues.

A disadvantage of the graph thumbnail visualization is that it needs to apply a circle packing algorithm\cite{collins2003circle}. It is impossible to pack circles efficiently inside other circles. In this specific example, both the 1- and the 2-core circles are larger than they should be given the number of nodes they contain. I needed to enlarge them, because otherwise they could not contain properly the two 3-cores of the network.

\subsection{Probabilistic Layout}
Following the same ``we can't visualize all nodes'' philosophy of graph thumbnails, we have probabilistic layouts\cite{schulz2016probabilistic}. This technique is handy when you have a generic guess of where the nodes \textit{should} be, but you cannot draw them all. You should use such layouts especially for very large graphs that you couldn't visualize otherwise, because they have too many nodes and/or edges.

The idea is as follows. First, you sample the nodes in your network, taking only a few of them. Then you calculate their positions using a deterministic force directed layout. You repeat the procedure multiple times, obtaining, for each node, a good approximation of where it should be. If you have nodes that you never sampled, you can reasonably assume that they are going to be in the area surrounding their neighbors. Since you're applying the algorithm to a sample, this won't take much time even if the original network was too large to be analyzed in its entirety.

Now each node is associated to a spatial probability distribution, much like elementary particles in quantum physics. You can assume that, if the node is anywhere, it'll be somewhere in the area where its probability is nonzero. At this point, you can merge nodes whose spatial probabilities overlap, by detecting and drawing a contour containing them. You should then bend edges and smudge them as well, to reflect the uncertainty of where their endpoints are.

\begin{figure}
\centering
\begin{subfigure}{.3\textwidth}
\includegraphics[width=\textwidth]{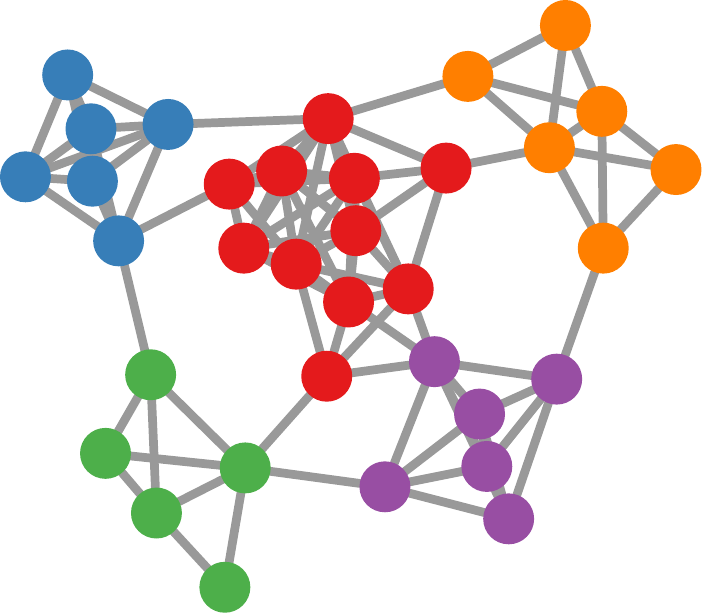}
\caption{Original}
\end{subfigure}\quad
\begin{subfigure}{.3\textwidth}
\includegraphics[width=\textwidth]{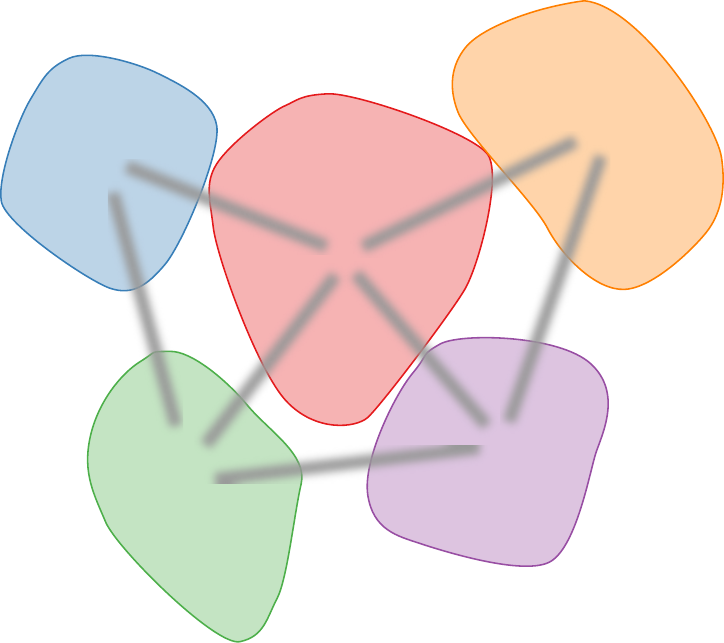}
\caption{$\alpha = 0.01$}
\end{subfigure}\quad
\begin{subfigure}{.3\textwidth}
\includegraphics[width=\textwidth]{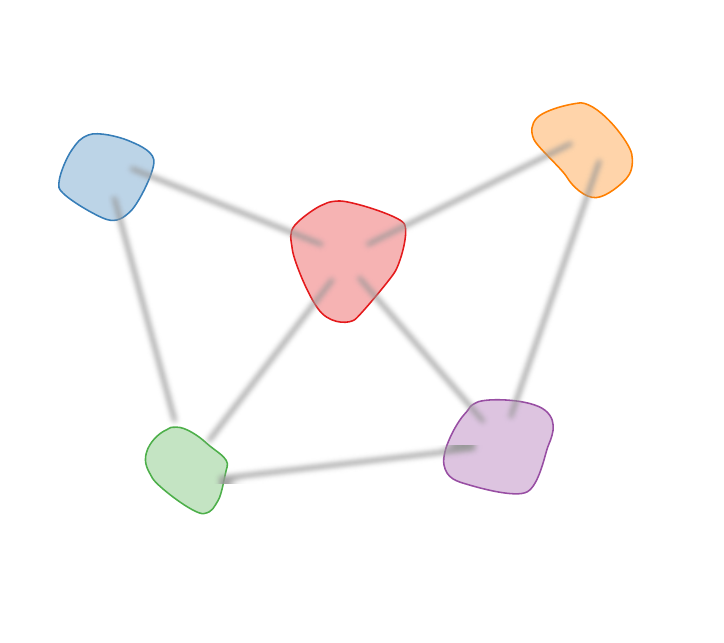}
\caption{$\alpha = 0.7$}
\end{subfigure}
\caption{A graph and its probabilistic layouts, for different levels of $\alpha$.}
\label{fig:problayout}
\end{figure}

You can specify a parameter $\alpha$ regulating how tight your smudges should be. For low values of $\alpha$, you get the quickest results at the price of large uncertainties. When $\alpha \sim 1$, your smudges become points, tightening up all nodes belonging to a smudge in the same area. Figure \ref{fig:problayout} shows a toy example, for two levels of $\alpha$.

\subsection{Revealing Matrices}
\begin{figure}[t]
\centering
\includegraphics[width=\columnwidth]{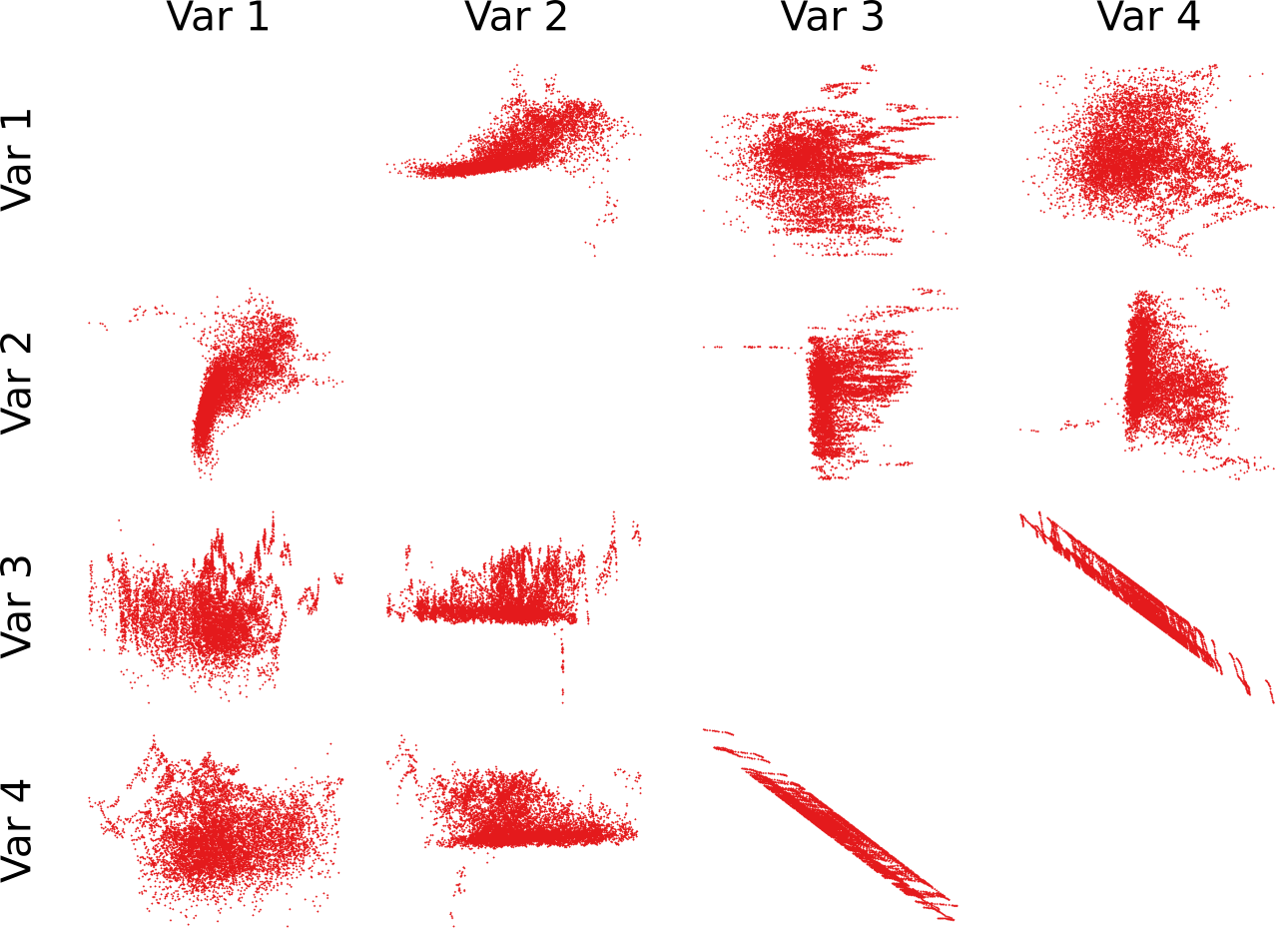}
\caption{An example of SPLOM visualization.}
\label{fig:splom}
\end{figure}

One key visualization technique is scatterplot matrices or SPLOMs. When you have multiple variables in your dataset, you might be interested in knowing which one correlates with which other. So you can create a matrix where each row/column is a variable, and each cell contains the scatter plot of the row variable against the column variable. Figure \ref{fig:splom} shows an example.

The same visualization technique can be applied to networks. In revealing matrices\cite{schich2010revealing}, each row/column of your matrix is an entity. Then, each cell of the matrix contains a bipartite network, where the nodes of one type are the row entity and the nodes of the other type are the column entity.

One defect of SPLOMs is that the main diagonal of the matrix is a bit awkward. In it, the row variable and the column variable are the same. Thus the scatter plot is meaningless, as it is the same variable on the $x$ and $y$ axes: a straight line. One could modify it by showing some sort of statistical distribution of the variable, but that would mean breaking the axis consistency of the SPLOM. For this reason, the main diagonal of a SPLOM is often omitted.

This defect does not apply to the revealing matrices visualization. The main diagonal in this case \textit{is} well defined: it is simply the direct relationship between entities of the same type. Thus, it contains a unipartite network per node type in your database.

\subsection{Timelines}
Temporally evolving networks are tricky to visualize. The most natural thing you can do is to use animations, showing a state of the network per frame, but that's unsatisfying. First because you can't put them on a piece of paper. Second because you cannot have an overview of the dynamics all at once; you need to wait for the animation to play out and you might have forgotten what was in the first frame by the time you get to the last.

\begin{figure}
\centering
\includegraphics[width=.8\columnwidth]{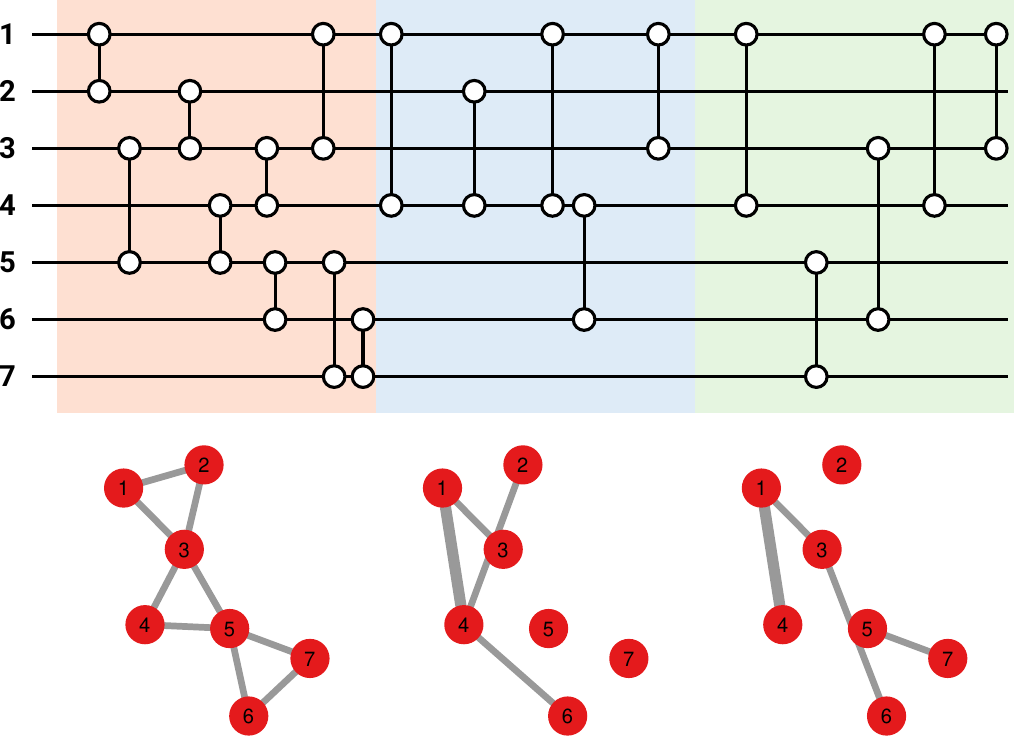}
\caption{A network timeline visualization: nodes as horizontal lines, edges as vertical lines. Each shaded area shows one snapshot and the corresponding classical node-link diagram visualization below it.}
\label{fig:temporal-timeline}
\end{figure}

There is one visualization technique I first saw in 2012\cite{holme2012temporal} (but it could be older) that changes fundamentally how we visualize a network to show time in a more natural way. Figure \ref{fig:temporal-timeline} shows an example.

In the figure a node becomes a horizontal timeline. Time flows from left to right, as we normally assume by convention. When two nodes are connected, we join the corresponding timelines with a vertical line. So, in this visualization, a node is a horizontal line and an edge is a vertical line. You can easily see in the figure that nodes $1$ and $4$ have a high level of activity and node $7$ is low activity, something that might be tricky to do otherwise. You may argue that this is a gimmick that only works for a handful of nodes and edges, but so are node-link diagrams, so checkmate there.

\begin{figure}
\centering
\includegraphics[width=.8\columnwidth]{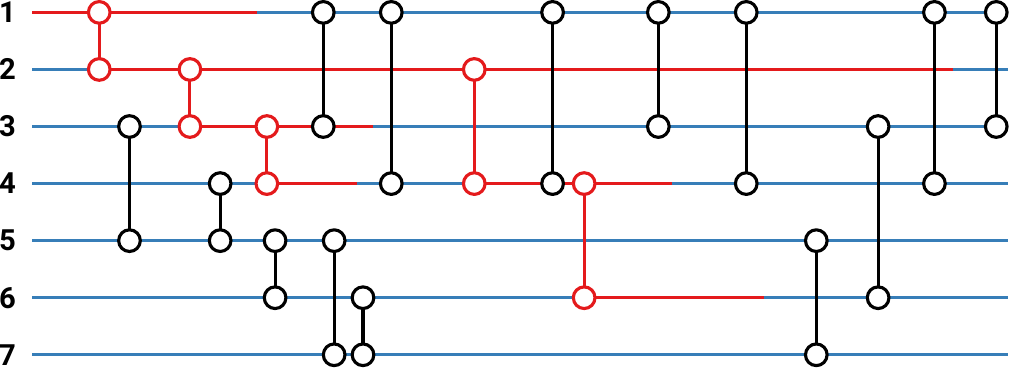}
\caption{Network timeline visualization of an SIS process. The node's line is red when the node is infected (I) and blue when it is susceptible (S). A contagion happens probabilistically when a red line joins a blue line, when it happens the connection is red. Note node $4$ that gets reinfected after recovering.}
\label{fig:temporal-timeline-sis}
\end{figure}

One neat thing that this visualization allows you to do is to keep track of events on the network. For instance, consider an SIS model -- which I show in Figure  \ref{fig:temporal-timeline-sis}. From the figure you can already infer that, once the disease is exclusively in node $2$ it is trapped and has nowhere to go in the future, because node $2$ won't interact with any other node. This kind of inference might be harder and/or slower to do visually with a different visualization.

\section{Case Studies}
As it often happens in data visualization, what you want to say with your network visualization might not be supported by any standard visualization technique out there. Sometimes, you need to craft a custom visualization, bending and breaking rules along the way. No one should really follow your workflow, because it applies only to your specific aim with your specific data. However, seeing some of these examples could be helpful in making you realize that you are not mad: sometimes you really do know better than everybody else. The aim of this section is to empower you in being daring: try to look at your data and your communication objective, and create your way to bringing them together.

I touch on two examples I worked on. These are custom ways of displaying a node-link diagram that I found useful. These node-link diagrams have special configurations given the need to highlight specific features of the networks they represent. Of course, there's much more out there, but these are two cases I'm familiar with.

\subsection{Product Space}
The Product Space\cite{hidalgo2007product}\cite{hausmann2014atlas} is a popular example. The Product Space is a network in which each node is a product that is traded among countries in the global market. Two products are connected if the sets of countries exporting them have a large overlap. The idea of this visualization is to show you which products are similar to each other, because if your country can make a given set of products, via the Product Space it can figure out which are the most similar products it should consider trying to export.

\begin{figure*}
\centering
\begin{subfigure}{.4\textwidth}
\includegraphics[width=\textwidth]{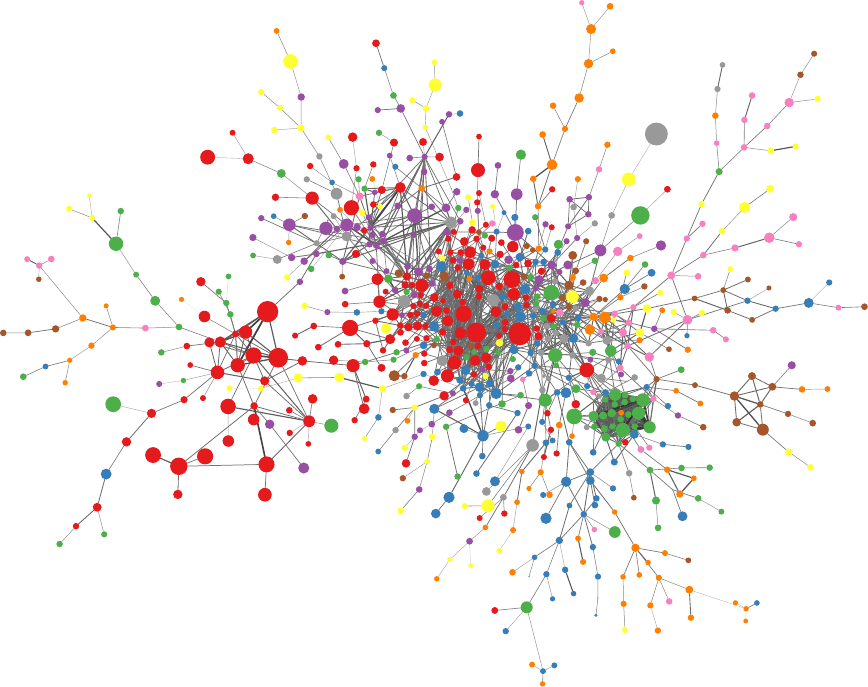}
\caption{}
\end{subfigure}\qquad
\begin{subfigure}{.5\textwidth}
\includegraphics[width=\textwidth]{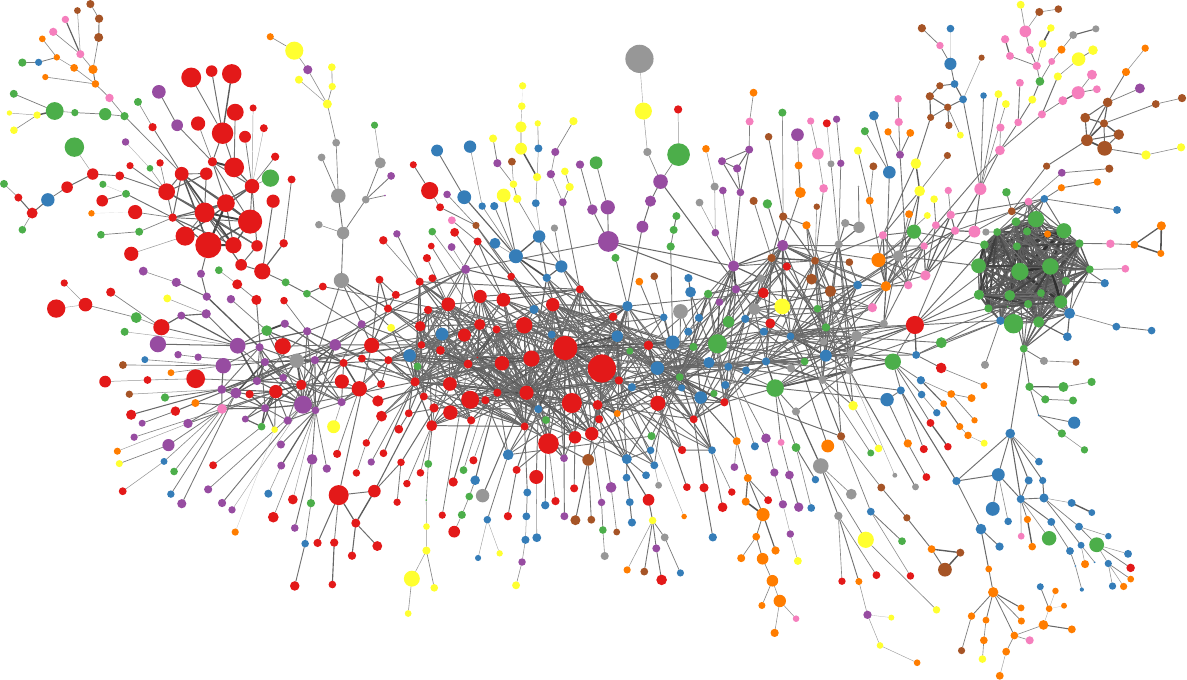}
\caption{}
\end{subfigure}
\caption{The Product Space. (a) Classical force directed layout. (b) Manually adjusted linear force directed. The node color is a product's Leamer category.}
\label{fig:prodspace}
\end{figure*}

The original way to try and visualize the Product Space was a simple force directed layout, as I show in Figure \ref{fig:prodspace}(a). However, as I mentioned previously, the force directed layouts have this tendency of forcing your networks on a sphere. This happens to work really poorly in the case of the Product Space. The reason is that not all products are the same. Some products are harder to export than others. This is a key concept in the original research, known as Economic Complexity.

This means that the Product Space has an inherent ``direction''. Countries want to move from simple to more complex products, as the latter is a more rewarding category to be able to export. However, the circle has no direction. It is a loop: you always get back to where you started. The shape of the Product Space in Figure \ref{fig:prodspace}(a) does not allow us to perceive the development path. That is why it is necessary to stretch out the visualization as I do in Figure \ref{fig:prodspace}(b): now the Product Space is a (complex, multidimensional) line\footnote[][3\baselineskip]{Eerily looking like an angel from Neon Genesis Evangelion, with that creepy head with multiple green eyes... Am I the only one seeing it?} and you can see that there is a clear direction going from right to left, from less to more complex products. It is still a type of force directed layout, but it needed to be customized to remove its inherent circularity.

\subsection{Cathedral}
In another paper of mine, I analyze government networks\cite{kosack2018functional}. My nodes are government agencies and I establish edges between them if the website of an agency has an hyperlink pointing to the website of another agency. One key question is verifying if this network has a hierarchical organization -- see Chapter \ref{cha:hier}. One obvious way to explore this question is visualizing the network and see if it looks like a hierarchy. Unfortunately, the network is relatively large and dense. So I need to come up with a custom layout. Such layout is useful to visualize dense hierarchical networks, and thus can be considered as an enhancement of the classical hierarchical layout presented earlier, that works only for tree-like structures.

The first step is to group nodes into a 2-level functional classification. This means to assign to each agency the function it performs in the government. For instance, a school is part of the education system (level $1$ function) and of the primary \& secondary education (level $2$ function). Or: a city government is part of general administration (level $1$ function) and of the municipal administration (level $2$ function).

\begin{figure}[b!]
\centering
\includegraphics[width=.9\columnwidth]{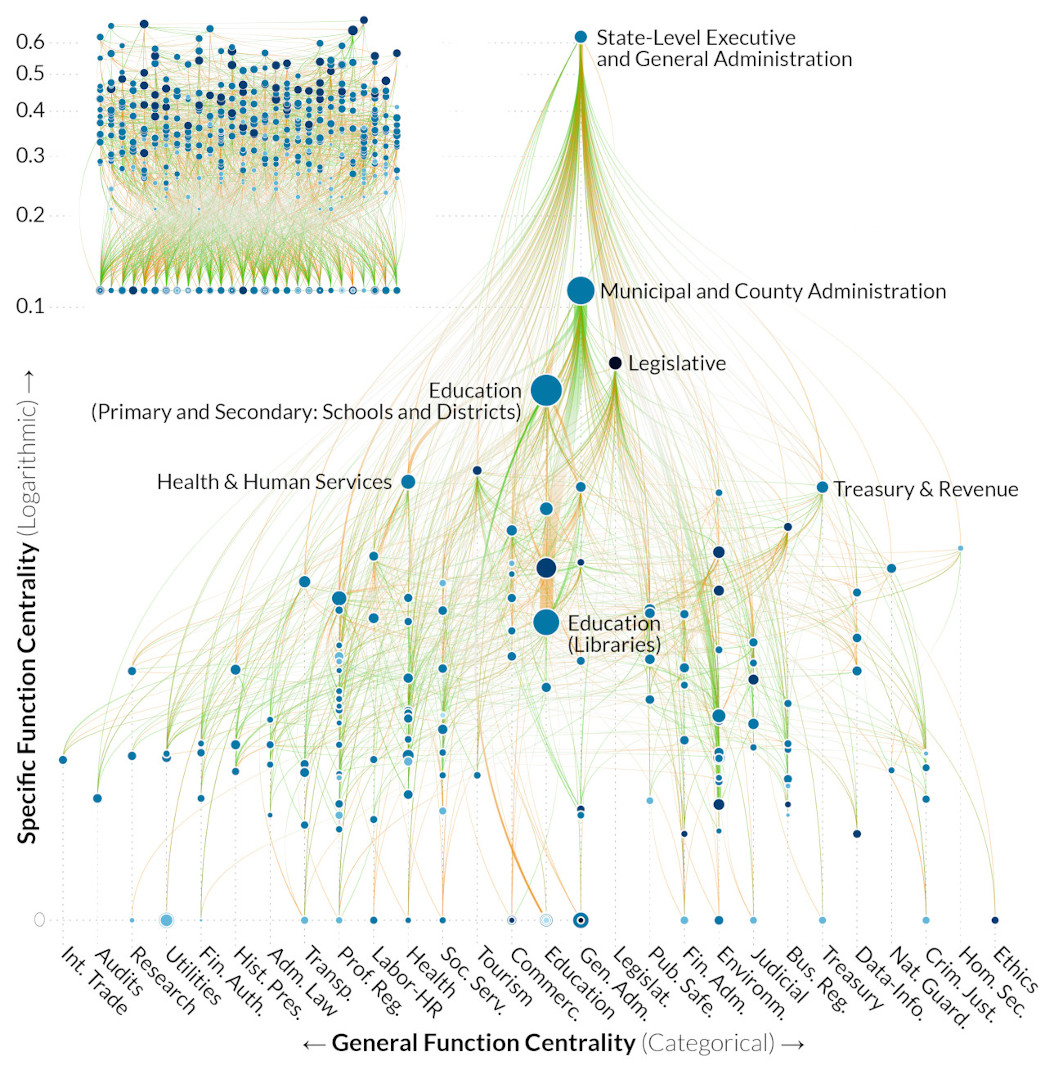}
\caption{The triangular two-level centrality plot I describe in the main text -- image by Kim Albrecht.}
\label{fig:cathedral}
\end{figure}

There are not many level $2$ functions so I can collapse all agencies into their level $2$ function. Then I display these functions in a scatter plot. On the x axis, I report the centrality of the level $1$ function. The most central level $1$ function is in the middle and, as we get to the edges of the visualization, we get to progressively less central level $1$ functions. Now, in this plot, each column contains all level $2$ functions belonging to a specific level $1$ function. On the y axis I report the centrality of the level $2$ function. Functions at the top are more central than functions at the bottom. The result is in Figure \ref{fig:cathedral}. It looks like a nice hierarchy!

The attack you could do to this visualization is that it might make any network look like a hierarchical network, no matter if it is actually hierarchical or not. That is why I embed a small inset in the top left corner. The network in the inset is the result of a configuration model version of the original network: it has the same number of nodes, edges, and the same degree distribution. The connections are rewired randomly, destroying the hierarchy -- if any is present. When I apply the same layout strategy, I obtain the visualization in the inset: there is not a trace of hierarchy any more! This proves that the layout is not showing hierarchies where there are none.

\section{Summary}

\begin{enumerate}
\item A network layout is an algorithm that determines where the nodes of your network visualization should be in a 2D space. The positions of the nodes are determined by the connections between them.
\item The most common principle is the one of the force directed layout. Nodes are charges of the same sign repelling each other and edges are springs trying to keep connected nodes together. This layout works for sparse networks with communities, whose topology fits on a circle.
\item Specialized layouts exists for even sparser networks with a hierarchical organization. Circular layouts can work for denser networks, provided that you use edge bends, bundling edges between nodes located in the same regions of the circle.
\item Edge bends help in many layouts, particularly avoiding the creation of ``ghost edges''. These happen when your network layout places a node on top of an edge that is not connected to it, giving the appearance that the node is part of a chain.
\item Node-link diagrams representing nodes as circles and edges as lines are not the only solution. For very dense networks you can show the network as a matrix, as a graph thumbnail via k-core decomposition, or with a probabilistic layout associating a node to a cloud of probability in space.
\item In many cases, your network will have a clear and unique message that has never been visualized before. In those cases, you need to bend rules and create a unique visualization serving your specific communication objective.
\end{enumerate}

\section{Exercises}

\begin{enumerate}
\item Which network layout is more suitable to visualize the network at \url{http://www.networkatlas.eu/exercises/51/1/data.txt}? Choose between hierarchical, force directed, and circular. Visualize it using all three alternatives and motivate your answer based on the result and the characteristics of the network.
\item Which network layout is more suitable to visualize the network at \url{http://www.networkatlas.eu/exercises/51/2/data.txt}? Choose between hierarchical, force directed, and circular. You might want to use the node attributes at \url{http://www.networkatlas.eu/exercises/51/2/nodes.txt} to enhance your visualization. Visualize it using all three alternatives and motivate your answer based on the result and the characteristics of the network.
\item Which network layout is more suitable to visualize the network at \url{http://www.networkatlas.eu/exercises/51/3/data.txt}? Choose between hierarchical, force directed, and circular. Visualize it using all three alternatives and motivate your answer based on the result and the characteristics of the network.
\end{enumerate}

\part{Useful Resources}

\chapter{Network Science Applications}\label{cha:history}
Network science is a vast field, exponentially expanding since the late nineties. There has been so much work on it. This book so far cites more than $1,000$ papers, and yet there still an incredible wealth of produced knowledge that did not fit in here. In this chapter I want to give you a taste of what network science can do. The idea is to briefly discuss the main contributions of a handful of classic papers that, for one reason or another, did not find space in the more pedagogical chapters that preceded this one.

Of course, the set of papers discussed here is subjective: it is my own perspective of the field, the papers and contributions on which I stumbled most often while working. Specifically, I am partial to the field of computational social science\cite{lazer2009life}: the use of computer science techniques to study social systems -- and humanities in general. I am still deeply embedded in the digital humanities tribe. It is also, ironically, a largely incomplete set. No matter how much effort I pour into this book to make it more exhaustive, it seems that each paper I add simply increases its surface area and makes it less thorough, not more. It's the fractal nature of complex systems, and it's something I will have to live with.

\section{Network Effects of Innovation}
Why is society nudging us to live in cities? Is there an invisible force gluing humans in larger and larger settlements? As a matter of fact, this might very well be true. A research collaboration\cite{bettencourt2007growth}\cite{gomez2012statistics} started investigating these questions by performing a deceptively simple analysis. They took data about as many cities in the world as possible. Then, they made a straightforward plot. They placed the population of the city on the x-axis, and plotted a bunch of other variables in the y-axis.

\begin{figure}
\centering
\begin{subfigure}{.475\textwidth}
\includegraphics[width=\textwidth]{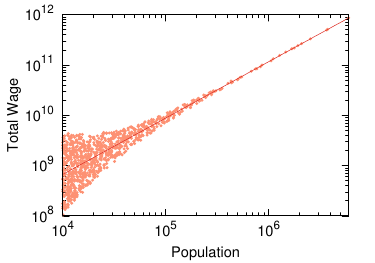}
\caption{$\alpha \sim 1.12$}
\end{subfigure}
\begin{subfigure}{.475\textwidth}
\includegraphics[width=\textwidth]{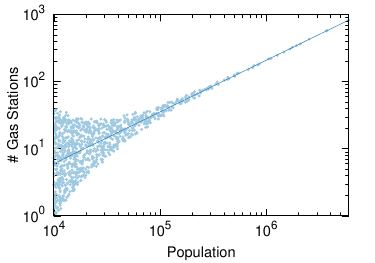}
\caption{$\alpha \sim 0.77$}
\end{subfigure}
\caption{(a) Total wage sum (y axis) as a function of a city's population (x-axis). (b) Number of gas stations (y axis) as a function of a city's population (x-axis).}
\label{fig:urban-scaling}
\end{figure}

Then they found a power relation between a city's population and its outcomes. Their plots looked like the ones I show in Figure \ref{fig:urban-scaling}. When they looked at their $\alpha$ exponents, they discovered something remarkable. The two plots in Figure \ref{fig:urban-scaling} might seem identical, but they differ in a crucial aspect: the value of the $\alpha$ exponent. Figure \ref{fig:urban-scaling}(a) has an $\alpha > 1$, while Figure \ref{fig:urban-scaling}(b) has an $\alpha < 1$. This is a much bigger deal than you might think.

The authors found a consistent higher-than-one $\alpha$ for all wealth creation and innovation activities in a city and, at the same time, a consistent lower-than-one $\alpha$ for all activities accounting for infrastructure management. $\alpha > 1$ means that each added individual to the city contributes more than its fair share to the total. If you have a city where each individual publishes a patent per year and you add a new inhabitant to the city, you don't get an additional patent that year: you get that now each individual publishes $1.12$ patents! Vice versa, $\alpha < 1$ is a classic ``economies of scale'' scenario: once you serve $100$ people, you can serve an additional person without increasing your effort by $1\%$.

So far, this is not a network paper: it's just a purely statistical observation. It is when trying to explain such phenomena that you find networks everywhere\cite[-0.5in]{bettencourt2013origins}\cite{arcaute2015constructing}\cite{gomez2018explaining}. Networks are a necessary ingredient to explain why an additional node enriches the network in a non linear way. Humans have limited resources, so they can only interact with what they can access. In network terms, these are the other nodes at a maximum distance $l$ from them. Every time you add a neighbor with new connections, lots of new nodes will get closer to you, some closer than $l$. 

\begin{figure}
\centering
\begin{subfigure}{.425\textwidth}
\includegraphics[width=\textwidth]{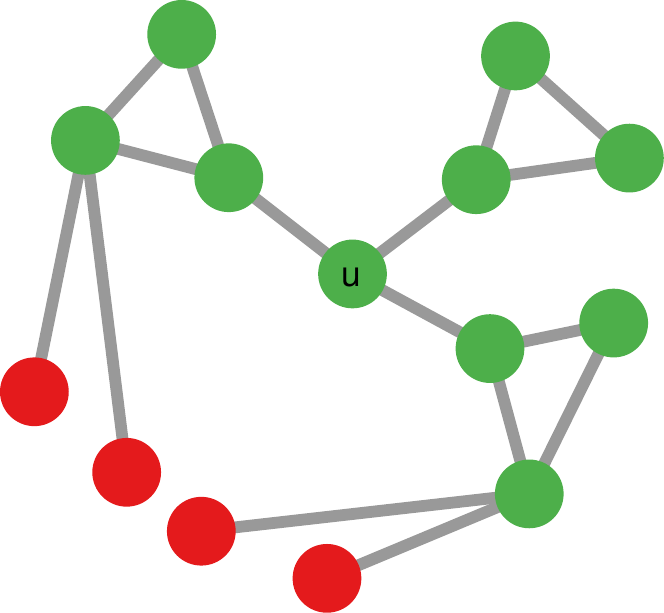}
\caption{}
\end{subfigure}\qquad
\begin{subfigure}{.425\textwidth}
\includegraphics[width=\textwidth]{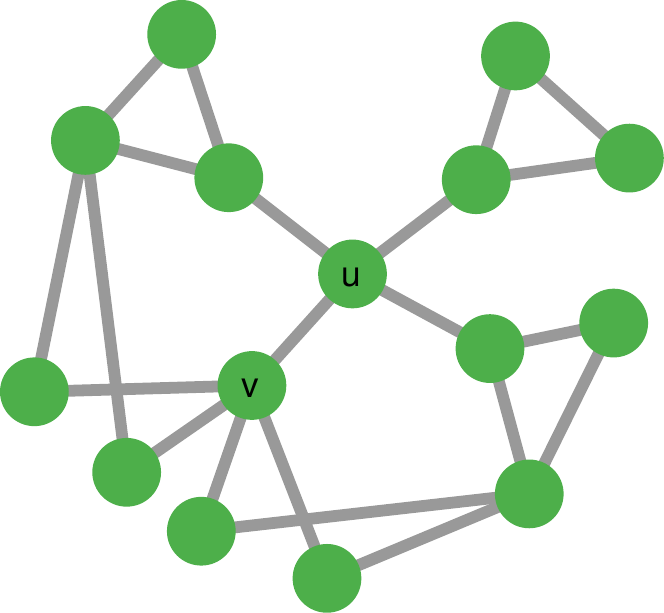}
\caption{}
\end{subfigure}
\caption{An explanation of non-linear node addition contribution. The node color encodes the reachability of a node from $u$: red = unreachable, green = reachable.}
\label{fig:urban-scaling2}
\end{figure}

Consider Figure \ref{fig:urban-scaling2} as an example. Let's assume that node $u$ is selling something, and it has a range: it can only serve up until its neighbors' neighbors. Its original productivity is then $10$. Then, node $v$ appears, and it connects to $u$. If $v$'s contribution were to be linear, $u$'s productivity would go up to $11$. But $v$ has other neighbors of its own, neighbors that were previously unreachable by $u$, as they were at distance $l = 3$. Thus, $u$'s productivity jumps to $15$!

This is a sort of combinatorial effect, where each node adds a new factor you can use and recombine with all the factors that were already present so far. This is easy to see especially in patents data\cite{youn2015invention}. Every time someone makes a new invention, that new invention can be combined with all the previous inventions to create a new one, and so on at infinity. Thus, the knowledge added by a new invention potentially \textit{multiplies} itself with the previously accumulated knowledge, rather than just \textit{adding} to it.

\section{Anonymity in the Age of Social Networks}
We live in troubling times when it comes to our privacy. Large organizations have an interest in gathering information about each individual, whether they do it for surveillance -- as highlighted by Snowden's leaks --, or for profit -- Facebook tracking is ubiquitous, but by no mean the exception in the private sector. The problem is that the simple usage of technology spreads an uncontrollable amount of information about us: the simple sequences of queries you ask a search engine might be enough to identify you\cite{backstrom2007wherefore}, and the combination of few elementary demographic pieces of data can de-anonymize $80\%$ of people\cite{sweeney2002k}.

\begin{figure}
\centering
\includegraphics[width=.425\textwidth]{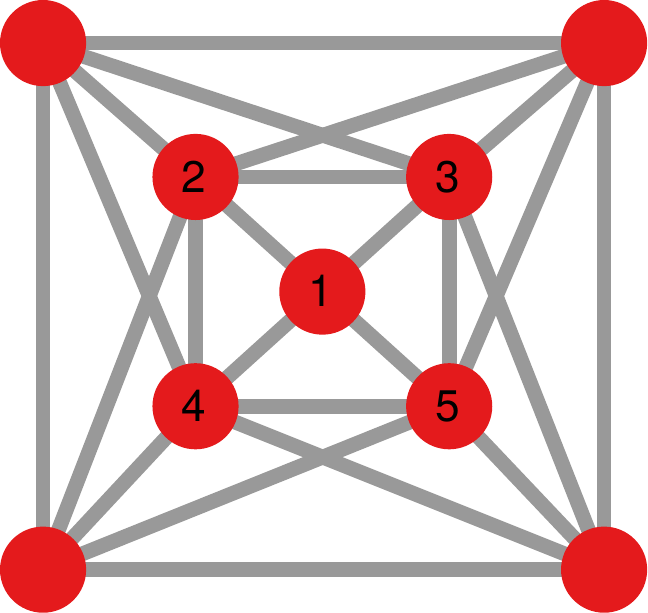}
\caption{A network in which I label the identified nodes.}
\label{fig:anonymity}
\end{figure}

If that worries you, consider that queries and demographics are simple unconnected data. When you talk about interconnected information, the problem is much worse. Consider what you see in Figure \ref{fig:anonymity}. The node $1$ in the center of this network might be you. If an attacker has identified some of your connections, they can say a lot about you\cite{nilizadeh2014community}: the only node in this network that has exactly $\{2, 3, 4, 5\}$ as the set of their friends. They might not be able to know your name, but under the assumption of homophily (Chapter \ref{cha:homophily}) they could infer what you like, your sexual orientation, and maybe even health issues. This is not solved by making your profile private\cite{zheleva2009join}. Even not having a profile at all on social media won't make you safe: a platform can create a shadow profile of you\cite{garcia2017leaking}.

De-anonymizing social networks\cite{narayanan2009anonymizing} is feasible, under a wide array of different scenarios -- whether the attacker is a government agency, a marketing campaign, or an individual stalker. This is usually done by creating a certain amount of auxiliary information that can then be used to recursively de-anonymize more and more nodes in the network. Counter-measures usually adopt the k-anonymity style: making sure that no individual can be identified by obfuscating enough data to make at least $k-1$ other individuals identical to her in some respect. For instance, a network is k-degree anonymous if there are at least $k$ nodes with any given degree value\cite{liu2008towards}.

Sometimes, the focus is preventing the disclosure of information about a relationship, i.e. to combat link re-identification\cite{zheleva2007preserving}. You might not want Facebook to know you are friend with someone, which they could do by performing some relatively trivial link prediction -- see Part \ref{par:lp}. In those cases, you might want to hide some of your relationships, and/or add a few fake connections, to throw off the score function of the link you want to hide.

\section{Human Connectome}
Quite likely, the most famous and studied network in human history is the brain. We have been studying neural networks of many animals, due to their limited size and ease of analysis: cats\cite{scannell1999connectional}, mices\cite{wang2012network}, and, of course, the superstar C. Elegans worm\cite{li2004map}. However, most of this is done with the big prize as the ultimate objective: the human brain. You might have heard of the Human Connectome Project. Proposed in $2005$\cite{sporns2005human}, its objective was to create a low-level network map of the human brain: a network where nodes are individual neurons and connections are the synapses between them.

\begin{figure}
\centering
\begin{subfigure}{.425\textwidth}
\includegraphics[width=\textwidth]{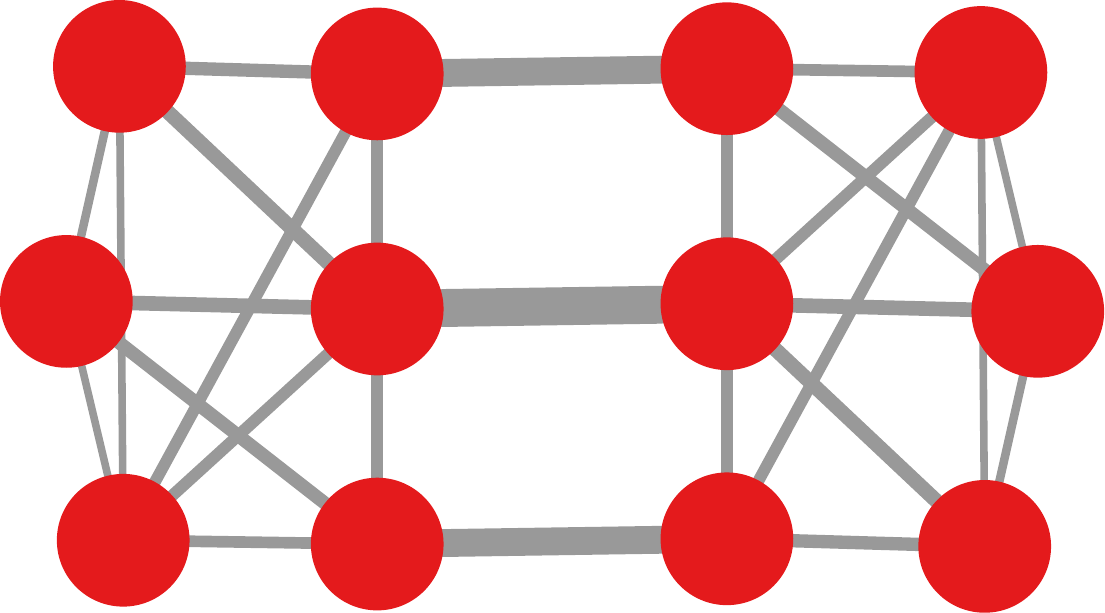}
\caption{}
\end{subfigure}\qquad
\begin{subfigure}{.425\textwidth}
\includegraphics[width=\textwidth]{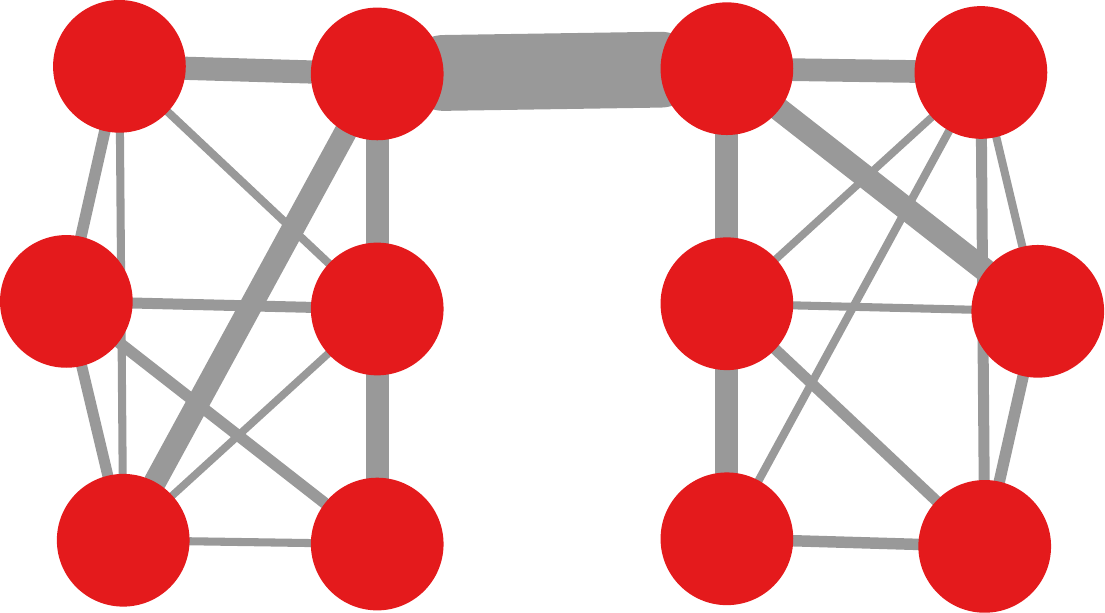}
\caption{}
\end{subfigure}
\caption{The comparison between (a) a healthy brain and (b) an unhealthy brain. Nodes are neurons, edges are synapses and their thickness is proportional to their edge betweenness centrality.}
\label{fig:brain-example1}
\end{figure}

The idea was that applying all the network science artillery to such a network would help us understanding better how our brains work\cite{bullmore2009complex} -- or don't, sometimes. In fact, one of the major lines of research is comparing the brain connection patterns between healthy and unhealthy individuals, because network analysis should be able to easily allow the identification of significant differences in the structures\cite{bassett2009human}. For instance, as Figure \ref{fig:brain-example1} shows, a simple edge betweenness centrality analysis could identify the overload on some synapses caused by structural differences.

There have been many studies of the human brain before the Human Connectome started. For instance, researchers have studied the effect of learning on the connections between brain areas\cite{bassett2011dynamic}. The reason why the Human Connectome is so revolutionary is the granularity of the data. Most brain network studies look at brain activity patterns: the high level difference in electrical potential of brain areas. In this case, the nodes are not individual neurons, but larger modules of the brain.

To be clear, one does not exclude the value of the other. In fact, the brain is an extremely complex organ: it is the most important organ of your body\footnote{According to the brain.}. This means that it operates at multiple scales\cite{betzel2017multi}, in a hierarchical fashion: neurons are part of modules\cite{bonifazi2009gabaergic}, and there are modules of modules, and so on -- check out Chapters \ref{cha:hier} and \ref{cha:hcd} for a few refreshers on hierarchies. In fact, one of the most appropriate models of the brain is multilayer networks\cite{de2017multilayer}.

\section{Science of Science and of Success}
Unsurprisingly, one of the things that interests scientists the most is... scientists. Network scientists are no exception to this rule. There is a large and healthy literature in analyzing networks of scientists. We already saw many examples of two types of science networks: co-authorship networks, where scientists are connected to each other if they collaborate on the same paper/project; and citation networks, connecting papers if one cites another.

The two can be combined to try and gather a general picture of how science gets done. Science is one of the most important human activities\footnote{According to scientists.}, because we rely on it to develop new and better ways to improve our everyday life. It's better to understand how it works, so that we can do it better. This is fundamentally the mission statement of the science of science field\cite{barabasi2012publishing}\cite{wang2013quantifying}\cite{fortunato2018science}, kickstarted by network scientists and making extensive use of network analysis tools.

One of the most peculiar findings is that the occurrence of the highest impact work of a scientist's career will happen at a random point in time\cite{sinatra2016quantifying}. In other words, there is no way to predict which of your papers will earn you a Nobel prize: it could be your first, it could be your last, or any in between. This is bad news if we want to predict the success of some research, but it's great news for me. The fact that I haven't come even close to making a groundbreaking discovery doesn't mean it won't happen eventually. I simply won't see coming if it does (it won't).

\begin{figure}
\centering
\begin{subfigure}{.475\textwidth}
\includegraphics[width=\textwidth]{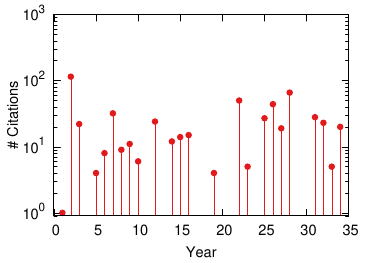}
\caption{}
\end{subfigure}\quad
\begin{subfigure}{.475\textwidth}
\includegraphics[width=\textwidth]{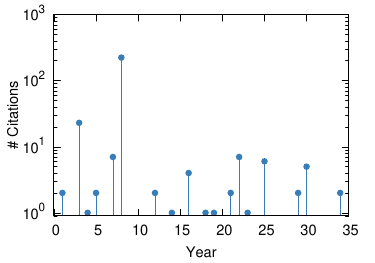}
\caption{}
\end{subfigure}
\caption{Two examples of career paths of scientists, showing the number of citations (y axis) gathered from papers published in a given year (x axis).}
\label{fig:science-of-science}
\end{figure}

Figure \ref{fig:science-of-science} shows an example of this concept. In both cases, the breakout paper arrived early, but there is no pattern in how citations come. Moreover, the red scientist (Figure \ref{fig:science-of-science}(a)) is a better scientist on average than the blue one (Figure \ref{fig:science-of-science}(b)), having $23.5$ citations per paper against blue's $16.2$. They're also more productive ($24$ vs $18$ papers). And yet, it is the blue scientists who published the best paper -- with $223$ citations, while red's best paper only has $115$ citations. Life is unfair this way.

Science of science ended up being a specialized niche of the more broad field of the science of success: the systematic investigation of the gap between one's performance and their success\cite{fraiberger2018quantifying}\cite{liu2018hot}. It is not always the best work of a person that ends up being the most successful. For instance, the Mona Lisa, the most famous painting of the world, is a masterpiece from one of the greatest intellectuals of all time -- Leonardo da Vinci -- but, among its other breathtaking creations, it is rather unremarkable. So unremarkable, in fact, that it was completely ignored and not even exposed until interest in it exploded after its theft.

This disconnect between performance and success is not an exclusive domain of art. It is a much more universal phenomenon. Another studied example is tennis\cite{yucesoy2016untangling}: it is not necessarily the tennis player at the top of the world ranking the one gathering the most Wikipedia page views -- or news articles about them, for that matter.

In fact, the performance-success disconnect can and should be applied to science as well. In this section, I equated ``success'' with citations: a successful paper gathers tons of citations. But is it the \textit{best} (read: highest performing) paper? Not at all! Citations and grant awards correlate with things that are independent of the science/performance itself (e.g., gender\cite{cole1979fair}, race\cite{ginther2011race} and how junior a person is\cite{blackburn1978research}). The world isn't a perfect meritocracy. Cumulative advantage is not just the pretty story of how you model broad degree distributions in networks (Section \ref{sec:physicsmodels-ba}): it is the real unfairness in front of everybody who does not start in the advantaged place/time/gender/race. We should investigate the performance-success disconnect in order to make the world suck a little less. One way to do it is to model science as the interaction between individual characteristics and systemic structures\cite{way2019productivity}.

\section{Human Mobility}
A significant portion of network scientists have also worked on issues of human mobility: describing and predicting how individuals and collectives move in the urban and global landscape\cite{gonzalez2008understanding}\cite{candia2008uncovering}. There are a few reasons for this. First, there is a strong connection between human mobility and many networked phenomena that network scientists investigate. Just to highlight the example from the previous sections: one can use the ``mobility'' of scientists between affiliations to predict their success\cite{deville2014career}. Alternatively, one can use mobility data to augment the de-anonymization process of people in social settings\cite{de2013unique}, or to better predict the spread of infectious diseases\cite{colizza2007modeling}\cite{balcan2009multiscale}\cite{tizzoni2014use}.

Second, complex networks are themselves useful tools to model and analyze mobility patterns. For instance, one can create a better synthetic model of human mobility by using an underlying social network to create realistic motivations for the simulated agents to move in space\cite{musolesi2006community}.

Classically, to predict the number of people moving from area $A$ to area $B$, one would use a ``gravity model''. This works just like Newton's gravity law: the mobility relation between two areas is directly proportional to how many people live in them (their ``mass'') and inversely proportional to their distance\cite{brockmann2006scaling}. In other words, there can be many people moving between New York and Chicago because they are huge cities, but Boston might attract more Newyorkers despite being less populous, simply because it's closer. The gravity model is overly simplistic: it's deterministic, it requires previous mobility data to fit parameters, it lacks theoretical grounding, and it simply doesn't predict observations that well. Network scientists have then developed a radiance model to fix these shortcomings\cite[9\baselineskip]{simini2012universal}.

\begin{figure}
\centering
\begin{subfigure}{.425\textwidth}
\includegraphics[width=\textwidth]{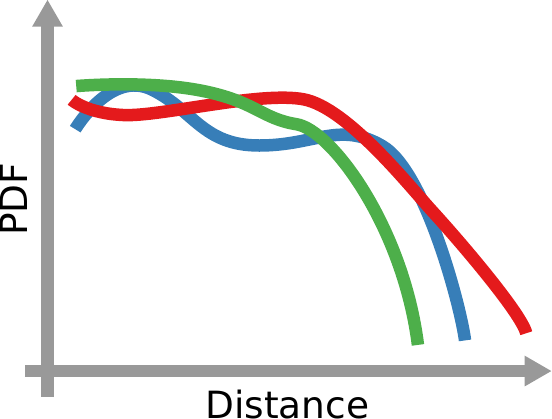}
\caption{}
\end{subfigure}\qquad
\begin{subfigure}{.425\textwidth}
\includegraphics[width=\textwidth]{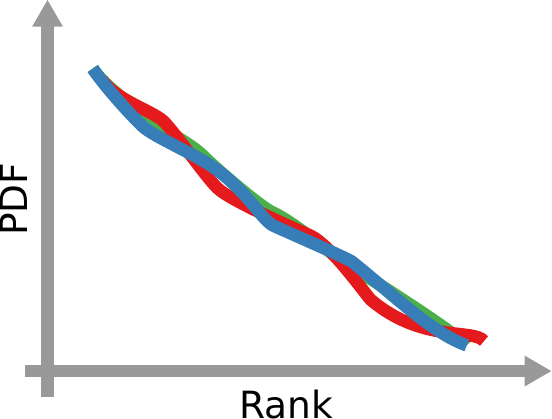}
\caption{}
\end{subfigure}
\caption{(a) Probability of a trip (y axis) as a function of the distance to the destination (x axis). (b) Probability of a trip (y axis) as a function of the destination's rank (x axis). In both cases, different colors report data from different cities.}
\label{fig:mobility-tale}
\end{figure}

What do we find? At a collective level, human mobility patterns are surprisingly universal, but \textit{not} when it comes to the covered distance\cite{noulas2012tale}. Figure \ref{fig:mobility-tale}(a) shows that the probability of making a trip is only mildly related to distance: there is no function properly approximating the likelihood of you visiting a place given its distance to you, and different cities have different scaling and cutoffs. In other words, it is not true that the farther apart a pizza place is, the least you go to eat there.

It is rather that how frequently you go to eat there is connected to the number -- and quality -- of the alternatives in between you and the pizza place. This is only correlated with, rather than being caused by, distance. What matters most, is the \textit{rank} of the place in your preferences. Figure \ref{fig:mobility-tale}(b) shows that there is a clear and universal function predicting a trip's probability given the popularity rank of the destination -- e.g. no matter the city, $20\%$ of trips go to the most popular destination.

The predictability of the collective dynamics, however, does not trickle down to predictability of individuals. Yes, we're animals of habit: we often commute between the same two places -- a property one can exploit to infer home and work locations by looking at incomplete mobility data\cite{iqbal2014development}\cite{alexander2015origin}. On average, one can confidently predict $93\%$ of individual mobility\cite{song2010limits}. However, there is a large variation between individuals: for some you could predict even better than that, while others are fundamentally unpredictable\cite{pappalardo2015returners}.

\section{Memetics}
Who around here doesn't like Internet memes? Cute and funny little pictures, perfect to waste time at work. Of course network scientists love them. However, we need to maintain appearances and pretend that the penguin image we have on our screens is there really for work. We're studying memes, you know? This is for science. There are a few angles with which network scientists attack the study of memes. Some of those already found space elsewhere in the book -- e.g. in Section \ref{sec:triggers-resistance}. 

The first is the relationship between the network structure and the probability of a rumor to spread -- or the fraction of nodes who will end up hearing a rumor. Theoretical calculations\cite{nekovee2007theory} show the impact of the network's topology: in a random graph, initial spread is slow but it will relentlessly cover the entire network; while for scale free networks the initial speed is fast but, in presence of degree correlations, it might fail to cover the entire network. All of this is very similar to simulations of diseases spreading on a network (see Chapter \ref{cha:epidemics}).

Other studies show how the large diversity in the meme success distribution -- few memes spread globally while most are immediately forgot -- are due to the limited capacity of brains to process information\cite{hodas2012visibility}\cite{weng2012competition}\cite[1.5in]{gleeson2014competition}. Another key question is whether memes spread following simple or complex contagion: is a single exposure sufficient or does reinforcement play a significant role? It seems that memes indeed obey the complex contagion rules\cite{monsted2017evidence}. For a refresher on the concepts, see Chapter \ref{cha:triggers}.

\begin{figure}[b]
\centering
\includegraphics[width=.425\textwidth]{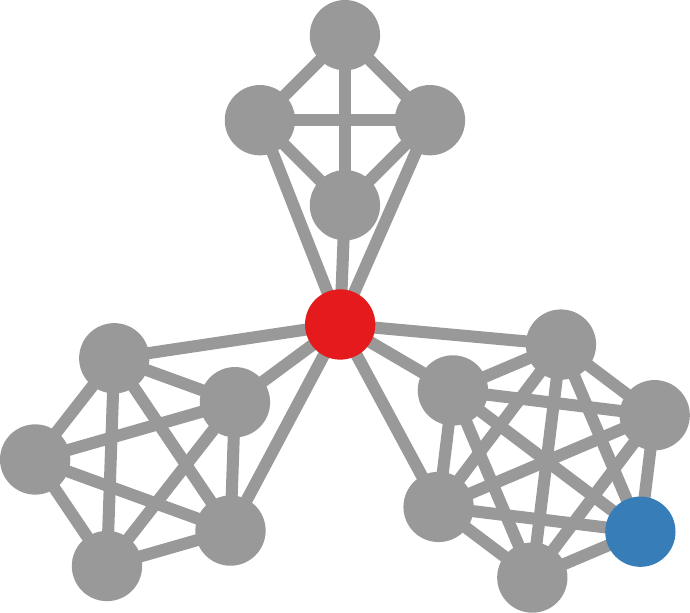}
\caption{A social network. The red and blue nodes are the origin points of two memes.}
\label{fig:community-meme}
\end{figure}

More complex topological features, such as communities, are difficult to treat mathematically, but their impact can be studied using real world data. Figure \ref{fig:community-meme} shows a toy example of the role of communities in meme propagation. Memes originating in the overlap between different communities -- in red in Figure \ref{fig:community-meme} -- have a better chance to go viral\cite{weng2013virality}\cite{weng2014predicting}. Being born well embedded in a community -- blue in Figure \ref{fig:community-meme} -- is bad for propagation, because there are not many paths leading the meme outside of the community.

In general, there are many empirical studies investigating how information propagates through a social network\cite{lerman2010information}, be it memes, rumors, news\cite{vosoughi2018spread}, videos\cite{cha2007tube}\cite{cha2009analyzing}, or photographs\cite{cha2008characterizing}\cite{cha2009measurement}. 

Specifically, one study focuses on the dynamics of ``following'' a content creator on social media\cite{cha2010measuring}. The common sense thing is that the more people are following you -- say on Twitter -- the better it is. You can be more influential if more people listen to you. However, experiments show that this is true only to a certain point. What matters most is the engagement of the followers. Just increasing the count is doing you no good if the people clicking on the \textit{Follow} button actually don't read what you produce. Your voice is diluted on the platform and you have less reach if you inflate those numbers.

The structure of the social network is not, however, the only thing that matters. I already mentioned elsewhere in the book (in Section \ref{sec:triggers-resistance}) that an important factor is also timing: when and how fast you get your appreciation matters a lot in determining whether you are going viral. Alternatively, one could look at the content itself of the meme: the specific image or text associated to it. Studies show how positive valence -- a happy meme -- are useful for propagation\cite{berger2012makes}. In my own research, I instead show how innovation is the key: you want to do something that is dissimilar from everything that has been done before\cite{coscia2014average}\cite{coscia2017popularity}.

\section{Digital Humanities}
Digital humanities is an umbrella term, covering a vast set of applications of computational tools to disciplines in the humanities. As a trained digital humanist myself, I cannot end this book without taking a closer look at this field. And, in a sense, I wasn't, because one could argue that the entire network analysis field is one of the largest subfield of digital humanities. In network analysis we have mathematical and computational models -- graphs and networks -- which are primarily applied to understand social systems. However, among the gigantic bazaar of network science applications, some stand out as poster children of digital humanities.

\begin{figure}[b]
\centering
\includegraphics[width=.55\textwidth]{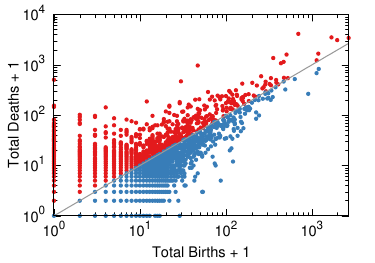}
\caption{The number of famous people who were born (x axis) and died (y axis) in a city. The identity line is in gray. Red points above the identity line and blue points below.}
\label{fig:birth-death}
\end{figure}

One is for sure the study of the birth-death network across cultural history\cite{schich2014network}. Figure \ref{fig:birth-death} shows an interesting historical pattern: each point is a city, and for each city we count the number of famous people who were born and who died in that city. In red we can see death attractors: cities who had more famous deaths than births. In blue we have the emitting cities. By analyzing the historical trajectories of cities, we can see how the cultural center of the world moved from Rome to Paris and then to New York, because more and more people die in the city where they work -- and notable people work where most notable people are. There are other interesting patterns, for instance the fact that the median distance between the birth and death place is increasing, reflecting technological advancements.

With a similar dataset, researchers built the ``notable people portfolio'' of cities and nations\cite{yu2016pantheon}. The idea is to classify all famous people in the area they contributed the most to humanity. Then, one can visualize in which areas places specialize\footnote{\url{https://pantheon.world/}}. For instance, the largest profession represented in the United States is actors, while it is politicians for Greece. But one could explore other dimensions. For instance, professions that are over-expressed in a country against the rest of the world, like chess players in Armenia ($6\%$ of all famous people!). Or explore gender divide: in Canada $27.4\%$ of male famous people were actors against $55.4\%$ female famous people. Finally, you can explore time as well. Before $1700$ AD, the most common way to become famous in Italy was to have a career in politics ($30.7\%$ of famous people did). Afterward? You're better off trying as a soccer player ($21.6\%$).

Other digital humanities applications of network science involve archaeology. This mostly involves the use of network visualization techniques to make sense of a complex, interconnected, and often largely incomplete set of evidence\cite{brughmans2013thinking}. However, it is not necessary to limit ourselves to this: network analysis can be used as a tool to explore evidence. For instance, there are studies of social networks in classical Rome\cite{brughmans2010connecting}. Other examples of prehistoric social network archaeology focus on pre-hispanic North America\cite{mills2013transformation}. Departing from archaeology, the field of social network analysis in a historic\cite{lemercier2015formal}, religious\cite{power2017discerning}, or anthropological\cite{flack2006policing} setting is alive and well.

And since network analysis endows us with powerful tools to study hidden preferences -- such as homophily and segregation, see Chapter \ref{cha:homophily} -- it is a natural instrument to use in other humanities fields, such as gender studies. In particular, there are studies showing unequal gender dynamics when it comes to power relations in online collaborative tools such as, e.g., Wikipedia. As you might expect, the majority of Wikipedia contributors are white men.

When it comes to female representation in the content\cite{wagner2015s}\cite{wagner2016women}, this gender gap shows. The researchers find that women are equally represented in article numbers -- at least in the main six language editions of Wikipedia. However, it is the way women are portrayed that is the problem. Women on Wikipedia tend to be more linked to men than vice versa. Moreover, romantic relationships and family-related issues are much more frequently discussed on Wikipedia articles about women than men.

\section{Political Polarization}\label{sec:app-polar}
Many researchers have used network analysis to study social aspects related to politics. Here I focus specifically on political polarization. Polarization is a complex phenomenon, which can be dissected in different ways from different perspectives. In general, one can say that polarization grows when people get more and more extreme in the political positions or ideologies they subscribe to. If everyone is a centrist, there is no polarization, but splitting camps in extreme left and extreme right indicates high polarization. Then there is the aspect of how you relate to people with different opinions than yours: are you civil, or is a disagreement enough to cause you to use toxic or aggressive language\cite{mejova2014controversy}? Finally, you could consider the polarization of the population at large as different from the one of elites.

It doesn't seem immediately obvious how this relates to networks. After all we're talking about opinions and how people use language. However, networks can be a powerful tool to study polarization. In fact, in the early times of the application of networks to polarization, researchers were mostly making a structural argument\cite{conover2011political}: homophily plays a strong role, people would connect only with people they agree, so we could identify echo chambers\cite{quattrociocchi2016echo}\cite{pariser2011filter}\cite{cinus2022effect}. We've seen this argument in Section \ref{sec:homophily-contagion}. It should be noted that the existence and the strength of echo chambers is a far from established fact, there are still questions to be answered\cite{abramowitz2008polarization}\cite{boxell2017greater}\cite{kubin2021role}.

\begin{figure}[h!]
\centering
\includegraphics[width=.6\textwidth]{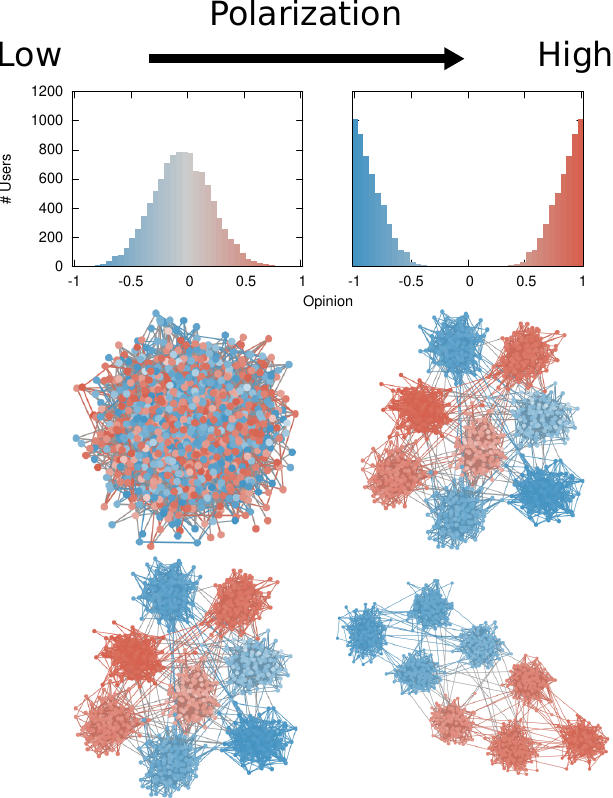}
\caption{The three factors we can consider for ideological polarization. From top to bottom: opinion extremity, echo chambers, and opinion homophily across echo chambers.}
\label{fig:app-polarization}
\end{figure}

To sum it up, looking at Figure \ref{fig:app-polarization} one could think of polarization as being three things: more extreme opinions, isolation in homogeneous groups known as echo chambers, and the tendency of those groups not to interact with groups that are too dissimilar from them. You could also add attributes on the edges recording the toxicity of the language used and correlate it with the opinion difference between the two people interacting. If you get more toxicity with a higher opinion difference, then you get higher polarization.

From there, researchers developed a few different ways to quantify how much it is difficult for people with a given opinion to reach people with a different opinion in a social network\cite{cota2019quantifying}\cite{cinelli2021echo}\cite{monsted2022characterizing}. The harder it is, the more society is polarized. Some popular approaches are based on partitioning the networks into two groups of opposite opinions and see how easy it is for random walks to cross the boundaries of this partition\cite{garimella2018quantifying}. I personally like to calculate the network distance between differing opinions\cite{hohmann2023quantifying} using the Generalized Euclidean measure I described in Section \ref{sec:nvd-ge-lapl}.

Finally, researchers have also created agent based models to see how polarization could arise -- and be counteracted by good policies\cite{de2022modelling}.

\chapter{Data \& Tools}\label{cha:utilities}
A professional worker is only as good as the tools they have and their mastery of them. Thus, knowing where to find the best tools is the first necessary step for being a good network scientist. The tools aren't going to do the job for you, but without them you're just a person armed with lots of good intentions. The aim of this chapter is to kickstart you to your career. I will give you a brief overview of the software libraries and programs one can use to analyze and visualize networks, point you to useful online resources -- especially to find new data sources --, and briefly discuss some of the most famous graph data you will find in the literature.

You might have already seen some of these resources here and there mentioned throughout the book. For instance, the vast majority of the exercises rely on you using Networkx, while in Part \ref{par:netviz} I heavily relied on the knowledge of what Cytoscape and Gephi can do for network visualization.

\section{Libraries}
It might be my bias as a computer scientist showing, but my opinion is that, if you want to have a career in network science, the first thing you have to look at is libraries that help you programming your custom network analyses. There are many fully fledged software programs, with their graphical interfaces and ready-to-use implemented analyses, but I don't think you're really going to be a complete network scientist if you only rely on them. At some point, you will find out something you cannot do with them, and you'll need to roll up your sleeves and get your hands dirty with programming. If you start by learning the libraries, instead, you can always have the option of lazily use another software for all the trivial tasks that don't need any specific customized contribution.

\subsection{Networkx}
I start by dealing with Networkx\cite{hagberg2008exploring}\footnote{\url{https://networkx.github.io/}}. Networkx is a Python library implementing a vast array of network algorithms and analyses. Networkx is -- as far as I can tell -- the most popular choice for students approaching network analysis tasks. I think it's a generalist tool that is not the best at anything specifically, but good enough at everything. One downside is that it is the library struggling with computational efficiency the most, but it is the most complete and popular. If this were a chapter about cinema, Networkx would be Steven Spielberg: everybody knows him, every movie he makes is good but not really great, but it's a bit of a boring choice as a favorite director.

\begin{figure}
\centering
\includegraphics[width=.6\textwidth]{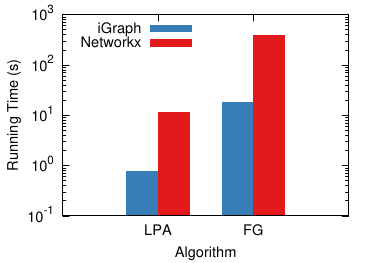}
\caption{The running times for different implementations of community discovery algorithms in Networkx (red) and iGraph (blue).}
\label{fig:networkx-time}
\end{figure}

As I mentioned, the biggest issue of Networkx is efficiency. Networkx is mostly implemented in Python, and it's so large it is impossible for the maintainers to guarantee high code standards throughout the library. Thus you might end up waiting for hours -- or days! -- for an operation that would take other libraries few seconds to complete. For instance, Figure \ref{fig:networkx-time} compares the running times -- on the same network with $\sim100k$ nodes and on the same machine -- of the label propagation and fast greedy modularity community discovery algorithms between Networkx and iGraph. Note the logarithmic scale on the y axis, and despair. Thus, for any analysis with a high time complexity -- for instance frequent pattern mining -- you should definitely look somewhere else.

Another downside I stumbled upon is buggyness. You will often find that the most obscure network functions are sometimes not implemented correctly. I found myself having to switch library because the graph isomorphism functions on labeled multigraphs were just returning the wrong results. This is to be expected for a vast library like this: bugs take time to be noticed, and their fixes to be incorporated.

That is not to say that things aren't improving. I hope that the bug I stumbled upon is now corrected. And other examples of slow implementations are actually on par with state of the art implementation. It used to take impossibly long to generate LFR benchmarks on Networkx, but when I checked on the current version at the time of writing this chapter, I noticed no runtime difference with the C binary provided by the original authors of the paper. Kudos to Networkx on this!

Networkx comes with big strengths as well. First, it is very \textit{pythonic}, which means intuitive and easy to use -- well, at least to me and to all who are comfortable with Python's style of doing stuff. Second, as mentioned, it has really a broad coverage. You can tell this was a tool made by network scientists. The array of functions included in Networkx has no peers in any other library that I know. Finally, it has a relatively decent ecosystem. Of course, not everything can be implemented in Networkx. The developers need to choose what to focus on. But it is a tool on which it is relatively easy to build. Thus you can easily find packages expanding Networkx's capabilities. For instance, Networkx doesn't have a way to find temporal communities, but researchers have built a library on top of Networkx to do so.

\subsection{graph-tool}
If you want to stay in the domain of Python, the obvious alternative to Networkx is graph-tool\cite{peixoto2014graph}\footnote{\url{https://graph-tool.skewed.de/}} by Tiago Peixoto. Graph-tool is, to some extent, the opposite of Networkx in almost every respect. For this reason, it represents a perfect complementary tool.

Graph-tool has several weaknesses. Many are connected to its strengths. For instance, one major strength of graph-tool is its efficiency: it is really fast in computing almost anything. This is due to the fact that it is one of the very few libraries I know that actually has parallel implementations of the network algorithms. This means that, if you are on a machine with multiple cores -- which is to say, your computer is not older than fifteen years --, your analyses are going to run much faster because each of your cores will be involved in the computation. This comes at the downside of requiring some non-trivial technical expertise when dealing with it. I have had students who had to give up on some parts of their projects because they could not install graph-tool.

The other weakness is its incompleteness. Graph-tool is nowhere near Networkx when it comes to offering network analysis tools. This is due to the fact that graph-tool is practically a one man show. Tiago has made a godly amount of work for one person, but he is still one human. On the other hand, this is linked to a strength as well. There might not be many functions implemented in graph-tool, but the ones that are there benefit from having a single mind behind them. The aforementioned issue I had with the bugs in labeled multigraph isomorphism was solved by simply using graph-tool.

Moreover, getting into Tiago's frame of mind is necessary to use graph-tool. You need to understand the way he does things in order to be able to do them as well. Things like function naming, object types, parameter passing -- what you call the interface of the library -- are not as pythonic and intuitive as in Networkx.

\begin{figure}
\centering
\includegraphics[width=.66\textwidth]{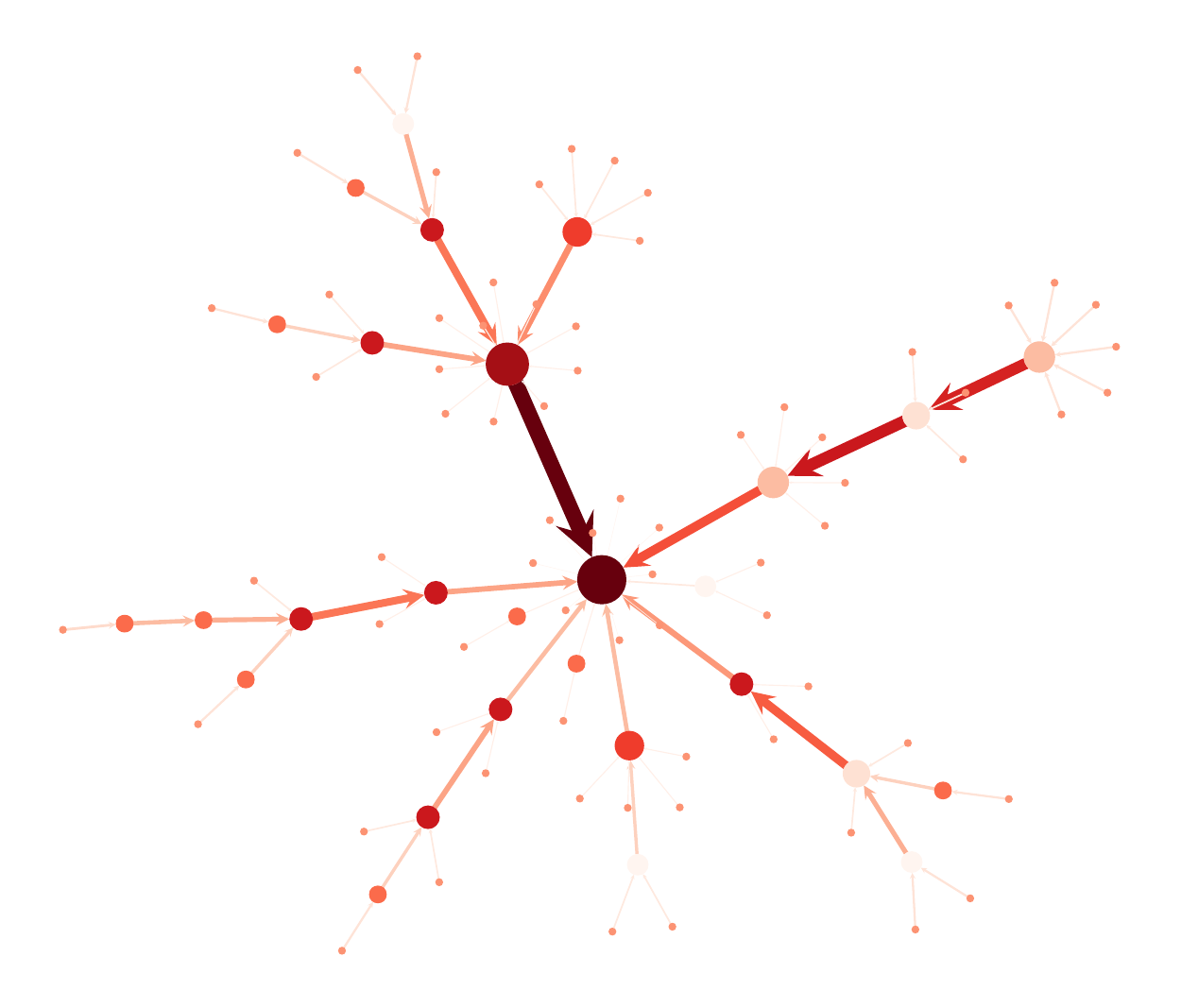}
\caption{A network visualization generated with graph-tool.}
\label{fig:graphtool-draw}
\end{figure}

The final strength of graph-tool is in visualization. I typically use other programs to visualize graphs, but it is undeniable that graph-tool is lightyears ahead any other library you can think of. Figure \ref{fig:graphtool-draw} is an example of what you can do with only minimal effort.

If you want to stay in Python and have performance and a bundle of graph utilities -- rather than a hand-holding library who thinks for you, but also limits you -- graph-tool is the way to go. Graph-tool is, in my movie director analogy, Werner Herzog. Extremely prolific and productive, but you already know that everything you're going to see will be heavily influenced by his charming accent.

\subsection{iGraph}
Among all the alternatives, iGraph\cite{csardi2006igraph}\footnote{\url{https://igraph.org/}} is certainly the most versatile tool. It combines the strengths -- and weaknesses -- of Networkx and graph-tool. On the one hand, it is a surprisingly complete tool with lots of implemented functions -- just like Networkx --, and it is pretty efficiently written -- like graph-tool. Other advantages reside in the fact that the library is available on a vast array of platforms: you can use it both in Python and in R. You can even import it directly as a C library. Thus, if you are capable of writing in C, you can probably cook up a customized analysis using the power of iGraph that cannot be beaten in terms of running time.

That said, iGraph is not the be all end all of network analysis. As I mentioned, it is available on R. In fact, I'd venture the guess that it was developed primarily for R. And I am personally incompatible with R -- some people love it, others like me cannot really understand it. To me, the interface of the library makes no sense. Function names, parameter passing, how things are stored and retrieved from objects: it is all in R style, which my brain unfortunately translates to ``incomprehensible randomness''. The fact that it is possible to import iGraph in Python should not fool you: it is not pythonic at all, and the Python code you end up writing while using iGraph doesn't even look like Python (for some it is a plus, but not for me). You've been warned.

An ironic note of merit to the documentation. Writing documentations is hard. Reading and understanding them is, at least for a dense researcher like me, even harder. Luckily, sometimes iGraph's documentation brightens your day with timeless comedic gems such as:

\begin{quote}
\texttt{layout\_on\_sphere} places the vertices (approximately) uniformly on the surface of a sphere, this is thus a 3d layout. It is not clear however what ``uniformly on a sphere'' means. 
\end{quote}

Gee, thank you, it's refreshing to see that not even who developed this function knows what the function is doing. Continuing my movie directors analogy, iGraph is David Lynch: probably the only one able to do what he is doing, but good luck knowing what's going on when you look at something made by him.

The fact that I'm badmouthing iGraph so hard and yet I am including it in the book and I use it should really convince you that it is a fundamental tool. If I could live without it -- trust me -- I would. But I can't, because sometimes it is the only thing that will save you.

\subsection{Julia}
Julia is a more recent alternative to Python. Like Python, it is a general purpose programming language, but it is particularly geared towards numerical analysis and so it is useful for data science. I call this section ``Julia'' rather than using the specific name of the library because Julia's library design is minimalist and modular. Each library will implement only a very specific set of things and you will find yourself having to import several libraries to make a complete network analysis pipeline. The libraries you'd find yourself using most often for your tasks are:

\begin{itemize}
\item Graphs.jl (\url{https://juliagraphs.org/Graphs.jl/stable/}) for basic graph models and operations;
\item GraphIO.jl (\url{https://github.com/JuliaGraphs/GraphIO.jl}) to read/write graphs to/from memory;
\item A bunch of libraries for plotting (\url{https://juliagraphs.org/Graphs.jl/stable/first_steps/plotting/});
\item A bunch of libraries if you want to have weighted or heterogeneous graphs (\url{https://juliagraphs.org/Graphs.jl/stable/ecosystem/graphtypes/});
\item Laplacians.jl (\url{https://danspielman.github.io/Laplacians.jl/dev/}) for some specific advanced linear algebra operations;
\item And others that you can find in the JuliaGraphs GitHub repository collection (\url{https://github.com/orgs/JuliaGraphs/repositories}).
\end{itemize}

The advantages of Julia are that, in general, code will run faster than Python, all things being equal -- I'll show an example later. Also, some of the code that has been implemented in Julia cannot be found anywhere else -- at least that I know of. For instance, I know of no other way to get Laplacian solvers than using Laplacians.jl (unless I were to implement them myself, obviously). Julia comes at the disadvantages that it compiles on the fly, so the first time you run a piece of code it might take a long time, while you wait for the compilation. Also, Julia's programming logic is different than Python. If you come from Python you will sometimes be surprised by the behavior of Julia, findings in your variables unexpected values because of how and where they were initialized. Julia for me is like Ari Aster: the hot new kid on the block who's doing amazing stuff, but you wonder whether you've become too old for that.

\subsection{Torch Geometric}
One thing you should always remember is that networks and graphs are, at the end of the day, matrices -- remember Chapter \ref{cha:mat}. I try to ignore this fact as much as I can, but it is an undeniable truth. Sometimes, the best thing you can do is to treat them as such, and to start doing some good old linear algebra. Which means that your toolbox can include specialized software like Matlab or Octave. There are in fact, network scientists who are able to do everything they need to do exclusively in these programming environments. I always look at them in awe, not knowing if I do so out of being fascinated or terrified by them.

If you are using Python, you cannot live without learning at least the basics of Numpy and Scipy\cite{virtanen2020scipy}\footnote{\url{https://www.scipy.org/}}, especially when it comes to use sparse matrices. Pandas\footnote{\url{https://pandas.pydata.org/}} is a good tool as well, because you can use it to pivot effortlessly between dealing with networks as edge lists and as matrices. Networkx can convert to and from its data structures into Numpy, Scipy, and Pandas.

However, the real game changes is the ability of performing matrix operations on GPUs. If the algorithm you're running can be expressed as a series of matrix multiplications, then you should load your graph onto your graphic card using PyTorch Geometric (\url{https://pytorch-geometric.readthedocs.io/en/latest/}) and get it done stupendously efficiently. The downside is that you need to get the whole GPU computing stack to work: get Torch to work, install your GPU compute libraries (such as CUDA) and so on. This can be daunting, but the rewards are incredible.

\begin{figure}
\centering
\includegraphics[width=\textwidth]{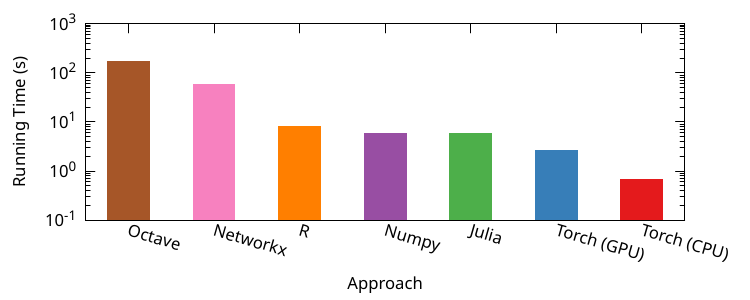}
\caption{The running times (y axis) to compute the effective resistance matrix in different frameworks (x axis and bar color).}
\label{fig:er-all-runtimes}
\end{figure}

To give you an idea, I tried to calculate the effective resistance on a graph of $3,000$ nodes in all the frameworks I could. Figure \ref{fig:er-all-runtimes} shows you how much time they took. I don't include graph-tool because it doesn't have a native way of calculating effective resistance and so you'd do it in numpy/scipy anyway. The runtimes I include here are only for the calculation of the matrix, and they do not take into account the read/write time nor the building of the basic data structure -- but it does count the calculation of the Laplacian.

Note that the figure has a logarithmic y axis. This shouldn't be taken as a formal benchmark, you know this operation might be a bit weird because the CPU version of torch geometric was faster than the GPU one. Another reason is that, normally, Octave is good for matrix operations, but its pseudoinverse is very slow, taking $3$ minutes. However, just to give an idea, Networkx takes $60$ seconds, but you can cut that by a factor of ten to $6$ second with a pure numpy/scipy operation -- or working with Julia, which has a slight edge over numpy. You can cut by another factor of ten to $0.6$ seconds (!!) using torch geometric.

So working with torch geometric is like watching an Alex Garland movie: you know you're going into the deep end, Ex Machina style.

\subsection{Other}
Networkx, graph-tool, iGraph, and Torch Geometric are the four horsemen of network analysis: most of the times, if you need a library, one of them will cover you. That is not to say they are the only things you should know. Here I group a bunch of miscellanea that could come in handy sooner of later.

There are a few competing libraries for doing network analysis. Stanford Network Analysis Project\cite{leskovec2016snap}\footnote{\url{https://snap.stanford.edu/}} (SNAP) comes to mind. This is another C++ library, thus it competes more directly with iGraph. It is owned by a research group at Stanford, which is both a blessing and a curse. It contains mostly implementations of the analyses developed in that research group, which are great and extremely useful. This, however, comes to the cost of completeness. It is very focused.

Specializing on multilayer networks, there's the Multinet package\cite{magnanianalysis}. This will help you deal with this complex data format, and it has versions for both R\footnote{\url{https://cran.r-project.org/web/packages/multinet/index.html}} and Python\footnote{\url{https://pypi.org/project/uunet/}}.

Then there are packages that provide specialized utilities. There are probably more out there than I can count, so I'm making only one example to make you aware of the wealth of useful things that you could find. The \texttt{powerlaw} package\cite{alstott2014powerlaw}\footnote{\url{https://github.com/jeffalstott/powerlaw}} allows you to perform statistical testing to verify whether your degree distribution is really a power law and cannot be explained by more mundane generating processes. It is what we rely on in Section \ref{sec:degree-fit}.

Other notable libraries include:

\begin{itemize}
\item CDlib\cite{rossetti2019cdlib}\footnote{\url{http://cdlib.readthedocs.io}}, specializing on implementing community discovery algorithms. It includes more than $50$ of them, plus around $30$ quality functions you can use to evaluate them (Chapter \ref{cha:cd-eval});
\item NDlib\cite{rossetti2018ndlib}\footnote{\url{http://ndlib.readthedocs.io}}, focusing on models of epidemics/spreading events on networks (Part \ref{par:sis});
\item DyNetX\footnote{\url{http://dynetx.readthedocs.io}}, which extends networkx adding to it the ability of dealing with temporal/dynamic networks.
\end{itemize}

\section{Software}
So far, we've seen software libraries: additional packages for programing languages that you can use to code your own solutions to network analysis problems. Sometimes, you instead want a fully-fledged software, possibly with a graphical user interface, to operate a set of standard operations. This is what we deal with in this section.

\subsection{For Visualization}
By far, the software I use the most is Cytoscape\cite{shannon2003cytoscape}\footnote{\url{https://cytoscape.org/}}. Mostly, I use it for network visualization. The visual style of Cytoscape is based on the Protovis Java library\cite{bostock2009protovis}, which is one of the ancestors of D3\cite{bostock2011d3}\footnote{\url{https://d3js.org/}} -- and it shows. The visual style of Cytoscape is really good, and you can customize a large quantity of visual attributes relatively easily.

Cytoscape supports some basic network analysis. You can calculate a bunch of node, edge, and network statistics, the ones you'd come up first in your exploratory data analysis phase -- nothing too fancy. This analytic capability is mostly there only to allow you to use node and edge statistical properties to augment your visualization. Since version $3.8$, Cytoscape shows fewer plots. For instance you cannot see any more the distribution of shortest path lengths. Moreover, I can't seem to be able to show them in a log-log scale -- it was possible before. However, the graphical quality of the plots greatly improved, and now they are interactive, allowing you to manipulate them to select nodes in the networks.

Another major strength of Cytoscape is its good selection of plugins. Just like Networkx, it supports a healthy ecosystems of contributors. The Cytoscape community skews heavily on the biological network crowd: protein-protein networks, gene interactions, and the like. Thus, if you're part of that community, you will likely find everything you need for Cytoscape. If you're not part of that community, the coverage is a bit more spotty.

Cytoscape comes with an API interface and a console. This means that you can write your customized piece of code that can use Cytoscape as a utility to augment your visualizations. I haven't tested it myself, but it is a nice option to have, if you're the tinkerer kind of analyst. Moreover, Cytoscape understands relatively advanced graph file formats such as GraphML\cite{brandes2013graph} and XGMML\cite{punin2001xgmml}. These are XML dialects specifically developed for graphs. They allow you to store a comprehensive list of node and edge attributes, which you can use to directly encode how your graph is supposed to look like. This means that you can transport your Cytoscape visualizations to any other software which understands them.

Finally, on the downsides, two more quick notes. First, it is a bit annoying to install, because at any given time it will rest on the previous long term support Java version (at the time of writing it is Java $11$, but for most of $2020$ you needed Java $8$, which came out in $2014$ -- i.e. you were running $6$ year old code). Second, I find Cytoscape to be a bit on the buggy side, with random mouse focus fails when selecting/editing text/nodes. This might be my personal experience using it on Linux, which is probably not as well supported as the versions for other major OS platforms. In any case, things have greatly improved over the years, so I like the directions in which it's going. Cytoscape is, in other words, Peter Jackson: he might not be perfect, he might have defects, but boy are his movies nice to look at!

The main alternative to Cytoscape is Gephi\cite{bastian2009gephi}\footnote{\url{https://gephi.org/}}. In fact, calling it an ``alternative'' might even be unfair: my sense is that Gephi is actually \textit{more} popular than Cytoscape among network scientists. However, that is not what I started using during my PhD and so I never ended up installing it. So I don't have any specific way to compare their relative strengths and weaknesses. Chances are that, for $99\%$ of visualization tasks you will find yourself doing, the two programs can be considered equivalent. Gephi is like Guillermo Del Toro: I am unable to tell him apart from Peter Jackson. Regarding the topic of file formats, Networkx can read the GEXF file format, which is the one Gephi uses to save your network visualizations.

Muxviz\cite{de2015muxviz}\footnote{\url{http://muxviz.net/}} is another great piece of software. Muxviz covers a slightly different angle from Cytoscape or Gephi. First, even if I classify it in the visualization subsection, it is much more analysis-oriented. In fact, it requires a lot of analytical power installed on your machine (Octave and R for instance). And you might find yourself using the command line interface more than the graphical interface. In this sense, it could have been listed as a library in the previous section.

More importantly, Muxviz is much more specialized. It has a specific focus on multilayer networks (Section \ref{sec:extended-multilayer}). Which is a good thing, because Cytoscape is not very good for visualizing them. As far as I know, with Cytoscape the only choice you have is to either visualize them with a multigraph, or visualizing one layer at a time and then use a lot of elbow grease to piece the layers together. Muxviz, instead, supports them natively, and it is thus a very complementary choice if you want to be prepared for all the layers life will throw at you.

Finally, there's NetLogo\cite{tisue2004netlogo}\footnote{\url{https://ccl.northwestern.edu/netlogo/}}. I frankly don't know where to classify it, because it is a weird mix of everything. First and foremost, NetLogo is a programming language. It is explicitly designed to facilitate the simulation of agent-based models. This includes all sorts of models, not necessarily the ones involving a network. Thus it is a more general tool, which allows you to do more than what this book focuses on.

Secondarily, you can use NetLogo for visualizing the effects of specific network processes. If you follow the link I provide, you can access NetLogo Web, a collection of simulations programmed in NetLogo that allows you to play with a bunch of different models. For instance, you can run a SIR model (Section \ref{sec:epidemics-sir}), modifying its parameters and tracking the effects of your actions. NetLogo is a godsend for all visual thinkers who need to \textit{see} things happening in front of them to really understand them.

\subsection{For Analysis}
There are many pieces of software out there that will allow you to perform network analysis and are commonly used by network professionals. They are far more than I can include here. So I will limit myself to those with which I had some personal experience.

The programs I talk about here are the ones that primarily provide analytic power. You can visualize networks with them, but you should not do that. Their visualization capabilities are not the main focus of the software, and are there mostly for you to get a quick sense of what sort of analyses you should ask the program to perform.

I think the program for network analysis I stumble the most upon in the literature is Pajek\cite{de2018exploratory}\footnote{\url{http://mrvar.fdv.uni-lj.si/pajek/}}. Pajek allows you to perform a vast array of network analysis, ranging from classical social science ones, to more computer science-y ones -- like community discovery. Pajek comes in different versions: Pajek, Pajek XXL, and Pajek 3XL. The main difference between the versions is the capability of handling larger and larger networks. The idea is that you would perform the memory-intense analyses on the XL versions of Pajek and then import the results for further investigation in the standard version of the program.

Pajek is such a popular program that its own specific file format is compatible with most of the software libraries I mentioned earlier. Both Networkx and iGraph have functions that will allow you to import networks saved in Pajek's file format. Pajek is Lars von Trier: perfect for geeking out every possible detail, but not the prettiest thing to look at.

A popular alternative from Pajek is UCINET\cite{borgatti2002ucinet}\footnote{\url{https://sites.google.com/site/ucinetsoftware/home}}. UCINET's strength is in its deep dive into the \textit{social} branch of social network analysis. It is possibly the most comprehensive tool for social scientists to use.

As a result, its coverage of the more computer science and physics branches is less than optimal. UCINET works best with small networks, it is not particularly well optimized for large scale analysis, and will lack some of the typical algorithms you might expect to find after reading this book. However, my biggest gripe with it is probably the fact that -- differently from almost everything I mentioned so far -- UCINET is not a free program. If you're a full time student, it will cost \$40. UCINET is George M\'{e}li\`{e}s: an immortal classic, but probably not the style you want to adopt in $2024$.

Finally, I should mention NodeXL\cite{hansen2010analyzing}\footnote{\url{http://nodexlgraphgallery.org/}}. NodeXL is a weird animal, which lives on the border between being a software analysis tool with a graphical user interace like Pajek, and a library like NetworkX. NodeXL is a graphical front-end that integrates network analysis into Microsoft Excel. Excel is a phenomenal tool, easily the best and most used software Microsoft has ever written. Excel allows you to perform powerful and sophisticated analysis tasks. There are people whose entire careers could be summed up by a handful of painstakingly crafted Excel spreadsheets. So it is no wonder that there is demand for integrating network analysis in Excel. NodeXL fills that niche.

\subsection{Community Discovery}
I create this special subsection to focus exclusively of implementations of algorithms solving the community discovery problem. This is easily the largest subfield of network analysis. Thus this subsection satisfies two needs. First, it gives you an idea about the immense wealth of code that cannot find space in generic libraries/software. Second, it contains the necessary references to the algorithms I consider in my algorithm similarity network that was included in Section \ref{sec:cd-partition-practical}.

The way this subsection works is as follows. Now I will list a bunch of labels that are consistent with Figure \ref{fig:community-asn}. For each label, I tell you where to find the implementation I used to build that figure. The general disclaimer is that, of course, some of these links are bound to break in the future. I accessed them last time around November $2018$, so the Internet Archive could help.

\begin{itemize}
\item edgebetween, fastgreedy, hrg, labelperc, leadeig, louvain, spinglass, walktrap: igraph implementation in R (\url{https://igraph.org/r/}).
\item mcl: \url{https://www.micans.org/mcl/\#source}.
\item tabu, extr: options \#5 and \#6 in \url{http://deim.urv.cat/~sergio.gomez/radatools.php}.
\item ganet, ganet+, moganet: \url{http://staff.icar.cnr.it/pizzuti/codes.html}.
\item ganxis: \url{https://sites.google.com/site/communitydetectionslpa/}.
\item conclude: \url{http://www.emilio.ferrara.name/code/conclude/}.
\item conga, copra, cliquemod, peacock: \url{http://gregory.org/research/networks/}.
\item mlrmcl: \url{https://sites.google.com/site/stochasticflowclustering/}.
\item metis: \url{http://glaros.dtc.umn.edu/gkhome/metis/hmetis/download}.
\item slpa, fluid, kerlin, kclique: networkx implementation in python (\url{https://networkx.github.io/documentation/stable/}).
\item pmm: \url{http://leitang.net/heterogeneous_network.html}.
\item crossass: \url{https://faculty.mccombs.utexas.edu/deepayan.chakrabarti/software.html}.
\item demon: \url{http://www.michelecoscia.com/?page_id=42}.
\item bigclam, agm: part of the SNAP library (\url{https://snap.stanford.edu/}).
\item hlc: \url{http://barabasilab.neu.edu/projects/linkcommunities/}.
\item tiles: \url{https://github.com/GiulioRossetti/TILES}.
\item oslom: \url{https://sites.google.com/site/andrealancichinetti/software}.
\item kmeans, dbscan, ward, agglomerative, spectral, meanshift, affinity, birch: \url{http://scikit-learn.org/stable/modules/classes.html\#module-sklearn.cluster}.
\item code-dense: \url{https://link.springer.com/article/10.1007/s10618-014-0373-y}.
\item moses, collapsed-sbm: \url{https://sites.google.com/site/aaronmcdaid/downloads}.
\item gce: \url{https://sites.google.com/site/greedycliqueexpansion/}.
\item ilcd: \url{http://cazabetremy.fr/rRessources/iLCD.html}.
\item svinet, mmsb: \url{https://github.com/premgopalan/svinet}.
\item bnmtf: \url{http://www.cse.ust.hk/~dyyeung/code/BNMTF.zip}.
\item rmcl: \url{https://rdrr.io/github/DavidGilgien/ML.RMCL/man/ML_RMCL.html}.
\item OLC: \url{http://www-personal.umich.edu/~mejn/OverlappingLinkCommunities.zip}.
\item cme-td, cme-bu: \url{https://github.com/linhongseba/ContentMapEquation}.
\item edgeclust: \url{http://homes.sice.indiana.edu/filiradi/Data/radetal_algorithm.tgz}.
\item infocentr: my own implementation of the algorithm described in the original paper\cite{fortunato2004method}.
\item msg, vm: \url{http://www.biochem-caflisch.uzh.ch/node/385}.
\item ocg: \url{http://tagc.univ-mrs.fr/tagc/index.php/software/ocg}.
\item savi: \url{http://dsec.pku.edu.cn/~tieli/}.
\item mixnet: \url{http://www.math-evry.cnrs.fr/logiciels/mixnet/mixnet}.
\item vbmod: \url{https://github.com/jhofman/vbmod_python}.
\item bridgebound, bagrowLocal, clausetLocal, lwplocal: \url{https://github.com/kleinmind/bridge-bounding}.
\item netcarto: \sloppy\url{http://seeslab.info/downloads/network-cartography-netcarto/}.
\item infomap, infomap-overlap: \url{http://www.mapequation.org/code.html\#Download-and-compile}.
\item graclus: \url{http://www.cs.utexas.edu/users/dml/Software/graclus.html}.
\item graclus2stage: my own implementation of the algorithm described in the original paper\cite{narasimhamurthy2008community}.
\item fuzzyclust: \url{https://github.com/ntamas/fuzzyclust}.
\item linecomms: \url{https://sites.google.com/site/linegraphs/} (to generate the line graph + igraph's implementation of Louvain).
\end{itemize}

It's now time to move to a section dedicated not to code, but to data. However, before doing so, I'll give you a small preview. In the paper building the algorithm similarity network, I test all these algorithms on $819$ real world networks that I use as a benchmark. These networks are taken from data kindly shared by the authors of a bunch of papers\cite[0.3in]{gabasova2015star}\cite{snijders2010introduction}\cite{van1999friendship}\cite{rhodes2015inferring}\cite{descormiers2011alliances}\cite{sundaresan2007network}\cite{schein1955social}\cite{gimenez2011illegal}\cite{beveridge2016network}\cite{de1999literary}\cite{lott1979dominance}\cite{kaminski2018moviegalaxies} and online webpages\footnote{\url{http://vlado.fmf.uni-lj.si/pub/networks/data/bio/foodweb/foodweb.htm}}\footnote{\url{http://wwwlovre.appspot.com/support.jsp}}.

\section{Data}
There's more to life than just the software running on your computer -- or so I'm told. Many great resources that can make you a better network analyst -- or even just a better person overall -- can be found online. Specifically, here I focus on online network resources concerning the first ingredient of every network paper: network data. You need data to test your models, to run your algorithm, to make a compelling case of why your paper is important. Often, your study will start from a dataset you already have, or you will collect one specially tailored for your purposes. In many other cases, you simply need any network you can put your hands on that fulfills some specific constraints. This section should help you with this task.

There are many places where you can find networks directly available for download, but I start with an index: the Colorado Index of Complex Networks\cite{clauset2016colorado}\footnote{\url{https://icon.colorado.edu/}} (ICON). This is quite possibly the most comprehensive index of network datasets from all domains of network science. Chances are that, if the network data is available somewhere, you can find it via ICON.

However, this is an index of network datasets, not a dataset repository, like the ones that will follow. This means that ICON is not hosting any network data itself. It rather contains the \textit{links} to those datasets. This has advantages and disadvantages. The advantage is completeness: not all datasets can be moved from their original source and hosted somewhere else. ICON can include those datasets, while the other repositories cannot. The other side of the coin is the dynamism of the Internet. Resources get moved all the time, and not everybody does it properly via HTTP redirects -- actually almost no one does it. Thus it is possible to find dead links in ICON, because the managers of the website cannot possibly constantly check that all links are working.

ICON will point you to tons of resources from which you can actually download your data, for instance Pajek's and UCINET's websites. There you can find a collection of network datasets you can download, which is a nice additional resource to the software. One issue you might have with this solution is that they distribute data in their own file formats, so you might need to convert them before you can use them with another software.

Also SNAP provides network data. While there is a large overlap between what you can find in Pajek and UCINET, SNAP's focus goes decisively more towards computer science. You will find \textit{very} large datasets there, sometimes larger than what you can handle -- at the time of writing this chapter, I believe the largest network is from Friendster, which contains more than $1.8$ billion edges. Just like with the implemented functions in SNAP, also the datasets are very much focused on the ones the Stanford research group used for their publications.

Another interesting resource is Konect\cite{kunegis2013konect}\footnote{\url{http://konect.cc/}}. Konect is also a Matlab package for network analysis. Since I do not use Matlab unless someone is pointing a gun at me, I have no experience with it as an analysis tool. However, I used to browse Konect daily to find and download some interesting network data. The list of available datasets, as far as I can tell, is a superset of what you can find in the websites of Pajek and UCINET, and more. The web interface is also well done, and you will be able to tell what are the main characteristics of a network before downloading it: if it's bipartite, what its degree distribution is, etc. I said ``used to browse'' because I believe the project might not be online any more, perhaps the Internet Archive could help.

Network Repository\footnote{\url{http://networkrepository.com/}} is probably the largest repository with direct network data download -- thus excluding ICON. The interface isn't as good as Konect, but it's free and it includes more network data. Tiago Peixoto of graph-tool fame also launched his own network data resource: Netzschleuder\footnote{\url{https://networks.skewed.de/}}, which rivals Network Repository in size. It mostly takes all the network data indexed by ICON or Konect and provides a direct download in different formats. The networks can also be directly imported in graph-tool via a function, without worrying about having the network file saved on your hard disk.

If you're specifically interested in multilayer network data, one cool resource is the Comune Lab\footnote{\url{https://comunelab.fbk.eu/}}. Comune is a project owned by the same people behind Muxviz and the two can be considered as closely integrated.

\section{Legendary Graphs}\label{sec:utilities-graphs}
There are some graphs that are so widely used that you don't really need to look for them in an online repository. These are the pillars on which the entire cathedral of network science is founded. They are often directly included in software and libraries, and all online self-respecting network data repositories have one or multiple copies of them. I include a few here.

\begin{figure}
\centering
\includegraphics[width=.8\textwidth]{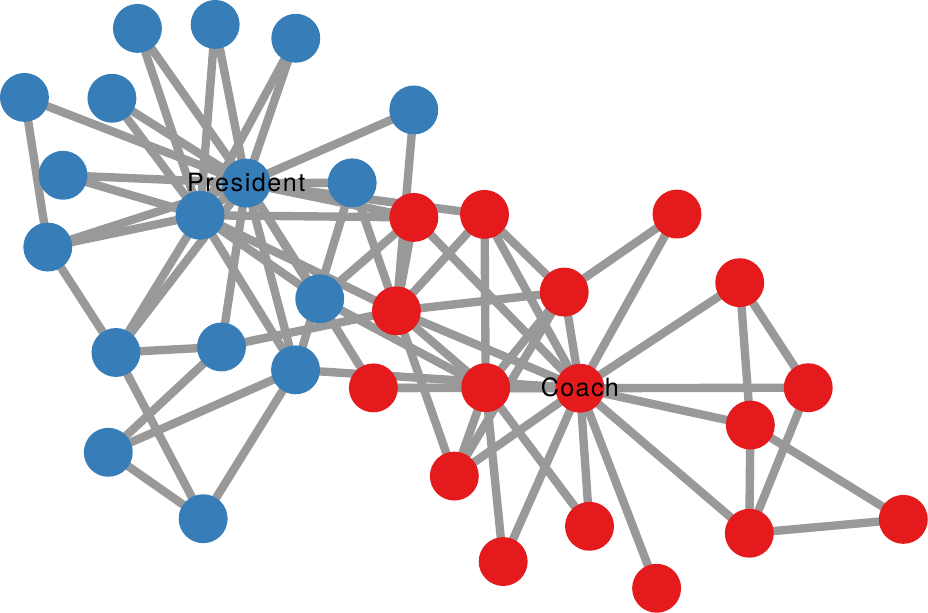}
\caption{The Zachary Karate Club network. I label the nodes representing the coach and the president. The node color indicates whom the member followed after the club split up.}
\label{fig:zachary}
\end{figure}

The first -- and by far most popular -- of these legendary graphs is the Zachary Karate Club\cite{zachary1977information}. This is a network of members of a karate club, connecting two members if they sparred against each other. It is often used because the network focuses on two main nodes: the coach and the president of the club. The club eventually split due to a disagreement between the two, and one can reconstruct on which side each member went by analyzing with whom they sparred. It is a classical example of community discovery. Figure \ref{fig:zachary} shows this beauty in all of its glory.

Aaron Clauset told me a fun fact about this network. Zachary's original paper contains a figure showing the undirected adjacency matrix of the Karate Club network, except that it's not fully undirected! One edge appears in one direction, but not in the other. This means that there are technically two Karate Club graphs, depending on whether this edge is a typo or not, one with $|E| = 77$ edges and one with $|E| = 78$ edges. The latter is the most common you'll find around, because it is the one that Mark Newman and Michelle Girvan used for their paper, which arguably launched the Karate Club network in the Olympus of network science.

Network scientists are obsessed with this network. It has its own t-shirt\footnote{\url{https://www.zazzle.co.uk/zachary_karate_club_with_label_t_shirt-235415254499870147}, the label says: ``If your method doesn't work on this network, then go home''.}. They even created the Zachary Karate Club Club\footnote{\url{https://networkkarate.tumblr.com/}}: the club of network scientists who are the first using the Zachary network as an example in their presentation at a network science conference. If you do so, you become the current holder of the Zachary Karate Club Trophy and you are responsible for handing it at the next conference you attend. This is fiercely competitive, and often you'll see this prize awarded at satellites events happening \textit{before} the conference itself, because people will use the network as an example as soon as they can, to get their hands on the trophy.

The Network Science Society hands many prestigious awards: the Erd\H{o}s-R\'{e}nyi prize\footnote{\url{https://netscisociety.net/award-prizes/er-prize}}, to the career of the most outstanding network scientist under the age of forty; or the Euler award\footnote{\url{https://netscisociety.net/award-prizes/euler-award}}, to the authors of paradigm-changing publications in network science. But don't get fooled. The Zachary Karate Club Trophy is where it's at.

\begin{figure}
\centering
\includegraphics[width=.66\textwidth]{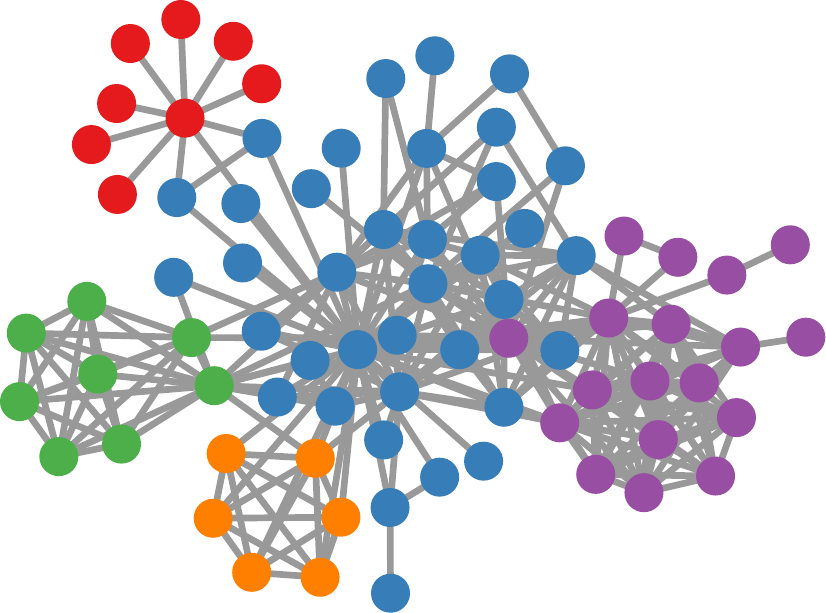}
\caption{The Les Miserables network. The node color follows the community assignment from a label percolation community discovery}
\label{fig:lesmis}
\end{figure}

Another commonly used network is the one obtained from Victor Hugo's novel \textit{Les Miserables}\cite{knuth1993stanford}. In the network, each node is a character, and two characters are connected together if they appear in the same chapter. Also in this case the classical application is for community discovery, given that there are sets of characters closely interacting with each other that never appear in chapters with other groups of characters. Figure \ref{fig:lesmis} shows an example. This is one of those graphs that even non-network scientist would use for examples related to other fields, for instance data visualization\footnote{\url{https://bost.ocks.org/mike/miserables/}}. The likely reason is the inclusion of this network in Knuth's popular book.

\begin{figure}
\centering
\includegraphics[width=.66\textwidth]{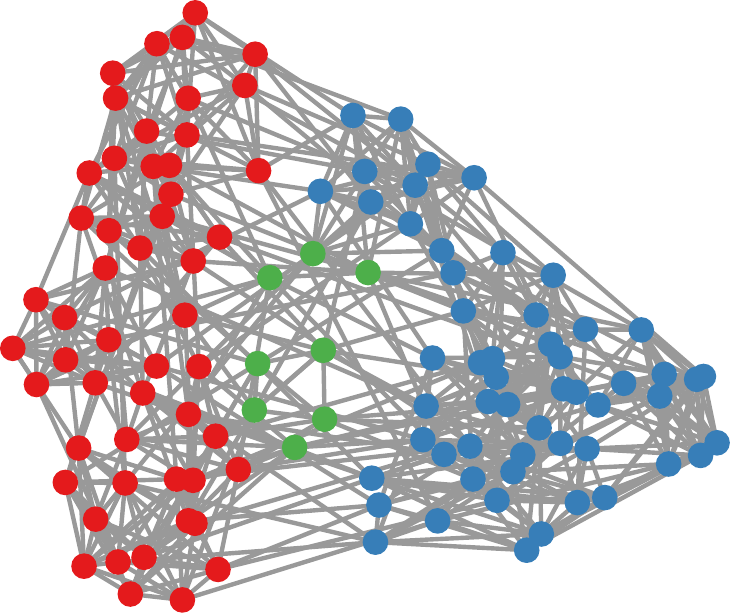}
\caption{The Football network. The node color follows the community assignment from a label percolation community discovery}
\label{fig:football}
\end{figure}

The college football network\footnote{\url{https://figshare.com/articles/American_College_Football_Network_Files/93179}} is another network commonly used for community discovery -- I'm sensing a pattern here. Figure \ref{fig:football} shows it. The reason it works well is due to the way sports are organized in the United States. Usually, teams are divided in conferences and divisions. A team will play with all other teams in their division, but only with a selected number of teams in the same conference and almost no team from the other conference. This creates a nice hierarchical community structure. There is also an overlap, as the most successful teams will then access to the finals and thus play a significant number of matches with teams from the other conference.

Other examples of networks I'm not going to discuss in details are:

\begin{itemize}
\item Florentine families\cite{breiger1986cumulated}: each node is a family from Renaissance Florence, and families are connected if there is a marriage tie between them (I used this network in Section \ref{seg:ergmodels-ergm} when talking about ERGMs);
\item Davis Southern women social network\cite{davis2009deep}: a bipartite network, connecting $18$ women to $14$ informal social events they attended;
\item C. Elegans: this is not a single graph, it is actually multiple. We have extracted all possible ways to represent this poor little worm in networks forms, from a neural network to protein-protein interaction networks.
\end{itemize}

\chapter{Glossary}

\section*{A}
Actor: The entity to which nodes in different layers in a multilayer network refer to. It can be considered as the connected component formed when using exclusively inter-layer couplings. \bigskip

\noindent Acyclic Graph: See Tree. \bigskip

\noindent Adjacency Matrix: A matrix where each row and column correspond to a node in the graph. The $A_{uv}$ entry of the matrix is one if nodes $u$ and $v$ are connected, zero otherwise. \bigskip

\noindent Adjusted Mutual Information: A uniform random variable will have some mutual information with a non-random variable. In AMI, by definition, if there is no relation between the two variables the result will be zero. Thus, AMI is mutual information adjusted for chance. \bigskip

\noindent Arborescence: A directed tree in which all nodes have in-degree of one, except the root, which has in-degree of zero. \bigskip

\noindent Arborescence Forest: A graph with multiple weakly connected components, each one of them being an arborescence. \bigskip

\noindent Assortativity: The tendency of nodes to connect with other nodes carrying similar attributes. Synonym of homophily. \bigskip

\noindent Arc: See Edge. \bigskip

\noindent Average Path Length: The sum of the lengths of all shortest paths in a network over the total number of such paths.

\section*{B}
Balanced Graph: A directed graph whose in- and out-degree sequences are the same. \bigskip

\noindent Betweenness Centrality: Normalized number of shortest paths passing through the node. \bigskip

\noindent Biclique: A clique in a bipartite network. \bigskip

\noindent Bipartite Network: A network with two types of nodes and whose edges can only connect two nodes of different type. \bigskip

\noindent Breadth First Search: The exploration of a graph by exploring all neighbors of a node before moving on the the neighbors of the next node.

\section*{C}
Chain: A set of nodes that can be ordered, and each node is connected only to its predecessor -- except the first node -- and its successor -- except the last node. \bigskip

\noindent Clique: A set of nodes where all possible edges are present, i.e. each node in a clique is connected with each other node in the same clique. \bigskip

\noindent Closeness Centrality: Normalized inverse of the average shortest path length from a node to all other nodes in the network (or connected component). \bigskip

\noindent Commute Time: the expected length of a random walk starting from node $u$ and coming back to $u$. \bigskip

\noindent Complement Graph: given a graph $G$, its complement is a graph where we remove all edges from $G$ and we connect all pairs of nodes that were not connected in $G$. \bigskip

\noindent Complement of the Cumulative Distribution: A plot telling you the fraction of points with value equal to or greater than $x$. \bigskip

\noindent Connected Component: The maximal (sub)set of nodes in a network that can all reach each other through walks. \bigskip

\noindent Connected Network: A network composed by a single connected component. \bigskip

\noindent Connection: see Edge. \bigskip

\noindent Convex Network: A network whose all connected subgraphs are convex, i.e. a tree of cliques. \bigskip

\noindent Convex Subgraph: A subgraphs that contains all shortest paths existing in the main network between its nodes. \bigskip

\noindent Coupling Strategy: The way nodes belonging to the same actor connects to each other across layers in a multilayer network. Example: clique, chain, star. \bigskip

\noindent Cumulative Distribution: A plot telling you the fraction of points with value lower than $x$. \bigskip

\noindent Cycle: A path in which the starting and ending node is the same. \bigskip

\noindent Cyclic Graph: A graph containing at least a cycle.

\section*{D}
Degree: The number of edges a node has. \bigskip

\noindent Degree Matrix: A matrix whose diagonal entries are the degrees of the corresponding nodes and the rest of the matrix is filled with zeros. \bigskip

\noindent Depth First Search: The exploration of a graph by exploring as far as possible along a branch before backtracking. \bigskip

\noindent Diameter: The length of the longest shortest path in a network. \bigskip

\noindent Digraph: See Directed Graph. \bigskip

\noindent Directed Acyclic Graph: A directed graph which does not contain a cycle. \bigskip

\noindent Directed Cyclic Graph: A directed graph containing a cycle. \bigskip

\noindent Directed Edge: A non-reciprocal edge, which implies a relationship that is not symmetric. \bigskip

\noindent Directed Graph: A graph containing directed edges. \bigskip

\noindent Directed Tree: A directed graph which would not contain a cycle even if we were to ignore edge directions. \bigskip

\noindent Disassortativity: The tendency of nodes to connect with nodes with unlike attributes. Opposite of homophily. \bigskip

\noindent Dynamic Network: A network whose edges can become active and/or inactive at different moments in time, usually represented as edge attributes.

\section*{E}
Edge: The interaction between two nodes, usually represented as a pair of nodes. \bigskip

\noindent Effective Resistance: the electric resistance between two points in an electric circuit (represented by nodes in $G$) assuming each edge in $G$ is a 1 Ohm resistor. \bigskip

\noindent Ego Network: A network focused on a node (ego). It contains the ego node, all his neighbors, and all the connections between these nodes. \bigskip

\noindent Eigenvalue: Given a matrix $A$, an eigenvalue of $A$ is the scaling factor of one of its eigenvectors, i.e. if $A v = \lambda v$ for some vector $v$, then $\lambda$ is an eigenvalue of $A$. \bigskip

\noindent Eigenvector: Given a matrix $A$, an eigenvector of $A$ is a special vector that only changes its length -- but not its direction -- when multiplied to $A$, i.e. if $A v = \lambda v$ for some value $\lambda$, then $v$ is an eigenvector of $A$.

\section*{F}
Fiedler Vector: The second smallest eigenvector of the Laplacian. \bigskip

\noindent Forest: A network composed by more than one connected component, each one of them being a tree.

\section*{G}
Giant Connected Component: In real world networks, the largest component which holds the majority of the nodes of a network. \bigskip

\noindent Graph: A set of nodes connected by a set of edges.

\section*{H}
Hairball: Incoherent ball of nodes and edges, typical result of a naive visualization of a network too large and dense to be spread out in a two dimensional plane. Also known as ridiculogram or spaghettigraph. \bigskip

\noindent Heterogeneous Network: A network with multiple node and edge types. \bigskip

\noindent Heterophily: The tendency of nodes to connect to nodes with unlike attributes. Synonym of disassortativity. \bigskip

\noindent Hitting Time: the expected length of a random walk starting from node $u$ and visiting $v$ for the first time. \bigskip

\noindent Homophily: The tendency of nodes to connect to other nodes with the same or similar attributes. Synonym of assortativity. \bigskip

\noindent Hub: A central node with many connections. \bigskip

\noindent Hyperedges: Edges that can connect more than two nodes at the same time. \bigskip

\noindent Hypergraph: A graph containing hyperedges.

\section*{I}
Identity Matrix: A matrix with ones on the diagonal and zeros everywhere else. \bigskip

\noindent In-Component: A weakly connected component in a directed graph whose paths can reach a strongly connected component but will never reach back. \bigskip

\noindent Incidence Matrix: A matrix with nodes on the rows, edges on the columns, and whose non-zero entries report to which edges a node is connected. \bigskip

\noindent Induced Subgraph: a subgraph of an original graph formed from a subset of the vertices of the graph and all of the edges connecting pairs of vertices in that subset. \bigskip

\noindent Interlayer Coupling: In a multilayer network, the special connections connecting the nodes belonging to the same actor. \bigskip

\noindent Isolated Node: a node with zero degree.

\section*{K}
$k$-Clique: A clique of $k$ nodes. \bigskip

\noindent $k$-core: Set of nodes that have a minimum degree of $k$, once you recursively remove from the network all nodes that have $k-1$ connections or fewer.

\section*{L}
Laplacian: The matrix obtained subtracting the adjacency matrix from the degree matrix. \bigskip

\noindent Lattice: A simple graph in which nodes are uniformly distributed in a n-dimensional space and they connect with a given number of their nearest neighbors. \bigskip

\noindent Leaf Node: A node with degree equal to one. \bigskip

\noindent Left Eigenvector: An eigenvector obtained multiplying the matrix from the left. If $v A = v \lambda$, the $v$ is a left eigenvector of $A$, in contrast with right eigenvectors. \bigskip

\noindent Line Graph: The graph that represents the adjacencies between edges of an undirected graph: each edge of the original graph is a node in the line graph, and two nodes in the line graph connect if they have a node in common in the original graph. \bigskip

\noindent Link: See Edge.

\section*{M}
Maximal Clique: A clique in a network to which you cannot add any nodes and still obtain a clique. \bigskip

\noindent Maximum Spanning Tree: A spanning tree of a weighted graph which has the highest edge weight sum of all spanning trees for that graph. \bigskip

\noindent Metapath: A path in a heterogeneous network, including nodes of different types. \bigskip

\noindent Minimum Spanning Tree: A spanning tree of a weighted graph which has the lowest edge weight sum of all spanning trees for that graph. \bigskip

\noindent Multidimensional Network: A network with multiple edge types. A subtype of multilayer networks. \bigskip

\noindent Multigraph: A graph in which there can be multiple parallel edges between the same two nodes. \bigskip

\noindent Multilayer Network: A network in which nodes can connect to each other with different types edges, and can have multiple identities. \bigskip

\noindent Multipartite Network: A network with two or more node types, and whose edges can only be established between nodes of unlike type. \bigskip

\noindent Multiplex Network: A network with multiple edge types. A subtype of multilayer networks. \bigskip

\noindent Mutual information: A relatedness measure between two random variables, namely the number of bits of information you obtain about a random variable if you know the other one.

\section*{N}
$n,m$-Clique: A biclique with $n$ nodes of type $1$ and $m$ nodes of type $2$. \bigskip

\noindent Neighbor: A node directly connected to your focus node by an edge. \bigskip

\noindent Node: The fundamental interacting unit of a graph. In a social network, it will be a person. In the Internet network, it will be a router. \bigskip

\noindent Normalized Mutual information: Equivalent to Mutual Information, normalized so that it takes values between zero and one.

\section*{O}
Out-Component: A weakly connected component in a directed graph which can receive paths from a strongly connected component but cannot reach it back.

\section*{P}
Parallel edges: Two (or more) edges established between the same pair of nodes. \bigskip

\noindent Path: A walk with no repeating nodes. \bigskip

\noindent Planar Graph: A graph you can draw on a 2D plane without intersecting any edges. \bigskip

\noindent Probabilistic Network: a network whose edges have probabilities of existing.

\section*{R}
Reverse Graph: the reverse graph of directed graph $G$ is another directed graph where we flip all edge directions. \bigskip

\noindent Ridiculogram: see Hairball. \bigskip

\noindent Right Eigenvector: An eigenvector obtained multiplying the matrix from the right. If $A v = \lambda v$, the $v$ is a right eigenvector of $A$, in contrast with left eigenvectors. \bigskip

\section*{S}
Self-loop: An edge connecting a node with itself. \bigskip

\noindent Simple Path: See Path. \bigskip

\noindent Simplicial Complex: A set of nodes whose connections to each other are part of a single high-order structure. Similar to hyperedges, with the constraint of being embedded in a geometric space. \bigskip

\noindent Singleton: see Isolated Node. \bigskip

\noindent Spaghettigraph: See Hairball. \bigskip

\noindent Spanning Tree: A subgraph that is a tree and includes all of the nodes of its parent graph. \bigskip

\noindent Square: A cycle of four nodes and four edges. \bigskip

\noindent Star: A set of nodes with one acting as a center connected to all other nodes in the star. All other nodes have only one connection, to the star's center. \bigskip

\noindent Stationary Distribution: The probability of ending in a node after a random walk of infinite length, equivalent to the degree for undirected networks. \bigskip

\noindent Stochastic Adjacency: A normalized adjacency matrix, whose rows have been divided by their sum. \bigskip

\noindent Strongly Connected Component: A component in a directed graph that contains paths from any node of the component to any other node of the component, respecting edge directions. \bigskip

\noindent Subgraph: A graph whose sets of nodes and edges are completely included in the node and edge sets of another graph.

\section*{T}
Temporal Network: See Dynamic Network. \bigskip

\noindent Tree: A graph containing no cycles. \bigskip

\noindent Triad: A connected graph with three nodes and two edges. \bigskip

\noindent Triangle: A connected graph with three nodes and three edges. \bigskip

\noindent Tripartite Network: A network with three node types and whose edges can only be established between nodes of unlike type.

\section*{U}
Undirected Network: A network whose edges are all without direction, i.e. all connections are symmetric. \bigskip

\noindent Uniform Hypergraph: A hypergraph containing hyperedges with the same cardinality -- i.e. each hyperedge contains the same number of nodes. \bigskip

\noindent Unipartite Network: A network with only one node type, without restrictions on how nodes connect -- in direct contrast with a bipartite network. \bigskip

\noindent Unweighted Network: A network whose edges have no weight -- or where all weights are equal to one. In direct contrast with a weighted network.

\section*{V}
Vertex: See Node.

\section*{W}
Walk: A sequence of nodes. Two consecutive nodes in the sequence must be adjacent. \bigskip

\noindent Weakly Connected Component: A component in a directed graph that contains paths from any node of the component to any other node of the component, but only if we ignore edge directions. \bigskip

\noindent Weighted Adjacency: An adjacency matrix which is not binary. Each cell contains the weight of its corresponding edge. \bigskip

\noindent Weighted Edge: An edge with quantitative information, determining its strength. \bigskip

\noindent Weighted Network: A network containing weighted edges.

\chapter{Most Common Abbreviations}

\section*{A}
$A$: Adjacency matrix. \bigskip

\noindent AMI: Adjusted Mutual Information. \bigskip

\noindent APL: Average Path Length. \bigskip

\noindent $APL_v$: The Average Path Length of paths starting from node $v$. \bigskip

\noindent AUC: Area Under the (ROC) Curve, a common way to estimate prediction performance.

\section*{B}
BFS: Breadth First Search.

\section*{C}
$C$: The commute time matrix, $C_{u,v} = H_{u,v} + H_{v,u}$, with $H$ being the hitting time matrix. \bigskip

\noindent CC: Global Clustering Coefficient. \bigskip

\noindent $CC_{avg}$: Average Clustering Coefficient of the network, the average of $CC_v$ for all $v$s in the network. \bigskip

\noindent $CC_v$: Local Clustering Coefficient of node $v$. \bigskip

\noindent CCDF: Complement of the Cumulative Distribution Function. \bigskip

\noindent CDF: Cumulative Distribution Function.

\section*{D}
$D$: Degree matrix. \bigskip

\noindent DFS: Depth First Search.

\section*{E}
$E$: Set of edges. \bigskip

\noindent ERGM: Exponential Random Graph Model, a technique to generate synthetic graphs for statistical testing.

\section*{F}
FN: False Negative. \bigskip

\noindent FP: False Positive. \bigskip

\noindent FPR: False Positive Rate, equal to $FP / (FP + TN)$.

\section*{G}
GCC: Giant Connected Component. \bigskip

\noindent GERM: Graph Evolution Rule Mining, a way to predict links in networks.

\section*{H}
$H$: The hitting time matrix, telling you how long it'll take for a random walker to visit one node when starting from another.

\section*{I}
$I$: The identity matrix.

\section*{K}
$\bar{k}$: The average degree of the network. \bigskip

\noindent $k_v$: The degree of node $v$.

\section*{L}
$L$: Laplacian matrix.

\section*{M}
MI: Mutual Information.

\section*{N}
NMF: Non-negative Matrix Factorization. \bigskip

\noindent NMI: Normalized Mutual Information. \bigskip

\noindent $N_u$: The set of neighbors of node $u$. \bigskip

\noindent $N_{u,l}$: In a multilayer network, the set of neighbors of node $u$ in layer $l$.

\section*{P}
$P_{uv}$: A path going from node $u$ to node $v$. \bigskip

\noindent PCA: Principal Component Analysis.

\section*{R}
ROC: Receiver Operating Characteristic, a common way to estimate prediction performance.

\section*{S}
SBM: Stochastic Block Model, a network generative model. \bigskip

\noindent SCC: Strongly Connected Component. \bigskip

\noindent SI: Susceptible-Infected, an compartmental epidemiology model with two states and one transition. \bigskip

\noindent SIR: Susceptible-Infected-Removed, an compartmental epidemiology model with two states and two irreversible transitions. \bigskip

\noindent SIS: Susceptible-Infected-Susceptible, an compartmental epidemiology model with two states and one reversible transition. \bigskip

\noindent SVD: Singular Value Decomposition, an operation to decompose an arbitrary matrix into a diagonal one. \bigskip

\noindent SVM: Support Vector Machine, a machine learning technique.

\section*{T}
TN: True Negative. \bigskip

\noindent TP: True Positive. \bigskip

\noindent TPR: True Positive Rate, equal to $TP / (TP + FN)$.

\section*{V}
$V$: Set of nodes.

\section*{W}
$W$: Set of possible weights in a weighted network. \bigskip

\noindent WCC: Weakly Connected Component.

\section*{Other Symbols}
$\Omega$: The effective resistance matrix of a graph.

\backmatter

\bibliography{biblio}
\bibliographystyle{plainnat}
\printindex
\end{document}